\documentclass[oneside, letterpaper, 11pt]{report}

\usepackage[utf8x]{inputenc}
\usepackage{amssymb,amsmath,amsfonts}
\usepackage[titletoc,title]{appendix}
\usepackage{array}
\usepackage{authblk}
\usepackage{booktabs}
\usepackage[font=small,labelfont=bf]{caption}
\usepackage{cite}
\usepackage{color}
\usepackage{comment}
\usepackage[utf8x]{inputenc}
\usepackage{csquotes}
\usepackage{fancyhdr}
\usepackage[text={6in,8.5in},headheight=14pt,asymmetric,centering]{geometry}
\usepackage{graphicx,epsfig}
\usepackage{graphicx}
\graphicspath{{./Figures}}
\usepackage[space]{grffile}
\usepackage{lscape}
\usepackage{ltablex}
\usepackage{makeidx}
\usepackage{mathpazo}
\usepackage{morefloats}   
\usepackage{multicol}
\usepackage{multirow}
\usepackage[final]{pdfpages}
\usepackage{placeins}
\usepackage{ragged2e}
\usepackage{rotating}
\usepackage{setspace}
\usepackage{siunitx}
\usepackage{tabto}
\usepackage{tabularx}
\usepackage[flushleft]{threeparttable}
\usepackage{titlefoot}
\usepackage[disable]{todonotes}   
\usepackage{url}

\allowdisplaybreaks[2]



\newcommand{\gev}{\operatorname{GeV}}

\newcommand{\fb}{\operatorname{fb}}

\newcommand{\gevcc}{\mbox{$\mathrm{GeV/}c^2$}}

\newcommand{\gevc}{\mbox{${\mathrm{GeV/}}c$} }
\newcommand{\mevc}{\mbox{${\mathrm{MeV/}}c$}}
\newcommand{\pT}{\mbox{$p_T$}}

\newcommand{\jpsi}{J\mskip -2mu/\mskip -0.5mu\Psi}

\newcommand{\sqrts}{\mbox{$\sqrt{s}$}}
\newcommand{\ep}{\textit{e}+\textit{p}}
\newcommand{\eD}{\textit{e}+\textit{D}}
\newcommand{\eA}{\mbox{\textit{e}+A}}

\newcommand{\pAu}{\textit{p}+Au}
\newcommand{\pPb}{\textit{p}+Pb}
\newcommand{\pA}{\textit{p}+A}

\newcommand{\eAu}{\mbox{\textit{e}+Au}}

\newcommand{\PbPb}{Pb+Pb}

\newcommand{\pp}{\mbox{\textit{p}+\textit{p}}}
\newcommand{\ee}{\mbox{$e^+e^-$}}
\renewcommand{\AA}{A+A}

\newcommand{\lumi}{\mbox{$\mathrm{cm}^{-2}\ \mathrm{sec}^{-1}$}}
\newcommand{\Qss}{\mbox{$Q_s^2$}}

\usepackage[plainpages=false,pdfpagelabels,pdfborder={0 0 0}]{hyperref}

\usepackage{wrapfig}
\usepackage[compat=1.1.0]{tikz-feynman} 
\usepackage{makecell}
\usepackage{bm}

\usepackage{caption}
\usepackage{subcaption}

\usepackage{afterpage}

\newcommand\blankpage{%
    \null
    \thispagestyle{empty}%
    \newpage}

\newcommand\semiblankpage{%
    \null
    \thispagestyle{plain}
    \newpage}

\makeatletter
\renewcommand{\l@section}{\@dottedtocline{1}{1.5em}{2.6em}}
\makeatother

\newcommand{\um}{$\mu$m}
\newcommand{\Xo}{X$_{0}$}

\begin{document}  \pagestyle{empty}


%
%
\newgeometry{textwidth=8.5in,textheight=11.0in}
\includegraphics[width=8.5in,height=10.99in]{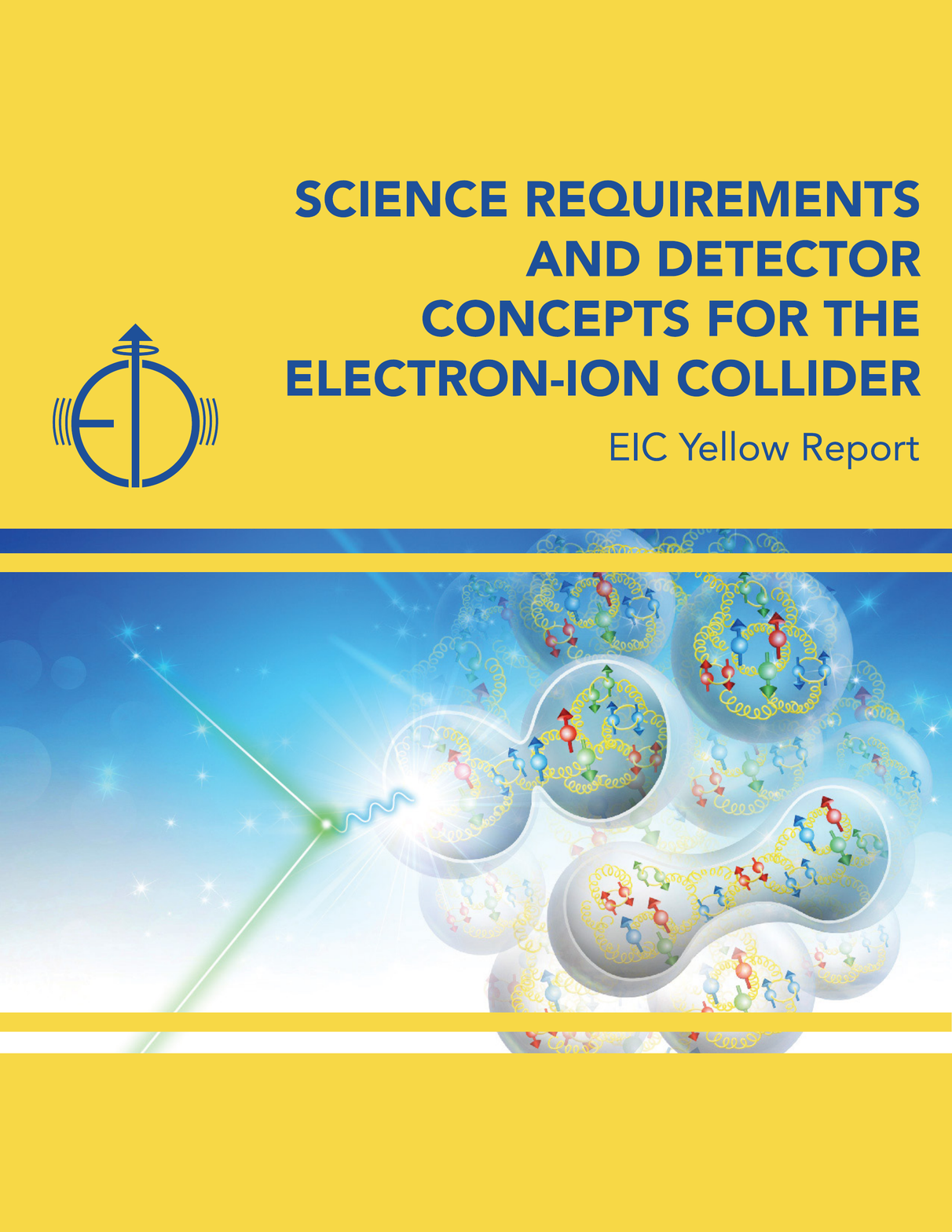}
\restoregeometry

%
%
\vspace*{\fill}
\large{Cover Art by Yulia Furletova and Shannon West}

\clearpage

\setcounter{page}{1}\pagenumbering{roman}  

\fancyhead{}
\fancyfoot{}
\pagestyle{plain}

\phantomsection
\addcontentsline{toc}{chapter}{Title Page}

\setcounter{page}{1} 
\renewcommand*{\thepage}{\roman{page}}

\hfill  \large{BNL-220990-2021-FORE}

\hfill  \large{JLAB-PHY-21-3198}

\hfill  \large{LA-UR-21-20953}

\vspace{2in}

\begin{center}
\textbf{\huge{Science Requirements and Detector Concepts for the Electron-Ion Collider\\}}
\vspace{0.5cm}
\LARGE{EIC Yellow Report}
\end{center}

\vspace*{\fill}
\hfill  \large{October 2021}

\cleardoublepage

\phantomsection
\addcontentsline{toc}{chapter}{Author List}
\begin{center}
   \LARGE{Author List}
\end{center}

{\small
\begin{center}
R.~Abdul~Khalek$^{1,2}$,
A.~Accardi$^{3,4}$,
J.~Adam$^{5}$,
D.~Adamiak$^{6}$,
W.~Akers$^{4}$,
M.~Albaladejo$^{4}$,
A.~Al-bataineh$^{7}$,
M.G.~Alexeev$^{8,9}$,
F.~Ameli$^{10}$,
P.~Antonioli$^{11}$
% ...
\end{center}
}

{\small
$^{1}${\it Vrije Universiteit Amsterdam, 1081 HV Amsterdam, The Netherlands}\\
$^{2}${\it Nikhef Theory Group, 1098 XG Amsterdam, The Netherlands}\\
$^{3}${\it Hampton University, Hampton, Virginia 23668, USA}\\
$^{4}${\it Thomas Jefferson National Accelerator Facility, Newport News, Virginia 23606, USA}\\
$^{5}${\it Brookhaven National Laboratory, Upton, New York 11973, USA}\\
$^{6}${\it Ohio State University, Columbus, Ohio 43210, USA}\\
$^{7}${\it Imam Abdulrahman Bin Faisal Univ., Dammam 34212, Saudi Arabia}\\
$^{8}${\it Universit\`a di Torino, I-10125 Torino, Italy}\\
$^{9}${\it INFN - Sezione di Torino, I-10125 Torino, Italy}\\
$^{10}${\it INFN - Sezione di Roma, I-00185 Roma, Italy}\\
$^{11}${\it INFN - Sezione di Bologna, I-40127 Bologna, Italy}\\
% ...
$^{21}${\it INFN - Sezione di Pavia, I-27100 Pavia, Italy}\\
% ...
}

\cleardoublepage

\phantomsection
\addcontentsline{toc}{chapter}{Abstract}
\thispagestyle{plain}
\begin{center}
    \LARGE{Abstract}
\end{center}

This report describes the physics case, the resulting detector requirements, and the evolving detector concepts for the experimental program at the Electron-Ion Collider (EIC). The EIC will be a powerful new high-luminosity facility in the United States with the capability to collide high-energy electron beams with high-energy proton and ion beams, providing access to those regions in the nucleon and nuclei where their structure is dominated by gluons. Moreover, polarized beams in the EIC will give unprecedented access to the spatial and spin structure of the proton, neutron, and light ions. The studies leading to this document were commissioned and organized by the EIC User Group with the objective of advancing the state and detail of the physics program and developing detector concepts that meet the emerging requirements in preparation for the realization of the EIC. The effort aims to provide the basis for further development of concepts for experimental equipment best suited for the science needs, including the importance of two complementary detectors and interaction regions. 

This report consists of three volumes. Volume I is an executive summary of our findings and developed concepts. In Volume II we describe studies of a wide range of physics measurements and the emerging requirements on detector acceptance and performance. Volume III discusses general-purpose detector concepts and the underlying technologies to meet the physics requirements. These considerations will form the basis for a world-class experimental program that aims to increase our understanding of the fundamental structure of all visible matter.

\cleardoublepage

\phantomsection
\addcontentsline{toc}{chapter}{Table of Contents}
\setcounter{tocdepth}{1}		
\tableofcontents 
\cleardoublepage

%
%

\fancyhead[RE]{\slshape \rightmark} 
\fancyhead[LE]{\thepage}
\fancyhead[LO]{\slshape \leftmark} 
\fancyhead[RO]{\thepage}
\fancyfoot{}
\pagestyle{fancy}

\setcounter{page}{1}\pagenumbering{arabic}                

%
%

\phantomsection
\addcontentsline{toc}{part}{\Large{\textbf{Volume I: Executive Summary}}}
\label{part1.ex.sum}
\newgeometry{textwidth=8.5in,textheight=11.0in}
\includegraphics[width=8.5in,height=10.99in]{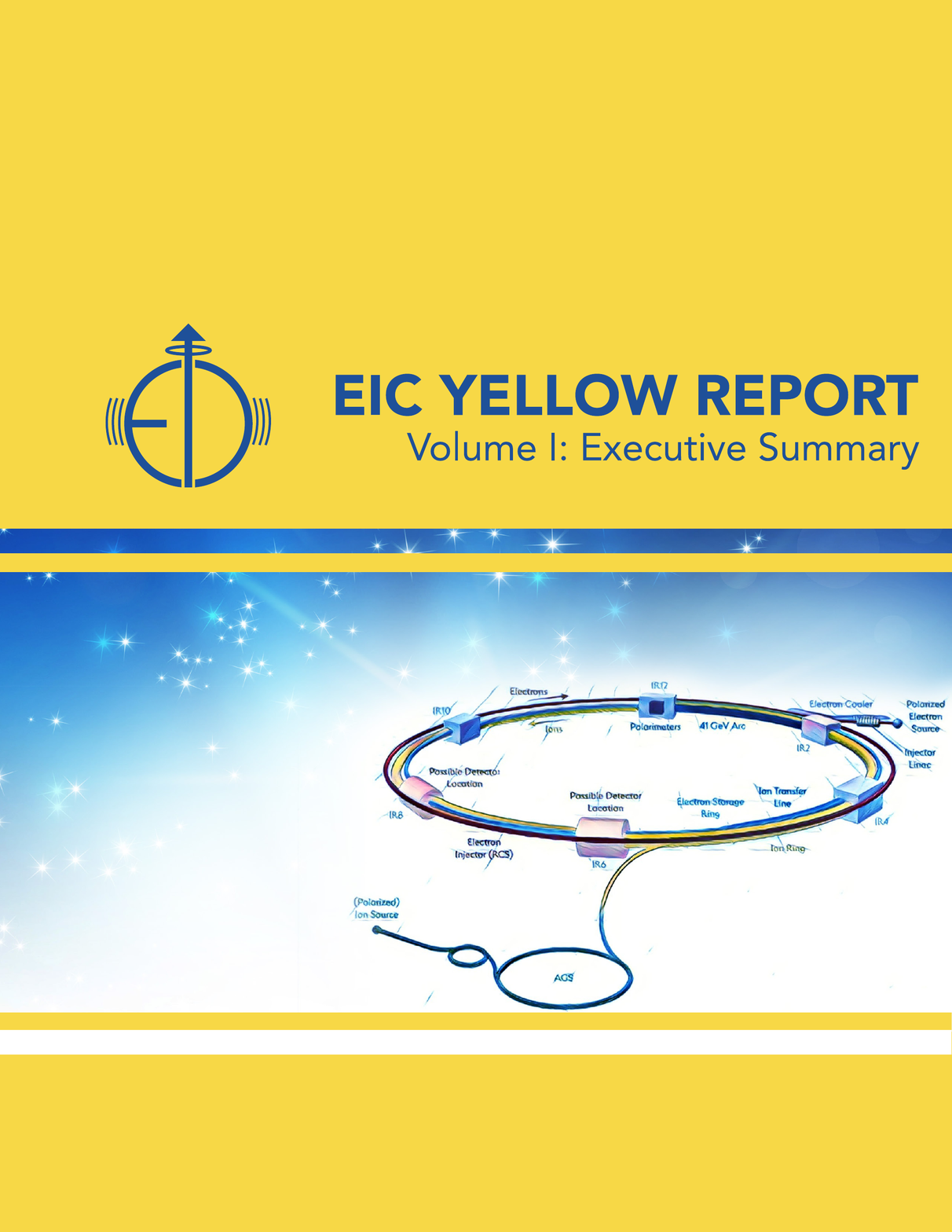}
\restoregeometry
\semiblankpage


\chapter{The Electron-Ion Collider}  
\label{part1-chap-EIC}

The Electron-Ion Collider (EIC) is a new, innovative, large-scale particle accelerator facility conceived by U.S. nuclear and accelerator physicists over two decades and planned for construction at Brookhaven National Laboratory on Long Island, New York by the U.S. Department of Energy in the 2020s. The EIC will study protons, neutrons and atomic nuclei with the most powerful electron microscope, in terms of versatility, resolving power and intensity, ever built. The resolution and intensity is achieved by colliding high-energy electrons with high-energy protons or (a range of different) ion beams. The EIC provides the capability of colliding beams of polarized electrons with polarized beams of light ions, and this all at high intensity.
The EIC was established as the highest priority for new construction in the 2015 US Nuclear Physics Long Range Plan, and was favorably endorsed by a committee established by the National Academy of Sciences in 2018 to assess the science case. In December 2019, the EIC was granted Critical Decision Zero (CD0) by the US Department of Energy, which launched the EIC as an official project of the US government.

\begin{figure}[ht]
\begin{center}
\vspace*{2mm}
\includegraphics[width=0.9\textwidth]{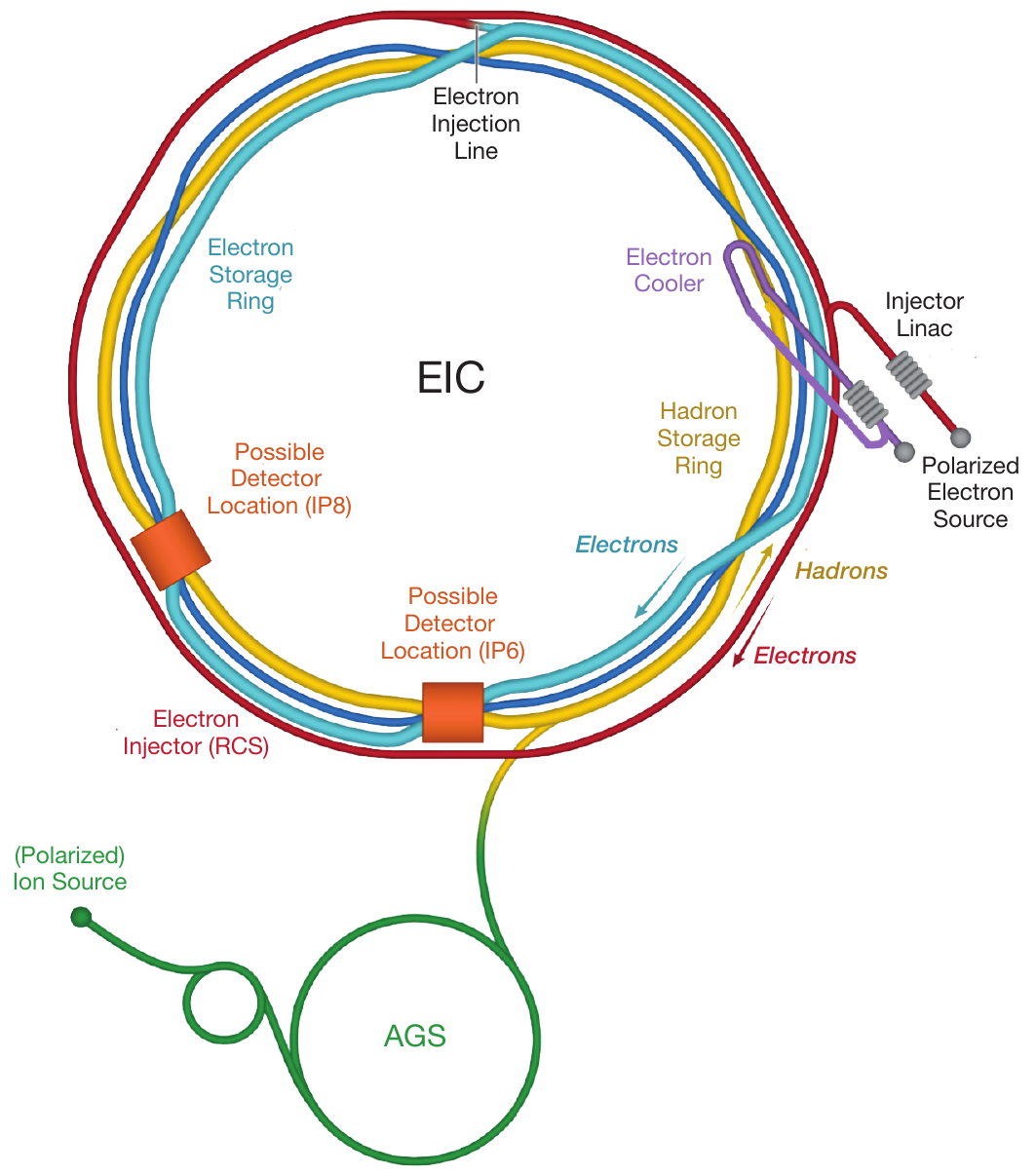}
\caption{Schematic layout of the planned EIC accelerator based on the existing RHIC complex at Brookhaven National Laboratory.}
\label{EIClayout}
\end{center}
\end{figure}

The main design requirements of the EIC are:
\begin{itemize}
\item Highly polarized electron ($\sim$70\%) and proton ($\sim$70\%) beams
\item Ion beams from deuterons to heavy nuclei such as gold, lead, or uranium
\item Variable \ep~center-of-mass energies from 20$-$100 GeV, upgradable to 140 GeV
\item High collision electron-nucleon luminosity 10$^{33}-$10$^{34}$  cm$^{-2}$ s$^{-1}$
\item Possibility to have more than one interaction region
\end{itemize}

Several of the above performance parameters will be realized for the first time at EIC in a collider mode, such as the availability of nuclear beams and polarized nucleon beams along with the operation at high collision luminosity. 
Shown schematically in Fig.~\ref{EIClayout}, the EIC will collide bright, intense counter circulating beams of electrons and ions and use sophisticated, large detectors to identify specific reactions whose precise measurement can yield previously unattainable insight into the structure of the nucleon and nucleus. The EIC will open a new window into the quantum world of the atomic nucleus and allow physicists access for the first time to key, elusive aspects of nuclear structure in terms of the fundamental quark and gluon constituents.  Nuclear processes fuel the universe. Past research has provided enormous benefit to society in terms of medicine, energy and other applications.
Particle accelerators and related technologies play a key role in the discovery sciences and it is estimated that about 30,000 worldwide are operating in industry.
The EIC will probe the frontiers of nuclear science well into the twenty-first century using one of the world’s most sophisticated particle accelerators and large detectors that will utilize cutting-edge technology.  

The realization of the EIC is led jointly by Brookhaven National Laboratory and Thomas Jefferson National Accelerator Facility at Newport News, Virginia. It will involve physicists and engineers from other laboratories and universities in the U.S. and from around the world. This realization is expected to roughly take a decade, with beam operations to start in the early 2030s.

The EIC Users Group (EICUG, www.eicug.org) was founded in 2016. It now contains over 1200 members from 245 institutions located in 33 countries around the world.  Late in 2019, the EICUG decided to organize an intensive, year-long consideration of the EIC physics measurements and scientific equipment by the members of the user group. This Yellow Report (YR) summarizes these studies and the conclusions that have been reached.  The purpose of the Yellow Report Initiative is to advance the state and detail of the documented community physics studies (EIC White Paper, Institute for Nuclear Theory program proceedings) and detector concepts (Detector and R\&D Handbook) in preparation for the realization of the EIC. The effort aims to provide the basis for further development of concepts for experimental equipment best suited for science needs, including complementarity of two detectors towards future Technical Design Reports. It is expected that this YR will be the cornerstone on which detector proposals will be developed by user collaborations beginning in 2021.

The work reported on here was organized by the EICUG at an in-person meeting in December 12-13, 2019 at the Massachusetts Institute of Technology, Cambridge, Massachusetts and was structured around four subsequent meetings in 2020: March 19-21, 2020 at Temple University, Philadelphia; May 20-22, 2020 at University of Pavia, Pavia, Italy; September 16-18, 2020 at the Catholic University of America, Washington, DC and November 19-21, 2020 at the University of California, Berkeley.  This was a massive, international, sustained effort through the year 2020 and was overseen by 8 conveners and 41 sub-conveners.  Because of the restrictions due to the pandemic, all of the EICUG meetings and interactions in 2020 were carried out remotely.

The EIC will be one of the largest and most sophisticated new accelerator projects worldwide in the next few decades, and the only planned for construction in the United States. 
It will address profound open questions in the fundamental structure of matter and attract new generations of young people into the pursuit of careers in science and technology.
Its high design luminosity and highly polarized beams are beyond state-of-the-art and its realization will likewise push the frontiers of particle accelerator science and technology.   


%
\chapter{Physics Measurements and Requirements}
\label{part1-chap-PhyMeasandReq}

\section{Introduction}

The Electron-Ion Collider (EIC) will address some of the most fundamental questions in science regarding the visible world, including the origin of the nucleon mass, the nucleon spin, and the emergent properties of a dense system of gluons. The science program has been reviewed by the National Academy of Sciences (NAS) which concluded that "the EIC science is compelling, fundamental, and timely."~\cite{NAP25171}. The NAS review was based on a series of workshops hosted by the Institute of Nuclear Theory (INT) culminating in a whitepaper in 2012 and an update in 2014 entitled "Understanding the glue that binds us all"~\cite{Accardi:2012qut}. The desire and need to construct a new collider facility were prominently featured in the 2015 US Long-Range Plan for Nuclear Science~\cite{Geesaman:2015fha}. 

\begin{table}[ht]
\caption{\label{DIS.processes} Different categories of processes measured at an EIC (Initial state: Colliding electron ($e$), proton ($p$), and nuclei ($A$). Final state: Scattered electron ($e'$), neutrino ($\nu$), photon ($\gamma$), hadron ($h$), and hadronic final state ($X$)).}
\begin{tabular}{p{9.2cm}p{4.5cm}} \toprule
 & \\
{\bf Neutral-current Inclusive DIS:} $e+p/\mathrm{A} \longrightarrow e'+X;$ for this process, it is essential to detect the 
scattered electron, $e'$, with high precision. All other final state particles ($X$) are ignored. The scattered electron is critical for all processes
to determine the event kinematics. &
\vspace{-15pt}
\includegraphics[width=4.5cm]{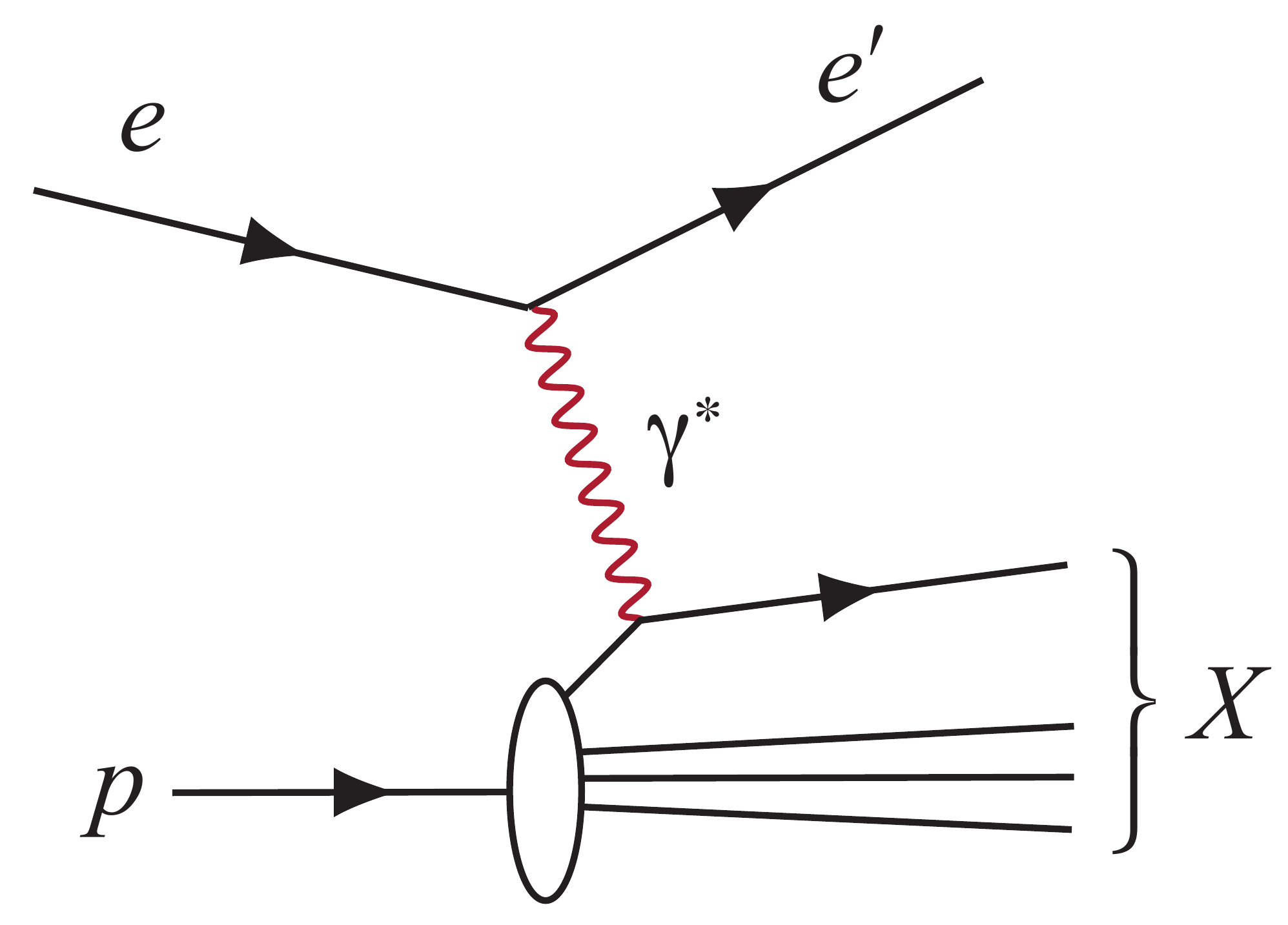}\\
& \\
{\bf Charged-current Inclusive DIS:} $e+p/\mathrm{A} \longrightarrow \nu + X;$ at high enough momentum transfer 
$Q^2$, the electron-quark interaction is mediated by the exchange of a $W^{\pm}$ gauge boson 
instead of 
the virtual photon. In this case the event kinematic cannot be reconstructed from the 
scattered electron, but needs to be reconstructed from the final state particles. & 
\vspace{-10pt}
\includegraphics[width=4.5cm]{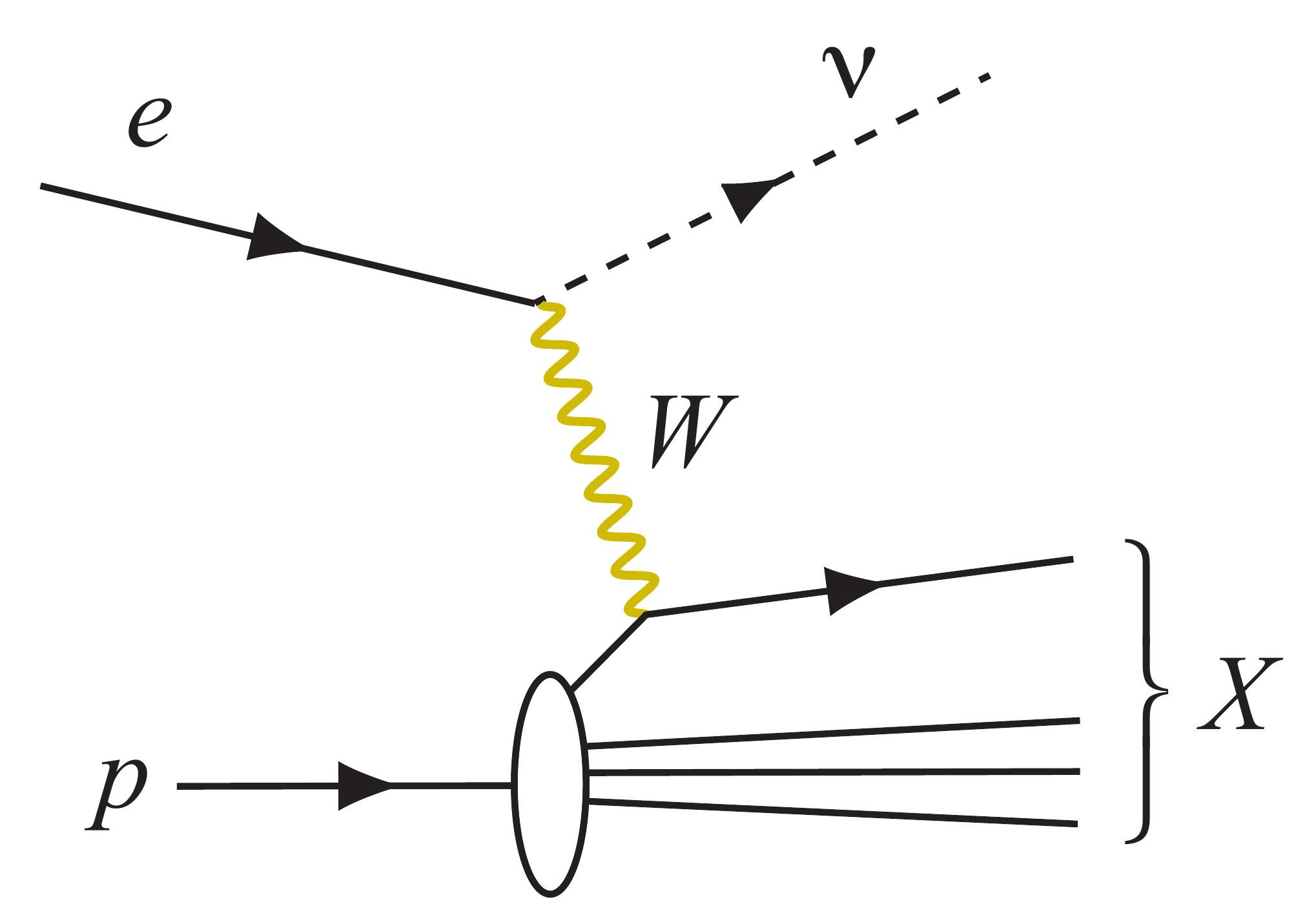} \\ 
& \\
{\bf Semi-inclusive DIS:} $e+p/\mathrm{A} \longrightarrow e'+h^{\pm,0}+X$, which requires measurement of  
{\it at least one} identified hadron in coincidence with the scattered electron. &
\vspace{-10pt}
\includegraphics[width=4.5cm]{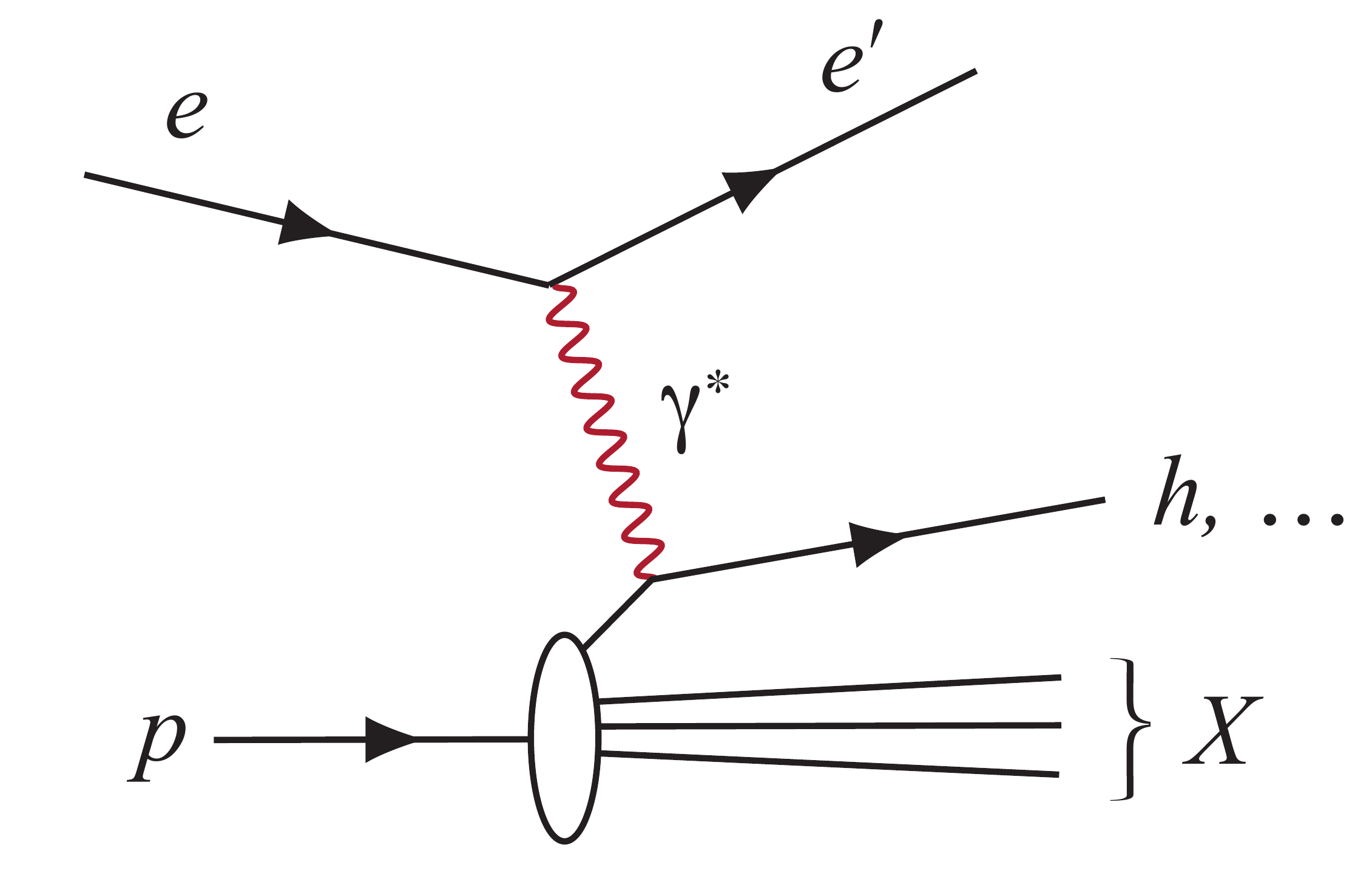} \\
& \\

{\bf Exclusive DIS:} $e+p/\mathrm{A} \longrightarrow e'+p'/A'+\gamma/h^{\pm,0}/VM$, which require the 
measurement of {\it all} particles in the event with high precision. &
\vspace{-15pt}
\includegraphics[width=4.5cm]{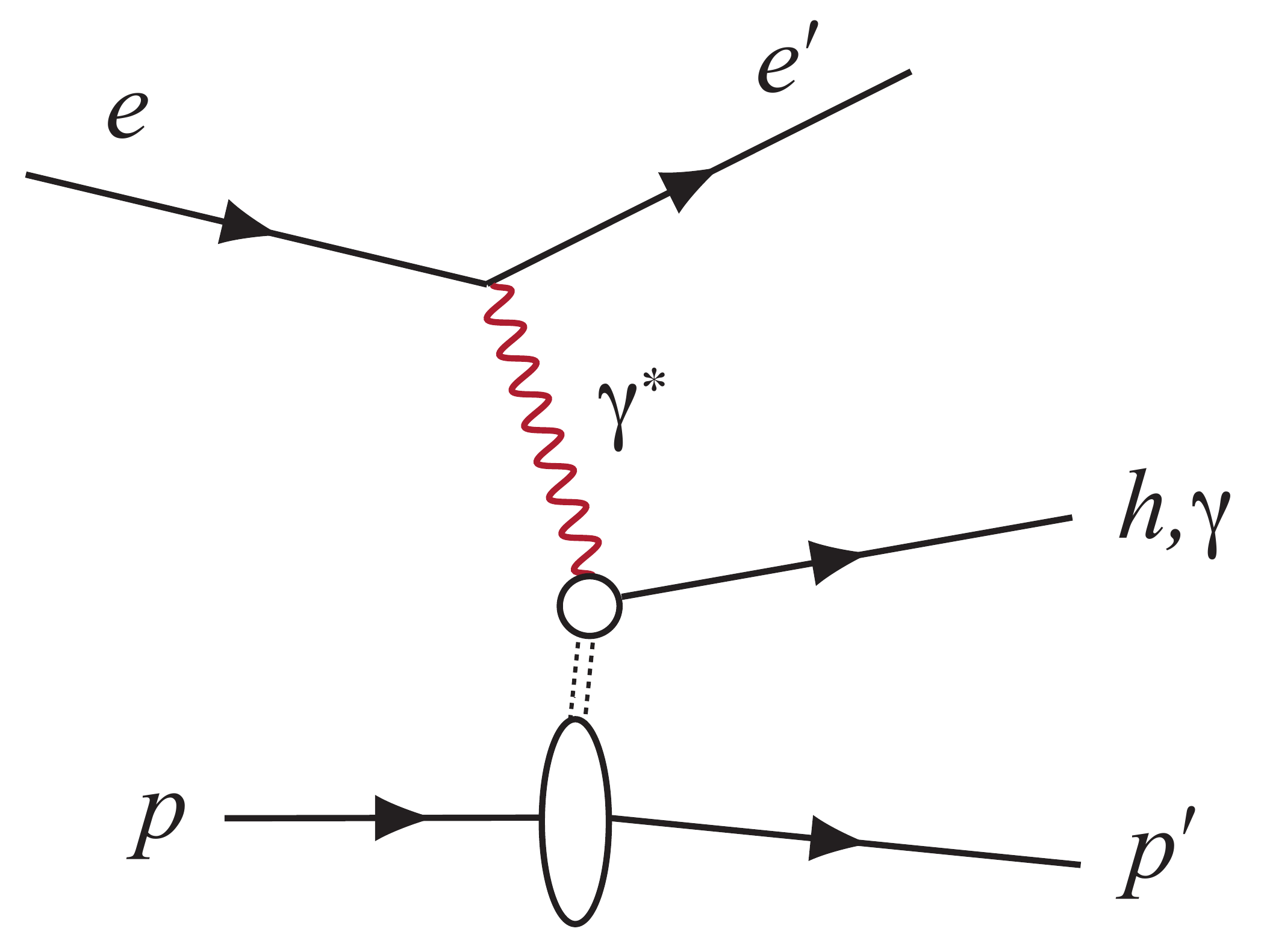} \\
& \\
\bottomrule
\end{tabular}
\end{table}

In this executive section, we present a selection of crucial physics topics that led
to the recommendation for the construction of an EIC, and summarize the machine parameters and detector requirements 
needed to address them.

Key science questions that the EIC will address are:
\begin{itemize}
\item How do the nucleonic properties such as mass and spin emerge from partons and their underlying interactions?
\item How are partons inside the nucleon distributed in both momentum and position space?
\item How do color-charged quarks and gluons, and jets, interact with a nuclear medium? How do the confined hadronic states emerge from these quarks and gluons? How do the quark-gluon interactions create nuclear binding?
\item How does a dense nuclear environment affect the dynamics of quarks and gluons, their correlations, and their interactions? What happens to the gluon density in nuclei? Does it saturate at high energy, 
giving rise to gluonic matter or a gluonic phase with universal properties in all nuclei and even in nucleons?
\end{itemize}

\begin{figure}[ht]
\begin{center}
\vspace*{2mm}
\includegraphics[width=1.0\textwidth]{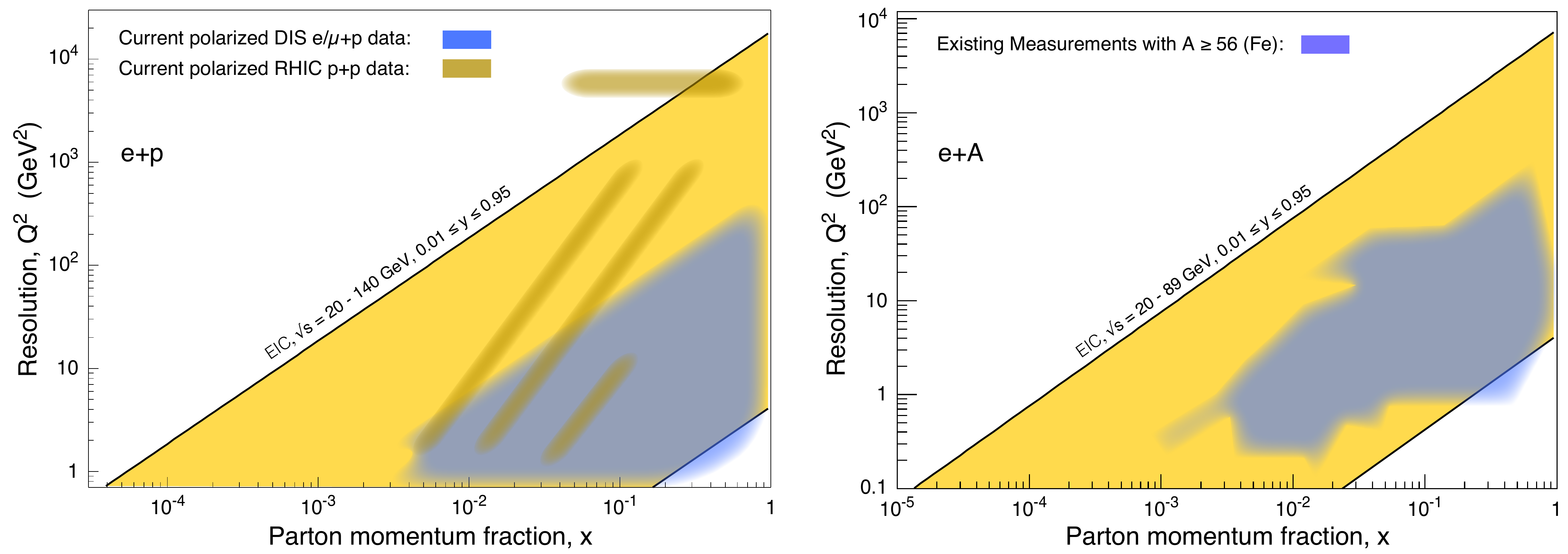}
\caption{\label{fig:xq2}Left: The $x$-$Q^{2}$ range covered by the EIC (yellow) in comparison with past and existing polarized $e/\mu$+$p$ experiments at CERN, DESY, JLab and SLAC, and \pp\ experiments at RHIC. 
Right: The $x$-$Q^{2}$ range for \eA\ collisions for ions larger than iron (yellow) compared to existing world data.}


\end{center}
\end{figure}

The EIC covers a center-of-mass energy range for \ep\ collisions of $\sqrt{s}$ of 20 to 140\,GeV. 
The kinematic reach in $x$ and $Q^{2}$ is shown in Figure~\ref{fig:xq2}. The quantities $x$, $y$, and $Q^{2}$ are obtained from measurements of energies and angles of final state objects, i.e. the scattered electron, the hadronic final-state or a combination of both. The quantity 
$x$ is a measure of the momentum fraction of the struck parton inside the parent-proton. $Q^{2}$ refers to the square of the momentum transfer between the electron and proton and is inversely proportional to the resolution. 
The diagonal lines in each plot represent lines of constant inelasticity $y$, which is the ratio of the virtual photon's energy to the electron's energy in the target rest frame. 
The variables $x,Q^{2},y$ and $s$ are related through the equation $Q^{2} \simeq s x y$. 
The left figure shows the kinematic coverage for polarized and unpolarized \ep\ collisions, and the right figure shows the coverage for \eA\ collisions. 
The EIC will allow in both collider modes an important overlap with present and past experiments. In addition, the EIC will provide access to entirely new regions in both $x$ and $Q^2$ in a polarized \ep~collider and \eA~collider mode, such as the low-$x$ region, providing critical information about the gluon-dominated regime.

Volume 2 of this Yellow Report provides a detailed overview of the EIC physics program, including several recent developments not addressed in the EIC White Paper.
In what follows, we focus on the most critical aspects of the scientific questions outlined above and motivate the machine and detector parameters needed to address these questions.

\section{Origin of Nucleon Spin}

Understanding the nucleon spin in terms of contributions from quark and gluon spin and angular momentum contributions has been an essential goal for nuclear scientists for several decades. 
The nucleon spin can be split into its components according to~\cite{Jaffe:1989jz}
\begin{equation}
\label{eq:spineq}
\frac{1}{2} = \frac{1}{2} \Delta \Sigma (\mu) + \Delta G(\mu) + L_{Q+G} (\mu) \,,  
\end{equation}
where $\Delta \Sigma$, $\Delta G$, and $L$ are the contributions from the quark plus antiquark spin, the gluon spin, and the parton angular momenta, respectively.
All terms of this spin decomposition depend on the renormalization scale $\mu$.
The parton spin contributions follow from the respective helicity distributions upon integration over the whole $x$-range from 0 to 1. 
The discovery by the EMC experiment at CERN in the 1980s that the $\Delta \Sigma$ term can only explain a small fraction of the nucleon spin brought this topic into the limelight. 
Numerous fixed-target polarized electron/muon DIS experiments and polarized \pp\  experiments at RHIC~\cite{Aschenauer:2015eha}, covering the range $0.005 \lesssim x \lesssim 0.6$, not only confirmed the general finding of the EMC experiment but also suggest that $\Delta G$ is not large enough to make up the missing contribution to the nucleon spin, thus providing a clear indication of a nonzero orbital angular momentum contribution.
However, the numerical values for $\Delta \Sigma$ and $\Delta G$ have large uncertainties because, thus far, we have no information at all about the parton helicity distributions for $x \lesssim 0.005$.
With measurements in this kinematic region combined with precision measurements over the full kinematic range accessible to EIC, the EIC will drastically reduce these uncertainties. Therefore, the EIC will put the nucleon spin decomposition's phenomenology on much firmer ground, and by inference well constrain the parton angular momenta contribution~\cite{Aschenauer:2015ata, Accardi:2012qut, Aschenauer:2017jsk}.

\paragraph{Machine and detector requirements for polarized DIS}

Obtaining information on $\Delta\Sigma$ and $\Delta G$ at the EIC requires measuring DIS with longitudinally polarized electrons and longitudinally polarized protons for a large range in $x$ and $Q^2$ and thus over a wide range in center-of-mass energy. 
One of the key detector requirements at low $x$ refers to the precision measurement of the scattered electron's energy ($E$) demanding good electromagnetic calorimetry at the level of $\sigma(E)/E \approx 10\%/\sqrt{E} \otimes (1-3)\%$ in the central detector region and superior performance at the level of $\sigma(E)/E \approx 2\%/\sqrt{E} \otimes (1-3)\%$ in the backward or rear direction. In addition, robust electron/hadron separation is essential. The reconstruction of kinematic variables at higher $x$-values including the hadronic final state requires good momentum resolution and calorimetric measurement, in particular in the forward direction. Radiative corrections need to be properly addressed, both in terms of theoretical treatment and experimental design. 

\section{Origin of Nucleon Mass}

More than 99\% of the mass of the visible universe resides in atomic nuclei, whose mass, in turn, is primarily determined by the masses of the proton and neutron.
Therefore, it is of utmost importance to understand the origin of the proton (and neutron) mass, particularly how it emerges from the strong interaction dynamics.
Interestingly, the proton mass is not even approximately given by summing the masses of its constituents, which can be attributed to the Higgs mechanism.
Just adding the masses of the proton's valence quarks provides merely about 1\% of the proton mass.
While a QCD analysis leads to a more considerable quark mass contribution to the proton mass, the qualitative picture that the Higgs mechanism is responsible for only a small fraction of the proton mass is not altered.
An essential role for a complete understanding of the proton mass is played by the trace anomaly of the QCD energy-momentum tensor~\cite{Ji:1994av, Lorce:2017xzd,Hatta:2018sqd, Metz:2020vxd}.
It is precisely this essential ingredient for which the EIC can deliver crucial input through dedicated measurements of quarkonia's exclusive production ($J/\psi$ and $\Upsilon$) close to the production threshold.

Another way to address the emergence of hadron mass is through chiral-symmetry features that manifest in the lightest mesons, the pion and kaon.
In this picture, the properties of the nearly massless pion are the cleanest expressions of the mechanism that is responsible for the emergence of the mass and have measurable implications for the pion form factor and meson structure functions~\cite{Aguilar:2019teb}. At variance with the pion, the effects of the Higgs mechanism, which gives a non-vanishing mass to the quarks, play a more substantial role for the kaon mass due to its strange quark content. Therefore, a comparison of the charged pion and charged kaon form factors over a wide range in $Q^2$ would provide unique information relevant to understanding the generation of hadronic mass.
The EIC can also open a vast landscape of structure function measurements constraining quark and gluon energy distributions in pions and kaons.


\paragraph{Machine and detector requirements for nucleon mass studies}

The main physics channel to study the nucleon mass's origin is the multi-dimensional measurement of the quarkonium production cross-section near threshold, which demands high luminosity.
A precise reconstruction of the scattered electron's energy at low-$Q^2$ is essential. To accurately measure the $t$-dependence for quarkonium production, 
recoil protons need to be detected, which demands a careful design of the interaction region to measure the forward-going protons scattered under small angles.

\section{Multi-Dimensional Imaging of the Nucleon}

Inclusive DIS provides a 1-dimensional picture of the nucleon as it reveals the $x$-distribution of (longitudinal) parton momenta in the direction of the nucleon momentum.
However, due to confinement, the partons also have nonzero momenta in the (transverse) plane perpendicular to the nucleon momentum.
The 3D parton structure of hadrons in momentum space is encoded in transverse momentum dependent parton distributions (TMDs). For both quarks and gluons inside a spin-$\frac{1}{2}$ hadron, a total of 8 leading-twist TMDs exist~\cite{Boer:1997nt}. These functions of different correlations between spins and transverse momenta reveal different insights into the dynamics of nucleons. 
TMDs can be measured via certain semi-inclusive processes, such as semi-inclusive DIS (SIDIS), where one detects an identified hadron in addition to the scattered lepton.

In the past, the unpolarized SIDIS cross-section and, in particular, the so-called Sivers asymmetry have been studied for different final-state hadrons.
The latter observable describes a single-spin asymmetry with a transversely polarized target, which gives direct access to the Sivers function $f_{1T}^\perp$~\cite{Sivers:1989cc, Sivers:1990fh}, one of the eight quark TMDs.
The data sets used to constrain TMDs are currently even more limited in $x$ and $Q^{2}$ than those shown in Fig.~\ref{fig:xq2} (left) used to constrain helicity parton distribution functions (PDFs).
With its polarized beams and the large energy range, the EIC will dramatically advance our knowledge of TMDs. 
The 3D momentum structure of the nucleon for the different quark flavors and the gluons will be mapped out over a wide range in $x$ and $Q^2$~\cite{Accardi:2012qut, Zheng:2018ssm}. 

An essential aspect of TMDs concerns their scale-dependence (evolution) as predicted in QCD, which is considerably more involved than the evolution of the 1D PDFs. 
There is, therefore, substantial interest in a quantitative understanding of the TMD evolution. 
The EIC will be ideal for such studies, complementing the high precision data becoming available from JLab at larger values of $x$.

It allows to explore SIDIS observables over an extensive range in $Q^2$ while covering transverse momenta of the final-state hadrons over a wide range from low (non-perturbative) to high (perturbative) values. 
 
While the current knowledge about the Sivers function in the valence region still has considerable uncertainties, the situation is even worse for sea quarks and gluons where hardly any experimental information exists.
At the EIC, the gluon Sivers function can be addressed through transverse single-spin asymmetries for the production of nearly back-to-back pairs of jets or heavy-flavor hadrons.
We note that, qualitatively, the above discussion of the Sivers function applies also to the other TMDs.
Generally, the EIC has transformative potential in the field of the nucleon's 3D structure in momentum space.

\paragraph{Machine and detector requirements for TMD measurements}

Measurements of TMDs require unpolarized as well as longitudinally and transversely polarized hadron beams colliding with (un)polarized electrons. 
The wide range in $x$ and $Q^2$ provided by the EIC is essential for mapping the TMDs. 
SIDIS requires the identification of final-state hadrons in coincidence with the scattered electrons. 
Identifying hadrons allows one to obtain information about the flavor of quarks, which have fragmented into the hadron(s).
Excellent particle identification (PID) is required to separate $\pi/K/p$ at the level of 3 $\sigma$ up to 50\,\gevc in the forward region, up to 10\,\gevc in the central detector region, and up to 7\,\gevc in the backward region. 
Mapping the TMDs in multiple dimensions will require larger data samples than for fully inclusive measurements. 
To disentangle the flavor dependence of the various TMDs, it is essential to collect data with neutron-rich transversely polarized beams of D or $^3$He under equivalent experimental conditions. 

\section{Imaging the Transverse Spatial Distributions of Partons}

As in the case of the transverse momentum distribution of partons inside a hadron, we know very little about their distribution in the transverse spatial dimensions, combined with the information about the longitudinal momentum fraction $x$. 
Those spatial distributions of partons yield a picture that is complementary to the one obtained from TMDs. 
So far, our level of knowledge of the spatial distributions for valence quarks, sea quarks, and gluons is still relatively low. 

It is possible to determine the transverse spatial distributions of quarks and gluons experimentally, where their study requires a particular category of measurements, that of exclusive reactions. 
Examples are deeply virtual Compton scattering (DVCS) and deeply virtual meson production. 
For those reactions, the proton remains intact, and a photon or a meson is produced. 
Exclusivity demands that all final-state products are detected, i.e., the scattered electron, the produced photon or meson, and the scattered proton. 
The spatial distributions of quarks and gluons in these measurements are extracted from the Fourier transform of the differential cross-section for the momentum transfer $t$ between the incoming and the scattered proton. 
The non-perturbative quantities that encode the spatial distributions are called generalized parton distributions (GPDs) \cite{Mueller:1998fv,Radyushkin:1996nd,Ji:1996nm}. 
In addition to the fundamental role of GPDs concerning the spatial distribution of partons inside hadrons~\cite{Burkardt:2002hr}, the second moment of particular sets of GPDs will provide more in-depth insight into the total angular momentum of quarks and gluons in the proton~\cite{Ji:1996ek}. 
GPDs offer a unique opportunity to probe the energy-momentum tensor and thus open the door to deepen our understanding of the nucleon mass.  
Moreover, GPDs contain information about the pressure and shear forces inside hadrons~\cite{Polyakov:2002yz}.

Our knowledge of GPDs from DVCS is currently limited and is based on fixed-target experiments at intermediate to high-$x$ or on the HERA collider measurements at low-$x$. 
The polarized beams and higher luminosity at EIC, along with forthcoming data from JLab at 12 GeV, will make a very significant impact on those measurements. 
It is anticipated that measurements made for protons in the range $0.04 \lesssim t < 1.5$\,GeV$^2$ will enable maps of parton distributions all the way down to $0.1 \, \textrm{fm}$~\cite{Aschenauer:2017jsk,Aschenauer:2013hhw}.
Such exclusive measurements performed on nuclei will enable us to understand the transverse quark and gluon distributions within. 

\paragraph{Machine and detector requirements for GPD measurements}

GPD physics is one of the most demanding aspects of the EIC program in terms of luminosity as it requires multi-dimensional binning of processes that have very low cross-sections. The collection of data at several center-of-mass energies to cover the physics program outlined in the EIC White Paper~\cite{Accardi:2012qut} is essential. The continuous measurement of the momentum transfer to the nucleon in the range $0.02 \, \textrm{GeV}^2 \lesssim |t| \lesssim 1.5 \, \textrm{GeV}^2$ demands a careful design of the interaction region to detect the forward-going protons scattered under small angles combined with a careful choice of the hadron beam parameters, {\it{i.e.}}, angular divergence, and large acceptance magnets. 

\section{Physics with High-Energy Nuclear Beams at the EIC}
\label{part1-sec-PhysMeasReq-nucl.phys}

The nucleus is a QCD molecule, with a complex structure corresponding to bound states
of nucleons. Understanding the formation of nuclei in terms of QCD degrees-of-freedom is an ultimate long-term goal of
nuclear physics. With its broad kinematic reach, as shown in Fig.~\ref{fig:xq2}, the capability
to probe a variety of nuclei in both inclusive and semi-inclusive DIS measurements, the
EIC will be the first experimental facility capable of exploring the internal 3-dimensional
sea quark and gluon structure of a nucleus at low $x$. Furthermore, the nucleus itself is
a unique QCD laboratory for discovering the collective behavior of gluonic matter
at an unprecedented occupation number of gluons, for studying the propagation of
fast-moving color charges in a nuclear medium to shed light on the mystery of the hadronization
process, and to explore the quark-gluon origin of short range nucleon-nucleon forces in the nuclei.

A key feature of gluon saturation is the emergence of a momentum scale $Q_{S}$, known as the saturation scale. 
When this scale significantly exceeds the QCD confinement scale $\Lambda_{QCD}$, the dynamics of strongly correlated gluons can be described
by weak coupling many-body methods. The framework that enables such computations is an effective field theory called the 
Color Glass Condensate (CGC)\cite{Gelis:2010nm}. The CGC predicts that $Q_{S}^{2} \propto A^{1/3}$; thus, the novel domain of saturated gluon fields can be accessed especially well in large nuclei. This regime of QCD is predicted to exist in all hadrons and nuclei when boosted to high energies where one can probe the low-$x$ region in full detail.
Unambiguously establishing this novel domain of QCD and its detailed study is one of the most critical EIC goals.

Multiple experimental signatures of saturation have been discussed in the literature \cite{Accardi:2012qut}. The EIC program follows a multi-pronged approach taking advantage of the versatility of the EIC facility. One of the key signatures concerns the suppression of
dihadron angular correlations in the process $e+\mathrm{A} \rightarrow e^{\prime}+h_{1}+h_{2}+X$. The angle between the two hadrons $h_{1}$ and $h_{2}$ in the azimuthal plane is sensitive to the transverse momentum of gluons to and their self-interaction \textemdash the mechanism that leads to saturation. 
The experimental signature of saturation is a progressive suppression of the away-side ($\Delta \Phi = \pi$)
correlations of hadrons with increasing atomic number A at a fixed value of $x$.
Diffraction and diffractive particle production in \eA\ scattering is another promising avenue to establish the existence of saturation and to study 
the underlying dynamics. 
Diffraction entails the exchange of a color-neutral object between the virtual photon and the proton remnant. As a consequence, there
is a rapidity gap between the scattered target and the diffractively produced system. At HERA, these types of diffractive events made up a large fraction of the total \ep\ cross-section 
(10--15\%).
Saturation models predict that at the EIC, more than $20\%$ of the cross-section will be diffractive. In simplified terms, since diffractive cross-sections are proportional to the square of the nuclear gluon distribution, $\sigma \propto g(x, Q^2)^2$, they are very sensitive
to the onset of non-linear dynamics in QCD.
An early measurement
of coherent diffraction in \eA\ collisions
at the EIC would provide the first unambiguous
evidence for gluon saturation.

\paragraph{Machine and detector requirements for studies of gluon saturation}

Operation of the EIC at the highest energies with the heaviest nuclei will be an essential requirement for
discovering gluon saturation. Good tracking performance and forward calorimetric measurements are important in addition to very forward instrumentation of measuring diffractive events using a specialized silicon detector system known as Roman pots. Studying diffractive processes poses stringent requirements on the hermeticity of a detector system. A detailed study of saturation beyond its discovery would require
a systematic variation of the nuclear size and of $\sqrt{s}$ to see where the saturation sets in. 

\section{Nuclear Modifications of Parton Distribution Functions}

When compared to our knowledge of parton distribution functions in the proton, our understanding of nuclear 
PDFs (nPDF) is significantly more limited. Most of it comes from fixed-target experiments in a region of intermediate to high-$x$ values. Recently available data from hadronic collisions at the LHC have had little impact on extracting
nuclear PDFs \cite{Eskola:2016oht}.  
High energy electron-nucleus collisions at the EIC will enable measurements of nuclear PDFs over a broad and continuous
range in $Q^{2}$, all the way from photo-production ($Q^{2} \sim 0$) to high $Q^{2}$ in the perturbative regime. This will lead to the study of the nPDFs 
with unprecedented precision and to the understanding
of the collective effects that lead to modifications of nuclear PDFs compared to a proton. How parton distributions in nuclei are modified can be quantified by plotting their ratio to parton distributions in the proton,
normalized by the nucleus's atomic number. The deviation of this ratio from unity is a clear demonstration
that the nuclear parton distributions are not simple convolutions of those in the proton. A ratio below unity is often called shadowing, while an enhancement is referred to as anti-shadowing.

Nuclear PDFs are determined through global fits to existing inclusive DIS data off nuclei.
These are the structure 
function $F_2$ and the longitudinal structure function $F_L$.
While $F_2$ is sensitive to the momentum distributions of (anti-)quarks and gluons mainly through scaling violations, $F_L$ has a more considerable direct contribution from gluons. Note that the measurement of F$_{L}$ requires one to operate the collider at several different center-of-mass energies.

An additional constraint on the gluon distribution at moderate to high-$x$ comes from charm production via photon-gluon fusion. The fraction of charm production grows with the energy, reaching about $\sim$15\% of the total cross-section at the highest $\sqrt{s}$, thus permitting one to set a robust and independent constraint on the gluon 
distribution in nuclei at high-$x$ \cite{Accardi:2012qut, Aschenauer:2017jsk}.

\paragraph{Machine and detector requirements for precision nuclear PDF measurements}

Based on recent studies for inclusive DIS and charm cross-section measurements \cite{Aschenauer:2017oxs}, large $\sqrt{s}$ provides access to a broader $x$-$Q^{2}$ coverage and reaches more in-depth into the small-$x$ regime of gluon dominance. Measurements involving charm-final states require good impact parameter resolution at the level of $\sigma_{xy} \sim 20/p_{T} \otimes 5~\mu\mathrm{m}$. The general detector requirements are similar to inclusive and semi-inclusive measurements. 

\section{Passage of Color Charge Through Cold QCD Matter}

In the standard regime of perturbative QCD at high $Q^2$ and moderate to high $x$, in \eA\ scattering events, the virtual photon transmits a large fraction of the electron's energy. It interacts with a quark from a nucleon in the nucleus. The struck quark will subsequently traverse the nucleus, interacting with the color charges within, and continually lose energy. At some point, this quark will hadronize and form a color-neutral hadron. Whether the hadronization process happens inside or outside the nucleus depends on the interplay between the quark's energy and the atomic number of the nucleus.
If the virtual photon energy (in the nuclear rest frame) is high, the quark kicked out of the nucleon will have considerable energy and produce a jet. Measuring the jets experimentally provides several advantages over studies of leading hadrons. Reconstructed from multiple (ideally all) final state particles produced by hadronization of the scattered parton, jets are much closer proxies for the parton kinematics than any single-particle observable. Using jets in many cases removes (or minimizes) hadronization uncertainties. On the other hand, jets are composite objects with rich internal substructure encoding shower evolution and hadronization details. 

At the EIC, the production of jets will be a useful tool to measure and study the hadronic component in high energy photon structure \cite{Chu:2017mnm} and gluon helicity in polarized protons~\cite{Aschenauer:2017jsk}. Jet measurements will also constrain polarized and unpolarized parton distribution functions, probing gluon transverse momentum dependent distributions, contribute to studies of QCD hadronization, shower evolution, and cold nuclear matter effects.
An energetic jet from a scattered parton encodes the history of multiple interactions with the target nucleus, which generate $p_T$-broadening. Thus, a comparison of the cross-section in \ep\ and \eA\ collisions is expected to be sensitive to in-medium broadening effects. Several key measurements relying on jets were identified for their sensitivity to parton energy loss in the nucleus~\cite{Arratia:2019vju}, together with the development of new tools for controlling hadronization effects. Among such measurements are several variables assessed via lepton-jet correlations, including the electron's ratio to jet transverse momenta and a relative azimuthal angle between the measured jet and electron. These measurements will constrain the parton transport coefficient in nuclei~\cite{Liu:2018trl}. 
It is expected that the variability of the collider's energy and the "dialing" of the nuclear size will allow us to study both the emergence of jets as a function of energy and the internal spatial structure of jets systematically as an additional topic of high interest. 

In addition to jet studies, identified hadron measurements will provide additional experimental avenues for a detailed understanding of cold-QCD effects of color-charge. Parton propagation through cold nuclear matter and its’ effects on hadronization have been previously studied by the HERMES collaboration in semi-inclusive deep-inelastic scattering on nuclei via relative hadron production cross-sections for various light-flavor particle species.

In addition to inclusive hadron and jet production measurements, mapping the modification of heavy flavor production in reactions with nuclei of different sizes will provide an experimental handle for understanding the transport properties of nuclear matter.

\paragraph{Machine and detector requirements for jets studies}

Jets can only be produced and identified cleanly at high enough center-of-mass energies.  
High momentum jets feature higher hadron multiplicity and a more complex internal structure.
As such, high center-of-mass energy is vital for jet studies. Nuclear size is an essential 
control variable in these experiments and a broad range from light to heavy nuclei is desired 
for systematic studies of energy loss in a nuclear medium. 
It is imperative to have matching beam energies for \ep\ and \eA\ collisions to avoid extrapolation-related uncertainties and deliver the most precise measurements of nuclear effects. One of the key detector specifications results from jet measurements requiring good tracking performance and good calorimetric resolution. At forward rapidity hadronic final state energies are very large and require good hadronic resolution at the level of $\sigma(E)/E \approx 50\%/\sqrt{E} \otimes 10\%$.

\section{Connections to Other Fields} 

EIC-based science is broad and diverse. It runs the gamut from detailed investigation of hadronic structure with unprecedented precision to explorations of new regimes of strongly interacting matter. EIC science can be characterized by a few distinguishing themes that reflect the major challenges facing modern science today, and that have deep links to cutting edge research in other subfields of physics.

A prominent example are the various opportunities for electro-weak (EW) and beyond the standard model (BSM) physics. Achieving heightened sensitivity to various BSM scenarios requires a variety of improvements including crucial constraints on the parton distributions. By recording copious high-precision data, the EIC has the potential to provide the much-needed precision at large-$x$, with further implications for precision QCD and EW theory in \pp\ collisions at the LHC. But the connections reach much further. Precision measurements at the EIC can provide new limits on various BSM couplings. For example, measurements at the EIC over a wide range of $Q^2$ will test the running of Weinberg’s weak mixing-angle. The availability of polarized electron (or positron) beams with proton or deuteron targets can scrutinize lepton flavor violation mechanisms in the charged lepton sector. Furthermore, the high energy and luminosity at the EIC offers opportunities for new particle searches such as a heavy photon or a heavy neutral lepton. 

Measurements at the EIC are also expected to deliver important input for several areas of astroparticle physics. Fields such as cosmic-ray air showers and neutrino astrophysics will benefit from better constrained models of hadronic interactions. Deeply inelastic scattering and photo-nuclear processes have natural ties with the physics of hadronic collisions. These relate to the issue of small-$x$ gluons and factorization in \ep\ and \eA\ versus \pp\ and \pA, and to the implications of the determination of parton distributions for \pA\ collision for an improved understanding of the initial conditions in heavy-ion collisions. Similar, the accurate characterization of parton distributions in nuclei provided by the EIC can directly benefit the neutrino physics program. In return, neutrino scattering can help better understand the parton structure of nucleons and nuclei, where the nucleon strangeness content is one example.

Even though the EIC is a high-energy collider with typical energy scales in the tens-to-hundred of GeV range, there are key measurements that are of relevance to nuclear physics at much lower energies in the tens-to-hundreds MeV range. In diffractive deep inelastic scattering at high energies, a clean separation develops between the fragmentation region of the electron and that of the nuclear target. Correlations amongst nucleons in the target fragmentation region have the potential to provide novel insight into the underlying quark-gluon correlations that generate short-range nuclear forces. Short-range nucleon-nucleon correlations dominate the high momentum tails of the many-body nuclear wave function and show signs of universal behavior in nuclei. On the other hand, accurate nuclear structure input is needed in coherent exclusive channels with light ions — enabling the study of nuclear tomography in partonic degrees of freedom — and in reactions with spectator tagging, which result in additional control over the initial nuclear configuration.

\section{Summary of Machine Design Parameters} \label{sec-SC-Param}

Here we summarize the machine requirements that were motivated in the previous sections through a set of critical measurements
that reflect the highlights of the EIC science program.
The successful scientific outcome of the EIC depends critically on: 
(a) the luminosity, 
(b) the center-of-mass energy, and its range, 
(c) the lepton and light-ion beam polarization, 
and (d) the availability of ion beams from deuterons to the heaviest nuclei. 
Two interaction regions are desired to ensure a robust physics program with complementary detector systems.

\paragraph{Luminosity}

The EIC is being designed to achieve peak luminosities 
between 
10$^{33}$cm$^{-2}$ s$^{-1}$
and 
10$^{34}$cm$^{-2}$ s$^{-1}$. 
To put these numbers into context, note that a luminosity of 
10$^{33}$cm$^{-2}$s$^{-1}$ with strong hadron cooling ($L_{\rm peak} = L_{\rm avg}$) 
yields an integrated luminosity of 1.5 fb$^{-1}$ per month. Here we assume a 60\% operation efficiency for the collider complex as routinely achieved by RHIC.
Without strong hadron cooling for the same operation's parameters, one would get a 30\% reduction, as the average 
luminosity $L_{\rm avg}$ per fill is reduced to 70\% of the peak luminosity $L_{peak}$. 
Most of the key physics topics discussed in the EIC White Paper~\cite{Accardi:2012qut} 
are achievable with an integrated luminosity of 10~fb$^{-1}$
 corresponding to 30 weeks of operations. One notable exception is studying the spatial distributions of quarks and gluons in the proton with polarized beams. These measurements require an integrated luminosity of up to 100~fb$^{-1}$ and would therefore benefit from an increased luminosity of 10$^{34}$cm$^{-2}$ sec$^{-1}$. 
It should be noted that many measurements can be performed simultaneously by judiciously choosing beam species and their spin orientation appropriately.

\paragraph{Center-of-Mass Energy} 

To ensure a wide kinematic reach and a large coverage of phase space, the EIC requires a variable center-of-mass energy $\sqrt{s}$ in the range of $\sim\!\!\!\,20-100$~GeV, upgradable to 140~GeV~\cite{Accardi:2012qut}. An energy of $\sqrt{s}_{eN} = 140$\,GeV is needed to provide sufficient kinematic reach into the gluon dominated regime. Some measurements require a variation in $\sqrt{s}$. The lower center-of-mass energy limit is driven by the ability to measure transverse quantities well, which are of the order of 10-100 MeV. This is important, for example, for the accurate determination of quark TMDs at high values of $Q^2$.

\paragraph{Polarization of beams} 

EIC Physics involves two types of asymmetries: ({\it i}) double-spin asymmetries, 
requiring both electron and hadron beams to be polarized, and ({\it ii}) single-spin asymmetries, requiring only one beam\textemdash typically the hadron beam\textemdash to be polarized.
The statistical uncertainties for spin asymmetries are strongly affected by the degree of polarization achieved. 
For double-spin asymmetries the dependence is $1/\left[ P_{e}P_{p}\sqrt{N}\right]$ and 
 for single-spin asymmetries it is $1/\left[P \sqrt{N} \right]$. Therefore, high beam polarizations are mandatory
to reduce the statistical uncertainties.
Measurements require longitudinal and transverse polarization orientation for protons, deuterons, $^3$He, and other polarizable light nuclei, as well as longitudinal polarization for the electron beam.

\paragraph{Nuclear Beams} 

Ion beams of heavy nuclei (Gold, Lead, or Uranium) combined with the highest $\sqrt{s}$, will provide access to the highest gluon densities and to an understanding of how colored particles propagate through nuclear matter. On the other hand, light ions are essential to study the A-dependence of gluon saturation and for precision studies 
 of short-range nuclear correlations.

\section{Summary of Detector Requirements} 
\label{sec-Det-SciReq}

\begin{figure}[ht]
\centering
\includegraphics[width=0.95\textwidth]{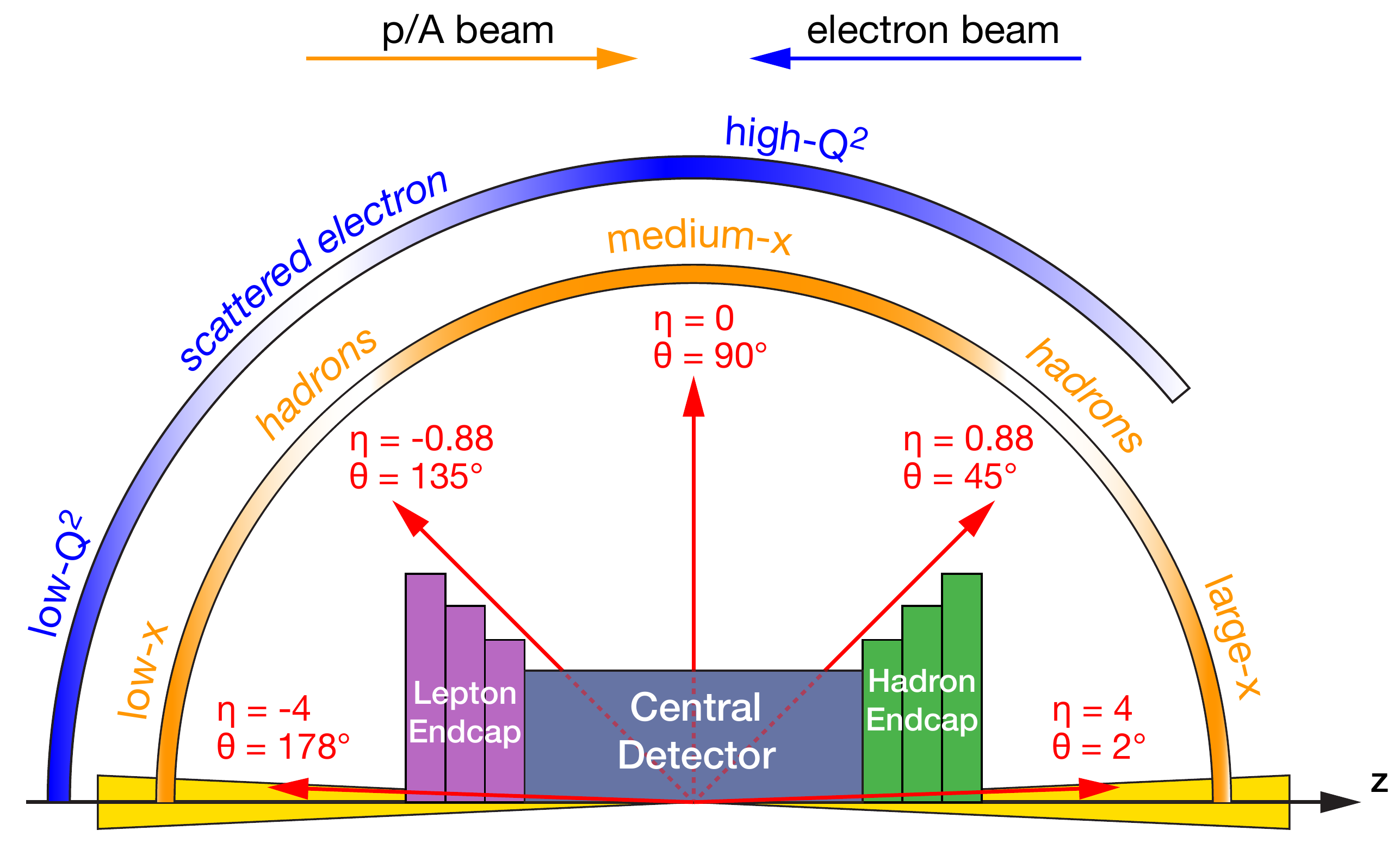}
\caption{Schematic showing the distribution of the scattered lepton and hadrons for different
$x-Q^2$ regions over the detector polar angle / pseudorapidity coverage.}
\label{Det_Schematics}
\vspace*{3mm}
\end{figure}

The diverse physics program promised by the new Electron-Ion Collider poses a technical and intellectual challenge for the detector design to accommodate multiple physics channels. Accommodating the needs of experimental measurements with different and, at times, competing requirements in one general-purpose detector design requires detailed consideration of the physics processes involved. Preliminary investigations on this topic were put forward in the EIC White Paper~\cite{Accardi:2012qut}. Further developing a more detailed set of physics-driven requirements for a future conceptual design of a general-purpose EIC detector
and considerations for a second complementary detector to overcome the technical and intellectual challenges were
the primary focus of the year-long Yellow Report EIC Users Group community effort. 

Figure~\ref{Det_Schematics} illustrates the correlation between 
polar angle ($\theta$) and pseudorapidity ($\eta=-\ln\tan(\theta/2)$)  
and the $x-Q^2$ phase space for the EIC physics program.
Recent studies of the physics-driven detector requirements were organized by three basic types DIS processes: Inclusive DIS both in neutral and charged current mode, semi-inclusive DIS, and exclusive DIS. Those basic processes are shown in Table~\ref{DIS.processes} 
For the following summary, and throughout this document, the beams' directions follow the convention used at the HERA collider at DESY: the hadron beam travels in the positive $z$-direction/pseudorapidity and is said to be going "forward." The electron beam travels in the negative $z$-direction/pseudorapidity and is said to be going "backward" or in the "rear" direction.  

All physics processes to be measured at an EIC require having the event and particle 
kinematics ($x, Q^2, y, W, p_t, z, \phi, \theta $) reconstructed with high precision. Kinematic variables such as $x$, $Q^2$, $y$, and $W$ can be determined from the scattered electron or the hadronic final state using the Jacquet-Blondel method \cite{Blumlein:2012bf} or a combination of both. The electron method provides superior resolution performance for $x$ and $y$ in the low $x$ region, while the Jacquet-Blondel method yields increased resolution performance for $x$ and $y$ towards large $x$ values.
To access the full $x-Q^2$ plane at different center-of-mass energies and for strongly asymmetric beam-energy combinations, the detector must be able to reconstruct events over a wide span in polar angle ($\theta$) and pseudorapidity ($\eta$). This imposes stringent requirements on both detector acceptance and the resolution of measured quantities such as the energy and polar angle in the electron-method case. 

\paragraph{Detector Requirements}
Below we summarize the critical detector requirements that are imposed by the rich physics program of an EIC.

\begin{itemize}
  \item The EIC requires a $4\pi$ hermetic detector with low mass inner tracking.
  \item The primary detector needs to cover the range of $-4 < \eta < 4$ for the measurement of electrons, photons, hadrons, and jets. It will need to be augmented by auxiliary detectors like low-$Q^2$ tagger in the far backward region and proton (Roman Pots) and neutron (ZDC) detection in the far forward region. 
  \item The components of an EIC detector will have moderate occupancy as the event multiplicities are low. However, specific components close to the beamline might see higher occupancies depending on the machine background level. 
  \item Compared to LHC detectors, the various subsystems of an EIC detector have moderate radiation hardness requirements.
  \item Excellent momentum resolution in the central detector ($\sigma_{p_T}/p_T (\%) = 0.05 p_T \otimes 0.5$).
  \item Good momentum resolution in the backward region with low multiple-scattering terms ($\sigma_{p_T}/p_T (\%) \approx 0.1 p_T \otimes 0.5$). 
  \item Good momentum resolution at forward rapidities ($\sigma_{p_T}/p_T (\%) \approx 0.1 p_T \otimes (1-2)$).
  \item Good impact parameter resolution for heavy flavor measurements ($\sigma_{xy} \sim 20/p_{T} \otimes 5~\mu\mathrm{m}$).
  \item Good electromagnetic calorimeter resolution in the central detector ($\sigma(E)/E \approx 10\%/\sqrt{E} \otimes (1-3)\%$ at midrapidity).
  \item Excellent electromagnetic calorimeter resolution at backward rapidities ($\sigma(E)/E \approx 2\%/\sqrt{E} \otimes (1-3)\%$).
  \item Good hadronic resolution in the forward region ($\sigma(E)/E \approx 50\%/\sqrt{E} \otimes 10\%$).
  \item Excellent PID for 3 $\sigma$ $\pi/K/p$ separation up to 50\,\gevc in the forward region, up to 10\,\gevc in the central detector region, and up to 7\,\gevc in the backward region.
\end{itemize}

Volume 2 of this Yellow Report provides further details on the detector requirements and a detailed discussion of the achievable precision for various observables.

The EIC physics program will allow us to deepen our understanding of the visible world around us, including the origin of the nucleon mass, the nucleon spin, and the emergent properties of a dense system of gluons.
This program requires a unique and versatile accelerator facility of colliding polarized electron and polarized proton and light-ion beams, and a range of light and heavy nuclei. The broad variety of processes required to address the diversity of the EIC science program similarly demands a unique $4\pi$ hermetic, asymmetric detector systems capable of identifying and reconstructing the energy and momentum of final-state particles with high precision.


\chapter{Detector Concepts}  
\label{part1-chap-DetCon}

EIC detectors are essential to make the detailed measurements described in the previous section to access the physical observables described by theoretical calculations. They will be large, sophisticated, and unique instruments which will be designed and constructed by multi-institutional collaborations of the EIC users from laboratories and universities around the world. This effort profits from a wealth of experience gained at the first \ep~collider facility HERA at DESY, Germany and the enormous development of novel detector concepts over the last several decades, since the first \ep~collisions at HERA in 1992. 
The detectors will be located at the interaction regions, where the electron and ion beams are brought into collision in a controlled way.  The 2015 US Nuclear Physics Long Range Plan recommended consideration of multiple EIC interaction regions and the EIC users have clearly stated their desire for two EIC detectors to effectively carry out the extensive scientific program.

The EIC detectors must be located in the interaction regions where space is constrained due to the requirements of high luminosity.  They must have strong integration of forward and backward detectors and multiple hermetic functionalities (precision energy measurement and particle tracking and identification) to determine the energy-momentum four-vector of final-state particles over a large range of energies: $\sim$10 MeV to $\sim$10 GeV. 

In addition to the major detector facilities, other sophisticated scientific instrumentation will be essential to carry out the scientific program. The collision luminosity must be determined using special purpose detectors close to the beams.  Furthermore, in spin-dependent measurements, precise knowledge of the polarizations of the electron and ion beams is essential.  This is obtained using special purpose polarimeters which are developed and operated by separate teams of physicists, engineers, and technicians.

The design for detector(s) at the EIC is centered around solenoidal superconducting magnets with bipolar fields, which can be achieved either through improvement modifications of the BABAR/sPHENIX magnet at 1.5\,T or with a new superconducting magnet at 3\,T. The solenoidal configuration naturally leads to tracking and vertexing, particle identification, and calorimetry systems organized in a configuration with barrel and endcap detectors. The detectors must be designed to operate with high efficiency in the presence of a large rate of background generated by the intense, circulating beams as they traverse the vicinity of the detectors.

In contrast to symmetric $ee$ and $pp$ colliders, the asymmetric nature of collisions at the EIC leads to unique detector requirements. The hadron endcap, barrel, and electron endcap detector systems see very different particle distributions, in terms of both momentum and particle types. Likewise, the performance requirements on these detector systems vary significantly between the detector regions. This is reflected by the critical detector requirements for the track, vertex, and energy resolution and particle identification separation summarized in the previous section.

The tracking and vertexing, particle identification, and calorimetry concepts described in this section do not identify a specific technology that can be used in all detector regions. Rather, the aim is to combine the best technology for each region in detector concepts that achieve the full set of requirements.

Multiple combined detector concepts for each detector functions are presented in this report. The complementarity between the detector technologies used in different detector concepts can be used to tailor detectors in the different interaction regions shown in Figure~\ref{EIClayout}, at the locations of the current STAR and sPHENIX detectors.

\section{Tracking and Vertexing Detector Systems}

The tracking and vertexing systems under consideration are based on semiconductor detector technologies and gaseous tracking detector technologies, with concept detectors combining both technologies.

Silicon semiconductor-based sensors collect electron/hole pairs caused by the passage of charged particles. The tracking and vertexing detector systems must have high granularity to satisfy the tracking (vertexing) resolution requirements of better than 5 $\mu$m (around 3 $\mu$m) while maintaining a low material budget below 0.8\% (0.1\%) of a radiation length in the barrel (endcap). Monolithic active pixel sensors (MAPS) have seen an evolution from the 180\,nm technology used by STAR and ALICE to the 150/180\,nm Depleted MAPS (D-MAPS). A third generation 65\,nm process is under development as a joint effort between the EIC and ALICE ITS3 vertex tracker upgrade.

In gaseous tracking chambers the ionization caused by tracks drifts to anode planes in endcaps where it is collected, potentially after additional amplification. A double-sided time projection chamber (TPC) with a central cathode plane and gas amplification modules at the endcaps is under construction for the sPHENIX experiment and may be modified for use at the EIC. Upgrades to the read-out pads for the EIC would be focused on micro-pattern gaseous detectors such as gas electron multipliers (GEMs), $\mu$MEGAs or $\mu$RWELL can provide electron amplification before read-out on high granularity anode printed circuit boards. Gaseous tracking detectors also aid in particle identification with ionization energy loss information. 

\begin{figure}
    \centering
    \includegraphics[width=\textwidth]{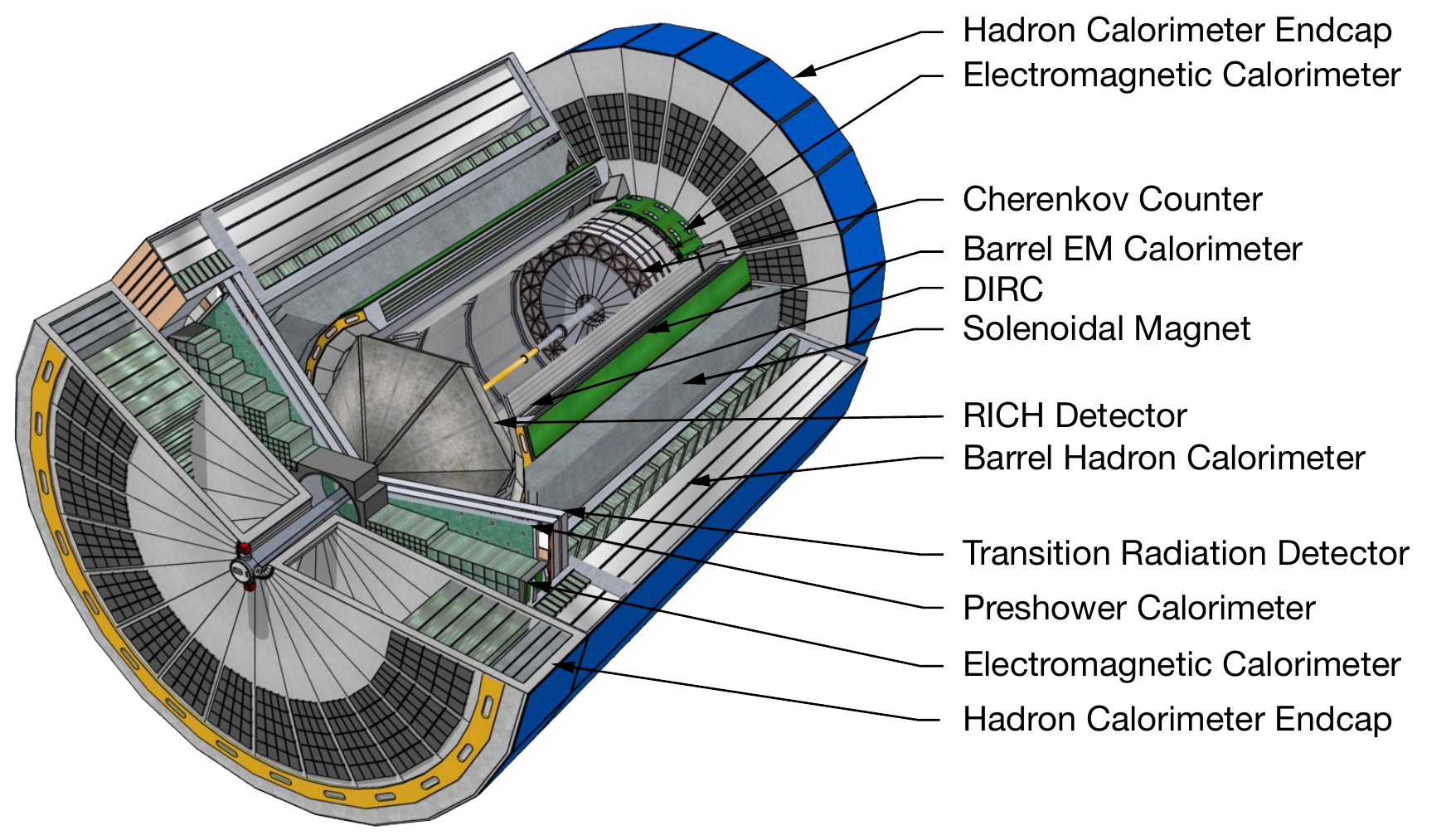}
    \caption{CAD model of a particular EIC detector concept, with the artistic rendering of the tracking, particle identification, and calorimetry subsystems.} 
\end{figure}

Two baseline tracking detector concepts are presented. An all-silicon tracking detector option with barrel and endcap silicon detector can be realized in a compact form. A hybrid tracking system combines a silicon vertex detector within a TPC and provides $dE/dx$ measurements that can aid particle identification. In both main options, alternative tracking options exist in the backward and forward tracking endcaps.

\section{Particle Identification Detector Systems}

The second major detector system, particle identification, separates electrons from pions, kaons, and protons, with significant pion/electron suppression and better than 3$\sigma$ pion/kaon/proton separation in all rapidity regions. Using the specific ionization ($dE/dx$) in time projection chambers with novel gas mixtures allows for improved resolution approaching the limit of Poisson statistics. However, dedicated particle identification detectors, based on Cerenkov light emission and time of flight measurements will be required.

In Cerenkov detectors light emitted by tracks faster than the speed of light in a gas, aerogel, or quartz radiator medium is detected. The tight integration with the tracking system exists here as well, since the originating track must be well known. Different radiator media are required in the electron endcap, barrel, and hadron endcap due to the different momentum ranges of particles in those regions. A hadron blind detector without focusing and with CsI photocathodes evaporated on GEMs can separate electrons from hadrons. A similar detection approach with focusing is used in the CsI ring imaging Cerenkov (CsI RICH) detector. Novel approaches that use nano diamond powder instead of CsI are also under consideration. A dual RICH (dRICH) with both a gas and aerogel radiator avoids the holes in the performance due to the Cerenkov thresholds. A modular RICH (mRICH) concept uses a Fresnel lens for focusing. Finally, detection of internally reflected Cerenkov light (DIRC) is under consideration for a high performance DIRC (hpDIRC), and would outperform the DIRC at BaBar and PANDA.

Other particle identification technologies are under development as well. Time of flight particle identification at low momentum can be reached through precision timing measurements in large area picosecond photon detectors (LAPPD) in the endcaps. GEM transition radiation detectors (GEM-TRD) combined with neural networks have also been shown to separate electrons and pions.

Based on these particle identification technologies, several combined concepts are presented. The forward direction includes a gas-based Cerenkov detector but requires another technology such as the dRICH. In the central region, a combination of the DIRC and TOF detectors must be augmented with, for example, the ionization loss measurements in the hybrid tracking detector system or other identification technologies in the more compact all-silicon tracking detector system. In the backward or rear direction several options satisfy the requirements, including the
mRICH with LAPPD.

\section{Calorimeter Detector Systems}

The third major detector system, calorimetry, measures particle energy and includes both electron and hadron calorimetry efforts. Only light-collecting calorimeters are considered here. Electromagnetic calorimetry (ECAL) requires excellent resolution to constrain the electron scattering kinematics, but also aids in separating electrons from hadrons, in the detection of neutral particles, and in the separation of the two photons from neutral pion decay. Hadron calorimetry (HCAL) is required for determination of the total energy in hadronic jets, in particular for neutral components which are not tracked.

Among possible ECAL technologies discussed are homogeneous detectors (PbWO$_4$, scintillating glass, and lead glass) and sampling calorimeters (scintillator fibers in tungsten powder and layered shashlyk detectors). Due to limited space available, short radiation length materials are favored. Likewise, silicon photomultipliers (SiPMs) are preferred since they take less space than regular photomultipliers and operate in the magnetic field. In the backward region, PbWO$_4$ appears to be the only option. In the central region, projective geometry is required. In the forward region, high granularity is required to resolve pion decay photons.

For the HCAL, existing technologies (e.g.  scintillating / depleted uranium sampling calorimeters as used at ZEUS) are considered sufficiently performant. Efforts are discussed to avoid lead in favor of steel, and achieve a design where the HCAL is the support structure for the ECAL. In the hadron endcap a denser material would be preferable, and the STAR Forward upgrade has allowed the construction of a small prototype of a compensating calorimeter with better resolution.

\section{Auxiliary Detector Systems}

In addition to the major central detector systems, specialized auxiliary detector systems will be necessary, all of which require close integration in the accelerator lattice. This is particularly true for the electron and hadron polarimeters, but also applies to the far-forward and far-backward regions of the detector.

In the far-forward region, silicon detectors in roman pots can detect very forward hadrons up to 5\,mrad with high timing resolution of low gain avalanche diodes (LGADs). Similar detector technologies will be used in the off-momentum detectors to tag nuclear breakup of Lambda decay products. Neutrons and low-energy photons in the forward direction will be detected in the Zero-Degree Calorimeter (ZDC), with both ECAL and HCAL components. Technologies from the ALICE FoCal and the LHC ZDC are considered.

In the far-backward region, bremsstrahlung photons detected in an electromagnetic zero-degree calorimeter or a pair spectrometer will be used to determine the luminosity, an important normalization quantity for many observables. Very low-$Q^2$ electrons will be tagged in far-backward position-sensitive detectors or segmentation in the zero-degree calorimeter.

In other sections of the EIC, electron and hadron polarimeters will non-destructively measure the polarization to a systematic precision better than 1\%. To allow timely feedback to accelerator operators, a statistical precision of similar size will be achieved on short time scales. 

For the electron beam, a Compton polarimeter can reach the needed luminosity using a diode laser with high repetition frequency and a fiber amplifier to reach powers up to 20\,W. To measure both longitudinal and transverse polarization, position sensitive detectors such as diamond strip or HV-MAPS detectors can be used.

For the hadron beam, the natural starting point is to use the existing polarimeters at RHIC: the atomic hydrogen jet for absolute measurements combined with a fast carbon ribbon for relative measurements. At the higher proton currents of the EIC, additional hydrogen jet detectors and alternative ribbon targets will be required. For $^3$He beams the hydrogen jet may be replaced by a polarized $^3$He target.

To manage the data acquisition bandwidth and associated selection of events of interest to the physics analyses, the EIC will use a streaming readout approach without trigger electronics that controls whether or not to record events, similar to the LHCb upgrade currently in progress. On the software side, new approaches in artificial intelligence are being explored.

\section{Two Complementary Detectors}

The EIC science program is diverse and broad. It requires a 4$\pi$ detector with strong integration of forward and backward detection capabilities. It requires multiple hermetic functionalities (precision energy measurement, and particle tracking and identification). It must cover a large and versatile range of energies spanning from nuclear energy scales to multi-GeV electron and ion beam energies.

The strong diversity of EIC science imposes the essential feature that the interaction region and the detector at the EIC are designed so all particles are identified and measured at as close to 100\% acceptance as possible and with the necessary resolutions. Variations of the interaction region design and beam line optics between the two interaction points can allow further optimization and enhancement of EIC science reach.

The broad science reach of the EIC is also reflected in the variety of detector technologies that are under consideration. Table 3.1 summarizes the high-level performance of different subdetectors based on a 3 T solenoid for a future EIC detector. The clear conclusion is that the best way to optimize the science output is through two detectors that differ in their basic features such as the Solenoid field and choices of sub-detector technologies. This will lead to complementarity in detector acceptance and systematic effects, and presents benefits due to technology redundancy. Studies performed to date already suggest the opportunity to optimize the overall physics output of the EIC in terms of precision and kinematic range through careful complementary choices of two general purpose detectors. 

In contrast to previous colliders, the complementarity between the two EIC detectors and their associated interaction regions will be built in from the start. Beyond maximizing EIC science promise, a further strong motivation for complementary detectors lies in the need for independent cross-checking of important results; the scientific community usually only becomes convinced of exciting new discoveries when two different experiments with different systematics arrive at the same conclusion.

\begin{sidewaystable}[thb]
\centering
\caption{This matrix summarizes the high level performance of the different subdetectors and a 3~T Solenoid. The interactive version of this matrix can be obtained through the Yellow Report Detector Working Group (https://physdiv.jlab.org/DetectorMatrix/).}
\label{tab:detector-matrix-exec}
\includegraphics[width=1.0\textwidth]{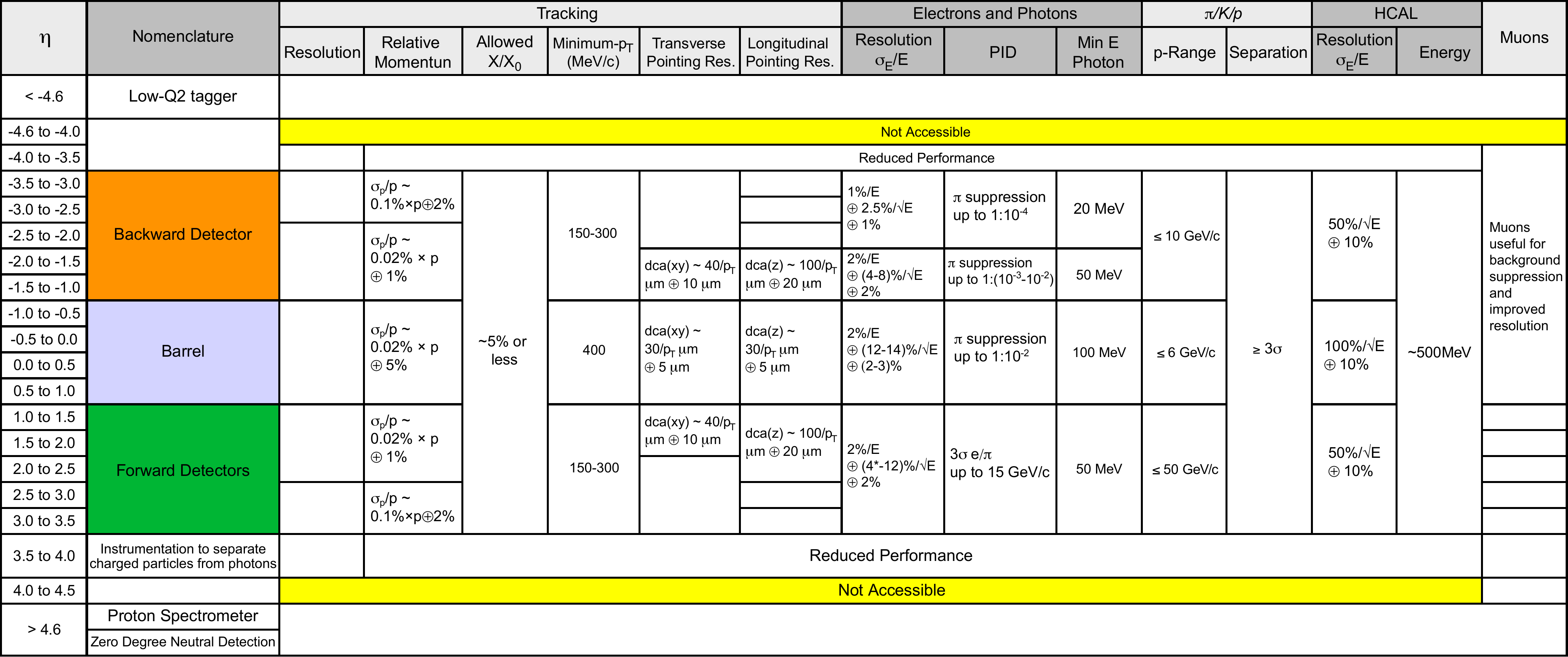} 
\vspace*{3mm}
\end{sidewaystable}
\FloatBarrier

\chapter{Opportunities for Detector Technology and Computing}
\label{part1-chap-OppDetTechComp}

In parallel with a nearly two-decade-long community effort of EIC science development and refinement, as well as experimental equipment conceptualization, BNL in association with TJNAF and the DOE Office of Nuclear Physics established in 2011 a highly successful generic EIC-related detector R\&D program. This program both built bridges between various domestic and international research groups and scientific communities, and was successful in its own right towards detector R\&D. Presently, 281 scientists are engaged in the generic EIC-related R\&D program, from 75 institutions in 10 countries. Most of the efforts have been organizationally merged in groups of topical consortia, which can provide the seeds for the EIC detector collaboration(s).

Many of the supported projects, ongoing or completed, developed technologies that are now integral parts of existing detector concepts or are regarded as potential alternatives. The vertex detector R\&D consortium aims to develop new improved Monolithic Active Pixel Sensors (MAPS) to meet the requirements demanded by the EIC requirements. Various Micro-Pattern Gas Detector (MPGD) technologies, such as Gaseous Electron Multiplier (GEM), Micromegas, and $\mu$RWELL, have been pursued for low material tracking in barrel and forward regions as well as Time-Projection Chamber (TPC) readouts. New concepts like miniTPCs and integrated Cherenkov-TPCs had been developed and tested. Many options for electromagnetic, and recently, hadronic calorimetry have received R\&D effort within the calorimetry consortium. From this grew the Tungsten-Scintillating Fiber (W-SciFi) calorimeter, scintillating fibers embedded in a W-powder composite absorber. In parallel, novel scintillating glasses (SciGlass) have been developed with unprecedented quality as cost-effective alternative to expensive lead-tungstate (PbWO4) crystals. The particle identification consortium is pursuing various technologies, such as Direct-Internally Reflected Cherenkov light (DIRC) detectors, modular and Dual Ring-Imaging Cherenkov (RICH) detectors, with Fresnel lens focalization in the former and with gas and aerogel radiators in the latter. New coating materials like nano-diamonds to replace Cesium-Iodide (CsI) for RICH photo sensors are also under investigation. Time-of-Flight detectors, as well as Roman Pots for forward proton detection, require highly segmented AC-coupled Low-Gas Avalanche Detector (AC-LGAD) sensors whose development has just started to get support from the program. Besides hardware R\&D the program has supported various vital projects such as machine background studies and simulation software developments to enable more accurate definition of the physics’ requirements. Sartre and Beagle are two examples of Monte-Carlo event generators whose development was substantially boosted by the program. Both were extensively used in the context of this report.

In general, due to this longstanding generic EIC-related detector R\&D program, and further support from Laboratory Directed Research \& Development (LDRD) Programs within the US national laboratories, and many university groups both inside and outside the US, the detector technologies to implement a successful comprehensive Day-One EIC Science program exist. For this reason the EIC User Group can continue to consider various technologies for many of the different detector functions to implement, with an eye also to possible detector complementarity for a second detector. The EIC also benefitted substantially from synergetic R\&D conducted for many high-energy and nuclear physics experiments, not only at BNL and TJNAF, but also for experiments such as ALICE and LHCb at CERN, PANDA at GSI and BELLE-II at KEK. 

On the other hand, further opportunities do remain. These are driven both by pursuing alternative detector technologies for a complementary second fully integrated EIC detector and Interaction Region, and to prepare for future cost-effective detector upgrades to enhance capabilities addressing new nuclear physics opportunities. Furthermore, the EIC will be a multi-decade nuclear physics facility after its construction is completed and will in this period likely require further detector upgrades driven by its science findings. It is expected that further physics opportunities enabled by new detector capabilities will already arise during the EIC design and construction phase.

Nuclear physics detection techniques typically need to cover a large range in energies. They can range from the MeV scale of nuclear binding energies and 100 MeV/c momentum scale below the Fermi momentum to isolate nuclear processes, all the way to the multiple tens of GeV scales to pinpoint the elementary sub-atomic quark-gluon processes and quark flavors. Due to this, nuclear physics drives detector technologies with often different demands than those in high-energy and particle physics. Examples are (i) particle identification techniques and their cost-effectiveness in readout (RICH, DIRC, ultra high-precision TOF, electro-magnetic and hadronic calorimetry), (ii) those driving detector material minimization to detect the lowest-momentum particles (inner tracking solutions, gaseous-based radial TPCs), (iii) those pushing for specific material radiation tolerances (electro-magnetic rather than hadronic, high-power target areas, and low-energy nuclear fragments), and (iv) those related to spin or polarization (beam, targets, polarimetry).

Further opportunities for detector technology within these overarching nuclear physics areas exist in the EIC design, construction, and science operations era. These can best be considered in detector functionality areas such as particle identification, calorimetry, tracking, and readout electronics, to address how one can enhance the performance of the EIC detector(s) with target R\&D projects in a year or more.

Examples of such detector opportunities include, but are not limited to, the following: material minimization in a possible all-Silicon tracker, particle identification reach at mid rapidity and at higher momenta, cost-effectiveness of readout of particle identification detectors by improvements to Silicon Photomultipliers (SiPMs) or to Large-Area Picosecond Photo-detectors (LAPPDs).
Furthermore, improvement of the achievable hadronic calorimetry resolutions, large-scale production and low-energy photon detection efficiency of possible glass-based electromagnetic calorimetry, new Application-Specific Integrated Circuit (ASIC) and front-end readout board needs required for streaming readout modes, or improved spatial and/or timing resolution of Zero-Degree Calorimeters driven by the imaging and diffractive science programs. It is crucial that some of this research for enhanced detector functionality continues and is recognized as driven by Nuclear Physics needs.

In parallel with these detector opportunities, unique opportunities exist to directly {\sl integrate} modern computing and data analysis methods in the experiment. Efforts are underway to develop methods and production systems to establish a quasi-instantaneous high-level nuclear physics analysis based on modern statistical methods. This requires a self-calibrated matrix of detector raw data synchronized to a reference time and would remove intermediate data storage requirements. This takes direct advantage of advances in micro-electronics and computing, and of artificial intelligence (AI) methods.

Micro-electronics and computing technologies have made order-of-magnitude advances in the last decades.  Combined with modern statistical methods, it is now possible to analyze scientific data to rapidly expose correlations of data patterns and compare with advanced theoretical expectations.  While many existing nuclear physics and high-energy physics experiments are taking advantage of these developments by upgrading their existing triggered data acquisition to a streaming readout model (where detectors are read out continuously), these experiments do not have the opportunity of integrated systems from data acquisition through analysis, such as the EIC has.  Hence, we aim to remove the separation of data readout and analysis altogether, taking advantage of modern electronics, computing and analysis techniques in order to build the next generation computing model that will be essential for probing the femto-scale science accessible at the EIC.

An integrated whole-experiment approach to detector readout and analysis towards scientific output will take advantage of multiple existing and emerging technologies.  Amongst these are: streaming readout, continuous data quality control and calibration, task-based high performance local computing, distributed bulk data processing at supercomputer centers, modern statistical methods that can detect differences among groups of data or associations among variables even under very small departures from normality, and systematic use of artificial intelligence (AI) methods at various stages.

To further elaborate on the latter, AI is becoming ubiquitous in all disciplines of Nuclear Physics.
EIC could be one of the first large-scale collider-based programs where AI is systematically employed from the start.
AI already plays an important role in existing experiments such as LHCb at CERN, where machine learning algorithms make already the majority of the near-real-time decisions what physics data should be written or proceed to a higher level analysis. 

Supported by the modern electronics able to continuously convert the analog detector signals, streaming readout can further the convergence of online and offline analysis: here the incorporation of high-level AI algorithms in the analysis pipeline can lead to better data quality control during data taking and shorter analysis cycles.  Indeed, AI could foster in the next years significant advances in the crucial area of fast calibration/alignment of detectors, greatly facilitating a data streaming readout approach.

For charged-particle tracking, where in nuclear physics experiments typically most of the computing cycles are spent in propagating the particles through inhomogeneous magnetic fields and material maps, AI can contribute to determine the optimal initial track parameters allowing to decrease the number of iterations needed. For particle identification, crucial for Nuclear Physics experiments, has recently seen a large growth of applications.

AI at the EIC is expected to play a role in high-level physics analysis such as searches for rare signatures which necessitates advanced techniques making strong use of machine learning to filter out events, the utilization of jets to empower taggers for boosted jets and quark flavors within the jets, and in the aid for construction of higher-level Wigner distributions from sparse and missing data. With the EIC detector design ongoing and opportunities for two detectors at the EIC, AI can be gainfully used for the design optimization process of the large and complex EIC detector systems that are based on computationally intensive simulations, for the optimization of the individual detector systems, and even the optimization of materials used within detectors for improved performance.

 
\cleardoublepage

%
%
\phantomsection
\addcontentsline{toc}{part}{\Large{\textbf{Volume II: Physics}}}

\newgeometry{textwidth=8.5in,textheight=11.0in}
\includegraphics[width=8.5in,height=10.99in]{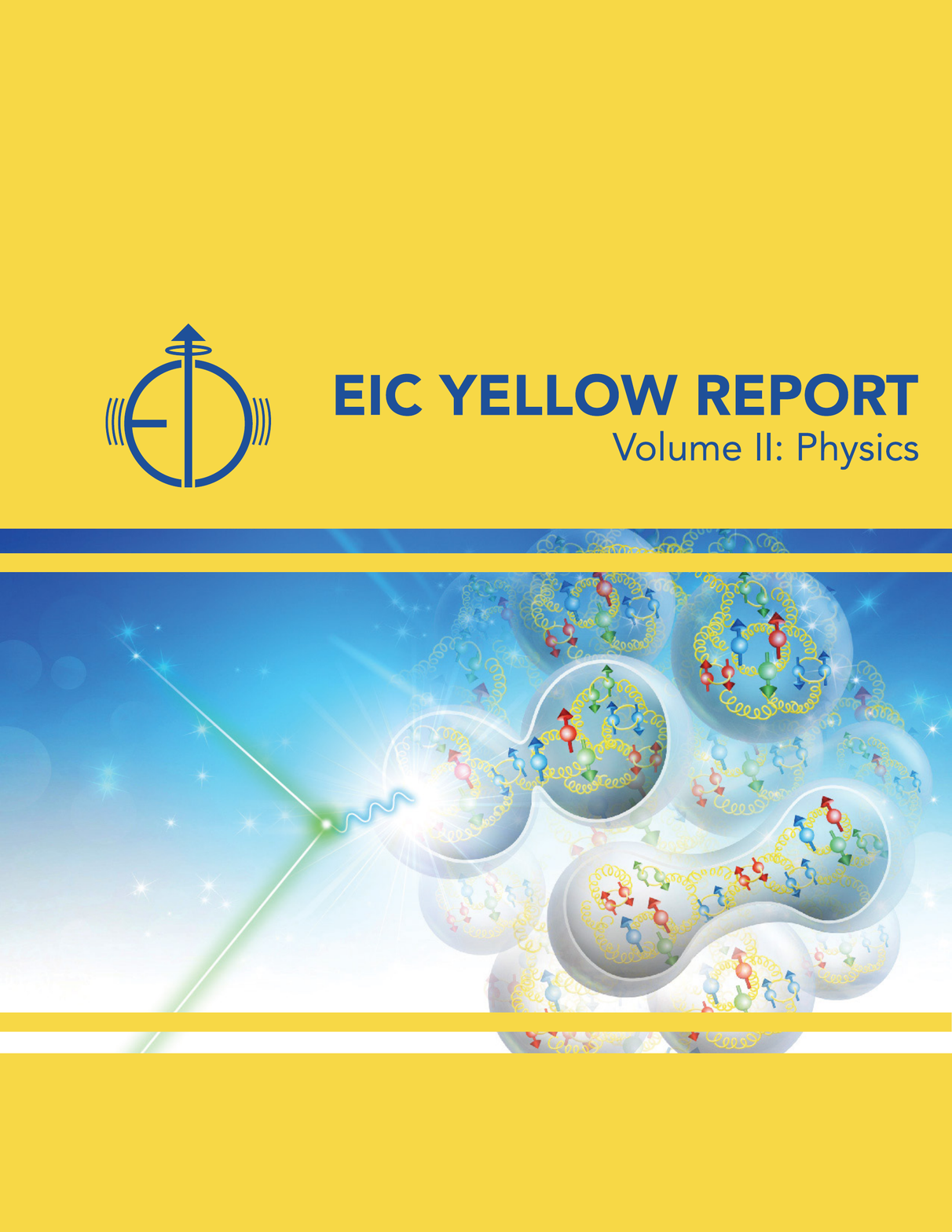}
\restoregeometry
\semiblankpage

%
%
\chapter{Introduction to Volume II}
\label{part2-chap-Intro}

For more than a decade, the physics community studying quantum chromodynamics (QCD) has gathered to come up with the next experimental facility that can answer the outstanding questions about the inner structure of matter. A consensus developed towards the need of a machine that can explore hadrons using an electromagnetic probe at high center-of-mass energies and high luminosity: an Electron-Ion Collider (EIC). Over the years, tremendous efforts have been devoted to making the physics case strong and defining the characteristics of such a collider. Important capabilities such as polarized beams, high luminosity, flexible center-of-mass energy, large variety of beam species are all essential for the success of the project. The physics case has been developed by the community in a White Paper~\cite{Accardi:2012qut} outlining the fundamental questions that an EIC would address. Crucial questions that could be addressed at an EIC include the origin of the mass of atomic nuclei, the origin of the spin of protons and neutrons, how gluons hold hadrons and nuclei together, and whether new emergent forms of matter made of gluons densely packed in phase space exist.

Another major milestone was the recommendation in the US 2015 Long Range Plan for Nuclear Science of a high-energy high-luminosity polarized EIC as the highest priority for new facility construction. Since then, the interest in the community has continued to grow. An assessment of a US-based electron-ion collider science program was carried out by the National Academy of Sciences, Engineering, and Medicine (NAS) in 2017, with a report produced in 2018~\cite{NAP25171} and findings which concluded that its physics case is ``compelling, fundamental, and timely". 

The NAS committee added that along with advancing nuclear science, an EIC would also benefit other areas such as astrophysics, particle physics, accelerator physics, and theoretical and computational modeling.  It would also play a valuable role in sustaining the U.S. nuclear physics workforce in the coming decades, incorporating new developments, including those in adjacent research domains.  Moreover, it would have a significant role in advancing more broadly the technologies that would as a result of the research and development undertaken in the implementation and construction of an EIC in the U.S.  The report emphasizes that an EIC is the only high-energy collider being planned for construction in the U.S. currently, and building such a facility would maintain U.S. leadership in accelerator collider science while benefiting the physical sciences.

Following the extremely positive assessment by the NAS, the US Department of Energy officially started the EIC project by establishing its CD-0 (mission need) in December 2019. In parallel to the DOE-driven activities, and in order to prepare the EIC construction, the physics community, organized around the EIC Users Group, started an initiative to define the detector requirements needed to deliver the science spelled out in the EIC White Paper and new topics highlighted in the NAS report and other publications. The goal is to advance the state and detail of the documented physics studies and detector concepts in preparation for the realization of the EIC. The effort aims to provide the basis for further development of concepts for experimental equipment best suited for science needs towards future Technical Design Reports (TDRs). These efforts were carried out during the year 2020 and are summarized in this ``Yellow Report''.

Since 1955, the CERN Yellow Reports series provides a medium for communicating CERN-related work where publication in a journal is not appropriate. Reports include material having a large impact on the future of CERN, as well as reports on new activities which do not yet have a natural platform. The series includes reports on detectors and technical papers, the criteria being that the audience should be large and the duration of interest long. The term Yellow Reports is now used frequently for documents with similar purpose in various physics communities unrelated to CERN.

To advance both the physics case and the detector concepts in preparation for the EIC, the EIC Yellow Report effort was initiated in December 2019 with a kick-off meeting hosted by the MIT. A total of 4 dedicated workshops, with an intermediate report to the community at the EIC Users Group meeting, were part of the program:
\begin{itemize}
    \item {\color{blue}{\href{https://indico.bnl.gov/event/7449/}{\underline{1st Workshop}}}}: March 19-21, 2020, Temple University, Philadelphia, PA
    \item {\color{blue}{\href{https://indico.bnl.gov/event/8231/}{\underline{2nd Workshop}}}}: May 22-24, 2020, University of Pavia, Pavia, Italy
    \item Status reports at {\color{blue}{\href{https://indico.bnl.gov/event/7352/}{\underline{Summer EICUG Meeting}}}}: August 3-7, 2020, FIU, Miami, FL
    \item {\color{blue}{\href{https://indico.bnl.gov/event/9080/}{\underline{3rd Workshop}}}}: September 17-19, 2020, CUA, Washington, DC
    \item {\color{blue}{\href{https://indico.bnl.gov/event/9913/}{\underline{4th Workshop}}}}: November 19-21, 2020, UC Berkeley, Berkeley, CA
\end{itemize}
All workshops and meetings and the entire effort were open to the participation of anyone in the community. The Yellow Report initiative was thus set to establish  a medium for broad community engagement, further the physics case, provide input to detector requirements, bring forward best available and emerging detector technologies and concepts, and document the progress towards EIC realization. This Yellow Report represents the current state of affairs and is intended to be used by the scientific community as the basis for further studies and developments. It is hoped that this intellectual investment into the future of the EIC will then guide the design and development of the actual EIC detectors.

The physics case and physics-driven detector requirement studies were organized within the Physics Working group. In parallel, detector R\&D efforts and detector technology choices were worked on within the Detector Working group. This Volume summarizes the Physics Working group efforts. It addresses the progress towards the main goal: carrying out a quantitative analysis of planned physics measurements for topics highlighted in the White Paper and for  physics topics developed more recently, and documenting emerging implications for detector design. To focus and further organize this effort, the Physics Working Group was divided into smaller subgroups, organized by physics processes, which provides a natural pathway for assessing detector requirements. These subgroups are: Inclusive Reactions, Semi-inclusive Reactions, Jets and Heavy Quarks, Exclusive Reactions, Diffractive Reactions \& Tagging.

The working groups were in charge of defining and studying the physics processes that fall into the particular categories. Through simulations, and by drawing from theoretical studies, the goal was to specify the requirements for the detector which would ensure that the physics outlined in the White Paper and the NAS report could be performed successfully. 
The working groups also addressed additional aspects that have gained  broader attention after the publication of the White Paper and the NAS report.

The physics studies that led to these detector requirements and novel physics ideas, as well as proposed measurements that may not yet provide input for the detector R\&D are documented in Chapter~\ref{part2-chap-EICMeasandStud}.The summary of the detector requirements delivered by each of the Physics subgroups is documented in the corresponding section of Chapter~\ref{part2-chap-DetRequirements}.  

\label{part2-ch.phy}

\chapter{The EIC Physics Case}
\label{part2-chap-EICPhyCase}

The physics case for the EIC has been presented in detail in previous documents such as the EIC White Paper~\cite{Accardi:2012qut}  and the Report of the NAS~\cite{NAP25171}.
Those documents spelled out a number of key questions, which the EIC would be able to answer --- see also the Executive Summary of this Yellow Report.
Questions that have been highlighted in the NAS Report are:
\begin{itemize}
\item How does the mass of the nucleon arise?
\item How does the spin of the nucleon arise?
\item What are the emergent properties of dense systems of gluons?
\end{itemize}
Beyond these questions, the EIC science case is, of course, ever evolving, and the following Chap.~\ref{part2-chap-EICMeasandStud} presents the current state of affairs.
In that chapter the various topics are grouped according to four major themes:
\begin{itemize}
\item Global properties and parton structure of hadrons
\item Multi-dimensional imaging of nucleons, nuclei and mesons
\item The nucleus: a laboratory for QCD
\item Understanding hadronization
\end{itemize}
Following this structure, we here give brief introductions and overviews for several key physics topics. 
More details are given in the White Paper~\cite{Accardi:2012qut} and the NAS Report~\cite{NAP25171}.
Further discussion as well as all the results of impact studies can be found in the various sections in Chap.~\ref{part2-chap-EICMeasandStud}.

\newpage

{\bf 1.~Global properties and parton structure of hadrons}

EIC measurements will reveal the quark and gluon structure of hadrons at the next level.
This, in particular, applies to global properties of the nucleon such as its spin and mass, that is, how those properties can be understood in terms of contributions from the partons.
\begin{figure}[t]
    \centering
    \includegraphics[width=0.65\textwidth]{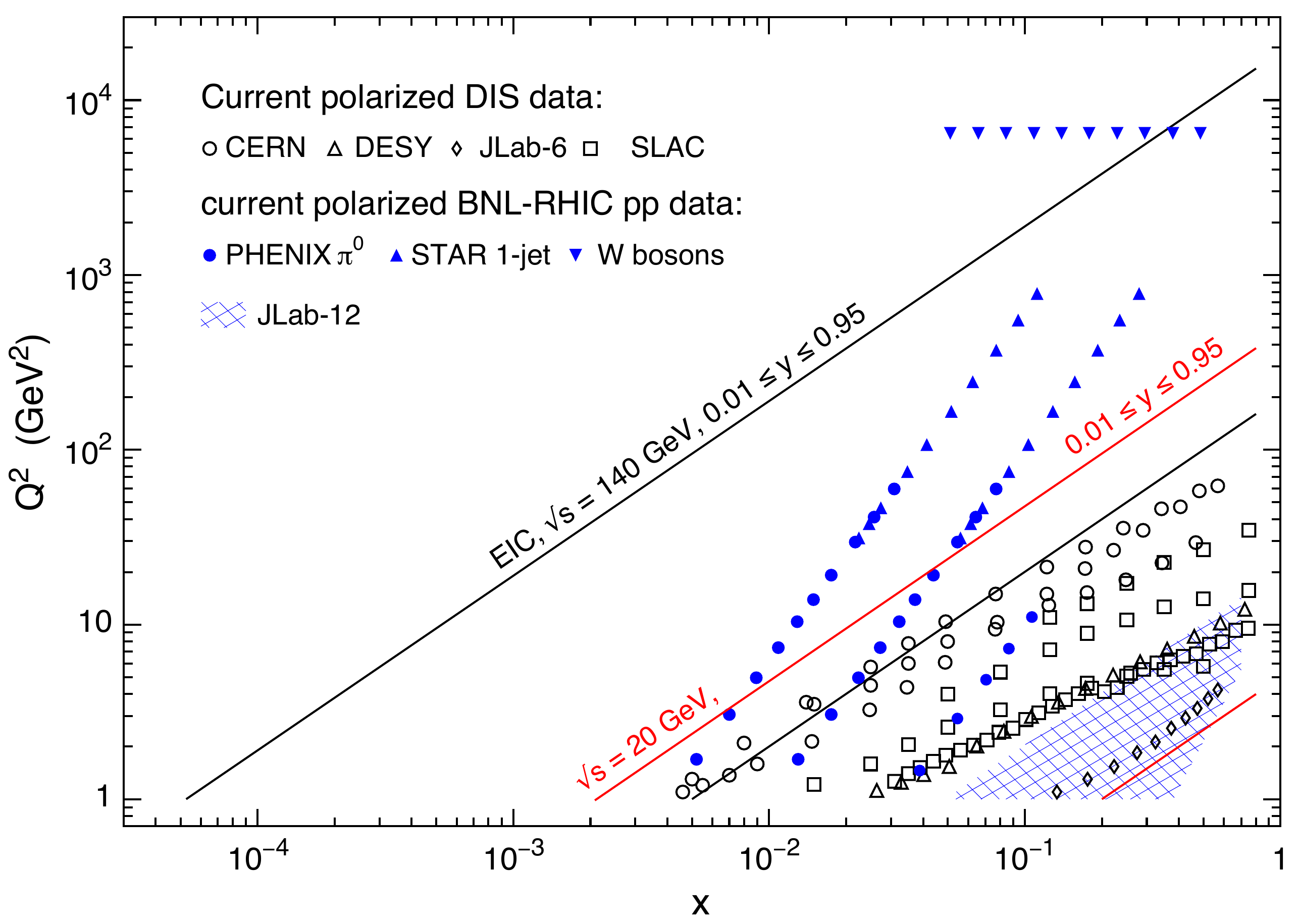}
    \caption{The $x$-$Q^{2}$ coverage of the EIC for two different center-of-mass energies, in comparison with polarized $ep$ experiments at CERN, DESY, Jefferson Lab and SLAC, as well as $pp$ experiments at RHIC.} 
    \label{fig:x_q2_poldis}
\end{figure}

{\bf Nucleon spin:} 
The spin and the mass are among the most important quantities characterizing any hadron.
Getting a deeper understanding of those quantities in QCD is a key mission of the EIC.
Starting with the spin, we recall that the spin of the nucleon can be decomposed according to~\cite{Jaffe:1989jz}
\begin{equation}
\frac{1}{2} = \frac{1}{2} \Delta \Sigma(\mu) + \Delta G(\mu) + L_q(\mu) + L_g(\mu)
\label{e:spin_sum_rule}
\end{equation}
into contributions from the quark plus antiquark (gluon) spin $\frac{1}{2} \Delta \Sigma$ ($\Delta G$), as well as quark (gluon) orbital angular momenta $L_q$ ($L_g$).
Each term of this decomposition depends on the renormalization scale $\mu$, where the scale dependence drops out upon summing over all contributions. 
For quite some time it was generally believed that the $\Delta \Sigma$ term is largely responsible for making up the nucleon spin.
It came therefore as a big surprise when in the late 1980s the EMC Collaboration reported a value for $\frac{1}{2} \Delta \Sigma$ which is only a small fraction of the nucleon spin~\cite{Ashman:1987hv}.
This ``nucleon spin crisis'' initiated a large number of further experimental and theoretical activities. 
Presently available results suggest that about $25\%$ of the nucleon spin is carried by the spins of the quarks and antiquarks~\cite{Aidala:2012mv}.
Mainly due to the RHIC spin program, we now also have clear evidence for a nonzero $\Delta G$~\cite{Aschenauer:2015eha}.
However, the values of $\Delta \Sigma$ and, in particular, $\Delta G$ still have very large uncertainties.
The main reason for this situation lies in the fact that, in order to determine the parton spin contributions in Eq.~(\ref{e:spin_sum_rule}), one in principle needs to know the corresponding helicity parton distributions (PDFs) for any parton momentum fraction $x$, since their integral defines $\frac{1}{2} \Delta \Sigma$ and $\Delta G$.
But the helicity distributions are presently only known for $x \gtrsim 0.01$ with good precision.
Through measurements of polarized DIS, the EIC will provide unprecedented detail of the parton helicity distributions down to $x \sim 10^{-4}$ --- see Fig.~\ref{fig:x_q2_poldis}.
This will not only result in a much better understanding of both $\Delta \Sigma$ and $\Delta G$, but also further constrain the sum $L_q + L_g$ in Eq.~(\ref{e:spin_sum_rule}) --- we refer to Sec.~\ref{part2-subS-SpinStruct.P.N} for more details.
The EIC may be able to also provide, for the first time, direct information on the parton orbital angular momenta through partonic Wigner functions, as discussed below in Sec.~\ref{part2-subS-SecImaging-Wigner}. 

{\bf Nucleon mass:} As with the spin of the nucleon, it is of fundamental importance to understand how the nucleon mass can be decomposed in QCD into contributions from the partons.
The mass of an atom is almost exactly equal to the sum of the masses of its constituents, that is, the nucleus and the electrons.
Likewise, the mass of an atomic nucleus is approximately given by the mass of the nucleons which make up the nucleus.
On the other hand, the nucleon mass cannot even be computed roughly by adding the masses of its constituents, which have their origin in the Higgs mechanism.
For instance, the sum of the masses of the valence quarks is just about 1\% of the nucleon mass.
While in a full QCD analysis, the quark mass contribution to the nucleon mass is larger,
studies show that the Higgs mechanism can only explain a small fraction of the nucleon mass. 
The bulk can be attributed to contributions from quark and gluon (kinetic and potential) energies.
Section~\ref{part2-subS-PartStruct-Mass} contains more details about the mass sum rule(s) of the nucleon.
That section also elaborates on how the EIC could significantly deepen our understanding of the nucleon mass in QCD by means of quarkonium production close to the production threshold.
\\

{\bf 2.~Multi-dimensional imaging of nucleons, nuclei and mesons}

Measurements of semi-inclusive and exclusive processes at the EIC will provide invaluable information about the multi-dimensional quark and gluon structure of nucleons, nuclei and even light mesons.

{\bf Imaging in position space --- form factors and generalized parton distributions:} 
The electromagnetic form factors are fundamental quantities containing information about the structure of strongly interacting systems.
They can be measured through elastic electron scattering, and they depend on the (squared) momentum transfer to the target, $t = - Q^2$. 
For the spin-$\frac{1}{2}$ nucleon two such form factors exist --- the Dirac form factor $F_1$ and the Pauli form factor $F_2$.
Alternatively, one often considers the electric and magnetic form factors $G_E$ and $G_M$, which are linear combinations of $F_1$ and $F_2$.
For a heavy target, the Fourier transforms of $G_E$ and $G_M$ can be interpreted as the 3D distributions of charge and magnetization, respectively.
But for the nucleon, and especially for the even lighter pions or kaons, such 3D distributions have no clean interpretation due to relativistic corrections.
(Some recent developments concerning 3D distributions can be found in Ref.~\cite{Lorce:2020onh}).
This problem does not arise for 2D distributions with the two dimensions being perpendicular to the (average) momentum of the incoming and outgoing nucleons~\cite{Miller:2007uy}.
For instance, the 2D electric charge distribution of the nucleon is given by the Dirac form factor through  
\begin{equation}
\rho(b_T) = \int \frac{d^2 \Delta_T}{(2 \pi)^2} \, F_1(Q^2 = \Delta_T^2) \, e^{-i \, \boldsymbol{\Delta}_T \, \cdot \, \boldsymbol{b}_T} \,,
\label{e:2D_dist}
\end{equation}
where $\boldsymbol{\Delta}_T$ denotes the total transverse momentum transfer to the target and $\boldsymbol{b}_T$ the transverse position (impact parameter).
Section~\ref{part2-subS-SecImaging-FF} outlines the prospects for measuring electromagnetic form factors of the proton, the deuteron and light mesons.
\begin{figure}[t]
    \centering
    \includegraphics[width=0.70 \textwidth]{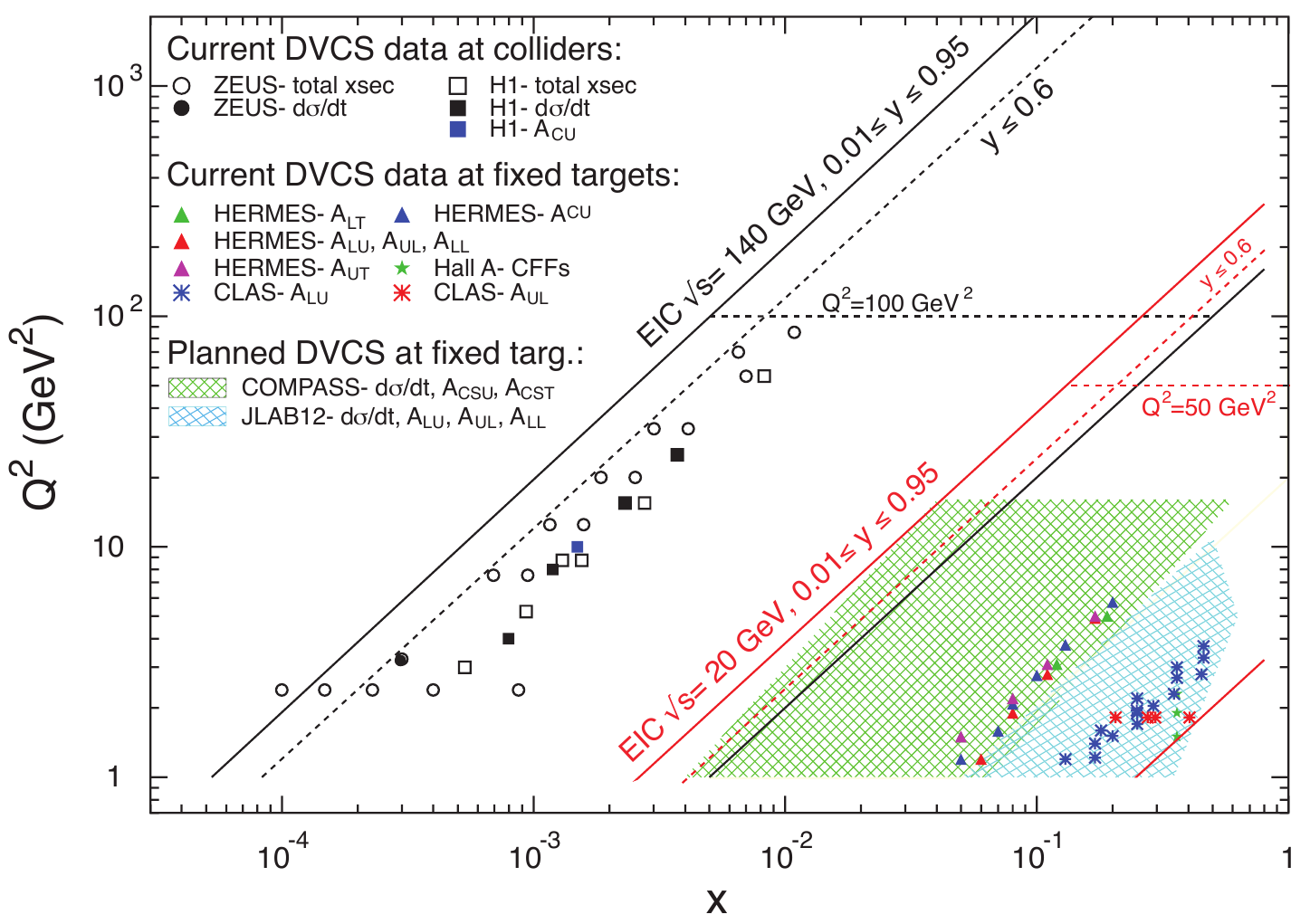}
    \caption{The kinematic coverage of the EIC for the DVCS process compared to other DVCS experiments.}
    \label{fig:x_q2_DVCS}
\end{figure}
The 2D charge distribution in Eq.~(\ref{e:2D_dist}) is related to the momentum-fraction-dependent ($x$-dependent) impact parameter distributions for individual quarks $q(x,b_T)$ according to  
\begin{equation}
\rho(b_T) = \sum_q e_q \int dx \, q(x, b_T) \,,
\end{equation}
where $e_q$ is the quark charge in units of the elementary charge.
The $q(x,b_T)$, and related quark (and gluon) distributions that exist due to the polarization degree of freedom of the nucleon and/or the partons, are the key quantities for the position-space imaging of hadrons.
It is very interesting that the $q(x, b_T)$ can be measured because of their relation to generalized parton distributions (GPDs) $H_q$~\cite{Burkardt:2000za},
\begin{equation}
q(x, b_T) = \int \frac{d^2 \Delta_T}{(2 \pi)^2} \, H_q(x, \xi = 0, \Delta_T^2) \, e^{-i \, \boldsymbol{\Delta}_T \, \cdot \, \boldsymbol{b}_T} \,,
\label{e:impact_GPD}
\end{equation}
with $\xi$ indicating the longitudinal momentum transfer to the target.
GPDs, which generalize the concept of ordinary PDFs~\cite{Mueller:1998fv}, appear in the QCD description of hard exclusive processes like deep-virtual Compton scattering (DVCS) and meson production~\cite{Ji:1996ek, Radyushkin:1996nd}.
The kinematic coverage of the EIC for the DVCS process is shown in Fig.~\ref{fig:x_q2_DVCS}.
The information encoded in GPDs is extraordinarily rich as they also allow for studies of the orbital angular momentum of partons, as well as the distribution of pressure and shear forces inside a hadron.
A more thorough discussion of GPDs and how the EIC will advance this crucial area of multi-dimensional imaging of hadrons can be found in Sec.~\ref{part2-subS-SecImaging-GPD3d}.
Imaging of the spatial distributions of quarks and gluons in nuclei is addressed in Sec.~\ref{part2-subS-LabQCD-Diffraction} via diffraction
and in Sec.~\ref{part2-subS-LabQCD-Photo} via coherent and incoherent vector meson production.

\begin{figure}[t]
    \centering
    \includegraphics[width=0.65\textwidth]{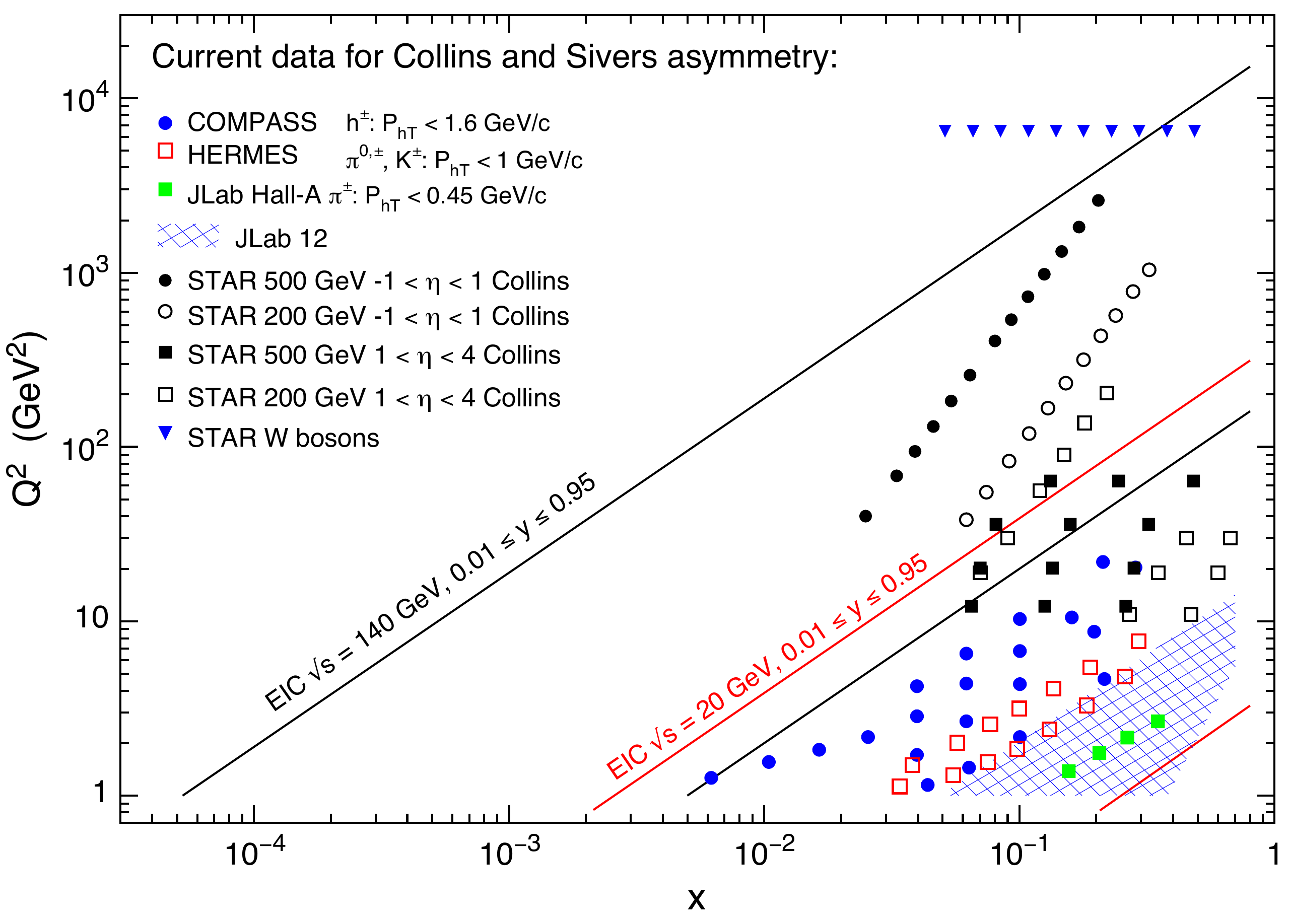}
    \caption{The kinematic coverage of the EIC for the Sivers and Collins effects in semi-inclusive DIS compared to other experiments for two exemplary energy configurations.}
    \label{fig:x_q2_tmd}
\end{figure}
{\bf Imaging in momentum space --- transverse momentum dependent parton distributions:}
Ordinary PDFs provide a 1D image of hadrons in momentum space.
Put differently, via PDFs we learn about the longitudinal motion of partons in a fast-moving hadron, that is, about their momentum distributions along the direction singled out by the hard momentum flow in the process.
However, the partons also have a nonzero transverse momentum relative to that direction. 
One can therefore define transverse momentum dependent parton distributions (TMDs), where for an unpolarized target and unpolarized quark typically the notation $q (x, k_T) = f_1^q (x, k_T)$ is used, with $\boldsymbol{k}_T$ indicating the transverse quark momentum.
A total of 8 leading-twist quark TMDs~\cite{Mulders:1995dh, Boer:1997nt} (and the same number of gluon TMDs~\cite{Mulders:2000sh, Meissner:2007rx}) can be identified for a spin-$\frac{1}{2}$ hadron. 
TMDs provide 3D images of hadrons in momentum space and as such they are complementary to GPDs.
In inclusive DIS, the information about the transverse parton motion is integrated out, and hence other reactions must be considered to address TMDs.
A flagship process for measuring TMDs is semi-inclusive DIS where at least one hadron is detected, in addition to the scattered lepton.
In Fig.~\ref{fig:x_q2_tmd} the kinematic coverage of the EIC is displayed for two very important TMD observables in semi-inclusive DIS, the Sivers effect~\cite{Sivers:1989cc} and the Collins effect~\cite{Collins:1992kk}.
It must be stressed that TMDs can also be studied via different final states in electron-nucleon collisions with di-hadrons or jets and, for instance, in reactions that are not lepton-induced such as the Drell-Yan process.
The fact that TMDs can be measured via a large number of reactions adds to their significance.
Like in the case of position space imaging, the EIC will tremendously enhance our knowledge about the momentum space image of the nucleon as discussed in much more detail in Sec.~\ref{part2-subS-SecImaging-TMD3d}.
\\

{\bf 3.~The nucleus: a laboratory for QCD}

The EIC will be the world’s first dedicated electron-nucleus $\eA$ collider and it will address a broad
program of fundamental physics with light and heavy nuclei.

{\bf Physics of non-linear color fields and gluon saturation:} Due to the rapid rise with energy of the gluon density in hadrons, gluons play a key role in our understanding of DIS and hadronic collisions at high energies. Gluons are responsible for much of the particle production in such collisions, as well as for the rise of total cross sections, related to the saturation of scattering amplitudes with energy as the unitarity limit is approached. This is a fundamental limit of nature on the maximal strength of color fields in hadrons and nuclei.
In particular, the scrutiny of nonlinear gluon dynamics will improve our insight into the strong
interaction rather profoundly, and help us to more
deeply understand this fundamental pillar of the standard model.

\begin{figure}
    \centering
    \includegraphics[width=0.65\textwidth]{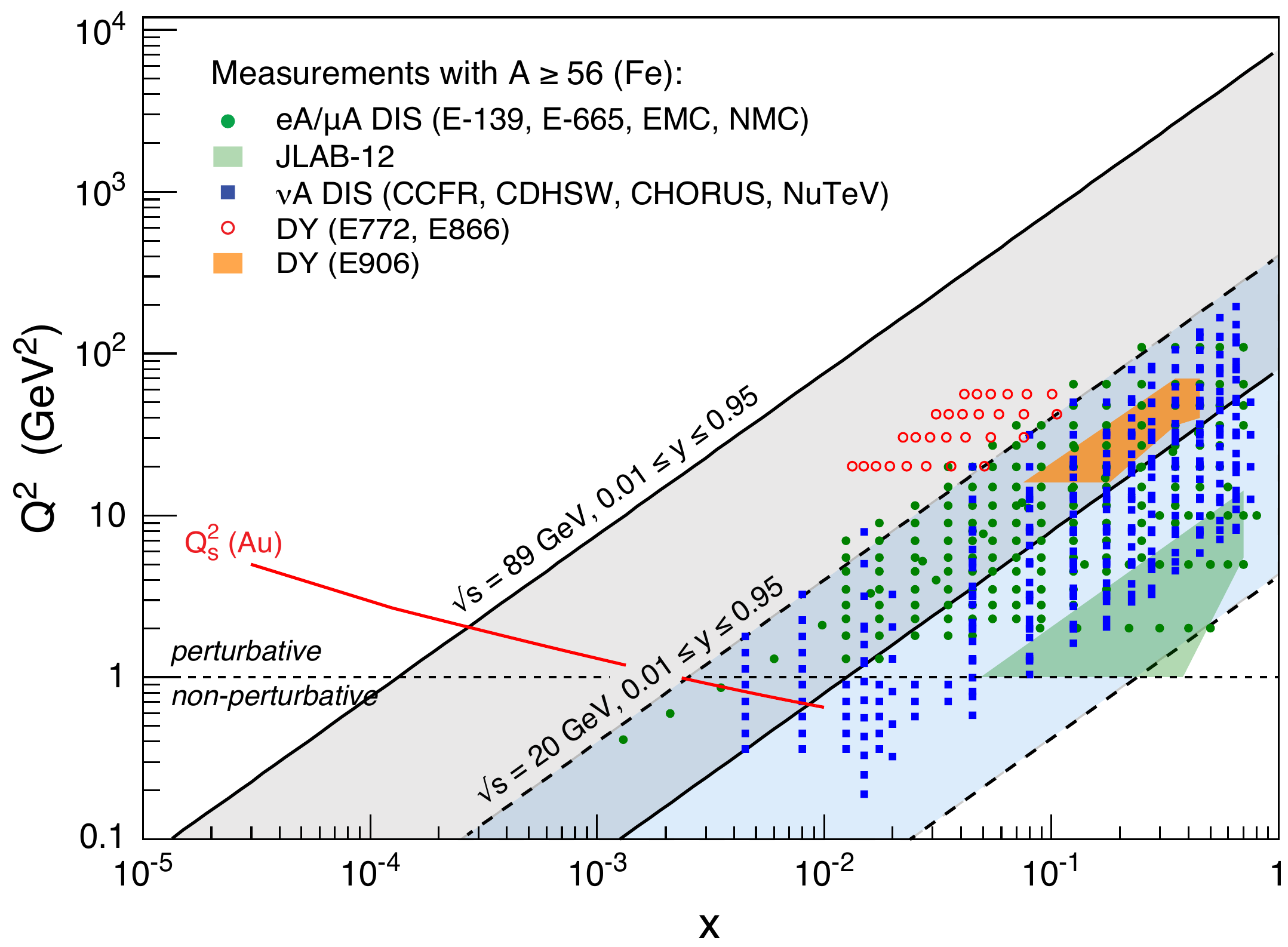}
    \caption{The kinematic coverage of the EIC for DIS on nuclei compared to that of previous experiments. The expected "saturation scale" $Q_s^2(x)$ for non-linear gluon dynamics in a large nucleus is indicated by a red line~\cite{Kowalski:2007rw,Kowalski:2003hm,Kowalski:2006hc}.}
    \label{fig:x-Q2-eA}
\end{figure}
DIS experiments on heavy nuclei at
high energies are ideally suited for the study of nonlinear
gluon dynamics. The projectile interacts coherently with a large number of stacked nucleons. This probes
very strong color fields at high energy, which is expected to lead to the phenomenon of gluon saturation,  described by an effective theory known as the Color Glass Condensate (CGC)~\cite{Gelis:2010nm}. In particular, the squared transverse
momentum (or inverse distance) scale where QCD becomes nonlinear is expected to grow in proportion to the average thickness of the target nucleus, and as a power $\lambda = 0.2 - 0.3$ of energy~\cite{Kowalski:2007rw,Kowalski:2003hm,Kowalski:2006hc}:
\begin{equation}
    Q_s^2 \sim \frac{A^{1/3}}{x^\lambda}~.
\end{equation}
Here, $x$ denotes the momentum fraction of gluons in the target which is probed in a particular process; it is inversely proportional to the energy.

One of the main goals of the physics program to be pursued at the EIC is to obtain clear evidence for nonlinear QCD dynamics at a perturbative scale, $Q_s > 1$~GeV, 
from the 
energy dependence of DIS cross-sections, structure functions, and other observables. This is predicted by the theory in the form of nonlinear evolution equations. Discovery of saturation requires unambiguous experimental evidence for these specific nonlinear equations.
While various features of the data from proton-nucleus and
heavy-ion collisions at RHIC and the LHC are consistent with
perturbative gluon saturation, there is nevertheless no consensus
in the field in favor of them providing unambiguous evidence for
nonlinear effects in the weak-coupling regime. The EIC is
expected to deliver a clean, direct measurement and
characterization of the gluon density in protons and nuclei, and how it depends on
energy and thickness of the target.
The high-energy aspects of DIS on nuclei have been presented more extensively in
the White Paper~\cite{Accardi:2012qut}, and are addressed  in
Secs.~\ref{part2-subS-LabQCD-Saturation}  and~\ref{part2-subS-LabQCD-Diffraction}. 

{\bf Nuclear PDFs:}
Nuclear parton distribution functions (nPDF) describe the behavior of bound partons in the nucleus.  Like their free-proton counterparts, nPDFs are assumed to be universal and are essential tools for understanding experimental data from collider experiments. To date, there is no compelling evidence for violation of the QCD factorization theorem~\cite{Collins:1989gx} or violation of universality. Thus, precise knowledge of PDFs in general, and nPDF in particular, becomes most relevant for advancing our understanding of strong interactions in a nuclear medium and for
interpreting results from collider experiments. 
Moreover, nPDFs provide an essential foundation for understanding the hot Quark-Gluon Plasma (QGP) medium produced in heavy-ion collisions at RHIC and the LHC, particularly for experimental measurements initiated by early-state hard scatterings. Proper characterization of the QGP dynamics also relies on adequate separation of initial and final state effects, the former encoded in the corresponding nPDFs.  Deep-inelastic neutrino scattering experiments with nuclear targets are also in critical need of precise nPDFs, which in turn impact the global analysis of proton PDFs. 

\begin{figure}
    \centering
    \includegraphics[width=0.55 \textwidth]{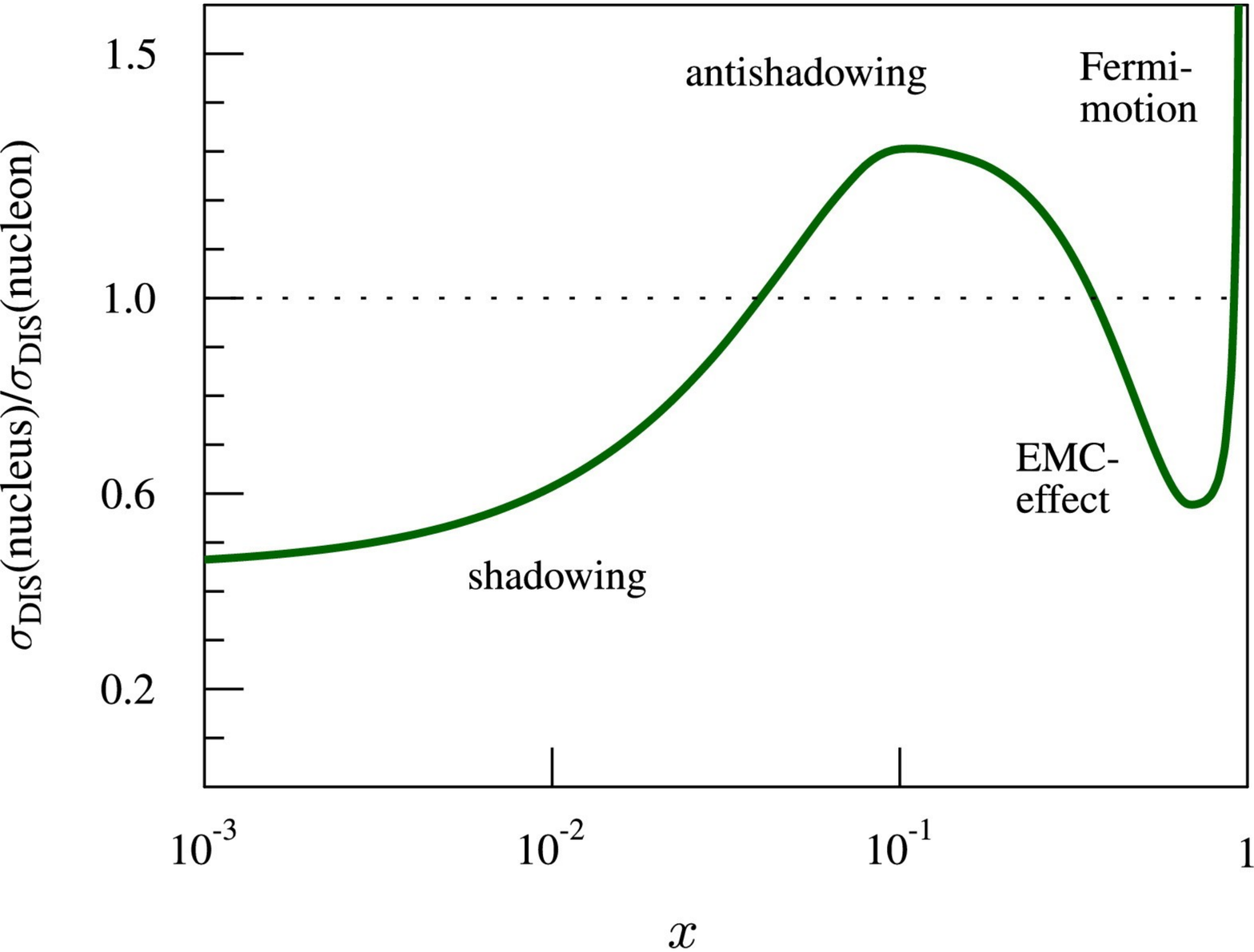}
    \caption{Typical nuclear effects seen in the DIS measurements. The figure is from~\cite{Paukkunen:2014nqa}.}
    \label{fig:nPDFcartoon}
\end{figure}

Experimentally, differences between PDF and nPDF have been firmly established by the deep-inelastic lepton-nucleus scattering data. The observed significant nuclear effects have ruled out a naive model of a nucleus as a superposition of quasi-free nucleons, and forced us to factor in modifications due to the nuclear environment.  These nuclear modifications are commonly described as shadowing, anti-shadowing, and the EMC effect ~\cite{ Arneodo:1996rv, Geesaman:1995yd, Gomez:1993ri, Norton:2003cb}. They are usually quantified in terms of the ratio to the free-nucleon PDFs, with $R<1$ indicating a suppression of the probability distribution compared to the free proton reference, and $R>1$ an enhancement. The approximate domains for these experimentally observed modifications, illustrated in Fig.~\ref{fig:nPDFcartoon}, are as follows: the shadowing regime ($R < 1$) is  prominent in the $x <0.1$ region; the anti-shadowing ($R>1$) effect is present between $0.1<x<0.3$, and the EMC effect refers to the slope of $R$ in the valence-quark dominated regime $0.3<x<0.7$. At higher $x$ there are effects due to Fermi motion in a nucleus.

Understanding how parton dynamics is modified in the nuclear medium, and the exact nature of the mechanisms that generate shadowing, anti-shadowing, and the EMC effects is a field actively pursued in both theory and experiment. 
It is commonly accepted that different physics processes contribute to different regions in $x$; however, there is no consensus on the exact nature of these contributions. The dependencies on nuclear size, impact parameter, and $x$ for these nuclear effects have not been derived from first-principles calculations but are instead inferred from fits to the existing experimental data. However, in contrast to the free-proton PDFs, the determination of nPDFs is severely limited by the kinematic coverage and the available precision of the data. The realization of the EIC with its intended versatility regarding available ion beam species and afforded phase space coverage will profoundly impact nPDF determination. Recent nPDF-related developments are discussed in sec.~\ref{part2-subS-LabQCD-NuclPDFs}.

{\bf Particle propagation through matter and transport properties of nuclei:}
In parallel with the qualitatively new constraints on nPDF, the EIC physics program will allow new advances in the related but intrinsically different modifications induced by cold nuclear matter – the energy loss of partons traversing the QCD medium. The energy loss is expected in both hot (QGP) and cold QCD matter through gluon radiation and collisional scattering losses. The quantitative assessments of the related processes in both types of media are central to relativistic heavy ion collisions and the field of nuclear physics in general. Cold nuclear matter has specific scales for gluon radiation that are different from the QGP medium.  Specifically, the (partially coherent) Landau-Pomeranchuk-Migdal (LPM) regime, with gluon formation times of the order of the length of the medium, and a fully coherent (or factorization) regime, dominating for significantly longer time frames~\cite{Peigne:2008wu}. While the fully coherent part could be evaluated via quarkonic measurements in hadronic collisions~\cite{Arleo:2018zjw}, the final-state energy loss in nuclei in the LPM regime will be probed  at the EIC most directly by hadron production measurements in semi-inclusive deep inelastic scattering events. 

The emerging EIC detector concepts suggest excellent capabilities for jet reconstruction and jet studies; relying on jets for extracting cold nuclear matter transport properties gives significant advantages over inclusive and semi-inclusive hadron measurements, as it allows to effectively reduce the role of nPDF modifications (with respect to the free-nucleon PDF) and enhances the effects due to final state interactions.  Medium-induced radiation resulting in broadening of the transverse profile of jet showers will be a discerning measurement for quantifying the final state gluon radiation and its angular dependence with the EIC data. Details of related studies are presented in Sec.~\ref{part2-subS-LabQCD-PropPartLoss}.
\\

{\bf 4.~Understanding hadronization} 

Intimately related to the prominent question of confinement is the one of hadron formation. How do the degrees of freedom of QCD, quarks and gluons, relate to the hadronic degrees of freedom we observe in nature? The EIC will not only address the many outstanding questions about hadron structure but also will hugely advance our understanding of hadron formation.

{\bf Parton fragmentation:}
The theoretical description of hadron formation usually involves factorization theorems where part of the production cross section can be calculated perturbatively, while the non-perturbative nature of hadronization is encoded in the so-called fragmentation function (FF). FFs describe how a parton transforms into the color-neutral hadrons that we observe. 

The unprecedented luminosity of the EIC will have a strong impact on the measurements of fragmentation functions for light mesons. Furthermore, it will be possible to study the dependence of
the hadronization process on polarization degrees of freedom via, for example, fragmentation into polarized $\Lambda$ hyperons.
 Much richer information yet can be obtained by measuring di-hadron FFs that appear in the description of semi-inclusive processes with two identified hadrons in the final state.
A partial-wave decomposition of polarized and unpolarized di-hadron FFs could be performed in order to obtain deeper insight into hadronization mechanisms.

Thanks to the availability of different beam species, the EIC will also be able to address hadronization in the nuclear medium. It is expected to provide the cleanest understanding yet from QCD of the energy loss of energetic partons traversing a nuclear medium, as measured via reconstructed leading jets.

\begin{table}[bth]
\centering
\includegraphics[width=1.0\textwidth]{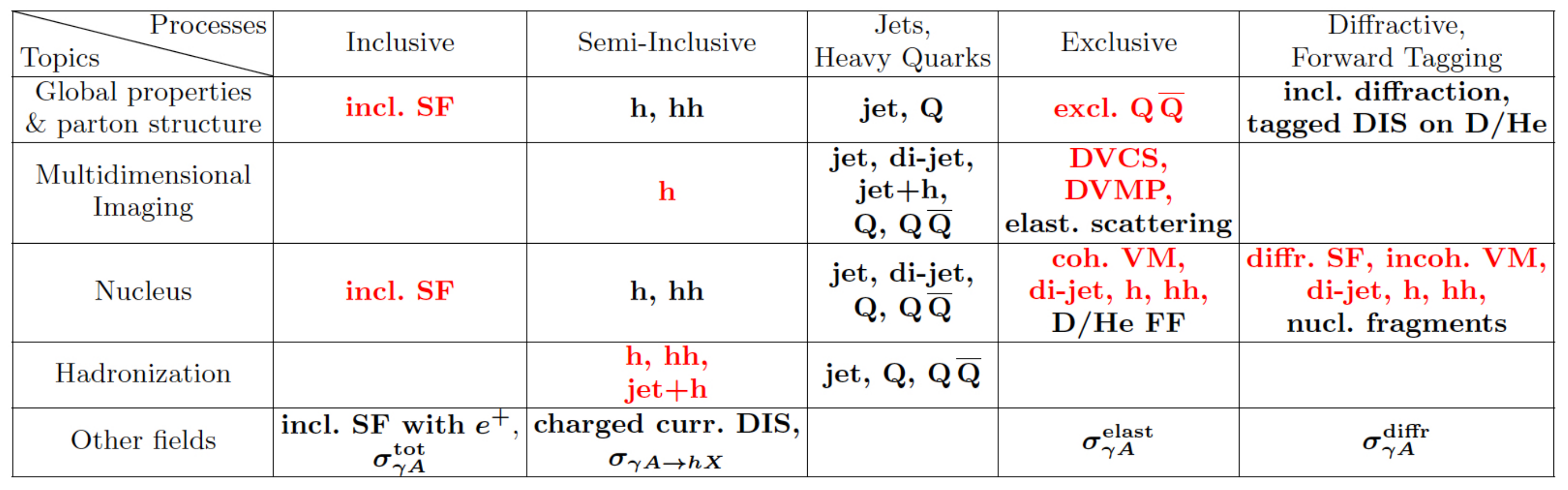}%
\caption{Relationship between the EIC science topics (rows) and the categories of measurements (columns). Measurements already discussed in the White Paper~\cite{Accardi:2012qut} or the NAS Report~\cite{NAP25171} are highlighted in red. Various additional measurements and physics ideas that have emerged since are also included in this table, but the table is not meant to be exhaustive. "Other fields" refers to neutrino, cosmic-ray and high-energy physics.
The acronym "SF" refers to structure function, "FF" to form factor, "h" to identified hadrons, $Q$ to heavy quarks; $Q \overline{Q}$ to heavy-quark bound states (quarkonium), and "VM" to vector mesons.}
\label{fig:MIT-topics-processes-matrix}
\end{table}

{\bf Jets and their substructure:} Jet substructure will be an important tool at the EIC to deeply scrutinize the process of hadronization.
In particular, jet substructure offers the opportunity to
study both the process of fragmentation, or parton radiation patterns, and hadronization,
or the formation of the parton shower into bound state hadrons. Hadrons-in-jets
will be used to study a variety of (un)polarized transverse-momentum-dependent fragmentation functions.
Furthermore, heavy flavor-tagged jet substructure
will also constitute an essential probe for parton propagation through a nuclear medium and
could be sensitive to both fragmentation and hadronization modifications.
The relatively low transverse momenta of reconstructed jets will enhance the role of the
heavy quark mass in measurements of jet splitting functions.

{\bf Production mechanism for quarkonia:} Studying the formation of heavy quark-antiquark bounds states will provide unique insight into hadronization. Due to the large mass of the quarks, their production involves both perturbative and non-perturbative processes. Quarkonium production will be studied in $\ep$ collisions, and also in $\eA$ collisions, where quarkonia hadronize inside the nucleus.

Most of the above topics have been identified in the past as pillars of the EIC science program.  
However, this YR also addresses a number of recent topics which had not been covered in depth or even at all in  the EIC White Paper~\cite{Accardi:2012qut} or the NAS Report~\cite{NAP25171}. 
These include the partonic structure of mesons in Sec.~\ref{part2-subS-PartStruct.M}, multi-parton correlations in Sec.~\ref{part2-subS-PartStruct-MultiPart}, jet-based TMD measurements in Sec.~\ref{part2-subS-SecImaging-TMD3d}, partonic Wigner functions in Sec.~\ref{part2-subS-SecImaging-Wigner}, particle propagation through matter and transport properties of nuclei in Sec.~\ref{part2-subS-LabQCD-PropPartLoss}, flavor-tagged jets and jet substructure in Sec.~\ref{part2-subS-LabQCD-Special}, short range correlations and the origin of the nuclear force in Sec.~\ref{part2-subS-LabQCD-ShortRange}, the structure of light (polarized) nuclei in Sec.~\ref{part2-subS-LabQCD-LightNuclei},  as well as hadronization and spectroscopy of exotic states in Sec.~\ref{part2-sec-Hadronization}. 
These exciting, timely topics further broaden the rich physics program to be pursued at the EIC.
Brief introductions to those topics are found at the beginning of each of Secs.~\ref{part2-sec-PartStruct} -- \ref{part2-sec-Hadronization}, while a more comprehensive account is left for the relevant subsections in Chap.~\ref{part2-chap-EICMeasandStud}.

The program pursued at the EIC is firmly focused on QCD in deeply-inelastic scattering, but it also connects to other fields.
Specific aspects of electroweak and beyond the Standard Model physics, neutrino, cosmic-ray \& astroparticle physics, the physics of proton-proton, proton-nucleus and nucleus-nucleus collisions studied at RHIC and the LHC, the physics of nuclear structure and of exotic nuclei and the general planning in the HEP community which can benefit from the insights provided by the EIC are outlined in Sec.~\ref{part2-sec-Connections}.

Finally, Sec.~\ref{part2-sec-TheoryEfforts} describes theory efforts tied to EIC science. 
They include lattice QCD which is the key tool for obtaining first-principles, non-perturbative results from the quantum field theory of the strong interactions.
Also added is a discussion of QED radiative corrections in electron scattering from a nucleon or nucleus.
A solid understanding of such corrections will be mandatory for obtaining the main information of interest from the EIC data.

At the Yellow Report kick-off meeting~\cite{MIT-YR-kick-off} it was discussed in detail how to best assess the detector requirements needed to carry out this broad and ambitious physics program.
It was decided that it would be most suitable to form into working groups that would study similar processes, since those would generally lead to similar detector requirements, independent of the physics topic addressed. 
The various physics measurements were grouped into the following categories which then also determined the structure of the physics working groups:
\begin{itemize}
    \item Inclusive reactions
    \item Semi-inclusive reactions
    \item Jets and heavy quarks
    \item Exclusive reactions
    \item Diffractive reactions and forward tagging
\end{itemize}
Table~\ref{fig:MIT-topics-processes-matrix} illustrates how the different categories of measurements above can address the different physics topics of the EIC. 
Chapter~\ref{part2-chap-DetRequirements} describes the detector requirements from each of the categories of processes, based on the studies performed by the five physics working groups.

\chapter{EIC Measurements and Studies}
\label{part2-chap-EICMeasandStud}

\section{Global Properties and Parton Structure of Hadrons}
\label{part2-sec-PartStruct}

The EIC will significantly enhance our knowledge of the parton structure of hadrons.  
This includes, in particular, the questions about how the global properties {\bf nucleon spin} and {\bf nucleon mass} which are discussed in detail in Sec.~\ref{part2-subS-SpinStruct.P.N} and Sec.~\ref{part2-subS-PartStruct-Mass}, respectively, can be understood in terms of contributions from quarks and gluons.
An introduction to these topics can be found in Chapter.~\ref{part2-chap-EICPhyCase}, while in the following a very brief overview of the other topics in this section is given.

Since the pioneering DIS experiments at SLAC in the late 1960s~\cite{Bloom:1969kc,Breidenbach:1969kd} it has been known that the nucleon has a partonic structure.
The simplest quantities describing how the partons are distributed inside the nucleon are the (one-dimensional) {\bf parton distribution functions} (PDFs), which depend on the fraction $x$ of the nucleon's momentum that is carried by the parton.
The most prominent ones are the twist-2 PDFs which have a density interpretation.
For a spin-$\frac{1}{2}$ hadron one can define the three quark PDFs
\begin{equation}
f_1^q(x) = q(x) \,, \qquad g_1^q(x) = \Delta q(x) \,, \qquad h_1^q(x) \,,
\label{e:PDF_def}
\end{equation}
where $f_1^q$ denotes the unpolarized quark PDF, while $g_1^q$ ($h_1^q$) denotes the helicity (transversity) PDF.
In~(\ref{e:PDF_def}) the most commonly used notations for the unpolarized and helicity PDFs are shown.
Even though the unpolarized PDFs are rather well known by now, the EIC can further this field as outlined in Sec.~\ref{part2-subS-UnpolPartStruct.P.N}.
The expected significant EIC potential for pinning down the helicity PDFs will be discussed in detail in Sec.~\ref{part2-subS-SpinStruct.P.N}, and prospects for the transversity distribution are presented in Sec.~\ref{part2-sec-Imaging} on multi-dimensional imaging; an important observable for measuring $h_1$ involves three-dimensional parton distributions.

Measurements at the EIC can also address the {\bf structure of mesons}.
Specifically, very detailed plans exist to explore pions, by far the lightest strongly interacting particles, as well as kaons.
Since mesons are unstable, they cannot be probed directly in a DIS experiment.
However, by considering suitable final states and kinematics in electron-proton scattering, one is able to largely single out lepton scattering off the meson of interest.
For instance, in order to explore DIS off a pion, the detection of a neutron is needed, in addition to the scattered lepton.
Studies of light mesons are very interesting in their own right, but may also offer deeper insights into the generation of hadron masses.
This aspect serves as an important driver behind those activities as explained in Sec.~\ref{part2-subS-PartStruct.M}.

The inclusive DIS process not only contains information about densities of single partons (twist-2 PDFs) but also about {\bf multi-parton correlations}, which characterize the structure of hadrons at a new level.
Specifically, the twist-3 structure function $g_T$, which is accessible in polarized DIS, is related to quark-gluon-quark correlations in the nucleon.
Additional quark-gluon-quark correlations can be studied in lepton-nucleon scattering by considering other final states beyond the fully inclusive one. 
It is timely to explore how the EIC can contribute to this important field, which in the past has received very little attention in documents articulating the EIC science case.
In Sec.~\ref{part2-subS-PartStruct-MultiPart} the prospects in that regard are briefly discussed.

About 10\% of the DIS events observed at HERA are {\bf (inclusive) diffractive}, that is, they show a large rapidity gap between the system $X$ and the target proton (or a low-mass excitation of the proton).
Therefore diffraction became a major research topic in the HERA community.
Since the proton is detected, diffractive events in DIS are characterized by additional kinematic variables beyond the standard variables $x$ and $Q^2$.
The EIC holds promise to significantly extend our understanding of inclusive diffraction.
In particular, as discussed in Sec.~\ref{part2-subS-PartStruct-InclDiff}, the kinematic range that can be explored at the EIC shows a considerable complementarity relative to the HERA measurements.

An important aspect of exploring QCD is the determination of the {\bf strong coupling $\alpha_s$} through precision measurements.
Key observables in that regard are {\bf event shapes} which by now can be computed very reliably in perturbative QCD.
In Sec.~\ref{part2-subS-PartStruct-GlobalEvent} the prospects for using event-shape measurements at the EIC to pin down $\alpha_s$ are presented in detail.

\subsection{Unpolarized parton structure of the proton and neutron}
\label{part2-subS-UnpolPartStruct.P.N}

\newcommand\IRFIGPATH{PART2/Figures.EICPhyCase/IR4YR20V2}

\subsubsection*{Inclusive neutral-current and charged-current DIS}

Historically, our knowledge of unpolarized collinear parton distribution functions (PDFs) has been driven by inclusive neutral-current (NC) and charged-current (CC) deep-inelastic scattering (DIS) cross sections from protons and deuterons, together with high-energy scattering data from proton-antiproton collisions at the Tevatron and more recent measurements from proton-proton collisions at the LHC. 
A detailed description of the data sets entering PDF determinations can be found in Refs.\cite{Gao:2017yyd,Ethier:2020way,Rojo:2015acz}.
The existing DIS data cover an impressive range in the outgoing lepton kinematics with $x$ down to $10^{-5}$  and $Q^2$ up to the order of $10^4{~\rm GeV^2}$. 
While there is a substantial kinematic overlap between the measurements at HERA and those in fixed-target experiments, they are complementary in accessing the small-$x$ and large-$x$ longitudinal hadron structure, respectively. 
On the other hand, the EIC covers an overlapping kinematic range between HERA and the fixed-target experiments, with an instantaneous luminosity potentially 3 orders of magnitude larger than at HERA. 
The EIC, together with other facilities and, in particular, the Jefferson Lab 12-GeV program, will allow for a new era in the exploration of the nucleon structure in high definition.   
\begin{figure}[ht]
    \centering
    \includegraphics[width=0.75\textwidth]{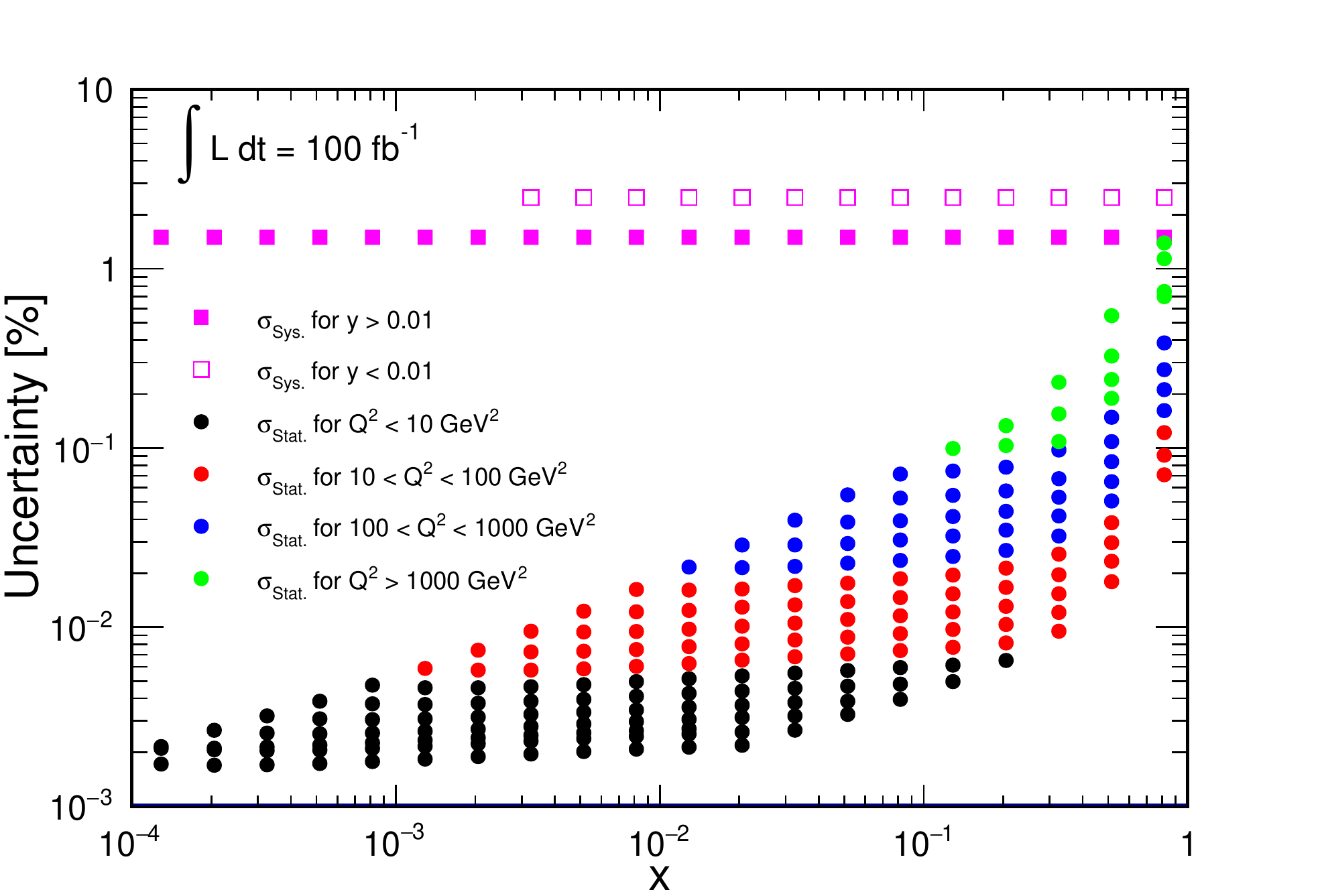}
    \caption{Simulated statistical and systematic uncertainties for electron-proton NC DIS at $\sqrt{s}=140.7\, {\rm GeV}$. The statistical uncertainties are calculated based on a two-dimensional binning with 5 bins per decade for both $Q^{2}$ and $x$. The determination of the displayed systematic uncertainties is discussed in Sec.~\ref{systematics_incl}.
    (The systematic uncertainties correspond to the ``Conservative Scenario'' discussed below.)}
    \label{fig:DIS-kinematics}
\end{figure}

In Fig.~\ref{fig:DIS-kinematics} we present statistical and systematic uncertainties for the EIC NC cross sections. While an integrated luminosity of $100~{\rm fb}^{-1}$ provides an impressively small statistical uncertainty at small $x$, the overall uncertainties are estimated at present to be limited by the systematic uncertainties.
Details for the projected uncertainties can be found in Sec.~\ref{part2-sec-DetReq.Incl}.

\begin{figure}[t!]
	\centering
	\includegraphics[width=\textwidth]{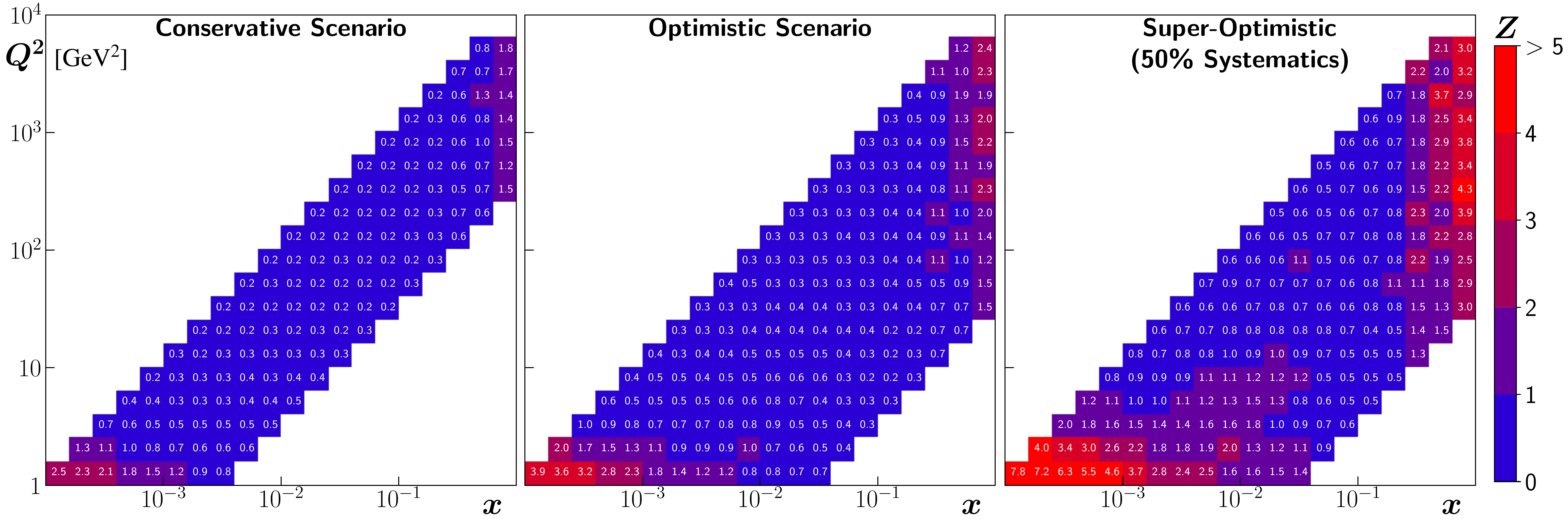}
 	\caption{Z-score analysis for differential cross sections using different scenarios for systematic uncertainties: conservative, optimistic and 50\% of optimistic (``super-optimistic'') as described in Sec.~\ref{part2-sec-DetReq.Incl}. The Z-score uses the NNPDF3.1 proton PDFs~\cite{Ball:2017nwa} to compare the cross sections generated from the central PDF replica (the null hypothesis) and non-central PDF replicas (the alternative hypothesis).}
	\label{fig:Z-impact}
\end{figure}

Fig. \ref{fig:Z-impact} shows the impact of the EIC NC DIS data on our current knowledge of the differential cross sections computed with the NNPDF3.1 PDF set~\cite{Ball:2017nwa,Bertone:2013vaa}. 
Using a $\chi^2$-based hypothesis test, we assess the EIC constraining power at the single-bin level with ${\cal L}$~=~100~${\rm fb}^{-1}$ of pseudodata and point-by-point systematic uncertainties as described in Sec.~\ref{part2-sec-DetReq.Incl} (left and central panels) and an additional optimized scenario reduced by a factor 2 (right panel). 
The impact of the EIC pseudodata is quantified in terms of a Z-score which measures the statistical separation in units of the standard deviation $\sigma$ between two hypotheses of cross sections. The figure shows that more than $5 \sigma$ (Z-score $\geq$ 5) average discrimination power between cross sections generated from the central PDF replica and non-central PDF replicas can be achieved across the entire EIC acceptance if the current projections for the  systematic uncertainties are reduced by a factor of 2. 
While the reduction of the uncertainties might not be achievable at the EIC, the Z-score profile indicates the need for designing detectors that maximize the purity and stability (see Figs.~\ref{fig:Purity_NC_18_275} -\ref{fig:Stability_5_100}) of the measurements in the phase space regions where the impacts are expected to be larger. 

%

\begin{figure}[th]
    \includegraphics[width=0.5125\textwidth]{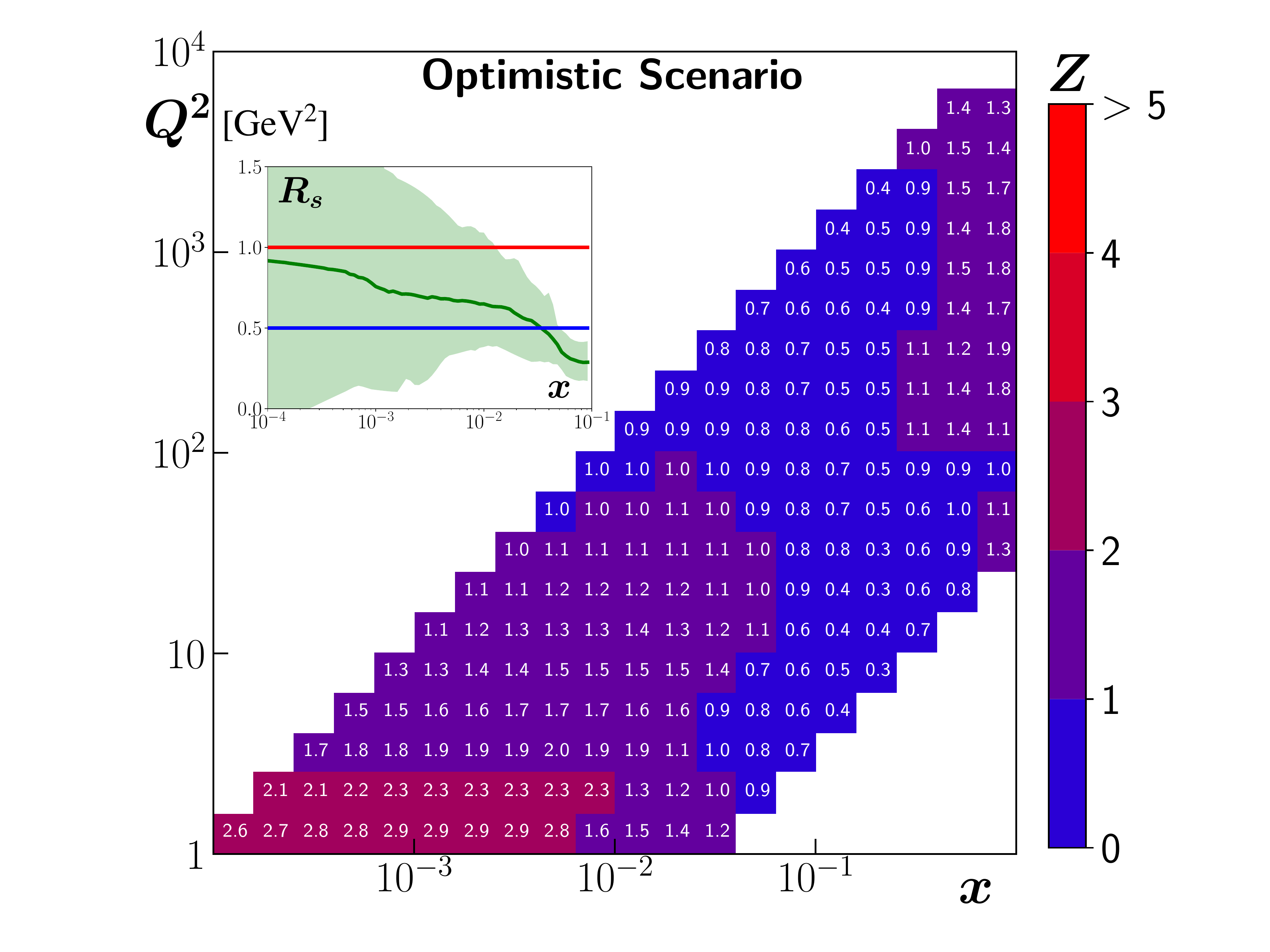}
     \includegraphics[width=0.4825\textwidth]{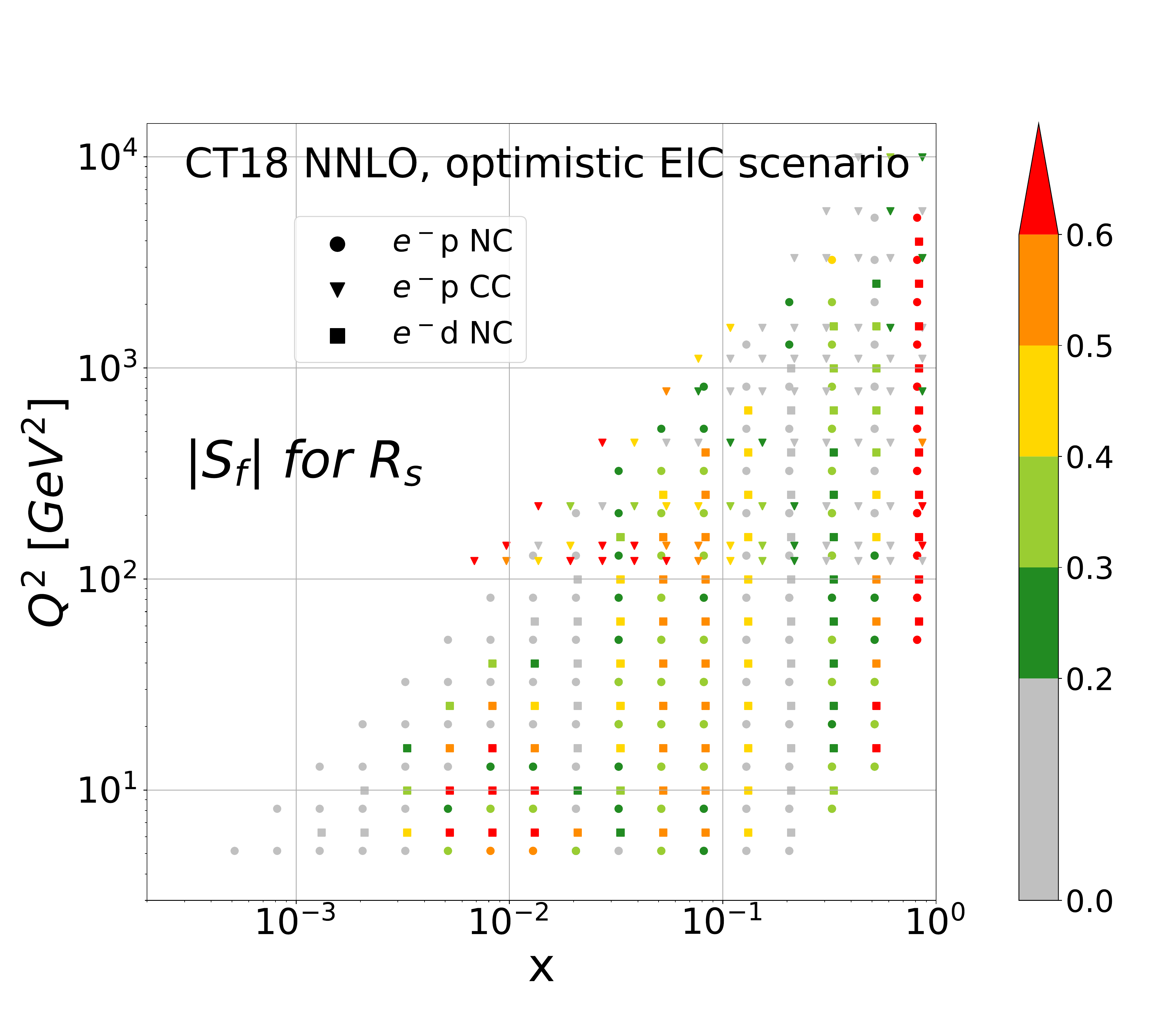}
    \caption{
    Left: $Z$-score analysis comparing the cross sections generated from PDF replicas satisfying $R_s=0.5$ (null hypothesis in blue) and $R_s=1$ (alternative hypothesis in red). These two hypotheses are built by modifying the NNPDF3.1 $s$, $\bar{s}$, $\bar{u}$, $\bar{d}$ distributions 
    in a way to conserve the sum rules.
    Right: The sensitivity $|S_f|$ of the EIC $e^-$
    pseudodata to the $R_s$ PDF ratio. Redder points indicate pseudodata with
    larger constraining power, as discussed in Ref.~\cite{Wang:2018heo}.
    }
    \label{fig:ZPDFmaps}
\end{figure}

In order to illustrate the impact from different bins in the $x$-$Q^2$ plane of 
the EIC data, we present in Fig.~\ref{fig:ZPDFmaps} (left) a parton-level analysis 
by focusing on the strange sector through the ratio $R_s=(s+\bar{s})\big/(\bar{u}+\bar{d})$ using the Z-score technique. 
We select the strange sector since it is one of the most difficult PDFs to be extracted from data from inclusive NC and CC reactions (see recent developments in Refs. \cite{Faura:2020oom,Sato:2019yez}) and therefore places stronger constraints on detector capabilities.
The analysis is carried out using the NNPDF3.1~\cite{Ball:2017nwa,Bertone:2013vaa} replicas and modifying the sea-quark PDFs requiring $R_s=0.5$ and $R_s=1$ in such a way that momentum conservation is not violated. 
Using the optimistic scenario, the Z-scores show that the strange sector can be discriminated up to $3 \sigma$ at low-$x$ and low-$Q^2$, and $2 \sigma$ in a narrow region of high $x$ and high $Q^2$. 
The sensitivity for moderate $x$ is found to be marginal. 

A complementary analysis done in the framework of PDFSense~\cite{Wang:2018heo,Hobbs:2019gob,Hobbs:2019sut} shown in Fig.~\ref{fig:ZPDFmaps} (right) also illustrates a potential for modest sensitivity of the EIC $e^-$ data to $R_s$.
This can be inferred based on the share of individual pseudodata with relatively larger values of the sensitivity parameter, $|S_f|$, which is defined and discussed in Sec. III of Ref.~\cite{Wang:2018heo}. 
The PDF-level impact of the EIC pseudodata can be judged in this context against the typical 
values of $|S_f|$~\cite{Wang:2018heo} for data fitted in the CT18 global analysis~\cite{Hou:2019efy}.

\begin{figure}[th]
	\centering
	\includegraphics[width=\textwidth]{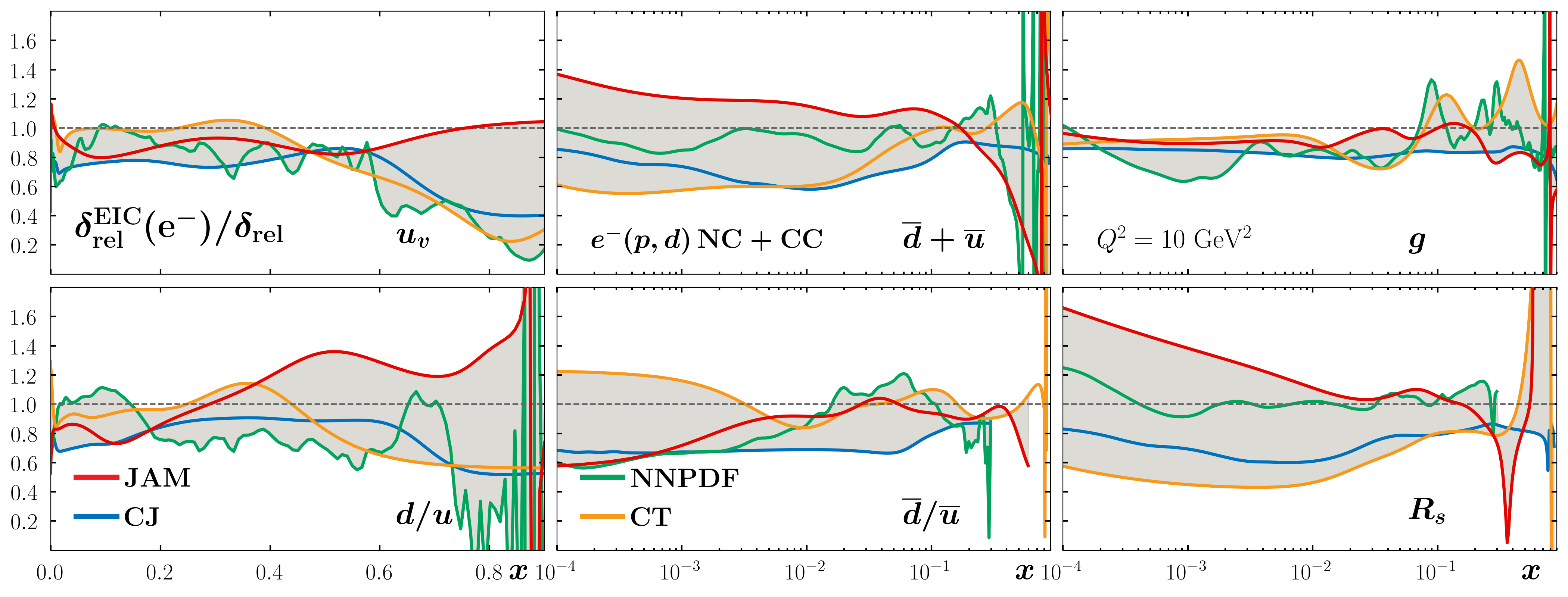}	
	\caption{Comparison of relative uncertainties for unpolarized PDFs $xf(x)$ for different partons, before and after the inclusion of EIC data, evaluated at $Q^2=10$~GeV$^2$. We include the analysis of different collaborations, limited to $e^-$ datasets.
	}
	\label{fig:e_minus}
\end{figure}

Using the optimistic scenario for the systematic uncertainties, in Fig.~\ref{fig:e_minus} we examine the potential impact of EIC NC and CC with incident electron beam colliding with proton and deuteron beams from a selection of PDF global analyzers (CJ\cite{Accardi:2016qay}, CT\cite{Hou:2019efy}, JAM\cite{cocuzza20,Sato:2019yez}, NNPDF\cite{Ball:2017nwa,Faura:2020oom}) that have incorporated the EIC pseudodata within their fitting framework.
For proton beams we use ${\cal L}=100 \, {\rm fb}^{-1}$ with  $\sqrt{s} = 28.6, ~44.7,~63.3,~140.7$~GeV for NC, and $140.7$~GeV for CC.
For deuteron beams we use ${\cal L}=10 \, {\rm fb}^{-1}$ and consider only NC at $\sqrt{s} = 28.6,~66.3,~89.0$~GeV. 
We stress that the various analyses are carried out under different conditions of data selection and PDF extraction methodologies.
Focusing on the DIS data sets, all groups use the bulk of the world DIS data from SLAC, BCDMS, NMC and HERA. 
While CT and NNPDF place strong cuts on $W$ ($W^2\geq 10 \, {\rm GeV^2}$) which exclude the region of very large $x$ and low $Q^2$, CJ and JAM use lower cuts ($W^2\geq 4 \, {\rm GeV^2}$) allowing to include PDF constraints from JLab. 
In terms of methodologies, the various groups have different approaches to carry out the Bayesian inference. 
CJ and CT use maximum likelihood augmented by the Hessian approach to estimate the confidence regions for the PDFs, while JAM and NNPDF utilize Monte Carlo approaches to sample the posterior distribution of the parameter space of the associated PDFs.
In order to reduce the systematic effects stemming from the Bayesian inference adopted by each group, in Fig.~\ref{fig:e_minus} we present relative uncertainties after EIC normalized to pre-EIC relative uncertainties for a selection of parton flavors.
The grey band built as an envelope from the various groups indicates the uncertainty on the impact from the projected EIC data. 
The impact of the EIC can be seen as the variations of the ratios away from unity, which occurs in most of the regions to be explored at the EIC.
Note that the ratios are not bound to be less than one since the inclusion of new data can change the relative strength of the flavor channels on the differential cross sections. 
However, the cross section uncertainties propagated from PDF uncertainties do decrease as expected. 
The results show that there is a potential strong impact on the valence sector where the uncertainties can decrease up to $80\%$ which should give new insights on the $d/u$ ratio. 
On the other hand, the sea sector is predominately modified in the small-$x$ region as expected, with a decrease of uncertainties up to $50\%$. 
Overall we find that the current detector setup, with systematic uncertainties as large as $2\%$, can induce significant constraints on the unpolarized PDFs. 
Those constraints will also raise the accuracy of information that can be obtained from the HL-LHC which includes studies of the Higgs boson --- see also the discussion in Sec.~\ref{part2-sec-Connections-Other}. 

\subsubsection*{Positron beam}

\begin{figure}[ht!] 
	\centering
	\includegraphics[width=\textwidth]{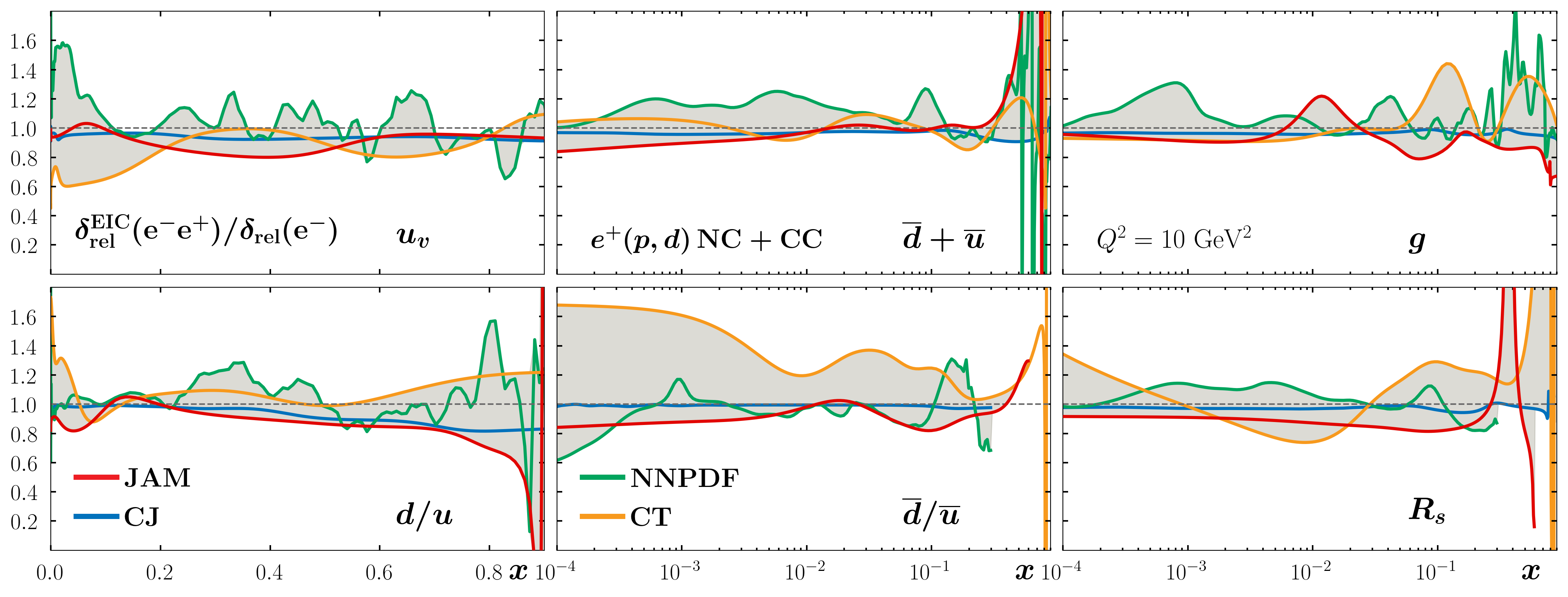}
    \caption{ PDF relative uncertainties after inclusion of NC and CC $e^+(p,d)$ data normalized to the electron-data-only case.}
	\label{fig:e_plus}
\end{figure}

While the EIC has the main focus on an incident electron beam, the possibility of having a positron beam to measure NC and CC is a relevant complementarity that boosts the exploration of the nucleon flavor structure. In particular, the different charge of
the exchanged $W^+$ boson is such that positron CC interactions are capable of probing
a unique combination of flavor currents inside the nucleon relative to the case of an electron beam. This potentially offers significant additional constraints on the $d$-type PDFs,
further constraining the $d/u$ ratio. Beyond this, positron beams may also allow for access to 
other effects, such as the breaking of the strange-antistrange symmetry, $(s=\bar{s})$,
or parton-level charge-symmetry violation~\cite{Hobbs:2011vy}.
In Fig.~\ref{fig:e_plus} we present the impact of positron data on top of all the electron data as ratios of the relative uncertainties. 
Furthermore, $e^+ d$ scattering can improve our understanding of nuclear effects for the simplest of all nuclei in a region free of contamination from $1/Q^2$ power corrections.

\subsubsection*{Parity-violating DIS} 

\begin{figure}[ht]
	\centering
	\includegraphics[width=0.45\textwidth]{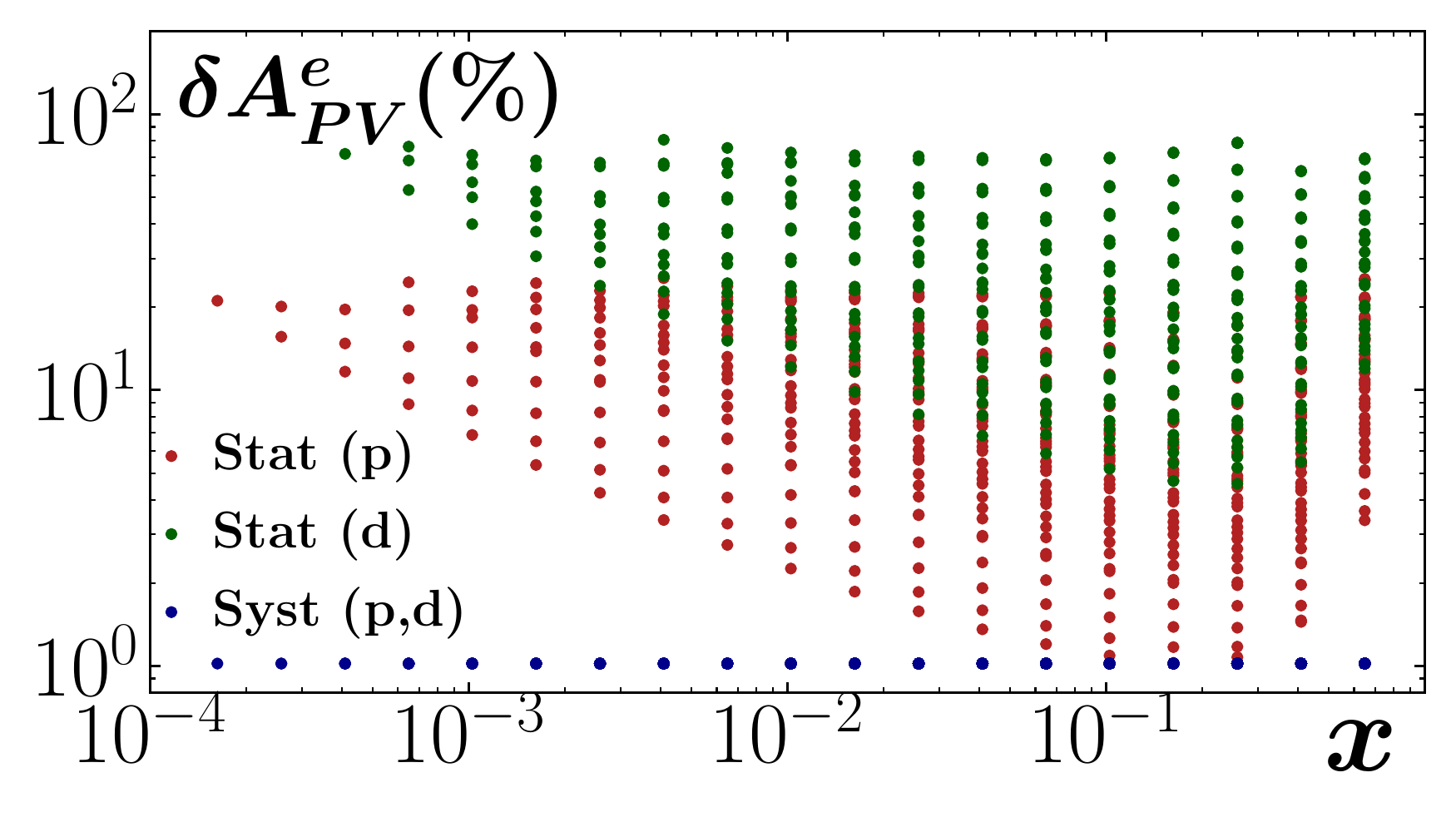}
	\includegraphics[width=0.45\textwidth]{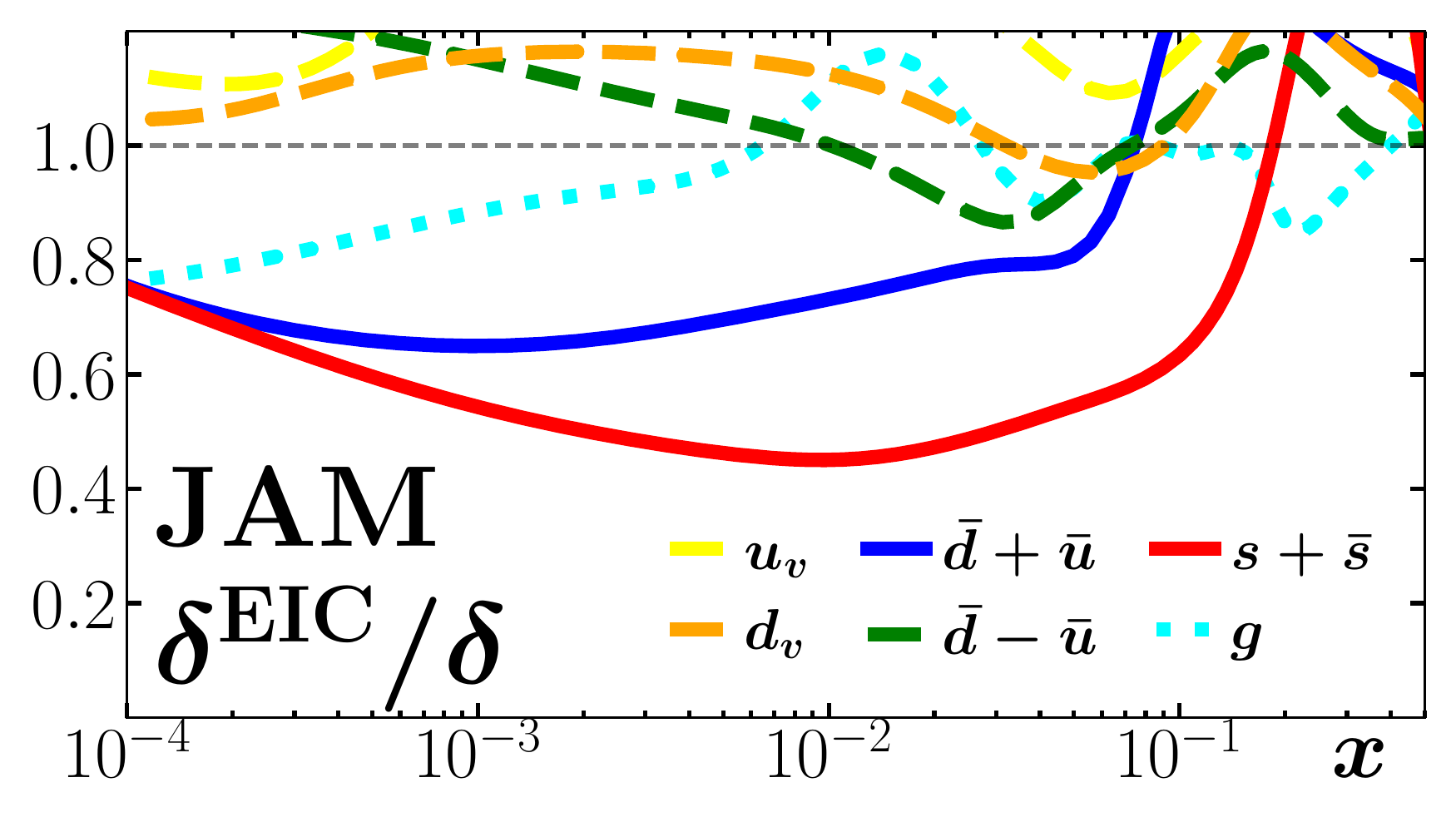}
	\caption{
	 Left:  $A^e_{\rm PV}$ uncorrelated statistical and systematic percent errors for proton and deuteron beams as a function of $x$.  Note that there are multiple values of $Q^2$ for each value of $x$. 
	Right: Ratio of uncertainties on the PDFs as functions of $x$, including EIC data on the parity-violating DIS asymmetry $A^e_{\rm PV}$ to those without EIC data, at $Q^2=10$~GeV$^2$.
	}
	\label{fig:APVe}
\end{figure}

The parity-violating asymmetry with (longitudinally) polarized leptons and unpolarized hadrons,
$A^e_{\rm PV} = A_{LU} = (\sigma^{\uparrow}-\sigma^{\downarrow}) / (\sigma^{\uparrow}+\sigma^{\downarrow})$,
is a unique observable accessible at the EIC, where the dominant proton structure function 
is given by a unique flavor combination, $F_1^{\gamma Z, p} \sim \frac{1}{9} (u+\bar{u}+d+\bar{d}+s+\bar{s}+c+\bar{c})$.
Figure~\ref{fig:APVe} displays the impact of this observable including both proton and deuteron beams with integrated luminosities of 100 fb$^{-1}$ and 10 fb$^{-1}$, respectively, at the energies $\sqrt{s} = 29, 45, 63$ and 141~GeV for proton and $\sqrt{s} = 29, 66$ and 89~GeV for deuteron.  The results show a strong impact on the strange-quark distribution $x(s+\bar s)$, particularly at low values of $x$.  We note that the inclusion of the EIC data guarantees smaller uncertainties for the observable but not necessarily for all PDF flavors at all values of $x$.

\subsubsection*{Tagged DIS}
\begin{figure}[ht]
    \centering
    \includegraphics[width=0.5\textwidth,trim={0 0 0 1 cm},clip ]{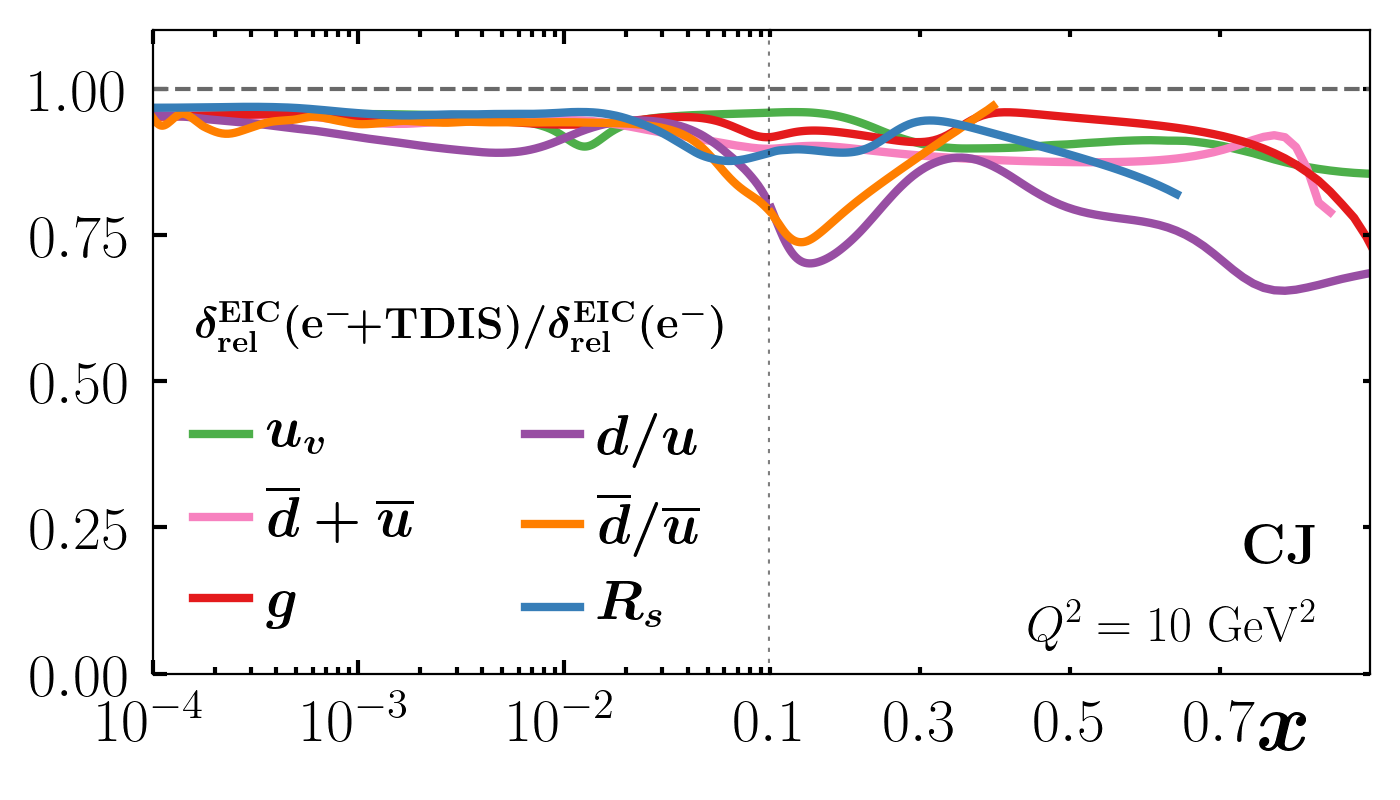}
    \caption{Impact of TDIS data on PDF determination within the CJ global fitting framework~\cite{Accardi:2016qay}. The vertical axis displays the ratio of relative PDF errors obtained with the TDIS-augmented EIC data set to those obtained with the baseline $e^-p$ EIC data.
    }
     \label{fig:CJ-tagged-D}
\end{figure}
Tagged DIS (TDIS) data offer a way to probe the structure of a barely off-shell neutron via semi-inclusive tagging of a slow  spectator proton in $e+d \rightarrow e'+p+X$ events. When these measurements are analyzed through the recently developed on-shell extrapolation technique, they provide one with an effective, free neutron DIS cross section \cite{Sargsian:2005rm,Strikman:2017koc}.
Figure~\ref{fig:CJ-tagged-D} shows the improvement
in the PDF relative uncertainty when using the EIC electron DIS data augmented with TDIS data~\cite{JLab_TDIS_LDRD}.
The addition of this data improves in general the determination of all flavors over the whole $x$-range, in particular the $d/u$ ratio at large $x$.
Generally, measurements at a collider will complement fixed-target tagging experiments planned at JLab in two ways. First, they will extend the kinematic reach to higher $Q^2$, where power corrections to the leading-twist factorized cross section are minimized. Second, they will provide a much cleaner way to detect the spectator nucleon, which propagates close to the beam direction and is naturally separated from the inelastic final state. In particular, this will make it much easier to tag a spectator neutron, and benchmark the on-shell extrapolation technique for a bound proton structure function against the already well-constrained free proton case.
Within a global analysis framework the combination of inclusive deuteron data and  ``free'' tagged neutrons will also provide one with new opportunities for understanding the dynamics of nuclear binding and Fermi motion, as well as measuring the nucleon off-shell quark and gluon structure~\cite{Accardi:2016qay,Alekhin:2017fpf,Tropiano:2018quk}.
Furthermore, the effective free-neutron data will allow for the first time to measure the $d/(p+n)$ ratio with data from the same machine, following the pioneering BONUS measurement~\cite{Griffioen:2015hxa}. This will eventually lead to a better understanding of the EMC effect~\cite{Malace:2014uea} starting from its very first manifestation in the nuclear modification of a bound proton-neutron system compared to a free one.

\subsubsection*{Sea quark PDFs via SIDIS measurements}
As fragmentation functions provide additional access to the flavor of the fragmenting parton via their dependence on fractional energy $z$ and the type of detected hadron, they are an excellent tool to gain further information on the PDF flavor structure of the nucleon. While inclusive cross section measurements only provide limited access to the parton flavor via isospin symmetry and the different weights between neutral and charged current interactions, the semi-inclusive DIS (SIDIS) cross sections add sensitivity via the fragmentation functions $D^{q,h}_1(z,Q^2)$ (in the case of unpolarized, single-hadron fragmentation).  
A detailed description of fragmentation functions and the prospects for constraining them at the EIC
can be found in Sec.~\ref{part2-subS-Hadronization-HadVacuum}. 
Here it should be stressed that the valence parton content of the detected hadron relates to the fragmenting parton flavor, particularly at high $z$. Kaons have a higher sensitivity to strange-quark fragmentation than pions,  while negative pions have a higher sensitivity to down-quark fragmentation compared to positive pions. In this way 
the combination of measurements of SIDIS cross sections for charged pions, kaons (and other hadrons) essentially allows one to disentangle the different valence, sea and gluon unpolarized PDFs. 

\begin{figure}[th]
    \centering
    \includegraphics[width=0.45\textwidth]{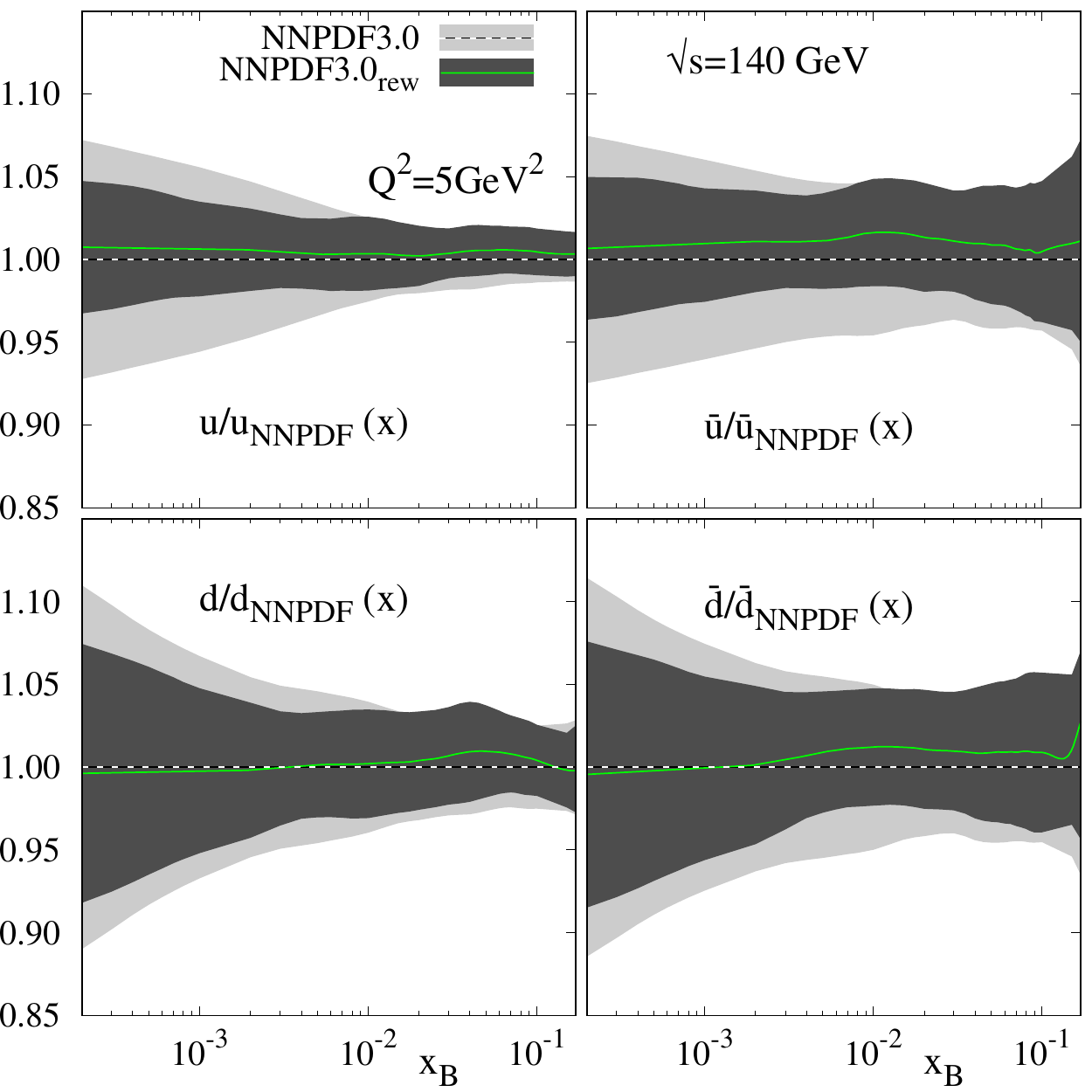}
    \includegraphics[width=0.45\textwidth]{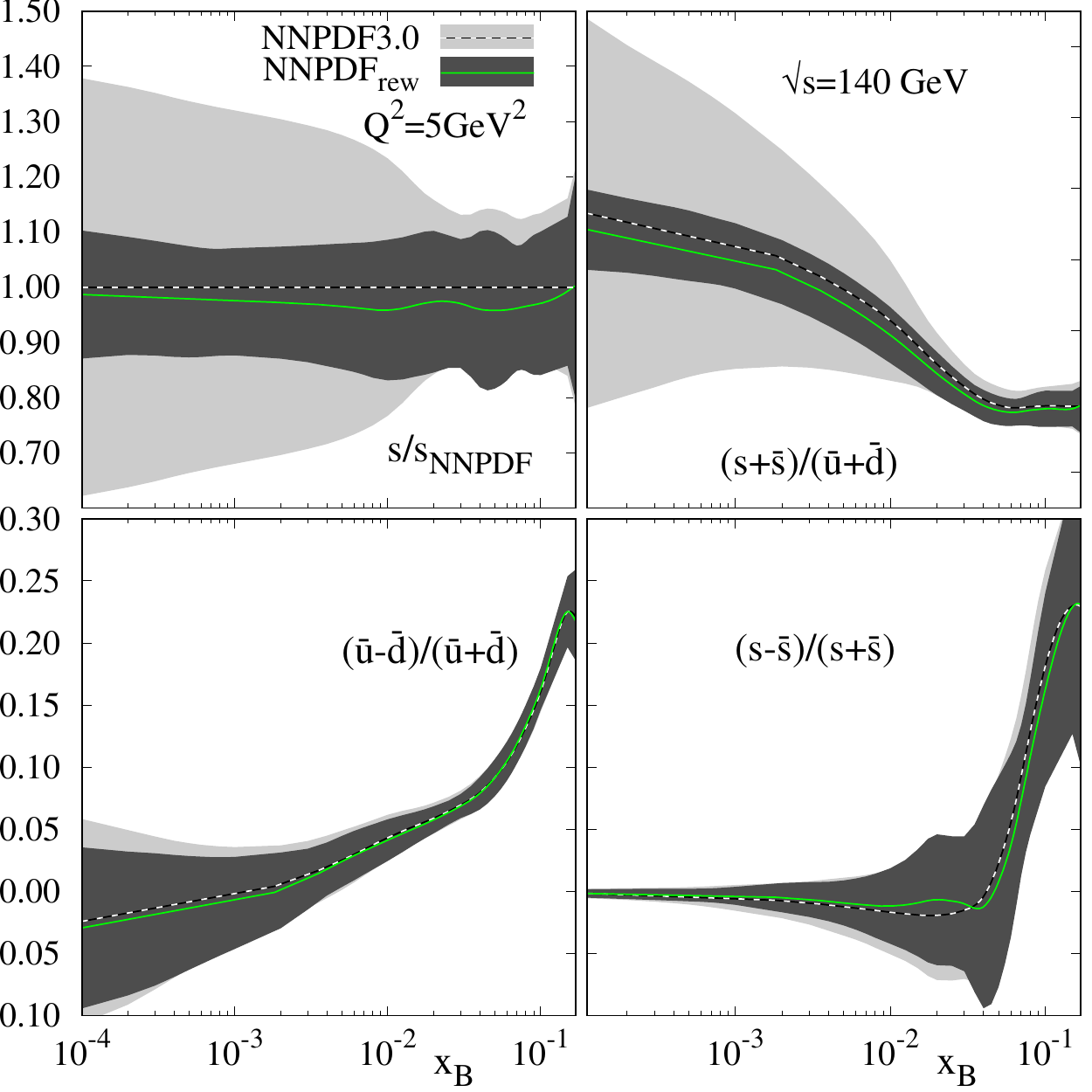}
    \caption{Expected impact on the unpolarized (sea) quark PDFs when adding SIDIS information from pions and kaons in {\it ep} collisions.
    The baseline NNPDFs were take from Ref.~\cite{Ball:2014uwa}.
    }
    \label{fig:part2-sec-PartStruct.unpolubar}
\end{figure}

A recent study of the expected impact on the unpolarized (sea) quark PDFs using simulated EIC pseudodata 
can be found in Ref.~\cite{Aschenauer:2019kzf}. 
In this work 
PYTHIA-6~\cite{Sjostrand:2006za} MC simulations were performed at the center-of-mass energies $\sqrt{s}=140$~GeV and $\sqrt{s}=45$~GeV and were extrapolated to 10~fb$^{-1}$ of integrated luminosity. The typical DIS selection criteria ($Q^2>1$~GeV$^{2}$, $0.01<y<0.95$ and $W^2>10$~GeV$^2$) were augmented by selecting charged pions and kaons that would end up in a main EIC detector, with momenta for which particle identification may be available.  
Using a reweighting technique~\cite{Ball:2010gb,Ball:2011gg} the impact was evaluated simultaneously for unpolarized PDFs and fragmentation functions. The effect on unpolarized (sea) quark PDFs can be seen in Fig.~\ref{fig:part2-sec-PartStruct.unpolubar}. While the impact on up, down, anti-up and anti-down quark PDFs is moderate, as they are already very well determined, the far less well-known strange PDFs will be constrained substantially, particularly at lower $x$. In addition, the presently heavily-debated strange to light sea-quark ratio will be determined well at lower $x$.


\subsubsection*{Nonperturbative charm}
\begin{figure*}[th]
\begin{tabular}{ll} 
\parbox[c]{0.5\textwidth}{\includegraphics[width=0.5\textwidth]{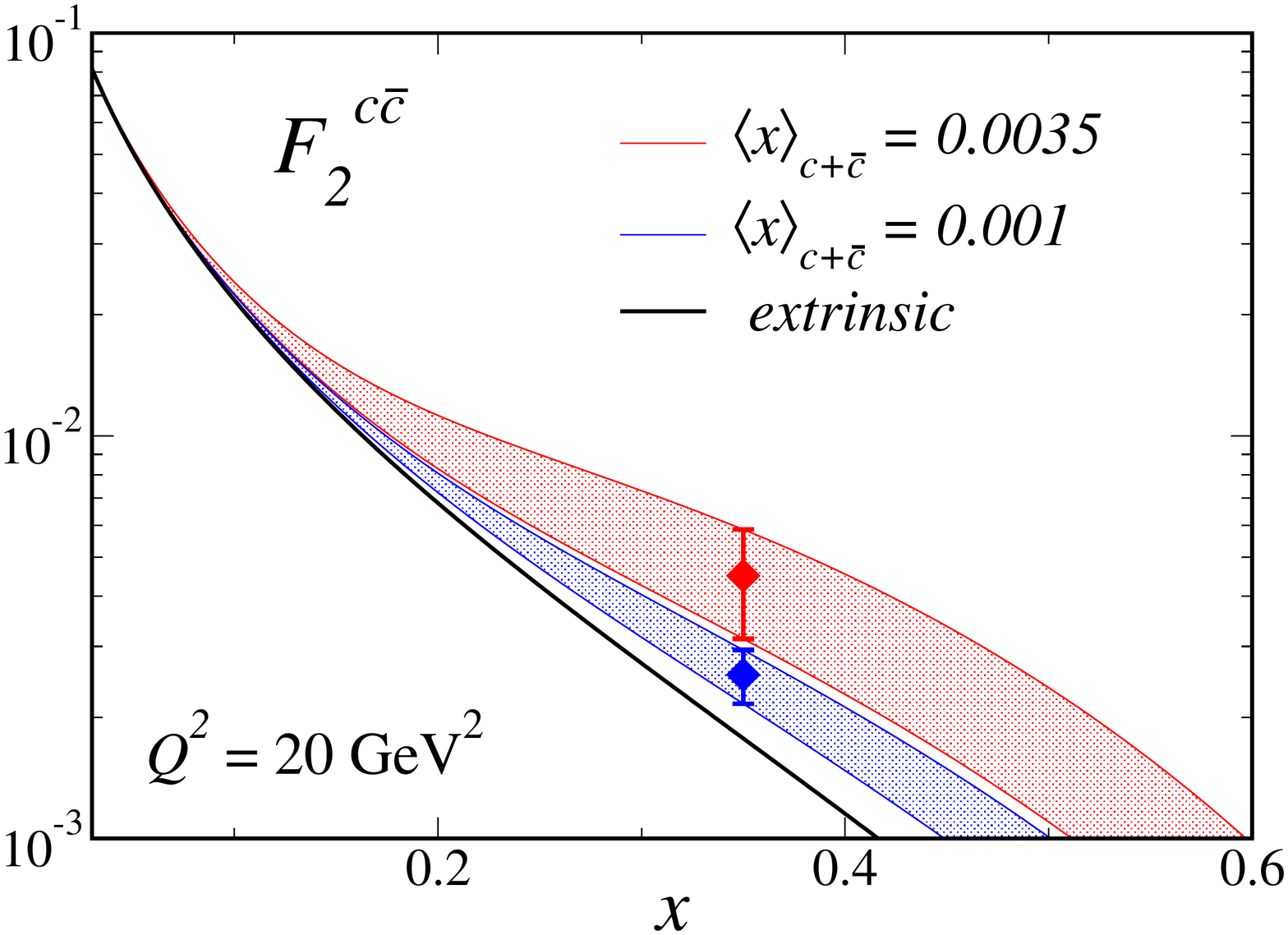}}
\hspace{0.01\textwidth}
\parbox[c]{0.4\textwidth}{
\caption[]{\label{fig:IC}
    Two scenarios~\cite{Hobbs:2017fom} for the potential magnitude of the nonperturbative charm contribution to the charm structure function $F^{c\bar{c}}_2$ of the proton.
    The presence of nonperturbative charm can be seen as a definite overhang in the charm structure function above the perturbative charm (black solid curve) result.
    The two points at high $x$ represent hypothetical measurements at $x\!\sim\!0.35$ with the minimal discrimination power needed to distinguish the model scenarios discussed in Ref.~\cite{Hobbs:2017fom}.
    }
}
\end{tabular}
\end{figure*}

The question of a possible nonperturbative charm component in the nucleon wave function~\cite{Brodsky:1980pb} has long challenged the field of hadronic physics. 
While numerous model calculations have been undertaken over the years, in addition to a significant number of QCD global analyses, a definitive signal has long been elusive, with most analyses~\cite{Jimenez-Delgado:2014zga,Hou:2017khm} generally placing upper limits on the total nonperturbative charm momentum, $\langle x \rangle_{c+\bar{c}}$, at the scale $Q\!=\!m_c$. 
The EMC charm structure function measurements of 1983~\cite{Aubert:1982tt} have been suggested as offering evidence for nonperturbative charm, but have been challenging to accommodate in a global fit.  
The kinematic region over which nonperturbative charm is expected to be visible in typical model calculations is high $x$ and low-to-moderate $Q^2$. 
In Fig.~\ref{fig:IC}, the size of the resulting effect in the charm structure function is plotted in a typical model calculation~\cite{Hobbs:2017fom} for two scenarios: highly suppressed [$\langle x \rangle_{c+\bar{c}} = 0.1\%$] and intermediate [$\langle x \rangle_{c+\bar{c}} = 0.35\%$]. 
Precision DIS data in this region, $x \gtrsim 0.3$ and $\langle Q^2 \rangle \sim 20$ GeV$^2$, would permit the direct measurement of the charm structure function and help resolve the proton charm content.
As discussed in Ref.~\cite{Hou:2017khm}, the nonperturbative charm contribution may be interpreted as involving twist-4 four-gluon correlator functions.
Measurements of nonperturbative charm may therefore constrain twist-4 gluon correlators in the same way that extrinsic charm is used to constrain the twist-2 gluon PDF.
A recent analysis carried out by the NNPDF Collaboration~\cite{Ball:2016neh} has demonstrated how measurements of $F_2^{c\bar{c}}$ at large $x$ have great potential to unravel intrinsic charm and that the constraints of the EIC on a nonperturbative charm component would complement those provided at the LHC, e.g., via weak boson production in the forward region. In addition, the charm-tagging abilities discussed briefly below and in greater detail in Sec. 8.3 will likely enhance the EIC's ability to disentangle a possible nonperturbative charm contribution to the structure of the nucleon.

%

\subsubsection*{Charm jets}
\begin{figure}[ht]
    \centering
    \includegraphics[width=0.48\columnwidth]{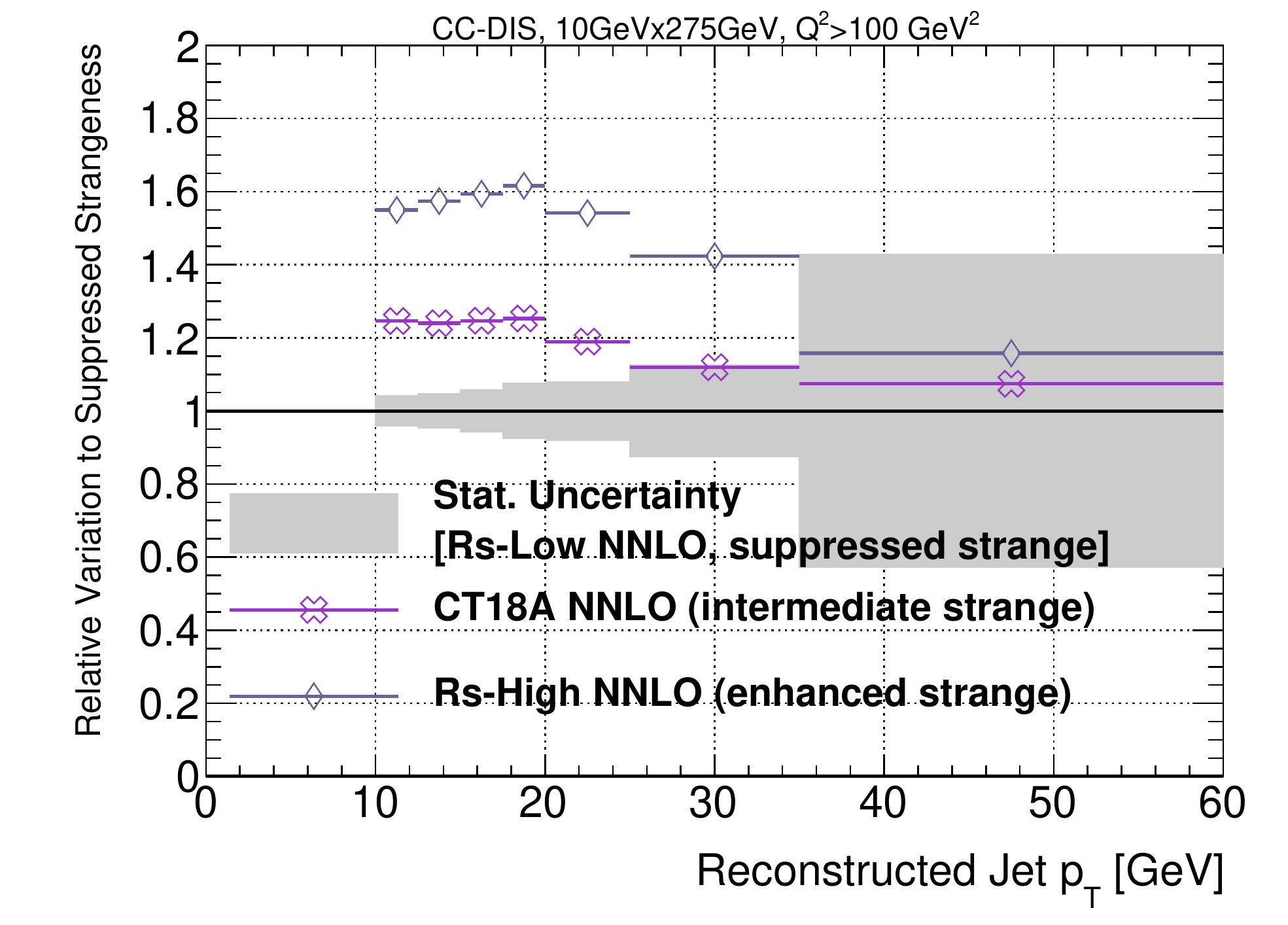} \ \
     \includegraphics[width=0.48\columnwidth]{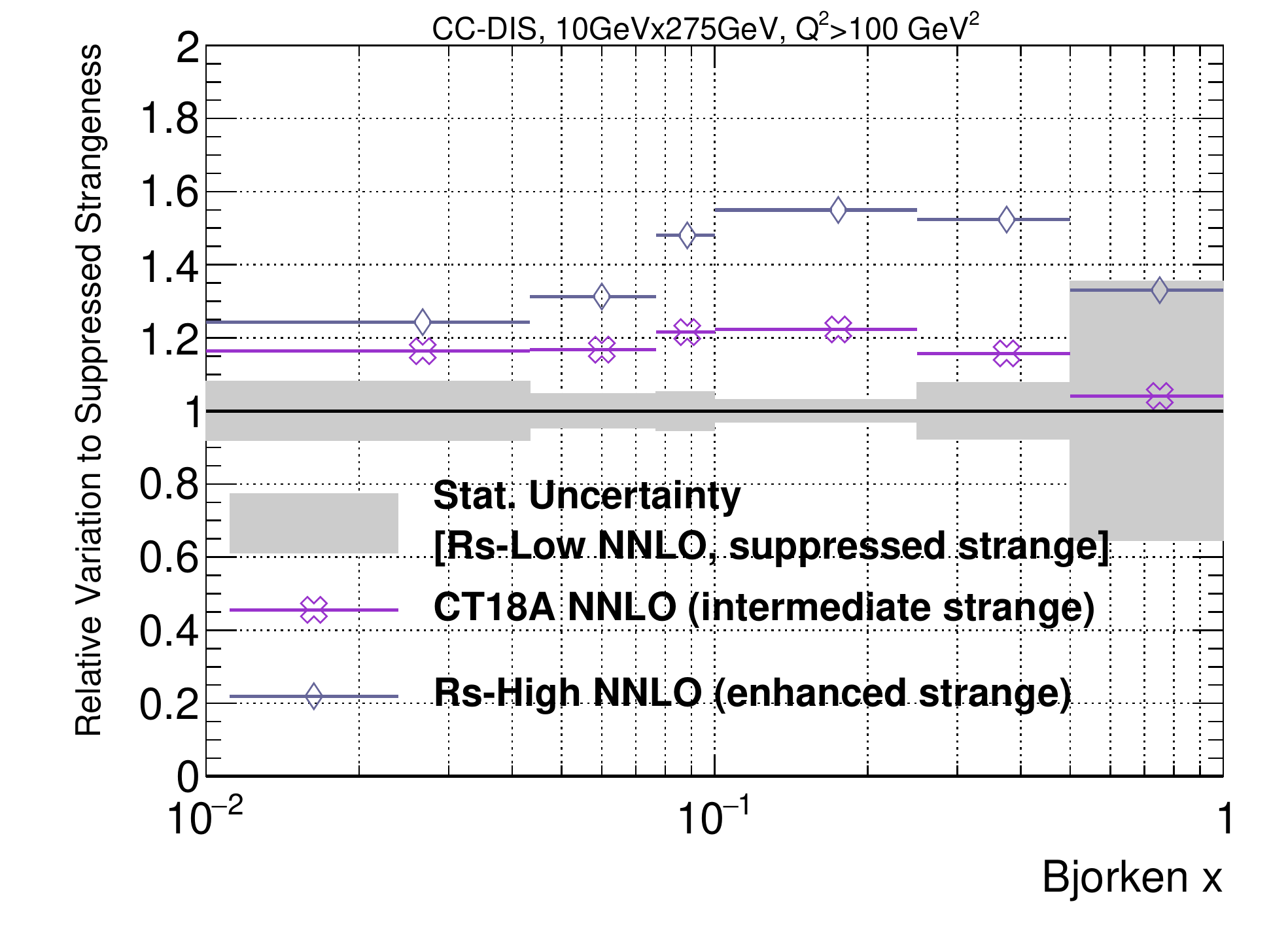}
    \caption{
        Comparison of charm-jet yields in electron-proton CC DIS under two scenarios for the behavior of nucleon strangeness and the light-quark sea: $R_s=2s/(\overline{u}+\overline{d})=0.325$ (``Rs-Low,'' CT18 NNLO with suppressed strangeness) and $R_s=0.863$ (``Rs-High,'' CT18Z NNLO with enhanced strangeness). An intermediate scenario, based on CT18A NNLO, is also shown. The gray band indicates the expected statistical error on the reconstructed and tagged charm jet $p_T$ (left) or Bjorken $x$ (right) spectrum for $\mathrm{100\,fb^{-1}}$ of data as simulated in Ref.~\cite{Arratia:2020azl}. The points indicate the relative difference in expected yields between the enhanced and suppressed strangeness cases, $1+(N_{0.863}-N_{0.325})/N_{0.325}$.  The relative magnitude of the blue points compared to the statistical uncertainty suggests that charm-jet measurements in CC DIS have strong sensitivity to the nucleon's unpolarized strange PDF.  While this calculation is from $ep$, similar discrimination power is expected for nuclear scattering as well.} 
\label{fig:CCDIS-charm}
\end{figure}

In addition to the moderate sensitivity to $R_s$ from inclusive EIC measurements, data involving final-state tagging of a produced charm quark may also help discriminate among scenarios for the strange sea, as demonstrated in a recent analysis \cite{Arratia:2020azl}. In Fig.~\ref{fig:CCDIS-charm}, we illustrate the event-level variation in CC DIS production of charm jets for $\sqrt{s}\! =\! 140$ GeV at the EIC, and find strong dependence on the input scenario for $R_s$ [$R_s = 0.325$ vs.~$R_s=0.863$, as obtained using extreme PDF sets in CT18(Z) NNLO].  This strong dependence suggests that charm-jet production may be a sensitive channel to constrain nucleon strangeness and disentangle patterns of $\mathrm{SU}(3)$ symmetry breaking in the light-quark sea.

\subsection{Spin structure of the proton and neutron}
\label{part2-subS-SpinStruct.P.N}

\subsubsection*{Inclusive $A_{LL}$} 

\begin{figure}[t]
    \centering
    \includegraphics[width=0.45\textwidth]{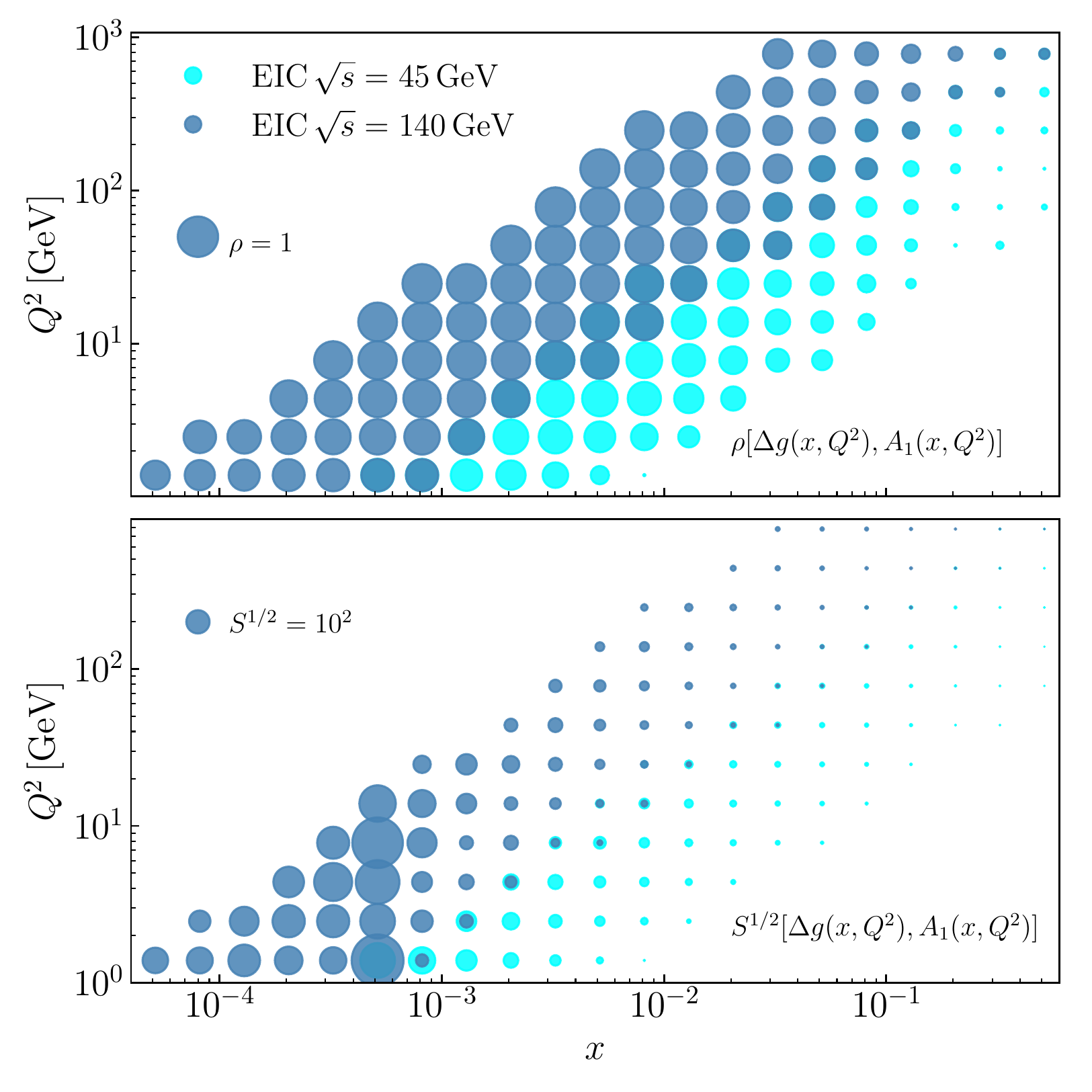} 
    \includegraphics[width = 0.45\textwidth]{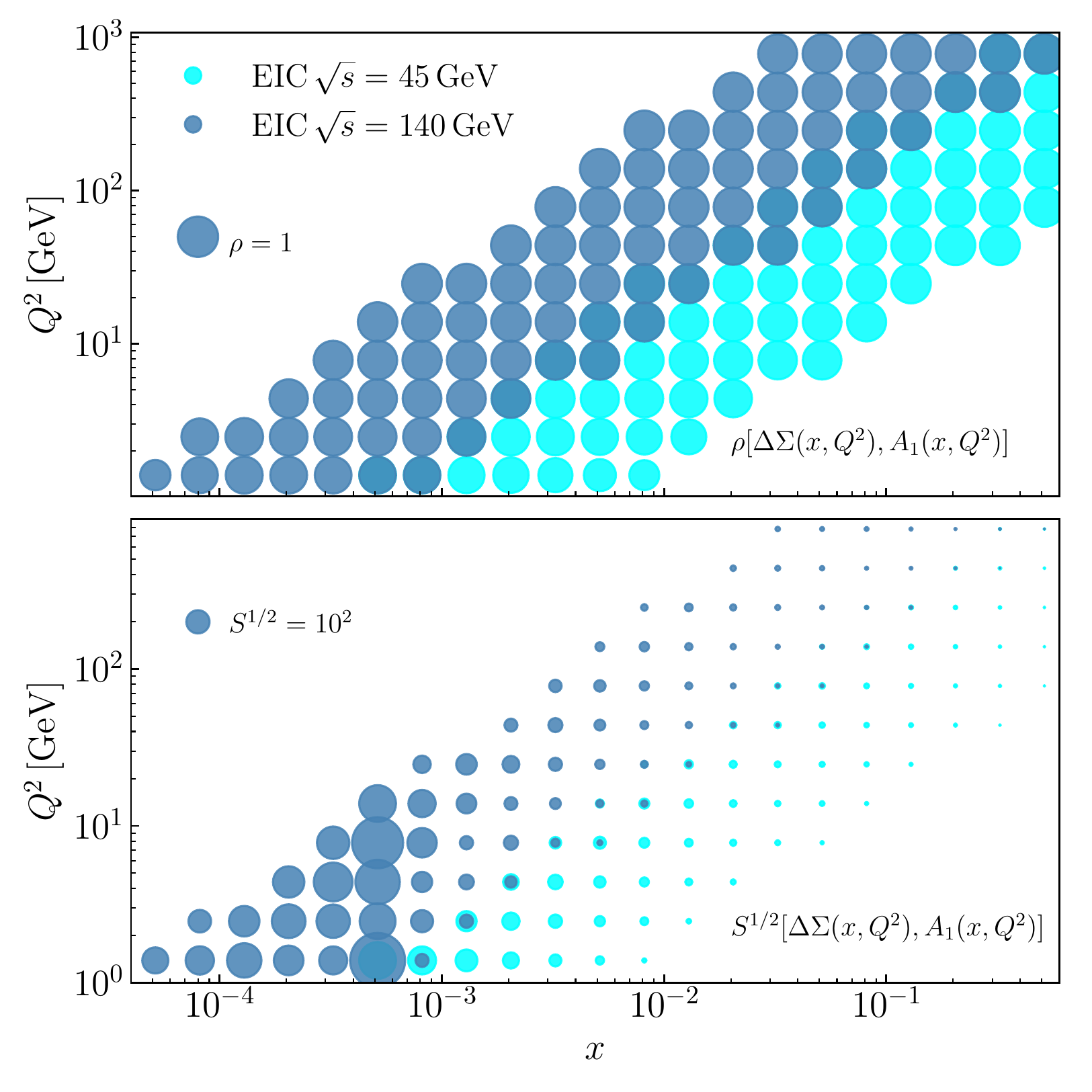}
    \caption{
        {\small Correlation (upper panel) and sensitivity (lower panel) coefficients between the gluon helicity distribution $\Delta g(x,Q^{2})$ and the (photon-nucleon) double-spin asymmetry $A_1$, as well as between the quark-singlet distribution $\Delta \Sigma(x,Q^{2})$ and $A_1$, as a function of $\{x,Q^2\}$. The lighter blue and darker blue circles represent the values of the correlation (sensitivity) coefficient for $\sqrt{s} = 45$ GeV and 140~GeV, respectively. In all the cases the size of the circles is proportional to the value of the correlation (sensitivity) coefficient.}
        }
    \label{fig:Correlation_DSSV}
\end{figure}

In studying the spin structure of the nucleon, the double spin asymmetry $A_{LL}$ has provided the bulk of the constraints on the spin-dependent collinear PDFs. 
In contrast to the unpolarized case, however, the existing $A_{LL}$ data have a much more limited kinematic coverage ($x \gtrsim 0.01$). 
As the world's first polarized lepton-hadron (and lepton-nucleus) collider, the EIC will explore uncharted territory in spin physics. 
In addition to the sensitivity to the quark sector, the wide $Q^2$-coverage of the EIC will probe scaling violations in the $g_1$ structure function, offering significant constraints on the gluon helicity PDF. This is demonstrated by the correlation maps
$\rho\left[f_i,{\cal O}\right]
=( \left<{\cal O}\cdot f_i\right>
  -\left<{\cal O}\right> \left<f_i\right>)
/\Delta {\cal O} \Delta f_i$
and sensitivity maps
$S\left[f_i,{\cal O}\right]
=( \left<{\cal O}\cdot f_i\right>
  -\left<{\cal O}\right> \left<f_i\right>)
/\delta {\cal O} \Delta f_i$
across kinematics 
between a given partonic structure and a physical observable $\cal O$ 
shown in Fig.~\ref{fig:Correlation_DSSV} \cite{Aschenauer:2020pdk}. 
In these metrics, $\Delta$ represent the uncertainties stemming from PDF uncertainties, while $\delta {\cal O}$ is the simulated observable uncertainty.

\begin{figure}[t]
    \centering
    \includegraphics[width=0.45\textwidth]{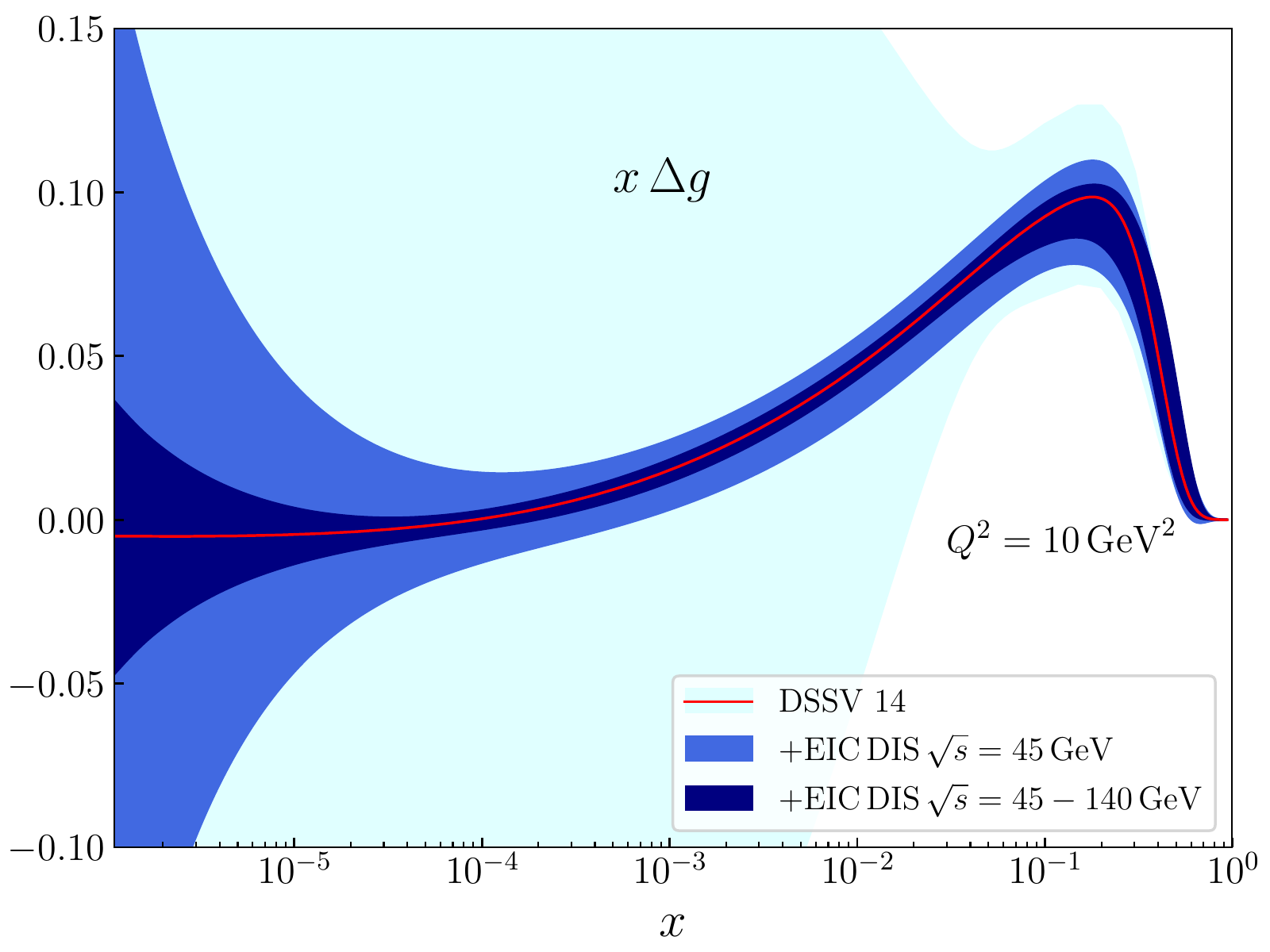} 
    \includegraphics[width = 0.45\textwidth]{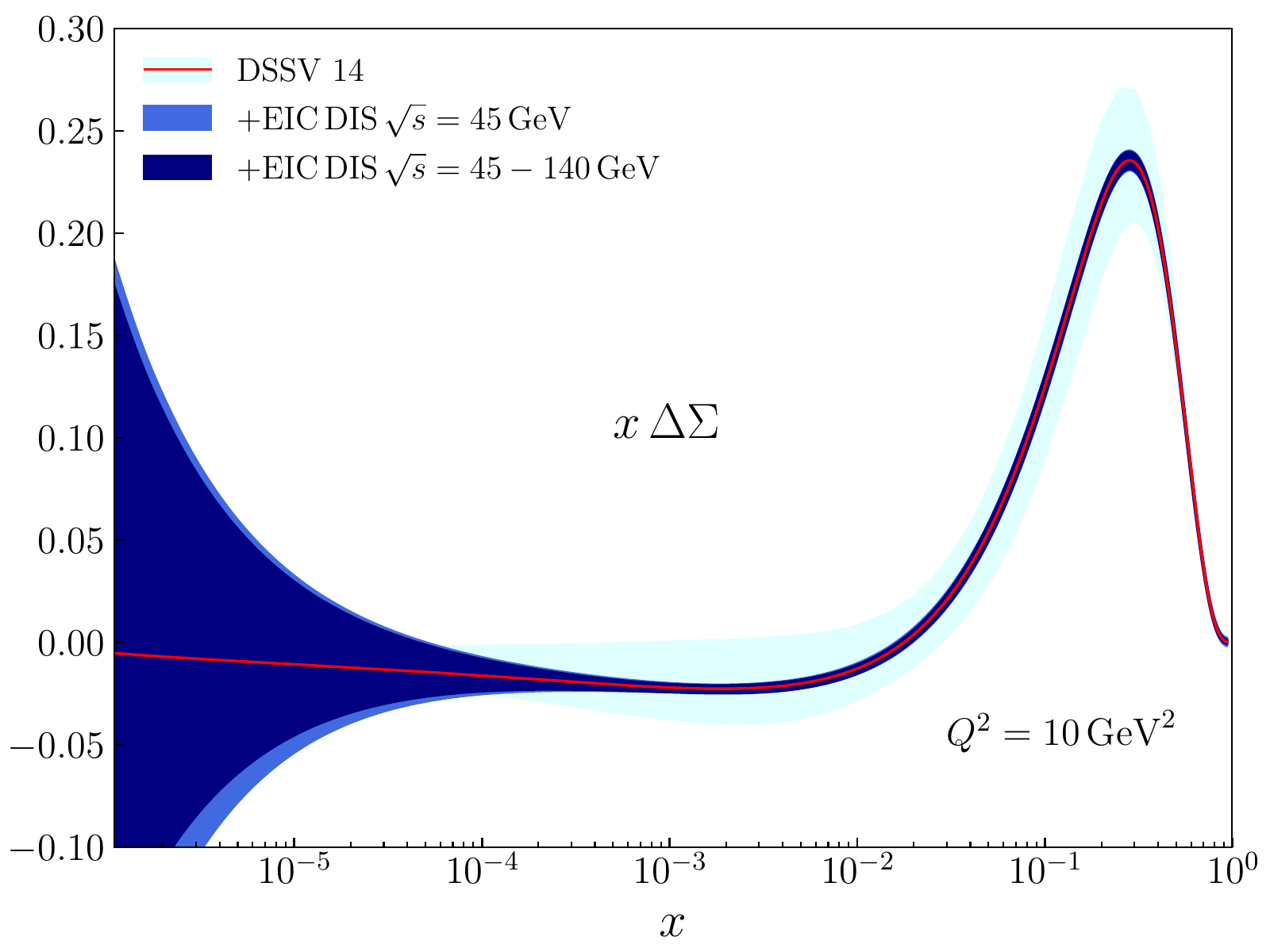}
    \caption{
        {
        \small 
        Impact of the projected EIC $A_{LL}$ pseudoda on the gluon helicity (left panel) and quark singlet helicity (right panel) distributions as a function of $x$ for $Q^2=10\,\mathrm{GeV}^2$. In addition to the DSSV14 estimate (light-blue), the uncertainty bands resulting from the fit including the $\sqrt{s}=45$~GeV DIS pseudodata (blue) and, subsequently,  the reweighting with $\sqrt{s}=140$~GeV pseudodata (dark blue), are also shown.}
        }
    \label{fig:Gl_Sigma_DISatEIC}
\end{figure}

The EIC impact on the helicity distributions is illustrated in Fig.~\ref{fig:Gl_Sigma_DISatEIC} where the $\sqrt{s}=45{\rm~GeV}$ pseudodata is included as part of a new global fit with an extended flexibility for the helicity PDFs parametrization. This extended version of the NLO DSSV14 baseline is then reweighted with the inclusion of the $\sqrt{s}=140{\rm~GeV}$ data. As indicated in the figure, the uncertainty on the gluon helicity is significantly reduced relative to the DSSV14~\cite{Aschenauer:2020pdk, deFlorian:2019zkl} baseline after the inclusion of the projected EIC pseudodata at ${\cal L}=10~{\rm fb}^{-1}$.
%

%

\begin{figure}[t]
    \centering
    \includegraphics[width = 0.45\textwidth]{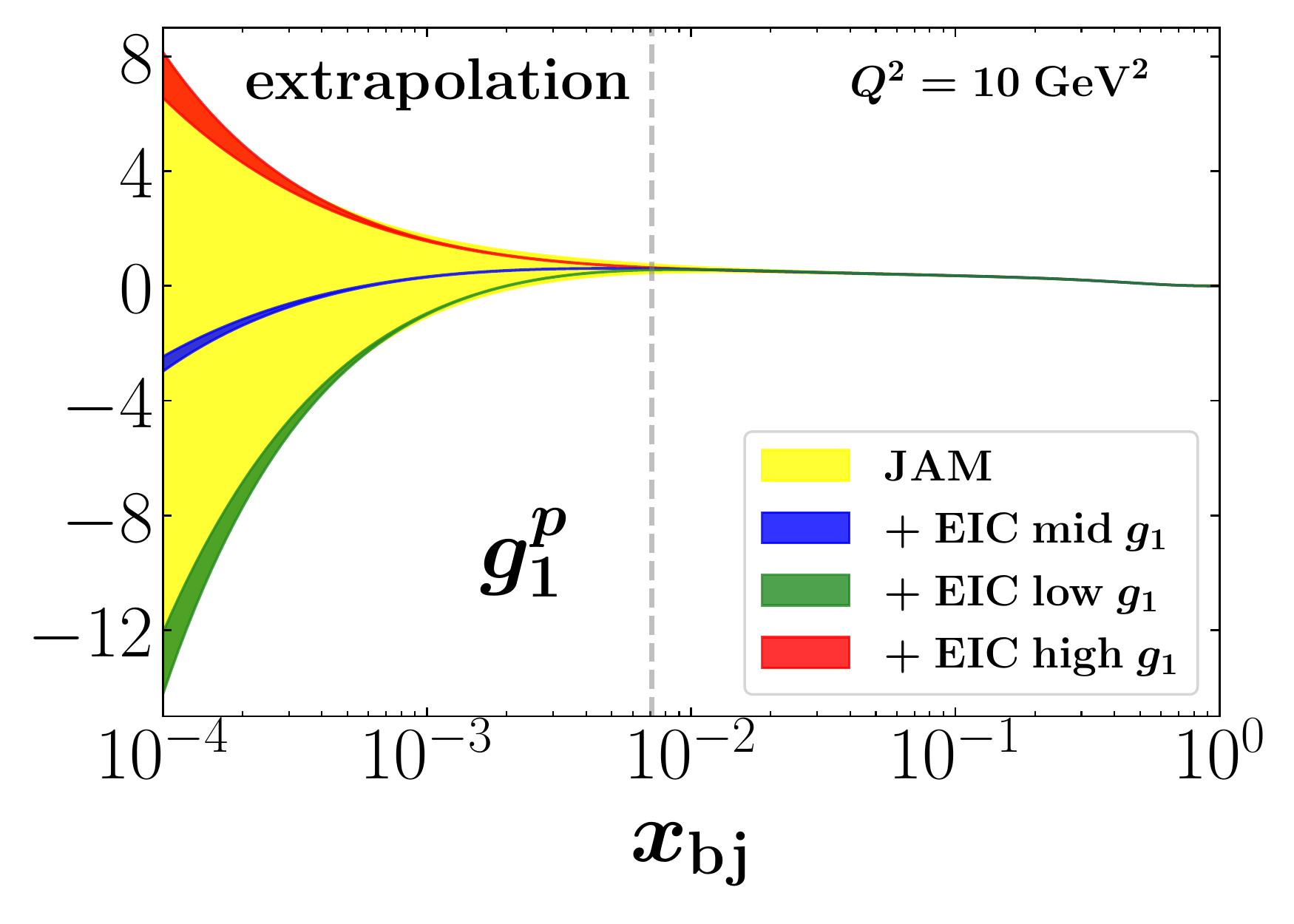}
    \includegraphics[width = 0.45\textwidth]{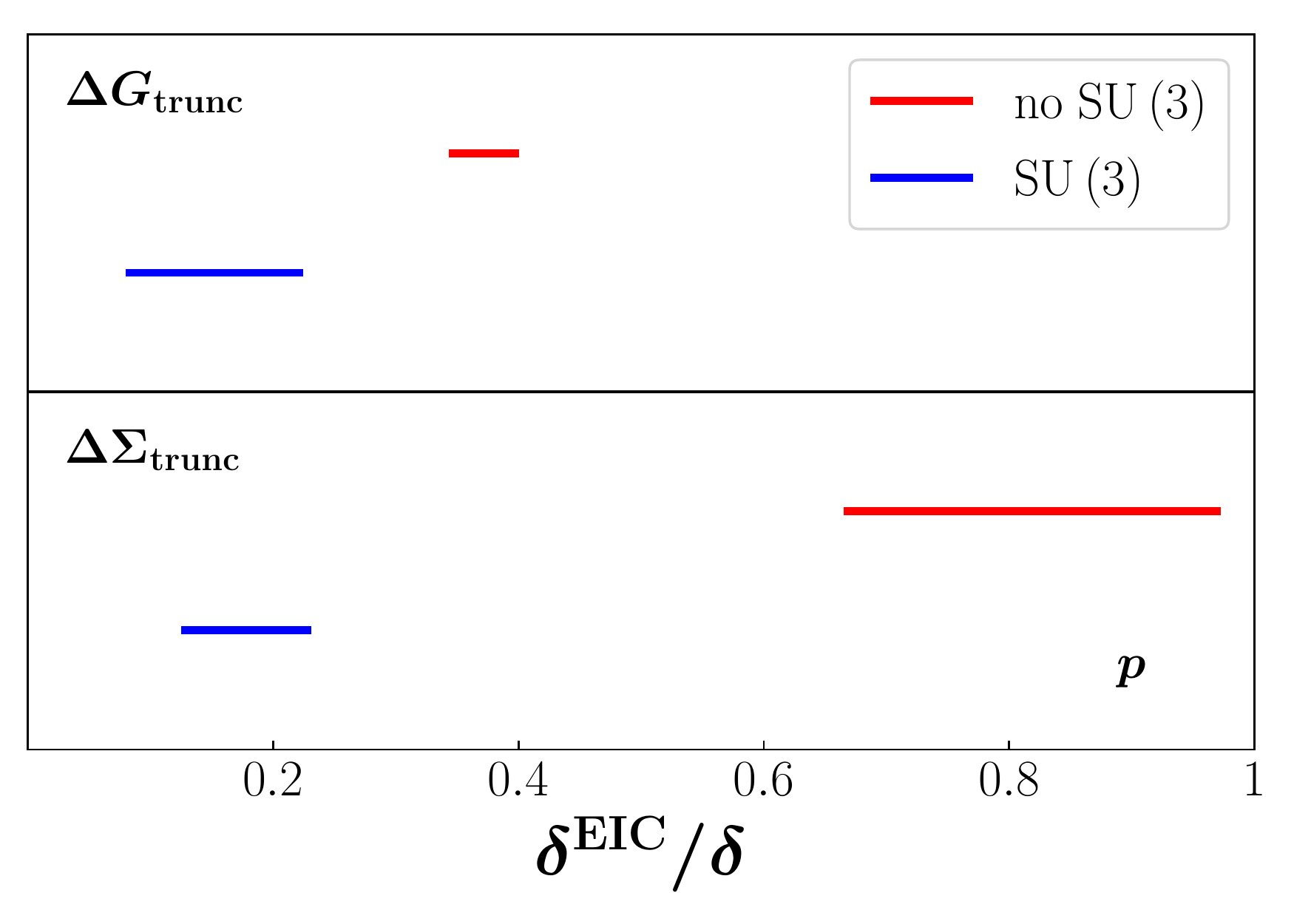}
    \caption{
        {\small 
        Left: Impact of projected $A_{LL}^p$ data at EIC kinematics on $g_1^p$, relative to the JAM global QCD analysis~\cite{Ethier:2017zbq, YiyuZhou20} (yellow band), taking $+1\sigma$ (``high $g_1$'', red band),  $-1\sigma$ (``low $g_1$'', green band) and central (``mid $g_1$'', blue band) uncertainties of $A_{LL}^p$.
        Right: Uncertainty on the gluon ($\Delta G_{\rm trunc}$) and quark-singlet ($\Delta \Sigma_{\rm trunc}$) truncated moments from $x_{\rm min}=10^{-4}$ to 1 with EIC data ($\delta^{\rm EIC}$) normalized to the baseline PDFs uncertainties ($\delta$)~\cite{Ethier:2017zbq, YiyuZhou20}, covering the ``low'', ``mid'' and ``high'' scenarios, for the case of no SU(3) symmetry (red lines) and with SU(3) symmetry (blue lines).
        }}
    \label{fig:A_LL_p}
\end{figure}

One of the challenges in reliably assessing the impact of the inclusive $A_{LL}$ measurements at the EIC is the fact that the predictions for the rates are based on extrapolation from existing measurements that only extend down to $x \sim 0.01$. 
A study exploring the uncertainty on the helicity distributions associated with the extrapolation of $A_{LL}$ for the EIC pseudodata is shown in Fig.~\ref{fig:A_LL_p}.
The analysis is carried out within the JAM global QCD analysis framework at NLO in pQCD, including all existing data on $A_{LL}$ and inclusive jet production from polarized $pp$ scattering at RHIC~\cite{YiyuZhou20}, along with $A_{LL}$ from EIC proton pseudodata simulated with ${\cal L} = 100~\mathrm{fb}^{-1}$, 2.3\% normalization uncertainty, and 2\% point-by-point uncorrelated systematic uncertainties.
The JAM analysis incorporates the projected EIC data along with the existing data using the full MC fitting framework.
To explore the impact of the extrapolation region, three sets of pseudodata were generated by shifting the unmeasured region at low $x$ with $\pm 1 \sigma$ confidence level, using existing helicity PDF uncertainties as well as the central predictions. 

In Fig.~\ref{fig:A_LL_p} the uncertainty bands for $g_1^p$ before and after the three scenarios ($\pm 1\sigma$ confidence level and central) at the EIC are shown, along with the ratios $\delta^{\rm EIC}/\delta$ of uncertainties on the truncated moments of the quark-singlet and gluon PDFs,  $\Delta\Sigma_{\rm trunc}$ and $\Delta G_{\rm trunc}$, integrated between $x_{\rm min}=10^{-4}$ and 1, with EIC data to the baseline JAM results with existing data.
The results show that, if one assumes SU(3) symmetry for the axial vector charges, the uncertainty on $\Delta G_{\rm trunc}$ can improve by $80 - 90\%$, depending on the behavior of the low-$x$ extrapolation of $g_1^p$, with an $\sim 80\%$ reduction in the uncertainty on $\Delta\Sigma_{\rm trunc}$. 
The reduction is more modest, however, if one does not impose SU(3) symmetry, in which case the gluon moment uncertainty decreases by $\sim 60\%$, but no clear reduction in the quark singlet uncertainty is apparent from proton EIC data alone.
%

%

\subsubsection*{Helicity and small-$x$ dipole formalism} 
%

\begin{figure}[t]
    \centering
    \includegraphics[width=0.6\textwidth]{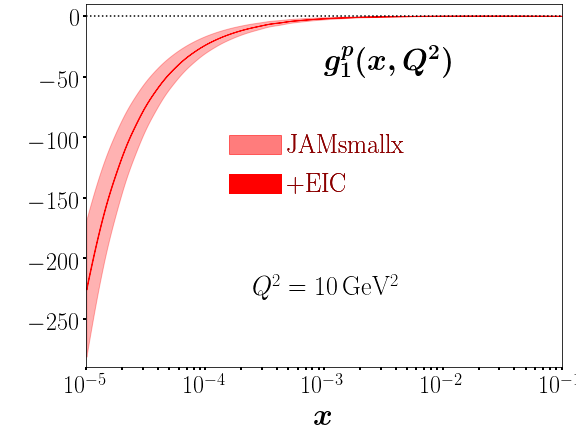}	
    \caption{
        {\small Plot of the $g_1(x)$ structure function obtained in Ref.~\cite{Adamiak:2021ppq} using     the small-$x$ helicity evolution formalism of Refs.~\cite{Kovchegov:2015pbl, Kovchegov:2016zex, Kovchegov:2016weo, Kovchegov:2018znm}.}}
    \label{fig:g1_smallx}
\end{figure}

A prediction for the $g_1$ structure function based on the novel small-$x$ evolution equations derived by Kovchegov, Pitonyak and Sievert (KPS) in Refs.~\cite{Kovchegov:2015pbl,Kovchegov:2016zex,Kovchegov:2016weo,Kovchegov:2018znm} is shown in Fig.~\ref{fig:g1_smallx}. 
The KPS equations evolve the polarized color-dipole scattering amplitude toward small values of $x$. 
At the leading order employed here, the KPS equations resum powers of $\alpha_s \, \ln^2 (1/x)$, generating pQCD predictions for the small-$x$ behavior of helicity PDFs and for the $g_1$ structure function. 
The curve in  Fig.~\ref{fig:g1_smallx} is obtained by using the large-$N_c$ LO KPS equations (along with their initial conditions) in the JAM framework~\cite{Sato:2016tuz,Ethier:2017zbq} to fit the existing world DIS data on $A_1$ and $A_{LL}$ for $x<0.1$ and extrapolate the resulting $g_1$ structure function to lower values of $x$ using the same KPS evolution. 
The plot was constructed in Ref.~\cite{Adamiak:2021ppq} for fixed $\alpha_s = 0.3$. 
Clearly the EIC data will significantly shrink the uncertainty of this prediction from the light red error band in  Fig.~\ref{fig:g1_smallx} to the solid red (very thin) one, allowing for a much better constraint on the proton spin coming from the small-$x$ quarks and moving the community closer to the resolution of the proton spin puzzle.
We stress that the $g_1$ extrapolation in Figs.~\ref{fig:A_LL_p} and \ref{fig:g1_smallx} are based on two different pQCD formalisms which give different uncertainty bands in the extrapolation region.

\subsubsection*{Neutron spin structure from inclusive and tagged DIS with polarized $^3$He and $^2$H}

Nucleon spin structure studies require measurements of polarized DIS on the neutron as well as the proton~\cite{Anselmino:1994gn,Kuhn:2008sy,Aidala:2012mv}. 
Neutron and proton data together are needed to determine the flavor composition of the $u$- and $d$-quark helicity distributions in the valence quark region $x \gtrsim 0.3$ (for sensitivity to strange-quark polarization additional observables are required), to separate singlet and nonsinglet structures in QCD evolution and the extraction of gluon polarization ($g_1^p - g_1^n$ and $g_1^p + g_1^n$ are generally of the same order at $x \gtrsim 10^{-3}$), and to evaluate the Bjorken sum rule. 
The extraction of neutron spin structure from DIS on polarized light nuclei must account
for nuclear effects (neutron polarization, Fermi motion, dynamical modifications),
which cause significant uncertainties and have been investigated
theoretically~\cite{Frankfurt:1988nt,Arneodo:1992wf,CiofidegliAtti:1993zs,Melnitchouk:1994tx,
Kulagin:1994cj,Piller:1995mf,Frankfurt:1996nf,Bissey:2001cw,Ethier:2013hna}.
The dynamical modifications of nucleon spin structure are themselves an object of study
and provide insight into the emergence of nuclear interactions from QCD --- see Sec.~\ref{part2-subS-LabQCD-LightNuclei}. 

\begin{figure}[b]
    \centering
    \includegraphics[width=0.4\textwidth]{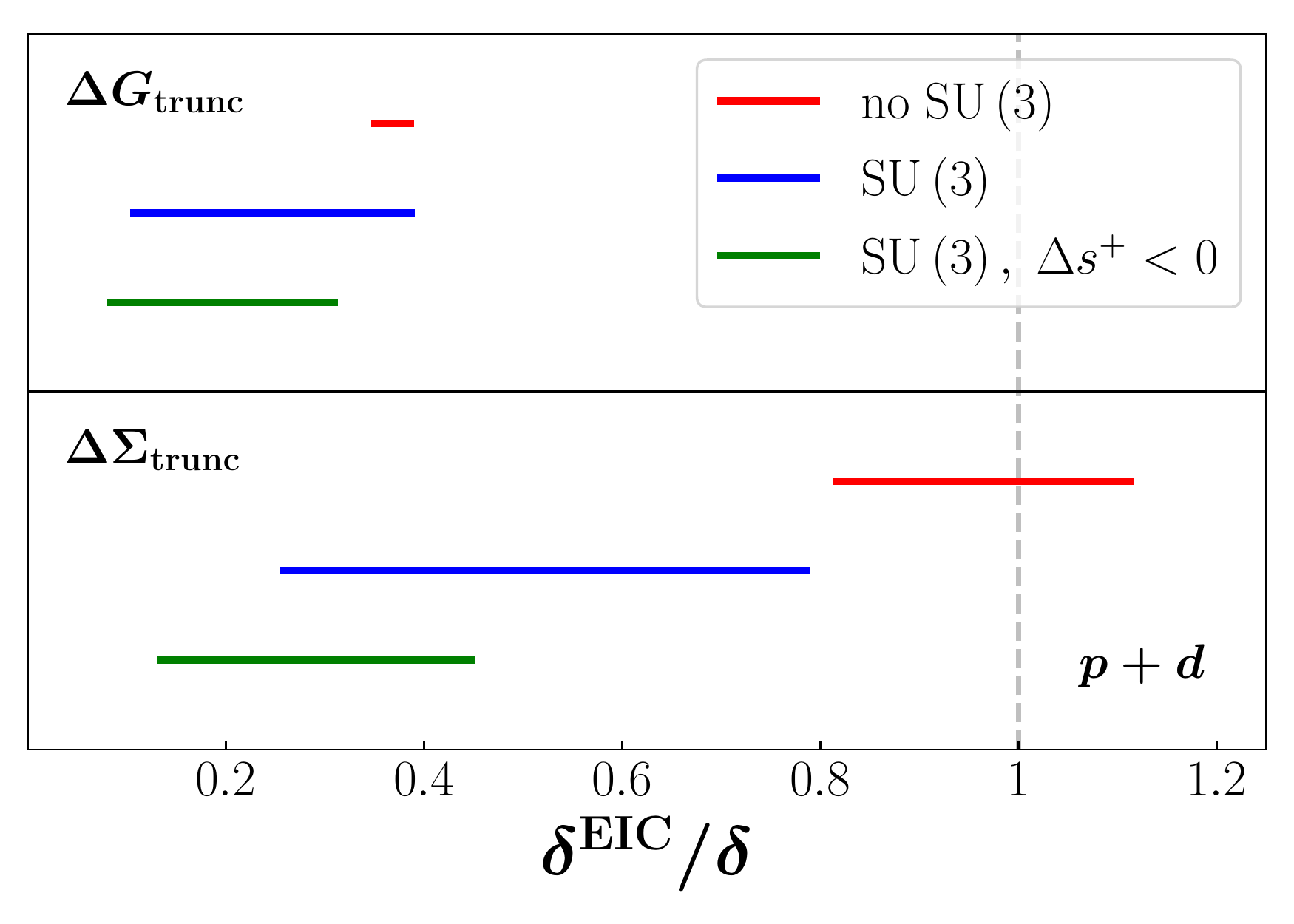}
    \includegraphics[width=0.4\textwidth]{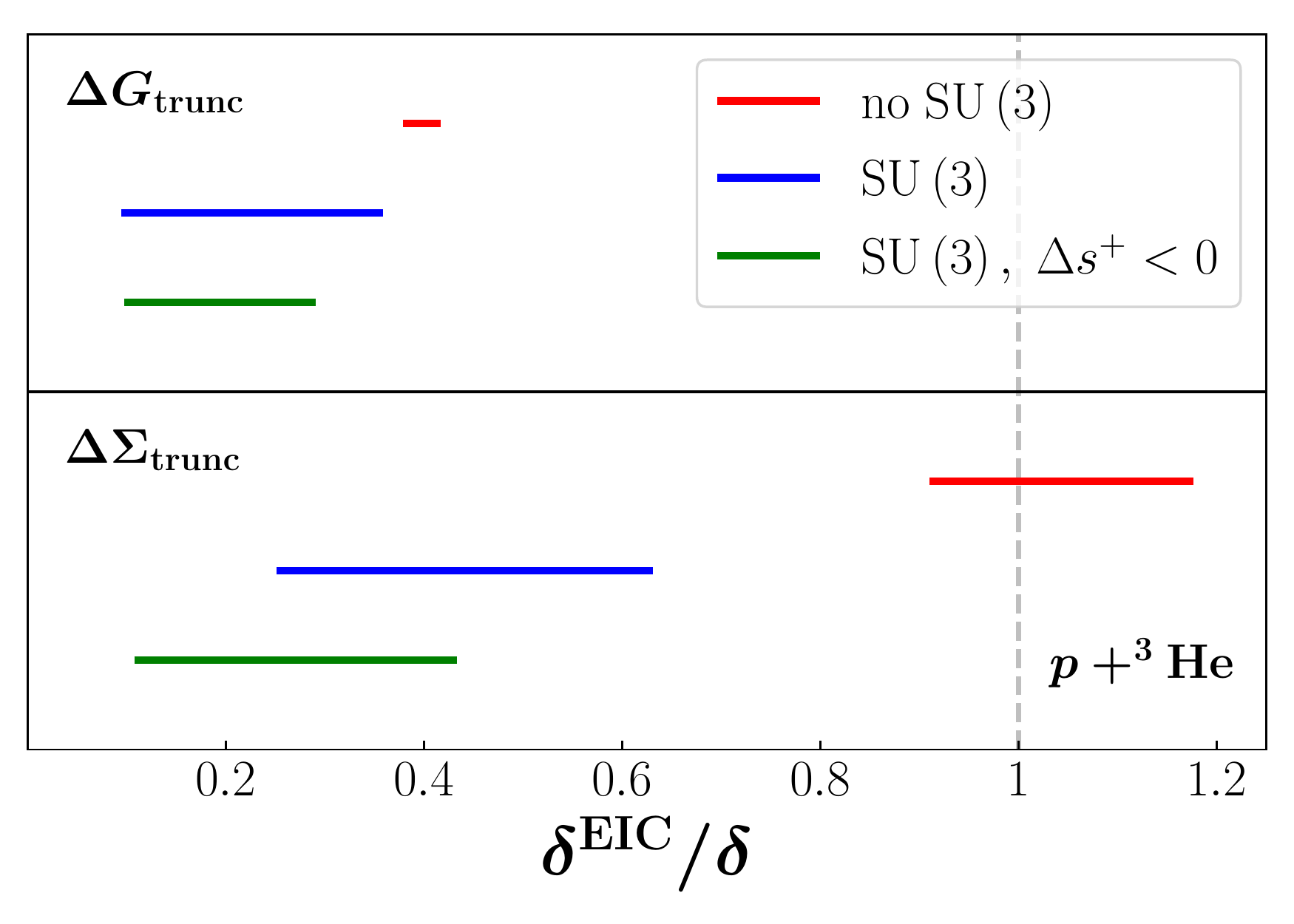}
    \caption{
        {\small Ratio of uncertainties $\delta^{\rm EIC}/\delta$ of the truncated quark-singlet and gluon moments with and without EIC data, for proton + deuteron (left), and proton + $^3$He (right), using the ``high'', ``low'' and ``mid'' extrapolations for $g_1$ in the unmeasured region. The scenario without SU(3) symmetry (red lines) is compared with those with SU(3) imposed (blue lines) and in addition restricting solutions to the ones with negative strangeness (green lines).}}
    \label{fig:A_LL_d_h}
\end{figure}

At the EIC, the neutron spin structure will be measured using DIS on polarized $^3$He, and possibly polarized deuteron $^2$H $\equiv d$. Measurements will be performed using both inclusive DIS (cross sections, spin asymmetries) and DIS with partial or full detection of the
nuclear breakup state (spectator tagging). Each of these methods brings unique
advantages and challenges to neutron structure extraction. Their combination offers
the prospect of substantial advances in the understanding of nuclear effects and
the precision of neutron structure extraction, making the theoretical uncertainties
commensurate with the projected experimental uncertainties of spin structure measurements at EIC.

The impact of data on polarized protons and neutrons, obtained from either deuteron or $^3$He polarization asymmetries, on the quark singlet and gluon truncated moments, $\Delta\Sigma_{\rm trunc}$ and $\Delta G_{\rm trunc}$, is illustrated in Fig.~\ref{fig:A_LL_d_h} for the ratio $\delta^{\rm EIC}/\delta$ of uncertainties with and without EIC data.
The $^3$He and deuteron EIC pseudodata are simulated with ${\cal L} = 10~\mathrm{fb}^{-1}$, 2.3\% normalization uncertainty, and 2\% point-by-point uncorrelated systematic uncertainty.
The impact of the EIC $p+d$ or $p+{}^3{\rm He}$ data is estimated by shifting the asymmetries in the unmeasured region at low $x$ by $\pm 1\sigma$ confidence level using existing helicity PDF errors.
%
%
Like for the proton data impact in Fig.~\ref{fig:A_LL_p}, the reduction of the uncertainties on the truncated moments depends strongly on the assumptions made about SU(3) symmetry for the axial charges, especially for the quark singlet $\Delta\Sigma$.
The gluon moment uncertainty reduction is relatively large, $\sim 60 - 90\%$, depending on the scenario, while the impact on the quark-singlet moment can range from $\sim 20\%$ to $ \sim 90\%$.
The strongest effect coincides with the most stringent constraints, such as removing solutions from the Monte Carlo samples that have positive strangeness.
For the least restrictive scenario, with no SU(3) assumptions, again there is no clear indication of a large impact, even when $d$ or $^3$He data are included, with generally similar behavior observed for both cases.
Deuteron and $^3\mathrm{He}$ data do not have significant impact in the uncertainty of the gluon truncated moment compared to proton data only, and additional observables, such as those from parity-violating DIS discussed below, will be needed to resolve the flavor decomposition of the quark helicity.

\begin{figure}[tbh]
\parbox[c]{0.48\textwidth}{\includegraphics[width=0.48\textwidth]{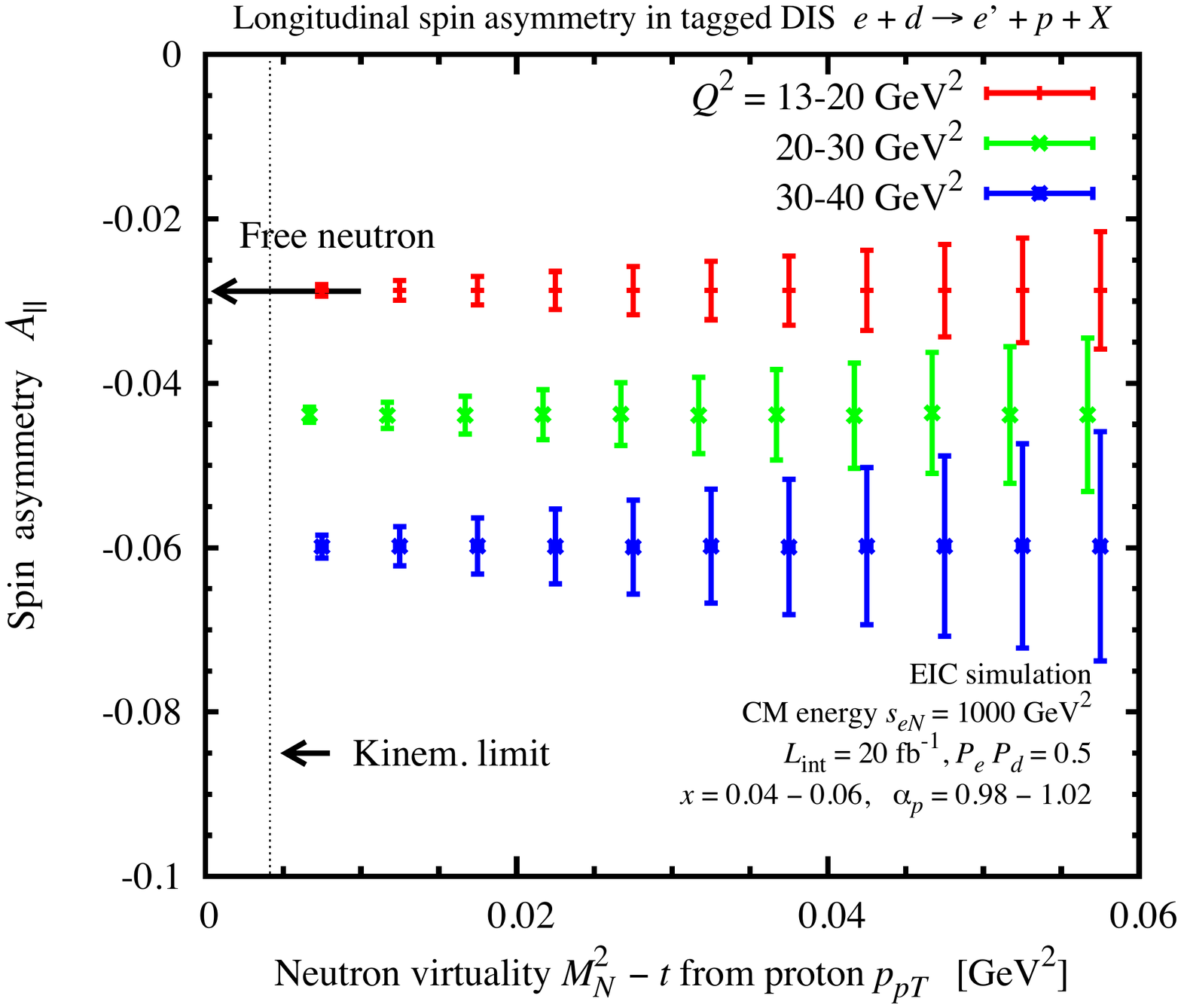}}
\hspace{0.02\textwidth} 
\parbox[c]{0.48\textwidth}{
\includegraphics[width=0.35\textwidth]{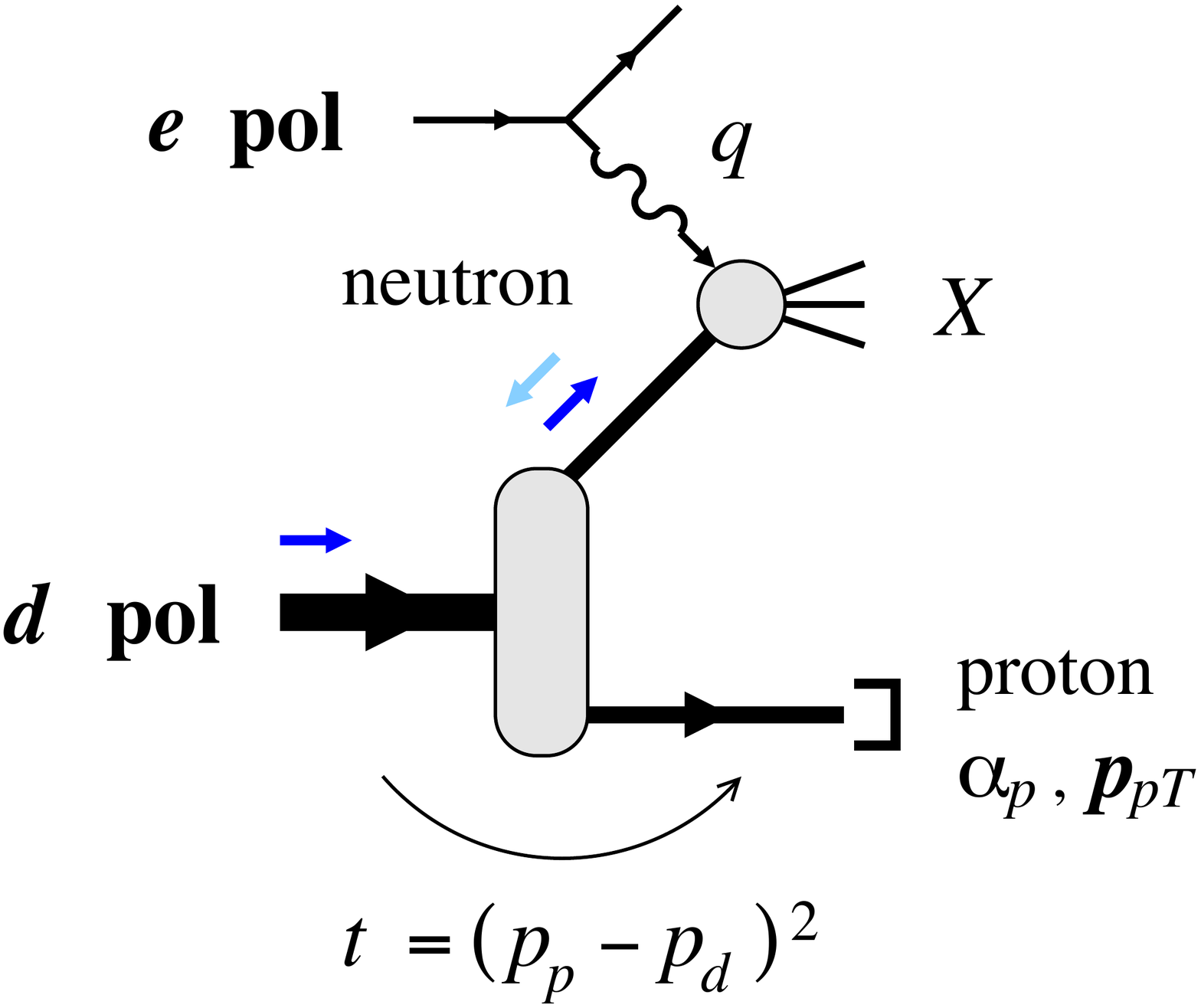}
}
\vspace*{-1cm}
\caption[]{\small Simulated EIC measurements of the longitudinal double-spin asymmetry
$A_{\parallel}$ in polarized deuteron DIS with proton tagging $e + d \rightarrow e' + X + p$.
The asymmetry is shown as a function of the neutron virtuality $t - M_N^2$, which is kinematically
fixed by the tagged proton momentum (light-cone momenta $\alpha_p$ and $\bm{p}_{pT}$).
In the limit $t - M_N^2 \rightarrow 0$ (on-shell extrapolation) the tagged spin asymmetry
coincides with the free neutron spin asymmetry $A_{\parallel n}$ \cite{Cosyn:2019hem,Cosyn:2020kwu}.
The uncertainties shown are statistical ($L_{\rm int}$ = 20 fb$^{-1}$, $P_e P_d$ = 0.5).
\label{fig:neutron_tagging}
}
\end{figure}

While the inclusive polarized DIS on $^3$He is the standard channel for neutron spin structure measurements at the EIC, their analysis relies on the effective neutron polarizations inferred from non-relativistic nuclear structure~\cite{Friar:1990vx,CiofidegliAtti:1993zs}.
Significant nuclear modifications arise from the presence of $\Delta$ isobars in the
$^3$He nucleus at $x \gtrsim 0.1$~\cite{Frankfurt:1996nf,Bissey:2001cw}, and from
spin-dependent nuclear antishadowing and shadowing at $x \lesssim 0.1$. The theoretical
uncertainty resulting from these effects is expected to be the dominant uncertainty and
should be reduced by further theoretical studies. DIS on $^3$He with spectator proton/neutron
tagging has been explored and appears feasible with the EIC forward detectors
(see Sec.~\ref{part2-subS-LabQCD-LightNuclei}).
The theoretical analysis of these measurements requires the modeling of nuclear final-state interactions,
for which corresponding methods have been developed~\cite{CiofidegliAtti:2002as,CiofidegliAtti:2003pb,%
  Palli:2009it,Kaptari:2013dma,Strikman:2017koc}.

DIS on the polarized deuteron complements the measurements on $^3$He and offers several
advantages \cite{Frankfurt:1983qs,Cosyn:2019hem,Cosyn:2020kwu}. In the deuteron $\Delta$
isobars and other non-nucleonic degrees of freedom are suppressed in average nuclear
configurations (nucleon momenta $\lesssim$ 300 MeV), so that the extraction of neutron spin structure
from inclusive DIS is generally simpler and more accurate than for $^3$He
\cite{Frankfurt:1981mk}.

In tagged DIS on the deuteron, the measured spectator momentum fixes the nuclear
configuration and permits a differential treatment of nuclear effects, significantly
improving the theoretical accuracy. The tagged proton momentum controls
the strength of $S$ and $D$ waves in the deuteron wave function and thus the
effective neutron polarization in DIS \cite{Cosyn:2019hem,Cosyn:2020kwu}.
On-shell extrapolation in the proton momentum eliminates nuclear modifications
and final-state interactions and permits the extraction
of the free neutron structure functions \cite{Sargsian:2005rm}.
Simulations show that an accurate determination of the neutron double-spin asymmetry
$A_{\parallel n}$ is feasible using polarized tagged DIS with on-shell extrapolation
(see Fig.~\ref{fig:neutron_tagging}). Further applications of tagged measurements
are discussed in Sec.~\ref{part2-subS-LabQCD-LightNuclei}.

\subsubsection*{Orbital angular momentum contribution to nucleon spin} 
\begin{figure}[t]
    \centering
	\includegraphics[width=0.80\textwidth]{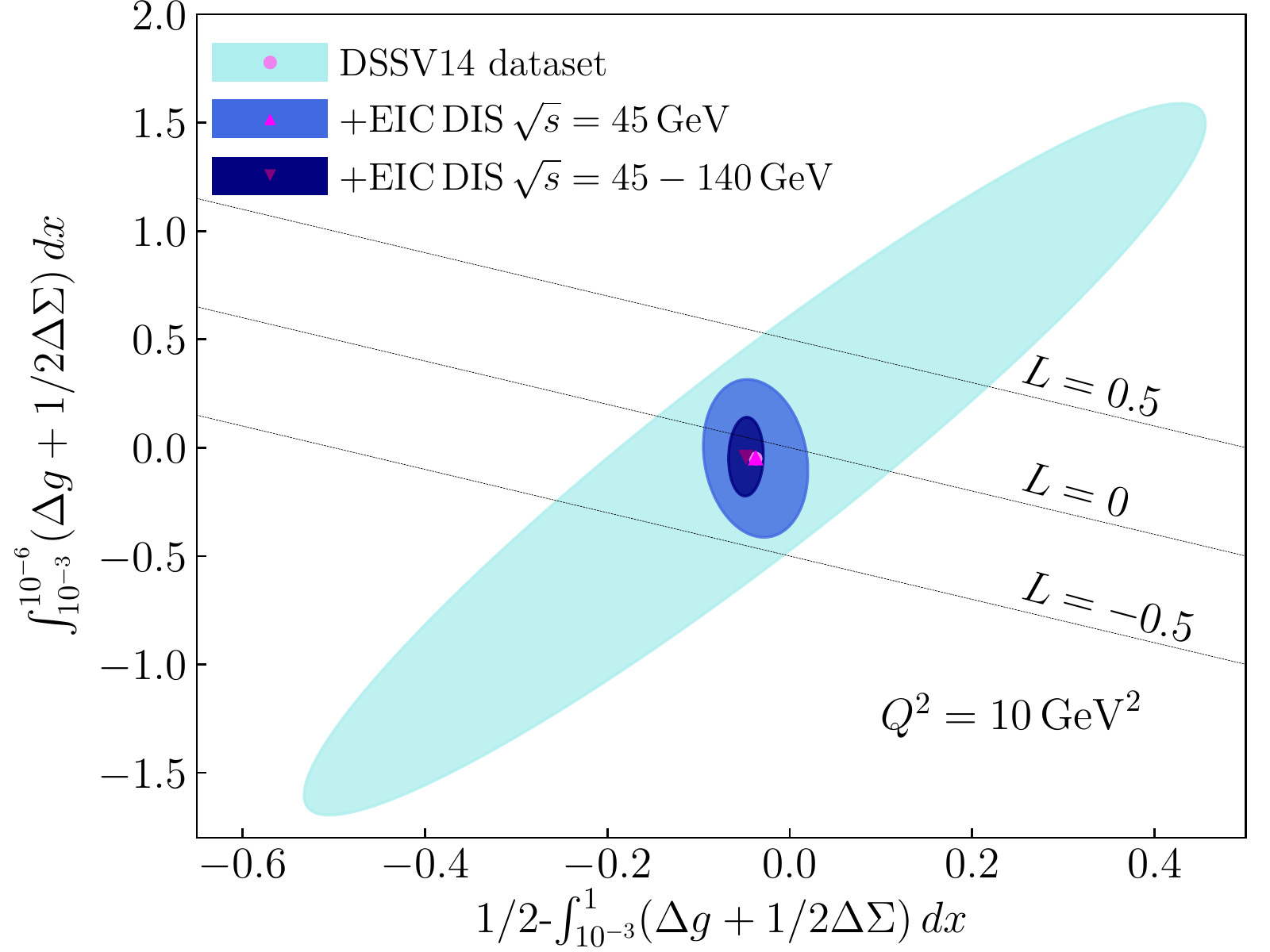}	
	\caption{
	{\small 
	Room left for potential OAM contributions to the proton spin at $Q^2=10\,\mathrm{GeV}^2$, using the existing data and future EIC measurements. The horizontal axis shows the difference between $\frac{1}{2}$ and the contribution from the spin of quarks and gluons for a momentum fraction down to $x=0.001$, which would be the room left for OAM if the spin contribution from partons with smaller momentum fractions was negligible.  The vertical axis presents the spin contribution from partons with momentum fractions between $10^{-6}$ and $10^{-3}$. The ellipses correspond to the $1 \sigma$ correlated uncertainty for the DSSV14 data set (light blue), the fit including EIC $\sqrt{s}=45$ GeV pseudodata (blue), and the reweighting with $\sqrt{s}=140$ GeV pseudodata.}
	}
	\label{fig:OAM_DSSV}
\end{figure}

%
The improved constraints on the spin of quarks and gluons allow for exploring the contribution to the proton spin due to the orbital angular momentum (OAM) of the partons.
Figure~\ref{fig:OAM_DSSV} presents the potential of the EIC to constrain this contribution,
which is identified with the difference between the quark and gluon spin contribution and the proton spin $\frac{1}{2}$. 
The horizontal axis shows the difference between $\frac{1}{2}$ and the contribution from the quark and gluon spins for a momentum fraction down to $x = 0.001$.  
The remaining contributions would be the room left for the parton OAM if the parton spin contribution with smaller momentum fractions is very small or even zero. But as the latter could actually be non-negligible, and is currently very uncertain, we represent on the vertical axis its potential contribution to the proton spin. The colored areas show the correlated $1 \sigma$ constraints on these values coming from present data, and those that one would expect from the projected EIC measurements. The diagonal lines represent the combinations of low-$x$ and high-$x$ contributions for which the resulting OAM would be as large as the proton spin and parallel to it, vanishing, or exactly opposite. The EIC data would be able to discard at least one of these extreme scenarios, and perhaps two of them.

The quark contribution of OAM to the nucleon spin can further be isolated via the extraction of generalized parton distributions (GPDs). These are functions which relate the longitudinal momentum fraction of a parton ($x$) to its position in the transverse plane (impact parameter)~\cite{Diehl:2003ny}. As such, they are connected to the OAM of partons, which is expressed in Ji's relation~\cite{Ji:1996ek} connecting the total angular momentum of quarks to the second Mellin moment of two GPDs, $H$ and $E$,
\begin{equation}
J^q = \frac{1}{2}-J^g = \frac{1}{2}\int_{-1}^{1} x\, dx \{H^q(x,\xi=0,t=0)+E^q(x,\xi=0,t=0)\},
\end{equation}
where a similar relation holds for gluons. GPDs, which are discussed in more detail in Sec~\ref{part2-subS-SecImaging-GPD3d}, are experimentally accessible in exclusive processes at low four-momentum square transfer to the nucleon, $|t|$, 
and high four-momentum square transfer to the struck parton, $Q^2$. This typically results in the production of a high-energy photon (in deeply-virtual Compton scattering, DVCS) or meson (in hard exclusive meson production), although other processes are possible --- kinematic studies of these are presented in Sec.~\ref{part2-sec-DetReq.Excl}. The variable $\xi$ encodes half of the parton's longitudinal momentum-fraction change, as a result of the scattering. 
Ji's spin decomposition allows one to identify the OAM contribution of quarks by subtracting the known contribution of the quark spin, $\frac{1}{2} \Delta \Sigma$, from $J^q$. 
Whether the OAM of gluons can be defined through $J^g - \Delta G$ is sometimes controversially discussed.
A comprehensive presentation of the spin decompsition of Ji and the one of Jaffe-Manohar can be found in Ref.~\cite{Leader:2013jra}.

The range of $x$ accessible at the EIC, combined with its high luminosity, will enable the GPDs $H$ and $E$ to be dramatically constrained --- an impact study for the GPD $E$ can be seen in Sec.~\ref{part2-subS-SecImaging-GPD3d}. While $H$ is fairly well-known in the valence region, determined mainly from the fixed-target experiments at JLab, $E$ is almost entirely unknown. Both GPDs are virtually unmapped in the low-$x$ region accessible at the EIC. Different observables have different sensitivity to the GPDs, and measurements from multiple processes are needed for their flavour separation. Thus DVCS on the proton and neutron will allow for the extraction of the GPDs for $u$- and $d$-quarks, which can also be obtained via the hard exclusive production of different mesons. Meson production is sensitive, at leading-twist and order, to gluon GPDs. The program of exclusive measurements at the EIC will enable, for the first time, OAM contributions from different quark flavors  as well as the total contribution of gluons to the nucleon spin to be determined, providing crucial insights into what has been long known as the ``proton spin puzzle''.

\subsubsection*{Parity-violating DIS} 

\begin{figure}[h]
    \centering
    \includegraphics[width=0.90\textwidth]{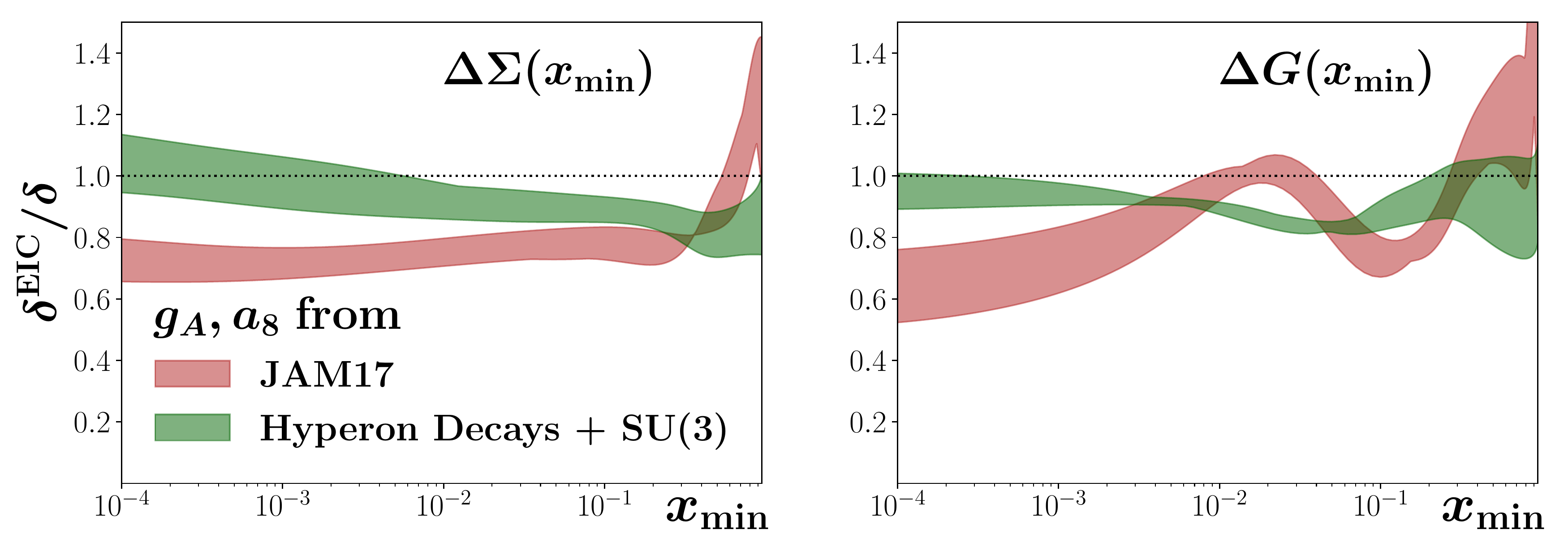}
    \caption{
        {\small 
        Ratio of uncertainties on the truncated moments of the quark singlet (left) and gluon (right) PDFs as functions of $x_{\rm min}$, including EIC data on the parity-violating DIS asymmetry $A_{\rm PV}^{\rm had}$ to those without EIC data, at $Q^2=10$~GeV$^2$.  Results with values of $g_A$ and $a_8$ taken from JAM17~\cite{Ethier:2017zbq} (red) are compared with those using values taken from hyperon decays and SU(3) (green).
        }}
    \label{fig:APVhad}
\end{figure}

Parity-violating DIS asymmetries with unpolarized electrons and polarized hadron beams can in principle provide additional constraints on the spin-dependent PDFs due to its unique flavor sub-processes. In Fig.~\ref{fig:APVhad} we present the impact of $A_{\rm PV}^{\rm had}$ at the EIC on the truncated moments of $\Delta\Sigma$ and $\Delta g$, assuming a proton beam with 100~fb$^{-1}$ integrated luminosity and uncorrelated systematic uncertainties from the pion background.  We find that the results depend upon the uncertainties of the triplet and octet axial charges, $g_A$ and $a_8$.  With values taken from the JAM17~\cite{Ethier:2017zbq} analysis, we see an impact of  $\sim 30\%$ for the quark singlet for low $x_{\rm min}$, which is significantly diminished, however, if one uses values from hyperon decays under SU(3) symmetry.  Similarly, we see a diminished impact on the gluon moment, especially at low $x_{\rm min}$.  This observable is limited by large statistical uncertainties, and thus higher luminosities would lead to a larger impact.  We note that the inclusion of the EIC data guarantees smaller uncertainties for the observable but not necessarily for all PDF flavors at all values of $x$.

\subsubsection*{Sea quark helicities via SIDIS }

\begin{figure}[t]
    \centering
    \includegraphics[width=0.9\textwidth]{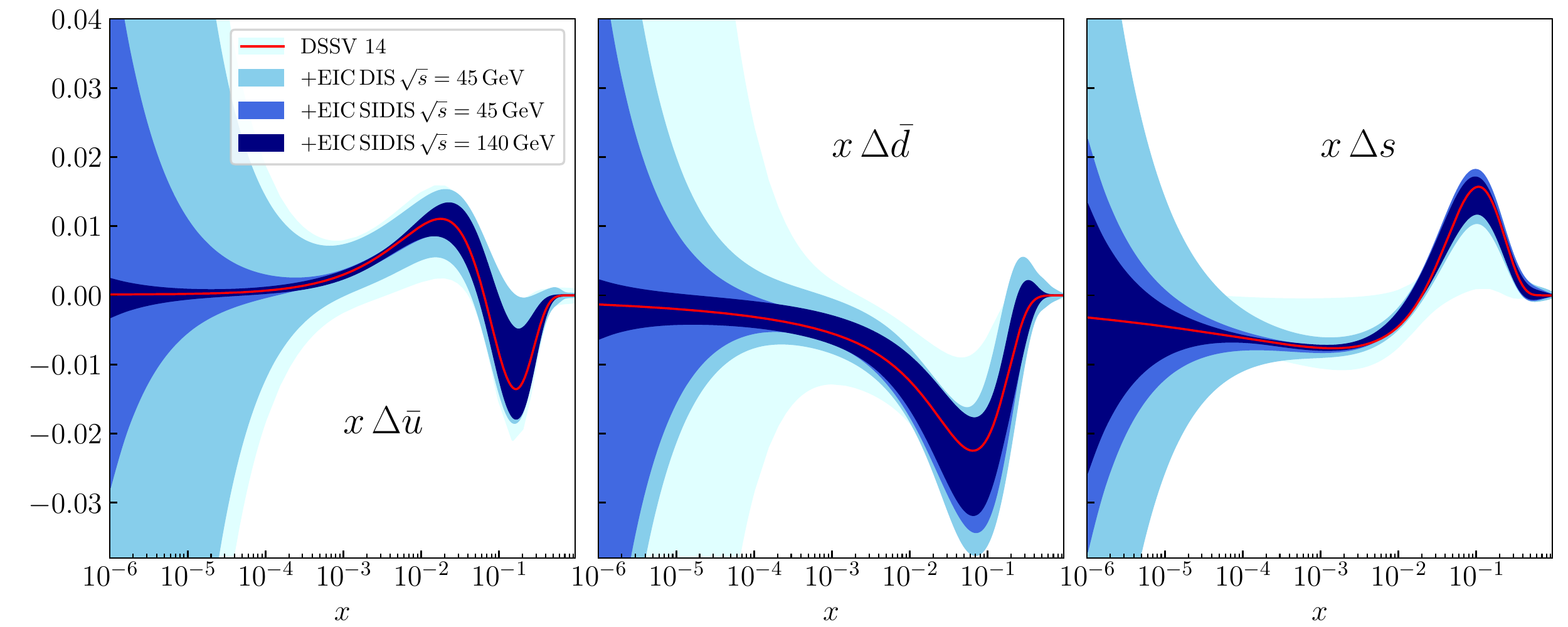}
    \caption{
        {\small Impact of SIDIS measurements at the EIC on the sea quark helicities $x \Delta \bar{u}$, $x \Delta \bar{d}$ and $x \Delta s$ as a function of $x$ at $Q^2 = 10$ GeV$^2$.}}
    \label{fig:part2-sec-PartStruct.deltaubar}
\end{figure}

The sensitivity on the struck parton that fragmentation functions provide can be used to leverage the understanding of the helicity structure of the nucleon --- see also Sec.~\ref{part2-subS-Hadronization-HadVacuum} concerning the fragmentation functions themselves.
In particular, the access to the sea quark helicities can be substantially improved over inclusive DIS measurements via SIDIS measurements that detect pions and kaons in addition to the scattered lepton.
Detailed impact studies that use {\sc PEPSI} as polarized MC generator and follow the previous DSSV~\cite{deFlorian:2009vb,deFlorian:2014yva,deFlorian:2019zkl} extractions have been performed on the expected EIC measurements using various collision energies and polarized proton as well as $^3$He beams~\cite{Aschenauer:2020pdk}. As can be seen in Fig.~\ref{fig:part2-sec-PartStruct.deltaubar}, the reduction in the uncertainties of all three sea quark helicities ($\Delta \bar{u}, \, \Delta \bar{d}, \, \Delta s$) in comparison to the current level of understanding is substantial. Similar to the gluon polarization, the highest impact at low $x$ relates to the data at the highest collision energies while intermediate to higher $x$ receive the biggest improvements already from the lower collision energies.  
One of the most important points that can be answered with the sea quark helicities are their contributions to the spin sum rule. In particular, the strange sea polarization is in current fits forced to negative values at lower $x$ due to the hyperon beta-decay constants and the assumption of $\mathrm{SU}(3)$-flavor symmetry in conjunction with no indication of a negative polarization in the $x$-range covered in the currently existing data~\cite{Airapetian:2004zf,Alekseev:2010ub}. The EIC SIDIS data will conclusively answer whether there is a nonzero strange polarization at $x > 0.5 \times 10^{-5}$.  
Further studies using similar pseudodata together with a re-weighting technique on the 
NNPDFpol~\cite{Nocera:2014gqa,Nocera:2017wep} replicas come to similar conclusions about the improvements to the sea quark helicities~\cite{Desai:2020}.

\subsubsection*{$\Delta G$ from dijet $A_{LL}$}

\begin{figure}[t!]
	\centering
	\includegraphics[trim={0 4cm 0 0},clip, width=\textwidth]{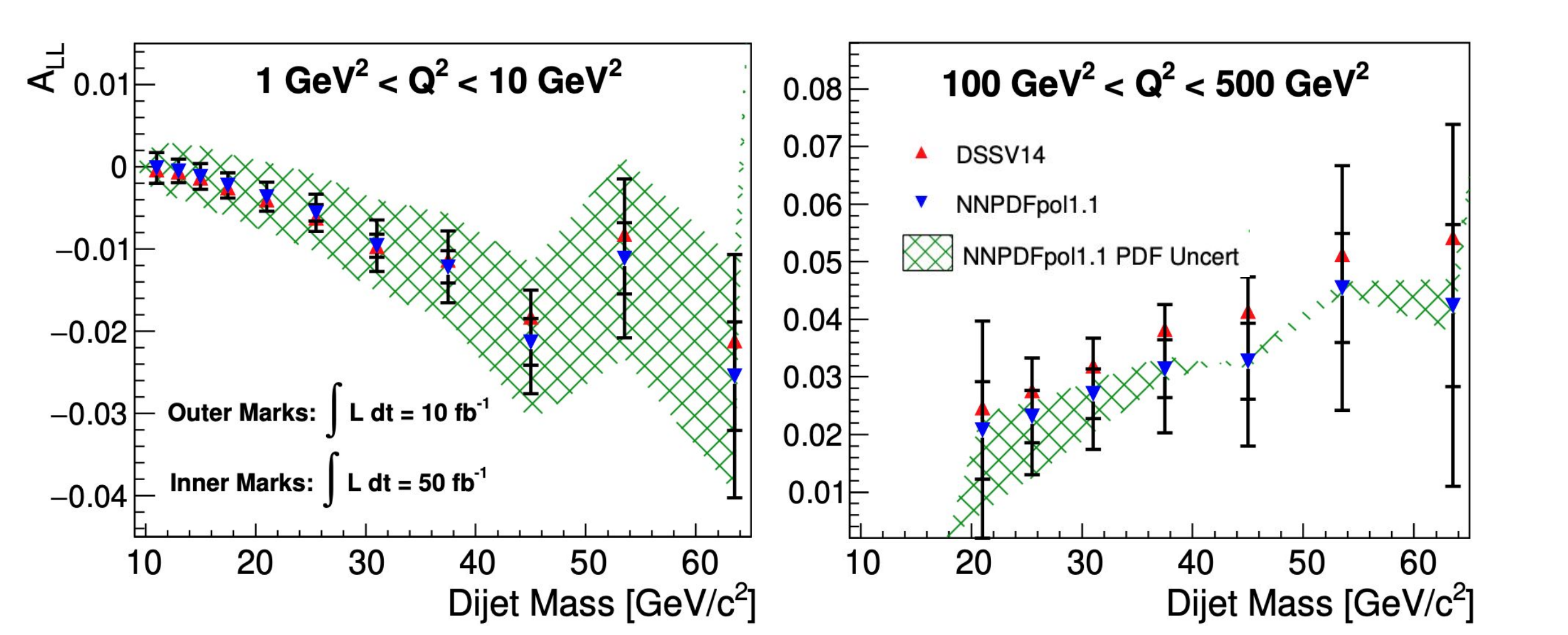}
	\caption[Dijet $A_{LL}$]{
	    {\small Dijet $A_{LL}$ as a function of dijet invariant mass for the combined QCD-Compton and PGF subprocesses using the DSSV14 and NNPDF1.1 polarized PDFs for $Q^2$ intervals of 1-10~GeV$^{2}$ (left) and 100-500~GeV$^{2}$ (right). Note that projected statistical uncertainties for the DSSV14 points are not shown for clarity, but are nearly identical to those from NNPDF1.1. (Figure from Ref.~\cite{Page:2019gbf}.)}}
	\label{fig:DIJETALL}
\end{figure}

As mentioned elsewhere in this section, the golden channel for the determination of $\Delta G$ at the  EIC will be the $Q^2$-variation of the inclusive $g_1$ structure function~\cite{Aschenauer:2015ata}. 
However, higher-order processes such as photon-gluon fusion (PGF) will provide direct access to the gluon and serve as an important cross-check to the inclusive result. 
A signature of the PGF process is the production of back-to-back partons with large momentum transverse to the virtual quark-parton axis. 
Therefore, detecting dijets in the Breit frame can be used to tag PGF events. 
A feasibility study was recently conducted~\cite{Page:2019gbf} which confirmed the viability of dijet reconstruction as a tag of PGF events, and also demonstrated the ability to use the dijet kinematics to reconstruct a number of relevant partonic quantities, such as the momentum fraction of the struck gluon. 
An estimation of the size of the expected dijet longitudinal double-spin asymmetry $A_{LL}$, which is sensitive to $\Delta G$, and associated statistical uncertainties was also performed following the procedure in Ref.~\cite{Chu:2017mnm} and compared to the uncertainties on the NNPDFPol1.1 polarized PDF~\cite{Nocera:2014gqa} as shown in Fig.~\ref{fig:DIJETALL}. 
While the expected statistical precision on $A_{LL}$ given a moderate amount of integrated luminosity would improve on our present knowledge of the polarized PDFs, it is likely that the inclusive $g_1$ measurement would provide superior constraining power. 
Nevertheless, the dijet $A_{LL}$ measurement will be important as a cross-check to the inclusive measurement as it arises from a different subprocess and will have different experimental uncertainties.

\subsubsection*{$\Delta G$ from heavy-quark $A_{LL}$} 
\begin{figure}
\centering
\begin{minipage}{.5\textwidth}
  \centering
  \includegraphics[width=1.0\linewidth]{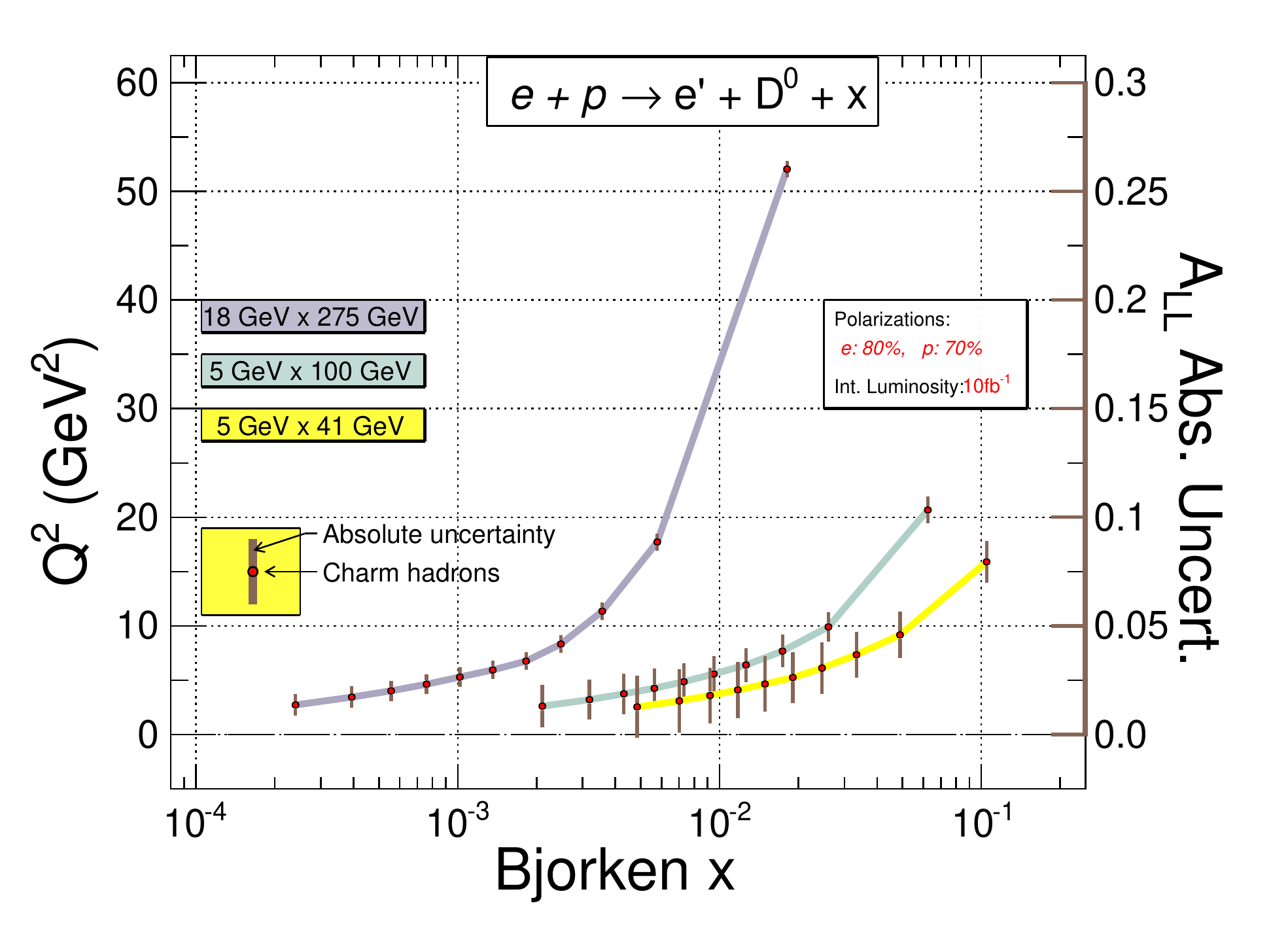}
\end{minipage}%
\begin{minipage}{.5\textwidth}
  \centering
  \includegraphics[width=1.0\linewidth]{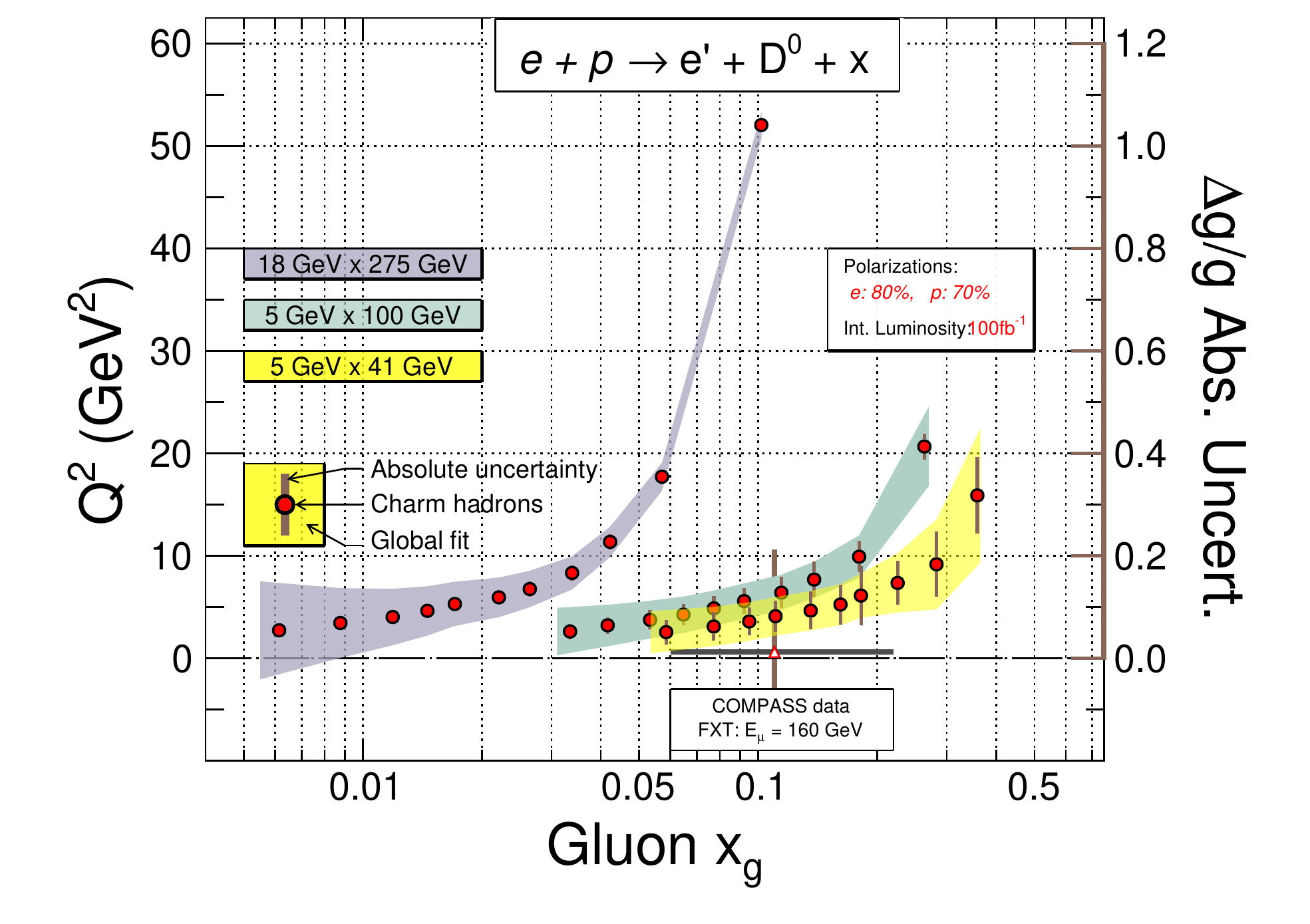}
\end{minipage}
\caption{
	    {\small Left: The longitudinal double-spin asymmetry $A_{LL}$ in semi-inclusive DIS production of $D^0$ mesons at $x$ and $Q^2$ of the measurements with the projected size of its statistical uncertainties. 
	    Right: Illustrated precision, kinematic coverage, and leading-order impact compared with the prior data point by the COMPASS collaboration and theory evaluation with their uncertainties as bands.}}
\label{fig:DGG}
\end{figure}

Heavy quarks are versatile probes for studying different aspects of QCD and nucleon structure. 
An example of their potential impact on EIC physics can be found in the measurement of the gluon polarization $\Delta g(x,Q^2)$.
The most precise insights on $\Delta g(x,Q^2)$ at the EIC will come indirectly from scaling violations of the inclusive structure function $g_1(x,Q^2)$.  Heavy-quark production forms a considerable contribution to the (polarized) cross section in lepton-nucleon DIS and hence to the corresponding structure functions.  Heavy quark production is dominated by gluon-induced processes already at the Born level in pQCD.  In the case of lepton-nucleon DIS, only the photon-gluon fusion process contributes at Born level, which makes heavy-quark production particularly sensitive to the gluon distributions.  The data on the charm contribution to the structure function $F_2(x,Q^2)$ from HERA, for example, are used in most global analyses of unpolarized PDFs.  Corresponding measurements of longitudinally polarized DIS at the EIC will provide insights on $\Delta g(x,Q^2)$.   The COMPASS collaboration at CERN has pioneered such a measurement.  To assess the potential and impact for the EIC, studies were performed of the semi-inclusive DIS production of single $D^0$ mesons and their subsequent decay into the charged $K\pi$ branch. These studies were based on PYTHIA-eRHIC and EIC fast simulations of detector response extended with vertex fitting and including selections based on topological reconstruction from the smeared (decay) tracks~\cite{Arrington:2021yeb}.
Figure~\ref{fig:DGG} (left) shows the contribution $g_1^{c\bar{c}}(x,Q^2)$ expressed as the asymmetry $A_{LL}$ at Bjorken-$x$ and $Q^2$ of the measurements with the size of their statistical uncertainties for the nominal EIC beam polarizations and an integrated luminosity of 10\,$\mathrm{fb}^{-1}$. The pQCD contributions to the inclusive heavy-quark production in polarized DIS are known~\cite{Hekhorn:2018ywm} at next-to-leading order.  Impact studies on equal footing with those of $g_1(x,Q^2)$ are thus in principle possible in global analyses.  However, they are not available as this report is being written.  Instead, the impact of future EIC data was assessed in a leading-order approach following that of the pioneering COMPASS determination~\cite{Adolph:2012ca}.
This impact is illustrated in Fig.~\ref{fig:DGG} (right), which shows the leading-order $A_{LL} = \Delta g(x,Q^2)/g(x,Q^2)$ versus gluon-$x$ for an integrated luminosity of 100\,$\mathrm{fb}^{-1}$ together with the prior COMPASS data and $\Delta g(x,Q^2)/g(x,Q^2)$ based on the NNPDF (polarized) parton distributions.

\subsection{Parton structure of mesons}
\label{part2-subS-PartStruct.M}



\subsubsection{Introduction}

The EIC, with its high luminosity and wide kinematic range, offers an extraordinary new opportunity to increase our knowledge of the pion and kaon structure~\cite{Aguilar:2019teb}. The properties of pions and kaons provide clear windows onto emergent hadronic mass (EHM) and its modulation by Higgs-boson interactions (see also Sec.~\ref{part2-subS-PartStruct-Mass}).  For an expanded discussion of the material contained in this section, we refer to~\cite{Arrington:2021biu}.

To facilitate this discussion, we translate current theoretical understanding of the light meson structure and the EHM (and structure) mechanisms into a set of critical science questions. 
They represent outstanding mysteries that require further experimental (and theoretical) examination, and illustrate the impact of a coherent study of pion and kaon structure yielding results similar to present studies of proton structure.
In Tab.~\ref{tab:MS_EIC_science_questions_table} we present the key science questions along with specific measurements and their experimental needs required to advance our understanding. 

\begin{figure}[ht]
\begin{center}
\includegraphics[width=0.25\textwidth]{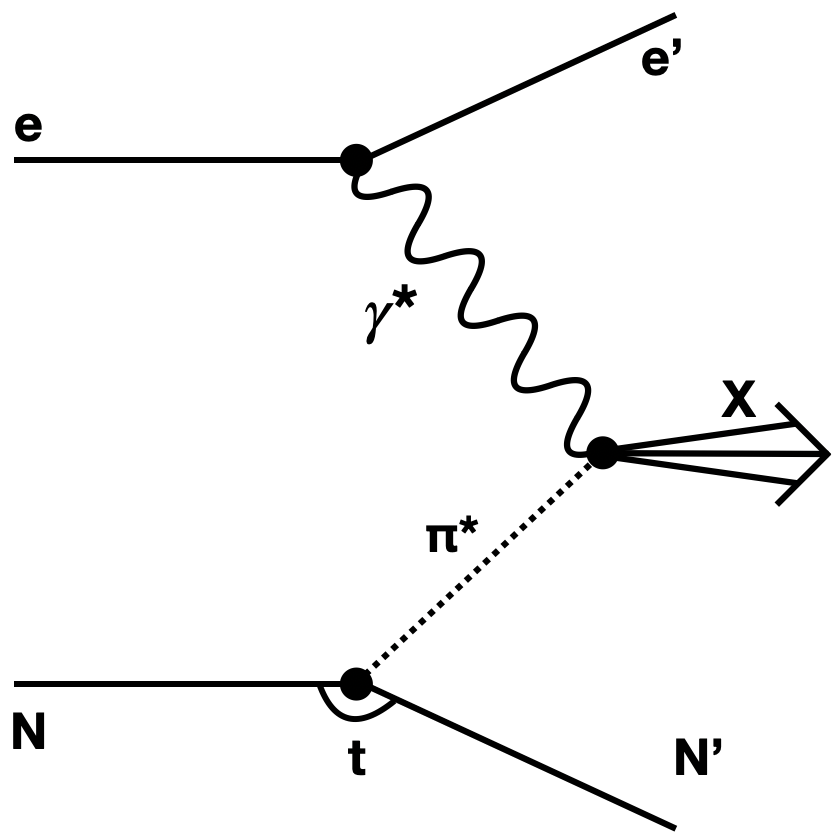}
    \caption{\label{fig:sullivan}
Diagram for the Sullivan process used to probe the structure of the pion.
}
\end{center}
\end{figure}

For all observables, a luminosity well above 10$^{33} \lumi$ is required to compensate for the (few times) 10$^{-3}$ fraction of the proton wave function related to the pion (kaon) Sullivan process --- see the diagram in Fig.~\ref{fig:sullivan}. Also, a large range in momentum fraction of the tagged nucleon $x_L=p^+_{N'}/p^+_N$ is required, up to $x_L \sim$ 1 for the $\ep$ reactions and $x_L$ at least $\sim$ 0.5 for the $\eD$ reactions. In addition to the $\pi^+$ channels listed in Tab.~\ref{tab:MS_EIC_science_questions_table}, data on $\pi^-$-channels (e.g. $\eD \to e^\prime + X + p + p$) and on  $\pi^0$-channels (e.g. $\ep \to e^\prime + X + p$) are crucial to constrain reaction mechanisms and theory backgrounds in extracting the physical pion (kaon) target information.


\begin{table}[htb]
\caption{Science questions related to pion and kaon structure and the understanding of the EHM mechanism accessible at the EIC, with the key measurements and some key requirements listed. Further requirements are addressed in the text.}
\tiny
\begin{tabular}{p{4.5cm} p{4cm} p{5cm}}   
\hline
\hline
\textbf{Science Question}  & \textbf{Key Measurement}    & \textbf{Key Requirements}   \\ \hline
\multirow{4}{\linewidth}{What are the quark and gluon energy contributions to the pion mass?}                                                   & \multirow{4}{\linewidth}{Pion structure function data over a range of $x$ and $Q^2$.}                      & \multirow{4}{\linewidth}{\begin{tabular}[c]{p{6.7cm}}$\bullet$ Need to uniquely determine\\\qquad $\ep \to e^\prime + X + n$ (low $-t$)\\$\bullet$ CM energy range $\sim$10-100 GeV\\ $\bullet$ Charged and neutral currents desirable\end{tabular}} \\
 &  & \\
 &  & \\
 &  & \\ \hline
\multirow{2}{\linewidth}{Is the pion full or empty of gluons as viewed at large $Q^2$?}                                                            & \multirow{2}{\linewidth}{Pion structure function data at large $Q^2$.}                                       & \multirow{2}{\linewidth}{\begin{tabular}[c]{p{4.3cm}}$\bullet$ CM energy $\sim$100 GeV\\$\bullet$ Inclusive and open-charm detection\end{tabular}}  \\
&  & \\ \hline
\multirow{3}{\linewidth}{What are the quark and gluon energy contributions to the kaon mass?}                                                   & \multirow{3}{\linewidth}{Kaon structure function data over a range of $x$ and $Q^2$.}                      & \multirow{3}{\linewidth}{\begin{tabular}[c]{p{5cm}}$\bullet$ Need to uniquely determine \\\qquad $\ep \to e^\prime + X + \Lambda /\Sigma^0$ (low $-t$)\\$\bullet$ CM energy range $\sim$10-100 GeV\end{tabular}}                                                                            \\
 &  & \\
 &  & \\ \hline
\multirow{2}{\linewidth}{Are there more or less gluons in kaons than in pions as viewed at large Q$^2$?}                                           & \multirow{2}{\linewidth}{Kaon structure function data at large $Q^2$.}                                        & \multirow{2}{\linewidth}{\begin{tabular}[c]{p{5cm}}$\bullet$ CM energy $\sim$100 GeV\\$\bullet$ Inclusive and open-charm detection\end{tabular}}                                                               \\
 &  & \\ \hline
\multirow{4}{\linewidth}{Can we get quantitative guidance on the emergent pion mass mechanism?}                                                    & \multirow{4}{\linewidth}{\makecell[l]{Pion form factor data \\for $Q^2$ = 10-40 (GeV/$c$)$^2$.}}                & \multirow{4}{\linewidth}{\begin{tabular}[c]{p{4.5cm}}$\bullet$ Need to uniquely determine exclusive process \\\qquad $\ep \to e^\prime + \pi^+ + n$ (low $-t$)\\$\bullet$ $\ep$ and $\eD$ at similar energies\\$\bullet$ CM energy $\sim$10-75 GeV\end{tabular}}                                          \\
 &  & \\
 &  & \\
 &  & \\ \hline
\multirow{4}{\linewidth}{What is the size and range of interference between emergent-mass and the Higgs-mass mechanism?}                        & \multirow{4}{\linewidth}{\makecell[l]{Kaon form factor data \\for $Q^2$ = 10-20 (GeV/$c$)$^2$.}}               & \multirow{4}{\linewidth}{\begin{tabular}[c]{p{4.5cm}}$\bullet$ Need to uniquely determine exclusive process \\\qquad $\ep \to e^\prime + K + \Lambda$ (low $-t$)\\$\bullet$  L/T separation at CM energy $\sim$10-20 GeV\\$\bullet$  $\Lambda /\Sigma^0$ ratios at CM energy $\sim$10-50 GeV\end{tabular}}            \\
 &  & \\
 &  & \\
 &  & \\ \hline
\multirow{3}{\linewidth}{What is the difference between the impacts of emergent- and Higgs-mass mechanisms on light-quark behavior?}            & \multirow{3}{\linewidth}{Behavior of (valence) up quarks in pion and kaon at large $x$.}                     & \multirow{3}{\linewidth}{\begin{tabular}[c]{p{4.5cm}}$\bullet$  CM energy $\sim$20 GeV (lowest CM energy to access large-x region)\\$\bullet$  Higher CM energy for range in $Q^2$ desirable\end{tabular}} \\
 &  & \\
 &  & \\ \hline
\multirow{3}{\linewidth}{What is the relationship between dynamically chiral symmetry breaking and confinement?}                     & \multirow{3}{\linewidth}{Transverse-momentum dependent Fragmentation Functions of quarks into pions and kaons.} 
                                    & \multirow{3}{\linewidth}{\begin{tabular}[c]{p{4.5cm}}$\bullet$ Collider kinematics desirable (as compared to fixed-target kinematics)\\$\bullet$  CM energy range $\sim$20-140 GeV\end{tabular}}\\                   
& & \\
& & \\ 
\specialrule{.1em}{.05em}{.05em}
\multicolumn{3}{l}{\textbf{More speculative observables}}                                                      \\ 
\hline
\multirow{4}{\linewidth}{What is the trace anomaly contribution to the pion mass?}                                                              & \multirow{4}{\linewidth}{Elastic $\jpsi$ production at low $W$ off the pion.}                                  & \multirow{4}{\linewidth}{\begin{tabular}[c]{p{4.5cm}}$\bullet$ Need to uniquely determine exclusive process\\\qquad  $\ep \to e^\prime + \jpsi + \pi^+ + n$ (low $-t$)\\$\bullet$  High luminosity ($\geq 10^{34} \lumi$)\\$\bullet$  CM energy $\sim$70 GeV\end{tabular}}                                               \\
& & \\
& & \\
& & \\ \hline
\multirow{4}{\linewidth}{Can we obtain tomographic snapshots of the pion in the transverse plane? What is the pressure distribution in a pion?} & \multirow{4}{\linewidth}{Measurement of DVCS off pion target as defined with Sullivan process.}                 & \multirow{4}{\linewidth}{\begin{tabular}[c]{p{4.5cm}}$\bullet$ Need to uniquely determine exclusive process \\\qquad $\ep  \to e^\prime + \gamma + \pi^+ + n$ (low $-t$)\\$\bullet$  High luminosity ($\geq 10^{34} \lumi$)\\$\bullet$  CM energy $\sim$10-100 GeV\end{tabular}}                                             \\
& & \\
& & \\
& & \\ \hline
\multirow{5}{\linewidth}{Are transverse momentum distributions universal in pions and protons?}                                                 & \multirow{5}{\linewidth}{Hadron multiplicities in SIDIS off a pion target as defined with Sullivan process.}    & \multirow{5}{\linewidth}{\begin{tabular}[c]{p{6.3cm}}$\bullet$ Need to uniquely determine SIDIS off pion\\\qquad  $\ep \to e^\prime + h + X + n$ (low $-t$)\\$\bullet$  High luminosity (10$^{34} \lumi$)\\$\bullet$ $\ep$ and $\eD$ at similar energies desirable\\$\bullet$  CM energy $\sim$10-100 GeV\end{tabular}} \\
& & \\
& & \\
& & \\
& & \\ 
\hline
\hline
\end{tabular}%

\label{tab:MS_EIC_science_questions_table}
\end{table}

\subsubsection{Sullivan process
\label{sec:713_sullivan}}
In specific kinematic regions, the observation of recoil nucleons ($N$) or hyperons ($Y$) in the tagged inclusive reaction $\ep \to e^\prime + X + (N\,{\rm or}\,Y)$ (see Fig.~\ref{fig:sullivan}) can reveal features associated with correlated quark-antiquark pairs in the nucleon, referred to as the ``meson cloud'' or ``five-quark component'' of the nucleon.  At low values of $-t$ (with $t$ the four-momentum transfer squared from the initial proton to the final nucleon or hyperon), the cross section displays behavior characteristic of meson-pole dominance and is referred to as the Sullivan process~\cite{Sullivan:1971kd}. For elastic scattering, this process carries information on the meson (pion or kaon) form factor. For DIS, the typical interpretation is that the nucleon PDFs contain a mesonic parton content through scattering from a meson target. An important development in establishing a formal link between the Sullivan process and QCD came with the realization that the moments of PDFs could be expanded in chiral effective field theory 
in terms of power series in the pion mass~\cite{Thomas:2000ny, Chen:2001nb, Detmold:2001jb}. 
While the total pion-cloud contribution contains short-distance contributions, which are model-dependent and must be fitted to data, the infrared behavior is model-independent and can only arise from a pionic component in QCD.

The Sullivan process can provide reliable access to a meson target in the space-like $t$-region, if the pole associated with the ground-state meson remains the dominant feature of the process and the structure of the related correlation evolves slowly and smoothly with virtuality. To check whether these conditions are satisfied empirically, one can take data covering a range in $t$, particularly low $-t$, and compare with phenomenological and theoretical expectations. A recent calculation explored the circumstances under which these conditions should be satisfied~\cite{Qin:2017lcd}. According to this study, for the pion (kaon) Sullivan process, low $-t$ equates to $-t <$ 0.6 (0.9) GeV$^2$ to be able to extract the pion (kaon) structure.  Substantial further theory input is required to solidify these numbers and data over a range of $-t$ down to the lowest accessible values are needed to verify the pion (kaon) structure extraction.

\subsubsection{Theoretical backgrounds in extracting the data}

The extraction of the mesonic structure of the nucleon from the tagged DIS cross section is inherently model-dependent. Therefore, it will be necessary to examine all available reasonable models 
to evaluate the theoretical uncertainty associated with extracting meson structure functions from the tagged DIS data. The measured cross section can be integrated over $t$ to obtain the leading-baryon
structure function introduced as $F_{2}^{LB(3)}$ in Ref.~\cite{Adloff:1998yg}. The pion structure function $F_{2}^{\pi}$ can then be extracted from $F_{2}^{LB(3)}$ using models, such as the Regge model of baryon production. 

The extraction of the pion structure function will have to be corrected for a number of effects beyond the simple Sullivan picture. These include non-pion-pole contributions, $\Delta$ and other $N^*$ resonances, absorptive effects, and uncertainties in the pion flux. 
While these corrections can be large and one cannot extract the pion structure function without including them, detailed calculations do exist~\cite{Kopeliovich:1996iw}. 
(A recent estimate of the absorptive effects was presented in Ref.~\cite{Carvalho:2020myz}.) 
Moreover, these corrections are minimized by measuring at the lowest $-t$, and having fine differential binning in $-t$.  A quantitative assessment of the desired resolution and binning in $-t$ needs future study. We note that the simulations of Fig.~\ref{fig:MC_fpi_t_10on135} result in a bin size in $-t$ of 0.1~$\gev^2$.
Having data from both protons and deuterons will provide essential cross checks for the models used in the extraction of the pion structure function, with different leading trajectories in the proton and neutron case, or the isospin 0 or 1 exchange~\cite{GolecBiernat:1997vy,Kazarinov:1975kw,Kopeliovich:1996iw}. 

The measured tagged cross sections and extracted tagged structure functions can be analyzed within a Regge framework, assuming the dominance of a single Regge exchange.
As pion exchange results in a different $x_L$-dependence of the cross section, it should be possible to determine the dominant exchange mechanism(s) by comparing the $x_L$-dependence from proton and neutron (deuteron) scattering.
The largest uncertainty in extracting the pion structure function will, however, likely arise from the (lack of) knowledge of the pion flux in the framework of the pion cloud model~\cite{DAlesio:1998uav,Stoks:1999bz}.
If we assume that all corrections can be performed with a 50\% uncertainty, and we assume a 20\% uncertainty in the pion flux factor, the overall theoretical, systematic uncertainty could approach $25\%$. The superior approach is to have a direct measurement of the pion flux factor by comparing to pionic Drell-Yan data --- see data from Refs.~\cite{Conway:1989fs,Drell:1969km} and future COMPASS data. 
On the other hand, we know that there must be a region at small $- t$ and large $x_L$ where the cross section should be dominated by soft pions and hence the dependence on the pion flux is minimal. In the context of a global QCD analysis (see below), one can fit the pion structure function at the same time as determining empirically the boundaries of the region of $- t$ and $x_L$ over which the pion exchange mechanism is the dominant one \cite{McKenney:2015xis, Barry:2018ort}.

\subsubsection{Kinematics of interest to address specific theory questions}

The science questions of interest of Tab.~\ref{tab:MS_EIC_science_questions_table} require a range of physics processes.
In general, a large range of center-of-mass energies is needed to access a wide range in $x$ and $Q^2$, as relevant for pion (kaon) structure function measurements or hadron-multiplicity measurements for a transverse-momentum dependent parton distribution program. This has to be balanced against the requirement to uniquely determine the remnant neutron (or $\Lambda$ or $\Sigma^0$) to ensure the scattering process occurs off a pion (kaon). The latter favors not-too-high center-of-mass energies to be able to determine the remnant $\Lambda$ (or $\Sigma^0$), both for missing-mass determination and to ensure that their decays occur before detection.  In addition, there is need for both $\ep$ and $\eD$ measurements at similar center-of-mass energies to validate the reaction mechanism. This drives the ``typical'' center-of-mass energies for pion and kaon structure function measurements to a $\sim 10-100$~GeV range. On the other hand, to access the largest-$x$ region, in order to address the valence quarks in pions (or kaons), the lowest center-of-mass energy to reach a sufficiently high and ``clean'' $Q^2$ level has the highest Figure-Of-Merit folding in all kinematic effects. 
Higher center-of-mass energies will increase the range in $Q^2$.

\subsubsection{Meson structure function projections}

\begin{figure}[ht]
\includegraphics[width=0.47\textwidth]{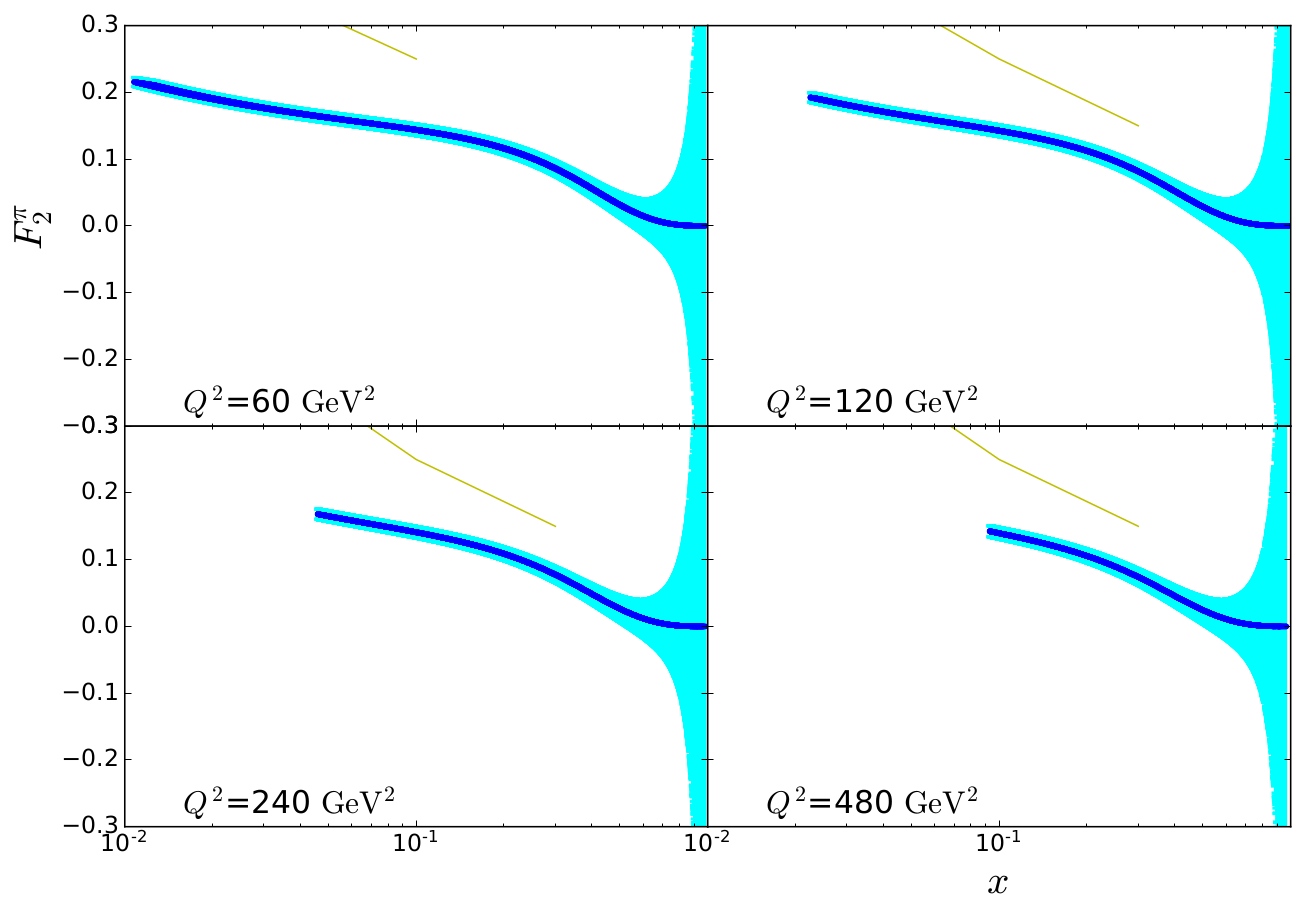}
\hspace*{0.20cm}
\includegraphics[width=0.47\textwidth]{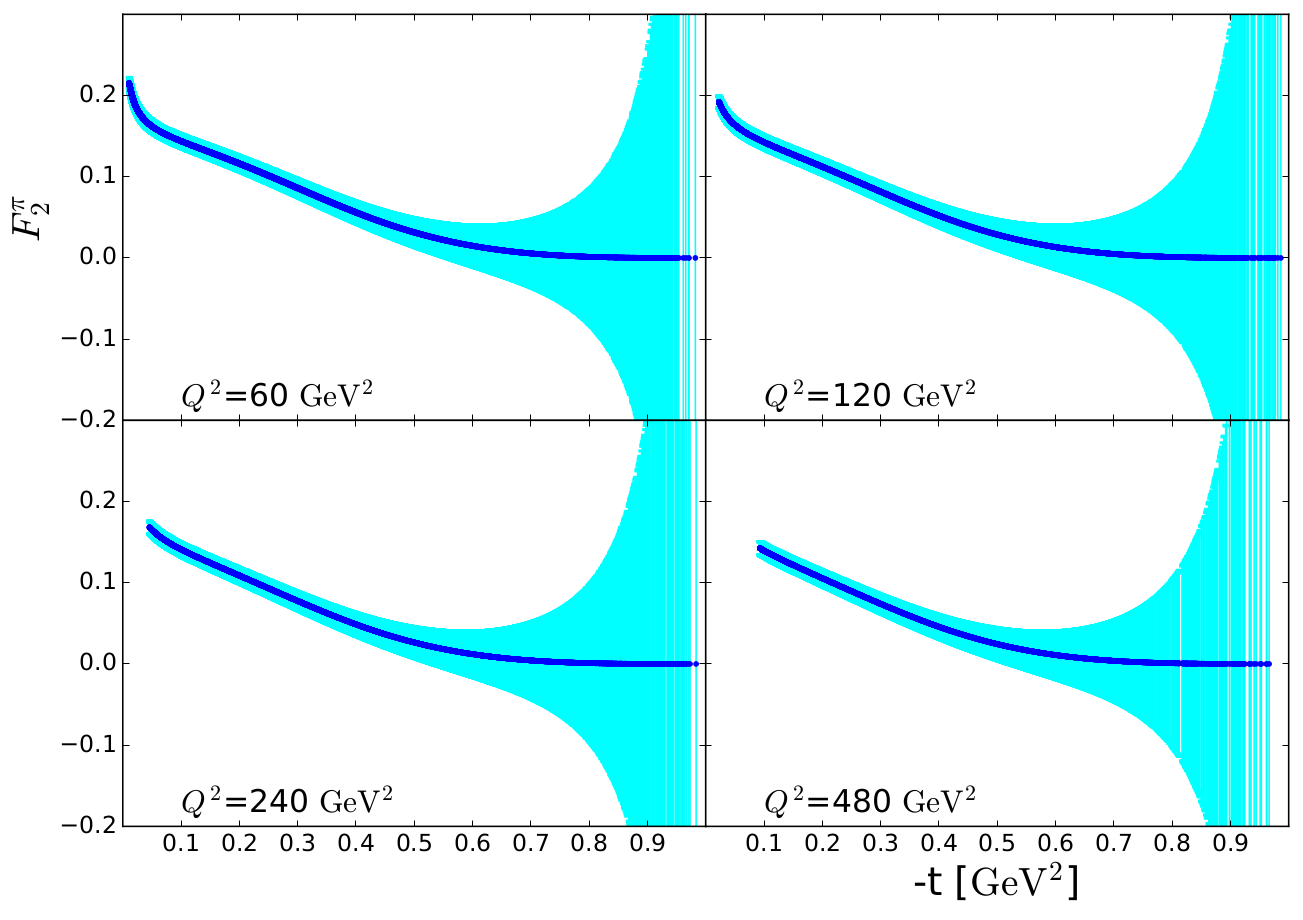}
    \caption{\label{fig:MC_fpi_t_10on135}
The Monte Carlo projections of the pion structure function vs $x$ (left) and $-t$ (right) for 10 $\gev$ electrons on 135 $\gev$ protons. The blue points are the Monte Carlo projections for $Q^2$ values of 60, 120, 240, 480~$\gev^2$. The projected data is binned in $x$ and $Q^2$ with bin sizes of 0.001 and 10 $\mathrm{GeV}^2$, respectively.  Each of these bins corresponds to one effective $t$-value on the right panel.  The blue shaded zones are the statistical uncertainties for an integrated luminosity of 100~$\fb^{-1}$. The brown line in the left figure is a GRV fit for similar pion structure function projections~\cite{Chekanov:2002pf}.
}
\end{figure}

A C++ and ROOT-based custom Monte Carlo event generator~\cite{EIC_mesonMC} was used for feasibility studies for pion and kaon structure function measurements. 
The generator calls various quantities such as CTEQ6 PDF tables, nucleon structure functions, and the tagged pion and kaon structure functions and splitting functions.
The pion structure function can be parametrized in a multitude of ways. Here, the parametrization outlined in Ref.~\cite{Chekanov:2002pf} was used. 
This parametrization is a scaled version of the proton structure function ($F^{\pi}_2=0.361*F_2$) which allowed an easy comparison to the available data from the H1 Collaboration and to the Gl\"{u}ck-Reya-Vogt (GRV) theoretical fit (shown on the left in Fig.~\ref{fig:MC_fpi_t_10on135}). The agreement with the HERA data validates the simulations in that regime, and one can also see good agreement with the GRV fit at higher $x$.

The left plot in Fig.~\ref{fig:MC_fpi_t_10on135} shows the reach in $x$ for four $Q^2$ bins. Statistical uncertainties with the addition of the leading neutron detection fraction were incorporated in the overall uncertainty using the integrated luminosity $\mathcal{L} = 100 \fb^{-1}$. The $x$-coverage is immediately apparent as the plot shows a reach from mid to high $x$. The uncertainties are reasonable for mid to mid-high $x$ but, as expected, increase rapidly as $x\to1$. Even with these restrictions the $x$-coverage 
is unprecedented and should allow for a detailed comparison between the pion and kaon structure.

As discussed above, theoretical calculations predict that the Sullivan process should provide clean access to the meson structure below a certain value of $-t$~\cite{Qin:2017lcd}. For the pion this is $-t\leq0.6~\gev^2$. The right panel of Fig.~\ref{fig:MC_fpi_t_10on135} shows the accessible range in $-t$ at the EIC for 10 $\gev$ electrons on 135 $\gev$ protons with reasonable uncertainties which would allow for an order-of-magnitude gain in statistics compared to HERA. The resulting access to a significant range in $Q^2$ and $-t$, including small $-t$, as well as significant $x$-coverage, 
will provide insights into the gluonic content of the pion.

\subsubsection{Impact on global QCD analysis}

\begin{figure}[t]
\includegraphics[width=0.95\textwidth]{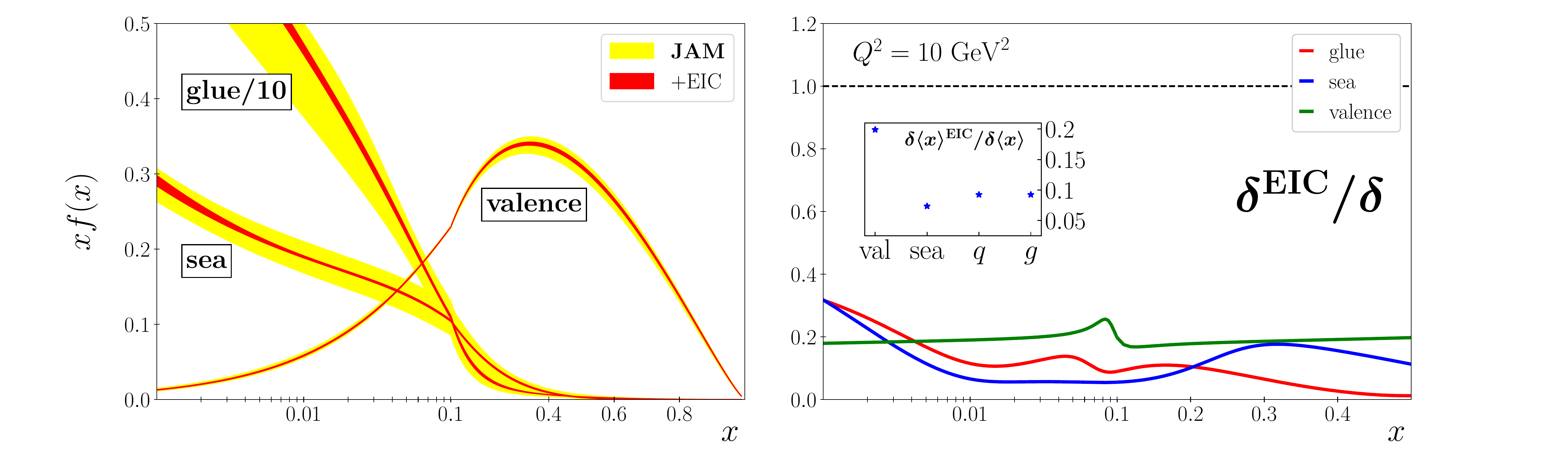}
\caption{
\label{fig:pion-pdf-impact}
    Left: Comparison of uncertainties on the pion valence, sea quark and gluon PDFs before (yellow bands) and after (red bands) inclusion of EIC data.
    Right: Ratio of uncertainties of the PDFs with EIC data to PDFs without EIC data, $\delta^{\rm EIC}/\delta$, for the valence (green line), sea quark (blue) and gluon (red) PDFs, assuming 1.2\% systematic uncertainty,
    and (inset) the corresponding ratios of the momentum fraction uncertainties, $\delta\langle x \rangle^{\rm EIC}/\delta\langle x \rangle$, for valence, sea, total quark and gluon PDFs~\cite{Barry-resummation-20}, at the scale $Q^2=10$~GeV$^2$.  Fits were obtained using a Monte Carlo procedure, using DGLAP at NLL with VFNS, NLL $\alpha_s$ and both Drell-Yan and $F_2$ for leading neutrons at NLO.}
\end{figure}

The potential impact of EIC neutron production data is illustrated in Fig.~\ref{fig:pion-pdf-impact}, which shows the valence, sea quark and gluon PDFs in the pion from the JAM global QCD analysis at the input scale $Q^2=10$~GeV$^2$~\cite{Barry:2018ort}, with current uncertainties compared with those expected with the addition of EIC data~\cite{Barry-resummation-20}.  At the moment this is the only impact study of its kind.
The analysis of the existing data includes pion-nucleus Drell-Yan cross sections, both $p_T$-differential and $p_T$-integrated, and the leading-neutron structure functions from HERA~\cite{Cao2020}, both treated at NLO.
The analysis assumes the center-of-mass energy $\sqrt{s}=73.5$~GeV, the integrated luminosity $\mathcal{L} = 100 \, \textrm{fb}^{-1}$ and a 1.2\% systematic uncertainty across all kinematics. 
This does not include an uncertainty coming from the model dependence of the extraction (see above).
For both the sea quark and gluon distributions, the PDF uncertainties are reduced by 
a factor $\sim 5-10$ for most of the range of $x$, with a (similar) factor $\sim 5$ reduction in the valence sector.
For the decomposition of the pion mass~\cite{Yang:2014xsa}, written in terms of matrix elements of the QCD energy momentum tensor (see Sec.~\ref{part2-subS-PartStruct-Mass}), the first moments, $\langle x \rangle_{q,g}$, are relevant.
For these quantities, the reduction in uncertainties is by a factor $\sim 10$ for both the total quark and gluon contributions, as can be seen in the inset of Fig.~\ref{fig:pion-pdf-impact} (right).
Note, however, that the errors do not include the aforementioned uncertainties associated with the model dependence of the pion flux, which may be of the order $10 - 20\%$~\cite{Aaron:2010ab, Chekanov:2002pf}, and 
could reduce the impact of the projected data on the uncertainties of the pion PDFs by several-fold.
A similar analysis may be performed for the PDFs in the kaon, which can be obtained from leading-hyperon production in the forward region.
In this case, the near-absence of empirical information on the parton structure of kaons will mean an even more striking impact of new EIC data. 

\subsubsection{Complementarity with other facilities}

The unique role of the EIC is its access to pion and kaon structure over a versatile large center-of-mass energy range, $\sim 29-141$~GeV. JLab will provide
tantalizing data for the pion form factor up to $Q^2 \sim$~10~GeV$^2$ ($Q^2 \sim$~5~GeV$^2$ for the kaon form factor), and measurements of the pion (kaon) structure functions at large $x$ ($> 0.5$) through the Sullivan process.

COMPASS++/AMBER~\cite{Denisov:2018unj} will play a crucial role as they can uniquely provide pion (kaon) Drell-Yan measurements in the center-of-mass energy region $\sim 10-20$~GeV. The phase 1 with pion Drell-Yan is planned in 2022 and later, phase 2 with kaon Drell-Yan no earlier than 2026. Some older pion and kaon Drell-Yan measurements exist, but for the kaon this is limited to less than 10 data points worldwide, so these measurements are absolutely important for a global effort of the pion structure function measurements 
(allowing a handle on the pion flux) 
and a \emph{sine qua non} for any kaon structure function data map. The COMPASS++/AMBER data in themselves will already give new fundamental insights in the EHM mechanism.

Lastly, an electron-ion collider in China (EicC) is under consideration with a similar center-of-mass energy range as COMPASS++/AMBER of $\sim 10-20$~GeV and bridging the energy range from JLab to EIC~\cite{Anderle:2021wcy}. The EicC on its own, and even more in combination with COMPASS++/AMBER, can provide good access to the region of $x \gtrsim$ 0.01 for pion, and especially kaon, structure function determination and the impact on EHM mechanisms on the valence quark and gluon structure. In addition, the EicC can extend the Rosenbluth L/T-separated cross section technique beyond JLab and access pion and kaon form factors to higher $Q^2$ values, roughly by a factor of 2-4.

In the end the EIC, with its larger center-of-mass energy range, will have the final word on the contributions of gluons in pions and kaons as compared to protons, settle how many gluons persist in pions and kaons as viewed with highest resolution, and vastly extend the range in $x$ and $Q^2$ of pion and kaon charts, and meson structure knowledge.

\subsubsection{Synergy with theory efforts}

\begin{figure}[htb]
\includegraphics[width=0.47\textwidth]{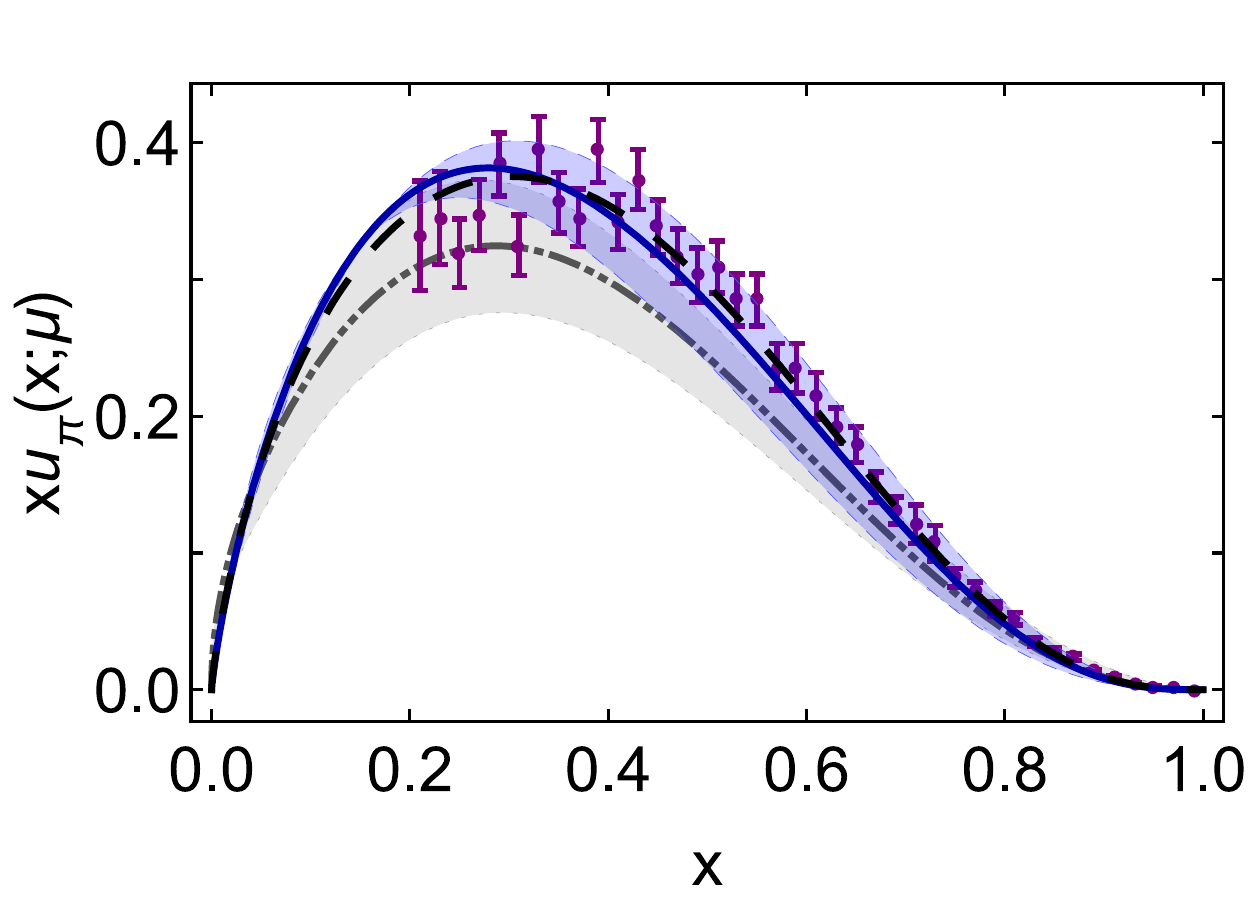}
\hspace*{0.20cm}
\includegraphics[width=0.47\textwidth]{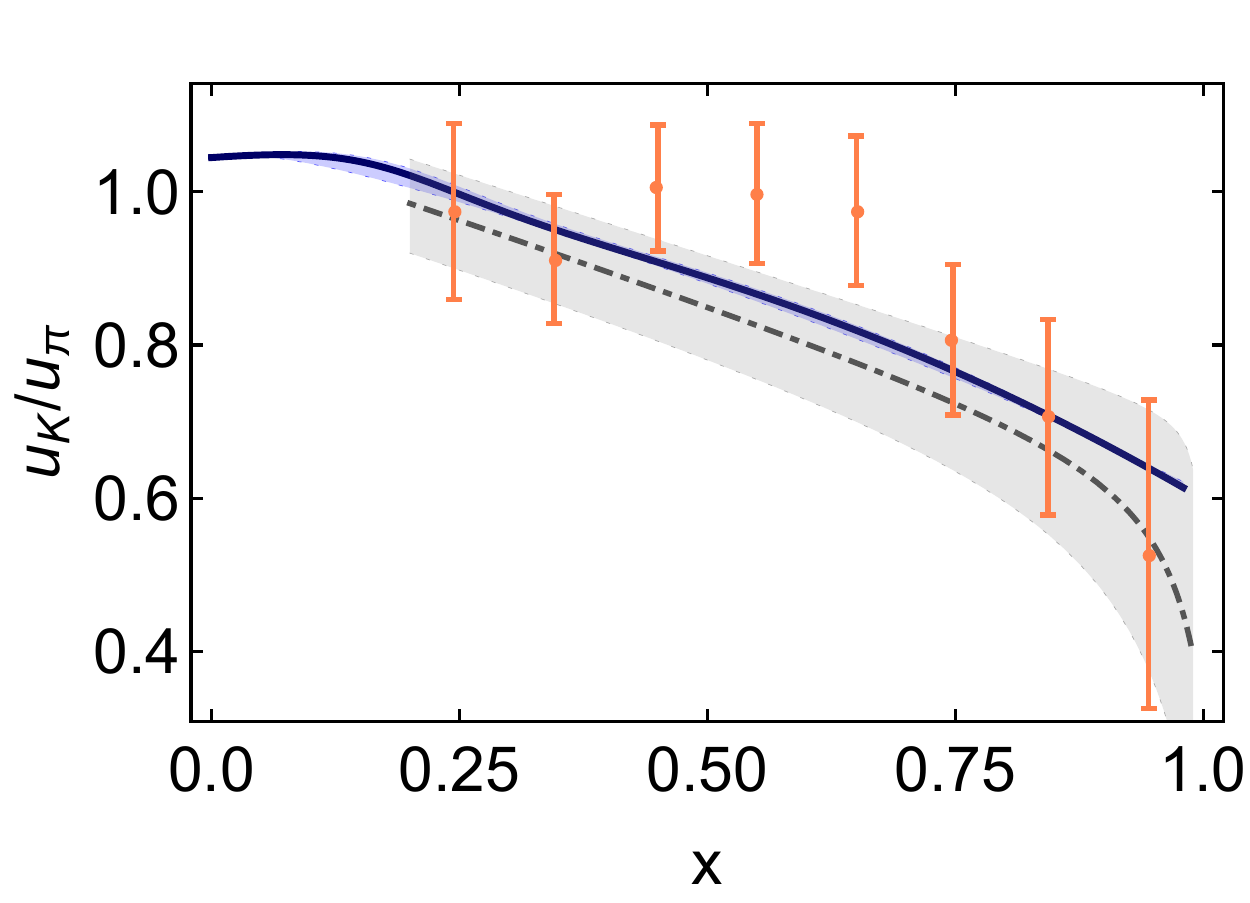}
\caption{\label{fig:F2JorSeg}
Left: Pion valence-quark momentum distribution $xu_\pi(x;\mu=5.2\,\text{GeV})$ 
from modern CSM calculation~\cite{Cui:2020dlm} (solid blue curve), early CSM analysis~\cite{Hecht:2000xa} (long-dashed black curve), and lattice QCD~\cite{Sufian:2019bol} (dash-dotted grey curve). 
The purple points are from a LO pQCD analysis of the data of Ref.~\cite{Conway:1989fs}, rescaled according to the analysis in Ref.~\cite{Aicher:2010cb}.
Right: Ratio $u_K(x;\mu)/u_\pi(x;\mu)$ 
from CSM calculation~\cite{Cui:2020dlm} (solid blue curve), and lattice QCD~\cite{Lin:2020ssv} (dash-dotted grey curve within grey band).
The data (orange) are from event distribution ratios of Ref.~\cite{Badier:1980jq}.
}
\end{figure}

Pion and kaon structure functions extracted from EIC data will be confronted with calculations from continuum Schwinger methods (CSM) and lattice QCD.
The CSM computation of Ref.~\cite{Cui:2020dlm} for the pion valence-quark PDF is depicted in the left panel of Fig.~\ref{fig:F2JorSeg}. It agrees with the predicted large-$x$ behaviour~\cite{Ezawa:1974wm, Farrar:1975yb, Berger:1979du, Brodsky:1994kg, Yuan:2003fs} and, moreover, its pointwise form matches that determined in Ref.~\cite{Aicher:2010cb}, which included the next-to-leading-logarithm resummation using the ``cosine method'' --- see Ref.~\cite{Chang:2014lva} for more details. Furthermore, Ref.~\cite{Cui:2020dlm} provides parameter-free predictions for pion glue and sea PDFs, in addition to all kaon PDFs.
Significantly, lattice QCD is now beginning to yield results for the pointwise behaviour of the pion valence-quark distribution~\cite{Sufian:2019bol, Chen:2018fwa}, with that delivered by the approach in Ref.~\cite{Sufian:2019bol} being in fair agreement with the CSM prediction.
%
%
Concerning the kaon valence-quark distributions from Ref.~\cite{Cui:2020dlm}, there are qualitative similarities between $u_K(x)$, $\bar s_K(x)$ and $u_\pi(x)$.
For instance, all three PDFs are consistent with the above-mentioned large-$x$ prediction, so that $\bar s_K(x)$ is much softer than the first lattice-QCD result~\cite{Lin:2020ssv}. There are also quantitative differences, as highlighted by the prediction for $u_K(x)/u_\pi(x)$ shown in the right panel of Fig.~\ref{fig:F2JorSeg} and compared with the result determined from a measurement of the $K^-/\pi^-$ event distribution ratio~\cite{Badier:1980jq}. 

\subsection{Origin of the hadron mass}
\label{part2-subS-PartStruct-Mass}

About $99\%$ of the mass of the visible universe come from all the nucleons that constitute it. The Higgs mechanism, which provides mass to the fundamental constituents of matter, can only explain  a small fraction of the nucleon mass. The rest finds its origin in the strong force that tightly binds quarks and gluons together.  
 Understanding how the hadron mass emerges in QCD is therefore of utmost importance.

One way to address the question is to determine how current quarks and gluons contribute to the hadron mass. There exist essentially two types of mass decomposition: one consists in a decomposition of the trace of the energy-momentum tensor (EMT)~\cite{Shifman:1978zn,Kharzeev:1995ij,Kharzeev:1998bz,Hatta:2018sqd,Tanaka:2018nae,Roberts:2020hiw}, and the other corresponds to an energy decomposition in the rest frame of the system~\cite{Ji:1994av,Ji:1995sv,Lorce:2017xzd,Lorce:2018egm,Rodini:2020pis,Metz:2020vxd}. In QCD, the EMT is given by the following rather simple-looking expression, \begin{equation}
T^{\mu\nu}=\overline\psi\gamma^\mu\,\tfrac{i}{2}\overset{\leftrightarrow}D\,\!^\nu\psi-G^{a\mu\lambda}G^{a\nu}_{\phantom{a\nu}\lambda}+\tfrac{1}{4}\,g^{\mu\nu} G^2 \,.
\end{equation} 
An essential feature related to the emergence of a mass scale is that the trace of the EMT receives anomalous contributions~\cite{Collins:1976yq, Nielsen:1977sy}, 
\begin{equation}
T^\mu_{\phantom{\mu}\mu}=\frac{\beta(g)}{2g}G^2+(1+\gamma_m)\overline\psi m\psi \,,
\end{equation}
where a summation over quark flavors is implied, $\beta(g)$ is the QCD beta function and $\gamma_m$ the anomalous dimension related to the renormalization of the quark mass.
Any mass decomposition starts with a particular split of the EMT into quark and gluon contributions, $T^{\mu\nu}=T^{\mu\nu}_q+T^{\mu\nu}_G$, which necessarily depends on the renormalization scheme and scale. For a spin-$0$ or spin-$\frac{1}{2}$ system, the forward matrix elements of these contributions can simply be parametrized in terms of two EMT form factors evaluated at vanishing momentum transfer~\cite{Ji:1994av,Ji:1995sv,Ji:1996ek}, i.e., 
 \begin{equation}
 \langle P|T^{\mu\nu}_{q,G}(0)|P\rangle=2P^\mu P^\nu A_{q,G}(0)+2M^2g^{\mu\nu}\bar C_{q,G}(0) \,,
 \end{equation}
while additional spin-dependent contributions are required for higher-spin systems~\cite{Cosyn:2019aio,Cotogno:2019vjb}. 
The trace decomposition takes the form $M=I_q+I_G$ with
\begin{equation}
    I_{q,G}\equiv g_{\mu\nu}\langle T^{\mu\nu}_{q,G}\rangle=\left[A_{q,G}(0)+4\bar C_{q,G}(0)\right]M \,,
\end{equation}
and the energy decomposition reads $M=U_q+U_G$ with
\begin{equation}
    U_{q,G}\equiv\langle T^{00}_{q,G}\rangle=\left[A_{q,G}(0)+\bar C_{q,G}(0)\right]M \,,
\end{equation}
where $\langle T^{\mu\nu}_{q,G}\rangle\equiv\frac{1}{2M}\langle P| T^{\mu\nu}_{q,G}(0)|P\rangle\big|_{\boldsymbol P=\boldsymbol 0}$ denotes the expectation value in the rest frame of the system. These two decompositions are consistent with each other since four-momentum conservation implies the constraints $A_q(0)+A_G(0)=1$ and $\bar C_q(0)+\bar C_G(0)=0$. Further decompositions of the quark and gluon energy contributions have also been discussed in the literature~\cite{Ji:1994av,Ji:1995sv,Lorce:2017xzd,Rodini:2020pis,Metz:2020vxd}, along with the case of massless systems~\cite{Lorce:2017xzd}.  Because of the constraints of four-momentum conservation,  only two independent inputs enter any mass decomposition. However,  an independent cross check of the mass sum rule (i.e., without imposing the four-momentum conservation) can be obtained only by  measuring  the four individual contributions $A_{q,G}(0)$ and $\bar C_{q,G}(0)$.

The various mass decompositions mentioned above become physically more meaningful if each component can be extracted from experimental observables. 
The first quark EMT form factor can be 
obtained as the second moment of the unpolarized PDF, i.e., $A_q(0)=\int\mathrm{d} x\,x\,f^q_1(x)$, and similarly for gluons. 
The other form factors $\bar C_i(0)$ are related to the hadron sigma term
$\sigma=\langle \overline\psi m\psi \rangle$ and the trace anomaly  $\langle\frac{\beta}{2g}G^2+\gamma_m\overline\psi m\psi\rangle$.
In the case of the nucleon,  the former is accessible through low-energy $\pi N$ phenomenology, like experimental
information on $\pi N$  scattering or $\pi$-atom spectroscopy measurements --- see, for instance, Refs.~\cite{Alarcon:2011zs,Alarcon:2012nr,Hoferichter:2015hva,Hoferichter:2015dsa}.
The missing term to directly test the mass sum rule of the proton is then the gluon contribution to the trace anomaly, i.e., the gluon condensate $\langle P|G^2|P\rangle$.
To probe this in experiments, the best way is to use heavy quarkonia  such as $J/\psi$ and $\Upsilon$ because they interact with hadrons primarily via gluon exchange. Besides, in order to maximize the sensitivity to the twist-four operator $G^2$, the center-of-mass energy must be as low as possible. These considerations have led to the proposed  near-threshold photo- or leptoproduction of $J/\psi$ or $\Upsilon$ in lepton-proton scattering~\cite{Kharzeev:1998bz,Hatta:2018ina,Boussarie:2020vmu} (see, also, Refs.~\cite{Brodsky:2000zc,Gryniuk:2016mpk,Du:2020bqj}). Recent studies have shown that this process is also sensitive to the so-called gluon D-term (or the gluonic ``pressure'' inside the nucleon)~\cite{Hatta:2018ina,Boussarie:2020vmu,Mamo:2019mka} which gives complementary information to the quark D-term measurable in DVCS (see, Sec.~\ref{part2-subS-SecImaging-GPD3d}). It is thus a unique process that can simultaneously address two important questions of the nucleon structure (mass, pressure), and is worth pursuing at the EIC. 

\begin{figure}[t]
\centering
\includegraphics[width=0.85\textwidth]{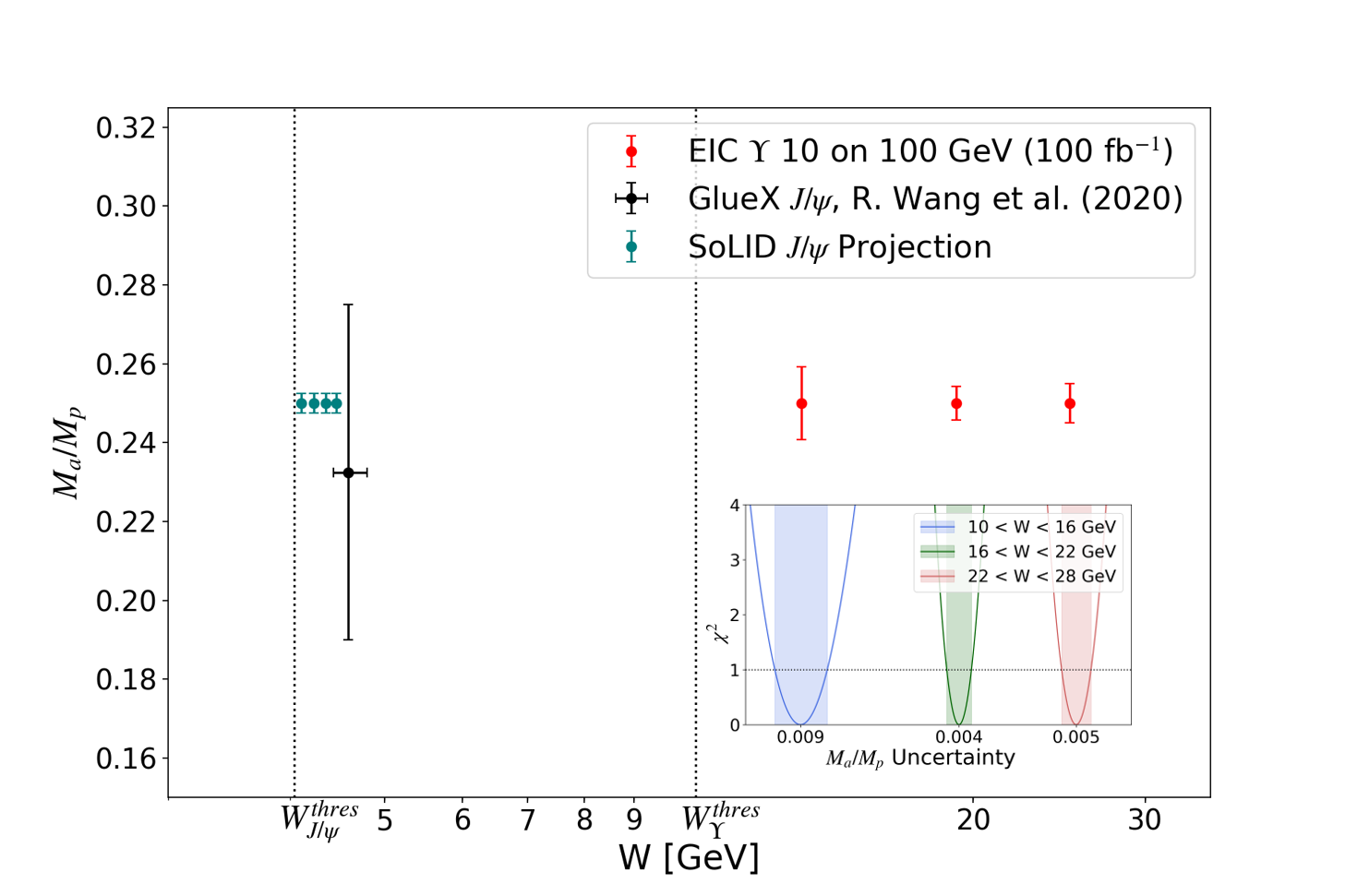}
\caption{Projection of the trace anomaly contribution to the proton mass ($M_a/M_p$) with $\Upsilon$ photoproduction on the proton at the EIC in 10 $\times$ 100 GeV electron/proton beam-energy configuration.
The insert panel illustrates the minimization used to determine the uncertainty for each data point.  
The black circles are the results from the analysis of the GlueX $J/\psi$ data~\cite{Wang:2019mza}, while the dark green circles correspond the JLab SoLID $J/\psi$ projections.
The $\Upsilon$ projections were generated following the approach from Ref.~\cite{Gryniuk:2020mlh}
with the lAger Monte Carlo generator~\cite{lager}.
\label{upsilon:proj_w_scan}}
\end{figure}

Currently, experiments are ongoing at JLab, and the first results for near-threshold $J/\psi$  exclusive  photoproduction have been reported recently~\cite{Ali:2019lzf}. However, the JLab energy is not sufficient to create an $\Upsilon$.
The impact of EIC measurements of $\Upsilon$ photoproduction on the proton is shown in Fig.~\ref{upsilon:proj_w_scan}. The projection of the trace anomaly contribution to the proton mass ($M_a/M_p$) is obtained by assuming the vector meson dominance model as described in Ref.~\cite{Wang:2019mza} and for a medium center-of-mass energy, using the nominal reference detector.  The uncertainties on the extraction of $M_a/M_p$ could be further improved with a larger acceptance for quasi-real events, and by combining the results from different energy configurations.
For comparison, the black and dark green circles show the results for $M_a/M_p$ from the  GlueX $J/\psi$ data~\cite{Wang:2019mza} and the JLab SoLID $J/\psi$ projections, respectively.
Besides, one could resort to leptoproduction at large photon virtualities $Q^2$~\cite{Boussarie:2020vmu}. Despite smaller cross sections, this process has certain advantages over photoproduction. The available region in $Q^2$ is rather limited at JLab ($Q^2 <10 \, \textrm{GeV}^2$), while these problems can be easily overcome at the EIC. 
On the other hand, studying a low-energy process at a high-energy collider inevitably entails new technical challenges. For example, one has to achieve high luminosity in lowest-energy runs. (It is important to distinguish the $ep$ center-of-mass energy from the  $\gamma^{(*)}p$ energy. The latter is constrained to be close to the threshold.) Moreover, the produced quarkonia and their decay products (lepton pairs) are typically in the very forward region, and this may require special detectors. Section~\ref{subsec:dvmp_ep} reports the results of detailed simulations which partly address these questions and indicate directions for future improvements.  

Another way to address the question of the origin of the hadron mass is 
through chiral symmetry. In this picture, different mechanisms due to  
dynamical chiral symmetry breaking (DCSB) are responsible for the 
emergent hadronic mass and should manifest themselves in observables 
that probe the shape and size of the hadron wave function~\cite{Zhang:2020ecj}. Five 
key measurements at the EIC expected to deliver far-reaching insights into
the dynamical generation of mass have been highlighted in Ref.~\cite{Aguilar:2019teb}. Among them, there are measurements of the meson structure 
functions as discussed in Sect.~\ref{part2-subS-PartStruct.M} (see Fig.~\ref{fig:pion-pdf-impact}) and of the pion 
electromagnetic form factor as reported in Secs.~\ref{part2-subS-SecImaging-FF} and \ref{diff-excl-mesons}. 
While the $\pi^+$ mass is barely influenced by the Higgs and is almost 
entirely generated by DCSB, the Higgs mechanism is expected to play a 
more relevant role for the $K^+$ mass due to its strange quark content.
Thus, the comparison of the charged pion
and charged kaon form factors over a wide range in $Q^2$ would provide unique information
relevant to understanding the generation of hadronic mass.  Planned simulation work for
2021-23 includes extensions of the pion form factor measurements to the case of the charged kaon, assuming that measurements at JLab-12 on exclusive $K^+$
electroproduction beyond the resonance region confirm the feasibility of this technique (see Sec.~\ref{part2-subS-SecImaging-FF}).

\subsection{Multi-parton correlations}
\label{part2-subS-PartStruct-MultiPart}

\subsubsection{Introduction}
Multi-parton correlations can be accessed through higher-twist observables with the underlying twist-classification that ``an observable is twist-$t$ if its effect is suppressed by  $(M/Q)^{t-2}$''~\cite{Jaffe:1996zw}. 
Despite the kinematical suppression, twist-3 observables are in principle not small and can even dominate for certain kinematics at moderate $Q^2$. This is illustrated by the fact that the first single-spin asymmetries (SSAs) in semi-inclusive DIS (SIDIS), $A^{\sin \phi_h}_{UL}$ and $A^{\sin \phi_h}_{LU}$, observed at HERMES~\cite{Airapetian:1999tv,Airapetian:2001eg,Airapetian:2002mf,Airapetian:2005jc,Airapetian:2006rx,Airapetian:2019mov} and CLAS~\cite{Gohn:2014zbz,Avakian:2003pk,Aghasyan:2011ha}, 
are twist-3 effects.  Observations of large transverse SSAs in single-inclusive hadron production in hadronic collisions (such as $p p\to hX$), dating back to the 1970s, are further evidence of the importance of twist-3 effects --- see, e.g., Refs.~\cite{Adams:1991rw,Krueger:1998hz,Allgower:2002qi,Adams:2003fx,Adler:2005in,Lee:2007zzh}.

Higher-twist distributions reflect the physics of the largely unexplored quark-gluon correlations which provide direct and unique insights into the dynamics inside hadrons including effects such as vacuum fluctuations --- see, e.g., Ref.~\cite{Ji:2020baz}.  They describe multi-parton distributions~\cite{Qiu:1991pp} corresponding to the interference between higher Fock components in the hadron wave functions, and as such have no probabilistic partonic interpretation. Yet they offer fascinating insights into the nucleon structure. A prominent example is the DIS structure function $g_2$~\cite{Jaffe:1989xx} related to the twist-3 PDF $g_T^q(x)$, which consists of a Wandzura-Wilczek (WW) part, that is given by the twist-2 helicity PDF $g_1^q(x)$~\cite{Wandzura:1977qf}, and the genuine twist-3 piece $\tilde{g}_T^q(x)$.
The Mellin moment $\int dx\,x^2\tilde{g}_T^q(x)$
describes the transverse impulse the active quark acquires after being struck by the
virtual photon due to the color Lorentz force~\cite{Burkardt:2008ps}.
The Mellin moment $\int dx\,x^2\tilde{e}^q(x)$ of the pure twist-3 piece in the scalar density $e^q(x)$ describes the average transverse force acting on a transversely polarized quark in an unpolarized target after interaction with the virtual photon~\cite{Burkardt:2008ps}.

The theoretical description of twist-3 SIDIS observables, like $A_{UL}^{\sin\phi_h}$ or $A_{LU}^{\sin\phi_h}$, is challenging due to open questions in the TMD factorization at the twist-3 level~\cite{Gamberg:2006ru, Bacchetta:2019qkv}. Under the assumption of factorization, 
twist-3 observables receive contributions from several unknown twist-3 TMD parton distributions and/or fragmentation functions~\cite{Bacchetta:2006tn}, typically requiring approximations in the data analyses~\cite{Efremov:2002ut,Courtoy:2014ixa}. An example is the WW(-type) approximation~\cite{Accardi:2009au, Bastami:2018xqd}, where contributions of genuine $\bar{q}gq$-correlators and mass corrections are neglected. The obvious disadvantage of such approximations is the neglect of exactly the new dynamics that enters at twist-3.
The situation simplifies in semi-inclusive jet production, a promising process  at EIC energies, which could provide valuable complementary information on twist-3 TMDs~\cite{Bacchetta:2004zf}.
The collinear twist-3 PDFs $e^q(x)$, $g_T^q(x)$, $h_L^q(x)$ are (also) accessible in di-hadron production~\cite{Bianconi:1999cd,Bacchetta:2003vn,Bacchetta:2002ux,Bacchetta:2006un,Jaffe:1997hf,Radici:2001na,Ceccopieri:2007ip}, a process which can be described using collinear factorization and for which a reduced number of terms contributes to the cross section. 
Higher-twist fragmentation functions can also be of interest in their own right~\cite{Accardi:2019luo}.

Furthermore, important connections between higher-twist parton correlators and twist-2 TMDs exist --- for example, in derivations of the evolution equations for transverse moments of TMDs~\cite{Zhou:2008mz,Vogelsang:2009pj,Kang:2008ey,Braun:2009mi,Kang:2010xv},
calculations of processes at high transverse momentum~\cite{Eguchi:2006mc},
or calculations of the high-transverse-momentum tails of TMDs~\cite{Ji:2006ub,Koike:2007dg, Bacchetta:2008xw}.
Ultimately, through global studies of all of these observables,
one will simultaneously obtain better knowledge of twist-3
collinear functions and twist-2 TMDs, and at the same time test
the validity of the formalism. Gathering as much information as
one can on the quark-gluon-quark correlator is essential to reach
this goal. One example of such a study is the global fit to twist-2 and twist-3 observables used to extract transversity and the tensor charge described in Sec.~\ref{sec:part2-subS-SecImaging-TMD3d.transversity}~\cite{Cammarota:2020qcw}.  In this respect, not only will SIDIS experiments play an important role at an EIC but also measurements of the transverse SSA in $eN^\uparrow\to hX$~\cite{Gamberg:2014eia,Airapetian:2013bim,Allada:2013nsw}.  This is the analogue of the corresponding measurements in $pp^{\uparrow}$ collisions at RHIC, which are sensitive to multi-parton correlators connected to the Sivers and Collins functions~\cite{Qiu:1998ia, Kouvaris:2006zy, Metz:2012ct, Kanazawa:2014dca,Gamberg:2017gle}.

The EIC spanning a large $Q$-range will be ideal to identify higher-twist effects.
The possibilities of extracting twist-3 observables  represent important doorways to 
study hadron structure in QCD.
Here we will describe two exemplary measurements that will be highly interesting at the EIC: The double-spin asymmetry $A_{LT}$ in inclusive DIS sensitive to $g_T^q(x)$ and the longitudinal beam-spin asymmetries in SIDIS to access $e^q(x)$.

\subsubsection{Twist-3 PDF \texorpdfstring{\boldmath{$g_T^q(x)$}}{gTq(x)} from inclusive DIS}
\begin{figure}[ht]
\centering
\includegraphics[width=\textwidth]{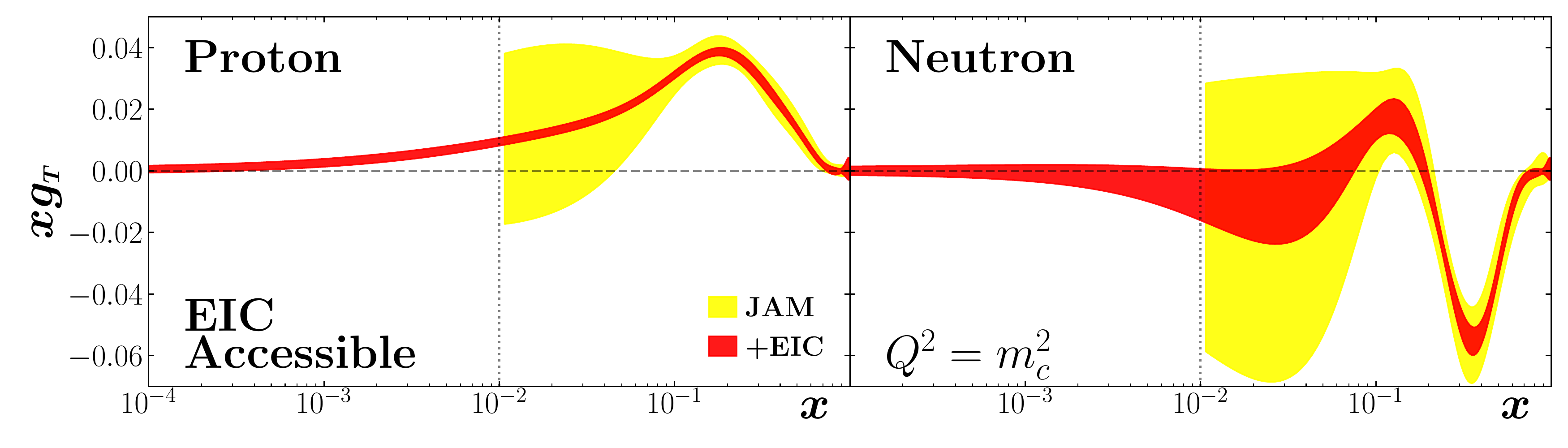}
\caption{\label{fig:part2-subS-PartStruct-MultiPart.gT}
Impact of EIC measurements on the structure function $g_T$, using proton, deuteron and helium targets at ${\cal L}=100$ fb$^{-1}$, with 1.6\% point-by-point uncorrelated systematic uncertainties.
The baseline structure functions $g_1$ and $g_T$ are extracted from the existing world data on $A_{LL}$ and $A_{LT}$ within the JAM framework. 
The extraction of $g_T$ was done using a suitable parametrization for the $x$-dependence at the input scale and assuming DGLAP for the $Q^2$-dependence.  
}
\end{figure}

The clearest example of higher twist, that is defined in a collinear framework and accessible in inclusive DIS, is the cross section for the collision of a longitudinally polarized electron beam and a transversely polarized target, $\vec{e} p^\uparrow\rightarrow e' X$. Factorization theorems lead to the introduction of the collinear twist-3 PDF $g_T^q(x)$~\cite{Jaffe:1989xx,Barone:2003fy}. Recent work on higher-twist contributions to the spin-dependent DIS structure functions $g_1$ and  $g_2$ was carried out by the JAM Collaboration~\cite{Sato:2016tuz}, which enables an extraction of $g_T^q(x)$ from the longitudinal-transverse double-spin asymmetry $A_{LT} \sim \left(g_1+g_2\right)=g_T$. 
The derivation of the inclusive cross section in collinear twist-3 factorization establishes the rigorous connection between $A_{LT}$ and the twist-3 PDF $g_T^q(x)$ --- see, for instance, Refs.~\cite{Qiu:1991pp,Metz:2012ct,Pitonyak:2013pt}. 
This in turn establishes the structure function $g_T$ as the most prominent observable to study multi-parton correlations at the EIC. Figure~\ref{fig:part2-subS-PartStruct-MultiPart.gT} shows the projected impact of the EIC data on $g_T$. 
The yellow band shows the uncertainty from DIS world data on the double-spin asymmetries $A_{LL}$ and $A_{LT}$ within the JAM analysis framework~\cite{Sato:2016tuz}, assuming DGLAP evolution for the $Q^2$-dependence. The impact from EIC pseudodata is reflected by the red band.
The EIC data will allow for entirely new insights on the PDF $g_T^q(x)$ and as such on multi-parton correlations and the interactions of the struck quark with the partons around it.

\subsubsection{Twist-3 PDF \texorpdfstring{\boldmath{$e^q(x)$}}{eq(x)} from semi-inclusive DIS}
\begin{figure}[ht]
\centering
\includegraphics[width=1\textwidth]{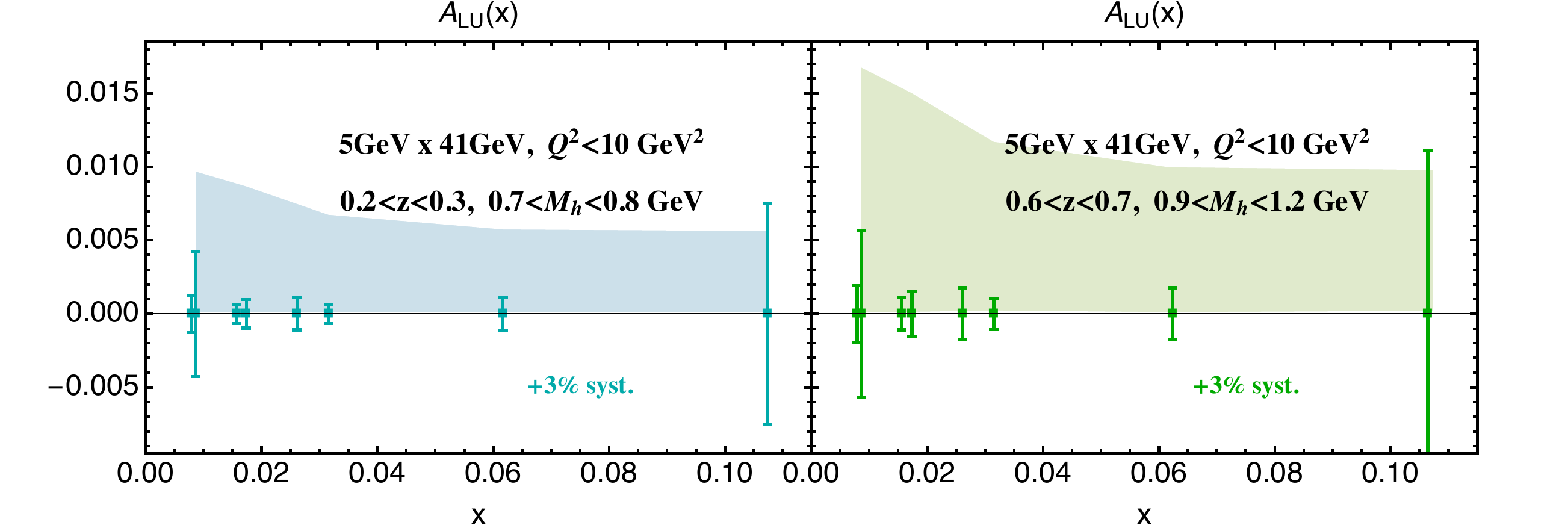}
\includegraphics[width=1\textwidth]{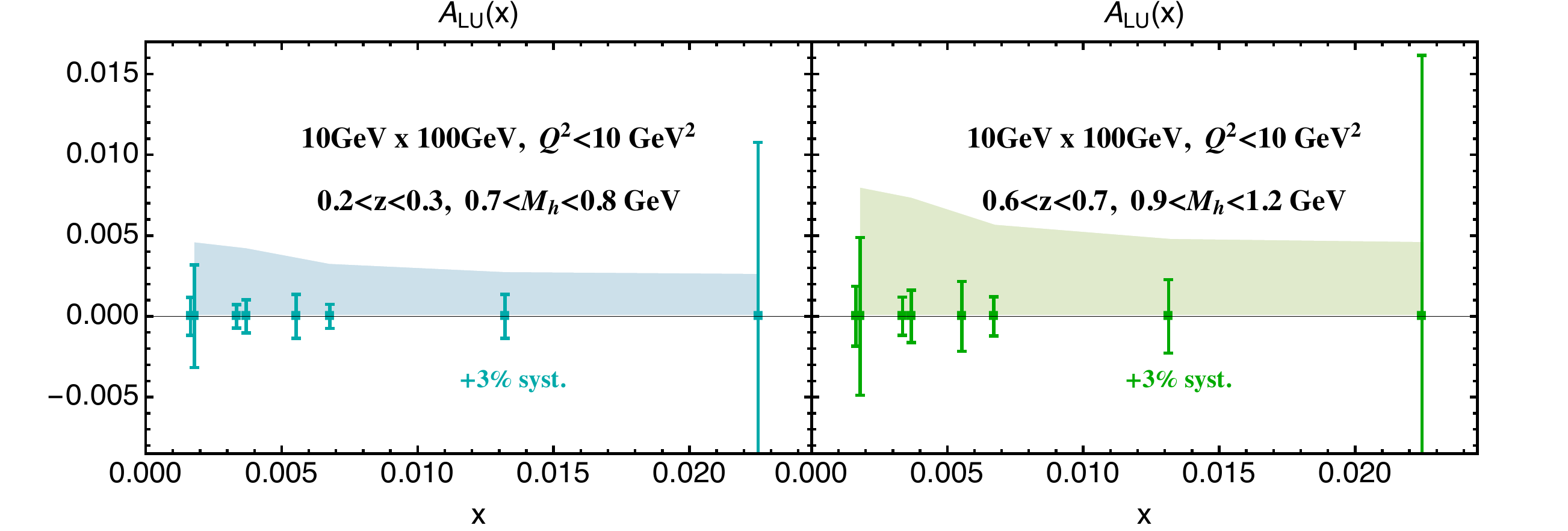}
\caption{Beam-spin asymmetry in semi-inclusive di-hadron production. Predictions corresponding to $Q^2=1$ GeV$^2$ based on the di-hadron fragmentation functions of Ref.~\cite{Radici:2015mwa}, low-energy models for the twist-$3$ PDF $e^q(x)$ and unpolarized PDFs from MSTW08 at leading order~\cite{Martin:2009iq} (see also text). The upper and lower panels show two different energy configuration. The left (blue) and right (green) plots correspond, respectively, to the fragmentation kinematics of ($0.2<z<0.3$,\, $0.7 \, \textrm{GeV} <M_h < 0.8 \, \textrm{GeV}$) and ($0.6<z<0.7$,\, $0.9 \, \textrm{GeV} < M_h < 1.2 \, \textrm{GeV}$), where $z$ is the longitudinal momentum fraction of the di-hadron pair and $M_h$ its invariant mass. 
The bands give the envelope of the model projections discussed in the text, folded with the uncertainty of the interference fragmentation function. The projected statistical uncertainties are plotted at zero.}
\label{fig:ex}
\end{figure}

The twist-3 PDF $e^q(x)$ can be decomposed into three  contributions~\cite{Efremov:2002qh} through QCD equations of motion. The first one is a $\delta(x)$-singularity related to the pion-nucleon sigma term and the non-trivial QCD vacuum structure~\cite{Burkardt:2001iy,Aslan:2018tff,Ma:2020kjz,Ji:2020baz,Hatta:2020iin,Bhattacharya:2020jfj}.
The second term $\tilde{e}^q(x)$ is related to a genuine $\bar q g q$ contribution with the aforementioned force interpretation~\cite{Burkardt:2008ps}. The third term is proportional to the quark mass and the unpolarized PDF $f_1^q(x)$. 
In the bag model, $e^q(x)$ is due to the bag boundary~\cite{Jaffe:1991ra,Avakian:2010br}, while constituent quark models feature the mass term as being dominant~\cite{Jakob:1997wg,Schweitzer:2003uy,Efremov:2002qh,Wakamatsu:2003uu,Ohnishi:2003mf,Lorce:2014hxa,Pasquini:2018oyz}. 
Phenomenologically, the chiral-odd PDF $e^q(x)$ must be paired to another chiral-odd object, associated with fragmentation for deep-inelastic processes. 
This is analogous to the extraction of the transversity described in Sec.~\ref{sec:part2-subS-SecImaging-TMD3d.transversity}.
In Fig.~\ref{fig:ex} the theoretical predictions are shown for the contribution of $e^q(x)$ to the beam-spin asymmetry in semi-inclusive di-hadron production, calculated in collinear factorization at leading order.
This asymmetry receives a contribution not only from $e^q(x)$ but also from a term involving a twist-3 di-hadron fragmentation function~\cite{Bacchetta:2003vn}. The latter has not been considered here~\cite{Courtoy:2014ixa}. 
The uncertainties in Fig.~\ref{fig:ex} come from the envelope of the uncertainties on the interference fragmentation function~\cite{Radici:2015mwa} and two models for $e^q(x)$, a light-front constituent quark model~\cite{Pasquini:2018oyz} and a model of the mass-term contribution to $e^q(x)$ with an assumed constituent-quark mass of 300 MeV and the unpolarized PDFs from MSTW08LO. All PDFs and fragmentation functions are taken at $Q^2=1$~GeV$^2$, and the projected uncertainties for the EIC are shown only for $Q^2$-values smaller than 10~GeV$^2$. 
A similar observable can be studied for single-hadron SIDIS in TMD factorization~\cite{Efremov:2002ut} which, however, is more complex as it involves 4 unknown contributions with one of them being related to the TMD $e^q(x,k_T)$. The evolution of the genuine twist-3 contribution $\tilde{e}^q(x)$ has been studied~\cite{Balitsky:1996uh,Belitsky:1997zw,Koike:1996bs} but never implemented in the PDF extractions. The EIC kinematics will especially allow for new insights in the low-$x$ region. This region is of high relevance due to the contribution of a local term related to the non-trivial QCD vacuum as well as the mass term whose $x$-behavior is enhanced like $f^q_1(x)/x$.


\def\p{I\!\!P}
\newcommand{\Pomeron}{I\!\!P}
\subsection{Inclusive and hard diffraction}
\label{part2-subS-PartStruct-InclDiff}
\subsubsection{Inclusive diffraction}

\begin{figure}[bth]
    \centering
   \includegraphics[width=0.45\textwidth]{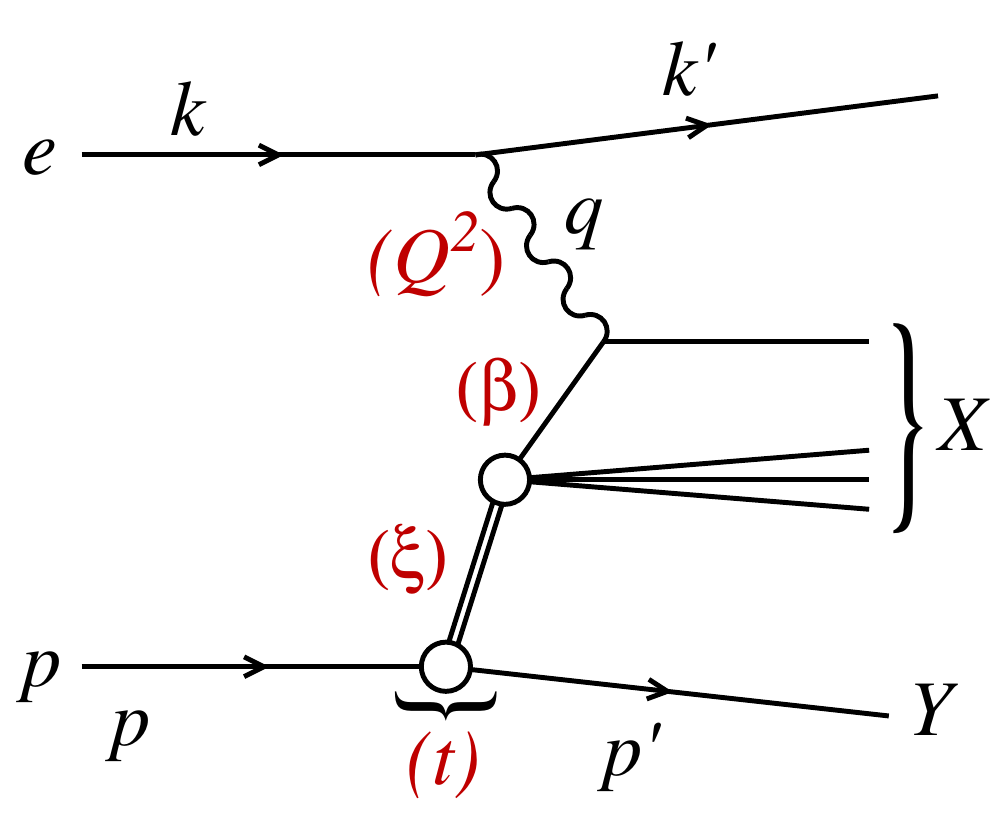}
    \caption{Diagram for diffractive event in DIS. 
    The final state includes the proton $Y$, the scattered electron, and the diffractive system $X$.
    (The four-momenta of the particles are indicated as well.)
    There is a rapidity gap between $X$ and the scattered proton. 
    The double-line indicates colorless diffractive exchange responsible for the presence of the rapidity gap.}
    \label{PWG-sec7.1.6-fig-diffevent}
\end{figure}

Inclusive diffraction has been extensively studied at the HERA collider~\cite{Adloff:1997sc,Breitweg:1997aa}. 
Diffractive events in DIS, ${ep\rightarrow eXY}$, are distinguished by the presence of a large rapidity gap between the diffractive system, characterized by the invariant mass $M_X$, and the final-state proton (or its low-mass excitation) $Y$.
In addition to the standard DIS variables $(x,Q^2)$, diffractive events (see diagram in Fig.~\ref{PWG-sec7.1.6-fig-diffevent}) are also characterized by a set of variables that are specific to diffraction and defined as
\begin{equation}
t=(p-p')^2\,, \;\;\; \xi=\frac{Q^2+M_X^2-t}{Q^2+W^2}\,, \;\;\; \beta = \frac{Q^2}{Q^2+M_X^2-t}\, .
\end{equation}
Here, $t$ is the squared four-momentum transfer at the proton vertex, $\xi$ (alternatively denoted by $x_{I\!P}$) can be interpreted as the momentum fraction of the ``diffractive exchange"  with respect to the hadron, and $\beta$ is the momentum fraction of the parton with respect to the diffractive exchange. 
The two momentum fractions combine to give the Bjorken variable, $x=\beta \xi$.
In analogy with unpolarized, non-diffractive inclusive DIS, the cross section for inclusive diffraction can be expressed in terms of the reduced cross section and the corresponding structure functions, which depend on the aforementioned additional variables specific to diffraction,
 \begin{equation}
\label{PWG-sec7.1.6-sred}
\sigma_{\rm red}^{D(3)} = F_2^{D(3)}(\beta,\xi,Q^2) - \frac{y^2}{Y_+} F_\mathrm{L}^{D(3)}(\beta,\xi,Q^2) \,,
\end{equation}
and, without integration over $t$,
\begin{equation}
\sigma_{\rm red}^{D(4)}= F_2^{D(4)}(\beta,\xi,Q^2,t) - \frac{y^2}{Y_+} F_\mathrm{L}^{D(4)}(\beta,\xi,Q^2,t) \,,
\end{equation}
where $Y_+= 1+(1-y)^2$.

\begin{figure}[ht]
    \centering
   \includegraphics[width=0.95\textwidth]{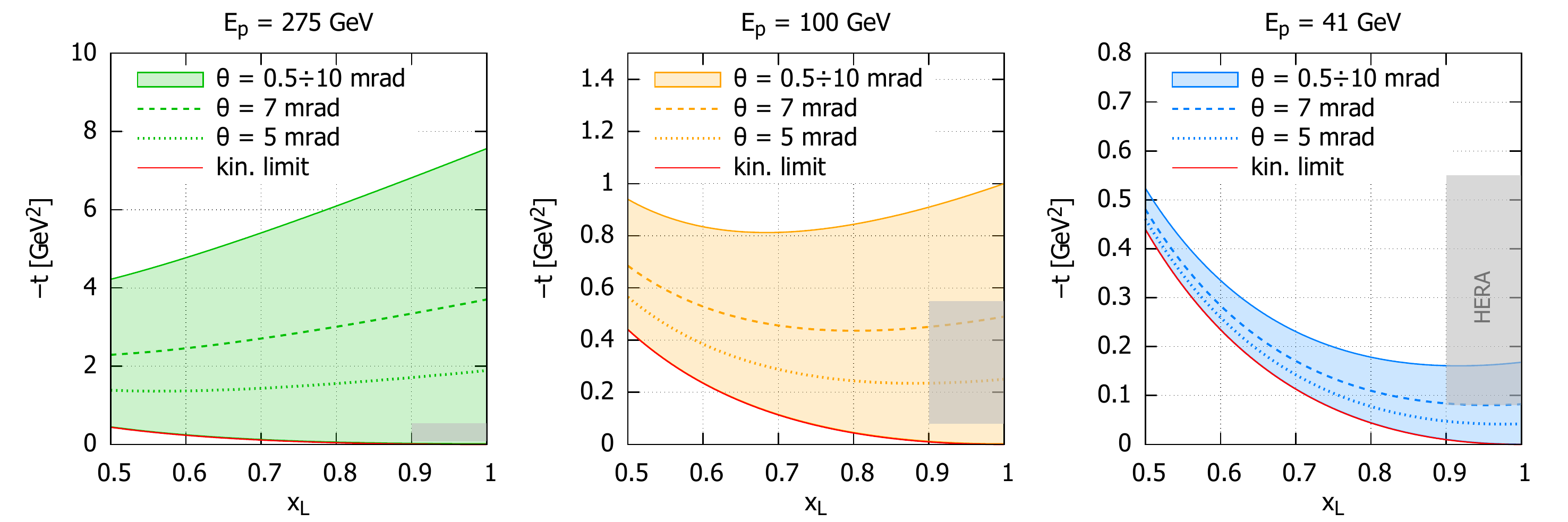}
    \caption{Accessible range in $t$ and $x_L$ for different small-angle $\theta$ acceptance of the final-state leading proton for three EIC energy scenarios: $E_p=275, 100, 41 \, \rm GeV$. The red line is the kinematic limit, and the grey area is the HERA range.}
    \label{PWG-sec7.1.6-fig-leadprot}
\end{figure}

The standard pQCD approach to inclusive diffraction is based on the collinear factorization~\cite{Collins:1997sr,Berera:1995fj,Trentadue:1993ka}.
The cross section is computed by the convolution of the perturbative partonic cross section and the diffractive parton distribution functions (DPDFs). 
The DPDFs are evolved using the DGLAP evolution equations with appropriately chosen initial conditions at some initial scale. 
At HERA fits to the diffractive structure functions were performed by H1~\cite{Aktas:2006hy} and ZEUS~\cite{Chekanov:2009aa}. 
They both parametrize the DPDFs in a two-component model, containing contributions 
from Pomeron and Reggeon exchange.
In both cases the proton-vertex factorization is assumed, meaning that the diffractive exchange can be interpreted as colourless objects called a ``Pomeron" or a ``Reggeon", with an appropriate parton distributions $f_i^{I\!P,I\!R}(\beta,Q^2)$ and factorized flux factor $f_p^{I\!P,I\!R}(\xi,t)$. 

There are number of areas where the EIC can significantly expand our knowledge of QCD diffraction. 
(We note that, since we discuss also the region of large $\xi$, we here talk about diffraction in the wider sense of leading proton and high-collision energies.)   
First, thanks to the instrumentation in the forward region, the EIC will be able to measure leading protons in a much wider range of $t$ and $x_L$ (fraction of the longitudinal momentum of the initial proton carried by the final proton) than at HERA. 
This is illustrated in Fig.~\ref{PWG-sec7.1.6-fig-leadprot} for different proton energies.
The red curves indicates the kinematic limit, and different curves indicate various angular cuts on the final-state proton. 
For example, for the highest proton energy, an angular acceptance extending to $7 \, \rm mrad$ translates into a range in $-t$ up to (at least) $2\, \rm GeV^2$.
This is well beyond the HERA range. 
\begin{figure}[th]
    \centering
   \includegraphics[width=0.75\textwidth]{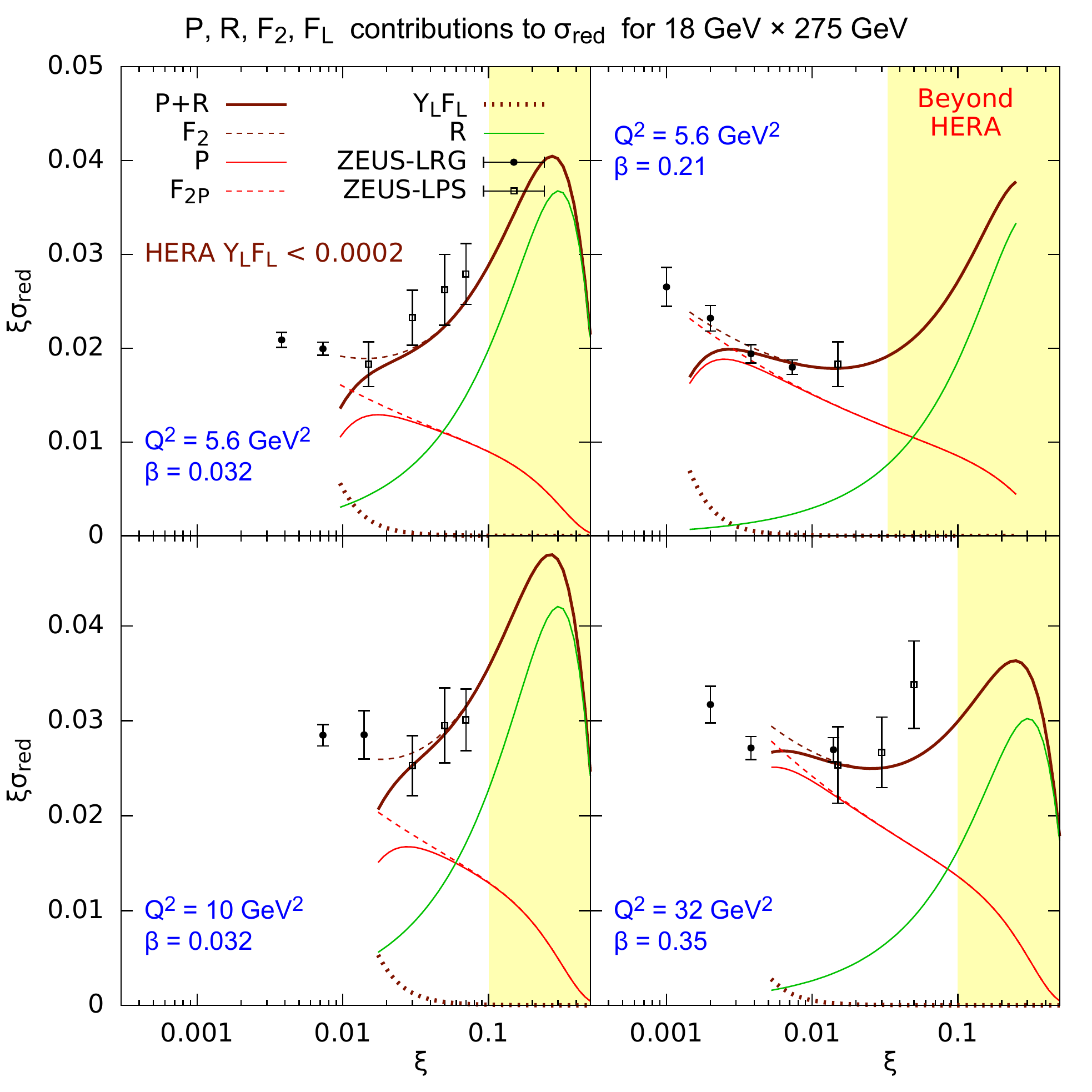}
    \caption{Reduced cross section as a function of $\xi$ in bins of $\beta$ and $Q^2$ for the EIC energy scenario $18 \times 275 \, {\rm GeV}$.
    Red solid curve: Pomeron contribution; green solid curve: Reggeon contribution; brown solid curve: sum of Pomeron and Reggeon contributions. 
    Dashed red curve: $F_2^D$ Pomeron contribution; dotted brown curve: $Y_L F_L$ contribution; dashed brown curve: $F_2$. 
    The data are from HERA, where the yellow region was not accessible. 
    The variable $Y_L$ is defined as $Y_L=y^2/Y_+$, and the calculations are based on the ZEUS SJ fit~\cite{Chekanov:2009aa}.}
    \label{PWG-sec7.1.6-fig-pomregg}
\end{figure}

The second area where the EIC could provide valuable information are the Pomeron and Reggeon contributions.
At HERA, the $t$-dependence of the Reggeon contribution could not be tested at all, as the range in $\xi$ was not sufficient to probe in detail the region where the Reggeon contribution is dominant. 
This is illustrated in Fig.~\ref{PWG-sec7.1.6-fig-pomregg}.
Here we show the reduced cross section as a function of $\xi$ in bins of $Q^2,\beta$ for the highest-energy scenario at the EIC, i.e., $18 \times 275 \, {\rm GeV}$.
The solid curves (red, green, brown) indicate the Pomeron contribution, Reggeon contribution, and their sum, respectively. 
It is clear that the region  $\xi>0.1$, which was not accessible at HERA, is where the Reggeon contribution starts to dominate. 
The EIC has the potential to explore that region to disentangle the two components. The same plot also illustrates the importance of the longitudinal structure function $F_L^D$ in the region of low $\xi$, indicated by the brown dotted curve. The superimposed data are from HERA, and they clearly follow the $F_2^D$ contribution only, since at HERA in this regime the contribution from $F_L^D$ was extremely small. 
On the contrary, at the EIC the contribution from $F_L^D$ for the same values of $Q^2,\beta,\xi$ is not negligible, as illustrated by the dotted brown curve in Fig.~\ref{PWG-sec7.1.6-fig-pomregg}.   
With its high luminosity and the variable center-of-mass energies, the EIC will provide  excellent opportunities to perform precise measurements of the longitudinal diffractive structure function. 

\begin{figure}[th]
    \centering
   \includegraphics[width=\textwidth]{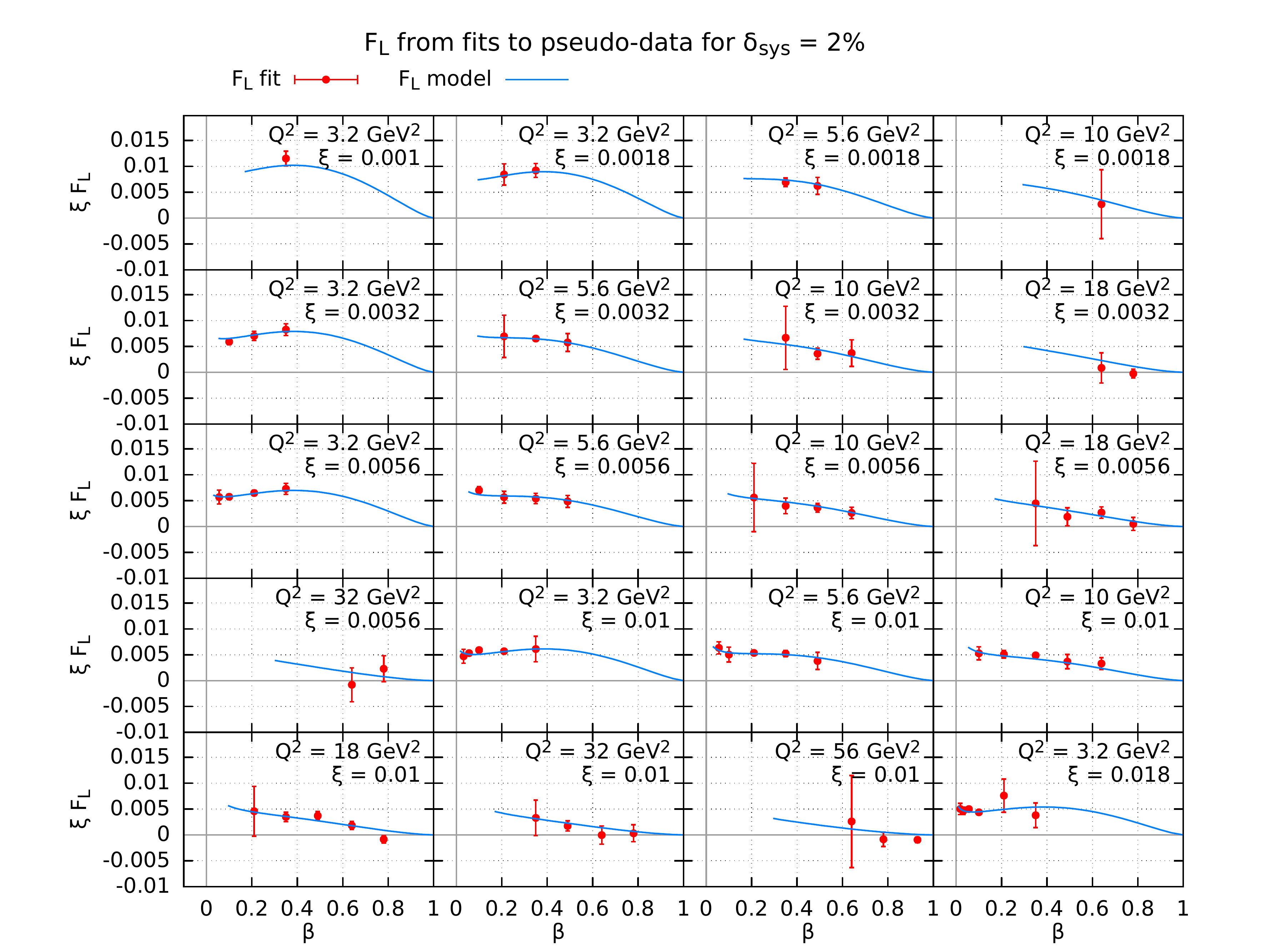}
    \caption{Longitudinal structure function $F_L^{D(3)}$ extracted from fits to the pseudodata as a function of $\beta$ in bins of $Q^2$ and $\xi$.  Red points indicate the data extracted from the analysis, and the curves show the prediction. The selected bins are ordered by $\xi$ and $Q^2$.}
    \label{PWG-sec7.1.6-fig-FL}
\end{figure}

In Fig.~\ref{PWG-sec7.1.6-fig-FL} the possibility of extracting $F_L^D$ is further explored. For this analysis, 18 energy setups were considered, 
$(5, 10, 18) × (41, 100, 120, 165, 180, 275) \, \rm GeV$, with $2 \, \rm fb^{-1}$ integrated luminosity for each setup.
There were 469 bins selected such that they are common to at least four beam setups, with the cuts $Q^2>3 \, \mathrm{GeV^2}, \, M_X>2 \, \rm GeV$. 
The measurement of this quantity is dominated by the systematic error, which for this analysis was assumed to be $2 \%$.
Figure~\ref{PWG-sec7.1.6-fig-FL} indicates that the potential is very good for measuring $F_L^D$ at the EIC in a wide range of kinematics.
It is worth noting, however, that it may be difficult to realize 18 energy setups in an experiment. 
More studies are thus needed to further scrutinize the potential for a measurement of $F_L^D$.

\begin{figure}[th]
    \centering
   \includegraphics[width=0.75\textwidth]{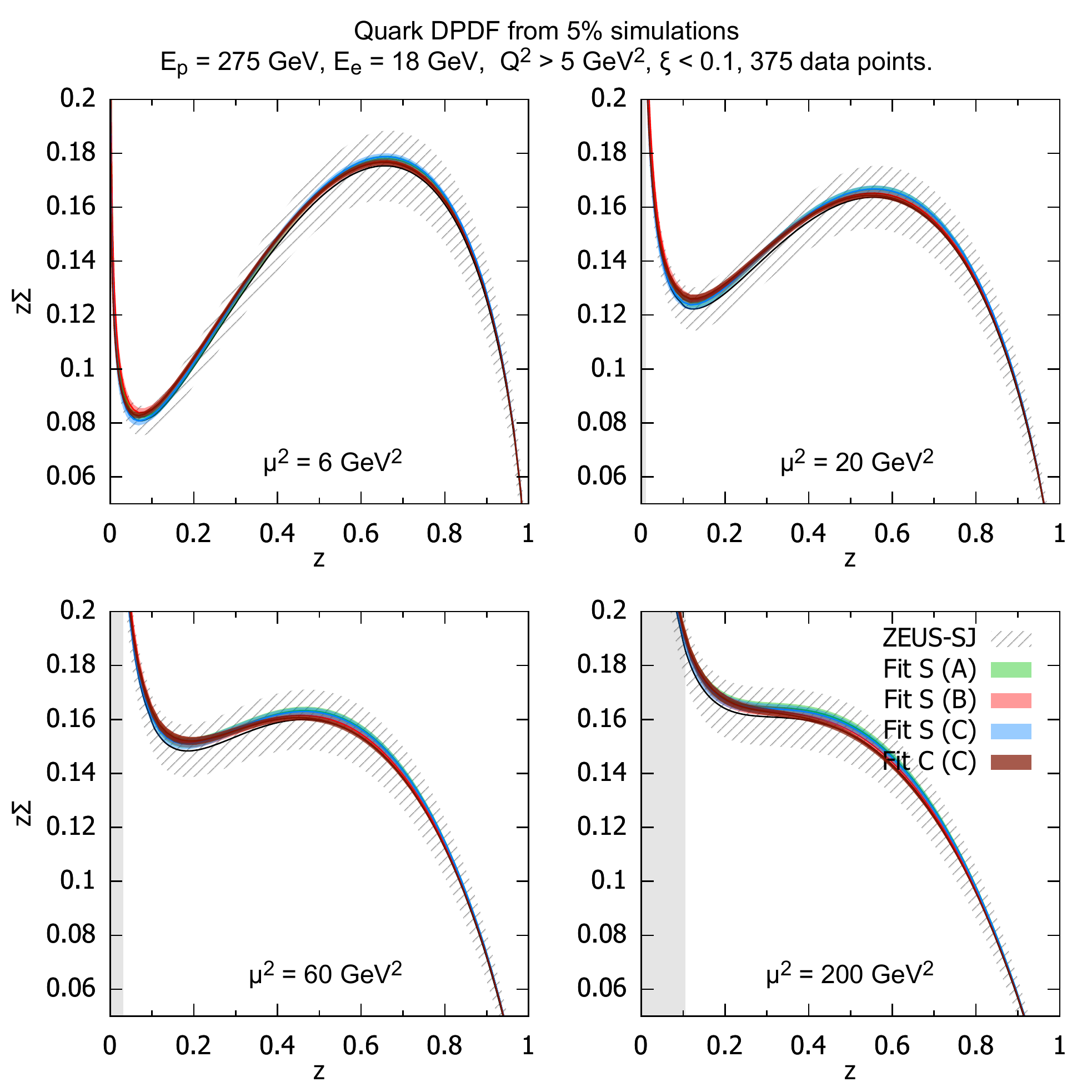}
    \caption{Diffractive quark distribution as a function of $z$ in bins of $Q^2$. The hatched bands indicate HERA uncertainty bands for the ZEUS SJ fit~\cite{Chekanov:2009aa}. The solid bands indicate the projected uncertainty after fitting to the EIC data.}
    \label{PWG-sec7.1.6-fig-DPDFs}
\end{figure}

Finally, the EIC could potentially improve the extraction of the DPDFs.
The DPDFs were for the first time extracted from data taken at HERA, and the success of the DGLAP fits to inclusive diffraction confirmed the applicability of the collinear factorization theorem to diffraction. Nevertheless, many open questions remained, since the fits were only valid for relatively large values of $Q^2$, that is, $Q^2 > 8.5 \, \rm GeV^2$ for H1 and $Q^2 > 5 \, \rm GeV^2$ for ZEUS, below which they failed, suggesting the need for additional corrections.  Studies indicated possible improvement of the description based on saturation models, which would incorporate higher-twist effects in the fits~\cite{Motyka:2017xgk}. In addition, the DPDFs from HERA were not very well constrained at large values of $z$, the longitudinal momentum fraction of the parton with respect to the diffractive exchange. 
(In the parton model $z = \beta$, but $\beta < z$ once higher orders are taken into account.) The EIC offers the unique opportunity to improve the extraction of the DPDFs at large values of $z$. In Fig.~\ref{PWG-sec7.1.6-fig-DPDFs} we show an example of an extraction of the quark DPDF in bins of $Q^2$ as a function of $z$. 
The DGLAP evolution using the HERA-type parametrization~\cite{Aktas:2006hy,Chekanov:2009aa} was fitted to EIC pseudodata generated assuming $5\%$ systematic error. The statistical error was negligible for the integrated luminosity of $2 \, \rm fb^{-1}$.
One finds that the projected uncertainty on the quark DPDF extracted from the EIC is reduced significantly with respect to HERA, in particular at large values of $z$.

\subsubsection{Diffractive dijets}

Studies of diffraction in high-energy electron-proton scattering is one of the highlights of the HERA heritage, which  
discovered that diffractive processes 
account for a substantial fraction ($10-15$\%) of all events. 
In DIS, taking advantage of the QCD factorization theorem~\cite{Collins:1997sr}, DPDFs of the proton have been determined~\cite{Aktas:2006hy,Chekanov:2009aa}, and their universality in diffractive dijet and open-charm production has been shown.
At the same time, in diffractive dijet photoproduction, next-to-leading order (NLO) perturbative QCD calculations~\cite{Klasen:1994bj} indicated that factorization is broken --- the theory agrees with the H1~\cite{Aktas:2007hn} and ZEUS~\cite{Chekanov:2007rh} data after assuming that either the resolved-photon contribution is scaled by a factor of $0.34$ or the entire pQCD cross section is scaled by a global factor of $0.4-0.7$ (with the values depending on the jet transverse momentum and having large theoretical uncertainties from scale variations and hadronization corrections).

To explore the EIC potential for diffractive dijet photoproduction, we performed detailed studies of this process in NLO QCD for \ep and \eA scattering~\cite{Guzey:2020gkk}. 
Using the framework developed in Ref.~\cite{Klasen:1994bj},
the cross section for the reaction $e+p \rightarrow e'+2~{\rm jets}+X+Y$ can be written as 
{\small{
\begin{equation}
d\sigma =
  \sum_{a,b} \int\!dy \int\!dx_{\gamma}\!\int\!dt\!\int\!dx_{\p}\!\int\!dz_{\p}
  f_{\gamma/e}(y) f_{a/\gamma }(x_\gamma,M^2_{\gamma }) f_{\p/p}(x_{\p},t)
  f_{b/\p}(z_{\p},M_{\p}^2) d\hat{\sigma}_{ab}^{(n)} \,.
  \label{eq:gk_1}
\end{equation}
}}
In Eq.~(\ref{eq:gk_1}), $f_{\gamma/e}(y)$ is the photon flux calculated in the improved Weizs\"acker-Williams approximation \cite{Budnev:1974de,Frixione:1993yw}, $y$ is the photon longitudinal momentum fraction, $f_{a/\gamma}(x_\gamma,M^2_{\gamma })$ is the PDF of the photon (for the resolved-photon contribution), and $x_{\gamma}$ the corresponding momentum fraction.
The diffractive PDF of the proton is written in the usually assumed form of Regge factorization as the product of the flux factor $f_{\p/p}(x_{\p},t)$, where $t$ is the invariant momentum transfer squared, and the PDFs of the Pomeron $f_{b/\p}(z_{\p},M_{\p}^2)$.
Finally, $d\hat{\sigma}_{ab}^{(n)}$ is the cross section for the production of an $n$-parton final state from two initial partons, $a$ and $b$.
In our analysis, 
we identified the
factorization scales $M_{\gamma}$, $M_{\p}$ and the renormalization scale
$\mu$ with the average transverse momentum $\bar{p}_T=(p_{T1}+p_{T2})/2$.
The longitudinal momentum fractions $x_\gamma$ and $z_{\p}$ can be experimentally determined from the two observed
leading jets through
\begin{equation}
x_{\gamma}^{\rm obs} = \frac{p_{T1}\,e^{-\eta_1}+p_{T2}\,e^{-\eta_2}}{ 2yE_e} \;\;\;\;\; {\rm and}\;\;\;\;\; 
z_{\p}^{\rm obs} = \frac{p_{T1}\,e^{\eta_1}+p_{T2}\,e^{\eta_2}}{ 2x_{\p}E_p} \,,
\label{eq:gk_2}
\end{equation}
where $p_{T}$ and $\eta$ is the transverse momentum and rapidity of jet-1 or jet-2, while $E_e$ and $E_p$ is the electron and proton beam energy, respectively.

Given the experience from HERA \cite{Aktas:2007hn,Chekanov:2007rh}, we defined jets with the anti-$k_T$ algorithm with the distance parameter $R=1$ and assumed that the detector(s)
can identify jets above the relatively low transverse energies of $p_{T1}>5$ GeV
(leading jet) and $p_{T2}>4.5$ GeV (subleading jet). 

Using the formalism outlined above, we made predictions for diffractive dijet photoproduction at the EIC as a function of  
the jet average transverse momentum $\bar{p}_T$, the observed longitudinal momentum fractions $x_{\gamma}^{\rm obs}$
and $z_{\p}^{\rm obs}$, the proton longitudinal momentum transfer $x_{\p}$, and
the jet rapidity difference $\Delta \eta$. Several features of the obtained results are important to emphasize.
The distribution in the jet average transverse momentum extends only to 8 GeV (it was 15 GeV at HERA) and, hence, 
the cross section is dominated by contributions from direct photons.
The $K$-factor, giving the ratio of the NLO and LO cross section, was found to be approximately constant, $K \approx 2$, and independent of kinematic variables. 
To address conclusively the mechanism responsible for factorization breaking requires a high proton-beam energy and wide ranges in $x_{\p}$ (where the subleading contribution at high $x_{\p}$ becomes important) and $x_{\gamma}^{\rm obs}$.
Replacing the proton by a heavy nucleus and using predictions for nuclear DPDFs from Ref.~\cite{Frankfurt:2011cs}, we also obtained results for these distributions in coherent diffractive dijet photoproduction on nuclei in the reaction $e+A \rightarrow e+2~{\rm jets}+X+A$.

\begin{figure}[ht]
\centering
 \includegraphics[width=0.69\textwidth]{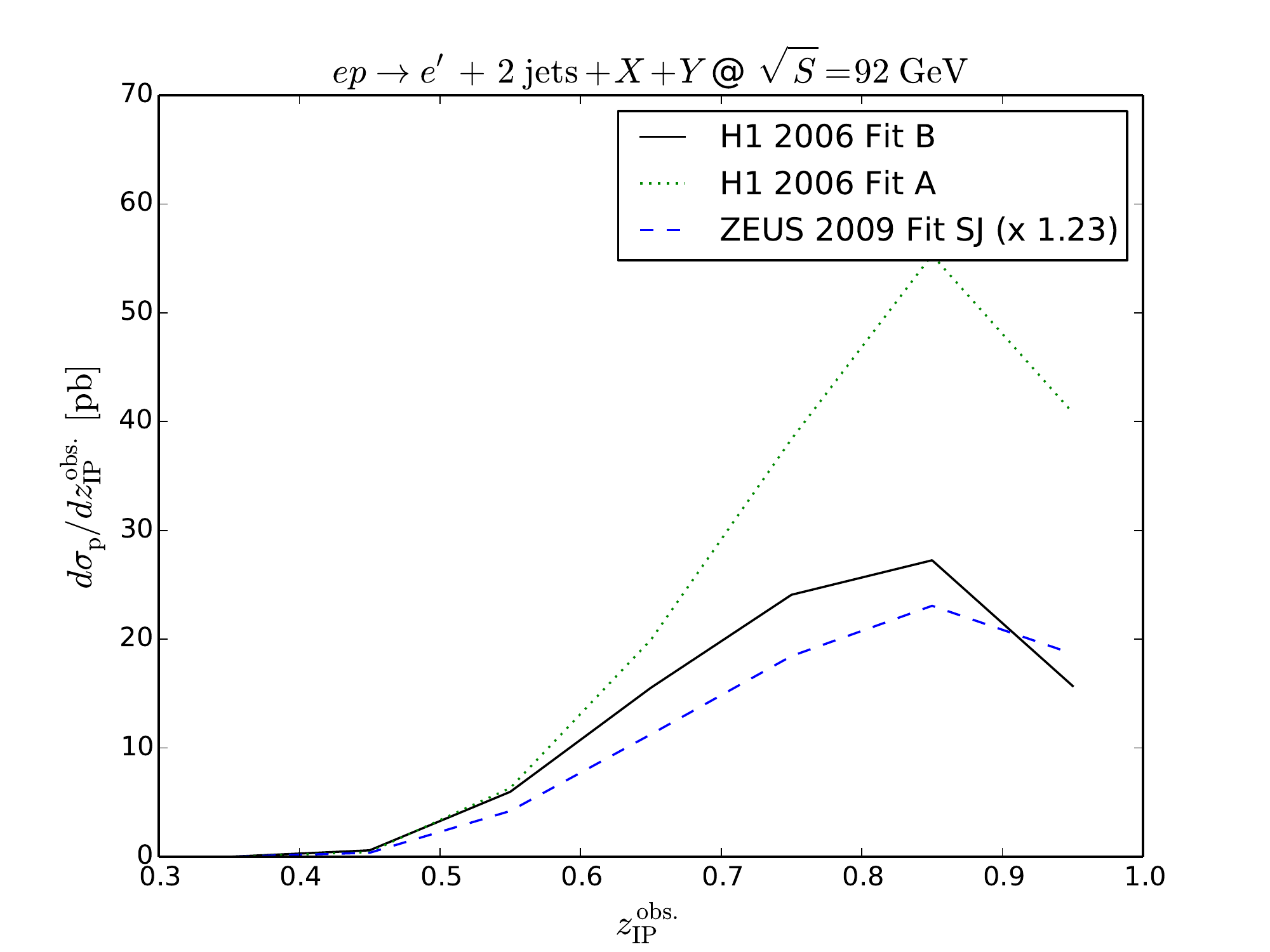}
 \caption{ NLO QCD predictions for the $z_{\p}^{\rm obs}$-dependence using three different sets of diffractive PDFs: 
 H1 2006 Fit B (full black), Fit A (dotted green), both from Ref.~\cite{Aktas:2006hy}, and ZEUS 2009 Fit SJ (dashed blue curves) from~ Ref.\cite{Chekanov:2009aa}.   The rescaling for the calculation using the ZEUS SJ fit is needed to take into account the
 contribution of proton dissociation, which has been included in the  H1 fits A and B.
 }
\label{PWG-sec7.1.6-fig-dijets}
\end{figure}

An example of our predictions is presented in Fig.~\ref{PWG-sec7.1.6-fig-dijets}, 
which shows the dijet cross section as a function of $z_{\p}^{\rm obs}$ for three different sets of diffractive PDFs, H1 2006 fits A and B~\cite{Aktas:2006hy} and ZEUS SJ fit~\cite{Chekanov:2009aa}.
We observe large  difference in the predictions due to the choice of the DPDFs, which indicates sensitivity and potential of diffractive dijet measurement at the EIC to DPDFs, particularly  at large values of $z_{\p}^{\rm obs}$.  

\subsubsection{Large-\texorpdfstring{$|t|$}{t} diffractive production of vector mesons}

Among the diffractive processes, the production of heavy vector mesons is particularly interesting.
Diffractive production of vector mesons is a great testing ground for details of the QCD dynamics, especially the interplay between soft and hard phenomena. 
This process is usually described in terms of a colorless exchange, with vacuum quantum numbers, which to lowest order is given by two-gluon exchange. 
At higher orders, this process is described by the $t$-channel exchange of a gluon ladder, which is often referred to as the perturbative Pomeron~\cite{Balitsky:1978ic,Kuraev:1977fs}.
At high energies, the exchange of the perturbative Pomeron leads to a significantly faster increase with energy of the scattering amplitude than in the soft regime.
Hence, 
a very important question in high-energy QCD is the energy dependence of the Pomeron on the size of the rapidity gap and its dependence on the momentum transfer $t$.

The exclusive channel, which dominates the region $|t| \le 1 \, {\rm GeV}^2$, is considered in detail elsewhere in this report. 
Here, we focus on the diffractive production of vector mesons at high $|t|$. 
In this case the proton usually does not stay intact but rather dissociates into a low mass excitation (which is however much larger than for low-$|t|$ diffraction). 
Such a process can be identified by the presence of a large rapidity gap between the heavy meson and the system produced in the fragmentation of partons knocked out from the target.   
An advantage of this class of processes is that there are two perturbative scales, of similar size,  which are present at both ends of the Pomeron, thus largely suppressing the diffusion of transverse momenta along the gluon ladder into the non-perturbative regime.

This is an excellent situation to investigate the energy dependence of the vacuum-exchange amplitude. In fact, the dependence of the cross section on the rapidity gap is directly converted into the intercept of the Pomeron exchange at a given $t$.
Roughly speaking, the dependence on the rapidity gap of the cross section should scale as $2(\alpha_{P}(t)-1)$, which should be about $0.4-0.5$ for the BFKL Pomeron as compared to $0.2$ for the soft regime.

The HERA detectors had a rather limited rapidity acceptance and therefore could not measure directly, for this process, the dependence of the cross section on the rapidity gap. 
As a result, the determination of the energy dependence of the Pomeron amplitude was sensitive to details of the $t$-dependence of the amplitude and also masked by the convolution with the $x$-dependence of the parton density in the target which was integrated over.

The EIC, with suitable detector(s), will have the potential to measure directly the rapidity gap, and thus the energy dependence of the cross section as a function of the gap size.
There are two different strategies that can be considered. In the first case, one can require the direct observation of the size of the rapidity gap. In this case one would require an activity in the central detector up to a certain angle. Detailed studies~\cite{Deak:2020mlz} show that an angle of about $4^{\circ}$, corresponding roughly to the rapidity $\eta = 3.3$, would already allow for a range of rapidity gaps. In the second scenario, the size of the gap is unknown since there is no activity in the forward part of the detector. 
In that scenario, only a lower limit on the gap size can be imposed.

Detailed estimates~\cite{Deak:2020mlz} found that the much higher luminosity of the EIC than of HERA compensates to some extent for the lower energy at the EIC. 
As a consequence, at the EIC one can test the cross section dependence on rapidity-gap intervals predicted by the BFKL model for rapidity gaps up to four units in rapidity.
A good detector acceptance in the nucleon fragmentation region for such studies is crucial. 
It was found~\cite{Deak:2020mlz} that an  acceptance up to $\eta=3.5$ is sufficient for this process, although acceptance up to higher rapidity (for example, $\eta=4.5$) would provide a longer lever arm allowing for more stringent tests of the small-$x$ dynamics and the Pomeron.
Apart from $J/\psi$ production, the rapidity-gap production of $\rho$-mesons maybe also very promising, perhaps even over a broader $|t|$-range.

\subsection{Global event shapes and the strong coupling constant}
\label{part2-subS-PartStruct-GlobalEvent}

%
%

\newcommand{\vect}[1]{\mathbf{#1}}
\newcommand{\abs}[1]{\left\lvert #1\right\rvert}
\newcommand{\eq}[1]{Eq.~\eqref{eq:#1}}
\newcommand{\eqs}[2]{Eqs.~\eqref{eq:#1} and \eqref{eq:#2}}
\newcommand{\nn}{\nonumber}
\newcommand{\fig}[1]{Fig.~\ref{fig:#1}}


\definecolor{darkgreen}{RGB}{50,150,50}
\newcommand{\redtext}[1]{\textcolor{red}{#1}}
\newcommand{\bluetext}[1]{\textcolor{blue}{#1}}
\newcommand{\greentext}[1]{\textcolor{darkgreen}{#1}}
\newcommand{\magentatext}[1]{\textcolor{magenta}{#1}}

\newcommand{\alphas}{\ensuremath{\alpha_\mathrm{s}}}
\newcommand{\tauonea}{\ensuremath{\tau_{1}^{a}}}


\newcommand{\Ee}{\ensuremath{\mathrm{E}_e}}
\newcommand{\Ep}{\ensuremath{\mathrm{E}_p}}

\subsubsection{Introduction}

Event shapes~\cite{Dasgupta:2003iq} are global measures of the momentum distribution of hadrons in the final state of a collision, using a single number to characterize how well collimated the hadrons are along certain axes. This simple and global nature makes them highly amenable to high-precision theoretical calculations and convenient for experimental measurements. They then become powerful probes of QCD predictions, the strong coupling $\alpha_s$, hadronization effects, etc. 

The classic example, for collisions $\ee\to X$ , is \emph{thrust}~\cite{Brandt:1964sa,Farhi:1977sg},
\begin{equation}
\label{eq:thrustee}
\tau = 1-T,\quad\text{where}\quad T = \frac{1}{Q}\max_{\hat{\boldsymbol{t}}}\sum_{i\in X}\abs{\hat{\boldsymbol{t}}\cdot \boldsymbol{p}_i} =  \frac{2}{Q}p_z^A\,,
\end{equation}
at a center-of-mass collision energy $Q$, summing the three-momenta $\boldsymbol{p}_i$ of all final-state hadrons $i\in X$ projected onto the thrust axis $\hat{\boldsymbol{t}}$, which is defined as the axis maximizing the sum. It is customary to use $\tau=1-T$, whose $\tau\to 0$ limit describes pencil-like back-to-back two-jet events, and which grows as the jets broaden, up to the limit $\tau=1/2$ for a spherically symmetric final state. Other examples of two-jet event shapes in $\ee$ are broadening $B$~\cite{Catani:1992jc}, $C$-parameter~\cite{Ellis:1980wv}, and angularities~\cite{Berger:2003iw,Berger:2004xf}.

The thrust axis $\hat{\boldsymbol{t}}$ determines two hemispheres $A$ and $B$ in the $\pm z$ directions, 
and $p_z^A$ in \eq{thrustee} is the total $z$ momentum in the $+z$ hemisphere. In DIS, a natural division of the final state into hemispheres occurs in the Breit frame, with the incoming current defining the $z$ direction. Accordingly, the DIS thrust $\tau_Q$ has been defined as~\cite{Antonelli:1999kx}
\begin{equation}
\label{eq:tauQ}
    \tau_Q \overset{\text{Breit}}{=}1 - \frac{2}{Q}\sum_{i\in\mathcal{H}_C}p_z^i\,,
\end{equation}
where $Q$ is now the DIS variable $Q$, and $\mathcal{H}_C$ is the ``current'' hemisphere in the Breit frame. This $\tau_Q$ does in fact have a Lorentz-invariant definition, in terms of the class of ``$N$-jettiness'' observables~\cite{Stewart:2010tn},
\begin{equation}
    \tau_N = \frac{2}{Q^2} \sum_{i\in X} \min\{q_B\cdot p_i,q_1\cdot p_i,\dots,q_N\cdot p_i\}\,,
    \label{eq:tauN}
\end{equation}
where $q_B$ is a a four-momentum vector in the proton beam direction (for DIS) and $q_{1,\dots, N}$ are four-momenta in $N$ ``jet'' directions in the final state along which one wishes to measure collimation of hadrons. 
The min operator in \eq{tauN} groups these hadrons into $N+1$ regions around the $q_i$.
One will find $\tau_N\to 0$ for $N+1$ perfectly collimated jets (and beam radiation).

We will focus on the simplest case for DIS, namely 1-jettiness~\cite{Kang:2012zr, Kang:2013wca, Kang:2013nha, Kang:2013lga}. There is a freedom to define the  vectors $q_B$ and $q_1$ (which we will now call $q_J$). Different choices of directions and normalizations give different measures of 1-jettiness. The choice
\begin{equation}
    q_B^b = xP\,,\quad q_J^b = q+xP\,,
\end{equation}
where $P$ is the incoming proton momentum, $x$ is the Bjorken variable, and $q$ is the current momentum, actually gives the same thing as $\tau_Q$,
\begin{equation}
\label{eq:tau1b}
    \tau_1^b \equiv \frac{2}{Q^2} \sum_{i\in X}\min\{q_B^b\cdot p_i\,, q_J^b\cdot p_i\} = \tau_Q\,.
\end{equation}
The label $b$ comes from notation used in Ref.~\cite{Kang:2013nha}. Momentum conservation leads to the last equality in \eq{tau1b} with $\tau_Q$ as defined in \eq{tauQ}. Note that $q_J^b = q+xP$ is the momentum the outgoing jet (quark) would have at Born (tree) level. Nonzero $\tau_1^b$ measures the deviations and broadening of the jet momentum and structure away from this Born limit. Another version of DIS 1-jettiness that we consider uses a $q_J$ that is adjusted to align with the physical jet momentum,
\begin{equation}
q_B^a = xP\,,\quad q_J^a = \abs{\boldsymbol{P}_J}(1,\hat{n}_J) \,,
    \label{eq:tauonea}
\end{equation}
where the jet momentum $\boldsymbol{P}_J$ and its direction $\hat n_J = \boldsymbol{P}_J/\abs{\boldsymbol{P}_J}$ may be found by a suitable algorithm, such as anti-$k_t$ \cite{Cacciari:2008gp}, or minimization over axes such as in $\ee$ thrust. 
For small enough $\tau_1^a$, the differences in $\tau_1^a$ measured using different infrared-collinear-safe algorithms will be power-suppressed, as long as they group  the same collinear, energetic particles into $P_J$. The difference between the true and Born-level jet axes used in $\tau_1^{a,b}$, however, is a leading-order effect \cite{Kang:2013nha}. Computing or measuring $\tau_1^a$ requires measuring particles in both beam and current hemispheres. Computing or measuring $\tau_1^b$, by contrast, according to \eq{tauQ} only requires a measurement of particles in the current hemisphere.


Here we shall consider the promise of 1-jettiness as a probe for the strong coupling $\alpha_s$ and of hadronization effects. Refs.~\cite{Kang:2012zr,Kang:2013wca} also explored 1-jettiness as a probe of nuclear PDFs and medium effects, another potentially powerful application of such observables at the EIC.

\subsubsection{Theoretical precision}
\label{Sect:TheoryPrecision}


{\bf Theoretical methods:} The global nature of event shapes such as those introduced above means that all collinear and soft radiation is probed with a single parameter. 1-jettiness distributions are sensitive to physics at three scales, hard $\mu_H = Q$, collinear $\mu_J = Q\sqrt{\tau_1}$, and soft $\mu_S=Q\tau_1$. In fixed-order perturbative predictions in QCD, logs of ratios of these scales appear at every order in $\alpha_s$.
For small $\tau_1$, these logs blow up at any fixed order in $\alpha_s$, and must be resummed to all orders. This is accomplished by factorization of the logs into pieces that depend on only one of the physical scales (hard, collinear, soft) at a time, and renormalization group (RG) evolution of each set of factorized contributions. 

Traditional methods in perturbative QCD (``direct'' QCD) have been used successfully to resum logs in event shapes (e.g., Refs.~\cite{Catani:1992ua,Dasgupta:2002dc,Almeida:2014uva}), but more recent methods use the technology of effective field theory to do so, namely, soft collinear effective theory (SCET)~\cite{Bauer:2000ew,Bauer:2000yr,Bauer:2001ct,Bauer:2001yt,Bauer:2002nz}. SCET has successfully been used to resum certain event shapes to N$^3$LL accuracy (e.g., Refs.~\cite{Abbate:2010xh,Hoang:2015gta}), including DIS 1-jettiness $\tau_1^a$ and $\tau_1^b$ (i.e., DIS thrust $\tau_Q$)~\cite{Kang:2015swk}. For these observables, SCET predicts the factorized cross sections, e.g., for $\tau_1^b$,
\begin{align}
    \label{eq:tau1bcs}
    \frac{d\sigma}{dx\,dQ^2\,d\tau_1^b} &= \frac{d\sigma_0}{dx\,dQ^2}\int dt_J dt_B dk_S d^2\boldsymbol{p}_T \,\delta\Bigl(\tau_1^b - \frac{t_J+t_B}{Q^2} - \frac{k_S}{Q}\Bigr) S(k_S,\mu)  \\
    &\quad\times J_q(t_J-\boldsymbol{p}_T^2)\sum_q H_q(y,Q^2,\mu)\mathcal{B}_q(t_B,x,\boldsymbol{p}_T^2,\mu)  + \sigma_{\text{ns}}(x,Q^2,\tau_1^b)\,,  \nn
\end{align}
where 
\begin{equation}
    \frac{d\sigma_0}{dx\,dQ^2} = \frac{2\pi\alpha_{\text{em}}^2}{Q^4} [(1-y)^2 +1]
\end{equation}
is the Born-level cross section, with $xys=Q^2$, and the sum over $q$ is over quark and antiquark flavors. (For $\tau_1^a$, $J_q$ depends only on $t_J$, and $\boldsymbol{p}_T$ is integrated over to turn the TMD beam function $\mathcal{B}_q$ into the ordinary beam function $B_q$~\cite{Kang:2013nha}.) The first factorized term in~\eq{tau1bcs} predicts the singular part of the $\tau_1^b$ distribution and will resum all the singular logs, while the final term $\sigma_\text{ns}$ is the nonsingular part of the distribution, predicted in fixed-order perturbation theory, to which the resummed prediction must be matched for large $\tau_1$. The factors $H_q,J_q,\mathcal{B}_q,S$ are universal factors that appear in predictions of many different observables. $H_q$ is the hard function coming from integrating out hard virtual fluctuations from QCD to match onto SCET and contains logs of $\mu/Q$. $J_q$ describes the collinear final-state radiation in the outgoing (current) jet giving it an invariant mass $t_J$ [and is independent of (light) quark flavor], and $\mathcal{B}_q$ the collinear radiation from the incoming proton beam with virtuality $t_B$ and transverse momentum $\boldsymbol{p}_T$. Both $J_q$ and $\mathcal{B}_q$ sum logs of $\mu/(Q\sqrt{\tau})$. The soft function $S$ sums the wide-angle radiation carrying momentum $k_S$ between the beam and jet, and depends on the smallest perturbative physical scale, summing logs of $\mu/(Q\tau)$.
The beam function 
satisfies a matching condition onto ordinary PDFs with a perturbatively calculable matching coefficient \cite{Stewart:2009yx,Stewart:2010pd,Stewart:2010qs,Gaunt:2014cfa,Gaunt:2014xga,Gaunt:2014xxa}.

The factorization of the singular parts of the $\tau_1$ cross section in \eq{tau1bcs} allows for resummation of the large logs of $\tau_1$. In terms of the integrated distributions $\sigma_c(\tau_1) = (1/\sigma_0)\int_0^{\tau_1} d\tau\,d\sigma/d\tau$,
these logs organize in the form
\begin{equation}
\label{eq:Laplaceexp}
    \sigma_c(\tau_1) = C(\alpha_s)e^{[Lg_\text{LL}(\alpha_s L) + g_\text{NLL}(\alpha_s L) + \alpha_s g_\text{NNLL}(\alpha_S L) + \cdots]} + D(\alpha_s,\tau_1)\,,
\end{equation}
where $L = \ln\tau_1$, $C$ is a constant coefficient, and  
$D$ contains the nonsingular terms.
Dependence on other kinematic variables is not shown in this schematic formula.
Resummation schemes determine the functions $g_\text{N$^k$LL}$ order by order, each summing an infinite tower of logs in the fixed-order expansion of $\sigma_c$.
This is achieved by solving the RG evolution equations for each piece of the factorized cross sections, each obeying
\begin{equation}
\label{eq:RGE}
   \mu\frac{d}{d\mu} F(x,\mu) = \gamma_F(\mu) F(x,\mu) \quad \Rightarrow\quad  F(x,\mu) = F(x,\mu_0) \exp\Bigl[ \int_{\mu_0}^\mu \frac{d\mu'}{\mu'}\gamma_F(\mu')\Bigr]\,,
\end{equation}
where $F = H$, $\mathcal{B}$, $J$,  $S$ (or, in this form, their Laplace/Fourier transforms), $x$ here is the natural variable for each function, and $\gamma_F$ is the anomalous dimension.
The solution for $F$ allows one to evaluate each function $F$ at a scale $\mu_0$ where the logs within each are minimized, and the large logs of $\mu/\mu_0$ at any other scale $\mu$ are summed into the exponential. The order to which these logs are resummed is determined by the accuracy to which the anomalous dimensions are known, each of which takes the form
\begin{equation}
    \gamma_F(\mu) = -\kappa_F\Gamma_\text{cusp}[\alpha_s]\ln\frac{\mu}{Q_F} + \gamma_F[\alpha_s]\,,
\end{equation}
where $\kappa_F$ is a constant, $\Gamma_\text{cusp}$ is a universal ``cusp'' anomalous dimension, $Q_F$ is the natural physical scale for the function $F$ [e.g., $Q_H=Q, Q_{J,B} = Q/(e^{\gamma_E}\nu)^{1/2}, Q_S = Q/(e^{\gamma_E}\nu)$, where $\nu$ is the Laplace transform variable for $\tau_1$], and $\gamma_F[\alpha_s]$ is the ``non-cusp'' part of the anomalous dimension. 
The order to which each piece is needed at a given accuracy is summarized in, e.g., Ref.~\cite{Almeida:2014uva}.

To date, the cusp anomalous dimension is known to four-loop ($\alpha_s^4$) accuracy~\cite{Korchemsky:1987wg,Korchemskaya:1992je,Moch:2004pa,Henn:2019swt}, and the non-cusp anomalous dimensions for the $\tau_1$ beam, jet, and soft functions to three-loop ($\alpha_s^3$) accuracy. The hard~\cite{Matsuura:1988cn,Matsuura:1988sm,Gehrmann:2005pd,Moch:2005id,Baikov:2009bg,Lee:2010cga} and jet functions~\cite{Lunghi:2002ju,Bauer:2003pi,Becher:2006qw,Moch:2004pa,Bruser:2018rad} themselves are known to $\alpha_s^3$, and the beam~\cite{Gaunt:2014xga,Gaunt:2014xxa} and soft functions to $\alpha_s^2$ accuracy~\cite{Kang:2015moa,Kelley:2011ng,Monni:2011gb,Hornig:2011iu}. 
This makes it possible to compute DIS 1-jettiness $\tau_1^{a,b}$ distributions currently to N$^3$LL resummed accuracy (see, e.g., Ref.~\cite{Kang:2015swk}). Furthermore, the non-singular part of the $\tau_1^b$ distribution has been computed analytically to first nontrivial order $\mathcal{O}(\alpha_s)$ in Ref.~\cite{Kang:2014qba}. The non-singular part of the 1-jettiness distribution in Ref.~\cite{Kang:2013lga} was also computed numerically to $\mathcal{O}(\alpha_s)$.

\begin{figure}
    \centering
    \includegraphics[width=.32\textwidth]{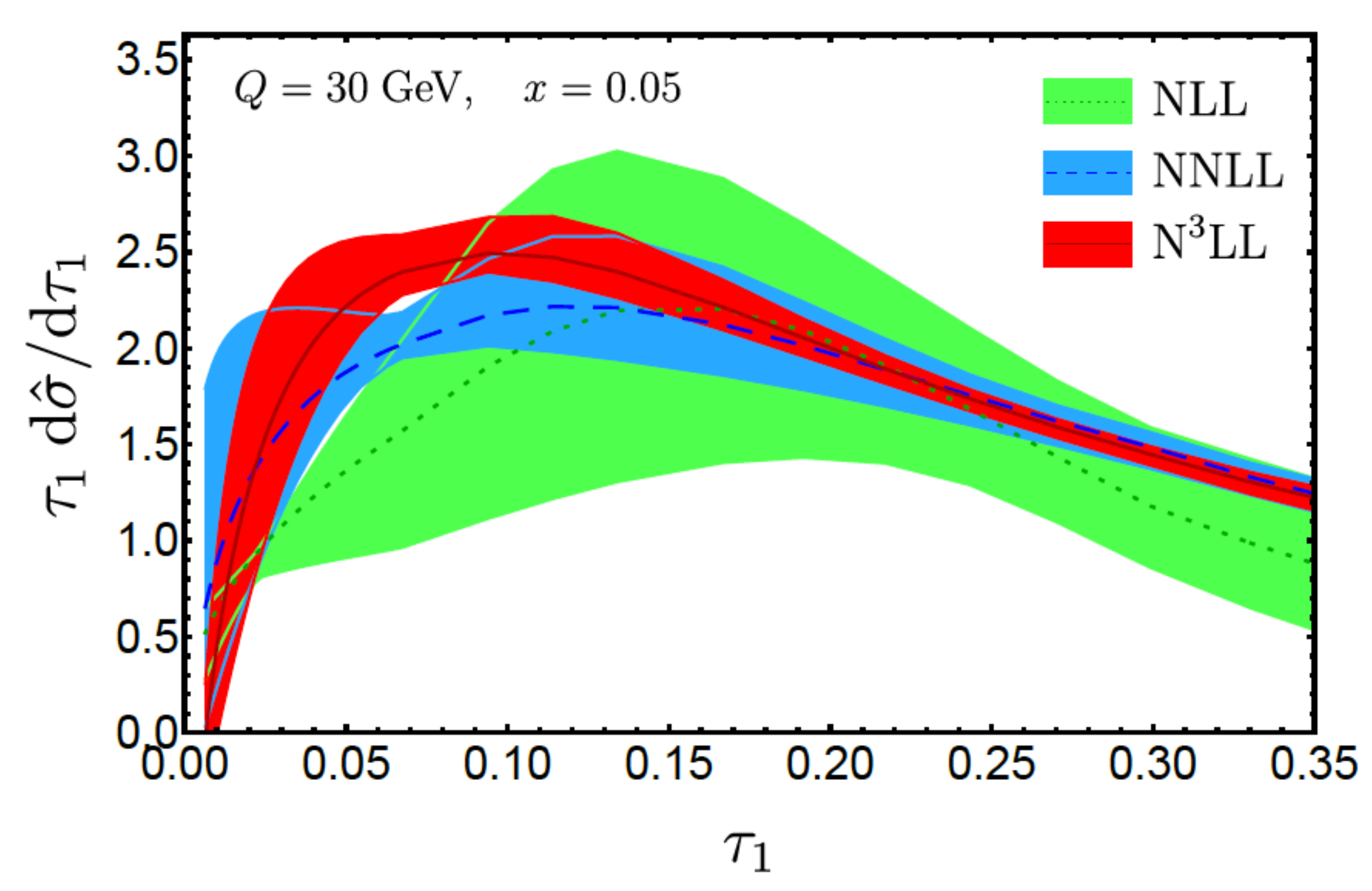}  \includegraphics[width=.33\textwidth]{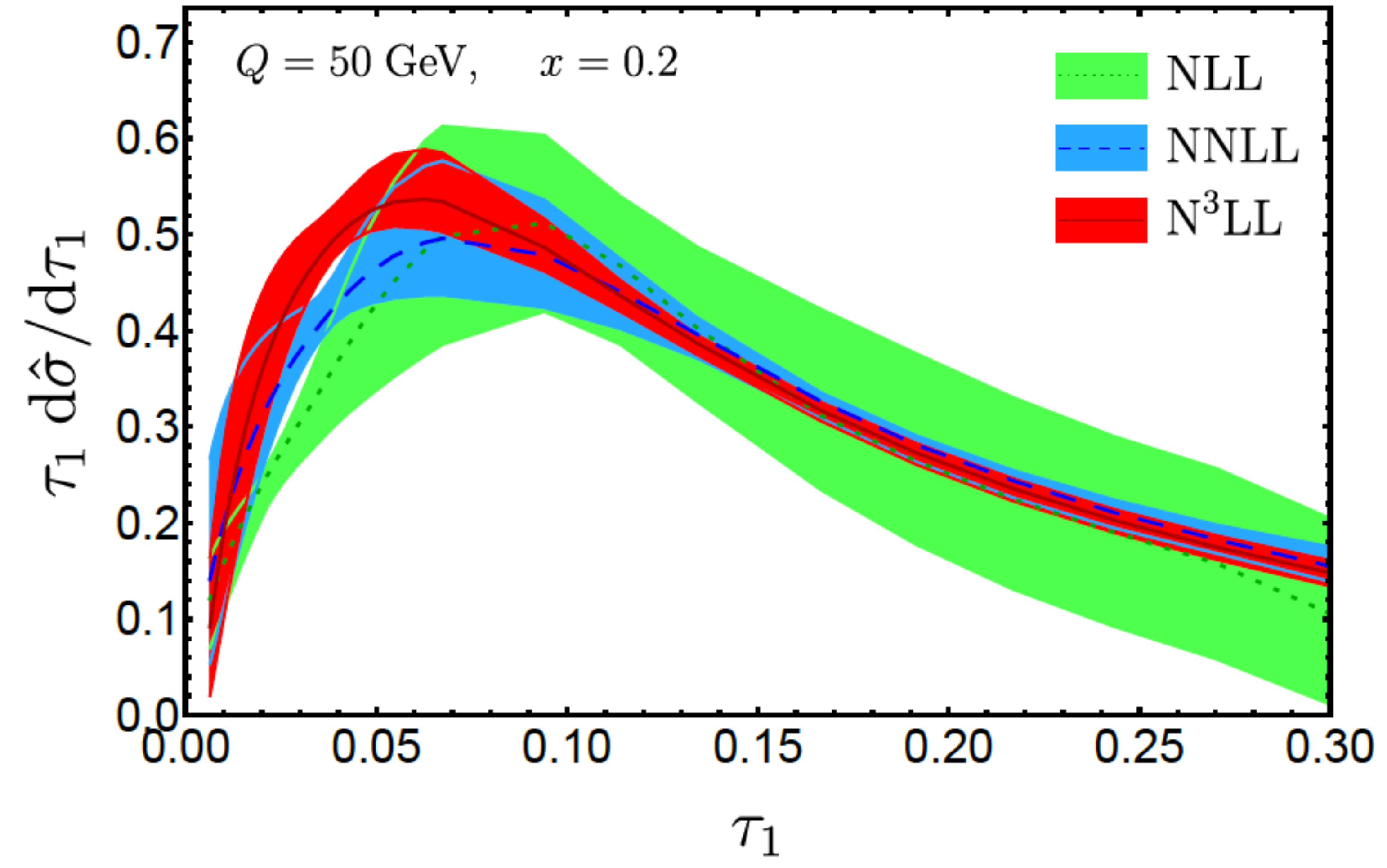}
    \includegraphics[width=.32\textwidth]{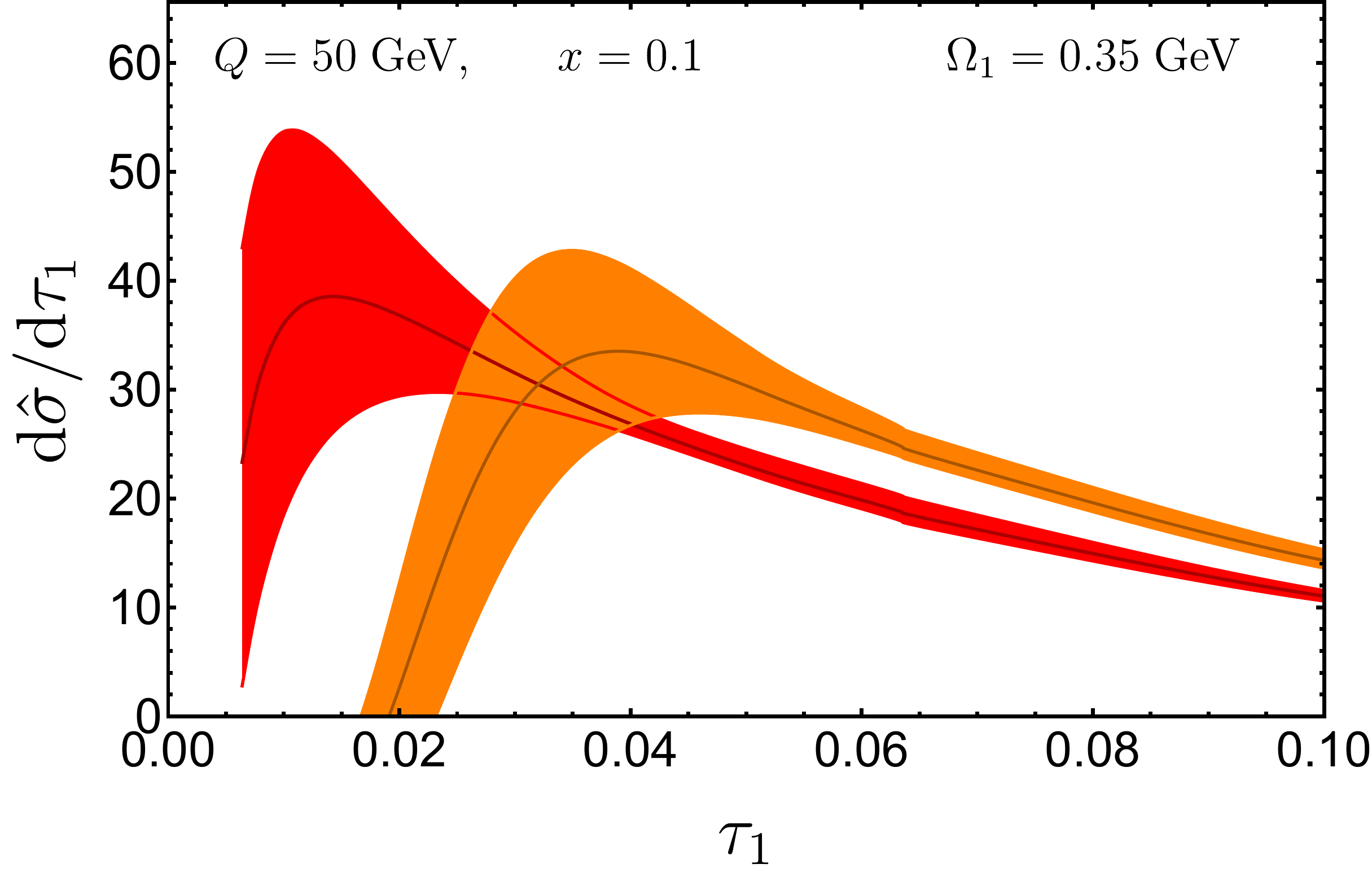}
    \caption{Theoretical predictions for 1-jettiness (or DIS thrust) $\tau_1^b$ distributions, from NLL to N$^3$LL accuracy at $Q=30$ GeV, $x=0.05$ (left) and $Q=50$ GeV, $x=0.2$ (center). The right-most panel shows a typical distribution before (red) and after (orange) convolution with a nonperturbative shape function, which for sufficiently large values of $\tau_1$ has the primary effect of shifting the perturbative distribution to the right, controlled by a universal shift (first moment) parameter $\Omega_1$, taken here to be $0.35$ GeV. A robust  determination of $\alpha_s$ from event shape measurements will also fit for $\Omega_1$ at the same time.}
    \label{fig:N3LL}
\end{figure}

In addition to the perturbative contibutions, the $\tau_1$ distributions receive nonperturbative corrections due to hadronization in the final state, growing more important for smaller $\tau_1$. These contribute to the soft function, which can be taken to have the form~\cite{Korchemsky:1999kt,Korchemsky:2000kp,Hoang:2007vb}
\begin{equation}
    S(k,\mu) = \int dk' S_\text{PT}(k-k',\mu) f_\text{NP}(k')\,,
\end{equation}
for a model function $f_\text{NP}$. For $k\sim\Lambda_\text{QCD}$ (i.e., $\tau_1\sim\Lambda_\text{QCD}/Q$), the full shape function $f_\text{NP}$ is needed to describe the distribution. For values $\tau_1\gg \Lambda_\text{QCD}/Q$, an OPE can be performed. For sufficiently large $\tau_1$, it can be shown that the first moment of the shape function is given by a universal quantity,
\begin{equation}
    \int_{-\infty}^\infty dk \,k\, f_\text{NP}(k) = \frac{2\Omega_1}{Q}\,,
\end{equation}
where $\Omega_1$ is defined through a universal vacuum matrix element of soft Wilson lines~\cite{Korchemsky:1999kt,Lee:2006nr}, and is the same for both versions of $\tau_1^{a,b}$ we have considered~\cite{Kang:2013nha}. 
For sufficiently large $\tau_1$, it simply shifts the distribution $d\sigma/d\tau_1$ to the right by $2\Omega_1/Q$ (see \fig{N3LL}) (cf.~Refs.~\cite{Korchemsky:1994is,Dokshitzer:1995zt,Dokshitzer:1995qm,Dokshitzer:1997ew}), and the value of $\Omega_1$ can be determined by a simultaneous fit for it and for $\alpha_s$ to the data. 

Further important corrections to these theoretical predictions can come from the effect of finite hadron masses on the nonperturbative corrections \cite{Salam:2001bd,Mateu:2012nk}, subtracting renormalon ambiguities between the perturbative and nonperturbative contributions \cite{Hoang:2007vb,Hoang:2008fs}, and computing the nonsingular corrections to higher order in $\alpha_s$. For direct comparison to experimental results, accounting for cuts on jet/hadron energies, rapidities, jet radii $R$, etc.~will also be necessary. 
Track-based observables should also be considered, for which theoretical technology has begun to be developed~\cite{Chang:2013iba,Chang:2013rca}, and which preliminary detector studies in Sec.~\ref{Sect:Experiment} suggest can be measured with greater experimental precision at the EIC.
These improvements will further enable the highest precision determination of $\alpha_s$ and $\Omega_1$ possible from DIS event shapes.


%
%

\begin{figure}
    \centering
    \includegraphics[width=.47\textwidth]{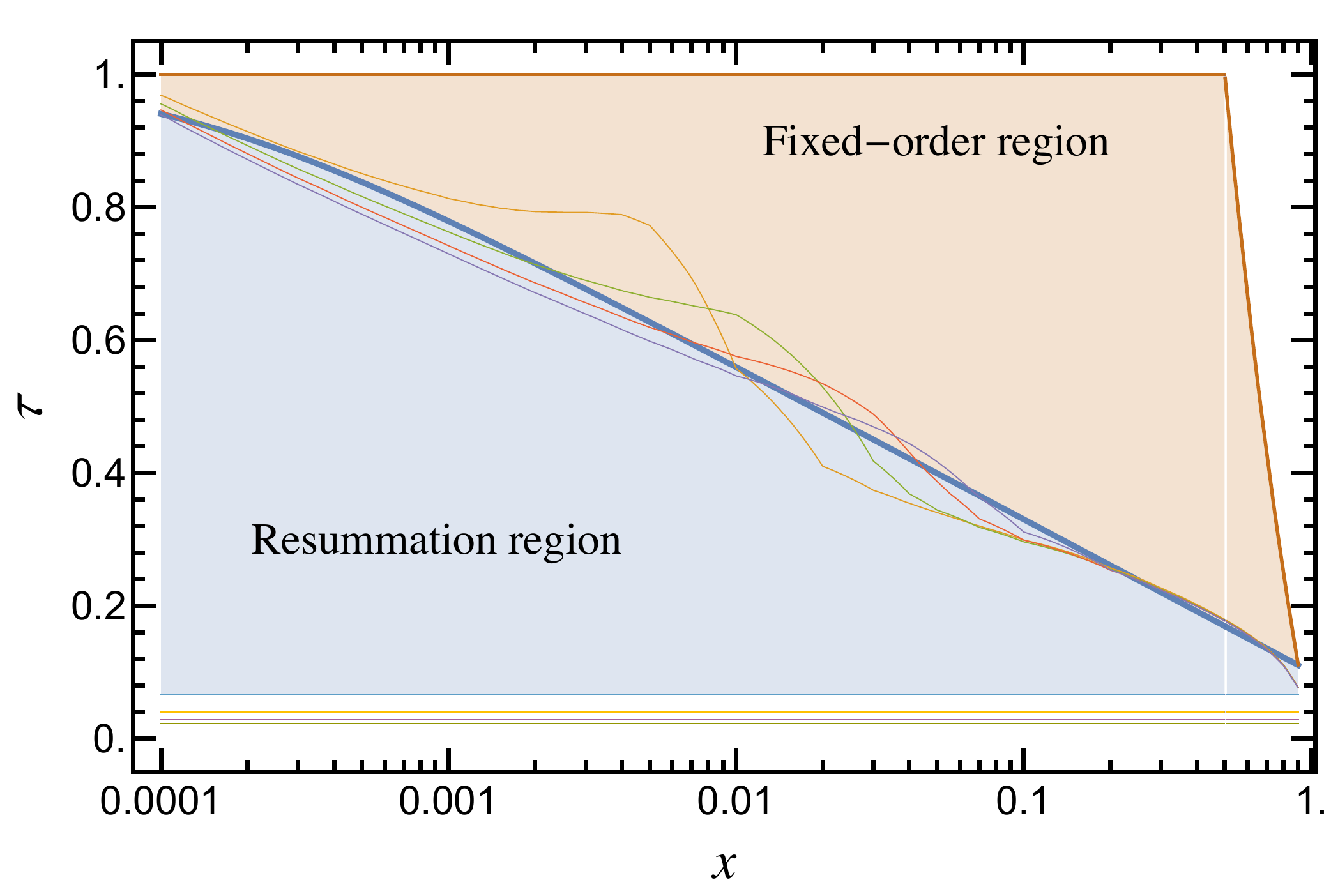} \ \includegraphics[width=.5\textwidth]{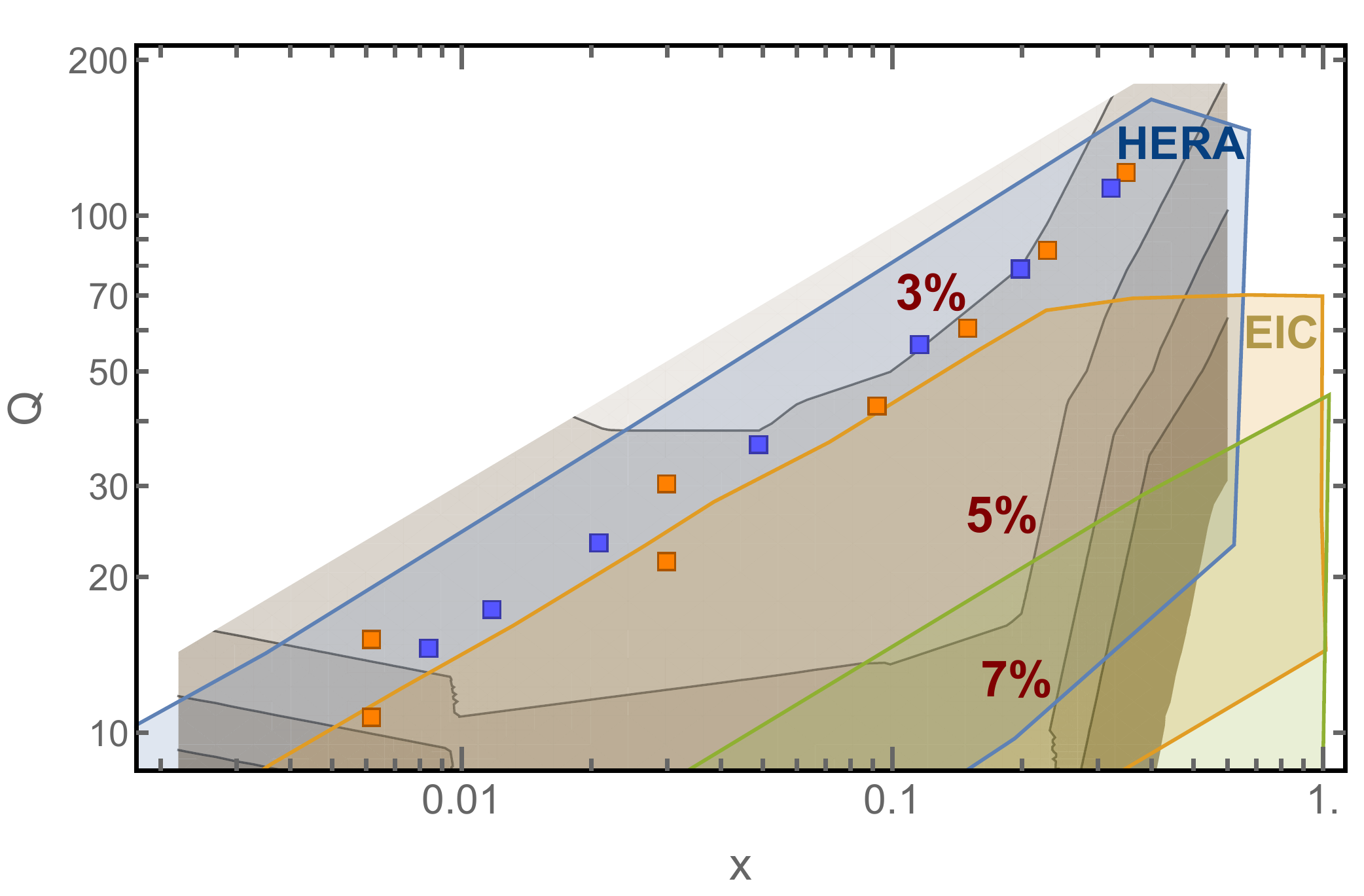} 
    \caption{Left: Region in $\tau=\tau_1^b$ over which resummed perturbation theory is expected to be more reliable than fixed-order perturbation theory, based on the value of $\tau$ at which singular logs and nonsingular terms become of comparable size \cite{Kang:2014qba}. Right: Contours of estimated theoretical uncertainty in $x,Q$ space, with coverage at HERA and (expected) EIC at high ($140 \, \textrm{GeV}$) and low ($45 \, \textrm{GeV}$) center-of-mass energies. Squares are values of $x,Q$ for previous HERA analyses of event shapes \cite{Aktas:2005tz,Chekanov:2006hv}.}
    \label{fig:resumregion}
\end{figure}

{\bf Predictions:} In \fig{N3LL} we show some of our predictions for $\tau_1^b$ distrbutions to N$^3$LL $+\mathcal{O}(\alpha_s)$ accuracy prepared for this Yellow Report. The uncertainties are estimated by varying the scales in resummed and fixed-order pieces in \eqs{tau1bcs}{RGE} \cite{Kang:2013nha,Kang:2014qba}.
The predictions also include the effect of a simple shape function, whose first moment is given by $2\Omega_1$, with $\Omega_1$ set to $0.35$ GeV.
One observes the good convergence in the perturbative region from one order to the next. In general, the theoretical uncertainties improve for larger $Q$
and, somewhat surprisingly, for smaller $x$. The resummation is turned off smoothly as $\tau_1$ grows large, and fixed-order predictions become more reliable than resummed.
This occurs around a value $\tau_1$ where the total contribution of the singular logs at fixed order in $\alpha_s$ becomes numerically comparable to the non-singular function, and based on $\mathcal{O}(\alpha_s)$ predictions \cite{Kang:2014qba}, this transition value turns out to be a function of $x$,  
see \fig{resumregion}.
This appears to be due to relative contributions of quark and gluon PDFs to the $\tau_1^b$ distribution as a function of $x$. This observation, however, is based on studies that do not yet include any resummation of logs of $x$ for very small $x$. 

In \fig{resumregion} we also plot contours of theoretical uncertainty in current N$^3$LL predictions for $\tau_1^b$ in $Q-x$ space, compared to coverages at HERA and the EIC, 
serving as a preliminary guide for where the best precision phenomenology might be expected. At present the best theoretical precision is achieved in a central region of $Q,x$ reflecting values that balance better perturbative behavior with smaller uncertainties from PDFs, as alluded to above. While the higher energy of HERA allowed coverage where our theoretical precision for $\tau_1$ is expected to be better, at EIC the opportunity exists to explore $x,Q$ values not analyzed by HERA, as well as to consider a larger set of event shapes under better theoretical control (for instance those, like $\tau_1^{a,b}$, that are free of non-global logs~\cite{Dasgupta:2002dc}), all of which will be important to reduce PDF uncertainties and to test universality of $\Omega_1$, not to mention higher statistics.

We have presented predictions for $\tau_1^b$ as this observable presently has the best theoretical accuracy available. Detector studies below will be presented for $\tau_1^a$ in Sec.~\ref{Sect:Experiment}. The theoretical predictions for $\tau_1^{a,b}$ would look substantially similar, with the small differences encoding effects of the transverse-momentum dependence of initial-state radiation (cf.~\eq{tau1bcs} and Ref.~\cite{Kang:2013nha}). In principle there is no obstacle to bringing predictions for the two observables to the same accuracy, and doing so is work in progress.

%
\begin{figure}
    \centering
    \includegraphics[width=.49\textwidth]{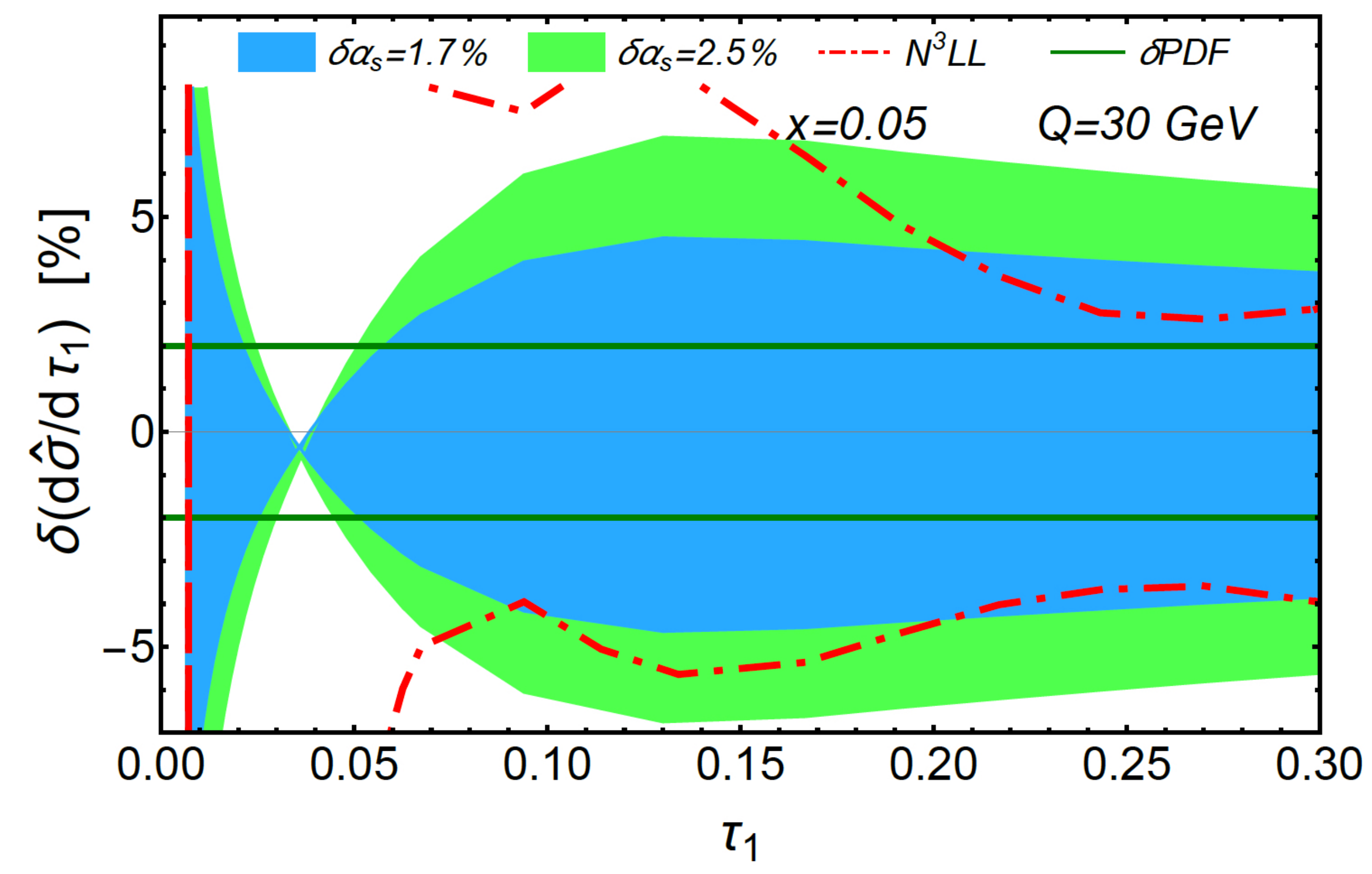}
    \includegraphics[width=.49\textwidth]{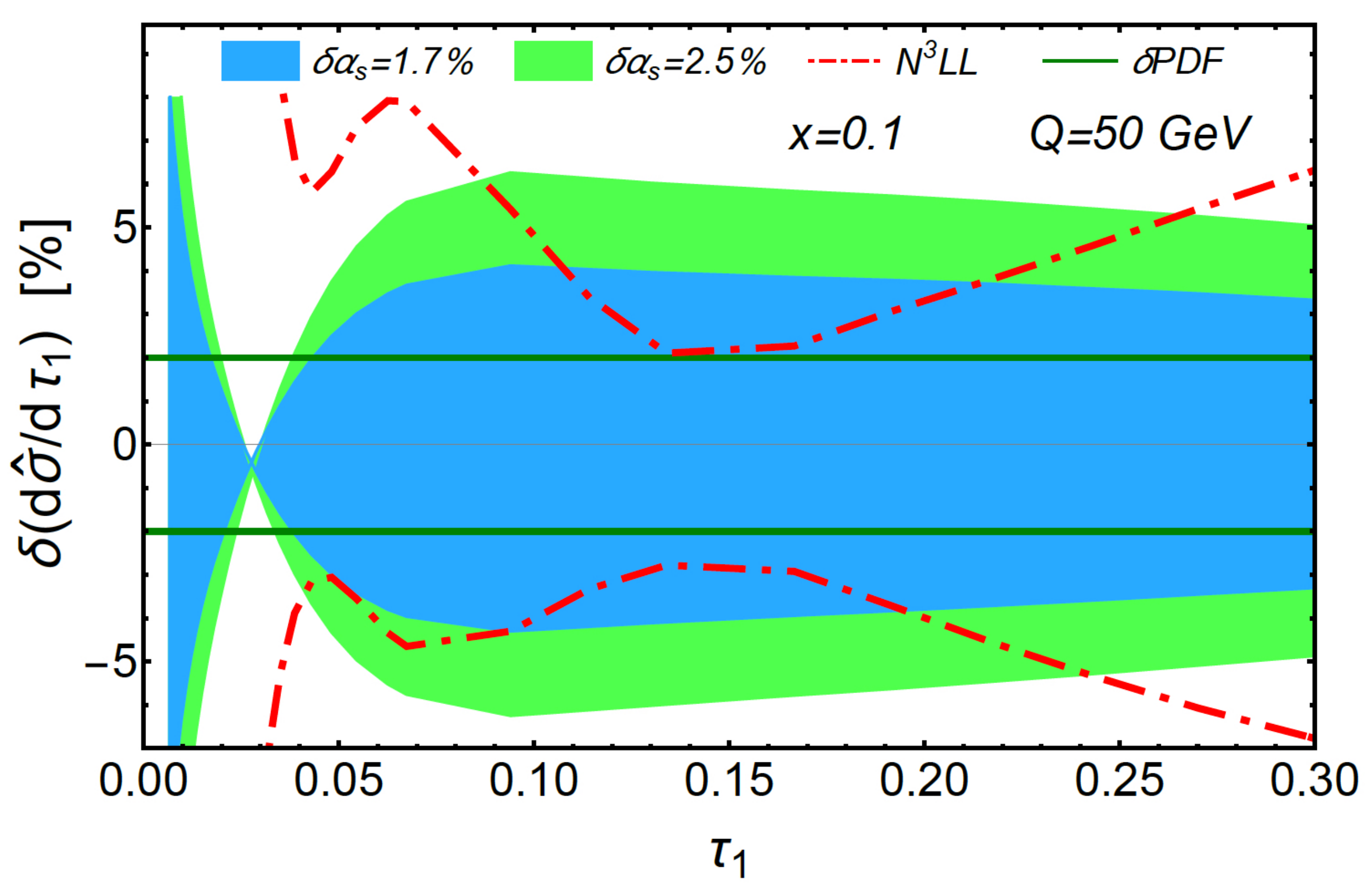}
    \caption{Current theoretical uncertainties in N$^3$LL$+\mathcal{O}(\alpha_s)$ predictions for $\tau_1^b$ (red dot-dashed) vs. the variations in the cross section from 1.7\% (blue) or 2.5\% (green) variations of $\alpha_s(M_Z)$ itself, along with the $1\sigma$ uncertainties in the MMHT 2014 PDF set used in these predictions (dark green), at two values $Q= 30$ GeV and $x=0.05$ (left) and $Q=50$ GeV and $x=0.1$ (right).}
    \label{fig:alphasvariation}
\end{figure}

{\bf Fitting for $\boldsymbol{\alpha_S}$:} In~\fig{alphasvariation} we illustrate the theoretical uncertainties in our N$^3$LL predictions for $\tau_1$ (red dot-dashed) vs.~the variations in the cross section from 1.7\% (blue) or 2.5\% (green) variations of $\alpha_s(M_Z)$ itself, along with the uncertainties in the MMHT 2014 PDF set~\cite{Harland-Lang:2014zoa} we used (dark green). These indicate that a single prediction and measurement of a $\tau_1$ distribution could yield a determination of $\alpha_S(M_Z)$ at the few percent level, with the prospect of using data from many $x,Q^2$ values only improving the ultimate sensitivity, competitive with current percent-level determinations from $\ee$ event shapes \cite{Zyla:2020zbs}. Using data from many $x,Q^2$ will also be important to combat PDF uncertainties.
The perturbative uncertainties in~\fig{alphasvariation} can be further significantly reduced by higher fixed-order calculations of the nonsingular contributions for large $\tau_1$.

The theoretical predictions depend sensitively on the value of $\alpha_s$ and, depending on the region of $\tau_1$, on nonperturbative corrections. As reviewed above, for large enough $\tau_1$, the dominant nonperturbative effect is a shift by $2\Omega_1/Q$. Thus many event-shape-based extractions~\cite{Davison:2008vx,Abbate:2012jh,Gehrmann:2012sc,Hoang:2015hka} of $\alpha_s$ involve  a two-parameter, $\{\alpha_s(Q_0), \Omega_1\}$, fitting procedure, using measurements in the relevant region of the distributions. The resulting analysis yields a correlation matrix that describes the degeneracy between the two parameters. Sampling a wide range in $x$-$Q$ is expected to play an important role in breaking this degeneracy. 

In practice, the theoretical predictions are integrated over a region of $x$-$Q$ and then binned in the event shape $\tau$. This way theoretical and experimental distributions are directly comparable. The fit can be performed within a single range of $x$-$Q$ or simultaneously for many/all ranges. This freedom is helpful for controlling the various systematics that may enter the theoretical distributions, such as PDF uncertainties and nonperturbative effects.  

Finally one needs to decide which region of the event shape spectrum (i.e., which bins in $\tau$) should be included in the analysis. This determination depends on many aspects, and the dependence of the fitting results on this  choice is a manifestation of the degeneracy of the two fitting parameters. 
Some deciding factors will be: i) the range of $x$-$Q$ which changes the location of the boundary between the resummation and fixed-order QCD regions (see~\fig{resumregion} left panel), ii) the treatment of non-perturbative corrections will play an important part in which bins are incorporated in the small-$\tau$ region. It is therefore important to have sufficiently fine binning in the observable  to allow for carefully choosing the region for which the most reliable extraction can be performed.

\section{Multi-dimensional Imaging of Nucleons, Nuclei, and Mesons}
\label{part2-sec-Imaging}

The multi-dimensional parton structure of nucleons, nuclei and mesons is a very important area of hadronic physics, and there is no doubt that the EIC can move this field to the next level.
Crucial in that regard will be (precision) measurements of certain exclusive and semi-inclusive processes.
Information about imaging in position space comes through form factors and, in particular, via generalized parton distributions (GPDs), whereas transverse momentum dependent parton distributions (TMDs) quantify the 3D parton structure of hadrons in momentum space.
Chapter~\ref{part2-chap-EICPhyCase} contains an introduction to those topics, while detailed discussions can be found below in Secs.~\ref{part2-subS-SecImaging-FF}, ~\ref{part2-subS-SecImaging-GPD3d} and~\ref{part2-subS-SecImaging-TMD3d}.
In what follows, a brief introduction is provided for the two additional topics of this section.
 
{\bf Wigner functions} can be considered the quantum-mechanical counterpart of classical phase space distributions.
In non-relativistic quantum mechanics they contain the same information as the wave function of a system.
Interestingly, Wigner functions can also be defined for partons in quantum field theory. 
A generic partonic Wigner function $W(x, \boldsymbol{k}_T, \boldsymbol{b}_T)$ depends on the longitudinal and transverse parton momenta as well as the impact parameter.
Therefore, Wigner functions not only contain all the physics encoded in TMDs and GPDs but also additional information.
For instance, they allow one to study spin-orbit correlations that are similar to the ones known from systems like the hydrogen atom.
The relation of the parton orbital angular momentum to a specific Wigner function is one example in that regard. 
For some time it was unclear if, even as a matter of principle, partonic Wigner functions can be measured.
But in the meantime some processes have been identified which are directly sensitive to those objects in a model-independent manner.
In relation to the EIC, at present the diffractive exclusive di-jet production is of particular interest, which holds promise to give access to Wigner functions for gluons.
More details about partonic Wigner functions and how the EIC can make significant contributions to this field are given in Sec.~\ref{part2-subS-SecImaging-Wigner}.

High-energy lepton scattering off {\bf light (polarized) nuclei} ($d$, $^3$He, $^4$He) typically serves a dual purpose.
First, both deuteron and $^3$He targets can be used to study the neutron, which is important for a complete picture of the nucleon.
This not only applies to PDFs, but also to GPDs and TMDs.
Nuclear corrections complicate the extraction of information about the neutron from light nuclei, but there exists decades-long expertise in this field to build on.
Second, obtaining information about the light nuclei is very interesting in its own right.
Topics include the investigation of the EMC effect in position space, exploring the pressure distributions in light nuclei, and exploiting the very unique opportunities which the spin-1 deuteron target offers through its possible tensor polarization. 
The prospects of this field at the EIC are summarized in Sec.~\ref{part2-subS-SecImaging-LpolNucl}.

\subsection{Nucleon and meson form factors}
\label{part2-subS-SecImaging-FF}

\subsubsection{Nucleon form factors}

The  electric and  magnetic form factors of the nucleon are measured  in  elastic electron-proton ($ep$) scattering experiments and provide a wealth of information about the radius and distributions of charge and magnetism 
in the nucleon~\cite{Punjabi:2005wq,Pacetti:2015iqa}.   
The future EIC will provide a unique opportunity to reach extremely high $Q^2$ and probe the transition from hadronic to partonic degrees of freedom.   At low $Q^2$, elastic scattering can serve as a straightforward reaction channel for experimentally determining luminosity by making use of the well known low-$Q^2$ electromagnetic form factors.
The cross section for the scattering process can be written as
\begin{equation}
    \sigma  =  \sigma_\mathrm{Mott} \times
     \left[\frac{G_E^2\left(Q^2\right)+\tau G_M^2\left(Q^2\right)}{1+\tau}+2\tau                           G_M^2\left(Q^2\right)\tan^2\left(\frac{\theta}{2}\right)\right] \text{,}
\label{elastic-gegm}
\end{equation}
where $G_E(Q^2)$ and $G_M(Q^2)$ are the charge and magnetic form factors, respectively, $Q^2$ is the four momentum transfer squared, 
and $\tau=\frac{Q^2}{4M^2}$ with the nucleon mass $M$.

\begin{figure}[htb]
\centering
\includegraphics[width=0.49\textwidth]{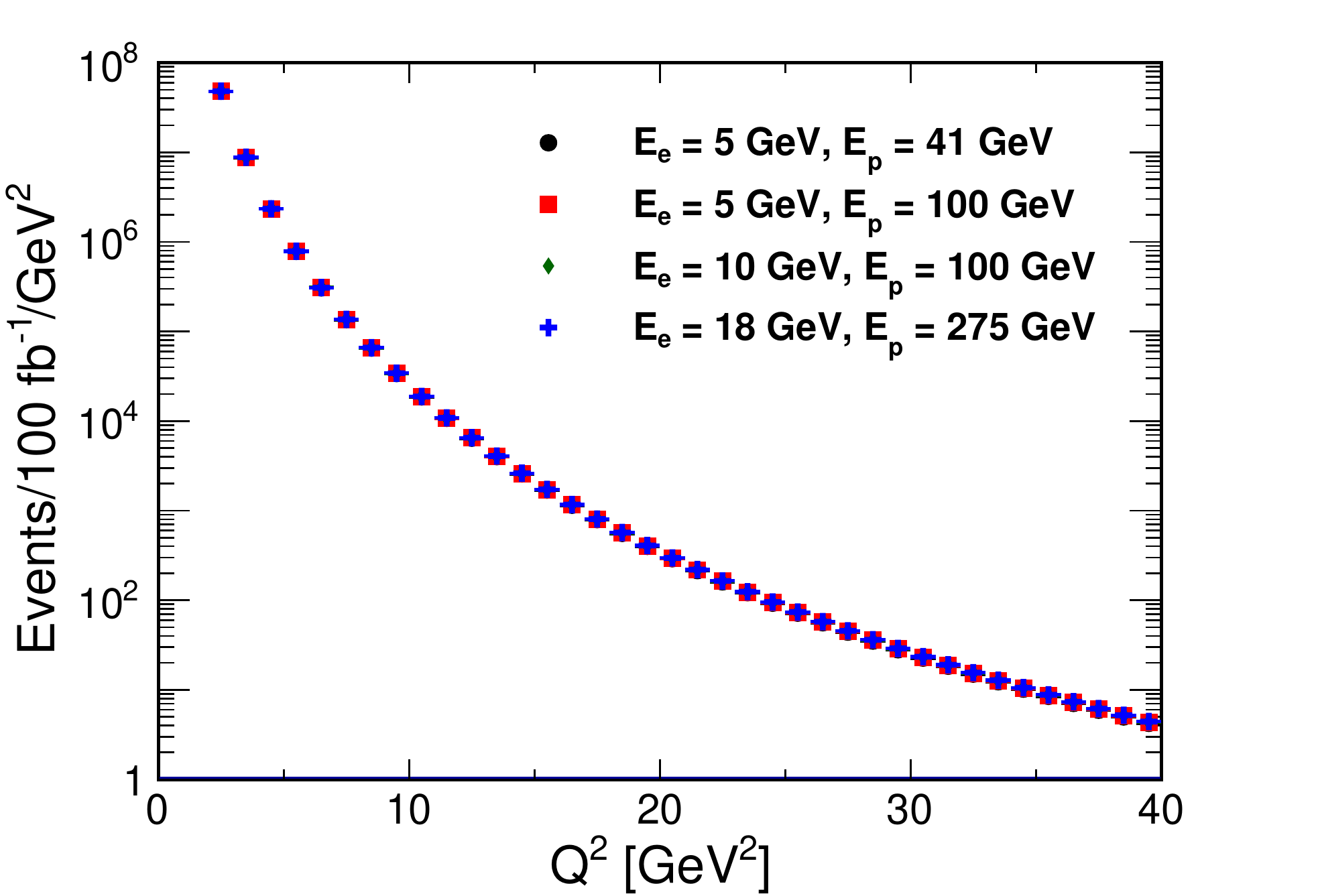}
\includegraphics[width=0.49\textwidth,page=1]{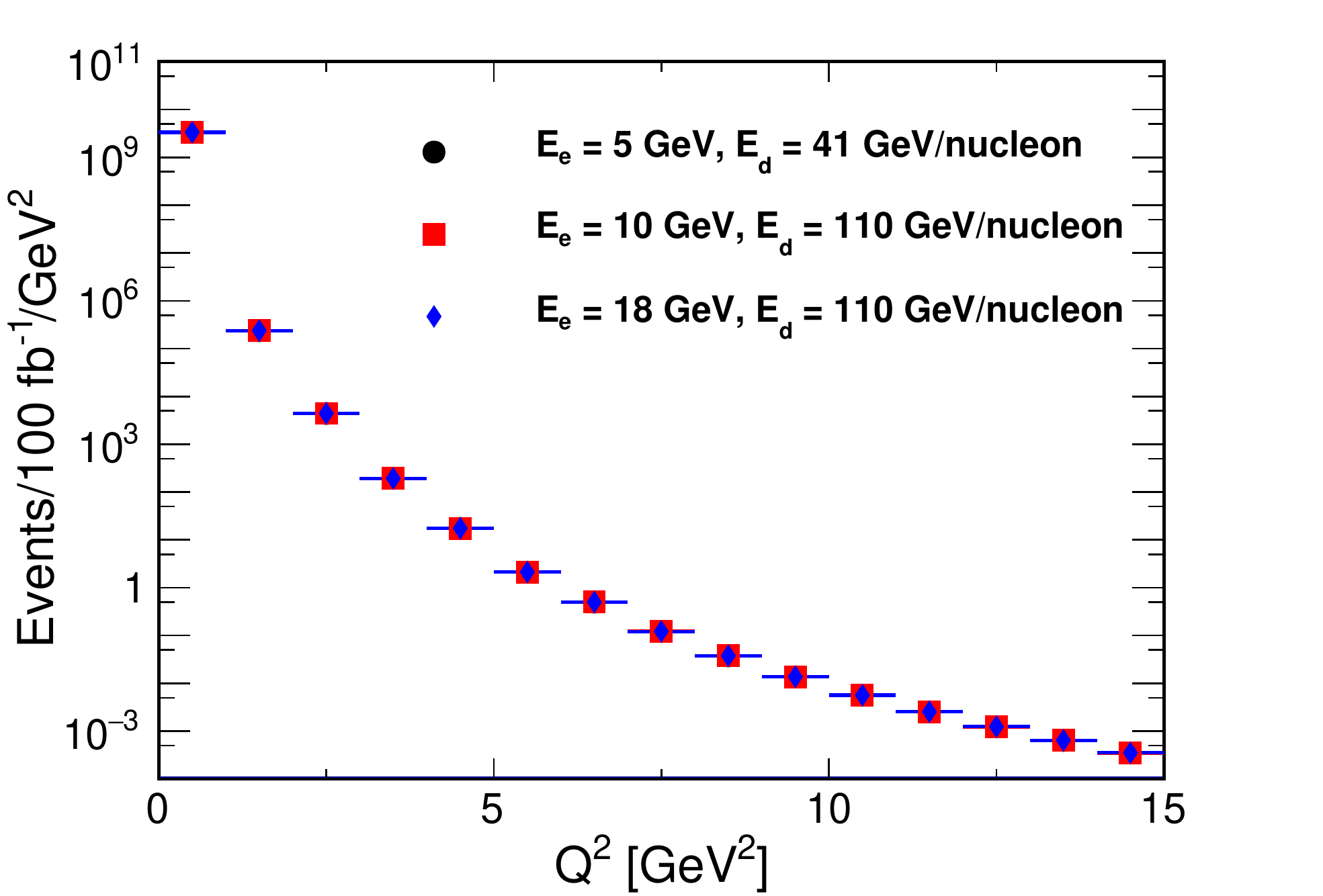}
\caption{The left panel shows the expected counts from elastic $ep$ scattering at the EIC.  The right panel shows the expected counts from elastic $ed$ scattering.  The lack of a large change in the rates for different beam energies implies that this data cannot be used for a Rosenbluth separation of the form factors.   On the other hand, as can be seen in Eq.~(\ref{elastic-gegm}), at large $Q^2$ the cross section is completely dominated by $G_M$; thus at large $Q^2$ the magnetic form factor can be extracted from the expected data.}
\label{fig:FormFactorRates}
\end{figure}

Measurements of the unpolarized $ep$ elastic cross section --- the cross section for the process where the final state consists of only an electron and proton with soft real-photon radiation --- will require both the scattered proton and scattered electron to be detected in order to separate these events from other processes.
As shown on the left side of Fig.~\ref{fig:FormFactorRates}, with the form factors parameterized by the form given in Ref.~\cite{Kelly:2004hm} and using the fit parameters in Ref.~\cite{Puckett:2017flj}, the EIC can potentially make measurements of the $ep$ elastic cross section up to $Q^2 \approx$ 40 $\gev^2$.  Though unlike fixed-target experiments, in collider kinematics, the data collected will be at values of the virtual photon longitudinal polarization $\epsilon \sim$ 1 where $\epsilon = [1+2(1+\tau )\tan^2(\theta/2)]^{-1}$.

\begin{figure}[htb]
    \centering
   \includegraphics[width=0.49\linewidth,page=1]{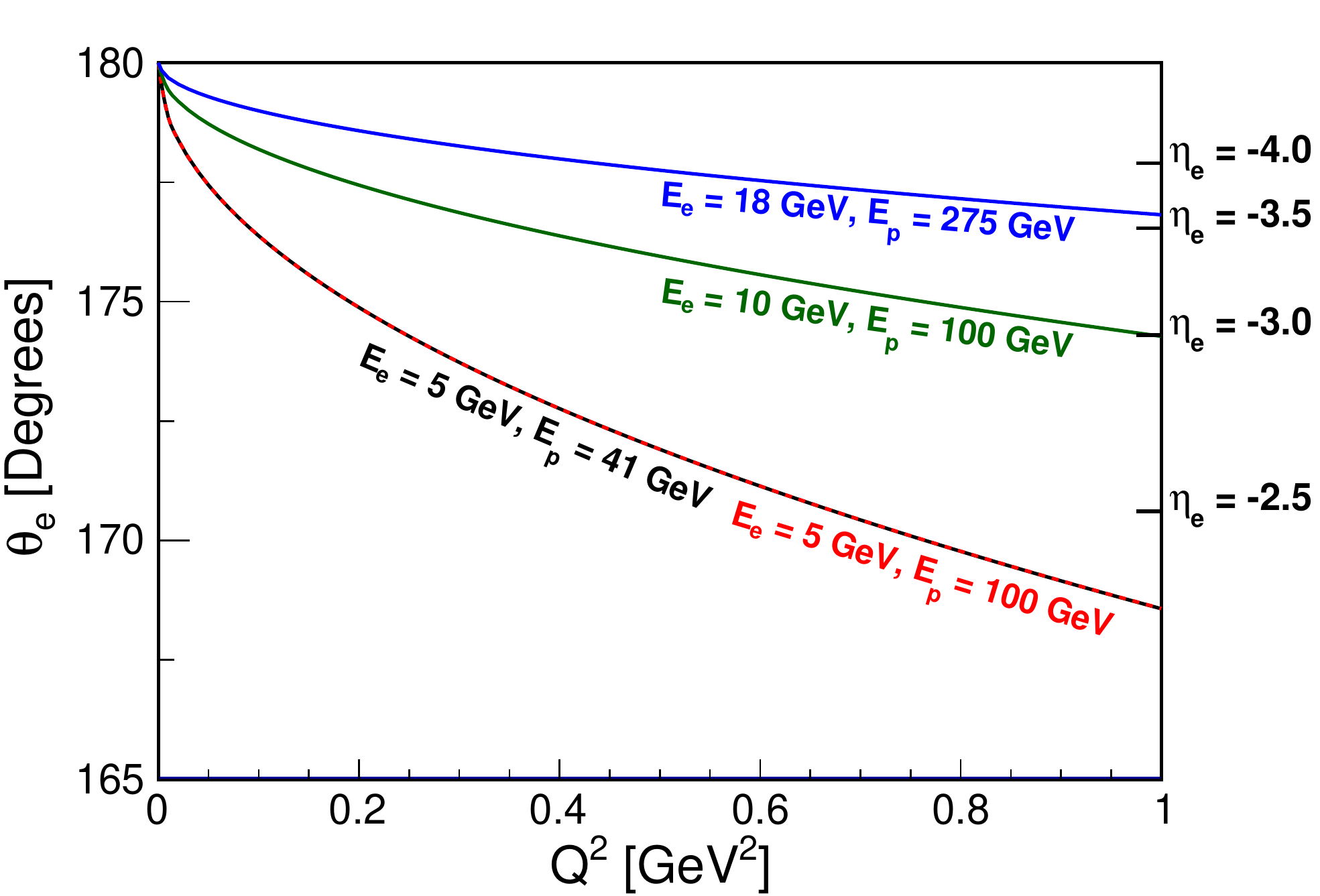}
    \includegraphics[width=0.49\linewidth,page=3]{PART2/Figures.EICMeasandStud/kin_yellow_deuteron.pdf}    
    \caption{Left figure shows the electron scattering angle and the right figure the proton scattering angle for different values of $Q^2$ and for different beam energy combinations.}
    \label{fig:barak-proton}
\end{figure}


If the central detector acceptance on the electron side extends down to $\eta = -3.5$, then the electron will enter the central acceptance at $Q^2$ below 1 $\gev^2$ (except for the highest $\sqrt{s}$ setting). 
In order to suppress inelastic backgrounds, 
it is necessary to detect the proton as well,
and far-forward detectors will be  needed for lower $Q^2$ (see Fig.~\ref{fig:barak-proton}).

Determination of the proton form-factor ratio at high $Q^2$ via double-spin asymmetry measurements will most likely not be possible due to the small expected asymmetry. However, the addition of a positron beam at the EIC would allow for the study of hard two-photon exchange effects at high $Q^2$~\cite{Accardi:2020swt}.

\begin{figure}[htb]
    \centering
    \includegraphics[width=0.49\textwidth,page=2]{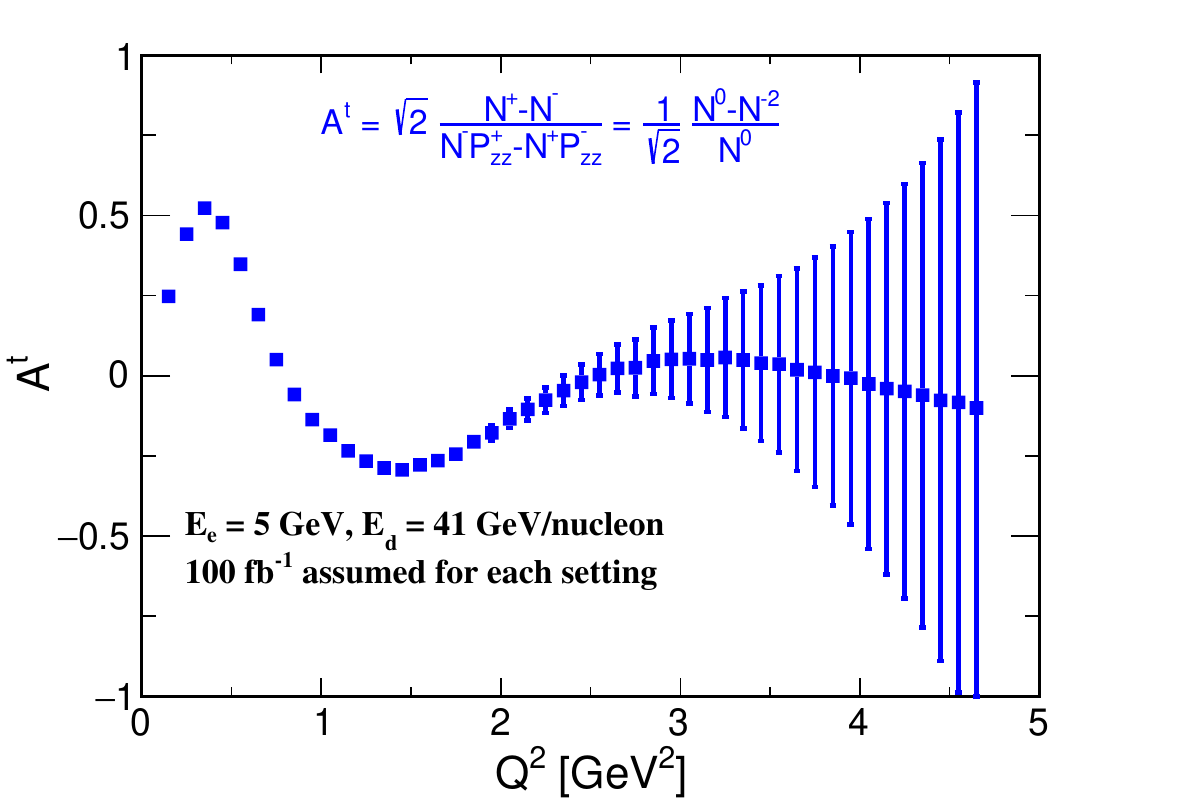}
    \caption{Shown is the tensor-polarized deuteron asymmetry for elastic $ed$ scattering.  The case of negative tensor polarization along the beam direction produces identical results as polarization perpendicular (positive vector, positive tensor) to the beam direction.}
    \label{fig:T20Deuteron}
\end{figure}

\begin{figure}[htb]
    \centering
     \includegraphics[width=0.49\linewidth,page=1]{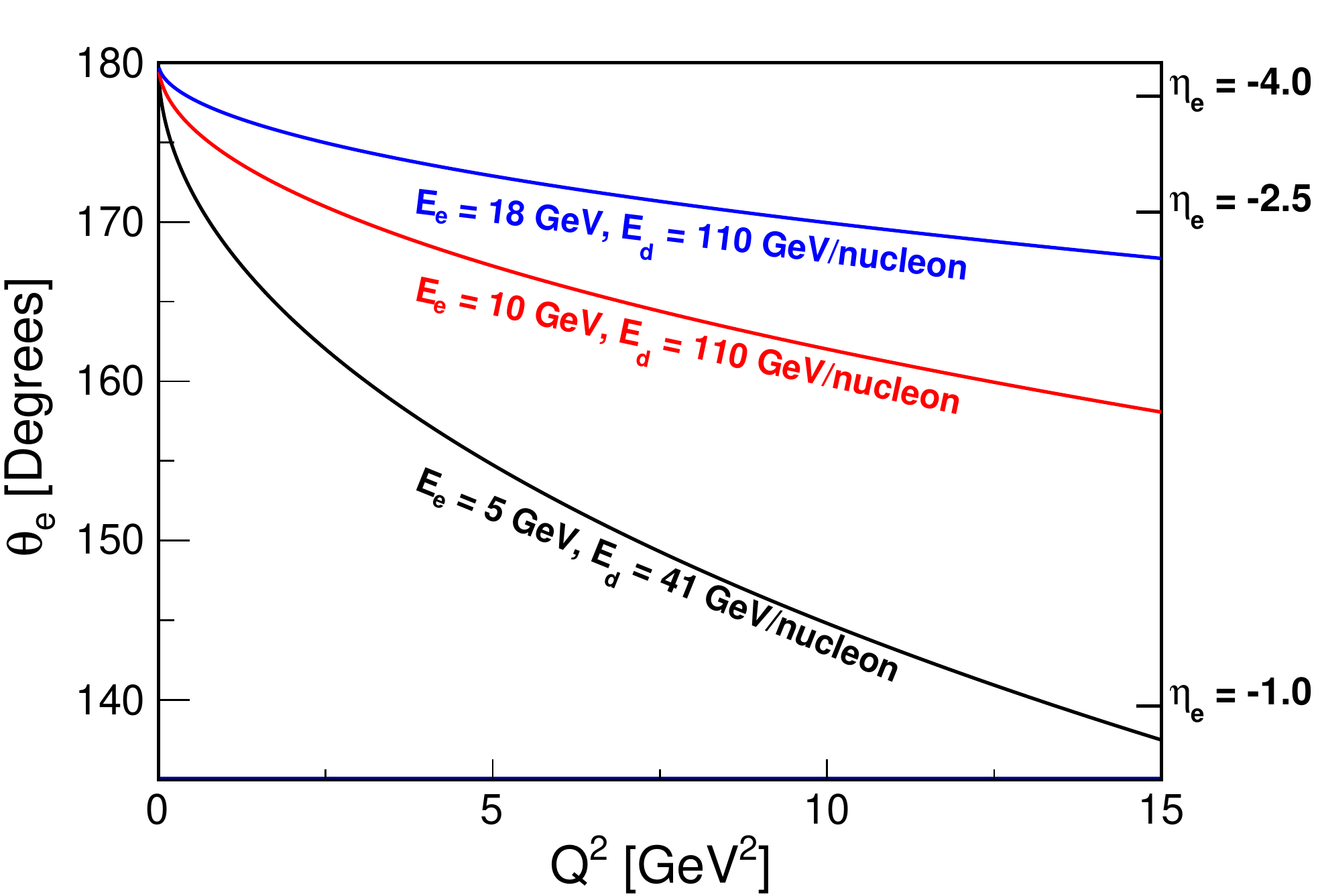}
    \includegraphics[width=0.49\linewidth,page=3]{PART2/Figures.EICMeasandStud/kin_yellow_deuteron.pdf}   
    \caption{Left figure shows the electron scattering angle and the right figure the deuteron scattering angle for different values of $Q^2$ and for different beam energy combinations.}
    \label{fig:barak-deuteron}
\end{figure}

Measurements of the unpolarized electron-deuteron ($ed$) cross section are possible at the EIC up to $Q^{2} \approx$ 5 $\gev^2$ (see right panel in Fig.~\ref{fig:FormFactorRates}), and measurements of the tensor-polarized asymmetry can be made up to $Q^{2} \approx$ 2.5 $\gev^2$ (see Fig.~\ref{fig:T20Deuteron}).
Here the deuteron form factors for the cross section calculation come from a refit of the Abbott experimental data~\cite{Abbott:2000ak,Abbott:2000fg,Parker:2020,Zhou:2020cdt}.
At low $Q^{2}$, this asymmetry is experimentally well known~\cite{Bouwhuis:1998jj,Nikolenko:2003zq} and can be used to help determine the polarization of a stored tensor-polarized deuteron beam.
Because only lower-$Q^2$ measurements are possible here, it is preferable to extend the electron acceptance down to $\eta = 4$ (see Fig.~\ref{fig:barak-deuteron}). 
It is also necessary to detect the scattered deuteron in order to suppress inelastic background, and far-forward detectors will be required for this. 
A positive tensor polarization for the deuterium beam along the beam direction can be achieved using high polarization in the  $m=+1$ or $m=-1$ state.  If the EIC is unable to create deuterons in the $m=0$ state (which has a negative tensor polarization), a possible way of creating a negative tensor polarization along the beam direction is shown in Fig.~\ref{fig:T20Deuteron}. Here, $ed$ elastic data is simulated first assuming a negative tensor polarization along the beam direction (blue points); then the simulation is repeated with a polarization perpendicular (positive vector, positive tensor) to the beam direction. 
The cross sections for these two cases are equivalent, and consequently coincide in the figure.

\subsubsection{Meson form factors}

Measuring meson form factors can help elucidate the interplay between emergent hadronic mass and the Higgs mechanism --- see also Sec.~\ref{part2-subS-PartStruct-Mass} and Tab.~\ref{tab:MS_EIC_science_questions_table}.
The experimental determination of the $\pi^+$ electric form factor ($F_{\pi}$) is challenging. In the timelike region, the form factor has been measured in $\ee \to \pi^+\pi^-$~\cite{Amendolia:1983di}. The best way to determine $F_{\pi}$ in the spacelike region would be elastic $e\pi$ scattering.
However, the lifetime of the $\pi^+$ is only 26.0~ns.
Since $\pi^+$ targets are not possible, and $\pi^+$ beams with the required properties are not yet available, one must employ high-energy exclusive electroproduction of pions, $p(e,e^\prime \pi^+)n$.  
This is best described as quasi-elastic ($t$-channel) scattering of the electron from the virtual $\pi^+$
cloud of the proton, where the Mandelstam variable $t$ is the momentum transfer squared to the target nucleon, $t=(p_{p}-p_{n})^2$.
As discussed in Sec.~\ref{part2-subS-PartStruct.M}, scattering from the $\pi^+$ cloud dominates the longitudinal photon cross section ($d\sigma_L/dt$) at sufficiently small $-t$.

To reduce background contributions, normally one separates the
components of the cross section due to longitudinal (L) and transverse (T)
virtual photons (and the LT, TT interference contributions), via a Rosenbluth
separation.  However, L/T separations are impractical at the EIC, due to the impossibility of acquiring low-$\epsilon$ data.  Below we propose an alternate technique to access $\sigma_L$ via a model, validated with exclusive $\pi^-/\pi^+$ ratios from deuterium.
Once $d\sigma_L/dt$ has been determined over a range of $-t$, from $-t_{min}$ to $-t\sim0.6$~GeV$^2$, 
the value of $F_{\pi}(Q^2)$ is determined by comparing the observed $d\sigma_L/dt$ values to the best available electroproduction model, incorporating off-shell pion and recoil nucleon effects.  The obtained $F_{\pi}$ values are in principle dependent upon the model used, 
but one anticipates this dependence to
be reduced at sufficiently small $-t$.  Measurements over a range of $-t$ are essential as part of the model validation process.
JLab-6 experiments were instrumental in establishing the reliability of this technique up to $Q^2=2.45$~GeV$^2$~\cite{Huber:2008id}, and extensive further tests are planned as part of the JLab E12-19-006~\cite{E12-19-006} experiment.

\paragraph{Requirements for separating exclusive and SIDIS events:}
The exclusive $\pi^+$ channel cross section is several orders of magnitude smaller than neighboring background from semi-inclusive DIS (SIDIS), but is distributed over a much narrower range of kinematics, and this is essential for the separation of the exclusive events from the background.  The exclusive $e+p \to e' + \pi^+ + n$ reaction is isolated by detecting the forward-going high-momentum neutron, that is, $e-\pi^+-n$ triple coincidences.
Since the neutron energy resolution is not very good, the neutron hit is used as a tag for exclusive events, and the neutron momentum is otherwise not used in the event reconstruction.  Thus, missing momentum is calculated as $p_\text{miss}=|\vec{p}_e + \vec{p}_p - \vec{p}_{e'} - \vec{p}_{\pi}|$.

The effectiveness of kinematic cuts to isolate the exclusive $\pi^+$ channel was evaluated by comparison to a simulation of $p(e,e'\pi^+)X$ SIDIS events, including both detector acceptance and resolution smearing effects.  
The most effective cuts are on the detected neutron angle ($\pm \, 0.7^{\circ}$ from the outgoing proton beam), on the reconstructed $-t<0.5$~GeV$^2$, and the missing momentum ($Q^2$-bin dependent cut).  The missing momentum cut ranges from $p_\text{miss} >95 \, \gevc$ at $Q^2=6 \, \gev^2$, to $p_\text{miss}>77 \, \gevc$ at $Q^2=35 \, \gev^2$;  i.e., all events above the cut value are removed, where the value is chosen to optimize the signal/background ratio for each $Q^2$ bin. After application of these cuts, the exclusive $p(e,e'\pi^+n)$ events are cleanly separated from the simulated SIDIS events.

\paragraph{Determining the longitudinal cross section \boldmath$d\sigma_L/dt$:}
After the exclusive $\pi^+$ event sample is identified, the next step is to separate the longitudinal cross section $d\sigma_L/dt$ from $d\sigma_T/dt$, needed for the extraction of the pion form factor.  However, a conventional Rosenbluth separation is impractical at the EIC due to the very low proton beam energy required to access $\epsilon<0.8$.
Fortunately, at the high $Q^2$, $W$ accessible at the EIC, phenomenological models predict $\sigma_L\gg\sigma_T$ at small $-t$.  For example, the Vrancx and Ryckebusch Regge-based model \cite{Vrancx:2013fra} predicts $R=\sigma_L/\sigma_T >10$ for $Q^2>10$~GeV$^2$ and $-t<0.06$~GeV$^2$, and $R>25$ for $Q^2>25$~GeV$^2$ and $-t<0.10$~GeV$^2$.  Thus, transverse cross section contributions are expected to be $1.3 - 14\%$.  The most practical choice appears to be to use a model to isolate the dominant $d\sigma_L/dt$ from the measured (un-separated) cross section $d\sigma_\text{uns}/dt$.

To control the systematic uncertainty associated with the theoretical correction to estimate $\sigma_L$ from the un-separated $\sigma_\text{uns}$, it is very important to confirm the validity of the model used.  This can also be done with EIC data, using exclusive $^2H(e,e'\pi^+n)n$ and $^2H(e,e'\pi^-p)p$ data for the same kinematics as the primary $p(e,e'\pi^+n)$ measurement.  The ratio of these cross sections is
$R=\frac{\sigma[n(e,e'\pi^-p)]}{\sigma[p(e,e'\pi^+n)]}=\frac{|A_V-A_S|^2}{|A_V+A_S|^2}$,
where $A_V$ is the isovector amplitude, and $A_S$ is the isoscalar amplitude.
Since the pion-pole $t$-channel process used to determine the pion form factor is purely isovector (due to $G$-parity conservation), the above ratio will be diluted if $\sigma_T$ is not small, or if there are significant non-pole contributions to $\sigma_L$.  The comparison of the measured $\pi^-/\pi^+$ ratio to model expectations, therefore, provides an effective means of validating the model used to determine $\sigma_L$.  The same model, now validated, can likely also be used to extract the pion form factor from the $\sigma_\text{uns}$ data.

\begin{figure}[htbp]
\begin{center}
\includegraphics[width=3.5in]{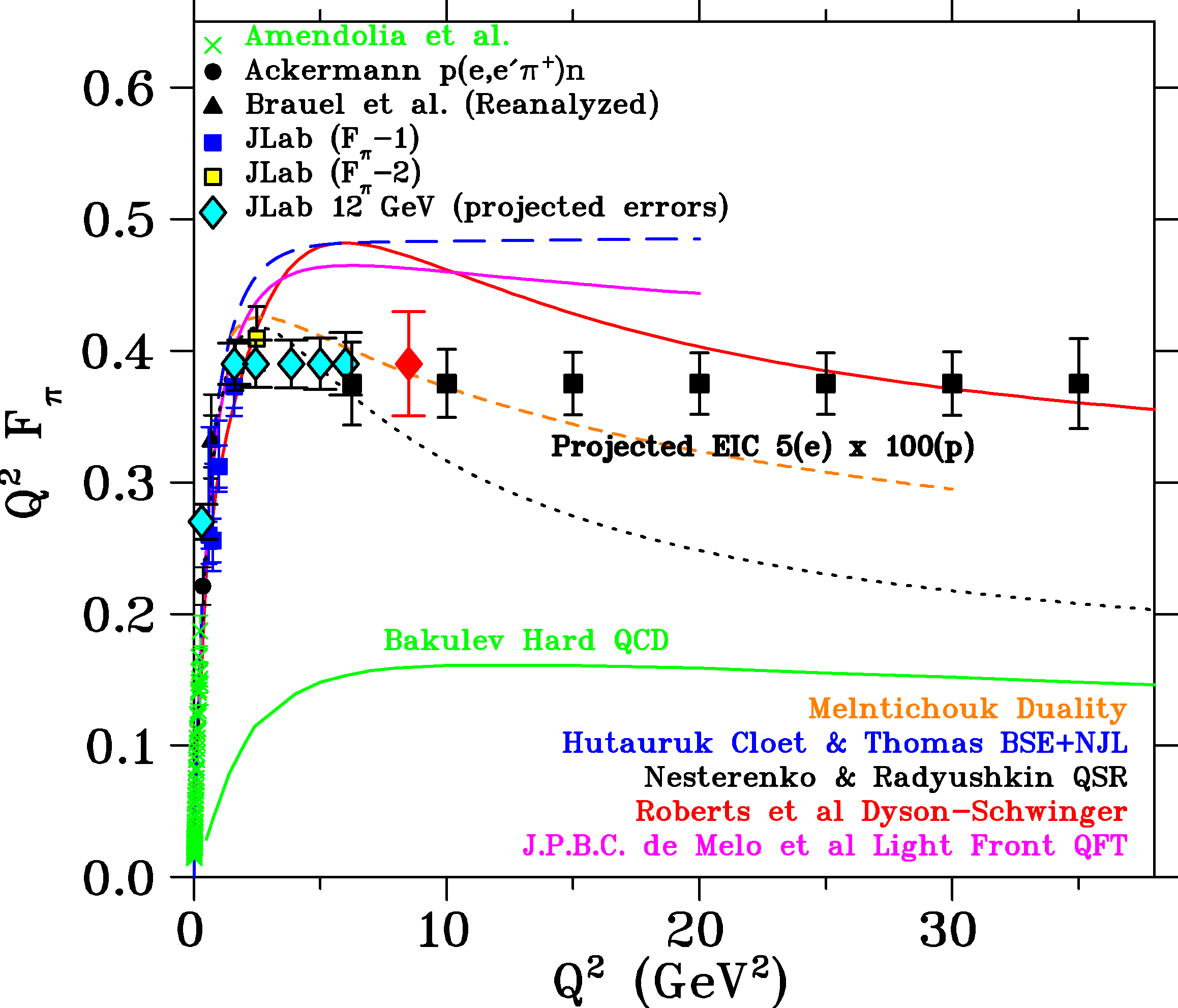}
\vspace{-0.2cm}
\caption{\label{fig:eic_fpi}
Existing data (green crosses~\cite{Amendolia:1984nz,Amendolia:1986wj}, black circles~\cite{Ackermann:1977rp} and triangles~\cite{Brauel:1979zk,Huber:2008id}, blue and yellow squares~\cite{Huber:2008id}) and projected uncertainties for 
future data on the pion form factor from JLab (cyan and red diamonds~\cite{E12-19-006}) and EIC (black squares), in
comparison to a variety of hadronic structure models~\cite{Bakulev:2004cu,Melnitchouk:2002gh,Hutauruk:2016sug,Nesterenko:1982gc,Chang:2013nia,Mello:2017mor}.  The EIC projections 
clearly cover a much larger $Q^2$ range than the JLab measurements.}
\end{center}
\end{figure}

\paragraph{Conclusions for pion form factor:}
The EIC can allow a pion form factor measurement up to $Q^2 =35$~GeV$^2$, as shown in Fig.~\ref{fig:eic_fpi}.  The error bars are based on the following assumptions: integrated luminosity of 20~fb$^{-1}$ for 5$\times$100 GeV measurement, clean identification of exclusive $p(e,e'\pi^+n)$ events by tagging the forward neutron, cross section systematic uncertainty of 2.5\% point-to-point, and 12\% scale, $R=\sigma_L/\sigma_T=0.013-0.14$ at $-t_{min}$, $\delta R=R$ systematic uncertainty in the model subtraction to isolate $\sigma_L$, pion pole channel dominance at small $-t$ confirmed in $^2$H $\pi^-/\pi^+$ ratios.
Also, a $-t<0.5$~GeV$^2$ cut was used.

\paragraph{\boldmath{$K^+$} form factor:} 
The reliability of the electroproduction method to determine the $K^+$ form factor $F_K$ has not been established yet. The JLab experiment E12-09-011~\cite{E12-09-011} has acquired data for the reactions
$p(e,e^\prime K^+)\Lambda$ and $p(e,e^\prime K^+)\Sigma^0$ at hadronic invariant mass $W=\sqrt{(p_{K}+p_{\Lambda,\Sigma})^2}>2.5$~GeV, to search for evidence of
scattering from the proton's ``kaon cloud''.  The data are still being analyzed, with L/T-separated cross sections expected in the next $\sim 2$ years.  If they confirm that
the scattering from the virtual $K^+$ in the nucleon dominates at low $|t|\ll m_p^2$, 
the experiment will yield the world's first quality data for $F_K$ above $Q^2>0.2$~GeV$^2$.  
This would then open up the possibility of using exclusive reactions to determine $F_K$ over a wide range of $Q^2$ at higher energies.  Studies are planned.  While the general technique will remain the same, 
the $\pi^-/\pi^+$ validation technique to confirm the $\sigma_L$ extraction cannot be used for the $K^+$.  We are optimistic that $\Lambda/\Sigma^0$ ratios can play a similar role, but conditions under which the clean separation of these two channels may be possible at the EIC requires further study and would only be possible at center-of-mass energies of $\sim 29 - 50$~GeV. 
Otherwise the hyperons decay too far down the beampipe for reconstruction using the far-forward detectors (see Sec.~\ref{subsec:DetReq.DT.meson}) and missing-mass resolution to separate the $\Sigma^0$ channel from the $\Lambda$ channel degrades at higher center-of-mass energies.

\subsection{Imaging of quarks and gluons in impact-parameter space}
\label{part2-subS-SecImaging-GPD3d}

A key challenge of nuclear physics is the tomographic imaging of the nucleon, encoded in the Generalized Parton Distributions (GPDs).
They provide a connection between ordinary PDFs and form factors and hence can describe the correlations between the longitudinal momentum of quarks and gluons and their position in the transverse spatial plane in a nucleon~\cite{Diehl:2003ny,Belitsky:2005qn}.

In the nucleon case, depending on the target and active-parton polarization, one can define four chiral-even GPDs ($H$, $E$, $\tilde{H}$ and $\tilde{E}$) and four chiral-odd GPDs ($H_T$, $E_T$, $\tilde{H}_T$ and $\tilde{E}_T$). 
They depend on three variables (considering the dependence on the factorization scale $Q^2$ to be known): $x$,  that is the average longitudinal momentum of the active quark as a fraction of the average target momentum; $\xi$ and $t$, that are, respectively, half the change in the fraction of longitudinal momentum carried by the struck parton and the squared four-momentum transferred to the target. However, one does not have complete direct experimental access to this multidimensional structure, since the dependence on $x$ enters observables in nontrivial convolutions with coefficient functions.

Since no single process is sufficient to determine GPDs fully, measurements of a variety of processes and observables are necessary to maximally constrain them. Fits of GPDs require educated choices of the fitting functions to incorporate the known theoretical constraints of GPDs, that is, polynomiality, sum rules, and positivity --- see, e.g., Refs.~\cite{Goeke:2001tz,Diehl:2003ny,Ji:2004gf,Belitsky:2005qn,Boffi:2007yc,Guidal:2013rya,Chouika:2017dhe} for a detailed account of the formalism and properties of GPDs.
Owing to QCD factorization theorems, a number of related processes are complementary to disentangle the various GPDs and their flavor dependence --- see, e.g., Refs.~\cite{Kumericki:2016ehc,Favart:2015umi,Kriesten:2020wcx} for recent works on the GPD phenomenology. 

\begin{figure}[t]
    \centering
    \includegraphics[width=1\textwidth]{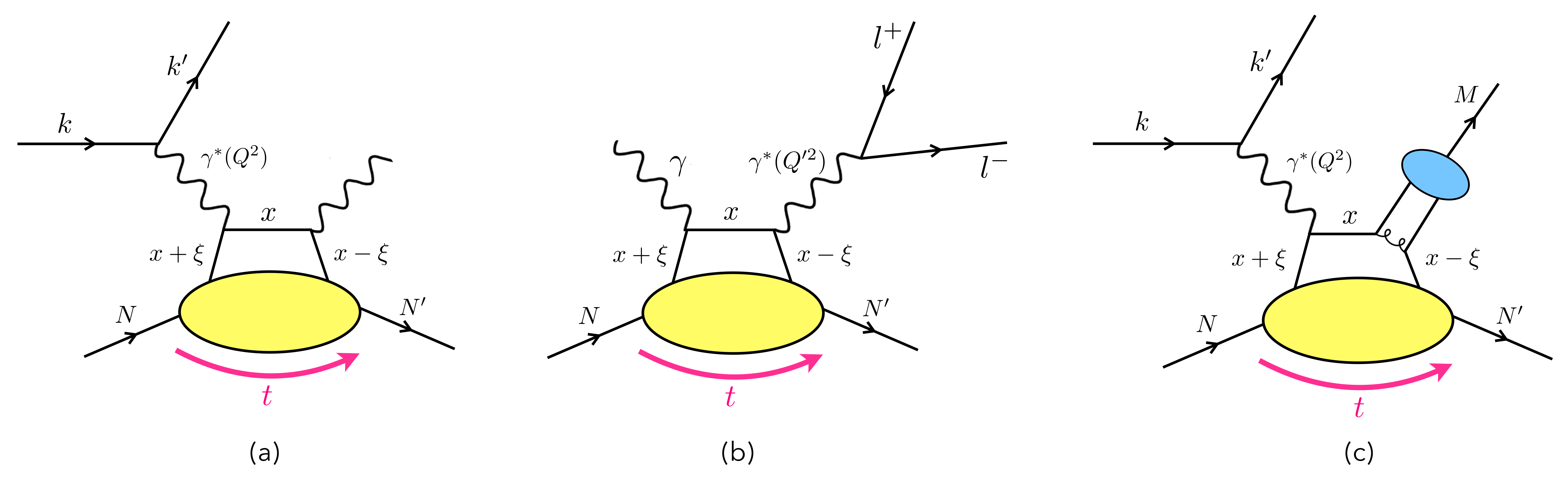}
    \caption{Illustrations of three main processes which are sensitive to GPDs: (a) exclusive electroproduction of a real photon, (b) TCS and (c) exclusive electroproduction of a meson.}
    \label{fig:GPD-processes}
\end{figure}

\subsubsection*{Phenomenology of GPDs}

The cleanest way to probe GPDs is via deeply-virtual Compton scattering (DVCS), i.e., $\gamma^* N \rightarrow \gamma N' $, at high photon virtuality ($Q^2 > 1$~GeV$^2$) and low (squared) momentum transfer ($|t|\ll Q^2$), where the scattering happens from a single parton~\cite{Belitsky:2001ns}. Experimentally, we access DVCS by measuring the exclusive electroproduction of a real photon (see diagram (a) in Fig.~\ref{fig:GPD-processes}). In this process the DVCS amplitude interfers with the so-called Bethe-Heitler (BH) process, that corresponds  to the emission of the photon by the incoming or the outgoing
electron and is exactly calculable in QED once the nucleon electromagnetic form factors are known.
DVCS is well described theoretically, including higher orders in $\alpha_{s}$, higher-twist and target mass corrections --- for a comprehensive review see, e.g., Ref.~\cite{Kumericki:2016ehc}. 

The DVCS cross section is parametrized in terms of Compton form factors (CFFs) through which, however, the dependence on $x$ is not directly accessible. CFFs are complex functions whose real and imaginary parts are convolutions over $x$ of the GPDs with a hard kernel, systematically computable in perturbative QCD. At leading order, the imaginary part of the CFFs gives the GPDs along the diagonals $x=\pm \xi$, while the real part of the CFFs probes a convoluted integral of
GPDs over the initial longitudinal momentum of the partons.
The interference between BH and DVCS provides a way to independently access the real and imaginary parts of the CFFs.
Beam and target single-spin asymmetries are proportional to the imaginary part of the DVCS-BH interference term.  
All three terms (pure BH, pure DVCS, and interference term) contribute to the unpolarized cross section. The DVCS and interference terms can be separated by exploiting their dependence on the incident-beam energy, which represents a generalized Rosenbluth separation.  
The real part of the DVCS amplitude also appears in double-spin asymmetries, but these can receive significant contributions from the pure BH process, making the extraction of the real part of the amplitude challenging. 
Beam-charge asymmetries (from measurements with both electron and positron beams), on the other hand, receive no direct contribution from the pure BH process and are also sensitive to the real part of the DVCS amplitude.  Therefore, an experimental program with positron beams can have a significant impact in accessing this crucial observable~\cite{Accardi:2020swt}.

Timelike Compton scattering (TCS) is a related process in which a real photon scatters off a parton to produce a virtual photon, detected through its lepton-pair decay (see diagram (b) in Fig.~\ref{fig:GPD-processes})~\cite{Berger:2001xd,Boer:2015fwa,Boer:2015cwa,Grocholski:2019pqj,Muller:2012yq,Pire:2011st,Moutarde:2013qs}. As such, this is an inverse process to DVCS, sensitive to the same set of GPDs. The complementarity of the DVCS and TCS processes relies mostly on the analyticity of the $Q^2$ behaviour of the scattering amplitudes~\cite{Muller:2012yq, Grocholski:2019pqj}. Confronting DVCS and TCS results together is a mandatory goal of the EIC to prove the consistency of the collinear QCD factorization framework and to test the universality of GPDs. The differences in the two processes also give experimental advantages in the extraction of CFFs --- for example, the asymmetries associated with the leptonic decay in TCS provide a more direct access to the real part of the dominant CFF~\cite{Grocholski:2019pqj,Boer:2015fwa,Boer:2015cwa}.  

Additional information on GPDs can be obtained from hard exclusive meson electroproduction (which is also called deeply-virtual meson production (DVMP)), where a meson, instead of the photon, is produced as a result of the scattering (see diagram (c) in Fig.~\ref{fig:GPD-processes})~\cite{Favart:2015umi}. These processes include: 
\begin{enumerate} 
\item heavy meson ($J/\psi$, Y)  electroproduction, which probes gluon GPDs and may also provide new information on the underlying mechanism of saturation by observing the change of the spatial gluon distribution from high to low $x_B$~\cite{Aschenauer:2017jsk};
\item light vector meson ($\rho^0$, $\rho^+$, $\omega$; $\phi$) electroproduction which, in addition, allows 
for a flavor separation of the GPDs;
\item light pseudoscalar meson ($\pi^+$, $\pi^0$, $\eta$) electroproduction which, at high $Q^2$, gives access to parity-odd GPDs ($\tilde H$ and $\tilde E$) and, at low $Q^2$, in a model-dependent way, to a group of chiral-odd GPDs that are inaccessible in DVCS at leading twist. Some of these GPDs are related to the transversity distributions extensively studied in semi-inclusive DIS and Drell-Yan processes.
\end{enumerate}

Hard exclusive production of $\pi^0$ mesons has a final state similar to that of DVCS. It consists of one scattered lepton in the DIS regime ($Q^{2} > 1~\mathrm{GeV}^{2}$), one scattered nucleon in a coherent state (i.e., no break-up of target particle in the interaction), and either one or two photons for DVCS and $\pi^0$ production, respectively. This similarity suggests that a common analysis of the detector requirements for both processes can be performed, as discussed in Sec.~\ref{subsec:dvcs_ep}. 

The information that can be extracted from a handful of DVCS measurements at low $x_{B}$ from HERA collider experiments, almost entirely consisting of cross sections in loose $Q^2-t$ bins, is very limited. GPD-based experiments at larger $x_{B}$ have been carried out at HERMES and COMPASS. Dedicated fixed-target experiments at JLab-12 will be addressing GPDs in the kinematic region dominated by valence quarks. 
More precise data mapping, with high granularity and a wider phase space, is required to fully constrain the entire set of GPDs for gluons and sea quarks. This will be provided by the EIC, which connects the domain typical of fixed-target experiments with that of collider measurements.
With its wide range in energy and high luminosity, the EIC will thus offer an unprecedented opportunity for a precise determination of GPDs. 

\begin{figure}[t]
  \centering
  \includegraphics[width=0.31\textwidth]{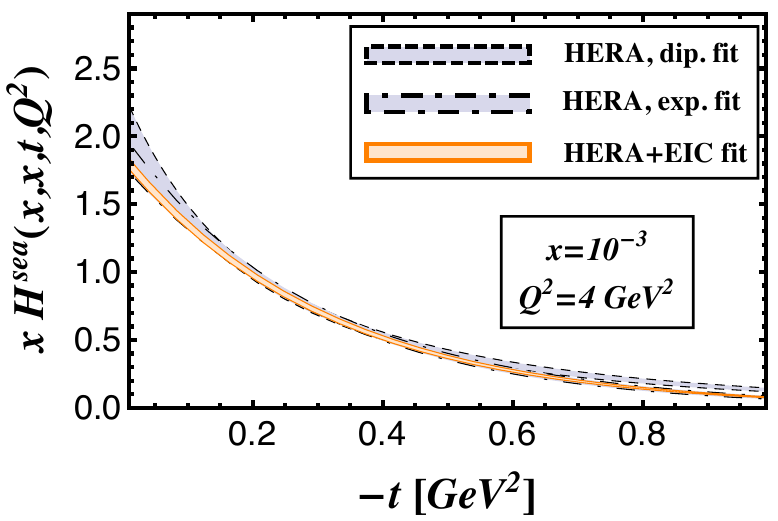}
  \includegraphics[width=0.31\textwidth]{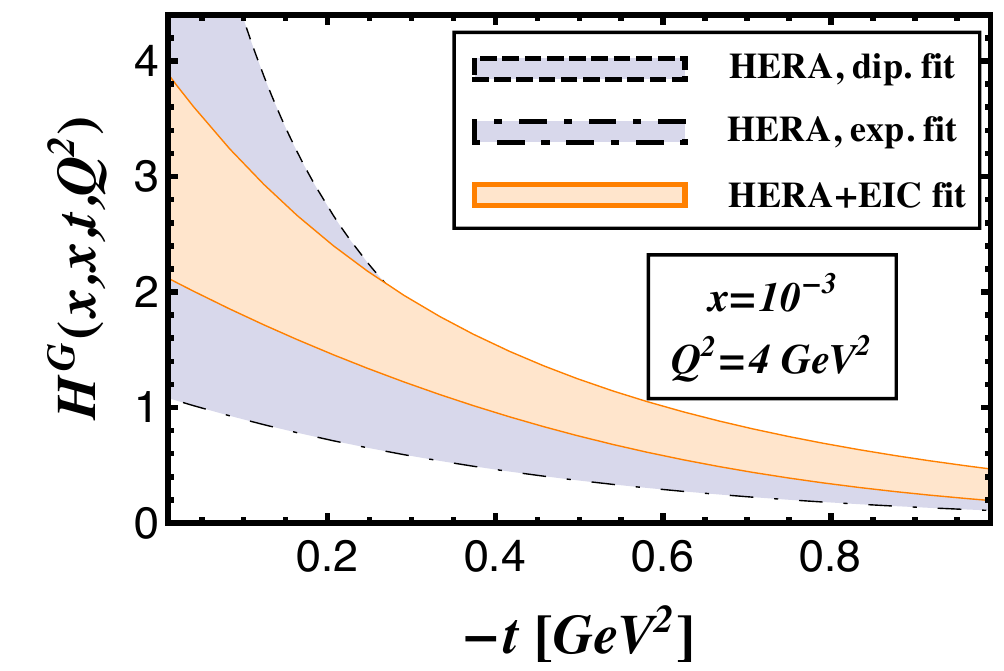}
  \includegraphics[width=0.31\textwidth]{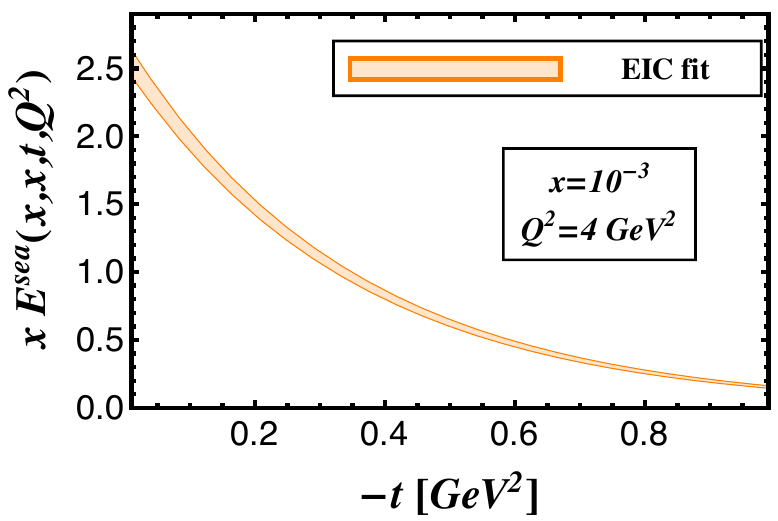}
  \caption{Extraction of the GPD $H$ for sea quarks (left) and gluons (center), and the GPD $E$ for sea quarks (right), at a particular $x$ and $Q^2$. The violet band is the uncertainty obtained excluding the EIC pseudodata from the global fit procedure~\cite{Aschenauer:2013hhw}.}
  \label{Fig:GPD}
\end{figure}

Simulation studies proved that the EIC can perform accurate measurements of DVCS cross sections and asymmetries in a very fine binning and with a very low statistical uncertainty~\cite{Aschenauer:2013hhw}. This pioneering assessment of the EIC capability to constrain GPDs solely relies on global fits of DVCS measurements. Figure~\ref{Fig:GPD} shows the uncertainties of GPDs extracted from current data (violet bands) and how they are constrained after including the EIC pseudodata into the fits (orange bands).  
This study demonstrated that the EIC can significantly improve our current knowledge of the GPD $H$ for gluons. 
Moreover, a precise measurement of the transverse target-spin asymmetry $A_{UT}$ 
leads to an accurate extraction of the GPD $E$ for sea quarks, which currently remains almost unconstrained~\cite{Aschenauer:2013hhw}.

\begin{figure}[t]
    \centering
     \includegraphics[width=0.5\textwidth]{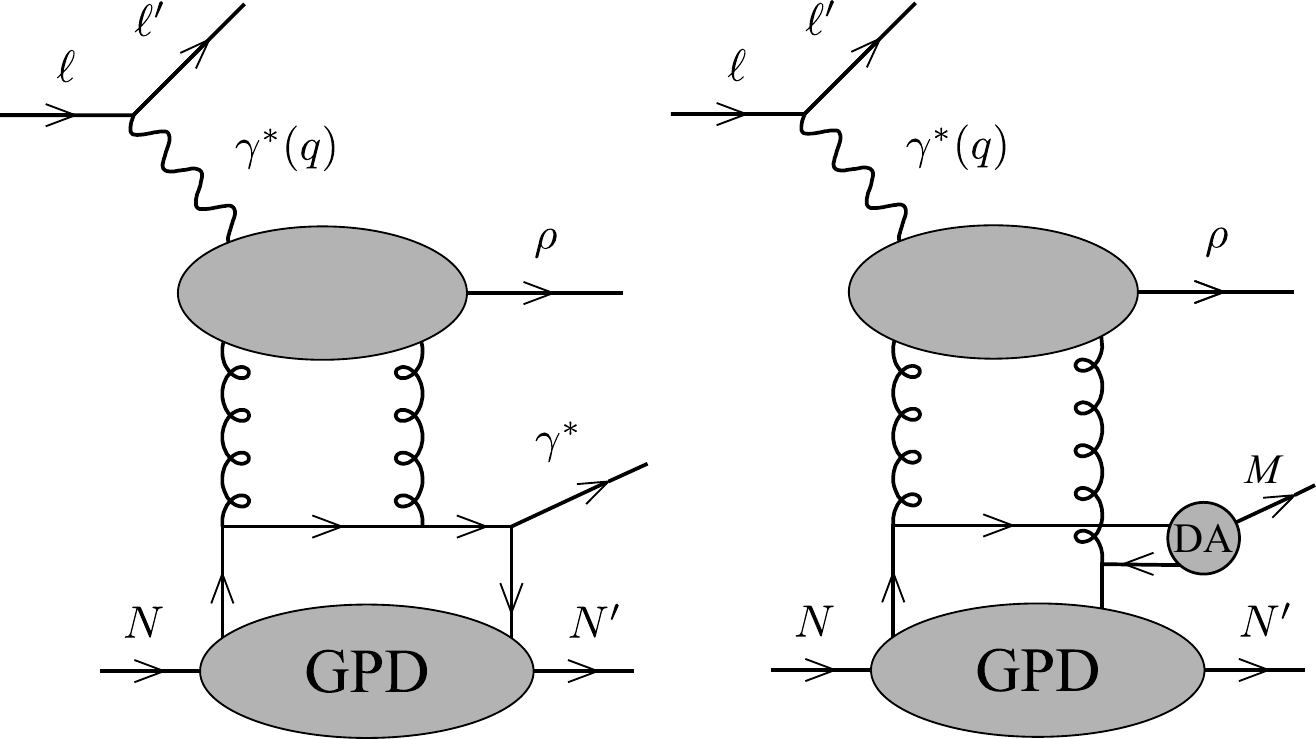}
    \caption{Leading-order diagram for the diffractive production of $\rho$ and a virtual photon (left panel) and diffractive two-meson production (right panel), with a large rapidity gap between the forward $\rho$ and the remaining $\gamma^*N'$ or $MN'$ final state~\cite{Pire:2019hos}.}
    \label{FigDiffex}
\end{figure}

Diffractive events are known to constitute a large part of the cross section in high-energy scattering. 
In Refs.~\cite{Pire:2019hos,Ivanov:2002jj,Cosyn:2020kfe}, access to GPDs is suggested in a diffractive process where a GPD-driven subprocess (${\cal P} N \to \gamma^*(Q'^2) N'$ or ${\cal P} N \to M N'$, with ${\cal P}$ a hard Pomeron and $M$ a meson) is triggered by a diffractive $\gamma^*(Q^2) \to \rho {\cal P}$ process, as shown in Fig.~\ref{FigDiffex}.
The kinematic domain is defined with a large rapidity gap separating the $\rho$ from the $\gamma^* N'$ or $ MN'$ final state, and a small momentum transfer between the initial and final nucleons.
Contrary to the usual DVCS and TCS processes, the integration over the quark momentum fraction in the amplitudes is restricted to a smaller domain ($-\xi < x < \xi$), with gluons not entering due to $C$-parity conservation. The skewness parameter $\xi$ is not related to $x_B$ as in DVCS, giving access to large $\xi$ even for high-energy processes~\cite{Ivanov:2002jj,Pire:2019hos}.
In the meson production process, as in DVMP, the nature of the meson and its polarization select specific GPDs (vector, axial vector, transversity).  The amplitudes are energy-independent at leading order, and would acquire a mild energy dependence when high-energy evolution is turned on. 
Cross section projections for the $\rho M$-production process at EIC kinematics were studied in Ref.~\cite{Cosyn:2020kfe}.  Detailed EIC simulations for both processes are in progress~\cite{Cepila:2016uku}.

In Ref.~\cite{Boussarie:2016qop, Pedrak:2017cpp, Duplancic:2018bum, Pedrak:2020mfm}, a new class of processes was proposed to access GPDs through $\gamma + N \rightarrow \gamma + M +N'$ and $\gamma + N \rightarrow \gamma + \gamma +N'$, focusing on the regime where $M_{\gamma M}^2$ or $M_{\gamma \gamma}^2$ provides a hard scale. The connection with GPDs relies on the fact that the subprocess $\gamma (q \bar{q}) \rightarrow \gamma + M$ or $\gamma (q \bar{q}) \rightarrow \gamma + \gamma$ factorizes from GPDs.
For photoproduction of $\gamma\gamma$, the hard subprocess gives access to GPDs at the special point $x=\pm\xi$~\cite{Pedrak:2017cpp}. On the other hand, the photoproduction of a photon-meson pair, for example $\gamma\rho$, is sensitive to chiral-odd (transversity) GPDs~\cite{Boussarie:2016qop}. Chirality constraints dictate that, for $\gamma \rho$ photoproduction, only chiral-odd GPDs contribute at leading twist (i.e., up to 1/$Q^2$ corrections) to the production of transversely-polarized $\rho$ mesons, while chiral-even GPDs enter for longitudinal meson polarization.

Selecting specific polarization states via measurements of the $\rho$ decay products enables the separation of the chiral-even and transversity GPDs. This process also benefits from a suppression of gluon GPDs in the amplitude, which typically introduce large NLO corrections. Simulations of the process at EIC kinematics are underway~\cite{Pire:2019hos}. 
The chiral-even sector ($M=\pi$ or $\rho_L$) can yield a large new set of observables, complementing the measurements of DVCS, TCS and DVMP which involve the same GPDs. In addition, the case $M = \pi^0$ provides a new way to access GPDs of gluons.

A novel method to extract GPDs has recently been proposed, which is based on the comparison of $\rho$- and $\pi$-meson production cross sections in charged-current processes~\cite{Siddikov:2019ahb}. The rates of these processes are suppressed compared to photoproduction and pose significant experimental challenges, yet are within reach of the EIC, as described in Sec.~\ref{subsec:charged_current}.

\subsubsection*{Impact parameter distributions}

Impact parameter distributions (IPDs) can be obtained by taking a Fourier transform of the GPDs in the variable $t$ at $\xi=0$. IPDs represent densities of partons with a given momentum fraction $x$ as a function of the position $\boldsymbol{b}_T$ (impact parameter) from the center of momentum of the nucleon in the transverse plane~\cite{Burkardt:2002hr}. 
A first attempt to obtain this information from DVCS measurements was illustrated in Refs.~\cite{Dupre:2016mai,Dupre:2017hfs}, using a model-dependent extrapolation to the point $\xi=0$ that is not accessible experimentally.
Recently, dispersion-relation techniques have been exploited to constrain the GPDs at $\xi=0$ from data~\cite{Moutarde:2018kwr}.
Both these analyses confirm that the width of the IPDs for unpolarized quarks in unpolarized protons has a very peaked transverse profile in the limit of $x\rightarrow 1$.
This behaviour comes from the fact that, in this limit, the active quark is always very close to the transverse center of momentum~\cite{Burkardt:2002hr,Pasquini:2007xz}. It suggests that the higher-$x$ valence quarks are localized closer to the center of the nucleon than the lower-$x$ sea quarks, which have a wider distribution in the transverse plane. 

\begin{figure}[t]
    \centering
     \includegraphics[width=0.9\textwidth]{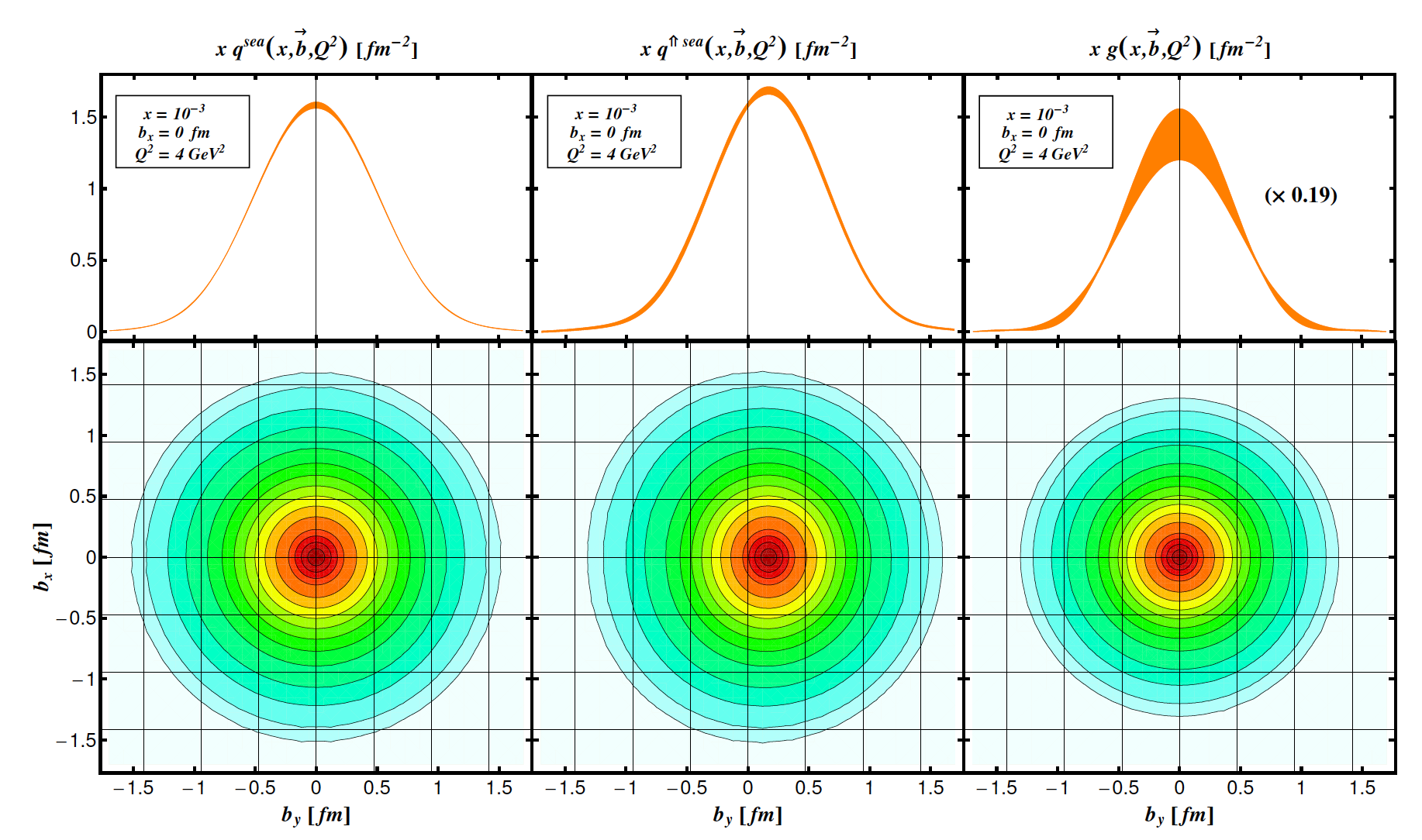}
    \caption{Impact parameter distributions at $x = 0.001$ and $Q^2 = 4 \, \textrm{GeV}^2$ for unpolarized sea quarks in an unpolarized proton (left), a transversely polarized proton (middle), and for unpolarized gluons in an unpolarized proton (right), obtained from a combined fit to the HERA collider data and EIC pseudodata~\cite{Aschenauer:2013hhw}. Top row: IPDs at fixed $b_x=0$ as a function of $b=b_y$.
    Bottom row: density plots of IPDs in the $(b_x,b_y)$-plane.}
    \label{Fig3dPartonic}
\end{figure}

\begin{figure}[tbp]
  \centering
  \includegraphics[width=0.8\textwidth]{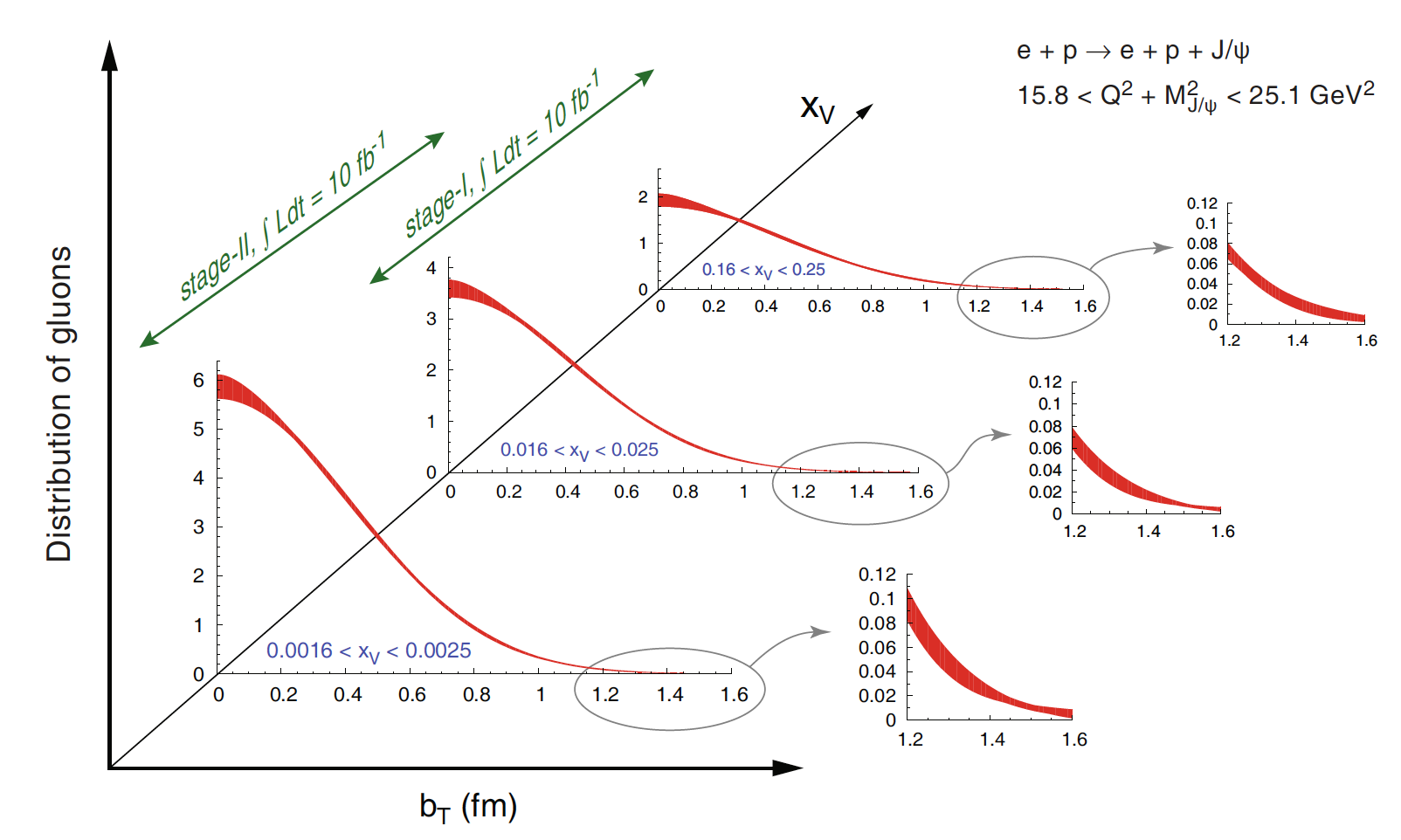}
   \includegraphics[width=0.5\textwidth]{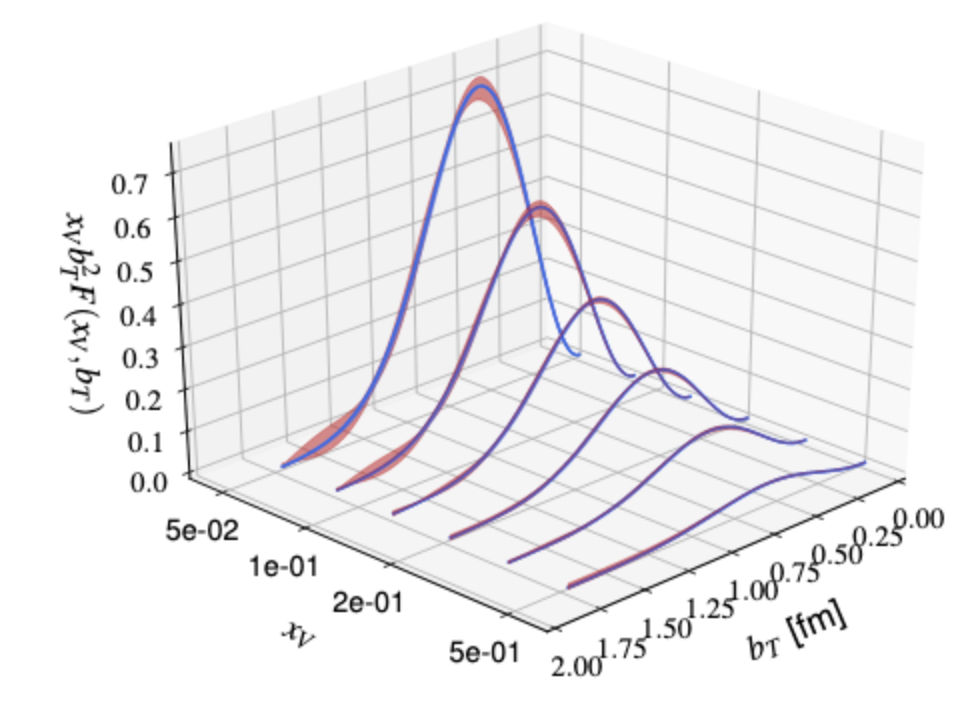}
  \caption{Top: Projected EIC uncertainties for the gluon IPD obtained from a
  Fourier transform  of the differential cross section for $J/\psi$ production for $15.8\,\textrm{GeV}^2 < Q^2+M_V^2 < 25.1\,\textrm{GeV}^2$, assuming a collection of $10\,\textrm{fb}^{-1}$ (from Ref.~\cite{Accardi:2012qut}). Bottom: Projected uncertainties for the gluon IPD multiplied with $b^2_T$, extracted by a Fourier transform of the differential cross section for $Y$ production for $89.5\,\textrm{GeV}^2 < Q^2+M_V^2 <91\,\textrm{GeV}^2$, assuming $100\,\textrm{fb}^{-1}$ (from Ref.~\cite{Joosten:2018gyo}).} 
  \label{Fig:jpsi-gluons}
\end{figure}

The DVCS-based EIC impact study of Ref.~\cite{Aschenauer:2013hhw} showed how by Fourier-transforming the GPDs constrained at the EIC, it is possible to extract the densities of quarks and gluons in the impact parameter space, as shown in Fig.~\ref{Fig3dPartonic}.
While this study is based only on measurements of DVCS, simulations have also shown how the EIC can provide high-precision measurements of the $|t|$-differential cross section of heavy vector mesons~\cite{Accardi:2012qut, Joosten:2018gyo}. More simulations for light and heavy mesons, performed in the context of this Yellow Report, are discussed in Sec.~\ref{subsec:dvmp_ep} and Sec.~\ref{diff-excl-mesons}. 
A Fourier transform of the $|t|$-differential cross section for the production of heavy vector mesons can help to visualize the uncertainty achievable for the gluon IPDs, though it still contains a contribution from the small but finite size of the meson, which needs to be disentangled in a full GPD analysis. For meson production, $Q^2+M^2_V$ becomes the relevant resolution scale. Therefore, $x_V= (Q^2+M^2_V)/(2P \cdot q)$ replaces the standard Bjorken variable $x_B=Q^2/(2P \cdot q)$. Figure~\ref{Fig:jpsi-gluons} shows the projected gluon IPDs measurable at the EIC, enabling us to accurately probe the spatial distribution of gluons over two orders of magnitude in $x_V$, up to the region where the valence quarks dominate.

\begin{figure}[th]
  \centering
  \includegraphics[width=0.95\textwidth]{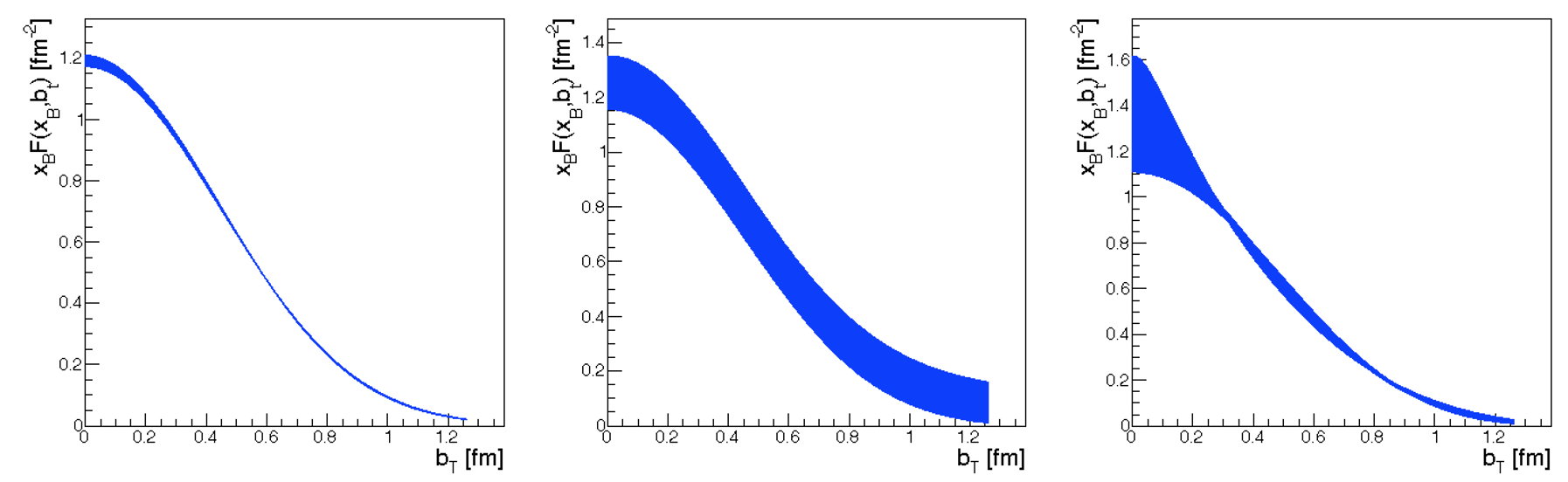}
  \caption{Fourier transform of the DVCS cross section as a function of the impact parameter $b_T$.   The cross sections are for different $|t|$ acceptance and an integrated luminosity of $10\,\textrm{fb}^{-1}$. The bands represent the parametric errors in the fit and the uncertainty from different extrapolations to the regions of unmeasured (very low and very high) $|t|$. Left: $0.03\,\textrm{GeV}^2 < |t| < 1.6\,\textrm{GeV}^2$, middle: $0.2\,\textrm{GeV}^2 < |t| < 1.6\,\textrm{GeV}^2$, right: $0.03\,\textrm{GeV}^2 < |t| < 0.65\,\textrm{GeV}^2$.
 }
  \label{Fig:t-range}
\end{figure}

The impact studies in Ref.~\cite{Aschenauer:2013hhw} assumed a measurement of $|t|$ in a very wide range, starting with the physical minimum $|t_{\rm min}| \sim 0.03 \, \textrm{GeV}^2$ up to $|t|= 1.6 \, \textrm{GeV}^2$ which, in a Fourier transform, corresponds to large values of the impact parameter. Studies by the same authors show that limiting the measured $|t|$-range would severely affect the precision of the extracted partonic densities, as shown in Fig.~\ref{Fig:t-range}. The bands represent the uncertainty from different extrapolations to the regions of unmeasured (very low and very high) values of $|t|$.

\subsubsection*{Form factors of the energy momentum tensor}

GPDs also offer the unique and practical opportunity to access the form factors of the energy momentum tensor (EMT), which are canonically probed through gravity~\cite{Ji:1998pc}. 
For a symmetric (Belinfante-improved) EMT, there are four form factors, usually referred as $A(t)$, $J(t)$, $D(t)$ and $\bar C(t)$, for each type of parton.
The first three form factors can be related to $x$-moments of the GPDs and, at $t=0$, the corresponding ``charges" for quarks and gluons give, respectively, the fraction of nucleon momentum carried by the partons, the quark and gluon contribution to the total angular momentum of the nucleon~\cite{Ji:1996ek} (see Sec.~\ref{part2-subS-SpinStruct.P.N}), and the $D$-term $D\equiv D(0)$, which is sometimes referred to as the ``last unknown global property" of the nucleon~\cite{Polyakov:2018zvc}.
Furthermore, the $\bar C(t)$ form factor is related to the EMT trace anomaly and plays an important role in the generation of the nucleon mass (see Sec.~\ref{part2-subS-PartStruct-Mass}).
The information encoded in the EMT form factors is revealed in the Breit 
frame~\cite{Polyakov:2002yz,Polyakov:2018zvc}, and has 
been discussed recently in other frames as well~\cite{Lorce:2018egm,Lorce:2017wkb}.
Working in the Breit frame, the $D(t)$ form factor 
can be related to the spatial distribution of shear 
forces $s(r)$ and pressure $p(r)$.
 
The relation for the shear forces holds also for quarks and gluons separately, while it is defined only for the total system in the case of pressure.
In this way, $D(t)$ provides the key to mechanical properties of the nucleon 
and reflects the internal dynamics of the system through the distribution of forces.
Requiring mechanical stability of the system, the corresponding force must be directed outwards so that one expects the local criterion
$2 s(r) + p(r) > 0$
to hold, which implies that the total $D$-term for any stable system must be negative,
$D<0$, as confirmed for the nucleon in models~\cite{Ji:1997gm,Goeke:2007fp,Cebulla:2007ei}, calculations from dispersion relations~\cite{Pasquini:2014vua} and lattice QCD~\cite{Hagler:2007xi,Shanahan:2018nnv}. 

Another consequence of the EMT conservation is the condition
$\int_0^\infty p(r)r^2{\rm d}r=0\label{stability}
$, which shows how the internal forces balance inside a composite particle.
This relation implies that the pressure must have at least one node. All models studied up to now show that the pressure is positive in the inner region and negative in the outer region, with the positive sign meaning repulsion towards the outside and the negative sign meaning attraction towards the center.
Recently, an analysis of JLab data taken at 6 GeV~\cite{Girod:2007aa,Jo:2015ema} has provided the first experimental information on the quark contribution to $D(t)$~\cite{Burkert:2018bqq}. The form factor parameters fitted to the JLab data, with the assumption of a negligible gluon contribution, were used to obtain the radial pressure distribution.
Within the uncertainties of the analysis, the distribution satisfies the stability condition, with a zero crossing near $r=0.6 \, \textrm{fm}$.
This analysis has been repeated in Ref.~\cite{Kumericki:2019ddg} using more flexible parametrizations by neural networks to improve the calculation of the uncertainties. The results show that presently available beam-spin asymmetry and cross-section measurements alone do not allow one to draw reliable conclusions. An independent study relying on neural-network-based global fits to existing DVCS data~\cite{Moutarde:2019tqa} also confirms that a reliable extraction of pressure forces from current experimental data is not achievable~\cite{Dutrieux:2021nlz}.

The method itself, however, appears valid and may provide a conclusive extraction of the quark contribution to $D(t)$ in the future, when used in combination with other observables, which are more sensitive to the real part of the CFFs and to $D(t)$ (such as the DVCS beam-charge asymmetry or the production of lepton pairs), and with forthcoming data from present facilities (JLab, COMPASS) and the EIC. Similarly, exploratory studies for the prospects of measuring the other EMT form factors at the EIC are in progress. Measuring beam-charge asymmetries, the most sensitive probe for $D(t)$, requires a positron beam, which can be unpolarized. 
While this is not envisioned in the EIC baseline, there seem no technical obstacles to this upgrade in the future.

\subsubsection*{Transition distribution amplitudes}
\begin{figure}[tbp]
\centering
\includegraphics[width=0.4\linewidth]{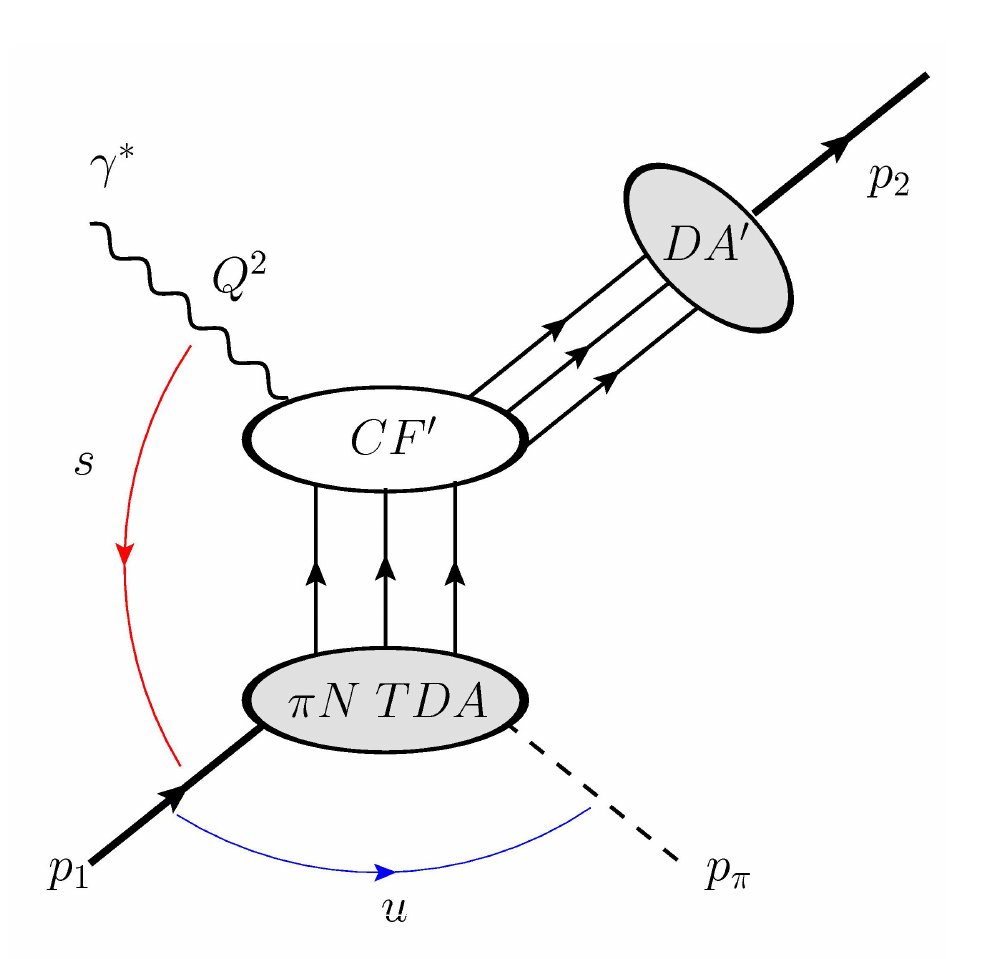}
\caption{The $\pi^0$ electroproduction process
in the
(backward-angle) TDA collinear factorization regime
(large $Q^2$, large $s$, fixed $x_B$, $u\sim
u_{\rm{min}}$). The $\pi N$ TDA (bottom grey oval)
corresponds to the transition distribution amplitude from a nucleon to a vector meson. The
forward-going nucleon is described by the DA
(top-right oval).\label{fig-tda}}
\end{figure}
New information on the parton composition of the nucleon can be accessed by studying hard exclusive meson production in the $u$-channel kinematics region ($t \sim t_{\textrm{max}}$ and $u \sim u_{\textrm{min}}$). 
This process is characterized by a nonzero baryon number exchange in the $u$-channel and can be studied in terms of non-perturbative objects known as nucleon-to-meson Transition Distribution Amplitudes (TDAs)~\cite{Pire:2004ie, Lansberg:2006uh,Pire:2005ax, Lansberg:2011aa}, as illustrated in Fig.~\ref{fig-tda}. 
TDAs describe the underlying physics mechanism of how the target
proton makes a transition into a $\pi$ meson in the final state.  One fundamental
difference between GPDs and TDAs is that the TDAs require three parton exchanges
between the TDA and the hard part.  At leading-twist, there are 8 independent TDAs that can be classified in terms of the light-cone helicity of the exchanged quarks~\cite{Pasquini:2009ki}.
This opens the way to specific and detailed analyses of the helicity content of correlated quarks in the nucleon.
Similarly to GPDs, after the Fourier transform in the transverse
plane, TDAs also carry valuable information on the transverse location of partons and, in particular, allow one to quantify the effect of diquark clustering in nucleons~\cite{Pire:2019nwa}. 
In order to advance in the exploration of the TDA physics, measurements at the EIC and at other facilities can play a crucial role --- see Sect.~\ref{subsec:bckw_pi0} for kinematic studies.

\subsection{Imaging of quarks and gluons in momentum space}
\label{part2-subS-SecImaging-TMD3d}

Transverse momentum dependent distribution and fragmentation functions (TMDs) describe not only the partons' longitudinal momentum, given by the variables $x$ and $z$ for distribution and fragmentation functions, respectively, but also their transverse momentum $k_T$. 
The study of partonic transverse momenta started already in the first years after the discovery of QCD~\cite{Close:1977mr,Parisi:1979se,Collins:1981uk}. 
Polarized TMDs were initially suggested as potential mechanisms for creating the unexpectedly large transverse single-spin asymmetries (SSAs) in hadronic collisions~\cite{Sivers:1989cc,Collins:1992kk}. 
Nowadays, TMDs are a widely-used tool for describing the 3D spin and momentum structure of the nucleon and other hadrons, and they provide access to previously elusive quantities~\cite{Diehl:2015uka, Bacchetta:2016ccz, Aschenauer:2015ndk, Boglione:2015zyc, Anselmino:2020vlp}.
At the EIC, the main access to TMDs comes from semi-inclusive DIS (SIDIS), where in addition to the standard DIS variables $x$, $Q^2$, and $y$, one also identifies final-state hadrons with a fractional energy $z$ and transverse momentum $P_T$ relative to the direction of the virtual photon. 
In several cases the azimuthal angles of the target spin $(\phi_S)$ and the fragmenting hadron momentum $(\phi_h)$ relative to the lepton scattering plane are also measured~\cite{Bacchetta:2004jz}.
Accounting for the transverse momentum degrees of freedom 
allows for the extraction of TMDs and, ultimately, the reconstruction of the 3D picture of hadrons in momentum space.
A general description of SIDIS for single-hadron observables can be found in Refs.~\cite{Mulders:1995dh,Bacchetta:2006tn} and their later extensions. 
Additional information of the momentum space image, including the flavor substructure and gluon TMDs, can be obtained also from other semi-inclusive processes, such as di-hadron production and jet-based measurements. 


The main theoretical tool for probing TMDs is the TMD factorization theorem. 
This theorem allows one to define universal TMD parton distributions~\cite{Collins:1981uk,Ji:2004wu,Collins:2004nx,Collins:2011zzd,GarciaEchevarria:2011rb} that are functions of $x$ and $k_T$. 
Similarly, one defines also TMD fragmentation functions that are functions of $z$ and $P_T$ and describe the hadronization of an outgoing parton~\cite{Metz:2016swz}.
TMD factorization has been intensively developed during the last decades, leading also to the discovery of new domains of applicability and deeper connections with fundamental properties of QCD. 
In TMD factorization, SIDIS structure functions have the following generic form, here for the example of $F_{UT}^{\sin(\phi_h-\phi_S)}$~\cite{Bacchetta:2006tn},
\begin{eqnarray}\label{7.2.3:TMD-generic}
F_{UT}^{\sin(\phi_h-\phi_S)}&=&\sum_{q}e_q^2 \, |C_V(Q)|^2 \, [R(Q,\mu_0)\otimes f_{1T}^{\perp q}(x;\mu_0)\otimes D_1^q(z;\mu_0)](P_T) \,,
\end{eqnarray}
where $\otimes$ indicates the convolution of transverse momenta, $|C_V|^2$ is the perturbative coefficient function, $R(Q,\mu_0)$ represents the evolution factor, while $f_{1T}^{\perp q}$ and $D_1^q$ are the Sivers TMD PDF and the unpolarized TMD FF, respectively. 
(For ease of notation, in Eq.~(\ref{7.2.3:TMD-generic}) the dependence of the nonperturbative functions on the transverse parton momenta has been suppressed.)
The reference scale $\mu_0$ depends on the details of the evolution implementation~\cite{Scimemi:2018xaf,Aybat:2011zv}.
The factorization formula (\ref{7.2.3:TMD-generic}) takes on a simpler form in $b$-space~\cite{Boer:2011xd},
{\small
\begin{eqnarray}\label{7.2.3:TMD-generic-inb}
F_{UT}^{\sin(\phi_h-\phi_S)}&=&\sum_{q}e_q^2|C_V(Q)|^2 \int 
\frac{d^2b}{(2\pi)^2}e^{i(b\cdot P_T)/z} R(Q,b,\mu_0) f_{1T}^{\perp q}(x,b;\mu_0)D_1^q(z,b;\mu_0),
\end{eqnarray}} 

\noindent
where the parameter $b$ is defined as the Fourier conjugate to $P_T/z$. 
In $b$-space, TMDs have a multiplicative evolution and simpler theoretical properties, and therefore this representation is often used in practice~\cite{Collins:1981va,Collins:1984kg}.

\begin{figure}[t]
\begin{center}
\includegraphics[width=1.\textwidth]{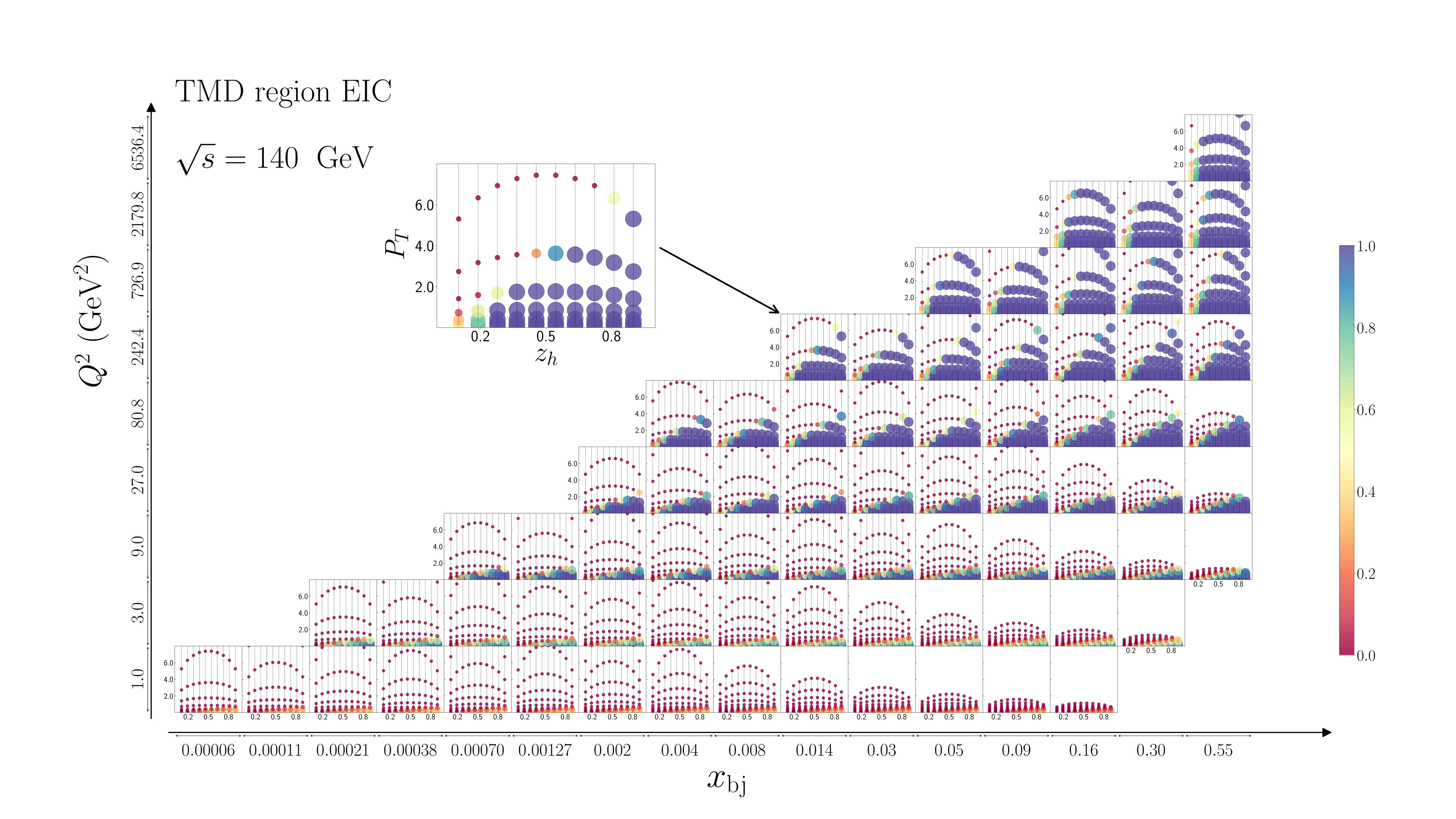}
\caption{\label{fig:affinity} The $x-Q^2$ plane with future EIC measurements at $\sqrt{s}=140$~GeV. Each panel displays $z_h \approx z$ vs.~$P_T$ ranges of measurements (shown by gray lines). The colored symbols represent the estimated affinity of the measurement to TMD factorization region. The color code is proportional to the affinity.
(See text for more details.)}
\end{center}
\end{figure}

The central feature of Eqs.~(\ref{7.2.3:TMD-generic},\,\ref{7.2.3:TMD-generic-inb}) is the presence of three non-perturbative functions: one TMD PDF, one TMD FF,
and the non-perturbative part of the evolution kernel, with the so-called Collins-Soper-kernel (CS-kernel) hidden in $R$. 
To clearly separate these three functions, measurements that are differential in $(Q,x,z)$ with large kinematic coverage are needed. 
The factorization formula~(\ref{7.2.3:TMD-generic}) receives corrections which enter in terms of powers of $\delta\sim P_T / (z Q)$. 
(We also refer to~\cite{Balitsky:2017flc, Balitsky:2017gis} where detailed investigations of power corrections for TMD factorization were presented.)
Identifying the domain of applicability of TMD factorization is nontrivial~\cite{Boglione:2016bph}. 
In recent analyses, usually the choice $\delta<0.25$ has been adopted, at least for high $Q$~\cite{Bacchetta:2017gcc,Scimemi:2017etj,Scimemi:2019cmh,Bacchetta:2019sam}. 
These restrictions reduce the significance of a large number of existing measurements. The EIC will provide measurements in an unprecedented large domain, which ultimately helps to pin down TMDs precisely. However, it also complicates the impact studies for such measurements since many features of TMDs are entirely unconstrained by current measurements, especially for $Q>5-10$~GeV.
In Fig.~\ref{fig:affinity} we show results of Ref.~\cite{Boglione2020} where the regions of pion production in SIDIS at the EIC are studied using results of Ref.~\cite{Boglione:2019nwk}.  The so-called affinity to TMD factorization region, that is, the probability (in the range from 0\% to 100\%) that the data can be described by TMD factorization, is calculated for each bin of the EIC measurements, and indicated by color and symbol size in the figure.
One can see that the bins with relatively high $z_h \approx z$ and $P_T$ (and relatively large $x$ and $Q^2$) are particularly important for the TMD factorization description. The rest of the data (or at least part of it) will be important for other descriptions
such as collinear factorization.

\subsubsection{Unpolarized TMDs and TMD evolution}
Presently, the unpolarized case is the best-studied part of the TMD program due to lots of measurements in many different kinematic ranges, starting from fixed-target experiments \cite{Derrick:1995xg,Adloff:1996dy,Asaturyan:2011mq,Airapetian:2012ki,Adolph:2013stb,Aghasyan:2017ctw} at low energies up to collider measurements at higher energies~\cite{Ito:1980ev,Moreno:1990sf,McGaughey:1994dx,Aidala:2018ajl,Affolder:1999jh,Aaltonen:2012fi,Abbott:1999wk,Abazov:2007ac,Abazov:2010kn,Aad:2014xaa,Aad:2015auj,Chatrchyan:2011wt,Khachatryan:2016nbe,Aaij:2015gna,Aaij:2015zlq,Aaij:2016mgv}. The precision and large span in $Q$ make unpolarized measurements ideal for the determination of the CS-kernel. The latest global analyses reach NNLO perturbative accuracy with N$^3$LL TMD evolution, demonstrate an excellent agreement between the theory and experiments, and provide extracted values of unpolarized TMDs with a good precision \cite{Bacchetta:2017gcc,Scimemi:2017etj,Scimemi:2019cmh,Bacchetta:2019sam} (named as Pavia17, SV17, SV19, Pavia19 for brevity). Nonetheless, the current overall status of the data (which includes extremely precise LHC measurements at $Q\sim M_Z$) does not allow for an accurate reconstruction of TMDs in the $b>1-2$~GeV$^{-1}$ region due to insufficient $P_T$-coverage. In this region, extractions accomplished by different groups can be very different --- see, for example, the comparison in Ref.~\cite{Vladimirov:2020umg}. Measurements at EIC energies will be able to fill in the gap between the low-energy fixed-target experiments and those at the LHC, and this will help pin down these functions at higher values of $b$, i.e., lower values of $k_T$. Additionally, unpolarized structure functions and unpolarized TMDs enter the definitions of other structure functions and spin asymmetries, and thus significantly influence the accuracy of polarized TMDs as well.

\begin{figure}[th]
\begin{center}
\includegraphics[width=0.45\textwidth]{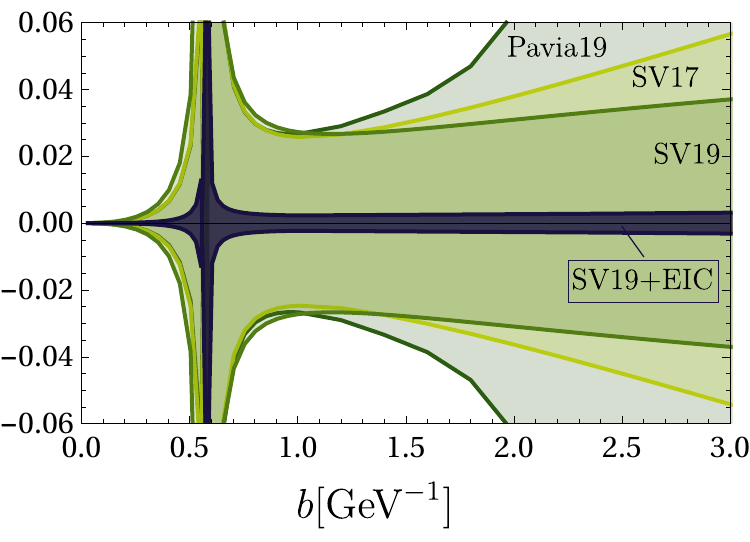}
\caption{\label{fig:RAD} Comparison of relative uncertainty bands (i.e.~uncertainties normalized by central value) for the CS-kernel at $\mu=2$ GeV.}
\end{center}
\end{figure}

\begin{figure}[hb]
\begin{center}
\includegraphics[width=0.95\textwidth]{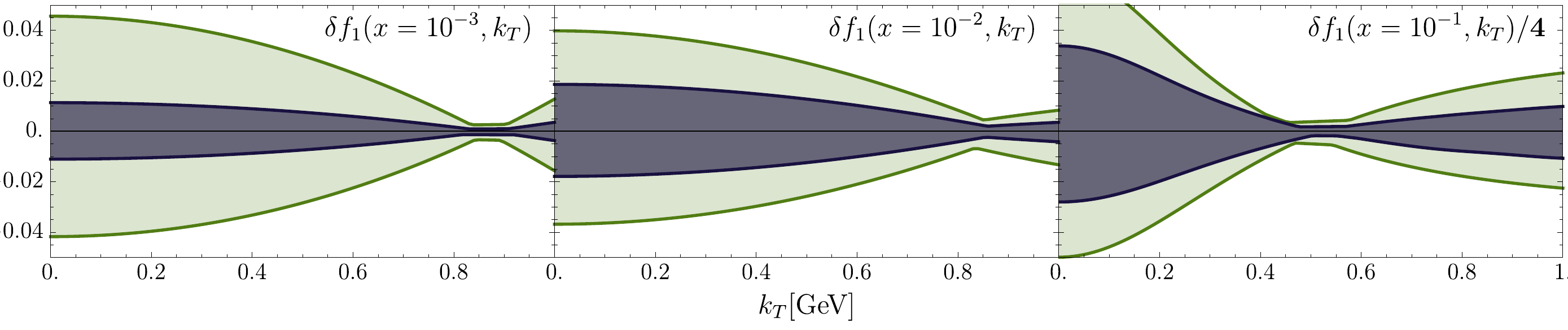}
\includegraphics[width=0.95\textwidth]{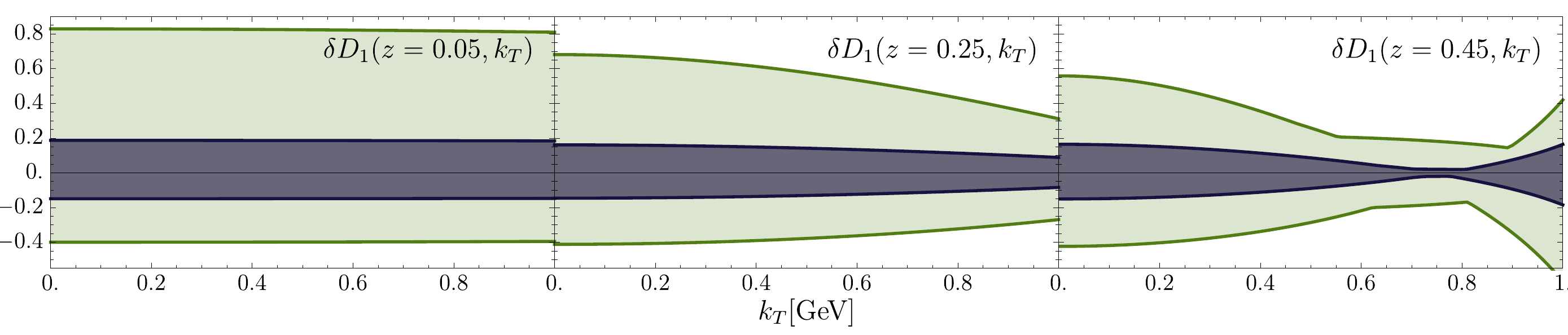}
\caption{\label{fig:unTMDimpact} Comparison of relative uncertainty bands (i.e.~uncertainties normalized by central value) for up-quark unpolarized TMD PDFs (upper panel) and $u \to \pi^+$ pion TMD FFs (lower panel), at different values of $x$ and $z$ as a function of $k_T$, for $\mu=2$ GeV. 
Lighter band is the SV19 extraction, darker is SV19 with EIC pseudodata.}
\end{center}
\end{figure}

To estimate the impact on the non-perturbative part of the CS-kernel and unpolarized TMDs, the SV19 fit was rerun with the inclusion of EIC pseudodata, in $5\times 41$~GeV, $5\times 100$~GeV, $10 \times 100$~GeV, $18\times 100$~GeV and $18\times 275$~GeV energy configurations, scaled to 10 fb$^{-1}$. 
The pseudodata, based on {\sc pythia}~\cite{Sjostrand:2006za} simulations, includes expected statistical and estimated systematic uncertainties, obtained for a hand-book detector design with moderate particle identification (PID) capabilities. 
The estimates for the expected uncertainty bands in comparison to existing ones are shown in Figs.~\ref{fig:RAD} and~\ref{fig:unTMDimpact}. The main impact on the unpolarized sector occurs in the CS-kernel, for which the uncertainty reduces by a factor $\sim 10$. This is due to the unprecedented and homogeneous coverage of the $(Q,x,z)$ domain, which can efficiently decorrelate the effects of soft-gluon evolution and internal transverse motion. Importantly, the current estimate is based on 
one-parameter
models (which likely explains the node-like structures seen in the figures), which are sufficient to describe the current data. Given the precision of the EIC measurements, one can expect to obtain a fine structure of the CS-kernel, which will help to explore properties of the QCD vacuum~\cite{Vladimirov:2020umg}. The unpolarized TMDs will also be significantly constrained through EIC data. The largest impact will be in the regions that are not covered by present data, i.e., for low $x$ and low $z$, where the size of the uncertainty bands can be reduced by a factor $\sim 4$. In other regions, the reduction of uncertainties is smaller, typically by a factor $\sim 2$. The EIC measurements will also play a key role in the study of the flavor structure of TMDs, which is currently almost unconstrained~\cite{Signori:2013mda}, making it difficult to estimate the impact of the EIC.

\subsubsection{Quark Sivers and Collins measurements }
\paragraph{Sivers function measurements:}
\begin{figure}[ht]
    \centering
    \includegraphics[width=0.8\textwidth]{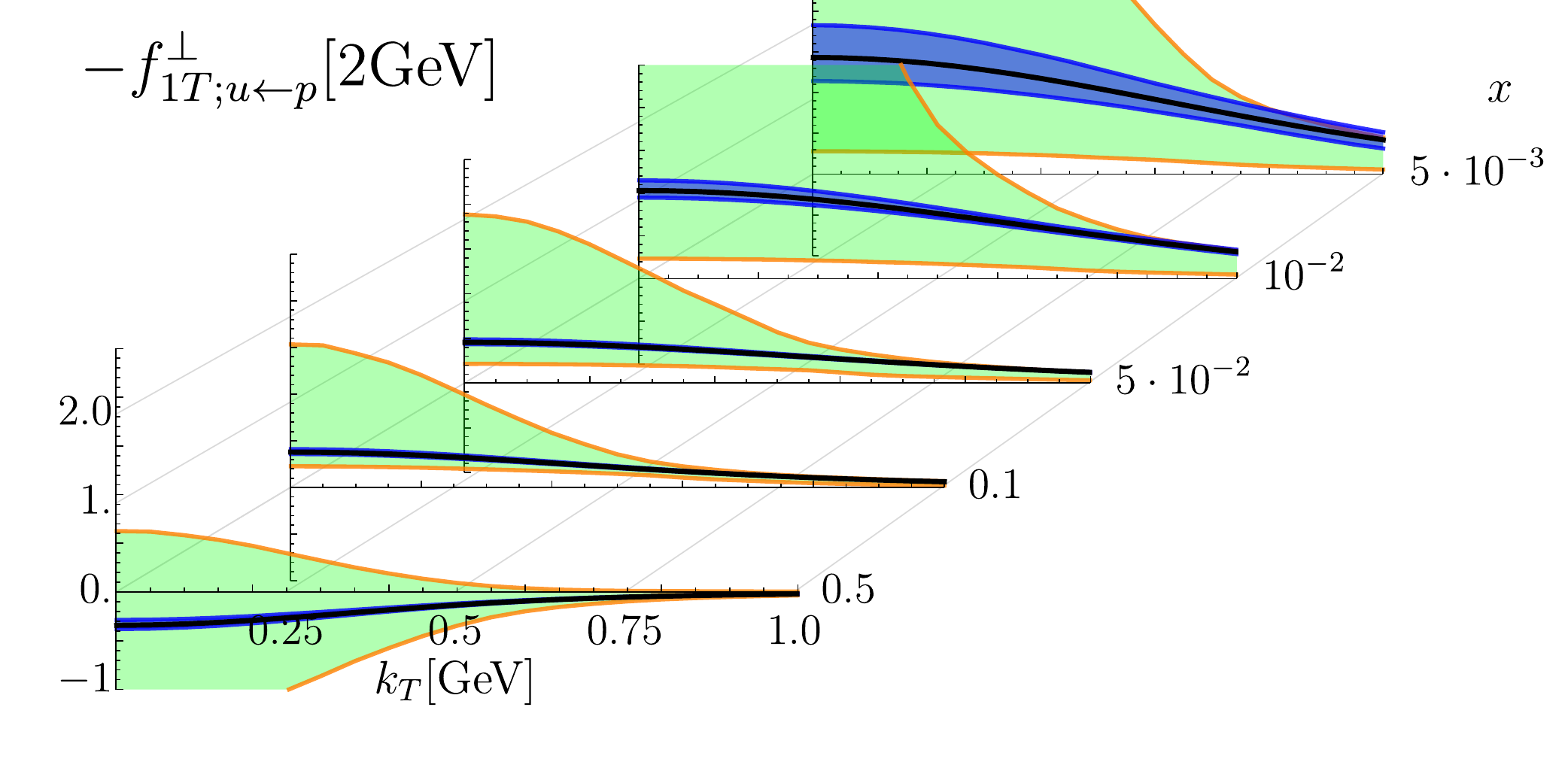}
    \includegraphics[width=0.8\textwidth]{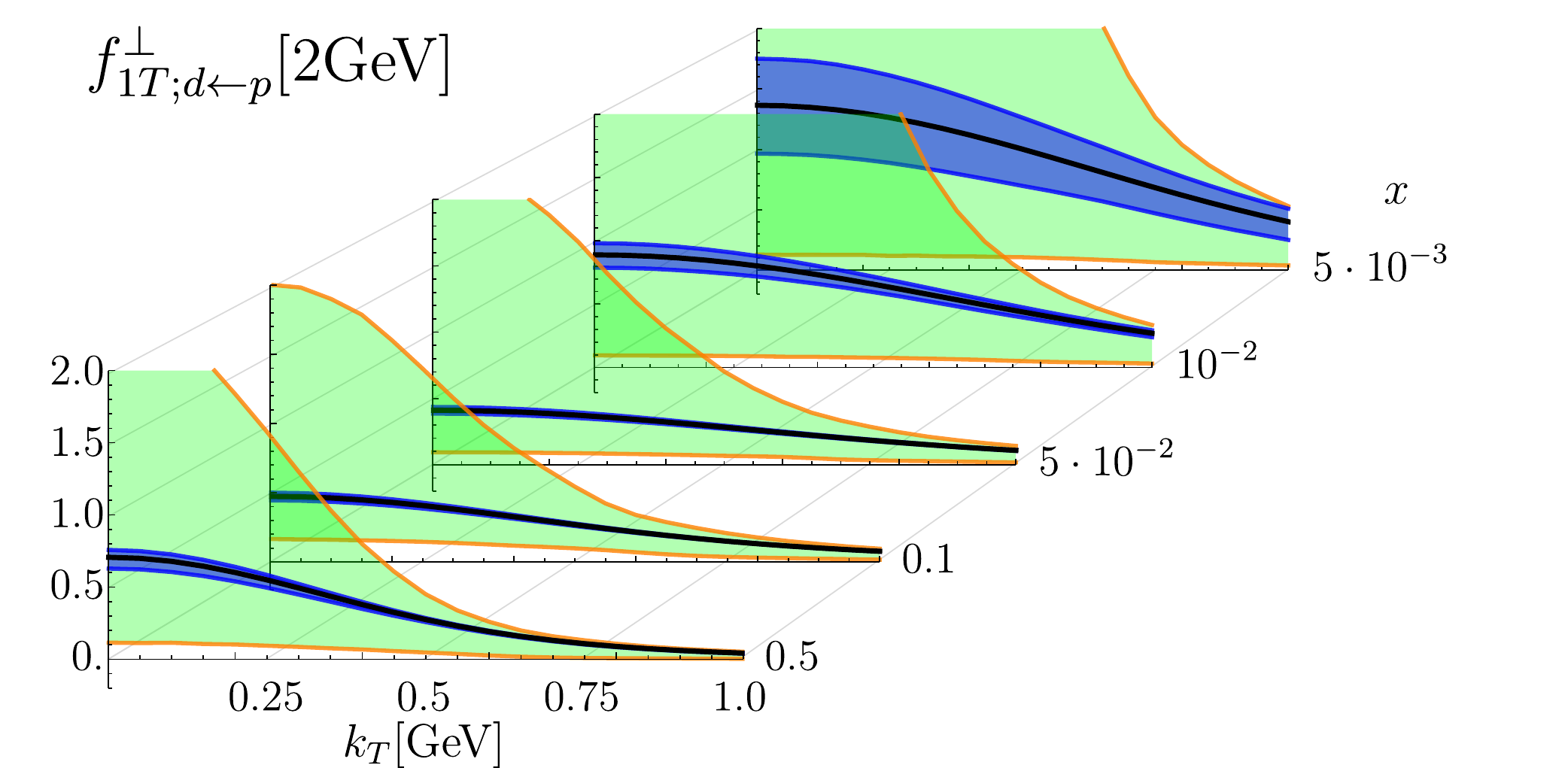}
    \caption{Expected impact on up and down quark Sivers distributions as a function of the transverse momentum $k_T$ for different values of $x$, obtained from SIDIS pion and kaon EIC pseudodata, at the scale of 2 GeV. The green-shaded areas represent the current uncertainty, while the blue-shaded areas are the uncertainties when including the EIC pseudodata.}
    \label{fig:part2-subS-PartStruct-MultiPart.udSivers}
\end{figure}
\label{sec:part2-subS-SecImaging-TMD3d.transversity}
The determination of the quark Sivers functions, $f_{1T}^{\perp q}(x,k_T)$, is one of the major goals for TMD physics. It can be extracted most directly from the transverse SSA proportional to the $\sin(\phi_h-\phi_S)$ modulation of the SIDIS cross section, which is expressed through the structure function $F_{UT}^{\sin(\phi_h-\phi_S)}$ (see Eq.~(\ref{7.2.3:TMD-generic})). 
The Sivers function is a T-odd TMD~\cite{Collins:2002kn}, that turns into the Qiu-Sterman matrix element~\cite{Efremov:1983eb,Qiu:1991pp} in the regime of small $b$~\cite{Boer:2003cm,Scimemi:2019gge}. The extraction of the Sivers TMD was performed by many groups \cite{Efremov:2004tp,Anselmino:2005ea,Collins:2005ie,Vogelsang:2005cs,Anselmino:2008sga,Bacchetta:2011gx,Sun:2013hua,Echevarria:2014xaa,Boglione:2018dqd,Luo:2020hki,Bacchetta:2020gko,Echevarria:2020hpy,Bury:2020vhj}. However, the global pool of Sivers asymmetry data currently has only a relatively small number of data points that satisfy the TMD factorization criterion. Consequently, the uncertainty bands on the Sivers function are very large. To determine the impact of EIC measurements on the Sivers function we used the pseudodata generated by {\sc pythia 6} with a successive reweighing by a phenomenological model for the Sivers and unpolarized structure functions from Ref.\cite{Anselmino:2008sga}. For the present impact study we used pseudodata for $\pi^\pm$ and $K^\pm$ production in $e+p$ collisions at the highest ($18\times 275$~GeV) and the lowest ($5\times 41$~GeV) energy configurations as well as $e+ {}^{3}\textrm{He}$ data at lower energies, scaled to 10 fb$^{-1}$. The systematic uncertainties were estimated as in the unpolarized case. The resulting pseudodata set 
has about two orders of magnitude more points than current data. Performing the fit of pseudodata with the initial setup of the Sivers function from the global analysis made in Ref.~\cite{Bury:2020vhj} based on  
SIDIS~\cite{Airapetian:2020zzo,Adolph:2014zba,Adolph:2016dvl,Alekseev:2008aa,Qian:2011py} and Drell-Yan~\cite{Aghasyan:2017jop,Adamczyk:2015gyk} data, we observe a drastic reduction of uncertainties, as shown in Fig.~\ref{fig:part2-subS-PartStruct-MultiPart.udSivers}. The uncertainty bands can be reduced by more than an order of magnitude, for all flavors.
Measuring the $P_T$-dependence of the cross section for a given $(x,z,Q)$ bin allows for the determination of the $k_T$-shape of the Sivers function, which is currently hardly constrained at all by experimental data.

\begin{figure}[th]
    \centering
    \includegraphics[width=0.97\textwidth]{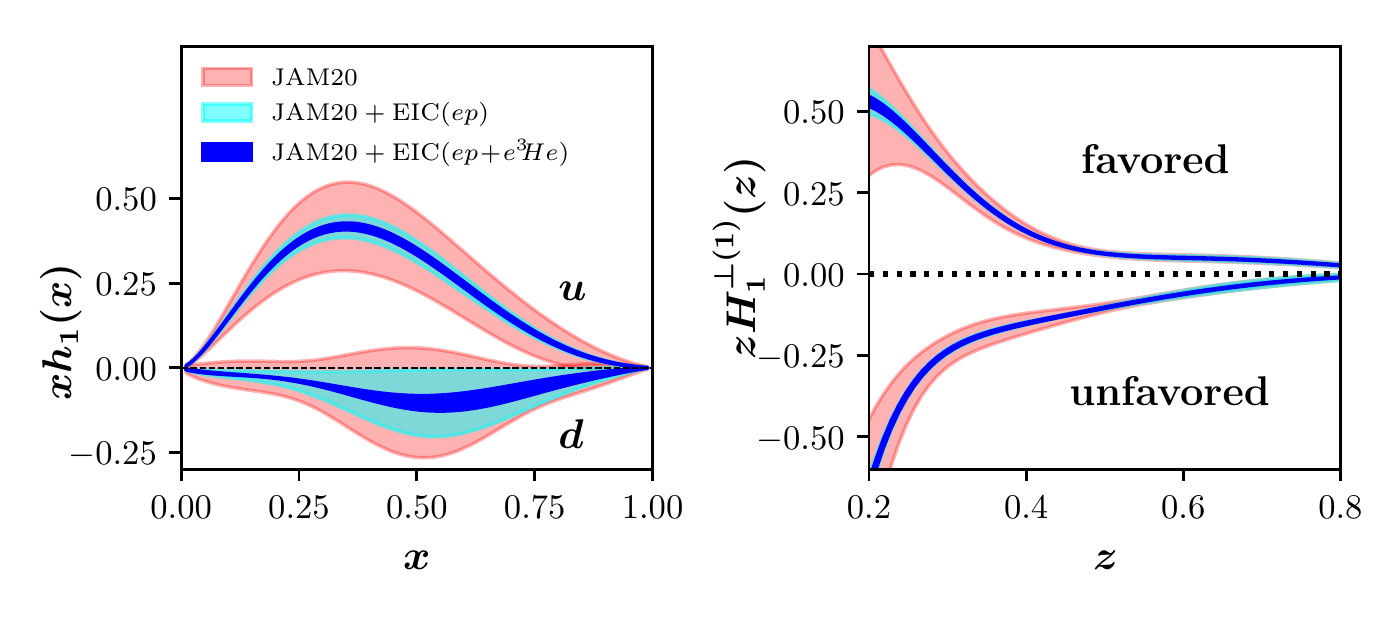}
    \includegraphics[width=0.94\textwidth]{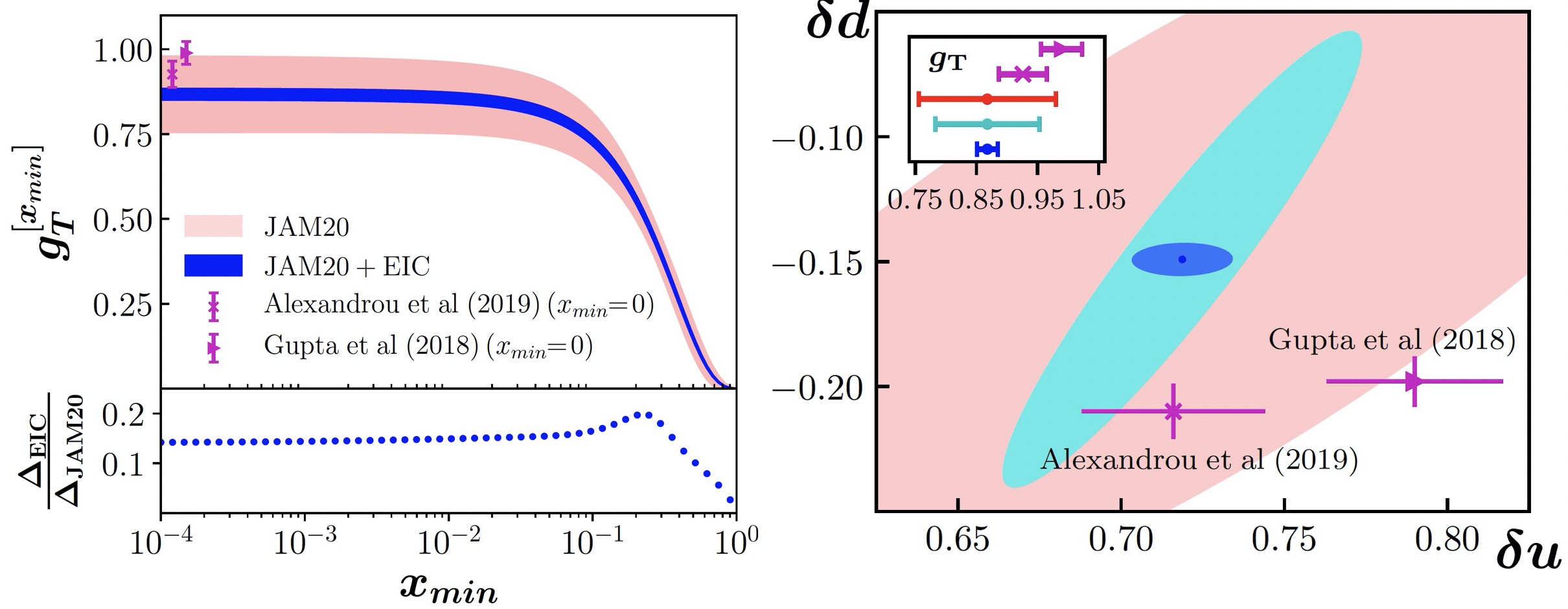}
    \caption{Top: Expected impact on the up and down quark transversity distributions and favored and unfavored Collins function first moment when including EIC Collins effect SIDIS pseudodata from \mbox{\textit{e}+p} and \mbox{\textit{e}+He} collisions~\cite{Gamberg:2021lgx}. 
    Bottom left: Plot of the truncated integral $g_T^{[x_{min}]}$ vs.~$x_{min}$.  Also shown is the ratio $\Delta_{\rm EIC}/\Delta_{\rm JAM20}$ of the uncertainty in $g_T^{[x_{min}]}$ for the re-fit that includes pseudodata from the EIC to that of the original JAM20 fit~\cite{Cammarota:2020qcw}. 
    Note that the results from two recent lattice QCD calculations~\cite{Gupta:2018qil,Alexandrou:2019brg} are for the full $g_T$ integral (i.e., $x_{min}=0$) and have been offset for clarity. 
    Bottom right: The impact on the up quark ($\delta u$), down quark ($\delta d$), and isovector ($g_T$) tensor charges and their comparison to the lattice data.}
    \label{fig:fig:part2-subS-PartStruct-MultiPart.singlehadtransversity}
\end{figure}

\paragraph{Collins-function-based transversity measurements:}
The tensor charges and the (chiral-odd) transversity distributions are the third leading-twist quantity of the nucleon.
They can only be accessed in combination with chiral-odd counter parts such as the Collins fragmentation function~\cite{Collins:1992kk}. Several results from fixed-target SIDIS measurements, $e^+ e^-$ annihilation and, recently, polarized proton-proton collisions, were included in various global fits~\cite{Anselmino:2007fs,Anselmino:2013vqa,Kang:2015msa,Kang:2017btw}, but the uncertainties are still very substantial.  
Using the framework of the QCD global analysis of transverse SSAs developed in Ref.~\cite{Cammarota:2020qcw} (JAM20), the impact of the EIC single-hadron Collins effect SIDIS data on the transversity distributions and the tensor charges was determined  (see Fig.~\ref{fig:fig:part2-subS-PartStruct-MultiPart.singlehadtransversity}). 
Again, {\sc pythia 6} was used to generate pseudodata for both proton and $^3$He beams at various energies and scaled to 10 fb$^{-1}$. The pseudodata were then re-weighted using structure functions based on the extraction presented in Ref.~\cite{Anselmino:2007fs}.
The Collins asymmetries are given by the $\sin(\phi_h+\phi_S)$ modulation of the cross section. 
The JAM20 fit utilizes the connection between TMDs and twist-3 multi-parton correlators to simultaneously fit data from SIDIS~\cite{Airapetian:2004tw,Airapetian:2010ds}, $e^+ e^-$
annihilation~\cite{Abe:2005zx,Seidl:2008xc,TheBABAR:2013yha,Aubert:2015hha,Ablikim:2015pta,Li:2019iyt}, Drell-Yan~\cite{Adamczyk:2015gyk,Aghasyan:2017jop}, and SSAs in proton-proton collisions~\cite{Adams:2003fx,Adler:2005in}.  The significant reduction in the uncertainties is apparent.  In particular, in going from JAM20 to JAM20+EIC($ep$) to JAM20+EIC($ep$+$e \,^3He$) the results for the tensor charges are $\delta u = 0.72(19) \to  0.72(6)\to 0.72(2), \delta d  = -0.15(16)\to -0.15(9)\to -0.149(7), g_T=0.87(11)\to 0.87(9)\to 0.87(2)$. 

The importance of the polarized $^3$He data is also manifest, especially for the down-quark transversity. Further details can be found in Ref.~\cite{Gamberg:2021lgx}. With the EIC data, uncertainties for phenomenological extractions of the tensor charges will become comparable to, and possibly smaller than, current lattice QCD calculations (see, e.g., Ref.~\cite{Gupta:2018qil,Alexandrou:2019brg}). 
As such, potential discrepancies may become relevant for searches of physics beyond the Standard Model~\cite{Courtoy:2015haa,Gao:2017ade}. 
We also mention that there is a significant reduction in the uncertainty for the Collins function (see Fig.~\ref{fig:fig:part2-subS-PartStruct-MultiPart.singlehadtransversity}), which will be an important test of universality with results from $e^+e^-$ annihilation.
(Theoretical considerations suggest that TMD fragmentation functions are universal, based on the specific kinematics of the fragmentation process --- see, for example, Refs.~\cite{Metz:2002iz, Collins:2004nx}.)
In addition, Fig.~\ref{fig:fig:part2-subS-PartStruct-MultiPart.singlehadtransversity} shows $g_T^{[x_{min}]}$ vs.~$x_{min}$, where $g_T^{[x_{min}]}$ is the following truncated integral:~$g_T^{[x_{min}]} \equiv \int_{x_{min}}^1 \! dx \big[(h_1^u(x)-h_1^{\bar{u}}(x))- (h_1^d(x)-h_1^{\bar{d}}(x))\big]$, as well as the reduction in the uncertainty of this quantity from the baseline fit of JAM20~\cite{Cammarota:2020qcw}. 
With the EIC the expectation is that the uncertainty will go down to $10\%$ of JAM20.

\subsubsection{Gluon TMD measurements}

Gluon TMDs encode different correlations between the momentum and spin of the gluon and its parent nucleon.
First classified in Ref.\cite{Mulders:2000sh}, they follow a TMD evolution analogous to that of quark TMDs~\cite{Echevarria:2015uaa}.
Apart from the unpolarized and linearly polarized gluon TMDs inside an unpolarized nucleon, $f_1^g$ and $h_1^{\perp g}$, respectively, of special interest for spin asymmetry measurements are the three naive T-odd gluon TMDs for a transversely polarized nucleon: the gluon Sivers function $f_{1T}^{\perp g}$ and two distributions of linearly polarized gluons, $h_1^g$ and $h_{1T}^{\perp g}$.

The operator structure of gluon TMDs is more involved than for quark TMDs.
Indeed, among the eight leading-twist gluon TMDs, four are naive T-odd, and thus expected to be process-dependent.
The underlying interpretation is that the gauge-link structures involved in the definition of gluon TMDs are different in DIS and hadronic collisions. 
As a result, it is predicted that these T-odd gluon TMDs accessed in $ep^{\uparrow}\rightarrow e' q\bar{q}X $ can be related by an overall sign change to those in $p^{\uparrow}p\rightarrow \gamma\gamma X$ (or any other color-singlet final state, such as di-$J/\psi$ or $J/\psi\,\gamma$).
Moreover, T-odd gluon TMDs can be cast into two types, namely the Weizs\"{a}cker-Williams (WW) type, also known as $f$-type, and the dipole type, also known as $d$-type, depending on the gauge-link structure involved in the scattering processes~\cite{Buffing:2013kca, Pisano:2013cya,Boer:2015pni,Bomhof:2006dp, Dominguez:2011wm, Boer:2015vso,Boer:2016xqr}.
This issue and its impact on EIC physics are discussed in detail in Sec.~\ref{part2-subS-LabQCD-Saturation}.  
The WW gluon TMDs appear exclusively in the $\gamma^{*}g\rightarrow q\bar{q}$ process in DIS, and they are generally difficult to extract in other hadronic collisions~\cite{Dominguez:2011wm}. 
Therefore, through the measurement of WW gluon TMDs, the EIC can provide a unique test of the non-universality of gluon TMDs, complementary to the proposed observables at hadron colliders~\cite{Brodsky:2012vg, Kikola:2017hnp,Hadjidakis:2018ifr}.

Currently, almost nothing is known from experiment about unpolarized or polarized gluon TMDs, except for the unpolarized gluon TMD at very small $x$.
Open-charm production is an ideal probe to study gluon TMDs~\cite{Boer:2016fqd,Boer:2011fh,Burton:2012ug}, but this process is statistically challenging~\cite{Zheng:2018awe} --- for some more details, see the discussion at the end of this section. 
One could also study charm jet pair production \cite{Kang:2020xgk} or measure (single, double or associated) quarkonium  production~\cite{Boer:2011fh,Boer:2012bt,Ma:2012hh,Zhang:2014vmh,Ma:2015vpt, Boer:2015uqa,Boer:2016fqd,Bain:2016rrv,Mukherjee:2015smo,Mukherjee:2016cjw,Lansberg:2017tlc,Lansberg:2017dzg,Bacchetta:2018ivt,DAlesio:2019qpk,Echevarria:2019ynx,Fleming:2019pzj,Scarpa:2019fol,Grewal:2020hoc,Boer:2020bbd}.
In this case, recent theoretical developments~\cite{Echevarria:2019ynx,Fleming:2019pzj} point to the need of new hadronic quantities, the TMD shape functions, which have not been experimentally constrained yet.

\begin{figure}[ht]
\centering
\includegraphics[width=0.47\textwidth]{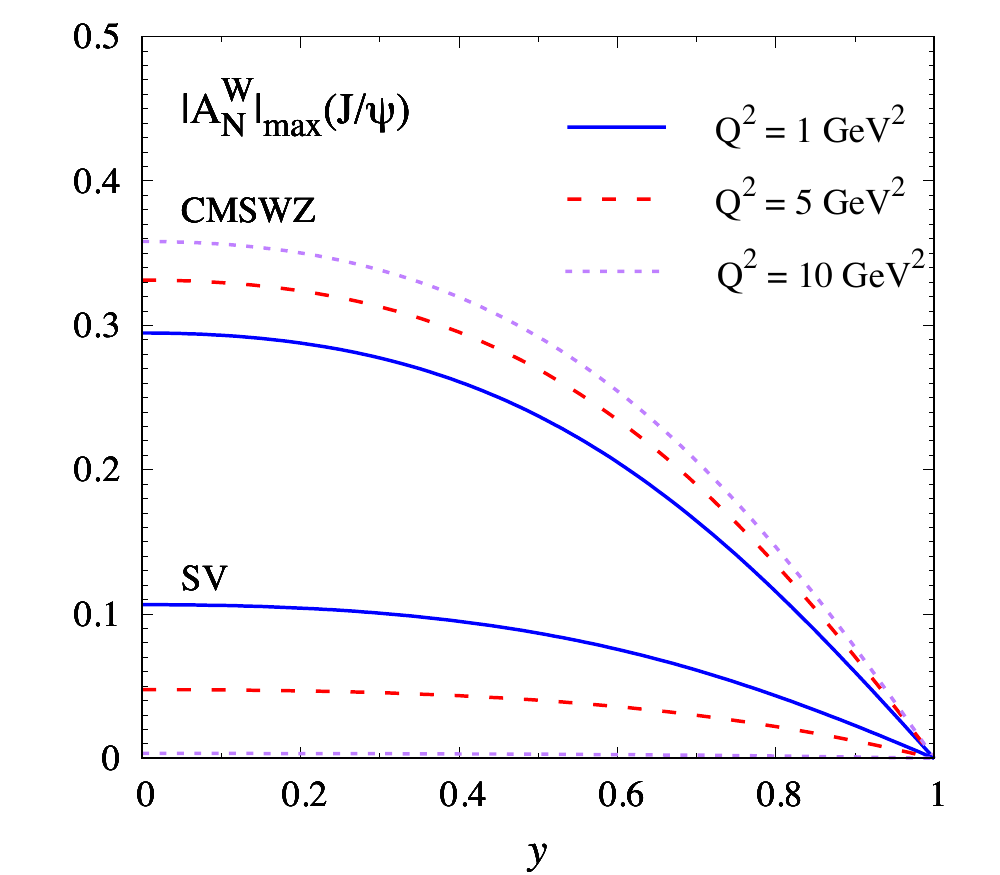}
\includegraphics[width=0.49\textwidth]{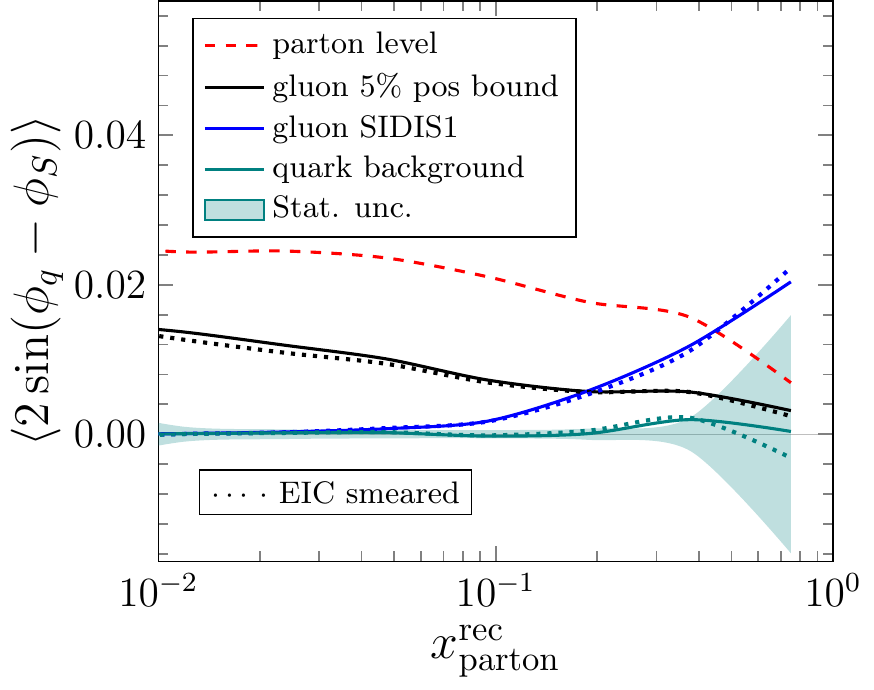}
\caption{
Left: Maximal $A_N^W$ asymmetries with $W =  \cos 2\phi_{\text{T}},\,\sin(\phi_S +\phi_T),\, \sin(\phi_S - 3 \phi_T)$, for $J/\psi$ production in SIDIS. These three asymmetries are sensitive to the linearly polarized gluon distribution $h_1^{\perp g}$ (in an unpolarized nucleon), and two linearly polarized gluon distributions (in a transversely polarized nucleon), $h_1^g$ and $h_{1 T}^{\perp g}$, respectively. The maximal asymmetries are calculated from the positivity bounds for the polarized gluon TMDs, and they become identical for all three weight functions.
The labels SV and CMSWZ refer to the implemented LDME sets~\cite{Chao:2012iv,Sharma:2012dy}. 
(Figure from Ref.~\cite{Bacchetta:2018ivt}.)
Right: projection of SSA $\langle 2\sin(\phi_q-\phi_S) \rangle$ modulation in the dijet channel as a function of the reconstructed parton momentum fraction $x_{\textrm{parton}}^{\textrm{rec}}$. 
The following kinematic cuts are used in the selection of events generated in the simulation at $\sqrt{s}=141$~GeV with an integrated luminosity of 10~$\fb^{-1}$: $0.01<y<0.95$, $1\,
\textrm{GeV}^{2}<Q^{2}<20 \, \textrm{GeV}^{2}$, trigger jet
$P_T^{\textrm{jet1}}>$~4.5~GeV and associated jet $P_T^{\textrm{jet2}}>$~4~GeV.
}
\label{fig:ANandDijet}
\end{figure}

In order to single out different azimuthal modulations of a measurement, which are related to different gluon TMDs, we introduce the azimuthal moments~\cite{Bacchetta:2018ivt}
\begin{align}
A^{W(\phi_S,\phi_T)} & \equiv 2\,\frac {\int   \mathrm{d} \phi_S \, \mathrm{d} \phi_T \, W(\phi_S,\phi_T)\,\mathrm{d}\sigma (\phi_S,\,\phi_T)}{\int  \mathrm{d} \phi_S\,  \mathrm{d} \phi_T \,\mathrm{d}\sigma (\phi_S, \phi_T)}\,,
\end{align}
where $\phi_S$ and $\phi_T$ denote the azimuthal angles of the transverse spin vector and the measured transverse momentum, respectively.
For instance, by taking $W= \cos2\phi_T$ we define $A^{\cos 2\phi_T} \equiv 2 \langle \cos 2\phi_T \rangle $.
The maximum values of such observables/asymmetries in $ep^\uparrow\to e J/\psi X$~\cite{Bacchetta:2018ivt}, obtained from the positivity bounds of the TMDs, are presented in Fig.~\ref{fig:ANandDijet} (left) in a kinematic region accessible at the EIC. They turn out to be measurable, but depend very strongly on the specific set of the adopted long-distance matrix elements (LDMEs).
Similar predictions are obtained for $\Upsilon$ production, and also for $J/\psi+ \textrm{jet}$ production~\cite{DAlesio:2019qpk}, which by varying the mass of the final state also allows one to test the evolution of gluon TMDs.

On the other hand, dijet or high-$\pT$ charged dihadron productions~\cite{delCastillo:2020omr} have also been recently proposed to access gluon TMDs.
In Fig.~\ref{fig:ANandDijet} (right), the projection~\cite{Zheng:2018awe} of the SSA $\langle
2\sin(\phi_q-\phi_S) \rangle $ via dijet production is shown as a function of 
$x_{\textrm{parton}}^{\textrm{rec}}=(p_{T}^{\textrm{jet1}}e^{-\eta^{\textrm{jet1}}}+p_{T}^{\textrm{jet2}}e^{-\eta^{\textrm{jet2}}})/\sqrts$
for $e+p^\uparrow$ collisions at $\sqrt{s}=141$ GeV. 
By defining $\vec{q}_{T}$ as the dijet momentum imbalance, we can compute the angle difference $\phi_{q}-\phi_{S}$ between $\vec{q}_{T}$ and the transverse spin vector $\vec{S}_T$ of the target. 
Using the anti-$k_{T}$ jet algorithm and the cone size $R=0.8$, jets are reconstructed from both charged and neutral particles with a minimum transverse momentum 
of $0.25\, \gevc$ within the pseudorapidity range $|\eta|<3.5$. 
First of all, since the gluon Sivers function is largely unconstrained, the dependence on $x_{\textrm{parton}}^{\textrm{rec}}$ of the resulting SSA strongly relies on the model inputs. 
Two different parametrizations of the gluon Sivers function are used as inputs in this projection. One is the SIDIS1 set extracted from the RHIC $A_N$ data fit of $\pi^0$~\cite{D'Alesio:2015uta}, and the other is an assumption based on 5\% of the positivity bound of the gluon Sivers function. 
The results of these two sets of parametrizations of the gluon Sivers function are shown as blue and black curves in Fig.~\ref{fig:ANandDijet} (right), respectively. 
In contrast, the background asymmetry arising from the quark Sivers function is labeled with the curve in teal. 
Second, the projected statistical uncertainty band as shown in Fig.~\ref{fig:ANandDijet} (right) is sufficient to resolve the signal of the gluon Sivers function down to 5\% of the positivity bound for a wide range of $x$. 
Last but not least, the red dashed curve represents the gluon Sivers asymmetry at the parton level. 
As expected, the jet-level SSAs inherit similar shapes as the parton level ones with a smaller magnitude. 
The dotted curves labeled as "EIC smeared" stand for the SSAs with the EIC detector response and smearing effects taken into account. 
The impact of detector responses becomes significant only in the large-$x$ region due to the limited statistics. 

In summary, the gluon Sivers function can be probed at the EIC down to 5\% of the positivity bound, which allows to explore the little-known correlation between the spin of the proton and the transverse orbital motion of the gluon inside. To measure the gluon Sivers effects (and also the gluon saturation as discussed in Sec.~\ref{part2-subS-LabQCD-Saturation}) via the dijet/dihadron process, a hermetic detector with good tracking (momentum and angular) resolutions will be required, since dijets are produced in the back-to-back azimuthal angle plane across a large range of rapidity, and their momentum imbalance is measured from the vector sum of the reconstructed jet momenta.

\subsubsection{Chiral-odd distribution functions via di-hadron measurements}
\label{sec:part2-subS-SecImaging-TMD3d.IFF}
Di-hadron correlations are sensitive to parton distribution functions via their coupling to di-hadron fragmentation functions (DiFFs)~\cite{Bianconi:1999cd,Radici:2001na,Bacchetta:2002ux,Bacchetta:2003vn}. Due to the extra degrees of freedom, they allow for a targeted access to the nucleon structure. 
One well-known example is the existence of transverse-polarization-sensitive FFs in the collinear framework. They have already been described in this report in their role to extract the twist-3 PDF $e(x)$ in Sec.~\ref{part2-subS-PartStruct-MultiPart}.
The partial-wave decomposition of DiFFs is addressed in Sec.~\ref{part2-subS-Hadronization-diHad-pw}.
Here, we focus on the access to transversity via the DiFF $H_1^\sphericalangle$ and to the Boer-Mulders function via the TMD DiFF $\bar{H}_1^\sphericalangle$. Note that here we follow the notation in Ref.~\cite{Bacchetta:2002ux}, while a notation that unifies the di-hadron and single-hadron FFs was proposed in Ref.~\cite{Gliske:2014wba}.

In SIDIS, transversity can be extracted from the $A^{\sin{(\phi_R+\phi_S)}}\propto h_1  H_1^\sphericalangle$ asymmetries, where $\phi_R$ is the azimuthal angle of the difference of the two hadron momenta~\cite{Bacchetta:2002ux,Gliske:2014wba}. 
The FF $H_1^\sphericalangle$ has been extracted from $e^+e^-$ data~\cite{Vossen:2011fk} by looking at correlations between the azimuthal orientations of two hadron pairs in back-to-back jets~\cite{Boer:2003ya,Bacchetta:2008wb,Courtoy:2012ry,Matevosyan:2018icf}, and it has been used with SIDIS data~\cite{Airapetian:2008sk,Adolph:2012nw,Adolph:2014fjw,Braun:2015baa} to extract the valence components of transversity~\cite{Bacchetta:2011ip,Bacchetta:2012ty,Radici:2015mwa,Benel:2019mcq}. 
The combination $h_1 H_1^\sphericalangle$ was predicted to be accessible also in $pp$ collisions~\cite{Bacchetta:2004it}. Using the STAR data~\cite{Adamczyk:2015hri}, transversity was recently extracted from a global analysis (Pavia18) including both SIDIS and $pp$ collision data~\cite{Radici:2018iag}. 
Similarly to the case of the single-hadron Collins effect, the impact of EIC di-hadron SIDIS data on transversity and the related tensor charges has been estimated by producing pseudodata for proton and $^3$He beams using the {\sc pythia 8} and {\sc DIRE} MC event generators, and re-weighing the pseudodata with structure functions based on the Pavia18 extraction, including a conservative estimate of the scaling of the $H_1^\sphericalangle$ error as $2/\sqrt{N_{\rm EIC}}$, with $N_{\rm EIC}$ the number of EIC pseudodata points. 
In this analysis, the error on $H_1^\sphericalangle$ is a major source of uncertainty. 
However, a significant reduction is expected beyond the current conservative estimate with future {\sc BELLE} II $e^+ e^-$ data and {\sc JLab12} SIDIS data. 

\begin{figure}[ht!]
    \centering
    \includegraphics[width=0.95\textwidth]{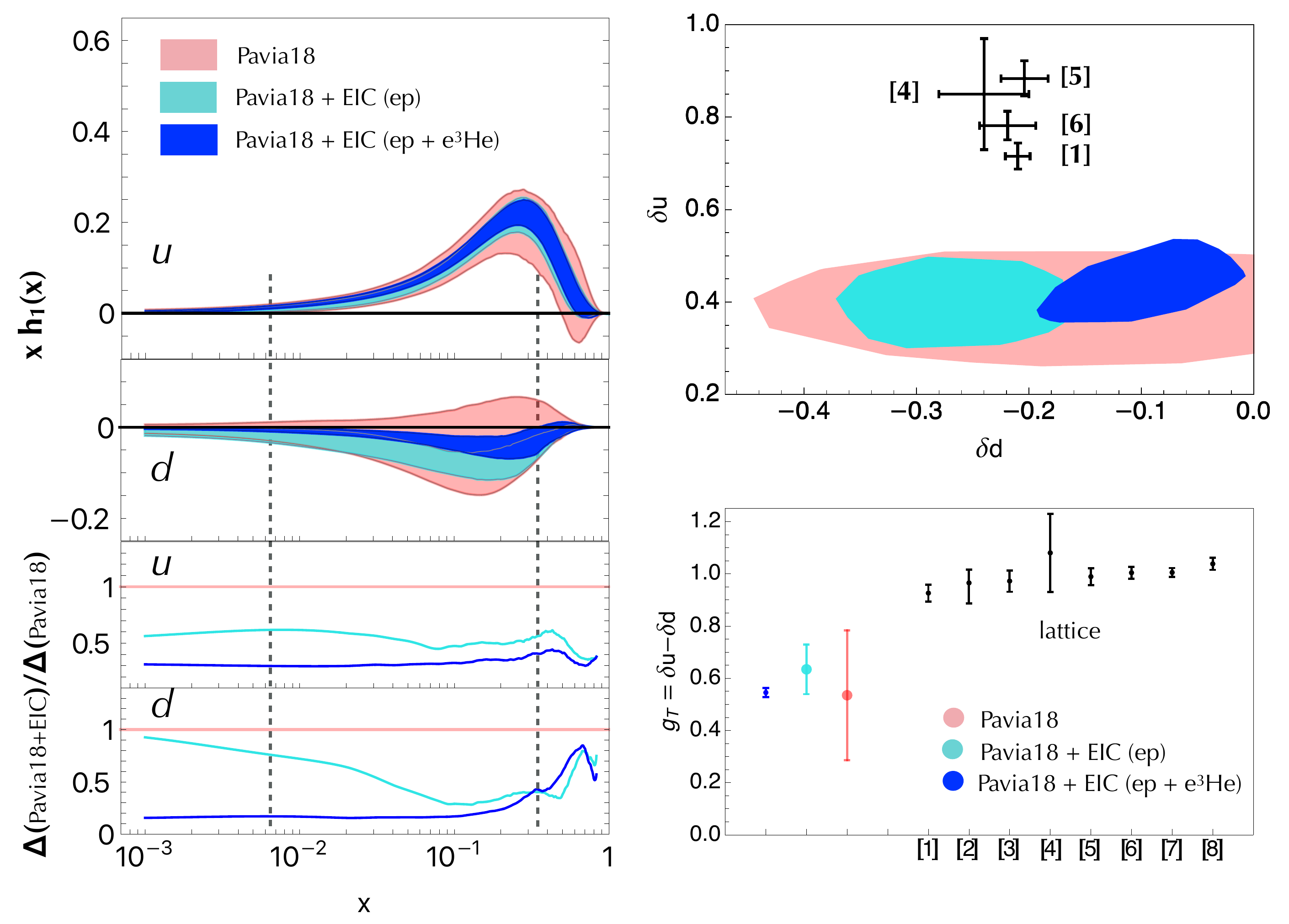}
    \vspace{-0.3cm}
        \caption{Left upper panel: The transversity $x h_1(x)$ as a function of $x$ at $Q^2 = 2.4$ GeV$^2$ for up and down valence quarks. Uncertainty bands for 68\% of all fitted replicas of data (see text). Pink band for the Pavia18 global extraction of Ref.~\cite{Radici:2018iag}, light-blue and blue bands when including EIC SIDIS di-hadron pseudodata from $ep$ and $e^3$He collisions, respectively, with electron/ion beam energy $10 \times 100$ GeV; vertical dashed lines indicate the $x$-range covered by existing data. Left lower panel: ratio of the size of uncertainties with respect to the Pavia18 extraction, with same color codes as before. Right panel: impact of EIC SIDIS di-hadron pseudodata on the up quark $(\delta u)$ vs. down quark $(\delta d)$ tensor charges, and on the isovector tensor charge $g_T$ (same color codes as before), in comparison with some recent lattice calculations, represented by black points and labeled as: [1] Ref.~\cite{Alexandrou:2019brg}, [2] Ref.~\cite{Harris:2019bih}, [3] Ref.~\cite{Hasan:2019noy}, [4] Ref.~\cite{Yamanaka:2018uud}, [5] Ref.~\cite{Gupta:2018qil}, [6] Ref.~\cite{Alexandrou:2017qyt}, [7] Ref.~\cite{Bali:2014nma}, [8] Ref.~\cite{Green:2012ej}. For more information on the EIC impact studies, see Ref.~\cite{Bacchetta20}}
    \label{fig:part2-subS-SecImaging-TMD3d.IFFImpact}
\end{figure}
Figure~\ref{fig:part2-subS-SecImaging-TMD3d.IFFImpact} shows the impact of EIC pseudodata for the electron-ion beam energy $10\times100$~GeV using 10~fb$^{-1}$ integrated luminosity --- see Ref.~\cite{Bacchetta20} for more details. 
The upper left panel shows $x h_1(x)$ at $Q^2 = 2.4$ GeV$^2$ for up and down valence quarks. The statistical error is estimated by making replicas of the pseudodata and by fitting them. 
The displayed uncertainty bands are built by taking the central 68\% of all replicas (if the statistical error has a Gaussian distribution, this is equivalent to $1\sigma$ standard deviation). In the lower left panel, the ratio of the uncertainty widths with respect to the Pavia18 extraction (pink curve) are shown when including only the EIC proton data (light-blue curve) and the EIC proton+$^3$He data (blue curve). Although the projections are made with one energy configuration only and with the conservative cut $0.1 \leq y \leq 0.85$, the impact is quite  evident with an average increase of precision by a factor $2$. The polarized $^3$He data are particularly important for the down quark transversity. They cause a reduction of the uncertainty width by almost an order of magnitude for $x \lesssim 0.01$ with respect to the Pavia18 extraction. Moreover, they shift up the minimum to higher values of $x$, as shown in the upper left panel. In going from Pavia18 to Pavia18+EIC($ep$) to Pavia18+EIC($ep$+$e^3$He), the isovector tensor charge changes as $g_T=0.53(25)\to 0.63(9)\to 0.54(2)$, respectively. The lower right panel of Fig.~\ref{fig:part2-subS-SecImaging-TMD3d.IFFImpact} clearly shows that with the EIC data the uncertainties for phenomenological extractions of $g_T$ can become comparable to, and very likely smaller than, current lattice-QCD calculations, which are indicated by black points and labelled by $[i]$, $i=1,\ldots,8$ (see corresponding references in the figure caption). In the upper right panel, the comparison is performed for the up tensor charge $\delta u$ vs. the down $\delta d$, involving only those lattice calculations that provide results for these flavor-diagonal components. The vertical dashed lines in the left plot indicate the $x$-range covered by current experimental data, with the mininum $x=0.0065$ attained by {\sc COMPASS}~\cite{Braun:2015baa}. It is important to note that no other existing or planned data covers the range $x < 0.0065$ whose impact on the full integral giving the tensor charge should not be neglected. 
Persisting potential discrepancies between phenomenology and lattice-QCD simulations, as in the upper right panel, would then become relevant for searches of physics beyond the Standard Model~\cite{Courtoy:2015haa,Gao:2017ade}.
As systematic effects are difficult to estimate from a fast simulation, consistent with experience from previous SIDIS measurements, a 3\% relative uncertainty and a 3\% scale uncertainty from the beam polarization were summed in quadrature to the statistical uncertainties.

Not shown here are projections for $\pi^\pm-\pi^0$ and $\pi-K$ pairs which will allow for improved flavor separation. The EIC will be able to make precision measurements also in these channels due to its excellent capabilities in PID and electromagnetic calorimetry. In the kinematic region where a $3\sigma$ separation between pion and kaons is possible, the background contribution to the $\pi-K$ sample will be less than 5\%.
The complementarity of extracting the tensor charges via three distinct methods (single, di-hadron and hadrons-in-jets FFs) will reduce the overall systematic uncertainties, both experimental as well as theoretical, significantly.

Like the Sivers function, the Boer-Mulders function, $h_1^\perp$, is naive T-odd~\cite{Boer:1997nt}, and as such allows for the study of similar aspects of QCD as $f_{1T}^\perp$. 
However, because it is chiral-odd, the present information on $h_1^\perp$ from experimental (SIDIS) data is sparse.
In addition to complementarity, extracting $h_1^\perp$ from di-hadron correlations, where it couples to the TMD DiFF $\bar{H}_1^\sphericalangle$, has the advantage that contributions from certain higher-twist contributions are expected to be significantly reduced.
Another less tangible advantage of using di-hadron asymmetries to extract modulations of the unpolarized cross section is that acceptance effects are averaged between the hadrons in the pair, contributing to the complementarity of the measurement and ideally leading to lower overall systematics. DiFFs can also be measured in jets allowing, e.g., access to the Boer-Mulders function with a collinear FF and a separation of the intrinsic transverse momenta of initial and final states in the measurement of TMD DiFFs analogues to in-jet measurements discussed in this section. There is some analogy between DiFFs and in-jet fragmentation, since both introduce an additional momentum vector, increasing the number of degrees of freedom. It will be interesting to explore opportunities that are given by the combination of these approaches.

\subsubsection{Jet-based TMD studies: electron-jet Sivers, hadron-in-jet Collins, and TMD evolution with substructure}
\label{part2-subS-SecImaging-TMD3d.jets}
\begin{figure}[ht]
    \centering
    \includegraphics[width=0.39\textwidth, height=0.32\textwidth]{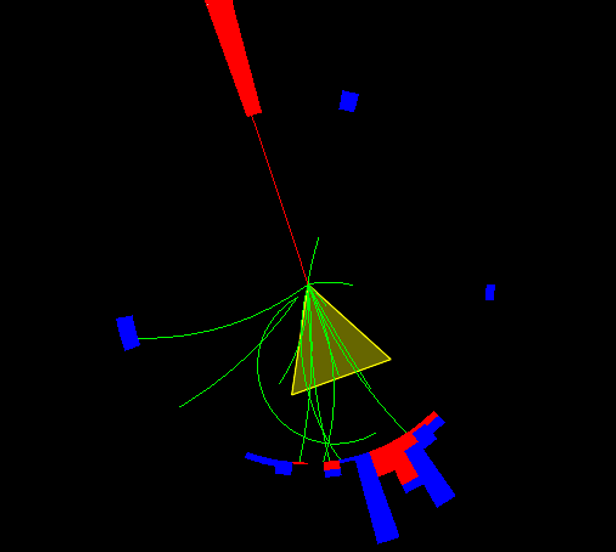}
    \includegraphics[width=0.56\textwidth]{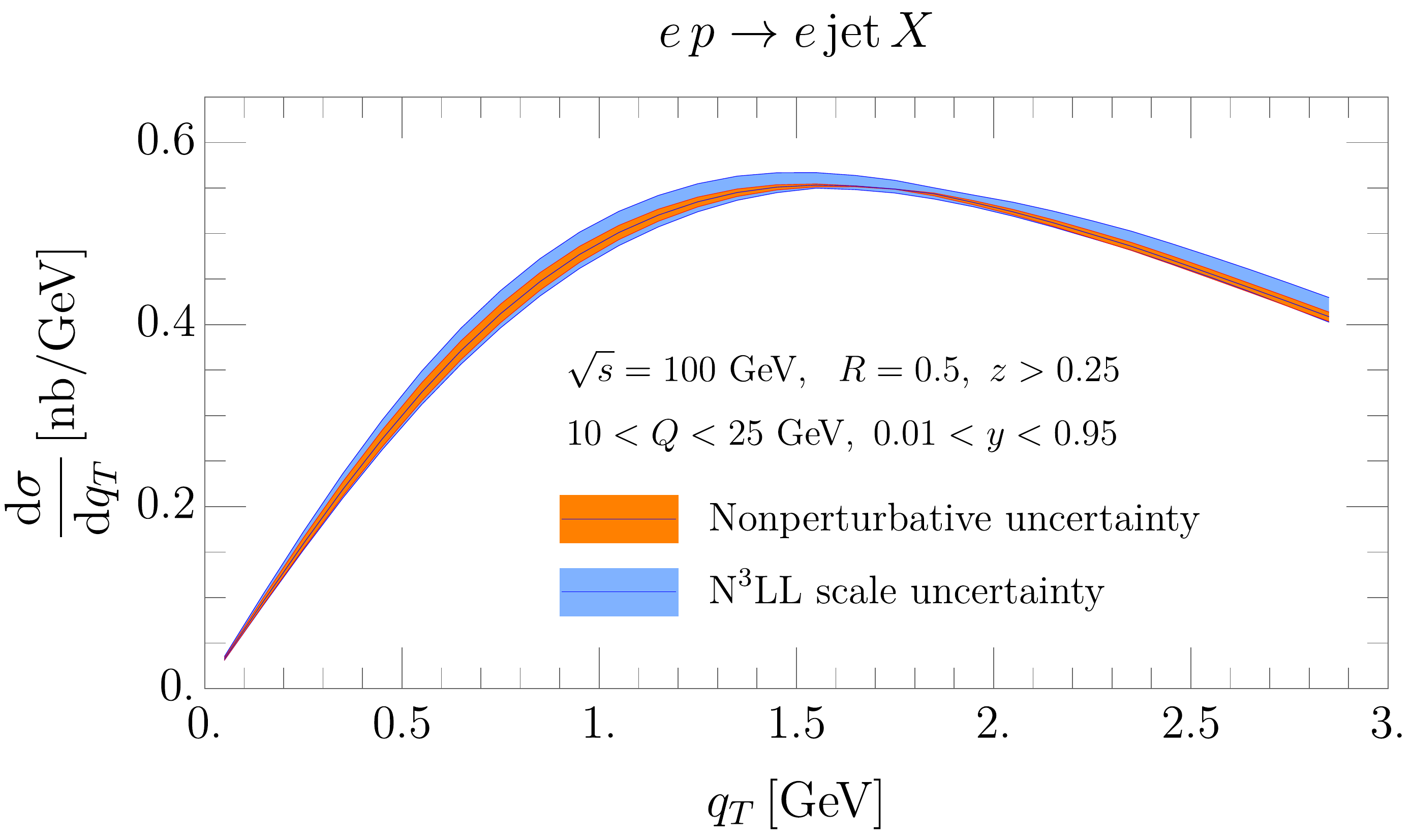}
    \caption{Left: Lepton-jet production close to the Born configuration in the laboratory frame. Right: predicted jet $q_{T}$ spectrum in the Breit frame, adapted from Ref.~\cite{Gutierrez-Reyes:2018qez,Gutierrez-Reyes:2019vbx} }
    \label{jetcartoons}
\end{figure}

\begin{figure}[th]
    \centering
    \includegraphics[width=0.49\textwidth]{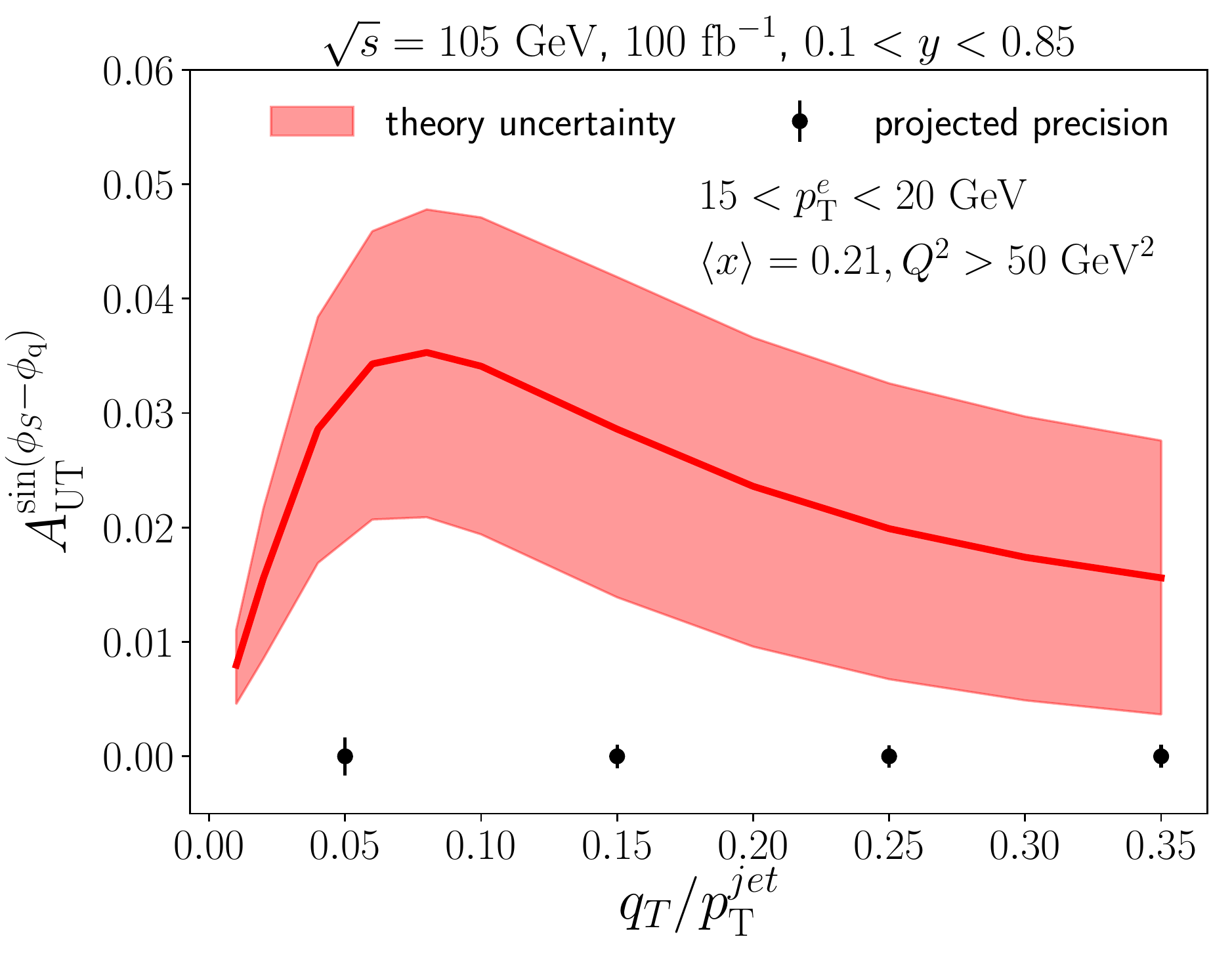}
    \includegraphics[width=0.49\textwidth]{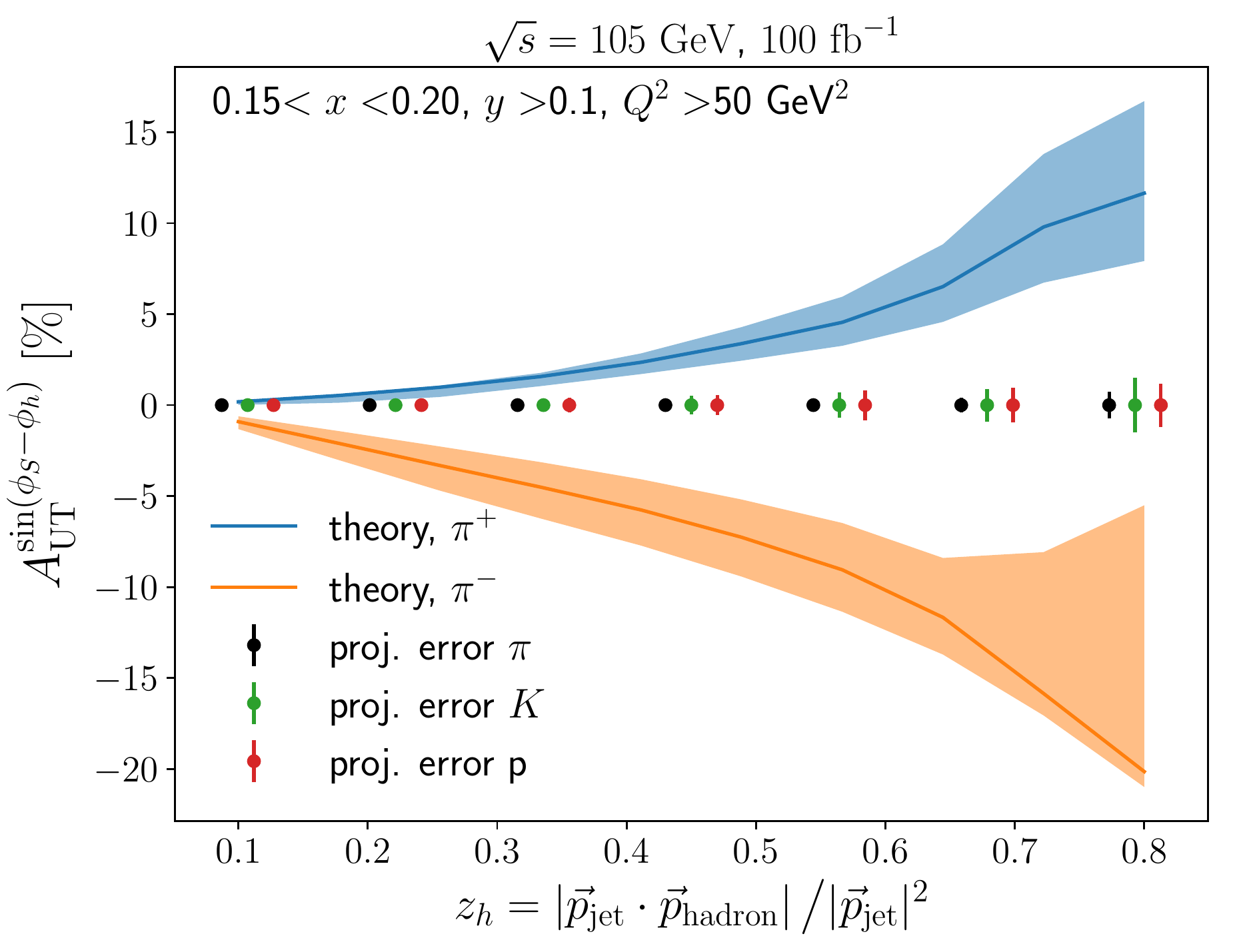}
    \caption{Left: Electron-jet Sivers asymmetry. Right: Hadron-in-jet Collins asymmetry. The error bars represent the expected precision, whereas the bands represent current uncertainties of the Sivers, transversity and Collins TMDs. Note that in this case the observables are calculated in the laboratory frame where $q_T$ corresponds to the transverse momentum imbalance between scattered electron and jet. Figures adapted from Ref.~\cite{Liu:2018trl,Arratia:2020nxw,Liu:2020dct}.}
    \label{JetCollinsSiver}
\end{figure}

Over the last few years, various studies~\cite{Page:2019gbf,Arratia:2019vju,Arratia:2020ssx,Gutierrez-Reyes:2018qez,Liu:2018trl,Gutierrez-Reyes:2019vbx,Gutierrez-Reyes:2019msa,Kang:2020fka,Liu:2020dct,Arratia:2020nxw,Kang:2020xyq,Yuan:2007nd,Bain:2016rrv,Neill:2016vbi,Kang:2017glf,Kang:2017mda,Kang:2017btw,Makris:2017arq,Neill:2018wtk,Cal:2019hjc,Cal:2019gxa} showed that jets offer a novel way to probe quark TMDs and TMD evolution at the EIC --- this possibility was not discussed in the INT Proceedings~\cite{Boer:2011fh} and the EIC White Paper~\cite{Accardi:2012qut}. Jets are excellent proxies for partons, cleanly separate current from target fragmentation~\cite{Page:2019gbf,Arratia:2019vju,Arratia:2020ssx}, and can deconvolve TMD PDFs from FFs~\cite{Gutierrez-Reyes:2018qez,Liu:2018trl,Gutierrez-Reyes:2019vbx,Gutierrez-Reyes:2019msa,Kang:2020fka,Liu:2020dct,Arratia:2020nxw,Arratia:2020ssx,Kang:2020xyq}. Moreover, jet substructure observables can probe TMD-evolution effects~\cite{Yuan:2007nd,Bain:2016rrv,Neill:2016vbi,Kang:2017glf,Kang:2017mda,Kang:2017btw,Makris:2017arq,Neill:2018wtk,Cal:2019hjc,Cal:2019gxa}. Jet physics is flourishing in the LHC era~\cite{Larkoski:2017jix} and transforming the heavy-ion field~\cite{Connors:2017ptx}, so is likely that the development of tailored jet techniques will also advance the field of 3D imaging at the EIC in synergy with traditional SIDIS studies. 
Studies of quark TMDs require jets close to Born kinematics ($\gamma^{*}q\to q$) as illustrated in the left panel of Fig.~\ref{jetcartoons}. There are two frameworks to access TMD PDFs with jets: One uses the Breit frame, which requires suitable jet algorithms~\cite{Arratia:2020ssx}, and defines the 
jet's energy fraction and transverse momentum 
in analogy to SIDIS~\cite{Gutierrez-Reyes:2018qez,Gutierrez-Reyes:2019vbx,Gutierrez-Reyes:2019msa}. Figure~\ref{jetcartoons} shows a jet $q_{T}$ prediction, which probes TMD PDFs independently of TMD FFs. The second possibility is to cluster jets with high 
transverse momenta in the lab frame, which provides another hard scale, in analogy to RHIC studies~\cite{Abelev:2007ii,Bland:2013pkt,Adamczyk:2017wld,Aschenauer:2016our}. The imbalance between the electron and jet probes TMD PDFs independently of TMD FFs~\cite{Liu:2018trl,Arratia:2020nxw,Liu:2020dct,Kang:2020xyq}. Figure~\ref{JetCollinsSiver} shows a prediction for the electron-jet Sivers asymmetry in transversely-polarized scattering. Flavor sensitivity can be achieved by tagging $u,d$ or strange-jets using the jet charge~\cite{Krohn:2012fg,Kang:2020fka}, identified leading hadrons inside the jet~\cite{Neill:2020mtc,Kang:2020fka}, or neutrino-jet correlations in charged-current DIS.

TMD fragmentation can be studied through jet substructure measurements~\cite{Yuan:2007nd,Bain:2016rrv,Kang:2017glf,Kang:2017btw,Kang:2020xyq,DAlesio:2017bvu,DAlesio:2010sag}. For example, in the transversely polarized case, hadron-in-jet measurements probe the quark transversity PDF and the Collins FF, as shown in Fig.~\ref{JetCollinsSiver}. Furthermore, novel techniques such as the winner-take-all scheme~\cite{Bertolini:2013iqa} and jet grooming~\cite{Dasgupta:2013ihk,Larkoski:2014wba} can boost the study of TMD evolution~\cite{Neill:2016vbi,Makris:2017arq,Kang:2017mda,Neill:2018wtk,Cal:2019hjc,Cal:2019gxa}. Future jet substructure studies will likely exploit the unprecedented combination of tracking, PID, and full calorimetry of the EIC detectors.

\subsubsection{Event shapes}
\begin{figure}[ht]
    \centering
    \includegraphics[width=0.495 \textwidth]{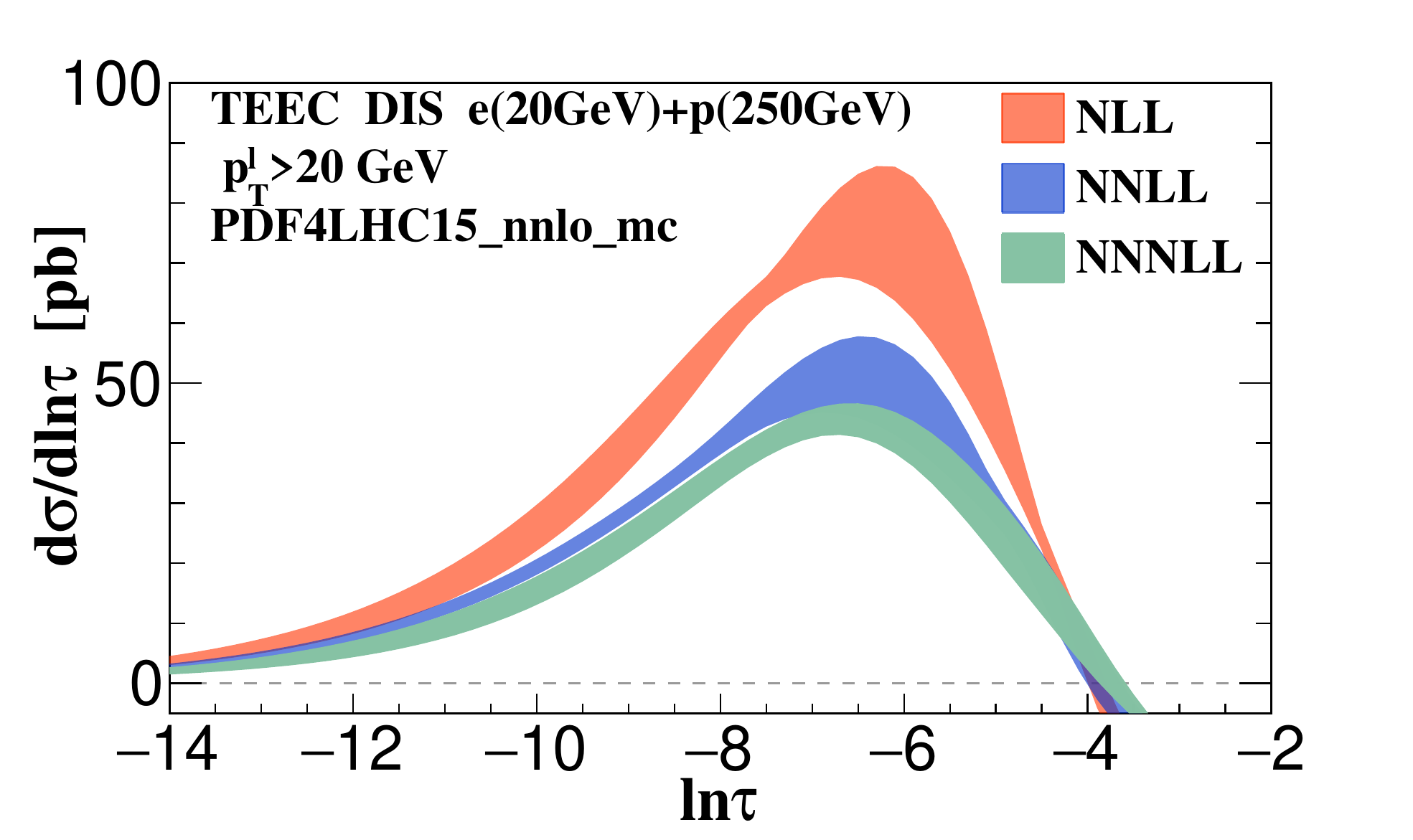}
    \includegraphics[width=0.495 \textwidth]{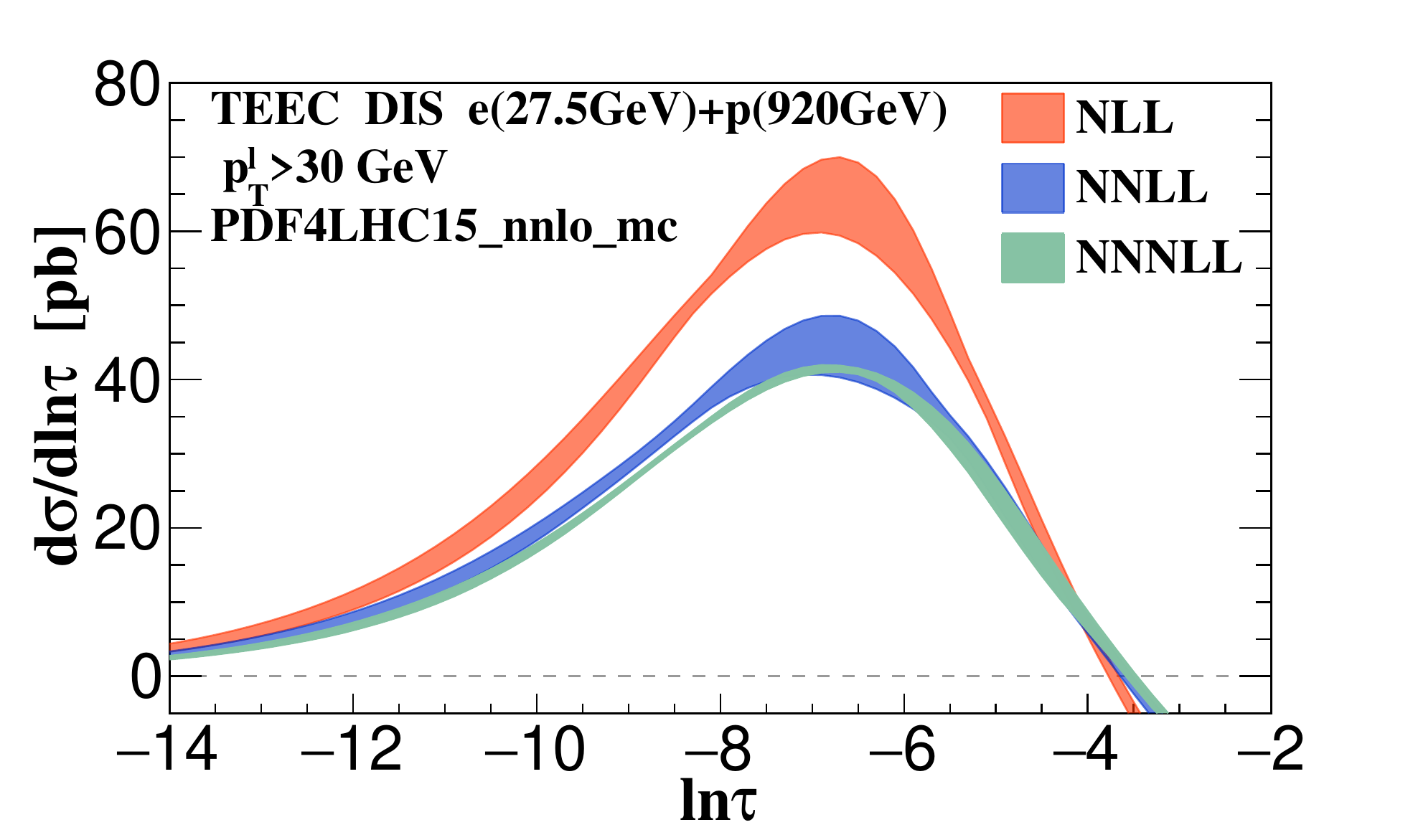}
    \caption{Resummed TEEC distributions in the back-to-back limit as a function of $\tau=(1+\cos\phi)/2$, which describes the deviation of the scattered lepton and the produced hadrons from being  back-to-back in the transverse plane. The orange, blue, and green bands are the predictions with scale uncertainties at NLL, NNLL and N$^3$LL, respectively. The left and right panels are for EIC and HERA energies, respectively.}
    \label{fig:resum}
\end{figure}

Event-shape observables have been widely used for precision QCD studies at various lepton and hadron colliders --- see also Sec.~\ref{part2-subS-PartStruct-GlobalEvent} above.
Transverse-energy-energy correlators (TEEC) have recently been calculated to high precision in hadronic collisions using techniques from soft-collinear effective theory~\cite{Gao:2019ojf}. In DIS, TEEC can be generalized by considering the transverse-energy-energy correlation between the lepton and hadrons in the final state, 
\begin{align}~\label{eq:teec_dis}
    \text{TEEC}(\phi) =&  \sum_{a} \int d\sigma_{lp\to l+a+X} \, \frac{ E_{T,l}  E_{T,a}}{E_{T,l} \sum_{i} E_{T,i}} \, \delta(\cos\phi_{la}-\cos\phi) 
    \nonumber \\ 
    =& \sum_{a} \int d\sigma_{lp\to l+a+X} \, \frac{  E_{T,a}}{\sum_{i} E_{T,i}} \, \delta(\cos\phi_{la}-\cos\phi) \, ,
\end{align}
where the sum runs over all the hadrons in the final states and $\phi_{la}$ is the azimuthal angle between the final-state lepton $l$ and hadron $a$ measured in a plane transverse to the collision axis in the lab frame. Recently, this observable has been evaluated to the highest resummed accuracy in DIS~\cite{Li:2020bub} --- N$^3$LL matched with the NLO cross section for the production of a lepton and two jets.  Figure~\ref{fig:resum} shows the precision of successive orders in the nearly back-to-back TEEC limit for EIC and HERA center-of-mass energies as a function of $\tau=(1+\cos\phi)/2$. 

The TEEC cross section can be factorized as the convolution of a hard function, beam function, jet function and soft function in the back-to-back limit. A close connection to  TMD factorization is established,  as the beam function, when combined with part of the soft function, is identical to the  conventional TMD parton distribution function,  and the jet function is the second moment of the TMD fragmentation function matching coefficient. As such, the generalization of TEEC to DIS~\cite{Li:2020bub}  provides a new way to precisely study TMD physics and non-perturbative effects at the future EIC. 

Recently, for DIS a new definition of energy-energy-correlations (EEC) adapted to the Breit frame has been introduced~\cite{Li:2021txc}. This observable can be calculated to the same next-to-leading order and next-to-next-to-next-to-leading logarithmic resummed accuracy as the TEEC. One of its important advantages is its insensitivity to experimental pseudorapidity cuts, often imposed in the lab frame due to detector acceptance limitations.

\subsubsection{Opportunities with heavy quarks}

Charm anti-charm hadron pair production in lepton-nucleon DIS proceeds at Born level via the photon-gluon-fusion process and thus offers attractive opportunities to study gluon TMDs~\cite{Boer:2011fh,Burton:2012ug,Boer:2015vso,Boer:2016fqd,Zheng:2018awe}.  
The gluon Sivers TMD, for example, and TMDs of linearly polarized gluons can be linked to azimuthal anisotropies of the produced charm anti-charm hadron pair. The Sivers asymmetry can be extracted from  measurements of the transverse SSA $A_{UT}$, as a function of the azimuthal angle of the $c\bar{c}$ hadron pair relative to the orientation of the proton spin. 
$A_{UT}(p_T)$ is defined in the standard way as $[\sigma_{L}(p_T) - \sigma_{R}(p_T)]/[\sigma_{L}(p_T) +\sigma_{R}(p_T)]$,
where $\sigma_{L(R)}$ are the cross sections for particle-of-interest production with spin polarized in the direction opposite to (same as) the spin of the proton, and $p_T$ is the transverse momentum of the heavy hadron pair. The SSA is directly related to the gluon Sivers effect, $A_{UT}(p_T) \propto f_{1T}^{\perp g}(x_g,k_T) / f_1^g(x_g,k_T)$, where $x_g$ is the gluon momentum fraction, $k_T$ is the transverse momentum of the incoming gluon,
and $f_{1T}^{\perp g}$ and $f_1^g$ are the gluon Sivers function and the unpolarized gluon TMD, respectively. 

\begin{figure}[ht]
\centering
\begin{minipage}{.5\textwidth}
  \centering
  \includegraphics[width=0.9\linewidth]{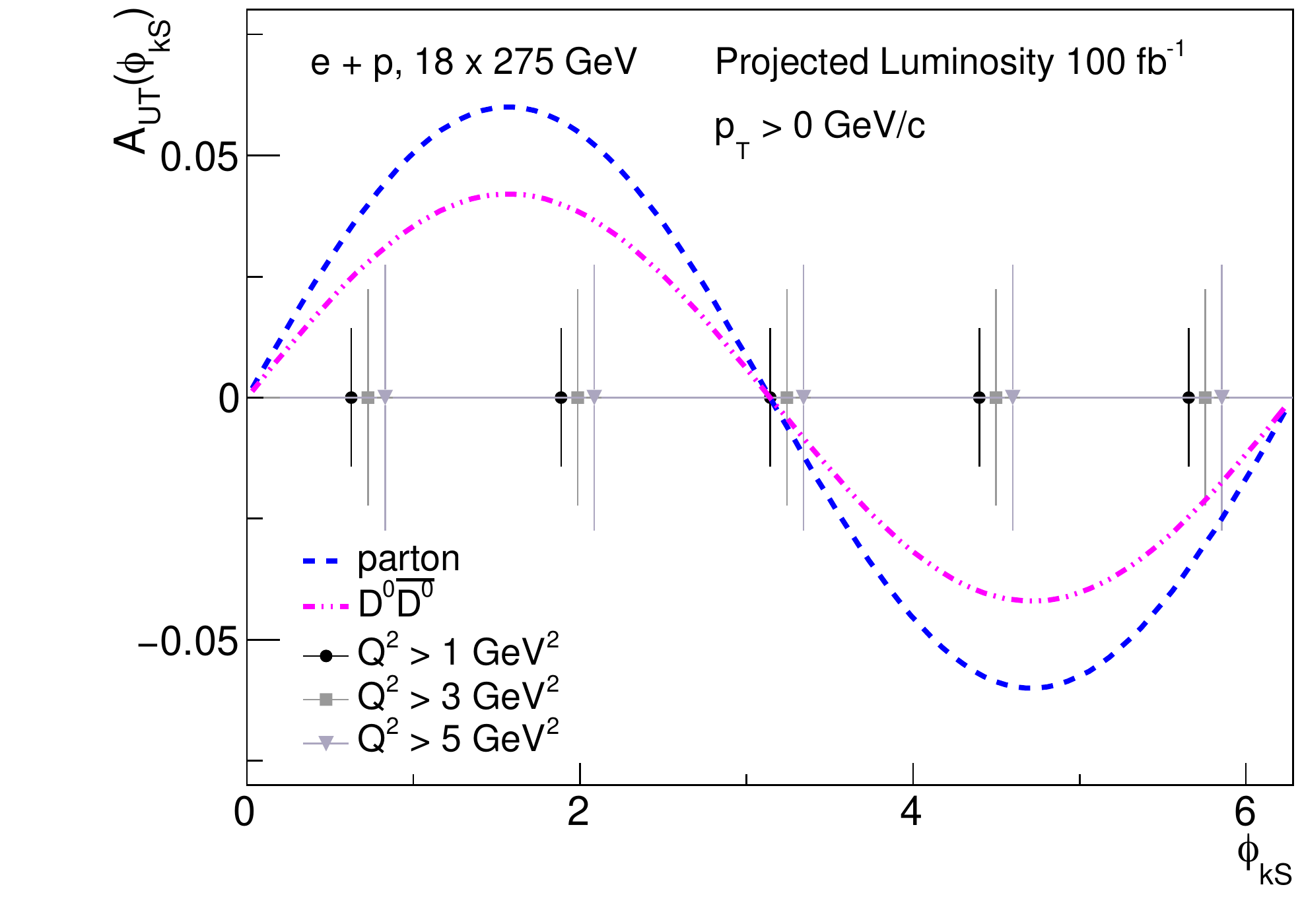}
\end{minipage}%
\begin{minipage}{.5\textwidth}
  \centering
  \includegraphics[width=0.9\linewidth]{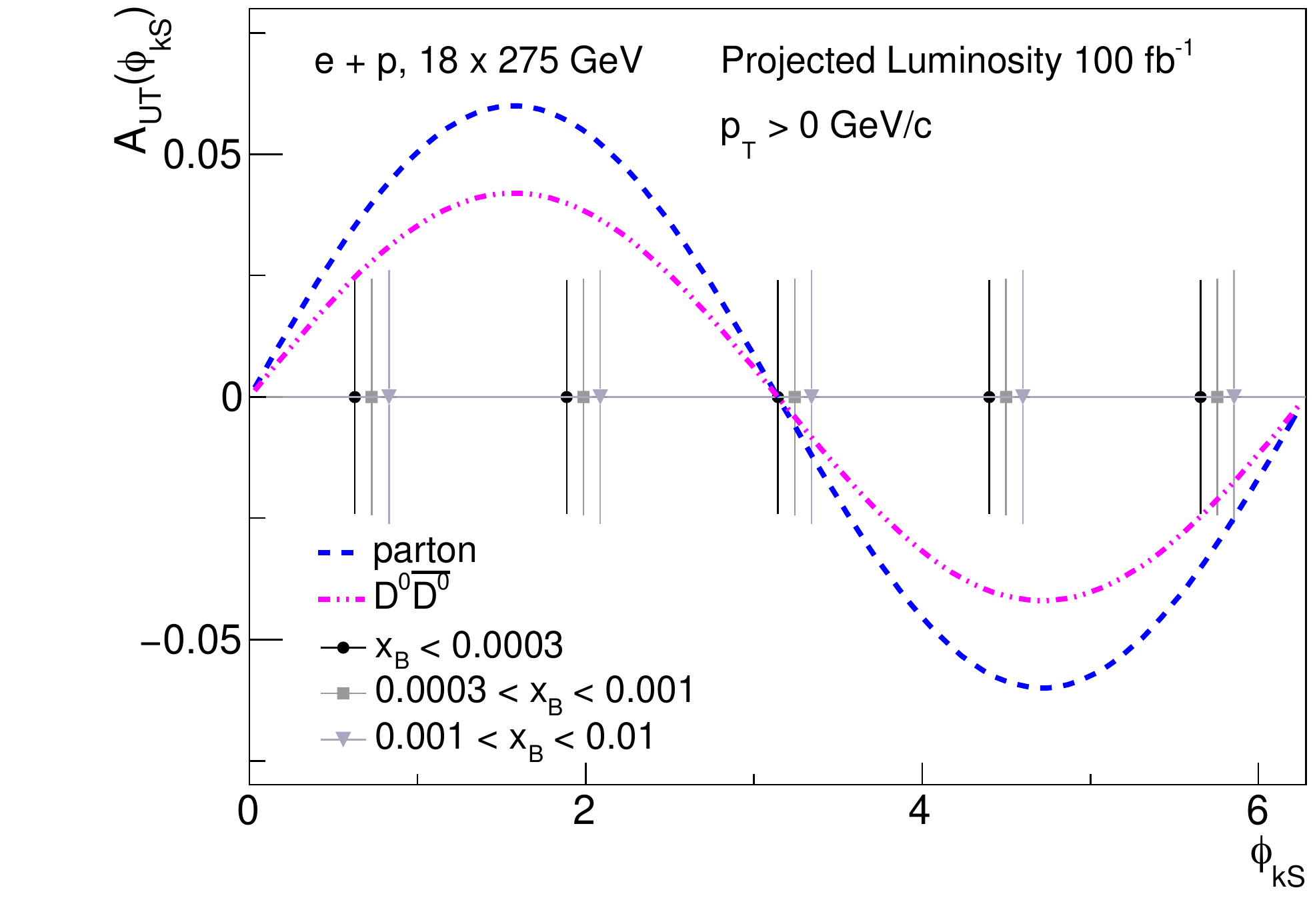}
\end{minipage}
\caption{
	    {\small Statistical uncertainty projections for $A_{UT}$ in bins of azimuthal angle of the momentum of the $D^0\overline{D}^0$ pair relative to the spin of the proton ($\phi_{kS}$), for different $Q^{2}$ (left) and $x_{B}$ (right) selections. The two curves indicate the signal strength at parton and $D^0\bar{D^0}$ levels. The above results are obtained by using the same inputs for the gluon TMDs as in Ref.~\cite{Zheng:2018awe} and assuming that the gluon Sivers function takes the magnitude of its $10\%$ positivity bound. }}
\label{fig:QQ_sivers}
\end{figure}

The correlation between the azimuthal angle of the $c\bar{c}$ pair momentum and that of the corresponding hadron pair momentum in the case of $D^0\overline{D}^0$ production was studied in PYTHIA 6.4 simulations and was found to be well-preserved during hadronization. 
This is because the $c\bar{c}$ pair momentum, which is identical to $k_T$ at leading order, is approximately equal to the $D^0\overline{D}^0$ pair momentum $p_T$. The signal strength, $A_{UT}$ at the partonic level, can be reduced by up to 30\% in the heavy-quark production and subsequent hadronization in these simulations. 
The effects of detector response were investigated using fast simulations that smeared the particle tracks. 
The analysis included topological selections entailing cuts on the secondary vertex fitted from the $D^0$ and $\overline{D}^0$ decay daughter particles to assess signal significance and backgrounds~\cite{Arrington:2021yeb}.
Figure~\ref{fig:QQ_sivers} shows uncertainty projections for $A_{\rm UT}$ for different values of $Q^2$ and $x_B$, in comparison with the possible signal size~\cite{Zheng:2018awe} of this thus far poorly constrained quantity.

\subsection{Wigner functions}
\label{part2-subS-SecImaging-Wigner}

Generalized transverse momentum dependent parton distributions (GTMDs) $G(x,\boldsymbol{k}_T,\boldsymbol{\Delta}_T,{\cal W})$ provide the most complete one-body information on the partons inside hadrons~\cite{Meissner:2009ww,Meissner:2008ay,Lorce:2011dv,Lorce:2013pza}.
They can be thought of as the mother distributions of TMDs and GPDs, since they reduce to these lower-dimensional distributions via appropriate projections.   
GTMDs contain richer physics than TMDs and GPDs combined, as they can describe nontrivial correlations between $\boldsymbol{k}_T$ and $\boldsymbol{\Delta}_T$ which are inaccessible from the studies of TMDs and GPDs separately~\cite{Lorce:2011kd,Lorce:2015sqe}. 
This is evident when considering Wigner distributions $W(x,\boldsymbol{k}_T,\boldsymbol{b}_T,{\cal W})$, which represent the counterpart of GTMDs in the phase-space of momentum ($k^+,\boldsymbol{k}_T$) and position ($\boldsymbol{b}_T$) coordinates~\cite{Ji:2003ak,Belitsky:2003nz}. 
GTMDs and Wigner distributions are related by a Fourier transformation in $\boldsymbol{\Delta}_T \leftrightarrow \boldsymbol{b}_T$, at vanishing longitudinal momentum transfer $\Delta^+=0$~\cite{Lorce:2011kd}. 

Originally introduced in the context of nucleon structure in 2003  \cite{Ji:2003ak,Belitsky:2003nz}, the Wigner distributions have long been thought of as purely theoretical constructs without experimental relevance. 
In the EIC White Paper~\cite{Accardi:2012qut} published in 2012, there was very little account of the Wigner distribution, let alone experimental probes of it. 
However, the situation has changed dramatically over the past several years. 
One of the remarkable findings is that the Wigner distribution for unpolarized $(U)$ partons in a longitudinally ($L$) polarized nucleon provides an intuitive, but rigorous and gauge-invariant, definition of the orbital angular momentum (OAM) of quarks and gluons~\cite{Lorce:2011kd,Hatta:2011ku,Lorce:2011ni},  
\begin{equation}
L^z_{q,g}=\int dx \, d^2 k_T \, d^2 b_T \, (\boldsymbol{b}_T \times \boldsymbol{k}_T)^z \, W^{q,g}_{LU}(x, \boldsymbol{k}_T, \boldsymbol{b}_T, {\cal W}) \,.
\label{oam}
\end{equation}
Depending on the path along which the Wilson line ${\cal W} $ is running, the relation~\eqref{oam} pertains to the two commonly used definitions of the quark OAM~\cite{Hatta:2011ku, Ji:2012sj, Lorce:2012ce}, i.e.,  the (canonical) one by Jaffe and Manohar ($L_{\rm JM}$)~\cite{Jaffe:1989jz}, and the (kinetic) one by Ji ($L_{\rm Ji}$)~\cite{Ji:1996ek}, and allows for an intuitive interpretation of the difference $L_{\rm JM} - L_{\rm Ji}$~\cite{Burkardt:2012sd}.
Therefore,  an experimental program to  extract the Wigner distributions will give the opportunity
to gain information on the kinetic OAM that is complementary to the study through Ji's relation from GPDs.  More remarkably, it will open the way to experimentally access the canonical OAM, for which we can obtain only indirect information from TMDs. While this is an unprecedented challenge, we believe that with sufficient theory efforts and experimental planning, one may pursue such measurements at the EIC. The first theoretical ideas for observables can be found in~\cite{Hatta:2016aoc,Ji:2016jgn,Rajan:2016tlg,Bhattacharya:2017bvs,Bhattacharya:2018lgm}, and their experimental feasibility tests have begun at the LHC. 
The first preliminary results from ATLAS~\cite{ATLAS:2017kwa} and CMS on dijet photoproduction~\cite{CMS:2020ekd} are now available for studies of gluon dynamics with implications for the future EIC. 

Another important development is the recognition that the gluon Wigner distribution at small $x$ is proportional to the so-called dipole S-matrix, which is a fundamental object in the physics of gluon saturation \cite{Hatta:2016dxp}. Almost all observables calculated in the Color Glass Condensate (CGC) framework involve the dipole S-matrix in one way or another. Moreover, the phase space distribution of gluons and their correlation have long been discussed in the small-$x$ literature, without calling it a Wigner distribution.  One can now put these efforts in a fresh context and integrate them into the general goal of studying multi-dimensional tomography at the EIC. This may also be a good starting point to explore the use of quark and gluon Wigner distributions at large $x$.

In this Yellow Report, we consider using exclusive dijet production in $ep$ collisions to access the gluon GTMD (Wigner) distribution at small $x$, as suggested in Refs.~\cite{Hatta:2016dxp,Hagiwara:2017fye}, see also \cite{Altinoluk:2015dpi}. 
This is possible because of the presence of two external momentum vectors, the proton recoil momentum $\boldsymbol{\Delta}_T$, which is approximately the negative total dijet momentum 
$-(\boldsymbol{p}_{T 1} + \boldsymbol{p}_{T 2})$, 
and the  dijet relative transverse momentum 
$\boldsymbol{P}_T = (\boldsymbol{p}_{T 1} - \boldsymbol{p}_{T 2})/2$
which is related to $\boldsymbol{k}_T$ in the GTMD.  
The angular correlations between 
$\boldsymbol{b}_T$ and $\boldsymbol{k}_T$ 
are translated into the azimuthal modulations of the dijet cross section in the angle between 
$\boldsymbol{\Delta}_T$ and $\boldsymbol{P}_T$,
which is measurable. 
First theoretical estimates in the CGC effective field theory suggest that the modulations can range from a few percents to some tens of percents, depending on the dijet kinematics~\cite{Salazar:2019ncp,Mantysaari:2019csc}. 
It is important to check whether such modulations survive after including higher-order corrections. The complete NLO calculation for this process in the CGC framework has already been performed~\cite{Boussarie:2016ogo}, but its numerical implementation is still in progress~\cite{Boussarie:2019ero}. In addition, the resummation of soft gluons in the final state should also be considered as it can strongly affect the total dijet momentum 
$\boldsymbol{p}_{T 1} + \boldsymbol{p}_{T 2}$~\cite{Hatta:2019ixj}. 
The connection between the dijet production cross section and the gluon Wigner distribution is established in the so-called correlation limit where 
$|\boldsymbol{P}_T|\gg |\boldsymbol{\Delta}_T|$~\cite{Hatta:2016dxp}.
It is also interesting to study dijet production away from this limit, where detailed information on multi-gluon correlations at small $x$ can be accessed~\cite{Mantysaari:2019hkq}.

First exploratory simulations of exclusive dijet production at EIC are presented in Sec.~\ref{subsec:diff_dijets}.
However, we emphasize that, despite the enormous progress in recent years, the study of GTMDs/Wigner distributions is still at an early stage.  Therefore, it is important to keep investigating if additional interesting physics is (exclusively) encoded in GTMDs, and whether GTMDs can be probed in processes other than diffractive dijet production. 
A single process has been identified so far to access information on the quark GTMDs, i.e., the exclusive pion-nucleon double Drell-Yan process $\pi N\rightarrow (l_1^-l_1^+)(l_2^-l_2^+)N'$~\cite{Bhattacharya:2017bvs}. 
This process is in principle sensitive to all leading-twist quark GTMDs by making use of suitable polarization observables. 
However, the count rate for the double Drell-Yan reaction is small since its cross section is proportional to $\alpha^4_{em}$. A pressing question is then whether one can identify a reaction that is sensitive to quark GTMDs, but which has a larger cross section than the double Drell-Yan process.

GTMDs also play an important role in exclusive $\pi^0$ production, $ep\to e'\pi^0 p'$~\cite{Boussarie:2019vmk}. At lower energies, this process is sensitive to chiral-odd GPDs as discussed in Sec.~\ref{part2-subS-SecImaging-GPD3d}. But at the top EIC energy where physics becomes gluon-dominated, the standard description in terms of quark GPDs ceases to be valid. Instead, the cross section in this channel will be dominated by a particular gluon GTMD called $F_{12}^{o}$~\cite{Meissner:2009ww,Lorce:2013pza} which does not reduce to any known GPD upon $k_T$-integration. Interestingly, the forward limit $\boldsymbol{\Delta}_T \to 0$ 
of $F_{12}^{o}$ is the gluon Sivers function, and is in fact equivalent to the QCD odderon at small $x$~\cite{Zhou:2013gsa}, which has evaded experimental detection for decades~\cite{Adloff:2002dw}.

\subsection{Light (polarized) nuclei}
\label{part2-subS-SecImaging-LpolNucl}

\subsubsection{Coherent DVCS on light nuclei}
\label{part2-subsubS-CohDVCS-LightNucl}

In hard photon electroproduction off nuclear targets (DVCS), two different channels are usually investigated: The fully exclusive coherent one, with detection of the recoiling nucleus, and the incoherent one, with detection of the struck proton.
An important recent achievement has been the first separation of the two channels in DVCS off $^4$He~\cite{Hattawy:2017woc,Hattawy:2018liu}, which opens the way to a new
series of measurements. The ones planned at the EIC, due to the
advantages offered by the collider setup in detecting recoiling systems with respect
to a fixed-target experiment, look very promising. 

We deal here mainly with the coherent channel, whose interest is described, for example, in Ref.~\cite{Dupre:2015jha}.
Here the main arguments are just listed, emphasizing the possible role of the EIC:
(i) nuclear tomography, along the lines of Ref.~\cite{Burkardt:2000za}, to obtain
a pictorial representation of the EMC effect in the transverse plane. 
This requires measurements at low $x_B$ in a wide range of $t$, a regime which is accessible at the EIC;
(ii) the comparison with realistic calculations using the impulse approximation (IA), which are possible for few-body nuclei, could expose non-nucleonic degrees of freedom, according to an idea proposed in Ref.~\cite{Berger:2001zb}. Indeed, the IA predicts a narrow light-cone momentum distribution for light systems. 
If measurements were performed in a wide enough range of $t$, which seems feasible at the EIC,
one could look for longitudinal momenta transferred to the struck nucleon in the target
larger than such a width, in a region forbidden by the conventional IA description,
pointing to possible contributions of non-nucleonic degrees of freedom, among other exotic effects;
(iii) access to information on the nuclear energy momentum tensor and the distribution of pressure and shear forces inside the nucleus, and the $D$-term --- see Ref.~\cite{Polyakov:2002yz} for the initial idea, Ref.~\cite{Polyakov:2018zvc} for a report, as well as Sec.~\ref{part2-subS-SecImaging-GPD3d}.
The necessity of accessing the real part of the nuclear Compton form factor makes this easier with an $e^+$ beam, whose use at the EIC is under study\cite{Accardi:2020swt}; 
(iv) gluon GPDs in nuclei, exposing possible gluon degrees of freedom in nuclei, planned
already at JLab \cite{Armstrong:2017wfw}, will be easier to study 
at the EIC, due to the accessible low values of $x_B$.
The same holds for all the low-$x$ phenomena, such as nuclear shadowing, for which a peculiar
behavior has been predicted \cite{Goeke:2009tu};
(v) a specific access to the information for the free neutron, possible by using specific nuclei and suitable polarization setups in the experiments~\cite{Rinaldi:2012ft}.

Light nuclei play a very important role in general: Their conventional structure is realistically known so that exotic effects in DVCS processes can be exposed.
In the following, the specific role of deuteron, $^3$He and $^4$He as beams at the EIC is summarized.

The deuteron is a spin-1 system with a rich spin structure and many GPDs
at leading twist~\cite{Berger:2001zb}. Being a two-body system, its relativistic description is possible.
The deuteron GPDs have been evaluated in a light-front framework in Ref.~\cite{Cano:2003ju} and, for the transversity sector, in Ref.~\cite{Cosyn:2018rdm}. 
Theoretically well studied, the measurements of coherent DVCS would allow the test of several predictions, concerning the energy momentum tensor~\cite{Cosyn:2018thq,Cosyn:2019aio} and the parton spin content, for which a specific sum rule has been proposed \cite{Taneja:2011sy}.
Nuclear effects are expected to be small and the deuteron is the obvious candidate
to extract neutron information, mainly in the unpolarized setup
and in the incoherent channel, where final-state interactions may be
relevant but, in principle, can be evaluated realistically.

Since $^3$He is a spin-$\frac{1}{2}$ system, the amplitude of DVCS off $^3$He has the same
GPD decomposition as that of the nucleon. 
The CFFs can be extracted measuring the same asymmetries defined for the proton target, and the GPDs can be obtained by performing the same analyses. 
Its binding energy is in between that of the deuteron and that
of $^4$He, making it the ideal target for studying the onset of nuclear
effects through the periodic table.
Among the light nuclei, it is the only one with nonzero isospin,
so that it is unique to study isospin-flavor dependence of nuclear effects \cite{Scopetta:2004kj,Scopetta:2009sn}, more easily seen through $^3$H beams, whose use at the EIC is presently under discussion.
Realistic conventional effects have been studied in IA calculations in terms of (spin-dependent) spectral functions \cite{Scopetta:2004kj,Scopetta:2009sn,Rinaldi:2012ft,Rinaldi:2012pj,Rinaldi:2014bba} and are under control, so that possible effects due to an exotic
nuclear parton structure can be safely exposed in forthcoming data.
Even the evaluation of relativistic effects in a light-front framework, although challenging, are under investigation along the lines of Ref.~\cite{DelDotto:2016vkh}.

The specific spin structure of $^3$He has been used 
extensively to extract information for the polarized neutron.
Also in this exclusive process specific CFFs in (polarized) coherent DVCS, where final-state interactions should be negligible with respect to the incoherent case, appear to be dominated by the neutron contribution~\cite{Rinaldi:2012ft,Rinaldi:2012pj,Rinaldi:2014bba}, so that 
the extraction of the neutron information, complementary to that
to be obtained from the deuteron, looks  promising.
The possible use of (polarized) $^3$He ($^3$H) beams, 
along with detection far from the interaction region,
makes the EIC the ideal machine for completely new studies.

\begin{figure}
\includegraphics[scale=0.21,angle=270]{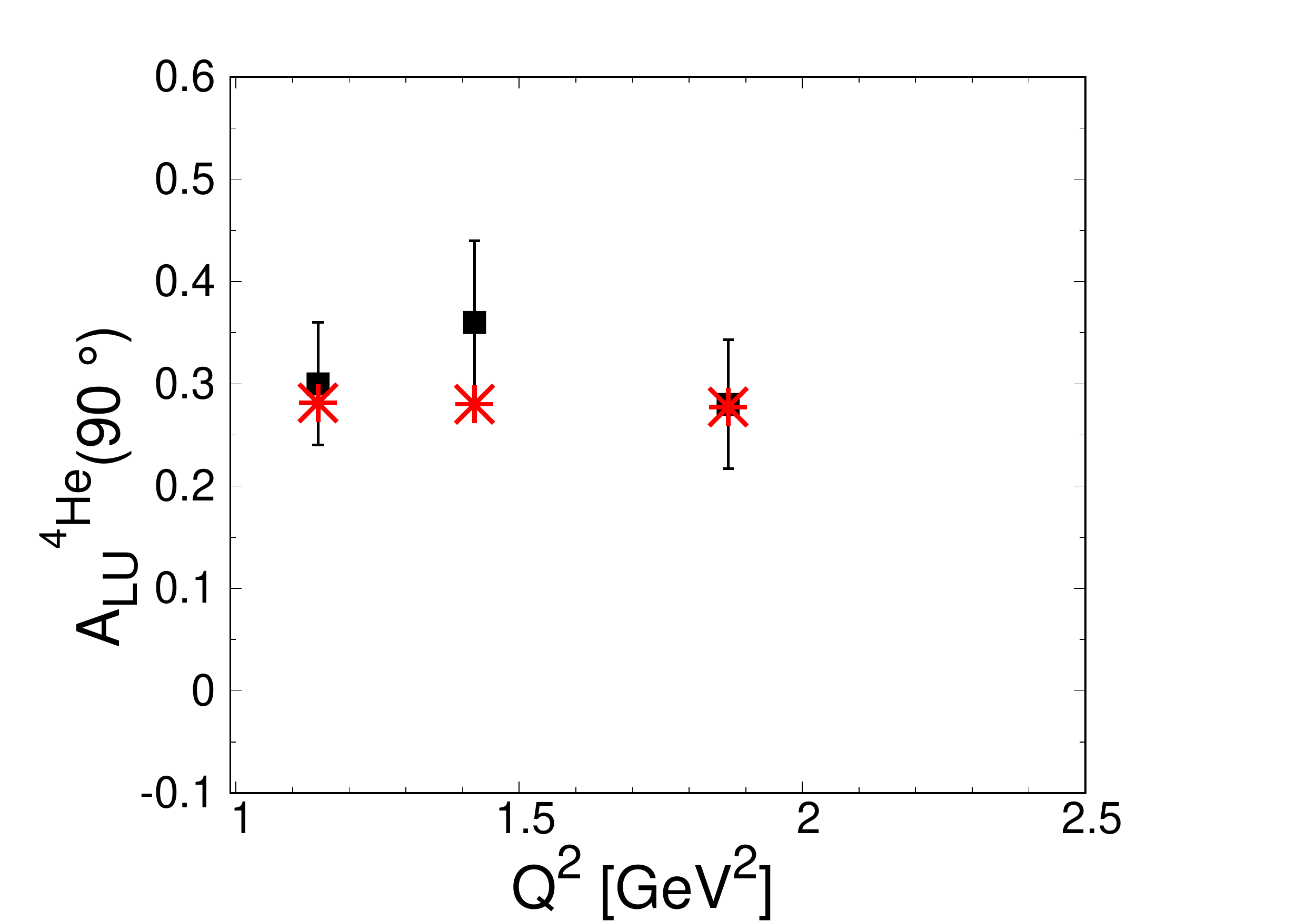}
\hspace{-0.6cm}
\includegraphics[scale=0.21,angle=270]{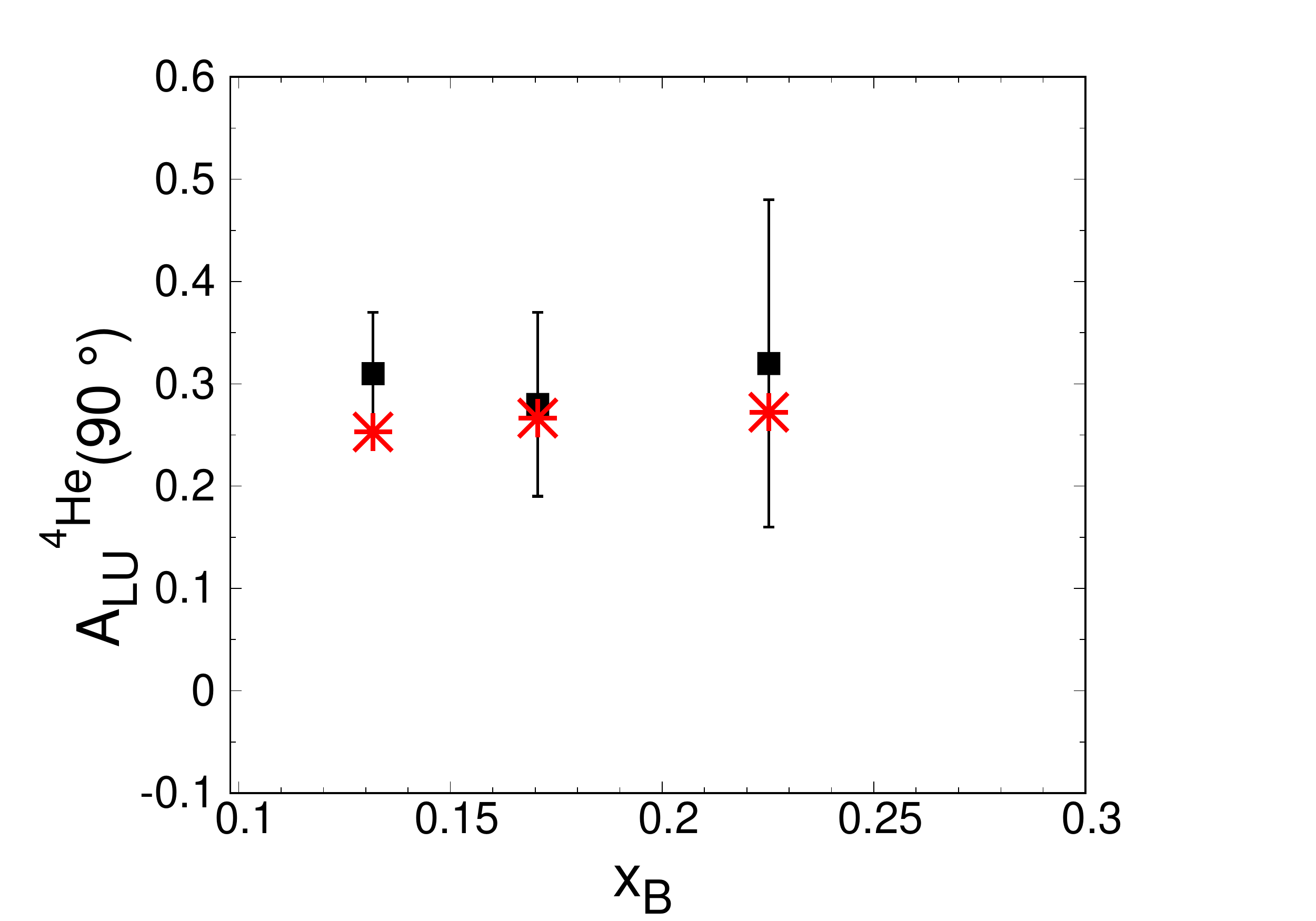}
\hspace{-0.6cm}
\includegraphics[scale=0.21,angle=270]{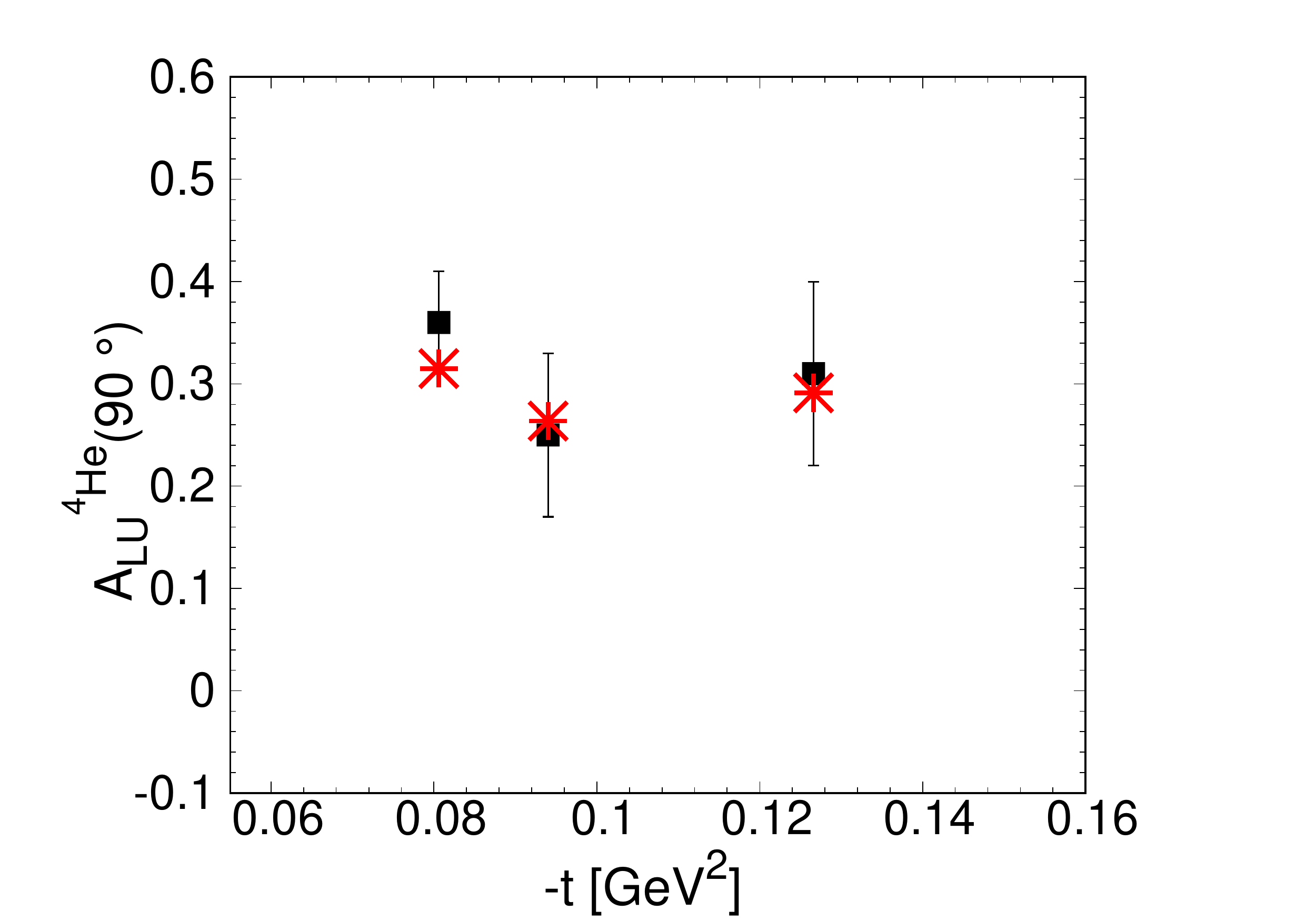}
\label{pg}
\caption{$^4$He azimuthal beam-spin asymmetry $A_{LU}(\phi)$,
for $\phi = 90^{\circ}$: results of Ref.~\cite{Fucini:2018gso} (red stars) compared with data (black squares) from the CLAS Collaboration at JLab~\cite{Hattawy:2017woc}.}
\label{coh4}
\end{figure}

The $^4$He nucleus is a spin-0 system and therefore the DVCS amplitude is described in terms of one leading-twist GPD in the chiral-even sector, so that only one CFF must be extracted in the experimental analysis, which is much easier than for the nucleon target.
From the point of view of nuclear dynamics, it is
a deeply-bound "real" nucleus and the nuclear effects are similar to those 
expected for heavier nuclei. For $^4$He,
realistic (non-relativistic) descriptions are challenging but possible, making it the ideal target to disentangle 
(novel) medium effects.
As already said, it is the only nucleus for which
data for the coherent channel have been released
\cite{Hattawy:2017woc}. Some calculations are also available since a long time \cite{Liuti:2004hd,Guzey:2003jh}. 
In the most recent IA theoretical analysis~\cite{Fucini:2018gso}, a semi-realistic nuclear description based on
the AV18 interaction~\cite{Wiringa:1994wb} and the UIX three-body forces~\cite{Pudliner:1995wk} was used, with the Goloskokov-Kroll model to parametrize the nucleon GPDs.
This framework, which was later extended to the incoherent channel~\cite{Fucini:2019xlc,Fucini:2020lxi}, proved successful in reproducing the data for the beam-spin asymmetry $A_{LU}$, as shown in Fig.~\ref{coh4}, and for the real and imaginary parts of the CFFs. 
This model is currently used in the TOPEG event generator and applied to simulate DVCS at the EIC kinematics. The region of low $x_B$, naturally accessible at the EIC and very important to study gluons in nuclei and expose exotic effects, is presently under investigation, for both DVCS and for the exclusive production of vector mesons. As for any coherent exclusive process, in DVCS off $^4$He the collider setup makes the detection of the recoiling intact nucleus easy,
which is very slow in a fixed-target experiment. This is even more important in the incoherent channel, for which the detection of other nuclear fragments could allow one to control effects from final-state interactions.

\subsubsection{Tensor polarized deuteron}

The deuteron, being a spin-1 particle, has additional spin degrees of freedom compared to the nucleon.  These can be probed using deuteron beams prepared in a spin ensemble with tensor polarization.  In inclusive DIS with a tensor polarized target, this gives rise to 4 additional structure functions, $b_{1-4}$, two of which are leading twist ($b_1,b_2$)~\cite{Hoodbhoy:1988am}. 
The $b_1$ structure function has a partonic density interpretation, which is also explicitly dependent on the nuclear magnetic state of the surrounding deuteron. This makes the observable unique because it directly probes nuclear interactions at the parton level.  As such, it can unravel novel information about nuclear structure, quark angular momentum, gluon transversity and the polarization of the quark sea that is not accessible in spin-$\frac{1}{2}$ targets~\cite{Efremov:1981vs,Jaffe:1989xy,Close:1990zw,Kumano:2010vz,Taneja:2011sy,Miller:2013hla}.

The $b_1$ structure function is experimentally extracted from the tensor asymmetry
\begin{equation}\label{eq:Azz}
    A_{zz} = 2\left( \frac{\sigma^+ - \sigma^-}{|P_{zz}^+|\sigma^- + |P_{zz}^-| \sigma^+} \right),
\end{equation}
where $P_{zz}$ is the amount of tensor polarization that ranges from $-2 \leq P_{zz} \leq 1$, and $\sigma^{+(-)}$ is the cross section when the target is polarized along (opposite to) the beam momentum.

Tensor-polarized deuteron is little explored in electron scattering. 
Elastic measurements are discussed in Sec.~\ref{part2-subS-SecImaging-FF}, quasi-elastic measurements were carried out at NIKHEF~\cite{Passchier:2001uc} and MIT Bates~\cite{DeGrush:2017azm}, and an inclusive-DIS measurement at HERMES~\cite{Airapetian:2005cb}. 
Two measurements are planned at JLab~\cite{C12-13-011, C12-15-005}. The JLab measurements can be extended to probe QCD effects at much higher energies and lower $x_B$ at the EIC.

Measurements of $A_{zz}$ in the quasi-elastic region illuminate QCD effects at short range and high momentum that depend on whether the deuteron wavefunction is hard or soft~\cite{Frankfurt:1988nt}. The discovery of short-range correlations (SRCs) in the quasi-elastic high-$x$ region, where all nuclear cross sections scale similarly to that of the deuteron, has led to a renewed interest in understanding these effects --- see also Sec.~\ref{part2-subS-LabQCD-ShortRange}.
The tensor asymmetry $A_{zz}$ provides a unique tool to experimentally constrain the ratio of the $S$ and $D$ wave functions at large momentum, which has been an ongoing theoretical issue for decades. 

Conventional nuclear physics models predict $b_1$ and $A_{zz}$ in inclusive DIS to be very small.  This is due to the averaging over initial deuteron configurations inherent to inclusive measurements and $b_1$ being proportional to the (small) deuteron $D$ wave.  In combination with tagging of a spectator nucleon (see Sec.~\ref{part2-subS-SpinStruct.P.N}), however, maximal $A_{zz}$ asymmetry values of $-2$ and $+1$ can be reached.  The tagged-DIS cross section also has several tensor-polarized structure functions which are zero in the impulse approximation and hence offer an opportunity to study spin-orbit effects specific to tensor polarization.  

Detailed rate estimates for the $A_{zz}$ asymmetry at the EIC are in progress.


\subsubsection{Medium modification of azimuthal modulations in SIDIS}

Measurements of medium modifications of spin-(in)dependent azimuthal asymmetries in semi-inclusive DIS (SIDIS) provide access to medium-modified parton distributions, and to the relative magnitude of the transverse momentum width of the nucleon TMDs.  Orbital motion of quarks is modified in the medium~\cite{Cloet:2009qs}. That makes a variety of physics observables which are sensitive to orbital motion, in particular several spin and azimuthal asymmetries, good candidates for providing important information on partonic distributions in bound nucleons.
Measurements of medium modifications of various spin and azimuthal asymmetries using unpolarized and longitudinally polarized leptons and nucleons will provide important information on the relative size of the transverse momentum in the proton, allowing to use nuclear targets as a microscope to study the proton in the medium. 

For SIDIS with nuclear targets, $e + \textrm{A} \to e+h+X$, the quark TMDs in a nucleus can be expressed as a product of nucleon TMDs and some functions depending on the total transverse broadening $\Delta_{2F}$, which itself depends on the quark transport parameter $\hat q$~\cite{Gao:2010mj,Song:2013sja},
\begin{align}
  f_1^{A}(x,k_T)&\approx\frac{A}{\pi\Delta_{2F}}\int d^2\ell_T \, e^{-(\boldsymbol{k}_T -\boldsymbol{\ell}_T)^2/\Delta_{2F}} \, f_1^{N}(x,\ell_T) \,,
 \nonumber\\
  k_T^2 \, g^{\perp A}(x,k_T)&\approx\frac{A}{\pi\Delta_{2F}}\int d^2\ell_T \, e^{-(\boldsymbol{k}_T -\boldsymbol{\ell}_T)^2/\Delta_{2F}} \, (\boldsymbol{k}_T \cdot \boldsymbol{\ell}_T) \, g^{\perp N}(x,\ell_T) \,,  
  \label{g1LAvsg1Lp} 
\end{align}
with $A$ the atomic number.
Using a Gaussian ansatz for the quark TMDs in a nucleon, the integrations in the leading-twist contributions can be carried out analytically~\cite{Gao:2010mj,Song:2013sja},
\begin{align}
f_1^{A}(x,k_T)&\approx\frac{A}{\pi\langle k_T ^2\rangle_A}\, f_1^{N}(x) \, e^{- k_T^2/\langle k_T ^2\rangle_A},\nonumber\\
g^{\perp A}(x,k_T)&\approx\frac{A}{\pi\langle k_T ^2\rangle^{g^\perp}_A} \, \frac{\langle k_T ^2\rangle^{g^\perp}}{\langle k_T ^2\rangle^{g^\perp}_A} \, g^{\perp N}(x) \, e^{-k_T^2/\langle k_T ^2\rangle^{g^\perp}_A},
\label{g-FLL}
\end{align}
where $\langle k_T ^2\rangle_A= \langle k_T ^2\rangle+\Delta_{2F}$ and  $\langle k_T ^2\rangle^{g^\perp}_A=\langle k_T ^2\rangle^{g^\perp}+\Delta_{2F}$.

\begin{figure}[htb!]
\begin{center}
\begin{tabular}{cc}
\includegraphics[width=.47\textwidth]{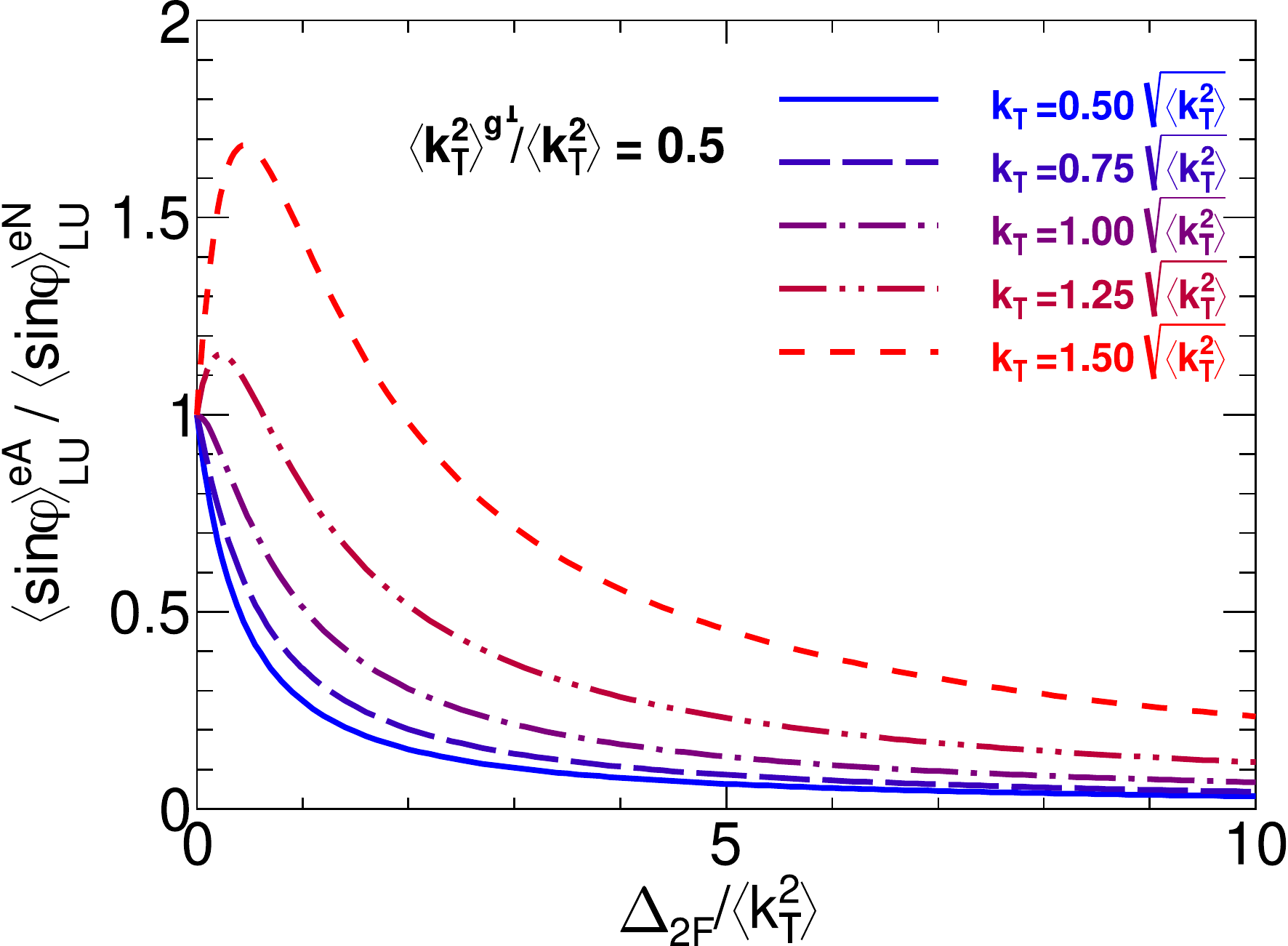}&
\includegraphics[width=.47\textwidth]{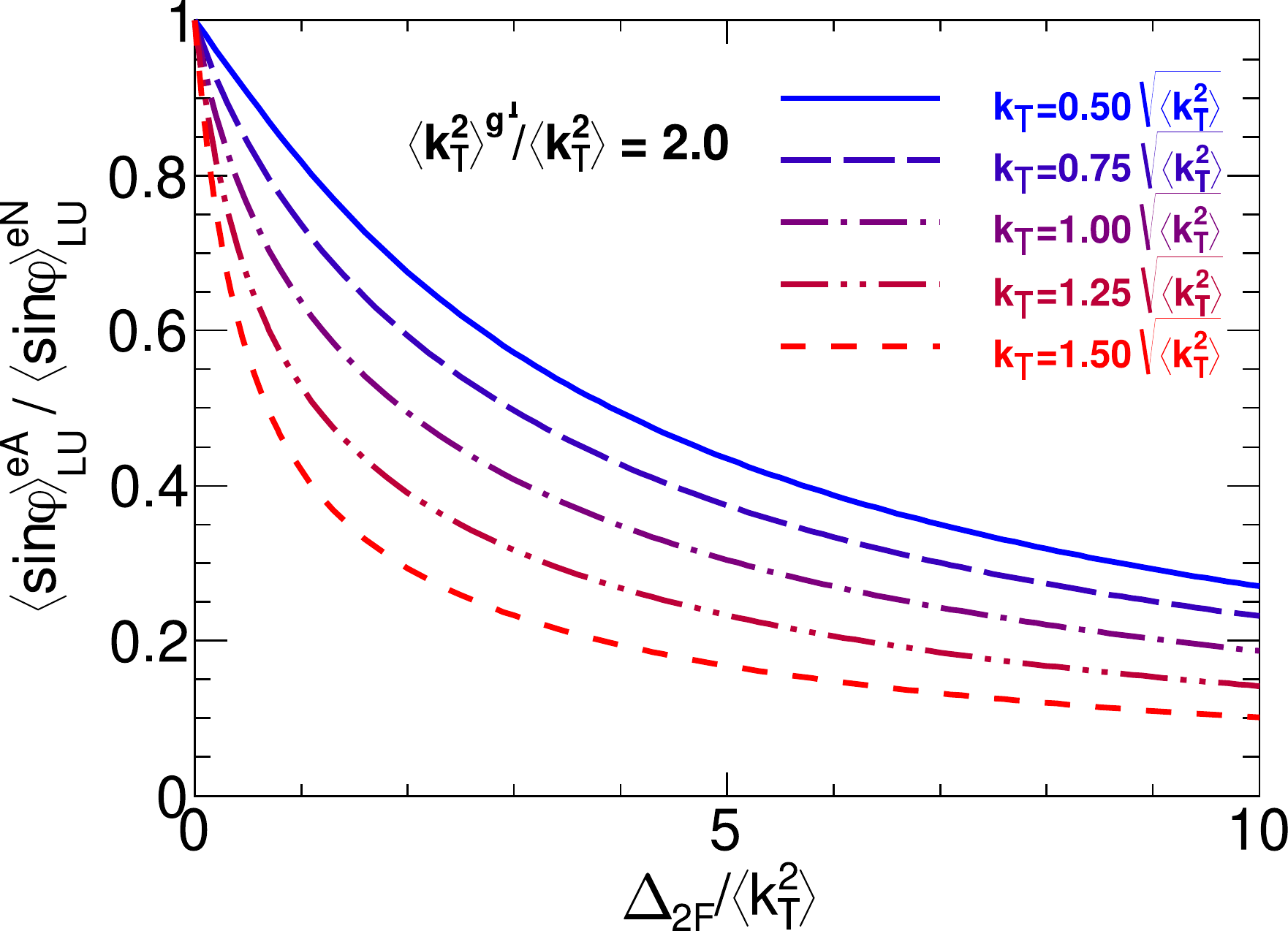}
\end{tabular}
\end{center}
\caption{Ratio $\langle\sin\phi\rangle_{LU}^{eA}/\langle\sin\phi\rangle_{LU}^{eN}$ as a function of $\Delta_{2F}/\langle k_T ^2\rangle$ for different $k_T$ values and $\langle k_T ^2\rangle^{g^\perp}/\langle k_T ^2\rangle$ ratios.}
\label{fig:tmd_medium_alu}
\end{figure}

Measuring higher-twist single-spin asymmetries (SSAs) in the medium will provide access to medium modifications of spin-orbit correlations, involving several higher-twist TMDs which are sensitive to 
final-state interactions ($f^\perp$, $g^\perp$, and $f_L^\perp$).
Those TMDs are also essential components of the 3D structure of the  nucleons. Detection of jets with the EIC would allow for studies of medium-modified TMDs, with completely different (compared to hadrons) systematics, viz., without involvement of unknown fragmentation functions, and their modification in the medium. 
Understanding jet formation in the medium and the correlation between jet momentum and its parent parton is, however, needed.
For example, medium modification of the SSA $A_{LU}^{\sin\phi}$ (where here $\phi$ is the azimuthal angle between the produced hadron and the leptoon plane) could be determined by the TMD $g^\perp$, modified in the medium~\cite{Gao:2010mj}, 
\begin{align}
  \frac{\langle\sin\phi\rangle_{LU}^{eA}}{\langle\sin\phi\rangle_{LU}^{eN}}\approx&\frac{\langle k_T ^2\rangle_A}{\langle k_T ^2\rangle}\left(\frac{\langle k_T ^2\rangle^{g^\perp}}{\langle k_T ^2\rangle^{g^\perp}_A}\right)^2
  \exp\Bigg[\Big(\frac{1}{\langle k_T ^2\rangle_A}-\frac{1}{\langle k_T ^2\rangle}-\frac{1}{\langle k_T ^2\rangle^{g^\perp}_A}+\frac{1}{\langle k_T ^2\rangle^{g^\perp}}\Big) \, k_T^2\Bigg] \,,
\end{align}
where $\langle k_T ^2\rangle^{g^\perp}$ and $\langle k_T ^2\rangle^{g^\perp}_A$ are the widths of $g^\perp$ in a free and bound nucleon, respectively. Nuclear modifications of this beam SSA, for example, appear to be very sensitive to the relative widths of the involved nucleon TMDs, and change significantly with transverse momentum (see Fig.~\ref{fig:tmd_medium_alu}). The overall magnitude of the effect is governed by the same transport parameter $\hat q$, or equivalently the transverse momentum broadening $\Delta_{2F}$, that controls nuclear suppression of hadron production in unpolarized and polarized electron-nucleus scattering. 

\newpage
\section{The Nucleus: A Laboratory for QCD}
\label{part2-sec-LabDenseQCD}

This section is devoted to an overview of the fundamental physics with light and heavy nuclei that will be performed at the EIC. Of course, nuclei are made of nucleons, which
in turn, are bound states of the fundamental
constituents probed at short distances, namely quarks and
gluons. 
The EIC will be the world’s first dedicated electron-nucleus $\eA$ collider. It will
explore the effect of the binding of nucleons on {\bf nuclear parton distributions}, the
momentum distribution of quarks and
gluons; but also, for the first time, determine
their spatial distributions in a nucleus via diffractive or exclusive
processes. In addition, the wealth of semi-inclusive probes at the EIC provides direct
and clean access to fluctuations of the
density of quarks and gluons in nuclei.

The high-energy aspects of DIS on nuclei such as the physics of non-linear color fields and gluon saturation are covered here in
Secs.~\ref{part2-subS-LabQCD-Saturation},~\ref{part2-subS-LabQCD-Diffraction}. The latter section also discusses the unique opportunity at the EIC to measure {\bf nuclear diffractive PDFs} and to understand their connection to shadowing. Section~\ref{part2-subS-LabQCD-Photo} covers {\bf coherent and incoherent photoproduction} on heavy ion targets. 

Additional topics addressed in this Sec.~\ref{part2-chap-EICMeasandStud} are as follows.

{\bf Collective effects} 

Until recently, "collective phenomena" were associated exclusively with large fireballs formed in nuclear collisions. These event-by-event multiparticle azimuthal correlations, particularly a long-range pseudorapidity feature termed "ridge"~\cite{Adams:2005ph}, were neither expected nor present in any modeling of "small" systems such as p+p or p+A collisions. This paradigm was shattered with the discovery of ridge "ridge"-correlation in (high multiplicity) proton-proton and proton-nucleus collisions~\cite{Khachatryan:2010gv,Aad:2015gqa,Khachatryan:2015lva,Khachatryan:2016txc,Abelev:2012ola,CMS:2012qk,Aad:2012gla,Aad:2013fja,Abelev:2014mda,Aad:2014lta,Khachatryan:2015waa}. These long-range correlations have two compelling theoretical explanations based on orthogonal premises: one interprets the observed effect in the small system data as a final-state phenomenon. The other is the initial state effect.  At the EIC, the selection of DIS-events at small-$x$ will provides us with a unique testing ground to test and understand in detail the physics mechanism behind the formation of these collective interactions, see Sec.~\ref{part2-subS-LabQCD-Collective}.

{\bf Special opportunities with jets and heavy quarks}

The technical capabilities of the EIC machine and detector concept under developments provide an excellent ground for extending jet and heavy quark measurements beyond the inclusive cross-sections and simple semi-inclusive correlations.   There is a palpable shift in the research efforts towards the jet substructure studies, including such studies with a required in-jet heavy flavor hadron presence, or "tagging." While the methods for exploring the jet constituent distributions began to develop in the 1990~\cite{Seymour:1994by}  and possibly earlier, this direction has exploded in recent years in the HEP community~\cite{Asquith:2018igt}.  At the same time, the jet substructure methods have been successfully adopted for studies of medium-induced jet modifications in the QGP by the nuclear physics community.

Jet substructure observables are indispensable tools for flavor-tagging, i.e., for statistical identification of the parton originating the jet. Flavor-tagging enables another dimension to explore the nuclear modification effects, of both initial and final state origin, for different hard-scattered partons. At the EIC, the jet substructure observables, specifically jet angularity, can image the nucleon/nuclei 3D structure and map out the hadronization process in a vacuum and nuclear medium. Another substructure tool that has been studied in connection with future EIC data is the jet charge, which showed promise to discern the contributions of quark and anti-quark jets and to, again, pinpoint their original parton flavor. A high precision jet charge measurements at the EIC will provide an excellent way to constrain isospin effects and the up/down quark PDFs in the nucleus. The details are supplied in Sec.~\ref{part2-subS-LabQCD-Special}. 

{\bf Short range correlations and the structure of light nuclei}

The EIC will also provide novel insight into the physics of short range correlations (SRC) in nuclei and how they relate to the mechanism by which QCD generates the nuclear force. The modification of the structure of bound nucleons as manifest, for example, in the EMC effect could be caused by short-range correlated nucleon pairs with high internal nucleon momentum. The new collider will investigate the underlying physics of SRC in kinematic regions that so far could not be reached.  A more detailed discussion is presented in Sec.~\ref{part2-subS-LabQCD-ShortRange}.

Also, the EIC will provide polarized $^3$He and $^3$H beams, possibly deuteron ($^2$H) beams and more.  This allows to probe the spin structure of the neutron, the tensor polarized deuteron, and measurements of the polarized EMC effect, in order to understand the interplay between partonic QCD phenomena and nuclear interactions; c.f.\ Sec.~\ref{part2-subS-LabQCD-LightNuclei}.
\subsection{High parton densities and saturation}
\label{part2-subS-LabQCD-Saturation}
The study of emergent properties of the ultra-dense gluonic matter is an important pillar of EIC physics. Since the emission of soft gluons is favored in QCD, a large number of low-momentum gluons exist inside high energy nucleons and heavy nuclei. These low momentum gluons are usually referred to as the low-$x$ gluons, where $x$ is the longitudinal momentum fraction of the gluon with respect to the parent hadron.  As a consequence, one observes a rapid increase in the gluon density towards smaller $x$ and a corresponding increase in the quark density coming from sea quarks via the $g\to q\bar q$ splitting process. On the other hand, the density of quarks and gluons in the small-$x$ limit will not become infinitely large due to the gluon saturation effect. When the gluon density is sufficiently high, the recombination of gluons via the $gg\to g$ process becomes important. Eventually, the gluon density is expected to saturate, as a balance is reached between gluon radiation and recombination.

The ``saturation'' of the densely populated gluonic system in nucleons and heavy nuclei~\cite{Gribov:1984tu,Mueller:1985wy}, is most commonly described in the theoretical framework provided by the Color Glass Condensate (CGC) formalism~\cite{McLerran:1993ni,Gelis:2010nm}. This framework can be viewed as an effective theory of high energy QCD in the low-$x$ limit. In the small-$x$ formalism, the emission of soft gluons is captured in the famous Balitsky-Fadin-Kuraev-Lipatov (BFKL) equation\cite{Balitsky:1978ic, Kuraev:1977fs}, while the gluon recombination manifests itself as the additional non-linear term in the extended evolution equations, which are known as the Balitsky-Kovchegov (BK) equation\cite{Balitsky:1995ub, Kovchegov:1999yj} and the Jalilian-Marian-Iancu-McLerran-Weigert-Leonidov-Kovner (JIMWLK) equation\cite{JalilianMarian:1997jx, JalilianMarian:1997dw, Iancu:2000hn, Ferreiro:2001qy}. These non-linear QCD evolution equations encode gluon saturation, which emerges at asymptotically small $x$ as the universal stable fixed point independent of initial conditions of color sources\cite{Munier:2003vc}. In practice, one often defines the so-called saturation momentum $\Qss (x)$ to separate the nonlinear saturated dense regime at low transverse momentum (or virtuality) from the linear dilute regime, and to characterize the strength of the saturation effects. Comparing to the saturation momentum in the proton, the saturation momentum $\Qss$ for a large nucleus with mass number $A$ is enhanced by a factor of $\sim A^{1/3}$ due to the overlap of nucleons at a given impact parameter~\cite{Kowalski:2007rw}. This nuclear enhancement factor is also known as the ``oomph" factor, which indicates that the saturation effect in \eA collisions is much stronger than that in $\ep$ collisions. 

The initial condition for small-$x$
evolution of a proton can be
related~\cite{Dumitru:2018vpr,Dumitru:2020fdh} to its light-front wave
function as constrained by imaging (see
sec.~\ref{part2-sec-Imaging}). This determines the dependence of the
dipole scattering amplitude on impact parameter and dipole size, on
their relative angle, and on the initial value of $x$ where the
resummation of soft gluon emissions is started~\cite{Dumitru:2020gla}.

\subsubsection*{Inclusive cross sections at small x}

At HERA, an extremely interesting phenomenon known as geometrical scaling\cite{Stasto:2000er}, has been discovered in the low-$x$ inclusive DIS data in $\ep$ collisions. In general, the inclusive cross section is a function of two independent variables, $x$ and $Q^2$. However, in the small-$x$ regime, the inclusive cross section can be cast into a single variable function which only depends on $\tau\equiv Q^2 /\Qss (x)$ with $\Qss (x)=Q_0^2\left(x/x_0\right)^{\lambda}$. Using $Q_0=1\gev$ and $x_0=3.04\times 10^{-4}$ and $\lambda=0.288$, one can show that all the inclusive data points within the range of $x<0.01$ and $Q^2 < 450\gev^2$ fall on a single curve. Later, it was demonstrated in Refs.~\cite{Munier:2003vc} that this remarkable geometrical scaling phenomenon can be elegantly derived from the traveling wave type solution of the non-linear BK equation. This has been reckoned as one of the striking pieces of evidence for the saturation formalism. At EIC, the inclusive cross section in $\eA$ collisions in the low-$x$ regime will provide important information about the nuclear shadowing effect and the saturation phenomenon. 

A systematic way of calculating inclusive cross sections in the CGC formalism is provided by the dipole factorization picture. Here one separates the process into an impact factor describing the fluctuation of the virtual photon into a partonic state, at leading order a dipole and at NLO also a $q\bar{q}g$ state, and the scattering amplitude of this state with the target. This dipole scattering amplitude generalizes the concept of a gluon distribution to include the possibility of nonlinear interactions with the target gluon field. Its dependence on $x$ (or the collision energy $W$) is described by the BK or JIMWLK equations.  Depending on the polarization state of the $\gamma^*$ one must consider separately transversely and longitudinally polarized photons. Measuring both cross sections will be a central part of the physics program at the EIC. In particular, the total virtual photon-target cross section for both protons and heavy nuclei are important ``day one'' measurements.

In a set of major theoretical advances in recent years both the BK~\cite{Balitsky:2008zza} and the JIMWLK equations~\cite{Balitsky:2013fea,Kovner:2013ona} are now known at NLO accuracy in the QCD coupling constant. An additional resummation of transverse (``collinear'') logarithms is required to stabilize the NLO equations, but by now robust practical methods to achieve this resummation have been achieved~\cite{Beuf:2014uia,Iancu:2015vea,Ducloue:2019ezk}. 
Also, the impact factor required for calculating the cross section is now known, both in momentum space~\cite{Balitsky:2012bs,Balitsky:2010ze} and, in a more practical form for use with the BK equation, in mixed transverse coordinate-longitudinal momentum space~\cite{Beuf:2016wdz,Beuf:2017bpd,Hanninen:2017ddy}.
Using this theoretical machinery, a very good description of existing HERA small-$x$ inclusive cross section data has been achieved using the collinearly resummed BK evolution equations both with LO~\cite{Ducloue:2019jmy} and NLO impact factors~\cite{Beuf:2020dxl}. 
An extension of the NLO calculation to heavy quarks is expected to appear soon. These calculations are straightforwardly generalizable from protons to nuclei without any parameters other than the Woods-Saxon nuclear density~\cite{Lappi:2013zma}. 
Together with EIC measurements of the total and longitudinal $\gamma^*$p and $\gamma^*$A cross sections, they will allow for a clear and precise window into the physics of gluon saturation.

\subsubsection*{Accessing low-$x$ gluons via di-jets or di-hadrons}
At the EIC, by going beyond inclusive measurements and studying SIDIS observables in the low-$x$ regime, we can obtain deeper insights into the gluon distributions. For many years, we have known that there are two gluon distributions in the CGC formalism. On the one hand, there is the dipole unintegrated gluon distribution (UGD), which is defined as the Fourier transform of dipole-target cross section\cite{Mueller:1993rr}. It often appears in the calculation of various inclusive processes. On the other hand, the so-called Weizs\"acker-Williams (WW) gluon distribution\cite{JalilianMarian:1996xn, Kovchegov:1998bi} has also been derived as the genuine number density of gluons in a target hadron by applying the well-known WW method of virtual quanta in QCD. In the McLerran-Venugopalan model~\cite{McLerran:1993ni} for a large nucleus, these two UGDs are found to have distinct $\pT$ behavior. The small-$x$ evolution of these two gluon distributions can be taken care of by applying the JIMWLK evolution equation to the corresponding correlators in coordinate space\cite{Dominguez:2011gc, Dumitru:2011vk}. Based on the gauge link structure of the WW gluon distribution, and calculations within the CGC formalism, it has been proposed~\cite{Dominguez:2010xd, Dominguez:2011wm} that the DIS back-to-back dijet/dihadron production at the EIC can be used to directly probe the WW distribution, which has not been measured before.

\begin{figure}
\begin{center}
		\includegraphics[width=0.48\textwidth]{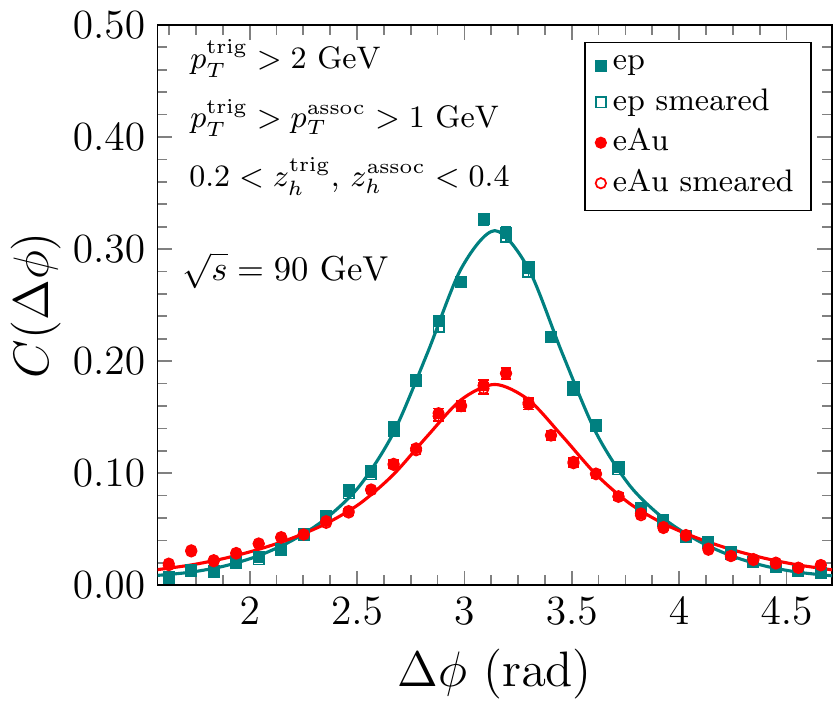}
\end{center}
\caption{Comparison between the dihadron azimuthal angle correlation in $\eAu$ collisions (labeled with filled red circles) and that in $\ep$ collisions (labeled with filled teal squares). The results with the detector smearing are shown in open markers. The solid lines represent the results obtained from the theoretical model calculations in the CGC formalism.}
\label{fig:dihadron_sat}
\end{figure}
To directly probe the WW gluon distribution and gluon saturation effects at low $x$, we can measure the azimuthal angle difference
($\Delta\phi$) between two back-to-back charged hadrons in $\eA$ collisions ($\eA\rightarrow e'h_1h_2X$). This azimuthal angle distribution can help us map the
transverse momentum dependence of the incoming gluon distribution. The
away-side peak of the dihadron azimuthal angle correlation is dominated by the back-to-back dijets produced in hard scatterings. 
Due to the saturation effect, the WW gluon TMD can provide additional transverse momentum broadening to the back-to-back correlation and cause the disappearance of the away-side peak when the saturation effect is overwhelming~\cite{Kharzeev:2004bw, Dominguez:2011wm}. A comparison of the heights and
widths of the coincidence probabilities
$C(\Delta\phi)=N_{pair}(\Delta\phi)/N_{trig}$ in $\ep$ and $\eA$ collisions
will be a clear experimental signature for the onset of the saturation effect. 

Furthermore, following the prescriptions in Ref.~\cite{Zheng:2014vka}, a Monte Carlo simulation has been carried out for the azimuthal angle correlations of two charged hadrons at $\sqrts=90 \, \gev $ in \ep and \eAu collisions. The results of the simulation are also compared with the prediction from the saturation formalism. To focus on the low-$x$ region, the events within the range of the virtuality $1<Q^{2}<2$ GeV$^{2}$ and inelasticity $0.6<y<0.8$ are selected. Events in nearby $Q^2$ and $y$ bins are expected to yield similar results. The hadron pairs are required to have an energy fraction
$0.2<z^{trig},z^{assc}<0.4$ within the pseudorapidity range $|\eta|<3.5$ with $p_{T}^{trig}>2$ GeV$/c$ and
1 GeV$/c$ $<p_{T}^{assc}<p_{T}^{trig}$. The away-side peak of the $C(\Delta\phi)$ distribution
is shown in Fig.~\ref{fig:dihadron_sat}. The $\ep$ reference is displayed by the squares in teal, while the results in $\eA$ collisions are shown in red. The results which include the detector response to this measurement are represented by the
open markers. Based on the current EIC tracking resolution design, the impact
of the detector smearing effect on this measurement is negligible. The theoretical results plotted in solid curves include the saturation effects together with the Sudakov resummation in
the CGC formalism\cite{Mueller:2012uf,Mueller:2013wwa, Stasto:2018rci}. A significant suppression of the away-side peak in \eAu collisions
compared to the \ep reference can be observed, given the projected statistical uncertainty corresponding to an integrated luminosity of 10 $\fb^{-1}$/A (the statistical error bars are too small to be visible.). 

At last, it should be pointed out that the production of di-jets/di-hadrons, and of heavy quark pairs, allows one to also probe the linearly polarized gluon distribution described by the TMD $h_1^{\perp g}$\cite{Boer:2010zf}. In the low-$x$ regime, the WW type gluon distribution $h_1^{\perp g}$\cite{Metz:2011wb,Dominguez:2011br, Dumitru:2015gaa, Hatta:2020bgy} can also be important, and generate sizable asymmetries in azimuthal angle distributions. Detailed numerical studies of the corresponding asymmetries at the EIC can be found in Ref.~\cite{Dumitru:2018kuw}.


\subsubsection*{Recent Progress in Probing Gluon Saturation with Jet Observables}

Jet observables provide an unprecedented opportunity to probe proton and nuclear structure at small Bjorken $x$ in a more differential fashion as compared to more inclusive observables. Several theoretical results have been obtained recently for inclusive dijet production in the forward direction, leading in particular to the introduction of new types of TMD distributions in the small $x$ regime that are sensitive to saturation effects\cite{Dumitru:2015gaa,Altinoluk:2019fui, Boussarie:2020vzf, Mantysaari:2019hkq}. Moreover, with $\eA$ collisions at the EIC, one will be able to probe the partonic content of dense nuclear targets. In particular, production of jets at moderate and low values of $x$ will offer a possibility to study the interplay of the
Sudakov effects related to hard scales present in the perturbative description, and the nonlinear effects due to gluon saturation phenomenon predicted in QCD~\cite{Mueller:2012uf,Mueller:2013wwa,Balitsky:2015qba,Balitsky:2016dgz,Altinoluk:2019fui}. Also, using a variety of nuclear targets one will have the opportunity to explore the so-called small-$x$ Improved TMD factorization framework (ITMD)\cite{Kotko:2015ura, vanHameren:2016ftb, Altinoluk:2019fui, Kotko:2017oxg}.

One of the most promising tools for studying the saturation regime is the measurement of multiparticle correlations
in \eA, in particular, the distribution of azimuthal angles between two hadrons in the diffractive process
$\eA → e' + h_1 + h_2 + X$. Furthermore, one will be able to measure similar processes with dijets.
These correlations are sensitive to the transverse momentum dependence of the gluon
distribution in a nucleus, as well as to gluon correlations, for which first principles computations are becoming available~\cite{Lappi:2012nh}.
The precise measurement of these dihadron and /or dijet correlations at an EIC would allow one not only to determine
whether the saturation regime has been reached, but to study as well the nonlinear evolution of spatial multigluon
correlations~\cite{Aschenauer:2017jsk}.


In addition, semi-inclusive photon + di-jet production is a highly differential measure of the many-body dynamics of gluon saturation in $\eA$ DIS at small $x$. This process can be computed systematically in the CGC effective field theory. At leading order, the cross-section is sensitive to both dipole and quadrupole correlators of lightlike Wilson lines~\cite{Roy:2018jxq}, and agrees with prior computations of the LO di-jet cross-section in the soft photon limit~\cite{Dominguez:2011wm}. This computation also allows one to extract the LO photon+jet/hadron cross-section; this channel has been 
shown recently~\cite{Kolbe:2020tlq} to have a clean and unique sensitivity to the saturation scale that is complementary to that of di-jet/di-hadron correlations.
In \cite{Roy:2019hwr,Roy:2019cux}, the $\eA$ photon+dijet cross-section was computed to next-to-leading-order. This result is sufficiently differential to encompass NLO computations
of inclusive photon production, photon+jet correlations, dijet, and single jet production, as well as the fully inclusive cross-section. When combined with the next-to-leading-log JIMWLK evolution~\cite{Balitsky:2013fea,Kovner:2013ona} for the dipole and quadrupole correlators, these NLO cross-sections can be computed to $O(\alpha_S^3 \ln(1/x))$ accuracy. Extracting concrete predictions for the semi-inclusive channels in EIC kinematics will be crucial for the clean and unambiguous 
characterization of gluon saturation.

\subsubsection*{Helicity TMDs and PDFs at small $x$}

In recent years much progress has also been made in extending the
discussion of quark and anti-quark helicity TMDs, PDFs, etc.\ (see
sec.~\ref{part2-subS-SpinStruct.P.N}) to the high-energy limit of
small $x$. This is a very interesting topic where the EIC is expected
to provide fundamental new insight.  Evolution equations for these
functions have been
derived~\cite{Kovchegov:2015pbl,Kovchegov:2016zex,Cougoulic:2019aja}
which resum powers of $\alpha_s \log^{2}(1/x)$ in the
polarization-dependent evolution, and powers of $\alpha_s \log (1/x)$
in the unpolarized evolution.  In the ladder approximation they reduce
to the Bartels-Ermolaev-Ryskin (BER) evolution equations~\cite{Bartels:1996wc} for the $g_{1}$ structure function. Initial conditions for these evolution equations
analogous to the McLerran-Venugopalan model for an unpolarized large
nucleus, have also been constructed~\cite{Cougoulic:2020tbc}.
\subsection{Diffraction}
\label{part2-subS-LabQCD-Diffraction}

\subsubsection* {Inclusive diffraction with nuclei}

Diffraction in \eA \ is a poorly studied subject, in particular inclusive diffraction, which has never been measured. Similar considerations apply to diffraction in \eA \ as to \ep \  collisions, Subsection~\ref{part2-subS-PartStruct-InclDiff}, the main difference being that for incoherent processes one must separately discuss processes where the nucleus breaks up into smaller nuclei, and nucleon-dissociative processes where an individual nucleon in the target dissociates. In terms of the typical $t$-dependence, the former is similar to coherent diffration in \ep, and the latter to proton dissociation in \ep. 

Coherent diffraction is mostly sensitive to the nuclear radius and global nuclear profile and structure, while incoherent diffraction is sensitive to nucleon degrees of freedom, specifically to nucleon and subnucleon fluctuations, see e.g. Refs.~\cite{Caldwell:2010zza,Mantysaari:2020axf} for reviews and Subsection~\ref{part2-subS-LabQCD-Photo}.

All of these cases are characterized by a rapidity gap between the target fragments and the photon fragment system. While detecting experimentally whether the nucleus has disintegrated or not might be challenging, the overall rapidity gap cross section that includes both coherent and incoherent processes should be more easily measurable.
In spite of the presence of more physically different sources of fluctuations in nuclei than in protons (fluctuating positions of the nucleons in the nucleus in addition to subnucleonic fluctuations), coherent diffraction is a larger part of the diffractive cross section in \eA \ than in \ep. This is due both to the fact that coherent diffraction grows parametrically as $A^{4/3}$ with the atomic mass number, and to the fact that nuclei are closer to the black disk limit, where there are no fluctuations and thus no incoherent processes.

\begin{figure}
\begin{center}%
	\includegraphics*[width=0.45\textwidth]{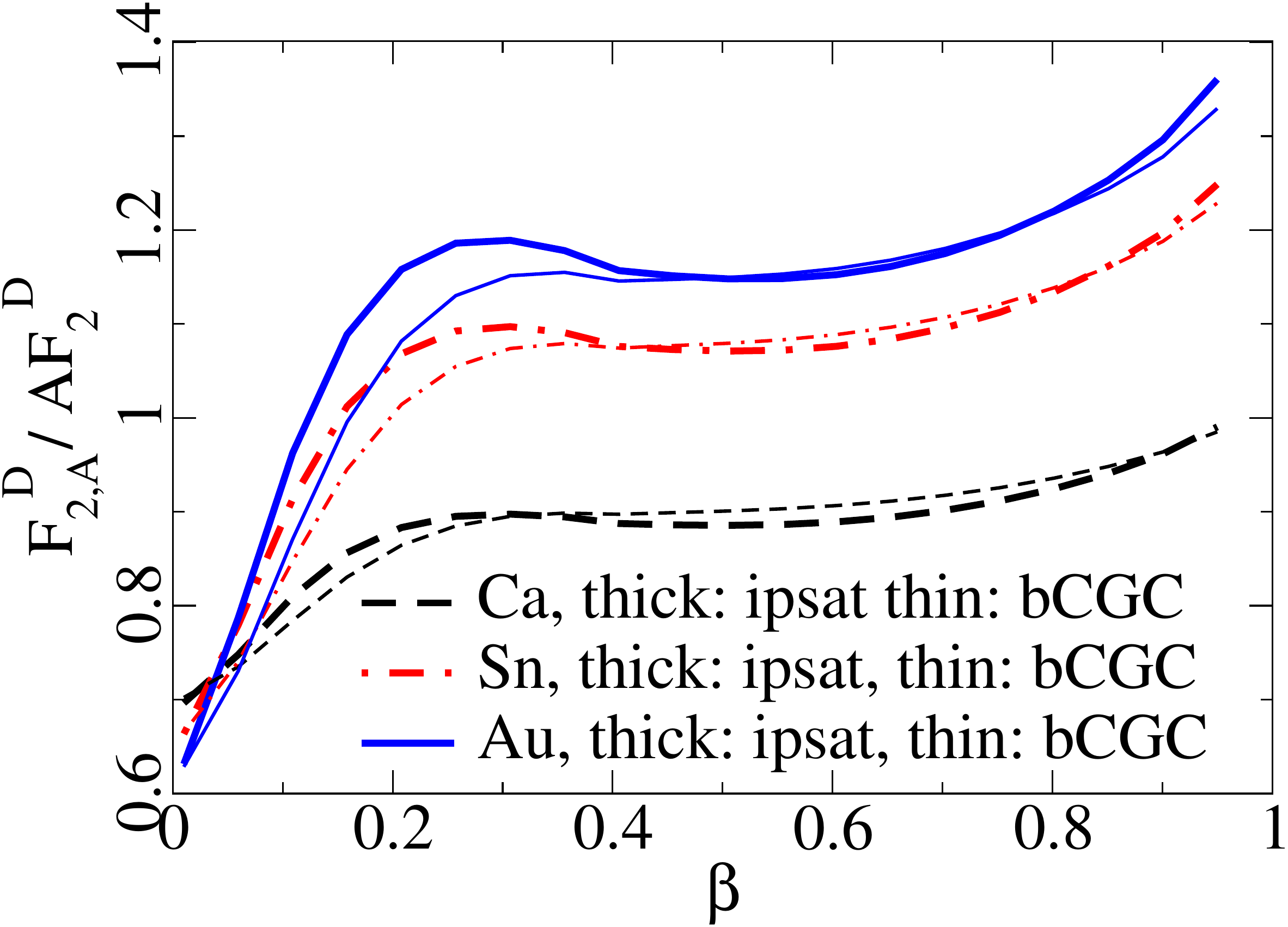}%
\hfill
	\includegraphics*[width=0.45\textwidth]{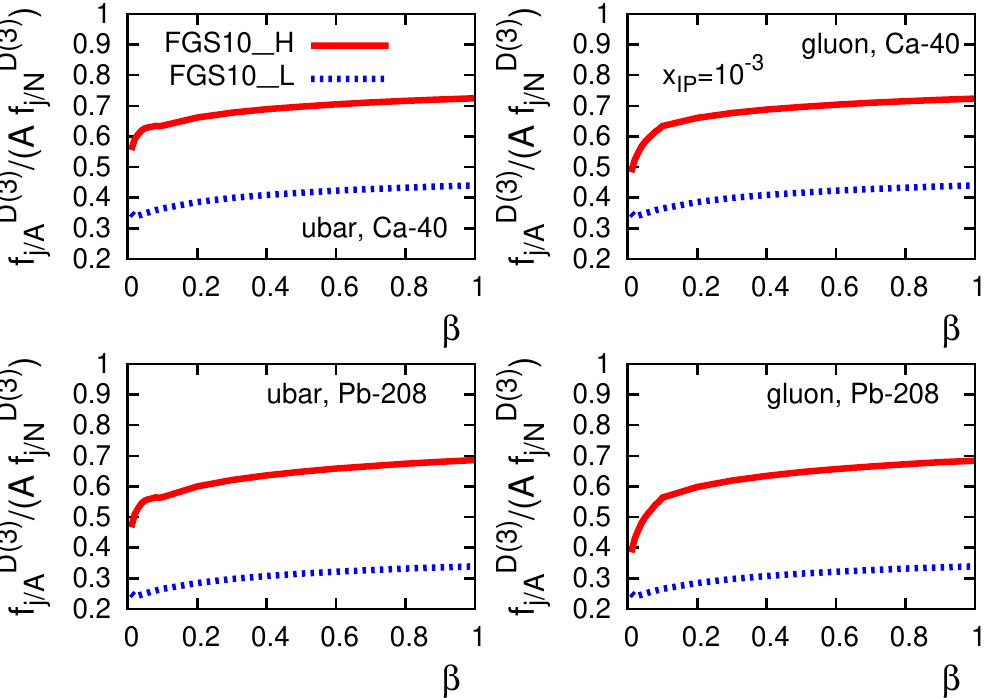}%
\end{center}
\caption{Left: Ratio of nuclear to proton diffractive structure functions, scaled by $A$, at $\xi=10^{-3}$ (also referred to as $x_\mathbb{P}$) as a function of  $\beta$ from dipole model calculations (Fig.~7 from Ref.~\cite{Kowalski:2008sa}). Right: ratios of nuclear to proton diffractive parton distributions, scaled by $A$, for sea quarks and gluons at the same $\xi$ (i.e. $x_\mathbb{P}$) from the Leading Twist Shadowing model (Fig.~72 from Ref.~\cite{Frankfurt:2011cs}).
}
\label{fig:dipole_vs_lts_diffraction}.
\end{figure}

Diffraction is generically more sensitive to gluon saturation than inclusive cross sections, since the diffractive cross section is proportional to the square of the gluon density. In hard diffraction, for instance, one should be able to distinguish predictions based on the strong field effects of
BK (or hard pomeron based approaches in general) from the soft pomeron physics associated with confinement~\cite{Deshpande:2005wd}. The ratio of the (coherent) diffractive cross section integrated over $t$ and some range  $M_X<M_{\textrm{max}}$ to the inclusive cross section is, in the dipole picture used in the saturation context, very generically enhanced in nuclei compared to protons, since in nuclei the dipole-target scattering amplitude at a fixed impact parameter is larger than in the proton~\cite{Nikolaev:1995xu,Kowalski:2008sa,Accardi:2012qut}. Some saturation models predict that hard diffractive events will constitute even up to 30-40\% of the cross-section~\cite{Levin:2002fj}.  The impact of nonlinear effects was recently investigated in Ref.~\cite{Bendova:2020hkp}, where predictions for diffractive structure functions and reduced cross sections  were obtained from solutions of the Balitsky–Kovchegov equation with the collinearly-improved kernel, and including impact-parameter dependence.

The gluon saturation computations can be compared with DGLAP predictions which match soft Pomeron physics with hard perturbative physics~\cite{Frankfurt:2003gx}. The latter result in a much smaller fraction of the cross-section and should therefore be easily distinguishable from CGC based ``strong field'' diffraction. Thus, in non-saturation parameterizations of nuclear diffractive  parton distribution functions one observes, in contrast to saturation calculations, a nuclear suppression rather than an enhancement. The difference can be traced back to  the interplay between multiple scattering and gap survival  probability. This leads to a very striking difference in the predictions from dipole models vs.\ leading twist shadowing, as illustrated in Fig.~\ref{fig:dipole_vs_lts_diffraction}.

In the dipole or CGC picture, the virtual photon interacts coherently with all nucleons at the same transverse coordinate. Parametrically, at large $A$, the amplitude for a fixed impact parameter is proportional to the number of overlapping nucleons $\sim A^{1/3}$. With a nuclear transverse area $\sim A^{2/3}$ this leads to a diffractive (or elastic) cross  section that parametrically depends on the mass number as $\sigma^{D}\sim A^{4/3}$, compared to $\sigma^{\mathrm{tot}}\sim A$. This growth saturates at the black disk limit, where the diffractive cross section is half of the total cross section. At small $\beta$, i.e.\ for large masses of the diffractive system, this saturation sets in much earlier because the partonic system is not a simple $q\bar{q}$ dipole, but a many-particle Fock state.

In the leading twist shadowing picture of Ref.~\cite{Frankfurt:2011cs}, a diffractive interaction also takes place coherently on all the overlapping nucleons of the nuclear target.  In contrast to the CGC dipole picture, one separates the interaction into a primary interaction with one nucleon and reinteractions with the other $A-1$ nucleons. The reinteractions with target nucleons are treated in the Gribov-Glauber approach to nuclear shadowing by including all possible diffractive intermediate states, which results in a strong nuclear suppression (shadowing) of nuclear diffractive structure functions and parton distributions compared to the proton case. In a different language, this suppression can be identified with the flavor-specific rapidity gap survival probability calculated in the leading twist nuclear shadowing model. Thus the question  of nuclear suppression vs enhancement in the diffractive/total cross section ratio is very sensitive probe of the role of coherence and saturation at a specific $x$ and $Q^2$.

In the following we report on a more detailed study~\cite{Armesto:2019gxy}  of coherent inclusive diffraction in the leading twist shadowing framework~\cite{Frankfurt:2011cs}. Here we assume that coherent events have been distinguished from the incoherent case 
using forward detectors, see Section~\ref{part2-sec-DetReq.Diff.Tag}. 

Assuming the same framework (collinear factorization for hard 
diffraction and Regge factorization) described for \ep in 
Subsection~\ref{part2-subS-PartStruct-InclDiff} to 
hold for \eA, nuclear diffractive PDFs (nDPDFs) can be 
extracted from the diffractive reduced cross sections, Eq.~\eqref{PWG-sec7.1.6-sred}. Such nDPDFs have never been measured. The kinematic coverage in $e$A at the EIC will be very similar to that shown for $ep$ in Fig. 3 in Ref.~\cite{Armesto:2019gxy}.

Due to the lack of previous measurements, there are no parameterizations for nDPDFs but models exist  for the nuclear effects on 
parton densities defined through the nuclear modification factor
\begin{equation}
R_k^A(\beta,\xi,Q^2) = \frac{f_{k/A}^{D(3)}(\beta,\xi,Q^2)}{A\,f_{k/p}^{D(3)}(\beta,\xi,Q^2)}\;,
\label{eq:ndmf}
\end{equation}
with 
diffractive parton densities in a nucleus $A$ denoted as $f_{k/A}^{D(3)}(\beta,\xi,Q^2)$.
We use the model proposed in~\cite{Frankfurt:2011cs}, where 
parameterizations for nuclear modification factors 
are provided at the scale $Q^2=4$ GeV$^2$. Then DGLAP evolution is employed to
evolve the ZEUS-SJ proton diffractive PDFs multiplied by $R_k^A$ from~\cite{Frankfurt:2011cs}
to obtain the nuclear diffractive PDFs, at any $Q^2$.
The structure functions and reduced cross sections are then
calculated in the same way as in the proton case, and these results are used to obtain the modification factors, analogous to Eq.~\eqref{eq:ndmf}, 
for these quantities.
We have also repeated the calculation in the Zero-Mass Variable Flavor Number Scheme in order to 
check that the resulting modification factors do not depend on the applied scheme.

The model in~\cite{Frankfurt:2011cs} employs 
Gribov inelastic shadowing~\cite{Gribov:1968jf} which relates diffraction in 
\ep to nuclear shadowing for total and diffractive 
\eA cross sections. It assumes that the nuclear wave function squared can be approximated by the product of one-nucleon densities, neglects the $t$-dependence of the diffractive $\gamma^*$-nucleon amplitude compared to the nuclear form factor, introduces a real part in the amplitudes~\cite{Gribov:1968uy}, and considers the colour fluctuation formalism for the inelastic intermediate nucleon states~\cite{Frankfurt:1994hf}. 
There are two variants of the model, named H and L, corresponding to different strengths of the colour fluctuations, giving rise to larger and smaller probabilities for diffraction in nuclei with respect to that in proton, respectively.
The corresponding nuclear modification factors, Eq.~\eqref{eq:ndmf}, 
for $F_{2}^{D(3)}$ and $F_{L}^{D(3)}$ in $^{208}$Pb, are shown in Fig. 13 in Ref.~\cite{Armesto:2019gxy}.

\begin{figure}
\begin{center}%
	\includegraphics*[width=0.5\textwidth]{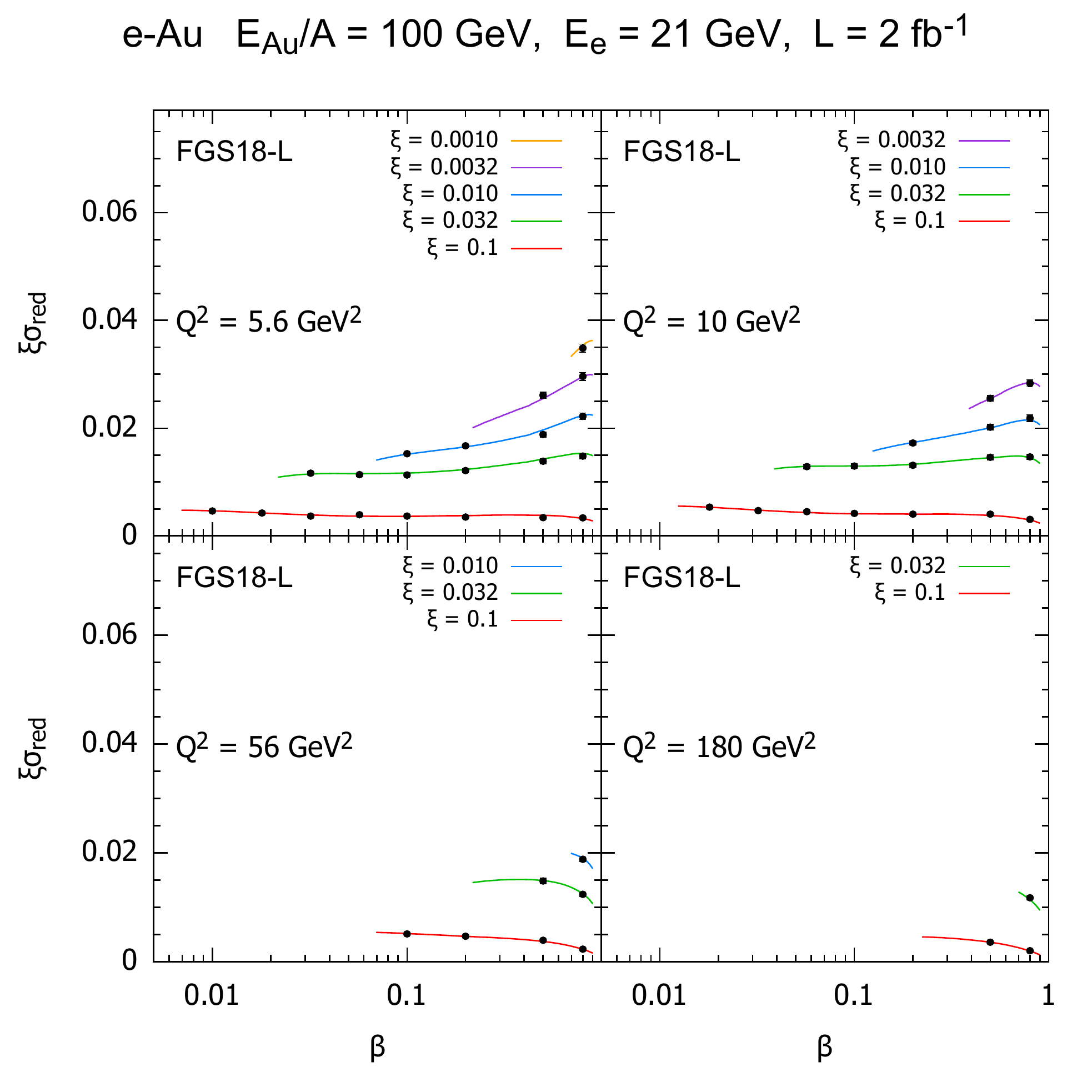}%
	\includegraphics*[width=0.5\textwidth]{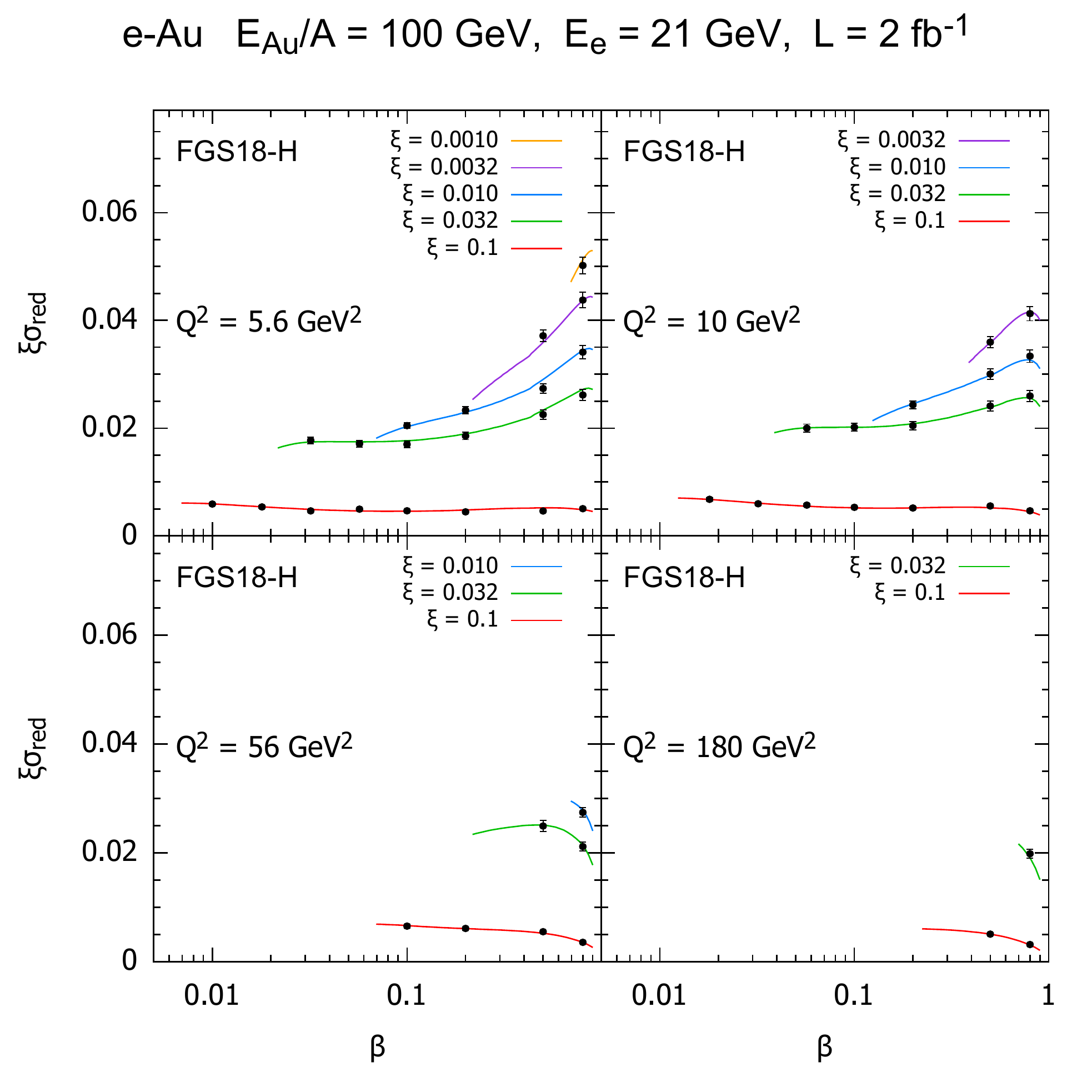}%
\end{center}
\caption{Simulated data for the diffractive reduced cross section as a function of $\beta$ in bins of $\xi$ and $Q^2$ for $e\,^{197}$Au collisions at the EIC, in the models L (left plot) and H (right plot) in~\cite{Frankfurt:2011cs}.
The curves for $\xi = 0.032, 0.01, 0.0032, 0.001$ are shifted up by 0.005, 0.01, 0.015, 0.02, respectively. Taken from~\cite{Armesto:2019gxy}.}
\label{fig:sigred_Au01e21}
\end{figure}

Pseudodata were generated for $e$Au collisions at the EIC using the same method, and taking the 
uncorrelated systematic error to be 5\%, as 
described for \ep in~\cite{Armesto:2019gxy}. We assumed $E_e=21$ GeV, $E_N=100$ GeV/nucleon and an integrated luminosity of 2 fb$^{-1}$.
The results are shown in Fig.~\ref{fig:sigred_Au01e21}. Studies performed for \ep at those energies show that the expected accuracy for the extraction of DPDFs at the EIC is comparable to that in existing DPDFs for the proton at HERA, with some improvements at large $\beta$, see Subsection~\ref{part2-subS-PartStruct-InclDiff}. Assuming a similar experimental uncertainty, integrated luminosity and kinematic coverage, the accuracy in the extraction of nDPDFs at the EIC would then be similar to that of existing HERA fits, see Subsection~\ref{part2-subS-PartStruct-InclDiff}. Improvements would crucially depend on a decrease of the systematic uncertainty, including the separation of coherent from incoherent diffraction.

Finally, the relation between diffraction and nuclear shadowing~\cite{Gribov:1968jf} could be tested at the EIC. The relation is a rigorous theoretical result for the deuteron case, while its extension to larger nuclei becomes model dependent~\cite{Frankfurt:2011cs,Armesto:2003fi}. The possibility of colliding electrons with different nuclear species, including deuterons, will therefore be very important.

A more differential observable giving a more detailed access to the parton level kinematics is provided by diffractive dijet measurements, since they also provide access to the angle of the dijet with respect to the proton, which otherwise is integrated over in inclusive diffraction.  In the CGC picture the theoretical description of diffractive dijet production has recently seen important theoretical advances with the calculations advancing to NLO accuracy~\cite{Boussarie:2014lxa,Altinoluk:2020qet}. Especially for the proton, this angular information is used to extract the Wigner distribution, as discussed in Sec.~\ref{part2-subS-SecImaging-Wigner}. Diffractive dijet production in nuclei can also  be addressed in the collinear diffractive parton distribution framework~\cite{Guzey:2020gkk}, in a similar fashion as in proton targets discussed in Sec.~\ref{part2-subS-PartStruct-InclDiff}.

\subsection{Nuclear PDFs}
\label{part2-subS-LabQCD-NuclPDFs}

Nuclear parton distribution functions (nPDFs) describe the behaviour of bound partons in the nuclear medium. Like free-proton PDFs they are assumed to be universal and are extracted through fits to existing data. To date, there is no compelling evidence of factorization breaking or violation of universality. 

The theoretical interpretation of {\AA} and {\pA} data from the LHC and RHIC also relies on precise knowledge of nPDFs. However, in contrast to the free-proton PDFs, the determination of nPDFs is severely limited by both the kinematic coverage and the precision of the available data.

\begin{figure}[th]
	\centering
	\includegraphics[width=12cm]{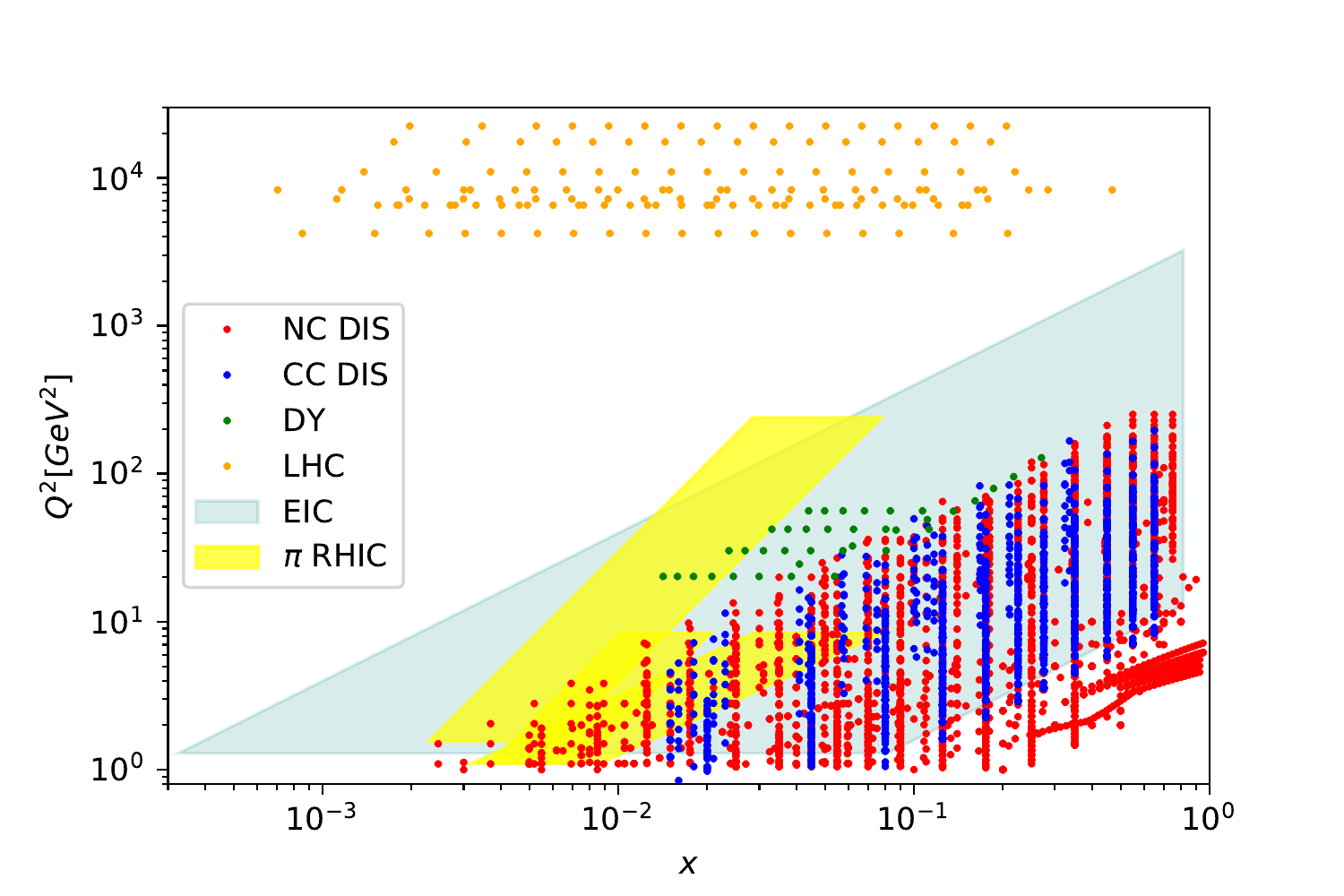}	
	\caption{Kinematic coverage of experimental data and EIC pseudo data used in nPDFs fits. The coverage corresponds to all measured nuclei together. Each nPDFs set has extra cuts that further reduce the explored space.}
	\label{fig:nuc_kin_cov}
\end{figure}

\begin{figure}[th]
	\centering
	\includegraphics[width=12cm]{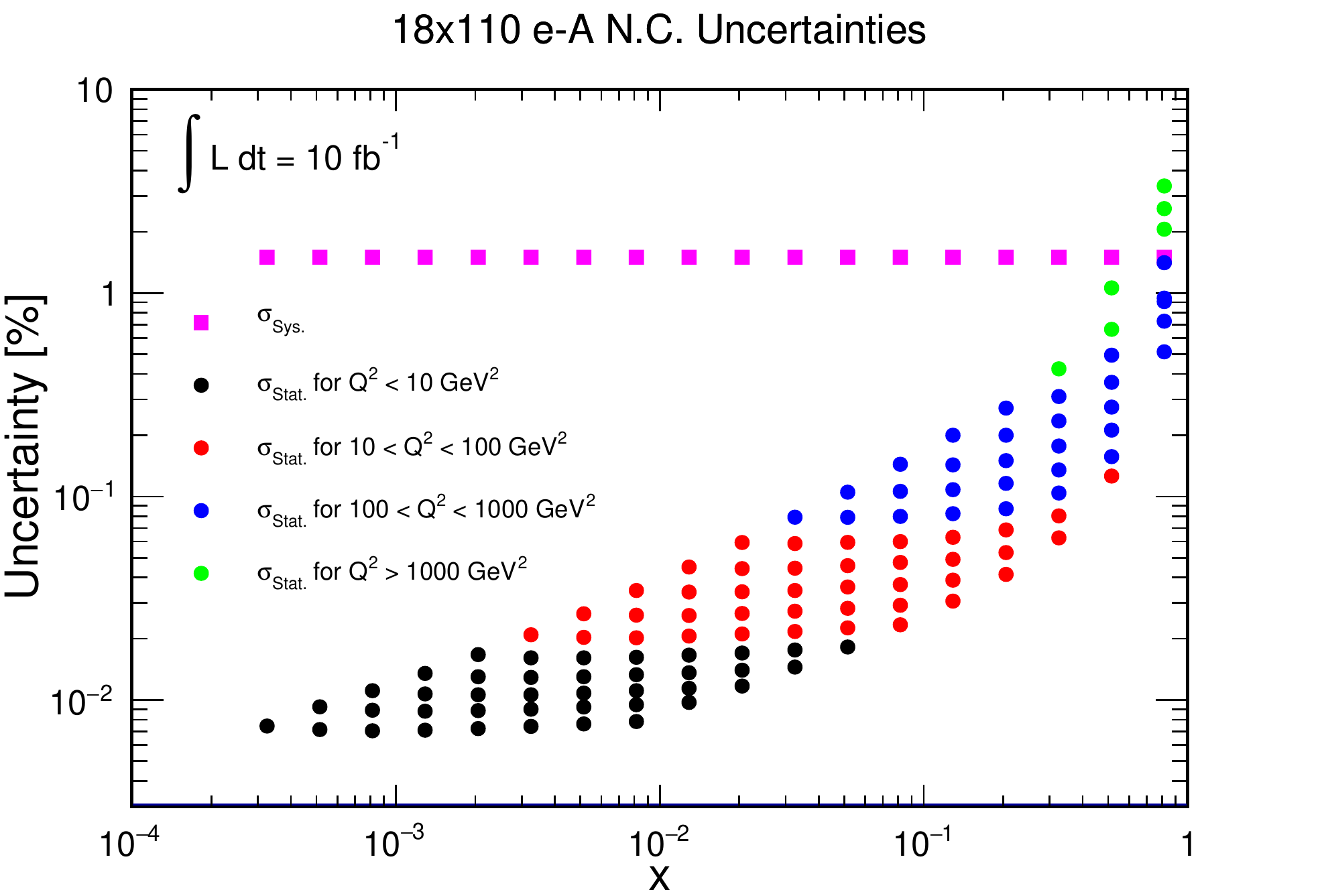}	
	\caption{Relative statistical and uncorrelated systematic uncertainties for inclusive cross section measurements in 18x110 GeV $\eA$ collisions expected at the EIC. Details of the systematic error estimate may be found in Section \ref{part2-sec-DetReq.Incl}.}
	\label{fig:relat_unc_nuc}
\end{figure}

The realization of the EIC will provide key constraints on nPDFs. Fig.~\ref{fig:nuc_kin_cov} shows the significant broadening of the kinematic coverage for all nuclei available at the EIC. Note that nPDFs sets make different selections and apply extra kinematic cuts that further reduce the explored space.
In contrast with previous experiments, the systematic uncertainties of the $e+A$ inclusive DIS cross section measurements at the EIC will be at most a few $\%$, as depicted in Fig. \ref{fig:relat_unc_nuc}. Additionally, the statistical uncertainties will be negligible for almost the whole $x$ coverage, gaining predominance only at the largest values of $x$. This broad kinematic coverage, almost doubling the one from existing data, will revolutionize our current understanding of partonic distributions in nuclei.

\subsubsection*{nPDFs via inclusive DIS}
The DIS cross section can be expressed in terms of the structure functions $F_{2}$ and $F_{L}$ 
\begin{equation}
\sigma \propto F_2(x,Q^2) - \frac{y^2}{1 + (1 - y)^2} F_L(x,Q^2) \, .
\label{Eq:SigmaRed}
\end{equation}
The former is mainly sensitive to the (anti-)quark content of the nucleon and dominates the cross-section at high values of $x$. The latter, relevant in the unexplored low $x$ region, has a direct contribution from the gluon density~\cite{Armesto:2010tg}. The large $Q^2$ lever arm of the EIC will allow us to precisely extract $F_L$ and further determine the nuclear gluon PDF. Longitudinal
and charm structure functions provide direct access to the magnitude of nuclear
effects on the gluon distribution~\cite{Cazaroto:2008qh}.

\begin{figure}[th]
	\centering
	\includegraphics[width=16cm]{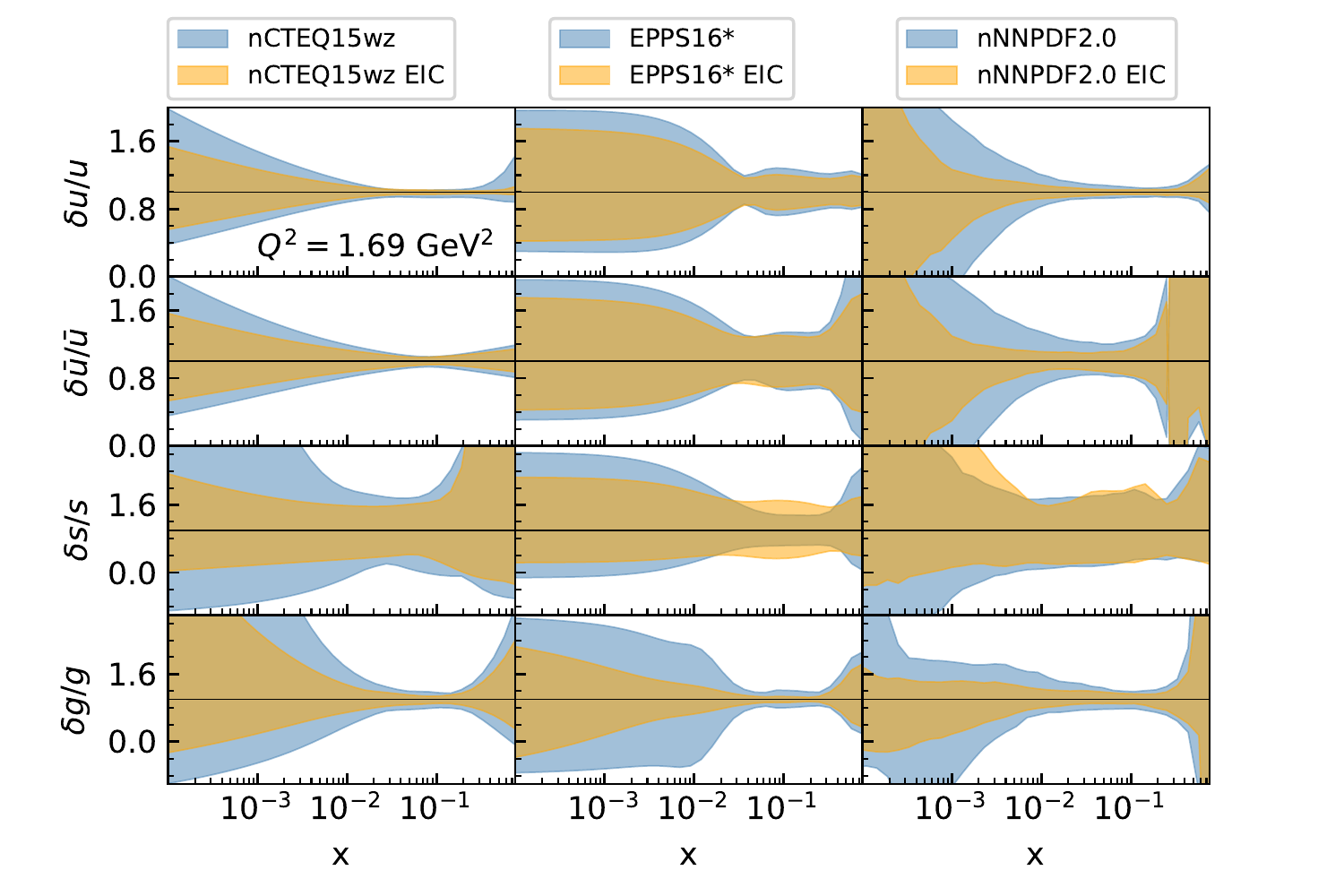}	
	\caption{Relative uncertainty bands for Au at $Q^{2}=1.69\text{ GeV}^{2}$ for $u$ (first row), $\bar{u}$ (second row), $s$ (third row) and gluon (lower row) for three different sets of nPDFs. The blue and orange bands correspond to before and after including the EIC pseudodata in the fit, respectively. }
	\label{fig:nuc_unc}
\end{figure}

The precision of the inclusive cross section measurements at the EIC at low values of $x$ ($x<10^{-2}$) and $Q^2$ will significantly reduce the current theoretical uncertainties. 
This is demonstrated in Fig.~\ref{fig:nuc_unc} which shows a comparison of the relative uncertainties of three modern sets of nPDFs~\cite{Aschenauer:2017oxs,Kusina:2020lyz,AbdulKhalek:2020yuc} in a gold nucleus (blue bands) and their modification when including EIC DIS pseudodata in the fits (orange bands). The overall effect is a significant reduction of the uncertainties in the low-$x$ region, where data is scarce or non-existent. The high-$x$, low $Q^{2}$ region is covered by fixed target experiments and will be further explored at CLAS.

 \subsubsection*{Probing nuclear gluons with heavy flavor production}

Heavy flavor (HF) production is a powerful observable that will complement inclusive DIS measurements in determining nuclear modifications of the PDFs, in particular for the gluon distribution. Recent results from ultraperipheral $A+A$ collisions \cite{Abelev:2012ba,Khachatryan:2016qhq, Khachatryan:2016qhq,Abbas:2013oua,Guzey:2013xba,Guzey:2020ntc} as well as HF and dijet production in $p+Pb$ \cite{Kusina:2017gkz,Eskola:2019bgf,Eskola:2019dui} at the LHC support nuclear suppression with respect to the proton gluon at $x \ll 0.1$ (shadowing). However, little is known about gluon enhancement (antishadowing) at $x \sim 0.1$ or a possible suppression at $x > 0.3$ (``gluonic EMC effect''). At the EIC it will be possible to obtain a direct constraint of the gluon density by measuring HF pairs which at LO are produced through the photon–gluon fusion process. This channel probes the gluon PDFs for $x > ax_{\rm B}$, where $a = 1 + 4 m_h^2/Q^2$ and $m_h$ is the heavy quark mass. This measurement will also permit the study of different heavy quark mass schemes and constrain the intrinsic HF components in the nPDFs~\cite{Accardi:2016ndt}. 

The feasibility and impact of nuclear gluon measurements with HF production at the EIC has been studied in dedicated efforts \cite{LD1601,Furletova:2020ejm,Aschenauer:2017oxs} by tagging, from the simulated DIS sample, the $K$ and/or $\pi$ decay products from the $D$ mesons produced in the charm fragmentation. The reconstruction methods used in this analysis \cite{LD1601} demonstrate the key role that particle identification (PID) will play. It was shown that the charm reconstruction is significantly increased \cite{Abramowicz:2013eja} when PID capabilities are included. 

\begin{figure}[ht]
  \centering
  \begin{minipage}[b]{0.47\textwidth}
	\includegraphics[width=8.3cm]{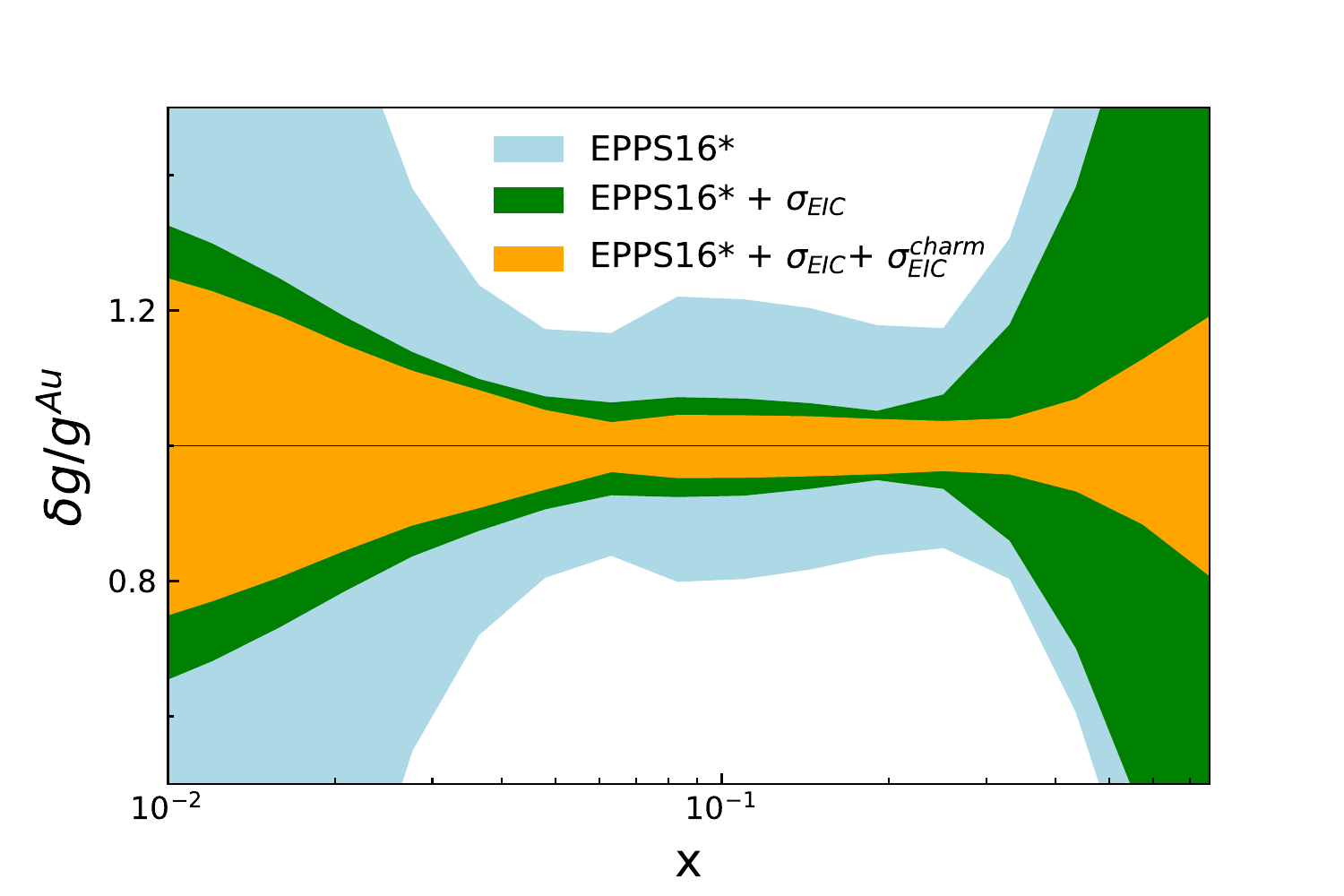}	
  \end{minipage}
  \begin{minipage}[b]{0.47\textwidth}
	\includegraphics[width=6.4cm]{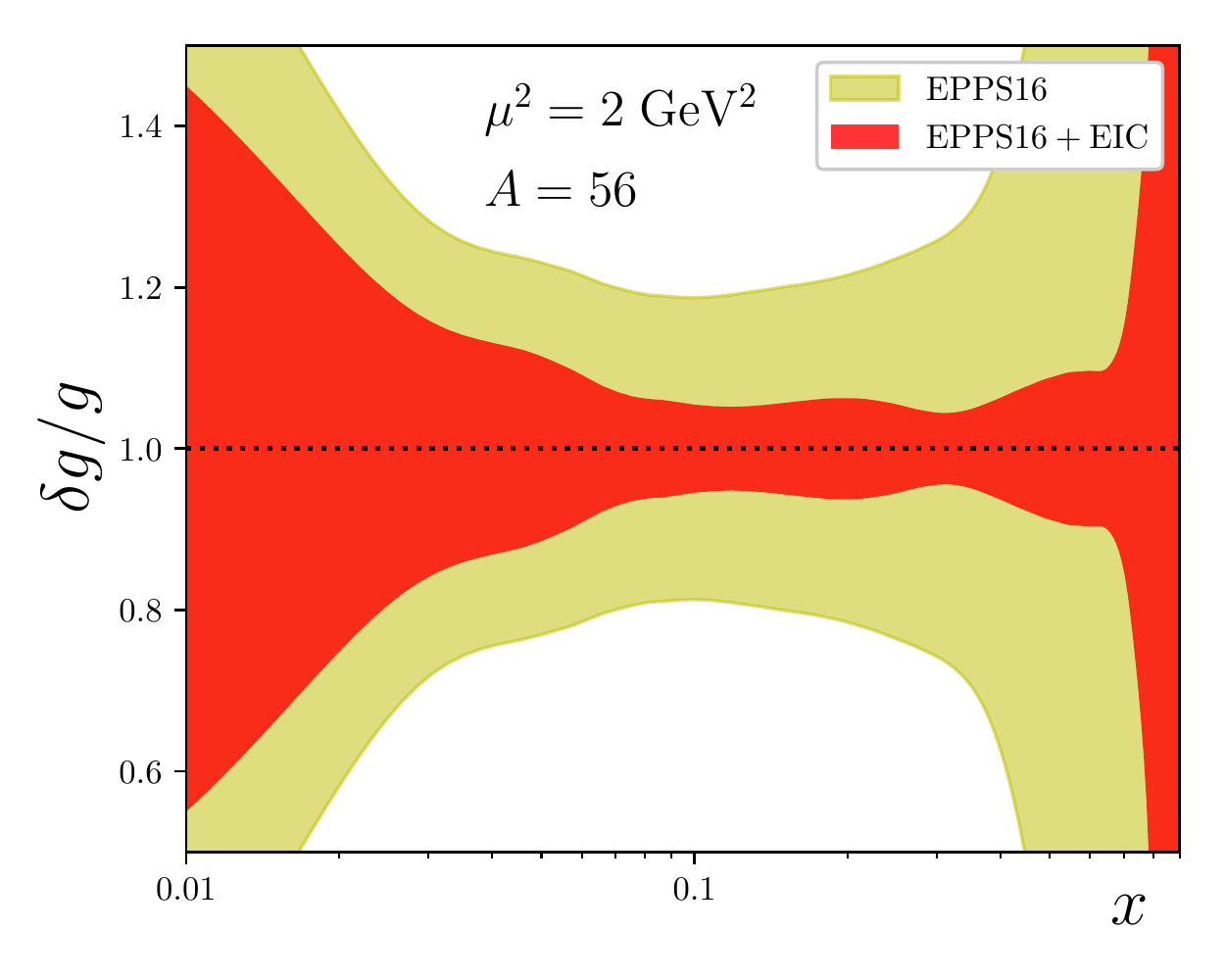}	
  \end{minipage}
    \caption{Left: Relative uncertainty bands of the gluon for $Au$ at $Q^{2}=1.69\text{ GeV}^{2}$ for EPPS16* (light blue), EPPS16*+EIC $\sigma$ (green) and EPPS16*+EIC $\sigma^{charm}$ (orange). Right: same as left panel but for Fe at $Q^{2}=2\text{ GeV}^{2}$ for EPPS16 (yellow) and EPPS16+EIC $\sigma^{charm}$ (red). }
    \label{fig:nuc_unc_hf1}
\end{figure}

In Ref.~\cite{Aschenauer:2017oxs} a full fit using the EIC pseudodata for the inclusive ($\sigma$) and the charm cross-section ($\sigma^{charm}$) has found a significant impact on the reduction of the gluon uncertainty band at high-$x$. This is illustrated in the left panel of Fig.~\ref{fig:nuc_unc_hf1}, where the blue band is the original EPPS16* fit, the green band incorporates $\sigma$ pseudodata and the orange one adds also $\sigma^{charm}$. A similar dedicated study using PDF reweighting with structure function $F_{2A}^{charm}$ was done in \cite{Sato:2016tuz}. In the right panel of Fig.~\ref{fig:nuc_unc_hf1} the impact of Fe pseudodata on the EPPS16 NLO gluon density \cite{Eskola:2016oht} is shown by the red band. The charm pseudodata substantially reduces the uncertainty at $x >\nobreak 0.1$, providing sensitivity to the presence of a gluonic EMC effect. Comparing the red band (only charm pseudodata) with the results of Fig.~\ref{fig:nuc_unc} one can see that the high-$x$ region can be equally studied considering inclusive or charm pseudodata. It is by combining both observables that a striking reduction is achieved (orange band, left panel of Fig.~\ref{fig:nuc_unc_hf1}). Moreover, the measurement will be complemented by jet studies that have already shown promising constraining power for gluons in $p+$Pb collisions \cite{Eskola:2019dui}.

\subsubsection*{Investigating the A dependence of nPDFs}
The EIC will have the capability to operate with a large variety of ion beams from protons to Pb in order to scrutinize the $A$-dependence of nuclear PDFs.
The different nuclei used in the nPDFs fits are usually connected through parameters for which an $A$ dependence is assumed. This allows one to use the whole set of available nuclei in a single fit instead of individually fitting nuclei, a task that, given the current data, would be difficult if not impossible. The EIC, by effectively being a \emph{nuclear} HERA will provide the opportunity to perform a full scan of the kinematic space for each nucleus individually, permitting a robust extraction of the $A$ dependence a posteriori. Studies have shown that the impact demonstrated in Fig.~\ref{fig:nuc_unc} is representative of the impact for all nuclei.

Finally, an EIC would allow, for the first time, a combined determination of proton, deuteron, and nuclear PDFs within an integrated global QCD analysis. This is a unique and invaluable benefit of this new facility, as it eliminates many of the biases and assumptions that plague current PDF determinations.

\subsection{Particle propagation through matter and transport properties of nuclei}
\label{part2-subS-LabQCD-PropPartLoss}

\subsubsection{Parton showers and energy loss in cold nuclear matter} 

Understanding particle propagation in matter and nuclear transport properties remain defining questions in the field. Theoretical efforts in the past decades have been focused on understanding the energy loss of partons as they propagate in strongly interacting 
environments~\cite{Baier:1994bd,Zakharov:1997uu,Gyulassy:2000er,Guo:2000nz,Arnold:2002ja,Guo:2006kz,Vitev:2007ve}. Applications to DIS have yielded a rather wide range of values for the transport coefficient of cold nuclear matter, defined as the mean transverse momentum squared transfer from the medium per unit length. Extracted $\hat{q}$ parameter values vary in the range of $0.02 - 0.14$    GeV$^2$/fm~\cite{Arleo:2003jz,Chang:2014fba}, thus varying by factor of seven between the low- and high-limits.  This spread in the possible $\hat{q}$ values could be addressed at the EIC.        

\begin{figure}[ht]
\begin{center}
\includegraphics[width= 0.8\textwidth]{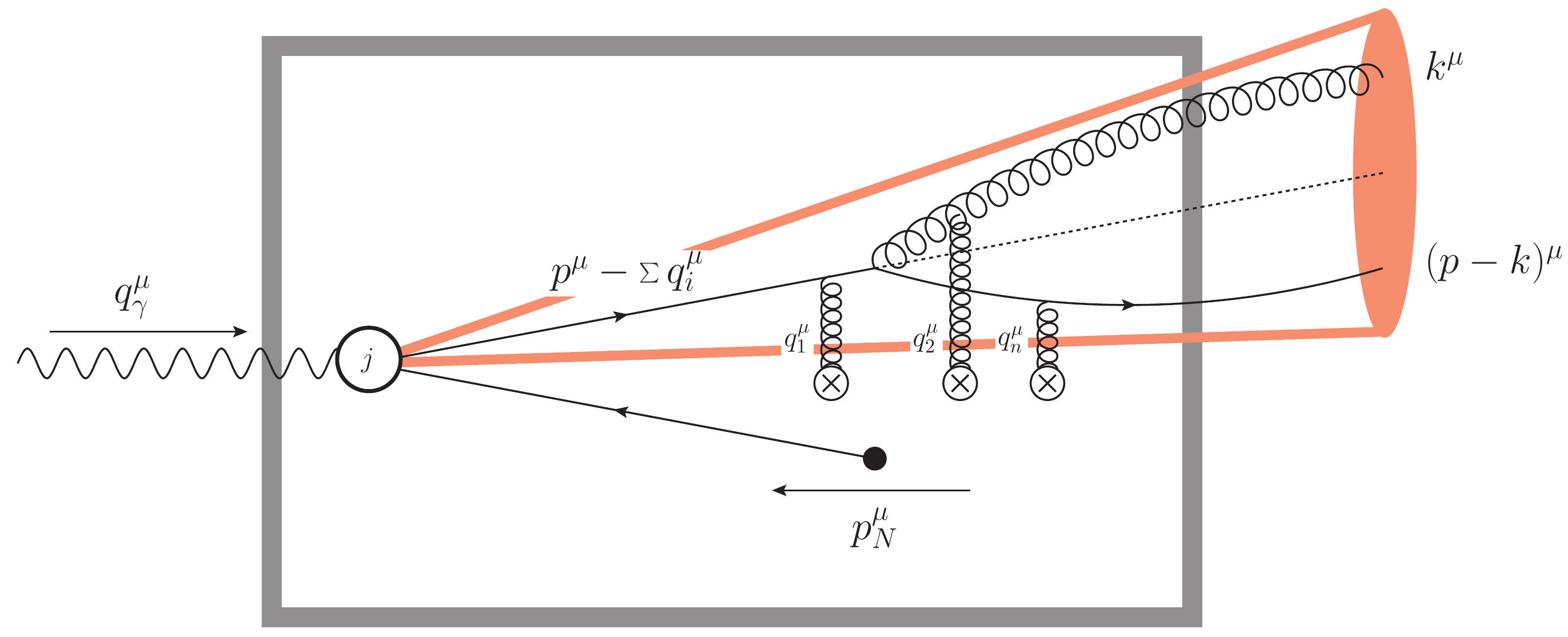} 
\caption{Illustration of the indicated jet kinematics for SIDIS in the Breit frame.  The dark box represents the medium (nucleus) and the red cone represents the jet.}
\label{f:Jet_Kinematics}
\end{center}
\end{figure}

More recently, a much more complete understanding of parton showers in matter, beyond the soft gluon approximation, has emerged~\cite{Wang:2009qb,Apolinario:2014csa,Fickinger:2013xwa,Sievert:2018imd,Sievert:2019cwq}. This allows techniques that bridge the gap between nuclear and particle physics, such as evolution and semi-inclusive jet functions to be applied to reactions with nuclei~\cite{Chang:2014fba,Kang:2014xsa,Li:2018xuv,Li:2020rqj}, see Fig.~\ref{f:Jet_Kinematics}.  
While theoretical approaches differ in  the way they treat final-state interactions, a universal feature is the fractional parton energy loss $\Delta E / E$, or more generally the contribution of medium-induced parton showers, becomes negligible  for very high parton energies $\nu$ in the rest frame of the nucleus owing to the non-abelian Landau-Pomeranchuk-Migdal effect. It is thus advantageous to minimize the rapidity gap between the measured jet and the target nucleus $\delta \eta = |\eta_{p/A}-\eta^{\rm jet}|$ and focus on the forward proton/nucleus going direction.  

\subsubsection{Formation of hadrons in matter}

The low center-of-mass energies for the HERMES collaboration measurements of DIS on nuclei~\cite{Airapetian:2003mi,Airapetian:2007vu} (fixed target and 27.6~GeV electron beam) have spawned phenomenology based upon the idea of early hadron formation and absorption in nuclear matter. 
If particle formation times are smaller than the time to traverse the nucleus, especially for small and large fragmentation fractions $z=p_{\rm had}/p_q$, they can be destroyed or transported away from the experimental acceptance~\cite{Accardi:2002tv,Kopeliovich:2003py}. This leads to suppression of hadron production cross section in \eA versus \ep reactions.

Hadron absorption phenomenology has not yet been developed for the EIC. The much larger $\sqrt{s}$ and the anticipated detector kinematic coverage relative to the nucleon/nucleus rapidity will provide a large boost to hadron formation times, making it unlikely for partons to hadronize in light pions and kaons at $\tau_f < 10$~fm. Uncertainty principle estimates \cite{Adil:2006ra,Sharma:2012dy} show that the formation time $\tau_f$ 
is inversely proportional to the mass of the quark/meson squared. Heavy mesons, such as the  D and the B, thus provide the best opportunity to study hadron absorption physics at the EIC  -  something that should be investigated further in the future.

\subsubsection{Jet production and modification in \texorpdfstring{\eA}{eA} collisions}

\begin{figure}[!t]
    \centering
    \includegraphics[width=0.49\textwidth]{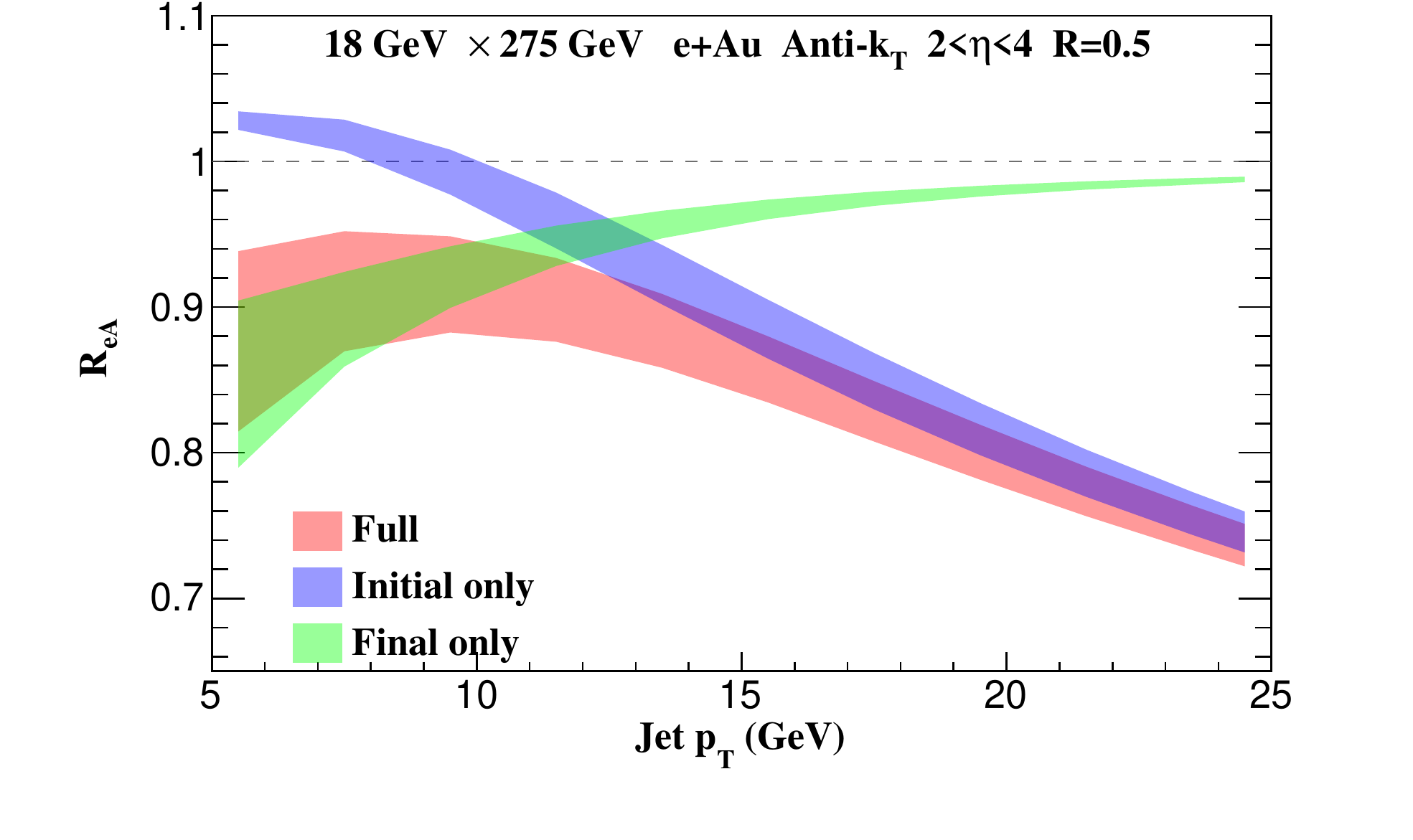}
    \includegraphics[width=0.49\textwidth]{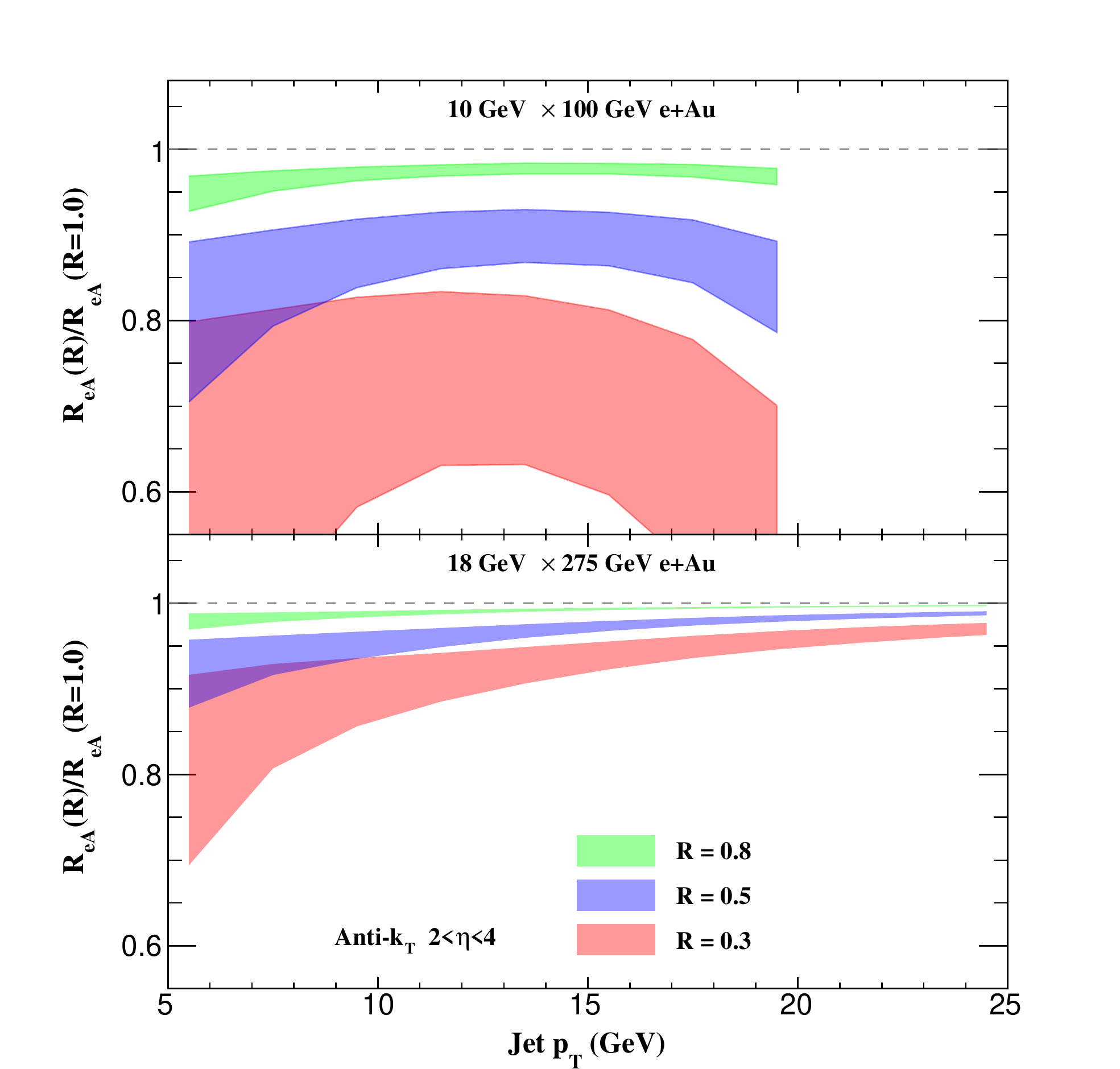}
    \vspace{0.2cm}
    \caption{Left: modifications of the inclusive jet cross section in  18 $\times$ 275 GeV  e+Au collisions for the rapidity interval $2< \eta < 4$.  The blue and green bands represent contributions from initial-state PDFs and final-state interaction between the jet and cold nuclear matter, while the red band is the full result. Right: ratio of jet  cross section modifications for different  radii $R_{\rm eA}(R)/R_{\rm eA}(R=1.0)$ in 10 $\times$ 100 GeV (upper) and  18 $\times$ 275 GeV (lower) e+Au collisions, where the smaller jet radius is R=0.3, 0.5, and 0.8,  and the jet rapidity  interval is $2<\eta<4$.   }
    \label{fig:ReAto1}
\end{figure}

Nuclear effects on  reconstructed jets in electron-nucleus collisions can be studied through the ratio 
\begin{align}
    R_{\rm eA}(R) = \frac{1}{A}   \frac{\int_{\eta1}^{\eta2} d\sigma/d\eta dp_T\big|_{e+A}}{   \int_{\eta1}^{\eta2}  d\sigma/d\eta dp_T\big|_{e+p}}\,.
\end{align}
In the semi-inclusive jet function approach $d\sigma \sim f_a(x,\mu) \otimes H_{ab}(x,z;p_T,\eta ) \otimes 
J_b(z,\mu,R)$. The effects of final-state in-medium parton shower that come at ${\cal O}(\alpha_s)$ are included as follows~\cite{Kang:2017frl,Li:2018xuv}, 
\begin{equation}
J_b(z,\mu,R) = J^{\rm vac}_b(z,\mu,R) + J^{\rm med}_b(z,\mu,R) \; , 
\end{equation}
and $J^{\rm med}_b(z,\mu,R)$ is calculated numerically from the medium-induced splitting functions. 
Predictions for the resulting jet modification are taken from Ref.~\cite{Li:2020rqj}. In the left panel of Fig.~\ref{fig:ReAto1}, for jets of radius $R=$0.5, bands correspond to scale uncertainties from varying the factorization scale and the jet scale by a factor of two independently. For the chosen kinematics 
the Bjorken-$x$ values corresponding to the so-called anti-shadowing and EMC regions of nuclear PDFs. As a result, there is an enhancement for small $p_T$ due to anti-shadowing  and an suppression for large $p_T$ due to  the EMC effect, which is  shown by the blue band.  The green band represents the final-state effects, which give rise to 10 - 20\% suppression when $p_T\sim 5$ GeV.  They are smaller for larger jet energy as expected, and going to backward rapidities further reduces the effect of medium-induced parton showers.  The predicted  full  $R_{\rm eA}(R=0.5)$ for 18 GeV (e) $\times$ 275 GeV (A) collisions  is given the red band.  
The measurements of  jet modification in future will improve our understanding of strong interactions inside nuclei and nuclear PDFs at moderate and large $x$. 

To study cold nuclear matter transport properties with jets at the EIC, it is essential to reduce the role of nPDFs and enhance the effects due to final-state interactions.  An efficient strategy  is to measure the ratio of the modifications with different jet radii, $R_{\rm eA}(R)/R_{\rm eA}(R=1)$,  which is also an observable  very sensitive to the details of  in-medium branching processes~\cite{Vitev:2008rz} and greatly discriminating with respect to theoretical models~\cite{CMS:2019btm}.  Furthermore, it is  very beneficial  to explore smaller center-of-mass energies.  Our  predictions for the ratio of jet cross section suppression for different radii at the EIC  is presented in the right panel of Fig.~\ref{fig:ReAto1},  where the upper and lower panels correspond to  results for 10 GeV (e) $\times$ 100 GeV (A) and 18 GeV (e) $\times$ 275 GeV (A) collisions, respectively.  The plot in the upper panel is truncated around $p_T \sim 20$ GeV because of phase space  constraints  in the lower energy  collisions.  By comparing the 18 GeV $\times$ 275 GeV e+Au collision results to the ones in Fig.~\ref{fig:ReAto1}  we see that $R_{\rm eA}(R)/R_{\rm eA}(R=1)$  indeed eliminates 
initial-state effects. The red, blue, and green bands denote ratios with $R=0.3\,,0.5\,,0.8$, respectively. 
Since medium-induced parton showers are broader than the ones in the vacuum,   for smaller jet radii the suppression from final-state interactions is more significant.   
Even though the scale uncertainties also grow,  the nuclear effect is clear an its magnitude is further enhanced  by the steeper $p_T$ spectra at lower $\sqrt{s}$.

\subsection{Collective effects}
\label{part2-subS-LabQCD-Collective}

For a long time it was believed that the collective multiparticle interactions in ``small'' collision systems (collisions of protons and electrons) are qualitatively  different from those in ``large'' systems, i.e. collisions of heavy nuclei. One of the most important discoveries in nuclear physics during the last decade has been that this is not necessarily the case. Instead, the kind of multiparticle correlations associated with hydrodynamical flow (see Sec.~\ref{part2-sec-Connections-Other}) of bulk, low $p_T$ particle  in nucleus-nucleus collisions have been observed in various systems where they were not anticipated. Some of the first indications of this were the azimuthally near-side structure elongated in the rapidity direction in the $\Delta \phi,\Delta \eta$-distribution of particles associated with a high-$p_T$ hadron, which became to be called the ``ridge''~\cite{Adams:2005ph}, or a similar correlation on the away-side once known as the ``Mach cone''~\cite{Adler:2005ee,Adams:2005ph}. These correlations were surprising since they involved high-$p_T$ particles that were not expected to form a part of the thermal medium. These discoveries were followed by the observation of the ``ridge''-correlation in (high multiplicity) proton-proton collisions at the LHC~\cite{Khachatryan:2010gv,Aad:2015gqa,Khachatryan:2015lva,Khachatryan:2016txc}, followed soon by  similar correlations in proton-nucleus collisions~\cite{Abelev:2012ola,CMS:2012qk,Aad:2012gla,Aad:2013fja,Abelev:2014mda,Aad:2014lta,Khachatryan:2015waa}, which were surprising because a collectively interacting medium was not expected to be present at all. 

Such multiparticle correlations are now typically analyzed in terms of Fourier harmonic coefficients of two, four, or higher multiparticle correlation functions, typically involving particles with a large rapidity separation to eliminate correlations from resonance decays. These coefficients are referred to in the field of heavy ion physics as ``flow coefficients,'' often even when they are calculated in models where they do not originate from any hydrodynamical interactions. Indeed, these correlations have 2 competing, ``final state'' and ``initial state'' explanations, that are in fact not mutually exclusive and may both contribute in varying degrees to the observed signals in different small collision systems. 

The explanation in terms of final state interactions assumes that the correlations result from interactions after the primary collision translating  transverse coordinate space structures into momentum space correlations (see Sec.~\ref{part2-sec-Connections-Other}). The interactions can be modeled by either hydrodynamics~\cite{Mantysaari:2017cni} or by kinetic theory~\cite{Koop:2015wea,Romatschke:2018wgi,Kurkela:2018qeb}. Many recent results, such as the ``geometry scan'' of collisions of proton, deuteron- and helium with gold ions reported by PHENIX~\cite{PHENIX:2018lia} seem to favor this interpretation. 

On the other hand, it is clear that there are also momentum correlations already present among the small-$x$ gluons in the colliding systems\cite{Dumitru:2010iy,Lappi:2015vta,Dusling:2017dqg}. Such correlations can naturally manifest themselves in the particles produced in the collision. The initial stage effects can be expected to be stronger in smaller collision systems~\cite{Dumitru:2008wn,Kovner:2010xk} where they are not diluted by a large number of uncorrelated domains elsewhere in the transverse plane. This is the opposite of the behavior of final state effects that would generically become smaller in smaller systems due to the shorter lifetime of the system. They could naturally explain e.g. observations of flow-like correlations for even heavy flavor hadrons~\cite{Sirunyan:2018toe,Zhang:2019dth}.

A DIS-event at small-$x$ provides us with an unique system to test and understand in detail the physics of these collective interactions. The virtual photon interacts as a hadronic system, whose size and lifetime can be tuned by varying $x$ and $Q^2$. Recently, the ATLAS collaboration reported collective phenomenon in the photo-nuclear ultra-peripheral \AA (this is equivalent to the $\gamma^\ast A$ collision with almost real photons) collisions as well \cite{Perepelitsa:2020pcf}. The CMS collaboration has also released a preliminary study of long-range azimuthal correlations in inclusive $\gamma p$ interactions~\cite{CMS:2020rfg}. There is an interesting and strong physical resemblance between the high multiplicity events in photo-nuclear collisions and those in \pA collisions.  On the other hand, in a reanalysis of HERA data a ridge-like signal was not found~\cite{ZEUS:2019jya}.

The wave function of a low-virtuality photon can in some event contain many active partons due to the rare QCD fluctuation and the dominant contribution to the high multiplicity events comes from such a partonic structure. Therefore, one can argue that the collective phenomenon could also be observed~\cite{Shi:2020djm} in certain kinematic region of the EIC where the incoming virtual photon has a sufficiently long lifetime. EIC can offer both \ep and \eA collisions with different values of virtuality $Q^2$, which allows one to change initial conditions for the target and the system size $\sim 1/Q$ of the collisional system. At the EIC one also expects a higher luminosity than at HERA, which increases the possibilities to collect a large sample of rare high multiplicity events, which has already been a condition to observe these correlations in proton-proton collisions. 
The future efforts in the era of the EIC can help us unravel the origin of the collectivity in high multiplicity events in small systems. 


\subsection{Special opportunities with jets and heavy quarks}
\label{part2-subS-LabQCD-Special}

\subsubsection{Flavor-tagged jets and the jet charge}

Different from inclusive jet cross sections, jet substructure measures the radiation pattern inside a given jet and is governed by a smaller intrinsic scale.  At  EIC energies, the phase space for radiation inside the jet cone is restricted, which makes it  more challenging to study  final-state cold nuclear effects. Even though the differences between the substructure of jets in $ep$ and \eA are expected to be smaller than in the case of heavy ion physics, the example of the jet charge~\cite{Field:1977fa} shows that  nuclear effects can indeed be  identified.  The average jet charge is defined as the transverse momentum $p_T^{i}$  weighted sum of the charges $Q_i $ of the jet constituents
\begin{align} \label{eq:charge}
    Q_{\kappa, {\rm jet}}  = \frac{1}{\left(p_T^{\rm jet}\right)^\kappa } \sum_{\rm i\in jet} Q_i \left(p_T^{i} \right)^{\kappa } \; ,  \quad \kappa > 0 \; .
\end{align}
Studies in proton and heavy-ion collisions~\cite{Krohn:2012fg, Li:2019dre,Chen:2019gqo,Sirunyan:2020qvi} have found that the jet charge is strongly correlated with the electric charge of the  parent parton and can be used to separate  quark jets from anti-quark jets and to pinpoint their flavor origin.

\begin{figure}
    \centering
    \includegraphics[width=0.7\textwidth]{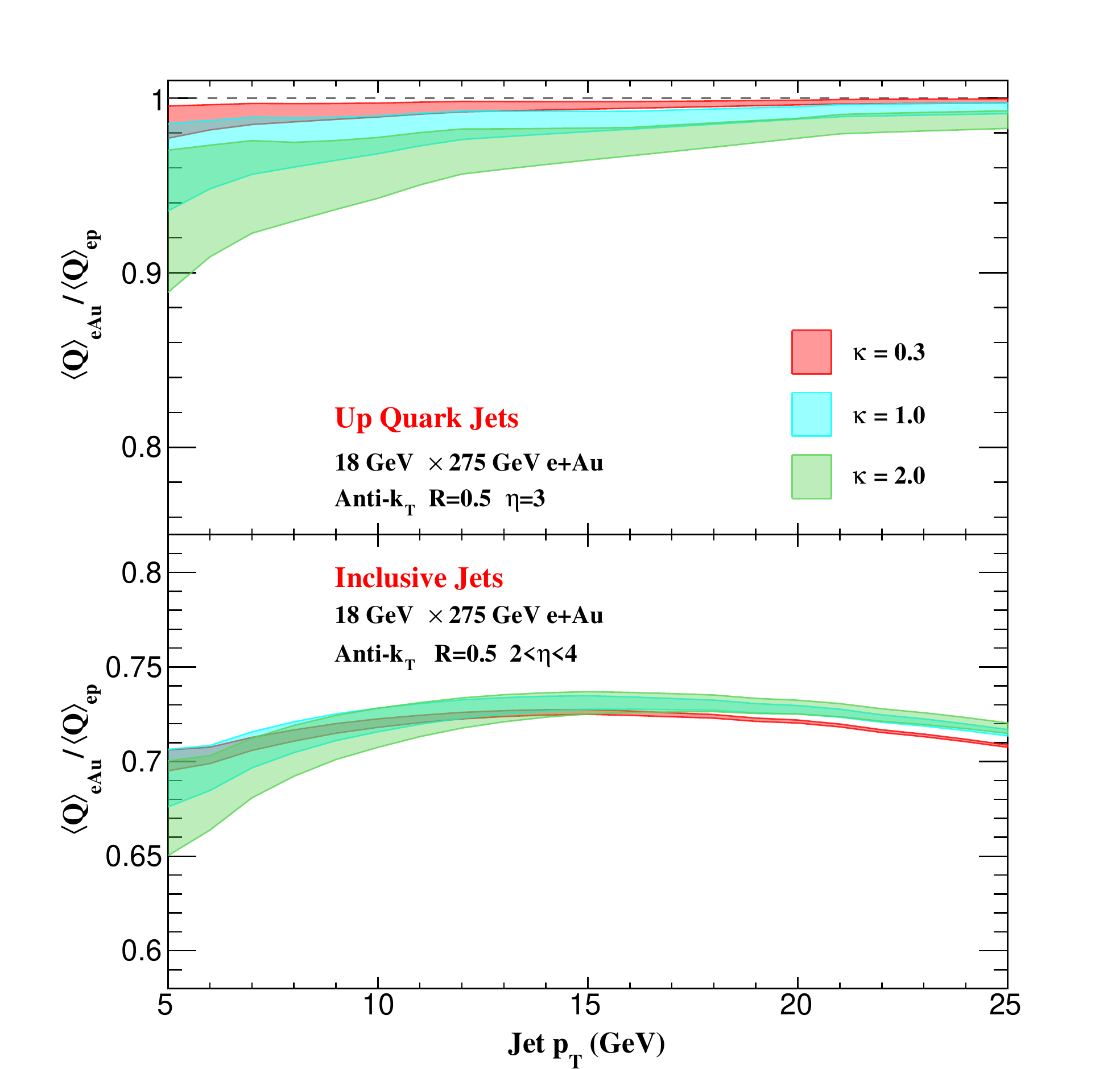}
    \caption{Modifications of the jet charge in $e+Au$ collisions. The upper panel is the modification for up-quark jet with $\eta=3$  
    in the lab frame. The lower panel is the results for inclusive jet with $2<\eta<4$ in 18 $\times$ 275 GeV $e+Au$ collisions.  }
    \label{fig:charge}
\end{figure}

Fig.~\ref{fig:charge} presents jet charge results at the EIC  in 18 GeV $\times$ 275~GeV \eAu collision and for radius parameter $R=0.5$.  
The red, blue and green bands correspond to the jet charge parameter $\kappa=0.3\,,1.0\,,2.0$, see Eq.~(\ref{eq:charge}),   respectively.  
The upper panel shows the modification for the average charge of up-quark initiated jets, where the rapidity is fixed to be  $\eta=3$. It  is defined as  $\langle Q_{q, \kappa}^{\rm eA} \rangle/\langle Q_{q, \kappa}^{\rm ep} \rangle$,  which is independent of the jet flavor and 
originates purely  from final-state interactions.  

Flavor separation for jets has  been accomplished at the LHC~\cite{Aad:2015cua} and  should be pursued at the EIC.
For a larger $\kappa$, the $(\kappa+1)$-th Mellin moment of the splitting function is more sensitive to  soft-gluon emission, this is the  $z\sim 1$ region in the splitting function where medium enhancement for soft-gluon radiation is the largest.  

As shown in the upper panel of Fig.~\ref{fig:charge}, the modification is more significant for larger $\kappa$. 
The modification of the average charge for inclusive jets behaves very differently because there is a  cancellation between contributions from jets initiated by  different flavor partons, in particular from up quarks and down quarks. The lower panel of Fig.~\ref{fig:charge} shows the ratio of average charges for inclusive jets with $R=0.5$ and $2<\eta<4$ for e+A and e+p collisions.  The modification is about 30\% and the $\kappa$ dependence is small due to the large difference between up/down quark density between proton and gold PDFs. Precision measurement of the charge for inclusive jets will be an excellent way to constrain isospin effects and  the up/down quark PDFs in the nucleus.

\subsubsection{Light and heavy-flavor tagged jet angularities}

In the clean EIC environment jet substructure studies can be extended to heavy flavor. In experimental simulations, initial jet reconstruction has been achieved based on true particle information. Inclusive jets are reconstructed with the anti-$k_{T}$ jet algorithm with cone radius at 1.0. Then jets are tagged with fully reconstructed heavy flavor mesons by requiring these reconstructed heavy flavor hadrons be within the associated jet cone \cite{Li:2020wyc}. If there is not a reconstructed heavy flavor hadron can be found within the jet cone, this jet is labeled as light flavor jet. 

\begin{figure}
\centering
\includegraphics[width=0.95\textwidth]{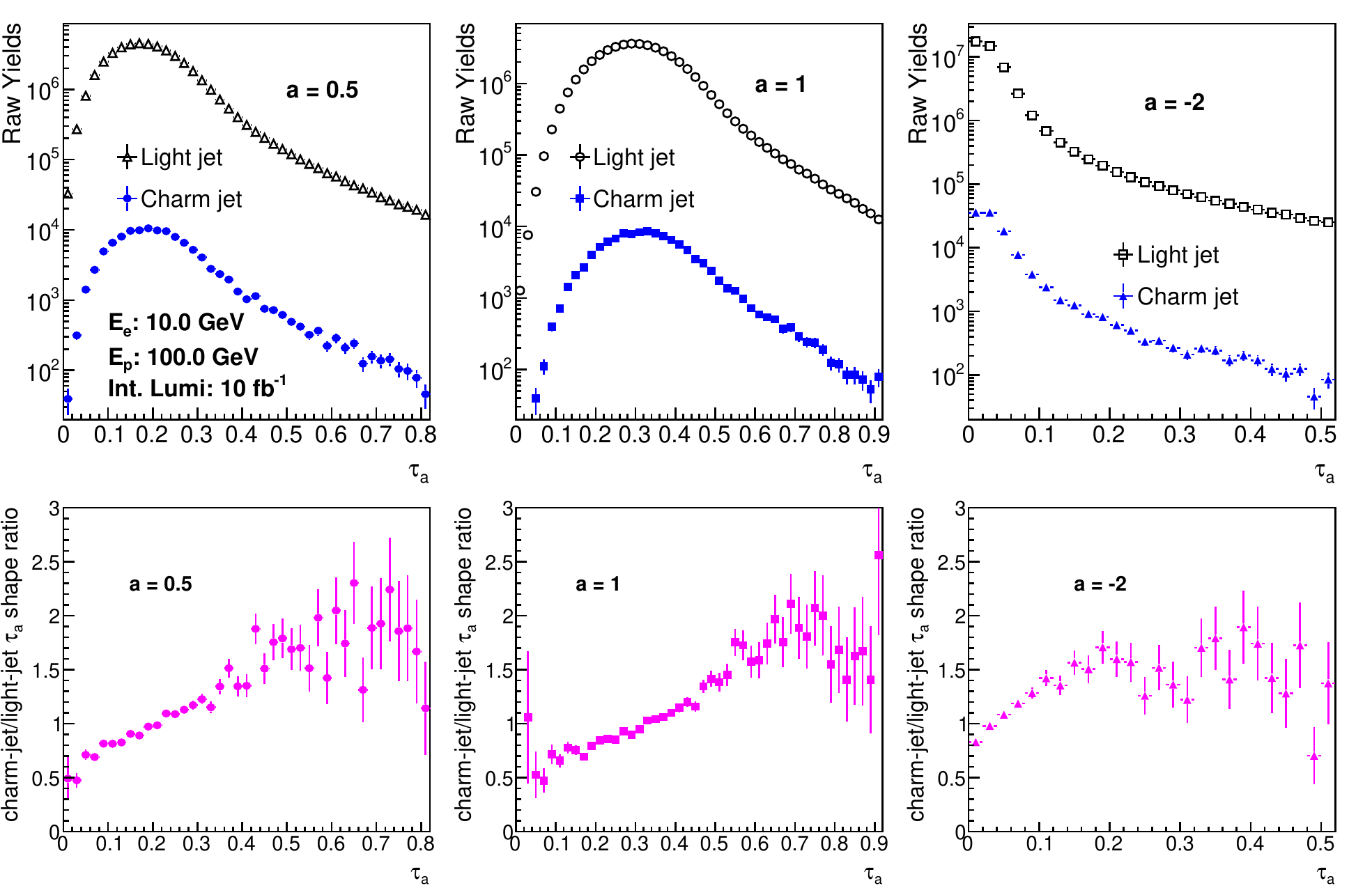}
\caption{\label{fig:Rjet_2} Jet angularity distributions for light flavor jets (black open symbols) and charm tagged jets (blue filled symbols) with different power order $a$ value selections are shown in the top panel. Distributions with $a = 0.5$ are shown on the left, with $a = 1$ are shown in the middle and $a = -2$ are shown on the right. Bottom panel shows the ratio of the normalized charm jet angularity distribution over the normalized light flavor jet angularity distributions in the top panel with the corresponding $a$ value selection. The statistical uncertainties are projected with 10 $fb^{-1}$ $e+p$ at $\sqrt{s} = 63$ GeV.}
\end{figure}

Jet substructure observables can image the nucleon/nuclei 3D structure and help map out the hadronization process in vacuum and nuclear medium. Recent theoretical developments suggest the jet angularity observable has  discriminating power to distinguish quark or gluon initiated jets. We have studied the jet angularity,
\begin{equation}
\tau_a=\frac{1}{p_T}\sum_{i\in J} p_T^i (\Delta R_{iJ})^{2-a} \, , 
\end{equation}
for light flavor jets and charm tagged jets with different power order $a$ value selections~\cite{Li:2020wyc,Wong:2020xtc}. Figure~\ref{fig:Rjet_2} shows the jet angularity distributions of light flavor jets and charm tagged jets and ratio distributions of their shapes in 10 $fb^{-1}$ $e+p$ at $\sqrt{s} = 63$ GeV. Charm jets have a broader jet shape which causes increasing trends in the jet angularity ratio distributions presented in the bottom panels of Figure~\ref{fig:Rjet_2}. Nuclear modification effects for different flavor jets are under study.

\subsubsection{Jets as precision probes in electron-nucleus collisions with electron-jet correlations and jet substructure measurements}

Jets produced in deep-inelastic scattering can be calibrated by a measurement of the scattered electron. Such “tag and probe” studies focus on Born kinematics and call for an approach orthogonal to most HERA jet measurements, which measured high $p_T$ jets in the Breit frame instead. The tag-and-probe approach has been discussed in Ref.~\cite{Liu:2020dct}. and Ref.~\cite{Arratia:2019vju}. Key observables include  electron-jet momentum balance, azimuthal correlations, and jet substructure, which can provide orthogonal constraints on the parton transport coefficient in nuclei and its kinematic dependence. Jet substructure of DIS jets offers a wealth of new opportunities, which have only recently been started; a calculation for the jet groomed radius is presented in Ref.~\cite{Arratia:2019vju}. Moreover, a comparison of standard jet reconstruction and the winner-take-all scheme for DIS jets and their correlation with the electron will help to gauge the modification of soft and collinear fragmentation in the nucleus~\cite{Arratia:2019vju}. Ref.~\cite{Arratia:2019vju} shows kinematic reach, projections of statistical uncertainty, and a discussion of detector requirements.

\subsubsection{Energy flow and quantum number correlations in jets}

An interesting opportunity is to study the correlations between energy flow and quantum numbers such as flavor, spin, and electric charge of the produced particles. The deconstruction of energy-energy correlations at a fixed angle into the contributions of specifically identified particles builds on classic observables such as jet charge. We have performed exploratory studies on samples generated from Pythia (in future we plan  to extend our studies to other event generators with different fragmentation assumptions). The observable we investigate focuses on leading and sub leading jet particles (for example, pion-pion) and the relative cross sections when their electric charges are of the same sign ($N_{CC}$) or opposite sign ($N_{C\overline{C}}$). Specifically, we consider the asymmetry observable $r$ defined as:
\begin{equation}
 r=(N_{CC}-N_{C\overline{C}})/(N_{CC}+N_{C\overline{C}})\, .
\end{equation}
We observe that in Pythia, opposite sign cross sections dominate. It is even more pronounced for kaon pairs and proton-antiproton pairs. Its dependence on particle kinematic phase space is being studied. We also propose to study the correlation between jet and forward particles which were conventionally hidden in the beam remnants. This direction will provide a new and essential window toward understanding hadronization through target fragmentation, global color neutralization and enhancing the precision for event tagging.

\subsection{Short-range correlations, origin of nuclear force}
\label{part2-subS-LabQCD-ShortRange}



\subsubsection{Diffractive J/psi production in electron-deuteron scattering at the EIC and its implication to short-range correlations}

\newcommand{\ed}{$\textit{ed}$}

Understanding the role of Quantum Chromodynamics in generating nuclear forces is important for uncovering the mechanism of short-ranged nuclear interactions and their manifestation in short range correlations (SRC). In recent years, experimental data from Jefferson Lab suggested a strong link between the SRC and the EMC effect~\cite{Schmookler:2019nvf,Hen:2016kwk,Weinstein:2010rt,Higinbotham:2012zz,Hen:2012fm,Hen:2013oha}. Specifically, they suggest that the underlying mechanism of nucleon modifications could be caused by short-range correlated nucleon pairs with high internal nucleon momentum, such as a quasi-deuteron inside the nucleus. However there are alternative phenomenological models that can explain the EMC effect without involving SRCs; see Ref.~\cite{Hen:2016kwk} for a recent review. 

The difficulty in drawing a definitive conclusion based on available experimental data is primarily due to the complexity of the nuclear environment. Given the differing structure and reaction dynamics of different nuclei, the nuclear mass ($A$) dependence could in principle be attributed to other underlying physical mechanisms. Nuclear effects that are driven by SRCs should be similar in light nuclei at extreme high internal nucleon momentum and in medium and heavy nuclei in a similar kinematic range. Therefore the observation of universal properties across a wide range of nuclei would suggest that the effect may be independent of the specifics of nuclear structure and reactions. A confirmation of such universal behavior would then provide a definitive explanation for the EMC puzzle. It may also provide insight into similarly universal dynamics, independent of microscopic details, in physical systems across varying energy scales. 

Besides the modifications in the valence quark region in the bounded nucleon, there are a number of  other outstanding questions:
\begin{itemize}
\item What role do gluons play in the short-range correlations of  nucleon pairs?
\item Are gluon modifications linked to the SRC, similar to that for  valence quarks?
\item What is the relation of SRCs to gluon shadowing?
Can this be related to the phenomenon of gluon saturation? 
\item What are the spatial and momentum distributions of partons in such high nucleon momentum configurations?
\end{itemize}
With regard to the last item, nucleon-nucleon elastic scattering experiments at high momentum transfer showed that the energy dependence of such reactions is quite sensitive to differing models of the internal spatial and momentum distributions of partons~\cite{Sterman:2010jv}. 

\begin{figure}[thb]
\centering
\includegraphics[width=0.4\linewidth]{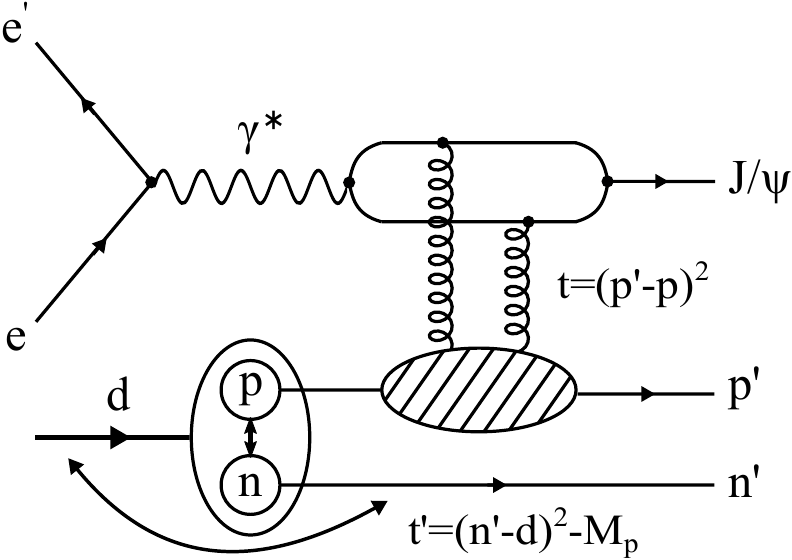}
  \caption{ \label{fig:figure_1} Diagram of incoherent diffractive $J/\psi$ productions in electron-deuteron scattering }
\end{figure}

The EIC will provide an unprecedented opportunity to systematically investigate the underlying physics of SRC for energies and kinematic regions that are otherwise impossible to reach. In this study, we propose to study the impact on gluon distributions inside of nucleons that are associated with a SRC pair in electron-deuteron ($ed$) scattering. Using the Monte Carlo event generator BeAGLE, we investigate the sensitivity of observables to high internal nucleon momentum in incoherent diffractive $J/\psi$ vector meson production. In a plane wave impulse approximation, the initial state deuteron wave function can be accessed directly from the four-momentum of the spectator nucleon (See Fig.~\ref{fig:figure_1} for the Feynman diagram). We use realistic physics estimates and a conceptual far-forward detector simulations of the EIC to fully reveal the potential of this exclusive process. In particular, we provide the luminosity and detector requirements necessary to study SRCs in the deuteron at an EIC. 

In Fig.~\ref{fig:figure_1}, kinematic variables are defined in the figure. In particular, the kinematic variable $t$ is defined between the four-momentum of the incoming and outgoing leading nucleon, while the incoming nucleon momentum inside of the deuteron is not known directly due to the internal nucleon momentum distribution. This is different from the process of electron-proton ($ep$) scattering where the incoming proton has the beam momentum. In an $ep$ collider experiment, the paradigmatic example thus far being the H1 and ZEUS experiments at HERA, the $t$ variable can in principle be reconstructed using different methods~\cite{Tu:2020ymk}, including a new method proposed in this study based on purely the spectator and the leading nucleon. 
The conclusion based on this study is that the best resolution of reconstruction of momentum transfer might come from a combination of different methods, i.e., the spectator tagging technique can be used for identifying the process while the method 3 in Ref.~\cite{Tu:2020ymk} can be used for the values of $t$.

\begin{figure}[thb]
\includegraphics[width=0.49\linewidth]{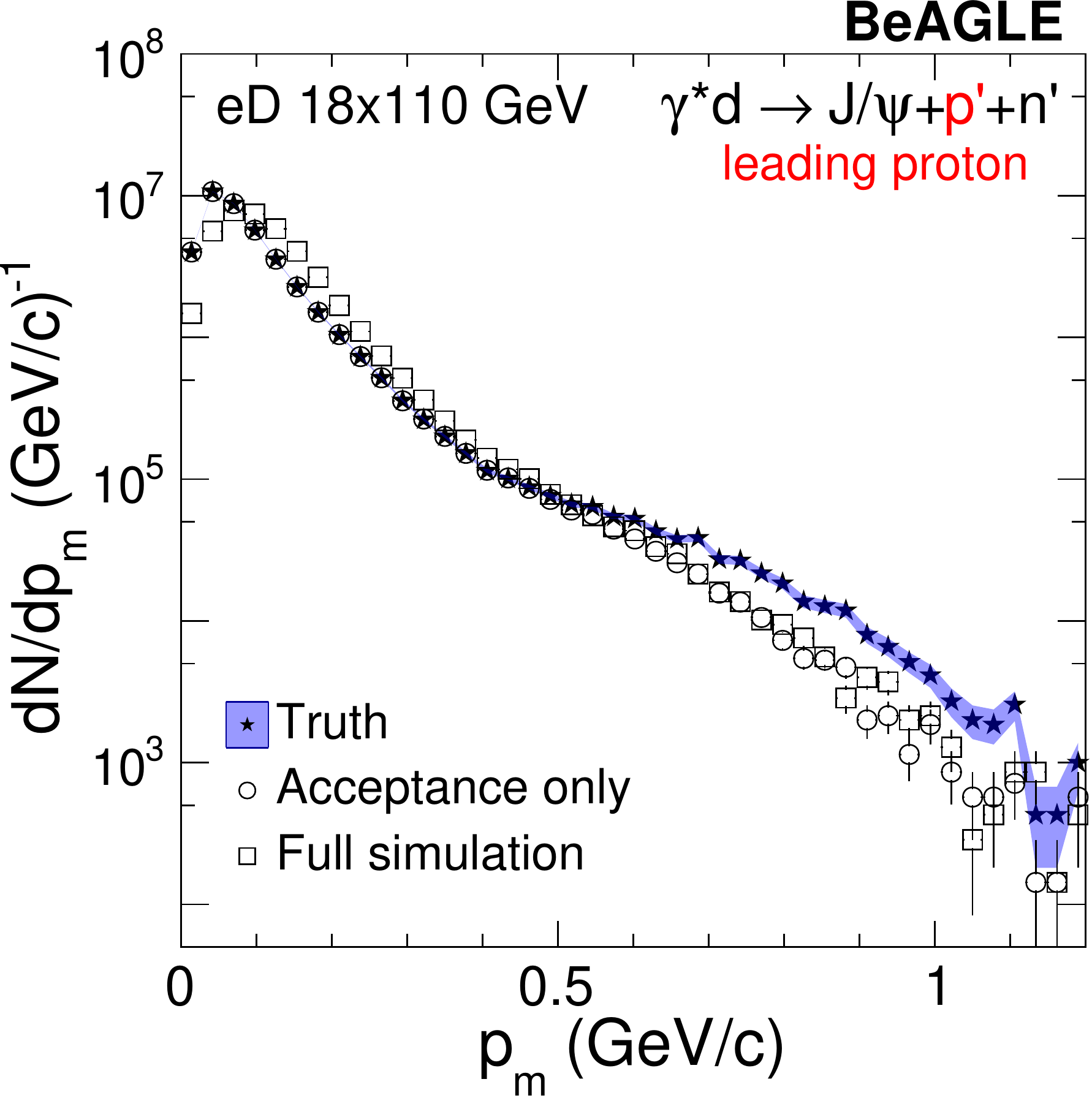}
\hspace{0.02in}
\includegraphics[width=0.49\linewidth]{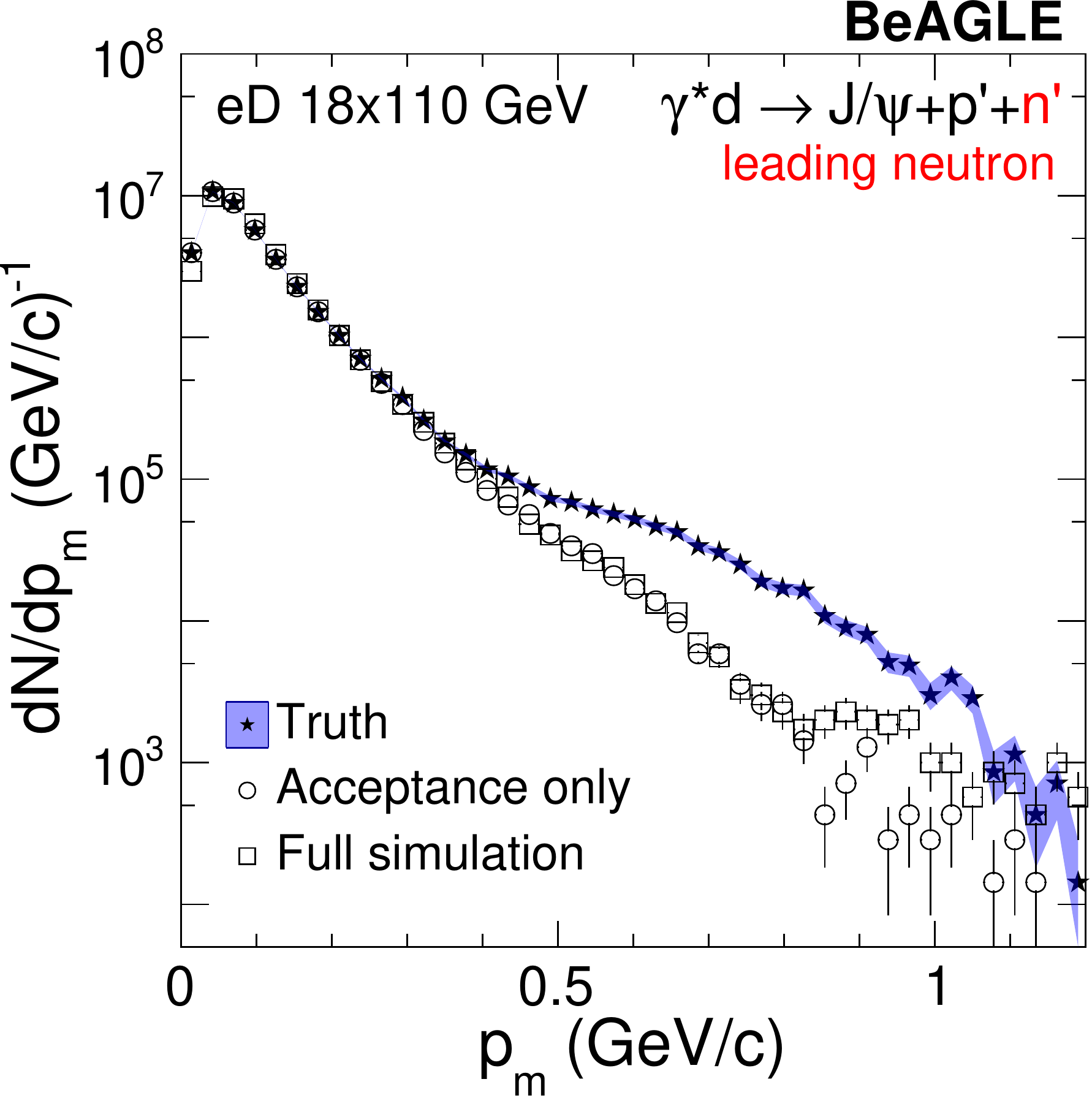}
  \caption{ \label{fig:figure_2} Distribution of the three-momentum of the spectator nucleon  in events associated with incoherent diffractive $J/\psi$ vector meson production in \ed\ collisions are shown for the BeAGLE event generator. The left panel is for the neutron spectator case, where the right panel is for the proton spectators. The simulations at the generator level, with acceptances effects only, and for the full simulations, are shown with solid, open circles, and open squared markers, respectively.}
\end{figure}

In BeAGLE simulations of incoherent diffractive $J/\psi$ meson production in \ed\ scattering, both cases where the spectator nucleon can be either a proton or a neutron are considered. In the simulations, the two cases are treated identically at the generator level, while in the reconstruction of the final state particles in the detector simulations, the spectator proton or neutron would experience different acceptances and detector smearing. In Fig.~\ref{fig:figure_2}, the three-momentum distributions of the spectator, $p_{\rm{m}}$, associated with incoherent diffractive $J/\psi$ production in \ed\ collisions, are shown for neutron  (left) and proton  (right) spectator, respectively. In each panel, the truth level simulation from BeAGLE is shown by solid star markers, where the open circles represent the results after the realistic simulation of the detector acceptance and forward instrumentation. The results of the full simulations (open square markers,)  include acceptances, smearing effects coming from intrinsic detector resolutions, and beam-related effects. With the capability of forward detectors, the access of high momentum configuration of the deuteron is experimentally possible.

For the detector and beam-related effect simulations, one sees that the measurements at low momentum would have a larger impact from detector resolutions but with almost 100\% acceptance; however, for the high momentum range, the impact is found to be opposite. 

Note even that at the generator level, proton and neutron spectator cases are identical, reflecting the assumptions on the deuteron wave function. However after acceptance effects and detector smearing are applied in the reconstruction, the resulting distributions are different. In the neutron spectator case, most of neutrons reconstructed by the ZDC are within a $\pm 4-6~\rm{mrad}$ cone varying with the azimuthal angle. The non-uniformity of the azimuthal acceptance is due to the aperture of  magnets and the other forward instrumentation. 
The neutron spectator acceptance is almost 100\%  up to 600~MeV, while about 80\% for $p_{m}\approx$1~GeV. The momentum smearing effect is noticeable for momenta up to 300~MeV. For a nominal beam momentum particle, e.g., 110~GeV, the resolution is typically 5\%, dominated by the constant energy resolution term of the ZDC. 

For the proton spectator case on the other hand, the $p_{\rm{m}}$ distributions are found to be different from the neutron. Since the proton has better overall resolution, the $p_{\rm{m}}$ distribution at low-momentum exhibits less bin migration in the tagged proton case, and a better acceptance for high nucleon momenta. Most of the proton spectators end up within the acceptance of the OMD instead of the RP due to the protons having less magnetic rigidity ($\sim 50\%$) compared to the deuteron beam. 

\begin{figure}[thb]
\includegraphics[width=0.49\linewidth]{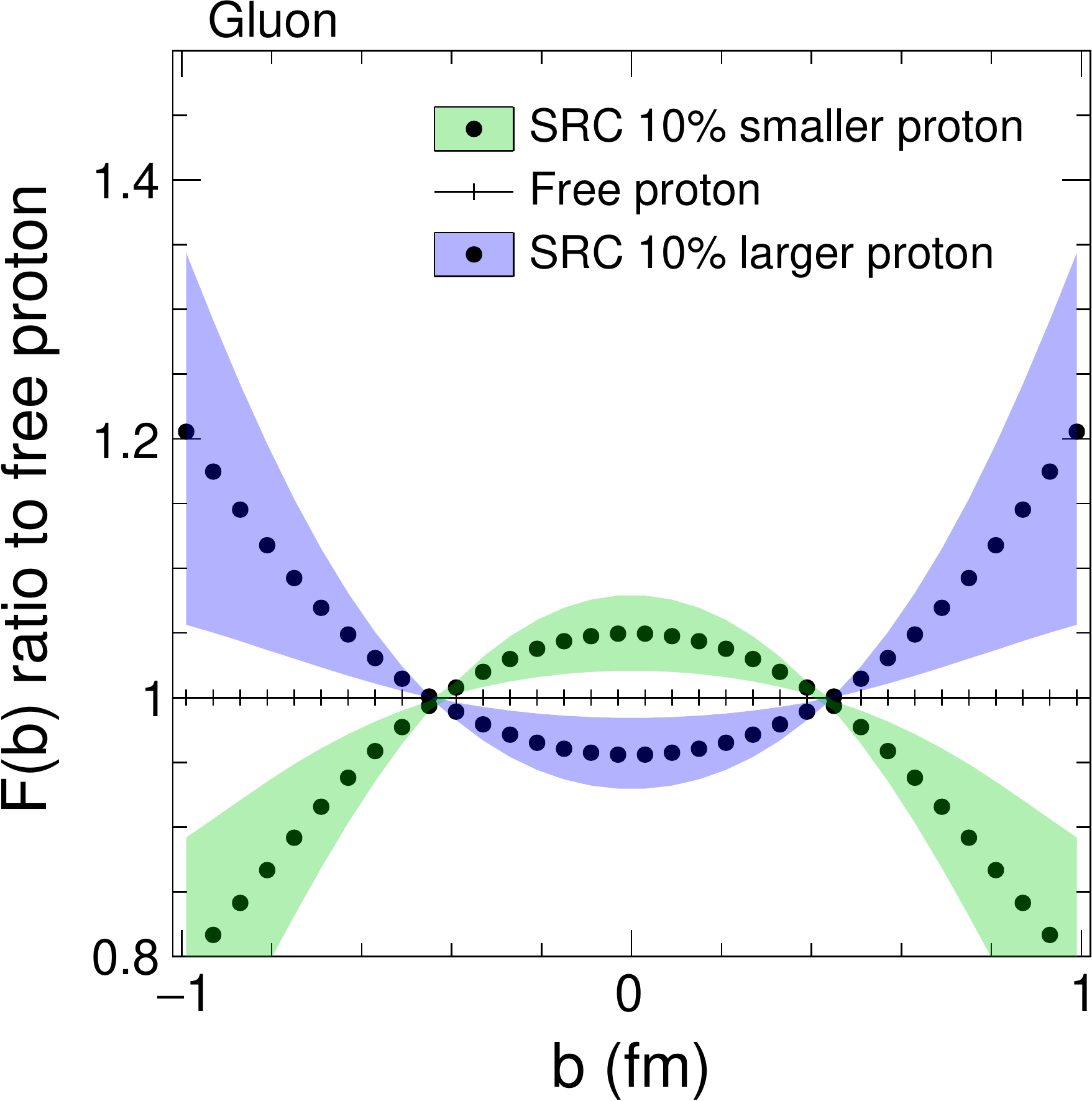}
\includegraphics[width=0.49\linewidth]{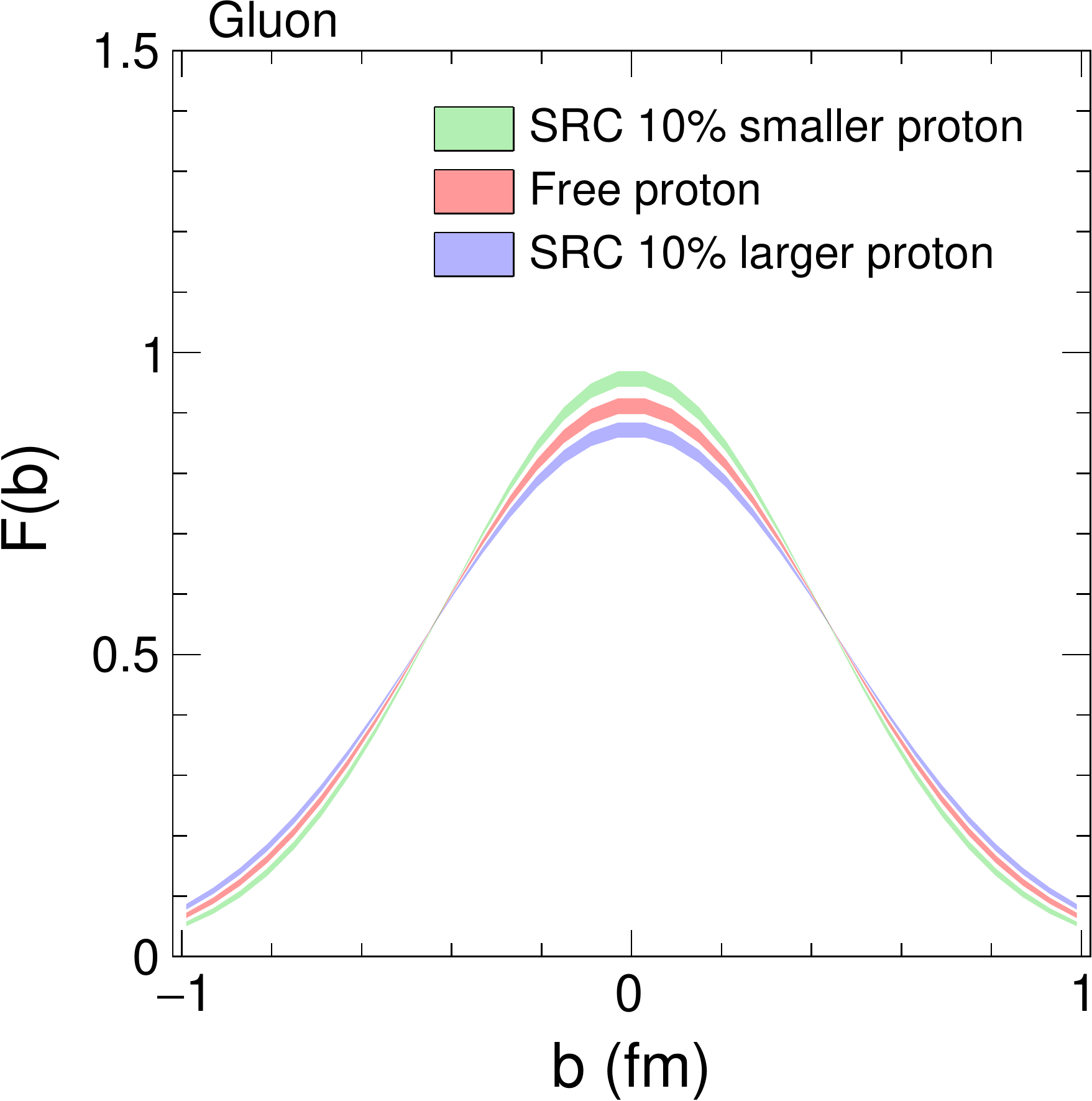}
  \caption{ \label{fig:figure_6} The gluon source distributions $F(b)$ (right) and their ratio between SRC protons and the free proton as a function of impact parameter $b$ (left), based on a Fourier transformation of the $t$ distributions of elastic $J/\psi$ production in $\gamma p$ collisions.  
  The color band indicates a 1$\sigma$ statistical uncertainty.  }
\end{figure}

In addition, by selecting different momentum range of the proton-neutron pair, the gluon density distributions can be compared. Based on Ref.~\cite{Tu:2020ymk}, a 10\% different size of proton in terms of gluon density is studied and is shown in Fig.~\ref{fig:figure_6}. The impact parameter distribution of the gluon density is based on the Fourier transformation on the momentum transfer $-t$ distributions, which can be measured up to high precision at the EIC. For details, see Ref.~\cite{Tu:2020ymk}. With the assumption of a similar statistical precision as obtained by the H1 result~\cite{Alexa:2013xxa}, 
the 10\% difference in the slope parameter of $-t$ will result in a 3$\sigma$ significant different source distribution. This difference will be dominated by the statistical uncertainty, while the systematic uncertainty will be largely, if not fully, canceled. Overall, the significance of the results depends on the signal strength and the statistical uncertainty. For a quantitative prediction, rigorous theoretical calculations are needed. 


\subsubsection{Studying short-range correlations with an EIC}

Understanding the modification of quarks in nucleons within nuclei (EMC effect) is a longstanding open question in nuclear physics~\cite{Higinbotham:2013hta}. Recent experimental results from electron scattering at Jefferson Lab strengthen the correlation between the EMC effect and nucleon-nucleon short-range correlated pairs (SRC) within nuclei~\cite{Weinstein:2010rt,Hen:2016kwk,Schmookler:2019nvf}. 
That means that the EMC effect is probably driven by the high-momentum highly-virtual nucleons of the SRC pairs. This connection can be tested experimentally by measuring electron deep inelastic scattering (DIS) from a nucleon and detecting its correlated SRC partner nucleon (tagging). 

The Electron-Ion-Collider (EIC) is an ideal machine for tagging measurements due to the unique capability of measuring recoil nucleons in a collider compared to fixed-target experiments. Furthermore, it will reach much higher $Q^{2}$ values than obtained in previous DIS measurements. The current design of the EIC detectors allows for a full acceptance for forward-going proton, neutrons and nuclear fragments besides the scattered electron. Ideally, it should be possible to measure the struck nucleon or its target-remnant jet, the SRC-partner, any spectators that were involved in final state interactions, and the nuclear remnant.

In the following figures~\ref{src-one} and \ref{src-two}, we show momentum distributions of the recoil nucleons determined by the electron and leading nucleon with the current IR design.   These results were generated using the Generalized Contact Formalism and then passed through EICROOT.   Presently head-to-head comparisons are being made between EICROOT and ESCalate's g4e codes.

\begin{figure}[htb]
    \centering
    \includegraphics[width =0.49\textwidth]{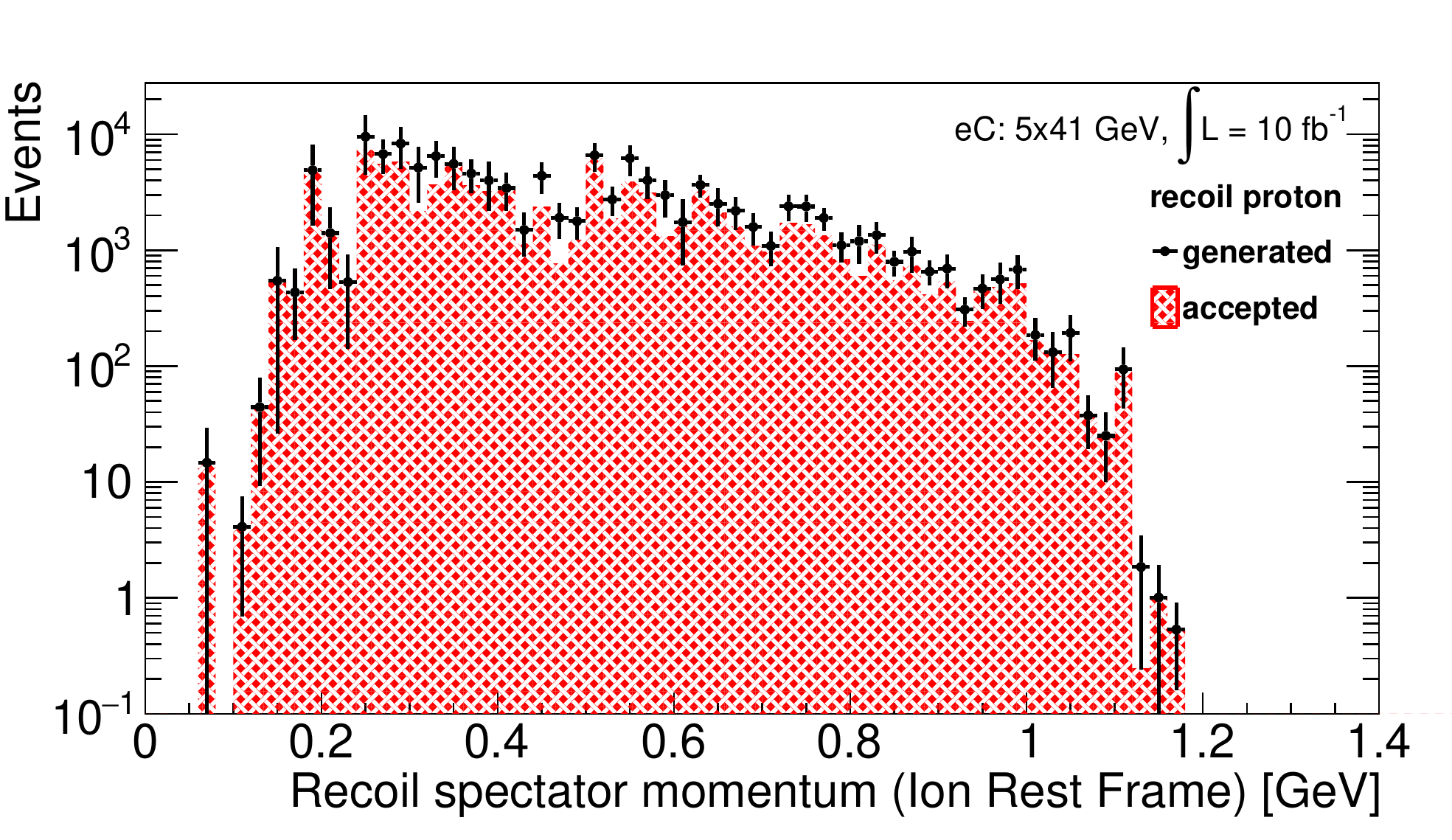}
    \includegraphics[width = 0.49\textwidth]{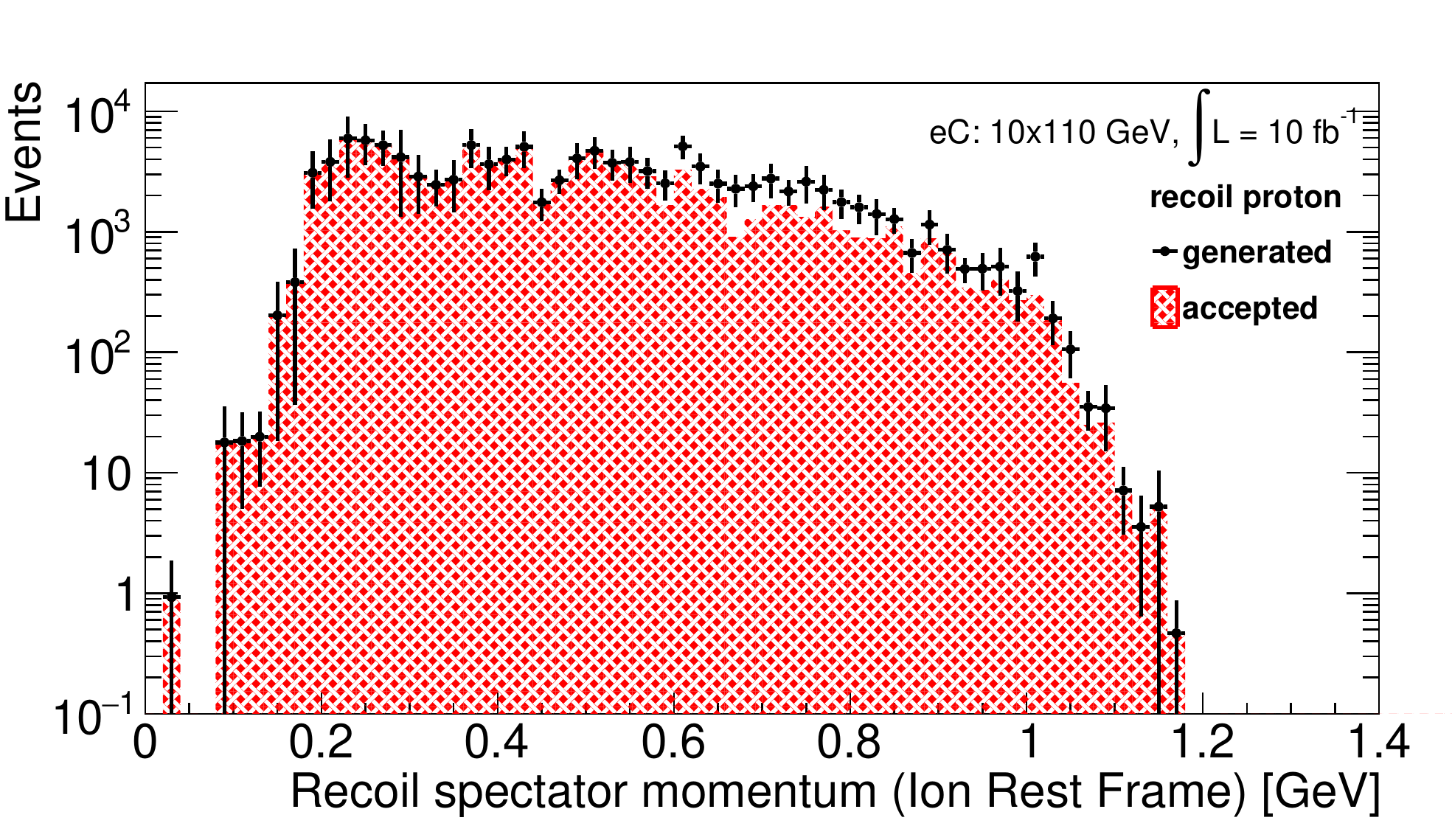}
    \caption{Momentum distributions for tagged recoil protons in quasi-elastic SRC breakup in $eC$ interactions. The black points show generated events for $10\,\mathrm{fb}^{-1}$ luminosity, and the accepted events are shown by the red-dashed histogram. (Left) Results for beam energies 5~GeV $e^{-}$ and 41~GeV ions. (Right) Results for beam energies 10~GeV $e^{-}$ and 110~GeV ions.}
    \label{src-one}
\end{figure}

\begin{figure}[htb]
    \centering
    \includegraphics[width =0.49\textwidth]{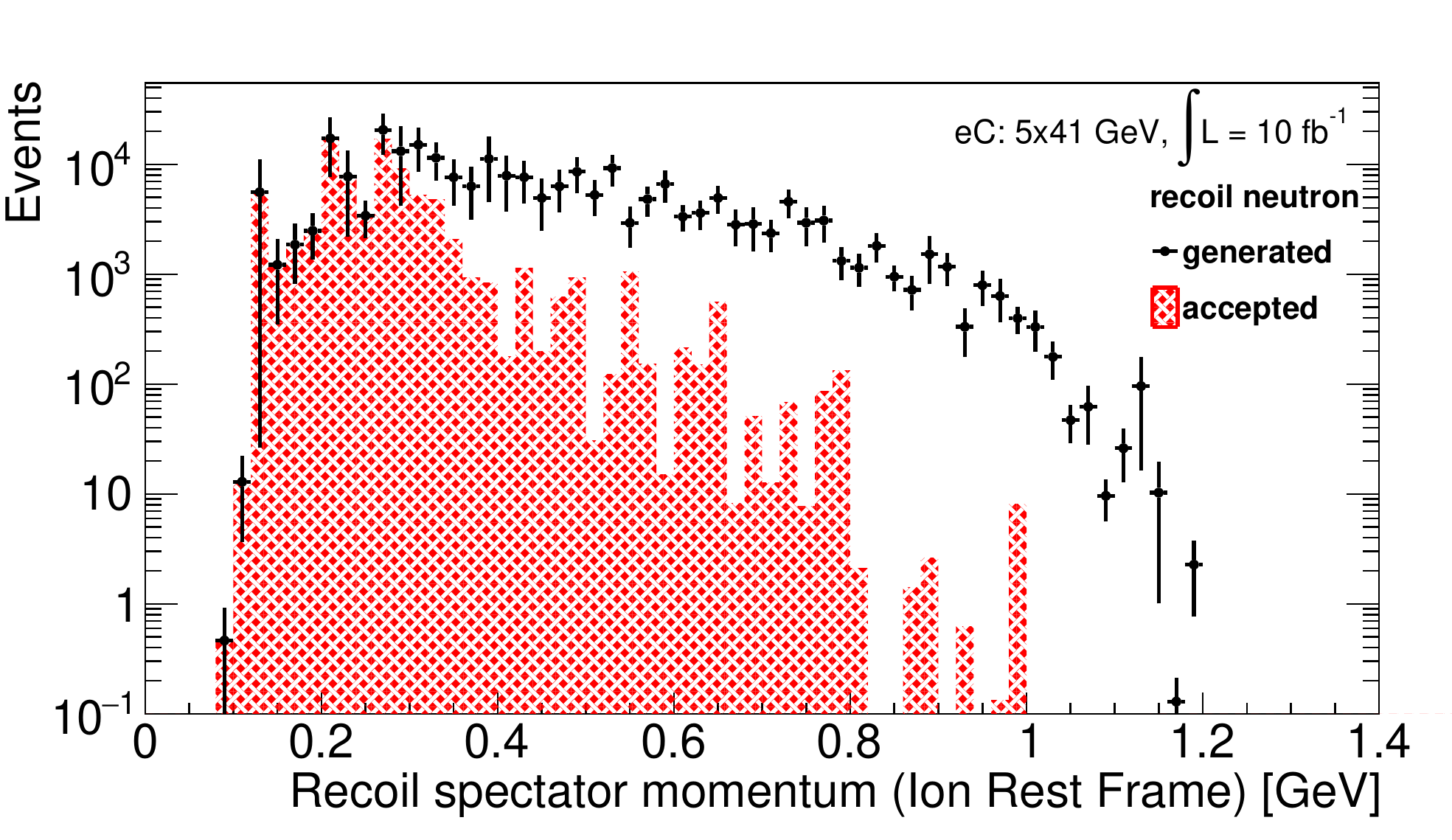}
    \includegraphics[width = 0.49\textwidth]{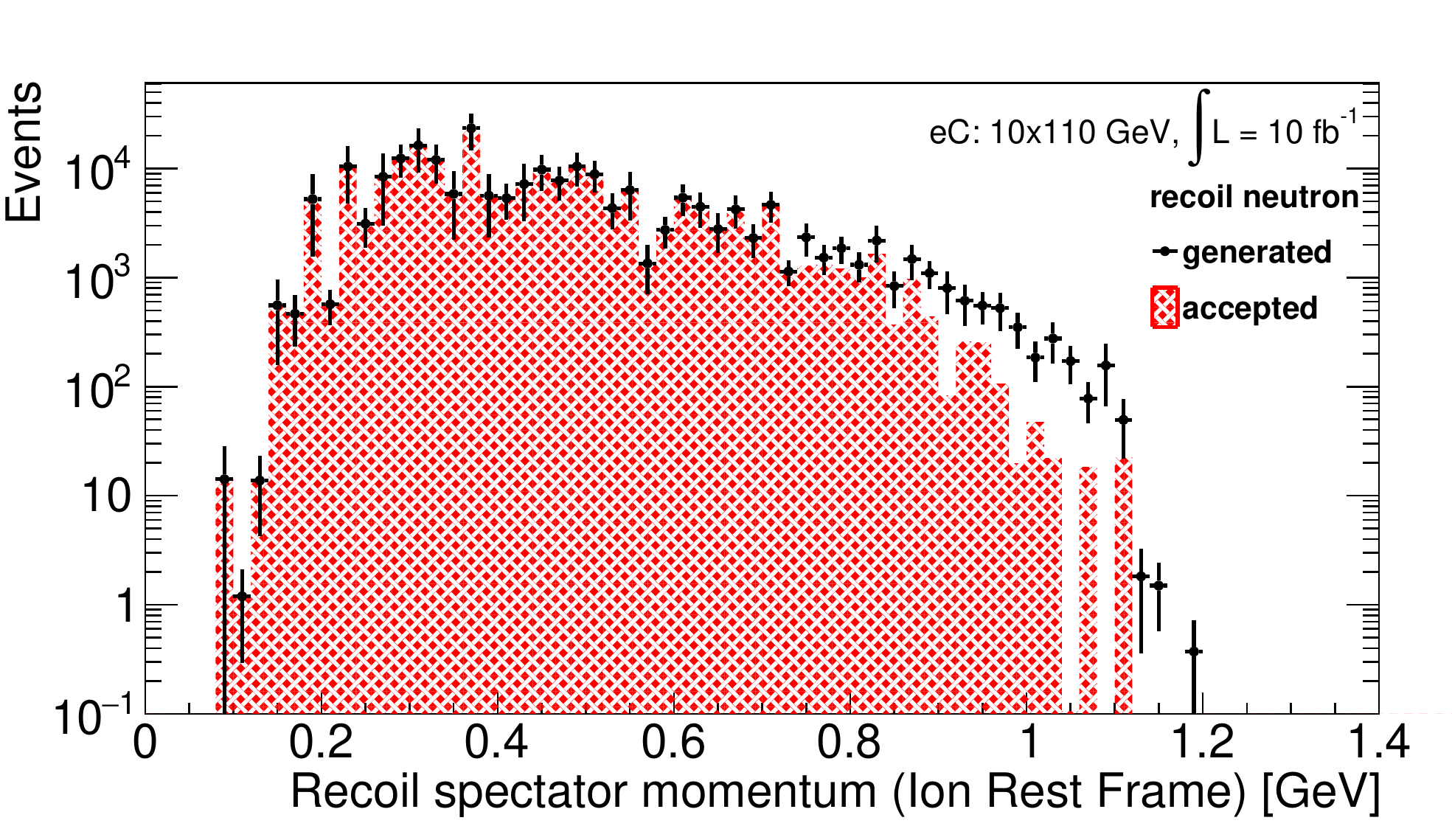}
    \caption{Momentum distributions for tagged recoil neutrons in quasi-elastic SRC breakup in $eC$ interactions. The black points show generated events for $10\,\mathrm{fb}^{-1}$ luminosity, and the accepted events are shown by the red-dashed histogram. (Left) Results for beam energies 5~GeV $e^{-}$ and 41~GeV ions. (Right) Results for beam energies 10~GeV $e^{-}$ and 110~GeV ions.}
    \label{src-two}
\end{figure}

The results clearly show very good acceptance for recoil spectator nucleons over a very large range of momentum.   For neutrons the acceptance is not as good for the lower energy setting though nearly complete at the highest energy.   This is simply the geometric effect of the size of the zero degree calorimeter along with kinematic focusing provided by the energy.


\subsection{Structure of light nuclei}
\label{part2-subS-LabQCD-LightNuclei}

The EIC, with its far forward detectors, provides a unique facility for studying light ions at high center of mass energies.
Light ions have several unique features that can be used to study the interplay between partonic QCD phenomena and nuclear interactions.
\begin{enumerate}
    \item Light ions can be polarized.  The EIC would allow for polarized $^3$He and $^3$H beams, possibly deuteron ($^2$H) beams and beyond.  This allows to probe neutron spin structure (see Sec.~\ref{part2-subS-SpinStruct.P.N}), tensor polarized deuteron (see Sec.~\ref{part2-subS-SecImaging-LpolNucl}) and measurements of the polarized EMC effect.
    \item The far forward detectors in the hadron going direction allow for the detection of specific nuclear breakup channels.  In inclusive processes scattering can take place on protons and neutrons, partonic structure can be modified by nuclear interactions and non-nucleonic d.o.f. play a role.  These effects can all be controlled by selecting particular break-up channels. Particularly effective breakup channels are the measurement of the so-called spectator nucleons, where one or more nucleons are detected in the target fragmentation region of the nucleus, see Fig.~\ref{fig:neutron_tagging} for the deuteron.  This allows to select the active nucleon in the reaction, suppress the contribution of non-nucleonic d.o.f. and select specific intra-nucleon distance scales in the initial nucleus.
    \item For light ions well-developed techniques exist to compute nonrelativistic nuclear wave functions from first principles, starting from microscopic NN interactions.  This makes it possible to describe the initial nuclear state and breakup into specific channels with high theoretical precision.
\end{enumerate}

In this subsection the focus is on the use of light ions in the study of nuclear interactions and their influence on medium modifications of parton distribution functions.  For free neutron structure, see Sec.~\ref{part2-subS-SpinStruct.P.N}; for 3D imaging of nuclear bound states, see Sec.~\ref{part2-subS-SecImaging-LpolNucl}.  

Nuclear interactions are effective interactions arising from QCD and describing the low-energy structure of nuclei using interacting nucleons has proven to be highly successful.  Several questions remain however.  How exactly do the effective NN forces arise from QCD? What are the short-distance properties of the nuclear interactions? Where do non-nucleonic degrees of freedom become manifest in nuclei?  Reactions where these questions can be addressed are quasi-elastic or diffractive knockout from nuclei (discussed in Sec.~\ref{part2-subS-LabQCD-ShortRange} in the context of nuclear short-range correlations) and DIS in the scaling regime.  

In the context of the influence of nuclear interactions on medium modifications, the wide $Q^2$-range available at the EIC means the $Q^2$-dependence of the EMC effect can be investigated.  DIS on polarized light ions results in measurements of the spin-dependent EMC effect, such as planned within the 12 GeV
program at Jefferson Lab. The magnitude of the spin-dependent contribution remains unknown so far.  In combination with spectator tagging the relevant distances in NN interactions that cause the EMC effect can be studied at the EIC.  Accurate descriptions of the reaction mechanisms are needed to disentangle the medium modification effects from nuclear final-state interactions~\cite{Kaptari:2013dma,Strikman:2017koc,Cosyn:2017ekf}.  This is especially important for spectators with momenta of a few hundred MeV (relative to the ion rest frame).  

In this way DIS on $d$ with neutron tagging, $^3$He with $d$ or $pn$ tagging and $^3$H with $nn$ tagging, in combination with free proton data can be used to study proton medium modifications and help to constrain reaction mechanism frameworks.  This in turn then will help in the disentanglement of medium modifications and final-state reactions for tagged DIS reactions on light nuclei accessing neutron structure ($d$ with proton tagging, $^3$He with $pp$ tagging and $^3$H with $d$ or $pn$ tagging).

For DIS at smaller values of Bjorken $x$, nuclear anti-shadowing and shadowing effects in light nuclei are of interest.  For the latter, light ions offer the advantage that the multiple scattering series is limited and thus controlled, see below for ${^4}$He.

\subsubsection{Studying nucleon structure in A=3 nuclei using double spectators tagging}

\begin{figure}[ht!]
    \centering
    \includegraphics[width = 0.49\textwidth ]{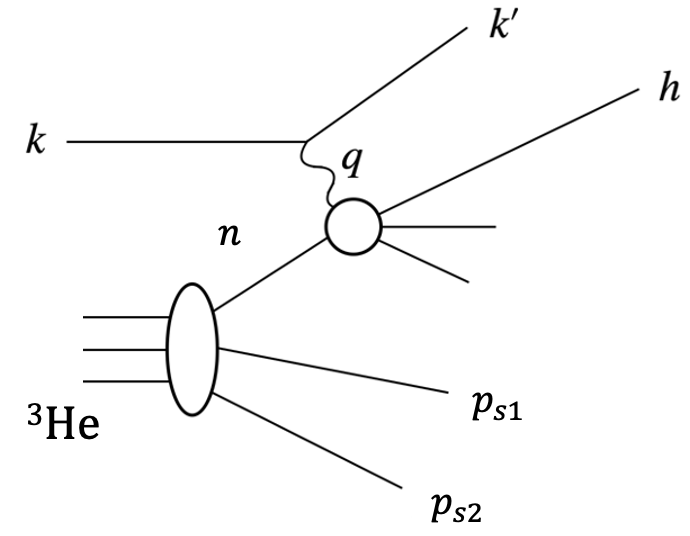}
    \includegraphics[width = 0.49\textwidth]{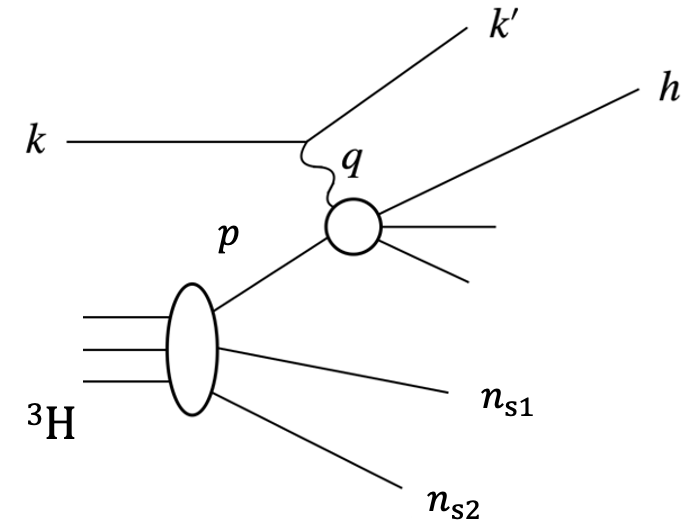}
    \caption{Shown are Feymann diagrams for double spectator nucleons tagging from $^3$He on the left and from $^3$H on the right.}
    \label{fig:3He_tagging_diagram}
\end{figure}

While there are highly accurate data for proton structure function, it is very hard to measure the neutron structure function due to the fact there is no free neutron target.   Neutron structure functions are determined using nuclear targets (deuterium or $^3$He) and often inferred using different theoretical models (see Secs.~\ref{part2-subS-UnpolPartStruct.P.N} and \ref{part2-subS-SpinStruct.P.N}).    Modern experiments, such as the Jefferson Lab BONuS experiment, tag the recoil proton from deuteron~\cite{Baillie:2011za,Fenker:2008zz}, 
but no such tagging experiment has been done with $^3$He (see Fig.~\ref{fig:3He_tagging_diagram} for a diagram).

Simulations of DIS and SIDIS scattering from $^3$He show that both spectator protons can be detected in the far forward region of the EIC. By ensuring that in the ion rest frame the
two spectator protons have a low momentum, the initial momentum of neutron can be constrained, which can provide an effective ``free neutron'' target with minimal model dependence, see Sec.~\ref{part2-subS-SpinStruct.P.N}. 
Taking this idea one step further, by using a tritium beam one can tag
two spectator neutrons.   Since in this case, the free proton's structure function is well known, one directly tests how well reaction mechanisms are understood (including final-state interactions with the detected spectators) and a unique way to test the validity of the $^3$He extractions of the neutron structure functions as well as offering a window onto nuclear medium modifications of proton partonic structure.  At higher spectator momenta spectator tagging yields information on the influence of nuclear interactions on DIS cross sections, both through medium modifications of partonic distributions (EMC effect, (anti)shadowing) and nuclear final-state interactions.

\begin{figure}[htb]
    \centering
    \includegraphics[width =0.49\textwidth]{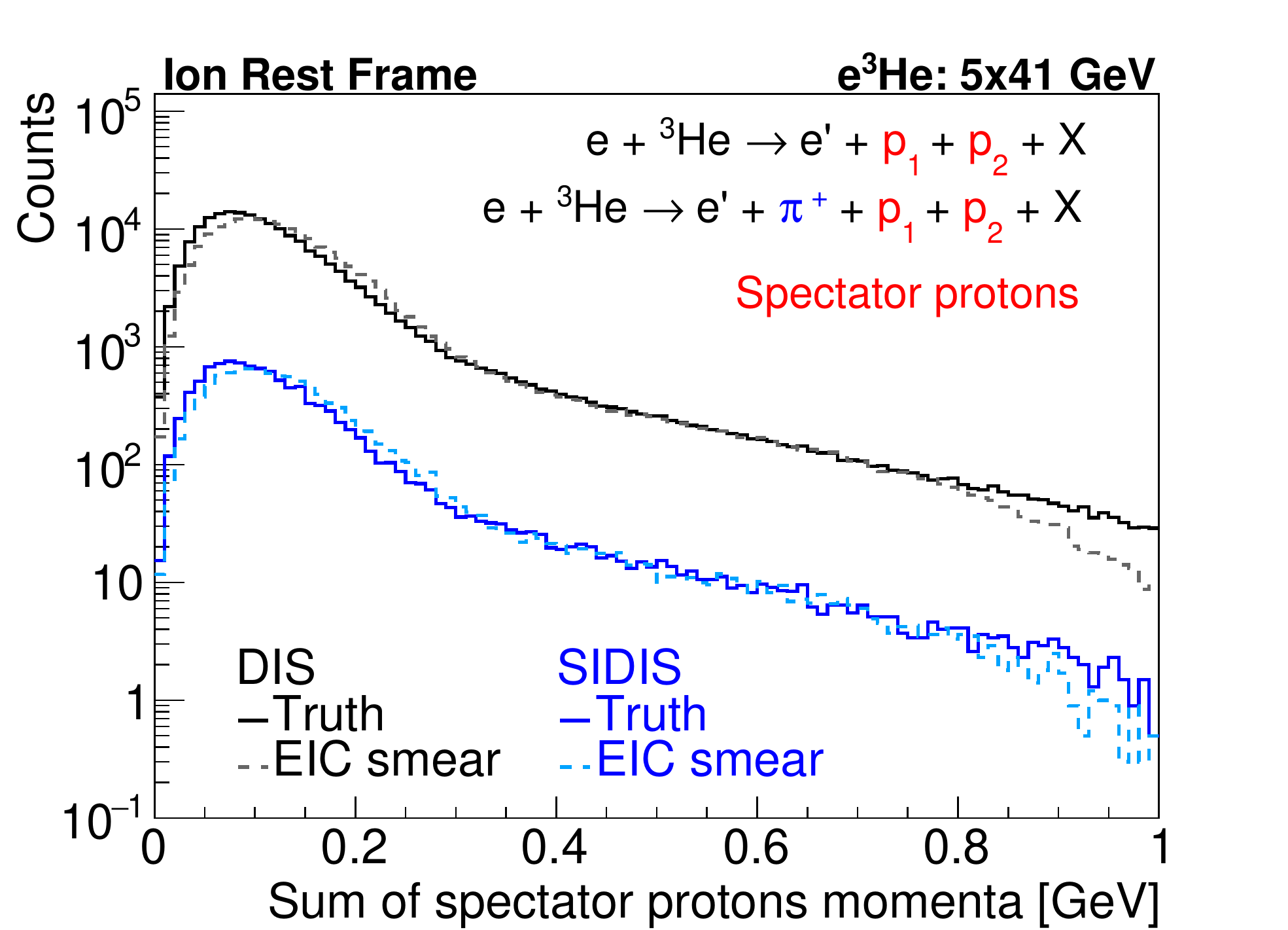}
    \includegraphics[width = 0.49\textwidth]{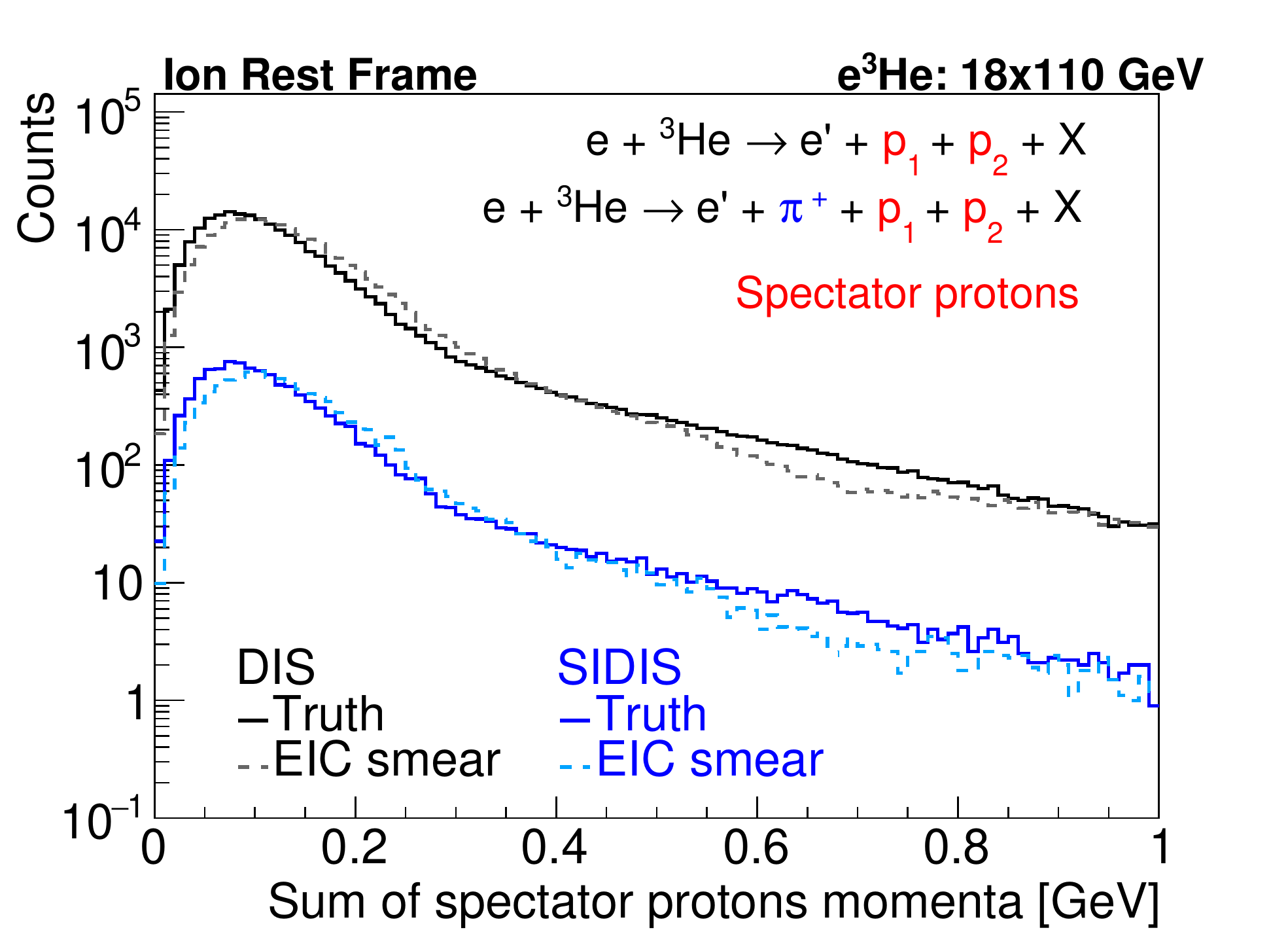}
    \includegraphics[width =0.49\textwidth]{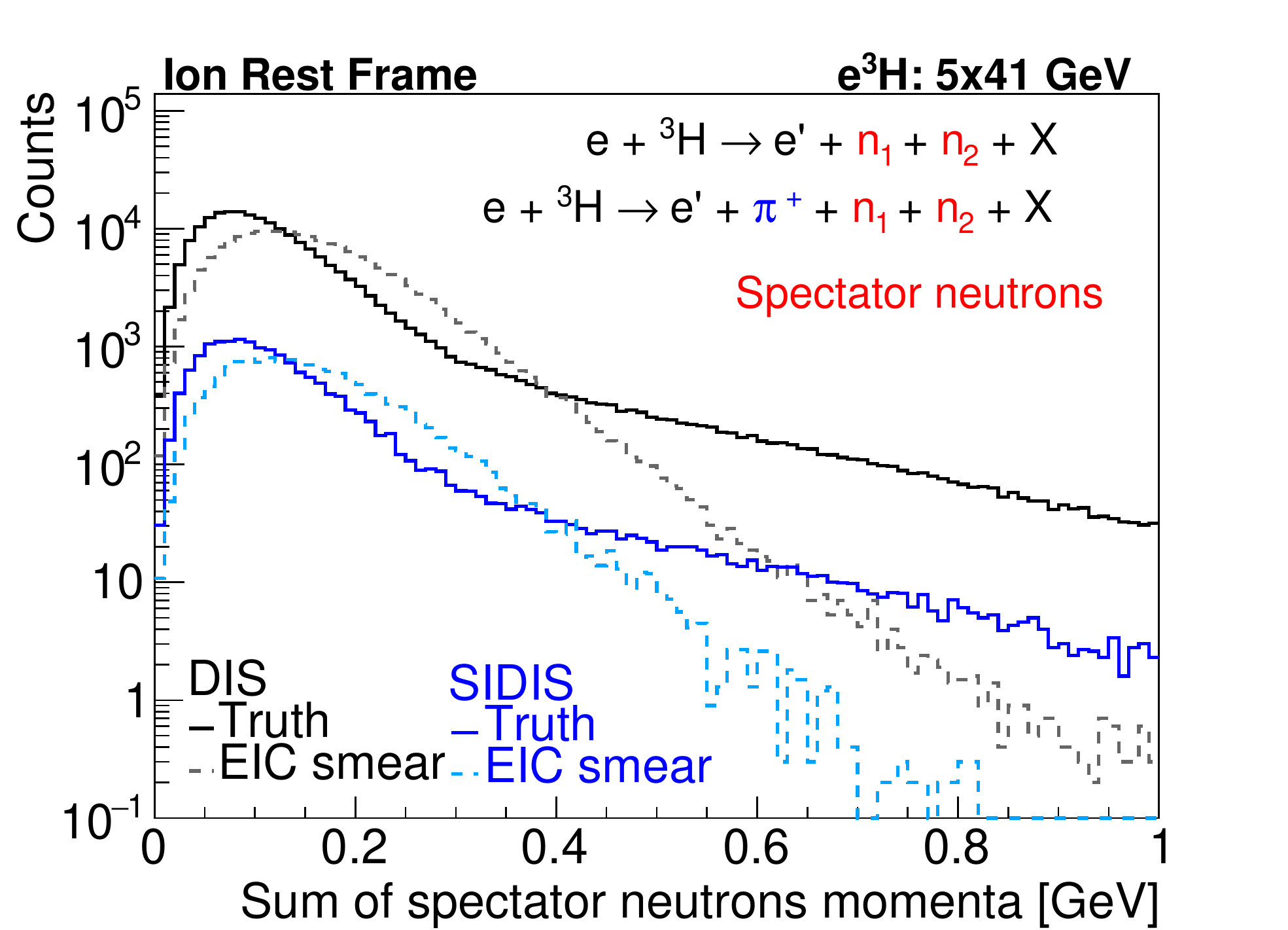}
    \includegraphics[width = 0.49\textwidth]{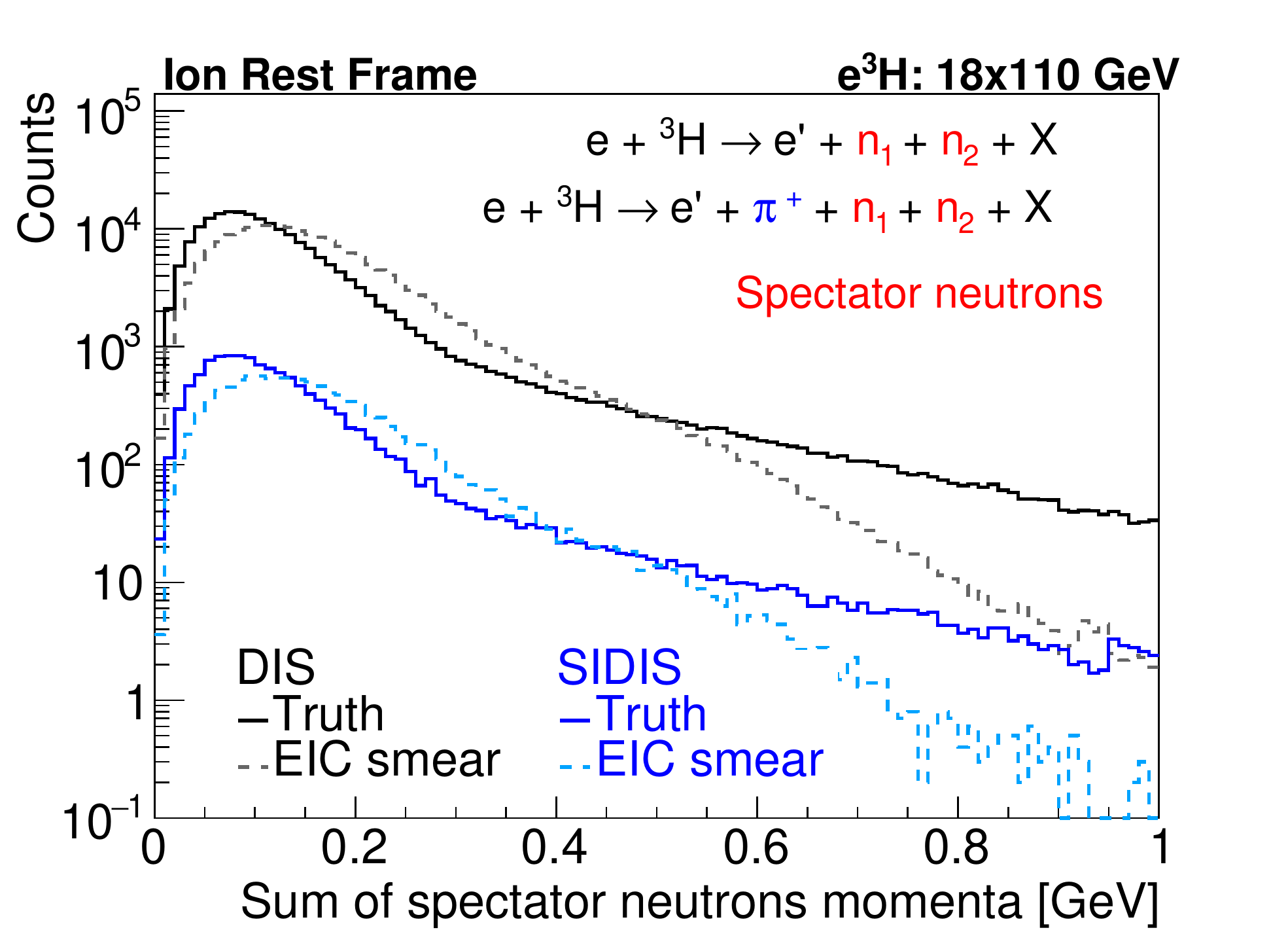}
    \caption{CLASDIS simulations for DIS and $\pi^+$ SIDIS on $^3$He with double proton spectator tagging (top) and on $^3$H with double neutron spectator tagging (bottom) at 5  on 41 GeV (left) and 18 on 110 GeV (right).}
\label{fig:He3-tagging}
\end{figure}

The DIS and SIDIS processes for both $^3$He and $^3$H targets were generated using the CLASDIS generator, a CLAS version of the PEPSI~\cite{Mankiewicz:1991dp} generator, which is based on LEPTO version 6.5 and JETSET version 7.410. CLASDIS was originally intended to be used for the fixed target event generation but has been extended for EIC kinematics. 
The generated events were modified to include the effects of nuclear motion. 
For a given event, the momentum of each nucleon was generated using the spectral function of Ref.~\cite{CiofidegliAtti:2004jg} (for leading nucleon momentum below 0.24 GeV) and the light-front Generalized Contact Formalism~\cite{Pybus:2020itv} (for higher-momentum leading nucleons). 
This provided distributions for the spectator nucleon momentum, as well as allowing adjustment of electron scattering quantities for Fermi smearing of the leading nucleon.
Results are shown in Fig.~\ref{fig:He3-tagging} and made use of a version of EICSMEAR which includes an approximation of the far forward region developed with EICROOT and GEANT4 for EIC (g4e). 
The results show that with the EIC one will be able to uniquely determine that the initial-state neutron was nearly at rest; minimizing model dependence for the extraction of quantities such as F$_2^n$ from unpolarized $^3$He and A$_1^n$ from polarized $^3$He.


Tagging of spectator nucleons in these reactions means final state interactions between the detected hadrons in the final state should be considered.   In Ref.~\cite{Kaptari:2013dma}, polarized DIS on $^3$He with deuteron tagging was considered.  This process would allow to study the onset of the spin-dependent EMC effect in three-body systems and can help to constrain FSI mechanisms in the deuteron tagged reaction on $^3$H, used to probe neutron structure.  Proper kinematical regions where the FSI effect is minimized were
identified in Ref.~\cite{Kaptari:2013dma}.  For spin dependent SIDIS off $^3$He and $^3$H  (important for the extraction of the neutron Sivers and Collins single-spin assymetry), a recent paper \cite{DelDotto:2017jub}  has shown that both at fixed target and the EIC,
the FSI described within a generalized eikonal approximation
using AV18~\cite{Wiringa:1994wb} wave functions are theoretically
under control in the experimental observables.  Neutron information can be safely extracted
using the same straightforward procedure proposed in a plane-wave impulse approximation analysis \cite{Scopetta:2006ww}.

\subsubsection{Coherent scattering off the lightest nuclei}


The leading twist theory of the gluon shadowing predicts shadowing both for the pdfs and for coherent production of heavy mesons like $\jpsi$. The theory predictions for coherent $\jpsi$ production off  lead  were tested in ultraperipheral collisions at the LHC  for $x$  down to $ 10^{-3}$. For the most recent comparison of the  leading twist theory of nuclear shadowing with the LHC data on coherent $\jpsi$ production of Pb and reference to the previous studies see Ref.~\cite{Guzey:2020ntc}.  So far the gluon shadowing was studied only in the case of heavy nuclei. At EIC studies of the coherent diffraction off heavy nuclei and parallel measurements of the gluon pdfs in the inclusive hard processes would provide further stringent tests of the gluon shadowing dynamics. A complementary set of measurements is possible at the EIC using the beams of the  lightest nuclei. In this case one can probe separately shadowing for scattering off two and of three nucleons. Note here that the average number of nucleons involved in the gluon shadowing at  $Q_0^2 \sim \text{few} \gev^2, x=10^{-3}$  is around two, so the measurements with heavy and light nuclei  would nicely complement each other.

\begin{figure}[ht]
\begin{center}
\includegraphics[width=0.45\textwidth]{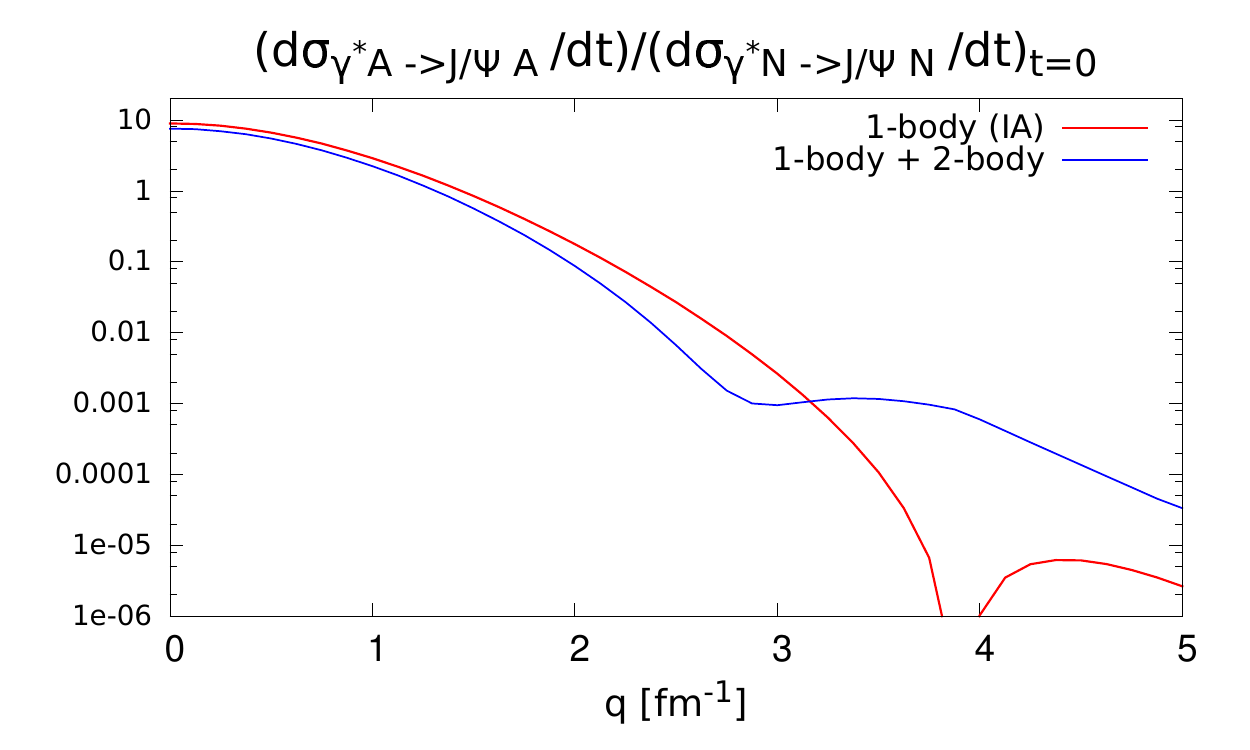}
\includegraphics[width=0.45\textwidth]{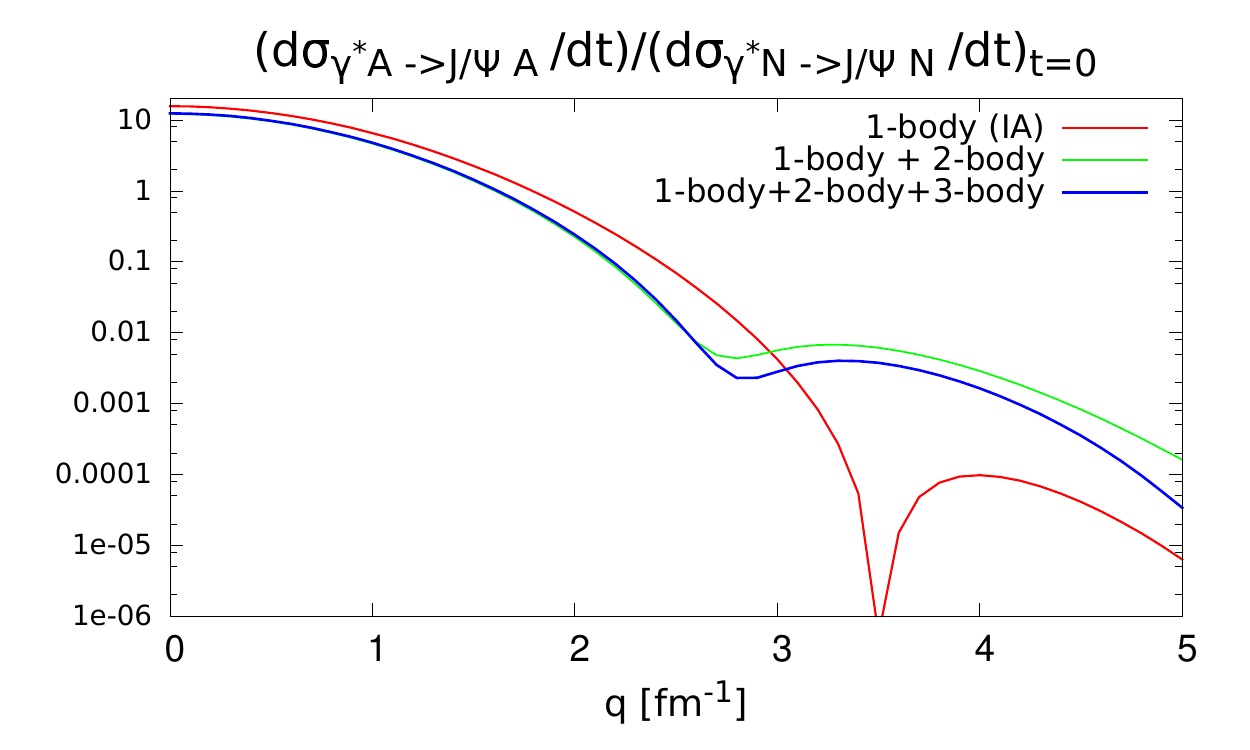}
    \caption{\label{Sec3.8_coh4He}
Coherent $\jpsi$ production on $^3$He (left panel) and $^4$He (right panel) at $x=10^{-3}$.  The cross section ratio to the $t=0$ value for production on the nucleon is shown, as a function of $q=\sqrt{-t}$.  Red curves do not include rescattering effects, green (blue) curves include double (triple) rescatterings.}
\end{center}
\end{figure}

An important advantage of the lightest  nuclei is that their wave functions are well known and so the impulse approximation term can be  reliably calculated. Also it would be possible to test the calculation  of the wave function by studying the cross section at $x \lesssim 0.03$ where rescattering effects are  small even for heavy nuclei and the impulse approximation dominates at all $t$.  An important advantage of $^4$He, $^3$He  is that the single scattering term (which is proportional to the nuclear form factor)  goes through zero at moderate $t$ ($-t\approx 0.3 \gev^2$ for $^4$He; see red line in Fig.~\ref{Sec3.8_coh4He}).
  This provides an opportunity to  separate the combination of double and triple rescattering amplitudes (interaction with all four nucleons is negligible) in a wide $t$ range.  Moreover combination of measurements off $^3$He and $^4$He   would allow to separate double and triple scattering amplitudes  in a practically model independent way due to the difference of the strengths of double and triple rescattering contributions in these two cases.

One can see from the result of the calculation at $x=10^{-3}$ the shift of the  position of the minimum is very significant and hence would be easy to measure provided  the detector has an acceptance in the discussed t-range ($-t\lesssim 0.5 \gev^2$)

In the case of the deuteron beams the rescattering effects are pretty small even for $-t\sim 0.5 \gev^2$ due to a large contribution of the quadrupole form factor. However, if the polarized deuteron beams would become available, one would be able to separate single and double rescattering contribution and consequently  further improve extraction of the amplitudes of scattering off two  and three nucleons.

There are many other interesting coherent reactions with  $^4$He, For example photo and electroproduction of  $\rho$-mesons. In the soft regime   screening should be stronger than for $\jpsi$  so the minimum of the differential cross section would be at smaller $-t$, up to $Q^2\sim 10 \div 15 \gev^2$  where the $t$-dependence of $\rho$-meson should approach that of $\jpsi$.

\subsection{Coherent and incoherent photoproduction on heavy targets}
\label{part2-subS-LabQCD-Photo}

Photoproduction and electroproduction are sensitive to the gluon content of the target nucleus, and provide information on how quark-antiquark dipoles interact in nuclei, providing information on the nuclear structure, as is discussed in Section 7.3.2.  This section will present the additional physics that can be probed by studying coherent and incoherent production separately. 

Exclusive production is an excellent probe of the gluon distributions in the nucleus, providing information on the overall gluon content through the overall cross-section, the spatial distribution of the gluons (through $d\sigma/dt$ for coherent production) and on event-by-event fluctuations in distributions, including gluon hot spots through inelastic production \cite{Mantysaari:2016ykx,Cepila:2016uku,Klein:2020fmr}.  

\begin{figure}
\centering
\includegraphics*[width=0.5\textwidth]{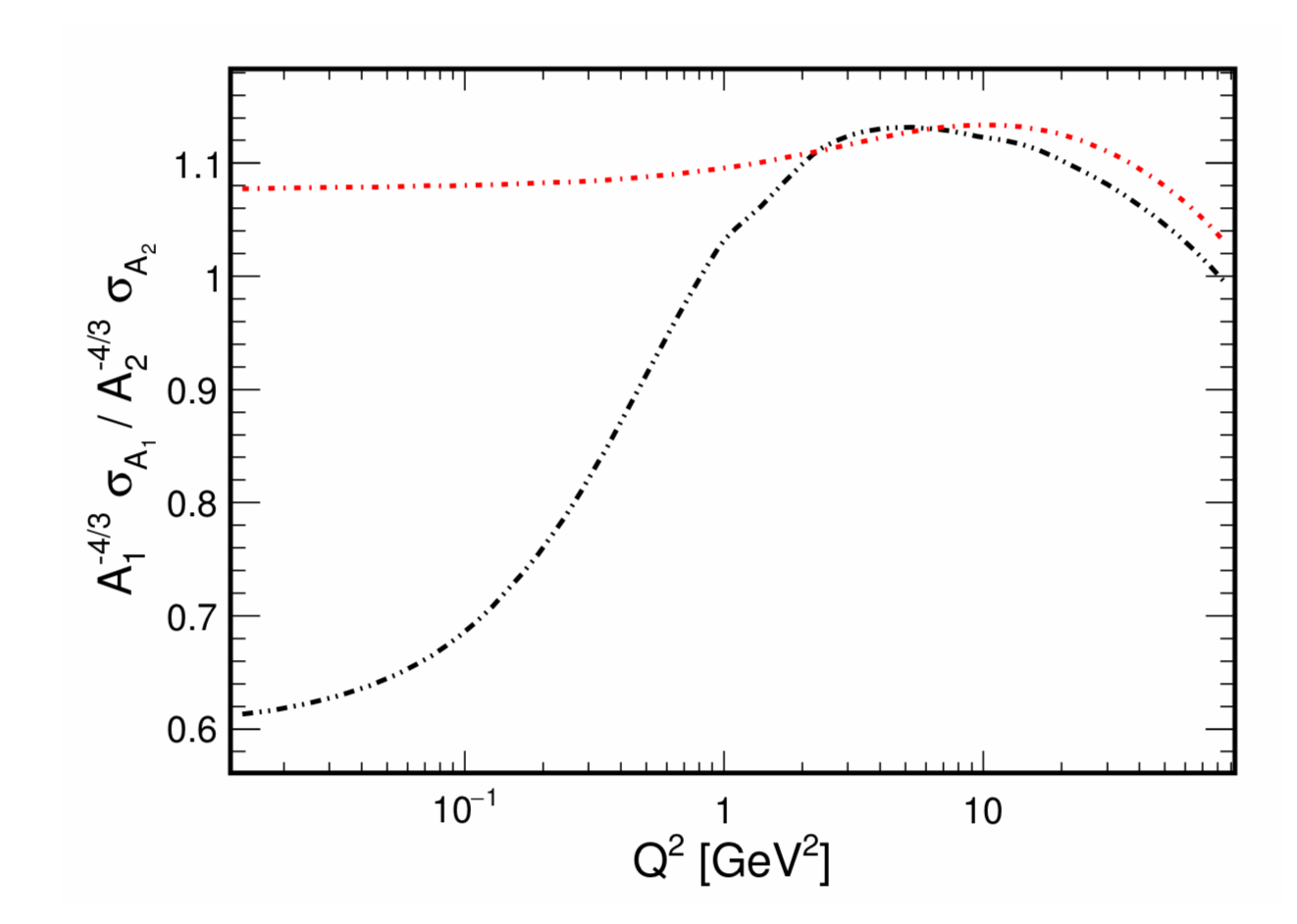}
\caption{The ratio of the coherent photoproduction cross-sections on lead and iron targets, scaled by $A^{-4/3}$, for  $\rho$ (black) and $J/\psi$ (red), as a function of $Q^2$. In the absence of nuclear effects, this ratio should be 1.  This ratio is roughly reached at high $Q^2$, when the nucleus is largely transparent, but, as the $Q^2$ drops, shadowing reduces the relative cross-section. From Ref. \cite{Lomnitz:2018juf}, but a very similar plot appears in Ref. \cite{Mantysaari:2017slo}.}
\label{fig:VMshadow}
\end{figure}

High-energy photoproduction and electroproduction cross-sections on proton targets were first studied extensively at HERA, where the cross-sections were measured for a variety of mesons, from the $\rho$ up to the $\Upsilon$ \cite{Crittenden:2001tn,Newman:2013ada}.  The photoproduction studies were extended to include heavy nuclear targets in ultra-peripheral collisions (UPCs) at RHIC \cite{Bertulani:2005ru} and the LHC \cite{Klein:2020fmr,Baltz:2007kq,Contreras:2015dqa}. The LHC UPC data \cite{TheALICE:2014dwa, Acharya:2018jua, Sirunyan:2019nog, Sirunyan:2018sav, Aaij:2018arx ,Aaij:2014iea} had a larger energy reach than HERA.   The LHC data exhibited a moderate suppression of the $J\psi$ cross-section, consistent with moderate gluon shadowing - roughly as predicted by leading order twist calculations \cite{Acharya:2019vlb,Khachatryan:2016qhq,Abelev:2012ba,Abbas:2013oua}.  Although the energy reach cannot match the LHC, the EIC will allow us to extend these measurements to large $Q^2$, and also, through vastly improved statistics, probe the production kinematics in detail, exploring the photon energy and $Q^2$ evolution of the nuclear targets.  As Fig. \ref{fig:VMshadow} shows, the ratio of photoproduction on light and heavy targets, scaled by $A^{-4/3}$ should be one in the absence of nuclear shadowing.  At large $Q^2$, a Glauber calculation (implemented in the eSTARlight Monte Carlo) predicts that the ratio should be close to one.  However, at lower $Q^2$, lighter mesons are expected to exhibit a lower ratio, reaching 0.6 for $\rho$ photoproduction.  In contrast, the heavier $J/\psi$ is predicted to exhibit little shadowing.  This plot shows the importance of studying light mesons, and covering a range of $Q^2$ extending down to zero. 

\subsubsection{Photoproduction and partonic structure}

At high energies, photoproduction occurs primarily via photon-Pomeron fusion.  In an optical approach, the Pomeron represents the absorptive part of the nuclear potential \cite{Sergeenko:2011kf}, so it has the same quantum numbers as the vacuum.  Therefore, the final state particles predominantly have the same quantum number as the photon, $J^{PC}= 1^{--}$, leading to vector meson dominance.  In lowest order pQCD, the Pomeron is treated as two-gluon exchange, thus the forward scattering cross-section for a vector meson with mass $M_V$ is \cite{Jones:2013pga}
\begin{equation}
\frac{d\sigma}{dt}\bigg|_{t=0} = \frac{\Gamma_{ll}M_V^3\pi^3}{48\alpha}
\bigg[\frac{\alpha_S({\overline Q}^2)}{\overline Q^4}xg(x,\overline Q^2\big)^2\big(\frac{1+Q^2}{M_V^2}\big)\bigg],
\label{eq:lopqcd}
\end{equation}
where $\Gamma_{ll}$ is width for the vector meson to decay to two leptons, $g(x,Q^2)$ is the gluon distribution and $\overline{Q}^2=(Q^2+M_V^2)/4$.  Because the vector meson mass provides a hard scale, for heavy quarkonium pQCD should hold, even for nearly real photons.  

There are several issues with this approach.  The most fundamental is the treatment of this 2-gluon exchange using parton distributions which, in rigorous theory, should be treated using a GPD \cite{Diehl:2009ut}.  Treatment using standard lowest-order pQCD introduces some theoretical issues:  the assumption of two gluons each carrying half the virtuality, the presence of two gluons with different $x$ values for the two gluons (fortunately asymmetric division, $x_1\gg x_2$ is preferred) and the choice of the mass scale \cite{Klein:2017vua,Flett:2019ept}.  

Similar difficulties apply in NLO pQCD.   Although there is not yet a complete NLO calculation. Ref. \cite{Jones:2016ldq} presents a partial NLO calculation.  This calculation has its own difficulties, but, with careful choice of scale, the accuracy seems good enough ($\pm15\%$ to $\pm 25\%$), good enough for use in determining gluon distributions \cite{Flett:2019pux}. 

Most of the interest in using vector meson production to probe gluon distributions has involved nuclear targets, since there is little data with nuclear targets at low $x$.  By using proton targets as a reference, it is possible to measure shadowing, with many of the theory uncertainties cancelling out in the ratio.  This has been done for $J/\psi$ photoproduction at the LHC, where the data clearly indicate moderate shadowing \cite{Abelev:2012ba,Khachatryan:2016qhq,Acharya:2021ugn}, with uncertainties that are much smaller than in current nuclear pdf parameterizations \cite{Eskola:2016oht}.  At the EIC, a comparison of vector meson production in \eA and \ep collisions able to probe gluons at $Q^2$ than other reactions, such as open charm and dijet production. 

The EIC should get large samples of the $J/\psi$ and $\psi'$, over a wide range of $Q^2$ and significant samples of photoproduction of the three $\Upsilon$ states (at small $Q^2$).  The rapidity $y$ of the final state vector meson depends on the photon energy $k$ and Bjorken-x of the struck gluon. In the lab frame, for low $Q^2$ and $x \ll 1$\cite{Klein:1999qj}
\begin{equation}
    k=\frac{M_V}{2}\exp{(y)}
\end{equation}
and 
\begin{equation}
    x=\frac{M_V}{4\gamma m_p}\exp{(-y)}.
    \label{eq:rapiditytox}
\end{equation}
Large $Q^2$ will shift the relationship between $y$ and $x$ these slightly \cite{Lomnitz:2018juf}.  These equations can be used to determine the $x$ of the struck gluons accurately, subject to the issues that come from the second gluon.

Other calculations use a dipole approach to determine the cross-section and $d\sigma/dt$ \cite{Kowalski:2006hc}.  They find the cross-section by integrating the interaction probability over $d^2b$.  In this approach, it is easy to introduce modified parton densities, such as those expected in a colored glass condensate \cite{Armesto:2014sma}. 

The rates for vector meson production are very high: 40-50 billion $\rho^0$ for 10 fb$^{-1}$/A luminosity, for both \ep and \eA 
\cite{Lomnitz:2018juf}.  The rates for the $\phi$ are smaller, about 2.5 billion per 10 fb$^{-1}$/A luminosity, while the $J/\psi$ rates are about 100 million events/year. Even the $\Upsilon$ states are accessible, with about 140,000/60,000 events expected for $ep$ and \eA respectively.   The rates for electroproduction $Q^2>1$ GeV$^2$ are lower, but, even for the $J/\psi$, about 5 million events events with $Q^2>1$ are expected per 10 fb$^{-1}$/A.   This statistics will permit detailed multi-dimensional studies, including studying the nuclear shape and fluctuations in narrow bins of $x$ and $Q^2$.

\subsubsection{Good-Walker and coherent photoproduction}

Further information can be extracted from vector meson production using  the Good-Walker paradigm \cite{Good:1960ba}.  It relates the coherent cross-section to the average wave function of the target, and the incoherent cross-section to wave function fluctuations \cite{Klein:2019qfb}.  In the Good-Walker approach, coherent production occurs when the final state nucleus remains in it's ground state, while incoherent production is when it is excited.  The total cross-section includes all possible final states. From the optical theorem:
\begin{equation}
\frac{d\sigma_{\rm tot}}{dt} = \frac{1}{16\pi} \big<|A(\Omega)|^2\big>
\label{eq:tot}    
\end{equation}
The cross-section is determined by summing the amplitudes for interacting on each nucleon.  Schematically,
\begin{equation}
\frac{d\sigma_{\mathrm{coh}}}{dt} = \frac{1}{16\pi} | \langle A(\Omega) \rangle|^2.
\label{eq:co}
\end{equation}
Here, $\Omega$ is the nuclear configuration (nucleon positions, subnucleonic fluctuations etc.).   This leads to an $A^2$ enhancement in $d\sigma_{\mathrm coh}/dt$ at small $t$.  This enhancement applies as long as coherence is maintained, with $|t| < (1/R_A)^2$ where $R_A$ is the nuclear radius.  In the absence of multiple interactions ({\it i. e.} for a small $\gamma p$ cross-section), the coherent cross-section scales as $A^{4/3}$.

The incoherent cross-section is just the difference between the total and coherent cross-sections, {\it i. e.} the difference between the cross-sections in Eqs. \eqref{eq:tot} and \eqref{eq:co}
\begin{equation}
\frac{\mathrm{d}\sigma_{\rm inc}}{\mathrm{d}t} = \frac{1}{16\pi} \bigg(\left\langle \big|\mathcal{A}(K,\Omega)\big|^2 \right\rangle - \big|\left\langle \mathcal{A}(K,\Omega)\right\rangle \big|^2\bigg)
\label{eq:inc}
\end{equation}
Because of the switched ordering of the squaring and averaging, the incoherent process is sensitive to fluctuations in the nuclear configuration. A more formal treatment of this approach is given in Ref.~\cite{Caldwell:2010zza}.

\subsubsection{Coherent photoproduction}

Measurements of $d\sigma/dt$ for coherent photoproduction can be used to image the nucleus; this is the gluonic nuclear equivalent of a GPD.
It maps out the transverse positions of photoproduction interactions within the target nucleus.  As was discussed in the EIC White Paper \cite{Accardi:2012qut}, this may be determined from the two-dimensional Fourier transform of $d\sigma/dt$ \cite{Toll:2012mb,Klein:2018grn}
\begin{equation}
F(b) \propto \int_0^\infty p_T dp_T J_0(bp_T)\sqrt{\frac{d\sigma}{dt}}
\label{eq:bessel}
\end{equation}
where $J_0$ is a Bessel function and $b$ is the impact parameter within the nucleus.   The $\sqrt{}$ converts  the cross-section into an amplitude.  The square root introduces a sign ambiguity.  The positive root is taken at $p_T=0$, and then the amplitude flips signs at each diffractive minimum.  This sign flip needs to be included when analyzing the data.  The detector resolution and photon $p_T$ limit how well the positions of these minima can be determined, limiting the accuracy of $F(b)$. 

\begin{figure}
\centering
\includegraphics*[width=0.5\textwidth]{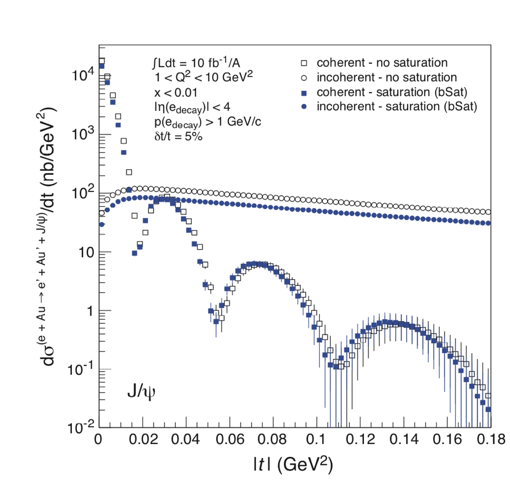}%
\includegraphics*[width=0.5\textwidth]{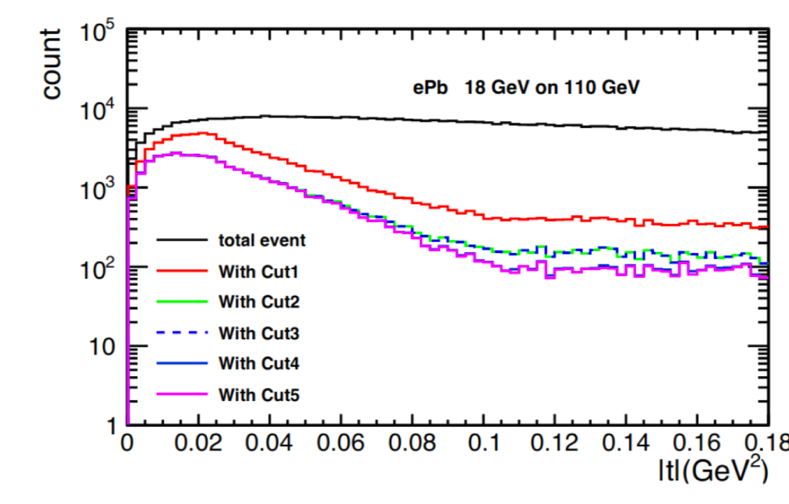}
\caption{(left) $d\sigma/dt$ for coherent and incoherent $J/\psi$ photoproduction with and without saturation.  From Ref. \cite{Accardi:2012qut}.  (right) $d\sigma/dt$ for incoherent $J/\psi$ photoproduction in the BeAGLE Monte Carlo (black).  The other lines show the effect of successive cuts that there are no neutrons (red), no photons with energy above 50 MeV (green), no protons in the draft Roman pot detector or off-energy detector, and no protons in the B0 forward detector, which is discussed in Section 8.5.2. }
\label{fig:dsdt}
\end{figure}

The STAR Collaboration has applied this approach to $\rho$ and direct $\pi^+\pi^-$ photoproduction in ultra-peripheral collisions and found some limitations in the method \cite{Adamczyk:2017vfu,Klein:2018grn}:

The $p_T$ integral runs from 0 to $\infty$, but the data is limited to a maximum $p_T$.   Unless the $p_T$ range encompasses several diffractive minima (up to $p_T\approx 15\hbar/R_A$), $F(b)$ will not fully capture the shape of the nucleus.  In the upper half of this $p_T$ range, incoherent production dominates over coherent, so good separating power is required.
This will be discussed in more detail later.

The measured $d\sigma/dt$ includes contributions from the photon $p_T$ and the experimental resolution.  Naively, one might expect to remove the photon $p_T$ by measuring it via the outgoing lepton.  However, this only works for large $Q^2$, where the scattered electron is detected.  Even if the electron is detected, the momentum spread of the electron beam limits the accuracy with which the electron $p_T$ can be determined.  The experimental resolution and remaining photon $p_T$ can be removed by an unfolding, but this is problematic near the diffractive minima.  Resolution and the photon $p_T$ will partially fill in these minima, and unfolding does not do a good job of restoring narrow spectral features.  This may limit how well the location of the minima can be found.  This will limit the accuracy of the placement of  the sign flips in Eq. \ref{eq:bessel}, reducing the accuracy of the measured $F(b)$.

Finally, $p_T$ is two-dimensional, but the $t$ in $d\sigma/dt$ is a four-vector, with three spatial components.  The former implies a cylindrical geometry (usually treated as a flat disk) while the latter implies spherical symmetry. Although the two cases are related by a Lorentz transform, they differ by the inclusion of (or neglect of) the longitudinal momentum transfer from the target.  This is unimportant for lighter mesons, where the longitudinal component is small, but it is a factor for heavier mesons like the $\Upsilon$.  The two geometries lead to slightly different predictions for impact-parameter dependent shadowing \cite{Emelyanov:1999pkc}.

More study is needed to understand the severity of all of these issues.

\subsubsection{Incoherent photoproduction}

The incoherent component of photoproduction is sensitive to event-by-event fluctuations in the target configuration, including variations in the positions of the individual nucleons, and partonic density fluctuations.
Equation \ref{eq:inc} can be used to test models of nucleon parton fluctuations, with small $t$ corresponding to long distance scales, and larger $t$ probing shorter range fluctuations.  However, $d\sigma/dt$ for incoherent production cannot be used to directly extract fluctuation measures.

The Good-Walker approach has been used to probe the proton using 
$J/\psi$ production data at HERA; predictions for incoherent photoproduction on a variety of nuclear targets at the EIC have been presented in ref.~\cite{Goncalves:2020ywm}. The analysis of HERA data found that the incoherent cross-section was compatible with a proton model where there were large event-by-event fluctuations in proton configuration, {\it i.e.} large variations in parton densities
\cite{Mantysaari:2016ykx,Cepila:2016uku}.   These fluctuations are expected to remain visible with ion targets.  Similar studies of ultra-peripheral collisions (UPCs) at the LHC found that the fluctuations should increase the incoherent cross-section by a factor of about 2 for heavy targets \cite{Mantysaari:2017dwh,Cepila:2016uku}.   Although the EIC has a smaller energy reach than HERA or LHC UPCs, it can study these fluctuations for a range of nuclear targets.

\subsubsection{Separating coherent and incoherent production}

A key problem for pursuing this physics involves separating coherent and incoherent production.  Although most incoherent interactions involve neutron (or sometimes proton) emission, this is not always the case, and nuclear excitations which decay via emission of MeV photons (in the nuclear rest frame) are not easy to detect.   The situation is further complicated because models of nuclear excitation in photoproduction are subject to large uncertainties; in most cases, there is no relevant data to constrain the models.

Figure \ref{fig:dsdt} shows the magnitude of the problem.  The left panel shows $d\sigma/dt$ for coherent and incoherent $J/\psi$ production, while the right panel shows the incoherent $J/\psi$ production as modelled in the BeaGLE Monte Carlo.  

At the position of the third diffractive minimum, the incoherent cross-section is about 400 times larger than the coherent one, so to determine the coherent cross-section, a rejection factor for incoherent photoproduction better than 400:1 must be achievable; this factor must also be known accurately.  In the right panel, BeAGLE simulations show that, by vetoing on neutrons, protons and high energy (above 50 MeV) photons only leads to a 100:1 rejection factor in that $|t|$ range. It should be noted that calculations of the frequency of nucleon-free incoherent production have large uncertainties, and one cannot use Monte Carlos to predict this fraction.

For $|t|<0.01$ GeV$^2$, the situation is reversed, with the coherent cross-section up to 100 times larger than the incoherent one.   Here, the major misidentification danger is the presence of random (uncorrelated) neutrons, forward protons or photons accompanying a coherent reaction.   This probability is not small. In full-energy eAu collisions at an eA luminosity of $8\times10^{31}$ cm$^{-2}$ s$^{-1}$, the rate for one background reaction, photo-excitation of a gold nucleus, followed by Giant Dipole Resonance decay, usually leading to a single neutron is about 2.5 MHz \cite{Klein:2014xoa}, leading to a significant rate of background coincidences.  Other sources will further increase the rate.  These rates can be measured using otherwise empty beam crossings, and statistically subtracted but, because of the large coherent to incoherent ratio, this will significantly increase the measurement uncertainty at low $|t|$.

Section 8.5.1 will present some experimental constraints in separating coherent and incoherent production, which will limit the use of exclusive production as a probe of nuclear targets.

\section{Understanding Hadronization}
\label{part2-sec-Hadronization}

How do the degrees of freedom of QCD, quarks and gluons, relate to the hadronic degrees of freedom we observe in nature? The EIC will not only address the many outstanding questions about hadron structure, as described in the previous sections, but also will make substantial progress in our understanding of hadron formation.
{\bf Hadronization in vacuum and in the nuclear medium} are
covered in Secs.~\ref{part2-subS-Hadronization-HadVacuum} and~\ref{part2-subS-Hadronization-HadNuclear}, respectively. 

Section~\ref{part2-subS-Hadronization-PartProd} is devoted to {\bf particle production for identified hadron species}. It is argued that
hadronization mechanisms other than parton fragmentation should also be considered for understanding particle production in lepton-hadron collisions: threshold production, string-breaking, and coalescence or recombination. The latter is arguably critical for particle production
in high density environments, such as in high multiplicity collisions involving hadrons or nuclei. Enhanced baryon production at intermediate transverse momenta in heavy-ion collisions has been considered for a long time to be one of the established signatures of the "Quark Gluon Plasma", and to provide evidence for coalescence contributions to hadron production. Could a similar effect occur in  $\eA$ collisions?
The EIC clearly is in a position to make groundbreaking
progress in our understanding of hadronization.

Production of {\bf quarkonia} in $\ep$ and $\eA$ collisions will provide unique insight into hadronization and is addressed in Sec.~\ref{part2-subS-Hadronization-Quarkonia}.

There is great potential also in studying {\bf new particle production mechanisms} such as exclusive backward $u$-channel production. Given its high luminosity the EIC may be able to discover fundamental QCD particle production processes with low cross sections such as via hard (perturbative) $C$-odd three gluon exchange.

This section also describes the impact the EIC will have on the study of hadron spectroscopy, in particular in the heavy quark sector. Here, too, the projected high luminosity of the EIC will enable detailed studies of {\bf exotic states} that have recently been observed at other facilities.

A complementary approach to insight into hadronization is via the study of {\bf target fragmentation} through the measurement of fracture functions. They describe the hadronization of the target after a parton with a given momentum fraction $x$ is removed from it. Factorization theorems apply and fracture functions are universal quantities, independent of the hard process.
\subsection{Hadronization in the vacuum}
\label{part2-subS-Hadronization-HadVacuum}

\subsubsection{Light meson fragmentation functions and flavor sensitivity}
Fragmentation functions, FFs, describe the formation of final-state hadrons off high-energetic, asymptotically free partons~\cite{Metz:2016swz}. As such, FFs directly connect to the confinement of the strong interaction. While data from electron-positron annihilation mostly constrain the singlet combination of the FFs and proton-proton collisions primarily constrain gluon FFs, the production of light mesons in semi-inclusive DIS is the primary channel for the differentiation between the fragmentation of light quarks and anti-quarks. In addition, semi-inclusive DIS data have a high sensitivity to the separation of quark flavors. Recent determination of spin-averaged single-hadron fragmentation functions from electron-positron annihilation data can be found in Refs.~\cite{Bertone:2017tyb,Sato:2016wqj} and global analyses based on the combination of data from electron-positron annihilation \cite{Braunschweig:1988hv,Abe:1998zs,Akers:1994ez,Abreu:1998vq,Buskulic:1994ft,Aihara:1986mv,Aihara:1988su,Abbiendi:1999ry,Lees:2013rqd,Leitgab:2013qh}, proton-proton collisions \cite{Adare:2007dg,Abelev:2006cs,Abelev:2014laa,Agakishiev:2011dc,Arsene:2007jd,Adamczyk:2013yvv} and semi-inclusive DIS \cite{Airapetian:2012ki,Adolph:2016bwc},  can be found in Refs.~\cite{Hernandez-Pinto:2016cnc,Hernandez-Pinto:2017jyv}. The latter find their origin in Ref.~\cite{deFlorian:2007aj}.

\begin{figure}[htp]
    \centering
    \includegraphics[width=0.45\textwidth]{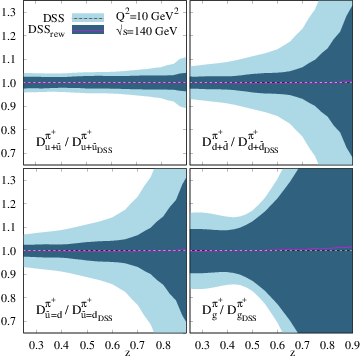}
    \includegraphics[width=0.45\textwidth]{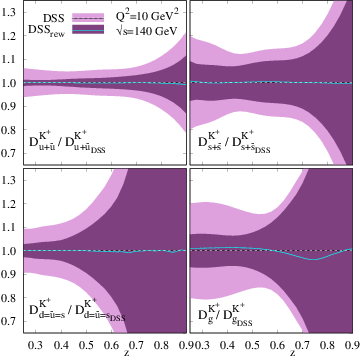}
    \caption{Pion (left) and kaon (right) FFs obtained from the global DSS analyses~\cite{Hernandez-Pinto:2016cnc,Hernandez-Pinto:2017jyv} (dashed line with light uncertainty bands) and obtained from the global DSS analyses with the inclusion of EIC pseudo-data, at a c.m.s. energy $\sqrt{s}=140$~GeV, (continuous line with dark uncertainty bands)~\cite{Aschenauer:2019kzf}, both normalized to the DSS best fit. The shaded bands reflect the statistical uncertainties from the pseudo-data (evaluated at a luminosity of 10 fb$^{-1}$) and the uncertainties from the PDFs. The upper (lower) row corresponds to the (un)favoured fragmentation.}
    \label{fig:part2-subS-Hadronization-HadVacuum.sidis}
\end{figure}

The impact of EIC pseudo-data can be seen in Fig.~\ref{fig:part2-subS-Hadronization-HadVacuum.sidis}~\cite{Aschenauer:2019kzf}, which shows the pion (left) and kaon (right) FFs obtained from the global analysis (DSS, \cite{Hernandez-Pinto:2016cnc,Hernandez-Pinto:2017jyv}) and those obtained by including the EIC pseudo-data, at a c.m.s. energy $\sqrt{s}=140$~GeV, to the DSS global analysis (DSS$_{rew}$, \cite{Aschenauer:2019kzf}), both normalised to the DSS best fit. The darkly shaded bands reflect the statistical uncertainties from the pseudo-data as well as those from the PDF set used.
It can be seen that the EIC has a strong potential to improve the determination of the light-meson FFs, both for favored (upper row) and for unfavored (lower row) fragmentation. The improvement is especially pronounced for the unfavored pion FFs and for the favored kaon FFs, while the results for the unfavored kaon FFs should be evaluated with some caution because of the more rigid functional form assumed.
Based on the results for the single-hadron, spin-averaged FFs it is reasonable to expect that EIC data will have also a significant effect on the extraction of di-hadron FFs and polarized FFs. In particular, the EIC will for the first time enable the measurement of spin-averaged and spin-dependent FFs in jets. 
This allows for a direct measurement of transverse-momentum-dependent FFs.
For more details on jet measurements at the EIC, see Sec.~\ref{part2-subS-SecImaging-TMD3d.jets}.

\subsubsection{Fragmentation into polarized
\texorpdfstring{$\Lambda$}{Lamdba} hyperons}
\label{part2-subS-Hadronization-lambda}


The detection of self-analyzing $\Lambda$ hyperons makes it possible to study the dependence of the hadronization process on polarization degrees of freedom in the final state. Here we will discuss polarized fragmentation functions that can be seen as the FF analogue to the polarized PDFs $g_1$ and $h_1$ as well as polarizing FFs that can be seen as the analogue to the Sivers TMD PDF or, in the twist-3 framework, the Qiu-Sterman matrix elements~\cite{Qiu:1998ia}.

\begin{figure}[htbp]
	\centering
\includegraphics[width=0.95\textwidth]{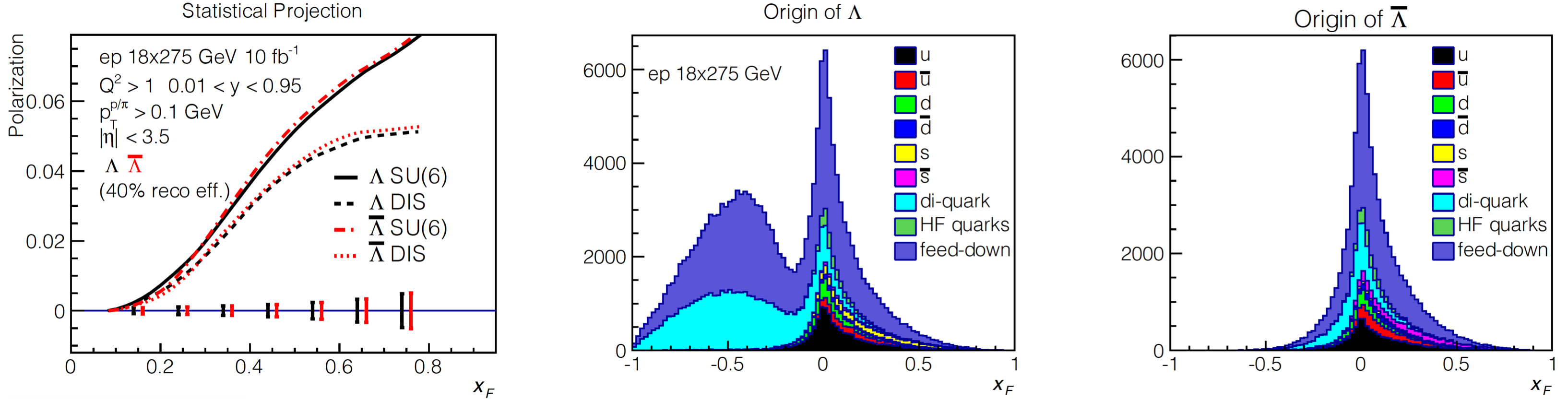}\\
	\caption{Left panel: Projection of longitudinal spin transfer for $\Lambda$ and $\bar \Lambda$ from proton beam at 18$\times$ 275 GeV at EIC. The curves are from model predictions~\cite{Zhou:2009mx}. The two right hand panels show the origin of the reconstructed $\Lambda/\bar{\Lambda}$. In the current fragmentation region a significant fraction originates from feed-down. A dominant part of the feed-down component is contributed by $\Sigma^0\rightarrow \Lambda \gamma$. For more information, see Ref.~\cite{Kang20}.
	}\label{fig:part2-subS-Hadronization-HadVacuum.lambdaPol}
\end{figure}

Polarized Fragmentation Functions (PFFs) describe the hadronization process of a polarized parton with a final state polarized hadron like $\Lambda$.  The PFF of $\Lambda$ and $\bar \Lambda$ have been widely studied in experiments including polarized lepton-nucleon DIS process \cite{Adams:1999px, Astier:2000ax, Airapetian:2006ee,Alekseev:2009ab} and polarized hadron-hadron collisions \cite{Bravar:1997fb,Abelev:2009xg,Adam:2018kzl,Adam:2018wce}.  
The first global analysis of longitudinally polarized fragmentation function $\Delta D$\cite{deFlorian:1997zj} has been performed with LEP data, but is not well constrained yet.
The polarized jet fragmentation function of $\Lambda$ within a fully reconstructed jet was also studied recently\cite{Kang:2020xyq}.
The spin transfer of $\Lambda$ and $\bar \Lambda$ from proton beam either longitudinally or transversely polarized in lepton-nucleon DIS process, may also provide valuable information on strange quark helicity or transversity distributions \cite{Lu:1995np,Anselmino:2000ga,Ma:2000uu,Ma:2001rm,Liu:2001yt,Ellis:2002zv,Ellis:2007ig,Zhou:2009mx} if the corresponding PFFs are reasonably determined. 
The unprecedented high precision data on hyperon spin transfer measurements at EIC will shed new light into the PFFs and the strange quark distributions.  
Fig. \ref{fig:part2-subS-Hadronization-HadVacuum.lambdaPol} shows a projection of longitudinal spin transfer to $\Lambda$ and $\bar \Lambda$ from polarized proton beam at EIC energy of 18$\times$ 275 GeV.

\begin{figure}[htp]
    \centering
    \includegraphics[width=0.95\textwidth]{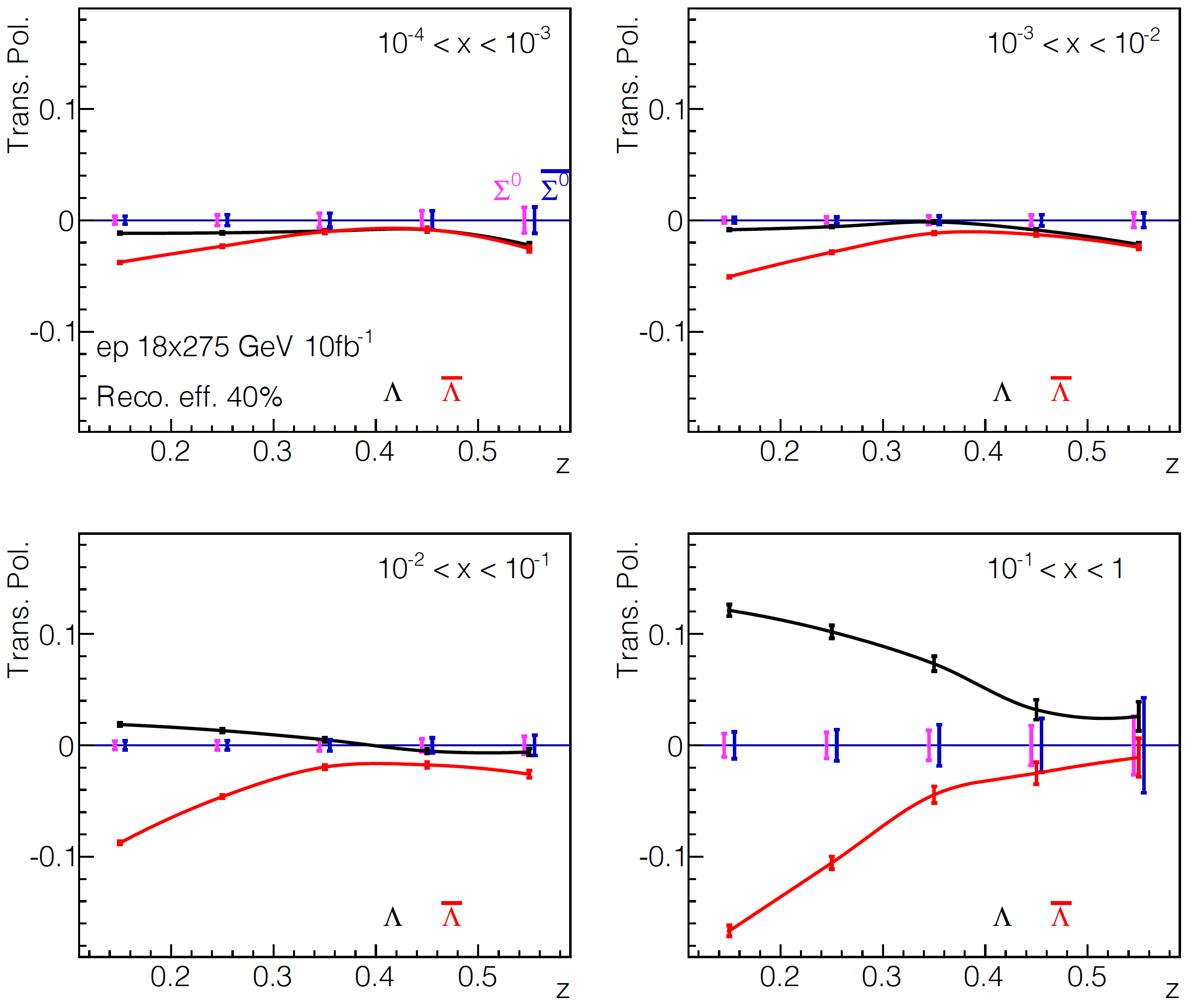}
  \caption{Projected transverse $\Lambda$ polarization using the extraction in Ref.~\cite{Callos:2020qtu} for the highest energy configuration. The projected uncertainty on $\Sigma^0$ polarization is also shown which is important by itself and to estimate the  polarization of the feed-down component. A 40\% reconstruction efficiency is assumed but not the effect from feed-down which most likely reduces the magnitude of the asymmetries.
    \label{fig:part2-subS-Hadronization-HadVacuum.lambda}}
\end{figure}

In the case, where the initial quark is not polarized, the final state $\Lambda$ can still carry polarization. In fact, it has been a long standing challenge to describe 
the transverse polarization of $\Lambda$ hyperons in unpolarized deep inelastic reactions from a
factorized framework in perturbative QCD. 
 Initiated  by the strikingly large 
transverse polarization asymmetries  of $\Lambda$ hyperons observed in early experiments at Fermilab (along with follow-up experiments) in $pA\to \Lambda X$ fixed target processes already 40 years ago  \cite{Bunce:1976yb,Schachinger:1978qs,Heller:1983ia,Lundberg:1989hw,Yuldashev:1990az,Ramberg:1994tk,Fanti:1998px,Abt:2006da}, experimental and theoretical investigations~\cite{Kane:1978nd,Panagiotou:1989sv,Dharmaratna:1996xd,Anselmino:2000vs,Anselmino:2001js,Boer:2010ya,Boer:2010yp,Gamberg:2018fwy} have spanned decades.
More recently, polarization of $\Lambda$ baryons were investigated at the LHC by the ATLAS collaboration \cite{ATLAS:2014ona}. While a small polarization was found in the ATLAS measurements-essentially consistent with zero-in the mid-rapidity region, such experiments demonstrate  that the polarization of $\Lambda$ baryons can be studied at the highest LHC energies and may be larger in different kinematical regions at forward rapidities. Experimentally, data on polarized $\Lambda$ fragmentation has been provided by the OPAL collaboration \cite{Ackerstaff:1997nh} at LEP. This measurement was performed on the $Z$-pole, i.e. at a center of mass energy equal to the mass of the $Z$-boson. While a substantial  {\it longitudinal} polarization of the $\Lambda$s was detected by OPAL, the {\it transverse} polarization was found to be zero within error bars.
Recently the BELLE collaboration  measured the production of
transverse polarization of  $\Lambda$-hyperons~\cite{Guan:2018ckx}    
in $e^+e^-$ - annihilation, where the hadron cross section is studied as a function of the event-shape variable called thrust $T$, fractional energy $z_{\Lambda}$, and the transverse momentum $j_\perp$ with respect to the thrust axis. 
They find a significant non-zero effect for  the  process $e^+e^-\rightarrow \Lambda^\uparrow {\rm (Thrust}{\rm )} X$ as well as for back to back production of $\Lambda + h$. In the TMD factorization  framework~\cite{Collins:1981uk,Boer:1997mf} for back to back production of $\Lambda + h$,  a chiral even, naively $T$-odd fragmentation function, the so-called polarizing fragmentation function $D_{1T}^\perp$ is predicted to be non-zero and universal~\cite{Collins:1992kk,Boer:2010ya}. Recent extractions of $D_{1T}^\perp$ from the $e^+e^-$ data can be found in Refs.~\cite{DAlesio:2020wjq,Callos:2020qtu}. The FF $D_{1T}^\perp$ can be seen as the fragmentation analogue to the Sivers function therefore a test of its universality compared to an extraction from SIDIS data is very interesting~\cite{Boer:2010ya}. As with other FF measurements, SIDIS data is needed to achieve better flavor separation. The EIC will be the first SIDIS facility where a high statistics sample of $\Lambda$'s in the current fragmentation region can be collected which will revolutionize the study of FFs with polarization degrees of freedom in the final state. 
Figure~\ref{fig:part2-subS-Hadronization-HadVacuum.lambda} shows the projected transverse $\Lambda$ polarization. For more information, see Ref.~\cite{Kang20}.

\subsubsection{Partial wave decomposition of polarized and unpolarized di-hadron FFs including TMDs}
\label{part2-subS-Hadronization-diHad-pw}
\begin{figure}[htp]
    \centering
    \includegraphics[width=0.95\textwidth]{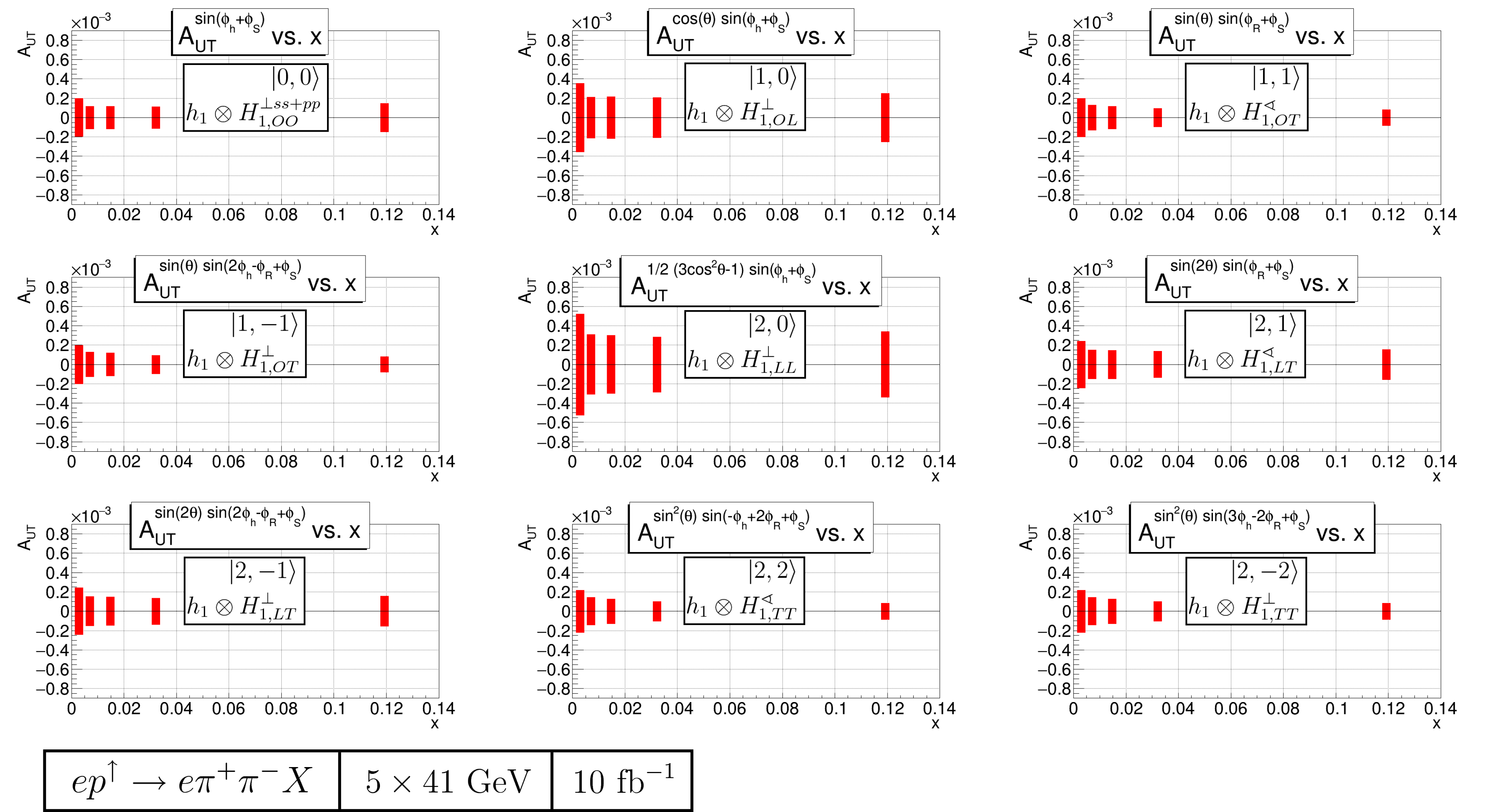}
    \caption{Projections for the nine partial waves contributing at twist-2 to $A_{UT}$ using $\mathcal{L}=10$~fb$^{-1}$ at $5\times41$ (in GeV).  The labels on the figure indicate the $m,l$ state and which PDF and FF the PW is sensitive to.
    \label{fig:part2-subS-Hadronization-HadVacuum.PW}}
\end{figure}

As discussed in Sec.~\ref{sec:part2-subS-SecImaging-TMD3d.IFF}, di-hadron FFs are more powerful than single-hadron FFs, due to the additional degrees of freedom. 
This allows FFs to exist that do not have a single-hadron analogue. One example that has been attracting recent interest is the helicity dependent FF $G_1^\perp$~\cite{Bacchetta:2002ux,Matevosyan:2017liq}. Recent models, similar to the ones for $H_1^\sphericalangle$~\cite{Bacchetta:2006un,Matevosyan:2017alv}, make projections for the magnitude of $G_1^\perp$ based on string fragmentation models~\cite{Matevosyan:2017uls} or on the interference of partial waves (PWs)~\cite{Luo:2020wsg}. 
However, another consequence is that the di-hadron cross-section contains an infinite series of angular modulations. In the context of a PW decomposition, this consequence can be interpreted as originating from the interference of waves with different quantum numbers. We use here the usual notation with the total angular momentum named $L$ and each PW characterized by angular momentum eigenvalues $l,m$. Already in the spin averaged cross-section one encounters 12 terms up to l=2 at leading twist, which is a limit following from angular momentum conservation~\cite{Bacchetta:2002ux,Gliske:2014wba}. 
The measurement of these structure functions will provide additional insight into hadronization mechanisms. 
Since these additional terms are essentially unknown and can interfere with the extraction of the asymmetries of interest, they are a dominant systematic effect for any di-hadron extraction (see, {\it e.g.}, discussion in Ref~\cite{Airapetian:2008sk, Schnell:2018xvv}), which is another strong motivation for their measurement. For a separation of the PWs, a fit not only in the azimuthal angles but also in the decay angle $\theta$ is necessary. A large acceptance in $\theta$ can only be achieved by low minimum momentum cutoff and this has been incorporated into the requirements for an EIC detector.     
Figure~\ref{fig:part2-subS-Hadronization-HadVacuum.PW} shows projections for the PWs that are contained in the azimuthal modulations of $A_{UT}$ sensitive to transversity coupled the chiral-odd, transverse polarization dependent, DiFF up to $l=2$, {\it i.e.}, taking only $s$- and $p$-waves into account. For more details, see the discussion around eq.~(52) in Ref.~\cite{Gliske:2014wba}. The projections are for a luminosity of $\mathcal{L}=10$~fb$^{-1}$ at the low CME configuration of $5\times41$ (in GeV), where the effect of the $p_T$ restriction is more severe. A reasonable precision can be achieved for all PWs. The larger uncertainties at low $x$ are due to the smaller depolarization factor $D(y)$ at these kinematics.

\subsubsection*{Leading jets}

\begin{figure}[th]
  \centerline{\includegraphics[width = .5\textwidth]{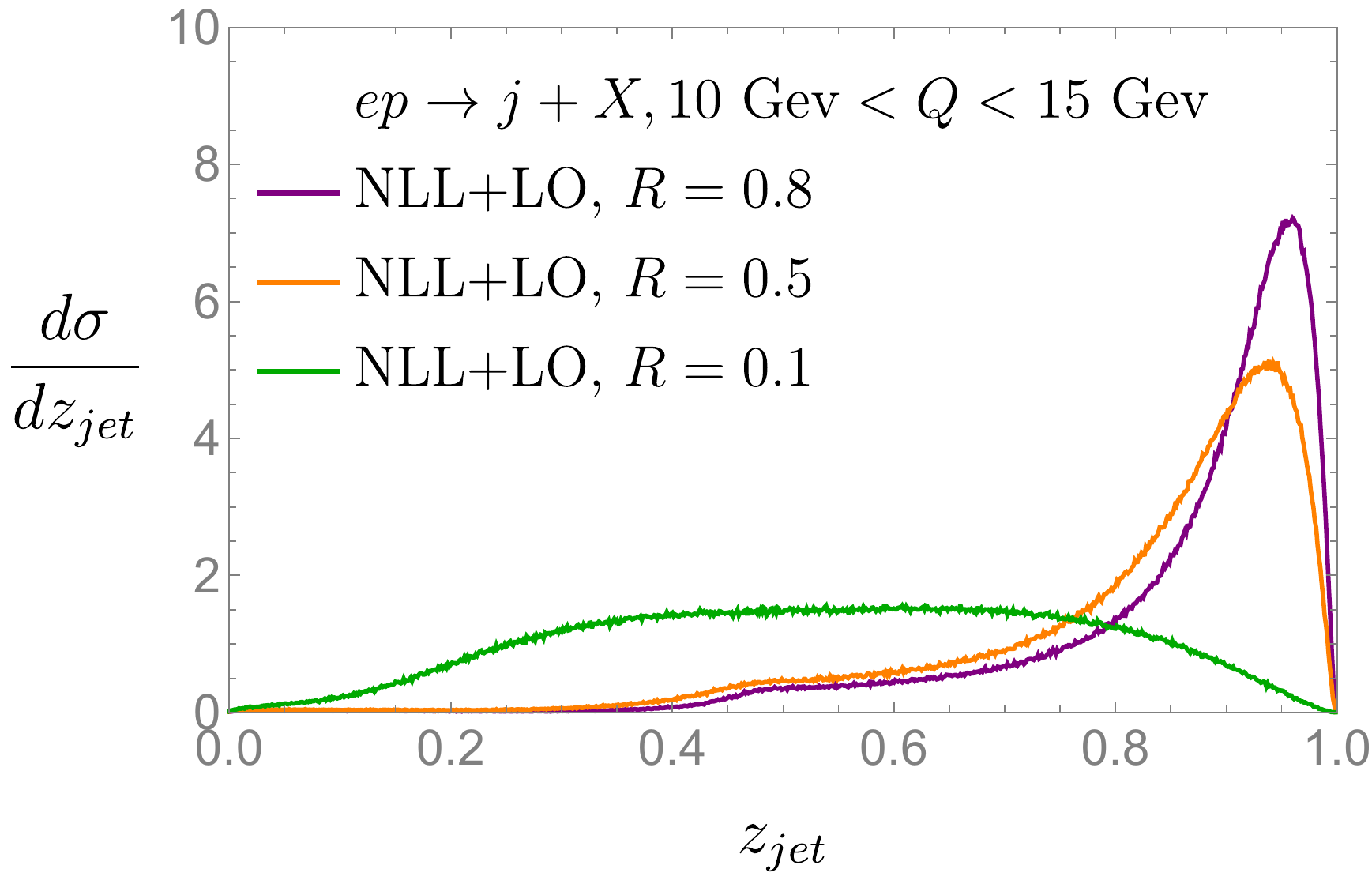}
  \includegraphics[width = .5\textwidth]{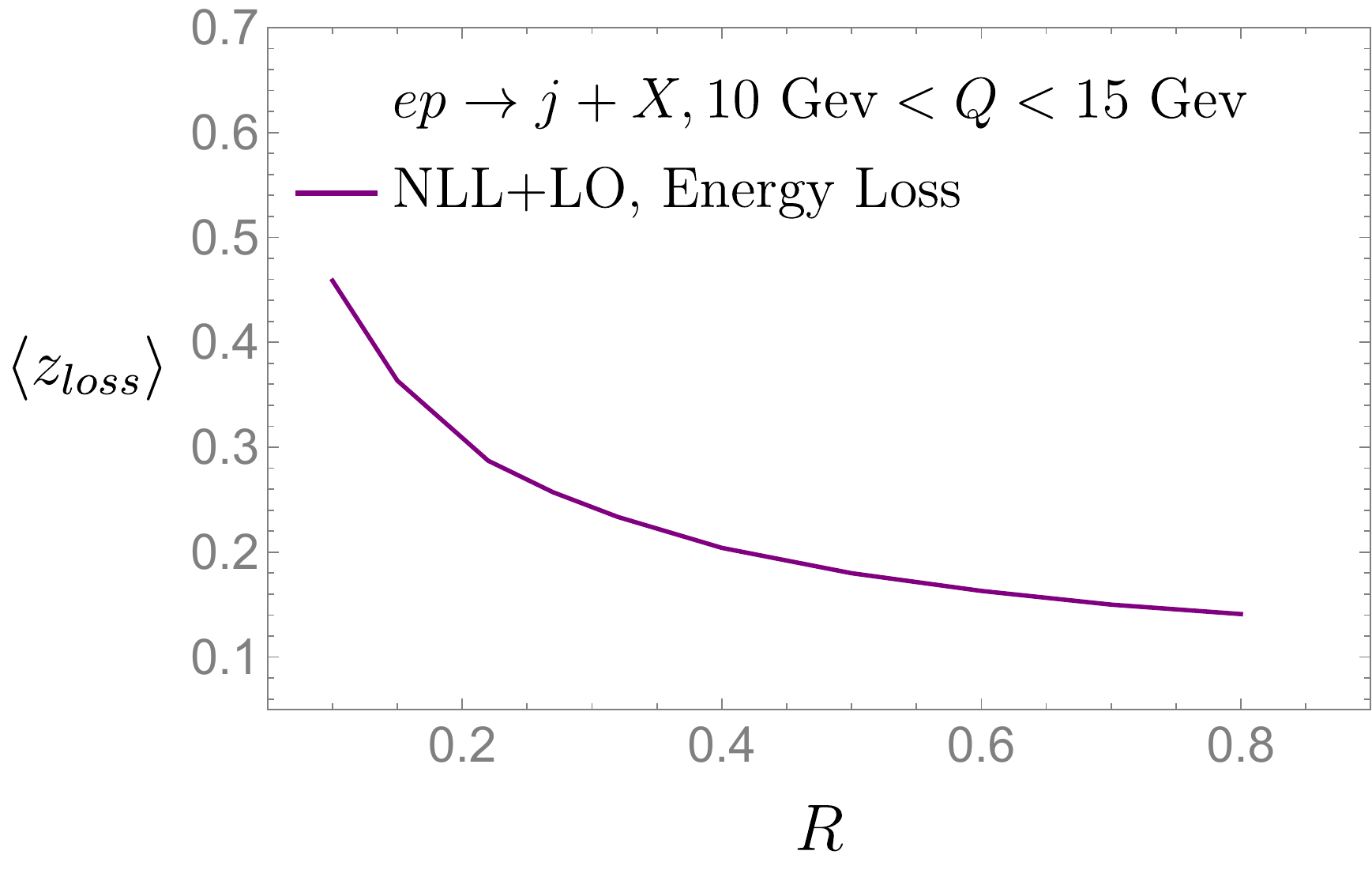}}
  \caption{Left: The leading jet cross section at the EIC at NLL$'$ as a function of the jet's longitudinal momentum fraction $z_{jet}$. Right: The average energy loss of leading jets $\langle z_{loss}\rangle$ as a function of the jet radius $R$.~\label{fig:eloss_qg}}
\end{figure}
Different than inclusive measurements, the reconstruction of leading jets or hadrons allows for a well defined notion of energy loss which can be directly measured at the EIC~\cite{Neill:2020mtc}. By identifying in addition a hard reference scale $Q^2=-q^2$, the photon virtuality, the average radiation outside the leading jet $\langle z_{loss} \rangle$ can be calculated from first principles in QCD which can be directly identified with parton energy loss. Semi-inclusive cross sections measurements with a suitable jet reconstruction algorithm in the Breit frame~\cite{Arratia:2020ssx}, allow for a unique opportunity to study the probability distribution of leading jets as well as their average energy loss, see Fig.~\ref{fig:eloss_qg}. In \ep collisions leading hadrons and jets probe non-linear QCD dynamics~\cite{Neill:2020mtc,Dasgupta:2014yra,Scott:2019wlk}. Additionally, in \eA collisions this provides a unique opportunity to quantify the interaction of energetic quarks and gluons with the cold nuclear matter environment which also allows for a connection to corresponding measurements in \pA and \AA collisions.

\subsubsection{Jet substructure}

Jets and their substructure will be an important tool for understanding hadronization. Since jets are closely related to scattered partons, they can be used to relate final-state hadrons to their parent parton. In particular, jet substructure offers the opportunity to study both the process of fragmentation, or parton radiation patterns, and hadronization, or the formation of the parton shower into bound state hadrons. Single hadrons-in-jets will be used to study a variety of (un)polarized transverse-momentum-dependent fragmentation functions. Fragmentation functions can also be measured for different parton flavors by, for example, tagging heavy quark mesons such as the $D^0$. Additionally, novel jet substructure techniques will be used to study parton radiation patterns in the theoretically clean environment of DIS that allows for better separation of the target and current fragmentation regions. Soft drop declustering techniques can also be used to suppress or enhance nonperturbative effects, which will be essential for better understanding the interplay between the perturbative and nonperturbative roles in the process of hadronization. 

\begin{figure}[ht!]
  \centerline{\includegraphics[width = 0.95\textwidth]{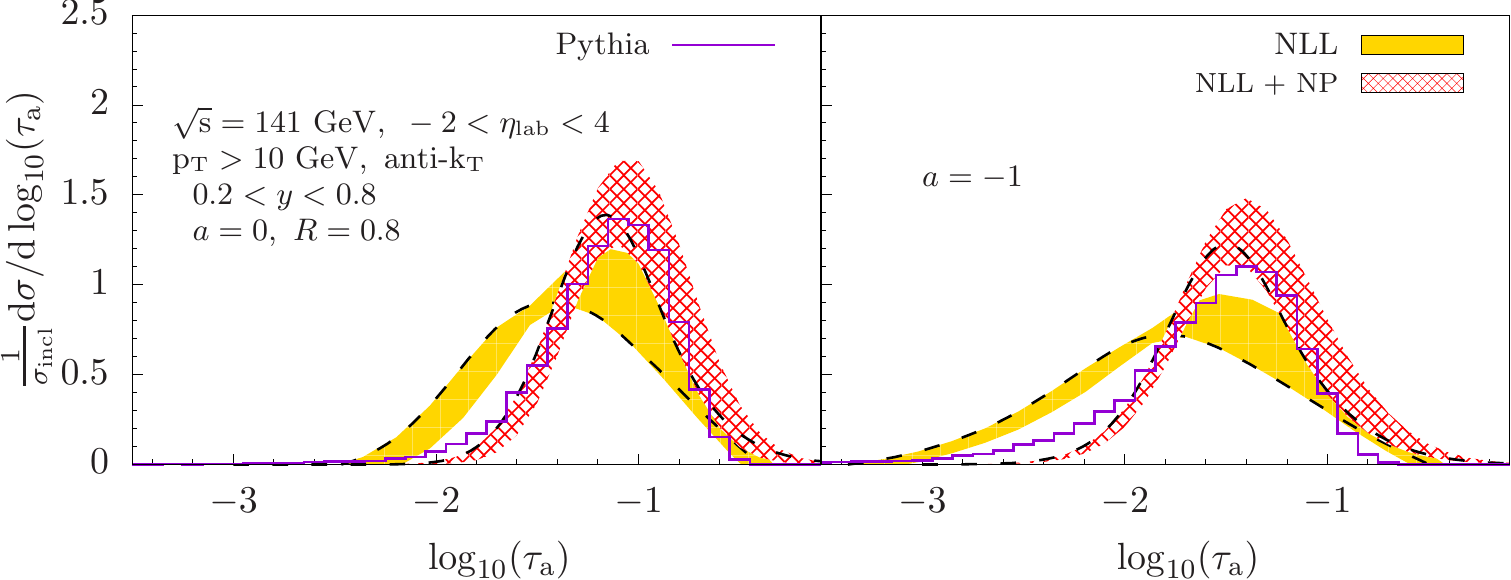}}
  \caption{Predictions for jet angularities at the EIC: Purely perturbative results (yellow band) and with a nonperturbative shape function (red band) compared to simulations from PYTHIA~6~\cite{Sjostrand:2006za}.~\label{fig:part2-subS-Hadronization-HadVacuum.angularity}}
\end{figure}

One set of substructure observables that have been explored in some detail for the EIC are the one-parameter family of constructs known as jet angularities. For jets with a given transverse momentum $p_T$, the observable is defined as~\cite{Berger:2003iw,Almeida:2008yp,Ellis:2010rwa,Hornig:2016ahz,Kang:2018qra}
\begin{equation}\label{eq:taua}
    \tau_a=\frac{1}{p_T}\sum_{i\in J}p_{Ti}\Delta R_{iJ}^{2-a} \,.
\end{equation}
Here $p_{Ti},\,\Delta R_{iJ}$ are the transverse momentum of each particle in the jet and their distance to the jet axis, respectively. Fig.~\ref{fig:part2-subS-Hadronization-HadVacuum.angularity} shows numerical results for the EIC~\cite{Aschenauer:2019uex} for two different values of $a$ and representative jet kinematics. Jet angularities and other jet substructure observables are of great interest at the EIC to study various physics aspects: Test of perturbative methods at low energies and particle multiplicities, universality aspects of nonperturbative shape functions which model hadronization effects, study power corrections, extractions of the QCD strong coupling constant, cold nuclear matter effects, and the tuning of parton showers.

\subsection{Hadronization in the nuclear environment}
\label{part2-subS-Hadronization-HadNuclear}

\subsubsection{Collinear nuclear fragmentation functions for light hadrons}
The modification of the final state in the presence of a nuclear medium is a well known but little understood phenomenon in cold nuclear matter physics, with different approaches proposed to describe the measured data \cite{Qiu:2019sfj,Burke:2013yra,Qin:2015srf,Blaizot:2015lma,Majumder:2010qh,KunnawalkamElayavalli:2017hxo,He:2018xjv}. In nucleus-light hadron collisions from both RHIC and the LHC the observables depend on nPDFs and in medium FFs simultaneously which leave very little sensitivity to the in medium modification of the FFs or jet functions in the global fits. Thus the cleanest way of exploring the final state effects is through SIDIS, pioneered by the HERMES collaboration \cite{Airapetian:2007vu}. In this case the use of multiplicities increases the sensitivity to the fragmentation functions. The first model independent extraction of nuclear fragmentation functions (nFFs) used the HERMES \cite{Airapetian:2007vu} and RHIC \cite{Adams:2003qm,Adler:2006wg} data \cite{Sassot:2009sh}. Unfortunately the former had a restricted kinematic coverage and no further exploration has been performed since. 

Assuming nuclear effects in SIDIS at the EIC will be similar to those seen at HERMES, the incredible precision expected will allow us to fully characterize the nFFs (as well as the nPDFs, see section \ref{part2-subS-LabQCD-NuclPDFs}). Using the latest pion FFs in vacuum from DEHSS \cite{Hernandez-Pinto:2016cnc,Hernandez-Pinto:2017jyv}, a new extraction of nFFs (LIKEn21) from the HERMES data was performed \cite{Zurita:2021kli}. 

\begin{figure}
   \centering
\includegraphics[width=0.9\textwidth]{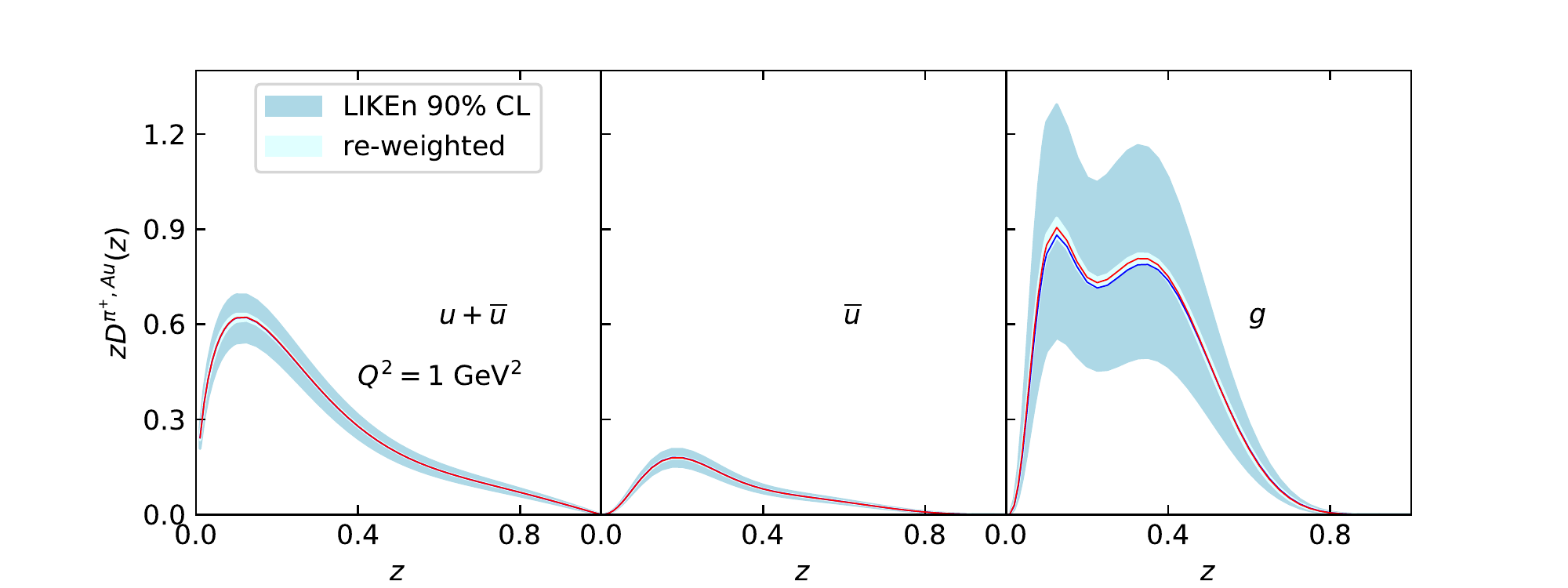}
\caption{\label{7.4.2:fig:nff_dist}Nuclear FFs LIKEn21 for Au and impact of EIC pseudodata at $\sqrt{s}\sim 30 \text{ GeV}$ for $u+\bar{u}$ (left), $\bar{u}$ (center) and gluon (right). Similar results are found for higher $\sqrt{s}$.
}
\end{figure}

\begin{figure}
   \centering
\includegraphics[width=0.9\textwidth]{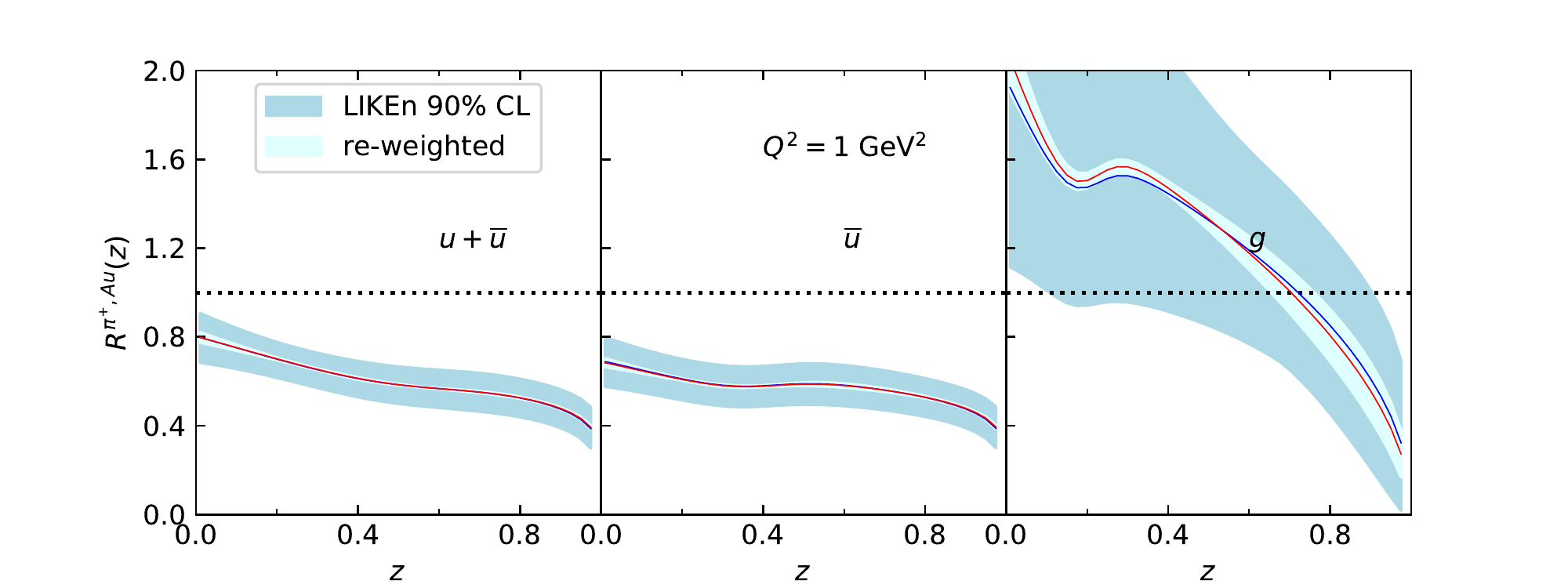}
\caption{\label{7.4.2:fig:nff_ratio}Ratio of the distributions in Fig.\ref{7.4.2:fig:nff_dist} to vacuum FFs from DEHSS \cite{Hernandez-Pinto:2016cnc,Hernandez-Pinto:2017jyv}.}
\end{figure}

With these novel nFFs an impact study for the EIC was performed using a re-weighting technique. Pseudo data was created using LIKEn21, with appropriate Gaussian noise from the estimated uncertainties that were obtained by {\sc pythia} simulations reweighted with the nuclear modification of \cite{Sassot:2009sh} for an accumulated luminosity of 10 fb$^{-1}$ for both collision species. Fig. \ref{7.4.2:fig:nff_dist} shows the distributions for $u+\bar{u}$, $\bar{u}$ and gluon densities at the initial scales. The corresponding ratios to DEHSS vacuum FFs can be seen in Fig. \ref{7.4.2:fig:nff_ratio}. The blue band corresponds to the $90\%$ CL of LIKEn21 and the (very narrow) light cyan band is the one resulting after re-weighting $\pi^{\pm}$ pseudo data using so far only the lowest collision energy of $\sqrt{s}\sim30$ GeV. Given the high precision expected at the EIC, about $10\%$ of the replicas remain. Higher values of $\sqrt{s}$ produce similar results. 

The $z\leq 0.2$ region is not covered by the data and therefore the shape of the modification should not be considered as anything other than an artificial outcome of the fit at this point. The EIC will definitely explore SIDIS in a much broader kinematic space, opening the way to a full characterization of the FFs and nFFs including flavour separation. Eventually SIDIS data could become an asset in the extraction of nPDFs as shown for the proton case in \ref{part2-subS-Hadronization-HadVacuum} for fragmentation functions in the vacuum.  

Studies that address the nuclear dependence of transverse momentum for final state hadrons either on the initial state or final state have been discussed in \ref{part2-subS-SecImaging-LpolNucl}. Further aspects of light hadron fragmentation, such as transverse momentum broadening, a possible $\nu$ dependence in the suppression of nFFs and other topics can also be studied in detail using the high precision of the EIC data.

\subsubsection{In medium evolution for light and heavy flavor mesons}

The effect of nuclear environment on hadronization is one of the key questions that the EIC will investigate. Fixed-target HERMES measurements with electron beam of energy $E_{\rm beam} = 27.6$~GeV~\cite{Airapetian:2003mi,Airapetian:2007vu} have clearly established attenuation of light particle production.  Different theoretical approaches have been proposed to explain the data that differ in the underlying assumptions and in the extracted transport properties of large nuclei~\cite{Wang:2002ri,Arleo:2003jz,Chang:2014fba,Accardi:2002tv,Kopeliovich:2003py,Brooks:2020fmf,Falter:2004uc}. With better understanding of in-medium parton showers, the traditional energy loss phenomenology can be generalized to  full  fragmentation function evolution in the presence of nuclear matter. It is  given by:
\begin{equation}
\label{eq:fullevol}
    \frac{d}{d \ln \mu^{2}} \tilde{D}^{h/i}\left(x, \mu\right)= 
    \sum_{j} \int_{x}^{1} \frac{d z}{z}  \tilde{D}^{h/j}\left(\frac{x}{z}, \mu\right)  
   \left( P_{j i}\left(z, \alpha_{s}\left(\mu\right)\right) +  P_{j i}^{\rm med}\left(z, \mu\right)  \right)  \, ,
\end{equation}
where in Eq.~(\ref{eq:fullevol})  $ P_{j i}^{\rm med}$ are the medium corrections to the splitting functions.
In addition to precision light flavor studies, the higher enter-of-mass energies at the EIC  provide new probes of hadronization - open heavy meson cross sections in  e+p and  e+A collisions~\cite{Li:2020zbk}.  

\begin{figure}
 	\centering
 	\includegraphics[width=0.48\textwidth]{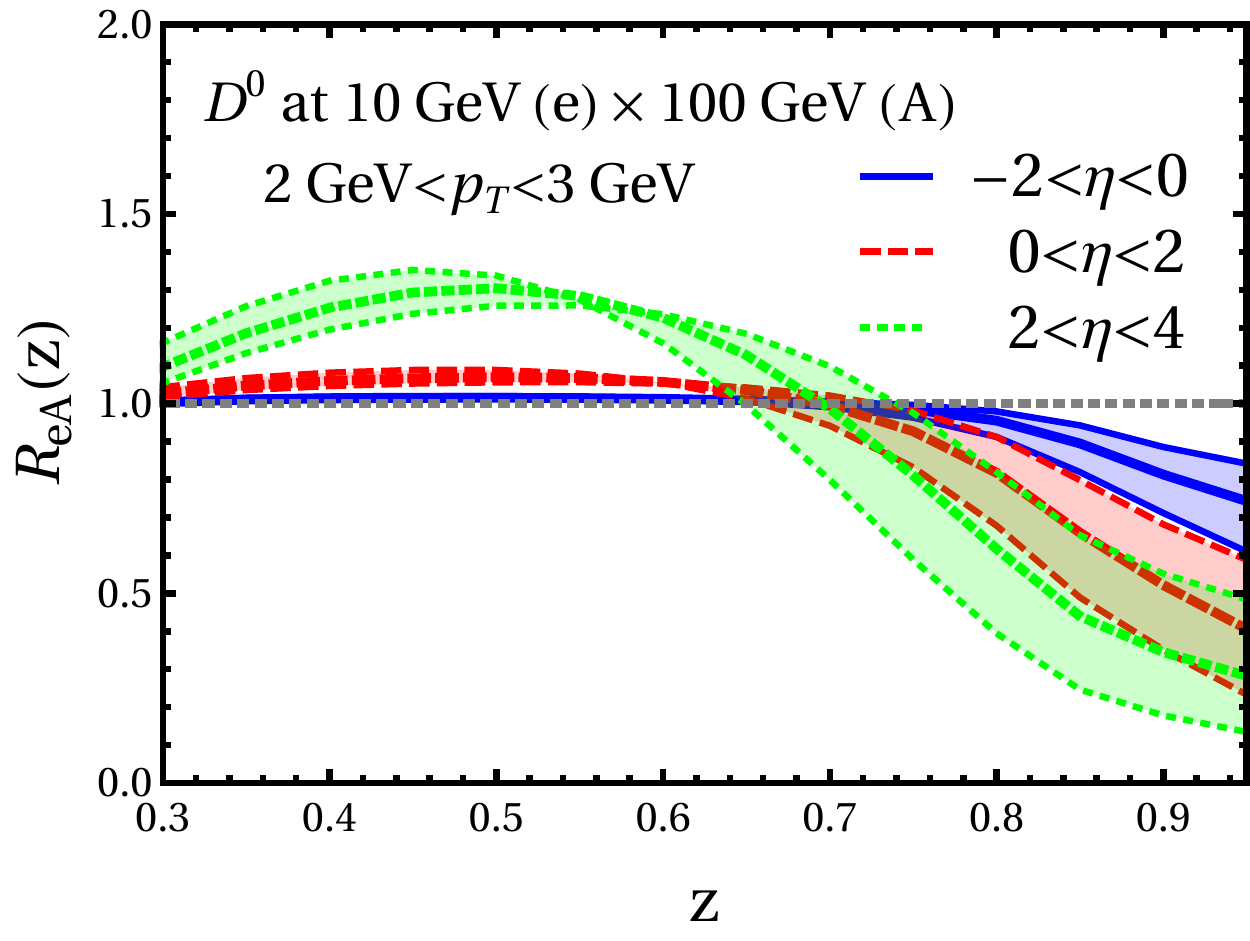}
 	\includegraphics[width=0.48\textwidth]{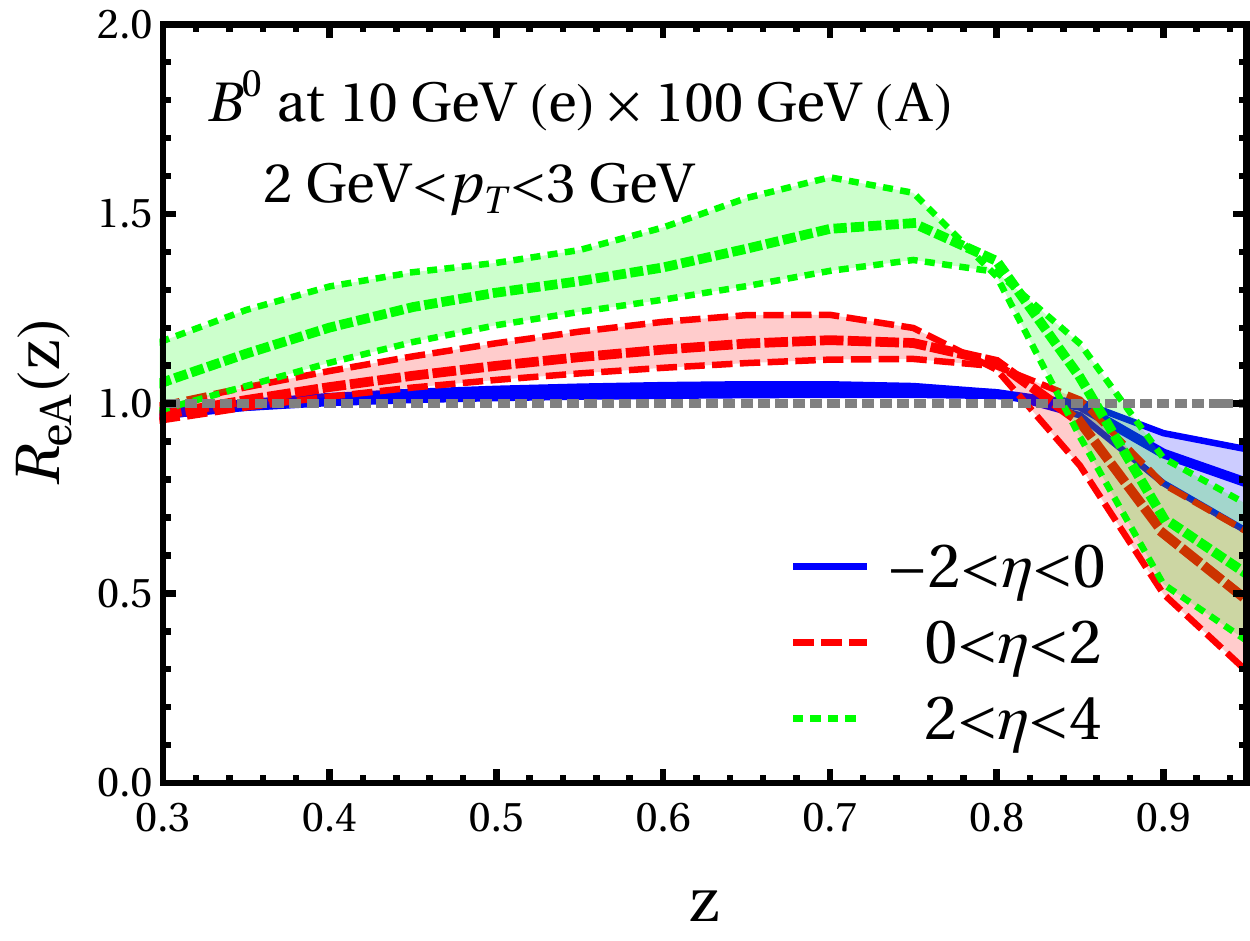}
 	\caption{ In-medium corrections for $D^0$ (left) and $B^0$ cross sections (right) as a function of the momentum fraction $z$ at the EIC in three rapidity regions. Left panel presents results for $D$-mesons and right panels is for $B$-mesons. The electron and proton/nucleus beam energies are 10 GeV$\times$100 GeV. }
 	\label{fig:zdisEIC_D0B0}
 \end{figure}
 
In contrast to light hadrons, the modification  of open heavy flavor in DIS reactions with nuclei,  such as the one for  $D^0$ mesons  and  $B^0$ mesons shown in  Fig.~\ref{fig:zdisEIC_D0B0},  is much more closely related to the  details of hadronization.  To investigate the nuclear medium effects, we study the ratio of the cross sections in electron-gold (e+Au) collision to the one in e+p collision.  We use the cross section of inclusive jet production for normalization that minimizes  the effect of nuclear PDFs.
\begin{equation}
R_{eA}^{h}(p_T,\eta,z)={\frac{N^{h}(p_T,\eta,z)}{N^{\rm inc}(p_T,\eta)}\Big|_{\rm e+ Au}}\Bigg / {\frac{N^{h}(p_T,\eta,z)}{N^{\rm inc}(p_T,\eta)}\Big|_{\rm e+p}} \, .
\label{RAatEIC}
\end{equation}
Here, $N^{\rm inc}(p_T,\eta)$ denotes the cross section of large radius jet production~\cite{Li:2020rqj}  with transverse momentum  $p_T$ and rapidity $\eta$. The observed  $R_{eA}(z)$ is qualitatively consistent with the  effective modification of fragmentation functions  even after their convolution with the PDFs and the perturbative hard part.  There is a significant suppression for large values of $z$, but it quickly evolves to enhancement for $z<0.65$ and $z<0.8$ for $D$-mesons and $B$-mesons, respectively.  The effect is most pronounced at forward rapidities and one finds that $R_{eA}^{h}$  as a function of $z$ is a more suitable observable  for  cold nuclear  matter tomography at the EIC than the  transverse momentum distributions' modification for hadrons in the laboratory frame alone. At smaller center-of-mass energies differential particle spectra fall faster with $p_T$, similar to what is observed in hadronic collisions \cite{Vitev:2004gn}, which further enhances the observed nuclear effects.   
Production of particles that contain strange quarks, such as kaons, can also be studied at the EIC~\cite{Chang:2014lla}.

 \subsubsection{Heavy meson reconstruction and physics projections}
 
 Due to the asymmetric nature of the collisions at the EIC, most of the final state hadrons are produced in the nucleon/nucleus beam going (forward) direction. A silicon vertex/tracking is critical to precisely measure these forward hadrons at the EIC. A LANL experimental team has produced conceptual designs of a Forward Silicon Tracker (FST) coupled to tracking in the central region to enable jet and heavy flavor physics at the EIC~\cite{Li:2020sru, Li:2020wyc, Wong:2020xtc}.  EIC Fun4all simulations were performed with both the Babar and BeAST magnets.  A $95$\% detection hit efficiency is used in both track and vertex reconstructions. In track reconstruction, the Kalman Filter algorithm is used and a $20$~$\mu$m vertex Gaussian smearing is applied to both $x$ and $y$ directions. 
The full simulation results, including momentum resolution and distance of closest approach resolution are applied to heavy meson reconstruction in physics simulation that are presented here. 

\begin{figure}
\centering
\includegraphics[width=0.99\textwidth]{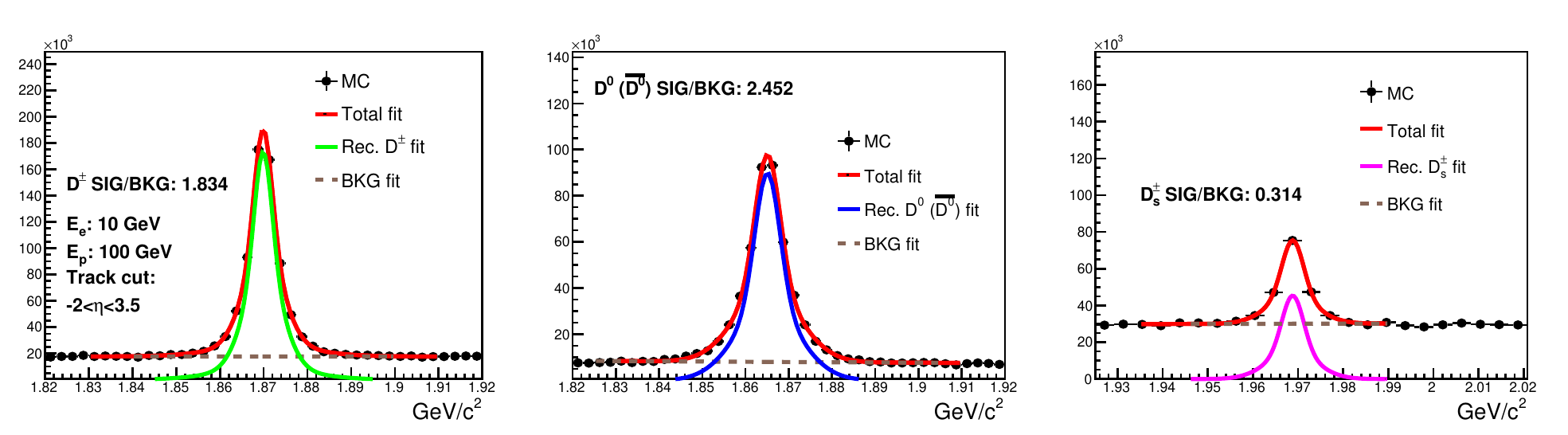}
\includegraphics[width=0.7\textwidth]{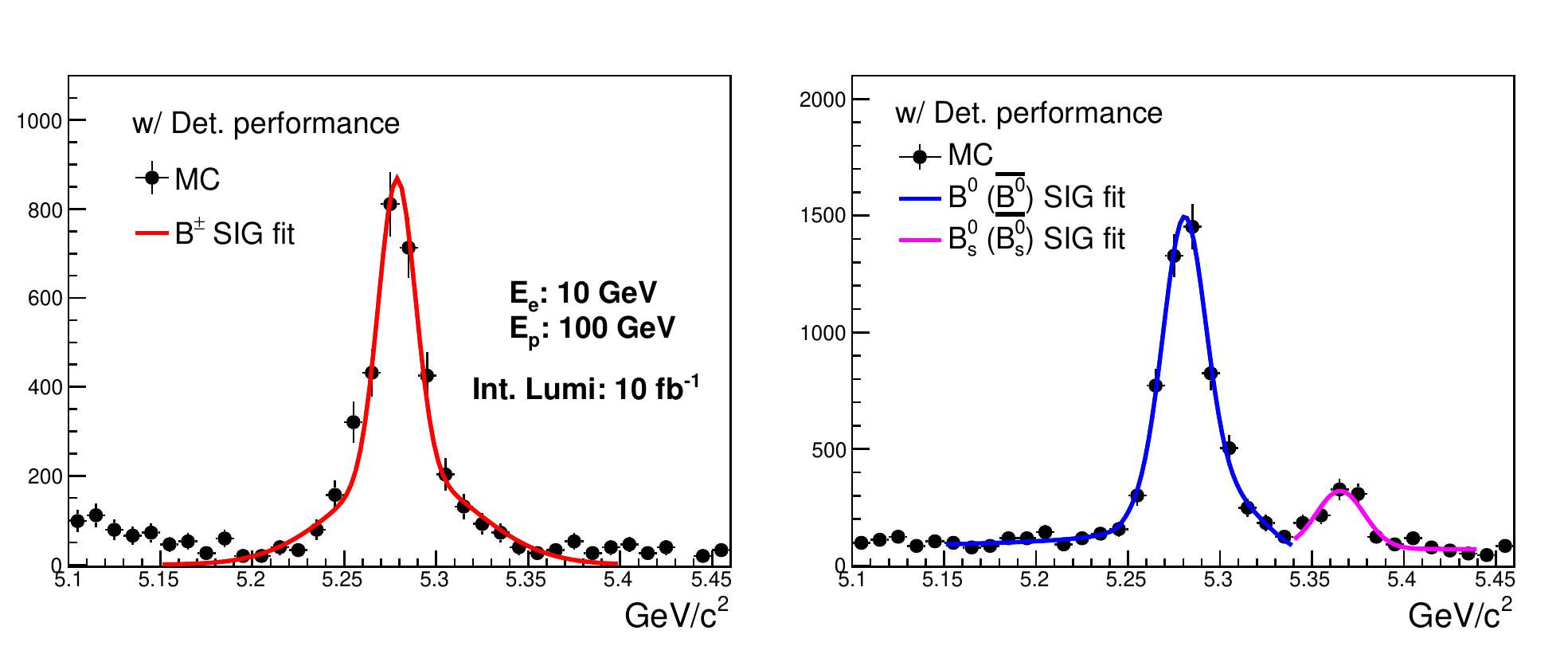}
\caption{\label{fig:HF_had_diff_1} Reconstructed D-meson and B-meson mass spectrum using the Forward Silicon Tracker  with the Beast magnetic field. Pixel pitch for both barrel layers and forward planes are selected at 20 $\mu m$. The integrated luminosity of $e+p$ collisions at $\sqrt{s} = 63$~GeV is 10~$fb^{-1}$.}
\label{fig:reconstructDB}
\end{figure}

Figure~\ref{fig:reconstructDB} shows the mass spectrum of fully reconstructed $D^{\pm}$, $D^{0}$ ($\bar{D^{0}}$), $D^{\pm}_{s}$, $B^{\pm}$, $B^{0}$ ($\bar{B^{0}}$) and $B^{0}_{s}$ ($\bar{B^{0}_{s}}$). For these heavy flavor hadron reconstructions charged tracks are required to have pseudorapidity within -2 to 4. Clear D-meson and B-meson signals have been obtained on top of the combinatorial backgrounds. The signal over background ratios and the reconstruction efficiency are listed in the associated panels. An integrated luminosity of  10~$fb^{-1}$ is assumed. 
In addition to heavy flavor meson reconstruction, we also looked for the heavy flavor baryon reconstruction (e.g. $\Lambda_{c}$). Although the combinatorial background is significantly higher than the D-meson mass spectrum, clear $\Lambda_{c}$ signal can be obtained. 

\begin{figure} 
\centering
\includegraphics[width=0.48\textwidth]{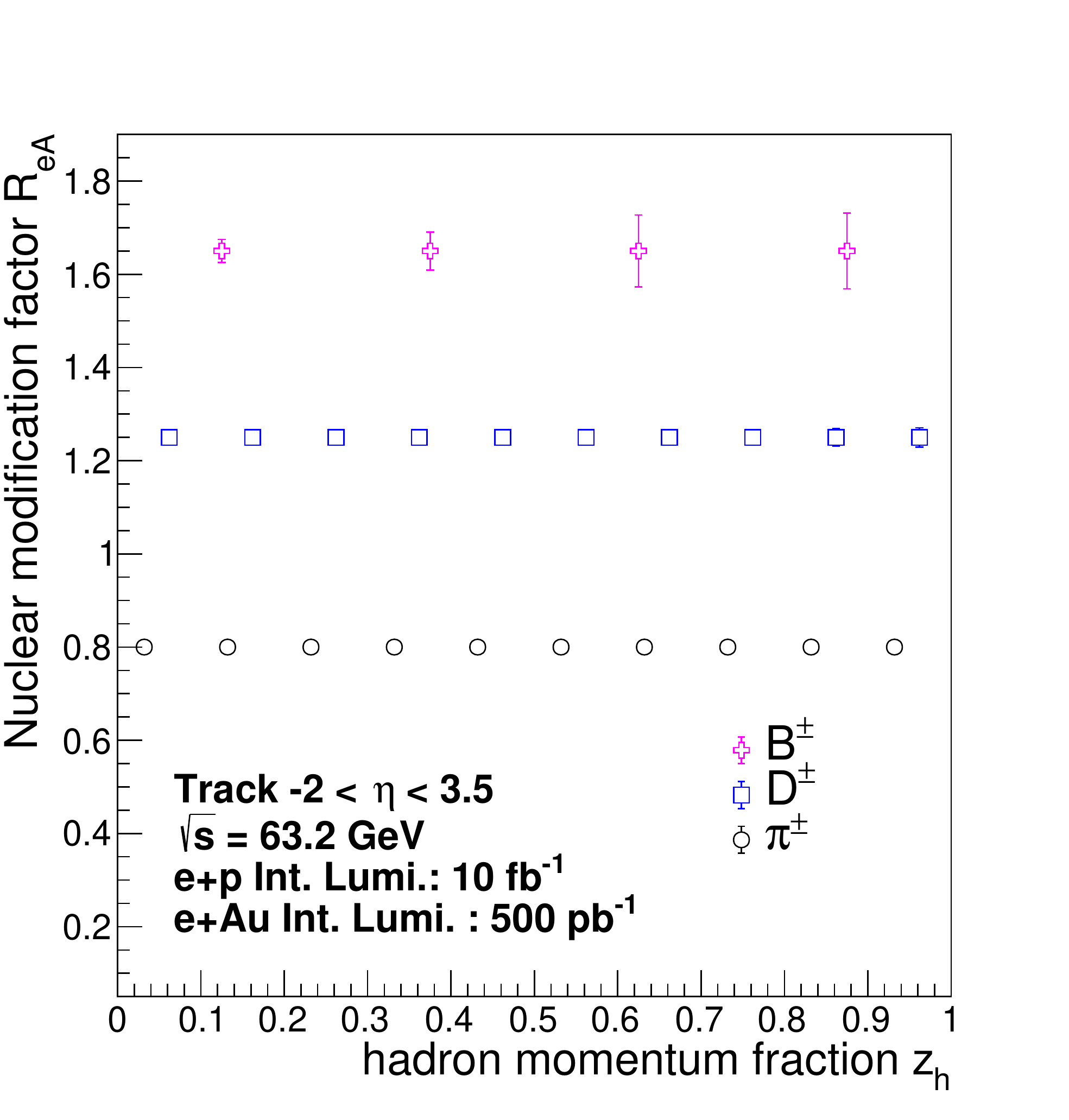}
\includegraphics[width=0.48\textwidth]{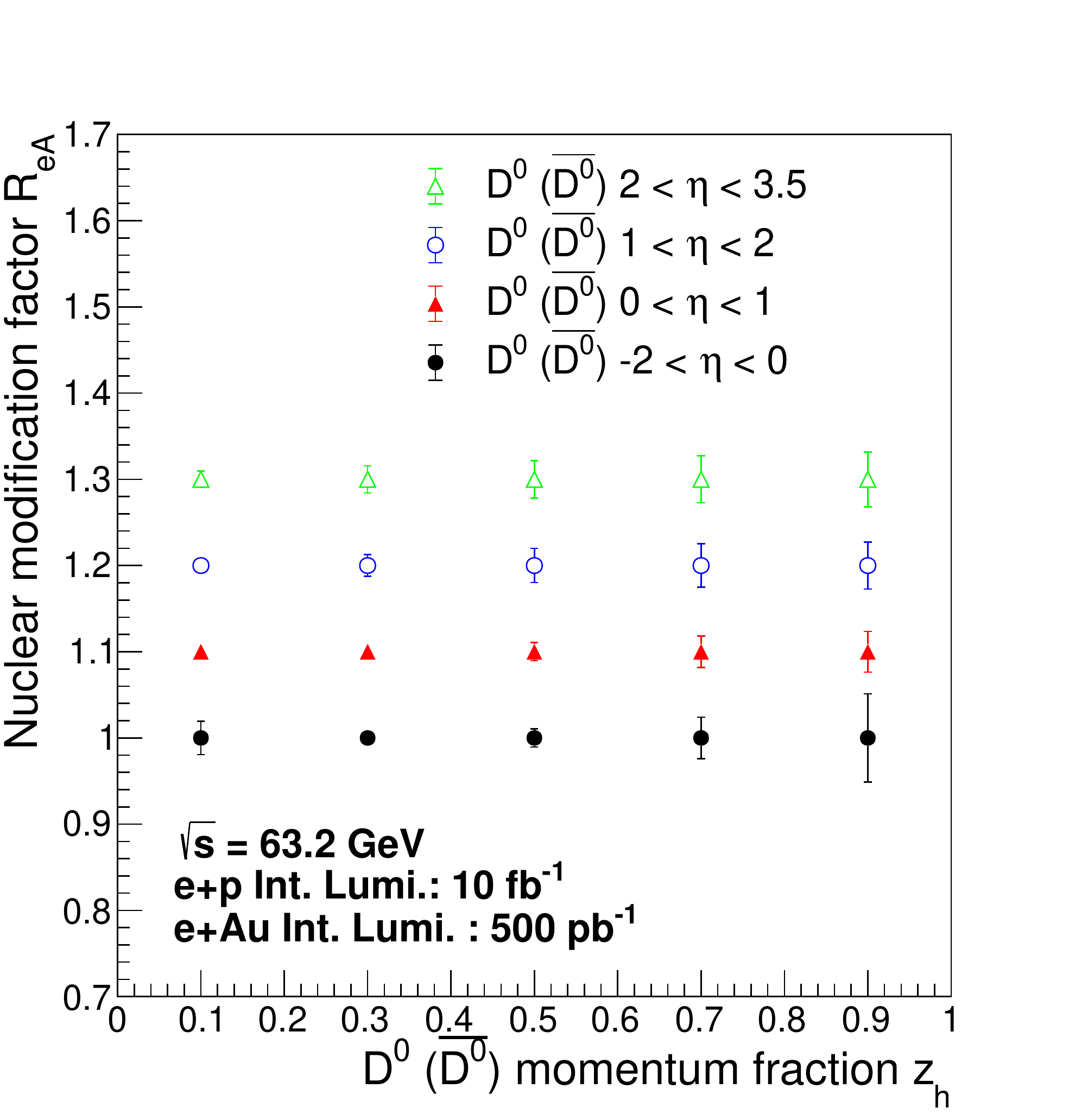}
\caption{Projected uncertainties for the nuclear modification factor $R_{eAu}$ for reconstructed flavor dependent hadrons versus the hadron momentum fraction $z_{h}$ for $\sqrt{s} = 63$~GeV.  (left panel). $R_{eAu}$ uncertainty projections for reconstructed $D^{0}$ ($\bar{D^{0}}$) in different pseudorapidity bins at the same center-of-mass energy (right panel).}
\label{fig:physicsReA}
\end{figure}

The nuclear modification factor $R_{eA}$ measurements for different flavor hadrons at the future EIC will not only explore both initial and final state effects on hadron production in nuclear medium, but also provide further information on hadronization process and its flavor dependence \cite{Li:2020zbk}. Figure~\ref{fig:physicsReA} gives the projected flavor-dependent nuclear modification factor for reconstructed flavor dependent hadron versus the hadron momentum fraction $z_{h}$ with detector performance derived from the FST design. The left panel shows that precise and differential measurements can be made for different hadron flavors. Even for B-mesons the modification due to final-state interactions can be clearly identified. The left panel shows the reconstructed  $D^{0}$($\bar{D^{0}}$)   $R_{eAu}$  in different pseudorapidity intervals which is essential to understand detector requirements as a function of $\eta$. Extensive studies with different magnetic field options, technology options and tracker designs have been performed~\cite{Li:2020sru, Li:2020wyc, Wong:2020xtc} to ensure that the physics can be delivered. It forms the basis of the reported tracking requirements.  

\subsubsection{Heavy flavor-tagged jet substructure at the EIC}

Jet substructure will also be an essential probe for studying a wide variety of QCD processes. Beyond the jet charge~\cite{Li:2020rqj}, observables such as jet shapes  and jet fragmentation functions can be used to study parton propagation through a nuclear environment and could be sensitive to both fragmentation and hadronization modification in a medium. The relatively low transverse momenta of reconstructed jets will enhance the role of the heavy quark mass in measurements of jet splitting functions~\cite{Li:2017wwc}.    Even though parton multiplicities at the EIC will be low, hadronization models, such as parton recombination, can also be tested by comparing jet substructure measurements in $e+p$ and $e+A$ collisions. This can be extended to a variety of mesons and baryons in jets, such as heavy flavor mesons or $\Lambda$ baryons. Studying the modification of hadronization and fragmentation with both light and heavy quark jets will further our understanding of mass dependent partonic interactions and energy loss with the medium. Techniques such as jet grooming can be used to tease out signals that are weaker than the ones observed in heavy-ion collisions.

\subsection{Particle production for identified hadron species}
\label{part2-subS-Hadronization-PartProd}

The ability to reliably identify different hadron species will be quintessential for many aspects of the physics program at the Electro-Ion Collider. Particle identification (PID) will be essential for determining the flavor of partons involved in the collision and elucidating how different quarks contribute to the integral properties of the nucleons and the nuclei.  PID will provide unique insights for TMD measurements through SIDIS studies, access to the strange sea quarks, and, of course, allows separating final state hadrons from leptons of the initial scatterings. But above all, PID capabilities will enable unprecedented access to the systematic studies of hadron formation processes, the hadronization.

While substantial efforts have been invested over past decades into studies of the nucleon structure, hadronization has received significantly less attention, with theoretical modeling of the process often not extended past in-vacuum collinear fragmentation of a single parton. The topic of hadronization was given renewed attention more recently, and received a more rigorous treatment, including  transverse momentum dependent FF, considerations for interferences between single parton and parton-gluon hadronizations, and di-hadron FF (see, for example, recent review~\cite{Metz:2016swz}).  However, other hadronization mechanisms have to be considered in understanding particle production in hadronic and lepton-hadron collisions. These include threshold production, string-breaking picture, and coalescence or recombination. The latter is arguably playing a critical role in particle production in higher-density environments, such as in collisions involving nuclei.  

It is known from the previous experimental observations that cold nuclear matter modifies the fragmentation patterns observed with identified particle species in a non-trivial way. Nuclear modification factors for pion and kaon fragmentation functions studied by HERMES in $\eA$   collisions show significant suppression with respect to “vacuum” reference, while protons suppression changes to enhancement in specific kinematic domain~\cite{ Airapetian:2007vu}. 

The enhancement in baryon production at intermediate transverse momenta, at some point termed “baryon-meson puzzle,” has been for years taken as one of the established QGP signatures and evidence for coalescence contributions in hadron production. This by itself begs the question: could the observed baryon enhancement in $\eA$   collision arise from a similar type of hadronization process?  Additionally, substantial baryon enhancements have been reported not only in $\AA$    collisions compared to $\pp$   across all flavor sectors (most recently in charm ~\cite{Adam:2019hpq, Sirunyan:2019fnc}), but in $\pp$  collisions over what is measured in $e^{+}e^{-}$ data. The enhancements in $\pp$  data are also found to be multiplicity dependent. These observations leave open questions about the details and balance of different hadronization contributions for future experiments to explore. 
The EIC is positioned to make groundbreaking progress in our understanding of hadronization. Identified particle corrections and ratios, particularly baryon to meson ratios across different flavors, are sensitive to hadronization details. Having broad kinematic coverage and energy lever-arm will allow constraining relative contributions from competing processes. Systematic studies of baryon-to-meson ratios with different ion species may offer sensitivity to density dependence of coalescence/recombination contributions.  Finally, semi-inclusive measurements of identified hadrons with or in jets will provide differentiating capabilities on light, strange and heavy quarks, as discussed in the following sections.

\subsection{Production mechanism for quarkonia and exotic states}
\label{part2-subS-Hadronization-Quarkonia}

\begin{figure}[ht!]
\begin{center}
\includegraphics[width= 0.9 \textwidth]{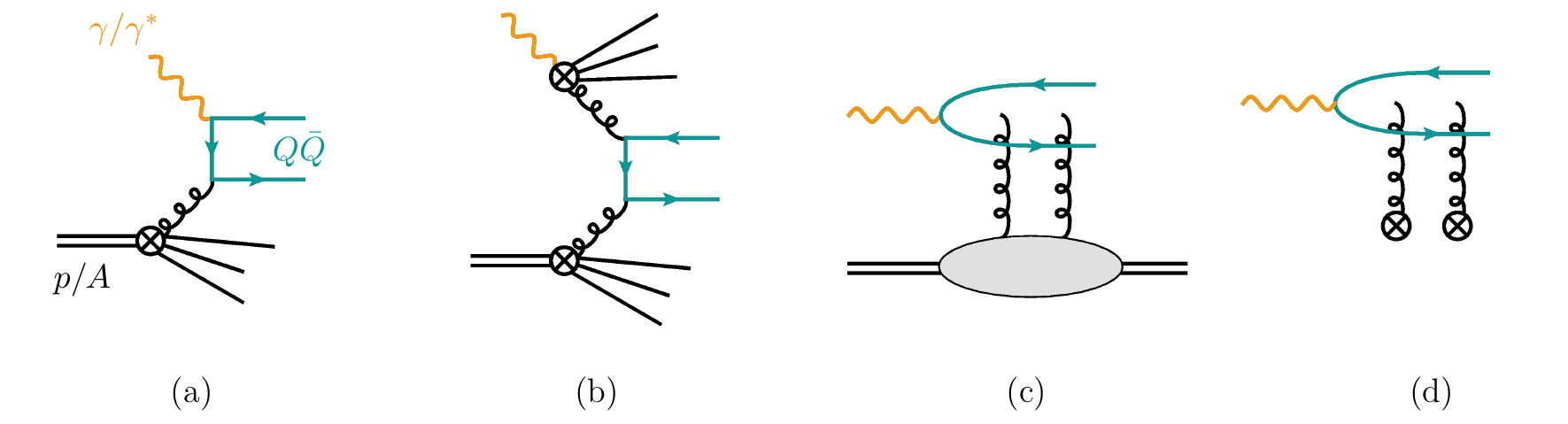}
\caption{\label{fig:quarkonium} Illustrative examples of quarkonium production mechanism in \ep and \eA colliders: (a) Direct photo/lepto-production, (b) resolved-photon quarkonium production, (c) exclusive quarkonium production, and (d) heavy quark pair production and subsequent Glauber/Coulomb gluon exchanges with nuclear matter. }
\end{center}
\end{figure}

Quarkonia, $\mathcal{Q}$, are the bound states of a heavy quark and the corresponding antiquark.
Due to the large mass of heavy quarks, quarkonium production entangles perturbative and non-perturbative QCD in a unique way. A quarkonium state is assumed to be produced in two steps. First, the perturbative generation of a heavy quark-antiquark pair with total momentum-squared near the bound state mass-squared, then the pair hadronize into the quarkonium state non-perturbatively. Since their discovery, three main production formalisms have been proposed: i) the color evaporation model (CEM)~\cite{Fritzsch:1977ay, Halzen:1977rs}, ii) color singlet model (CSM)~\cite{Chang:1979nn}, and iii) the effective theory of non-relativistic-QCD (NRQCD)~\cite{Bodwin:1994jh} with the
Lagrangian
\begin{eqnarray}
  \label{eq:L-vNRQCD}
  \mathcal{L}_{\text{vNRQCD}} &=& \sum_{{p}} \psi^{\dag}_{{p}} \left ( iD^0 - \frac{({\mathcal{P}}-i{D})^2}{2m} \right ) \psi_{{p}} +\mathcal{L}^{(2)} + (\psi \to \chi, T \to \bar{T}) \nonumber  \\
 && +  \mathcal{L}_s(\phi,\bar{\phi},A_q^{\mu})  +  \mathcal{L}^{V}(\psi,\chi,A_q^{\mu}) \;,
\end{eqnarray}
where $\psi$ denotes the heavy quark field and $\chi$ the corresponding antiquark. The Lagrangian terms $\mathcal{L}^{(2)}$ are higher order terms, $\mathcal{L}_s$ is the soft gluon and ghost part of the Lagrangian, and $ \mathcal{L}^{V}$ contains the potential terms.
While all three approaches assume that quarkonia are produced from the hadronization of a heavy quark-antiquark pair, they differ in how the probability of this happening depends on other quantum numbers. In the CEM one assumes uniform probability for all other quantum numbers, where in NRQCD the probability depends on the angular momentum and color configuration of the pair. In CSM the only non-zero probability is assigned to the leading color singlet combination. While all these frameworks enjoyed partial success, the theory of quarkonium production still remains an open question. For recent reviews of quarkonium physics see Refs.~\cite{Lansberg:2019adr,Chapon:2020heu}. 

\subsubsection{Precision quarkonium physics at the EIC}

Lepton-nucleon/nucleus collisions  constitute an excellent laboratory for studies of quarkonium production. They provide a simpler and cleaner environment than hadronic collisions, yet are far richer than electron-positron annihilation. Quarkonia can be produced  either through photo-production ($Q^2 \simeq 0$) or lepto-production ($Q^2> 1$ GeV) processes. In these two cases the direct, resolved, and diffractive/exclusive productions (see fig.~\ref{fig:quarkonium}(a), (b), and (c) respectively) can be relevant depending on the kinematic regimes  considered.  At HERA both photo-production~\cite{Aid:1996dn, Breitweg:1997we, Aaron:2010gz, Chekanov:2002at, Adloff:2002ex, Abramowicz:2012dh,Chekanov:2009ad} and lepto-production~\cite{Aaron:2010gz, Adloff:1999zs, Adloff:2002ey, Chekanov:2005cf,Sun:2020stm} have been studied. Some of the variables typically  used in these studies are: the inelasticity, $y = P\cdot p_{\mathcal{Q}} / P\cdot q$, the quarkonia transverse momenta $p_T$ and $p_T^{\star}$ in the laboratory and the $\gamma/\gamma^*$-proton c.m.\ frame respectively, as well as the corresponding rapidities. 
Although the data collected at HERA had a major impact on our current view of quarkonium production, the interpretation of these data remains a subject of debate until this day~\cite{Kramer:1995nb, Butenschoen:2009zy, Flore:2020jau, Fleming:2006cd}.  At the EIC the hadronization of quarkonia can be studied in both the laboratory frame and the $\gamma/\gamma^*$-proton center-of-mass frame, and in various kinematic regimes. We thus foresee that with the high luminosities available at the EIC, we could also obtain multi-differential distributions, which would help establish a global picture of quarkonium hadronization.  

The various production channels can be disentangled by considering different kinematic regimes, establishing this way DIS as a prime framework for the study of quarkonium production channels. While at small values of inelasticity ($y \lesssim 0.3$) the resolved process dominates, at intermediate and large values of $y$ direct production is paramount. At the same time, diffractive processes are expected to be enhanced at the kinematic endpoint $y \simeq1$, but drop quickly toward smaller $y$.   Quarkonium photo-production near threshold is also related to the trace anomaly and the origin of the proton mass~\cite{Gryniuk:2016mpk,  Hatta:2019lxo, Gittelman:1975ix, Kharzeev:1998bz, Frankfurt:2002ka}. This is one of the fundamental  questions to be answered at the EIC as discussed in more detail in sec.~\ref{part2-subS-PartStruct-Mass}. 

At small transverse momentum, $p_T^{\star}$, quarkonium  production can be approached from the TMD factorization perspective. Several studies have considered both polarized and unpolarized proton beams~\cite{Bacchetta:2018ivt, DAlesio:2019qpk, Kishore:2019fzb, Boer:2020bbd, Echevarria:2020qjk}.  Recent theoretical developments in NRQCD~\cite{Beneke:1997qw,  Fleming:2002rv,  Fleming:2003gt, Fleming:2006cd,  Leibovich:2007vr, Echevarria:2019ynx, Fleming:2019pzj} incorporate the leading perturbative effects from soft radiation to all orders in the strong coupling expansion -- allowing us this way to safely study the non-perturbative effects, which can be accessed in the  semi-inclusive process in the $y\to1$ and/or $p_T^{\star} \to 0$ limits. While the $ y\to1$ limit involves the quarkonium fragmentation shape functions, the $p_T^{\star} \to 0$ limit involves the recently introduced TMD shape functions. There has been no phenomenological extraction of the TMD shape functions. Meanwhile, it has been proposed that exclusive quarkonium production can be understood through the formalism of GPDs and Wigner functions~\cite{Cui:2018jha, Chen:2019uit}.  Exclusive photoproduction of exotic states, such as the so-called $XYZ$ mesons, is described in Section~\ref{part2-subS-Hadronization-Spectroscopy}.

\subsubsection{Production of quarkonia and exotics in \texorpdfstring{\eA}{eA} collisions}

The EIC will also offer the opportunity to observe quarkonium production in \eA collisions where one can study the interactions with nuclear matter and the formation of quarkonia in a nuclear medium. The study of nuclear effects in quarkonium production is an emerging  field of nuclear physics where EIC measurements are expected to play a major role in our understudying of these effects.  Most of the recent field-theoretic developments rely on the effective theory of NRQCD. In a recent formulations of NRQCD~\cite{Makris:2019ttx, Rothstein:2018dzq} the Glauber/Coulomb gluon interactions with heavy quarks, Fig.~\ref{fig:quarkonium} (d), in the non-relativistic limit, are incorporated into the effective theory. The Lagrangian of NRQCD$_{\rm G}$ is constructed by adding to the vNRQCD Lagrangian the additional terms that encode such interactions with quark and gluon sources through (virtual) Glauber/Coulomb gluons exchanges. We may then write,
\begin{equation}
  \mathcal{L}_{\text{NRQCD}_{\rm G}} = \mathcal{L}_{\text{vNRQCD}} + \mathcal{L}_{Q-G/C} (\psi,A_{G/C}^{\mu,a}) + \mathcal{L}_{\bar{Q}-G/C} (\chi,A_{G/C}^{\mu,a})\;,
\end{equation}
where the effective fields $A_{G/C}^{\mu,a}$ incorporate the information about the nuclear medium. 
This provides a systematic and formal approach to the inclusion of nuclear effects on quarkonium propagation~\cite{Sharma:2012dy, Aronson:2017ymv}. A similar approach has been applied successfully to the propagation of jets through nuclear medium~\cite{Ovanesyan:2011xy, Chien:2014nsa}.  In a different direction, recently the formalism of open quantum systems  has been gaining a significant attention~\cite{Akamatsu:2014qsa, Brambilla:2017zei, Yao:2018nmy, Akamatsu:2020ypb, Yao:2020eqy, Das:2018xel}. Although primarily formulated in the context of quark-gluon plasma, these formalisms can also be applied to cold nuclear matter effects.

\begin{figure}[ht]
\begin{center}
\includegraphics[width= 0.9 \textwidth]{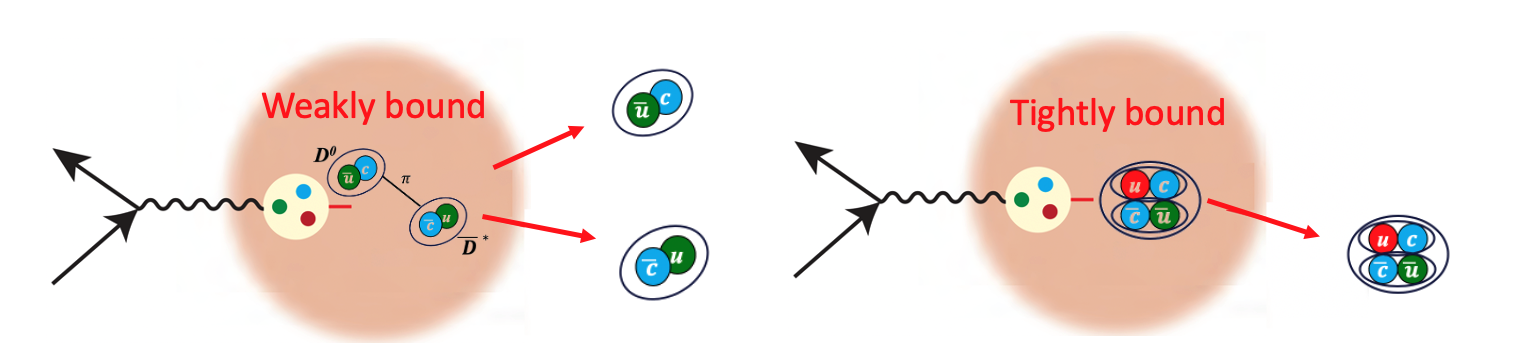}
\includegraphics[width= 0.7 \textwidth]{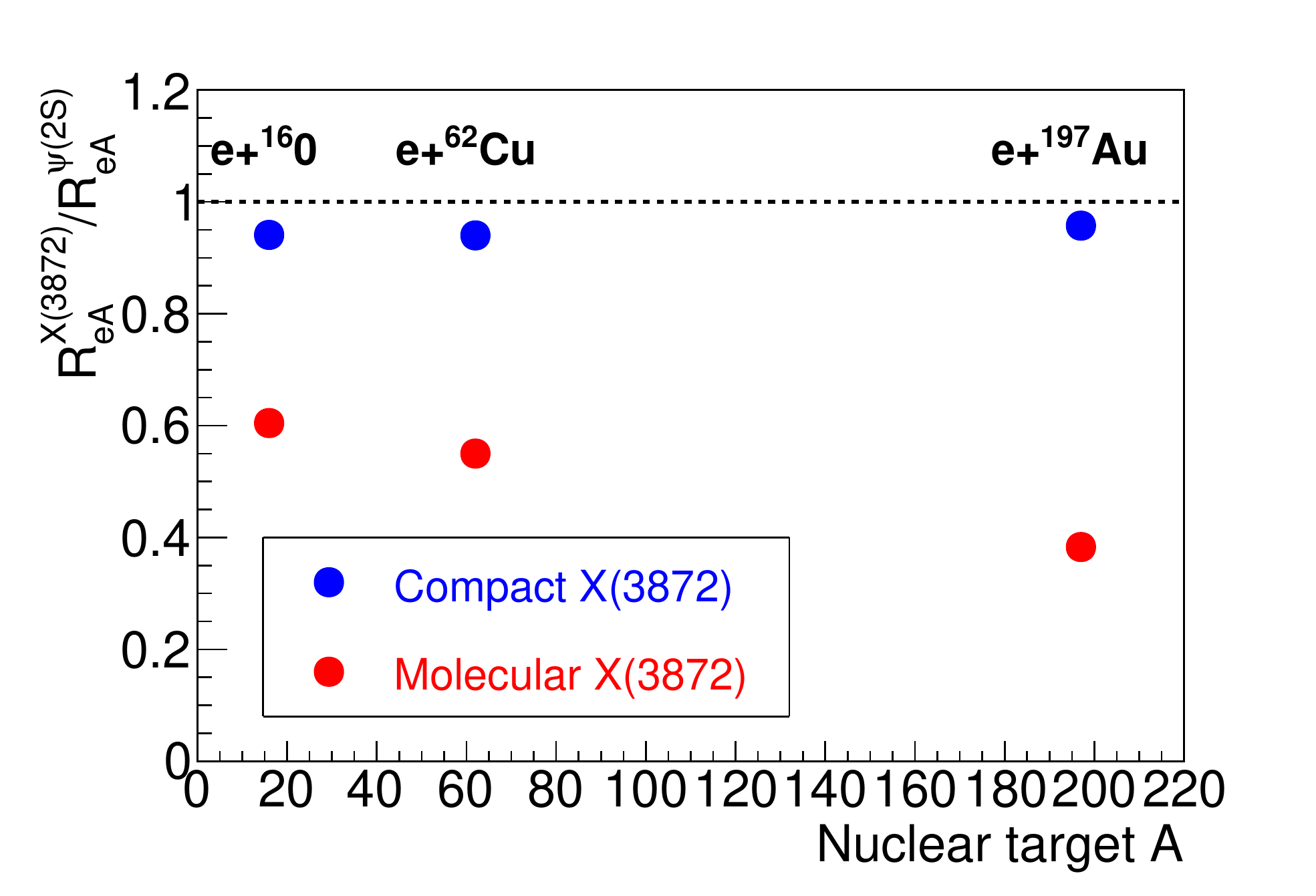}
\caption{\label{fig:exotics} Top panel: illustration of weakly and tightly bound tetraquark state propagation in a large nucleus.  Bottom panel: the ratio of nuclear modification factors $R_{eA}$ for X(3872) to $\psi(2S)$, for two different assumptions of the X(3872) structure.}
\end{center}
\end{figure}

Quarkonia produced in \eA collisions at the EIC can hadronize inside the nucleus.  As the heavy $Q\bar{Q}$ pairs propagate through cold nuclear matter, they will be subject to disruption via interactions with partons in the nucleus, which can lead to suppression with respect to \ep collisions.  These effects are distinctly different from those measured in \pA collisions at RHIC and the LHC, since at those colliders the crossing time is shorter than the charmonium formation time. Charmonium production inside the nucleus has been previously studied in fixed-target \pA collisions, where the crossing time is sufficiently long that hadronization also occurs inside the nucleus.  These measurements showed that the relatively weakly bound $\psi(2S)$ is suppressed more than the $J/\psi(1S)$ state~\cite{Leitch:1999ea, Alessandro:2006jt}.  This is understood phenomenologically in terms of the size of the state: weakly bound charmonia with a larger radius will effectively sample a larger volume of the nucleus as they propagate outwards, and therefore have a higher probability of interacting and being disrupted~\cite{Arleo:1999af}.  The low backgrounds at the EIC will allow these models to be tested on higher charmonium states which are difficult to reconstruct at hadron-hadron colliders.

These effects can also be used to discriminate between models of exotic hadron structure.  Multiple candidates for tetra- and pentaquark states have been identified, such as the X(3872) and the $P_{c}^{+}$ states, but there is no consensus on whether these states are hadronic molecules or compact multiquark states~\cite{Olsen:2017bmm}.  Embedding these resonances in the nuclear medium provides a new environment to study their properties.  From previous experience with conventional charmonium states, one would expect that large, weakly bound hadronic molecules would undergo significantly more disruption while traversing the nucleus than a compact state, as shown in Fig.~\ref{fig:exotics}.  

\subsection{New particle production mechanisms}
\label{part2-subS-Hadronization-NewPPMech}

\begin{figure}[t]
\includegraphics[width=0.91\textwidth]{
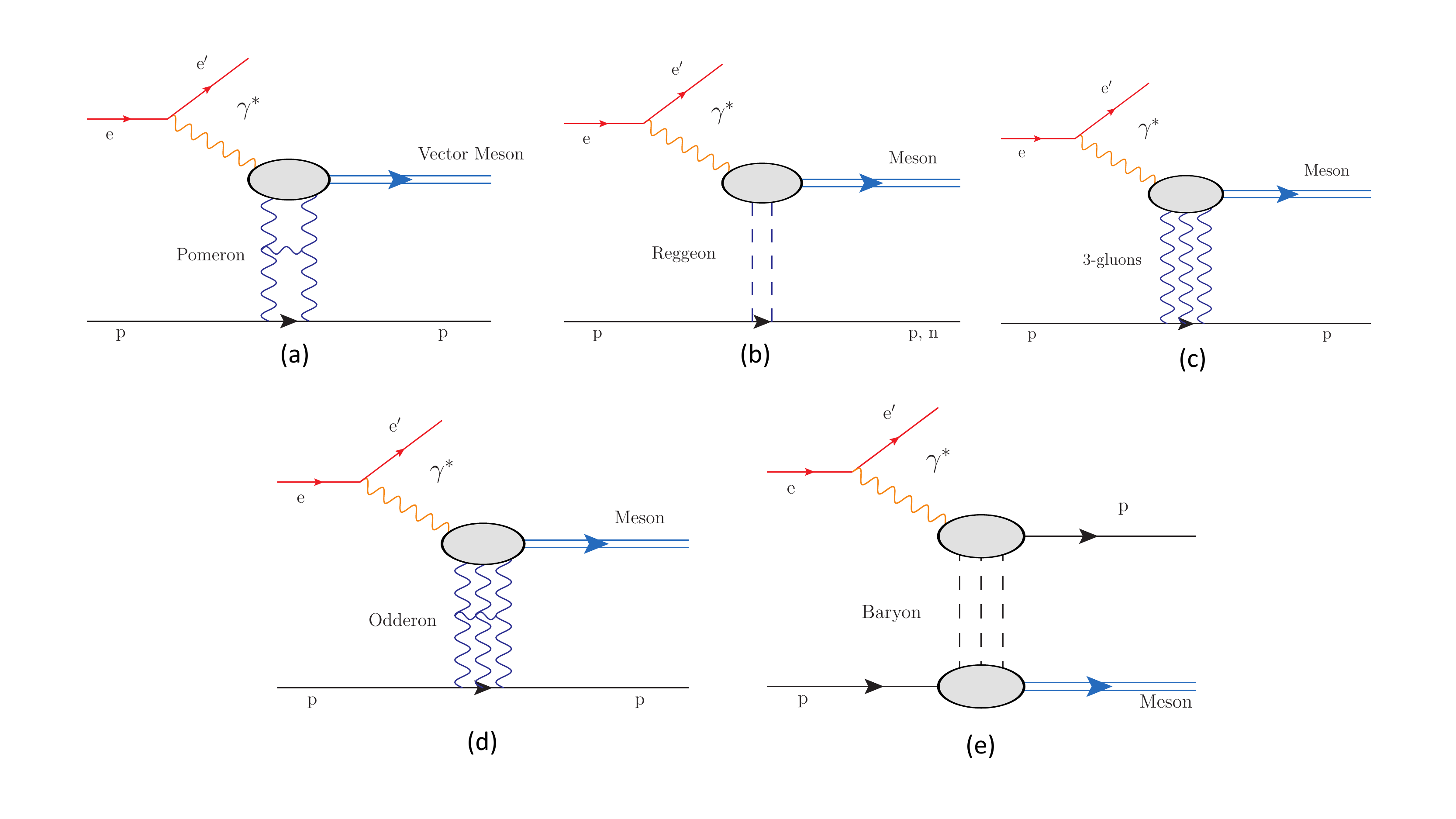}
\caption{The exclusive photoproduction mechanisms discussed in this section. They include (a) Pomeron exchange, (b) Reggeon exchange,  (c) 3-gluon exchange, (d) Odderon exchange  and (e) baryon exchange.}
\label{fig:diagrams}
\end{figure}

Exclusive vector meson production is usually modelled as occurring via the exchange of a Reggeized particle, either a Pomeron or a Reggeon.  Pomeron and Reggeon exchange models have provided excellent fits to a wide range of fixed target and HERA data \cite{Crittenden:1997yz}.  However, other types of exchange are possible; this section will explore some of the possibilities, which are shown in Fig. \ref{fig:diagrams}. 

Pomeron exchange  represents the absorptive part of the cross-section, so the Pomeron has the same quantum numbers as the vacuum, $J^{PC}=0^{++}$.  This naturally explains why vector mesons are predominantly produced; the final state has the same quantum numbers as the incident photon.  In pQCD Pomerons can be treated as a gluonic ladder built on two gluon exchange, and obeying the BFKL evolution equations \cite{Balitsky:1978ic, Kuraev:1977fs}.  Almost uniquely, the cross-section for Pomeron-mediated reactions rises with increasing energy.

Reggeons normally represent summed meson trajectories (baryon trajectories will be discussed below), so carry a wider range of quantum numbers, including charge.  Thus, they can lead to a wide range of final states; this makes photon-Reggeon fusion an attractive venue for meson spectroscopy.  The cross-section for Reggeon-mediated reactions drops with increasing energy, so these reactions are best studied at energies that are not too high.

This section will explore alternate exchange mechanisms for particle production, involving the Odderon (the three-gluon analog of the Pomeron, with negative charge-parity) and backward production involving baryon trajectories.   It will also consider some unique facets of near-threshold production. 

\subsubsection{Odderon exchange}

\begin{figure}[t]
\includegraphics[width=0.91\textwidth]{
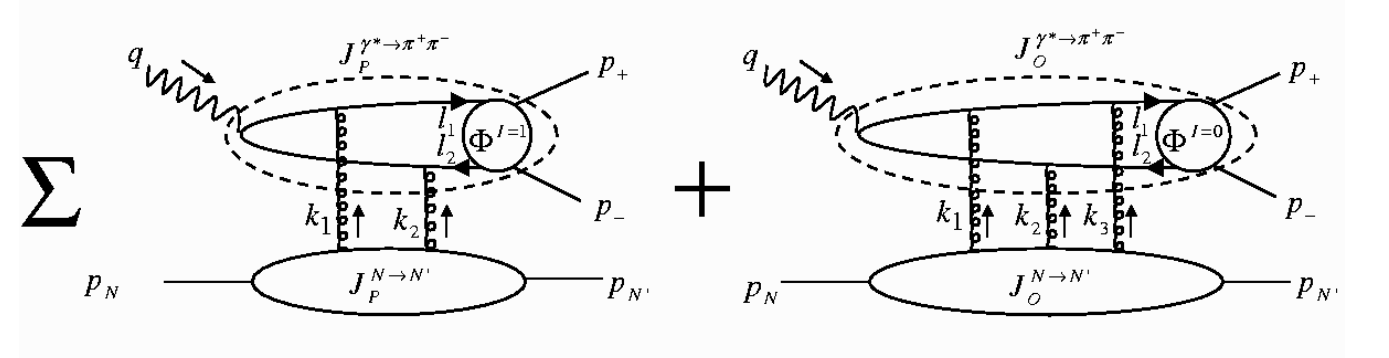}
\caption{$\pi^+ \pi^-$ pairs may be diffractively produced through Pomeron and Odderon exchange. The interference of the two amplitudes leads to a characteristic signature of a charge asymmetric observable.}
\label{fig:interfere}
\end{figure}
Hadronic reactions at low momentum transfer and high energies for charge-odd exchange are described in Regge language in terms of Odderon exchange, Fig. \ref{fig:diagrams}(d), which is in QCD and at Born level a three gluon color singlet. Although mandatory to explain the difference between $pp$ and $\overline{p}p$ scattering, and a natural object in QCD~\cite{Bartels:1980pe,Jaroszewicz:1980mq,Kwiecinski:1980wb}, its properties remain quite elusive.  At an EIC, the Odderon might manifest itself via the production of the $f_2(1270)$ via photon-Odderon fusion (the same state may also be produced in two-photon interactions).

Exclusive $\pi^0$ electroproduction data at HERA indicate a small magnitude of the hard Odderon - exchange cross-section. However, with the high luminosity of the EIC one may be able to observe hard $C$-odd three gluon exchange in large-$|t|$ exclusive production of a pseudo-scalar $J^{PC} = 0^{-+}$
meson. Examples include the production of light $\pi^0, \eta,
\eta\prime$ at high $Q^2$ or of heavy $\eta_c, \eta_b$ at low
$Q^2$~\cite{Czyzewski:1996bv, Engel:1997cga, Dumitru:2019qec}. The
latter would be rather challenging to observe, not only due to the
low cross section but also because of the large background of $\eta_c$ from radiative $J/\Psi \to \eta_c + \gamma$
decays~\cite{Klein:2018ypk}. These decays involve a soft and easy to miss photon. $\eta_c$ from the photon-Odderon process should have a harder $|t|$ spectrum, so the large $|t|$ region should be a good place to search for this process \cite{Dumitru:2019qec,Klein:2018ypk}.  Analogous processes with a heavy ion
target are largely unexplored.

The study of observables where Odderon effects are present at the amplitude level would therefore promise increased sensitivity to the rather small normalization of this contribution. One such observable involves charge asymmetries in open charm production \cite{Brodsky:1999mz} or two-pion electroproduction with the exchange of a soft Odderon \cite{Ginzburg:2002zd,Ginzburg:2004qc} (shown in Fig. \ref{fig:interfere}) or a hard Odderon \cite{Hagler:2002nh, Hagler:2002nf}. In this latter process
\begin{equation}
eN\rightarrow e\pi^+\pi^- N
\end{equation}
at high energy, large $Q^2$, modest pion pair invariant mass $ M_{\pi \pi}^2 =O(1 $GeV$^2$) and large rapidity gap between the pion pair and the final nucleon, one may factorize the Pomeron-Odderon proton impact factor from a perturbatively calculable hard subprocess where the $\pi^+ \pi^-$ pair  is described by a generalized distribution amplitude (GDA) \cite{Diehl:1998dk,Polyakov:1998td,Diehl:2000uv}. The interference of the C-even and C-odd amplitudes leads to measurable charge asymmetries with a characteristic  $ M_{\pi \pi}$ dependence, with a magnitude  large enough to be detected if the  relative strength of the Odderon to Pomeron couplings is not unexpectedly small.

Lastly, we note that $C$-odd three gluon exchange has also been related to the dipole gluon Sivers TMD function of a transversely polarized proton~\cite{Yao:2018vcg}. These functions have been discussed in sec.~\ref{part2-subS-SecImaging-Wigner}.

\subsubsection{Exclusive backward (\texorpdfstring{$u$-channel}{u-channel}) production}

\begin{figure}[thb]
\includegraphics[width=0.91\textwidth]{
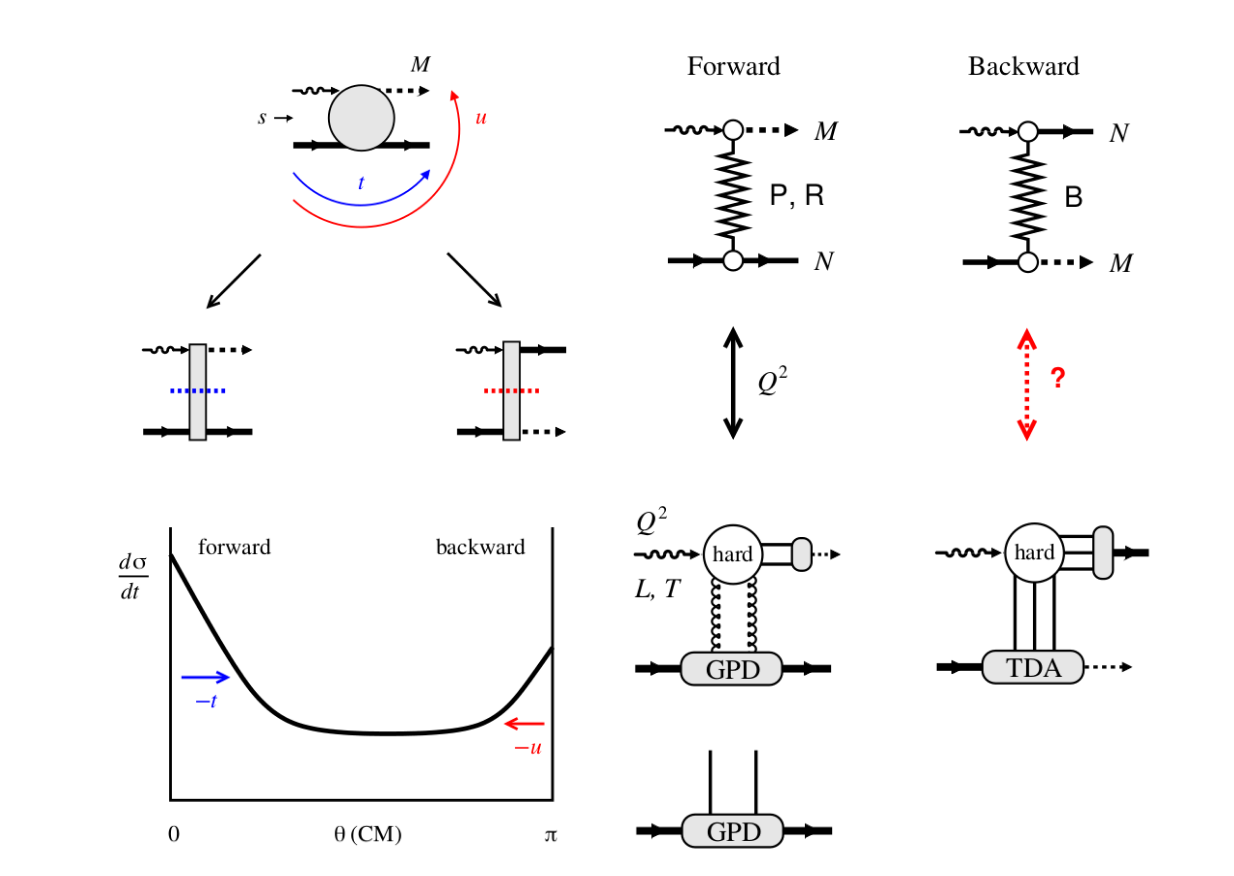}
\caption{(left) Soft-hard-soft structure transition.  (right) Forward-backward factorization scheme.}
\label{fig:transition}
\end{figure}

In Reggeon exchange events, the cross-section is usually largest at small $|t|$, with $d\sigma/dt\propto\exp(-b|t|)$, where $b$ is related to the square of the transverse interaction radius.  In the 1970s fixed-target experiments made a surprising discovery that, at very large $|t|$,  there is an increase in cross-section near the maximum  possible $|t|$, corresponding to the region of small $u$, as is shown in the lower-left part of Fig. \ref{fig:transition}.  This is known as backward production, because, in the center of mass frame, the produced meson recoils, while the struck nucleon recoils.  In very high energy collisions, such as at the EIC, the produced meson will have a rapidity near that of the incident ion beam, while the final state nucleon will be more central.

This can be easily accommodated in a Reggeon framework by allowing the exchange of baryon number, {\it i. e.} baryon trajectories.  This effectively swaps $u$ and $t$, and may explain the photoproduction data on backward production.    A comprehensive review article on the Regge model and its success in describing photoproduction of mesons can be found in Ref.~\cite{Laget:2019tou}.



In this paradigm, it is relatively easy to extrapolate fixed-target data on backward photoproduction upward to EIC energies.  This is easiest for the $\omega$ meson, for which there is data at sufficiently different photon energies to be able to fit for the energy dependence.  The photon-Reggeon component for $\omega$  production (with the photon-Pomeron contribution removed) can be parameterized as
\cite{Crittenden:1997yz,Klein:1999qj}: 
\begin{equation}
\frac{d\sigma}{dt}= A(s/{\rm 1 GeV})^B \exp(-Ct)
\end{equation}
where  $A \approx 18\ \mu b/{\rm GeV}^2$ and $B=-1.92$ for the $\omega$ and, from HERA data \cite{Derrick:1996yt}
$C\approx 10$ GeV$^{-2}$.  
In contrast, the backward-production data can be parameterized
\begin{equation}
    \frac{d\sigma}{du}= A(s/{\rm 1 GeV})^B  \exp(-Cu)
\end{equation}
where $A\approx 4.4 \mu b$/GeV$^2$, $B=-2.7$ and $C=-21$ GeV$^{-2}$.  Compared to forward photoproduction by Regge exchange, this formula has a constant about four times smaller, a faster drop-off with energy, and a somewhat larger $C$.   The $C$ values may not be directly comparable, because the fixed-target backward-production experiments were at much lower photon energies, so $t_{\rm min}$ was much larger; this could have led to a larger $C$ than would be seen at larger beam energies.

The relative $A$, $B$ and $C$ all lead to smaller backward to forward cross-section ratios at EIC energies.  However, the cross-sections are still large enough  for backward production to be easily observable. 

The final state consists of a proton at mid-rapidity, and an $\omega$ that is typically near the beam rapidity.  The decay $\omega\rightarrow\pi^0\gamma$ could be studied with far-forward calorimetry.  At lower proton beam energies, the $\omega$ moves further away from the beam, so may be easier to detect. More importantly, if one chooses interactions where the proton is produced closer to the incident hadron direction, the $\omega$ is shifted further from the beam.  In all cases, backward production studies require good far-forward instrumentation. 

\subsubsection{\texorpdfstring{$u$-Channel}{u-channel} exclusive meson electroproduction}

Exclusive electroproduction of mesons from the photoproduction to a large $Q^2$ above the resonance region, is another good handle to study the Regge exchange reactions.  The pioneer experimental and phenomenological effort from JLab \cite{Park:2017irz, Li:2019xyp} raise further questions: what are forward-backward cross section ratios in other $u$-channel electroproduction interactions such as $\pi^0$, $\pi^\pm$, $\rho$, $\eta$, $\eta$ and $\phi$?  Could the $t$-channel phenomenology recipe for mapping out the $W$ and $x$ dependence, be applied to the $u$-channel interactions? How would the $u$-channel interactions factorize? These important questions form the core bases for the future studies.

The large acceptance, wide kinematics (in $Q^2$ and $W$) and forward tagging capability at the EIC provide a great opportunity to study the near-forward and near-backward electroproduction of all mesons simultaneously.

Combining the data collected at JLab 12 GeV and EIC, we aim to accomplish the following objectives to unveil the complete physics meaning of $u$-channel interactions:
\begin{itemize}
\item At low $Q^2$ limit: $Q^2 < 2$ GeV$^2$, mapping out the $W$ dependence for electroproduction of all mesons at near-backward kinematics.
\item Extracting the $u$-dependence ($\sigma \propto e^{-b \cdot u}$) as a function of $Q^2$. This could be used to study the transition from a “soft” Regge-exchange type picture (transverse size of interaction is of order of the hadronic size) to the “hard” QCD regime.
\item Studying the model effectiveness between the hadronic Regge based (exchanges of mesons and baryons) and the partonic description through Transition Distribution Amplitudes (exchanges of quarks and gluons), is equivalent to studying the non-perturbative to perturbative QCD transition.

\end{itemize}

\subsubsection{Three-gluon exchange and near-threshold production}

Near threshold, other quarkonium production mechanisms may visibly contribute.   One involves three-gluon exchange with a target, as in Fig. \ref{fig:diagrams}(c)
Three-gluon exchange is expected to be subdominant at high energies, but might be visible near threshold. The GlueX experiment at JLab has studied $J/\psi$ production near the energy threshold, and found that the cross-section was above that expected in a two-gluon exchange model, but consistent with the sum of two-gluon plus three-gluon exchange \cite{Ali:2019lzf}.  GlueX also searched for narrow peaks in the $J/\psi p$ cross-section with increasing photon energy, and found none.  From this, they set limits on pentaquark production, eliminating some (not all) pentaquark models. 

Near-threshold production of quarkonia is sensitive to the quarkonium-nucleon potential\cite{Strakovsky:2020uqs}. The quarkonium may be treated as a dipole, with length inversely proportional to mass.  The potential is itself sensitive to the internal structure of the nucleon target. The potential is often quantified in terms of the meson-nucleon scattering length. This scattering length should decrease with increasing vector meson mass, if, as expected, the potential decreases for small dipoles. The proton - vector meson coupling involves the strong coupling constant $\alpha_s$ and the separation of the corresponding quarks. This separation (in first approximation) is proportional to $1/m_V$.

Although there is an active program on near-threshold $J/\psi$ production at  JLab, its energy reach does not extend to the $\psi'$ or $\Upsilon$ states, so this should be new territory for the EIC.  The EIC will be able to study electroproduction in addition to photoproduction.  The threshold region involves relatively low-energy photons, so good detector acceptance in the hadron-going direction is required to study it. The $J/\psi$-N and $\Upsilon$-N cross sections measured via their re-scattering/absorption inside a nucleus are anomalously small for low energy photoproduction. This can be explained because we deal with “young” $J/\psi$ and $\Upsilon$ \cite{Feinberg:1980yu} that are still small dipoles. 
In $J/\psi$ (or $\Upsilon$) electroproduction, the “young” $J/\psi$ and $\Upsilon$ produced at larger $Q^2$ should have smaller formation times and correspondingly smaller radii of the heavy quarkonium.

\subsection{Spectroscopy}
\label{part2-subS-Hadronization-Spectroscopy}

Considerable progress in hadron spectroscopy has been made in recent years through many unexpected observations in the heavy quark sector including the proliferation of so-called $XYZ$ states, charmed pentaquark $P_c$ candidates and more (see reviews Ref.~\cite{Lebed:2016hpi, Esposito:2016noz, Olsen:2017bmm, Brambilla:2019esw}).  Relatively early in the $XYZ$ discoveries in $e^+e^-$ colliders and $b-$hadron decays it was recognized that they could be studied in alternative processes, such as photoproduction, with many calculations for individual reactions over the years, for example Refs.~\cite{Liu:2008qx, He:2009yda, Lin:2013mka, Klein:2019avl}.  Photoproduction can also be used to study conventional mesons as well as exotica \cite{Klein:2019avl}.  

Fixed target experiments using the Jefferson Lab 12 GeV electron beam, such as GlueX~\cite{Adhikari:2020cvz} or CLAS12~\cite{Burkert:2020akg}, provide access to the light-quark regime and $s$-channel production of $P_c$ states discovered by LHCb~\cite{Ali:2019lzf}, however they do not have sufficient energies to produce $XYZ$ states \textit{via} $t$-channel exchange.   Previous measurements in $ep$ collisions at HERA, however, have demonstrated the ability to study heavy quarkonia through photoproduction, particularly the well known vector $c\bar{c}$ and $b\bar{b}$ states~\cite{Adloff:2002re,Alexa:2013xxa,Adloff:2000vm,Chekanov:2009zz}.  The COMPASS collaboration has studied muonproduction of the $J/\psi \pi^+\pi^- p$ final state finding an indication of a new state $\widetilde{X}(3872)$~\cite{Aghasyan:2017utv} and also set limits on $Z_c$ photoproduction in the $J/\psi \pi^+ n$ final state~\cite{Adolph:2014hba}.  
The integrated luminosities expected for the EIC, provide the opportunity to study rare exclusive processes not accessible at HERA.

\begin{figure}
\begin{center}
\includegraphics[width= 0.37 \textwidth]{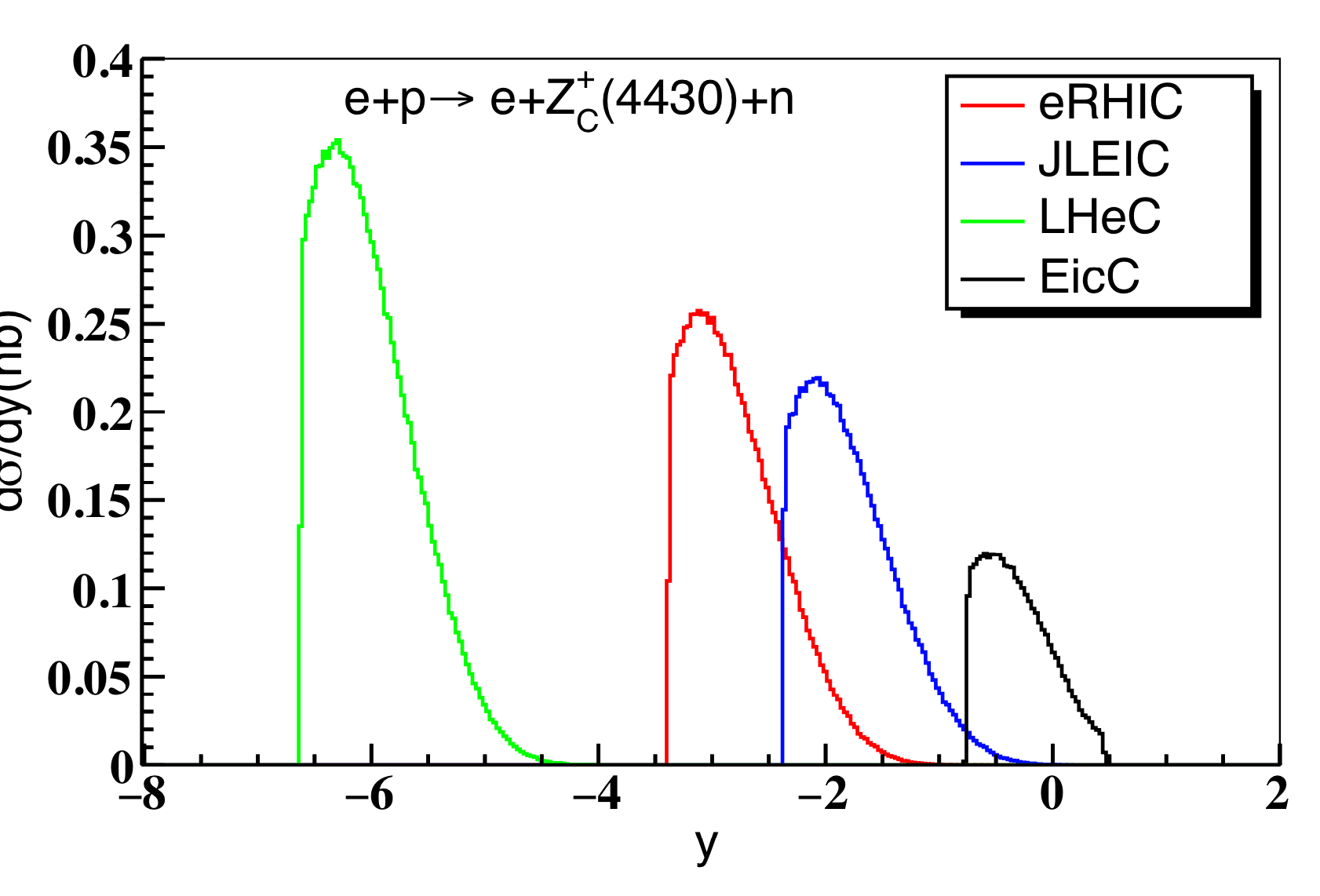}
\includegraphics[width= 0.28 \textwidth]{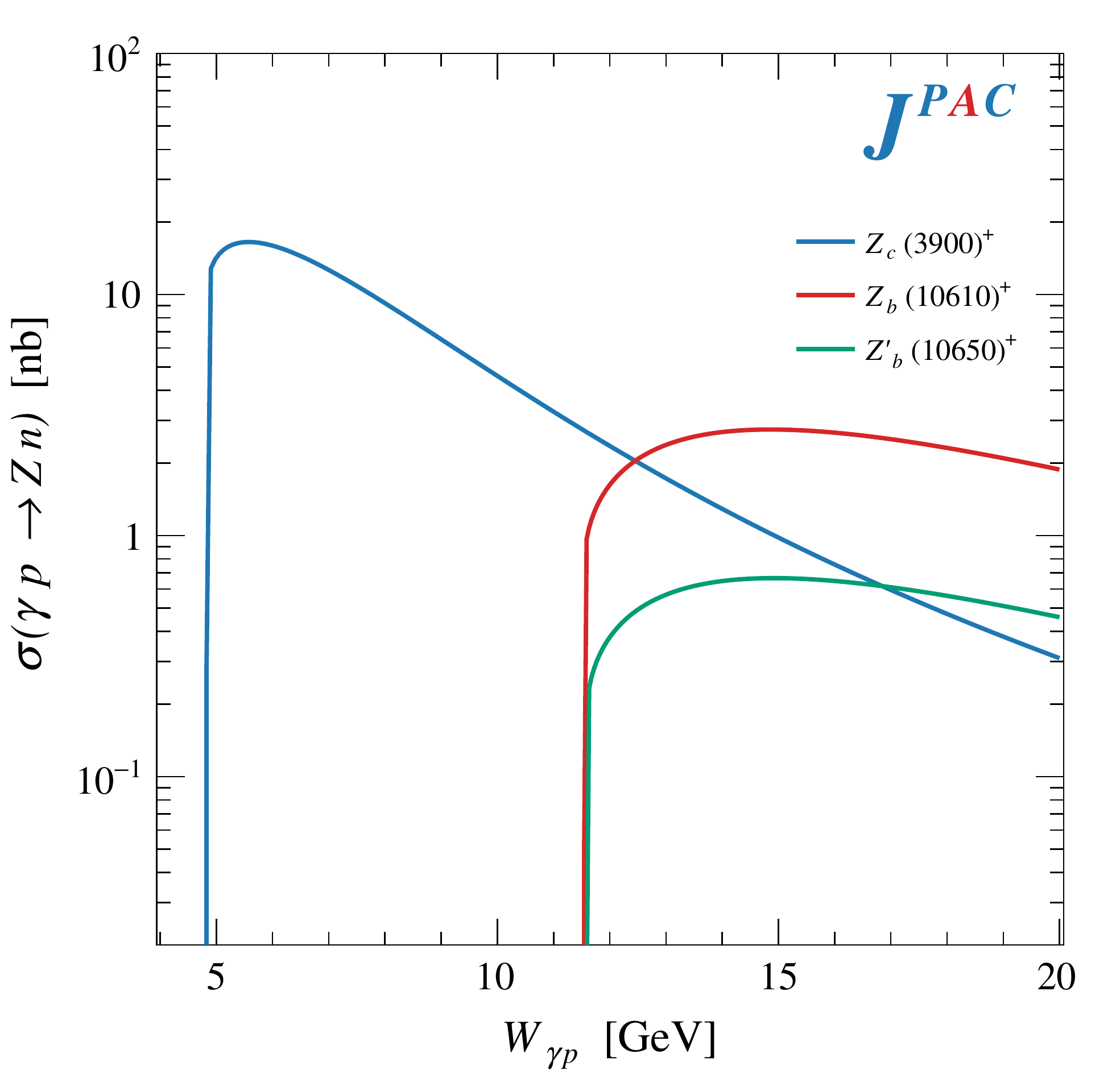}
\includegraphics[width= 0.28 \textwidth]{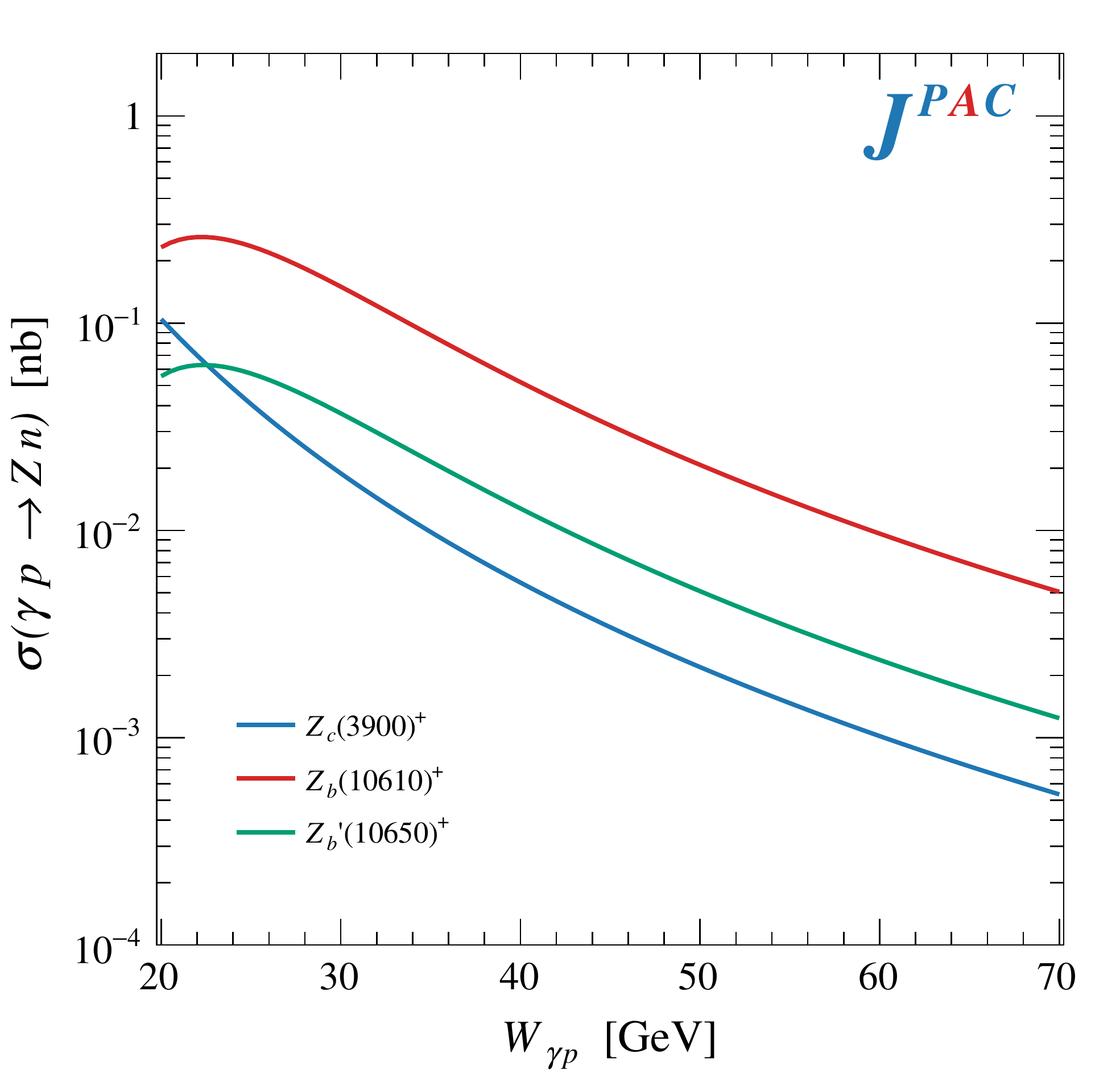}
\caption{\label{fig:jpac} (left) $Z^+$ photoproduction rapidity distributions for colliders with varying energies from \cite{Klein:2019avl} and (right) integrated cross section predictions~\cite{Albaladejo:2020tzt} for fixed-spin exchange, valid at low energies (center), and for Regge exchange, valid at high energies (right).}
\end{center}
\end{figure}

Photoproduction through photon-Pomeron fusion lead predominantly to $J^{PC}=1^{--}$ states like the $J/\psi$, etc., so is only sensitive to exotic with those quantum numbers.   Photon-Reggeon fusion leads to states with a wider range of spin, parity and even charge, so can be used to search for a much wider range of both conventional mesons and exotica.  Since the photon-Reggeon fusion cross-sections are typically peaked at low photon-nucleon center of mass energies (a few times the threshold energy) (Fig. \ref{fig:jpac} center-right) these reaction products are typically produced in the forward direction, requiring good detector acceptance in that region.   The left panel of Fig. \ref{fig:jpac} shows the predictions for a typical exotica model,  where the $Z_c^+$ is a spin-1 tetraquark candidate.  The different curves correspond to different proposed colliders, with the eRHIC curves very close to the current design.  It can be seen that lower-energy collisions, \textit{e.g.} at the Electron-Ion Collider in China (EicC) lead to more central production; lower beam energies at the EIC may therefore be beneficial for studying exotica.

Recent predictions from the JPAC Collaboration~\cite{Albaladejo:2020tzt} provide a comprehensive assessment of the exclusive cross sections for several $XYZ$ states.  Figure~\ref{fig:jpac} shows the predicted photoproduction cross section as a function of the center-of-mass energy for three $Z$ states: $Z_c(3900)^+, Z_b(10610)^+$ and $Z_c(10610)^+$ all previously observed in produced in $e^+e^-$ collisions.  In the low energy region near threshold (left) fixed-spin charge exchange is expected to provide a valid description, while high energies are described by Regge exchange.  With expected cross sections at the $\sim$1-10~nb level the statistics available for some of these reactions are comparable to current measurements of similar states in $e^+e^-$ machines and heavy flavor decays.  More details on the simulation and detector requirements are provided in Sec.~\ref{part2-sec-DetReq.SIDIS.Spectroscopy}.  
\subsection{Target fragmentation}
\label{part2-subS-Hadronization-TargFrag}

Target fragmentation in $ep/eA$ DIS (hadron production in the target rapidity region)
offers new ways of exploring hadronization dynamics and nucleon structure in QCD.
The QCD factorization theorem for single-inclusive hadron production
$e + p \rightarrow e' + X + h(x_{\rm F}, p_T)$ permits separation of hard
and soft contributions, including QCD radiation, and enables a description of target fragmentation
in close analogy to the total DIS cross section \cite{Collins:1997sr,Trentadue:1993ka}.  
The fracture functions (or conditional PDFs) depend on $x$ and $Q^2$ as well as on the
hadronic variables $x_{\rm F}$ and $p_T$ and combine aspects of parton distribution
and fragmentation functions. They obey standard DGLAP evolution
and are independent of the hard process (universal). Physically, the fracture functions
describe the hadronization of the target after removal of a parton (quark, gluon) with given $x$
at the scale $Q^2$. As such they contain rich information about hadronization dynamics
(confinement, chiral vacuum structure) and nucleon structure (multiparton correlations).

Present experimental knowledge of target fragmentation in DIS is very limited;
see \cite{Arneodo:1984rm,Arneodo:1984nf,Arneodo:1986yb,%
Adams:1993wv,Derrick:1977zi,Allen:1980ip,Allen:1982jp}
for fixed-target results.
The HERA experiments measured $p$ production in the diffractive peak $x_{\rm F} \approx 1$,
and $p$ and $n$ production in the region $x_{\rm F} \gtrsim 0.3$
\cite{Chekanov:2008tn,Aaron:2010ab,Andreev:2014zka,Alexa:2013vkv,Ceccopieri:2014rpa}.
The results indicate strong baryon number flow in DIS at small $x$ (up to 50\%
is moved to $x_{\rm F} < 0.3$), which raises interesting questions about multiparton
dynamics that cannot be answered with the HERA data alone.

EIC could transform the knowledge of target fragmentation and open this area up to systematic
study~\cite{CFNS2020_Target_Fragmentation}.
Measurements should focus on the following features connected with specific questions of
dynamics and structure:

{\it (a) $x$-dependence of target fragmentation:} Theoretical arguments predict qualitative
changes of the $x_{\rm F}$ distributions of $p$ and $n$ depending on the $x$ of
the removed parton: $\propto (1 - x_{\rm F})$ at $x > 0.2$; constant in $x_{\rm F}$ at $x \sim 0.2$; 
$\propto 1/x_{\rm F}$ at $x \ll 0.1$ \cite{Strikman:1977he}.
Observing these changes would provide direct evidence
of the nucleon's multiparton structure and enable quantitative understanding.

{\it (b) Spin dependence and polarization transfer:} Polarized $ep$ DIS removes a parton
with definite spin from the nucleon wave function. Measuring the spin dependence of the
$x_{\rm F}$ distributions in targer fragmentation provides insight into the role of spin-dependent
forces in fragmentation, a major open question with broad implications (string fragmentation,
chiral vacuum structure, spin-orbit effects). Fragmentation into self-analyzing $\Lambda$
baryons \cite{Ceccopieri:2012rm} or use of the Collins variable \cite{Collins:1992kk}
would allow one to study the polarization transfer to the produced system.

{\it (c) Quark vs.\ gluon fracture functions:} Another interesting question is how the hadronization
process changes depending on whether a quark or gluon is removed from the nucleon wave function.
This could be studied by measuring target fragmentation induced by quark- or gluon-sensitive
hard processes (e.g. heavy flavor production).
At $x \ll 0.1$, this comparison will probe the multiparton structure of configurations
building up small-$x$ parton densities. At $x > 0.1$, it will reveal the coupling of large-$x$
gluons to valence quarks (e.g., if the leading Fock component in the nucleon dominates,
gluon fracture functions should be strongly suppressed).

{\it (d) Correlations of target and current fragmentation:} Measurements of hadron correlations
between the current and target fragmentation regions could directly probe the multiparton
structure of the nucleon. Correlations between sea quarks are induced by the short-range
non-perturbative forces causing the dynamical breaking of chiral symmetry -- the phenomenon
responsible for hadron mass generation in QCD; these correlations could be revealed in
back-to-back pion correlations with $p_T \approx 0.5$ GeV and moderate rapidity
separations $\Delta y \approx 4$ \cite{Schweitzer:2012hh}.
Generally, such measurements could elucidate the
dynamical origin of intrinsic transverse momentum in the nucleon.

\begin{figure}[t]
{\includegraphics[width=0.35\textwidth]{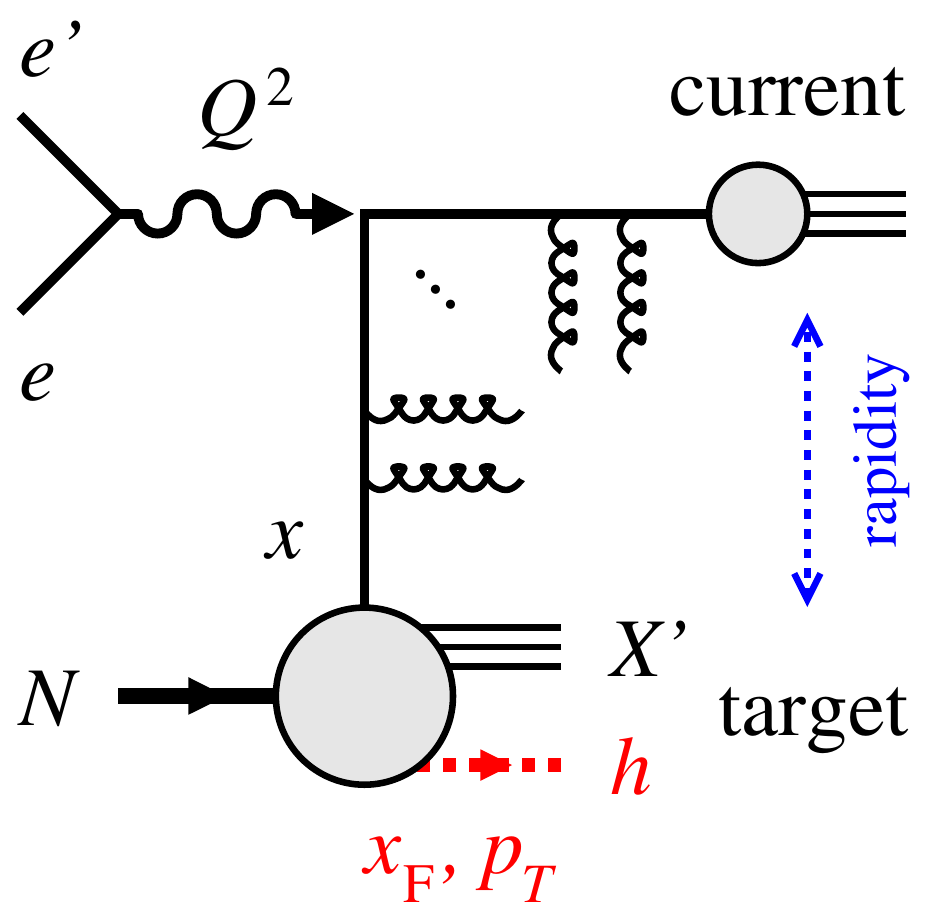}}
\hspace{0.02\textwidth} 
\includegraphics[width=0.6\textwidth]{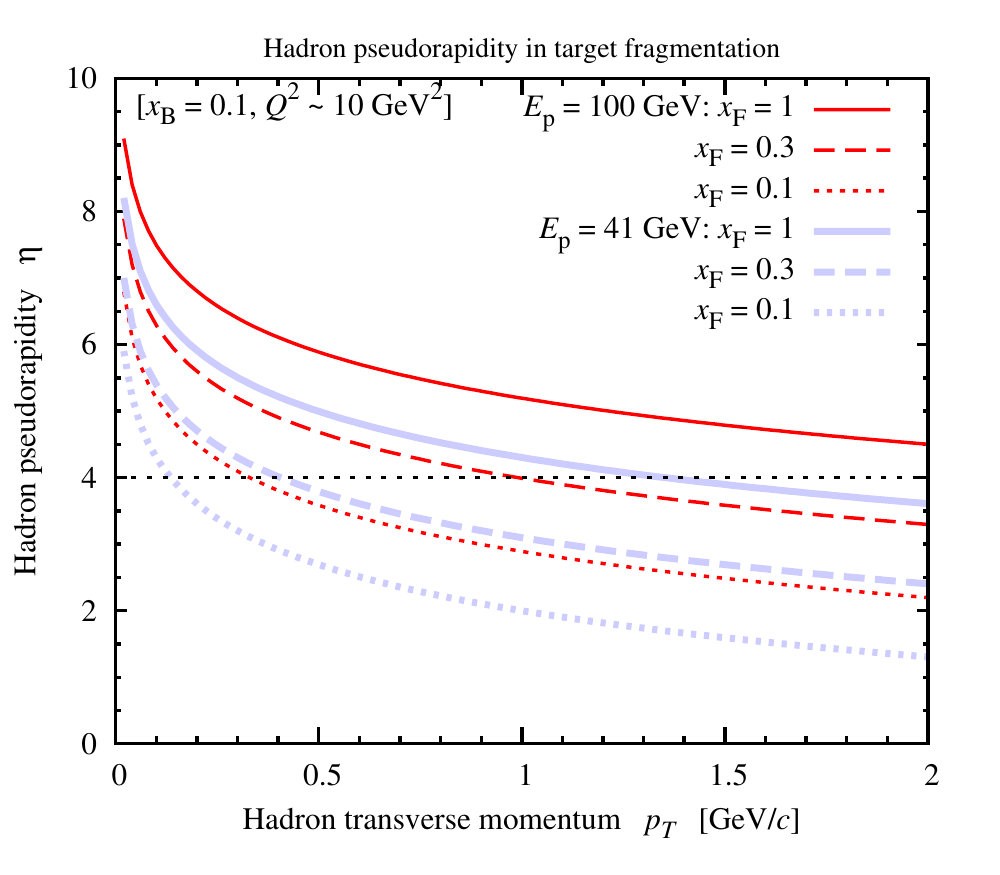}
\caption{\small Left: QCD factorization of target fragmentation in DIS.
Right: Pseudorapidity distribution of hadrons in target fragmentation in DIS as a function of $p_T$
for fixed values of $x_F$, for two different values proton beam energies, $E_p =$ 100 and 41 GeV.
The horizontal line at $\eta = 4$ is the approximate boundary of the central detector. The distributions shown here do not depend explicitly on the electron beam energy;
the information on the electron scattering process enters only through the variables $x_B$ and$ Q^2$.}
\label{fig:target_fragmentation}
\end{figure}

The target fragmentation measurements described here could largely be performed with the baseline
EIC detector design. An important requirement is
continuous coverage in $x_{\rm F}$ from $\sim 1$ down to $\sim 0.1$, without gaps between
the central ($\eta < 4$) and forward detectors. Fig.~\ref{fig:target_fragmentation}
shows the pseudorapidity distribution of hadrons in target fragmentation in DIS as a function of $p_T$
for fixed values of $x_F$, for two different values proton beam energies, $E_p =$ 100 and 41 GeV.
Simulations are in progress.
%
%
%
%
%
%
%

\newpage
\section{Connections with Other Fields}
\label{part2-sec-Connections}

While the principal focus of the physics program at the EIC is QCD, there are nevertheless important and unique points of contact with other fields. This is mainly due to the high luminosity, the availability of polarized lepton and hadron beams, and the wide kinematic coverage of the EIC. Section~\ref{part2-sec-Connections-EW} presents the main opportunities for {\bf electroweak (EW)} and {\bf beyond the standard model (BSM)} physics.

Precision measurements at the EIC can provide new limits on various BSM couplings.  For example, a polarized positron beam would provide access to the parity-conserving but lepton charge-conjugation-violating
couplings $C_{3q}$ via a measurement of the charge conjugation asymmetry of cross sections of polarized leptons and anti-leptons scattering off a nuclear target. Precision measurements of the $C_{iq}$ couplings at the EIC over a wide range of $Q^2$ can also test the running of Weinberg's weak mixing angle. Furthermore, the availability of
polarized electron or positron beams with proton or deuteron targets, over a wide range of kinematics, can scrutinize lepton flavor violation mechanisms in the charged lepton sector. 

The high energy and luminosity at the EIC in principle offers opportunities for new particle searches such as a heavy photon or a heavy neutral lepton. The ability to polarize both electron and proton beams at an EIC may lead to much stronger  constraints on heavy new physics operators in the standard model effective field theory (SMEFT) than exist today.

There is considerable overlap between the EIC science program and {\bf neutrino physics}, as outlined in some detail in Sec.~\ref{part2-sec-Connections-Neutrino}.
In fact, those two research areas are of mutual benefit.
Measurements at the EIC could provide important input for future experiments in neutrino physics, such as a more accurate estimate of nuclear effects.
In return, neutrino scattering can help to better understand the parton structure of both nucleons and nuclei, where the nucleon strangeness content is one example.

Measurements at the EIC are also expected to
deliver important input for several areas of astroparticle physics.  Fields such as
{\bf cosmic-ray  air  showers  and  neutrino astrophysics}
will benefit from better constrained models of hadronic interactions, see discussion in Sec.~\ref{part2-sec-Connections-Cosmic}.

Deeply-inelastic scattering and photo-nuclear processes of course have close ties with the physics of hadronic collisions. Sec.~\ref{part2-sec-Connections-Other} focuses on a few particular aspects which have not been addressed in other parts of this Yellow Report. These relate to the issue of {\bf small-$x$ gluons and factorization} in DIS vs.\ {\textit{p}+\textit{p}} and {\textit{p}+A}, to the implications of the determination of PDFs for {\textit{p}+\textit{p}} and {\textit{p}+A} collisions, to {\bf initial conditions for hydrodynamics} of heavy-ion collisions, and to {\bf parton interactions in nuclear matter}.

Remarkably, the EIC could also contribute to fundamental insights into {\bf nuclear structure} and the physics of {\bf exotic nuclei}. Secs.~\ref{part2-sec-Connections-NuclStruc} and~\ref{part2-sec-Connections-ExoNuc} provide an overview of these interesting aspects.

Finally, Sec.~\ref{part2-sec-Connections-HEP} presents the interface to current and future efforts in the {\bf High-Energy Physics} community. In particular, in that section several points of overlap with the ongoing Snowmass process are highlighted.
\subsection{Electroweak and BSM physics}
\label{part2-sec-Connections-EW}
\subsubsection{Introduction}

The high luminosity, polarized lepton and hadron beams, variety of nuclear targets,  and wide kinematic range at the EIC open the door to searches that go beyond the typical nuclear physics "boundaries". While the electroweak interaction, through charged-current interactions, can be used as a clean probe to separate quark flavors \cite{Aschenauer:2013iia}, it can also be used to access new observables such as $\gamma-Z$ interference structure functions~\cite{Zhao:2016rfu} through neutral current interactions. Further details on spin structure observables can be found in Sec.~\ref{part2-subS-SpinStruct.P.N}. This section will explore how the EIC will contribute to searches of physics Beyond the Standard Model (BSM) and where possible will compare to the expected reach of other experimental efforts.

\subsubsection{Weak neutral-current measurements}

For electron-hadron scattering, in the region where the virtuality of the exchanged boson satisfies $Q^2\ll M_Z^2$, the weak neutral current can be  parametrized in terms of contact interactions
\begin{equation}
\label{eq:contactCiq}
{\cal L} = \frac{G_F}{\sqrt{2}} \sum_{\ell, q} \Big [ C_{1q} \>\bar{\ell}
\gamma^{\mu}\gamma_5 \ell \, \bar{q}\gamma_\mu q + C_{2q} \> \>\bar{\ell} \gamma^{\mu}
\ell \, \bar{q}\gamma_\mu \gamma_5 q + C_{3q} \>\bar{\ell} \gamma^{\mu}\gamma_5 \ell \,
\bar{q}\gamma_\mu \gamma_5 q \Big ] \,,
\end{equation}
where the $C_{iq}$ are perturbatively calculable coefficients. The $C_{1q}$ and $C_{2q}$ depend sensitively on the weak mixing angle, $\theta_W$. 
A comparison of the experimentally extracted values of the $C_{iq}$ couplings with the SM predictions can be used to set limits of the scale $\Lambda$ at which new interactions may arise. At low energies, well below the scale $\Lambda$, these new interactions can be parametrized by the effective Lagrangian
\begin{eqnarray}
\delta{\cal L} &=& \frac{g^2}{\Lambda^2}\sum_{\ell,q} \Big \{  \eta^{\ell q}_{LL} \>\bar{\ell}_L \gamma_\mu \ell_L \, \bar{q}_L \gamma_\mu q_L 
+ \eta^{\ell q}_{LR}\> \bar{\ell}_L \gamma_\mu \ell_L \, \bar{q}_R \gamma_\mu q_R  \nonumber \\
&+& \eta^{\ell q}_{RL}\>\bar{\ell}_R \gamma_\mu \ell_R \, \bar{q}_L \gamma_\mu q_L   
+ \eta^{\ell q}_{RR} \> \bar{\ell}_R \gamma_\mu \ell_R \, \bar{q}_R \gamma_\mu q_R \Big \} \,,
\end{eqnarray}
where the mass limit for $\Lambda$ is defined with the convention $g^2=4\pi$. The coefficients $\eta_{ij}^{\ell q}$ take on the values of  $+1$, $0$, or $-1$, allowing for the possibility of constructive or destructive interference with the SM contributions. The $C_{iq}$ coefficients can now be written as $C_{iq}=C_{iq}({\rm SM}) + \Delta C_{iq}$ , corresponding to the sum of the SM and new-physics contributions. For example, the new-physics contribution to the $C_{2q}$ couplings takes the form $\Delta C_{2q}= [g^2/(2\sqrt{2} G_F\Lambda^2)](\eta_{LL}^{\ell q} - \eta_{LR}^{\ell q}+\eta_{RL}^{\ell q}-\eta_{RR}^{\ell q})$, depending on a specific combination of chiral structures. Similar expressions can be obtained for the $C_{1q}$ and $C_{3q}$ couplings. A new-physics scenario with the chiral structure  $\eta_{LL}^{\ell q}=1$ and $\eta_{RR}^{\ell q}=\eta_{RL}^{\ell q}=\eta_{LR}^{\ell q}=0$, results in a specific pattern of shifts $\Delta C_{1q}=\Delta C_{2q}=-\Delta C_{3q} =g^2/(2\sqrt{2}\Lambda^2 G_F)$  relative to the SM values which can be used to set limits on $\Lambda$. 

Different flavor combinations of the $C_{1q}$ coefficients have best been measured  through atomic parity violation \cite{Wood:1997zq} and elastic parity-violating electron scattering \cite{Androic:2013rhu}. The $C_{2q}$ couplings are more challenging due to their relatively small values in the SM. They can be accessed through parity-violating DIS on a deuteron target by measuring the cross section asymmetry between left-handed and right-handed electrons, 
\begin{equation}
\label{eq:Apv}
A_{PV}^e = \frac{d\sigma_L-d\sigma_R }{d\sigma_L +d\sigma_R} \,.
\end{equation}
Recently~\cite{Wang:2014guo,Wang:2014bba} at JLab, 6-GeV polarized electrons incident on a unpolarized deuteron target were used to extract the combination $2C_{2u}-C_{2d}=-0.145\pm 0.068$ at $Q^2=0$, showing for the first time that this combination is nonzero at the $95\%$ confidence level. The SoLID spectrometer~\cite{Chen:2014psa} as part of the JLab-12 program is expected to further improve the precision of this measurement. The EIC can provide complementary high-precision measurements on both proton and deuteron due to its high luminosity and  wide kinematic range, with access to different linear combinations of the $C_{iq}$ and extraction of  the combination $2C_{2u}-C_{2d}$ over a different (higher) $Q^2$-range.

On the other hand, experimental data on the $C_{3q}$ couplings are quite sparse. They are parity-conserving but charge-conjugation-violating (charge conjugation for the lepton).  A positron beam would provide a unique opportunity to access these couplings
via a measurement of the charge conjugation asymmetries \cite{Berman:1973pt,Accardi:2020swt}
\begin{equation}
A^{e^- -e^+} = \frac{d\sigma(e^- N ) -d\sigma(e^+ N) }{d\sigma(e^-N ) +d\sigma(e^+ N) } \,, \qquad A^{e^-_L -e^+_R} = \frac{d\sigma(e_L^- N ) -d\sigma(e_R^+ N) }{d\sigma(e_L^-N ) +d\sigma(e_R^+ N) } \,,
\end{equation}
through a comparison of cross sections of unpolarized and polarized leptons and anti-leptons scattering off a nuclear target, respectively. Such a measurement has been carried out only once before at CERN~\cite{Argento:1982tq}, using polarized muon and anti-muon beams scattering off a carbon target, resulting in the extraction $0.81(2C_{2u}-C_{2d}) + 2C_{3u}-C_{3d}=1.53\pm 0.45$. Using the current experimental value~\cite{Wang:2014guo} of $2C_{2u}-C_{2d}$, yields the result $2C_{3u}-C_{3d}= 1.65\pm0.453$.  A positron beam at the EIC can improve upon this measurement~\cite{Furletova:2018nci,Accardi:2020swt}. The isoscalar deuteron target is preferred over the proton target since it provides access to the combination $2C_{3u}-C_{3d}$ and minimizes the uncertainty from the d/u PDF ratio. However, both the proton and deuteron targets can be used in a complementary manner to extract the $C_{3q}$ couplings.

The importance of precision measurements at the EIC of the $C_{iq}$ couplings is well illustrated by BSM scenarios that involve heavy leptophobic $Z'$-bosons~\cite{Buckley:2012tc,GonzalezAlonso:2012jb}. Since they couple very weakly to leptons, their primary signature at hadron colliders corresponds to dijets with invariant mass  $m_{j_1j_2} \sim M_{Z'}$. The large dijet background at hadron colliders makes it difficult to constrain such a scenario. However, at the EIC, the existence of a leptophobic $Z'$-boson could introduce a deviation from the SM values in the $C_{2q}$ couplings without affecting the $C_{1q}$ couplings. This occurs via a vacuum polarization quark loop that connects a photon vector coupling to the electron current and the $Z'$ axial-vector coupling to the quark current, leading to a shift only in the $C_{2q}$ couplings.

Precision measurements of the $C_{iq}$ couplings at the EIC for a wide $Q^2$-range can also test the running~\cite{Zhao:2016rfu} of $\sin^2\theta_W(\mu)$ in the previously unexplored region $10 \, \textrm{GeV} < \mu < 70 \, \textrm{GeV}$. We denote the $\overline{\rm MS}$-scheme value  by $\sin^2\theta_W(\mu)_{\overline{\rm MS}}$.  
Note that in the region $Q^2 < M_Z^2$, close to the Z-pole, the effective contact interaction parametrization in Eq.~(\ref{eq:contactCiq}) is no longer applicable and the full $Q^2$-dependence in the Z-boson propagator must be included for a proper interpretation of tests of the running of the weak mixing angle. It is useful to define an effective running weak mixing angle~\cite{Czarnecki:1995fw,Czarnecki:1998xc, Czarnecki:2000ic,Ferroglia:2003wa,Kumar:2013yoa} by $\sin^2 \theta_W (Q^2)=\kappa(Q^2) \sin^2 \theta_W (m_Z)_{\overline{\rm MS}}$, where $\kappa(Q^2)$ incorporates $\gamma-Z$ vacuum polarization mixing and other universal corrections that appear in the low-energy parity-violating observables. 

\begin{figure}[ht]
	\centering
	\includegraphics[width=0.5\textwidth]{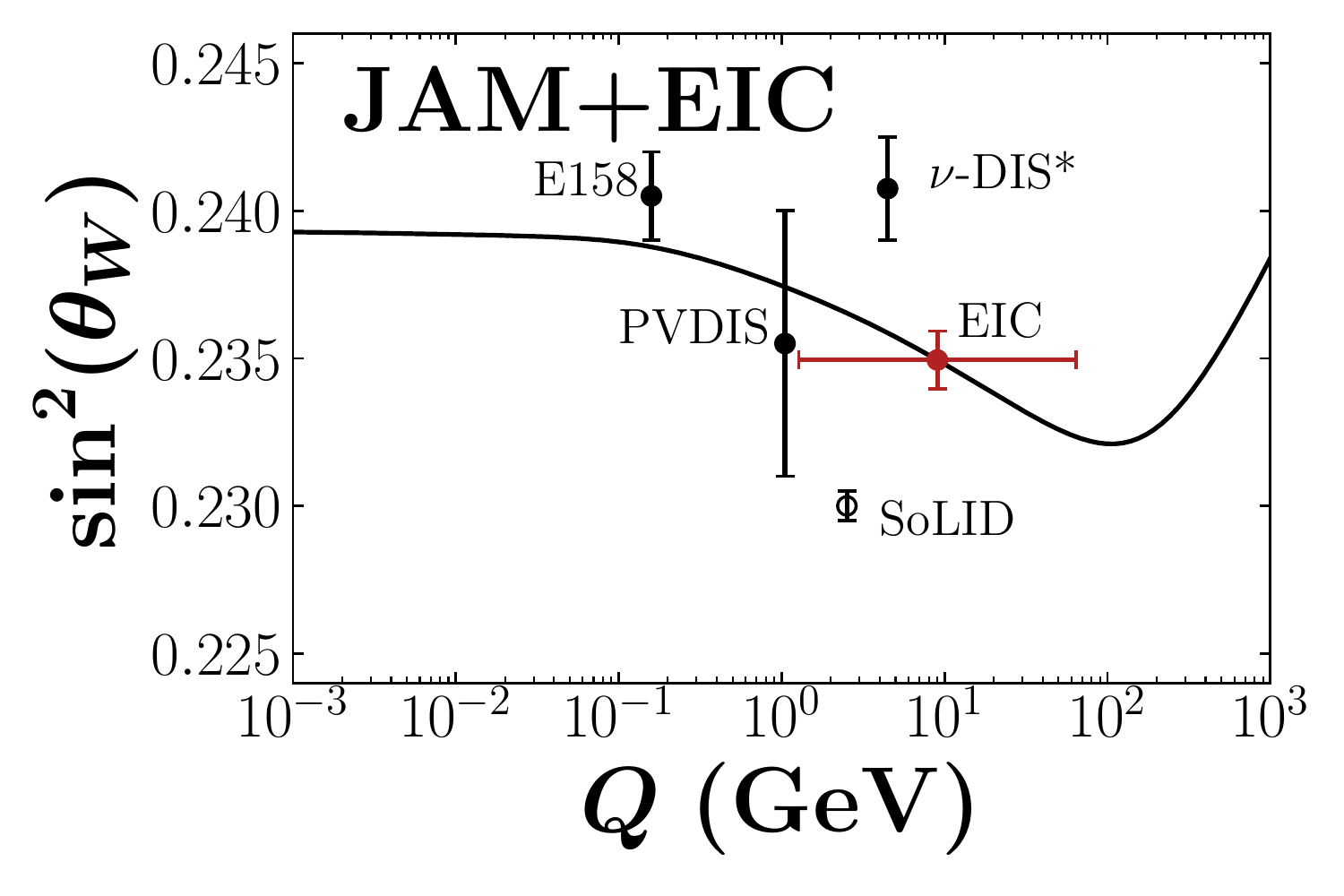}
	\caption{Impact of entire electron-proton and electron-deuteron EIC data on $\sin^2(\theta_W)$. The horizontal uncertainty depicts the extent of the kinematic range.}
	\label{fig:sin2w}
\end{figure}

The parity-violating electron-scattering asymmetry in Eq.~(\ref{eq:Apv}), for the extraction of the weak mixing angle $\sin^2 \theta_W (Q^2)$ below the Z-pole, is typically seriously considered only for the isoscalar deuteron target where structure function effects largely cancel at leading twist facilitating improved precision and sensitivity to the weak mixing angle. However, the high precision of the PDF data obtained by the EIC may also allow for extractions to be made using a proton target. 
Figure~\ref{fig:sin2w} shows the impact of the observable $A_{PV}^{e}$ on $\sin^2 \theta_W (Q^2)$, assuming a proton and deuteron target with 100 fb$^{-1}$ and 10 fb$^{-1}$ luminosity, respectively, and an uncorrelated systematic uncertainty of 1\% from the pion background --- for statistics details as a function of Bjorken variable $x$ see left part of Fig.~\ref{fig:APVe}. Within the JAM Monte Carlo framework, the normalization of $\sin^2\theta_W (Q^2)$, along with the spin-averaged PDFs, were simultaneously fit to obtain the result. 
The functional form of $\sin^2\theta_W (Q^2)$ was fixed in this study and the anticipated EIC statistics were used only to constrain the overall normalization. While the EIC will be able to make measurements over a wide kinematic range (as depicted by the extent of the horizontal uncertainty) this study determined the constraints from the entire data set (as seen from the vertical uncertainty). The $Q^2$ of this point was selected to be in the middle of the EIC kinematic range.
The wide EIC kinematic region, where very little data exist, will provide significant constraints on this fundamental parameter of the SM. 

Such precision tests of the $Q^2$-dependence of $\sin^2 \theta_W (Q^2)$ at the EIC, can also be an effective probe of physics associated with new light degrees of freedom that cannot be parametrized as contact interactions. One such scenario, dark parity violation~\cite{Davoudiasl:2012ag,Davoudiasl:2012qa}, arises through a light dark boson $Z_d$ corresponding to a spontaneously broken ${\rm U}(1)_d$ gauge symmetry in the dark sector. The dark-Z boson is a generalization of the standard dark photon which only couples to the electromagnetic current by mixing with the photon, a result of kinetic mixing with the ${\rm U}(1)_Y$ hypercharge sector of the SM before the electroweak symmetry breaking. 

The dark photon is phenomenologically motivated to explain the observed gamma ray~\cite{Jean:2003ci} and positron~\cite{Adriani:2008zr,Barwick:1997ig,Beatty:2004cy,Aguilar:2007yf} excesses through dark matter annihilation near the galactic center.  An extended Higgs sector can generalize the dark photon to the dark-Z boson which couples to the SM via both kinetic and mass mixing with the photon and the Z-boson, with couplings $\epsilon$ and $\epsilon_Z=m_{Z_d}/M_Z \delta$ respectively. Here $\delta$ is a model-dependent parameter arising from the extended Higgs sector. 

The dark-Z gives rise to an additional source of parity violation through its coupling to the weak neutral current by mixing with the Z-boson. In parity-violating DIS, its effects can be absorbed into a shift in the measured weak mixing angle,
\begin{equation}
\Delta \sin^2 \theta_W (Q^2) \simeq -0.42\> \varepsilon\> \delta \> \frac{M_Z}{m_{Z_d}} \frac{m_{Z_d}^2}{Q^2+m_{Z_d}^2} \,.  
\end{equation}
For $m_{Z_d} \ll M_Z$, this shift is negligible near the Z-pole $Q^2\sim M_Z^2$. However, at low $Q^2$, below the Z-pole, the shift can be significant. In the region explored by the EIC, $10<$ GeV $Q < 70$ GeV, the mass range $m_{Z_d}\sim 10 - 30 \, \textrm{GeV}$ could result in deviations in the running of the weak mixing angle, large enough to be within reach of the projected EIC sensitivities.

\subsubsection{Charged lepton flavor violation}
The discovery of neutrino oscillations have now firmly established lepton flavor violation (LFV) in the neutrino sector, confirming that neutrinos have nonzero mass, pointing to new physics beyond the SM. By contrast, there has been no experimental observation of flavor violation in the charged lepton sector. LFV in the neutrino sector implies charged-lepton flavor violation (CLFV) in processes such as $\mu \to e\gamma$. However, it is mediated at one loop and suppressed by the smallness of the neutrino masses, yielding Br$(\mu \to e \gamma) < 10^{-54}$, well beyond the reach of any current or planned experiments. 

However, many BSM scenarios~\cite{Tanabashi:2018oca} predict significantly higher CLFV rates that are within reach of current or future planned experiments. A variety of experiments~\cite{Litchfield:2014qea,WIEDNER:2013nwa,Wintz:1998rp, Lee:2018wcx,Baldini:2018nnn,vanderSchaaf:2003ti,Aaron:2011zz,Chekanov:2005au} across the energy spectrum have searched for and set limits on CLFV processes that involve transitions between the electron and the muon. 

By contrast, the limits  on CLFV involving the $e \leftrightarrow\tau$ transition are worse by several orders of magnitude. Extensive searches for this CLFV transition  $e^\pm + p \to \tau^\pm +X$ were conducted at HERA \cite{Chekanov:2005au,Aktas:2007ji}. Theoretical and simulation studies \cite{Gonderinger:2010yn,Boer:2011fh} have been performed for the EIC and indicate that
significant improvement over the HERA limits can be achieved. A polarized positron beam would complement these planned studies at the EIC, providing an independent probe  that can help distinguish between different CLFV mechanisms. 

It is convenient to study CLFV within the Leptoquark (LQ) framework. LQs are color triplet particles that couple to leptons and quarks
and mediate CLFV processes at tree-level allowing for larger cross sections.  The Buchm\"uller-R\"uckl-Wyler (BRW) parameterization classifies the LQs into 14 different types according to their spin (scalar or vector), fermion number $F = 3B + L$ (0 or $\pm2$), chiral couplings to leptons (left-handed or right-handed),  SU(2)$_L$ representation (singlet, doublet, or triplet), and ${\rm U(1)}_Y$ hypercharge. 
\begin{figure}[ht]
\label{Fig:LQz}
    \centering
    \includegraphics[width=0.6\textwidth]{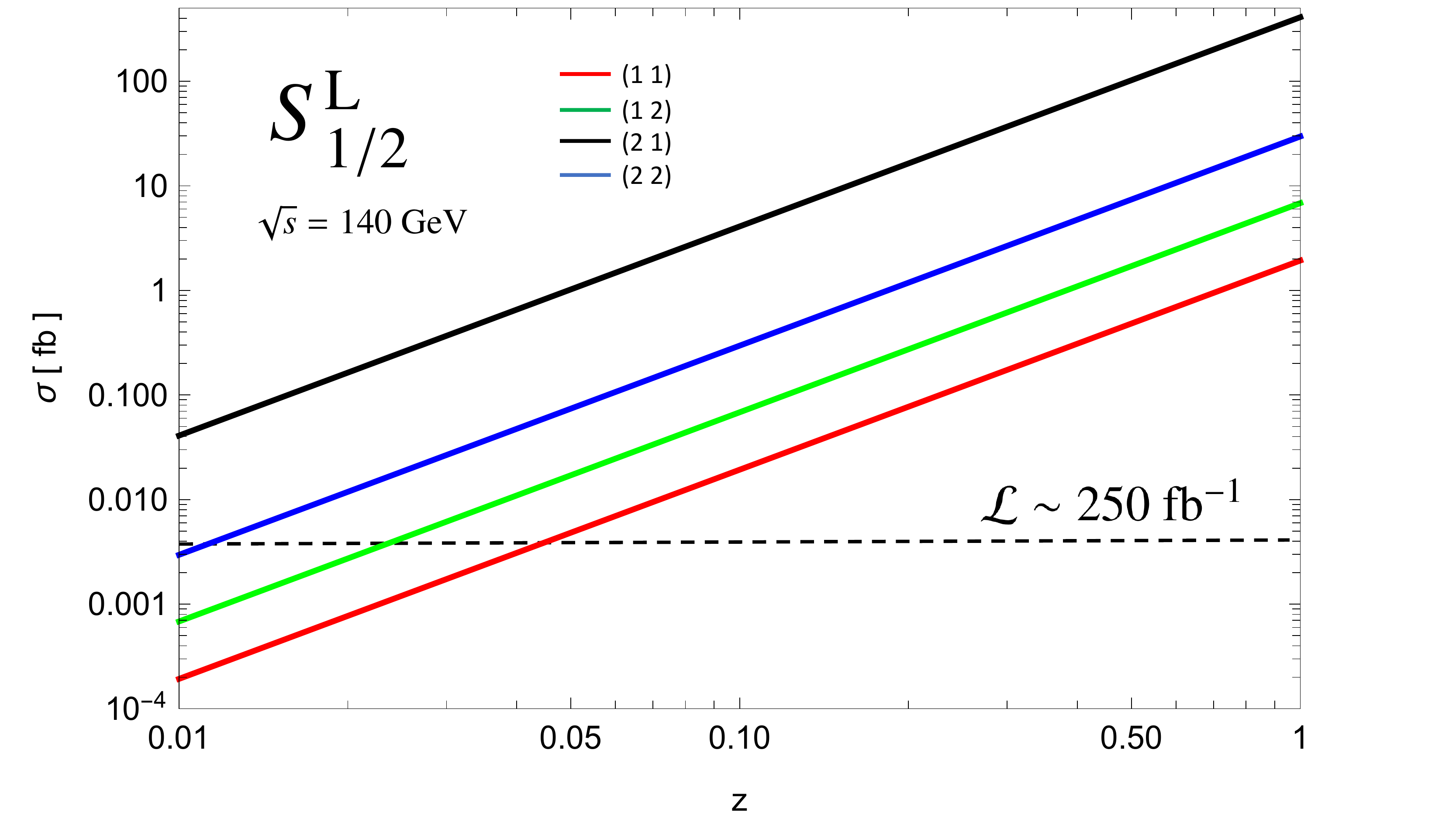}
    \caption{Improvement over limits set by HERA on CLFV. The different color lines indicate different leptoquark states, while the dashed line indicates the expected reach of the EIC with 250~fb$^{-1}$}.
    \label{fig:LQ}
\end{figure}

In the region of LQ mass $M_{LQ}\gg \sqrt{s}$, the CLFV process is mediated via a contact interaction and the cross section for $e^- + N \to \tau^-+X$ is proportional to the combination $\kappa_{\alpha \beta} \equiv \lambda_{1\alpha}\lambda_{3\beta}/M_{LQ}^2$. Here $\lambda_{1\alpha}$ denotes the coupling of the LQ to the electron and quark of generation $\alpha$, $\lambda_{3\beta}$ denotes the coupling of the LQ to the $\tau$-lepton and quark of generation $\beta$. 
Limits on $\kappa_{\alpha \beta}$ have been set at HERA and through low-energy experiments \cite{Chekanov:2005au,Aktas:2007ji}. In Fig.~\ref{fig:LQ},  we plot the cross section (for $\sqrt{s}= 140$ GeV) for the production of the $F=0$ scalar LQ state,  $S_{1/2}^{\rm L}$, as a function the variable $z\equiv \kappa_{\alpha \beta}/\kappa_{\alpha \beta}^{\rm limit}$, where $\kappa_{\alpha \beta}^{\rm limit}$ denotes the maximum value that saturates the existing limit. Thus, $z=1$ corresponds to the current limit $\kappa_{\alpha \beta} = \kappa_{\alpha \beta}^{\rm limit}$, corresponding to the largest allowed cross section. The factor of $10^2-10^3$ increase in luminosity of the EIC compared to HERA, will allow sensitivity to smaller cross sections and correspondingly smaller values of $z <1$. As seen in Fig.~\ref{fig:LQ}, with an integrated luminosity of ${\cal L} \sim 250$ fb$^{-1}$, the EIC can improve the bound on $z$ by one to two orders of magnitude depending on the specific LQ state considered. See Ref.~\cite{Cirigliano:2021img} for a recent study evaluating the reach of EIC to probe CLFV physics in the context of SMEFT.

Simulation studies are ongoing to evaluate the discovery sensitivity to $e-\tau$ conversion given the conceptual design of the EIC detector. The challenge is to identify the $\tau$-lepton amidst the hadron remnants of the DIS event. The goal is to reach a sensitivity to an $e-\tau$ appearance cross section at the level of 0.1 fb. Beyond increased luminosity the EIC will take advantage of the improvements to detector technology. In particular, improved reconstruction of jets, better tracking resolution and most importantly a vertex detector will be hugely beneficial for this search. Early studies indicate that the potential exists --- though this must be confirmed with many detailed studies --- to achieve an efficiency approaching 10\% while being background-free for an integrated luminosity of 100 fb${^{-1}}$.  

The use of different combinations of polarized electron or positron beams with proton or deuteron targets, over a wide range of kinematics, can allow for distinguishing between specific LQ states or CLFV mechanisms. The positron and electron beams can be used to separate the contributions of the $F=0$ and $|F| = 2$ LQ states. The lepton-beam polarization can be used to distinguish between left-handed and right-handed LQ states. A wide kinematic range allows distinguishing between scalar and vector LQs through the difference in the $y$-dependence of the corresponding cross sections. Finally, proton vs deuteron targets can be used to distinguish between ``eu" and ``ed" LQs, corresponding to LQs with different electroweak quantum numbers~\cite{Taxil:1999pf}.

\subsubsection{Charged current chiral structure}
The chiral structure of electroweak interactions allows only left-handed electrons and right-handed positrons to couple to the $W$-boson. Thus, the SM predicts a linear dependence on the lepton beam polarization for the charged-current processes $e^- + p \rightarrow \nu _e  + X $, $e^+ + p \rightarrow \bar{\nu}_e  + X$.
Precision measurements of this polarization dependence can test the chiral structure of the charged-current interactions. A right-handed W-boson ($W_R$), arising in Left-Right Symmetric models with spontaneous symmetry breaking ${\rm SU(2)}_L\otimes {\rm SU(2)}_R\otimes {\rm U}(1)_{B-L} \to {\rm SU(2)}_L\otimes {\rm U(1)}_Y$, could lead to a deviation from a linear dependence on the lepton-beam polarization. The higher luminosity and degree of lepton-beam polarization at the EIC can allow for modest improvements~\cite{Furletova:2018nci} over the HERA limits~\cite{Aktas:2005ju} on the $W_R$-boson mass, $M_R$. While the Tevatron and the LHC have already set more stringent limits on $M_R$, in the TeV range, by looking for deviations in the transverse mass distribution of the Drell-Yan process $pp \rightarrow  W \rightarrow l \nu _l$, the observed distribution is sensitive to a time-like charged boson and in general can be affected by physics involving different chiral and flavor structures. In this way, the EIC measurements will complement the limits that will be set by the LHC.

\subsubsection{Heavy-photon and neutral-lepton searches}
The high energy and luminosity at the EIC offers interesting opportunities for searches for new particles including heavy photons as well as heavy neutral leptons (HNL).  
At the EIC, these searches could take advantage of the unique kinematics, which make it equivalent to a multi-TeV lepton beam on a fixed-proton or heavier target.  In the nuclear rest frame, radiative production prefers to give a substantial fraction of the beam energy to the radiated particle, producing highly-boosted final states.  Compared to that hypothetical fixed-target experiment, however, electron-going final states at the EIC have significantly lower boost and hence wider opening angles in the laboratory frame, allowing access to kinematics that are otherwise difficult to capture. 

For the heavy-photon searches, current limits beyond 1~GeV are set primarily by BaBar, LHCb, and CMS~\cite{Filippi:2020kii,Fabbrichesi:2020wbt}. The center-of-mass energy of the EIC reaches above the $Z^0$ threshold, competitive with the CMS dimuon result~\cite{CMS:2019kiy}, the highest mass range currently probed by a collider experiment.  In particular, the presence of an initial-state lepton with large center-of-mass energy may also make it possible to substantially expand probing of the parameter space in models with new force mediators with leptonic couplings. 

Dilepton searches for radiative production of a dark photon via $e+p\rightarrow e + p + A' \rightarrow e+p+l^++l^-$ would directly test electronic and muon couplings to this new mediator with minimal model dependence, and could also be performed in $e+A$ collisions, where the (additional) charge in the nucleus is expected to enhance radiative production. While detailed studies of the impact of these model-independent searches are still in their early stages, they have the potential to provide a significant contribution to the field.

The HNL searches~\cite{HNL-LoI} are motivated by models suggesting that they could contribute to the neutrino mass generation through the Type-I Seesaw mechanism~\cite{Minkowski:1977sc,Yanagida:1979as,GellMann:1980vs,Glashow:1979nm,Mohapatra:1979ia,Schechter:1980gr} as well as the matter-antimatter asymmetry in the Universe~\cite{Asaka:2005pn,Akhmedov:1998qx}. At the EIC the primary search will focus on the $e + p \rightarrow N + X$ production channel with a particular focus on the events with a displaced vertex. 
Preliminary studies suggest that searches at the EIC for such particles would offer sensitivities beyond existing limits for mass ranges around 5~GeV. This mass range can also be probed by other proposed experiments~\cite{Beacham:2019nyx}.

\subsubsection{General BSM searches}
\paragraph{SMEFT:}
The SM effective field theory (SMEFT) provides a convenient theoretical framework for investigating indirect signatures of heavy new physics without associated new particles at
low energies. Considerable effort has been devoted to performing global
analyses of the available data within the SMEFT and other frameworks.  An issue 
that arises in such global fits is the appearance of ``flat directions" that occur when 
the available experimental measurements cannot disentangle the contributions from different EFT operators.

The flat directions that appear when studying 2-lepton, 2-quark four-fermion operators can be
resolved with the inclusion of high-precision measurements using polarized beams. 
Although the naive expectation is that these operators are well probed by high invariant-mass
Drell-Yan distributions at the LHC, only a very limited number of combinations of Wilson 
coefficients can be probed by such measurements.  The ability to polarize both the electron and the proton beam at the EIC allows for probes of Wilson coefficient combinations not accessible at the LHC~\cite{Boughezal:2020uwq}. Combined fits of LHC and projected EIC data lead to much stronger constraints than either experiment alone.  Moreover, the addition of polarized positrons to the EIC would provide constraints on further flat directions in the SMEFT framework leading us closer to a fully constrained system.

\paragraph{Lorentz- and CPT-violating effects:}
Lorentz and CPT symmetry are among the most well established symmetries in physics. However, many BSM
theories admit regimes where one or both of these symmetries can be spontaneously broken. Low-energy
tests of Lorentz and CPT symmetry can be performed using the effective field theory known
as the Standard-Model Extension~(SME)~\cite{Colladay:1998fq,Kostelecky:2003fs,Colladay:1996iz}. To date, SME operators describing Lorentz- and CPT-violating effects on QCD degrees of freedom are largely unconstrained.

Recent studies suggest that differential cross section measurements at the EIC will allow 
for precision tests of Lorentz and CPT symmetry in the quark sector~\cite{Lunghi:2016sgk,Lunghi:2018uwj,Kostelecky:2019fse}. Data for unpolarized inclusive DIS at 100~fb$^{-1}$ integrated luminosity can increase bounds on quark-sector coefficients by two orders of magnitude compared to HERA data. Symmetry violations would be visible as variations of the cross section as a function of sidereal time. Additional processes that can be measured at the EIC, including those with polarization effects, charged-current exchange, and QCD corrections, can place first constraints on a number of completely unexplored effects stemming from Lorentz and CPT violation.

\subsection{Neutrino physics}
\label{part2-sec-Connections-Neutrino}

The accurate characterization of the structure of nucleons and nuclei 
provided by the EIC program can directly benefit neutrino physics. 
Massive nuclear targets are typically required in neutrino experiments 
to collect sizable statistics, but they also introduce uncertainties related to 
nuclear effects~\cite{Alvarez-Ruso:2017oui}.
Since the energy of the incoming neutrino is unknown on an event-by-event basis, 
it must be inferred from the detected final-state particles, which are affected by a substantial nuclear smearing. 
The latter is present even for an ideal detector since the initial momentum 
of the bound nucleon is not known and hadrons produced in the primary 
interactions can be absorbed or re-interact within the nucleus. 
Target nuclei commonly used in neutrino experiments include C, O, Ar, Fe, Pb. 
Understanding the impact of nuclear effects on the measured cross sections 
and event distributions is particularly critical for the next-generation 
long-baseline neutrino oscillation experiments like DUNE~\cite{Abi:2020evt} 
and Hyper-Kamiokande~\cite{Abe:2018uyc}, which are looking for CP violation via tiny differences between neutrino and antineutrino interactions off Ar and H$_2$O targets, respectively. 
The kinematic coverage of the EIC 
(see, for instance, Figs.~\ref{fig:xQ2_18_275} - \ref{fig:xQ2_5_41})
has overlap with the region accessible at the Long-Baseline 
Neutrino Facility (LBNF), which is dominated by inelastic interactions. 
In addition to the default energy spectrum optimized 
for the neutrino oscillation measurements in DUNE --- 
in which more than 54\% of the events have $W>1.4 \, \textrm{GeV}$ --- 
a high-energy beam option is available with energies 
in the $10-20 \, \textrm{GeV}$ range. 

Neutrinos and antineutrinos have many desirable properties for a probe 
of the structure of nucleons and nuclei, including a complete flavor separation 
($d/u, s/ \bar s, \bar d / \bar u$, valence/sea) through the charged-current (CC) 
process, the presence of an axial-vector component of the weak current, 
and the natural spin polarization. The possibility to address the main 
limitations of (anti)neutrino experiments at future facilities, allowing 
the collection of high statistics samples combined with an accurate control 
of the targets and fluxes~\cite{Petti:2019asx}, can provide valuable information complementary 
to the EIC program. In particular, precision measurements of $\nu(\bar \nu)$ 
interactions on both hydrogen and various nuclear targets with 
the high intensity and the energy spectra of the planned LBNF beams can offer a broad mixture of measurements of electroweak parameters, parton and hadron structure of nucleons and nuclei, 
nuclear physics, form factors, structure functions and cross-sections, 
as well as searches for new physics or verification of existing outstanding
inconsistencies~\cite{ESGprop,Snowmass21}. Exploring possible synergies and complementarities between the EIC and future neutrino programs could potentially further enhance 
their physics reach by combining the unique features of the electron 
and (anti)neutrino probes~\cite{petti_poetic9}. 
For example, the $\nu$-p CC DIS at LBNF and the e-p CC DIS at the EIC are characterized by similar 
hadronic tensors and trivially-related leptonic tensors, allowing to elucidate the complete flavor structure of the nucleon. 
In the following, we review the neutrino-nucleon cross section 
and the various kinematic regions in scattering with nuclei, and 
we list a few topics of particular interest. 


\subsubsection{Cross sections and kinematical regions} 
\label{sec:cross-section-kinematics} 

The CC neutrino- or antineutrino-proton scattering 
cross section is given by three structure functions $F_1$, $F_2$, 
and $F_3$ 
\cite{Winter:2000xh,CooperSarkar:2011pa,kumano_1988_kobe,Kumano:2018bwh},
\begin{align}
\! \!
\frac{d\sigma_{CC}^{\nu/\bar\nu }}{dx \, dy}
 =  \frac{G_F^2 \, s }{2 \pi \, (1 + Q^2 /M_W^2 )^2} 
     \, \bigg [  \, F_1^{\, CC} \, x \, y^2
     + F_2^{\, CC} \left ( 1 -y -\frac{Mxy}{2E} \right )
     \pm F_3^{\, CC} \, xy \, 
         \left( 1 - \frac{y}{2} \right ) \,
        \, \bigg ] ,
\label{eqn:nu-p-cross}
\end{align}
where $\pm$ indicates $+$ and $-$ for neutrino and antineutrino, respectively, $G_F$ is the Fermi coupling constant, $s$ is the square of the center-of-mass energy, and $M_W$ is the $W$ boson mass.
The structure function $F_1$ is related to $F_2$ 
by the Callan-Gross relation $2x F_1 = F_2$
in the parton model, and the structure functions
are expressed by the PDFs for the proton as
\begin{align}
F_2^{\nu p \, (CC)} 
    & = 2x \, \big ( \, d + s + \bar u + \bar c \, \big ) , \ \ 
x F_3^{\nu p \, (CC)} 
      = 2x \, \big ( \, d + s - \bar u - \bar c \, \big ) ,
\nonumber \\
F_2^{\bar\nu p \, (CC)} 
    & = 2x \, \big ( \, u + c + \bar d + \bar s \, \big ) , \ \ 
x F_3^{\bar\nu p \, (CC)} 
      = 2x \, \big ( \, u + c - \bar d - \bar s \, \big ) ,
\label{eqn:f123-parton}
\end{align}
in the leading order of $\alpha_s$.
The neutral-current (NC) cross section is given
in the same way by the replacements
$M_W \to M_Z$, $G_F^2 \to \rho G_F^2$ where
$\rho = M_W^2 / ( M_Z^2 \, \cos^2\theta_W )$
with the weak-mixing angle $\theta_W$,
and $F_{1,2,3}^{(CC)} \to F_{1,2,3}^{(NC)}$.
The NC structure functions are expressed by the PDFs 
defined by $q^\pm = q \pm \bar q$ as
\begin{align}
F_2^{\nu/\bar\nu\, p \, (NC)} 
  & \! = \! 2x \, \left [ \, (u_L^2+u_R^2) \, \big ( u^+ + c^+ \big )
      + (d_L^2+d_R^2) \, \big ( d^+ + s^+ \big ) \, \right ] ,
\nonumber \\
x F_3^{\nu/\bar\nu\, p \, (NC)} 
  & \! = \! 2x \, \left [ \, (u_L^2-u_R^2) \, \big ( u^- + c^- \big )
      + (d_L^2-d_R^2) \, \big ( d^- + s^- \big ) \, \right ] .
\end{align}
The left- and right-handed couplings for a quark are 
expressed by the third component of isospin $T_a^{\, 3}$,
charge $e_q$, and the weak-mixing angle $\theta_W$ as
$q_{_L} =  T_q^{\, 3} - e_q  \sin ^2  \theta _W$, 
$q_{_R} =  - e_q  \sin ^2  \theta _W$,
with $T_q^{\, 3},\ e_q  = +1/2, \ +2/3 \ (-1/2, \ -1/3)$ 
for $q=u,c \ (d,s)$.

\begin{wrapfigure}[12]{r}{0.40\textwidth}
\begin{center}
\begin{minipage}[c]{0.40\textwidth}
    \vspace{-0.85cm}
    \hspace{+0.10cm}
    \includegraphics[width=6.0cm]{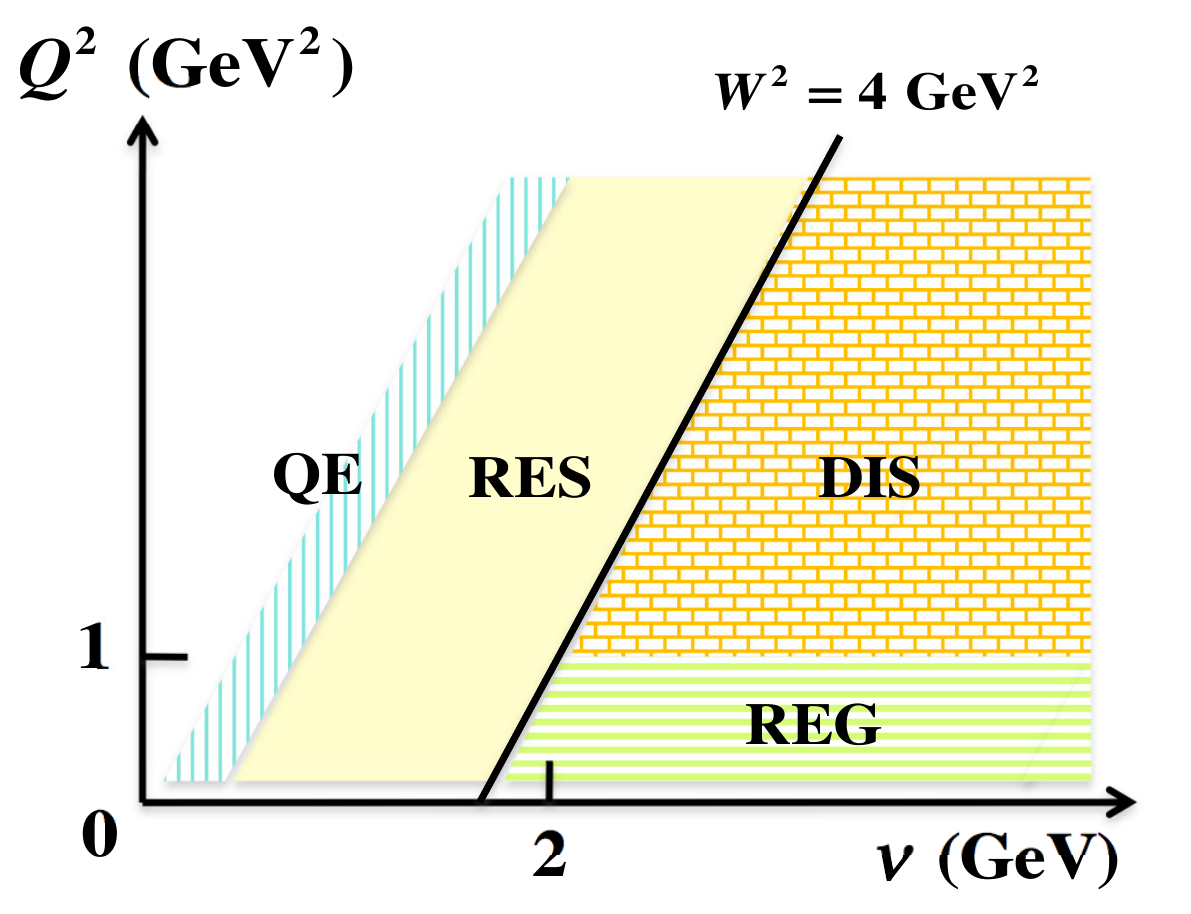}
\vspace{-0.80cm}
\caption{Kinematical regions of neutrino-nucleus scattering.}
\label{fig:nu-nucleus-kinematics}
\vspace{-0.5cm}
\end{minipage}
\end{center}
\end{wrapfigure}

\noindent
In lepton-nucleus scattering, the kinematical regions of
quasi-elastic (QE), resonance (RES), 
deep-inelastic scattering (DIS), and Regge-exchange (REG)
are shown in Fig.~\ref{fig:nu-nucleus-kinematics}
as functions of the energy transfer $\nu$ and $Q^2$.
For the current neutrino oscillation experiments with
neutrino energies ranging from a few hundred MeV to a few GeV,
all of these kinematical regions should be understood accurately~\cite{Nakamura:2016cnn}.
At low energies, the lepton interacts with nucleons almost elastically,
and nucleon resonances appear as the lepton energy increases.
In the DIS region characterized by $W^2 \ge 4 \, \textrm{GeV}^2$ and $Q^2 \ge 1 \, \textrm{GeV}^2$, the nucleon breaks up for the majority of the events. 
The region $W^2 \ge 4 \, \textrm{GeV}^2$ and $Q^2 < 1 \, \textrm{GeV}^2$
is described by the Regge theory.
For calculating neutrino cross sections in all of those regions,
theoretical descriptions should be tested by charged-lepton data.
Then axial-vector components should be added, so that
EIC measurements are valuable.


\subsubsection{Studies of (anti)neutrino-nucleus interactions} 
\label{sec:nuclear} 

The EIC program will provide important information on the nuclear 
modifications of the nucleon properties (see also Sec.~\ref{part2-sec-LabDenseQCD}) 
which are relevant to understand (anti)neutrino-nucleus interactions~\cite{Alvarez-Ruso:2017oui}. As a result of the structure 
of the weak current, significant differences are expected for 
nuclear effects in charged leptons and (anti)neutrino DIS. 
In general, nuclear modifications of structure functions 
and parton distributions depend on the isospin of the target and 
on the $C$-parity, and can therefore differ for neutrino and 
antineutrino interactions~\cite{Kulagin:2007ju}. At the typical 
$Q^2$ values accessible in $\nu(\bar \nu)$ inelastic scattering, 
higher-twist contributions play an important role, both at the nucleon 
and at the nuclear level. Using both neutrino and antineutrino DIS 
we can access the different structure functions $F_2, \, xF_3, \, F_T$, 
as well as $R=F_L/F_T$, which is expected to have a different behavior 
at small $Q^2$ with respect to electromagnetic interactions~\cite{Kulagin:2007ju}. 

The flavor separation of the weak current and the availability of 
precision (anti)neutrino measurements off different nuclear targets 
would allow for exploring the flavor dependence of nuclear effects on 
parton distributions, structure functions, and form 
factors~\cite{Hirai:2007sx,Eskola:2016oht,Kovarik:2015cma,deFlorian:2011fp,Khanpour:2016pph,Kulagin:2014vsa}. 
To this end, a comparison between interactions on hydrogen~\cite{Duyang:2018lpe,Duyang:2019prb,Snowmass21h}
and on nuclear targets is particularly relevant. 
The isospin symmetry can provide a determination of the free neutron 
structure function ($F_{2,3}^{\bar \nu p} = F_{2,3}^{\nu n}$) and hence the one of the average isoscalar nucleon using interactions on hydrogen  
$F_{2,3}^{\nu N} \equiv (F_{2,3}^{\nu p} + F_{2,3}^{\bar \nu p})/2$. 
We can then obtain a direct measurement of the nuclear ratios 
$R_A \equiv F_{2,3}^{\nu A} / F_{2,3}^{\nu N}$. An interesting point 
to study, in addition to the flavor dependence of nuclear effects, 
is the role of the axial-vector current and the corresponding 
differences with electromagnetic interactions. 
All of those nuclear physics measurements offer complementary 
information to the EIC program and the corresponding 
CC DIS measurements. 

\subsubsection{Measurements of the strangeness content of the nucleon} 
\label{sec:strange} 

Sensitivity to the strange-quark content of the nucleon can be achieved 
at the EIC from measurements of identified hadrons in semi-inclusive 
DIS~\cite{Aschenauer:2019kzf} and with charm jets in CC DIS~\cite{Arratia:2020azl}. Neutrino and antineutrino scattering provide a direct access to the strangeness of the nucleon 
via charm production. The intensity of LBNF beams will allow for
precision measurements of exclusive decay modes of charmed hadrons 
(e.g., $D^{*+}, \, D_s, \, \Lambda_c$) and of charm fragmentation 
and production parameters. In addition, the strange sea quark distributions
$s(x)$ and $\bar s(x)$ can be probed with the $\mu \mu$ and $\mu e$ 
inclusive semi-leptonic charm-decay channels with a statistics more than 
one order of magnitude higher than the largest samples currently available~\cite{Samoylov:2013xoa,Goncharov:2001qe}. 
The analysis of both neutrino- and antineutrino-induced charm production, 
in combination with the EIC measurements (Sec.~\ref{part2-subS-UnpolPartStruct.P.N}), 
can provide an accurate 
determination of the strange-quark content of the nucleon and of 
the corresponding $s-\bar s$ asymmetry~\cite{Alekhin:2018dbs,Alekhin:2017olj,Alekhin:2014sya}. 

While the elastic form factors of the strange quark vector current have been measured 
with good accuracy in parity-violating electron scattering (PVES)~\cite{Young:2006jc}, the strange axial-vector form factors are still poorly constrained by experiment. 
Neutrino and antineutrino measurements can 
accurately determine the latter from NC elastic scattering off protons, 
$\nu_\mu (\bar \nu_\mu) p \to \nu_\mu (\bar \nu_\mu) p$~\cite{Ahrens:1986xe,Garvey:1992cg,Alberico:1998qw,AguilarArevalo:2010cx}. 
In the limit $Q^2 \to 0$, the NC differential cross section is proportional 
to the axial-vector form factor, 
$d \sigma / d Q^2 \propto G_1^2=(-G_A/2+G_A^s/2)^2$, where $G_A$ 
is the known axial form factor and $G_A^s$ is the strange form factor. 
This process provides a direct measurement of the strange-quark contribution 
to the nucleon spin, $\Delta s$, by extrapolating the NC differential 
cross section to $Q^2=0$ since in this limit $G_A^s \to \Delta s$. 
A combined analysis with PVES data would allow for
an accurate determination of all three strange form factors 
$G_E^s, \, G_M^s, \, G_A^s$~\cite{Pate:2003rk,Pate:2008va}. 

\subsubsection{Isospin physics and sum rules} 
\label{sec:isospin} 

Isospin physics is a compelling topic for future neutrino experiments 
looking for differences between neutrino and antineutrino interactions. 
The EIC can provide accurate measurements of the $d/u$ content of 
the nucleons with proton and deuteron data (Sec.~\ref{part2-subS-UnpolPartStruct.P.N}), 
as well as the corresponding nuclear 
modifications in nuclei (Sec.~\ref{part2-sec-LabDenseQCD}). Complementary measurements can be obtained 
using both $\nu$ and $\bar \nu$ interactions on hydrogen~\cite{Duyang:2018lpe,Duyang:2019prb,Snowmass21h}. 
In particular, the isospin symmetry allows for a direct measurement of the free neutron 
structure functions $F_{2,3}^{\nu n} \equiv F_{2,3}^{\bar \nu p}$ 
and $F_{2,3}^{\bar \nu n} \equiv F_{2,3}^{\nu p}$. 
This measurement provides, in turn, a precise determination of 
the $d/u$ quark ratio up to values of $x$ close to 1~\cite{Alekhin:2017fpf,Accardi:2016qay}.  

The Adler sum rule~\cite{Adler:1964yx,Allasia:1985hw}, 
$S_A=0.5 \int^1_0  dx/x ( F_2^{\bar \nu p} - F_2^{\nu p} ) = I_p$, 
gives the isospin of the target and can be measured as a function 
of $Q^2$ using $\nu(\bar \nu)$ interactions 
on the proton and heavier nuclei~\cite{Kulagin:2007ju}. 
The value of $S_A$ is sensitive to possible 
violations of the isospin (charge) symmetry, heavy-quark (charm) production, 
and strange sea asymmetries $s-\bar s$. 
The Gross-Llewellyn-Smith (GLS) sum rule~\cite{Gross:1969jf,Kim:1998kia}, 
$S_{GLS} = 0.5 \int^1_0  dx/x ( xF_3^{\bar \nu p} + xF_3^{\nu p} )$, 
can also be measured in $\nu$ and $\bar \nu$ interactions. The value 
of $S_{GLS}$ receives both perturbative and non-perturbative 
QCD corrections and its $Q^2$ dependence can be used to extract 
the strong coupling constant $\alpha_s$~\cite{Larin:1991tj,Kataev:1994rj}. 
Measurements with both proton and heavier nuclei~\cite{Kulagin:2007ju} would allow for an investigation of the isovector and nuclear corrections. 
The EIC measurements can constrain the small-$x$ behavior of the structure 
functions, reducing the uncertainties on both the Adler and GLS sum rules. 

We repeat that isospin symmetry implies the relation $F_{2,3}^{\bar \nu p} = F_{2,3}^{\nu n}$ 
and that for an isoscalar target $F_{2,3}^{\bar \nu} = F_{2,3}^{\nu}$.
These relations as a function of $x$ and $Q^2$ can be exploited for precision 
tests of isospin (charge) symmetry using a combination of proton and
isoscalar nuclear targets. 
The EIC data can provide valuable constraints to improve the accuracy of such measurements --- see Sec.~\ref{part2-subS-UnpolPartStruct.P.N} for more details.

\subsubsection{Electroweak measurements and the NuTeV anomaly} 
\label{sec:weak-mixing} 

Neutral currents not only contain isospin currents,
which act on left-handed components, but also 
electromagnetic currents which act on both
left- and right-handed ones.
The mixing fraction of those currents is determined by the weak-mixing angle $\theta_W$.
This angle was accurately measured by collider experiments;
however, the NuTeV Collaboration reported an anomalously large 
weak-mixing angle~\cite{Zeller:2001hh},
$\sin^2 \theta_W = 0.2277 \pm 0.0013 \, \rm{(stat)} 
                         \pm 0.0009 \, \rm{(syst)}$.
It is different from a global analysis of other data,
$\sin^2 \theta_W= 0.2227 \pm 0.0004$,
in the year 2002.
This is called the NuTeV anomaly, which has not been fully understood until now.
Since $\theta_W$ it is one of the fundamental parameters of the SM, 
it is important to find the cause for this difference.
Neutrino and antineutrino charged- and neutral-current events were
analyzed for extracting $\sin^2 \theta_W$ in the NuTeV experiment.
The Paschos-Wolfenstein relation
$ R^-  = ( \sigma_{NC}^{\nu N}  - \sigma_{NC}^{\bar\nu N} ) /
         ( \sigma_{CC}^{\nu N}  - \sigma_{CC}^{\bar\nu N} )
        =  1/2 - \sin^2 \theta_W $,
which is supposed to hold for the isoscalar nucleon,
was used for its determination.

Since the NuTeV target was iron instead of the isoscalar nucleon, 
various correction factors needed to be considered to this
Paschos-Wolfenstein relation. 
In addition, this relation was obtained by assuming isospin symmetry which provides a relation between PDFs of the neutron and PDFs of the proton.
There are correction terms to the Paschos-Wolfenstein relation
from isospin breaking in the PDFs,
different nuclear corrections for $u_v$ and $d_v$,
finite distributions of $s(x)-\bar s(x)$ and $c(x)-\bar c(x)$,
and neutron-excess effects~\cite{Kumano:2002ra,Hirai:2004ba,Martin:2004dh,Bentz:2009yy,Kulagin:2003wz}.
These factors can be accurately measured from the $\gamma$-$Z$ interference at the EIC via parity-violating asymmetries in polarized electron scattering, providing an independent 
check of the NuTeV anomaly~\cite{Zhao:2016rfu}. It must be emphasized that 
$\nu$ and $\bar \nu$, contrary to charged leptons, give a direct access to both the weak-mixing angle $\theta_W$ and the $Z^0$ coupling to (anti)neutrinos. Therefore, combined electroweak 
measurements using electrons and (anti)neutrinos are required. 
Precision measurements of electroweak parameters can be performed at LBNF using various independent channels including DIS, $\nu e^-$ elastic scattering, elastic scattering 
off protons, coherent $\rho$ production, etc. These processes are characterized by different 
scales of momentum transfer, providing a tool to test the running of $\sin^2 \theta_W$, in 
addition to the EIC measurements. 
The range of accessible scales covers a region from $0.01 \, \textrm{GeV}$ to a few GeV, partially overlapping the EIC coverage (Sec.~\ref{part2-sec-Connections-EW}) and extending it to lower scales.

\subsubsection{Possible GPD measurements in neutrino scattering} 
\label{sec:neutrino-GPD} 

\begin{wrapfigure}[12]{r}{0.40\textwidth}
\begin{center}
\begin{minipage}[c]{0.40\textwidth}
    \vspace{-0.80cm}
    \hspace{+0.37cm}
    \includegraphics[width=4.5cm]{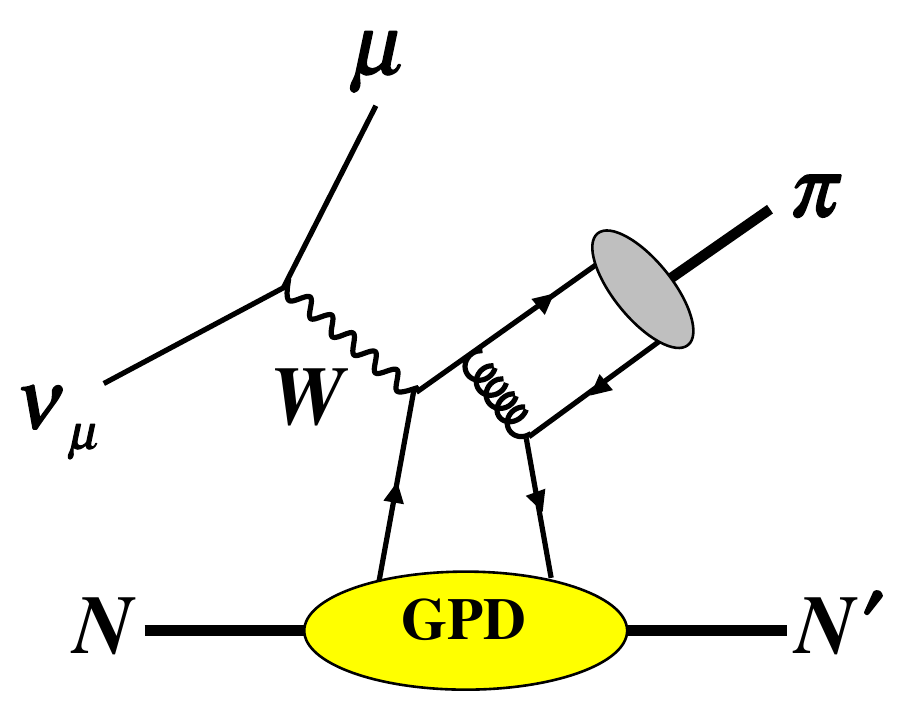}
\vspace{-0.10cm}
\caption{Sample leading-order diagram for the reaction 
         $\nu_\mu + N \to \mu +\pi+ N^\prime$ involving GPDs.}
\label{fig:nu-GPD}
\vspace{-0.2cm}
\end{minipage}
\end{center}
\end{wrapfigure}

The detailed understanding of the GPDs~\cite{Diehl:2003ny} is one of major goals of the EIC science program --- see Sec.~\ref{part2-subS-SecImaging-GPD3d}. 
For instance, GPDs can provide crucial new insights into the origin of the nucleon spin through quantifying the contributions due to partonic orbital angular momenta.
However, there is another important motivation to study GPDs, namely to determine gravitational form factors to find the origin of hadron masses and pressure distributions inside hadrons in terms of quark and gluon degrees of freedom, as obtained in the timelike GPDs 
(or generalized distribution amplitudes)~\cite{Kumano:2017lhr}.
The LBNF can supply a $5-10 \, \textrm{GeV}$ neutrino beam, so that it is possible to measure the GPDs, for example, in the pion-production reaction 
$\nu_\mu + N \to \mu +\pi+ N^\prime$ as shown in Fig.\,\ref{fig:nu-GPD}~\cite{Pire:2017tvv,Kumano:2018bwh,Kopeliovich:2012dr}.
Since the neutrino reactions are sensitive to the quark flavor, their measurements are complementary to the EIC project, and the flavor separation of the quark GPDs will be facilitated using measurements from both EIC and LBNF.

\subsection{Cosmic ray/astro-particle physics}
\label{part2-sec-Connections-Cosmic}

Measurements at the EIC will provide important input into several areas of astro-particle physics.  These areas require more precise models of hadronic interactions to be able to interpret astrophysical data.  These areas include cosmic-ray air showers and neutrino astrophysics.

\subsubsection{Cosmic ray air showers}

Cosmic-ray air showers occur when a high-energy proton or heavier nucleus strikes the atmosphere, producing a shower of millions to trillions of particles.   Cosmic rays with energies above about $10^{15}$ eV are rare enough so that they can only be studied with ground-based detectors.  These detectors sample the shower particles that reach the ground, measuring their density and lateral spread.  Cosmic-ray physicists  use these indirect data to determine the energy spectrum and nuclear composition of cosmic-rays. The energy can be determined largely from the overall particle density.  The composition is often inferred from the muonic content of the shower.  The muons are mostly from the decays of charged pions and kaons and neutral kaons, while photons and electrons come from photons from $\pi^0$ decays.  Strangeness production models thus play a large role in inferring the composition from muon data.  Very few hadrons reach the ground, so they are only a useful observable below about 1 PeV, where showers are more copious.

Air fluorescence detectors image the shower as it develops in the atmosphere to find the depth of shower-maximum, $X_{\rm max}$, the point along the shower trajectory containing the most particles.  For a given cosmic-ray energy, heavier particles, like iron, have a larger interaction cross-section and a lower per-nucleon energy, so reach $X_{\rm max}$ at a higher altitude than for lighter primaries.  Because more of the interactions occur at higher energies, they also produce more muons than incident protons. A hadronic interaction model is required to quantify this relationship, and to infer event energies and composition. Cosmic-ray physicists use a number of different models, for this, with SIBYLL, QGSJet and EPOS being the most common.  These models use pQCD to model hard interactions, with a Pomeron inspired phenomenology to simulate the soft interactions that account for most of the produced particles.  They are tied, to varying degrees, to RHIC and LHC data, but still vary significantly in their predictions \cite{Dembinski:2019uta}. 

Since cosmic-rays essentially follow a fixed-target geometry, measurements in the far forward region are critical to track energy flow downward through the atmosphere.  Although the TOTEM \cite{Royon:2020pom} 
and LHCf \& RHICf \cite{Sako:2020dom} experiments have made some cross-section and forward multiplicity measurements, this phase space has not been well studied at colliders. The EIC will have excellent forward and far-forward instrumentation, allowing for accurate studies in the target fragmentation region.  
In particular, a knowledge of the inelasticity of struck protons in hadronic collisions is a vital input to hadronic models. Electron-proton collisions are not the same as $pp$, but they will help constrain the models. 

These models are receiving attention because the energy spectra measured by the two very large (area more than 1,000 km$^2$) experiments, Auger \cite{Fenu:2017hlc} and Telescope Array (TA) \cite{Ivanov:2020rqn} are in tension.  A joint working group could not resolve this disagreement \cite{Ivanov:2017juh}.  The difference may be due to physically different cosmic-ray spectra in the Northern and Southern hemispheres \cite{diMatteo:2020dlo}.  This would be a very important discovery, pointing to the existence of a few local cosmic-ray sources.  However, before reaching that conclusion, we need to exclude other possibilities.  The two experiments use somewhat different detection techniques - water Cherenkov detectors for Auger, scintillator for TA, so inaccuracies in hadronic models could lead to differences in energy calibration.

Unfortunately, there is considerable tension between different composition measurements, both between the two experiments, and, for Auger, between composition measurements using air fluorescence and that using muon detectors. Composition is often quantified using the log of the mean atomic number, $\langle \ln(A)\rangle$. This is a single number, easy to quantify, but it misses the intricacies of the nuclear mass spectrum.

The Auger $X_{\rm max}$ analysis finds a significant composition shift with increasing energy, from an apparent mostly-proton composition, with $\langle \ln(A)\rangle\approx 0.3$ at $10^{18.5}$ eV to a mixed composition at $10^{19.5}$ eV, with $\langle \ln(A)\rangle\approx 1.3$ \cite{Aab:2017cgk}.  This is with the QGSJET-II-04 hadronic model; with EPOS-LHC, $\langle \ln(A)\rangle$ is about 0.5 larger - a very significant shift.   As was discussed in the EIC White Paper, this inflection point in $dX_{\rm max}/dE$ could also come from a change in the character of the hadronic interaction, such as the onset of saturation.  In contrast, the TA study of $X_{\rm max}$ finds a rather small change in  $\langle \ln(A)\rangle$ with increasing energy \cite{Bergman:2020swo}.  For QGSJET-II-04, their data is consistent with mostly protons, while other hadronic models imply somewhat heavier compositions.

Other Auger analyses, using muons, have found somewhat different results.  A study using detector rise time as an indicator of muon content found a somewhat larger composition shift with energy, with the composition at $10^{19.5}$ eV consistent with pure iron (for the EPOS-LHC model) \cite{Mezek:2018rpo}.  A newer, still preliminary measurement using dedicated muon counters, found an even larger muon excess, pointing to, if naively interpreted, a composition heavier than iron \cite{Muller:2019dvf}. The muon content is 40-50\% higher than is expected with a lighter (consistent with the $X_{\rm max}$ analysis) composition.  These muon mismatches are a rather clear sign that there is a issue with the air shower simulations, likely in the hadronic model.  

At medium energies ($10^{15}$ - $10^{17.5}$ eV), the situation appears somewhat better, in that there are fewer obvious inconsistencies.  However, recent studies, such as a new IceCube  measurement \cite{IceCube:2019hmk} show that uncertainties to hadronic models are a major contributor to systematic uncertainty on the cosmic-ray composition.  Complementary studies, using high $p_T$ muons in air shower, should allow for composition extraction in a pQCD framework, are less advanced \cite{Abbasi:2012kza,Soldin:2018vak}.

EIC data could significantly help to reduce these uncertainties, by providing high-accuracy measurements of hadronic particle production to tune the models, especially in the forward region, and, for high $p_T$ muon analyses, by pinning down parton distributions at low $x$.  If it is observed, saturation could also explain some of the inflection points seen in the composition distributions.   Data from oxygen and/or nitrogen targets is of particular value, to match the air-shower targets. 

\subsubsection{Astrophysical neutrinos}

The discovery of high-energy (up to at least $10^{16}$ eV) cosmic neutrinos \cite{Aartsen:2014gkd} is one of the most exciting recent developments in high-energy astrophysics.  EIC data can offer important information for future astrophysical neutrino studies, by helping constrain the background from atmospheric neutrinos, and by better predicting the absorption of high-energy neutrinos in the Earth.

Atmospheric neutrinos from cosmic-ray air showers are a significant background to astrophysical neutrinos. A better understanding of atmospheric neutrinos would lead to smaller systematic uncertainties on cosmic neutrino fluxes, especially at lower energies, where atmospheric neutrinos are dominant.   One important question is how often downward-going atmospheric neutrinos are self-vetoed by being accompanied by high-energy muons \cite{Gaisser:2014bja}, either by a muon produced in the same weak decay as the neutrino, or elsewhere in the hadronic interaction or elsewhere in the air shower.

Prompt neutrinos are of particular interest, since they have a harder energy spectrum than neutrinos from pions/kaons, so they are closer to astrophysical neutrinos.  A standard calculation \cite{Bhattacharya:2015jpa} has large uncertainties, due to uncertainties on the cosmic-ray flux and composition, on the pQCD cross-sections, and on the low-x gluon distributions in nitrogen and oxygen.  Currently, the only measurements of gluon distributions in medium nuclei come from fixed target experiments, which can only probe the region $x\gtrapprox 10^{-2}$ (and that at low $Q^2$).  EIC data will extend these results downward in $x$ and cover a wide range in $Q^2$, reducing the uncertainties on prompt neutrino production.  Beyond this, it will be another place to compare charm production cross-sections with theoretical calculations; this comparison should lead to more accurate treatments of charm production in air showers.

Looking ahead, future detectors, like IceCube Gen2 will instrument much larger areas (order 100 km$^3$) with radio-detection stations and thereby study astrophysical neutrinos at higher energies, up to $10^{20}$~eV \cite{Aartsen:2020fgd}.  It is necessary to know the interaction cross-section to measure the flux accurately.
At $10^{20}$ eV, neutrino interactions probe quark distributions with a typical $x\approx 10^{-7}$, at $Q^2 \approx M_W^2$.  One recent calculations, using NLO pQCD with DGLAP evolution found roughly 15\% \cite{CooperSarkar:2011pa} uncertainty in the DIS cross-sections at $10^{20}$ eV, while another calculation, with different assumptions about parton evolution, found a considerably larger, 50\% uncertainty at the same energy \cite{Connolly:2011vc}.  A newer calculation used data from LHCb on D meson production to constrain the low-$x$ parton distributions, with a consequent decrease in uncertainty \cite{Ball:2016neh}.  This study used a NNLO calculation found a cross-section about 10\% below the previous calculations, except at neutrino energies above $10^{19}$ eV. The stated uncertainties were under 10\%.

These results were for isoscalar targets, without nuclear effects.  Water has more protons than neutrons, so there are deviations from the expectations for targets with $Z=N$.  Also, shadowing may be significant for very high energy neutrinos.    Ref. \cite{Bertone:2018dse} finds that nuclear corrections are small at low energies, but increase to up to 10\% (with large uncertainties) at energies above $10^{20}$ eV.  The calculations in Ref. \cite{Klein:2020nuk} find smaller overall changes, with anti-shadowing increasing the cross-section in some ranges.  Further, in the region where large-$x$ parton distributions are significant, the proton excess in water can have a very large effect, with large uncertainties.  This change is particularly visible in the $\nu$ inelasticity (fraction of neutrino energy transferred to the nuclear target) distribution.  Since, for most analyses, acceptance varies with inelasticity \cite{Aartsen:2018vez}, assuming the correct inelasticity distribution is critical in determining the neutrino flux.   Measurements of the inelasticity distribution are also used to determine the $\nu/\overline\nu$ ratio in atmospheric neutrinos, using the different distributions for $\nu$ and $\overline\nu$. 

Knowing the cross-section is critical for estimating interaction rates and zenith angle distribution in a detector \cite{Garcia:2020jwr,Klein:2019nbu}. A larger cross-section increases the number of downward-going neutrinos that interact, but, because neutrinos interact in the Earth, reduces the number of upward-going events.  Neutrino absorption in the Earth has been observed for energies above 6.3 TeV, at a level consistent with the standard model \cite{Aartsen:2017kpd}.  The inelasticity distribution is important because, for charged-current $\nu_\mu$ and most $\nu_\tau$ interactions, only the energy transferred to the struck target is visible in radio-based detectors \cite{Aartsen:2018vez}. 

Beyond rate estimates, a good understanding of the standard-model cross-section is critical in searching for BSM contributions to the total cross-section.  A variety of BSM models predict an increase in the neutrino-nucleon cross-section, including those positing sphalerons, leptoquarks, and extra rolled-up dimensions  \cite{Klein:2019nbu}.  EIC measurements of quark distributions in nuclei should reduce the uncertainties on the cross-section and inelasticity distributions significantly.  Although EIC measurements cannot cover the full range of Bjorken-$x$ needed for neutrino astrophysics, EIC data should be able to significantly constrain the non-linear evolution to lower $x$ values, reducing the uncertainties at all energies. 

Finally, some recent studies have applied pQCD to calculations of prompt neutrino production in astrophysical accelerators \cite{Bhattacharya:2014sta}. Improved measurements of gluon distributions in protons would reduce the uncertainties on the predicted neutrino flux and energy spectrum.  The spectra are of particular importance because different spectra can be characteristic of different source classes.
\subsection{Other connections to \pp, \pA, \AA}
\label{part2-sec-Connections-Other}

\subsubsection*{Low-x gluons and factorization in $\textrm{eA}$ ($\textrm{ep}$) vs $\textrm{pA}$ and $\textrm{AA}$}
A few years ago, an interesting connection between the CGC formalism and the TMD factorization has been established, and a lot of progress in both fields is made\cite{Angeles-Martinez:2015sea}. In particular, this has led to a fundamental understanding of two different gluon distributions, namely the Weizs\"{a}cker-Williams (WW) gluon distribution and the dipole gluon distribution, from the perspective of the operator definition in the CGC formalism~\cite{Bomhof:2006dp, Dominguez:2011wm}. The progress and its impact on EIC physics are discussed in detail in Sec. \ref{part2-subS-LabQCD-Saturation}. This part is devoted to the small-$x$ gluon and its factorization in \eA vs \pA and \AA collisions. 

First, based on the small-$x$ factorization in DIS and \pA collisions, the WW and dipole gluon distributions can be viewed as the fundamental building blocks of all gluon distributions in the CGC formalism, and they can be used to construct other more complicated gluon distributions appearing in the dijet production in \pA collisions in the large $N_c$ limit. Since the gauge links associated with the gluon distributions depend on the details of the scattering process, the low-$x$ gluons are then process-dependent as shown in Table~\ref{twog}. The table also shows that many processes are sensitive to the dipole gluon distribution, while the back-to-back dijet production in DIS can provide the direct measurement of the WW gluon distribution. Meanwhile, $\textrm{pA}$ collisions can serve as a gateway to the EIC as far as saturation physics is concerned, and it also plays an important and complementary role in the study of these two fundamental gluon distributions. Furthermore, in this factorization at the low-$x$ limit, the virtual photon and the proton are treated as dilute probes for the dense gluons in the target. The corresponding cross section for a certain process in DIS and $\textrm{pA}$ collisions can still be expressed in terms of the convolution of the relevant gluon distributions and the short distance hard part. The small-$x$ factorization in DIS and $\textrm{pA}$ collisions is expected to hold at higher order\cite{Dominguez:2012ad}, since the higher-order corrections do not generate genuine new correlators in the large $N_c$ limit. Last but not least, the production of color-neutral particles in hadron-hadron collisions, such as the Higgs production\cite{Sun:2011iw, Mueller:2013wwa, Liou:2012xy} in \pA and \AA, are also sensitive to the WW gluon distribution in heavy nuclei. However, for the productions of hadronic final states in \AA collisions, the issue of the factorization\cite{Collins:2007nk, Rogers:2010dm} becomes more complicated due to the color entanglement of initial and final state interactions, and one needs to use a more complicated form\cite{Gelis:2008ad, Gelis:2008rw, Gelis:2008sz} of small-$x$ factorization in the CGC formalism and resort to numerical methods to obtain results in this case. 

\begin{table}[bth]
\begin{tabular}{c|c|c|c|c|c|c }
 \hline
 &\textrm{Inclusive DIS} & SIDIS & DIS dijet & Inclusive in \pA & $\gamma$+jet in \pA & dijet in \pA  \\
 \hline $xG_{\textrm{WW}}$& $- $ & $-$ & $+$ & $-$ & $- $& $+$ \\
 \hline $xG_{\textrm{DP}}$ &$+$ &$+$ &$-$ &$+$ & $+$& $+$ \\
\hline
 \end{tabular}
\caption{The process dependence of two gluon distributions (i.e., the Weizs\"{a}cker-Williams (WW for short) and dipole (DP for short) distributions) in \eA (\ep) and \pA collisions. Here the $+$ and $-$ signs indicate that the corresponding gluon distributions appear and do not appear in certain processes, respectively.}
 \label{twog}
 \end{table}
 

\subsubsection*{Implications  of PDF determinations for proton-proton collisions}

\begin{figure}[tbh!]
\includegraphics[clip,width=0.5025\textwidth]{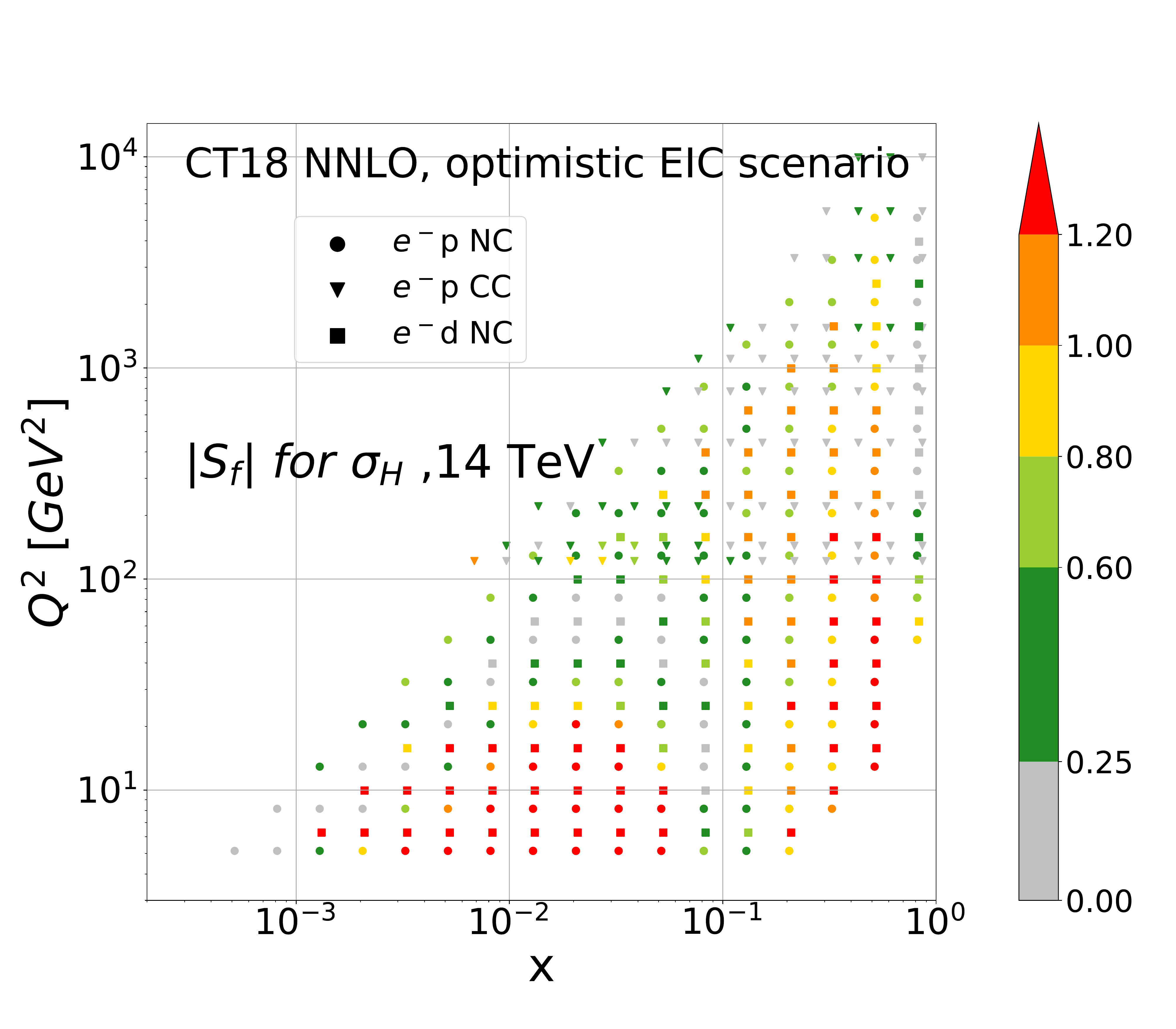}
\includegraphics[clip,width=0.5025\textwidth]{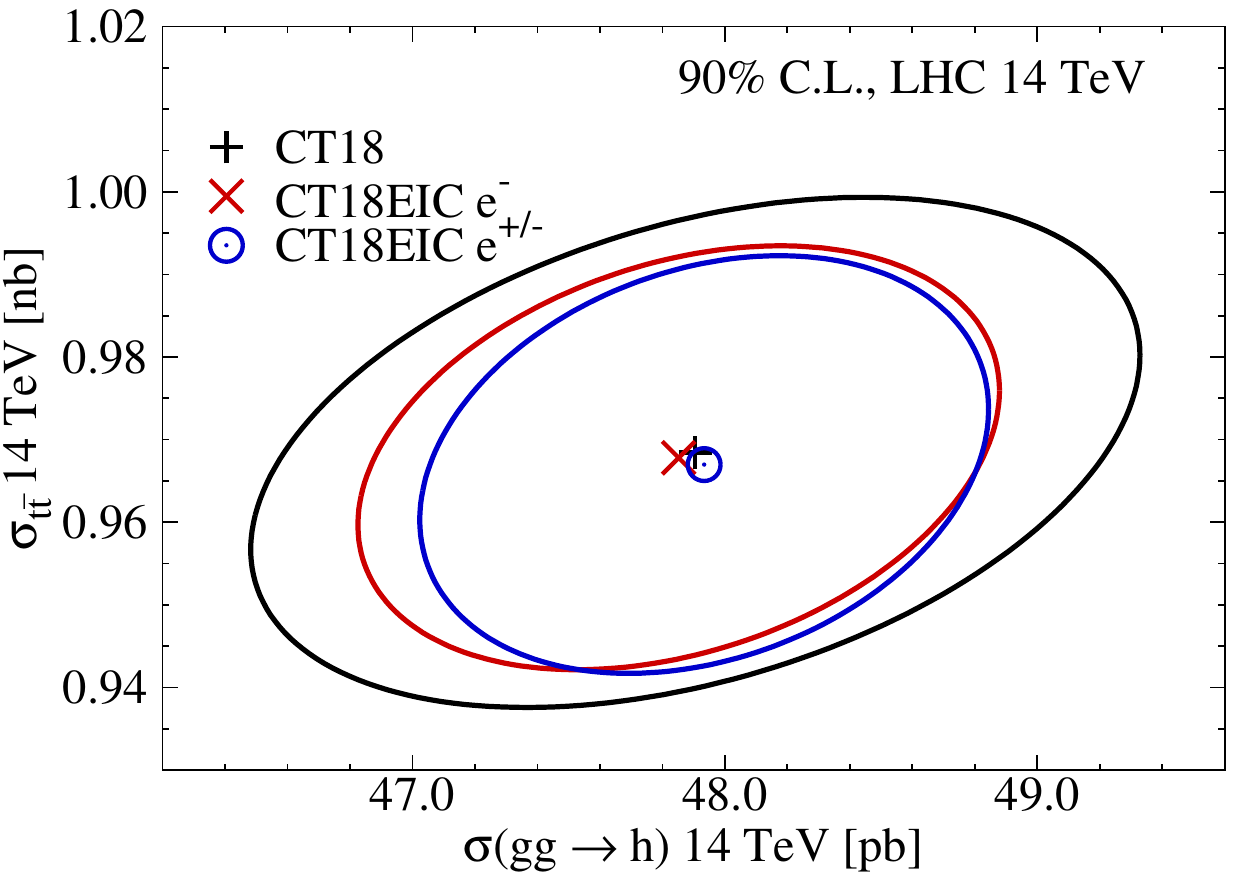}
	\caption{
		(Left) The PDF {\it sensitivity}, $|S_f|$, of the main EIC $e^-$ pseudodata
		of Sec.~7.1.1 to $\sigma_H(\mathrm{14\,TeV})$, the total $gg \to H$ cross
		section for Higgs production at 14 TeV. Redder points indicate points with stronger
		pull as described in Refs.~\cite{Wang:2018heo,Hobbs:2019gob,Hobbs:2019sut},
		implying significant impact of the $e^-$ information on $\sigma_H(\mathrm{14\,TeV})$.
		(Right) Computed total Higgs cross sections at 14 TeV cross sections, with
		Higgs-$t\bar{t}$ uncertainty ellipses for the calculations based on the CT18
		NNLO \cite{Hou:2019efy} baseline shown as the outer black and EIC post-$e^-$ uncertainties shown by the inner red and post-$e^-$/$e^+$ (inner blue) fits.
	}
\label{fig:EIC-Higgs}
\end{figure}

The realization of ultimate precision at hadron colliders like the (HL-)LHC
remains limited by uncertainties in the unpolarized proton PDFs. Achieving
hightened sensitivity to various BSM scenarios requires a variety of 
improvements, including next-generation theoretical accuracy (such as N$^3$LO
hard cross sections) and additional constraints to the PDFs themselves.
By recording copious high-precision DIS data, the EIC has the potential to
impose important constraints to the PDFs as illustrated in Sec.~7.1.1 with
implications for observables in $pp$ scattering at the LHC and precision
QCD and EW theory. We show here two representative examples. 

Figure~\ref{fig:EIC-Higgs},
concentrating on potential EIC impacts in the Higgs sector, shows the PDF sensitivity of the total $gg\to H$ cross section. The plot uses EIC pseudodata assuming an integrated luminosity of $\mathcal{L}=100\,\mathrm{fb}^{-1}$, corresponding to the optimistic scenario for detector performance and systematic uncertainties. 
The significant per-point sensitivities translate into a pronounced total impact on the computed 14~TeV cross sections. 

\begin{figure}[tbh!]
\includegraphics[clip,width=0.485\textwidth]{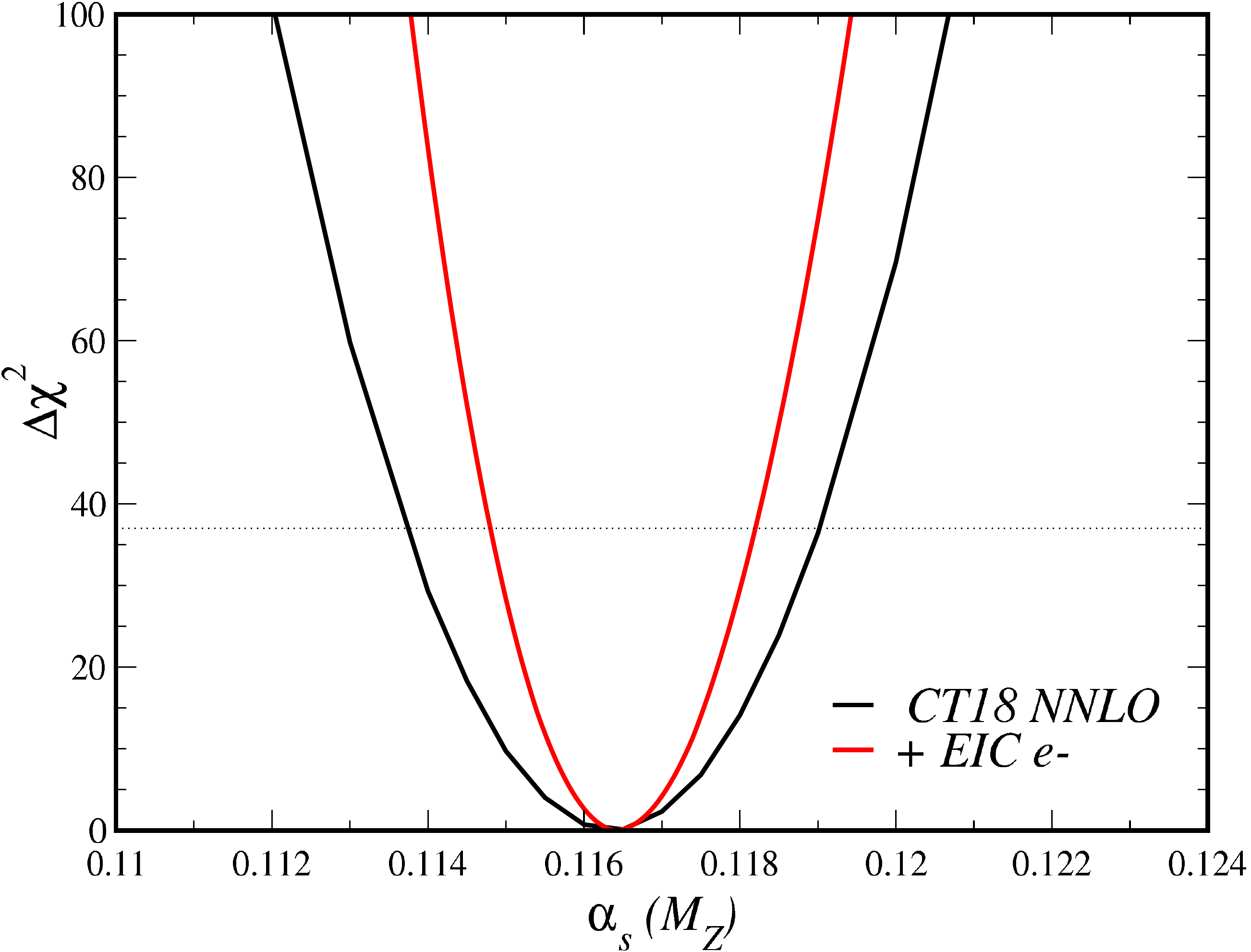}
\includegraphics[clip,width=0.51\textwidth]{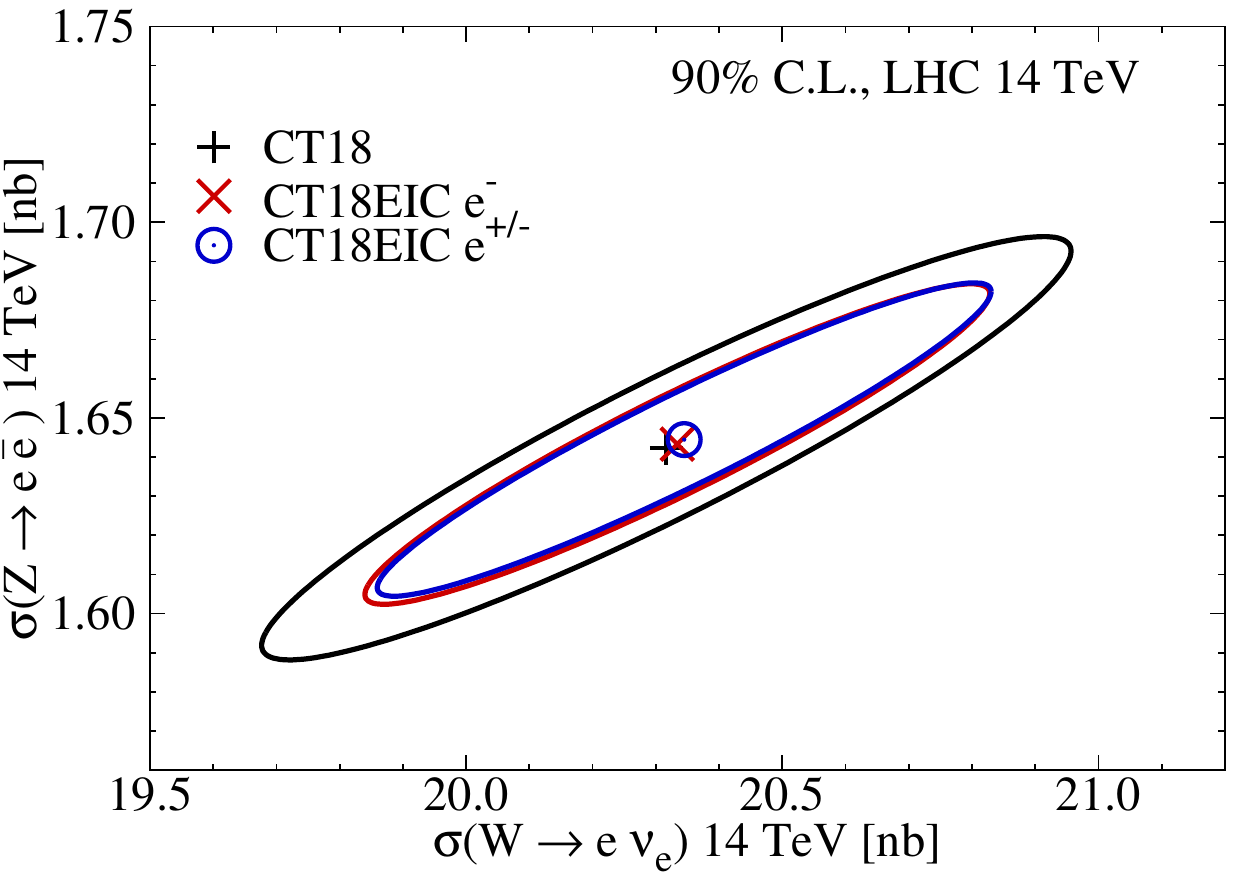}
	\caption{
		(Left) The likelihood function, $\chi^2$, for $\alpha_s$  in a fit with the
		EIC $e^-$ data (red) relative to the  CT18 NNLO global analysis baseline (black).  The narrowing corresponds to a $\sim\!40\%$ reduction of the $68\%$ C.L.~uncertainty of the strong coupling. 
		(Right) The uncertainty ellipses for total $W$ vs.~$Z$ cross sections  in $pp$ collisions at 14 TeV once the EIC $e^-$ (inner red ellipse) or combined $e^-/e^+$ (inner blue) pseudodata are included on top of the baseline CT18 prediction (outer black).
	}
\label{fig:alpha-EW}
\end{figure}

Figure~\ref{fig:alpha-EW} focuses on  precision QCD and EW physics, where precise DIS data over a wide range of $x$ and $Q^2$ are expected to impose significant constraints to QCD-sector Standard Model inputs, including $\alpha_s$ and heavy-quark masses. The figure demonstrates the effect of the main EIC $e^-$ optimistic scenario pseudodata on the  likelihood function, $\chi^2$, for  $\alpha_s$  in the CT18 NNLO global analysis. The effect of the EIC pseudodata corresponds  to a $\sim\!40\%$	reduction of the $68\%$ C.L.~uncertainty of the strong coupling. Also shown is the impact the uncertainties  for total $W$ vs.~$Z$ production at 14 TeV once the EIC $e^-$ pseudodata are included on top of the baseline CT18 prediction.

\subsubsection*{Implications  of PDF determinations for proton-nucleus collisions}

It has been demonstrated \cite{Aschenauer:2017oxs,AbdulKhalek:2019mzd} that measurements at EIC are bound to significantly improve our knowledge of the nuclear PDFs. In particular, it should be possible to tightly constrain the gluon distribution at $x \gtrsim 10^{-2}$ at scales comparable to the charm-quark mass, $Q \sim m_{\rm charm}$. Due to DGLAP dynamics, this translates to well constrained gluons even at $x \ll 10^{-2}$ at higher interaction scales

Such an improved description will lead to more precise predictions for \pA collisions at the LHC and thereby allow for stringent tests of factorization. For example, the forward D-meson production measured by LHCb \cite{Aaij:2017gcy} can probe the nuclear structure down to $x \sim 10^{-6}$ at perturbative scales. It thus serves as an ideal place to search e.g.\ for non-linear dynamics beyond DGLAP. At the moment, the nuclear PDFs are not particularly well constrained at small $x$ and can be easily fitted to reproduce the LHCb D-meson data without conflicts with the other existing data \cite{Kusina:2017gkz,Eskola:2019bgf}. However, since the EIC constrains nuclear PDFs at much lower $x$ than existing DIS data, there should be significantly less room for additional tuning to fit LHC data. This increases the chances for discovering e.g.\ the onset of non-linear evolution at the LHC. Similar conclusions hold in the case of other observables such as the direct photon production at forward direction \cite{Helenius:2014qla} possibly to be measured by ALICE \cite{ALICECollaboration:2020rog}. In fact, even the inclusive Z and W production at the CMS and ATLAS acceptances carry a significant sensitivity on the gluon PDFs \cite{Citron:2018lsq} (Sect.~10.4.2) as the gluons at low $Q$ dictate the behaviour of sea quarks at the electroweak interaction scale. On the other hand, phenomenological studies of J/Psi production in ultra-peripheral Pb-Pb collisions will profit from precise nuclear PDFs when modeling the generalized nuclear PDFs. In this sense an extraction of nuclear PDFs in a clean environment such as EIC will allow for precision searches of new phenomena in a broad range of observables in \pA collisions.


\subsubsection*{Initial conditions for hydrodynamics in \AA collisions}
Heavy ion collision experiments at RHIC and LHC aim to produce  deconfined quark-gluon matter and study its properties~\cite{Akiba:2015jwa}. The standard model of the ``little bang'' of a heavy ion collision consists of several stages. The initial particle production is followed by a phase of thermalization and equilibration leading to the creation of a droplet of quark-gluon plasma. The plasma then cools and expands in a process that is usually modeled by relativistic hydrodynamics, before undergoing a phase transition to ordinary hadronic matter which then decouples into the hadronic final states that are observed by the detectors. Extracting properties of the quark gluon plasma from measurements requires a simultaneous analysis of different experimental observables in a common theoretical framework. Recent years have seen significant process performing such analyses in a systematical statistical framework (see e.g.~\cite{Bernhard:2019bmu}).

The standard little bang framework is able to predict heavy ion collision observables starting from a given initial condition at the time of equilibration, and from a given set of transport coefficients describing the evolving matter. Inverting this process to infer both the matter properties and the initial conditions is a daunting task, and can in many cases be an ill-posed problem. This is where the physics program of the EIC is relevant in several ways. Firstly, exclusive and diffractive measurements of protons and nuclei at the EIC, discussed e.g. in Secs.~\ref{part2-subS-SecImaging-GPD3d}, \ref{part2-subS-SecImaging-Wigner} and \ref{part2-subS-LabQCD-Photo}, will provide accurate information on the spatial distribution (and its fluctuations) of quarks and gluons in protons and nuclei~\cite{Mantysaari:2020axf}. This spatial structure is one of the most important inputs into hydrodynamical calculations of the quark-gluon plasma, because collective interactions can transform spatial structures in the initial condition into momentum space correlations among produced particles, i.e. hydrodynamical flow generated as a response to pressure gradients in the matter.

As discussed in Sec.~\ref{part2-subS-LabQCD-Collective}, multiparticle correlations that are present in the wavefunctions of the colliding systems can have effects that are very similar to ones resulting from hydrodynamical flow~\cite{Schlichting:2016sqo,Altinoluk:2020wpf}. Especially in what are referred in the heavy ion context to as ``small systems'', i.e. proton-proton and proton-nucleus collisions, it can be difficult to disentangle the effects of hydrodynamical correlations (i.e. ``flow'') from such multigluon ``initial state'' correlations. The ``initial state'' correlations can be studied very precisely at the EIC, and analyzed in terms of concepts like linear gluon polarization~\cite{Dumitru:2015gaa,Marquet:2016cgx}, or Wigner functions.  A precise understanding of the CGC wavefunction from the EIC will also help constrain our understanding of the pre-equlibrium thermalization stage of the heavy ion collision.

\subsubsection*{Parton interactions in matter}

\begin{figure}[t]
\includegraphics[width=0.9\textwidth]{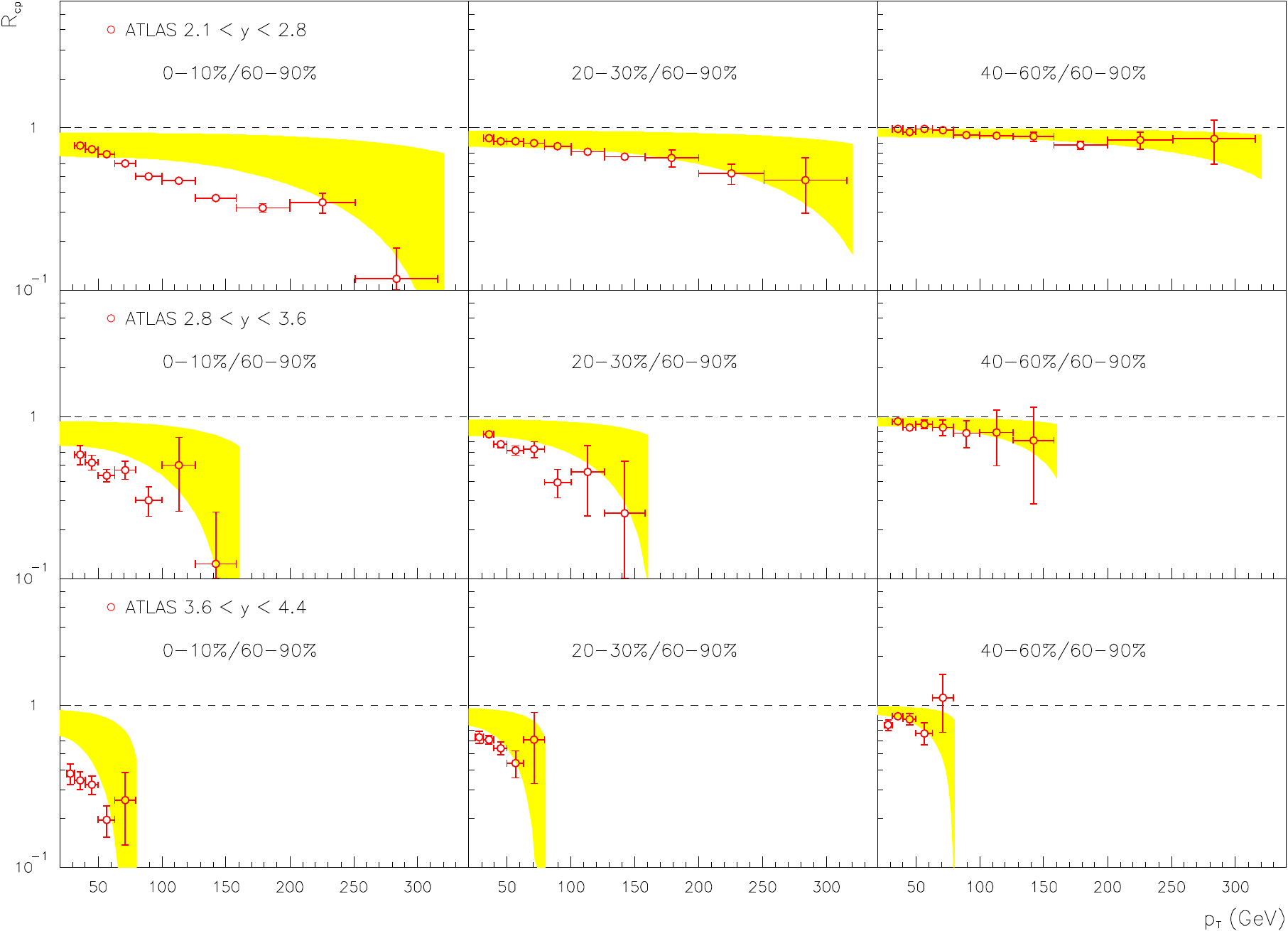}
\caption{$R_{\rm cp}$ for inclusive jet production in p+Pb collisions at $\sqrt{s}=5.02$ TeV in central (left), mid-central (middle) and mid-peripheral (right) events. Four different rapidity intervals ($2.1<y<2.8$, $2.8<y<3.6$ and $3.6<y<4.4$) are presented. Data are from the ATLAS collaboration at the LHC~\cite{ATLAS:2014cpa}.}
\label{fig-Rcp_pt2}
\end{figure}

 The importance of understanding parton and particle  propagation and parton energy loss in DIS has been well articulated in Secs.~\ref{part2-subS-Hadronization-HadNuclear}, \ref{part2-subS-Hadronization-Quarkonia}, \ref{part2-subS-LabQCD-PropPartLoss}. Advances in this direction will facilitate the interpretation of the data from $\textrm{pA}$ and $\textrm{AA}$ reactions. In proton-nucleus collisions, final-state effects associated with the quark-gluon plasma (QGP) are expected to be absent/suppressed. However, experimental results on the centrality dependence of high energy jet cross sections in $\textrm{pPb}$ collisions at LHC~\cite{ATLAS:2014cpa} and in $\textrm{dAu}$ collisions at RHIC~\cite{Adare:2015gla} show highly nontrivial and large nuclear effects from different centrality selections. These are observed at all transverse momenta $p_T$ at forward (in the direction of the proton beam) rapidities and for large $p_T$ at mid-rapidity, and they are manifest as suppression of the jet yield in central events and enhancement in peripheral collisions~\cite{ATLAS:2014cpa}, see Fig.~\ref{fig-Rcp_pt2}. Theoretical work on hadron, jet, Drell-Yan, and $\textrm{J}/\psi$  production in \pA reactions has emphasized the importance of cold nuclear matter (CNM) energy loss~\cite{Vitev:2007ve,Neufeld:2010dz,Xing:2011fb,Arleo:2012hn}. Calculations that incorporate this physics are qualitatively consistent with the central to peripheral cross section ratio denoted $R_{\rm cp}$ in Fig.~\ref{fig-Rcp_pt2}~\cite{Kang:2015mta}. Away from kinematic bounds, CNM energy loss effects are small but can still contribute to the observed quenching in 
 $\textrm{AA}$~\cite{Chien:2015hda}. The impact on cross sections and particle correlation is amplified at smaller center-of-mass energies~\cite{Vitev:2004gn,Qiu:2003pm}.

\subsection{The EIC and nuclear structure physics}
\label{part2-sec-Connections-NuclStruc}

Even though the EIC is a high-energy collider with typical energy scales in the tens-to-hundred GeV range, there are key measurements that are of relevance to nuclear physics at much lower energies in the tens-to-hundreds MeV range. This section outlines the possibilities for measurements at the EIC that have the potential to provide insights into certain aspects of nuclear structure physics. Conversely, guidance from nuclear structure physics is also important in realizing key aspects of the EIC program. 

Short-range nucleon-nucleon correlations (SRCs)~\cite{Arrington:2011xs,Atti:2015eda,Hen:2016kwk} 
dominate the high-momentum tails of the many-body nuclear wave function and show signs of universal behavior in nuclei from deuterium to the heavy nuclei~\cite{Tang:2002ww, Piasetzky:2006ai,Subedi:2008zz,Fomin:2011ng,Korover:2014dma,Hen:2014nza,Duer:2018sby,Duer:2018sxh}. 
Data from JLab indicate that SRCs may provide novel insights into the EMC effect, and one expects that future Jlab data will help to distinguish between competing explanations~\cite{Hen:2016kwk,Arrington:2019wky,Wang:2020uhj}.  The dynamics of these SRCs can also be studied at the EIC using quasi-elastic two-nucleon knockout --- see Sec.~\ref{part2-subS-LabQCD-ShortRange} for more details. 
    
\begin{figure}[t]
    \centering
    \includegraphics[width=0.7\linewidth]{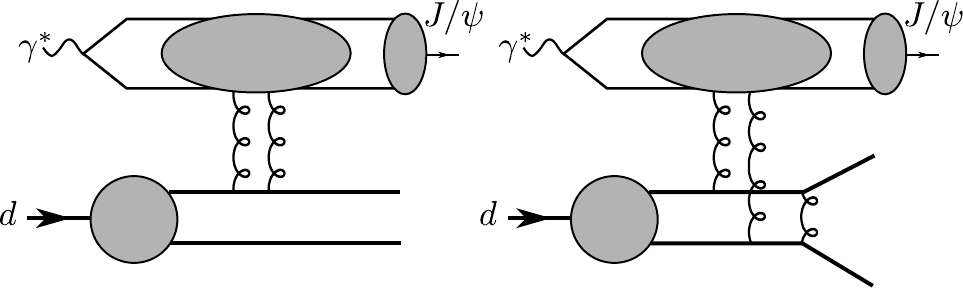}
    \caption{Diagrams for two mechanisms in exclusive quarkonium production on the deuteron with diffractive breakup.  The virtual photon fluctuates into a color dipole and exchanges two gluons with the deuteron.  Left: Two gluons attach to the same nucleon (color-singlet SRC).  Right: Two gluons attach to different nucleons (color-octet SRC).}
    \label{fig:src_diagram}
\end{figure}

A key question is the role of gluons in SRCs, and in the EMC effect. At small $x$, gluon shadowing is expected to be important for nucleon modification in nuclei, as confirmed now in ultraperipheral heavy-ion collisions~\cite{Khachatryan:2016qhq,Abelev:2012ba}.
Also, what is the role of gluons at large $x$? At the EIC, the higher energies and the large lever arm in $Q^2$ will allow one to probe the role of gluons in short-range forces. 
A novel way to explore gluon degrees of freedom in the nucleon-nucleon potential is by looking at exclusive heavy quarkonium (``onium") production in coincidence with knock-out reactions of protons and neutrons. The simplest example to consider is exclusive scattering off the deuteron\footnote{Other light nuclei (such as $^3$He) will also be interesting to consider but will require to reconstruct a more complicated kinematics.}. 
In this context, there exist two interesting possibilities as shown in Fig.~\ref{fig:src_diagram}. The first is when there is a color-singlet gluon exchange between the onium and a ``leading" proton or neutron with a spectator counterpart. In this case, at large relative momentum transfer between the two knock-out nucleons, the scattering is sensitive to color-singlet SRCs in the deuteron wave function~\cite{Strikman:2017koc}. The second case also corresponds to color-singlet exchange between the onium and the deuteron, but in this instance one gluon attaches to the proton and the other to the neutron. The color-singlet structure of the interaction is then necessarily sensitive to color-octet SRCs in the deuteron wave function~\cite{Miller:2015tjf}. 
The first of these gluon-dominated SRCs has been simulated recently using the Monte Carlo event generator BeAGLE~\cite{Beagle} and a novel idea for extracting the $t$-dependence of the knock-out reaction by simultaneously tagging the leading and spectator nucleons~\cite{Tu:2020ymk}. This study suggests that deuteron configurations with typical internal momenta of up to $1 \, \textrm{GeV}$ are accessible --- see Sec.~\ref{part2-subS-LabQCD-ShortRange} for more details. 

An important challenge for both the outlined processes and nuclear breakup channels is to identify and isolate final-state interactions (FSIs).  In exclusive measurements with nuclear breakup, the dominant FSI will be between the slow-moving (relative to the nucleus center-of-mass) breakup products.  In nuclear DIS with tagging, re-interactions of slow hadrons in the DIS target fragmentation region with the spectator fragments also contribute~\cite{Strikman:2017koc}.  Given the low relative momenta in these FSIs, the dynamics have a lot in common with low- and medium-energy nuclear breakup reactions~\cite{Frankfurt:1996xx,Sargsian:2001ax,Laget:2004sm,Egiyan:2007qj,Sargsian:2009hf,Boeglin:2011mt,Strikman:2017koc,Cruz-Torres:2020uke}.

Because of the collider environment, the potential for clean measurements of nuclear fragments with its far-foward detectors, and the high luminosity, the EIC can provide novel insights into low-energy nuclear reactions and correlations. The most relevant measurements are diffractive observables corresponding to a low-momentum transfer color-singlet exchange (of momenta larger than the nuclear Fermi momentum) and a large rapidity gap separating nuclear fragments from the current fragments for a wide range of invariant masses $M_X$. At high energies, there is a very clean separation of time scales between the hard QCD physics of the current fragmentation region and the soft physics of nuclear fragments.
On the other hand, accurate nuclear structure input is needed in coherent exclusive channels with light ions --- enabling the study of nuclear tomography in partonic degrees of freedom --- and in reactions with spectator tagging, which result in additional control over the initial nuclear configuration.  The latter is important for studies of the structure of free neutrons, the short-range nature of the nuclear force and medium modifications of partonic structure.
    
Since the nuclei are strongly boosted at the EIC, the light-front framework for nuclear structure and correlations is appropriate. Because the nucleus is probed at fixed light-front time in a high-energy reaction, boost-invariant wave functions can be constructed and off-shell effects remain finite.  In the reaction frameworks these enter as light-front momentum distributions and spectral functions~\cite{Frankfurt:1988nt,Frankfurt:1981mk,Miller:1997xh,Miller:1999ap,Kaptari:2013dma,Strikman:2017koc,Cosyn:2020kwu}. High-precision phenomenological or effective-field-theory-derived potentials~\cite{Wiringa:1994wb,Machleidt:2000ge,Epelbaum:2008ga,Machleidt:2011zz} enter in this description or can be reformulated starting from light-front quantization~\cite{Cooke:2000ef, Miller:1999ap}. The similarity between light-front approaches like the ``basis light-front quantization''~\cite{Vary:2009gt} and many-body approaches to nuclear structure and correlations such as the no-core shell model~\cite{Barrett:2013nh} has been noted previously. For a review of such {\it ab initio} methods, see Ref.\cite{Hiller:2016itl}. 
    
In particular, considerable insight can be gained into understanding how the high-momentum tails of nuclear wave functions that we noted in the context of SRCs scale with nuclear size~\cite{Chen:2016bde}. Furthermore, similarity renormalization group methods~\cite{Bogner:2006pc}, pioneered originally in the light-front framework~\cite{Glazek:1993rc}, allow for extending the study of such ultraviolet correlations in nuclei to lower momenta~\cite{Bogner:2012zm}. 
    
Polarized light ion beams will be available at the EIC. Measurements with the deuteron's tensor polarization allow one to probe the interplay between high-energy QCD dynamics and low-energy nuclear interactions --- see Sec.~\ref{part2-subS-SecImaging-LpolNucl} for more details.  Tensor-polarized observables are proportional to the deuteron's radial D-wave ($L=2$) component and therefore place constraints on the size and momentum dependence of the D-wave~\cite{Frankfurt:1983qs}. This is important in the context of the universality of SRCs.  With spin-1 nuclei such as the deuteron one can also probe the distribution of linearly polarized gluons in inclusive DIS~\cite{Jaffe:1989xy}, which is impossible for the nucleon.

Precision nuclear-structure input is essential to extract the full potential of the (high-luminosity) EIC for many of the channels considered in this Yellow Report.    For inclusive DIS, the dominant neutron structure uncertainty in the high-$x$ region arises from nuclear structure corrections.  Pinning down the link between medium modifications and nuclear interactions or inferring the size of non-nucleonic components in nuclei (like $\Delta$ isobars) cannot be done without a baseline nuclear structure calculation.  Observables such as the deuteron tensor $b_1$ structure function encode the difference of the unpolarized quark PDF between a deuteron in the polarization state $M=\pm 1$ (``dumbbell'') and $M=0$ (``donut''), and are thus inherently sensitive to nuclear interactions.

Another aspect of the EIC program where nuclear structure input can play a decisive role is the separation of the coherent and incoherent part of the cross section for diffractive and exclusive processes on heavy nuclei.  This is critical for nuclear imaging and studies of gluonic fluctuations. In the incoherent signal, the reaction can involve nuclear excitations that decay with the emission of MeV photons in the nucleus rest frame.  Detecting these photons with the EIC detectors would be very hard, and is further complicated because of large uncertainties (due to lack of data) in the models of nuclear excitations and differences in the level structure between different heavy nuclei (spectra, decay times) which have an impact on the detection.  Other contributions to the incoherent signal have nucleons evaporating from the nucleus.  The models for nucleon evaporation in heavy nuclei used in \eA Monte Carlo generators for the EIC (such as BeAGLE) can be better constrained.  All this complicates the vetoing of incoherent events at larger momentum transfer (direct detection of a coherent scattering event with a heavy nucleus being impossible) --- see Secs.~\ref{part2-subS-LabQCD-Photo} and \ref{part2-sec-DetReq.Diff.Tag} for more details. 

Further input from and collaboration with the nuclear structure community will help resolve those as well as other outstanding issues.  In recent years there have been several workshops with specific input from the nuclear structure community~\cite{Ghent2018,CFNS2018_EICSRC}, and a stronger interest is expected with the EIC project fully underway now.

\subsection{Exotic Nuclei at the EIC}
\label{part2-sec-Connections-ExoNuc}

Seventy years after the introduction of the nuclear shell model, low-energy nuclear structure is still a vibrant field of research. Ever more capable rare isotope facilities have made it possible to create new super-heavy elements and reveal the properties of nuclei far from stability. The latter is important for understanding basic nuclear properties such as modifications to the shell structure, and provides input for solving outstanding problems in other fields, an example of which is nucleosynthesis in astrophysics. Elements that are heavier than iron are typically created in stellar events (supernovae, neutron star mergers) where high fluxes of neutrons lead to captures intermixed with $\beta$-decays. The rapid neutron-capture process (r-process) defines a path towards the heavier elements going through the most neutron-rich nuclei. Measuring the properties of these is therefore one of the key goals of the Facility for Rare Isotope Beams (FRIB), where rare isotopes are produced in low-energy nuclear reactions. A more detailed overview of the FRIB program can, for instance, be found in Ref.~\cite{NAP11796}. However, recent studies suggest that rare isotopes could also be measured at the EIC in parallel with other planned experiments using heavy-ion beams. In addition, the unique capabilities offered by the EIC would make these data complementary to those collected by FRIB.

Short-lived nuclei will decay in flight between the production and detection points (at the IP and Roman pots, respectively). But at the EIC, where the relativistic $\gamma$ is 100, a flight time in the lab frame of 100 ns corresponds to only 1 ns in the rest frame of the ion. As a consequence, only nuclei with half-lives below 1 ns will experience significant in-flight decay. The high survival probability would allow the EIC to study the shortest-lived isotopes very far from stability.

The high-energy kinematics of the EIC are helpful for photon detection. In \eA scattering at the EIC, the initial interaction causes an intra-nuclear cascade that typically knocks out a few nucleons, leaving the residual nucleus in an excited state. The de-excitation initially proceeds through fission and/or emission of nucleons and light nuclei ($d$, $\alpha$). Once below the neutron separation energy, the decay proceeds through $\gamma$ emission. Measuring these photons is of particular interest since the transitions reveal the level structure of the final (rare) isotope. 
The photons are emitted isotropically in the rest frame, but in the lab frame they are preferentially moving in the ion beam direction, where they can be detected close to zero degrees with an energy up to a factor 100 higher than in the rest frame. For spectroscopy, this shift in photon energy means that naturally occurring sources (as well as activation in the vicinity of the beamlines), will not contribute to the background. Since the photons will be measured in coincidence with the ion, the resulting spectrum should be much cleaner than in traditional $\gamma$ spectroscopy measurements. On a practical side, it also means that it should be possible to replace the commonly used HPGe detectors with, for instance, LYSO crystals that do not require cryogenics. One should also note that in contrast to other applications such as vetoing of incoherent backgrounds in coherent diffraction on heavy ions, spectroscopy would benefit from a good photon acceptance, but it would not be a critical requirement.

\begin{figure}[thp]	
	\centering
	\subfloat[]{\label{fig:NZ_a}\includegraphics[width=0.45\linewidth,page=1]{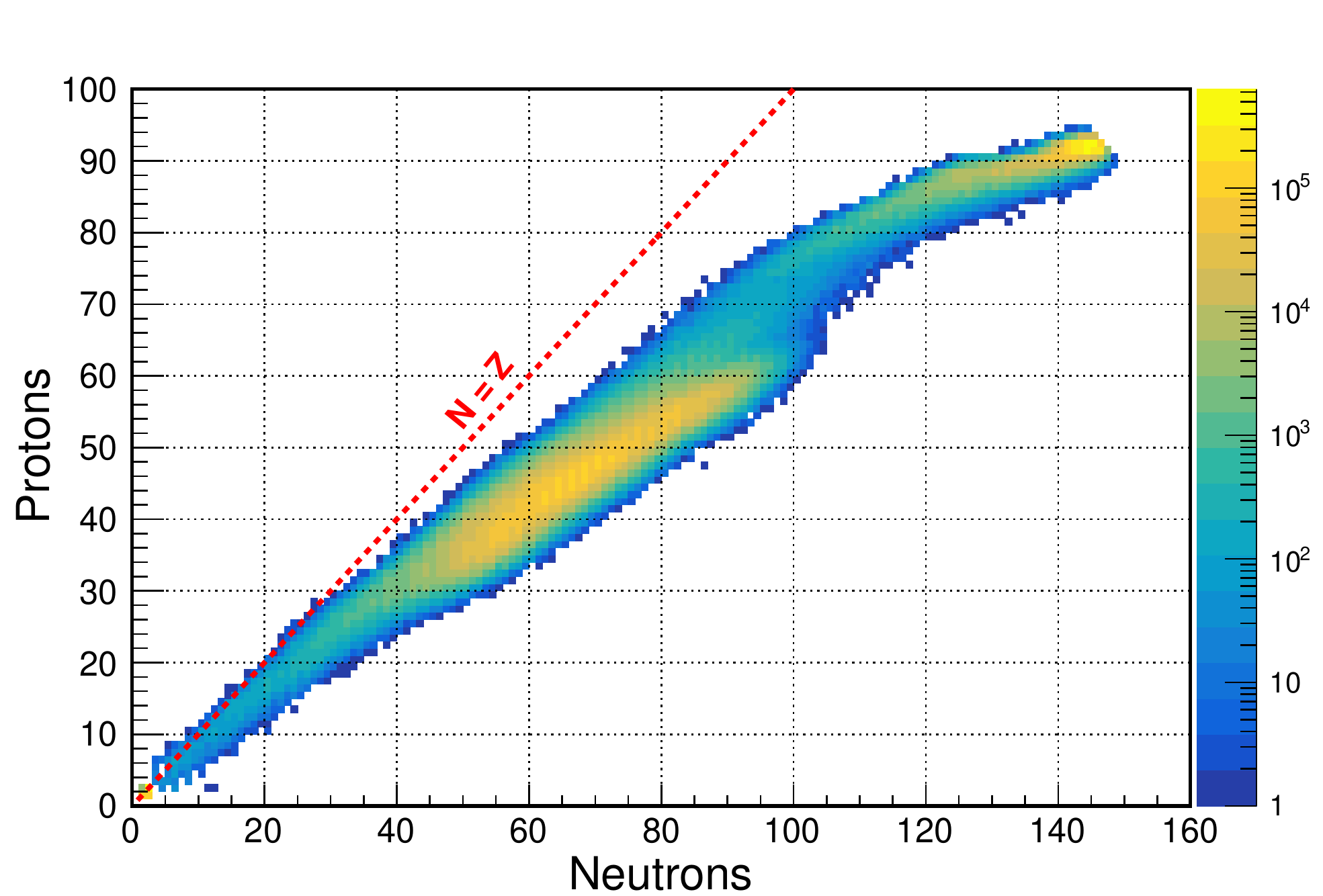}}\qquad%
	\subfloat[]{\label{fig:NZ_b}\includegraphics[width=0.45\linewidth,page=1]{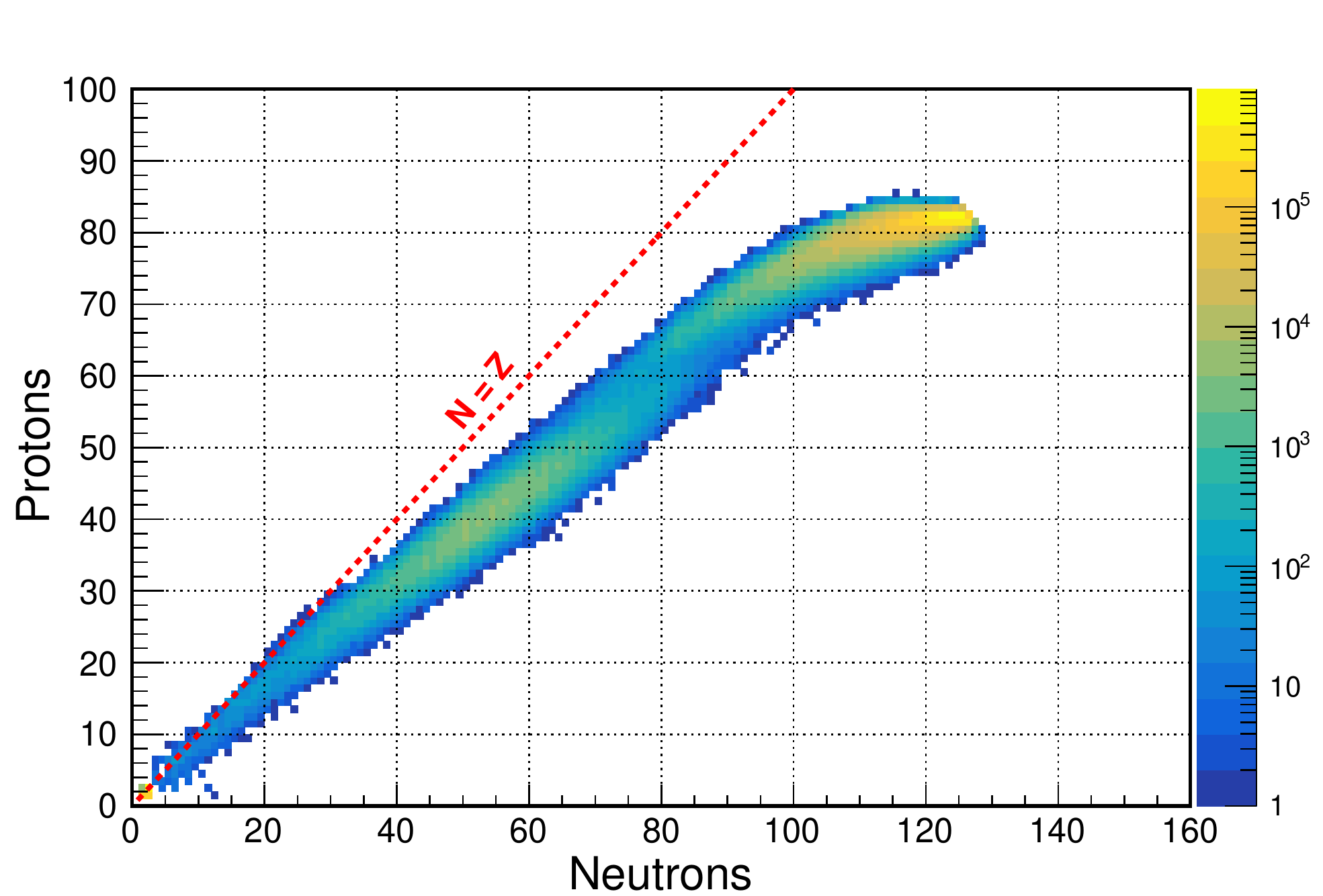}}\qquad%
	\subfloat[]{\label{fig:NZ_c}\includegraphics[width=0.45\linewidth,page=2]{PART2/Figures.EICMeasandStud/U238_NZ_V1.pdf}}\qquad%
	\subfloat[]{\label{fig:NZ_d}\includegraphics[width=0.45\linewidth,page=3]{PART2/Figures.EICMeasandStud/Pb208_NZ_V1.pdf}}
	\caption{Final-state nuclei produced in 4 minutes of beam time in collisions at $10^{33} \thinspace {\rm cm}^{-2} {\rm s}^{-1}$ luminosity between 18-GeV electrons and 110-GeV/A ions. The upper panels show the full distribution for (a) $^{238}$U and (b) $^{208}$Pb, while the lower panels zoom in on the (c) fission region for $^{238}$U and (d) evaporation region for $^{208}$Pb, respectively.}\label{Fig:rare_isotopes_yields}
\end{figure}

The production rates of rare isotopes in \eA collisions have been studied using the BeAGLE event generator. Figure~\ref{Fig:rare_isotopes_yields} shows the distribution of final nuclei with statistics corresponding to about 4 minutes of running at the EIC. The left panels show results for $^{238}$U and the right panels for $^{208}$Pb targets. In the former case, medium-mass fragments are produced abundantly through fission. While such fragments are already on average more neutron-rich than the parent nucleus, the tails of the mass distribution reach very far from stability. This is already evident in the simulated sample. For $^{208}$Pb, fission is less prevalent, but heavier exotic nuclei are produced through evaporation following the intra-nuclear cascade. This process may be intrinsically less efficient than multi-step fragmentation reactions at FRIB, but the EIC production rates seem sufficiently high to provide opportunities for complementary measurements.

Our initial BeAGLE simulations use Fluka for the de-excitation process. Detailed comparisons between Fluka and codes used at FRIB~\cite{TARASOV20084657, TARASOV2016185} and GSI~\cite{Kelic:2009yg} are in progress. The goal is to use the latter to make more reliable estimates for the EIC. Nevertheless, preliminary comparisons suggest a good qualitative agreement.

Experimentally, identification of the produced ions would use the far-forward detectors located 30-50 m downstream of the collision point. The forward spectrometer measures the magnetic rigidity of the ion, which is equivalent to measuring A/Z. However, in order to uniquely identify an isotope one would need an additional independent measurement of Z. The most straightforward way to do this would be to use a small Cherenkov detector behind the tracker in the Roman pot(s). The number of Cherenkov photons produced by an ion is proportional to $Z^2$, and a heavy ion passing through a few mm of fused silica can produce $10^5$ photons. Adding such a “mini-DIRC” should be relatively straightforward. The associated R\&D challenge would be to find an optimal readout which could count the photons with desirable precision. Alternative solutions exist, but would be more bulky (other Cherenkov radiators) or introduce more mass ($dE/dx$). High-resolution timing could also provide a helpful constraint.

A rare-isotope program at the EIC would greatly benefit from an excellent forward acceptance for isotopes that undergo small changes in rigidity compared with the ion beam (which is equivalent to small changes in $p_T$ for light ions). A second focus on the Roman pots, which has part of the original IR design for the JLab EIC concept, would thus be highly desirable. However, it is also essential to retain the highest possible luminosity at the maximum ion beam energies, where most of the data on heavy ions will be taken at the EIC.

\subsection{Interface to High-Energy Physics efforts}
\label{part2-sec-Connections-HEP}

A new generation of high-energy physics experiments will deepen our knowledge of the subatomic matter and evolution of the Universe. 
Particle interactions through the fundamental strong force are of crucial importance in their own right and play the key role across many studies, from electroweak precision tests and Beyond-Standard-Model (BSM) searches at the Large Hadron Collider (LHC) to high-intensity experiments and neutrino and cosmic ray physics. 
There is significant potential for cross-fertilization between studies of hadronic matter in particle and nuclear physics experiments. While the EIC will provide essential new inputs about the structure of nucleons and nuclei for particle experiments, multi-decade experience of the particle community in QCD studies at colliders can benefit various aspects of the EIC program.

The Particle Physics Community Planning Exercise ("Snowmass'2021")~\cite{Snowmass:2021} is a study organized by the Division of Particles and Fields (DPF) of the American Physical Society to consider all options of interest to the US particle physics community and to identify a path forward. The Snowmass study is taking place concurrently with the EIC Yellow Report study. It will inform the Particle Physics Project Prioritization Panel (P5) and High Energy Physics Advisory Panel (HEPAP) with recommendations and research priorities for the US Department of Energy and the US National Science Foundation to pursue.

The EIC will be a unique facility  to study electron-hadron/nucleus collisions in a broad range of center-of-mass energies, $\sqrt{s}= 29-141 \, \textrm{GeV}$, and with high luminosity --- about 100-1000 times larger than that achieved by HERA. 
The versatility of the machine performance, required by the EIC science, pushes the frontiers of accelerator science on many fronts~\cite{NAP25171}. It is hence natural that the EIC Yellow Report and the Snowmass'2021 process cooperatively explore the opportunities that the EIC and its integrated detector will provide to the HEP community and the broader scientific community. It is fortuitous that both the European and the US particle physics communities are engaged in their future planning activities exactly at the time when the US EIC crossed a major milestone of a formal approval (CD0) by the US Department of Energy. The European Particle Physics Strategy Update (EPPSU)~\cite{EPPSU-Report} concluded in Spring of 2020, and the US particle physics community is conducting the Snowmass study as this report goes into print. Both studies proactively engage the high-energy physics communities in exploring opportunities for their own interests at the EIC.
Community inputs pertinent to the EIC studies are submitted to the Snowmass process in the form of Letters Of Interest, Snowmass Proceedings, and Snowmass Frontier reports (see website for details~\cite{Snowmass:2021}. 

{\bf Hadron tomography:} The EIC is expected to have a significant impact on the reach in precision of hadron scattering experiments at future hadron colliders. The program of the High-Luminosity LHC (HL-LHC) is premised on achieving the next-generation sensitivity to a wide variety of SM and BSM processes. The success of this program in testing the SM and performing impact measurements at the TeV scale is critically dependent upon advancements in knowledge of the internal structure of hadrons within QCD.
As developed in this Yellow Report, the EIC will undertake a dedicated tomography program to measure the 2+1-dimensional (dependent on two transverse and one longitudinal direction) structure of the nucleons and a broad range of nuclei~\cite{LOIEICProtonTomography}. This program envisions measurement of observables sensitive to various parton distributions in the proton and other QCD bound states, including TMDs and GPDs, in addition to (un)polarized collinear (longitudinal) PDFs. By facilitating controlled extractions of these various parton distributions and testing relations among them predicted by QCD, the EIC will provide unique data that will clarify detailed mechanisms of formation of QCD bound states. The EIC measurements will be confronted with advanced predictions from multi-loop QCD and lattice QCD.  
While currently fixed-target DIS experiments provide leading constraints on the spin-independent nucleon PDFs at large $x$~\cite{Hou:2019efy,Hobbs:2019gob}  relevant for new-physics searches at the HL-LHC, the EIC will significantly advance in constraining these PDFs and separating parton flavors in the same kinematic region. 

Precise determinations of PDFs and TMDs at the EIC will elevate accuracy of the HL-LHC measurements of electroweak parameters: weak mixing angle and weak boson mass. 
For example, precision measurements of the weak boson mass at the LHC rely on the theoretical formalism of TMD factorization to model the  transverse recoil of weak bosons against QCD radiation. The EIC will constrain TMD PDFs associated with the nonperturbative radiation from up- and down-type quarks. The knowledge of the flavor dependence of TMD PDFs will reduce an important theoretical uncertainty in the LHC $W$ boson mass measurement~\cite{Bacchetta:2018lna,Bozzi:2019vnl,Bacchetta:2019sam}. 
These measurements will stimulate theoretical developments to accurately compute QCD and electroweak radiative contributions, as well as their interplay, in a consistent framework applying to both the EIC and LHC. 

{\bf Semi-inclusive DIS, hadron fragmentation, and jet formation:} The large range of beam energies available at the EIC, combined with the fine resolution and particle identification of the EIC detector, opens a unique venue for exploring formation of hadronic jets, especially the interplay of perturbative QCD radiation and nonperturbative hadronization. The process of semi-inclusive production of hadronic states in DIS will measure in detail the flavor composition of the initial hadronic states as a function of the parton's momentum fraction $x$ and fragmentation of partons into various hadrons as a function of the momentum fraction $z$. At the EIC, it will be possible to study the multiplicity and angular distributions of final hadronic states as a function of the variable center-of-mass energy of lepton-hadron scattering events~\cite{LOIJets}. These observations will offer unique insights about the formation of final-state jets, jet substructure and jet angularity, and they will test universality of underlying perturbative and nonperturbative QCD mechanisms. In turn, production of hadronic jets accompanied by the relevant theoretical advancements will offer novel channels to probe the flavor and spin composition of the EIC initial states ranging from nucleons to heavy nuclei.  EIC studies of jet formation and jet properties go hand-in-hand with the LHC jet physics program, by focusing on aspects of nonperturbative hadronization that are difficult to access in the complex LHC environment.
These observations will guide the development of advanced parton shower algorithms for event generators used by LHC experiments.  

{\bf Heavy-flavor production:} At the EIC, heavy-flavor production will play an important role and will elucidate QCD factorization formalisms ("factorization schemes") for processes with massive quarks, as well as the nonperturbative aspects of heavy-quark scattering dynamics~\cite{LOIHQ}. Advanced capabilities for detection of jets containing charmed particles will open avenues for unique measurements, like the determination of the strangeness content of the (polarized) nucleons and nuclei at momentum fractions $x>0.1$~\cite{Arratia:2020azl}. Hypothetical dynamical mechanisms for massive quark scattering such as "intrinsic charm"~\cite{Brodsky:1980pb} will be constrained.
As an example of unique synergistic capabilities, the construction of a multi-purpose Forward Physics Facility (FPF) is proposed in the next 5-10 years in a cavern in a far-forward region of ATLAS to carry out diverse searches for long-lived particles such as neutrinos and dark bosons~\cite{LOIFPF}. Production of forward neutrinos in the ATLAS collision point and their detection in the FPF will proceed through interactions of weak bosons with charm quarks in a proton or a heavy nucleus, in a similar kinematic region as the one accessed at the EIC. Studies of charm-quark production at large momentum fractions at the EIC thus can provide essential theoretical inputs for the LHC FPF program, by measuring the "intrinsic charm" and other production mechanisms. 

{\bf Electroweak precision and BSM physics:} The combination of high luminosity, the range of accessible energies, and beam polarization at the EIC opens unique opportunities for precision tests of the SM and searches for new BSM physics~\cite{LOIEWBSM}. A whole series of observables will be extracted from the measurement of the parity-violating asymmetry $A_{PV}$ in DIS, the unique EIC process that can be studied using a combination of initial beams and spin polarizations. By tagging on either a final-state electron or missing transverse energy, the EIC will distinguish between DIS events mediated by neutral ($\gamma^*, Z$) and charged ($W^\pm$) vector bosons. In neutral-current DIS, the EIC will measure the less constrained parity-violating structure function $F_3^{\gamma^*,Z}$ that would access the EW coupling constants $g_{1,5}^{\gamma,Z}$ of the fundamental SM Lagrangian and look for their possible deviations due to BSM interations. 

The difference between neutral-current weak couplings of leptons to up- and down-type quarks can be measured in the previously unexplored region of boson virtualities $Q=10-70 \, \textrm{GeV}$, providing information about the energy dependence of the weak mixing angle $\theta_W$. These measurements will impose constraints on BSM models based on "dark parity violation" and a leptophobic $Z'$ boson.  

The ability to register various leptonic final states --- $e,\tau$, missing transverse energy --- in DIS with either electron or positron beams will be used to look for evidence of heavy elementary particles such as dark photons, extra $Z'$ bosons and new fermionic states. A search for charged-lepton flavor violation in the $e+N\to \tau+X$ channel at the EIC will surpass the HERA collider in searches for massive leptoquarks generically predicted by grand unified theories. The kinematic reach of the EIC will allow one to distinguish between scalar and vector leptoquarks. The polarization of the EIC beams, alternations between electron and positron beams, and between hadronic targets will enable versatile tests of leptoquark electroweak couplings as a function of spin and flavor.

We expect synergy between the EIC and HL-LHC programs of new-physics searches, especially if periods of operation of the two colliders overlap. The EIC has the potential to significantly constrain the PDFs and their flavor composition in the region of large partonic momentum fractions, $x>0.01$, where the main constraints on the nucleon structure are currently provided by the fixed-target experiments.
The measurements of the PDFs at the EIC will not be affected by possible new physics contributions that may be present in the relevant kinematic region at the LHC. In addition, EIC searches for indirect signatures of new physics in the framework of the Standard Model Effective Field Theory (SMEFT) will be complementary to the LHC. The EIC, with its polarizable beams, will likely constrain dimension-6 SMEFT operators from new physics that cannot be easily accessed at the LHC. "Combined fits of LHC and projected EIC data can lead to much stronger constraints than either experiment alone."~\cite{LOIEWBSM}

{\bf Saturation and diffractive effects:}
The intermediate energy of the EIC will allow one to examine power-suppressed hadronic contributions and their dependence on the nuclear target. A large part of the EIC program will be dedicated to the structure of nuclei probed in high-energy collisions, including shadowing or saturation effects predicted by QCD for the scattering of high-density partonic systems. 

To summarize, the numerous instances in which the studies of precision QCD and hadronic structure at the EIC will impact activities at the Snowmass Energy Frontier include, but are not limited to 
\begin{itemize}
\item high-energy QCD measurements with accompanying improvements in PDF precision; measurements of the QCD coupling and heavy-quark masses;
\item new knowledge on the gluonic structure of the proton affecting Higgs phenomenology at the HL-LHC;
\item TMDs for precision electroweak physics, including determination of the $W$ boson mass at the LHC;
\item in-depth studies of formation and structure of hadronic jets, and of scattering processes with heavy-quark states;
\item improved resolution of nuclear structure and nuclear-medium effects, with
connections to phenomena like ultra-peripheral photonuclear collisions at hadron colliders;
\item accurate measurements of unpolarized and polarized parton distributions with large momentum fractions that can be confronted with predictions from lattice QCD. \end{itemize}

These developments will depend on various technical and methodological advances, including
\begin{itemize}
\item next-generation perturbative QCD developments, such as multi-loop QCD computations, explorations of QCD factorization theorems and resummation formalisms;
\item phenomenological studies of TMDs/GPDs, including QCD fits and model-based calculations;
\item advancements in precise calculations of electroweak radiative effects;
\item novel insights from lattice QCD; 
\item application of artificial intelligence and machine-learning techniques to proton tomography.
\end{itemize}

\section{Connected Theory Efforts}
\label{part2-sec-TheoryEfforts}

In order to maximize the output of the EIC program, dedicated theory efforts in various directions are of utmost importance.
One example is the calculation of higher-order QCD corrections, including the resummation of large logarithmic corrections.
The status of this field depends on the final state under consideration.
It is also important that data analyses include higher-order corrections as much as possible. 
For a number of reasons, also non-perturbative approaches/models will continue to be essential in the EIC era.
They can reveal general, model-independent results, where the ``re-surrection'' of the Sivers function, based on a model calculation~\cite{Brodsky:2002cx}, is just one striking example in that regard.
Such approaches can as well provide intuition about non-perturbative quantities and, in particular, allow one to compute (new) observables which can help to guide the experiments.
Also the synergy with {\bf lattice QCD} will be absolutely critical.
This field has become remarkably mature by now, thanks to improvements in algorithms, increased computer power and new conceptual breakthroughs.
An example for the latter are new space-like parton correlators through which, for the first time, the $x$-dependence of PDFs and related quantities can be computed directly in lattice QCD.
Lattice calculations can be used to interpret data from the EIC.
Moreover, combining information from lattice QCD with EIC data will (considerably) increase our knowledge about the structure of strongly interacting systems.
An overview of this field in relation to the EIC science is given in Sec.~\ref{part2-sec-Connections-Lattice}.
Another very important area is {\bf QED radiative corrections}, as discussed in more detail in Sec.~\ref{part2-sec-Connections-RadCor}. 
Extraction of precision information about nucleons, nuclei and mesons will be impossible without having a very good understanding of such perturbative effects.
\subsection{Lattice QCD}
\label{part2-sec-Connections-Lattice}

Understanding the internal quark-gluon structure of hadrons from first principles remains a long-term goal of Nuclear Physics, and has been emphasized, e.g., in the Nuclear Science Advisory Committee's Long Range Plan~\cite{Geesaman:2015fha}. 
At the typical energy scales associated with hadron physics, QCD is not amenable to a perturbative expansion in the strong coupling. 
The only known approach that captures the full non-perturbative QCD dynamics involves a discretization of the continuum theory on a 4-dimensional Euclidean lattice, which allows to study QCD via numerical simulations. 
This approach, known as lattice QCD (LQCD), provides a rigorous framework for understanding the non-perturbative aspects of QCD directly from the underlying fundamental theory.
LQCD has advanced significantly since the first numerical explorations about four decades ago, and has successfully reproduced many measurable quantities, such as hadron masses. The progress has enabled lattice calculations to predict new excited and exotic states, some yet to be discovered experimentally. 
LQCD also enables many studies of the structure of hadrons, for example elucidating the decomposition of the proton's spin among its constituents. 
Thermodynamics calculations on a lattice have significantly shaped our ideas about properties of QCD matter at high temperatures. 
The lattice formulation of QCD is also used for high-precision calculations of standard-model parameters, studies of confinement, and weak decays. In many of these areas, LQCD follows a scientific program that strongly aligns with the EIC physics. Moreover, the synergy between perturbative QCD and LQCD will help the EIC program reach its full potential and highest impact. 

LQCD calculations currently provide the most precise determination of the strong coupling constant $\alpha_s$.
Both phenomenological and lattice calculations have systematic uncertainties associated with 
perturbative series truncation.
In lattice calculations, however, it is possible to follow the evolution of
$\alpha_s(\mu)$ to high scales where the perturbative expansion becomes more precise and non-perturbative corrections are negligible\footnote{
  Current LQCD determination of $\alpha_s^{\overline{\rm MS}}$ is done from non-perturbative calculations with up to $N_f=4$ dynamical flavors, with perturbative matching at the quark mass thresholds.}.
The only experimental input needed in the LQCD determinations are the quark masses and the overall scale which are determined from matching to the hadron spectrum and have a negligible impact on precision.
The other important LQCD uncertainty is due to discretization effects, and it can be progressively improved with more computing.
The most recent lattice average value~\cite{Aoki:2019cca} has 30\% improved precision
compared to the 2016 value, and is already more precise than non-lattice determinations~\cite{Zyla:2020zbs}.
Non-lattice determinations combine multiple experimental inputs, in particular $\ell N$
deep-inelastic scattering data, which will be further improved at the EIC.
Since lattice calculations allow for an entirely independent determination, they will continue to
provide an excellent check for the theory, phenomenology and experiments combined to determine 
$\alpha_s$.

LQCD calculations have also provided key insights into thermodynamics and phases of QCD.
Among them are precise calculations of the equation of state at zero baryon density, demonstration that the phase transition is a cross-over, and determination of its temperature to 1\% accuracy. 
While calculations at nonzero density are not possible due to the notorious sign problem, significant progress has been made by Taylor expansion and/or analytic continuation from imaginary chemical potential, which have yielded lower bounds on the location of the critical point. 
Improving these results to full maturity will require expensive computations of the exaflop scale in the years to come. 
Novel ideas as well as quantum computing are currently explored to reach larger baryon densities that correspond to the lowest-energy RHIC collisions and densities above the critical point.

As discussed extensively above in this Yellow Report, one can obtain information on the partonic structure of hadrons through the PDFs and their generalizations (GPDs and TMDs). 
These parton distributions are defined through operators on the light-cone, which is inaccessible in LQCD as it is formulated in Euclidean space. 
Limited information on those quantities may be accessed through their Mellin moments, which have been extensively studied in LQCD for the PDFs and GPDs. However, a systematic calculation of moments beyond the third nontrivial moment is obstructed due to the decaying signal and power-law mixing between operators. A new field has emerged in recent years, most notably the so-called quasi-PDFs approach~\cite{Ji:2013dva}, which connects lattice-calculable matrix elements to  light-cone PDFs via a perturbative matching procedure in the so-called Large Momentum Effective Theory (LaMET)~\cite{Ji:2014gla,Ji:2020ect}. Other ways to extract the $x$-dependence of distribution functions have been proposed earlier that are based on the hadronic tensor~\cite{Liu:1993cv,Liu:1998um,Liu:1999ak}, as well as auxiliary quark field approaches~\cite{Detmold:2005gg,Braun:2007wv}. Following the work on the quasi-PDFs, a number of other methods have been developed, such as the current-current correlator approach~\cite{Ma:2014jla,Ma:2014jga,Ma:2017pxb}, the pseudo-PDFs~\cite{Radyushkin:2016hsy}, and a method based on the operator product expansion~\cite{Chambers:2017dov}. These approaches are now widely applied in LQCD, for the study of proton PDFs, GPDs and more recently TMDs. They have also been extended to other particles, such as the pion and kaon, mostly for the distribution amplitudes. It is expected that more studies of the pion and kaon PDFs will follow using the aforementioned methods to access their $x$-dependence. Improvements in these calculations will complement the experimental effort, as discussed in Sec.~\ref{part2-subS-PartStruct.M}. Finally, the study of the $x$-dependence of the PDF of the $\Delta$ resonance may provide useful insights into hadron structure that cannot be obtained experimentally. For recent reviews of the aforementioned approaches and their implementation in LQCD, see Refs.~\cite{Monahan:2018euv,Cichy:2018mum,Ji:2020ect,Constantinou:2020pek}. The technical advances in LQCD for the calculation of momentum-boosted hadrons can find application in other areas. For instance, the momentum-smearing method~\cite{Bali:2016lva} can be useful for the study of nucleon and meson form factors at high momentum transfer. Progress in this direction will complement the experimental program outlined in Sec.~\ref{part2-subS-SecImaging-FF}.

\begin{figure}[t]
\centering
\includegraphics[width=0.4\textwidth]{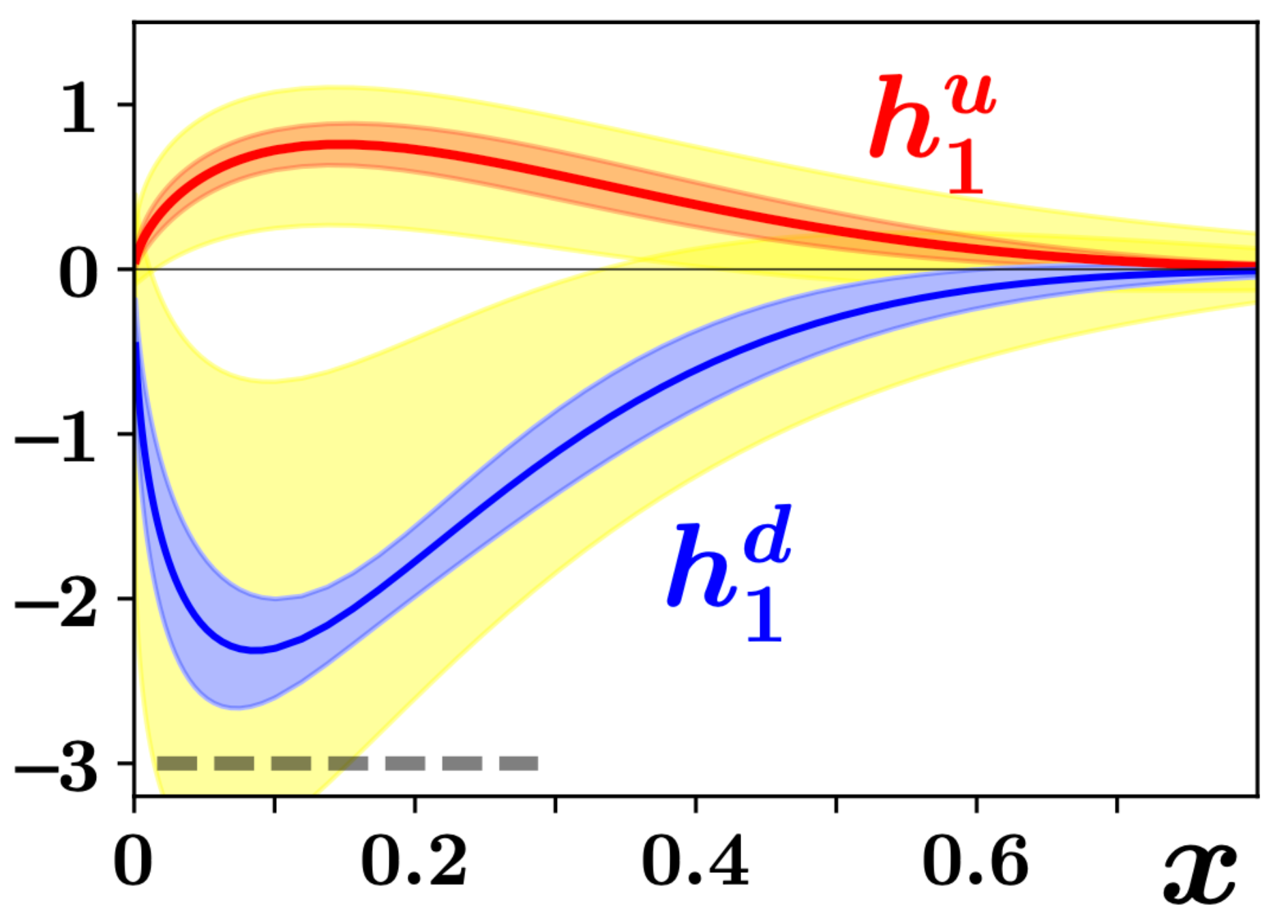}
\hspace*{1cm}
\includegraphics[width=0.35\textwidth]{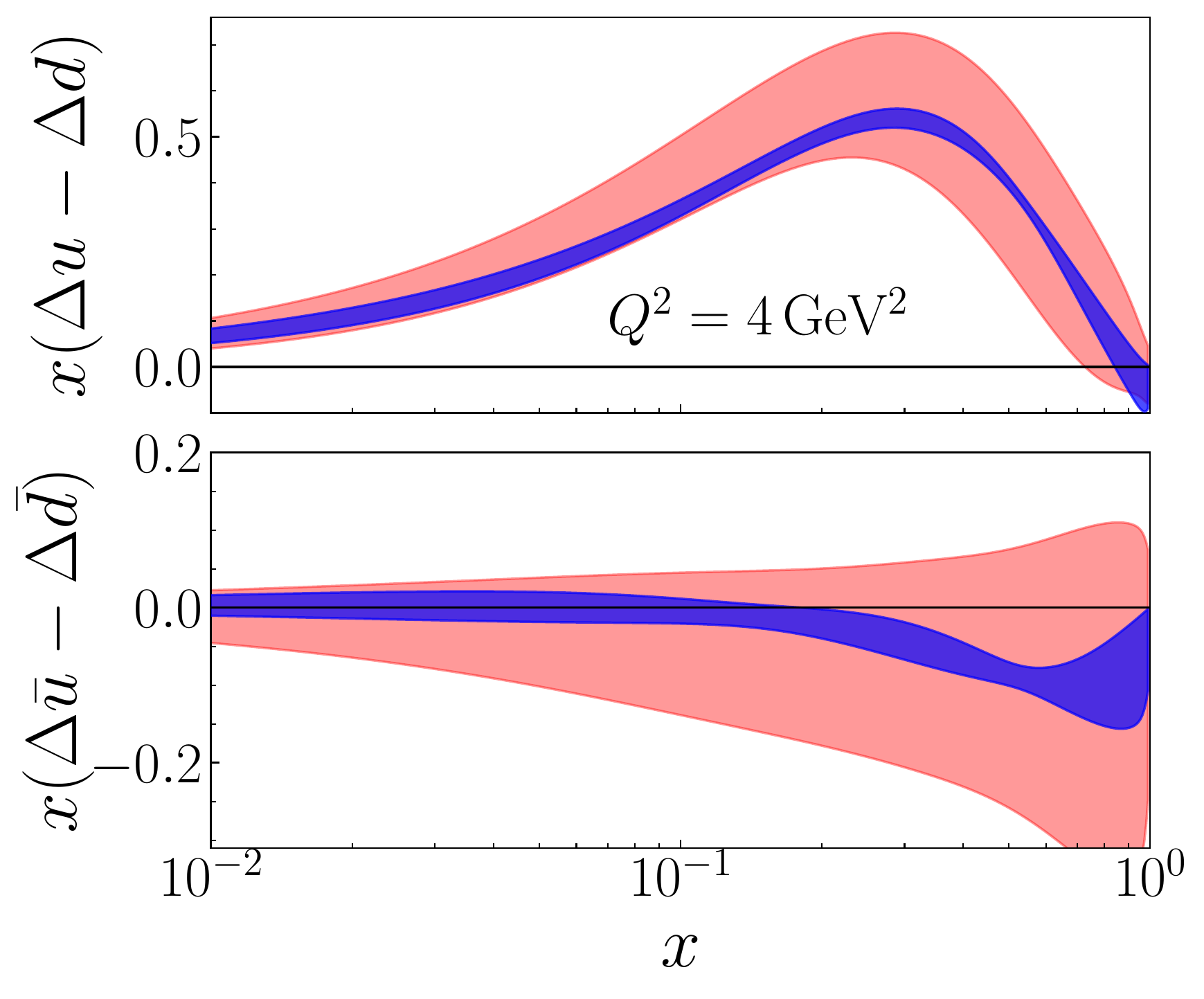}
\caption{Left: The transversity PDF from the JAM global analysis without constraints from LQCD (yellow band) and with constraints using the lattice estimate for the tensor charge (red band for the up quark and blue for the down quark)~\cite{Lin:2017stx}. 
Right: The helicity PDF from the JAM global analysis obtained from experimental data sets only (red bands), or combined with lattice data (blue bands)~\cite{Bringewatt:2020ixn}.}
\label{fig:pheno_lat}
\end{figure}

The range of beam energies and kinematic coverage of the EIC is ideally suited for measuring PDFs, and the corresponding Mellin moments and compare with LQCD~\cite{Hobbs:2019gob}. 
Comparisons of accurate EIC measurements can serve as benchmark quantities for lattice results. Understanding the sources of systematic uncertainties and reproducing well-known experimental results validates the lattice methods. Having confidence in the lattice methodologies, promotes the exploration of a variety of quantities, such as spin-dependent PDFs (for instance, the transversity distribution) which are difficult to access experimentally.
Along those lines, the progress of lattice calculations in obtaining Mellin moments of PDFs and $x$-dependent distribution functions, is of interest to the phenomenological community analyzing experimental data sets to extract the distribution functions. In particular, lattice data are now beginning to be incorporated into global analyses on similar footing as the experimental data sets. This leads to improved estimates of PDFs, particularly in regions where the experimental data are either sparse, imprecise, or non-existing. Synergy between phenomenology and LQCD already lead to better estimates of the transversity PDFs by using lattice results for the tensor charge~\cite{Lin:2017stx}. The left panel of Fig.~\ref{fig:pheno_lat} shows how the lattice data constrain the JAM Collaboration global fits, based on SIDIS data, of the up-quark and down-quark transversity PDFs. 
As can be seen, the accuracy of the PDFs is improved by a factor of 3--4, demonstrating the constraining power of LQCD results. Lattice data on the helicity PDFs were also included within the JAM global analysis framework, and the combined helicity PDFs is shown in the right panel of Fig.~\ref{fig:pheno_lat}~\cite{Bringewatt:2020ixn}. Here, compared to experiment alone, the combined analysis reduces the uncertainties by a factor of 3--6 depending on the $x$ region. 
This holds for both the quark and antiquark contributions. 
Along similar lines, further possibilities for synergy between LQCD and global QCD fits exists for a variety of other quantities, such as the ($x$-dependent) transversity PDF, twist-3 PDFs, GPDs and TMDs. 
The extraction of those functions from the experimental data depends, to some extent, on models, and the LQCD data can therefore be very valuable. 
Exploratory studies within LQCD exist for the aforementioned quantities~\cite{Alexandrou:2018eet,Bhattacharya:2020cen,Chen:2019lcm,Alexandrou:2020zbe,Shanahan:2020zxr,Zhang:2020dbb}. 
Other possibilities for synergy are better constraints in the low- and intermediate-$x$ regions. It should be noted that, currently, the small-$x$ region, as well as the large-$x$ region, cannot be accessed reliably through LQCD calculations, as extremely fine lattice discretizations and large lattice volumes are required. 
Nevertheless, combining the lattice data and the experimental data helps to constrain a wide range of the kinematic regions. 
It is expected that, within the next few years, more combined analyses will be available, utilizing lattice results from the various approaches mentioned above. 

The EIC provides a unique tool with which to probe the modification of the partonic structure of the nucleon in nuclei. 
Along with precise studies of nuclear modification of the unpolarized PDFs through the $F_2$ structure function, the famous EMC effect~\cite{Aubert:1983xm,Arneodo:1992wf,Geesaman:1995yd,Hen:2013oha}, it will also provide access to a polarized analog~\cite{Chen:2004zx,Smith:2005ra,Cloet:2005rt,Chen:2016bde}. LQCD calculations of moments of parton distributions in light nuclei are just beginning~\cite{Detmold:2020snb} and in the coming years will improve significantly. It is expected that LQCD predictions for the spin and flavor dependence of EMC-like effects will be available before EIC begins taking data. Additionally, the EIC will enable studies of double-helicity-flip structure functions of nuclei with spin $J > \frac{1}{2}$; these distributions isolate contributions of exotic nuclear gluons that cannot be localized to the individual constituent nucleons. First attempts to access moments of this distribution for the deuteron have been made~\cite{Winter:2017bfs} and will be improved upon in the coming years.

LQCD can also play an important role in tests of electroweak and beyond-Standard-Model physics at the EIC. As one example, the polarization asymmetry in $eD$ scattering provides a method for extracting the weak mixing angle, $\theta_W$~\cite{Zhao:2016rfu}. In the limit that charge symmetry violation is neglected (up quarks in the proton are the same as down quarks in the neutron) and sea-quark effects are negligible ($\overline{s}-s=0$), the asymmetry is independent of hadron structure for $Q^2\to\infty$~\cite{Shanahan:2013vla}. However, in reality these approximations limit the precision with which $\theta_W$ can be extracted and even rudimentary LQCD calculations of $u_p-d_n$ or $\overline{s} - s$ or their moments will enable better determinations.

\subsection{Radiative corrections}
\label{part2-sec-Connections-RadCor}
In many nuclear physics experiments, radiative corrections quickly become a dominant source of systematics. 
In fact, the uncertainty on the corrections might be the dominant source for high-statistics experiments. 
It is therefore important to have a good understanding, both experimentally and theoretically, of this important issue.  
The following discussion mainly focuses on QED radiative corrections.

{\bf Introduction:}
Radiative processes can roughly be divided into several groups --- see the diagrams in Fig.~\ref{fig:PWG-sec7-rcfeyn} as an example for DIS. 
The first-order radiative diagrams relative to the base process (a) include an additional photon. 
The effect of the self-energy diagrams of type (b) and the vertex correction in diagram (c), can be combined using the QED Ward identity, and absorbed in the electron wave function renormalization. 
The dominant QED contribution to the vacuum polarization (d) can be calculated exactly in QED. 
Hadronic vacuum polarization can best be extracted from measurements --- see e.g.~Ref.~\cite{Keshavarzi:2018mgv} for a recent analysis. 
Corrections described by Feynman diagrams with photons attached on the hadron side are harder to handle. 
Even in the easiest case of elastic scattering, they depend on the hadronic substructure. 
In the case of DIS, photons can be radiated by one of the quarks in the hard scattering process. 
Two-photon exchange, very topical currently in the discussion of the puzzle of the proton form factor ratio, requires additional information about the hadron structure, and so far eludes accurate theoretical treatment even for elastic electron-proton scattering --- see e.g.~Refs.~\cite{Afanasev:2017gsk, Henderson:2016dea, Tomalak:2017shs}.

\begin{figure}[t]
    \centering
    \begin{tikzpicture}   
    \begin{feynman}   
      \vertex (ev);
      \vertex[above left= 0.5cm and 1.5cm of ev] (ei) {$e$};
      \vertex[above right=0.5cm and 1.5cm of ev] (ef) {$e^\prime$};
      \vertex[below=1.5cm of ev,blob,pattern=none] (pv) {};
      \vertex[below left= 0.5cm and 1.5cm of pv] (pi) {$p/A$};
      \vertex[above right=0.2cm and 1.5cm of pv](x1) ;
      \vertex[                right=1.5cm of pv](x2) {X};
      \vertex[below right=0.2cm and 1.5cm of pv](x3) ;
      \vertex[below left=0.75cm and 1.5cm of ev](label){(a)};            
    \diagram*{
      (ei) -- [fermion] (ev)-- [fermion] (ef);
      (pi) -- [double distance=2pt, with arrow=0.5cm] (pv) ;
      (pv) --[plain] (x1);
      (pv) --[plain] (x2);
      (pv) --[plain] (x3);
      (ev) --[photon] (pv);
        };
    \end{feynman}
    \end{tikzpicture}    
    
    \begin{tikzpicture}   
    
    \begin{feynman}   
    
      \vertex (ev);
      \vertex[above left= 0.5cm and 1.5cm of ev] (ei) {$e$};
      \vertex (ei1) at ($(ei)!0.25!(ev)$);
      \vertex (ei2) at ($(ei)!0.75!(ev)$);
      \vertex[above right=0.5cm and 1.5cm of ev] (ef) {$e^\prime$};
      \vertex[below=1.5cm of ev,blob,pattern=none] (pv) {};
      \vertex[below left= 0.5cm and 1.5cm of pv] (pi) {$p/A$};
      \vertex[above right=0.2cm and 1.5cm of pv](x1) ;
      \vertex[                right=1.5cm of pv](x2) {X};
      \vertex[below right=0.2cm and 1.5cm of pv](x3) ;
      \vertex[below left=0.75cm and 1.5cm of ev](label){(b)};       
    \diagram*{
      (ei) -- [plain] (ei1) --[fermion] (ei2) --[plain] (ev)-- [fermion] (ef);
      (ei1) --[photon,out=72,in=72,looseness=2.0] (ei2);
      (pi) -- [double distance=2pt, with arrow=0.5cm] (pv) ;
      (pv) --[plain] (x1);
      (pv) --[plain] (x2);
      (pv) --[plain] (x3);
      (ev) --[photon] (pv);
        };
    \end{feynman}
    \end{tikzpicture}
    \begin{tikzpicture}   
    \begin{feynman}   
      \vertex (ev);
      \vertex[above left= 0.5cm and 1.5cm of ev] (ei) {$e$};
      
      \vertex[above right=0.5cm and 1.5cm of ev] (ef) {$e^\prime$};
      \vertex[below=1.5cm of ev,blob,pattern=none] (pv) {};
      \vertex[below left= 0.5cm and 1.5cm of pv] (pi) {$p/A$};
      \vertex[above right=0.2cm and 1.5cm of pv](x1) ;
      \vertex[                right=1.5cm of pv](x2) {X};
      \vertex[below right=0.2cm and 1.5cm of pv](x3) ;
      \vertex (ei2) at ($(ef)!0.75!(ev)$);
      \vertex (ei1) at ($(ei)!0.75!(ev)$);    
      \vertex[below left=0.75cm and 1.5cm of ev](label){(c)};       
    \diagram*{
      (ei) -- [fermion] (ei1) -- [plain] (ev) --[plain] (ei2)-- [fermion] (ef);
      (ei1) --[photon,out=72,in=108,looseness=1.5] (ei2);
      (pi) -- [double distance=2pt, with arrow=0.5cm] (pv) ;
      (pv) --[plain] (x1);
      (pv) --[plain] (x2);
      (pv) --[plain] (x3);
      (ev) --[photon] (pv);
        };
    \end{feynman}
    \end{tikzpicture}
    \begin{tikzpicture}   
    \begin{feynman}   
      \vertex (ev);
      \vertex[above left= 0.5cm and 1.5cm of ev] (ei) {$e$};
      
      \vertex[above right=0.5cm and 1.5cm of ev] (ef) {$e^\prime$};
      \vertex[below=1.5cm of ev,blob,pattern=none] (pv) {};
      \vertex[below left= 0.5cm and 1.5cm of pv] (pi) {$p/A$};
      \vertex[above right=0.2cm and 1.5cm of pv](x1) ;
      \vertex[                right=1.5cm of pv](x2) {X};
      \vertex[below right=0.2cm and 1.5cm of pv](x3) ;
      \vertex (ei2) at ($(ev)!0.6!(pv)$);
      \vertex (ei1) at ($(ev)!0.2!(pv)$);  
      \vertex[below left=0.75cm and 1.5cm of ev](label){(d)};       
    \diagram*{
      (ei) -- [fermion]  (ev) -- [fermion] (ef);
      (ev) --[photon] (ei1);
      (ei2) --[photon] (pv);      
      (ei1) --[fermion,out=0,in=0,looseness=1.6] (ei2);
      (ei2) --[fermion,out=180,in=180,looseness=1.6] (ei1);
      (pi) -- [double distance=2pt, with arrow=0.5cm] (pv) ;
      (pv) --[plain] (x1);
      (pv) --[plain] (x2);
      (pv) --[plain] (x3);
      
        };
    \end{feynman}
    \end{tikzpicture}
    \begin{tikzpicture}   
    \begin{feynman}   
      \vertex (ev1);
      \vertex[right=1cm of ev1] (ev2);
      \vertex[above left= 0.5cm and 1cm of ev1] (ei) {$e$};
      
      \vertex[above right=0.5cm and 1cm of ev2] (ef) {$e^\prime$};
      \vertex[below right=1.5cm and 0.25cm of ev1,blob,pattern=none] (pv) {};
      \vertex[below left= 0.5cm and 1.5cm of pv] (pi) {$p/A$};
      \vertex[above right=0.2cm and 1.5cm of pv](x1) ;
      \vertex[                right=1.5cm of pv](x2) {X};
      \vertex[below right=0.2cm and 1.5cm of pv](x3) ;
      \vertex[below left=0.75cm and 1.5cm of ev](label){(e)};       
    \diagram*{
      (ei) -- [fermion]  (ev1) --[fermion] (ev2) -- [fermion] (ef);
      (ev1) --[photon] (pv);
      (ev2) --[photon] (pv);
      (pi) -- [double distance=2pt, with arrow=0.5cm] (pv) ;
      (pv) --[plain] (x1);
      (pv) --[plain] (x2);
      (pv) --[plain] (x3);
      
        };
    \end{feynman}
    \end{tikzpicture}  
    \begin{tikzpicture}   
    \begin{feynman}   
      \vertex (ev);
      \vertex[above left= 0.5cm and 1.5cm of ev] (ei) {$e$};
      
      \vertex[above right=0.5cm and 1.5cm of ev] (ef) {$e^\prime$};
      \vertex[below=1.5cm of ev,blob,pattern=none] (pv) {};
      \vertex[below left= 0.5cm and 1.5cm of pv] (pi) {$p/A$};
      \vertex[above right=0.2cm and 1.5cm of pv](x1) ;
      \vertex[                right=1.5cm of pv](x2) {X};
      \vertex[below right=0.2cm and 1.5cm of pv](x3) ;
      \vertex (ei1) at ($(ei)!0.5!(ev)$);     
      \vertex[above right=0.5 and 1.5 cm of ei1] (ei2);
\vertex[below left=0.75cm and 1.5cm of ev](label){(f)};       
    \diagram*{
      (ei) -- [fermion] (ei1) --[fermion] (ev) -- [fermion] (ef);
      
      (ev) --[photon] (pv);
     (ei1) --[photon] (ei2);      
      (pi) -- [double distance=2pt, with arrow=0.5cm] (pv) ;
      (pv) --[plain] (x1);
      (pv) --[plain] (x2);
      (pv) --[plain] (x3);
      
        };
    \end{feynman}
    \end{tikzpicture}
    \begin{tikzpicture}   
    \begin{feynman}   
      \vertex (ev);
      \vertex[above left= 0.5cm and 1.5cm of ev] (ei) {$e$};
      
      \vertex[above right=0.5cm and 1.5cm of ev] (ef) {$e^\prime$};
      \vertex[below=1.5cm of ev,blob,pattern=none] (pv) {};
      \vertex[below left= 0.5cm and 1.5cm of pv] (pi) {$p/A$};
      \vertex[above right=0.2cm and 1.5cm of pv](x1) ;
      \vertex[                right=1.5cm of pv](x2) {X};
      \vertex[below right=0.2cm and 1.5cm of pv](x3) ;
      \vertex (ei1) at ($(ef)!0.5!(ev)$);     
      \vertex[below right=0.6 and 0.75 cm of ei1] (ei2);
\vertex[below left=0.75cm and 1.5cm of ev](label){(g)};       
    \diagram*{
      (ei) -- [fermion] (ev) --[fermion] (ei1) -- [fermion] (ef);
      
      (ev) --[photon] (pv);
     (ei1) --[photon] (ei2);      
      (pi) -- [double distance=2pt, with arrow=0.5cm] (pv) ;
      (pv) --[plain] (x1);
      (pv) --[plain] (x2);
      (pv) --[plain] (x3);
      
        };
    \end{feynman}
    \end{tikzpicture}

    \caption{Typical Feynman diagrams describing first-order radiative corrections for DIS, $ep/A \rightarrow eX$.}
    \label{fig:PWG-sec7-rcfeyn}
\end{figure}

Correction terms described by Feynman diagrams with loops of the type (b)--(e) do not change the observed kinematics. 
In the lowest order, they contribute via the interference term of (a) with the sum of (b)--(e), and are naturally suppressed by a factor of the fine structure constant $\alpha$, which stems from the two additional photon vertices. 
However, due to the loop integrals their contribution can be enhanced beyond this level. 
These graphs (b)--(e) contain ultraviolet and infrared divergences. 
The former can be treated using various schemes, for example dimensional regularization, and are absorbed in renormalization constants. 
The latter remain and cancel at the level of the cross section with the graphs of type (f) and (g), where an additional photon exists in the final state. 
These real radiation corrections have to be taken into account since scattering accompanied by soft photons cannot be distinguished experimentally from the elastic scattering. 
The cancellation of infrared divergences between virtual and real radiative corrections is proven to be correct to all orders~\cite{Yennie:1961ad}.

Since the final state is different, diagrams (f) and (g) do not interfere with those of types (b)--(e). 
In contrast to those, for finite photon momenta, they change the observable final-state kinematics. 
This makes corrections dependent on the experimental acceptance. 
The cross section drops typically inversely proportional to the photon energy. 
However, for radiation from the lepton line, there is an additional enhancement. 
Since the photon carries away energy, reducing $Q^2$ for the same electron scattering angle, the cross section is enhanced. 
In extreme cases, where the photon carries away almost all the lepton energy, the radiative cross section can be (much) larger than the elastic one.

{\bf Peaking approximation:}
Photon radiation is dominated by emission that is (almost) collinear with the emitting particle, but with a tail extending to large angles. 
Assuming strictly collinear radiation leads to the common peaking approximation~\cite{Mo:1968cg}, where the particle scattering angles are unmodified. 
However, this is not exact, and for high-precision predictions for cross sections one has to take into account non-collinear photon radiation. 

{\bf Expected size, higher orders and uncertainty:}
Because of the acceptance and resolution dependence of radiative corrections, absolute numbers for the effect and, more importantly, its uncertainty, strongly depend on the details of the experimental conditions. 
Numerical effects of a couple of tens of a percent are easily possible. 
Calculations for DIS~(see Ref.~\cite{Liu:2020rvc} for a recent study), specifically, show that radiative corrections can be very large, in particular at small $x$ and at large inelasticity $y$, with a strong sensitivity to the exact treatment of the radiative corrections. 

Especially when first-order corrections are large, one should ask whether unknown higher-order effects will induce large uncertainties for the cross section predictions. 
In the soft-photon approximation, one can estimate such higher-order corrections since they are known to exponentiate to all orders~\cite{Bloch:1937pw,Yennie:1961ad}. 
In practice, this exponentiation is often used even for non-vanishing photon momenta, but this approach does not capture all important higher-order corrections. In particular, there are logarithmically enhanced contributions due to hard collinear photons. 
Those can also be estimated~\cite{Kripfganz:1990vm,Blumlein:2007kx}, and recent work provides a practical way to include them based on an approach which factorizes leading logarithms into structure functions~\cite{Liu:2020rvc}. 
Faithful error estimates require an accurate calculation of the second order, which recently became available~\cite{Bucoveanu:2018soy,Banerjee:2020rww}. 
Results so far show that second-order QED corrections are comparatively small. 

While classically, radiative corrections were applied post-hoc to the data as a correction factor (see Ref.~\cite{Badelek:1994uq} for a discussion of the so-called Dubna scheme~\cite{Akhundov:1994my} and a comparison with the Mo and Tsai scheme~\cite{Mo:1968cg}), this approach quickly reaches uncertainty limits, especially when complicated acceptances come into play. 
Most modern experiments therefore use Monte Carlo integration with generators including radiative effects~\cite{Kwiatkowski:1990es, Charchula:1994kf} (see also Ref.~\cite{Aschenauer:2013iia} for a recent update of the Monte Carlo program DJANGOH) to extract physics in their analysis, as has been done in the HERA experiments H1 and ZEUS.

It is important to note that the size of the corrections and the size of their uncertainty are generally unrelated.
This means that, depending on the situation, a reliable calculation of large corrections may be feasible while, on the other hand, numerically small corrections may have large (absolute) uncertainties.
Particularly relevant are acceptance-dependent contributions to the corrections since they depend on experimental conditions which may not be known with good precision. 
It may be possible that acceptance cuts, for example on the accepted energy deficit, can be chosen such that the total radiative corrections are small, but uncertainties from neglected higher orders are large, whereas a different cut would produce larger corrections, but minimize the influence of neglected terms. 
Additionally, an uncertainty on the acceptance cut may induce uncertainties on the corrections which may be difficult to estimate. 

{\bf Polarization degrees of freedom:}
For measurements using polarization degrees of freedom, the situation is even more complex. 
On the one hand, two-photon-exchange corrections on extractions of the proton form factor ratio from experiments using polarization are generally expected to be small compared to the effect on Rosenbluth-type measurements --- similar to common normalization uncertainties, they are suppressed in asymmetries. 
On the other hand, measurements of polarization transfer observables exhibit so far theoretically unexplained deviations from their Born prediction~\cite{Meziane:2010xc}. 
For semi-inclusive DIS, calculations for arbitrary polarization of the initial-state nucleon can be found in Ref.~\cite{Akushevich:2019mbz}.  

{\bf Monte Carlo methods:}
The structure of initial-state and final-state radiation makes the development of efficient MC generators a particularly hard problem. 
In convenient parametrizations of the scattering kinematics, cross sections vary by many orders of magnitude. 
A MC generation with weighted events is then very inefficient. 
Hence, also naive rejection sampling methods to un-weight events do not work, as the ratio of accepted events is then approaching zero. 
Approaches using automatic volume reweighting are challenged by the high derivatives near the peak cross section. 
The analytic structure has therefore to be exploited in order to generate events suitably and efficiently. 
For the fixed-target elastic case, multiple generators for first-order (partly including soft-photon exponentiation) exist (see, for instance, Refs.~\cite{Bernauer:2010zga,Akushevich:2011zy,Gramolin:2014pva,Henderson:2016dea}). 
They use different ways to keep event weights constant or low-variance, improving MC efficiency.
Recently, higher-order MC generators became available~\cite{Bucoveanu:2018soy,Banerjee:2020rww}.
It should be straightforward to adapt these generators to the situation at a collider. 
However, extra work is needed to validate the calculations since numerical precision and efficiencies strongly depend on the kinematic conditions. 

{\bf QCD radiation and new directions:} 
Treating radiative effects as corrections to some Born-level base process becomes increasingly difficult for reactions beyond DIS. 
On the other hand, QCD higher-order graphs bear a close resemblance to those QED diagrams, and a unified approach handling both QCD and QED effects is possible. 
Based on the factorization theorem, it is possible to describe resummed leading logarithmic higher-order corrections with the help of distribution and fragmentation functions~\cite{Liu:2020rvc}. 
Partons and photons can be treated democratically in the event generation, allowing for the resummation of higher-order corrections in the form of parton showers~\cite{Hoeche:2009xc}. 
This approach is currently implemented in the main HEP generators~\cite{Buckley:2011ms}.  

{\bf The way forward:}
While the theoretical treatment of QED radiative corrections is well established, some questions remain, especially when hadronic effects come into play. 
Experimental validation is then required, and allows us to extract new information about these processes, transforming radiative corrections from a nuisance in the extraction of physics to a physics goal. 
For example, an exclusive measurement where the photon is detected, can give additional important information about the internal structure of the target. 
Deeply-virtual Compton scattering is an example. 

The ability to change the lepton charge, that is, to collide positron as well as electron beams with nucleons and ions, offers a powerful tool to control radiative corrections~\cite{Bernauer:2020vue}.  
Most importantly, the next term beyond the leading single-photon exchange, is directly accessed via the ratio of positron to electron scattering, as has been pursued in experiments at DESY~\cite{Henderson:2016dea}, Jefferson Lab~\cite{Rimal:2016toz} and VEPP-3~\cite{Rachek:2014fam}.  
Positron beams also offer unique opportunities to access flavor separation of parton distributions and new electroweak structure functions, using the charged-current electroweak interaction, as well as access to beam charge asymmetries to determine GPDs via deeply-virtual Compton scattering. 
A possible program for JLab-12 is described in Ref.~\cite{Accardi:2020swt}; similar physics opportunities would exist for an EIC. 
Realization of a positron beam at Jefferson Lab before the advent of EIC would be very desirable in terms of achieving a better understanding of the radiative corrections to electron scattering.

Experiment design can be instrumental in reducing uncertainties. 
Here, progress thus far has been limited by the lack of suitable MC generators that apply in collider kinematics, a deficit which is currently being addressed.
In parallel, it seems prudent to consider a systematic experimental test of radiative correction procedures once the EIC becomes operational.  
This will likely require a dedicated detector/beam configuration to enhance sensitivity to particular kinematics, where such corrections are large --- see, for instance, Ref.~\cite{Sofiatti:2011yi}.
It seems important to organize a dedicated effort focused on radiative correction generator development and experimental validation in the EIC era.

{\bf Summary:}
Radiative corrections introduce an important uncertainty, and experiment design should take them into account, not only to minimize their effect, but also as a physics goal itself. 
While significant progress in the theoretical treatment has been made, there is still some way to go, especially in the description of electron-nucleus scattering.
Also, more work must be done to either include radiative MC generators, or include radiative effects into existing generators.
There appear to be no real fundamental problems, but the amount of work required is substantial, and such efforts must be supported by the EIC community.
Furthermore, a whole class of new theoretical developments is needed to obtain radiative corrections for higher-twist contributions that can be measured at the EIC.
A more detailed snapshot of the current state is reflected by the contributions to a recent ad-hoc workshop at the Center for Frontiers in Nuclear Science~\cite{adhocrcwhitepaper,Afanasev:2020hwg}, the first in a series of meetings on this topic, with the next meeting planned for the first half of 2021.

\chapter{Detector Requirements}
\label{part2-chap-DetRequirements}

The detector requirements best suited for the EIC physics opportunities have been under discussion for many years. The new 
collider is planned to accelerate a variety of ion beams, polarized electron and light nuclei species over a broad range of center of mass energies. 
Building on the experience from HERA at DESY, the first complete set of requirements for the EIC was documented in the EIC White Paper~\cite{Accardi:2012qut}.  
These developments were further advanced by the Detector Requirements and R\&D Handbook~\cite{EIC:RDHandbook}. 
The  EIC physics program scope was considered in terms of several physics processes. 
Specifically, both White Paper and the R\&D Handbook studies considered inclusive, semi-inclusive, exclusive reactions and diffractive processes, much like the Yellow Report working groups (apart from processes with jets and heavy quarks in the final state).

The EIC ``general-purpose detector" requirements outlined in the Handbook were based on expected event geometries at the EIC and the kinematic coverage in the context of each of these processes. Some of the experimental apparatus capabilities are essential for the entire physics program. Examples include high precision luminosity monitors and polarimeters. The other universal requirement is  high precision/high resolution scattered electron reconstruction (and identification). For all the processes involving final state particles in the central region, magnet design considerations become inseparable from charged particle tracking, when ensuring desired momentum resolution.  In addition to the above requirements, it was already known that studies involving SIDIS reactions and processes with jets and heavy quarks would require extended hadron identification over broad momentum and rapidity ranges, full azimuthal coverage, and excellent vertexing resolution.  The jet performance is known to be coupled to the precision of the hadronic and electromagnetic calorimeters. Exclusive reactions and diffractive processes imposed even more constraints on calorimetry. It is essential for such reactions to accurately reconstruct all the particles in the event; achieving such reconstruction with sufficient accuracy requires  using numerous detector components. Thus exclusive and diffractive reactions demand hermeticity of the setup while also  
adding the additional requirement of precision far-forward detectors.

\begin{figure*}[ht]
\centering
\includegraphics[width=0.9\textwidth]{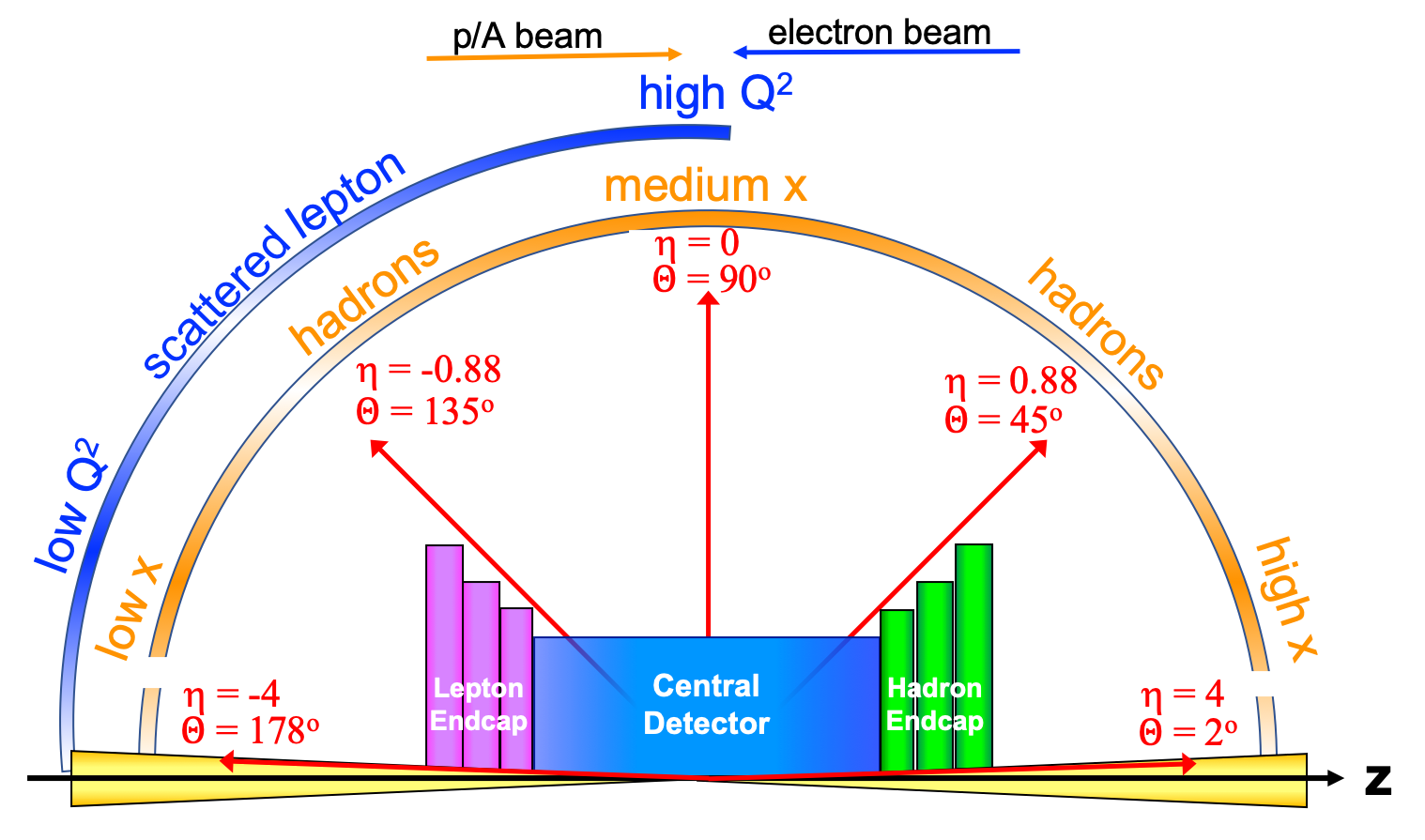}
\caption{A schematic showing how hadrons and the scattered lepton for different
$x-Q^2$ are distributed over the detector rapidity coverage.}
\label{fig:Det.Schematics}
\vspace*{3mm}
\end{figure*}

The summary of the previously developed detector requirements from the Detector R\&D Handbook~\cite{EIC:RDHandbook} was used as a starting point for the Yellow Report studies.  A sketch of the EIC generic detector outline is shown in Fig. \ref{fig:Det.Schematics}. The coordinate system for all the plots in this chapter is defined as indicated in the sketch: the hadron beam goes in the positive z-direction ($\theta \sim 0^\circ$) while the electron beam moves in the negative z-direction ($\theta \sim 180^\circ$).  Positive pseudorapidities $\eta>0$ are associated with angles $\theta<90^\circ$, and negative pseudorapidities $\eta<0$  are for particles with $\theta>90^\circ$.
For the central rapidity region, the starting generic detector configuration assumes near hermetic coverage. Initially, the pseudo-rapidity range of $\pm 4$ was taken (later in the process, a $\pm 3.5$ units with reduced performance of up to 4 was found more realistic). A very low material budget for the central tracking region not exceeding 5\% of $X/X_D$ is needed with a tracking momentum resolution in under the 5\% range.  Reliable electron identification (and thus significant pion suppression of $10^{-4}$) was one of the program's critical requirements. The spatial resolution of the primary vertex was assumed to be $\sim$~20 micrometers level. Efficient hadron identification at the $3\sigma$ level was accepted for pions, kaons, and protons, reaching to about 7 and 50~GeV/$c$ at mid-to forward rapidities, respectively.

\begin{figure}[!ht]
  \centering
  \includegraphics[width=0.8\textwidth]{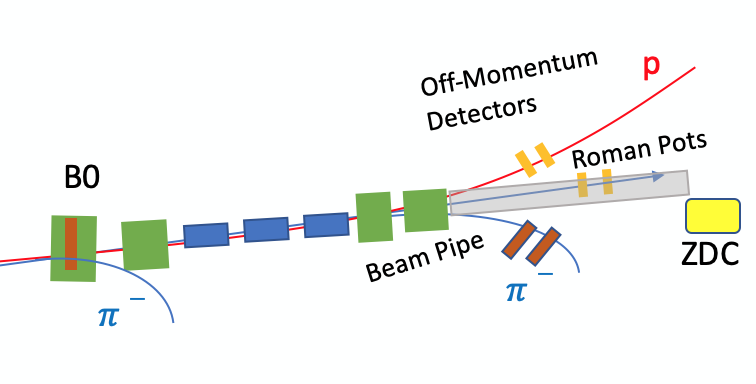}
  \caption{Geometric layout of the far-forward area setup in the hadron going direction at very small $\theta$.  The green boxes denote the dipole magnets; the blue rectangles denote the quadrupole focusing magnets; the gray tube is a simple representation of the beam pipe in the drift region where many of the far-forward protons and neutrons are detected.  A detailed engineering design of the beam pipe is currently in progress. Blue $\pi^-$ and red proton tracks originating from $\Lambda$ decays are shown, of importance to the meson structure studies, see Sec.~\ref{subsec:DetReq.DT.meson}.}
  \label{fig:far-forw}
\end{figure}


In addition to the central region instrumentation, the region most relevant from  requirements standpoint  for exclusive and diffractive processes is situated beyond the main detector. This region is known as the far-forward area.   This unique area allows particles that have extremely small scattering angles to be detected.   In the hadron going direction, this area includes Roman Pots (RP), off-momentum detectors (OMD), a B0 tracker, and a Zero Degree Calorimeter (ZDC), see Sec.~\ref{part3-sec-Det.Aspects.FFDet}  for more details. In Fig.~\ref{fig:far-forw}, the geometric layout of the far-forward region of the IR is shown, together with a tentative conceptual design of the far-forward particle detectors, see caption for details. The details of the emerging requirements for this far-forward area are discussed in the following.

The table, depicted in Fig.~\ref{fig:HandbookT2} below, is  taken directly from~\cite{EIC:RDHandbook}, and summarizes all the requirements and the outlined detector performance expectations for the central rapidity region as a function of pseudo-rapidity. These requirements were parameterized for fast simulation studies in the first official release of EICSmear, a Monte Carlo package allowing fast smearing of simulated events to study the effects of detector resolution for EIC~\cite{git:eicsmeardetectors}. They were used for subsequent developments of this report as discussed in the following subsections.
\begin{figure}[htbp]
\centering
{\includegraphics[width=0.99\linewidth]{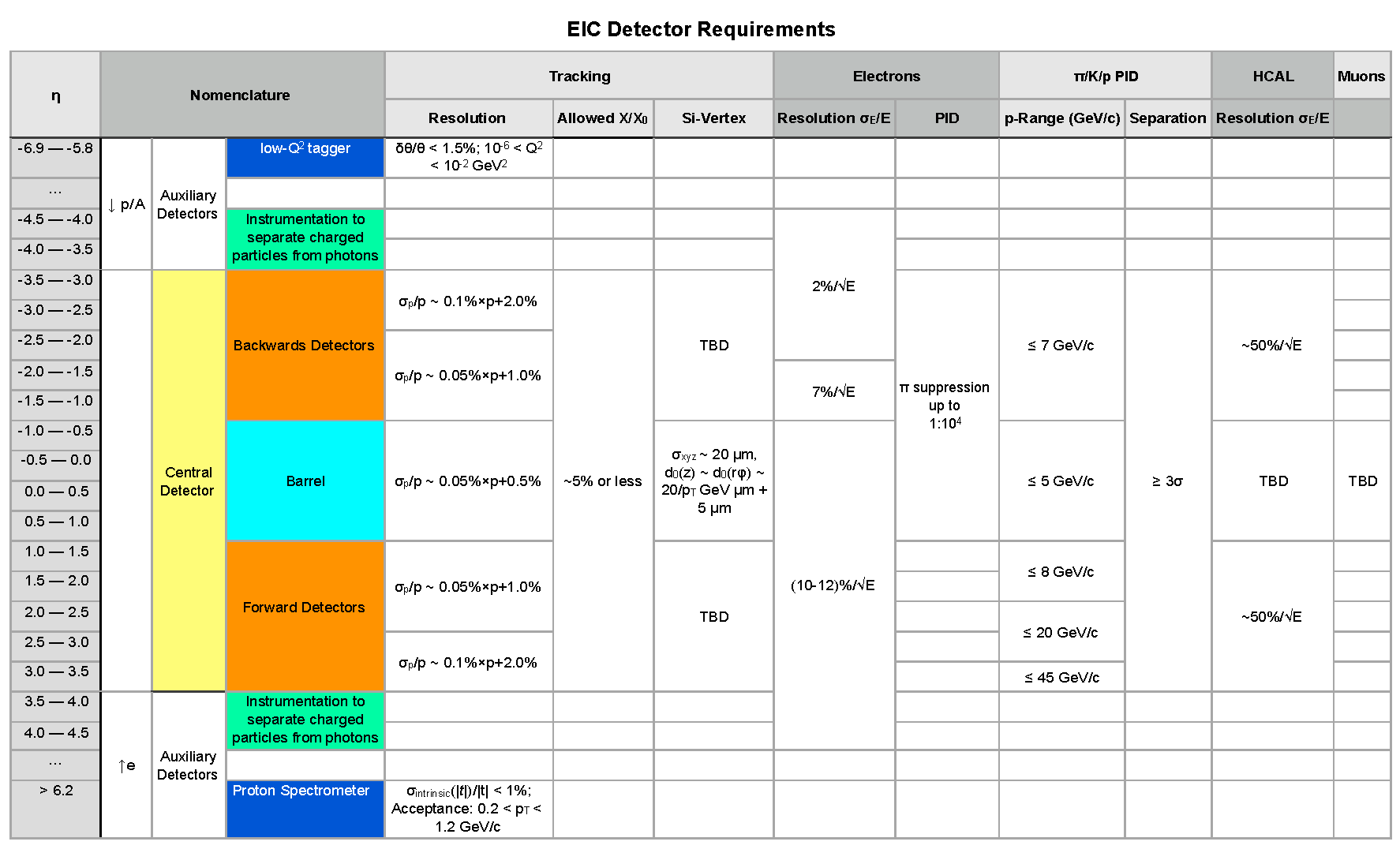}}
\caption{Summary of the EIC Detector requirements from the R\&D Handbook~\ref{fig:HandbookT2}.}
\label{fig:HandbookT2}
\end{figure}

%
%
%
%

\section{Inclusive Measurements}
\label{part2-sec-DetReq.Incl}

Inclusive reactions may be divided into two types of interactions -- those that proceed via the exchange of a virtual photon or Z boson, referred to as neutral current (NC) events, and those that proceed via the exchange of a charged W boson, referred to as charged current (CC) events.  By definition, inclusive reactions do not place any constraints on the flavor of the interaction that occurs at the boson-quark vertex or the type of particles produced in the final state. As a result, inclusive channels are sensitive to a range of QCD, electroweak and beyond-the-standard-model (BSM) processes and provide a wealth of physics opportunities. Examples include spin-averaged and spin-dependent nucleon and nuclear parton distributions functions, non-linear QCD and higher twist effects as well as CPT and Lorentz symmetry violating measurements. 

\subsection{Reconstruction of kinematic variables}

For inclusive reactions the kinematics of the interaction are reconstructed by either detecting the scattered beam electron, or by reconstructing the hadronic recoil. In the case of electron reconstruction the following definitions for $x$, $Q^2$ and $y$ are used:

\begin{equation}
Q^2 = 4EE'\cos^2{(\theta^{e'}_{p}/2)}\qquad
y = 1 -\frac{E'(1-\cos{\theta^{e'}_{p}})}{2E}\qquad
x = \frac{Q^2}{sy} 
\end{equation}

\noindent where $s$ is the center-of-mass energy, $E$ and $E'$ are the energy of the incoming and scattered electron and $\theta^{e'}_{p}$ is the angle of the scattered beam electron measured with respect to the incoming proton axis (+z). If the hadronic recoil is used to reconstruct the event kinematics, then the Jacquet-Blondel definitions are used instead:

\begin{equation}
\label{eq:eReco}
Q^2_{JB} = \frac{p^2_T}{1-y_{JB}} \qquad
y_{JB} = \frac{(E-p^z)}{2E}\qquad
x_{JB} = \frac{Q^2_{JB}}{sy_{JB}} 
\vspace{0.4cm}
\end{equation}

\noindent where $p^2_T = (\sum_hP^x_h)^2 + (\sum_hP^y_h)^2 $ and $(E-p^z) = \sum_h(E_h - p^z_h)$ are summed over the all of the final state hadrons in the event. For CC channels, JB reconstruction is the only option, while for NC channels it is possible to use electron, JB or a mixture of electron and hadronic reconstruction techniques. Typically, JB reconstruction is used for events with small $y$, as the $x$ and $y$ resolutions decrease rapidly in this region for the electron reconstruction variables.

\begin{figure}[htbp]
\centering
\subfloat[e+p 18x275 GeV]{\label{fig:xQ2_18_275}\includegraphics[width=0.45\linewidth]{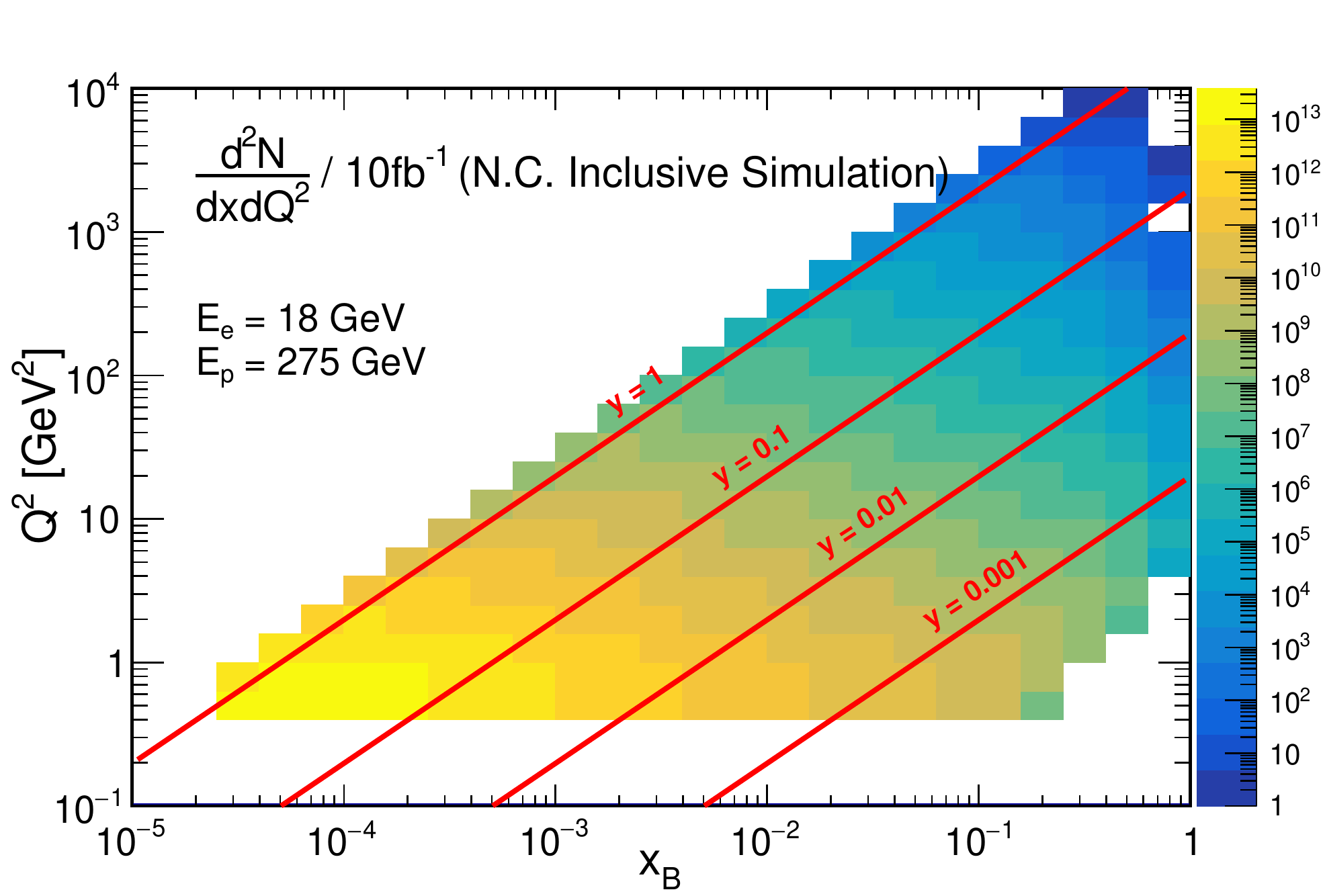}}\qquad%
\subfloat[e+p 10x100 GeV]{\label{fig:xQ2_10_100}\includegraphics[width=0.45\linewidth]{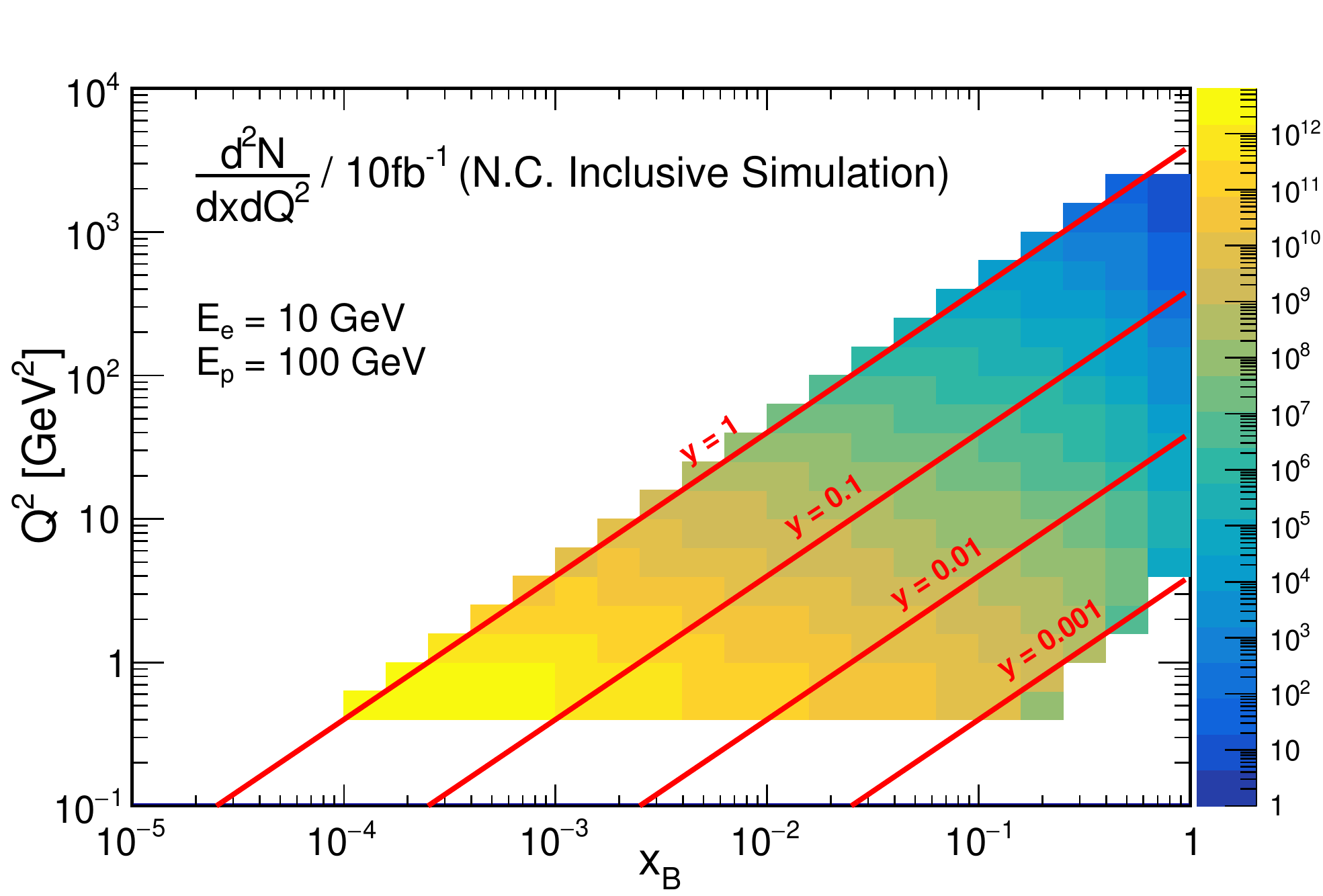}}\qquad%
\subfloat[e+p 5x100 GeV]{\label{fig:xQ2_5_100}\includegraphics[width=0.45\textwidth]{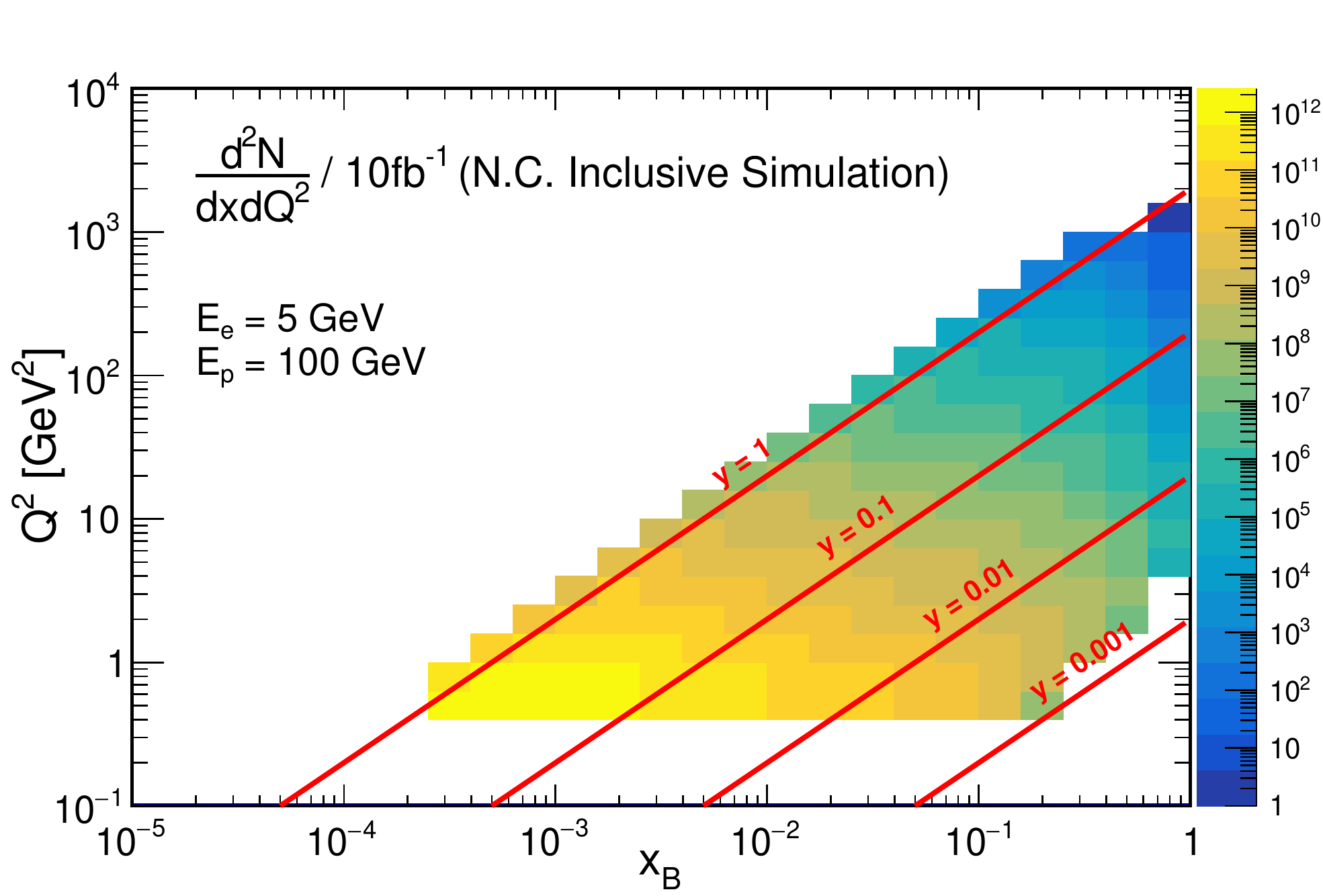}}\qquad%
\subfloat[e+p 5x41 GeV]{\label{fig:xQ2_5_41}\includegraphics[width=0.45\textwidth]{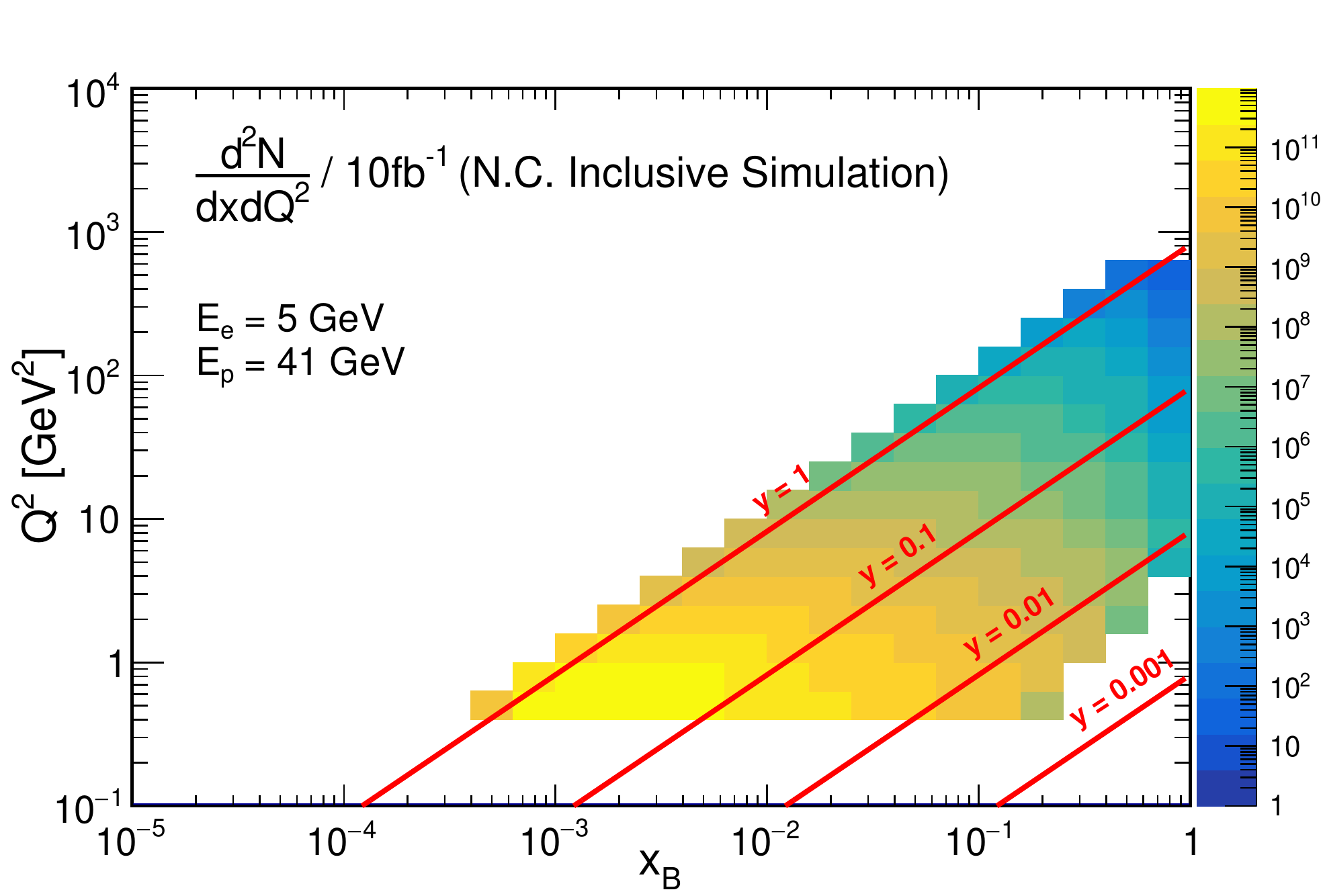}}%
\caption{Differential yields for neutral-current interactions binned in $x$ and $Q^2$ for four proposed  center-of-mass energies. These plots were created using the PYTHIA6 event generator with cuts of $Q^2 >$ 0.5 GeV$^2$ applied on the generated particles. Radiative effects are not included. }
\end{figure}

The $x$ vs $Q^2$ coverage for NC electron reconstruction for four e+p beam configurations, 18x275, 10x100, 5x100, 5x41 GeV are shown in Figures \ref{fig:xQ2_18_275} - \ref{fig:xQ2_5_41}.


\subsection{Kinematic phase space}

\begin{figure}[htbp]
\centering
\subfloat[e+p 18x275 GeV]{\label{fig:PS_18_275}\includegraphics[width=0.45\linewidth]{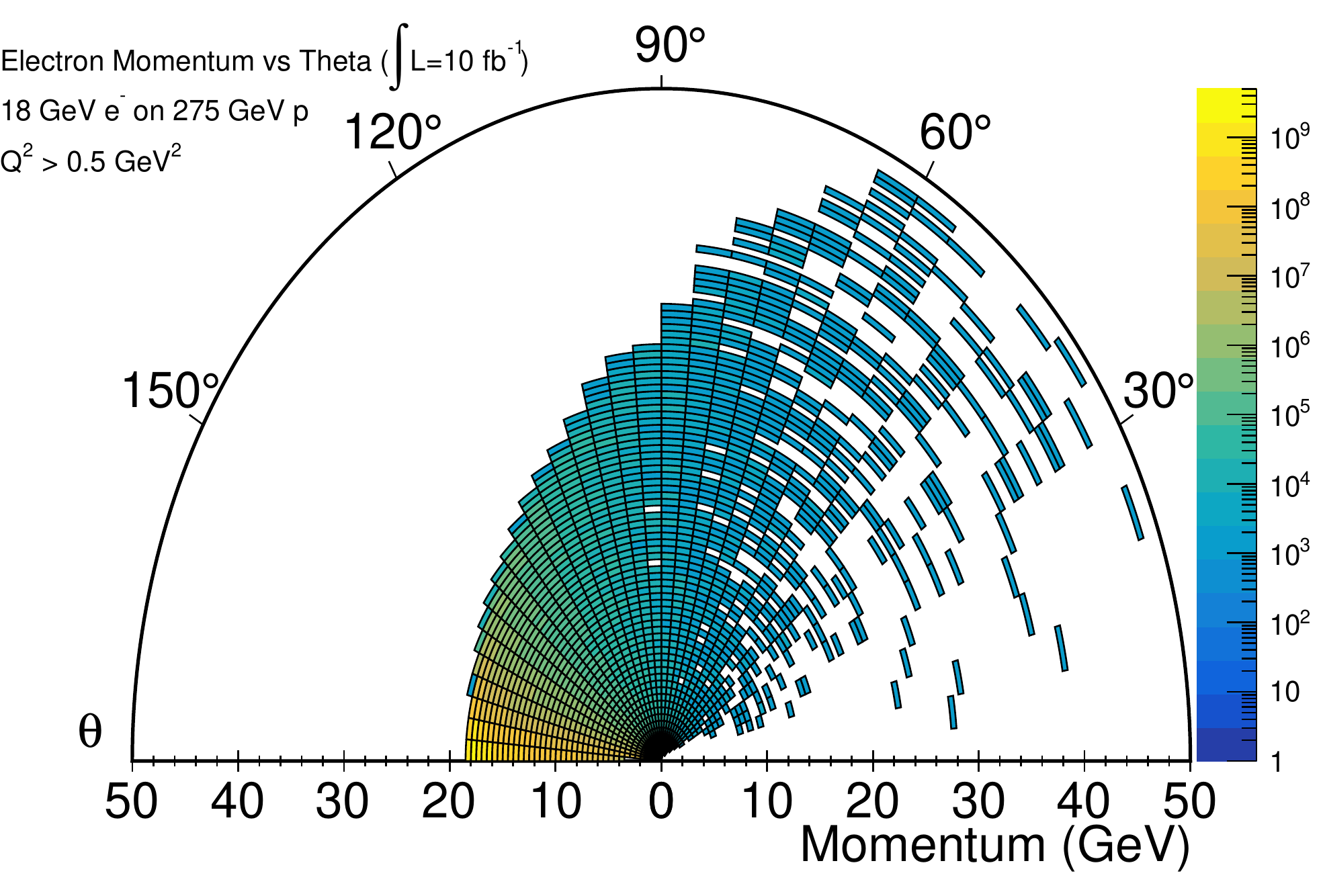}}\qquad%
\subfloat[e+p 10x100 GeV]{\label{fig:PS_10_100}\includegraphics[width=0.45\linewidth]{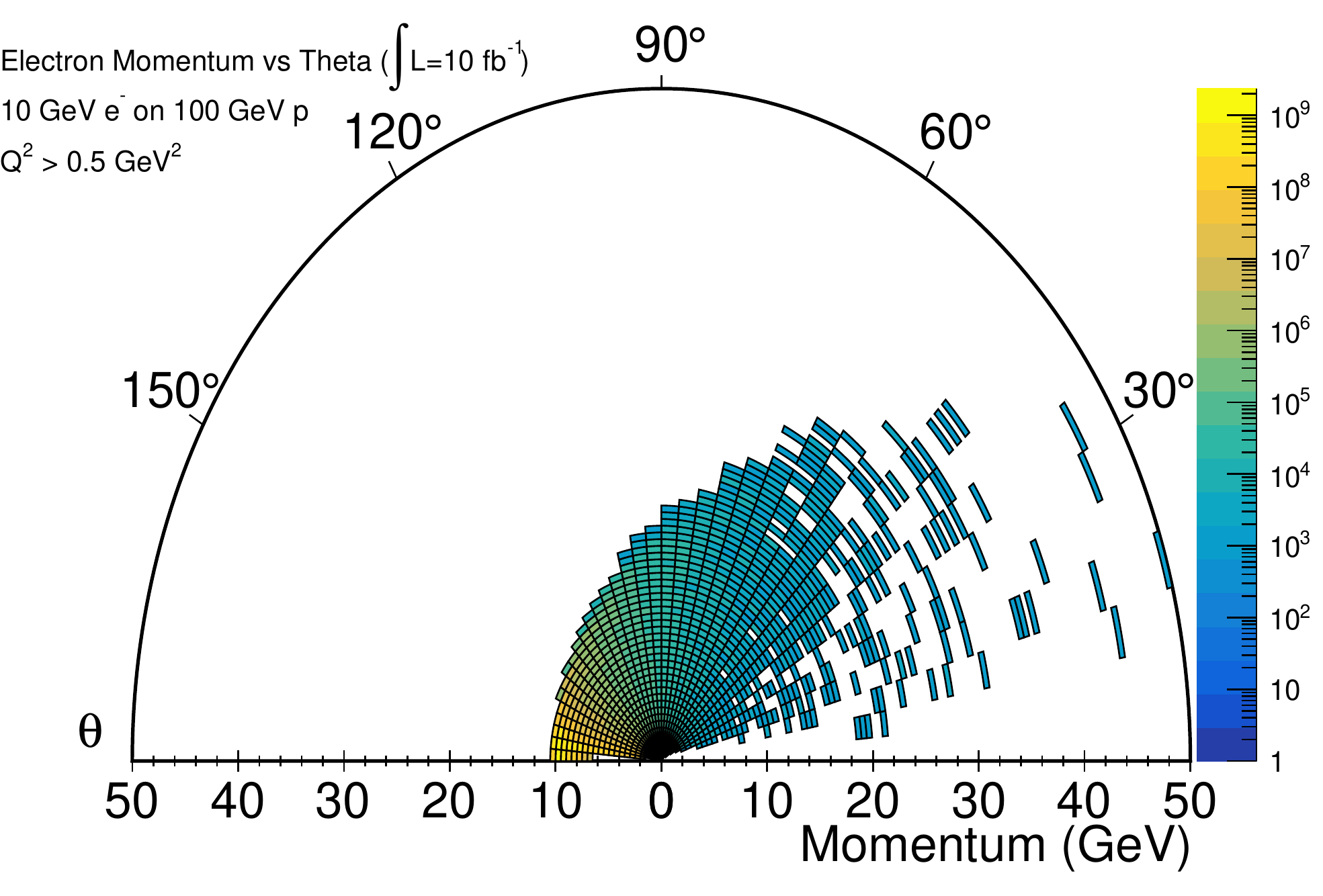}}\qquad%
\subfloat[e+p 5x100 GeV]{\label{fig:PS_5_100}\includegraphics[width=0.45\textwidth]{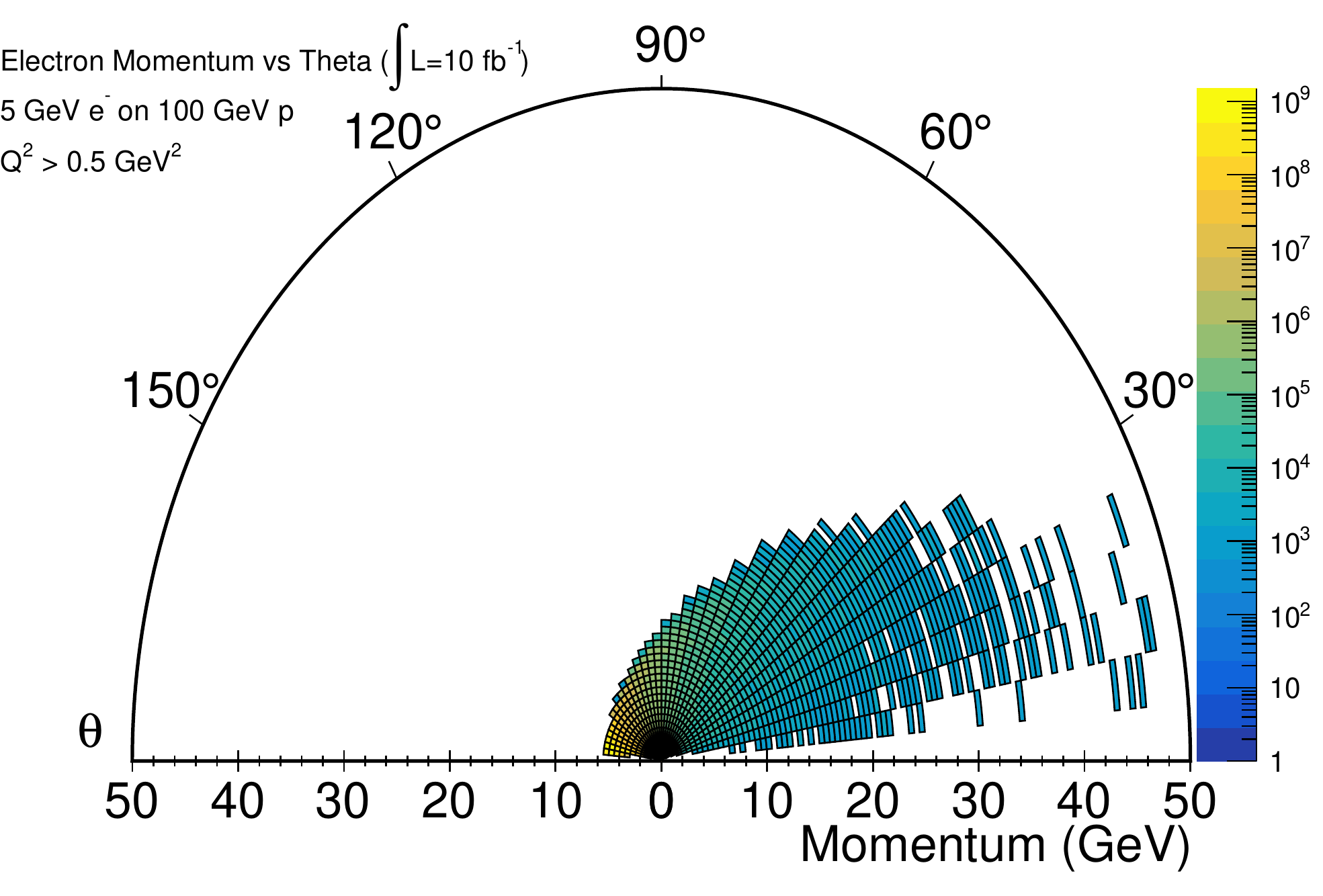}}\qquad%
\subfloat[e+p 5x41 GeV]{\label{fig:PS_5_41}\includegraphics[width=0.45\textwidth]{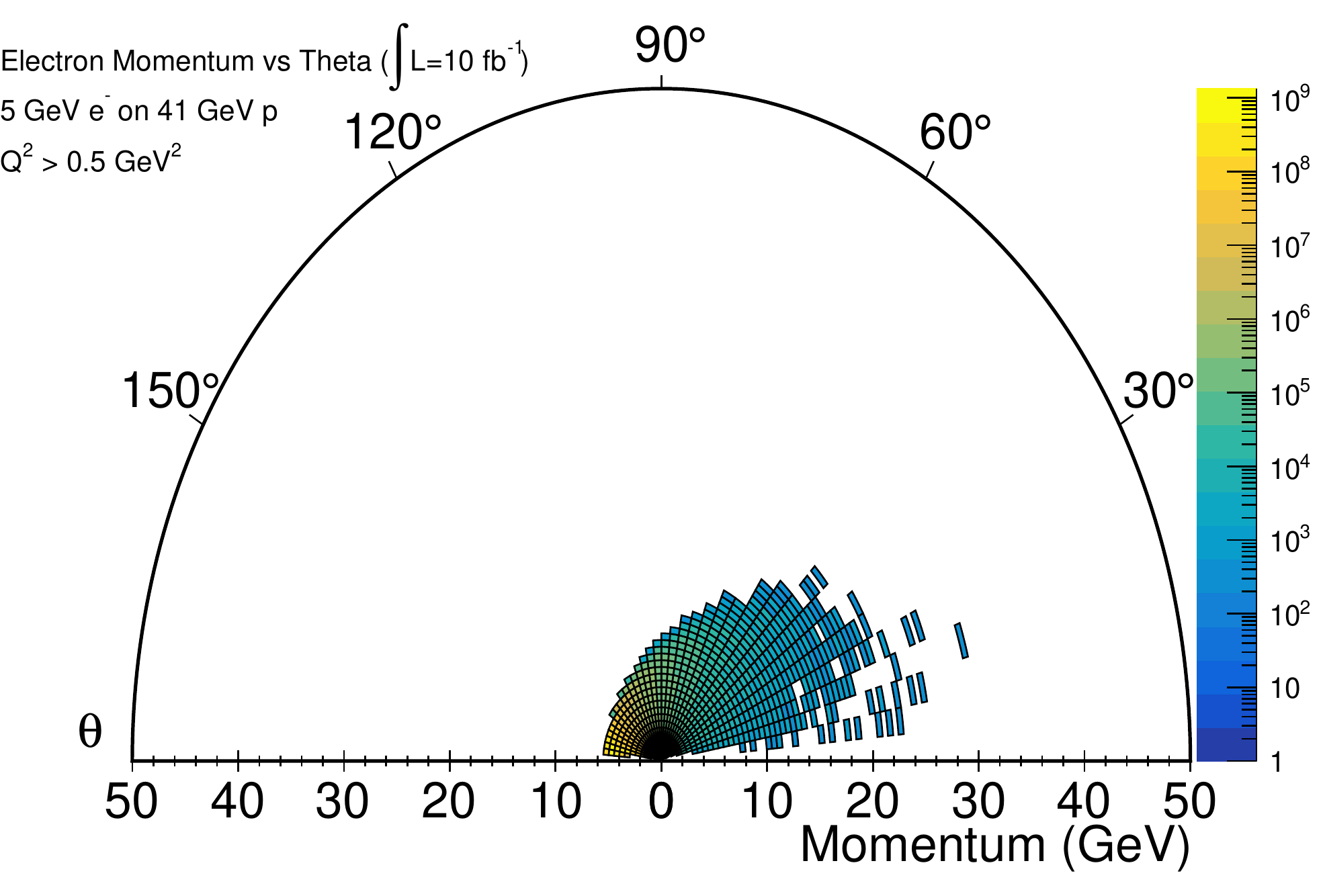}}%
\caption{Polar plots of yields for scattered electrons in NC interactions binned in $\theta^{e'}_p$ and $p$ for four proposed  center-of-mass energies. These plots were created using the Pythia6 event generator with cuts of $Q^2 >$ 0.5 GeV$^2$ applied at the vertex level. Radiative effects are not included.}
\end{figure}

The first step in defining detector requirements for the inclusive channels is to identify the kinematic regions in scattering angle $\theta^{e'}_p$ and momentum $p^{e'}$ for the final state electrons, photons and hadrons. Figures \ref{fig:PS_18_275}-\ref{fig:PS_5_41} show the momentum-$\theta^{e'}_p$ distributions of the scattered beam electron for four beam e+p beam configurations, 18x275, 10x100, 5x100, 5x41 GeV. As expected, the electron typically scatters in the backward direction, between $-3.5<\eta < 1$. The average momentum of the scattered electron increases with $\sqrt{s}$ and $Q^2$, peaking at mid-to-forward rapidities. 

\begin{figure}[htbp]
\centering
\subfloat[e+p 18x275 GeV]{\label{fig:PSall_18_275}\includegraphics[width=0.45\linewidth]{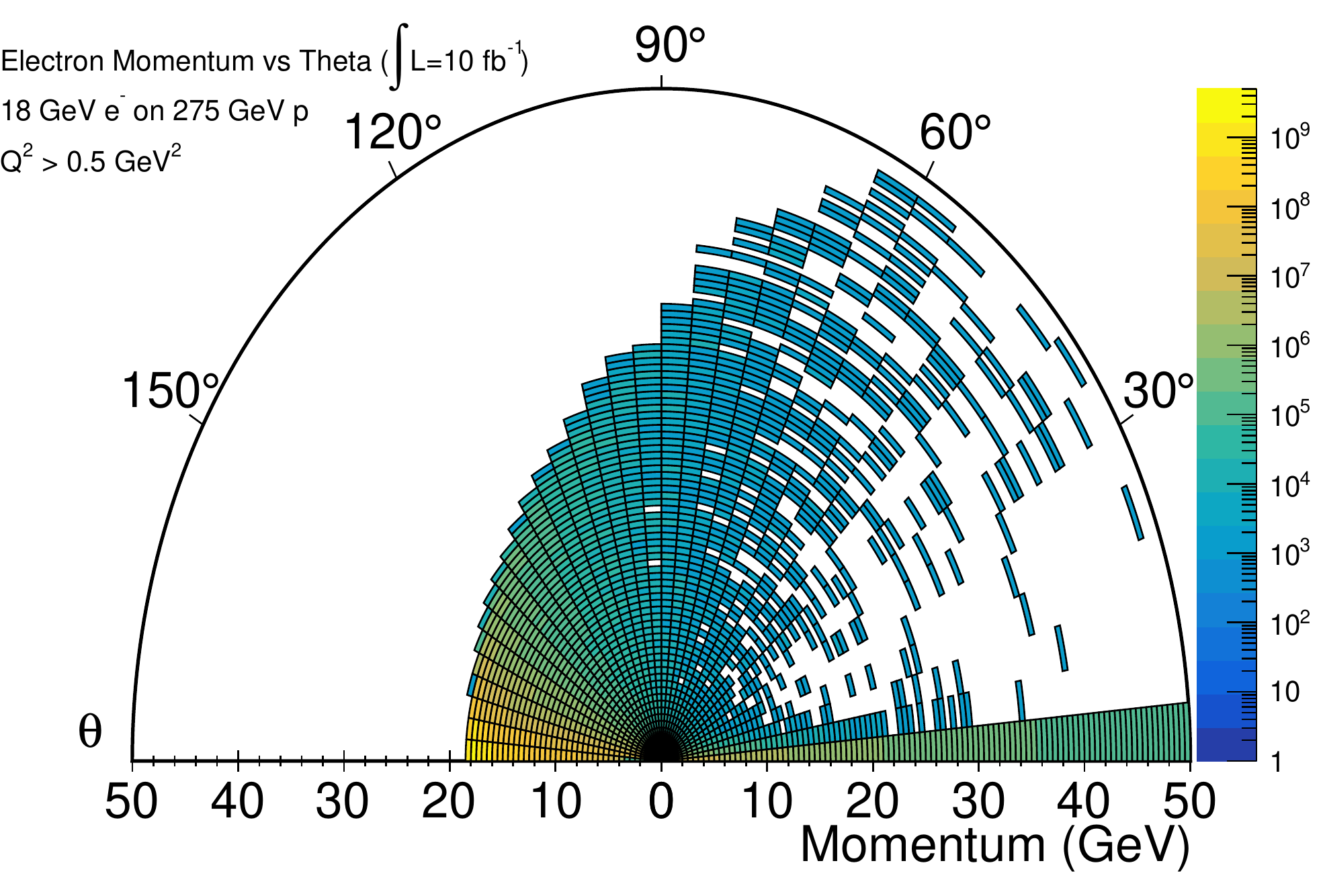}}\qquad%
\subfloat[e+p 18x275 GeV]{\label{fig:PSall_Q2cut_18_275}\includegraphics[width=0.45\linewidth]{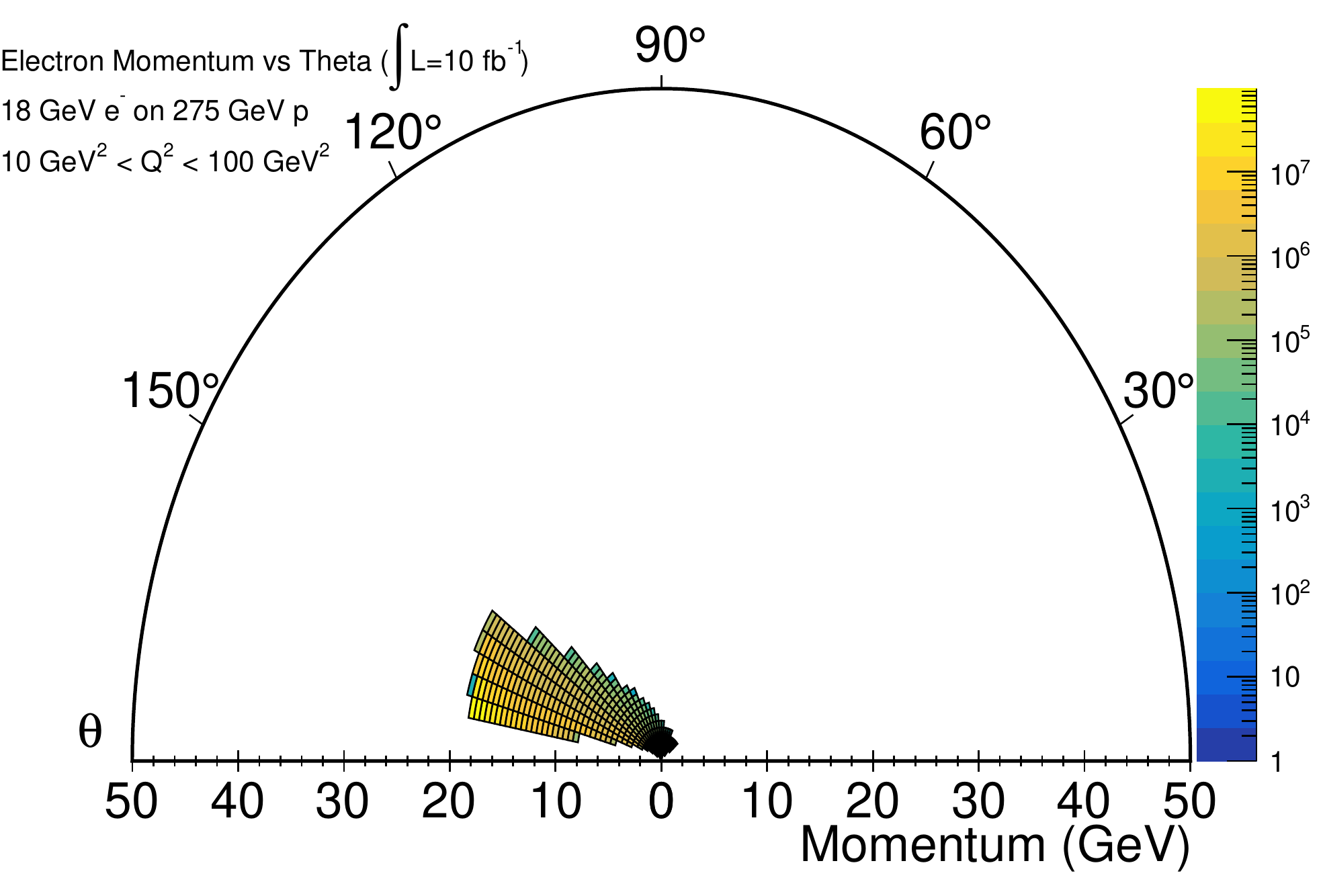}}\qquad%
\caption{Polar plots of yields for detected electrons in NC interactions binned in $\theta^{e'}_p$ and $p$ for 18x275 beam configuration. The left plot includes scattered and decay electrons. The right plot is the scattered electron distribution but with a $10< Q^2 < 100$~\gevcc  cut applied.}
\end{figure}

The distributions in \ref{fig:PS_18_275}-\ref{fig:PS_5_41} only show the kinematics of the scattered beam electron with a minimal cut of $Q^2>0.5$ GeV$^2$. Figure \ref{fig:PSall_18_275} shows the enhanced electron yield in the far forward regions, due largely to hadronic and leptonic decays. Figure  \ref{fig:PSall_Q2cut_18_275} shows the same distribution for the scattered electron with (10$<Q^2<$100) GeV$^2$ cuts applied.

\begin{figure}[htbp]
\centering
\subfloat[e+p 18x275 GeV]{\label{fig:PS_18_275_hadrons}\includegraphics[width=0.45\linewidth]{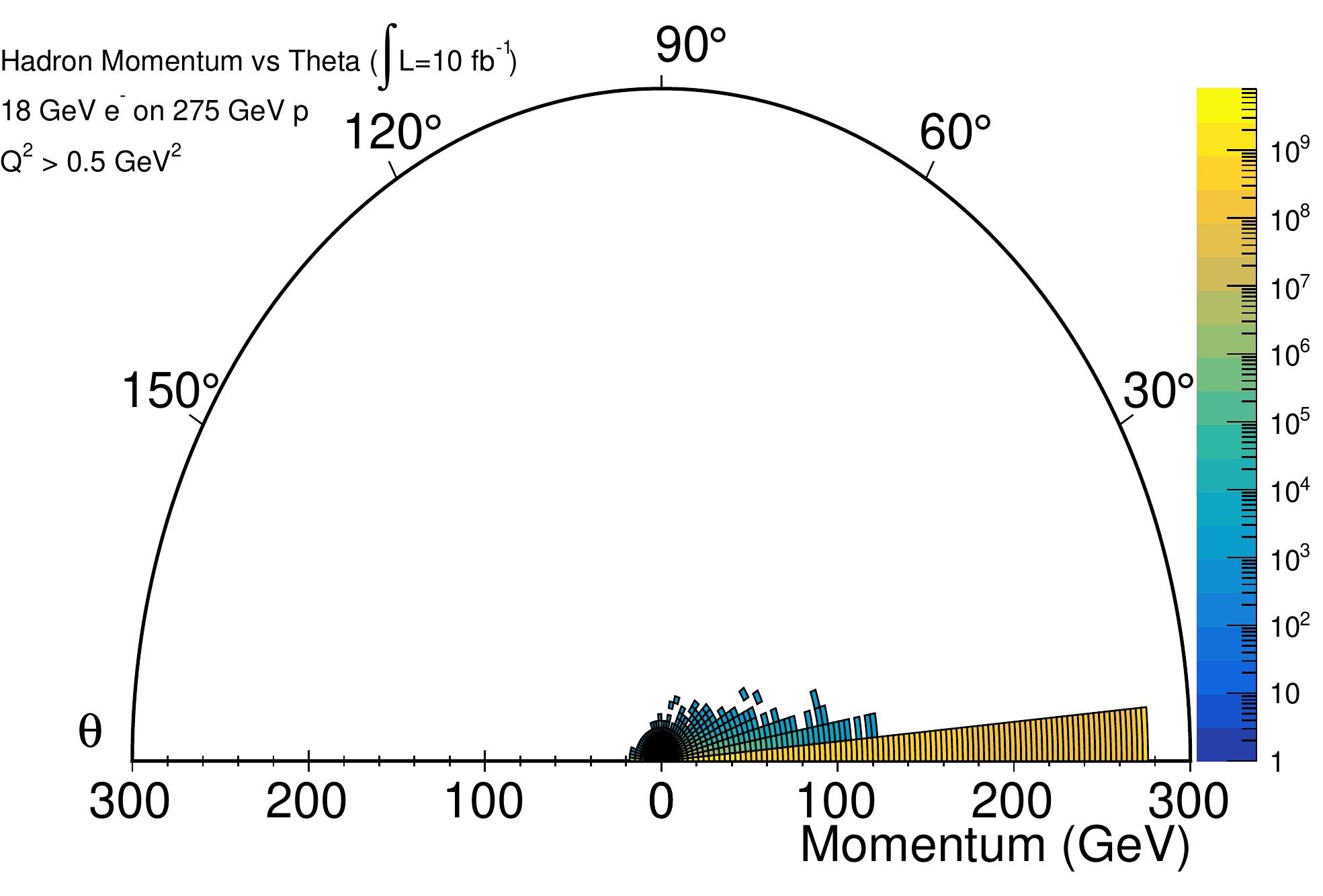}}\qquad%
\subfloat[e+p 10x100 GeV]{\label{fig:PS_10_100_hadrons}\includegraphics[width=0.45\linewidth]{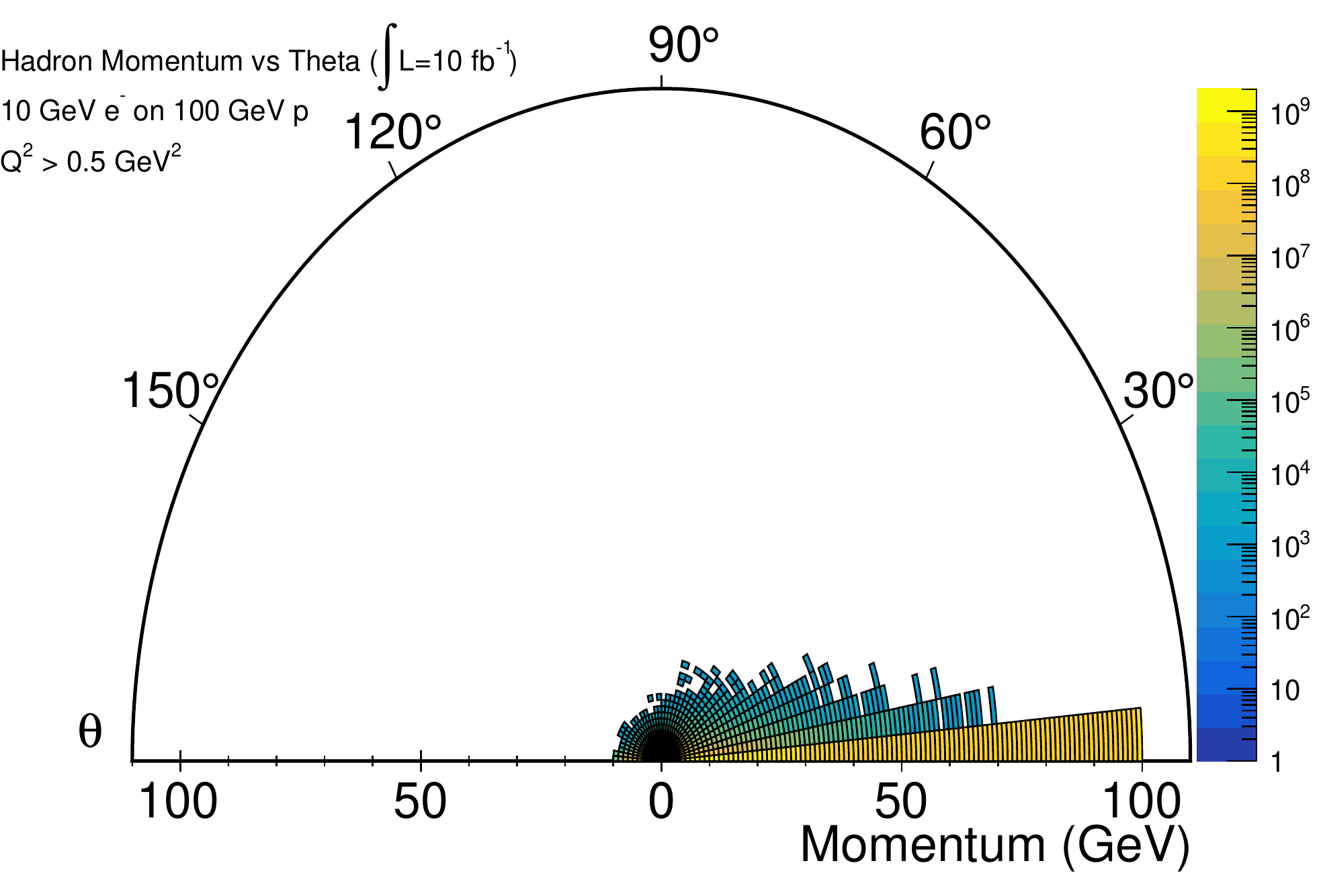}}\qquad%
\subfloat[e+p 5x100 GeV]{\label{fig:PS_5_100_hadrons}\includegraphics[width=0.45\textwidth]{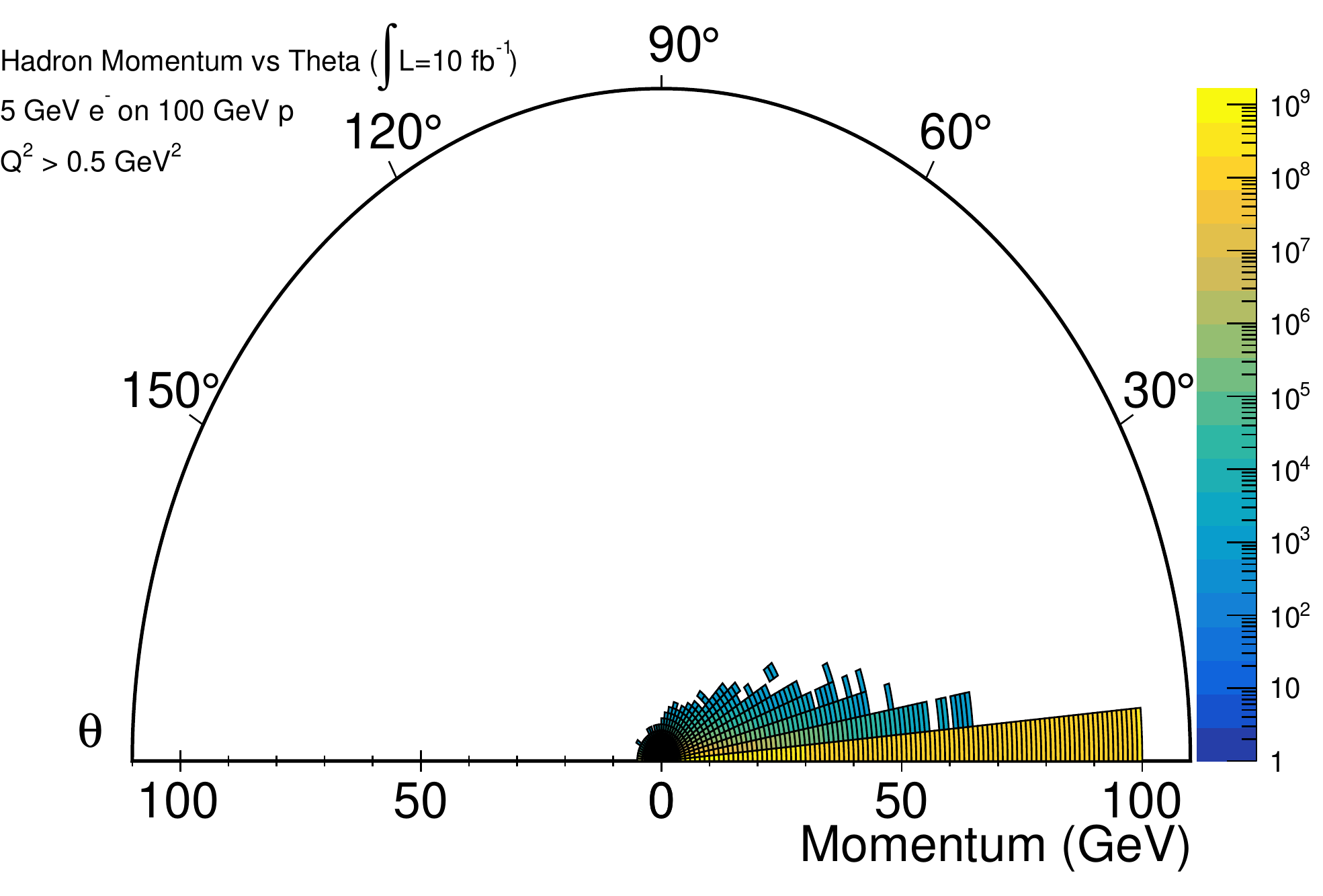}}\qquad%
\subfloat[e+p 5x41 GeV]{\label{fig:PS_5_41_hadrons}\includegraphics[width=0.45\textwidth]{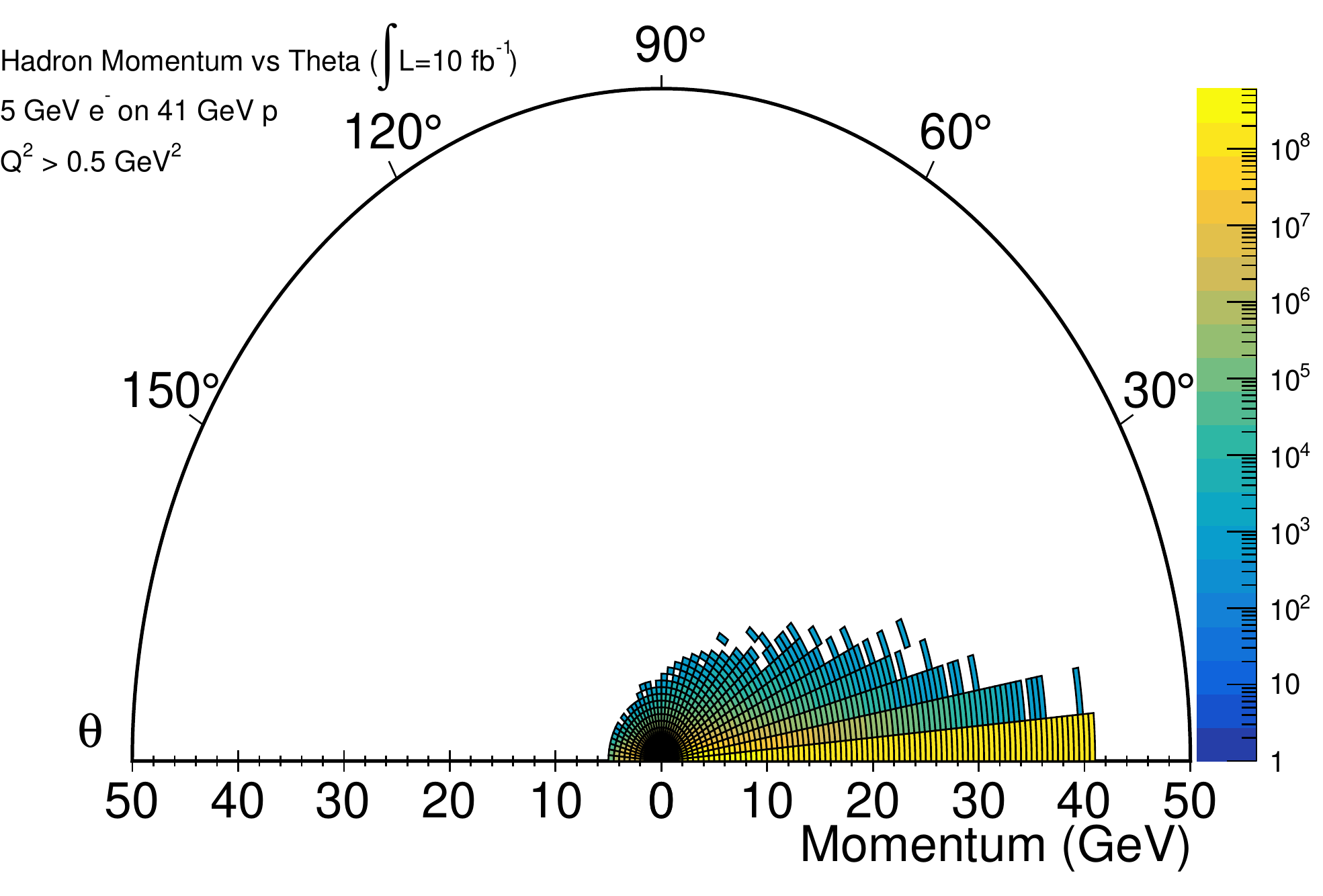}}%
\caption{Polar plots of yields for scattered hadrons in NC interactions binned in $\theta^{e'}_p$ and $p$ for four proposed  center-of-mass energies. These plots were created using the Pythia6 event generator with a cut cut of $Q^2>0.5$ GeV$^2$ applied at the vertex level. }
\end{figure}

Figures \ref{fig:PS_18_275_hadrons} - \ref{fig:PS_5_41_hadrons} show the momentum-$\theta^h_p$ distributions of the hadronic recoil in NC events for the four canonical beam configurations. The yield and average momentum are highly peaked in the proton beam direction. The accuracy of the JB reconstruction method relies on detecting as much of this hadronic recoil as possible, motivating forward coverage for both electromagnetic (for photons) and hadronic calorimeters. Distributions for neutrons, protons and photons are documented in the inclusive reactions wiki\footnote{https://wiki.bnl.gov/eicug/index.php/Yellow$\_$Report$\_$Physics$\_$Inclusive$\_$Reactions}. The same distributions produced by the DJANGOH event generator are nearly identical to those generated by Pythia6 and may be found on the inclusive reactions wiki as well.

\begin{figure}[htbp]
\centering
\subfloat[e+p CC 18x275 GeV]{\label{fig:PS_CC_photons}\includegraphics[width=0.45\linewidth]{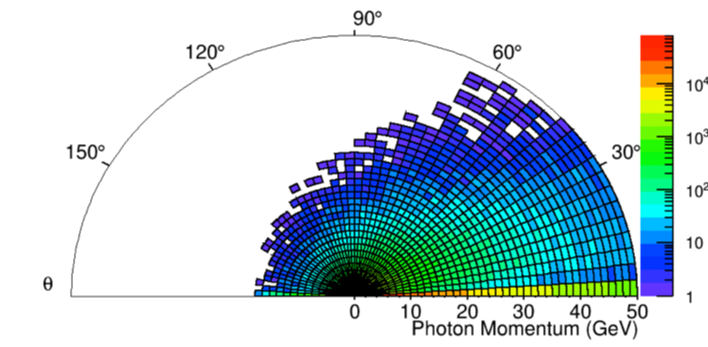}}\qquad%
\subfloat[e+p CC 18x275 GeV]{\label{fig:PS_CC_hadrons}\includegraphics[width=0.45\linewidth]{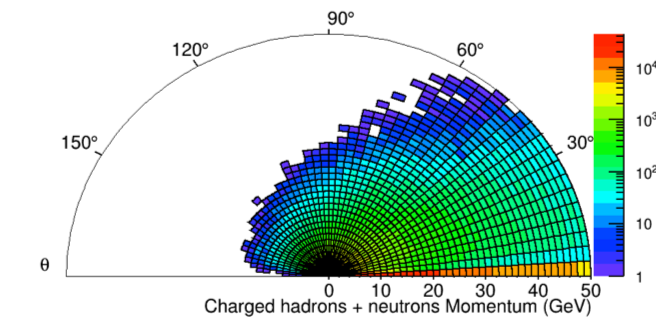}}\qquad%
\caption{Polar plots of yields for photons (left) and hadrons (right) in CC interactions binned in $\theta^{h/\gamma}_p$ and $p$ for 18x275 GeV beam configuration.}
\end{figure}

Charged Current events also rely on the JB reconstruction method so it is critical to investigate the electromagnetic and hadronic recoil in CC events. Similar to the NC case, Figures \ref{fig:PS_CC_photons} and \ref{fig:PS_CC_hadrons} show a highly peaked distribution in the far forward region. However, in constrast to NC events, there is a much larger, higher momentum tail extending into the mid-forward rapidity region. The photon distributions are very similar to the hadron distributions, reinforcing the need for continuous electromagnetic and hadronic calorimeter coverage through the mid-rapidity region and as far possible into the forward region.

\subsection{Electron acceptance and particle identification}
\label{sec:ePID}

The detection and identification of the scattered electron is critical for nearly all inclusive reaction channels. The exception is the CC channel, where the resulting neutrino escapes undetected, and the kinematic reconstruction relies completely on the detection of the hadronic recoil. In those cases no particle identification (PID) is necessary, only energy and momentum reconstruction. 

\begin{figure}[htbp]
\centering
\includegraphics[width=0.8\linewidth]{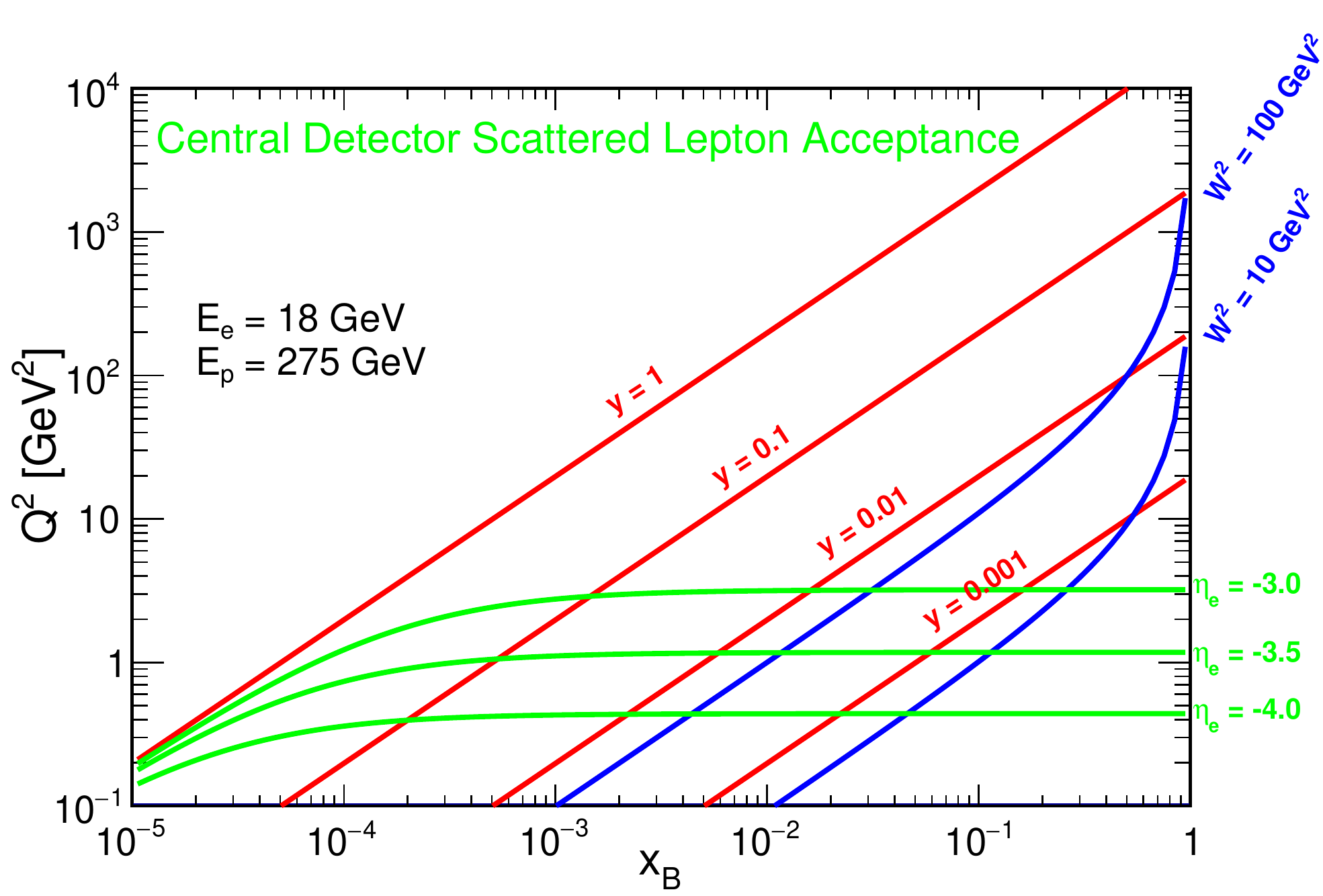}
\caption{Scattered lepton acceptance as a function of $x_B$ and $Q^2$ for 18x275 GeV beam collisions. Green curves denote possible fiducial edges of the proposed central arm detector. Red and blue curves indicate the lines of constant $y$ and $W^2$.\label{fig:Acceptance}}
\end{figure}

The minimum electron momentum that can be detected is set by the proposed acceptance of $-3.5 < \eta < 3.5 $ and the magnetic field of the detector. Figure \ref{fig:Acceptance} shows the limits placed on $Q^2$, $x$ and $y$ due to the detector acceptance for 18x275 GeV NC events. The vast majority of the inclusive channels require a $Q^2 > 1$ GeV$^2$ and W$^2$ > 4 GeV$^2$ cut for interpretation within a pQCD framework. These requirements alone exclude nearly the entire available phase space for $\eta < -3.5$. The conclusion is that inclusive reactions with these kinematic cuts do not require detection capabilities beyond $\eta = -3.5$. By their nature, gluon saturation, color glass condensate and low $Q^2$ photo-production studies do not have these requirements and would likely utilize detectors in the far backward region. 

\begin{wrapfigure}{R}{0.45\linewidth}
\centering{} 
\includegraphics[width=0.95\linewidth]{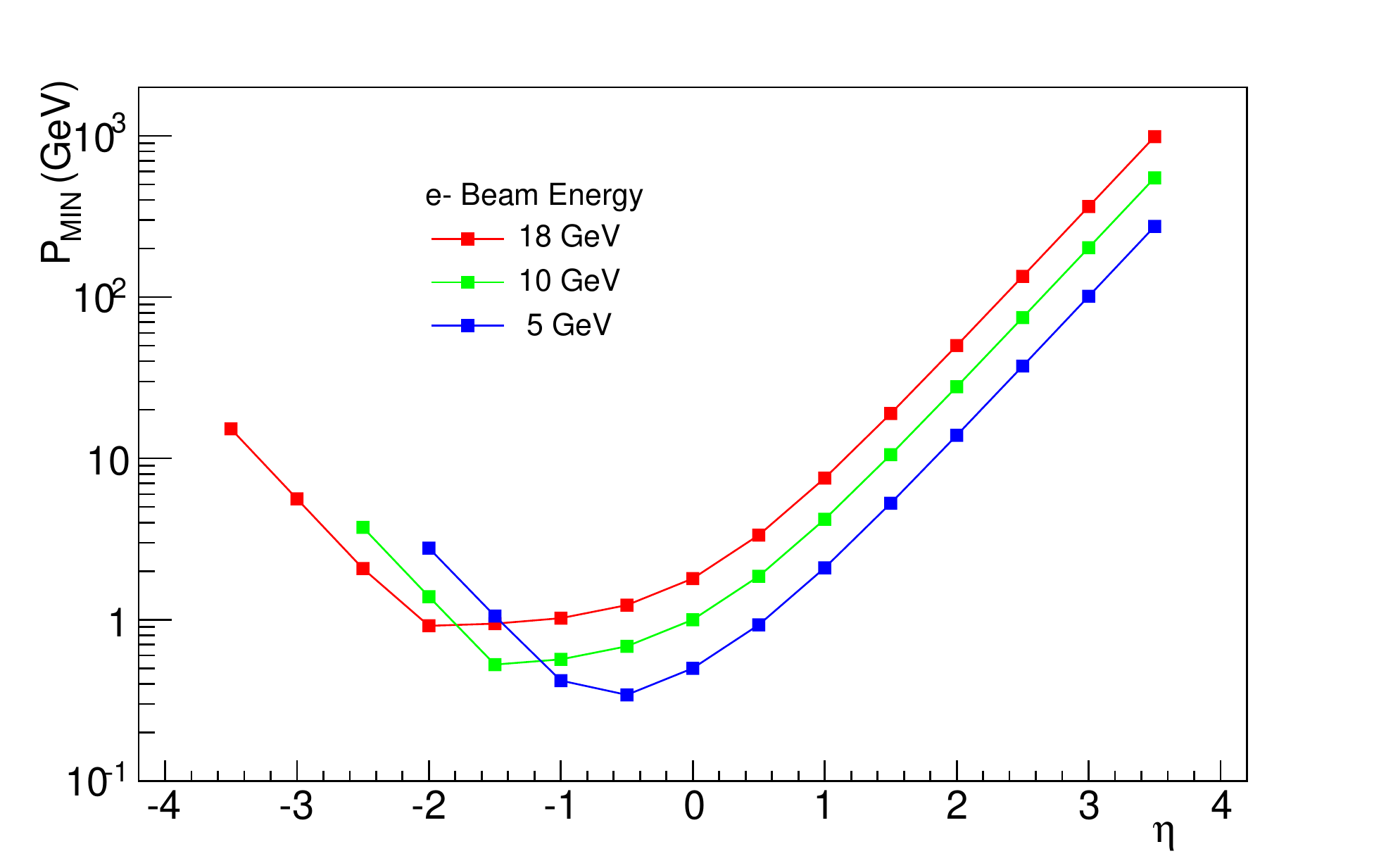}
\caption{The $P_{MIN}$ of the scattered $e^-$ vs detector $\eta$. In the central, central-backward region, the scattered $e^-$ momenta extend down to 0.3-0.5 GeV/$c$. These calculations do not include magnetic field effects.}
\label{fig:EtaMom}
\end{wrapfigure}
In addition to the minimum $Q^2$ requirement, a $y < 0.95$ cut is typically applied in order to maintain a reasonable $Q^2$ resolution. Figure \ref{fig:EtaMom} shows the minimum momentum, $P_{MIN}$, of the detected electron as a function of detector $\eta$. This figure does not account for any acceptance losses at low momentum due to curvature in the magnetic field. Once included, the minimum momentum is likely to be $\sim{500}$ MeV/$c$.

\begin{figure}[!ht]
\centering
\includegraphics[width=0.8\textwidth]{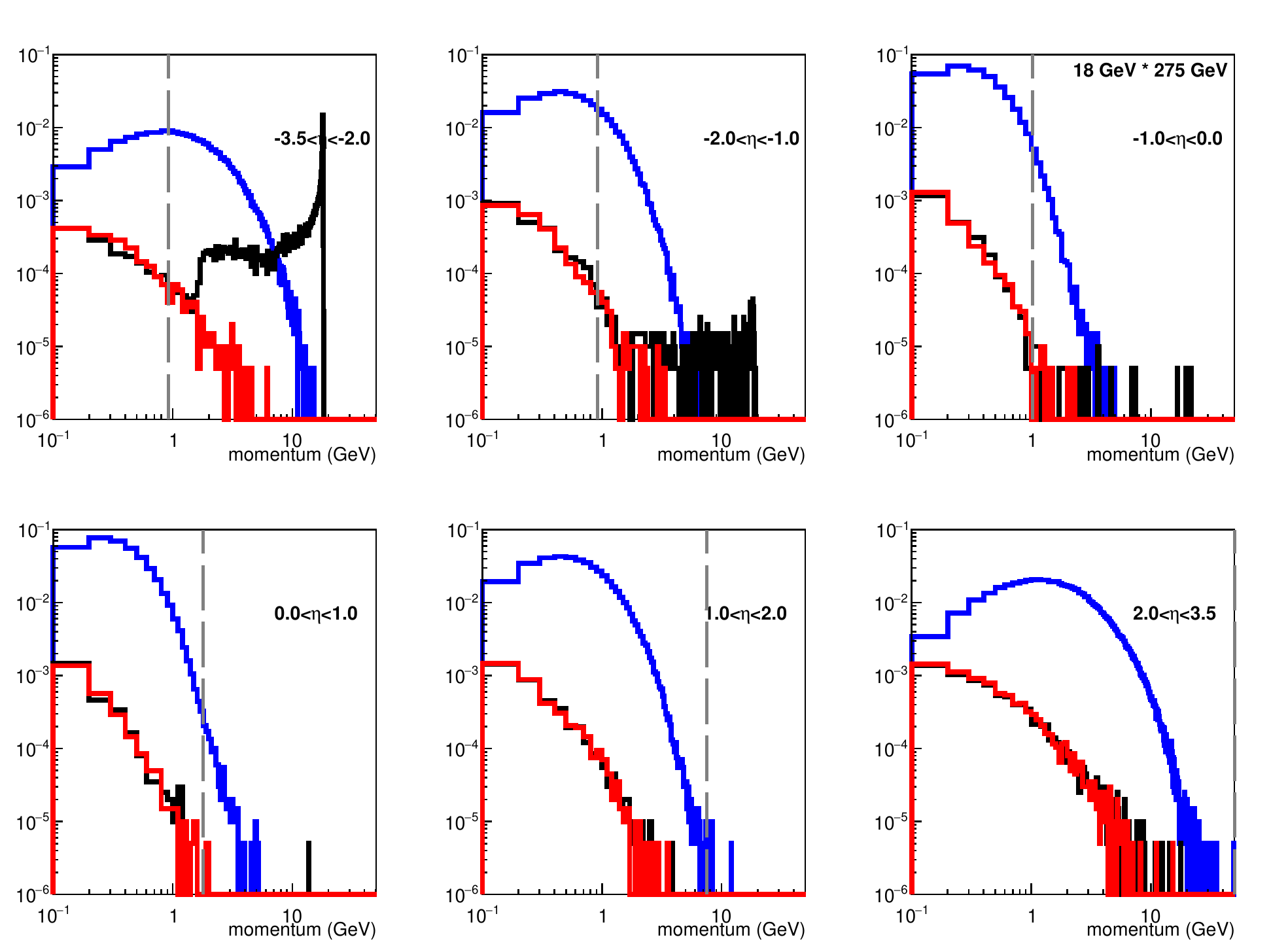}
\caption{Event normalized $e^-$(black), $e^+$(red) and $\pi^-$(blue) yields for the 18x275 GeV beam configuration. The dashed line marks the $P_{MIN}$ for electrons detected in that $\eta$ bin.}
\label{fig:PID_18_275}
\end{figure}

\begin{figure}[!ht]
\centering
\includegraphics[width=0.8\textwidth]{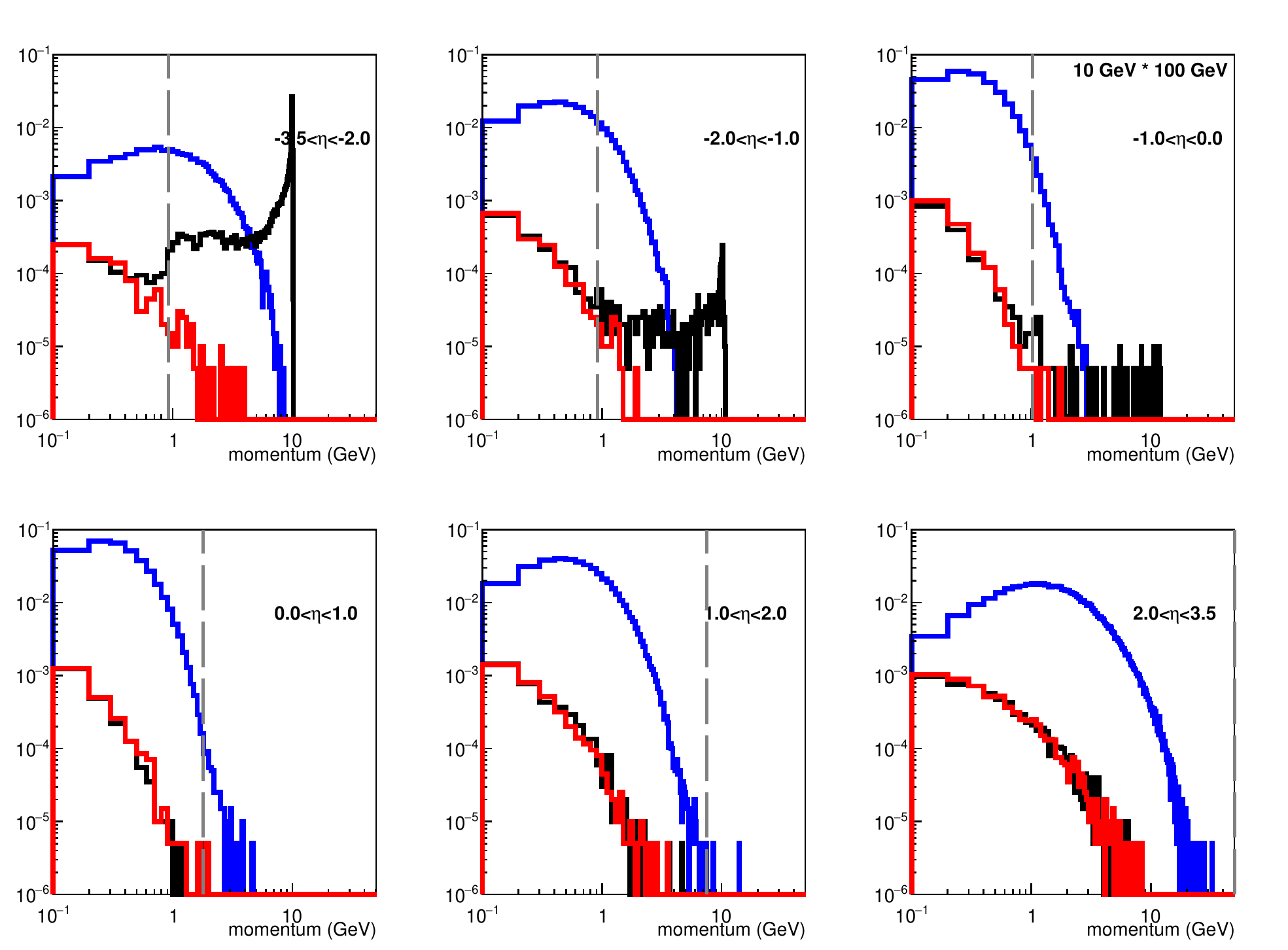}
\caption{Event normalized $e^-$(black), $e^+$(red) and $\pi^-$(blue) yields for the 10x100 GeV beam configuration. The dashed line marks the $P_{MIN}$ for $e^-$ detected in that $\eta$ bin.}
\label{fig:PID_10_100}
\end{figure}

\begin{figure}[!ht]
\centering
\includegraphics[width=0.8\textwidth]{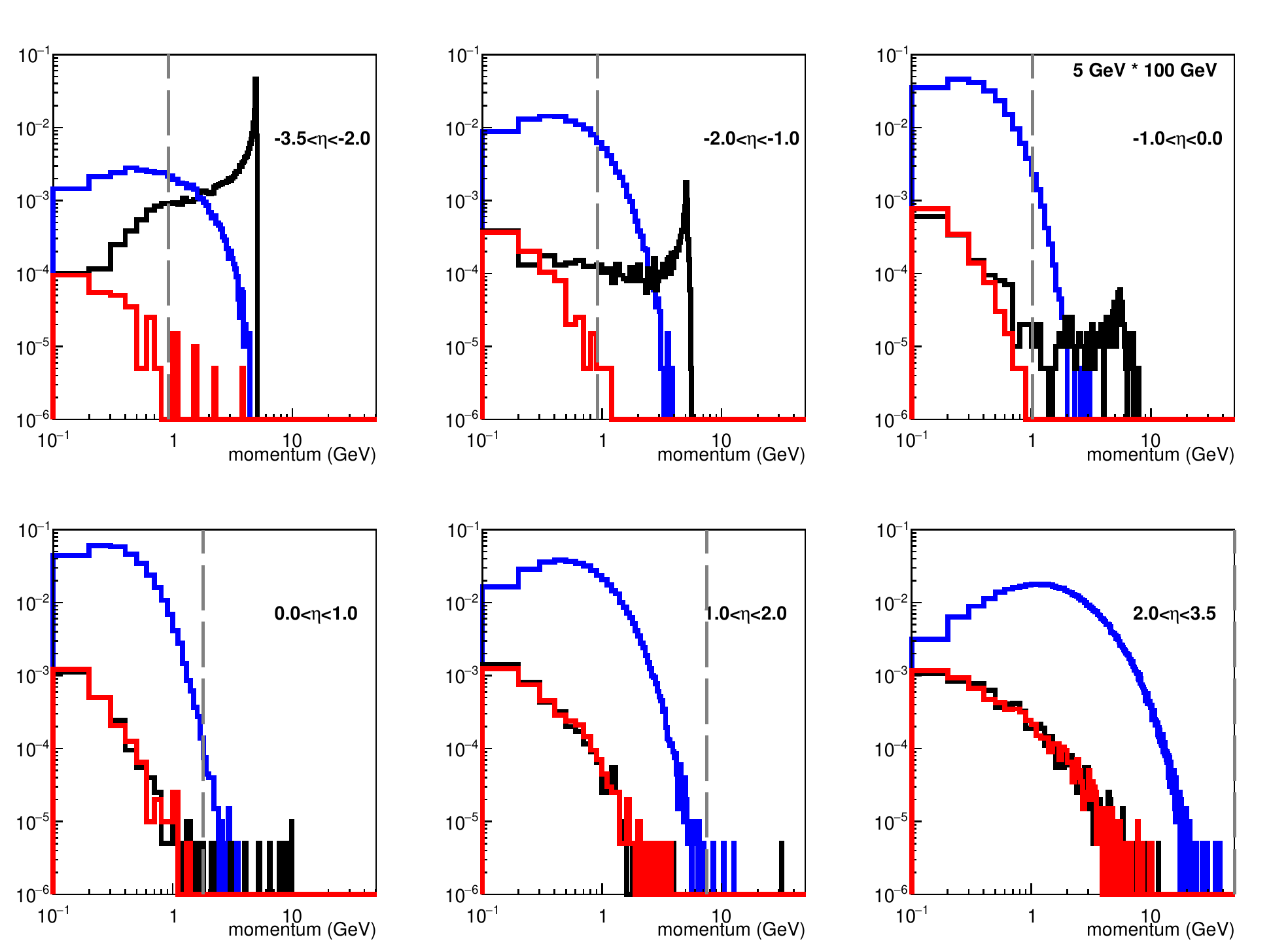}
\caption{Event normalized $e^-$(black), $e^+$(red) and $\pi^-$(blue) yields for the 5x100 GeV beam configuration. The dashed line marks the $P_{MIN}$ for $e^-$ detected in that $\eta$ bin.}
\label{fig:PID_5_100}
\end{figure}

\begin{figure}[!ht]
\centering
\includegraphics[width=0.8\textwidth]{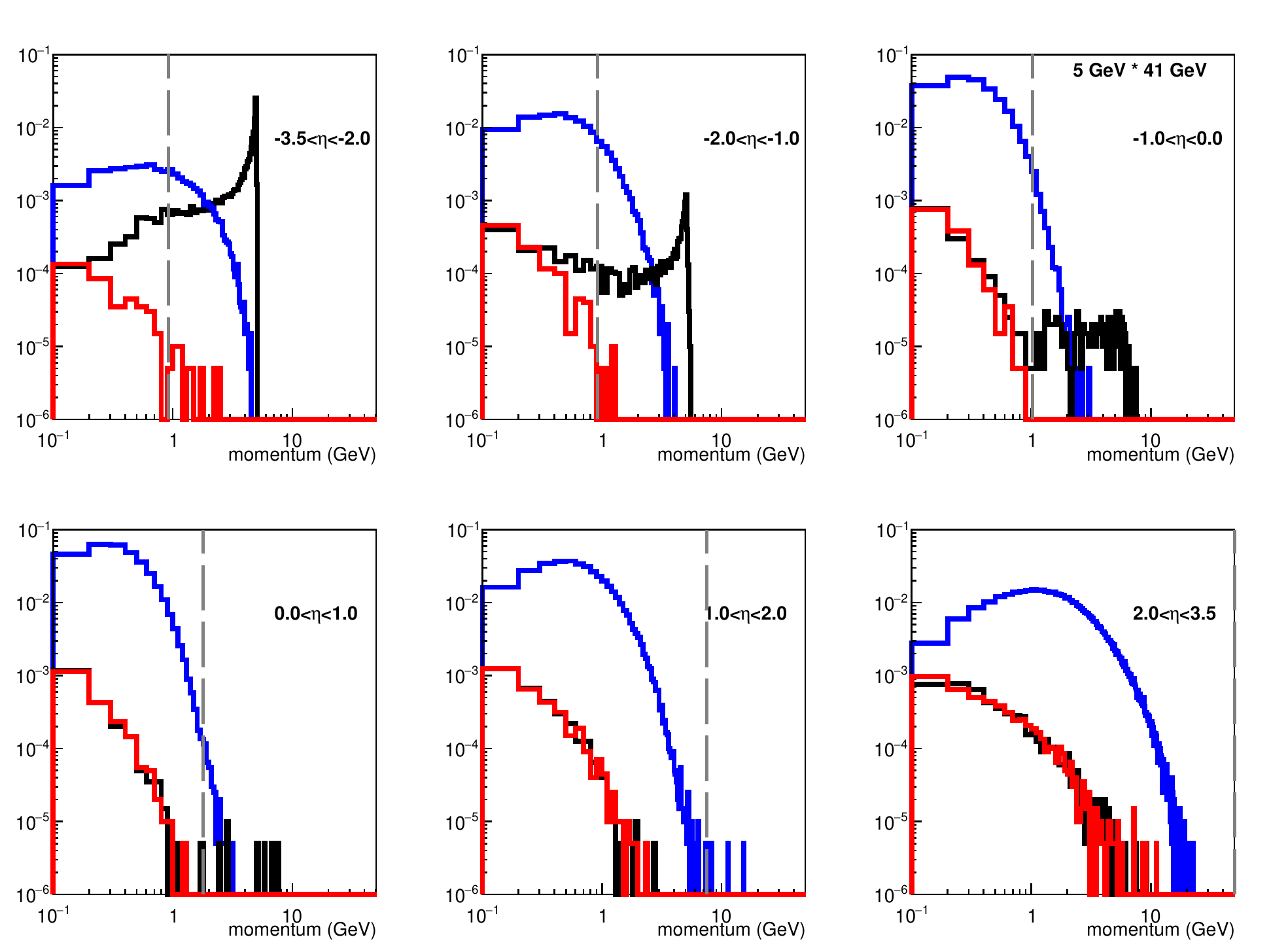}
\caption{Event normalized $e^-$(black), $e^+$(red) and $\pi^-$(blue) yields for the 5x41 GeV beam configuration. The dashed line marks the $P_{MIN}$ for electrons detected in that $\eta$ bin.}
\label{fig:PID_5_41}
\end{figure}

Electron PID is required to suppress two types of backgrounds. The largest background comes from the significant rate of same-sign charged pion production. Figures \ref{fig:PID_18_275}-\ref{fig:PID_5_41} show the scattered $e^-$ and $\pi^-$ yields, normalized to the number of thrown events, for each center-of-mass configuration and as a function of particle momentum for six bins in the range $-3.5<\eta<3.5$. The dashed grey lines mark the $P_{MIN}$ of the detected electron for the given $\eta$ bin. Table \ref{tab:ePID} documents the maximum $\pi^-/e^-$ ratio, for $p > P_{MIN}$, in each beam configuration and $\eta$ bin. Note, bins with negligible $e^-$ rates above $P_{MIN}$ are omitted from the table.


\newpage
The most stringent constraints on the detector electron PID capabilities come from the longitudinal double spin and the electron parity violating asymmetries, $A_{LL}$ and $A^e_{PV}$. The pion contamination has two effects; it inflates the statistical errors and incurs an associated systematic error. In the case of $A_{LL}$ the raw asymmetry is typically corrected for the measured pion $A^\pi_{LL}$ and the statistical errors from both asymmetries combined appropriately. The PID systematic error for $A_{LL}$ is then proportional to the {\it uncertainty} on the pion contamination. Assuming the uncertainty is an order of magnitude less than the contamination, the $A_{LL}$ measurement can tolerate contamination of $\sim{0.01}$ $\pi^-/e^-$  before becoming systematically limited. For $A^e_{PV}$ the pion asymmetry $A^\pi_{PV}$ is typically consistent with zero, so the pion contamination acts as a pure dilution. In this case the associated PID systematic error incorporates the statistical uncertainty on the $A^{\pi}_{PV}$ measurement and is dominated by a term that is proportional to the fraction of pions in the reconstructed electron sample. A conservative upper limit of $1\times 10^{-3}$ $\pi^-/e^-$ pion contamination is set by requiring the associated systematic error to be less than  $10\%$ of the statistical error. 
This estimate is based on the statistical errors shown in Figure \ref{fig:APVhad} and the systematic error formulation outlined in Eq. 3.12.2 in Ref.\cite{Wang:2013iiv}. 

\begin{table}[htbp]
\centering
\vspace{0.2cm}
\begin{tabular}{|c|c|c|c|c|}
\hline
$E^{e^-}_{beam}$ (GeV) & $\eta$ bin & $P^{e^-}_{min}$ (GeV) &  Max $\pi^-/e^-$ & final $\pi^-/e^-$ ratio\\
\hline
\hline
 18 &  (-3.5,-2) & 0.9 & 200 & 0.02\\ 
 18 &  (-2,-1)  & 0.9 & 800 & 0.08\\ 
 18 &  (-1, 0)  & 1.0 & 1000 & 0.1\\ 
 18 &  (0, 1)  & 1.8 & 100 & 0.01\\ 
 \hline
 10 &  (-3.5,-2) & 1.4 & 10 & 0.001\\ 
 10 &  (-2,-1)  & 0.5 & 400 & 0.04\\ 
 10 &  (-1, 0)  & 0.6 & 800 & 0.08\\ 
 10 &  (0, 1)  & 1.0 & 1000 & 0.1 \\ 
 \hline
 5 &  (-3.5,-2) & 2.8 & 0.1 & 0.00001 \\ 
 5 &  (-2,-1)  & 0.4 & 100 & 0.01\\ 
 5 &  (-1, 0)  & 0.3 & 500 & 0.05\\ 
 5 &  (0, 1)  & 0.5 & 1000 & 0.1\\ 
 \hline
\end{tabular}
\caption{The minimum detected $e^-$ momentum (column 3), the maximum $\pi^-/e^-$ ratio for electrons with $p^{e-} > P^{e-}_{min}$ (column 4) and the final $\pi^-/e-$ ratio after the $10^4$ suppression determined for the original baseline detector (column 5) for each $e^-$ beam energy and scattered electron $\eta$ bin. The calculation of $P^{e^-}_{min}$ includes a $Q^2 > 1$~GeV$^2$ and $y< 0.95$ requirement.}
\label{tab:ePID}
\end{table}

The original estimate of the $\pi^-$ suppression at the detector level (Fig.~\ref{fig:HandbookT2}) was $10^{4}$ for $-3.5 < \eta < 1$. Applying this factor to the  maximum $\pi^-/e^-$ rates (column 4) for electrons with momentum greater than $P^{e-}_{min}$ (column 3) gives the expected fraction of pions in the reconstruction electron sample (column 5) indicated in Table $\ref{tab:ePID}$. The goal of $1\%$ contamination for $A_{LL}$ is never met for the highest beam energy. The situation improves for lower energy beams but is still not sufficient for $\eta>-1 (-2)$ for the 5 (10) GeV $e^-$ beams. 

\begin{figure}[!ht]
\centering
\includegraphics[width=0.45\textwidth]{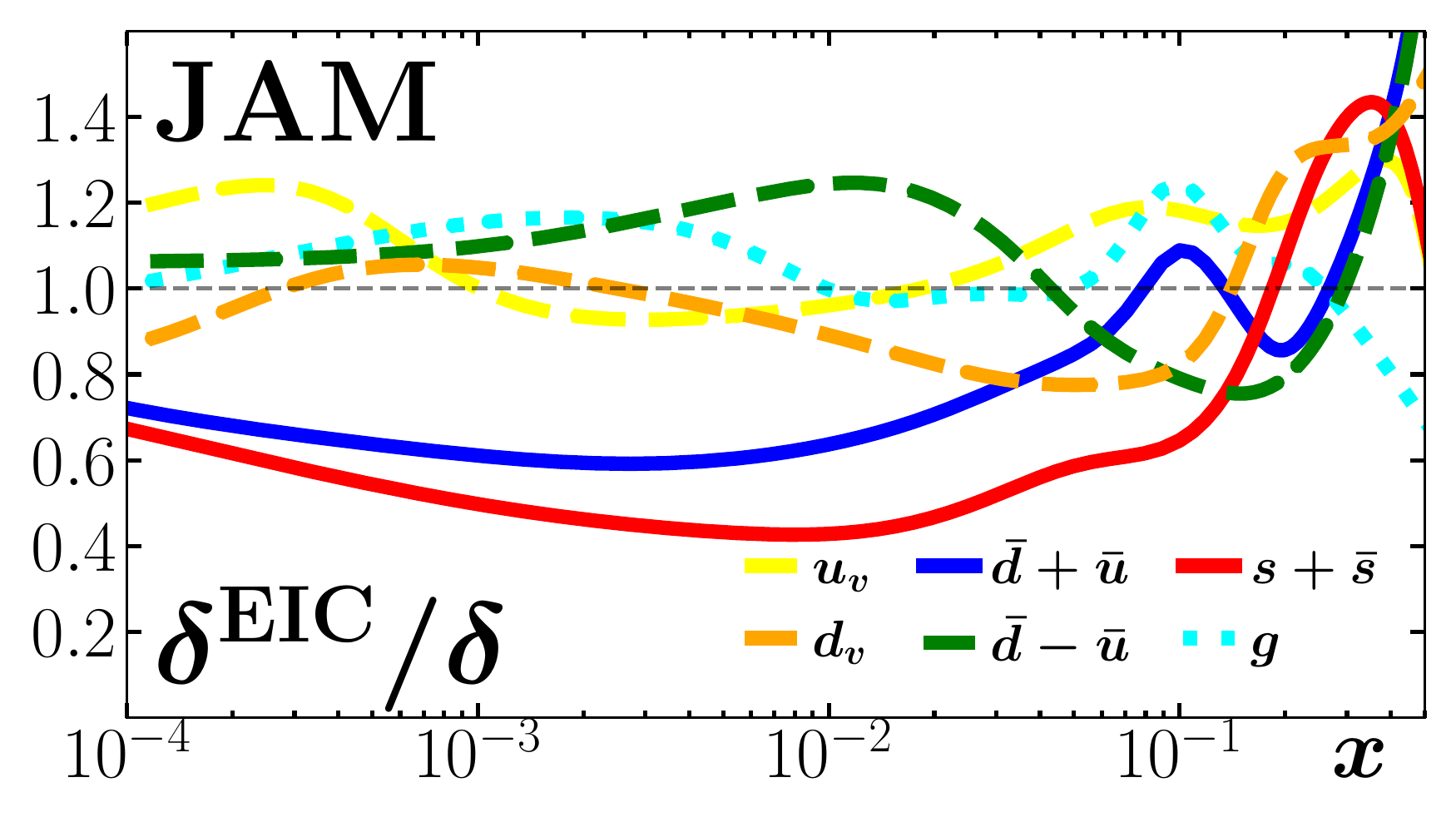}
\hspace{0.5cm}
\includegraphics[width=0.45\textwidth]{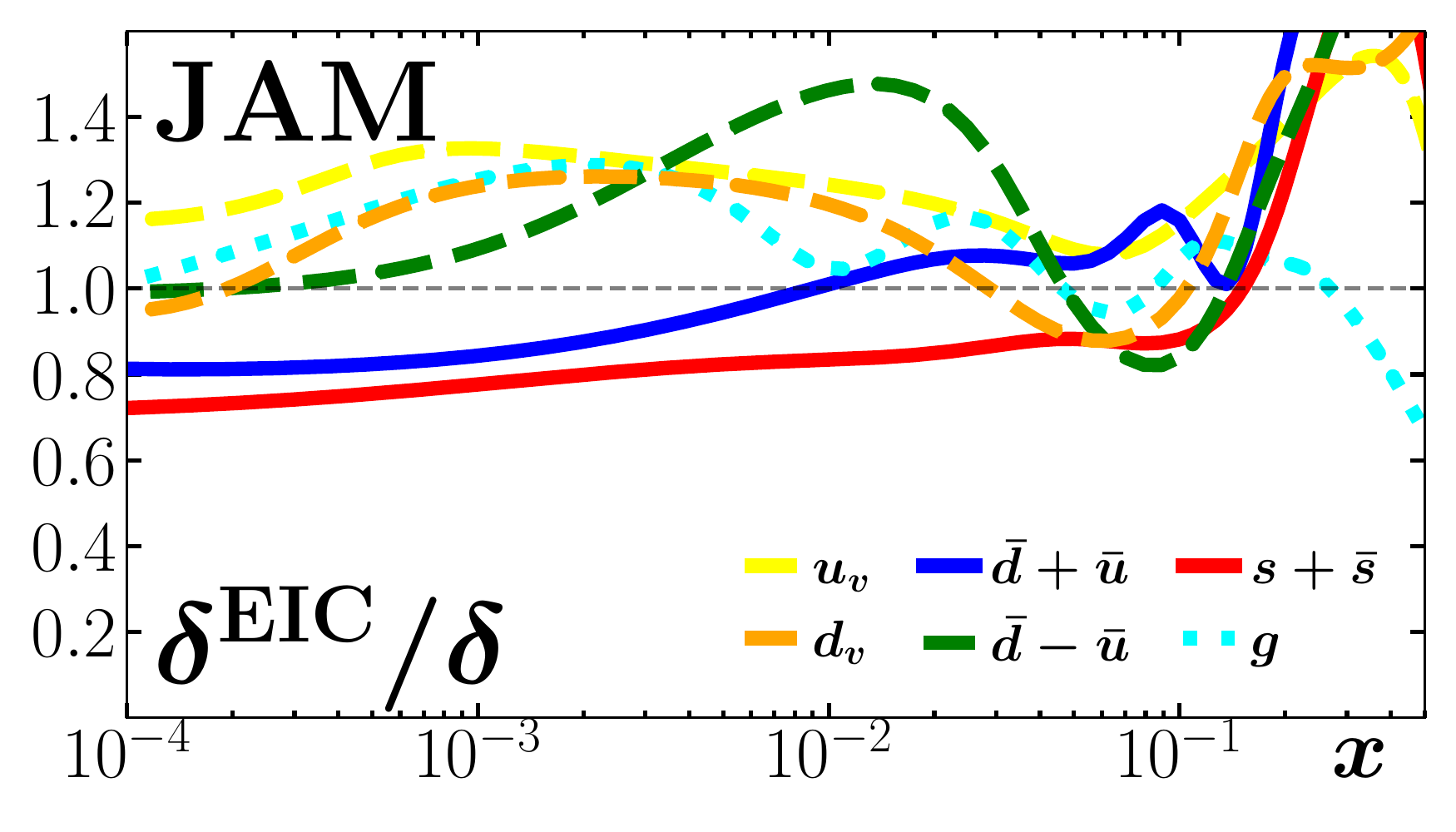}
\caption{Constraints on quark and gluon PDFs due to the inclusion of $A^{e-}_{PV}$ as measured at the EIC. The left/right side incorporates systematic errors from the original/updated EIC Handbook detector matrix. }
\label{fig:APVe_sys}
\end{figure}

The situation for $A^{e-}_{PV}$ is even more dire, with only the far backward region for the two lowest energy beams meeting the desired $ 1\times 10^{-3}$ $\pi^-/e^-$ contamination. The updated estimate for the $\pi^-$ suppression at the detector level maintains the $10^{4}$ suppression for only the $-3.5<\eta<-2$ bin, but decreases it to $10^{3}$ for $-2<\eta<-1$ and then $10^{2}$ for $-1<\eta<1$. The effect is a significant reduction in the constraints on the quark and gluon PDFs. This is demonstrated by comparing Fig.~\ref{fig:APVe}, which assumes a flat $1\%$ systematic error, to Fig.~\ref{fig:APVe_sys} which incorporates the $\eta$ dependent systematic errors associated with the original uniform $10^4$ suppression (left) and updated estimate that decreases to $10^2$ at midrapidity (right).

It is worth noting that $\pi^-$ suppression may be enhanced through software pion ID algorithms that were not included in these studies and are expected to increase pion suppression by factors of 2-4. Further studies that incorporate realistic detector materials and response to hadronic interactions are necessary to robustly evaluate $\pi^{-}$ suppression capabilities. 

Pair-symmetric production of $e^-$ constitutes the second most significant background contribution to inclusive channels. These electrons are the result of pair-production via interactions of the scattered electron with detector material and Dalitz decays. The pair-symmetric contribution to the total $e^-$ yield is represented by the red $e^+$ distribution in Figures \ref{fig:PID_18_275}-\ref{fig:PID_5_41}. Robust evaluations of the size of the pair-symmetric background require simulation of the full material budget. Corrections are typically based on a combination of dedicated systematic runs (with a reversed magnetic field for example) and analysis level-algorithms that discriminate between the scattered electron and pair-symmetric backgrounds.
 
\subsection{Resolution and bin migration effects in electron reconstruction.}

The resolution on the reconstruction of $x$, $Q^2$ and $y$ via electron detection in NC events was evaluated by passing 10 $fb^{-1}$ of 18x275, 10x100 and 5x100 GeV neutral current pseudo-data through the EICSmear fast simulation package. The pseudo-data was produced using the DJANGOH event generator with full radiative effects turned on. The EICSmear package implements a one mRad smearing in $\theta$ and $\phi$ and an energy resolution of $2\% \oplus 1\%$, $7\% \oplus 1\%$, $12\% \oplus 2\%$  for the backward, mid and forward electromagnetic calorimeters.

\begin{wrapfigure}{R}{0.5\linewidth}
\centering{}
\vspace{-0.4cm}
\includegraphics[width=0.95\linewidth]{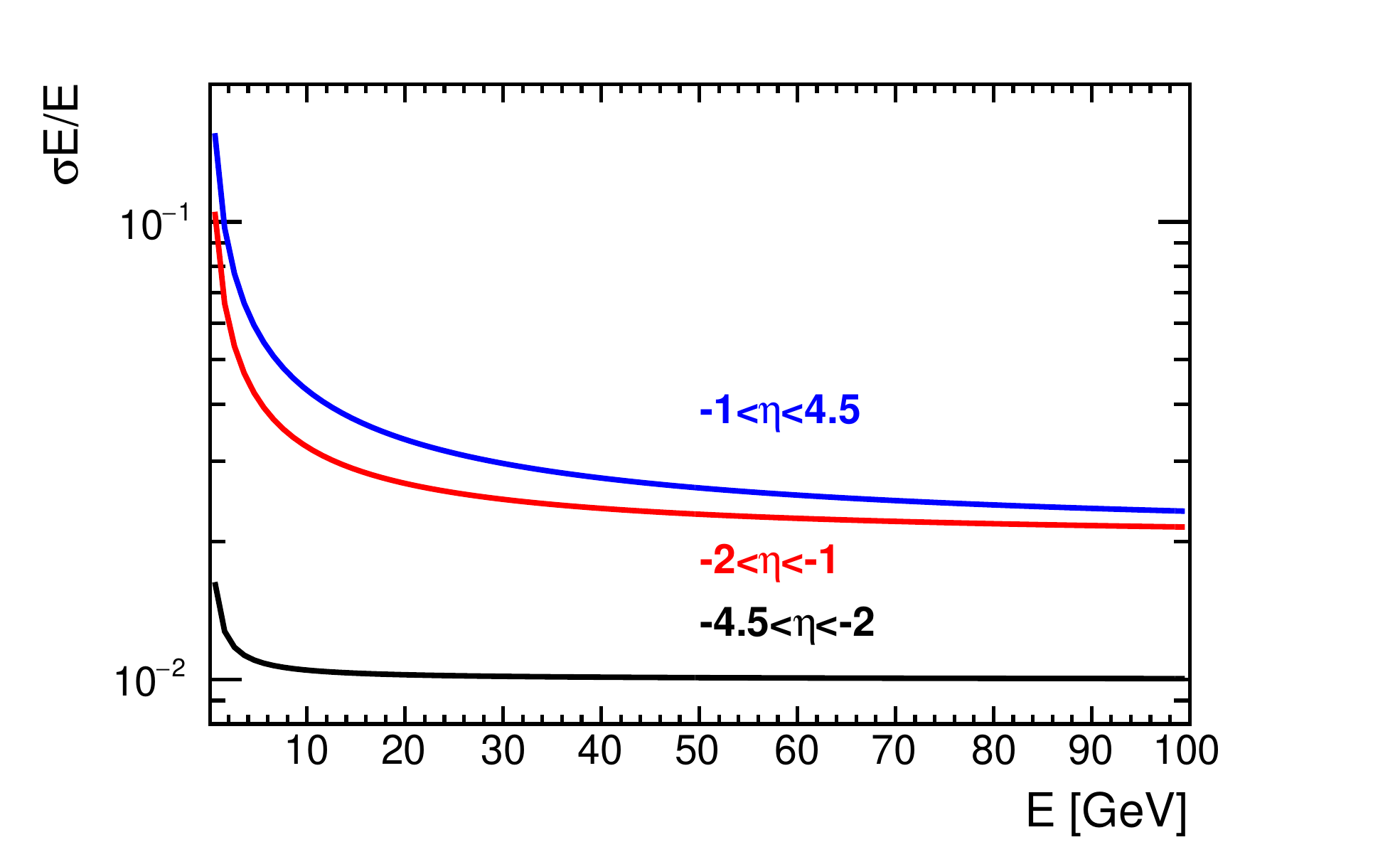}
\caption{Energy resolution as a function of $e^-$ energy for the backward, central and forward electromagnetic calorimeters.}
\label{fig:dEeta}
\end{wrapfigure}
\begin{figure}[!ht]
\centering
\includegraphics[width=0.95\linewidth]{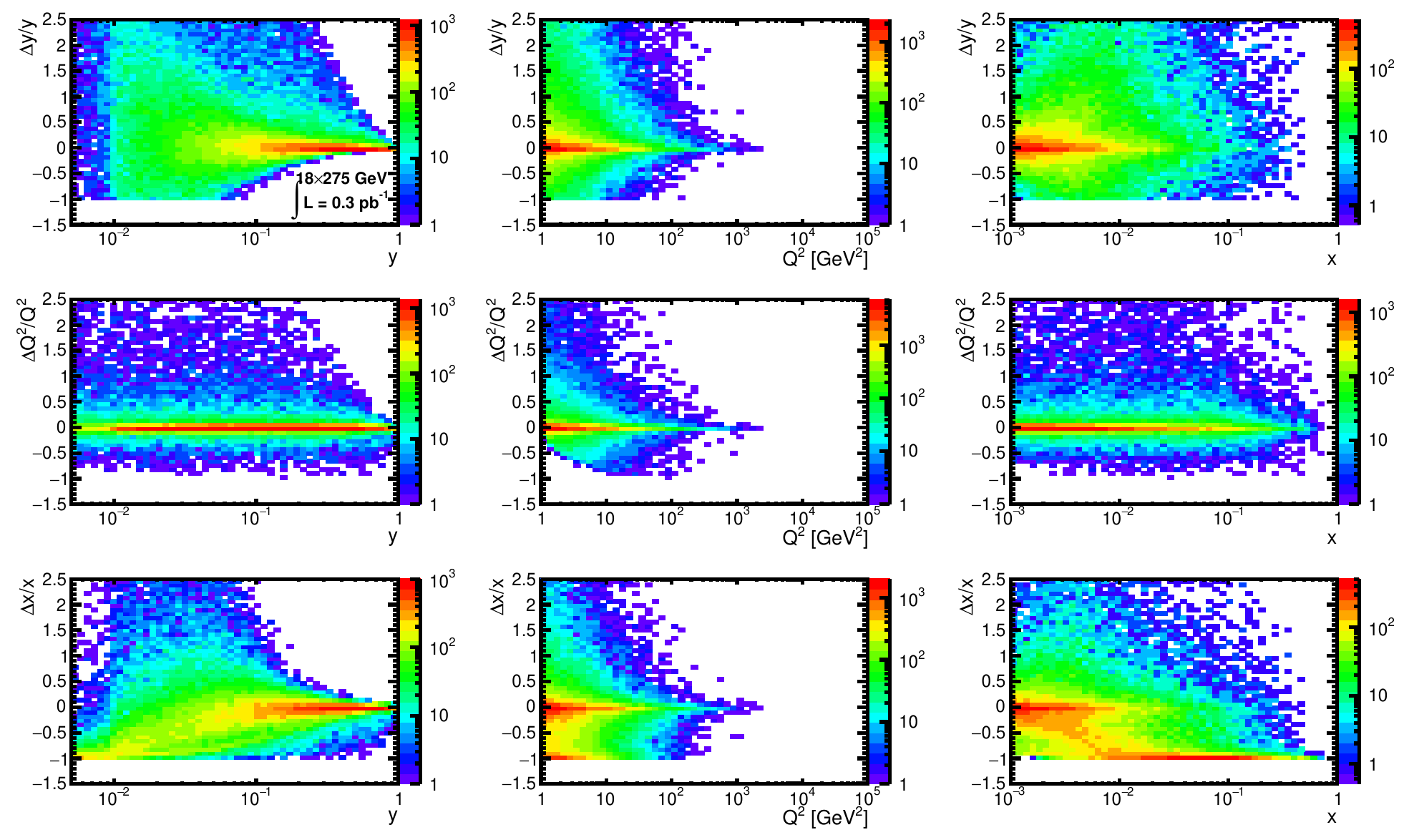}
\caption{Resolutions, defined as (reconstructed - true)/true, for kinematic variables in NC 18x275 GeV events. The ineleasticity is require to be $y < 0.95$}
\label{fig:eRecoReso_18_275_noLowYcut}
\end{figure}
\begin{figure}[!ht]
\centering
\includegraphics[width=\linewidth]{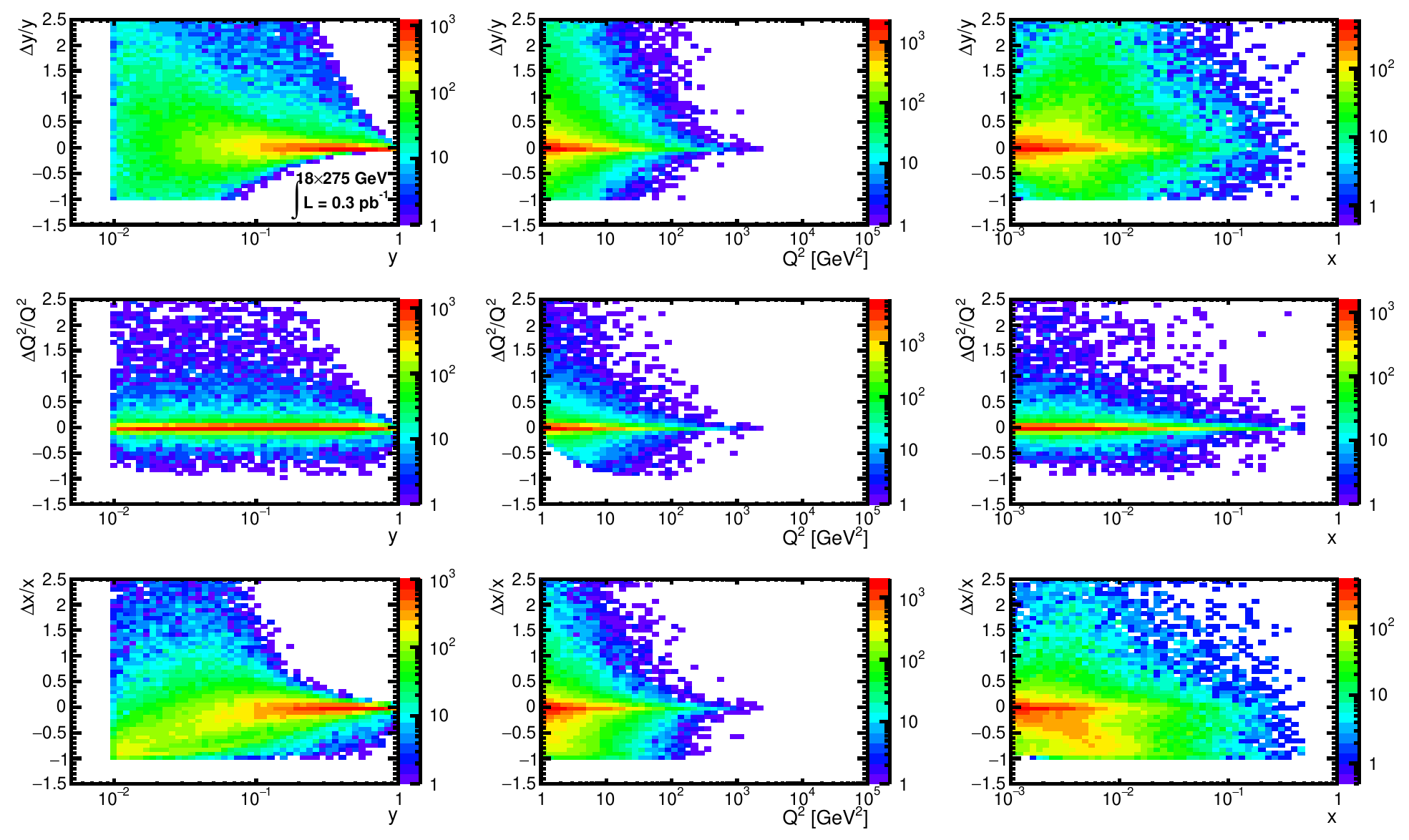}
\caption{Resolutions, defined as (reconstructed - true)/true, for kinematic variables in NC 18x275 GeV events. The inelasticity is required to be $ 0.01 < y < 0.95$}.
\label{fig:eRecoReso_18_275_lowYcut}
\end{figure}

The resolution $\Delta{x}/x$ and $\Delta{y}/y$ can be shown from Eqs.~\ref{eq:eReco} to diverge as $y\rightarrow0$. Indeed the plots in Figure \ref{fig:eRecoReso_18_275_noLowYcut} show that $\Delta{y}/y$ and $\Delta{x}/x$ diverge at small y. The resolution on the reconstruction of $x$ also develops a systematic offset with a subset of high $x$ events being reconstructed with very low $x$. This offset originates from the large positive fluctuations of $y$, due to the increasing poor resolution at low $y$, which then lead to the suppression in $x$. The plots in figure \ref{fig:eRecoReso_18_275_lowYcut} show that this offset effect is mitigated once a $y>0.01$ cut is applied. Similarly, as $y\rightarrow{1}$ the scattered electron energy and $\theta^{e'}_p$ become small. Figure \ref{fig:dEeta} shows that the resolution on the reconstructed energy degrades as the the scattering goes more forward, motivating a $y < 0.95$ cut as well.

\begin{figure}[htb]
\centering
\subfloat[NC 18x275 GeV]{\label{fig:Purity_NC_18_275}\includegraphics[width=0.45\linewidth]{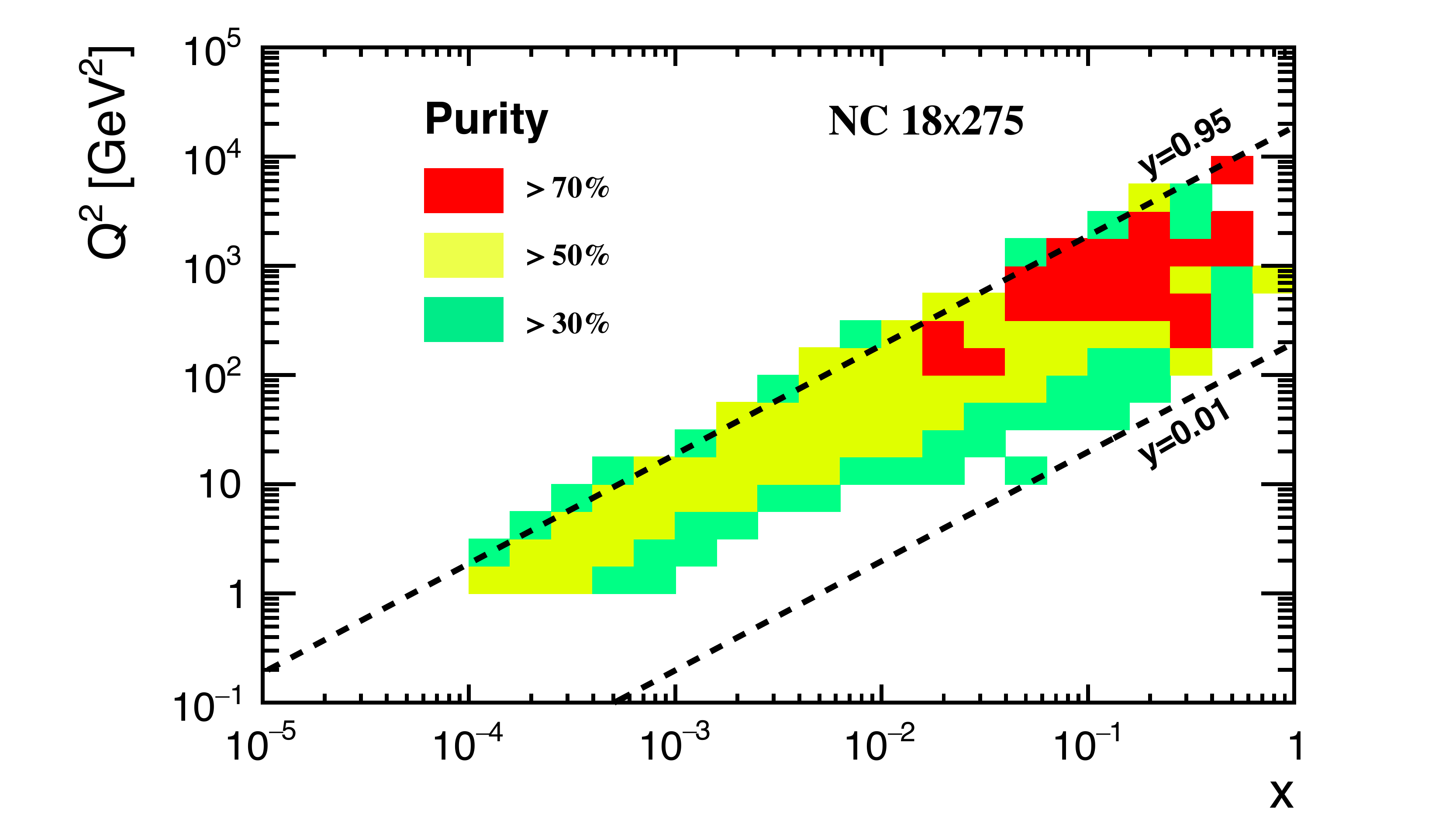}}\qquad%
\subfloat[NC 18x275 GeV]{\label{fig:Stability_18_275}\includegraphics[width=0.45\linewidth]{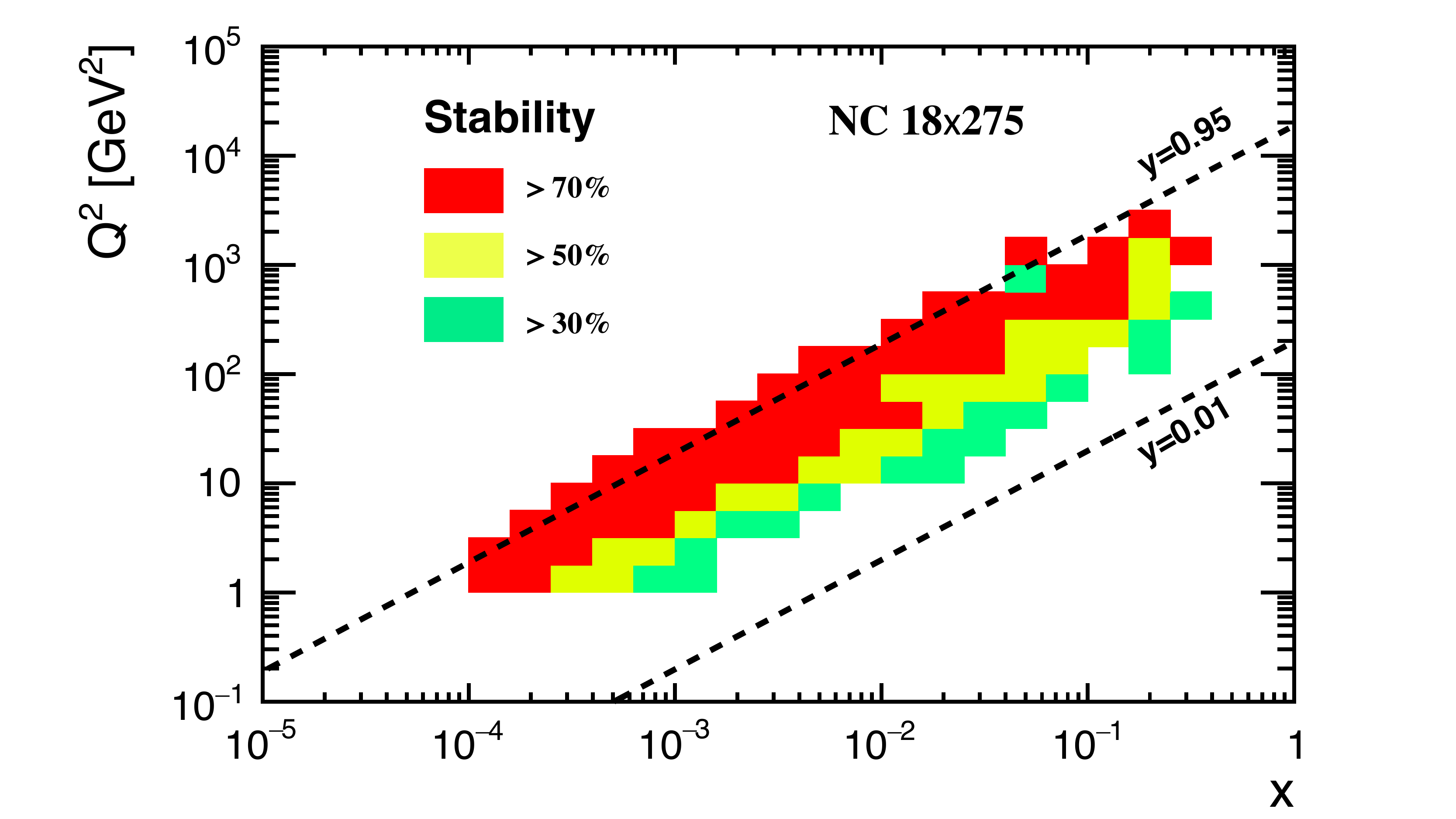}}\qquad%
\subfloat[NC 10x100 GeV]{\label{fig:Purity_NC_10_100}\includegraphics[width=0.45\linewidth]{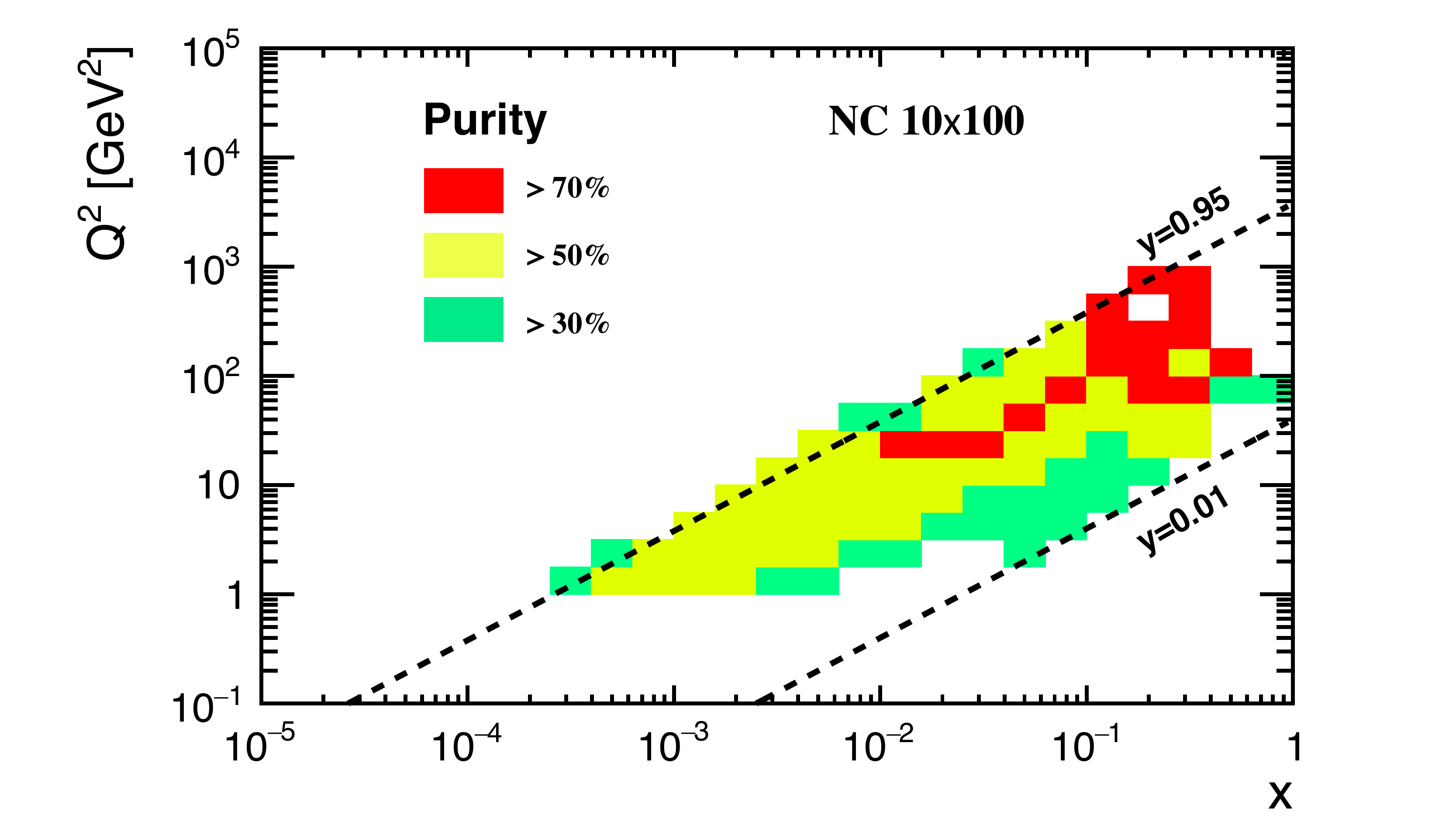}}\qquad%
\subfloat[NC 10x100 GeV]{\label{fig:Stability_10_100}\includegraphics[width=0.45\linewidth]{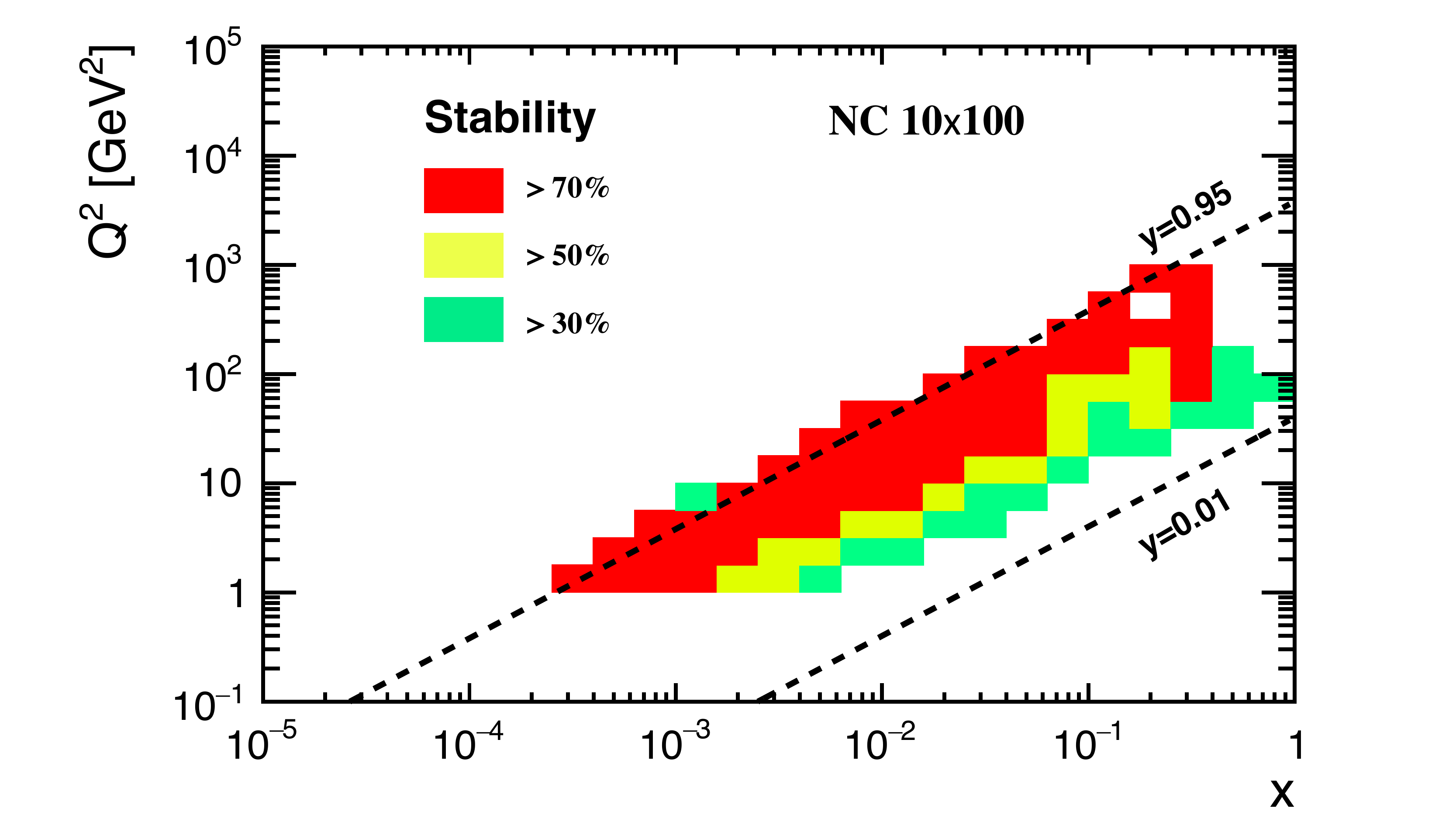}}\qquad%
\subfloat[NC 5x100 GeV]{\label{fig:Purity_NC_5_100}\includegraphics[width=0.45\linewidth]{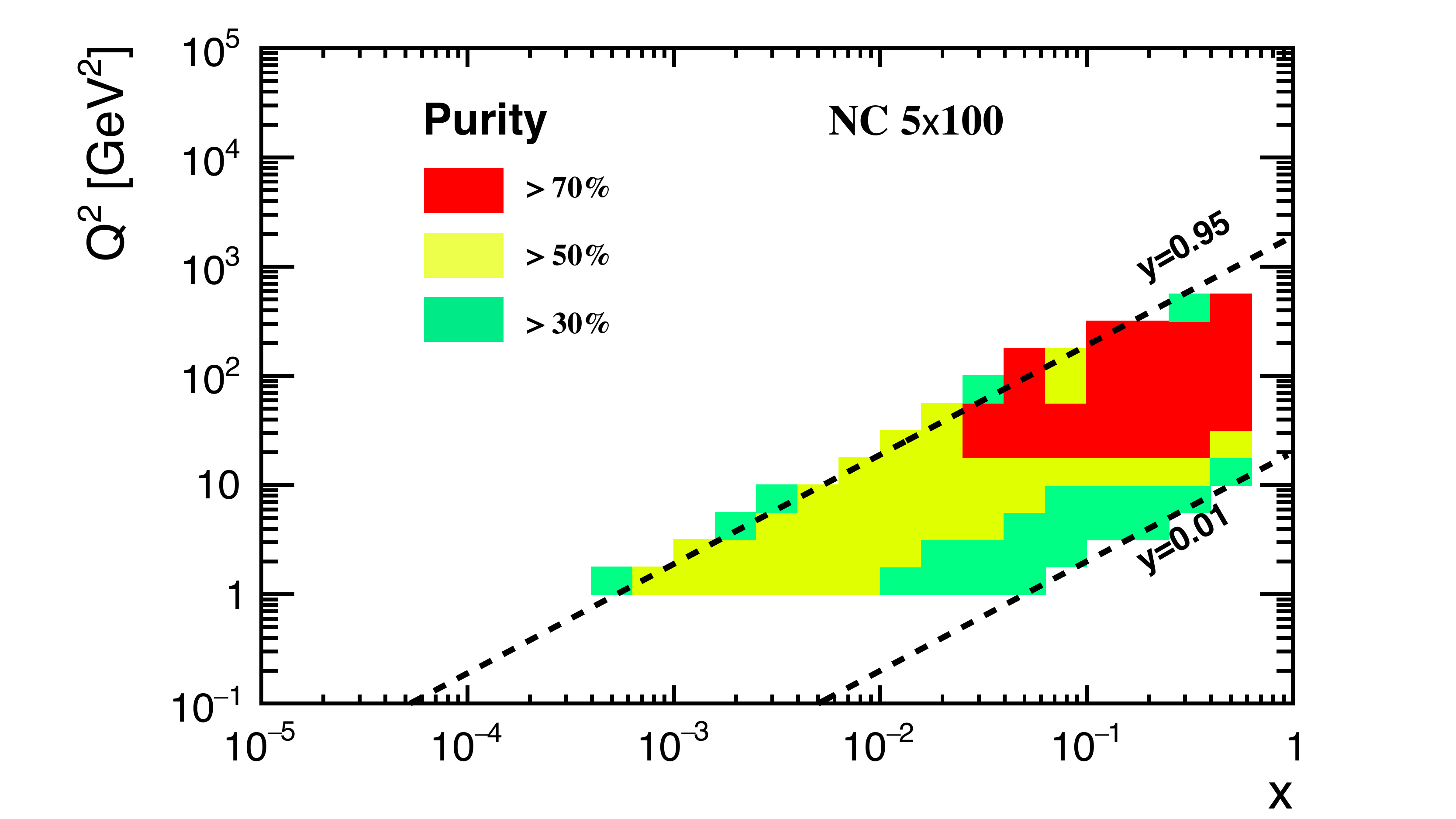}}\qquad%
\subfloat[NC 5x100 GeV]{\label{fig:Stability_5_100}\includegraphics[width=0.45\linewidth]{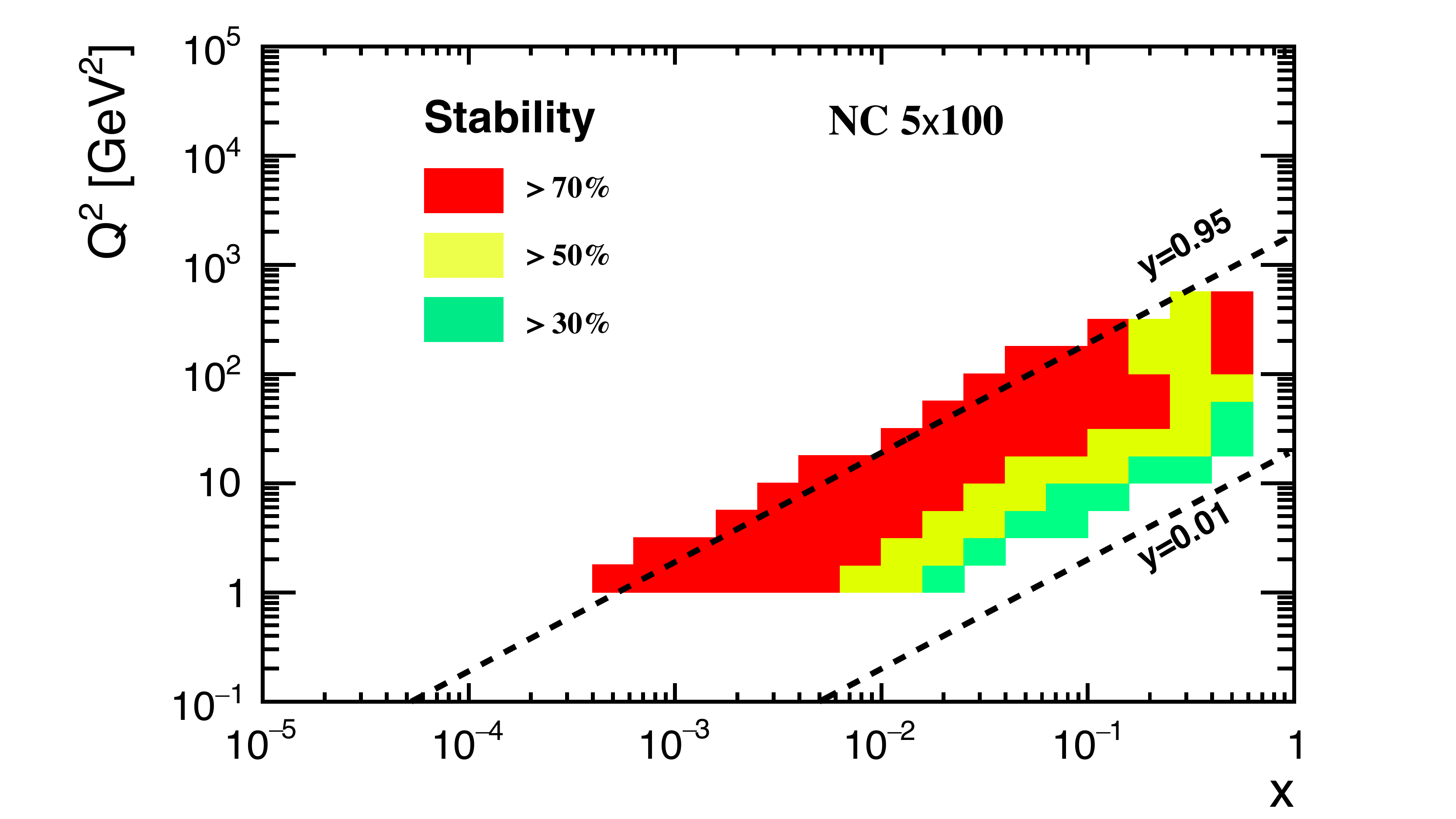}}\qquad%
\caption{The purity (left) and stability (right) for 18x275 GeV (top), 10x100 GeV (middle) and 5x100 GeV (bottom) beam configurations.}
\end{figure}

With the appropriate inelasticity cuts established it is important to evaluate the purity and stability for the detector resolutions proposed in the detector matrix. The purity is defined as the fraction of events reconstructed in a given bin that were generated in the same bin. It reflects the bin migration into a reconstructed kinematic bin $(x_R, y_R, Q^2_R)$.  The stability is defined as the fraction of events generated in a given bin that were reconstructed in the same bin. It reflects the migration of events outside of a generated kinematic bin $(x_G, y_G, Q^2_G)$. For a given detector configuration, the $x$ and $Q^2$ binning should be optimized in order to maximize both purity and stability and therefore minimize the size and systematic errors associated with kinematic corrections. In an effort to test the proposed calorimeter resolutions the process was reversed, with the binning first chosen to be five bins in $x$ and 4 bins in $Q^2$ per decade. Figures  \ref{fig:Purity_NC_18_275} - \ref{fig:Stability_18_275} show that the purity and stability are both well above $30\%$ for all bins, the typical lower limit deemed acceptable in HERA analyses. The conclusion is that, with the $y > 0.01$ cut applied, the current detector resolutions are sufficient. 

\clearpage

\subsection{Acceptance, resolution, and bin migration effects in Jacquet-Blondel reconstruction}
\label{sec:J-B_rec}
The resolution on the reconstruction of $x$, $Q^2$ and $y$ via hadronic reconstruction in either CC or NC events was evaluated by passing 10 $fb^{-1}$ of 18x275 GeV NC and CC pseudo-data through the EICSmear fast simulation package. The pseudo-data was produced using the DJANGOH event generator with full radiative effects turned on. In addition to the EICSmear settings described in the previous section, the barrel hadronic calorimeter resolution was set at $85\%\oplus7\%$ and the back/forward hadronic calorimeters at $45\%\oplus6\%$. Here the $\oplus$ symbol represents a quadrature sum.
\clearpage

\begin{wrapfigure}{R}{0.5\linewidth}
\centering{} 
\vspace{-0.4cm}
\includegraphics[width=0.95\linewidth]{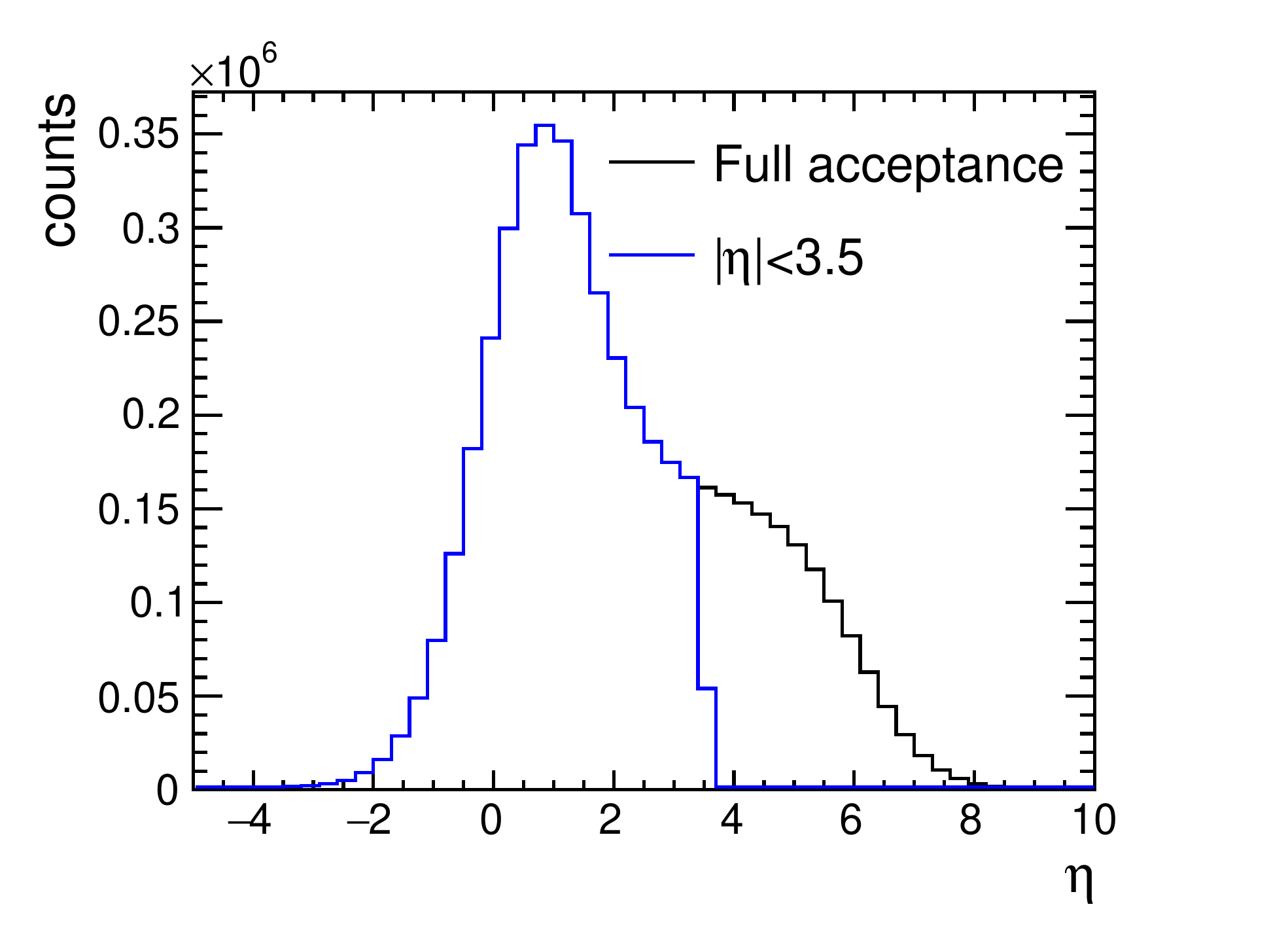}
\caption{The $\eta$ distribution of all final state hadrons and photons with (blue) and without (black) the nominal acceptance cut for 18x275 GeV beam configuration.}
\label{fig:CC_beyondEta_3.5}
\end{wrapfigure}
\begin{figure}[!ht]
\centering
\includegraphics[width=0.95\linewidth]{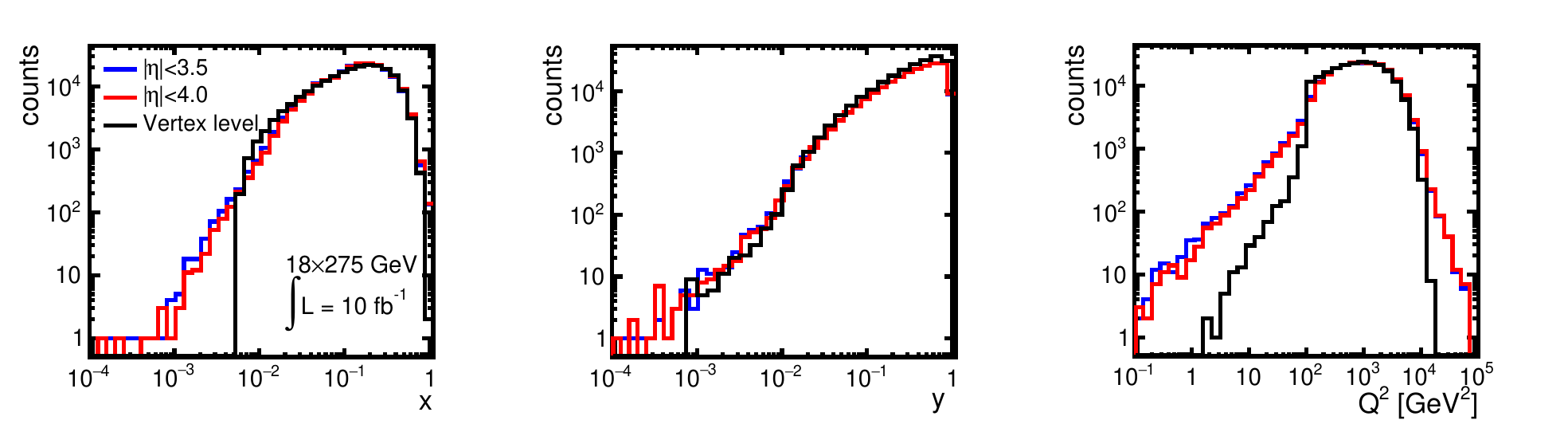}
\caption{JB reconstruction of $x$, $y$ and $Q^2$ for the vertex level (black), nominal (blue) and expanded (red) reconstruction.}
\label{fig:CC_x_Q2_y_reco}
\end{figure}

As demonstrated in Figures \ref{fig:PS_CC_photons} and \ref{fig:PS_CC_hadrons} the hadronic recoil of CC and NC events extends past the proposed nominal coverage of $-3.5<\eta<3.5$. The accuracy of the JB reconstruction method relies, in part, on the fraction of particles that are reconstructed in the detector acceptance. The blue and black curves in Figure \ref{fig:CC_beyondEta_3.5} show the  distribution of the final state kaons, pions, neutrons, protons and photons, with and without the acceptance cut applied. Approximately $30\%$ of all final state particles fall outside the $-3.5 < \eta < 3.5$ acceptance. Despite losing nearly a third of the final state particles in the CC event, the changes in the reconstructed $x$, $y$, and $Q^2$ are minimal. Figure \ref{fig:CC_x_Q2_y_reco}  shows the kinematic variables reconstructed at the true, or vertex level, compared with the reconstructed variables with nominal $( -3.5 < \eta < 3.5)$ or expanded $( -4.0 < \eta < 4.0)$ acceptance. The difference between the nominal and expanded reconstruction is negligible for $x$, $y$ and $Q^2$. 

\begin{figure}[htbp]
\centering
\subfloat[Purity High Resolution]{\label{fig:Purity_JB_res40}\includegraphics[width=0.45\linewidth]{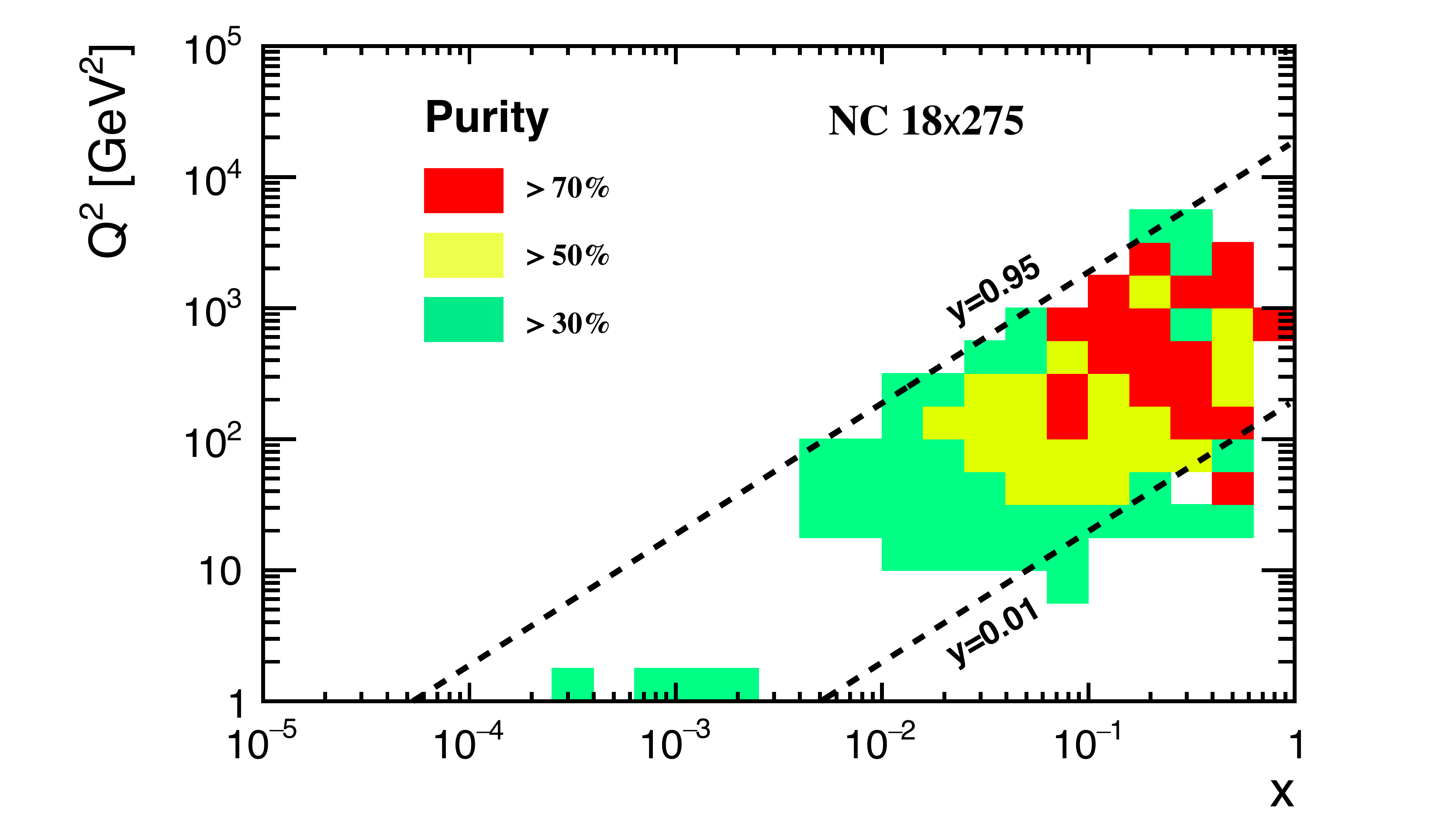}}\qquad%
\subfloat[Stability High Resolution]{\label{fig:Stability_JB_res40}\includegraphics[width=0.45\linewidth]{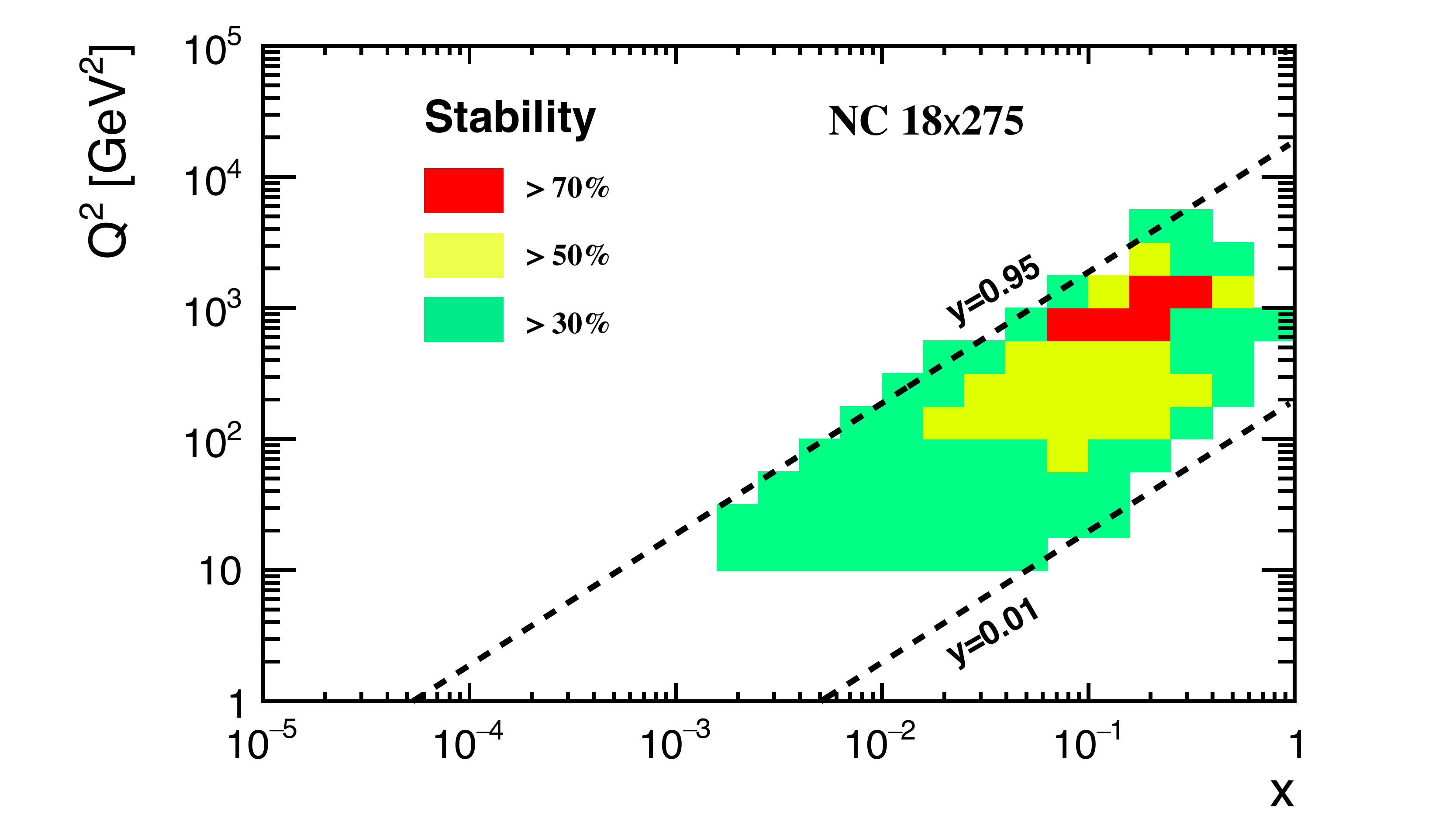}}\qquad%

\subfloat[Purity Mid Resolution]{\label{fig:Purity_JB_res45}\includegraphics[width=0.45\linewidth]{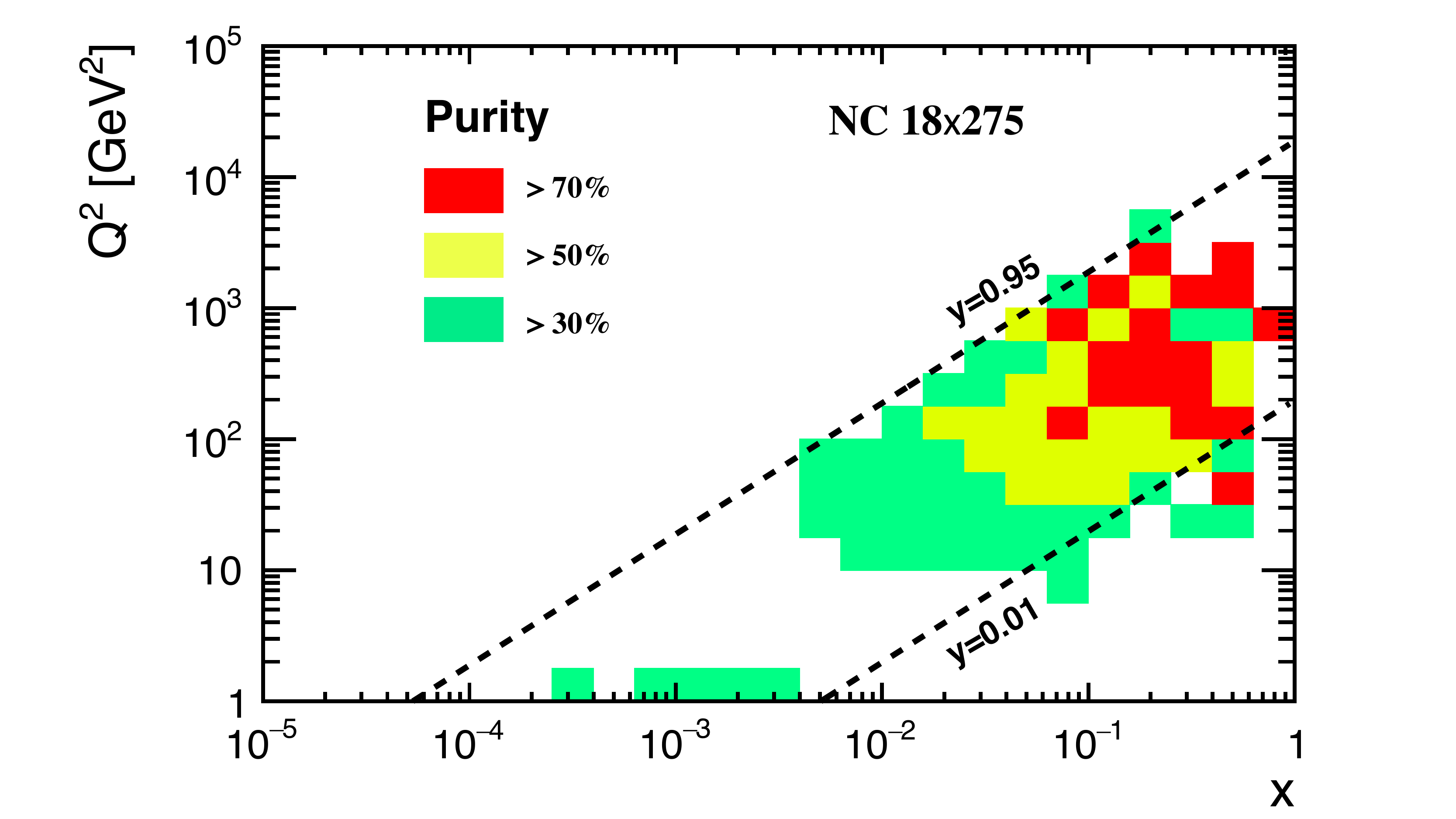}}\qquad%
\subfloat[Stability Mid Resolution]{\label{fig:Stability_JB_res45}\includegraphics[width=0.45\linewidth]{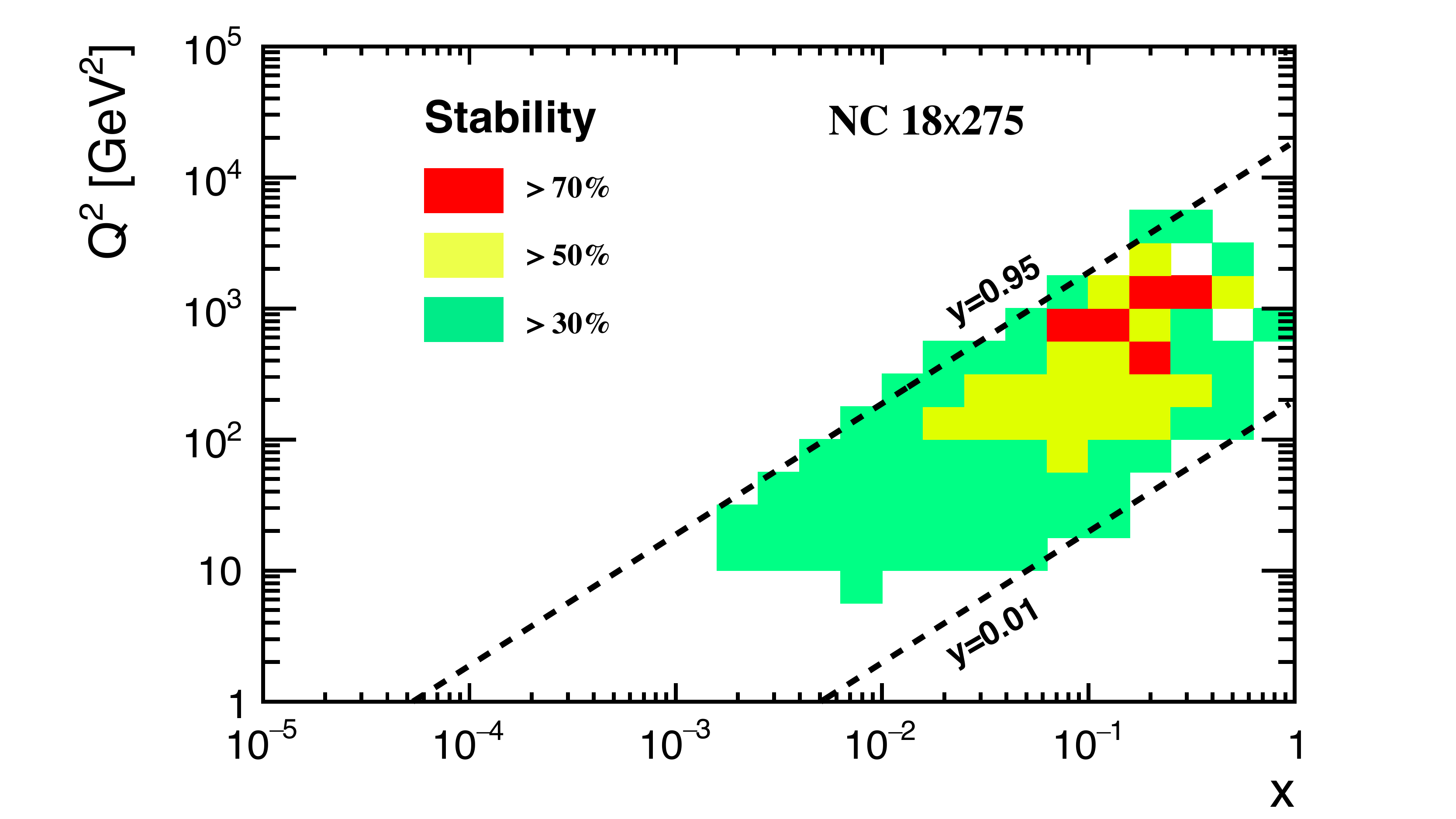}}\qquad%

\subfloat[Purity Low Resolution]{\label{fig:Purity_JB_res50}\includegraphics[width=0.45\linewidth]{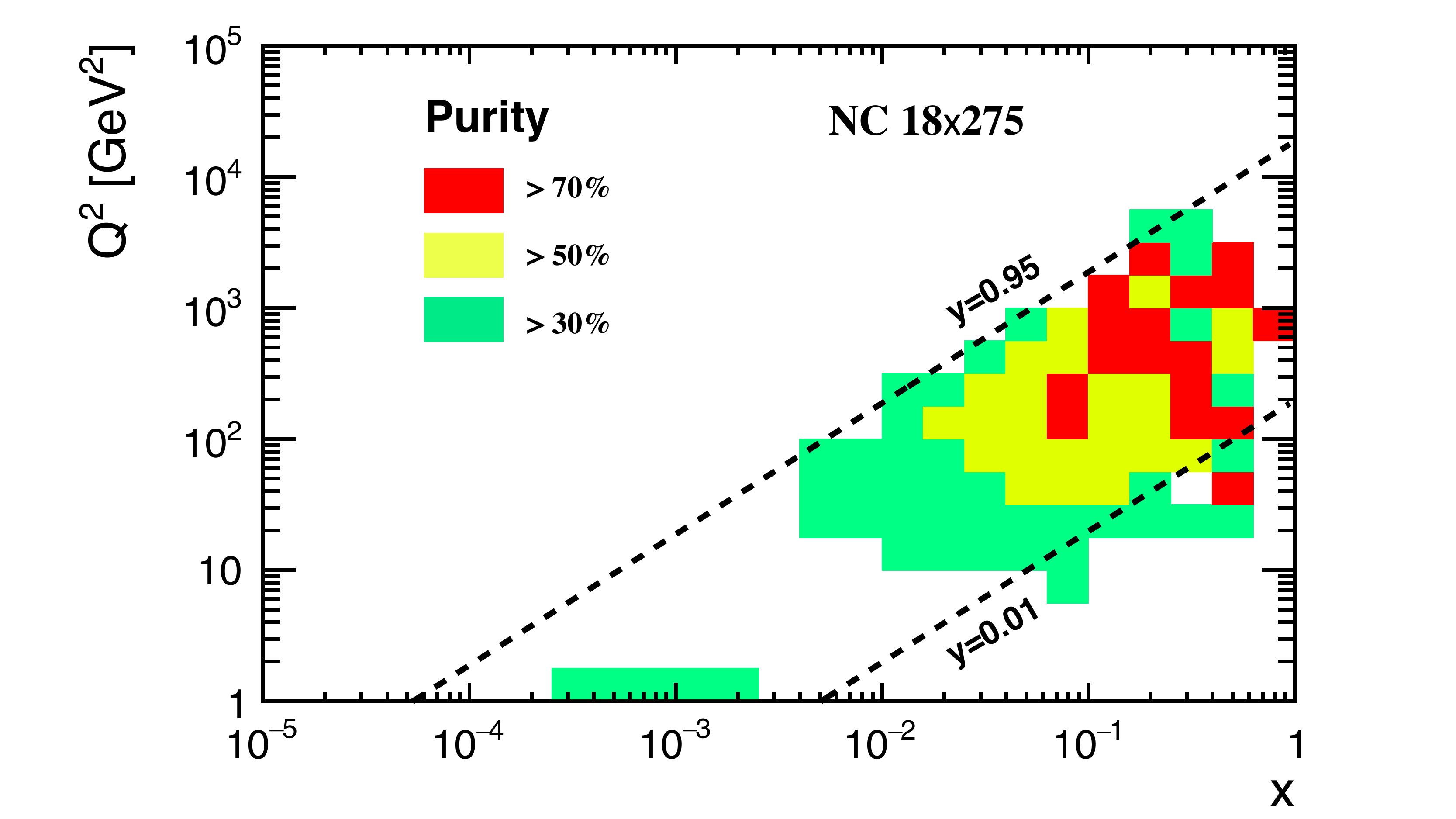}}\qquad%
\subfloat[Stability Low Resolution]{\label{fig:Stability_JB_res50}\includegraphics[width=0.45\linewidth]{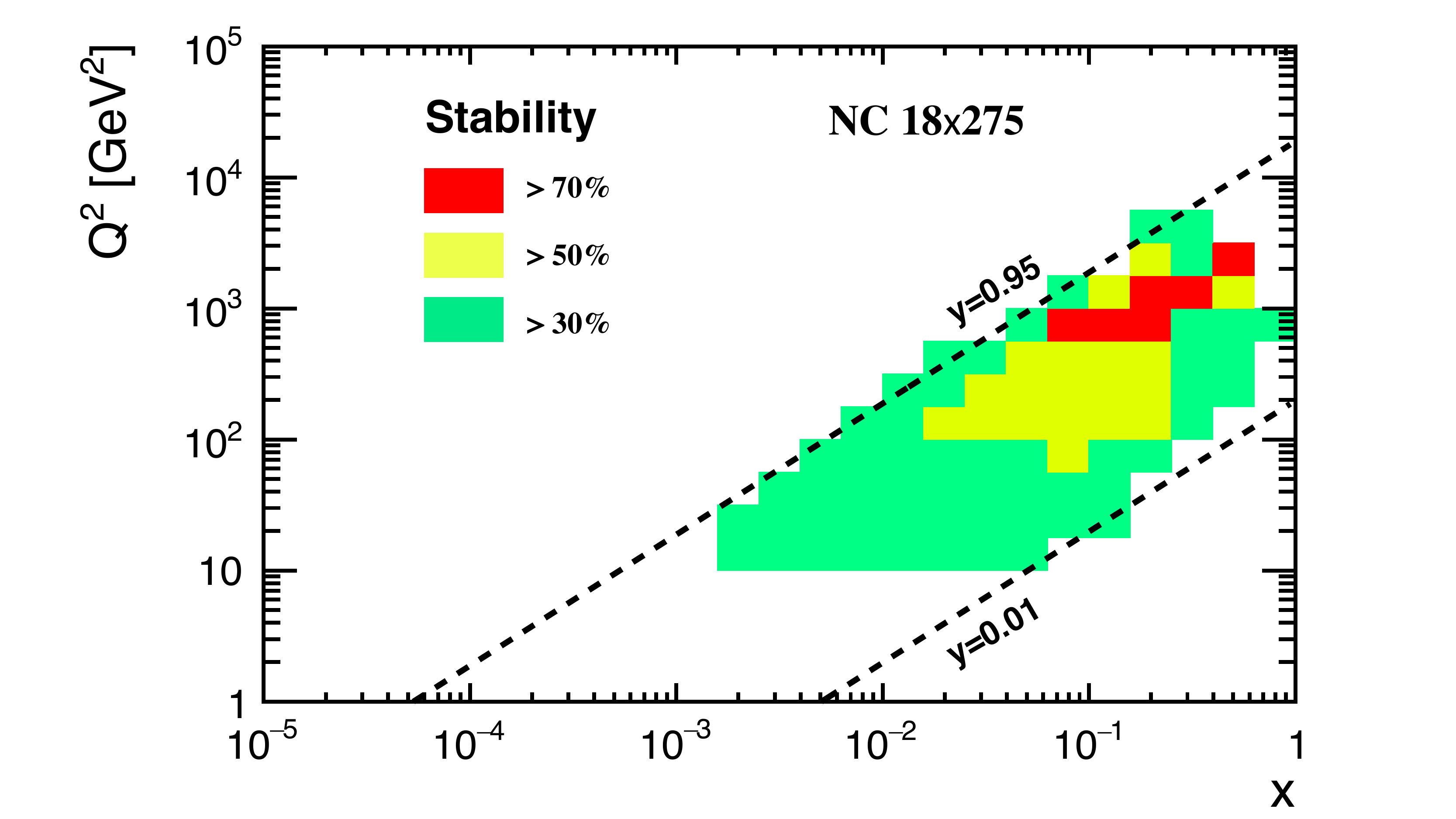}}\qquad%

\caption{The purity (left) and stability (right) for JB reconstruction of NC events in 18x275 GeV collisions. The resolution of the back/forward hadronic calorimeter is varied from $40\% \oplus 5\%$ (top), $45\% \oplus 6\%$ (middle) to $50\% \oplus 10\%$ (bottom). The resolution of the barrel hadronic calorimeter is fixed at $85\% \oplus 6\%$.}
\end{figure}

As in the case of the electron reconstruction, it is important to investigate bin migration effects in JB reconstruction. Specifically, the detector group requested an investigation into several resolution performances for the hadronic calorimeter in the forward region.  Figures \ref{fig:Purity_JB_res40} - \ref{fig:Stability_JB_res50} show the purity and stability for $40\% \oplus 5\%$, $45\% \oplus 6\%$ and $50\% \oplus 10\%$ resolution. In all cases the stability and purity are worse than in the electron reconstruction case, but the differences are not large, suggesting the bin migration effects not driven by differences at this level in the hadronic calorimeter resolution.

\subsection{Generator verification}
The PYTHIA6 and DJANGOH Monte-Carlo event generators were used extensively to determine the above detector requirements, as well as provide estimates of the statistical and systematic uncertainties used in the inclusive pseudo-data. For these reasons, it was necessary to validate that the Monte-Carlo generators give reasonable results when compared to data and theory calculations.

\begin{figure}[htbp]
\centering
\subfloat[e+p HERA: Lower $Q^2$]{\label{fig:ep_HERA_sim_low}\includegraphics[page=1, width=0.85\linewidth]{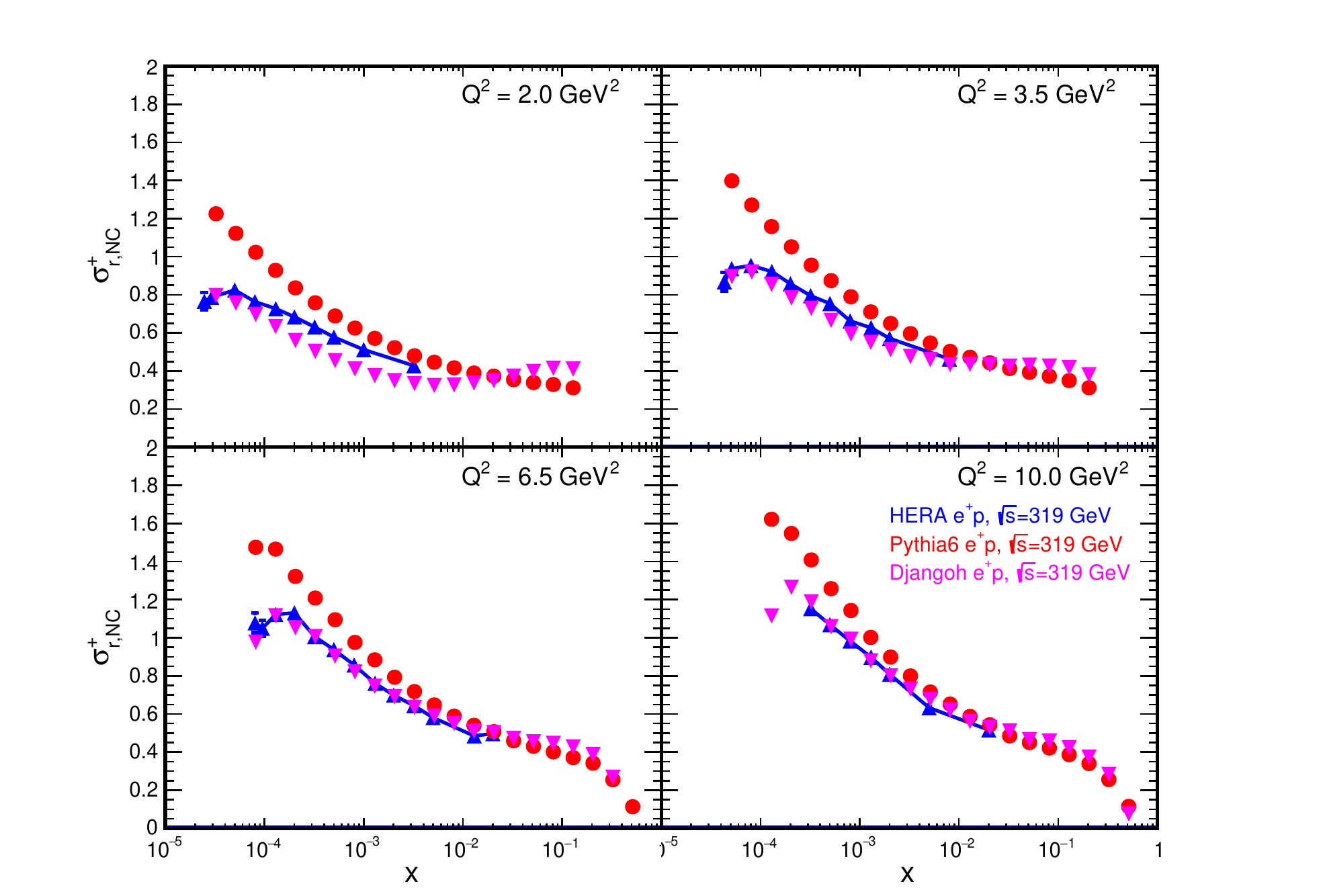}}\qquad%
\subfloat[e+p HERA: Higher $Q^2$]{\label{fig:ep_HERA_sim_hi}\includegraphics[page=2, width=0.85\linewidth]{PART2/Figures.DetRequirements/Inclusive/hera_compare.pdf}}\qquad%
\caption{Comparison of PYTHIA6 and DJANGOH Monte-Carlo event generators at HERA energy with the EIC tune to HERA NC inclusive cross section measurements. The simulation cross sections are calculated from the generators at the vertex level with QED radiative effects turned OFF. Approximately 10 $pb^{-1}$ of pseudo-data was created with each generator. }
\label{fig:Hera_sim}
\end{figure}

\begin{figure}[htbp]
\centering
\subfloat[e+p HERA: Lower $Q^2$]{\label{fig:ep_HERA_fits_low}\includegraphics[page=1, width=0.85\linewidth]{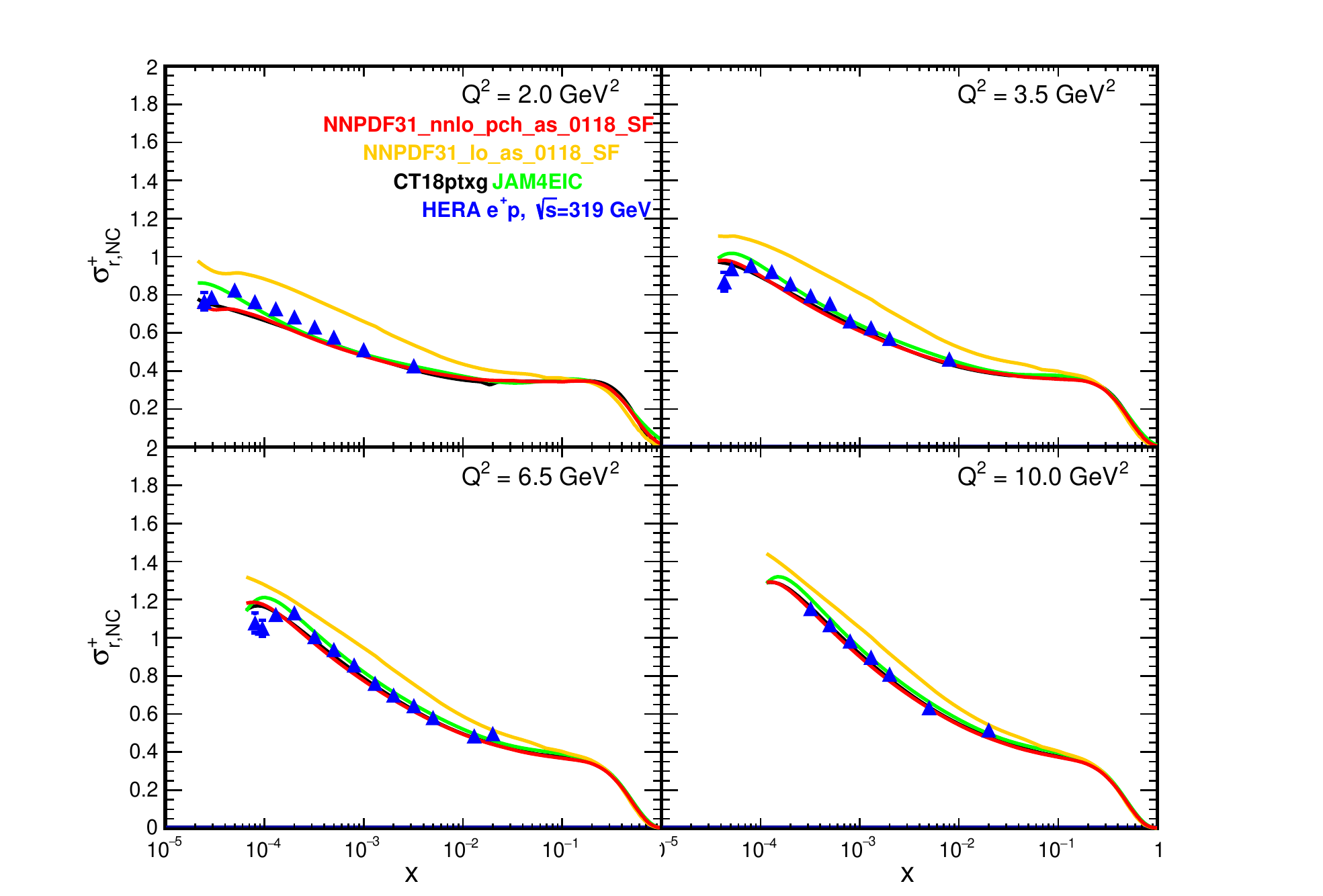}}\qquad%
\subfloat[e+p HERA: Higher $Q^2$]{\label{fig:ep_HERA_fits_hi}\includegraphics[page=2, width=0.85\linewidth]{PART2/Figures.DetRequirements/Inclusive/hera_theory.pdf}}\qquad%
\caption{Comparison of modern PDF fit results to HERA NC inclusive cross section measurements. The cross section is calculated at LO for the orange curve; at NLO for the green curve; and NNLO for the red and black curves.}
\label{fig:Hera_fits}
\end{figure}

Comparisons of the reduced inclusive positron-proton NC cross section for both event generators to HERA data are shown in figure~\ref{fig:Hera_sim}. The DJANGOH simulation was performed using the cteq61.LHgrid (10150) LHAPDF5 grid as a PDF input, while the PYTHIA6 simulation was performed using the cteq6ll.LH pdf (10042) LHAPDF5 grid. As can be seen in the figure, the DJANGOH simulation agrees better with the HERA data at lower $Q^2$, and both simulation programs agree well with the data at higher $Q^2$. In addition, modern PDF fits at LO, NLO, and NNLO are compared to the HERA data in figure~\ref{fig:Hera_fits}. These theory calculations then served as the baseline for validating the Monte-Carlo simulations at \textit{EIC} energies. 

\begin{figure}[htbp]
\centering
\subfloat[e-p 18x275 GeV: Lower $Q^2$, \textit{cteq6l1} PDF set used for PYTHIA6]{\label{fig:ep_18275_fits_lo1}\includegraphics[page=1, width=0.85\linewidth]{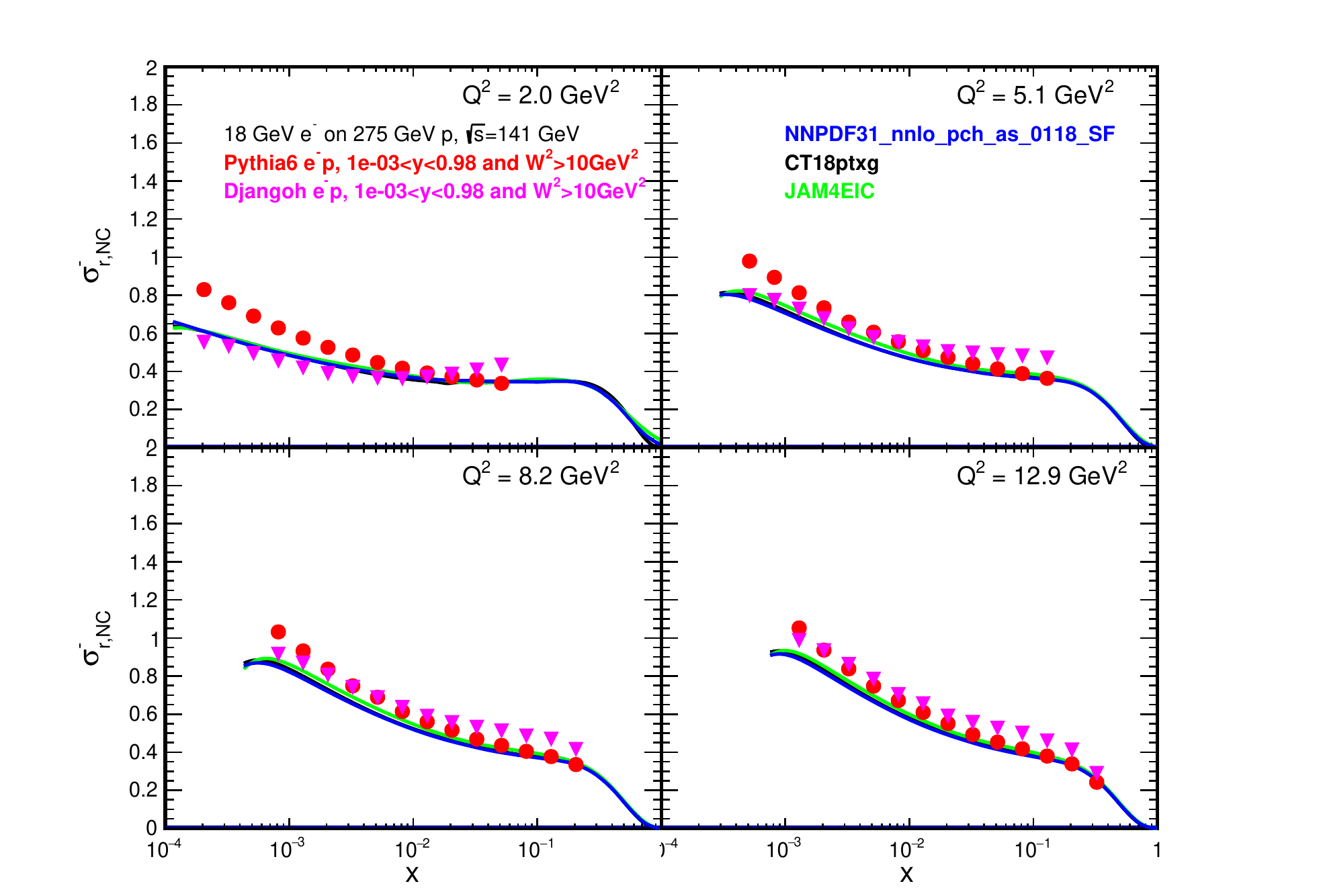}}\qquad%
\subfloat[e-p 18x275 GeV: Higher $Q^2$, \textit{cteq6l1} PDF set used for PYTHIA6]{\label{fig:ep_18275_fits_hi1}\includegraphics[page=2, width=0.85\linewidth]{PART2/Figures.DetRequirements/Inclusive/cs_compare_18275_1.pdf}}\qquad%
\caption{Comparison of PYTHIA6 (red) and DJANGOH (pink) reduced cross-sections for $e^-+p$ scattering at 18x275 to NNPDF\cite{Ball:2017nwa}\cite{Faura:2020oom}, CT18\cite{Hou:2019efy} and JAM\cite{cocuzza20} global fits. The \textit{cteq6l1} PDF was used for PYTHIA6, while the \textit{cteq61} PDFs were used for the DJANGOH simulations. Approximately 10 $pb^{-1}$ of pseudo-data was created with each generator. }
\label{fig:18275_sim_fits_A}
\end{figure}

\begin{figure}[htbp]
\centering
\subfloat[e-p 18x275 GeV: Lower $Q^2$, \textit{cteq61} PDF set used for PYTHIA6]{\label{fig:ep_18275_fits_lo2}\includegraphics[page=1, width=0.85\linewidth]{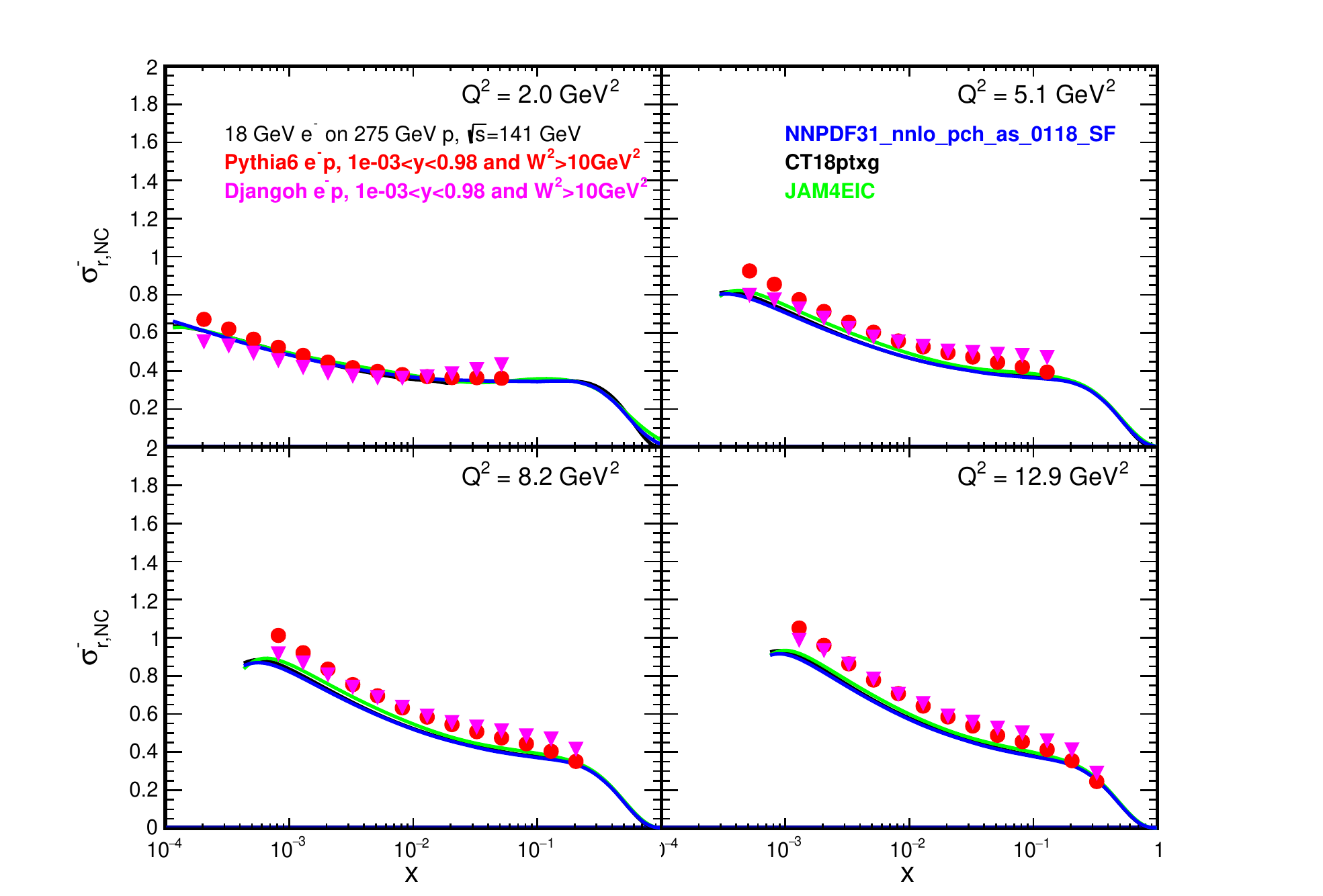}}\qquad%
\subfloat[e-p 18x275 GeV: Higher $Q^2$, \textit{cteq61} PDF set used for PYTHIA6]{\label{fig:ep_18275_fits_hi2}\includegraphics[page=2, width=0.85\linewidth]{PART2/Figures.DetRequirements/Inclusive/cs_compare_18275_2.pdf}}\qquad%
\caption{Comparison of PYTHIA6 (red circles) and DJANGOH (pink triangles) reduced cross-sections for $e^-+p$ scattering at 18x275 to NNPDF\cite{Ball:2017nwa}\cite{Faura:2020oom}, CT18\cite{Hou:2019efy} and JAM\cite{cocuzza20} global fits. The \textit{cteq61} PDF was used for both PYTHIA6 and DJANGOH simulations. Approximately 10 $pb^{-1}$ of pseudo-data was created with each generator. }
\label{fig:18275_sim_fits_B}
\end{figure}

Comparisons of the electron-proton NC reduced cross sections at the EIC energy setting of 18x275 GeV extracted from the Monte-Carlo simulations to those calculated from the PDF fits are shown in Figures ~\ref{fig:18275_sim_fits_A} and ~\ref{fig:18275_sim_fits_B}. Results are similar for the other studied beam energy configurations -- 5x41 GeV, 5x100 GeV, and 10x100 GeV. The binning chosen is equivalent to the one shown in figure~\ref{fig:xQ2_18_275}. The simulation events have been corrected for bin-centering effects (which are on the order of 5-8\% for the chosen binning) using a cross section model in order to quote the reduced cross sections at the center of each $x$-$Q^2$ bin.

In figures~\ref{fig:ep_18275_fits_lo1} and \ref{fig:ep_18275_fits_hi1}, the PYTHIA6 simulation was performed using the \textit{cteq6ll} PDF set; while in figures~\ref{fig:ep_18275_fits_lo2} and \ref{fig:ep_18275_fits_hi2}, the PYTHIA6 simulation was run using the \textit{cteq61} PDF set. In both cases, the DJANGOH simulation was run using the \textit{cteq61} PDF set. As can be seen, the simulation performed using the \textit{cteq61} PDF set more accurately reproduces the low $x$, low $Q^2$ fit cross sections results; and the simulation performed using the \textit{cteq6ll} PDF set more accurately reproduces the higher $x$, higher $Q^2$ fit cross sections results. The simulation results agree with the fit results to the 10\% level over most of the kinematic phase space.

\begin{figure}[htbp]
\centering
\includegraphics[width=0.95\linewidth]{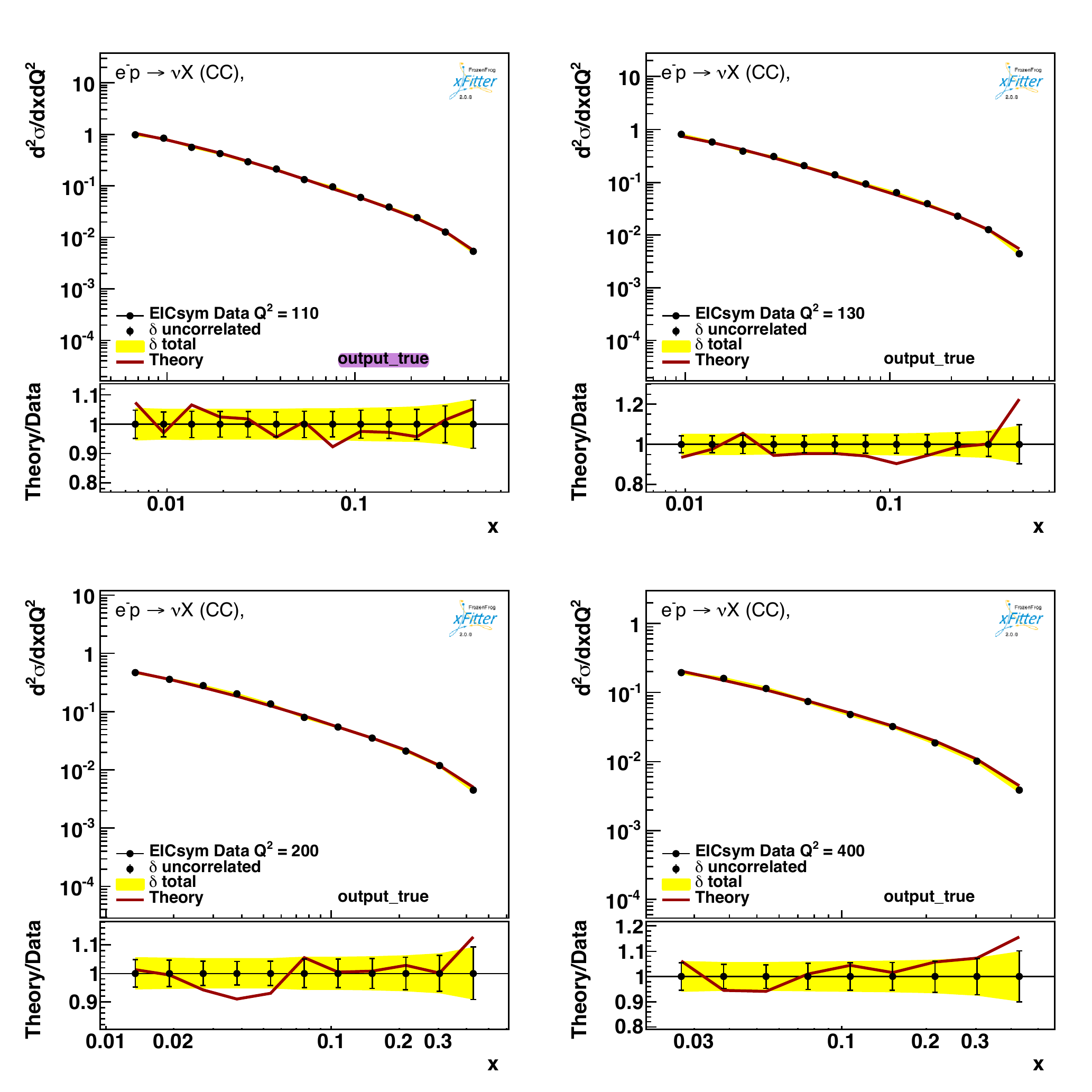}
\caption{Comparison of CC differential cross sections reconstructed from 10 $fb^{-1}$ of pseudo-data created with DJANGOH and NLO theory curves from xFitter\cite{Alekhin:2014irh}.}
\label{fig:CC_xsection}
\end{figure}

Figure \ref{fig:CC_xsection} shows excellent agreement, for Q$^2$ of 110, 130, 200 and 400 $GeV^2$, between the differential charged current cross-sections reconstructed at the vertex level with NLO theory curves produced with xFitter. The 10 $fb^{-1}$ of CC pseudo data was simulated with the DJANGOH generator. 

\subsubsection{Systematic uncertainties for the inclusive pseudo-data} \label{systematics_incl}
As the uncertainties on the NC inclusive cross section measurements at the EIC will be dominated by systematic errors for much of the probed kinematic phase space, it is necessary to make estimates of those errors for the generated pseudo-data. For the CC inclusive cross section measurements the systematic uncertainty will most likely be at a similar level as the statistical uncertainty for much of the measured kinematic phase space. Although it is very difficult to determine systematic uncertainties for an accelerator and detector which have not yet been constructed, estimates of these uncertainties can be made based on the experience of previous experiments (primarily those at HERA) as well as simulation studies using the EIC Handbook detector and the current EIC detector matrix.

The systematic uncertainties on the pseudo-data were divided into uncorrelated uncertainties (i.e. point-to-point or bin-by-bin uncertainties) and scale uncertainties (i.e. normalization uncertainties). No attempt was made to estimate partially correlated systematic uncertainties. Two sets of systematic uncertainties were constructed: an optimistic set and a conservative set.

For the unpolarized NC electron(positron)-proton cross section measurements, the estimate of the uncorrelated uncertainty was 1.5\% (2.3\%) in the optimistic (conservative) scenario. These uncertainties came from a 1\% uncertainty on the radiative corrections; and a 1-2\% uncertainty due to detector effects. An additional uncertainty of 2\% was added for the pseudo-data with y $<$ 0.01, as hadronic reconstruction methods are required in that kinematic region. The normalization uncertainty was set at 2.5\% (4.3\%) in the optimistic (conservative) scenario. This included a 1\% uncertainty on the integrated luminosity; and 2-4\% uncertainty due to detector effects. During the fits of the pseudo-data, the normalization uncertainty was treated as fully correlated between different beam energy settings.

The same uncertainties were used for the unpolarized NC electron-nucleus cross section measurements, with the exception that data at y $<$ 0.01 was not included for the e-A pseudo-data sets.

For the unpolarized CC electron(positron)-proton cross section measurements, an uncorrelated uncertainty of 2\% and a normalization uncertainty of 2.3/5.8\% was used for the optimistic/conservative scenarios of the pseudo-data. The point-to-point uncertainties came from a 2$\%$ on the background subtraction and a 0.5\% uncertainty on acceptance and bin-centering effects. The normalization uncertainties include optimistic/conservative contributions from luminosity, radiative corrections and simulation errors. 

The systematic errors for the asymmetry measurements, $A_{LL}$ and $A_{PV}$, vary depending on the analysis group and the channel. In Chapter~\ref{part2-sec-PartStruct}, Figures~\ref{fig:A_LL_p} and \ref{fig:A_LL_d_h} implemented a flat $2\%$ point-to-point systematic, while Figures~\ref{fig:APVe} and \ref{fig:APVhad} implemented a flat $1\%$ point-to-point error. Note that the $1\%$ systematic errors for the parity violating asymmetries are dominated by the pion background contribution and were based on estimates from previous fixed-target measurements that covered different kinematic regions. Studies discussed in Section \ref{sec:ePID} found this $1\%$ estimate to be optimistic for the majority of the EIC kinematics, especially in the regions $\eta>-2$. 

\subsection{Summary of the inclusive detector requirements}

In summary, the detector coverage and capabilities outlined in the EIC Handbook, and implemented into EICSmear, are sufficient for nearly all of the inclusive reaction channels. The only point of tension is the limited $e^-/\pi^-$ discrimination for regions of $\eta > -2$. The reduction in $e^-$ PID capabilities at mid-rapidity will result in $A_{PV}$ and $A_{LL}$ being systematically limited in this region.

\section{Semi-Inclusive Measurements}
\label{part2-sec-DetReq.SIDIS}
\subsection{General SIDIS kinematics and requirements}
\label{sec:generalSIDIS}
\begin{figure}[ht]
    \centering
    \includegraphics[width=\textwidth]{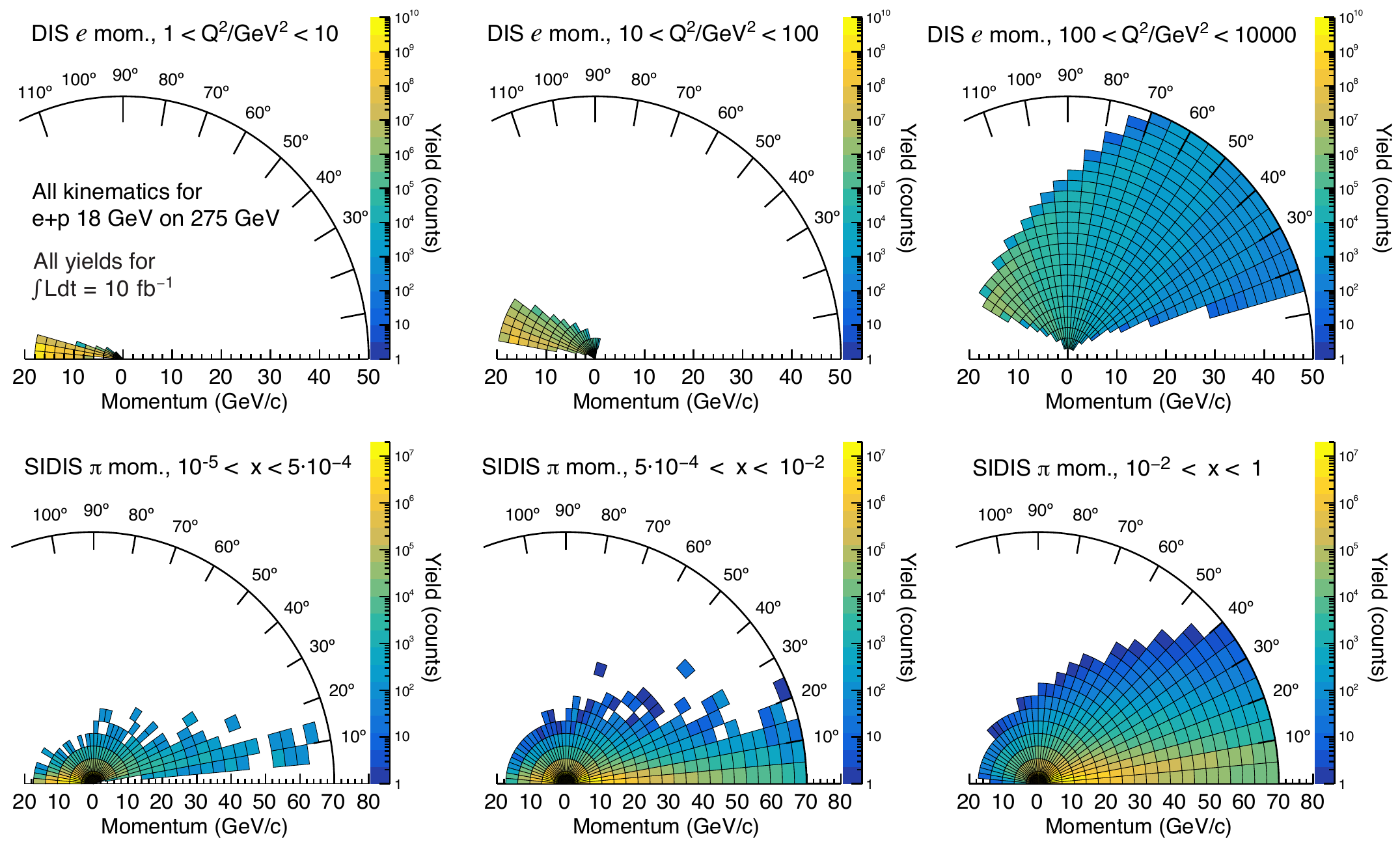}
    \caption{Top panel: Polar figures of the scattered DIS lepton momentum distributions for the highest collision energies for small to high momentum transfers from left to right. 
    Bottom panel: Polar figures of SIDIS hadron momentum distributions for the highest collision energies for small to high $x$ from left to right. All yields are extrapolated to 10 fb$^{-1}$ of accumulated luminosity.}
    \label{fig:part2-sec-DetReq.SIDIS.letponhadronkine}
\end{figure}

Semi-inclusive DIS (SIDIS) mostly uses the information of the hadronic final state to obtain additional information about the nucleon or nuclei with the help of fragmentation functions. Fragmentation functions can inform about the spin, momentum (transverse and longitudinal) and, in particular, flavor of an outgoing parton. As such, in addition to the main DIS kinematics as obtained by the scattered lepton or the total hadronic final state, one or several final-state hadrons have to be detected as well. 
For the applicability of perturbative QCD for the hard processes and factorization into non-perturbative distribution and fragmentation functions to be valid, the DIS kinematics require a $Q^2$ larger than 1 $\gev^2$. Together with a selecion of $0.01<y<0.95$ and $W^2>10 \gev^2$ they will be referred to as DIS cuts in the following.
For the final state hadrons, the fractional momentum $z=(P_{in}\cdot P_h)/(P_{in}\cdot q)$, where $P_{in}$ is the incoming nucleon's 4-momentum, $q$ is the momentum transfer and $P_h$ is the 4-momentum of the outgoing hadron, ranges typically between 0.1 and 1. Particularly higher $z$ values are of interest, since the correlation with the fragmenting parton's flavor, spin, etc is higher. Typically, hadrons that are part of the outgoing nucleon remnant (so-called target fragmentation) are found at low $z$. Such hadrons are not as important for the main SIDIS measurements. 
In addition to the $z$ of the hadron, also the type of detected hadrons needs to be identified as it is strongly correlated with the fragmenting flavor, particularly for higher $z$. 
These characteristics define the general range of scattered leptons and detected hadrons in terms of energy and rapidity. They are summarized in Fig.~\ref{fig:part2-sec-DetReq.SIDIS.letponhadronkine} for the highest collision energies. 
As can be seen, the scattered lepton rapidity and momentum moves with increasing $Q^2$ from backward rapidities and being bounded by the electron beam energy to forward rapidities and energies that are governed by the hadron beam energy. 
In contrast, the hadron kinematics are more closely correlated with the $x$ of the event. At low $x$ the hadron gets predominantly emitted into the backward region and the momentum follows that of the electron beam. With increasing $x$ the fragmenting hadrons move into the forward region and can take up substantial fractions of the hadron beam momentum.   

Within these rapidity and momentum ranges, a good tracking resolution is required not only for the hadrons but even more so for the scattered leptons. The reason for this lies in the fact that hadron transverse momenta and azimuthal angles, that are generally necessary for all TMD related measurements, are typically defined relative to the virtual photon axis in the frame where the incoming nucleon is at rest. The scattered lepton momenta provide the boost into that frame and can substantially distort those distributions. 

\paragraph{Requirements for the detector}
A homogeneous electron and hadron coverage from at least $-3<\eta<3$ is needed, preferably extended to $\pm3.5$. Good momentum resolution of scattered lepton and final-state hadrons is important in the correct reconstruction of SIDIS kinematics.

\subsection{Hadron PID impact on 4D TMD measurements}
\label{sidis_TMD}
As described in the previous part, the hadron momenta can reach up to 18 \gevc in the backward region ($-4<\eta<-1$), occasionally more than 10 \gevc  at central rapidities ($-1<\eta<1$), and more than 50 \gevc  at forward rapidities ($1<\eta<4$). For all SIDIS measurements that is also the maximum momentum for which the hadron type needs to be well identified. The lowest momenta of interest in all rapidities are of several hundred \mevc.

\begin{figure}[ht]
    \centering
    \includegraphics[width=0.9\textwidth]{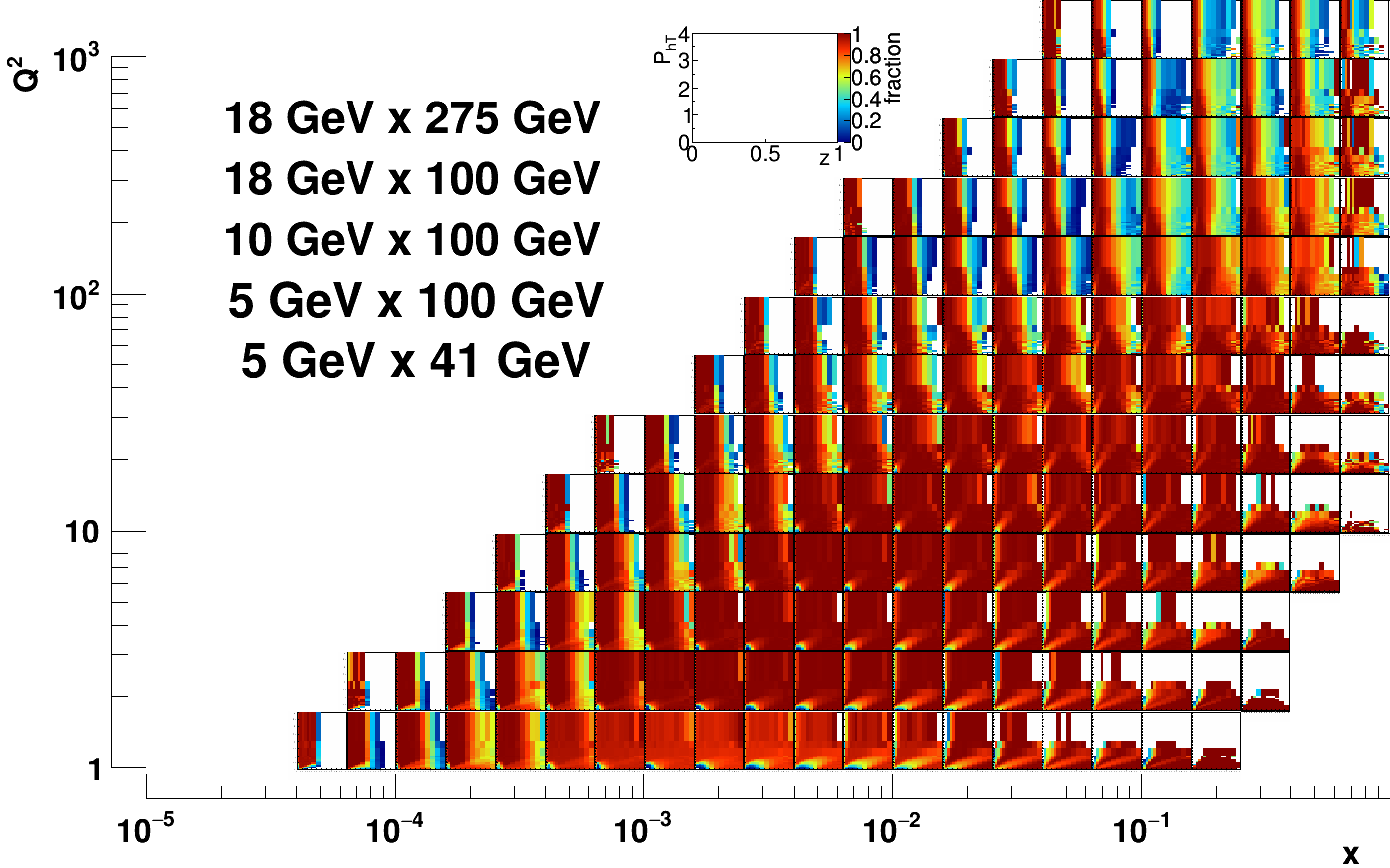}
    \caption{PID acceptance fractions as a function of pion fractional energy $z$ and transverse momentum $P_{hT}$ in bins of $x$ and $Q^2$, taking into account all collision energies. The fractions are evaluated by calculating the yield of accepted pions within the PID momentum ranges described in the text normalized by all pions.The standard DIS event selection criteria are applied.}
    \label{fig:part2-sec-DetReq.SIDIS.4dacc}
\end{figure}

As the high momenta are usually also related to high fractional momenta $z$, where the correlation to the fragmenting parton flavor and spin is largest, the high momentum hadron PID requirements are most important. 
To illustrate the relevance of the PID requirements, the impact of PID ranges are displayed as a function of $x$, $Q^2$, $z$ and $P_T$ assuming perfect tracking using a realistic PID range normalized by the hadron yields in these variables with perfect tracking and PID information. Losses in some regions, particularly at intermediate $x$ and $Q^2$, can be compensated between different beam energy combinations, but particularly high $x$ at low and high $Q^2$ can only be obtained by the lowest and highest beam energy combinations, respectively. It is however important to also keep in mind that in addition to the bins in the DIS kinematics the coverage of SIDIS variables $z$ and $P_T$ is important. 
The coverage is displayed in Fig.~\ref{fig:part2-sec-DetReq.SIDIS.4dacc} assuming PID coverage of up to 7 \gevc  from $-3.5<\eta<-1$, 6 \gevc  from $-1<\eta<1$ and 50 \gevc  from $1<\eta<3.5$ for pions after combining all beam energy options. For the most part this PID selection covers most kinematic regions, but the rather low maximum momentum in the barrel part of the detector significantly cuts into the intermediate-$x$, high-$Q^2$ and higher-$z$ range that would be particularly important for any TMD evolution studies. 

\begin{figure}[ht]
    \centering
    \includegraphics[width=0.9\textwidth]{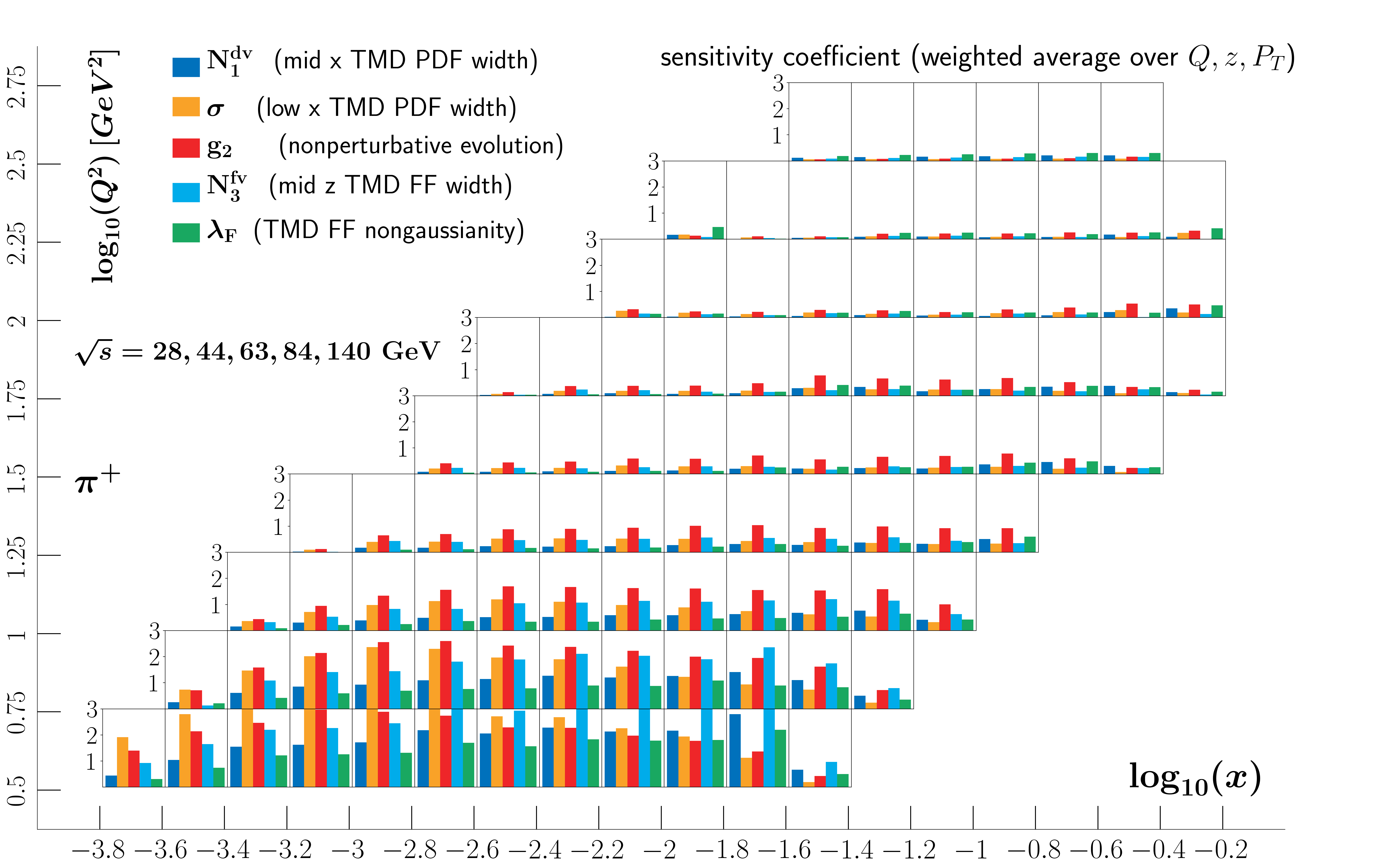}
    \caption{Expected sensitivities to various TMD PDF and FF parameters, as well as the TMD evolution as shown for the verious collision energy options and for detected final-state positive pions. The impact has been averaged over final state hadron transverse momentum and fractional energy for better visibility.}
    \label{fig:part2-sec-DetReq.SIDIS.4dsensitivity}
\end{figure}

In order to see why all kinematic regions in the previous figure are relevant, a simplified version of the expected sensitivities from the EIC data to the unpolarized TMDs are shown in Fig.~\ref{fig:part2-sec-DetReq.SIDIS.4dsensitivity} based on Ref.~\cite{Bacchetta:2019sam}.
The bars represent the overall impact to a set of TMD PDF and FF parameters, as well as the TMD evolution itself. Naturally at lower $x$ the sensitivity to the low-$x$ dependence of transverse momentum width of the PDFs is the largest. For the TMD evolution, both low and high $Q^2$ data are important, although, for unpolarized TMDs, some LHC data that is available somewhat reduces the need for very high $Q^2$, at least at intermediate $x$. It is interesting to see, that the non-Gaussian tails of the transverse momentum dependent fragmentation functions are very relevant at both low-$x$ and higher-$x$ and $Q^2$. Not shown in this figure is that at lower collision energies naturally the higher $x$ regions become more relevant as well as the fragmentation-related parameters are getting better constrained at those $x$. 

\paragraph{Requirements for the detector}
Three $\sigma$ separation of pions from kaons is needed over a large area of the central detector. Due to the hadron energy ranges at the various collision energies 7 GeV/$c$ is sufficiently high in the $-3.5<\eta<-1$ region, 8 to 10 GeV/$c$ would be preferable in the central region ($-1<\eta<1$) and up to 50 GeV/$c$ is needed in the more forward regions ($1<\eta<3.5)$.     

\subsection{Using the hadronic final state to reconstruct SIDIS variables}
\label{sec:SIDIS_final-state}
The JB method discussed in Sec.~\ref{part2-sec-DetReq.Incl} can also be used to reconstruct $x$ and $Q^2$ in SIDIS. When considering neutral current events with a reconstructed electron, one can also use methods that use information from both, the scattered electron and the hadronic final state to increase the precision of the reconstructed kinematic variables. Two of those methods are the so-called "mixed" method and the double-angle method~\cite{Blumlein:2012bf}.
In the mixed method, the exchanged 4-momentum $q$ is calculated from the electron, and the energy transfer $y$ is calculated from the hadronic final state, whereas the double angle method uses only information about the angles of the scattered lepton and hadronic final state. This method is therefore less affected by the loss of relative energy resolution of the scattered lepton which make measurements at low values of $y$ unfeasible.

\begin{figure}[ht]
    \centering
    \includegraphics[width=0.49\textwidth]{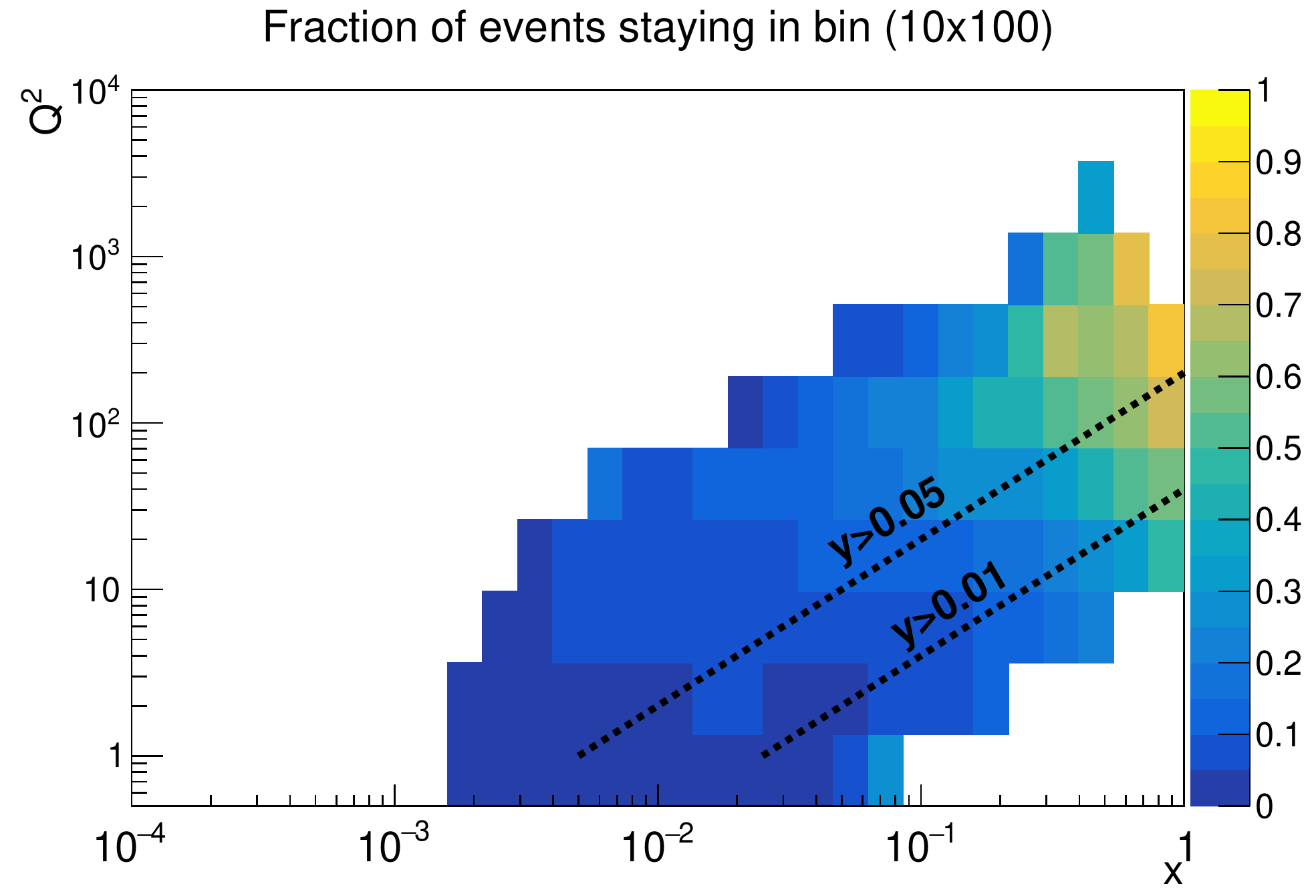}\includegraphics[width=0.49\textwidth]{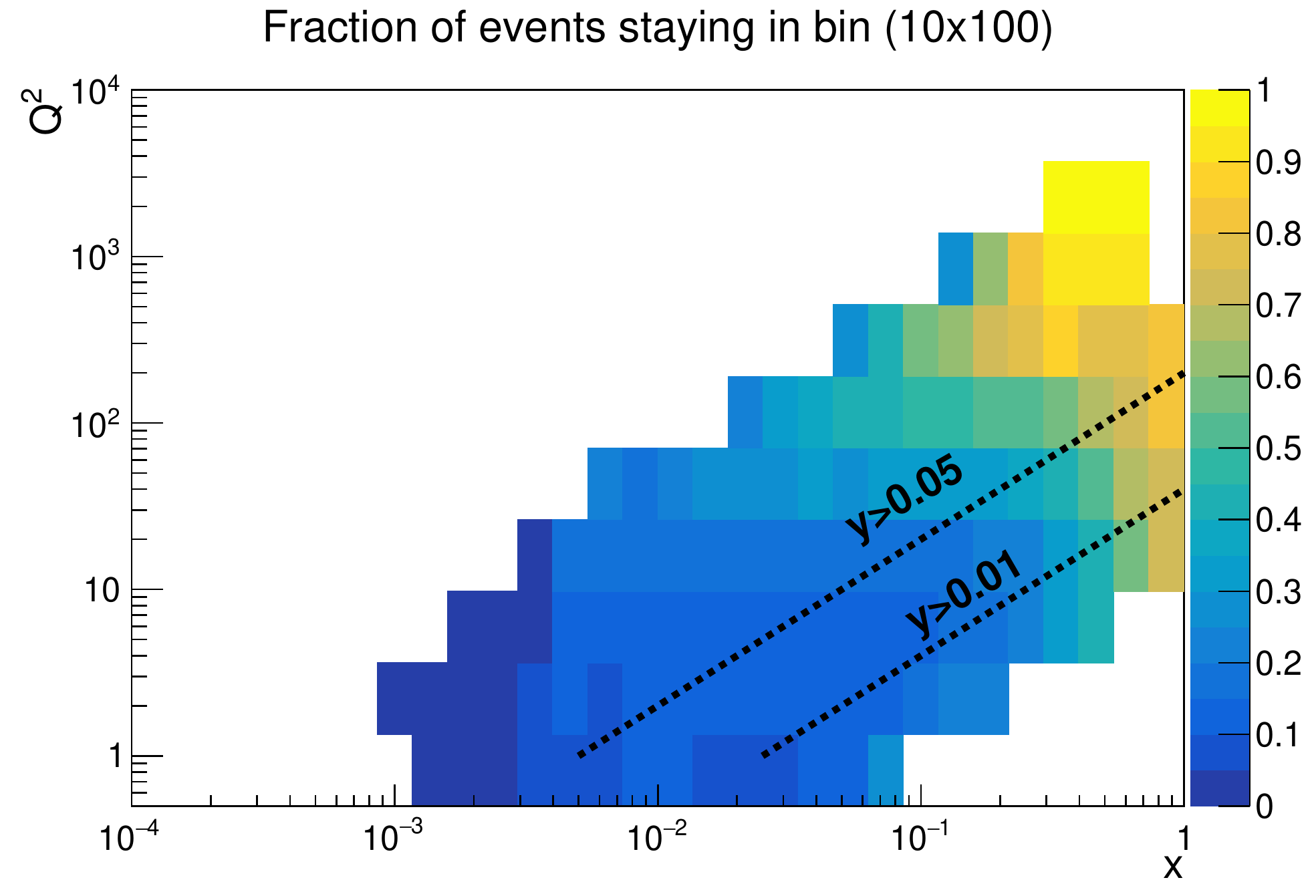}
    \caption{Fraction of events staying in the respective $x/Q^2$ bin for the $10\times100$ GeV$^{2}$ configuration. JB method is shown on the left, whereas the double angle method is shown on the right. The double angle method improves on the JB method at high $x$ and moderate $Q^2$.
    \label{fig:part2-sec-DetReq.SIDIS.recMethods}}
\end{figure}

\begin{figure}[ht]
    \centering
  \includegraphics[width=0.49\textwidth]{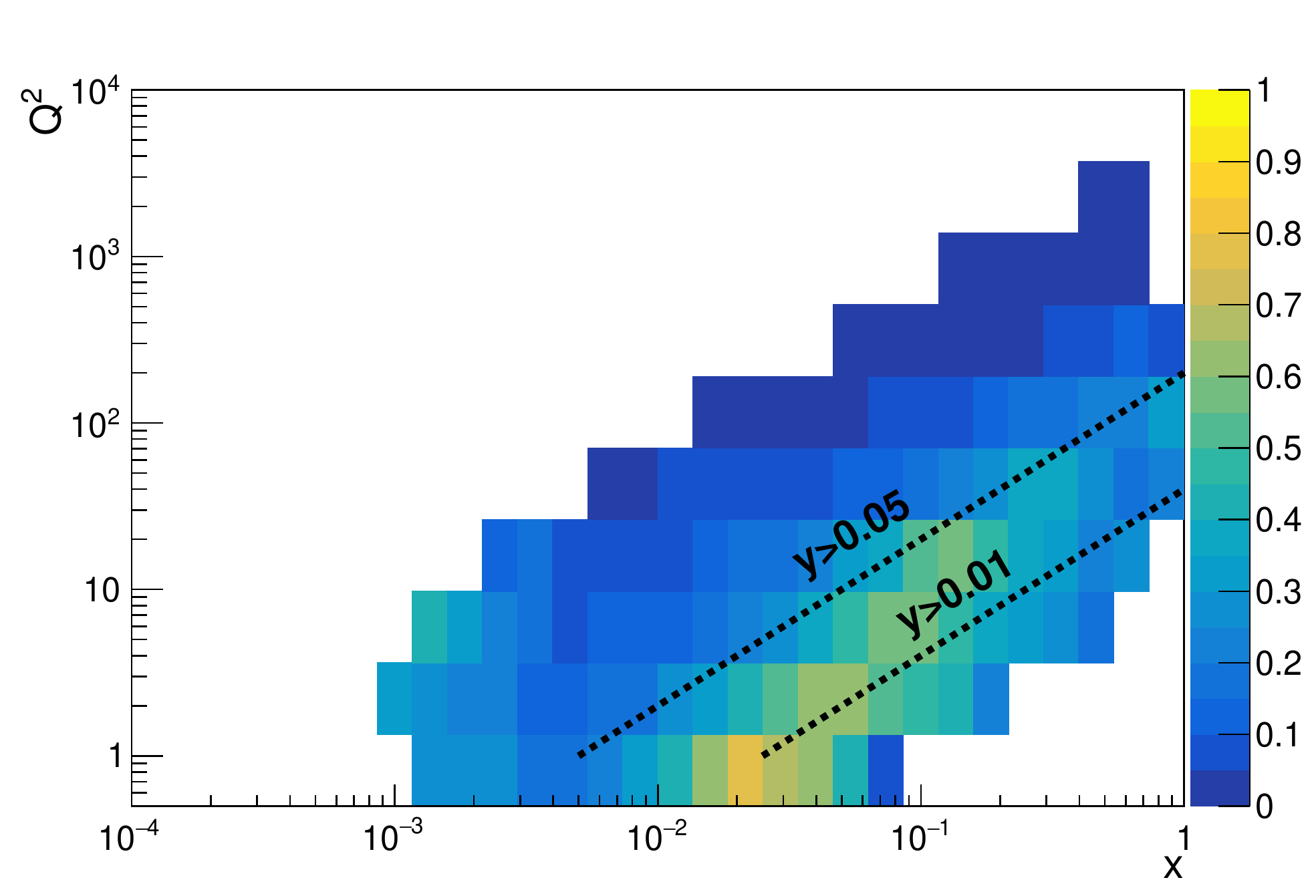}\includegraphics[width=0.49\textwidth]{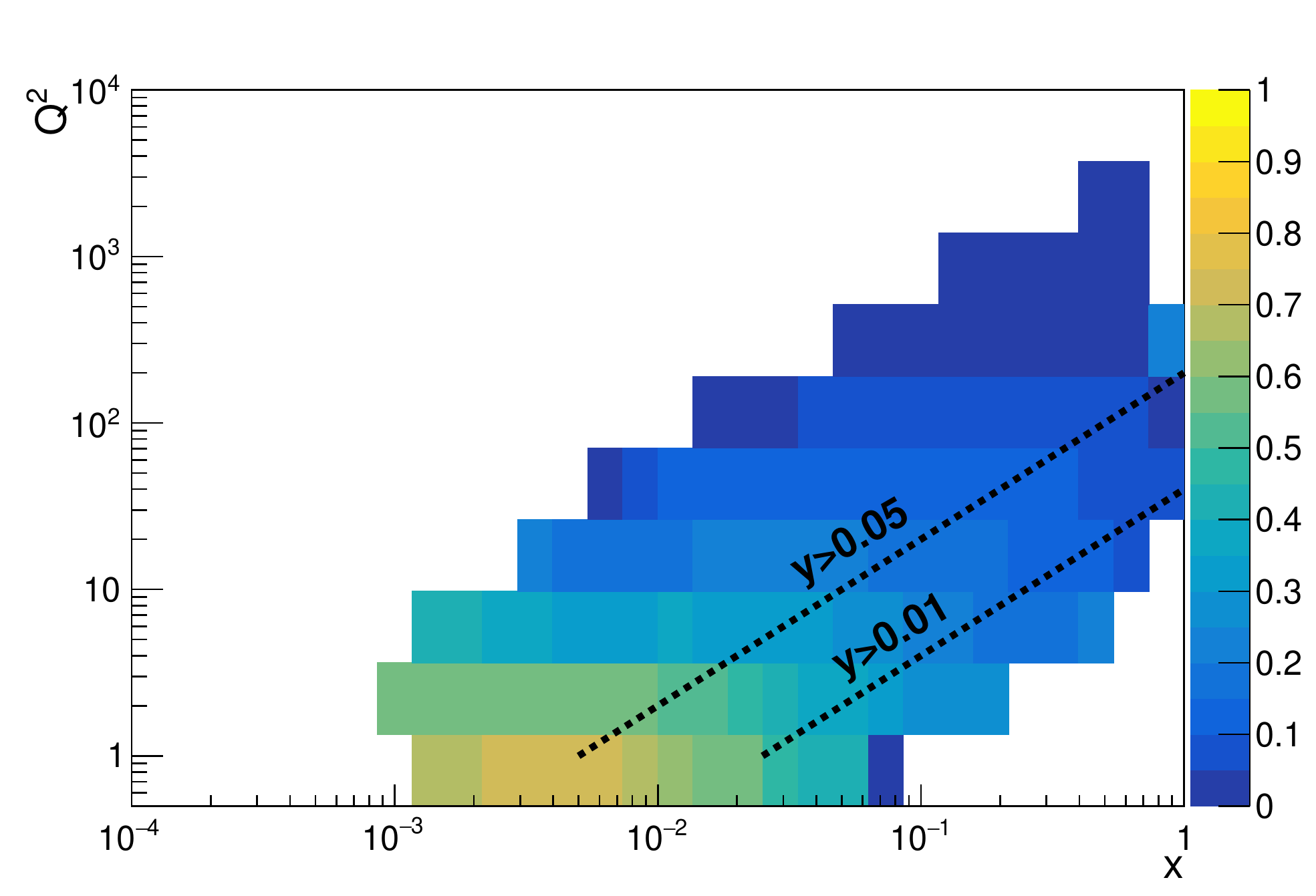}
    \caption{Mean relative error in reconstructing $z$ using just the scattered electron (left) and the double-angle method (right). The double-angle method shows improved resolution at low $y$. The $10\times 100$ GeV $^2$ configuration was simulated for these results.}
    \label{fig:part2-sec-DetReq.SIDIS.recHadronicSIDIS}
\end{figure}

We studied JB, mixed and double-angle method in a simulation using {\sc pythia}8+DIRE and Delphes. 
Consistent with the finding in Sec~\ref{part2-sec-DetReq.Incl} the JB method increases the resolution of $x$-$Q^2$ reconstruction at high $x$ and $Q^2$. We also observed that the methods that include the final state perform better than the JB method as shown in Fig.~\ref{fig:part2-sec-DetReq.SIDIS.recMethods}.
While the reconstruction of the DIS variables from the hadronic final state is well known, to our knowledge the reconstruction of the SIDIS variable, in particular $z$, $p_T$ and azimuthal angles in the Breit frame from the hadronic final state has not been studied so far. 
Similar to the reconstruction of the DIS variables, the importance of the extension of these methods to SIDIS variables lies in the loss of resolution in the measurement of Breit frame variables due to the poorly reconstructed 4-vector $q$. The transverse component of $q$ can be calculated from the transverse momentum of the hadronic final state which only leaves two components that can be determined by solving two equations
\begin{align}
    Q^2&=-q^\mu q_\mu\\
    y&=\frac{P^\mu q_\mu}{P^\mu l_\mu}.
\end{align}
Here $P$ is the four-vector of the nucleon, $l$ the four-vector of the lepton.
The values for $Q^2$ and $y$ are determined consistently with the method used (JB, mixed or double-angle).
A complication arises, since the quadratic equation has two solutions. From simulations it is determined that the smaller solution is most often the correct one, but more studies should be done to improve this criterion.
In our simulation studies, we observed a similar improvement for the SIDIS variables as for the DIS variables. Figure~\ref{fig:part2-sec-DetReq.SIDIS.recHadronicSIDIS} shows the comparison between the reconstruction of $z$ using just the scattered electron or the double-angle method. The improvement at low $y$ is evident.
It is worth emphasizing that using these methods also allows the reconstruction of jets in the Breit frame. We studied this as well using the recent Centauro algorithm~\cite{Arratia:2020ssx} and found that at low $y$ and low ${p_T}_{\textrm{jet}}$ the hadronic methods perform better than using just the electron for the reconstruction of the particle momenta in the Breit frame.
\paragraph{Requirements for the detector}
The requirements on the detector to be able to reconstruct the final state are obviously similar to the requirements for the JB methods described in Sec.~\ref{part2-sec-DetReq.Incl}. In particular it is observed that expanding the coverage from $\eta=3.5$ to $\eta=4.0$ on the hadron going side extends the region where DIS/SIDIS variables can be reconstructed in the highest $x$ region at low $y$. We note that, unlike for the JB method, a hadronic calorimeter does not lead to a significant improvement of the mixed or double angle methods.

\subsection{Requirements for di-hadron measurements}
\label{part2-sec-DetReq.SIDIS.DiHads}
The di-hadron channels naturally share most requirements with the single hadron TMD requirements discussed above. There are two additional aspects though, which need special consideration. 
The first is the dependence of the acceptance on the decay angle $\theta$ and the second the PID performance for hadron pairs. We will first discuss these requirements in order.
As discussed in Sec.~\ref{part2-subS-Hadronization-diHad-pw}, one physics objective is the decomposition of the di-hadron cross-section in terms of partial waves (PWs). The different PWs can be distinguished by their dependence on azimuthal angles in the Breit-frame and the decay angle $\theta$ defined in the di-hadron CMS as the angle between one hadron and the direction of the pair in the lab system~\cite{Bacchetta:2002ux}. 

While the azimuthal coverage requirement is similar to the single-hadrons and therefore has been on criterium of optimization for the EIC detector design, the coverage in $\theta$ is not that obvious. Small values of $\theta$ are correlated with an asymmetric distribution of the energy between the hadrons and thus with one hadron having a small momentum relative to the other. Therefore, restrictions on the minimum accepted hadron momentum, mainly due to the curling up of low $p_T$ tracks, lead to restrictions on the accessible $\theta$ range and thus on our ability to do a PW separation of the cross-section. 

\begin{figure}[ht]
    \centering
    \includegraphics[width=0.95\textwidth]{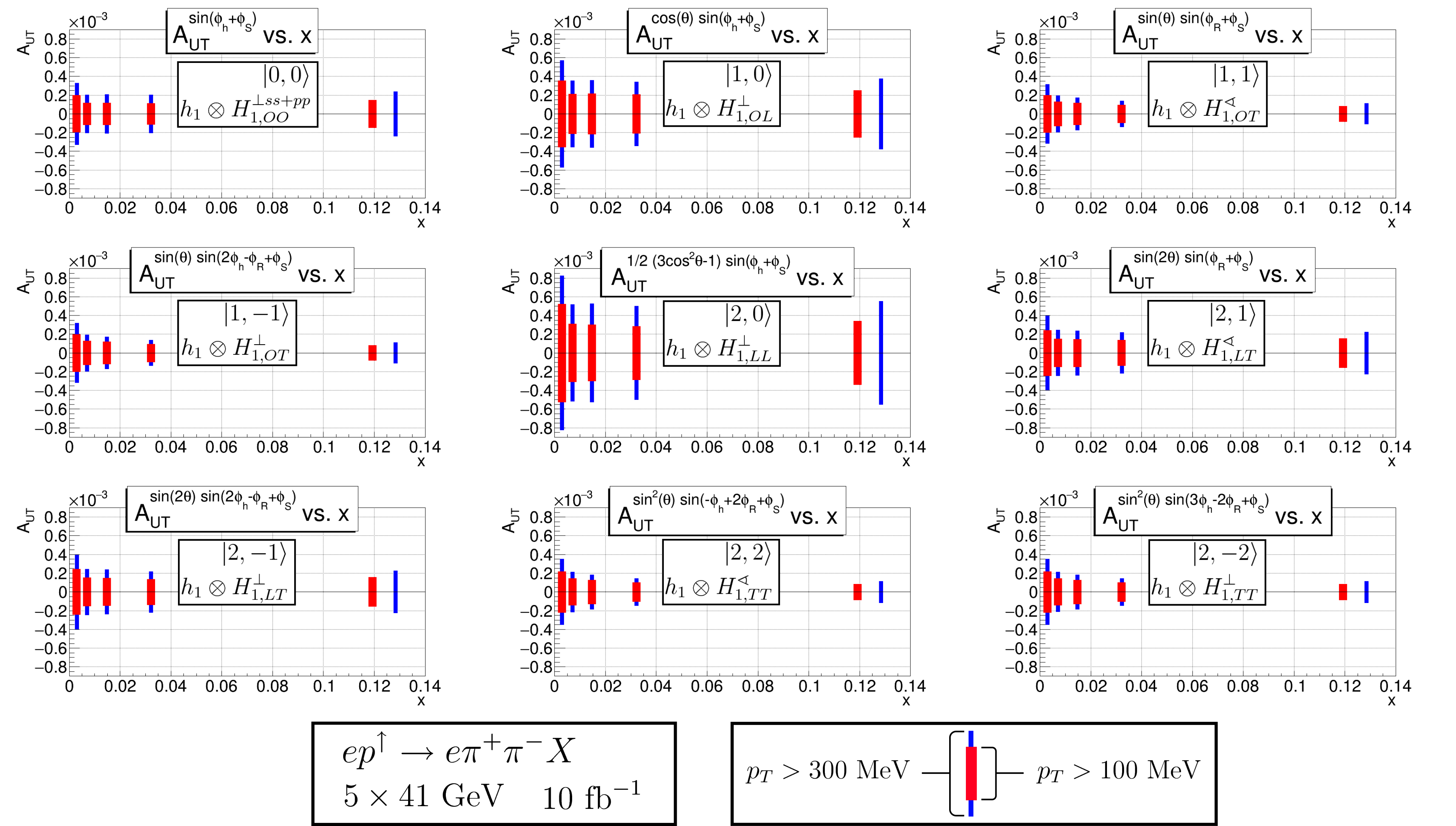}
    \caption{Statistical uncertainties estimated for the PW decomposition up to L=2 for $H_1^\sphericalangle$ for 10~fb$^{-1}$ at 5~$GeV\times$~41~GeV. Narrow blue bands correspond to a  requirement of $p_T>$300~MeV and wide, red bands to a requirement of $p_T>$100~MeV on the pion tracks. The labels on the figure indicate the $m,l$ state and which PDF and FF the PW is sensitive to.}
    \label{fig:part2-sec-DetReq.SIDIS.DiHadPW}
\end{figure}

We investigated the impact of several minimum $p_T$ values corresponding to several proposed detector layouts on the projected statistical uncertainties of the PW separation.  Fig.~\ref{fig:part2-sec-DetReq.SIDIS.DiHadPW} shows the statistical uncertainties for $p_T>100$ and 300~MeV. It is observed that lowering the limit below 100~MeV, which corresponds to the requirement imposed by the $\Lambda$ reconstruction discussed in Sec.~\ref{part2-sec-DetReq.SIDIS.Lambda}, does not improve the uncertainties significantly.

Concerning PID requirements, again, the di-hadron requirements are similar to the single hadron requirements discussed above. However, since two hadrons are required, the dilution of the samples for a given significance in separation can be more severe. A study was performed to asses the impact of two and three $\sigma$ separation between pions and kaons. This is mainly motivated by the PID in the central region for which a three~$\sigma$ separation at high momenta is more difficult to achieve.
For di-hadrons, the $p_T$ dependence of the dilution is flat. As expected, with two~$\sigma$ separation, the $\pi\pi$ samples are still very pure, above $95$\%. However, the purity of $\pi K$ samples drops to below 70~\% and $KK$ pairs to about 75~\%. With three~$\sigma$ separation, the purity of the samples is above 95~\% for all cases. Therefore having $\pi-K$ separation at three $\sigma$ is very important for precision measurements involving kaons. 

\paragraph{Requirements for the detector}
In addition to the single hadron requirements, coverage down to low hadron momenta of $p_T > 100$ MeV/$c$ is required for the partial-wave analysis of di-hadron final states.

\subsection{Requirements for \texorpdfstring{$\Lambda$}{Lambda} measurements}
\label{part2-sec-DetReq.SIDIS.Lambda}
\begin{figure}[ht]
    \centering
    \includegraphics[width=0.95\textwidth]{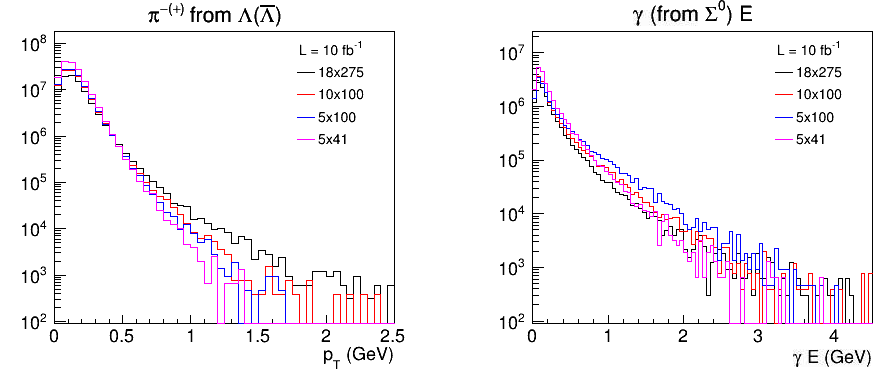}
    \caption{Transverse momentum and energy spectrum of pions from $\Lambda\rightarrow p \pi$ and $\Sigma^0\rightarrow \Lambda \gamma$ respectively.}
    \label{fig:part2-sec-DetReq.SIDIS.LambdaPionSpec}
\end{figure}

\begin{figure}[ht]
    \centering
    \includegraphics[width=0.48\textwidth]{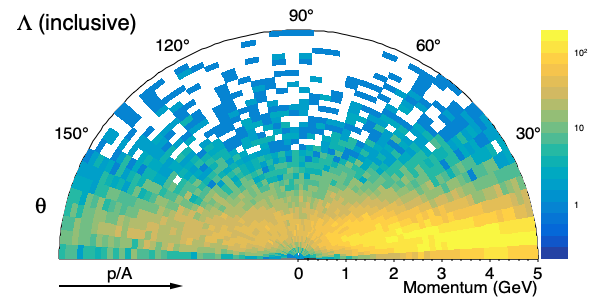}
    \includegraphics[width=0.48\textwidth]{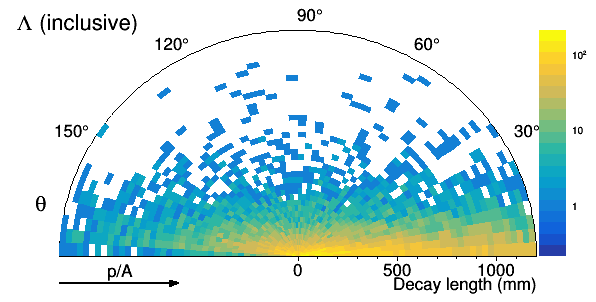}\\
    \includegraphics[width=0.48\textwidth]{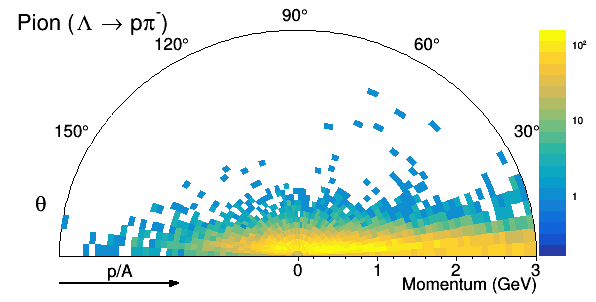}
    \includegraphics[width=0.48\textwidth]{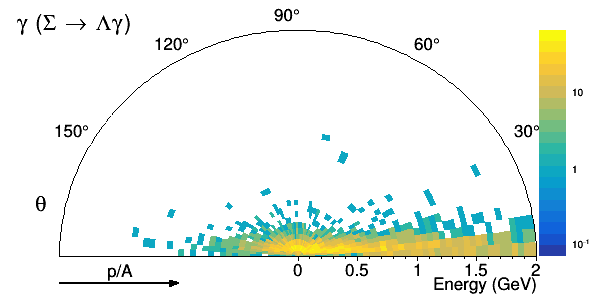}
    \caption{From top left to bottom right: kinematics of $\Lambda$, decay length of $\Lambda$, kinematics of decay $\pi^-$ and kinematics of decay $\gamma$ from the non-prompt $\Lambda$ production chain $\Sigma^0\rightarrow \Lambda+\gamma$. All plots from fast simulation of the $18\times 275$ configuration.}
    \label{fig:part2-sec-DetReq.SIDIS.LambdaKins}
\end{figure}

The efficient detection of $\Lambda$ hyperons poses specific requirements to a potential detector. Experimentally the channels $\Lambda\rightarrow p \pi^-$ and $\bar{\Lambda}\rightarrow \bar{p} \pi^+$ are the most promising. In these channels the $p/\bar{p}$ carries the majority of the momentum of the parent $\Lambda$, leading to a very soft spectrum of the decay pions. 
Fig~\ref{fig:part2-sec-DetReq.SIDIS.LambdaKins} shows the kinematics of the produced $\Lambda$ and its decay products. The left panel in Fig.~\ref{fig:part2-sec-DetReq.SIDIS.LambdaPionSpec} shows the projected spectrum for the decay pion for a representative energy and $x_F>0$. It is evident that any minimum $p_T$ restriction will cut significantly into the $\Lambda$ spectrum. For example, requiring $p_T>300$~MeV leaves less than 5\% of $\Lambda$s even at the highest energies. From these studies we deduced a requirement of $p_{T\pi}>100$~MeV.

The $\Lambda$ measurements discussed in this report are assuming that the hyperon is promptly produced by the fragmenting quark. However, a significant fraction of $\Lambda$s are coming from non-strong decays and thus have to be corrected for.
The right panel in Fig.~\ref{fig:part2-sec-DetReq.SIDIS.LambdaPionSpec} shows the spectrum for the decay photons. For $\Sigma^0$ boosted in the forward direction the spectrum is somewhat harder but additional material in that region is still a concern. We concluded that a requirement of $\gamma$ detection with the nominal resolution in that region for $\gamma_E>200$~MeV up to $\eta=3.0$ and $\gamma_E>400$~MeV for $3.0<\eta<4.0$ is sufficient to maintain a reasonable acceptance for $\Sigma^0$s.

\paragraph{Requirements for the detector}
$\Lambda$ decay pions need to be detected from transverse momenta as low as $P_T > 100$ MeV/$c$ in order to recover most $\Lambda$s. Similarly, photon energies need to be also well reconstructed down to 200 MeV and 400 MeV in the central and forward regions, respectively, to distinguish $\Lambda$ feed-down from $\Sigma^0$.

\subsection{Spectroscopy requirements for forward electron identification}
\label{part2-sec-DetReq.SIDIS.Spectroscopy}

The spectroscopy of unconventional quarkonia, referred to as $XYZ$ mesons, in photoproduction relies on the efficient detection of all the meson decay products and adequate resolutions to identify and study the produced resonances.  One particular reaction of interest $\gamma p \rightarrow Z_c(3900)^+ n$, introduced in Sec.~\ref{part2-subS-Hadronization-Spectroscopy}, provides useful insight to the detector requirements in $ep$ collisions.  Here we will focus on the $Z_c(3900)^+ \rightarrow J/\psi\pi^+$ decay mode with subsequent $J/\psi \rightarrow e^+e^-$, however the $Z_c^+$ has been observed in open charm decay modes as well.  This reaction (and other $XYZ$ final states) are simulated using the amplitudes provided by the JPAC Collaboration~\cite{Albaladejo:2020tzt} integrated into the \textsc{elSpectro} event generator ~\cite{GitHub:elspectro}.

\begin{figure}[ht]
\begin{center}
\includegraphics[width= 0.9 \textwidth]{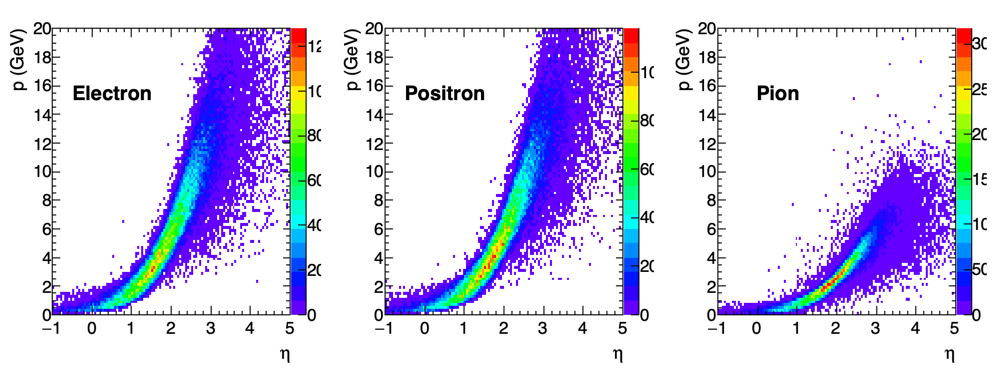}
\caption{\label{fig:zc_5_41} Simulated kinematics p vs $\eta$ for $Z_c^+ \rightarrow J/\psi\pi^+, J/\psi\rightarrow e^+e^-$ decay products for $5 \times 41$ GeV beam energies}
\end{center}
\end{figure}

\begin{figure}[ht]
\begin{center}
\includegraphics[width= 0.9 \textwidth]{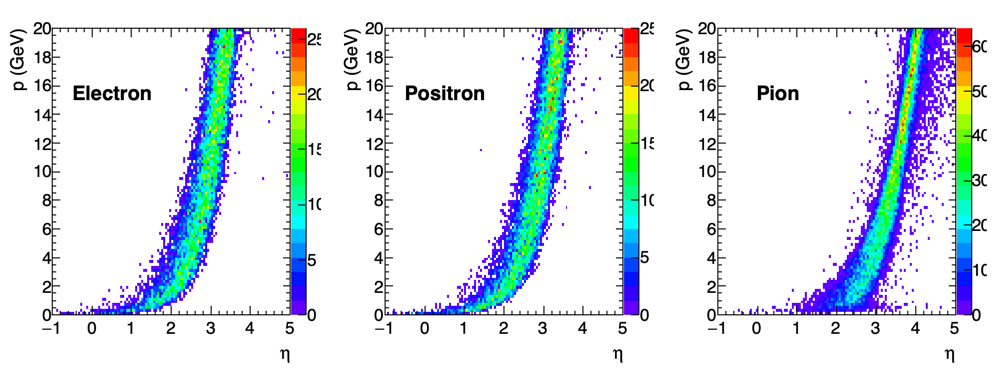}
\caption{\label{fig:zc_18_275} Simulated kinematics p vs $\eta$ for $Z_c^+ \rightarrow J/\psi\pi^+, J/\psi\rightarrow e^+e^-$ decay products for $18 \times 275$ GeV beam energies}
\end{center}
\end{figure}

The lab frame kinematics ($p$ vs. $\eta$) of the meson decay particles $e^-, e^+$ and $\pi^+$ are shown in Fig.~\ref{fig:zc_5_41} and~\ref{fig:zc_18_275} for $5\times41$ and $18\times275$ GeV beam energies, respectively.  Due to the large kinematic boost from the proton beam, the decay particles populate most of the forward hadron acceptance.  In particular, for the highest proton beam energies many of the tracks have $\eta > 3.5$ which will limit the acceptance for these reactions.  Therefore, we expect the low beam energy (high acceptance) and high beam energy (high luminsoity) to be complementary in studying the photoproduction of $XYZ$ states.

\begin{figure}[ht]
\begin{center}
\includegraphics[width= 0.9 \textwidth]{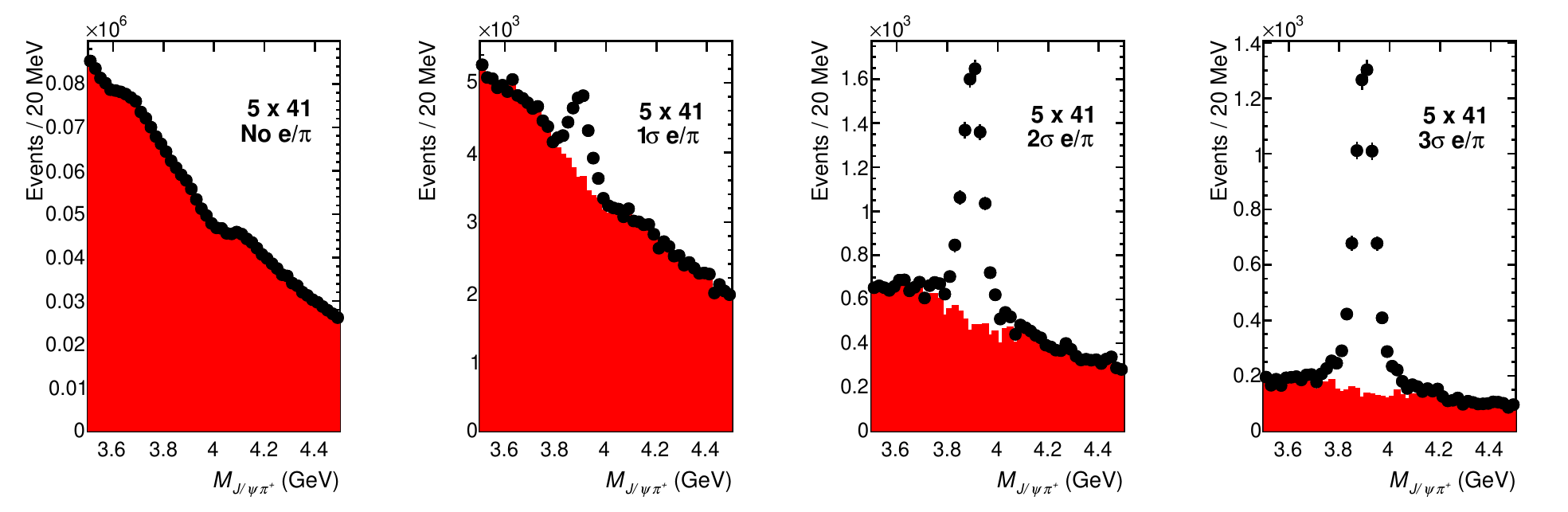}
\caption{\label{fig:zc_epi} Simulated $Z_c^+ \rightarrow J/\psi\pi^+$ mass distribution from smeared simulation for $5 \times 41$ GeV beam energies with increasing (left-to-right) $e/\pi$ separation.  The PYTHIA background (red) is significantly reduced relative to the signal by more stringent $e/\pi$ separation.}
\end{center}
\end{figure}

Additional requirements on the particle identification are also necessary to identify the $J/\psi \rightarrow e^+e^-$ decay particles in this forward region of the detector.  Background from inclusive hadron production was studied using PYTHIA minimum bias events (\textit{i.e.} without any $Q^2$ selection), where charged hadrons were nominally assumed to be mis-identified as $e^\pm$ candidates.  The $J\psi\pi^+$ mass distribution in Fig.~\ref{fig:zc_epi} illustrates the level of background (red) and expected $Z_c(3900)^+$ signal (black) for electron-pion separations from none to $3\sigma$.  For an $e/\pi$ separation of $3\sigma$, a purity of $\sim90\%$ is achieved which would allow further study of the observed resonance, including the decay angular distributions.  The relative normalization of the signal and PYTHIA background requires further study and validation of the current $XYZ$ models with data from light mesons may provide some important benchmarks.

\paragraph{Requirements for the detector}
Unlike most SIDIS measurements, spectroscopy requires electron-pion separation ($>3\sigma$) in the forward region for J/$\psi$ identification. 
Tracking and hadron identification exceeding $\eta > 3.5$ would be preferrable due to the boost of produced states of interest.

\subsection{Summary of SIDIS-related Detector requirements}
The summary of all SIDIS related detector requirements are summarized in Table \ref{tab:SIDIS_requirements}. As discussed in the previous sections, particle identification over as large a momentum range as possible and low minimum momentum/energy reconstruction are the leading detector requirements. Forward calorimetry and tracking coverage up to $\eta < 4$ becomes important when using the hadronic final state to determine the DIS kinematics.

\begin{table}[ht]
    \centering
    \caption{SIDIS related maximum PID momentum ranges that need to be covered as well as minimum tracking momentum and EM energy ranges.}
    \label{tab:SIDIS_requirements}
    \begin{tabular}{c c c c c }
    \hline\hline
rapidity &  hadron ID ($\pi$/K/p) & e ID (e/h separation) & minimum $P_T$ & minimum E \\ \hline
-3.5 -- -1.0 & 7 GeV/$c$ & 18 GeV/$c$ & 100 MeV/$c$ & 100 MeV \\
-1.0 -- \hphantom{-}1.0 & 8-10 GeV/$c$ & 8 GeV/$c$ & 100 MeV/$c$ & 100 MeV \\
\hphantom{-}1.0 -- \hphantom{-}3.5 & 50 GeV/$c$ & 20 GeV/$c$ & 100 MeV/$c$ & 100 MeV \\ \hline\hline
    \end{tabular}

\end{table}

\section{Jets and Heavy Quarks}
\label{part2-sec-DetReq.Jets.HQ}

An impressive amount of information on the partonic structure of the nucleon can be extracted from 
inclusive DIS reactions, where only the scattered lepton is measured. However, including information 
from the hadronic final state as in semi-inclusive DIS (SIDIS) measurements (see Sec.~\ref{part2-sec-DetReq.SIDIS}) 
can provide further 
insight that is not possible with purely inclusive analyses. The observables evaluated by the Jets 
and Heavy Quarks working group characterize more fully the total final state by taking into 
account correlations between the produced particles or by identifying specific final states. They 
provide complimentary, and often times, unique information compared to standard SIDIS measurements. 

Jets cluster many of the final state particles arising from a hadronizing parton into a single object, 
which gives a better representation of the kinematics of that parton than a single hadron measurement 
could. In addition, many well defined techniques exist to systematically explore the distribution of 
energy within a jet, providing a unique way to study the hadronization process. 

Heavy quark production 
at the EIC will provide a clean probe of gluon dynamics in the nucleon/nucleus while the large mass 
of charm and bottom mesons will provide many advantages in the study of hadronization and cold nuclear 
matter effects. Finally, global event shapes aim to classify the energy flow of 
the entire final state and can yield very high precision measurements of fundamental quantities such 
as the strong coupling constant. 

The precision reconstruction of jets, heavy quark states, and global event shapes places performance 
constraints on tracking, calorimetry, and particle identification in the central detector region. Specific 
requirements for each of these components, along with the driving physics considerations, will be enumerated 
in the sections below. We also discuss the simulation and detector smearing frameworks used to evaluate detector 
effects and the basic kinematics of the observables of interest. Finally, we discuss the topics of unfolding and 
evaluation of systematic uncertainties using the 1-jettiness observable as an example.

\subsection{Simulation and detector modeling} \label{part2-sec-DetReq.Jets.HQ.SIMU}

The detector performance requirements listed in the sections below were derived from the analysis of a 
number of representative jet, heavy flavor, and event shape observables. For each relevant observable, the 
appropriate process was simulated using a Monte Carlo event generator  
and the output was run through a smearing framework, or a more detailed detector 
simulator, to evaluate the effects of finite resolutions and acceptances and determine the detector qualities 
needed to measure the observable with the necessary precision to extract the desired physics.

\begin{figure}[ht!]
\centering
\subfloat[\label{PWG-sec-JHQ-fig-PYTHIA8DATA}]{\includegraphics[width=0.45\textwidth]{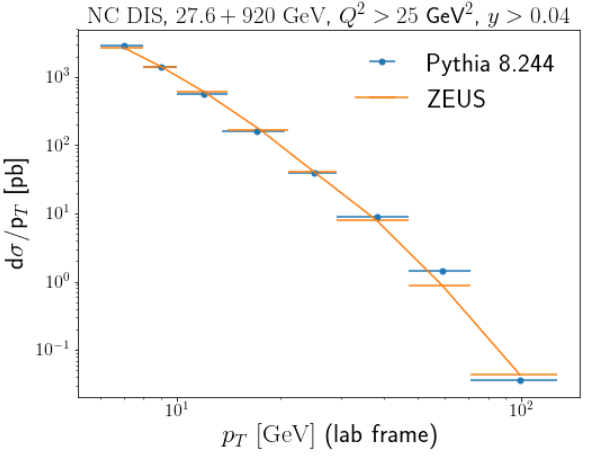}} \quad
\subfloat[\label{PWG-sec-JHQ-fig-PYTHIA8THEORY}]{\includegraphics[width=0.45\textwidth]{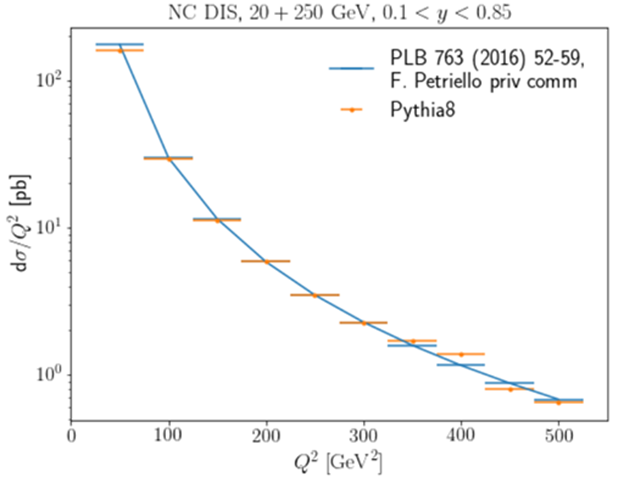}}
\caption[PYTHIA8 Validation]{Comparison of neutral current DIS jet cross section as a function of \pT\ for inclusive lab frame jets from ZEUS data and PYTHIA8 (left). Comparison of the lab frame inclusive jet cross section as a function of $Q^2$ between PYTHIA8 and theory (right).}
\label{PWG-sec-JHQ-fig-PYTHIA8VALIDATION}   
\end{figure}

All studies carried out by this group utilized either the PYTHIA6~\cite{Sjostrand:2006za} or 
PYTHIA8~\cite{Sjostrand:2007gs} event generator. PYTHIA6 was 
implemented within the pythiaeRHIC program, which has been tuned to reproduce \ep\ data from HERA.
The pythiaeRHIC Monte Carlo has been used in numerous EIC studies and has been shown to reproduce HERA (di)jet cross 
sections and jet profiles~\cite{Chu:2017mnm,Aschenauer:2019uex}. 
The PYTHIA8 generator 
was used primarily for simulations of jet production in neutral and charged current DIS at moderate to large $Q^2$ 
values as well as heavy flavor production. As PYTHIA8 has not been specifically tuned to reproduce \ep\ jet data 
(beyond the development and setting of default parameters by the PYTHIA authors), several checks were made to 
ensure the Monte Carlo was giving reasonable output. Figure~\ref{PWG-sec-JHQ-fig-PYTHIA8DATA} compares 
$\mathrm{d}\sigma/\mathrm{d}p_{T}$ for lab frame jets as measured by the ZEUS experiment to that obtained 
from PYTHIA8. A comparison is also made between PYTHIA8 and a theoretical calculation of the jet cross 
section as a function of $Q^2$ for beam energies relevant for the EIC in Fig.~\ref{PWG-sec-JHQ-fig-PYTHIA8THEORY}. In both cases, the agreement with the 
simulation is seen to be excellent, giving confidence that at least for the relevant kinematic regions, 
PYTHIA8 produces reasonable results.

The second step in the evaluation of detector requirements involves distorting the output from the event 
generators to mimic the effects of finite detector resolutions and acceptances. For jet observables, this 
was done using a fast smearing framework, either eic-smear~\cite{git:eicsmeardetectors} or 
DELPHES~\cite{deFavereau:2013fsa}. 
Eic-smear is a relatively 
light-weight framework that allows for the definition of different `detector volumes', 
which will smear either the momentum or energy of particles that traverse these regions. The energy 
or momentum is smeared according to a Gaussian distribution whose sigma is set by the user and 
should correspond to the proposed energy or momentum resolution of a given subsystem. DELPHES is a 
more sophisticated tool that takes into account the bending of charged particles in a solenoidal 
magnetic field and gives access to higher level observables such as particle flow objects and missing 
transverse energy. As with eic-smear, the resolutions of the detector components are supplied in 
parameterized form (see \cite{delphesCard} for settings) and incident particles are smeared according 
to a log-normal distribution. Because of 
the event generator / smearing frameworks in place during most of the Yellow Report exercise, eic-smear 
was used with pythiaeRHIC and DELPHES was used with PYTHIA8 for all analyses. 

While fast smearing based on parameterized descriptions of tracker and calorimeter responses are largely 
adequate for jet observables, a more detailed accounting of detector characteristics is needed to evaluate 
requirements for heavy flavor reconstruction. Therefore, many of the heavy flavor analyses utilized 
more complete detector geometries implemented in the eic-root and Fun4All frameworks.

\subsection{Kinematics summary} \label{part2-sec-DetReq.Jets.HQ.KINE}

Plots of relevant kinematic quantities for jets and heavy flavor states are presented below for several beam 
energy combinations and $Q^2$ ranges.

\begin{figure}[ht!]
\centering
\subfloat{\label{PWG-sec-JHQ-fig-jetPtKinPhotoLo}
 \includegraphics[width=0.45\textwidth]{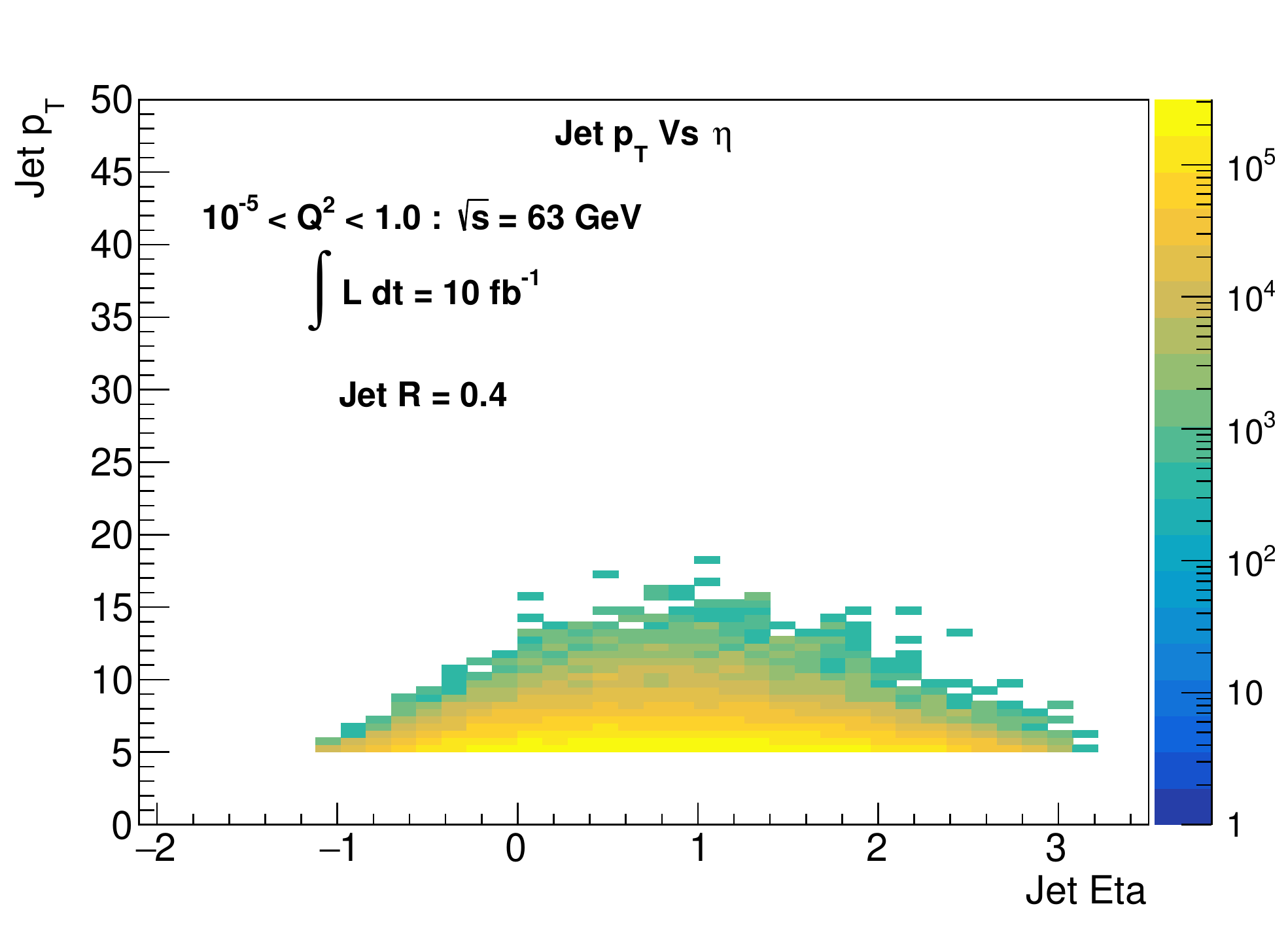}} \quad
\subfloat{\label{PWG-sec-JHQ-fig-jetPtKinDISLo}
\includegraphics[width=0.45\textwidth]{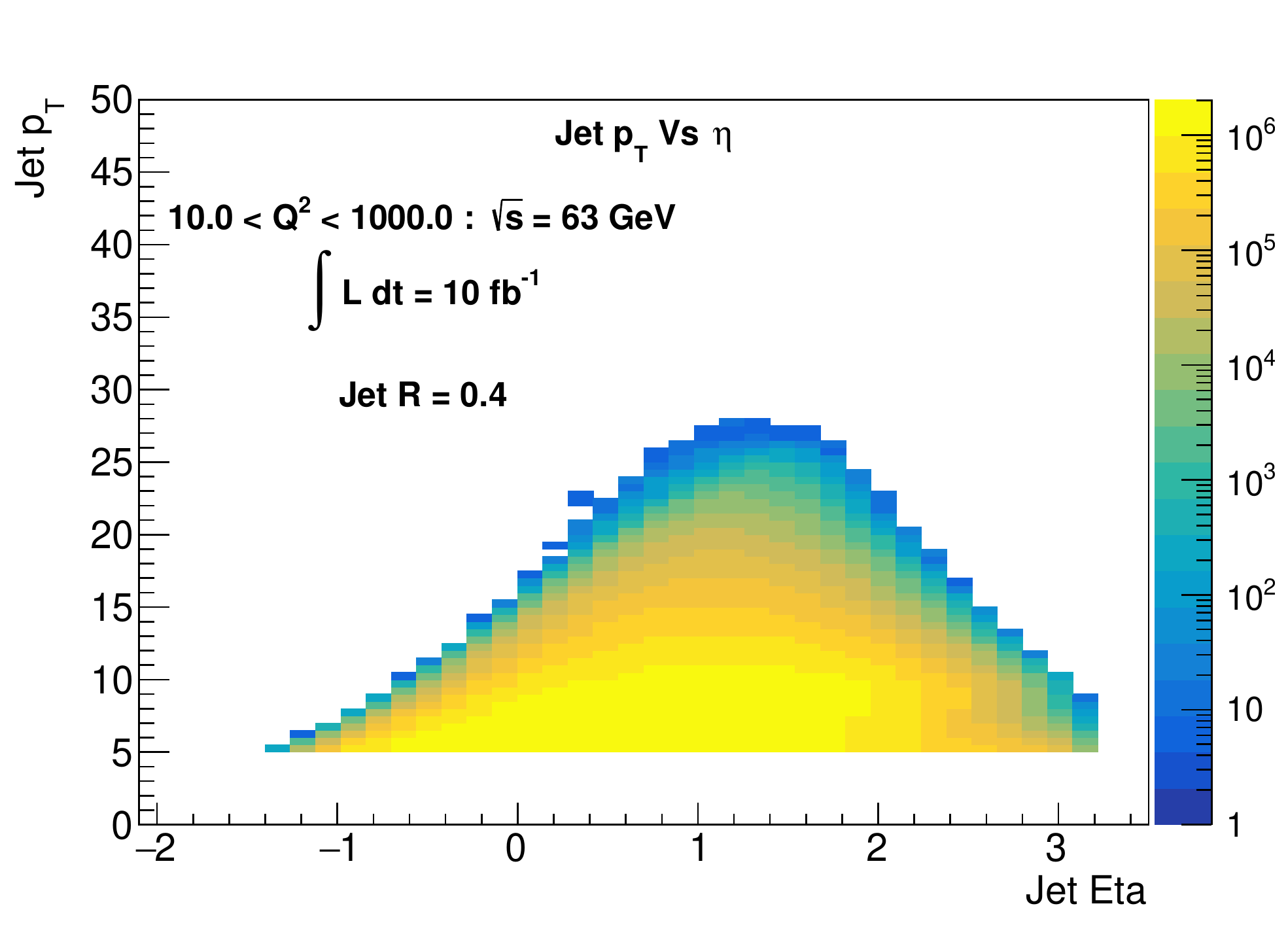}} \quad
\subfloat{\label{PWG-sec-JHQ-fig-jetPtKinPhotoHi}
\includegraphics[width=0.45\textwidth]{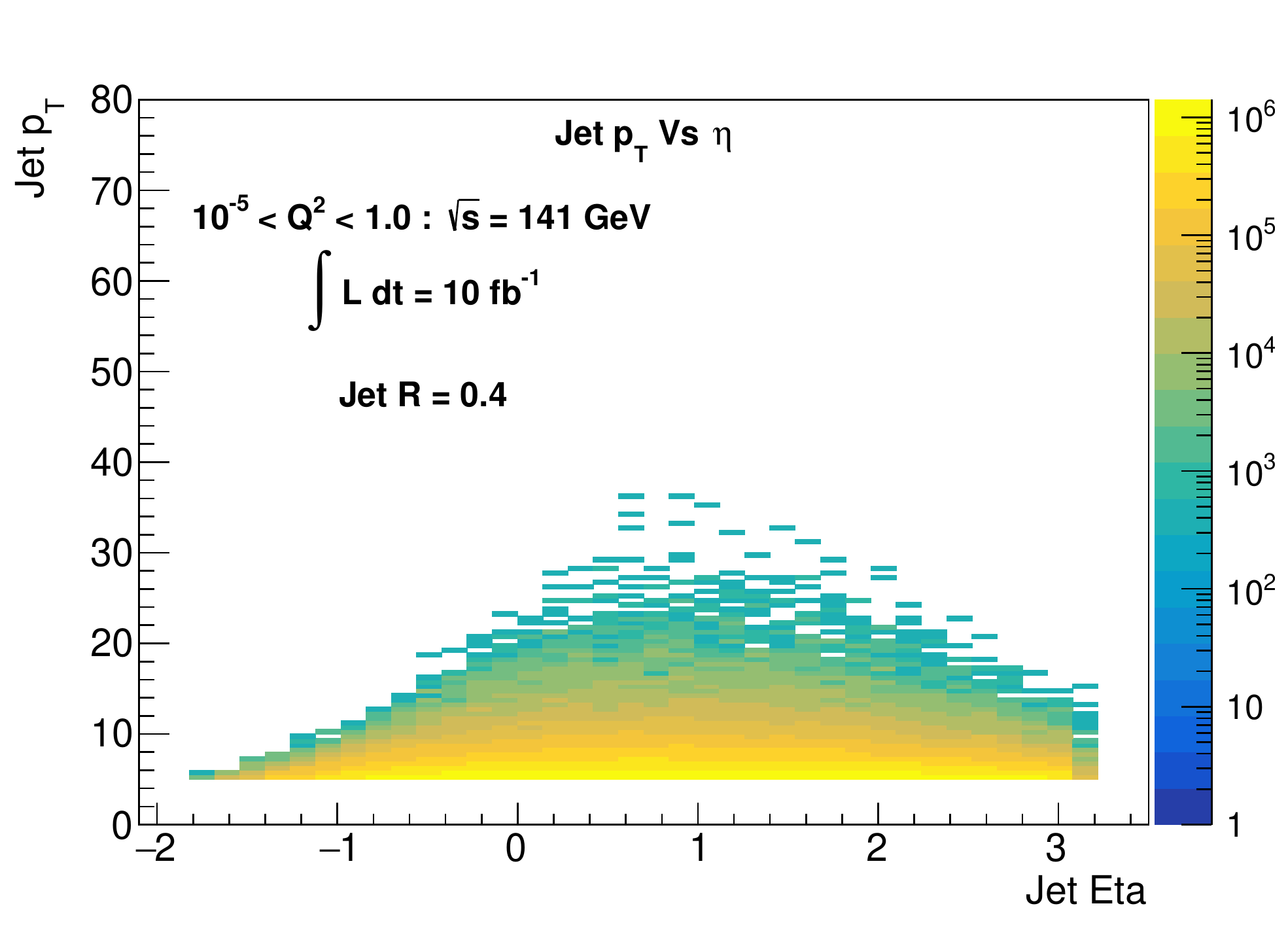}} \quad
\subfloat{\label{PWG-sec-JHQ-fig-jetPtKinDISHi}
\includegraphics[width=0.45\textwidth]{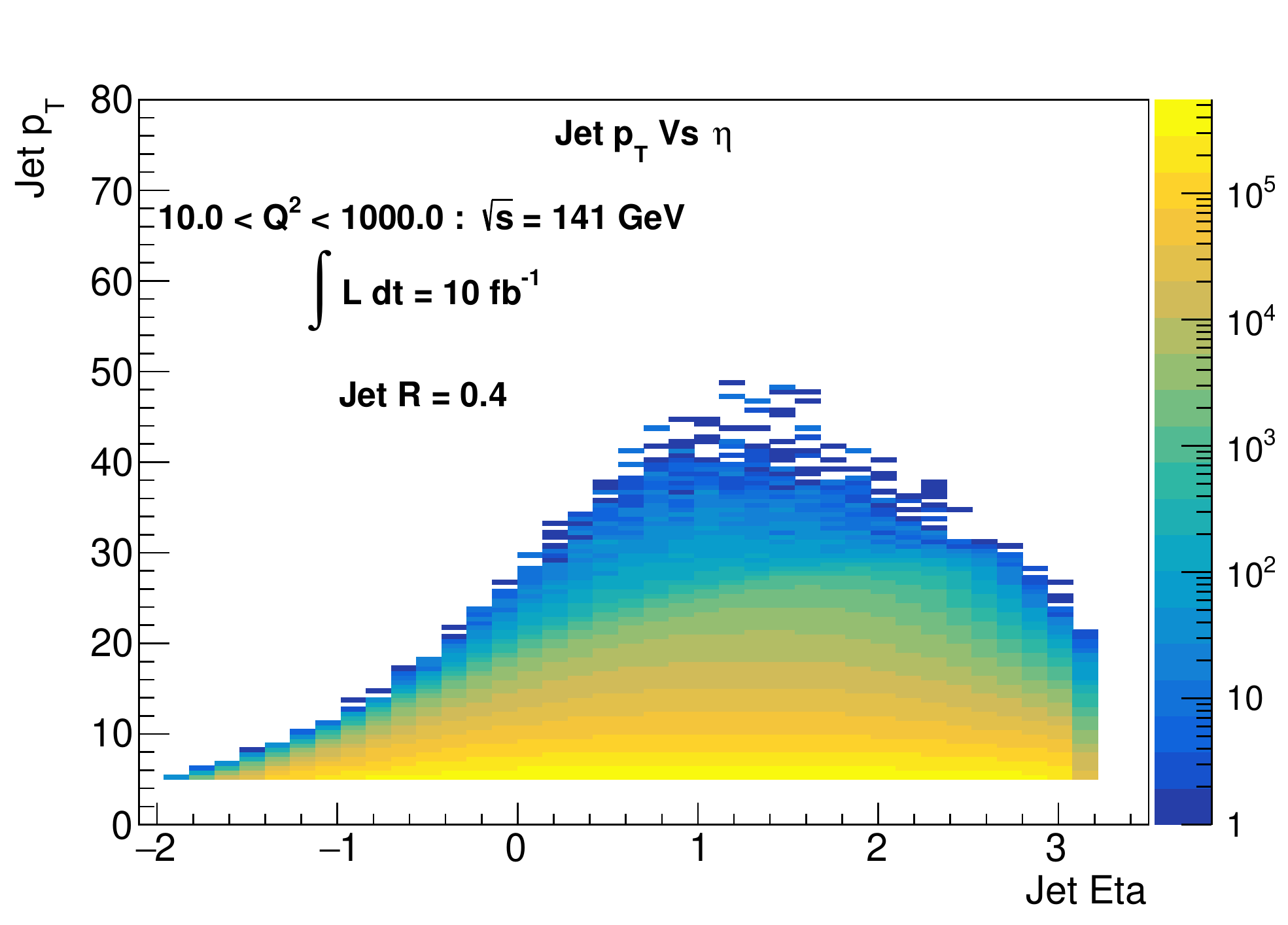}}
\caption[Jet Kinematics pT]{Jet $p_T$ vs pseudorapidity for beam energies of 10x100~GeV (top row) and 18x275~GeV (bottom row) and $10^{-5} < Q^2 < 1.0$~GeV$^2$ (left column) and $10 < Q^2 < 1000$~GeV$^2$ (right column). The jet resolution parameter used is 0.4. Counts have been scaled to correspond to an integrated luminosity of 10~fb$^{-1}$.}
\label{PWG-sec-JHQ-fig-jetPtVsEta}   
\end{figure}

\begin{figure}[ht!]
\centering
\subfloat{\label{PWG-sec-JHQ-fig-jetEKinPhotoLo}
\includegraphics[width=0.45\textwidth]{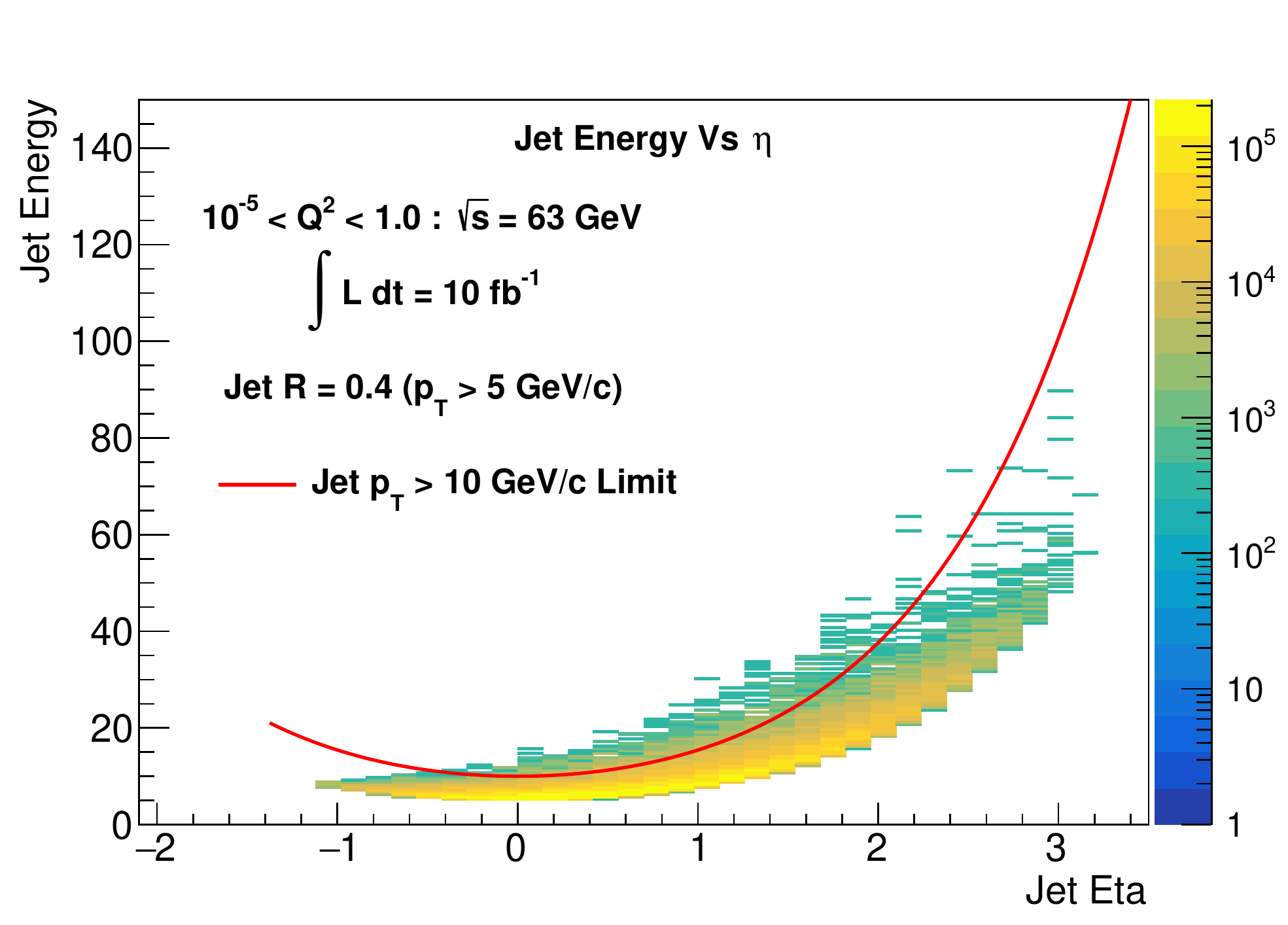}} \quad
\subfloat{\label{PWG-sec-JHQ-fig-jetEKinDISLo}
\includegraphics[width=0.45\textwidth]{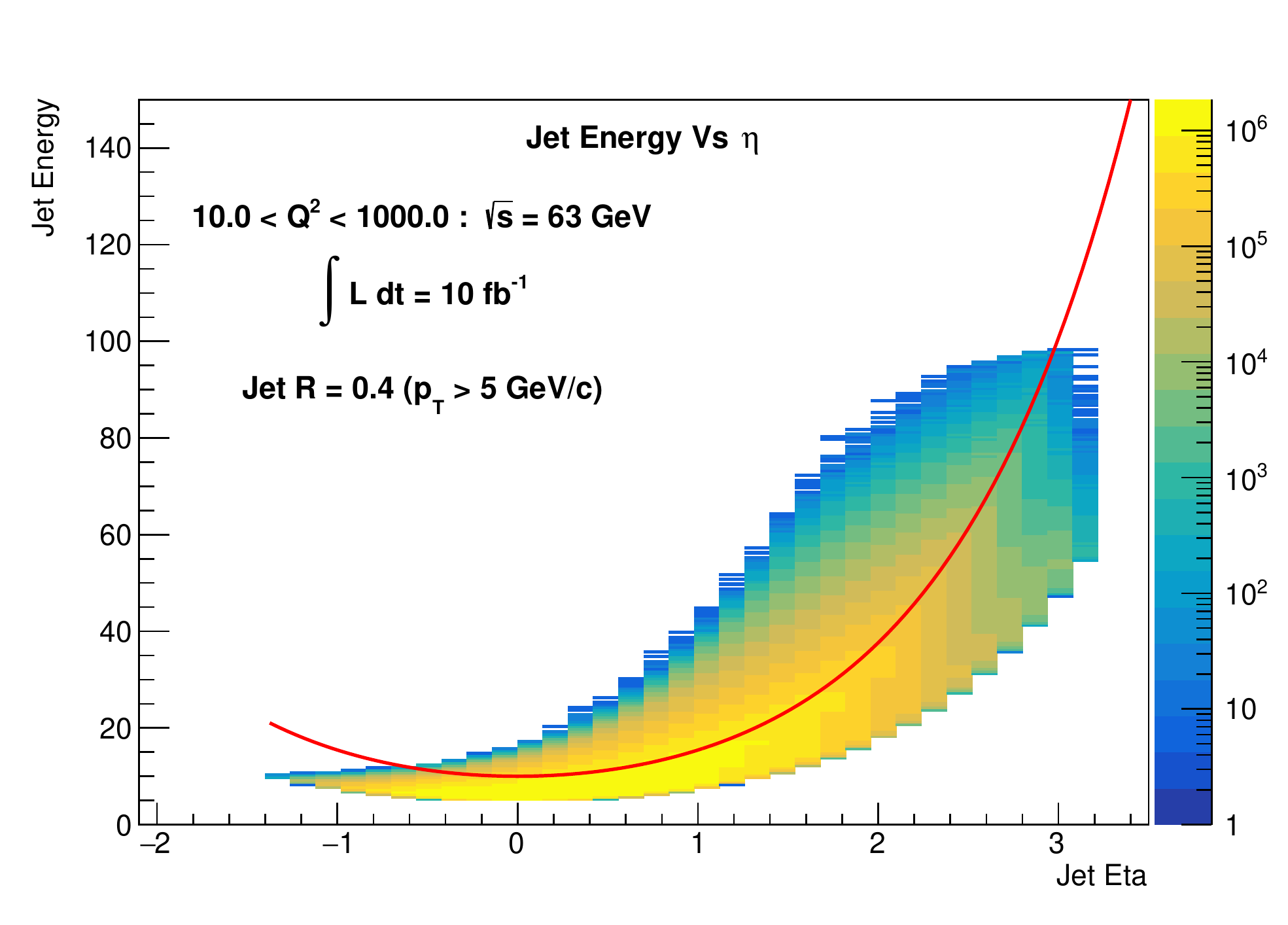}} \quad
\subfloat{\label{PWG-sec-JHQ-fig-jetEKinPhotoHi}
\includegraphics[width=0.45\textwidth]{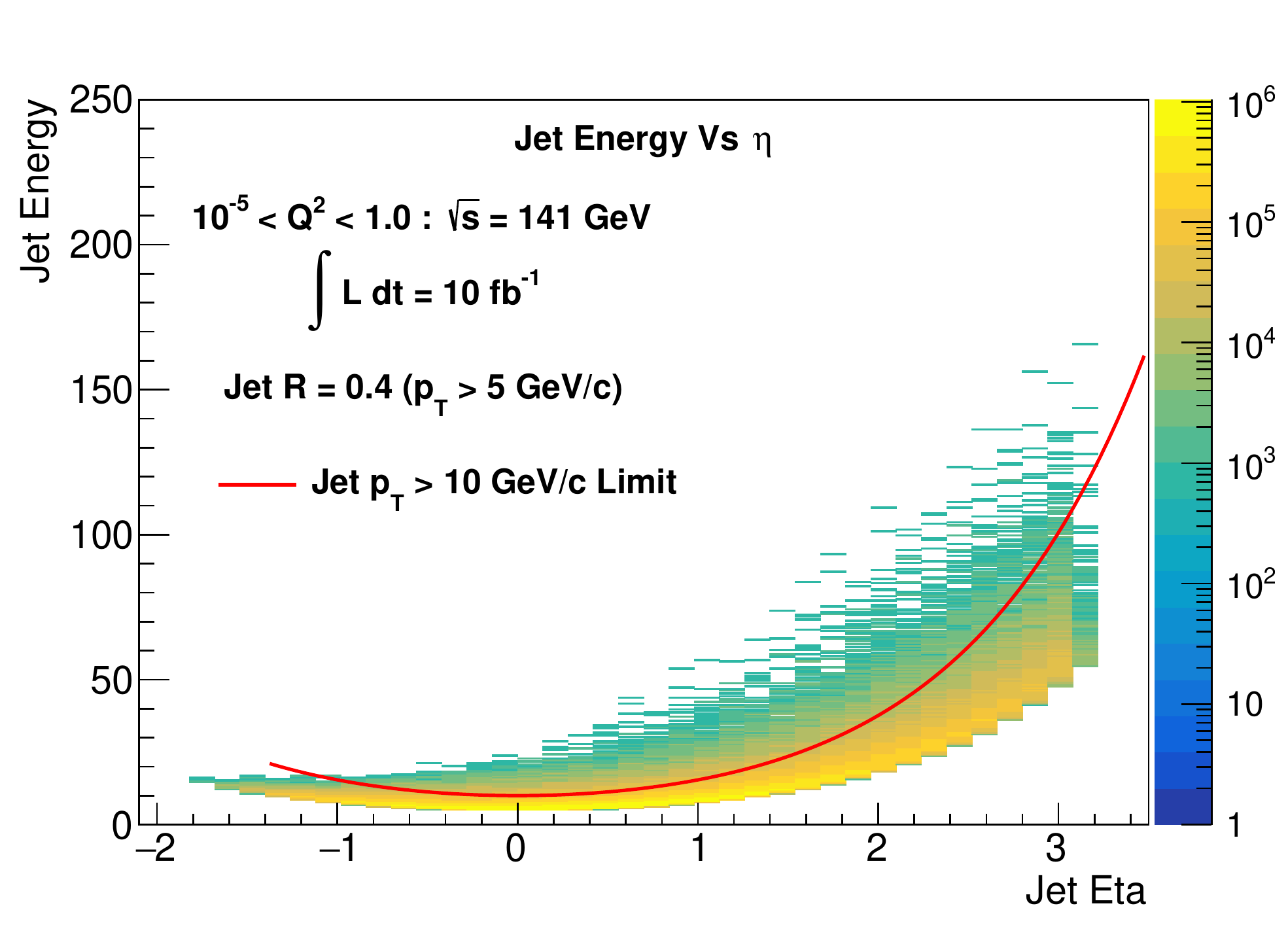}} \quad
\subfloat{\label{PWG-sec-JHQ-fig-jetEKinDISHi}
\includegraphics[width=0.45\textwidth]{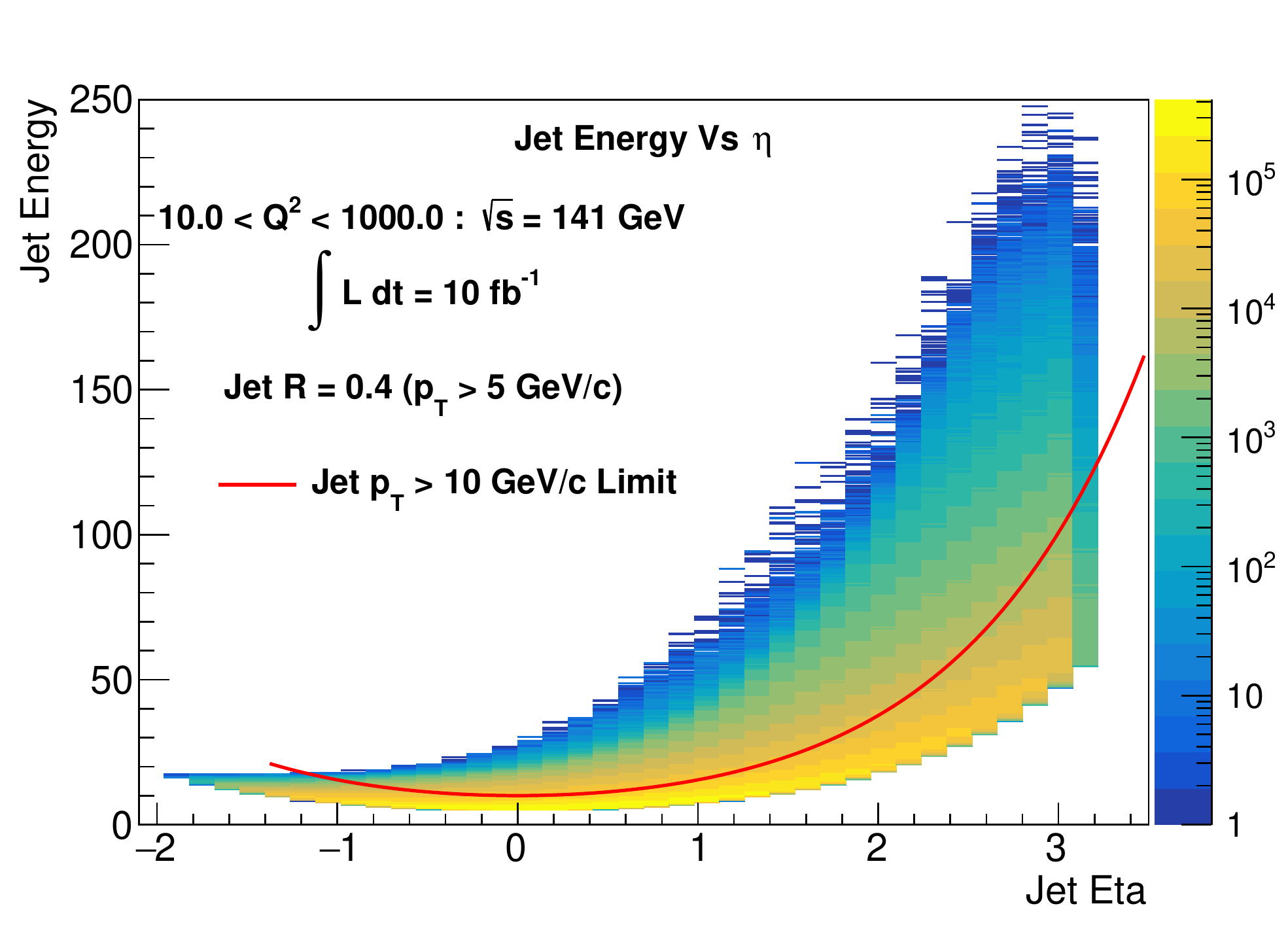}}
\caption[Jet Kinematics Energy]{Jet energy vs pseudorapidity for beam energies of 10x100~GeV (top row) and 18x275~GeV (bottom row), and $10^{-5} < Q^2 < 1.0$~GeV$^2$ (left column) and $10 < Q^2 < 1000$~GeV$^2$ (right column). The jet resolution parameter used is 0.4. Counts have been scaled to correspond to an integrated luminosity of 10~fb$^{-1}$. The distributions are shown with a minimum jet \pT\ of 5~GeV; the red line shows the lower limit of the distribution assuming a minimum jet \pT\ of 10~GeV.}
\label{PWG-sec-JHQ-fig-jetEVsEta}   
\end{figure}

For jets, both the transverse momentum (Fig.~\ref{PWG-sec-JHQ-fig-jetPtVsEta}) 
and energy (Fig.~\ref{PWG-sec-JHQ-fig-jetEVsEta}) are plotted as a function of pseudorapidity. Diffractive 
jet production is not considered here. Jets were found 
in the laboratory frame using the Anti-k$_T$ algorithm~\cite{Cacciari:2008gp} with a resolution parameter of 0.4. 
The radius used here is 
somewhat smaller than the $\mathcal{O}(1.0)$ radius jets that were used in most analyses in order to show 
more of the behavior at large pseudorapidity while keeping the jet fully contained in the range 
$|\eta| < 3.5$. The counts have been scaled to the equivalent of 10~fb$^{-1}$. Note that because of the 
large cross section, the number of generated 
events for the $10^{-5} < Q^2 < 1.0$ sample amount to only 0.02~fb$^{-1}$ and so the available phase space 
for the corresponding plots is not fully populated. Further jet kinematic distributions can be found in~\cite{Page:2019gbf,Arratia:2019vju}.

\begin{figure}[ht!]
\centering
\subfloat{\label{PWG-sec-JHQ-fig-D063KIN}
\includegraphics[width=0.90\textwidth]{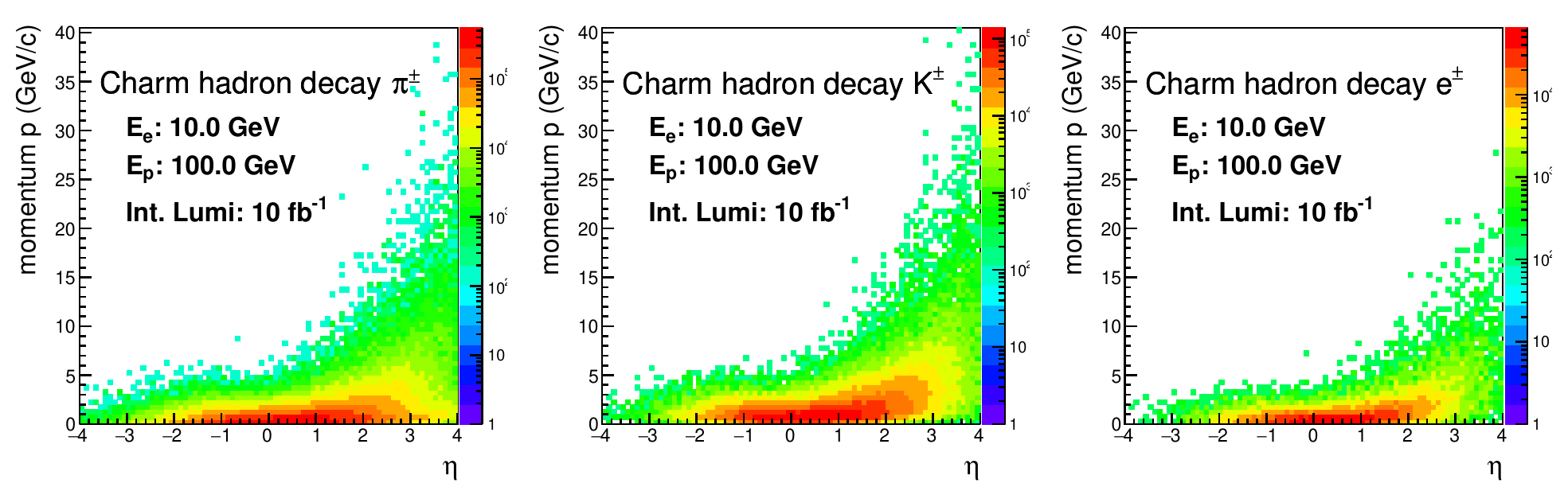}} \quad
\subfloat{\label{PWG-sec-JHQ-fig-D085KIN}
\includegraphics[width=0.90\textwidth]{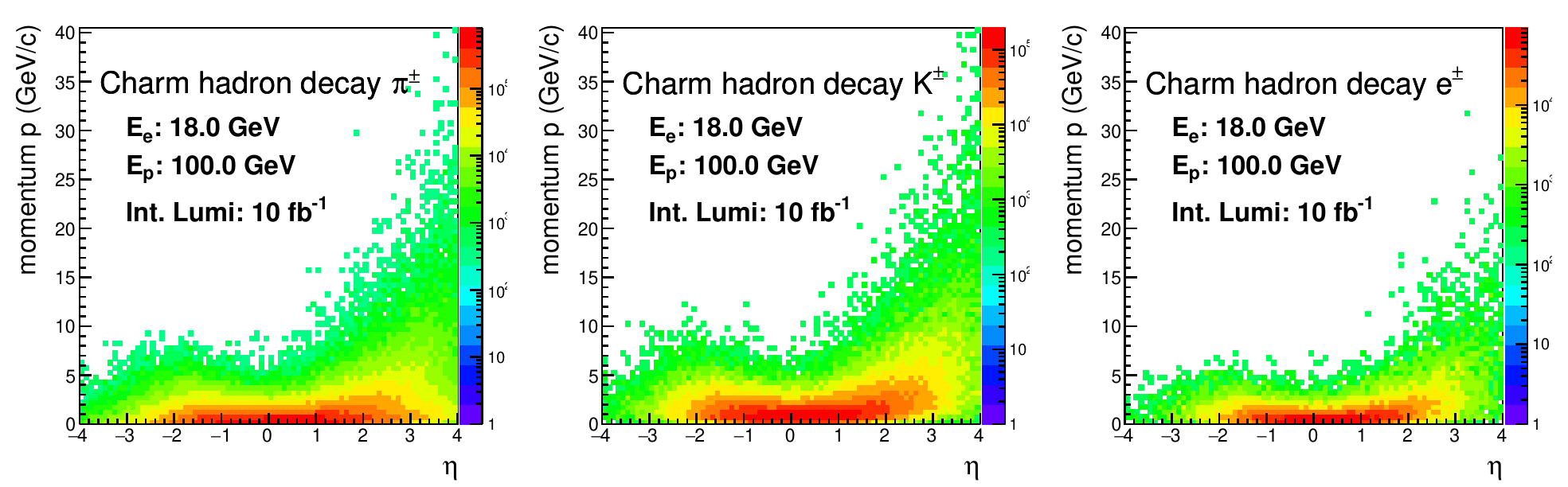}} \quad
\subfloat{\label{PWG-sec-JHQ-fig-D0141KIN}
\includegraphics[width=0.90\textwidth]{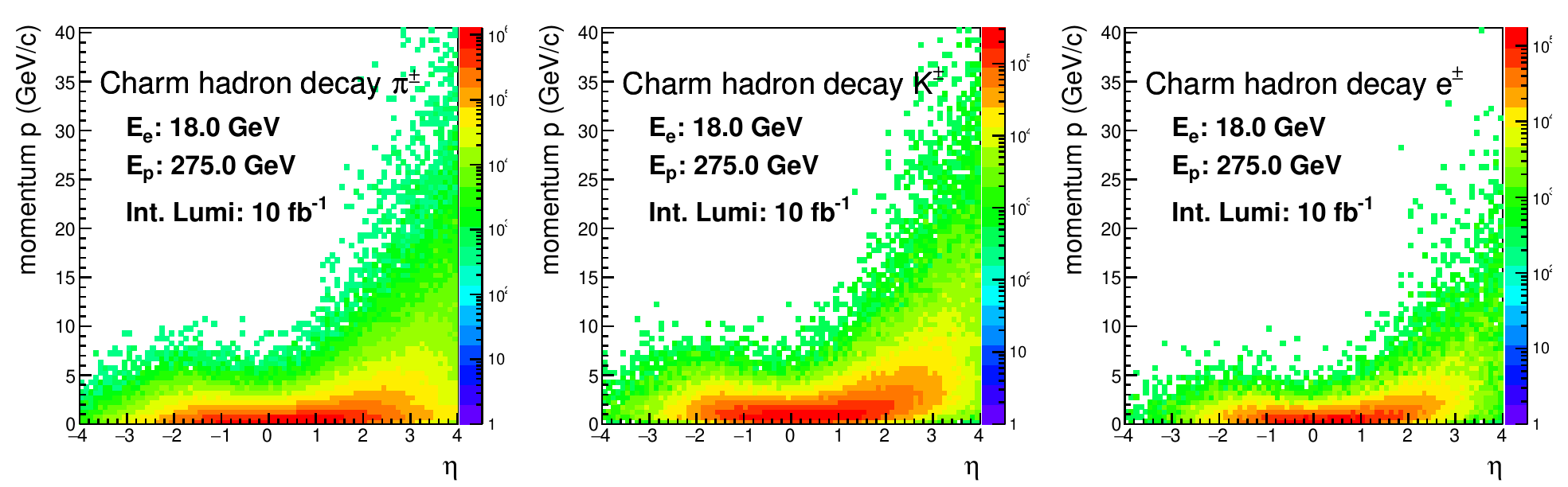}}
\caption[D0 Kinematics]{Momentum vs pseudorapidity for the decay products of $D^0$ mesons for beam energies of 10x100~GeV (top row), 18x100~GeV (middle row), and 18x275~GeV (bottom row). Charged pions are in the left column, charged kaons in the middle column, and electrons/positrons in the right column. Counts have been scaled to correspond to an integrated luminosity of 10~fb$^{-1}$.}
\label{PWG-sec-JHQ-fig-d0Kin}
\end{figure}

The momentum vs pseudorapidity distributions for pions and kaons emitted in $D^0$ decays can be found in 
Fig.~\ref{PWG-sec-JHQ-fig-d0Kin} for beam energies of 10x100~$\gev$, 18x100~$\gev$, and 18x275~$\gev$. As with the jet 
plots, the counts have been scaled to the equivalent of 10~fb$^{-1}$. Further heavy flavor kinematic distributions 
can be found in~\cite{Wong:2020xtc}.

\subsection{Tracking} \label{part2-sec-DetReq.Jets.HQ.TRACKING}

A high-resolution, large acceptance tracker will be essential for all analyses considered by this 
working group, including jet and heavy meson reconstruction and event shape measurements. Due 
to the generally low energies of produced hadrons at the EIC, the tracking system will 
provide better resolution than hadron calorimetry at all but the most forward 
rapidities, where particle energies can be large and tracking resolution is expected to 
degrade. It is therefore anticipated that charged tracks (which comprise roughly 
two-thirds of the energy contained in an average jet) will be the dominant input to the 
jet-finding algorithms. The superior pointing resolution for tracks compared to 
calorimeter clusters will also be critical for jet substructure and event shape 
measurements where the spatial distribution of energy is an explicit aspect of the 
observable. Charged tracks are also indispensable to the study of heavy flavor states, 
where the invariant mass of (identified) track combinations is used to tag heavy hadrons. 
The ability to accurately reconstruct track trajectories also allows for the identification 
of the secondary vertex associated with the decaying heavy particles which aids in 
background suppression.

\subsubsection{Momentum resolution} \label{part2-sec-DetReq.Jets.HQ.PRES}

Over the course of the Yellow Report effort, several sets of track momentum resolution 
parameters, representing reasonable assessments of potential tracker performances, were 
made available and were evaluated by the Jets and Heavy Quarks group. These resolutions 
were parameterized in the form $A\% \times \mathrm{P} \oplus B\%$, where 
P is the track 3-momentum, in six pseudorapidity intervals between $\pm 3.5$ and can be 
found in Tab.~\ref{PWG-sec-JHQ-tab-TRACKRES}. The resolutions in the column labeled 
`Handbook' were based on values from the EIC R\&D Handbook~\cite{EIC:RDHandbook}, which 
predates the Yellow Report, and were used in a number of jet analyses, including studies of 
neutral and charged current jet production for TMD extractions (see discussion in Sec.~\ref{part2-subS-SecImaging-TMD3d.jets} and ~\cite{Arratia:2020nxw}) and jet substructure studies (see Sec.~\ref{part2-subS-Hadronization-HadVacuum}).
Based on the work done for the Yellow Report, momentum resolution parameters were also released 
for magnetic field strengths of 3~T and 1.5~T (see Tab.~\ref{tab:solenoid_magnet} and related discussion).
The two field 
strengths also imply different minimum track transverse momenta thresholds, which are discussed 
in Sec.~\ref{part2-sec-DetReq.Jets.HQ.ADDCONS}. In the following, perfect efficiency is assumed for tracks above the minimum \pT\ threshold.

\begin{table}[ht!]
	\setlength{\tabcolsep}{9.0pt}
	\begin{center}
	    \caption[Track Momentum Resolution]{Charged particle momentum resolution parameterizations for different pseudorapidity bins that were evaluated by the Jets and Heavy Quarks working group. See text for discussion.}
		\scalebox{0.9}{
			\begin{tabular}{ c  c  c  c }
				\hline \hline
				Pseudorapidity Range & Handbook ($\sigma$P/P\%) & 3~T ($\sigma$P/P\%) & 1.5~T ($\sigma$P/P\%) \\ \hline
				$-3.5 < \eta < -2.5$ & $0.1\%*\mathrm{P} \oplus 2\%$ & $0.1\%*\mathrm{P} \oplus 2\%$ & $0.2\%*\mathrm{P} \oplus 5\%$ \\ 
				$-2.5 < \eta < -1.0$ & $0.05\%*\mathrm{P} \oplus 1\%$ & $0.02\%*\mathrm{P} \oplus 1\%$ & $0.04\%*\mathrm{P} \oplus 2\%$\\ 
				$-1.0 < \eta < 1.0$ & $0.05\%*\mathrm{P} \oplus 0.5\%$ & $0.02\%*\mathrm{P} \oplus 0.5\%$ & $0.04\%*\mathrm{P} \oplus 1\%$ \\ 
				$1.0 < \eta < 2.5$ & $0.05\%*\mathrm{P} \oplus 1\%$ & $0.02\%*\mathrm{P} \oplus 1\%$ & $0.04\%*\mathrm{P} \oplus 2\%$\\ 
				$2.5 < \eta < 3.5$ & $0.1\%*\mathrm{P} \oplus 2.0\%$ & $0.1\%*\mathrm{P} \oplus 2\%$ & $0.2\%*\mathrm{P} \oplus 5\%$\\ \hline \hline
			\end{tabular}
		}
		\label{PWG-sec-JHQ-tab-TRACKRES}
	\end{center}
\end{table}

\begin{figure}[ht!]
\centering
\includegraphics[width=0.90\textwidth]{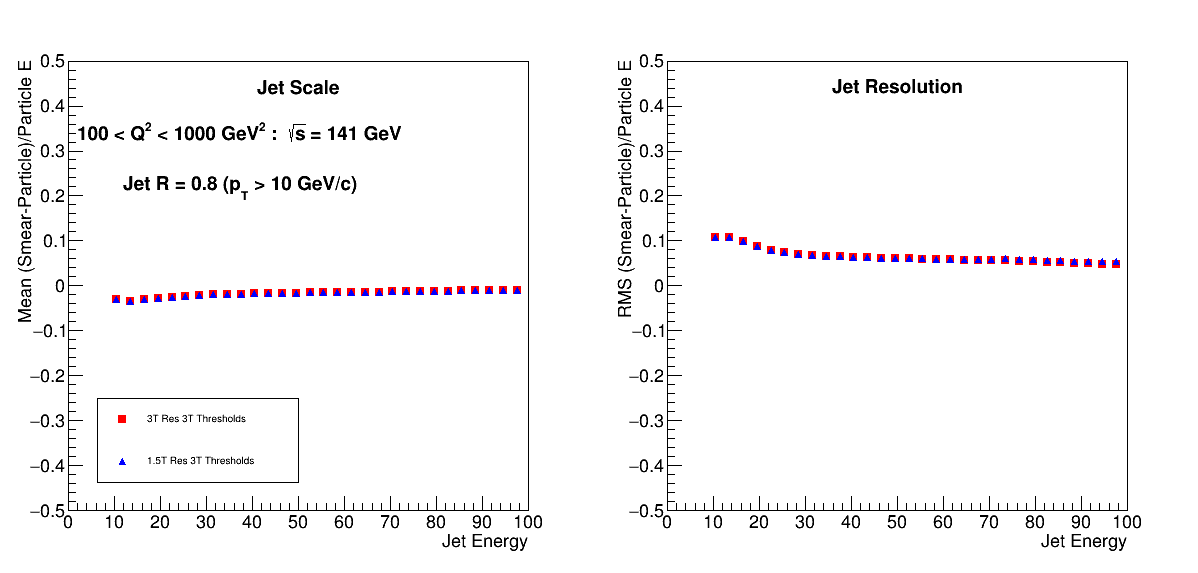}
\caption[Jet Energy Resolution]{Comparison between jet energy scale (left) and jet energy resolution (right) for the 3~T (red squares) and 1.5~T (blue triangles) track momentum resolutions listed in Tab.~\ref{PWG-sec-JHQ-tab-TRACKRES} as a function of jet energy. The track $p_T$ thresholds associated with the 3~T field were used and smearing was done in the eic-smear framework.}
\label{PWG-sec-JHQ-fig-JETENERGYRESPONSE}   
\end{figure}

For jet analyses, the primary metrics used to evaluate the suitability of track momenta 
resolutions are the jet energy resolution (JER) and jet energy scale (JES). A 
comparison of JES and JER for the 3~T and 1.5~T resolution configurations can be seen in 
Fig.~\ref{PWG-sec-JHQ-fig-JETENERGYRESPONSE} as a function of jet energy for $R = 0.8$ jets found in the laboratory frame 
using the Anti-k$_T$ algorithm with transverse momentum greater than 10~GeV/$c$. The event 
$Q^2$ range was limited to between 100 and 1000~GeV$^2$, but jets in the photoproduction 
region show the same behavior. The JES is taken as the mean of the smeared jet energy 
minus the true jet energy divided by the true jet energy distribution while the JER is 
the RMS. Note that for this comparison, the same set of minimum \pT\ thresholds were used 
in order to isolate the variation due to track momentum resolution. Also note that the matching 
procedure used was to select each truth level jet above the \pT\ threshold of 10~GeV/$c$ and 
then loop through all smeared jets in the event to find the one closest in $\eta - \phi$ space. 
This shows the extent that a truth level jet will be distorted by detector effects. This 
can also be inverted such that smeared jets above threshold are selected and all truth level 
jets are looped over to find the closest. In this case, the JES and JER will reflect biases which 
can arise when truth jets below threshold are smeared to higher energy. This will be discussed 
further in Sec.~\ref{part2-sec-DetReq.Jets.HQ.HCAL}.

As can be seen in Fig.~\ref{PWG-sec-JHQ-fig-JETENERGYRESPONSE}, there is practically no change to JES or JER between 
the 3~T and 1.5~T track momentum resolution settings, despite the roughly factor of two difference between the 
resolution parameters. The lack of impact on JER indicates that the contribution from the track momentum resolution 
constitutes a negligible amount to the overall jet energy resolution budget compared to other sources. This implies 
that bulk jet quantities will likely be insensitive to variations in tracking resolution at the scale expected for 
an EIC detector. Also, the 
magnitudes of the JES (several percent) and JER (less than 10\% for jet energies above roughly 20~GeV) 
should be amenable to unfolding corrections. This level of resolution has also been shown to 
be adequate for measurements for the lepton-jet Sivers asymmetry differential in 
$q_T = \overrightarrow{p_{T}}^{\mathrm{Lepton}} +  \overrightarrow{p_{T}}^{\mathrm{Jet}}$ 
(see Fig.~\ref{JetCollinsSiver} and \cite{Arratia:2020nxw}). While the magnitudes of the JES and JER also depend on the 
calorimeter resolutions, we can conclude that tracking resolutions of the order of those 
presented in Tab.~\ref{PWG-sec-JHQ-tab-TRACKRES} should be sufficient for our jet measurements. 

\subsubsection{Minimum \texorpdfstring{$p_T$}{pT} threshold and efficiency} \label{part2-sec-DetReq.Jets.HQ.ADDCONS}

In addition to momentum resolution, the minimum track transverse momentum threshold and 
tracking efficiency will be important parameters when determining the suitability of a 
detector design. As mentioned in the previous section, the assumed strength of the solenoidal 
magnetic field will affect the efficiency for reconstructing low transverse momentum particles. 
Table~\ref{PWG-sec-JHQ-tab-TRACKTHRESH} shows the assumed minimum \pT\ threshold corresponding to the 3~T and 1.5~T 
magnetic field strengths (see Tab.~\ref{tab:solenoid_magnet}). Note that these thresholds represent the points at which at least 
90\% of tracks could be reconstructed using a simple Kalman filter algorithm and more sophisticated 
reconstruction techniques could improve these thresholds. Reconstruction of lower momentum 
particles will also be possible, albeit at lower efficiencies.

\begin{table}[ht!]
	\setlength{\tabcolsep}{9.0pt}
	\begin{center}
	    \caption[Track Transverse Momentum Threshold]{Transverse momentum thresholds assumed for magnetic field strengths of 3~T and 1.5~T.}
		\scalebox{0.9}{
			\begin{tabular}{ c  c  c }
				\hline \hline
				Pseudorapidity Range & Min $p_T$ (3~T) [MeV/$c$] & Min $p_T$ (1.5~T) [MeV/$c$]\\ \hline
				$0.0 < |\eta| < 1.0$ & 400 & 200 \\
				$1.0 < |\eta| < 1.5$ & 300 & 150 \\
				$1.5 < |\eta| < 2.0$ & 160 & 70 \\
				$2.0 < |\eta| < 2.5$ & 220 & 130 \\
				$2.5 < |\eta| < 3.5$ & 150 & 100 \\ \hline \hline
			\end{tabular}
		}
		\label{PWG-sec-JHQ-tab-TRACKTHRESH}
	\end{center}
\end{table}

\begin{figure}[ht!]
\centering
\includegraphics[width=0.90\textwidth]{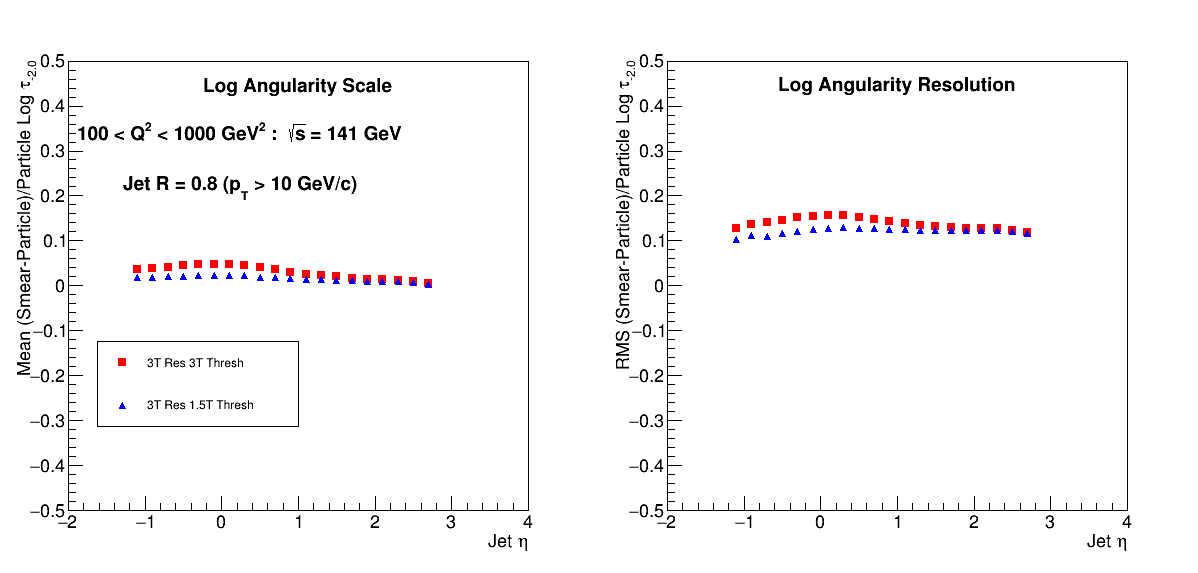}
\caption[Jet Angularity Resolution]{Comparison between jet angularity scale (left) and angularity resolution (right) for the 3~T (red squares) and 1.5~T (blue triangles) transverse momentum thresholds listed in Tab.~\ref{PWG-sec-JHQ-tab-TRACKTHRESH} as a function of jet pseudorapidity. The angularity parameter $a$ was set to -2.0. The track momentum resolutions associated with the 3~T field were used for both theshold settings and smearing was done in the eic-smear framework.}
\label{PWG-sec-JHQ-fig-JETANGULARITYRESPONSE}   
\end{figure}

As was the case with the momentum resolutions, the \pT\ thresholds have very little effect 
on the observed JES and JER, with slight improvement in JES for the lowest jet energies. 
While bulk jet quantities such as energy seem relatively unaffected by changes in the tracking 
resolution and threshold, more differential observables such as jet substructure should also 
be evaluated. Figure~\ref{PWG-sec-JHQ-fig-JETANGULARITYRESPONSE} shows the offset and RMS of the smeared vs particle level jet 
angularity (see Sec.~\ref{part2-subS-Hadronization-HadVacuum} and \cite{Aschenauer:2019uex}), defined analogously to the JES and JER. In order 
to isolate threshold effects, the 3~T momentum resolution parameters were used for both sets 
of threshold values. A 
preference can be seen for lower thresholds in both the scale and resolution. While the 
effect is not large, this indicates that substructure measurements will benefit from lower 
track \pT\ thresholds.

Lower track \pT\ thresholds will also be useful in the tagging and reconstruction of certain 
heavy hadrons. The $D^{*}$ for example decays to a $D^{0}$ via the emission of a soft charged 
pion. GEANT simulations utilizing an all silicon tracker concept show that for a minimum \pT\ threshold 
of roughly 200~MeV/$c$, the soft pion acceptance is 60\% at mid-rapidity and drops to 
a low of 20\% at $|\eta| = 3$. However, decreasing the tracking threshold to 100~MeV/$c$ increases 
the acceptance to 90\% at mid-rapidity with a falloff to 70\% at $|\eta| = 3$.

The effects of track \pT\ thresholds on jet substructure measurements and heavy hadron reconstruction 
imply that the bulk tracking efficiency will be an important consideration for jet and heavy flavor 
physics. Given the relatively low energy/transverse momentum of jets at the EIC, missing even a 
small fraction of tracks with moderate momentum could have a large effect on jet scale and resolution. 
In the absence of a recommended tracking efficiency function, many analyses performed by this group 
assumed an efficiency of 100\% above the minimum \pT\ threshold. The 1-jettiness analysis did look at 
the effect of variations in tracking efficiency and found a modest degradation of resolution when 
decreasing the tracking efficiency by 2\% below the nominal assumed value of 95\% when doing a 
track based evaluation of 1-jettiness (see Fig.~\ref{PWG-sec-JHQ-fig-JETENERGYRESPONSEECAL}). 

\subsubsection{Vertex resolution} \label{part2-sec-DetReq.Jets.HQ.VRES}

When considering heavy flavor reconstruction, there are two vertices of interest: the primary 
collision vertex where the initial scattering occurs, and a secondary vertex which is the 
origin of tracks arising from the decaying heavy meson. The reduction of background associated 
with the heavy meson reconstruction can benefit greatly from cuts on a number of topological 
relationships between the two vertices, such as the decay length (distance between vertices) and 
the distance of closest approach (DCA) between the primary vertex and reconstructed meson 
trajectory. One of the most relevant parameters for determining the 
vertices and the relationships 
between them is the transverse DCA and its resolution, $\sigma_{xy}$. This 
resolution is parameterized for different pseudorapidity ranges in the 
form $A/p_{T} \oplus B$ where $A$ and $B$ are in microns and $p_{T}$ is the track transverse 
momentum in GeV/$c$. The requirements on $\sigma_{xy}$ requested by our working group can be seen in 
Tab.~\ref{PWG-sec-JHQ-tab-VERTEXRES}. We currently do not request independent limits on the longitudinal 
DCA or primary vertex resolution.

\begin{table}[ht!]
	\setlength{\tabcolsep}{9.0pt}
	\begin{center}
	    \caption[Vertex Pointing Resolution]{Requested vertex position resolution.}
		\scalebox{0.9}{
			\begin{tabular}{ c  c }
				\hline \hline
				Pseudorapidity Range & Resolution \\ \hline
				$-3.5 < \eta < -3.0$ & N/A \\ 
				$-3.0 < \eta < -2.5$ & $\sigma_{xy} \sim 30/p_{T} \oplus 40~\mu\mathrm{m}$ \\ 
				$-2.5 < \eta < -1.0$ & $\sigma_{xy} \sim 30/p_{T} \oplus 20~\mu\mathrm{m}$ \\
				$-1.0 < \eta < 1.0$ & $\sigma_{xy} \sim 20/p_{T} \oplus 5~\mu\mathrm{m}$ \\ 
				$1.0 < \eta < 2.5$ & $\sigma_{xy} \sim 30/p_{T} \oplus 20~\mu\mathrm{m}$ \\ 
				$2.5 < \eta < 3.0$ & $\sigma_{xy} \sim 30/p_{T} \oplus 40~\mu\mathrm{m}$ \\
				$3.0 < \eta < 3.5$ & $\sigma_{xy} \sim 30/p_{T} \oplus 60~\mu\mathrm{m}$ \\ \hline \hline
			\end{tabular}
		}
		\label{PWG-sec-JHQ-tab-VERTEXRES}
	\end{center}
\end{table}

\begin{figure}[ht!]
\centering
\subfloat{\label{PWG-sec-JHQ-fig-DMESONS}
\includegraphics[width=0.90\textwidth]{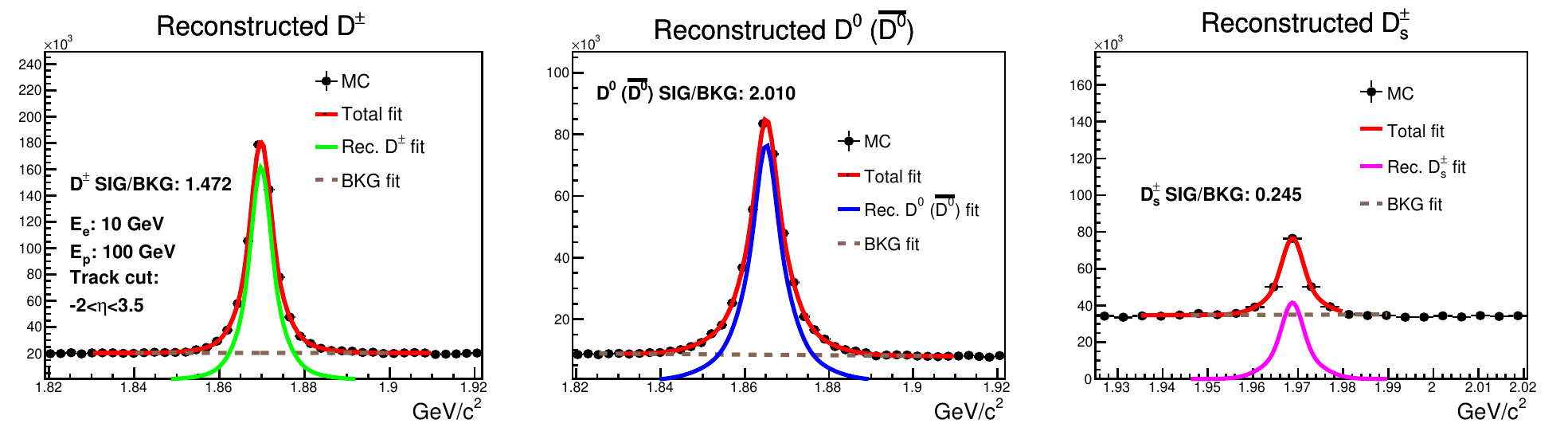}} \quad
\subfloat{\label{PWG-sec-JHQ-fig-BMESONS}
\includegraphics[width=0.90\textwidth]{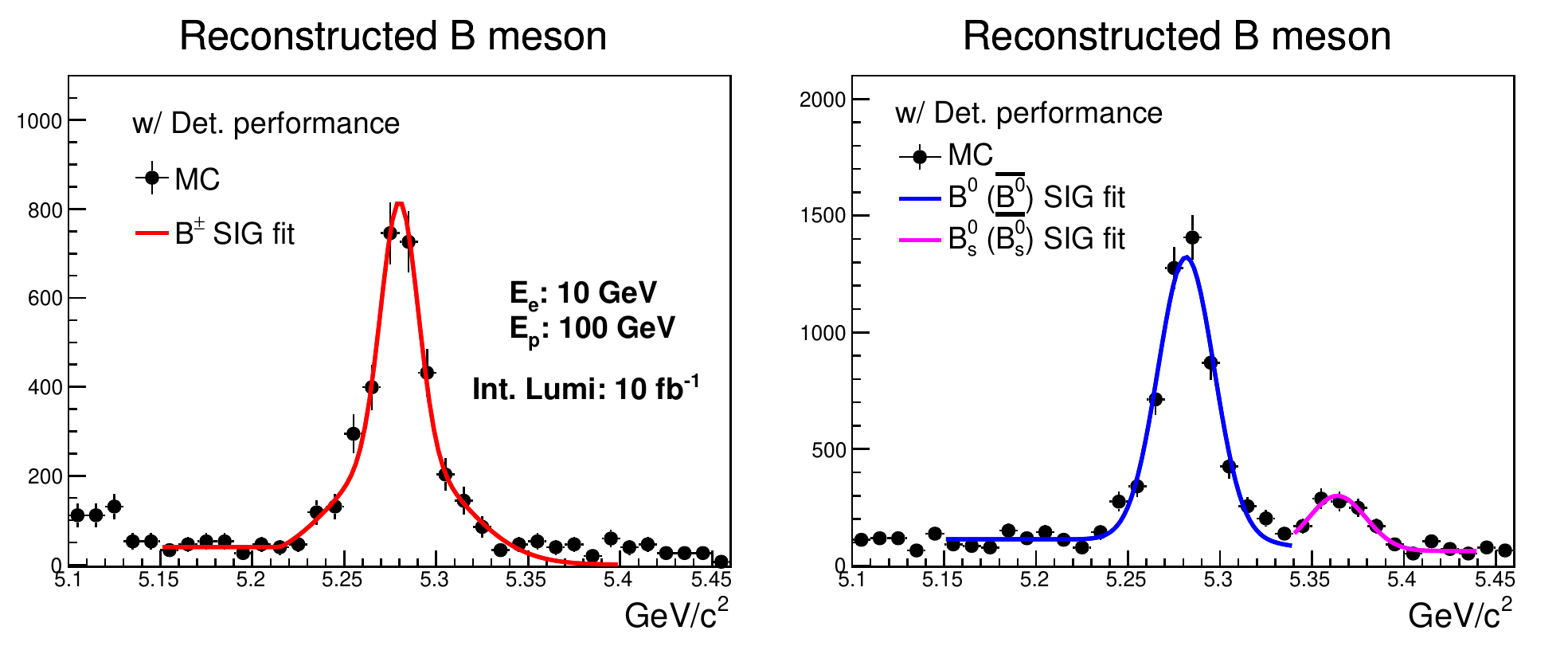}} 
\caption[Heavy Meson Reconstruction]{Reconstructed mass spectra for charm (top panels) and bottom (bottom panels) mesons achievable using the forward silicon tracker design described in~\cite{Wong:2020xtc} and assuming the transverse pointing resolutions listed in Tab.~\ref{PWG-sec-JHQ-tab-VERTEXRES}. Counts have been scaled to correspond to an integrated luminosity of 10~fb$^{-1}$.}
\label{PWG-sec-JHQ-fig-HEAVYMESONS}   
\end{figure}

\begin{figure}[t!]
	\centering \includegraphics[keepaspectratio=true, width=0.65\linewidth]{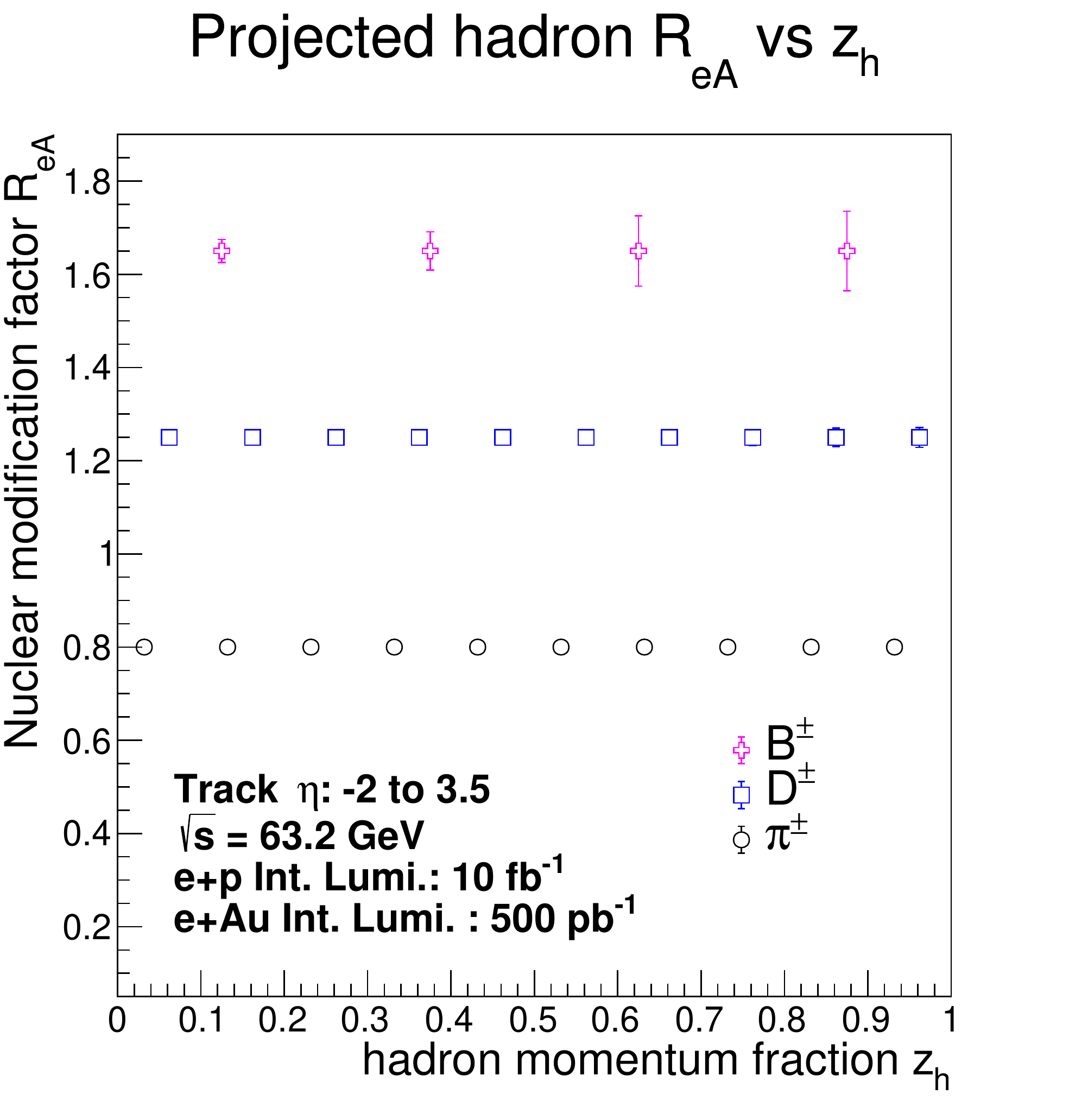}
	\caption[$R_{eA}$]{Achievable precision on the flavor dependent $R_{eA}$, shown as statistical error bars, as a function of hadron mometum fraction $z_h$ given the mass reconstruction shown in Fig.~\ref{PWG-sec-JHQ-fig-HEAVYMESONS}.}
	\label{PWG-sec-JHQ-fig-HFRECOREA}
\end{figure}

The requirements listed in Tab.~\ref{PWG-sec-JHQ-tab-VERTEXRES} were largely derived from feasibility studies 
of using open charm and bottom mesons in high precision measurements of 
the nuclear modification factor $R_{eA}$ over a wide pseudorapidity range in order to discriminate 
between different models of parton energy loss and hadronization. Further details on the models and 
required precision on $R_{eA}$ can be found in~\cite{Li:2020zbk}. 
The requested $\sigma_{xy}$ resolutions lead to the $D$ and $B$ reconstructed mass spectra seen in Fig.~\ref{PWG-sec-JHQ-fig-HEAVYMESONS}, 
which in turn lead to the precision extractions of $R_{eA}$ in Fig.~\ref{PWG-sec-JHQ-fig-HFRECOREA}. A detailed technical 
note describing designs of a silicon vertex detector and forward tracking system leading to the 
performance listed above can be found here: \cite{Wong:2020xtc}.

While the parameters above have been shown to be sufficient for measurements of $R_{eA}$, it should 
be noted that more differential measurements may benefit from higher vertex 
resolutions. The advantage would come in the form of better signal to noise ratios which would 
lead to higher significance measurements for a given sampled luminosity. Studies based on an all 
silicon tracking system~\cite{Arrington:2021yeb} have shown that an improvement in $\sigma_{xy}$ 
from 20~$\mu\mathrm{m}$ to 10~$\mu\mathrm{m}$ at a track $p_T$ of 1~GeV/$c$ can increase $D^0$ significance 
by 20\%. In addition, other heavy mesons of interest, such as the $\Lambda_{c}^{+}$ have transverse displacements 
from the primary vertex of on the order 20~$\mu\mathrm{m}$, making tighter resolutions advantageous~\cite{Arrington:2021yeb}. 

The effects of vertex resolution have also been studied in the context of charm jet tagging via (among other 
parameters) the presence of a certain number of `high impact parameter' tracks \cite{Arratia:2020azl}. Here, the (unsigned) 
impact parameter is defined as 
$\sqrt{(d_{0}/\sigma_{d_{0}})^{2} + (z_{0}/\sigma_{z_{0}})^{2}}$ 
where $d_0$ 
and $z_0$ are the distances of closest approach of a track in the transverse and longitudinal direction, 
respectively and $\sigma_{d_{0}} = \sigma_{xy}$ and $\sigma_{z_{0}}$ are the corresponding uncertainties. The tagging 
procedure was optimized assuming $\sigma_{d_{0}} = \sigma_{z_{0}} = 20$~$\mu\mathrm{m}$. To test the impact of 
vertexing performance, both DCA resolutions were set to 100~$\mu\mathrm{m}$ and the tagging procedure re-optimized. 
The degraded resolutions led to a 60\% loss in charm jet efficiency. A more optimistic scenario was also 
tested with both resolutions being improved to 10~$\mu\mathrm{m}$ leading to a 30\% increase in charm yield for 
the same light jet contamination. While it is clear that better $\sigma_{xy}$ resolution will benefit this 
measurement, it has not been demonstrated that the requested resolutions are inadequate. Therefore, we did 
not elevate these considerations to the level of requirements.

\begin{figure}
	\centering \includegraphics[keepaspectratio=true, width=0.95\linewidth]{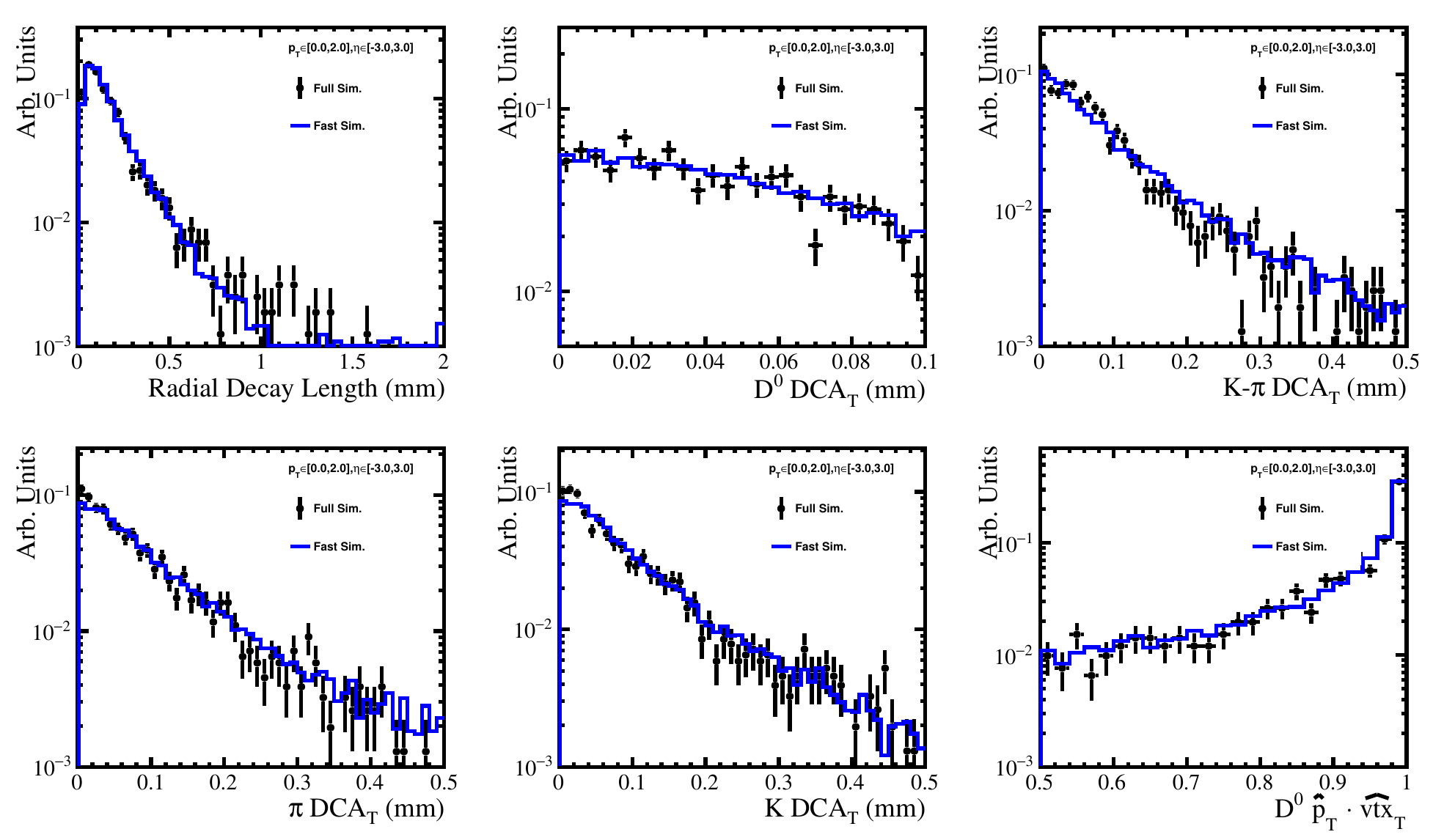}
	\caption{Comparison of the reconstructed $D^{0}$ topological variables in the GEANT4-based all silicon simulation (data points) and the fast simulation (blue histograms)~\cite{Arrington:2021yeb}. All distributions are normalized to have unit area. The $D^{0}$ candidates shown here are required to have a $|\eta|<3$ and $p_{T}<2$ GeV/$c$.}
	\label{fig:PWG-sec-JHQ-fig-fastfull}
\end{figure}

It should be noted that eic-smear incorporates fast simulation of track parameters, but does not incorporate vertex fitting.  This presents a conundrum for several heavy-quark studies in this report where they rely on displaced vertex techniques.
An extensive set of GEANT4-based simulations have been performed to extend fast simulations of the detector response for topological observables~\cite{Arrington:2021yeb}. 
This response, illustrated in Figure~\ref{fig:PWG-sec-JHQ-fig-fastfull} for $D^0$ mesons, has been propagated in the studies of gluon polarization and the gluon Sivers' TMD using heavy quark probes discussed in the preceding chapter.

\subsection{Particle identification} \label{part2-sec-DetReq.Jets.HQ.PID}

The ability to positively identify charged hadron species (topics such as electron and heavy meson 
identification are not covered here) will enable a suite of measurements which are sensitive to 
flavor dependencies in the final state. Chief among these will be analyses of the unpolarized 
identified hadron-in-jet fragmentation functions and the related polarized Collins asymmetry 
(Sec.~\ref{part2-subS-SecImaging-TMD3d} and \cite{Arratia:2020nxw}). It is also expected that 
certain grooming and substructure techniques, as well as novel observables which track 
correlations between leading and sub-leading jet particles (see Sec.~\ref{part2-subS-LabQCD-Special}), will utilize 
PID to trace flavor evolution through the jet shower. PID will also be an asset to heavy flavor 
tagging as charm and bottom mesons will often contain a kaon in their decay chain, while 
lambda particles will emit a proton, so being able to tag these particles will help reduce 
combinatoric background.

\begin{figure}[ht!]
\centering
\subfloat{\label{PWG-sec-JHQ-fig-PARTETAVSPLOW}
\includegraphics[width=0.45\textwidth]{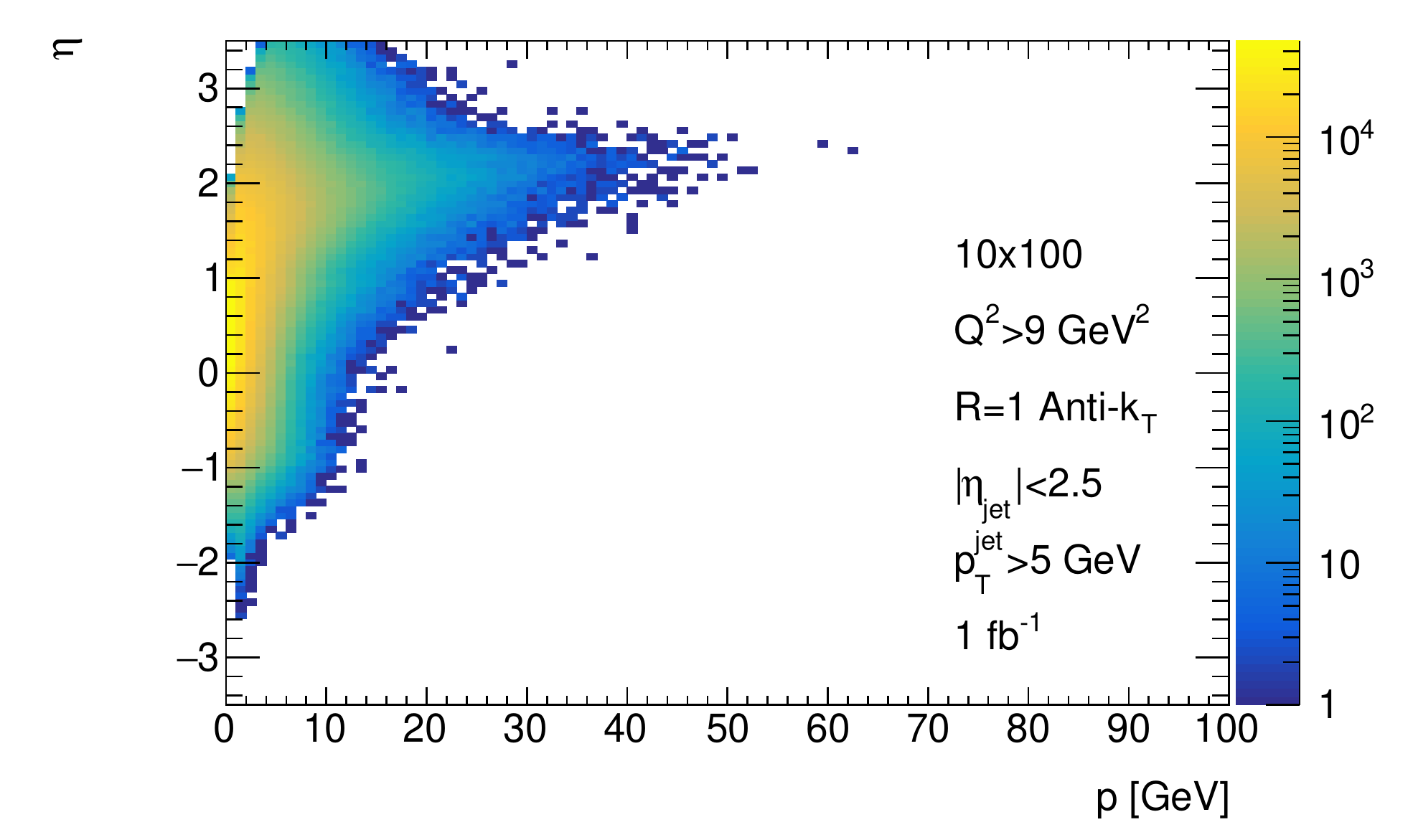}} \quad
\subfloat{\label{PWG-sec-JHQ-fig-PARTETAVSPHIGH}
\includegraphics[width=0.45\textwidth]{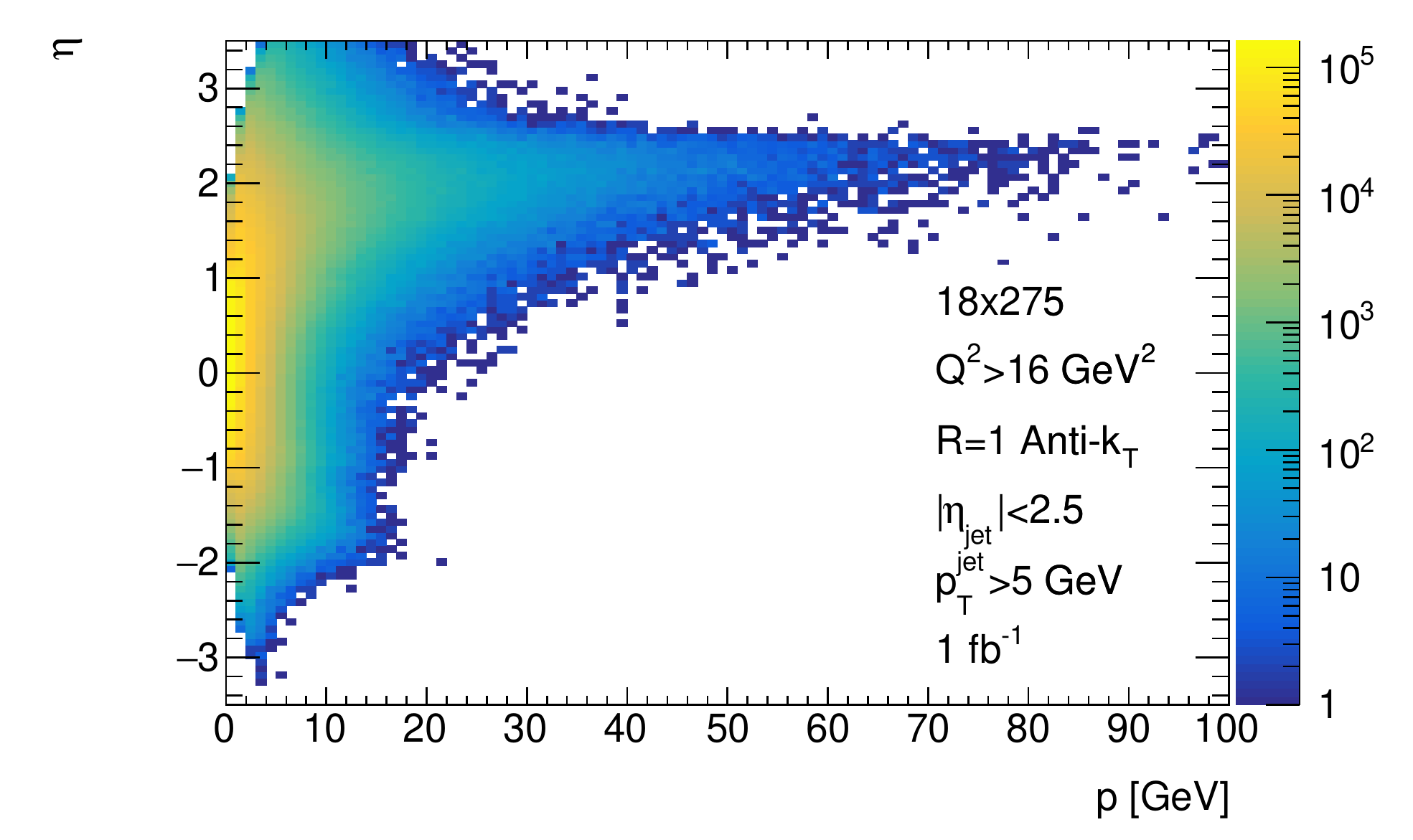}}
\caption[Particle in Jet Momentum]{Pseudorapidity vs momentum for charged hadrons found inside jets with a radius 1.0 at beam energies of 10x100~GeV (left) and 18x275~GeV (right). Counts have been scaled to correspond to an integrated luminosity of 10~fb$^{-1}$.}
\label{PWG-sec-JHQ-fig-PARTETAVSP}   
\end{figure}

When evaluating the PID needs for an analysis, there are two aspects to consider: how well the charged 
hadron can be identified and the momentum range over which a given identification power is needed. The 
requirements on momentum range are informed simply by the momentum spectrum of charged hadrons within 
reconstructed jets and the desire to measure as much of the $z = p^{\mathrm{had}}/p^{\mathrm{jet}}$ 
spectrum as possible. The pseudorapidity vs momentum for charged hadrons inside of jets at moderate to 
large $Q^2$ can be see in Fig.~\ref{PWG-sec-JHQ-fig-PARTETAVSP} for beam energy combinations of 10x100~GeV and 18x275~GeV. As expected, 
the maximum particle momentum increases with pseudorapidity until roughly $\eta = 2.2$, where the particle 
momentum begins to drop due to an interplay between fiducial volume cuts and the jet radius. 

The hadron momenta spectra shown in Fig.~\ref{PWG-sec-JHQ-fig-PARTETAVSP} inform our requested PID momentum coverage 
presented in Tab.~\ref{PWG-sec-JHQ-tab-PIDMOM}, where the momentum limit represents the particle momentum up to which 
at least $3\sigma$ separation between pions and kaons would be possible. This request is more 
ambitious in terms of momentum reach than the initially proposed values of 5~GeV/$c$ in the barrel and between 
8~GeV/$c$ and 45~GeV/$c$ over the hadronic endcap. The barrel request in particular will necessitate 
additional research and development to realize a technical path to separation at momenta above 5~GeV/$c$. 
Despite this, the requested PID capabilities will enable a host of measurements which would not be possible 
otherwise. For example, measurements with light and Heavy Ions, which have maximum beam energies of 100-110~GeV, will 
see significant fraction of jets produced in the barrel, and the phase space which is not covered due to low PID limits 
will not be recoverable by moving to different detector regions or center-of-mass energies.

\begin{table}[ht!]
	\setlength{\tabcolsep}{9.0pt}
	\begin{center}
	    \caption[PID Momentum Range]{Requested PID momentum coverage for 3$\sigma$ pion/kaon separation.}
		\scalebox{0.9}{
			\begin{tabular}{ c  c }
				\hline \hline
				Pseudorapidity Range & Momentum Range \\ \hline
				$-3.5 < \eta < -1.0$ & $\leq 7~\mathrm{GeV}/c$ \\ 
				$-1.0 < \eta < 0.5$ & $\leq 10~\mathrm{GeV}/c$ \\ 
				$0.5 < \eta < 1.0$ & $\leq 15~\mathrm{GeV}/c$ \\
				$1.0 < \eta < 1.5$ & $\leq 30~\mathrm{GeV}/c$ \\ 
				$1.5 < \eta < 2.5$ & $\leq 50~\mathrm{GeV}/c$ \\ 
				$2.5 < \eta < 3.0$ & $\leq 30~\mathrm{GeV}/c$ \\
				$3.0 < \eta < 3.5$ & $\leq 20~\mathrm{GeV}/c$ \\ \hline \hline
			\end{tabular}
		}
		\label{PWG-sec-JHQ-tab-PIDMOM}
	\end{center}
\end{table}

\begin{figure}[ht!]
\centering
\subfloat[\label{PWG-sec-JHQ-fig-JTVSZ}]{
\includegraphics[width=0.45\textwidth]{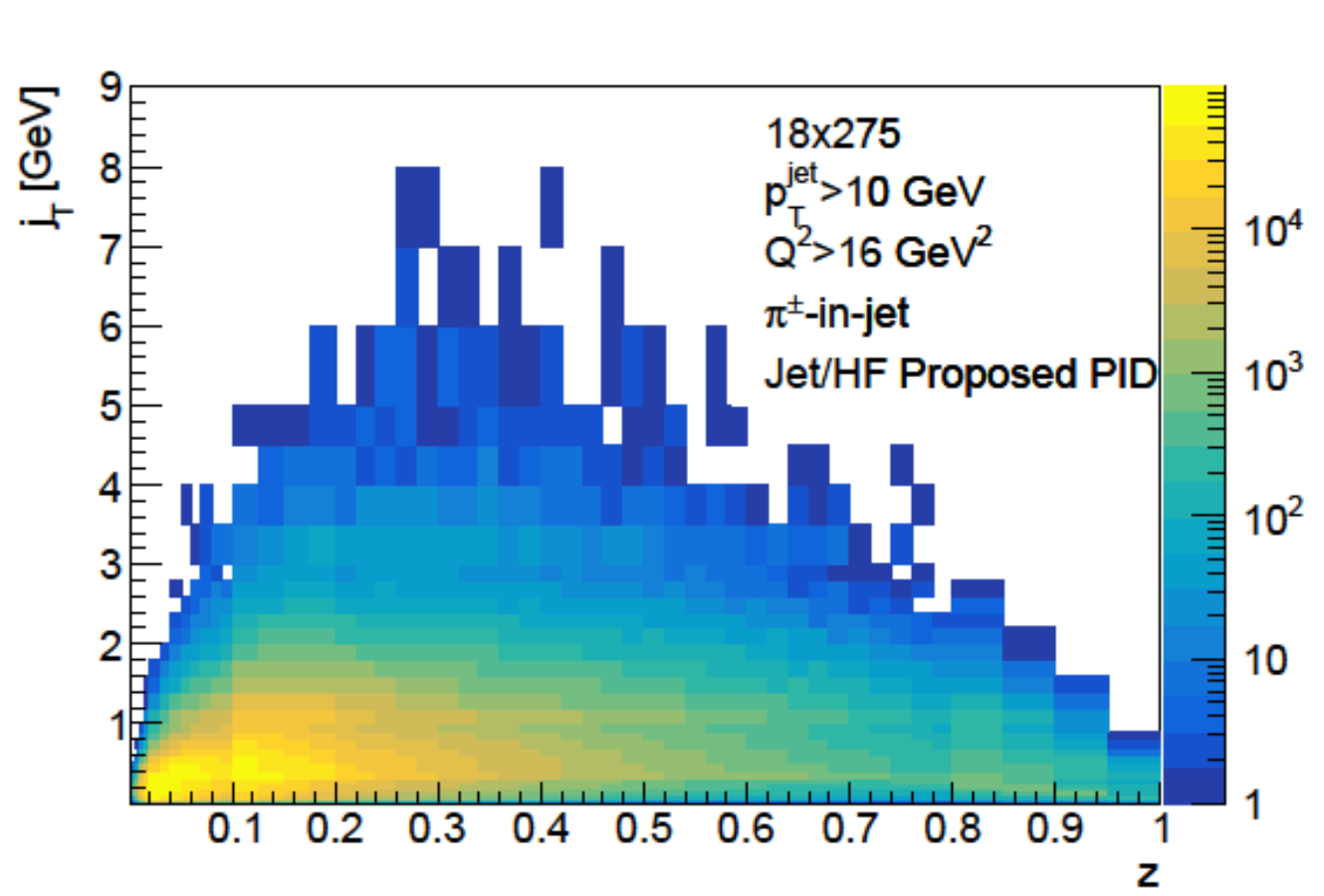}} \quad
\subfloat[\label{PWG-sec-JHQ-fig-JTVSZRATIO}]{
\includegraphics[width=0.45\textwidth]{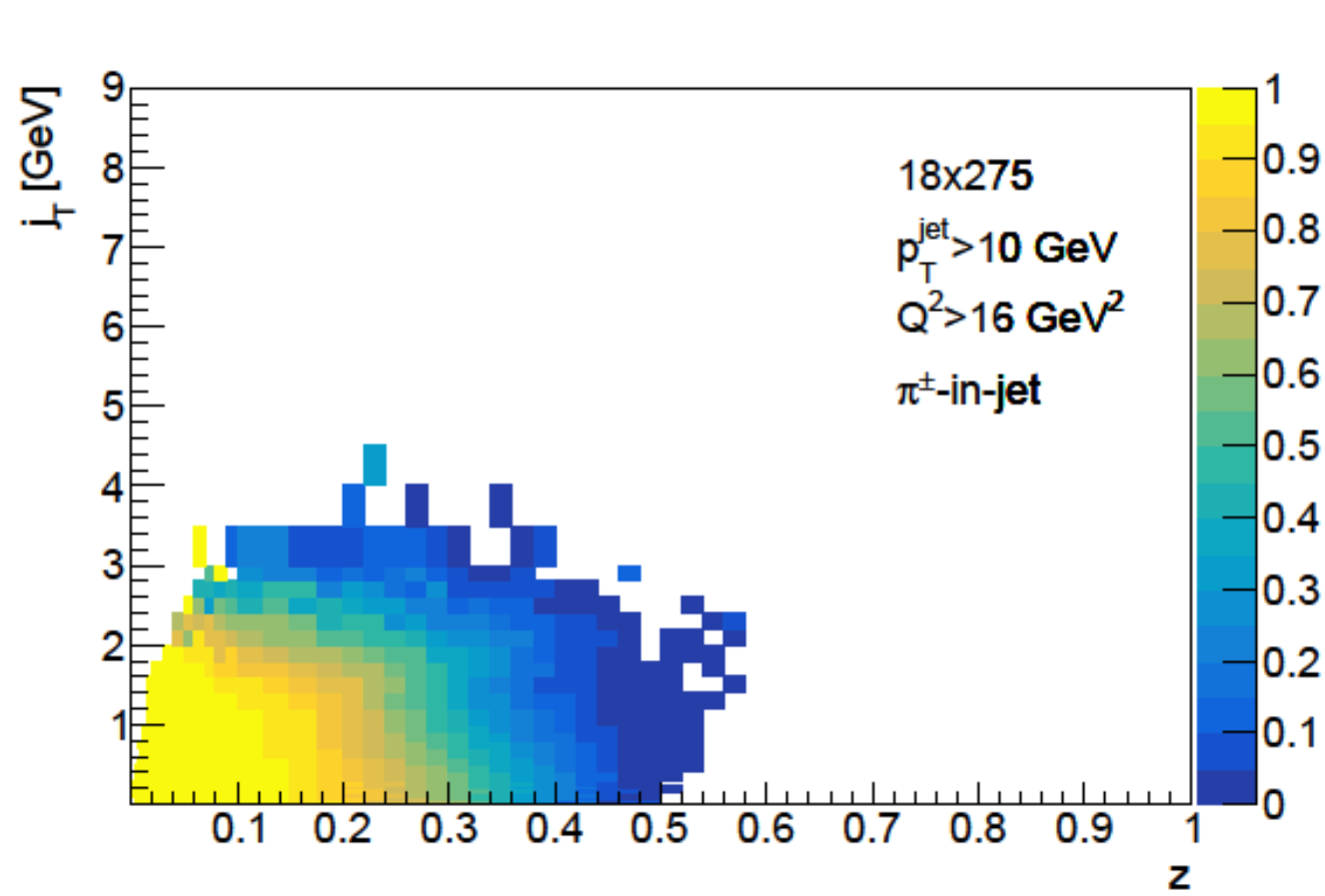}}
\caption[PID Phase Space]{(Left) $j_T$ vs $z$ distribution for identified hadrons inside jets assuming the PID capability listed in Tab.~\ref{PWG-sec-JHQ-tab-PIDMOM}. (Right) Ratio of the accessible $j_T$ vs $z$ phase space assuming restricted PID momentum limits enumerated above over the phase space assuming the momentum limits in Tab.~\ref{PWG-sec-JHQ-tab-PIDMOM}.}
\label{PWG-sec-JHQ-fig-PIDJTVSZ}   
\end{figure}

One way of characterizing the kinematic distribution of hadrons within 
a jet is via the particle $j_T$ and $z$, which are the transverse momentum of the particle 
with respect to the jet axis and the ratio of the total particle momentum over the jet 
momentum, respectively. Figure~\ref{PWG-sec-JHQ-fig-PIDJTVSZ} demonstrates the loss in $j_T$ and $z$ 
coverage for identified particles which would result from lowering the PID momentum reach 
with respect to what we have requested. Figure~\ref{PWG-sec-JHQ-fig-JTVSZ} shows the available phase space for 
identified hadrons assuming the PID momentum ranges in Tab.~\ref{PWG-sec-JHQ-tab-PIDMOM} while Fig.~\ref{PWG-sec-JHQ-fig-JTVSZRATIO} 
shows the ratio of the restricted momentum coverage phase space to the requested momentum 
coverage phase space. The restricted momentum coverage values were taken as $\leq 5$~GeV/$c$ 
for $-1.0 < \eta < 1.0$, $\leq 8$~GeV/$c$ for $1.0 < \eta < 2.0$, $\leq 20$~GeV/$c$ for 
$2.0 < \eta < 3.0$, and $\leq 45$~GeV/$c$ for $3.0 < \eta < 3.5$. For the highest $\eta$ bin, 
track momenta within a jet are restricted to approximately 20~GeV because the jet thrust axis 
needs to be at least one jet radius distance away from the detector fiducial volume edge, so the 
`restricted' value was simply taken as the expected performance range given by the detector group. 
It is readily apparent that 
a large fraction of $j_T$ coverage is lost with the more restrictive PID reach and $z$ values 
above 0.5 become inaccessible.

\begin{figure}[ht!]
\centering
\subfloat{\label{PWG-sec-JHQ-fig-JTVSZBARREL}
\includegraphics[width=0.80\textwidth, clip=true, trim = 100bp 100bp 100bp 90bp]{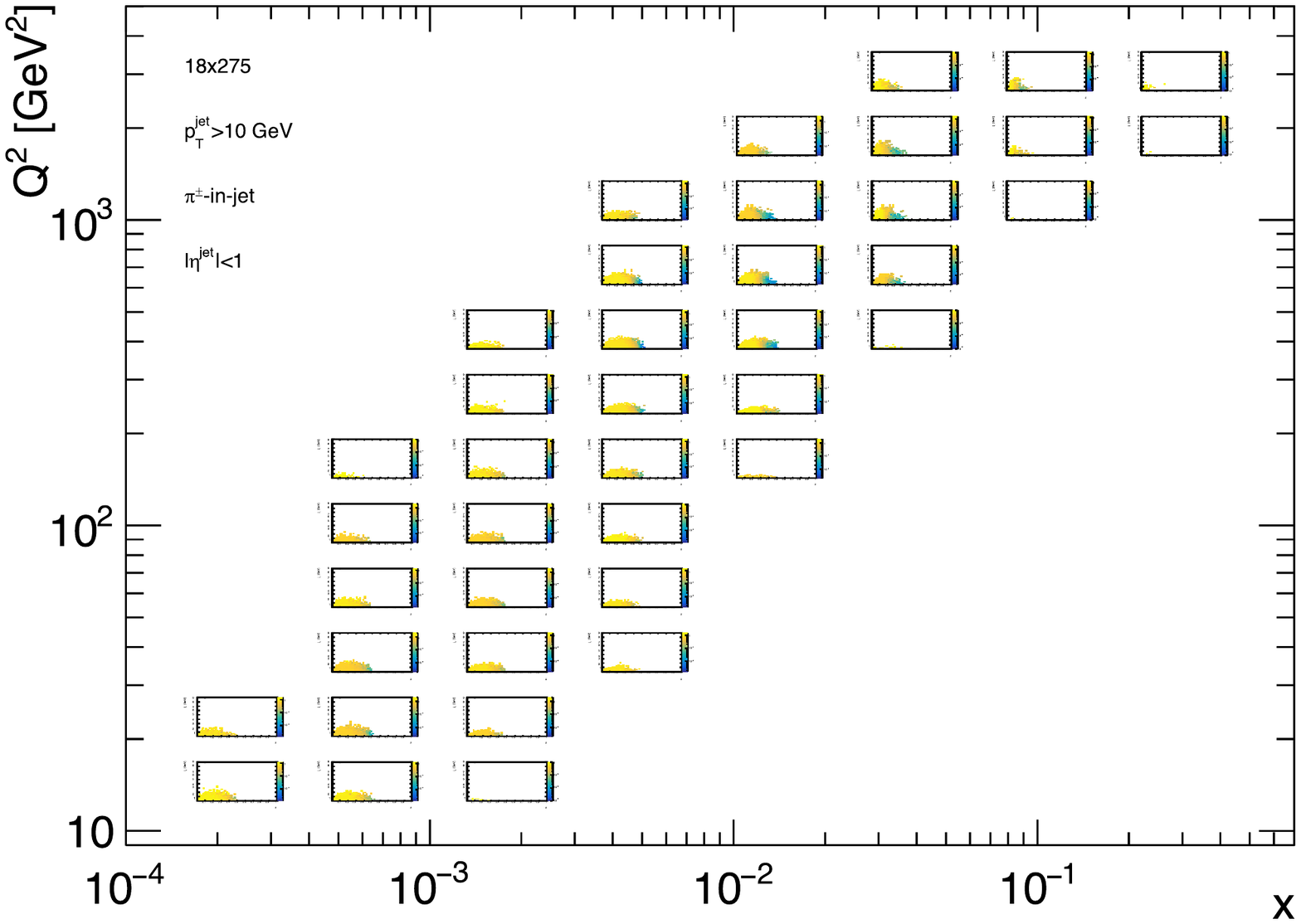}} \quad
\subfloat{\label{PWG-sec-JHQ-fig-JTVSZENDCAP}
\includegraphics[width=0.80\textwidth, clip=true, trim = 100bp 100bp 100bp 90bp]{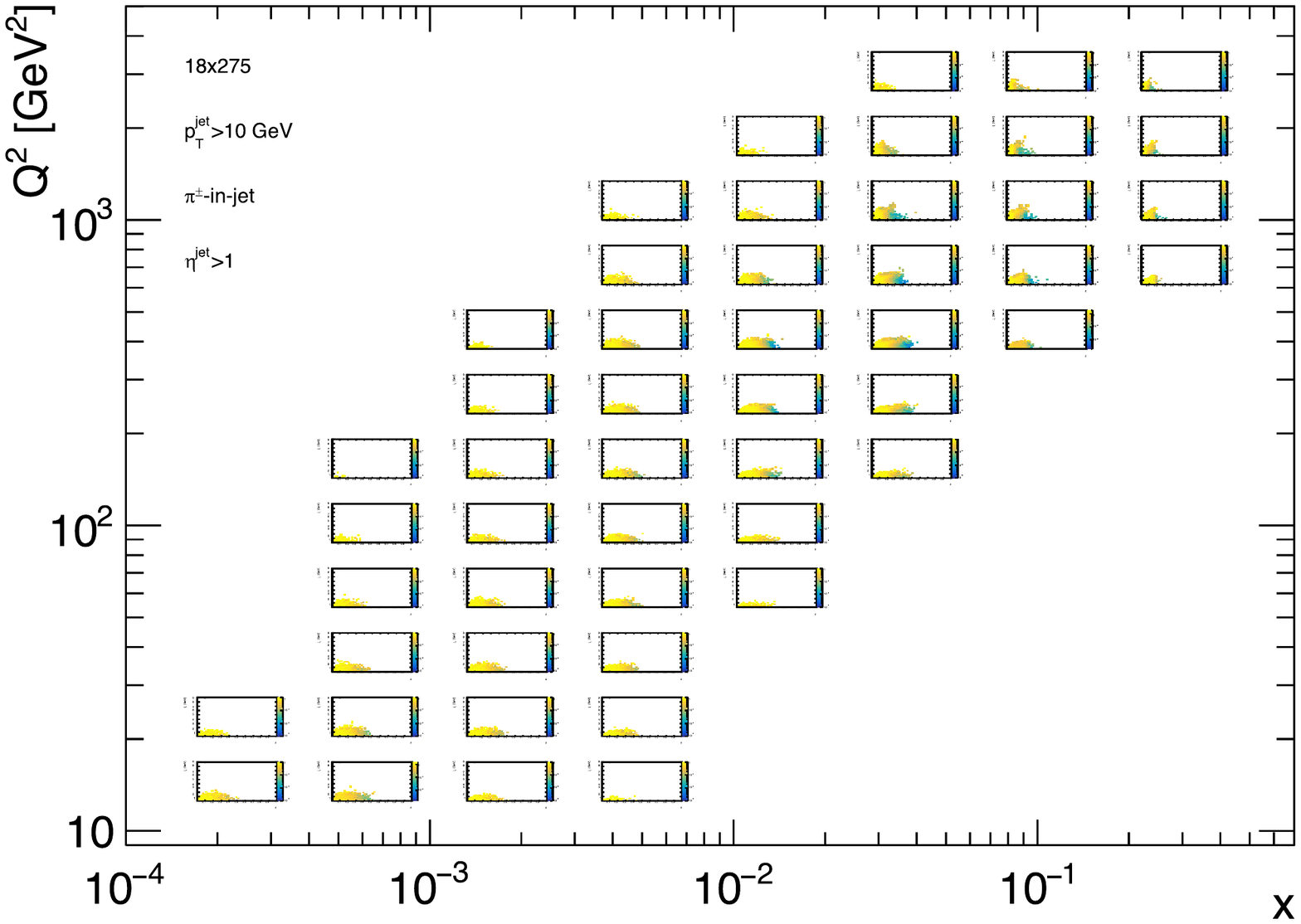}}
\caption[PID Phase Space $x$ vs $Q^2$]{Ratio of the accessible $j_T$ vs $z$ phase space assuming restricted PID momentum limits enumerated above over the phase space assuming the momentum limits in Tab.~\ref{PWG-sec-JHQ-tab-PIDMOM} in bins of $x$ and $Q^2$ for jets with $-1 < \eta < 1$ (top panel) and $1 < \eta < 3.5$ (bottom panel).}
\label{PWG-sec-JHQ-fig-PIDJTVSZXQ2}   
\end{figure}

Figure~\ref{PWG-sec-JHQ-fig-PIDJTVSZ} shows the effect of restricted PID coverage integrated over $x$ and $Q^2$, as well 
as over the full pseudorapidity range of the central detector. To get a better idea of where coverage losses occur, 
Fig.~\ref{PWG-sec-JHQ-fig-PIDJTVSZXQ2} displays the restricted over requested PID coverage ratio as a function of $x$ 
and $Q^2$. Each sub-panel is equivalent to Fig.~\ref{PWG-sec-JHQ-fig-JTVSZRATIO}, and yellow regions correspond to 
small losses while blue and blank regions indicate areas where much of the coverage is lost. Because the technical 
feasibility for extending PID momentum reach is different in the barrel and endcap regions, we show the phase space 
loss for the $-1 < \eta < 1$ and $1 < \eta < 3.5$ regions separately.

\begin{figure}[ht!]
\centering
\subfloat[\label{PWG-sec-JHQ-fig-COLL1}]{
\includegraphics[width=0.80\textwidth]{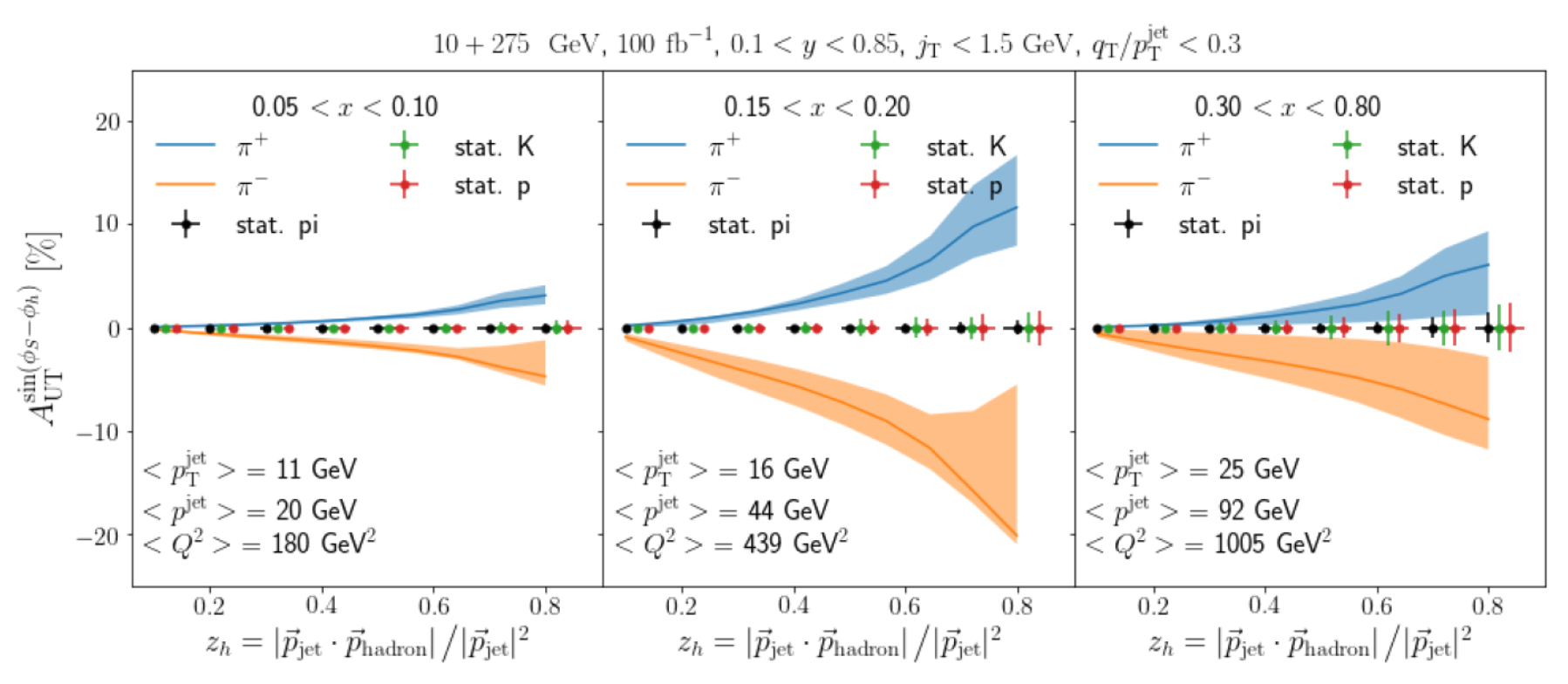}} \quad
\subfloat[\label{PWG-sec-JHQ-fig-COLL2}]{
\includegraphics[width=0.80\textwidth]{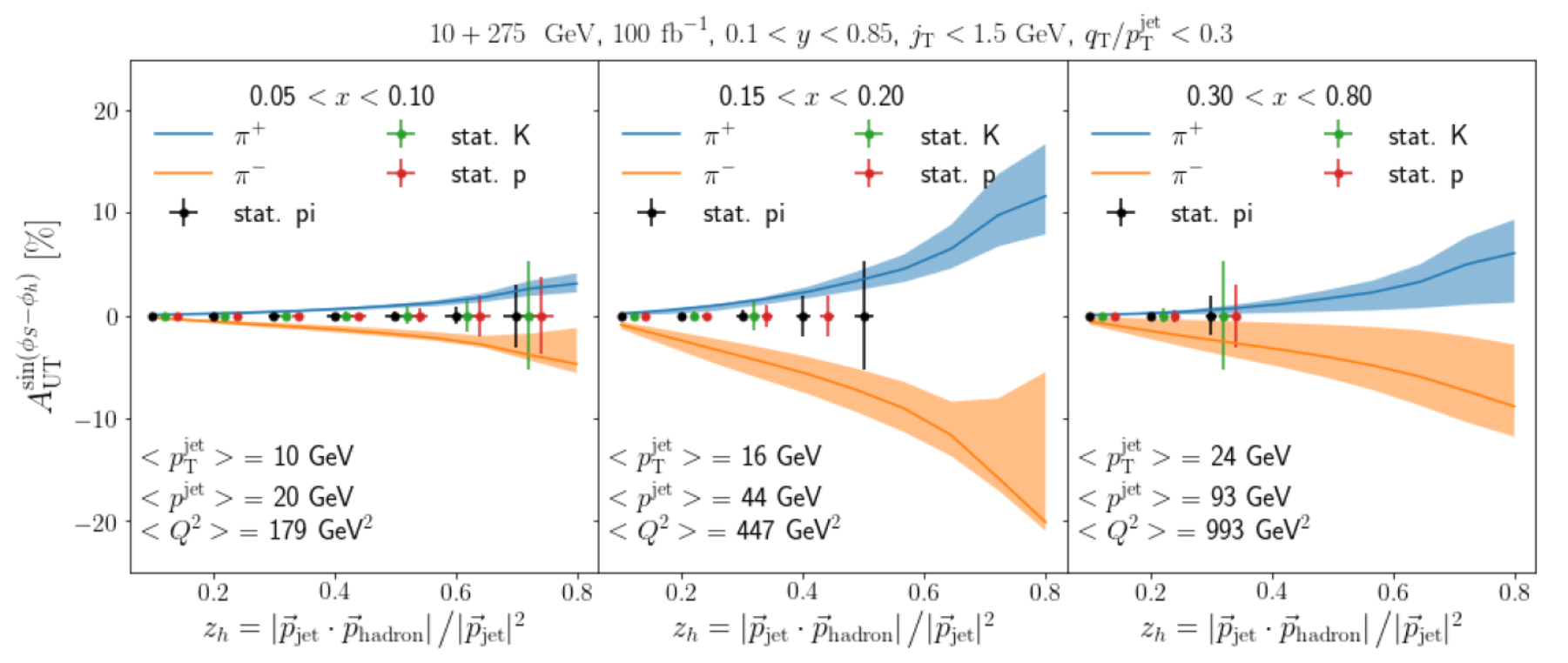}} \quad
\subfloat[\label{PWG-sec-JHQ-fig-COLL3}]{
\includegraphics[width=0.80\textwidth]{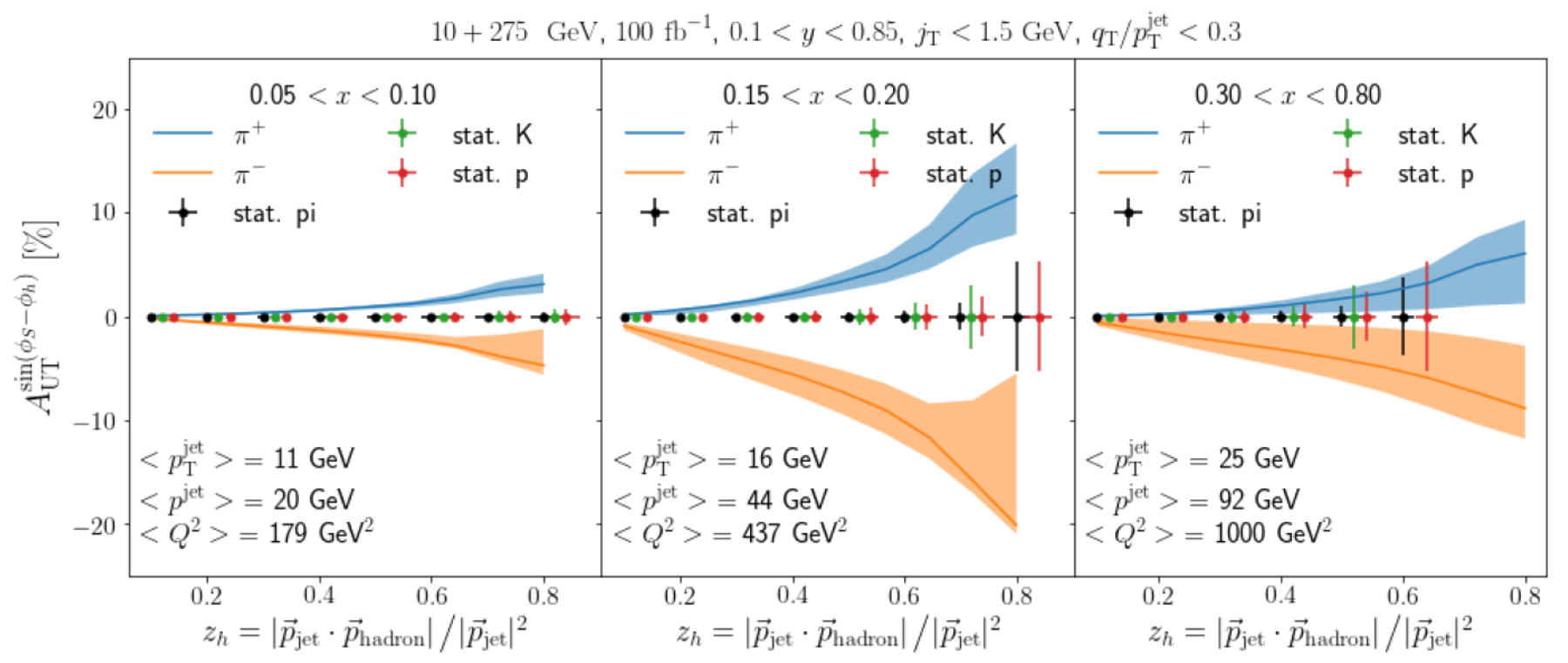}}
\caption[Collins Asymmetry]{Expected precision of Collins asymmetry measurement for identified hadrons as a function of $z$ in three $x$ bins assuming perfect PID (top panel), PID with the restricted momentum reach (middle panel), and PID with the requested momentum reach (bottom panel). The blue and orange curves (bands) correspond to theoretical predictions (uncertainties) for the $\pi^+$ and $\pi^-$ Collins asymmetries. See \cite{Arratia:2020nxw} and references therein for further details.}
\label{PWG-sec-JHQ-fig-COLL}   
\end{figure}

The impact of this phase space loss on the Collins asymmetry measurement can be seen in 
Fig.~\ref{PWG-sec-JHQ-fig-COLL}, which shows the statistical precision of the extracted Collins asymmetry for 
identified particles as a function of $z$ for three $x$ ranges assuming perfect PID (\ref{PWG-sec-JHQ-fig-COLL1}), 
the restricted PID momentum ranges listed above (\ref{PWG-sec-JHQ-fig-COLL2}), and the requested PID momentum ranges 
listed in Tab.~\ref{PWG-sec-JHQ-tab-PIDMOM} (\ref{PWG-sec-JHQ-fig-COLL3}). It is evident that restricting the PID coverage 
below our requested values, especially in the region $1.0 < \eta < 2.5$, will severely restrict the ability to measure the Collins asymmetry in 
the high $z$ region where both the expected asymmetry and theoretical uncertainties are largest.

Another observable for which PID will be necessary is the relative charge asymmetry between leading and sub-leading 
identified particles within a jet, $r_{asy}=(N_{CC}-N_{C\overline{C}})/(N_{CC}+N_{C\overline{C}})$ (Sec.~\ref{part2-subS-LabQCD-Special}). Here $N_{CC}$ 
and $N_{C\overline{C}}$ represent the yields when the leading and sub-leading particles have the same and opposite 
electric charge, respectively. Plotting this observable as a function of various kinematic variables like 
$z = p_{NL}/(p_{L} + p_{NL})$ will allow studies 
of correlations between energy flow and quantum numbers such as flavor, spin, electric and color charges among the 
particles in jet and shed light on jet formation and evolution. Here $p_{NL}$ and $p_L$ denote the momenta of the 
sub-leading and leading particles, respectively.

\begin{figure}[ht!]
\centering
\includegraphics[width=0.9\textwidth]{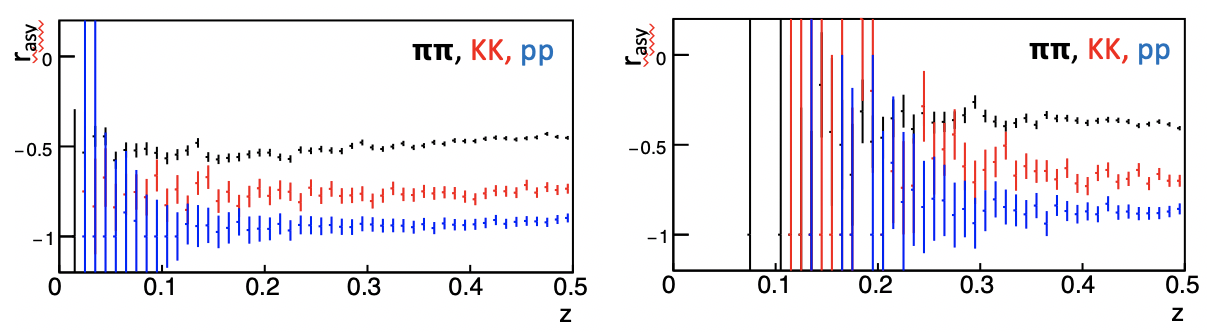}
\caption[Subleading Particle Asymmetry]{Left: $r_{asy}$ with z for PID limit up to  10~GeV/$c$. Right: $r_{asy}$ with z for  PID limit up to  5~GeV/$c$.}
\label{PWG-sec-JHQ-fig-RASYM}
\end{figure}

Figure~\ref{PWG-sec-JHQ-fig-RASYM} demonstrates the effect of limited PID momentum coverage on the $z$ reach of 
$r_{asy}$. The left panel shows the expected statistical precision on $r_{asy}$ as a function of $z$ for identified 
proton-proton, kaon-kaon, and pion-pion pairs assuming PID capabilities up to 10~GeV/$c$ for $-1 < \eta < 1$. Results 
are for 18~GeV electrons on 275~GeV protons and $Q^2 > 65$~GeV$^2$. The right panel shows the increase in statistical 
uncertainty when barrel PID is restricted to less than 5~GeV. It is evident extracting useful data from the $z<0.4$ 
region will be difficult under the 5~GeV limit scenario.

\subsection{Calorimetry} \label{part2-sec-DetReq.Jets.HQ.CALO}

For analyses considered by the Jets and Heavy Quarks working group, electromagnetic and 
hadron calorimetry will be utilized 
primarily in jet and event shape reconstruction where it will compliment information from the 
tracking detector to measure the full energy of the event. The calorimeters will 
provide the only means of measuring neutral energy, such as photons from (predominantly) $\pi^0$ 
decay or long-lived neutral hadrons such as neutrons and $K^{0}_{\mathrm{L}}$s. The electromagnetic 
and neutral hadron energy fractions in a typical jet are roughly 25\% and 10\%, respectively. As 
discussed in Sec.~\ref{part2-sec-DetReq.Jets.HQ.TRACKING}, the superior momentum resolution of the tracker will 
make it the primary subsystem for detecting charged hadrons, yet at the most forward rapidities, 
jet energies are expected to be large enough for the calorimeter resolution to be competitive or 
possibly superior to that of the tracker.

\subsubsection{Electromagnetic calorimetry} \label{part2-sec-DetReq.Jets.HQ.ECAL}

The primary performance criteria for the electromagnetic calorimeters (ECals) is energy resolution, 
which directly affects the jet energy resolution as well as the accuracy of the reconstructed 
event shape. The resolution of the ECals, especially in the lepton-going direction, are primarily 
driven by the need to accurately reconstruct the scattered beam electron in order to reconstruct 
the event kinematics. As with the track momentum resolution, several sets of `reasonable' ECal 
resolutions were provided and assessed by this working group. The bulk of our analyses, such as jet 
TMD studies (see Sec.~\ref{part2-subS-SecImaging-TMD3d.jets} and \cite{Arratia:2020nxw}) used values 
based off the EIC R\&D Handbook~\cite{EIC:RDHandbook}, which were parameterized in the form $A\%/\sqrt{E} \oplus B\%$. 
Recently, new values were released based on work performed during the Yellow Report effort and 
parameterized in the form $A\%/\sqrt{E} \oplus B\% \oplus C/E$. In addition, a set of improved resolutions 
based on the most optimistic technology choices was used to investigate variations in jet performance 
with changes in ECal resolution. All resolution values investigated are summarized in Tab.~\ref{PWG-sec-JHQ-tab-ECALRES}.

\begin{table}[ht!]
	\setlength{\tabcolsep}{9.0pt}
	\begin{center}
	    \caption[ECal Resolution]{Electromagnetic calorimeter energy resolutions investigated by the Jets and Heavy Quarks group. See text for details of resolution sets.}
		\scalebox{0.8}{
			\begin{tabular}{ c  c  c  c }
				\hline \hline
				Pseudorapidity range & HBK resolution & Standard resolution & Ideal resolution \\ \hline
				$-3.5 < \eta < -2.0$ & $1\%/\sqrt{E} \oplus 1\%$ & $2.5\%/\sqrt{E} \oplus 1\% \oplus 1\%/E$ & $2.5\%/\sqrt{E} \oplus 1\% \oplus 1\%/E$ \\ 
				$-2.0 < \eta < -1.0$ & $8\%/\sqrt{E} \oplus 2\%$ & $8\%/\sqrt{E} \oplus 2\% \oplus 2\%/E$ & $4\%/\sqrt{E} \oplus 2\% \oplus 2\%/E$ \\ 
				$-1.0 < \eta < 1.0$ & $12\%/\sqrt{E} \oplus 2\%$ & $14\%/\sqrt{E} \oplus 3\% \oplus 2\%/E$ & $4\%/\sqrt{E} \oplus 1.5\% \oplus 2\%/E$ \\
				$-1.0 < \eta < 3.5$ & $12\%/\sqrt{E} \oplus 2\%$ & $12\%/\sqrt{E} \oplus 2\% \oplus 2\%/E$ & $4\%/\sqrt{E} \oplus 1.5\% \oplus 2\%/E$ \\ \hline \hline 
			\end{tabular}
		}
		\label{PWG-sec-JHQ-tab-ECALRES}
	\end{center}
\end{table}

\begin{figure}[ht!]
\centering
\includegraphics[width=0.90\textwidth]{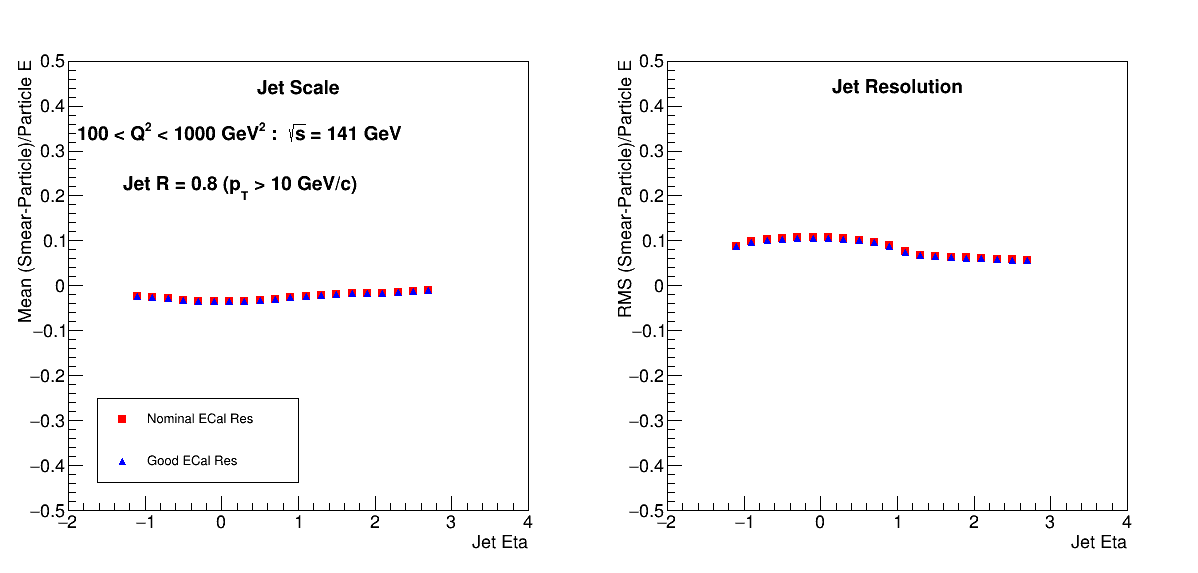}
\caption[Jet Energy Resolution ECal Comp]{Comparison between jet energy scale (left) and jet energy resolution (right) for the standard and ideal ECal energy resolution settings listed in Tab.~\ref{PWG-sec-JHQ-tab-ECALRES} as a function of jet pseudorapidity. Smearing was done in the eic-smear framework.}
\label{PWG-sec-JHQ-fig-JETENERGYRESPONSEECAL}   
\end{figure}

Figure~\ref{PWG-sec-JHQ-fig-JETENERGYRESPONSEECAL} compares the JES and JER for the standard and 
ideal ECal resolution cases as a function 
of jet pseudorapidity. As was the case with the track momentum resolution, very little change in JER 
is seen between the two scenarios, indicating that the jet resolution budget is dominated by other 
contributions, such as energy fluctuations in the hadron calorimeter.

In addition to the energy resolution, there are several other performance aspects of the ECals which 
should be considered, such as detection thresholds, cluster separation, and position resolutions. 
The threshold for detection of a `hit' in the ECal will influence the amount of energy from a 
hadronizing parton is included in a jet, or accounted for in event shape and missing energy 
measurements. Several analyses~\cite{Arratia:2020azl,Arratia:2020nxw} have assumed minimum energy thresholds of 
200~MeV/$c$ and see good jet energy scales and missing transverse energy resolutions. The ECal energy 
thresholds assumed for the standard resolution scenario range between 20 and 100~MeV and so should 
be sufficient. The ability to isolate and accurately 
determine the position of clusters in the ECals will be important for analyses which consider the 
spatial distribution of energy such as (groomed) jet substructure and event shape measurements. 
The impact of ECal cluster position resolution was not studied in detail, but it is our expectation 
that resolutions arising from tower segmentation on the order of the Moli\`{e}re radius would be 
adequate. 

\subsubsection{Hadron calorimetry} \label{part2-sec-DetReq.Jets.HQ.HCAL}

As with the electromagnetic calorimeters, the main performance criteria for the hadron calorimeters 
(HCals) will be energy resolution. Unlike with tracking or the ECals, however, the combination of low particle 
energies and poor resolution will make the relationship between HCal energy resolution and JER/JES 
non-trivial. For low energy jets, primarily at mid to negative rapidity, the poor HCal resolution 
can introduce large biases in the detector level jet sample due to the misreconstruction of low 
energy hadrons to much higher energies. Thus, in some regions, the capability of the HCal to 
identify and isolate neutral hadrons could have just as much impact on jet reconstruction as 
the intrinsic energy resolution. 

\begin{figure}[ht!]
\centering
\subfloat{\label{PWG-sec-JHQ-fig-HIEPHASESPACE}
\includegraphics[width=0.45\textwidth]{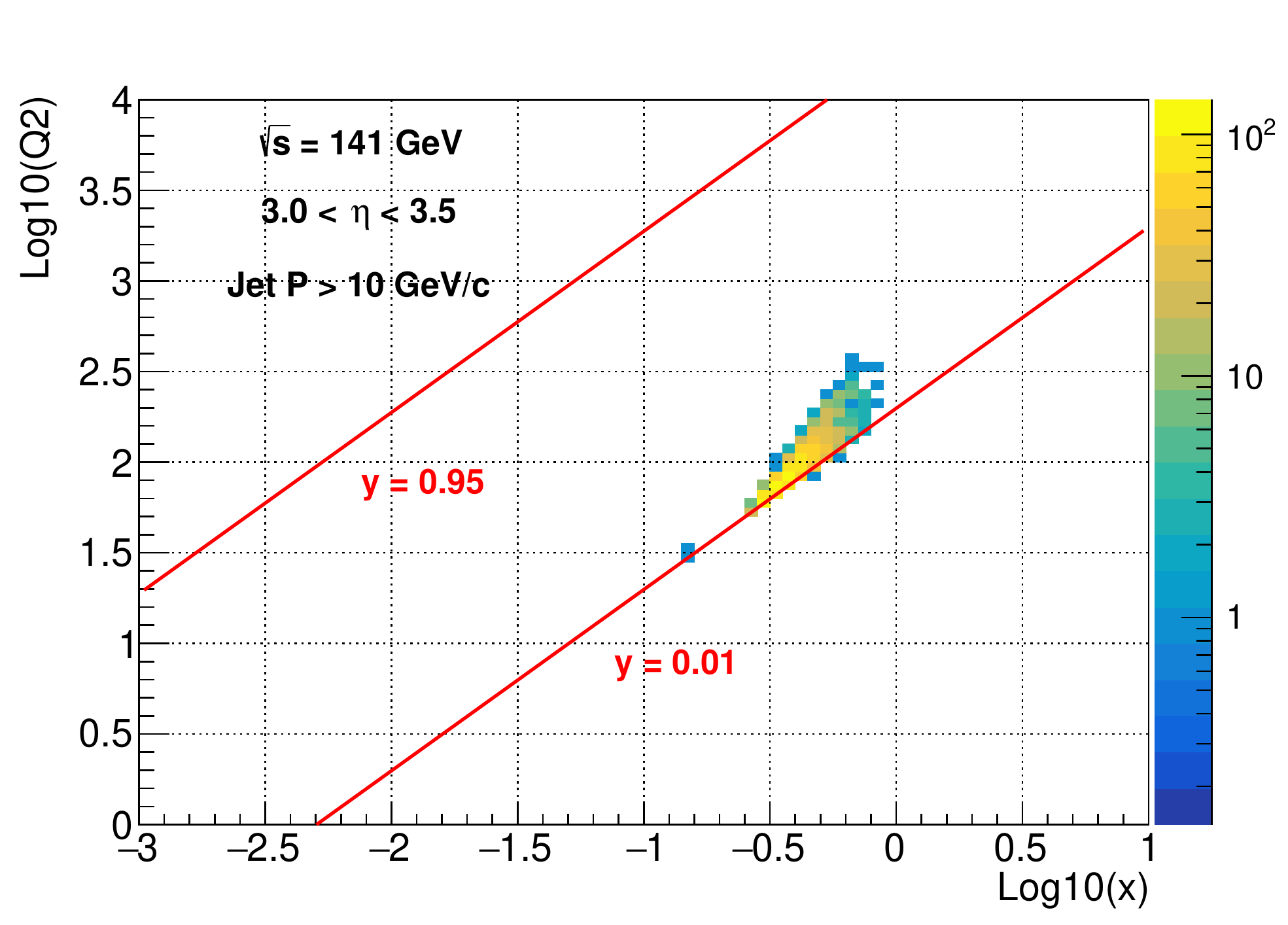}} \quad
\subfloat{\label{PWG-sec-JHQ-fig-LOWEPHASESPACE}
\includegraphics[width=0.45\textwidth]{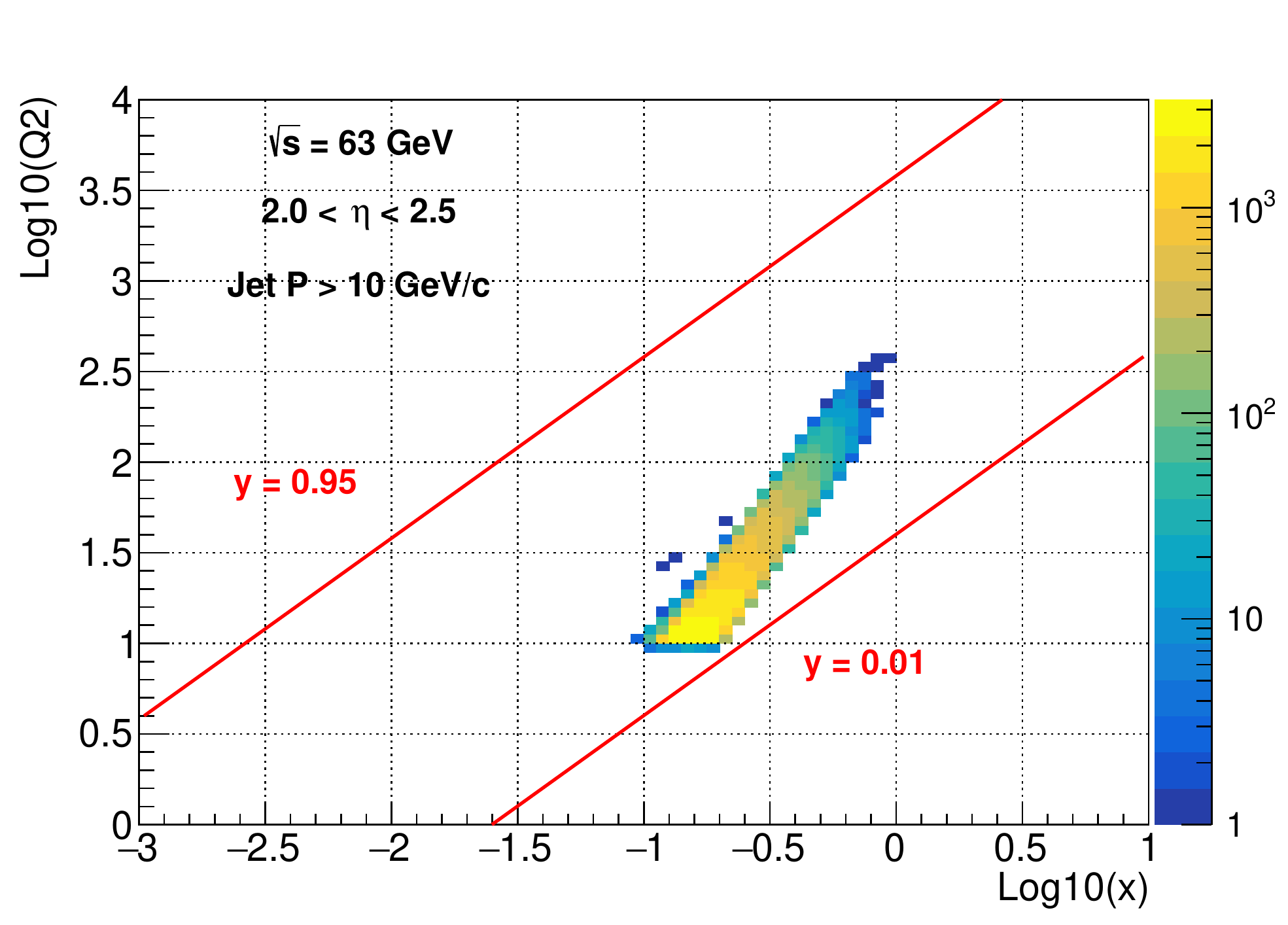}} 
\caption[Jet $x-Q^2$ Phasespace]{$x-Q^2$ phase space covered by the struck quark (jet) from LO DIS at an energy combination of $18x275$~GeV in the pseuorapidity range 3.0 to 3.5 (left panel) and at an energy combination of $10x100$~GeV in the pseudorapidity range 2.0 to 2.5 (right panel). The $x-Q^2$ range covered by high rapidity jets at the largest beam energy is also covered by lower rapidity jets at the lower beam energy.}
\label{PWG-sec-JHQ-fig-JETENERGYPHASESPACE}   
\end{figure}

The Jets and Heavy Quarks working group requests hadron calorimeter coverage over the full extent of the 
central detector $(-3.5 < \eta < 3.5)$ with energy resolutions specified in 
Tab.~\ref{PWG-sec-JHQ-tab-HCALRES}. The baseline 
performance guidance indicated that an energy dependent resolution term of $50\%/\sqrt{E}$ was realistic 
for the end caps and it was determined that this, plus a 10\% constant term, gave adequate jet energy resolutions 
in the lepton end cap $(\eta < -1.0)$. Likewise, a resolution of 
$50\%/\sqrt{E} \oplus 10\%$ was found to lead to acceptable jet energy resolutions in the forward region 
$(1.0 < \eta < 3.5)$. A minimum energy threshold of 500~MeV/$c$ was assumed for all pseudorapidities. It should be noted that for the highest energy (most forward) jets, the constant 
term dominates the resolution and reducing this term from 10\% to 5\% would dramatically improve jet energy 
resolution in this region. We acknowledge that achieving this constant term for $\eta > 3$ will be 
difficult due to shower leakage and thus keep 10\% as our official request. The issue of shower leakage at the 
highest pseudorapidities, and thus the highest $x$ values can be alleviated by lowering the hadron beam energy, 
which will shift jets at a given $x - Q^2$ point to lower pseudorapidity as seen in 
Fig.~\ref{PWG-sec-JHQ-fig-JETENERGYPHASESPACE}. The ability to shift jet pseudorapidities to lower 
values by reducing the beam energy also relieves the requirement that the calorimeter resolution 
performance extend beyond $\eta > 3.5$, which has been requested by other working groups. 

The resolution requirements 
in the barrel region $(|\eta| < 1.0)$ were explored in detail in two analyses: charged current DIS tagging and 
photoproduction jet reconstruction. One signature of charged current events is a large missing transverse 
energy $(E_{T}^{\mathrm{miss}})$, which is defined as the magnitude of the vector sum of all energy deposits 
in the detector. A study of charged current tagging was carried out in the Delphes framework and good 
resolution in $(E_{T}^{\mathrm{miss}})$ was seen even assuming a relatively poor barrel HCal resolution of 
$100\%/\sqrt{E} \oplus 10\%$ \cite{Arratia:2020azl}. However, the absence of a hadron calorimeter in the barrel led to large 
asymmetric tails in the $(E_{T}^{\mathrm{miss}})$ distribution that would complicate unfolding procedures 
and reduce photoproduction and NC DIS background rejection ability.

\begin{table}[ht!]
	\setlength{\tabcolsep}{9.0pt}
	\begin{center}
	    \caption[HCal Resolution]{Requested hadron calorimeter energy resolution.}
		\scalebox{0.9}{
			\begin{tabular}{ c  c }
				\hline \hline
				Pseudorapidity range & Energy resolution ($\sigma E/E\%)$ \\ \hline
				$-3.5 < \eta < -1.0$ & $50\%/\sqrt{E} \oplus 10\%$ \\ 
				$-1.0 < \eta < 1.0$ & $100\%/\sqrt{E} \oplus 10\%$ \\ 
				$1.0 < \eta < 3.5$ & $50\%/\sqrt{E} \oplus 10\%$ \\
			\end{tabular}
		}
		\label{PWG-sec-JHQ-tab-HCALRES}
	\end{center}
\end{table}

\begin{figure}[t!]
	\centering \includegraphics[keepaspectratio=true, width=0.95\linewidth]{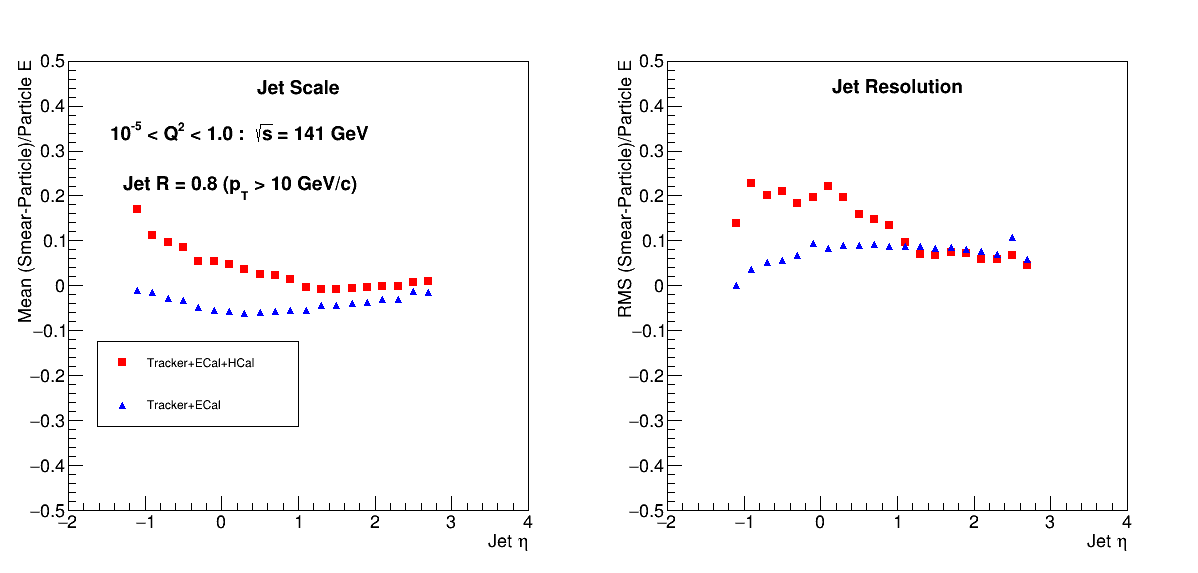}
	\caption[HCal Bias]{Jet energy scale (left) and resolution (right) as a function of jet pseudorapidity when selecting smeared jets with $p_T > 10$~GeV/$c$. Results for jets found with the tracker, ECal, and HCal are shown in red while jets found without including HCal information are shown in blue.}
	\label{PWG-sec-JHQ-fig-HCALBIAS}
\end{figure}

\begin{figure}[t!]
	\centering \includegraphics[keepaspectratio=true, width=0.95\linewidth]{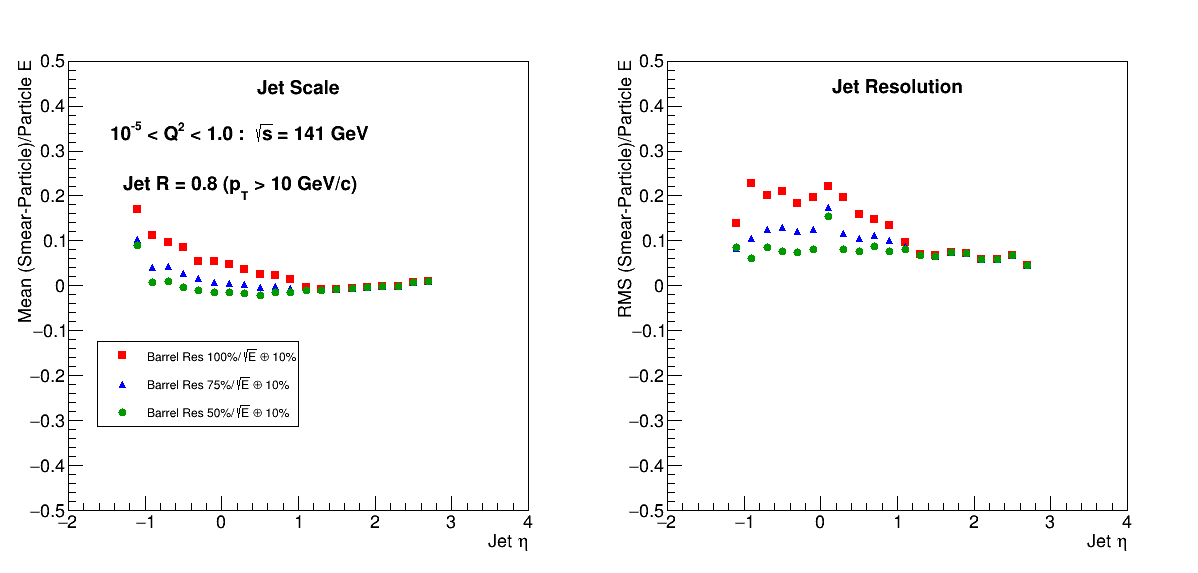}
	\caption[HCal Energy Resolution]{Jet energy scale (left) and resolution (right) as a function of jet pseudorapidity when selecting smeared jets with $p_T > 10$~GeV/$c$ for different values of HCal energy resolution: $100\%/\sqrt{E} \oplus 10\%$ (red squares), $75\%/\sqrt{E} \oplus 10\%$ (blue triangles), and $50\%/\sqrt{E} \oplus 10\%$ (green circles).}
	\label{PWG-sec-JHQ-fig-HCALRES}
\end{figure}

\begin{figure}[t!]
	\centering \includegraphics[keepaspectratio=true, width=0.95\linewidth]{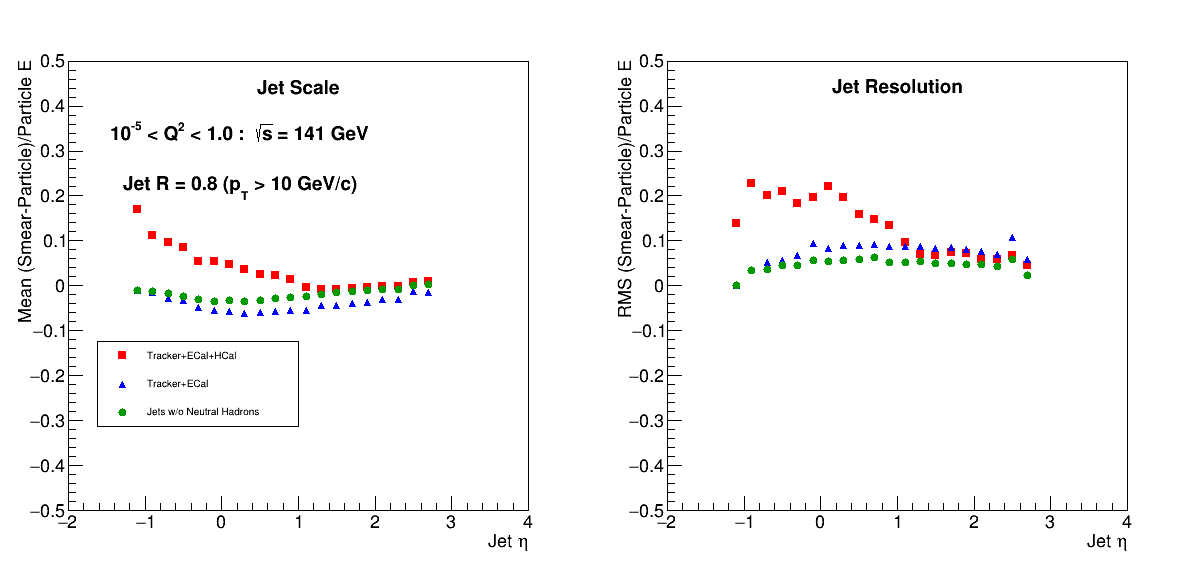}
	\caption[HCal Neutral Hadron Veto]{Demonstration of the effect of selecting only jets which do not contain a neutral hadron (green circles) on the jet energy scale (left) and resolution (right) as compared to the cases when all subsystems are used in jet finding (red squares) and when HCal information is excluded (blue triangles).}
	\label{PWG-sec-JHQ-fig-HCALVETO}
\end{figure}

As was done with the tracker momentum and ECal energy resolutions, an evaluation of the barrel HCal energy resolution 
on the jet energy scale and resolution was carried out using the eic-smear framework. In this case, however, the order 
of the matching between truth 
and smeared jets was reversed such that for each smeared jet above the $p_T$ threshold of 10~GeV/$c$, the closest truth 
level jet in $\eta - \phi$ space was found. This ordering allows for the assessment of biases to the JES and JER 
caused by sub-threshold truth level jets that get smeared above threshold. Figure~\ref{PWG-sec-JHQ-fig-HCALBIAS} shows the JES and JER for jets with $p_T$ above 10~GeV/$c$ from the 
photoproduction region, $10^{-5} < Q^2 < 1.0$~GeV$^2$. When using all detector subsystems, a large bias is seen in 
the JES, however, when the HCal information is excluded from the jet-finding, the bias (as well as the energy resolution) is substantially reduced. 
This effect arises because a certain fraction of the jets in a smeared energy bin actually arose from a truth level 
jet with lower energy that contained a neutral hadron whose energy was smeared to a much higher value. 
Because of the steeply falling energy spectrum, even small smearing at low jet energy can contribute to higher smeared bins.
This effect becomes more pronounced at lower eta where jets 
have smaller energies. The effect is also present at higher $Q^2$, but the bias is not as severe. This bias can be reduced by improving the resolution of the barrel HCal as shown in 
Fig.~\ref{PWG-sec-JHQ-fig-HCALRES}. However, a more effective method would be to select only those jets which do not contain a neutral 
hadron, and thus do not suffer from the large energy distortion, by using the HCal as a neutral hadron veto. This 
is illustrated in Fig.\ref{PWG-sec-JHQ-fig-HCALVETO} where the green circles, which are for jets which do not contain 
a neutral hadron, show little bias and better resolution than jets found with tracker+ECal+HCal information or with the HCal excluded. This neutral hadron veto capability depends critically 
on the ability to physically isolate individual showers within the calorimeter and match them to a charged particle 
track to select clusters arising from long-lived neutral hadrons. The feasibility of this method will need to be 
studied in more detail using full detector models to determine what detector granularity and clustering performance 
is required to effectively tag neutral hadrons. 

\subsubsection{Coverage continuity} \label{part2-sec-DetReq.Jets.HQ.HERM}

All analyses carried out by our working group assumed uninterrupted calorimeter coverage over the full 
range of the central detector. In a real detector, however, there is the possibility of a coverage 
gap(s) (likely near the barrel-endcap interface) to accommodate services to/from the inner detectors 
or due to interference between barrel and endcap detector components. Without a full detector model, 
it is impossible to know the exact size and location of any gap(s) and determine the effect one 
would have on our observables. However, a qualitative assessment of the impact of a gap can be achieved 
in fast simulation simply by zeroing out the calorimeter response in the desired region. 

\begin{figure}[t!]
	\centering \includegraphics[keepaspectratio=true, width=0.95\linewidth]{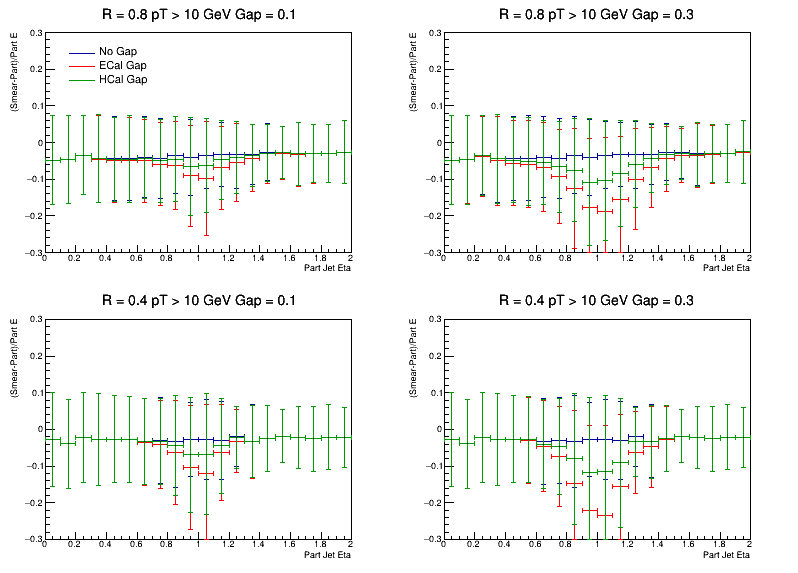}
	\caption[Calorimeter Gap]{Effect of electromagnetic and hadron calorimeter coverage gaps of 0.1 (left column) and 0.3 (right column) units of pseudorapidity on particle level jets of $R = 0.8$ (upper row) and $R = 0.4$ (lower row) and $p_T >$~10 GeV/$c$. The gap is centered at $\eta = 1$.}
	\label{PWG-sec-JHQ-fig-CALOGAP}
\end{figure}

Figure~\ref{PWG-sec-JHQ-fig-CALOGAP} shows the effect of a gap in calorimeter coverage on the smearing of 
particle level jets. As should be expected, greater deviations are seen for the larger gap size, while jets with 
larger radii show a less pronounced dip than their smaller radii counterparts due to the fact that 
they always cover more of the pseudorapidity range unaffected by the coverage gap. It is also seen 
that a gap in electromagnetic calorimetry will have a larger effect than a break in hadron calorimetry. 

While the simple analysis above does not set a requirement on the tolerable size of a gap for any given 
analysis, it should give a qualitative picture as to what the effects on a jet would be. Given that 
jets are extended objects and event shape analyses aim to characterize the energy distribution of the 
entire event, the Jets and Heavy Quarks group requests that any calorimeter gap in the main detector 
volume be kept as small as possible.

\subsection{Systematic uncertainties and unfolding: 1-jettiness}
\label{Sect:Experiment}

This section assesses the experimental issues in the measurement of 1-jettiness at the EIC, for a high-precision determination of the strong coupling \alphas. In this first assessment of experimental capabilities we focus on the measurement of \tauonea\ (Eq.~\ref{eq:tauonea}) in kinematic region $Q>30$ GeV, where the calculational precision is expected to be greatest (Sec.~\ref{part2-subS-PartStruct-GlobalEvent}).

The axes used for minimization in \tauonea\ are the beam direction and the jet centroid axis (Eqs.~\ref{eq:tauN} and \ref{eq:tauonea}). The jet is determined using the Anti-k$_T$ algorithm \cite{Cacciari:2008gp} with $R=1$, and we assume the pion mass for all tracks and clusters. The jet centroid is defined using the \pT-weighted average of jet constituents. In each event we utilize the highest-\pT\ jet whose centroid lies within $|\eta|<2.5$.

Once the hardest jet in the event is found, we loop over all particles in the acceptance and compute the scalar product of each particle four-vector and the jet four-vector as well as the product of the particle four-vector and that of the beam, according to Eqs.~\ref{eq:tauN} and \ref{eq:tauonea}. The minimum of the two products is chosen and accumulated in a loop to obtain \tauonea\ after normalization by $Q^{2}$.

We then calculate \tauonea\ at both particle and detector-level. At detector level we consider two cases: the observable is computed using only tracks as input and the observable is computed using the Delphes particle-flow components as input.

\begin{figure}[ht]
\begin{center}
    \includegraphics[width=0.49\textwidth]{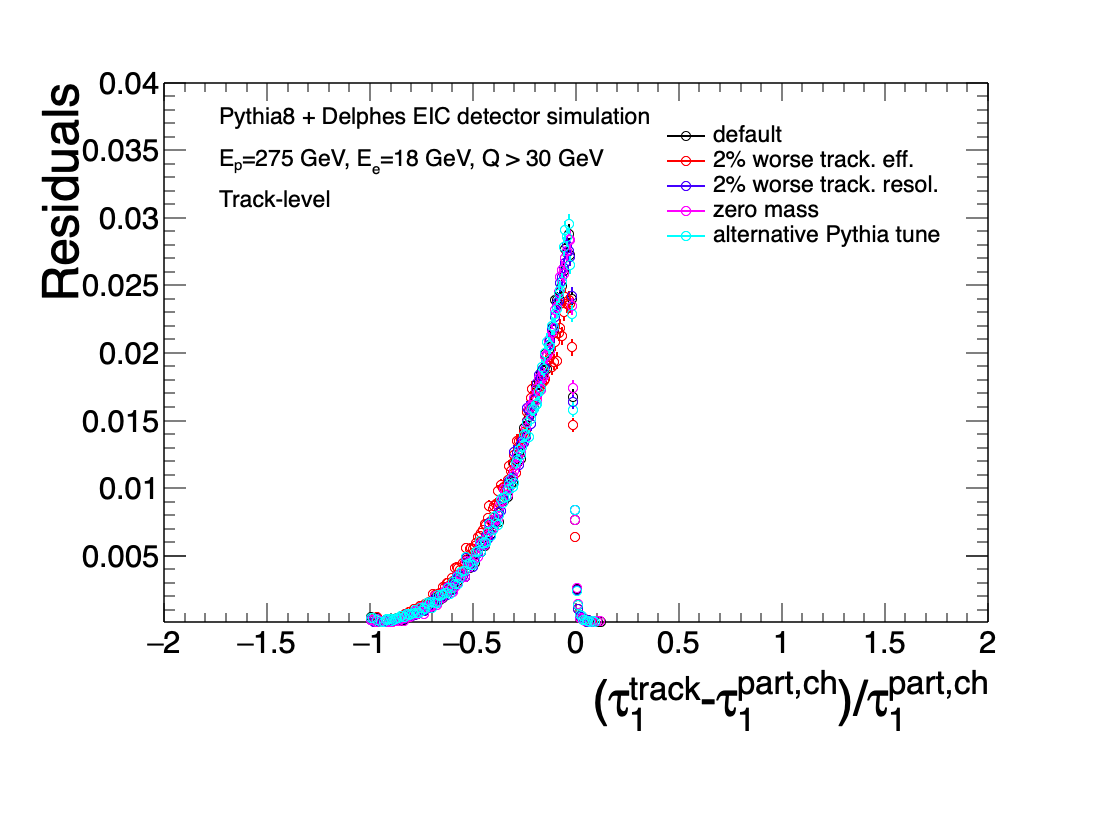}
    \includegraphics[width=0.49\textwidth]{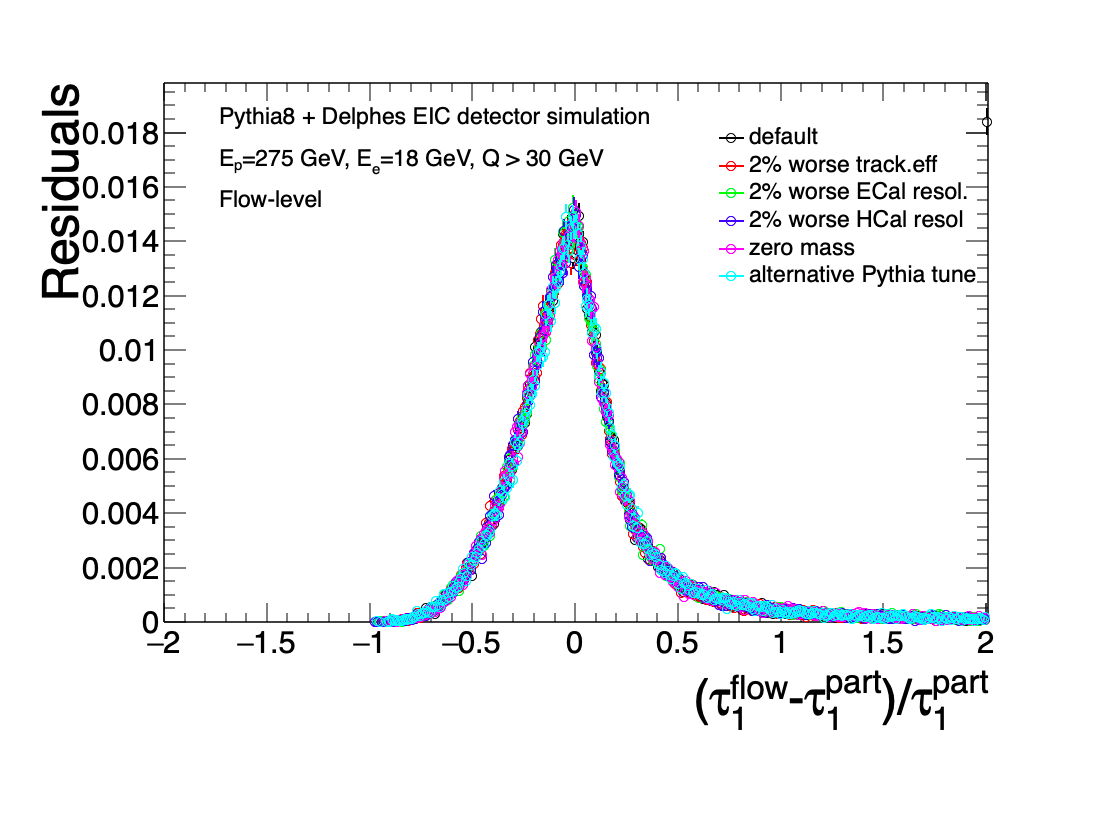}
    \includegraphics[width=0.8\textwidth]{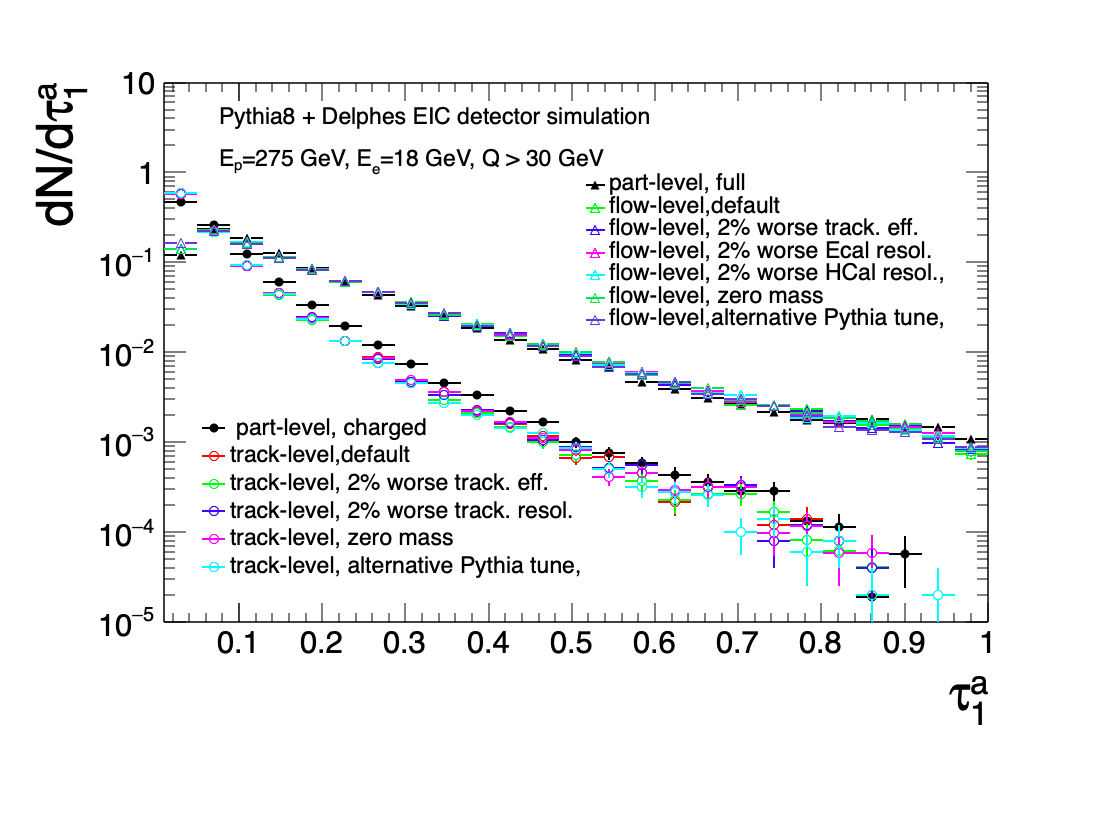}
    \caption{Upper plots: Residual distributions of \tauonea\ at particle and detector-levels, for variations choices of detector performance. Lower plots: distribution of \tauonea\ at particle and detector-levels, for variations choices of detector performance.}
    \label{fig:tau1resol}
\end{center}
\end{figure}

The relative difference of \tauonea\ at detector and particle level, at both track-level and particle-flow level, is shown in the upper panels of Fig. \ref{fig:tau1resol}. The residuals distributions are significantly non-Gaussian, especially at track-level, with RMS of $18\%$ and $36\%$ for track and flow-level respectively. The residuals distribution is largely insensitive to the variations in detector performance considered here.

The distributions of \tauonea\ at flow and track-level, again for various choices of detector performance, are shown directly in the bottom plot of Fig. \ref{fig:tau1resol}. The most notable feature of the figure is the significant change in the distribution from flow (all particles) to track-level (charged tracks only). 
The systematic shift to lower value of \tauonea\ for the track-level measurement is due to both lower jet \pT\ and fewer terms in the sum in the numerator of Eq.~\ref{eq:tauN}. Variations in detector performance give much smaller variation in the \tauonea\ distribution than the shift in the overall distribution from flow to track-level.

Our goal is a high-precision measurement of \tauonea. If the instrumental response due to detector effects is known quantitatively, this information can be utilized to correct such effects and maximize the precision of the measurement, using the approach of regularized unfolding. In this section we estimate the precision achievable for measuring \tauonea\ at EIC, using unfolding.

The precision of an unfolding procedure is dependent upon the statistical precision of the raw data. We utilize the differential cross section for \tauonea\ given by PYTHIA8, scaled by the assumed integrated luminosity of 10 fb$^{-1}$, to calculate the statistical uncertainty in the ``raw'' \tauonea\ distributions. We then vary the content of each bin using a Poisson distribution with mean of the number of projected counts.

The unfolding procedure uses the Bayesian approach, as implemented in RooUnfold\cite{RooUnfold}. The response matrix is constructed by calculating \tauonea\ at the particle and detector level for each event. 

We consider the following contributions to the systematic uncertainty of the \tauonea\ measurement:

\begin{enumerate}

\item Tracking efficiency is degraded by (absolute) 2$\%$. A few percent uncertainty is motivated by the magnitude of he discrepancy between the MC and data in describing the tracking efficiency in LHCb\cite{Aaij:2014pwa} and ALICE.

\item 2$\%$ poorer energy resolution of the ECal and the HCal calorimeters. 

\item Variation in PYTHIA8 tune: Monash tune. Further studies will utilize Herwig.  

\item Mass assumption: compare assumption of pion mass for all tracks (default) and perfect PID at detector level.  

\end{enumerate}

For each systematic variation, a new response matrix is constructed and the unfolding procedure is repeated.

\begin{figure}[ht]
\begin{center}
    \includegraphics[width=0.8\textwidth]{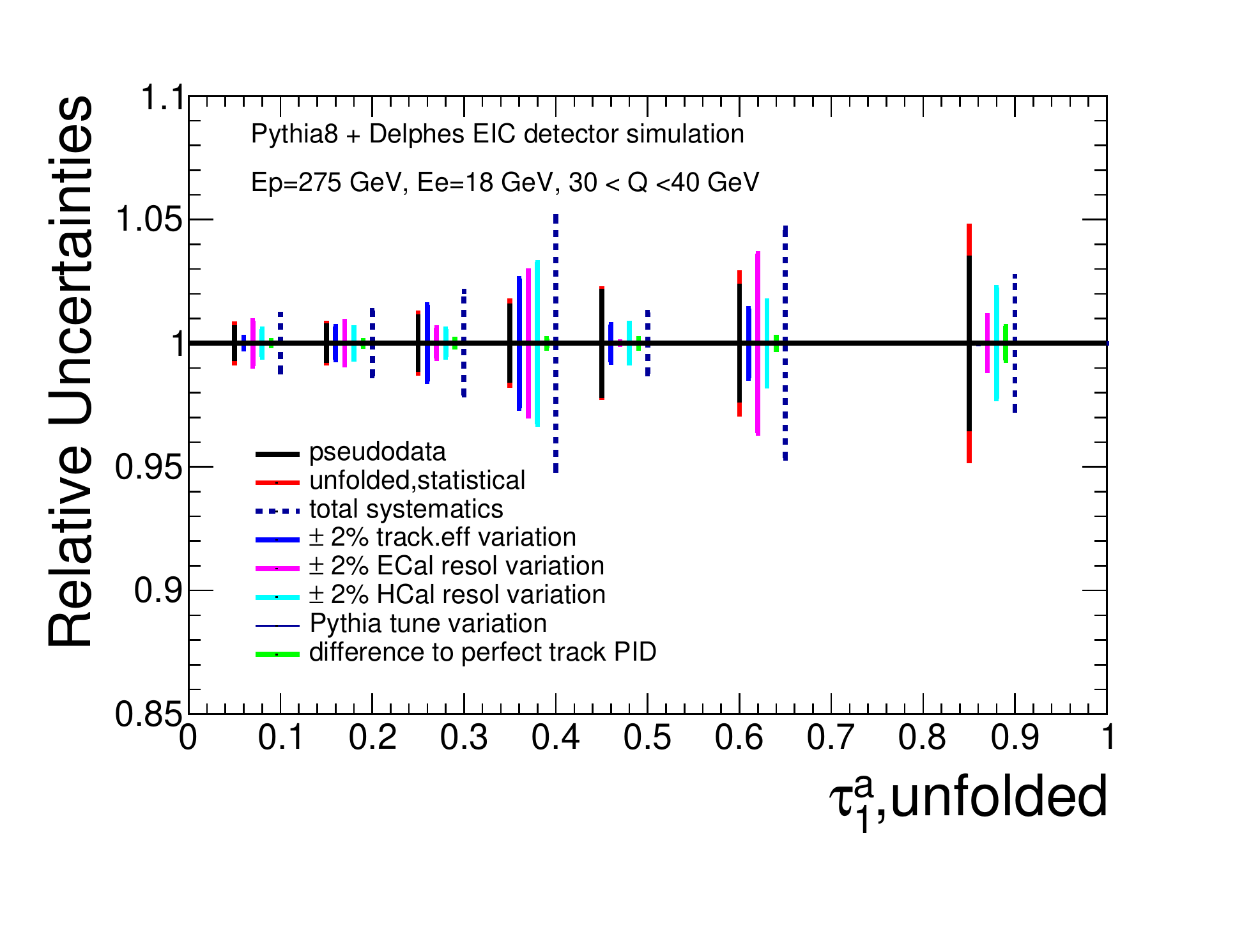}
      \caption{Relative statistical and systematic uncertainties of the unfolded \tauonea\ distribution (flow-level) for events with $30<Q<40$ GeV.}
    \label{fig:RelErrorsTau1}
\end{center}
\end{figure}

Figure \ref{fig:RelErrorsTau1} shows the results of this unfolding procedure for \tauonea\ at the particle flow level, for events with $30 < Q < 40$ GeV. Statistical uncertainties of the raw distribution and the unfolded solution in the range $0.1< \tauonea <0.7$ are within 2\%. The systematic variations in detector performance, described above, give systematic uncertainty in that region on the order of 4\%, as indicated by the dashed lines. 

\paragraph{Discussion}
\label{Sect:Discussion}

Figure \ref{fig:RelErrorsTau1} indicates that statistical and systematic uncertainties of flow-level \tauonea\ $30<Q<40$ GeV are expected to be of the order of 2 to 4 $\%$ over the full range of the  distribution, for this specific choice of binning in virtuality $Q$. Assessing the correlation between the statistical and systematic uncertainties is beyond the scope of this report. 

Note that the largest contributions to the systematic uncertainties in 
Fig.~\ref{fig:RelErrorsTau1} are due to the calorimeter performance. As the next step in this analysis, calorimeter performance based upon specific EIC designs should be considered, include the projected effects and uncertainties due to calorimeter non-linearity. 

In addition, Fig.~\ref{fig:RelErrorsTau1} indicates that track-level measurements may have greater relative precision experimentally. As noted in Sec.~\ref{part2-subS-PartStruct-GlobalEvent}, theoretical calculations of track-level event shape observables with controlled and improvable precision may be possible, though such an approach still requires development in both theory and experiment. We will continue to consider the measurement of \tauonea\ and other event shape obseravbles at both the flow and track-levels.

As discussed in Sec.~\ref{part2-subS-PartStruct-GlobalEvent}, fundamental parameters of QCD will be extracted by comparing analytic calculations to the EIC data in bins of Q and x.

In this section, we have performed a first evaluation of the expected measurement precision of \tauonea. Uncertainties of few percent are to be compared to systematic uncertainties of typically $10\%$ in $\alpha_{S}$ extractions at HERA using inclusive jet cross sections\cite{Aaron:2010ac} for instance. Further investigation requires MC pseudo-data based on specific EIC detector designs, together with the application of Bayesian Inference tools~\cite{RevModPhys.83.943}.

\subsection{Summary} \label{part2-sec-DetReq.Jets.HQ.SUMM}

The preceding section discussed the detector requirements necessary to carry out the physics program covered by the 
Jets and Heavy Quark working group. These requirements are summarized below:

\begin{description}
\item[Track Momentum Resolution] Track momentum resolution parameters were listed in Tab.~\ref{PWG-sec-JHQ-tab-TRACKRES} for two magnetic field strengths. It was concluded that either the 3T or 1.5T settings would be adequate.
\item[Minimum Track Transverse Momentum] Minimum track $p_T$ was also enumerated for two magnetic field settings and are found in Tab.~\ref{PWG-sec-JHQ-tab-TRACKTHRESH}. A preference for the lower thresholds associated with the 1.5T field was found.
\item[Transverse Pointing Resolution] The required track transverse pointing resolution necessary for the Heavy Flavor program can be found in Tab.~\ref{PWG-sec-JHQ-tab-VERTEXRES}.
\item[PID] The Jets and Heavy Quarks working group requests $\pi/K/P$ separation at the $3\sigma$ level over a wide momentum range, specific numbers can be found in Tab.~\ref{PWG-sec-JHQ-tab-PIDMOM}.
\item[Electromagnetic Calorimetry] The default ECal energy resolution put forward by the detector group and listed under Standard resolution in Tab.~\ref{PWG-sec-JHQ-tab-ECALRES} was determined to be adequate. An energy threshold of 100~MeV or better is also requested.
\item[Hadron Calorimetry] The requested HCal energy resolution can be found in Tab.~\ref{PWG-sec-JHQ-tab-HCALRES}. It was found that neutral hadron isolation could also be important for jet energy scale and resolution.
\end{description}

\section{Exclusive Measurements}
\label{part2-sec-DetReq.Excl}


\subsection{Deeply virtual Compton scattering and exclusive production of \texorpdfstring{$\pi^0$ in \ep}{pi0 in ep}}
\label{subsec:dvcs_ep}
Deeply virtual Compton scattering (DVCS) and the hard exclusive production of $\pi^0$ mesons off a nucleon play a prominent role in the studies of GPDs. DVCS gives access to chiral-even GPDs, which are important for the extraction of information on both the nucleon tomography and the energy-momentum tensor (EMT), including access to the total angular momentum of partons, as discussed in Sec.~\ref{part2-subS-SecImaging-GPD3d}. 

Hard exclusive production of $\pi^0$ mesons, on the other hand, gives access to chiral-odd GPDs, some of which are related to transversity distributions, which are extensively studied in SIDIS and Drell-Yan processes.

The similarity between the final states of DVCS and hard exclusive production of $\pi^0$, see Sec.~\ref{part2-subS-SecImaging-GPD3d}, suggests that the detectability of both processes in an apparatus equipped with electromagnetic calorimeters (ECALs) can be studied together. Another reason for a joined study of this type is that DVMP $\pi^0$ may become a background to DVCS. This happens when one low-energy photon coming from $\pi^0$ decay misses ECALs, its energy is too low for a detection, or if two electromagnetic cascades induced by photons can not be distinguished from each other. A common analysis of DVCS and DVMP $\pi^0$ therefore allows to stress three main aspects of ECALs design: geometrical acceptance, energy thresholds and granularity.

The study presented in this subsection is based on two Monte Carlo generators. The first one is MILOU 3D, a recently updated version of MILOU \cite{Perez:2004ig}, used to generate DVCS events. The original version of MILOU is supplied with two-dimensional $(x_{\mathrm{B}}, Q^{2})$ lookup tables of DVCS sub-amplitudes, refereed to as Compton form factors (CFFs), while $t$-dependance of those factors is factorised out and modeled with either exponential or dipole Ansatz. This way of modeling of CFFs has been modified for the purpose of this study. Namely, MILOU 3D can now be supplied with three-dimensional $(x_{\mathrm{B}}, Q^{2}, t)$ tables, allowing to account for  an interplay between all three variables, which is important to describe data at energies lower than those available at HERA \cite{Akhunzyanov:2018nut}. This modification allowed using two realistic GPD models to generate the lookup tables: KM \cite{Kumericki:2007sa, Kumericki:2009uq, Kumericki:2013br, Cuic:2020iwt} implemented in GeParD and GK \cite{Goloskokov:2005sd,Goloskokov:2007nt,Goloskokov:2009ia} implemented in PARTONS \cite{Berthou:2015oaw}. These two models significantly differ by construction, i.e. they are based on different schemes of GPD modelling, and they are constrained by different experimental data. The second generator is toyMC, which was developed for the purpose of this study. It assigns a weight to each generated event, which corresponds to either DVCS or DVMP $\pi^0$ cross section. For this generator the lookup tables of cross sections were generated with the GK model, which includes chiral-odd GPDs crucial for the description of exclusive $\pi^0$ production. The amplitudes for DVMP $\pi^0$ were evaluated using the GK formalism~\cite{Goloskokov:2005sd}, which is based on the modified perturbative approach \cite{Botts:1989kf}, allowing one to overcome the problem of infrared divergences that appear for transversely polarised virtual photons. Among many available versions of chiral-odd GK GPDs we have chosen the one that successfully describes cross sections measured by COMPASS \cite{Alexeev:2019qvd}. The kinematic domain covered by this measurement significantly overlaps that of EIC, particularly at its lowest beam energy configuration: $5~\mathrm{GeV} \times 41~\mathrm{GeV}$. Both generators can be interfaced with EIC-smear.

\begin{figure}[!ht]
    \begin{center}
       \includegraphics[width=0.85\textwidth]{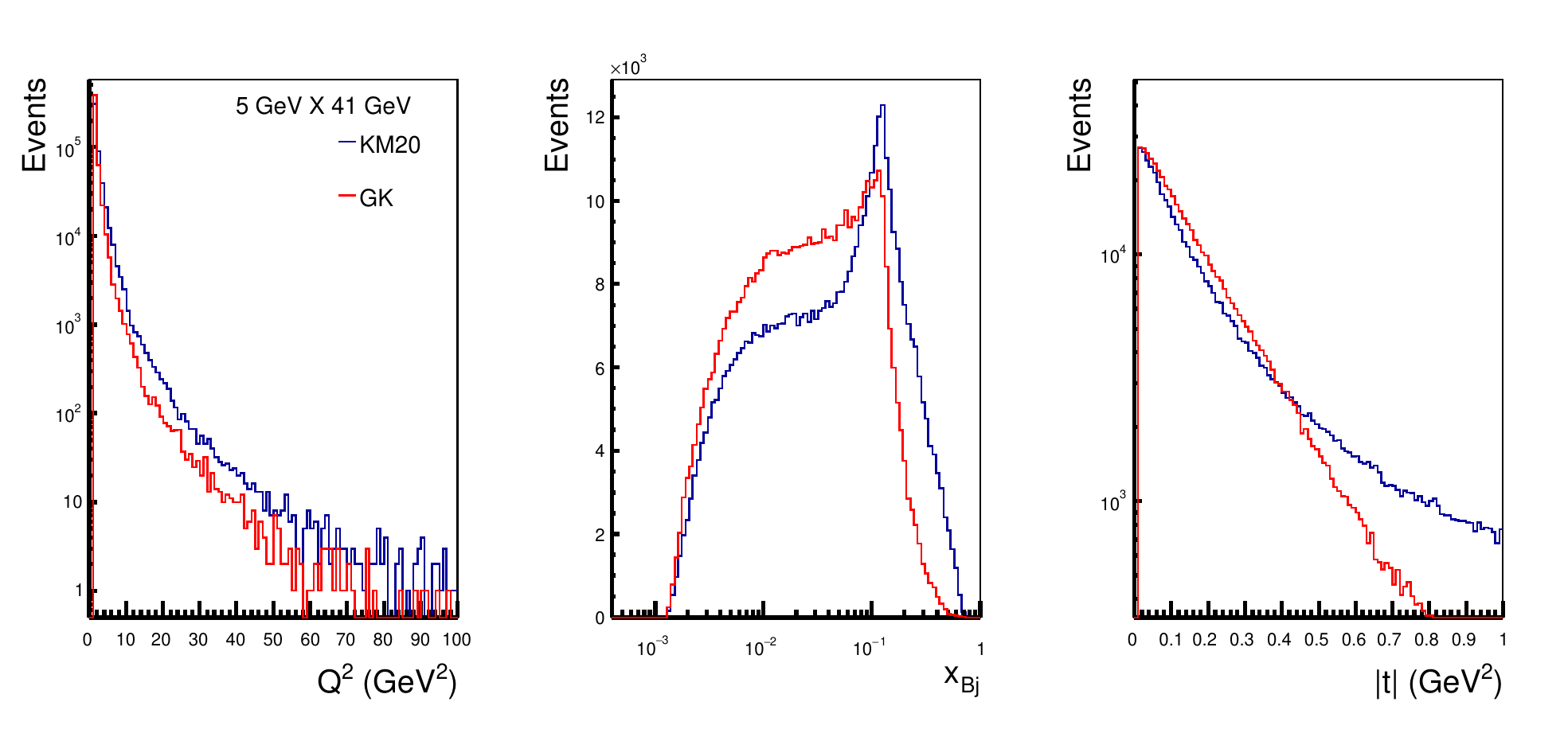}
       \includegraphics[width=0.85\textwidth]{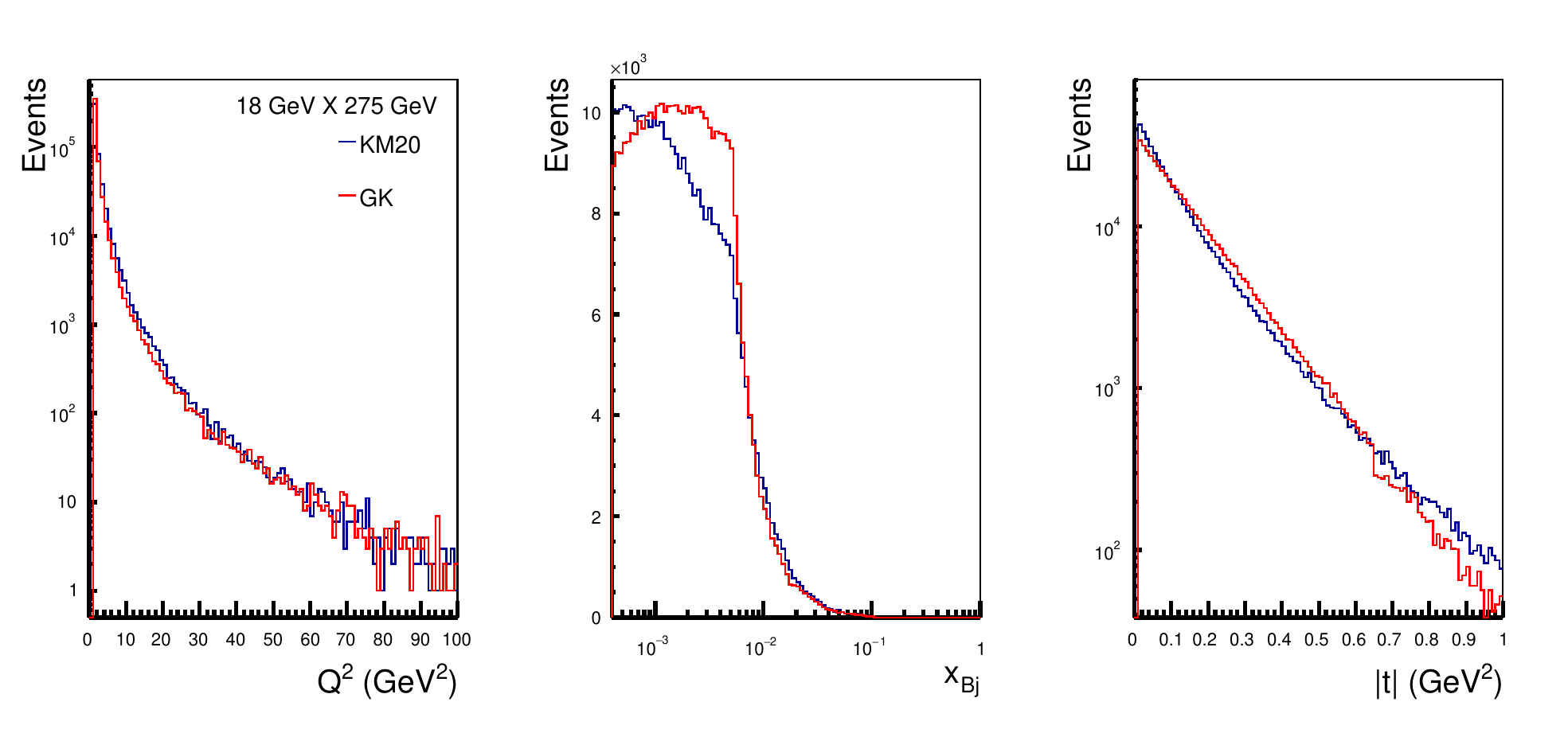}
    \end{center}
    \caption{Kinematic distributions of DVCS events generated according to KM20~\cite{Cuic:2020iwt}  (blue) and GK~\cite{Goloskokov:2009ia} (red) models.}
    \label{fig:exclusive_dvcs_EventKine}
\end{figure}

Figure~\ref{fig:exclusive_dvcs_EventKine}, based on 500k events simulated with with MILOU 3D, compares the distributions of events generated according to KM20~\cite{Cuic:2020iwt} (blue) and~GK~\cite{Goloskokov:2009ia} (red) for the lowest and highest beam energy configurations. The following cuts have been applied at generation level: $Q^2 > 1~\mathrm{GeV}^{2}$, $0.01 < y < 0.95$, and $0.01 < |t|~\mathrm{GeV}^{2} < 1.6$. Both models predict a significant drop of the cross section with $Q^2$.The different $|t|$ distributions for GK (exponential) and KM20 (dipole) are also evident.   

\begin{figure}[!ht]
    \begin{center}
       \includegraphics[width=0.45\textwidth]{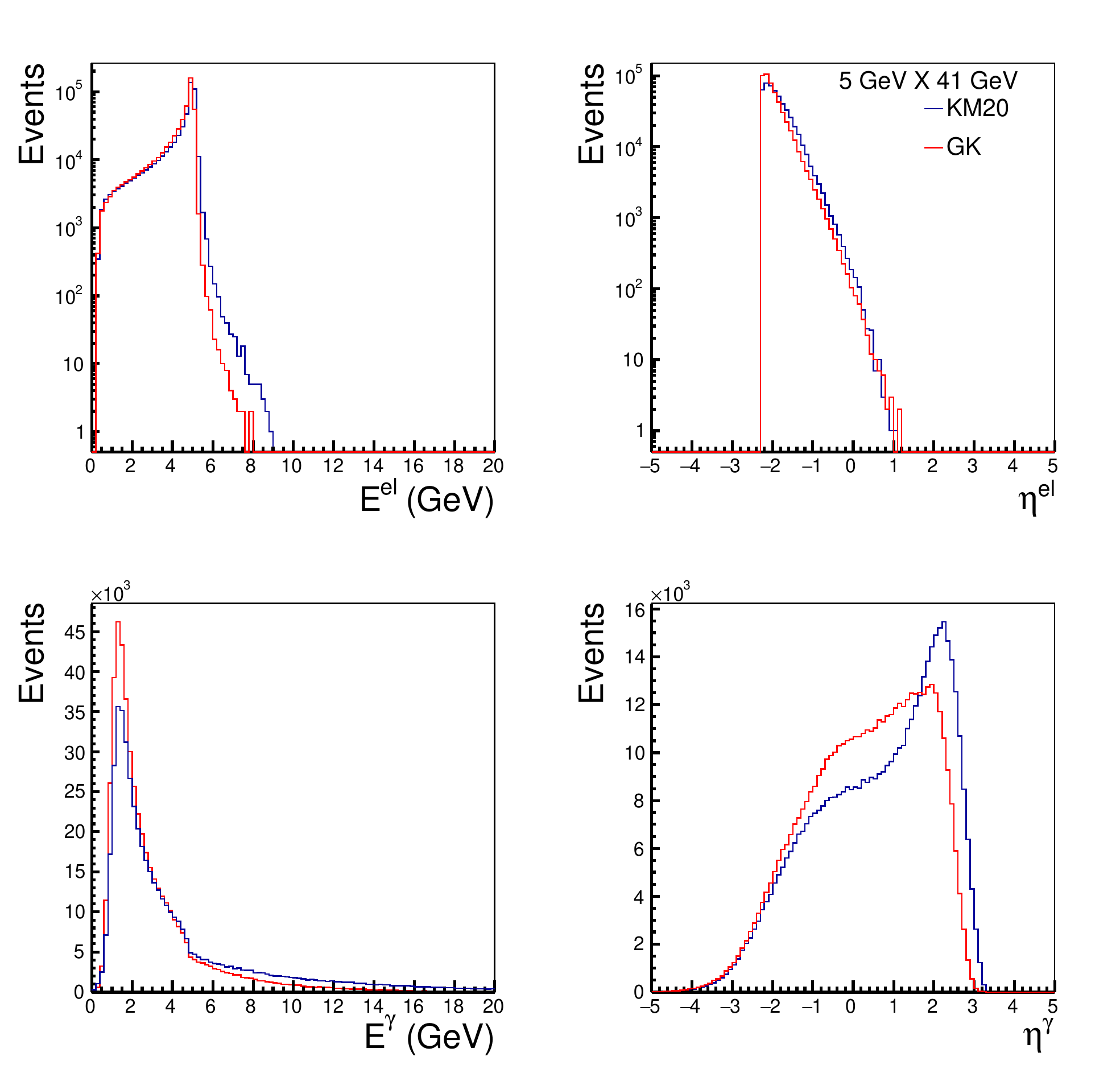}
       \includegraphics[width=0.45\textwidth]{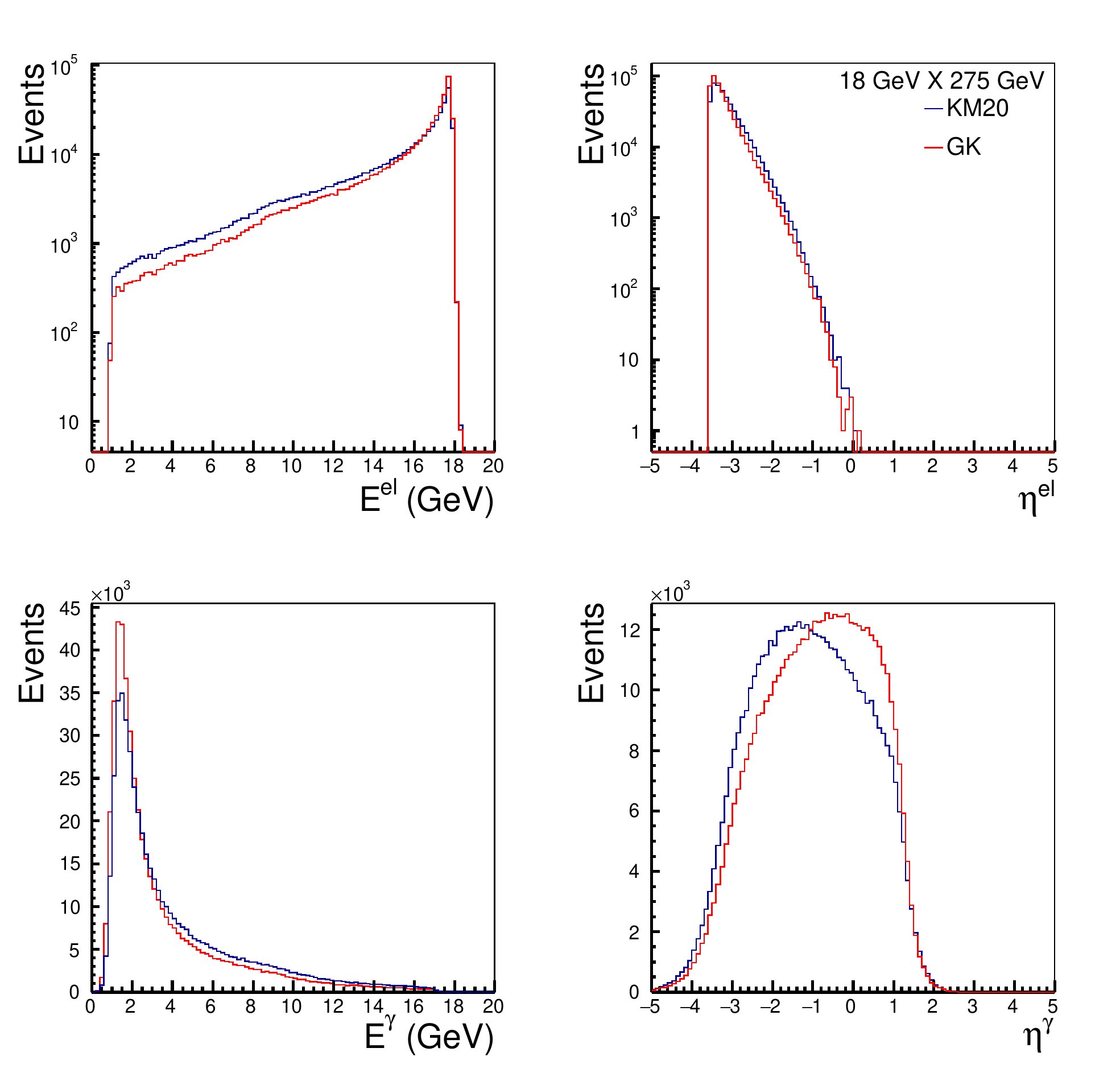}
    \end{center}
    \caption{Energy and pseudo-rapidity spectra of DVCS events generated according to KM20 (blue) and GK (red) models.}
    \label{fig:exclusive_dvcs_LeptonGamma}
\end{figure}

Using the same sample of generated DVCS events, we have also simulated the energy and pseudo-rapidity distributions of the scattered electron and produced real photon in a DVCS process. Figure~\ref{fig:exclusive_dvcs_LeptonGamma} visualizes these spectra for the lowest and highest beam energy configurations. At lower energies (left) the electron is predicted to be scattered within the nominal combined EMCAL+Tracker acceptance of $|\eta| < 3.5$. At the highest electron-beam energy (right), the peak in the scattered electron distribution is predicted to be at $\eta \sim -3.6$. This is expected to be valid for other exclusive processes, making a slightly extended acceptance at backwards pseudo-rapidities beneficial for detection efficiency. Both models predict that the nominal $|\eta| < 3.5$ acceptance should be enough for detecting most of the produced photons, with a nearly perfect situation at smaller beam energies and a slight loss in efficiency at top beam energy, notably at very low values of $x_{B}$.     

\begin{figure}[!ht]
    \begin{center}
        \includegraphics[width=0.45\textwidth]{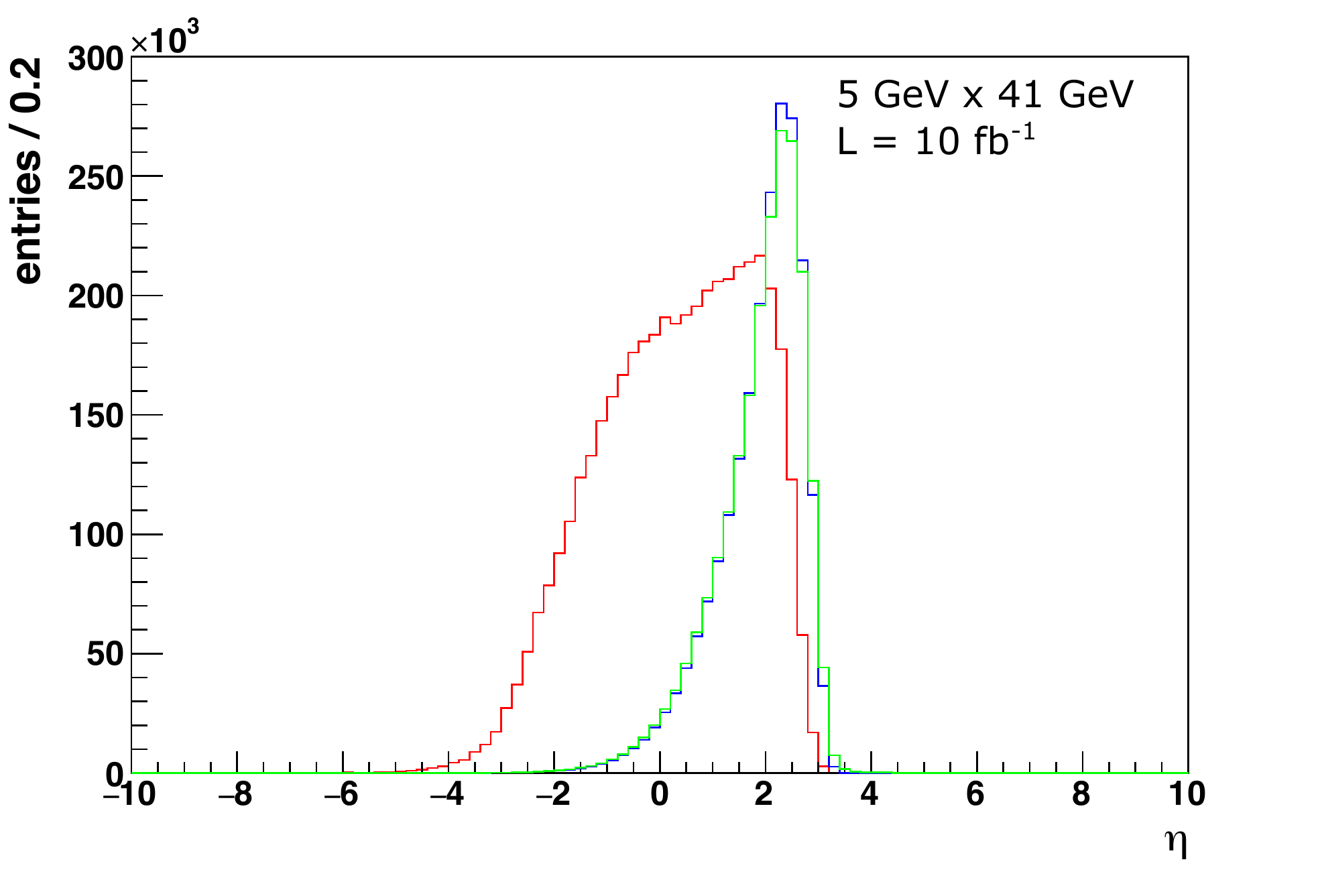}
        \includegraphics[width=0.45\textwidth]{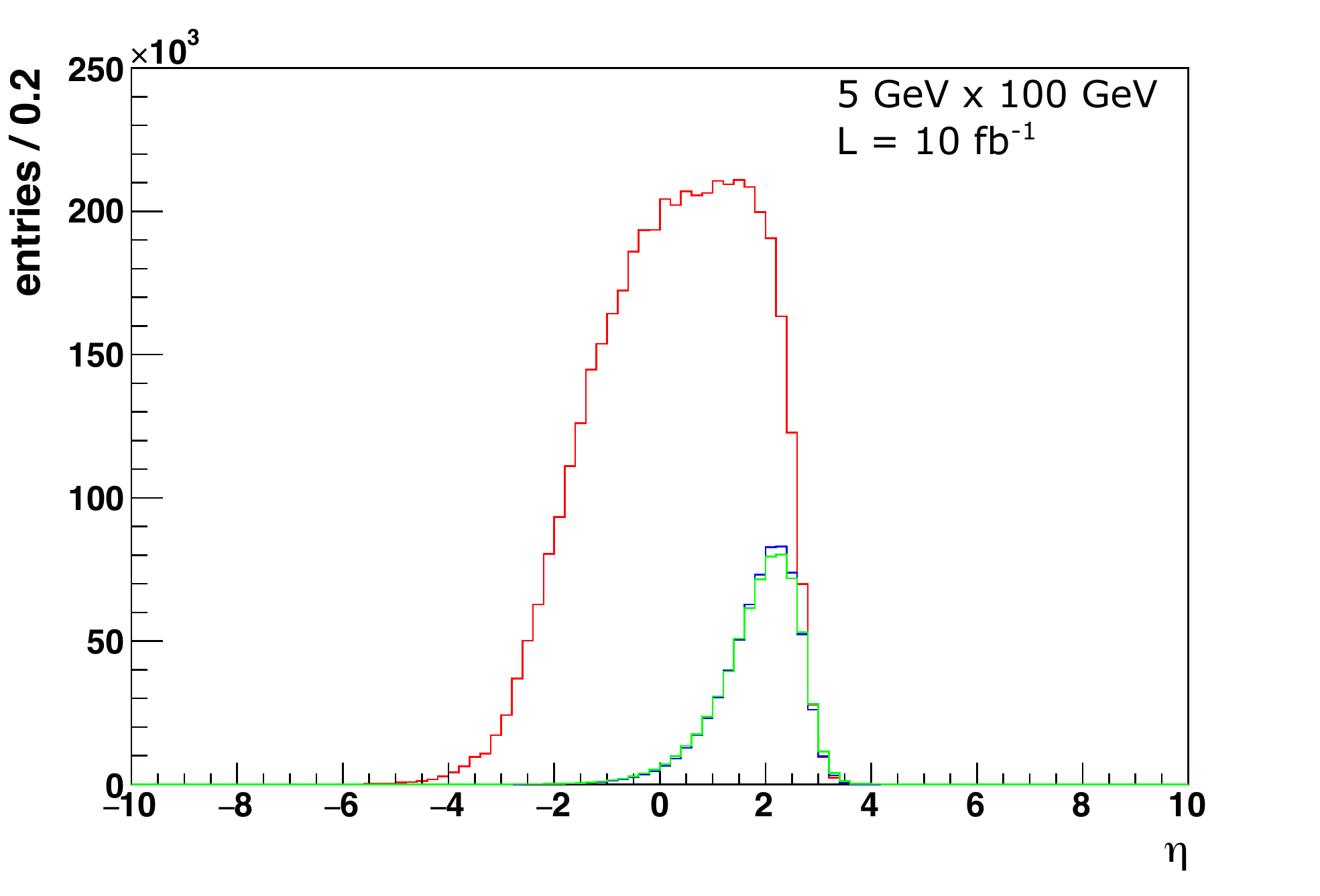}
        \includegraphics[width=0.45\textwidth]{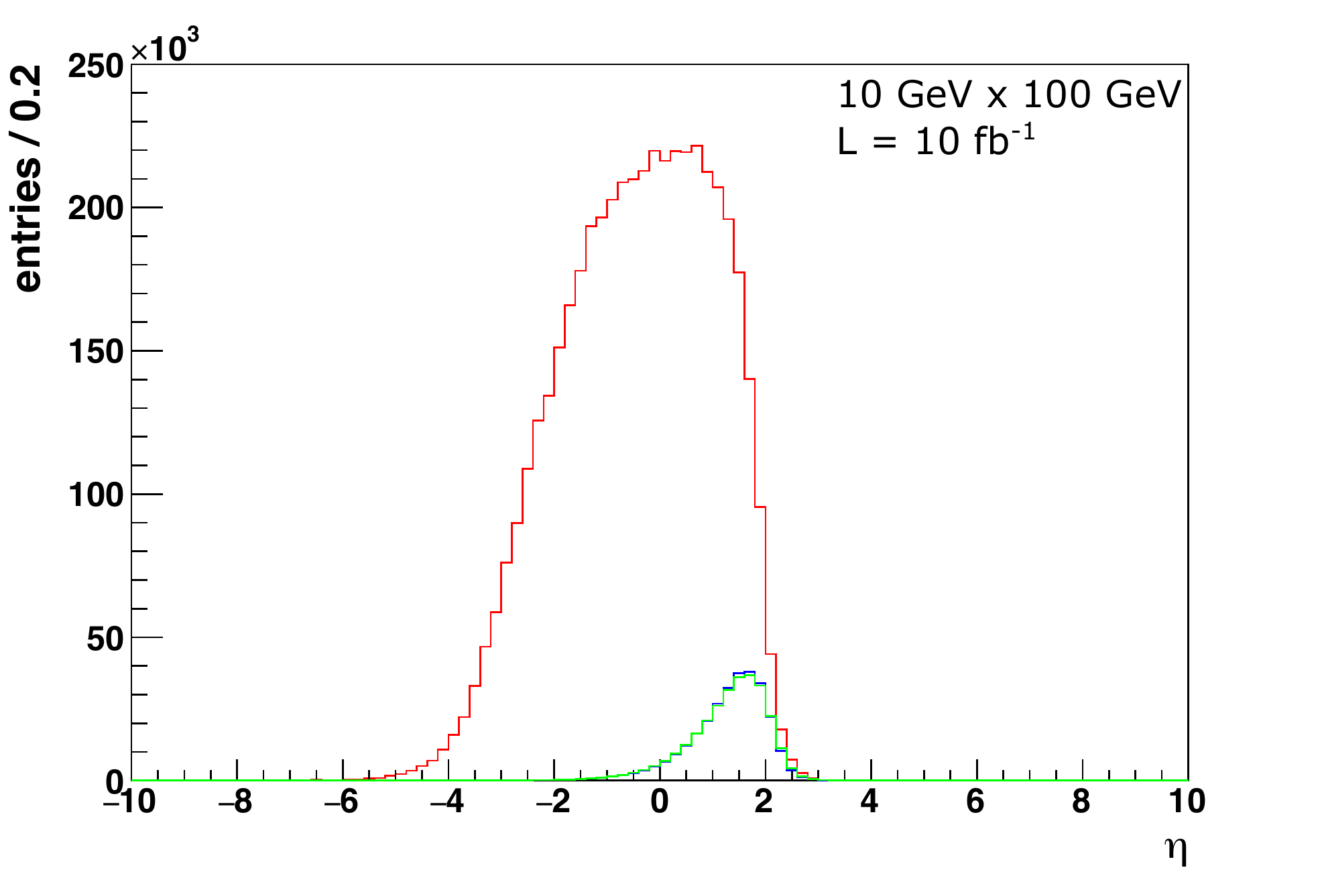}
     \includegraphics[width=0.45\textwidth]{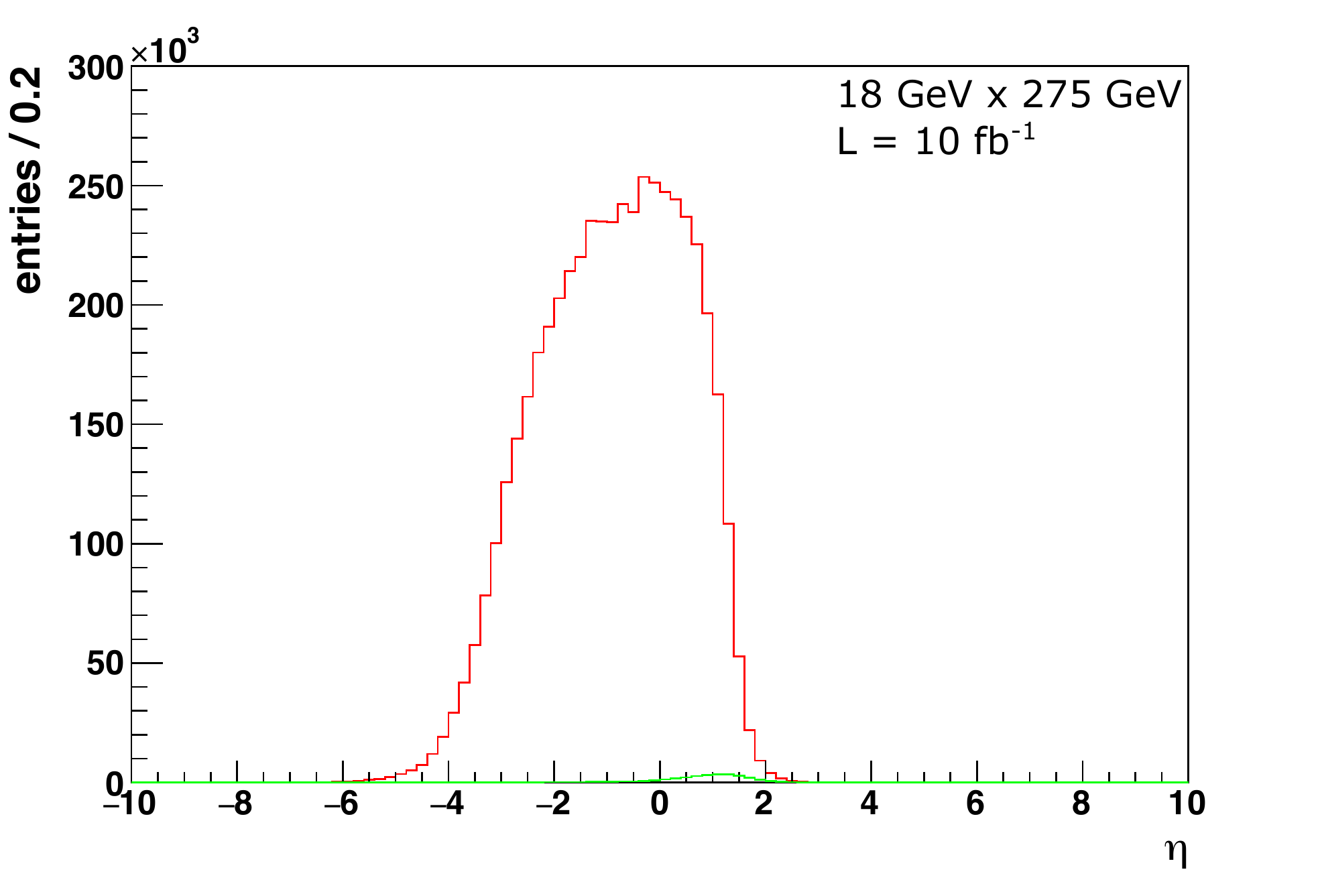}
    \end{center}
    \caption{Distributions of pseudo-rapidity for DVCS photons (red), exclusive $\pi^0$ mesons (blue) 
    and photons coming from decays 
    of 
    $\pi^0\rightarrow \gamma\gamma$ 
    (green, histogram scaled by $\frac{1}{2}$) 
    for four beam energy configurations (see the insert labels) and $10~\mathrm{fb}^{-1}$ of integrated luminosity.}
    \label{fig:exclusive_dvcs_pi0_pseudorapidity}
\end{figure}

In order to assess the significance of $\pi^0$ background, the pseudo-rapidity distribution of photons from DVCS and exclusive $\pi^0$ production before applying the smearing is compared in Fig.~\ref{fig:exclusive_dvcs_pi0_pseudorapidity}. These histograms represent the sample of events generated with toyMC for the four beam energy configurations considered in this report and assume an integrated luminosity of $\mathcal{L} = 10~\mathrm{fb}^{-1}$ per each configuration. Two cuts were applied to the sample before making the histograms: $Q^{2} > 1~\mathrm{GeV}^{2}$ and $0.01 < y < 0.95$. The validity of the $y$-cut was checked with a sample of events after applying the smearing. This study shows that at $y = 0.01$ one may expect the resolution of this variable at the order of $dy/y \approx 0.5$ for $5~\mathrm{GeV} \times 41~\mathrm{GeV}$ and $dy/y \approx 1$ for $18~\mathrm{GeV} \times 275~\mathrm{GeV}$ beam energy configuration. We conclude that for  $18~\mathrm{GeV} \times 275~\mathrm{GeV}$ and the assumed acceptance of $|\eta| < 3.5$ for both electrons and photons one may expect to loose $14\%$/$17\%$ of DVCS events and $11\%$/$12\%$ of exclusive $\pi^0$ events, where the first number is due to the acceptance on electrons, while the second one is due to the acceptance on both electrons and photons. The loss is mainly seen for $Q^{2} \approx 1~\mathrm{GeV}^2$ events. The loss of DVCS events can be almost entirely recovered by slightly extending coverage in backwards pseudo-rapidity to $\eta < -3.7$ from the currently assumed value of $\eta < -3.5$. For lower beam energies the loss is smaller, in particular for $5~\mathrm{GeV} \times 41~\mathrm{GeV}$ it is of the order of $1\%$.

\begin{figure}[!ht]
    \begin{center}
        \includegraphics[width=0.45\textwidth]{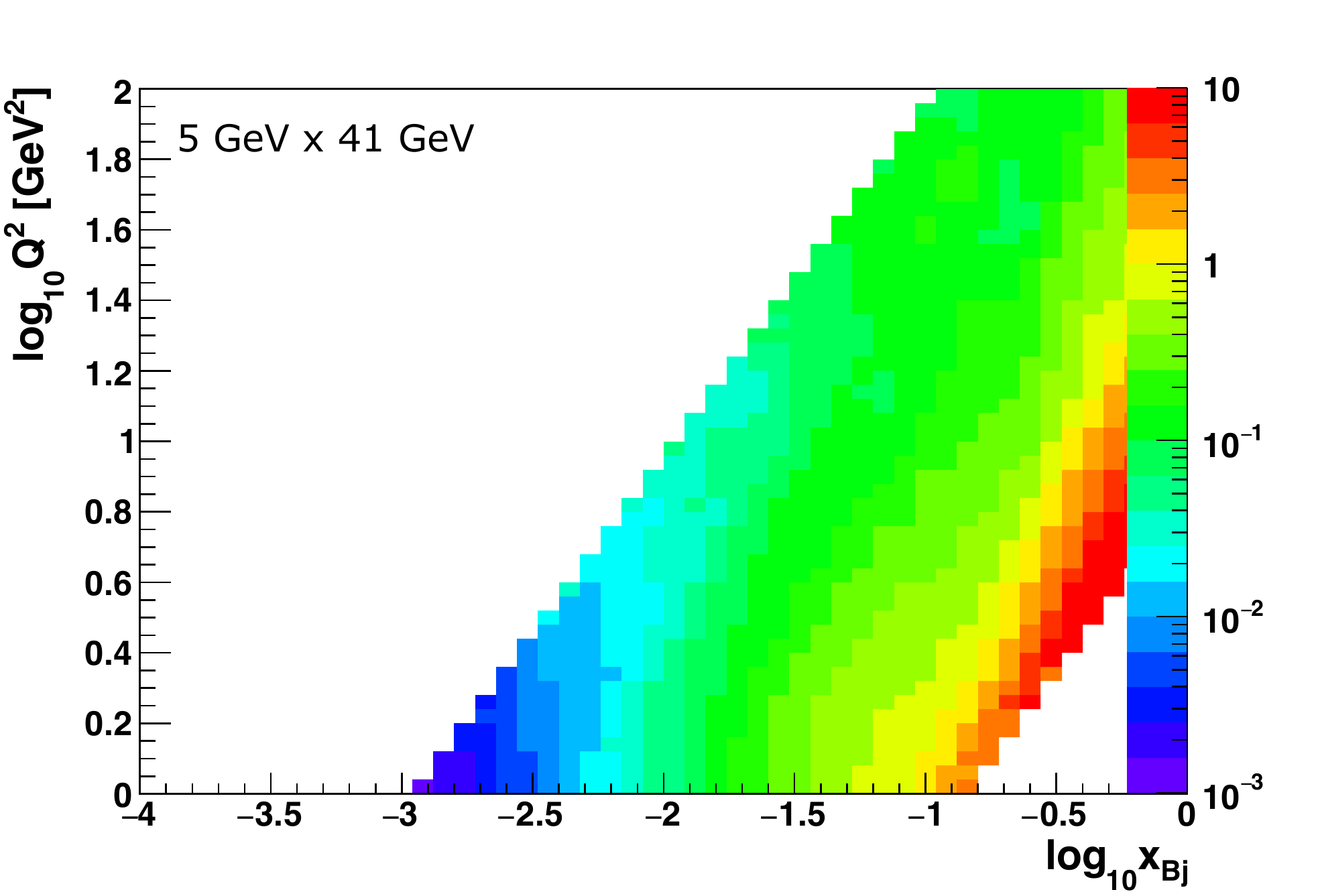}
    \end{center}
  \caption{Ratio of DVMP $\pi^0$ to DVCS event yields in phase-space of $(x_{\mathrm{B}}, Q^{2})$ for $5~\mathrm{GeV} \times 41~\mathrm{GeV}$ beam energies. For more details see the text.}
    \label{fig:exclusive_dvcs_pi0_ratio}
\end{figure}

\begin{figure}[!ht]
    \begin{center}
        \includegraphics[width=0.45\textwidth]{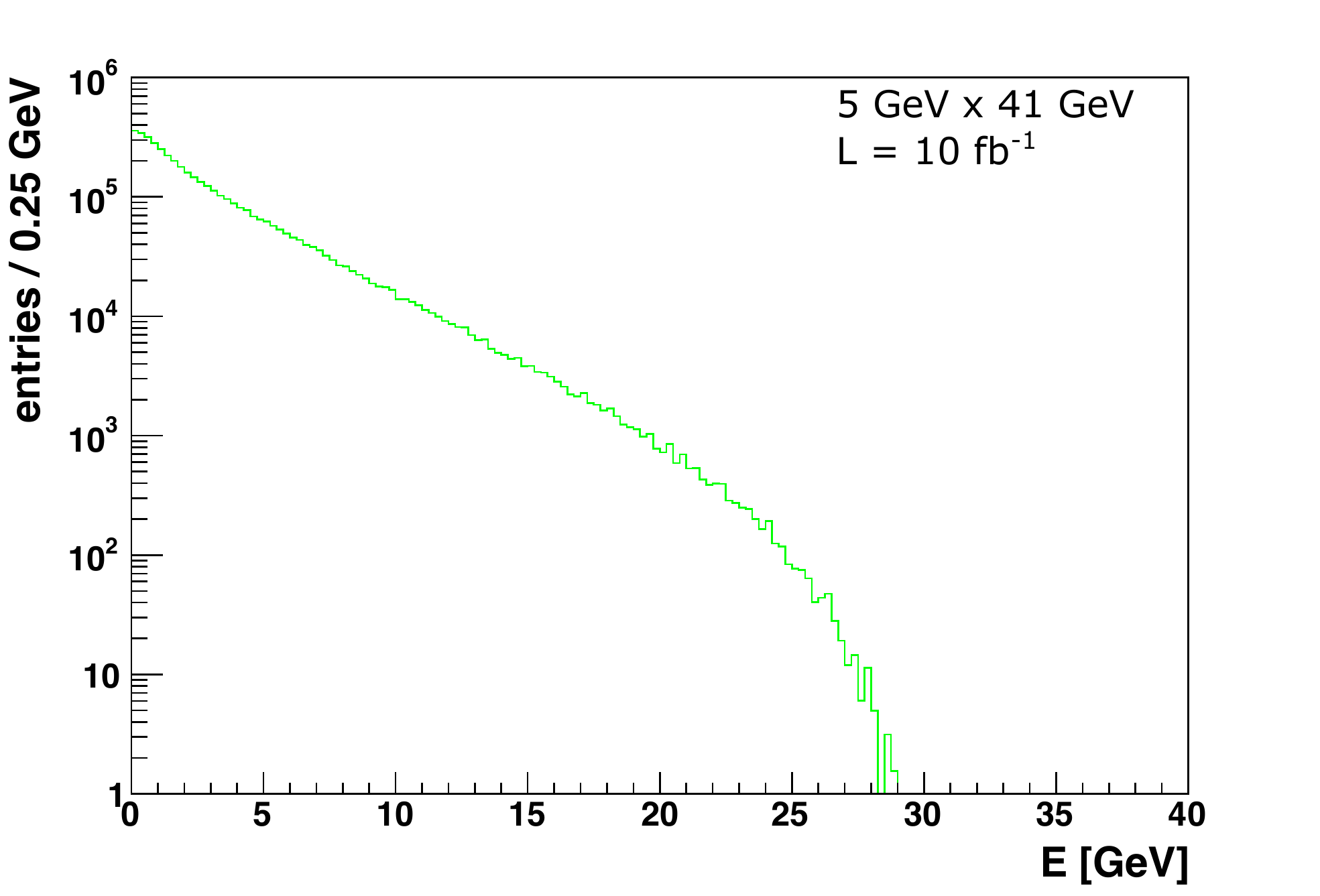}
        \includegraphics[width=0.45\textwidth]{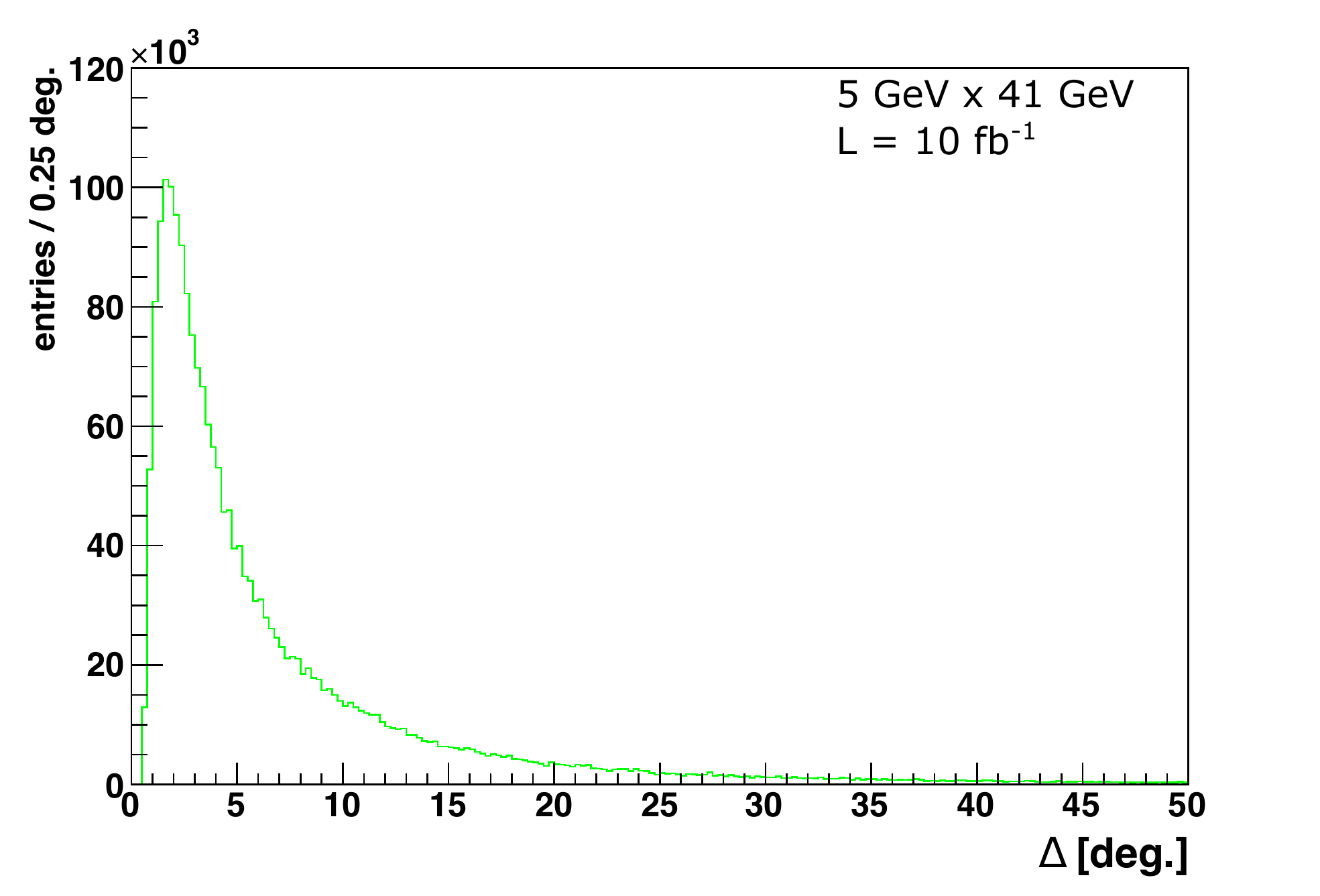}
    \end{center}
    \caption{Energy distribution of photons coming from decays of exclusively produced $\pi^0\rightarrow \gamma\gamma$ (left) and distribution of opening angle in the lab frame between those photons (right). Plots are for $5~\mathrm{GeV} \times 41~\mathrm{GeV}$ beam energies and $10~\mathrm{fb}^{-1}$ of integrated luminosity}
    \label{fig:exclusive_dvcs_pi0_energy_angle}
\end{figure}

The contamination of DVCS sample by misinterpreted exclusive $\pi^0$ events is demonstrated with Fig. \ref{fig:exclusive_dvcs_pi0_ratio}, where the ratio of events in bins of $(x_{\mathrm{B}}, Q^{2})$ is shown for $5~\mathrm{GeV} \times 41~\mathrm{GeV}$ beam energies. This energy configuration is chosen because the relative event yield (cf. Fig. \ref{fig:exclusive_dvcs_pi0_pseudorapidity}) is the largest.  The plot is made for the sample of events generated with toyMC after applying  $Q^{2} > 1~\mathrm{GeV}^{2}$ and $0.01 < y < 0.95$ cuts, requiring both electrons and photons to be reconstructed assuming $|\eta| < 3.5$ acceptance. With no additional cuts on energy thresholds for the detection of photons in ECALs and cuts on the spacial separation of $\pi^0$ decay photons, we may estimate that in the domain of high-$x_{\mathrm{B}}$ one may expect a significant yield of DVMP $\pi^0$ events with respect to DVCS. The effect of such cuts can be deduced from plots like those shown in Fig. \ref{fig:exclusive_dvcs_pi0_energy_angle}, where spectra of energy and opening angles of photons in the lab frame are shown. 

\subsection{Neutron target}
\label{subsec:neutron_gpd}
Neutron GPDs can also be studied at the EIC, where the unique collider experiment at high energy with fully reconstructed final-state particles can provide insights into the neutron substructure. Since there is no easy source of a free neutron target at collider facilities, one of the experimental challenges of constraining the neutron GPDs is to separate the background from nucleus-induced effect when colliding with nuclei. Hereby, the newly proposed experimental technique of spectator tagging in light nuclei~\cite{Strikman:2017koc}, e.g. deuteron, will provide a clean and unambiguous way to measure the neutron GPD. The idea is to measure an exclusive reaction, e.g. DVCS, on the deuteron, where the final-state particles are exclusively identified, including the real photon and the spectator nucleon. Generally the real photon goes into the main detector at mid-rapidity, while the spectator nucleon has a rapidity close to the deuteron beam that goes to the far-forward region. For the case of studying the neutron GPDs, the spectator nucleon would be a proton such that the exclusive process can be unambiguously measured on the neutron. This will be almost no different than the DVCS measurement on a free neutron because the deuteron is very loosely bound. The detector requirements to perform DVCS on neutrons  are i very similar to those needed to  perform  diffractive $J/\psi$ measurements in electron-deuteron scattering at the EIC, which can be found in Ref.~\cite{Tu:2020ymk} and also discussed in Sec.~\ref{subsec:deuteronSpectatorTag} of this Yellow Report. The conclusion based on  detailed $J/\Psi$ studies is that with  a reasonable design of forward proton and neutron detectors, the momentum transfer distributions of the DVCS process can be measured very precisely for a wide range of $t$. This can be achieved by combining two different methods described in Sec.~\ref{subsec:dvmp_ea} for $t$ reconstruction with spectator tagging to identify the events. The spectator proton acceptance is almost 100\% for the general case of deuteron breakups. In addition, different methods of reconstructing the $t$ distribution can provide a better handle on the  systematic uncertainties of the measurement, including sources arising from beam momentum spread and angular divergence. 

\subsection{Deeply virtual Compton scattering off helium}
\label{subsec:dvcs_he4}

In this section we discuss the feasibility of making measurements of coherent  exclusive reactions on light nuclei to study the physics described in section \ref{part2-subsubS-CohDVCS-LightNucl}. The expected main limitation is the detection range in $t$, for which the most challenging situation is helium, thus we will concentrate here on that reaction. Many processes are of great interest, but here we will focus on coherent DVCS. For reactions such as tagged DVCS or DVMP, the detector needs combine the ones we identify in this section, with those identified as needed to study the similar processes of  tagged DIS and DVMP on the proton.

\subsubsection{The Orsay-Perugia Event Generator (\texttt{TOPEG})}

For this study, a new Monte Carlo event generator for the coherent DVCS off the $^4$He nucleus has been developed. This tool, called the Orsay-Perugia event generator (\texttt{TOPEG}), is based on the \texttt{Foam} ROOT library \cite{Jadach:2005ex}. The generated cross section exploits at LO the model for the chiral even GPD describing the $^4$He parton structure presented in Ref. \cite{Fucini:2018gso}. Checks at the JLab kinematics with an electron beam energy of 6 GeV have been successfully performed. 
\texttt{TOPEG} works in two stages: the exploration and the generation. During the exploration, a \textit{foam} of cells is generated and filled with an approximated cross section constant in each cell. The weight, i.e. the ratio between the true distribution and the approximated one, is calculated. The generation is then based on the approximated distribution. For each generated event, a call of the function is done to check that the cross section has a reasonable value at the exact generated kinematic. 
In the model calculation, the real part of the Compton form factor, involving a principal value integral, is most time consuming. Besides, since this term affects the magnitude of the cross section only with a little impact at the kinematics considered, we neglected this term in the generation of the event presented in the following.

For the three scenarios of EIC, we generated 1 million events with $Q^2 > 4 $ GeV$^2$ and $|t_{min}| < |t| < |t_{min}+0.5| $ GeV$^2$. The corresponding luminosity ranges between 100 and 250 nb$^{-1}$ going from the high energy configuration to the lower ones.
The obtained total cross section ranges between 4 and 11$\, \mu$b and grows with the beam energies since it is dominated by the form factor of the $^4$He strongly dependent on $t$. This value, however, is significantly reduced (around 95$\%$ or more) when accounting for the $t$ acceptance of the helium nuclei in the far forward detectors.

\subsubsection{Central detector}

\begin{figure}[ht]
\includegraphics[scale=0.21]{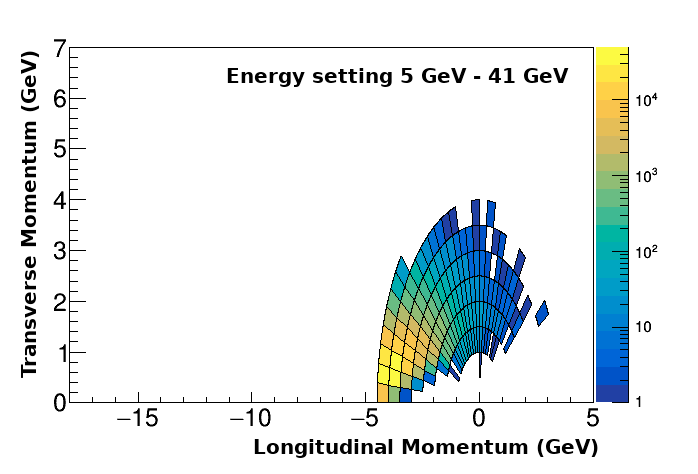}
\hspace{-0.6cm}
\includegraphics[scale=0.21]{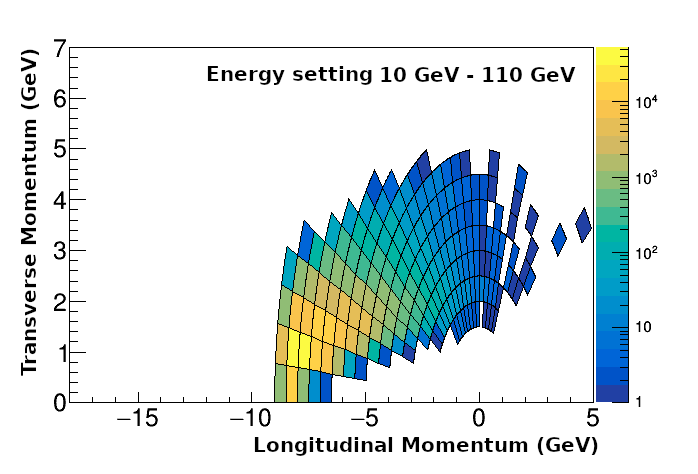}
\hspace{-0.6cm}
\includegraphics[scale=0.21]{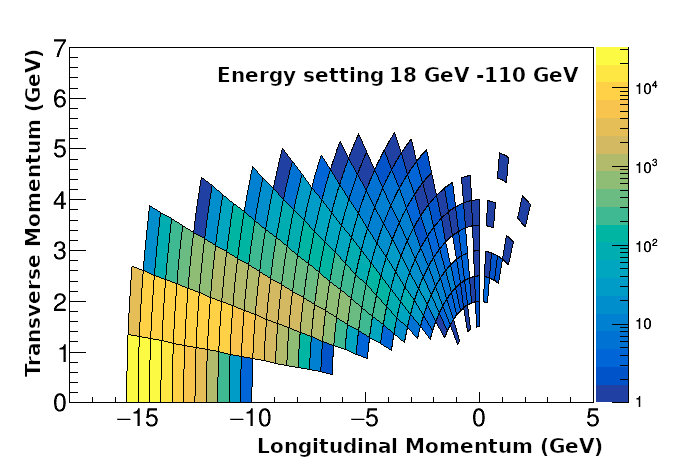}
\label{pg-two} 
\caption{Kinematic distribution of the photons produced in coherent DVCS on helium-4 as generated with \texttt{TOPEG} for the three energy configurations envisioned for the EIC.}
\label{pho-kine-coh-dvcs}
\end{figure}

From these events, we evaluated the acceptance of electrons and photons to be detected in the central detector. The DVCS photons are shown in Fig.~\ref{pho-kine-coh-dvcs} for all energy configurations. 
We observe that they are mostly in easily accessible kinematics within the acceptance described by the detector matrix. 
The limit is reached only for some low angle photons in the highest energy configuration. This concerns mostly the lowest $x_B$ events and thus the highest energy settings for $\phi \approx 0$ kinematics, where $\phi$ is the angle between the hadronic and leptonic planes. This is not too concerning as this kinematic region concentrates mostly the BH contribution to the process and has little importance for the extraction of the Compton Form Factor (CFF) and the physics goals in general. However, the proportion of lost events rises to 20\% for the highest energy setting, which shows that the pseudorapidity limit of 3.5 is critical and starts to significantly affect the data collected. If this limit was to be modified, it could affect strongly the physics reach at low $x_B$ and the interest to run light nuclei at the highest energy settings.

\subsubsection{Forward detector}

In the far forward region, where the scattered helium nuclei are detected, the situation is more complicated. At low $x_B$, $t_{min}$ becomes very small leading to kinematics impossible to access,meaning the minimum $t$ reported will be  set by the detector acceptance. While for  high $t$ the limitation will be the luminosity. These two values are the critical limits that need to be evaluated for the light nuclei coherent processes. 

In the detector matrix, the limit at low $t$ is given by the Roman pot capabilities. It is expected to detect recoil nuclei at transverse momenta as low as 0.2 GeV. This corresponds to $-t \approx 0.04$ GeV$^2$. In a similar fashion to the photon acceptance, this appears to leave plenty of room to study the $t$ dependence between $t_{min}$ and the first minimum (at $-t \approx 0.7, 0.42, 0.48$ GeV$^2$ for $d$, $^3$He,
$^4$He, respectively). However, since $|t| \sim \pT^2$, a degradation of the lower momentum reach proposed would significantly affect our capability to study $^3$He and $^4$He coherent DVCS.

\subsubsection{Overall performance}

Overall, it appears that the detector capabilities proposed in this report have a wide enough kinematical range to study the tomography and other possible elusive nuclear parton dynamics around the critical first diffraction minimum of the electromagnetic form factor. We identify two key points as critical for these studies, in the sense that a degradation would directly affect the accessible physics. These are the minimum angle of photon detection in the backward detector and the minimum transverse momentum accessible in the Roman Pots for recoil nuclei.

\begin{figure}[ht]
\includegraphics[scale=0.118]{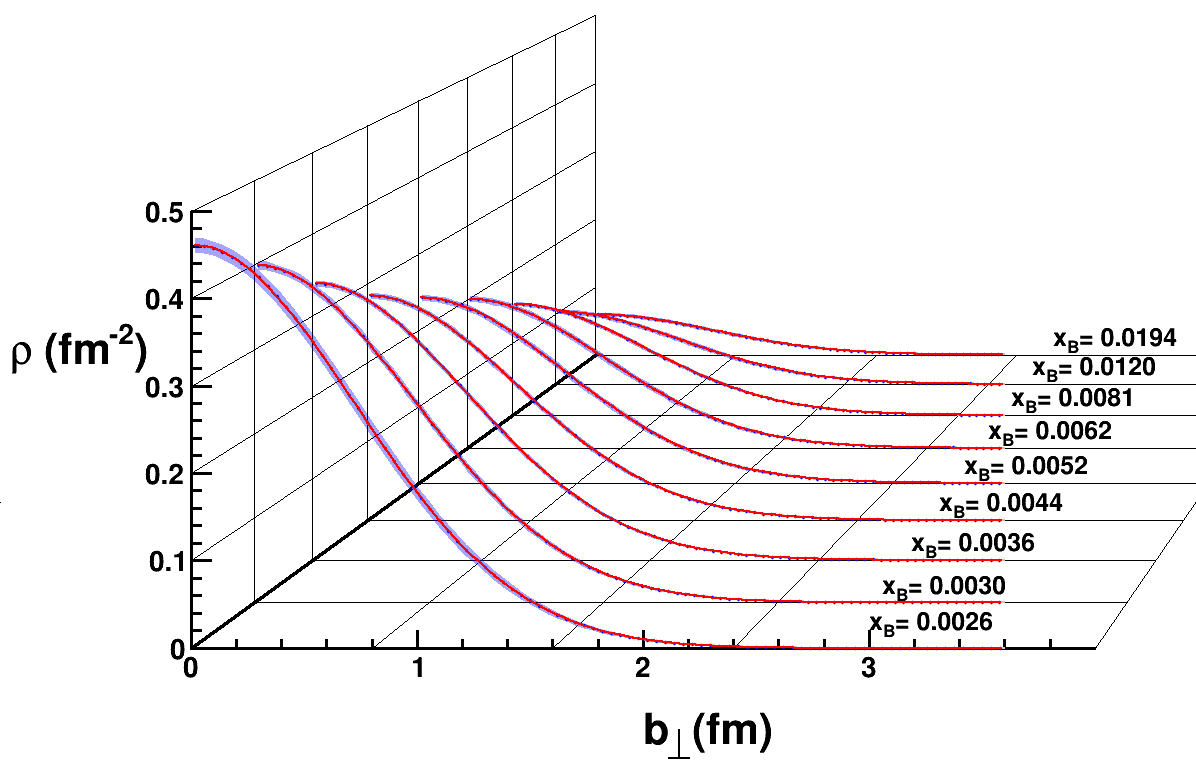}
\includegraphics[scale=0.118]{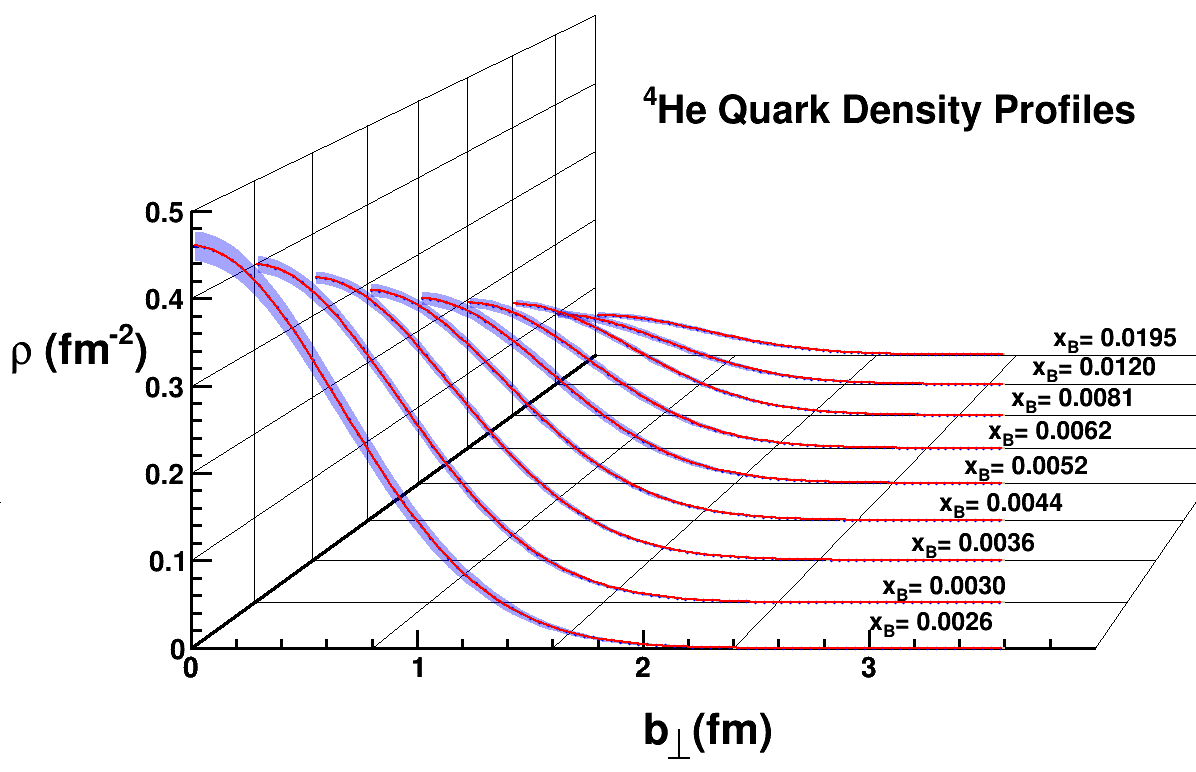}
\includegraphics[scale=0.118]{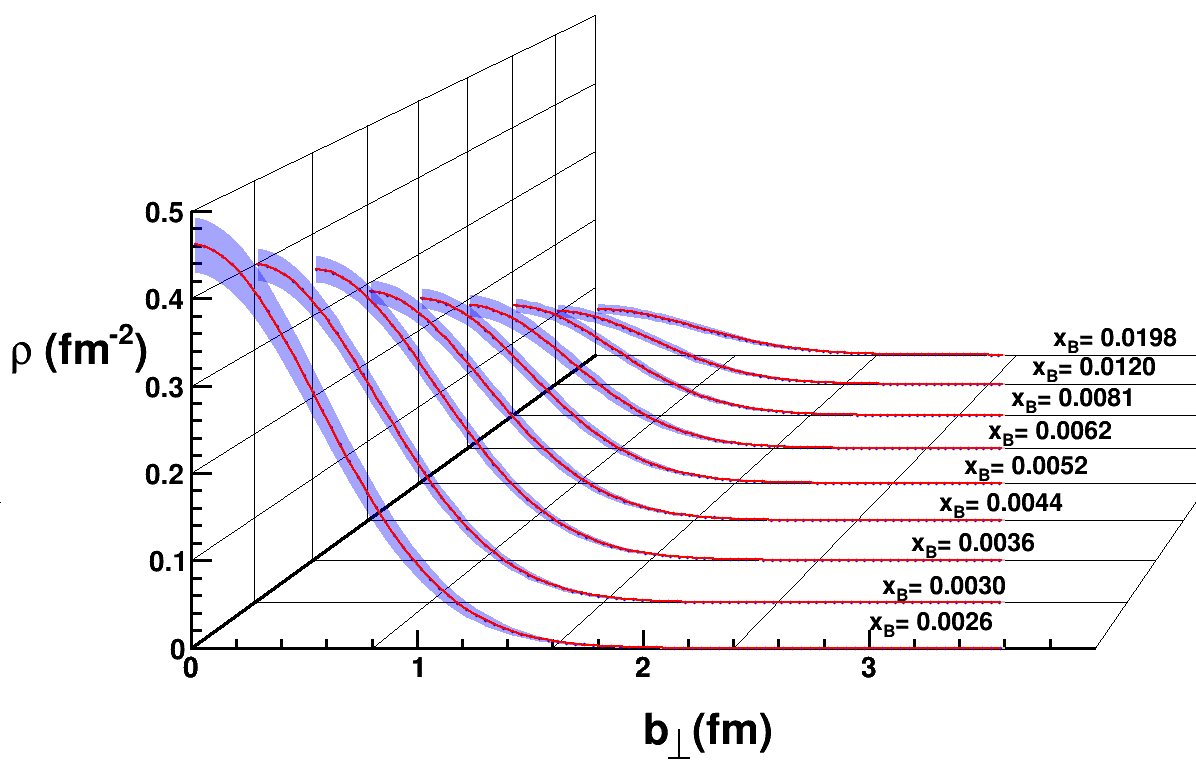}
\caption{Quark density profiles extracted from pseudo-data corresponding to a 10 fb$^{-1}$ integrated luminosity and generated with the \texttt{TOPEG} software. The extraction was performed using a fit based on the leading order formalism \cite{Kirchner:2003wt,Belitsky:2008bz} and a Roman pots detection threshold set to $\pT = 0.1$ (left), $0.2$ (center) and $0.3$~GeV (right).}
\label{fig-coh-dvcs-profile}
\end{figure}

In order to quantitatively assess the effect of different recoil nuclei minimum transverse momentum, we performed a fit of pseudo-data generated with the \texttt{TOPEG} software using different assumptions. We show in Fig. \ref{fig-coh-dvcs-profile} the quark density profiles extracted using the leading order formalism \cite{Kirchner:2003wt,Belitsky:2008bz}, assuming three different minimum transverse momenta for the Roman pots and a 10~fb$^{-1}$ integrated luminosity. We chose to present here the figure of merit for the nominal design value of the Roman Pots in this report $\pT = 0.2$~GeV, as well as a higher and lower value ($\pT = 0.1 , 0.3$~GeV) to illustrate the criticality of this parameter for the measurement. As we can see, the density profile extraction remains doable for all the assumptions used here, however we notice that the error is highly correlated to the measurement threshold of the Roman Pots. This highlights the importance to optimize the Roman Pot threshold to maximize the phenomenological exploitation of the nuclear DVCS data.

\subsection{Timelike Compton scattering}
\label{subsec:tcs}

\begin{figure}[htb]
    \begin{center}
        \includegraphics[width=0.49\textwidth]{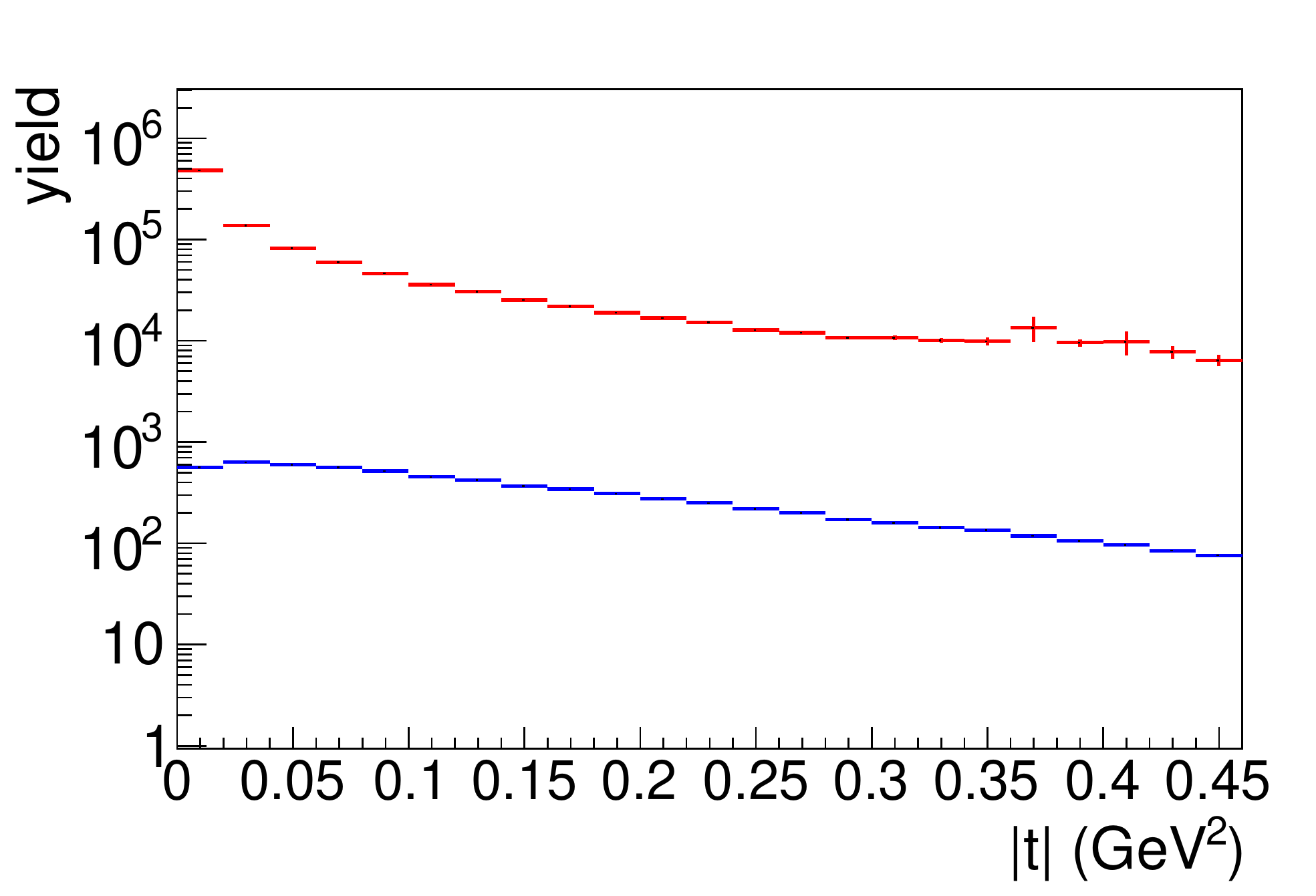}
        \includegraphics[width=0.49\textwidth]{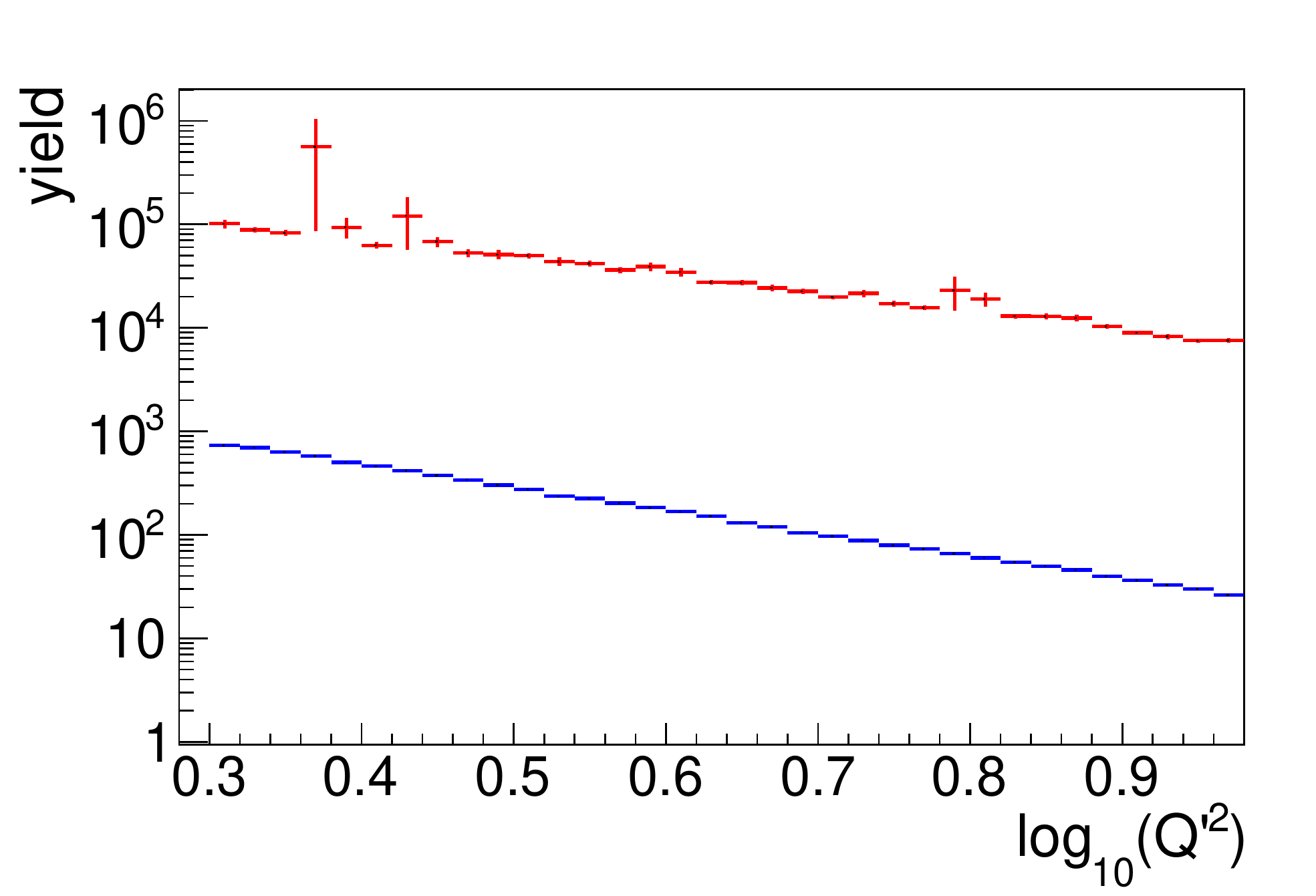}
        \includegraphics[width=0.49\textwidth]{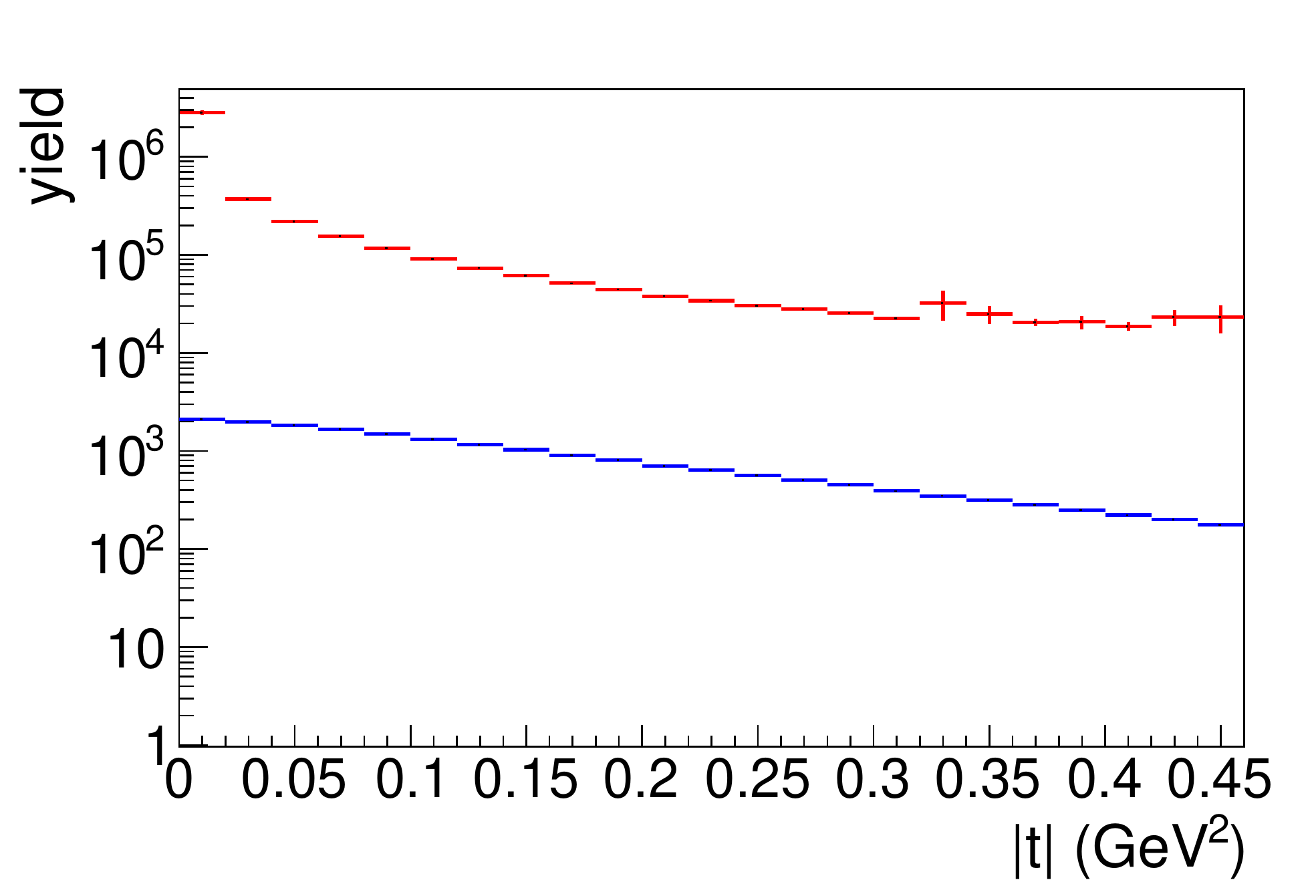}
        \includegraphics[width=0.49\textwidth]{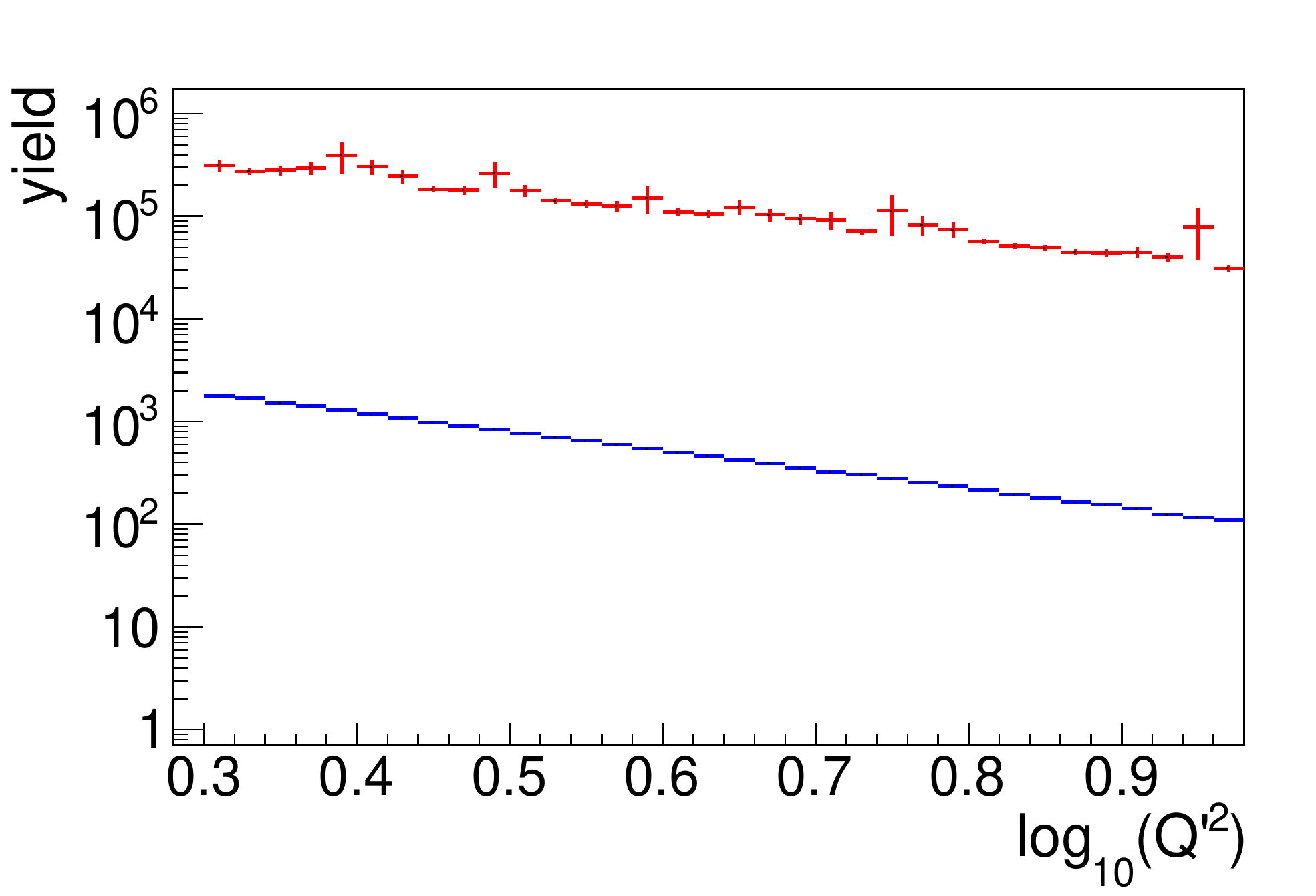}
    \end{center}
    \caption{Comparison of the kinematics between simulated pure BH (red) and pure TCS (blue) at two collision energies: 5~$\times$~41~GeV (top) and 18~$\times$~275~GeV (bottom). Left column: $|t|$ distribution, right column: $Q'^2$. The yield is quoted for an integrated luminosity of 10~fb$^{-1}$.}
    \label{fig:tcs_kin}
\end{figure}

\begin{figure}[htb]
    \begin{center}
        \includegraphics[width=0.49\textwidth]{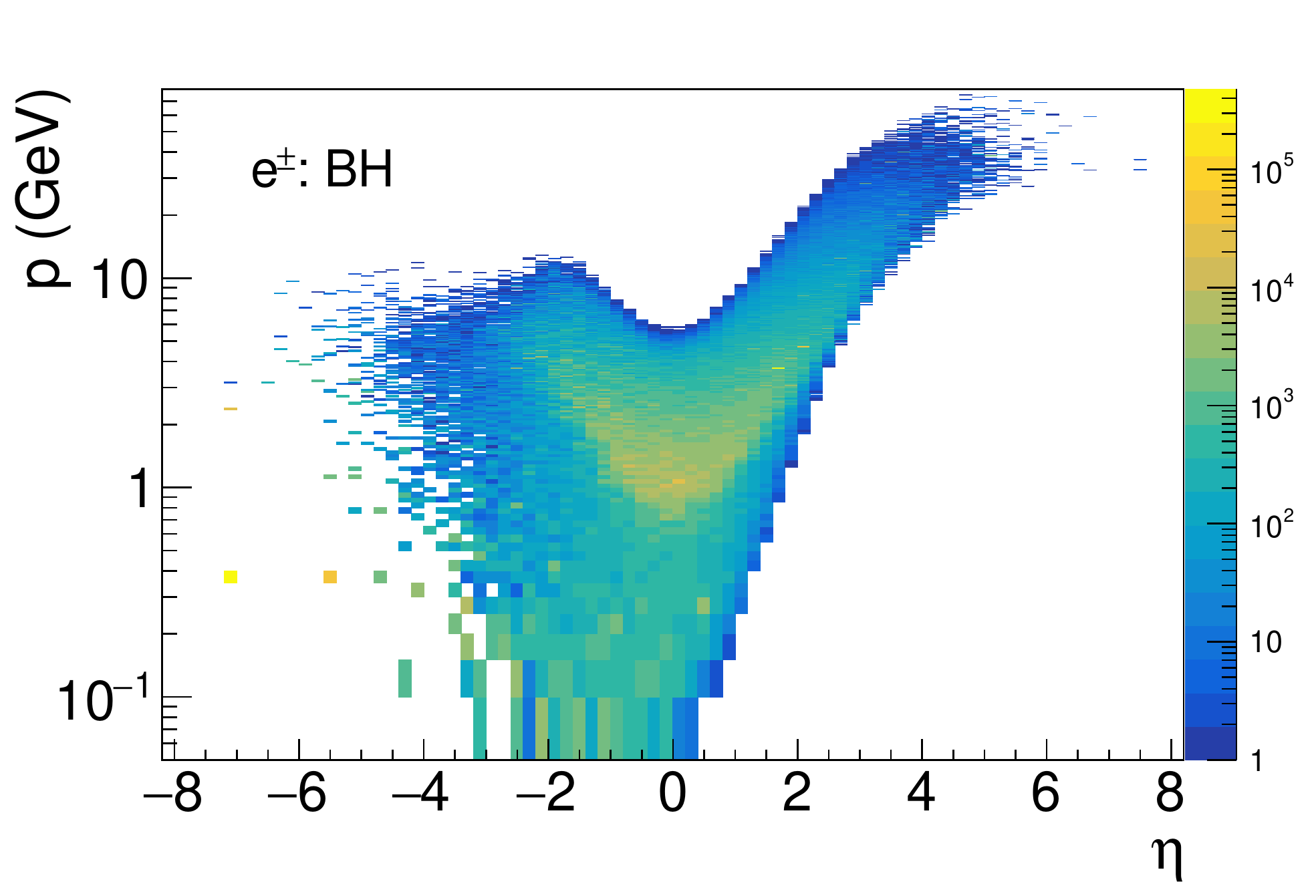}
        \includegraphics[width=0.49\textwidth]{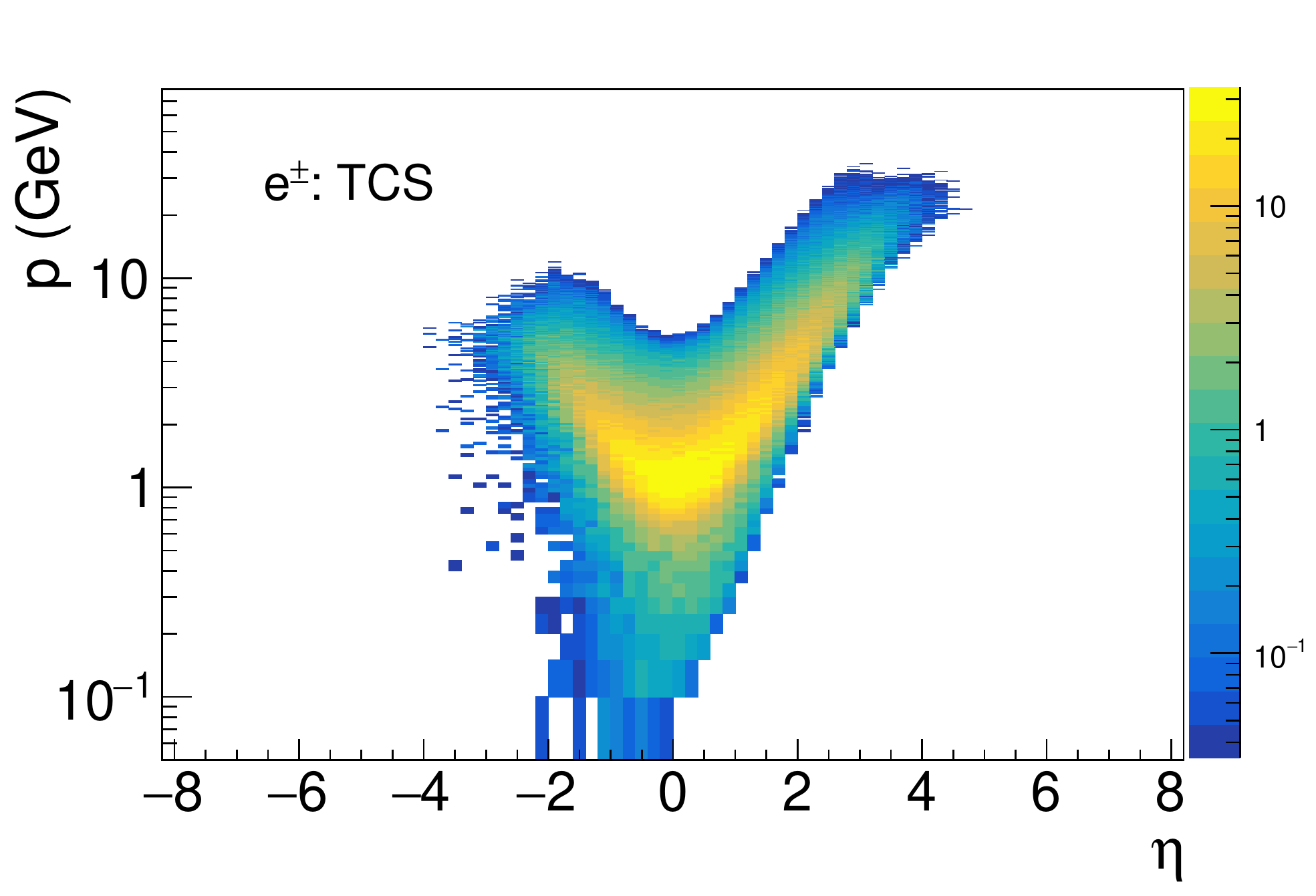}
        \includegraphics[width=0.49\textwidth]{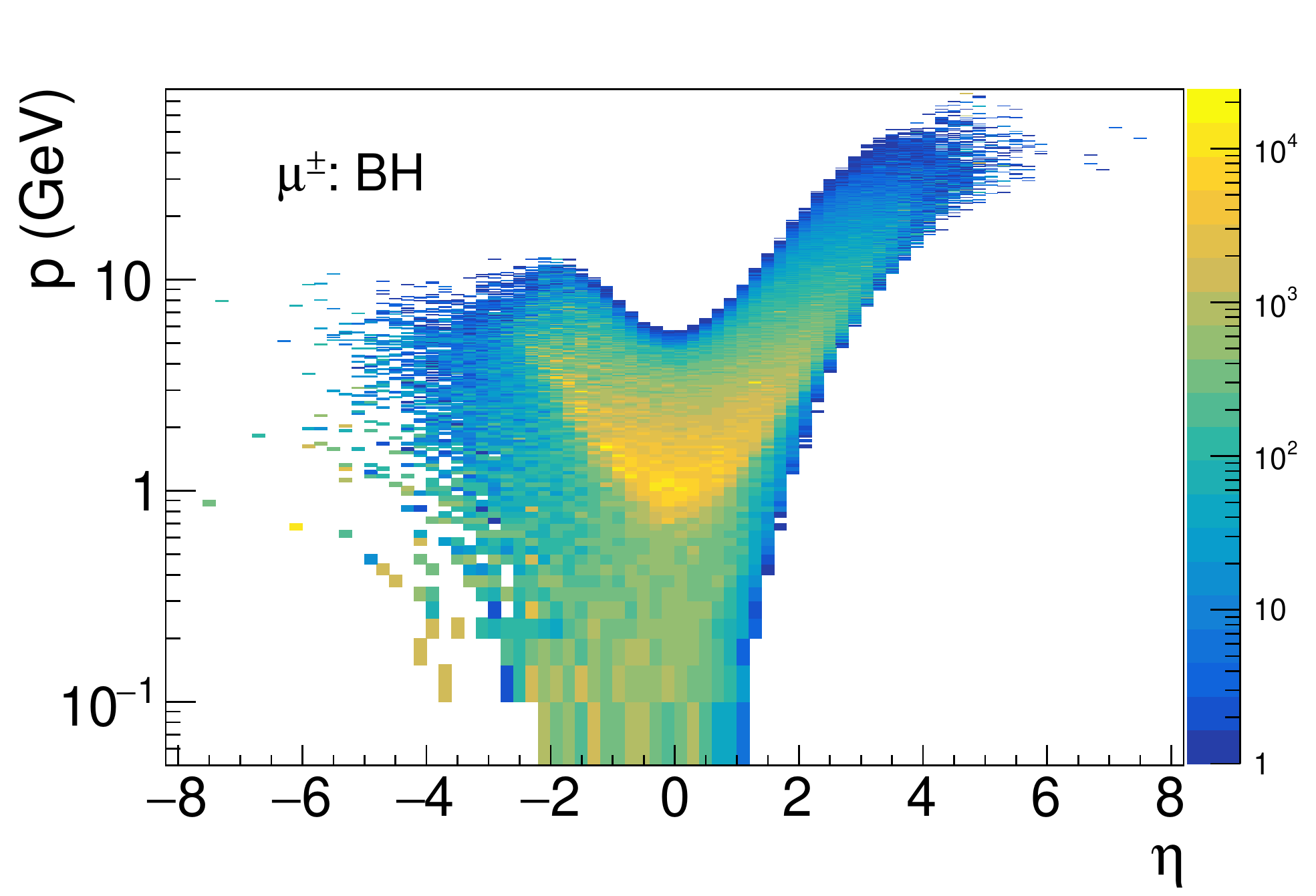}
        \includegraphics[width=0.49\textwidth]{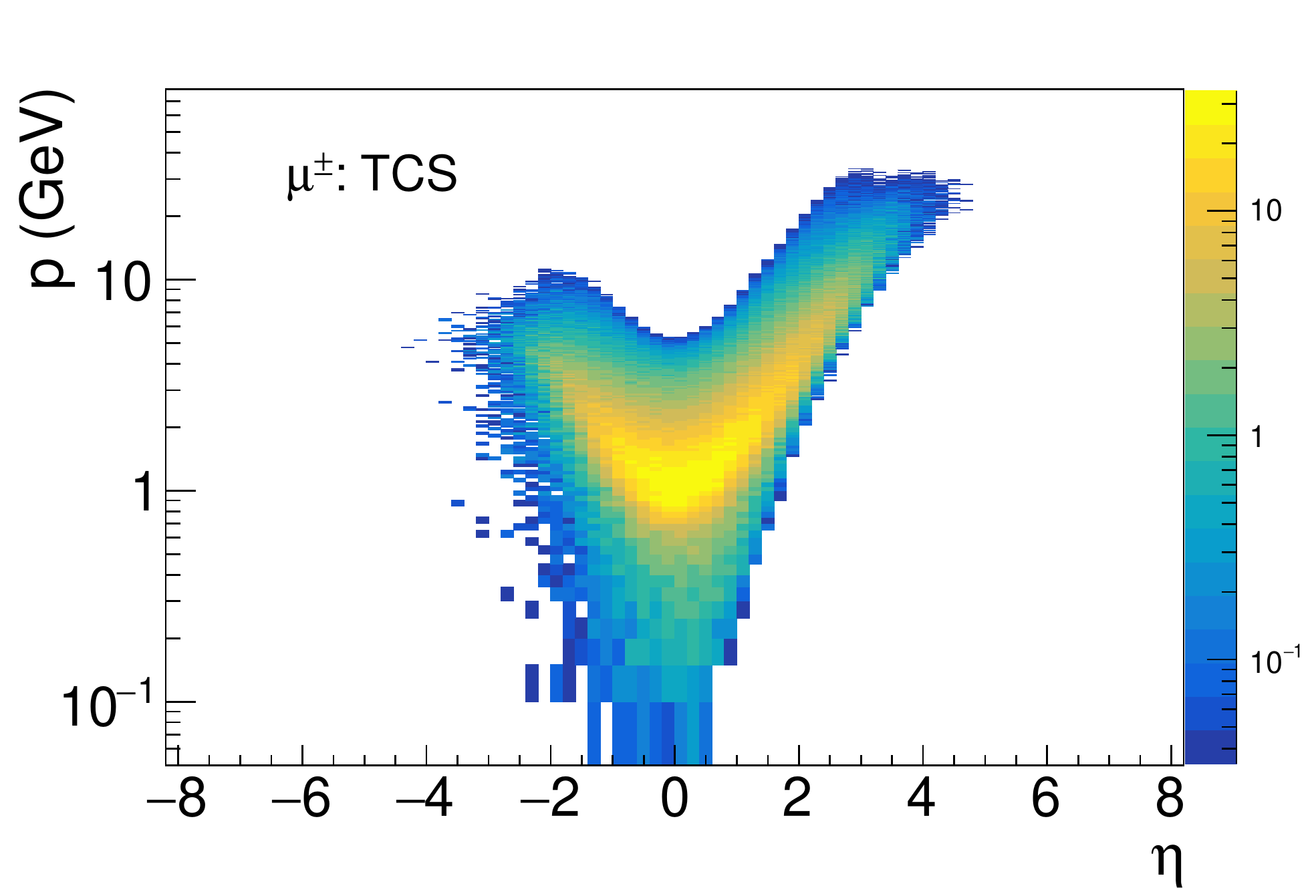}
    \end{center}
    \caption{Momentum vs pseudorapidity for $e^+e^-$ (top) and $\mu^+\mu^-$ (bottom) at the 5~$\times$~41~GeV collision energy, for an integrated luminosity of 10~fb$^{-1}$. Left: BH, right: TCS.}
    \label{fig:leptons_5x41}
\end{figure}

\begin{figure}[htb]
    \begin{center}
        \includegraphics[width=0.49\textwidth]{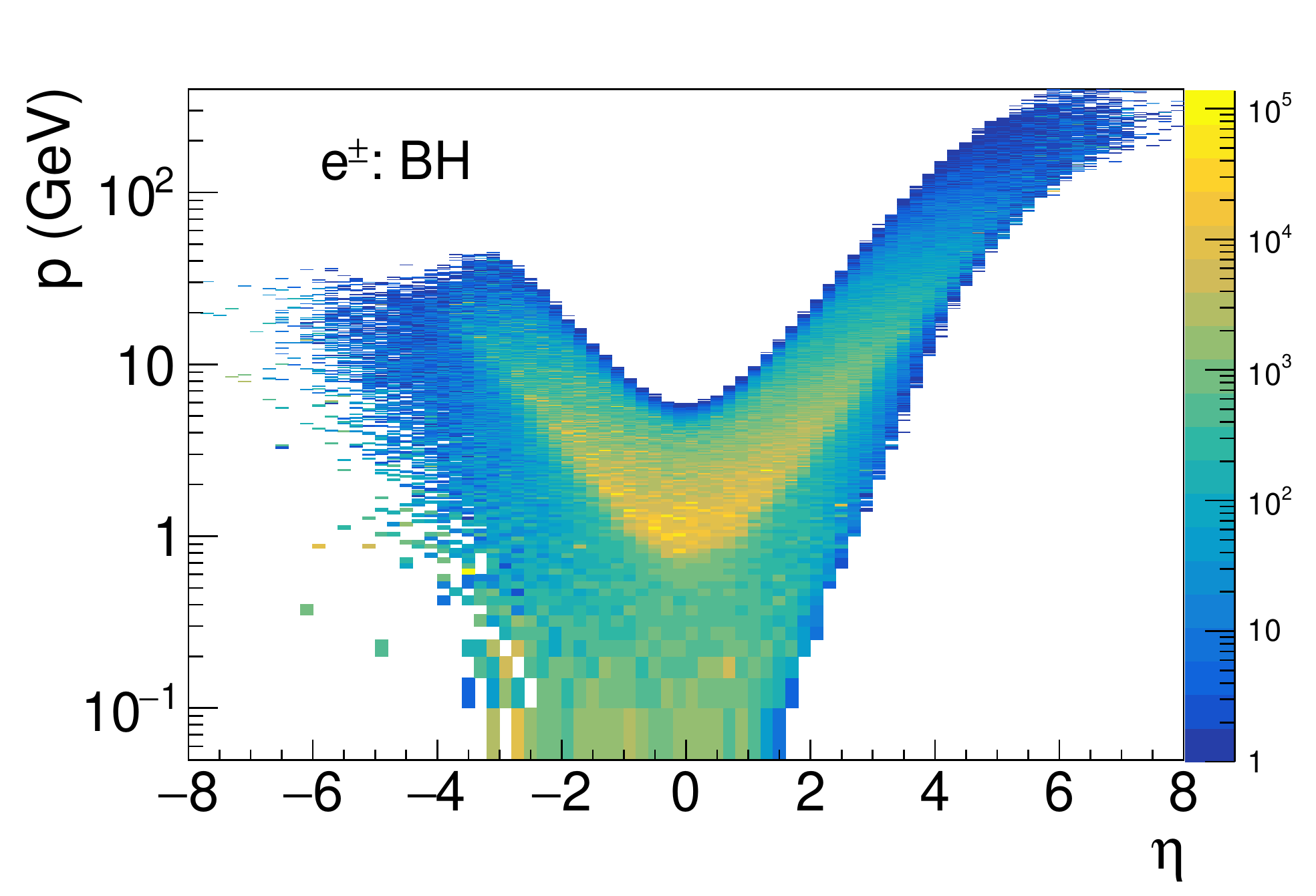}
        \includegraphics[width=0.49\textwidth]{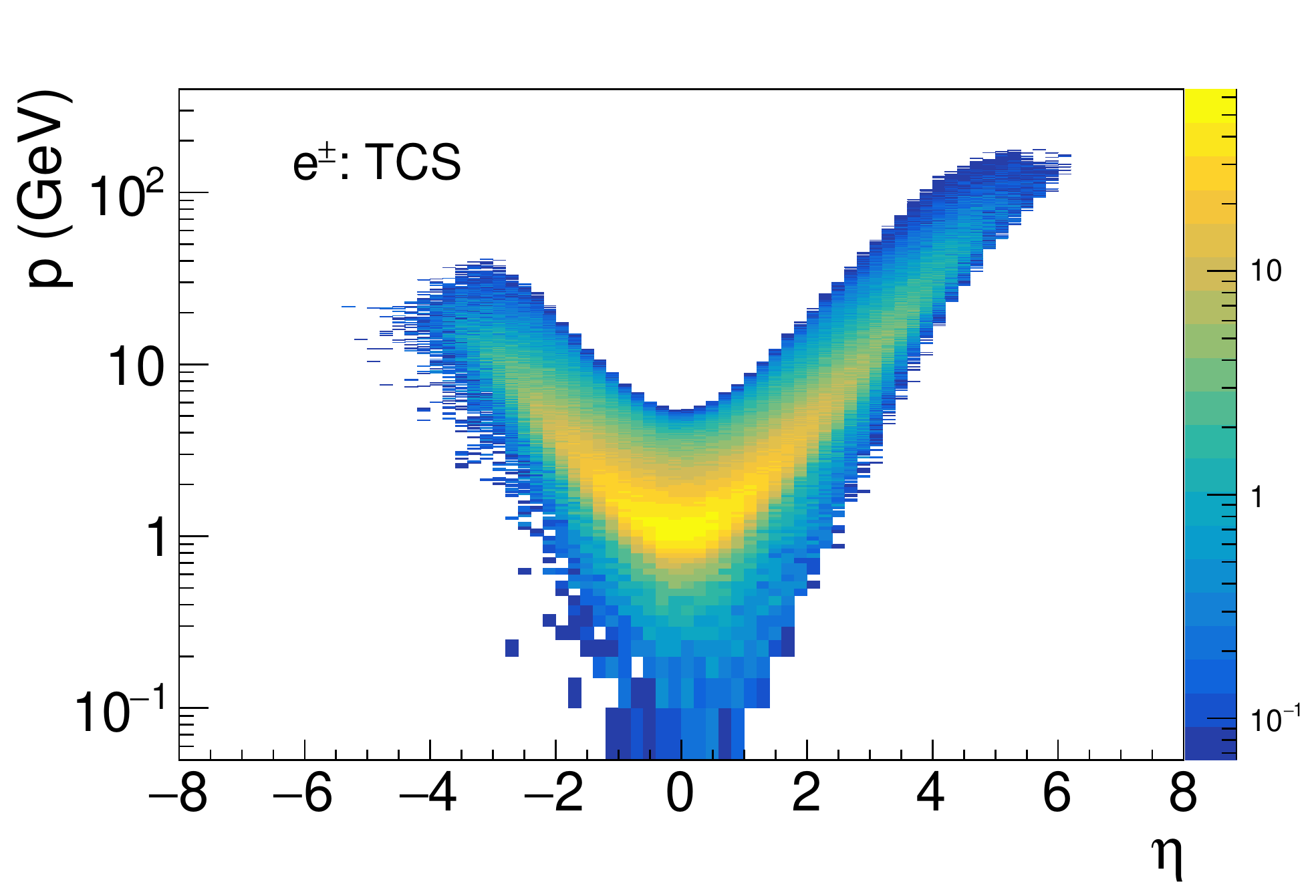}
        \includegraphics[width=0.49\textwidth]{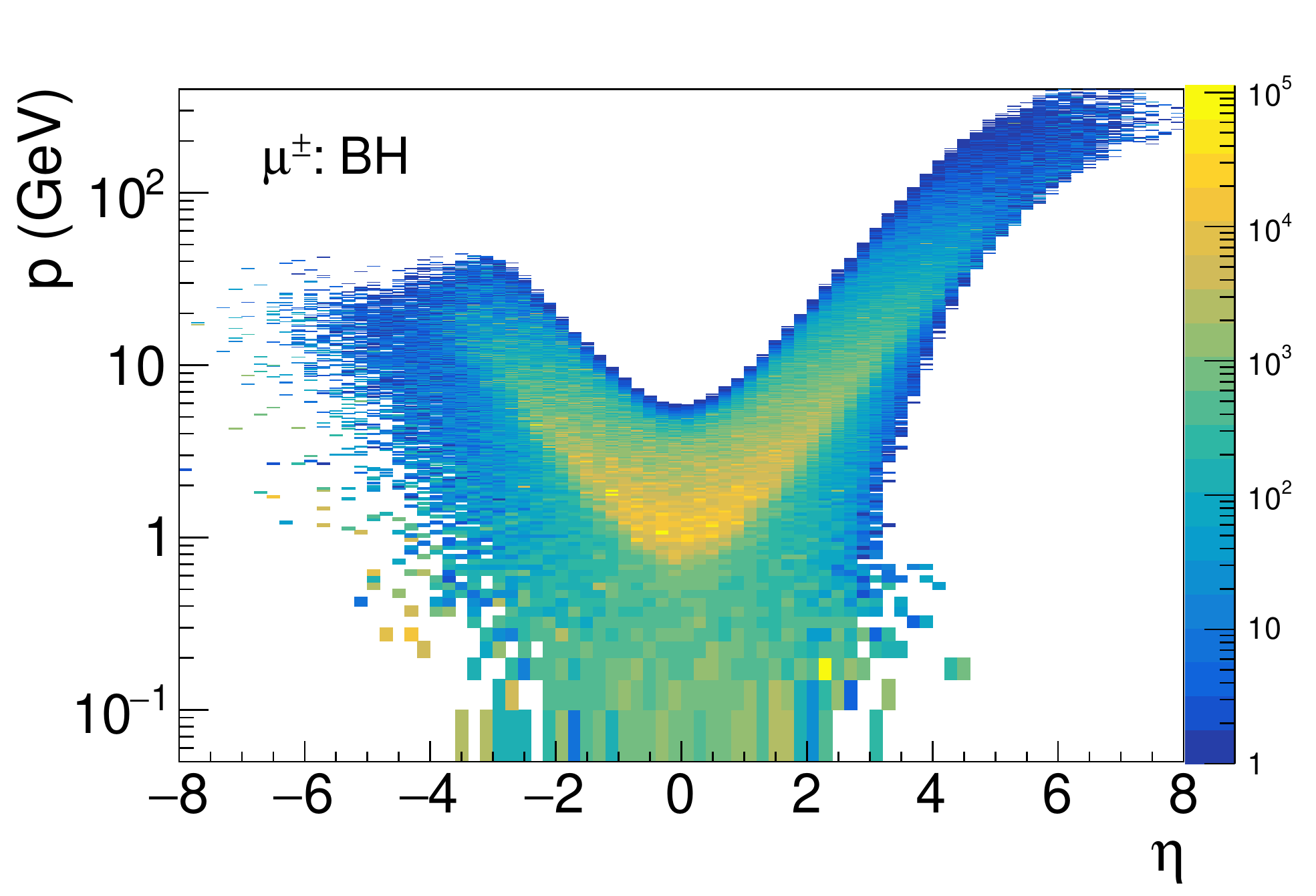}
        \includegraphics[width=0.49\textwidth]{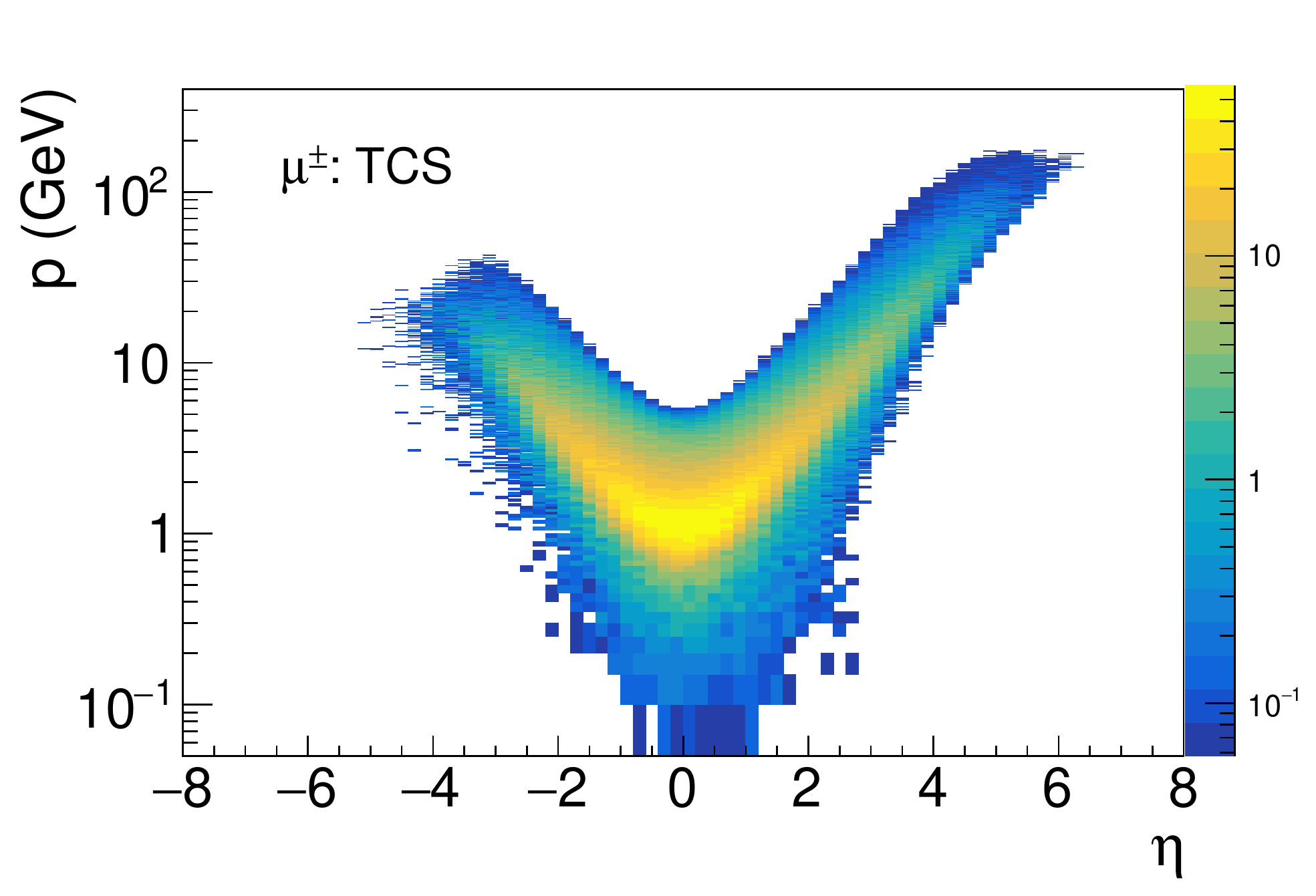}
    \end{center}
    \caption{Momentum vs pseudorapidity for $e^+e^-$ (top) and $\mu^+\mu^-$ (bottom) at the 18~$\times$~275~GeV collision energy, for an integrated luminosity of 10~fb$^{-1}$. Left: BH, right: TCS.}
    \label{fig:leptons_18x275}
\end{figure}

\begin{figure}[htb]
    \begin{center}
        \includegraphics[width=0.49\textwidth]{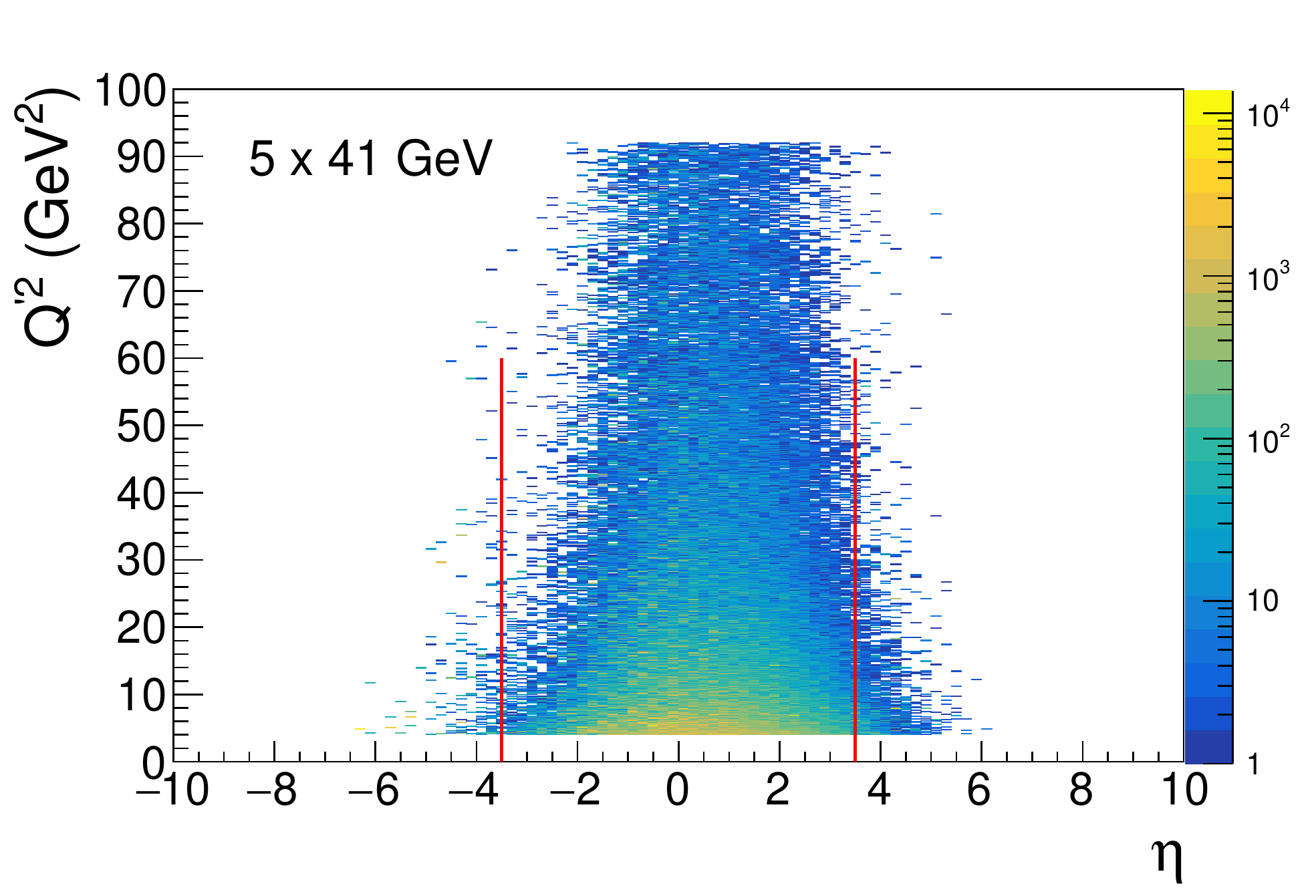}
        \includegraphics[width=0.49\textwidth]{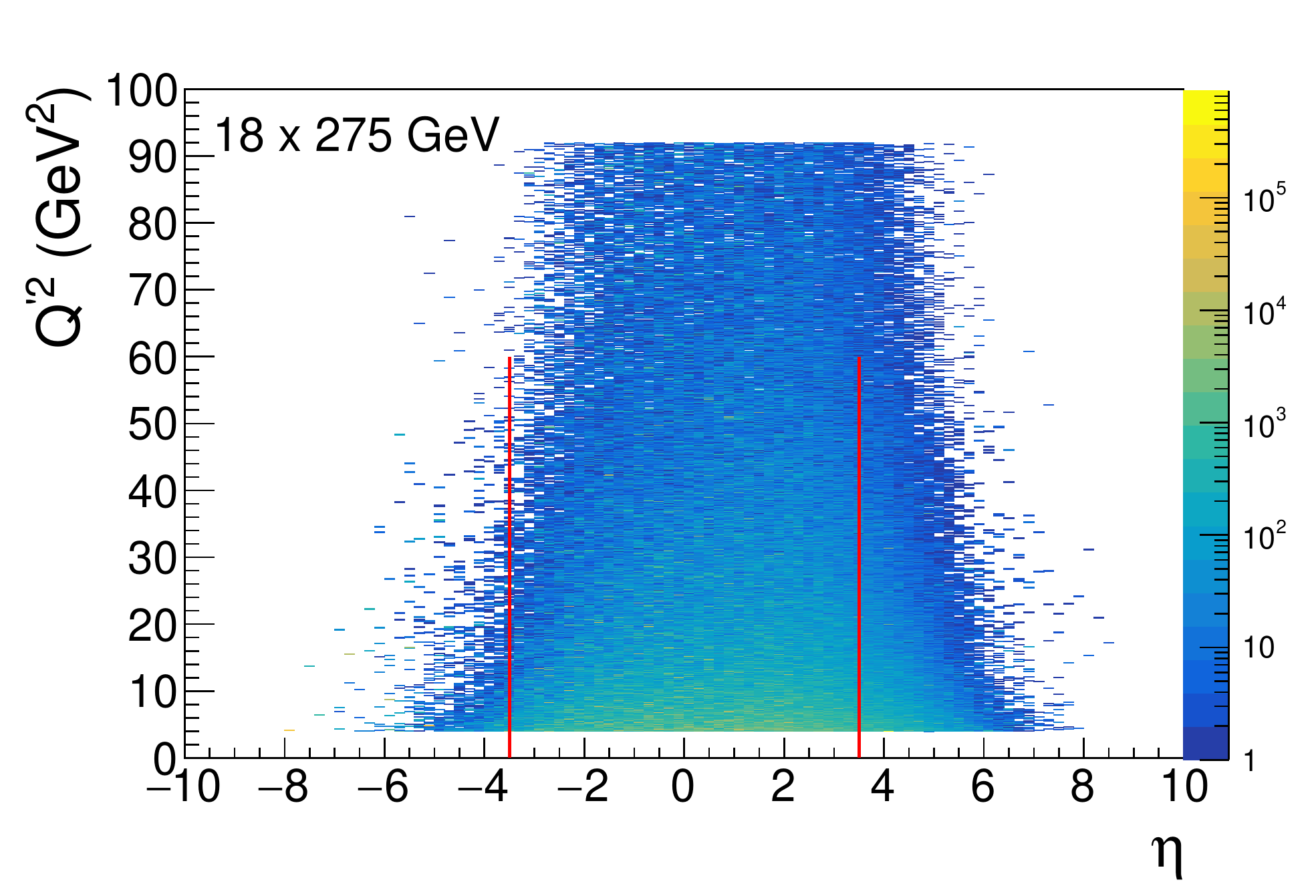}
    \end{center}
    \vspace*{-5mm}
    \caption{$Q'^2$ vs pseudorapidity of the produced electron for the 5~$\times$~41~GeV (left) and 18~$\times$~275~GeV (right) collision energies, BH events for an integrated luminosity of 10~fb$^{-1}$. Red, vertical lines indicate the edges of the nominal central detector acceptance: $|\eta| < 3.5$.}
    \label{fig:tcs_qprim2_eta}
\end{figure}

Timelike Compton Scattering (TCS) is an inverse process to DVCS, where a real photon scatters from a nucleon to produce a large $Q'^2$ virtual photon which decays into a pair of leptons. As such, TCS has a final state identical to the exclusive production of $J/\Psi$, but without the advantage of a well-defined invariant mass for the lepton pair.  Exclusive reconstruction of TCS is, therefore, experimentally more challenging. We have simulated TCS using a toy Monte Carlo which generates a spectrum of quasi-real photons in a head-on collision of electrons and protons and interpolates through tables of CFFs to calculate TCS cross sections, which are applied as weights to each event \cite{tcs_mc}. The CFF tables were produced with the PARTONS framework and used the Goloskokov-Kroll parametrization~\cite{Goloskokov:2009ia} -- the same CFFs were also employed in the DVCS simulations in Sec. \ref{subsec:dvcs_ep}. Similarly to DVCS, the TCS amplitude interferes with the Bethe-Heitler process, which can produce the same final state. Both the pure TCS and the pure BH distributions were simulated, each for $e^+e^-$ and $\mu^+\mu^-$ decay leptons. To suppress the BH contribution where it particularly dominates the kinematics, a cut was applied on $\pi/4 < \theta < 3\pi/4$, which is the angle between the positive lepton momentum and scattered proton in the lepton centre of mass frame. An additional cut on the virtuality of the produced photon, $Q'^2 > 2$~GeV$^2$, ensured a hard scale in the scattering, while requiring $Q'^2 < M_{J/\Psi}^2$ suppressed the resonant background. All quasi-real photon virtualities up to $Q^2 = 0.1$~GeV$^2$ were included. The integrated kinematics of the simulation, for both the BH and the TCS signal, are shown in Fig.~\ref{fig:tcs_kin}. The resulting distributions of the decay leptons, for TCS and BH separataly and for the lowest and highest collision energy settings, are shown in Figs. \ref{fig:leptons_5x41} - \ref{fig:leptons_18x275}. The distributions are very similar for both processes, except for the difference in the yield. The nominal acceptance of the central detector, $|\eta| < 3.5$, would result in the loss of only the highest momentum leptons. An extended far-forward acceptance of $\eta < 4.5$ would catch the majority of the lepton pairs even at the highest collision energy. Any loss due to the acceptance would not have a significant effect on the distribution of $Q'^2$, which provides the hard scale in the process (Fig.~\ref{fig:tcs_qprim2_eta}). 

The BH contribution dominates over the TCS cross section by approximately two orders of magnitude. Similarly to DVCS, access to the TCS amplitude is usually obtained via the BH-TCS interference term in the overall cross section. This is expected to provide a boost to the TCS signal on the order of 10-15\%.  

\begin{figure}[htb]
    \begin{center}
        \includegraphics[width=0.49\textwidth]{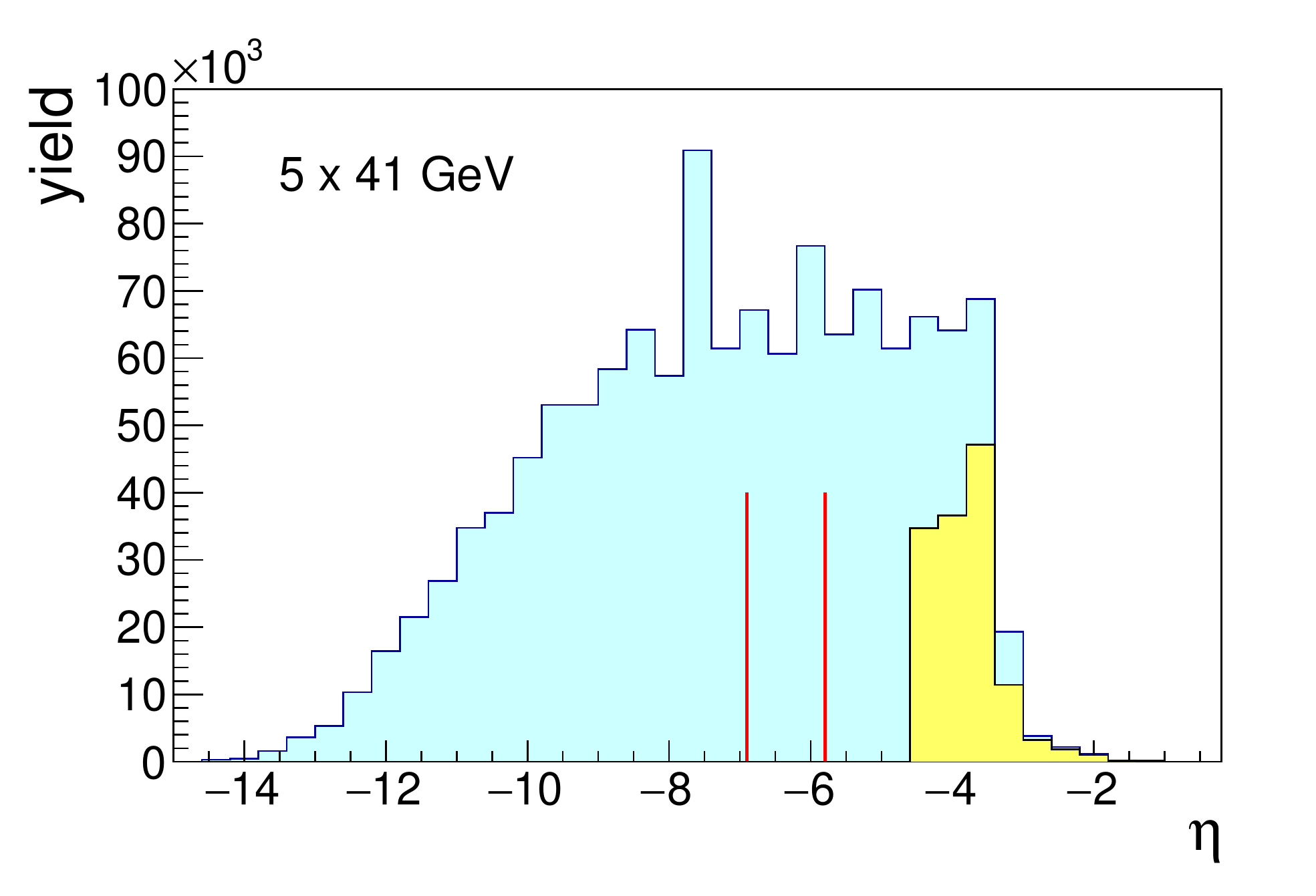}
        \includegraphics[width=0.49\textwidth]{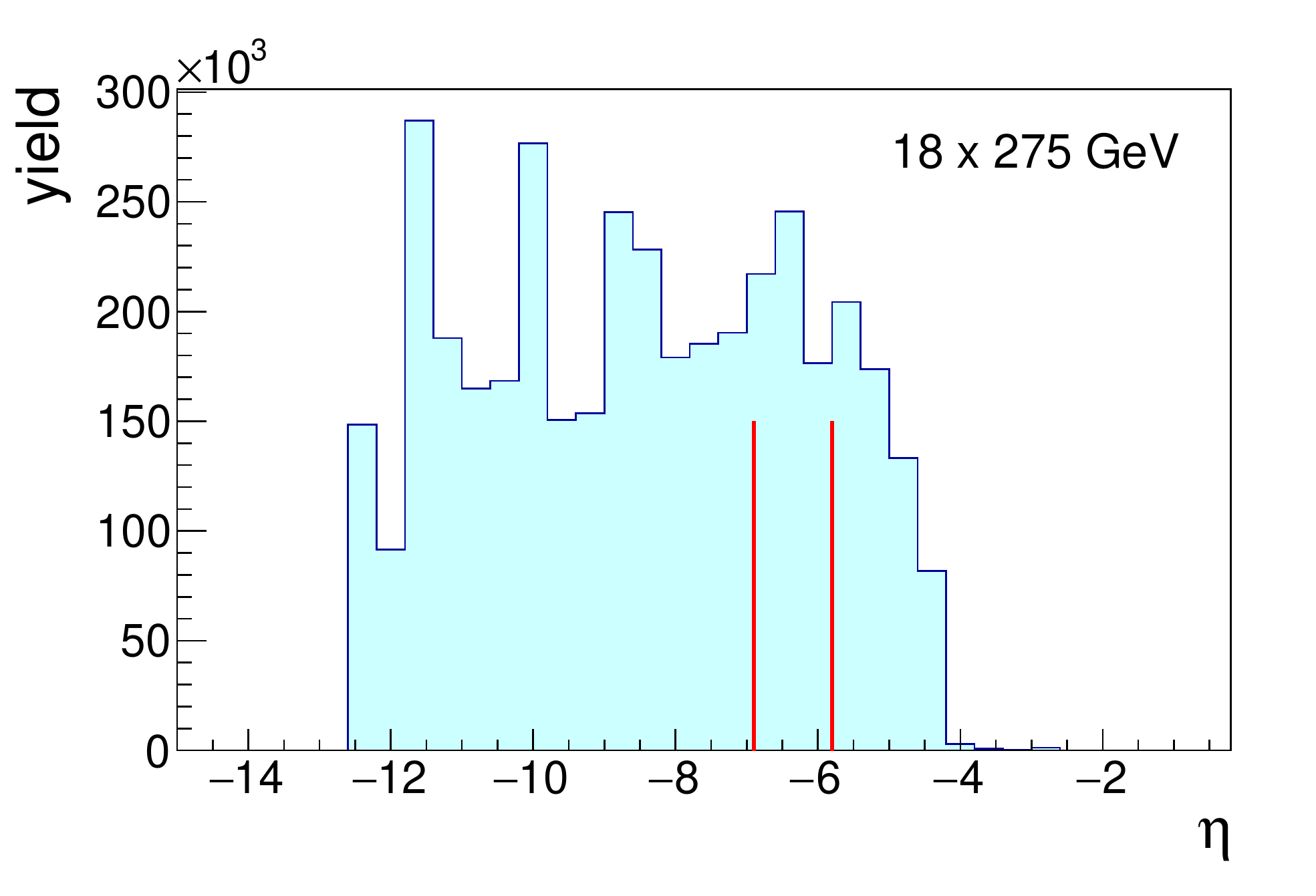}
    \end{center}
     \vspace*{-5mm}
    \caption{The generated pseudorapidity of the scattered electron for the 5~$\times$~41~GeV (left) and 18~$\times$~275~GeV (right) collision energies, shown in cyan. The acceptance range of the low-$Q^2$ tagger is indicated by red lines, while the yellow distribution shows the case where all four final state particles are reconstructed. BH events at an integrated luminosity of 10~fb$^{-1}$.}
    \label{fig:tcs_scattered_el}
\end{figure}

\begin{figure}[htb]
    \begin{center}
        \includegraphics[width=0.49\textwidth]{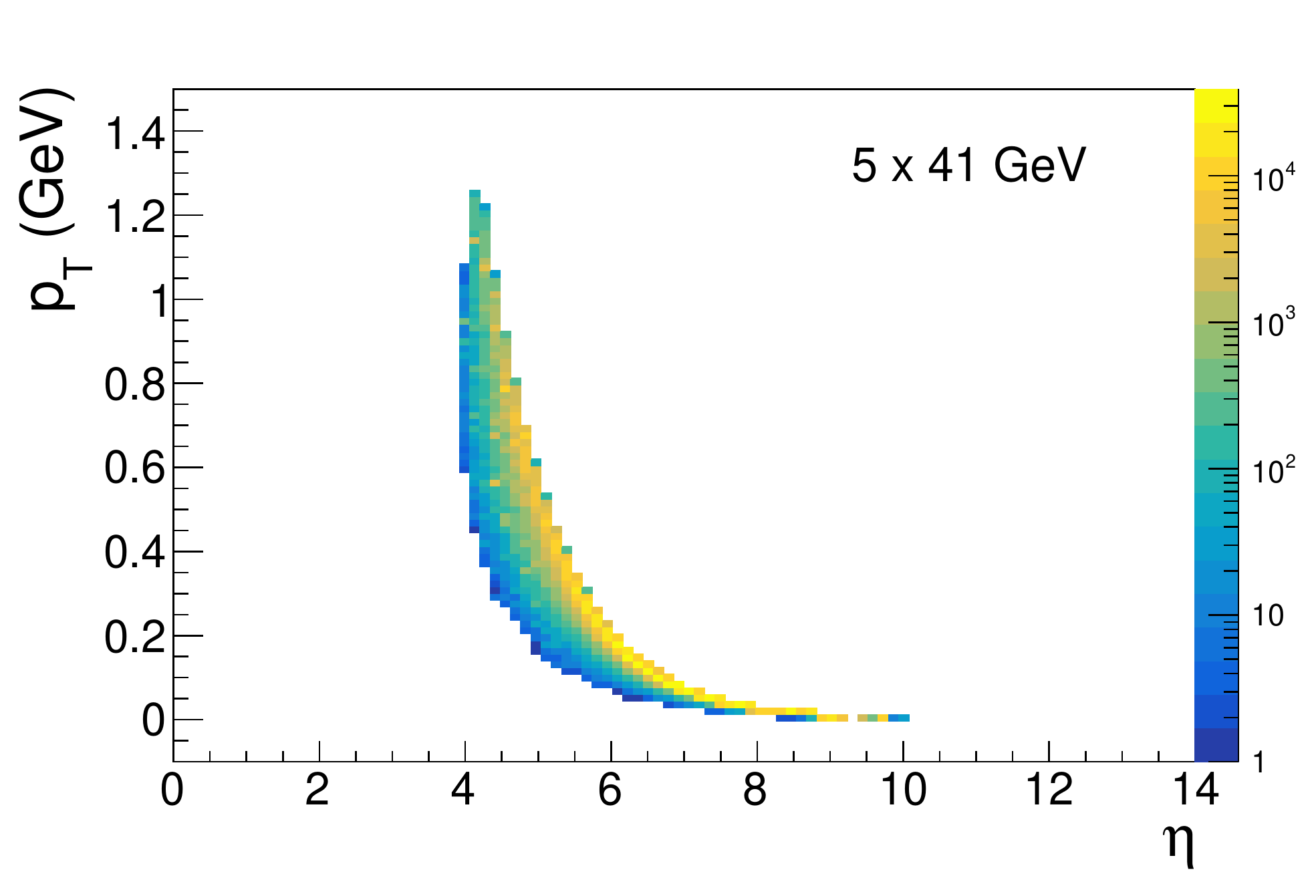}
        \includegraphics[width=0.49\textwidth]{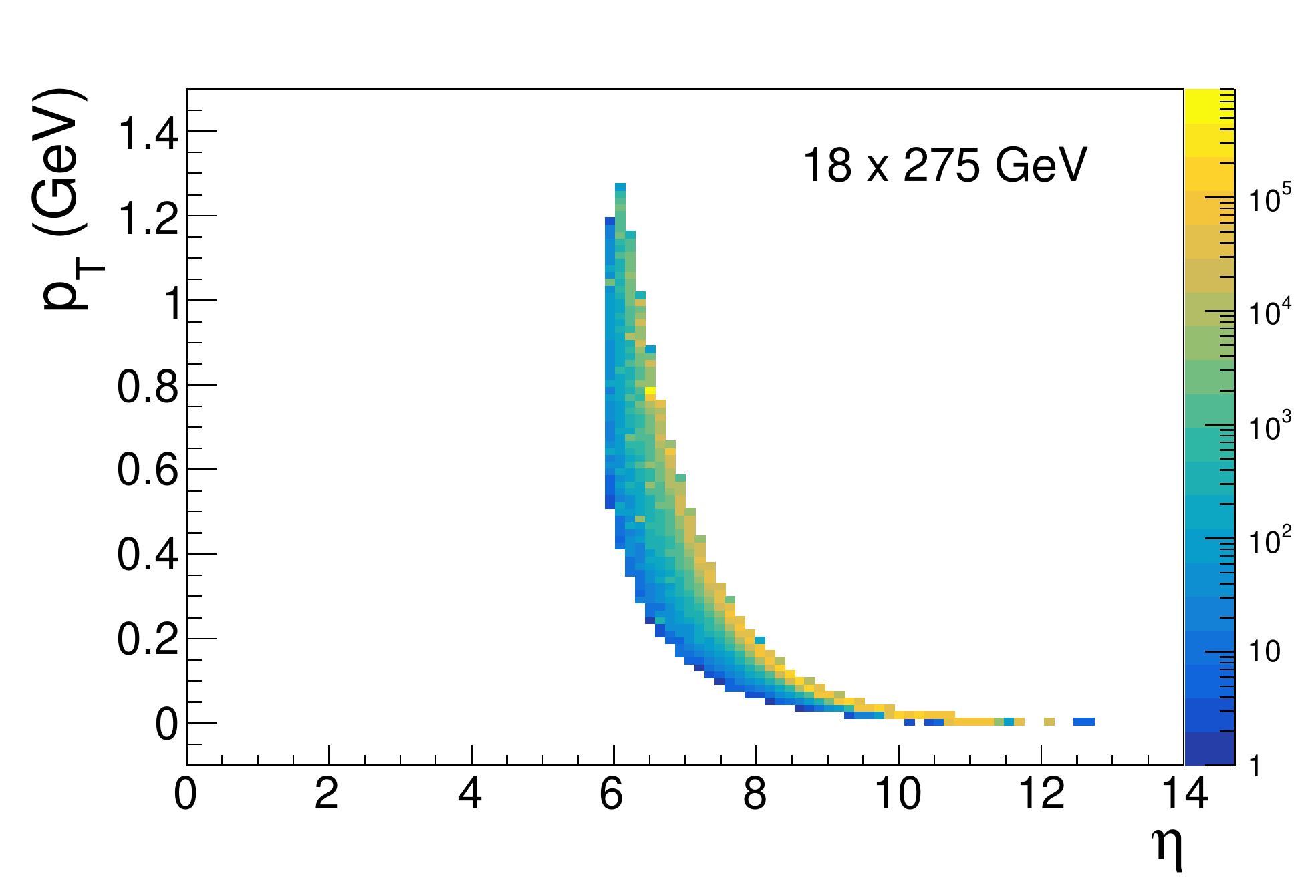}
    \end{center}
     \vspace*{-5mm}
    \caption{The generated $\pT$ vs pseudorapidity of the recoil protons for the 5~$\times$~41~GeV (left) and 18~$\times$~275~GeV (right) collision energies, for BH events at an integrated luminosity of 10~fb$^{-1}$.}
    \label{fig:tcs_scattered_p}
\end{figure}

The generated events were passed through EIC-smear to determine the effect of detector resolutions. Requiring $Q^2 < 0.1$GeV$^2$ for photoproduction results in scattered electron angular distributions shown, for the lowest and highest collision energies, in Fig.~\ref{fig:tcs_scattered_el}. While at the lowest energy some electrons may be detected in the far backward detectors, at the highest energy the low-$Q^2$ tagger is necessary for fully exclusive reconstruction. Distributions of generated $\pT$ vs. $\eta$ for the recoil protons are presented in Fig.~\ref{fig:tcs_scattered_p} for both collider energies -- their detection hinges on Roman Pots.   

\begin{figure}[htb]
    \begin{center}
        \includegraphics[width=0.32\textwidth]{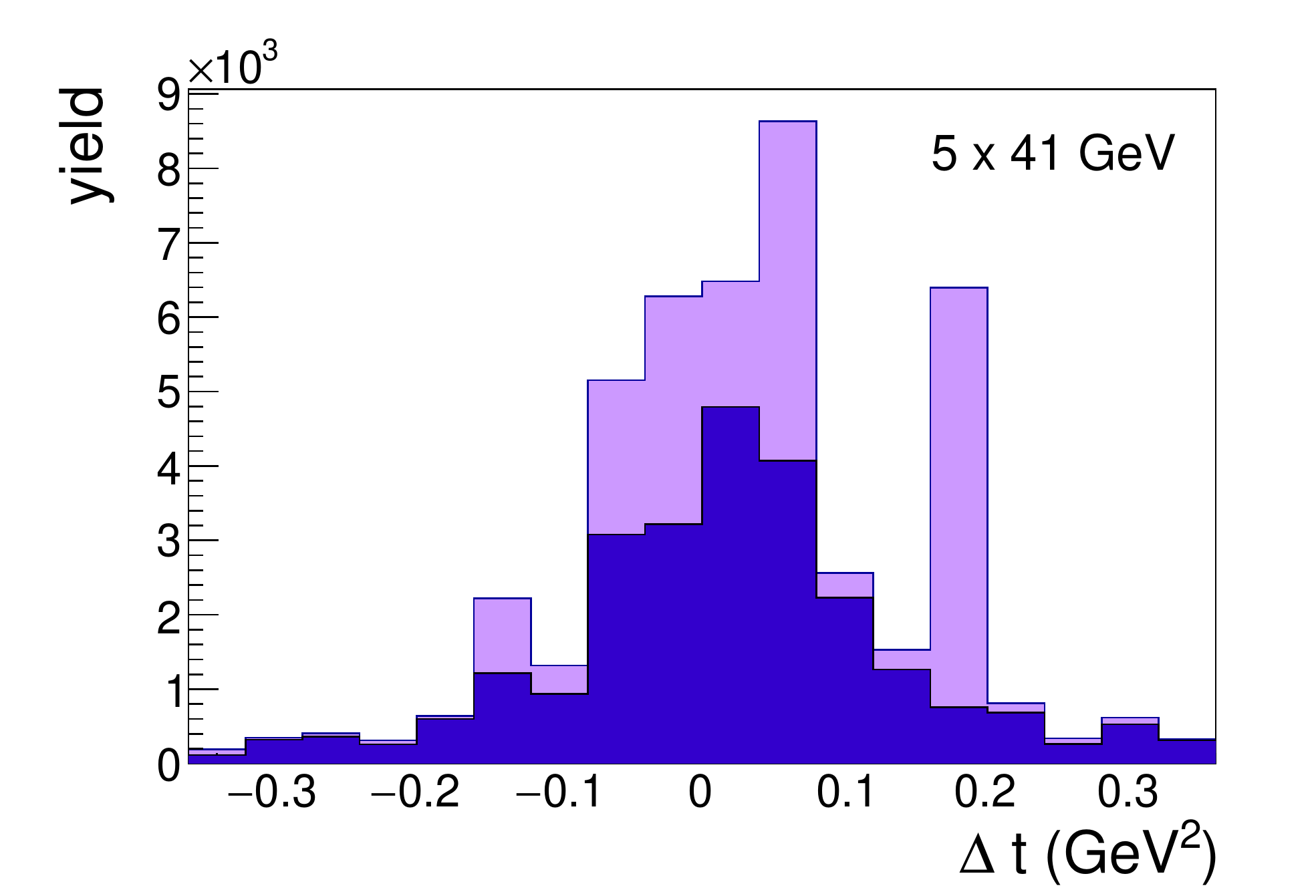}
        \includegraphics[width=0.32\textwidth]{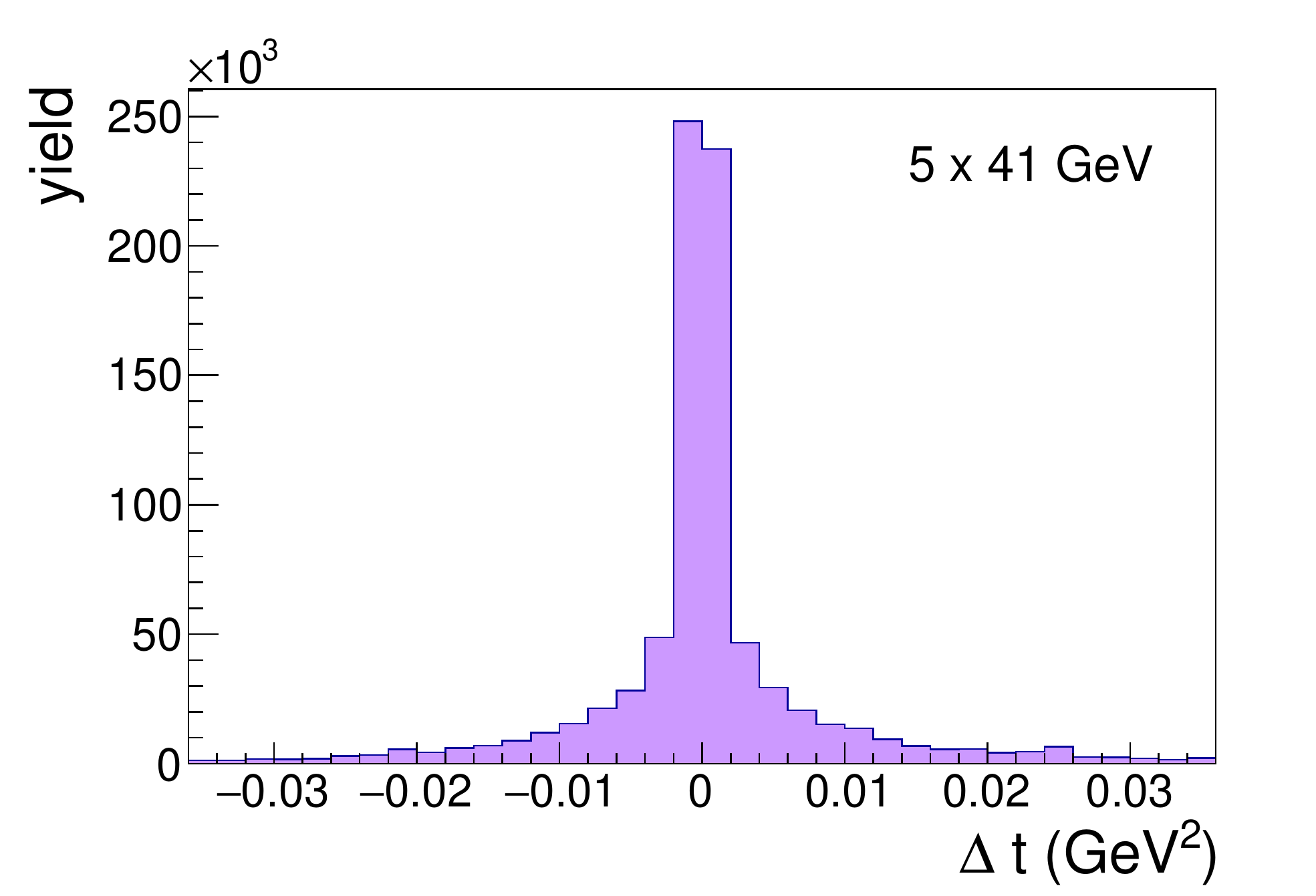}
        \includegraphics[width=0.32\textwidth]{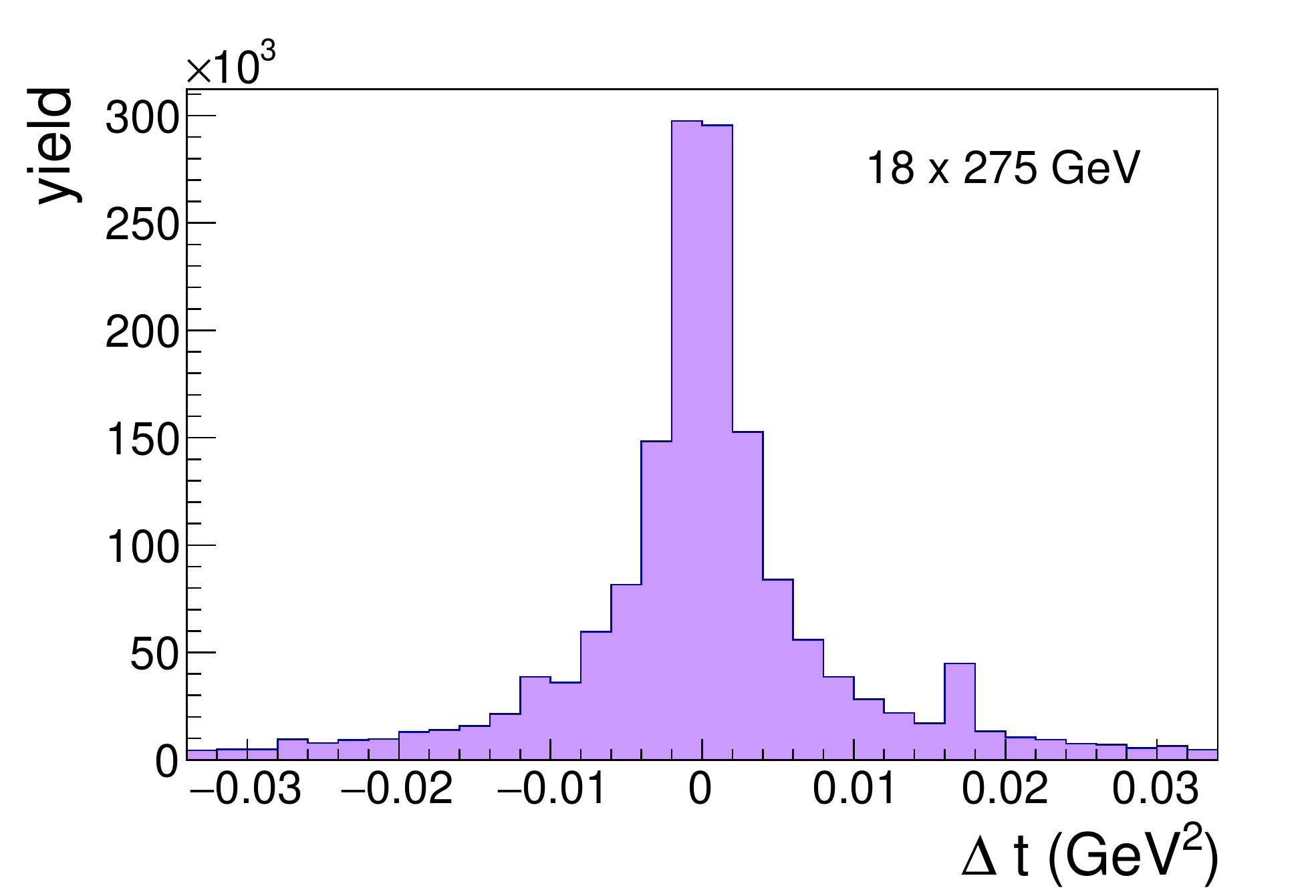}
    \end{center}
     \vspace*{-5mm}
    \caption{$\Delta t$ (generated - reconstructed) distributions, where $t = (q' - q)^{2}$ (left, showing in light purple all events and in dark blue those where a proton had also been reconstructed) and $t = (p' - p)^{2}$ (middle) for the 5~$\times$~41~GeV and 18~$\times$~275~GeV (right) collision energies. At the highest collision energy, $t$ can only be reconstructed using the proton. BH events at an integrated luminosity of 10~fb$^{-1}$.}
    \label{fig:tcs_delta_t}
\end{figure}

Fig.~\ref{fig:tcs_delta_t} shows the expected resolution on $t$, where the difference between the generated and reconstructed (using EIC-smear) $t$ is plotted. At the lowest collider energy, where some scattered electrons are reconstructed, it is possible to obtain $t$ from the quasi-real and virtual photons, but the resolution is significantly worse than calculating $t$ on the basis of the beam and recoil protons.

While the electron-positron final state may be the obvious experimental choice, measurement through muon decays has two advantages: it avoids any combinatorial background from $e^{+}e^{−}$ pairs where the $e^{-}$ is the scattered electron, and muons provide a considerably better mass resolution due to the absence of bremsstrahlung, and thus a better signal-to-background ratio. 
The cross sections for both decay channels are equal and the kinematic distributions of the leptons are very similar. 
A measurement of both channels would therefore allow for systematic cross-checks and a doubling of statistics. The scattered electron follows the same distribution as for all quasi-real photoproduction processes -- detection of the electron in a low-$Q^2$ tagger would further help to suppress the background and ensure exclusivity.

\subsection{Exclusive vector meson production in \texorpdfstring{\ep}{ep}}
\label{subsec:dvmp_ep}
Exclusive production of vector mesons is complementary to DVCS. As discussed in Sec.~\ref{part2-sec-Imaging}, production of light vector mesons allows one to separate quark flavor GPDs whereas heavy mesons probe gluon-GPDs. Studies of light vector meson production are separately discussed in Sec.~\ref{diff-excl-mesons}.  
We used the lAger event generator~\cite{lager} to obtain samples of $J/\psi$ and $\Upsilon$ events, where the $J/\psi$ and $\Upsilon$ photoproduction cross sections are those from fits in a vector-meson dominance model to the world data from Refs.~\cite{Gryniuk:2016mpk,Gryniuk:2020mlh}.
This photoproduction cross section is then used to obtain the electroproduction 
cross section as described in Ref.~\cite{Gryniuk:2020mlh} Appendix A.
For the decay into $e^\pm$ and $\mu^\pm$ we assumed s-channel helicity conservation.
We used the PHOTOS~\cite{Davidson:2010ew} package to account for the radiative effects on the vector meson decay.
Finally, we used the GRAPE-DILEPTON~\cite{Abe:2000cv} program to simulate the dilepton background to the detected exclusive final state.
To simulate detector effects, we used EIC-smear with matrix detector~\cite{git:eicsmeardetectors}, including far-forward elements.
Finally, for our studies, we evaluated the nominal beam configurations: 5 GeV electrons on 41 GeV protons, 5 GeV electrons on 100 GeV protons, 10 GeV electrons on 100 GeV protons, and 18 GeV electrons on 275 GeV protons. For brevity, we only include results for the lowest and highest energy configuration in this document, as they form a realistic envelope for the other configurations.

\subsubsection{Detector requirements}
\begin{figure}[htb]
    \begin{center}
        \includegraphics[width=0.49\textwidth]{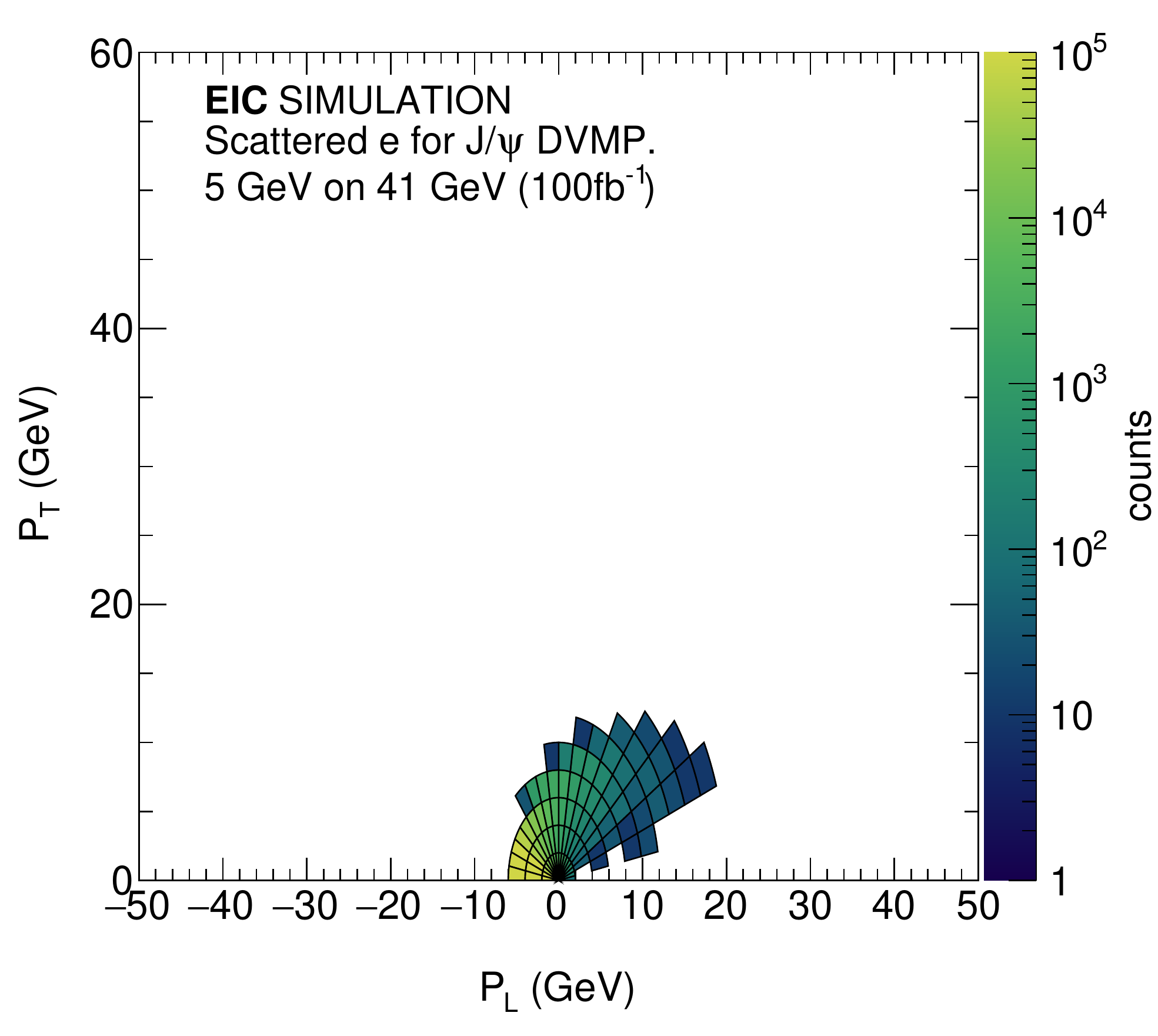}
        \includegraphics[width=0.49\textwidth]{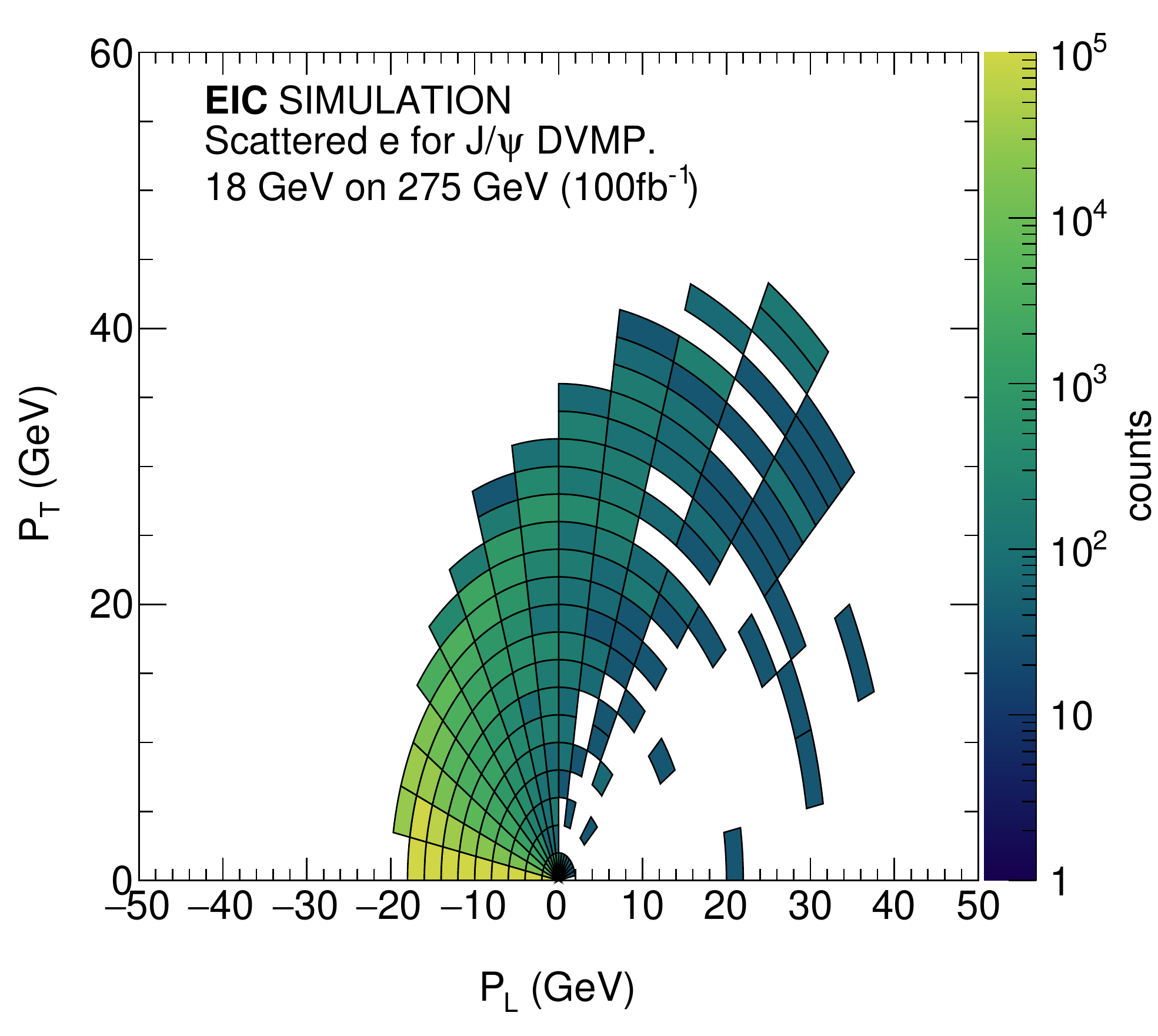}\\
        \includegraphics[width=0.49\textwidth]{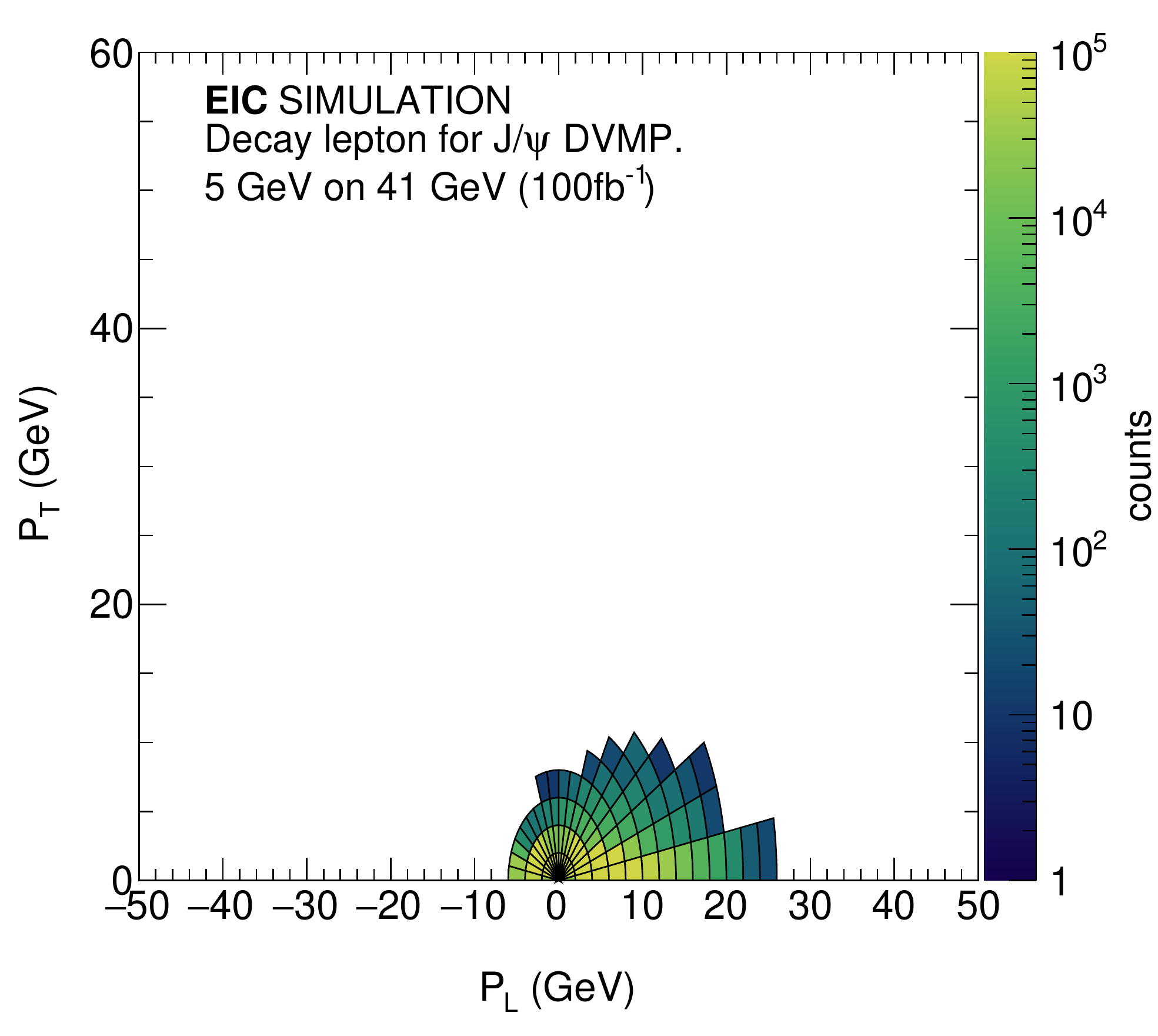}
        \includegraphics[width=0.49\textwidth]{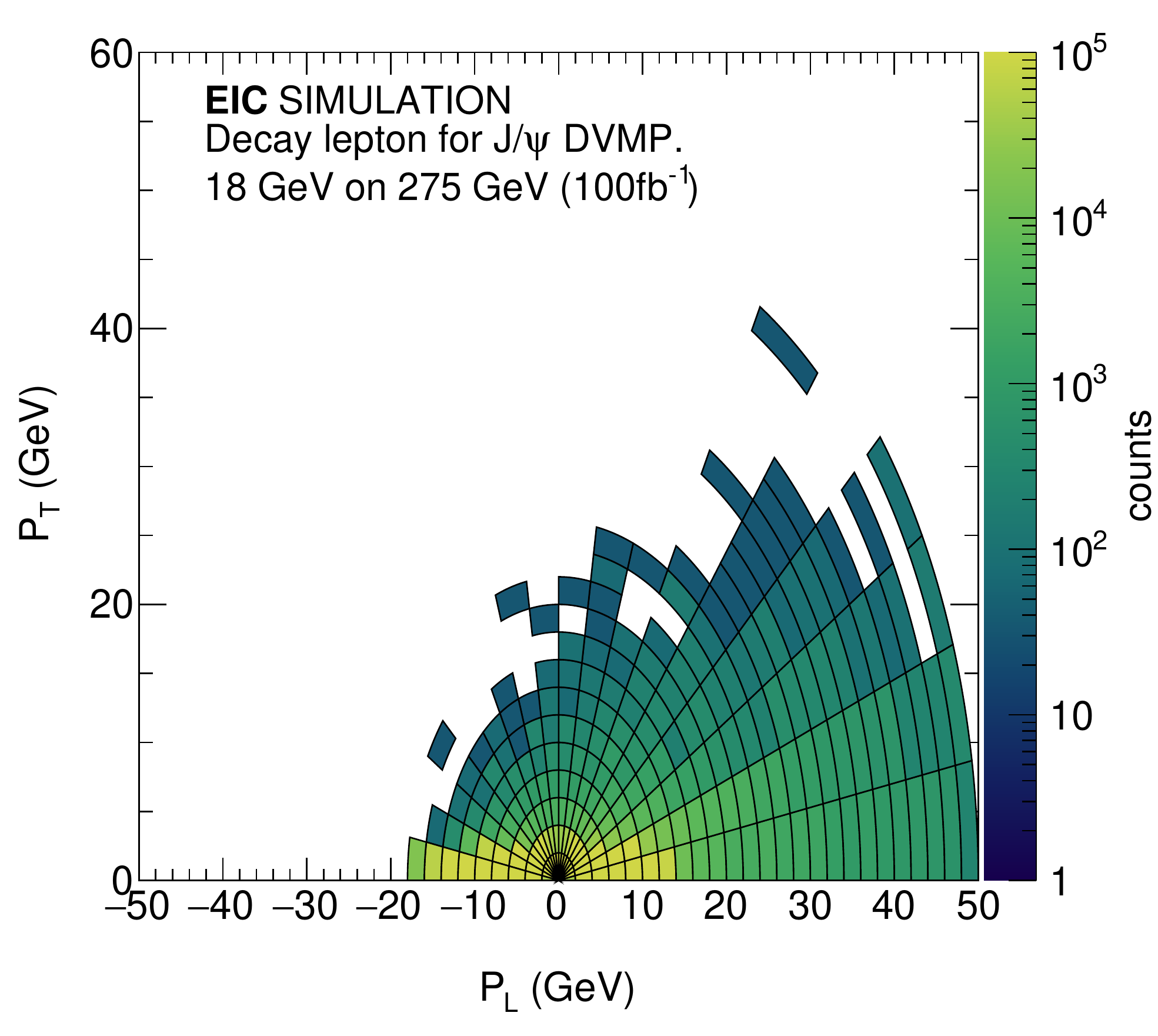}
    \end{center}
    \caption{Polar figures of the momentum distribution for the scattered electron (top panels) and decay leptons (bottom panels) for 
    $J/\psi$ DVMP at the lowest and highest energy configurations.
    }
    \label{fig:exclusive_dvmp_polar_lepton}
\end{figure}
To fully detect the exclusive reaction, we need to detect the scattered lepton, recoil proton, and both decay leptons. Note that the event geometry concerning the scattered lepton and recoil proton is very similar to that of DVCS. 
A summary of the polar distributions of the leptons in DVMP for the lowest and highest collision energies are shown in Figure~\ref{fig:exclusive_dvmp_polar_lepton}.

\begin{figure}[htb]
    \begin{center}
        \includegraphics[width=0.49\textwidth]{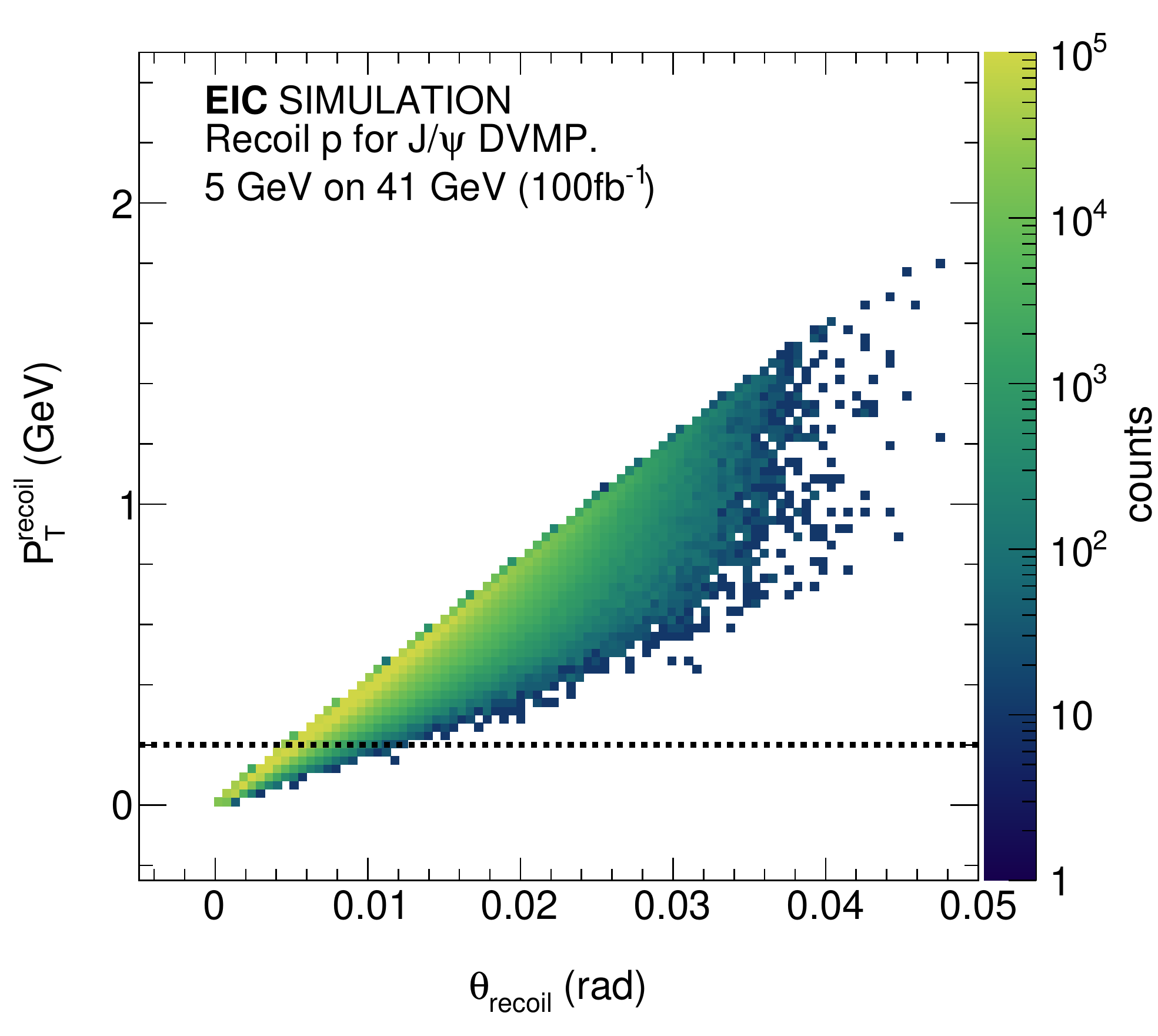}
        \includegraphics[width=0.49\textwidth]{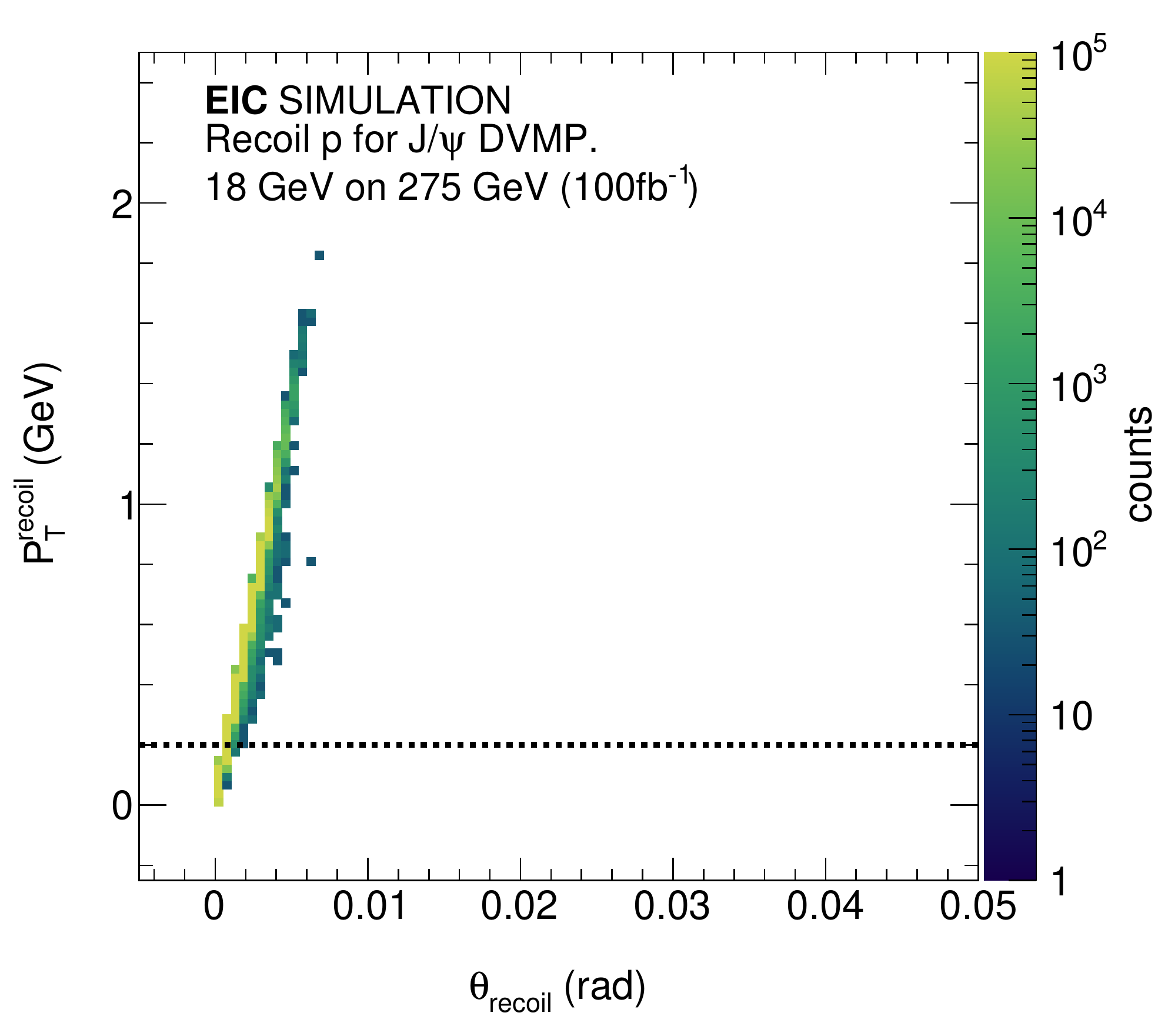}
    \end{center}
    \caption{Transverse momentum $\pT$ versus polar angle $\theta$ for recoil protons in exclusive $J/\psi$ DVMP for the lowest and highest energy configurations. 
    The dashed line corresponds to a lower $\pT$ cut of 200 MeV.
    }
    \label{fig:exclusive_dvmp_recoil_pT_eta}
\end{figure}

The recoil proton detection occurs through a combination of forward and far-forward detector elements. 
For increasing collision energy, the recoil becomes increasingly forward.
Figure \ref{fig:exclusive_dvmp_recoil_pT_eta} shows recoil $\pT$ as a function of the recoil polar angle $\theta$. 
The dashed line shows a nominal $\pT$ cutoff below 200 MeV, which does not substantially impact the detector's physics reach.
Comparing the lower and higher beam setting in this figure, we will need a smooth transition between a forward B0-style detector into a Roman Pot-like system.

\begin{figure}[htb]
    \begin{center}
        \includegraphics[width=0.49\textwidth]{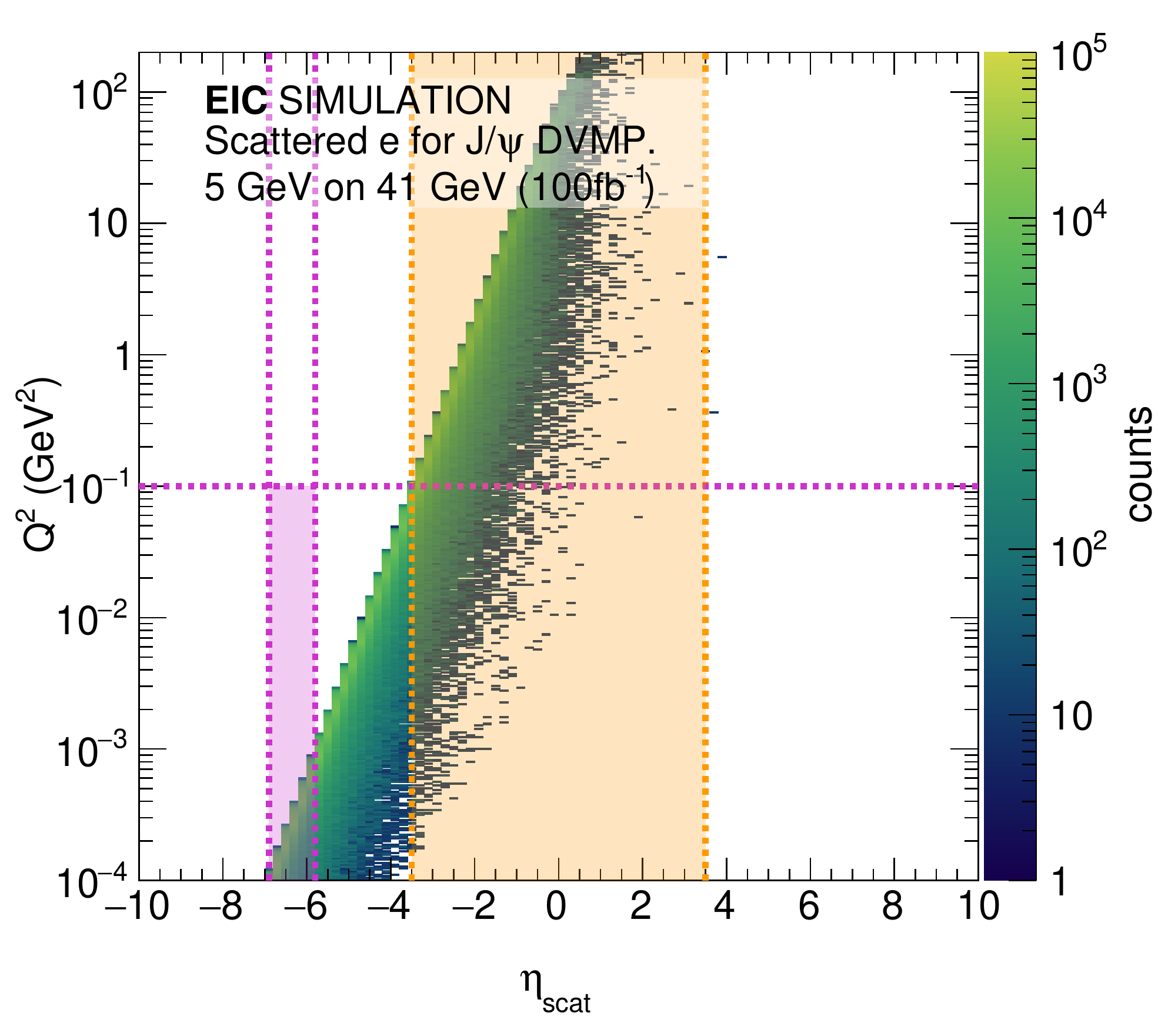}
        \includegraphics[width=0.49\textwidth]{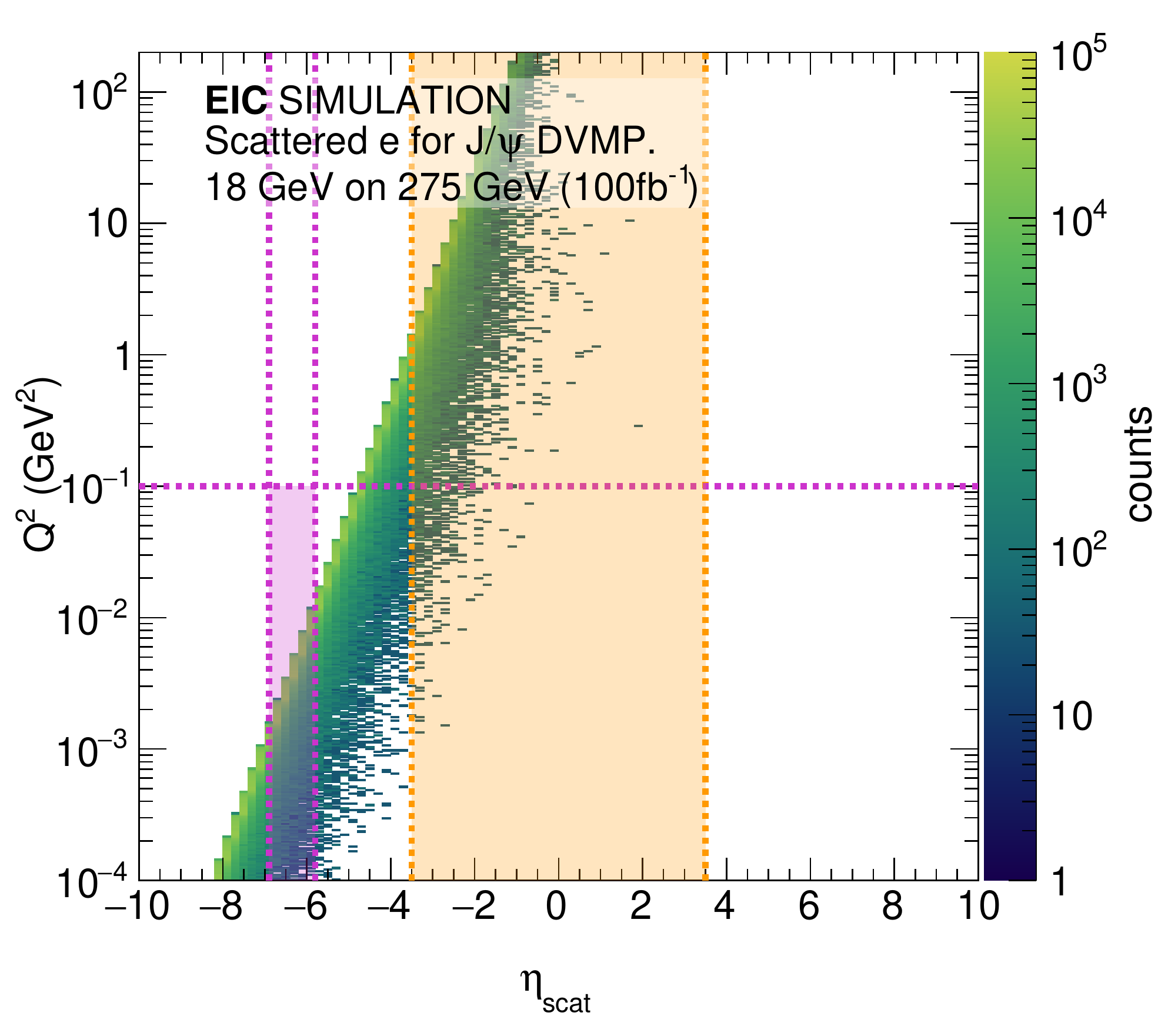}
    \end{center}
    \caption{$Q^2$ as a function of the scattered electron pseudo-rapidity for exclusive $J/\psi$ DVMP with the lowest and highest energy configurations. 
    The orange box indicates scattered electrons detected in the nominal central detector, while the magenta box corresponds to events detected with the low-$Q^2$ tagger.}
    \label{fig:exclusive_dvmp_scat_Q2_eta}
\end{figure}

Fig. \ref{fig:exclusive_dvmp_scat_Q2_eta} shows $Q^2$ as a function of pseudorapidity $\eta$.
The orange box corresponds to a nominal central detector covering $|\eta| < 3.5$, while the magenta box corresponds to a nominal low-$Q^2$ tagger accepting $-6.9 < \eta < -5.8$.
The central detector acceptance is sufficient for all configurations to accept events from $Q^2>0.1$GeV$^2$ to large values of $Q^2$.
The lower limit of $\eta > -3.5$ is restrictive for photoproduction events in the main detector, especially for higher collision energies.
The photoproduction of DVMP at higher energies will completely depend on the low-$Q^2$ tagger unless a significant enhancement of the backward region's electron acceptance is possible.
More acceptance in the low $Q^2$ tagger would directly translate into more measured photoproduction events.
A better acceptance for low $Q^2$ events would benefit Upsilon photoproduction near threshold, where the projected statistical precision is particularly low (cf. Fig.~\ref{fig:exclusive_dvmp_W_thresh}). 
For a summary graph of the polar distributions for the scattered electron, see Figure~\ref{fig:exclusive_dvmp_polar_lepton} (top panels).

\begin{figure}[htb]
    \begin{center}
        \includegraphics[width=0.49\textwidth]{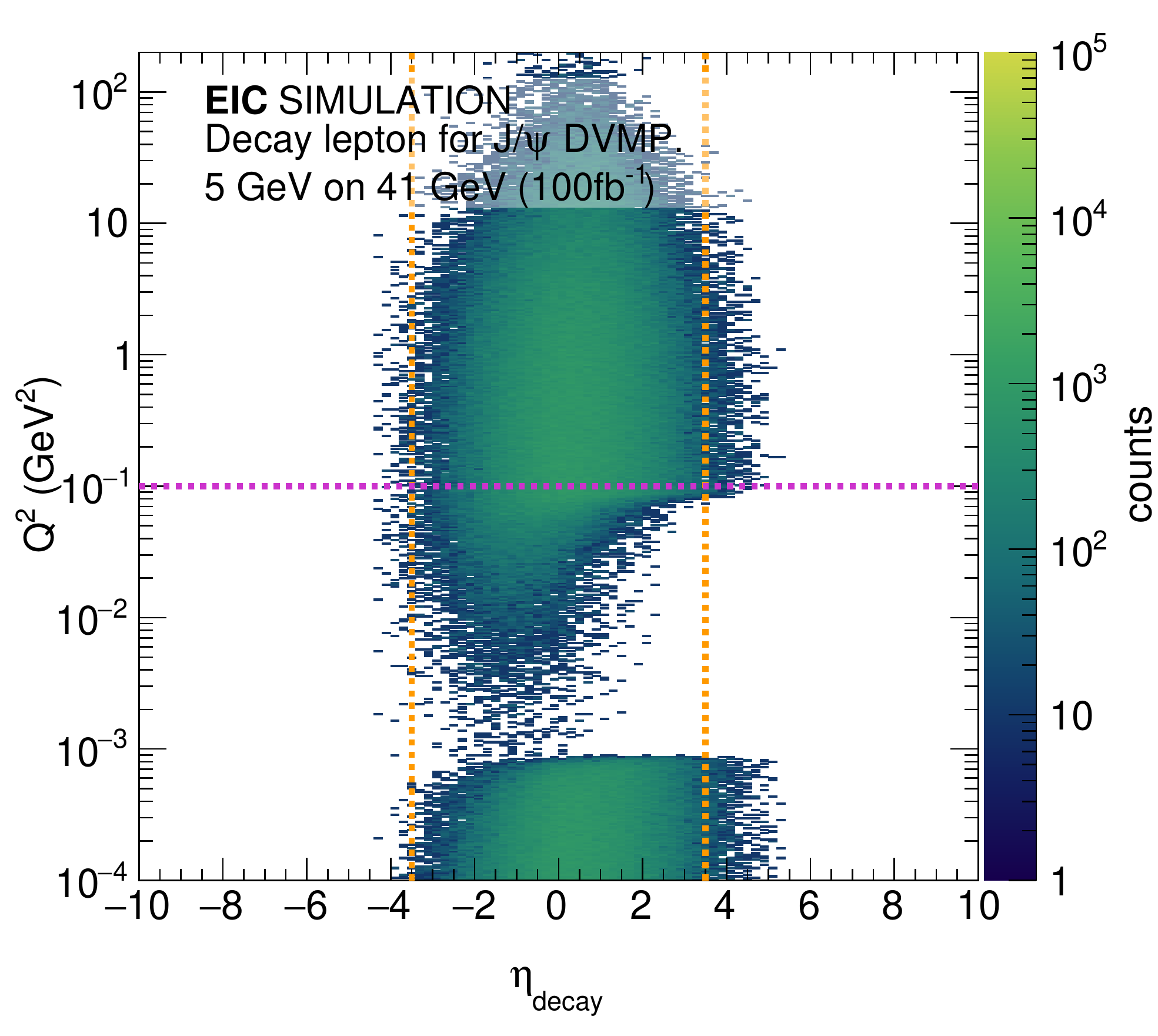}
        \includegraphics[width=0.49\textwidth]{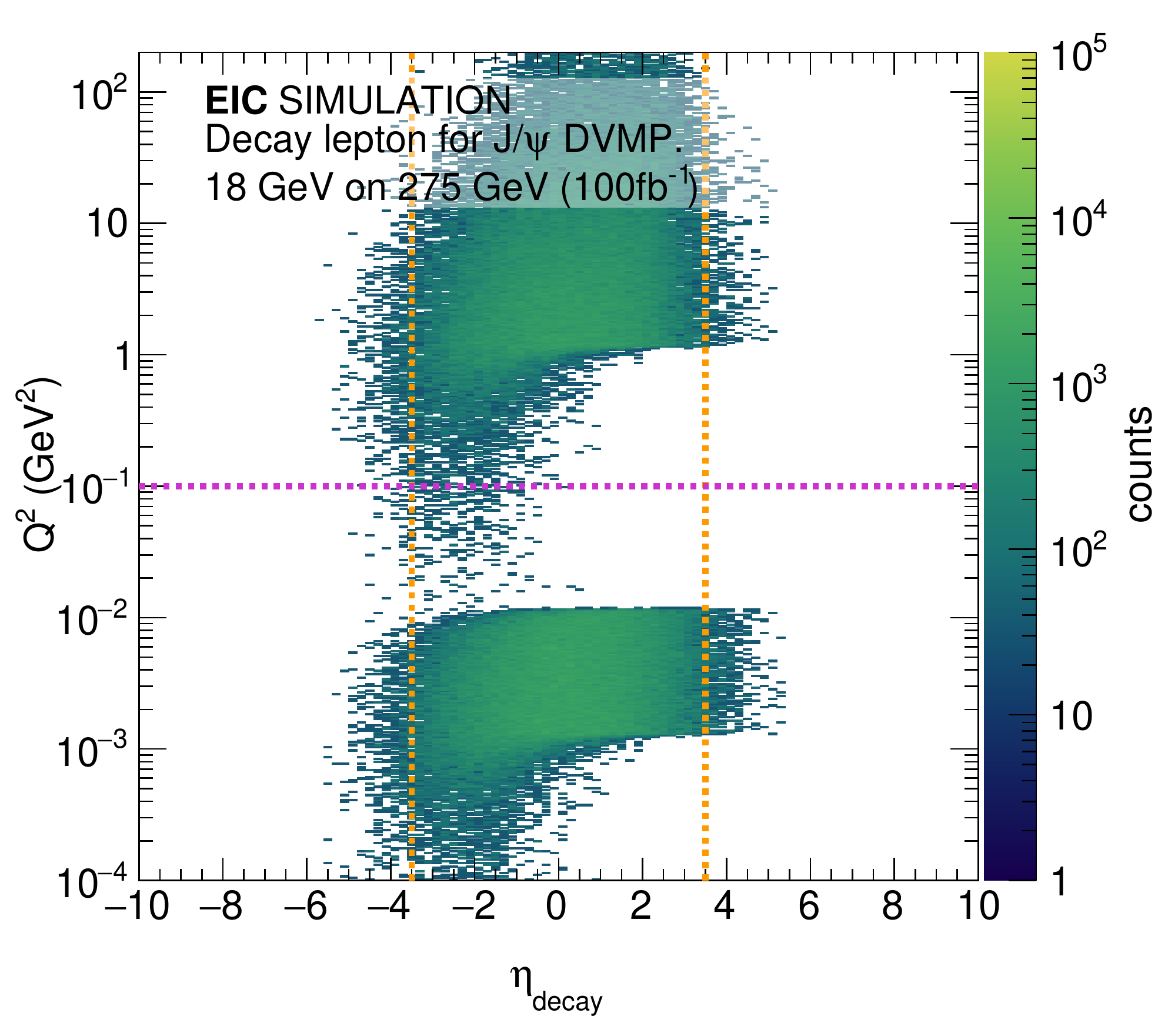}\\
        \includegraphics[width=0.49\textwidth]{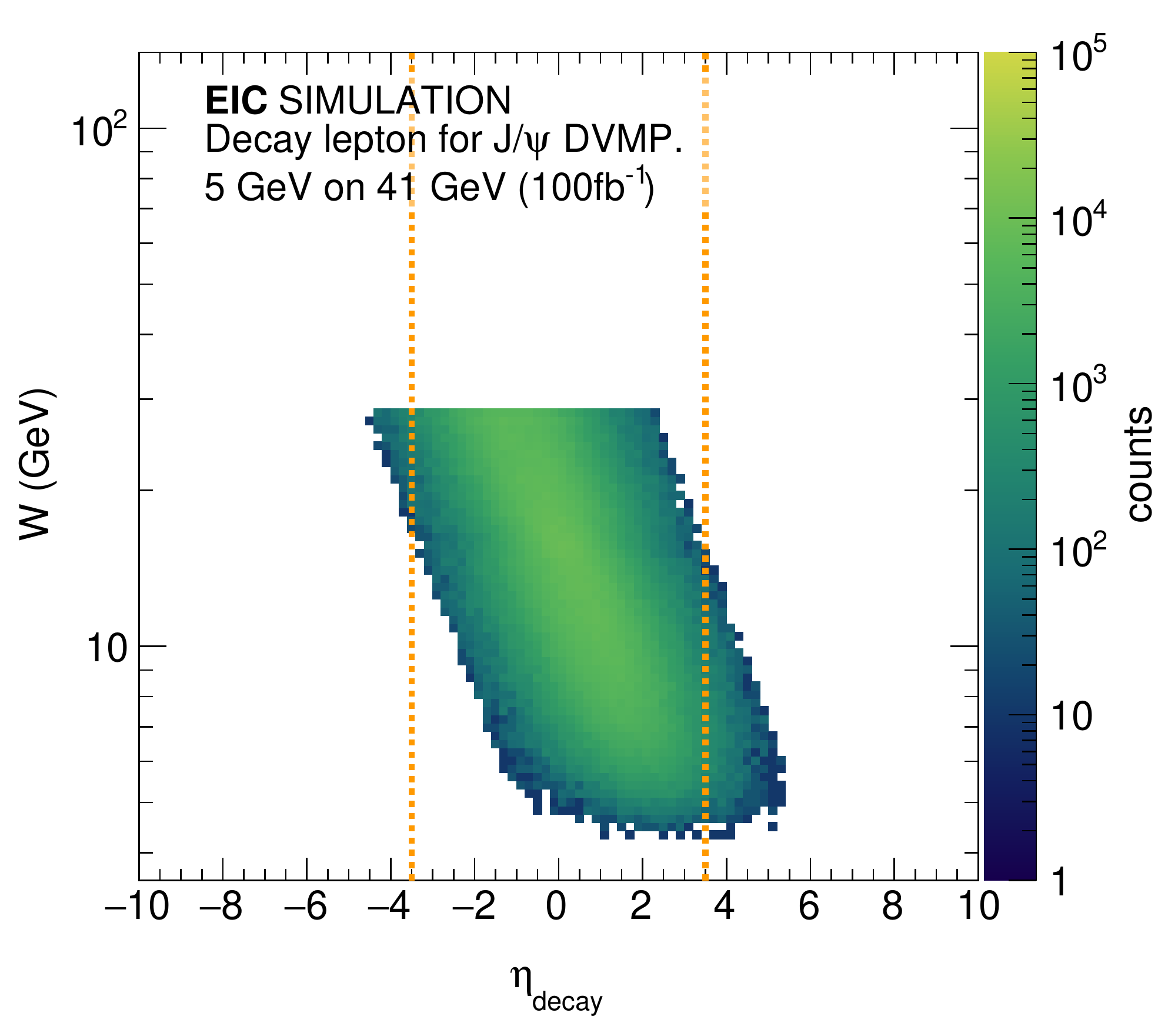}
        \includegraphics[width=0.49\textwidth]{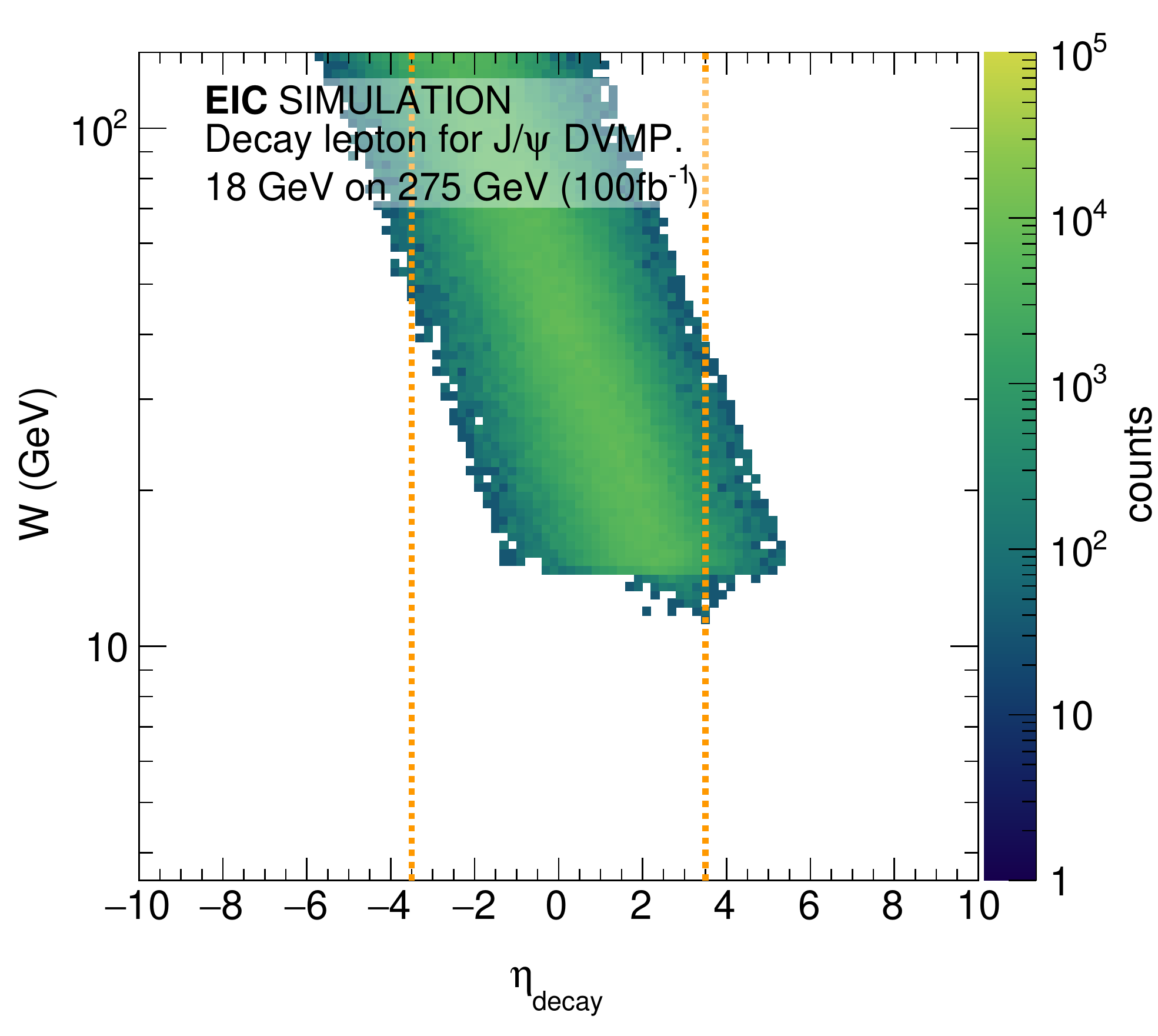}\\
    \end{center}
    \caption{$Q^{2}$ (top) and $W$ (bottom) versus decay lepton rapidity for exclusive $J/\psi$ DVMP for the lowest and highest energy configurations (left and right).
    The two structures on the top graphs are due to the discontinuity between scattered electrons detected in the main detector and those detected in a nominal low-$Q^2$ tagger.
    The orange lines on both graphs show a nominal main detector acceptance for $|\eta| < 3.5$, and the magenta line on the top graph shows a upper limit for photoproduction events for $Q^2 < 0.1\text{GeV}^2$ .
    }
    \label{fig:exclusive_dvmp_decay_kin}
\end{figure}
The main detector will measure the $e^\pm$ and $\mu^\pm$ pair from $J/\psi$ and $\Upsilon$ decay.
Figure \ref{fig:exclusive_dvmp_decay_kin} shows $Q^2$ (top) and $W$ (bottom) versus decay lepton rapidity.
The nominal central detector range of $|\eta| < 3.5$ envelopes the majority of events.
The bottom panels show a slight $W$ dependence of the decay particle $\eta$, but the loss of acceptance at lower $W$ is relatively minor.
The top panels show a discontinuity between the higher and lower $Q^2$ points, caused by the rapidity gap between the electron endcap and the low-$Q^2$ tagger.
From a detector point of view, the limiting factor for DVMP is the low $Q^2$ acceptance, limiting the event count in the $\Upsilon$ threshold region important for the physics of the proton mass.
Expanding the acceptance in the low-$Q^2$ tagger would dramatically improve the statistics in these regions.
For a summary graph of the polar distributions for the decay leptons, see Figure~\ref{fig:exclusive_dvmp_polar_lepton} (bottom panels).

\begin{figure}[htb]
    \begin{center}
        \includegraphics[width=0.49\textwidth]{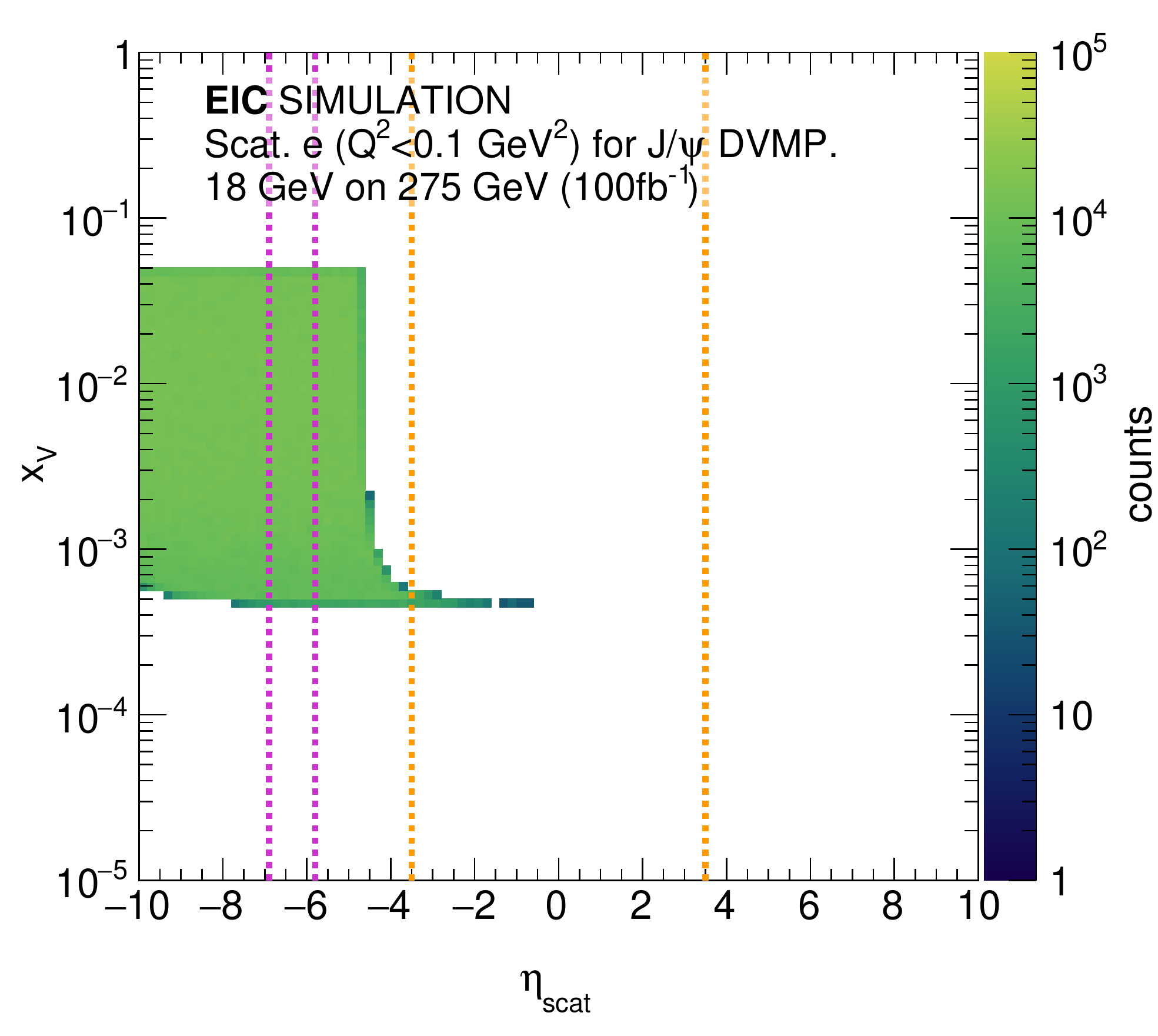}
        \includegraphics[width=0.49\textwidth]{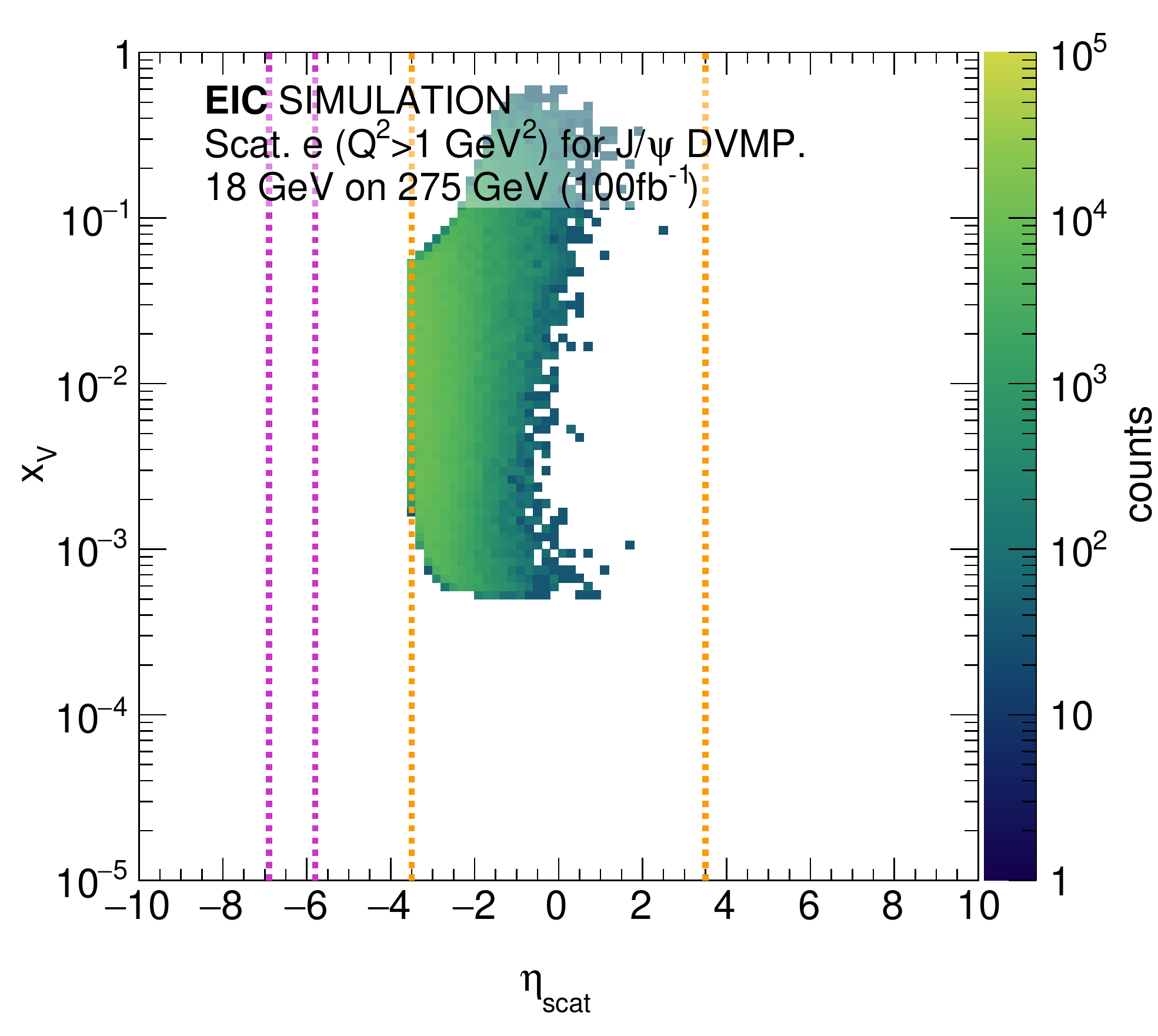}\\
        \includegraphics[width=0.49\textwidth]{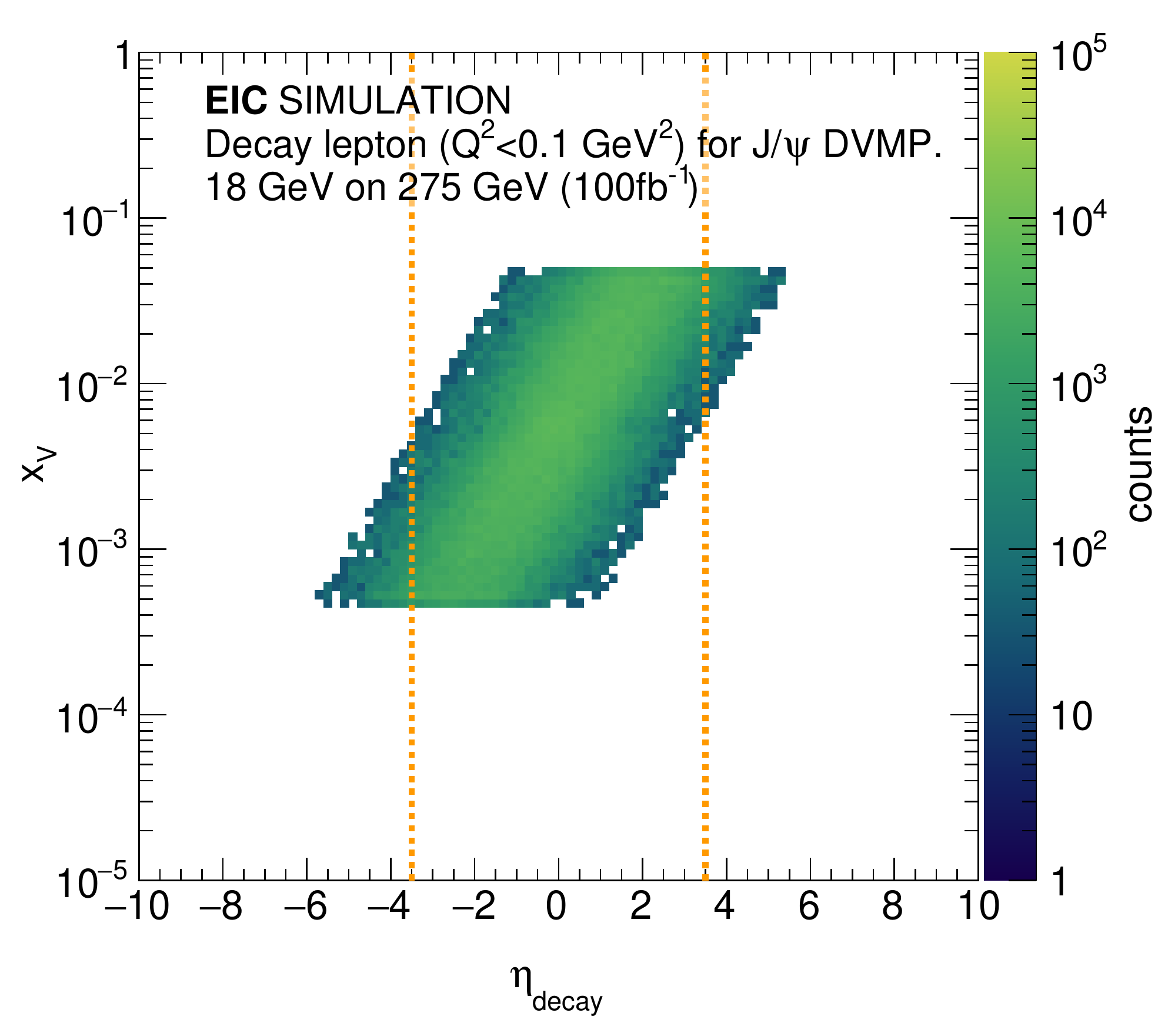}
        \includegraphics[width=0.49\textwidth]{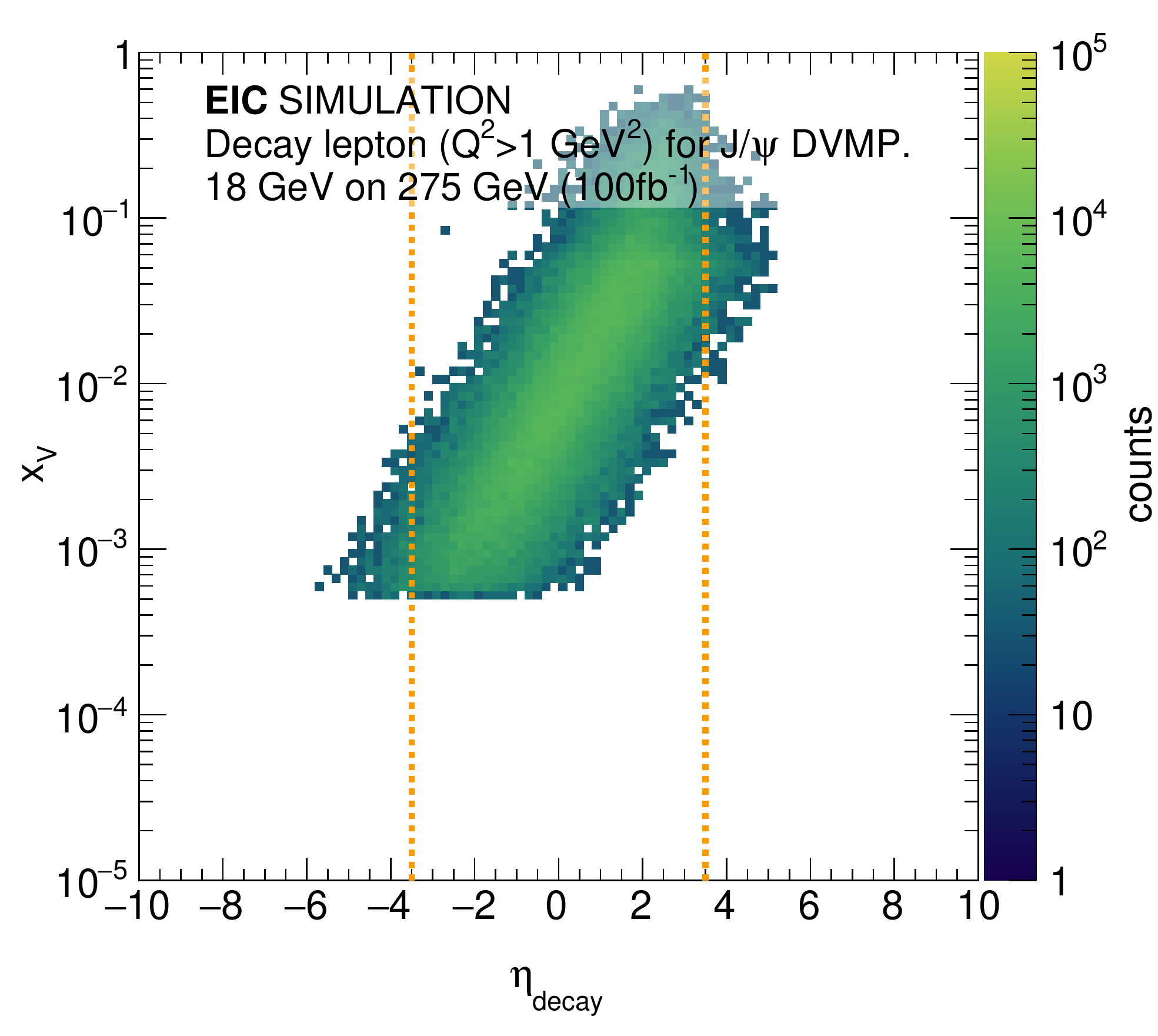}
    \end{center}
    \caption{$x_V$ versus the pseudo-rapidity of the scattered electron (top panels) and decay lepton (bottom panels) 
    in photoproduction (left) and electroproduction (right) for $J/\psi$ DVMP at the highest energy configuration.
    The dashed orange lines show the nominal central detector range of $|\eta| < 3.5$,
    and the dashed magenta line shows a nominal low-$Q^2$ tagger at $-6.9 < \eta < -5.8$.
    There are no detector cuts in the top figures, while the bottom figures have a nominal acceptance applied to only the 
    scattered electron and recoil proton.
    The straight edge in the top right figure is caused by the $Q^2 > 1\text{ GeV}^2$ requirement for electroproduction.
    }
    \label{fig:exclusive_dvmp_x_vs_eta}
\end{figure}
The impact of the nominal fiducial volume of $|\eta| < 3.5$ in the main detector on the $x_V$ coverage at the highest collision energy
for $J/\psi$ DVMP
is explored in Figure~\ref{fig:exclusive_dvmp_x_vs_eta} for electroproduction and photoproduction.
The top panels show $x_V$ versus the scattered electron $\eta$ without detector cuts. Here too, it is clear that for photoproduction 
we are fully dependent on the low-$Q^2$ tagger.
For electroproduction, we have good coverage of the full $x_V$ range within the main detector. Note that the lower bound of $Q^2 > 1$GeV$^2$
coincides almost exactly with the $\eta > -3.5$ cutoff in the backward region.
The case for the $J/\psi$ decay leptons is shown in the bottom panels. While there is a clear relation
between $\eta_\text{decay}$ and $x_V$, we do not loose access to any kinematic region due to the
main detector acceptance. This is similar to what we concluded for $W$ (see Fig.~\ref{fig:exclusive_dvmp_decay_kin}).

\begin{figure}[htb]
    \begin{center}
        \includegraphics[width=0.49\textwidth]{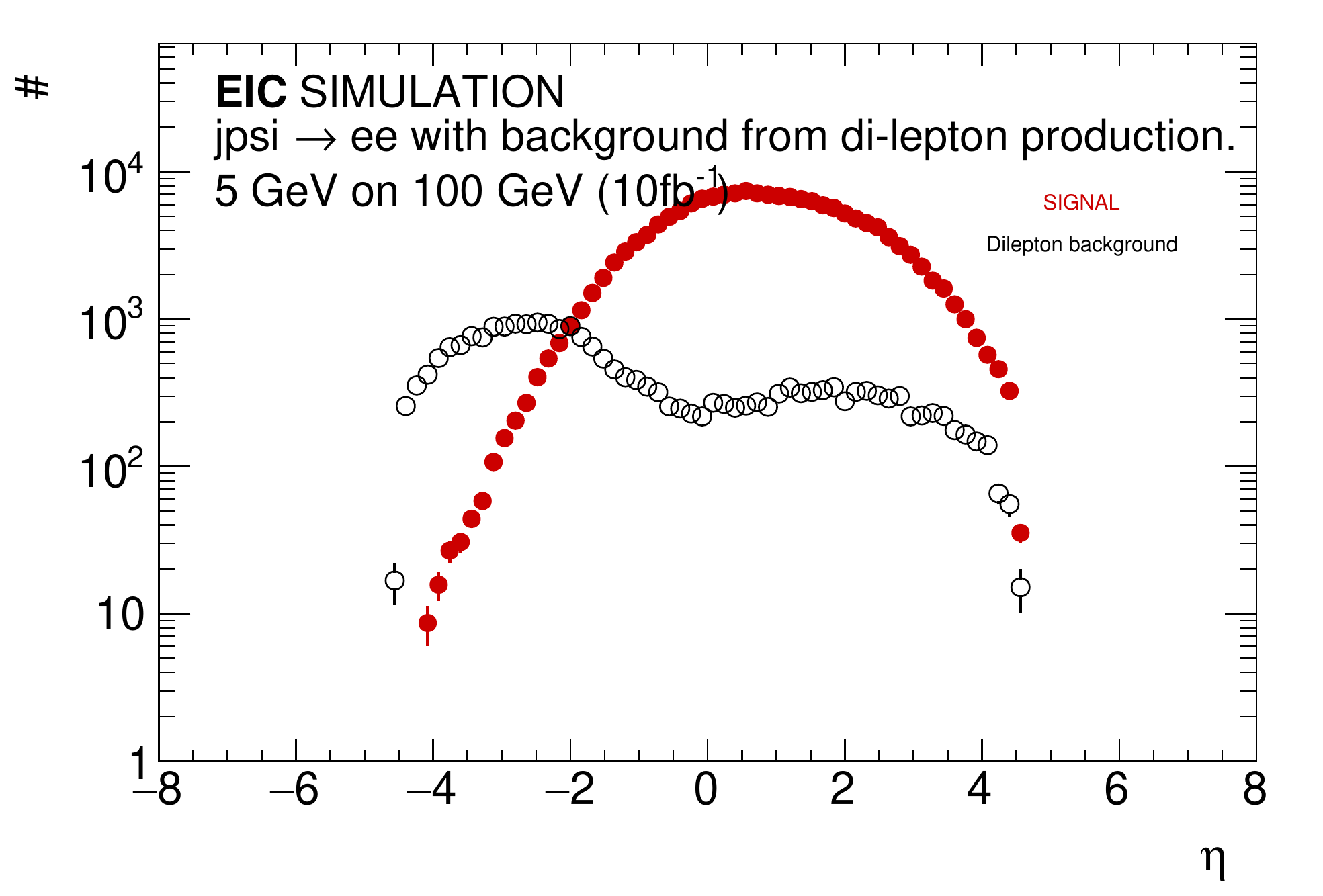}
        \includegraphics[width=0.49\textwidth]{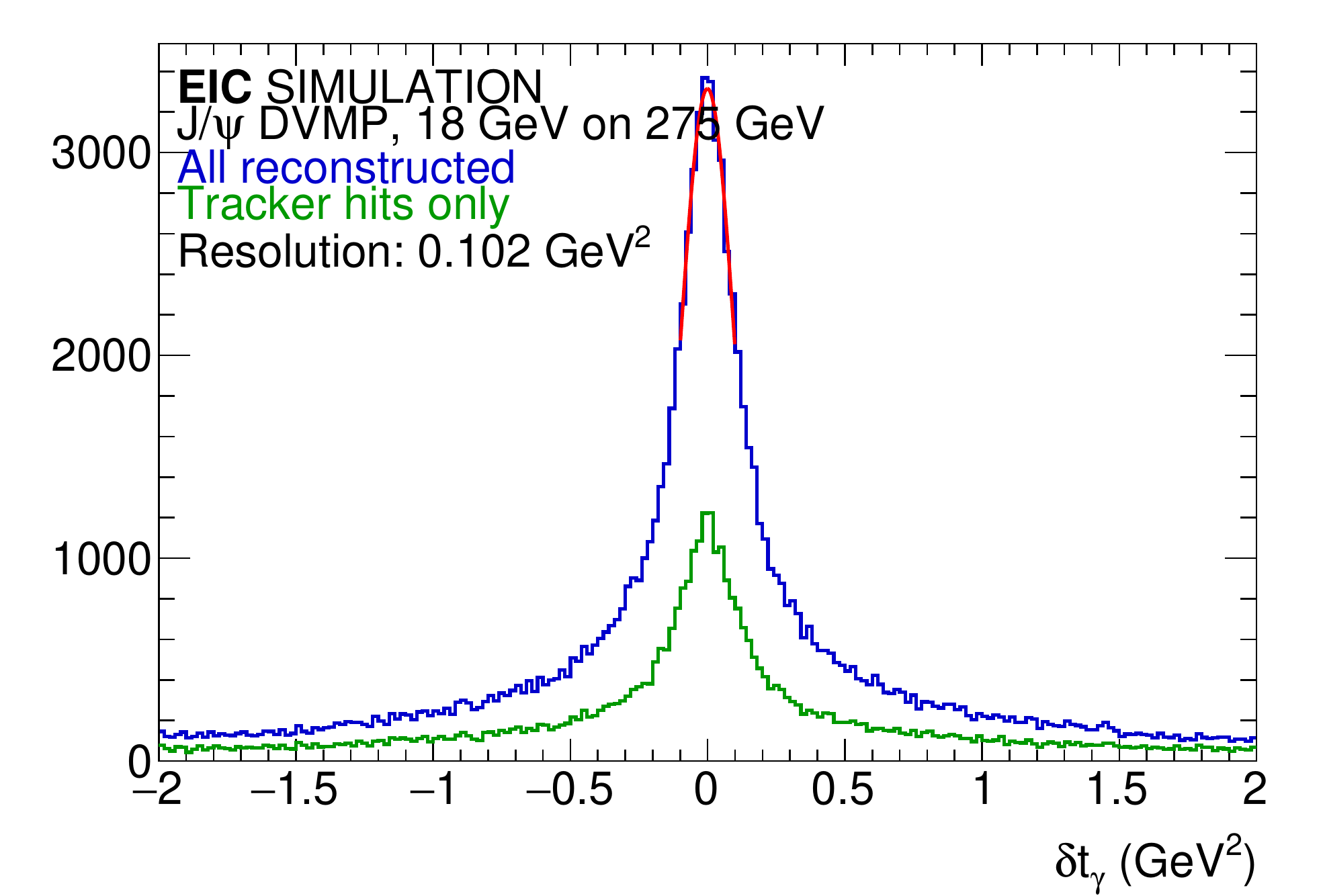}
    \end{center}
    \caption{Left: Signal and dilepton background for photoproduction. Right: Resolution on effects on the reconstructed vector meson kinematics. The red curve is a fit, yielding the resolution 0.102~GeV$^2$. }
    \label{fig:exclusive_dvmp_need_muon}
\end{figure}
Figure \ref{fig:exclusive_dvmp_need_muon} makes the case for muon PID for exclusive DVMP.
The left panel compares the $J/\psi\rightarrow e^\pm$ projected count rate with the dilepton background count for $J/\psi$ photoproduction events for an intermediate energy configuration.
The background becomes non-negligible for events with decay leptons in the backward region.
While this background issue is only significant for certain kinematic corners for DVMP, it will be much more severe for TCS.
Muon PID will be important to control for this.

The right panel in Fig.~\ref{fig:exclusive_dvmp_need_muon} shows the difference $\delta t$ between generated and measured $t$ based on the reconstructed scattered electron and vector meson kinematics.
Note that beam divergence effects and beam energy spread, which will further complicate this reconstruction, are not accounted for in this study.
The long tails on this resolution originate from limited resolution effects and radiative effects in the vector meson decay.
These heavy tails are present in all reconstructed kinematics, particular at lower $W$.
More sophisticated tracking reconstruction algorithms should make this situation a bit better.
The improved resolution for muon events and the smaller impact of radiative effects will make this channel crucial to study and control for these resolution effects.

Finally, the muon channel's availability will double the available statistics, vital for threshold physics where a typical bin in $ W $ and $t$ may have single-digit counts, illustrated in Figure \ref{fig:exclusive_dvmp_W_thresh}.

\begin{figure}[htb]
    \begin{center}
        \includegraphics[width=0.49\textwidth]{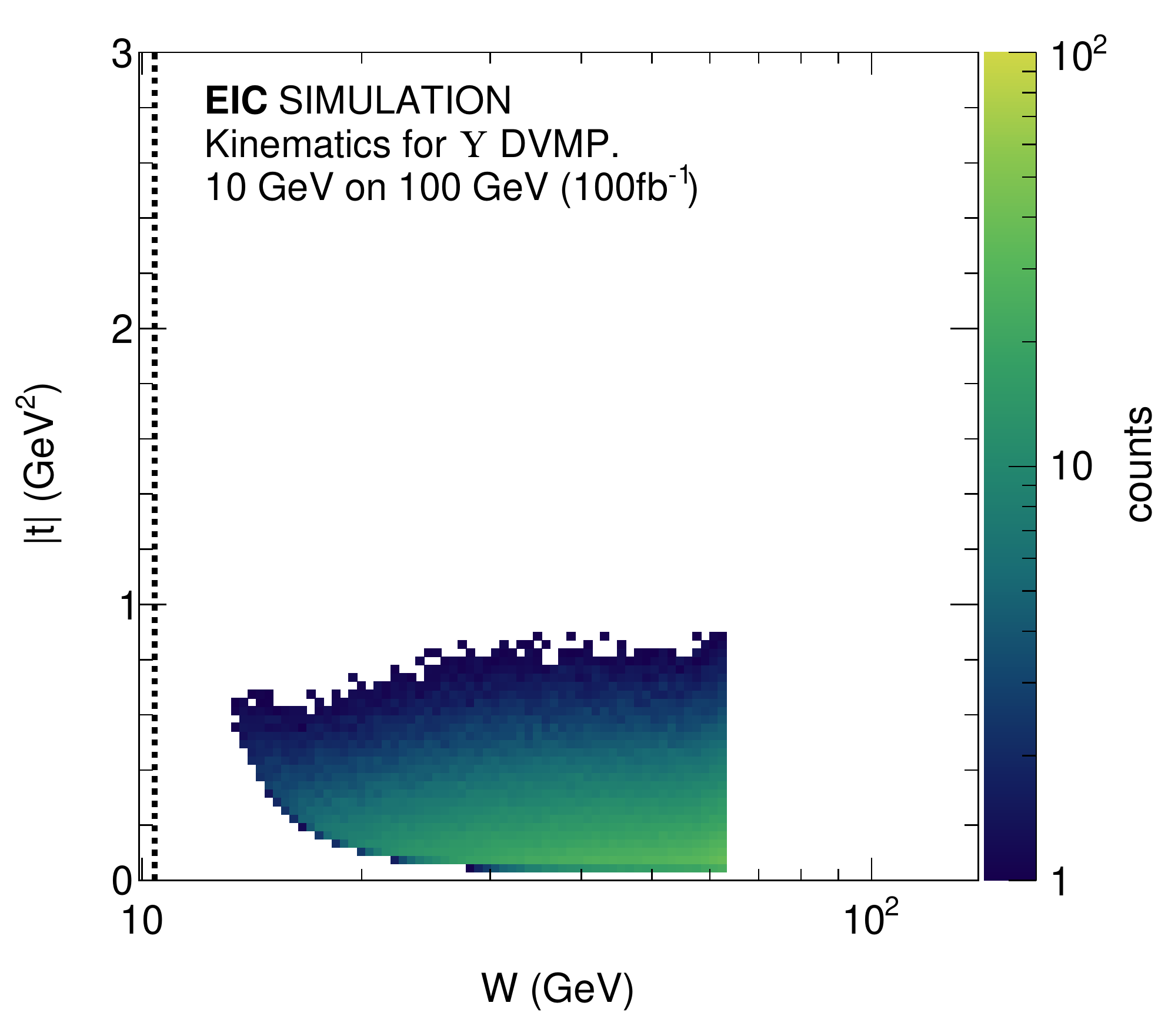}
        \includegraphics[width=0.49\textwidth]{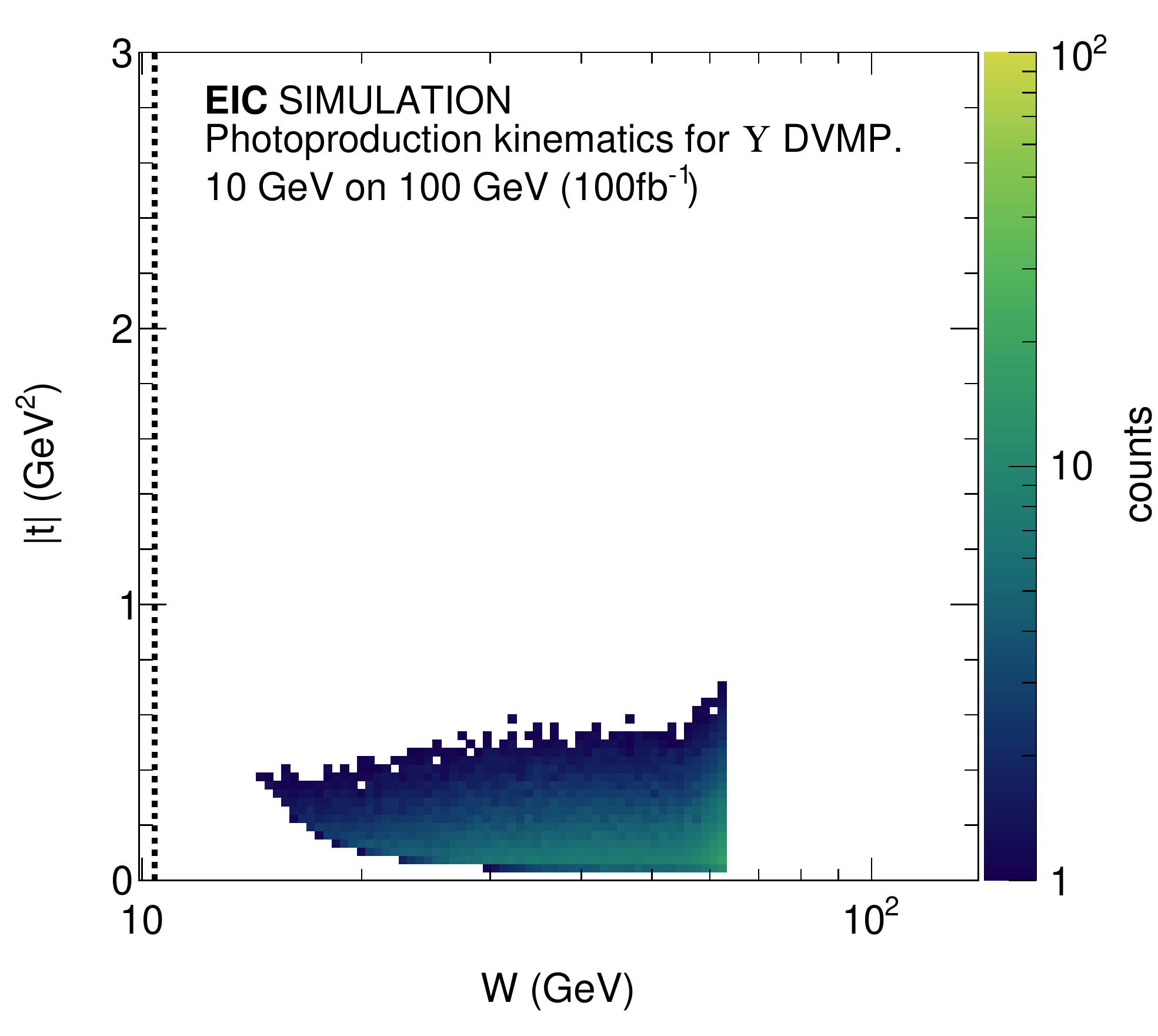}
    \end{center}
    \caption{Coverage in $W$ and $t$ for $\Upsilon$ DVMP for an intermediate energy configuration. The left panel shows all events, and the right panel shows photoproduction events only. Note that the photoproduction events result in relatively low statistics.
    They would directly benefit from either more acceptance in the backward region of the main detector, or an expanded $\eta$-coverage for the low-$Q^2$ tagger.}
    \label{fig:exclusive_dvmp_W_thresh}
\end{figure}

\subsubsection{Kinematic coverage}
\begin{figure}[htb]
    \begin{center}
        \includegraphics[width=0.49\textwidth]{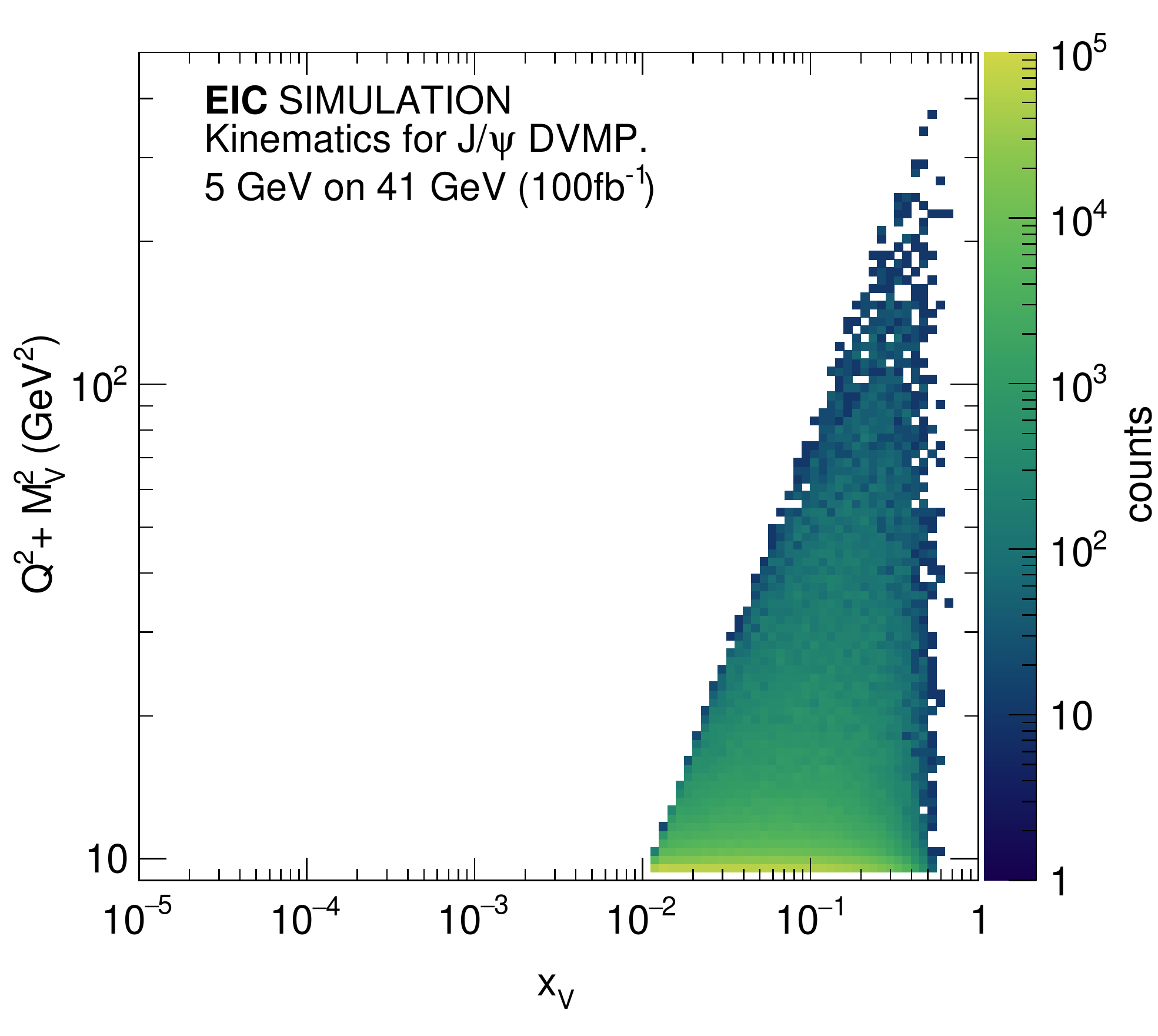}
        \includegraphics[width=0.49\textwidth]{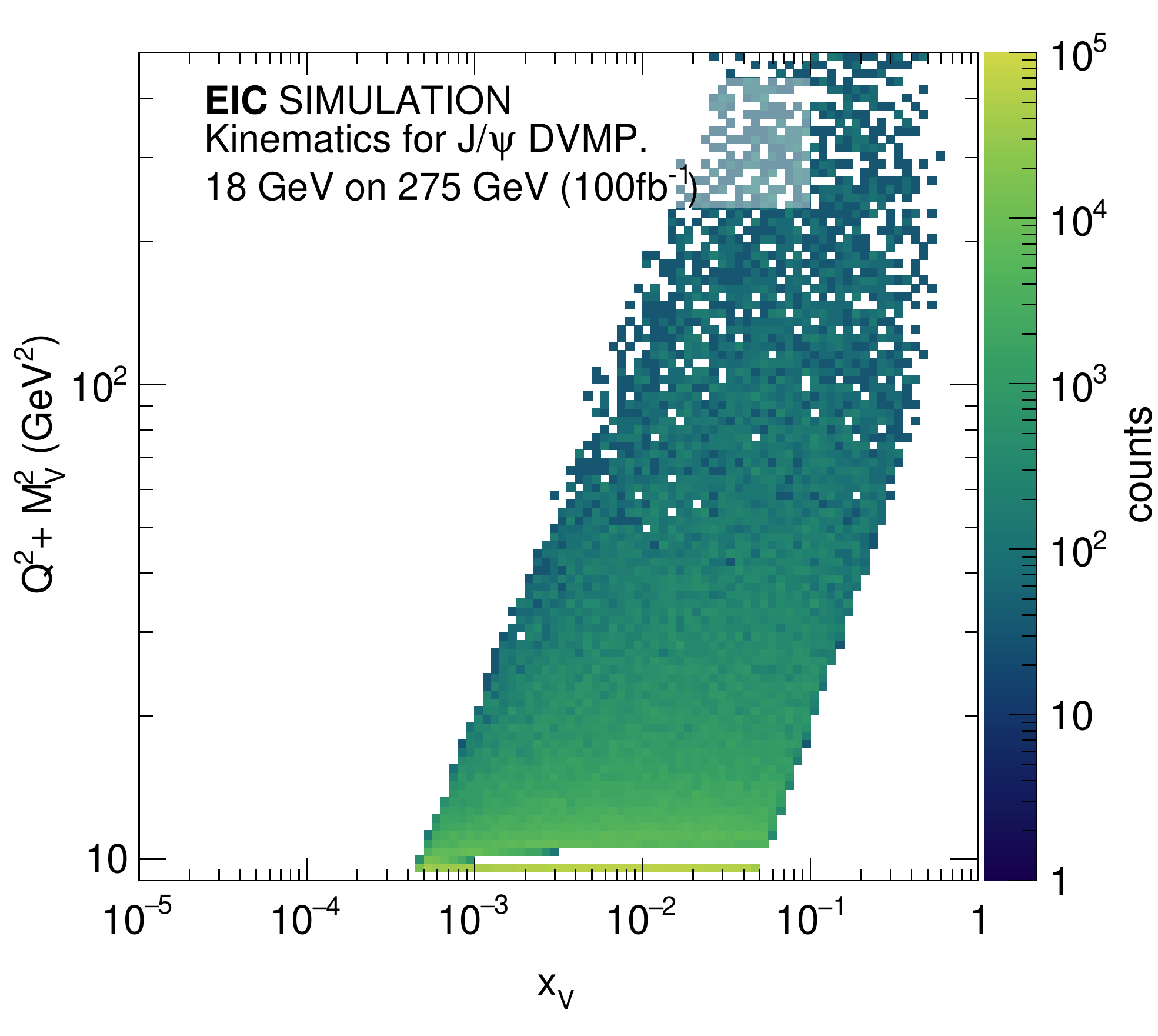}\\
        \includegraphics[width=0.49\textwidth]{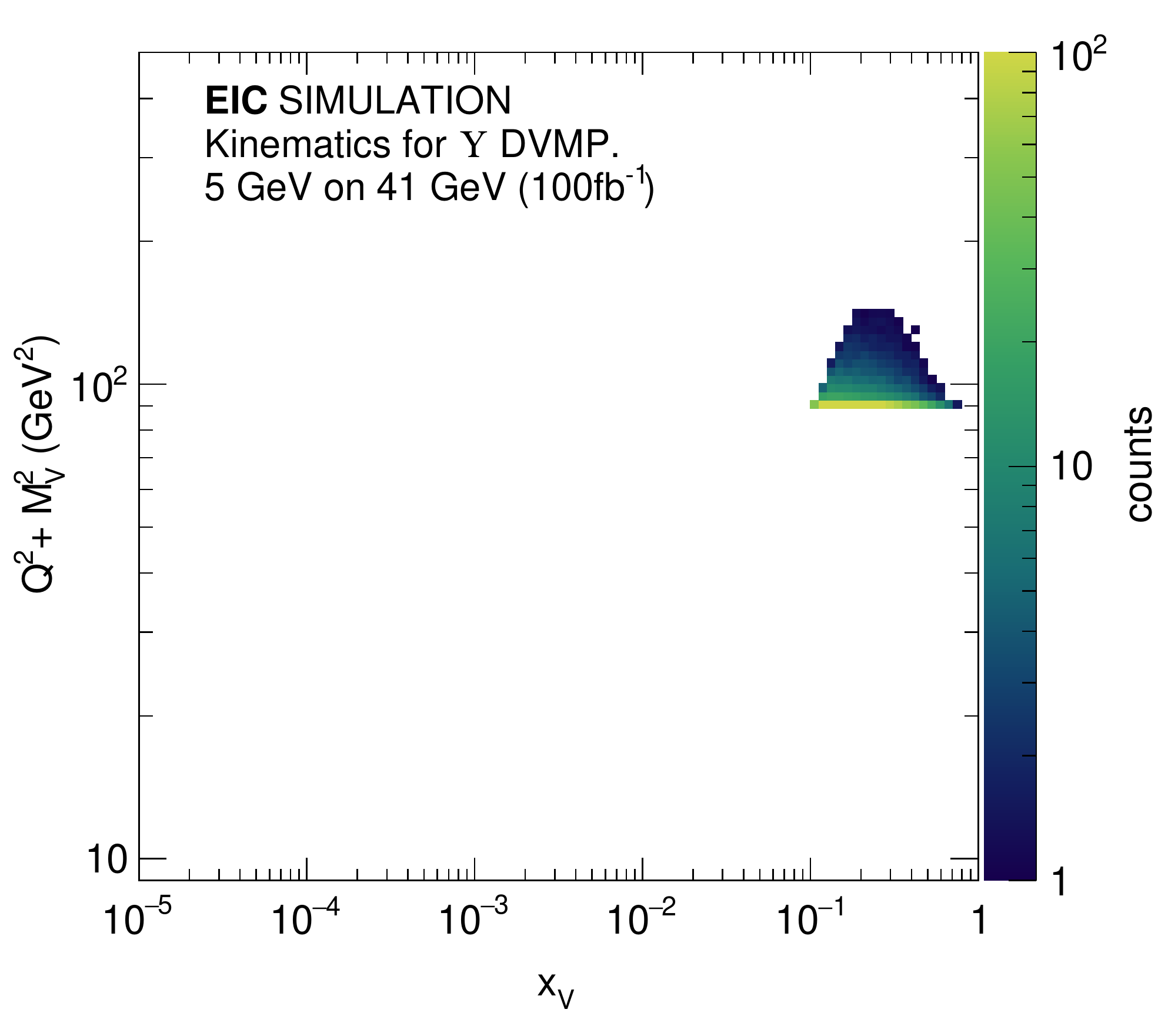}
        \includegraphics[width=0.49\textwidth]{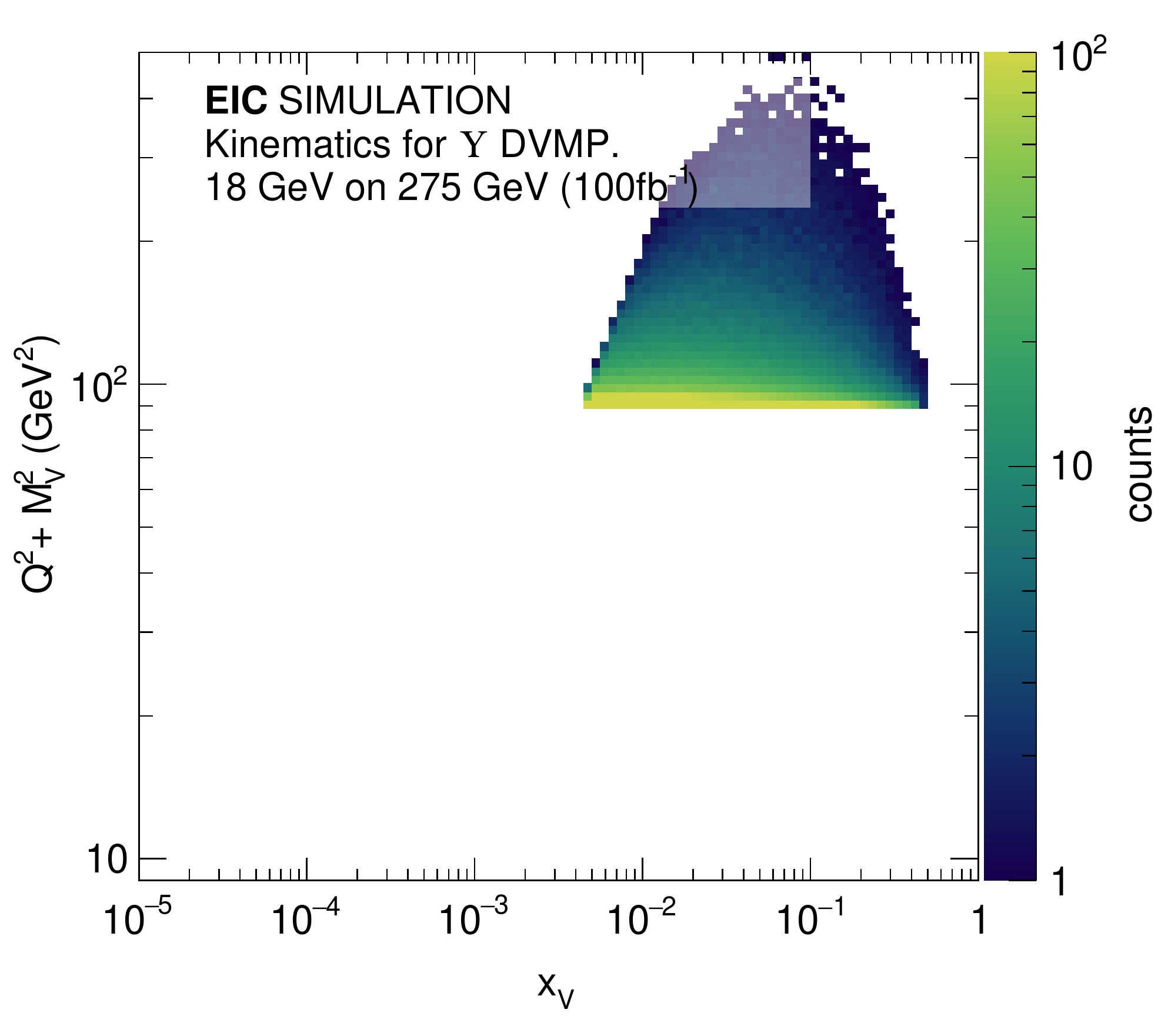}\\
    \end{center}
    \caption{$Q^2 + M_V^2$ versus $x_V$ for DVMP of $J/\psi$ (top) and $\Upsilon$ (bottom). The left panels show the lowest energy configuration, and the right panels show the highest energy configuration.
    The discontinuity at lower $Q^2+M_V^2$ in the top right graph is due to events with scattered electrons in either the low $Q^2$ tagger or the main detector.
    }
    \label{fig:exclusive_dvmp_x_Q2}
\end{figure}
Figure \ref{fig:exclusive_dvmp_x_Q2} shows the phase space in $Q^2 + M_V^2$ versus $x_V$ for $J/\psi$ (top) and $\Upsilon$ (bottom) DVMP at EIC.
Varying the collision energy will provide sensitivity to the gluon GPD from the valence region to the sea region.
Due to its much larger mass, $\Upsilon$ DVMP will access this gluonic structure at a much larger scale than $J/\psi$ production, providing for an important handle on the evolution and factorization of the formalisms used to extract the GPDs.
A nominal luminosity of 100fb$^{-1}$ will allow for a precise determination of the $Q^2$ dependence of $J/\psi$ production, while being sufficient to study $\Upsilon$ production in several bins of $Q^2 + M_V^2$.

Figure \ref{fig:exclusive_dvmp_W_thresh} shows $|t|$ as a function of $W$ for $\Upsilon$ DVMP for all detected events (left), and photoproduction events where $Q^2 < 0.1\text{~GeV}^2$.
Measuring DVMP near the threshold is challenging due to the steeply dropping cross section as the production phase space closes.
Furthermore, going to the threshold region is intrinsically limited due to finite detector resolution.
The nominal lower limit of $y > 0.01$ driven by resolution effects translates to a lower limit on $W$.
The measurement of the $\Upsilon$ photo-production cross section near-threshold would greatly benefit from increased statistics, achievable through either an extension of the acceptance in the backward region or through an increased acceptance in the low-$Q^2$ tagger.
The situation is better for electroproduction, and in both cases a nominal luminosity of 100 fb$^{-1}$ is sufficient as long as the low-$Q^2$ can be improved from the nominal values used for this study.

\subsection{Exclusive vector meson production in \eA}
\label{subsec:dvmp_ea}

Measuring exclusive cross sections for vector mesons in heavy nucleus targets,  $e + A \rightarrow e^\prime + A^\prime + V$, where $V = \rho, \phi, J/\psi, \Upsilon$,  played a prominent role in the EIC White Paper~\cite{Accardi:2012qut} and is considered as one of the key measurements of the \eA  program at the EIC.  The cross section for these exclusive processes, especially for lighter mesons~\cite{Toll:2012mb} is generically more sensitive to saturation (or shadowing) than inclusive cross sections.

In addition to the integrated  cross section, one is particularly interested in the $t$-distribution $d \sigma/dt$ and the separation of coherent from incoherent events, where the target proton or nucleus stays intact or breaks up into color neutral fragments, respectively.  These give access to the transverse spatial structure and fluctuations of the gluons in the target, see Sec.~\ref{part2-subS-LabQCD-Photo}.
For typical values of $t$ for coherent and incoherent processes see the discussion in Sec.~\ref{part2-subS-LabQCD-Photo} and Fig.~\ref{fig:combo_incoherent_new} below.
 For coherent events with heavy nuclei, unlike for proton targets, the nucleus does not leave the beam pipe, and $t$ must be reconstructed within the central detector. This is the procedure discussed in this section. Here we discuss specifically production of $\rho$, $\phi$, and $J/\psi$ in coherent $e$+A events, but the ability to reconstruct $t$ in the central detector will also be useful for light ions, incoherent events with either protons or nuclei, and also as a cross check for events using the far forward spectrometers. However coherent \eA processes typically involve the smallest values of $|t|$ and thus pose the most stringent requirements. 

Here we follow earlier studies by \cite{Caldwell:2010zza} and focus on the $\pT$-resolution as the dominating factor that determines the precision with which we can measure the momentum transfer $t$. The resolution is parametrized as
$\sigma_{\pT}/{\pT} =  (\sigma_{\pT}/{\pT})_\mathrm{meas}  \oplus (\sigma_{\pT}/{\pT})_\mathrm{MS}$, where MS refers to the multiple scattering term. We use as  start values the ones  listed in the EIC Detector Requirements and  R\&D Handbook \cite{EIC:RDHandbook} for the respective pseudorapidity interval.

\subsubsection{Event generation and decay channels}
 
 Simulations were carried out with the Sartre event generator \cite{Toll:2012mb,Toll:2013gda,Sambasivam:2019gdd} version 1.34, based on the bSat~\cite{Kowalski:2006hc} dipole model.  The generator describes vector meson production at HERA (\ep) and data from ultra-peripheral collisions at the LHC in \pPb{} and \PbPb~\cite{Sambasivam:2019gdd} quite well.   To speed up simulations, skewedness corrections as well as corrections of the real part of the amplitude were not applied.  This affects the magnitude of the cross section but has little impact on the kinematics, which is the focus of this study.
 
 Data sets for photoproduction, defined here as $Q^2 < 0.01$ GeV$^2$ and moderate $Q^2$, $1 < Q^2 < 10$ GeV$^2$  were generated for $\rho$, $\phi$, and $J/\psi$ mesons at maximum energy, i.e., $E_e = 18$ GeV and $E_\mathrm{Au} = 110$ GeV. For each meson we generated between 160 and 430 million events. Decays of the vector mesons were conducted with tools from the Sartre library taking the polarization of the virtual photon into account.  We assume that the particle  identification will be sufficient to reconstruct the vector mesons via the invariant mass of the respective state. A continuous background from the Bethe-Heitler process $\eAu \rightarrow e^\prime + \mathrm{Au} ^\prime + e^+ + e^- ( \mu^+ + \mu^-)$ will have to be handled in the analysis of the data. Since most event generators assume colliding beams with a zero crossing angle,  one cannot directly generate events in the detector frame from beams that have a momentum spread and divergence. Instead, we generate events from nominal head-on beams and only after this smear the incoming electron and ion 4-momenta.
 
For the $J/\psi$ we use here the $e^+e^-$ decay channel. 
 The muon decay channel would have similar kinematics, with  two main advantages: ({\it i}) avoiding  combinatorial background from  the scattered electron and ({\it ii}) the absence of bremsstrahlung.  For the $\phi$ we use here the decay $\phi \rightarrow K^+ K^-$, with a branching ratio of 49.2\%. The decay into kaons has a serious disadvantage that since $m_\phi - 2 m_K = 32.11$ MeV at    low $Q^2$ as the $\pT$ of the decay kaons can  remain below the cut-off values of any EIC detector. A possible remedy would be tracking of curled tracks in the vertex detectors tolerating the absence of particle identification, or runs with small magnetic field settings. The $\rho$ is measured through $\rho \rightarrow \pi^+ \pi^-$.  Since $m_\rho- 2 m_\pi = 496.35$ MeV the measurement at low $Q^2$ is feasible, but also requires a low $\pT$ threshold for tracking.

 \subsubsection{Kinematic coverage}
\label{sec:kineVM_new}

\begin{figure}[tbh!]
	\begin{center}
		\centerline{\includegraphics[width=0.7\linewidth]{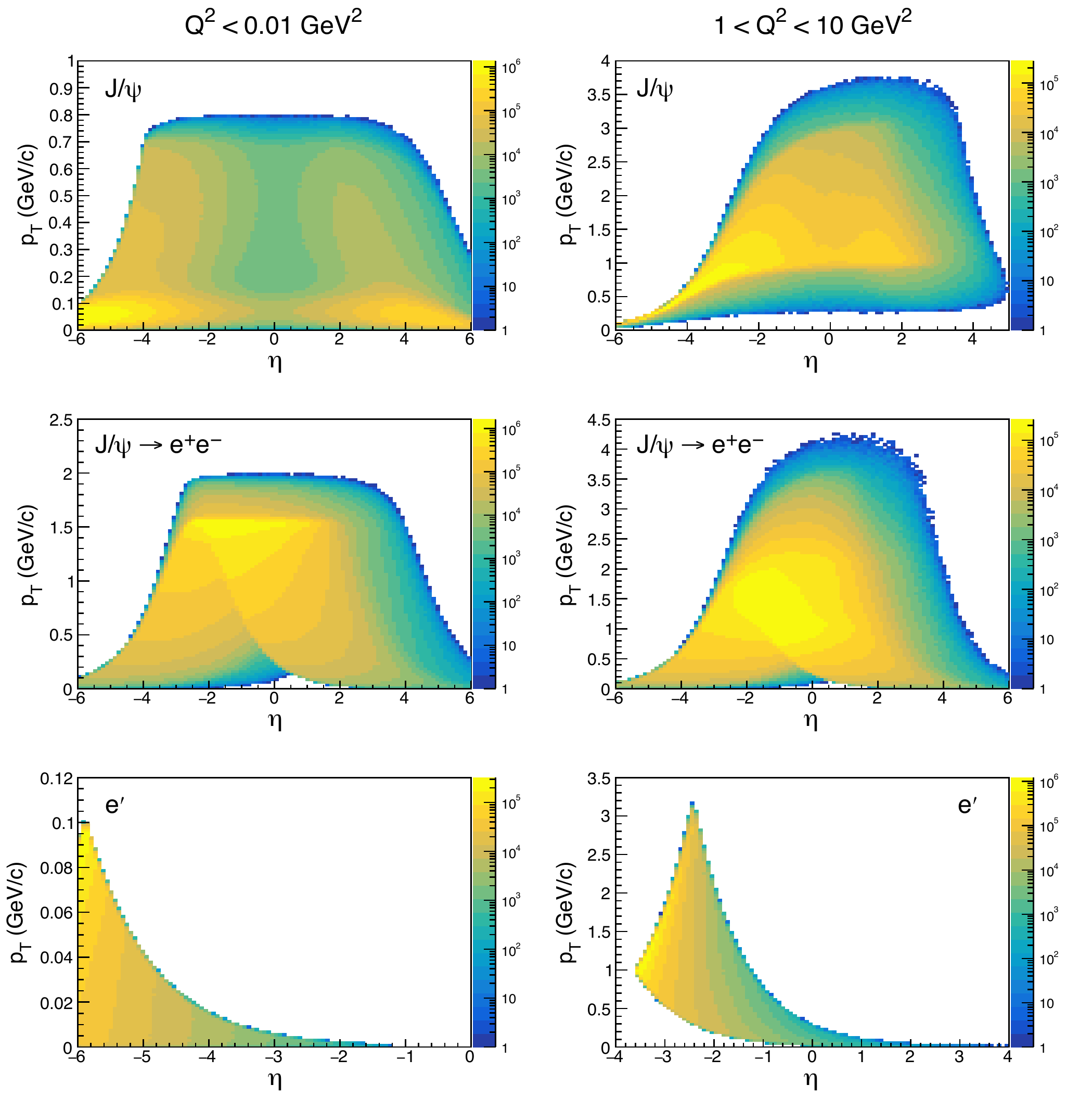}}
		\caption{\label{fig:jpsiKine}  Kinematics for diffractive $ e + \mathrm{Au} \rightarrow e^\prime + \mathrm{Au}^\prime + J/\psi$ with $J/\psi$ decaying into $e^+ e^-$. The left column is for photoproduction and the right for $1< Q^2 < 10$ GeV$^2$. Shown, from top to bottom are $\pT$ versus pseudorapidity ($\eta$)  for $J/\psi$,  electrons from the $J/\psi$ decay, and the scattered electron.}
	\end{center}
\end{figure}

\begin{figure}[tbh!]
	\centering
		\includegraphics[width=0.48\linewidth]{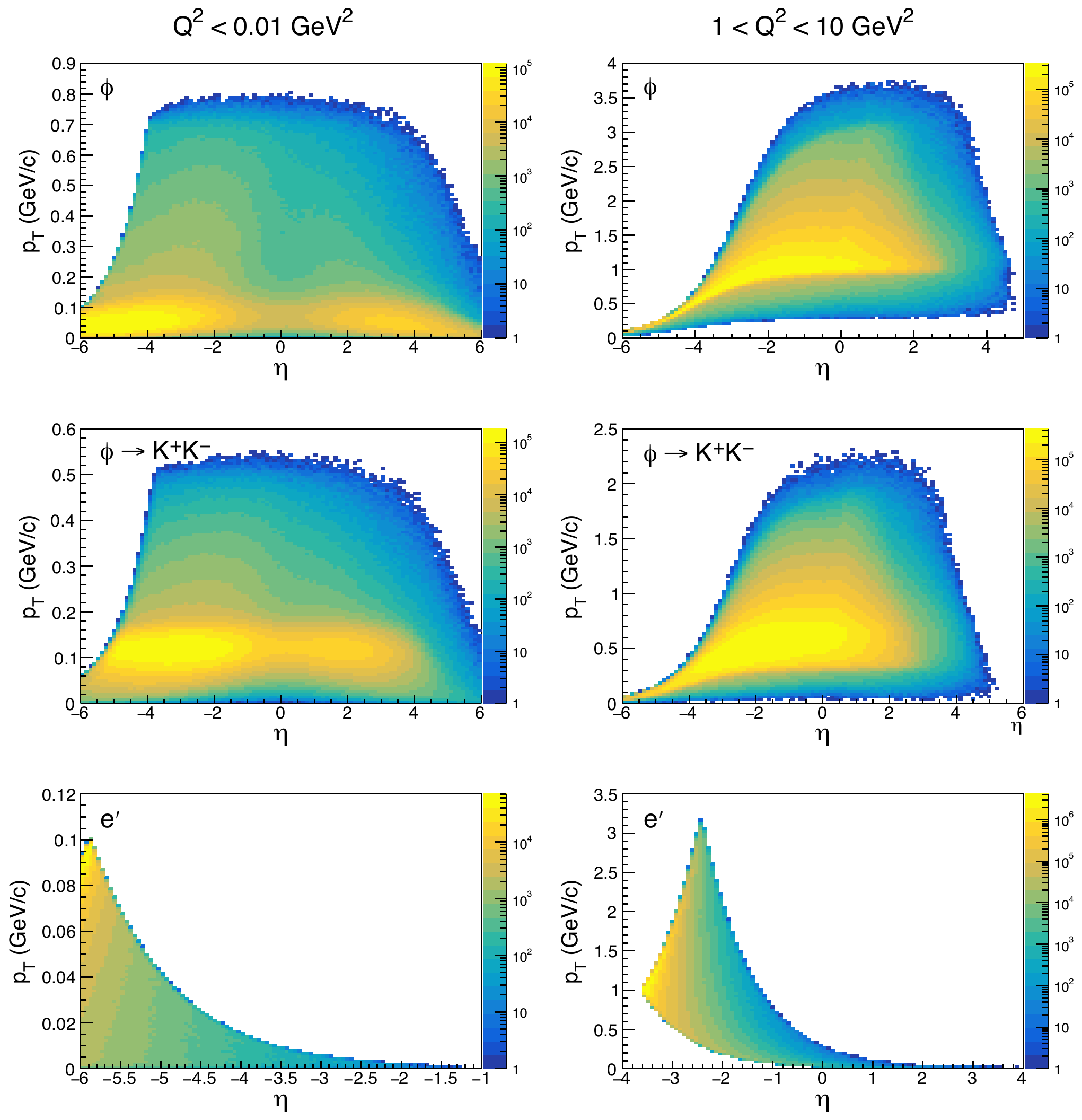}
		\hspace{2mm}
		\includegraphics[width=0.48\linewidth]{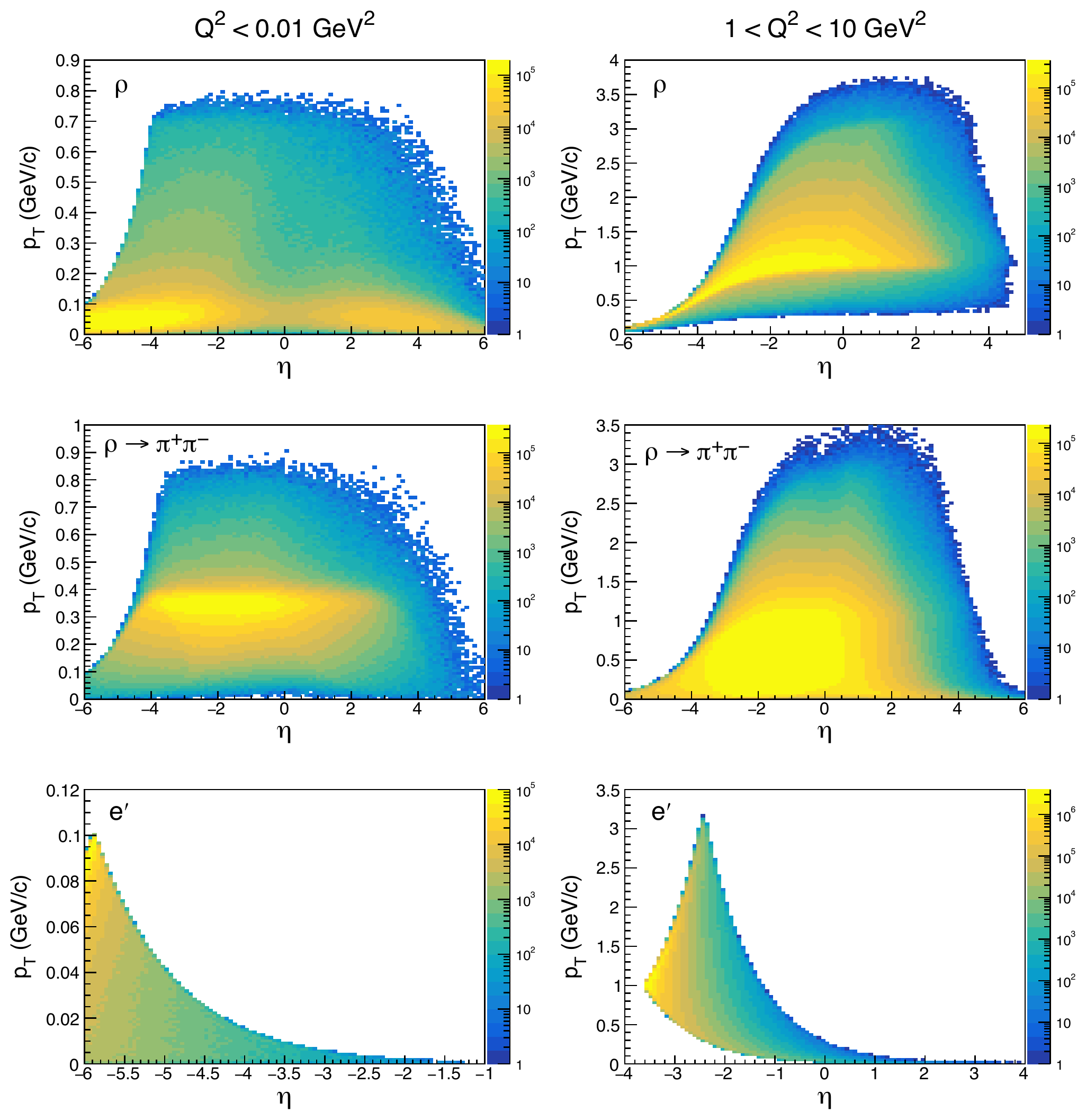}
		\caption{\label{fig:phirhoKine}  {\bf Left:} Kinematics for diffractive $ e + \mathrm{Au} \rightarrow e^\prime + \mathrm{Au}^\prime + \phi$ with $\phi$ decaying into $K^+ K^-$. The left column is for photoproduction and the right for $1< Q^2 < 10$ GeV$^2$. Shown, from top to bottom are $\pT$ versus pseudorapidity ($\eta$)  for $\phi$,  kaons from the $\phi$ decay, and the scattered electron. {\bf Right:} Same for $ e + \mathrm{Au} \rightarrow e^\prime + \mathrm{Au}^\prime + \rho$ with $\rho$ decaying into $\pi^+ \pi^-$. Note the different scale on the vertical axis for photoproduction and electroproduction.}
\end{figure}

Figures \ref{fig:jpsiKine} and \ref{fig:phirhoKine} show the kinematics in $\pT$ and pseudorapidity for $J/\psi$, $\phi$, and $\rho$, respectively.  From these figures we can determine the pseudorapidity range we want to focus our studies on. For $1< Q^2 < 10$ GeV$^2$ the decay daughters of the vector meson are spread over a wide range with the bulk sitting at midrapidity for the $J/\psi$, at $\eta \sim -1$ for the $\phi$ and at $\eta \sim -2 $ for the $\rho$. In all cases, measurements in the barrel region of $-1 < \eta < 1$ would yield sufficient statistics for a successful measurement. For all vector mesons the scattered electron falls dominantly in the  backward range of $-3.5 < \eta \lesssim -2.5$.
For photoproduction ($Q^2 < 0.01$ GeV$^2$ )  the vector meson decay products tend to drift further backward peaking at $\eta \sim -1.5$ for the $J/\psi$, at $\eta \sim -3$ for the $\phi$ and $\eta \sim -4$ for the $\rho$.  The scattered electron is pushed into the far backward region, $\eta < -5$ and can only be detected with a low-$Q^2$ tagger. 


The \pT range of the vector meson decay particles is from the tracking threshold (few hundred MeV) to 2-3 GeV.
From this we can already conclude that the multiple-scattering term in the $\pT$ resolution will play a dominant role. In our simulations we use a lower $\pT$ cut of 300 MeV unless otherwise noted.

\subsubsection{Beam effects}
\label{sec:beamsigmatovert}

 \begin{table} [tbh!]
 	\footnotesize
 	\begin{center}
 			\begin{tabular}{|l|cc|cc|cc|cc|}
 				\hline 
 				Species                                               & Au & e & Au & e &  Au & e &  Au & e  \\
 				Energy (GeV)                                      & 110 & 18 &   110  &  10  & 110    & 5  & 41 & 5 \\ \hline \hline
 				\multicolumn{9}{|l|}{Strong hadron cooling:} \\ \hline
 				RMS $\Delta \theta$, h/v ($\mu$rad) & 218/379 & 101/37 & 216/274 & 102/92 & 215/275 & 102/185 & 275/377 & 81/136 \\
 				RMS $\Delta p/p$ ($10^{-4}$)          & 6.2 & 10.9 & 6.2 & 5.8 & 6.2 & 6.8 & 10 & 6.8 \\ \hline
 				\multicolumn{9}{|l|}{Stochastic Cooling:} \\ \hline
 				RMS $\Delta \theta$, h/v ($\mu$rad)  &77/380 &109/38 &136/376& 161/116 &108/380 &127/144& 174/302 &77/77 \\
               RMS $\Delta p/p$ ($10^{-4}$)         &10 &10.9& 10 &5.8 &10 &6.8 &13 &6.8  \\
 				\hline
 			\end{tabular}
 		\caption{Horizontal and vertical beam divergence and beam momentum spread and for various energies for e+Au running used in this study. The values vary depending on the beam cooling option.}
 		\label{tab:beameffects}
 	\end{center}
 \end{table}
 
\begin{table} [tbh!]
 		\footnotesize
 		\begin{center}
 			\begin{tabular}{|l|l||c|c|c|c|c|c|}
 				\hline 
 				{} & {} & \multicolumn{6}{|c|}{$t$-range (GeV$^2$)}  \\ 
 				method &  effect & 0-0.1 & 0.1-0.4 & 0.04 - 0.07 & 0.07 - 0.10 & 0.10 - 0.13 & 0.13 - 0.18 \\ \hline \hline
 				E &  beam divergence   & 0.061 & 0.015 & 0.008 & 0.007 & 0.006 & 0.005 \\ \hline   
				E & beam mom.~spread & 149.61  & 10.36 & 3.03 & 1.86 & 1.37 & 1.03 \\ \hline     
				L &  divergence \& mom.~spread & 0.048 & 0.016 & 0.009 & 0.007 & 0.006 & 0.005 \\ 
 				\hline
 			\end{tabular}
 			\caption{Effect of beam momentum spread and beam divergence on $t$-resolution, $\sigma_t/t$,  with method E and L  for $J/\psi$ production in $1 < Q^2 < 10$ GeV$^2$. Shown is the relative difference between smeared and actual $t$ for 6 ranges in $t$. The quoted $\sigma_t/t$ is the r.m.s of the respective distribution calculated in the full range.}
 			\label{tab:beamEffectsOnRes}
 		\end{center}
 	\end{table}

There are two beam effects that potentially can affect the $t$ resolution, the spread of the beam momentum $\mathrm{d}p/p$ and the horizontal and vertical beam divergence $\sigma_h$ and $\sigma_v$. None of these effects can be corrected on an event-by-event basis. Note that this is different from the effect of a finite crossing angle that is well defined. In this study we neglect therefore crossing-angle  effects assuming that  they can be fully corrected for. We assume the values in Table~\ref{tab:beameffects}.

The default method for reconstructing $t$ is based on using only the transverse momenta of the vector meson and the scattered electron ignoring all longitudinal momenta. This method was extensively used at HERA in diffractive vector meson studies. 
We take
\begin{equation}
	 t = \left[\vec{p}_T(e^\prime) + \vec{p}_T(V)\right]^2,
	\label{eq:methodA}
\end{equation}
which we refer to as method A here. 
We compared this method with the actual $t$ in our generated events without any smearing due to beam and detector effects and made the following observations:
	\begin{itemize}
		\item In $J/\psi$ production method A underestimates the actual $t$. The offset is largest at $Q^2 =$ 1-2 GeV$^2$ with around 2\% and decreases towards larger $Q^2$ to 1\% at $Q^2 = $ 9-10 GeV$^2$. The offset is absent for photoproduction ($Q^2 < 0.01$ GeV$^2$). For $1< Q^2 < 10$ GeV$^2$ and including the offset we obtain $\sigma_t$ resolutions (r.m.s.) of 10\% ~for $t < 0.01$ GeV$^2$, 1.8\% at $t  = 0.10$ GeV$^2$, and 1.6\% at $t  = 0.16$ GeV$^2$. In photoproduction we observe no $t$ smearing except at the lowest $t$ ($t < 0.01$ GeV$^2$) of 1.3\%.
		
		\item In $\phi$ production method A shows similar issues as for the $J/\psi$ but to a lesser degree. Except at the lowest $t$, the offset is $\sim$0.5\%. The  $\sigma_t/t$ resolutions (r.m.s.) is 6.3\% for $t < 0.01$ GeV$^2$ and 0.45\% for $t  > 0.10$ GeV$^2$, both for $1< Q^2 < 10$ GeV$^2$.
		
		\item The trend continues in $\rho$ production. Including a minimal offset in the range of $1< Q^2 < 10$ GeV$^2$ we find a  $\sigma_t/t$ resolutions (r.m.s.) is 6.3\% ~for $t < 0.01$ GeV$^2$ and 0.3\% for $t  > 0.10$ GeV$^2$.	
	\end{itemize}

In general the approximation of method A leads to small to negligible $t$ resolution effects due to the beam divergence and momentum spread, except for very small $t$ (close to $t_\textrm{min.}$), where the longitudinal component of the  momentum exchange dominates.   At small $t$ one can improve the resolution by using the full measured 4-momenta. In principle one would use the conservation of 4-momentum to calculate the momentum transfer as $t  = (p_V + p_{e^\prime} - p_e)^2$, which we call here method E. Calculating this inner product involves the accurate subtraction of two large numbers: the incoming and outgoing electron longitudinal momenta. This makes it very sensitive to the momentum spread in the electron beam, rendering this exact reconstruction impractical. 
    
The poor resolution at small $t$ can be significantly improved with the very nontrivial assumption that it is possible to assure that the reaction is inclusive by vetoing all decay products of the target nucleus. In this case one can use one additional constraint, namely that the invariant mass of the outgoing nucleus must be $M_A^2$,  to determine the actual longitudinal momentum of the incoming electron, instead of assuming the nominal electron beam momentum. We refer to this procedure as method L.  As an additional check in method L (that has not been performed here) one could verify that the inferred $e$ energy is within the expected spread of electron energies in the beam.
The impact of beam effects on the $t$ resolution is summarized in Table~\ref{tab:beamEffectsOnRes}.

 \subsubsection{Impact of momentum resolution}
\label{sec:momres} 

In order to study the impact of momentum resolution effects on the $t$-resolution, we look at processes with $1 < Q^2 < 10$ GeV$^2$ since the resolution of the scattered electron plays a more important role than is the case in photoproduction. We focus on the region discussed in Sec.~\ref{sec:kineVM_new} where the vector meson is detected in the barrel  ($|\eta|<1|$)and the scattered electron at $-3.5 < \eta < -2.5$). 

 Tables \ref{tab:jpsiRes_new} , \ref{tab:phiRes_new}, and  \ref{tab:rhoRes_new} show our results for $\sigma_t/t$ for $J/\psi$, $\phi$, and $\rho$ mesons, respectively, in 6 $t$ bins between 0 and 0.18 GeV$^2$.
  The first line in each table is the Detector and R\&D Handbook \cite{EIC:RDHandbook} value, and the subsequent lines show the effect of an improved resolution. The lowest $t$ bin has a poor resolution, but as discussed above in Sec.~\ref{sec:beamsigmatovert} for low $t$ a different analysis method is needed that takes into account the longitudinal component of the momentum exchange.
  
    \begin{figure}[tbh!]
 	\begin{center}
 		\includegraphics[width=0.5\linewidth]{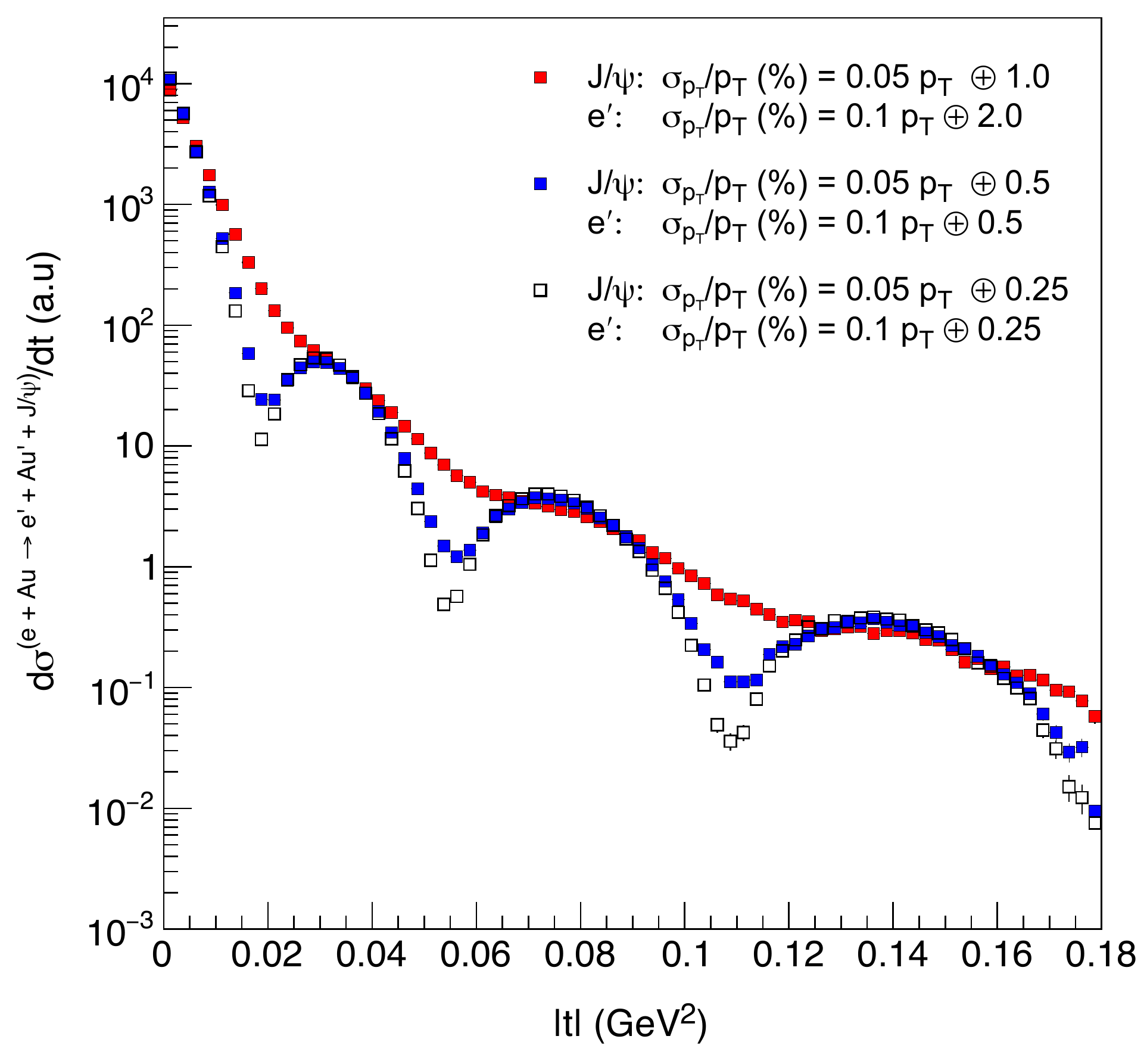}
 		\caption{\label{fig:dsdtCompare_new}  Illustration of the impact of different $\pT$ resolutions on the coherent $J/\psi$ production cross section, $d\sigma/dt$, for $1 < Q^2 < 10$ GeV$^2$. }
		\vspace{-6mm}
 	\end{center}
 \end{figure}

 \begin{table} [htb!]
 	\footnotesize
 	\begin{center}
 		\begin{tabular}{|l|l||c|c|c|c|c|c|}
 			\hline 
 			measurement & MS & \multicolumn{6}{|c|}{$t$-range (GeV$^2$)}  \\ 
 			precision term for & term for barrel & \multicolumn{6}{|c|}{}  \\  
 			barrel (backward) (\%) &  (backward) (\%) & 0-0.1 & 0.1-0.4 & 0.04 - 0.07 & 0.07 - 0.10 & 0.10 - 0.13 & 0.13 - 0.18 \\ \hline \hline
 				0.05 (0.1) & 1.0 (2.0)   & 4.58 & 0.45 & 0.25 & 0.19 & 0.16 & 0.14 \\ \hline   
				0.1 (0.2) & 1.0 (2.0)   & 4.71  & 0.46 & 0.25 & 0.20 & 0.17 & 0.14 \\ \hline     
				0.025 (0.05) & 1.0 (2.0)  & 4.54 & 0.45 & 0.24 & 0.19& 0.16 & 0.14 \\ \hline  
				0.05 (0.1) & 0.5 (2.0)  & 3.53 & 0.38 & 0.21 & 0.17 & 0.14 & 0.12 \\ \hline    
				0.05 (0.1) & 0.5 (1.0) & 1.29 & 0.22 & 0.12 & 0.10 & 0.08 & 0.07 \\ \hline    
				0.05 (0.1) & 0.5 (0.5)  & 0.78 & 0.16 & 0.09 & 0.07 & 0.06 & 0.05\\ \hline    
 				0.05 (0.1) & 0.25 (0.5)   & 0.49 & 0.12& 0.07 & 0.05 & 0.05 & 0.04\\ \hline    
 				0.05 (0.1) & 0.25 (0.25)   & 0.36 & 0.09 & 0.05 & 0.04 & 0.04& 0.03\\    
 				\hline
 			\end{tabular}
 			\caption{$\sigma_t/t$ for $J/\psi$ production in $1 < Q^2 < 10$ GeV$^2$ in 6 different $t$ bins. Each row shows the $t$ resolution for the two different terms that make up the $\pT$ resolution of the $J/\psi$ decay particles. The measurement precision term and the MS term are shown for the two different regions studied, the barrel region for the $J/\psi$ detection and the backward region for the measurement of the scattered electron. See text for details.
 			\label{tab:jpsiRes_new}}
 		\end{center}
 	\end{table}
 	 	
 \begin{table} [htb!]
 		\footnotesize
 		\begin{center}
 			\begin{tabular}{|l|l||c|c|c|c|c|c|}
 				\hline 
				measurement & MS & \multicolumn{6}{|c|}{$t$-range (GeV$^2$)}  \\ 
precision term for & term for barrel & \multicolumn{6}{|c|}{}  \\  
barrel (backward) (\%) &  (backward) (\%) & 0-0.1 & 0.1-0.4 & 0.04 - 0.07 & 0.07 - 0.10 & 0.10 - 0.13 & 0.13 - 0.18 \\ \hline \hline
  				0.05 (0.1) & 1.0 (2.0)   & 5.91 & 0.31  & 0.18  & 0.14  & 0.12 & 0.11  \\ \hline   
 				0.1 (0.2) & 1.0 (2.0)     & 6.00  & 0.32 & 0.18 & 0.14 & 0.13 & 0.11 \\ \hline     
 				0.025 (0.05) & 1.0 (2.0)  & 5.88  & 0.31  & 0.18  & 0.14  & 0.12  & 0.11 \\ \hline  
 				0.05 (0.1) & 0.5 (2.0)    & 5.41  & 0.30 & 0.17  & 0.14 & 0.12 & 0.10  \\ \hline    
 				0.05 (0.1) & 0.5 (1.0)    & 1.59 & 0.15  & 0.09 & 0.07  & 0.06  & 0.05 \\ \hline    
 				0.05 (0.1) & 0.5 (0.5)    & 0.63 & 0.09  & 0.05  & 0.04  & 0.04& 0.03 \\ \hline    
 				0.05 (0.1) & 0.25 (0.5)    & 0.51 & 0.08 & 0.05 & 0.04 & 0.03& 0.03 \\ \hline    
 				0.05 (0.1) & 0.25 (0.25)  & 0.26 & 0.05 & 0.03 & 0.02 & 0.02  & 0.02  \\    
 				\hline
 			\end{tabular}
			\caption{$\sigma_t/t$ for $\phi$ production in $1 < Q^2 < 10$ GeV$^2$ in 6 different $t$ bins. Each row shows the $t$ resolution for the two different terms that make up the $\pT$ resolution of the $\phi$ decay kaons. The measurement precision term and the MS term are shown for the two different regions studied, the barrel region for the $\phi$ detection and the backward region for the measurement of the scattered electron. See text for details.
	       \label{tab:phiRes_new}}
 		\end{center}
 	\end{table}
 	
 		 	 \begin{table} [htb!]
 		\scriptsize
 		\begin{center}
 			\begin{tabular}{|l|l||c|c|c|c|c|c|}
 				\hline 
				measurement & MS & \multicolumn{6}{|c|}{$t$-range (GeV$^2$)}  \\ 
precision term for & term for barrel & \multicolumn{6}{|c|}{}  \\  
barrel (backward) (\%) &  (backward) (\%) & 0-0.1 & 0.1-0.4 & 0.04 - 0.07 & 0.07 - 0.10 & 0.10 - 0.13 & 0.13 - 0.18 \\ \hline \hline
  				0.05 (0.1) & 1.0 (2.0)    & 6.87 & 0.33  & 0.19 & 0.15 & 0.13 & 0.11   \\ \hline   
 				0.1 (0.2) & 1.0 (2.0)       & 6.99 & 0.33 & 0.19 & 0.15 & 0.13 & 0.11    \\ \hline     
 				0.025 (0.05) & 1.0 (2.0)    & 6.99 & 0.33 & 0.19 & 0.15 & 0.13 & 0.11  \\ \hline  
 				0.05 (0.1) & 0.5 (2.0)      & 6.17  & 0.31  & 0.18 & 0.14 & 0.12 & 0.10    \\ \hline    
 				0.05 (0.1) & 0.5 (1.0)       & 1.83 & 0.16 & 0.09 & 0.07  & 0.06  & 0.06  \\ \hline    
 				0.05 (0.1) & 0.5 (0.5)      & 0.74 & 0.10 & 0.06 & 0.05 & 0.04  & 0.04    \\ \hline    
 				0.05 (0.1) & 0.25 (0.5)     & 0.57 & 0.08  & 0.05 & 0.04 & 0.03 & 0.03   \\ \hline    
 				0.05 (0.1) & 0.25 (0.25)    & 0.29  & 0.05  & 0.03  & 0.03 & 0.02  & 0.02  \\     
 				\hline
 			\end{tabular}
			\caption{$\sigma_t/t$ for $\rho$ production in $1 < Q^2 < 10$ GeV$^2$ in 6 different $t$ bins. Each row shows the $t$ resolution for the two different terms that make up the $\pT$ resolution of the $\rho$ decay pions. The measurement precision term and the MS term are shown for the two different regions studied, the barrel region for the $\rho$ detection and the backward region for the measurement of the scattered electron. See text for details.
        	\label{tab:rhoRes_new}}
 		\end{center}
 	\end{table}

The most important finding is that the measurement precision term in $\sigma_{\pT}/{\pT}$ has little impact on the overall $t$ resolution in the barrel as can be seen in columns 1-3. This holds for all studies of vector mesons. We conclude that a precision term of 0.05\% for the barrel and 0.1\% for the backward region seems adequate. 
   The case is different for the MS term, especially in the backward region, which appears to have a substantial impact on $\sigma_t/t$ as seen in columns 4-8.  Figure \ref{fig:dsdtCompare_new} illustrates this effect, demonstrating that  the Handbook settings completely wash out the diffractive structure of the distribution.  Note that unfolding procedures will not improve the situation as the positions of the minima are a priori unknown.
   The $t$ resolution for the $\phi$ and the $\rho$ appears to be better than that of the $J/\psi$ at lower $t$ due to the different $\pT$ range of the decay particles.
   
  The evaluation of the $t$-resolution on photoproduction is  more involved since much will depend on the performance of a low-$Q^2$ tagger and its potential $\pT$ resolution, which is currently unknown. However, $t$ can also be calculated in photoproduction by ignoring the scattered electron $\pT$ in method A, taking  $t \approx - \pT^2$, with an error less than $Q^2$ \cite{Chekanov:2002xi}, which is acceptable e.g. for a selection of events with $Q^2 \lesssim 10^{-4}$. For $\phi$ meson  photoproduction one needs to detect decay kaons with momenta of 100-150 MeV/$c$ (see Fig. \ref{fig:phirhoKine}), that can only be captured in either low-field runs or with the inner layer of a vertex tracker. The situation is only slightly better for the $\rho$ where the decay pions have $\pT$ around 300-400 MeV/$c$ and optimal for the $J/\psi$ with decay electrons of 1-1.5 GeV/$c$.
 
 	\FloatBarrier


 	  \begin{figure}[tbh!]
	\begin{center}
		\includegraphics[width=0.3\linewidth,clip=true,bb=192 5 365 140]{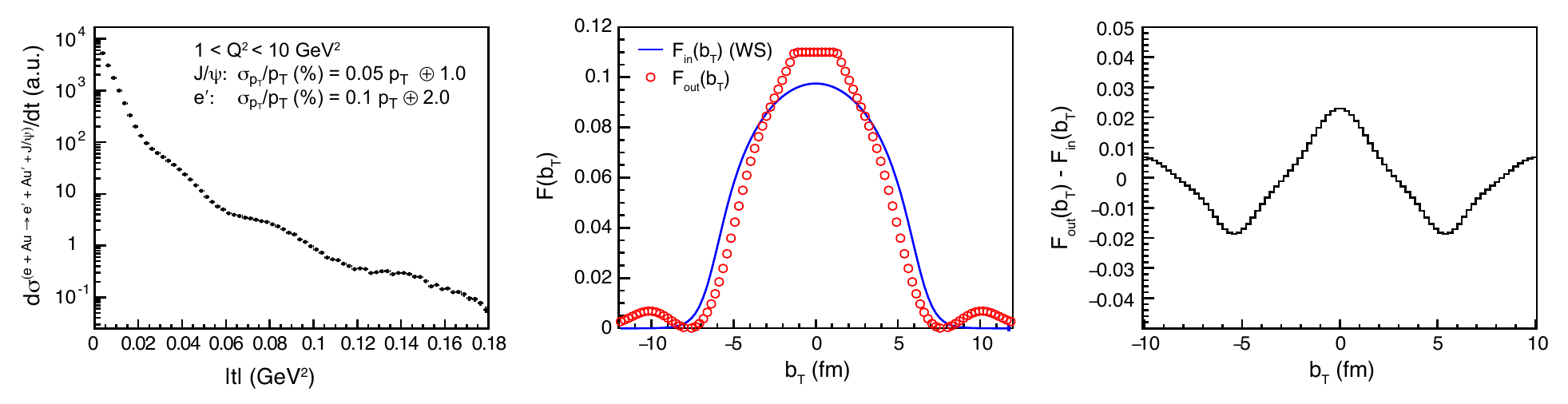}
		\includegraphics[width=0.3\linewidth,clip=true,bb=192 5 365 140]{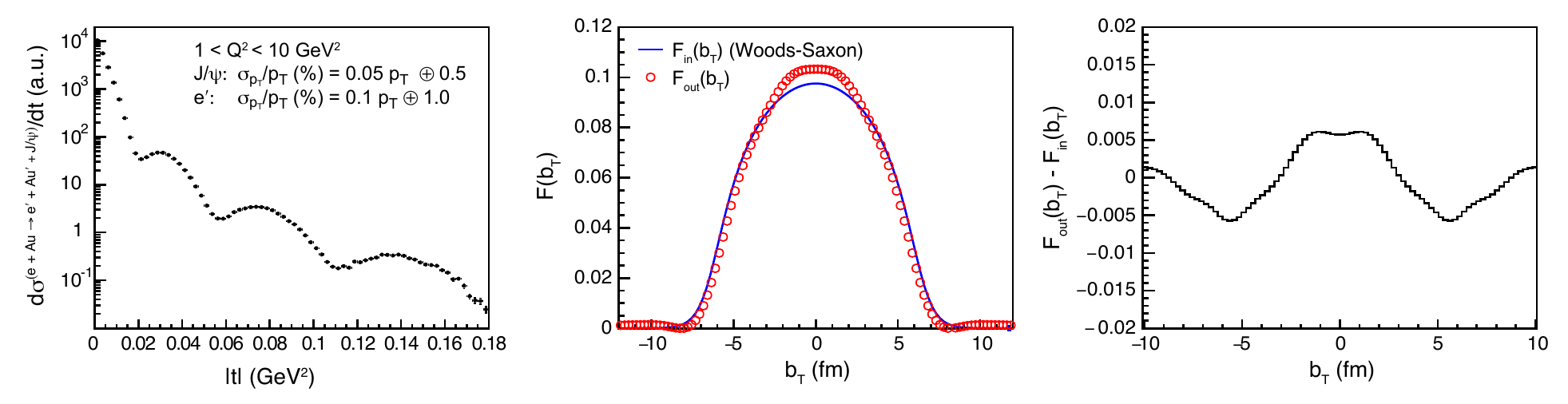}
		\includegraphics[width=0.3\linewidth,clip=true,bb=192 5 365 140]{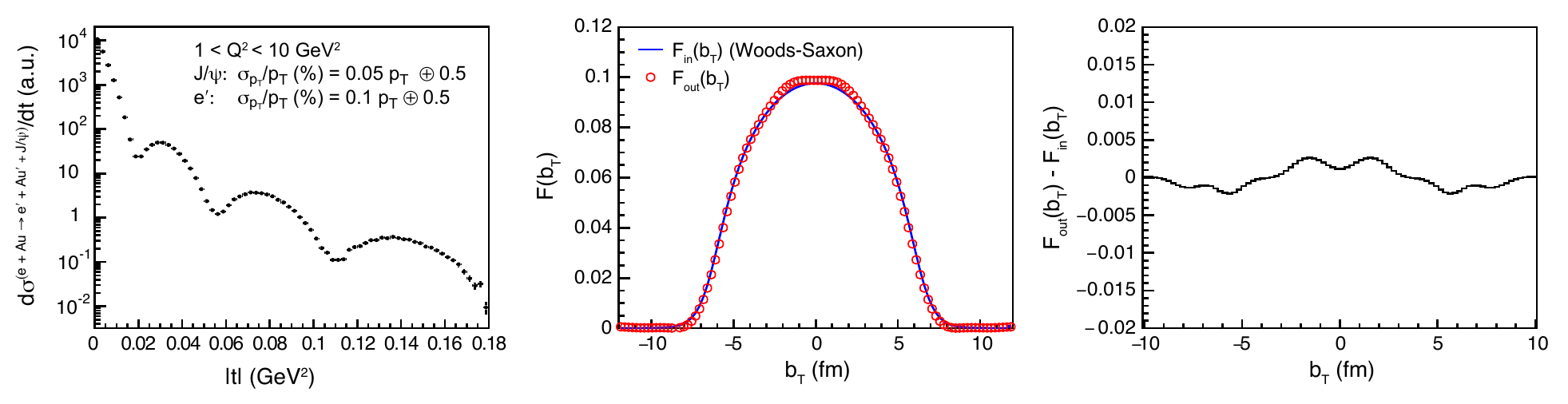}
		
		\caption{\label{fig:Taextraction}   Extracted $F(b_T)$ (red circles) compared to the Wood-Saxons input distribution (blue line). 
		Left: handbook detector, with $\sigma_{\pT}/\pT = 0.05\pT \oplus 1.0 \%$ for $J/\Psi$ and $\sigma_{\pT}/\pT = 0.1\pT \oplus 2.0$ for $e'$. Center: a MS-term resolution improved by a factor 2: $\sigma_{\pT}/\pT = 0.05 \pT \oplus 0.5 \%$ for $J/\Psi$ and $\sigma_{\pT}/\pT = 0.1 \pT \oplus 1.0$ for $e'$.  Right: resolution on the scattered electron improved by a further factor 2: $\sigma_{\pT}/\pT = 0.05\pT \oplus 0.5 \%$ for $J/\Psi$ and $\sigma_{\pT}/\pT = 0.1\pT \oplus 0.5$ for $e'$, corresponding to our ``nominal'' resolution.
		}
		\vspace{-3mm}
	\end{center}
\end{figure}

\begin{figure}[tbh!]
	\begin{center}
		\includegraphics[width=0.3\linewidth,clip=true,bb=192 5 365 140]{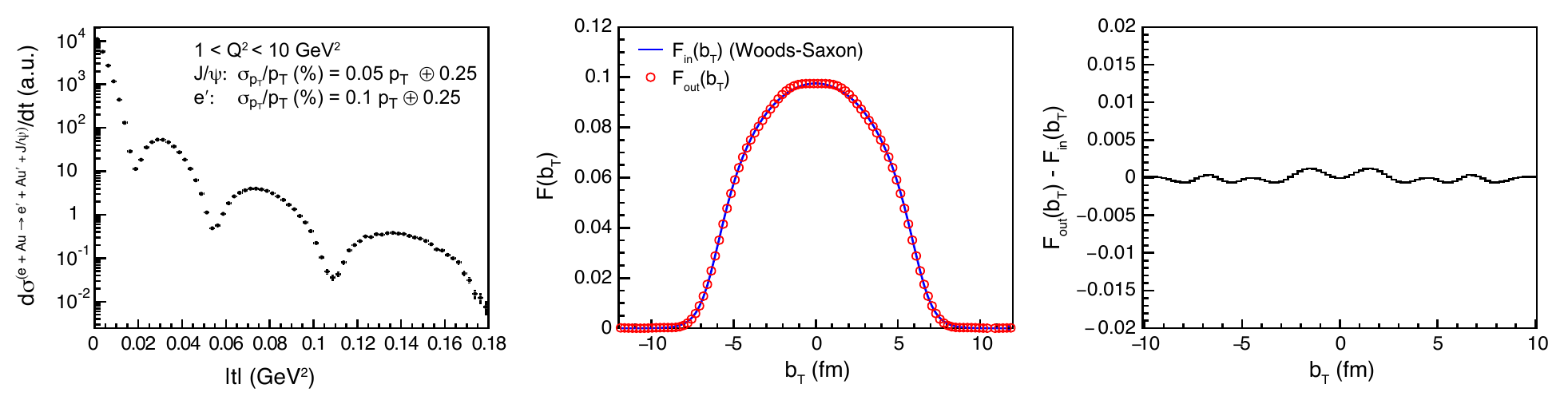}
		\includegraphics[width=0.3\linewidth,clip=true,bb=192 5 365 140]{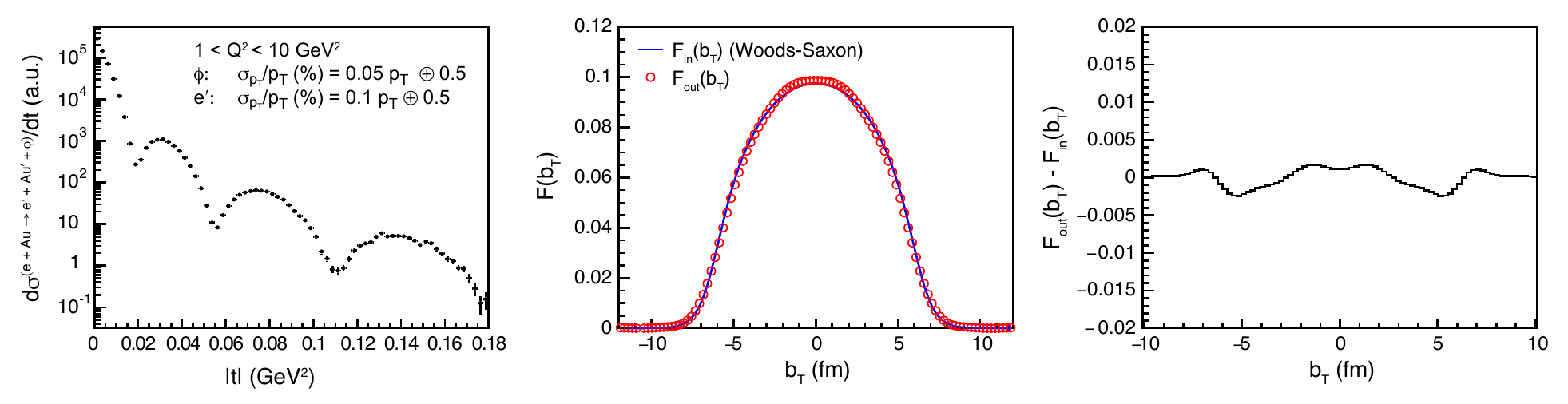}
		\includegraphics[width=0.3\linewidth,clip=true,bb=192 5 365 140]{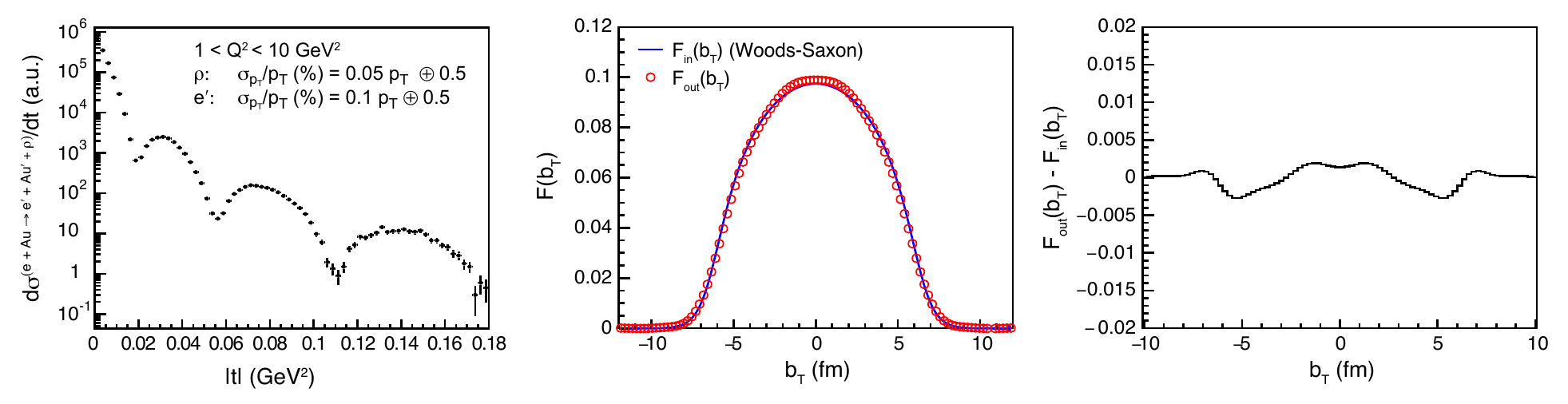}
		
		\caption{\label{fig:moreTaextraction}   Extracted $F(b_T)$ (red circles) compared to the Wood-Saxons input distribution (blue line). 
		Left: $J/\Psi$ source extraction with a further improvement compared to the nominal resolution of Fig~\ref{fig:Taextraction} (right), with  $\sigma_{\pT}/\pT = 0.05\pT \oplus 0.25 \%$ for $J/\Psi$ and $\sigma_{\pT}/\pT = 0.1\pT \oplus 0.25$ for $e'$. Center and right: $\phi$  and $\rho$ production with the nominal resolution of $\sigma_{\pT}/\pT = 0.05\pT \oplus 0.5 \%$ for $J/\Psi$ and $\sigma_{\pT}/\pT = 0.1\pT \oplus 0.5$ for $e'$. 
		}
		\vspace{-3mm}
	\end{center}
\end{figure}

The coherent distribution $d\sigma/dt$ allows, as discussed in Secs.~\ref{part2-subS-SecImaging-GPD3d} and~\ref{part2-subS-LabQCD-Photo}, one to
obtain information about the gluon distribution in impact-parameter
space through a Fourier transform \cite{Toll:2012mb}. This is regarded as one of the 
key studies in the \eA program.   Successfully  extracting the source
distribution is essential and will be used in the following  to establish the requirements on $\sigma_t/t$ and thus on $\sigma_{\pT}/\pT$.
  
Assuming here for simplicity that a complex phase of the amplitude does not  depend on $t$ or $b$,   we can regain the impact-parameter dependent  amplitude $F(b)$ as a Fourier transform of the square root of the cross section.  In order to maintain the oscillatory structure of the amplitude we have to switch its sign in every second minimum.  

The Sartre generator starts from  an explicit transverse density function  $T_A(b)$. Gluon saturation results in a deviation of the $b$-dependence of the amplitude from the input density. Here we investigate the accuracy of extracting $F(b)$  as the difference $(F_\mathrm{out}(b) - F_\mathrm{in}(b))$ between the input and extracted amplitudes. Figure \ref{fig:Taextraction} shows a comparison between the input and extracted source densities. This figure uses the bNonSat model~\cite{Toll:2012mb}, where the amplitude is exactly proportional to the input distributions, and thus $(F_\mathrm{out}(b)$ would be equal to $F_\mathrm{in}(b))$ for an ideal detector and in the absence of beam effects, and including the longitudinal component of the momentum transfer. We see that a reduction of the MS term to $0.5$~GeV for both the meson decay products and the scattered electron is required for a reconstruction of the impact parameter profile, representing a factor 2 improvement with respect to the handbook detector for the barrel and a factor 4 for the scattered electron. \emph{This is our nominal detector requirement resulting from this study.} Figure \ref{fig:moreTaextraction} (left) shows the result of an even further improvement by a factor 2.  A closer look at the Fourier-transforms reveals that what is crucial is to resolve the minima up to the third one, as discussed in the next subsection. 

From studies discussed in Sec.~\ref{sec:momres} we already observed that the $\sigma_t/t$ resolution for a given $\pT$-resolution is smaller for the $\rho$ and $\phi$ than for the $J/\psi$.   Figure \ref{fig:moreTaextraction} (center, right) shows the source extraction accuracy for $\phi$ ad $\rho$ with the nominal resolution  $\sigma_{\pT}/\pT = 0.05\pT \oplus 0.5 \%$ in the barrel and $\sigma_{\pT}/\pT = 0.1\pT \oplus 0.5$ for $e'$

\subsubsection{Separating coherent and incoherent processes}
\FloatBarrier

\begin{figure}[tbh!]
	\begin{center}
		\includegraphics[width=0.5\linewidth]{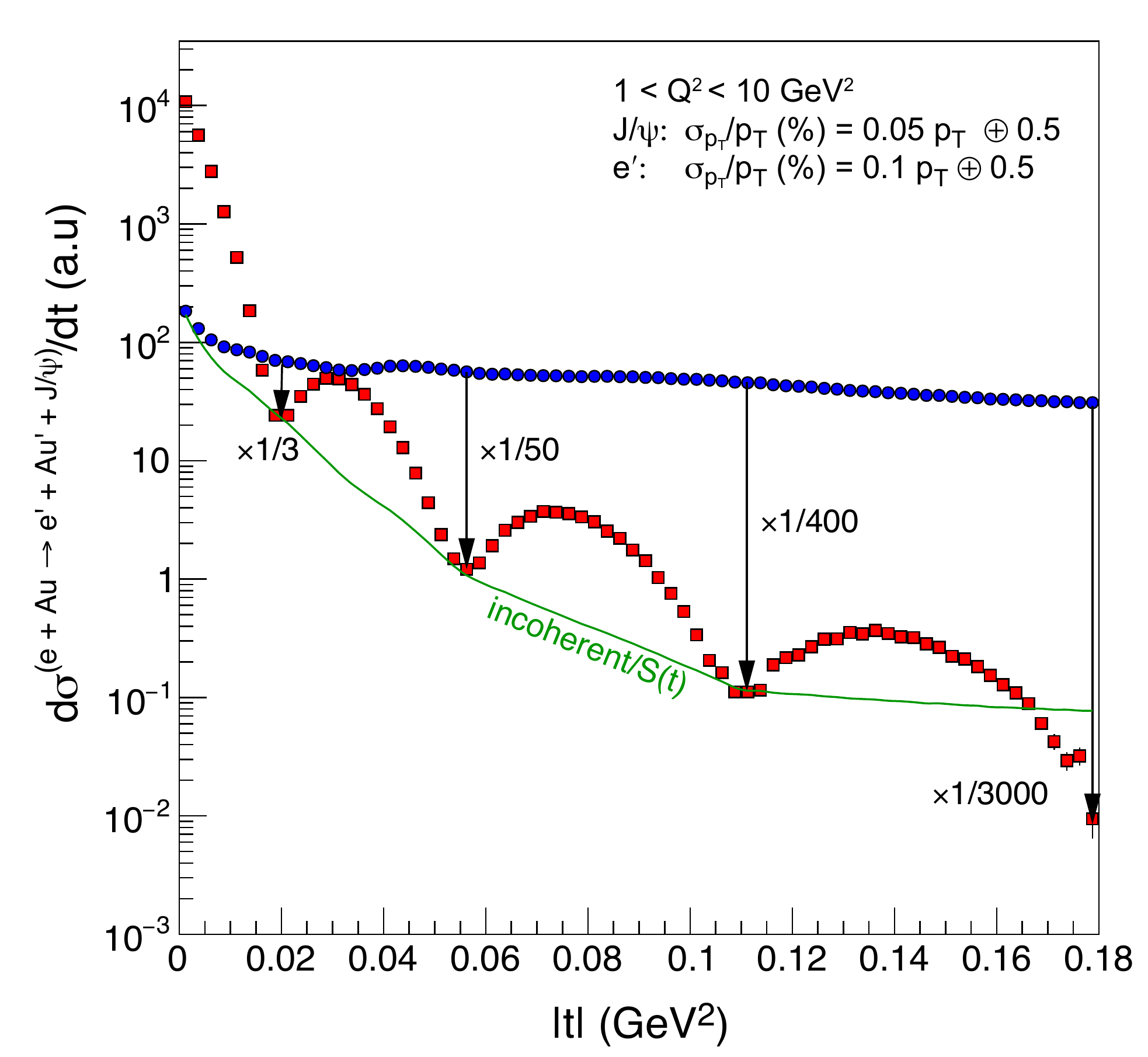}
		\caption{\label{fig:coh_incoh_eAu_new}   Coherent (red) and incoherent (blue) cross section $d\sigma/dt$ for diffractive $J/\psi$ production in $1 < Q^2 < 10$ GeV$^2$. }
	\end{center}
\end{figure}

\begin{figure}[tbh!]
	\begin{center}
		\includegraphics[width=0.32\linewidth]{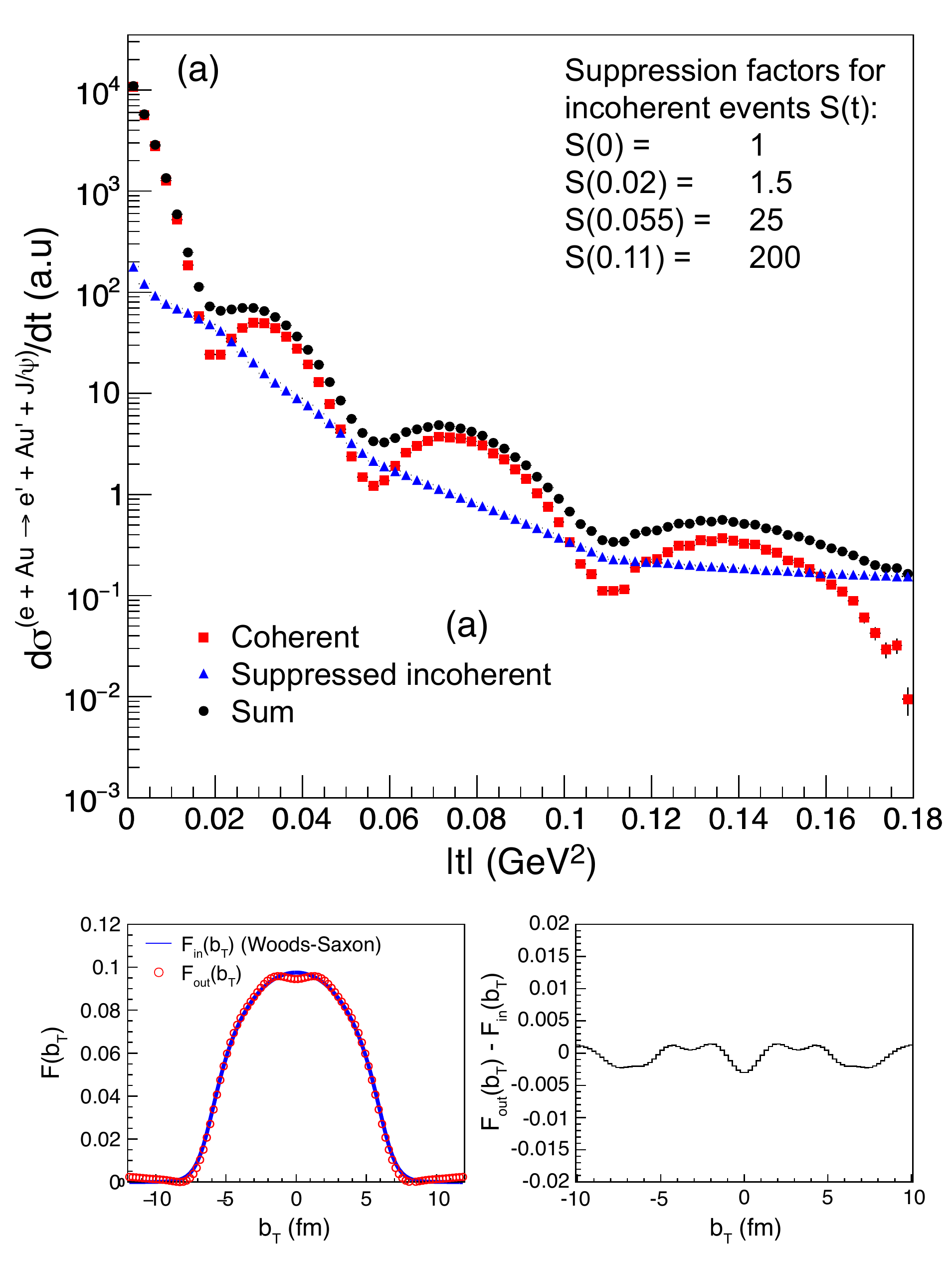}
		\includegraphics[width=0.32\linewidth]{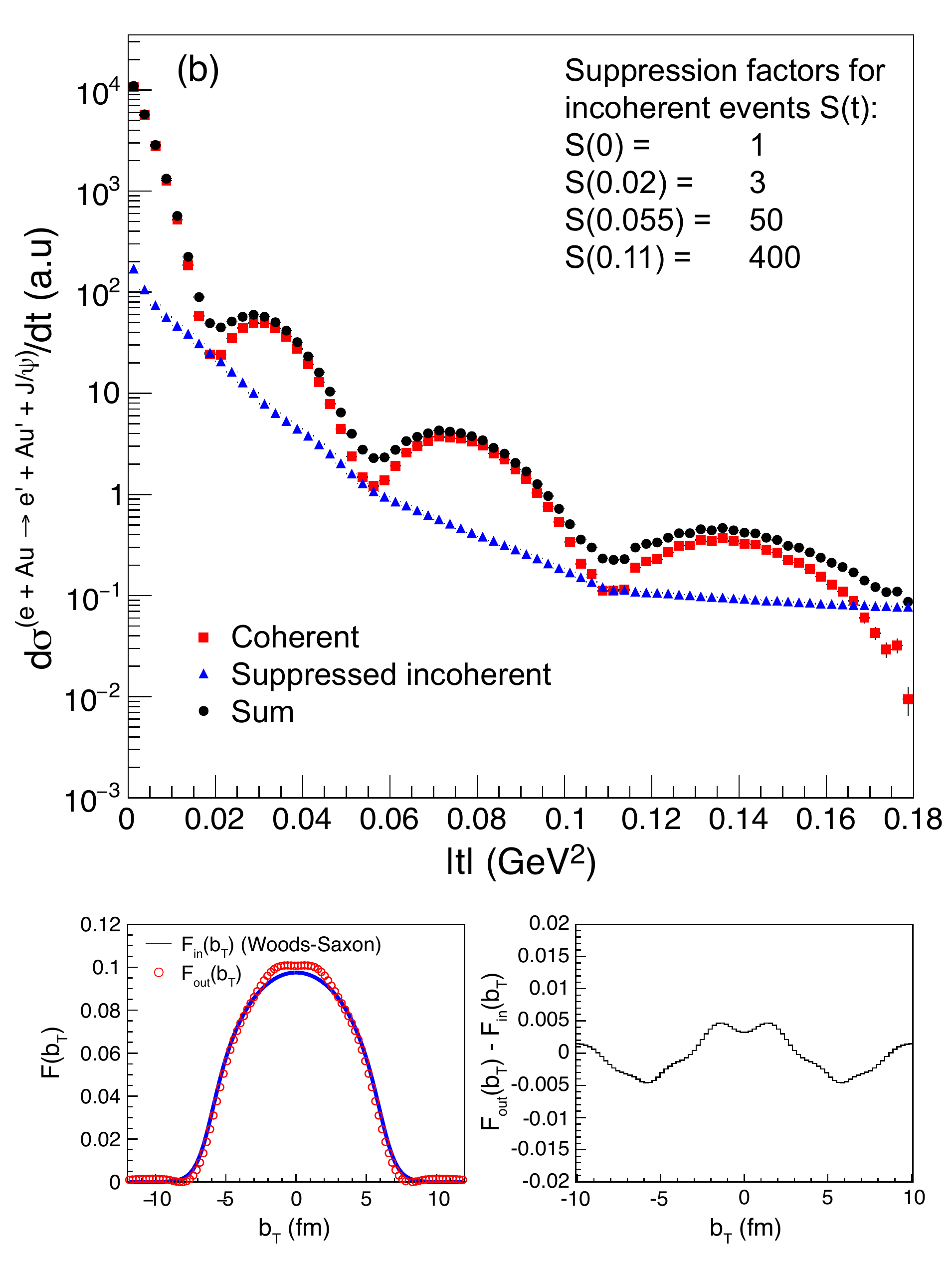}
		\includegraphics[width=0.32\linewidth]{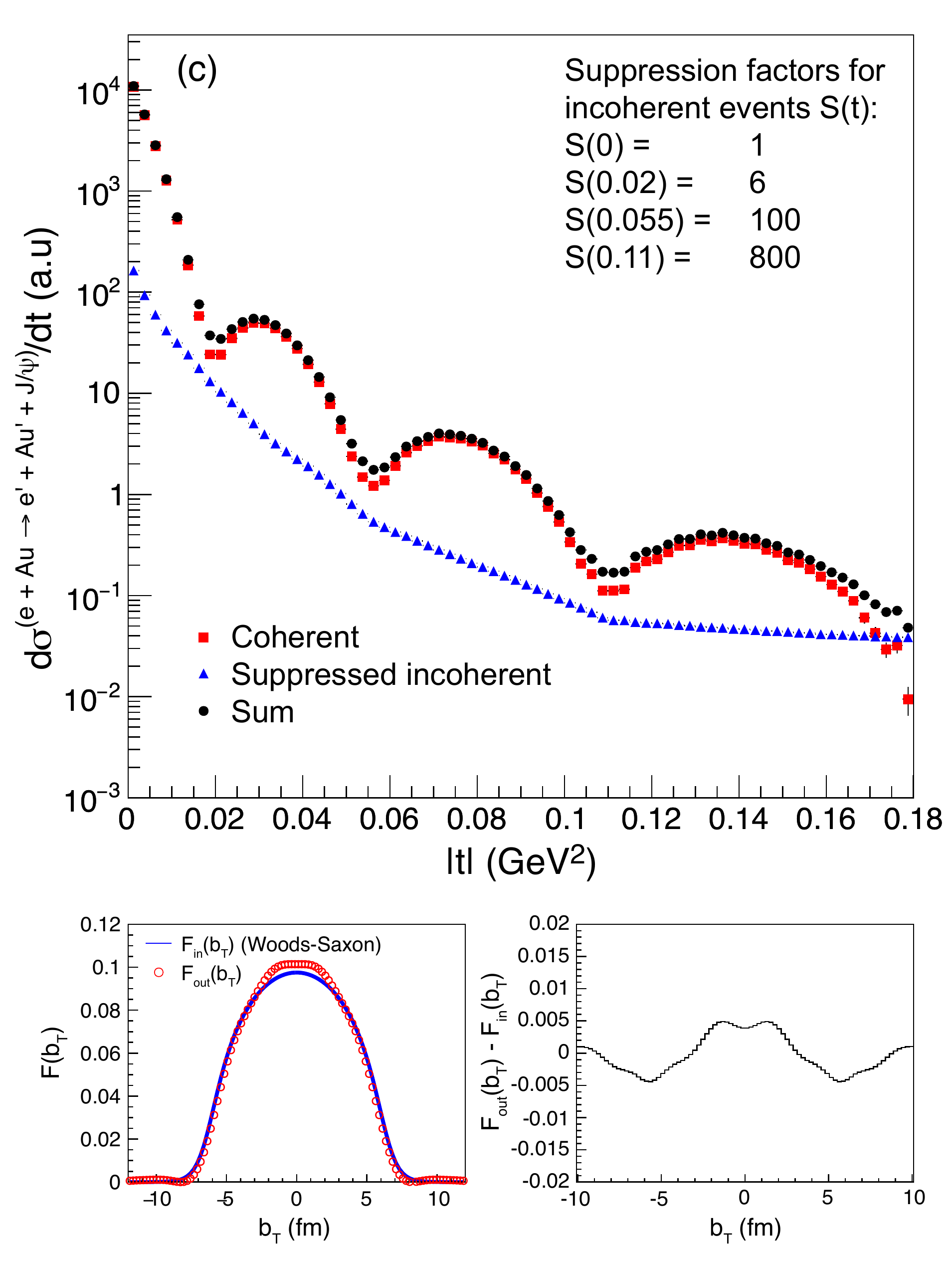}
			\end{center}
		\caption{\label{fig:combo_incoherent_new}   Each panel depicts $d\sigma/dt$ for coherent, suppressed (detected and rejected) incoherent events and their sum. The sum is used to extract the source $F(b_T)$ which is shown in the bottom left of each panel and compared to the input Woods-Saxon distribution. The plots on the bottom right illustrate the  difference between extracted and true $F(b_T)$.  The suppression of the incoherent background increases from the left to right panel in steps of a factor 2.  }
\end{figure}

Experimentally, the measured spectra in diffractive vector meson production contain the sum of coherent and incoherent processes (see Fig.~\ref{fig:coh_incoh_eAu_new}). At low $t$, coherent production dominates the cross section while already at around $|t| > 0.02$ GeV$^2$ the incoherent process starts to take over. Both processes are of substantial interest in their own rights as discussed in Sec.~\ref{part2-subS-LabQCD-Photo}.  While it is relatively easy to select a clean sample of incoherent events by requiring a breakup neutron in the Zero Degree Calorimeter (ZDC) or a charged fragment in the Roman Pots, the inverse is not true. The coherent spectra will thus be contaminated by a fraction of incoherent events that passed all cuts. Our purpose here is to assess the degree of suppression of the incoherent background required for the physics. Here we should emphasize, that the  ratio $\sigma_\mathrm{incoh}/\sigma_\mathrm{coh} (t)$ is strongly model dependent, and while it is kinematically accessible in UPC events at the LHC, models are currently not calibrated to LHC experimental data.

Figure~\ref{fig:coh_incoh_eAu_new} depicts the coherent (red) and incoherent (blue) cross section with the nominal resolution defined above in Sec.~\ref{sec:momres}.
In order to vary the level of suppression to define reliable requirements we 
construct first a template suppression curve $S(t)$ (green in Fig.~\ref{fig:coh_incoh_eAu_new}), by linear  interpolation  in the logarithm of the suppression factor, between  fix points at the minima  and leveling at larger $t$.  With this template at hand, we now can vary the suppression values (except at $t=0$) by a common factor, $c_s$, to study the effects of different background levels on the extraction. 
  
Figures \ref{fig:combo_incoherent_new}(a)-(c) depict 3 scenarios with (from left to right) increasing levels of suppression of the incoherent contribution.  The extraction
of $F(b_T)$ is surprisingly  robust in a considerable range around the nominal scenario shown in Fig.~ \ref{fig:coh_incoh_eAu_new} and  \ref{fig:combo_incoherent_new}(b). Further studies, not displayed here, show that significant distortions start to affect $F(b_T)$ for suppressions that are a factor 4 less than the nominal values and that they are fully negligible for suppression level of 4 times larger than nominal. Conducting the same studies for $\phi$ and $\rho$ production yields the same conclusions as for the $J/\psi$, in fact even slightly better.

   	 	  \begin{figure}[tbh!]
	\begin{center}
		\includegraphics[width=0.5\linewidth]{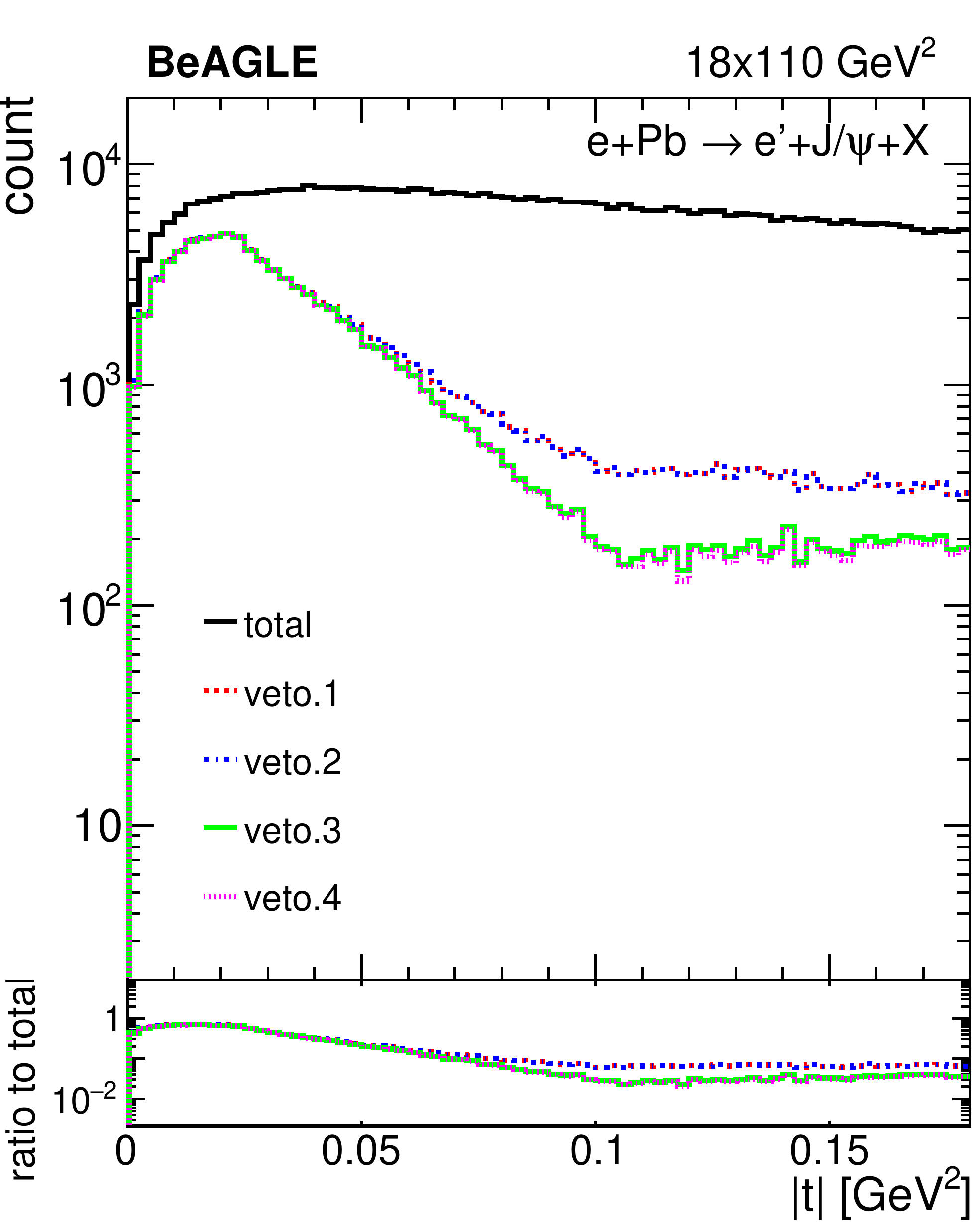}
		\caption{\label{fig:incohsuppression} 
		Suppression of incoherent background events by successive cuts: see text for explanation. 
		}
		\vspace{-3mm}
	\end{center}
\end{figure}
A separate study using the BeAGLE generator has been initiated to assess the veto inefficiency for incoherent events. Preliminary studies use the following  combination of cuts: 
\begin{description}
\item[Veto 1] no neutrons in the ZDC
\item[Veto 2] Veto 1 and no protons in the Roman Pots
\item[Veto 3] Veto 2 and no proton in the off-energy detector
\item[Veto 4] Veto 3 and no proton in the B0.
\end{description}
These cuts alone, as shown in Fig.~\ref{fig:incohsuppression}, are not yet enough to suppress the incoherent contribution to measure the diffractive pattern of the coherent contribution down to the required level. There is an additional prospect of improving the vetoing efficiency by detecting decay photons from the nuclear $\gamma$-decay, ideally  in both the ZDC and the B0. This additional rejection factor is, however, still in the process of being quantified.

\subsection{\texorpdfstring{$u$-channel}{u-channel} exclusive electroproduction of \texorpdfstring{$\pi^0$}{pi0}} 
\label{subsec:bckw_pi0}

\begin{figure}[htb]
\centering
\includegraphics[width=0.56\linewidth]{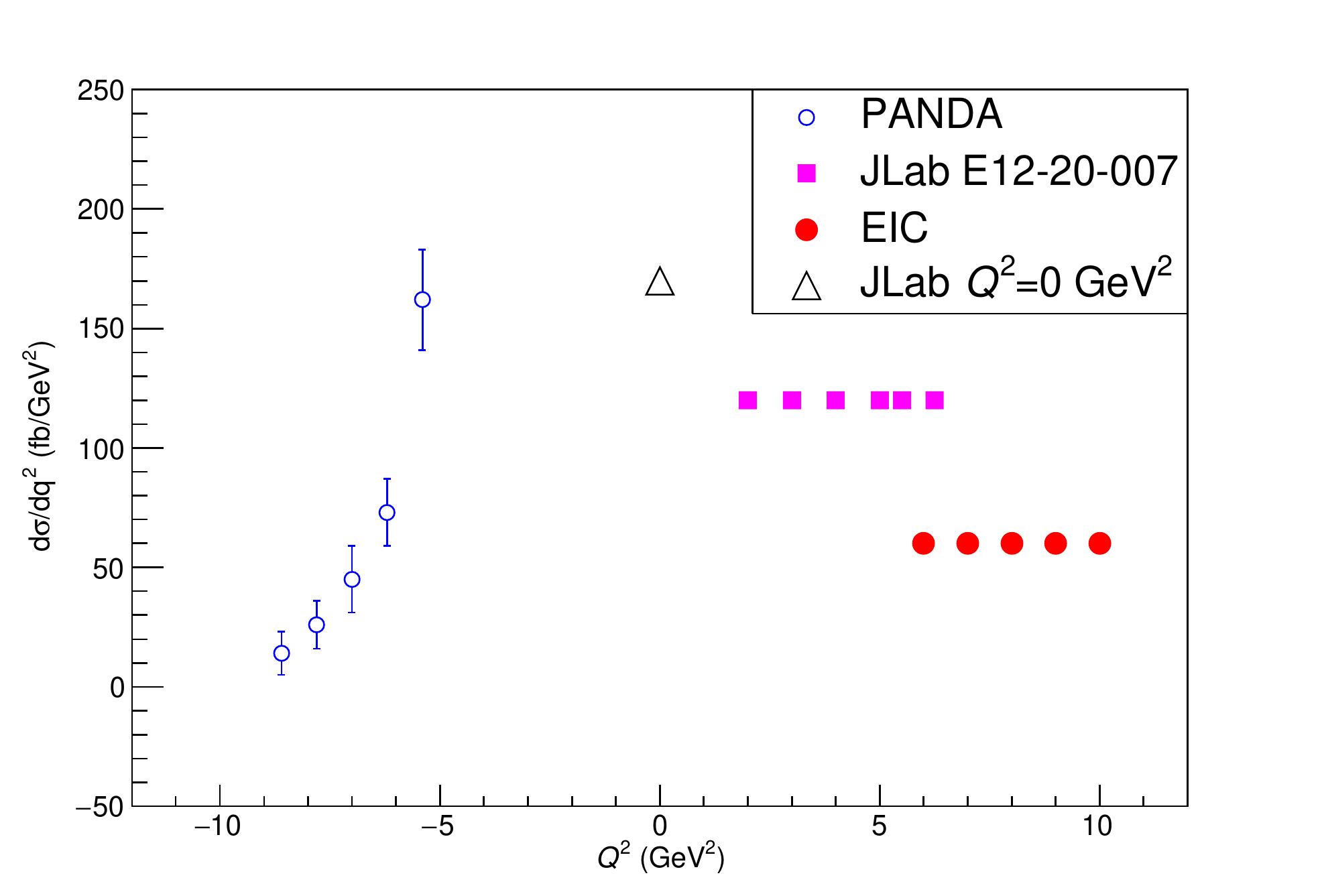}
\vspace{-0.2cm}
\caption{The anticipated global data set of $d\sigma/dq^2 (\gamma^*p \rightarrow p\pi^0)$ vs $Q^2$ at fixed $s=10$ GeV$^2$. The projected data points are: open blue circles, results  
 from PANDA (FAIR)~\cite{Singh:2014pfv}; 
 magenta squares,  results from the 
JLab E12-20-007 experiment~\cite{Li:2020nsk}; red full circles, results at EIC.  A potential JLab 12 measurement at real-photon point $Q^2 = 0$ GeV$^2$ is indicated by the open triangle. 
\label{fig:u_channel_pi0-chp8}}
\end{figure}

\begin{figure}[hbtp]
\center
\includegraphics[width=0.325\textwidth]{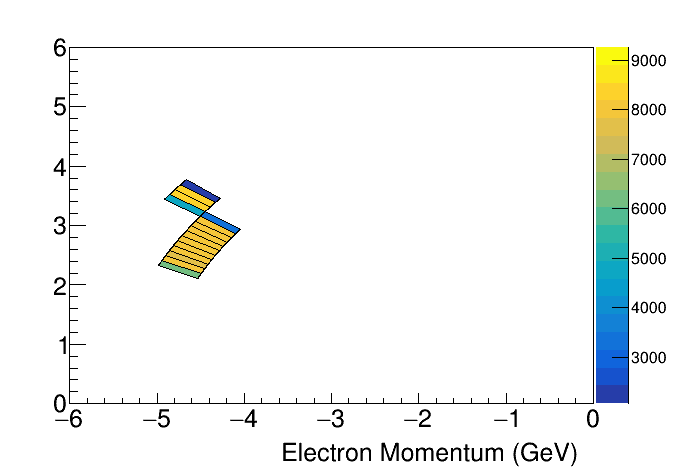}
\includegraphics[width=0.325\textwidth]{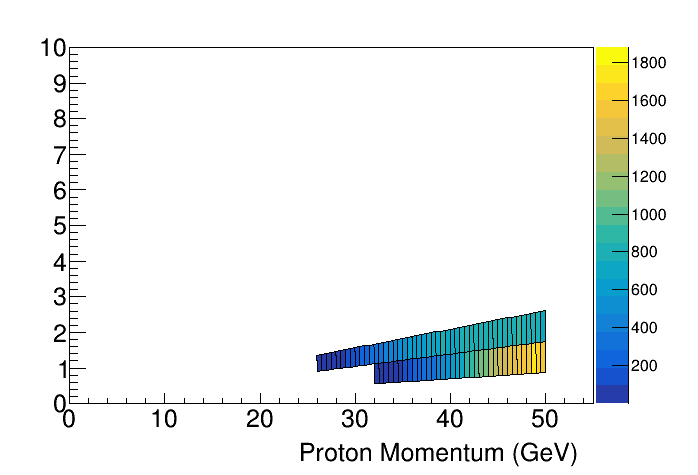}
\includegraphics[width=0.325\textwidth]{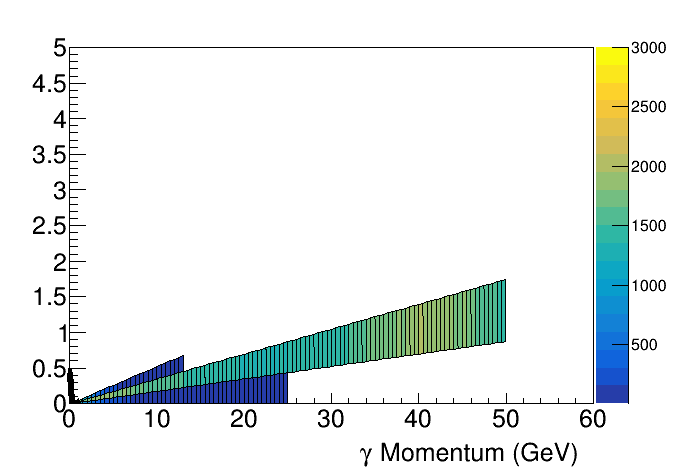}
\caption{Momentum distributions for scattered electrons (left), recoiled proton (middle) and decayed photons (right) in the exclusive $e+p\rightarrow e^\prime+ p^\prime + \pi^0$ process, for integrated luminosity of 10 fb$^{-1}$.}
   \label{Fig:u_chan_pi0_angular_mom-chp8}
\end{figure}

 The  backward ($u \sim u_{\textrm {min}}$) exclusive $\pi^0$ process $e+p\rightarrow e^\prime+ p^\prime + \pi^0$  can be studied at EIC in kinematics beyond the reach of existing facilities, providing new information on the transition distribution amplitudes (TDAs) introduced in Sect.~\ref{part2-subS-SecImaging-GPD3d}.
The Fig.~\ref{fig:u_channel_pi0-chp8}  illustrates 
the perspective of $Q^2$ evolution from 2 to 10 GeV$^2$ from the combination of planned measurements  at JLAB~\cite{Li:2020nsk}, PANDA~\cite{Singh:2014pfv} and EIC, at fixed $s=10$ GeV$^2$.
All the detection parameters required for the measurement at EIC with a 5 GeV electron beam colliding with a 100 GeV proton beam and 
 in the range from  $Q^2=6.2$ to $10.5$ GeV$^2$ are summarized 
in Table~\ref{tab:u-channel_kinematics-chp8} and Fig.~\ref{Fig:u_chan_pi0_angular_mom-chp8}.
For the experimental setup, we intend to use the ZDC to detect the decayed photons from $\pi^0$ with momentum from 40 to 60 GeV. 
Exploring also different collision energies, we found that a lower proton energy would produce a lower momentum $\pi^0$, and the decayed photon will not reach the ZDC due to acceptance.

 In  summary, the pseudorapidity and momentum  are, respectively,  $|\eta| \sim 4.1$ and $P_{p'} \sim 50$ GeV for the recoiled proton and   $|\eta| < 1.5$ and  $P_{e'}\sim 5.4$ GeV for the scattered electron, while the $\pi^0$ momentum is $P_{\pi^0} \sim 50$ GeV. 
 
According to the latest detector study, there may be problems to reach the pseudorapidity  values $|\eta| \ge 4$ at the Hadron End Cap.   In this case, a dedicated detector is required to tag the recoiled proton at $\eta \sim 4.1$. Otherwise, we should apply  the missing mass reconstruction technique  to resolve the proton.

\begin{table}[bth]
\centering\caption{The nominal values for the particle momentum and $\eta$ of scattered electrons, recoiled protons and produced $\pi^0$ in the exclusive  $e+p\rightarrow e^\prime+ p^\prime + \pi^0$ process. 
\label{tab:u-channel_kinematics-chp8}}
\begin{tabular}{ccccccccc}
\hline
$Q^2$& $\eta_{e^{\prime}}$ & $P_{e^\prime}$  & $\eta_{p^{\prime}} $ & $P_{p^{\prime}}$  & $\eta_{\pi^0}$& $P_{\pi^0}$  & $P_{\gamma}$  \\
(GeV$^2)$   &    & (GeV)            &           & (GeV) & & (GeV$) $   & (GeV$) $        \\ 
\hline
6.2 & -1.39 & 5.31 &  4.13 & 43.40 &  4.38  &  56.29 &   28.24 \\
7.0 & -1.32 & 5.35 &  4.09 & 45.50 &  4.38  &  54.12 &   27.06 \\
8.2 & -1.24 & 5.40 &  4.12 & 49.74 &  4.38  &  49.84 &   24.72  \\
9.3 & -1.19 & 5.46 &  4.09 & 51.90 &  4.38  &  47.60 &   23.80  \\
10.5& -1.12 & 5.52 &  4.07 & 54.96 &  4.38  &  44.50 &   22.25 \\
\hline
\end{tabular}

\end{table}

\subsection{Exclusive meson production by charged currents}
\label{subsec:charged_current}

\begin{figure}[htb]
\begin{center}
\includegraphics[width=0.7\columnwidth]{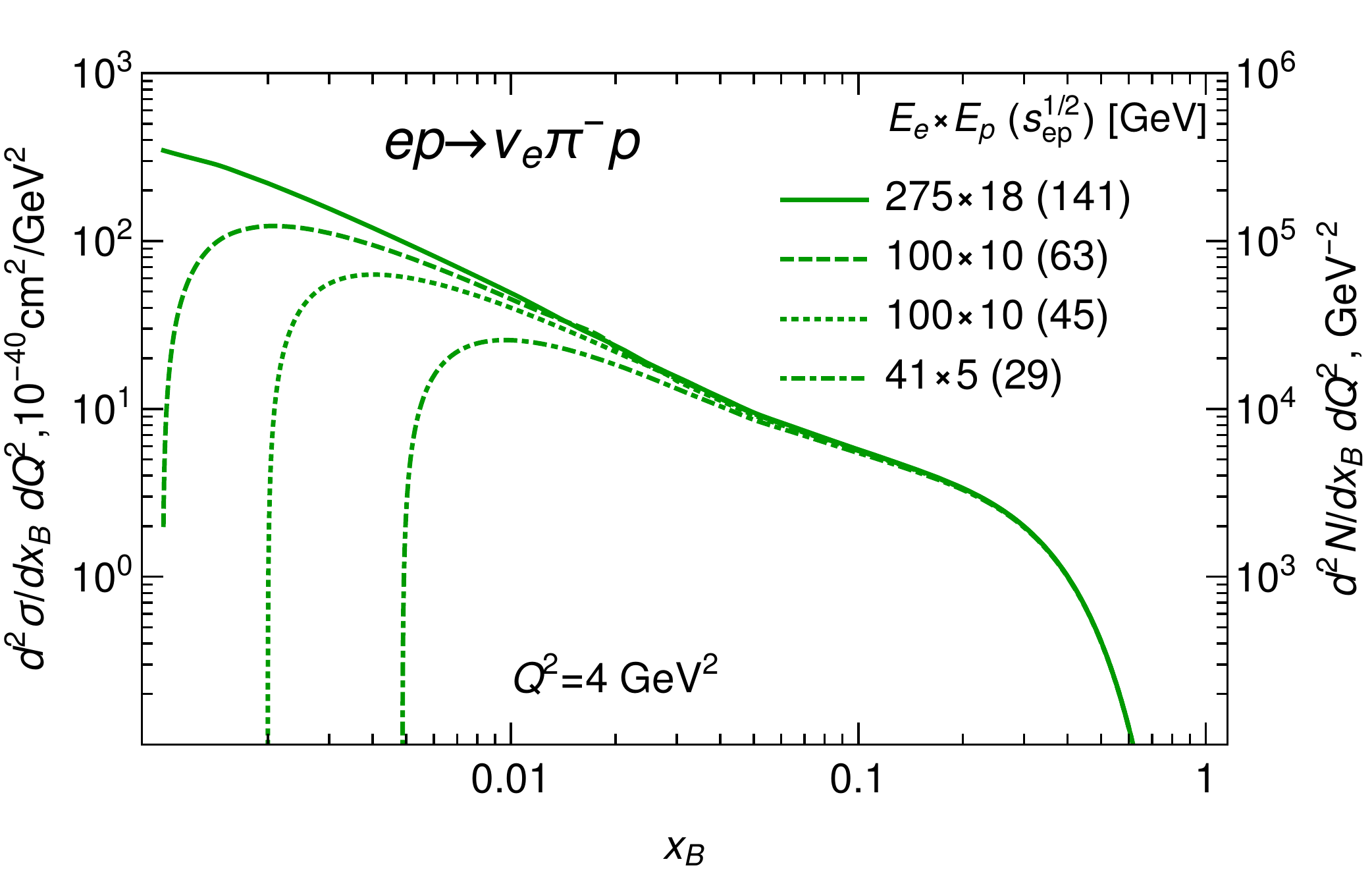} 
\end{center}
\caption{\label{fig:CCDVMP}The cross section of charged current pion production
in EIC kinematics evaluated in the framework of~\cite{Siddikov:2019ahb}. The right vertical axis is rescaled by the integrated luminosity  $\int dt\,\mathcal{L}=100\,{\rm fb}^{-1}$ to give
the expected number of events per bin as in Eq.~(\ref{eq:CCDVMP_N-chp8}). }
\end{figure}

The Charged Current (CC) DVMP processes
are  suppressed compared to photoproduction, yet are within
the reach of the Electron Ion Collider. 
Fig.~\ref{fig:CCDVMP} shows the results for the cross section of exclusive $\pi^{-}$production in
CCDVMP process \ep$\to\nu_{e}\,\pi^{-}p\,$ in the framework of~\cite{Siddikov:2019ahb,Pire:2017lfj},
as function of the  Bjorken variable $x_{B}=Q^{2}/(2p\cdot q)$, with 
 $Q^{2}=-q^{2}$ the virtuality of the charged boson and $p$ the incoming proton momentum.
 The details of the evaluation of the cross section and its relation to proton GPDs can be found in \cite{Siddikov:2019ahb}.
For the sake of convenience, we rescaled the right vertical axis of
the figure to show the values of the product
\begin{equation}
\frac{d^{2}N}{dx_{B}dQ^{2}}=\frac{d^{2}\sigma}{dx_{B}dQ^{2}}\times\int dt\,\mathcal{L}\,\label{eq:CCDVMP_N-chp8}
\end{equation}
which facilitates estimates of the expected number of events per bin
(we used an integrated luminosity $\int dt\,\mathcal{L}=100\,{\rm fb}^{-1}$
for estimates). For other channels (like exclusive CC production of
strangeness or charm) we expect that the cross section is of
the same order of magnitude. For nuclei we expect that the exclusive
cross section should scale with atomic number as $\sim Z^{4}$ (modulo
a factor of $\sim$2 due to model-dependent nuclear corrections).

However, there are certain challenges which must be addressed in order
to make possible measurements of charged current exclusive processes,
ensure their exclusiveness, as well as suppress backgrounds from quasi-real
photoproduction.\emph{ }For nuclei additional backgrounds stem from
the subprocess on neutrons. In the following, 
we will focus on the \ep$\to\nu_{e}\,\pi^{-}p\,$ channel, yet many
challenges are common to all other final states.

\subsubsection{Kinematics reconstruction}
The CCDVMP processes contain an undetected neutrino in the final state, and
no recoil electron which is conventionally used as a trigger in experimental
setup. For this reason, there are certain challenges for reconstruction
of the kinematics of such processes. Since in the \ep$\to\nu_{e}\,\pi^{-}p\,$ channel
both the final state hadrons ($\pi^{-}$ and $p$) are charged, their kinematics might be reconstructed with good precision.
From energy-momentum conservation,  the neutrino momentum is
\begin{equation}
p_{\nu_{e}}=p_{e}+p_{i}-p_{f}-p_{\pi},
\end{equation}
where $p_{i}\equiv(E_{i},\,\boldsymbol{p}_{i})$ and $p_{f}\equiv(E_{f},\,\boldsymbol{p}_{f})$
are, respectively, the initial and final four-momenta of the nucleon, $p_{e}$ is the four-momentum
of the incident electron, and $p_{\pi}$ is the four-momentum of the
produced pion. If we detect the four-momenta of recoil proton and produced
pion, we may reconstruct the kinematics of the process. For the Bjorken
variables $Q^{2}$, $x_{B}$ and $t$, in the massless limit  ($m_{N}\approx m_{\pi}\approx0$) valid 
 in high energy kinematics, we have
\begin{align}
Q^{2} & \approx4E_{e}\left(E_{p}-p_{f}\cos^{2}\left(\frac{\theta_{f}}{2}\right)-p_{\pi}\cos^{2}\left(\frac{\theta_{\pi}}{2}\right)\right),\\
x_{B} & \approx \frac{2E_{e}}{E_{p}}\left(1+\frac{E_{p}-p_{f}-p_{\pi}}{p_{f}\sin^{2}\left(\frac{\theta_{f}}{2}\right)+p_{\pi}\sin^{2}\left(\frac{\theta_{\pi}}{2}\right)}\right),\nonumber \\
t & \approx-4E_{p}p_{f}\sin^{2}\left(\frac{\theta_{f}}{2}\right)\label{eq:tExp},
\end{align}
where we used the notation $(p_{i},\,\theta_{i})$, with  $i\equiv\pi,f$, for the absolute value
of the three-momentum and polar angle (w.r.t. incident proton) of the
pion and recoil proton. Since the cross section is
exponentially suppressed as a function of $t$, from~Eq.~(\ref{eq:tExp})
we may expect that the dominant contribution comes from configurations
with small scattering angle $\theta_{f}$ of recoil proton and small
momenta $p_{f}$. 

\begin{figure}[thp]
\begin{center}
\includegraphics[width=0.48\textwidth]{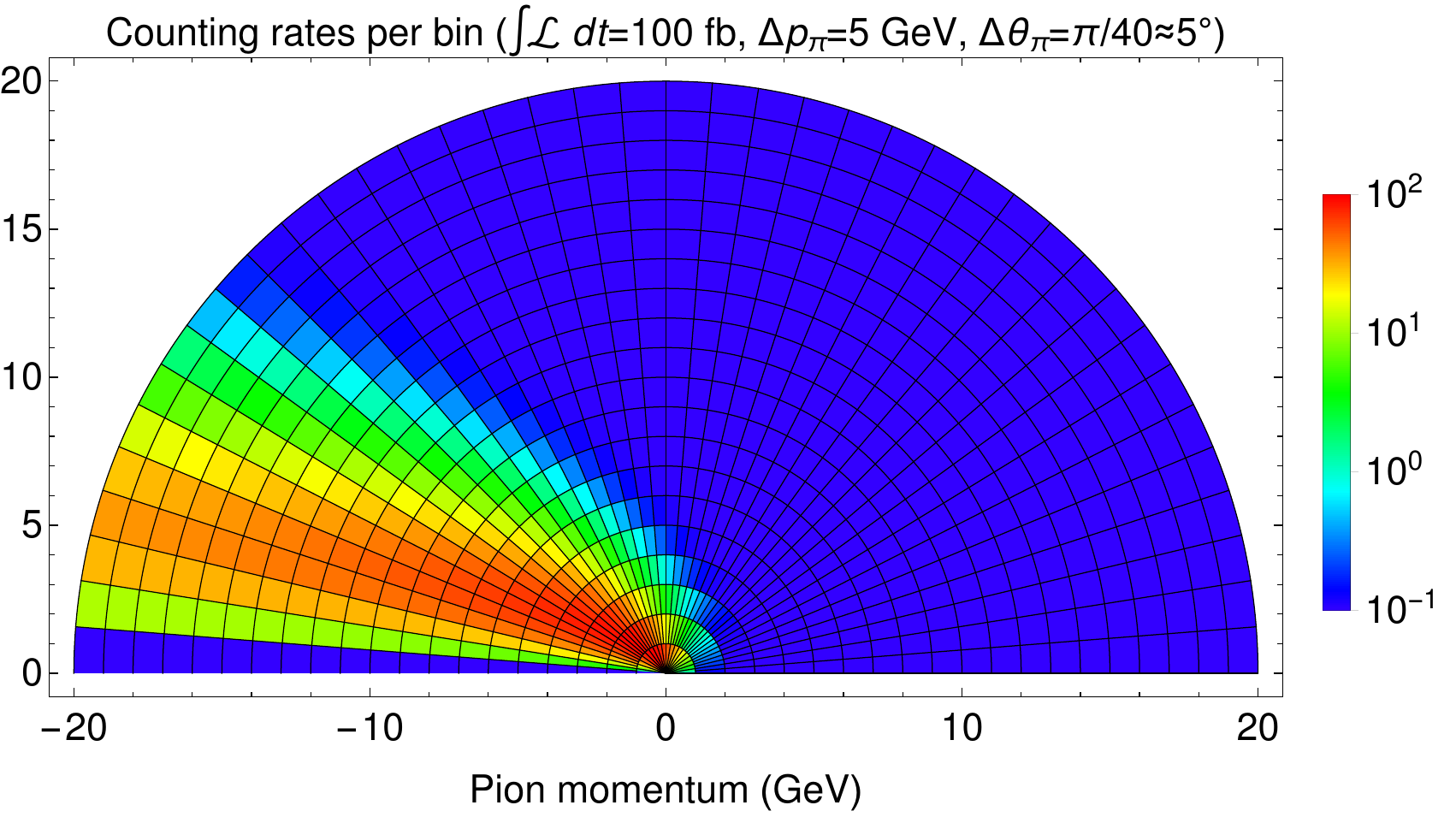}\includegraphics[width=0.48\textwidth]{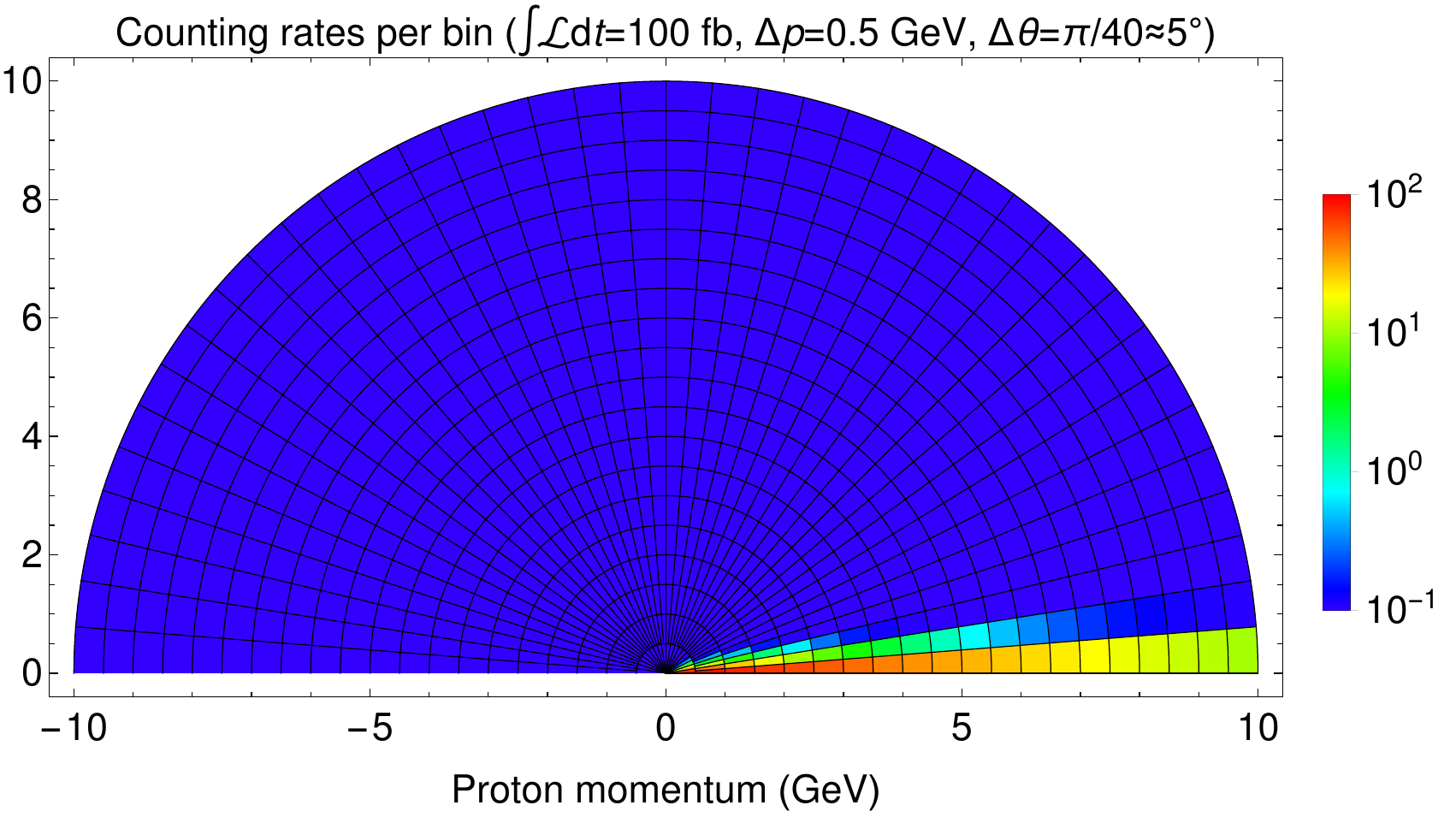}
\end{center}
\caption{\label{fig:Angular} 
Kinematic plots for the \ep$\rightarrow \nu_e\,p\,\pi^-$ process. Left plot: Angular distributions of produced
pions (integrated over the proton momentum in the kinematically allowed domain). Right plot: Angular distributions of recoil protons (integrated
over the pion momentum in the kinematically allowed domain). The number of
events was estimated assuming integrated luminosity $\int\mathcal{L}dt=100\,{\rm fb}^{-1}$
and the size of the bins $\Delta p_{i}\times\Delta\theta_{i}$ given
in the title of each Figure. In both cases, the angle $\theta$ is measured with
respect to direction of incident proton beam. }
\end{figure}

In Figure~\ref{fig:Angular} we show the angular distributions
of $\pi^{-}$ and recoil proton (integrated over the momenta
of the spectator particles in the kinematically allowed domain). As
expected, the recoil protons predominantly scatter in the forward
direction, whereas pions are produced mostly in the backward direction.
It is expected that the dominant contribution will come from the pions
with momenta $\lesssim20$~GeV and protons with momenta $\lesssim10$~GeV.

\begin{figure}[thb]
\begin{center}
\includegraphics[width=0.95\textwidth]{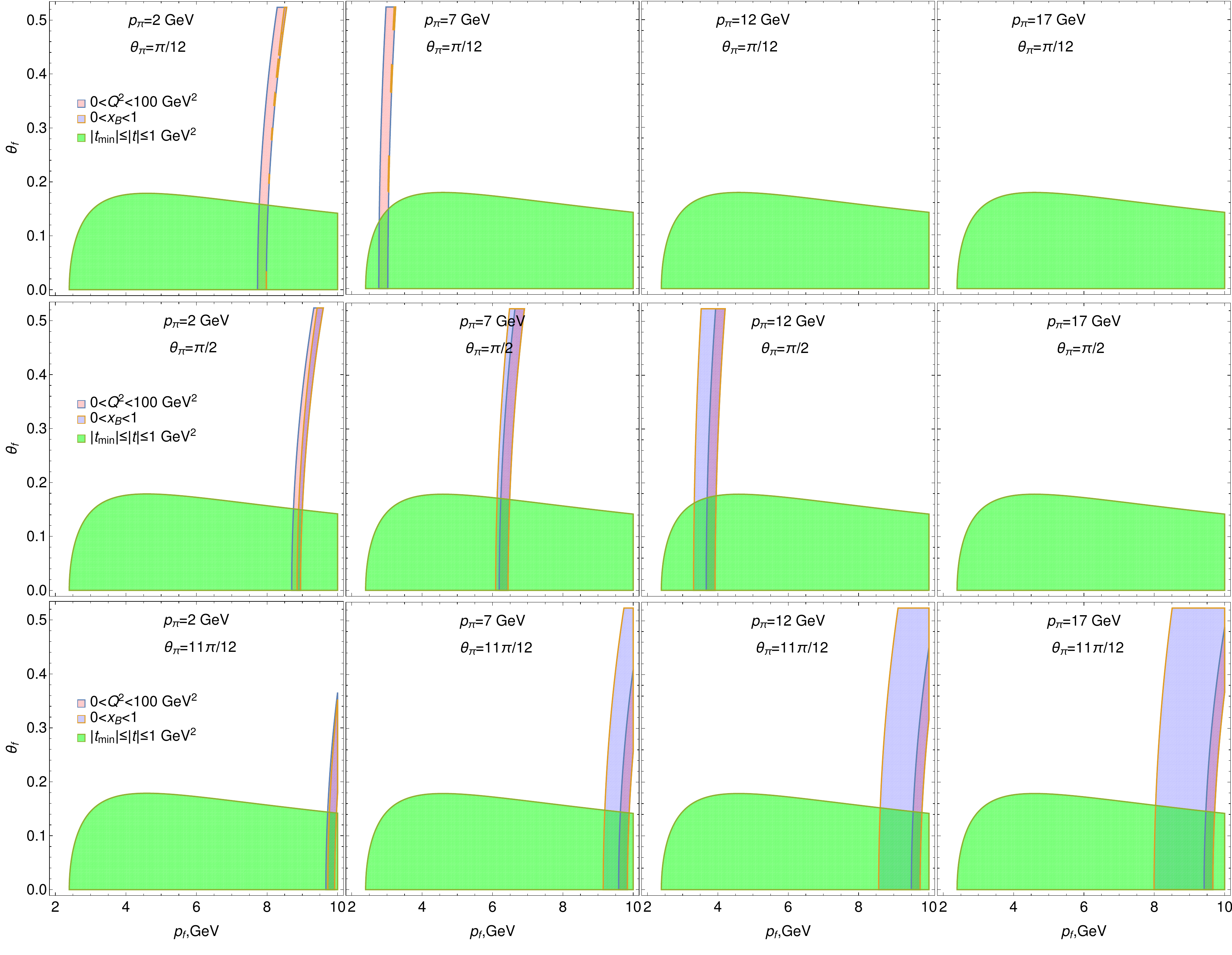}
\end{center}
\caption{\label{fig:Angular-1}Kinematic restrictions in the \ep$\rightarrow \nu_e\,p\,\pi^-$ process on the  momentum $p_{f}$
and scattering angle $\theta_{f}$ of the recoil proton in the case when the three-momentum
of the pion is fixed. The blue band corresponds to the physical region
$0<x_{B}<1$. The intersection of the subdomains $0<Q^{2}\lesssim100\,{\rm GeV}^{2}$
(pink band) and $|t|\lesssim1\,{\rm GeV}^{2}$ (green area) gives
the dominant contribution to the DVMP cross section, then limiting significantly the momentum of the proton. The angles
$\theta_{f},\,\theta_{\pi}$ are measured w.r.t. direction
of the incident proton beam (the first, second and third rows correspond
to a pion produced in forward, central and backward directions, respectively).}
\end{figure}

The distributions shown in the Fig.~\ref{fig:Angular} do not reflect
the kinematic constraints which impose limits on possible mutual variations
of the momenta of the particles. In order to illustrate such constraints,
in Fig.~\ref{fig:Angular-1} we show that the three-momentum of
the recoil proton may change in a very limited range when the momentum
of the produced pion is fixed. 

In summary, it is possible
to reconstruct the kinematics of the process using only the momenta
of the pion and recoil proton.

\subsubsection{Photoproduction backgrounds}
We expect that sizeable backgrounds to the charged current processes might
come from the processes mediated by virtual photon, \ep$\to e\pi^{-}p\,X$,
where the scattered electron and some remnants remain undetected.
The cross sections of photon-mediated processes are enhanced by the kinematic
factor $\sim(Q^{2}+M_{W}^{2})^{2}/Q^{4}$ compared to the charged current
channel, and for the values of $Q^{2}$ available at EIC the quasi-real
photoproduction will exceed by several orders of magnitude the contributions
of charged current exclusive processes. For this reason, the possibility
to measure the charged current processes will depend crucially on the
possibility to reject events which contain anything except $\pi^{-}$
and $p$ in the final state. Detectors with $4\pi$ coverage are needed
for this purpose. Since this is experimentally very challenging,
 we suggest some additional checks (``cutoffs'')
which might be used in order to ensure there is no undetected remnants
of the photoproduction processes.
\begin{itemize}
\item From charge conservation, we deduce that the photoproduction of $\pi^{-}p$
should include at least a charged pion, e.g. via the subprocess the $e\,p\to e\,\pi^{-}\pi^{+}p$.
 In order to ensure that such backgrounds are missing,
we suggest to check that the value of the missing mass  (the mass of the  undetected
neutrino) is below the  threshold of the mass of a pion, i.e. 
\begin{equation}
m_{\nu_{e}}^{2}\equiv p_{\nu}^{2} =\left(p_{e}+p_{i}-p_{f}-p_{\pi}\right)^{2}<\,m_{\pi}^{2}.\label{eq:cut_M2}
\end{equation}
Such a cutoff is sufficient to exclude contributions of any  process with photoproduction of additional (undetected) hadrons. However, the strict implementation
of the cut defined by Eq.~(\ref{eq:cut_M2}) is experimentally challenging: the quantities in the
l.h.s. of Eq.~(\ref{eq:cut_M2}) are of order of dozens of GeV, for this
reason the cut requires measurements of the momenta
of the scattered particles with outstanding precision (with a relative error
$\lesssim10^{-5}$). Yet we still believe that even a relaxed cut~(\ref{eq:cut_M2}),
with a higher upper limit, might be useful to improve signal/noise ratio.
\item Another important background comes from quasi-elastic scattering \ep$\to$\ep,
which might give an  important contribution due to the misidentification of
electrons as pions in the EIC detectors. For this reason, we suggest to
use an additional missing energy cutoff 
\begin{equation}
\Delta E=E_{e}+E_{i}-E_{(\pi/e')}-E_{f}\gtrsim0.5\,{\rm GeV}\label{cut_E},
\end{equation}
where $E_{(\pi/e')}$ is the energy of the final pion or misidentified
electron produced in the collision. Imposing this cut, it will completely
eliminate the quasi-elastic background. 
\item In case of nuclear targets, there are additional contributions from
photoproduction processes on neutrons, e.g., the  $en\to e\pi^{-}p$ subprocess
which will give the dominant contribution, both for inclusive and exclusive
production\footnote{In case of exclusive production on heavy nuclei we expect that CCDVMP
will be overshadowed by photoproduction of long-living nuclides via
$en\to e\pi^{-}p$ subprocess, e.g. $e\,^{63}{\rm Cu}\to e\,\pi^{-}\,^{63}{\rm Zn}$.
Experimentally it is very challenging to distinguish its final state
from that of CCDVMP process \eA$\to\nu\pi^{-}A$.}. For this reason, we find that nuclear targets cannot be used for
CCDVMP studies. 
\end{itemize}
To summarize, we believe that the CCDVMP processes have sufficiently large
cross sections to be measured at EIC kinematics.  However, there are
huge backgrounds from photoproduction processes. The ability to exclude these backgrounds will be crucial  if charged current exclusive processes are to be studied at  the EIC. We checked that a combination of the cuts~(\ref{eq:cut_M2}) and (\ref{cut_E}) allows us to get rid of different photon-induced background processes, though the cut (\ref{eq:cut_M2}) might be difficult to implement.

\subsection{Diffractive dijets}
\label{subsec:diff_dijets}

\begin{figure}[htbp]
\includegraphics[width=0.49\textwidth]{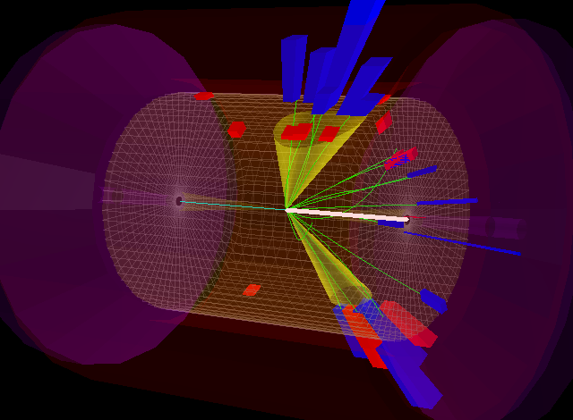}
\includegraphics[width=0.49\columnwidth]{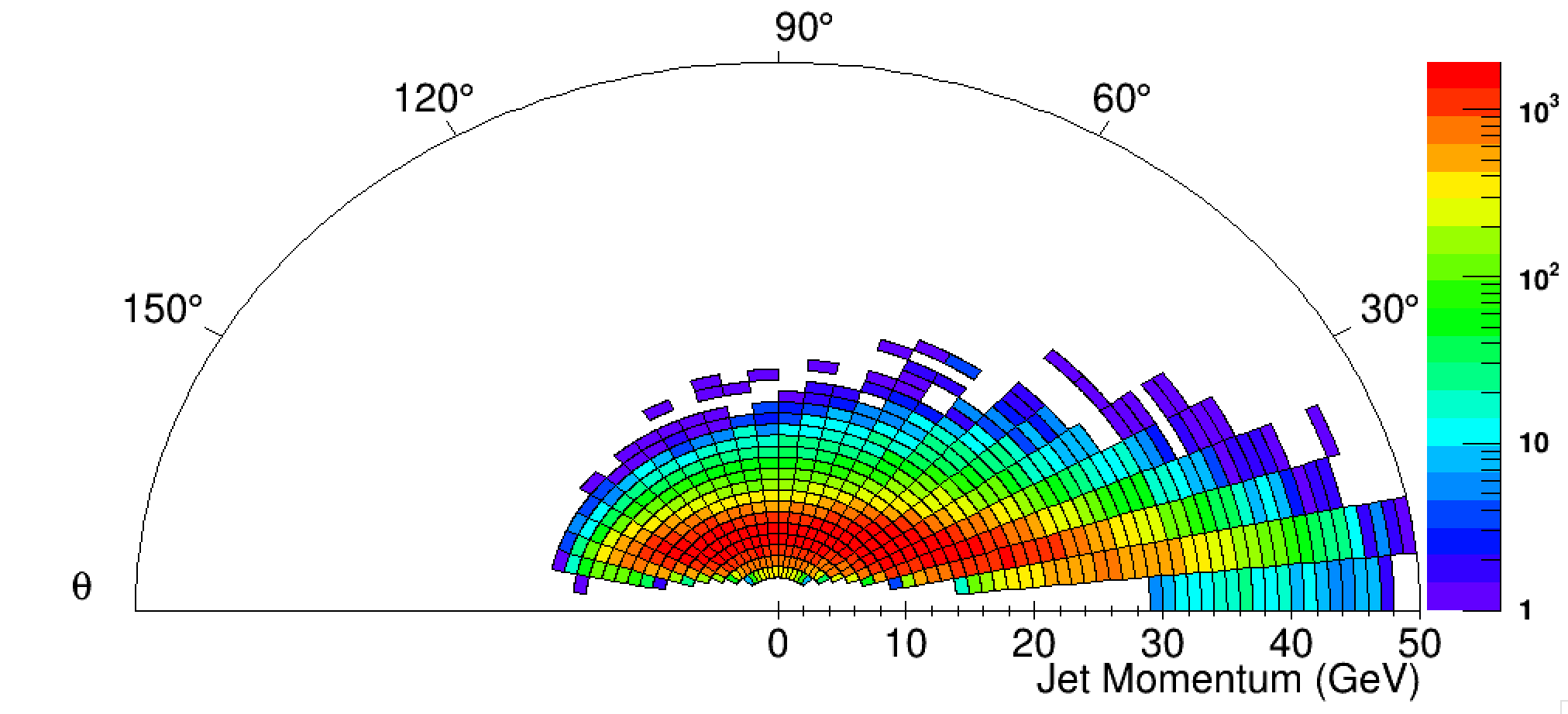}

\caption{Left: event display of diffractive dijet event at 140~GeV center-of-mass energy.  PYTHIA~8.244 simulation with 18$\times$275 GeV collision energy.
Right: Distribution of jets in angle and momentum.
}
 \label{fig:diffractivejets}
\end{figure}

\begin{figure}[ht]
\centering
\includegraphics[width=0.9\columnwidth]{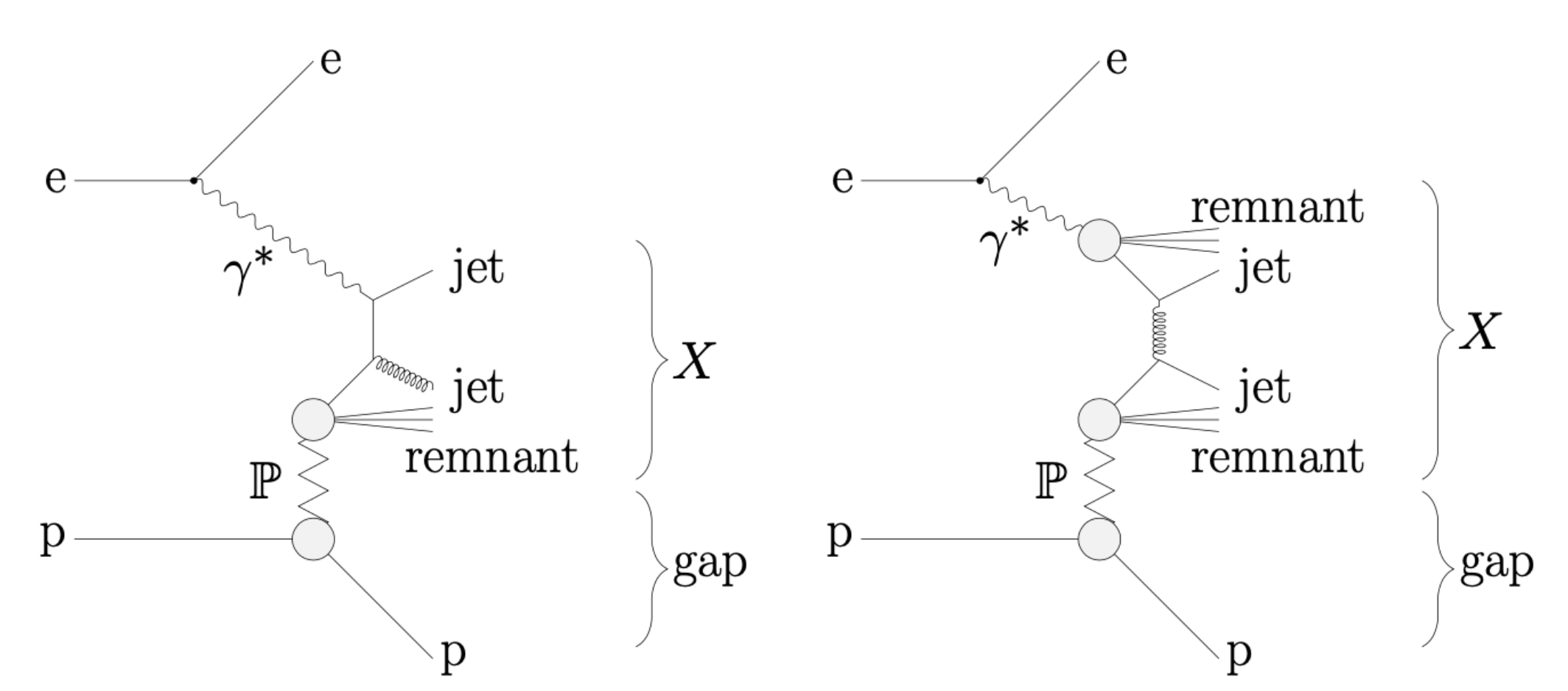}
\caption{Leading order Feynman diagrams for diffractive dijet photoproduction in ep collisions. In the left part, the photon participates directly in the hard scattering matrix element. In the right part,  a parton from the resolved photon participates.}
\label{fig:dijetPhotoproduction}
\end{figure}
In this section we focus on the photo-produced diffractive dijets in ep collisions. A typical event display is shown in Fig.~\ref{fig:diffractivejets}, showing a back-to-back dijet separated by a rapidity gap from the beam.  The initial state consists of an electron and a proton, with the former radiating off a (virtual) photon. If the photon is highly virtual, we are in the range of deep inelastic scattering (DIS) while a photon with low enough virtuality can be considered (quasi-)real. This is the photoproduction regime. No clear distinction between the two regimes exists, however, and photons of intermediate virtuality require careful consideration to avoid double-counting. A special feature in the photoproduction regime is that the  process can be separated into different ``resolved'' and ``unresolved'' contributions~\cite{Helenius:2019gbd} depending on whether the partonic structure of the photon is resolved or whether it directly participates in the hard subprocess, see Fig.~\ref{fig:dijetPhotoproduction}. These resolved photons open up for all possible hadron-hadron processes, including diffractive ones~\cite{Helenius:2019gbd}.

Several complementary experimental methods have been developed to identify diffractive events in \ep collisions. Each method exploits a specific signature characteristic of diffraction. The diffractive events can be directly detected by means of a forward spectrometer~\cite{Adloff:1998yg,Derrick:1995tw}. Because of the low $t$ of the process, the outgoing p or nucleus is scattered at very low angles with respect to the initial direction and one needs to place the spectrometer very far from the interaction point and very close to the beam axis. The other common technique to tag on diffraction is to require a "rapidity gap" in the detector. This means that there is a region in the detector from the hadron beam towards the center of the detector in which there is no activity from the hadronic final state \cite{Derrick:1993xh,Derrick:1995tw,Ahmed:1994nw,Aktas:2006hy}. The efficiency for detecting, and the purity of, diffractive events therefore depends strongly on the rapidity coverage of the detector. 

\begin{figure}[tbh!]
\centering
\includegraphics[width=1\columnwidth]{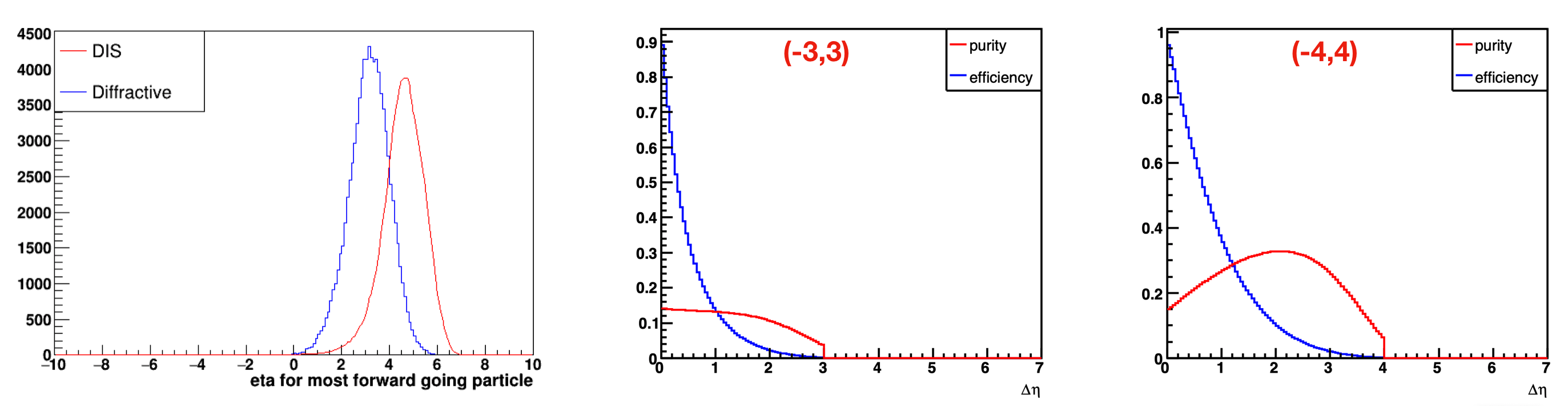}
\caption{Left figure: The $\eta$ distribution of the most forward particle  in the event, for both inclusive DIS and diffractive event samples; The middle and right  figures: the efficiency and purity distribution in ep collisions for different $\eta$ coverage (-3,3) and (-4,4). Here we assume the inclusive DIS to diffractive cross section ratio is 7:1. }
\label{fig:purity_efficiency_eta}
\end{figure}

Experimentally one measures the pseudorapidity of the most forward particle in the detector ($\eta_{max}$) and requires that the $\eta$ range between it and the edge of the forward detector instrumentation is large enough. The diffractive events concentrate therefore at low values of $\eta_{max}$, corresponding to large values of $\Delta\eta$ (The $\eta$ gap between the most forward particle of the event and the edge of the forward detector instrumentation). The large rapidity gap method has the advantage of a much higher statistics compared to the forward spectrometer method. The model for photoproduced diffractive dijets for 18$\times$275~GeV beam energy presented here is based on the general-purpose event generator PYTHIA~8, and the DIS events are simulated by PYTHIA~6. We select $Q^2<1$~GeV${}^2$ events. The $\pT$ cut for the leading jet is 5~GeV and for the associated jet is 4~GeV. The jet kinematic distribution in diffractive events is shown in Fig.~\ref{fig:diffractivejets} (right). We assume the ratio of inclusive DIS events and diffractive events is 7:1. In the left figure in Fig.~\ref{fig:purity_efficiency_eta}, we show the $\eta$ distribution of the most forward particle in the event, for both inclusive DIS and diffractive event samples. So we can obtain the purity and efficiency distribution as shown in Fig.~\ref{fig:purity_efficiency_eta}. The larger $\eta$ coverage would give us better purity of diffractive events.

\subsection{Summary}
Limitations of the nominal detector design identified through the study of exclusive processes outlined in this section are summarised below. The required constraints are presented in Tables~\ref{tab:Excl_summary_tracking} - \ref{tab:Excl_summary_PID}.

\begin{description}
\item[Acceptance in the backward region] The lowest edge of the angular acceptance of the electron endcap in the central detector is critical for most of the exclusive processes studied. At the highest collision energy, the nominal detector edge at $\eta = -3.5$ cuts into the distributions in DVCS, the exclusive production of $\pi^0$s, vector mesons and diffractive jets. In DVCS, the electron distribution peaks at $\eta = -3.6$. Cutting at $\eta > -3.5$ results in a loss of efficiency of 14\% (due to electron acceptance) and 17\% (due to electron and photon acceptance) for DVCS and 11\%/12\%, (due to electron/electron and photon acceptance), for $\pi^0$-production. In the case of DVCS off $^4$He, the loss is estimated at 20\%. This cuts into the lowest-$x_B$ reach. The loss can be recovered by extending $\eta > -3.7$.
For the case of vector meson production, exclusive photoproduction depends entirely on extending acceptance beyond $\eta = -3.5$: either with the use of a low-$Q^2$ tagger or with far-backward detectors beyond the electron endcap. This is particularly important for $\Upsilon$ photoproduction near threshold. Extending the coverage to $\eta > -4$ additionally increases the purity and efficiency for difractive jet reconstruction. 
\item[Acceptance of the Far-Forward detectors] At low $x_B$, the physical $t_{min}$ for DVCS in $^4$He cannot be reached by the detectors, therefore detector acceptances directly define the minimal $t$ which can be experimentally accessed, which translates into the uncertainty in quark density profiles. At the high $t$ end, the limitation comes from luminosity. The $t$-acceptance of the nominal detector is sufficient to map out the first minimum but a lower $\pT$ reach of 200 MeV is critical and cannot be degraded. For the case of \eA collisions, suppression of the incoherent background up to the necessary third minimum in $t$ cannot be achieved with the cuts studied, but may be possible with a veto based on detection of nuclear decay photons in ZDC and B0. The $u$-channel exclusive electroproduction of $\pi^0$ also relies on proton detection at $\eta \sim 4.1$ and a detection of the $\pi^0$ decay photons with momenta 40 - 60 GeV in the ZDC. For the lower proton beam energies, acceptance reaching lower angles than the ZDC is necessary to detect the decay photons. Extending the coverage to $\eta < 4$ also increases the purity and efficiency for difractive jet reconstruction.    
\item[Muon detection] The ability to detect muons in the central detector would be of great benefit for vector meson production and TCS. This would not only double statistics, but help to suppress backgrounds and improve the resolution in $t$ due to the smaller impact of radiative effects.
\item[Tracking resolution] Good momentum resolutions in the central detector are crucial for vector-meson production in \eA collisions as they have a direct effect on the $t$-resolution. The minimum requirements are listed in table~\ref{tab:Excl_summary_tracking}. Momentum resolution also plays a critical role in charged current meson production, where it is needed to suppress photoproduction backgrounds.   
\end{description}

\begin{table}[ht]
    \scriptsize
    \renewcommand\theadfont{}
    \centering
    \begin{tabular}{c c c c c c}
    \hline\hline
pseudorapidity & \thead{tracking\\ resolution} & \thead{vertex \\ resolution} & \thead{material \\ budget} & detector & comments \\ \hline
-6.9 -- -5.8 & $\sigma_{\theta}/\theta = 1.5\%$ &    &    &  low-$Q^2$ tagger & \thead{$10^{-6} < Q^2 < 10^{-2}$ \\ GeV$^2$}\\
\hline
-4.5 -- -3.5 &  &    &    &  \thead{instrumentation \\ to separate $\gamma$ and \\ charged particles} & \thead{need coverage for \\ DVMP at highest \\ energy settings} \\
\hline
-3.5 -- -2.0 & \thead{$\sigma_{\pT}/\pT \sim$ \\ $0.1\pT + 0.5\%$} & TBD & $X/X_0 \leq 5\%$ & electron endcap & \\
-2.0 -- -1.0 & \thead{$\sigma_{\pT}/\pT \sim$ \\ $0.05\pT + 0.5\%$} & TBD & $X/X_0 \leq 5\%$ & electron endcap & \\
\hline
-1.0 -- 1.0 & \thead{$\sigma_{\pT}/\pT \sim$ \\ $0.05\pT + 0.5\%$} & $\sigma_{xyz} \sim 20 \mu m$ & $X/X_0 \leq 5\%$ & barrel & \\
\hline
1.0 -- 2.5 & \thead{$\sigma_{\pT}/\pT \sim$ \\ $0.05\pT + 1\%$} & TBD & $X/X_0 \leq 5\%$ & hadron endcap & \\
2.5 -- 3.5 & \thead{$\sigma_{\pT}/\pT \sim$ \\ $0.1\pT + 2\%$} & TBD & $X/X_0 \leq 5\%$ & hadron endcap & \\
\hline
3.5 -- 4.0 &  &    &    &  \thead{instrumentation \\ to separate $\gamma$ and \\ charged particles} & \thead{$\pi/K$ minimum $\pT$ \\ (see D+T section)} \\
\hline
$>  6.2$ & $\sigma_t/t < 1\%$ &    &    &  \thead{proton \\ spectrometer} & \thead{$0.2 < \pT < 1.2$ GeV \\ for protons, TBD \\ for light ions} \\

 \hline\hline
    \end{tabular}
    \caption{Summary of tracking constraints from exclusive processes.}
    \label{tab:Excl_summary_tracking}
\end{table}

\begin{table}[ht]
    \footnotesize
    \renewcommand\theadfont{}
    \centering
    \begin{tabular}{c c c c c}
    \hline\hline
pseudorapidity & \thead{ECal \\ energy resolution \\ $\sigma_E/E$} & PID in ECal & \thead{HCal \\ energy resolution \\ $\sigma_E/E$} & detector \\ \hline
-4.5 -- -4.0 & $2\%/\sqrt{E}$ &  &  & \thead{instrumentation \\ to separate $\gamma$ and \\ charged particles} \\
\hline
-4.0 -- -3.5 & $2\%/\sqrt{E}$ &  & \thead{$50\%/\sqrt{E} + 6\%$ \\ for di-jet studies}  & \thead{instrumentation \\ to separate $\gamma$ and \\ charged particles} \\
\hline
-3.5 -- -2.0 & $2\%/\sqrt{E}$ & \thead{$\pi$ suppression \\ up to 1:104} & \thead{$50\%/\sqrt{E}$ \\ constant term TBD} & electron endcap\\
\hline
-2.0 -- -1.0 & $7\%/\sqrt{E}$ & \thead{$\pi$ suppression \\ up to 1:104}  & \thead{$50\%/\sqrt{E}$ \\ constant term TBD} & electron endcap\\
\hline
-1.0 -- 1.0 & $(10-12)\%/\sqrt{E}$ & \thead{$\pi$ suppression \\ up to 1:104} & \thead{HCal needed, \\ resolution TBD} & barrel\\
\hline
1.0 -- 3.5 & $(10-12)\%/\sqrt{E}$ & & \thead{$50\%/\sqrt{E}$ \\ constant term TBD} & hadron endcap\\
\hline
3.5 -- 4.0 & $(10-12)\%/\sqrt{E}$ & & \thead{$50\%/\sqrt{E} + 6\%$ \\ for di-jet studies} & \thead{instrumentation \\ to separate $\gamma$ and \\ charged particles} \\
\hline
4.0 -- 4.5 & $(10-12)\%/\sqrt{E}$ & & & \thead{instrumentation \\ to separate $\gamma$ and \\ charged particles} \\
\hline
$> 4.5$ & \thead{$4.5\%/\sqrt{E}$ \\ for $E_{\gamma} > 20$ GeV} & $\leq 3$ cm granularity & & \thead{neutral particle \\ detection} \\

 \hline\hline
    \end{tabular}
    \caption{Summary of electromagnetic and hadronic calorimeter constraints from exclusive processes.}
    \label{tab:Excl_summary_Cal}
\end{table}

\begin{table}[ht]
    \small
    \renewcommand\theadfont{}
    \centering
    \begin{tabular}{c c c c c}
    \hline\hline
pseudorapidity & momentum range & \thead{$\pi / K / p$ \\ separation}  & \thead{muon \\ detection} & detector \\ \hline
-4.0 -- -3.5 &  &  & \thead{required \\ for background \\ suppression and \\improved resolution} & \thead{instrumentation \\ to separate $\gamma$ and \\ charged particles} \\
\hline
-3.5 -- -1.0 & $\leq 7 $ GeV/$c$ & $\geq 3 \sigma$  & \thead{required \\ for background \\ suppression and \\improved resolution} & electron endcap \\
\hline
-1.0 -- 1.0 & $\leq 5 $ GeV/$c$ & $\geq 3 \sigma$  & \thead{required \\ for background \\ suppression and \\improved resolution} & barrel \\
\hline
1.0 -- 2.0 & $\leq 8 $ GeV/$c$ & $\geq 3 \sigma$  & & hadron endcap  \\
\hline
2.0 -- 3.0 & $\leq 20 $ GeV/$c$ & $\geq 3 \sigma$ & & hadron endcap  \\
\hline
3.0 -- 3.5 & $\leq 45 $ GeV/$c$ & $\geq 3 \sigma$ & & hadron endcap  \\
 \hline\hline
    \end{tabular}
    \caption{Summary of $\pi / K / p$ separation and muon detection constraints from exclusive processes.}
    \label{tab:Excl_summary_PID}
\end{table}

\section{Diffractive Measurements and Tagging}
\label{part2-sec-DetReq.Diff.Tag}

\subsection{Requirements for exclusive vector meson production}
\label{diff-excl-mesons}

Although exclusive vector meson production, $\ep/A\rightarrow e+V+X$ is a simple reaction, with a final state that is typically the scattered electron, two charged mesons or leptons from a vector meson decay, and, for incoherent photoproduction, the products of nuclear breakup, it does impose some significant requirements on the detector.  Here, we discuss requirements related to pseudorapidity coverage of tracking detectors, ability to track soft kaons, momentum resolution and detecting nuclear breakup.  More details are given in Ref.~\cite{Arrington:2021yeb}.  

\begin{figure}
\centering
\includegraphics*[width=0.5\textwidth]{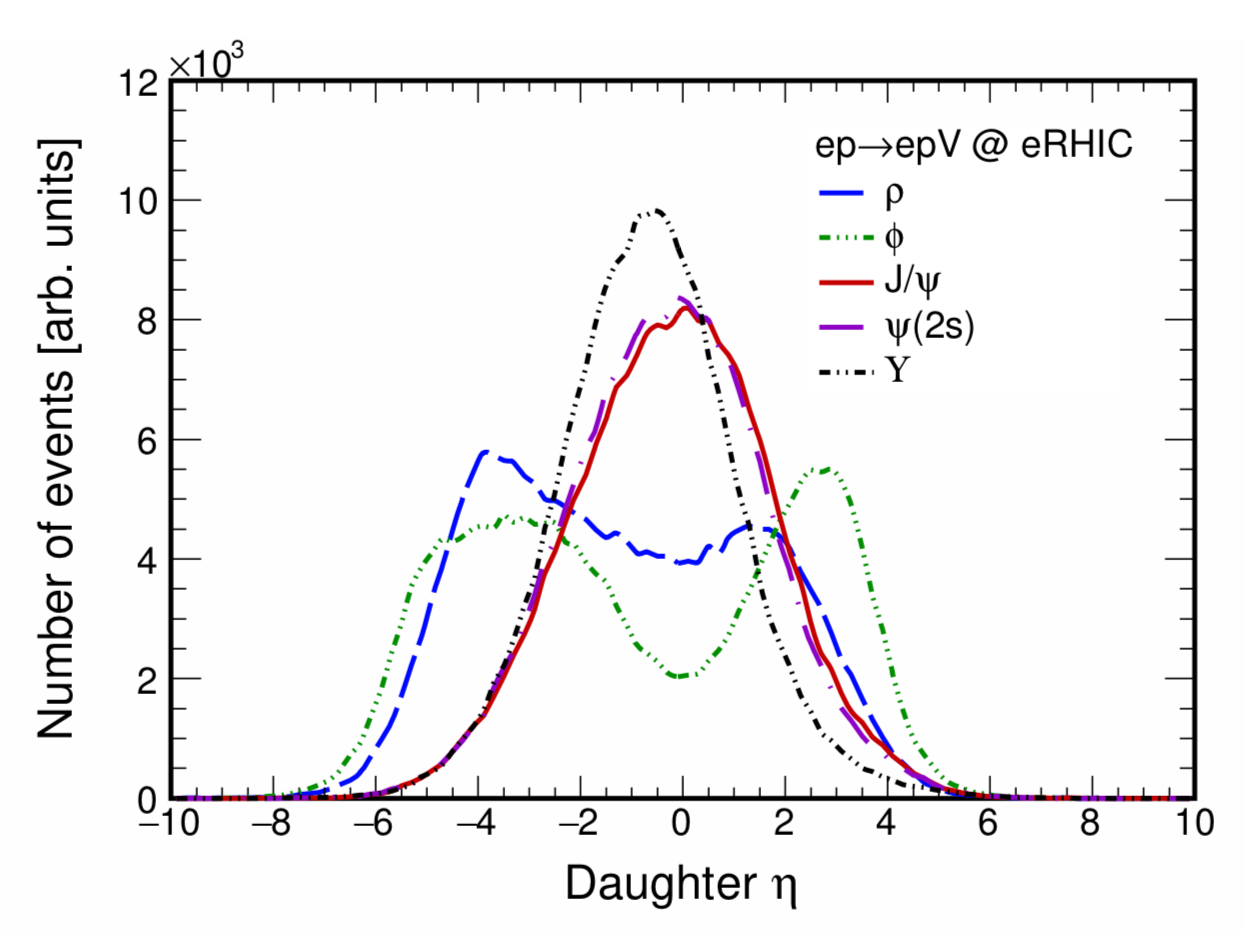}%
\caption{The pseudorapidity distribution for the daughter particles from the decay of different vector mesons at the EIC: $\rho\rightarrow\pi^+\pi^-$, $\phi\rightarrow K^+K^-$, $J/\psi\rightarrow e^+e^-$, $\psi'\rightarrow e^+e^-$ and  $\Upsilon(1S)\rightarrow e^+e^-$.  The lighter mesons have a broader pseudorapidity distribution because the Clebsch-Gordon coeffients for a spin 1 particle decaying to two spin-0 particles is very different than from a decay to two spin-1/2 particles. 
From Ref. \cite{Lomnitz:2018juf}.
}
\label{fig:clebsch}
\end{figure}

The simulations that are shown were done with the eSTARlight Monte Carlo generator \cite{Lomnitz:2018juf} which accurately reproduces the essential features of the vector meson production and decay.  eSTARlight is based on parameterized HERA data, and has been benchmarked against many HERA reactions.  The ratio of longitudinal to transverse production as a function of $Q^2$ is also based on HERA data, with judicious extrapolations where needed.  The vector meson decays depend on the vector meson polarization and on the Clebsch-Gordon coefficients related the vector meson to its daughters, as is shown in Fig. \ref{fig:clebsch}.

The main requirements presented touch on the pseudorapidity acceptance, momentum resolution and ability to track low $p_T$ particles.  A fourth detector concern is thickness, to minimize bremsstrahlung by $e^\pm$ traversing the beampipe or detector \cite{Arrington:2021yeb}.  When $e^\pm$ from vector meson decays radiate photons, the pair will be reconstructed with a lower mass, but higher $p_T$. The same thing happens for radiative $J/\psi$ decays to $e^+e^-\gamma$. The higher $p_T$ is problematic for studies of $d\sigma/dt$.  Some of these events can be saved if the photon is reconstructed in a calorimeter, and some can be removed with a cut on pair mass, but many soft bremsstrahlung events will remain. The detector momentum resolution will determine the width of the pair mass cut; this is another place where improved momentum resolution can improve vector meson reconstruction.

\subsubsection{Pseudorapidity acceptance}

\begin{figure}
\centering
\includegraphics*[width=0.9\textwidth]{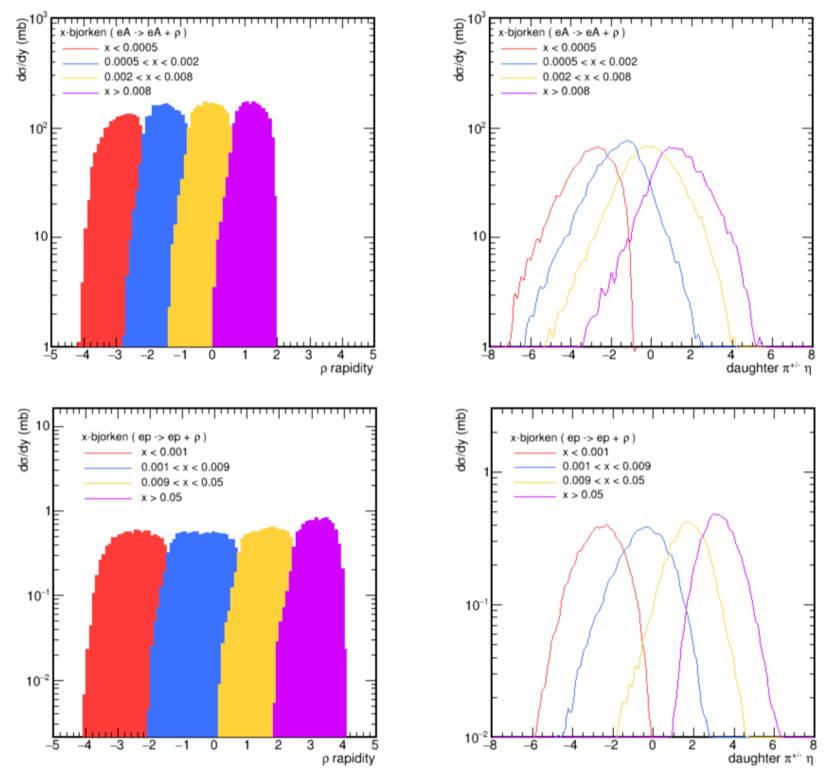}%
\caption{(left) $d\sigma/dy$ for coherent $\rho$ photoproduction using the eSTARlight Monte Carlo \cite{Lomnitz:2018juf}.  The top plots are for 18 GeV electrons colliding with 100 GeV ions, while the bottom plots are for 18 GeV electrons on 275 GeV protons.  The total production is divided up in terms of $x$ of the struck gluon; rapidity increases smoothly with $x$.  (right) Daughter pion pseudorapidity for coherent $\rho$ photoproduction under the same conditions, also divided up by $x$.}
\label{fig:rhorapidity}
\end{figure}

From Eq. \ref{eq:rapiditytox}, it should be immediately clear that a detector with broad acceptance in pseudorapidity is required. Figure \ref{fig:rhorapidity} shows the rapidity distribution for photoproduced $\rho$ for both \ep and \eA collisions, along with the pseudorapidity distribution for the daughter pions. Vector mesons from $ep$ collisions cover a wider range in rapidity, for a couple of reasons.  First, the ion Lorentz boost is larger, allowing collisions down to lower Bjorken-$x$ values.  Second, the coherent requirement for photoproduction on an ion is roughly $x< \hbar/m_p R_A$.  Therefore, for heavy ions, coherence is only possible for roughly $x<0.03$.  Incoherent photoproduction is possible for all $x$ values, so incoherent \eA photoproduction looks more like the $ep$ coherent distribution. It is important to note that, for ions, Fermi motion allows interactions to occur with $x>1$.  That is not included in these simulations, but could lead to final states with an even larger rapidity.

Very roughly, the decay of a $\rho$ with rapidity $y$ leads to pions in the pseudorapidity range $y-1$ to $y+1$. It is important to have good acceptance in at least this broad a range around the $\rho$ rapidity to be able to reconstruct the spin-density matrix of the $\rho$, and, from that, determine the mixture of longitudinal to transverse polarization.

If the acceptance is reduced for $y>-4$, sensitivity to partons with the lowest $x$ values will be lost.  This region is also critical for probing backward production of mesons, as discussed in Sec. 7.4.5.   From Eq. \ref{eq:rapiditytox}, each unit of rapidity in Fig. \ref{fig:rhorapidity} that is lost raises the minimum accessible $x$ by a factor of $e$. 

Similarly, if acceptance is cut off on the other side of the detector, sensitivity to near-threshold production will be lost.  This may be a particular issue for nuclear targets, where the study of partons with $x>1$ (possible because of Fermi motion) is important to understand nuclear correlations~\cite{Fomin:2017ydn}.  This region is also important for studying the production of exotics like pentaquarks, discussed in Sec. 7.4.6.

\subsubsection{Soft kaons from \texorpdfstring{$\phi$}{phi} decays}

Exclusive production of the $\phi$ was one of the featured reactions in the EIC White Paper \cite{Accardi:2012qut}.  The White Paper considered only electroproduction, but it is very important to study the $Q^2$ evolution of exclusive production, to see how saturation turns on as the $Q^2$ is reduced \cite{Mantysaari:2017slo,Lomnitz:2018juf}.  Among the different $\phi$ final states, only ${\rm K}^+{\rm K}^-$ seems feasible.  The ${\rm K_SK_L}$ final state is problematic because of the long ${\rm K_L}$ lifetime; it is too soft to be easily reconstructible in hadronic calorimeters.  The dilepton final states would be easy to reconstruct, but the branching ratios are too low to allow for adequate statistics; their small signals would also challenge particle identification systems.  

$\phi\rightarrow K^+K^-$ is challenging because the kaons are so soft. In the $\phi$ rest frame, the kaon momenta are only 135 MeV, or $v \approx 0.2c$. Besides the low velocity, they are heavily ionizing, so are easily stopped in detector elements.  Multiple scattering of the kaons is likely to dominate the $K^+K^-$ mass resolution.  Combining timing information with track points could help improve this mass resolution \cite{Klein:2020sts}. 

\begin{figure}
\centering
\includegraphics*[width=0.9\textwidth]{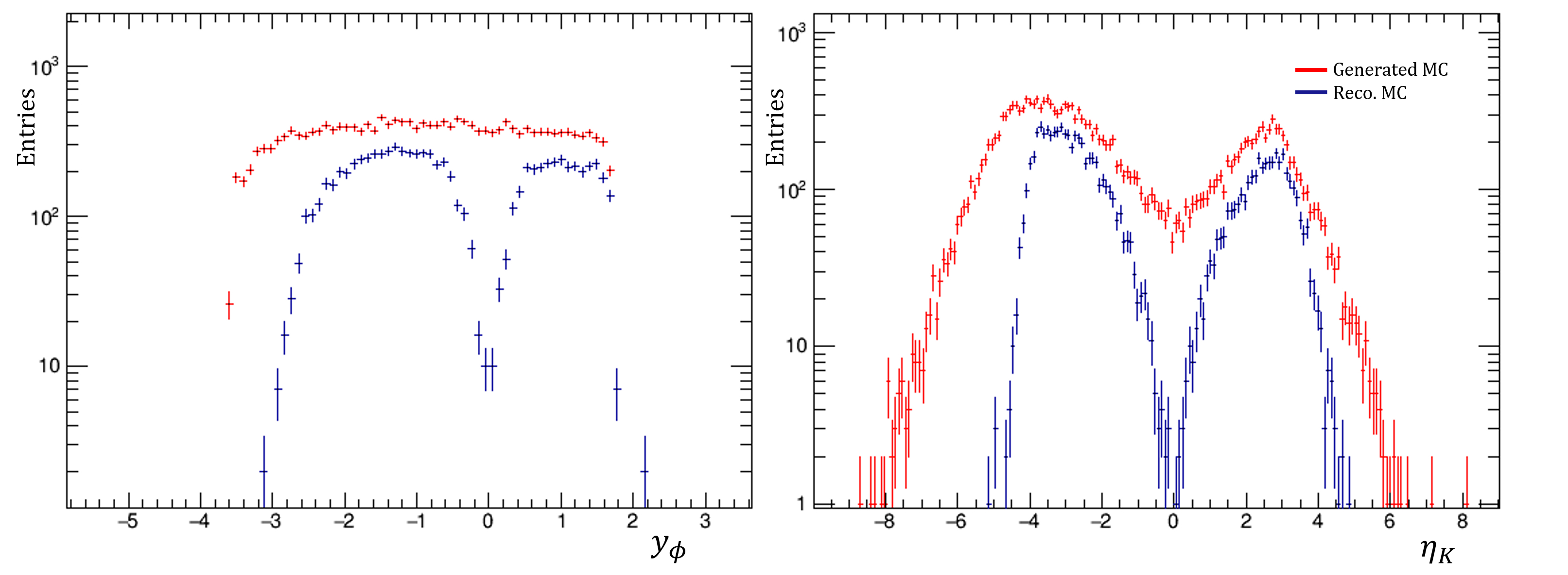}%
\caption{(left) $d\sigma/dy$ for coherent $\phi$ production using the eSTARlight Monte Carlo \cite{Lomnitz:2018juf} for 18 GeV electrons on 275 GeV protons. The red histograms show the total production while the blue histograms show the acceptance in an all-silicon tracker with a 1.5 T solenoidal field.  The drop-off around $y=0$ is because the tracker cannot reconstruct the low-momentum tracks from kaons that decay nearly at rest.  (right) Pseudorapidity distribution for charged kaons from coherent $\phi$ production under the same conditions, with the same red and blue curves. The gaps around $y=0$ and $\eta=0$ are because the tracker cannot reconstruct the soft kaons from $\phi$ decays near rest.
}
\label{fig:kaonrapidity}
\end{figure}

Figure \ref{fig:kaonrapidity} shows the rapidity distribution for coherently produced $\phi$, along with the pseudorapidity distribution of its charged kaon decay products.  In addition to showing all generated $\phi$ and daughters, it also shows the $\phi$ and kaons that are reconstructed in a model of an all-silicon detector \cite{Arrington:2021yeb} in a 1.5 T solenoidal magnetic field.  There is a large drop-off for kaons with pseudorapidity near zero, and a corresponding fall for $\phi$ with rapidity near zero.   The reason is that this detector cannot reconstruct soft kaons from $\phi$ decays near rest.  Although this is only one example detector, it has thinned silicon monolithic active pixel sensors in a precision vertex chamber, and it would not be easy to do significantly better than this.     Away from $y=0$, the kaons are Lorentz boosted.  The higher velocity kaons are more penetrating, and the $\phi$ can be reconstructed. At still higher $|y|$, the typical kaon pseudorapidity is larger than the $\phi$ rapidity, reducing the acceptance at large $|y|$ \cite{Arrington:2021yeb}. 

At higher $Q^2$ the $\phi$ should have a significant transverse momentum, leading to increased acceptance.  The acceptance limitations near $y=0$ and $p_T=0$ will create a hole in acceptance at  $Q^2=0$ and will preclude $\phi$ measurements around $x=1/2\gamma$ ($x=0.005$ for ions).  The importance of this hole will depend on its size; this should be studied in future detector designs.

\subsubsection{Momentum resolution}

The need to be able to separate $\Upsilon(1S)\rightarrow ll$, $\Upsilon(2S) \rightarrow ll$ and $\Upsilon(3S) \rightarrow ll$ requires a detector with good momentum resolution; separating the $J/\psi$ and $\psi'$ is much easier.  The mass difference between the $\Upsilon(2S)$ and $\Upsilon(3S)$ is only 334 MeV, or 3\% of their mass. This separation is a signature requirement for the sPHENIX collaboration, who found that a mass resolution of 100 MeV was required giving a bit over $3\sigma$ separation \cite{Osborn:2020soo}. The same requirement should be applicable for an EIC detector.

\begin{figure}
\centering
\includegraphics*[width=0.8\textwidth]{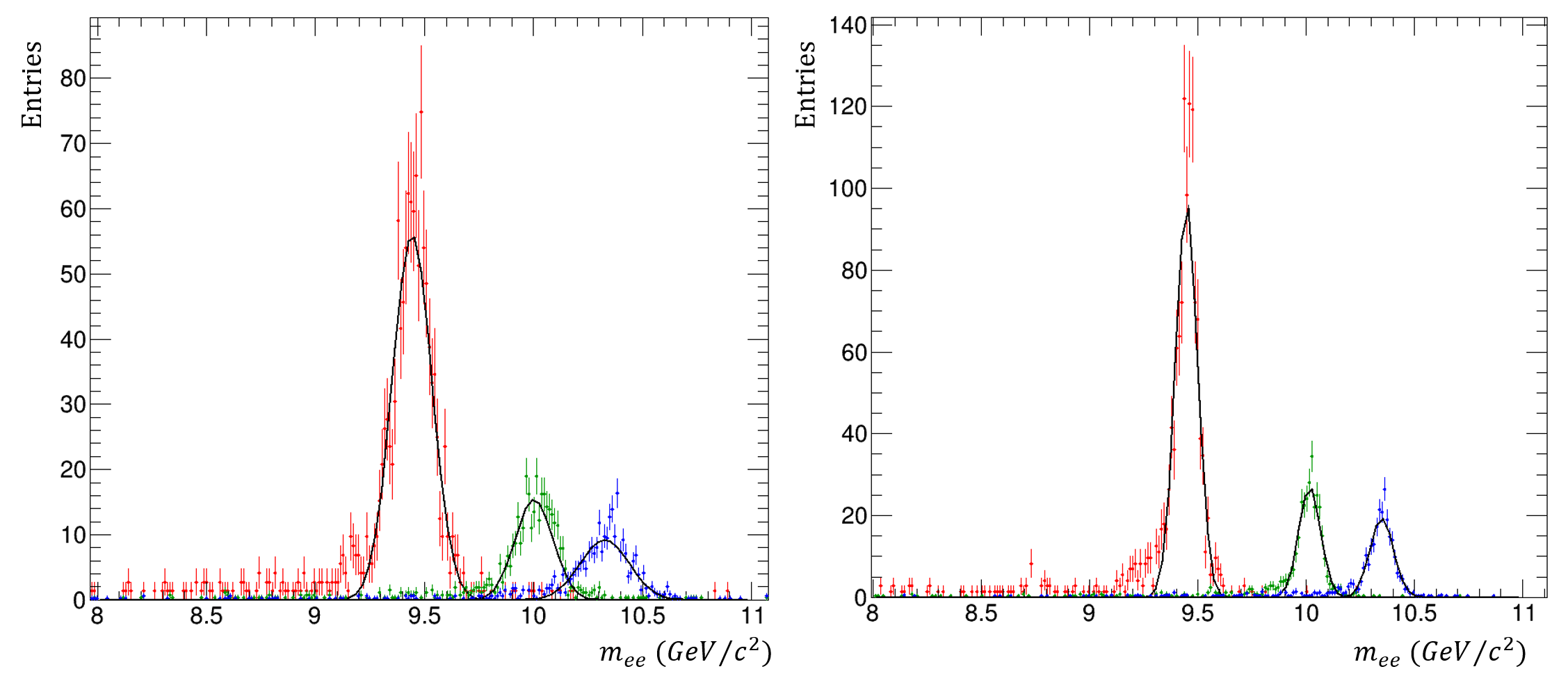}%
\caption{Dielectron mass spectra for combined $\Upsilon(1S)$ plus $\Upsilon(2S)$ plus $\Upsilon(1S)$ production in an all-silicon detector in (left) a 1.5 T magnetic field and (right) a 3.0 T magnetic field.  The histograms are for an integrated luminosity of 10 fb$^{-1}/A$, where $A=197$ is the atomic number.
}
\label{fig:upsilon}
\end{figure}

It is possible to use this mass resolution to estimate the required momentum resolution, in a simple case:  An $\Upsilon$ at  $y=0$ and $p_T=0$ has two back-to-back tracks each with energy $M_{ll}/2$ and momenta close to that value.  Then, neglecting the small lepton masses, $M_{ll}^2=4p_1p_2$.  Assuming that the two track momenta and their resolutions are roughly equal, then $\sigma_M/M=2\sigma_p/p$.  So, for tracks with momentum around 5 GeV, the required momentum resolution is about 0.5\%.  Away from $y=0$, $p_T=0$, an analytic analysis is more difficult, but it seems that the Lorentz boosts lead to slightly looser requirements, when expressed in terms of $\sigma_p/p$.  Going further requires simulations; Fig. \ref{fig:upsilon} shows the dielectron mass spectrum for a simulation combining the three $\Upsilon$ states in an all-silicon detector in 1.5 and 3.0 T magnetic fields.  Adequate separation is seen at 1.5 T, while at 3.0 T the separation is almost complete.

\subsubsection{Separating coherent and incoherent interactions}

As was discussed in Section 7.3.9, separating coherent and incoherent production is critical for using vector meson production for nuclear imaging and studies of gluonic fluctuations.  For moderate/large $|t|$, rejection factors of more than 400:1 are required.  This poses extreme requirements on the forward detection elements at an EIC detector.  The nuclear excitation typically occurs via neutron, proton or photon emission.   Neutrons and protons can be detected with zero degree calorimeters (ZDCs) and forward proton spectrometers (Roman Pots) respectively, but it is critical that these both have excellent acceptance out to transverse momenta of several times the Fermi momentum.  Photon detectors must be able to detect photons with energies considerably less than 1 MeV in the nuclear rest frame. Gold has detectable (short-lived) excited states at 269 and 279 keV, for example. Different nuclei will impose different requirements.

Lead is doubly magic, so its lowest energy excited state is at 2.6 MeV \cite{Nlevels}, leading to lab-frame photon energies of hundreds of MeV.  These photons are likely isotropic in the nuclear frame.  Backward-going photons will be Lorentz downshifted or not sufficienly boosted to be detectable.  Allowance must be made for these missed photon. 

In contrast, gold has an excitation with an energy of 77 keV, and a lifetime of 1.9 nsec \cite{Nlevels}. The long lifetime means that the excited gold nucleus may travel tens of meters before decaying, making the decay products essentially impossible to observe.  There are additional low-lying states with energies of 269 and 279 keV, which translate to maximum detector-frame energies below 60 MeV.  Although we do not have good models to predict which levels are excited in exclusive vector meson production, it seems unlikely that the required separation can be achieved with gold nuclei.

In short, some nuclear deexcitations will involve very soft photons; for gold, some of these photons are emitted after the excited nucleus has left the interaction region.  These photons are probably undetectable.  Separating coherent and incoherent production is likely to be considerably easier with lead beams, but more study is required to determine what rejection factor is achievable, and how it will impact the physics.  


\subsection{Meson structure}
\label{subsec:DetReq.DT.meson}

For the detection of particles of relevance to meson structure studies all sub-components of the far-forward area play an important role, the detection in the B0 area, detection of decay products with the off-momentum detectors, and detection of forward-going protons and neutrons with the Roman Pots and ZDC.

\subsubsection{Sullivan process for pion structure: \texorpdfstring{$\bm{e+p \rightarrow e^\prime + X + n}$}{}}

\begin{figure}[htb]
\includegraphics[width=0.47\textwidth]{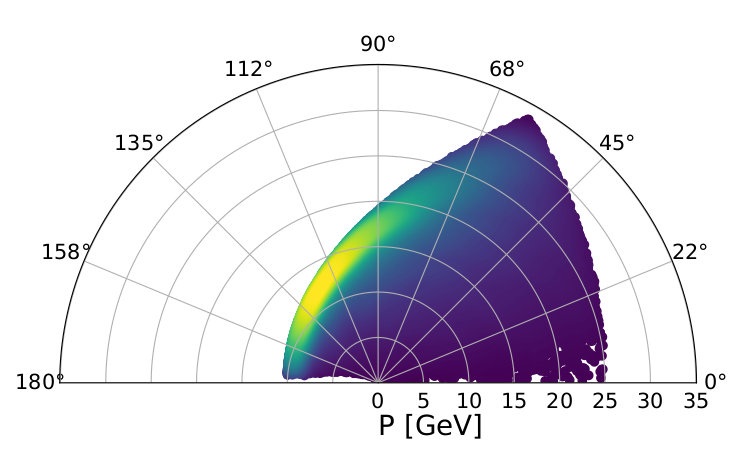}
\includegraphics[width=0.47\textwidth]{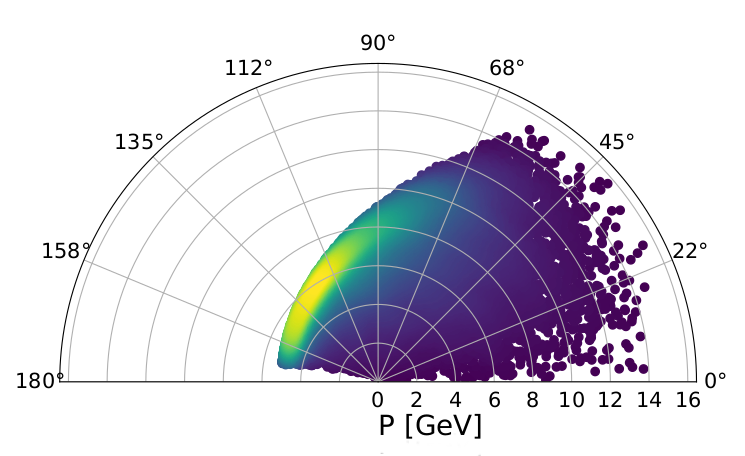}
\hspace*{0.20cm}
\includegraphics[width=0.47\textwidth]{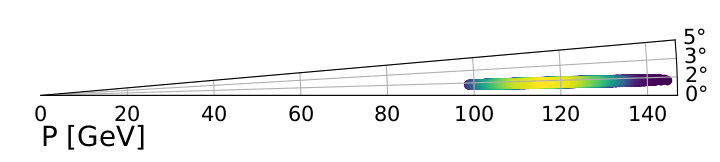}
\includegraphics[width=0.47\textwidth]{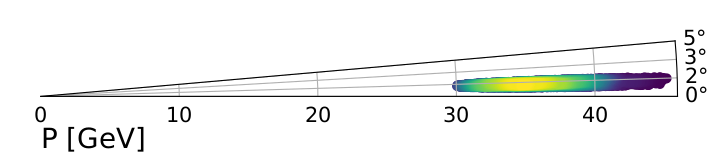}
\caption{\label{fig:kin_scat_n_10on135}
A comparison of the scattered electron (top) and leading neutron (bottom) kinematics in $e+p \rightarrow e^\prime + X + n$ for two energy settings - 10 GeV on 135 GeV (left) and 5 GeV on 41 GeV(right) with a luminosity of 100~$\fb^{-1}$. The momentum, P and angle, $\theta$ are defined in the lab frame. The scattered electrons, in both cases, are within the acceptance of the central detector and the leading neutrons are at small forward angles and carry most of the proton beam energy after the scattering process.
}
\end{figure}

The initial pion structure studies were conducted at the highest energy of 18 GeV on 275 GeV (corresponding to the electron and proton beam energy, respectively, both in GeV) to maximize the kinematics coverage. However, to improve access to the high $x_{\pi}$ region (see Sec.~\ref{part2-subS-PartStruct.M}), alternate lower beam energies 10 GeV on 135 GeV and 5 GeV on 41 GeV were also selected. These lower beam energies allow access to this high $x_{\pi}$ regime over a wider range of $Q^2$. For a comparison, the 18 GeV on 275 GeV energies allow access to high $x_{\pi}$ data over a $Q^2$ range of $\sim$200-1000~GeV$^{2}$, while with the 10 GeV on 135 GeV energies that range was increased to $\sim$30-1000~GeV$^{2}$, and with the 5 GeV on 41 GeV energies to $\sim$5-1000~GeV$^{2}$. The lower-energy combination of 5 GeV on 41 GeV is even more beneficial to tag kaon structure by allowing detection of the leading $\Lambda$ events (see below).

The kinematics for the more advantageous lower energy settings, 10 GeV on 135 GeV and 5 GeV on 41 GeV, are shown in  Fig.\,\ref{fig:kin_scat_n_10on135}. While the scattered electrons are within the acceptance of the central detector, the leading neutrons for these two energy settings are at a very small forward angle while carrying nearly all of the proton beam momentum. These leading neutrons will be detected by the ZDC.

\begin{figure}[htb]
  \centering
   \includegraphics[width=0.3\textwidth]{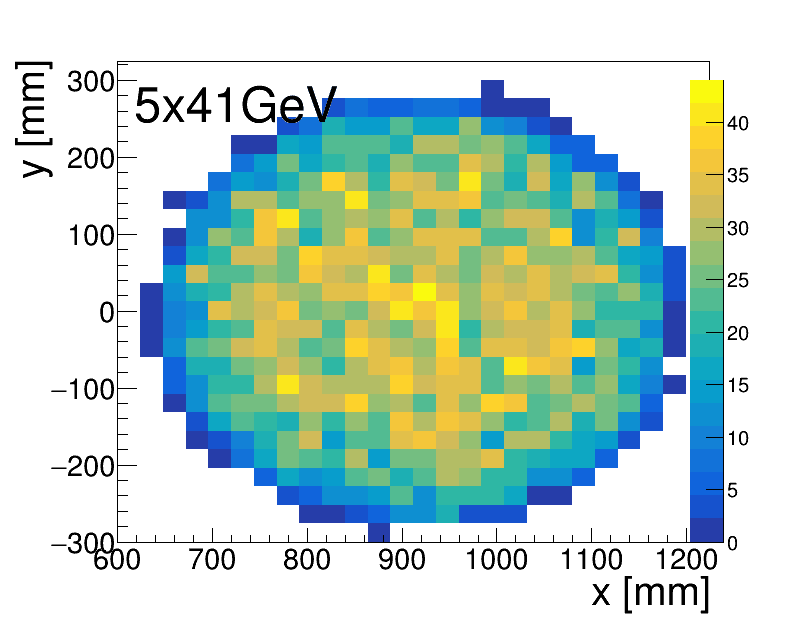}
  \includegraphics[width=0.3\textwidth]{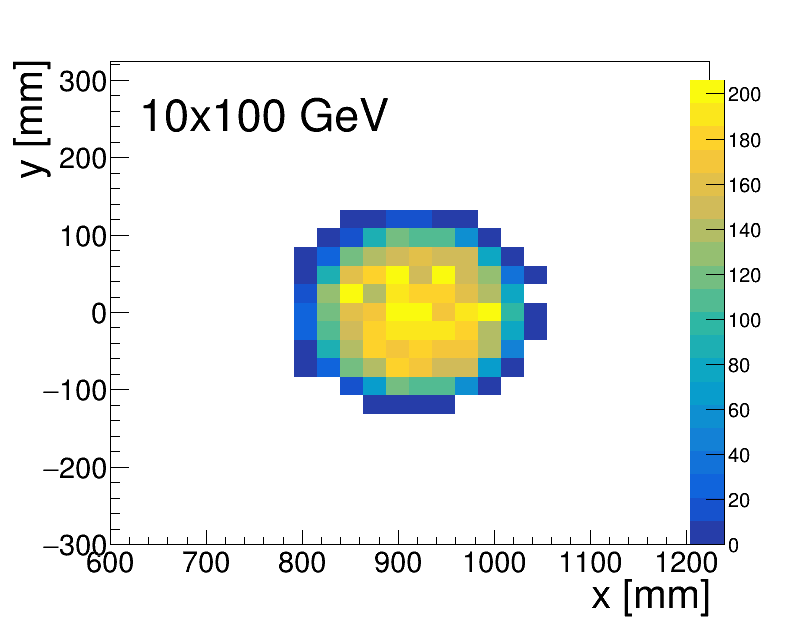}
  \includegraphics[width=0.3\textwidth]{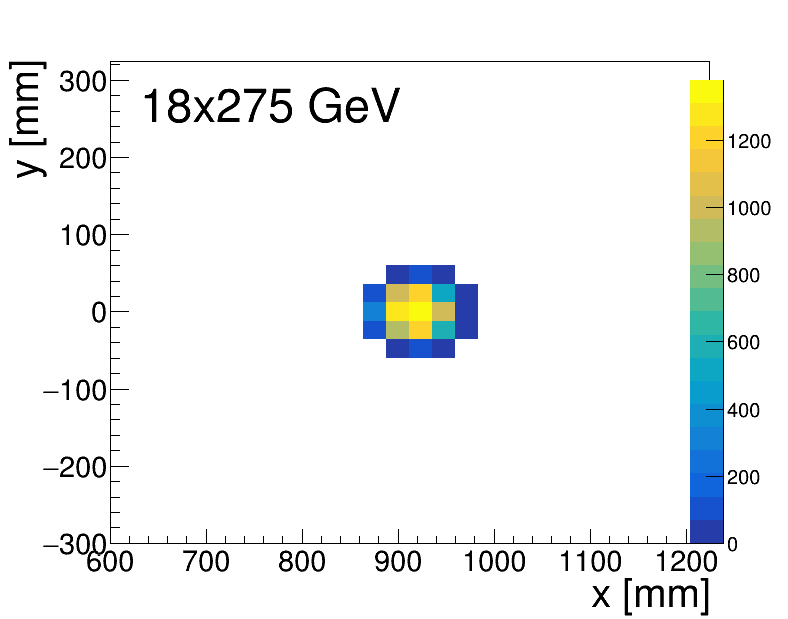}
  


  \includegraphics[width=0.3\textwidth]{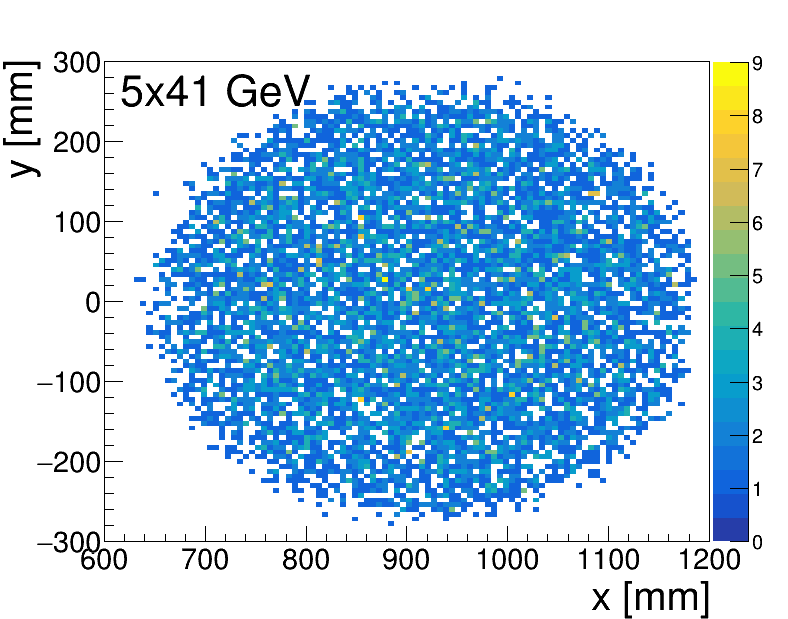}
  \includegraphics[width=0.3\textwidth]{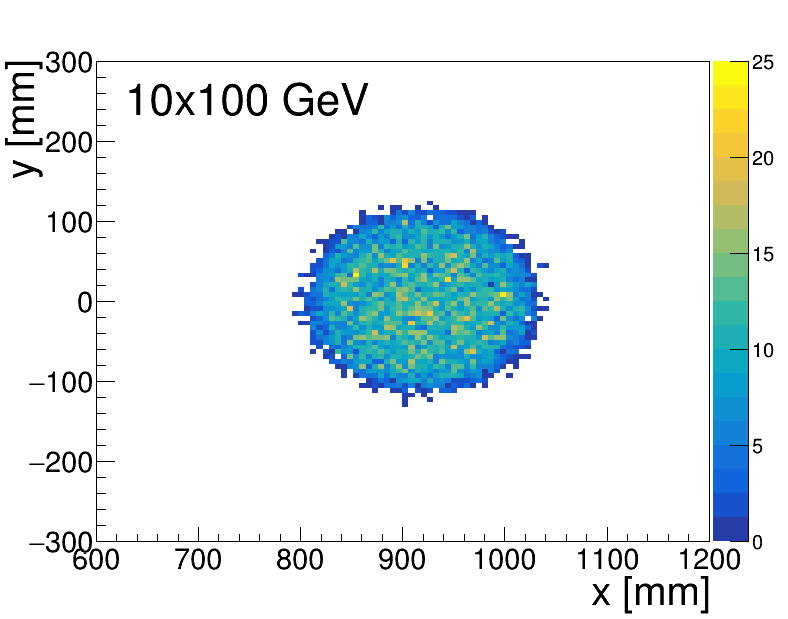}
  \includegraphics[width=0.3\textwidth]{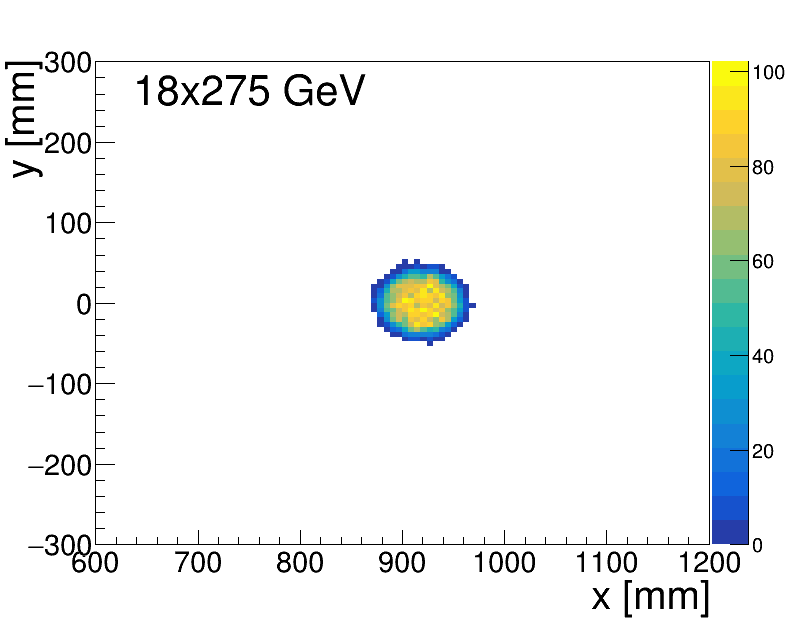}
  \caption{ Sullivan process $e+p \rightarrow e^\prime + X + n$: acceptance plot for neutrons in 60$\times$60~cm$^2$  ZDC, with a low spatial resolution of 3~cm (upper panels) and with a high spatial resolution of 0.6~cm (lower panels),  for different beam energy settings, from left to right 5 GeV on 41 GeV, 10 GeV on 100 GeV, and 18 GeV on 275 GeV, all with a luminosity of 100~$\fb^{-1}$. The acceptance plot for 5 GeV on 100 GeV would be similar as shown for 10 GeV on 100 GeV. The lower proton (ion) energies set the requirement for the size of the ZDC, whereas the higher proton (ion) energies drive the spatial resolution requirement.}
  \label{fig:n-ZDC-res}
\end{figure}

Figs.~\ref{fig:n-ZDC-res} shows the acceptance plots for neutrons in the ZDC for different beam energy settings. As one can see, the spatial resolution of ZDC plays an important role for the higher beam energy setting  (18 GeV on 275 GeV), since it is directly related to the measurements of  $p_T$ or $t$. For the lower beam energy setting (5 GeV on 41 GeV), the total acceptance coverage of the ZDC is important. This sets a  requirement  for the total size of ZDC to be a minimum of 60$\times$60~cm$^2$. Such a configuration of the ZDC allows to achieve close to  $100\%$ neutron detection efficiency for this channel. 

\subsubsection{Sullivan process for kaon structure: \texorpdfstring{$\bm{e+p \rightarrow e^\prime + X + \Lambda}$}{}}

For the case of a leading $\Lambda$ event, to tag the DIS process on a kaon, both its decay products are detected at small forward angles due to the nature of two-body decay kinematics. The detection of these $\Lambda$ decay products requires additional high-resolution and granularity due to the small angle of separation of decay products. 

Detection of the decay channel $\Lambda\rightarrow n+ \pi^0$ is feasible but will require a means for EM Calorimetry before the ZDC, in order to distinguish the neutron and the two photons coming from $\pi^0$ decay. 
Detection of the other decay channel, $\Lambda\rightarrow p+\pi^-$, poses a more challenging measurement due to its requirement of additional charged-particle trackers or a veto trigger on the path to ZDC.

The reconstruction of the $\Lambda$ event in the far-forward detection area is one of the most challenging tasks. This comes mainly from the fact that these leading $\Lambda$'s have energy close to the initial beam energy, and thus their decay lengths can be tens of meters along the $Z$-axis (or beam-line). This complicates detection of the decay products, and thus the final $\Lambda$ mass reconstruction.

\begin{figure}[hbt]
\centering
    \includegraphics[width=0.4\textwidth]{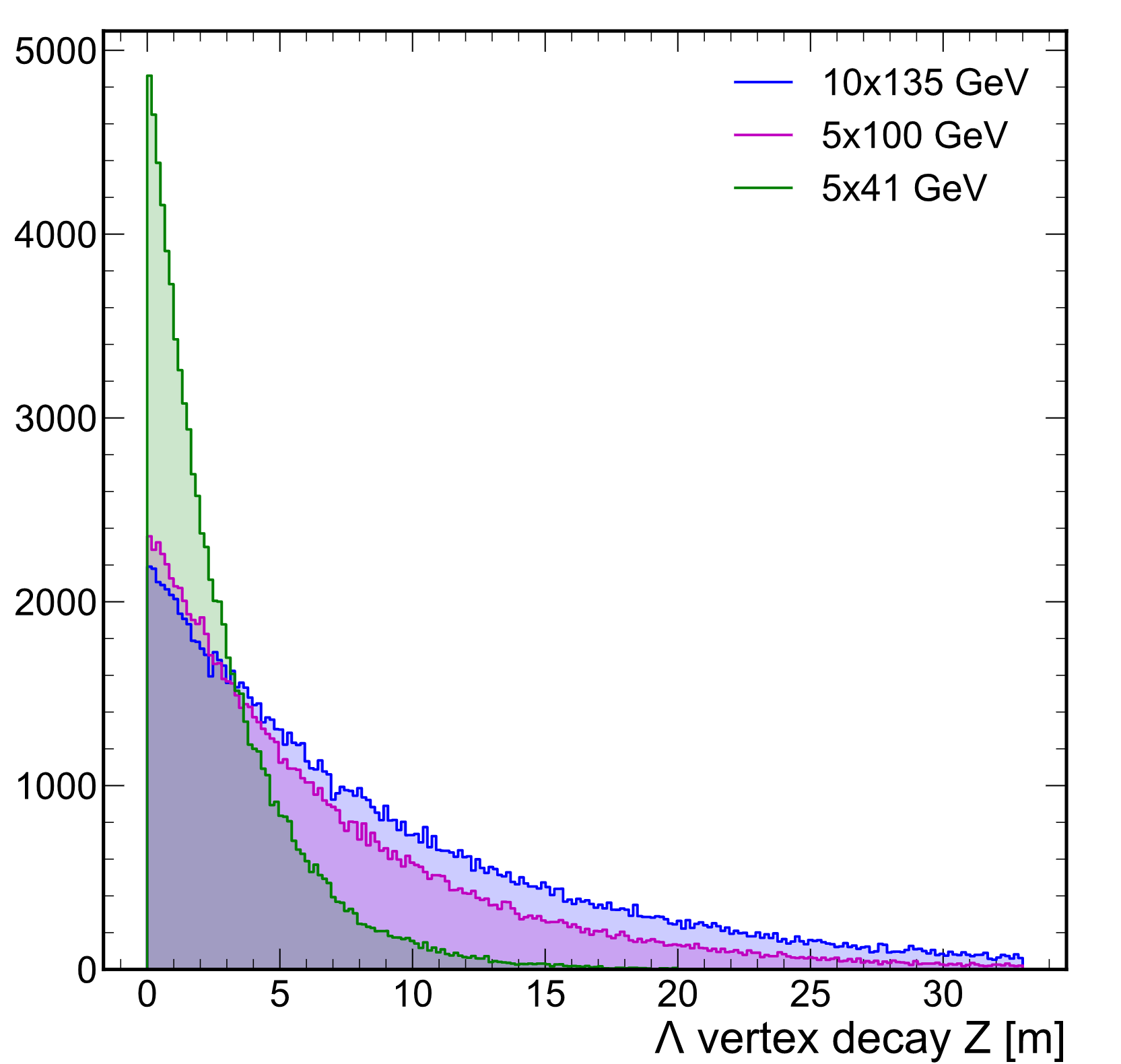}

    \caption[]{ 
        ${e+p \rightarrow e^\prime + X + \Lambda}$: the $\Lambda$-decay spectrum along the beam line for different beam energies with a luminosity of 100~$\fb^{-1}$. Vertical axis shows unnormalized events.
        \label{fig:lam_z_decay}
    }
\end{figure}

Figure~\ref{fig:lam_z_decay} illustrates this further, and shows the $Z$-coordinate of where the $\Lambda$-decay occurs for different beam energies. For the lower beam energy settings (5 GeV on 41 GeV) most $\Lambda$ decays are within the central detector region, but at the higher proton (ion) beam energies the $\Lambda$ decays happen more in the forward-detection area, with tails of the decay process to near the ZDC location. Table ~\ref{tab:lambda_decay} shows the percentage of decayed $\Lambda$ for different energies and different Z ranges.

\begin{table}[hbt]
    \caption{ $e+p \rightarrow e^\prime + X + \Lambda$: Percentage of decayed $\Lambda$'s in different detection ranges. }
    \centering
    \begin{tabular}  {  l c c c c c c    }
   \hline
   \hline
    $E_\text{beams}$ & & $Z_\text{vtx}<5$\,m & & 5\,m\,$< Z_\text{vtx}<30$\,m  & &$Z_text{vtx}>30$\,m\\ 
    \hline
    5 GeV on 41 GeV   & &83.0\% & & 16.6\% & & 0.4\% \\  
    10 GeV on 100 GeV  & &52.1\% & & 46.7\% & & 1.2\% \\
    10 GeV on 130 GeV & &41.8\% & & 54.2\% & & 4\% \\  
    18 GeV on 275 GeV & & 23.3\% & & 56.2\% & & 20.5 \% \\
    \hline
    \hline
    \end{tabular}
\label{tab:lambda_decay}
\end{table}

To study the possibility of $\Lambda$ mass reconstruction further, both main decay modes were looked into: $\Lambda \rightarrow p + \pi ^{-}$ with a branching ratio of $63.9\%$, and $\Lambda \rightarrow n + \pi ^{0}$ with a branching ratio of $35.8\%$. Both channels can be  clearly separated by the different charge of the final-state particles, and thus the different detector components which will play a role for their detection. 


\paragraph{$\bm{\Lambda \rightarrow p + \pi ^{-}}$}

For this process we only  have charged particles in the final state.  Therefore, for detection,  we have to rely on sub-components along the far-forward area such as the B0 tracker, the Off-Momentum trackers, and Roman Pots. 

\begin{figure}[htbp]
\includegraphics[width=0.3\textwidth]{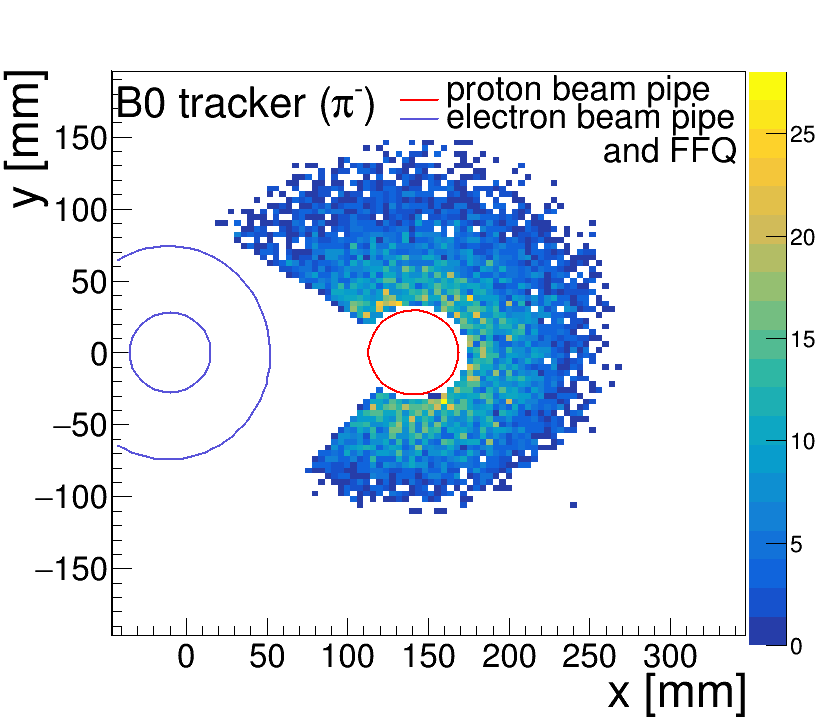}
\includegraphics[width=0.3\textwidth]{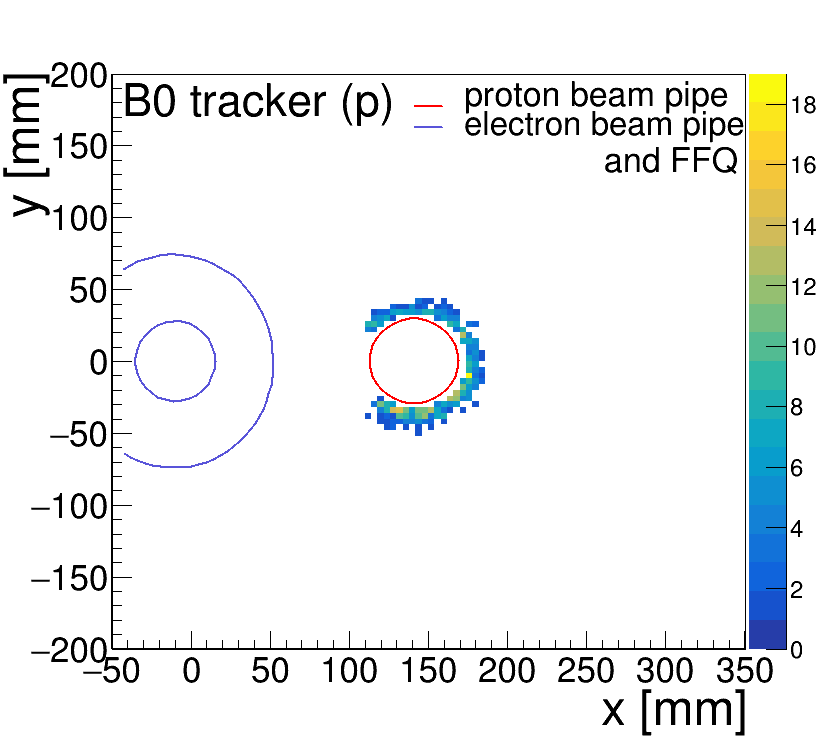}
\includegraphics[width=0.3\textwidth]{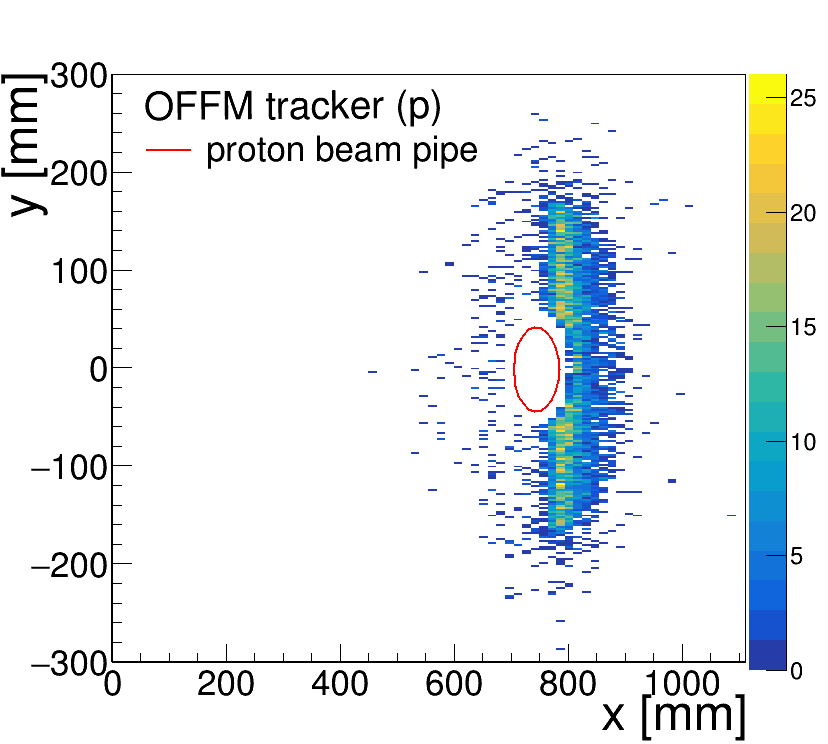}
\caption[]{$\Lambda \rightarrow p + \pi ^{-}$ decay in ${e+p \rightarrow e^\prime + X + \Lambda}$:  Occupancy plots for energy setting 5 GeV on 41 GeV with a luminosity of 100~$\fb^{-1}$.  For  $\pi ^-$ in the B0 tracker (left panel). For protons in the B0 tracker (middle panel) and in Off-Momentum detectors (right panel). The red circle shows the beam pipe position and the blue circle shows the electron Final-Focus Quadrupole (FFQ) aperture inside the B0 dipole. 

\label{fig:occup_L_p_41}
}
\end{figure}

\begin{figure}[htbp]
%
\centering
  \includegraphics[width=0.45\textwidth]{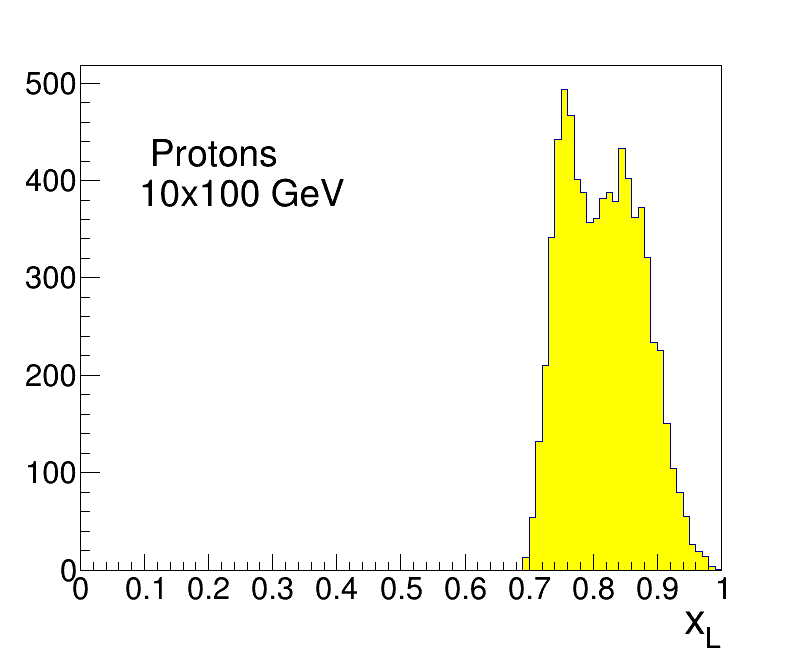}
  \includegraphics[width=0.45\textwidth]{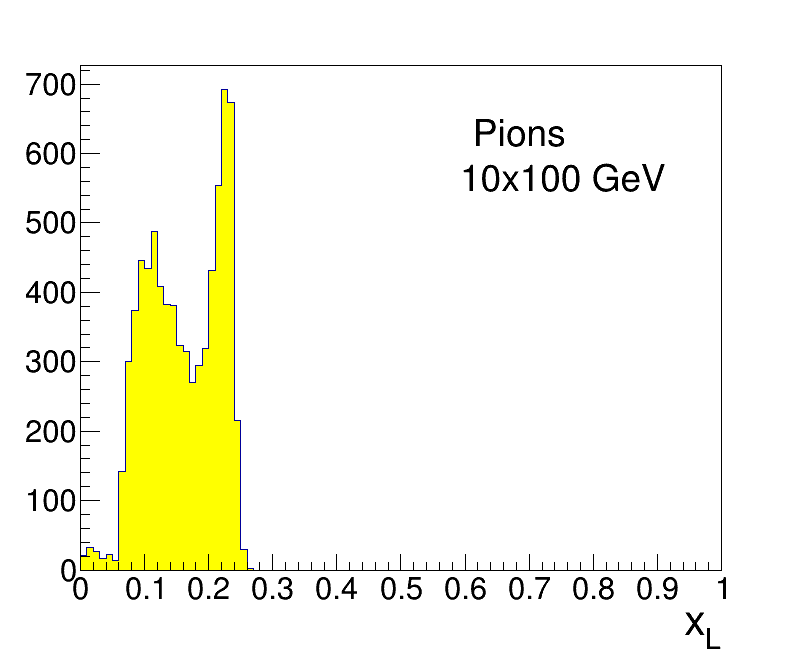}
  \includegraphics[width=0.45\textwidth]{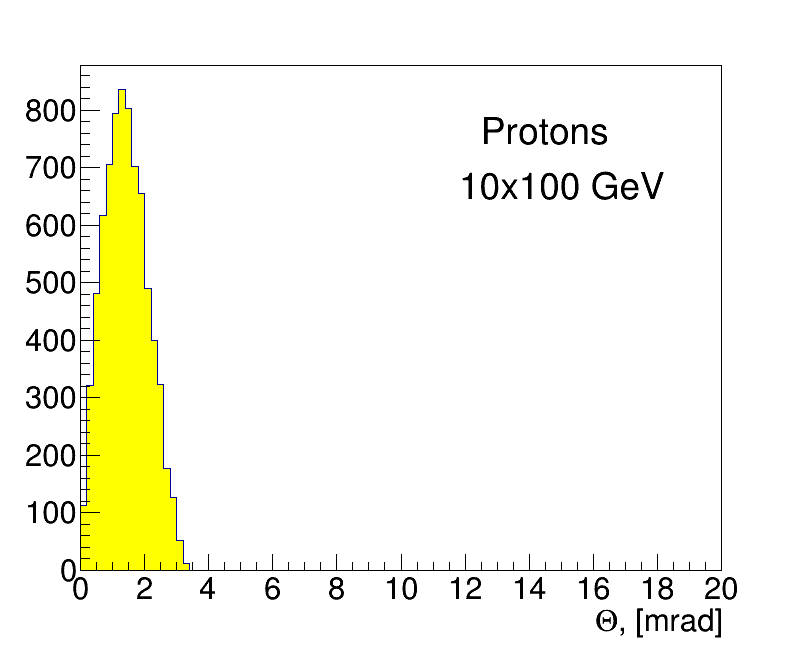}
  \includegraphics[width=0.45\textwidth]{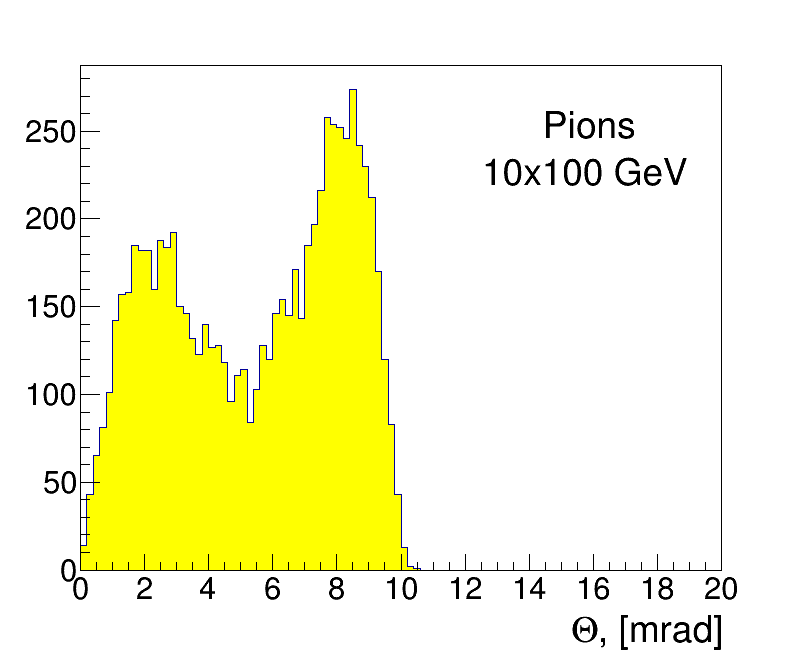}

\begin{tabular}{lll}
(a) ~~~~~~~~~~~~~~~~~~~~~~~~              & ~~~~~~~~~~~~~(b) 
\end{tabular}
\caption[]{ $\Lambda \rightarrow p + \pi ^{-}$ decay in ${e+p \rightarrow e^\prime + X + \Lambda}$: $x_L$ (top row) and theta (bottom row) distributions for detected decay products of $\Lambda$ particles for the 10 GeV on 110 GeV beam energy combination with a luminosity of 100~$\fb^{-1}$: protons (left panels)  and pions (right panels).  Vertical axes show unnormalized events.

\label{fig:XL}
}
\end{figure}
As an example, occupancy plots for the beam energy setting of 5 GeV on 41 GeV are shown in Fig.~\ref{fig:occup_L_p_41}. Since this is the lowest beam energy setting, most of the lambdas would decay in the first meter (before the B0 magnet), and the decay products of lambda are expected to have low momenta. Therefore, as expected, protons coming from the $\Lambda$ decays will mostly be detected, due to their lower rigidity, in the off-momentum detectors (c) and partially in a B0 tracker (b). While for pions, the tracker inside the B0 dipole will be the only detecting element (a). As one can also see from this figure, the proton-beam-pipe aperture inside the B0-dipole plays an important role and sets the detection efficiency for pions, as well as the azimuthal angle $\phi$-coverage of the detecting elements around the proton beam-pipe. Further information on the distributions for detected decay products at these lower beam energies of 10 GeV on 110 GeV are given in Fig.~\ref{fig:XL}.

\begin{figure}[htbp]
\includegraphics[width=0.3\textwidth]{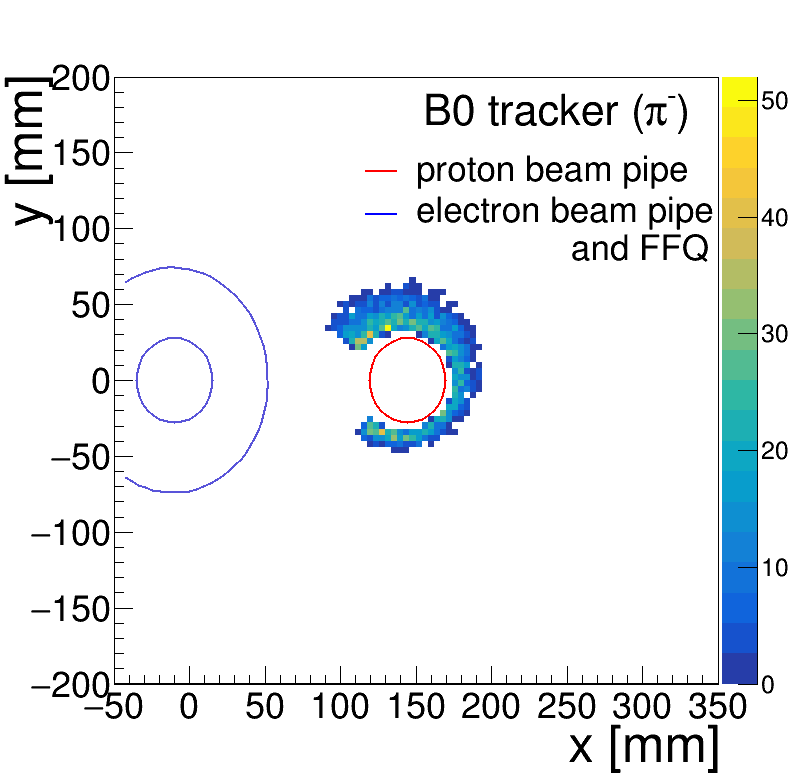}
\includegraphics[width=0.3\textwidth]{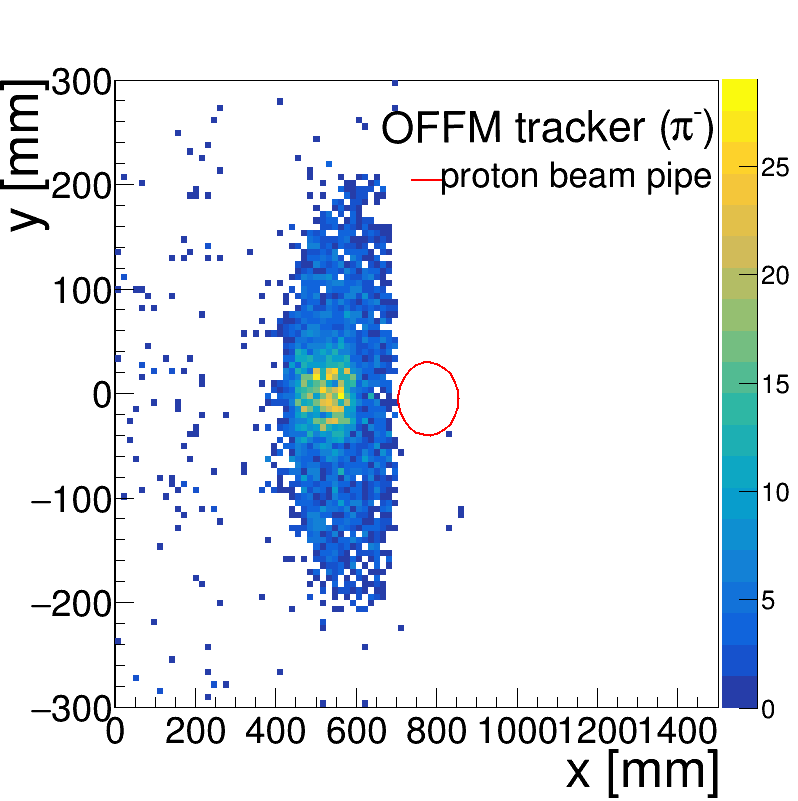}
\includegraphics[width=0.3\textwidth]{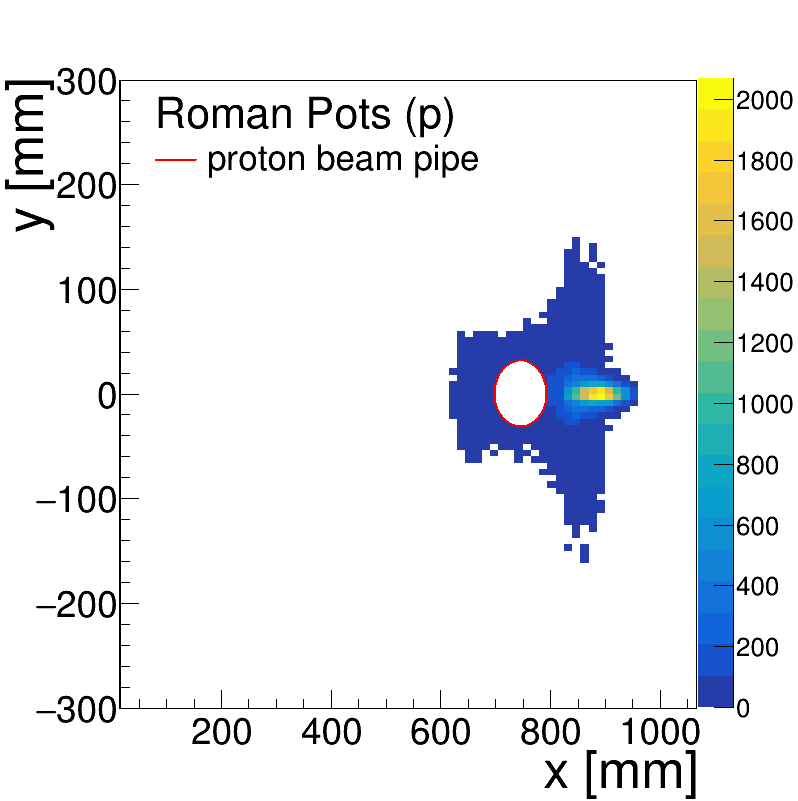}
\caption[]{$\Lambda \rightarrow p + \pi ^{-}$ decay in ${e+p \rightarrow e^\prime + X + \Lambda}$: Occupancy plots for energy setting 10 GeV on 100 GeV with a luminosity of 100~$\fb^{-1}$. For $\pi ^-$ in the B0 tracker (left panel) and the off-momentum tracker (middle panel). For protons in the Roman Pots detectors (right panel). The red circle shows the beam pipe position and the blue circle shows the electron FFQ aperture inside the B0 dipole. 

\label{fig:occup_L_p_100}
}
\end{figure}

For the higher beam-energy settings, for example 10 GeV on 100 GeV, the protons are to be detected in the Roman Pots (and partially in Off-Momentum) detectors, see Fig.~\ref{fig:occup_L_p_100}. Pions originating from a $\Lambda$-decay with $Z_{vxt} < 4$m will only partially be detected in the B0-area, while most of them will go undetected  through the proton beam pipe. Pions with higher momentum and lower angles ($p_t$ or $theta$) can pass through the bores of the Final-Focusing Quadrupole magnets (FFQs) and be detected in the Off-Momentum detectors. Their detection represents the denser (light) area of detection in the Off-Momentum detectors (Fig.~\ref{fig:occup_L_p_100}(b)). Note that due to the negative charge of the pions, they will experience an opposite bending in dipoles, as compared to protons (compare with the protons in the Off-Momentum detectors on Fig.~\ref{fig:occup_L_p_41}(b)). Therefore, in order to detect the $\Lambda$-decays in this channel the Off-Momentum detectors need to provide a full azimuthal coverage, to establish a proper detection for the negatively-charged particles.

\begin{figure}[htb]
  \centering
 \includegraphics[width=0.4\textwidth]{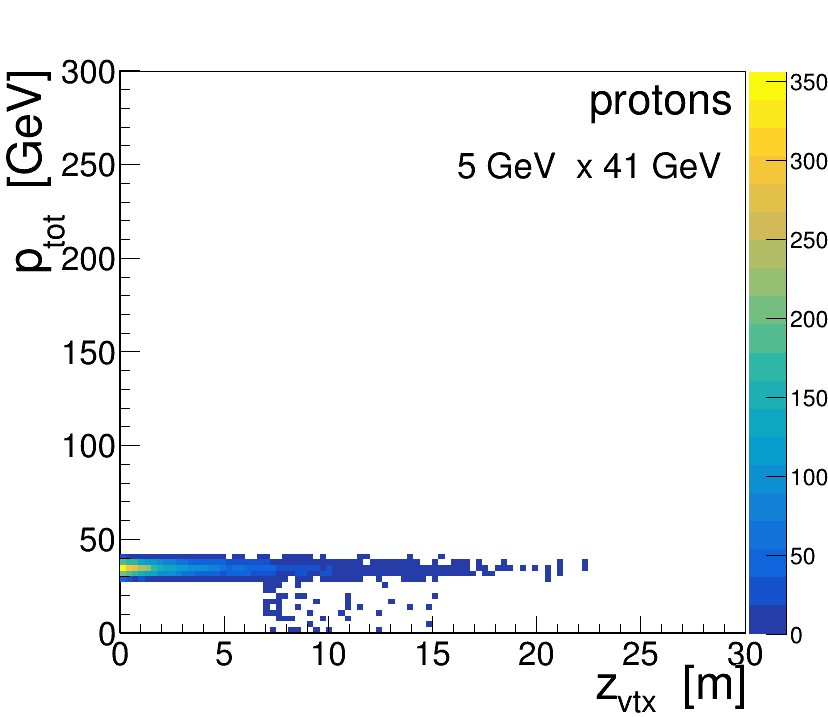}  
  \includegraphics[width=0.4\textwidth]{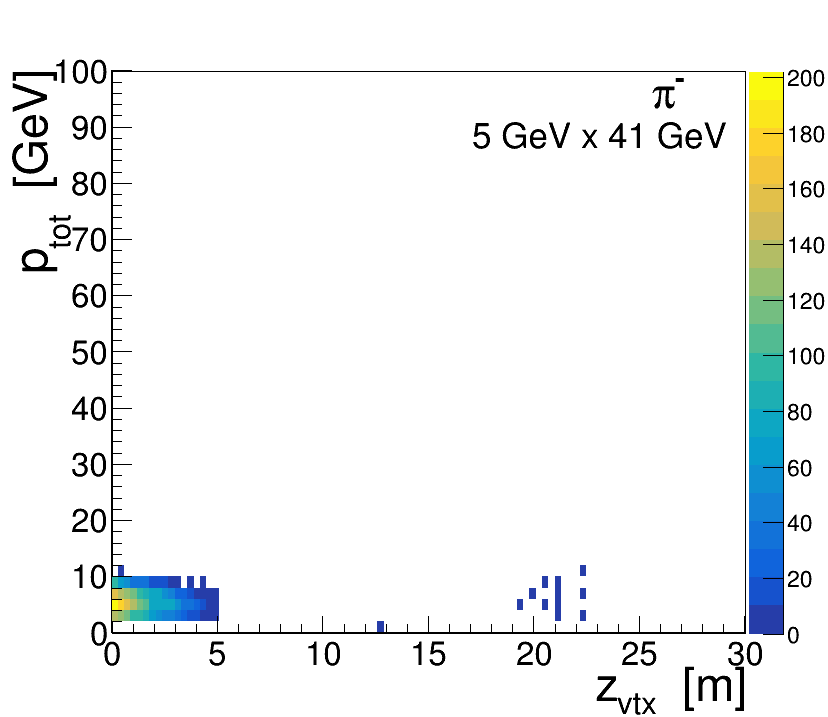}  
  
  \includegraphics[width=0.4\textwidth]{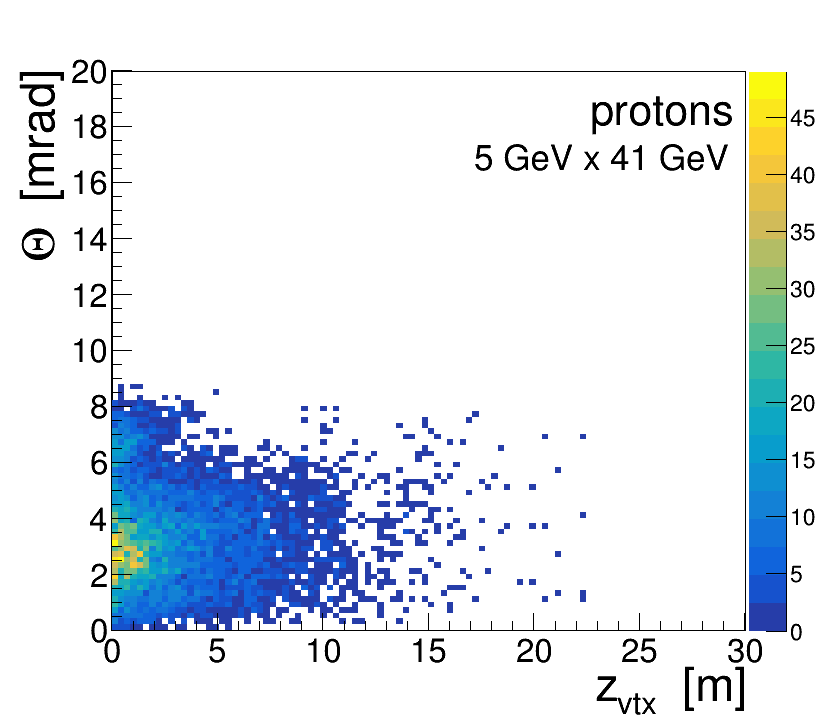}  
  \includegraphics[width=0.4\textwidth]{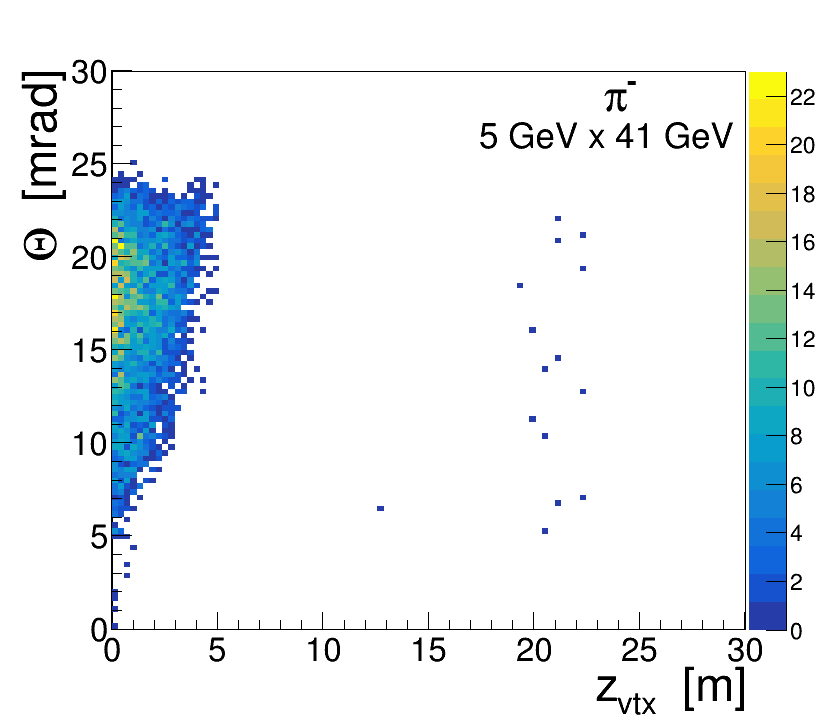}  
  
  \caption{$\Lambda \rightarrow p + \pi ^{-}$ decay in ${e+p \rightarrow e^\prime + X + \Lambda}$: Momentum (top) and angular (bottom) distributions of protons (left) and $\pi^-$ (right) at beam energy setting 5 GeV on 41 GeV with a luminosity of 100~$\fb^{-1}$, as registered in the far-forward detectors as a function of their origination (the decay vertex).  }
  \label{fig:L-p_Zvtx_41_f4}
\end{figure}

For the 5 GeV on 41 GeV beam energy combination, Fig.~\ref{fig:L-p_Zvtx_41_f4} shows the momentum (top panels) and angular (bottom panels) distributions of protons (left panels) and pions (right panels) from $\Lambda$-decay as a function of distance of the $\Lambda$-decay point, as detected in one of the beam line sub-detectors. This then in turn illustrates which of the sub-detectors along the beam line detect the decay products. The protons carry most of the initial proton beam momentum and extend over the far-forward direction, with angles less than 8 mrad. On the other hand, as one can clearly see from the high density of hits, the $\Lambda$-reconstruction efficiency will mainly depend on the efficiency for the detection of pions in the B0 area, with angles in the 5-25 mrad range.  

\begin{figure}[htb]
  \centering
 \includegraphics[width=0.4\textwidth]{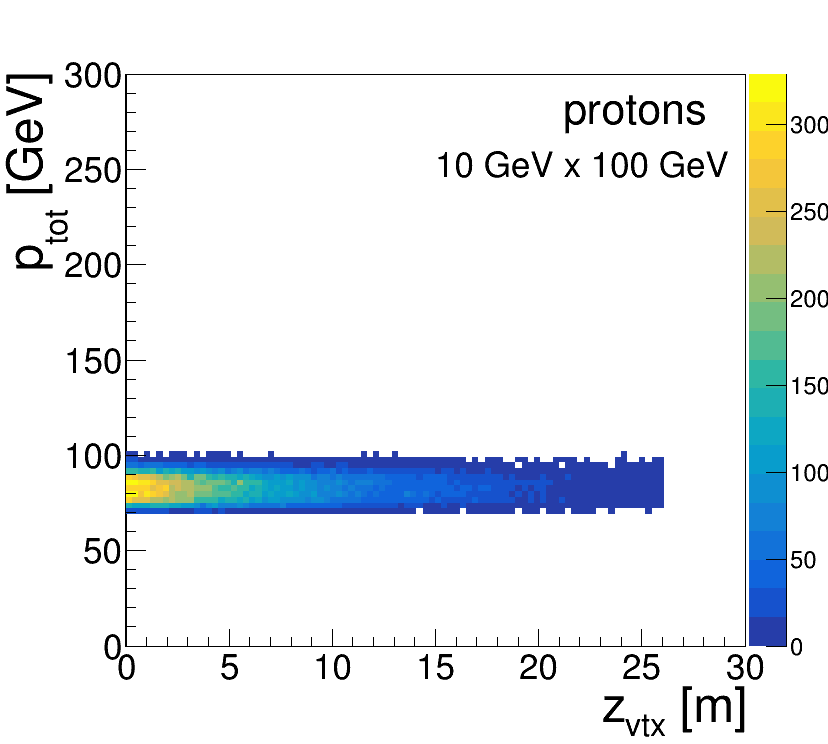}  
  \includegraphics[width=0.4\textwidth]{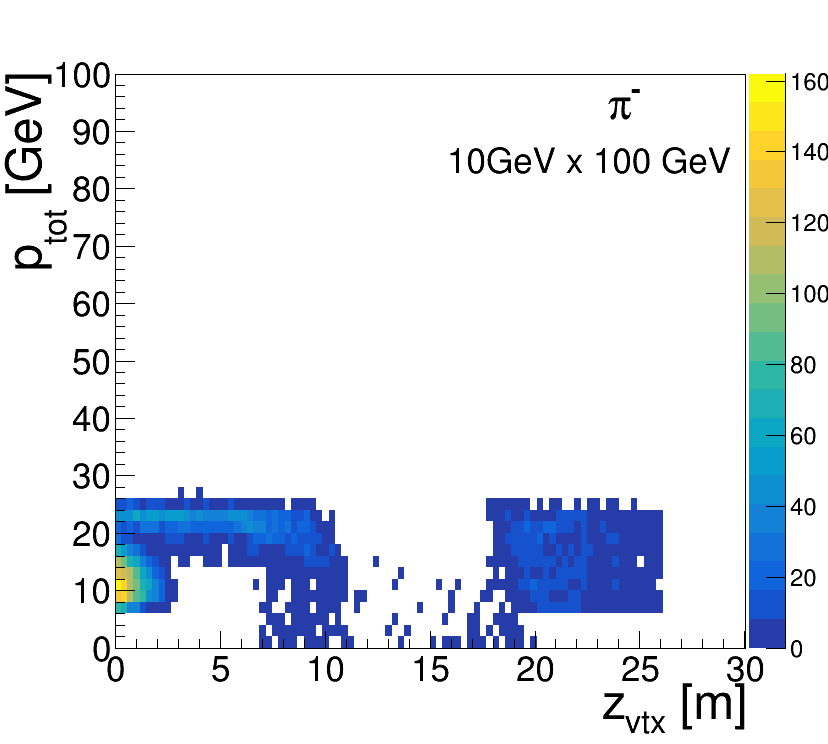}  
  
  \includegraphics[width=0.4\textwidth]{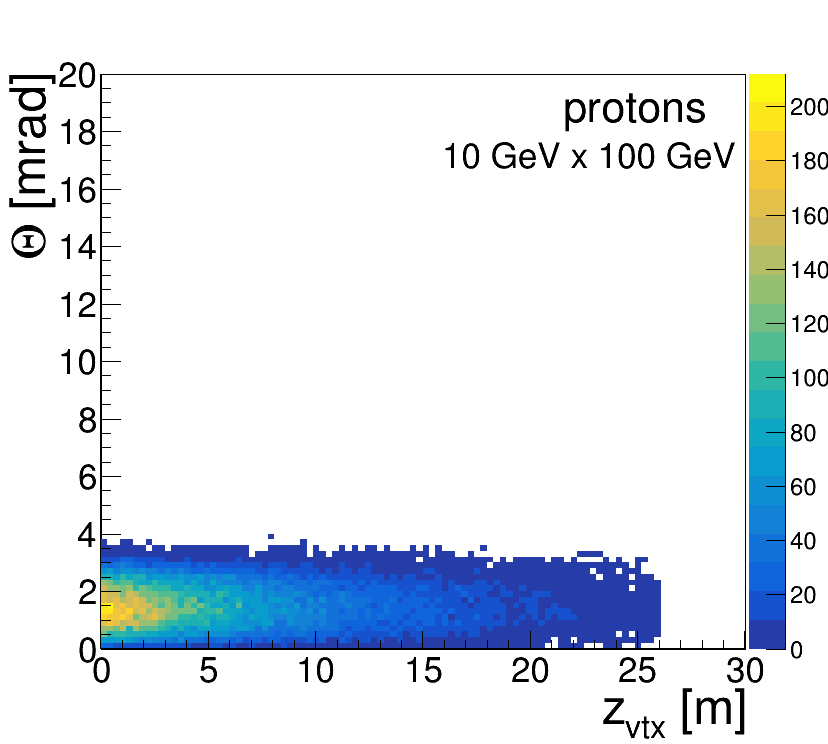}  
  \includegraphics[width=0.4\textwidth]{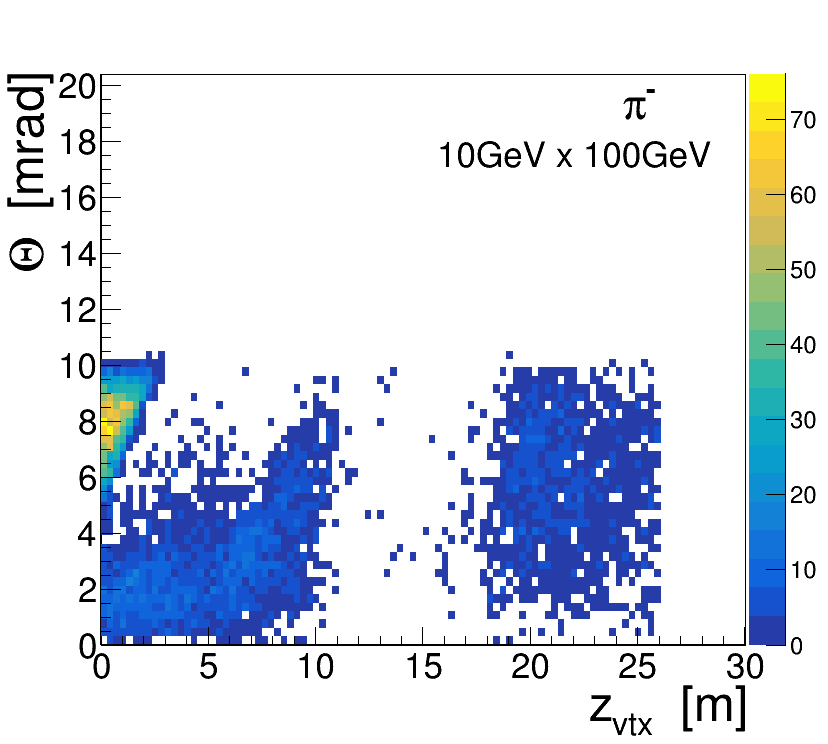}  
   \caption{$\Lambda \rightarrow p + \pi ^{-}$ decay in ${e+p \rightarrow e^\prime + X + \Lambda}$: Momentum (top) and angular (bottom) distributions of protons (left) and $\pi^-$ (right) at beam energy setting 10 GeV on 100 GeV with a luminosity of 100~$\fb^{-1}$, as registered in the far-forward detectors as a function of their origination (the decay vertex). For the $\pi^-$, the "dead" area in the Final-Focusing Quadrupole magnet region where placement of detectors is impossible is apparent.}
  \label{fig:L-p_Zvtx_100}
\end{figure}

\begin{figure}[!ht]
\includegraphics[width=0.45\textwidth]{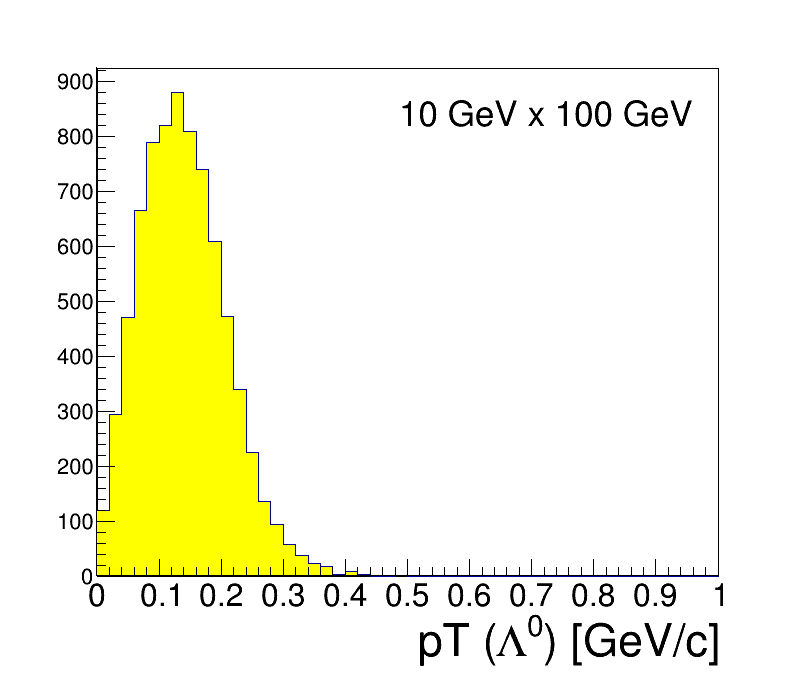}
\includegraphics[width=0.45\textwidth]{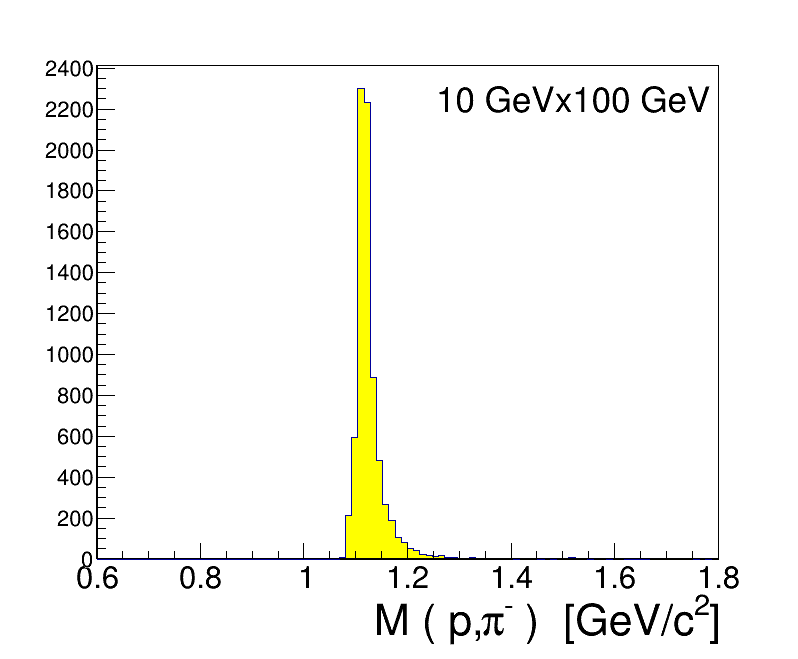}
\caption{ $\Lambda \rightarrow p + \pi ^{-}$ decay in ${e+p \rightarrow e^\prime + X + \Lambda}$: The $p_T$ ( left) and invariant mass (right) of the reconstructed $\Lambda$ particles for the beam energy settings of 10 GeV on 100 GeV with a luminosity of 100~$\fb^{-1}$. Vertical axes display unnormalized events.}
\label{fig:Mass}
\end{figure}

For the higher beam energy combination, for example 10 GeV on 100 GeV, the situation will be much different. Fig.~\ref{fig:L-p_Zvtx_100} shows the momentum and angular distributions for protons and $\pi^-$. For the latter, one can clearly see a ``dead" area appear along the beam line, where the FFQ beam elements are located, prohibiting placement of detectors and thus $\pi^-$ detection. This comes from the fact that these pions have significantly low momentum, and the beam magnet optics settings does not allow them to pass through this area, they get swept into the magnets and beam line. Those $\Lambda$s which decay after the set of FFQs will be tagged  by the off-momentum detector, but since the $Z_{vtx}$ is unknown, it will be hard (or impossible) to make a one-to-one correlation between the tagged position and the particle's momentum or angle. Therefore, for the final reconstruction of the $\Lambda$ invariant mass, one has to use only events with $ Z_{vtx} < 3-5$ meters, to make this correlation possible. That this indeed remains possible is shown in Fig~\ref{fig:Mass} (right panel), which shows the invariant mass spectra of the $\Lambda (p,\pi ^-) $ channel for this 10 GeV on 100 GeV beam energy setting. The corresponding $p_T$ spectrum of the $\Lambda$ particles is shown on the left panel of Fig~\ref{fig:Mass}.  Distributions for 5 GeV on 41 GeV are very similar, for 18 GeV on 275 GeV distributions were not considered as too few $\Lambda$ decays survive the $Z_{vtx}$ cut (see Table~\ref{tab:lambda_decay}).

\begin{table}[hbt]
    \caption{ $e+p \rightarrow e^\prime + X + \Lambda$: Final $\Lambda$ detection efficiency, as a function of beam energy combinations, for $\Lambda$ detection with a cut on decay applied of $Z_{vtx} < 4$\,m to ensure $\Lambda$-mass reconstruction. }
    \centering
    \begin{tabular}  {   l c c c c  c   }
    \hline
    \hline
 Beam energies  & 5 GeV on 41 GeV &  & 10 GeV on 100 GeV &  & 18 GeV on 275 GeV\\ 
      \hline 
Lambda Efficiency & 20\% & & 15\% & & 1\% \\  
      \hline
      \hline
    \end{tabular}
\label{tab:lambda}
\end{table}

We summarize this result in Table ~\ref{tab:lambda}, which shows the expected $\Lambda$ detection efficiency for the decay $\Lambda \rightarrow p + \pi^-$. A cut on decay within 4 meters, $Z_{vtx} < 4$m has been applied for this selection. The decrease in detection efficiency for the higher-energy settings comes mainly from this $Z_{vtx}$ cut, but is necessary to ensure $\Lambda$ mass reconstruction.


\paragraph{ $\bm{\Lambda \rightarrow n + \pi ^\circ}$}

\begin{figure}[htbp]
    \includegraphics[width=\textwidth]{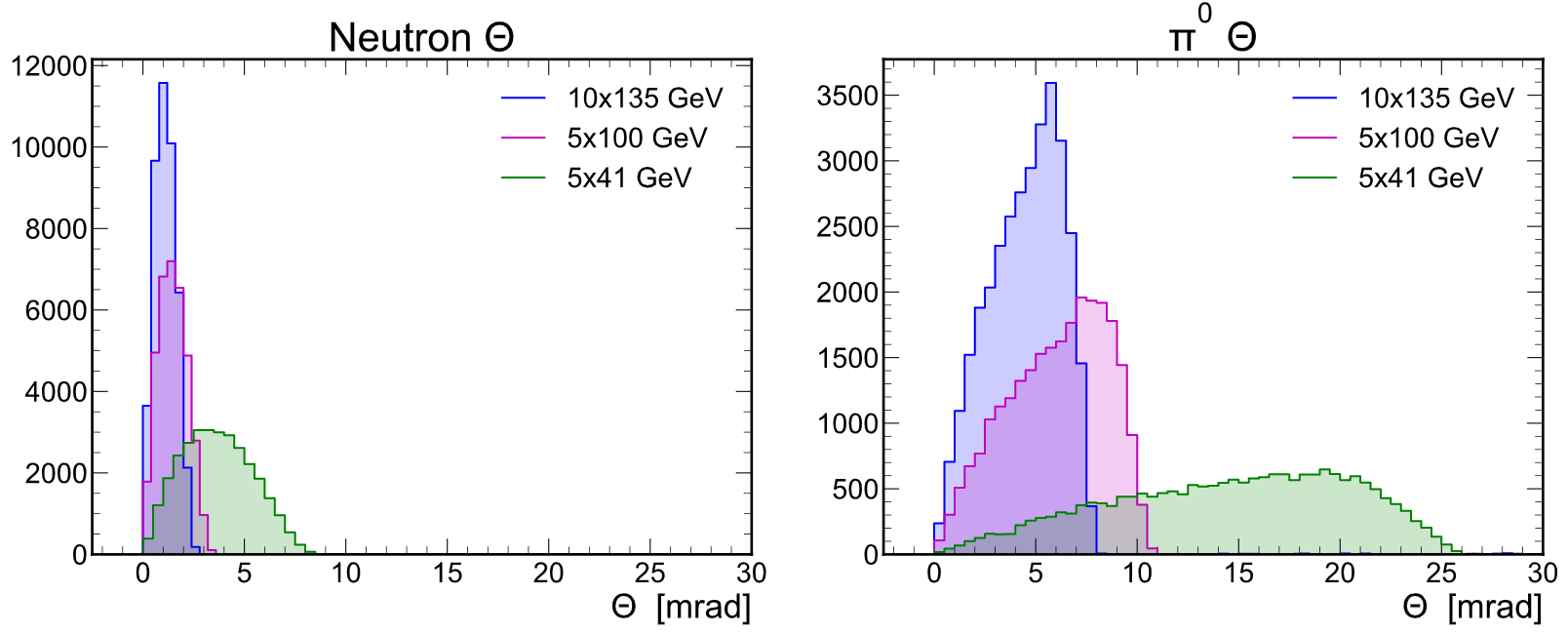}
    \caption[]{ 
        $\Lambda \rightarrow n + \pi ^{0}$ decay in ${e+p \rightarrow e^\prime + X + \Lambda}$: Angular distributions for neutrons (a)  and  $\pi^{0}$ (b), for beam energy settings 10 GeV on 135 GeV, 5 GeV on 100 GeV, and 5 GeV on 41 GeV with a luminosity of 100~$\fb^{-1}$. Vertical axes show unnormalized events.
        \label{fig:n_pi0_theta}
    }
\end{figure}

For this process we only have neutral particles in the final state. The main scheme of detection for these particles will be the ZDC and/or some kind of electromagnetic calorimeter/photon detector in the B0 area. Similar as for the $p + \pi^{-}$ decay mode, with lower beam energies more particles can be detected in the central detector region. Fig.~\ref{fig:n_pi0_theta} shows the angular ($\Theta$) distributions for $n$ and $\pi^{0}$ for different beam energies. It is furthermore assumed that the $\pi^{0}$ is reconstructed from $\pi^{0} \rightarrow \gamma\gamma$, where the photons are deposited in one of the corresponding detectors.


\begin{figure}[htbp]
    \includegraphics[width=\textwidth]{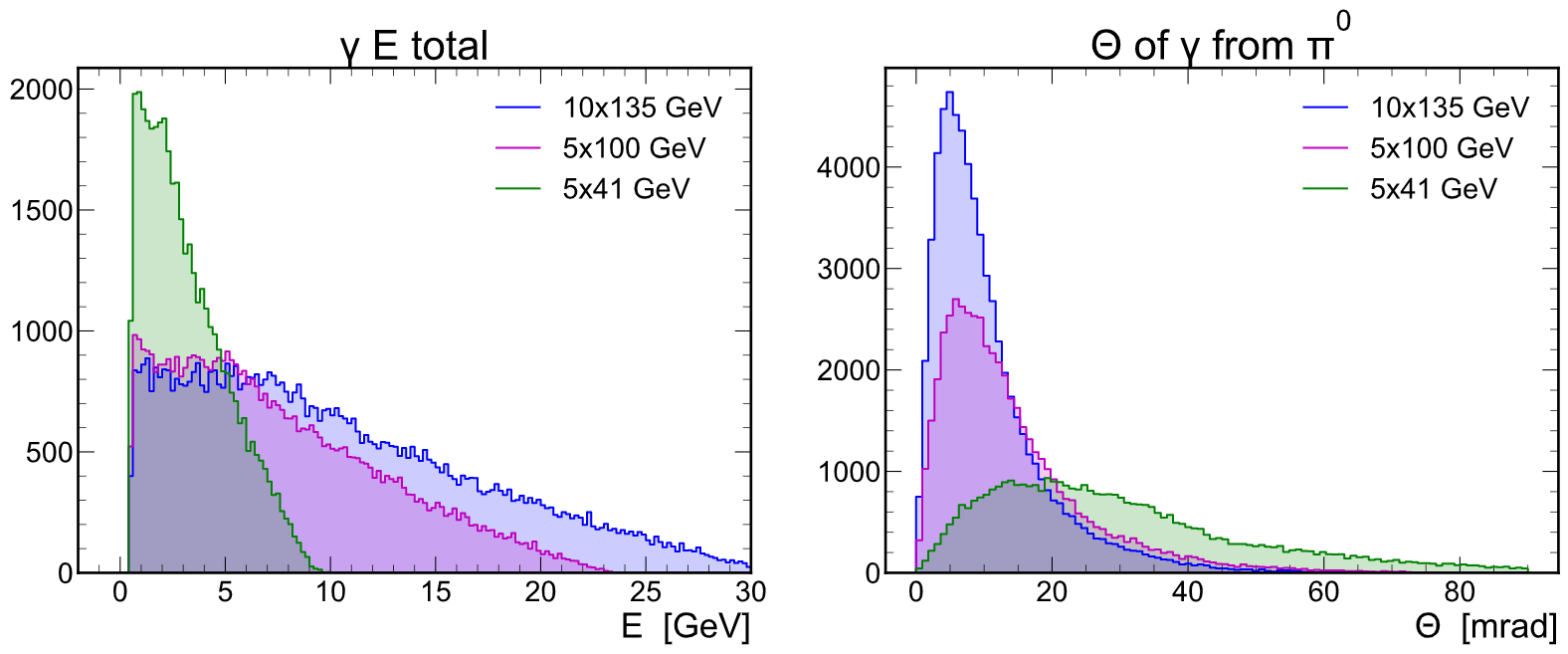}
    \caption[]{ $\Lambda \rightarrow n + \pi ^{0}$ decay in ${e+p \rightarrow e^\prime + X + \Lambda}$:
        Energy and angular $\Theta$ distributions for detected $\gamma\gamma$ from $\pi^{0}$ of $\Lambda$ decay with a luminosity of 100~$\fb^{-1}$. Vertical axes show unnormalized events.
        \label{fig:gam_etot_theta}
    }
\end{figure}

\begin{figure}[htbp]
    \includegraphics[width=\textwidth]{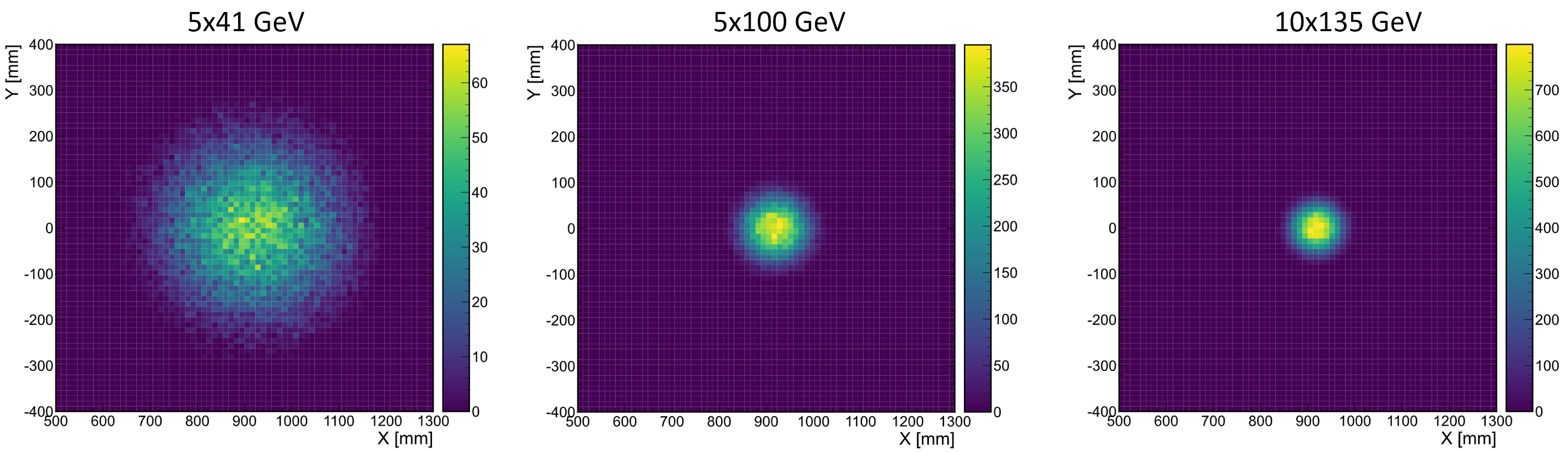}
    
    \vspace*{\floatsep}
    
    \includegraphics[width=\textwidth]{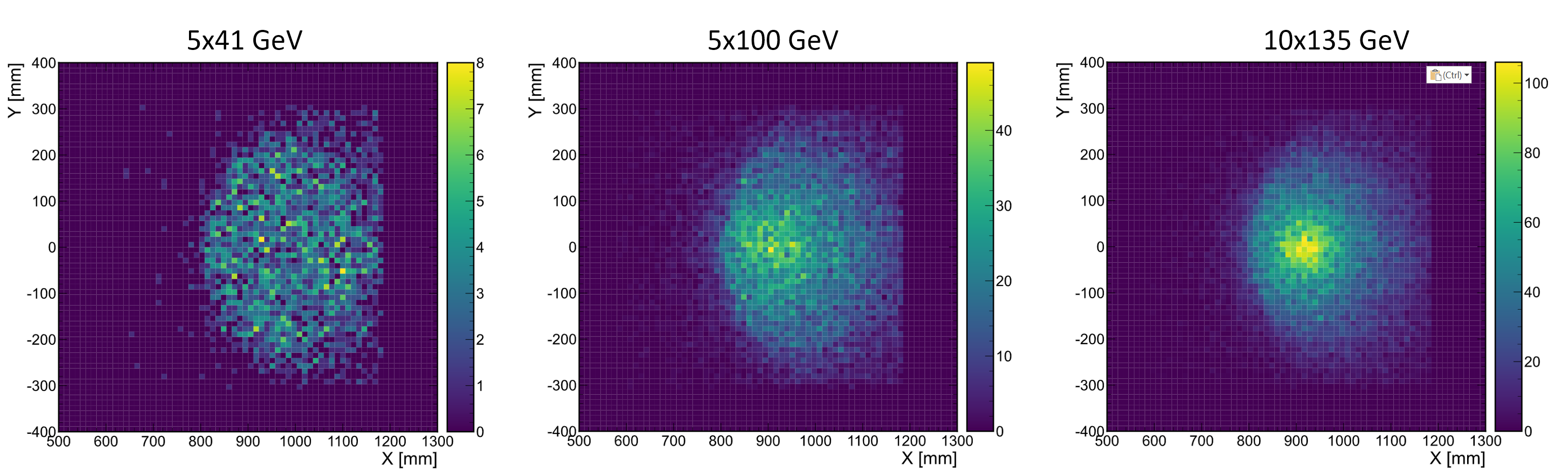}
    \caption[]{ 
        $\Lambda \rightarrow n + \pi ^{0}$ decay in ${e+p \rightarrow e^\prime + X + \Lambda}$: Occupancy distribution for neutrons (top panels) and $\gamma\gamma$ from $\pi^{0}$ decay (bottom panels) as detected in the ZDC for different beam energy settings with a luminosity of 100~$\fb^{-1}$.
        \label{fig:gam_zdc_diff}
    }
\end{figure}

The energy and angular distributions of the two photons from the $\pi^{0}$ decay are shown in Fig.~\ref{fig:gam_etot_theta}, for various beam energy settings. At lower beam energy settings, like 5 GeV on 41 GeV, some measurements to detect the larger-angle photons in the B0 area is required to recapture efficiency. As the beam energy increases, the ZDC starts playing the main role for detection of both neutrons and neutral-pions.
Fig ~\ref{fig:gam_zdc_diff} shows occupancy plots of $n$ and $\gamma\gamma$ used for $\pi^{0}$ reconstruction for different energy settings. 

\begin{figure}[htbp]
    \centering
\includegraphics[width=0.34\linewidth]{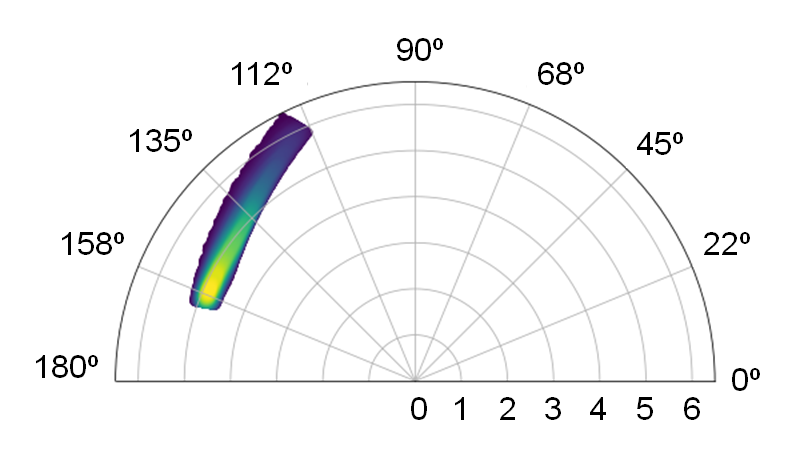}
\includegraphics[width=0.32\linewidth]{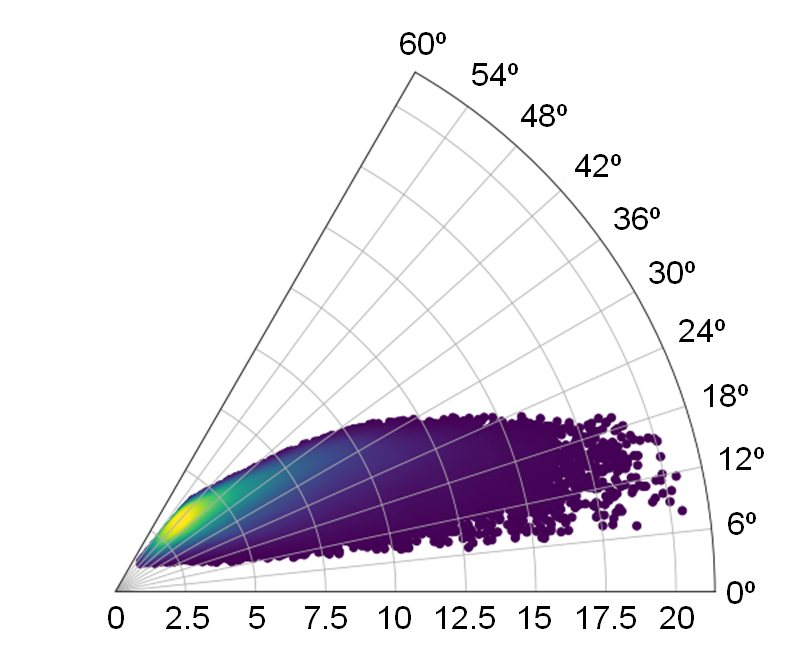}
\includegraphics[width=0.32\linewidth]{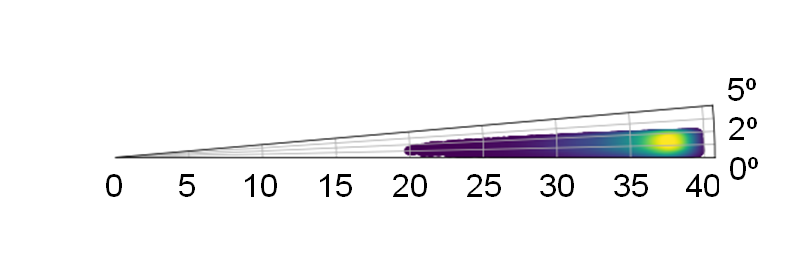}

\includegraphics[width=0.34\linewidth]{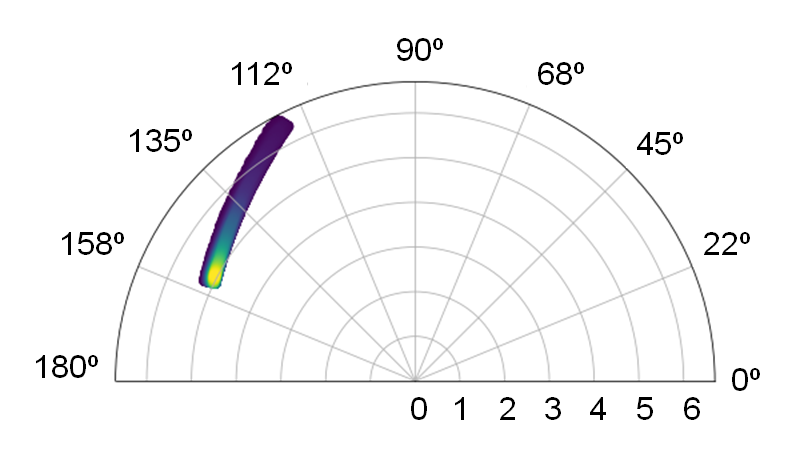}
\includegraphics[width=0.32\linewidth]{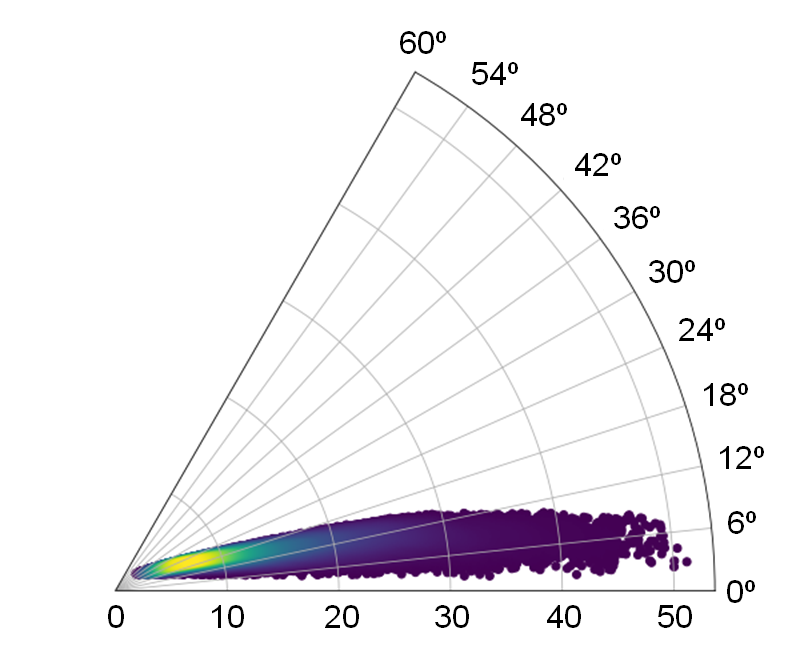}
\includegraphics[width=0.32\linewidth]{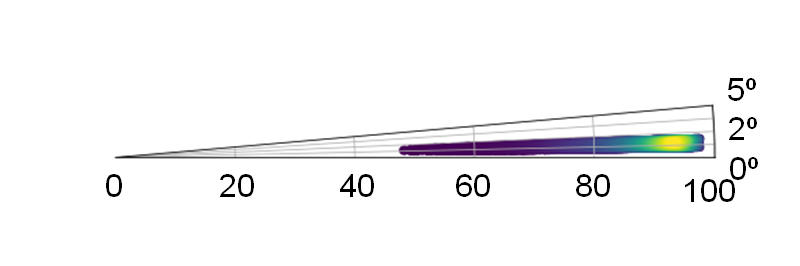}

\includegraphics[width=0.34\linewidth]{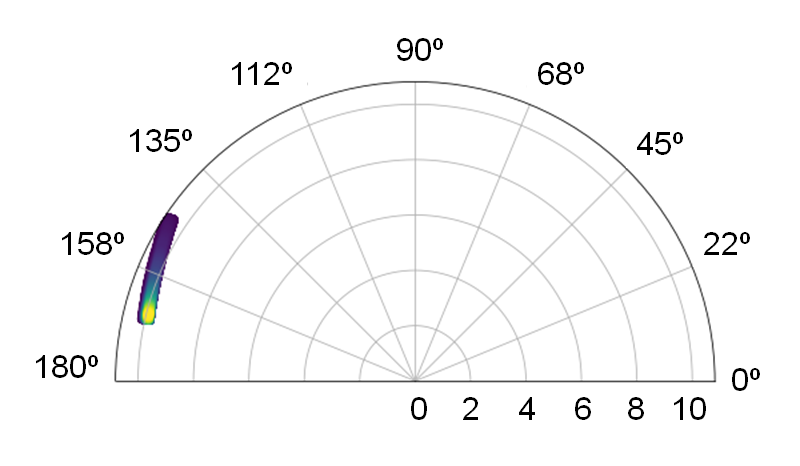}
\includegraphics[width=0.32\linewidth]{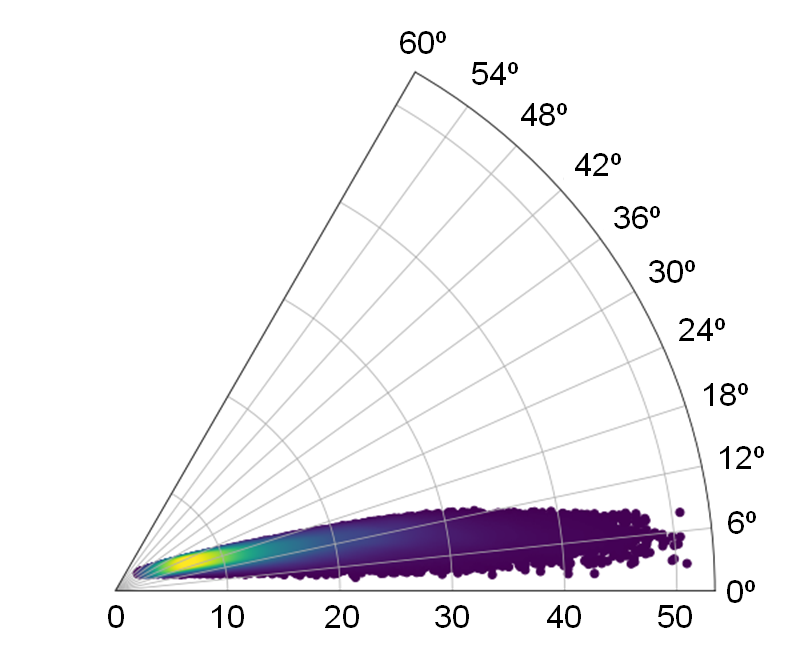}
\includegraphics[width=0.32\linewidth]{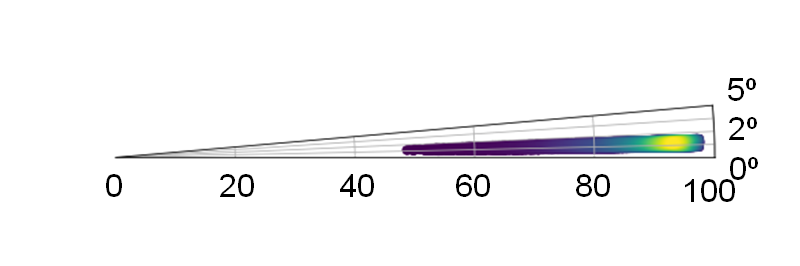}

    \caption{Kinematic distributions for exclusive $p(e,e'\pi^+n)$ events for $e'$ (left), $\pi^+$ (center), and $n$ (right), at beam energies of 5 GeV on 41 GeV (top), 5 GeV on 100 GeV (middle), and 10 GeV on 100 GeV (bottom) with a luminosity of 100~$\fb^{-1}$. The radial component is momentum [GeV], and the polar coordinate is the scattering angle, with the proton (electron) beam direction pointing to the right (left).}
    \label{fig:fpi_kin}
\end{figure}

\subsubsection{Exclusive \texorpdfstring{$p(e,e'\pi^+n)$}{} events}

Simulations demonstrating the feasibility of pion electric form factor measurements at the EIC have been performed using 
 a Deep Exclusive Meson Production (DEMP) $p(e,e'\pi^+n)$ event generator based upon the Regge model of Ref.~\cite{Choi:2015yia}. This model provides an excellent description of the existing JLab data up to $-t$=4.3 GeV$^2$ \cite{Basnet:2019cpg} and is well-behaved over a wide kinematic range. The kinematic distributions for exclusive $p(e,e'\pi^+n)$ events used to extract pion form factors (Sec.~\ref{part2-subS-SecImaging-FF}) are shown in Fig.~\ref{fig:fpi_kin}. As for tagged DIS events, the neutrons assume nearly all of the proton  beam  momentum, and need to be detected at very forward angles in the ZDC. The scattered electrons and pions have also similar momenta as in the tagged DIS case, except that here the electrons are distributed over a wider range of angles. E.g., for the 5 GeV on 100 GeV beam energy setting, the 5-6~GeV/$c$ electrons are primarily scattered 25$^\circ$-45$^\circ$ from the electron beam, while the 5-12~GeV/$c$ $\pi^+$ are at 7$^\circ$-30$^\circ$ from the proton beam.

\subsubsection{Accelerator and instrumentation requirements}
As one can see from the detector simulation examples shown, access to meson structure physics greatly benefits from EIC operations at the lower center-of-mass energies. Apart from that there is need for both $ep$ and $ed$ measurements at similar center-of-mass energies. Lower energies enhance the range of $Q^2$ at large $x_\pi$.  The detection needs to uniquely tag kaon structure require lower energies to enhance $\Lambda$ decay probability at short distances.  This allows $\Lambda$-mass reconstruction from the detected decay products. To tag the pion and kaon structure, proper instrumentation of B0 tracking detectors is needed, requiring full azimuthal coverage and perhaps pushing a smaller proton-beam pipe diameter. Off-momentum detectors have to also provide this full azimuthal coverage for detection of negatively-charged decay particles.

\subsection{Deuteron DIS with spectator tagging: Free neutron structure and nuclear modifications}

DIS on the deuteron with detection of the spectator nucleon (``spectator tagging''),
\begin{equation}
e + d \rightarrow e' + X + N, \hspace{2em} N = \textrm{$p$ or $n$}
\end{equation}
offers a unique
method for extracting the free neutron structure functions and studying the nuclear modifications of
proton and neutron structure (EMC effect, antishadowing, shadowing). Detection of the spectator nucleon
identifies the active nucleon in the DIS process and eliminates dilution. Measurement of the
spectator nucleon momentum (typically $p_N \lesssim$ 300 MeV/$c$ in the deuteron rest frame) controls
the nuclear configuration in the deuteron initial state and permits a differential analysis of
nuclear effects, enhancing the experimental reach and theoretical accuracy of the analysis.
In DIS with proton tagging, on-shell nucleon extrapolation in the spectator proton momentum selects large-size
$pn$ configurations in the deuteron where nuclear interactions are absent and provides a
model-independent method for extracting the free neutron structure functions \cite{Sargsian:2005rm}.
In DIS with proton or neutron tagging, measurements with spectator momenta $p \sim$ 200-600 MeV/$c$
(rest frame) select small-size $pn$ configurations where the EMC effect is enhanced and can
be studied systematically as a function of the size of the $pn$ configuration (tagged EMC effect).
Measurements of bound proton structure in deuteron DIS with neutron tagging can be compared with
free proton structure measured in proton DIS, transforming the analysis of nuclear modifications
(independent normalization, size of modifications). The theoretical framework for tagged DIS on the
deuteron is being developed, including final-state interactions and polarization
\cite{JLab_TDIS_LDRD,Strikman:2017koc,Frankfurt:1983qs,Cosyn:2019hem,Cosyn:2020kwu}.
%
%
\begin{figure}[b]
\parbox[c]{0.25\textwidth}{\includegraphics[width=0.25\textwidth]{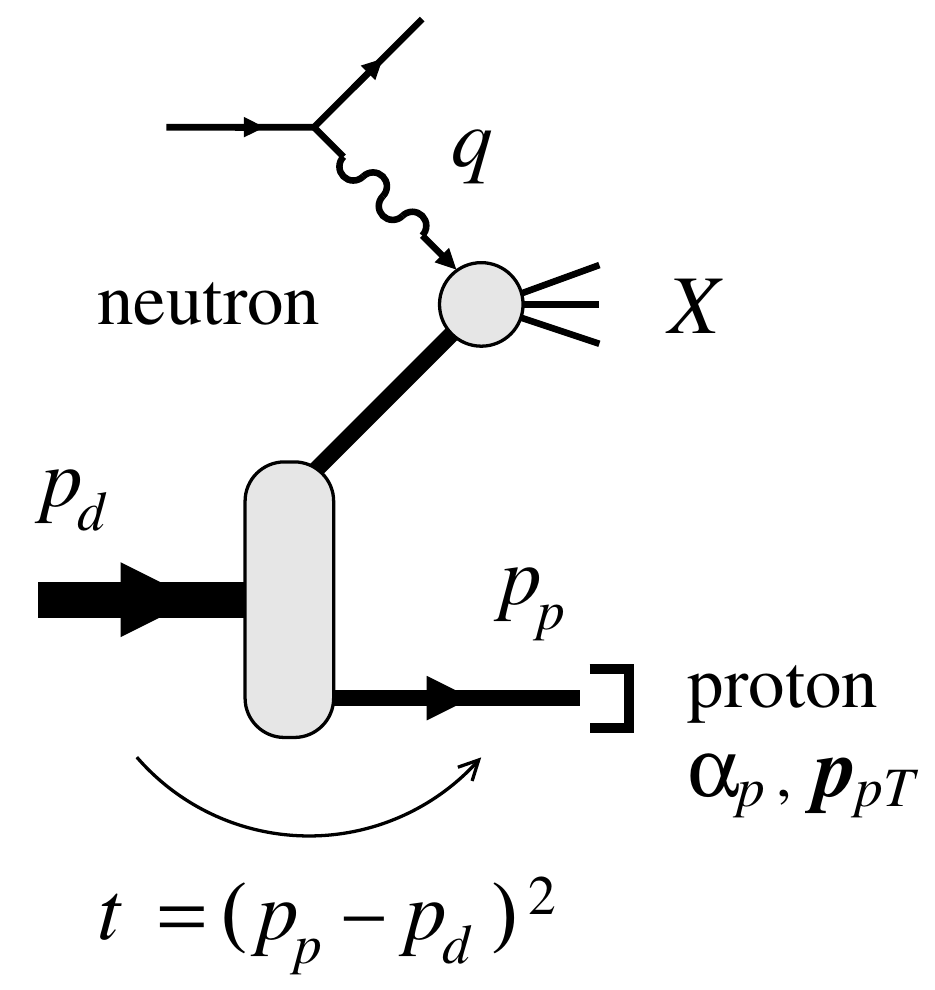}}
\hspace{0.1\textwidth} 
\parbox[c]{0.6\textwidth}{
\caption[]{\small Deuteron DIS with spectator proton tagging, $e + d \rightarrow e' + X + p$. 
The measured proton momentum is described by the light-cone fraction $\alpha_p = 2 p_p^+/p_d^+$
and transverse momentum $p_T$ in the deuteron-photon collinear frame. The proton controls the
neutron virtuality $t - m^2$ in the deuteron (off-shellness). The on-shell point can be reached
by extrapolation to $p_T^2 \rightarrow -a_T^2$, where $a_T^2 (\alpha_p)$ is the pole position in
the deuteron light-cone wave function and is of the order $\sim \epsilon_d m$ ($\epsilon_d = 2.2$ MeV
is the deuteron binding energy, $m$ is the nucleon mass). The graph shows the cross section in
the impulse approximation; final-state interactions and polarization effects are discussed in
Refs.~\cite{Strikman:2017koc,Frankfurt:1983qs,Cosyn:2019hem,Cosyn:2020kwu}.
}
\label{fig:tagging_diagram}
}
\end{figure}

In tagged DIS at EIC, the spectator nucleon moves in the forward ion direction with $\approx 1/2$
the deuteron beam momentum, with an offset that is determined by its boosted rest-frame momentum
(its light-cone momentum fraction is conserved): $p_{\parallel, N}[\textrm{collider}]
= x_L p_{\parallel, d}$ with $x_L \approx \frac{1}{2} (1 + p_{\parallel, N}[\textrm{restframe}]/m) =$ 0.35--0.65,
$p_{\perp, N}[\textrm{collider}] = p_{\perp, N}[\textrm{restframe}] \lesssim 300$ MeV/$c$.
Proton spectators are detected with the forward spectrometer; neutron spectators are detected
with the ZDC. Generic detector requirements are: (a) Acceptance for proton and neutron spectators
in the given $x_L$ and $p_\perp$ range; (b) Proton longitudinal momentum resolution
$\delta x_L/x_L \ll 10^{-2}$ and transverse momentum resolution $\delta p_\perp/p_\perp \approx$ 20--30 MeV/$c$;
(c) neutron momentum resolution of $9-11$ GeV/$c$ for $p > 110$ GeV/$c$ and $p_{T}$-resolution of
40--80 MeV/$c$, assuming a ZDC with energy resolution $\sigma_E < {50\%}/{\sqrt{E}}$ for hadrons,
and an angular resolution of 3 mrad/$\sqrt{E}$ as assumed for other e+d studies
in this report 
(see Sec.~\ref{subsec:deuteronSpectatorTag} for details).

\begin{figure}[th]
\includegraphics[width=\textwidth]{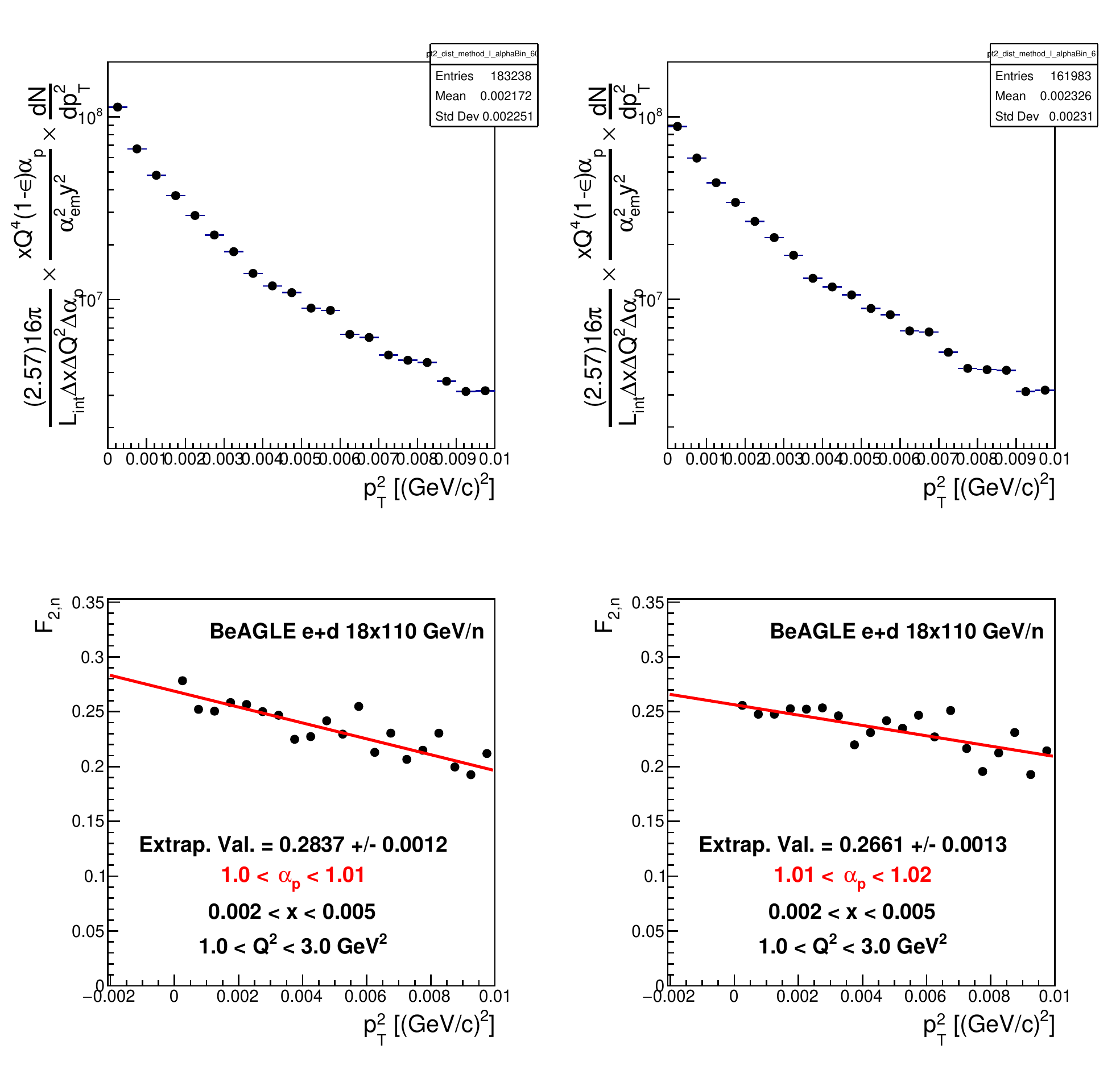}
\caption[]{Top row: $p_{T}^{2}$ distributions for two proton
spectator $\alpha_{p}$ bins in the same $x$ and $Q^{2}$ bins (cuts are
shown in the text on the bottom row of plots). The distributions in
the top row include the bin-weighting by the flux factor, as well as
all constants included to achieve the final, scaled histograms. Bottom row:
The same distribution multiplied by the inverse spectral function pole factor
$(p_{T}^{2} + a_{T}^{2})^{2} /[2(2\pi)^{3} R]$ \cite{Cosyn:2020kwu}. The resulting dependence is then
used to perform the on-shell extrapolation to calculate the neutron
structure functions. The red line shows the first-degree polynomial
used to fit the data and perform the extrapolation to $p_{T}^{2} \rightarrow -a_{T}^{2}$.}
\label{fig:tagging_onshell_simulation}
\end{figure}

Simulations have been performed of neutron structure extraction with deuteron DIS with spectator
proton tagging and pole extrapolation in the proton momentum
(see Fig.~\ref{fig:tagging_onshell_simulation}); see Ref.~\cite{JLab_TDIS_LDRD} for
earlier studies. The kinematic variables and
theoretical method are described in Refs.~\cite{Strikman:2017koc,Cosyn:2020kwu}.
The spectator proton momentum is specified by the
light-cone fraction $\alpha_p = 2 p_p^+/p_d^+$ and transverse momentum $p_T$ in the deuteron-photon
collinear frame. Electron-deuteron DIS events are generated with BeAGLE \cite{Beagle}.
The top plots in Fig.~\ref{fig:tagging_onshell_simulation} show the kinematic distribution of events
in $p_T^2$ in two bins of $\alpha_p$, as determined by the strong momentum dependence of the deuteron
spectral function. The bottom plots show the kinematic distributions after removal of the deuteron
pole factor $(p_T^2 + a_T^2)^2 / [2 (2\pi)^3 R]$, which removes most of the momentum dependence,
and the pole extrapolation to unphysical $p_T^2 \rightarrow - a_T^2$, which gives the free neutron
structure function $F_{2n}$ \cite{Cosyn:2020kwu}.
The plots show the procedure in a typical $(x, Q^2)$ bin in EIC kinematics;
similar results are obtained in other bins. The procedure has been verified by comparing the extrapolation
results with the free nucleon structure functions in the physics model. The study of detector and
beam smearing effects is in progress.
%
%
%

%
%
\subsection{Diffractive \texorpdfstring{$\jpsi$}{} production on the deuteron with spectator tagging}
\label{subsec:deuteronSpectatorTag}

Diffractive $\jpsi$ production on the deuteron with the goal of probing short-range correlations requires that the spectator nucleon is tagged, see Sec.~\ref{part2-subS-LabQCD-ShortRange}.   In the EIC, these spectators are moving with the hadron beam will end up in the far-forward region of the EIC.
Some general considerations used to establish baseline particle 
acceptance and detector resolutions via full simulations in Geant4~\cite{Agostinelli:2002hh} are presented here and  further details can be found online~\cite{Tu:2020ymk}.

Four different forward detectors are considered in the current study:
the $B0$ silicon tracker, off-momentum detectors, Roman pots and ZDC. The location, size, resolution of transverse and total momentum, energy, and scattering angle of these four detectors used in the simulation are summarized in Ref.~\cite{Tu:2020ymk}. In addition to the intrinsic detector related effects, the beam related effects, smearing of the three-momentum components of the nominal deuteron beam are carried out using Gaussian smearing with a width proportional to the values of angular divergence and beam energy spread. The modified deuteron beam four-vector is used to calculate a Lorentz boost vector. The final state protons and neutrons are then boosted from the lab-frame to the deuteron rest frame using the original, unsmeared deuteron boost vector, and then boosted back to the lab-frame using the smeared boost.

With the current baseline design of the far-forward detectors  and the consideration of beam-related effects, the measurement of diffractive $\jpsi$ production in electron-deuteron collision is experimentally possible with good precision up to 800 MeV in internal nucleon momentum. The gluon density distributions can be obtained via measurement of the momentum transfer distributions in different bins of nucleon momentum, where the underlying mechanism of gluon dynamics can be directly studied in terms of short-range nuclear correlations. Depending on the $Q^{2}$ and nucleon momentum range, the required integrated luminosity is between 30 to 500 $\rm{fb^{-1}}$.

%
%
\subsection{Double tagging in the far forward region}

As discussed in section~\ref{part2-subS-LabQCD-ShortRange}, one can do very interesting and unique physics with the A=3 nuclei if it is possible to tag the spectator protons from $^3$He or the spectator neutrons from $^3$H.

\begin{figure}[htb]
    \centering
    \includegraphics[width =0.49\textwidth]{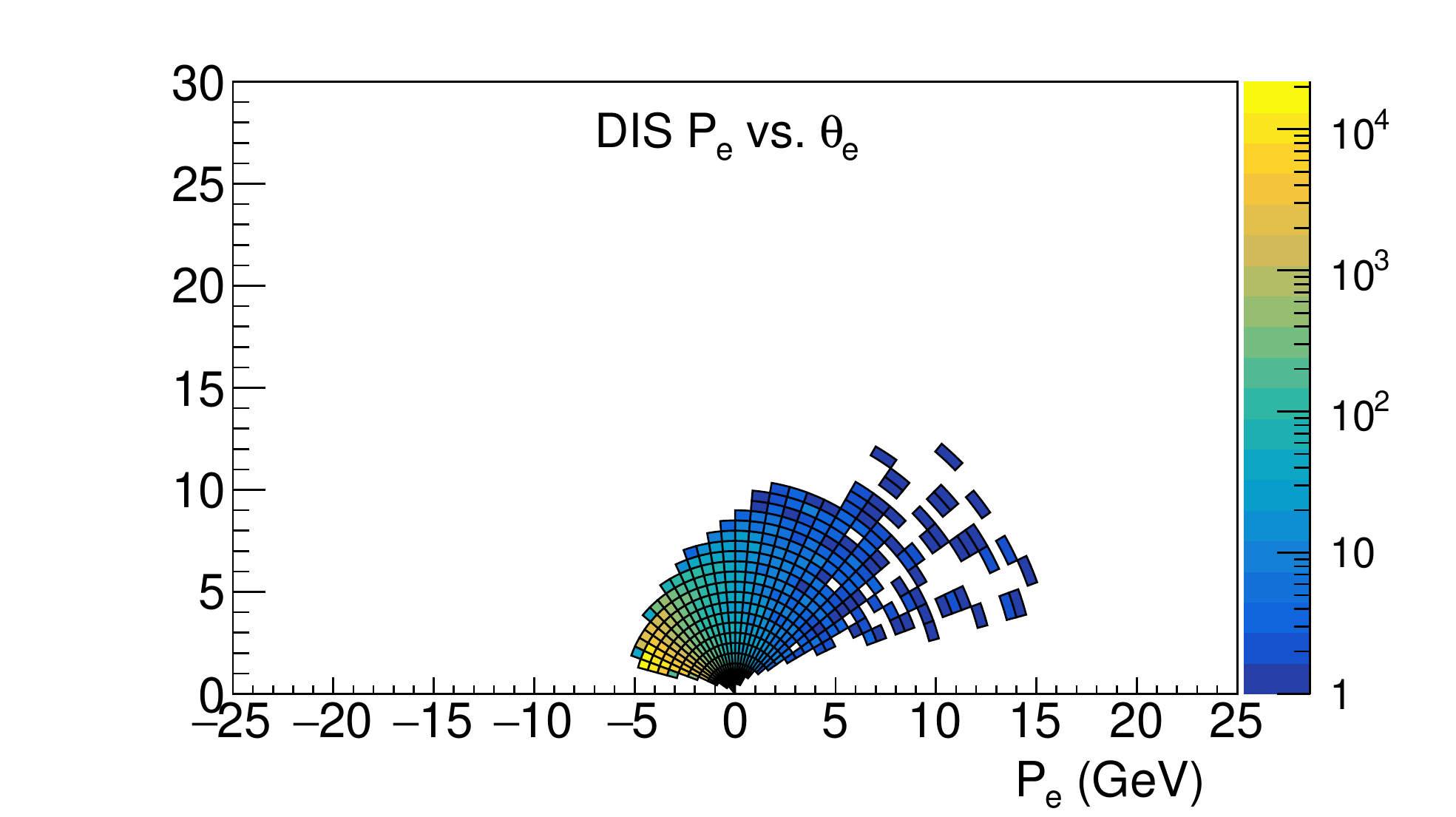}
    \includegraphics[width = 0.49\textwidth]{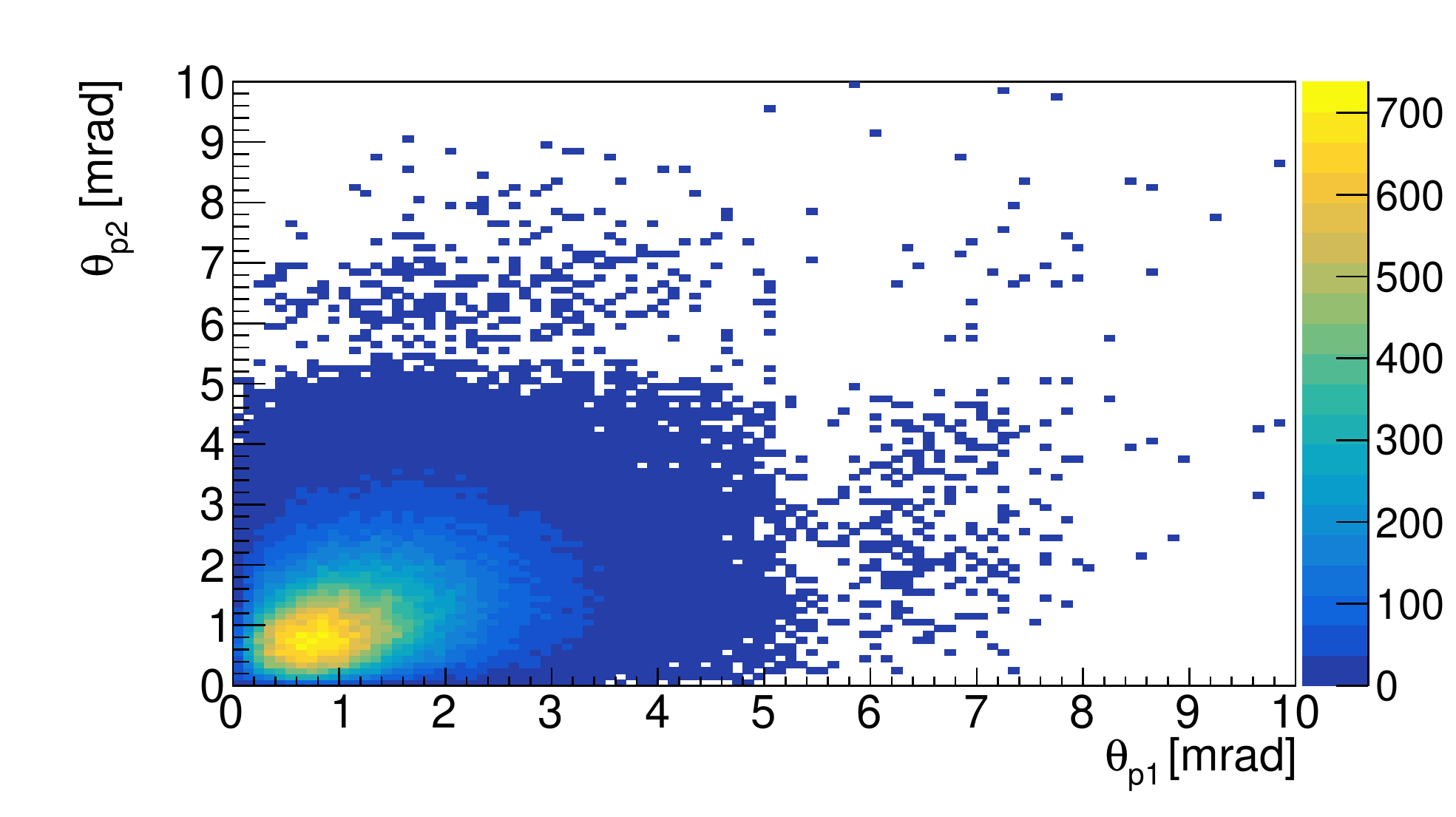}
    \caption{DIS kinematics for the $^3$He(e,e'pp)X reaction at 5x41 GeV with the protons detected in the far forward region.}
    \label{qe-41x5_pp}
\end{figure}

\begin{figure}[htb]
    \centering
    \includegraphics[width =0.49\textwidth]{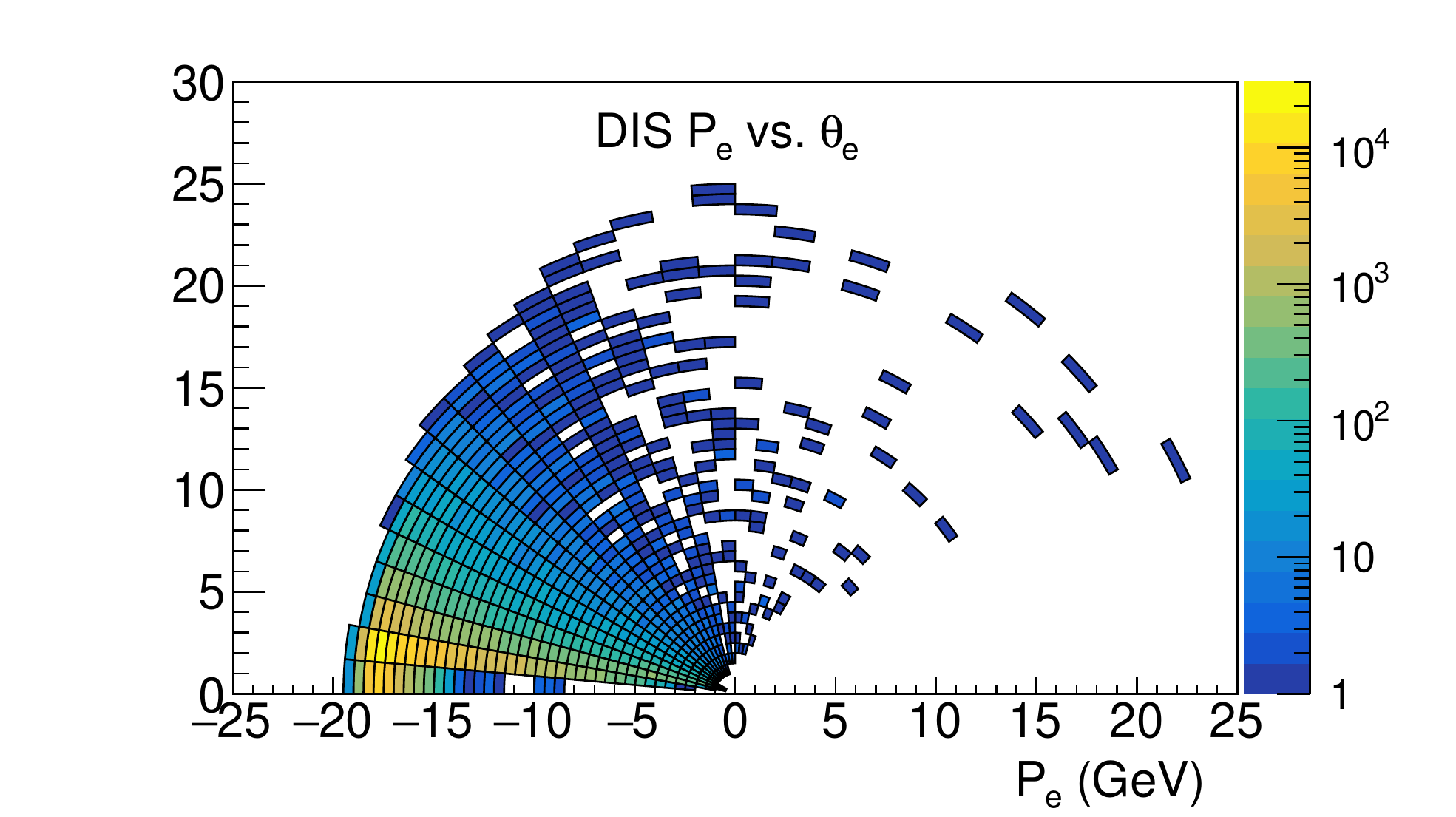}
    \includegraphics[width = 0.49\textwidth]{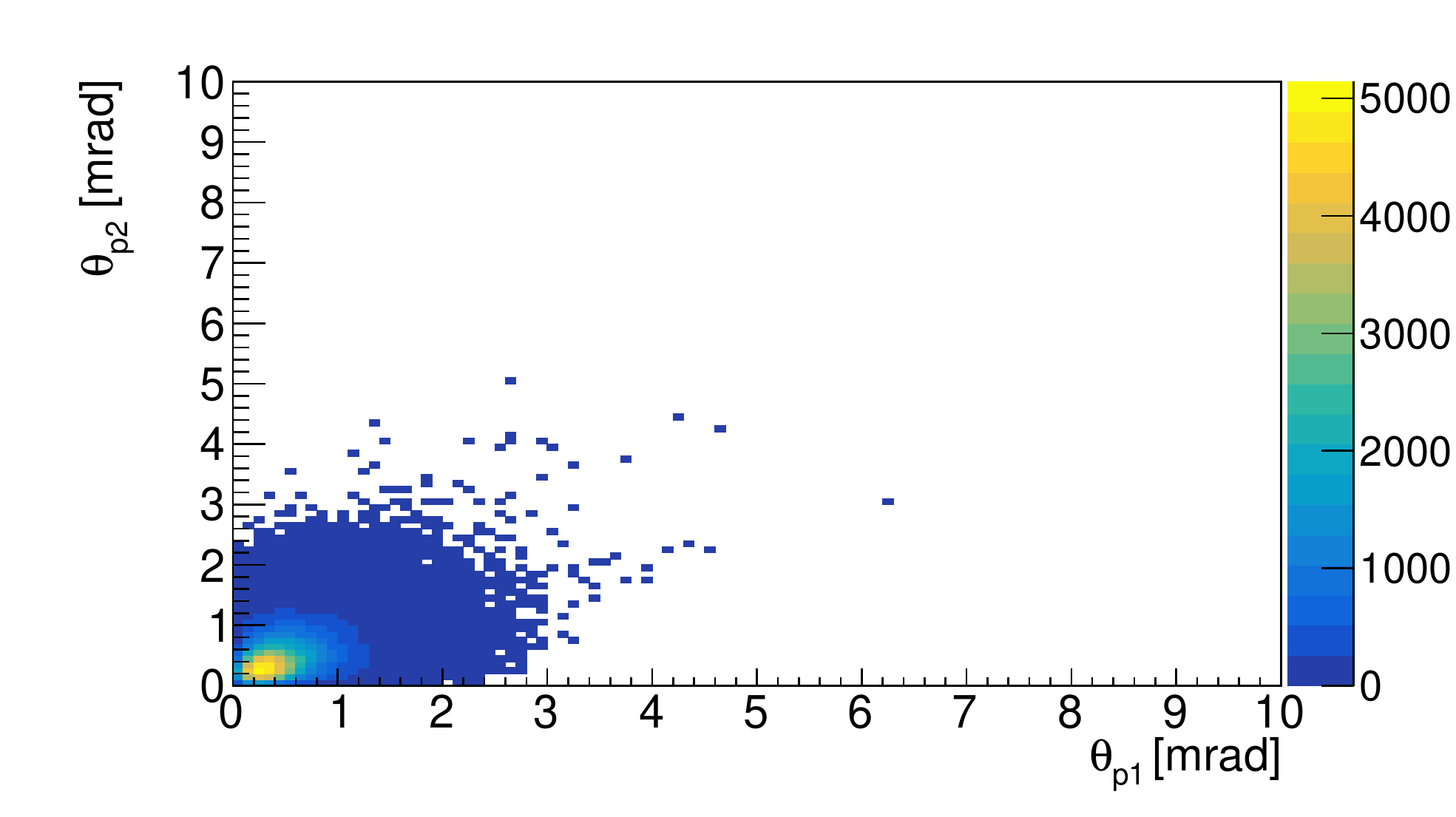}
    \caption{DIS kinematics for the $^3$He(e,e'pp)X reaction at 18x110 GeV.   Here the protons are more kinematically focused then in the lower energy case and will be better detected in the Roman pots.}
    \label{qe-110x18_pp}
\end{figure}

\begin{figure}[htb]
    \centering
    \includegraphics[width =0.49\textwidth]{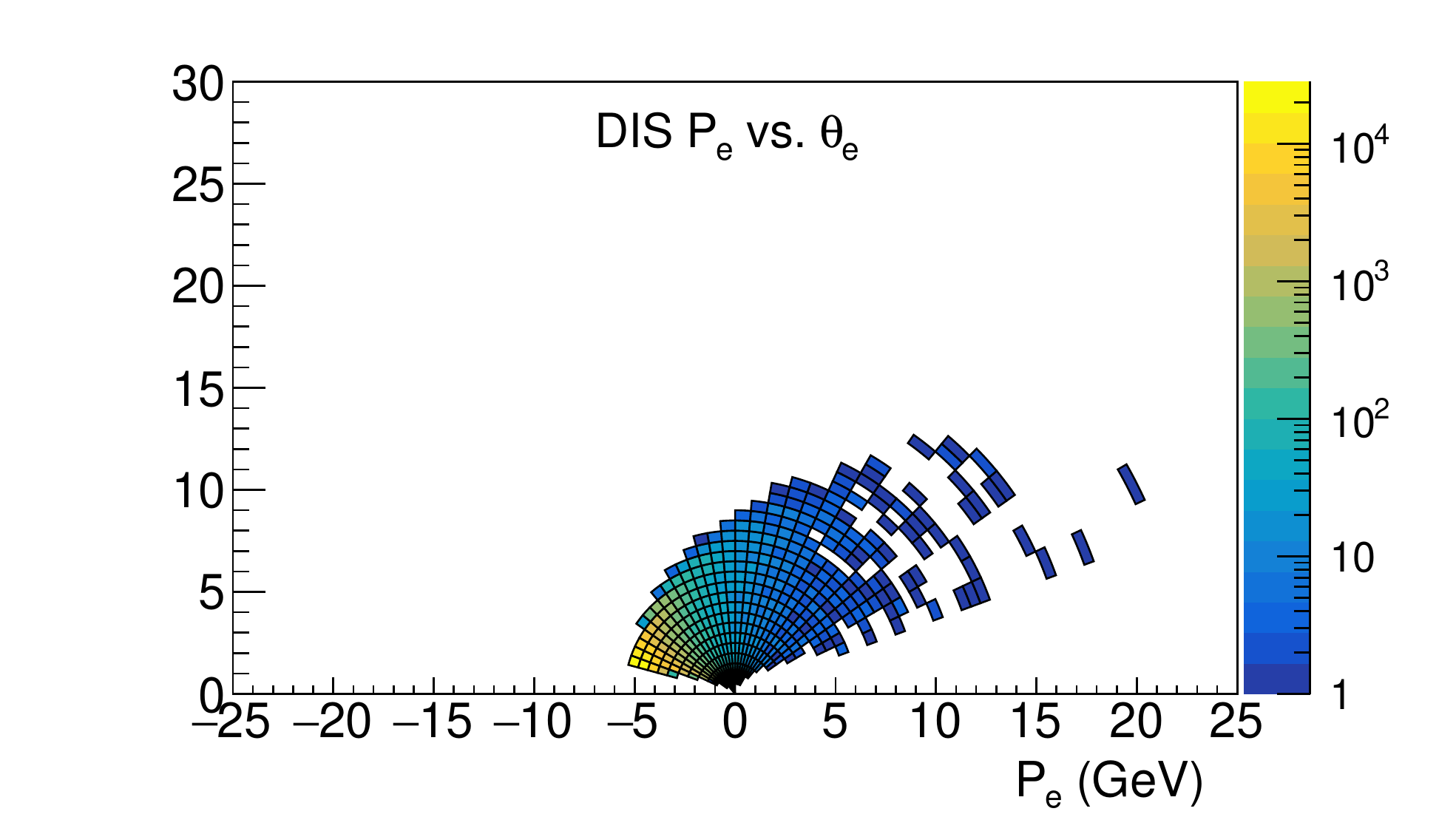}
    \includegraphics[width = 0.49\textwidth]{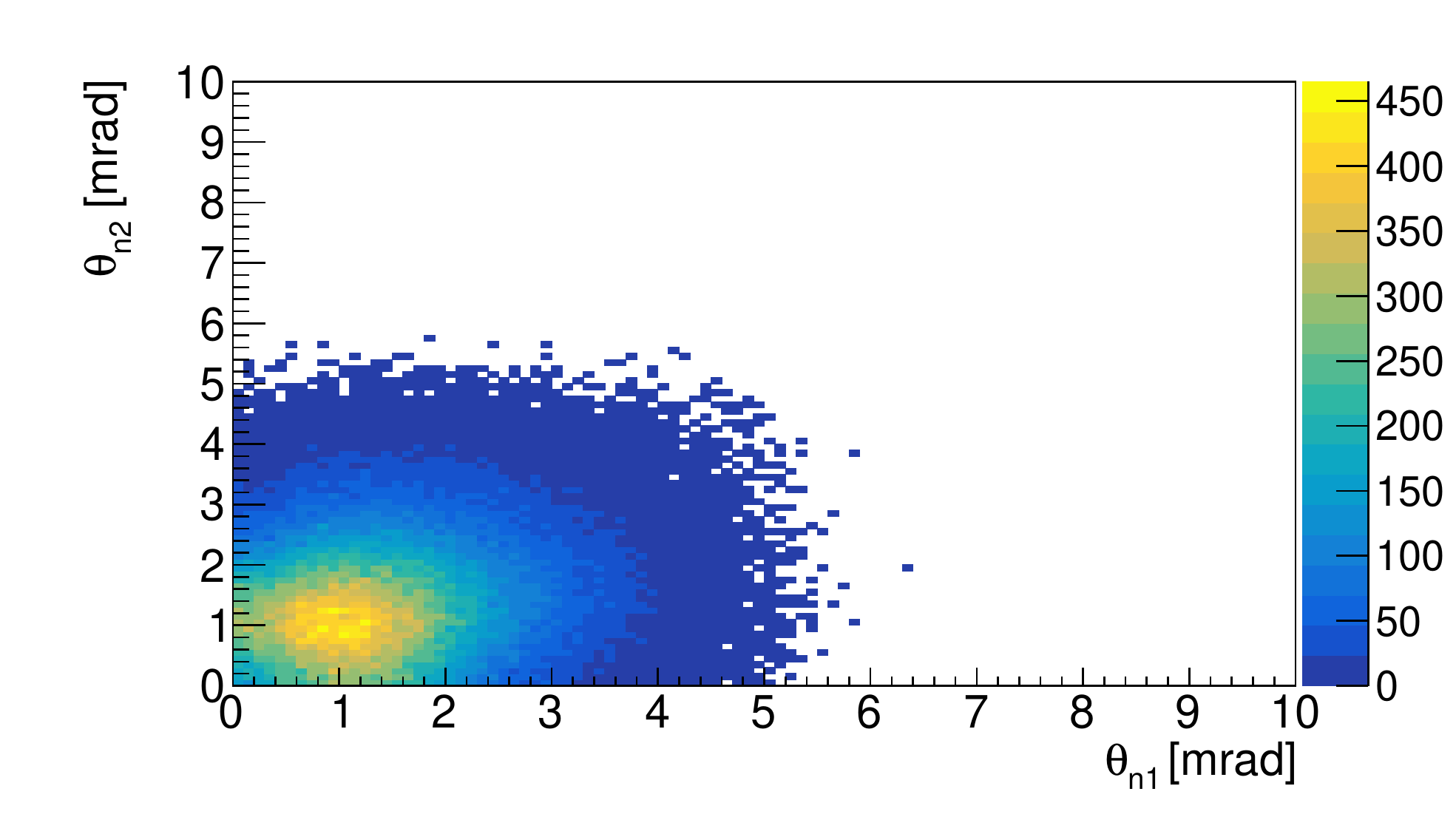}
    \caption{DIS kinematics for the $^3$H(e,e'nn)X reaction at 5x41 GeV with the neutrons detected in the far forward region.}
    \label{qe-41x5_nn}
\end{figure}

\begin{figure}[htb]
    \centering
    \includegraphics[width =0.49\textwidth]{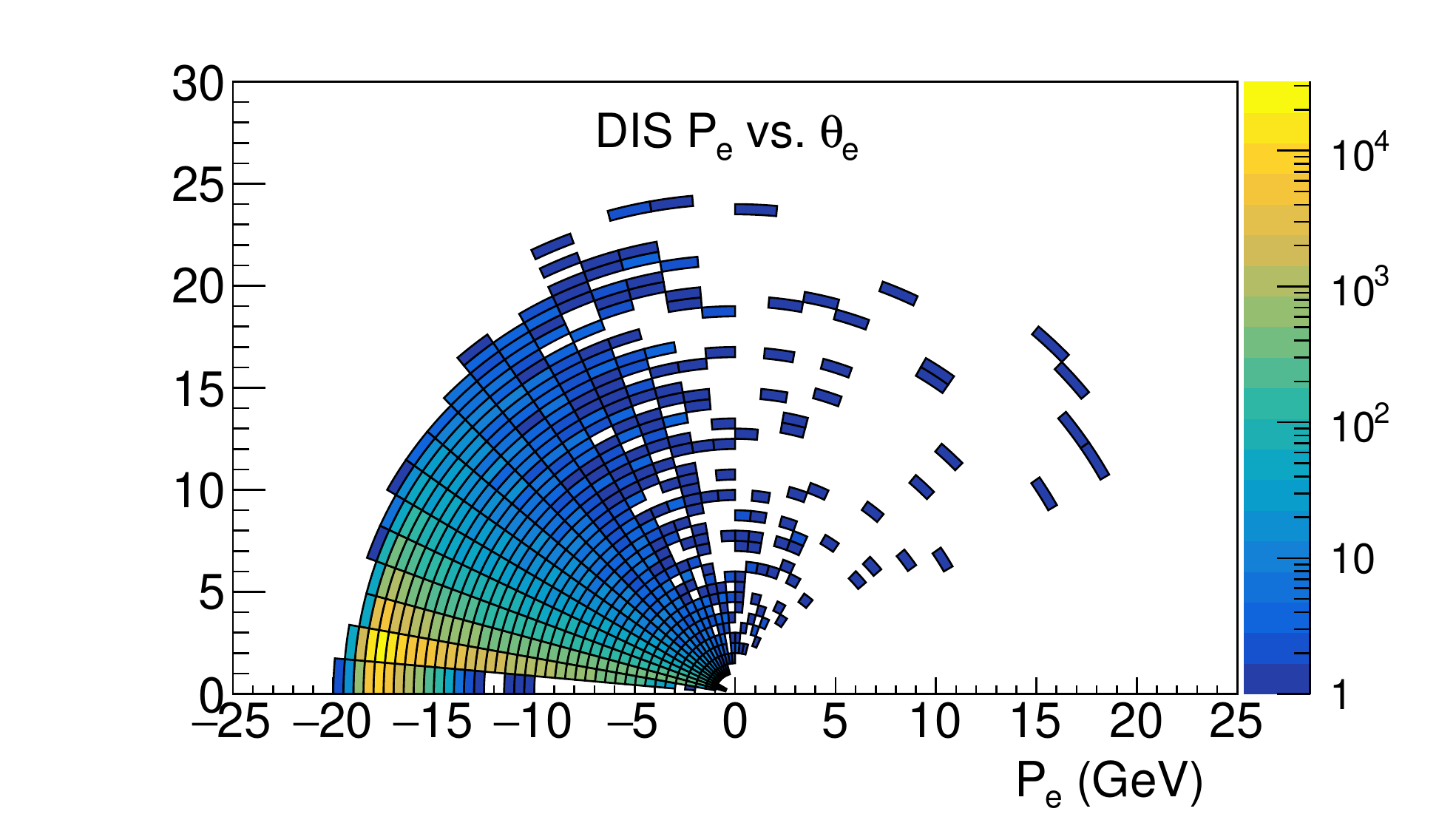}
    \includegraphics[width = 0.49\textwidth]{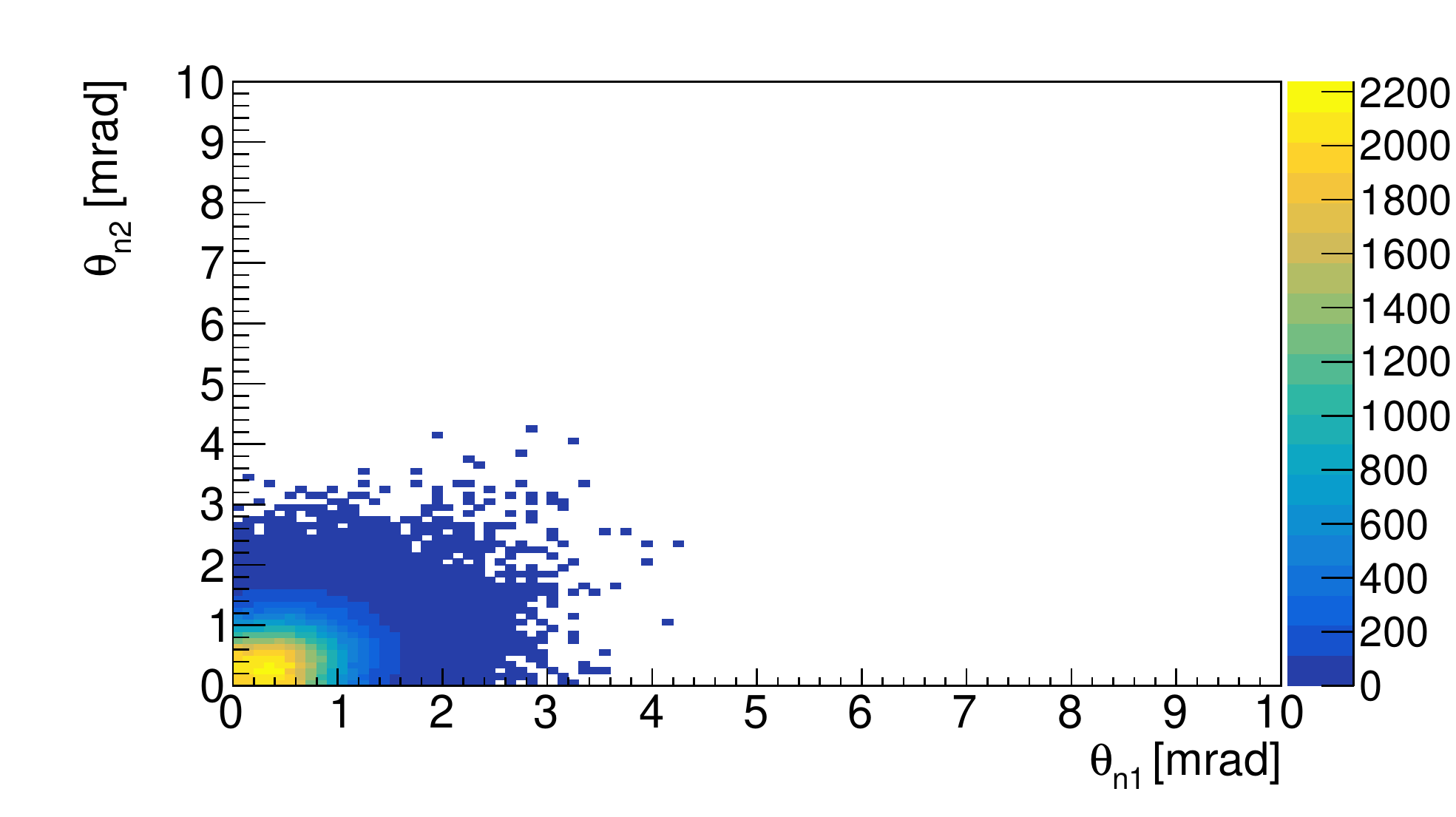}
    \caption{DIS kinematics for the $^3$H(e,e'nn)X reaction at 18x110 GeV with the neutrons detected in the far forward region.}
    \label{qe-110x18_nn}
\end{figure}

Thus following the requirements as detailed in the Meson production section of the far forward region, one will be able to do double spectator tagging from A=3 nuclei, opening new ways to extract nucleon information as well as study reaction mechanism effects.

\subsection{Short-range correlations and EMC effect studies}
\label{sec:diff_short-range}
Going beyond just using nuclei as an effective free nucleon target by tagging spectator particles, one can also make use of tagging to determine when the system was in a highly offset state.   This is critical for studying short-range nucleon-nucleon correction and could be the key to finally fully understanding the EMC effect.

\begin{figure}[htb]
    \centering
    \includegraphics[width =0.49\textwidth]{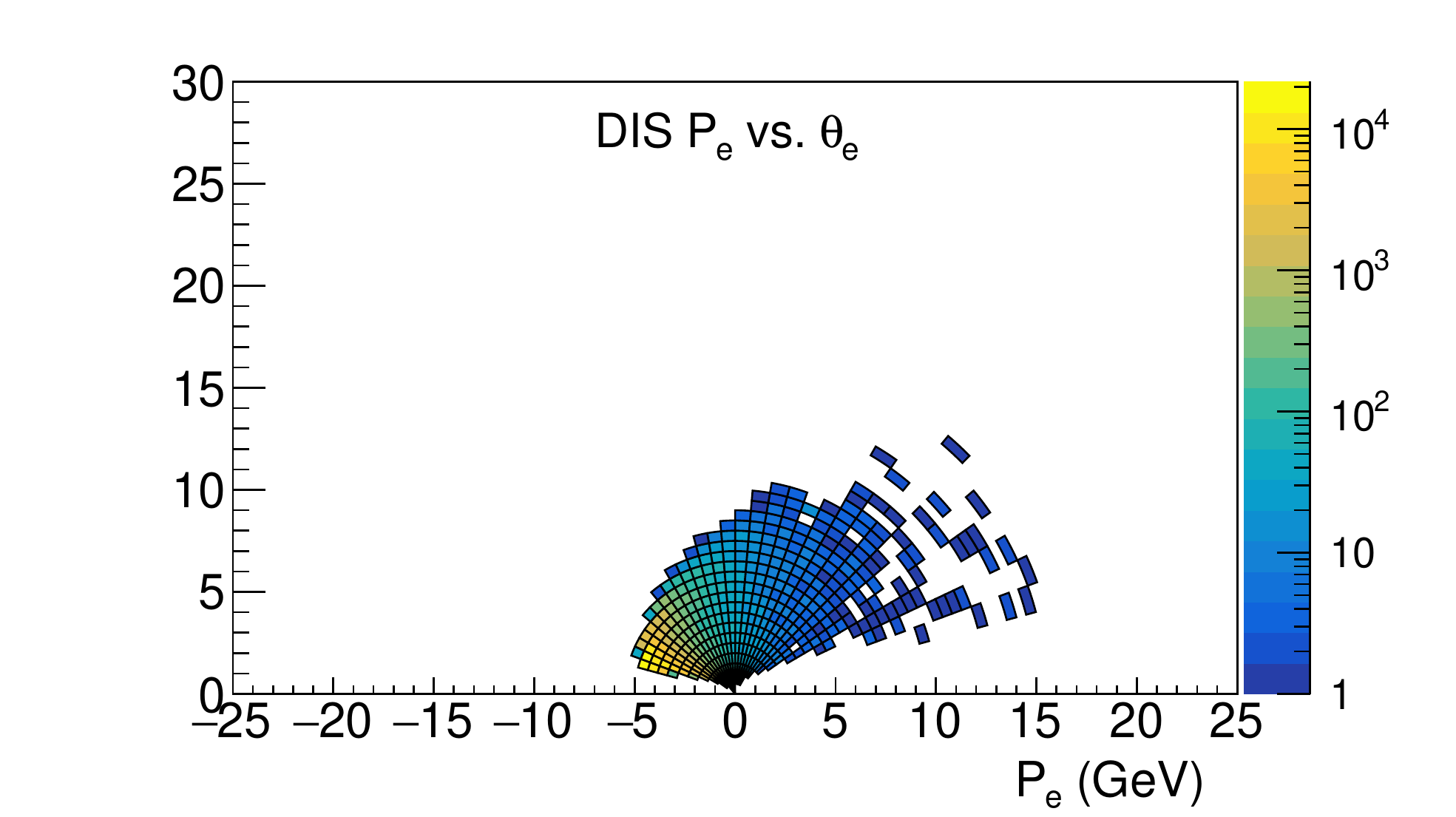}
    \includegraphics[width = 0.49\textwidth]{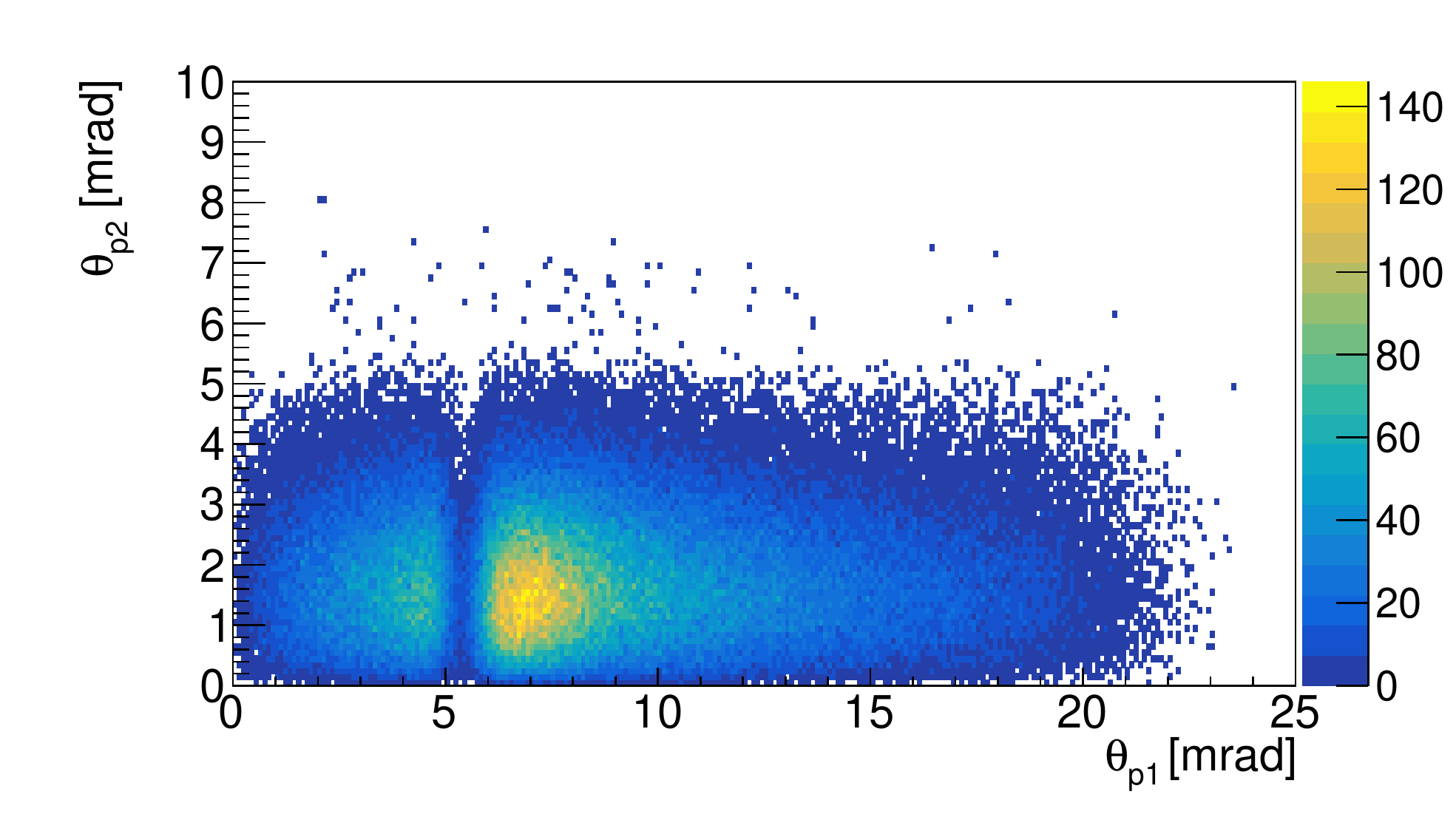}
    \caption{DIS kinematics for the $^3$He(e,e'pp)X reaction at 5x41 GeV but now for initial-state SRC proton pairs with the protons detected in the far forward region.}
    \label{src-41x5}
\end{figure}

\begin{figure}[htb]
    \centering
    \includegraphics[width =0.49\textwidth]{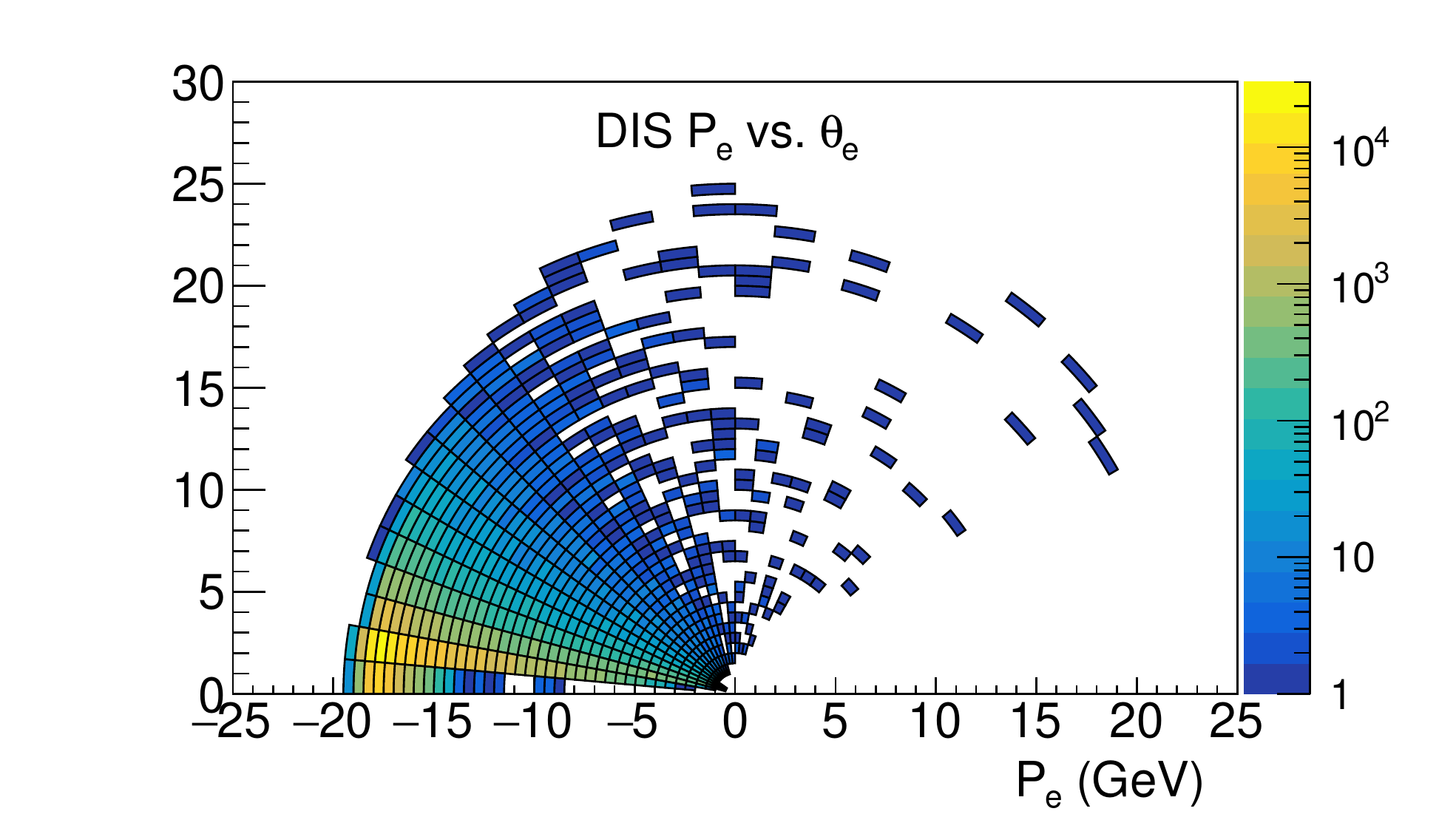}
    \includegraphics[width = 0.49\textwidth]{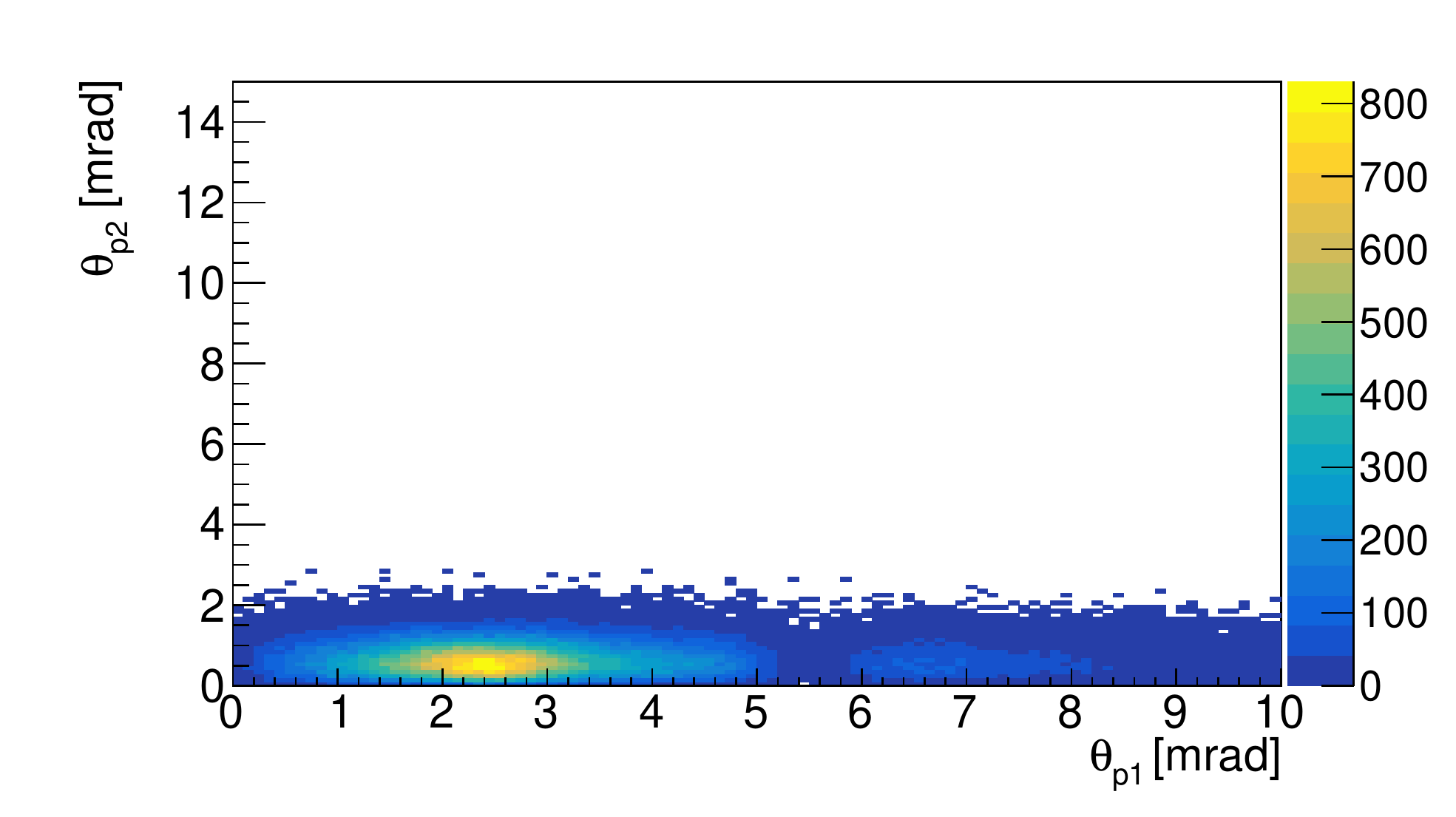}
    \caption{DIS kinematics for the $^3$He(e,e'pp)X reaction at 18x110 GeV but now for initial-state SRC proton pairs with the protons detected in the far forward region.}
    \label{src-18x110}
\end{figure}

To model the effect of initial-state corrections, the generalized contact formalization was used and study preformed with EICROOT, g4e as well as EICSMEAR.   All the studies show a significant fraction of the highly-correlated nucleons from pairing can be detected in the far forward region. 
As a representative example, in Fig.~\ref{src-41x5} and Fig.~\ref{src-18x110} are shown the angles where a proton-proton SRC pair from $^3$He and a neutron-neutron SRC pair from $^3$H go.   This is an extreme example, nevertheless, a significant fraction of the events would be detected, especially at the highest center-of-mass energies.

\subsection{Inclusive diffraction}

The studies of inclusive 
diffractive DIS ($ep \rightarrow eXp$) presented in sections~\ref{part2-subS-PartStruct-InclDiff} 
and~\ref{part2-subS-LabQCD-Diffraction} by default assume 
that diffractive final states can be identified with perfect efficiency. In this section, we 
discuss methods of selecting diffractive processes in which the proton remains intact and 
summarise the challenges in achieving a high level of experimental performance. 

\begin{figure}
\begin{center}%
        \includegraphics*[width=0.48\textwidth]{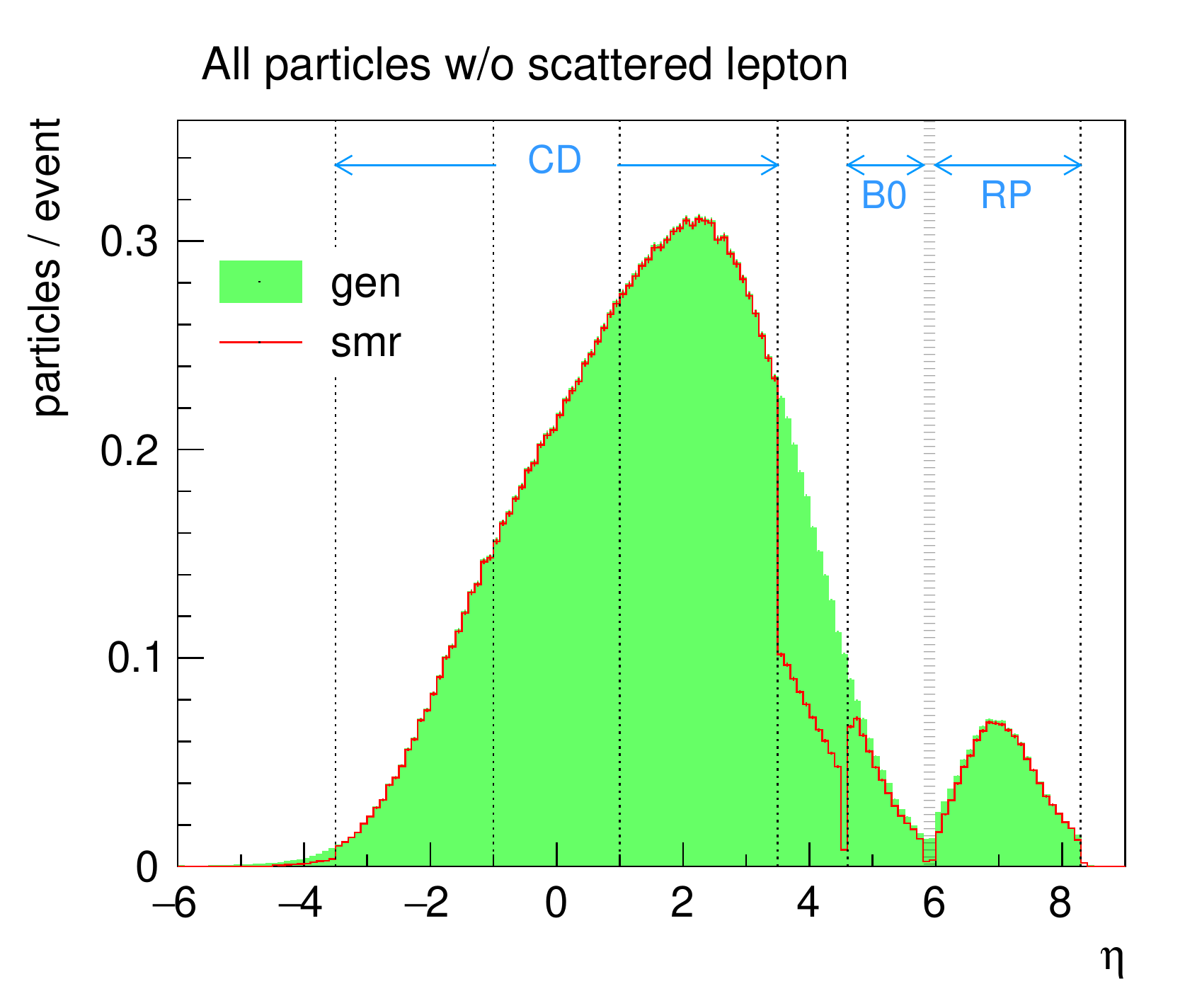}%
        \includegraphics*[width=0.48\textwidth]{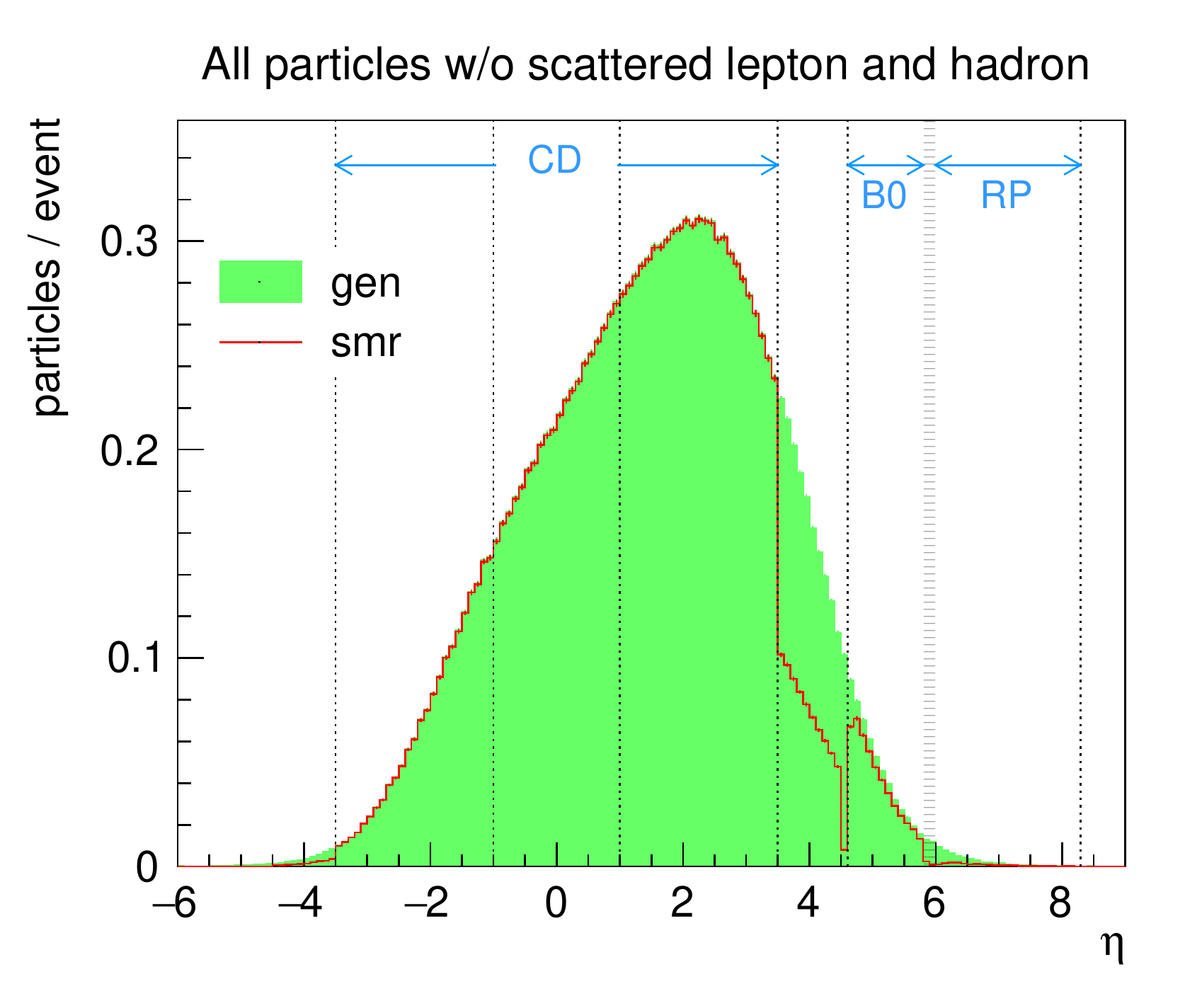}%
\end{center}
\caption{Pseudorapidity distribution of stable final state particles in a RAPGAP simulation
of diffractive DIS with $x_L > 0.6$
at the EIC (electrons with energy $18 \ {\rm GeV}$ and protons of $275 \ {\rm GeV}$).
The left panel corresponds to all final state particles whilst the right excludes the final state proton. 
The full set of generated particles is shown in green ('gen').
Those remaining after passing through an EIC detector simulation incorporating losses due to 
incomplete acceptance and migration due to imperfect detector resolution are shown with the red curve ('smr'). 
The vertical lines delineate the central detector region with full coverage (CD), the region in
which only electromagnetic calorimetry is currently envisaged and the regions covered by the 
B0 and Roman Pot ('RP') detectors. The normalisation is to the average number of particles per bin per event.}
\label{fig:difacceptance}
\end{figure}

At HERA, the inclusive diffraction process was studied successfully using a 'rapidity gap' 
method \cite{Aktas:2006hy,Breitweg:1997aa,Newman:2013ada}
in which events 
were identified on the basis of an absence of activity in 
a large forward region of pseudorapidity extending as close to the outgoing proton direction as 
possible. The gap size is strongly correlated with the fractional energy loss of the
proton $\sim \xi$, with an approximate dependence of the form $\Delta \eta \sim - \ln \xi$. 
Events in which the proton did not remain intact were vetoed on the basis
of particles observed in `forward detectors', either directly from the primary interaction, 
or as a result of secondary interactions with the beam-pipe or collimators. 
For  example in the  H1 experiment \cite{Aktas:2006hy}, the most forward
acceptance for veto detectors was obtained using scintillators surrounding 
the beam pipe $26$~m downstream, which were able to detect charged particles in the range 
$6.0 < \eta < 7.5$. Whilst it led to high acceptance at sufficiently small $\xi$, this 
method suffered from a number of drawbacks. Most importantly, the non-observation of the
scattered proton leads to contamination from proton dissociation sources (${\rm ep \rightarrow eXY}$) where $Y$ 
is a low mass proton excitation) and even from non-diffractive processes with
naturally occurring fluctuations in the hadronisation process, generating rapidity gaps that are exponentially
suppressed as a function of gap size. These unobserved backgrounds were ultimately the largest source
of systematic uncertainty.  
Furthermore, 
the use of the most forward part of the calorimeter 
as part of the veto (using an `$\eta_{\rm max}$' cut) limited the range in $\xi$ and $\beta$ that
could be accessed. 
The reduced center-of-mass energy at EIC leads to smaller
gap sizes for fixed dissociation system mass $M_X$, implying that a gap based selection would at best
be applicable only at small $M_X$. Nonetheless, rapidity gap
identification remains a powerful tool, even if it is only used as a veto against background contamination.
It is correspondingly advisable to ensure the fullest possible acceptance for forward-going particles.

Figure~\ref{fig:difacceptance} illustrates final state particle flow in a sample of diffractive events 
simulated using the RAPGAP Monte Carlo model \cite{Jung:1993gf}.
The potential for vetoing
forward activity is clear from the comparison between the distribution at the generator level ('gen') and
that after accounting for experimental effects ('smr'). Most interesting from this point of view is
the most forward part of the central detector, where there is almost full acceptance 
for both electromagnetic and hadronic calorimetry up to 
$\eta = 3.5$ and continued electromagnetic calorimetry up to $\eta = 4.5$ (hence approximately half of 
final state
particles are considered to be observed in the region $3.5 < \eta < 4.5$ according to the simple simulation). 
The use of the B0 detectors (integrated into the beampipe around $6$~m downstream and covering an angular
region $5.5 < \theta < 20 \ {\rm mrad}$) or the addition of dedicated veto detectors 
using for example scintillating tiles (the 
'off-momentum' detectors envisaged in the current EIC
design may be useful in this context) could extend this
considerably.
Nonetheless, the rapidity gap method doesn't reach the level of precision 
to which we aspire for diffractive studies at EIC and limits the acceptance in $\xi$ and $\beta$. 

The second method of studying diffractive processes with intact protons
is through the direct observation and measurement
of the scattered proton. This has been achieved at various colliders by using `Roman Pot' insertions
to the beam pipe which house sensitive detectors that are able to approach the beam to within a few mm
without compromising the vacuum. Roman Pot technologies, first applied at ISR and more recently  at the LHC, with the TOTEM
experiment for example operating 14 separate stations. Detection and tracking have benefited
from the use of radiation-hard silicon pixel detectors, whilst Time-of-Flight techniques have been
introduced in the case where both protons are detected, using either 
multiple layers of `ultra-fast' silicon or diamond-
or quartz-based Cerenkov radiators to achieve precisions approaching $20 \ {\rm ps}$ per proton. 

The current version of the EIC design incorporates Roman pot detectors at $26$~m 
and $28$~m, with an 
angular acceptance extending up to $5 \ {\rm mrad}$. 
Full coverage in azimuthal angle may be possible, exploiting the spectroscopic
effects of the machine magnets to ensure that all protons scattered with lower energy than the beam
pass inside the arc of the main beam. 
As can be seen from a comparison between the left and right panels of Fig.~\ref{fig:difacceptance},
the Roman pot detector acceptance is well-matched to the detection of scattered protons over a wide
range of kinematic phase space.
The B0 silicon detectors incorporated in the current EIC design add acceptance at larger scattering angles.
This angular acceptance range can be mapped directly onto a kinematic plane in $\xi$ and $t$. 
As illustrated in Fig.~\ref{PWG-sec7.1.6-fig-leadprot}, the upper limit of the Roman pot
angular coverage is adequate for the study of diffractive DIS across the full region of 
interest in $\xi$ and also across a wide range in $t$ for high energy EIC configurations. The addition
of the kinematic range covered by the B0 further improves matters at very large $|t|$ such that there
is almost full coverage at least to $|t| = 1 \ {\rm GeV^2}$ at $E_p = 275 \ {\rm GeV}$ and 
$E_p = 100 \ {\rm GeV}$. Whilst the situation is more challenging at $E_p = 41 \ {\rm GeV}$,
highly interesting measurements could still be made. 

It should be noted that the choice of lower limit of  
scattering angle $\theta = 0.5 \ {\rm mrad}$ in Fig.~\ref{PWG-sec7.1.6-fig-leadprot} 
is somewhat ad hoc. In practice this limit is determined by the proximity to the beam-line that
can be tolerated without causing disruption to the beam. Experience at previous colliders
suggests that this lower limit is likely to be decreased slowly and steadily over time and
is a matter for constant discussion between experiments and machine experts. A natural way to
decrease the lower limit in $\theta$ is to locate the Roman Pots further from the interaction point.
However, it is not possible to instrument the beam pipe further downstream than around $30$~m in the 
current EIC design due to the crab cavities. 

\begin{figure}
\begin{center}%
        \includegraphics*[width=\textwidth]{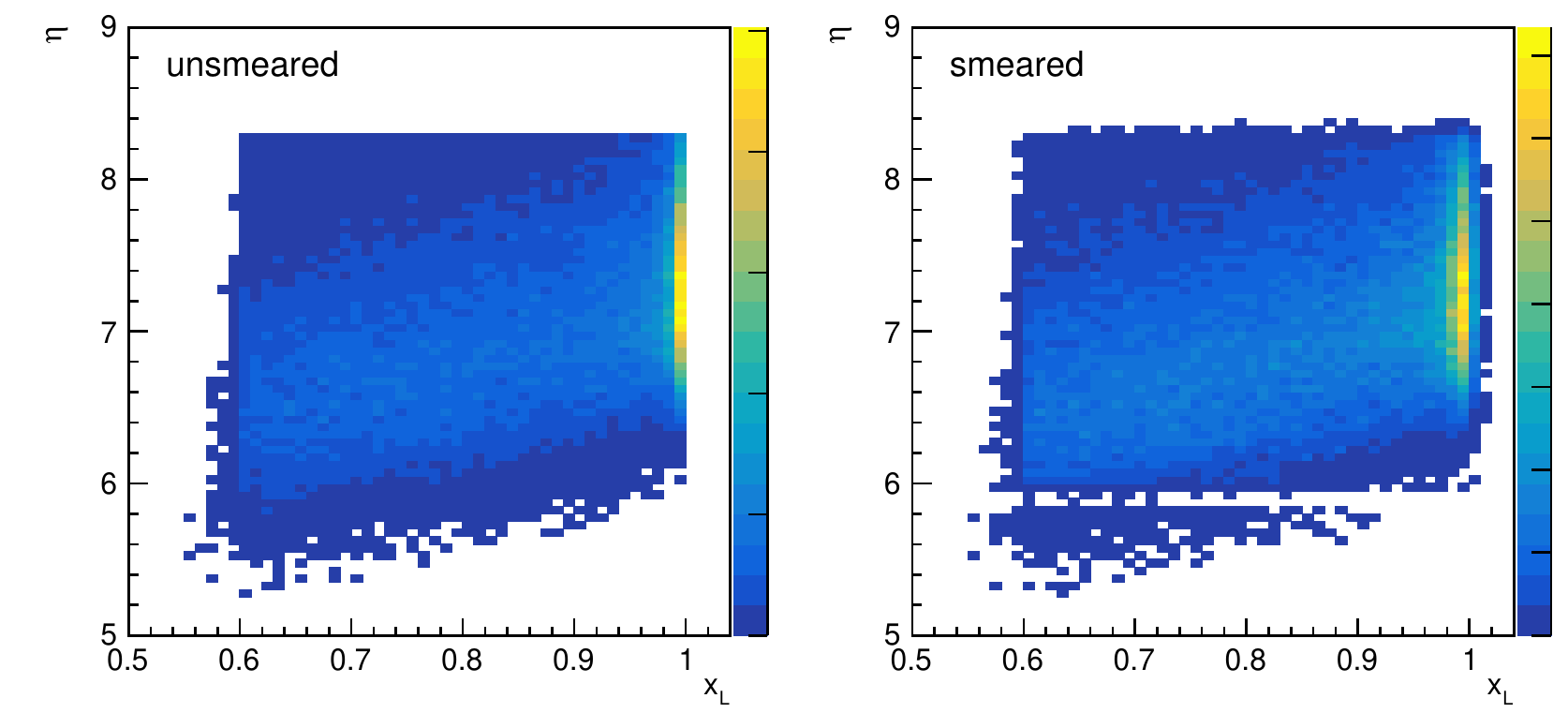} \\
\vspace*{1cm}
        \includegraphics*[width=0.49\textwidth]{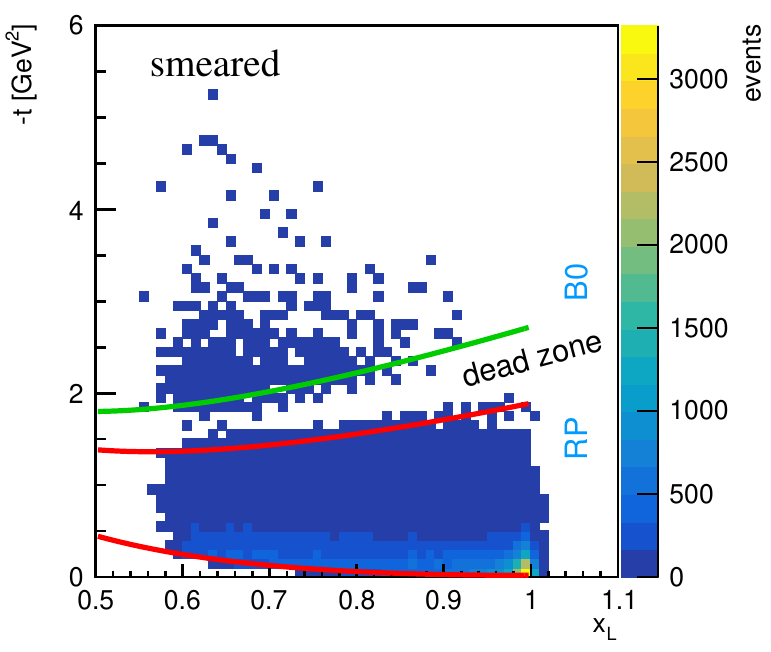} 
        \includegraphics*[width=0.49\textwidth]{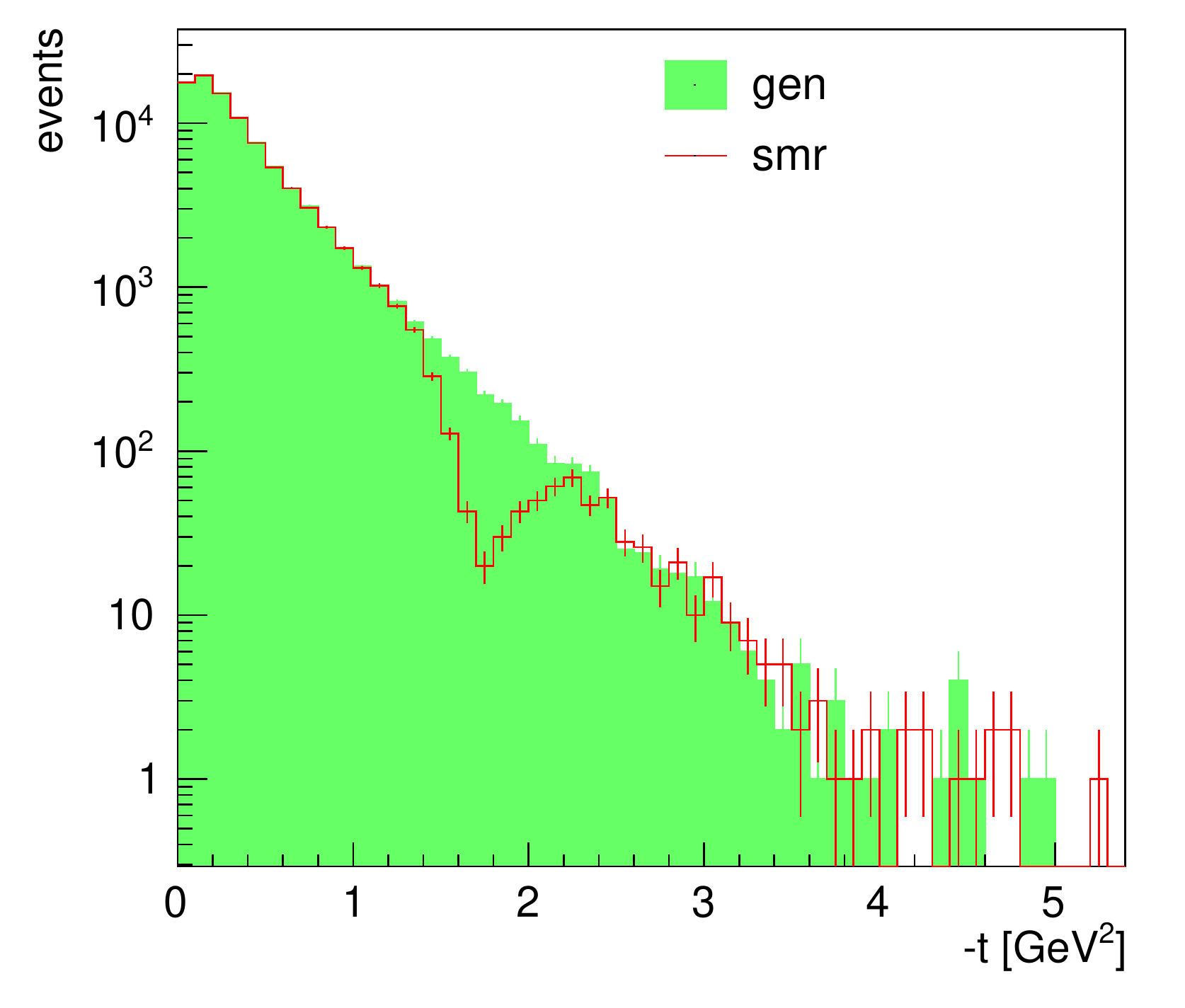} 
\end{center}
\caption{Distribution of final state protons in the RAPGAP simulation of diffractive DIS at the EIC
(electrons with energy $18 \ {\rm GeV}$ and protons of $275 \ {\rm GeV}$). 
Top left: generator level distribution in the plane of $\eta$ versus 
$x_L = 1 - \xi$. Top right: similar distribution after passing through a 
simple model of instrumental effects (migrations and acceptance). 
Bottom left: distribution mapped onto the pane of $x_L$ versus $t$ after 
passing through the simulation of migration and acceptance effects.
Bottom right: one dimensional projection onto the $t$ axis after passing through the
simulation of migration and acceptance effects.  
No lower limit on the acceptance in $\theta$ is applied in these figures.}
\label{fig:difdeadzone}
\end{figure}

One final consideration in proton tagging is to avoid a large gap in acceptance between the B0 and Roman
pot detectors. The current design achieves this 
for the most part, as illustrated in figure~\ref{fig:difdeadzone}(top), where
protons from the RAPGAP simulation are shown in the plane of $\eta$ versus $x_L = 1 - \xi$. The plot
after passing through a simple simulation of EIC instrumental effects exhibits a small band in which
acceptance is lost around $\eta = 6$, corresponding to the gap between the B0 and the Roman pot acceptances.
Mapping these distributions onto the plane in $x_L$ and $t$, as shown in 
figure~\ref{fig:difdeadzone}(bottom), shows that, at least for the highest $\sqrt{s}$, 
the 'dead zone' band of
lost acceptance is at relatively large $|t|$ values and is unlikely to seriously compromise 
measurements. On the other hand, the band migrates to smaller $|t|$ vales at lower $\sqrt{s}$.
It may be possible to mitigate this if there are two EIC detectors, by adjusting the design of the 
forward detectors in the second interaction region such that the gap falls at a different angle.

\subsection{Summary of far forward region physics requirements}



For the far forward region, we have seen that numerous physics channels can be studied with the current design and layout of the various detectors.  For the B0 sensors, 3.4 cm inner radius and 20 cm outer radius with a 50 x 50 $\mu$m$^2$ pixel is needed.  For the off momentum tracker an 10 cm inner radius is assumed along with a 10 x 30 cm$^2$ sensor with a 500 x 500 $\mu$m$^2$ pixel pitch.   For the Roman Pots, it is assume that they will be 10 $\sigma$ from the beam halo size with a 20 x 10 cm$^2$ sensor with 500 x 500 $\mu$m$^2$ pixel pitch.

For the zero degree calorimeter, it is assumed that the device will be at least 60 x 60 cm$^2$ made up of low and high granularity electromagnetic calorimeter as well as a 10 x 10 cm$^2$ hadronic calorimeter.
Ideally, the calorimeter would have an energy resolution of 35\%/$\sqrt(E)$, but less than 50\%/$\sqrt(E)$ is acceptable for carrying out the physics goals that have been laid out by the diffractive and tagging physics working group.

In addition to these detector requirements,  the spin physics studies with the light nuclei require that the accelerator be able to store polarized spin-1/2 and spin-1 particles (e.g. polarized $^3$He and Deuterium).   It is understood that preserving the polarization of polarized deuterons in a circular storage ring is particularly challenging and at a glance would seem to be beyond the scope of the planned spin rotation capabilities of the EIC.    A clever solution to this problem is to use natural preserving deuteron energies, as known as magic energies, where only a limited amount of spin rotator and/or Siberian snakes are required to maintain the polarization.   
At per nucleon momentum of 104.9, 111.5, 124.6 and 131.2 GeV, full polarization can be provided at one interaction region while at 39.3 GeV or 118.0 GeV full polarization could be provided at both of the planned interaction regions.   

Finally, the measurement of the inclusive, diffraction puts additional requirements on the forward and far-forward detectors. The rapidity gap method for inclusive diffraction requires vetoing forward activity, thus hermiticity of the detector setup is crucial. The B0 detectors are very useful for this measurement, but it would be also necessary to have additional detectors to veto the activity. The method of the tagged proton puts another requirement on the far forward detectors. While the current setup is already very well suited for this measurement there exists a gap in the proton acceptance between B0 and Roman pot detectors. While it would be difficult to completely close the gap within the current detector setup, perhaps the complementary detector design could be modified in such a way that the gap would appear in the different regime of angles and thus at different values of momentum transfer for the extracted cross section.

\section{Summary of Requirements}
\label{part2-sec-DetReq.Sum.Req}
Table~\ref{fig:finalmatrix} summarizes the combined detector requirements that emerged from the studies performed by the working groups. Several requests for improved performance as compared to the previously assumed values have been identified in order to deliver key measurements at the future EIC:

\begin{table}[hbt]
\includegraphics[width=\linewidth]{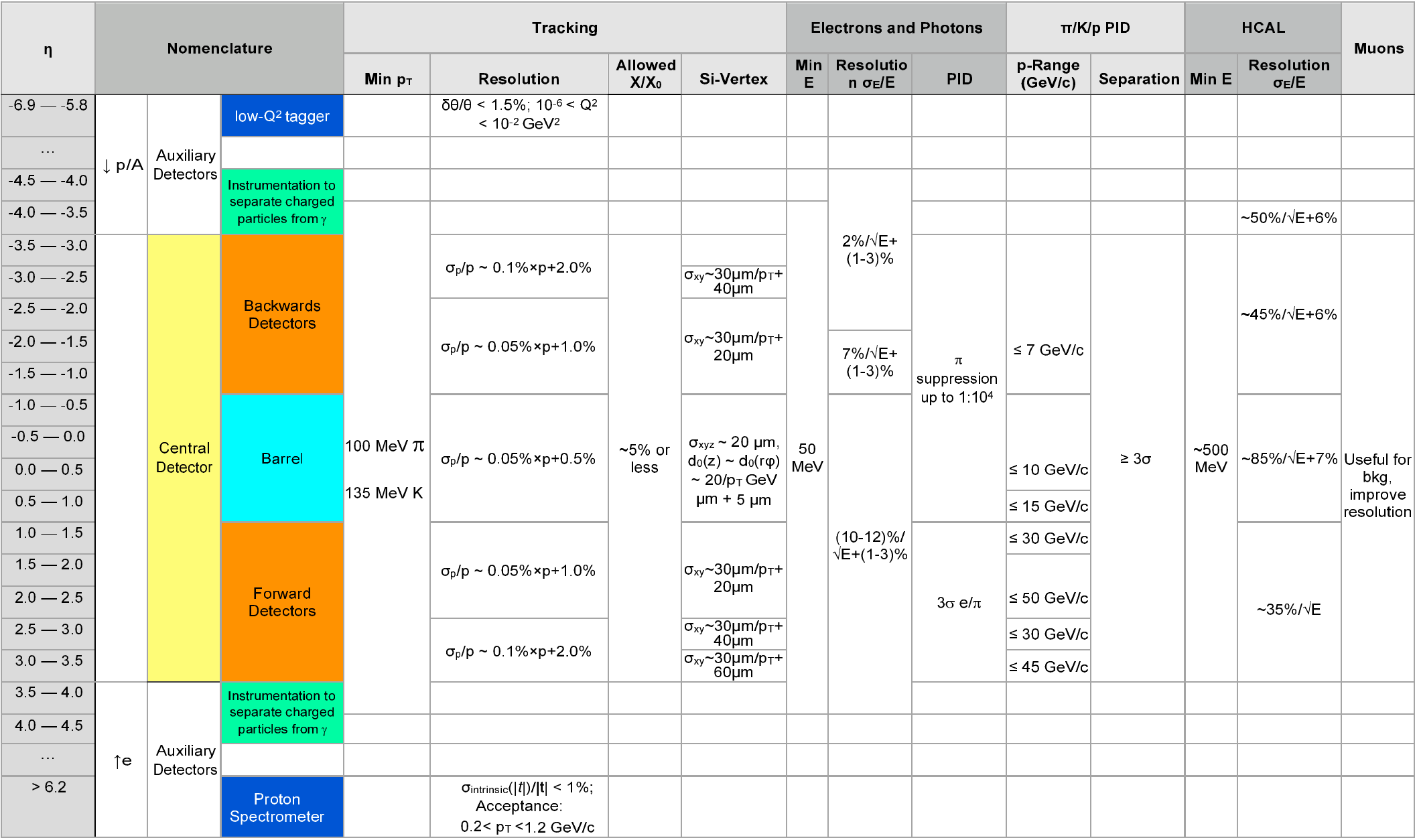}
\caption{Summary of the Physics Working Group detector requirements}
\label{fig:finalmatrix}
\end{table}

\begin{enumerate}
\item {\bf Hermiticity} A pseudo-rapidity coverage of $-4<\eta<4$ in the central detector is crucial for many flagship measurements, in the exclusive and diffractive channels. The highest $\eta$ coverage should be strongly pursued in order to reduce the pseudo-rapidity gap between the central and the far-forward region detectors. An asymmetric configuration where the acceptance in $\eta$ is increased in a limited region in azimuth could potentially overcome the pseudo-rapidity limitation enforced by the beams crossing angle. Indeed, the measurement of inclusive diffractive channels using the rapidity gap method requires an excellent hermiticity of the overall detector.
  
The increased pseudo-rapidity coverage in the forward region is also essential to reconstruct the hadronic state in inclusive and semi-inclusive reactions using the JB method, extending the accessible area in the highest $x$ at low $y$.

\item {\bf Momentum resolution} Excellent momentum resolution in the central region is required for a wide variety of physics studies. In particular, for those inclusive and semi-inclusive reactions that use the Jacquet-Blondel method for reconstructing event kinematics, for jet reconstruction, jet correlations, and jet substructure studies, and others.

\item {\bf Minimum $p_T$} A low detection threshold for pions and kaons is required. In order to measure soft pions from $D^*$ and $\Lambda$ decays, as well as for partial-wave analysis of di-hadron final states, a minimum detection $p_T$ of 100~MeV is needed. Soft kaons from $\phi$ decays should be measured up to $p_T$ of 135~MeV.

\item {\bf Vertex resolution} The requirements on the vertex resolution listed in Tab.~\ref{fig:finalmatrix} are driven by the heavy flavor reconstruction where the reduction of the background relies strongly on analyses-related selections performed in different combinations of the primary scattering vertex and the secondary vertex of the decaying heavy meson. The required impact parameter resolution for the heavy flavor measurements is of the level of $\sigma_{xy} \sim 20/p_{T} \oplus 5~\mu\mathrm{m}$ at mid-rapidity.

\item {\bf Electron ID} At mid-rapidity inclusive measurements will be limited by the electron/pion discrimination, which will drastically impact the measurements of longitudinal double spin and parity violating asymmetries in the highest $Q^2$ region. A pion suppression capability at the level of $10^{-4}$ is deemed critical for several important measurements in studies of inclusive reactions. In addition, spectroscopy measurements require $3\sigma$ electron/pion separation in the forward region for $J/\psi$ identification.

\item {\bf Photon detection threshold} Soft photon detection is driven by requirements to separate coherent and incoherent production of vector mesons. Some nuclear deexcitations will produce very low energy photons, depending on the nucleus involved.

\item {\bf Hadron ID} PID at the level of $3\sigma$ is required over a large momentum range and is primarily driven by mid-to-high $z$ measurements of jet fragmentation functions and the related polarized Collins asymmetry. TMD measurements also require  $3\sigma$ separation of pions from kaons up to large values of their momenta, and are driving the requirements in the forward-rapidity region, too. Particularly, 3 $\sigma$ $\pi/K/p$ separation up to 50\,GeV/$c$ in the forward region,  up to 10\,GeV/$c$ in the central detector, and up to 7\,GeV/$c$ in the backward region are required.

\item{\bf Electromagnetic calorimetry} Good electromagnetic calorimeter resolution in the central detector, at the level of $\sigma(E)/E \approx 10-12\%/\sqrt{E} \oplus 1-3\%$ at midrapidity, would be sufficient for jet physics, as jet reconstruction could take advantage of expected excellent tracking performance. However, at forward rapidities, excellent electromagnetic calorimeter resolution  ($\sigma(E)/E \approx 2\%/\sqrt{E} \oplus 1-3\%$) will become an essential driver of jet reconstruction performance.

\item{\bf Hadron calorimetry} In the mid-rapidity region, the energy resolution of hadron calorimeters is driven by single jet measurements.  Neutral hadron isolation could also be important for jet energy scale and resolution. In the forward and backward rapidity region diffractive di-jets need a good hadron energy measurement, with a resolution of the level of $\sigma(E)/E \approx 50\%/\sqrt{E} \oplus 10\%$. The requirement on the constant factor at the highest rapidities is driven by the need for good energy resolution where tracking dies out. A minimum energy threshold of 500 MeV/$c$ was assumed for all the studies performed.
\end{enumerate}


In addition to the requirements described above and included in Tab.~\ref{fig:finalmatrix}, requirements
for the far-forward region have
originated primarily from the exclusive and diffractive working groups, as follows.

B0 sensors with an acceptance between 3.4 and 20 cm from the beamline and 50$\times$50 $\mu$m pixels are required. The off-momentum tracker needs a 10-cm inner radius and an active area of 10$\times$30 cm$^2$. Its position resolution should be at least 0.5$\times$0.5 mm$^2$. Similar position resolution is required in the Roman Pots that would need a sensitive area of 20$\times$10 cm$^2$, assuming they are placed at 10 $\sigma$ from the beam envelope. 

A zero degree calorimeter at least 60$\times$60 cm$^2$ in size should include both electromagnetic and hadronic calorimetry, with low and high granularity. Its energy resolution should approach 50\%/$\sqrt{E}$, but physics would benefit from an enhanced performance of 35\%/$\sqrt{E}$.

Finally, for the spin physics program with light nuclei, such as deuterons, high polarization is key. In order to preserve deuteron polarization, which is very challenging in a circular storage ring, it is suggested to use particular beam energies that only require a limited number of spin rotators and/or Siberian snakes to maintain polarization. At per nucleon momentum of 104.9, 111.5, 124.6 and 131.2 GeV/$c$, full polarization can be provided at one interaction region while at 39.3 GeV/$c$ or 118.0 GeV/$c$ full polarization could be provided at both of the planned interaction regions.

\FloatBarrier

\cleardoublepage

%
%
\phantomsection
\addcontentsline{toc}{part}{\Large{\textbf{Volume III: Detector}}}
\label{part3.det}

\newgeometry{textwidth=8.5in,textheight=11.0in}
\includegraphics[width=8.5in,height=10.99in]{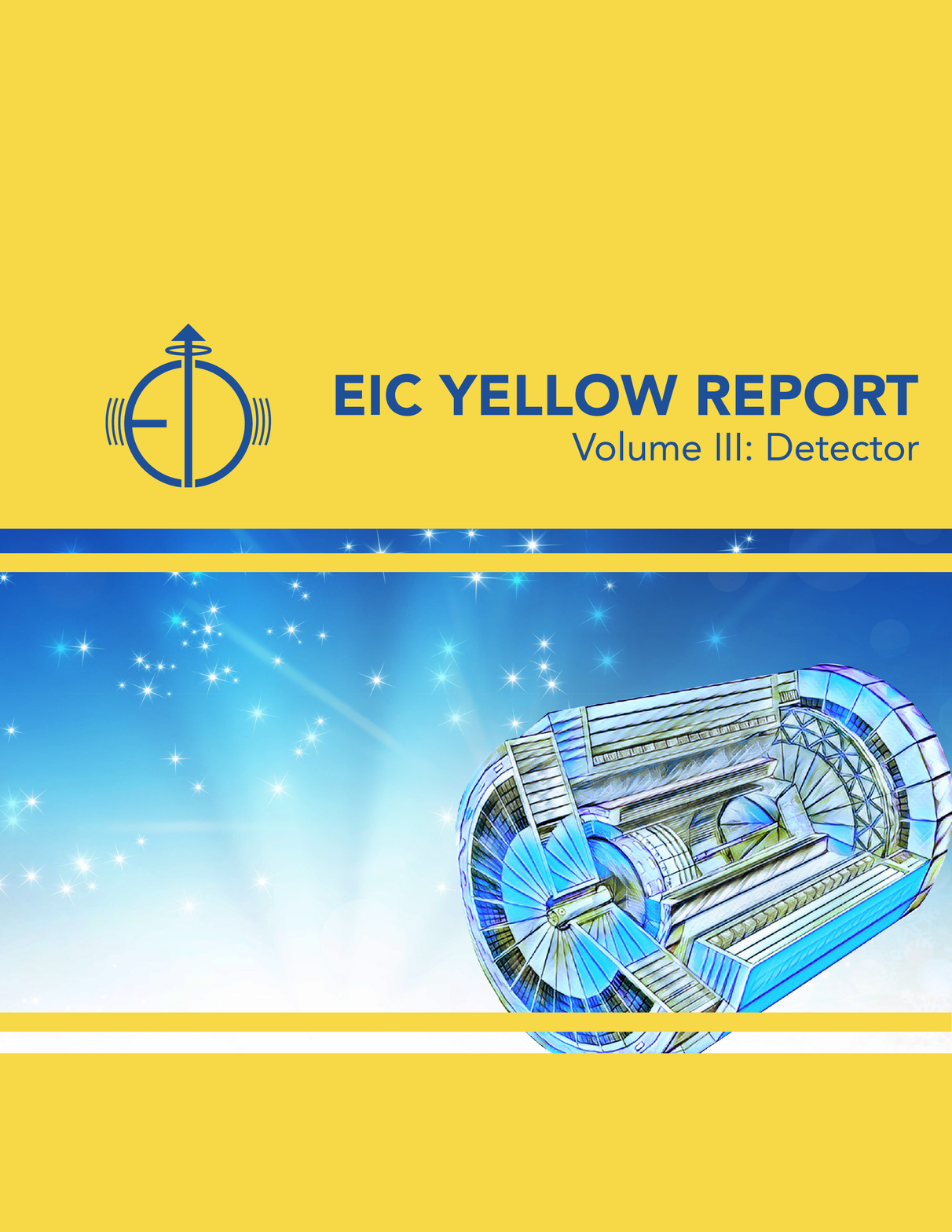}
\restoregeometry
\semiblankpage

%
%

\chapter{Introduction to Volume III}
\label{part3-chap-Intro}

The third volume of the EICUG Yellow Report is dedicated to the EIC detector(s) with reference to the EIC studies (Vol. II, Chapter 7) and the detector requirements (Vol. II, Chapter 8), which result from the analysis of the physics scope. Volume III assumes as its starting point, the performance requirements and the corresponding detector challenges (Chapter 10). On this basis, the detector technologies adequate to answer the EIC requests and the performance these technologies can provide are surveyed (Chapter 11). Considerations in favor of instrumenting the two interaction regions foreseen in the overall design of the accelerator complex are also presented (Chapter 12). These studies are complemented with an initial assessment of the EIC detector integration (Chapter 13). The on-going R\&D effort within the program “Generic R\&D for the Electron Ion Collider” is presented in Chapter 14.
\par
The matter discussed in Volume III results from a rich variety of inputs, including:
\begin{itemize}
    \item 
    Previous studies~\cite{Accardi:2012qut, EIC:RDHandbook};
    \item 
    Central detector concepts proposed over years, namely BEAST~\cite{pCDR}, JLEIC~\cite{Abeyratne:2012ah, Morozov:2019uza},  EIC-sPHENIX~\cite{Adare:2014aaa}  and TOP-SIDE~\cite{Repond:2019hth};
    \item
    The development efforts ongoing within the program “Generic R\&D for the Electron Ion Collider” (\url{https://wiki.bnl.gov/conferences/index.php/EIC_R\%25D}), which is summarized in Chapter~\ref{part3-chap-DetTechnology};
    \item
    Dedicated simulation studies performed in the context of the “Generic Detector R\&D for an Electron Ion Collider” and, more, during the preparation of the present Yellow Report;
     \item
    Novel ideas and options proposed by the community during the one-year long effort resulting in the present Yellow Report.
\end{itemize}
The present level of assessment of the EIC detector matter could not have been achieved without the Yellow Report initiative. In fact, this activity has created an unique opportunity for the EIC community at large: the synergistic effort of theorists, experimentalists, accelerator experts and computer science experts has made possible the relevant progress obtained on a time-scale of one year.

\section{General EIC Detector  Considerations}

\begin{figure*}[ht]
\centering
\includegraphics[width=0.9\textwidth]{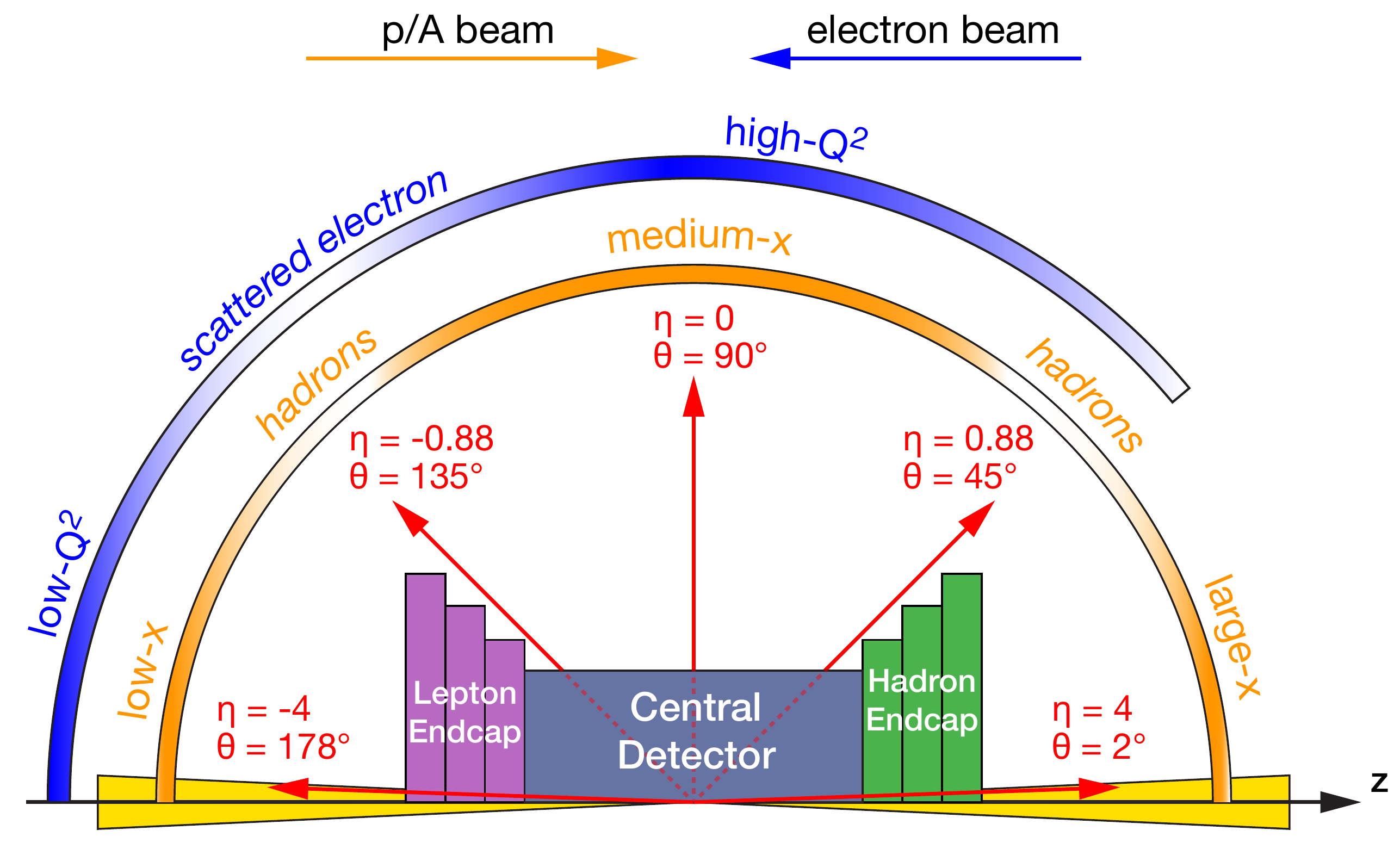}
\caption{A schematic showing how hadrons and the scattered lepton for different
$x-Q^2$ are distributed over the detector rapidity coverage.}
\label{Det.Schematics}
\vspace*{3mm}
\end{figure*}

All the different physics processes to be measured at the EIC require having the event and particle kinematics (x, Q$^2$, y, W, p$_t$, z, $\Phi$, $\theta$, defined in Appendix \ref{appendix:kinematics}) reconstructed with high precision. The key variables x, Q$^2$, y, and W are either determined from the scattered electron or from the hadronic final state. In order to access the full x--Q$^2$ plane at different center-of-mass energies and for strongly asymmetric beam energy combinations, the detector must be able to reconstruct events over a very wide range in rapidity. This imposes requirements on both detector acceptance and resolution. 
At EIC, without good coverage of the rapidity range $\lvert {\eta} \rvert>$2, 
a significant fraction of the x--Q$^2$ phase space will be missed. This puts strong emphasis on the lepton and hadron end-caps. Figure ~\ref{Det.Schematics} illustrates, for the detector elements in the interaction region, the correlation between the pseudo-rapidity and scattering angle and the x--Q$^2$ phase space. The central detector, approximately covering the range of $\lvert {\eta} \rvert<$1 is also referred to as the barrel detector, while the hadron end-cap and the electron end-cap are often indicated as forward and backward end-cap, respectively. The setup is completed by very-small-angle counters situated at a larger distance from the interaction point, forming the set of the very forward and very backward detectors. 
The geometrical acceptance is not the only potential limiting parameter. Other constraints that are more critical for forward high energy particles come from the minimum detectable
particle momentum, the acceptance in transverse momentum and the resolution in the momentum measurement. These limitations can be mitigated by data taking at different settings of the central solenoid resulting in different magnetic field values. 
Ancillary systems, namely the luminosity monitor and the lepton and hadron polarimeters, complete the equipment needed for the EIC physics program.
\par
A reference general purpose detector is discussed in the following: it is still a concept and consists of a set of equipment technologies that can meet the majority of the EIC requirements. The included technologies correspond to the ones most studied so far and can form a detector design, that can be used as a reference. Nevertheless, no technology ranking between the reference and the alternative options has yet been established. The fresh and stimulating possibilities coming from more recent proposals are useful for establishing a rich basis for the future detector choices. Therefore, all the technologies presented in the Yellow Report Volume III have the same applicability and relevance, while they attest the fecundity of the overall EIC user group community. The reference general purpose detector concept is suggested by the EIC project. The EIC facility is able to support two interaction points, but the EIC project foresees only the instrumentation of a single one, 
able to answer to the needs posed by the whole EIC physics program. 
Nevertheless, the EICUG community is confident that the EIC science output can be increased by instrumenting also the second interaction region, exploiting complementary features, where the complementary aspects can be related both to the detector and the interaction region design. In this context, the benefit of having different options for the detector technologies appears in its full capacity: different detector approaches can substantially contribute to the definition of the two-detector scenario, which is the EIC user group favored scenario.
\par
The EIC physics program poses challenges to the detector performance, which becomes more severe when the constraints coming from the collider design are also taken into account. The challenges can be understood by identifying the points of tension between the requirements and the expected performance of the reference detector. 
The existence of these critical aspects encourages the community to make progress in the preparatory detector R\&D effort, to explore novel options, and to reconsider the overall detector design. 
\par
The simulation efforts have been started for various detector elements mostly as standalone exercises with various levels of maturity: analytical calculations, simplified Monte Carlo exercises (fast simulations), and Geant4-based Monte Carlo approaches (full simulations). The path towards more advanced and comprehensive approaches is presently evolving within three simulation tools, EicRoot, ESCalate, and Fun4All. One of them will be selected and  will eventually include all detectors in the central, far-forward electron, and far-forward ion regions. Fast simulations in Eic-Smear and Delphes are being used for physics studies. 
The detector studies in this Yellow Report represent an initial assessment of the detector technologies for the EIC and their capabilities, but need to be further refined and in particular complemented in an integrated approach, where the information from various detector elements and also the interaction region, support structures and other dead material is taken into account. This is essential for understanding, e.g., the performance at the edges of the detector system or the effect of combining different technologies. 
An integrated approach will benefit from one detector simulation toolkit that is maintained, and supported by the whole community.
All this considered, the results reported are subject to future improvements and updates, even if they represent a solid initial assessment.
\section{Reference EIC Detector}

\label{part3-chap-Intro-section-reference-detector}
The physics opportunities at EIC are intimately connected to the overall design of the experiments and to the performance of the required detectors. From the experimental point of view, the broad physics EIC program can be accomplished by the study of (i) inclusive, (ii) semi-inclusive and (iii) exclusive processes, all of them with an initial state of electrons and light or heavy nuclei, with polarized electron and light nuclei beams and spanning a wide range of center-of-mass energies. The main requirements for the experimental apparatus are based on these processes and the requirements of the wide kinematic coverage,  adding more and more complexity moving from reactions (i) to (iii):
\begin{itemize}
    \item 
    Precise identification of the scattered electron and extremely fine resolution in the measurement of its angle and energy are a key requirement for all experimental channels; other essential tools for the whole physics scope are the central magnet and the tracking system required for momentum measurements and full rapidity coverage with electromagnetic and hadronic calorimetry;
    \item
    More is needed to access the semi-inclusive processes (ii): excellent hadron identification over a wide momentum and rapidity range, full 2$\pi$ acceptance for tracking and momentum analysis and excellent vertex resolution by a low-mass vertex detector; 
    \item
    Exclusive reactions (iii) impose the necessity to accurately reconstruct all particles in the event using a tracker with excellent space-point resolution and momentum determination, electromagnetic calorimetry with excellent energy resolution, hadronic calorimetry in the end-caps, 
    the complete hermeticity of the setup with the additional requirement of very forward detectors such as Roman pots, and large acceptance zero-degree calorimetry to effectively detect neutrons from the breakup of nuclei or neutral decay products from tagged DIS processes;
    \item 
    For the entire experimental program a precise determination and monitoring of the luminosity will be essential;
    \item
    Measurements with polarized beams require the use of electron, proton, and light nucleus polarimeters;
    \item
    The strategy for detector read-out and data acquisition has to be defined taking into account the data rate of the experiment, as well as the rapid developments in the field of digital electronics and computing power, suggesting a integrated approach to both the read-out and data acquisition and software and computing.
\end{itemize}
\par
\begin{figure*}[ht]
\centering
\includegraphics[width=0.95\textwidth]{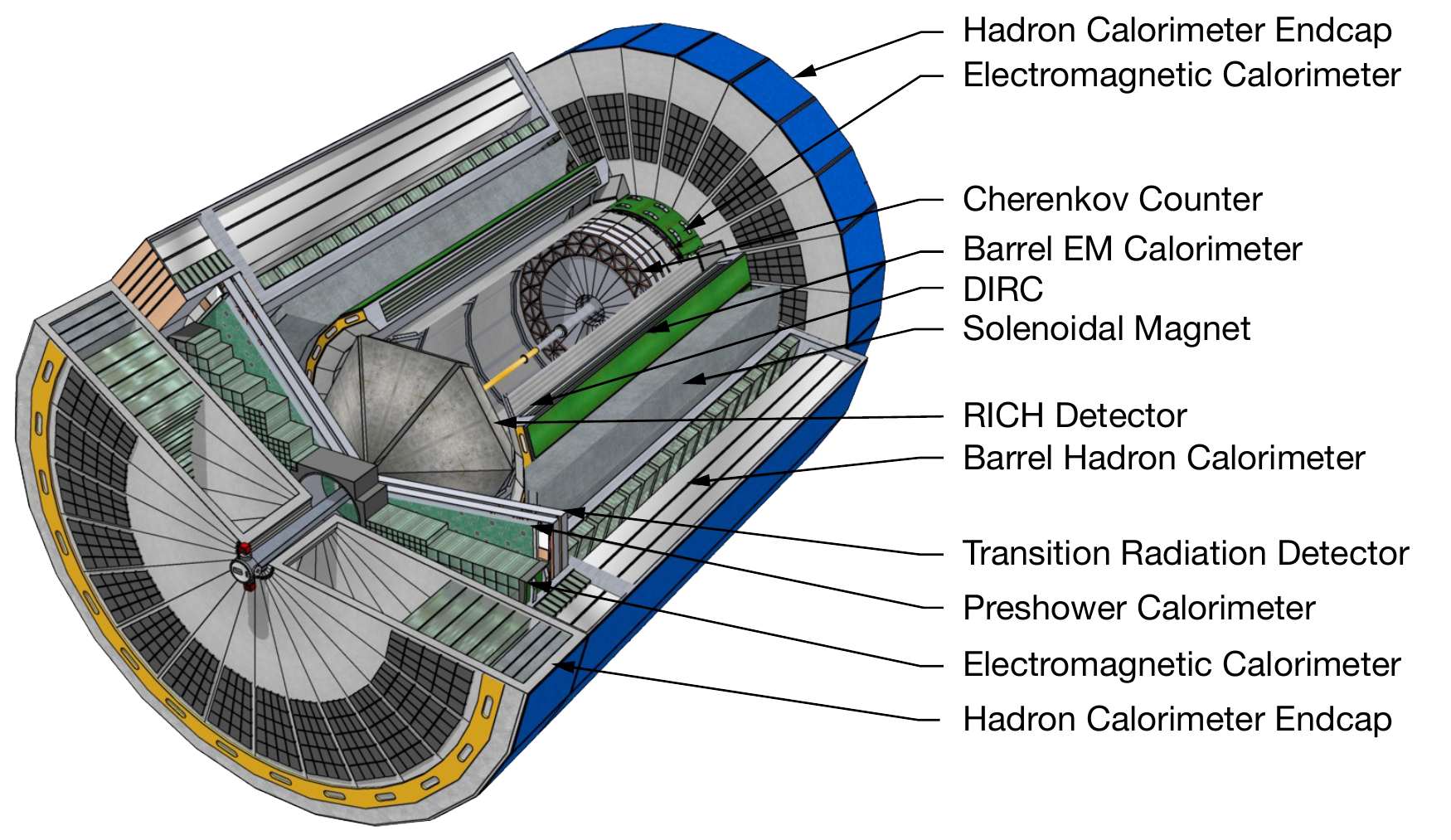}
\vspace*{3mm}
\caption{A cutaway illustration of a generic EIC concept detector. } 
\label{central-detector-CAD}
\vspace*{3mm}
\end{figure*}
\begin{figure*}[ht]
\centering
\includegraphics[width=0.85\textwidth]{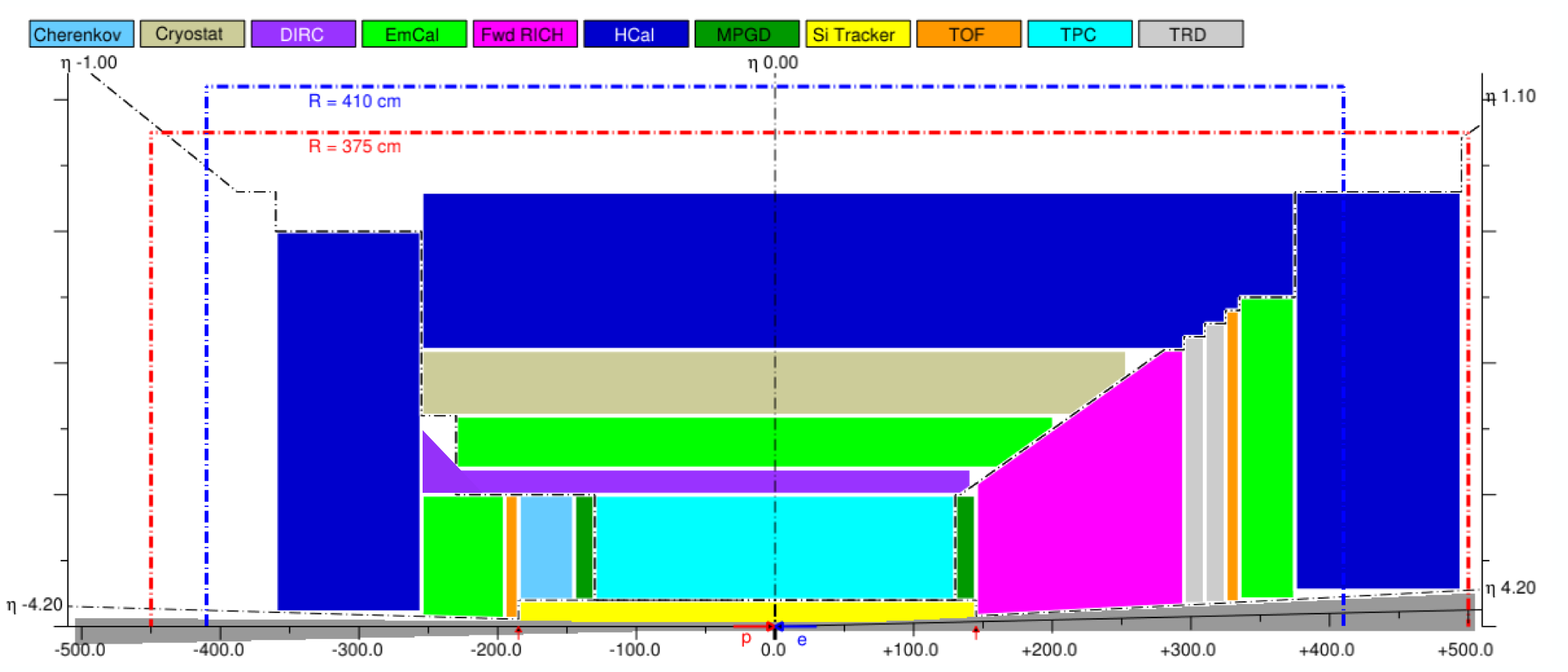}
\vspace*{3mm}
\caption{A 2D sketch of an EIC detector layout, horizontal cross cut. Only one half of the detector is shown, with the other half being mirror-symmetric in this view, up to the crossing angle. The beam pipe footprint (in dark gray) is to scale. The blue dashed line shows the doorway size between the assembly and the installation halls. The red dashed line shows the realistic central detector envelope with the available [-4.5, 5.0] m space along the beam line.} 
\label{central-detector-cartoon}
\vspace*{3mm}
\end{figure*}
\begin{figure}[ht]
    \centering
    \includegraphics[width=.98\textwidth]{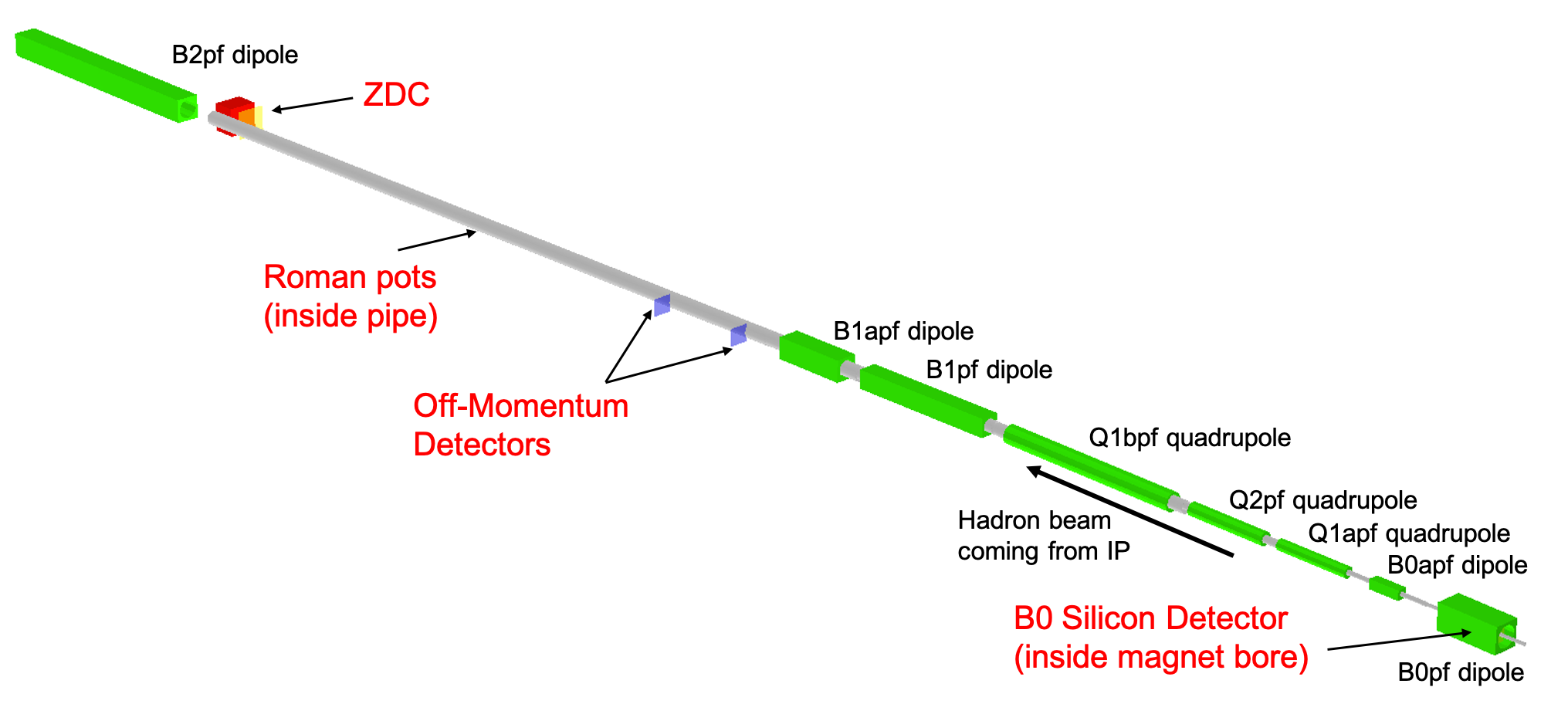}
    \caption{\textsc{Geant4} rendering of the far-forward hadron beam magnets shown in green, a simple sketch of a beam pipe, and the four detector subsystems currently included in the reference detector.}
    \label{fig:IR_FF_layout}
\end{figure}

A reference central detector design, largely matching the physics requirements, is presented as a 3D model in Fig.~\ref{central-detector-CAD} and in 2D schematic form in Fig.~\ref{central-detector-cartoon}. Figure~\ref{fig:IR_FF_layout} illustrates the very forward detectors. 
The following characteristics are assumed. The central detector instruments the pseudo-rapidity region $-4 < \eta < 4$ with full coverage of the range $\lvert \eta \rvert <$~3.5 (details are provided in Sec.~\ref{part3-sec-Det.Aspects.challenges}). This acceptance range matches the needs of the inclusive, semi-inclusive, jet physics and spectroscopy studies. It is complemented by the very forward and backward detectors ensuring the hermeticity and the forward tagging required by specific topics of the physics program, in particular exclusive reactions and diffractive channels. The main requirements of the central detector are dictated by the event geometry and the physics program, as illustrated
in detail in Vol.~II, chapter~8. They are related to {\bf (1) tracking and momentum measurements, (2) electron identification, (3) hadron identification} and {\bf (4) jet energy measurements}, while {\bf (5) the overall detector size} is imposed by collider design considerations:
\begin{enumerate}
\item{Very fine vertex resolution, at the 20~\si{\micro}m level for the three coordinates, is needed, while a moderate momentum resolution around 2\% for p$_T>$0.1~GeV/$c$ matches the physics requirements; in the reference detector this is obtained with Si vertexing surrounded by a TPC and completed by disc-shaped detectors in the forward and backward regions;}
\item{The purity requirements for electron/hadron separation are at the 10$^{-4}$ level in the backward and barrel region and, for this reason, the figures for the electron energy resolution are very demanding, in particular in the backward region where an r.m.s. of 2\%/$\sqrt(E)$ is needed, in the reference detector this is realized by PbWO$_4$ (PWO) crystals; in the same direction, provided the requested light collection system, and the material budget should not exceed 5\% X$_0$ in front of the electromagnetic calorimeter;}
\item{The identification of the different hadron species in the whole central detector coverage, namely for hadrons with momenta up to 50\,GeV/$c$, is  3$\sigma$ $\pi$/K separation over the whole range as a reference figure. In the reference detector this is obtained with various technologies: a focusing aerogel RICH in the backward arm, a high performance Detection of Internally Reflected Cherenkov light (DIRC) in the barrel, 
a dual RICH with aerogel and fluoro-carbon gas in the forward arm;}
\item{The measurement of the jet energy in the forward direction is a necessity, a resolution of the order of 50\%/$\sqrt(E)$ is required to match the needs. Sampling calorimeters with ion converter are assumed in the reference detector;}
\item{The detector extension along the beam lines impacts on the required length around the IP that has to be kept free of machine elements, typically referred to as $L^{\ast}$. The reference figure is [-4.5, 5.0] m space along the beam line, assuming the 
iteration point at zero.}
\end{enumerate}

\chapter{Detector Challenges \texorpdfstring{\&}{and} Performance Requirements}
\label{part3-chap-DetChalReq}

\section{Beam Energies, Polarization, Versatility, Luminosities}
\label{part3-sec-DetChalReq.BEPVL}

The science mission of the Electron-Ion Collider is to provide us with an understanding of the internal structure of the proton and more complex atomic nuclei that is comparable to our knowledge of the electronic structure of atoms. Unlike the more familiar molecular and atomic structure, the interactions and structures are not well separated in protons and other forms of nuclear matter, but are inextricably mixed up, and the observed properties of nucleons and nuclei, such as mass and spin, emerge out of this complex system.

A consensus study report of the National Academies of Sciences, published in 2018, on an {\bf Assessment of U.S.-based Electron-Ion Collider Science}~\cite{NAP25171}, recognized this and concluded “EIC science is compelling, timely and fundamental”. The NAS study further found that “An EIC can uniquely address three profound questions about nucleons — neutrons and protons — and how they are assembled to form the nuclei of atoms:
\begin{itemize}
\item How does the mass of the nucleon arise?
\item How does the spin of the nucleon arise?
\item What are the emergent properties of dense systems of gluons?”
\end{itemize}
They also concluded that “These three high-priority science questions can be answered by an EIC with highly polarized beams of electrons and ions, with sufficiently high luminosity and sufficient, and variable, center-of-mass energy.”

This reinforces the unique accelerator requirements of the Electron-Ion Collider, requiring a large luminosity, 10$^{33-34}$~cm$^{-2}$s$^{-1}$ over a large and variable range of center-of-mass energies, between 20 and 140 GeV, high electron and (light) ion beam polarizations of above 70\%, and a large range of accessible ion beams, from deuterium to the heaviest nuclei (uranium or lead). The electron polarization is longitudinal, whereas the polarization requirement for protons and light ions is for both longitudinal and transverse. Due to the broad science program foreseen at the Electron-Ion Collider, the possibility to have a second interaction region and associated detector is emphasized in the report.

As stated in the National Academy of Sciences Committee recommendations, “a central goal of modern nuclear physics is to understand the structure of the proton and neutron directly from the dynamics of their quarks and gluons, governed by the theory of their interactions, QCD, and how nuclear interactions between protons and neutrons emerge from these dynamics.” The scientific program of the EIC is designed to make unprecedented progress towards this goal. The large versatility of the EIC can be at the highest level understood by the requirements from the various studies driving the EIC science, for details see chapter \ref{part2-chap-EICMeasandStud} and \ref{part2-chap-DetRequirements}.

Thus, the successful scientific outcome of the EIC depends critically on:
(a) the luminosity, 
(b) the center-of-mass energy and its range, 
(c) the lepton and light ion beam polarization, 
and (d) the availability of ion beams from deuteron to the very heavy nuclei.
Two interaction regions are desired to ensure an overall optimized physics output of the EIC in terms of precision and kinematic range through careful complementary choices of basic features of the two general purpose detectors such as the Solenoid, as well as sub-detector technologies, leading to different acceptances, technology redundancy and cross calibration.

The design of the EIC was guided by the described requirements and originally expressed in the White Paper and reinforced by the recommendations of National Academy of Sciences. In its current design the machine provides collisions of polarized electrons and polarized protons in the center-of-mass energy region from  29 to 141\,GeV, and polarized electron-heavy ion collisions up to 
89\,GeV/nucleon. This is accomplished by an electron storage ring with up to 18\,GeV beam energy and colliding those electrons with polarized protons or heavy ions stored in the hadron storage ring  operating at energies 41\,GeV and 100 to 275\,GeV (protons), or 41\,GeV and 100 to 110\,GeV/nucleon (ions). 
Electron-proton luminosities reach $1.0\times 10^{34}\,{\rm cm}^{-2}{\rm sec}^{-1}.$
The two beams collide at a crossing angle of 25\,mrad, which allows to separate beams quickly, to bring the focusing beam elements close to the interaction point and at the same moment keeping synchrotron radiation background low. The loss of luminosity from the crossing angle is compensated by sets of crab cavities in each beam. Both the electron and the proton beam in the EIC will reach 70\% polarization. The same high level of polarization is also expected for $^3$He beams.

In collision the electrons and hadrons must circulate their respective storage rings with the same revolution period, which defines the  collision energy modes that will be available. 
The beam parameters and energies relevant for our detector studies including the corresponding luminosities are listed in Table~\ref{table:epBeamParameter} for \ep\ and \ref{table:eABeamParameter} for \eAu\ with and without strong hadron cooling to improve emittance of the proton/ion beams. Shown in each table are also the respective spread of the beam momentum $\mathrm{d}p/p$ and the horizontal and vertical beam divergence $\sigma_h$ and $\sigma_v$, two effects that can potentially impact the precision of various measurements and that cannot be corrected on an event-by-event basis. The former is a beam effect while the latter is caused by the machine optics. The divergence adds a $p_T$ to the beams. It can be reduced by increasing $\beta^*$ since $\sigma \propto 1/\sqrt{\beta^*}$ but at the expense of a substantially lower luminosity since $\beta^* \propto 1/{\cal L}$.   

\begin{sidewaystable}[p]
\begin{footnotesize}
		\begin{tabular}{l|cc|cc|cc|cc|cc} 
 			Species & $p$ & $e$  & $p$  & $e$  & $p$  & $e$  & $p$  & $e$  & $p$  & $e$ \\  \hline \hline
			Beam energy [GeV] & 275 & 18 & 275 & 10 & 100 & 10 & 100 & 5 & 41 & 5 \\ 
			$\sqrt{s}$ [GeV] & \multicolumn{2}{c|}{140.7} & \multicolumn{2}{c|}{104.9} & \multicolumn{2}{c|}{63.2} & \multicolumn{2}{c|}{44.7} & \multicolumn{2}{c}{28.6} \\
			No. of bunches & \multicolumn{2}{c|}{290} & \multicolumn{2}{c|}{1160} & \multicolumn{2}{c|}{1160} & \multicolumn{2}{c|}{1160} &\multicolumn{2}{c}{1160}  \\  \hline 
			{} & \multicolumn{10}{c}{High divergence configuration} \\ \hline
			RMS $\Delta \theta$, h/v [$\mu$rad] &  150/150 & 202/187 & 119/119 & 211/152 & 220/220 & 145/105 & 206/206 & 160/160 & 220/380 & 101/129 \\ 
			RMS $\Delta p/p$ [$10^{-4}$]  &  6.8 & 10.9 & 6.8 & 5.8 & 9.7 & 5.8 & 9.7 & 6.8 & 10.3 & 6.8 \\  
			Luminosity [$10^{33}{\rm cm}^{-2}{\rm s}^{-1}$] & \multicolumn{2}{c|}{1.54} & \multicolumn{2}{c|}{10.00} & \multicolumn{2}{c|}{4.48} & \multicolumn{2}{c|}{3.68} & \multicolumn{2}{c}{0.44} \\  \hline
			{} & \multicolumn{9}{c}{High acceptance configuration} \\ \hline
			RMS $\Delta \theta$, h/v [$\mu$rad] &  65/65 & 89/82 & 65/65 & 116/84 & 180/180 & 118/86 & 180/180 & 140/140 & 220/380 & 101/129 \\ 
			RMS $\Delta p/p$ [$10^{-4}$]  &  6.8 & 10.9 & 6.8 & 5.8 & 9.7 & 5.8 & 9.7 & 6.8 & 10.3 & 6.8 \\ 
			Luminosity [$10^{33}{\rm cm}^{-2}{\rm s}^{-1}$] & \multicolumn{2}{c|}{0.32} & \multicolumn{2}{c|}{3.14} & \multicolumn{2}{c|}{3.14} & \multicolumn{2}{c|}{2.92} & \multicolumn{2}{c}{0.44} \\ \hline
		\end{tabular}
\end{footnotesize}
		\caption{Beam parameters for $e$+$p$ collisions for the available
	center-of-mass energies $\sqrt{s}$ with strong hadron cooling. Luminosities and beam effects depend on the configuration. Values for high divergence and high acceptance configurations are shown.}
\label{table:epBeamParameter} 
\vspace{5mm}
\begin{footnotesize}
		\begin{tabular}{l|cc|cc|cc|cc} 
			Species & Au & $e$ & Au & $e$ & Au & $e$ & Au & $e$\\  \hline \hline	
			Beam energy [GeV] & 110 & 18 & 110 & 10 & 110 & 5 & 41 & 5  \\ 
			$\sqrt{s}$  [GeV] & \multicolumn{2}{c|}{89.0} & \multicolumn{2}{c|}{66.3} & \multicolumn{2}{c|}{46.9} & \multicolumn{2}{c}{28.6} \\ 
			No. of bunches & \multicolumn{2}{c|}{290} & \multicolumn{2}{c|}{1160} & \multicolumn{2}{c|}{1160} & \multicolumn{2}{c}{1160} \\ \hline
			{} & \multicolumn{8}{c}{Strong hadron cooling} \\ \hline
			RMS $\Delta \theta$, h/v [$\mu$rad] &  218/379 & 101/37 & 216/274 & 102/92 & 215/275 & 102/185 & 275/377 & 81/136  \\ 
			RMS $\Delta p/p$ [$10^{-4}$]  &  6.2 & 10.9 & 6.2 & 5.8 & 6.2 & 6.8 & 10 & 6.8 \\ 
			Luminosity [$10^{33}{\rm cm}^{-2}{\rm s}^{-1}$] & \multicolumn{2}{c|}{0.59} & \multicolumn{2}{c|}{4.76} & \multicolumn{2}{c|}{4.77} & \multicolumn{2}{c}{1.67} \\  \hline
			{} & \multicolumn{8}{c}{Stochastic  cooling} \\ \hline
			RMS $\Delta \theta$, h/v [$\mu$rad] &  77/380 & 109/38 & 136/376 & 161/116 & 108/380 & 127/144 & 174/302 & 77/77  \\ 
			RMS $\Delta p/p$ [$10^{-4}$]  &  10 & 10.9 & 10 & 5.8 & 10 & 6.8 & 13 & 6.8 \\ 			
			Luminosity [$10^{33}{\rm cm}^{-2}{\rm s}^{-1}$] & \multicolumn{2}{c|}{0.14} & \multicolumn{2}{c|}{2.06} & \multicolumn{2}{c|}{1.27} & \multicolumn{2}{c}{0.31} \\  \hline
		\end{tabular} 
\end{footnotesize}
		\caption{Beam parameters for $e$+Au collisions for the available
		center-of-mass energies $\sqrt{s}$. Luminosities and beam effects depend on the cooling technique. Values for strong hadronic and stochastic cooling are shown. }
	\label{table:eABeamParameter}
\end{sidewaystable}

The physics working groups had selected a set of beam species and energy modes as benchmarks for the studies conducted for this report. They are listed in Table \ref{table:YRCollisionSystemMatrix}. This set could be also considered as a good example for the operation modes of a future scientific program at the EIC.  
Most studies considered integrated luminosities of 10 fb$^{-1}$ and 100 fb$^{-1}$ and assumed a polarization of 70\% for electrons, protons, and light ions.

\begin{table}[tbh]
	\centering
			\begin{tabular}{l|cc|cc|cc|cc} 
				Beams & \multicolumn{8}{c}{Collision energy modes (GeV)} \\ \hline 
				{} & $E_e$  & $E_h$ & $E_e$  & $E_h$ & $E_e$  & $E_h$ & $E_e$  & $E_h$ \\ \hline \hline
                $e$+$p$        & 18 & 275 & 10 & 100 & 5  & 100 &  5 & 41 \\
                $e$+$d$        & 18 & 135 & 10 & 100 & 5  & 100 &  5 & 41 \\
                $e$+$^3$He     & 18 & 110 & 10 & 110 & {} &  {} &  5 & 41 \\
                $e$+$^4$He     & 18 & 110 & 10 & 110 & {} &  {} &  5 & 41 \\
                $e$+C          & 18 & 110 & 10 & 110 & {} &  {} &  5 & 41 \\
                $e$+$^{40}$Ca  & 18 & 110 & 10 & 110 & {} &  {} &  5 & 41 \\
                $e$+Cu         & 18 & 110 & 10 & 110 & {} &  {} &  5 & 41 \\
                $e$+Au         & 18 & 110 & 10 & 110 & {} &  {} &  5 & 41 \\ \hline
		\end{tabular}
		\caption{Collision species and energy modes considered for physics simulations for this report.  Energies of ion beams ($E_h$) are always quoted per nucleon.}
\label{table:YRCollisionSystemMatrix}
				
\end{table}

\FloatBarrier

\section{Integrated Detector and Interaction Region}
\label{part3-sec-DetChalReq.ID.IR}

Beyond the unique accelerator requirements, EIC science also leads to a unique set of detector requirements, and to integrate the main and all the detectors along the hadron ans electron beam from the beginning into the interaction region layout. All final state particles carry information 
about the 3D QCD structure of nuclear matter and the emergent phenomena. Therefore, it is essential that the interaction region and the detector 
at the EIC are designed so all particles are identified and measured at as close to 100\% acceptance as possible and with the necessary resolutions.

\begin{figure}[ht]
\begin{center}
\includegraphics[width=0.50\textwidth]{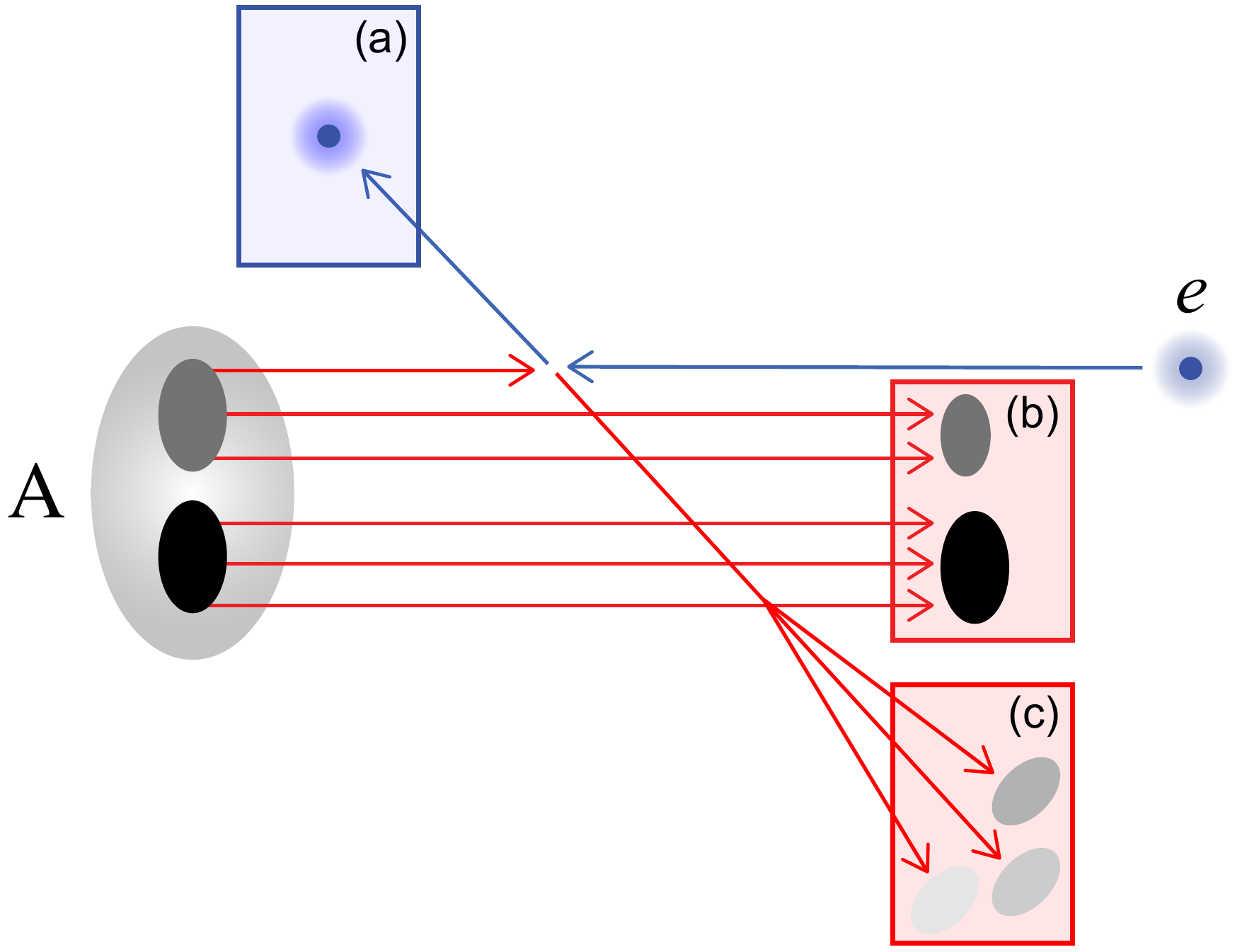}
\end{center}
\caption {\label{fig:finalstateparticles} Classification of the final-state particles of a DIS process at the EIC: (a) scattered electron, (b) the particles associated with the initial state ion, and (c) the struck parton.}
\end{figure}

The basic physics process at the EIC is Deep Inelastic Scattering (DIS), which is represented in Fig.~\ref{fig:finalstateparticles}. An ion, composed of nucleons, which are in turn composed of partons (quarks and gluons), moves in one direction and collides with an electron moving in the other direction. The electron collides with a parton within the ion in a hard collision.

We qualitatively define three classes of particles in the final state:
\begin{itemize}
\item The scattered electron,
\item Particles associated with the initial state ion, and
\item Particles associated with the struck parton.
\end{itemize}

The difficulty in achieving good acceptance in the forward regions at a collider has to do with the accelerator elements needed to deliver the colliding beams. To first order, the luminosity at the interaction point is inversely proportional to the distance between the nearest quadrupole magnets. On the other hand, the closer the beam elements are to the interaction point, the more they obstruct the acceptance at shallow angles with respect to the beam axis, and restrict the acceptance for forward particles. This complicates achieving close to 100\% acceptance for all three types of particles, but especially the particles associated with the initial state ion.

This leads to a unique and novel integration of the detector in the interaction region, extending over a large region ($\pm$ 40 m) beyond the main detector. It also should be pointed out that any change has impact on a wide variety of systems, from beam dynamics to accelerator performance to magnet engineering requirements to detection capability – the latter in terms of (gaps in) acceptance, particle identification and resolutions.

Further integration is required for ancillary measurements necessary to deliver on the EIC science program, such as those to absolutely determine the longitudinal electron and longitudinal or transverse proton/light-ion polarizations with beam polarimeters, with good systematic (order 1\%) understanding. For the electrons, a transverse polarization measurement is not a requirement but is often useful to underpin the systematic understanding of the spin direction. Further, the exact frequency of electron-ion collisions per second must be experimentally determined with luminosity monitors with a goal of better than $\sim$1\% understanding. Both of these goals are non-trivial, 
with beam dynamics potentially impacting long-time averaging methods.

The central detector region of the EIC is designed to measure those final state particles from the hard collision between the electron and 
the parton in the ion (particles of types (a) and (c) in Fig.~\ref{fig:finalstateparticles}) and is very much like the traditional collider detectors. The EIC central detector needs to provide the measurements to determine $Q^2$ and $x$ variables of the electron scattering kinematics (related to the spatial resolving power $\sim$ $1/Q^2$ and the quark/gluon density $\sim$ $1/x$ of the QCD landscape). The central detector is divided into three sections, the Electron-endcap, the Hadron-endcap and the Barrel. The three different central detector sections correspond to different $x$ and $Q^2$ regions for the scattered electron.

Beyond determination of $x$ and $Q^2$, measurement of two transverse kinematics degrees of freedom (transverse momentum and impact parameter), as well as flavor identification of the partonic collision is central to the 3D QCD nucleon and nuclear structure program planned for the EIC. The energy scale of the transverse kinematics is $\sim$200 MeV/$c$. This means that identification and precise measurements of single hadrons among the particles associated with the scattered partons (Particles of type (c) in Fig.~\ref{fig:finalstateparticles}) are also needed in the central detector.

Crucial information on hadron structure is carried by particles that do not emerge from the beam envelope within the coverage of the central detector. Broadly speaking, there are two types of forward final state particles that need to be reconstructed. The first type of forward particles comes from interactions in which the beam particle receives a large transverse momentum kick and fragments into many parts. These particles typically retain a velocity similar in magnitude, but with significantly different kinematics, from that of the incident beam particle and may have very different charge-to-mass ratios. Such particles will separate relatively rapidly from the beam. An example of such a particle is a forward proton from a deuteron-electron DIS, a process that can give information on QCD neutron structure (inside a deuteron) comparable to QCD proton structure. Another example are forward neutrons.

The second type is a (hadron) beam particle that stays intact during the collision, only loses a small fraction of its longitudinal momentum, and acquires a small transverse momentum. These particles are for example protons or ions in non-dissociative diffractive interactions, and will have a trajectory that is close to the proton (ion) beam. To map these types of forward final state particles, a highly-integrated extended (“far-forward”) detector region (see fig. \ref{fig:IR_FF_layout}) is defined downstream of the ion beam, and after the beam final-focusing elements, covering about 30 m. The far-forward detector region together with the central detector provides essential near-complete coverage for final state particles associated with the incident ion-beam particle.

Similarly, the “far-backward” detector region, (detectors along the outgoing lepton beam) is highly integrated to capture a third type of measurements, those close to the beam line in the electron-beam direction. This allows monitoring of the luminosity and an increase of the low-$Q^2$ coverage of the detector. Electron-ion collisions, where the electron is scattered through a very shallow angle, correspond to the case where the exchanged photon is almost real. Such photoproduction processes are of interest in their own right, but also can enable a program of hadron spectroscopy.

The science program at the EIC has the potential to revolutionize our understanding of 3D QCD nuclear and nucleon structure. It will also explore new states of QCD. In order to maximize the potential of an EIC, it is important to have a large (near-100\%) acceptance not only in the central region, but also in the region that is close to both the ion-beam and electron-beam direction — i.e. a total acceptance detector is needed. There has never been a collider detector that has both the central and forward (or backward) acceptances maximized in tandem, and this design is uniquely suited to the EIC physics program.

\section{Rate and Multiplicities}
\label{part3-sec-DetChalReq.RatMul}


\begin{table*}[hbt]
\vspace*{2mm}
\caption{Total \ep ~cross-section ($Q^2 > 10^{-9}, 10^{-9} < y < 0.99$ as a function of electron and proton beam energies. The cross-sections were calculated using PYTHIA6 event generator and might change slightly depending on the settings.}
\label{Collision.table1}
\centering
\begin{tabular}{l|l|llll}
\toprule
\multicolumn{2}{c|}{\multirow{2}{*}{$\sigma_{tot}$(\si{\micro}b)}} & \multicolumn{3}{c}{$E_e$~[GeV]} \\ \cline{3-5} 
\multicolumn{2}{c|}{}                                         & 5       & 10     & 18      \\ \hline
\multirow{3}{*}{$E_p$~[GeV]}              & 41                & 25.9    & 30.1   & 35.0   \\ 
                                          & 100               & 32.1    & 37.1   & 41.6    \\ 		
                                          & 275               & 39.4    & 44.6   & 49.3    \\ \bottomrule
\end{tabular}
\end{table*}

\begin{figure}[tb]
\includegraphics[width=0.99\textwidth]{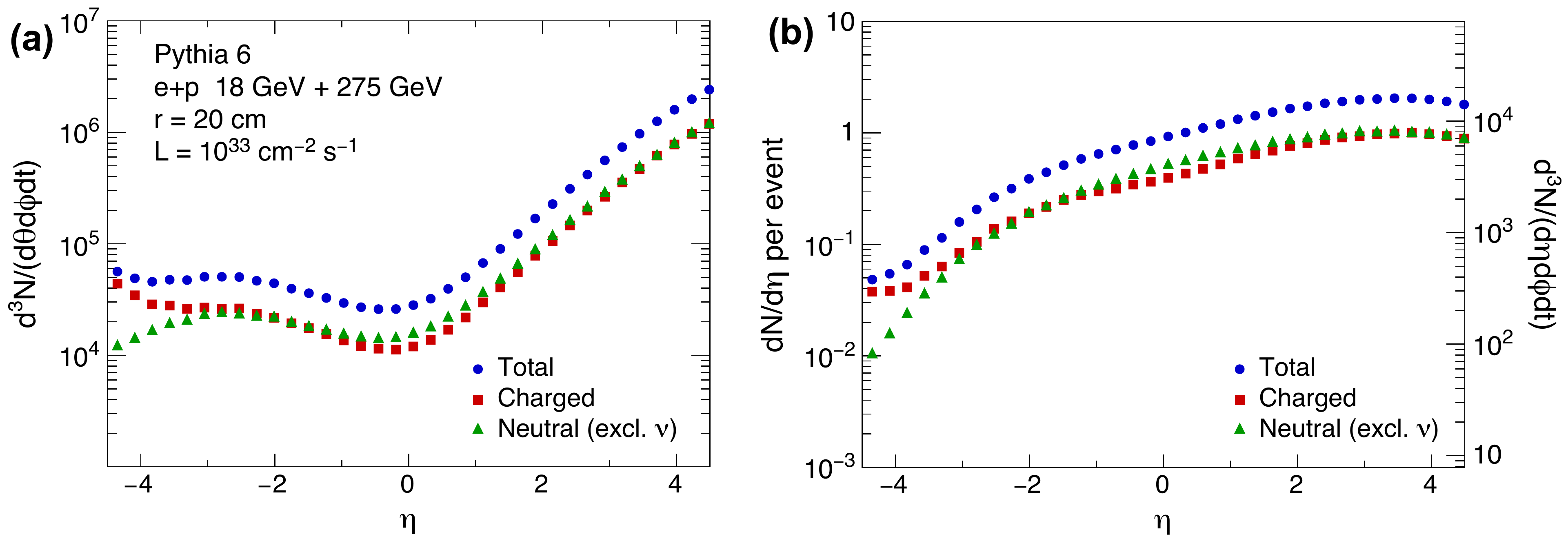}
\caption{Particle production rates as a function of pseudo-rapidity at EIC for 18\,GeV on 275\,GeV \ep\ collisions and a luminosity of $10^{33} cm^{-2} s^{-1}$. (a) particles per second per unit ($\phi, \theta$), i.e., the $\eta$-dependent flux at a distance of 1 m from the interaction point. (b) mean numbers of particles per event (left axis) and particles per second per unit ($\eta,\phi$) (right axis).}
\label{IR.rates}
\end{figure}

The EIC total \ep\ cross-section is estimated using the PYTHIA6 event generator as listed in Table~\ref{Collision.table1}. 
For each collision, Figure~\ref{IR.rates} shows the particle production rates for the 20~GeV on 250~GeV beam energy configuration. Events were simulated using PYTHIA6, and the total cross section reported by PYTHIA6 was used to scale event counts to rates. No cuts, for example on event $Q^2$ or particle momentum, were applied. The $\eta$-range spans the expected acceptance of the main EIC detector. The term "charged" particles refers to electrons, positrons, and charged long-lived hadrons, while "neutrals" refers to photons, neutrons, and $K^0_L$.

\begin{figure}[tb]
\centering
\includegraphics[width=1.0\linewidth]{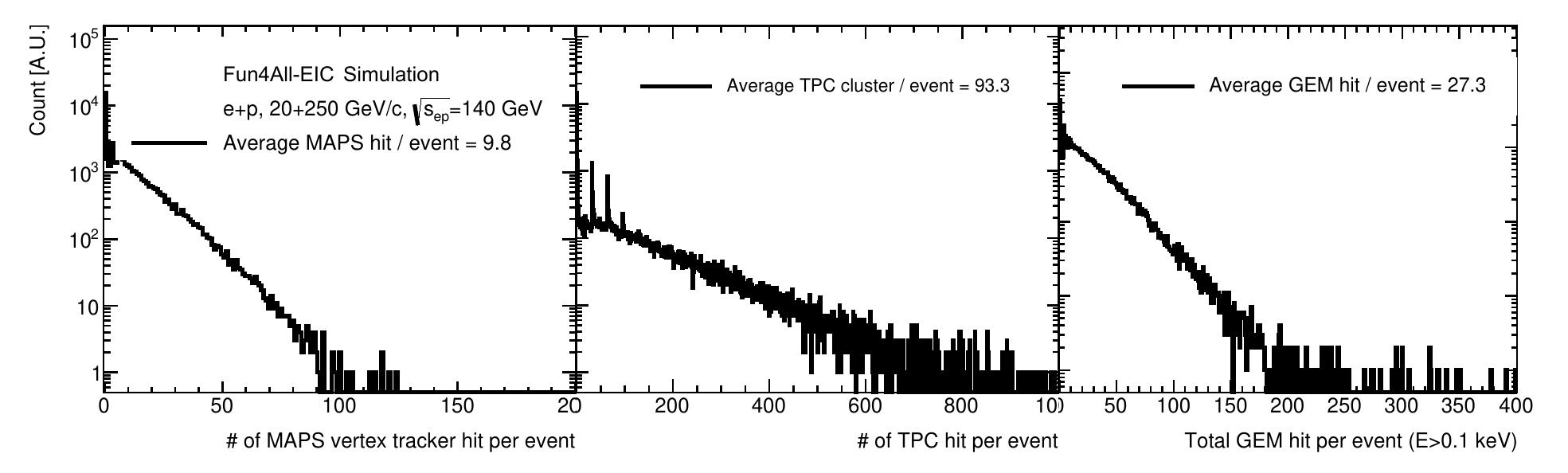}
\caption{Distribution of number of hits in the tracking detectors that originated from a single \ep~collision at $\sqrt{s_{ep}}=140$~GeV.}
\label{fig:daq-rate-tracker}
\end{figure}

\begin{figure}[tb]
\centering
\includegraphics[width=1.0\linewidth]{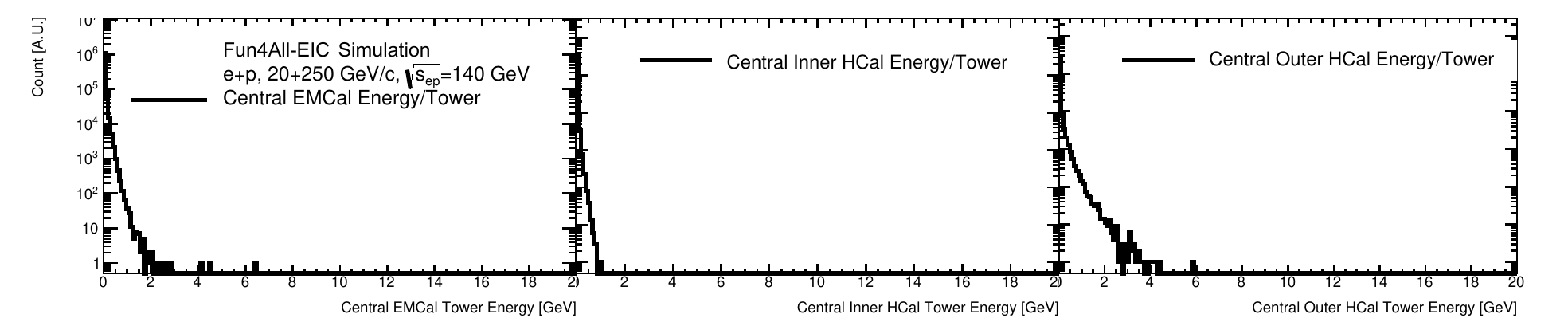}
\includegraphics[width=1.0\linewidth]{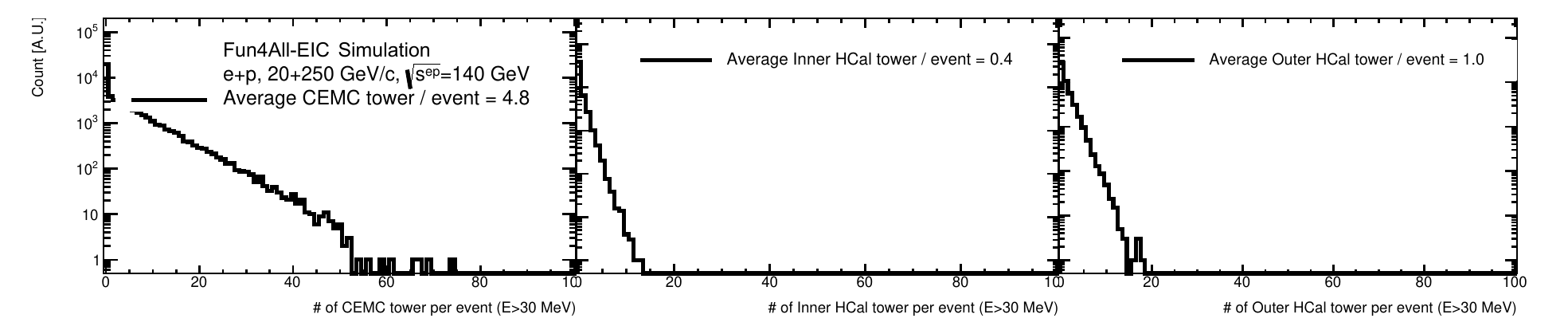}
\caption{Distribution of per-tower energy and the number of active towers in the central calorimeters that originated from a single  \ep~collision at $\sqrt{s_{ep}}=140$~GeV.}
\label{fig:daq-rate-ccal}
\end{figure}

\begin{figure}[tb]
\centering
\includegraphics[width=1.0\linewidth]{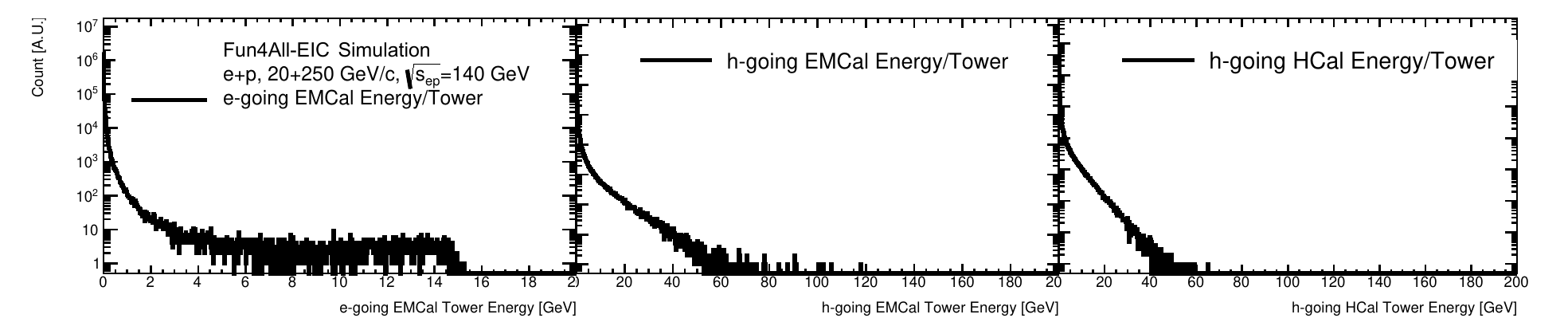}
\includegraphics[width=1.0\linewidth]{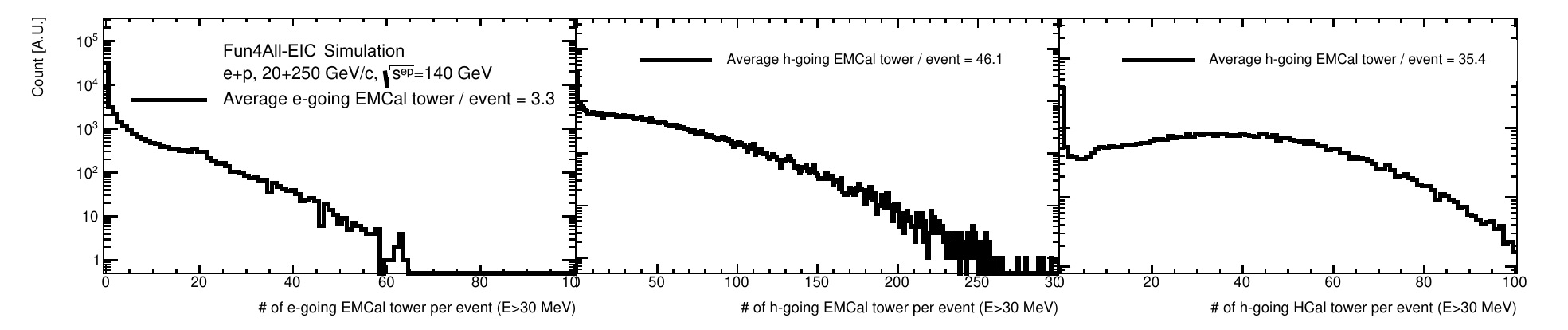}
\caption{Distribution of per-tower energy and the number of active towers in the forward calorimeters that originated from a single \ep~collision at $\sqrt{s_{ep}}=140$~GeV.}
\label{fig:daq-rate-fcal}
\end{figure}

\begin{figure}[tb]
\centering
\includegraphics[viewport=125bp 0bp 1300bp 450bp,clip,width=1.0\linewidth]{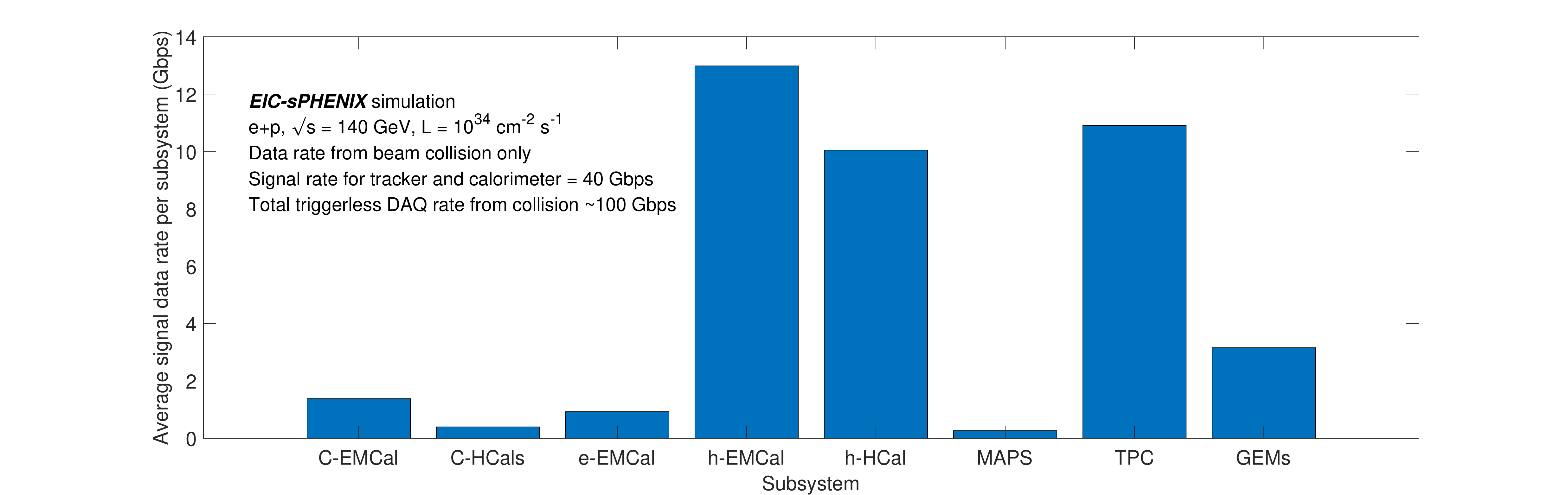}
\caption{
Signal data rates from tracking and calorimetric detectors from EIC collisions via full detector \textsc{Geant4} simulation of an EIC detector concept based on sPHENIX~\cite{Adare:2014aaa,sPH-cQCD-2018-001}, which also applies to the reference EIC detector as in this report. Zero-suppression and realistic data format based on sPHENIX prototyping are assumed in this estimation.  The overall tracker data rate is 40~Gbps. The estimated rate with PID detector and moderate detector noise would reach 100 Gbps for full experiment. Please note that the backgrounds, e.g. beam gas interaction, excessive detector noise, synchrotron photon hit rate, may dramatically increase the data rate in some detectors, but they are not included in this plot and they will be discussed in
Section~\ref{part3-sec-DetChalReq.Backgrounds}.
}
\label{fig:daq-rate-collision}
\end{figure}

The EIC detector response to the collisions and the data rate were studied using full detector \textsc{Geant4} simulations of 
a generic EIC detector model~\cite{Adare:2014aaa,sPH-cQCD-2018-001}.
The subsystem multiplicity distributions (Figures~\ref{fig:daq-rate-tracker},
\ref{fig:daq-rate-ccal}, and~\ref{fig:daq-rate-fcal}) and the average data rate
(Figure~\ref{fig:daq-rate-collision}) are studied in a simulation combining the
EIC tune of PYTHIA6, which samples $\sim50$~\si{\micro}b of the \ep~collision cross
section, and the full detector \textsc{Geant} simulation. At the top instantaneous luminosity of $10^{34}$~cm$^{-1}$s$^{-1}$, the collision-induced zero-suppressed streaming data rate from EIC collisions is around 100~Gbps, which is the minimal amount of raw data that has to be recorded to disk in order to record all minimum-bias EIC collisions in the central detector without the assumption of online reconstruction and reduction.

\FloatBarrier

\section{Backgrounds}
\label{part3-sec-DetChalReq.Backgrounds}




The combination of the relatively low signal rate of the EIC collisions and the requirement for stringent systematic control for 
EIC measurements calls for low background and detector noise at an EIC experiment. In turn, the types and levels of backgrounds are one of the main considerations on the detector design and it is a major consideration for IR integration, such as the arrangement of the beam magnets as well as 
other beam parameters and optics. 
The experience at earlier accelerator facilities, especially the previous HERA electron-proton collider, indicates the importance of background studies in the early design phase. Primary sources of machine-induced background are discussed in this subsection. One of the problems at HERA the accumulation of positively charged ions or dust when running with electrons has not yet been considered.


\subsection {Ionization radiation dose and neutron flux from the EIC collisions}
    
\begin{figure}[th]
\centering
\includegraphics[width=.82\linewidth]{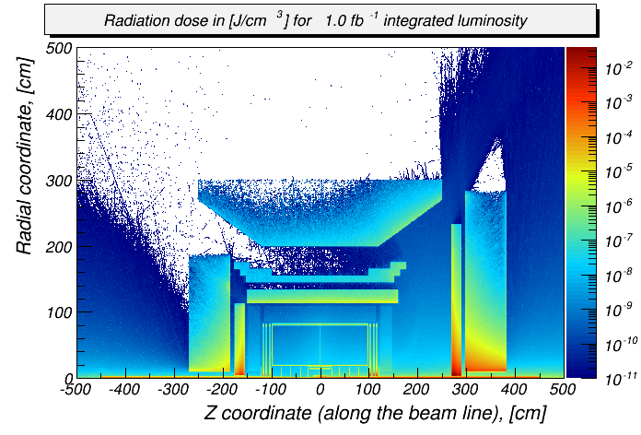}
\caption{Ionizing radiation energy deposition from \ep~collision at $\sqrt{s_{ep}}=140$~GeV studied using the BeAST detector concept, which also applies to the reference EIC detector as in this report.}
\label{fig:TID-collision}
\end{figure}

\begin{figure}[bh]
\centering
\includegraphics[width=.82\linewidth]{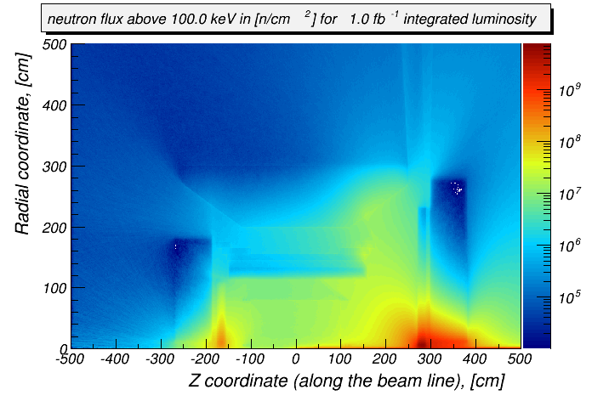}
\caption{Neutron flux from the \ep~collision at $\sqrt{s_{ep}}=140$~GeV studied using the BeAST detector concept with the assumed location in the RHIC, located/placed in the RHIC IP6 experimental hall, which also applies to the reference EIC detector as in this report.}
\label{fig:neutron-fix-collision}
\end{figure}

The ionization radiation dose and neutron flux from the \ep~collisions are studied using EICROOT and a generic EIC detector model in the RHIC IP6 experimental hall. The simulation is generated with the EIC tune of PYTHIA6 with $20\times250$\,GeV beam energy and is based on the 
\textsc{Geant3} package with the $HADR=5$ option. As shown in Figure~\ref{fig:TID-collision}, the near-beam-line regions experience relatively high ionizing radiation. For example, the crystal calorimeters in the backward arm show approximately 2.5\,kRad/year max ionizing radiation dose from the \ep~collisions at the top luminosity ($10^{34}$\,cm$^{-2}$s$^{-1}$). The above-100\,keV neutron flux is shown in Figure~\ref{fig:neutron-fix-collision}. The near-beam-line regions, in particular the vertex tracker and the forward-backward calorimeters also experience relatively high neutron flux, exceeding $10^{10}$~neutrons/cm$^2$ per year from the \ep~collisions at the top luminosity ($10^{34}$~cm$^{-2}$s$^{-1}$). 




\subsubsection {Synchrotron radiation}

Various sources of synchrotron radiation could have an impact on the background level at the IP.
When the trajectory of a charged particle is bent, synchrotron photons are emitted that are tangential to the particle’s path. 
Bending and focusing of the electron beam is the main cause of synchrotron radiation within the IR. It is important to place the IP far away 
from strong bending magnets in the arcs to minimize synchrotron radiation. The tracking detectors in the central detector as well as the 
calorimeter have to be properly shielded against synchrotron radiation, therefore a number of absorbers and masking must be applied along the electron beam direction. Synchrotron radiation also deposits several kilowatts of power into the beam pipe in the central detector region, which must then be cooled. Additionally, synchrotron radiation can degrade vacuum quality by causing material desorption from vacuum
chamber walls and/or heating residual gas. Synchrotron radiation is also a direct and indirect source of background in  the luminosity monitor, and low-Q$^2$ tagger located on the downstream electron side of the IR.
However, background from the contribution from the upstream electron beam scattering off residual gas must still be assessed.

\begin{figure}[htbp]
\includegraphics[width=\textwidth]{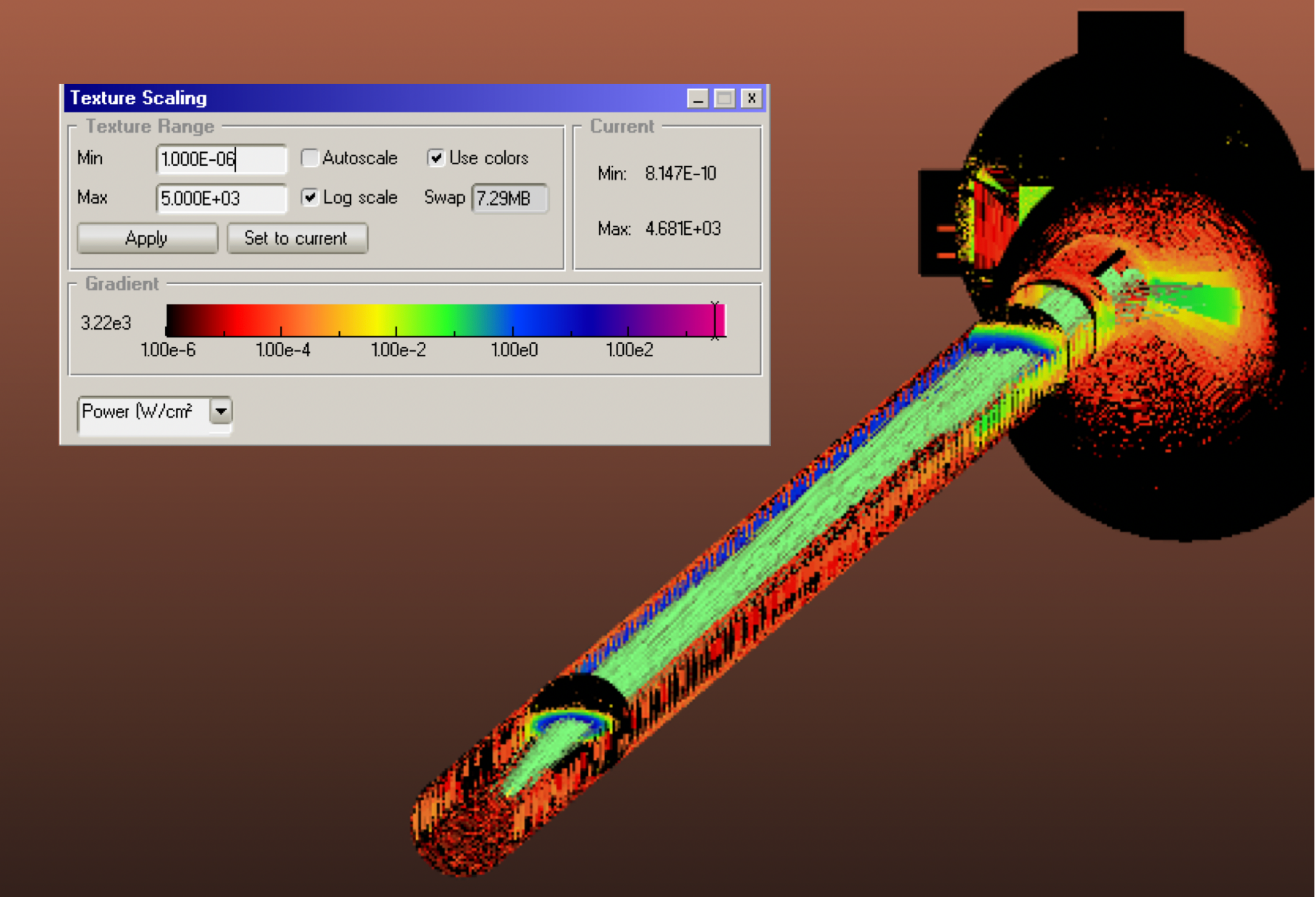}
\caption{SYNRAD generation of synchrotron radiation from 0.260\,A of 18\,GeV electrons. The color scale is logarithmic, with blue approximately 1\,W/cm$^2$.  Electrons enter from the lower left in the figure, the initial radiation fan is generated from the last dipole, at approximately 40 m upstream of the IP. Individual photons are traced by the green lines.  The vertical striations on the beam pipe result from the sawtooth inner
profile of the pipe, which ensures photons hit the wall locally head-on.  }
\label{fig:SYNRAD}
\end{figure}

\begin{figure}[htbp]
\begin{center}
\includegraphics[width=1.0\linewidth]{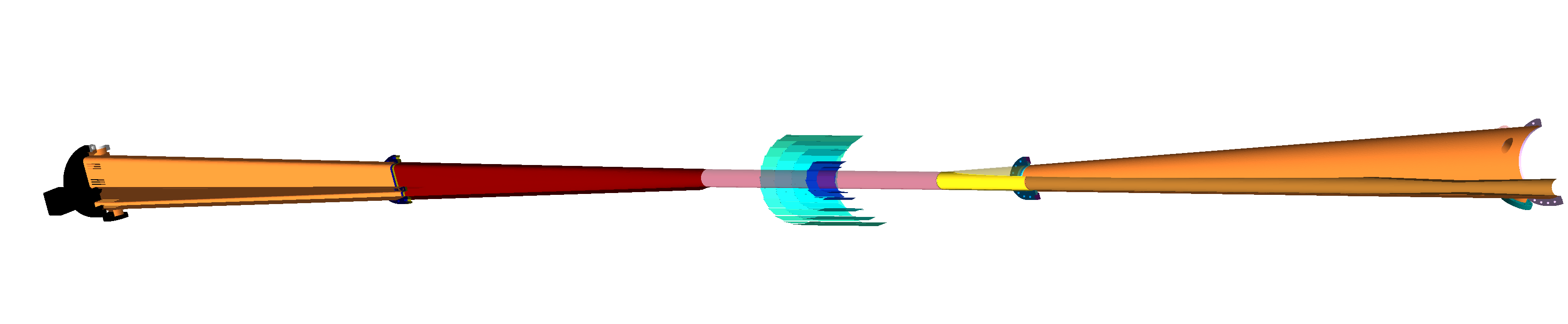}
\caption{\textsc{Geant4} model of IR1 Beam Pipe, with Si Vertex Tracker.  The electron beam enters horizontally from the right, and exits in the rectangular beam channel to the left.  The ion beam enters in the small tube on the lower left, and exits via the large cone on the upper right.}
\label{fig:IR1-beampipe}
\end{center}
\end{figure}

A model of the electron beamline (Fig.~\ref{fig:IR1-beampipe}) has been used in SynRad~\cite{Kersevan:2019vme}, where 
synchrotron radiation for the 18\,GeV electron beam at the maximal design value
of 0.260~A, including 26 mA in a broad tail distribution, has been generated. 
Figure~\ref{fig:SYNRAD} shows a view of
the  upstream electron beamline and IP, with synchrotron radiation generated by the
last upstream dipole and FFQ quadrupoles.  
Electrons enter from the lower left on the figure, at the location of the last dipole, $\approx 40$ m from the IP. The IP itself is obscured by the hourglass shape of the central region of the beam pipe. 

\begin{figure}
\centering
\includegraphics[width=\linewidth]{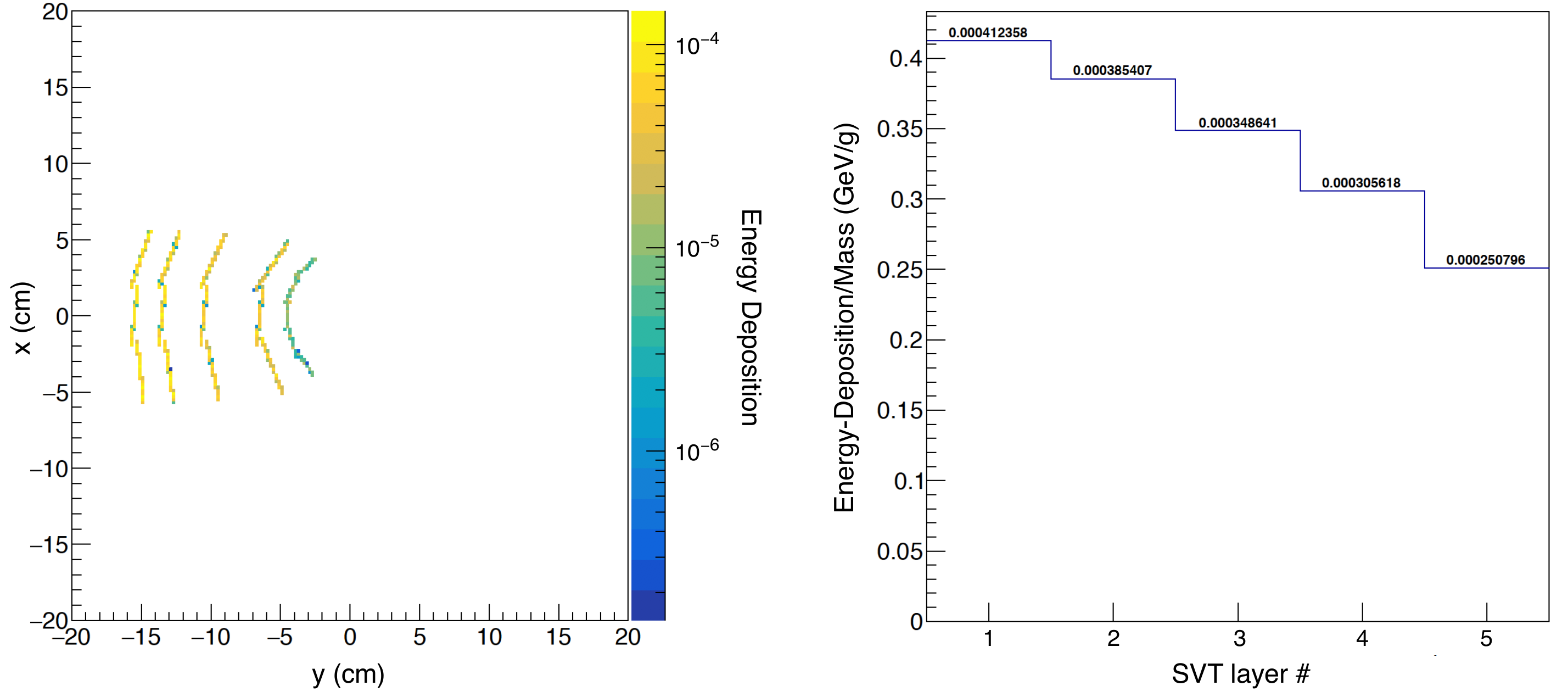}
\caption{Left: synchrotron energy deposition in central Be pipe and 5 layers of SiVT.
Energy is integrated over the length of each element.
Right: synchrotron radiation dose (GeV/gram) in each of 5 layers of SiVT, which is averaged over the length of each element. The photon flux in these figures is integrated over $0.465~\mu$sec of an 18\,GeV electron beam at the design current of 0.26 A, including a beam tail.}
\label{fig:E-SiVT}
\end{figure}

Figure~\ref{fig:E-SiVT} left panel illustrates the energy deposition in the Be beam pipe and Si Vertex Tracker layers and the right panel
the dose (energy per mass) in the Si layers. The photon flux in these figures is integrated over 0.465\,\si{\micro}s of an 18\,GeV electron 
beam at the design current of 0.26\,A, including a beam tail.

\begin{figure}
\vspace*{2mm}
\centering
\includegraphics[viewport=25bp 100bp 800bp 550bp,clip,width=.9\linewidth]{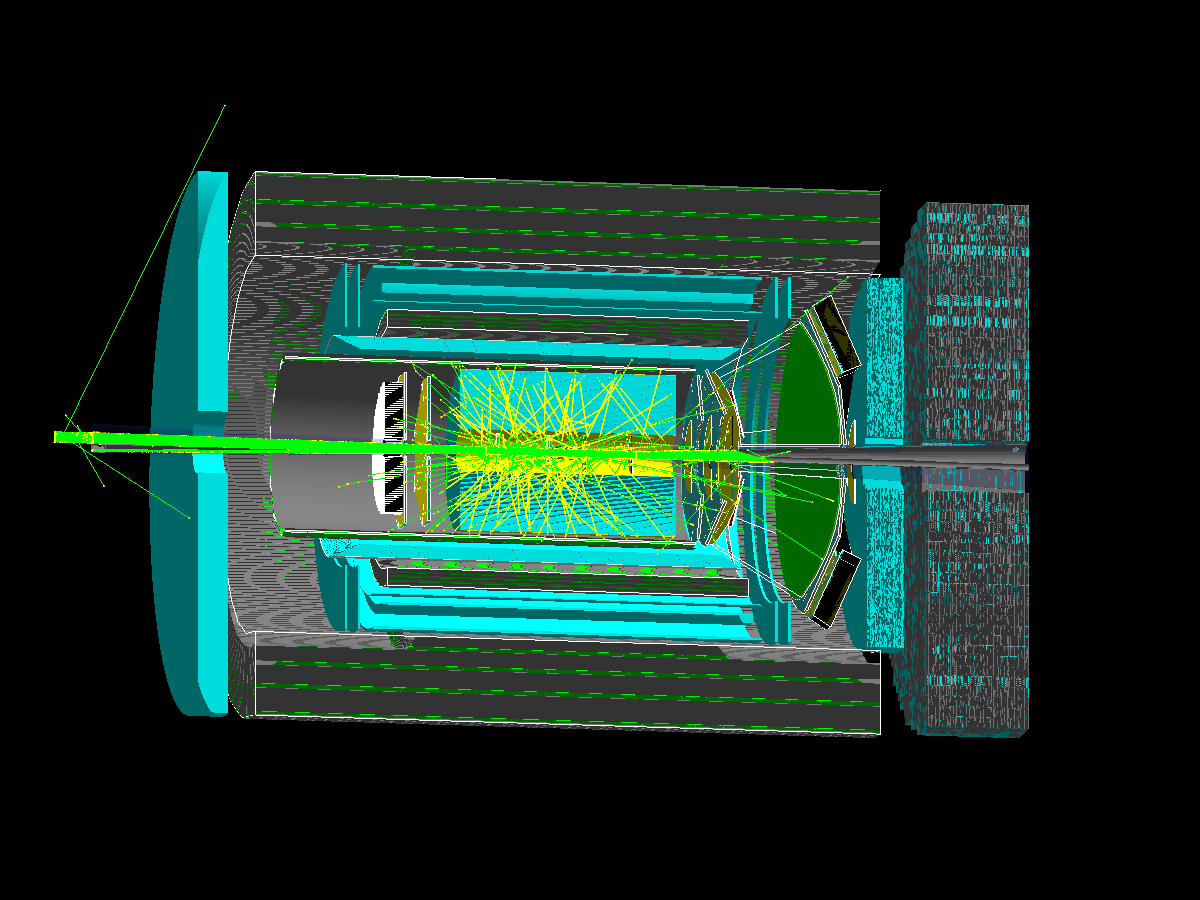}
\caption{\textsc{Geant4} simulation of the synchrotron photon background in an EIC detector model~\cite{Adare:2014aaa,sPH-cQCD-2018-001}. The photons (green lines) are generated in SynRad simulation~\cite{Kersevan:2019vme} and the photons are interfaced to the \textsc{Geant4} simulation after passing the Final Photon Absorber around $z>3$~m. Although the inner detectors have the highest flux of synchrotron photon background, the background affects tracking and PID detectors in much higher radii too, due to the scattering and secondary interactions.}
\label{fig:daq-sim-synrad-display}
\end{figure}

Furthermore, a \textsc{Geant4}-based tool-set is being prepared to examine the hit rate that originates from the synchrotron radiation background 
in the full experiment apparatus. The SynRad synchrotron radiation simulation~\cite{Kersevan:2019vme} as previously discussed is interfaced with the detector response as modeled in full detector \textsc{Geant4} simulations and the digitization model of a generic EIC detector model
~\cite{Adare:2014aaa,sPH-cQCD-2018-001}, as illustrated in Figure~\ref{fig:daq-sim-synrad-display}. The detector hit rate results are pending to be updated with the July-2020 beam chamber and optics adjustment. 

\subsubsection {Beam-gas interactions}

Beam-gas interactions occur when proton or ion beam particles collide with residual gas. Ion beam interactions with gas cause beam particle losses and halo, which reach detectors. This is an important source of neutrons that thermalize within the detector hall. The large synchrotron radiation load could heat the beam pipe and residual gas particles from the beam pipe walls could be released, which would lead to a degradation of the vacuum. A crossing angle and short section of shared beam pipe in the EIC design minimize the beam-gas problem.

\begin{figure}[htbp]
\begin{center}
\includegraphics[width=.9\linewidth]{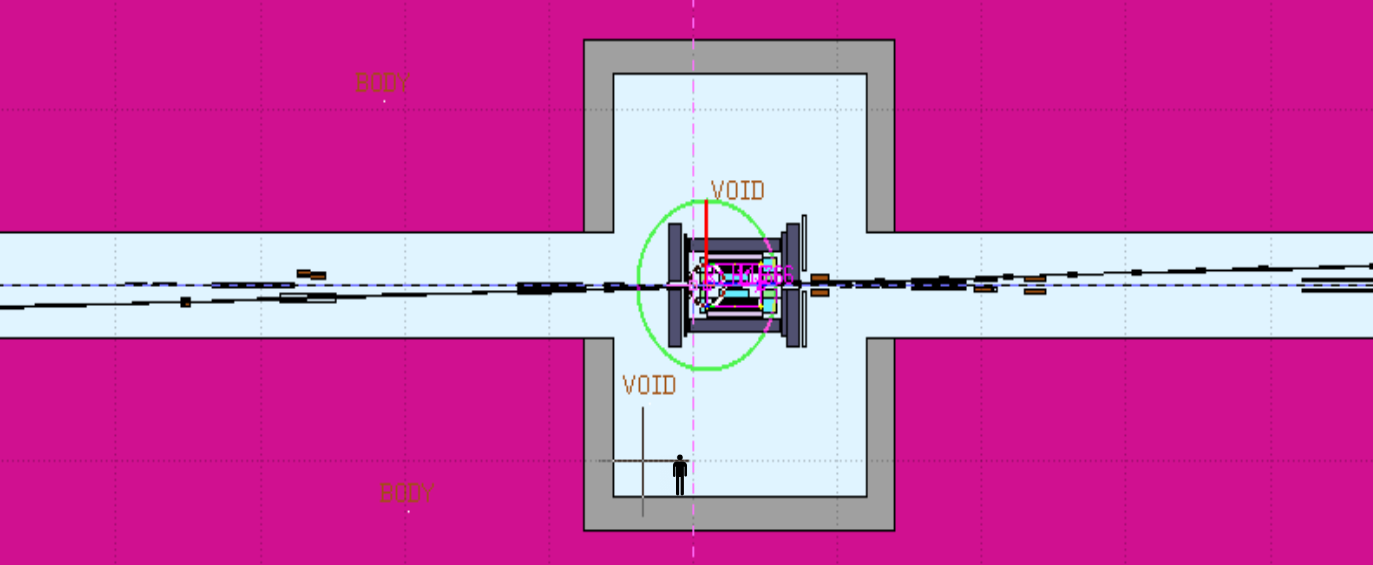}
\caption{Top view of a FLUKA model of the EIC Interaction Region 1 (person for scale comparison is shown at the bottom).
Ions enter from lower left, electrons enter on the solenoid axis from the right.}
\label{fig:IR1-FLUKA}
\end{center}
\end{figure}

A model of the  interaction region-1 (IR1), $\pm\ 30$ m, including all magnets, the tunnel walls, the detector cavern, and a simplified representation of the detector  have been created in FLUKA.   This is illustrated in Fig.~\ref{fig:IR1-FLUKA}.  
A more detailed view of the detector model is presented in Fig.~\ref{fig:Det1-FLUKA}.

\begin{figure}[htbp]
   \begin{center}
    \includegraphics[width=.9\textwidth]{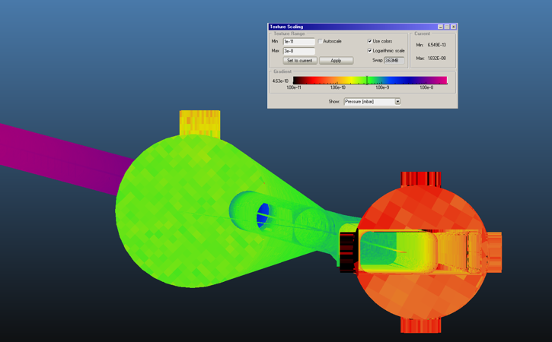}
\caption{MOLFLOW calculation  of the static vacuum in the IR.  The beam pipe layout is the same as Fig.~\ref{fig:SYNRAD}.
In this view, the electron beam enters from the upper left and exits through the large horizontal aperture on the right.  The incident ions enter from the right at $z=-4.5$~m via the smaller upright rectangular aperture. The light green color in the central region indicates a vacuum of $\approx 5\cdot 10^{-9}$ mbar. The downstream ion beam pipe is not shown beyond the flange at $z = 4.5$~m.}
\label{fig:StaticVacuum}
\end{center}
\end{figure}

\begin{figure}[htbp]
\begin{center}
\includegraphics[width=.9\linewidth]{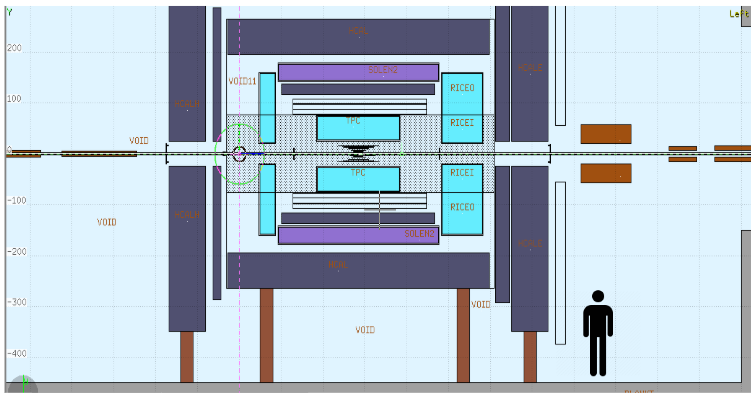}
\caption{Elevation view of the FLUKA model of the EIC Detector.  The Si Vertex Tracker (SiVT) in this rendition includes six layers.}
\label{fig:Det1-FLUKA}
\end{center}
\end{figure}

The studies of the dynamic vacuum in the IR are directly linked to the synchrotron radiation flux impacting the beam pipe.  
Figure~\ref{fig:StaticVacuum} illustrates the static vacuum (without synchrotron radiation) in IR1, based on nominal out-gassing rates, the molecular flow conductance of the beam pipe,  and the pumping speed of the NEG pumps at $\pm 4.5$ m.

In order to efficiently simulate the interactions of the ion beam with the residual gas in the beam-line vacuum, we 
artificially create a thin ``pencil'' (diameter 3mm) of air at pressure $P_F = 100$ mbar along the beam-line.  A global view of
the neutron fluence is presented in the top panel of Figure~\ref{fig:nflmp}.  
The simulation includes the full cascading and thermalization of secondaries from
the primary beam-gas interactions.  The figure illustrates the fact that the detector itself, especially the iron flux return, serves as both
a neutron sink and neutron source.

\begin{figure}[!htb]
\center{\includegraphics[clip, trim=1.9cm 0.2cm 0.7cm 0.9cm, width=.65\linewidth]{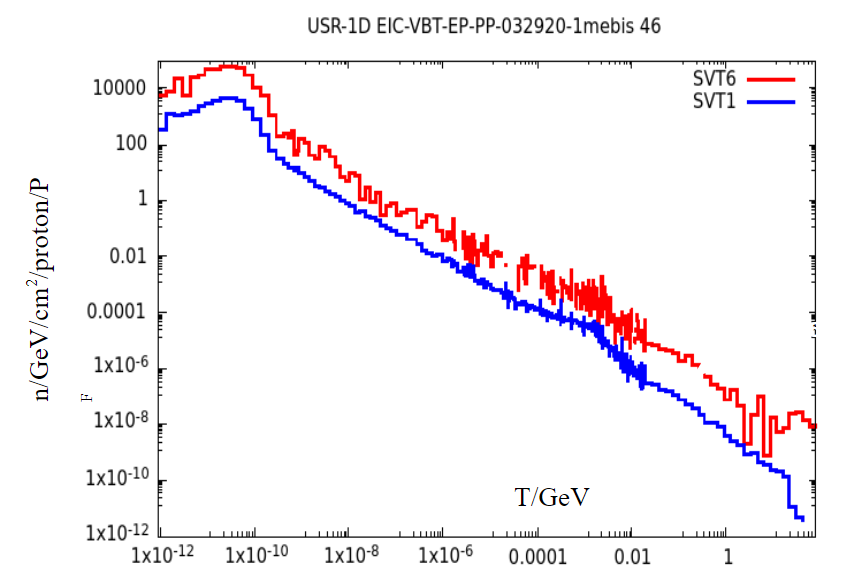}}
\caption{Neutron energy spectra from FLUKA simulations in two layers of the  SiVT : (1) Outer-most Si layer (SVT1) and (2) Inner-most Si layer (SVT6). The vertical scale is   fluence in units of $neutrons/GeV/sr/cm^2/proton$ at pressure $P_F=100$ mbar. The horizontal scale is neutron 
energy in $GeV$.  Absolute realistic flux in neutrons/s/sr/cm$^2$/GeV is obtained by multiplying the vertical axis by  
$\approx 6.25\cdot 10^7\text{ protons/s}$ (see Fig.~\ref{fig:nflmp} caption). }
\label{fig:nSVT}
\end{figure}

\begin{figure}[htbp]
\begin{center}
\includegraphics[width=\linewidth]{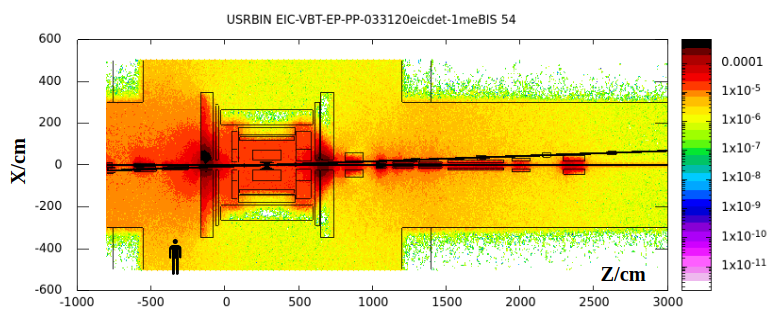}
\includegraphics[width=\linewidth]{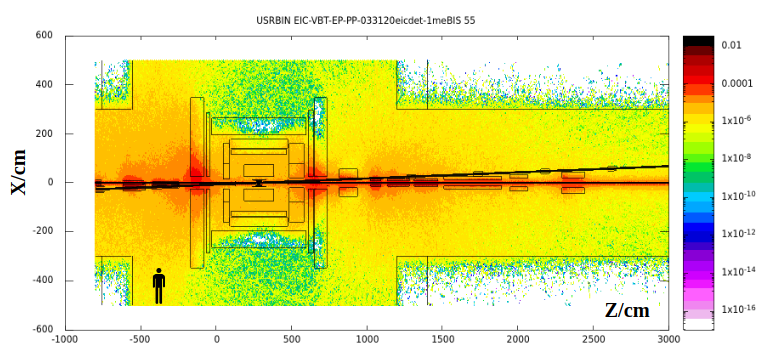}
\caption{
Maps for Neutron fluence (top) and 1-MeV-equivalent Neutron fluence (bottom) from 
$p+Air$ interactions in the beam pipe at proton energy $E_p=275\,\text{GeV}$ and an artificial pressure $P_F$ (``P-FLUKA'') in a thin
cylinder along the beam line. The IP  is located at 
$Z=285$~cm. Neutron fluence is given by the color chart at the right side of the plot  in units of  neutrons/cm$^2$/proton at $P_F=100$~mbar. 
Normalized rates for current $I=1$ A and a realistic average beam-line vacuum $P=10^{-9} \text{ mbar}$ are obtained by multiplying
the color values by $(I/e)(P/P_F) = 6.25\cdot 10^7\text{ protons/s}$.  Thus dark red regions (almost yielding to black) correspond to a realistic fluence of $\approx 6\cdot 10^4 \text{ neutrons/s/cm}^2$.
}
\label{fig:nflmp}
\end{center}
\end{figure}

The energy spectrum of beam-gas induced neutron at the central Si Vertex Tracker (SiVT) is illustrated in Fig.~\ref{fig:nSVT}.  
The energy distribution
shows a clear peak of fully thermalized neutrons below 1 eV, as well as a knee around 10 MeV from evaporation neutrons.  
Neutron damage to Si sensors
occurs primarily via displacement of nuclei from their ideal lattice positions.  
This can happen both by direct $n$Si scattering, and also by recoil
from Si$(n,\gamma)$ reactions.  
The latter can dislodge nuclei, even for neutron energies well below 1~eV.  

\begin{figure}[!htb]
\center
\includegraphics[width=.95\linewidth]{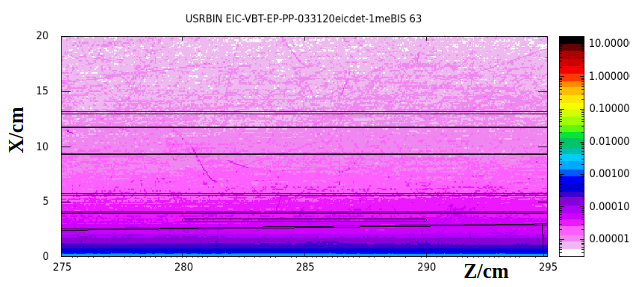}
\caption{
One-$MeV$ equivalent neutron fluence map in the area of SiVT; $p+Air$ interactions in the beam pipe at proton energy $E_p=275$\,GeV. 
IP  is located at 
$Z=285$\,cm. Fluence is given by the color chart at the right side of the plot  in units of  
neutrons/cm$^2$/proton/$P_F$, where $P_F=0.1$\,bar is the pressure used in the FLUKA model. }
\label{fig:svt1meveq}
\end{figure}

The damage induced by neutrons is
frequently quantified by an equivalent flux of 1 MeV neutrons.  
This is shown in the lower panel of  Figure~\ref{fig:nflmp}.
From Fig.~\ref{fig:svt1meveq}, we
obtain an annual dose of 
$6\cdot 10^{10} \text{n/cm}^2$ 
(1 MeV equivalent) in the SiVT.  This is more than
three orders of magnitude less than the suggested tolerance of 
$10^{14} \text{n/cm}^2$. 

\begin{figure}
\centering
\includegraphics[width=.9\linewidth]{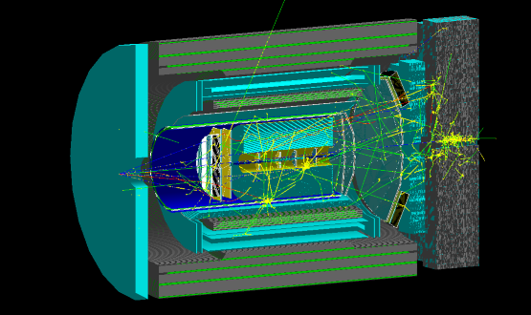}
\caption{\textsc{Geant4} simulation of a beam gas interaction background in an EIC detector model~\cite{Adare:2014aaa,sPH-cQCD-2018-001}. The interaction originate after the last focusing magnet at $z=-4$~m. The produced particle shower will cascade through the central detector stack and induce high multiplicity background throughout the forward and backward spectrometers.}
\label{fig:daq-sim-beamgas-display}
\end{figure}

\begin{figure}
\centering
\includegraphics[viewport=125bp 0bp 1300bp 450bp,clip,width=.9\linewidth]{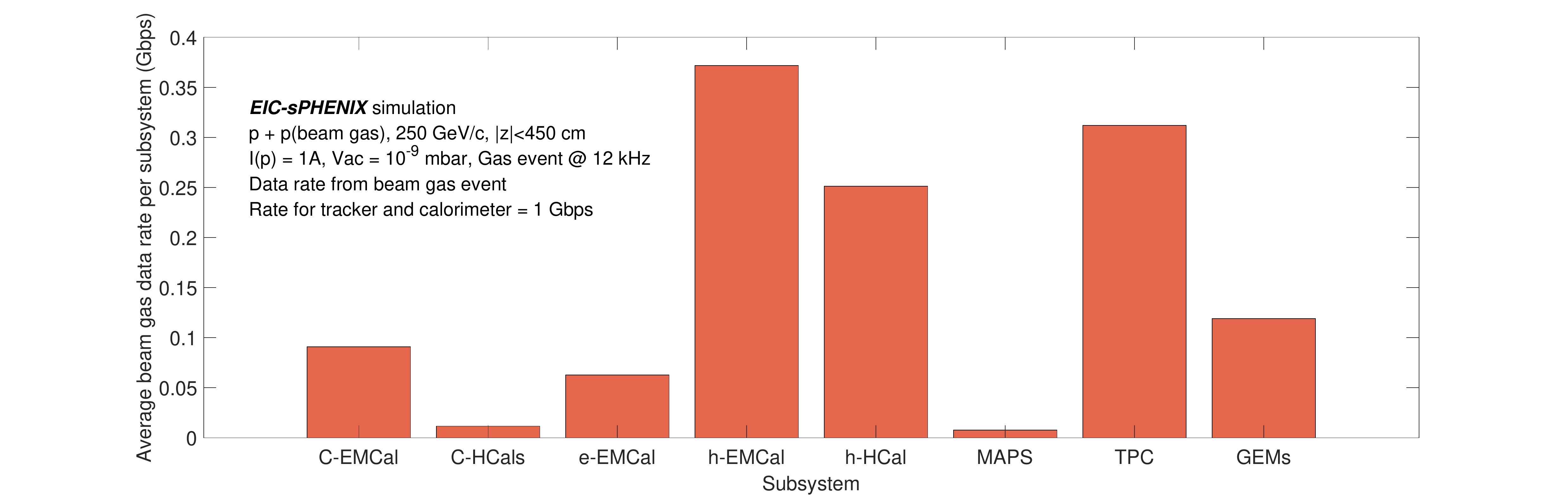}
\caption{
Signal data rates from tracking and calorimetric detectors from beam gas collisions via full detector \textsc{Geant4} simulation of a detector concept based on sPHENIX~\cite{Adare:2014aaa,sPH-cQCD-2018-001}, which also apply to the reference EIC detector as in this report. This simulation assumes constant $10^{-9}$~mbar vacuum in the experimental region of $|z|<450$~cm, which would be modified with a dynamic vacuum profile in the future. }
\label{fig:daq-rate-beamgas}
\end{figure}

Next we we further estimated the data rate across the whole experiment that originated from the beam-gas interactions. The full detector simulation model 
is used to simulate the proton beam hydrogen gas interaction generated with PYTHIA8 in the $p+p$ fixed-target configuration. The hydrogen gas pressure is assumed to be a constant $10^{-9}$~mbar across the experimental region  $|z|<450$~cm, which leads to approximately 10~kHz inelastic beam gas interaction rate. The result collision is propagated through the detector model as illustrated in Figure~\ref{fig:daq-sim-beamgas-display}. The result data rate is summarized in Figure~\ref{fig:daq-rate-beamgas}.

\subsubsection {Beam halo}

Particles produced in elastic collisions of both electron and proton beams with residual gas or beam-beam interactions can form a 
halo distribution around the beam. There also can be halo muons produced in inelastic p-A collisions. Often the result is an
on-momentum electron or ion with large scattering angle. These particles can then generate additional background by interacting with the 
beam pipe and can impact the stability of the beam. Beam halos are being studied to determine whether ”scraping” the halo with collimators
is required, as well as proper placement of those collimators.


\subsubsection {Summary and outlook}

Although multiple background sources are discussed in this section, at the time of this report, many aspects of these background estimation can still vary considerably as the accelerator and experiment conceptual designs proceed. Nonetheless, we feel it crucial to point out a few key experimental regions which are susceptible to high background. 

\begin{itemize}
\item 
A silicon vertex tracker is expected to be installed with a minimal clearance outside the beam pipe. The overall collision charged particle flux is relatively low (Section~\ref{part3-sec-DetChalReq.RatMul}). However, the proximity to the beam pipe exposes this detector to the background such as the synchrotron radiation, beam gas interactions, and beam halo. The current estimation shows an annual neutron fluence reaches $O(10^{11})n/$cm$^2$. The synchrotron hit rate ranges from $O(10^8)$ to $O(10^{12})$ pixels per second, depending on the choices of the beam chamber coating. Optimization of the machine and detector design is ongoing to reduce and refine the background rate, and to protect this key detector from unexpected beam conditions. 
\item 
In addition to the silicon vertex tracker, many tracking and PID detectors would observe a considerable hit rate from the
synchrotron photons, in particular the main barrel tracker and forward-backward silicon tracker, as illustrated in Figure~\ref{fig:daq-sim-synrad-display}. Optimization and detailed estimation are still ongoing. 
\item 
Hadronic showers lead to enhanced neutron fluence near the first few hadron interaction lengths of the calorimeters at the vicinity of the beam pipe. Current estimations are at the orders of $O(10^{10})n/$cm$^2$. Further study and refinement will be carried out. 
\end{itemize}

\FloatBarrier

\section{Systematic Uncertainties}
\label{part3-sec-DetChalReq.Sys.Ancillary}

This sections gives a summary of physics and detector systematic uncertainties critical at the EIC.
Systematic effects will limit the precision of many EIC measurements. This puts very high requirements on the statistical precision of the luminosity and polarization measurement as well as on their systematic uncertainties. The bremsstrahlung process $ep\rightarrow ep\gamma$ 
was used successfully for the measurement of the luminosity by the HERA collider experiments. It has the features of a precisely known QED cross-section, and a high rate, which allowed negligible statistical uncertainty.

Different from HERA, where only the lepton beam was polarized, in an EIC both the lepton and proton/light ion beams will be polarized. Then the bremsstrahlung rate is sensitive to the polarization dependent term $a$ in the cross section: $\sigma_{brems}=\sigma_0(1+aP_eP_h)$.
Thus, the polarization ($P_e,P_h$) and luminosity measurements are coupled, and the precision of the luminosity measurement is limited by the precision of the polarization measurement. This also limits the precision of the measurement of double spin asymmetries 
$A_{||} = \frac{1}{ P_e,P_h}[\frac{N^{++/--} - RN^{+-/-+}}{ N^{++/--} + RN^{+-/-+}}]$
through the determination of the relative luminosity $R=L^{++/--}/L^{+-/-+}$, with $N$ being the measured  for a specific beam polarisation alignment (parallel ++ and antiparallel -+) and the spin sorted luminosity $L$

\begin{figure}[ht]
\centering
\includegraphics[width=0.65\textwidth]{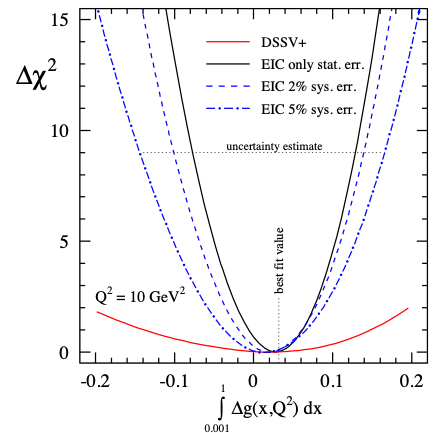}
\caption{\label{fig:lumi.systematics.fig1} The $\chi^2$ profiles for the first moments of the gluon truncated to the 
region $0.001 <x <1$. 
The results are based on using current data (DSSV+) 
and sets of projected EIC data with two different c.m.s.-energies (5GeVx100GeV and 5GeVx250GeV). 
The $\chi^2$  profiles assuming only statistical uncertainties 
and adding 2\% and 5\%
systematic uncertainty in the fit, solid, dotted and dot-dashed
lines, respectively.}
\end{figure}

As an example, we show the impact of the systematic uncertainties on the determination of the gluon polarization $\Delta g(x,Q^2)$~\cite{Aschenauer:2012ve} by measuring the double spin asymmetry $A_{||}(x,Q^2)$, which is proportional to the ratio of the polarized and unpolarized structure function $g_1(x,Q^2)$, and $F_1(x,Q^2)$, respectively. A full QCD analysis has been performed using EIC pseudo data for inclusive and semi-inclusive double spin asymmetries.  Figure \ref{fig:lumi.systematics.fig1} shows the impact of the systematic and statistical uncertainties vs. only statistical uncertainties in the determination of the integral contribution of the gluons to the total spin of the proton. A total systematic uncertainty on the order of 2\% would be important to achieve to profit from the statistical precision available at an EIC. With the low-x double `spin asymmetries being on the level of $10^{-4}$, this will require measuring the relative luminosity to $\sim 10^{-5}$.

For absolute cross sections the precision on the absolute luminosity measurement is critical; the current understanding is that $\delta L/L$ should be $\sim$1\%. To achieve this, it is critical to control and monitor the following parameters known to impact the systematics of the luminosity measurement the most:
\begin{itemize}
\item beam parameters, like position and divergence, to not impact the acceptance of the bremsstrahlung photons
\item radiation damage of the electromagnetic calorimeter system, as this would influence the energy threshold and energy scale 
measurement of the calorimeters 
\item reduce impact of synchrotron radiation on the calorimeter system
\item keep the homogeneity of  conversion window in thickness and material composition
\end{itemize}

{\bf Radiative Corrections} – Radiative effects are inescapably present in relativistic electron scattering from hadrons. They are essential 
to the interpretation of all electron scattering data and indeed can be central to the determination of important aspects of hadron structure. 
The uncertainty on the radiative correction can often be a significant contribution, and in certain cases may dominate the total measurement uncertainty.  

While there is a mature understanding of radiative corrections, EIC measurements are likely systematics limited and radiative corrections systematics might be dominant. EIC will involve polarized hadrons and nuclear beams in collider kinematics for the first time. Radiative corrections are an important uncertainty~\cite{Afanasev:2020hwg}, and EIC experiment design should take them into account, not only to minimize their effect, but also as a physics goal itself. While progress has been made on the theoretical development, there are still important open issues, especially for the theory for e+A collisions. Event generators used in physics simulations, more work must be done to either include radiative generators, or extend the existing generators with radiative effects.

Radiative effects are conventionally treated by introducing correction factors to map the measured quantity back to the pure, radiation-effects-free base quantity. This procedure becomes increasingly difficult beyond inclusive scattering, when more and more scales and complex multi-dimensional quantities are probed. Recently, a new approach was proposed to unify the (QED) radiative effects with the QCD radiation~\cite{Liu:2020rvc}. This would allow a uniform treatment of radiative corrections for the extraction of parton distribution functions, transverse momentum dependent distributions, and other partonic correlation functions from electron scattering data, including polarization and nuclear effects.

In parallel, it seems prudent to consider a systematic experimental test of radiative correction procedures once the EIC becomes operational.  This will likely require a dedicated detector/beam configuration to enhance the sensitivity to particular kinematics, where such corrections are large.  It seems reasonable to organize a dedicated effort focused on radiative correction generator development and experimental validation in the EIC era.  There appear to be no fundamental problems, but the amount of work required to execute a precise measurement program at EIC is still substantial, and corresponding efforts must be supported by the EIC community.

{\bf Inclusive Processes Systematics} – Uncertainties for inclusive structure functions at the EIC were guided by experience from the ZEUS neutral current experiment at HERA~\cite{Chekanov:2001qu}, where the normalization (scale) uncertainty was $\approx 2\%$ on the reduced cross section with an uncorrelated uncertainty under $3\%$ for most bins. As the individual sources of uncertainty are similar, we anticipate $\approx 1.5\%$ uncorrelated (point-to-point) and $\approx 2.5\%$ normalization uncertainties in the optimistic EIC case (see Table~\ref{table:epsystematics}). This gives a total uncertainty on each point of $\approx 3\%$. In the Table one can see the normalization uncertainty related to the luminosity measurement, and the point-to-point estimate of the radiative corrections effect in the HERA data. The latter may well differ at the EIC and is to be estimated process by process. Of direct relevance is also the estimated point-to-point uncertainty of 1-2\% on acceptance, bin migration, tracking and trigger efficiencies, and backgrounds. This 1-2\% point-to-point uncertainty sets the scale for many EIC measurements.

\begin{table}[htb]
\centering
\begin{tabular}{|c|c|c|}

\hline
                          & point-to-point (\%) & normalization (\%) \\ \hline
Statistics (10 fb$^{-1}$) & 0.01-0.35           & \\
\hline

Luminosity &
  &
  1 \\
\hline

Electron purity &
  &
  1 (for 90\% purity) \\
\hline

Bin-centering &
$<0.5$  &
$<0.5$  \\
\hline

Radiative corrections (HERA) &
  1 &
    \\
\hline
Acceptance / Bin-migration, & & \\ 
Trigger and tracking efficiency, & 1-2& 2-4 \\
Charge symmetric background &
   &
   \\
\hline

Additional uncertainty for y$<0.01$ bins &
2 &
  \\
\hline
\hline

Total &
1.5-2.3  (2.5-3 for y$<0.01$) &
2.5-4.3 \\
\hline

\end{tabular}
\caption{Projected inclusive neutral current uncertainty sources and values}
\label{table:epsystematics}
\end{table}

{\bf Exclusive Processes Systematics} – For exclusive processes, there is a further impact on the knowledge of point-to-point systematics due to the dependence of the cross sections on $-t$. Thus, the uncertainty goes beyond the precision of the absolute luminosity measurement of 1\% and the illustrated point-to-point uncertainties for (semi-)inclusive measurements. For tagged structure function measurements at HERA, a 2.5\% point-to-point systematic uncertainty was quoted. We believe we can improve on this at the EIC as the design is optimized for forward detection. Based on scaling from experience of exclusive measurements at Jefferson Lab, and folding in alignment expectation, the best point-to-point {\sl $-t$-dependent} systematic uncertainty that can likely be achieved for such reactions is 1.0-1.2\%. This can be further validated by actual EIC measurements and comparing measured and calculated dependencies on $-t$.

{\bf Electron Polarization Systematics} – One of the primary goals of the reference Compton polarimeter concept presented in Section~\ref{sec-Spin} is the ability to measure the electron beam polarization with a systematic uncertainty of $\frac{dP}{P}=1\%$ or better.  This systematic uncertainty is driven primarily by knowledge of the degree of circular polarization (DOCP) of the laser at the interaction point, as well as the detector response.

The EIC Compton polarimeter will make use of a technique that employs optical reversibility theorems to constrain the laser DOCP inside the vacuum, at the interaction point, via measurements of back-reflected light~\cite{Narayan:2015aua, Vansteenkiste:93}. This technique is expected to determine the laser polarization to a precision of $\approx$0.1\%.

Extraction of the transverse electron beam polarization is accomplished via measurement of an up-down asymmetry. In this case, a key requirement is to have a detector of sufficient resolution in the vertical direction to be able to fit the shape of the asymmetry with minimal distortion.  Monte Carlo studies show that this can be accomplished with strip detectors of $\approx$~100$\mu$m (50$\mu$m) for the backscattered photons (scattered electrons).

Measurement of the longitudinal beam polarization depends on knowledge of the backscattered photon and scattered electron energy spectrum. For the electron detector, this is accomplished by measuring the horizontal position of the scattered electrons after being momentum-analyzed in the dipole after the laser-electron interaction point.  A detector that measures both the Compton kinematic end-point and the asymmetry zero-crossing is self-calibrating (with respect to energy response) and about 30-40 channels between those two points are sufficient to determine the longitudinal beam polarization with high precision.  At the same time, such an electron detector can be used to provide in-situ energy response calibration for the photon calorimeter.  The two detectors together are expected to constrain the longitudinal electron beam polarization to better than 1\%.

{\bf Hadron Polarization Systematics} – High energy hadron beam polarimetry is utilizing recoil particles from elastic scattering at small momentum transfer (with resulting recoil energies on the order of a few MeV). The methods have been established at proton beam energies up to 255 GeV/$c$ at RHIC over two decades. The absolute beam polarization is measured through a polarized hydrogen jet target which is calibrated with a Breit-Rabi analyzer. The potential polarization decay during a few-hour long store is tracked with fast measurements using ultra-thin Carbon targets, which also allow to scan the transverse profile of the beam bunches.

The dominant source of systematic uncertainties until recently was due to the molecular fraction in the atomic hydrogen target. Recent studies have been able to essentially eliminate this uncertainty, but at a cost to an increased background in the detector. So far, the best achieved uncertainty for the polarization scale uncertainty has been below 1.5\% (relative).

Comparison of store-to-store differences between the hydrogen jet and Carbon measurements have in the past exhibited additional uncertainties beyond purely statistical fluctuations and need to be monitored. Polarization decay and the bunch polarization profile will need to be folded into the time-dependent polarization in collision at the EIC experiments, each of which introduces their own uncertainties.

The reduced polarized proton/ion bunch spacing at the EIC as compared to RHIC operations is potentially magnifying background in the polarimeters, where the time-of-flight of the recoil particles is on the order of or larger than consecutive beam bunches. This is of particular concern when bunches have opposite polarization direction as the background may carry its own spin-dependent asymmetry. If pile-up of elastic reactions is small, this can be mitigated by modified detectors, which allow for an efficient rejection of high-energetic background particles. However, since the particle energy is directly related to the time-of-flight, the kinematic coverage of the detectors will suffer with the reduced bunch spacing.

Work is on-going to further study the sources of backgrounds in proton and light-ion beam measurements and the necessary steps to reach the required uncertainty for the EIC physics program. An uncertainty like the best achieved to date, better than 1.5\%, may be anticipated, with the goal to achieve 1\% with time and gain of EIC beam experience.

\section{Physics Requirements}
\label{part3-sec-DetChalReq.PhysReq}
The physics program of an EIC imposes several challenges on the design of a general purpose detector, and more globally the extended interaction region, as it spans center-of-mass energies from 29\,GeV to 141\,GeV, different combinations of both beam energy and particle species, and several distinct physics processes.
The various physics processes encompass inclusive measurements 
$e + p/A \rightarrow e'+X$;
semi-inclusive processes 
$e + p/A \rightarrow e'+h+X$;
which require detection of at least one hadron in coincidence with the scattered lepton;
and exclusive processes
$e + p/A \rightarrow e'+N'A'+\gamma/h$,
which require the detection of all particles in the reaction with high precision. 
\par
The high level requirements for the EIC general purpose detector are:
\begin{itemize}
    \item The EIC requires a \emph {hermetic} detector with \emph {low mass} inner tracking.
    \item The main detector needs to cover the range of $-4 < \eta < 4$ for the measurement of electrons, photons, hadrons, and jets. It will need to be augmented by auxiliary detectors like low-$Q^2$ tagger in the far-backward region and proton (Roman Pots) and neutron (ZDC) detection in the far-forward region. 
    \item The components of an EIC detector will have moderate occupancy as the event multiplicities are low (Sec.~\ref{part3-sec-DetChalReq.RatMul}). 
    However, depending on the machine background level certain components close to the beamline might experience higher occupancy (Sec.~\ref{part3-sec-DetChalReq.Backgrounds}). 
    \item An EIC detector will have to cope with a data rate of up to $\sim 500$ kHz at full luminosity.
    \item Compared to LHC detectors, the various subsystems of an EIC detector have moderate radiation hardness requirements, e.g. at the calorimeters up to $\sim$3 krad/year electromagnetic and 10$^{11}$ n/cm$^2$ hadronic at the top luminosity.
\end{itemize}
\par
The intensive work done during the Yellow Report initiative has resulted in detailed requirements for the different subdetectors forming the central detector, and the individual detectors along the beamline (Chapter~\ref{part2-chap-DetRequirements}). The current status of these requirements is shown in Table~\ref{SubDetReq}. The interactive version of the Table at (https://wiki.bnl.gov/eicug/index.php/Yellow\_Report\_Physics\_Common) provides for every requirement a link to all the physics simulations and the study, which sets the most stringent requirement. 
\begin{sidewaystable}[thb]
\centering
\caption{This matrix summarizes the high level requirements for the detector performance. The interactive version of this matrix can be obtained through the Yellow Report Physics Working Group WIKI page\\ (https://wiki.bnl.gov/eicug/index.php/Yellow\_Report\_Physics\_Common).}
\label{SubDetReq}
\includegraphics[width=1.0\textwidth]{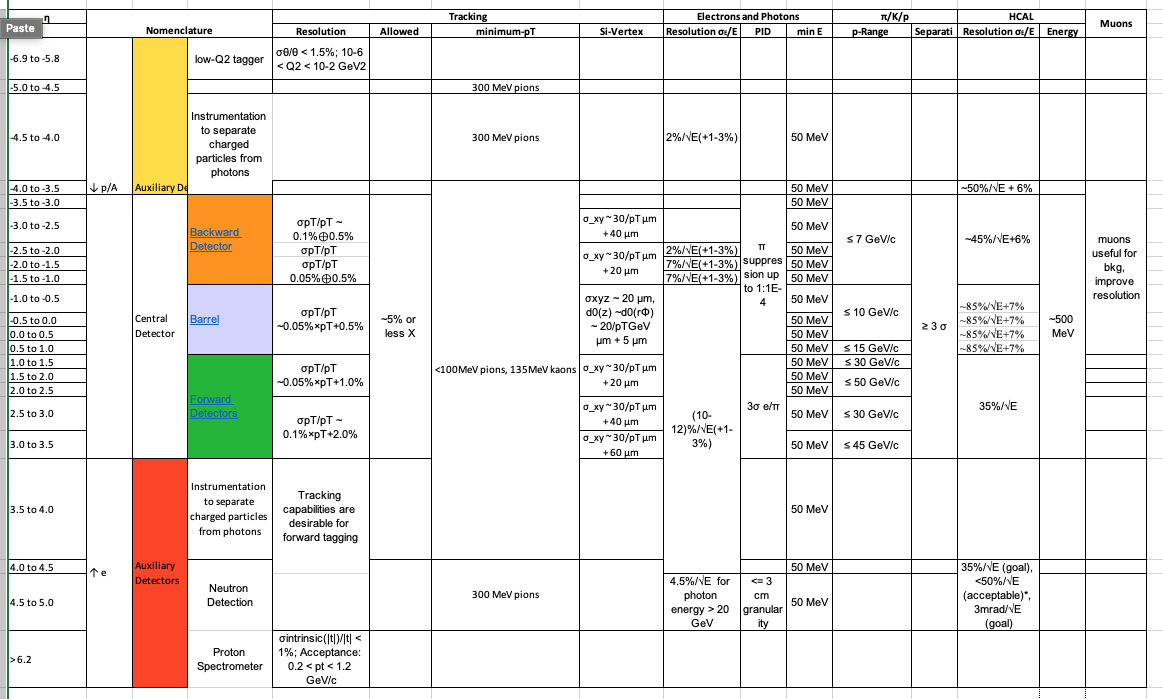} 
\vspace*{3mm}
\end{sidewaystable}
\FloatBarrier

\chapter{Detector Aspects}
\label{part3-chap-Det.Aspects}

\section{Magnet}
\label{part3-sec-Det.Aspects.Magnet}

We consider the case of an EIC detector built around a
central superconducting solenoid, possibly with trim coils to adjust the solenoidal field. Tracking resolutions in the central pseudo-rapidity range suggest a nominal field of 1.5~T, but a range between 1.5~T and 3~T would make a wider range of physics measurements accessible. A central field as high as 3~T would maximize the effective $|B|\cdot$dl integral for particles scattered at small polar angles, both in forward and backward directions. However, high magnetic fields come at the cost of reducing the low-$p_T$ acceptance of charged tracks. The acceptance for low-$p_T$ particles down to the momenta $\sim$100 MeV/$c$ requires that a fraction of physics data are taken at a substantially lower field. Field polarity flip is a standard measure to address systematic effects due to a different acceptance for the positively and negatively charged particles, hence a bipolar magnet operation with a polarity switch is one of the magnet requirements. 

Physics studies to date suggest a solenoid with a
bore diameter of 2.5-3.5~m would be able to accommodate the detectors necessary for EIC measurements. 
Specifications on coil length, presently assumed to be able to provide a $\sim$3.0~m magnetic length as a reference figure, cryostat radial space, and coil configuration require an optimization integrated with the overall detector design. The solenoid design is characterized by three regions: the barrel and backward/foreward endcaps.  In the barrel and the backward endcap regions, the field should be parallel to the magnet axis. In the forward direction, the RICH detector extends from +100~cm to +240~cm with respect to the magnet centre, where the field lines should be projective with respect to the nominal IP location. A flux return path could be provided through the hadronic calorimeter assemblies in the forward and backward directions. Correction coils in the hadron end-cap may be required to meet the RICH detector readout on field projectivity. The need for these coils should be avoided as they will adversely affect the hadron calorimeter performance, but if needed, should be allowed to occupy a maximum of 10~cm of the available linear space.

Alternative detector integrated designs, where a dipole or toroidal field are superimposed with the solenoid field in the central region of the detector, have been considered to improve the $|B|\cdot$dl integral at small scattering angles. These integrated designs could be an option if an acceptance that meets the physics requirements can be demonstrated.

Re-use of the existing BABAR/sPHENIX magnet is an alternative to the realization of a new solenoid with optimized design. Whereas the new solenoid main specifications are an up-to 3T magnetic field, a 2.5-3.5~m diameter bore, and a magnetic length of $\sim$3~m, the BABAR/sPHENIX magnet provides an up-to 1.5~T field, a 2.8~m diameter bore, and similar magnetic length. The magnet for the BABAR experiment at PEP-II at SLAC, CA was manufactured by Ansaldo, Italy in 1997 and was commissioned in 1998. It was then transferred to BNL, NY in 2015 for use in the sPHENIX experiment where it still resides today. It received a high-field test (up to 1.3~T) in 2018. 
The prolonged use of the BABAR/sPHENIX magnet may require the implementation of several maintenance and improvement modifications, including new protection circuits such as voltage taps, inspection of and as needed reinforcement of the internal mechanical support, including new strain gauges, and replacement of control instrumentation sensors. Several of these implementations involve the delicate operation of disassembly of the magnet. To repair an existing small leak in the valve box for the cryogenic cooling system requires a replacement of the valve box or disassembly to inspect cooling pipework and to repair leaks. Moreover, additional changes may be required for re-using the magnet, for example those needed to match the requirements of projective field lines in the RICH region.

The main parameters of both a new superconducting solenoid magnet, at the present stage of magnet optimization integrated with the overall detector design, and the existing BABAR/sPHENIX magnet are shown in Table~\ref{tab:solenoid_magnet}. For a new magnet, a slightly larger bore of 3.2 meter is chosen as compromise between, on one hand, magnet complexity and mechanical hall space considerations, and on the other hand providing some much-needed space in the bore to permit more detector technology choices to ensure functionality of tracking, hermetic electromagnetic calorimetry and particle identification (both e/$\pi$ and $\pi$/K/p) over a large range of particle momenta. No detection beyond hadronic calorimetry is foreseen outside the magnet, alleviating any requirement for low radiation length materials in the mechanical design and permitting the choice of a NbTi conductor in a Cu matrix for the new magnet. 

The coil length is driven by the present definition of the barrel region as between pseudo-rapidity of -1 and 1. This corresponds to an angle of $\sim$40 degrees. This means that for a certain bore size, the space for the mechanical length of the magnet cryostat is roughly 20\% larger, or 3.84 meter for a 3.2-meter bore. Folding in an approximate need of 12 cm additional space on each side of the magnet coil for inner vacuum and helium vessels, and multi-layer isolation, determines the coil length requirement to be 3.6 meter. A somewhat larger coil length of 3.8 meter would not be a major issue, but likely not much more as the edge of the cryostat is one of the regions where detector infrastructure (support, cabling, etc.) will reside. There will also be a trade-off between the need for equal coverage of tracking and electromagnetic calorimeter and particle identification detector readout.

\begin{table}[ht]
    \centering
    \begin{tabular}{|c|c|c|}
    \hline
         Parameter & New Magnet & BABAR/sPHENIX Magnet \\
         \hline
         Maximum Central Field (T) & 3 & 1.5 \\
         \hline
         Coil length (mm) & 3600 & 3512 \\
         \hline
         Warm bore diameter (m) & 3.2 & 2.8 \\
         \hline
         Uniformity in tracking region &  & \\
         (z = 0, r $<$ 80 cm) (\%) & 3 & 3 \\
         \hline
         Conductor & NbTi in Cu Matrix & Al stabilized NbTi \\
         \hline
         Operating Temperature (K) & 4.5 & 4.5 \\
         \hline
    \end{tabular}
    \caption{Summary of some of the main requirements of the EIC detector solenoid magnet.}
    \label{tab:solenoid_magnet}
\end{table}

The main advantage of accessibility of low central solenoid fields (down to $\sim$0.5 T) is the low-$p_T$ acceptance of charged-particle tracks. A central field of 0.5 T roughly equates to a detection capability of charged particles down to transverse momenta of below $\sim$ 0.1 GeV/$c$. This is relevant for mapping the decay products of heavy-flavor mesons. The main advantage of a 3 T versus a 1.5 T central solenoid field is for the momentum resolution of charged particles as function of pseudo-rapidity. Doubling the magnetic field can lead to an improvement of the momentum resolution by a factor of $\sim$~2 from a leading order $\sim 1/B$ dependence. This is relevant in the central region, but even more so in the forward pseudo-rapidity regions, $\eta >$ 2, where the momentum resolutions rapidly worsen. For example, for $\eta \sim$ 3, a momentum resolution of $\sim$2-3\% is achievable for pions with momenta up to about 30 GeV/$c$ with a 3 T central field, and only double that resolution for a 1.5 T central field.

\section{Tracking}
\label{part3-sec-Det.Aspects.Tracking}
\subsection{Introduction}
This section represents an attempt to combine the requirements from the physics working groups and  tracking technologies and detector design into viable detector concepts that can meet these requirements. These concepts contain assessments of the current state of the art in both the technologies, services, mechanical support and other components to deliver a design that is deemed to be consistent with what can reasonably be expected to be deployed at the EIC in the needed timescales. In order to reduce risk and ensure that development proceeds apace with the construction schedule, a list of areas in need of targeted R\&D has been compiled and is presented in Chapter 14.
\subsection{Main requirements and acceptance coverage}
\begin{table}[h!bt]
\centering
\caption{Requirements for the
tracking system from the physics groups.}
\label{tab:reqTable}
\includegraphics[width=1.\columnwidth,trim={0pt 0mm 0pt 0mm},clip]{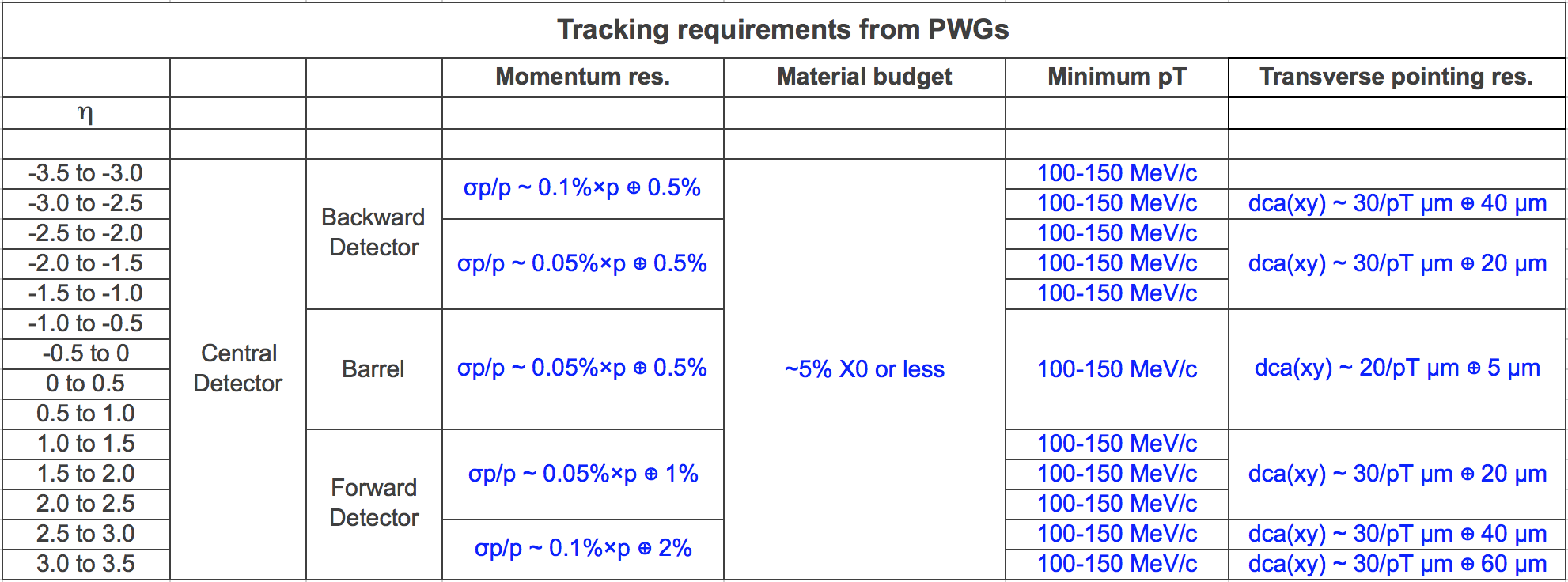}
\end{table}
The requirements for the tracking in an EIC detector are derived from the physics simulations and are represented by the detector requirements table shown in Table~\ref{tab:reqTable}. The ranges in pseudorapidity are accompanied with requirements for relative momentum resolution, allowed material budget in terms of radiation length, minimum p$_T$ cutoff, transverse and longitudinal pointing resolutions. These requirements form the basis of the designs and concepts that are presented.
\subsection{Silicon Detector Technologies for EIC}
\label{Sec:SiTrackingTech}
To satisfy the requirements detailed above, the EIC silicon vertex and tracking (SVT) detector needs to have high granularity and very low material budget. Performance simulations of the detector concepts presented in Section \ref{sec:hybrid-tpc} 
highlight the need for a spatial resolution $\leq 5 \mu m$ in tracking layers and disks, and around 3 $\mu m$ in the vertex layers, combined with a material budget $\leq 0.1\% X_{0}$ in the vertex layers, $\leq 0.8\% X_{0}$ in the tracking layers and $\leq 0.3\% X_{0}$ in the disks.

A broad survey of silicon detector technologies was presented and discussed at the first EIC Yellow Report Workshop in March 2020~\cite{temple} covering hybrid pixel detectors, strip detectors, Low Gain Avalanche Detectors (LGAD), the DEPFET sensor, and Monolithic Active Pixel Sensors (MAPS). The survey considered existing examples of these detectors as well as the silicon technologies used for their development to understand their potential for application at the EIC. MAPS have been identified as the best detector technology to satisfy the requirements of the EIC SVT and are discussed below. These detectors provide the highest granularity, have lower power consumption and consequently lower material budget, as well as the required readout speed in one device. Recent developments in MAPS at the 65~nm node offer some attractive features and satisfy the requirements of the EIC vertex layers, as shown in Section~\ref{65nm}. The integration of charge collection and readout capabilities into one silicon substrate is well suited for the required level of integration and acceptance coverage of the EIC SVT. Silicon technologies such as LGAD and SOI, whose developments in the next few years could produce a viable alternative for the EIC SVT, are presented in Chapter~\ref{part3-chap-DetTechnology}.

\begin{figure}[ht!]
	\centering
    \includegraphics[width=0.6\columnwidth]{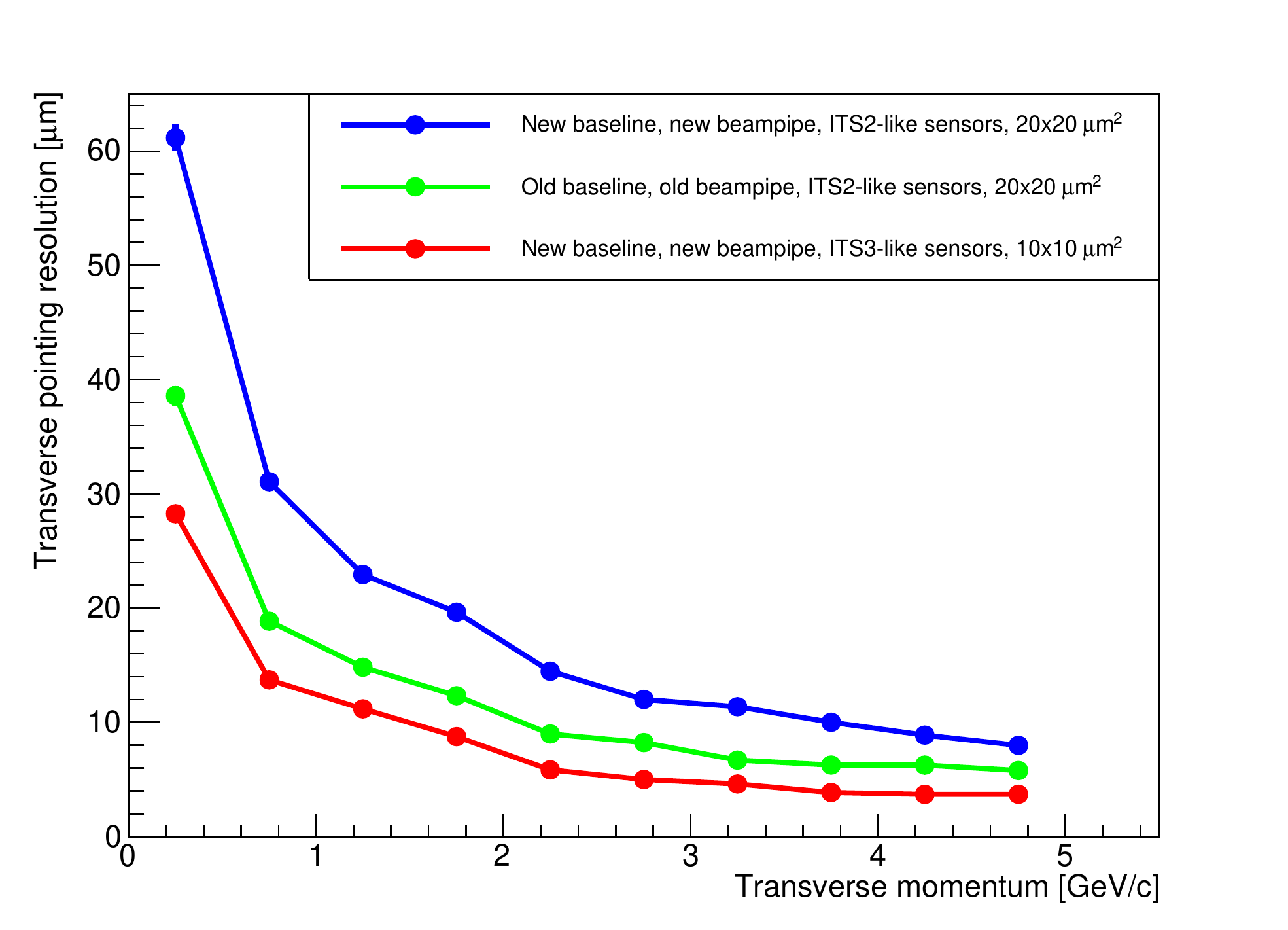}
	\caption{Transverse pointing resolution versus transverse momentum, comparing the ALICE ITS2 based detector configuration with old (green) and new (blue) beam pipe, and the ALICE ITS3 based detector configuration with new beam pipe (red).}
	\label{fig::newBPtest_forSVTEICwriteup_transvPointRes}
\end{figure}

The stringent requirements for the vertex layers are driven by the rather large beam pipe radius and are necessary to obtain the required vertex reconstruction performance. This is shown in Figure~\ref{fig::newBPtest_forSVTEICwriteup_transvPointRes}. Pre-CD0 simulations assumed a beam pipe radius of 18 mm and an upgraded ALICE Inner Tracker (ITS2)~\cite{Abelevetal:2014dna} derived SVT detector where vertexing layers and disks had a material budget of 0.3\%~X/X$_0$ per layer, and the tracking layers had a material budget of 0.8\%~X/X$_0$ per layer. The pixel size was $20 \times 20$~$\mu$m$^2$. This configuration gives the transverse pointing resolution described by the green curve in Figure~\ref{fig::newBPtest_forSVTEICwriteup_transvPointRes}. With the updated beam pipe radius of 31 mm, this configuration would lead to a severe decrease in tracking performance (blue curve). The transverse pointing resolution can be recovered, and even improved, with higher granularity and lower material budget. The result in the red curve assumes a configuration based on the ALICE ITS3~\cite{its3det} technology explained below, where the vertexing layers have a material budget of 0.05\%~X/X$_0$ per layer, the tracking layers 0.55\%~X/X$_0$ per layer, and the disks each have a material budget of 0.24\%~X/X$_0$. The pixel size is $10 \times 10$~$\mu$m$^2$.

In addition to these requirements, an EIC SVT detector needs to be designed with an integration time below 2 $\mu$s to cope with the interaction frequency expected at the highest luminosity, i.e. 500 kHz at $10^{34}$~cm$^{-2}$~s$^{-1}$. These requirements drive the choice of the silicon detector technology. 
%

\subsubsection{MAPS}
\label{MAPS}

\begin{figure}[hbt]
	\centering
        \includegraphics[width=0.6\columnwidth]{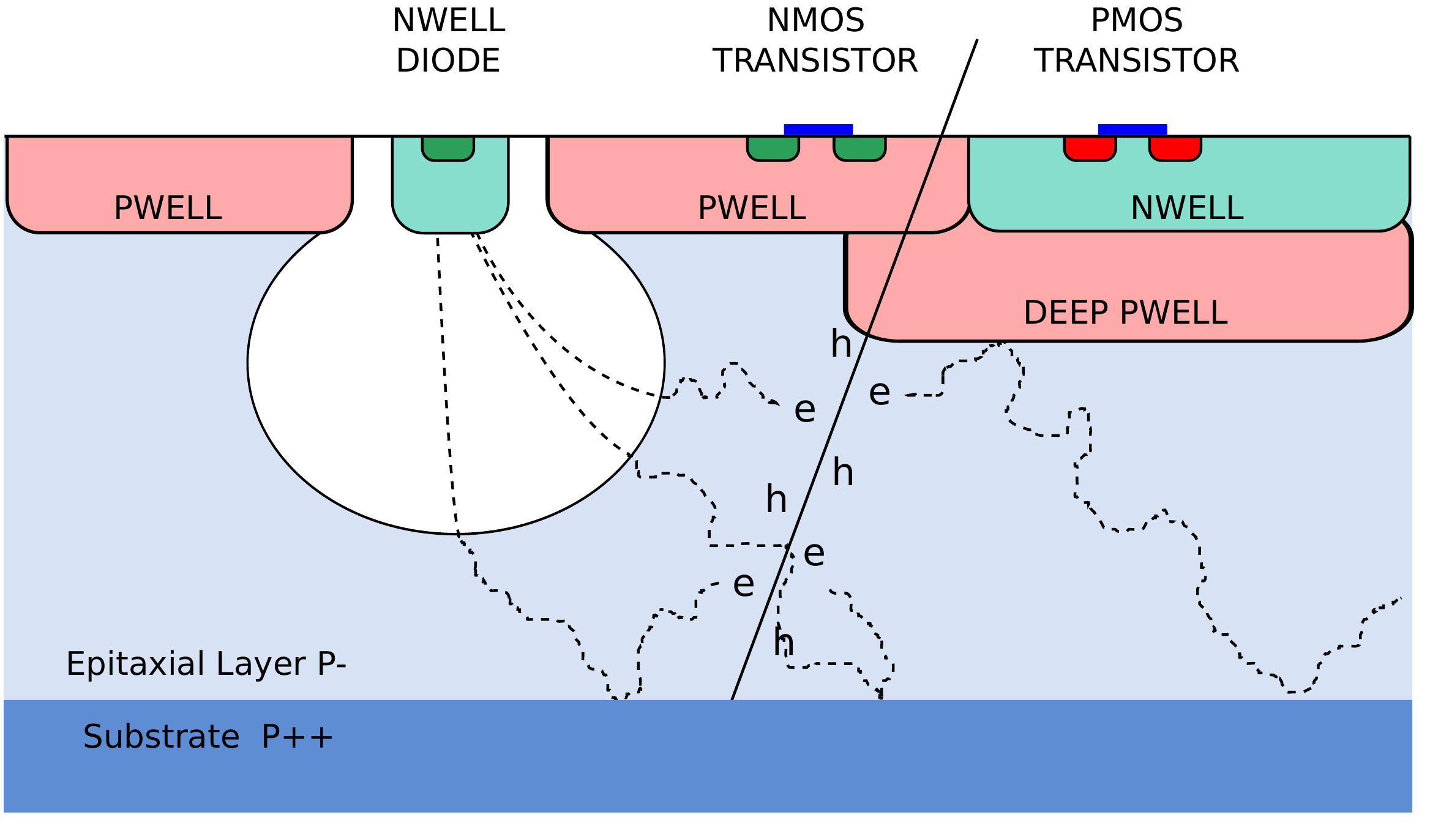}
	\caption{Cross-section of an ALPIDE pixel, showing the small n-type collection electrode and the p-wells containing the electronics in a p-type epitaxial layer. A small depletion region develops around the collection electrode for an applied reverse bias voltage of a few volts \cite{Abelevetal:2014dna}.}
	\label{alpide}
\end{figure}

MAPS are currently used as vertex detectors in the STAR Heavy Flavour Tracker~\cite{Contin:2017mck} and in the ALICE ITS2. The latter deploys the ALPIDE sensor \cite{AglieriRinella:2017lym}. The sPHENIX experiment also uses the ALPIDE sensor. This sensor represents a breakthrough with respect to traditional MAPS such as the MIMOSA used by the STAR experiment. ALPIDE is fabricated in a commercial 180 nm CMOS imaging process provided by Tower Jazz (TJ). The main novelty of this device is the possibility to partially deplete the substrate and thus collect part of the charge by drift, and to integrate both PMOS and NMOS transistors. These features have improved MAPS charge collection properties, radiation hardness, and signal processing capabilities. Figure \ref{alpide} shows the cross-section of an ALPIDE pixel. This design retains as in previous MAPS generations a small collection electrode and thus a small sensor capacitance of a few fF that is key to low power, low noise, fast sensor readout, and compact front-end electronics design for small pixel pitch.

Following on from the ALPIDE, a new generation of MAPS sensors has been developed in the past ten years with the goal of reaching the rate and radiation tolerance capability typically required by high luminosity particle physics experiments. These so-called Depleted MAPS (DMAPS) are fabricated in High Voltage or High Resistivity commercial 150/180 nm CMOS imaging technologies and can be fully depleted. A number of prototypes have been fabricated targeting the upgrades of the ATLAS pixel detector for the HL-LHC in different CMOS technologies. The ATLASPix sensor in the AMS/TSI technology \cite{atlaspix3} and the LF-MONOPIX in the  LFoundry technology \cite{Moustakas:2018ure} feature a large collection electrode that contains the electronics.  This results in a uniform electric field in the sensor substrate needed to achieve the required speed and radiation hardness but comes at the price of high sensor capacitance. The MALTA and TJ-MONOPIX prototypes in the TJ 180 nm technology \cite{Cardella:2019ksc, Dyndal:2019nxt, Moustakas:2018ure} keep the small collection electrode and achieve full depletion with a modification of the process by adding a deep n-implant so that the depletion region grows from below the collection electrode and electronics implant~\cite{Snoeys:2017hjn, Munker:2019vdo}. These sensors have demonstrated that they fulfil the requirements of operation at the HL-LHC, but use at the EIC SVT would have to be demonstrated as they have been designed to match very different requirements. An application of the MALTA sensor for tracking at large $z$ is described in \ref{LANL-concept}.  

It is however important to note that the CMOS imaging technologies in which existing DMAPS prototypes have been fabricated could be used to design a dedicated MAPS sensor for the EIC SVT. In particular, the TJ 180 nm modified CMOS imaging process is very interesting because of the benefit of the small sensor capacitance leading to low power and fine pitch. This technology has been positively evaluated for use at the EIC by the eRD18 project, a collaboration between the University of Birmingham and the RAL CMOS Sensor Design group (CSDG), in the framework of the EIC Generic Detector R\&D program \cite{eRD18}.

Recently, an effort is emerging to develop a third generation MAPS in a 65 nm CMOS imaging technology. A large community is gathering to develop this process for future experiments through the ALICE ITS3 project and the CERN EP R\&D program. This path is more attractive for the development of an EIC MAPS as the 65 nm technology offers improved performance in terms of granularity and power consumption that are key for precision measurements at the EIC, as well as process availability on the EIC project timescale. The drawbacks with respect to older technology nodes are higher non recurring engineering (NRE) costs and complexity.

A joint EIC SVT sensor development has started with the ALICE ITS3 group. The ALICE ITS3 project aims at developing a new generation MAPS at the 65 nm node to build an extremely low mass detector for the HL-LHC. The ITS3 sensor specifications and development timescale are largely compatible with those of the EIC. Furthermore, non-ALICE members are welcome to contribute to the R\&D to develop and use the technology for other applications. Through the ITS3 collaboration, the EIC can benefit from shared development costs to design an innovative sensor solution at the 65 nm node for an experiment starting in approximately 10 years and that will demonstrate the capabilities of this technology for future collider experiments.
\subsubsection{65 nm MAPS SVT detector}
\label{65nm} 

An EIC SVT concept is being developed based on the proposed 65 nm MAPS sensor and ITS3 detector concept \cite{eRD25}. Both baseline configurations presented under investigation (\ref{all-Si-Rey}, \ref{sec:hybrid-tpc}) assume the use of this technology to define pixel pitch and realistic estimates of material budget for services and support structure \cite{leo}, and configuration of the vertex layers. In addition to the advantages discussed in \ref{MAPS}, joining the ITS3 development has additional benefits.

Table \ref{spec_its3} compares the specifications for the proposed ITS3 sensor to the ones of the existing ALPIDE. Table \ref{spec_svt} shows preliminary specifications for an EIC sensor. From these it is clear that the ITS3 fully satisfies and even exceeds the requirements of the EIC SVT with higher granularity, lower power consumption, shorter integration time and lower fake hit rate. In particular, the 10 $\mu$m pixel pitch is key to the design of the vertex layers (Figure~\ref{fig::newBPtest_forSVTEICwriteup_transvPointRes}).
\begin{table}[hbt]
	\caption{Specifications for the ALICE ITS2 ALPIDE sensor and the proposed sensor for the ITS3 upgrade.}
	\label{spec_its3}
	\centering
        \includegraphics[trim=0 100 0 140, clip, width=0.8\columnwidth]{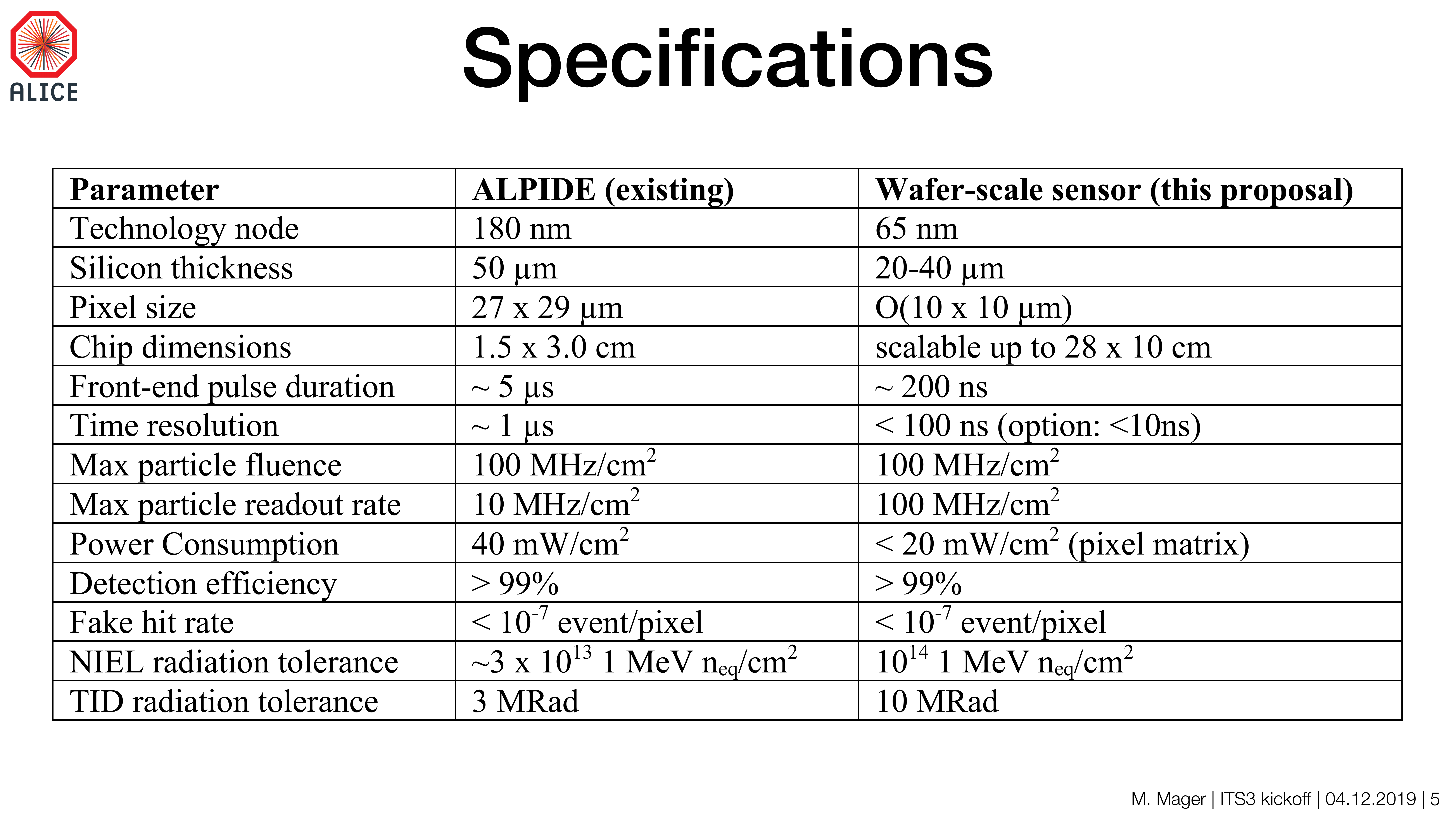}
\end{table}

\begin{table}[hbt]
	\caption{Preliminary specifications for an EIC SVT MAPS sensor based on simulations by the eRD18 (Birmingham/RAL CSDG) and eRD16 (LBNL) projects of the EIC Generic Detector R\&D program.}
	\label{spec_svt}
	\centering
        \includegraphics[width=0.6\columnwidth]{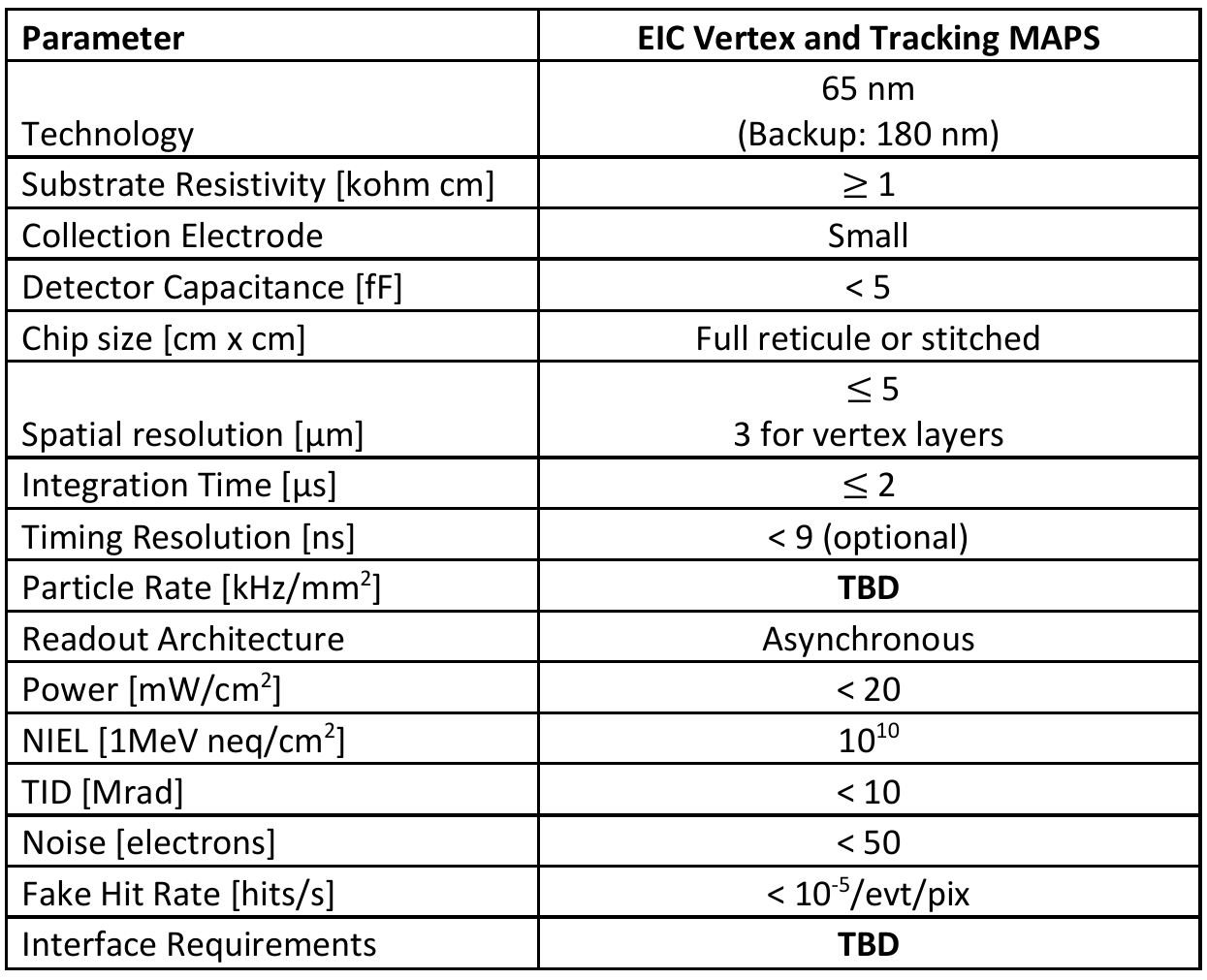}
\end{table}

\begin{figure}[hbt]
	\centering
        \includegraphics[width=0.6\columnwidth]{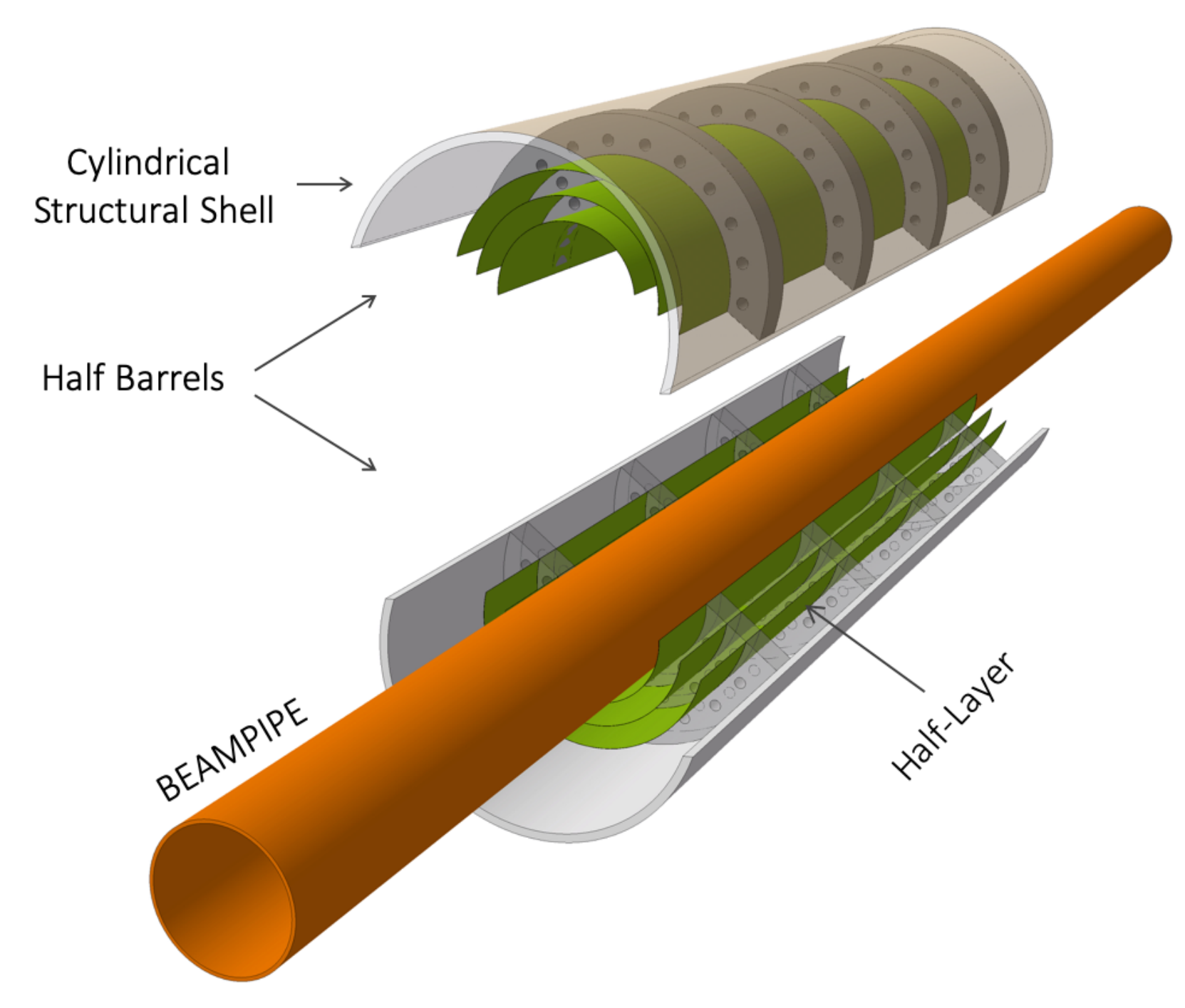}
	\caption{Layout of the ITS3 Inner Barrel. The figure shows the two half-barrels mounted around the beampipe \cite{its3det}.}
	\label{its3}
\end{figure}

The ITS3 project is taking an integrated approach where design and post-processing techniques are combined to develop a three-layer vertex detector with an extremely low material budget (Figure \ref{its3}).  The use of low power design techniques, large area, 2D stitched sensors thinned below 50 $\mu$m and bent around the beam pipe minimises cooling, support structure and services in the active area, enabling a material budget of only 0.05\% $X_{0}$. Such a detector concept is a very attractive solution for the EIC vertex layers where extremely low material budget coupled with the sensor's high granularity will deliver the required vertex resolution (Figure~\ref{fig::newBPtest_forSVTEICwriteup_transvPointRes}). The implementation of the ITS3 detector concept into the EIC vertex layers is currently being worked out by the EIC Silicon Consortium.

Despite the large overlap, the EIC and ITS3 detectors have some significant differences, most notably the size. The ITS3 is a 0.12 m$^2$, three layers vertex detector. The EIC SVT baseline configurations presented in (\ref{all-Si-Rey}, \ref{sec:hybrid-tpc}) have an area of approximately 12 m$^2$ and 15 m$^2$, for hybrid and all-silicon respectively. Cost and yield of stitched wafer-scale sensors will not be compatible with use in the EIC detector outside the vertex layers. For the tracking layers and disks the EIC sensor development will fork off the ITS3 sensor design path to develop a reticule-size version of the ITS3 sensor (no changes in other aspects of the sensor design are foreseen a part from its size) as well as a more conventional design of support structures (classical staves and disks), where dedicated engineering solutions will be deployed to meet the material budget constraints \cite{leo}.

\subsection{Gaseous Detector Technologies}
\label{gaseous_tracking}
\begin{figure}[h!bt]
\centering
\includegraphics[width=\columnwidth,trim={0pt 0mm 0pt 0mm},clip]{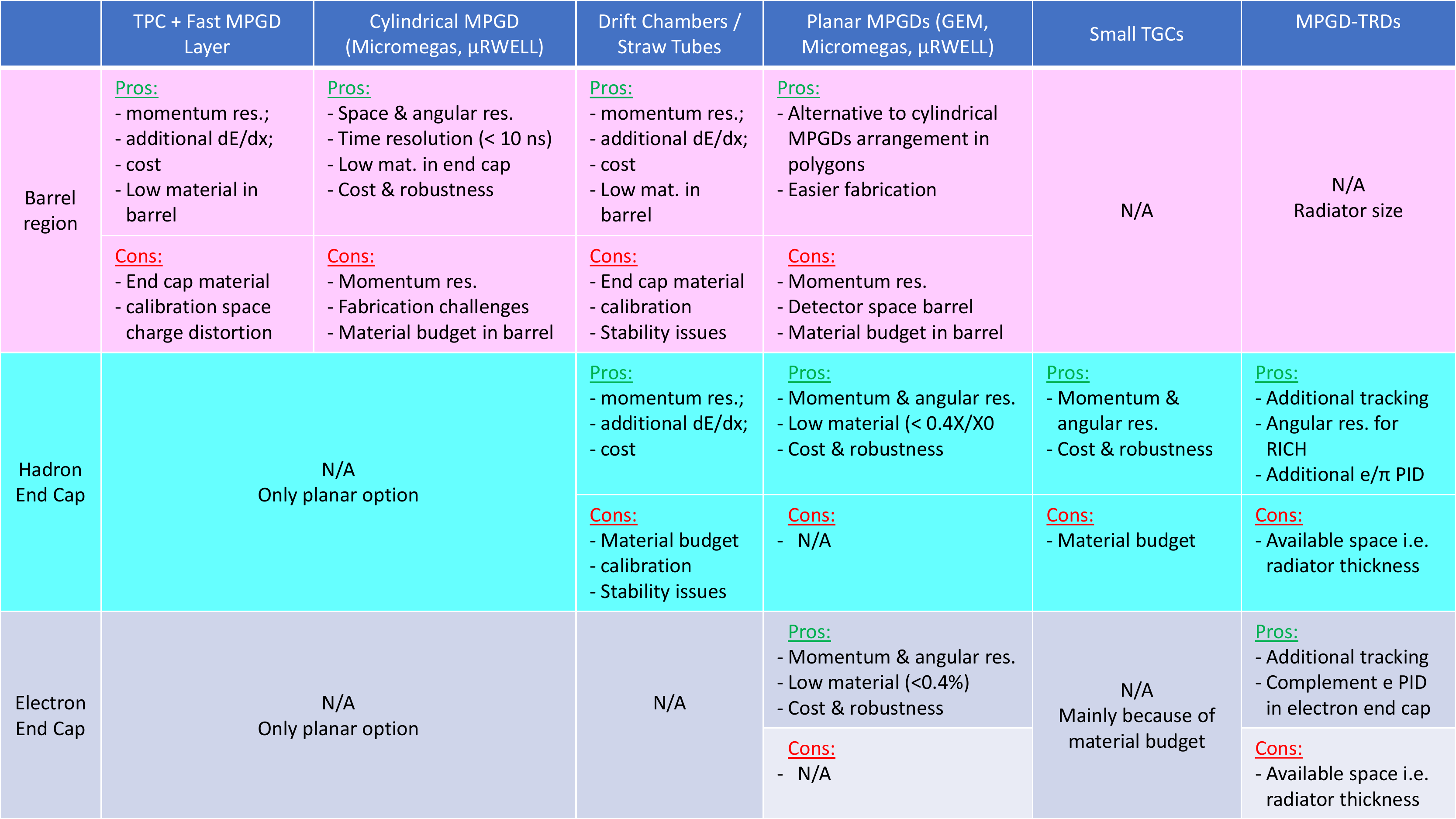}
\caption{Comparison of different gaseous detectors technologies for tracking in EIC.}
\label{fig:gaseous_compare}
\end{figure}

In this section, we describe various gaseous detector technologies under consideration for the tracking in the central region of an EIC detector. Figure~\ref{fig:gaseous_compare} shows a comparative study of each of these technologies and where in the EIC detector, each of these technologies is best suited. 
\subsubsection{Time Projection Chambers (TPC)}
\label{sec:tpc} 
\begin{figure}[thb!]
	\centering
        \includegraphics[width=\columnwidth]{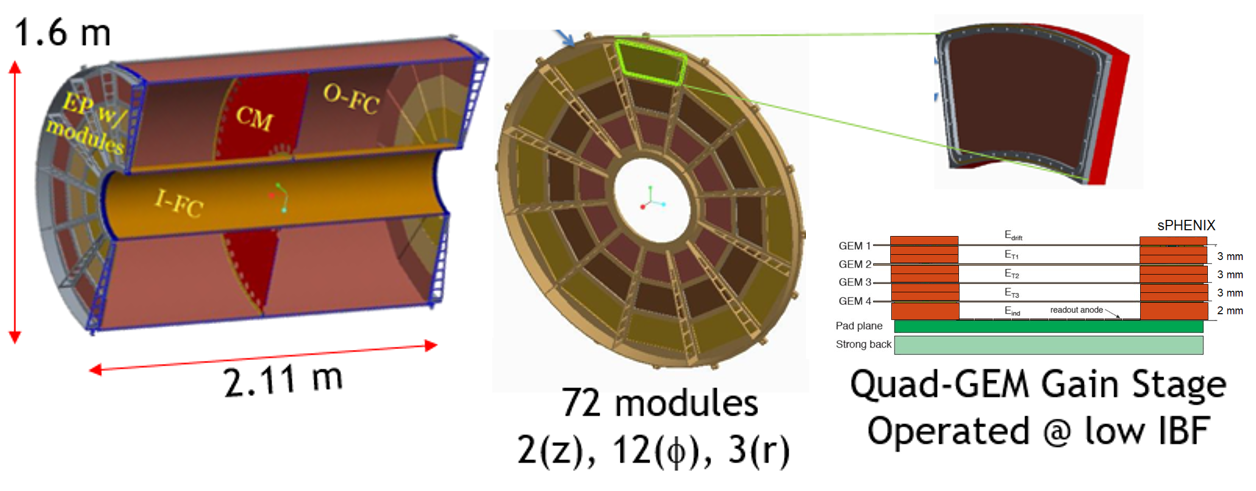}
	\caption{Pictorial diagram of the sPHENIX TPC. The gas volume and thus the active registration volume for charged particle tracks is between the inner field-cage (I-FC) and the outer field-cage (O-FC). The cathode consist of a thin metallized membrane.}
	\label{fig:stpc1}
\end{figure}

\begin{figure}[h!bt]
	\centering
        \includegraphics[width=\columnwidth]{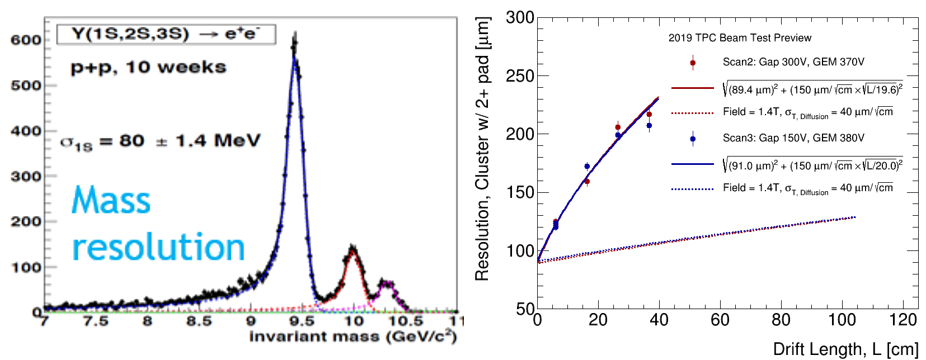}
	\caption{Left: simulation for the mass resolution sufficient to separate Upsilon states. Right: test-beam results extrapolated to sPHENIX conditions.}
	\label{fig:performancesTPC}
\end{figure}

A TPC is an option for the central detector in an EIC detector. It will provide required momentum resolution for the physics program at an EIC and is also a detector that can deliver PID by means of dE/dx.\newline
A TPC is presently under construction for the sPHENIX experiment which is expected to start taking data in IP8 of RHIC, in 2023. The sPHENIX-TPC is a compact detector with a minimum material budget in the central region. It has been also designed with an eye toward the use in an EIC detector. This concerns the minimization of the material budget in the forward region which takes into account not only the front-end electronics but also necessary infrastructure, like mounting structure and cooling.\newline
The TPC design follows the classical cylindrical double-sided TPC layout, with a cathode located at the middle of the interaction region dividing the TPC into two mirror-symmetric volumes.  The end-caps of the TPC accommodate gas-amplification modules in a subdivided arrangement; 12 sectors in azimuth and 3 sectors in radial extension. This results in a total of 72 readout modules for both end-caps. An illustration can be seen in Fig.~\ref{fig:stpc1}.
\
The physics program with the sPHENIX detector requires excellent pattern recognition as well as excellent momentum resolution. One of the performance parameters to be fulfilled for the sPHENIX program is the separation of the $Y$-states which requires a momentum resolution from the TPC in the order of $\Delta p/p\approx 1.2\%$ in the range of $4~\gevc<p<10~\gevc$ for the e$^\pm$ daughter particles. This translates to a required position resolution $\sigma_{r\phi}\lesssim 300\mu m$ with 40 track points in the sPHENIX TPC. This requirement is relaxing with more space points. A test-beam campaign with a TPC prototype verified that this resolution goal more than achievable, see Fig.~\ref{fig:performancesTPC}. 
\newline
The TPC has to be operated in a gate-less configuration such that the readout is not limited due a severe dead-time. This requires in turn the use of Micro Pattern Gas Detectors (MPGDs). For the sPHENIX TPC the choice was made to use a quadruple-GEM avalanche structure, similar to the solution that has been implemented in the ALICE-TPC at the LHC. The operating point of the GEM-stack has been adapted to the sPHENIX environment.
\paragraph{Gas Amplification}
The goal to limit space charge effects requires a low ion-back flow from the amplification device into the main tracker gas volume. A vast R\&D program to this extent has been performed by the ALICE collaboration and the experience gained there directly affected the design choices for the sPHENIX TPC.
\begin{figure}[hbt]
	\centering
        \includegraphics[width=\columnwidth]{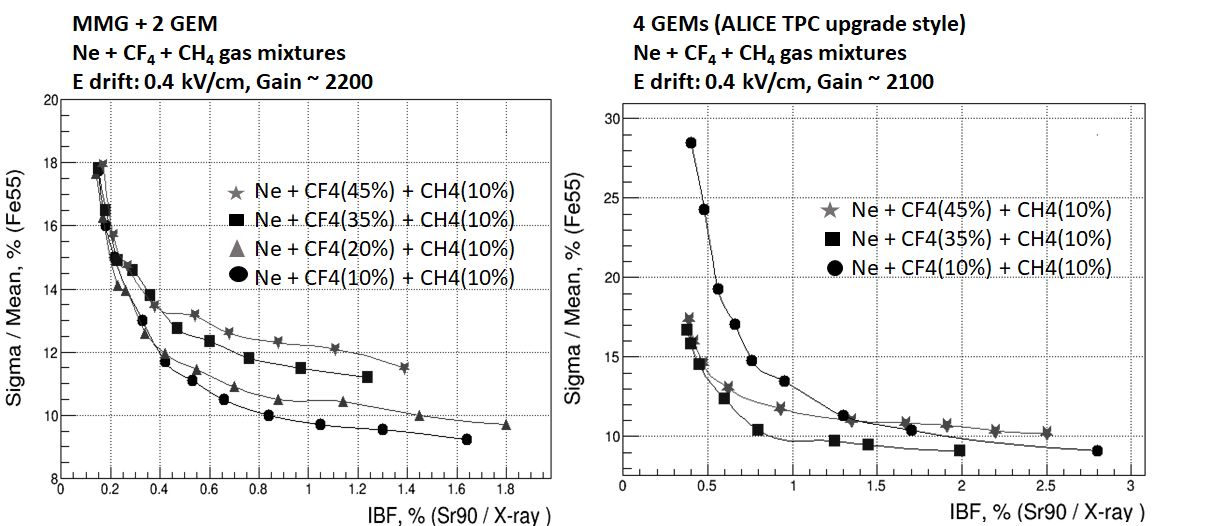}
	\caption{Comparison of two amplification structures regarding their energy resolution and ion-back flow performance with various gas mixtures. Left: operation regime of a MM2G. Right: operation regime of a quadruple-GEM amplification device.}
	\label{fig:ibfGEM1}
\end{figure}
For the sPHENIX program the energy resolution does not play a role, hence the operating point for the readout has been chosen around the minimum IBF ($\sim$ 0.3\%). For the EIC program this choice can be modified to gain back good energy resolution: the space charge effects in an EIC TPC might be less severe. Studies on the effect of space are ongoing. In principle, there are already solutions in the prototype stage if it turns out that IBF will play a similar role in an EIC environment (see next paragraph). The gas choice for the sPHENIX TPC is based on Ne-CF$_4$ because of its advantageous properties: 1) high drift velocity, 2) low transverse diffusion and 3) comparatively fast ion drift velocity. The Neon component could be exchanged with Argon which provides a higher ionization yield and therefore improves dE/dx performance. Other gas components can be added to the gas mixture which is under consideration for optimizing the TPC for EIC purposes. It is worth the mentioning that the sPHENIX configuration has been investigated in a test-beam environment with a modified operating point and promising dE/dx performance has been measured.
An alternative to the quadruple GEM readout option is the MM2G option. It consists of a double-GEM layer on top of a MicroMegas as the main amplification device, hence the term MM2G. The double-GEM structure provides the necessary field ratios to maintain a low IBF and act as a pre-amplifier. It has been shown that it is possible to obtain a low IBF while maintaining an energy resolution of better than 12\% (Fig.~\ref{fig:ibfGEM1}, left).
\paragraph{Modifications to the sPHENIX TPC}
A major modification of the TPC presently under construction for sPHENIX will be the recovery of about 10 cm vertical track length. The design for sPHENIX was chosen such that the first 10 cm in radial extension will not be read out electronically. This choice has its origin in that space charge distortions, i.e., deflections from the ideal electron trajectory are largest in the vicinity of the field cage. Therefore, the space charge distortions will be still real within the vicinity of the field-cage, however, the track information from this part will not be considered and therefore not electronically read out. This can be easily reverted in the EIC era.\newline
A modified readout pad-geometry with perhaps a modified readout electronics might improve the performance for the TPC in the EIC era. However, these are topics which are discussed in the Section \ref{part3-sec-DetTechnology.Tracking}. 

\subsubsection{Micro Pattern Gaseous Detectors (MPGDs)}
%
\begin{figure}[h!bt]
\centering
\includegraphics[width=0.95\columnwidth,trim={0pt 0mm 0pt 0mm},clip]{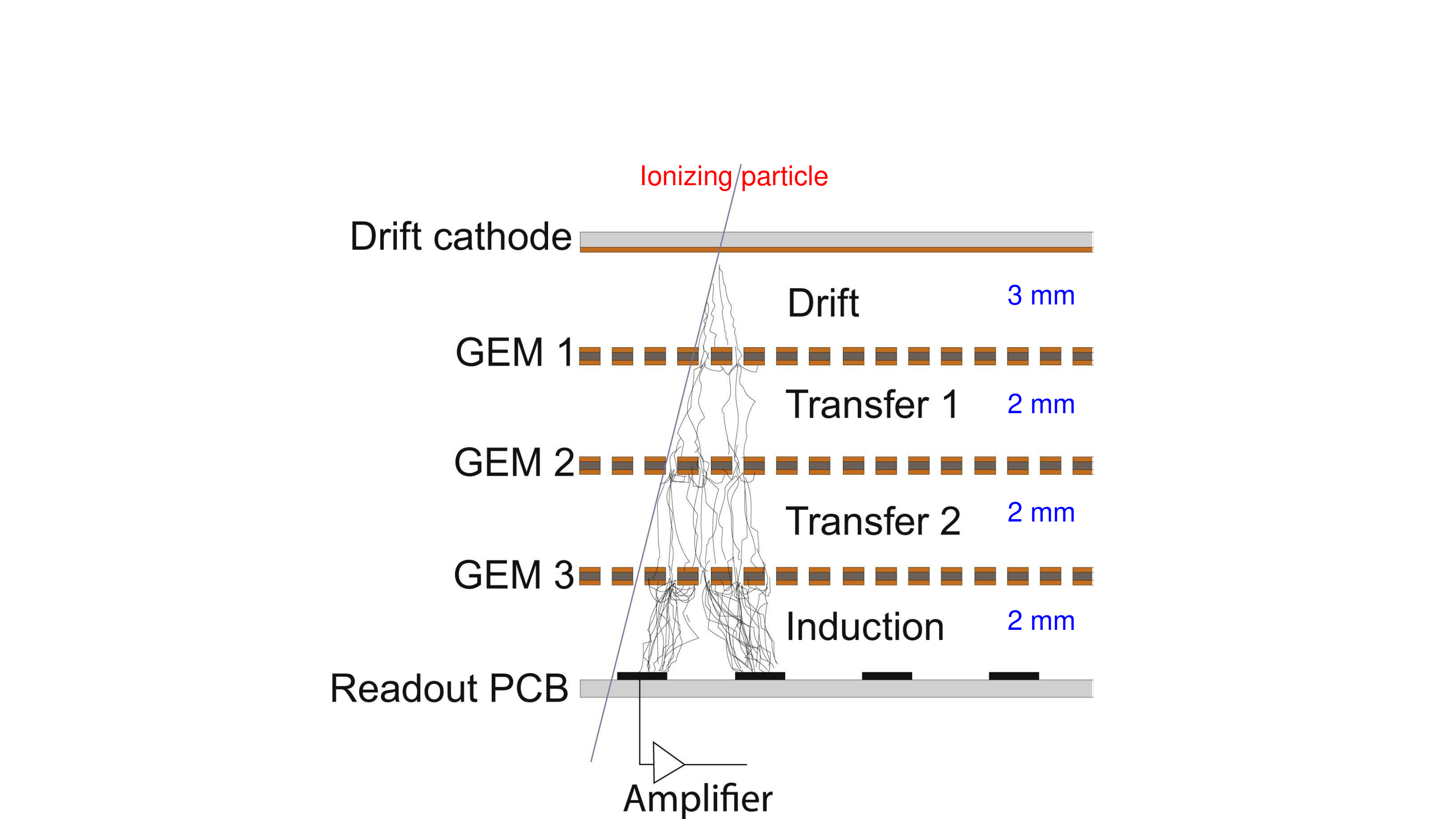}
\caption{Cross sectional view of Triple-GEM detector~\cite{Sauli:1997qp}.}
\label{fig:gem_tech}
%
\includegraphics[width=0.95\columnwidth,trim={0pt 0mm 0pt 0mm},clip]{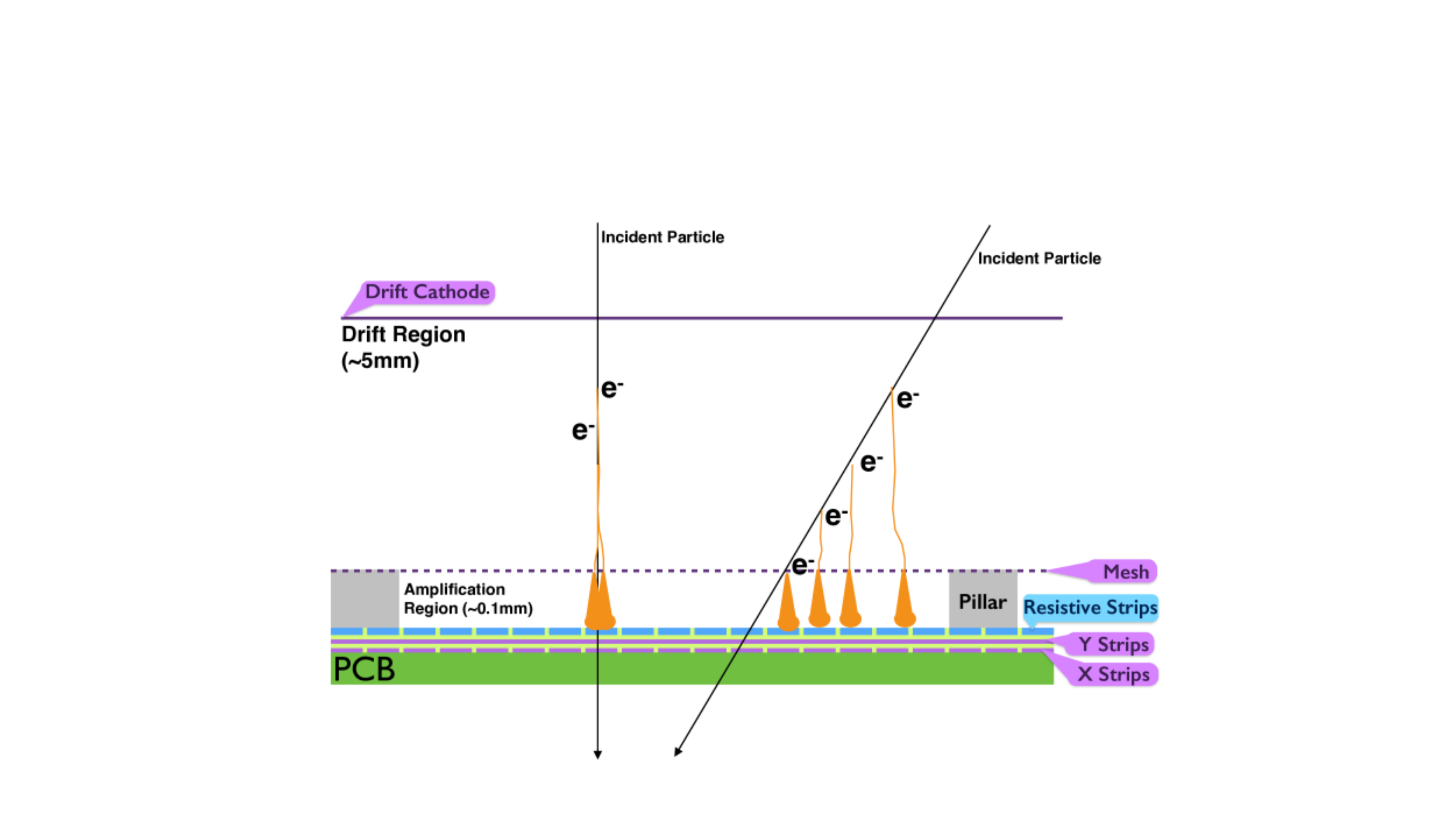}
\caption{Cross sectional view of micromegas detector~\cite{Giomataris:1995fq}.} 
\label{fig:micromegas_tech}
\end{figure}
\begin{figure}[h!bt]
\centering
\includegraphics[width=0.95\columnwidth,trim={0pt 0mm 0pt 0mm},clip]{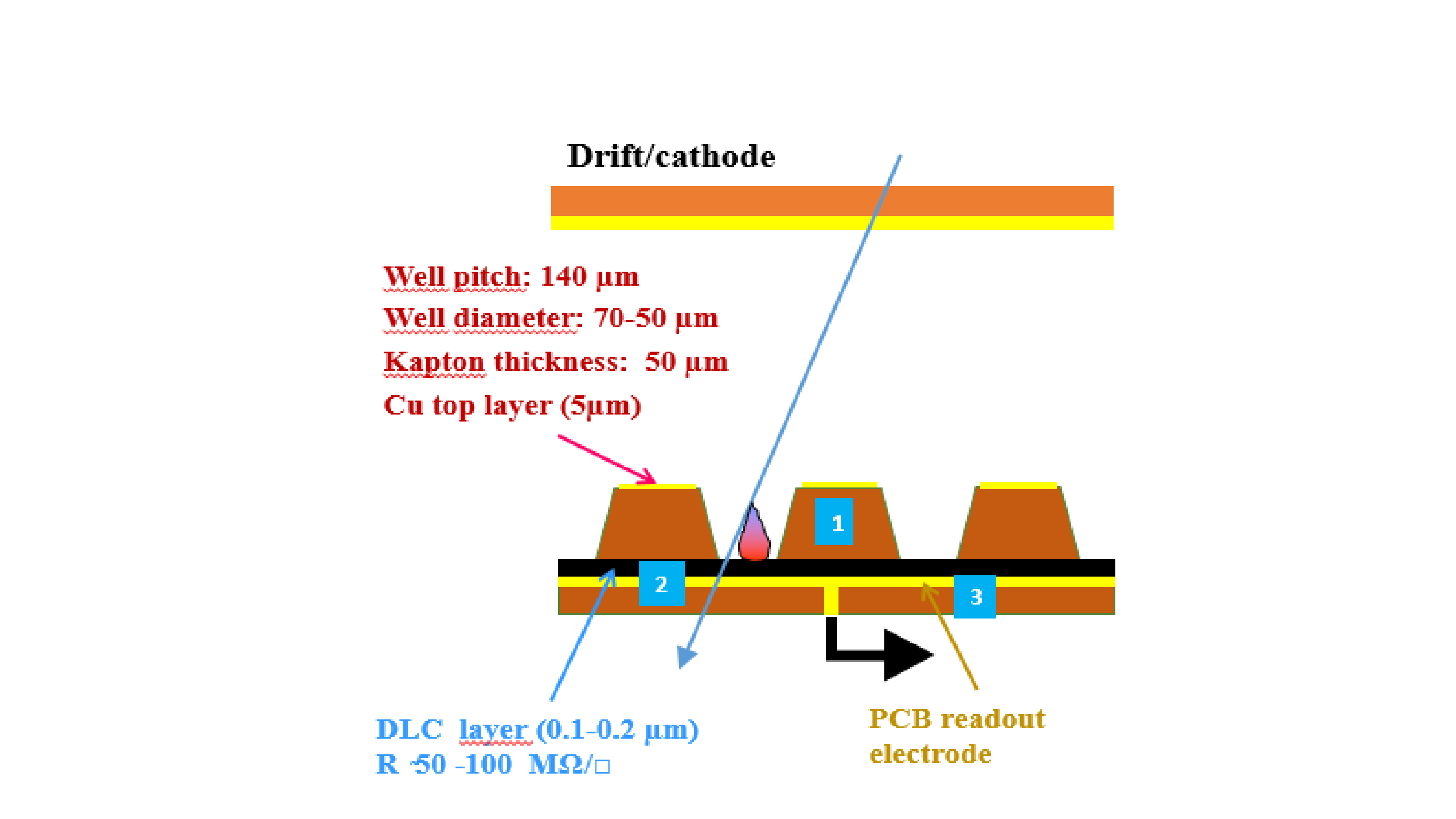}
\caption{Cross sectional view of $\mu$RWELL detector~\cite{Bencivenni:2014exa}} 
\label{fig:urwell_tech}
\end{figure}

%
%
MPGD  technologies such as Gas Electron Multiplier (GEM)~\cite{Sauli:1997qp}, Micro Mesh Gaseous Structures (Micromegas)~\cite{Giomataris:1995fq}, Resistive Micro Well ($\mu$RWELL)~\cite{Bencivenni:2014exa} are widely used for tracking in various particle physics experiment across the world such as the COMPASS~\cite{Altunbas:2002ds}, LHC main detectors upgrade (ATLAS~\cite{Wotschack:2013ola}, CMS~\cite{cmsGem}, ALICE~\cite{Gasik:2014sga} \& LHCB~\cite{Alfonsi:2006ww}) at CERN, SBS~\cite{Gnanvo:2014hpa}, CLAS12~\cite{Acker:2020qkv}, and PRad~\cite{Xiong:2019umf} 
at Jefferson Lab, STAR FGT and PHENIX HDB at BNL. These technologies typically combine a gaseous device for electron amplification with high granularity strips or pads anode readout PCB (see Fig. \ref{fig:gem_tech}, \ref{fig:micromegas_tech}, \ref{fig:urwell_tech}) to provide a combined excellent 2D space point resolution $(\approx 50\, \mu \rm{m})$, fast signal ($\approx \rm {5ns}$), high rate capabilities $(\approx\rm {MHz/cm^2)}$, low material budget $(\approx 0.5\% \rm{X_0})$ per layer, radiation hardness and large area capabilities at a significantly lower cost compared to silicon trackers.  \\ An extensive R\&D program conducted by the eRD6 Consortium \cite{eRD6cons} within the EIC Generic Detector R\&D program is dedicated to the development and optimization of MPGD technologies \cite{Zhang:2017dqw, Gnanvo:2015xda, Posik:2015gha, Aiola:2016rld} as main tracker in the central region of an baseline EIC hybrid tracker as descried in section~\ref{sec:hybrid}. In this hybrid configuration, two options, both of them involving MPGD detectors, are under study for the barrel tracker. The first option has a TPC detector (see  section~\ref{sec:tpc}) for the main tracker with a MPGD device or a combination of two MPGD devices for electron amplification and readout in the TPC end cap. The alternative to the TPC in the barrel region explores large cylindrical Micromegas or $\mu$RWELL layers for the main tracker. Both TPC and and cylindrical MPGDs options are complemented in the hadron end electron end caps by planar MPGD discs. Performance studies for various geometrical configurations of the planar MPGD layers in the end cap regions are reported in section~\ref{sec:hybrid-endcap} 

\paragraph{Drift Chambers \& Straw Tubes (DCs)}

An ultra-light drift chamber with particle identification capabilities represents a sound and a well-founded proposal for the tracker candidate of a general-purpose detector at EIC.\\
The recently built and under commissioning drift chamber, CDCH~\cite{Tassielli:2020wap} for the experiment MEG II~\cite{Baldini:2018nnn} at PSI can serve as a guide for such a hypothesis.\\
The MEG II experiments aims at searching for the decay of a positive muon at rest into a positron and a photon with a sensitivity of 6${\times}$10$^{-14}$. The prerequisites for accurate and efficient tracking of the 52.8 MeV/c momentum positrons rely on transparency, to minimize the contribution from multiple Coulomb scattering to the momentum measurement, and on high spatial resolution.
The MEG II drift chamber~\cite{Tassielli:2020wap} derives directly from the drift chamber~\cite{Andryakov:1996gn} of the KLOE experiment~\cite{Tortora:1999tg}, run for over 20 years at the 1 GeV center of mass e$^+$e$^-$ accelerator DAFNE, the environment of which is not too different from the one at EIC, although at a much lower energy and lower luminosity. In MEG II, a continuous beam of 7$\times10^7$ muon/s is stopped on a thin target, placed at the center of the drift chamber, and all the muon decay products are efficiently tracked inside the active volume, at a rate of 30 kHz/cm$^2$. A resolution of better than 100 KeV/c is expected for the 52.8 MeV/c positron momentum.\\
Furthermore, CDCH will adopt the cluster counting and timing techniques, which allow for particle identification with unprecedented resolutions, typically a factor two better than the traditional dE/dx technique, and for improving, well below 100 ${\mu}$m for short drift cells, the spatial resolution obtainable with the conventional measurement of the fastest drifting electron.
\begin{figure}[ht]
\includegraphics[width=1\textwidth]{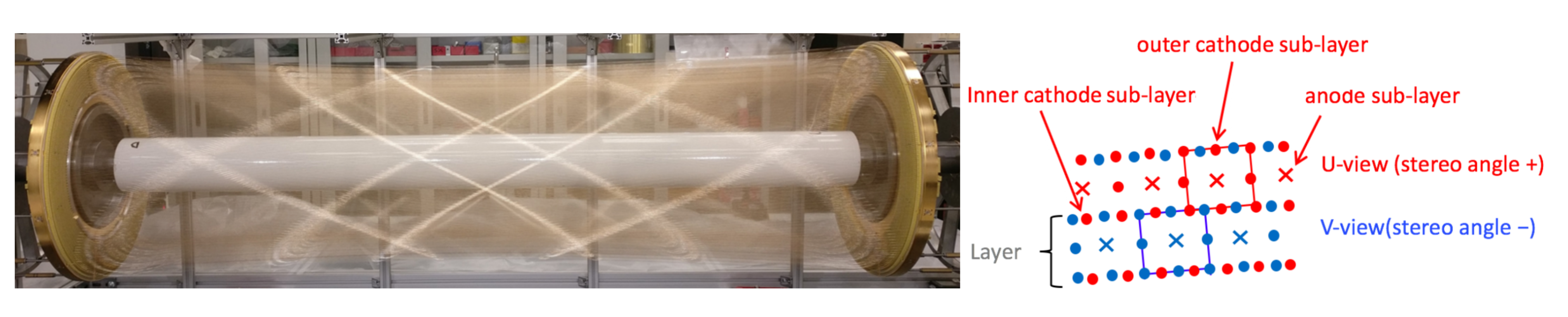}
\caption{The fully wired CDCH on the assembly structure (left) and a sketch of the drift cells within two alternating sign stereo layers (right).\label{fig:DC:cdchview}}
\end{figure}
The CDCH (Fig. \ref{fig:DC:cdchview}-left) is a unique volume, high granularity, all stereo, low mass cylindrical drift chamber, extending from 19.6 cm inner radius to 28.4 cm outer radius, for 193 cm length. It consists of 9 co-axial layers, at alternating sign stereo angles divided in 12 identical 30$^{\circ}$ sectors. Each layer has 192 square cells, with a ratio of field to sense wires equal to 5:1 to ensure the proper electrostatic configuration, and is composed by one anode and two cathode sub-layers, as sketched in Fig. \ref{fig:DC:cdchview}-right. The cell size varies linearly between 6.7 mm, at the innermost radius, and 8.7 mm at the outermost radius. The stereo angles range from 5.8$^{\circ}$ to 8.5$^{\circ}$.\\
The anodes are 20 ${\mu}$m diameter tungsten wires, while the cathodes are 40 and 50 ${\mu}$m light aluminum alloy wires. In total, the CDCH is made with 1728 sense wires, 9408 cathode wires and 768 guard wires to equalize the gain of the innermost and outermost layers. The CDCH active volume is confined within a 2 mm thick cylindrical carbon fiber shell at the outer radius and a thin (20 ${\mu}$m) aluminized Mylar foil at the inner radius. The CDCH gas mixture is 90/10 He/i-C$_{4}$H$_{10}$, chosen for the low radiation length (${\sim}$1400 m), a fast enough average drift velocity (${\sim}$2 cm/${\mu}$s) and a good spatial resolution (110 ${\mu}$m, \cite{Baldini:2016rrk}).\\
\\For a generic detector at EIC, assuming an available cylindrical volume of 320 cm length, extending from 10 cm to 90 cm in radius, one could think of a similar drift chamber with 270 cm active length (320 cm, including services at the end-plates) co-axial to the beam line and asymmetrically placed with respect to the interaction point with 2/3 of the length (180 cm) in the forward hadron direction and 1/3 (90 cm) in the forward electron direction. Extrapolating from the electrostatic structure of the MEG II drift chamber, one could fit in this volume 80 co-axial layers, at alternating sign stereo angles, arranged in 24 identical azimuthal sectors, of approximately 1 cm$^{2}$ size square cells for a total of approximately 25,000 cells and 150,000 wires. The challenges potentially arising from such a large number of wires are overcome by the peculiar design of the modular wiring procedure successfully adopted for the MEG II CDCH (here one would have 6 wires/cm$^{2}$ as opposed to 12 wires/cm$^{2}$ in the case of the MEG II CDCH). The sense wires can be read from both ends for charge division and time difference measurement in order to ease track finding seeding. The angular coverage would extend down to 12$^{\circ}$ in the forward hadron direction, corresponding to 99\% of the full solid angle and to 23$^{\circ}$ in the forward electron direction for a 97\% full coverage. The chamber can be operated with the same CDCH gas mixture, 90/10 He/i-C$_{4}$H$_{10}$, corresponding to a maximum drift time of ${\sim}$250 ns. The number of ionisation clusters generated by a minimum ionizing particle (m.i.p.) is about 12.5 cm$^{{-}1}$, allowing for efficiently exploiting the cluster counting/timing techniques to improve both spatial resolution (${\sigma}_{r\phi}$ {\textless} 100 ${\mu}$m) and particle identification (${\sigma}_{(dNcl/dx)}$/(dN$_{cl}$/dx) {\textless} 3.6\%). Making use of the cluster counting technique for particle identification, the proposed drift chamber for the EIC detector possesses the capability of separating (> 4 sigma) pions from kaons over the full range of momenta, up to 100 GeV/c, except for a narrow window around 1 GeV/c, which can easily be recovered with a modest 100 ps resolution time-of-flight system (see below for more details).\\
A very similar drift chamber~\cite{Chiarello:2019qvm,Tassielli:2021D1}, proposed for the detectors at the future e$^+$e$^-$ accelerators FCCee~\cite{Abada:2019zxq} and CEPC \cite{DC::CEPC}, has been extensively studied with full simulations at center of mass energies from 90 GeV to 365 GeV and luminosities up to 2$\times10^{36}$ cm$^{-2}$s$^{-1}$, corresponding to trigger rates of 100 kHz and charged track multiplicities in excess of 20/event.\\
The total amount of material in the radial direction towards the barrel electromagnetic calorimeter would be of the order of 1.5\% X$_{0}$, (10$^{-4}$ for the inner Mylar foil, 2.5${\times}$10$^{-3 }$ for gas and wires, 1.2${\times}$10$^{-2}$ for the outer carbon fiber shell) whereas, in the forward direction, it would be about 4.0\% X$_{0}$, including frontend electronics and cables (see Fig. \ref{fig:DC:trnsConf}-bottom left). The analytically calculated expected performance of such a drift chamber, illustrated in Fig. \ref{fig:DC:momres} and Fig. \ref{fig:DC:momresrel}, would be:\\
\indent ${\Delta}$p$_{t}$/p$_{t}$ = (0.34p$_{t}$ + 1.1)${\times}$10$^{-3}$; \hspace{1cm} ${\Delta \phi}$ = (0.23 + 0.9/p$_{t}$)${\times}$10$^{-3}$ rad.;\\
\indent ${\Delta \theta}$ = (0.24 + 0.6/p$_{t}$)${\times}$10$^{-3}$ rad.; \hspace{0.75cm} ${\sigma}_{dN/dx}$/dN/dx = 3.6\% \\
where the sums in parenthesis are intended in quadrature.
\begin{figure}[ht]
\includegraphics[width=0.49\textwidth]{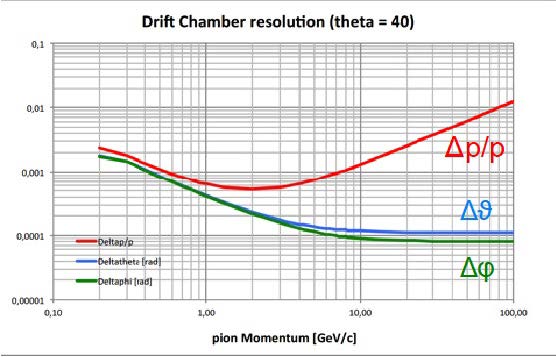}
\includegraphics[width=0.49\textwidth]{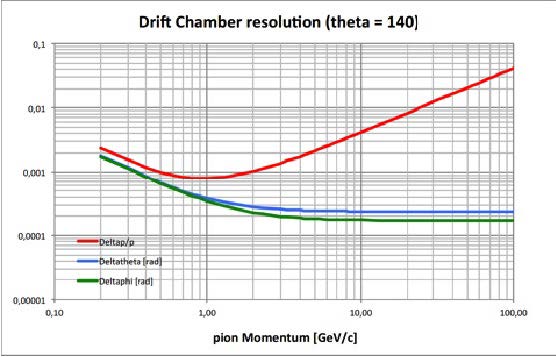}
\caption{Relative transverse momentum and angular resolutions (in radiant) from analytical calculations, as a function of momentum, in the forward hadron direction (at left, for ${\theta}$ = 40$^{\circ}$) and in the forward electron direction (at right, for ${\theta}$ = 140$^{\circ}$).\label{fig:DC:momres}}
\end{figure}
\begin{figure}[ht]
\centering
\includegraphics[width=0.5\textwidth]{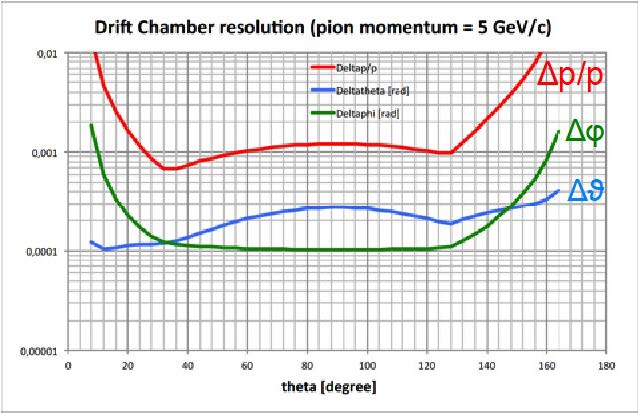}
\caption{Relative transverse momentum and angular resolutions (in radiant) from analytical calculations, as a function of polar angle for a transverse momentum of 5 GeV/c.\label{fig:DC:momresrel}}
\end{figure}
In Fig. \ref{fig:DC:pid}-left the expected particle separation (${\mu}$/${\pi}$ in red, ${\pi}$/K in blue, K/p in green) performance in terms of numbers of standard deviation is illustrated as a function of momentum (for ${\theta}$ = 40$^{\circ}$). Solid curves refer to separation with the cluster counting technique and dashed curves refer to the optimal energy loss truncated mean technique. Fig. \ref{fig:DC:pid}-right shows how the ${\pi}$/K separation varies as a function of the polar angle for a momentum of 2.8 GeV/c, both for cluster counting (solid curve) and for dE/dx (dashed curve). A cluster counting efficiency of 80\% is assumed in the calculations. To be noticed the relative gain of a factor 2, in terms of particle separation, of the cluster counting technique with respect to dE/dx.
\begin{figure}[ht]
\includegraphics[width=0.49\textwidth]{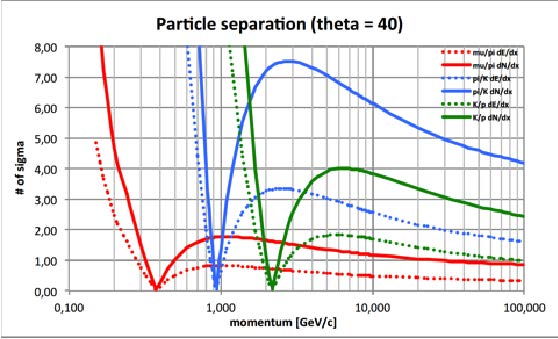}
\includegraphics[width=0.49\textwidth]{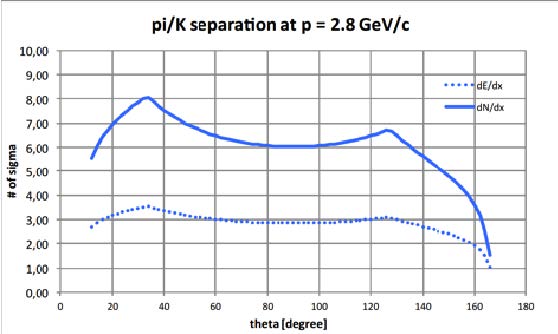}
\caption{Left: The expected particle separation (${\mu}$/${\pi}$ in red, ${\pi}$/K in blue, K/p in green) performance in terms of numbers of standard deviation as a function of momentum (for ${\theta}$ = 40$^{\circ}$). Right: ${\pi}$/K separation as a function of the polar angle for a momentum of 2.8 GeV/c. Solid curves refer to separation with the cluster counting technique and dashed curves refer to the optimal energy loss truncated mean technique.\label{fig:DC:pid}}
\end{figure}
The excellent performance of the drift chamber, not only in terms of momentum and angular resolutions, as it is shown in Fig. \ref{fig:DC:momresrel}, but also in terms of particle separation (Fig. \ref{fig:DC:pid}-right), degrades for ${\theta}$ {\textless} 32$^{\circ}$ and ${\theta}$ {\textgreater} 124$^{\circ}$, because of the limited longitudinal extension of the chamber. \\
This could be remedied with a different geometrical configuration by shortening the cylindrical drift chamber to an active length of 120 cm and extending the active volume longitudinally as follows. By using a planar drift chamber unit module as the one sketched in Fig. \ref{fig:DC:trnsConf}-top left, one can assemble two transverse planar drift chambers at its extremities, as shown in Fig. \ref{fig:DC:trnsConf}-bottom right: the one in the forward electron direction made of a set of 27 triplets of planes, as those illustrated in Fig. \ref{fig:DC:trnsConf}-top right, for a total of 162 unit modules, 10,000 drift cells and a longitudinal extension of approximately 80 cm and the one in the forward hadron direction made of a set of 40 triplets of planes, for a total of 240 modules, 15,000 drift cells and a longitudinal extension of approximately 120 cm.
\begin{figure}[ht]
\includegraphics[width=1\textwidth]{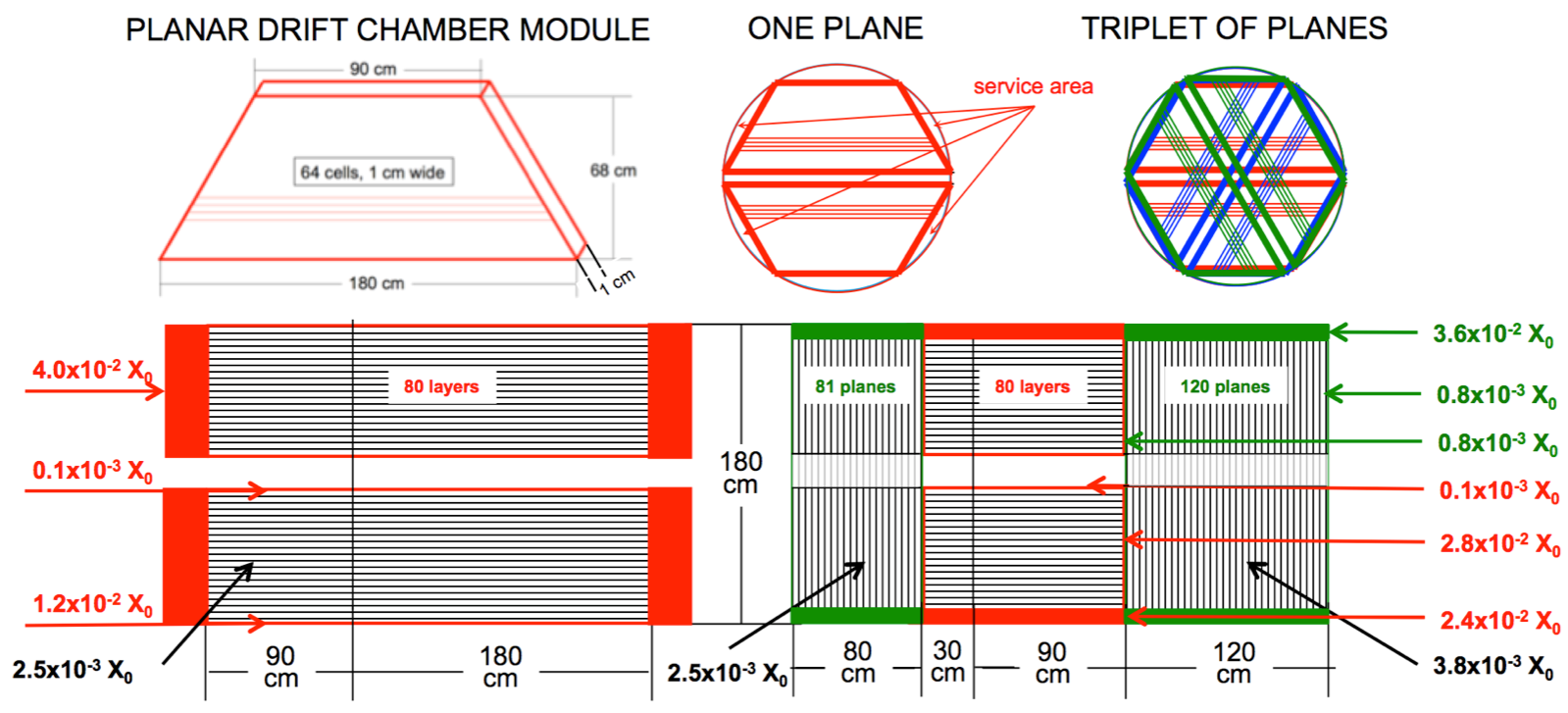}
\caption{Top left: A unit module of a planar drift chamber hosting 64 cells in an electrostatic configuration analogous to the one sketched in Fig. \ref{fig:DC:cdchview}-right. Top center: One plane constituted by two facing unit modules. Top right: A triplet of planes assembled after a ${\pm}$60$^{\circ}$ rotation. Bottom left: Schematic view of the cylindrical drift chamber discussed in the text, with indication of dimensions and of material budget in terms of radiation length. Bottom right: Different configuration with a short cylindrical drift chamber and two transverse planar drift chamber at the two extremities, as described in the text. Solid red and green boxes represent the service areas where the front-end electronics is placed.\label{fig:DC:trnsConf}}
\end{figure}
In this configuration, the tracking system will occupy the same volume, although with twice as many channels. A different arrangement will be needed for the services of the cylindrical drift chamber. However, given its short radial extension, 80 cm, one could, with proper cables, distribute azimuthally at the outer radius the front-end electronics, thus marginally reducing the momentum lever arm and slightly increasing the material budget towards the barrel electromagnetic calorimeter from 1.5\% X$_{0}$ to approximately 4\% X$_{0}$. This will be compensated by a substantial increase to greater than 99.5\% of the solid angle coverage and in the flattening of the resolution functions in the forward and backward directions. Particle identification will also greatly benefit from the increase in the average number of samples with a resulting improvement of the ${\sigma}_{dN/dx}$/dN/dx from 3.6\% to better than 3.0\% in the forward electron direction and to less than 2.4\% in the forward hadron direction.

\subsubsection{Small-strip Thin Gap Chambers (sTGCs)}

\begin{figure}[hbt]
	\centering
        \includegraphics[width=0.45\columnwidth]{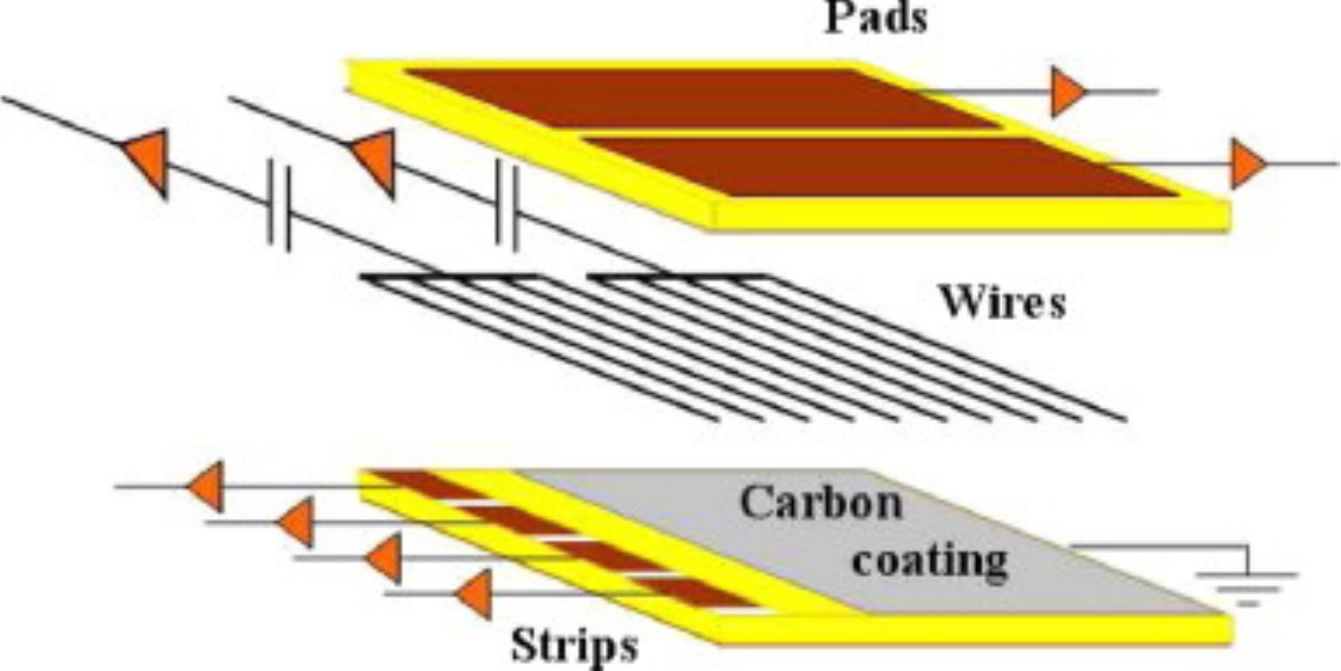}
	\caption{Schematic diagram of the basic sTGC structure reproduced from \cite{Abusleme:2015yja}.}
	\label{fig:stgc_tech}
\end{figure}

Small-strip Thin Gap Chamber (sTGC) detector technology was developed for the ATLAS new small wheel upgrade~\cite{Stelzer:2016mky}. ATLAS stGCs are used for muon detectors to provide trigger capabilities.
A modified version of the sTGC tracker, based on the ATLAS design, is being used for the STAR forward rapidity upgrade~\cite{SN0648STARForward}. 
The small-strip thin gap chamber detector technology offers a reasonably good space-point resolution ($\approx100\mu$m) and low material budget $\sim0.5\% X_0$ per layer, for a relatively low cost compared to various other technologies.  
The sTGC as designed by ATLAS for the new small wheel upgrade consists of a grid of 50$\mu$m diameter gold-plated tungsten wires with a 1.8mm pitch sandwiched between two cathode planes 1.4mm from the wire plane. The sTGC wires operate at 2.9 kV in a gas mixture of 55\% $CO_2$ and 45\% n-pentane. 
The sTGC modules feature both strip and pad readout. Copper strips with a pitch of 3.2 mm are located on one of the anode planes and run perpendicular to the wires. Large rectangular readout pads, useful for fast triggering, are located on the other anode plane. 
An illustration of the basic design of an sTGC is shown in Fig.~\ref{fig:stgc_tech}. 

While position resolution better than 50$ \mu$m has been achieved in test beam studies~\cite{Abusleme:2015yja}, in practice, the sTGC strip readout is expected to provide position resolution on the order of $100-150 \mu$m, depending on the charged track's incident angle. 
The ATLAS new small wheel setup employs sTGC modules with strips aligned to provide precise position measurement in the bending coordinate, with measurement of the azimuthal information provided by wire readout.  
The STAR forward upgrade application employs sandwiches of two layers of sTGC modules with one layer providing precise x-position measurements and the other layer providing precise y-position measurements\cite{SN0648STARForward,Yang:2020prn}. In addition, the design used by STAR replaces pads on one of the two layers with diagonal strips to help improve space point reconstruction.

Since the sTGC detectors are highly cost-effective with a low material budget and robust up to single hit rates of 100 kHz/cm$^2$, they are a suitable technology choice for large area planar regions of tracking. Specifically, sTGC layers could be employed for tracking in the hadron-going (forward) direction at a $z\approx300$ cm  beyond the RICH detector. The sTGC may be a good choice for tracking in this region, beyond the central tracking and PID detectors, where the magnitude of the multiple scattering effects will be larger making precise space point resolution less important.  Similarly, sTGC planes may be a viable cost-effective option for the regions that require large area trackers in the electron-going (backward) direction.  

\subsection{Detector concepts and performance studies}
\label{tracking_concepts}
In this section we present two alternative detector concepts. The first is an all-silicon set of tracking layers and discs. The second concept is a hybrid design that contains silicon tracking layers and discs with a gaseous tracking detector surrounding the silicon based barrel layers. Each of these designs has particular strengths. In the all-silicon case, the full tracking detector can be realized in a comparatively compact form while retaining excellent tracking capabilities. In the hybrid case, the gaseous detector can provide dE/dx measurements that can add to the PID capabilities while maintaining tracking that meets the EIC requirements. A set of attributes for each configuration is shown in Table~\ref{tab:det_compare}.
\begin{table}[h!bt]
\centering
\caption{Comparison of attributes for the two simulated detector configurations.}
\label{tab:det_compare}
\includegraphics[width=0.975\columnwidth,trim={0pt 35mm 0pt 35mm},clip]{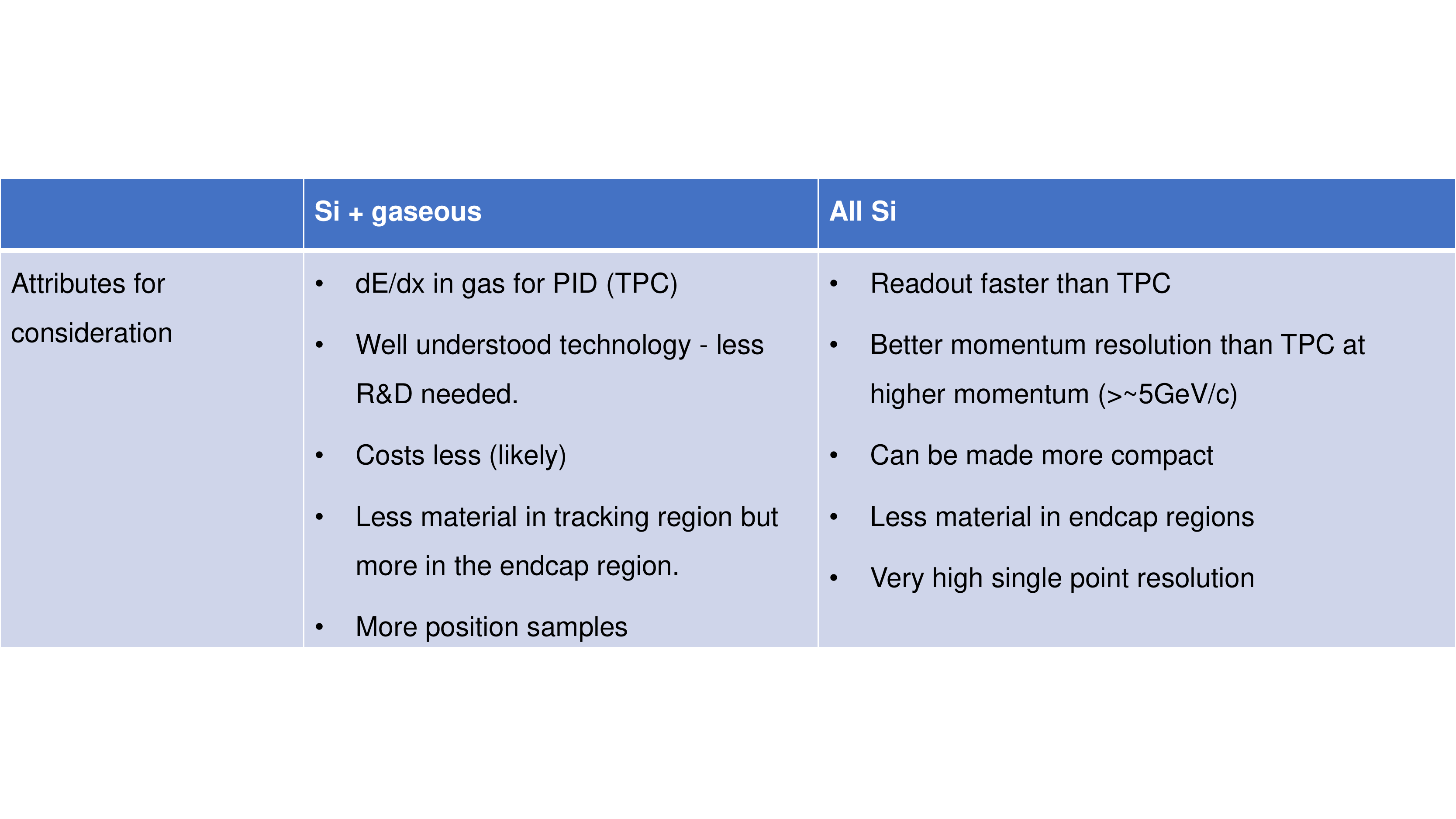}
\end{table}
It is hoped that the inclusion of two alternative configurations will aid in the selection of a detector optimized with respect to the full set of overall detector requirements. In addition, these options may aid in the formulation of complementary detector configurations for the second interaction point. The performance of each detector option in comparison to the physics derived requirements can be found in summary tables in the concluding Summary section of the tracking chapter.
\subsubsection{Baseline All-Silicon Tracking Option (Barrel \& End Caps)}
\label{all-Si-Rey}

\begin{figure}
    \centering
    \includegraphics[width=0.49\textwidth]{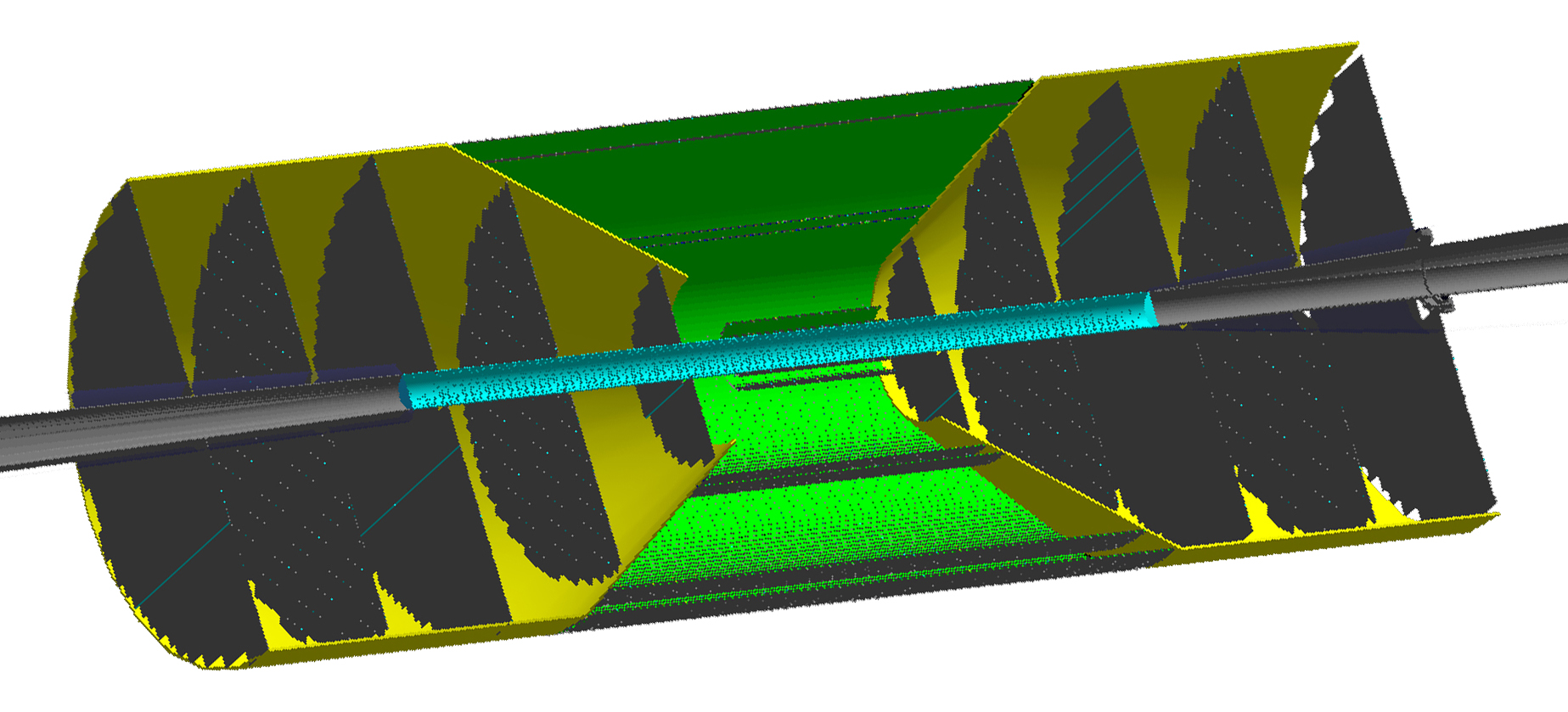}
    \includegraphics[width=0.49\textwidth]{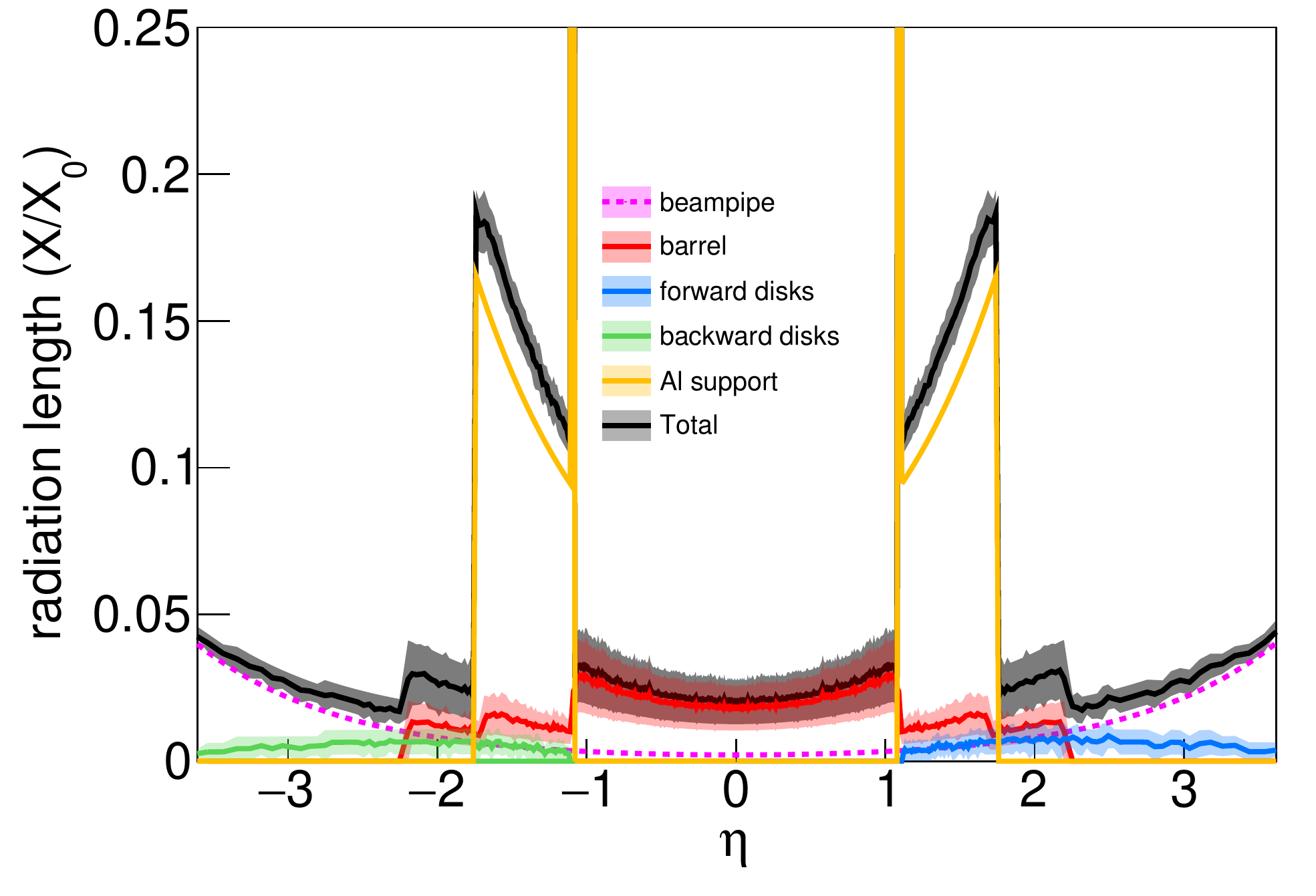}
    \caption{All-silicon tracker geometry. Left: 
    Geant-4 model showing half of the detector. The barrel, disks, and support structure correspond to the green, dark-gray, and yellow components, respectively. The beryllium section of the beam pipe is shown in cyan. The rest of the beam pipe, which takes into account the expected electron-hadron-beam crossing angle is shown in light-gray.
    Right: Detector material scan. The dashed line describes the baseline material budget from the beam pipe. The red, blue, and green curves correspond to the barrel, forward, and backward components of the detector, respectively. The uncertainty band defines the minimum and maximum amounts of material found in a given $\eta$ as the material is scanned around $\phi$. The yellow curve describes an aluminum structure that is used as a mass equivalent for support structure and services. See text for details.
    }
    \label{fig:all_si:det_geom}
\end{figure}

A pixelated all-silicon tracker prototype for the EIC is shown in Fig.~\ref{fig:all_si:det_geom} (left).
The detector is cylindrically symmetric and has three main regions: a 6-layer barrel in the mid-rapidity region, 5 disks in the forward region, and 5 disks in the backward region.
The extent of the tracker along the beam axis is identical in both directions, a constraint consistent with the current choice to have the nominal beam collision point coincide with the geometric center of the overall general purpose detector concepts.
In the barrel region, the trade-off from pairing layers to gain momentum-resolution performance is primarily with the momentum measurement threshold, $2 p_T \simeq 0.3 B\cdot r$ (about 0.2\,GeV/$c$ for a representative $B = 3\,\mathrm{T}$ and $r \simeq 0.4\,\mathrm{m}$).
Pairing of layers also reduces the number of stave designs and associated tooling.
In the all-silicon concept under consideration, the layers that constitute the barrel are thus paired with the outermost pair at $\simeq 0.4\,\mathrm{m}$ and the intermediate pair near the mid-point to the beam axes to best capture the sagitta.
The transition between the outer barrel layers and the disks is near $|\eta| \simeq 1.1$ to minimize the amount of traversed material.
Further details on the barrel and disk geometries are presented in tables~\ref{tab:all_si:barrel}~and~\ref{tab:all_si:disks}, respectively.
In this concept, the innermost barrel layers drive the vertexing performance.
Their length (well) exceeds the extent of the $\simeq 8\,\mathrm{cm}$ beam-collision region and is chosen to accept (displaced) tracks for $|\eta| \lesssim 2$ without relying on track-pointing with the disks, which will near-inevitably involve tracking across inactive material from services and supports in this region of the detector.
The dominant parts of the services and supports are thought to be guided out in a projective way along the transition angle between the barrel and the disks.
While the actual support structure will likely be made of carbon fiber composite, the material is modeled in a simplified form as an effective 5-mm-thick aluminum cone in the performance simulations thus far; engineering evaluations remain to be done.
This geometry is wrapped around the EIC beam pipe, which in the region $-79.8 < z < 66.8$ cm corresponds to a 3.17-cm-radius beryllium cylinder of thickness of 760 \um.

In this configuration, the detector is made up of ALICE-ITS3-like staves, each having an average material budget of $X/X_0 = 0.3 \%$. These staves, assembled into the detector geometry, contribute the amount of material  shown in Fig.~\ref{fig:all_si:det_geom} (right). 
Since the staves form a periodic but changing material budget, the azimuth ($\phi$) is swept for each pseudorapidity
direction, and the minimum and maximum found $X/X_0$ define the width of the uncertainty band.
Overall, the active areas of the detector provide a material budget of $X/X_0<5\%$. The support structure adds a significant amount of material. The projective design of this structure ensures that most of this material is concentrated in a small pseudorapidity range, at $|\eta|\approx1.1$.

\begin{table}
    \parbox{.45\linewidth}{
        \centering
\caption{Main barrel-layer characteristics.}
\label{tab:all_si:barrel}
\begin{tabular}{c|cc}
Barrel  & radius    & length along z        \\
layer   & {[}cm{]}  & {[}cm{]}              \\
  \hline
1       & 3.30      & 30                    \\
2       & 5.70      & 30                    \\
3       & 21.00     & 54                    \\
4       & 22.68     & 60                    \\
5       & 39.30     & 105                   \\
6       & 43.23     & 114                    
\end{tabular}
}
\parbox{.45\linewidth}{
\centering
\caption{Main disk characteristics.}
\label{tab:all_si:disks}
\begin{tabular}{c|ccc}
Disk  		& z position	& outer 			& inner 			\\
number		& {[}cm{]}      & radius {[}cm{]}	& radius {[}cm{]}	\\
\hline
-5			& -121		    & 43.23			    & 4.41			    \\
-4			& -97		    & 43.23			    & 3.70			    \\
-3			& -73		    & 43.23			    & 3.18			    \\
-2			& -49		    & 36.26			    & 3.18			    \\
-1			& -25		    & 18.50			    & 3.18			    \\
\hline
1           & 25            & 18.50			    & 3.18 			    \\
2           & 49            & 36.26			    & 3.18 			    \\
3           & 73            & 43.23			    & 3.50 			    \\
4           & 97            & 43.23			    & 4.70 			    \\
5           & 121           & 43.23             & 5.91
\end{tabular}
}
\end{table}

This configuration was studied and optimized using the Geant-4-based Fun4All framework~\cite{Fun4All,Pinkenburg:2011zza,Pinkenburg:2005zza}.
Momentum, pointing, and angular resolutions at the vertex were studied by populating the detector over the entire acceptance with single particles
(charged pions, electrons, and protons) generated in the momentum range of $0 < p < 30$ GeV$/c$ with a fixed vertex at $(x, y, z) = (0, 0, 0)$, and reconstructing their tracks with the detector.
The simulated silicon-pixel size corresponds to 10 \um\  (point resolution $=10/\sqrt{12}$ \um).
The studies were carried out with magnetic-field maps describing the BaBar (1.4 T) \cite{Babar} and Beast (3.0 T) \cite{Beast1,Beast2} solenoids.

\begin{figure}
    \centering
    \includegraphics[width=\textwidth]{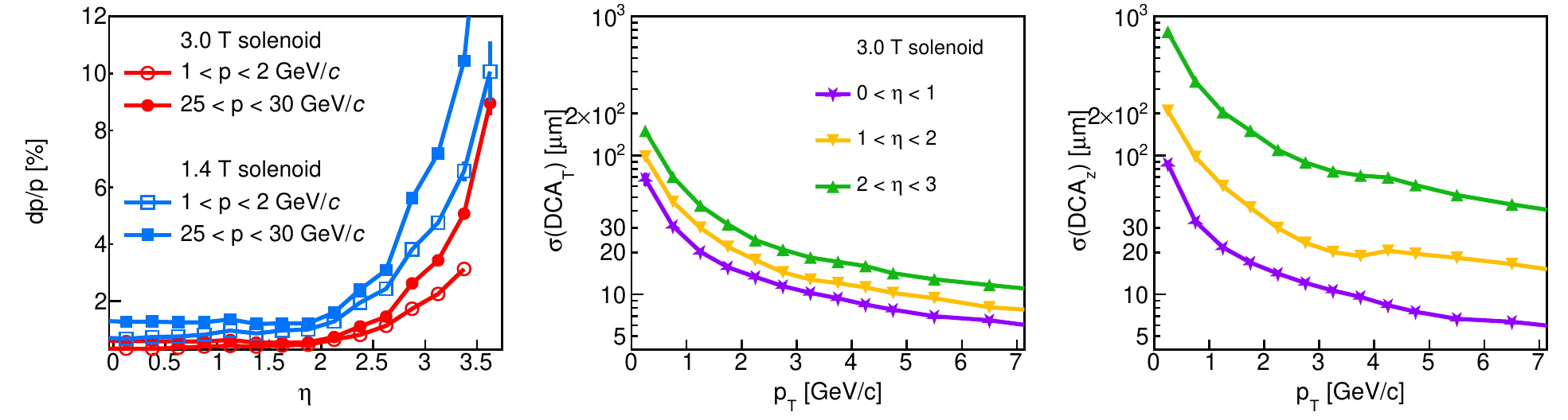}
    \caption{Detector resolutions. Left: Momentum resolution as a function of pseudorapidity for pions for two magnetic-field configurations for representative momentum bins.
    Center: Transverse Distance-of-Closest-Approach (DCA$_T$) resolution as a
    function of transverse momentum for several pseudorapidity bins.
    Right: Longitudinal Distance-of-Closest-Approach (DCA$_z$) resolution as a
    function of transverse momentum for several pseudorapidity bins.
    }
    \label{fig:all_si:resolutions}
\end{figure}

The fractional momentum resolution is determined as
the standard deviation of a normal function fitted to the $\Delta p/p \equiv (p_{truth} - p_{reco})/p_{truth}$ distribution. Here, the labels `truth' and `reco' represent generated and reconstructed variables, respectively.
Momentum-resolution results for pions are shown as a function of pseudorapidity in Fig.~\ref{fig:all_si:resolutions} (left).
As expected from the leading-order $\sim 1/B$ dependence of the momentum resolution, doubling the magnetic field improves the momentum resolution by a factor of $\approx$2.
The resulting distributions were characterized using fits with the functional form

\begin{equation}
    dp/p=Ap\oplus B,
    \label{eq:dpp_v_p_param}
\end{equation}
where $\oplus$ is shorthand notation for sum in quadrature. The $A$ and $B$ fit parameters are presented in Table~\ref{tab:param_AllSi}.

\begin{table}
\small
\caption{All-silicon tracker momentum and pointing resolution parametrizations.}
\label{tab:param_AllSi}
\begin{tabular}{cc|cc|cc|cc}
&          		& \multicolumn{2}{c}{$\delta p/p = Ap\oplus B$} & \multicolumn{2}{c}{DCA$_z = A/p_T\oplus B$} & \multicolumn{2}{c}{DCA$_T = A/p_T\oplus B$} \\
                                    			&          		& A[\%/GeV]	& B[\%]                & A[$\mu$m GeV]       	& B[$\mu$m]      	& A[$\mu$m GeV]                    & B[$\mu$m]                     \\
\hline \hline
\multirow{2}{*}{$0.0 < |\eta| < 0.5$} 	& B = 3.0T 	& 0.018		& 0.382                & 27	            		& 3.2         		& 25     		& 4.9                 \\
                                    			& B = 1.4T   	& 0.041		& 0.773                & 27                 		& 3.3               		& 26                 	& 3.9                 \\
\hline
\multirow{2}{*}{$0.5 < |\eta| < 1.0$} 	& B = 3.0T 	& 0.016             & 0.431		   & 37				& 3.8        		       	& 28           	& 4.5                 \\
                                    			& B = 1.4T   	& 0.034             & 0.906                & 35           			& 3.8		               	& 31               	& 4.0                 \\
\hline
\multirow{2}{*}{$1.0 < |\eta| < 1.5$} 	& B = 3.0T 	& 0.016             & 0.424                & 56              			& 5.9		               	& 33 	              	& 5.5                 \\
                                    			& B = 1.4T   	& 0.034             & 0.922                & 56               		& 5.4        		       	& 35	               	& 5.1                 \\
\hline
\multirow{2}{*}{$1.5 < |\eta| < 2.0$} 	& B = 3.0T 	& 0.012             & 0.462                & 111              		& 7.0               		& 40               	& 5.1                 \\
                                    			& B = 1.4T   	& 0.026             & 1.000                & 112             		& 7.1               		& 41               	& 4.9                 \\
\hline
\multirow{2}{*}{$2.0 < |\eta| < 2.5$} 	& B = 3.0T 	& 0.018             & 0.721                & 213				& 13.8             		& 47               	& 7.1                \\
                                    			& B = 1.4T   	& 0.041             & 1.551                & 212				& 16.0             		& 48               	& 7.7               \\
\hline
\multirow{2}{*}{$2.5 < |\eta| < 3.0$} 	& B = 3.0T 	& 0.039             & 1.331                & 347                     	& 40.5                     	& 52                  & 11.9              \\
                                    			& B = 1.4T   	& 0.085             & 2.853                & 373                     	& 37.9                     	& 59                  & 11.2              \\
\hline	
\multirow{2}{*}{$3.0 < |\eta| < 3.5$} 	& B = 3.0T 	& 0.103             & 2.441                & 719                     	& 87.6                      	& 59                  & 26.0               \\
                                    			& B = 1.4T   	& 0.215             & 5.254                & 732                     	& 87.7                     	& 66                  & 25.3               \\
\hline
\multirow{2}{*}{$3.5 < |\eta| < 4.0$} 	& B = 3.0T 	& 0.281             & 4.716                & 1182                     	& 206                     	& 69                  & 65.9               \\
                                    			& B = 1.4T   	& 0.642             & 9.657                & 1057                     	& 221                     	& 69                  & 72.1                    
\end{tabular}
\end{table}

The Distance of Closest Approach (DCA) is defined as the spatial separation between the primary vertex and the reconstructed track projected back to the $z$ axis (DCA$_{z}$) or to the $x-y$ plane (DCA$_{T}$). The DCA resolutions were determined as the standard deviation of
normal functions fitted to the DCA$_{z}$ and DCA$_{T}$ distributions.
DCA-resolution results as a function of transverse momentum ($p_T$) for pions are shown in Fig.~\ref{fig:all_si:resolutions} (center and right). The resulting distributions were characterized via fits with the functional form

\begin{equation}
\sigma({\rm DCA})=A/p_T\oplus B.
\label{eq:dca_v_pT_param}
\end{equation}
The $A$ and $B$ fit parameters are presented in Table~\ref{tab:param_AllSi}.

\begin{figure}
    \centering
    \includegraphics[width=0.6\textwidth]{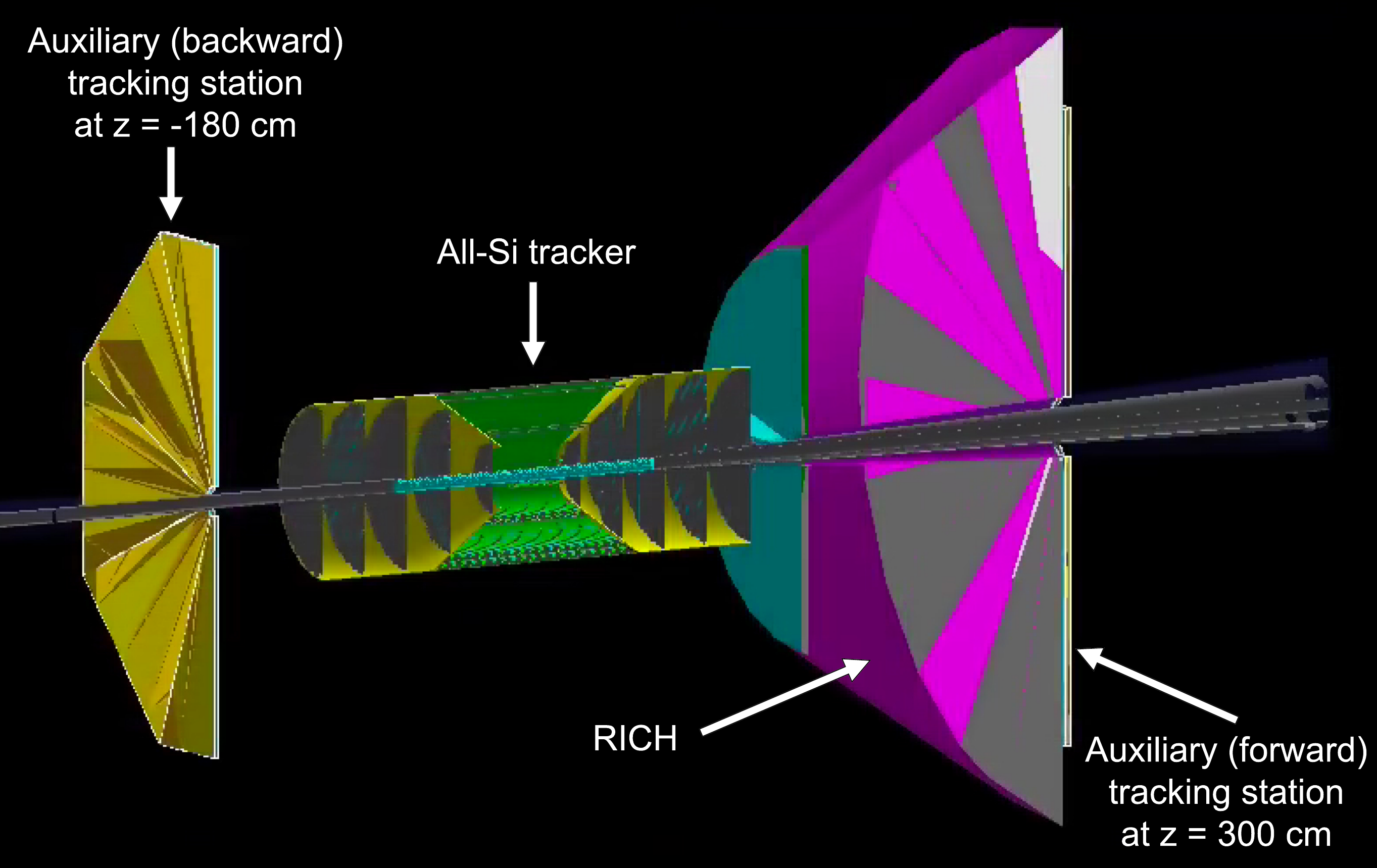}
    \caption{Event display showing the all-silicon tracker complemented with additional tracking stations in the available space~\cite{ayk}. In the backward region, the tracking station is installed at $z = -180$ cm with no significant amount of material expected between the all-silicon tracker and the complementary tracking station. In the forward region, the auxiliary tracking station is installed at $z=300$ cm, behind the Ring Imaging Cherenkov (RICH) detector. The RICH material parameters were provided by the PID detector working group~\cite{cisbani}.}
    \label{fig:GEM_RICH_schematic}
\end{figure}

In closing, we have discussed several of the considerations for an instrument-performance driven integration of barrel tracking and vertexing layers with backward and forward disk arrays into an all-silicon tracking concept based on MAPS technology~\cite{eRD25_2}.
This all-silicon concept offers similar or better momentum and angular performance than the hybrid TPC-silicon concept  of BeAST~\cite{Aschenauer:2014cki} with identical vertexing performance.  It is radially more compact, $r = 43.23 \,\mathrm{cm}$ versus $r = 80.0\,\mathrm{cm}$, thereby freeing $36.77\,\mathrm{cm}$ that could be used for alternate purposes such as PID and offering opportunities for complementary baseline EIC general purpose central detector concepts.
\FloatBarrier
\subsubsection{Alternative Forward and Backward Tracking Options}
\label{LANL-concept}

As seen in Fig.~\ref{fig:all_si:resolutions} (left), the momentum resolution is overall constant as a function of pseudorapidity up to $\eta\sim2$, and then rapidly worsens.
We studied the possibility of improving the momentum resolution at forward and backward pseudorapidities by complementing the all-silicon tracker with auxiliary tracking stations, including
Gas Electron Multiplier (GEM) detectors with 50-\um\ resolution in the radial and azimuthal directions and additional 10-\um-pixel silicon disks in the available space away from the interaction point~\cite{ayk}. The available space for such additional detectors is different in the forward and backward directions, as shown in Fig.~\ref{fig:GEM_RICH_schematic}. In the electron-going (backward) direction, a complementary tracking station can be installed at $z\sim -180$ cm, and no significant amount of material is projected to be placed between said detector and the all-silicon tracker. In the hadron-going (forward) direction, the additional station can be installed at $z\sim300$ cm, behind the RICH detector.

\begin{figure}[htb]
    \centering
    \includegraphics[width=0.95\textwidth,page=1]{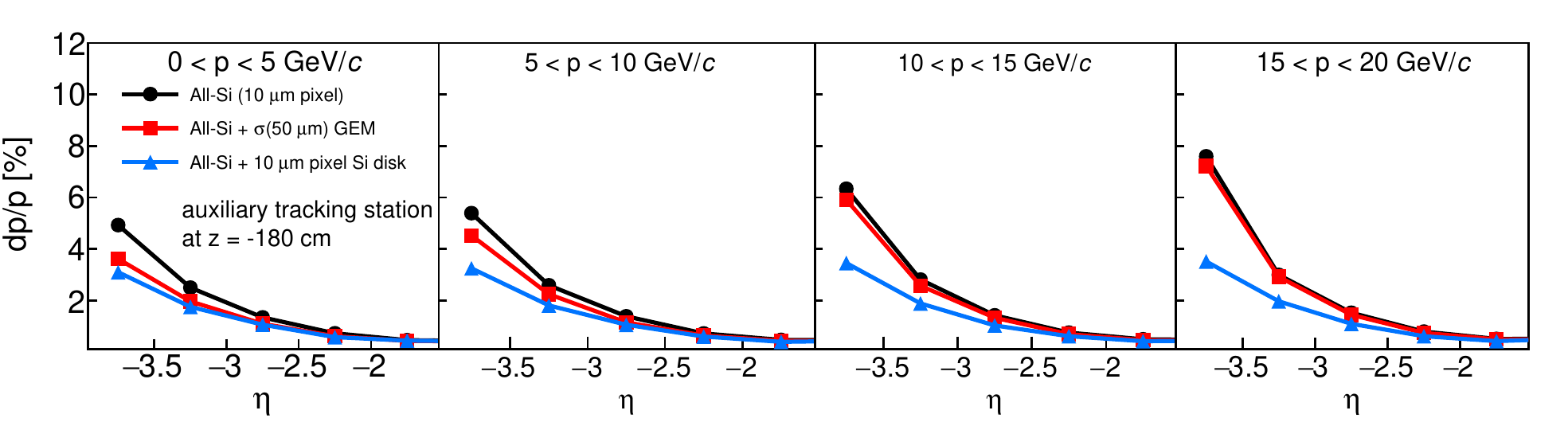}
    \caption{Momentum resolution as a function of pseudorapidity demonstrating the effect of complementing the all-silicon tracker in the electron-going (backward) direction. Each panel corresponds to a different momentum bin, from 0 to 20 GeV/$c$. The black circles correspond to the standalone all-silicon tracker (for a 10 \um\ $\times$ 10 \um\ pixel size). The red squares and blue triangles correspond to the all-silicon tracker complemented with a 50-\um-resolution GEM detector and a 10-\um-pixel silicon disk, respectively.}
    \label{fig:backward_GEM}
\end{figure}

\begin{figure}[htb]
    \centering
    \includegraphics[width=0.95\textwidth,page=3]{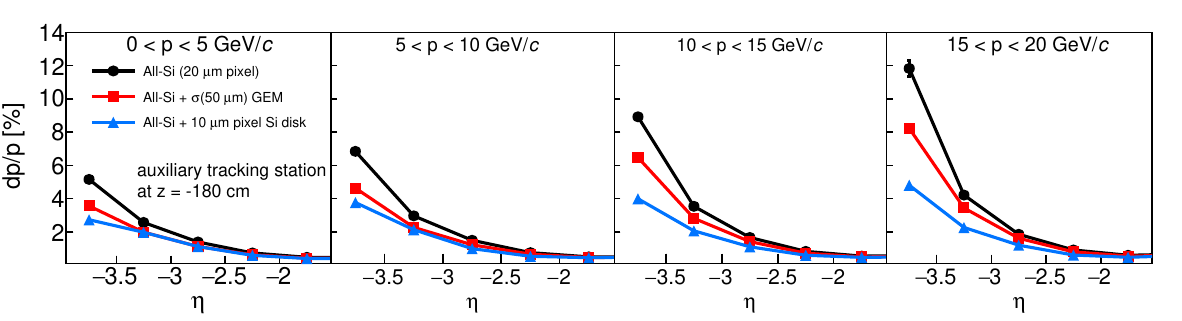}
    \caption{Same as Fig.~\ref{fig:backward_GEM}, but for a 20 \um\ $\times$ 20 \um\ all-silicon-tracker pixel size.}
    \label{fig:backward_GEM_20}
\end{figure}

The effect of complementing the all-silicon tracker in the electron-going direction is shown in Figs.~\ref{fig:backward_GEM}~and~\ref{fig:backward_GEM_20} for a 10 \um\ and 20 \um\ all-silicon-tracker pixel sizes, respectively. In the backward region, where the available space is closer to the all-silicon tracker, 
an auxiliary 10-$\mu$m-pixel detector provides a significantly better momentum resolution, mainly in the higher momentum region.

\begin{figure}[htb]
    \centering
    \includegraphics[width=0.95\textwidth,page=2]{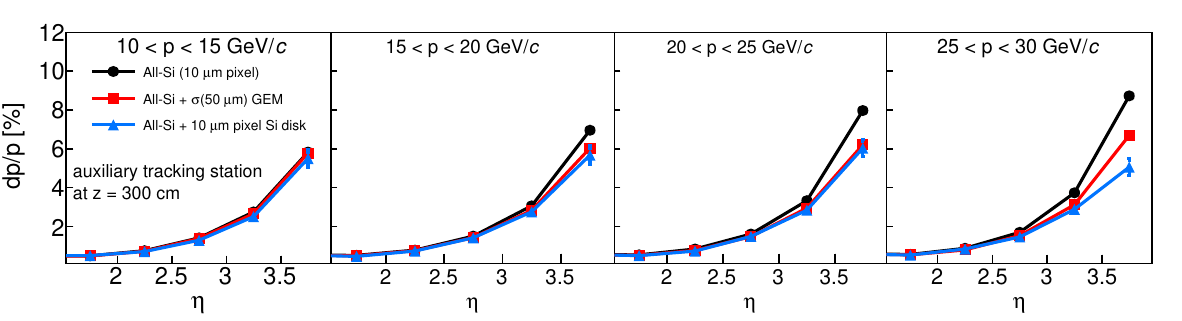}
    \caption{Momentum resolution as a function of pseudorapidity demonstrating the effect of complementing the all-silicon tracker in the hadron-going (forward) direction. Each panel corresponds to a different momentum bin, from 10 to 30 GeV/$c$. The black circles correspond to the standalone all-silicon tracker (for a 10 \um\ $\times$ 10 \um\ pixel size). The red squares and blue triangles correspond to the all-silicon tracker complemented with a 50-\um-resolution GEM detector and a 10-\um-pixel silicon disk, respectively.}
    \label{fig:forward_GEM}
\end{figure}

\begin{figure}[htb]
    \centering
    \includegraphics[width=0.95\textwidth,page=4]{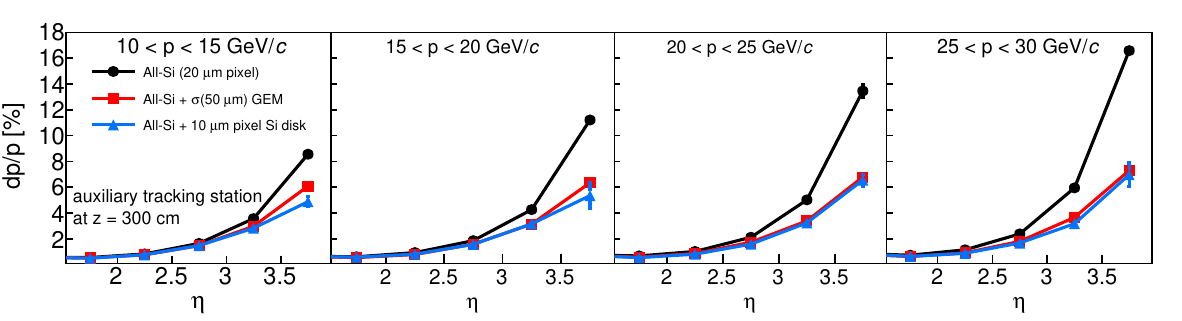}
    \caption{Same as Fig.~\ref{fig:forward_GEM}, but for a 20 \um\ $\times$ 20 \um\ all-silicon-tracker pixel size.}
    \label{fig:forward_GEM_20}
\end{figure}

Results in the forward region and with a 3 T magnetic field are shown in Figs.~\ref{fig:forward_GEM}~and~\ref{fig:forward_GEM_20} for a 10 \um\ and 20 \um\ all-silicon-tracker pixel sizes, respectively. The auxiliary station is placed behind the RICH detector. The RICH material parameters were provided by the PID detector working group~\cite{cisbani}. Since in the forward region the available space is farther away from the all-silicon tracker, the path traversed by a charged particle through the magnetic field in the tracking region ($\int{\bf B}\cdot d {\bf l}$) is larger.
As a result, the resolution is less sensitive to the complementary detector resolution, and while the silicon disk provides the best performance, the GEM detectors considered provide a comparable enhancement to the momentum resolution.
The effect of these auxiliary tracking stations depends on the EIC magnetic-field details. In these simulations, solenoidal fields were used. Likely, the magnetic field lines will be shaped to minimize bending inside the RICH detector, which will lower the $\int{\bf B}\cdot d {\bf l}$.
While the auxiliary tracking stations in these simulations cover pseudorapidities $|\eta|>1.2$, they have a larger impact at higher pseudorapidities ($|\eta|\gtrsim 2.5$). Consequently, smaller tracking stations can be used to complement the all-silicon tracker.

\paragraph{Forward Silicon Tracker (FST)}

Another forward silicon-tracker configuration (hereon referred to as FST), designed for heavy flavor and jet measurements in the EIC~\cite{Li:2020sru,Wong:2020xtc}, is presented here. The proposed FST covers pseudorapidities between $1$--$3.5$ and momenta up to $30$~GeV. An integrated detector design with the use of both FST and GEM tracker, which could be a cost effective option, are also studied in detector simulation.

\paragraph{Detector Design} The FST, which is implemented in Fun4All simulation, consists of six planes of silicon sensor as shown in Figure~\ref{fig:LANL_FST_geom}. The FST detector design parameters are listed in Table~\ref{tab:LANL_FST_geom}. The FST is placed between $35$~cm and $300$~cm along the $z$~axis. The inner radius of each plane changes with the $z$~position to fit the ion beam pipe geometry. Studies of the detector performance with different pixel pitch and silicon thickness are documented in ref~\cite{Wong:2020xtc}. In the latest FST detector design, the first three planes (plane~0-2) use a pixel pitch of $20$~\um\ and a silicon thickness of $50$~\um\ that are close to the ALICE ITS-3 type sensor~\cite{Abelevetal:2014dna,its3det} while the last three planes (plane~3-5) apply MALTA sensor properties~\cite{Pernegger:2017trh,Berdalovic:2018dlt,Hiti:2019ujr}. With both sensor technologies, the FST can provide excellent spatial and timing resolutions.

\begin{minipage}{\textwidth}
  \begin{minipage}[c]{0.46\textwidth}
    \centering
    \includegraphics[width=\textwidth]{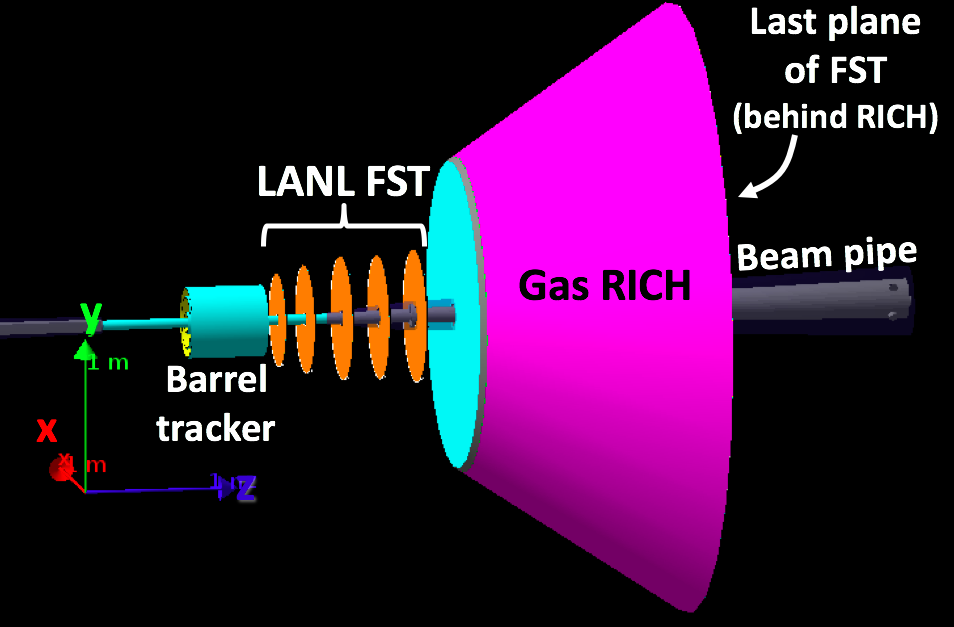}
    \captionof{figure}{\label{fig:LANL_FST_geom}FST setup in Fun4All simulation.}
  \end{minipage}
  \begin{minipage}[c]{0.49\textwidth}
    \centering
    \small
    \begin{tabular}{c c c c c c}
    \hline
    \multirow{2}{*}{Plane} & z & r\textsubscript{in} & r\textsubscript{out} &  pixel & silicon\\
    & (cm)& (cm)& (cm)& Pitch (\um) & thickness (\um)\\ 
    \hline
    0 & 35  & 4   & 25   & 20   & 50\\
    1 & 62.3  & 4.5 & 42   & 20   & 50\\
    2 & 90  & 5.2   & 43   & 20   & 50\\
    3 & 115 & 6   & 44 & 36.4   & 100\\
    4 & 125 & 6.5 & 45   & 36.4 & 100\\
    5 & 300 & 15  & 45   & 36.4 & 100\\
    \hline
    \end{tabular}
    \captionof{table}{\label{tab:LANL_FST_geom}FST geometry parameters}
    \end{minipage}
\end{minipage}

\begin{figure}[htb]
    \centering
    \includegraphics[width=0.4\textwidth]{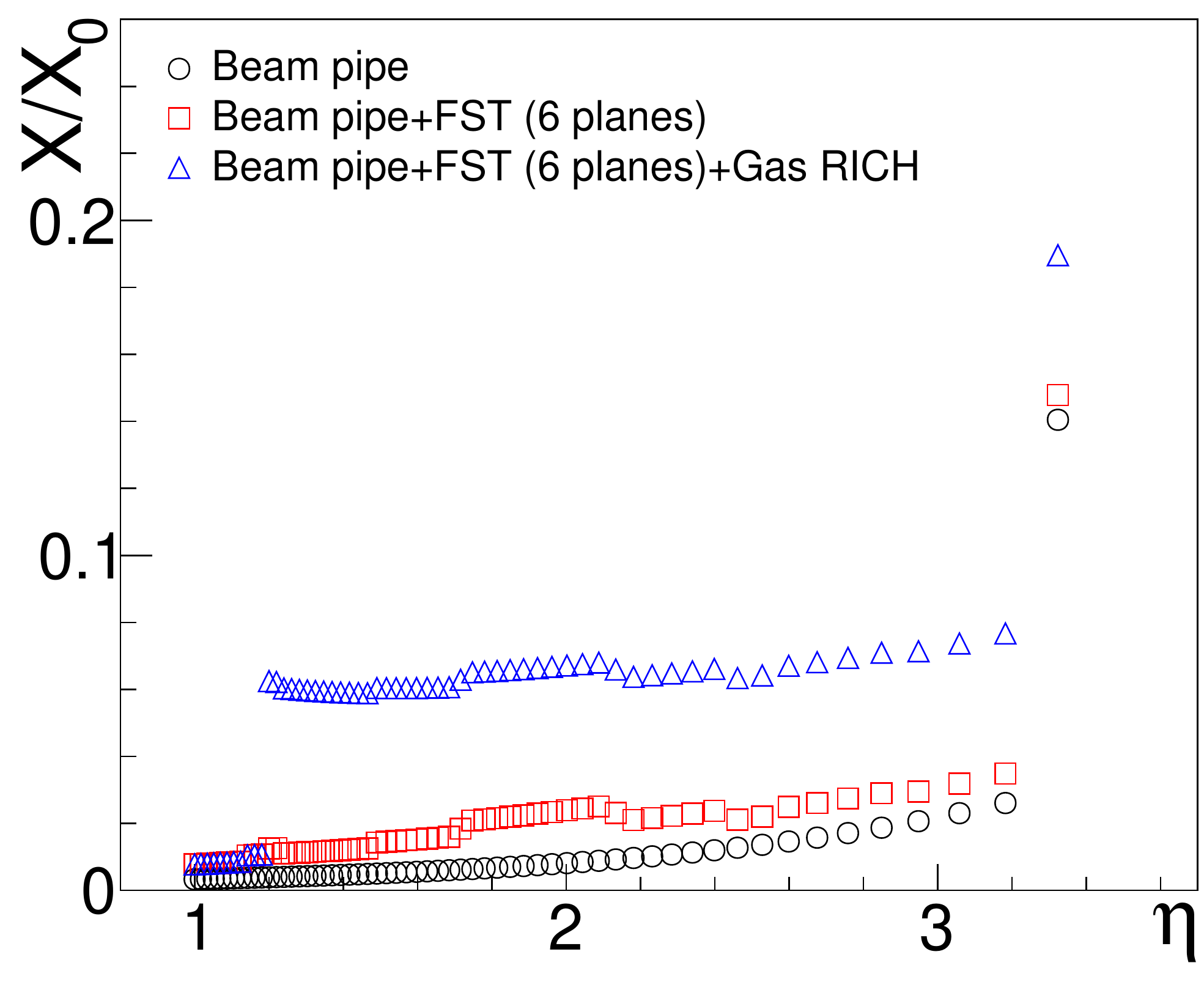}
    \includegraphics[width=0.4\textwidth]{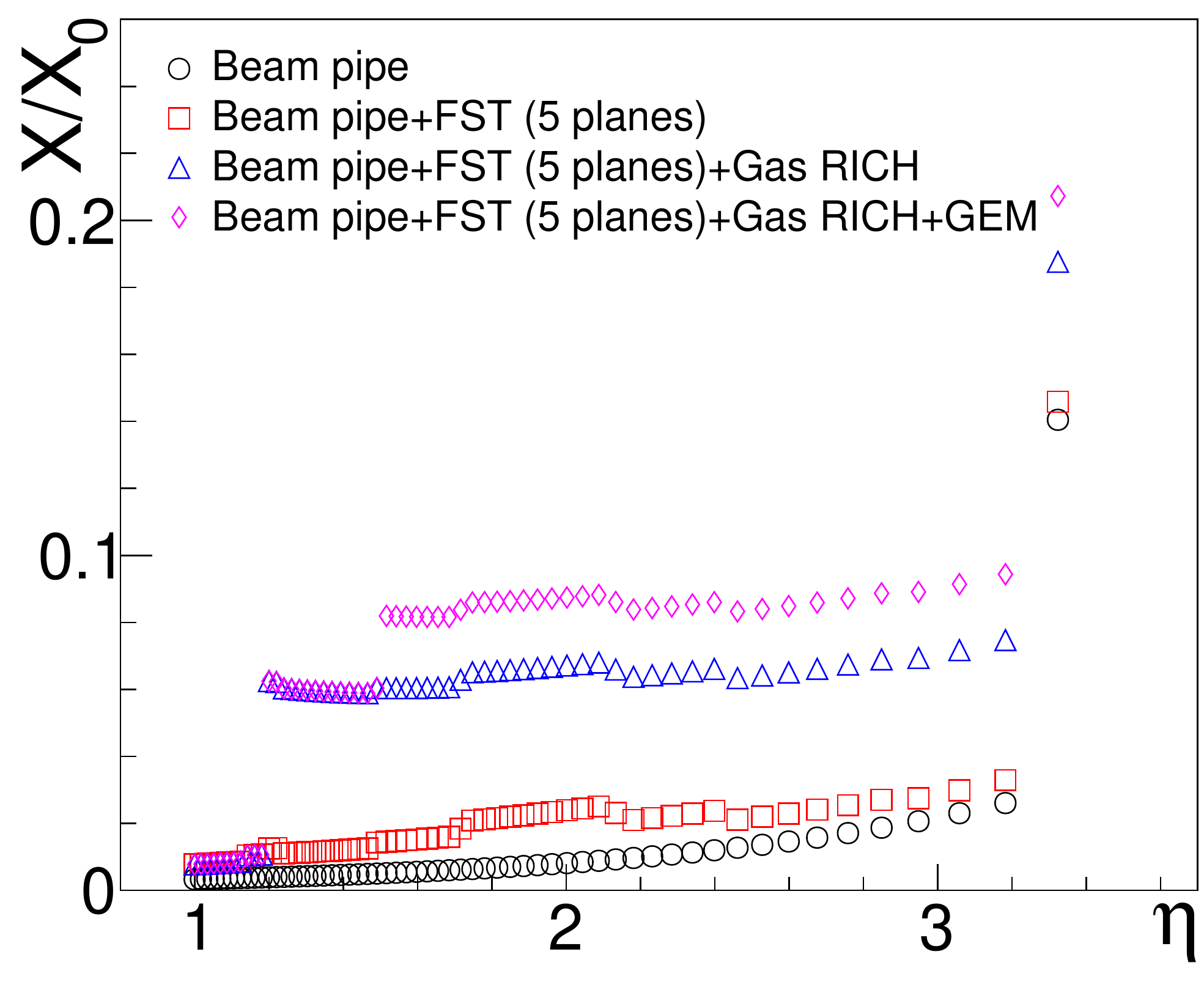}
    \vspace{-4mm}
    \caption{\label{fig:LANL_FST_matscan}Material budgets of different integrated detector setups.}
\end{figure}

\paragraph{Detector Integration} Integrated detector setups are also implemented in the simulation. The first setup, which is shown in Fig.~\ref{fig:LANL_FST_geom}, includes an additional gas RICH with aerogel and C\textsubscript{2}F\textsubscript{6} gas as radiator. The second setup replaces the last plane (plane~5) of FST with a mockup GEM tracker. The GEM tracker, which consists of three planes filled with methane, covers $1.5<\eta<3.5$. The material budgets of the integrated setups are shown in Fig.~\ref{fig:LANL_FST_matscan}. The material budgets of the first and the second integrated setup are $<8$\%~\Xo\ and $<10$\%~\Xo\ at $\eta<3.3$, respectively.

\begin{figure}[htb]
    \centering
    \includegraphics[width=\textwidth]{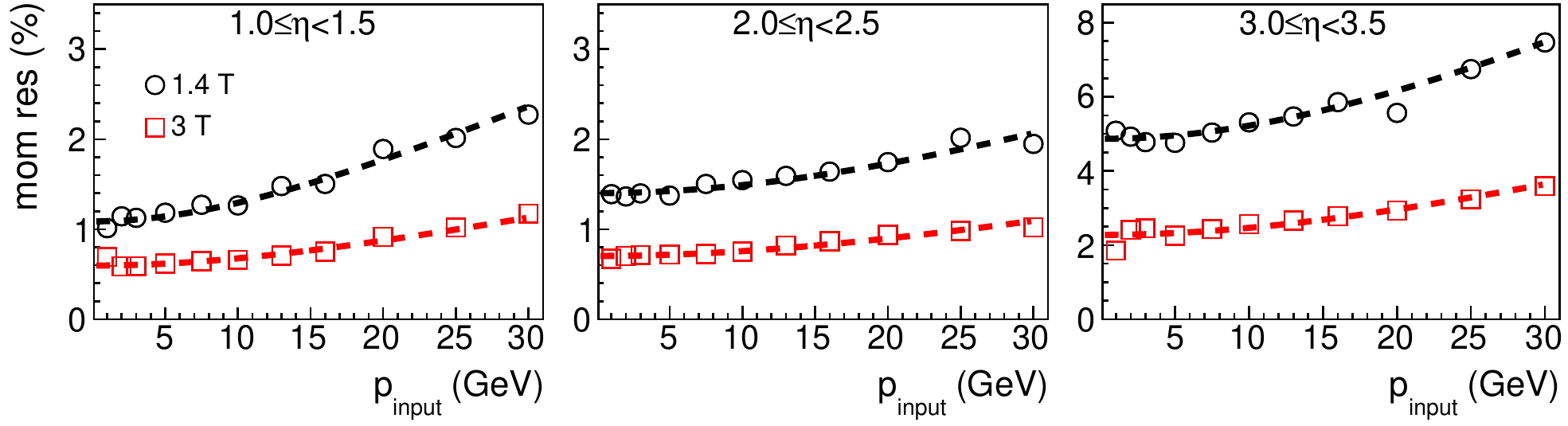}
    \vspace{-6mm}
    \caption{\label{fig:LANL_FST_momRes}Momentum resolutions as a function of input momentum of the integrated detector setup with the beam pipe, the barrel tracker, the five-plane FST, the gas RICH and the GEM tracker. The dash lines are the fits using Equation~\eqref{eq:dpp_v_p_param}. The fit results are shown in Table~\ref{table:LANL_FST_momRes}.}
\end{figure}

\begin{figure}[htb]
    \centering
    \includegraphics[width=\textwidth]{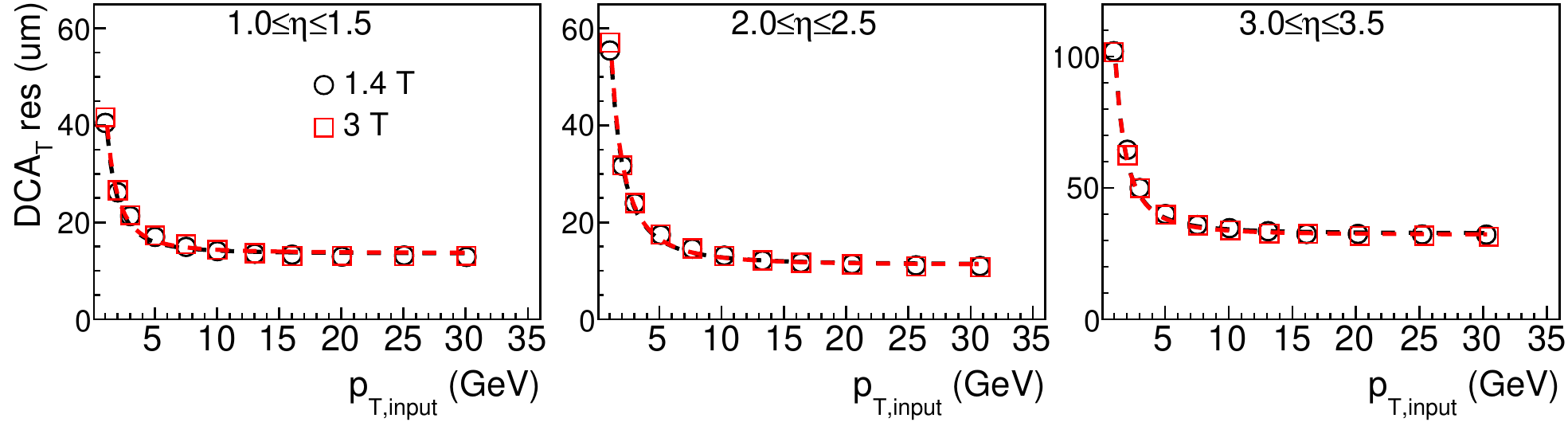}
    \vspace{-6mm}
    \caption{\label{fig:LANL_FST_DCA2D}$DCA_{T}$ resolutions as a function of input transverse momentum of the integrated detector setup with the beam pipe, the barrel tracker, the five-plane FST, the gas RICH and the GEM tracker. The dash lines are the fits using Equation~\eqref{eq:dca_v_pT_param}. The fit results are shown in Table~\ref{table:LANL_FST_DCA2DRes}.}
\end{figure}

\paragraph{Detector performance} The momentum resolutions of the integrated detector setup with the GEM are shown in Fig.~\ref{fig:LANL_FST_momRes}. The results are fitted using Equation~\eqref{eq:dpp_v_p_param}. The fit results of different detector setups are listed in Table~\ref{table:LANL_FST_momRes}. The momentum resolutions of the integrated detector setups with the $3$~T ($1.4$~T) magnet are $<10$\% ($18$\%) and $<4$\% ($8$\%) at $\eta<1$ and $\eta>1$, respectively. Comparing results of different detector setups as shown in Table~\ref{table:LANL_FST_momRes}, the additional gas RICH worsens the momentum resolutions by about $1$\% at $\eta>2.5$. Furthermore, Table~\ref{table:LANL_FST_momRes} shows that replacing the last plane of FST with a GEM tracker does not significantly change the momentum resolution. The $DCA_{T}$ resolutions of the integrated detector setups with the GEM tracker are shown in Fig.~\ref{fig:LANL_FST_DCA2D}. The $DCA_{T}$ resolutions are fitted using Equation~\eqref{eq:dca_v_pT_param}. The fit results of different detector setups with the use of the $3$~T magnetic field are listed in Table~\ref{table:LANL_FST_DCA2DRes}. The fit results of $DCA_T$ resolutions with the use of the $1.4$~T magnetic fields are not shown in Table~\ref{table:LANL_FST_DCA2DRes} as the $DCA_T$ resolutions show a weak dependence on the magnetic fields. Table~\ref{table:LANL_FST_DCA2DRes} shows that the $DCA_{T}$ resolutions are $<50$~\um\ and $<110$~\um\ at $\eta<2$ and $\eta>2$, respectively. Furthermore, Table~\ref{table:LANL_FST_DCA2DRes} shows that the replacement of the last plane of FST with the GEM tracker gives no significant differences in $DCA_{T}$ resolution. The results of momentum and $DCA_{T}$ resolutions, which show that replacing the last plane of FST by the GEM tracker does not give significant differences in detector performance, make the integrated detector setup with the GEM tracker an possible option considering the potential of lower cost of a GEM tracker compared to a silicon detector with a large pseudorapidity coverage needed for the RICH detector.

\begin{table}[ht]
\footnotesize
\begin{center}
\caption{\label{table:LANL_FST_momRes}Fit parameters of the momentum resolutions of different detector integration setups.}
\begin{tabular}{|c |c | c c | c c | c c| }
\hline
\multirow{2}{*}{$\eta$} & \multirow{2}{*}{B field} & \multicolumn{2}{c|}{FST (6 planes)} & \multicolumn{2}{c|}{FST (6 planes) + RICH} & \multicolumn{2}{c|}{FST (5 planes) + RICH + GEM}\\ \cline{3-8}
&  & A (\%/GeV) & B (\%) & A (\%/GeV) & B (\%) & A (\%/GeV) & B (\%)\\ \hline
\multirow{2}{*}{$1.0$--$1.5$} & $3$~T & 0.039 & 0.568 & 0.040 & 0.551 & 0.032 & 0.597\\
& $1.4$~T & 0.076 & 1.039 & 0.077 & 1.120 & 0.070 & 1.088\\
\hline
\multirow{2}{*}{$1.5$--$2.0$} & $3$~T & 0.019 & 0.454 & 0.018 & 0.448 & 0.013 & 0.445\\
& $1.4$~T & 0.039 & 0.839 & 0.039 & 0.882 & 0.026 & 0.876\\
\hline
\multirow{2}{*}{$2.0$--$2.5$} & $3$~T & 0.032 & 0.687 & 0.035 & 0.682 & 0.028 & 0.704\\
& $1.4$~T & 0.068 & 1.346 & 0.070 & 1.374 & 0.051 & 1.402\\
\hline
\multirow{2}{*}{$2.5$--$3.0$} & $3$~T & 0.037 & 1.190 & 0.062 & 1.306 & 0.062 & 1.336\\
& $1.4$~T & 0.086 & 2.362 & 0.127 & 2.607 & 0.123 & 2.629\\
\hline
\multirow{2}{*}{$3.0$--$3.5$} & $3$~T & 0.063 & 1.746 & 0.095 & 2.069 & 0.095 & 2.278\\
& $1.4$~T & 0.124 & 3.378 & 0.189 & 4.305 & 0.189 & 4.868\\
\hline
\end{tabular}
\end{center}
\end{table}

\begin{table}[ht]
\footnotesize
\begin{center}
\caption{\label{table:LANL_FST_DCA2DRes}Fit parameters of the $DCA_{T}$ resolutions of different detector setup with the use of the $3$~T magnetic field. }
\begin{tabular}{|c | c c| c c| c c|}
\hline
\multirow{2}{*}{$\eta$}  & \multicolumn{2}{c|}{FST (6 planes)} & \multicolumn{2}{c|}{FST (6 planes) + RICH} & \multicolumn{2}{c|}{FST (5 planes) + RICH + GEM}\\ \cline{2-7}
 & A (\um$\cdot$GeV) & B (\um) & A (\um$\cdot$GeV) & B (\um) & A (\um$\cdot$GeV) & B (\um)\\ \hline
$1.0$--$1.5$ & 41.54 & 14.19 & 39.47 & 14.39 & 40.73 & 14.06 \\
\hline
$1.5$--$2.0$ & 49.57 & 8.24 & 48.49 & 8.43 & 51.56 & 7.36 \\
\hline
$2.0$--$2.5$ & 57.87 & 13.73 & 54.79 & 14.16 & 59.58 & 11.48 \\
\hline
$2.5$--$3.0$ & 76.78 & 20.42 & 81.63 & 21.13 & 83.90 & 20.35 \\
\hline
$3.0$--$3.5$ & 77.79 & 29.71 & 95.90 & 30.01 & 104.95 & 31.55 \\
\hline
\end{tabular}
\end{center}
\end{table}


\subsubsection{Hybrid Tracking Option (Si-Vertex + Gaseous Detector Trackers)}\label{sec:hybrid}
\begin{figure}[ht]
	\centering
    \includegraphics[width=0.6\columnwidth]{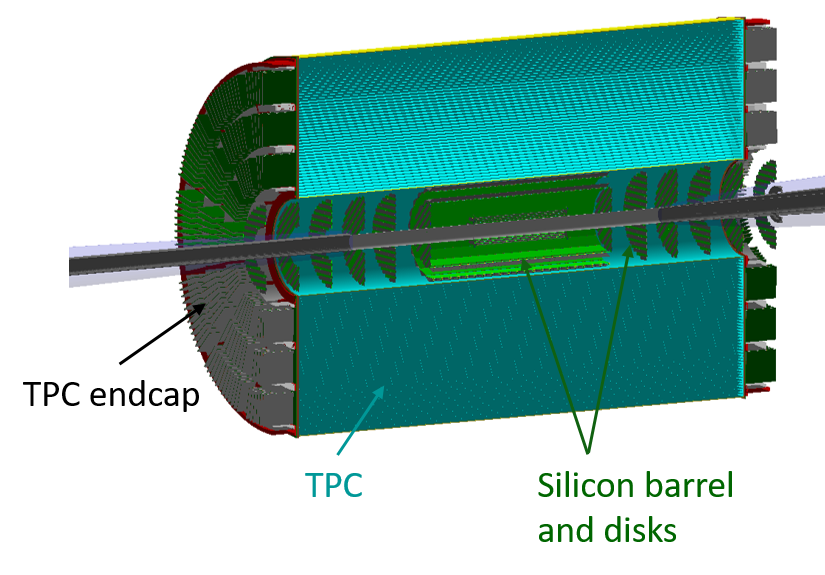}
	\caption{New hybrid baseline layout. The silicon layers and disks are shown in green, and the TPC in light blue.}
	\label{fig::newBaselineLayout_1}
\end{figure}

\paragraph{Barrel: Silicon Vertex + TPC}
\label{sec:hybrid-tpc} 
Figure~\ref{fig::newBaselineLayout_1} shows the simulated layout of this hybrid configuration. The silicon part is made of three layers close to the beampipe (vertexing layers) and two layers at larger radii (tracking layers) in the central region, and seven disks in the forward and backward regions. A TPC surrounds the central region and two TPC endcaps are placed after the silicon disks in both forward and backward regions.

The silicon detector parameters are based on the ALICE ITS3 technology. The vertexing layers have a material budget of 0.05\%~X/X$_0$ each, the tracking layers 0.55\%~X/X$_0$, and the disks each have a material budget of 0.24\%~X/X$_0$. The pixel size is $10 \times 10$~$\mu$m$^2$. 
The placements and parameters of barrel layers and disks are described in detail in Tables~\ref{tab::siPlusTPCbaseline_barrel} and~\ref{tab::siPlusTPCbaseline_disks}.
\begin{table}[ht]
\footnotesize
\centering
\subfloat[Barrel region][Barrel region]{
\begin{tabular}{|l|l|l|}
\hline
\textbf{Layer} & \textbf{Length} & \textbf{Radial position} \\ \hline
Layer 1 & 420 mm & 36.4 mm \\ \hline
Layer 2 & 420 mm & 44.5 mm \\ \hline
Layer 3 & 420 mm & 52.6 mm \\ \hline
Layer 4 & 840 mm & 133.8 mm \\ \hline
Layer 5 & 840 mm & 180.0 mm \\ \hline
TPC start & 2110 mm & 200.0 mm \\ \hline
TPC end & 2110 mm & 780.0 mm \\ \hline
\end{tabular}
\label{tab::siPlusTPCbaseline_barrel}
}
\subfloat[Disk region][Disk region]{
\begin{tabular}{|l|l|l|l|l|}
\hline
\textbf{Disk} & \textbf{$z$ position} & \textbf{Inner radius} & \textbf{Outer radius} \\ \hline
Disk 1 & 220 mm & 36.4 mm & 71.3 mm  \\ \hline
Disk 2 & 430 mm & 36.4 mm & 139.4 mm  \\ \hline
Disk 3 & 586 mm & 36.4 mm & 190.0 mm  \\ \hline
Disk 4 & 742 mm & 49.9 mm & 190.0 mm  \\ \hline
Disk 5 & 898 mm & 66.7 mm & 190.0 mm  \\ \hline
Disk 6 & 1054 mm & 83.5 mm & 190.0 mm  \\ \hline
Disk 7 & 1210 mm & 99.3 mm & 190.0 mm \\ \hline
\end{tabular}
\label{tab::siPlusTPCbaseline_disks}
}
\caption{Positions and lengths of detector parts in the barrel region and the disk region. In the disk region, the seven disks in the forward region are shown, but this layout is symmetric so it is the same with reversed sign on the $z$ position in the backward region.}
\end{table}
The table for the disks only shows the forward region, since this detector layout is symmetric in $z$. The radial positions for the barrel layers are based on the minimum distance between layers used in the ALICE ITS2 system \cite{aliceTDR}. While it may be possible to put layers closer together, using these distances give a detector that is plausible to build with currently existing technologies and structure solutions.
Each detector layer is built up of overlapping staves, consisting of several chips along with material representing cables, cooling pipes, and simple support structures.

\paragraph{Momentum and pointing resolutions;}

Studies for the resolutions are made in the following parameter space:
\begin{itemize}
    \item Transverse momentum range: 0 to 30~\gevc
    \item Pseudorapidity: $-1.0 \leq \eta \leq 1.0$, $1.0 \leq \eta \leq 2.5$, $2.5 \leq \eta \leq 3.5$
    \item Magnetic field: 1.5 T and 3.0 T
\end{itemize}
Since this detector layout is symmetric, negative pseudorapidities will have the same resolutions as the positive ones. Positive pions are used, with 1,000,000 events in each pseudorapidity range. 

\begin{figure}[htb]
    \centering
    \includegraphics[width=0.7\textwidth]{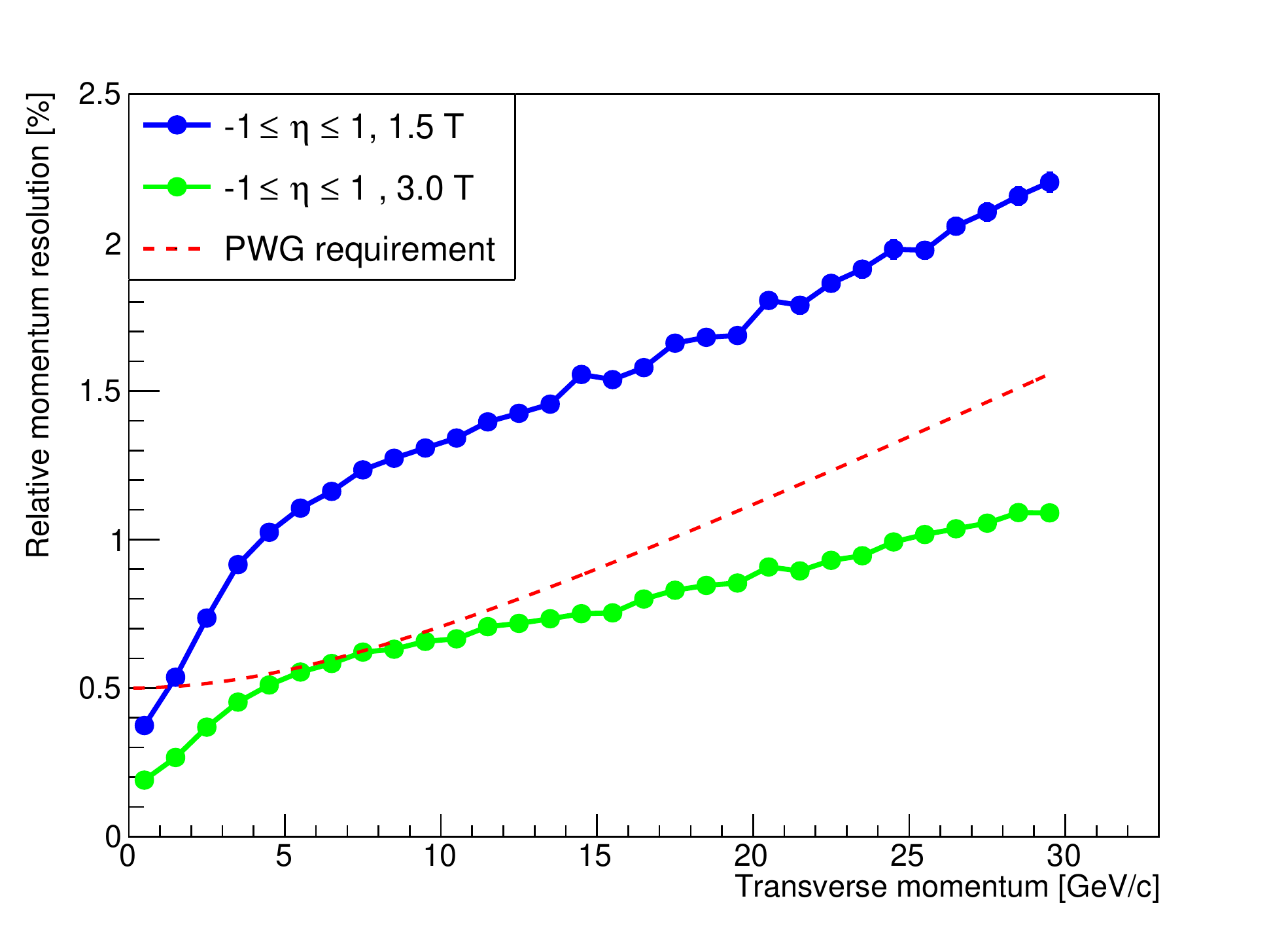}
    \caption{Relative transverse momentum resolution versus transverse momentum, for the baseline hybrid silicon plus TPC layout. The data are for the central region ($-1 \leq \eta \leq 1$). The blue curve shows the resolution for a 1.5 T field, and the green curve shows the resolution for a 3.0 T field. The red line shows the relative momentum resolution parametrisation requirement as given by the Physics Working Group (see Equation~\ref{eq::relMomResFit_transvPointResFit}).}
    \label{fig::hybridWithTPC_central_relMomRes_0to30GeV_withPWG_1}
\end{figure}

The formulae for resolution parametrisation are given in Equation~\ref{eq::relMomResFit_transvPointResFit}, where $A$ and $B$ indicate constants.
\begin{equation}
    \frac{\sigma_p}{p} = A \cdot p \oplus B = \sqrt{(A \cdot p)^2 + B^2}, \qquad \frac{\sigma_{xy}}{p_\text{T}} = \frac{A}{p_\text{T}} \oplus B = \sqrt{\left( \frac{A}{p_\text{T}} \right)^2 + B^2}
    \label{eq::relMomResFit_transvPointResFit}
\end{equation}
This parametrisation works well for the pointing resolution, but it has limitations for the relative transverse momentum resolution when using a gas TPC. In this case, as can be seen from Figure~\ref{fig::hybridWithTPC_central_relMomRes_0to30GeV_withPWG_1}, the parametrisation works well for $p_\text{T}$ between 0 and 4~\gevc, but the resolution value goes into a less steep linear increase after this point. The figure shows the relative transverse momentum resolution versus transverse momentum for both a 1.5~T field and a 3.0~T field, and the dashed line shown is the parametrisation provided by the Physics Working Group.
Fits to these data will be split up in momentum intervals to characterise the two clear regions (above and below 4~\gevc) separately. The pseudorapidity interval $1 \leq \eta \leq 2.5$ receives similar treatment. The final results from the relative transverse momentum fits, with parameters taken from Equation~\ref{eq::relMomResFit_transvPointResFit} can be seen in Table~\ref{tab::relMomResFitResults_transv_1} for a 1.5~T field and a 3.0~T field. The same can be seen for the full momentum resolution (i.e. $p$ rather than $p_\text{T}$) in Table~\ref{tab::relMomResFitResults_full_1}.

\begin{table}[ht]
\caption{Relative transverse momentum resolution fit parameters for a 1.5~T magnetic field and a 3.0~T magnetic field, using the fit presented in Equation~\ref{eq::relMomResFit_transvPointResFit}.}
\label{tab::relMomResFitResults_transv_1}
\centering
\begin{footnotesize}
\begin{tabular}{|l|l|l|l|}
\hline
\textbf{Interval} & \textbf{$p_\text{T}$ interval} & \textbf{Fit 1.5 T [\%]} & \textbf{Fit 3.0 T [\%]} \\ \hline
$-3.5 \leq \eta \leq -2.5$ & 0 to 30 GeV/$c$ 	& $A = 0.6 \pm 0.01$, $B = 4.2 \pm 0.03$   & $A = 0.3 \pm 0.01$, $B = 2.1 \pm 0.01$	\\ \hline
$-2.5 \leq \eta \leq -1.0$ & 0 to 4 GeV/$c$ 	& $A = 0.5 \pm 0.01$, $B = 0.9 \pm 0.01$ & $A = 0.2 \pm 0.01$, $B = 0.5 \pm 0.01$ 	\\
 & 4 to 30 GeV/$c$ 							& $A = 0.1 \pm 0.01$, $B = 2.2 \pm 0.01$	& $A = 0.06 \pm 0.001$, $B = 1.1 \pm 0.01$ 	\\ \hline
$-1.0 \leq \eta \leq 1.0$ & 0 to 4 GeV/$c$ 	& $A = 0.2 \pm 0.01$, $B = 0.4 \pm 0.01$ 	& $A = 0.1 \pm 0.01$, $B = 0.2 \pm 0.01$	\\
 & 4 to 30 GeV/$c$ 							& $A = 0.07 \pm 0.001$, $B = 1.1 \pm 0.01$ 	& $A = 0.03 \pm 0.001$, $B = 0.5 \pm 0.01$ 	\\ \hline
$1.0 \leq \eta \leq 2.5$ & 0 to 4 GeV/$c$ 	& $A = 0.5 \pm 0.01$, $B = 0.9 \pm 0.01$ 	& $A = 0.2 \pm 0.01$, $B = 0.5 \pm 0.01$ 	\\
 & 4 to 30 GeV/$c$ 							& $A = 0.1 \pm 0.01$, $B = 2.2 \pm 0.01$ 	& $A = 0.06 \pm 0.001$, $B = 1.1 \pm 0.01$ 	\\ \hline
$2.5 \leq \eta \leq 3.5$ & 0 to 30 GeV/$c$ 	& $A = 0.6 \pm 0.01$, $B = 4.2 \pm 0.03$ 	& $A = 0.3 \pm 0.01$, $B = 2.1 \pm 0.01$	\\ \hline
\end{tabular}
\end{footnotesize}
\end{table}
\begin{table}[hb]
\scriptsize
\caption{Relative momentum resolution fit parameters for a 1.5~T magnetic field and a 3.0~T magnetic field, using the fit presented in Equation \ref{eq::relMomResFit_transvPointResFit}.}
\label{tab::relMomResFitResults_full_1}
\begin{tabular}{|l|l|l|l|}
\hline
\textbf{Interval} & \textbf{Mom. interval} & \textbf{Fit 1.5 T [\%]} & \textbf{Fit 3.0 T [\%]} \\ \hline
$-3.5 \leq \eta \leq -2.5$ & 0 to 30 GeV/$c$ & $A = 0.09 \pm 0.005$, $B = 3.71 \pm 0.046$ & $A = 0.05 \pm 0.003$, $B = 1.90 \pm 0.024$  \\ \hline
$-2.5 \leq \eta \leq -1.0$ & 0 to 8 GeV/$c$ & $A = 0.24 \pm 0.003$, $B = 0.67 \pm 0.011$ & $A = 0.11 \pm 0.002$, $B = 0.33 \pm 0.006$ \\
 & 8 to 30 GeV/$c$ & $A = 0.07 \pm 0.001$, $B = 1.81 \pm 0.020$ & $A = 0.04 \pm 0.001$, $B = 0.88 \pm 0.010$\\ \hline
$-1.0 \leq \eta \leq 1.0$ & 0 to 5 GeV/$c$ & $A = 0.21 \pm 0.002$, $B = 0.34 \pm 0.004$ & $A = 0.11 \pm 0.001$, $B = 0.18 \pm 0.003$ \\  
 & 5 to 30 GeV/$c$ & $A = 0.06 \pm 0.0004$, $B = 1.09 \pm 0.007$ & $A = 0.03 \pm 0.0002$, $B = 0.54 \pm 0.003$ \\ \hline
$1.0 \leq \eta \leq 2.5$ & 0 to 8 GeV/$c$ & $A = 0.24 \pm 0.003$, $B = 0.67 \pm 0.011$ & $A = 0.11 \pm 0.002$, $B = 0.33 \pm 0.006$ \\  
 & 8 to 30 GeV/$c$ & $A = 0.07 \pm 0.001$, $B = 1.81 \pm 0.020$ & $A = 0.04 \pm 0.001$, $B = 0.88 \pm 0.010$ \\ \hline
$2.5 \leq \eta \leq 3.5$ & 0 to 30 GeV/$c$ & $A = 0.09 \pm 0.005$, $B = 3.71 \pm 0.046$ & $A = 0.05 \pm 0.003$, $B = 1.90 \pm 0.024$\\ \hline
\end{tabular}
\end{table}

Table~\ref{tab::transvPointFitResults_1} shows the fit values for the transverse pointing resolution data from simulations, using the silicon plus TPC hybrid baseline detector.
\begin{table}[ht!]
\caption{Transverse pointing resolution fit parameters, using the fit presented in Equation~\ref{eq::relMomResFit_transvPointResFit}.}
\label{tab::transvPointFitResults_1}
\centering
\begin{tabular}{|l|l|l|}
\hline
\textbf{Interval} & \textbf{Fit 1.5 T [$\mu$m]} & \textbf{Fit 3.0 T [$\mu$m]} \\ \hline
$-3.5 \leq \eta \leq -2.5$ & $A = 49.3 \pm 0.2$, $B = 9.64 \pm 0.02$ & $A = 48.5 \pm 0.2$, $B = 9.58 \pm 0.02$ \\ \hline
$-2.5 \leq \eta \leq -1.0$ & $A = 23.3 \pm 0.1$, $B = 3.32 \pm 0.01$ & $A = 23.1 \pm 0.1$, $B = 3.31 \pm 0.01$ \\ \hline
$-1.0 \leq \eta \leq 1.0$ & $A = 14.1 \pm 0.1$, $B = 2.11 \pm 0.01$ & $A = 13.7 \pm 0.1$, $B = 2.14 \pm 0.01$ \\ \hline
$1.0 \leq \eta \leq 2.5$ & $A = 23.3 \pm 0.1$, $B = 3.32 \pm 0.01$ & $A = 23.1 \pm 0.1$, $B = 3.31 \pm 0.01$ \\ \hline
$2.5 \leq \eta \leq 3.5$ & $A = 49.3 \pm 0.2$, $B = 9.64 \pm 0.02$ & $A = 48.5 \pm 0.2$, $B = 9.58 \pm 0.02$ \\ \hline
\end{tabular}
\end{table}
Table~\ref{tab::longPointFitResults_1} shows the same for the longitudinal pointing resolution.
\begin{table}[ht]
\small
\caption{Longitudinal pointing resolution fit parameters, using the fit presented in Equation~\ref{eq::relMomResFit_transvPointResFit}.}
\label{tab::longPointFitResults_1}
\centering
\begin{tabular}{|l|l|l|}
\hline
\textbf{Interval} & \textbf{Fit 1.5 T [$\mu$m]} & \textbf{Fit 3.0 T [$\mu$m]} \\ \hline
$-3.5 \leq \eta \leq -2.5$ & $A = 596.9 \pm 1.5$, $B = 41.05 \pm 0.12$ & $A = 596.5 \pm 1.5$, $B = 40.79 \pm 0.12$ \\ \hline
$-2.5 \leq \eta \leq -1.0$ & $A = 78.3 \pm 0.2$, $B = 3.11 \pm 0.02$ & $A = 78.1 \pm 0.2$, $B = 3.12 \pm 0.02$ \\ \hline
$-1.0 \leq \eta \leq 1.0$ & $A = 23.2 \pm 0.1$, $B = 2.64 \pm 0.01$ & $A = 22.9 \pm 0.1$, $B = 2.64 \pm 0.01$ \\ \hline
$1.0 \leq \eta \leq 2.5$ & $A = 78.3 \pm 0.2$, $B = 3.11 \pm 0.02$ & $A = 78.1 \pm 0.2$, $B = 3.12 \pm 0.02$ \\ \hline
$2.5 \leq \eta \leq 3.5$ & $A = 596.9 \pm 1.5$, $B = 41.05 \pm 0.12$ & $A = 596.5 \pm 1.5$, $B = 40.79 \pm 0.12$ \\ \hline
\end{tabular}
\end{table}

These results show that the requirements on pointing resolutions can be met with this layout and the ITS3-like technology, in all regions. The relative momentum resolution does not meet the requirements however, especially with a 1.5~T magnetic field. With a 3.0~T magnetic field the requirements are met apart from at $|\eta| \geq 2.5$.

\paragraph{Minimum-$p_\text{T}$ limit;}

The minimum reconstructable $p_\text{T}$ is investigated in the full pseudorapidity range, by sending out low-momentum (0 to 0.5 GeV/$c$ in  $p_\text{T}$) kaons and pions from the vertex, and seeing what fraction of total tracks in a region can be reconstructed, using a simple fast Kalman filter reconstruction algorithm. Improved reconstruction methods may fare better, but as an approximation of the highest limit of the minimum-$p_\text{T}$ that can be reconstructed, this method is deemed feasible. Table~\ref{tab::siPlusTPC_minPtvalues} contains minimum reconstructable $p_\text{T}$ values for different pseudorapidity regions. Results for pions and kaons are similar, and thus only one value is presented, representing the cutoff point where 90\% of events are reconstructed. This cutoff point is important to keep in mind; lower $p_\text{T}$ tracks can also be reconstructed up to a point, but less efficiently.

\begin{table}[ht]
\caption{Minimum reconstructable $p_\text{T}$, using simple Kalman filter reconstruction algorithm, for different pseudorapidity intervals. This study is done using the silicon plus TPC baseline layout.}
\label{tab::siPlusTPC_minPtvalues}
\centering
\begin{tabular}{|l|l|l|}
\hline
\textbf{$\eta$ interval} & \textbf{Min-$p_\text{T}$, 1.5 T} & \textbf{Min-$p_\text{T}$, 3.0 T} \\ \hline
$-3.0 \leq \eta \leq -2.5$ & 100 MeV/$c$ & 150 MeV/$c$ \\ \hline
$-2.5 \leq \eta \leq -2.0$ & 130 MeV/$c$ & 220 MeV/$c$ \\ \hline
$-2.0 \leq \eta \leq -1.5$ & 70 MeV/$c$ & 160 MeV/$c$ \\ \hline
$-1.5 \leq \eta \leq -1.0$ & 150 MeV/$c$ & 300 MeV/$c$ \\ \hline
$-1.0 \leq \eta \leq 1.0$ & 200 MeV/$c$ & 400 MeV/$c$ \\ \hline
$1.0 \leq \eta \leq 1.5$ & 150 MeV/$c$ & 300 MeV/$c$ \\ \hline
$1.5 \leq \eta \leq 2.0$ & 70 MeV/$c$ & 160 MeV/$c$ \\ \hline
$2.0 \leq \eta \leq 2.5$ & 130 MeV/$c$ & 220 MeV/$c$ \\ \hline
$2.5 \leq \eta \leq 3.0$ & 100 MeV/$c$ & 150 MeV/$c$ \\ \hline
\end{tabular}
\end{table}

\subsubsection{Barrel: Silicon Vertex + Cylindrical MPGDs}
\label{sec:hybrid-barrelmpgd} 

In the barrel, the silicon vertex tracker can be complemented by several layers of MPGDs, i.e. a barrel MPGD tracker (BMT).
Each cylindrical layer of the MPGD tracker consists of curved detector elements (tiles) of about 50~cm in width and long enough to cover the range $|\eta|<1$. Each detector element is considered to have a 2D readout and the spatial resolutions both in the z and the $r\cdot\varphi$ directions are assumed to be 150 $\mu$m. The detailed implementation in simulation of each tile is based of the technology developed for the CLAS12 barrel Micromegas tracker~\cite{Acker:2020qkv}: the material budget in the active area of each detector is about 0.3\%~$X/X_0$. The tiles in each layer are separated by a gap of about 2 cm of printed circuit board with a copper layer that mimic the routing of the readout lines back to the end caps. 

\begin{figure}[ht!]
  \centering
  \includegraphics[width=0.40\columnwidth]{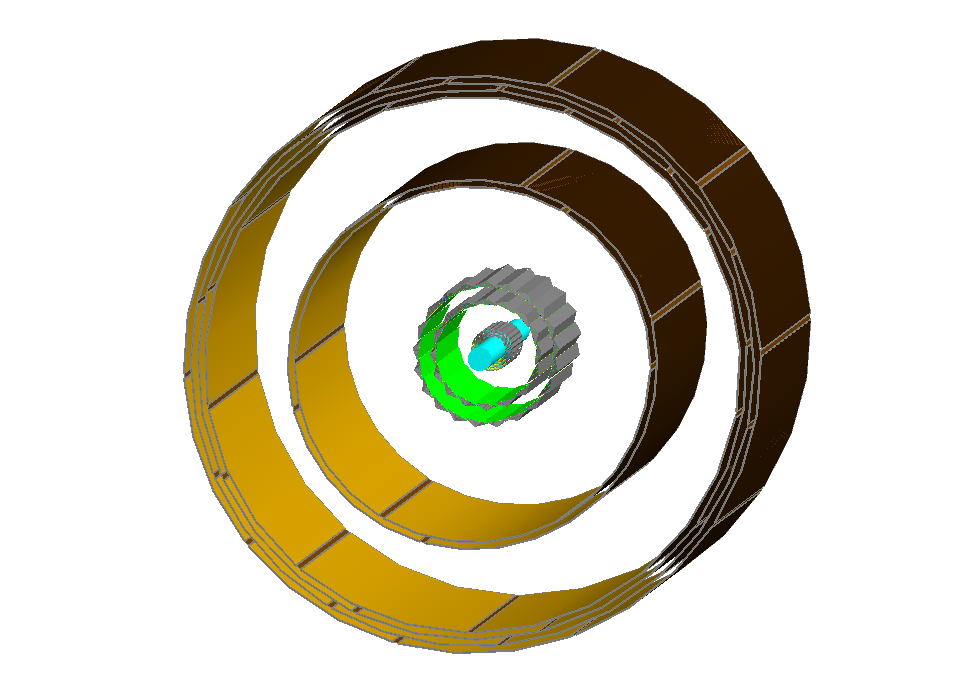}
  \includegraphics[width=0.28\columnwidth]{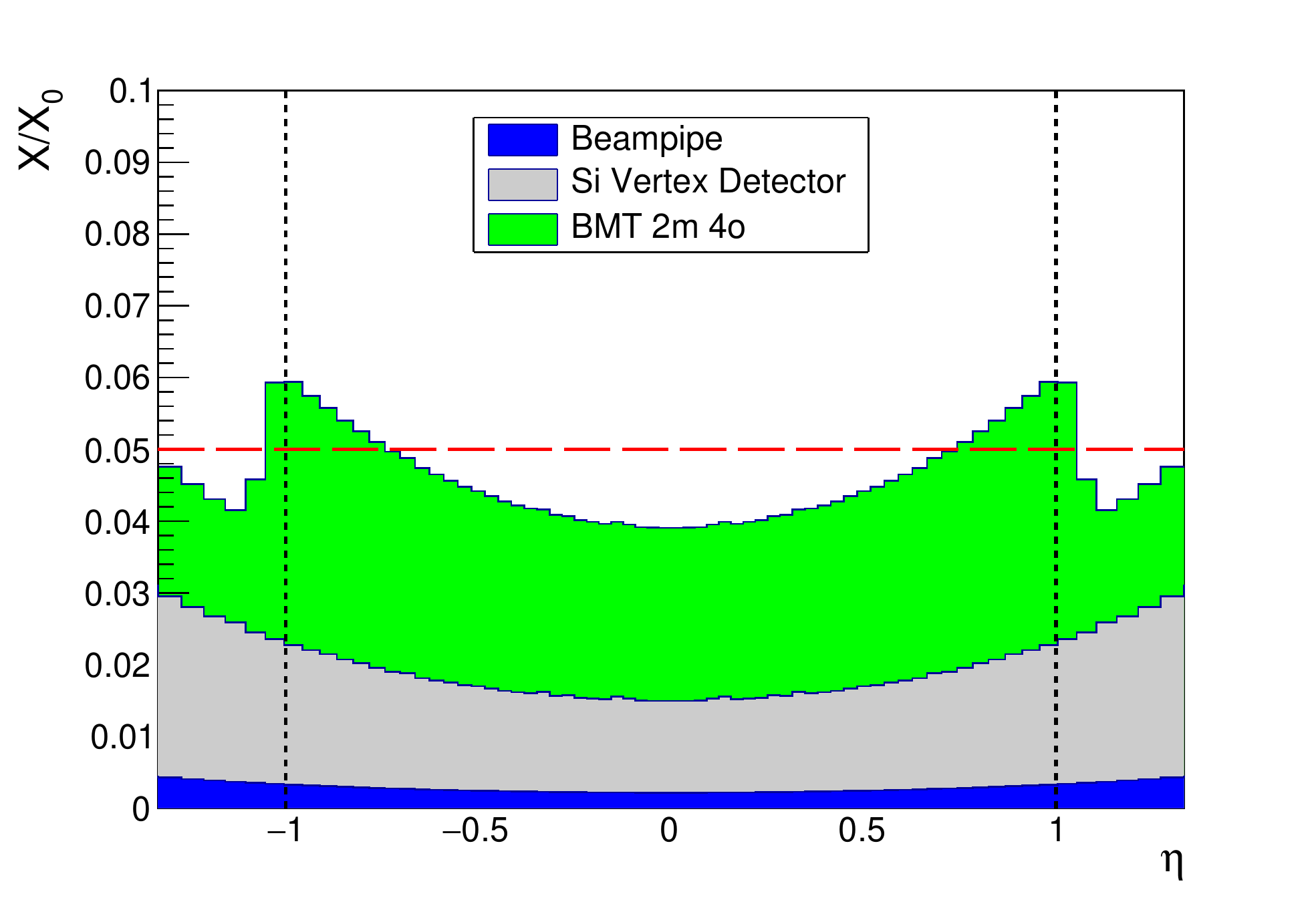}
  \includegraphics[width=0.28\columnwidth]{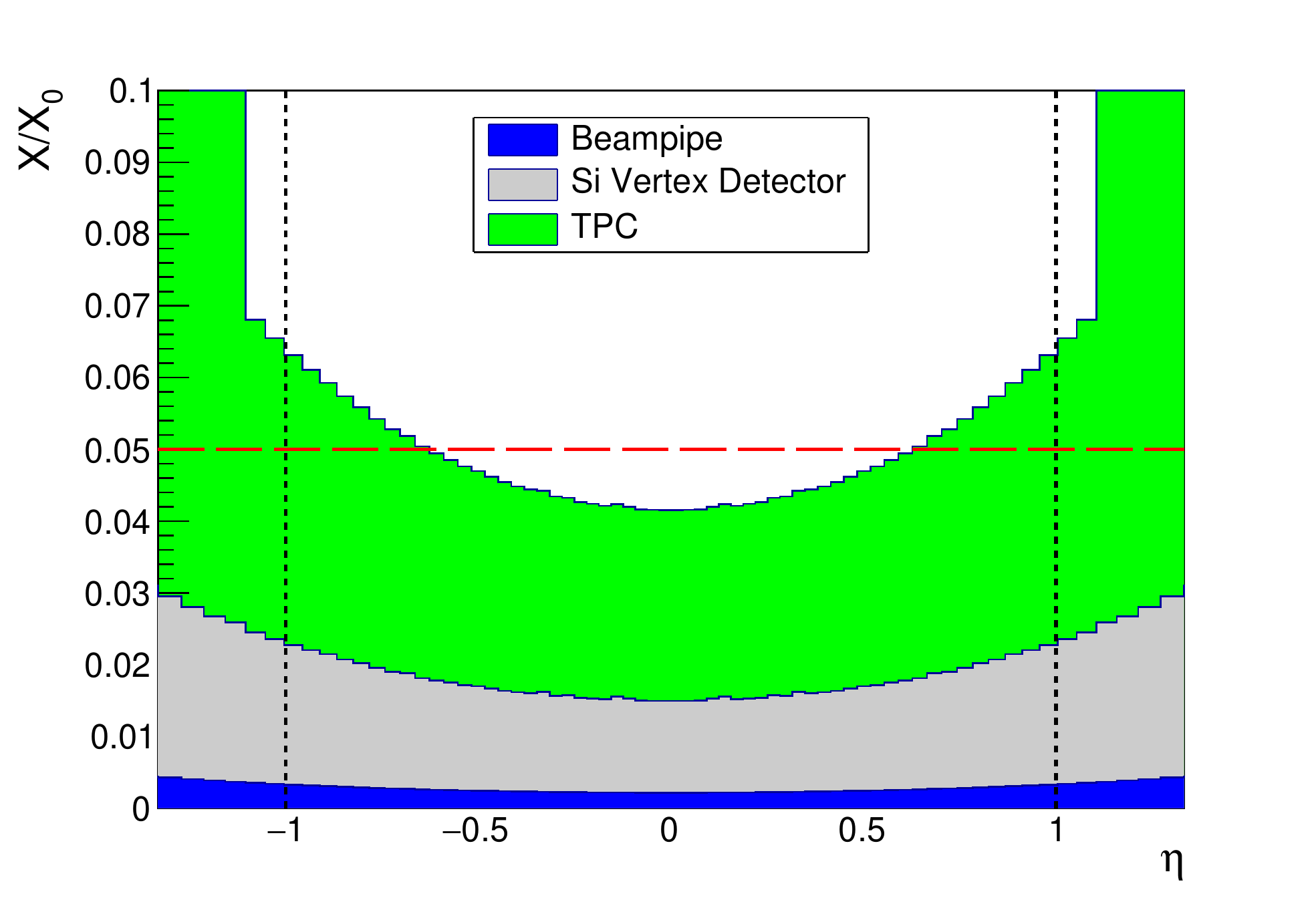}
  \caption{(left) A possible configuration of the cylindrical MPGD tracker with two pairs of layers at mid way between the SVT and the four outer detector layers. The material budget of the hybrid detector with MPGD layers (center) is comparable with the TPC solution (right). In the stack plots, the contribution of the beam pipe in blue, in gray the one of the silicon vertex detector and in green the MPGD tracker (or TPC) contribution.}
  \label{fig::hybrid_barrel_mpgd_geom}
\end{figure}

Figure~\ref{fig::hybrid_barrel_mpgd_geom} shows a possible configuration of the MPGD tracker with six layers: two layers are placed at a radial distance from the beam pipe of about 50 cm and four layers are placed at about 80 cm. 
Several configurations have been investigated: one configuration with six layers equally spaced at regular radial intervals, one with three pairs of layers (inner, middle and outer pairs) and a configuration with two layers in the middle and four layers in the outer part of the barrel. 
Table~\ref{tab::siPlusMPGD_barrel} shows the radial position of the layers for the last two configurations.

\begin{table}[htpb]
\caption{Radial position of MPGD tracker layers used in the hybrid detector simulation}
\label{tab::siPlusMPGD_barrel}
\centering
\begin{tabular}{|l|l|}
\hline
\textbf{Layer} & \textbf{Radial position} \\ \hline
0 inner  & 198 mm \\ \hline
1 inner  & 217 mm \\ \hline
2 middle & 477 mm \\ \hline
3 middle & 496 mm \\ \hline
4 outer  & 719 mm \\ \hline
5 outer  & 736 mm \\ \hline
6 outer  & 756 mm \\ \hline
7 outer  & 775 mm \\ \hline
\end{tabular}
\end{table}

\begin{figure}[ht!]
  \footnotesize
  \centering
  \includegraphics[width=0.4\columnwidth]{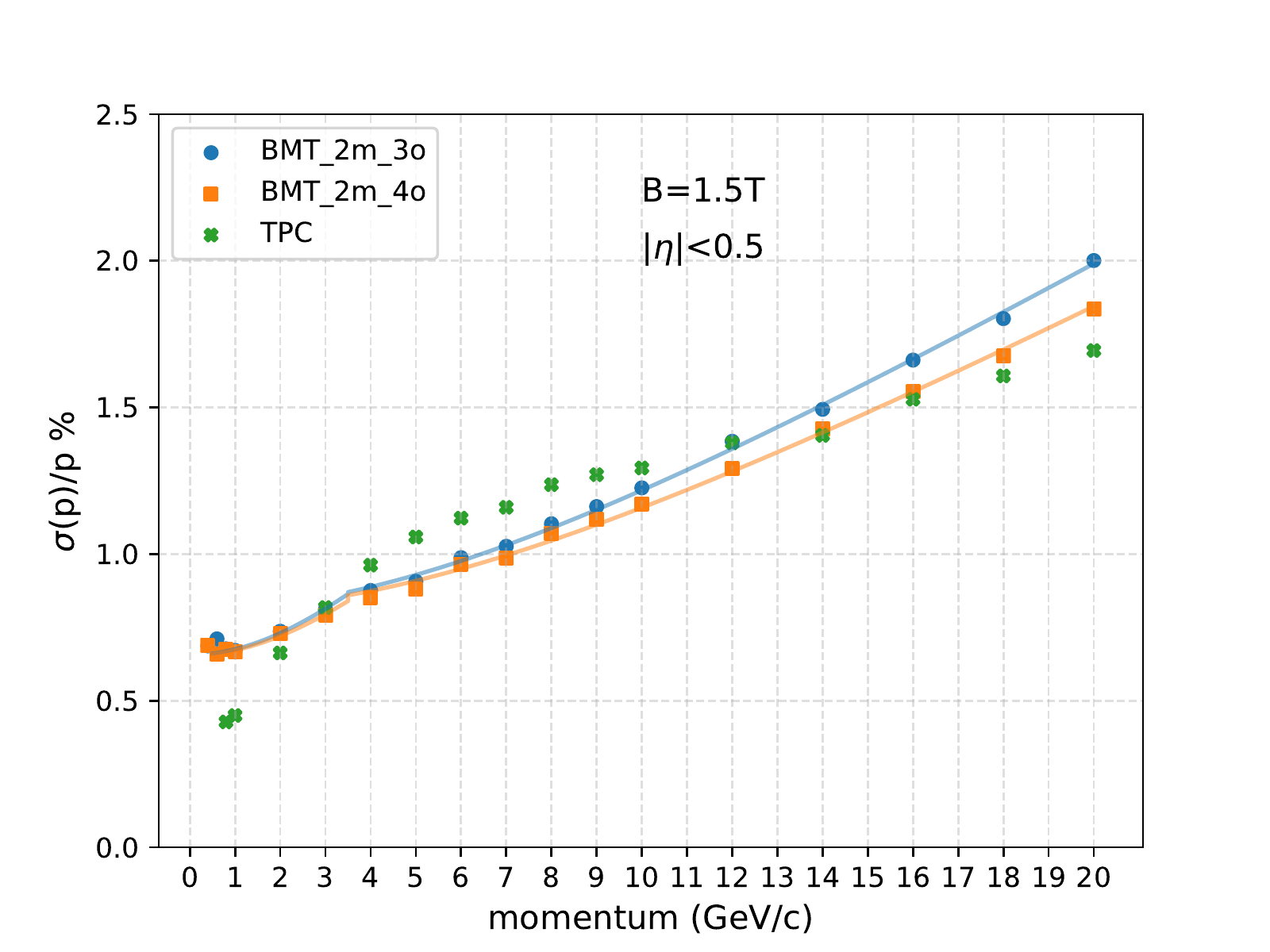}
  \includegraphics[width=0.4\columnwidth]{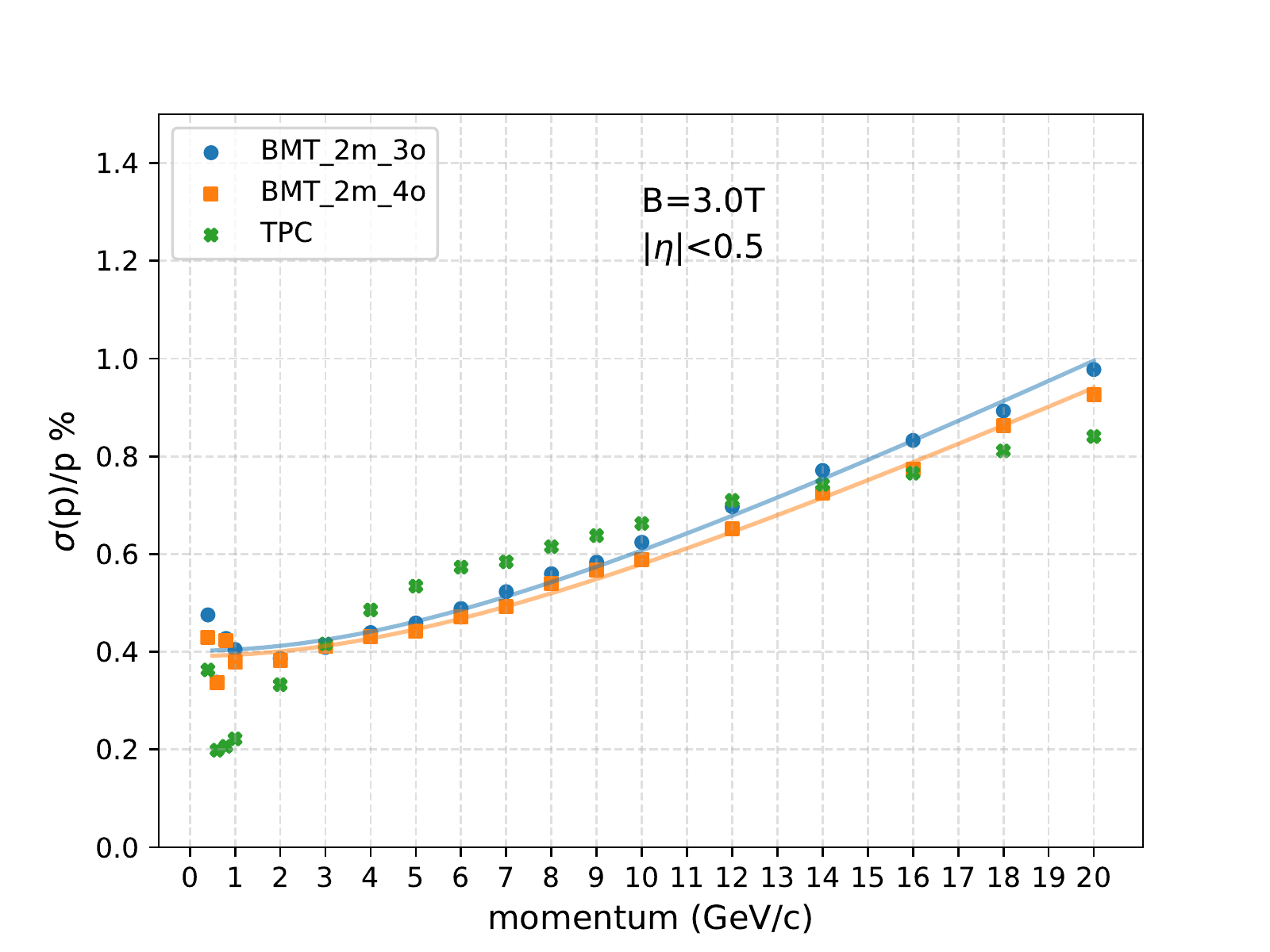}
  \caption{Relative transverse momentum resolutions of two hybrid detector configurations with MPGD cylindrical layers compared toe the configuration with the TPC: (left) with a magnetic field of 1.5~T and (right) 3~T. The configuration ``2m\_4o'' is depicted in Figure~\ref{fig::hybrid_barrel_mpgd_geom}. The ``2m\_3o'' only has three layers in the outer region. }
  \label{fig::hybrid_barrel_mpgd_resolutions}
\end{figure}

Studies of the relative momentum resolution have been performed by simulating five thousand $\pi^-$ per momentum bin in the range $|\eta|<0.5$ with a solenoid magnetic field of 1.5~T and 3~T. The beam pipe and the silicon vertex tracker are the same as in Section~\ref{sec:hybrid-tpc}.
The results for the configurations with two middle layers and three or four layers in the outer region are shown in Figure~\ref{fig::hybrid_barrel_mpgd_resolutions} together with resolutions obtained with the hybrid TPC detector of Section~\ref{sec:hybrid-tpc}.
The hybrid detectors studied (TPC and BMT) show similar relative momentum resolutions overall, with the hybrid TPC solution performing better at very low momenta and at higher momenta, while the hybrid BMT solution being better in the momentum range $3<p<12$~GeV/$c$. The results have been interpolated with the equation shown in Equation~\ref{eq::relMomResFit_transvPointResFit} and are reported in Table~\ref{tab::siPlusMPGD_barrel_p_res_fit}.
When compared to the PWG requirements, these results echo those presented in Section~\ref{sec:hybrid-tpc}, in particular the results at 3~T magnetic field exceed the PWG requirements.

\begin{table}[ht!]
\footnotesize
\caption{Results of the fitting of the relative momentum resolutions for two hybrid BMT configurations. }
\label{tab::siPlusMPGD_barrel_p_res_fit}
\centering
\begin{footnotesize}
\begin{tabular}{|l|cc|cc|cc|}
\hline
            & \multicolumn{4}{c|}{$B=1.5$ T}   & \multicolumn{2}{c|}{$B=3.0$ T} \\ \hline
            & \multicolumn{2}{c|}{$p<4$ GeV/$c$} & \multicolumn{2}{c|}{$p>4$ GeV/$c$}  &  \multicolumn{2}{c|}{$p>0.5$ GeV/$c$}  \\ \hline
	    &     A  (c/GeV) &     B      &     A  (c/GeV) &     B          &     A (c/GeV)   & B  \\ \hline
BMT\_2m\_3o &  0.161$\pm$0.007    &  0.656 $\pm$ 0.008 &  0.091$\pm$0.007    &  0.81 $\pm$ 0.01  & 0.0456 $\pm$ 0.0005   &  0.402 $\pm$ 0.007  \\ \hline
BMT\_2m\_4o &  0.15$\pm$0.01    &  0.65 $\pm$ 0.01 &  0.083$\pm$0.0008    &  0.81 $\pm$ 0.01 & 0.0427 $\pm$ 0.0004   &  0.391 $\pm$ 0.005  \\ \hline

\end{tabular}
\end{footnotesize}
\end{table}


\subsubsection{Hadron \& Electron End Cap: MPGDs}
\label{sec:hybrid-endcap} 
\begin{figure}[!ht]
\begin{center}
\includegraphics[width=0.95\columnwidth,trim={0pt 40mm 0pt 30mm},clip]{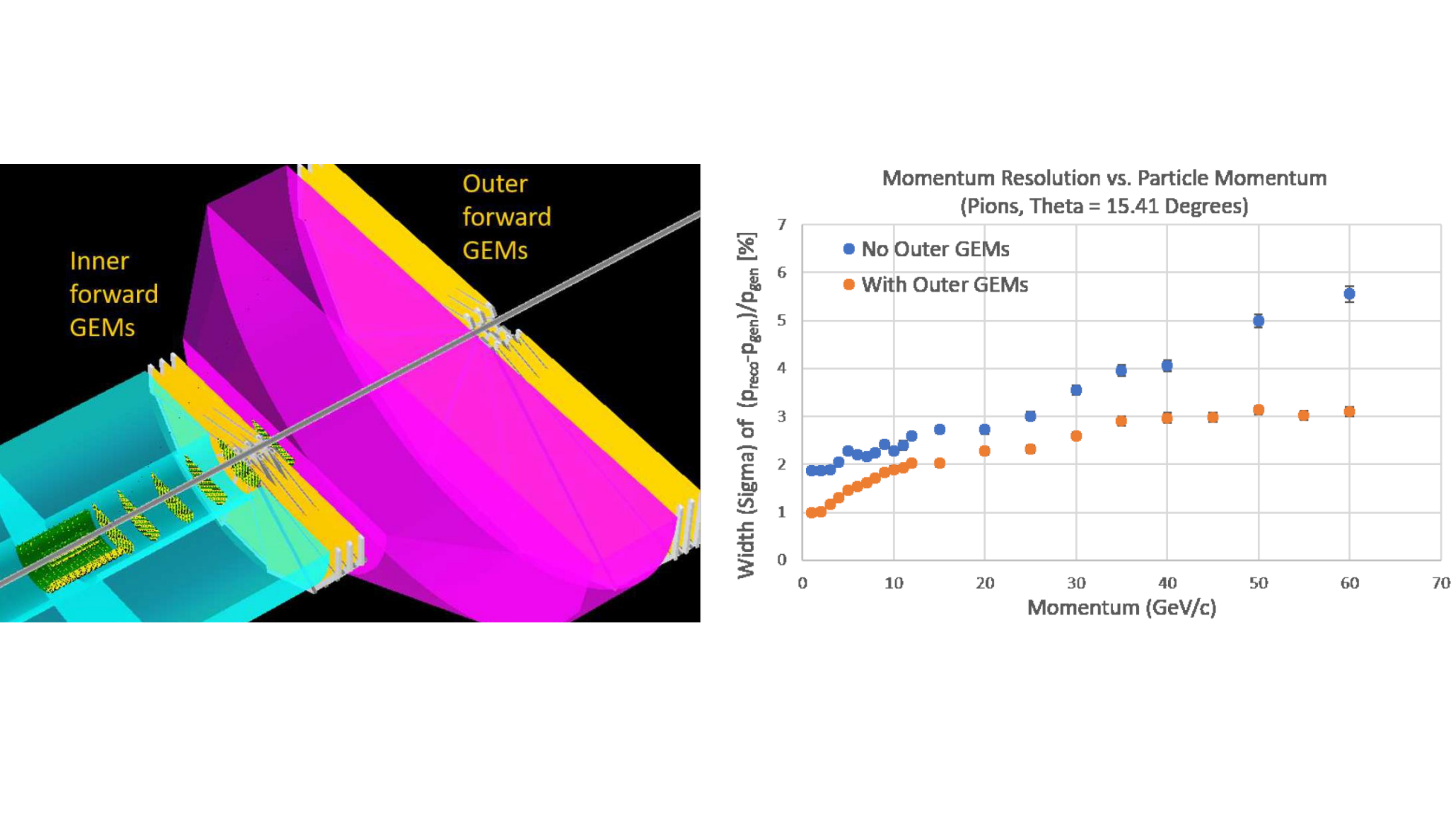}
\caption{\label{fig:geom}\textit{Left:} Simulated BeAST geometry with outer forward GEM detectors. \textit{Right:} Momentum resolution vs.\ momentum for pions at fixed scattering angle $\theta = 15.41^{\circ}$ ($\eta = 2.0$) with (orange) and without (blue) outer forward GEMs.}
\end{center}
\end{figure}

The tracking in the forward region of the hybrid configuration is composed of two large--area GEM stations, the inner forward GEMs and the outer forward GEMs, with each station made of three disks of triple-GEM detectors as shown in a standard BEAST detector geometry in Fig.~\ref{fig:geom} (left). Using the EicRoot simulation framework, we have studied the impact of inner and outer GEMs on the momentum resolution and the number of hits available for track fitting as a function of particle scattering angle and particle momentum. 
The simulated detector components include the beam pipe, the vertex and forward silicon trackers, the time projection chamber (TPC), the inner forward GEM station, the ring imaging Cerenkov (RICH) detector gas volume, and the outer forward GEM station behind the RICH. Specifically, the impact of the outer GEM detector on the tracking performance is studied by comparing the performance of the BeAST detector in the standard configuration with only the inner GEMs against the configuration including the outer GEMs while varying the particle parameters (scattering angle and momentum). Here, it is assumed that the detector would operate with a 1.5~T B-field. The scattering angle $\theta$ is varied from 5$^{\circ}$ to 75$^{\circ}$. The outer GEMs impact performance within their angular acceptance of $5^\circ < \theta < 35^{\circ}$ ( $3.1 > \eta > 1.15$). The dimensions of the outer GEMs in these simulations are chosen to closely match the acceptance of the inner GEMs. 

\begin{figure}[h]
\begin{center}
\includegraphics[width=0.95\columnwidth,trim={0pt 30mm 0pt 55mm},clip]{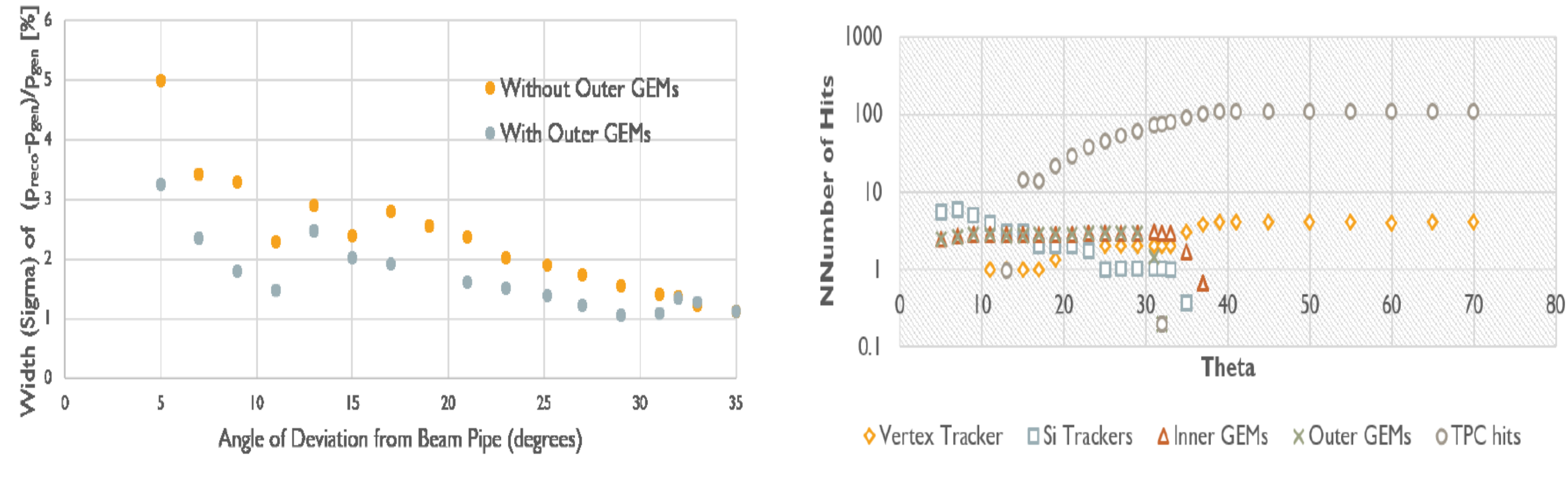}
\caption{\label{fig:numhits} \textit{Left}: Momentum resolution vs.\ scattering angle $\theta$ for 10 GeV/c pion tracks in a 1.5~T magnetic field from simulation of the standard BeAST detector with (gray) and without (orange) outer forward GEMs added. \textit{Right}: Average number of hits in each tracking subdetector (vertex, TPC, silicon, inner and outer GEMs) vs.\ scattering angle $\theta$ for this configuration.}		
\end{center}
\end{figure}

Fig.~\ref{fig:geom} (right) shows the momentum resolution as a function of momentum while keeping the scattering angle fixed at \mbox{\( \theta  \) = 15.41$^{\circ}$ (\( \eta  \) = 2.00).} It demonstrates that the significant improvement from outer GEMs holds over a large momentum range from 1 - 60 GeV/c. From the results shown in Fig.~\ref{fig:numhits} (left), it is clear that the outer GEMs significantly improve the momentum resolution, particularly for small scattering angles where the improvement reaches a factor of two. The particular structure of the graph is presumably due to the varying number of hits on the individual detectors. In order to verify this, we plot the average number of hits in each tracking subdetector as a function of $\theta$ in Fig.~\ref{fig:numhits} (right). Over the full $5^\circ < \theta < 35^{\circ}$ acceptance region of the outer forward GEM, both inner and outer GEM subdetectors provide a constant number of hits while the number of TPC hits drops rapidly below $\theta = 15^{\circ}$ and the number of vertex hits is down to one hit below $\theta = 18^{\circ}$ . In this angular range, the number of forward Si hits is comparable to the number of hits in each GEM subdetector. The design of the two GEM subdetector is very similar, so adding the outer forward GEM doubles the total number of GEM hits in this region. The forward Si detector, inner GEMs, and outer GEMs each contribute roughly a third to the total number of track hits in this region. This explains the significant impact of the outer forward GEM in the angular range below $\theta = 15^{\circ}$ ($\eta>2$).

\paragraph{Transverse Momentum Resolution Study}

\begin{figure}[ht!]
	\centering
    \includegraphics[width=0.6\columnwidth]{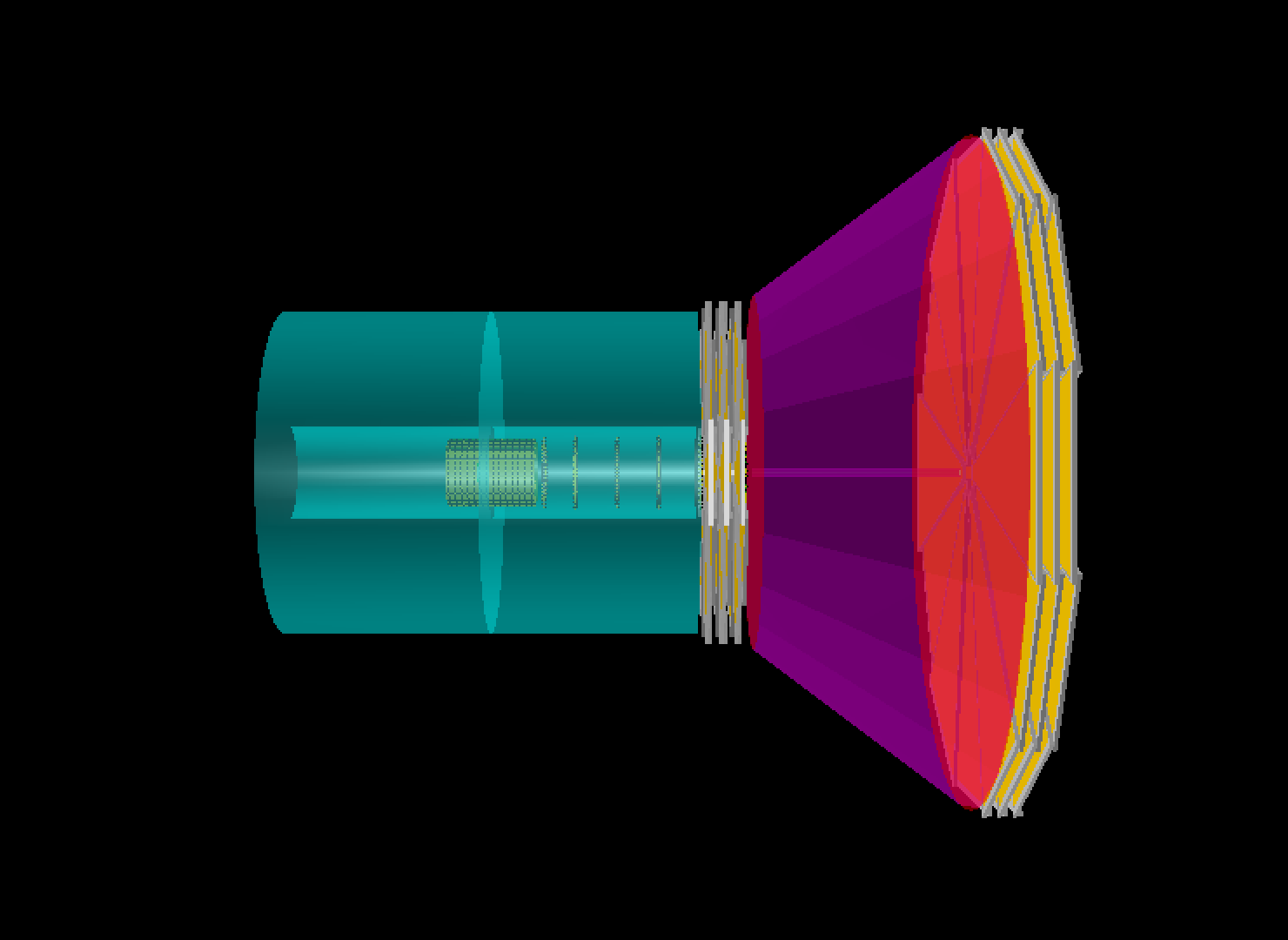}
	\caption{Hybrid barrel with forward GEM layers. TPC is shown in teal, the silicon layers within the TPC are shown in yellow, the forward GEM layers are shown in yellow, and the RICH is shown in purple.}
	\label{fig::BeASTModel_EICRoot}
\end{figure}

A study on the impact of the magnetic field strength on the transverse momentum resolution was performed using a model of the hybrid detector including TPC with a longitudinal hit point resolution given by Equation~\ref{eq::TPCResParam} where $D$ is the drift distance, $ A = 100\mu\mathrm{m}/\sqrt{\mathrm{cm}}$ and $B=500 \mu\mathrm{m}$, and a transverse hit point resolution also given by Equation~\ref{eq::TPCResParam} with $A = 15\mu\mathrm{m}/\sqrt{\mathrm{cm}} $ and $ B =200 \mu\mathrm{m}$, and vertical pad size of $0.5 \mathrm{cm}$; Silicon Vertex Tracker with hit point resolution of $5.8 \mu\mathrm{m} \times 5.8 \mu\mathrm{m}$; Forward Silicon Tracker with hit point resolution of $5.8 \mu\mathrm{m} \times 5.8 \mu\mathrm{m}$; forward GEM trackers with hit point resolution of $50 \mu\mathrm{m} \times 50 \mu\mathrm{m}$; far-forward GEM trackers with hit point resolution of $100 \mu\mathrm{m} \times 100 \mu\mathrm{m}$; material for the gas RICH between the inner forward GEM trackers and outer forward GEM trackers is also included. This model is shown in Figure~\ref{fig::BeASTModel_EICRoot}.

The placements and parameters of barrel layers and disks are described in detail in Tables~\ref{tab::hybridSiTPC_VST} and~\ref{tab::hybridSiTPC_FST}.

\begin{equation}
    \sigma = A \cdot \sqrt{D\mathrm{[cm]}} \oplus B = \sqrt{(A \cdot \sqrt{D\mathrm{[cm]}})^2 + B^2}
    \label{eq::TPCResParam}
\end{equation}

\begin{table}[ht]
\footnotesize
\centering
\caption{Positions and lengths of detector parts in the barrel region and the disk region.}
\subfloat[Barrel region][Barrel region]{
\begin{tabular}{|l|l|l|}
\hline
\textbf{Layer} & \textbf{Length} & \textbf{Radial position} \\ \hline
Layer 1 & 270 mm & 23.4 mm \\ \hline
Layer 2 & 270 mm & 46.8 mm \\ \hline
Layer 3 & 420 mm & 140.4 mm \\ \hline
Layer 4 & 420 mm & 157.2 mm \\ \hline
TPC start & 1960 mm & 200.0 mm \\ \hline
TPC end & 1960 mm & 800.0 mm \\ \hline
\end{tabular}
\label{tab::hybridSiTPC_VST}
}
\subfloat[Disk region][Disk region]{
\begin{tabular}{|l|l|l|l|l|}
\hline
\textbf{Disk} & \textbf{$z$ position} & \textbf{Inner radius} & \textbf{Outer radius} \\ \hline
Disk 1 & 250 mm & 18.0 mm & 185.0 mm  \\ \hline
Disk 2 & 400 mm & 18.0 mm & 185.0 mm  \\ \hline
Disk 3 & 600 mm & 18.0 mm & 185.0 mm  \\ \hline
Disk 4 & 800 mm & 18.0 mm & 185.0 mm  \\ \hline
Disk 5 & 1000 mm & 18.0 mm & 185.0 mm  \\ \hline
Disk 6 & 1210 mm & 18.0 mm & 185.0 mm  \\ \hline
\end{tabular}
\label{tab::hybridSiTPC_FST}
}
\end{table}

Studies for the resolutions are made in the following parameter space:
\begin{itemize}
    \item Transverse momentum range: 0 to 30 GeV/c
    \item Pseudorapidity: $\eta = 1.25, 1.5, 1.75, 2.0, 2.5,$ and $3.0$
    \item Magnetic field: 1.5~T and 3.0~T
\end{itemize}
In the simulation positive pions are used, with 1000 events for each transverse momentum value at each pseudorapidity value. 

The formula for resolution parameterization is given in Equation~\ref{eq::relPtResFit}, where $A$ and $B$ indicate constants.
\begin{equation}
    \frac{\sigma_{p_{T}}}{p_{T}} = A \cdot p_{T} \oplus B = \sqrt{(A \cdot p_{T})^2 + B^2}
    \label{eq::relPtResFit}
\end{equation}

\begin{figure}[ht]
\centering
\subfloat[$\eta = 1.25$][$\eta = 1.25$]{

    \includegraphics[width=0.48\textwidth]{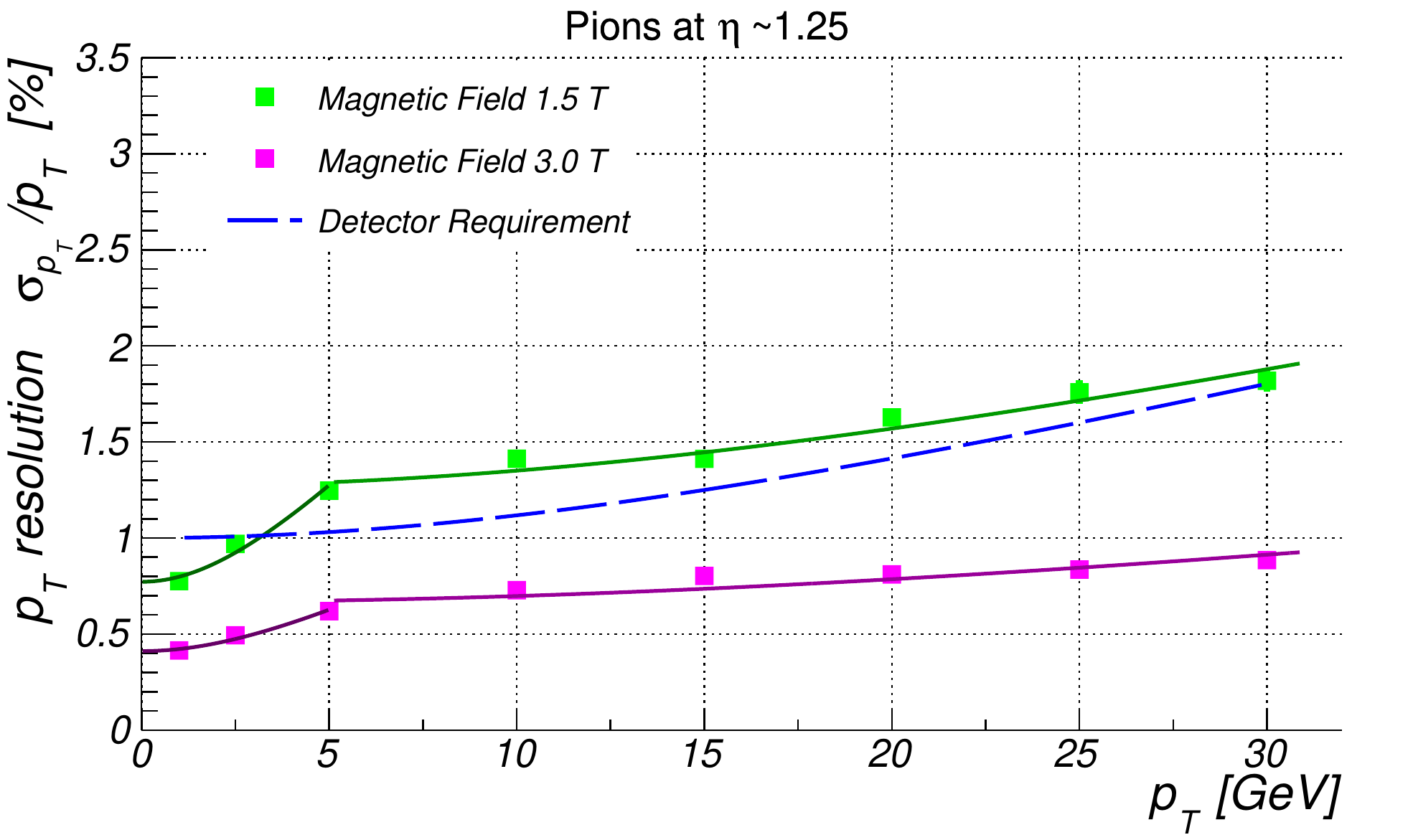}
    \label{fig::hybridWithTPC_ForwardGEM_PTRes1p25}

}
\subfloat[$\eta = 2.5$][$\eta = 2.5$]{

    \includegraphics[width=0.48\textwidth]{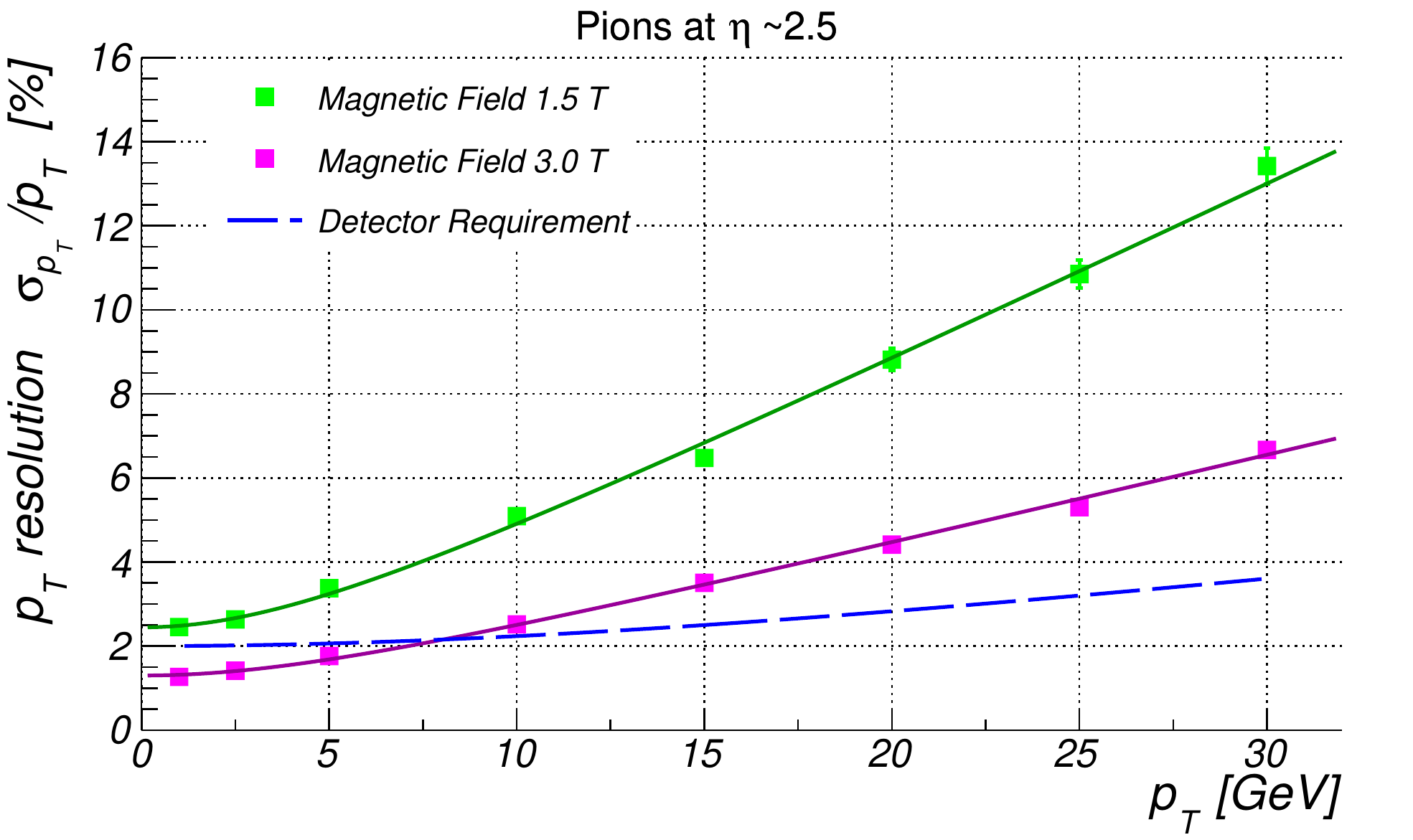}
    \label{fig::hybridWithTPC_ForwardGEM_PTRes2p5}

}
\caption{Relative transverse momentum resolution versus transverse momentum. The data are for the pions at $\eta = 1.25$ (left) and at $\eta = 2.5$ (right). The green points shows the resolution for a 1.5 T field, and the magenta points show the resolution for a 3.0 T field. The dashed blue line represents the physics momentum tracking requirement, which is calculated from values in Table~\ref{tab::relPtResFitResults} (assuming $p = p_T$), and Equation~\ref{eq::relPtResFit} for the eta range: $1.0 < |\eta| < 1.5$ (left) and $2.5 < |\eta| < 3.0$ (right).}
\end{figure}

As noted in the studies from Section~\ref{sec:hybrid-tpc}, this parameterization has limitations for the relative transverse momentum resolution when using a gas TPC. In this case, as can be seen from Figure~\ref{fig::hybridWithTPC_ForwardGEM_PTRes1p25}, the trend of the resolution values changes, going into a less steep linear increase after 5 GeV/c. So the parameterization is done in two regions. The figure shows the relative transverse momentum resolution versus transverse momentum for both a 1.5~T field in green and a 3.0~T field in magenta.

Fits to these data will be split up in momentum intervals to characterise the two clear regions (above and below 5 GeV/c) separately. The evaluation of resolutions at pseudorapidity values $\eta \leq 2.0$ are treated this way. The final results from the relative transverse momentum fits, with parameters taken from Equation~\ref{eq::relPtResFit}, can be seen in Table~\ref{tab::relPtResFitResults} for a 1.5~T field and a 3.0~T field. With a TPC that reflects the material budget and resolution of the sPHENIX TPC, a 3T magnetic field would be needed for the current gas-silicon hybrid detector to meet the current physics tracking momentum resolution requirements in the central region. However, it should be noted that an EIC optimized TPC (material budget, gas choice, etc.) still needs to be investigated and could yield different conclusions. Additionally, as the TPC resolution deteriorates at larger rapidities and the endcap trackers kick in, within this detector setup, a 3T field would not meet the required physics requirements for high momenta particles at large rapidity. This shows that additional tracking elements will be needed. 

\begin{table}[ht]
\footnotesize
\centering
\caption{Relative transverse momentum resolution fit parameters for a 1.5~T magnetic field and a 3.0~T magnetic field, using the fit presented in Equation~\ref{eq::relPtResFit}.}
\begin{tabular}{| c | c | c c | c c |}
\hline
\multirow{2}{*}{\textbf{$\eta$ Interval} } & \multirow{2}{*}{\textbf{$p_\text{T}$ Interval} } & \multicolumn{2}{c|}{\textbf{Fit 1.5 Tesla}} & \multicolumn{2}{c|}{\textbf{Fit 3.0 Tesla}} \\ \cline{3-6}
&  & \textbf{A} [\%/(GeV/c)] & \textbf{B} [\%] & \textbf{A} [\%/(GeV/c)] & \textbf{B} [\%] \\ \hline

\multirow{2}{*}{$\eta = 1.25$} &0 to 5 GeV/c & $ 0.203 \pm 0.009$ & $ 0.772 \pm 0.019$ 	& $ 0.095 \pm 0.005$ & $ 0.411 \pm 0.010$	\\ 
& 5 to 30 GeV/c 	        & $ 0.046 \pm 0.002$ & $ 1.269 \pm 0.025$ 	& $ 0.021 \pm 0.001$ & $ 0.666 \pm 0.012$	\\ \hline
\multirow{2}{*}{$\eta = 1.5$} & 0 to 5 GeV/c & $ 0.228 \pm 0.019$ & $ 1.361 \pm 0.034$ 	& $ 0.128 \pm 0.009$ & $ 0.721 \pm 0.017$	\\ 
& 5 to 30 GeV/c 	        & $ 0.068 \pm 0.003$ & $ 1.819 \pm 0.037$ 	& $ 0.034 \pm 0.002$ & $ 0.931 \pm 0.019$	\\ \hline
\multirow{2}{*}{$\eta = 1.75$}& 0 to 5 GeV/c & $ 0.356 \pm 0.023$ & $ 1.745 \pm 0.048$ 	& $ 0.183 \pm 0.011$ & $ 0.905 \pm 0.022$	\\  
& 5 to 30 GeV/c 	        & $ 0.097 \pm 0.004$ & $ 2.524 \pm 0.057$ 	& $ 0.048 \pm 0.002$ & $ 1.309 \pm 0.027$	\\ \hline
\multirow{2}{*}{$\eta = 2.0$} & 0 to 5 GeV/c & $ 0.459 \pm 0.025$ & $ 1.882 \pm 0.051$ 	& $ 0.210 \pm 0.013$ & $ 1.029 \pm 0.027$	\\ 
& 5 to 30 GeV/c         	& $ 0.196 \pm 0.006$ & $ 2.959 \pm 0.085$ 	& $ 0.090 \pm 0.002$ & $ 1.506 \pm 0.036$	\\ \hline
$\eta = 2.5$ &0 to 30 GeV/c & $ 0.426 \pm 0.006$ & $ 2.445 \pm 0.046$ 	& $ 0.214 \pm 0.003$ & $ 1.299 \pm 0.026$	\\ \hline
$\eta = 3.0$ &0 to 30 GeV/c & $ 1.077 \pm 0.020$ & $ 4.668 \pm 0.109$ 	& $ 0.520 \pm 0.007$ & $ 2.143 \pm 0.048$	\\ \hline
\end{tabular}
\label{tab::relPtResFitResults}
\end{table}

\subsubsection{Fast tracking Layers \& Additional PID detectors}
\paragraph{Fast Signal \& High Resolution MPGDs for DIRC in the Barrel Region}
For the scenario where a TPC is chosen as the central tracker option for the EIC detector and MAPS technology is adopted as the vertex tracker, we have identified three strong motivations for the need of a high-precision and fast-signal tracking detector to complement the inherent limitations of the TPC + MAPS as main tracking detectors in the barrel region. 

\textbf{High angular resolution tracking layer for the barrel PID detector:} Particle identification at an EIC is going to be critical. High angular tracking resolution will improve the effectiveness of the PID detectors, in particular the DIRC and RICH detectors. We have studied the impact that our fast cylindrical $\mu$RWELL trackers would have on the angular tracking resolution in the central region. 

We simulated a detector setup within the EicRoot framework, which implemented a silicon vertex tracker, TPC, and cylindrical $\mu$RWELL trackers~\cite{eRD6-July20}. The study was performed with $\pi^{-}$ particles in a $1.5$ T magnetic field for scattering angles of $43^o, 66^{o}$, and $89^{o}$ over a momentum range of $1$ to $7$ GeV. We find an improvement in the angular resolution of tracks entering and exiting the DIRC when cylindrical $\mu$RWELL layers are located in front and behind the DIRC~\cite{eRD6-July20}. The simulation studies demonstrate that the two layers configuration surrounding the DIRC detector will improve the PID detector performances, and help aid in achieving the required 3$\sigma$ $\pi$/K separation at 6 GeV.

\textbf{High space point resolution tracking layer for TPC field distortions correction/calibration:} In addition to providing the angular resolution information to the DIRC detector, cylindrical $\mu$RWELL layers will also provide precision tracking to calibrate the TPC tracks and help correct for well known "scale distortions" of TPC tracks. For this case, the optimal configuration will be two  cylindrical $\mu$RWELL layers, the first inside the TPC inner field cage and the second outside. We are performing simulation studies for the two-layers configurations to evaluate the performances.

\textbf{Fast tracking layer to complement slow TPC and MAPS detector} Both TPC and MAPS technologies are slow detectors and having an additional fast tracker with a timing resolution of a few ns will be required to provide the bunch crossing timing information to the reconstructed vertex as well as central tracks.  $\mu$RWELL detector technologies provides the timing resolution better than 9 ns~\cite{Bencivenni:2017wee, urwellTiming, urwellTiming2} needed to satisfy these requirements
%

\paragraph{MPGD-based-TRDs for Electron PID and Tracking in the End Caps}\label{TRK-TRD}
Identification of secondary electrons plays a very important role for physics at the Electron-Ion Collider (EIC). J/$\psi$ has a significant branching ratio for decays into leptons (the branching ratio into $e^+e^-$ is  $\sim 6\%$). The branching ratio of D-mesons is Br($D^+ \rightarrow e +X$) $\sim 16 \%$ and the branching ratio of B-mesons is Br($B^{\pm} \rightarrow e  + \nu +X_c$ ) $\sim 10\%$ , and also not negligible contributions from  $B \rightarrow  D \rightarrow e +X $.  Electron identification is also important for many other physics topics, such as spectroscopy, beyond the standard model physics, etc. By using more sophisticated electron identification the efficiency of those channels could be increased. A high granularity tracker combined with a transition radiation option for particle identification could provide additional information necessary for electron identification or hadron suppression in the energy range from 2-100 GeV (note, that  pions only start to emit TR-photons above $\sim$150~GeV, which are  beyond the EIC kinematic range). Due to asymmetric beam energies and boosted kinematics, it is important to provide such additional instrumentation in the hadron endcap. %
\begin{figure}[hbt]
\includegraphics[width=0.325\textwidth]{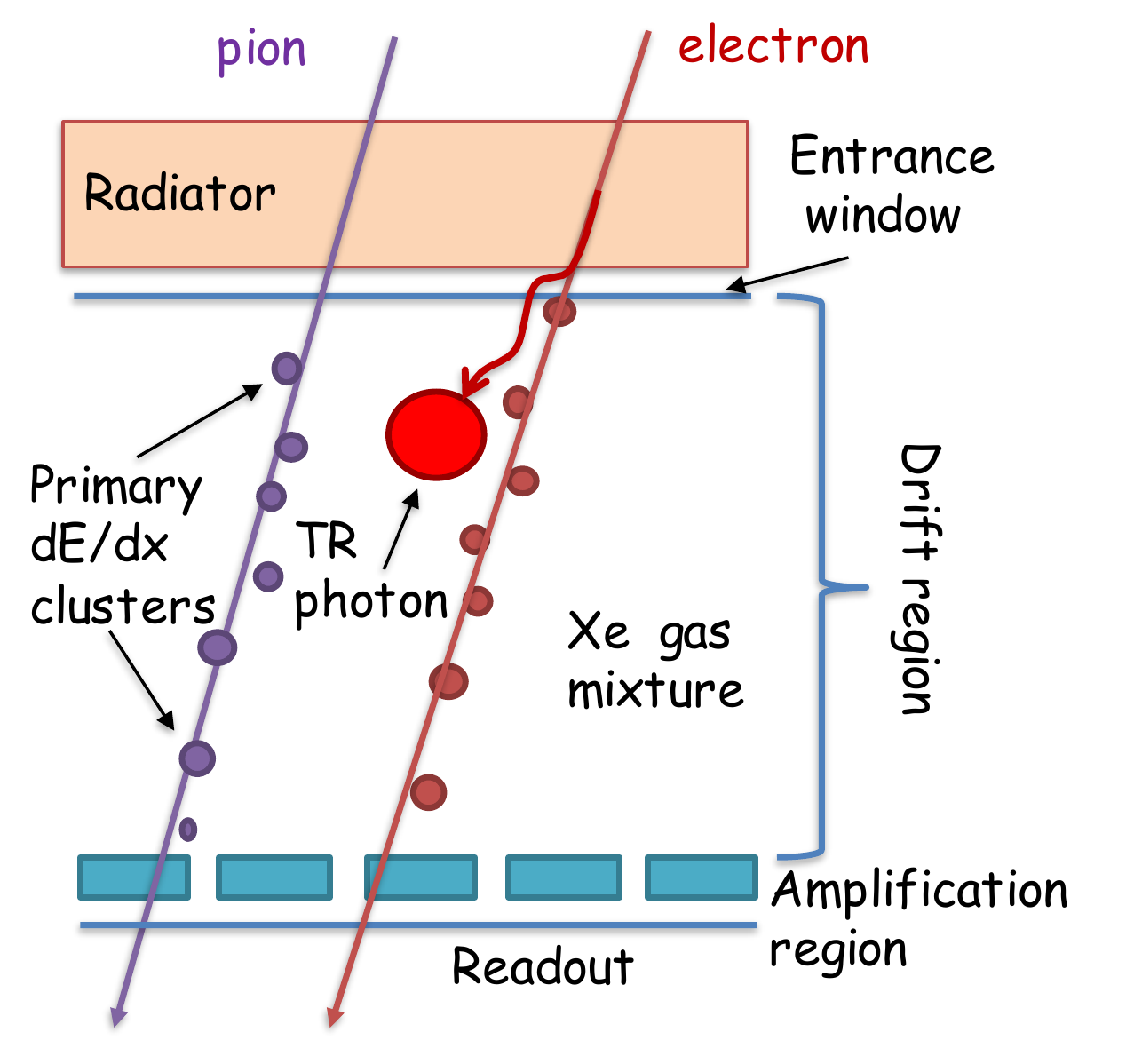}
\includegraphics[width=0.4\textwidth]{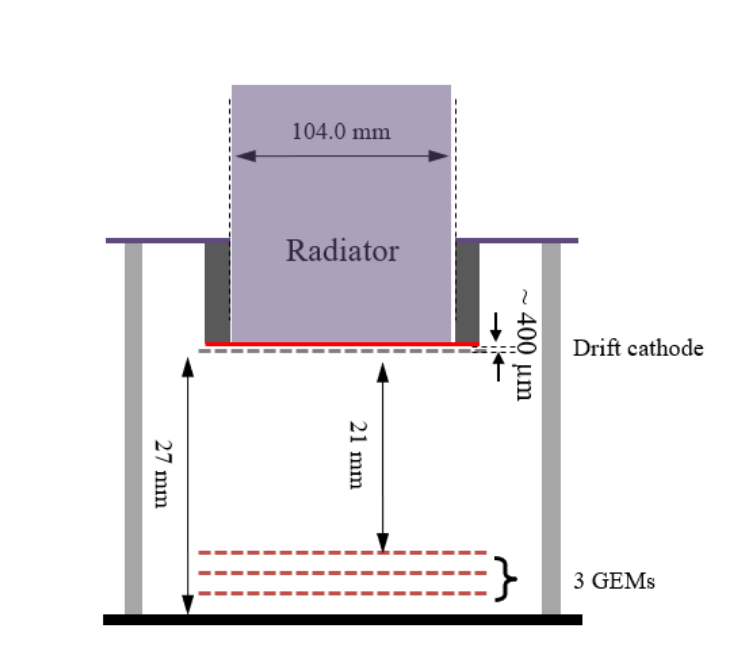}
\includegraphics[width=0.25\textwidth]{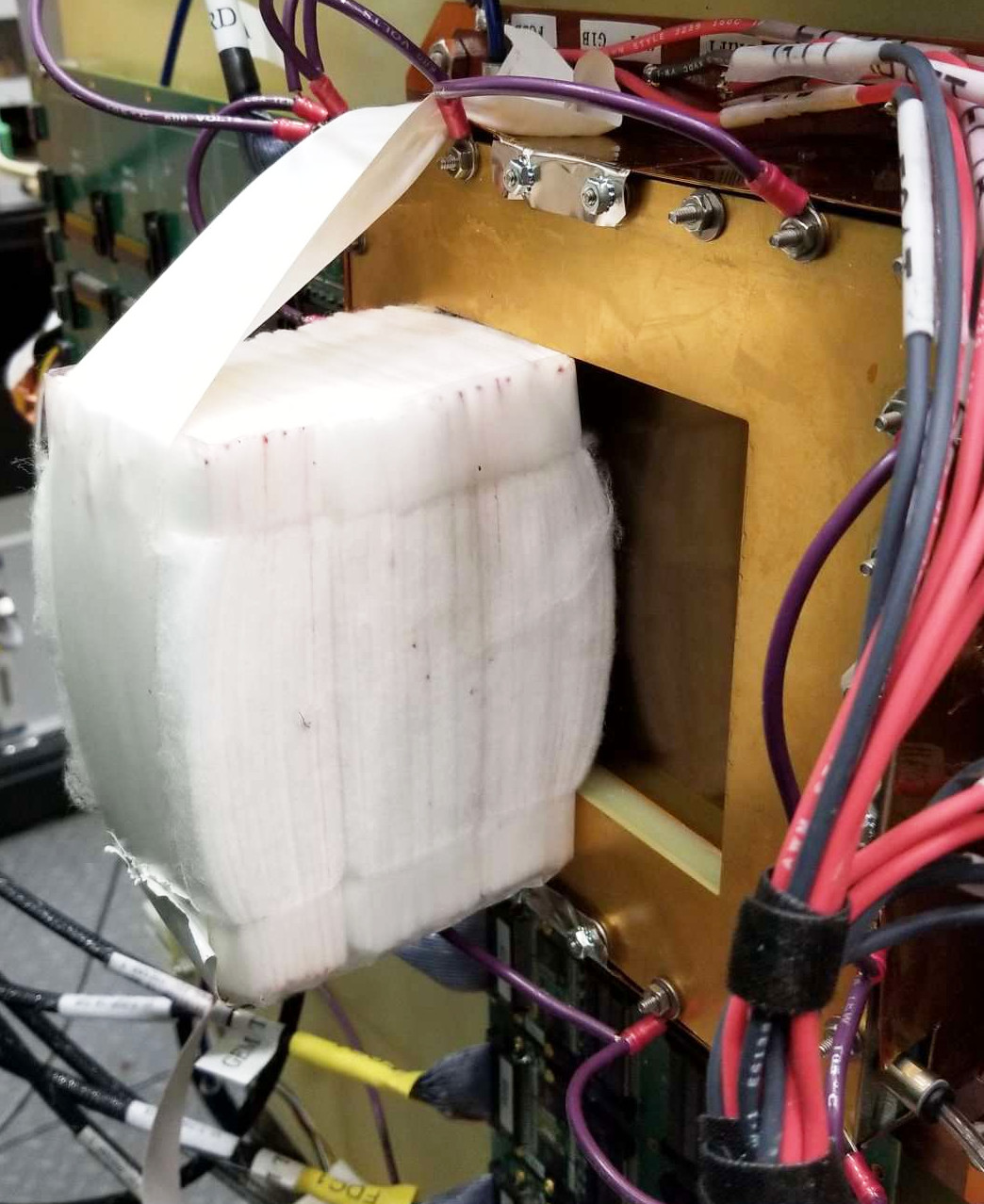}
\caption{ The concept of GEM-based TRD (left), the prototype scheme (middle), prototype in testbeam setup (right) \label{fig:TRD1}}
\end{figure}
The basic concept of GEM-based TRD is shown on the Fig.~\ref{fig:TRD1}. A standard triple-GEM detector~\cite{Sauli:1997qp} with high granularity strip pitch (400~$\mu$m) capable of providing high resolution space point position information was converted into a transition radiation detector and tracker (GEM-TRD/T)~\cite{Barbosa:2019hux}. This was achieved by making several modifications to the standard GEM tracker. First, since heavy gases are required for efficient absorption of X-rays, the operational gas mixture has been changed from an Argon based mixture to a Xenon based mixture. Secondly, the drift region needed to be increased from $\sim$3~mm to 20-30~mm in order to detect more energetic TR photons. To produce the TR photons, a TR radiator was installed in front of the GEM entrance window. Finally, the standard APV25 GEM readout electronics was replaced with faster electronics based on FADCs~\cite{FADC125} developed for the JLab GlueX Drift Chambers.
A GEANT4 simulation was performed to optimize the radiator and detector thicknesses for a single chamber (Fig.~\ref{fig:TRD1}). The G4XTRGammaRadModel model was used for the fleece radiator, which was implemented in GEANT4 as an irregular type of radiator with a given  density and two parameters ($\alpha_1$, $\alpha_2$), which define the spread of materials and air-gaps within a radiator.  Due to the self-absorbing property of the radiator, soft photons (3-6 keV) generated within the first few centimeters of the TR-radiator will be absorbed, leading to an increase in the hard X-ray photon spectrum at the exit from the radiator. A thin layer of gas in Xe-based detector will not be effective at detecting hard X-ray photons.  As one could see in Fig.~\ref{fig:TRD_Rad_scan1} (left), rejection power is saturated after 22cm of radiator for our GEM detector with 21mm gas thickness, including 400$\mu$m of dead gas layer in front. Experimental data points (stars) show a good agreement with MC projections. 
A TRD needs information about the ionization along the track, to discriminate TR photons from the ionization of the charged particle. The GEM-TRD/T prototype used a precise (125~MHz, 12~bit) FADC~\cite{FADC125} coupled with fast shaper pre-amplifiers, developed at JLAB, with a VME-based readout. The FADCs have a pipeline readout window  of up to 8~$\mu$s, which covers the entire drift time (500~ns) of the GEM-TRD/T prototype and gives a room for HV scan. The pre-amplifiers used GAS-II ASIC chips to provide 2.6~mV/fC amplification with a peaking time of 10~ns. 
\begin{figure}[hbt]
\resizebox{0.5\textwidth}{!}{\includegraphics{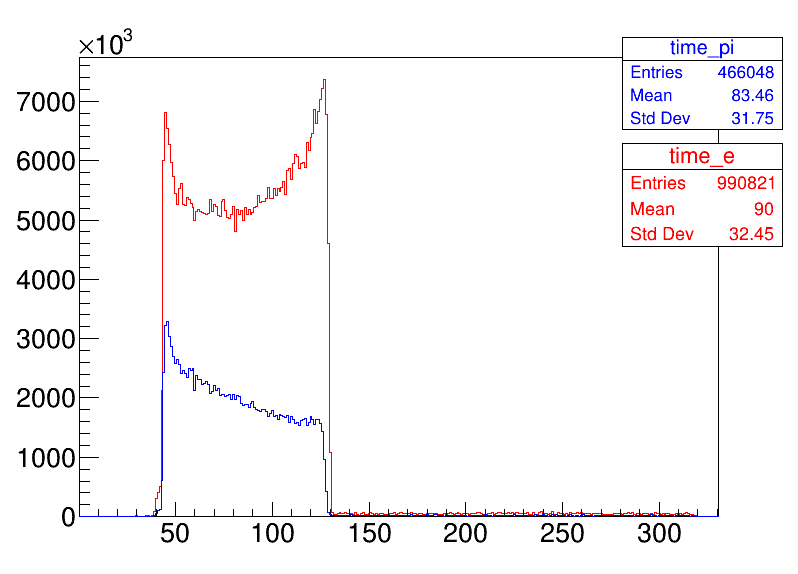}}
\resizebox{0.475\textwidth}{!}{\includegraphics{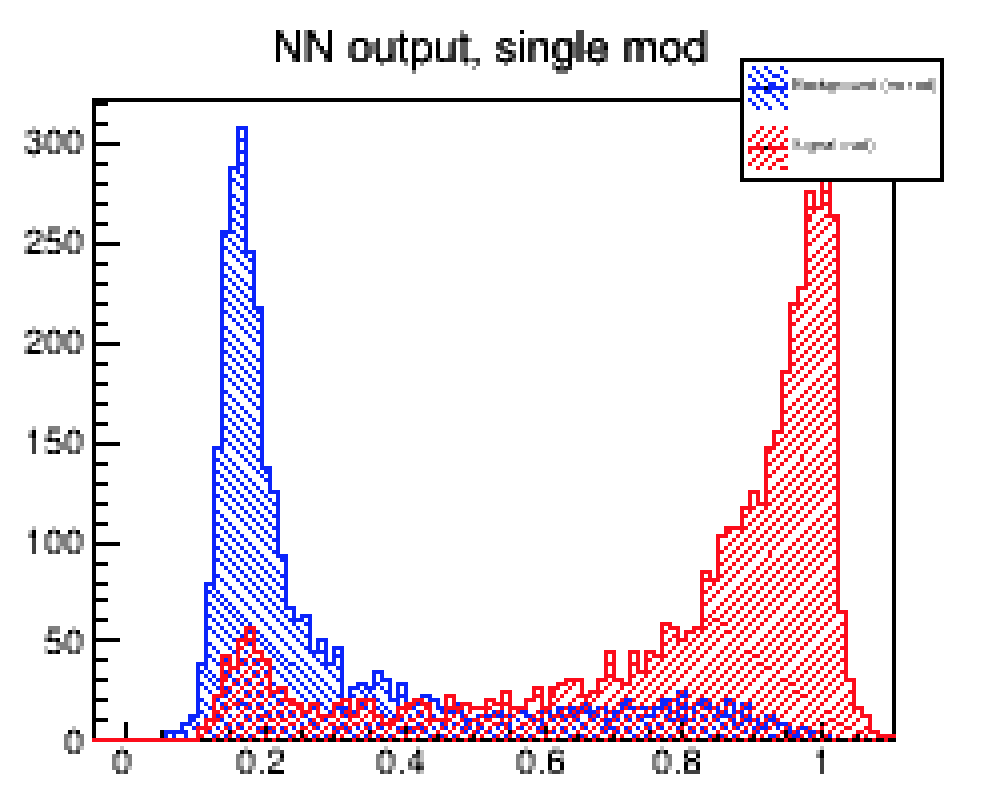}}\\
\caption{ Left plot shows average energy deposition along the drift time (x-axis in fADC time-bins). Right plot is output from Neural Network, showing the separation between electrons and pions.  \label{fig:BestRun1}}
\end{figure}
%
\begin{figure}[ht]
\centering 
 \begin{minipage}{0.475\textwidth}
    \includegraphics[width = .95\textwidth,keepaspectratio]{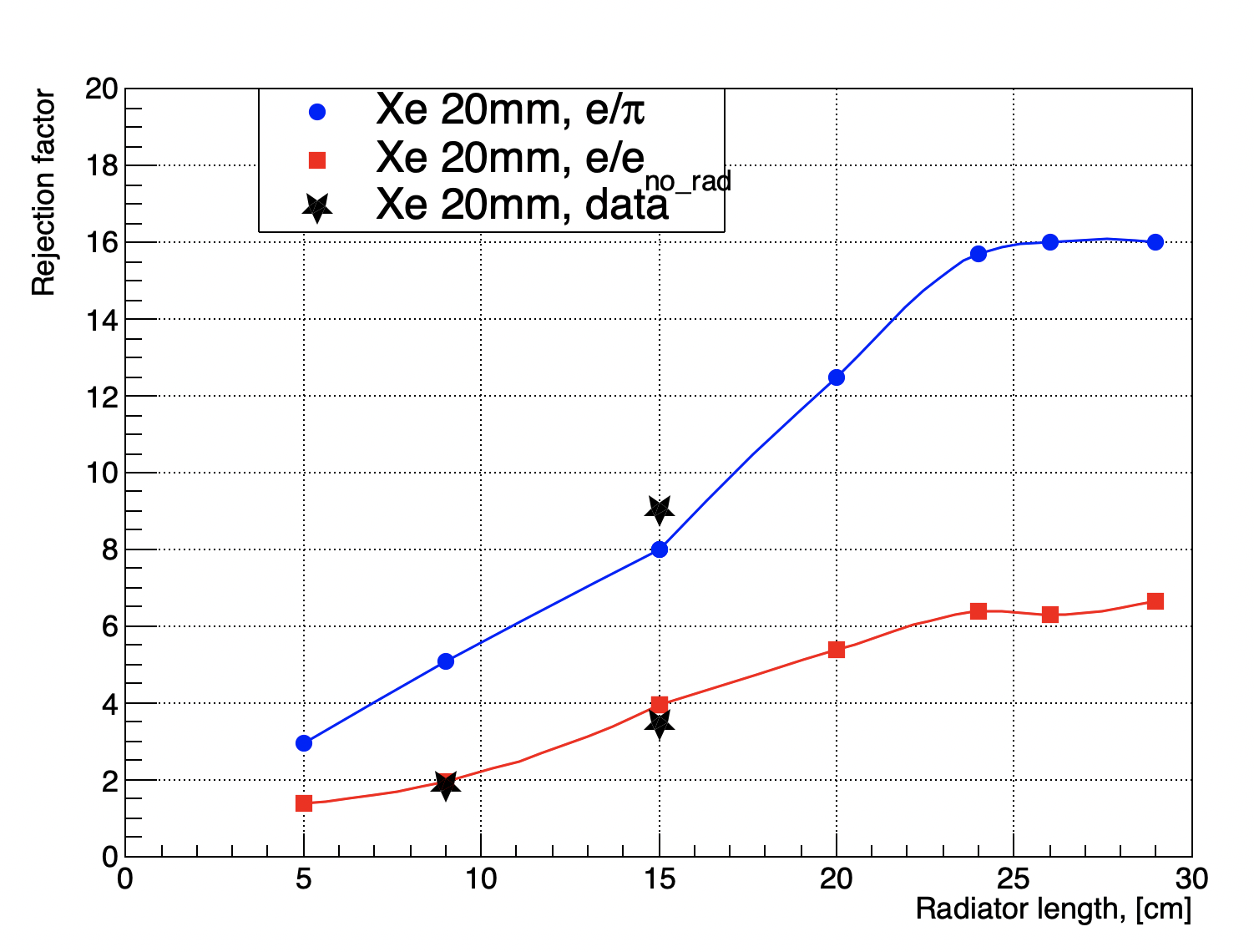}
 \caption{Rejection  vs. TR-radiator thickness. }
   \label{fig:TRD_Rad_scan1}
  \end{minipage}
  \hspace{0.1cm}
  \begin{minipage}{0.5\textwidth}
   \includegraphics[width = .95\textwidth,keepaspectratio]{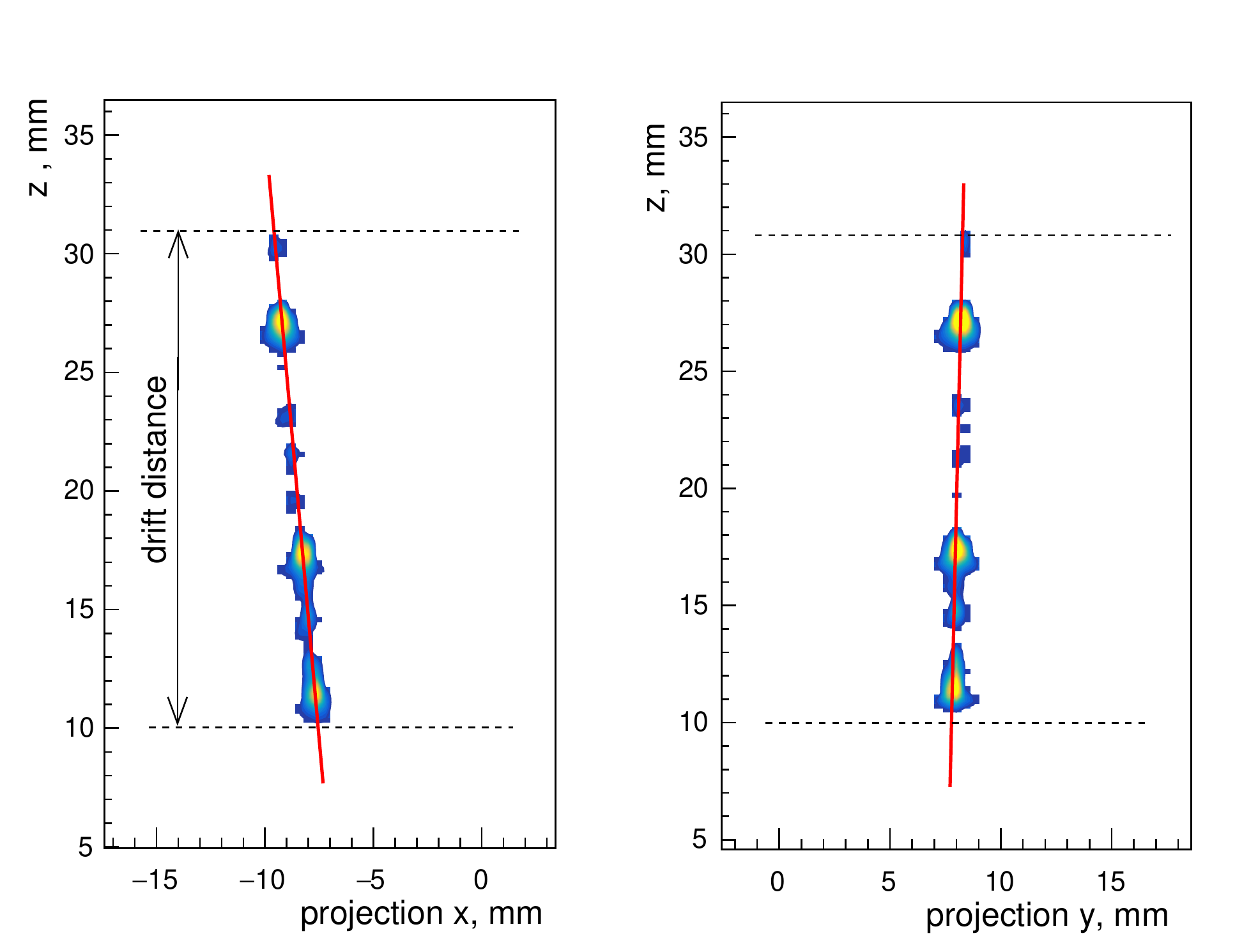}
    \caption{ Single track reconstruction. }
    \label{fig:gem_track}
  \end{minipage} 
\end{figure}
For the e/$\pi$ rejection factor the amplitude and arrival time of each individual cluster along the drift time were analyzed.  All this information (up to 20 variables) was used as input for likelihood and artificial neural network (ANN) programs, such as JETNET or ROOT-based Multi-layer Perceptron. The ANN system was trained with MC or data samples of incident electron and pions. Then an independent sample was used to evaluate the performance. An example of such a training procedure is shown in Fig.~\ref{fig:BestRun1}.  A 90\% efficiency for our electron identification was required. The neural network output for e/$\pi$ rejection is shown  Fig~\ref{fig:TRD_Rad_scan1}, demonstrating that a rejection factor of around 9 could be achieved with a single module and $ ~\sim$15~cm radiator. A single GEM-TRD/T module has  $\sim 3\%X_0$ depending on a TR-radiator length.  \\ 
As for tracking aspects, a standard GEM plane can only provide the 2D X-Y position of a track, while the GEM-TRD/T with increased drift volume and with  Flash ADC readout allows for 3D track segments to be reconstructed as in $\mu-$TPC configuration. In the hadron end cap region, in addition to the e/$\pi$ rejection capabilities, GEM-TRD track segment behind dRICH could be used to:
\begin{itemize}
\item measure a track angular resolution and therefore help to improve dRICH performance;
\item correct for a multiple scattering before EMCAL and improve tracking performance for charged particles. 
\item improve pointing track resolution and cluster-seed position measurements for EMCAL
\item could be used as a seed-element for track finding algorithms.
\end{itemize}
Figure~\ref{fig:gem_track} shows projections of a typical 3D reconstructed track from the GEM-TRD/T prototype. The left panel shows the track projection in XZ plane with $Z$ the drift time as a function of the cluster position in the  $X$ direction. The right panel shows corresponding projection in YZ plane. 



\paragraph{Readout structures for MPGDs}

\textbf{Zigzag Shaped Charge Collection Anodes}
\newline
The segmentation of the readout plane for MPGD-based detectors can play a critical role for the detector performance, especially for the spatial and angular resolution and should be seriously considered for future experiments. To improve the resolution, a typical strategy is to simply reduce the pitch of the anodes, but this comes at the cost of greater instrumentation. As an alternative, highly interleaved anode patterns, such as zigzags offer relatively coarse segmentation, while preserving performance~\cite{ZZmpgd4eic_tns2020} \cite{ZZmpgd4eic_tns2018}. By optimizing the three main operant geometric parameters of the zigzag (including the pitch, the periodicity of the zigzag, and the degree of interleaving, here referred to as the “stretch” parameter), charge sharing among neighboring pads or strips may be finely tuned for specific avalanche schemes.
\begin{figure}[hbt]
	\centering
        \includegraphics[width=1.0\columnwidth]{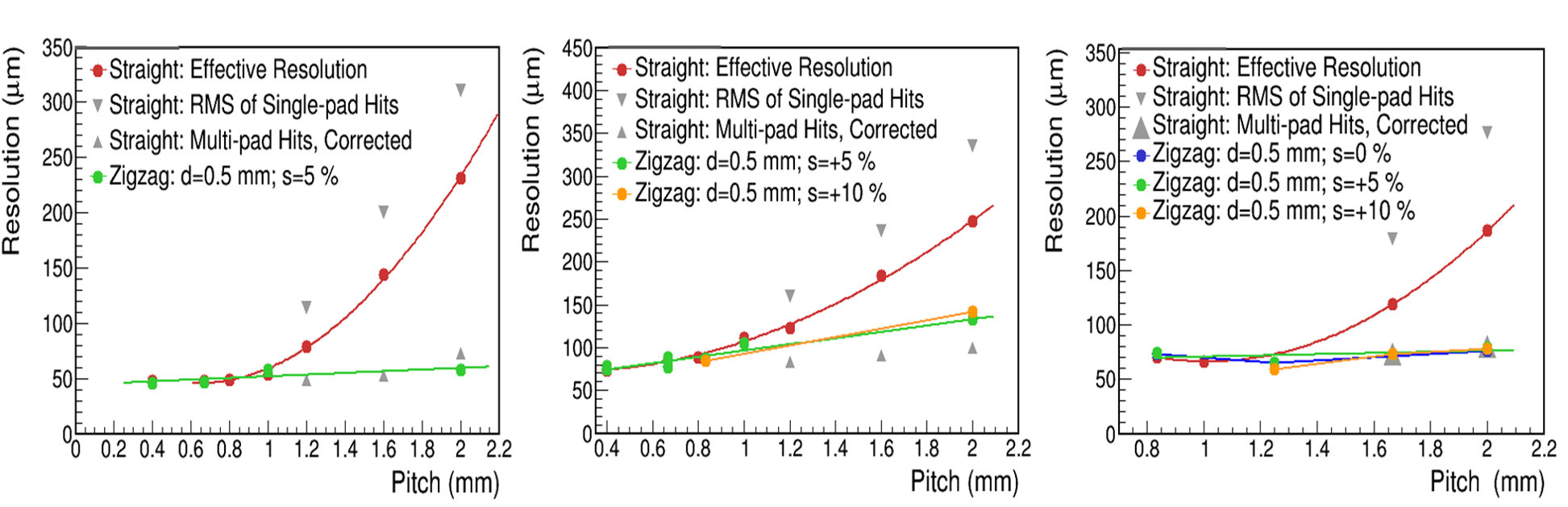}
	\caption{\label{fig:ZZ_summary_1.png}
Position resolution vs. pitch for straight strip and zigzag shaped anodes in GEM, Micromegas and $\mu$RWELL detectors respectively, measured in a 120 GeV proton beam. The resolution is shown for a zigzag period (\emph{d}) of 0.5~mm and various stretch parameters (\emph{s}). The resolution for straight strips is corrected using pad response functions, however the raw resolutions are quoted for the zigzags. The resolution for the straight strips is broken down into regions of the readout dominated by single and multi-pad clusters (grey points), where the red points denote the weighted average. 
 }
\end{figure}

\begin{figure}[hbt]
	\centering
        \includegraphics[width=0.75\columnwidth]{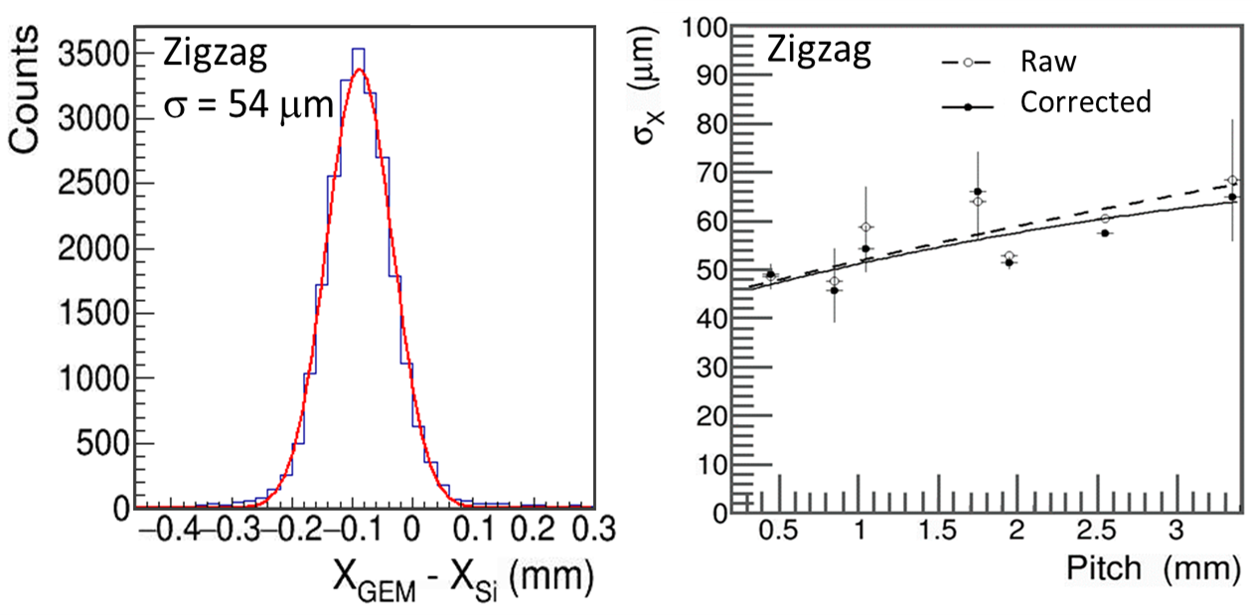}
	\caption{\label{fig:ZZ_summary_2.png}
Raw residual distribution for zigzag anodes with pitch = 2~mm, period = 0.4~mm, and stretch = 0~\% and a plot of the position resolution vs. pitch in the case of a 4-GEM detector, respectively. 
 }
\end{figure}

Fig.~\ref{fig:ZZ_summary_1.png} compares the resolution as a function of the pitch for standard straight strips and various zigzag parameters for GEM, Micromegas, and $\mu$RWELL detectors. In all cases, the position resolution is comparable below a pitch of 1mm, but the resolution quickly degrades for straight strips at larger pitch. This is mainly due to poor charge sharing, where the majority of charge is collected by a single pad. An equally beneficial feature of zigzags is the ability to maintain a highly uniform and linear response across the full detector acceptance.  The “out of the box” detector response of optimized zigzag anodes is shown in Fig.~\ref{fig:ZZ_summary_2.png}, which includes a purely Gaussian raw residual distribution, without the need for pad response functions, as in the case of straight strips.  Ultimately, in situations where the detector occupancy is fairly low and a relatively coarse readout segmentation is acceptable, zigzag shaped charge collection anodes provide a very efficient means of encoding high resolution positional information, with values remaining below 65~$\mu$m for a pitch as large as 3.3~mm as indicated in the right-hand plot.
\subsubsection{Comparison: requirements versus performance}
The tracking performances of the two baseline detector concepts (“all-silicon” and “hybrid” systems) have been compared with the corresponding requirements coming from physics simulations reported in Table~\ref{tab:reqTable}. While both reference systems correspond to a total material budget well below the 5\% X$_0$ limit in the tracking region, the two options exhibit slightly different performances, in particular in terms of relative momentum resolution as a function of total momentum and pointing resolution in the transverse plane as a function of p$_T$. Results coming from fit description of the corresponding distributions according to the functional forms (~\ref{eq:dpp_v_p_param}) and (~\ref{eq:dca_v_pT_param}), have been reported in table~\ref{tab:param_AllSi} and tables~\ref{tab::relMomResFitResults_full_1}, \ref{tab::transvPointFitResults_1}, for the all-silicon and hybrid baseline respectively. They have been also extracted for the two different values of the magnetic field currently under consideration, namely 1.5 T and 3 T. 
Tables~\ref{tab:allSiPerf1T} and ~\ref{tab:allSiPerf3T} illustrate the comparison of the all-silicon tracking performance for both fields with the corresponding requirements. The transverse pointing resolution stays unchanged with varying magnetic field and satisfies pretty well the requirements both at 1.5 T and 3 T field. The relative momentum resolution better matches the requirements for the higher field value and in the central pseudo-rapidity region.  

\begin{table}[h!bt]
\centering
\caption{Comparison of performance and requirements for the all-silicon concept at 1.5 T magnetic field.}
\includegraphics[width=0.95\columnwidth,trim={0pt 0mm 0pt 0mm},clip]{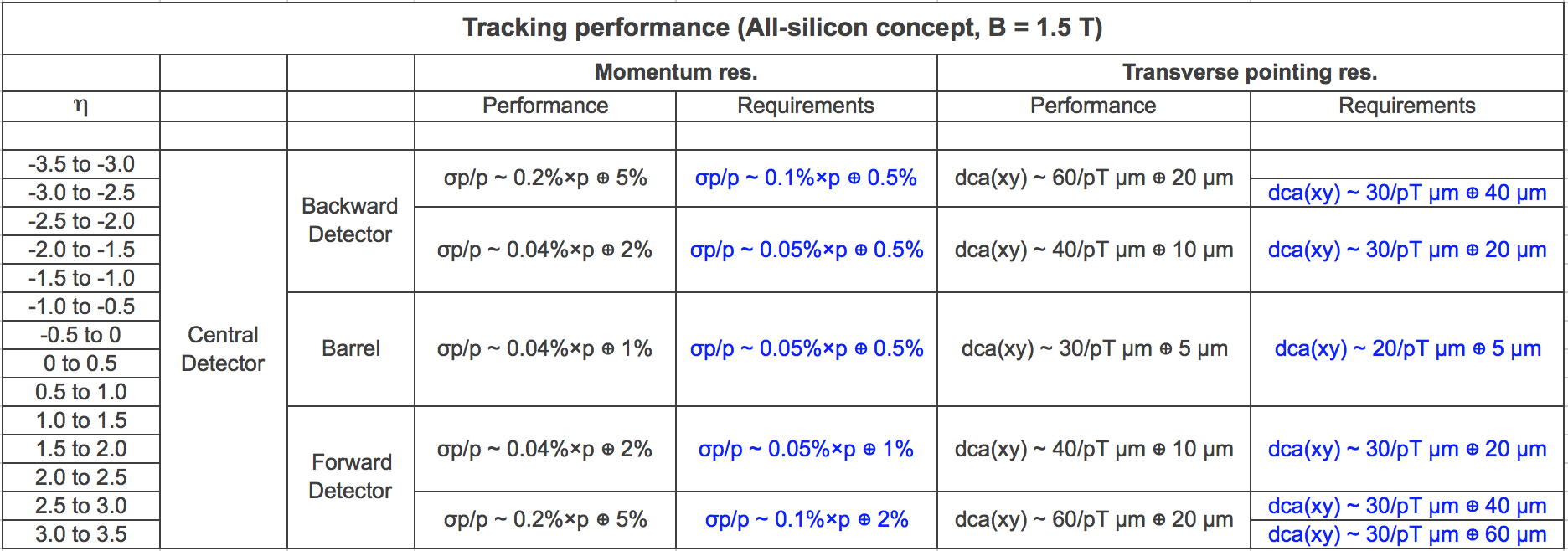}
\label{tab:allSiPerf1T}
\end{table}
\begin{table}[h!bt]
\centering
\caption{Comparison of performance and requirements for the all-silicon concept at 3 T magnetic field.}
\includegraphics[width=0.95\columnwidth,trim={0pt 0mm 0pt 0mm},clip]{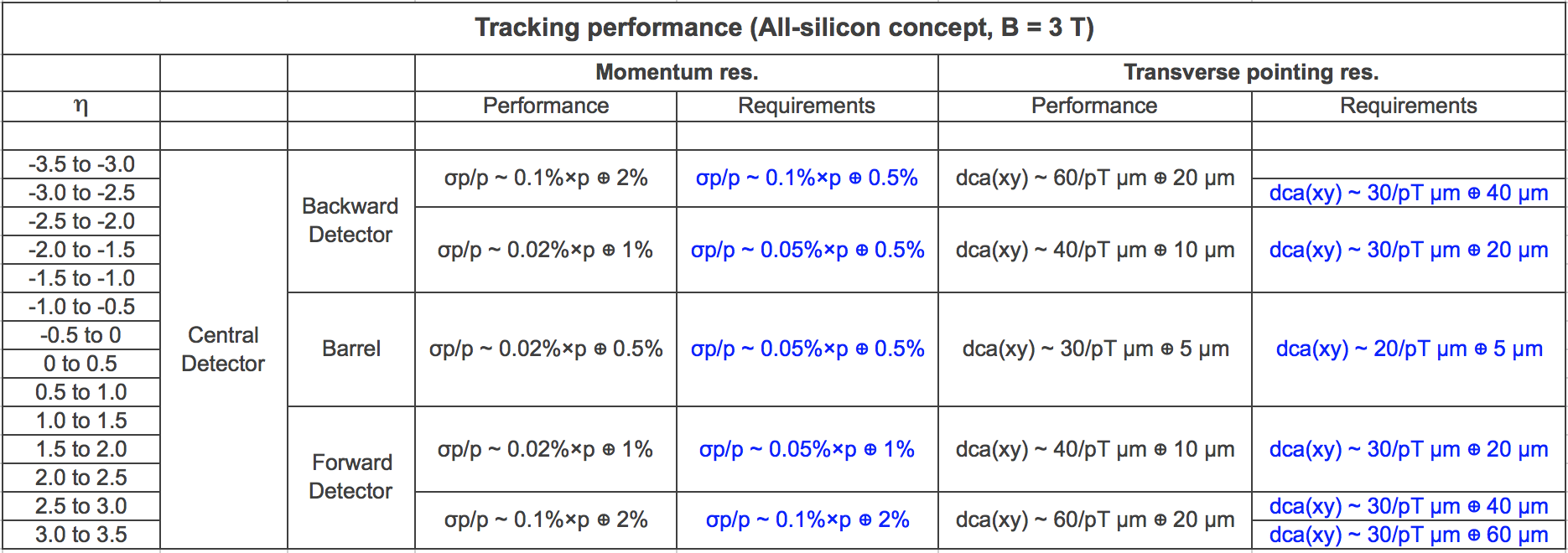}
\label{tab:allSiPerf3T}
\end{table}

The same comparison can be done for the performances worked out with the hybrid baseline concept, including also the preliminary estimate of the track reconstruction efficiency at low transverse momenta reported in table~\ref{tab::siPlusTPC_minPtvalues}. The corresponding comparison is shown in Tables~\ref{tab:hybridPerf1T} and ~\ref{tab:hybridPerf3T} hereafter, again for both the magnetic field values. Also for the hybrid tracker the transverse pointing resolution performance does not depend on the magnetic field and satisfies the requirements, while with the low magnetic field option the relative momentum resolution degradation brings the performance below the level set by requirements. However, a more efficient reconstruction of the tracks at the lower transverse momenta would better fit the physics requirements in this case.

\begin{table}[h!bt]
\centering
\caption{Comparison of performance and requirements for the hybrid concept at 1.5 T magnetic field.}
\includegraphics[width=0.95\columnwidth,trim={0pt 0mm 0pt 0mm},clip]{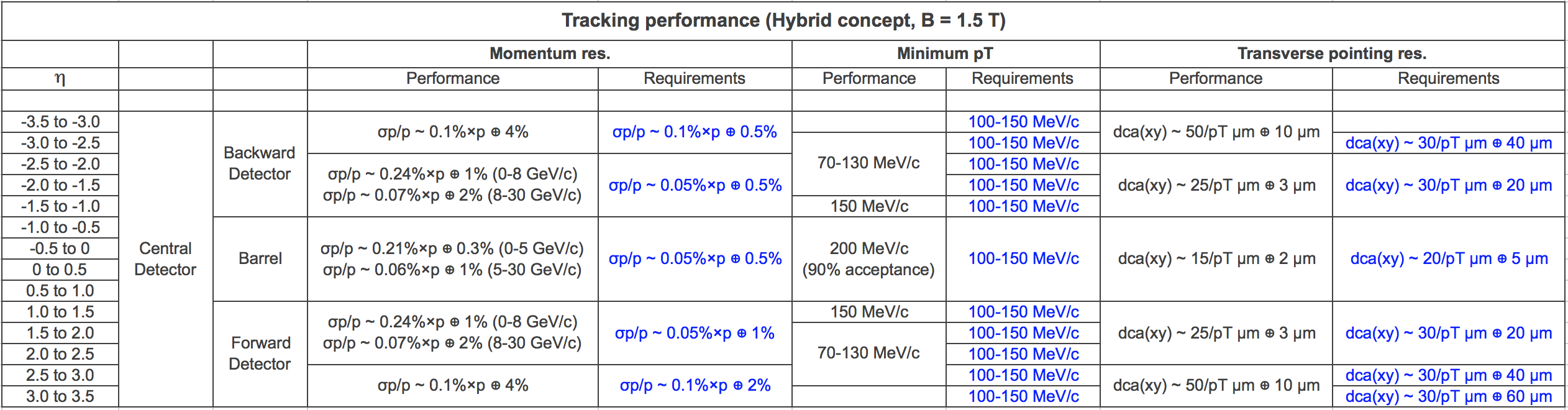}
\label{tab:hybridPerf1T}
\end{table}
\begin{table}[h!bt]
\centering
\caption{Comparison of performance and requirements for the hybrid concept at 3 T magnetic field.}
\includegraphics[width=0.95\columnwidth,trim={0pt 0mm 0pt 0mm},clip]{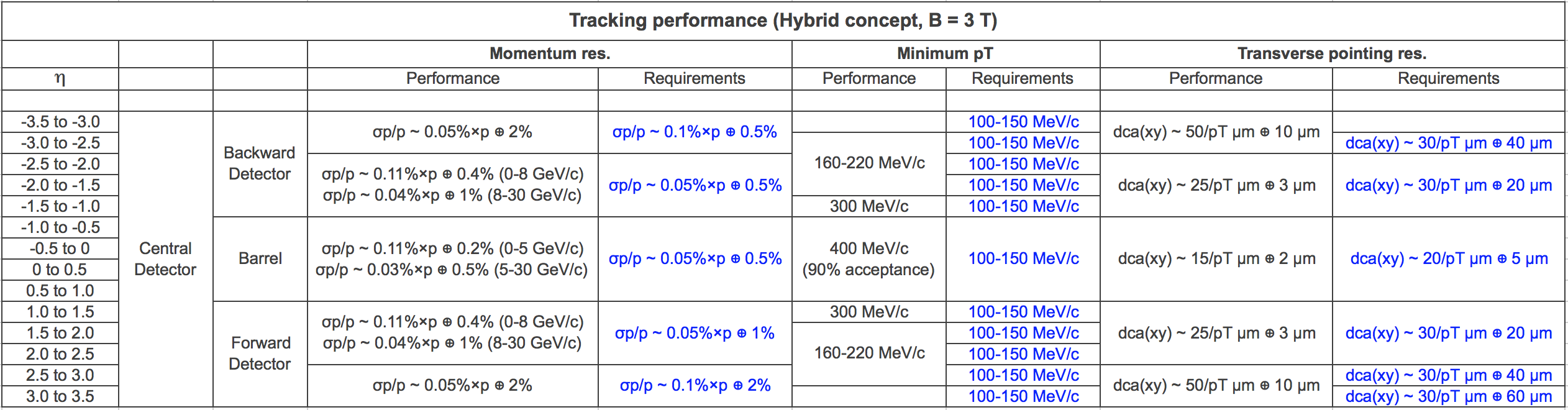}
\label{tab:hybridPerf3T}
\end{table}
\subsection{Material Budget Considerations}
As is clear from the requirements, the silicon tracking layers require a very low material budget per layer/disc and this need for low mass material budget in the acceptance extends to the surrounding detectors. In order to assess the balance of mass that contributes to the overall load, it is necessary to make an estimate of the additional material in the fiducial volume that is associated with the tracking detectors.
The material budget for the tracking detector is dependent on the parameters of the silicon sensors used, the architecture of the services (powering, readout, cooling, monitoring, safety interlocks, etc.) employed in the deployed detector design and the design and composition of the mechanical support structures used for precisely locating the tracking detector in the main detector volume. A reasonable starting point for estimating the services load is to start with existing technology and powering/readout architectures and project what could be expected should we adopt what has been accomplished. The current state of the art tracking detector of similar characteristics (MAPS sensors, $10/m^{2}$ of silicon area) is the recently upgraded ALICE ITS. As part of the EIC User Group Yellow Report activities, the service loads have been estimated and parameterized \cite{leo}. These estimates have been scaled for what can be expected for a detector system based on the ITS3 sensor which is currently under development \cite{its3det}. These parameterizations are currently being added to the simulation efforts for the EIC silicon detector baseline detector configurations so that the effects of these mass loads on the physics measurements can be assessed.
The largest mass in the services, by far, is the power supply and return cabling. This can be addressed in multiple ways. The most obvious avenue to explore is reducing the power required by the sensors. This is under investigation. An EIC sensor based on the ITS3 type development is expected to reduce the power needed by half to a dissipation of 20 mW/$cm^{2}$. This helps, but as the voltage supplied to the sensors is also reduced from 1.8V to 1.2V, to maintain the cable voltage drops to manageable levels, a significant fraction of the conductor is still required. It is possible to reduce the radiation length of the power cabling by moving to copper clad aluminum conductors. This can help significantly since the $X_{0}$ of aluminum is a factor of $\sim$ 6 lower than the $X_{0}$ of copper. Using aluminum conductors unfortunately  comes at a cost in space required by the services since the conductivity of aluminum is ~65\% that of copper. Other options would include significantly reducing the number of required conductors to power the detector. This could be addressed by either serial powering of detector staves, or the integration of radiation tolerant DC-DC converters at the stave ends \cite{leo2}. Both of these options require exploration and R\&D to become viable.
The readout cabling is also a significant load. It could be possible to combine stave outputs and multiplex the data from multiple staves on detector for readout over high speed fiber optical connections. The multiplexing circuitry and fiber optic drivers would need to be radiation tolerant. In addition, this reduction in the readout granularity would lead to larger portions of the detector becoming inactive in the case of single point failures in the multiplexing and fiber circuits. Clearly an optimization using these factors will need to be carried out. This is also an area for targeted R\&D.
The reduction in the sensor power dissipation using ITS3 like sensors would significantly help the cooling requirements so smaller and possibly fewer lines could be used. Air cooling is also a possibility, but the envisioned detector is very compact and arranging proper flow and ducting would require careful study. 
For the detector safety system sensors and environmental monitors, it is likely that the level of services would be similar to what is seen in the ALICE ITS.


\clearpage
\section{Electromagnetic Calorimetry}
\label{part3-sec-Det.Aspects.ECAL}


The EIC is a collider with diverse physics topics that impose
unique requirements on the electromagnetic calorimeter (ECAL)
design. Nearly all physics processes require the detection of the
scattered electron for the momentum or energy reconstruction and 
particle identification. The 3-momentum is measured with the tracker
and the energy is measured with ECAL. An excellent energy
resolution of ECAL is essential at small scattering angles where
the momentum resolution of the tracker is not expected to be sufficient. 
ECAL also must detect and identify single photons from DVCS 
and photon pairs from $\pi^0$ decays.
The main tasks of the ECAL can be summarized as:
\begin{itemize}
  \item{} Detect the scattered electrons in order to separate them from pions and also
          improve the energy/momentum resolution at large $|\eta|$.
  \item{} Detect neutral particles - photons, and measure the energy and the coordinates of the impact.
  \item{} PID: separate secondary electrons and positrons from charged hadrons.
  \item{} Provide a spacial resolution of two photons sufficient to identify 
          decays $\pi^0\rightarrow{}\gamma\gamma$ at high energies. 
\end{itemize}

\subsection{Requirements and Overview}
\label{sec:part3-Det.Aspects.ECAL-req}

\begin{figure}[htb]
\begin{center}
  \begin{minipage}[t]{0.32\linewidth}
    \includegraphics[angle=0,width=0.97\linewidth]{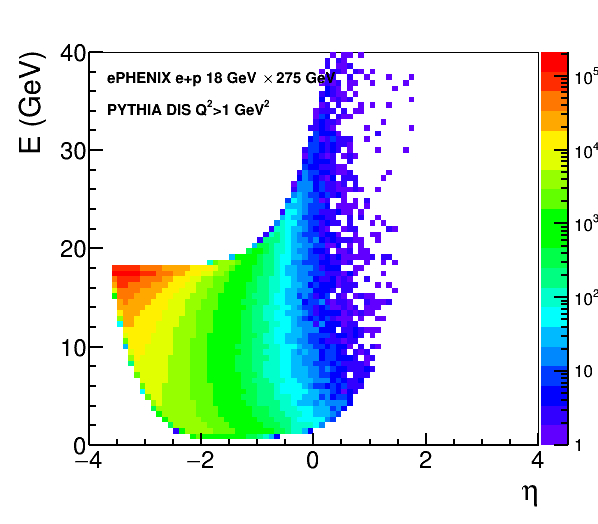}
  \end{minipage}
   \begin{minipage}[t]{0.32\linewidth}
    \includegraphics[angle=0,width=0.90\linewidth]{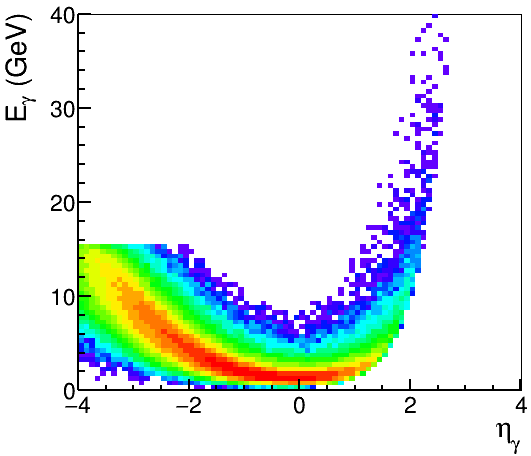}    
  \end{minipage}
  \begin{minipage}[t]{0.32\linewidth}
    \includegraphics[angle=0,width=0.97\linewidth]{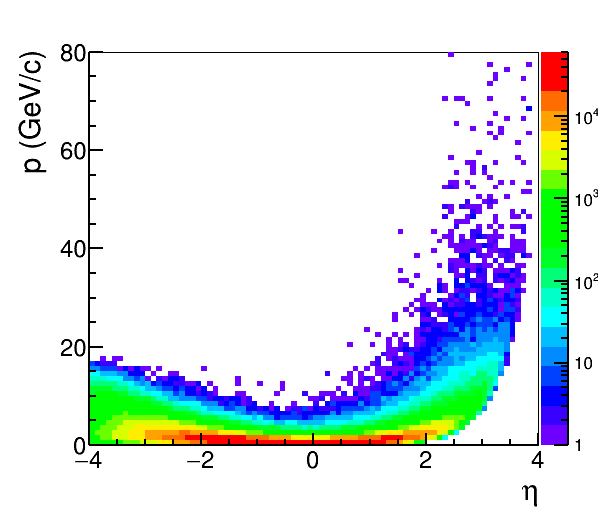}
  \end{minipage}
\end{center}
\caption{Calculated momentum spectra of particles in $e+p~ 18\times{}275$ GeV collisions\cite{Bazilevsky:tmplmar21}. 
         Left: DIS $e^-$ from PYTHIA~\cite{Sjostrand:2007gs}; Middle: DVCS $\gamma$ from MILOU~\cite{Perez:2004ig};
         Right: $\pi^0$ from PYTHIA.
        }
\label{fig:part3-Det.Aspects.ECAL-3spectra} 
\end{figure}

\begin{figure}[htb]
\begin{center}
  \includegraphics[angle=0,width=0.97\linewidth]{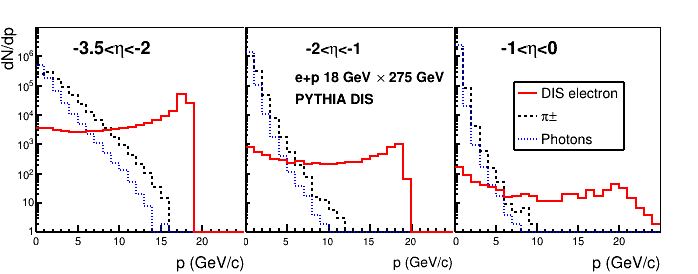}
\end{center}
\caption{Calculated momentum spectra of DIS electrons, photons and pions 
          in $e+p~ 18\times{}275$ GeV collisions\cite{Bazilevsky:tmplmar21}. 
        }
\label{fig:part3-Det.Aspects.ECAL-DIS-background} 
\end{figure}

\begin{figure}[htb]
\begin{center}
  \begin{minipage}{0.35\linewidth}
      \includegraphics[angle=0,width=1.00\linewidth]{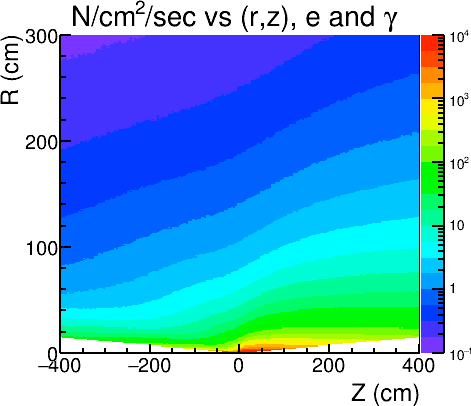} \\ 
      \includegraphics[angle=0,width=1.00\linewidth]{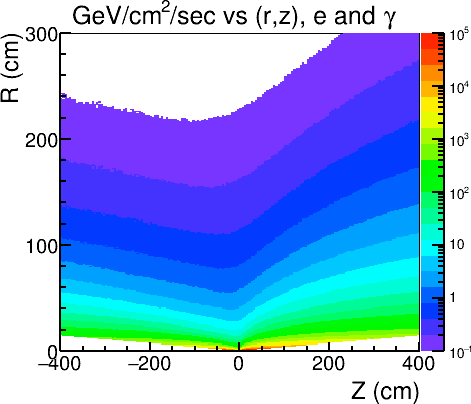} 
  \end{minipage}
  \begin{minipage}{0.35\linewidth}
      \includegraphics[angle=0,width=1.00\linewidth]{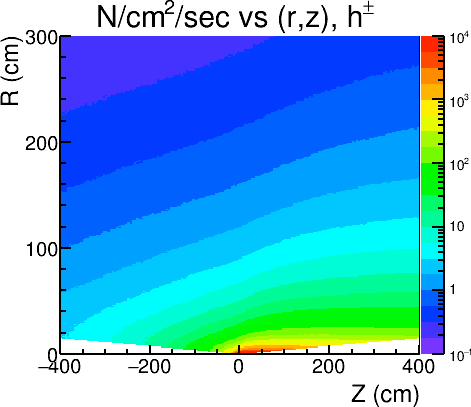} \\
      \includegraphics[angle=0,width=1.00\linewidth]{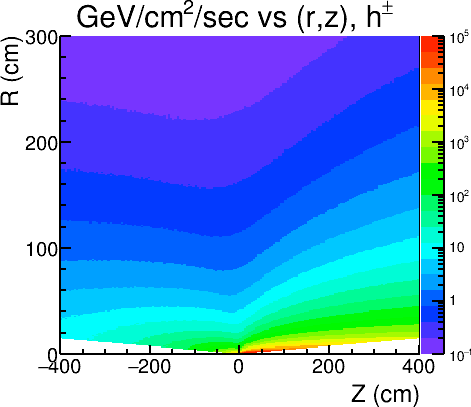} 
  \end{minipage}
\end{center}
\caption{Calculated particle and energy flux from DIS in $e+p~ 10\times{}100$~GeV collisions at a luminosity of
         $10^{34}$~cm$^{-2}$s$^{-1}$, for $\eta{}<4$. The top row shows the particle flux for electrons plus photons (left) and
         for hadrons (right). The bottom row shows the energy flux.  
        }
\label{fig:part3-Det.Aspects.ECAL-rates} 
\end{figure}

The physics requirements on the EIC detectors are shown in Table~\ref{SubDetReq} (also outlined in Ref.~\cite[p.~25]{EIC:RDHandbook}). The requirements on ECAL are summarized in Table~\ref{tab:part3-Det.Aspects.ECAL-req}.
The kinematic range and the requirements for the electron detection in
ECAL was discussed at length in presentations \cite{%
Bazilevsky:dwgcalfeb25,
Bazilevsky:tmplmar21,
Bazilevsky:dwgcalmay21,
Bazilevsky:dwgcaljun2,
Bazilevsky:dwgcaljul14
}
 (see Fig.\ref{fig:part3-Det.Aspects.ECAL-3spectra}). 
The background to DIS electrons is shown in Fig.\ref{fig:part3-Det.Aspects.ECAL-DIS-background}.
The expected particle flux at the full luminosity is shown in
Fig.~\ref{fig:part3-Det.Aspects.ECAL-rates}. 

\begin{table}[htb]
 \begin{center}
   \begin{tabular}{l|cccc}
     \hline
     \hline
     $\eta$   & -4 to -2  & -2 to -1  & -1 to 1 & 1 to 4 \\
     \hline
     $\sigma_E/E\cdot{}\sqrt{E/1~\mathrm{GeV}}$   & 2\%  & 7\%  & 10-12\% & 10-12\% \\
     \hline
   \end{tabular}
 \end{center}
 \caption{The physics requirements on the ECAL energy resolution from Table~\ref{SubDetReq}.
         }
 \label{tab:part3-Det.Aspects.ECAL-req}
\end{table}

The best energy resolution is required at $\eta{}<-2$. Such a resolution can be achieved with 
heavy scintillating crystals. 
The best two-photon resolution is required at $\eta{}>2$, which can be achieved 
with a fine granularity of a detector made of heavy materials, 
or by using a preshower detector with a fine granularity. 
The physics goals favor a reasonably hermetic detector, covering a range of 
about $-4<\eta<4$.

The space available for the entire EIC detector system is finite. This
poses a practical constraint on the technologies for the EIC
calorimeters. At the writing of this Yellow Report no final decisions
on the space have been made, but one can consider general drivers of
space constraints, e.g. tightness of the space in the endcaps is
driven by the luminosity requirements while the barrel calorimeter
space depends on the magnet design\footnote{The barrel ECAL is assumed to be located inside of the magnet's bore. Placing ECAL outside of the magnet cryostat would significantly compromise the performance. For example, the cryostat of the BaBar magnet is about 1.4$X_0$ thick\cite{Aubert:2001tu} which would add $\approx{}10\%/\sqrt{E}$\cite{Fabjan:2003aq} to the energy resolution for $\eta=0$ particles, and more at larger impact angles. Furthermore, it would strongly compromise the position resolution, $e/\pi$ separation and the reconstruction of high-energy $\pi^0$s.}.
For example, with the BaBar magnet
the outer diameter of ECAL can go up to 140 cm, while the minimal
radial thickness of ECAL is about 30~cm (based on the sPHENIX
experience).
The space in the magnet barrel is valuable for the momentum measurements, the PID etc. 
The general purpose detector illustrated in Fig.~\ref{central-detector-cartoon} allocates:
\begin{itemize}
  \item{} $\Delta{}Z<60$~cm in the electron endcap; 
  \item{} $\Delta{}R<30$~cm in the barrel; 
  \item{} $\Delta{}Z<40$~cm in the hadron endcap; 
\end{itemize}
Limited space generally favors calorimeter materials with a short radiation length ($X_0$).

The expected particle flux (Fig.~\ref{fig:part3-Det.Aspects.ECAL-rates}) is relatively low. In the electron arm
the highest flux from DIS and photoproduction at small angles is below 0.5~kHz for a $2\times{}2$~cm$^2$ module. Taking into account the leakage from the adjacent modules the signal rate per module is expected to be $<5$~kHz in the electron arm. In the hadron arm the signal rate is expected to be $<20$~kHz per similar module. The contribution from various accelerator backgrounds remains to be evaluated. 
The calculated energy flow in this area (Fig.~\ref{fig:part3-Det.Aspects.ECAL-rates}) allows to evaluate a dose rate
of about 1.2~krad per full year.
Several specifications for ECAL follow from the particle spectra and flux: 
\begin{itemize}
  \item{} The energy range depends on $\eta$ (Fig.\ref{fig:part3-Det.Aspects.ECAL-3spectra}).
 \item{} The expected particle flux from the $ep$ collisions is below a few kHz per module. The expected signal rate is below a few dozens of kHz per module.   
  \item{} The timing resolution should be good enough to separate the bunches ($\approx{}$10~ns apart). 
  \item{} Moderate radiation hardness up to 3 krad/year (30 Gy/year) electromagnetic and 10$^{10}$~n/cm$^2$ hadronic at the top luminosity.
\end{itemize}


Only light-collecting calorimeters for the EIC have been considered in this 
report~\footnote{An alternative approach to the EIC spectrometer - {\it TOPSiDE}~\cite{Repond:2018kap} features most of the detectors based on silicon, including calorimeters with very fine granularity.
   More details can be found in Section~\ref{part3-sec-Det.Aspects.CAL_TOPSIDE}}
ECAL will be located in a strong magnetic field - in the bore of the solenoid, 
or in the stray field of $>0.1~$T. Therefore, regular PMTs cannot be used. Silicon photomultipliers (SiPM) are 
considered the most promising photosensor for ECAL. Compared to PMTs, SiPMs take much less longitudinal space,
mostly for the readout electronics, which is an advantage at EIC. Based on the current experience the SiPM readout
and the services (cables, cooling pipes etc) may take about 15~cm of the longitudinal space.

\subsection{ECAL: Requirements, Options and Features}
\label{sec:part3-Det.Aspects.ECAL-overviewtech}

The important parameters of calorimeters are:
\begin{itemize}
  \item{} {\bf Energy resolution}. The commonly used approximation for a
    particle of energy E is: 
    \begin{equation}
      \label{eq:part3-Det.Aspects.ECAL-overviewtech-eres}
      \sigma{}/E = \alpha \oplus \beta{}/\sqrt{E} \oplus \gamma{}/E .  
    \end{equation}
    The term $\gamma$ depends on the noise level and is typically small 
    for photosensors with high gains. The constant term $\alpha$ depends on 
    a number of factors, including the calorimeter thickness
    (on the leakage of showers outside of the calorimeter active area),
    and also on the quality of the detector calibration. For ECALs with
    hundreds of channels or more, typically $\alpha{}>1\%$\cite{Aubert:2001tu,TheBABAR:2013jta}.
    The stochastic term $\beta$ depends on the technology used 
    (the sampling ratio, the size of the signal observed etc.).  
  \item{} {\bf Position resolution} of the
    particle impact. An approximation is used: 
    \begin{equation}
      \label{eq:part3-Det.Aspects.ECAL-overviewtech-xres}
      \sigma_X = \delta{} \oplus{} \epsilon{}/\sqrt{E} \oplus{} \Delta\cdot{}\sin{\theta_I} . 
    \end{equation}
    The resolution depends on the granularity (for
    ECAL limited by the Moli\`{e}re radius) and the energy
    resolution. The coefficients $\delta$ and $\epsilon$ are approximately 
    proportional to the cell size. The third term describes the dependence on the
    angle $\theta_I$ between the incoming particle direction and 
    the longitudinal axis of the calorimeter cell. The coefficient $\Delta$ is approximately $X_0$, where 
    $X_0$ is the average radiation length of the calorimeter material~\cite[p.~527]{Aphecetche:2003zr}.
  \item{} {\bf Lowest detectable energy} depends on the signal
    size versus the noise and low-energy background.  
  \item{} {\bf Electron/pion separation}
    mostly depends on the energy resolution and the
    longitudinal segmentation (if any).  
  \item{} {\bf Two-photon separation}. Two photons not hitting adjacent cells can be separated at the clustering level. 
          An analysis of the shower profile allows to separate photons hitting adjacent cells, provided the hits 
          are at least one cell-size apart.
  \item{} {\bf Detector longitudinal size}. A denser material allows to make the detector shorter for
       the given thickness in radiation lengths. The resolution may depend on the thickness.
  \item{} {\bf Signal timing}. A long signal may affect the
       signal/noise ratio and the pattern recognition.

\end{itemize}

The energy resolution of any calorimeter depends on:
\begin{itemize}
   \item{} Uniformity of the measured response across the volume of the
           detector. The effect may be important both in high-resolution homogeneous
           calorimeters, in particular of a trapezoidal shape\cite{Britton:2005bt, Diehl:2017mmg}, 
           and in medium-resolution sampling calorimeters\cite{Aidala:2020toz, Woody:dwgcaljul}. 
   \item{} Shower containment. 
           In a shorter calorimeter the fluctuations of the shower leakage 
           lead to a higher constant term $\alpha$ and a worse
           resolution at high energies. 
           The dependence of the energy resolution of the calorimeter on its depth 
           in radiation length was calculated in Ref.~\cite{Bazilevsky:dwgcalaug11}.
           For the expected energy range of $E<20$~GeV the impact of the downstream 
           leakage would not significantly change the resolution, for a thickness:\\
           \centerline{
           \begin{tabular}{c|ccc}
             $\beta$            & 2.5\%   & 7\%     &  12\%  \\ \hline
             thickness in $X_0$ & $>$22   & $>$20   & $>$ 18 \\
           \end{tabular}
           }
           \\
           The dependence of the constant term of a sampling calorimeter with a $0.25X_0$ layer thickness
on the overall thickness $x=X/X_0$ has been calculated~\cite[p.~12]{eRD1:talkjul20} in a range of 18-24      (Fig.~\ref{fig:part3-Det.Aspects.ECAL-shashlyk_resol_simul}). The result is well fit using a polynomial $\alpha\approx{}(1.31-0.251(x-20)+0.0144(x-20)^2)\%$ and can be extrapolated  
           to a wider range as 14-28.
   \item{} Signal size. More photoelectrons/GeV lead to
       smaller relative fluctuations and a lower impact of noise. A typical
       yield of a classic lead glass calorimeter is about 1000~p.e./GeV
       providing fluctuations of RMS=3\% at 1~GeV, to be compared with the
       factor $\beta$. For high resolution calorimeters of $\beta < 3$\% the
       yield should be higher.
   \item{} The readout threshold may be important since a shower splits 
         between several cells. It is selected depending on the noise and background.
\end{itemize}

Numerous ECAL technologies have been developed for the field and the development is still ongoing.
A number of technologies have been studied and developed in the
framework of the EIC R\&D, project eRD1\cite{eRD1:jul20}. The results have
been used in this report. The technologies considered are discussed in more details in
in Section~\ref{sec:part3-Det.Aspects.ECAL-technologies}.

\subsubsection{Homogeneous Calorimeters}
\label{sec:part3-Det.Aspects.ECAL-homegeneous}

Typically, the best energy resolution is obtained with homogeneous detectors not affected by 
the sampling fluctuations. Heavy scintillating materials produce large signals per MeV absorbed,
leading to a good resolution. The best results have been achieved so far with scintillating crystals.
Detectors using the Cherenkov light in heavy glass provide a medium resolution. 
\begin{itemize}
  \item{} {\bf PbWO$_4$}.  
      A combination of the requirements for the resolution,
     compactness, radiation hardness, the signal length, as well as the cost and availability
     considerations led to one candidate among the scintillating crystals: lead tungstate PbWO$_4$ (see Sec.~\ref{sec:part3-Det.Aspects.ECAL-pbwo})
     - a mature technology used in many experiments (Tab.~\ref{tab:part3-Det.Aspects.ECAL-tech_pbwo}).
     It typically provides $\beta \approx{}2.5$\%. 
  \item{} {\bf Scintillating glass}.  
     A search for a new, cheaper material - scintillating glass (see Sec.~\ref{sec:part3-Det.Aspects.ECAL-scglass}) - 
     is being pursued   
     in the framework of eRD1~\cite{eRD1:jul20}. Such a material may provide a resolution comparable with the lead tungstate. 
     The material is less dense than lead tungstate and would require more space for the same thickness in $X_0$. A potential
     advantage with respect to lead tungstate would be a lower cost and higher availability. 
  \item{} {\bf Lead glass}.
     This technology uses the Cherenkov light produced in glass containing lead oxide (see Sec.~\ref{sec:part3-Det.Aspects.ECAL-LG})
     and provides a medium resolution of $\beta\approx{}6$\%. Lead glass is less dense than lead tungstate and would require more space. 
     It has been widely used in experiments 
     since the 1960's, and some of those detectors may become available for re-use at EIC.
\end{itemize}

\subsubsection{Sampling Calorimeters}
\label{sec:part3-Det.Aspects.ECAL-sampling}

The resolution of sampling detectors may vary $\beta{}\sim{} 5-15\%$ depending on the sampling fraction 
and the granularity of the active and passive material: 
\begin{itemize}
   \item{} Sampling fraction $f_{samp}$ is the fraction of the total energy released in the active material, 
           evaluated typically for MIPs.
           For a better resolution one needs a larger sampling fraction, 
           which typically increases the detector length for the same thickness in $X_0$.
   \item{} Sampling frequency is related to the thickness of one "layer" of the absorber 
           and the active material (scintillator). This parameter is well defined for the "sandwich"-type geometry. 
\end{itemize}
It has been argued \cite[p.~119]{Wigmans:2018fua} that the stochastic coefficient is approximately proportional to 
$\sqrt{d[mm]/f_{samp}}$, where $d$ is the thickness of the active material layer (or the fiber's diameter). 
This subject is discussed in 
Section~\ref{sec:part3-Det.Aspects.ECAL-shashlyk} ({\it shashlyk} subsection), 
Figure~\ref{fig:part3-Det.Aspects.ECAL-shashlyk_resol} and Equation~\ref{eq:part3-Det.ECAL_resol_shashlyk}.

The requirements for the resolution and radiation hardness favor the absorber-scintillator combination. The popular 
technologies are:
\begin{itemize}
  \item{} {\bf Absorber/Scintillating Fibers: Pb/ScFi or W/ScFi}. The fibers are embedded into 
         a heavy material as lead or tungsten 
         (see Sec.~\ref{sec:part3-Det.Aspects.ECAL-scfibers}). In one implementation the fibers are glued between lead sheets. 
         Such {\it SPACAL-type} detectors have been used in a number of 
         experiments~\cite{Acosta:1990qs, Antonelli:1994kf, Beattie:2018xsk}. 
         In another implementation tungsten powder is used 
         for the absorber. This technology~\cite{Aidala:2020toz} has been developed for the sPHENIX experiment. The resolution
         depends of the fiber density and the absorber material and may vary in a range of $\beta = 6-15$\%. A better resolution
         is provided by a less dense detector. 
  \item{} {\bf Shashlyk} - a stack of absorber and scintillator plates (see Sec.~\ref{sec:part3-Det.Aspects.ECAL-shashlyk}).
       The light is collected with the help of WLS fibers passing through the plates. For the absorber lead or tungsten are used. 
       The technology is widely used and allows detectors of various resolutions and sizes 
       (see Tab.~\ref{tab:part3-Det.Aspects.ECAL-tech_shashlyk}). The resolution depends on the thickness of the plates
       and may vary between $\beta = 5-15$\%. Tungsten for the absorber material provides a high density and a short
       length of the calorimeter.  
\end{itemize}

\subsubsection{ECAL technologies considered for EIC}
\label{sec:part3-Det.Aspects.ECAL-considered}

Technologies which may fit the EIC requirements are listed in Table~\ref{tab:part3-Det.Aspects.ECAL-considered}.
\small
\begin{table}[htb]
 \small
 \begin{center}
  \scalebox{1.0}{
   \begin{tabular}{rl|cccccccc|cc}
     \hline
     \hline
     \# & Type     & samp- & $f_{samp}$ & $X_0$ & $R_M$ & $\lambda_I$ & cell           & $\frac{X}{X_0}$ & $\Delta{}Z$   & 
           \multicolumn{2}{|c}{$\sigma_E/E$, \%} \\ \cline{11-12}
        &          & ling, mm  &   &  mm   & mm   &  mm         & mm$^2$   &    &   cm   & $\alpha$ & $\beta$ \\
     \hline
     1 & W/ScFi$^{**}$    & $\oslash{}$0.47 ScFi & 2\% & 7.0  & 19  & 200 & $25^2$  & 20  & 30 & 2.5 & 13     \\
       &                 & W powd.              &     &      &     &     &         &     &     &    &        \\
     \hline
     2 & PbWO$_4^{***}$   & -                    & -   & 8.9  & 19.6 & 203 & $20^2$ & 22.5 & 35 & 1.0 & 2.5  \\
     \hline
     3 & Shashlyk$^{***}$ & 0.75 W/Cu$^a$       & 16\% & 12.4 & 26   & 250 & $25^2$ & 20  & 40 & 1.6 & 6.3  \\
       &                 & 1.5 Sc              &     &      &       &     &       &     &    &     &      \\
     \hline
     4 & W/ScFi$^{**}$    & 0.59$^2$ ScFi       & 12\% & 13  &  28   & 280 & $25^2$  & 20  & 43 & 1.7 & 7.1     \\
       & with PMT        & W powd.             &      &     &       &     &        &     &    &     &      \\
     \hline
     5 & Shashlyk$^{***}$ & 0.8  Pb             & 20\% & 16.4 & 35   & 520 & $40^2$  & 20  & 48 & 1.5 & 6  \\
       &                 & 1.55 Sc             &      &      &      &     &        &     &    &     &      \\
     \hline
     6 & TF1 Pb glass$^{***}$ & -               & -    & 28   & 37   & 380 & $40^2$ & 20  & 71  & 1.0 & 5-6   \\
     \hline
     7 & Sc. glass$^{*b}$ & -                   & -    & 26   & 35   & 400 & $40^2$ & 20  & 67  & 1.0 & 3-4   \\
     \hline
     \multicolumn{11}{l}{
        \footnotesize
        \begin{tabular}{@{}l@{}}
          *** ~ Mature technology, well understood. used in several experiments \\
          **~ ~ New technology, proven in test beams , in production for experiments \\
          *~~ ~ Technology under development, not fully proven in test beams \\
          a~~ ~ Material 80\% W + 20\% Cu by volume, $X_0$=4.1~mm \\
          b~~ ~ The parameters of scintillating glass are tentative, see Section~\ref{sec:part3-Det.Aspects.ECAL-scglass}. \\
        \end{tabular}
        \small
     }\\
   \end{tabular}
  }
 \end{center}
\normalsize
 \caption{The technologies promising for ECAL, ordered by the radiation length of the material. 
         The Moli\`{e}re radius $R_M$ is defined as $R_M=X_0\cdot{}21~$MeV$/E_{crit}$ and calculated for
         mixtures according to Ref.~\cite{Zyla:2020zbs} (Eq.~34.37--34.38).
         $X/X_0$ is the thickness of the active area measured in radiation lengths, selected to provide
         the resolution presented in the table. 
         A shorter active area would increase the constant term $\alpha$.   
         $\Delta{}Z$ denotes the full length of the module calculated as $X+15$~cm, where 15~cm is reserved for 
         everything but the active area and includes the photosensors, the readout electronics, the cables and services,
         and the support structure. The resolution is parametrized 
         using Equation~\ref{eq:part3-Det.Aspects.ECAL-overviewtech-eres}. The ``noise'' factor $\gamma$ depends 
         on the type of the photosensor, for SiPM $\gamma{}\approx{}0.01$~GeV is expected. 
 }
 \label{tab:part3-Det.Aspects.ECAL-considered}
\end{table}
\normalsize

Comments to Table~\ref{tab:part3-Det.Aspects.ECAL-considered}:
\begin{enumerate}
  \item{} Such a W/ScFi detector is being built for sPHENIX~\cite{Aidala:2020toz}. The properties have been measured in test beams.
  \item{} PbWO$_4$ crystals have been used in a number of experiments (Tab.~\ref{tab:part3-Det.Aspects.ECAL-tech_pbwo}) and 
          typically provide such properties.
  \item{} Such a 20$X_0$ calorimeter would fit into 40~cm space. The W/Sc sampling is similar to 
          the Pb/Sc sampling of \#5. The resolution coefficients $\alpha$ and $\beta$ have been
          evaluated using Eq.~\ref{eq:part3-Det.ECAL_resol_shashlyk}. In order to account for calibration uncertainties
          1\% was added to the constant term: $\alpha \rightarrow \alpha \oplus{} 0.01$. 
  \item{} Such a W/ScFi prototype has been built and the properties measured in a test beam~\cite{eRD1:jul16}. 
          It used a long light guide and a PMT. The sampling can be adjusted to fit into a shorter space, as 40~cm.
  \item{} Such a Pb/Sc {\it shashlyk} calorimeter (but 23$X_0$) is used in the COMPASS experiment~\cite{Abbon:2014aex}. 
          The constant term $\alpha$ is scaled to a shorter calorimeter of 20$X_0$. See also Table~\ref{tab:part3-Det.Aspects.ECAL-tech_pbwo}.
  \item{} TF1 glass has been used in many experiments (see Ref.~\cite{Abbon:2014aex, Aphecetche:2003zr} for example). 
          Cherenkov light is detected. 
          For details see Section~\ref{sec:part3-Det.Aspects.ECAL-LG}.
  \item{} Several types of Scintillating glass are being tested~\cite{eRD1:jul20}. 
          For details see Section~\ref{sec:part3-Det.Aspects.ECAL-scglass}.
\end{enumerate}

The technologies listed can provide the energy resolution close to the
initial requirements (Table~\ref{tab:part3-Det.Aspects.ECAL-req}). 
The PbWO$_4$ crystals nearly fit the requirements for the $-4<\eta<-2$ area. 
The $1<\eta<4$ area requires a medium resolution, and a high
granularity, which implies a dense material. The choice of the technologies for the $-2<\eta<4$ areas 
will depend on the geometrical constraints of the spectrometer and the space allocated. 

All the described technologies are considered radiation hard for the radiation 
levels expected at the EIC. 

\subsubsection{Impact of the material in front of ECAL}
\label{sec:part3-Det.Aspects.ECAL-material-impact}

\begin{figure}[htb]
\begin{center}
  \begin{minipage}{0.325\linewidth}
    \centerline{
      \tikzstyle{background grid}=[draw, black!50,step=.5cm]
      \begin{tikzpicture}[x=0.1\linewidth, y=0.1\linewidth]
        \node [inner sep=0pt,above right]
            {\includegraphics[viewport=25 11 512 540,clip,angle=0,width=0.95\linewidth]{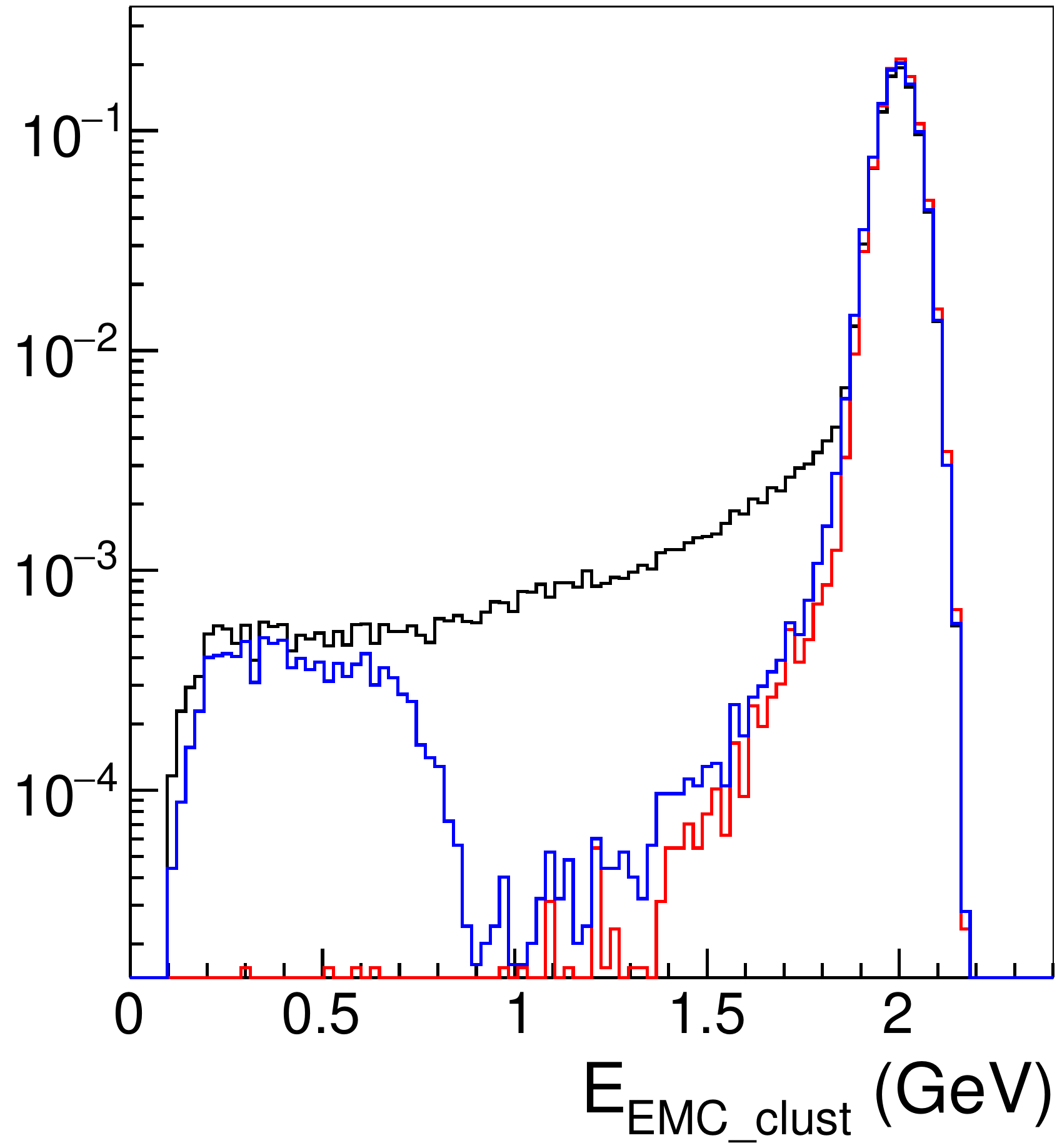}};
        \node (p0) at (1.3,9.6 ) [right, rotate=0] {\scalebox{0.9}{\color{black}{$\eta=-3$~~~ $e^-$ 2~GeV}}};  
        \draw [red,thick] (1.5,8.8) -- +(0.8,0);
        \node (p1) at (2.4,8.8 ) [right, rotate=0] {\scalebox{0.9}{\color{red}{ no radiation}}};  
        \draw [black,thick] (1.5,8.0) -- +(0.8,0);
        \node (p2) at (2.4,8.0 ) [right, rotate=0] {\scalebox{0.9}{\color{black}{ radiation}}};  
        \draw [blue,thick] (1.5,7.2) -- +(0.8,0);
        \node (p3) at (2.4,7.2 ) [right, rotate=0] {\scalebox{0.9}{\color{blue}{ radiation, corr.}}};  
      \end{tikzpicture}
    }
  \end{minipage}
  \begin{minipage}{0.650\linewidth}
    \centerline{
      \tikzstyle{background grid}=[draw, black!50,step=.5cm]
      \begin{tikzpicture}[x=0.1\linewidth, y=0.1\linewidth]
        \node (p0) at (0.3,2.0) [right, rotate=90] {\scalebox{1.0}{\color{black}{Losses}}};  
        \node at (0.5,0) [inner sep=0pt,above right]
           {\includegraphics[viewport=49 11 514 490,clip,angle=0,width=0.47\linewidth]{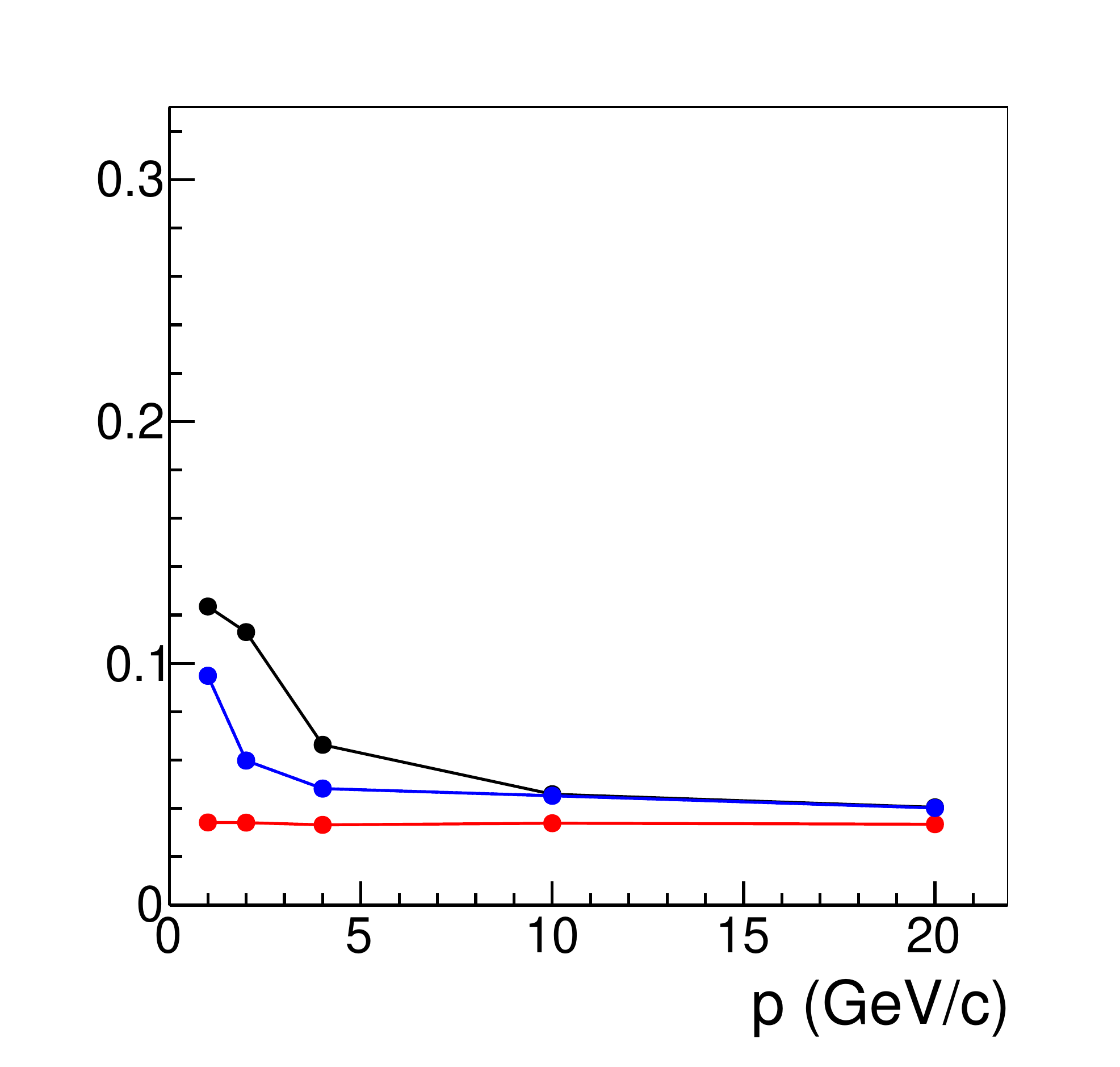}};
        \node at (5.3,0) [inner sep=0pt,above right]
           {\includegraphics[viewport=49 11 514 490,clip,angle=0,width=0.47\linewidth]{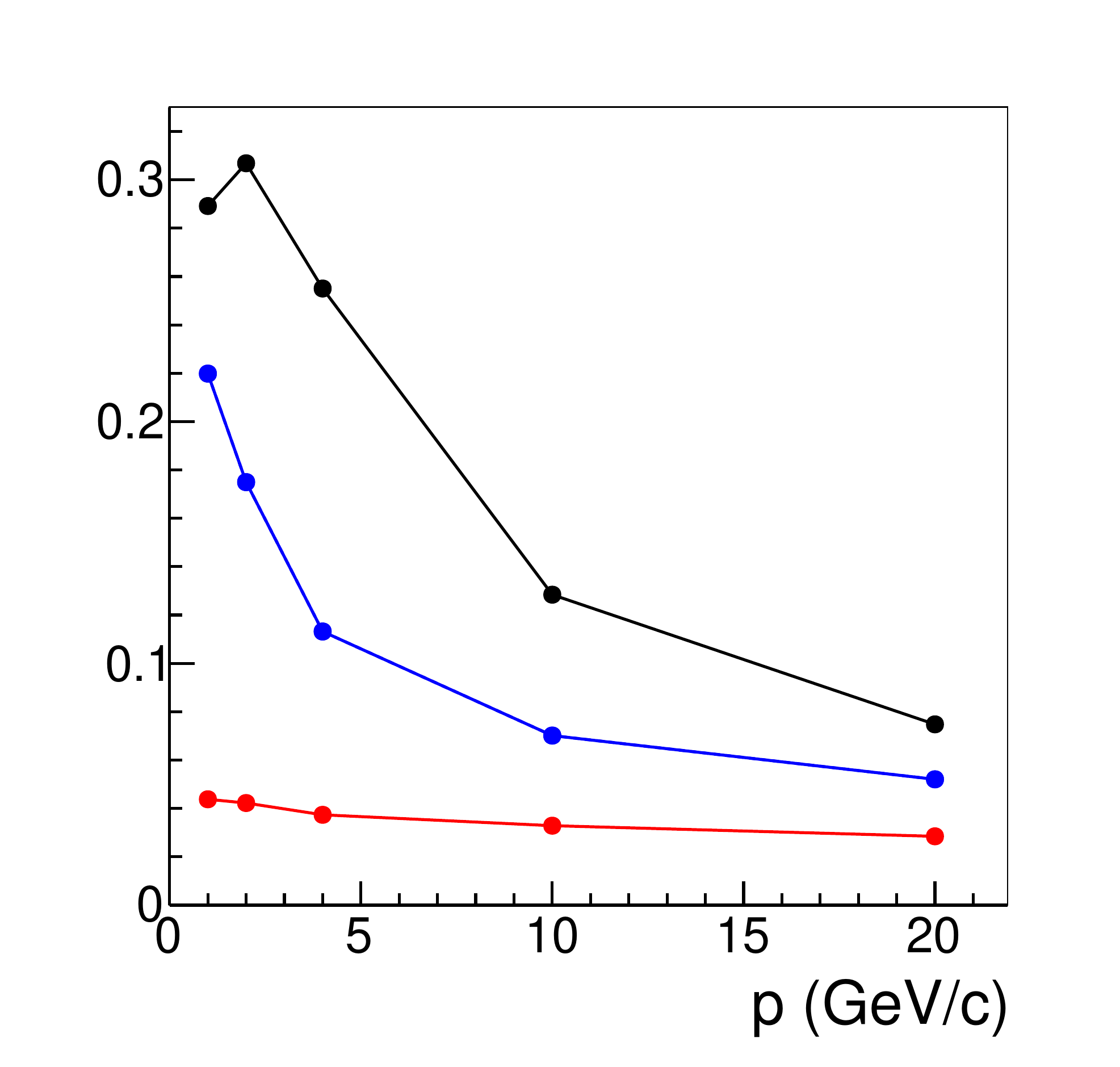}};
        \node (p1) at (1.5,4.4 ) [right, rotate=0] {\scalebox{1.0}{\color{black}{$\eta=-3$, $\approx{}4.5\%X_0$}}};  
        \node (p1) at (6.5,4.4 ) [right, rotate=0] {\scalebox{1.0}{\color{black}{$\eta=-2$, $\approx{}9.\%X_0$}}};  
      \end{tikzpicture}
    }
  \end{minipage}

\end{center}
\caption{Simulated impact of material in front of a PbWO$_4$-crystal ECAL on detection of 2~GeV electrons~\cite{Bazilevsky:dwgcaljul14}. 
         The amount of material depends on $\eta$.  
         Three cases are considered: a) electron does not radiate in the material;
         b) electron radiates and only the cluster associated with electron track is considered; c) attempt made to recover the photons radiated in
         the thick objects upstream, whose positions can be predicted.  Practically all the energy has been recovered up to
         a loss of a half of the initial energy, which was an arbitrary cutoff. The real cutoff will depend on the background,
         the tracking quality etc.\\ 
         {\bf Left}: The energy spectrum of the cluster in ECAL from GEANT4~\cite{Agostinelli:2002hh} simulation.\\
         {\bf Middle} and {\bf Right}: the losses of electrons, selected using $E/p > 1 - 2\sigma_E$.   
        }
\label{fig:part3-Det.Aspects.ECAL-material-impact} 
\end{figure}

A certain amount of material will be distributed along the path of particles from the interaction
point to the face of ECAL. The electrons radiate and the photons convert to pairs. 
Because of the magnetic field the radiated photons may hit the calorimeter at a distance from
the impact of the electron. Simulated signals are shown in 
Figure~\ref{fig:part3-Det.Aspects.ECAL-material-impact}. The detected energy distribution has a tail
to lower energies. A typical identification criteria for electrons $E/p>1-2\sigma_E$ may lead
to losses of 5-30\%, in particular at low momenta 
(Fig.~\ref{fig:part3-Det.Aspects.ECAL-material-impact}). The losses can be partly recovered,
since the material is expected to be concentrated at certain places, allowing 
to predict the impact position of radiated photons for a given particle trajectory. 
Still losses of 10-20\% are expected for certain areas at $p<10$~GeV.

\subsubsection{Impact of the Cell Size and the Projective Geometry}
\label{sec:part3-Det.Aspects.ECAL-cellsize}

In order to have the best coordinate resolution while minimizing the number of the readout channels the cell transverse size
is usually selected close to the Moli\`{e}re radius of the calorimeter material.
The coordinate resolution depends on the position of the hit and is the best at the boundary between two cell. The average
resolution depends on the particle energy and the incident angle 
(see Equation~\ref{eq:part3-Det.Aspects.ECAL-overviewtech-xres}).  
Based on experience (see Table~\ref{tab:part3-Det.Aspects.ECAL-xres}) we may expect a resolution for the normal
incident angle $\theta_I$ of about $(1\oplus{}3/\sqrt{E/1\mathrm{GeV}})$~mm for the cell size 20-25~mm, and
$(1\oplus{}6/\sqrt{E/1\mathrm{GeV}})$~mm for the cell size of about 40~mm. Let us consider a dense detector
with $X_0\approx{}10$~mm and the cell size of 25~mm.
In the non-projective geometry, at $\theta_I=45^\circ$,
the additional term $X_0\sin{\theta_I}\approx{}7$~mm will dominate the coordinate resolution.
The relative deterioration of the resolution does not depend strongly on the density of the material. 

\begin{table}[htb]
 \begin{center}
  \scalebox{0.9}{
   \begin{tabular}{l|rcccrc}
     \hline
     \hline
        Type     & $R_M$, & cell size, & $\sigma_E/E$ & $\delta$ & $\epsilon$, mm  & Ref \\
                 &  mm    &   \multicolumn{1}{c}{mm}       
                 &  \multicolumn{1}{c}{at 1~GeV}   &   mm     
                 & \multicolumn{1}{c}{GeV$^{0.5}$} &     \\
     \hline
       PbWO$_4$  & 20   & ~20         &  2.9\%      &  0.4     & 2.6         & \cite{Shimizu:2000jk} \\
       PbWO$_4$  & 20   & ~22         &  3.9\%      &  0.3     & 2.6         & \cite{Conesa:2005wj} \\
       TF1       & 37   & ~38         &  5.7\%      &  0.5     & 6.0         & \cite{Abbon:2007pq} \\
       Shashlyk  & 41   & ~55         &  8.4\%      &  1.6     & 5.7         & \cite{Aphecetche:2003zr} \\
       Shashlyk  & 59   & 110         &  4.7\%      &  3.3     & 15.4        & \cite{Kharlov:2008tw}\\
     \hline
   \end{tabular}
  }
 \end{center}
 \caption{The coordinate resolutions observed with several detectors for the normal incident angle $\theta_I$. 
         The resolution is parametrized using
         Equation~\ref{eq:part3-Det.Aspects.ECAL-overviewtech-xres}. The stochastic factor $\epsilon$ appears
         to be approximately proportional to the cell size.   
 }
 \label{tab:part3-Det.Aspects.ECAL-xres}
\end{table}

\begin{figure}[htb]
\begin{center}
  \begin{minipage}{0.49\linewidth}
    \includegraphics[angle=0,width=0.97\linewidth]{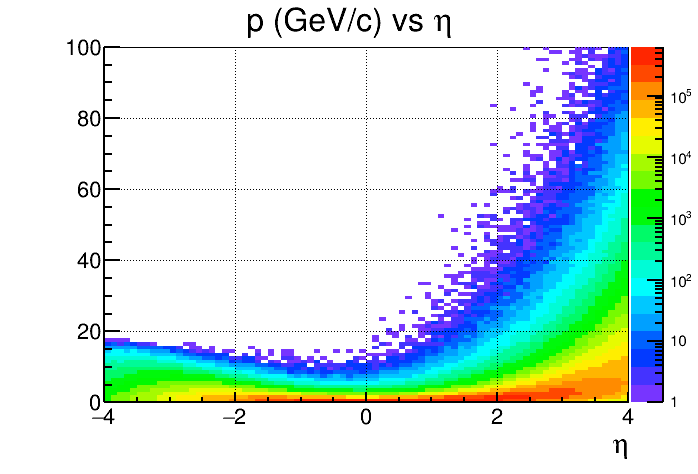}
  \end{minipage}
  \begin{minipage}{0.49\linewidth}
    \includegraphics[angle=0,width=0.97\linewidth]{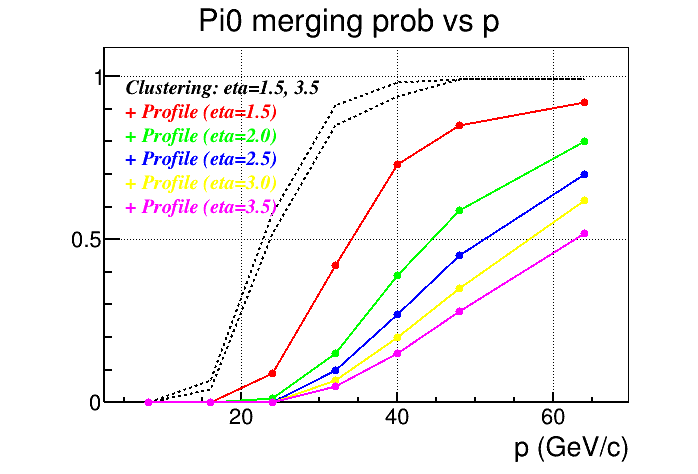}
  \end{minipage}
\end{center}
\caption{{\bf Left}: The calculated $\pi^0$ momentum spectrum for SiDIS at $e+p~18\times{}275$~GeV collisions,
                using PYTHIA~\cite{Sjostrand:2007gs}.
         {\bf Right}: The probability of two photons to merge, calculated~\cite{Bazilevsky:dwgcalaug18} using
                GEANT4~\cite{Agostinelli:2002hh}
                for the cell size of $25\times{}25$~mm$^2$ located at 3~m from
                the interaction point, for the non-projective geometry. For the projective geometry the
                results for $\eta{}>3.5$ would be close to the non-projective curve at for $\eta$=3.5.  
        }
\label{fig:part3-Det.Aspects.ECAL-pi0detection} 
\end{figure}

 
Another important characteristics of ECAL is the ability to discriminate a single photon from a
merged photon pair from a high momentum $\pi^0$ meson decay. For a high momentum
$\pi^0$ the minimal angle between two photons in the Lab frame is
$\approx{}2m_{\pi^0}/p_{\pi^0}$ and most of the decays produce two photons
at angles close to the minimal angle.
At high enough momentum
two photons appear in the ECAL in a close proximity to each
other, so that the ECAL response to a pair of decay photons becomes
indistinguishable from the response to a single photon with the energy
equal to a sum of decay photon energies. ECAL granularity defines the
highest momentum at which ECAL can discriminate single photon from
merged photons from $\pi^0$ meson decay. Usually, two photons are
easily distinguishable in the ECAL when they are separated at least
by a distance equal to twice of cell size. In this case two photons
produce two clusters in ECAL, or a single cluster with two distinct
local maxima. With smaller distance between two photons, they produce
a single cluster with one local maximum. Even in this case, different
mathematical techniques to analyze the energy distribution among the
cluster cells still can discriminate a single photon cluster from a
merged photon cluster, down to a distance between two photons equal
to the cell size, or even down to a half of the cell size, though with
limited efficiency. Figure~\ref{fig:part3-Det.Aspects.ECAL-pi0detection}
 illustrates such a capability for the
hadron endcap ECAL with the cell transverse size of 2.5~cm, located at 3~m
from the collision point. The performance deteriorates for a
non-orthogonal impact (here at lower $\eta$), due to a wider shower
profile and its larger fluctuations in the ECAL transverse plane. For
a transverse size $d$ and the distance
to the collision point $Z_{ECAL}$, the momentum reach for $\pi^0 /
\gamma$ discrimination scales roughly as $Z_{ECAL}/d$.

The requirements to the hadron endcap strongly favor a calorimeter material with a short radiation length and 
a small Moli\`{e}re radius, allowing a fine segmentation of $\leq{}25$~mm. 

Using the projective geometry for ECAL in the barrel is standard for solenoid-based spectrometers. 
For the endcaps it is geometrically more complex. The projective geometry would provide
a significantly better coordinate resolution at large radii.

\subsubsection{Electron/pion separation}
\label{sec:part3-Det.Aspects.ECAL-e-pi-separ}

\begin{figure}[htb]
\begin{center}
  \begin{minipage}{0.38\linewidth}
    \centerline{
      \includegraphics[angle=0,width=1.00\linewidth]{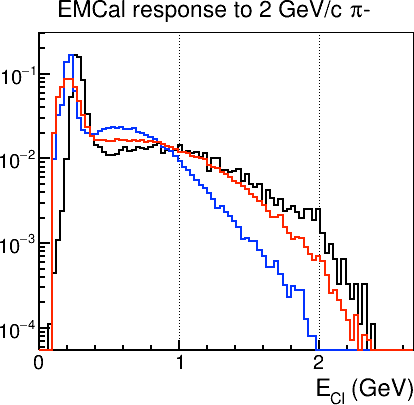}
    }
  \end{minipage}
  \begin{minipage}{0.42\linewidth}
    \centerline{
      \tikzstyle{background grid}=[draw, black!50,step=.5cm]
      \begin{tikzpicture}[x=0.1\linewidth, y=0.1\linewidth]
        \node [inner sep=0pt,above right]
            {\includegraphics[angle=0,width=0.90\linewidth]{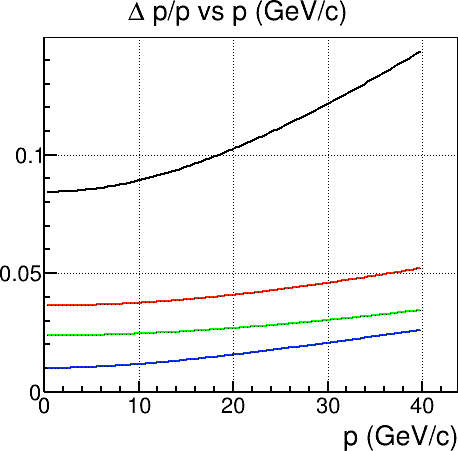}};
         \node (p0) at (8.0,7.0 ) [rotate=0] {\scalebox{1.0}{\color{black}{$\eta$}}};  
         \node (p1) at (8.0,6.0 ) [rotate=0] {\scalebox{1.0}{\color{black}{-3.5}}};  
         \node (p2) at (8.0,5.4 ) [rotate=0] {\scalebox{1.0}{\color{red}{-3.0}}};  
         \node (p3) at (8.0,4.7 ) [rotate=0] {\scalebox{1.0}{\color{green}{-2.0}}};  
         \node (p3) at (8.0,4.1 ) [rotate=0] {\scalebox{1.0}{\color{blue}{-1.0}}};  
      \end{tikzpicture}
    }
  \end{minipage}
\end{center}
\caption{
         Ref.~\cite{Bazilevsky:dwgcalmay21, Bazilevsky:dwgcaljun2}. 
         {\bf Left}: The response to 2~GeV pions for: black - measured for Pb/Sc $8\%/\sqrt{E}\oplus{}2\%$ (PHENIX);
               red - simulated (GEANT4~\cite{Agostinelli:2002hh}) for W/ScFi $13\%/\sqrt{E}\oplus{}3\%$; blue - simulated for PbWO$_4$ $2.5\%/\sqrt{E}\oplus{}1\%$.\\  
         {\bf Right}: Assumed momentum resolution (RMS) for charged particles.
        }
\label{fig:part3-Det.Aspects.ECAL-e-pi-separ0} 
\end{figure}

The DIS momentum spectra of the DIS electrons and pions are shown in
Fig.~\ref{fig:part3-Det.Aspects.ECAL-DIS-background}. At lower momenta
the pion flux dominates the flux of scattered electrons by orders of
magnitudes. ECAL is expected to be the main tool for the electron identification.
Pions produce smaller signals in ECAL than electrons of the same momentum
(Fig.~\ref{fig:part3-Det.Aspects.ECAL-e-pi-separ0}, left).
Using the measured momentum of the charged track $p$ and the energy deposited by
this track in ECAL one can select electrons requiring  
$E/p>1-\Delta$. The fluctuations of the $E/p$ value are characterized by 
$\sigma(E/p)=E/p(\sigma_E/E\oplus{}\sigma_p/p)$, where $\sigma_p/p$ 
are expected to be significant at $|\eta|>2$.
In this review we use typically $\Delta{}=1.6 \sigma_E/E$,
using only the Gaussian width of the calorimeter signal. For the 
Gaussian calorimeter response the efficiency for electrons would
be 95\%. However, the response typically has a tail extending to lower energies, 
increased by material in front (Fig.~\ref{fig:part3-Det.Aspects.ECAL-material-impact}),
which reduces the efficiency for electrons. 
Larger $\sigma_E$ and $\sigma_p$ lead to a lower efficiency for electrons 
and a smaller rejection factor for pions for a given $\Delta$.  
 
In general, one expects a better electron-PID performance for a better 
energy resolution of the calorimeter and the momentum resolution of the spectrometer.
Analysis of the shower
profile can provide an additional pion suppression. However, the effect depends on
the impact angles, and therefore, on the geometry of the calorimeter
(projective or not). 

The pion suppression performance of calorimeters has been measured in test beams and  
also evaluated using simulation. One should note that it is challenging to measure
or calculate large rejection factors $R_\pi>1000$ because of beam contamination,
or uncertainties in simulation of hadronic processes.
The pion rejection factor may be limited by charge exchange processes as $\pi^-+p\rightarrow{}\pi^0+n$ that
would produce signals similar to electrons at the same energy (noted in Ref.~\cite{Allen:2009aa} for example).
The cross section for such processes typically falls with energy.
 
\begin{table}[htb]
 \begin{center}
  \scalebox{0.85}{
   \begin{tabular}{ll|rrrrcrrrr}
     \hline
     \hline
        Type  & Experi-    & \multicolumn{3}{c}{$\sigma_E/E$,~\%} & E,     & \multicolumn{2}{c}{$\varepsilon_e$} & par-     & $R_\pi$ & Ref \\
              \cline{3-5} \cline{7-8}
              &       ment & $\alpha$ & $\beta$ & $\gamma$       &  GeV   &  meas.       &  calc.    &    ticle &         &     \\
     \hline
     PbWO$_4$    & -        &  0.1 &  3.1 &    &  1.0-2.5 &        &  98\% & $\pi^-$ &   500 & \cite{Inaba:1994jd} \\
     \hline
     PbWO$_4$    & -        &  0.5 &  4.0 &    &  80.     &  90\%  &       & $\pi^-$ &  6000 & \cite{Alexeev:1996be} \\
     \hline
     TF1         & PHENIX   &  0.8 &  6.0 &    &  1.5-4.0 &  80\%  &  98\% & $\pi^-$ &   250 & \cite{Aphecetche:2003zr} \\
                 &          &      &      &    &          &  90\%  &       &         &   160 &                          \\
                 &          &      &      &    &          &  95\%  &       &         &   100 &                          \\
                 &          &      &      &    &      1.0 &  80\%  &  98\% &         &    80 &                          \\
                 &          &      &      &    &     0.75 &  80\%  &  98\% &         &    45 &                          \\
                 &          &      &      &    &     0.50 &  80\%  &  98\% &         &     7 &                          \\
     \hline
     TF1         & Hall C   &  1.0 &  6.0 &    &   3.2    &  95\%  &       & $\pi^-$ &$^*$200 & \cite{Mkrtchyan:2012zn} \\
     \hline
     Pb/Sc       & PHENIX   &  2.1 &  8.1 &    &  40      &  77\%  &  84\% & $\pi^+$ &   430 & \cite{Awes:2002sb} \\
                 &          &      &      &    &          &  88\%  &  95\% &         &   350 &                    \\
                 &          &      &      &    &          &  92\%  &  98\% &         &   300 &                    \\
                 &          &      &      &    &          &  95\%  & 100\% &         &   200 &                    \\
                 &          &      &      &    &   4      &        &  95\% &         &   100 & unpub              \\
                 &          &      &      &    &   3      &        &  95\% &         &    80 & unpub              \\
                 &          &      &      &    &   2      &        &  95\% &         &    43 & unpub              \\
                 &          &      &      &    &   1      &        &  95\% &         &    12 & unpub              \\
                 &          &      &      &    &   0.5    &        &  95\% &         &   3.4 & unpub              \\
     \hline
     Pb/Sc       & ALICE    &  1.7 & 11.1 & 5.0 & 100     &  90\%  &       & $\pi^-$ &  2000 & \cite{Allen:2009aa} \\
                 &          &      &      &    &  100     &  95\%  &       &         &  1100 &                     \\
                 &          &      &      &    &   40     &  90\%  &       &         &   700 &                     \\
                 &          &      &      &    &   40     &  95\%  &       &         &   400 &                     \\
     \hline
     W/ScFi      & sPHENIX  &  2.8 & 15.5 &    &    8     &        &  50\% & $\pi^-$ &   710 & \cite{Aidala:2017rvg} \\
                 &          &      &      &    &          &        &  84\% &         &   330 &                       \\
                 &          &      &      &    &          &        &  95\% &         &   210 &                       \\
                 &          &      &      &    &          &        &  98\% &         &   160 &                       \\
                 &          &      &      &    &          &       & 99.9\% &         &    90 &                       \\
     \hline
     \multicolumn{11}{l}{\footnotesize * The longitudinal segmentation not used}\\
   \end{tabular}
  }
 \end{center}
 \caption{Measured pion suppression factor $R_\pi$ and the associated efficiency $\varepsilon_e$ to electrons. 
          The shower shape has not been taken into account, except for Ref.~\cite{Alexeev:1996be} (PbWO$_4$).
          In several
          studies the $\varepsilon_e$ was measured and from the data reported it was possible to calculate the ``Gaussian'' 
          efficiency, that is considerably higher than the measured one, as expected. 
          For other studies only the calculated ``Gaussian'' efficiency is available. The measurements marked ``unpub''
          come from the authors of the paper, but have not been included into the paper.
 }
 \label{tab:part3-Det.Aspects.ECAL-e-pi-sep}
\end{table}

Several examples of the measured pion suppression in various calorimeters are 
shown in Table~\ref{tab:part3-Det.Aspects.ECAL-e-pi-sep}. For the sampling calorimeters
the largest reported rejection factor of $R_\pi$=2000, at the measured $\varepsilon_e=90$\%,
was obtained at 100~GeV, where the energy resolution was about 2\%. A rejection of $R_\pi=500$
was measured for a PbWO$_4$ calorimeter at 2.5~GeV, where the energy resolution was 2\%.
In this test a cut $\Delta>2\cdot{}\sigma_E$ was applied (98\% ``Gaussian'' efficiency),
which may translate to a $\varepsilon_e\approx{}90$\% of the real efficiency.
 
\begin{figure}[htb]
\begin{center}
  \begin{minipage}{0.40\linewidth}
    \centerline{
      \tikzstyle{background grid}=[draw, black!50,step=.5cm]
      \begin{tikzpicture}[x=0.1\linewidth, y=0.1\linewidth]
        \node [inner sep=0pt,above right]
            {\includegraphics[angle=0,width=0.90\linewidth]{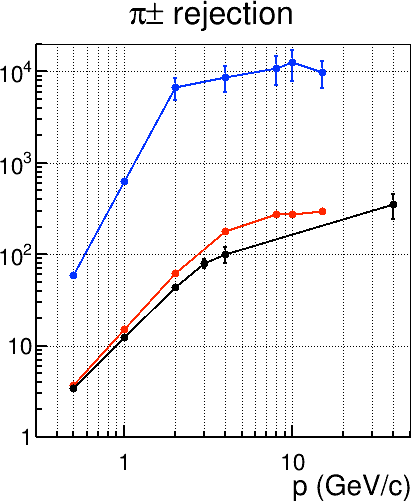}};
         \node (p0) at (6.0,4.5 ) [rotate=0] {\scalebox{0.7}{\color{black}{Pb/Sc~\cite{Awes:2002sb}}}};  
         \node (p1) at (6.2,6.7) [rotate=0] {\scalebox{0.7}{\color{red}{W/ScFi simul}}};  
         \node (pa) at (8.0,5.0) [rotate=0] {\scalebox{0.7}{\color{red}{\cite{Aidala:2017rvg}}}};
         \draw[red] (6.1,6.0) circle (0.7mm);
         \draw[->,red,thick] (pa) edge [] (6.1,6.0);   
         \node (p2) at (6.5,8.5 ) [rotate=0] {\scalebox{0.7}{\color{blue}{PbWO$_4$ simulation}}};  
         \draw[blue] (3.6,6.8) circle (0.7mm);
         \node (p3) at (4.0,7.3 ) [rotate=0] {\scalebox{0.7}{\color{blue}{PbWO$_4$}}};  
         \node (p4) at (3.5,6.3 ) [rotate=0] {\scalebox{0.7}{\color{blue}{\cite{Inaba:1994jd}}}};  
         \draw [blue,very thick] (2.8,6.8) -- +(1.4,0);
      \end{tikzpicture}
    }
  \end{minipage}
  \begin{minipage}{0.40\linewidth}
    \centerline{
      \tikzstyle{background grid}=[draw, black!50,step=.5cm]
      \begin{tikzpicture}[x=0.1\linewidth, y=0.1\linewidth]
        \node [inner sep=0pt,above right]
            {\includegraphics[angle=0,width=0.90\linewidth]{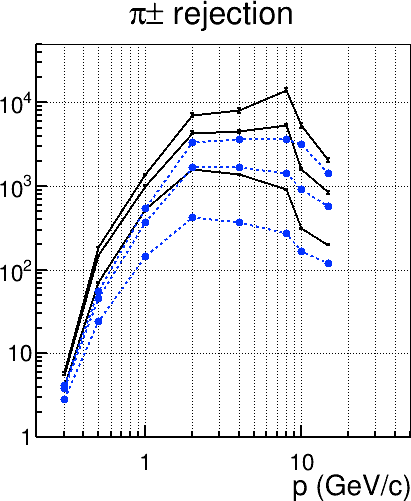}};
         \node (p0) at (8.0,8.0 ) [rotate=0] {\scalebox{0.9}{\color{black}{$\eta$}}};  
         \node (p1) at (8.0,5.0 ) [rotate=0] {\scalebox{0.7}{\color{black}{-3.5}}};  
         \node (p2) at (8.0,6.3 ) [rotate=0] {\scalebox{0.7}{\color{black}{-3.0}}};  
         \node (p3) at (8.0,7.0 ) [rotate=0] {\scalebox{0.7}{\color{black}{-2.5}}};  
         \node (p4) at (4.0,9.5 ) [rotate=0] {\scalebox{0.7}{\color{blue}{PbWO$_4$ simulation}}};  
      \end{tikzpicture}
    }
  \end{minipage}
\end{center}
\caption{Measured and simulated (GEANT4~\cite{Agostinelli:2002hh}) pion suppression~\cite{Bazilevsky:dwgcalmay21, Bazilevsky:dwgcaljun2}, evaluated
         with a $\Delta=1.6\cdot\sigma_E$ cut (``Gaussian'' $\varepsilon_e=95\%$). \\ 
         {\bf Left}: pion suppression, momentum resolution neglected; $E/p$ cut only;\\
           black - measured for Pb/Sc $8.1\%/\sqrt{E}\oplus{}2.1\%$ (PHENIX~\cite{Awes:2002sb}, see also Table~\ref{tab:part3-Det.Aspects.ECAL-e-pi-sep});\\
           red - simulation for W/ScFi $12\%/\sqrt{E}\oplus{}3\%$, calculation compared with a measurement at 8~GeV~(Ref.~\cite{Aidala:2017rvg} and Tab.~\ref{tab:part3-Det.Aspects.ECAL-e-pi-sep});\\
           blue - simulation for PbWO$_4$ $2.5\%/\sqrt{E}\oplus{}1\%$. The calculation, exceeds a measurement at 1-2.5~GeV 
                   (Table.~\ref{tab:part3-Det.Aspects.ECAL-e-pi-sep}) by a factor of $<$10. \\
         {\bf Right}: simulated pion suppression for PbWO$_4$ $2.5\%/\sqrt{E}\oplus{}1\%$ , 
           at $\eta=-3.5, -3.0, -2.5$, the momentum resolution taken into account. The dependence 
           on $\eta$ is caused by the momentum resolution. \\           
           blue - $E/p$ cut, $\varepsilon_e=95$\%; \\
           black -$E/p$ and shape cuts, $\varepsilon_e=92$\%. 
        }
\label{fig:part3-Det.Aspects.ECAL-e-pi-separ1} 
\end{figure}

Figures~\ref{fig:part3-Det.Aspects.ECAL-e-pi-separ0}, \ref{fig:part3-Det.Aspects.ECAL-e-pi-separ1} show
the calculated suppression dependence on energy, the calorimeter resolution
and the track momentum resolution (Fig.~\ref{fig:part3-Det.Aspects.ECAL-e-pi-separ0}, right). 
The simulated pion-produced signals 
in PbWO$_4$ and in sampling detectors are compared -
the former have a shorter tail to high values. A stronger 
response to neutrons by the plastic scintillator than by an inorganic one
may contribute to the effect. While the results of calculations for sampling calorimeters
are consistent with the measurements, the calculated $R_\pi$ for PbWO$_4$ is more than
an order of magnitude higher than a measurement at 2.5~GeV. It may be caused by systematic
uncertainties of both the measurement and simulation. At this time we can not claim 
that a rejection power higher than 1000 is achievable at moderate energies
even with the relatively high-resolution PbWO$_4$ detector. 

Figure~\ref{fig:part3-Det.Aspects.ECAL-e-pi-separ1} (right) demonstrates how the 
momentum resolution affects the performance at small angles $|\eta|>3$.

Calculations also show that taking the shower shape into account can improve the
pion rejection by a factor of 2 even in non-projective geometry, or by a factor of 3-4
at small incident angles and in projective geometry.   

In summary,
in the energy range 4-20~GeV sampling calorimeters and lead glass calorimeters
can provide a pion rejection factor
from a hundred to a thousand. PbWO$_4$ crystals may be able to provide factors 3-5
higher, but factors $>1000$ need to be confirmed by measurements.
 
\begin{figure}[htb]
\begin{center}
  \small
  \begin{minipage}{0.32\linewidth}
\centerline{\small~~ ~~~~~~$\sigma_E/E=2.5\%/\sqrt{E}\oplus{}1\%$}    \end{minipage}
  \begin{minipage}{0.32\linewidth}
\centerline{\small $\sigma_E/E=7\%/\sqrt{E}\oplus{}2\%$}    
  \end{minipage}
  \begin{minipage}{0.32\linewidth}
 \centerline{\small $\sigma_E/E=12\%/\sqrt{E}\oplus{}2\%$~~~~~~~~~~}    
  \end{minipage}
  \includegraphics[viewport=0 0 540 210,clip,angle=0,width=0.90\linewidth]{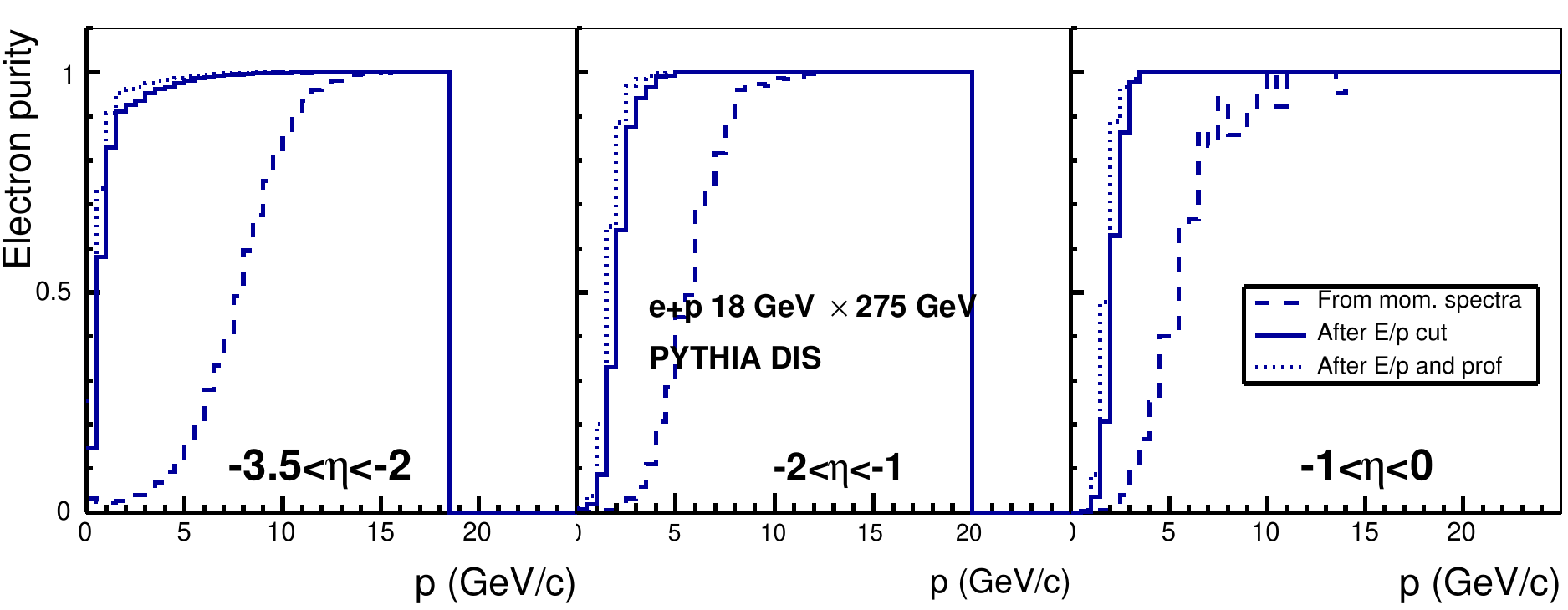}\\
  \includegraphics[viewport=0 0 540 210,clip,angle=0,width=0.90\linewidth]{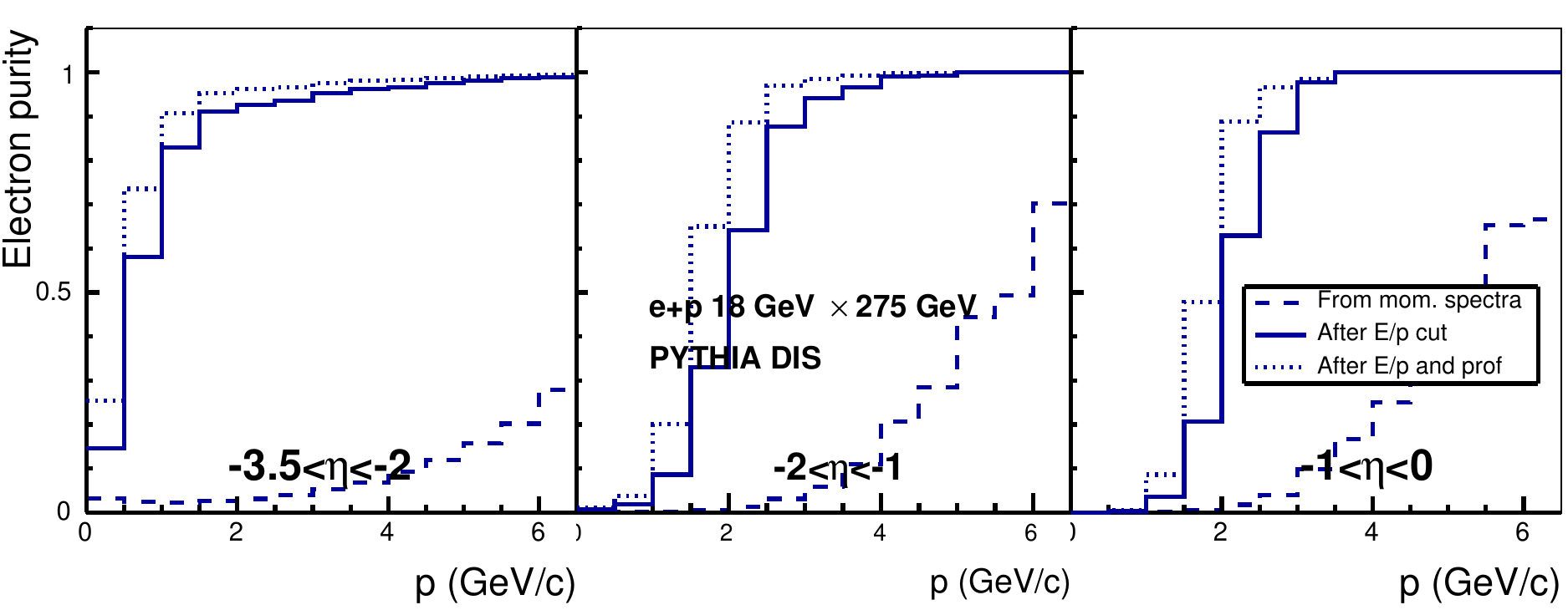}  
\end{center}
\caption{Calculated purity of the DIS electron sample and the effect of the pion 
         suppression~\cite{Bazilevsky:dwgcaljun2} for e+p~18$\times$275~GeV.
         The pion suppression was evaluated using
         a $\Delta=1.6\cdot\sigma_E$ cut (``Gaussian'' $\varepsilon_e=95\%$).
         The columns present
         three areas of $\eta$ with the assigned $\sigma_E/E$ for each area. The bottom
         panel presents the zoomed in plots of the top panel. Dashed lines - no cuts,
         solid lines - $E/p$ cut, dotted lines - $E/p$ and shower shape cuts. A cap of 1000 on 
         the calculated pion rejection was set in order to address the existing uncertainties.
        }
\label{fig:part3-Det.Aspects.ECAL-e-pi-separ2} 
\end{figure}

Figure~\ref{fig:part3-Det.Aspects.ECAL-e-pi-separ2} shows the
calculated purity of electrons in the DIS sample, in 3 areas of
$\eta$, each equipped with ECAL of a different resolution, close to
the specifications in Table~\ref{tab:part3-Det.Aspects.ECAL-req}. The
levels of the pion background are also different in these areas
(Fig.~\ref{fig:part3-Det.Aspects.ECAL-DIS-background}).  
In all areas a $>$95\% purity is reached at $p>4$~GeV. 
The
$-3.5<\eta<-2$ area is supposed to be covered with a high-resolution
ECAL ($\beta=2.5\%$), and the purity $>$90\% is reached at
$E>2$~GeV.  
The other areas are envisioned to be covered with
$\beta=7\%$ and $\beta=12\%$-resolution calorimeters and at $p<4$~GeV
the pion contamination remains high.  

Pion rejection can be improved by using a ``preshower'' detector. 
Also, the calorimeter itself may be equipped with a second
readout from the front part of the modules, which does not require an extra space
for another detector. A factor of 2 improvement in $e/\pi$ 
separation was achieved equipping a {\it shashlyk} detector with scintillator
plates with different emission times~\cite{Benvenuti:2001ss}. 

Since $X_0/\lambda_I$ is smaller for heavier materials, they should provide
a better pion suppression. One should note that the material passed by the electron track
in front of the calorimeter will reduce the energy reaching the calorimeter, 
affecting the $E/p$ ratio and the $e/\pi$ PID. For the same pion rejection factor
the efficiency to electrons will be reduced. A lower efficiency typically causes  
a higher uncertainty of the measurement.

\subsubsection{Lowest detectable energy}
\label{sec:part3-Det.Aspects.ECAL-minenergy}

The lowest detectable energy depends on the amount of light detected versus noise of various origin
and low-energy background.
With PbWO$_4$ as low as 20~MeV photons can be detected, provided low-noise sensors and electronics, although
with a 30-50\% energy resolution. For sampling detectors one may expect the lowest detectable energy of 
50-100~MeV. 

\subsubsection{Readout Considerations}
\label{sec:part3-Det.Aspects.ECAL-readout}

Only detectors with optical readout have been considered. In the
current scenarios the endcap ECAL photosensors will be located in a
magnetic field of $>$0.1~T, which precludes the usage of regular
PMTs. The barrel ECAL is located in a $>$1~T field. At the moment the
sensor of choice is SiPM, which provides a high gain (about $10^6$) and a
medium photodetection efficiency of about 20\%. The drawbacks are
small surface, noise, susceptibility to radiation,
in particular to neutron/proton radiation~\cite{Qiang:2012zh, Biro:2018yxc},
sensitivity to temperature, a small dynamic range, and the intrinsic
nonlinearity~\cite{Chirikov-Zorin:2019pjd}. Radiation leads to a higher noise.
Additionally, the performance degrades with the charge flown through the SiPM~\cite{Tsai:tmplmar21}. 
For the same
amount of light a SiPM can fire a number of pixels comparable
to a PMT photoelectron count~\cite{Anfimov:2011zz}. However, a fraction of the pixels
fire due to the cross talk, not improving the statistical
fluctuations. While a SiPM readout is natural for the fiber
technologies as {\it Shashlyk}, it does not look optimal for a large-surface - 16~cm$^2$
- glass blocks. Such a readout has not been tested yet with 4~cm$^2$ crystals. 

The effect of non-linearity for SiPMs depends on the desired dynamic
range and the calorimeter resolution. Let us consider the requirements
of a 2\% energy resolution at 1~GeV and the maximum energy of 20~GeV (at the
center of the electron endcap), and find the total number of pixels
needed for one calorimeter cell. With the optimal cell size
about 80\% of a shower energy on average goes to one cell, but with considerable 
fluctuations.
The p.e. (or pixel) count at 1~GeV should be ≳10k
(1\% statistical fluctuations). Then, at 20~GeV the pixel count with no
saturation would be about 200k. 
The saturation effects for MPPC S12572-010P 90k-pixel, 3$\times$3~mm$^2$
device have been shown to be as large~\cite{Chirikov-Zorin:2019pjd} 
as would be expected for a 30k-pixel device. The nonlinearity correction has to be calibrated rather than calculated, and may
contain large uncertainties. This may require to limit the number
of fired pixels to $<$20\% of the total. Therefore, per one crystal
one would need a device(s) with about 1M pixels in total. The technology
of SiPM is still developing and the linearity might be improved in the
future. Another factor to consider is the number of photons coming from the 
crystal's face per mm$^2$ compared with the PDE and the density of pixels. 

A {\it shashlyk} module made for MPD~\cite{NICA-MPD:TDR-ECAL-2019}
has been tested with a Hamamatsu MPPC S13360-6025 which contains 57k pixels
$25\times{}25~\mu$m$^2$. With the yield of about 5000~pxs/GeV the loss to
non-linearity at 2~GeV was about 10\%. 

It is expected that for the electronics readout special ASIC chips will be developed
(see Section~\ref{part3_DetAspects.DaqElectronics.Constraints}),
which will provide the bias voltage to the SiPM, read out the signals using fADC,
and process the signals producing the timing and the integral and/or maximum amplitude.
Since both the detectors considered and the SiPM sensors are fast, one may expect
a timing resolution of $<$1~ns. At least 12-bit fADCs are needed. In order to provide 
a 0.02-20~GeV dynamic range and the $2\%/\sqrt{E}$ energy resolution at $\eta{}<-2$
(with PbWO$_4$ crystals) a 14-bit fADC is needed.

It will be also important to
be able to send out not only the processed, but also the raw signals as waveforms. 
Without an ASIC chip the power consumption of the on-detector electronics will be considerably higher
and its functions may be limited.

The full number of readout channels depends on the geometry and the light detection technology. 
Assuming the geometry presented in Fig.~\ref{central-detector-cartoon} and the average cell size of
$25\times{}25$~mm one comes to an approximate number of 60-70k cells. Several photosensors 
per cell may be required. It remains to be decided whether a combined or separate readout
should be used in such a case.  

ECAL must be equipped with a monitoring system, which distributes light flashes to the
photosensors. The on-board electronics, additionally to the readout, must also operate
the monitoring system. 

\subsubsection{Discussion}
\label{sec:part3-Det.Aspects.ECAL-discussion}

The EIC resolution requirements to the electromagnetic calorimetry system can be met or nearly
met by using developed technologies. For the area $\eta{}<-2$ the PbWO$_4$ crystals
appear to be the only practical choice providing a performance close to the required, and also being
compact enough to meet the expected geometrical constraints. For the other areas several options
exist. The choice strongly depends on the geometrical constraints. A better performance may be
achieved with more space, which is a subject for a global optimization of the experiment.    
Other considerations to be mentioned:
\begin{itemize}
  \item{} The area $\eta>1$ requires a high granularity of ECAL in order to resolve photons from $\pi^0$ decays. It favors
      a small cell size and high-A materials, which would also allow a shorter space.
  \item{} The projective geometry allows a better coordinate resolution and $e/\pi$ separation. It also improves the $\pi^0/\gamma$ identification at high energy. The barrel part
          is supposed to be projective. A decision has to be made about the endcaps. 
  \item{} The $e/\pi$ separation provided by the ``basic'' ECAL with the required resolution
          will be sufficient to study the e+p~18$\times$275~GeV DIS at $p>4$~GeV. At $\eta<-2$ the high-resolution ECAL
          will extend the coverage to about 2~GeV. With the electron beam energy of 10~GeV the signal to background
          ratio is different and a similar purity can be reached at momenta of about 1~GeV lower.
          The ePD can be improved  
          either by using calorimeters with a much better resolution, or by providing a ``preshower'' capability,
          or by using different detectors as a Cherenkov or TRD.
  \item{} At this time a SiPM is the photosensor of choice. However, such a sensor may bring limitations to 
           the performance of high-resolution detectors, as PbWO$_4$, 
          that have to cover a relatively large dynamic range. Large-surface sensors with a 
          high pixel density are needed for this application.
          
\end{itemize}

The eRD1 ``EIC Calorimeter R\&D Consortium''~\cite{eRD1:jul20} is expected
to continue the development of a number of technologies, including PbWO$_4$ crystals,
scintillating glass, W/ScFi and {\it shashlyk} detectors.    
\clearpage

\subsection{ECAL technologies}
\label{sec:part3-Det.Aspects.ECAL-technologies}

\subsubsection{PbWO\texorpdfstring{$_4$}{}  crystals}
\label{sec:part3-Det.Aspects.ECAL-pbwo}

The basic requirements on the EIC high-resolution EM calorimeters (see section~\ref{part3-sec-Det.Aspects.ECAL}) rule out most of the well-known scintillator materials. Finally, even a compact geometrical design requires, due to a minimum granularity, a large quantity of crystal modules, which rely on existing technology for mass production to guarantee the necessary homogeneity of the whole calorimeter. For hadron physics measurements with electromagnetic reactions, such as at multiple setups at Jefferson Lab and also at PANDA/GSI, the most common precision calorimeter of choice has been lead tungstate, PbWO${_4}$ (PWO). This is mostly driven by the requirement of good energy resolution and high granularity to detect and identify electrons, photons and pions. Good energy resolution aids in electron-pion separation and to determine the electron scattering kinematics, compactness and high granularity is driven by the need for position resolution and separation of single-photons from neutral-pion decays. PWO meets the requirements of an extremely fast, compact, and radiation hard scintillator material providing sufficient luminescence yield to achieve good energy resolution. 

Crystalline scintillators like NaI(Tl), CsI(Tl), and CsI used in detectors at electron-positron colliders like Crystal Ball at SPEAR, Crystal Barrel at LEAR, BaBar at PEPII, BELLE at KEK, or KTeV at FNAL have high light output, but cannot provide the required granularity and have relatively slow decay time (except CsI). 
These materials (except CsI)  have a relatively low radiation resistance, which makes them not suitable for the EIC operating at top luminosity. 
BaF$_2$ used at Crystal Barrel/TAPS experiment at ELSA has similar limitations for the granularity and scintillation kinetics. Although BaF$_2$ has a very fast component its separation from the slow component is nontrivial. BGO is a slow scintillator. CeF$_3$ is the closest candidate crystalline scintillator to PWO. 

\small
\begin{table}[htb]
\small
\begin{center}
\resizebox{\columnwidth}{!}{%
 \setlength{\tabcolsep}{3pt}
 \begin{tabular}{||c c c c c c c c c c||} 
 \hline
 Material & ${NaI(Tl)}$ & ${CsI(Tl)}$ & ${CsI}$ & ${BaF_2}$ & ${CeF_3}$ & ${BGO}$ & ${PbWO_4}$ & ${LSO(Ce)}$ & SciGlass \\ [0.5ex] 
 \hline\hline
 Density (${g/cm^3}$) & 3.67 & 4.51 & 4.51 & 4.89 & 6.16 & 7.13 & 8.3 & 7.4 & 3.7-5.4 \\[0.75ex]
 
 Melting Point ($^{\circ}$C)& 651 & 621 & 621 & 1280 & 1460 & 1050 & 1123 & 2050 & 1200-1300* \\[0.75ex]
 
 Radiation Length (cm) & 2.59 & 1.86 & 1.86 & 2.03 & 1.70 & 1.12 & 0.89 & 1.14 & 2.2-2.8 \\[0.75ex]

 Moliere Radius (cm) & 4.13 & 3.57 & 3.57 & 3.10 & 2.41 & 2.23 & 2.00 & 2.07 & 2-3 \\[0.75ex]

 Interaction Length (cm) & 42.9 & 39.3 & 39.3 & 30.7 & 23.2 & 22.7 & 20.7 & 20.9 & ~40 \\ [0.75ex]

 Refractive index${^a}$ & 1.85 & 1.79 & 1.95 & 1.50 & 1.62 & 2.15 & 2.20 & 1.82 & ~2 \\ [0.75ex]

 Hygroscopicity & Yes & Slight & Slight & No & No & No & No & No & No \\ [0.75ex]

 Luminescence${^b}$ (nm) & 410 & 560 & 420 & 300 & 340 & 480 & 425 & 420 & 440  \\
  (at Peak)              &     &     & 310 & 220 & 300 &     & 425 &     & 460  \\ [0.75ex]          

 Decay Time${^b}$ (ns)   & 245 & 1220 & 30 & 650 & 30 & 300 & 30 & 40 & 450 (40) \\
                         &     &      &  6 &  0.9 &    &     & 10 &    & 10-20 \\[0.75ex]  
 
                                   
Light Yield (${\gamma/MeV}$) & 41k  & 60k & 1.3k & 16k  & 2.8k & 8k & 240 & 35k & (0.5-2)k \\[0.75ex]      

 d(LY)/dT ${^{b,c}}$(${\%/^\circ C}$ )       & -0.2 & 0.4 & -0.6  & -1.9 & 0 & -0.9 & -2.5 & -0.2 & 0 \\[0.75ex]            

  Radiation Hardness & 1-2 & 1 & 10 & 1 & ${>50}$ & ${>1000}$ & ${>1000}$ & ${>1000}$ &${>1000}$ \\
  (krad)             &     &   &   &   &      &  recovery    &       &       &          \\[1.5ex]

 Experiment        & Crystal & CLEO   & KTeV  &  TAPS & - & L3    & CMS     & SuperB &\\
                   & Ball    & BaBar  &       &       &   & BELLE & ALICE   & KLOE   &\\
                   &         & BELLE  &       &       &   &       & PrimEx  &        &\\
                   &         & BESIII &       &       &   &       & PANDA   &        &\\
                   &         &        &       &       &   &       & HPS     &        &\\
                   &         &        &       &       &   &       & NPS     &        &\\
  
 \hline   
\end{tabular}%
}
\end{center}
a At the wavelength of the emission maximum.\\
b Top line: slow (intermediate) component, bottom line: fast component.\\
c At room temperature.\\
* Melting temperature for glass
\normalsize
\caption{Properties of Heavy Scintillator materials with Mass production Capability}
\label{table:part3-Det.Aspects.ECAL-tech_PWO_scintillators}
\end{table}
 \normalsize 

The original PWO specifications developed for applications in experiments at the LHC at CERN such as the Electromagnetic CALorimeter (ECAL) of CMS and the PHOton Spectrometer (PHOS) of ALICE have been optimimized over the last decades. PWO produced for the CMS/ECal and used for many other applications until 2008, e.g. for the Primex/HyCal, have fast scintillation kinetics and, for full size crystals of 23 cm length (28 ${X_0}$), a light output of 8-12 phe/MeV (measured with a bi-alkali photocathode at room temperature (RT)). The light output was significantly improved for application in experiments at PANDA/GSI. The improved material is called PWO-II and features a relative light output of 0.6 at RT (2.5 at -25${^\circ}$C), as compared to 0.3 (0.8 at -25${^\circ}$C) for PWO from CMS. The steps to achieve PWO-II revealed that radiation hardness becomes a very sensitive parameter, when operating temperatures are below T=0${^\circ}$C. Much effort has thus been devoted to understanding the radiation hardness of PWO-II crystals. The general characteristics of rectangular PWO-II crystals produced between 2014 and 2020 for the Neutral Particle Spectrometer (NPS) at JLab are listed in Table ~\ref{table:part3-Det.Aspects.ECAL-tech_PWO_twovendors}. The measurement details can be found in Ref. \cite{Horn:2019beh}. A summary of measurement details on and results of characterization tests at PANDA/GSI can be found in the PANDA TDR update. 

 \small
 \begin{table}[htb]
 \small
\begin{center}
\resizebox{\columnwidth}{!}{%
 \begin{tabular}{||c c c c c c c c ||} 
 \hline
       &           &         & Longitudinal  & Longitudinal   & Light       & Radiation         & Mass prod. \\      
Vendor & Sides     & Length  & Transmittance & Transmittance  & Yield${^a}$ & hardness          & quality   \\
       & Dimension &         & (at ${\lambda}$=360nm)   &  (at ${\lambda}$=420nm)    &    & coefficient${^b}$ &issues      \\
       &  (mm)     & (mm)    &   ${\%}$            &     ${\%}$           &  (phe/MeV)           &(${m^{-1}}$)       &  ${\%}$           \\[0.75ex]
\hline \hline
 CRYTUR & 20.460${\pm}$0.018 & 200.0${\pm}$0.1 & 45.46${\pm}$2.71 & 69.27${\pm}$1.35 & 16.05${\pm}$0.86  & ${\leq}$1.1  & No        \\[1.0ex] 
 SICCAS & 20.550${\pm}$0.028 & 200.0${\pm}$0.2 & 29.23${\pm}$4.73 & 63.77${\pm}$2.43 & 16.39${\pm}$2.55  & ${\leq}$1.5  & (20-30)   \\
\hline
\end{tabular}%
}
\end{center}
a Measured at Room Temperature(25${^\circ}$C).\\
b Induced radiation absorption coefficient dk at ${\lambda}$ = 420 nm for integral dose ${\leq100 Gy}$.\\ 
\normalsize
\caption{The properties of PWO crystals vendor with Mass production Capability}
\label{table:part3-Det.Aspects.ECAL-tech_PWO_twovendors}
\end{table}
\normalsize

PWO is available from two commercial vendors with established mass production capability, the Shanghai Institure of Ceramics, Chinese Academy of Science (SICCAS) (Shanghai, China) and the company Crystal Turnov (CRYTUR), Turnov, Czech Republic. SICCAS uses a modified Bridgeman method to grow the crystals. Full size crystals PWO-II have been evaluated by JLab and PANDA collaborations with typical failure rates of 20-30\% (see Table  ~\ref{table:part3-Det.Aspects.ECAL-tech_PWO_twovendors}). Aside from limitations of the technology that impact the optical properties, part of the failure might be attributed to large variations in the quality of the raw material. Nevertheless, SICCAS crystals can be used in EMCals, in particular if quality parameters can be relaxed. CRYTUR has gained much experience in the development and production of different types of inorganic oxide crystals for a long time and has entered the mass production of PWO-II based on the Czochralski method in 2018. CRYTUR has been producing the tapered PANDA geometry type PWO crystals and the rectangular crystals for the NPS. In August 2020, the total number of crystals produced for both PANDA and NPS passed 1000 crystals. None of the mass-produced crystals had to be rejected so far. 

The parameters of PbWO${_4}$ calorimeters used in various experiments and results of beam tests are summarized in Table~\ref{tab:part3-Det.Aspects.ECAL-tech_pbwo}. Those with dates before 2008 use PWO-I and those with dates after 2008 use PWO-II.

\small
\begin{table}[htb]
 \small
 \begin{center}
  \scalebox{0.88}{
   \setlength{\tabcolsep}{3pt}
   \begin{tabular}{lrrccccccl|rrr}
     \hline
     \hline
       Experi- & Ref & \#        & cell      & $\frac{X}{X_0}$ & Photo-       & Tempe- & Test      & mat-   & p.e./MeV 
             & \multicolumn{3}{c}{$\sigma_E/E$[GeV],\%} \\ \cline{11-13}
       ment    &     &           &  size     &                 & sensor       & rature &  beam     & rix    & $E_{min}$, 
             & \multicolumn{1}{c}{$\alpha$}  & \multicolumn{1}{c}{$\beta$} & \multicolumn{1}{c}{$\gamma$} \\ 
               &     &           &  mm$^3$   &                 &  mm$^2$      & $^\circ$C & GeV     &        & MeV 
             &             & & \\ 
     \hline
      GAMS   & \cite{Alexeev:1995bh}     & 35   & 20$^2$      & 20~~ & XP1911$^b$ &  14        & 10    & 5$\times$5 & 6 p.e. &   0.47      &   2.8      &  \\
      1995   &                           &      &$\times{}180$&      & 176~mm$^2$ & $\pm{}0.2$ &  70   &            &        & $\pm{}0.06$ & $\pm{}0.2$ &  \\
     \hline
      KEK    & \cite{Shimizu:2000jk}     & 9    & 20$^2$      & 22.5 &  R4125$^b$ &  13        & 0.2   & 3$\times$3 &        &   0.0       &   2.5      & 1.4 \\
      2000   &                           &    &$\times{}200$  &      & 25~mm$^2$  &            &  1.0  &            &        & $\pm{}2.7$  & $\pm{}0.1$ & $\pm{}0.1$ \\
     \hline
      ALICE  & \cite{Aleksandrov:2005yu} & 18k$^N$ & 22$^2$     & 20~~ &  S8148$^a$ & -25        & 0.6   & 3$\times$3 & 7.5 p.e.&  1.1       &   3.6      & 1.1 \\
      2005   &                           &    &$\times{}180$  &      & 25~mm$^2$  & $\pm{}0.1$ &  150  &            &        & $\pm{}0.3$  & $\pm{}0.2$ & $\pm{}0.3$ \\
     \hline
      CMS    & \cite{Adzic:2006za}       & 76k$^B$ & 22$^2-$27$^2$ & 26~~ & S8148$^a$ &  18           & 25   & 3$\times$3 & 10 p.e.$^f$  &  0.4       &   2.9      & 12.9  \\
      2006   &                           &    &$\times{}230$  &   & 2$\times$25~mm$^2$ & $\pm{}0.1$ &  100  &            &   & $\pm{}0.3$  & $\pm{}0.2$ & $\pm{}0.2$ \\
     \hline
      PRIMEX & \cite{Kubantsev:2006uf}   &  1k$^S$ & 20.5$^2$   & 20~~ & R4125A$^b$ &  14        & 25    & 5$\times$5 &    &  0.9        &   2.5      & 1.0  \\
      2006   &                           &         &$\times{}180$  &      & 176~mm$^2$ & $\pm{}0.1$ &  100  &            &    &             &            &            \\
     \hline
      PANDA  & \cite{Rosenbaum:2016kyk}  & 11k$^C$ & $21^2-27^2$ &22.5 & LAAPD$^{ac}$  &  -25     & 0.05  & 3$\times$3 & 16 p.e. &  0.5   &   2.3      & 0.27  \\
      2011   & \cite{Kavatsyuk:2011zz}   &  5k$^B$ &$\times{}200$  &   & 190~mm$^2$   &          &  0.75  &            &  10    &        &             &      \\
     \hline
      HPS    & \cite{Balossino:2016nly}  & 442$^B$ & $13.3^2-16^2$ &18~~ & S8664-1010$^a$ &  -17 & 0.35  & 3$\times$3 &       &  2.5   &   2.87     & 1.62  \\
      2017   &                           &         &$\times{}160$  &   & 100~mm$^2$   & $\pm{}0.3$ &  2.35 &          &       &        &            &       \\
     \hline
     CLAS12& \cite{Acker:2020brf}        & 332$^S$ & $15^2$        &22.5 & S8664-1010$^a$ &  0.0       & 2.2  & 3$\times$3 & 230 p.e.$^e$ &  3.3$^d$   &         &    \\
     FT 2020  &                          &         &$\times{}200$  &     & 100~mm$^2$     & $\pm{}0.1$ &      &           &           &        &            &       \\
     \hline
      NPS   & \cite{Horn:2019beh}       & 670$^S$ & $20.5^2$      &22.5 & R4125$^b$       &  18.0       & 4.7  & 3$\times$3 & 14 p.e.  &  1.6$^d$   &         &    \\
      2019  &                           & 350$^C$ &$\times{}200$  &     & 176~mm$^2$      &             &      &           &           &        &            &       \\
     \hline
      CCAL-NPS  & \cite{CCAL-NPSprot:2020nov}              & 140$^S$ & $20.5^2$      &22.5 & R4125$^b$       &  17.0       & 4.7  & 3$\times$3 & 14 p.e.  &  0.4   &   2.6      &    1.9\\
      2019  &                           &       &$\times{}200$  &     & 176~mm$^2$      &             &      &           &           &        &            &       \\
     
     \hline

     \multicolumn{13}{l}{
         \footnotesize {{\it Manufacturer:}  \bf B} ~~~ BTCP ; \hspace{2em} {\bf N} ~~~NCC-RSS ; \hspace{2em} {\bf S} ~~~SICCAS ; \hspace{2em} {\bf C} ~~~CRYTUR.
     }\\
     \multicolumn{13}{l}{
         \footnotesize {\bf a} ~~~ - APD; \hspace{3em} {\bf b} ~~~- PMT;  \hspace{3em} {\bf c} ~~~- Signal shaping $1~\mu$s. 
                                          \hspace{3em} {\bf d} ~~~- The full resolution at the given energy                        
     }\\
     \multicolumn{13}{l}{
         \footnotesize {\bf e} ~~~ It is unclear why the yield is much higher than the yield from PANDA. \hspace{3em} {\bf f} ~~~from Ref.~\cite{Horn:2019beh}. 
     }\\
     
   \end{tabular}
  }
 \end{center}
\normalsize
 \caption{
         List of parameters of PbWO$_4$ EM calorimeters.
         }
 \label{tab:part3-Det.Aspects.ECAL-tech_pbwo}
\end{table}
\normalsize

The installation of crystal modules in the ECAL frame will require a frame structure for supporting the weight and allowing to service detector modules. Several examples of such support structures exist. It should be noted that unlike PANDA/GSI a PWO-based ECAL at EIC is not envisioned to be cooled to sub-zero temperatures. The impact of the support structure on the desired resolution has to be balanced with the mechanical aspects. An initial study suggests that a carbon fiber structure of thickness of the order of 1mm does not significantly deteriorate the energy resolution.

\subsubsection{Scintillating glass}
\label{sec:part3-Det.Aspects.ECAL-scglass}

%
A bridge between PWO and less stringent resolution requirements could be provided by SciGlass, a novel scintillating glass developed by Scintilex in collaboration with the Vitreous State Laboratory (VSL) at The Catholic University of America. The VSL is a premier glass facility with extensive expertise in materials development and glass composition-property development and optimization. 

In the past, production of glass scintillators has been limited to small samples due to difficulties with scale-up while maintaining the needed quality. Some of the most promising materials investigated include cerium doped hafnate glasses and doped and undoped silicate glasses and nanocomposite scintillators. All of these have major shortcomings including lack of uniformity and macro defects, as well as limitations in sensitivity to electromagnetic probes. One of the most promising recent efforts is the development of DSB:Ce, which is a cerium-doped barium silicate glass nanocomposite. However, lack of uniformity and macro defects persist in this type of glass and need to be resolved before scale up would be possible.

\begin{figure}
	\centering
	\includegraphics[width=0.75\linewidth]{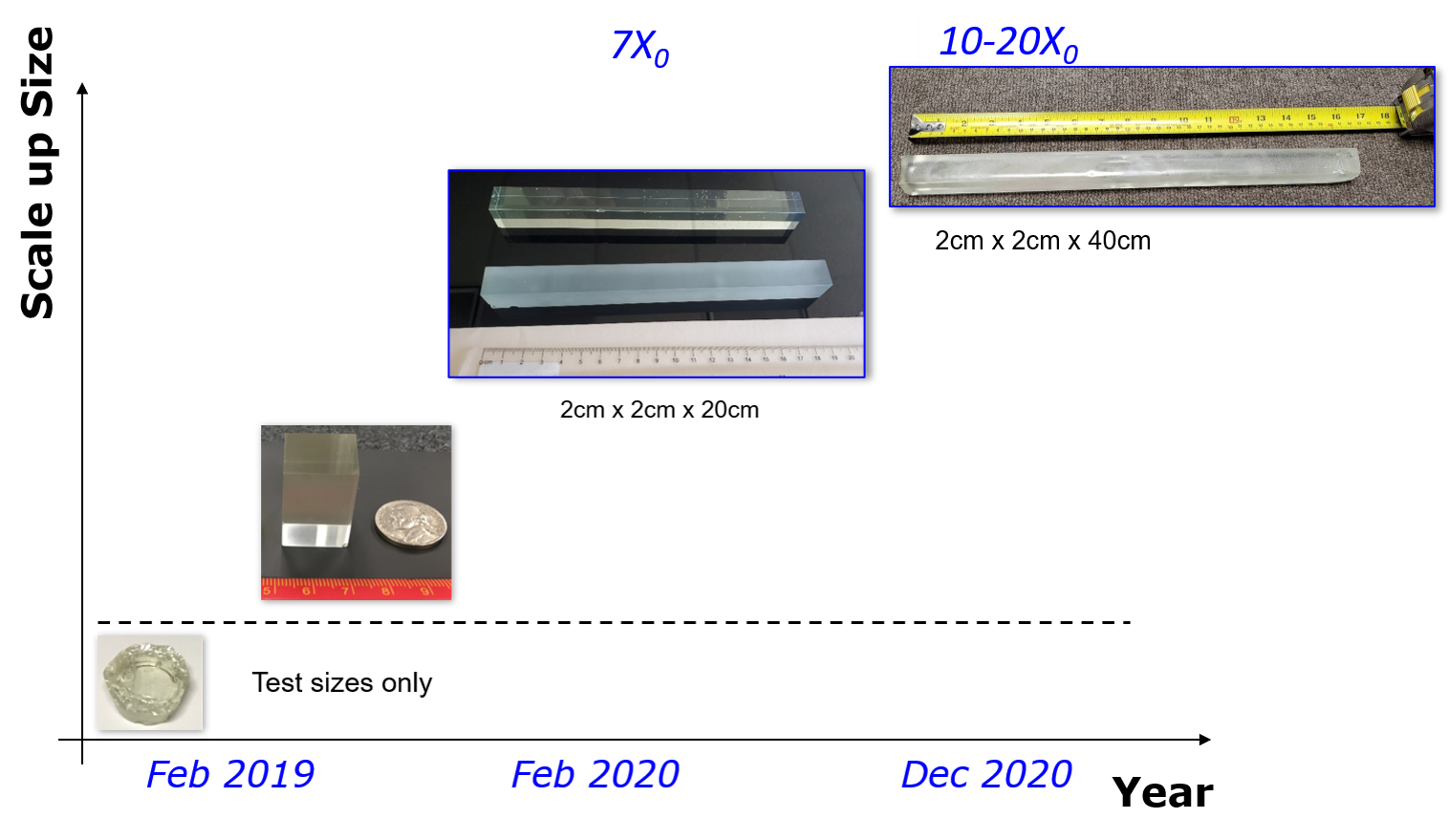}	
	\caption{Progress with SciGlass scaleup.
	}
	\label{fig:part3-Det.Aspects.ECAL-Scintilex-SciGlass-scaleup}
\end{figure}

Scintilex has developed a new family of glass scintillators (SciGlass) that have comparable or better performance to current nanocrystalline glass ceramic scintillators but have considerable advantages in terms of simplified manufacturing processes and ease of scale up. Scintilex has demonstrated a successful scaleup method and can now reliably produce glass samples of sizes up to ${\sim}$10${X_0}$ and scale-up to ${\sim}$20${X_0}$ was demonstrated with production of the first 40 cm long block. The scale-up progress of SciGlass over the last year is shown in Fig.~\ref{fig:part3-Det.Aspects.ECAL-Scintilex-SciGlass-scaleup}. After establishing the formulation and fabrication techniques for producing small batches (10-20 blocks) of SciGlass, Scintilex has initiated a research program towards larger scale production, which has started in 2020.

\begin{figure}
	\centering
	\includegraphics[width=0.3\linewidth]{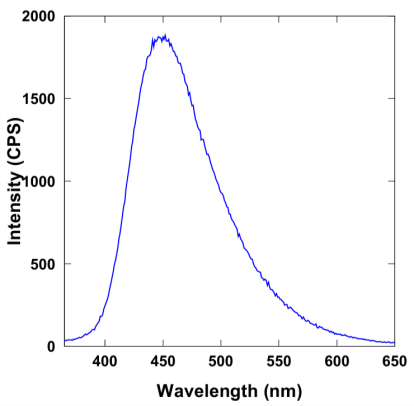}
	\includegraphics[width=0.3\linewidth]{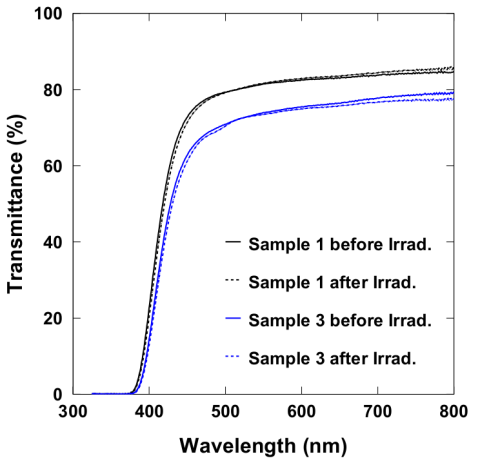}
	\includegraphics[width=0.3\linewidth]{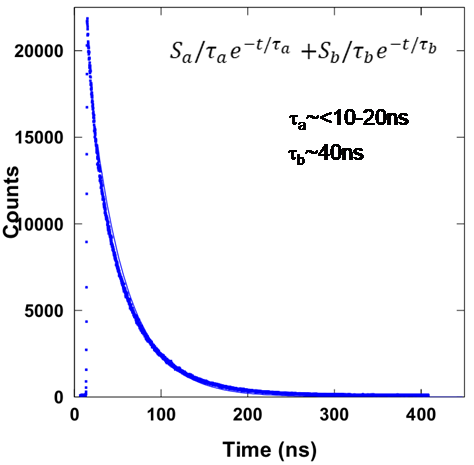}
	\caption{Emission spectrum, transmittance before/after EM irradiation, and scintillation kinetics of SciGlass.
	}
	\label{fig:part3-Det.Aspects.ECAL-SciGlass-characteristics}	
\end{figure}

%

The properties of SciGlass in comparison to PbWO$_4$ are shown in Table ~\ref{table:part3-Det.Aspects.ECAL-tech_PWO_scintillators}. 
Initial measurements with particle energies of 4-5 GeV together with simulation indicate that SciGlass has an energy resolution comparable to PWO for block sizes of a comparable number of radiation lengths. 
\begin{figure}
	\centering
	\includegraphics[width=0.5\linewidth]{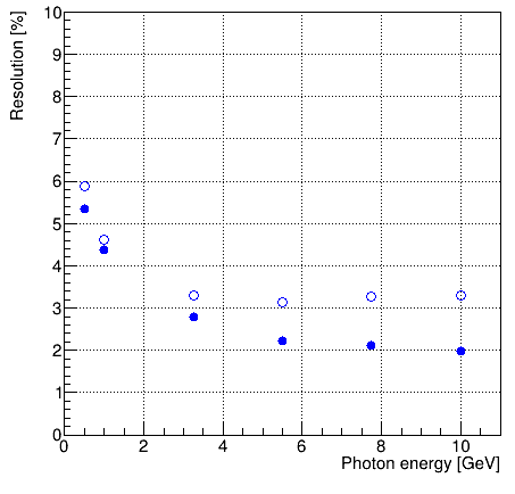}
	\caption{Projected energy resolution for SciGlass of 40~cm (open symbols) and 50~cm (filled symbols) lengths.}
	\label{fig:part3-Det.Aspects.ECAL-Scintilex-SciGlass-simulations}
\end{figure}

SciGlass blocks of sizes up up to ${\sim}$10${X_0}$ were characterized on the test bench using the same methods as used for PWO. Radiation hardness tests were carried out at IJCLab-Orsay with Co-60 sources (EM probes) and at the U. Birmingham MC40 cyclotron with hadron probes. SciGlass has excellent radiation resistance - no damage up to 1000 Gy electromagnetic and ${10^{15} n/cm^2}$ hadron irradiation, the highest doses tested to date, response time of 20-50 ns, and good transmittance in the near UV domain (78${\%}$ at 440 nm). The SciGlass insensitivity to temperature is another clear advantage over PbWO$_4$, which has a dependence of about 2-3${\%/^\circ C}$ and has to be continuously monitored. The present samples have densities up to 5.4 g/cm$^3$, ${X_0}$=2.2-2.8 cm, and a Moli\`{e}re radius of 2-3 cm.


\subsubsection{Lead glass}
\label{sec:part3-Det.Aspects.ECAL-LG}

The lead-glass electromagnetic calorimeters for EIC can be analogous
to VENUS at Tristan~\cite{Sumiyoshi:1987fi}, OPAL at
LEP~\cite{Ahmet:1990eg}, JLab GlueX forward calorimeter
~\cite{GlueX-HallD:FCAL-2008}, or PHENIX~\cite{Aphecetche:2003zr}. Several
types of lead glass, of different parameters have been used as
radiator. The light generation mechanism for the lead-glass is
dominantly Cherenkov radiation (scintillation is below 1-2\%).  

\begin{table}[htb]
 \small
 \begin{center}
  \scalebox{1.00}{
   \begin{tabular}{lccccc}
     \hline
     \hline
     Material & $X_0$  & $R_M$  & $E_{crit}$ & Refrac. & $\rho$   \\
              &  cm    & cm     &  MeV     & index   & g/cm$^3$ \\
     \hline
       TF1    & 2.74   & 3.70   &  15      & 1.647   & 3.86 \\
       F8-00  & 3.10   & 4.20   &   -      & 1.620   & 3.60 \\
       SF57   & 1.55   & 2.61   &  12      & 1.890   & 5.51 \\
     \hline
   \end{tabular}
  }
 \end{center}
 \normalsize
 \caption{
         Parameters of several types of lead glass.
         }
 \label{tab:part3-Det.Aspects.ECAL-tech_lg}
\end{table}

The
fraction of PbO$_2$ in chemical composition may vary from 45\% to 75\% by
weight, and density from 3.6~g/cm$^3$ to 
5.5~g/cm$^3$~\cite{Ahmet:1990eg, Mkrtchyan:2012zn, GlueX-HallD:FCAL-2008, Bruckner:1991de} 
(see Table~\ref{tab:part3-Det.Aspects.ECAL-tech_lg}). 
The radiation length is within 1.5~cm to 3.1~cm. The Moli\`{e}re radius is 2.6~–~3.7~cm, typically.
In practice, homogeneous calorimeters must be 20
radiation length deep to contain electromagnetic shower. For the lead
glass radiator this implies 30~–~50~cm length. For a hodoscopic
construction, the optimal granularity size is $\approx$ 4~cm, which depends on
the Moli\`{e}re radius.  The refractive index ranges from
1.62 to 1.89 (1.65 typically) ~\cite{Ahmet:1990eg,
  GlueX-HallD:FCAL-2008, Mkrtchyan:2012zn}. The transparency window
starts from ~350~nm, except for the Ce doped radiation resistant lead
glass for which it starts from 400~nm~\cite{Mkrtchyan:2012zn}. PMTs
with bialcali photcathode (sensitivity range from 300~nm to 600~nm,
peak quantum efficiency ~20\% at 400~nm) are well suited for Cherenkov
light detection from electromagnetic showers in lead glasses.  The
lead glass calorimeters have modular construction. The glass blocks
are wrapped in thin reflector (usually aluminized Mylar), then by
light tight Tedlar film. It is important to have a thin layer of air
between the block and Mylar, for full internal reflection of light at
oblique incident angles. The PMTs are optically coupled to the blocks
by means of optical glue or grease of suitable refractive index. In
the moderate magnetic field the PMTs can be shielded by layers of
$\mu$-metal. In stronger fields (1~–10~mT) additional shielding of
photocathode by soft iron can be implemented. A light guide between
the block and PMT, no shorter than diameter of photocathode shall be
placed between the block and PMT. Such design of modules has been
effectively used in many lead glass 
calorimeters~\cite{Sumiyoshi:1987fi,Ahmet:1990eg,GlueX-HallD:FCAL-2008,Prokoshkin:1995rd}.
Radiation hardness of lead-glass crystals is ∼10~krad integral dose
for TF1, and 50~krad for F101 type radiation hard glass. The
glass blocks recover from damage on their own within 1 to 3
months~\cite{Mkrtchyan:2012zn}. They can be cured in situ by exposing
to UV radiation. A 30\% reduction in transparency of 4~cm glass
thickness can be recovered within 8 hours. Alternatively, off-line
gradual heating, up to 260~$^{\circ}$C and cooling may be implemented.
The energy resolution of lead-glass calorimeters strongly depends on the optical
quality, the light yield, the light detection efficiency,
and the electronic noise, and may vary from $\approx 5\%/\sqrt{E}\oplus{}1\%$
~\cite{Prokoshkin:1995rd} to $\approx 8\%/\sqrt{E}\oplus{}3\%$
~\cite{Prokoshkin:1995rd, Avakian:1996ge, Mkrtchyan:2012zn}. A
coordinate resolution of $6.4/\sqrt{E}$~mm for incoming photons was
obtained in a hodoscopic construction of a $4\times{}4$~cm$^2$ 
granularity~\cite{GlueX-HallD:FCAL-2008}.

A large number of lead glass blocks from older experiments 
may become available for applications at EIC. For example,
PHENIX~\cite{Aphecetche:2003zr} has used
about 9000 TF1 blocks $40\times{}40\times{}400$~mm$^3$ (14.4$X_0$). 
COMPASS~\cite{Abbon:2007pq} has used about 3000 TF1 blocks  
$38\times{}38\times{}450$~mm$^3$ (16$X_0$). 

\subsubsection{Scintillating fibers embedded in absorber}
\label{sec:part3-Det.Aspects.ECAL-scfibers}

Scintillating fiber calorimeters ({\it SPACAL}s) have been built and used in many experiments 
in both High Energy and Nuclear Physics and have been used for both electromagnetic 
and hadronic calorimeters 
\cite{Acosta:1990qs, Antonelli:1994kf, Beattie:2018xsk, Sedykh:2000ex, Armstrong:1998qs}. 
They consist of many scintillating 
fibers embedded in an absorber material which are then gathered at the front or 
the back (or both) and read out with photosensors. The sampling fraction and sampling 
frequency can be adjusted by changing the number of fibers and their spacing to 
provide a range of energy resolutions and other properties. In addition, the absorber 
material can be selected for a specific application in order to achieve a variety 
of requirements.

One of the requirements for any ECAL at EIC is that it be compact, i.e., that 
it has a short radiation length and small Moli\`{e}re radius so that the total length 
of the calorimeter can be minimized and that the lateral extent of the shower can 
be contained to provide good separation of neighboring showers. This can best be 
achieved with a high Z absorber such as tungsten. The sPHENIX barrel ECAL utilizes 
a tungsten {\it SPACAL} (W/SciFi) design where an array of scintillating fibers is embedded 
in a matrix of tungsten powder and epoxy. Some of the properties of this design 
are listed in Table~\ref{tab:part3-Det.Aspects.ECAL-considered}. 
This design was originally developed at UCLA \cite{Tsai:2012cpa} 
and then later adopted by the sPHENIX Experiment \cite{Aidala:2017rvg} 
which then further developed the technology into an industrialized  process to 
produce more than 6000 2D projective absorber blocks. These blocks are read out 
using SiPMs that are coupled to the blocks using short light guides, which keeps 
the total radial length of the calorimeter to 26~cm inside the BaBar solenoid magnet, 
including the readout and supporting structure. This calorimeter is currently under 
construction and is expected to be completed by the end of 2021.

One of the issues with the W/SciFi design is that the boundaries between the blocks 
and the light guides introduce certain non-uniformities in the energy response. 
These can be measured using the position information provided by the calorimeter 
itself and/or the tracking system and used to correct the energy response. For 
the sPHENIX design, this leads to an energy resolution 
$\sim 13\%/\sqrt{E} \oplus 2.5\%$. 

For any future W/SciFi calorimeter for EIC, it would be advantageous to minimize 
the number of boundaries produced by the blocks and the light guides, which is 
possible by greatly reducing the length of the light guides to just a few mm (which 
is necessary to act as mixer for the light coming out of the fibers) and then covering 
nearly all of the readout area with SiPMs. This is now also possible with the availability 
of large area (6 $\times$ 6 mm$^2$) SiPMs at 
an affordable cost. 

\subsubsection{Shashlyk}
\label{sec:part3-Det.Aspects.ECAL-shashlyk}

{\it Shashlik} calorimeters have been used in many High Energy and Nuclear Physics 
experiments (Table~\ref{tab:part3-Det.Aspects.ECAL-tech_shashlyk}). The light produced in 
an alternating stack of absorber plates and scintillating tiles is collected with the help of 
wavelength shifting (WLS) fibers passed through the stack 
and is detected on one or both of the fiber's ends. 
The outgoing light is concentrated on a surface much smaller than the cell size - an advantage for 
using the relatively small semiconductor photosensors.

The plate thickness can be selected in order to obtain the required sampling fraction and 
the sampling frequency. It should be noted that, as with other sampling calorimeters, 
a larger sampling fraction leads to a larger radiation length and the Moli\`{e}re radius, 
which then increases the length of the total stack and allows the shower to spread out 
more laterally. 

Most of the {\it shashlyk} calorimeters to date have used lead for the absorber plates.
Using tungsten helps to reduce the lateral overlap of showers 
(Table~\ref{tab:part3-Det.Aspects.ECAL-tech_shashlyk}: HERA-B inner calorimeter),
and to reduce the total length of the calorimeter.

\begin{figure}[htb]
\begin{center}
  \begin{minipage}{0.32\linewidth}
    \includegraphics[angle=0,width=0.97\linewidth]{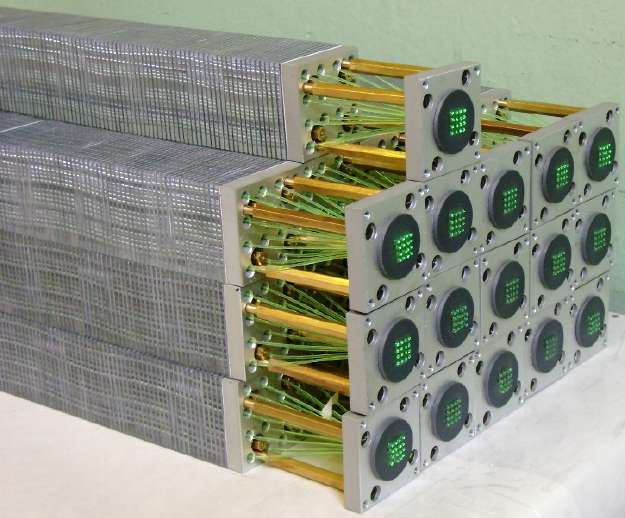}
  \end{minipage}
  \begin{minipage}{0.40\linewidth}
    \includegraphics[angle=0,width=0.86\linewidth]{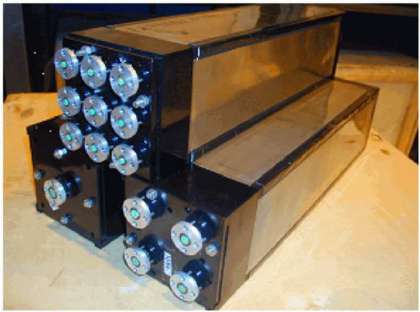}
  \end{minipage}
  \begin{minipage}{0.24\linewidth}
    \includegraphics[angle=0,width=0.97\linewidth]{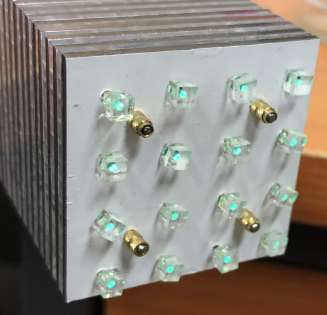}
  \end{minipage}
\end{center}
\caption{
         Left: COMPASS Pb/Sc ``spiral'' $4\times{}4$~cm$^2$ modules~\cite{Abbon:2014aex};
         Middle: LHC-B Pb/Sc 3 module types, with a single $12\times{}12$~cm$^2$ cell, with four $6\times{}6$~cm$^2$ cells, 
          and with nine $4\times{}4$~cm$^2$ cells~\cite{Barsuk:2010dlp}; 
         Right: eRD1 W/Sc prototype $4\times{}4$~cm$^2$ cell, readout: 16 small SiPM per 
                cell~\cite{Kuleshov:dwgcalmar30, eRD1:talkjul20}.
        }
\label{fig:part3-Det.Aspects.ECAL-shashlyk_photo} 
\end{figure}

\begin{figure}[htb]
  \begin{minipage}{0.49\linewidth}
    \includegraphics[angle=0,width=0.97\linewidth]{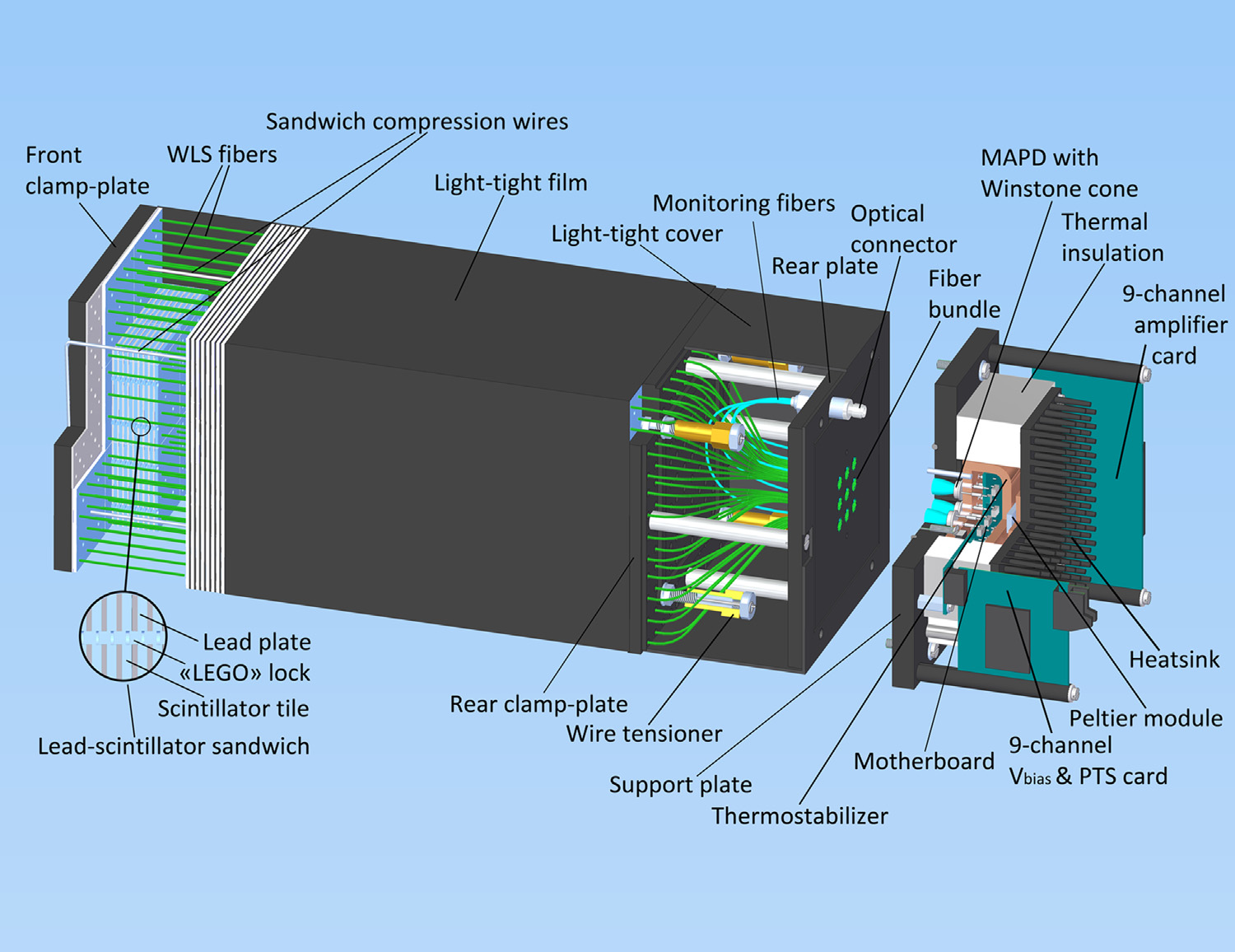}
  \end{minipage}
  \begin{minipage}{0.49\linewidth}
    \includegraphics[angle=0,width=0.97\linewidth]{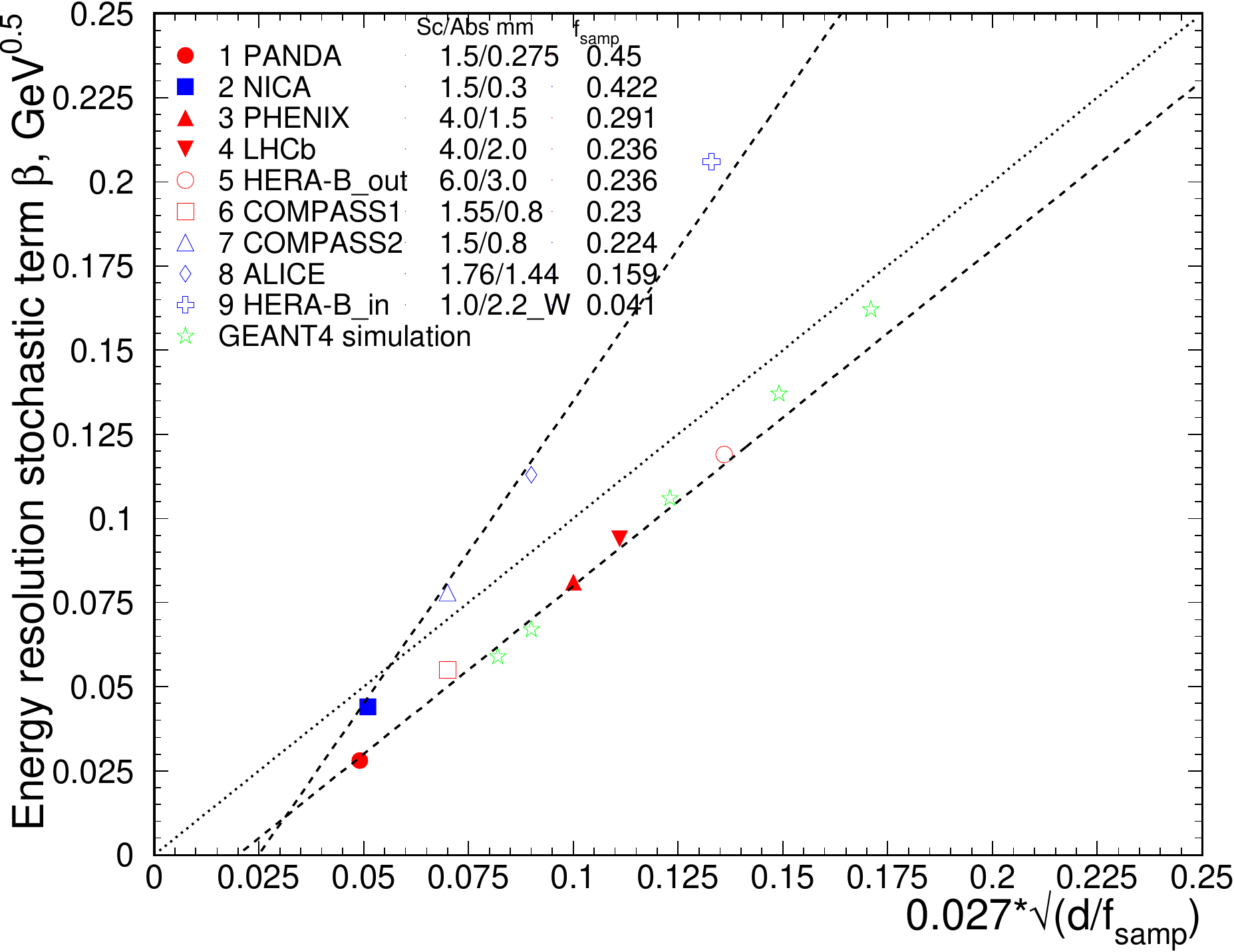}
  \end{minipage}
\caption{
         Left: A 3$\times$3 module design~\cite{Chirikov-Zorin:2016yoa} with a SiPM readout.\\  
         Right: The measured stochastic term $\beta$ of the resolution of {\it shashlyk} calorimeters 
         (Tab.~\ref{tab:part3-Det.Aspects.ECAL-tech_shashlyk})
         against a predicted value of $\beta=0.027\sqrt{d/f_{samp}}$~\cite[p.~119]{Wigmans:2018fua}, where $d$ is the thickness
         of the scintillator tile in mm. The ``scaling variable'' $\sqrt{d/f_{samp}}$ at the first order does not depend
         on the thickness of the scintillator tile. The plot indicates that the data can be split in two groups. In each group 
         the dependence on the ``scaling variable'' is nearly linear: 
         $\beta\approx{}0.027(\sqrt{d/f_{samp}}-0.74)$ and $\beta\approx{}0.049(\sqrt{d/f_{samp}}-0.92)$.  
         The origin of the difference between the groups is unclear 
         at this moment. The results of GEANT4 simulation (Fig.~\ref{fig:part3-Det.Aspects.ECAL-shashlyk_resol_simul}) match
         the lower curve well at moderate layer thickness.
        }
\label{fig:part3-Det.Aspects.ECAL-shashlyk_resol} 
\end{figure}

Figure~\ref{fig:part3-Det.Aspects.ECAL-shashlyk_photo} shows several examples of the {\it shashlyk} detector.
Typically, the WLS fibers are 1~cm apart. 
In most {\it shashlyk} calorimeters, the WLS fibers are bundled at the back of the detector 
and read out with a single photosensor.
In a new eRD1 W/Sc prototype (Fig.~\ref{fig:part3-Det.Aspects.ECAL-shashlyk_photo}, right) each fiber 
is readout by a small SiPM.  
One module is often split into several readout cells in order to reduce the effect of the edges between cells
(Fig.~\ref{fig:part3-Det.Aspects.ECAL-shashlyk_photo}, middle).
The grid of fibers leads to variations of the response across the cell surface. The best uniformity has been
achieved with a ``spiral'' geometry of the fibers
(Fig.~\ref{fig:part3-Det.Aspects.ECAL-shashlyk_photo}, left). An example of a 3$\times$3 module design is 
shown in Figure~\ref{fig:part3-Det.Aspects.ECAL-shashlyk_resol}~(left). 

The scintillator and WLS fibers are selected in order to match their spectral properties.

\begin{table}[htb]
 \footnotesize
 \begin{center}
  \scalebox{0.89}{
   \setlength{\tabcolsep}{3pt}
   \begin{tabular}{lrccccccrccc|cc}
     \hline
     \hline
       Experi- & Ref & sampling & $f_{samp}$ & $\rho$   & $X_0$ & $R_M$ & $\frac{X}{X_0}$ & cell & read- & Yield/ & Beam 
           & \multicolumn{2}{c}{$\sigma_E/E$[GeV]} \\ \cline{13-14}
       ment    &     & mm       &   \%      & $\frac{\mathrm{g}}{\mathrm{cm}^3}$ 
                                                       &  mm   & mm    &                 & mm   & out   &  MeV   & GeV 
           & $\alpha$ \% & $\beta$ \% \\ \cline{4-7}
       \# ch   &     & \# layers& \multicolumn{4}{c}{\it 2$^{nd}$ line: calculation}      &                 & {\scriptsize WLSF}  
                                                                                                & mm$^2$ &       &       
           &  \multicolumn{2}{c}{$\gamma$ \%} \\ 
     \hline
       KOPIO & \cite{Atoian:2007up} & Pb/Sc 0.275/1.5 &    & 2.75 & 35. & 60. & 16.& 110.& APD  & 50~p.e. & 0.2-0.4 & 2.0 & 2.74 \\ 
        few  &                      & $\times$300     & 45.& 2.60 & 35. & 57. &    & 144 & 200. &         &           &     &      \\
     \hline
       PANDA & \cite{Kharlov:2008tw}& Pb/Sc 0.275/1.5 &    &      & 34. & 59. & 20.& 110.& PMT  &  5~p.e. & 1-19      & 1.3 & 2.8  \\ 
$\sim$ 2000  &                      & $\times$380     & 45.& 2.60 & 35. & 57. &    & 144 & 200. &         &           & 
              \multicolumn{2}{c}{3.5} \\
     \hline
    MPD\hspace{2pt}NICA  & 
               \cite{Semenov:2020glg} & Pb/Sc 0.3/1.5 &    &      & 32. & 62. & 12.& 110.& SiPM &         & 0.5-3.0   & 1.0 & 4.4  \\ 
     38000   &                      & $\times$220     & 43.& 2.70 & 32. & 55. &    & 144 & 36.  &         &           &     &      \\
     \hline
    PHENIX   & \cite{Aphecetche:2003zr} 
                                    & Pb/Sc 1.5/4.0   &    &      & 20. &     & 18.&  55.& PMT  & 1.5 p.e.& 5-80      & 2.1 & 8.1  \\ 
     15500   &                      &  $\times$66     & 29.& 3.81 & 20. & 42. &    &  36 & 200. &         &           &     &      \\
     \hline
    LHCb     & \cite{Barsuk:2010dlp}& Pb/Sc 2.0/4.0   &    &      &     & 37. & 24.&  40.& PMT  & 3.0 p.e.& 5-100     & 0.8 & 9.4  \\ 
     6000    &                      &  $\times$66     & 24.& 4.44 & 17. & 35. & 25.&  16 &      &         &           &            
              \multicolumn{2}{c}{14.} \\
     \hline
    HERA-B   & \cite{Avoni:2007zza} 
                                    & Pb/Sc 3.0/6.0   &    &      & 17. & 37. & 20.&  56.& PMT  & 0.8 p.e.& 5-28      & 1.4 & 11.9 \\ 
     4000    &                      &  $\times$37     & 24.& 4.45 & 17. & 42. &    &  36 & 490. &         &           &     &      \\
     \hline
    COMPASS  & \cite{Abbon:2014aex} & Pb/Sc 0.8/1.55  &    &      &     &     & 23.&  38.& PMT  &         & 1 - 7     &     & 5.5  \\ 
      888    &                 & spiral $\times$156   & 23.& 4.50 & 16. & 37. &    &  16 & 490. &         &           &     &      \\
     \hline
    COMPASS & \cite{Anfimov:2013kka} 
                                    & Pb/Sc 0.8/1.5   &    &      & 16.4& 35. & 16.&  40.& SiPM &         & 1 - 7     & 2.3 & 7.8  \\ 
     $\approx${2000} 
            & \cite{Anfimov:2015jwa}& $\times$109     & 22.& 4.60 & 16. & 36. &    &  16 &  9.  &         &           &     &      \\
     \hline
    ALICE   & \cite{Allen:2009aa}                
                                    & Pb/Sc 1.44/1.76 & 9.5& 5.68 & 12.3& 32. & 20.&  60.& APD  & 4.4 p.e.& 0.5-100   & 1.7 & 11.3 \\ 
     12288           
            &                       & $\times$77      & 16.& 5.63 & 12.4& 30. &    &  36 & 25.  &         &           &    
              \multicolumn{2}{c}{5. } \\
     \hline
    HERA-B   & \cite{Avoni:2007zza} 
                                    &  W$^a$/Sc 2.2/1.0   
                                                      &    &      & 5.6 & 14. & 23.&  22.& PMT  & 0.13 p.e.& 5-28     & 1.2 & 20.6 \\ 
     2100    &                      &  $\times$37     & 4.1& 12.5 & 5.7 & 13.9&    &   9 & 490. &          &          &     &      \\
     \hline
     \hline
    eRD1     & \cite{Kuleshov:dwgcalmar30}
                                    &  W$^b$/Sc 1.58/1.63   
                                                      &    &      &     &     & 31.& 38. & SiPM &   &           &    &   \\ 
             & \cite{eRD1:talkjul20}                     
                                    &  $\times$79     & 9  & 8.9  & 8.4 & 19. &    & 16  &      &          &   & 1. & 7.7~$^c$     \\
     \hline
     \multicolumn{14}{l}{
         \footnotesize {\bf a} ~~~- W/Fe alloy 90/10~\% by volume; \hspace{3em} {\bf b} ~~~- W/Cu alloy 80/20~\% by volume 
     }\\
     \multicolumn{14}{l}{
         \footnotesize {\bf c} ~~~- Results of GEANT simulation;   
     }\\
   \end{tabular}
  }
 \end{center}
\normalsize
 \caption{
         List of parameters of {\it shashlyk} EM calorimeters used
         in experiments. The values of the average properties of the
         calorimeter material ($f_{samp}$, $\rho$, $X_0$, and $R_M$),
         if published, are presented in the top lines of the proper
         cells.  The values calculated using the published sampling
         structure are presented in the bottom line. The calculation
         is simplified, but done in a standard way for all the
         entries facilitating the comparison between the entries.  The
         results of the calculations are usually close to the published values, except
         the only one published value of $f_{samp}$.  The resolution
         is parametrized using
         Equation~\ref{eq:part3-Det.Aspects.ECAL-overviewtech-eres}. The
         resolution was measured in test beams in the energy range
         specified. The size of the readout cell is shown, along with
         the number of WLS fibers per cell.}
 \label{tab:part3-Det.Aspects.ECAL-tech_shashlyk}
\end{table}
\normalsize

Table~\ref{tab:part3-Det.Aspects.ECAL-tech_shashlyk} shows the parameters of several large-scale {\it shashlyk} detectors
as well as two prototypes. These detectors are built both in rectangular and trapezoidal shapes, the latter provides the projective
geometry (ALICE~\cite{Allen:2009aa} and MPD~\cite{Semenov:2020glg} for example). 
Various photosensors have been used: conventional PMTs, avalanche photodiods (APD), and
SiPMs. 

\begin{figure}[htb]
  \begin{minipage}{0.49\linewidth}
    \includegraphics[viewport=95 75 434 406,clip,angle=0,width=0.97\linewidth]{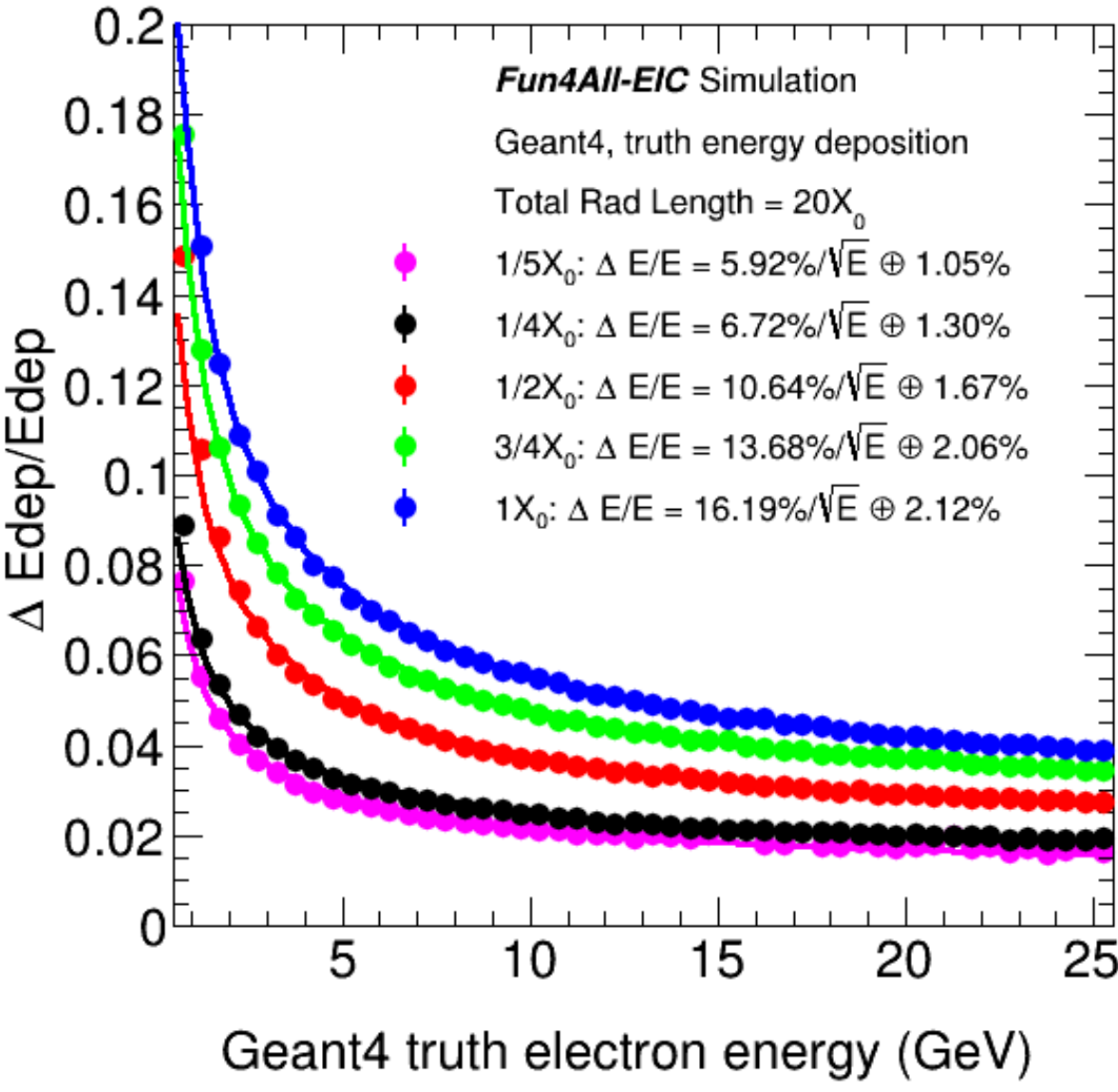}
  \end{minipage}
  \begin{minipage}{0.49\linewidth}
    \includegraphics[viewport=515 75 856 406,clip,angle=0,width=0.97\linewidth]{PART3/Figures.Det.ECAL/W-Shashlik_GEANT_Simulations_eRD1_jul_2020.pdf}
  \end{minipage}
\caption{
         GEANT4 calculation of the {\it shashlyk} W/Sc detector resolution~\cite{eRD1:talkjul20}. The scintillator tile is 1.5~mm thick.
         Left: Dependence on the absorber plate thickness for $20X_0$ total thickness. The results are compared with data in 
        Fig.~\ref{fig:part3-Det.Aspects.ECAL-shashlyk_resol}.
        Right: Dependence on the total thickness $x=X/X_0$ for a $0.25X_0$ thick layer. The constant term is described by a polynomial: $\alpha=(1.31-0.251(x-20)+0.0144(x-20)^2)\%$. 
        }
\label{fig:part3-Det.Aspects.ECAL-shashlyk_resol_simul} 
\end{figure}

It has been argued \cite[p.~119]{Wigmans:2018fua} that the stochastic coefficient is approximately proportional to 
$\sqrt{d/f_{samp}}$, where $d$ is the thickness of the active material layer (or the fiber's diameter). For a number of 
sampling calorimeters of various types (LAr, LKr, Pb/Sc {\it shashlyk}, SPACAL etc) it was found that 
$\beta{}\approx{}2.7\%\sqrt{(d/1~\mathrm{mm})/f_{samp}}$. The data from Tab.~\ref{tab:part3-Det.Aspects.ECAL-tech_shashlyk}
is shown in Fig.~\ref{fig:part3-Det.Aspects.ECAL-shashlyk_resol}~(right). It is not clear what causes the data to split in
two groups with different slopes. The lower group is described by the predicted slope of 0.027, while the higher group
is described by a larger slope of 0.049. The offsets of the linear functions are not physical ($\beta$ can not be negative
at any layer thickness) but indicate that at a smaller layer thickness some other processes must dominate the resolution. 
Results of GEANT4 simulation of {\it shashlyk} calorimeters are shown in Fig.~\ref{fig:part3-Det.Aspects.ECAL-shashlyk_resol_simul}.
The stochastic term describes the data well (Fig.~\ref{fig:part3-Det.Aspects.ECAL-shashlyk_resol}). The 
constant term simulated for $0.25\cdot{}X_0$ is well described by a parabolic function. In summary, the resolution of a {\it shashlyk} calorimeter
can be approximated by:
 \begin{eqnarray}
    \label{eq:part3-Det.ECAL_resol_shashlyk}
     \beta{} \approx 2.7\cdot{}(\sqrt{d/f_{samp}} - 0.74)\% \\
     \alpha{} \approx (1.31-0.251(x-20)+0.0144(x-20)^2)\% \nonumber ,
 \end{eqnarray}
where $d$ is the thickness of the scintillator tile in mm, $f_{samp}$ is the sampling ratio; $x=X/X_0$ is the full thickness of the calorimeter.
The constant term depends on the layer thickness as well. In a real experiment it also depends on the calibration quality and other factors.

The expected spacial constraints of the EIC favor the use of tungsten alloys 
for the absorber. One can select the sampling structure in order to 
be able to fit the detector into 40~cm of space (see Tab.~\ref{tab:part3-Det.Aspects.ECAL-considered}). Assuming
the approximation of Eq.~\ref{eq:part3-Det.ECAL_resol_shashlyk}
one may expect that such a structure of W/Sc 0.75/1.5~mm would provide a stochastic term of 6.3\%~GeV$^{0.5}$.

The eRD1 Consortium is planning to study in a test
beam a 3$\times$3 W/Sc detector prototype~\cite{Kuleshov:dwgcalmar30,eRD1:talkjul20}
(see Fig.~\ref{fig:part3-Det.Aspects.ECAL-shashlyk_photo}, right and
Tab.~\ref{tab:part3-Det.Aspects.ECAL-tech_shashlyk}). Instrumenting
each individual fiber with its own 
small SiPM may provide more detailed information about the position 
of the shower inside the stack, thus providing better position resolution, and 
also allowing this information to be used to correct for any non-uniformity in 
either the light collection or energy response. One may
also consider adding the signals from several of those SiPMs electrically, reducing
the number of readout channels.
Such a design reduces the length of the module, saving a few cm of space needed for bundling
of the fibers.


\section{Hadron Calorimetry}
\label{part3-sec-Det.Aspects.HCAL}

\subsection{General consideration for Hadron Calorimeters}

The major point for the design of hadron calorimeters (HCAL) at EIC is the capability of the whole detector to provide a precise reconstruction of the jet energy (see Section~\ref{part2-sec-DetReq.Jets.HQ} and Ref.~\cite{Page:2019gbf,Arratia:2019vju}). Additionally, the application of the Jaquet-Blondel method (Section~\ref{sec:J-B_rec}, Ref.~\cite{Jacquet:1979jb,Tsai:tmplmar21}) requires the detection of all the final state hadrons, including the proton fragmentation products which are mostly concentrated in the very forward area.  

\begin{figure}[htb]
\begin{center}
  \begin{minipage}[t]{0.40\linewidth}
    \includegraphics[angle=0,width=0.97\linewidth]{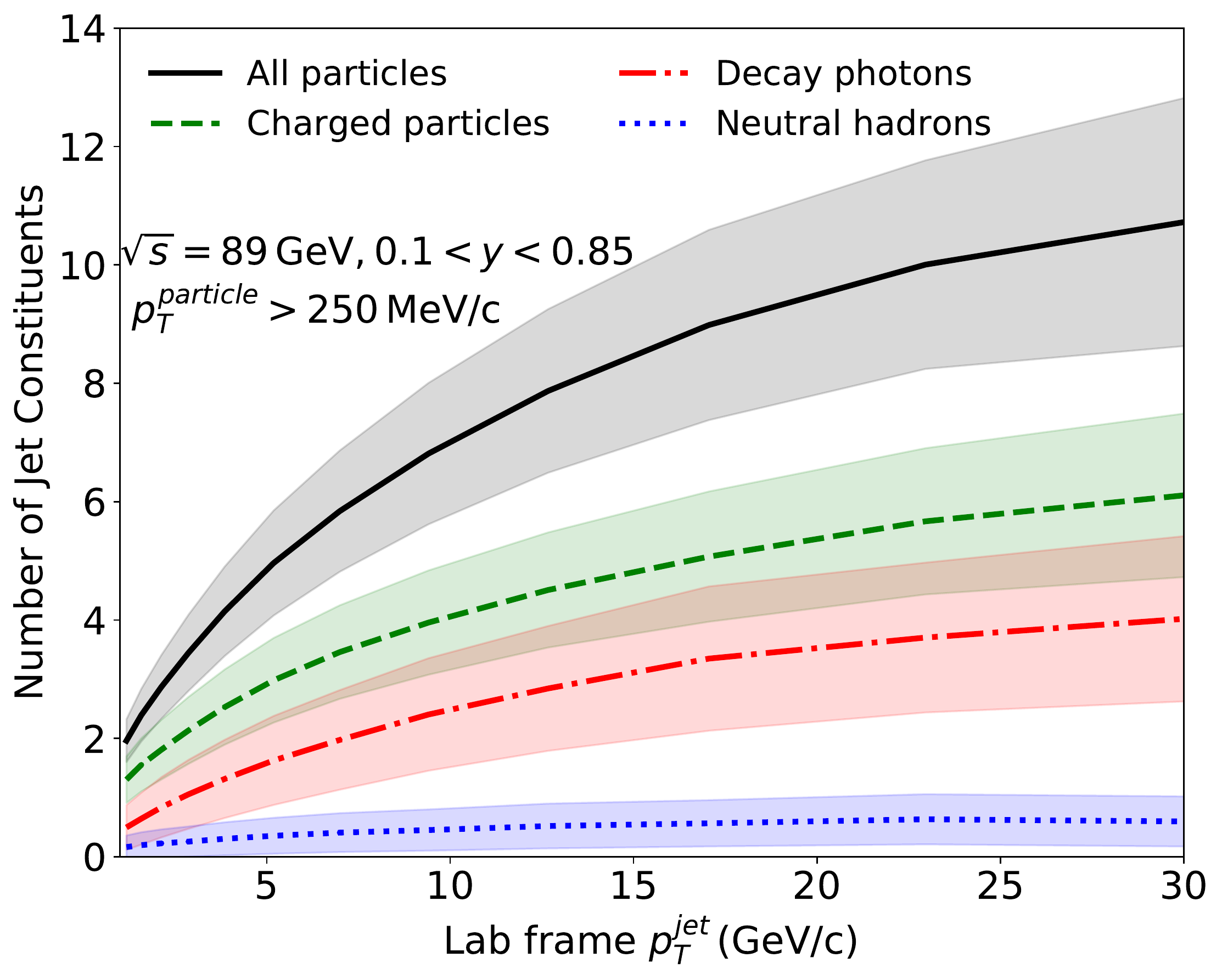}
  \end{minipage}
  \begin{minipage}[t]{0.29\linewidth}
    \includegraphics[angle=0,width=0.97\linewidth]{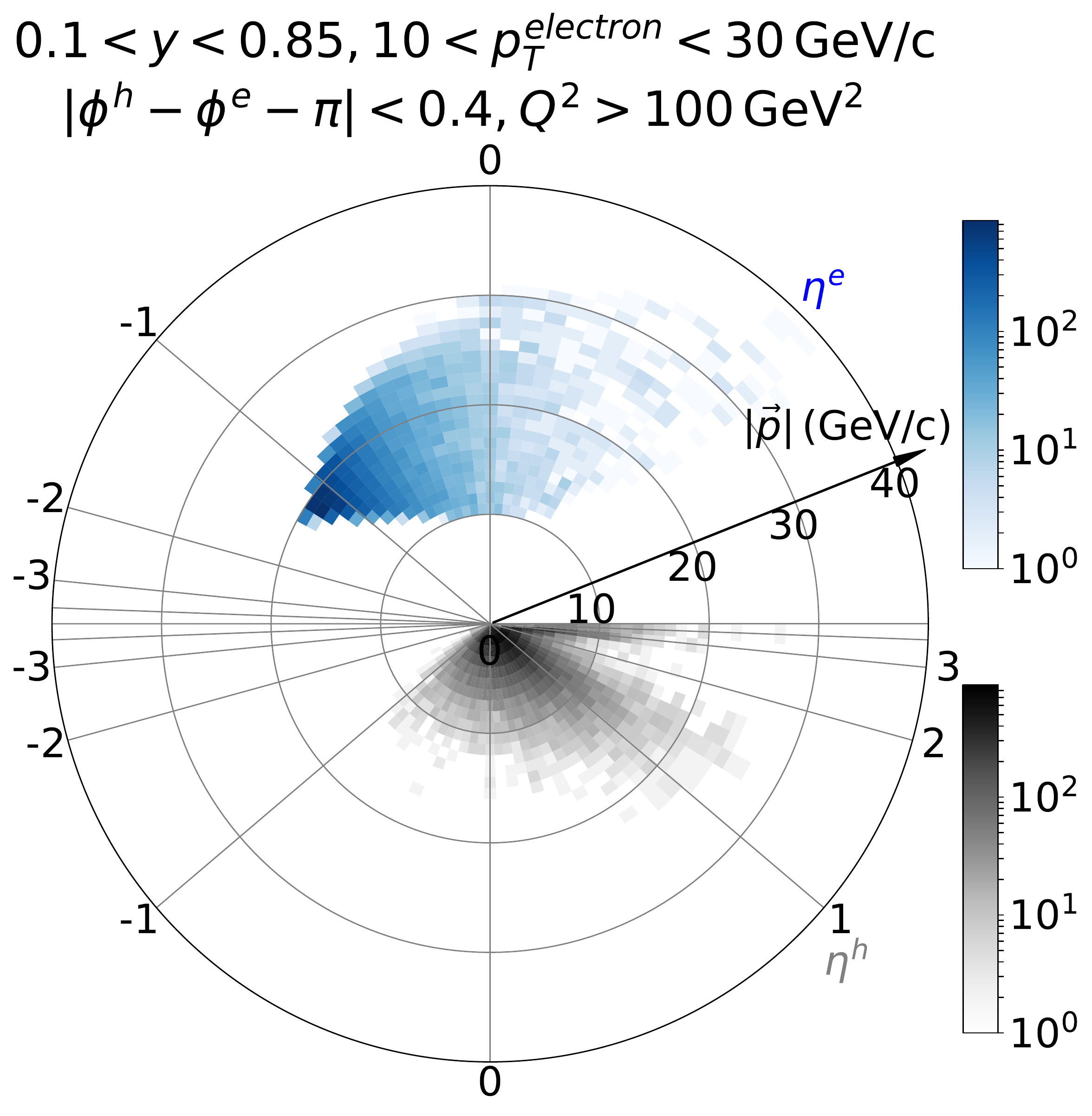}
  \end{minipage}
  \begin{minipage}[t]{0.29\linewidth}
    \includegraphics[angle=0,width=0.97\linewidth]{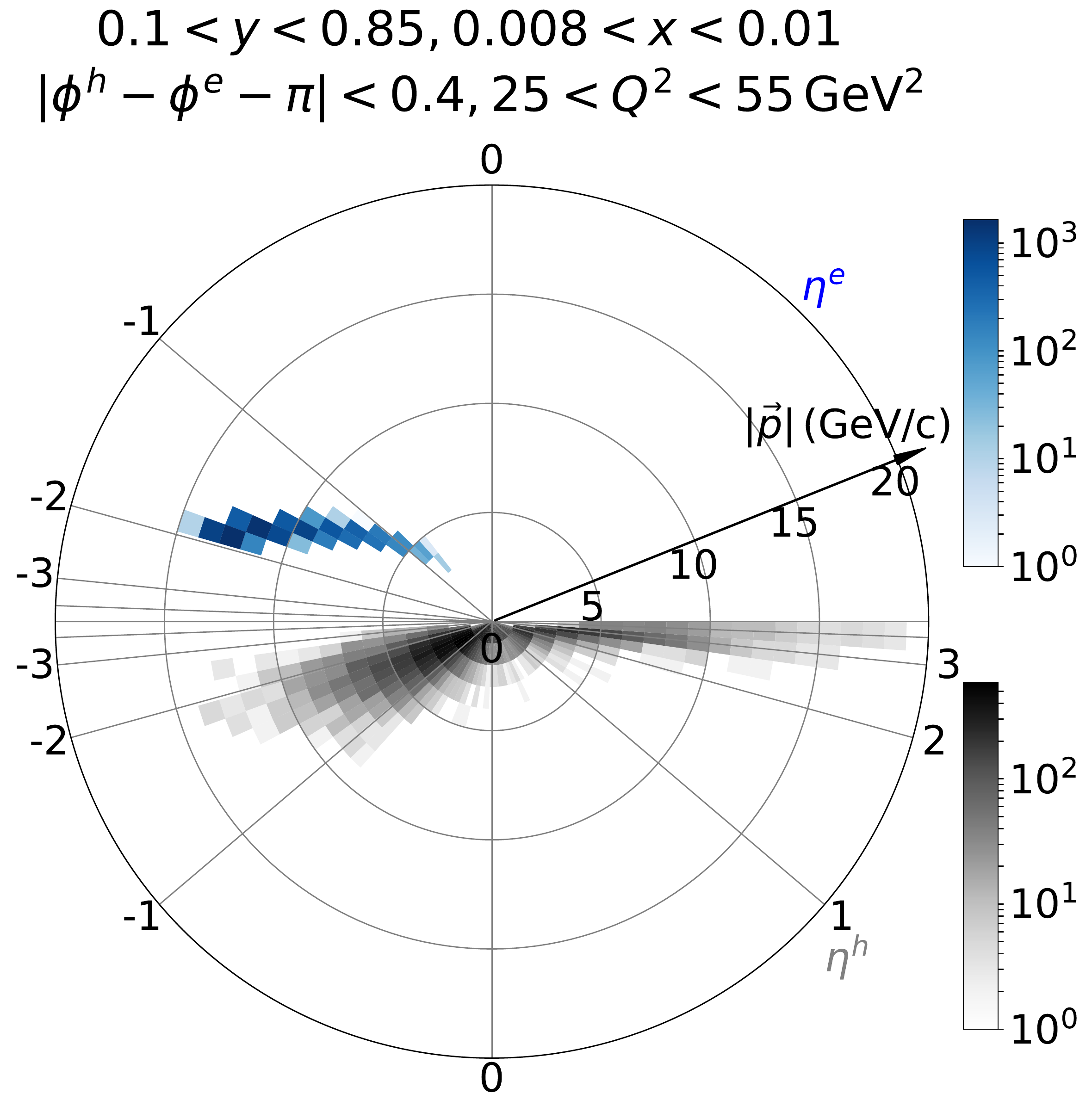}
  \end{minipage}
\end{center}
\caption{Jet properties from Ref.~\cite{Arratia:2019vju}.
         Left: Number of particles inside the jets as a function of the transverse momentum $p_T^{jet}$ in the lab frame. 
         Middle and Right: Polar plots of the kinematic distributions of the particles from jets produced in DIS (in gray) and
         of the scattered electrons (in blue) for two kinematic ranges. 
        }
\label{fig:part3-Det.Aspects.HCAL-multipl} 
\end{figure}

A jet in the final state consists of charged particles, photons, and neutral hadrons as neutrons and kaons. The energy fractions carried by these components are about 65\%/25\%/10\%~\cite[p.~4]{Arratia:dwgcalmay26} on average, with considerable fluctuations. The multiplicity depends on $p_T^{jet}$ (see Fig.~\ref{fig:part3-Det.Aspects.HCAL-multipl}, left) and is moderate. The jet fragments in the final state populate the whole range of the central detector $-3.5<\eta{}<3.5$ (see Fig.~\ref{fig:part3-Det.Aspects.HCAL-multipl}, middle and right). A large fraction of these particles have momenta below 10-20~\gevc.  

Jet measurements can be performed using purely calorimetric information, or alternatively using a combination of tracking and calorimetric info. The former method requires a high resolution hadronic calorimetry systems as was used at HERA in the ZEUS experiment~\cite{Hilger:1986xj}. The latter is known as ``energy-flow method'', which was first used in the ALEPH experiment at LEP~\cite{Buskulic:1994wz}, and then in several collider experiments  including H1 at HERA~\cite{Peez:energyflow} and CMS at the LHC~\cite{Sirunyan:2017ulk}, and is planned to be used in sPHENIX. The modern implementations of this method are using machine-learning techniques~\cite{Komiske:2018cqr}.    
  
 At EIC the energy-flow approach is envisioned for the jet reconstruction. 
The tracker and ECAL will measure about 90\% of the jet energy with a precision much higher than any hadron calorimeters built in the past, except may be for the very forward region in the hadron endcap, where tracking performance starts to deteriorate (depends on the magnet design). Energy-flow methods rely on precision measurements of the charged fragments of jets using the tracker instead of the calorimeters (ECAL + HCAL). However, calorimeter information is still needed to account for contributions from neutral hadrons for which a sufficient high granularity may be important to disentangle the different contributions, i.e. proper assignment of a signal to the neutral components of the jet. The physics requirements on the EIC detector system are discussed in Chapter~\ref{part2-sec-DetReq.Sum.Req} and are outlined in Table~\ref{SubDetReq}. The requirements on HCAL are summarized in Table~\ref{tab:EIC_HCAL_energy_resolution}.
The energy resolution in the table is referred to a single particle resolution rather than jet.  

%
%
%
\begin{table}[ht!]
\centering
  \small
  \begin{tabular}{|c|c|c|c|c|} 
  \hline \hline
   $\eta$  & \multicolumn{2}{c|}{EIC Specifications} & \multicolumn{2}{c|}{Conservative option}  \\ 
           \cline{2-5}
           & $\sigma_E/E$, \% & $E_{min}$, MeV & $\sigma_E/E$, \% & $E_{min}$, MeV \\
  \hline
  -3.5 to -1.0 &  $45/\sqrt{E}+7$ & 500 &  $~50/\sqrt{E}+10$ & 500  \\
  -1.0 to +1.0 &  $85/\sqrt{E}+7$ & 500 &  $100/\sqrt{E}+10$ & 500  \\
  +1.0 to +3.5 &  $35/\sqrt{E}$   & 500 &  $~50/\sqrt{E}+10$ & 500  \\
  \hline
 \end{tabular}
 \normalsize
 \caption{HCAL parameters from the EIC specifications (Table~\ref{SubDetReq}) and for a technically conservative option. 
 Several ways to improve the energy resolution are described in the text.}
 \label{tab:EIC_HCAL_energy_resolution}
 \end{table}

The requirements and options for technical implementation were discussed in presentations~\cite{
Page:dwgcalmar19,
Arratia:dwgcalmar30,
Arratia:dwgcalmay26,
Tsai:dwgfeb25,
Tsai:tmplmar21,
Tsai:dwgapr7
}.
Only light-collecting HCAL options have been considered in this section. Some alternative approaches are discussed in Section~\ref{part3-sec-Det.Aspects.HCAL.altern}. At EIC all envisioned calorimetry systems are sequential, i.e. ECAL followed by HCAL. 
This is driven by the relatively high EM energy resolution requirement which will be difficult to achieve with a single device serving simultaneously as ECAL and HCAL. A possible exception may be the very forward hadron endcap where the stochastic term for EM energy resolution might be relaxed due to the higher energy of incoming particles. Achieving high resolution for both EM particles and hadrons is a very difficult task, and there are no precedents from past experiments. For instance, the ZEUS collaboration at HERA operated a very high resolution hadron calorimeter, but paid a price in the form of a rather mediocre performance for EM shower detection $18\%/\sqrt{E}$, while the situation was vice versa for the H1 detector~\cite{Abt:1996hi}. An excellent EM resolution typically leads to a poor hadronic shower detection. As an example pointed out in Ref.~\cite{Wigmans:2002je}, once the choice is made for a crystal ECAL, it essentially does not matter what one installs behind it. The hadronic energy resolution will be poor. It will be completely determined by fluctuations in the energy sharing between the EM and hadronic calorimeter sections, which in this case have very different e/h values. This results in a typical hadronic resolution of approximately $100\%/\sqrt{E}$. Even the most sophisticated compensating hadronic sections cannot alter this conclusion. The challenge of balancing EM and hadronic calorimeter performance is a common problem for any calorimetry system. Other detector/collider specific limitations such as available space, dead material between the EM and hadronic sections, choice of the readout etc., also affect the hadronic resolution of any system. 

The total hadronic resolution of three high-resolution calorimeters (approximately compensated) and the various factors contributing to it are listed in Table~\ref{table:HCAL_energy_resolution_diff_cal}, where $\sigma_p $, $\sigma_s$, $\sigma_i$ are the fluctuations of the number of signal quanta, the sampling fluctuations and the intrinsic fluctuations, respectively.

\begin{table}[ht!]
\centering

 \begin{tabular}{||c||c||c||c||} 
 \hline
          & ZEUS $U^{238}$ & ZEUS Pb & SPACAL       \\ [0.5ex]
 \hline\hline
  $\sigma_p$  & $6\%/\sqrt{E}$  &  $10\%/\sqrt{E}$     &   $5\%/\sqrt{E}$ \\
 \hline
 $ \sigma_s$  & $31\%/\sqrt{E}$  &  $42\%/\sqrt{E}$     &   $27\%/\sqrt{E}$ \\
 \hline
  $\sigma_i$  & $19\%/\sqrt{E}$  &  $11\%/\sqrt{E}$     &   $11\%/\sqrt{E}$ \\
 \hline
  $\sigma_h$  & $37\%/\sqrt{E}$  &  $44\%/\sqrt{E}$     &   $30\%/\sqrt{E}$ \\
 
 \hline
\end{tabular}

\caption{Hadronic energy resolution of various calorimeters. Data taken from ~\cite{Drews:1989vr,Bernardi:1987xu,Acosta:1991ap,Wigmans:1997px}}
\label{table:HCAL_energy_resolution_diff_cal}
\end{table}

In all three detectors, the hadronic resolution is dominated by the sampling fluctuations. This is a direct consequence of compensation (e/h=1), which requires small sampling fractions, for example, 2.3\% for lead/plastic detectors and 5.1\% for uranium/plastic devices.

Much effort went into understanding of the mechanism of compensation in the past ~\cite{Wigmans:2000vf}, upon which the high-resolution ZEUS calorimetry system was build. However, one aspect of compensation was not immediately clear at that time, namely, the energy dependence which affects the precision of jet reconstruction. Data from ZEUS showed that, for particles below 10 GeV, the e/h ratio of the ZEUS calorimeter gradually decreases by ~30\% with decreasing energy ~\cite{Andresen:1989qr}. There is no known solution to this problem. For the EIC central detector, with exception of the very forward region in the hadron endcap (at ~ $\eta>2.5$), most hadrons will have energies below
10 GeV, and thus there is little value to pursue compensation (such as using depleted uranium) for the hadronic calorimeter section in these regions. In the very forward region of the hadron endcap the hadron energy will be above 10 GeV and the compensation technique is very relevant.  

One should point out that a fine granularity of a non-compensating calorimeter allows to improve the resolution by assigning weights to the detected signals (``off-line compensation''). The method was first used~\cite{Abramowicz:1980iv} at the CDHS experiment, and later applied and improved at H1~\cite{Abt:1996xv, Issever:2004qh}. H1 used a 45k-channel Pb/Fe Liquid Argon calorimeter. At EIC such a method can be considered where a longitudinal segmentation of the ECAL and HCAL readout appears practical. 

\subsection{Central detector consideration}

Precise measurements of the hadron energy with calorimeters requires sufficient containment of hadronic showers. Unlike the compact electromagnetic showers hadronic showers are very broad. The longitudinal and radial containment $L_{95\%}$ and  $R_{95\%}$, the required length and radius of the calorimeter for a 95\% hadronic energy deposition containment (Ref.~\cite{Leroy:2000mj}), scales as:

   \begin{equation}
      \label{eq:part3-Det.Aspects.HCAL-hadr_en_deposition}
       L_{95\%} \approx t_{max} + 2.5 \lambda_a , ~~~~R_{95\%} \approx 1 \lambda_{int},
    \end{equation}
    
 
where $t_{max} \approx 0.2\log[e]{E(GeV)} + 0.7$ is the shower maximum depth, and $\lambda_a$ (in units of $\lambda_{int}$ ) describes the exponential decay of the cascade beyond $t_{max}$ and varies with hadron energy as $\lambda_a=[E(GeV)]^{0.13}$.
For the EIC central detector a calorimeter system of approximately 5 $\lambda_{int}$ depth seems sufficient for most regions,
except for the forward region of the hadron endcap where it should be of order 6-7 $\lambda_{int}$. Table 
~\ref{table:HCAL_absorber_material} lists absorber materials typically used in HCALs and the thickness needed for the 95\% containment.

\begin{table}[ht!]
\centering
\begin{tabular}{|l|c|c|}
\hline
Absorber material & ${L_{95}}$ 15~GeV & ${L_{95}}$ 30~GeV \\
\hline
Fe & 80~cm & 94~cm \\
Pb & 83~cm & 99~cm \\
Cu & 72~cm & 86~cm \\
W  & 47~cm & 56~cm \\
U  & 52~cm & 61~cm \\
\hline
\end{tabular}
\caption{Absorber materials used in HCALs}
\label{table:HCAL_absorber_material}
\end{table}

The choice of the absorber material is often driven (apart from the energy resolution) by costs, engineering constraints, the magnet design, desire for a compensated calorimeter system, and, in case of a SiPM readout, by an acceptable level of the neutron fluence. 
Lower-$Z$ absorbers generate fewer neutrons. In this regard steel absorbers are preferable. 

\subsection{HCAL Energy resolution}

Precise measurements of hadron energy with sampling calorimeters require sufficiently high sampling fraction and sampling frequency to keep sampling fluctuations and number of signal quanta fluctuations below the acceptable threshold (see Table~\ref{table:HCAL_energy_resolution_diff_cal}). Increasing the sampling fraction leads to a significant reduction of the final calorimeter density. In addition, calorimeters with large sampling fraction require significant additional space for mechanical stability, as they are usually not self supporting. For example, $\lambda_{abs}$ for DU is 10.5 cm. However, the effective $\lambda_{abs}$ of the ZEUS calorimeter is 24 cm, about a factor of two larger. Collider central detectors are generally large-volume detectors, and the cost is an important factor in the calorimeter, in particular hadron calorimeter, design. As a consequence, compromises are usually necessary. 
As an example, the ZEUS and SPACAL HCAL systems listed in Table ~\ref{table:HCAL_energy_resolution_diff_cal} can give an idea of the space requirement for high resolution calorimeters. The ZEUS calorimeter system (hadron endcap) extended over almost 4 meters, of which about half the space was occupied by the high resolution DU/Sc calorimeter. The remaining space was occupied by the backing calorimeter whose purpose was to control longitudinal leakages. The SPACAL system required about 2 meters for the Pb/ScFi structure and additional 0.7 meters for the readout, which is similar for the E864 calorimeter based on the SPACAL 
design~\cite{Bernardi:1987xu,Acosta:1991ca,Armstrong:1998qs}.

The space available for all EIC detectors including the calorimeter systems is finite. 
Desired properties for the EIC calorimeters, beyond the requirement on energy resolution, are the following: compactness and mechanical sturdiness, which allows for building self-supporting structures and minimizing the space required for passive mechanical support structures. This for example, makes lead a non-ideal choice for the HCAL absorber as it would require a significant passive reinforcement in order to keep the mechanical stability of the detector plus an additional space to support the ECAL section. Ideally, it would be preferable to use the HCAL structure as a support for the ECAL. This is possible to achieve with a steel absorber. This choice of material would also eliminate dead material between ECAL and HCAL sections. Such dead material degrades the overall system performance 
as it is located almost at the shower maximum position~\cite{eRD1:rj2}.
To control longitudinal leakage one usually employs 
tail catchers, or backing calorimeters, as in case of ZEUS. At EIC the tail catcher would have to be integrated with the main calorimeter due to the lack of space for a separate device. Such an approach is 
described in Ref.~\cite{eRD1:rj2}, where the last few layers of the HCAL section would have additional independent readout. The information from the tail catcher allows for a clean identification of the showers without a longitudinal leakage.
There is a desire to have a higher resolution (better than $ \approx 40\%/\sqrt{E}$ with a constant term of $\approx 5\%$) calorimeter in the forward hadron endcap ($\eta >$ 2.5), i.e. the region where the  calorimeter performance is anticipated to exceed that of the tracker. This interplay of the calorimeter and tracker performance is similar to that in the electron endcap, where the inner part requires the highest resolution ECAL and the outer part has more relaxed requirements (see Table~\ref{tab:EIC_HCAL_energy_resolution}). In the hadron endcap, taking into account the limited available space, a very dense calorimeter that minimizes leakage and ideally serves as both ECAL and HCAL with a single readout would be preferable. Such a calorimeter should have a small sampling fraction and a sufficiently high sampling frequency (to keep the EM energy resolution at an acceptable level), which is currently only possible with the fiber calorimeter technology.

\begin{figure}[htb]
\begin{center}
  \begin{minipage}[t]{0.495\linewidth}
    \includegraphics[angle=0,width=0.99\linewidth]{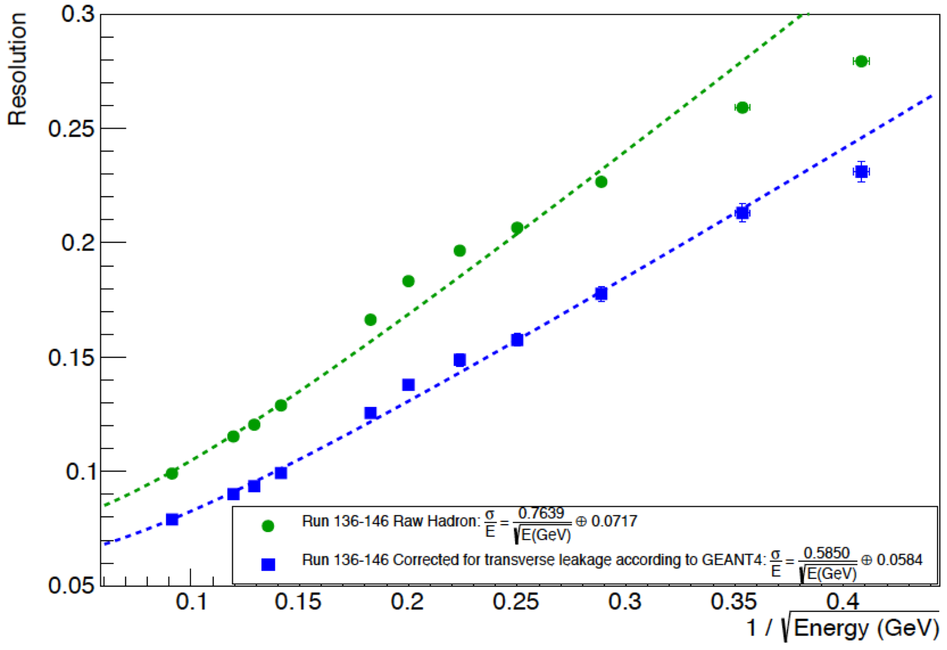}
  \end{minipage}
  \begin{minipage}[t]{0.495\linewidth}
    \includegraphics[angle=0,width=0.99\linewidth]{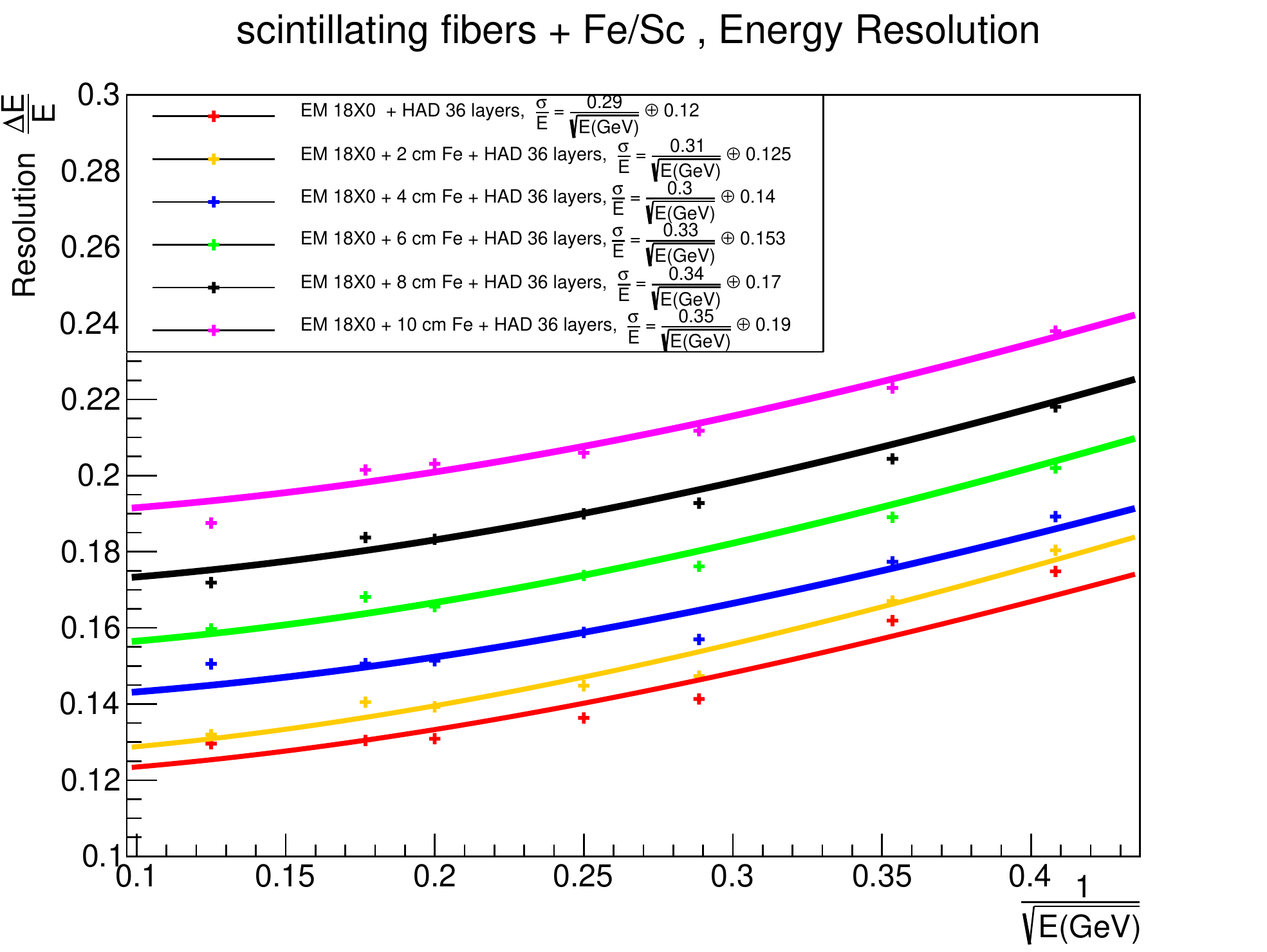}
  \end{minipage}
\end{center}
\caption{Energy resolution from Ref.~\cite{Tsai:tmplmar21}.
         Left: Measured resolution for ECAL (shashlyk) and HCAL (Fe/Sc). The top points are the measurements. 
         The lower points and the lower curve $\sigma(E)/E\approx{}0.06 \oplus 0.59/\sqrt{E}$
         were obtained using a correction (based on simulation) for the lateral leakage.
         Right: A calculated impact of material between ECAL (W/ScFi) and HCAL (Fe/Sc) on the energy resolution. The resolution changes from
         $0.12 \oplus 0.29/\sqrt{E}$ (no material) to $0.19 \oplus 0.35/\sqrt{E}$ (10~cm of iron).
        }
\label{fig:part3-Det.Aspects.HCAL-resol} 
\end{figure}

The ``conservative option'' for the endcaps shown in Table~\ref{tab:EIC_HCAL_energy_resolution} is based on measurements of prototypes (see Fig.~\ref{fig:part3-Det.Aspects.HCAL-resol}, left). The barrel HCAL parameters can be modeled from the sPHENIX barrel HCAL, which has a similar size and geometry. At sPHENIX the ECAL+HCAL have a resolution for pions of about $13\% \oplus{} 85\%/\sqrt{E}$~\cite{Nattrass:2018bdt}. There, the HCAL consists of a short section inside the bore of the BaBar magnet and the main Fe/Sc section outside of the cryostat. The outside part also serves as the magnet's flux-return yoke. At EIC no HCAL inner section is envisaged. In order to scale the sPHENIX results to the EIC configuration in a conservative way, let us assume that at sPHENIX there is no material in the cryostat, while at EIC the cryostat contains material, and the outside HCAL has the same thickness as both parts of the sPHENIX HCAL.  
The estimated impact of material (up to 10~cm or iron) between ECAL and HCAL is shown in Fig.~\ref{fig:part3-Det.Aspects.HCAL-resol}, right. In comparison, the BaBar cryostat is about $1.4X_0$, which is equivalent to 2.5~cm of iron. Such amount of material should cause only a moderate degradation and one may expect that a resolution of $10\% \oplus{} 100\%/\sqrt{E}$ in Table~\ref{tab:EIC_HCAL_energy_resolution} is a conservative estimate and is achievable. 

\subsection{eRD1 EIC R\&D and STAR forward developments}

To date R\&D efforts towards high-resolution hadron calorimetry at the EIC have been limited since the existing technologies have been considered sufficient.  The very first eRD1 calorimetry consortium proposal~\cite{eRD1:pr1}
aimed at developing a new W-powder/ScFi technology for both EM and HCAL sections to help to balance the requirements on the EM and hadron energy resolutions. In particular, the technique was aimed at simplifying the construction of EM calorimeters with high sampling frequency and small sampling fraction (approximately being compensated) providing a $\approx{}12\%/\sqrt{E}$ energy resolution. With a support from the STAR Forward upgrade project the eRD1 consortium built a small prototype of a compensated calorimeter system with the new W/ScFi technology in the EM section and an HCAL section copying the ZEUS Pb/Sc prototype, listed in Table~\ref{table:HCAL_energy_resolution_diff_cal}. This system was tested at FNAL in 2014~\cite{Tsai:2015bna} and was modeled in the BEAST EIC detector model
as the hadron endcap. Such a compensated system can meet the requirements for the EIC hadron calorimeters listed in Table~\ref{tab:EIC_HCAL_energy_resolution}. However, the non-compensated variant was considered as well. This originated from a
budgetary constraint for the STAR forward upgrade that eventually led to the development of a non-compensated calorimetry system consisting of Pb/Sc shashlyk for the EM section (utilizing the existing EM blocks from the PHENIX experiment) and Fe/Sc for the hadronic section. A small prototype of this system was built and tested at FNAL in 2019.

With accounting for the transverse leakages in the test beam prototype, the  energy resolution for STAR FCS system is close to ${60\%/\sqrt{E}+6\%}$ (see Fig.~\ref{fig:part3-Det.Aspects.HCAL-resol}, left). An earlier tested compensated prototype had a $\approx$30\% better hadronic energy resolution compared to the non-compensated version.   

Additional R\&D efforts have been carried out to demonstrate a similar system with W/ScFi for the ECAL section that could meet the EIC physics requirements~\cite{BNL-Prop:2011mar}.

\subsection{Alternative methods for high resolution HCAL}
\label{part3-sec-Det.Aspects.HCAL.altern}

Over the past two decades there were attempts to significantly improve the energy resolution of hadron calorimeters using the dual readout method. This method uses an observable which correlates with the number of neutrons released in the hadronic shower, which correlates with the "invisible" energy ($\approx 40\%$ in the hadronic shower~\cite{Wigmans:1998hkz}).

By comparing the signals produced by scintillation light and Cherenkov light in the same detector, and considering the timing and spatial characteristics of the showers, the EM shower fraction can be determined for individual events. The EM shower fluctuations are the main culprit for problems encountered with hadronic calorimetry, The validity of this principle has been demonstrated with the DREAM fiber calorimeter~\cite{Benaglia:2016adr}. A realization of the dual readout at EIC would have to take into account the relatively low energy of hadrons and the spacial constraint. eRD1 took an opportunity to look at the timing characteristic of showers using the STAR Forward calorimeter prototype with steel absorber during the 2019 test run at FNAL. No meaningful correlation of the fast component of the hadronic shower with the total energy has been observed. Accounting for the EM fraction of the shower on an event-by-event basis using this method does not look promising (at least with the steel absorber).

At the end of this section we should also mention alternative concepts of designing the whole detector in which the role of calorimeters is quite different compared to what has been traditionally used. These concepts were initially driven by the HEP community for the future linear collider program, which requires an extremely high energy resolution for jets. Tungsten-silicon sampling calorimeters allow for very fine granularity required for detection of dense high-energy jets, and typically combine the electromagnetic and hadron parts into a uniform structure. Such a calorimeter, which also provides a very strong radiation hardness, is being build for the CMS endcaps upgrade~\cite{Barney:2020qbq}.
Calorimeters in these concepts are essentially digital devices with hundreds of millions of channels to track every single particle in hadronic showers, as required by particle flow algorithms. 
The {\it TOPSiDE}~\cite{Repond:2018kap,Armstrong:dwgaug18} concept of the EIC detector, discussed in more details in Section~\ref{part3-sec-Det.Aspects.CAL_TOPSIDE}, is an example of such an approach.

In summary, the set of parameters for the technically conservative option listed in Table~\ref{tab:EIC_HCAL_energy_resolution} should be achievable with existing technologies as demonstrated by the eRD1 consortium and the STAR Forward upgrade, with some additional R\&D efforts to improve the performance of a STAR-like forward calorimeter system. Higher resolution hadron calorimetry will require additional R\&D efforts, e.g. to develop a high density fiber calorimeter with SiPM readout or another suitable technology.

\subsection{TOPSiDE Calorimetry}
\label{part3-sec-Det.Aspects.CAL_TOPSIDE}

The TOPSiDE (timing optimized PID silicon detector for the EIC) concept leverages recent developments in ultra-fast silicon sensor technology to simplify tracking and particle identification for the barrel and endcap regions.
At its core, TOPSiDE consists of a high-precision silicon vertex detector surrounded by a tracker composed of ultra-fast silicon detectors (UFSD), such as low-gain avalanche-diode (LGAD) detectors, which is then surrounded by calorimetry.
With the excellent time resolution of the tracking detector, particle identification ($\pi-K-p$ separation) is done with time-of-flight alone, removing the need for a dedicated particle-identification system for the majority of the central detector. This minimizes material thickness, number of detector subsystems, and distinct technologies required for the central detector.

The TOPSiDE concept does not put any special requirements on the calorimetry system,
and the reference detector calorimetry requirements could be met by a wide variety of designs and technologies, all capable of integrating within the TOPSiDE concept.

In the spirit of minimizing the distinct detector technologies throughout, a sampling (or imaging) electromagnetic calorimeter of silicon sensors interleaved by tungsten plates forms a natural choice --- a SiW calorimeter.
In an earlier incarnation of the TOPSiDE concept, the ambitious use of high-resolution LGAD sensors throught the electromagnatic calorimeter was evaluated, in the spirit of high precision digital calorimetry~\cite{Breton:2020xel}.
In this case, it would be possible track every single particle in hadronic showers, as required by particle-flow algorithms~\cite{Repond:2019fbz}.
The drawbacks to this approach lie in the demanding power and cooling requirements for such a calorimeter, as well as its cost and large amount of readout channels. 
In particular as the performance of such a system may very well surpass what is required for the EIC science program.

A less ambitious, more grounded approach to SiW-based electromagnetic calorimetry is to use a different type of silicon sensor which better matches the calorimetry requirements at the EIC.
A natural choice in this direction would be to leverage the recent progress in monolithic silicon sensors, e.g.,  ATLASPix~\cite{Schoning:2020zed},
a low-power pixel detector that was built for the CERN experiment ATLAS, and further optimized as AstroPix \cite{Brewer:2021djn}, developed for a future space-based gamma-ray telescope.
This family of sensors has demonstrated an excellent energy resolution at low energies ($\sim$7\% at 30keV) and does not have stringent power and cooling requirements (it will be used for space-based detectors).
Its timing resolution of 25ns is sufficient for a medium-granularity electromagnetic calorimeter.
This would result in an imaging calorimeter with potentially great resolution for low energy photons from $\pi^0$-decay, while having excellent electron-pion separation.
Furthermore, to optimize the resolution on scattered electron, TOPSiDE uses a PbWO$_{4}$ crystal calorimeter at the inside of the backward endcap, similar to the reference detector. 

For hadron calorimetry, the TOPSiDE baseline design uses the same approach as the reference detector.
Other options being considered beyond the baseline design are based around  a granular (imaging or semi-digital) approach using interleaved layers of steel with scintillator pads \cite{Adloff:2012gv} or resistive plate chambers (RPC)~\cite{Baulieu:2015pfa}. 
Even a single RPC layer can provide additional timing measurements and enhance TOPSiDE's 4D track fitting, leading to better hadron separation. 
This type of system will have a very competitive energy resolution with the additional feature of robust muon particle identification.

\section{Particle Identification}
All multi-purpose detectors, for example as illustrated in 
Figure~\ref{central-detector-CAD}, contain systems that work symbiotically toward achieving the physics goals.  Among these detector systems is the subset that identifies the species of collision ejectiles commonly known as Particle Identification Detectors or PID.  
Typically, the tracking systems provide a momentum measurement ($\vec{p} = m \gamma \vec{\beta}$) which when combined with information on velocity ($\vec{\beta}$) is sufficient to distinguish the various particle species.  Most often "PID" refers to the separation of $\pi$, $K$, and $proton$ whereas eID refers to the identification of electrons.  This section discusses each of these two topics, the requirements for EIC, and possible technological implementations necessary to achieve the physics goals.

\begin{figure}[hbt]
	\centering
        \includegraphics[width=1.0\columnwidth]{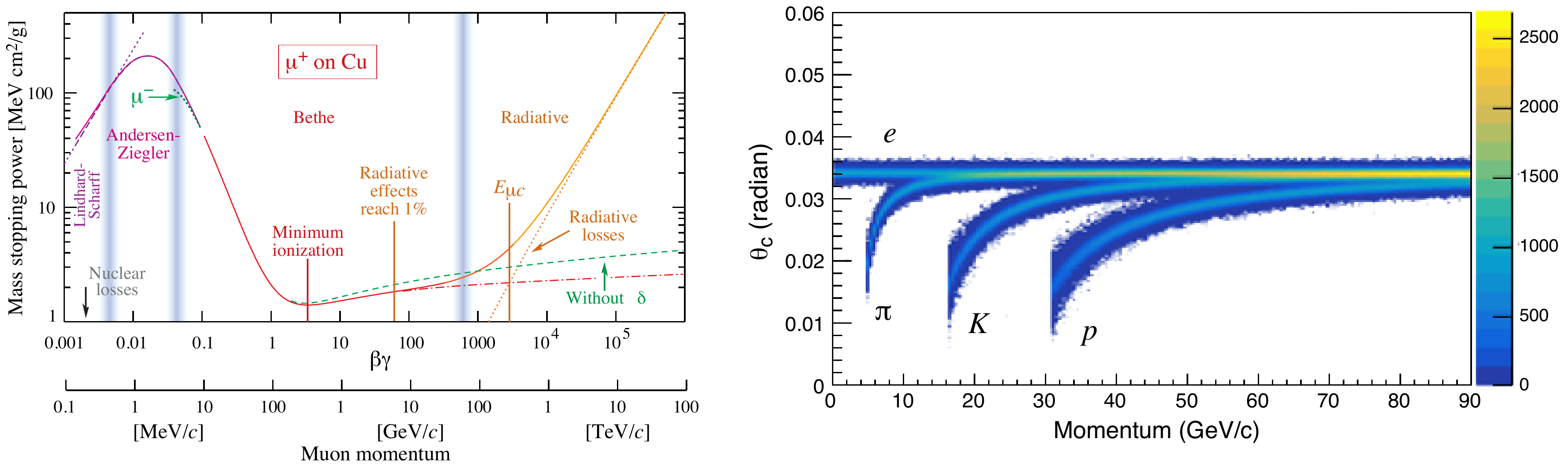}
	\caption{Examples of the velocity-dependent detector interactions used for PID devices.  The left panel denotes the rate of energy loss of a charged particle passing through matter.  The right panel denotes the angle of Cherenkov radiation.}
	\label{Fig:01-PID-Physics}
\end{figure}

The two basic approaches to PID are the direct measurement of the particle's velocity (known as Time-of-Flight or "TOF") and the measurement of velocity dependent interactions of the particle with the detector.  Four common velocity-dependent detector interactions are \cite{Zyla:2020zbs}:
\begin{itemize}
    \item Specific Ionization (aka dE/dx), wherein the rate of energy deposit (typically left in a gasseous medium) is measured precisely.
    \item Cherenkov Radiation, wherein the angle of Cherenkov photon production depends upon velocity as 
    cos($\theta$)~=~1/(n$\dot \beta$).
    \item Bremsstrahlung, wherein the power dissipated to braking radiation goes as $P=\frac{q^2\gamma^4}{6\pi\epsilon_0c}\left( \dot{\beta} + \frac{\left(\vec{\beta}\cdot\dot{\vec{\beta}} \right)^2}{1-\beta^2}\right)$
    \item Transition Radiation (TR), wherein the intensity of transition radiation goes as $I=\frac{Z^2e^2\gamma\omega_p}{3c}$.
\end{itemize}
Bremsstrahlung is the effect by which eID is accomplished in an electromagnetic calorimeter.  The calorimetry requirements for EIC are discussed in Section~\ref{part3-sec-Det.Aspects.ECAL} and will not be additionally discussed here. The velocity necessary to produce sufficient transition radiation is high enough that at EIC a Transition Radiation Detector (TRD) should be considered specifically as an eID device.  The velocity dependence of $\frac{dE}{dx}$ and the Cherenkov Effect, as highlighted in Figure~\ref{Fig:01-PID-Physics} \cite{Zyla:2020zbs}, are suitable for PID and eID applications.

\subsection{Physics Requirements}
\begin{figure}[htb!]
\centering
\includegraphics[width=\textwidth]{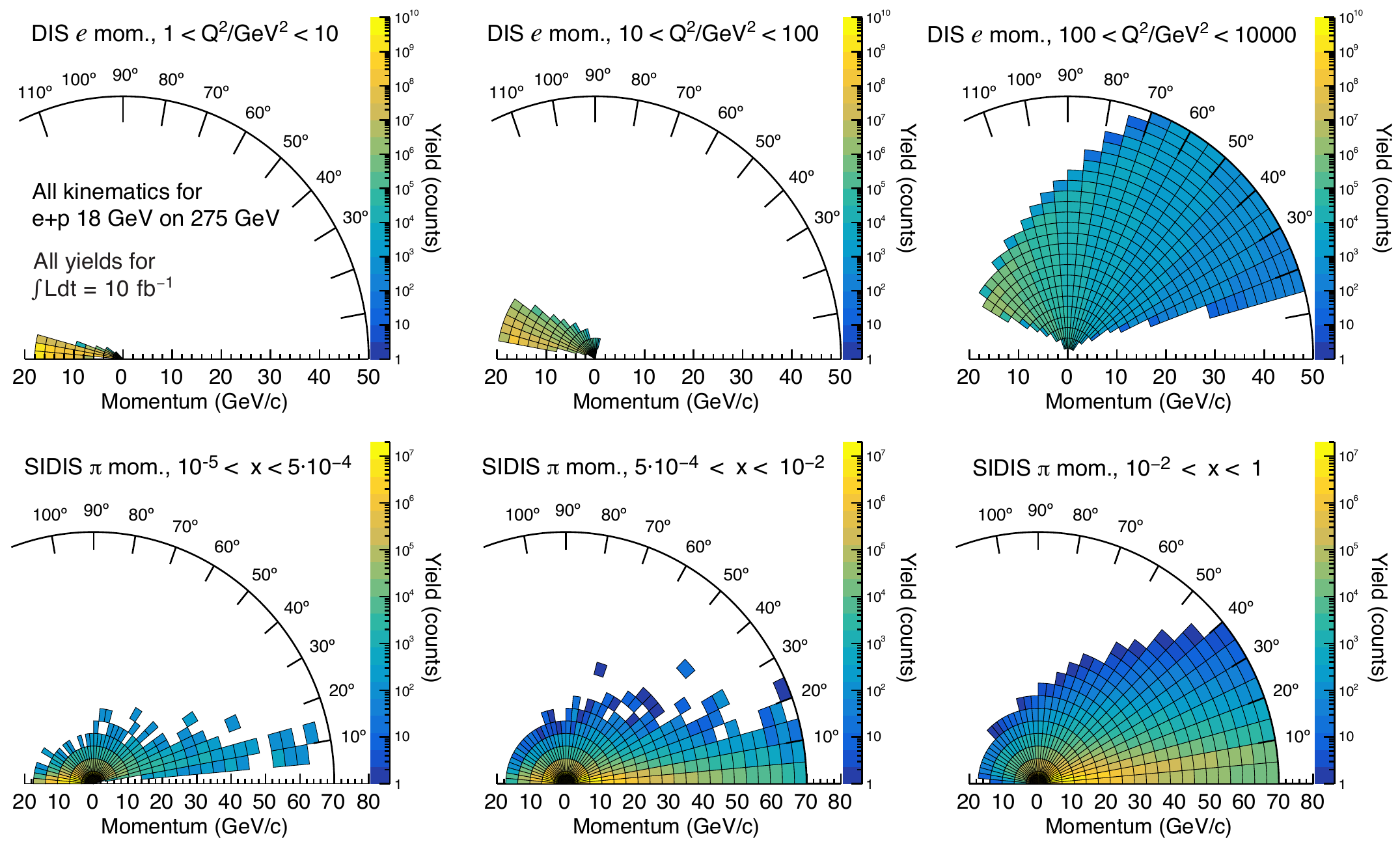}
\caption{\label{fig:Kinematics-in-PID-section}
DIS electron and SIDIS pion simulated yield for 18 GeV electron on 275 GeV proton collisions.  Yield (color scale) is plotted in polar coordinates with the radial coordinate indicating momentum and the azimuthal coordinate indicating ejectile polar angle. The top row is for electrons, the bottom row is for pions.  The columns are selected ranges of $Q^2$ as indicated.  The asymmetry of the initial state is reflected as a momentum asymmetry in the ejectiles.
}
\end{figure}

As described in Volume II in this report, simulations of collisions for an extensive list of physics processes, each spanning the $\sqrt{s}$ anticipated at EIC have been performed.  As an example, Figure ~\ref{fig:Kinematics-in-PID-section} displays an overview of electron and hadron production as a function of particle lab momentum and polar production angle.
The full suite of such calculations was considered and used to formulate the so-called "Requirements Matrix" that specifies relevant detector performance parameters as a function of $\eta$.  A successful detector design is any that satisfies the detector performance requirements.  The PID-relevant subset of the detector matrix is shown in Table~\ref{table:PID-Matrix}.

In the following sections we discuss the performance characteristics of multiple suitable detector technologies for the final EIC detector.  Following that we discuss how these technologies can be arrayed to best address the requirements matrix.


\begin{table}[htb!]
\centering
\caption{\label{table:PID-Matrix}
Detector performance matrix.  This truncated form of the matrix selects only the requirements for Particle Identification.
}
\includegraphics[width=0.8\linewidth]{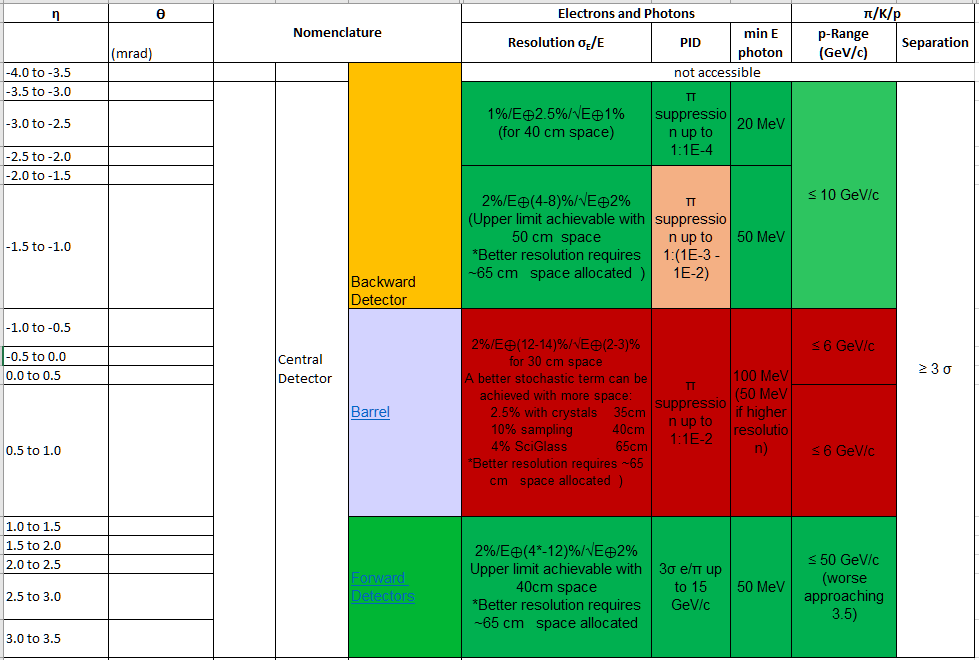}
\end{table}

\subsection{PID via Energy Loss}
\label{dE/dx}
Many tracking detector configurations are possible (as described in section\ref{part3-sec-Det.Aspects.Tracking}). The ”hybrid” option includes a Time Projection Chamber (TPC) as its outer layer, which may provide PID information via dE/dx . It is thus important to understand the limits of such devices.
Available space is at a premium, partly due to the longitudinal limit of $\pm$ 4.5 meters in Z. Given available space, tracking is generally limited to a radial extend of roughly 1~meter, which is significantly smaller than common TPCs such as STAR (2m)~\cite{Anderson:2003ur} and ALICE (2.5m)~\cite{Alme:2010ke}.  It is thus, important to work to achieve excellent $\frac{dE}{dx}$ performance in a small distance.

The primary challenge in any $\frac{dE}{dx}$ measurement comes from the process of energy loss being two steps.  Each locus of ionization is independent of its neighbors and therefore the rate of primary ionization follows Poisson statistics.  This rate is typically captured by the parameter $N_p$ which counts the primary ionization sites per unit length (usually expressed as $\frac{primary}{cm}$).  Unfortunately, primary electrons are often released with sufficient energy to generate several secondaries making a so-called "cluster" of ionization.  The total ionization is characterized by $N_t\frac{total}{cm}$ and follows a probability distribution with a long "Landau tail".  Battling the resolution loss due to the Landau tail is the primary challenge for any PID detector.

\begin{figure}[hbt]
	\centering
      \includegraphics[width=0.9\columnwidth]{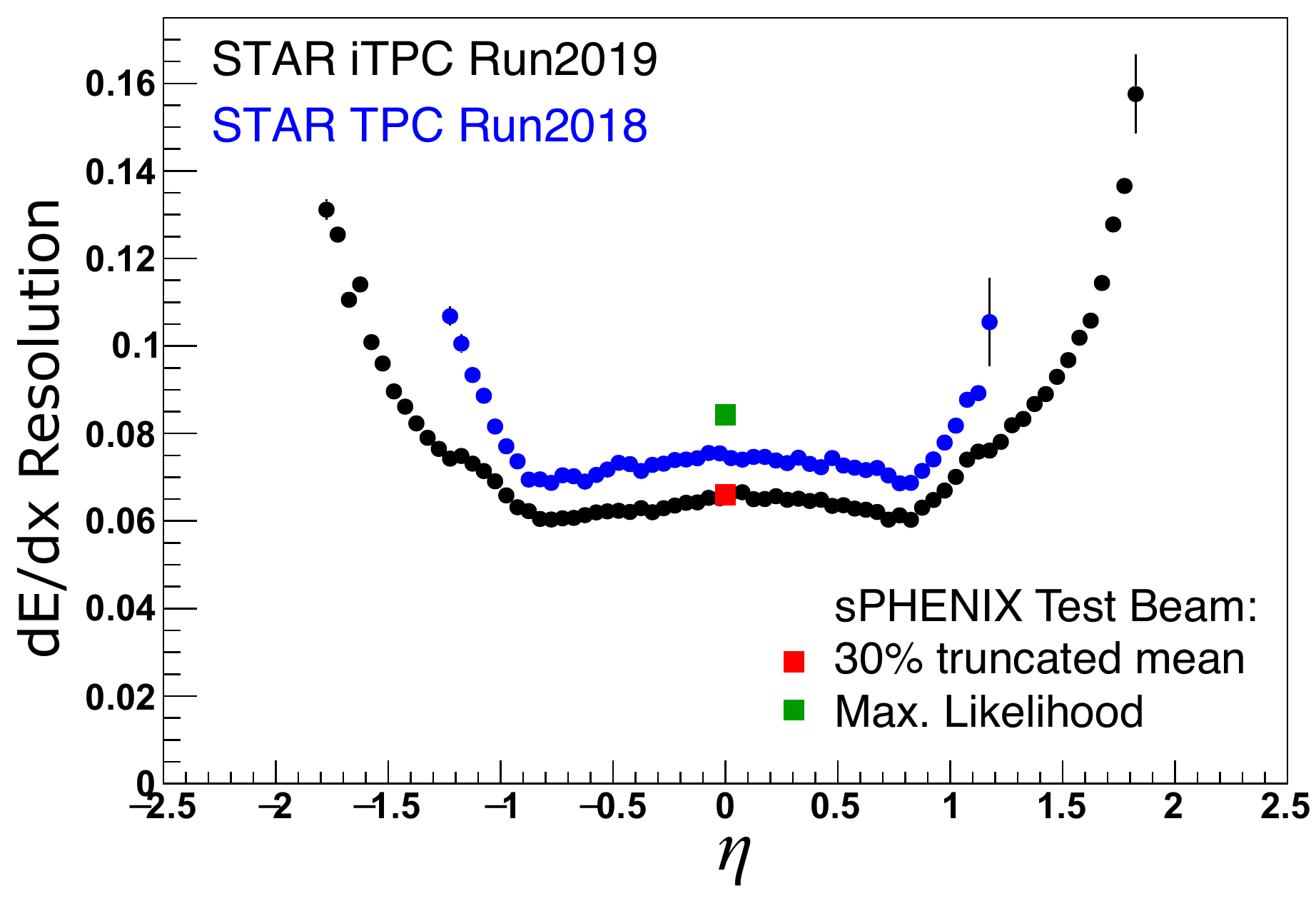}
	\caption{A comparison of STAR $\frac{dE}{dx}$ resolution (round points) with sPHENIX Test Beam results (squares) is shown.  The sPHENIX results are comparable by a suitable gas choice: using a gas with a limited number of ionization secondary electrons  per primary ionization event, the relevance of the tail in the Landau distribution can be mitigated. 
	\label{Fig:03a-Dedx-Measured-a}}
\end{figure}

\begin{figure}[hbt]
	\centering
      \includegraphics[width=0.8\columnwidth]{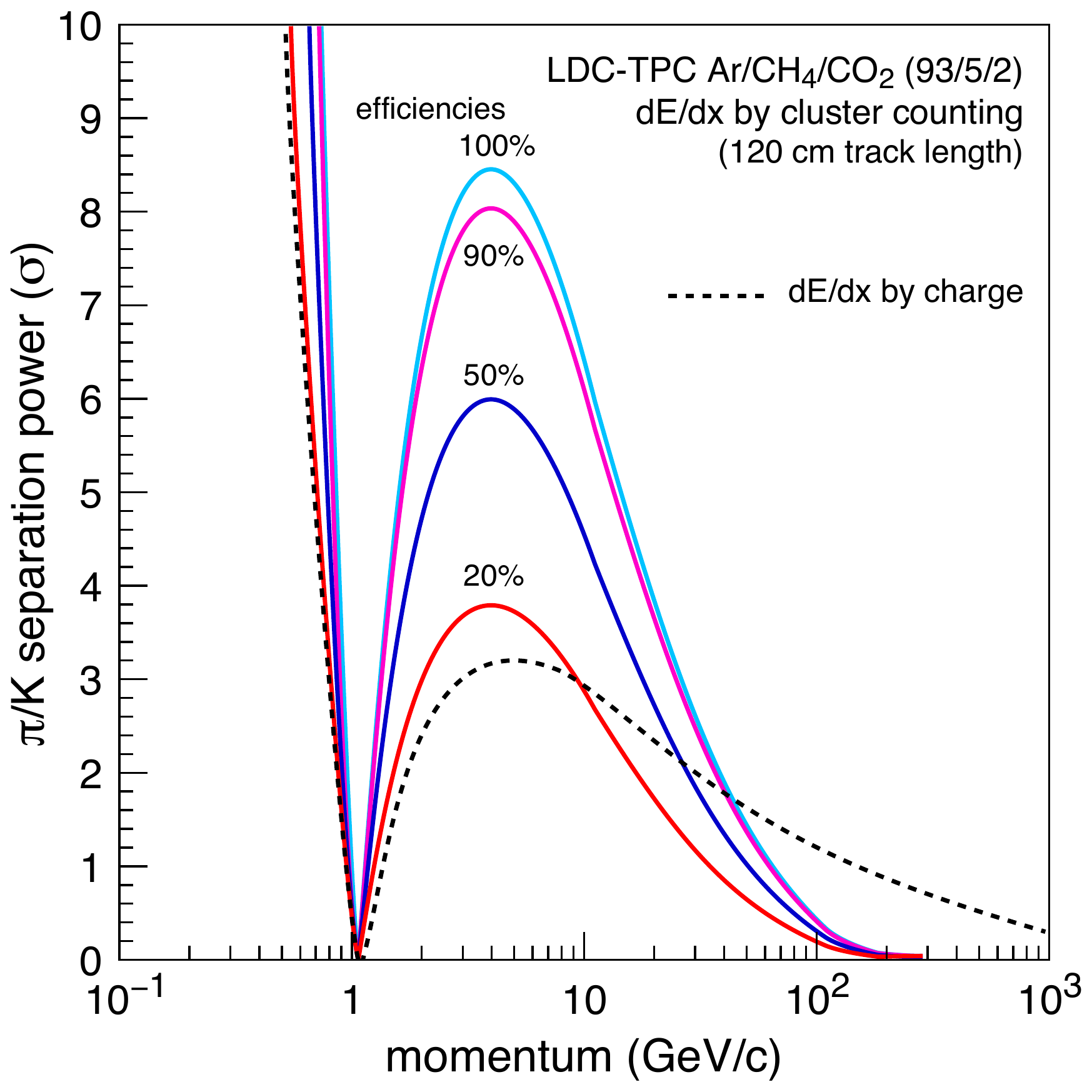}
	\caption{A simulation of $\pi$-K separation in a "cluster-counting" detector, for which the fluctuations are reduced from Landau to Poisson, thereby extending the effective momentum range.
	\label{Fig:03a-Dedx-Measured-b}}
\end{figure}

The traditional method of addressing the Landau is to make many independent samples of the ionization and perform either a fit to the $\frac{dE}{dx}$ probability distribution or via a so-called "truncated mean" calculation.  An improvement recently suggested and tested by sPHENIX for EIC applications is to use a gas that has an intrinsically small ratio of total electrons to primary electrons, $\frac{N_{t}}{N_{p}}$, so that the fundamental ionization statistics are closer to Poisson.  Figure~\ref{Fig:03a-Dedx-Measured-a} shows a comparison of STAR $\frac{dE}{dx}$ resolution (Ar:$CH_4$[90:10], 72 samples, 150 cm, $\frac{N_t}{N_p}=3.9$) to a small sPHENIX prototype (Ne:$CF_4$[50:50], 48 samples, 60 cm, $\frac{N_t}{N_p}=2.3$) as measured in test beam\cite{PID:Jhuang}.  Despite fewer samples and shorter detector length, similar performance is indeed achieved.  Figure~\ref{Fig:03a-Dedx-Measured-b}  shows a simulation of performance assuming that one can count explicitly the individual clusters of ionization and thereby approach the limit of Poisson statistics\cite{PID:Timepix}.  Such a device might be rather attractive for EIC, but requires further R\&D to demonstrate its efficacy in short length applications such as required for EIC. A drift chamber option where cluster counting is implemented is presented in Sec.~\ref{gaseous_tracking}.


\subsection{Cherenkov}
\label{cherenkov}
The measurement of the emission angle of Cherenkov photons is a powerful PID technique with a tunable dynamic range.  Because the Cherenkov angle depends upon velocity as $\cos\left(\theta_C \right)=\frac{1}{n\beta}$, one is able to accomplish PID at the highest momentum using the lowest index of refraction, $n$.  There are two penalties for choosing low $n$.  First, with lower index, the Cherenkov threshold $\beta = \frac{1}{n}$ goes up, resulting in non-detection of low momentum particles.  Second, with lower index, the photon yield per unit length $\frac{dN_{\gamma}}{dL}=2\pi\alpha\sin^2 \left( \theta_c \right)\int\frac{d\lambda}{\lambda^2}$,  goes down resulting in the need of long radiators.  As a result, Cherenkov detectors must be carefully tuned to the required physics.  Because the momentum range needs at EIC 
vary significantly with $\eta$ it is necessary to tune the radiator index differently in three regions called "electron endcap", "barrel", and "hadron endcap".

\begin{figure}[hbt]
	\centering
        \includegraphics[width=1.0\columnwidth]{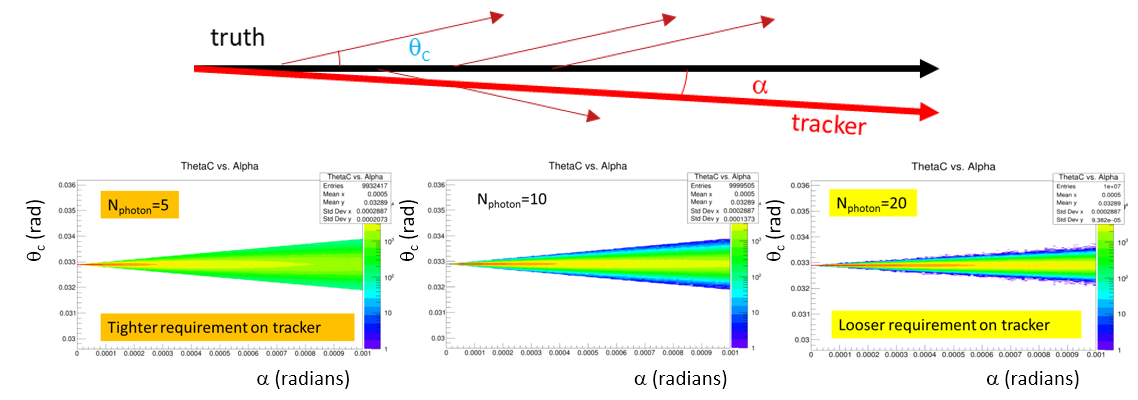}
	\caption{Simulations of the effect of tracking resolution on apparent Cherenkov angle.  Reconstructed $\theta_C$ distributions are plotted as a function of tracking error $\alpha$ under three assumptions for mean Cherenkov photon yield:  5, 10, and 20 photoelectrons per ring.}
	\label{Fig:03b-TrackingResolutionEffect}
\end{figure}

A subtle coupling between Cherenkov measurement and tracking resolution is illustrated in Figure ~\ref{Fig:03b-TrackingResolutionEffect}.  Because a Cherenkov detector rarely measures the trajectory of the track, it is reliant upon the tracker to provide a direction vector of the track itself \underline{while the track passes through the radiator}.  Figure~\ref{Fig:03b-TrackingResolutionEffect} shows a toy Monte Carlo simulation of this effect.  Here, a track with it's true direction indicated by the black arrow releases photons with angle $\theta_C$.  The tracker is presumed to report the particle direction as indicated by the red arrow; different from truth by the angle $\alpha$.  A mis-measurement by $\alpha$ generates photon-by-photon errors in the apparent $\theta_C$ in the range $\left(\theta_C-\alpha\right)$ to $\left(\theta_C+\alpha\right)$. The three bottom panels show reconstructed $\theta_C$ as a function $\alpha$ under different assumptions for mean photoelectron yield:  5 photons/ring, 10 photons/ring, and 20 photons/ring. These plots demonstrate that the impact of finite track direction errors of size $\alpha$ can be mitigated to some degree by higher Cherenkov photon statistics and that the goals for track pointing resolution will be tightly coupled to the chosen PID technology and performance.

A variety of effects limit the precision of measurement of Cherenkov angle in any realistic device.  These are listed and discussed here:
\begin{itemize}
    \item {\bf Chromaticity}\\All materials suffer from an index of refraction that varies with wavelength ($n(\lambda)$) thereby creating a photon-by-photon chromatic smearing of the Cherenkov angle.  This effect is particularly acute near the transmission cutoff of the radiating medium.
    \item {\bf Optical Aberration (aka "Emission Point Error")}\\Even at normal particle-to-mirror incidence, a spherical mirror is just an approximation to a parabolic reflector.  Furthermore, as the angle of incidence strays from the normal, optical aberrations increase.  This effectively means that the location at which a photon is detected picks up a dependence on the place within the radiator at which the photon was emitted.  It is therefore most often termed as an "Emission Point" Error.
    \item {\bf Pixelation}\\Cherenkov photons are detected individually and the finite pixel density of the focal plane readout detector thereby generates an uncertainty in the initial emission angle.
    \item {\bf Magnetic Field}\\Ideally the radiator medium for a Cherenkov radiator is free of magnetic field so that the particle direction is not changing as it propagates through the radiator medium.  In a compact application like EIC this is often difficult to arrange and is sometimes approximated by attempting to minimize $\vec{v} \times \vec{B}$ through careful adjustment of the magnetic field orientation.  Imperfections necessarily generate uncertainty in the Cherenkov angle.
    \item {\bf Tracking}\\Finally, the Cherenkov angle resolution can be limited by the knowledge of the track direction as it traverses the radiator medium.
\end{itemize}

In the following sections, we'll discuss in detail several options for Cherenkov detector configurations that have been studied in the EIC context.

\subsubsection{Hadron Blind Detector (HBD)}
\begin{figure}[hbt]
	\centering
        \includegraphics[width=1.0\columnwidth]{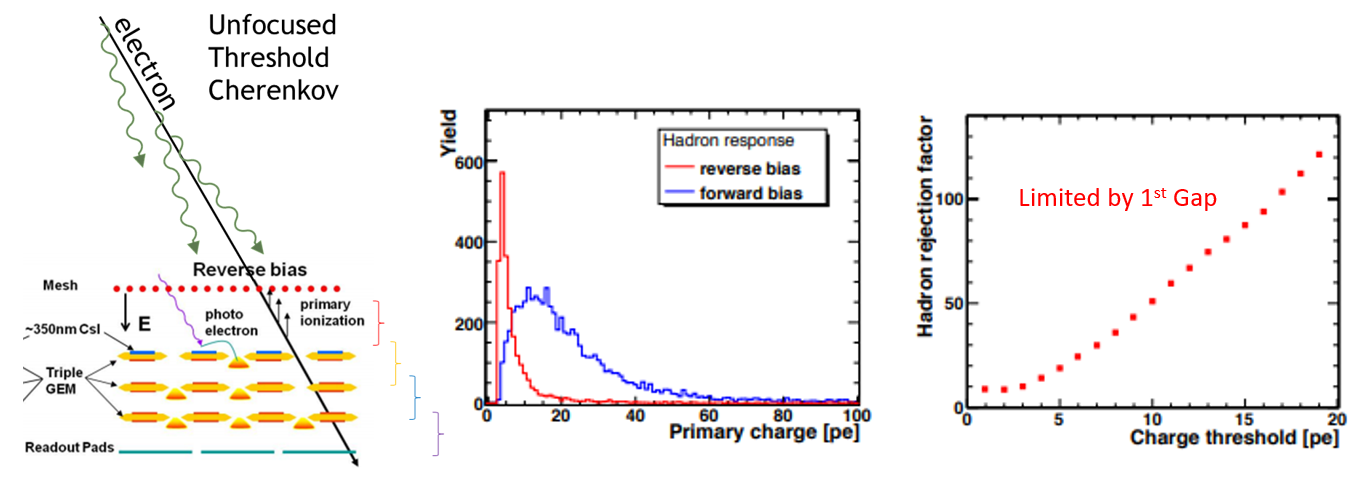}
	\caption{The left upper panel shows the triple-GEM with CsI coating used in PHENIX.  The center and right panels show the response to pions and electrons as well as the rejection.  Rejection of the device is limited by response to ionization in the gap between the first two GEMs.}
	\label{Fig:HBD-01-Configuration}
\end{figure}

An HBD device collects unfocused Cherenkov light and makes no attempt to focus the light so as to determine the Cherenkov angle.  It is instead operated in a "Threshold Mode" wherein the fastest particles will radiate, making it suitable only for eID and not for PID.  The PHENIX experiment was the first implementation of such a device\cite{Anderson:2011jw, Fraenkel:2005wx}.  That implementation is shown in Figure~\ref{Fig:HBD-01-Configuration}.  Pure $CF_4$ gas (n=1.00056) is used as a radiator.  The transparency at low wavelength is leveraged to take advantage of the $\frac{1}{\lambda^2}$ photon yield.  As measured by the "$N_0$" parameter (325), this is the brightest Cherenkov detector ever built.

A CsI photocathode is evaporated onto Gas Electron Multipliers (GEMs) and provides sensitivity to $\lambda < 200 nm$ and has a yield of 20 photoelectrons in 50 cm.  In PHENIX, the device was optimized for distinguishing closed Dalitz pairs (40 p.e.) from isolated electrons (20 p.e.).  It was not optimized for e/$\pi$ separation and suffers from an ionization signal generated by any charged particle passing through the focal plane.

\begin{figure}[hbt]
	\centering
        \includegraphics[width=1.0\columnwidth]{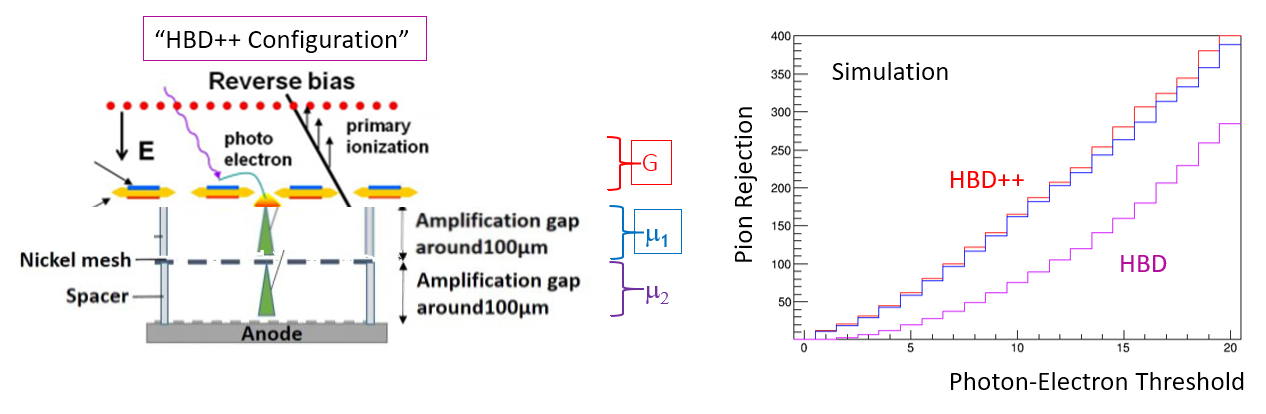}
	\caption{Simulated performance of an HBD configured with $\mu$MEGAS amplification (left panel) immediately following the photocathode bearing GEM.  A roughly 2 times improvement of pion rejection might be achieved in this configuration (right panel).}
	\label{Fig:HBD-02-Simulation}
\end{figure}

Simulations have been done on an alternative HBD implementation (HBD++) as is shown in Figure~\ref{Fig:HBD-02-Simulation}\cite{PID:BiweeklyTKH20200731}.  Here the later GEM gain stages are replaced by MICROMEGAS detector(s) thereby minimizing the ionization signal from the charged particles.  This results in a near doubling of the pion rejection provided by the device, but has never been proven in an actual implementation.

\subsubsection{CsI RICH}
A corollary to the HBD design can be achieved by focusing the Cherenkov light into the focal plane and thereby enabling a measurement of the Cherenkov angle\cite{Blatnik:2015bka}.  This configuration mostly retains the brightness of the original HBD although there is additional light loss due to both the increased gas path length (round trip including the mirror).  The concept benefits from the fact that the low material budget of the photon detector (GEMs) can be placed directly in the path of the particles at the entrance of the device, minimizing the emission term.  The design suffers, in two critical aspects:
\begin{itemize}
    \item The low refractive index, n=1.00056, results in a rather high threshold for pion and kaon radiation (4.17 and 14.75 GeV/$c$ respectively).  The CsI RICH must therefore be supplemented by an additional PID device to match the physics requirements at lower momenta.
    \item Use of the radiation gas down to the transparency cutoff results in a high distortion due to chromaticity.
\end{itemize}

\begin{figure}[hbt]
	\centering
        \includegraphics[width=0.5\columnwidth]{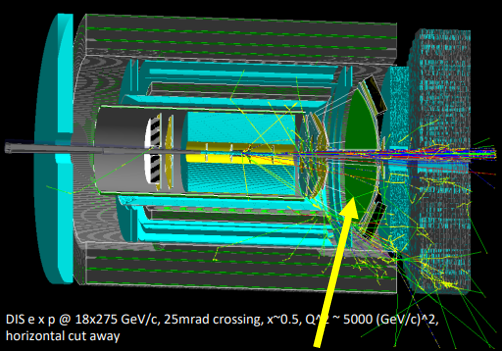}
	\caption{EIC-sPHENIX configuration of a gas Cherenkov in the style of the prototype tested for EIC.}
	\label{Fig:GRICH-01-Configuration}
\end{figure}

\begin{figure}[hbt]
	\centering
        \includegraphics[width=1.0\columnwidth]{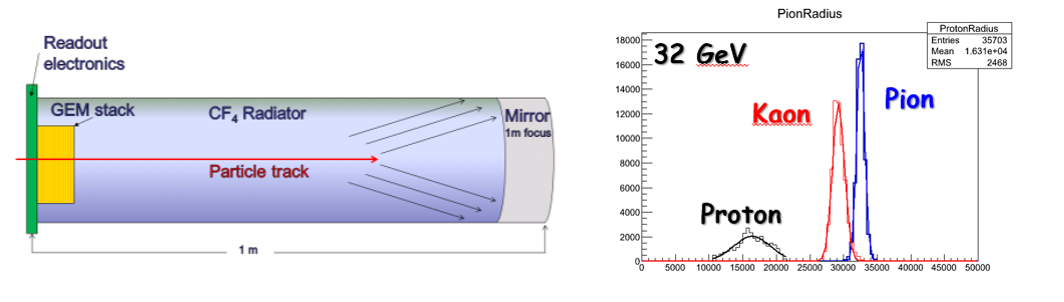}
	\caption{The left panel shows the test beam configuration with 1~m $CF_4$ radiator and quintuple GEMstack.  The right panel shows the response of the test beam detector to protons, kaons, and pions at 32 GeV/$c$ momentum.}
	\label{Fig:GRICH-03-TestBeam}
\end{figure}

A detector concept called "EIC-sPHENIX" (Figure~\ref{Fig:GRICH-01-Configuration}) is what results from maximal reuse of sPHENIX detectors and accompanying devices placed in both end caps~\cite{Seele:2013aua}.  The default configuration of this device uses a Cherenkov radiator with $CF_4$ gas as described above.  A prototype of this device was tested at Fermilab (Figure~\ref{Fig:GRICH-03-TestBeam}) with excellent $\pi$-K separation measured at 32~GeV/$c$ and extrapolated 3-$\sigma$ performance to 60~GeV/$c$. 

\begin{figure}[hbt]
	\centering
        \includegraphics[width=1.0\columnwidth]{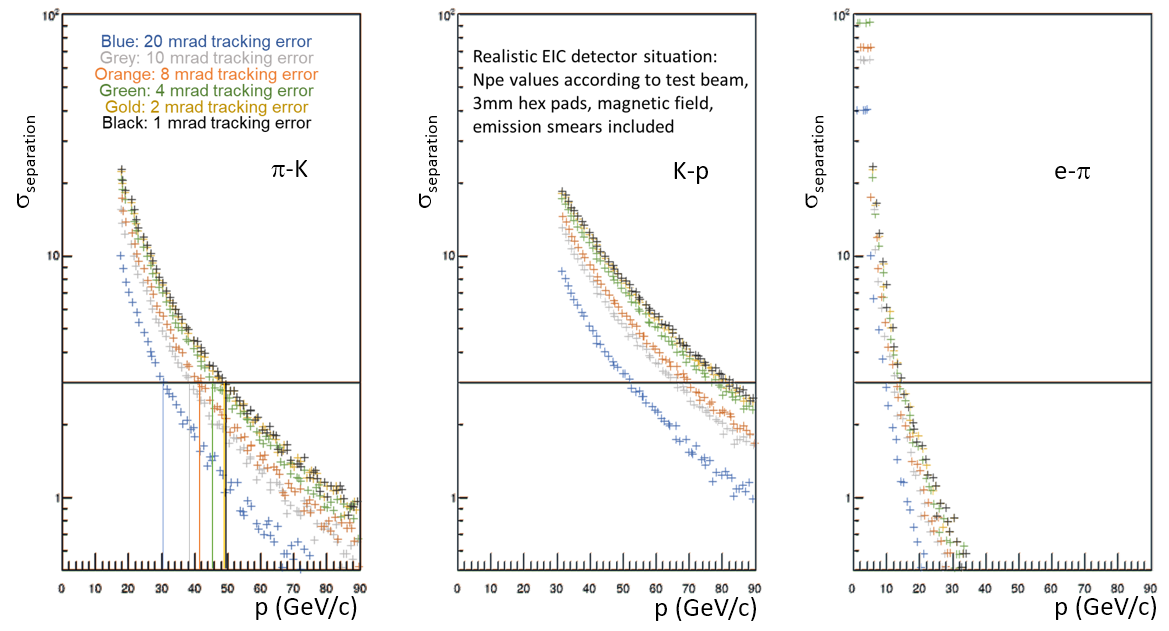}
	\caption{Fast simulations show the $\pi$-K, K-p, and e-$\pi$ separation anticipated for a 1 meter CsI RICH assuming various tracking precisions.  Because of the large chromatic term, the tracking does not degrade PID until the pointing resolution becomes worse than 2 mrad.}
	\label{Fig:GRICH-02-Simulation}
\end{figure}

Figure~\ref{Fig:GRICH-02-Simulation} shows the result of a simulation in which the measured detector performance is parameterized and subjected to varying errors in track direction\cite{PID:BiweeklyHK20200508}.  Because the device is severely limited by "chromaticity" (wavelength-dependent index), this device has comparatively lax requirements on the tracking and is unaffected by track pointing errors of roughly 2~mrad or less.

In the ePHENIX implementation, the CsI RICH is complemented by mRICH detectors (see below) that compensate for the high Cherenkov threshold over some of the aperture.  Nonetheless, there is a gap in PID coverage between the maximum achieved by the mRICH and the minimum achieved by the GEM RICH (gRICH).

\begin{figure}[hbt]
	\centering

\includegraphics[width=.99\textwidth]{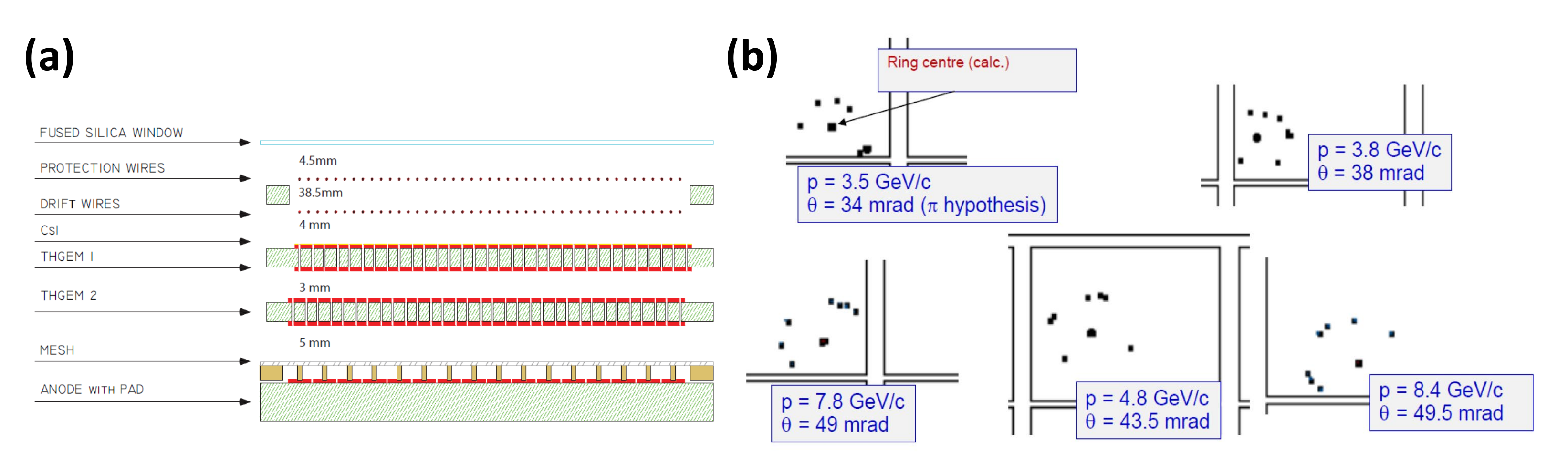}
	\caption{(a) Sketch of the hybrid single photon detector: 
two staggered THGEM
layers are coupled to a resistive bulk MM; image not to scale. (b) Ring images
detected with the hybrid single photon detector of COMPASS RICH; ring centres calculated from the reconstructed trajectory; no image filtering applied.}
	\label{Fig:compass-hybrid}
\end{figure}
\begin{figure}[hbt]
	\centering
\includegraphics[width=.9\textwidth]{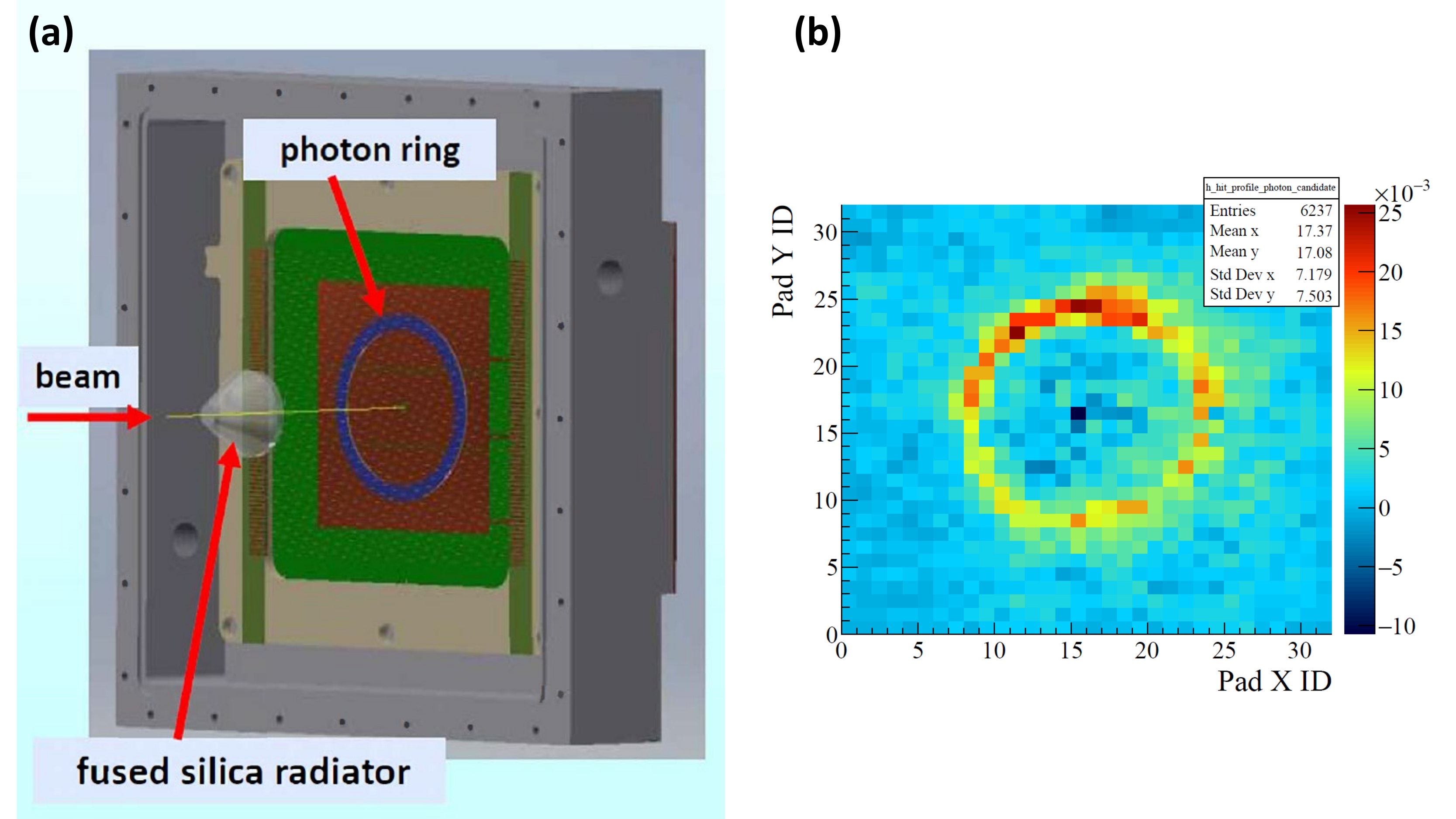}
	\caption{(a) Formation of the ring image on the photon detector prototype   by Cherenkov photons generated in a quartz radiator crossed by beam particles (principle). (b) 2-D histogram of the hits produced by the Cherenkov photons in the small pad-size prototype}
	\label{Fig:small-pad_hybrid_prototype}
\end{figure}

An alternative to the GEM-based photon detector, is represented by the hybrid MPGD photon detector in use since 2016 in COMPASS RICH~\cite{Agarwala:2018wba}: two THick GEM (THGEM) multiplication layers, the first one coated with a CsI film and acting as photocathode are followed by a resistive MICROMEGAS stage (Fig.~\ref{Fig:compass-hybrid}). 
A reduced pad size is needed to match the compact configuration  at EIC, where the gaseous RICH focal length is of the order of 1~m. A prototype with  pad-size reduced from 8~mm to 3~mm has been designed, built and successfully tested in a beam (Fig.~\ref{Fig:small-pad_hybrid_prototype}). Its operation in a window-less configuration such as that in the  EIC-sPHENIX one has yet to be confirmed. 

So far, CsI is the only photocathode material  that has been successfully used in gaseous detectors, although its usage is affected by several difficulties and limitations. In applications at high luminosity, its quantum efficiency is reduced over time by the bombardment of ions produced in the multiplication process when the integrated charge exceeds 1~mC/cm$^2$. Therefore, these detectors must be used at limited gain, reducing their overall efficiency. CsI is chemically fragile, in particular if exposed to water in excess of ~20ppm. This requires tedious manipulation techniques to maintain a dry, inert atmosphere. A novel, more robust option is offered by Hydrogenated Nano Diamond (HND) powder\cite{
Brunbauer:2020tpl,Chatterjee:2019vpm,Agarwala:2019wcy,Agarwala:2018qdm}.
It has a similar quantum efficiency to CsI, spanning the same range of UV wavelengths.  Hydrogenation requires high temperatures. The powder hydrogenation before forming the converting coating makes this 
approach compatible with the components of gaseous detectors. HND exhibits good chemical stability and the coating layer is mechanically robust.

\begin{figure}[hbt]
	\centering
\includegraphics[width=1\textwidth]{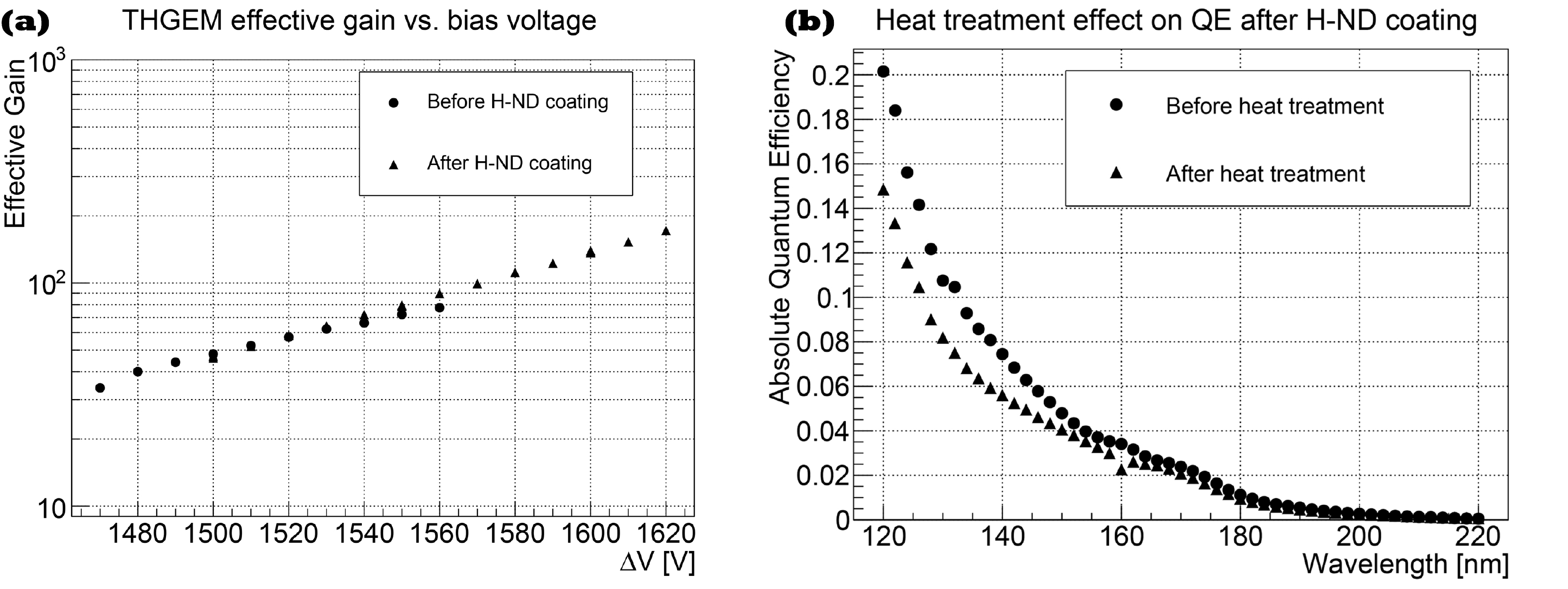}
	\caption{(a) Effective gain versus applied biasing voltage for a THGEM measured with the bare device and with HND coating after applying the heating protocol. (b) QE versus wavelength of a NHD-coated sample measured before and after the heating protocol.}
	\label{Fig:thgem-hnd}
\end{figure}

The avalanche performance of THGEMs with HND coating is unchanged, when an appropriate 
post-coating heating protocol is applied (Fig. ~\ref{Fig:thgem-hnd}, (a) ).  The preservation of the quantum efficiency when the protocol is applied is under study  Fig. ~\ref{Fig:thgem-hnd}, (b) ); presently, the protocol is being optimized to obtain a complete preservation of QE. The goal of this R\&D is to obtain a valid alternative to CsI suitable for gaseous single photon detection in high luminosity applications like EIC. 

\subsubsection{Dual RICH (dRICH)}
\begin{figure}[hbt]
	\centering
        \includegraphics[width=1.0\columnwidth]{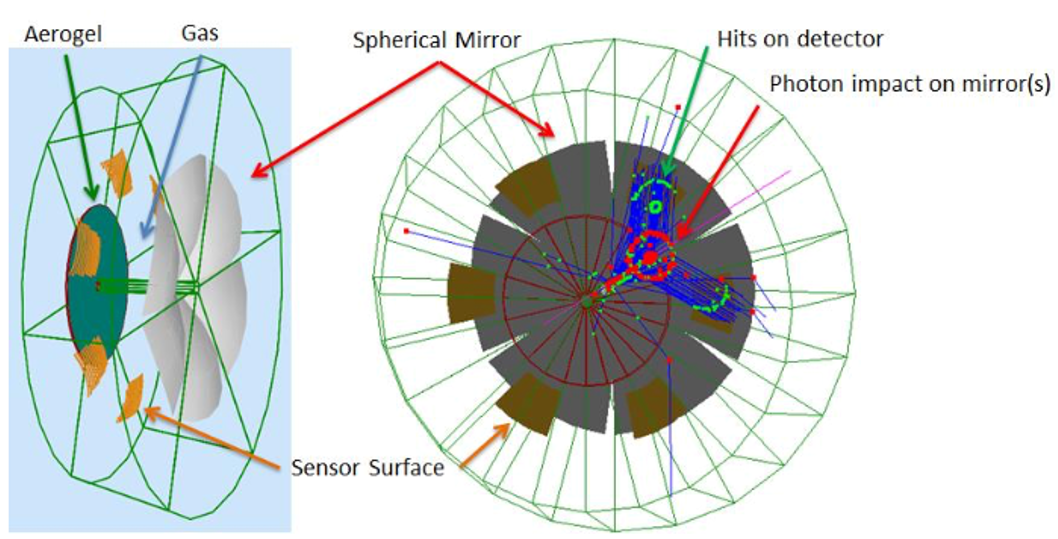}
	\caption{Dual RICH detector configuration after AI-driven optimization.  Multiple mirror panels (gray) focus rings from both aerogel and $C_2F_6$ perfluorocarbon radiators onto the same focal plane.}
	\label{Fig:DRICH-01-Configuration}
\end{figure}

A so-called "Dual RICH" utilizes two different radiator indices and thereby is able to cover the full momentum range without penalty owing to the Cherenkov threshold of the gas section\cite{PID:drich1, PID:drich2, PID:drich3}.  The design optimized for EIC is shown in Figure~\ref{Fig:DRICH-01-Configuration} and uses both an aerogel radiator and a gas radiator ($C_2F_6$) to cover the full momentum range in a single device.  In the current design, 
the photo sensors are located outside the acceptance. This has multiple effects that drive the device performance:

\begin{itemize}
    \item The optics is less ideal and therefore the emission term becomes dominant in the resolution.
    \item The focal plane is moved to a lower radiation zone.  This helps not only in the level of background hits that can interfere with the photon ring, but also may allow the use of emerging technology such as SiPM detectors to be used for the readout.
\end{itemize}

Shifting the focal plane to one side widens and complicates the parameter space for detector design, making optimization a daunting task.  The present design of the detector was optimized using AI-based optimization techniques to investigate a wide space of detector configurations. The design shown here is the result of that exhaustive investigation~\cite{Cisbani:2019xta}.

\begin{figure}[hbt]
	\centering
        \includegraphics[width=1.0\columnwidth]{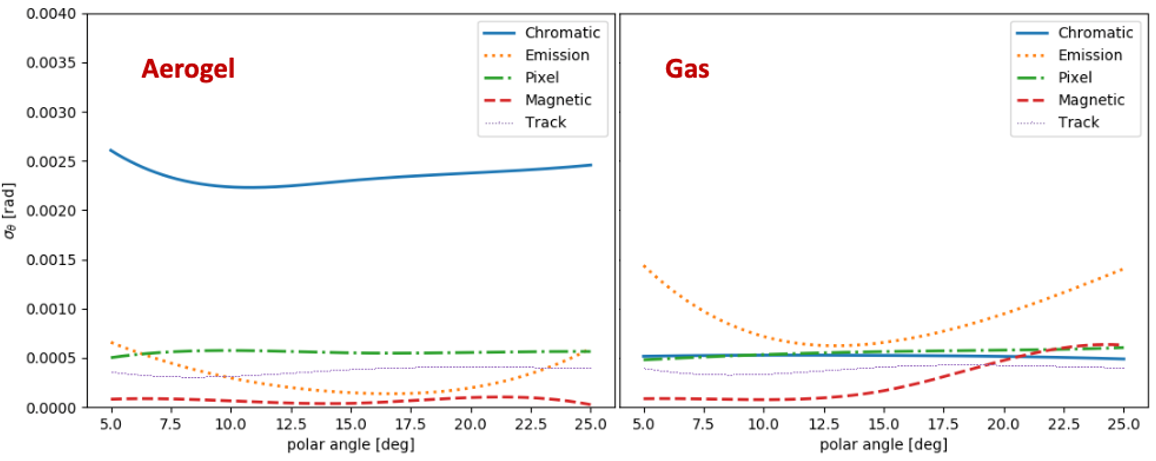}
	\caption{Resolution contributions for the Dual RICH.  As is true for most aerogel implementations (left panel), the chromatic dispersion of the radiator itself is the limiting factor in the resolution.  Conversely, for the gas detector, the emission term dominates due to off-axis focusing.}
	\label{Fig:DRICH-02-Resolution}
\end{figure}

Figure~\ref{Fig:DRICH-02-Resolution} shows the converged solution for the detector performance optimization in both the aerogel and the gas sections.  Each term in the final resolution is isolated by its contribution of the Cherenkov angle resolution.  The aerogel performance is dominated by the natural chromaticity of the radiator medium itself.  All other contributing factors to the aerogel performance are negligible as compared to chromaticity which represents a fully optimal performance.

The angular resolution of the gas section is more complex.  As referenced previously, emission terms (aberration) are dominant and peak at the edges of the segmented RICH mirrors.  The optimization of this factor is evident by the fact that the Emission resolution term is of equal height at the two extremes of the polar angle acceptance.  

\begin{figure}[hbt]
	\centering
        \includegraphics[width=1.0\columnwidth]{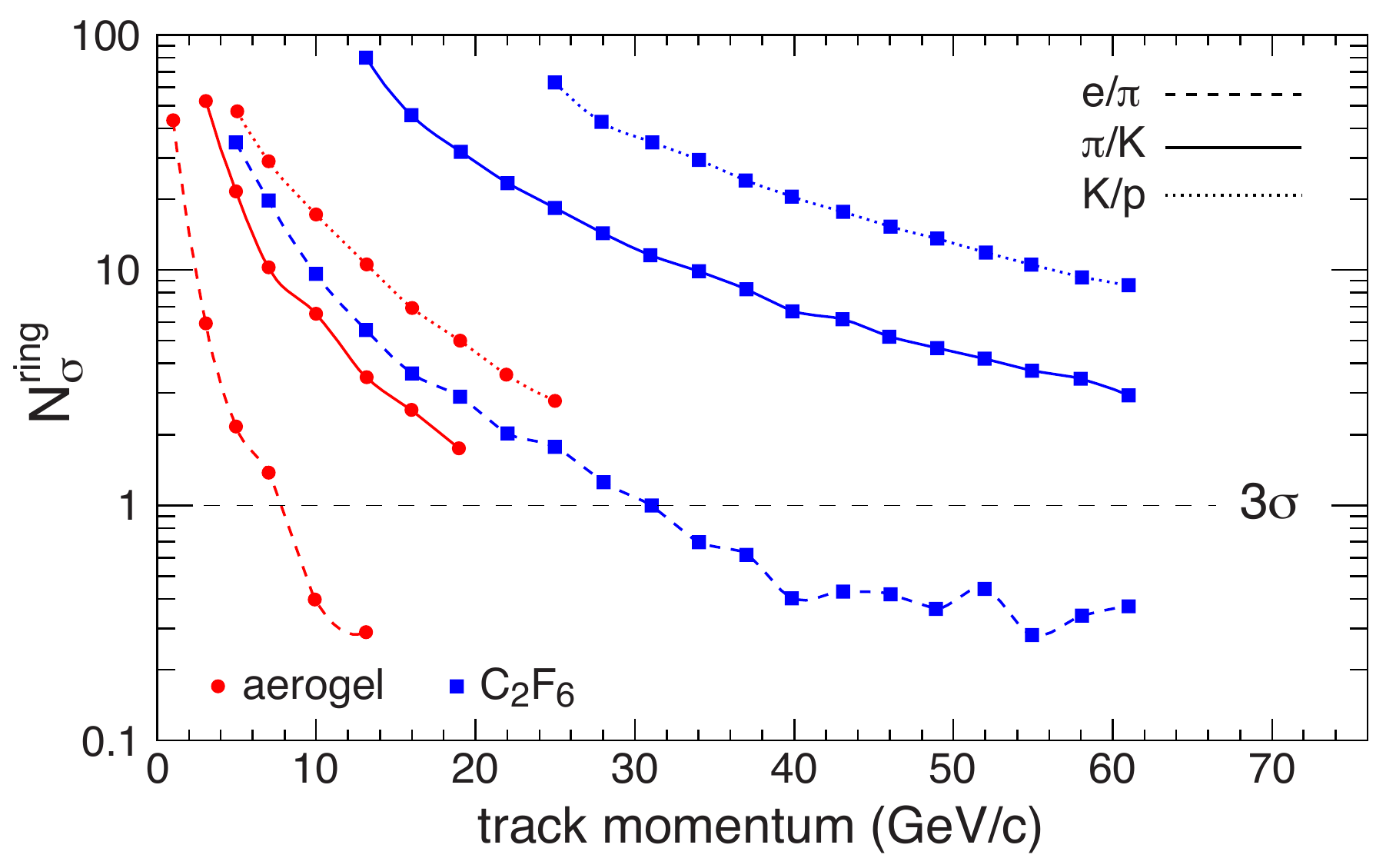}
	\caption{Performance of the Dual RICH for a variety of particle species.  In each case, the combination of aerogel and gas provides uninterrupted PID across the full range.  The device also serves for eID across more than the required momentum range.}
	\label{Fig:DRICH-03-Performance}
\end{figure}

Figure~\ref{Fig:DRICH-03-Performance} indicates the calculated performance of the dRICH detector for e-$\pi$, $\pi$-K, and K-p separation.  Several features are worth noting.  First, the dRICH is not merely limited to PID application, but also provides excellent eID out to roughly 20 GeV/$c$ momentum.  Second, the dRICH does not have "holes" in the performance either at low momentum (due to aerogel) nor at intermediate momentum due to the index match of the aerogel and gas radiator performance.  Finally, the $\pi$-K performance achieves the full goals of the requirements matrix.

As is true for most modern gas Cherenkov detectors, the dRICH design utilizes the superior performance of perfluorocarbon radiator gas ($C_2F_6$).  Future environmental concerns can have two kinds of impact:

\begin{itemize}
    \item It may be required to recover and purify the radiator gas to avoid release to the environment, which is a significant cost and complexity.
    \item Environmental concerns in the worst case could drive the cost and availability of the gas beyond tolerable levels.
\end{itemize}

Current calculations demonstrate that these issues could be avoided by running an environmentally friendly gas at high pressure.  Indeed, current calculations indicate that the dRICH performance would be insignificantly affected by a switch to Ar gas at 3 atm~\cite{PID:hpRICH1, PID:hpRICH2}.  This will nonetheless impose an engineering challenge to maintain a low material budget.

One final note is that the external requirement on the tracking systems was modeled to be a limiting resolution of 1 mrad on track inclination while the track passes through the whole length of the gas radiator.  Due to the large lever arm (1.5 meters) and possible scattering internal to the detector itself (entrance window in the high pressure version), it is likely wise to supplement the tracking prior to the dRICH with a detector that provides an additional space point beyond the radiation volume.  This latter point can be rather low resolution as compared to the rest of the tracking while still providing the necessary 1 mrad uncertainty in track direction.

At the time of this writing, the dRICH is the best known approach to EIC particle at the highest possible momenta due to its full coverage of the dynamic range in momenta desired for the hadron arm.  

\subsubsection{Modular RICH (mRICH)}
\begin{figure}[hbt]
	\centering
        \includegraphics[width=1.0\columnwidth]{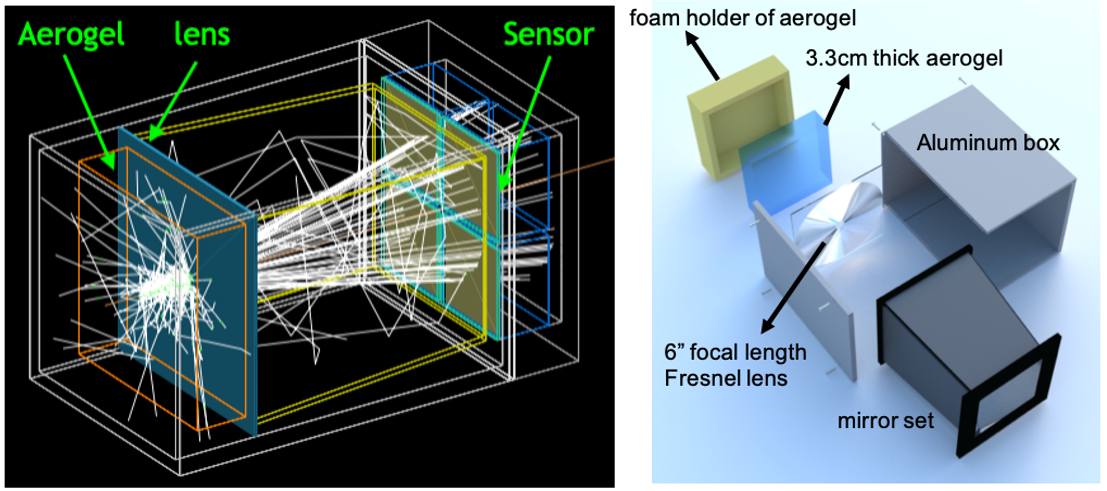}
	\caption{Configuration of the second mRICH prototype. An event display from a GEANT4 simulation with realistic optical properties is shown on left. The mRICH components are shown on right from a 3D design model.}
	\label{Fig:MRICH-01-Configuration}
\end{figure}

A so-called "Modular RICH" is an aerogel-based RICH\cite{PID:mrich1,PID:mrich2,PID:mrich3,PID:mrich4}.  A unique feature of this device is the use of a Fresnel lens to make a focused ring, thereby significantly improving the performance as compared to a ``proximity focused" detector which is more common in aerogel applications.  Figure~\ref{Fig:MRICH-01-Configuration} shows the key components of the second mRICH prototype which was tested at Fermilab in 2018. Also shown in Figure~\ref{Fig:MRICH-01-Configuration} is an event display from a realistic GEANT4-based simulation with proper optical properties implemented. The mirrors along the sides of the device allow it to collect light which is not initially directed to the photocathode found at the detector exit. Several aspects of the design of this device allow it to outperform conventional aerogel-based RICH detectors:
\begin{itemize}
    \item The Fresnel lens acts to generate a lens-focused rather than a proximity-focused ring.
    \item The Fresnel lens imposes a wavelength cutoff on the transmitted light limiting the Rayleigh scattering effect.
    \item The focusing aspect somewhat relaxes the mechanical tolerances on the exit surface of the aerogel.
    \item The modular, compact and projective features of mRICH provide a flexible instrumentation and installation of array of mRICH modules with uniform performance.
    \item The mRICH can possibly be configured with a photodetector that exhibits precision timing.  
\end{itemize}

This device is useful both in the electron arm performing both eID and PID functions and also in the hadron arm (under the presumption of a gas RICH instead of a Dual RICH).  

\begin{figure}[hbt]
	\centering
        \includegraphics[width=1.0\columnwidth]{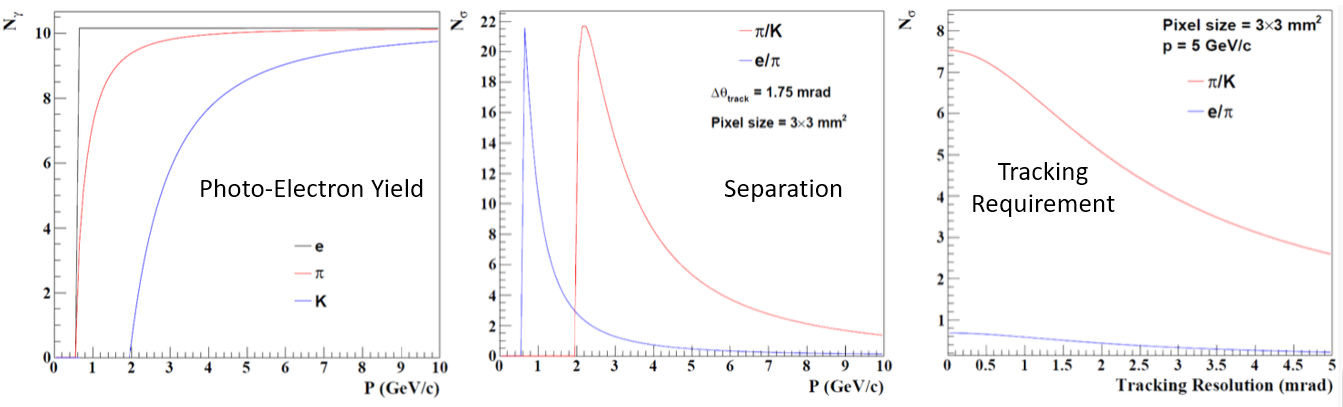}
	\caption{GEANT4 simulations show the expected photon yield (left panel) and separation performance (center panel) of the Modular RICH.  The right panel shows the degradation of performance with less than perfect track pointing resolution.}
	\label{Fig:MRICH-02-Performance}
\end{figure}
The limit to the resolution of the mRICH detector is the chromaticity term (as was true for the aerogel section of the dRICH), indicating the design is presently optimal.  The simulated performance of the mRICH is shown in Figure~\ref{Fig:MRICH-02-Performance}.  The saturation yield of Cherenkov photons is 10 per ring and is shown as a function of momentum for $\pi$ and K.  The center panel shows that the e-$\pi$ rejection extends until roughly 2 GeV/$c$ and $\pi$-K until roughly 6-7 GeV/$c$.  These are well, but not perfectly matched to the requirements matrix in the electron arm. Finally, as with all precision Cherenkov devices, the mRICH has strict requirements on the tracking resolution provided.  The third panel shows the degradation in separation as the tracking resolution worsens indicating a tolerance of roughly 1 mrad as supplied by the external system.

\begin{figure}[hbt]
	\centering
        \includegraphics[width=1.0\columnwidth]{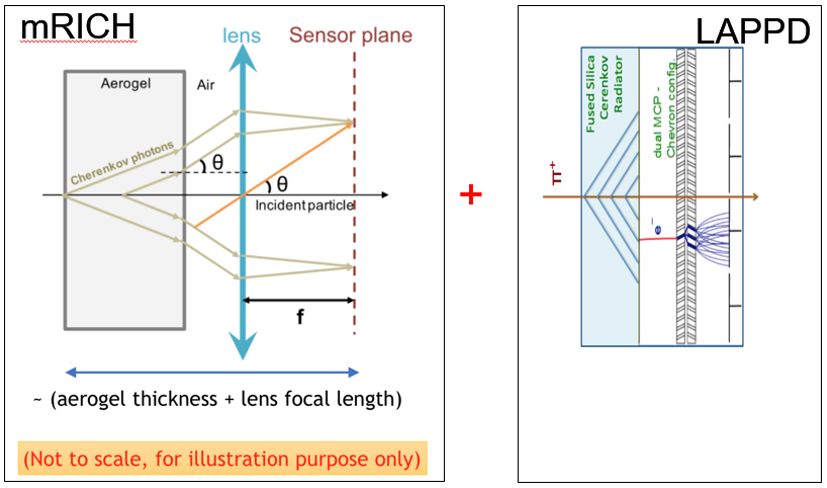}
	\caption{Timing Option for the Modular RICH.  If LAPPD sensors are used in conjunction with the mRICH this allows a high resolution TOF determination to extend the mRICH performance to negligible momentum.}
	\label{Fig:MRICH-03-TimingOption}
\end{figure}

Finally, Fig.~\ref{Fig:MRICH-03-TimingOption} shows an option for configuring the output detection stage of the mRICH with a high precision timing detector so that it can additionally serve as a TOF tag, thereby improving its PID capability.

\subsubsection{Detection of Internally Reflected Cherenkov (DIRC)}
\label{hpDIRC}
An interesting aspect of Cherenkov detectors emerges at high refraction index.  Since both the saturation Cherenkov angle ($\beta=1$) and the angle for total internal reflection are solely dependent upon refractive index, one finds that at normal incidence, Cherenkov light will be totally internally reflected by any material whose index satisfies the condition $n > \sqrt{2}$.  This technique offers the unique advantage that, so long as the sides and corners of the radiator are made with high precision, the light can be propagated to the end of the radiator while preserving the Cherenkov angle.  The result is a geometrically thin device that allows light detection only at the end(s).  Furthermore, due to the in-medium light propagation length depending upon the Cherenkov angle, timing can also be used to aid the refining the Cherenkov angle determination.

\begin{figure}[hbt]
	\centering
        \includegraphics[width=0.9\columnwidth]{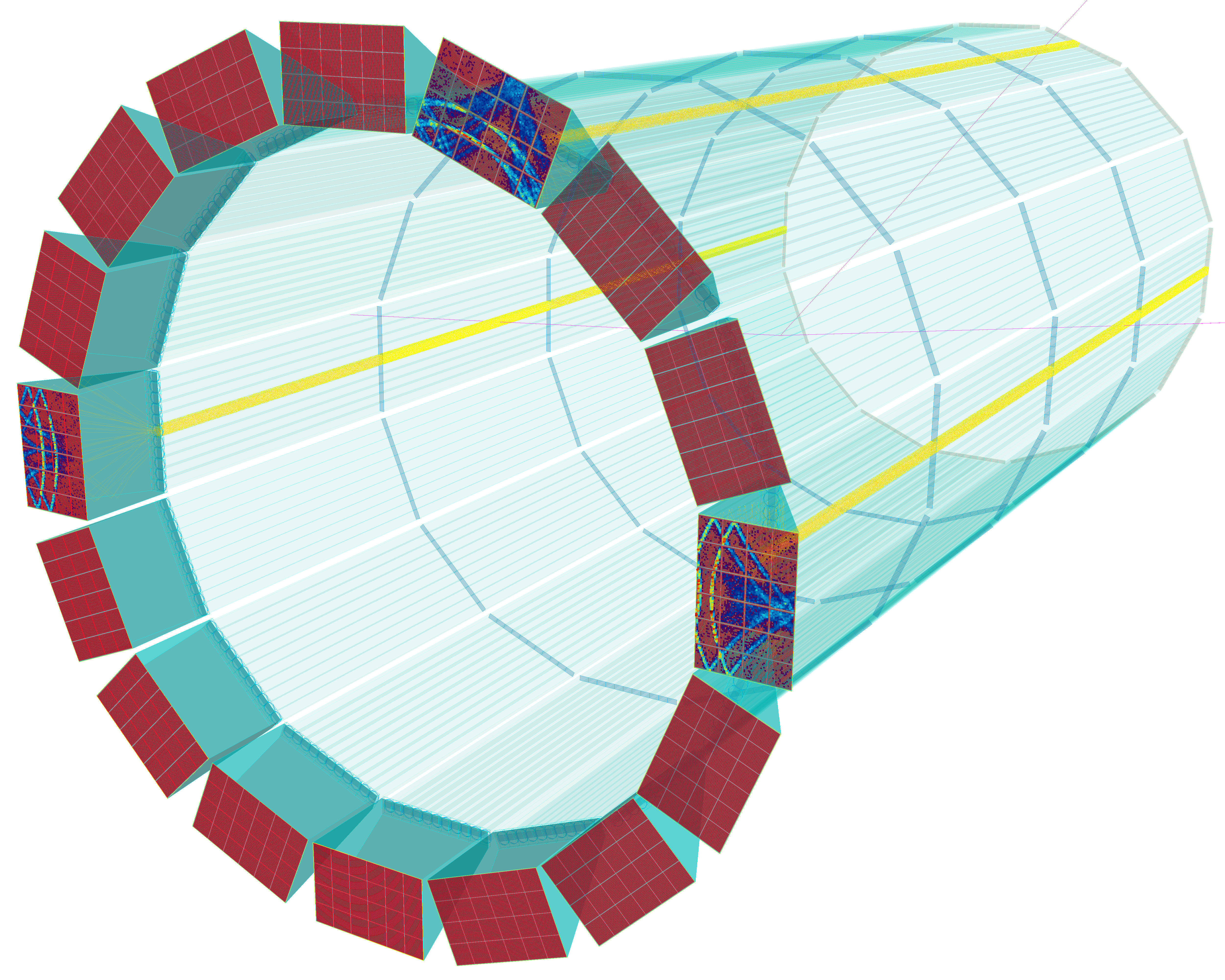}
	\caption{Configuration of the DIRC.  Newly developed compact "expansion boxes" at one end of the device focus the light into segmented rings.  The DIRC is indicated in the reference detector (Fig.~\ref{central-detector-cartoon}) as the light blue barrel detector immediately following the TPC.}
	\label{Fig:DIRC-01-Configuration}
\end{figure}

The original application of DIRC was in the BaBar experiment~\cite{Schwiening:1997uu} at SLAC wherein the barrel section of the detector was surrounded by a series of radiator bars made of synthetic fused silica (colloquially referred to as "quartz"). Rings were imaged by a so-called "expansion volume" that effectively made for a "proximity focus".  In the years that followed many advances of DIRC technology have been accomplished to effectively replace the proximity focus with an actual focus.  The result is that it is conservatively anticipated that an EIC application of DIRC technology can be made that far outperforms the BaBar application while dramatically reducing the size of the expansion volume\cite{PID:dirc1,PID:dirc2,PID:dirc3,PID:dirc4,Kalicy:2020ogn,Kalicy:2020kws}.  A picture of this High Performance or hpDIRC is shown in Figure~\ref{Fig:DIRC-01-Configuration}.  The left panel shows the quartz bars and the expansion volume isolated from the rest of the EIC detector.  The right panel shows one possible geometry by which the DIRC could be realized in an EIC detector.  Here the expansion volume is terminated with the photon detectors as indicated in red.  Because the photon detectors prefer to be normal to the spectrometer's magnetic field, their explicit locations will be tightly coupled to the edge field orientation.  It has been demonstrated that all plausible magnetic field orientations can be accommodated with little or no degradation in overall performance.

The DIRC application in many ways represents exquisite precision in all the geometric aspects of Cherenkov.  As a result, the DIRC's precision must be similarly reflected in the tracking.  The contribution of tracking resolution falls into the category of "correlated terms" in the analysis of the DIRC performance.

\begin{figure}[hbt]
	\centering
        \includegraphics[width=1.0\columnwidth]{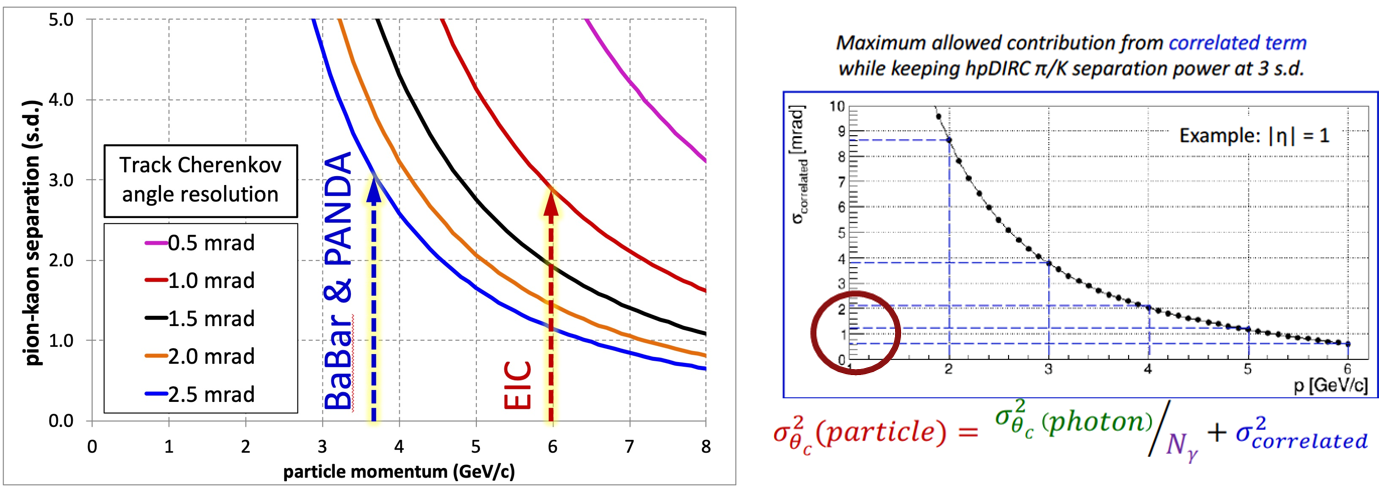}
	\caption{Performance of the DIRC in simulation.  The left panel shows the anticipated performance of the hpDIRC as compared to predecessors.  Improved focusing leads to improved momentum reach for $\pi$-K separation.  The right panel shows the influence of external factors (the "correlated term") on the resolution of the DIRC setting a stringent limit in pointing precision.}
	\label{Fig:DIRC-02-Performance}
\end{figure}

\begin{figure}[hbt]
	\centering
        \includegraphics[width=1.0\columnwidth]{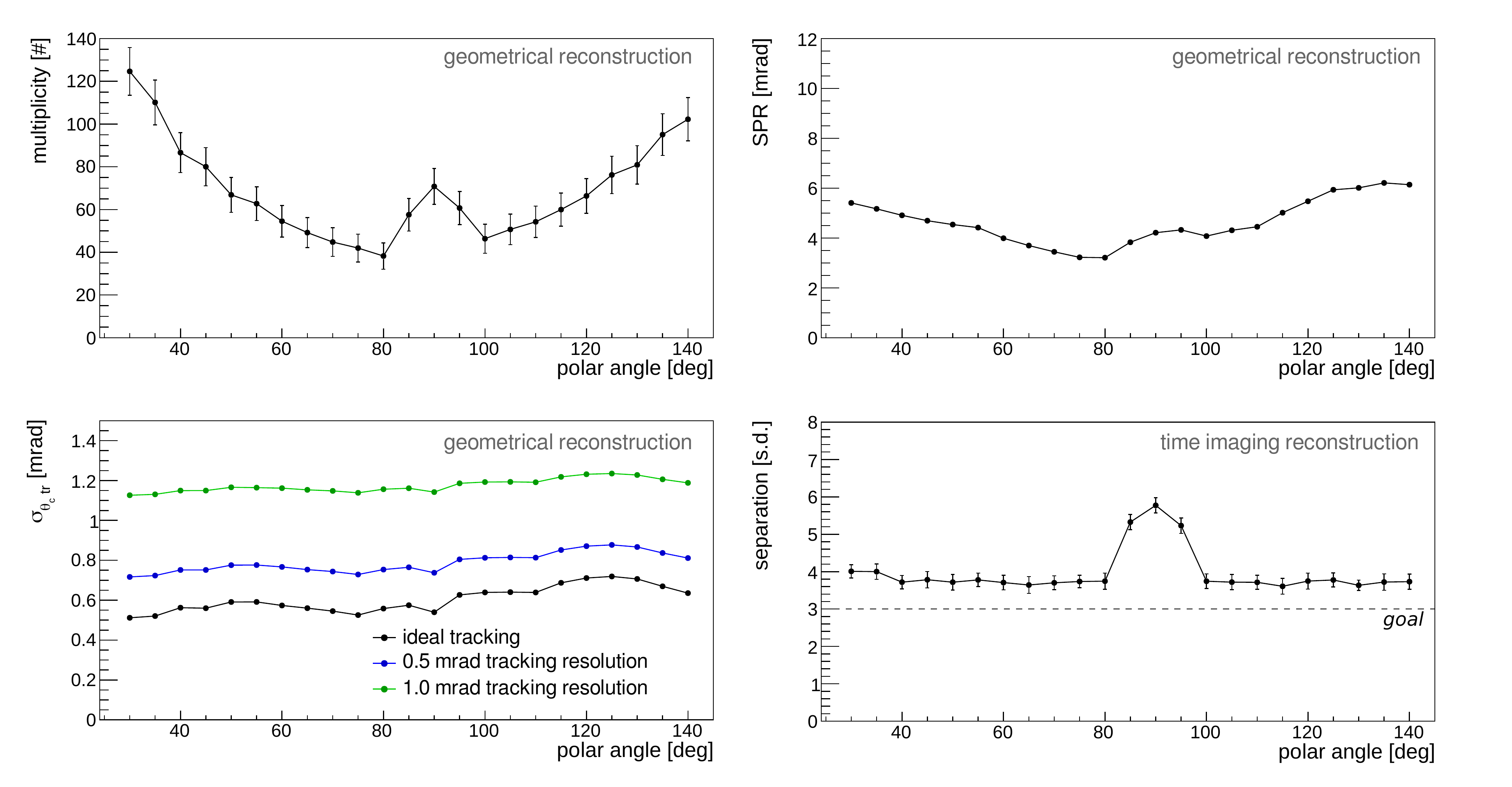}
	\caption{Dependence of the DIRC performance on $\eta$.  Photon yield and resolution per photon are simulated for the hpDIRC as a function of polar angle.  Despite variations in these terms, the separation power is reasonably flat with polar angle (lower right panel).}
	\label{Fig:DIRC-03-Simulation}
\end{figure}

\begin{figure}[hbt]
	\centering
        \includegraphics[width=1.0\columnwidth]{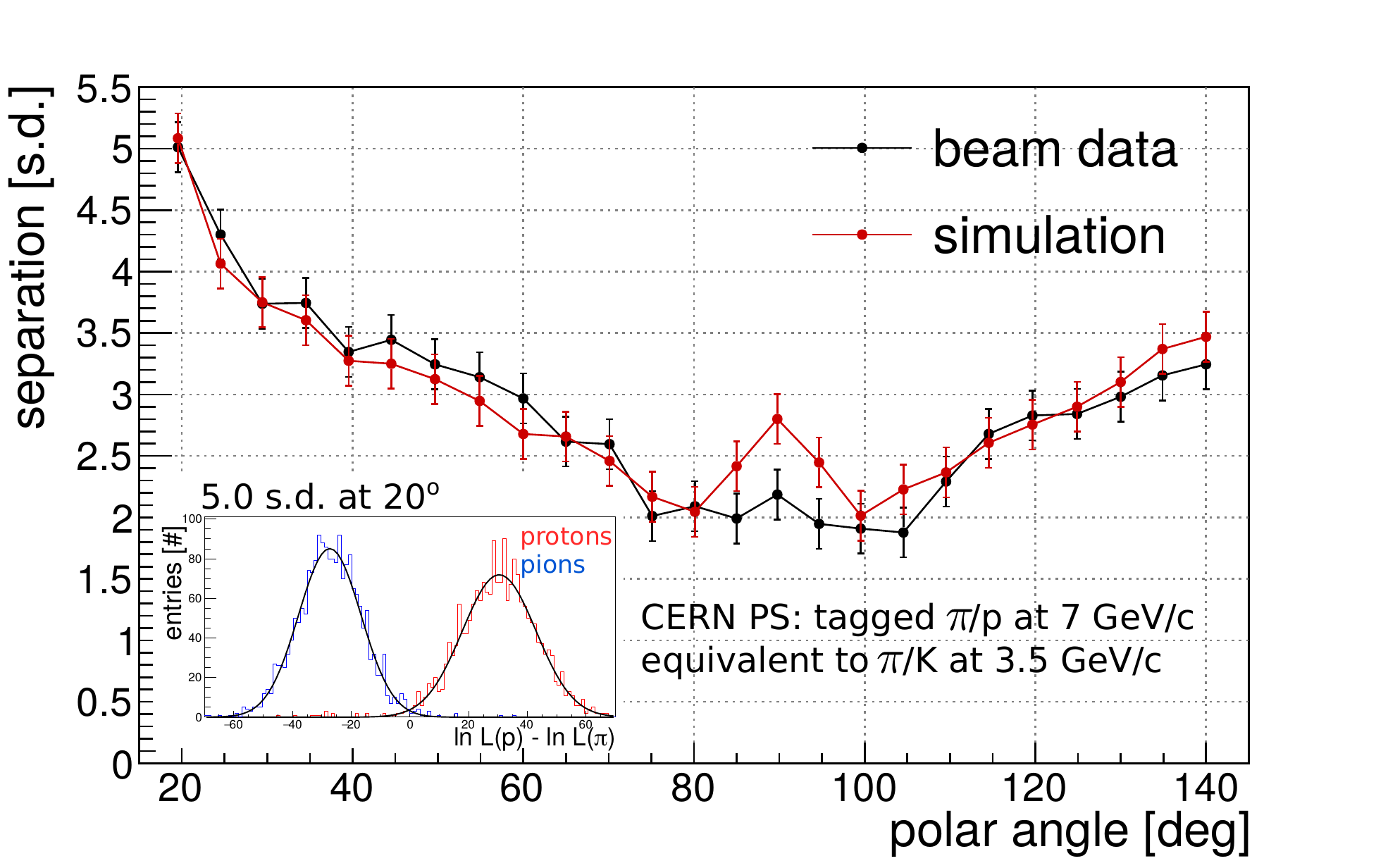}
	\caption{Comparison of the DIRC performance to test beam measurements.  The agreement is excellent.}
	\label{Fig:DIRC-04-TestBeam}
\end{figure}

Figure~\ref{Fig:DIRC-02-Performance} summarizes the anticipated DIRC performance in the so-called "hpDIRC" configuration and also compares the performance to the BaBar and PANDA applications of this technology.  The improvement in $\pi$-K separation in moving from the Babar to the hpDIRC design is close to a factor of two and reaches 6 GeV/$c$.  It is important to note that as with all PID detector technologies, various assumptions about the performance of other detector systems is vital to estimate the efficacy of the device.  These factors can be combined into a single so-called "correlated term", the effect of which is indicated by the right half of Fig.~\ref{Fig:DIRC-02-Performance}.  In particular, this figure denotes the limit applied to the convolution of all sources of correlated term as a function of desired 3-$\sigma$ $\pi$-K separation goal.  To reach the required performance for EIC, it is clear that the correlated term must not exceed 0.8~mrad and this places a restriction on the tracking performance at the level of 0.5~rad.

The nature of using internally reflected Cherenkov light makes the DIRC performance sensitive to the inclination of the particle and thereby the polar angle of the emission.  These effects have been simulated in detail and are summarized in Fig.~\ref{Fig:DIRC-03-Simulation}.  Although specific small features exist (such as the improved resolution at a polar angle of $90^o$), the overall performance of the DIRC with polar angle is rather uniform.  Fig.~\ref{Fig:DIRC-04-TestBeam} demonstrates this for a test beam run with remarkable agreement between measurement and simulation.

\subsection{Time Of Flight (TOF)}
\label{TOF}
Recent years have seen major advancements in the precision by which detector devices can measure the time of passage of a particle\cite{Hattawy:2018qbv,Xie:2020yat,PID:lappd3,Apresyan:2018oln,Pellegrini:2014lki,Cartiglia:2015iua,Breton:2016zoz}.  Such time, whether compared to a reference time for the collision as a whole (aka "Start time") or whether measured at multiple points along the trajectory of a particle as it passes through the spectrometer allow for a direct measurement of the particle's velocity and hence are useful forms of particle identification.  An intrinsic advantage of measurements is that they contain no limiting threshold in performance ({\it e.g.} Cherenkov radiation is only produced for $\beta > \frac{1}{n}$) and thereby produce signals for charged particle of any momentum.  These detectors are most often rather thin measured both by radiation length and by physical dimension.

One can divide modern TOF technologies into two categories depending upon whether the technology converts light into photo-electrons (which subsequently avalanche) or whether they produce and detect ionization directly.  The former case (as discussed in more detail in Section~\ref{PID-PhotonDetection}) is most often sensitive both to the strength and orientation of the external magnetic field.

\begin{figure}[hbt]
	\centering
        \includegraphics[width=1.0\columnwidth]{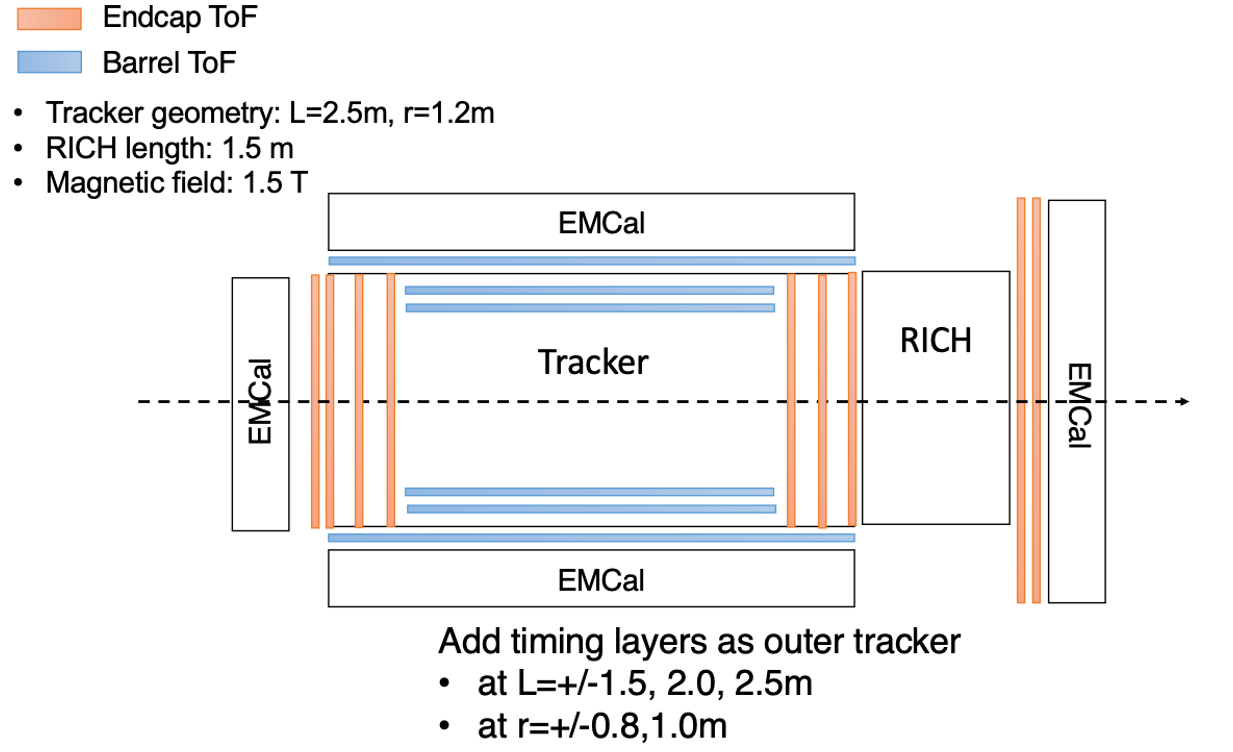}
	\caption{One possible configuration of Time of Flight for EIC.  Differing technologies may chosen in the barrel and endcap respecting the sensitivity to photon detectors wrt external magnetic fields.}
	\label{Fig:TOF-01-Configuration}
\end{figure}

Figure~\ref{Fig:TOF-01-Configuration} displays one possible configuration of TOF detectors as arrayed into the typical EIC detector geometry.  This particular geometry makes the assumption that the ToF measurements would be achieved with a silicon-based technology such as LGAD that is intrinsically insensitive to magnetic fields.  The technology is layered in each direction so that several measurements of time are performed on every track and that these measurements additionally contribute to the tracking system by virtue of providing precise space points as well as precise timing ("4D" tracking).  

\begin{figure}[hbt]
	\centering
        \includegraphics[width=1.0\columnwidth]{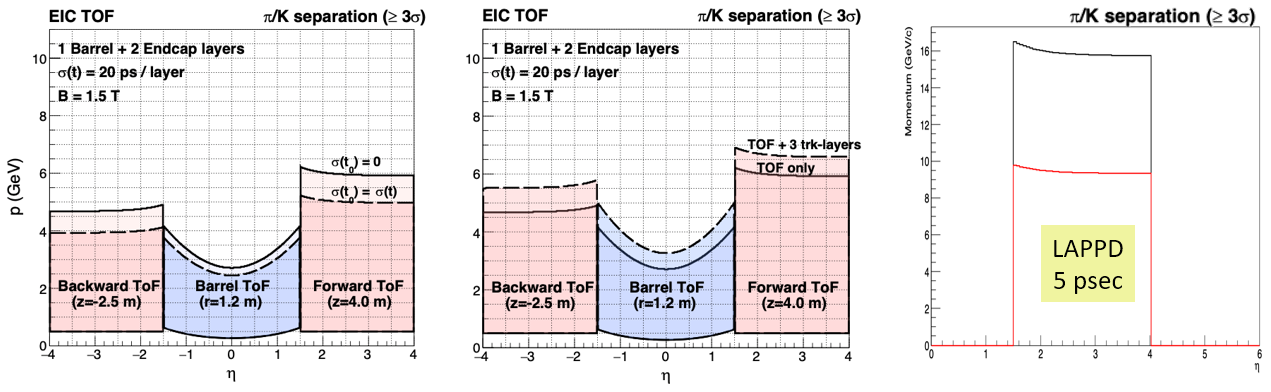}
	\caption{Performance of Time of Flight for EIC.  The left panels show the separarion power results for multiple layers of LDAG silicon sensors, here taken to benefit by $\sqrt{N_{layer}}$ scaling.  The right panel shows the performance of LAPPD for two different flight paths 3m (red) and 4 m (black).}
	\label{Fig:TOF-02-Performance}
\end{figure}

The performance of the all-silicon TOF system shown previously is summarized in the left two panels of Figure~\ref{Fig:TOF-02-Performance}\cite{PID:tof1,PID:tof2,PID:tof3,PID:tof4}.  We note several aspects of this calculation.  First, the calculation assumes that the overall time measurement scales with the number of measurements as $\frac{1}{\sqrt{N_{meas}}}$.  This requires that common issues such as clock jitter are small compared to the intrinsic detector resolution.  Appropriate R\&D is ongoing to ensure that this will be the case by the time of EIC.  Second, the calculation assumes the absence of HCAL detectors in the endcaps so that the flight path of the particles can be maximized.  The existence or not of HCAL is thus one of the issues that can addressed in the design of complementary EIC designs.

At the time of this writing, the best TOF performance is supplied by LAPPD (Large Area Picosecond Photon Detector) devices providing timing resolution of roughly 5~ps $\sigma$.  The performance of that detector is summarized in the right-most panel of Figure~\ref{Fig:TOF-02-Performance}.  Because these devices utilize the avalanche of photo-electrons to generate their signal they are sensitive to the magnetic field.  The current implementations of the technology are therefore limited to end cap implementations.

\subsection{Transition Radiation Detection}\label{PID=-TRD}
As discussed previously, the transition radiation (TR) photon rate
preliminary depends on the Lorentz-$\gamma$ and thereby serves as an effective means of performing particle identification.  The momentum range of produced particles at the EIC effectively makes TRD an electron identification device since heavier particles will not produce enough TR photons for effective measurement.  Most TRD applications are configured to simultaneously measure particle trajectory and TR radiation.  The details of the TRD studies specifically for the EIC have been discussed previously in the tracking section and specifically in Section~\ref{TRK-TRD} and are not repeated here.

%
%

\subsection{Photon Detection Technology Options}\label{PID-PhotonDetection}

Many of the devices discussed previously involve the detection of visible or UV photons, frequently with an accompanying requirement of being able to discriminate between noise and the signal resulting from a single photo-electron. Furthermore, the photo-sensor must maintain much of its efficiency and gain while immersed in the magnetic field of the spectrometer. Many traditional devices for single photo-electron detection fail the final criterion of operation when immersed in a magnetic field. Several suitable technologies exist or are under development and have been studied in the context of EIC applicability~\cite{PID:photondetection}.

The MCP PMT uses micro-channel plate technology to replace the traditional dynode structure for achieving gain in a photomultiplier tube. These devices are intrinsically more tolerant to an external field but are not entirely immune. Several options have been studied, one of which is summarized in Fig.~\ref{fig:Updated-field-effect}.  Here, the gain performance of a 10-$\mu$m pore-size MCP PMT, namely PHOTONIS\footnote{PHOTONIS FRANCE S.A.S, Avenue Roger Roncier, 19100 Brive B.P. 520, 19106 BRIVE Cedex France. URL: http://www.photonis.com} Planacon XP85112 is shown as a function of magnetic field magnitude and for several orientations of the PMT relative to the direction of the magnetic field (B-field). The value of the high voltage is 96.4\% of the maximum recommended value by the manufacturer. For all studied angles, the device maintains sufficient gain up to at least 1~T to be suitable for use in a RICH or DIRC detector. The smaller the angle of incidence of the B-field to the plane of the sensor, the wider the B-field range over which the gain is acceptable. For example, at 10$^\circ$, the device has sufficient gain up to 1.2~T. At higher fields, the gain exhibits a continuous drop. 
The angle between the field and the MCP PMT is a critical parameter as the rate of gain drop increases as the angle increases: for an angle of 20$^\circ$, gain drop to about 1\% of the reference gain at 0~T is observed at 1.5~T, whereas the same drop of 1\% occurs at 2.2~T for an angle of 0$^\circ$. It is imperative to carry out further studies to understand the extent to which the rate of the gain decrease can be mitigated using a different voltage distribution across the MCP PMT stages than the nominal established by the manufacturer, which has not been fully optimized for such high-B field applications. These gain studies must also include a survey of the collection efficiency of the device as at high fields one would expect not only loss of gain, but also loss of photoelectrons. Such studies are needed because tracking considerations imagine the central value of the field to be as high as 3~T. At such a field, it may be difficult to position the photon detectors of all Cerenkov devices (dRICH, mRICH, DIRC) in B-fields smaller than 1 – 1.2~T. Conversely, it has been shown that for a 3~T central field, 1~T in the region of photon detection is plausible, but requires careful design.

\begin{figure}[hbt]
	\centering
        \includegraphics[width=1.0\columnwidth]{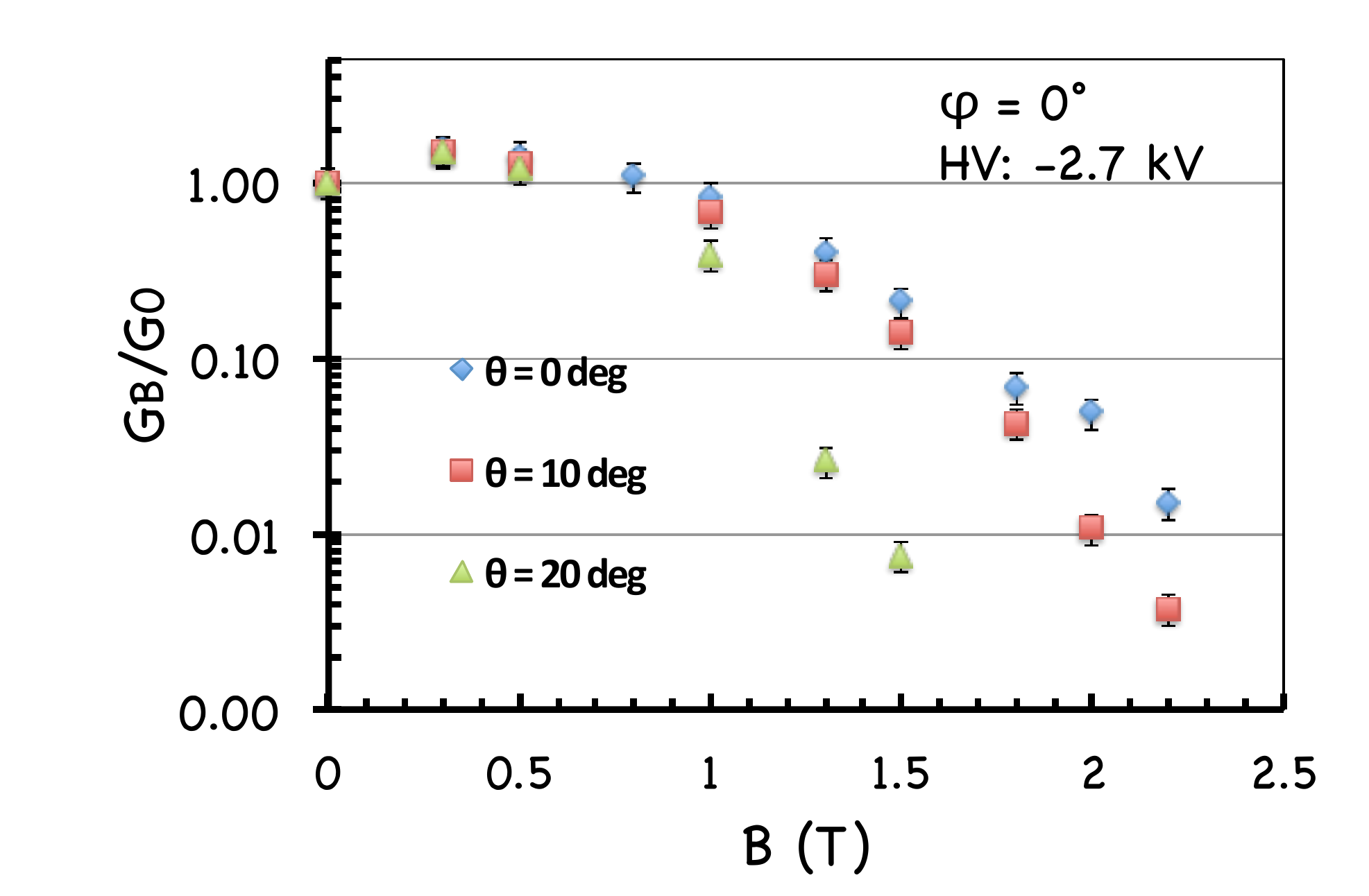}
	\caption{Magnetic-field effect on the gain of a 10-$\mu m$ pore-size Planacon XP85112 MCP PMT. The B-field magnitude is shown on the X-axis, whereas the gain (at a given B-field) relative to the gain at 0 T is shown on the Y-axis. The error bars show 20\% total uncertainty, which contains systematic and statistical uncertainties added in quadrature. The value of the high voltage is chosen to be as high as possible, but not too high as to cause unacceptable ion feedback. For the purpose of assessing the gain performance of the PMT, a gain drop of up to 50\% (relative to the gain at 0T, which is typically of the order of 10$^6$ at this high voltage) is considered acceptable. This value is used to determine the upper limit of the feasible operational B-field range of the device for EIC Cherenkov-detector application.}
	\label{Fig:Updated-field-effect}
\end{figure}

\begin{figure}[hbt]
	\centering
        \includegraphics[width=1.0\columnwidth]{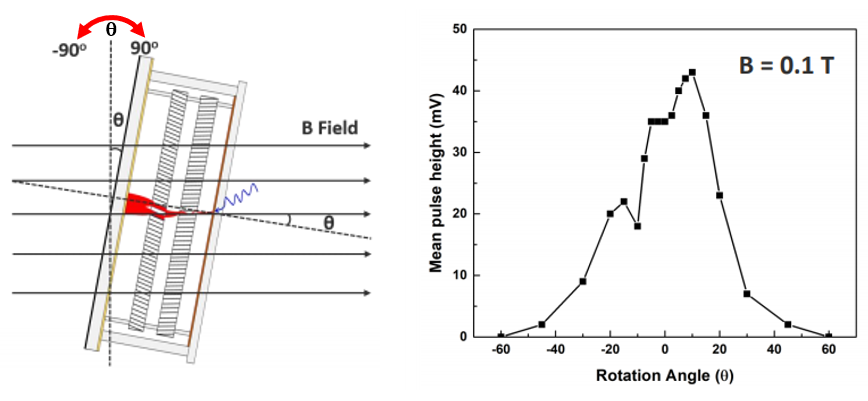}
	\caption{Magnetic Field Effects on LAPPD Devices which are quite similar to the MCP-PMT performance.}
	\label{Fig:PHOT-02-LAPPD}
\end{figure}

Another developing photon detection technology is that of LAPPD.  These devices also use micro channel plates as their basic of avalanche.  These can be used both for Cherenkov readout ({\it e.g.} in an mRICH configuration to add timing) or directly as a TOF detector.  As shown in Figure~\ref{Fig:PHOT-02-LAPPD}, these devices also suffer a significant loss in signal strength which is a combination of gain loss (somewhat tolerable) and efficiency loss~\cite{Hattawy:2018qbv}.  The QE loss is a second order impact when the LAPPD is used as a TOF detector since the primary signal already consists of multiple photoelectrons.  However, this loss is critical to the use of LAPPD as a Cherenkov detector readout.

Finally, we note the developments in recent years of silicon photo-multipliers or SiPMs.  Initially, these devices (which operate in a Geiger avalanche mode in each pixel) were highly susceptible to radiation damage.  Much work has been done to improve their performance intrinsically and it is now known as well that operation at low temperatures (below -30$^o$C) and post-annealing processes have been effective means to maintain and restore operation.  For this reason, SiPM technology seems a leading choice for readout of light signals from calorimeter devices at the EIC.  That said, more work is required to demonstrate the efficacy and long term viability of SiPM technology for use in a Cherenkov detector.  The basic distinction is signal size.  A well designed and high performance calorimeter will register many photons into a single pixel, making the presence or absence of "several" photo-electrons a mere shift and widening of the pedestal.  RICH detectors, on the other hand, must distinguish zero from one photo-electron and thereby are much more vulnerable to radiation damage of an SiPM.  It is therefore a clear priority to continually develop and evaluate the performance of SiPM detectors for RICH applications in the coming years.
Dedicated studies to establish SiPMs as single photon detectors adequate for the dRICH are starting. The goals consists in proving that they can be used up to integrated fluences of 10$^{11}$ n-equivalent per cm$^2$, a figure adequate to ensure several years of operation at maximum EIC luminosity. This possibility is suggested by recent results~\cite{Calvi:2018ulw} indicating that the goal is reachable when operating SiPMs at low temperature (in the range -30~-~-40 ~$^o$C) and undertaking annealing cycles at high temperature (170~$^o$C). SiPMs from different producer will be characterized by laboratory measurements and in test beam. Samples of SiPMs not exposed to irradiation, irradiated and recovered with an annealing cycle after irradiation will be tested. The study includes the coupling with the front-end ALCOR chip~\cite{Kugathasan:2019tgu}, developed for the SiPMs of the DarkSide experiment.

\subsection{Configuration for EIC}
Based upon the characteristics of known detector technologies as described above, it is possible to assert solutions for the detector performance matrix in each of the pseudo-rapidity regions.  These possibilities are outlined in the table~\ref{Table:PID-Assignments} and discussed in more detail in the sections that follow.  

\begin{table}[ht]
\centering

\begin{tabular}{|c|c|c|}
\hline
    region &  eID-only Technologies & eID \& PID Technologies\\
    \hline
    Electron & HBD, TRD & mRICH, LAPPD, LGAD\\
        Central & - & dE/dx, DIRC, LGAD \\
        Hadron & TRD & dRICH, gRICH/LAPPD \\ 
     \hline
\end{tabular}
\caption{\label{Table:PID-Assignments} 
Configurations of the PID detectors capable of meeting the performance requirements.}
\end{table}

It is necessary to establish simple criteria for what can be considered as acceptable performance and in particular the definition of dynamic range.  The PID/eID task can be simplified to the identification of four particle species which in mass order are the electron, pion, kaon, and proton.  For Cherenkov technologies, both the threshold and "imaged" mode of operation can be utilized as part of the ID process.  Positive ID is defined as follows:
\begin{itemize}
    \item Positive eID for a threshold device is valid up to the momentum at which the pion begins to radiate.
    \item Positive PID for an imaging device begins at the momentum where the kaon starts to radiate.
\end{itemize}
While careful analysis shifts these limits somewhat, they are nonetheless useful in comparison across detector technology options.  To this end, we list the Cherenkov thresholds for each radiator considered in any of our detector systems in Table~\ref{Tab:PID_indices}.

\begin{table}[ht]
    \centering
    \begin{tabular}{|c|c|c|c|c|c|}
    \hline
      &  & \multicolumn{4}{|c|}{Threshold (GeV/$c$)}\\ 
         radiator & index & e  & $\pi$  & K  & p \\
         \hline
         quartz (DIRC)    & 1.473   & 0.00048 & 0.13 &  0.47 & 0.88 \\
         aerogel (mRICH)  & 1.03    & 0.00207 & 0.57 &  2.00 & 3.80\\
         aerogel (dRICH)  & 1.02    & 0.00245 & 0.69 &  2.46 & 4.67\\
         $C_2F_6$ (dRICH) & 1.0008  & 0.01277 & 3.49 & 12.34 & 23.45\\
         $CF_4$ (gRICH)   & 1.00056 & 0.01527 & 4.17 & 14.75 & 28.03\\
         \hline
    \end{tabular}
    \caption{Table of Cherenkov thresholds for various media.}
    \label{Tab:PID_indices}
\end{table}

These thresholds, along with the detailed calculations shown in the prior sections, are summarized for application to each detector arm in the sections that follow.

\subsubsection{Forward Region}
The PID requirements in the hadron-going direction are naturally the most stringent in the spectrometer owing to the broad momentum range required for hadron identification.  The various technologies considered have been accumulated into a table distinguishing their range in e-$\pi$ separation and also in $\pi$-K separation.  For Cherenkov devices, the highest momentum for e-$\pi$ is put at the pion threshold and the lowest momentum for $\pi$-K is placed at the kaon threshold.  The results are summarized in Table~\ref{Tab:PID-Hadron arm sensitivity ranges.}.

\begin{table}[ht]
    \centering
    \begin{tabular}{|c|c|c|}
     \hline
      forward region  & \multicolumn{2}{|c|}{Range (GeV/$c$)}\\ 
         Technology & e - $\pi$  & $\pi$ - K \\
         \hline
         CsI RICH         & 0.0150 - 20   & 14.75 - 50 \\
         dRICH (aerogel)  & 0.0025 -  5   &  2.46 - 16\\
         dRICH (gas)      & 0.0127 - 18   & 12.34 - 60\\
         dRICH (overall)  & 0.0025 - 18   &  2.46 - 60\\
         TOF (LGAD)       &  0 - 1          &  0.00 -  5\\
         TOF (LAPPD 4m 5ps)      &  0 - 2.5   &  0.00 - 16\\
         TRD              & 1.0 -- 270.0 & -- \\
         \hline
    \end{tabular}
    \caption{Performance ranges for possible forward region detector technologies.}
    \label{Tab:PID-Hadron arm sensitivity ranges.}
\end{table}

Among the various options it becomes evident that there is a clear need for gas-based Cherenkov to reach the high end momentum requirements of the EIC.  IsT is also immediately clear that owing to the high threshold imposed by a low-index radiator choice necessary to reach the high momentum range, there must be an additional technology.  The dRICH presents an elegant solution to the issue by incorporating aerogel.  The gRICH option must be augmented by the addition of technology like aerogel-base mRICH or by high resolution TOF in order to cover the full dynamic range.

\subsubsection{Barrel}
\label{pid_central-arm}
The principle challenge of the barrel is the lack of space provided therein.  As a result the DIRC technology and TOF technology become leading options in most designs.  There exist, however, two significant issues with a DIRC-only solution.  These are:
\begin{itemize}
    \item The DIRC provides a threshold for kaon radiation at 0.47 GeV/$c$.
    \item There is a need for eID (e-$\pi$) that may not be fully met.
\end{itemize}

\begin{table}[ht]
    \centering
    \begin{tabular}{|c|c|c|}
     \hline
       barrel & \multicolumn{2}{|c|}{Range (GeV/$c$)}\\ 
         Technology & e - $\pi$  & $\pi$ - K \\
         \hline
         $\frac{dE}{dx}$  & 0 - 2           & 0 - 3 \\
         $\frac{dE}{dx}$ (Cluster Count) & 0 - 10      & 0 - 15 \\
         DIRC             & 0.00048 -  1  &  0.47 -  6\\
         TOF (LGAD)       &  0 - 1        &  0.00 -  5\\
         HBD              & 0.0150 - 4.17   & N/A \\
         \hline
    \end{tabular}
    \caption{Performance ranges for possible central barrel detector technologies.}
    \label{Tab:PID-Central barrel sensitivity ranges.}
\end{table}

It is therefore likely that a complementary technology in addition to the DIRC is required for the central barrel.  The use of dE/dx follows naturally when one assumes that the tracking system would contain a hybrid of silicon and TPC.  However, one much be cautious.  Because of the so-called "band crossings" in any dE/dx measurement, it is absolutely necessary to have a "tag" of low velocity particles to eliminate these from any eID system (wherein the electron is well into the high beta plateau).  TOF provided either by the DIRC system or by the inclusion of timing layers in the silicon tracker will be a must for such systems.

In a non-hybrid tracking system (internal silicon layers to 50 cm radius), one can imagine utilizing the additional space for a new PID device to complement the DIRC and TOF options.  In this case, one can even imagine exceeding the TPC dE/dx performance by a significant factor utilizing cluster counting rather than merely energy loss measurements.  Cluster counting devices require further R\&D in the coming time to demonstrates that this capability can be reached.  RICH systems modeled similar to the Delphi Barrel RICH can also be imagined, however the viability of SiPM devices as readouts for RICH detectors long term in the face of a high radiation environment must be demonstrated.

\subsubsection{Backward Arm}
 In the backward direction, several possibilities exist.  One of these possibilities is that despite the asymmetry of the collision itself, one could choose to place a device such as dRICH in the backward region as well.  This creates a challenge since the dRICH technology requires significant space.  Nonetheless, dRICH in the backward end-cap would over-perform all the requirements of the electron ion collider and provide a singular solution for both endcaps.
 
 More conventional thinking would attempt to fulfill the less stringent needs in the backward region by instead using one or several layers of a more compact PID technology.  The ideal requirement of 4 GeV/$c$ eID capability is well matched to the HBD-style technology.  In the sPHENIX application, a 50 cm radiation of gas with a 4.17~GeV/$c$ pion threshold.  A limitation of this technology is that its original design is optimized for separation of 2e from 1e and not for e-$\pi$.  Calculations exist as shown above for a new avalanche stage that promises to produce a pion rejection factor of roughly 100.  An alternative, is to split the HBD volume into two halves and square a lesser pion rejection factor.  Both these concepts are unproven at the time of the Yellow Report and would require further R\&D to prove their validity.

\begin{table}[ht]
    \centering
    \begin{tabular}{|c|c|c|}
     \hline
      Backward Region & \multicolumn{2}{|c|}{Range (GeV/$c$)}\\ 
         Technology & e - $\pi$  & $\pi$ - K \\
         \hline
         dRICH (aerogel)  & 0.0025 -  5   &  2.46 - 16\\
         dRICH (gas)      & 0.0127 - 18   & 12.34 - 60\\
         dRICH (overall)  & 0.0025 - 18   &  2.46 - 60\\
         HBD              & 0.0150 - 4.17   & - \\
         mRICH            & 0.0025 -  2   &  2.00 - 6\\
         TOF (LAPPD 4m, 5ps)      &  0 - 3    &  0.00 - 16\\
         TOF (LAPPD 3m, 10ps)      &  0 - 1.8 &  0.00 - 10\\
         TRD              & 1.0 -- 270.0 & -- \\
         \hline
    \end{tabular}
    \caption{Performance ranges for possible backward end-cap detector technologies.}
    \label{Tab:PID-Electron arm sensitivity ranges.}
\end{table}

A more conventional approach is to use one or more compact PID technologies.  The mRICH is reasonably well suited to the task for providing additional eID and also PID.  An option is being considered for augmenting the readout of mRICH with LAPPD which adds high resolution TOF to the mix.  Two improvements occur.  First, the TOF tag does not need to exceed the Cherenkov threshold for aerogel (instead it need to exceed the threshold in the LAPPD window).  This enhances the capability at the lowest momenta.  Second, the TOF information will augment the performance so long as the mRICH would be placed with a long enough flight path (not a restriction for the ring-based mRICH mode of operation).  

TRD is also a possibility in the backward end-cap.  TRD, like HBD, can be thought of as a threshold technology in that only the electrons radiate while the pions do not.  The threshold is at roughly 1 GeV which makes the TRD technology an excellent complement to the mRICH in providing the necessary eID in the backward region.

\section{Far-Forward Detectors}
\label{part3-sec-Det.Aspects.FFDet}

\subsection{Introduction}

The EIC physics program includes a very broad domain for diffractive physics measurements. Experimentally, this means that robust far-forward ($\eta > 4.5$) hadron and photon detection, and far-rear electron detection ($\eta < -4.5$) is required. These regions of the IR require multiple detector concepts to meet the needs of the physics program, including calorimetry for electrons, neutrons, and photons, silicon sensors for charged particle tracking, timing, and detector concepts such as Roman Pots for detecting protons or nuclear remnants that are very close to the beam. The subsequent sections will introduce the various detectors and technologies, and discuss the results of simulations (including realistic acceptance, beam effects, and detector resolutions) and the associated impact on the physics.

\subsubsection {General Layout of Far-Forward IR Region}

\begin{figure}[th]
\includegraphics[width=.99\textwidth]{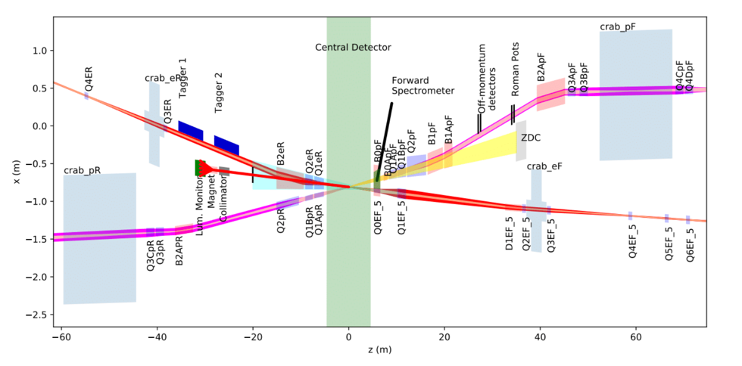}
\caption{Image of the full EIC baseline IR layout.}
\label{fig:overallIRLayout}
\end{figure}

\begin{figure}[th]
\includegraphics[width=.9\textwidth]{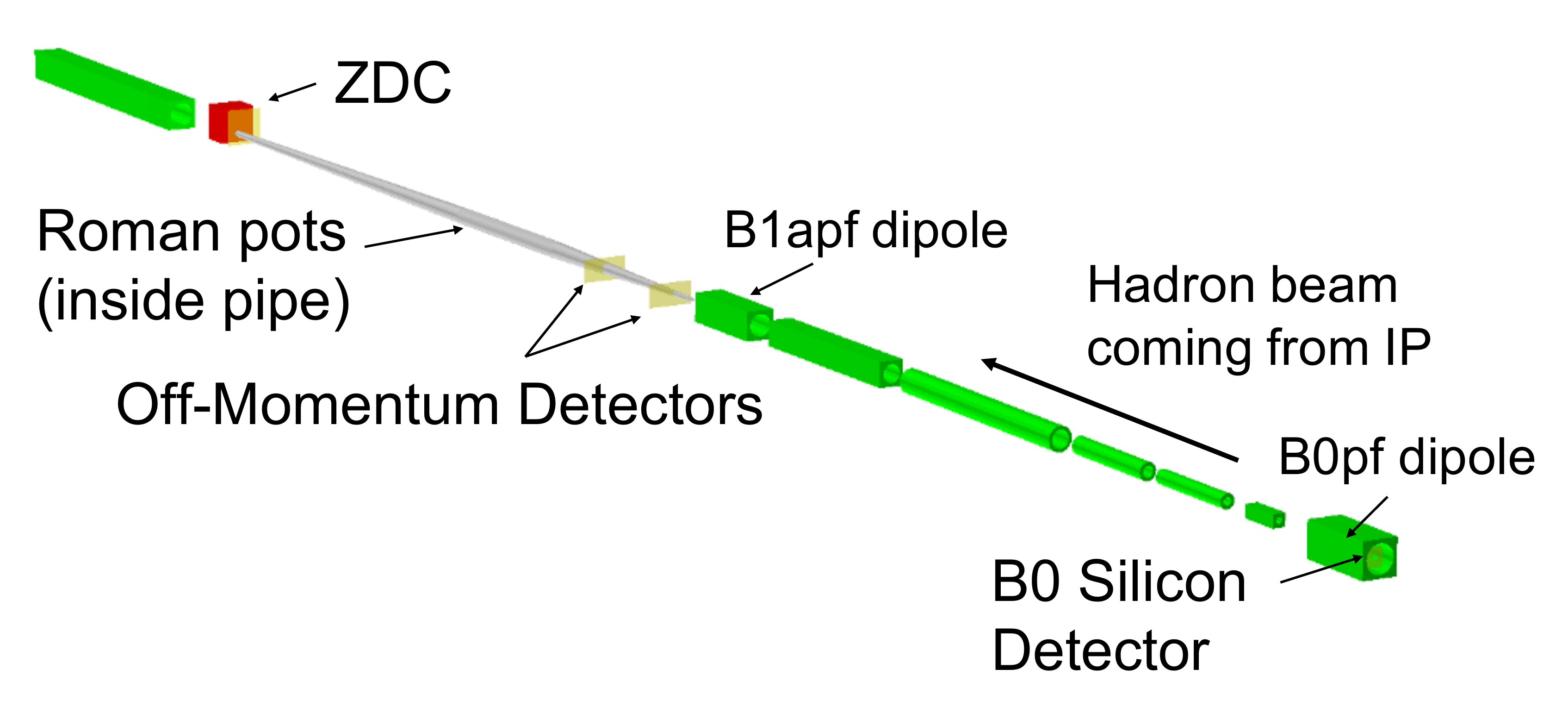}
\caption{Image of the Far-Forward IR and the associated detector components. \\Image generated using Geant4+EicRoot.}
\label{fig:FFHadronDetectors}
\end{figure}

The far-forward region of the interaction region at the baseline EIC detector is complex and requires novel ideas for covering a broad acceptance for charged and neutral particles from a long list of interactions. Fig. \ref{fig:overallIRLayout} shows a plan view of the full baseline EIC IR region. Fig. \ref{fig:FFHadronDetectors} shows the layout of the far-forward region used in the GEANT4 simulations. The image shows the various magnets for the hadron beam that create a unique engineering problem for placement of particle detectors and for allowing passage of particles scattered away from the beam.

The various subsystems involved in the far-forward region are summarized in Table \ref{tab:FF_accep} and depict the challenge of building a suite of detectors to cover the full acceptance for the various interaction channels.

\begin{table}[ht]
    \centering
    \resizebox{0.98\textwidth}{!}{
    \begin{tabular}{| l | c | c | c | c |}
    \hline
        \textbf{Detector}\ & \textbf{(x,z) Position [m]} & \textbf{Dimensions}& \textbf{$\theta$ [mrad]} & \textbf{Notes} \\
         \hline
         ZDC & (0.96, 37.5) & (60cm, 60cm, 2m) & $\theta < $ 5.5 & $\sim$4.0 mrad at $\phi = \pi$ \\
         \hline
         Roman Pots (2 stations) & (0.85, 26.0) (0.94, 28.0) & (25cm, 10cm, n/a) & $0.0 < \theta$ $< 5.5$ & 10$\,\sigma$ cut. \\
         \hline
         Off-Momentum Detector & (0.8, 22.5), (0.85, 24.5) & (30cm, 30cm, n/a) & $0.0 < \theta < 5.0$ & $0.4 < x_{L} < 0.6$ \\
         \hline
         B0 Spectrometer & (x = 0.19, $5.4<$\,z\,$< 6.4$) & (26cm, 27cm, n/a) & $5.5 < \theta < 13.0$ & $\sim$20\,mrad at $\phi$=0 \\
         \hline
    \end{tabular}
    }
    \caption{Summary of far-forward detector locations and angular acceptances for charged hadrons, neutrons, photons, and light nuclei or nuclear fragments. In some cases, the angular acceptance is not uniform in $\phi$, as noted in the table. For the three silicon detectors (Roman Pots, Off-Momentum Detectors, and B0 spectrometer) a depth is not given, just the 2D size of the silicon plane. For the Roman Pots and Off-Momentum Detectors, the simulations have two silicon planes spaced 2m apart, while the B0 detectors have four silicon planes evenly spaced along the 1.2m length of the B0pf dipole magnet bore. The planes have a "hole" for the passage of the hadron beam pipe that has a radius of 3.2cm. }
    
    \label{tab:FF_accep}
\end{table}

\subsection {Roman Pots}

\begin{figure}[ht]
\centering 
\includegraphics[width=.9\textwidth]{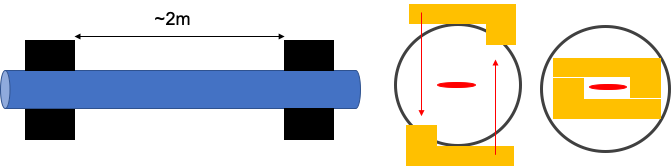}
\caption{Cartoon sketch of the Roman Pots concept. This Yellow Report study assumes two Roman Pots stations, separated by 2 meters, for all of the simulations. The right side of the cartoon shows a potential shape of the sensors and how the pots could be inserted into the beam line. Given the small amount of space between the hadron and electron beams at this spot in the IR, horizontal movement of the pots is challenging.}
\label{fig:RPCartoon}
\end{figure}

Roman Pots (RP) are vessels with a thin window in which silicon detectors are placed. The pot vessel is inserted into the beam pipe vacuum, allowing detection of scattered charged particles that are very close to the beam, These detectors can measure scattered protons or light nuclei which are separated from the hadron beam by up to 5 mrad. The windows on the pots through which protons or light nuclei can enter to be measured by the silicon detectors are generally placed within 1 mm or so of the beam (depending on the beam optics and hence the transverse beam size at the RP location), with safe distance being defined as the ``10 $\sigma_{x,y}$" region, where $\sigma_{x,y}$ is the transverse size of the beam in x and y. Fig. \ref{fig:RPCartoon} shows a cartoon sketch of the basic concept being considered, but note that the stainless steel pots themselves are not shown in the cartoon. In this section, basic requirements for the sensors will be discussed first, and technology appropriate for use in the EIC diffractive physics program will be discussed at the end. 

\subsubsection{Basic Requirements for Roman Pots}

In general, the Roman Pots need to have both the necessary acceptance and resolution to carry out the diffractive physics program at the EIC. The acceptance is driven by the machine optics (i.e. the transverse beam size at RP location) and active sensitive region of the detector (sensor size). The various tables of optics configurations used in these studies can be found in the pre-Conceptual Design Report for the EIC ~\cite{pCDR}. From studies of Deeply Virtual Compton Scattering in e+p collisions, the sensitive area of the sensors needs to be about 25cm x 10cm to capture the majority of the protons within the required 5 mrad acceptance. Protons at larger scattering angles can be measured with the B0 detector (see Sec. \ref{B0Sec}). At the highest proton beam energy (275 GeV), the protons are within the 5 mrad acceptance, with the lower cutoff on the scattering angle acceptance being driven by the size of the beam at the  location. For protons at the lower beam energies (100 GeV and 41 GeV), the B0 and RP detectors are both required to fully cover the acceptance range. 
The details and potential physics impact of the different acceptances across beam energy configurations are discussed in Sec. \ref{sec:DVCSStudyDetails}. 

The $p_{T}$ resolution of the RP is dictated by both beam effects and detector effects. They are listed below in general order of the size of the effect, with the first being the largest contribution.

\begin{itemize}
    \item Beam angular divergence
    \item Crab cavity rotation
    \item Silicon pixel pitch
    \item Transfer matrix uncertainty
\end{itemize}

The beam angular divergence sets the lower bound of the achievable resolution, so the goal is to mitigate the other effects such that their impacts are less than the angular divergence contribution. The contribution due to the crab cavity rotation manifests itself as an effective vertex smearing, since the crab cavity rotates the bunch horizontally such that the electron and hadron bunches arrive at the IP head-on. The effective vertex smearing is approximately $(0.5*\theta_{crossing}*L_{bunch})$. This contribution can be mitigated with fast timing (with resolution $\sim$35ps), allowing for precise measurement of the location of the collision within the bunch. Table \ref{tab:DVCSSmearing} summarizes the smearing contributions with reference to the study that generated the quantitative assessment of these values and their relative impact.

\subsubsection{Silicon Sensors for Roman Pots}

The development of high spatial resolution pixel detectors with high per-pixel time resolution has been one of the major technological drivers in collider physics in recent years in order to meet some of the challenges posed by future collider experiments. Current particle trackers in collider experiments are based on silicon technology with a spatial resolution of few tens of microns, while novel silicon technologies have recently allowed timing resolution of few tens of ps, for instance with the Low Gain Avalanche Diodes (LGADs)~\cite{Giacomini:2018tbt,Moffat:2018kxw}. For example, the ATLAS and CMS experiments~\cite{Collaboration:2296612,Collaboration:2623663} at the High Luminosity LHC (HL-LHC)~\cite{CERN-ACC-2015-0140,CERN-ATS-2012-236} have developed fast-timing detectors based on LGAD sensors.

The LGAD is based on a simple $p$--$n$ diode concept, where the diode is fabricated on a thin high-resistivity $p$-type silicon substrate. A highly-doped $p$--layer (the ``gain" layer) is implanted under the $n^+$: application of a reverse bias voltage creates an intense electric field in this superficial region of the sensor, able to start an avalanche multiplication for the electrons. The gain is limited to a factor of typically 10-100, such that the noise is low compared to the case of avalanche photodiodes. The drift of the multiplied carriers through the thin substrate generates a fast signal with a time resolution of few tens of ps. However, there is a severe limit on the spatial resolution this detector can achieve. Importantly, the dead areas exist at the edges of the pixels and in-between the pixels, so that large-pitch pixel only are possible lest a low fill-factor is introduced. For example, the LGAD sensors developed for the ATLAS and the CMS timing-detectors have relatively large pads of about 1.3 x 1.3 mm$^2$ size.   

\begin{figure}[ht]
\centering 
\includegraphics[width=.45\textwidth]{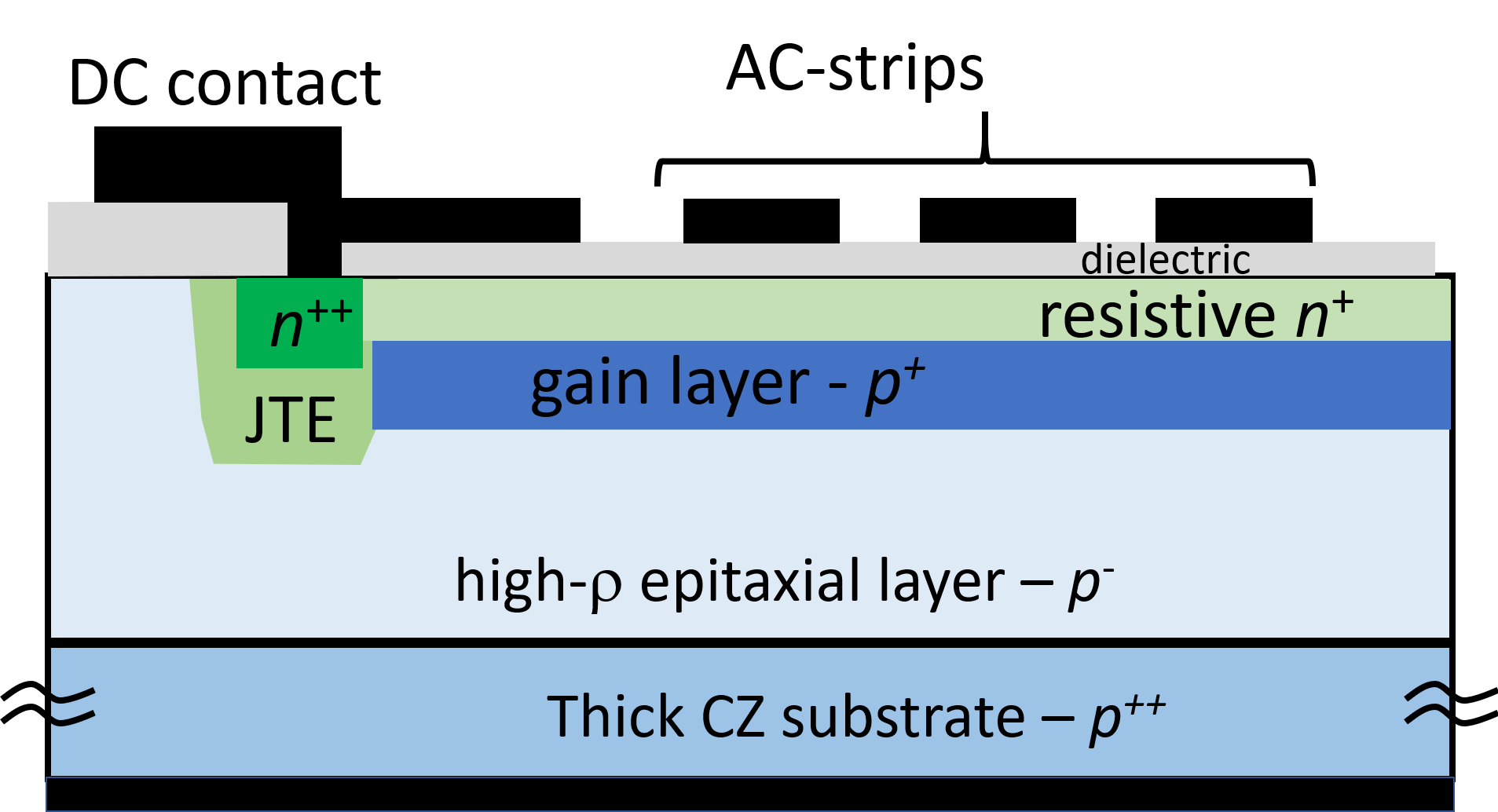}
\includegraphics[width=.45\textwidth]{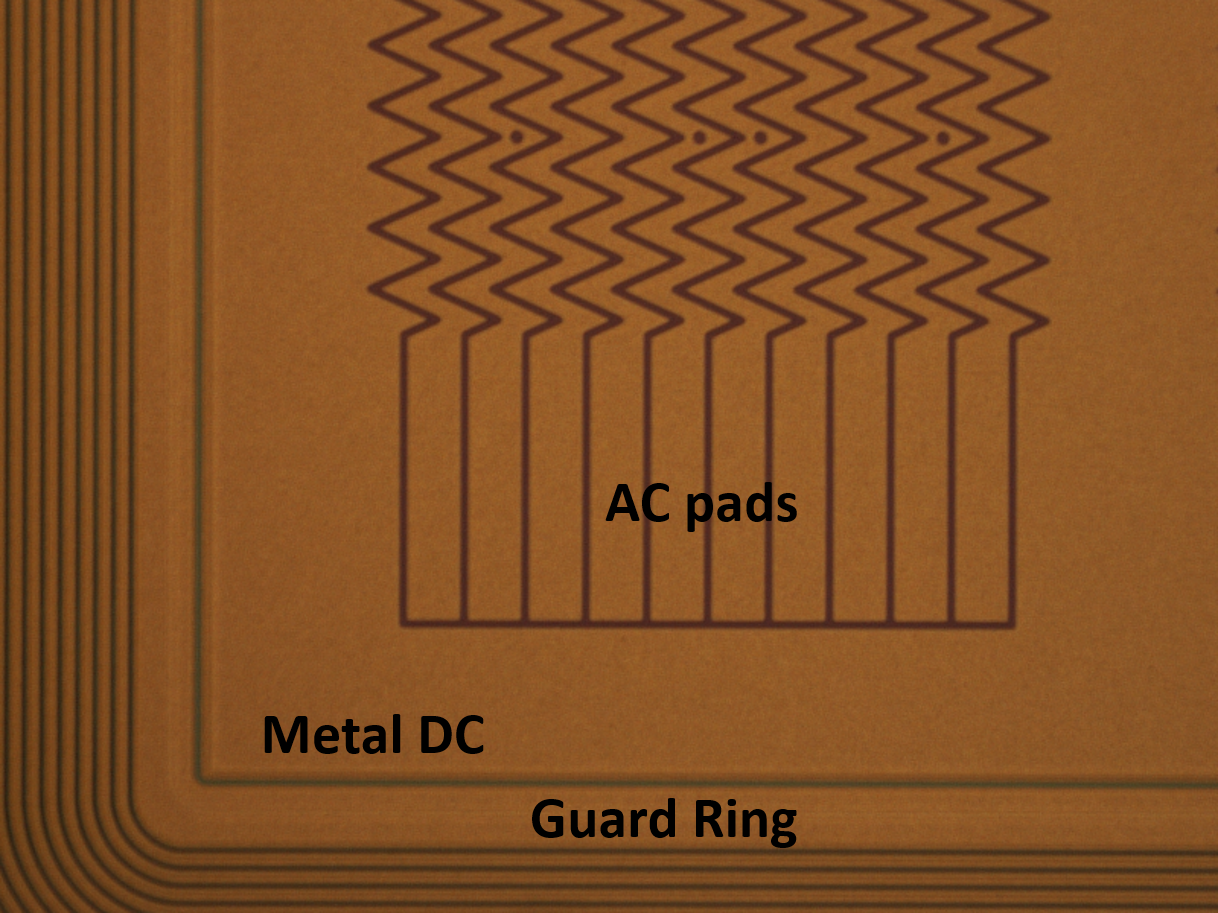}
\caption{\label{fig:sketch} Left. sketch of the cross section of a segmented AC-LGAD (not to scale). For simplicity, only three AC electrodes are shown. Right: microscope image of an AC-LGAD, fabricated at BNL.}
\end{figure}

Recent research has studied how to segment LGAD sensors~\cite{Mandurrino:2020ukm}, e.g. with pixels or strips with pitches in the tens of microns,  in order to achieve fine spatial resolution while maintaining the fine LGAD time resolution. It was demonstrated~\cite{Giacomini:2019kqz,Mandurrino:2019csy} that the new technology of AC-coupled LGADs 
(AC-LGADs~\cite{Mandurrino:2020ukm}) is a good candidate for a 4-dimensional (4-D) silicon detector to provide time resolution in the few tens of ps and segmentation of few tens of microns.  Figure~\ref{fig:sketch} shows a schematic section of a segmented AC-LGAD sensor. 

Differently from a standard DC-coupled LGAD, see Ref.~\cite{Giacomini:2018tbt}, the $n^+$ layer is more resistive than in the standard LGAD. Above the active area, a thin dielectric layer is deposited and, on top of it, metal pads are placed to define the AC-couple electrodes of the structure. Signals are induced on these pads,  which  are connected to the read-out electronics. If the pads are close enough, there is an important cross-talk between them that can be used for interpolation. Since the geometry of these pads can be arbitrary, by patterning the pads as  zigzag (Figure~\ref{fig:sketch}) it is possible to use the cross-talk among strips to enhance the spatial resolution and, at the same time, to keep the number of the read-out channels low.

Since they are fabricated on thin  substrates,
the LGAD sensors,intrinsically have a very limited dead area external to the active region. One floating guard ring is sufficient to sustain the high voltage and scribelines  at a distance smaller than 100 $\mu$m are possible.

Another sensor option for the Roman Pots  is the  3D pixel technology that has been used, for example,  in the ATLAS IBL. 3D sensors are intrinsically fast and are lacking the "Landau" noise, which constitutes the ultimate limit of the timing resolution of the LGAD. On the other hand, 3D sensors do not have intrinsic gain and have a capacitance which is 4-5 times higher than that of an LGAD of the same area. Their fast timing properties cannot therefore be exploited by a power budget-limited readout electronics.

In fact, a critical aspect for the development of a Roman Pot pixel detector with fast-timing capabilities is the readout. The front-end electronics must have timing and feature size compatible with those of the sensor.  Current ASICs for ATLAS
(ALTIROC) and CMS (ETROC) are designed in the CMOS TSMC 130 nm and CMOS 65 nm
technologies respectively, and they use TDCs to measure the Time of Arrival and Time over Threshold, as well as RAM for data buffering. In the ALTIROC, for example, the maximum jitter is of the order of 25 ps for 10 fC charge, and the ALTIROC and ETROC total power consumption per unit area is about 200-300 mW / ${\rm cm^2}$. As a comparison, the RD53 readout chip for pixel detectors for tracking (i.e. no timing) at the HL-LHC with 50 x 50 $\mu{\rm m^2}$ and 25 x 100 $\mu{\rm m^2}$ feature sizes is estimated to have a power density of about 1 W / ${\rm cm^2}$ or less.  Small pixels complicate the design due to limited space to accommodate TDCs and RAM and increased preamp and TDC power density. However, it seems reasonable  to reach 500 x 500 $\mu{\rm m^2}$ feature size by rearranging blocks and removing components that are likely unnecessary in a Roman Pot detector (e.g. a large RAM), while maintaining the same timing performance. In addition, by using Time-Over-Threshold (TOT) features in the ASICs, the charge sensed by pixel can be measured and in turn the charge sharing among pixels estimated. Therefore, using the TOT information the spatial resolution may improve beyond the fixed pixel pitch.

\subsubsection{Summary of the Current Design Constraints}

Based on the requirements listed above, and the results of the studies detailed in Secs. \ref{sec:DVCSStudyDetails} and \ref{sec:He3SimulationStudy}, the overall optimized Roman Pots requirements can be summarized in the following way. In order to fully cover the $p_{t}$ range of scattered protons and ions from the various physics channels covered by the Roman Pots, a total active sensor area of 25cm x 10cm will be required. This can be achieved with various different arrangements of the sensors, but the total area covered must be preserved in order to maximize the kinematic coverage. 

The studies to date, along with the expected improvement of the proposed silicon sensor technology (AC-LGADs), indicate that a 500 $\mu$m x 500 $\mu$m pixel size will properly balance the smearing contribution and R\&D efforts. The simulations detailed in Secs. \ref{sec:DVCSStudyDetails} and \ref{sec:He3SimulationStudy} assumed two RP stations with one sensor plane each. However, in actual operation, anywhere between 2 and 5 sensor planes per station would likely be used for redundancy and background rejection. With the assumed active area per plane and pixel size, this leads to 100k channels (pixels) per plane. 

Finally, in order to meet the needs of both background rejection and reduction of vertex smearing from the crab cavity rotation, a timing resolution per plane of $\sim$35ps will be required. 
\subsection{Off-Momentum Detectors} \label{OFFMSec}

\subsubsection{Basic Design Considerations}

\begin{figure}[hb]
\centering 
\includegraphics[width=.9\textwidth]{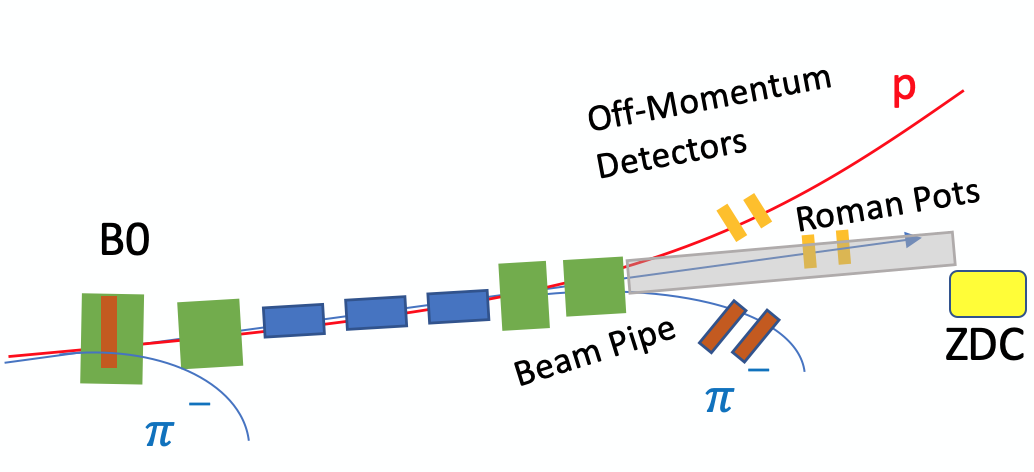}
\caption{Cartoon schematic of the operation of the off-momentum detectors. In the schematic a nuclear beam is being used and the final state particles shown are from various potential collision events, such as from nuclear breakup or lambda decay.}
\label{fig:OFFM-drawings}
\end{figure}

In any e+A collision event, protons and other charged particles can appear in the final state with very small scattering angles (e.g. proton spectators in nuclear breakup). In this scenario, the resulting charged particles will be directed toward the far-forward (FF) detectors, but will have a significantly different magnetic rigidity compared to the nuclear beam in question. For example, a proton with 100 GeV/$c$ of total momentum arising from an e+d collision where the deuteron beam has 200 GeV/n of energy would mean that the proton has an $x_{L} \sim$0.5, and half the rigidity of the deuteron beam, causing it to experience more severe magnetic deflections in the lattice. In this case, the protons will not stay in the beam pipe all the way down to the Roman Pots, and will instead be bent out of the beam pipe after the B1apf dipole magnet, as shown in Fig. \ref{fig:OFFM-drawings}. Measuring these so-called ``off-momentum" protons (or other charged particles) will require additional sensor planes outside the beam pipe - the so-called ``off-momentum detectors" (OMD). These detectors will cover $0.25 < x_{L} < 0.6$ for protons, with the azimuthal symmetry of the acceptance degrading at $x_{L} < 0.4$ due to losses in the quadrupole magnets.

The technology employed in these detectors can be the same used for the Roman Pots since the reconstruction approach using a transfer matrix will be similar. The main difference aside from the detection of off-momentum particles is that there will be no need for a $10\sigma$ cut that limits low-$p_{T}$ acceptance since the detectors sit outside of the beam pipe. 

Results of the simulations obtained using the OMD system are presented in Secs. \ref{sec:deuteronSpecTagSim}, \ref{sec:He3SimulationStudy}, and \ref{sec:lambdaDecaySimulations}. The spectator proton studies included only detectors on one side of the beam pipe, with two stations and the reconstruction approach as with the Roman Pots. The studies of $\Lambda$ decay indicate the need for detectors on the other side as well for detection of negative pions, and also a more complicated reconstruction method to account for the highly displaced $\Lambda$ decay vertex.

\subsubsection{Summary of the Off-Momentum Detector Considerations}

The Off-Momentum Detectors will be important for tagging final state charged particles from nuclear breakup and $\Lambda$ decay. The reconstruction of charged particles with this subsystem will be carried out in the same way as for the Roman Pots, using a lattice transfer matrix to reconstruct the IP coordinates from the hits at the sensor planes. It should be noted, however, that significant non-linear contributions to this transfer matrix approach will need to be considered for the Off-Momentum Detectors compared to the Roman Pots due to the off-momentum particles interactions with the quadrupole fields. More sophisticated reconstruction methods should be considered for future development. In the simulations, the sensors were assumed to be 30cm x 30cm, covering both sides of the beam pipe after the B1apf dipole magnet. This assumption will need to be refined when more up-to-date beam pipe designs are finalized. 

\subsection{B0-spectrometer} \label{B0Sec}

\subsubsection{Basic requirements for B0}
\begin{figure} [!ht]
\centering 
\includegraphics[width=.48\textwidth]{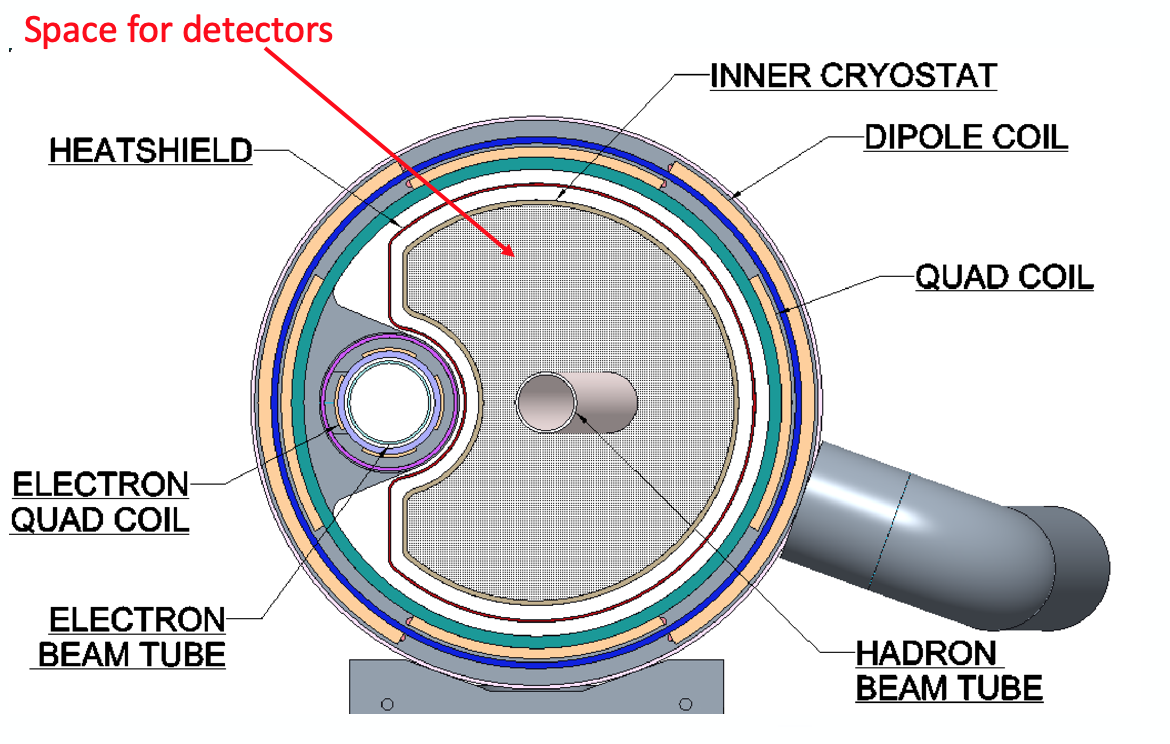}
\includegraphics[width=.48\textwidth]{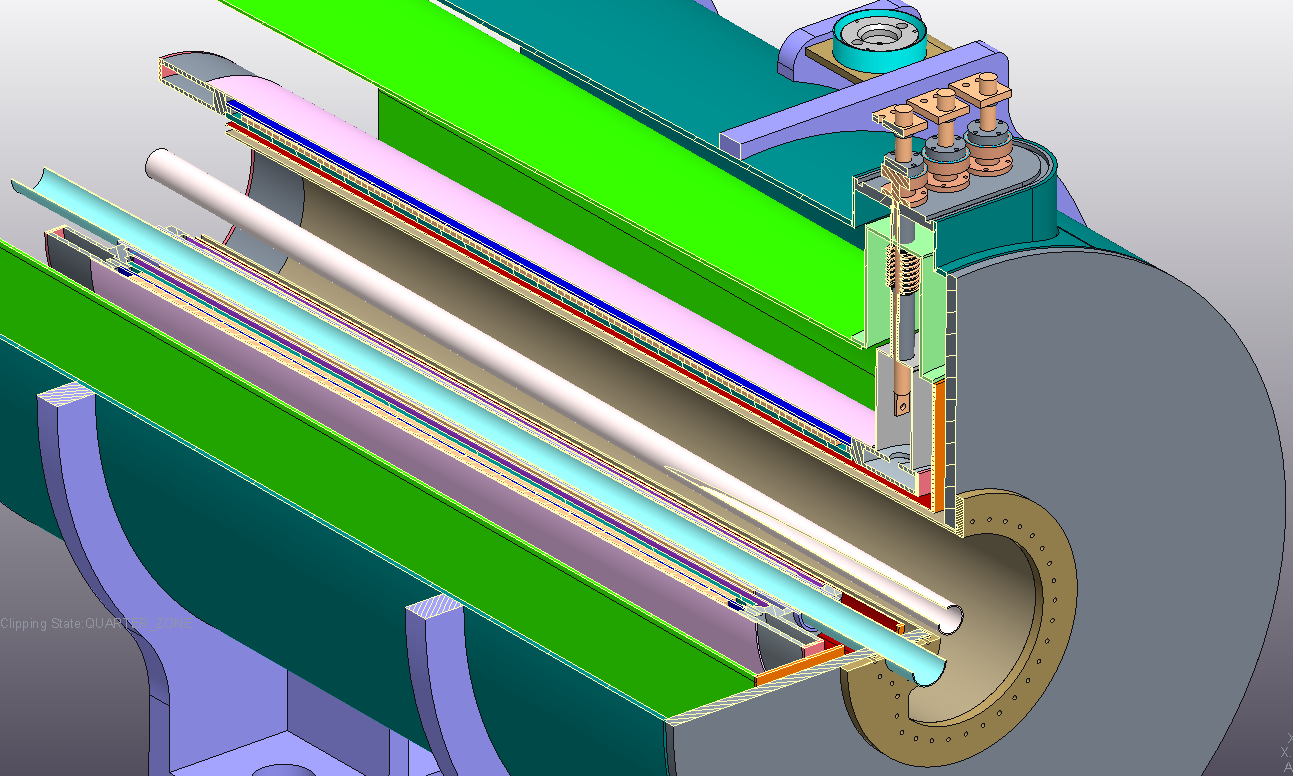}
\caption{Closeup image of the B0pf magnet bore with both hadron and electron beampipes and warm area for placement of detectors in 2D (left) and 3D (right).  The detector area allows for up to $\sim$13 mrad of angular coverage between the two beam pipes, and up to $\sim$20 mrad of angular coverage between the hadron beam pipe and the inner wall of the bore.  Both silicon tracking detectors as well as compact electromagnetic calorimetry are under consideration for integration into the open space in the bore.}
\label{fig:b0Images}
\end{figure}

The B0 tracker can help to provide very forward tracking capability for charged tracks. Such capability  is important for forward ($\eta > 3$) particle measurements as well as  event characterization and separation. Figure \ref{fig:b0Images} shows some conceptual drawings of the B0 bore with the sensors included. There has also been discussion on including electromagnetic calorimetry into the B0pf magnet bore, but simulations have not been carried out at this point.

\subsubsection{Silicon Sensors for B0-tracker}
To meet the radiation tolerance, spatial and timing resolutions in this kinematic region, several silicon sensor candidates are considered. 

\begin{figure} [!ht]
\centering 
\includegraphics[width=.88\textwidth]{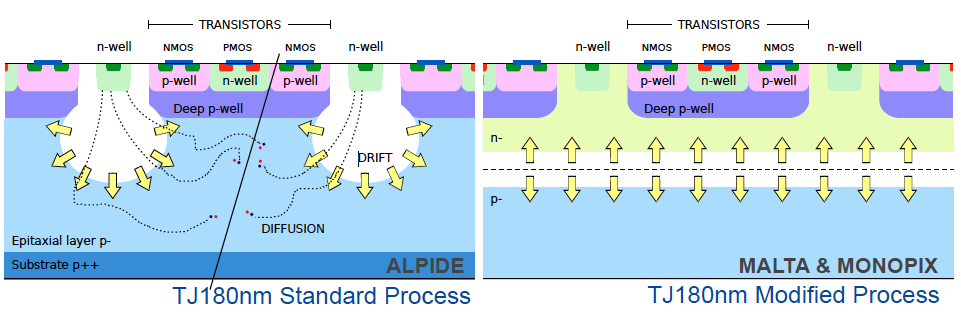}
\caption{\label{fig:maps} Comparison of the charge ionization process between a normal Tower Jazz 180 nm Monolithic Active Pixel Sensor (left panel) and a depleted Tower Jazz 180 nm Monolithic Active Pixel Sensor (right panel). the Figure from H. Pernegger presentation in the HSTD2019 Hiroshima conference.}
\end{figure}

One candidate is the Low Gain Avalanche Detector (LGAD) which has better than 30 ps timing resolution. Technical details are described in the previous chapter. Another top candidate is the radiation hard Monolithic Active Pixel Sensor technique: MALTA \cite{Pernegger:2017trh, Pernegger:2019qkb, Berdalovic:2018dlt}. This technique utilizes the Depleted Monolithic Active Pixel Sensor (or High Voltage Monolithic Active Pixel Sensor) to meet high granularity, low cost and low material budgets. The existing MALTA sensors based on the Tower Jazz 180 nm design, contain 36.4$\mu m$ by 36.4$\mu m$ pixels with the average silicon thickness of around 300 $\mu$m. 
Recent developments in the Monolithic Active Pixel Sensor (MAPS) technology with updated charge ionization process from diffusion to full depletion allow for achieving fast readout speeds. One type of this advanced technology is the MALTA sensor, which has a readout speed of about 5 ns.
High radiation tolerance ($>10^{15} n_{eq}/cm^{2} $) has been demonstrated at shaping time of 25\,ns \cite{Sharma:2019cdd}. Table\ref{tab:silicon} summarizes the performance of the LGAD and the MALTA technique. Ongoing R$\&$D for different silicon sensor techniques will improve their radiation tolerance, achieve better timing and finer spatial resolutions and get low material budgets.

\begin{table}[hbt]
\begin{center} 
\begin{tabular}{ lcc} 
 \hline
  \hline
 \textbf{Parameter} & \textbf{LGAD or AC-LGAD} & \textbf{MALTA} \\ \hline
 Technique &  Low Gain Avalanche Diode & 180\,nm Tower Jazz HV-MAPS \\ \hline
 Pixel size & current 1.3\,mm$\times$1.3\,mm  & 36.4\,$\mu$m $\times$ 36.4\,$\mu$m \\ 
  & towards 100\,$\mu$m $\times$ 100 $\mu$m & \\ \hline
 Integration time & $<$ 100\,ps  & $<$ 5\,ns \\ \hline
 Thickness per layer & $<$ $1\% X_{0}$ & $<$ $0.5\% X_{0}$ \\ \hline
 Radiation tolerance & $ \sim 10^{14} n_{eq}/$\,cm$^{2}$ &  $ > 10^{15} n_{eq}/$cm$^{2}$ \\ 
 \hline
 \hline
\end{tabular}
 \caption{Comparison of the LGAD and MALTA sensor performance}
    \label{tab:silicon}
\end{center}
\end{table}

Some things that will need to be considered for the future design of the B0 detector system are listed below.

\begin{itemize}
    \item Radiation background ( in particular a  synchrotron radiation and a radiation coming from the primary collisions) in the proposed very forward pseudorapidity region.
    \item The need for higher resolution sensors for reconstruction compared to the Roman Pots (pixels size $\sim$50$\mu$m)
    \item Available space in bore for sensors, support structure, and cabling.
\end{itemize}

\subsubsection{Pre-shower or EMCAL in the B0 spectrometer}
In order to provide a detection of low-energy photons and to provide a coverage in the transition area between central detector and ZDC calorimeters, a pre-shower detector or  electro-magnetic calorimeter might be considered in this area. Taking into account a limited amount of free space available along a  Z-axis,  and difficulties with integration a pre-shower might be a better option.

\subsubsection{Summary of the Current Design Constraints}

For the current design, 26x27cm$^2$ planes were used with 50x50$\mu$ m$^2$ pitch size. At least 4 layers will be needed, as a combination of high granularity and fast-timing detectors, to provide proper charged particle detection/tracking, momentum reconstruction, and to deal with the high-background expected in this area. Currently, the final layout of the sensors is still under consideration as the design of hadron and electron beam pipes inside the B0pf magnet bore are still being finalized.

\subsection { Zero-Degree Calorimeter (ZDC) }
\subsubsection{Basic requirements for the ZDC}
The Zero Degree Calorimeter (ZDC) will serve critical roles for a number
of important physics topics at EIC, such as distinguishing between coherent 
diffractive scattering in which the nucleus remains intact, and incoherent 
scattering in which the nucleus breaks up; measuring geometry of $e + A$ 
collisions, spectator tagging in $e + d / ^3He$, asymmetries of leading 
baryons, and spectroscopy. 
These physics goals require that the ZDCs have high efficiency for neutrons 
and for low-energy photons, excellent energy, $p_T$ and position resolutions, 
large acceptance and sufficient radiation hardness. 

\begin{figure} [!ht]
\centering 
\includegraphics[width=.48\textwidth]{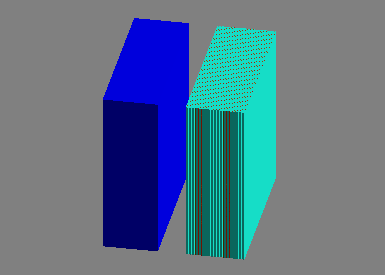}
\caption{\label{fig:ZDCschematic} Schematic ZDC in Geant4 simulation.}
\end{figure}

\begin{figure}[!ht]
  \centering
  \begin{subfigure}[h]{0.49\textwidth}
    \includegraphics[width=\textwidth]{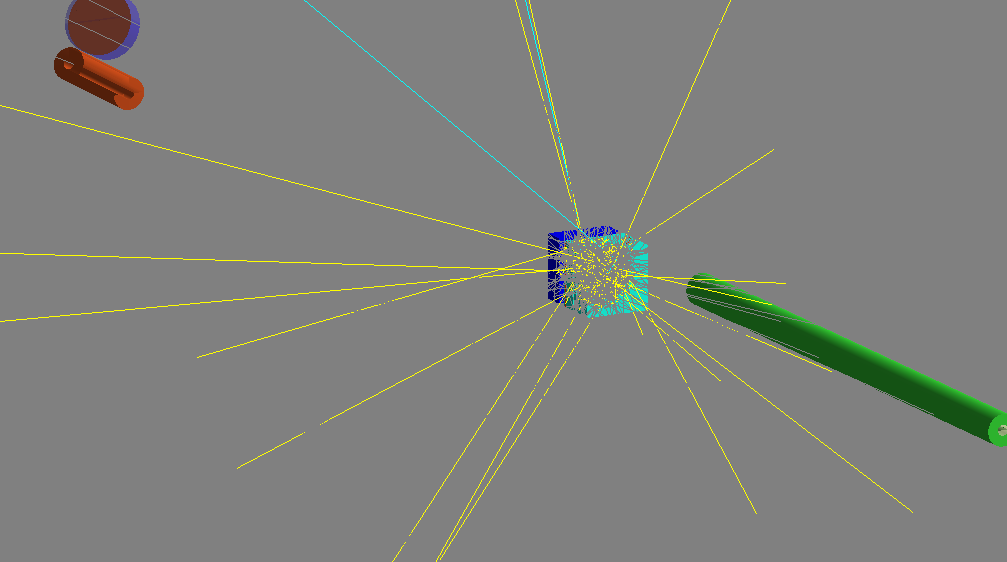}
    \caption{ZDC 20 GeV neutron event display.}
    \label{fig:ZDC20GeVN}
  \end{subfigure}
  \begin{subfigure}[h]{0.49\textwidth}
    \includegraphics[width=\textwidth]{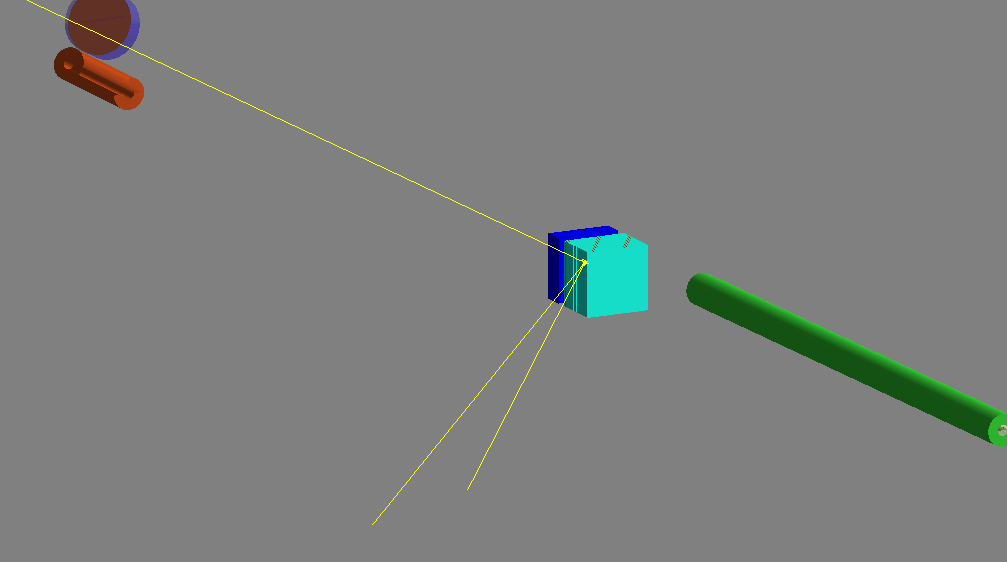}
    \caption{ZDC 500 MeV photon event display}
    \label{fig:ZDC500MeVGamma}
  \end{subfigure}
  \caption{Event displays for (Left) a 20 GeV neutron and (Right) a 500 MeV photon interacting with the ZDC.}
  \label{fig:ZDCEventDisplay}
\end{figure}

The ZDC schematic representation is shown in Fig.~\ref{fig:ZDCschematic}.
A 10 cm lead tungstate absorber is placed in front of 20 layers in the ALICE FoCal.
Figure~\ref{fig:ZDCEventDisplay} shows the event display for a 20 GeV neutron and a 500 MeV photon interacting with the ZDC.

\subsubsection{EMCAL technologies for ZDC}

\begin{figure*}[ht]
    \centering
    \includegraphics[width=\textwidth]{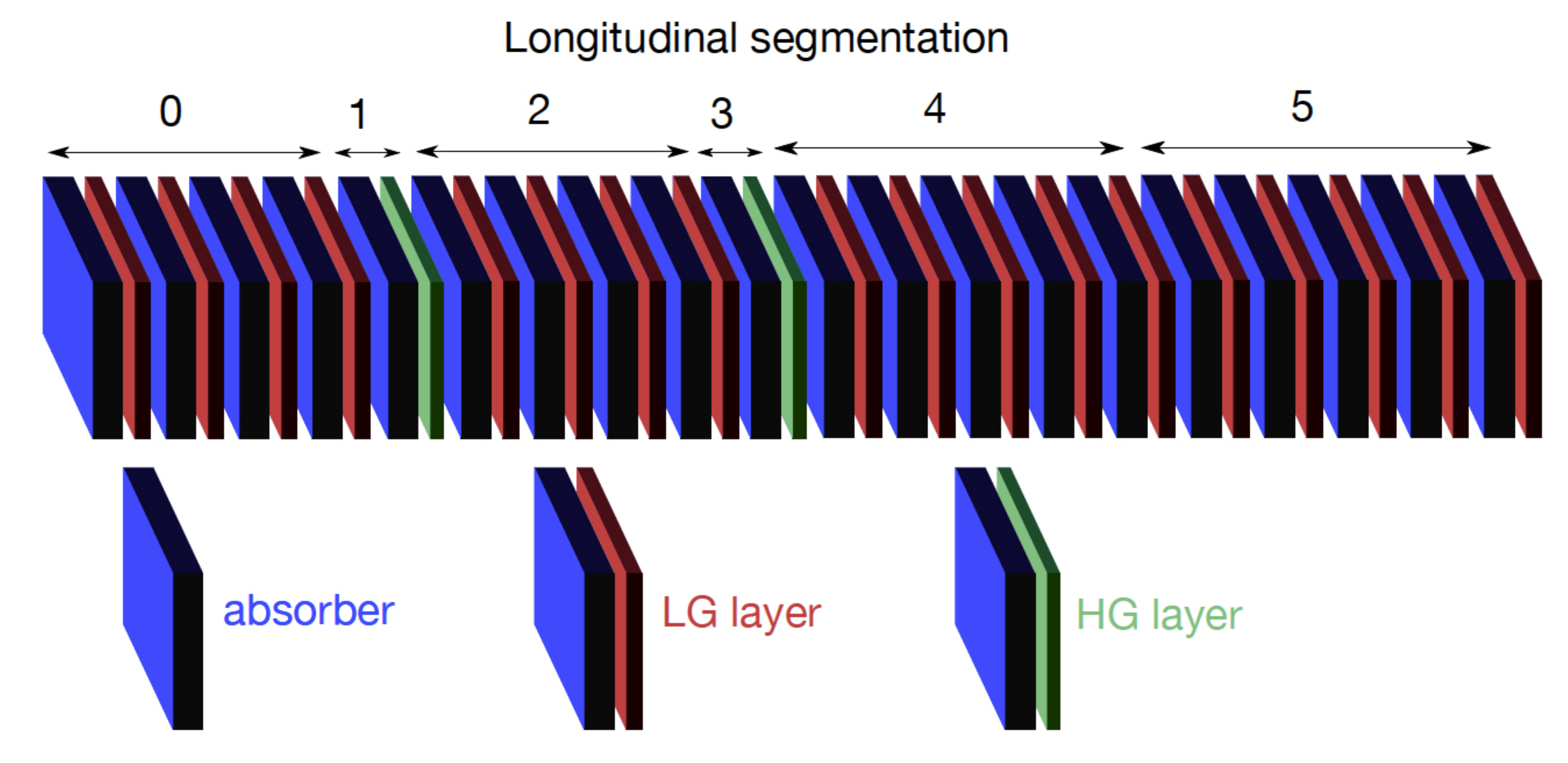}
    \caption{Schematic of the FoCal electromagnetic calorimeter. 
The blue absorber is tungsten, the red low granularity silicon layers 
are used for energy measurement while the green high granularity layers 
give precise position information \cite{ALICE-PUBLIC-2019-005}.}
    \label{fig:FoCal}
\end{figure*}

There are several possible approaches to achieve high energy and 
position resolution in an electromagnetic calorimeter. 
As an example, the ALICE FoCal \cite{ALICE-PUBLIC-2019-005}, is  
silicon-tungsten (Si+W) sampling calorimeter with longitudinal 
segmentation. 
Low granularity layers are used for the energy measurement while higher 
granularity layers provide accurate position information. A schematic 
of FoCal is shown in Fig.~\ref{fig:FoCal}. 

From simulations the photon energy resolution for FoCal is estimated to 
be $\sigma_E = 25\% / \sqrt{E} \oplus 2\%$. 
This is comparable to that expected for the sPHENIX W/SciFi calorimeter.
Other technologies that would provide suitable resolution include 
crystals (PbWO$_4$, LYSO, GSO, LSO), SciGlass glass, and W/SciFi. 
PbWO$_4$ crystals and SciGlass glass have been developed and 
characterized by the eRD1 Consortium and the Neutral Particle 
Spectrometer project at Jefferson Lab. 
Tests have shown energy resolutions of $\sim 2\%/\sqrt{E}$ for photon 
energies $\sim 4$ GeV \cite{Horn:2015yma}. 
The orbiting Fermi Gamma Ray Telescope uses a CsI crystal array and 
tracker to achieve very high spatial and energy resolution~\cite{Atwood:2009ez}. 

\subsubsection{HCAL technologies for ZDC}
The hadronic part of the ZDC is needed for neutron identification. To tag spectrator neutrons from both heavy and  light nuclei 
an energy resolution of $\sigma_E < {50\%}/{\sqrt{E}}$ and an angular 
resolution of at least 3 mrad / $\sqrt{E}$ are required. 
Cerenkov calorimeters, which measure only the high energy component of 
the showers, give excellent position resolution  and tight containment 
but are non-compensating and so somewhat non-linear. 
Sampling all charged particles 
produces a better energy resolution 
at the cost of worse lateral containment. 
We seek to exploit both techniques to maximize 
both the energy and 
position resolution of the ZDC. 
This could be done by using the quartz fibers developed for the LHC 
ZDCs, \cite{JZCap}, with traditional scintillators. 

\subsubsection{Soft photon detection}

In order to detect coherent collisions it is necessary to veto events in which 
 soft photons are emitted from an excited nucleus. 
In general, the photon decay chain of a heavy nucleus is dominated
by photons of energy of the order of 10 keV. 
These photons may be indistinguishable from background. 
However, for a doubly magic nucleus such as ${}^{208}$Pb, every bound-state 
decay sequence has at least one photon with an energy of at least 2.6 MeV. 
After accounting for the boost of the nucleus with momentum 
110 GeV/$c$ per nucleon, 20\% of these decay photons (with minimum energy 
455 MeV) are detectable within the ZDC aperture of $\sim 4.5$ mrad.
In order to detect such  photons from nuclear excitation 
it is important that the ZDC have the largest possible 
aperture. 
It is possible that a 2nd IR design would allow a larger ZDC acceptance.
Resolving nuclear decay photons from background will require a full 
absorption EM calorimeter with  excellent energy resolution, e.g. made 
with crystal scintillator (LYSO, PWO, ...). 

\subsubsection{Scintillator Tracker Detector} 

The meson structure research for the EIC has shown the need of a tracker, in combination with the ZDC, to be used as a veto detector for $\pi^-$ for an efficient measurement of the $\Lambda \rightarrow n + \pi^0$ channel~\ref{subsec:DetReq.DT.meson}. Besides this main purpose, adding a tracker could improve the reconstruction of charged particles in the ZDC for other different channels. An inexpensive and feasible option is the use of scintillating fibers (SciFi) as a tracker detector.

\begin{figure}[ht]
	\centering
	\includegraphics[width=0.85\linewidth]{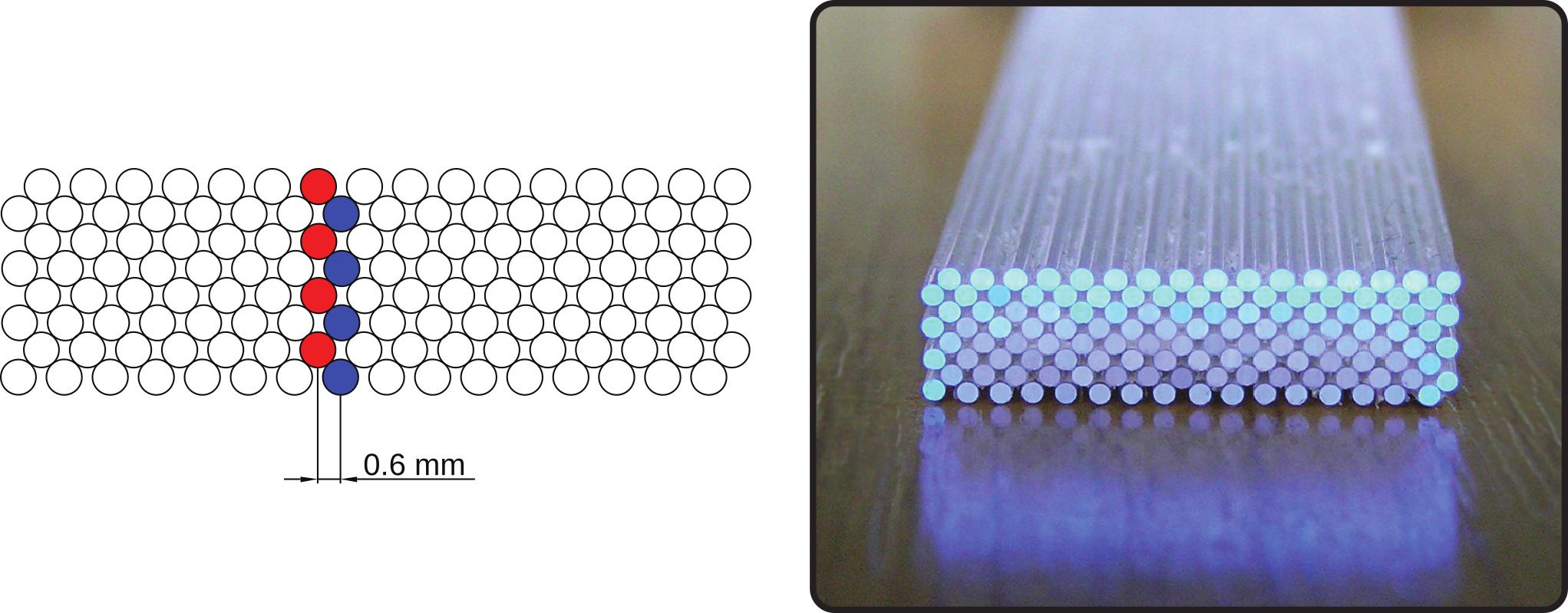}
	\caption{
	(Left) Schematic layout of a 4 layers SciFi bundle showing, in red and blue, two corresponding read-out channels for particles coming in the vertical direction. (Right) Picture of an assembled 4 layers, 32 channels SciFi bundle. 	}
	\label{fig:0degBundle}
\end{figure}

SciFi trackers combine the fast response of scintillator detectors with the flexibility and granularity that fibers can provide. A high efficiency fiber is made of a core of polystyrene-based scintillator surrounded by a cladding of PMMA, and some fibers by another cladding of fluorinated PMMA.  A SciFi tracker can handle high rates and is highly tolerant to radiation\,\cite{Kharzheev:2019tfk}, but on the other hand, the photon yield is quite low due to the small photon capture fraction, about 5\% for the double cladding fibers\footnote{one end output of the fiber}. Detection efficiency is increased adding extra fiber layers (Fig.\ref{fig:0degBundle}). Scintillating light can be read-out by several pixel devices like Avalanche Photo-diodes, Silicon photo-multipliers or multi-anode photo-multipliers. A SciFi tracker with a layout as in Figure \ref{fig:0degBundle} can achieve a spatial resolution of $\approx\,300\,\mu$m and a time resolution of $\approx\,500-220$\,ps \cite{Achenbach:2008ub}\cite{Gayoso:2012wga}, but different fiber diameter and overlap between channels results in similar spatial resolutions\,\cite{Blanc:1603129}.

\subsubsection{Summary of the current design} 

The number of spectator neutrons is predicted to be correlated with the collision geometry.  
The required performance of the detector to identify the coherence of 
the collision (via tagging of photons with $E_{\gamma}\sim50-150$ MeV) is under development using simulations with BeAGLE
\cite{Beagle}. Some of the performance parameters are still under study, but reasonable assumptions for the technology and detector performance were used to evaluate the impact on physics observables.

\paragraph{Acceptance} 

A large acceptance (e.g. 60$\times$60 cm$^2$) to establish good 
identification efficiency between coherent and incoherent collisions is 
necessary for vetoing spectator neutrons from nuclear breakup. 
This large acceptance is also required to determine the collision 
geometry~\cite{EIC:RDHandbook}. 
Additionally, studying very forward production and asymmetry of hadrons and 
photons also requires a high-acceptance detector. 
The EIC aperture of $\pm$4 mrad gives $p_T < 1$GeV/$c$ coverage for 
275 GeV hadrons and photons, which covers the transition from 
elastic/diffraction to the incoherent regime; for the low-energy hadron beam 
the acceptance in terms of $p_T$ is more limited (e.g. 
$p_T < 0.4$GeV/$c$ coverage for a 100\,GeV beam). 

\paragraph{Energy, position, and $p_T$ resolutions} 

Due to the strong $\beta$ squeeze to $<$ 1 meter to achieve the required high luminosity, a beam momentum spread of $\sim$20 MeV and a hadron beam angular divergence of $\sim$80-200 $\mu$rad is induced.
Thus a position resolution below the centimeter range for neutrons will not be useful.  A 
1 cm position resolution provides 300 $\mu$rad angular resolution, which 
can be translated to transverse momentum resolution $p_{\rm T} \sim 30$ 
MeV/$c$ for a 100 GeV spectator neutron. 

The collision geometry can be deduced from the number of spectator neutrons. Counting these neutrons implies a 
 minimum energy resolution of 
 $\Delta E/E \sim 50\%/\sqrt{E(GeV)}$.   In order 
 to accommodate signals ranging from single MIP track to 30 spectator neutrons, a wide dynamic energy range in the readout electronics is required. 

The ZDC at the EIC is anticipated to be a sampling type calorimeter with a sufficient longitudinal size of $\sim$10 interaction lengths\cite{EIC:RDHandbook}. 
It is also required to have a sufficient transverse size of $\sim$2 
interaction length to avoid transverse leakage of the hadron shower and 
to achieve good hadron energy resolution.

\subsection{Integration with accelerator}
\subsubsection{Beam parameters and lattice}

The integration of Far-forward detector components with the accelerator plays an important role for emerging EIC physic program. It is important to start it at the earliest stage of the design, since it could have an impact on both parties: it could affect accelerator impedance, or, on the other hand, inappropriate placement of accelerator elements  could have an impact on the detector acceptance by blocking or obscuring incident particles. 
The current studies were done with the accelerator lattices described  in the Table~\ref{tab:ion_ir} for an ion beam  and  Table~\ref{tab:el_ir} for an electron beam. Note, that sets of the quadrupoles will be placed in the common cryostat volume, therefore there will be no possibility to place any detecting elements there.

{\tiny {
\begin{table}[hbt]
    \resizebox{0.98\textwidth}{!}{
    \begin{tabular}{c| c| c | c | c | c | c| c | c | c  | c  }
    \hline
    
       \scriptsize{Name}&   \scriptsize{Type }  & \scriptsize{L}  &  \scriptsize{$R_{in}$}&   \scriptsize{$R_{out}$}&  \scriptsize{Dipole} &  \scriptsize{Quadrupole}    &  \scriptsize{$X_c$} &  \scriptsize{$Y_c$} &  \scriptsize{$Z_c$}&  \scriptsize{ $Theta_c$}  \\
       &   & [m]&  [m] & [m] & [T] & [T/m] &  [m] & [m] &  [m]  &  [rad]  \\ 
       \hline 
       \multicolumn{11}{c}{ \small {Rear elements}}\\
       \hline
\small{iYI6\_HB2} &  \small{SBEND} &  5.69   & 0.05  &  0.3  &  0 / -4.64    & 0      & -1.18   &  0     &  -48.96 &   0.011 \\ 
\hline 
\small{iYI6\_HQ3} & \small{QUAD}   &    1.2  &  0.05 &  0.3  &  0 / 0 & 47.8 & -0.52  & 0       & -20.7  &  0.025  \\
\hline 
\small{iYI6\_HQ2} & \small{QUAD}   &    2.57 &  0.05 &  0.3  &  0 /  0       & 47.1   & -0.323  & 0   &    -12.9  &  0.025   \\
\hline 
\small{iYI6\_HQ1} & \small{QUAD}   &    3.42 &  0.05 &  0.3  &  0 /  0       & -67.45 & -0.2046 & 0   &   -8.18 &   0.025  \\
        \hline 
       \multicolumn{11}{c}{ \small {Forward elements}}\\
       \hline  
\small {iB0PF}  & \small{SBEND}  & 1.2  & 0.2   & 0.5   & 0/-1.3   & 0      & 0.148  & 0 & 5.9    & 0.0259 \\
\small {iB0APF} & \small{SBEND}  & 0.6  & 0.043 & 0.256 & 0/-3.47  & 0      & 0.2    & 0 & 7.7    & 0.0278 \\ 
\small {iQ1APF} & \small{QUAD}   & 1.46 & 0.056 & 0.28  & 0/0      & -72.61 & 0.24   & 0 & 9.23   & 0.0289  \\
\small {iQ1BPF} & \small{QUAD}   & 1.61 & 0.078 & 0.34  & 0/0      & -66.18 & 0.293  & 0 & 11.06  & 0.0289\\
\small {iQ2PF}  &\small{QUAD}    & 3.8  & 0.131 & 0.58  & 0/0      & 39.45  & 0.383  & 0 & 14.16  & 0.0289 \\
\small {iB1PF}  & \small{SBEND}  & 2.99 & 0.135 & 0.5   & 0/-3.79  & 0      & 0.505  & 0 & 18.06  & 0.035  \\ 
\small {iB1APF} & \small {SBEND} & 1.5  & 0.168 & 0.4   & 0 /-2.70 & 0      & 0.6113 & 0 & 20.81  & 0.0436 \\ 
\small {iB2APF} &\small{SBEND}   & 5.7  &  0.05 & 0.3   & 0 /6.00  & 0      & 1.5221 & 0 & 41.890 & 0.02713 \\
       \hline
      \hline
    \end{tabular}
    }
    \caption{ Ion  beam lattice  for 275 GeV }
    \label{tab:ion_ir}
\end{table}
}
}

{\tiny {
\begin{table}[hbt]
    \resizebox{0.98\textwidth}{!}{
    \begin{tabular}{c| c| c | c | c | c | c| c | c | c  | c  }
    \hline
    
       \scriptsize{Name}&   \scriptsize{Type }  & \scriptsize{L}  &  \scriptsize{$R_{in}$}&   \scriptsize{$R_{out}$}&  \scriptsize{Dipole} &  \scriptsize{Quadrupole}    &  \scriptsize{$X_c$} &  \scriptsize{$Y_c$} &  \scriptsize{$Z_c$}&  \scriptsize{ $Theta_c$}  \\
       &   & [m]&  [m] & [m] & [T] & [T/m] &  [m] & [m] &  [m]  &  [rad]  \\ 
       \hline 
       \multicolumn{11}{c}{ \small {Rear elements}}\\
       \hline

\small{eQ5ER} &	\small{QUAD} &	1.2	& 0.05	& 	0.3	 & 0/0 & 	7.481 &	0.4131 & 	0 & 	-46.8267 &	0   \\ 
\small{eQ4ER} & \small{QUAD} &	0.6	 & 0.05	 & 	0.3	& 0/0 &	8.85796 &	0.4131 & 	0 & 	-37.99667 & 	0  \\ 
\small{eDB3ER}& \small{RBEND} & 5.199 &	0.05 &	0.3	 & 0/0.2115  0& & 0.39525 &	0 &	-34.79671 &	-0.00916  \\
\small{eQ3ER} &\small{QUAD}	& 0.6 &	0.05  &	0.3	 &0/0 &	-22.7971 &	0.354 & 	0	 & -31.597 &	-0.01832  \\ 
\small{eDB2ER} & \small{RBEND} & 5.5 &  0.05 & 	0.3	 & 0/-0.1999 &  0& 	0.01889 &	0 &	-12.249 &	-0.00916 \\
\small{eQ2ER}& \small{QUAD} &	1.4	 & 	0.05 &	0.3	& 0/0 &	14.1466 & 0 & 	0 &	-8.3 &	0 \\ 
\small{eQ1ER} & \small{QUAD} &	1.8	 &  0.05 &	0.3	& 0/0 & -14.478	& 0 &	0&	-6.2&	0  \\

        \hline 
       \multicolumn{11}{c}{ \small {Forward elements}}\\
       \hline  
\small{eQ0EF} &\small{QUAD} &1.2	&0.0031  & 0.007 &	0/0 &	-13.54 & 0 &	0 &	5.9	 &0	\\
\small{eQ1EF} &\small{QUAD} &	1.61& 0.05	 & 0.3 &	0/0 &	7.4612 &	0&	0  &	11.065	 & 0 \\	
\small{eQ2EF} &\small{QUAD} &	3.8	& 	0.05&	0.3	 &0/0 &	0	&0	&	0 &	14.17	&0	\\
\small{eQ3EF} &\small{QUAD}& 1.2 & 	0.05& 0.3 &	0/0	&-5.5461 &	0 &	0 	&20.82 &	0	\\
\small{eQ4EF} &\small{QUAD}&	1.2 &0.05	&0.3	&0/0&5.85445 &	0	& 0&	29.95	& 0	\\

      \hline
      \hline
    \end{tabular}
    }
    \caption{ Electron  beam lattice  for 18 GeV }
    \label{tab:el_ir}
    
\end{table}
}
}

\subsubsection{Beam Pipe, Vacuum, and Background }

At this point we do not have a mature engineering design of the beampipe in the far-forward area. In this section we just formulate some requirements for it, and discuss implications of the design.

\begin{figure}[!ht]
  \centering
  \includegraphics[width=0.55\textwidth]{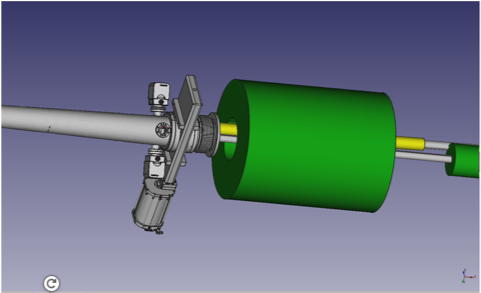}
  \includegraphics[width=0.25\textwidth]{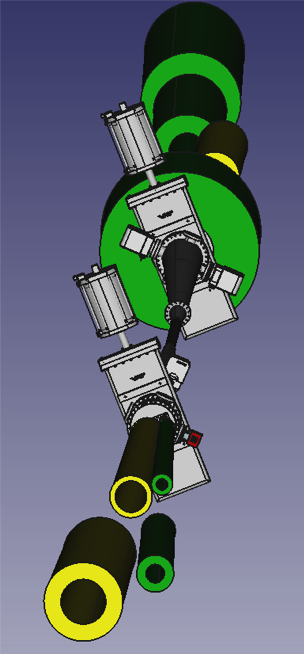}
  \caption{Beampipe at the B0 location. To give an idea of scale B0 is 1m long.  The left hand panel shows a side view of B0, represented as cylinder,  with vacuum pumps visible on the left and the separate electron and hadron beam pipes on the right. The right hand panel shows the same elements from a vantage point close to the beams, looking away from the interaction point. }
  \label{fig:B0-beampipe}
\end{figure}

\begin{figure}[!ht]
  \centering
  \includegraphics[width=0.8\textwidth]{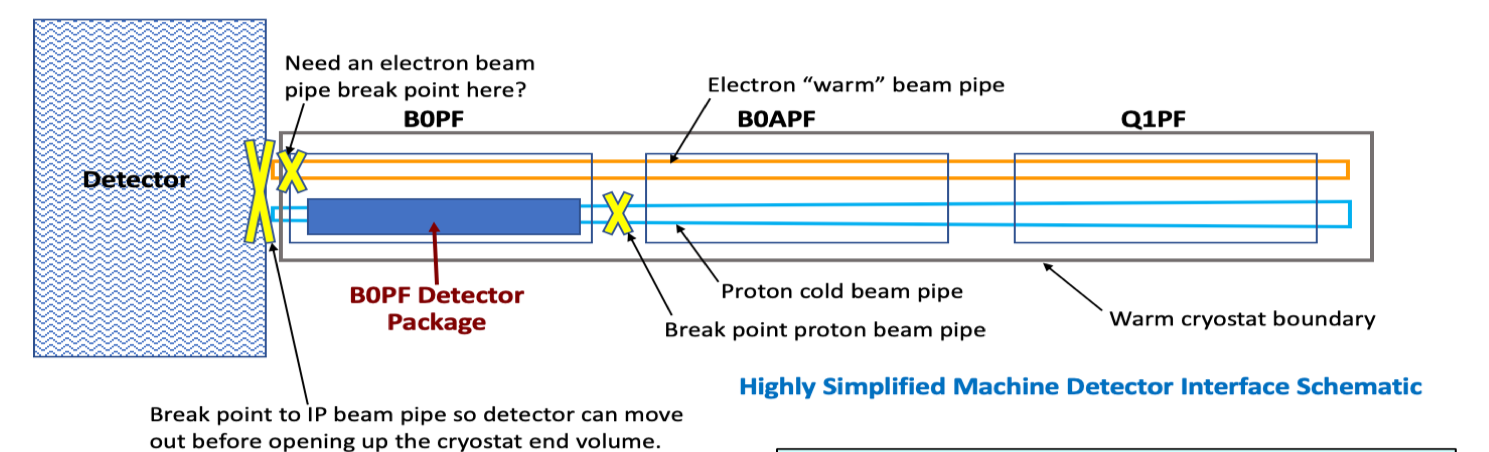}
  \caption{Schematic of the integration of B0-dipole, showing the electron and proton beam break points.}
  \label{fig:B0-int}
\end{figure}

One of the important areas to pay attention to the material budget while designing the beampipe is the B0-dipole location. First of all we have to minimize the amount of material at the exit window - this will be the area where the conical beampipe shared by both beams transfers to the two separate beampipes for the incoming electrons and outgoing ions. Vacuum pumps in front of the B0 dipole, shown on Fig.~\ref{fig:B0-beampipe}, will be large sources of beam+machine background where incident particles could start to develop showers, increasing the occupancy in the B0-tracker. Also, shower-tails from the central detector HCAL or cryo-module around the B0- dipole could potentially give an additional source of background for the B0-tracker.
The overall integration, assembly, and maintenance of detectors in this area will require a significant engineering effort. Preliminary locations of the beampipe/vacuum breaking points are shown on Fig.~\ref{fig:B0-int}, which would allow proper integration of accelerator elements, two separate beampipes, as well as allowing proper detector assembly and maintenance.

The exit window from the beampipe after the B1apf dipole magnet for the Zero Degree Calorimeter needs to be properly designed, due to the impact on the detection efficiency for low-energy photons used to veto incoherent heavy-nuclear breakup in the far-forward direction. A small section of beryllium beam pipe could also be considered as an option to minimally impact low-energy photons.

The beampipe material near the off-momentum detectors needs to be minimized in order to minimize impact on the momentum resolution due to multiple scattering. This same section of beam pipe is also shared by particles (neutron, photons) exiting to go toward the ZDC, so the design choices must accommodate both needs.

The scattering chamber and movable Roman Pots which house the silicon detectors need to be developed
and the RP impact on the accelerator impedance needs to be evaluated and minimized. In order to protect the sensors from incident beam losses, a proper collimation scheme needs to be designed together with a beam-loss monitor system. 
The former will play a crucial role in the machine and the detector protection. The latter, designed as part of the machine safety system, can successfully be used in the Beam Based Alignment procedure of the RP detectors.

In the backward direction a proper collimating scheme against synchrotron radiation from the electron beam needs further development to protect the low-Q2 tagger (see Sec.~\ref{sec:FB_electrons_main}). The beampipe material also needs to be optimized to minimize the impact on the multiple scattering in this area. An exit window for the bremsstrahlung photons is needed for the luminosity monitor, as discussed in Sec.~\ref{sec:FB_photons_main}. 

In addition to backgrounds from beam+machine sources, beam+gas background could also be a potential challenge for the far-forward and backward regions. The synchrotron radiation from the electron beam causes heating of the beam pipe, which will in-turn cause out-gassing and reduce the vacuum quality and increase the possibility for particles in the ion beams to collide with residual gas molecules. This will be a particularly challenging problem for the increase in neutron flux, which will have impact on especially the B0 detector system. This has been studied for a central silicon vertex tracker at mid-rapidity, and the annual dose of radiation has been found to be $\sim$3 orders of magnitude less than the suggested tolerances. More studies are needed to asses the annual radiation doses experienced by the B0 detectors, but the impact will be similar.

\subsection{Physics impact }
\subsubsection{Simulation Details} \label{sec:simDetails}

The simulations presented for the far-forward region of the IR were carried out using Geant4 implemented in either EicRoot or ESCalate. The simulations include the most-recently available layout of the IR magnets and engineering components (e.g. beampipe) and additionally include beam effects such as the smearing of the vertex due to rotation of the bunch by the crab cavity and beam angular divergence, unless noted otherwise for a particular study (e.g. studies of acceptance only). The parameters for the various beam effects can be found in the CDR.

\subsubsection{Deeply Virtual Compton Scattering (DVCS)} \label{sec:DVCSStudyDetails}

In e+p DVCS, the initial proton is scattered by very small angles ($\sim$ few mrads), and therefore is within the far-forward acceptance - specifically in the Roman Pots or the B0 spectrometer. Using only the tagged final-state proton, one has access to the momentum transfer, t, in the interaction. The precise measurement of this t-distribution yields access to the impact parameter distribution related to the gluon GPD. 

\begin{figure}[ht]
\includegraphics[width=.98\textwidth]{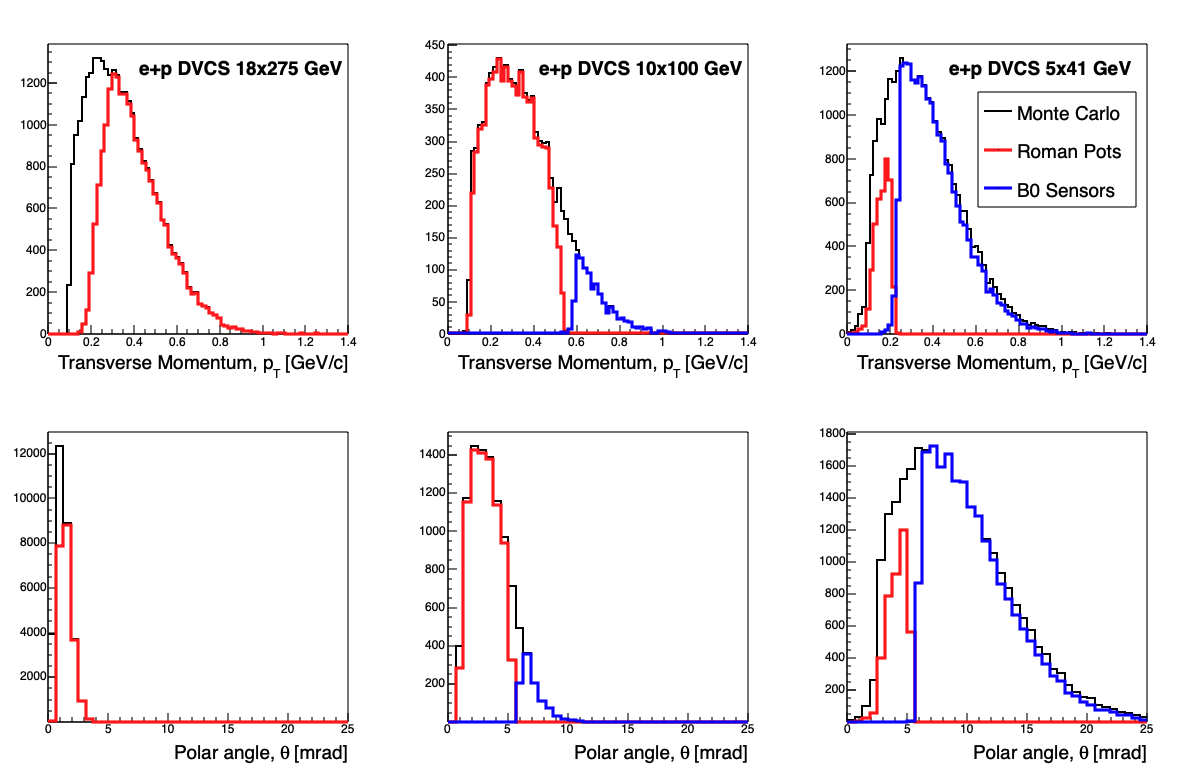}

\caption{$p_{T}$ (top row) and polar angle (bottom row) acceptance for three different beam energy configurations: 18x275 GeV (left), 10x100 GeV (middle), and 5x41 GeV (right). The black data in each figure represent the MC information from MILOU, the red lines are the accepted particles in the Roman Pots, and the blue lines are particles accepted in the B0 sensors. }
\label{fig:dvcsPtAcceptance}
\end{figure}

This simulation study was carried out using the MILOU MC generator to produce the simulated DVCS events, which were then passed through EicRoot and GEANT4 to simulate detector responses. These full simulations were then used to evaluate the DVCS proton acceptance and detector smearing. The study was conducted using three beam energy combinations, and included all of the smearing effects noted in the above description of the Roman Pots, namely, the effects of angular divergences, crab cavity rotation (which effectively smears the primary vertex), and detector reconstruction smearing. Fig. \ref{fig:dvcsPtAcceptance} shows the $p_{T}$-acceptance for three different beam energy configurations. 

\begin{figure}[ht]
\includegraphics[width=.95\textwidth]{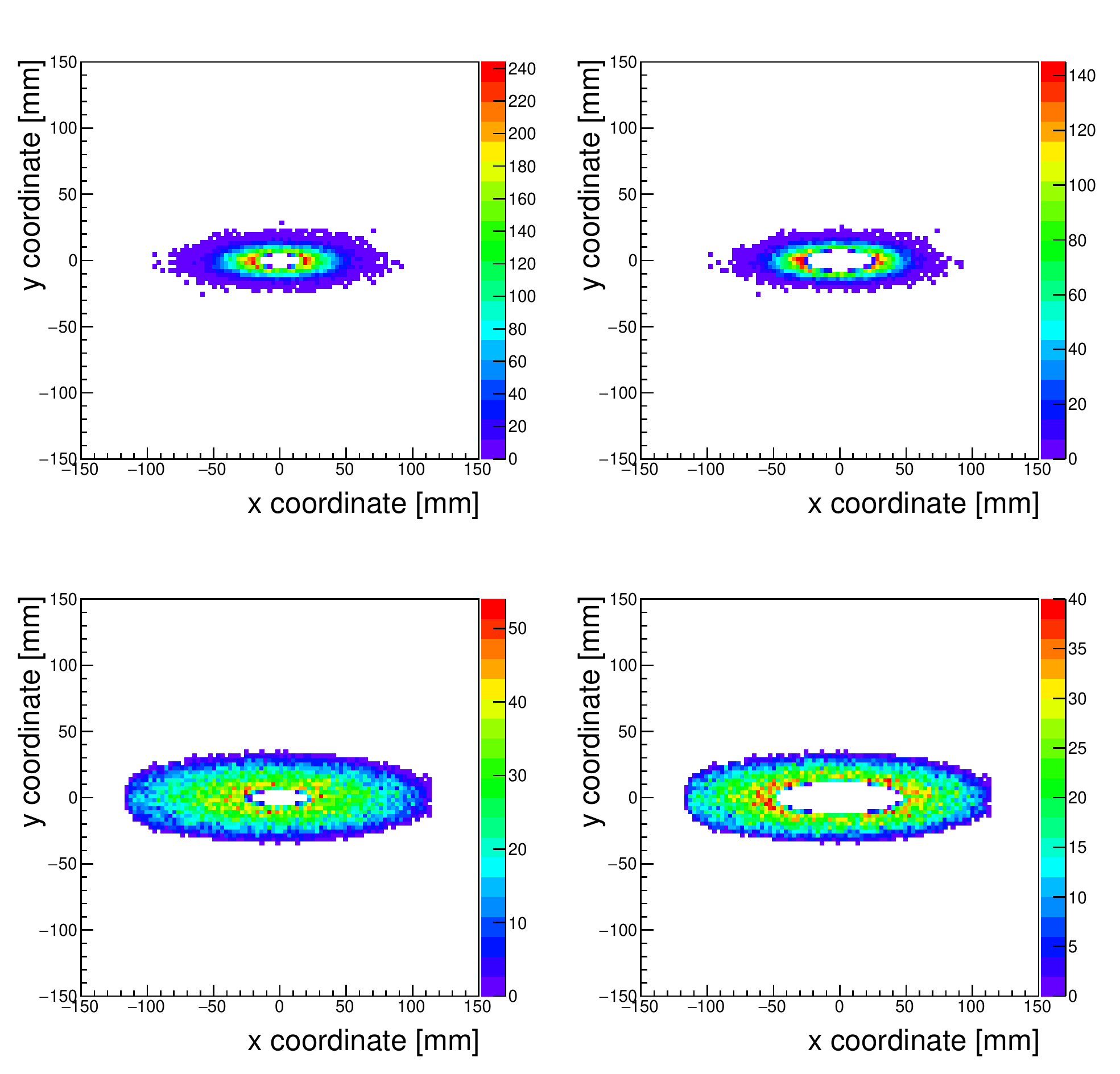}
\caption{Acceptance images for protons incident on the first Roman Pots sensor plane. The top row is for the 18x275 GeV beam energy configuration, with the left plot and right plot being the high acceptance and high divergence optics configurations, respectively. The bottom row is for the 10x100 GeV beam energy configuration. Note the decrease in the size of the $10\sigma$ region (iris in the center of the plots) when we use the high acceptance optics, with left and right plots being the high acceptance and high divergence optics configurations, respectively. As noted previously, the trade off for more acceptance is a drop in luminosity at the IP.}
\label{fig:RPDVCSAccepImage}
\end{figure}

The acceptance is driven by the aperture size (which affects high-$p_{T}$ acceptance) and the beam optics choice, which determines the transverse beam size at the Roman Pots location, and provides the low-$p_{T}$ acceptance cutoff. When using the optics configurations optimized for Roman Pots acceptance, there is a reduction in the overall luminosity up-to approximately a factor of 2. Fig. \ref{fig:RPDVCSAccepImage} shows the impact of the optics choices for the $10\sigma$ safe distance for two different beam energies. 

Another important conclusion to be drawn from these acceptance plots is the need for a large active sensor area to maximize the high-$p_{T}$ acceptance. Fig. \ref{fig:RPDVCSAccepImage} implies the need for sensors to cover an active area of approximately 25cm $\times$ 10cm.

Table \ref{tab:DVCSSmearing} summarizes the smearing contributions from this study. Based on this study and discussions ongoing in the EIC R\&D effort, a 500 $\mu$m x 500 $\mu$m pixel size gives the necessary resolution while still keeping the cost and design constraints reasonable. 

\begin{table}[ht]
\small
    \centering
    \begin{tabular}{|c|c|c|c|c|c|c|}
    \hline
         $\Delta p_{T}$ & Ang. Div. (HD) & Ang. Div. (HA) & Crab Cavity & 250 um & 500 um & 1.3 mm \\
         \hline
          18$\times$275 GeV & 40 & 28 & 20 & 6 & 11 & 26 \\
         \hline
         10$\times$100 GeV & 22 & 11 & 9 & 9 & 11 & 16 \\
         \hline
         5$\times$41 GeV & 14 & - & 10 & 9 & 10 & 12\\
         \hline
    \end{tabular}
    \caption{Summary of smearing contributions from angular divergence, crab cavity rotation, various pixel size choices (for the Roman Pots). HD and HA refer to the ``high divergence" and ``high acceptance" optics configurations, respectively.}
    \label{tab:DVCSSmearing}
\end{table}

\subsubsection{Spectator Tagging in e+D Interactions} \label{sec:deuteronSpecTagSim} 

In diffractive e+D interactions, either the proton or the neutron acts as a spectator, while the other nucleon is active. For this study, only the p+n final state for each spectator case was considered. The major difference here in proton detection is due to the proton having a different magnetic rigidity compared to the deuteron beam, requiring use of the off-momentum detector system for tagging these breakup protons.

\begin{figure}[ht]
\includegraphics[width=.95\textwidth]{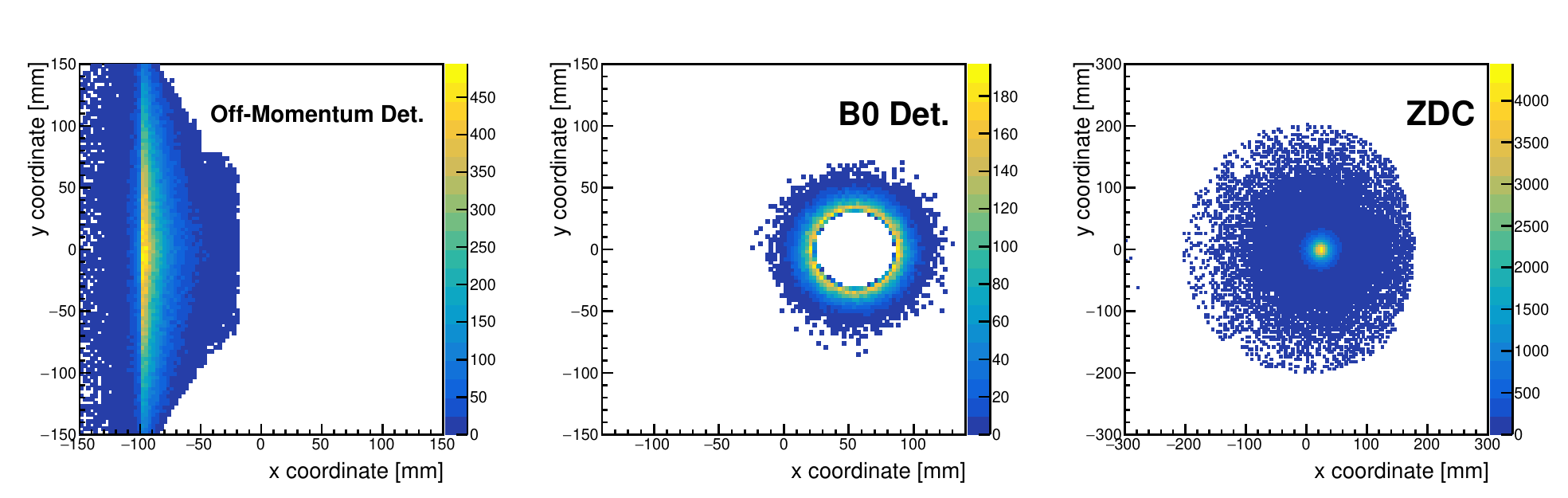}
\caption{Acceptance images for protons and neutrons in the case where the neutron acts as a spectator. In this case, the protons have a larger range of scattering angles, and detection requires both the off-momentum detectors and the B0 detector. The plots show the protons incident on the off-momentum detectors (left), the B0 detector (middle), and the neutrons incident on the ZDC (right). All coordinates are local to the sensor plane. }
\label{fig:specNeutronAccepImage}
\end{figure}

\begin{figure}[ht]
\includegraphics[width=.95\textwidth]{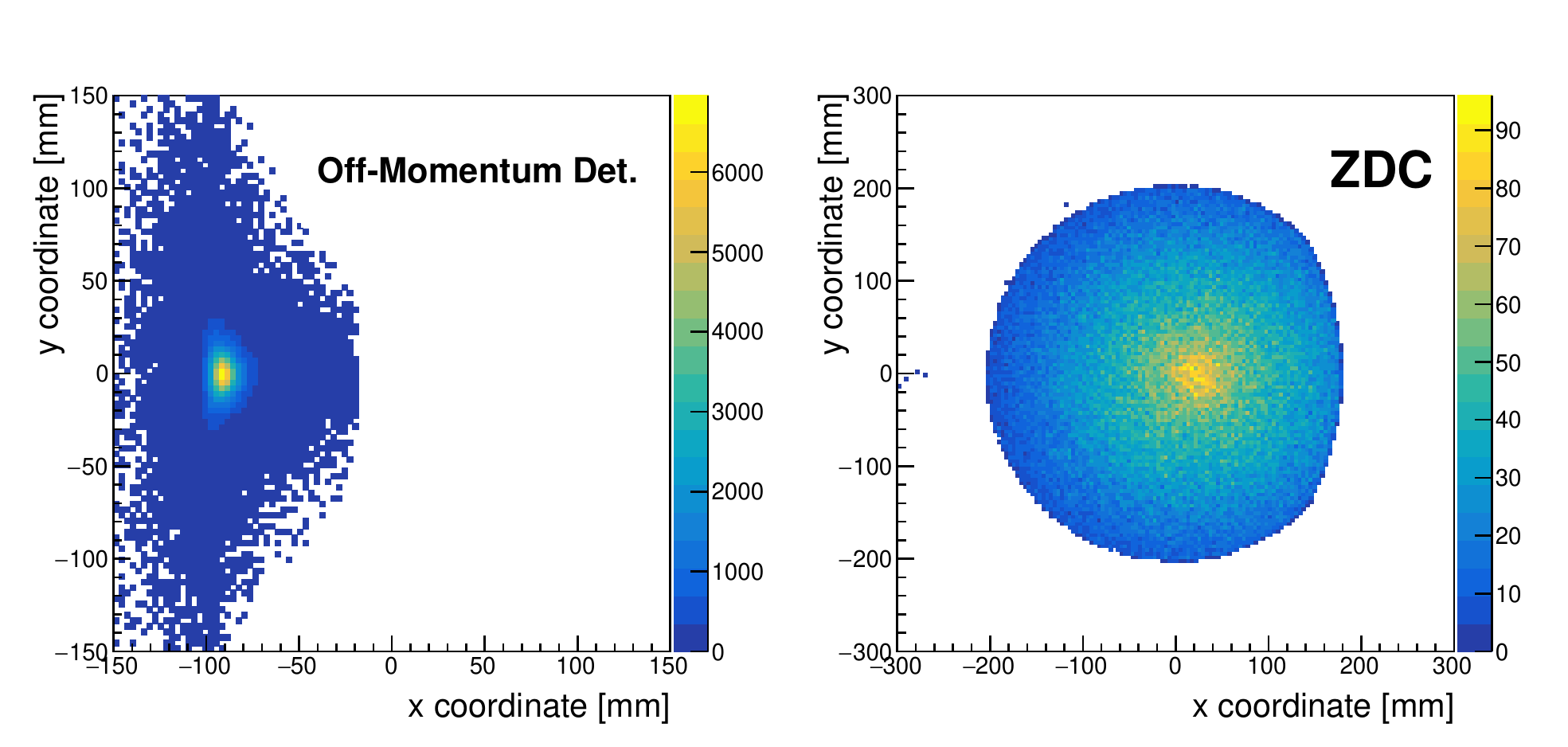}
\caption{Acceptance images for protons and neutrons in the case where the proton acts as a spectator. The plots show the protons incident on the off-momentum detectors (left), and the neutrons incident on the ZDC (right). All coordinates are local to the sensor plane. }
\label{fig:specProtonAccepImage}
\end{figure}

\begin{figure}[ht]
\includegraphics[width=.95\textwidth]{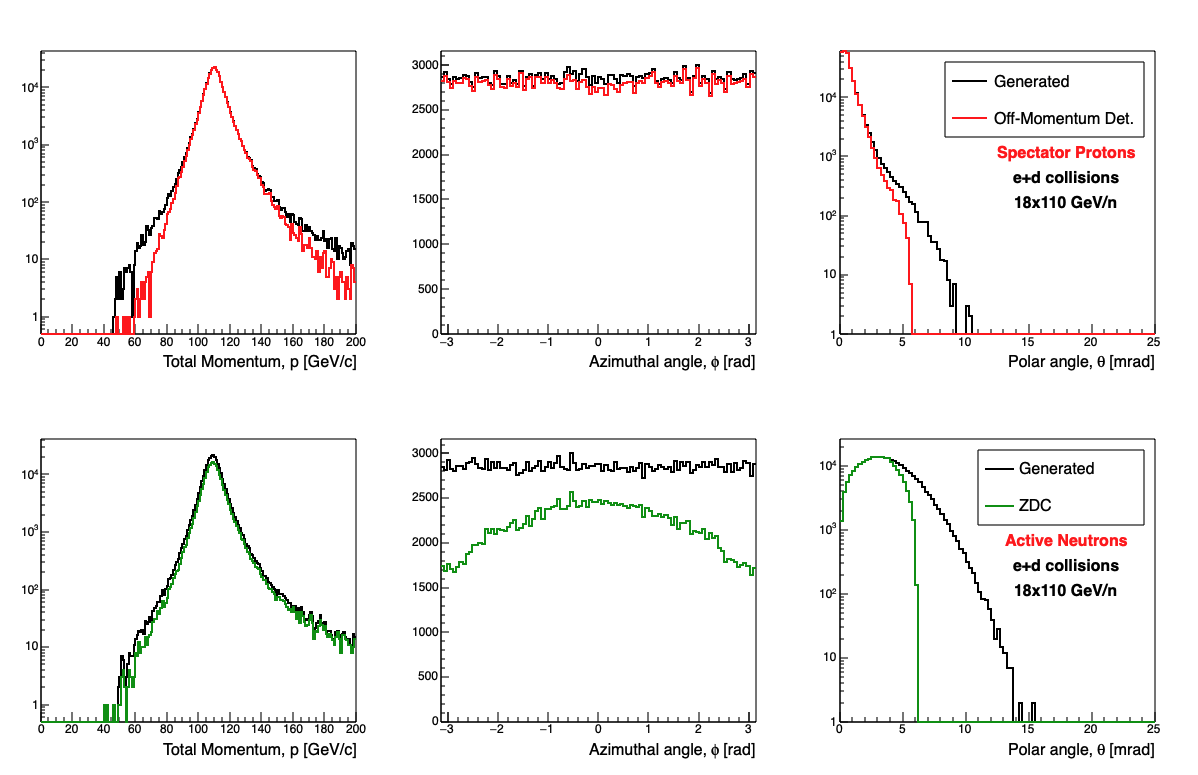}
\caption{3-momentum $p$, azimuthal angle ($\phi$), and polar angle ($\theta$) acceptance for protons (top) and neutrons (bottom) for the proton spectator case. For both rows, the black lines show the BeAGLE MC particles. For the protons, the red and blue lines show the accepted protons in the off-momentum and B0 detectors, respectively. For the neutrons, the green line shows the ZDC acceptance.  }
\label{fig:specProtonAccep}
\end{figure}

\begin{figure}[ht]
\includegraphics[width=.95\textwidth]{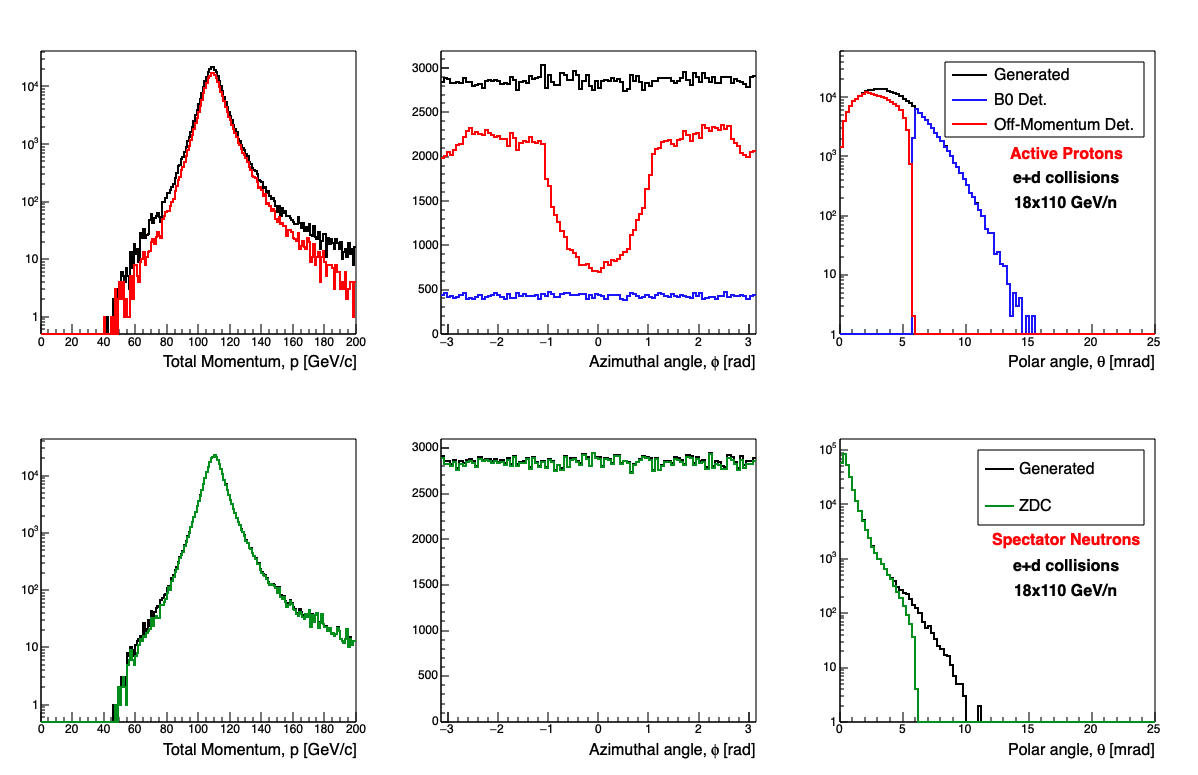}
\caption{3-momentum $p$, azimuthal angle ($\phi$), and polar angle ($\theta$) acceptance for protons (top) and neutrons (bottom) for the neutron spectator case. For both rows, the black lines show the BeAGLE MC particles. For the protons, the red and blue lines show the accepted protons in the off-momentum and B0 detectors, respectively. For the neutrons, the green line shows the ZDC acceptance. }
\label{fig:specNeutronAccep}
\end{figure}

Figures \ref{fig:specNeutronAccepImage} and \ref{fig:specProtonAccepImage} show the kinematic acceptances for the protons and neutrons for the active neutron case, while Figs. \ref{fig:specProtonAccep} and \ref{fig:specNeutronAccep} show the kinematic acceptances for the protons in neutrons for the active proton case.

From these figures it is clear that when a particle acts as a spectator, its acceptance is optimized because its distribution of scattering angles is narrower. For the neutrons, this helps the acceptance because the aperture size limits the neutron acceptance to $<$ 5 mrad. For the protons, larger scattering angles are acceptable in the case of angles $>$ 5 mrad, since many of these enter the acceptance of the B0 spectrometer. However, at smaller scattering angles, the larger spread in transverse momentum imparted to the protons in the neutron spectator cases causes many protons to be lost in the lattice before making it to the off-momentum detectors. 

In addition to the acceptances, the resolutions were studied in detail, and their effect on various physics observables evaluated. These results can be found in~\cite{Tu:2020ymk}.

\subsubsection{Spectator Proton and Neutron Tagging in \texorpdfstring{e+$^{3}\rm{He}$ and e+$^{3}\rm{H}$}{e+3He and e+3H} Collisions} \label{sec:He3SimulationStudy}

\begin{figure}[ht]
\includegraphics[width=.98\textwidth]{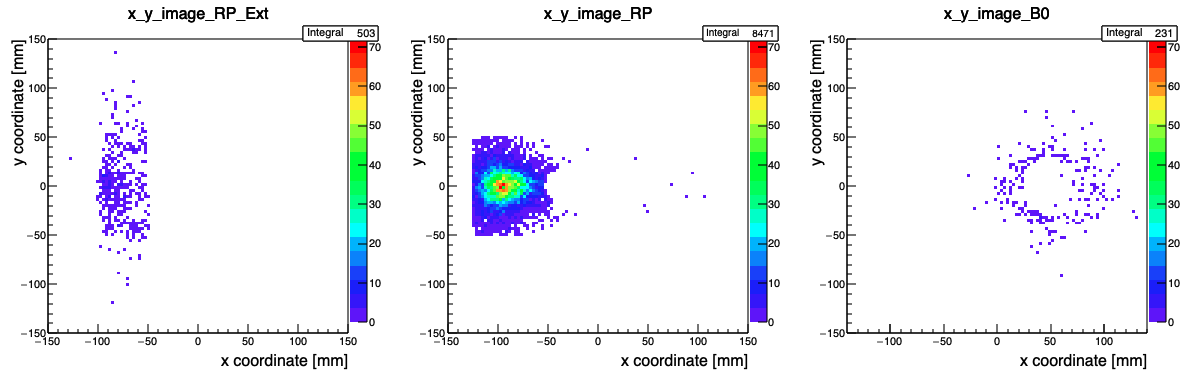}
\includegraphics[width=.98\textwidth]{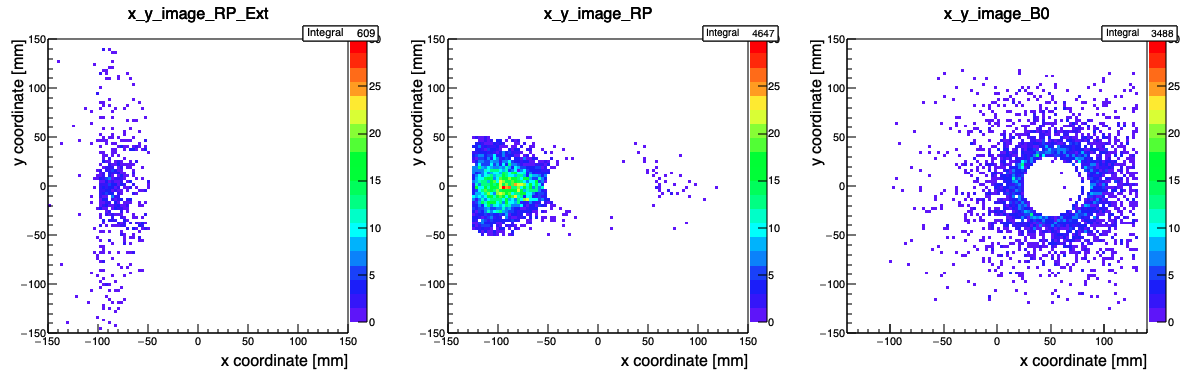}
\caption{Occupancy plots of protons incident on the various FF detectors. The top row is 5x41 GeV BeAGLE DIS events, while the bottom row is 5x41 GeV SRC events. The left column is protons incident on the off-momentum detectors, the middle column is the Roman Pots, and the right column is the sum of the 4 individual planes of the B0 detector used in this simulation. 
All plots show the local coordinate system for the particular detector.}
\label{fig:he3AcceptanceImages}
\end{figure}

Studying short-range correlations (SRC) and the polarized neutron structure can be accomplished by studying e+$^{3}\rm{He}$ (e+$^{3}\rm{H}$) collision events in which the neutron (proton) is the active nucleon in the collision and the protons (neutrons) act as spectators. In order to do this type of study, the prospects of tagging both spectator protons or neutrons in the far-forward region needs to be assessed. A full-simulation study was carried out to this end using e+$^{3}\rm{He}$ DIS events from BeAGLE, as well as SRC events using a spectral function approach. These two paradigms allow for the study of the double-tagging of the final state spectator protons in two very different kinematic regimes. In the DIS case, the two protons end up with very similar final state kinematics, while in the SRC case, one of the protons is in an SRC pair with the active neutron and therefore has a very different initial $p_{T}$ distribution than the other spectator proton. Fig. \ref{fig:he3AcceptanceImages} shows the occupancy of protons incident on the various detector subsystems. These plots show the repeated need for multiple subsystems to cover the acceptance, as well as the need for a large active area for the Roman Pots subsystem. Fig. \ref{fig:he3AcceptanceImages} only shows the lowest beam energy configuration since it is the most demanding on the acceptance. 

\begin{figure}[ht]
\includegraphics[width=.98\textwidth]{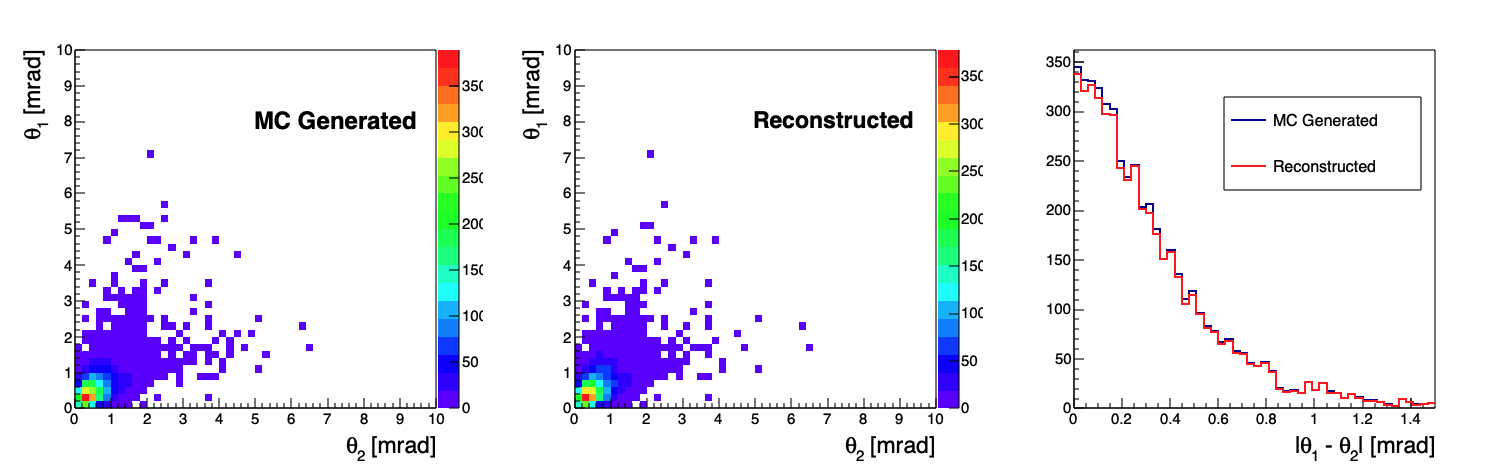}
\includegraphics[width=.98\textwidth]{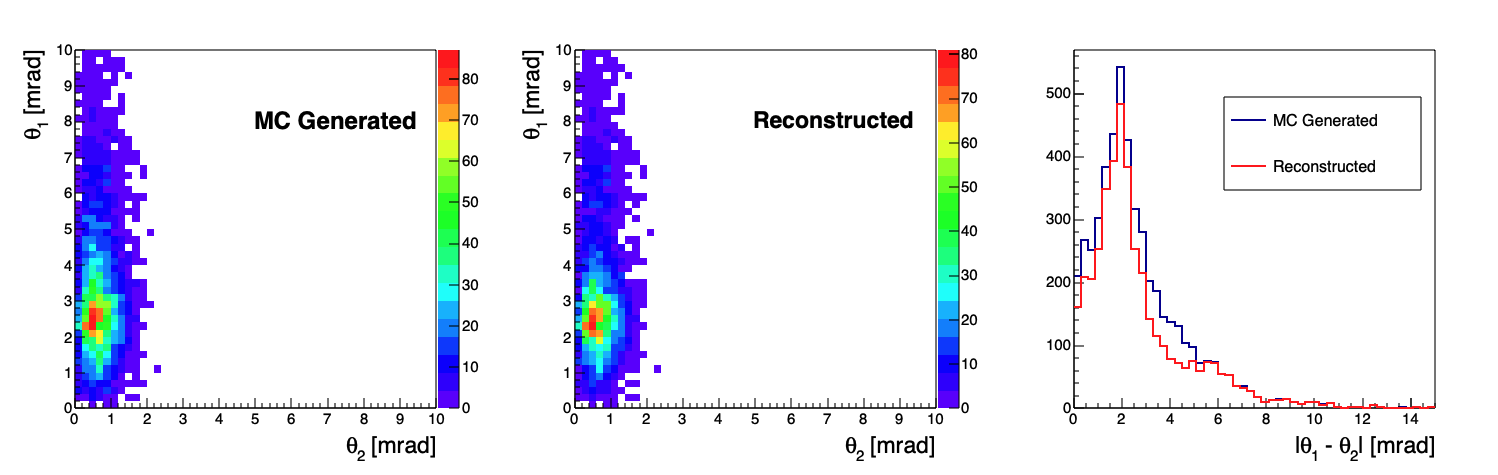}
\caption{Scattering angle plots for spectator protons from e+$^{3}\rm{He}$ collisions using BeAGLE DIS events at 10x110 GeV (top row) and SRC events at 18x110 GeV (bottom row). The left panel in both rows shows the scattering angle of proton one vs. proton two from the MC generator, the middle plots shows what is reconstructed in the EicRoot GEANT simulation, and the right panel shows the absolute value of the difference between the angles, which tells us how close together they are when they arrive at the detector.}
\label{fig:he3Acceptance110GeV}
\end{figure}

\begin{figure}[ht]
\includegraphics[width=.98\textwidth]{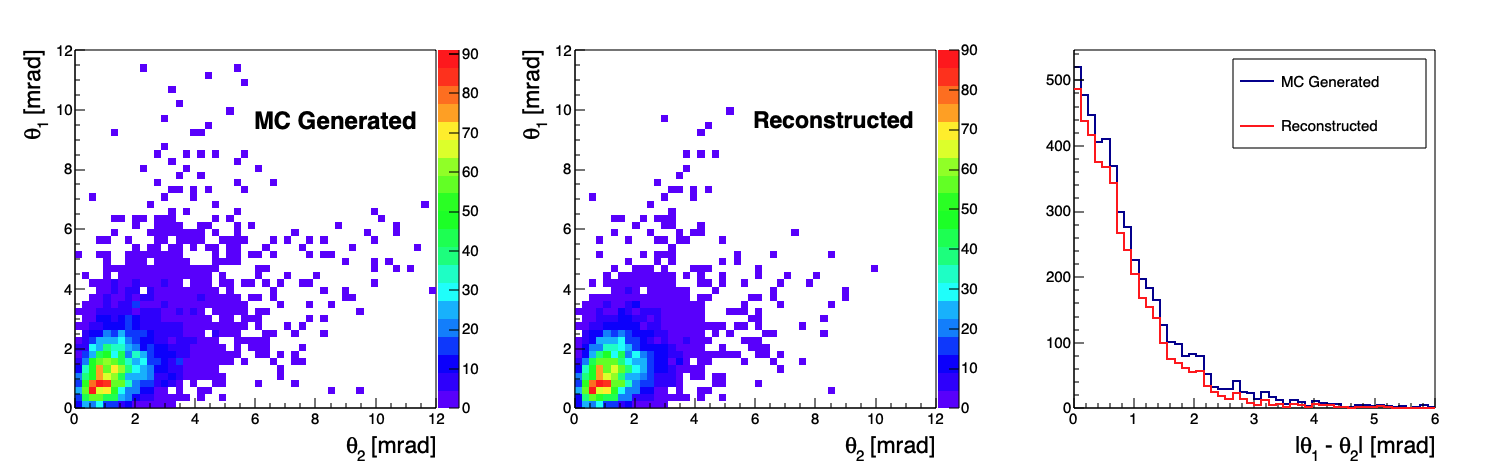}
\includegraphics[width=.98\textwidth]{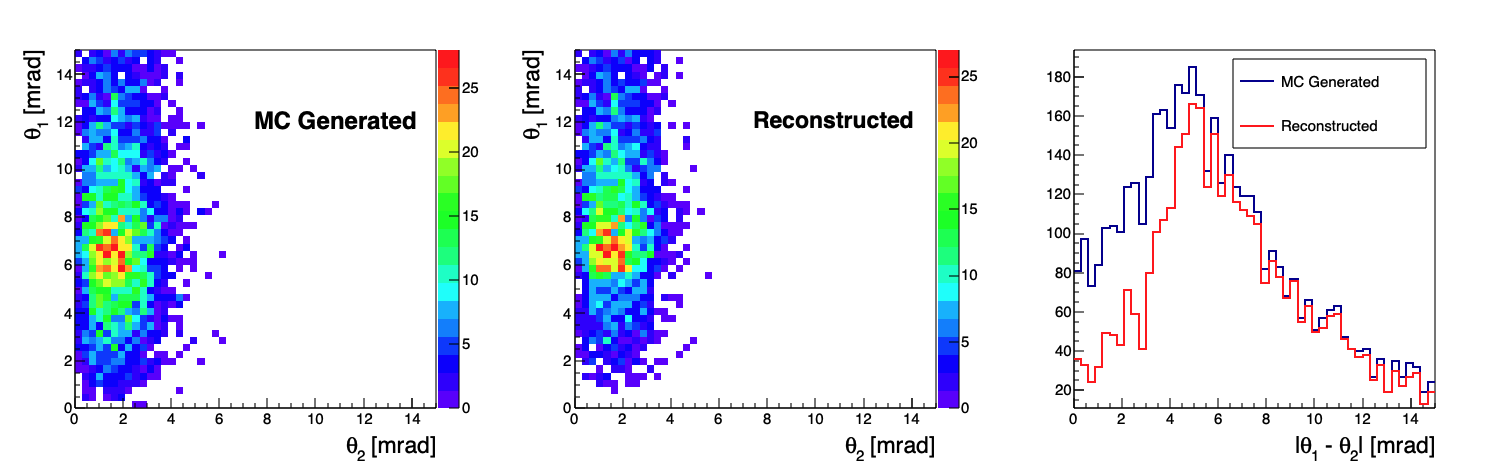}
\caption{Scattering angle plots for spectator protons from e+$^{3}\rm{He}$ collisions at 5x41 GeV using BeAGLE DIS events  (top row) and SRC events (bottom row). The left panel in both rows shows the scattering angle of proton one vs. proton two from the MC generator, the middle plots shows what is reconstructed in the EicRoot GEANT simulation, and the right panel shows the absolute value of the difference between the angles, which tells us how close together they are when they arrive at the detector.}
\label{fig:he3Acceptance41GeV}
\end{figure}

Figs. \ref{fig:he3Acceptance110GeV} and \ref{fig:he3Acceptance41GeV} show the results of the study for two different energy configurations. The results indicate that the double-tagging efficiency for the spectator protons look very promising for the baseline interaction region, with most cases having a double-tagging efficiency above 85\% (above 90\% for the higher energy configuration), except for the lower energy SRC case which has an efficiency above 75\%. Most of the losses in the double-tagging efficiency comes from a single proton being lost between the B0 detector and Roman Pots, or between the off-momentum detectors and Roman Pots. These acceptance gaps are to some point unavoidable due to the finite thickness of the beam pipe, which is the main driver of that gap between the detectors. 

A study of the neutron double-tagging efficiency in e+$^{3}\rm{H}$ events was also carried out using fast simulations in eic-smear. The results indicate that the neutron double-tagging efficiency is also quite good, with most of the acceptance losses being in the SRC case when one of the neutrons has a larger scattering angle that may cause it to be lost in the 4.5 mrad aperture.

\subsubsection {Far-forward tagging ions}

At the time of writing this document, no MC samples for light-nuclei tagging in the FF direction were available for validation in our simulation framework. However, based on the numerous other studies, some basic conclusions can be drawn. Light-ion tagging (e.g. $^{4}\rm{He}$) should have similar constraints as those seen for the tagging of protons in the FF direction (e.g. proton DVCS). The machine optics can be tuned similarly to maximize the low-$p_{T}$ acceptance at the Roman Pots. From this, the main limitation will be the shape of the $p_{T}$ distribution given by the coherent light nuclear scattering process. If the $p_{T}$ distributions are similar as for the e+p case, than the acceptance of these light nuclei at the Roman Pots will also be similar. More studies should be carried out in the future to asses the impact of the various choices of machine optics on the FF light nuclei acceptance.

\subsubsection{Meson Structure and FF \texorpdfstring{$\Lambda$}{Lambda} Decay} \label{sec:lambdaDecaySimulations}

The reconstruction of $\Lambda$s in the target fragmentation regime is one of the most challenging tasks in the FF region of the IR. It comes from the fact that the decay vertex of such $\Lambda$s is spread by tens of meters along the Z-axis (along the beam-line) which makes detection of the decay products and mass reconstruction very difficult. 

\begin{figure*}[hb]
\begin{tabular}{lll}
\parbox[c]{0.37\textwidth}{\includegraphics[width=0.38\textwidth]{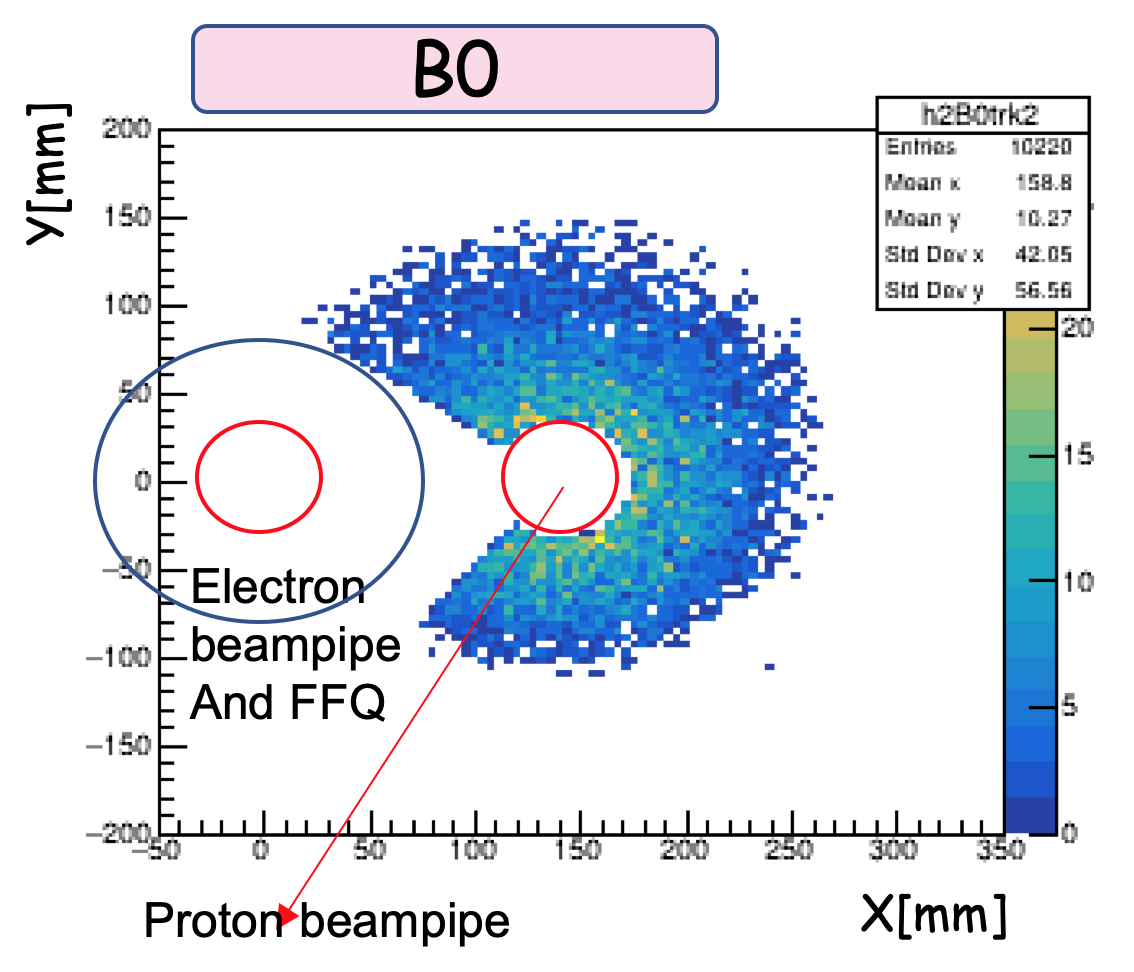}}
&
\parbox[c]{0.6\textwidth}{\includegraphics[width=0.6\textwidth,height=0.33\textwidth]{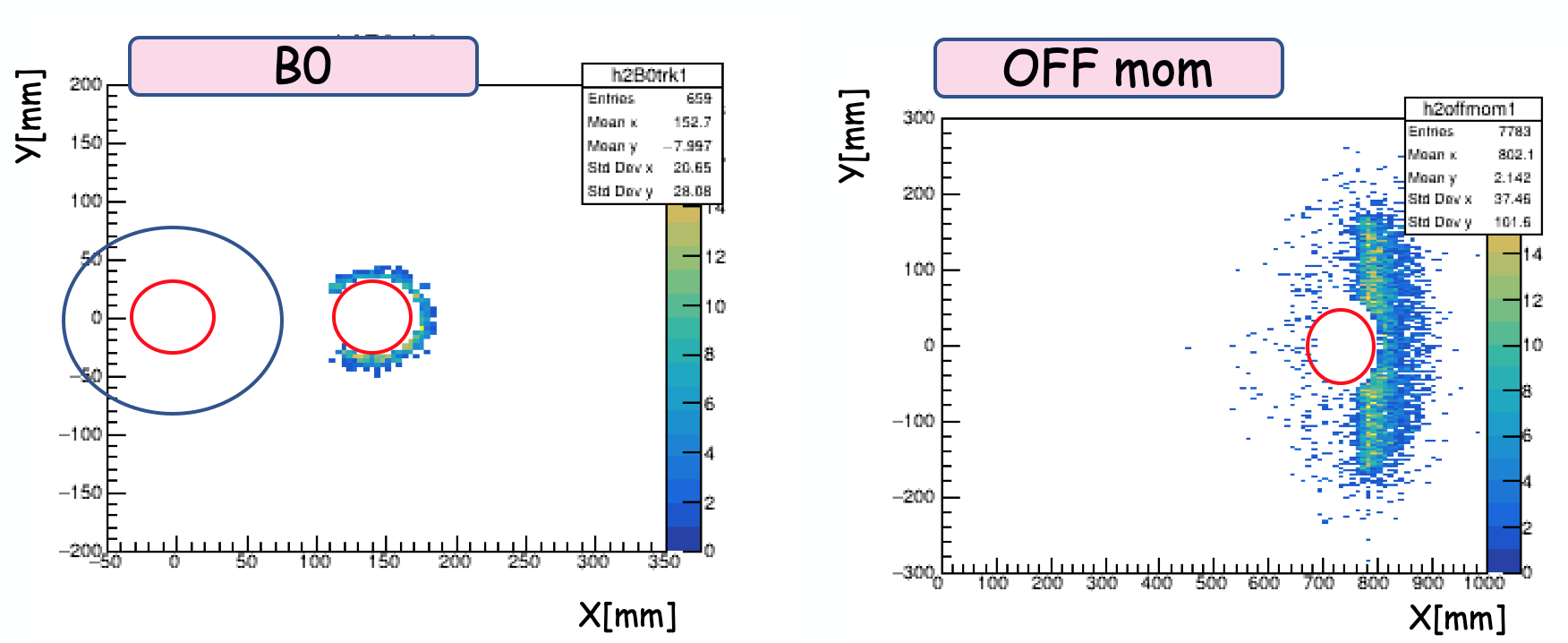}}
\\[-.2ex]
(a) & (b) 
\end{tabular}
\caption[]{Occupancy plots for energy setting 5x41 GeV  (a) for $\pi ^-$ in B0 tracker  (b) for protons in B0 and Off-Momentum detectors.The red circle shows the beampipe position and the blue circle shows electron FFQ aperture inside B0 dipole. 
}
\label{fig:Lambda-B0_p_41}
\end{figure*}

Occupancy plots for the beam energy setting of 5$\times$41 GeV for pions and protons from $\Lambda$ decays is shown in Figure~\ref{fig:Lambda-B0_p_41}. Since this is the lowest beam energy setting, most of the $\Lambda$s would decay in the first meter (before the B0 magnet), and the decay products of $\Lambda$ are expected to have low momenta and larger angle.  Therefore, as expected, protons coming from the $\Lambda$ decays will mostly be detected, due to their lower rigidity, in the off-momentum detectors and partially in a B0 tracker, while  the B0 tracker  will be the only detecting element for pions. As one can also see from this Figure, the proton-beam-pipe aperture inside the B0-dipole plays an important role and sets the detection efficiency for pions. Also a full azimuthal angle $\phi$-coverage of the detecting elements around the proton beam-pipe is important: outer radius of electron FFQ needs to be minimized  to provide enough space for tracking detectors.

\begin{figure}[!ht]
  \centering
  \includegraphics[width=0.95\textwidth]{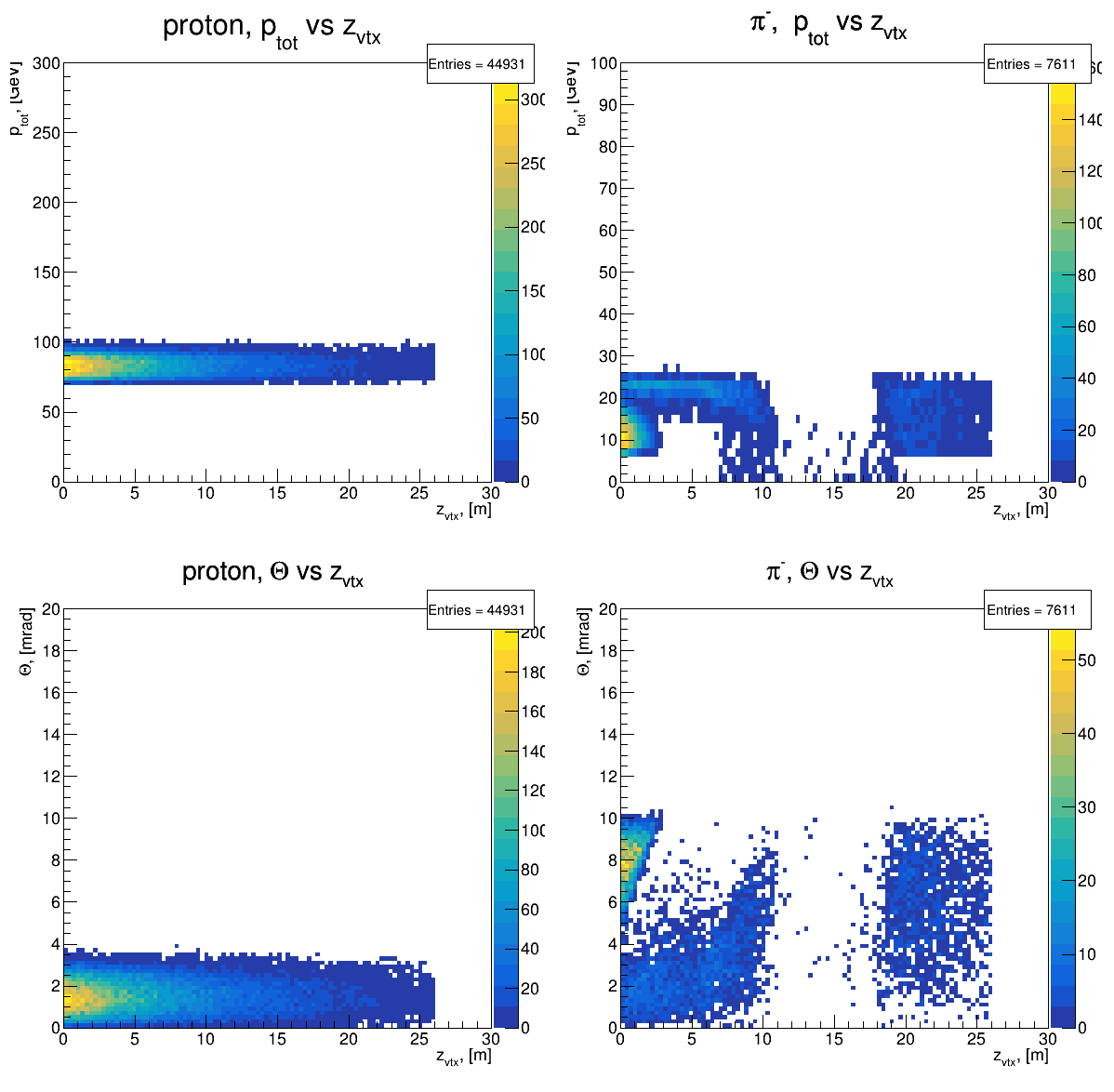}  \caption{ Beam energy 10x100 GeV. Momentum and Theta distributions for Lambda decay particles, protons (left) and  $\pi^-$(right), registered in far-forward detectors  vs their origination (decay vertex).  }
  \label{fig:L-p_Zvtx_41}
\end{figure}

For another beam energy setting, for example 10~GeV$\times$100~GeV (Fig.~\ref{fig:L-p_Zvtx_41}), one could clearly see, for charged pions, the ``dead" area along the beamline, where the beam elements (quadrupoles)  are located.  This comes from the fact that those pions have significantly lower momentum than the beam, and very small $x_{L}$, causing the pions to be lost in the lattice before they can be detected. 
It is also important to point out that negative charged particles ( pions) will bend into opposite direction, compared to protons, as shown on the Fig.~\ref{fig:OFFM-drawings}, therefore a proper coverage of off-momentum detectors would be required  to to provide an efficient detection for those particles.

\subsection{Conclusions}

The far-forward region of the EIC baseline IR has been studied extensively throughout this entire Yellow Report process. The main conclusions from these studies are that several detector subsystems are needed to cover the entire far-forward region including Roman Pots, a high resolution zero-degree calorimeter, a silicon-based spectrometer in the first dipole magnet after the IP, and various planes of silicon on either side of the beam pipe after the B1apf dipole to capture charged particles with $x_{L} < 0.6$, so-called ``off-momentum particles". The technology choices detailed in this chapter reflect the R\&D efforts of numerous people and represent our recommendations to meet the needs of the FF physics programs at the EIC. As can be seen throughout the document, the IR design has undergone some revisions (especially the B0 magnet) that have led to different considerations for the detector geometry, as seen in the difference in the B0 coverage between the e+D study and the $\Lambda$ study. These details are not yet final, and the different assumptions should make it clear what kinds of design difficulties could be faced in the B0 detector planning. As the IR design progresses and the community moves toward the formation of an experiment collaboration, more detailed simulations will need to be carried out in addition to what has been provided by these studies. Nevertheless we hope these studies provide a strong foundation for validation of the future detector simulation and design efforts. All details related to IR design, optics parameters, etc. used in these studies can be found in the pre-Conceptual Design Report for the EIC \cite{pCDR}.

%

%

%
\section{Far-Backward Detectors}

\begin{figure*}[ht]
  \centering
  \includegraphics[width=0.8\textwidth]{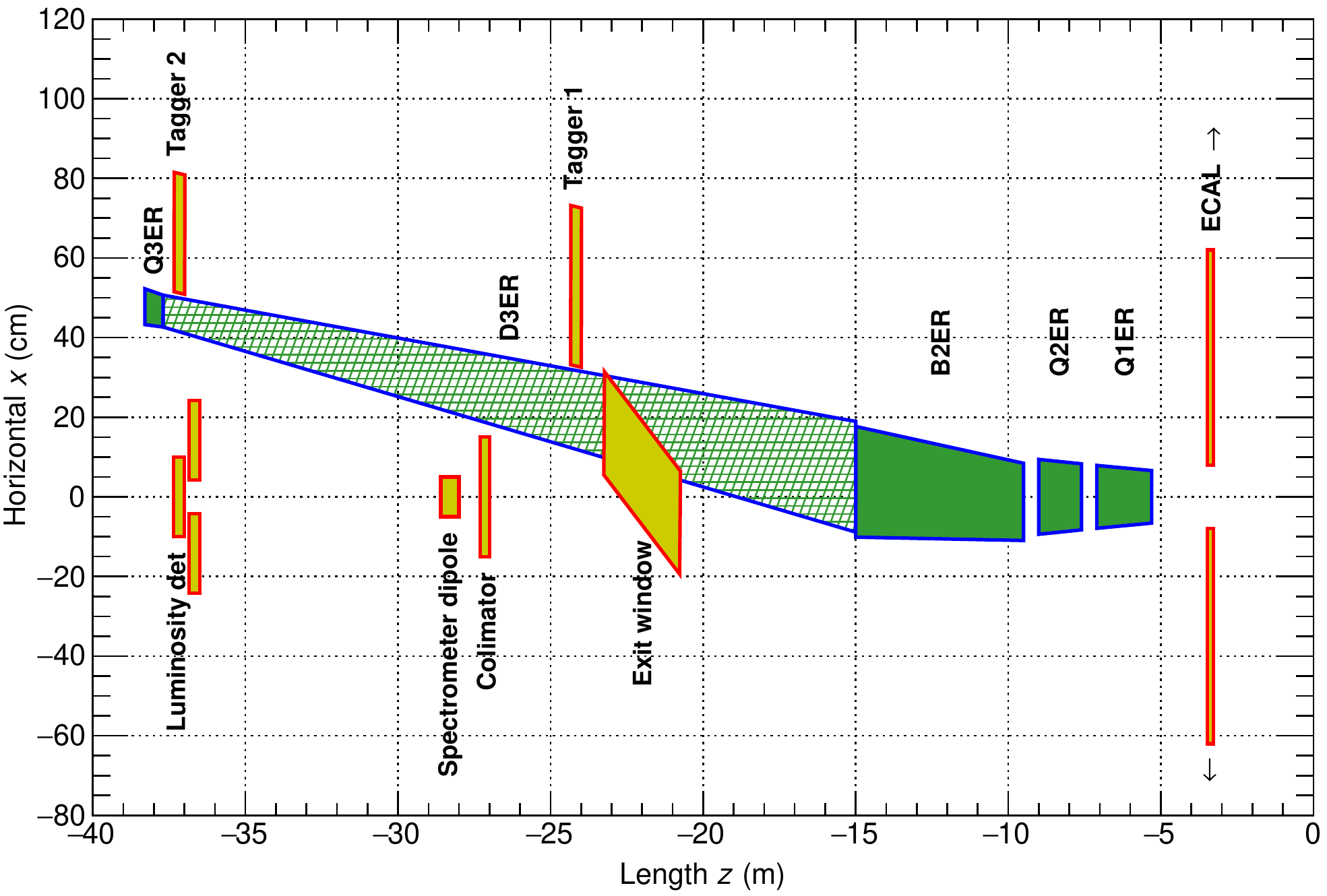}
  \caption{The region downstream of the interaction point in the
    electron direction.}
  \label{fig:rear_layout}
\end{figure*}

The path of the electron beam downstream of the interaction point is
shown in Fig.~\ref{fig:rear_layout}.
Beam magnets are shown in full green, drift space in hatched green and detectors and components in red and yellow.
The horizontal axis is aligned with the direction of the beam at the
collision point, along which photons from \ep\ and \eA\ interactions
will travel.
These photons come predominantly from the bremsstrahlung process
used for luminosity determination.
The lower left of the figure shows possible instrumentation for the
luminosity measurement.
Bremsstrahlung and low-$Q^2$ processes also produce electrons with momenta slightly below
the  beam energy.
After being bent out of the beam by lattice dipoles they may be
measured by taggers as shown in the top left of the figure.
This section will detail studies of a luminosity monitor as well a tagger for electrons from bremsstrahlung and low-$Q^2$ events. The technology considerations and machine-driven acceptances are both addressed as well.

\subsection{Far-Backward Photons} \label{sec:FB_photons_main}

\subsubsection{Luminosity Measurement and Bremsstrahlung Photons}
The luminosity measurement provides the required normalization for
all physics studies. At the broadest scale it determines absolute cross sections,
such as needed for the structure function $F_2$ and derived PDFs.
On an intermediate scale, it is also required to combine different running periods,
such as runs with different beam energies needed to measure $F_L$, or runs with
different beam species to study $A$ dependencies.
Asymmetry measurements are conducted using beams with bunches of both
spin states. On the finest scale,
the relative luminosity of the different bunch crossings is needed to
normalize the event rates for the different states; the uncertainty on
the relative bunch luminosity is a limiting factor for asymmetry
measurements.

The bremsstrahlung process $e+p \longrightarrow e+p+\gamma$
was used successfully for the measurement
of luminosity by the HERA collider experiments~\cite{Amaldi:1979qp,Adamczyk:2013ewk,Andruszkow:2001jy,Andruszkow:1992rz,Helbich:2005qf,Frisson:2011zz,Aaron:2012kn}.
It has a precisely known QED cross-section~\cite{Haas:2010bq} which is large, minimizing
theoretical uncertainty and providing negligible statistical uncertainty.
Thus the scale uncertainty of the luminosity is determined by the
systematic uncertainties of the counting of bremsstrahlung events.
The ZEUS collaboration at HERA-II measured luminosity with a 1.7\% scale
uncertainty. Experiences there indicate directions for improvement, such as improved understanding of the photon acceptance. The electron taggers discussed in Section~\ref{sec:FB_electrons_main} will provide a direct measurement of this acceptance.
Such improvements at the EIC should be able to reduce
the scale uncertainty to $\sim$1\%.

In contrast to HERA, where only the electron beam was polarized, both
the electron and proton/light ion beams will be polarized in the EIC.
In this case the bremsstrahlung rate is sensitive to the polarization dependent
term $a$ in the cross section $\sigma_{\rm brems}= \sigma_0(1+ aP_eP_h)$.
Thus, the polarizations $P_e, P_h$ and luminosity  measurements are
coupled, and the precision of the luminosity measurement is limited by
the precision of the polarization measurement.
This is especially important for relative luminosities for asymmetry
measurements, where the bremsstrahlung process used for normalization
has different cross sections for different spin states.
The precision needed for the relative luminosity measurement is driven
by the magnitude of the physics asymmetries which can be as low as
$10^{-4}$; the uncertainty on relative bunch luminosities must exceed
this level of precision.

\begin{figure*}[ht]
  \centering
  \begin{tabular}{cc}
    \includegraphics[width=0.5\textwidth]{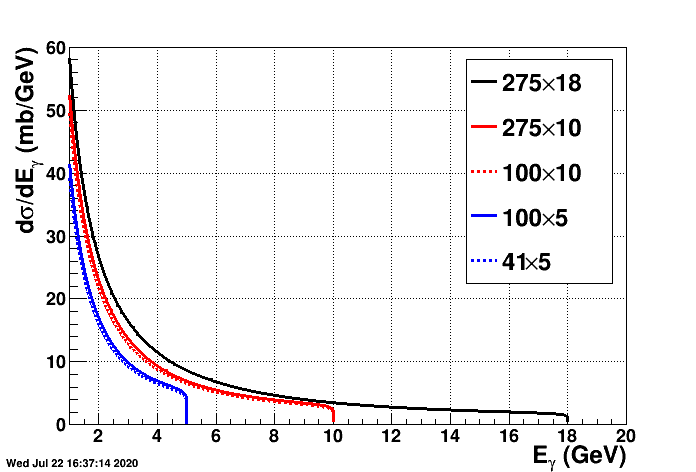}
    &
    \includegraphics[width=0.5\textwidth]{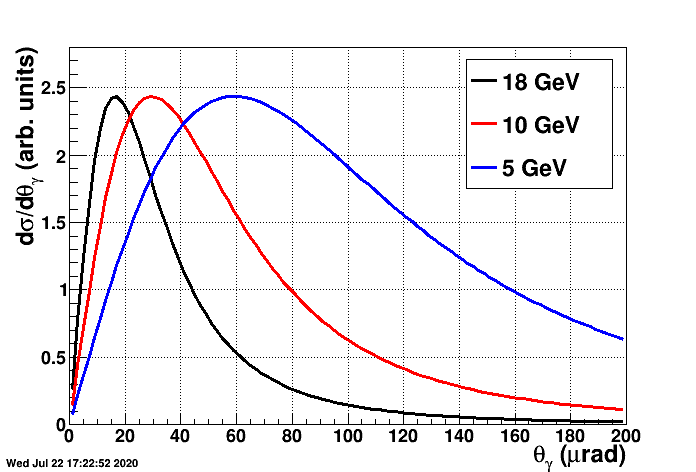}
  \end{tabular}
  \caption{Bremsstrahlung photon energy (left) and angular (right)
    distributions for EIC beam energies.}
  \label{fig:brems_Egthg}
\end{figure*}

\begin{figure}[ht]
  \centering
    \includegraphics[width=0.5\textwidth]{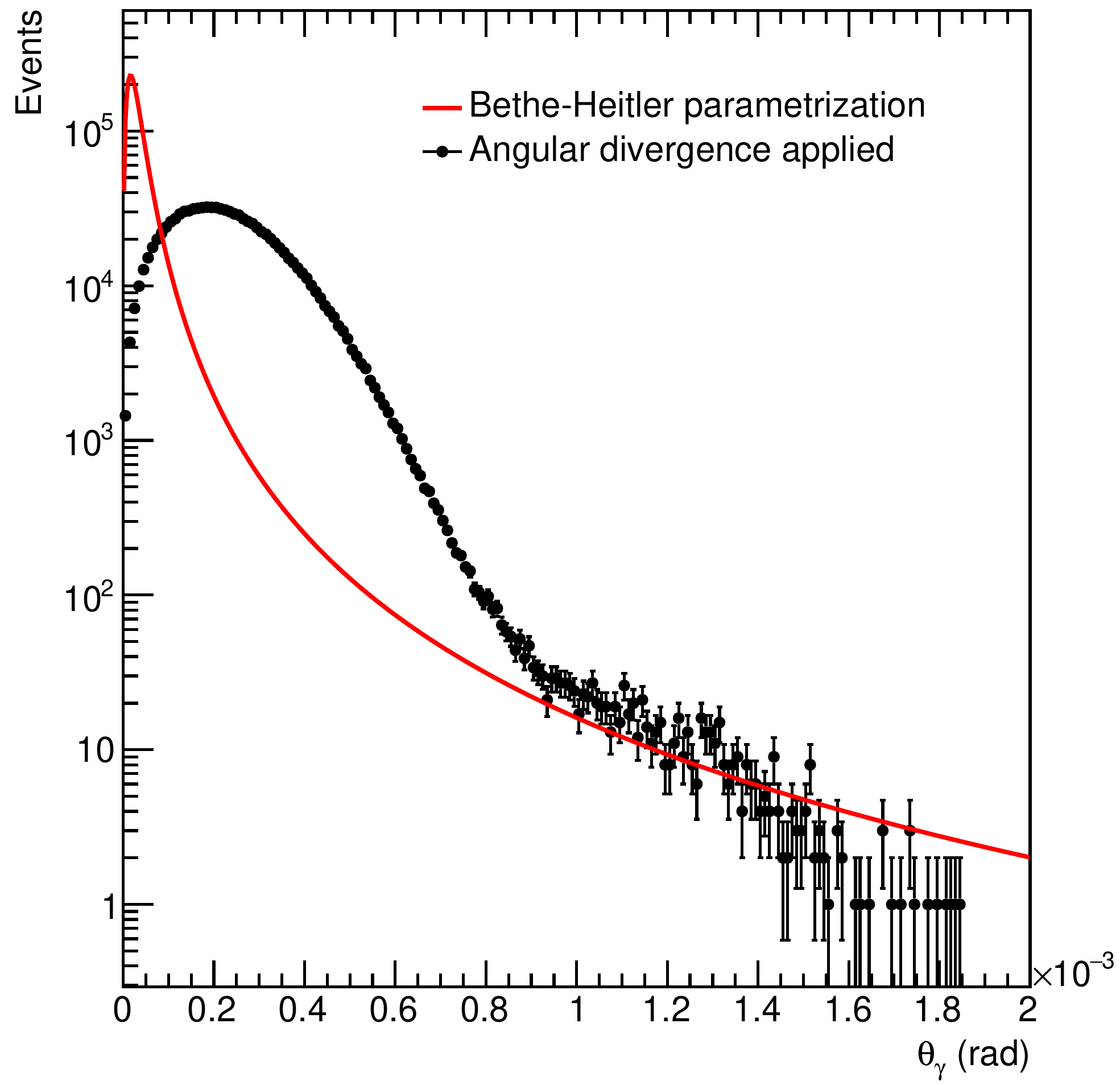}
    \caption{Angular dependence of bremsstrahlung cross section. The effect of beam
    angular divergence is shown.}
    \label{fig:lumi-sigma-smeartheta}
\end{figure}

The bremsstrahlung photon energy $E_{\gamma}$ distributions for EIC beam energies are shown in left of Fig.~\ref{fig:brems_Egthg}.
They diverge as $E_{\gamma}\rightarrow0$ and have sharp cutoffs at the
electron beam energies.
As shown in the right of Fig.~\ref{fig:brems_Egthg},
the bremsstrahlung photons are strongly peaked in the forward
direction with typical values of
 $\theta_{\gamma} \approx m_e/E_e$,
with values of 20-60 $\mu$rad at the EIC.
The RMS angular divergence of the electron beam is significantly
larger than these values and will dominate the angular distribution of
bremsstrahlung photons
as shown in Fig.~\ref{fig:lumi-sigma-smeartheta}.

\begin{figure}[ht]
  \centering
  \includegraphics[width=\textwidth]{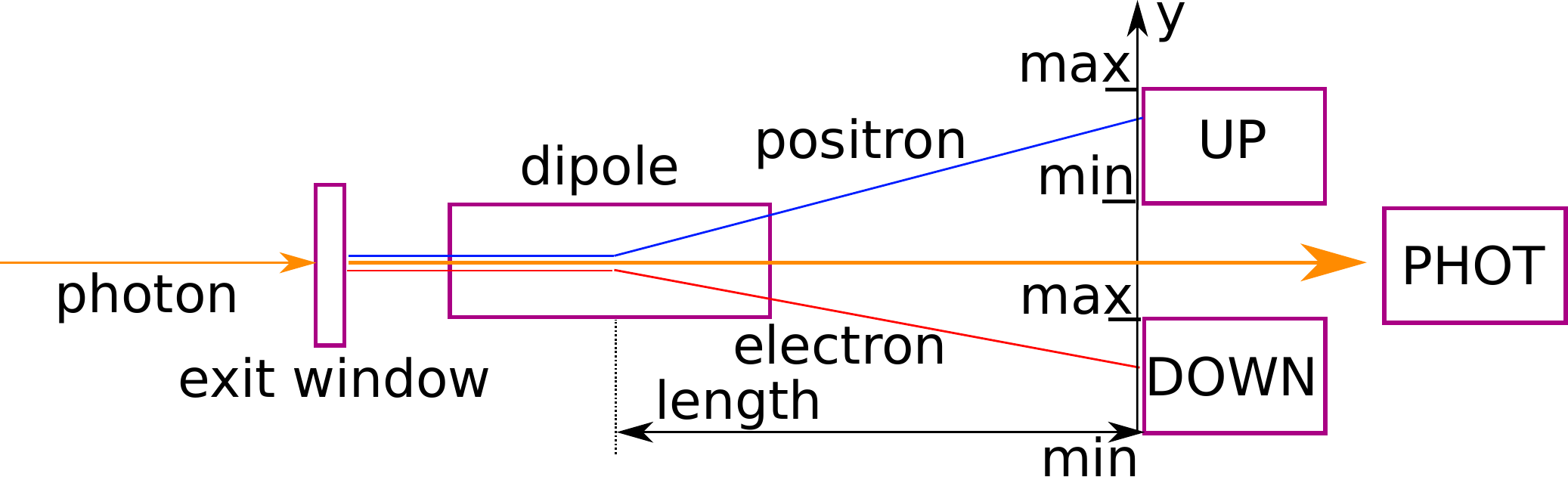}
  \caption{Principle of the luminosity measurement. Bremsstrahlung photons are incident on an aluminum exit window.
  Converted electron-positron pairs are split in the spectrometer dipole magnet and detected in the UP and DOWN
  detectors. Non-converted photons reach the photon calorimeter PHOT.}
  \label{fig:lumi-zeus}
\end{figure}

\subsubsection{Bremsstrahlung Photon Detection Principles and Requirements}

Figure~\ref{fig:lumi-zeus} shows a side view of detector components along the photon zero-degree line in the backward direction.
The straightforward method for measuring bremsstrahlung situates a
calorimeter at zero degrees in the electron direction, PHOT in the figure,
and counting the resulting photons.
The calorimeter is also exposed to the direct synchrotron radiation
fan and must be shielded, thus degrading the energy resolution.
This also imposes a rough low energy cutoff on photons typically
\mbox{$\approx$ 0.1-1 GeV} below which the calorimeter is
insensitive.
At peak HERA luminosities, the photon calorimeters were sensitive to 1-2
photons per HERA bunch crossing.
At an EIC luminosity of $10^{33}\,{\rm cm}^{-2}\,{\rm s}^{-1}$, the
mean number of such photons per bunch crossing is over 20 for
electron-proton scattering and increases with $Z^2$ of the target for
nuclear beams.
The per bunch energy distributions are broad, with a mean proportional
to the number of photons per bunch crossing. The counting of
bremsstrahlung  photons thus is effectively an energy measurement
in the photon calorimeter with all of the related systematic
uncertainties  (e.g. gain stability) of such a measurement.

An alternative method to directly counting bremsstrahlung photons,
used effectively by the ZEUS collaboration at HERA~\cite{Andruszkow:2001jy}, employs a pair
spectrometer.
A small fraction of photons is converted into
$e^+e^-$ pairs in the vacuum chamber exit window. A dipole magnet
splits the pairs vertically and each particle
hits a separate calorimeter adjacent to the unconverted photon path,
UP and DOWN in
Fig.~\ref{fig:lumi-zeus}.
This has several  advantages over a zero-degree photon calorimeter:
\begin{itemize}
\item The calorimeters are outside of the primary synchrotron radiation fan.
\item The exit window conversion fraction reduces the overall rate.
\item The spectrometer geometry imposes a  well defined low energy cutoff in the photon spectrum, which depends on
the magnitude of the dipole field and the location of the calorimeters.
\end{itemize}
The variable parameters of the last two points (conversion fraction, dipole field and calorimeter
locations) may be chosen to reduce the rate to less than or of order one $e^+e^-$ coincidence per bunch
crossing even at nominal EIC luminosities. Thus, counting of bremsstrahlung photons is simply
counting of $e^+e^-$ coincidences in a pair spectrometer with only
small corrections for pileup effects. 

The locations of a zero-degree calorimeter and pair spectrometer are
shown in the bottom left of Fig.~\ref{fig:rear_layout}.
Careful integration into the machine lattice is required, not only to
allow for enough space for the detectors,
but also to accommodate the angular distribution of the photons.
This is dominated by the angular divergence of the electron beam,
with RMS values as high 0.2 mrad.
Thus a clear aperture up to a few mrad is required to measure the
angular distribution and minimize the acceptance correction.
The spectrometer rate is directly proportional to the fraction of
photons which convert into $e^+e^-$ pairs, placing stringent
requirements on the photon exit window.
It must have a precisely known material composition, and a precisely
measured and uniform thickness along the photon direction.

Calorimeters are required for both luminosity devices, for triggering
and energy measurements. The high rates dictate a radiation hard
design, especially for the zero-degree calorimeter, which must also
have shielding against synchrotron radiation.
The spectrometer must also have precise position detectors to measure
the $e^{\pm}$.
Combined with the calorimeter energy measurement this allows
reconstruction of the converted photon positions.
The distribution of photon positions is required to correct for
the lost photons falling outside the photon aperture and detector
acceptances.

\subsubsection{Bremsstrahlung Photon Detector Implementation}

\begin{figure}[!ht]
  \centering
  \includegraphics[width=0.9\textwidth]
  {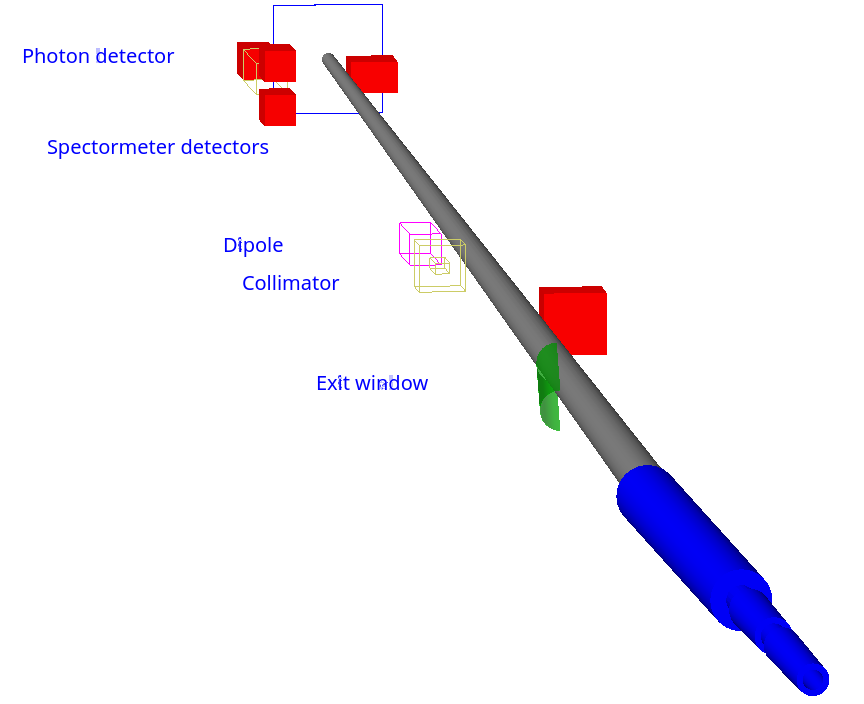}
  \caption{Geant4 model of luminosity detectors.}
  \label{fig:lumi-geant}
\end{figure}

The luminosity system is shown in plan view in the bottom left of Fig.~\ref{fig:rear_layout}.
A Geant4 model of all essential components for the luminosity measurement is shown in Fig.~\ref{fig:lumi-geant}.
A photon exit window is placed at $z$ = -20.75\,m. It is tilted by 100\,mrad relative to the axis of the electron beam
(and of the photons), to achieve an acceptable heat load from synchrotron radiation. A collimator
at $z$\,=\,-27\,m will prevent synchrotron radiation at larger angles from entering the luminosity system.
A dipole spectrometer magnet at $z$\,=\,-28\,m will split converted electron-positron pairs into the spectrometer
detectors. A direct photon detector is placed at $z$\,=\,-37.8\,m, after a graphite filter of 5\,$X_0$ length.
The pair spectrometer detectors are at $z$\,=\,-36.5\,m.
The spectrometer detectors are displaced vertically so their nearest edges are 42\,mm above and below $y=0$.
The detectors are implemented in the model as boxes which register hits by all incoming particles.

The spectrometer acceptance as a function of bremsstrahlung photon energy depends on the layout as shown in Fig.~\ref{fig:lumi-zeus}. It depends on the distance \textit{length} from the dipole magnet
to the spectrometer detectors, the magnetic field of the dipole and the detector positions \textit{min} and \textit{max}
of the UP and DOWN detectors along the vertical \textit{y} axis.

\begin{figure}[!ht]
  \centering
  \includegraphics[width=0.6\textwidth]{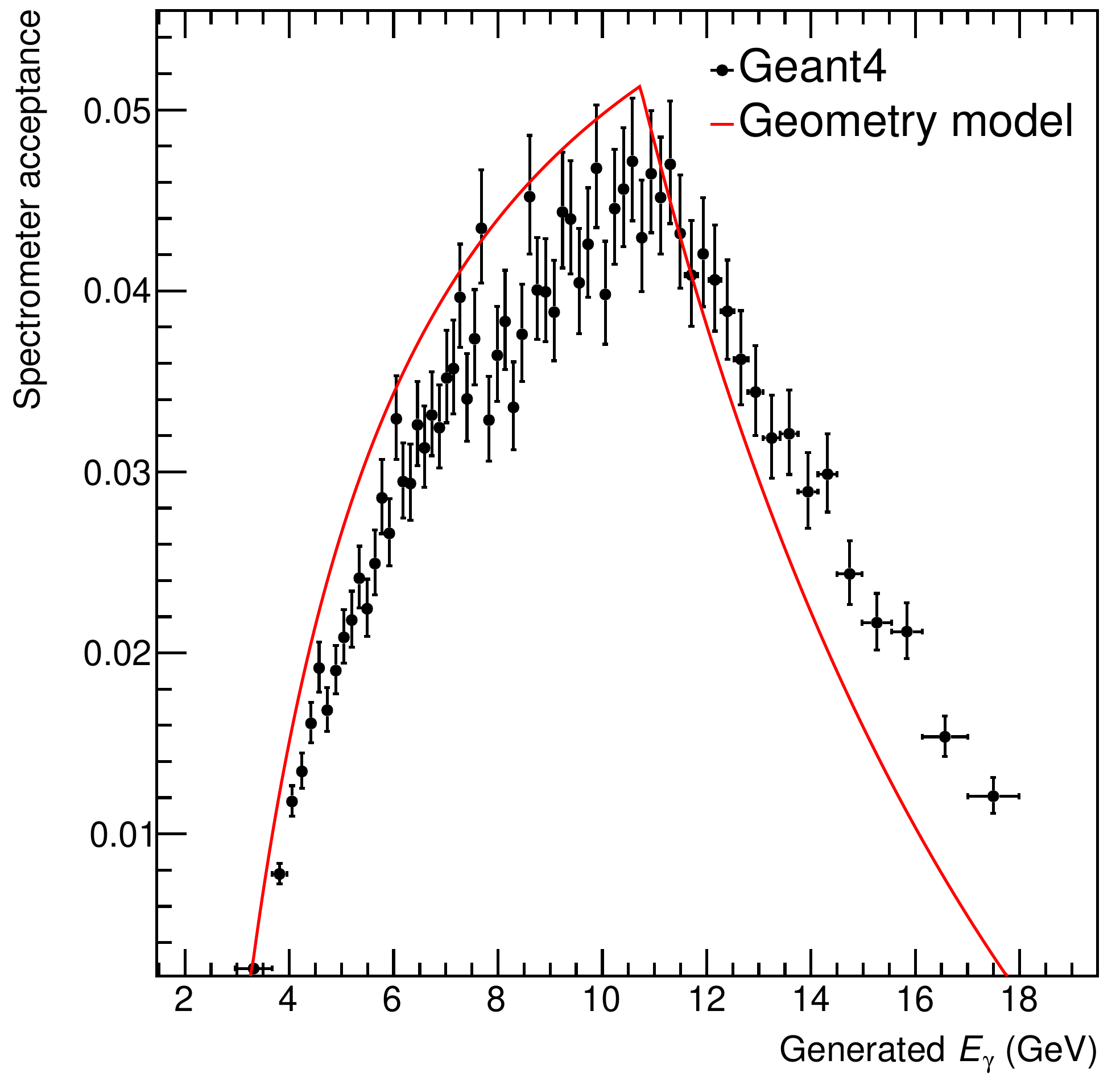}
  \caption{Luminosity spectrometer acceptance as a function of the bremsstrahlung photon energy $E_{\gamma}$.
  The acceptance includes a photon to pair conversion probability of $\sim$8\%.}
  \label{fig:lumi-spec-acc}
\end{figure}

The acceptance is shown in Fig.~\ref{fig:lumi-spec-acc} for
$18\times 275$~GeV beams,
as a function of generated bremsstrahlung photon energy $E_{\gamma}$. The Geant4 distribution is a result
of simulation of 1M bremsstrahlung events generated by the eic-lgen event generator \cite{eic-lgen}
and passed through the layout of Fig~\ref{fig:lumi-geant}. The acceptance is constructed as a fraction of events
with at least 1~GeV of energy hitting both the UP and DOWN detectors.

A geometric model for the acceptance, shown in Fig.~\ref{fig:lumi-spec-acc} as a solid line, is based
on the formula for deflection of a charged particle in a magnetic field and the coincidence requirement
of both pair electrons hitting the detectors. A unit-charge particle moving along the $z$ direction through a magnetic field
$B_x$ oriented along $x$ axis adds a transverse momentum $p_T \propto \int B_x\mathrm{d}z$ along the vertical
$y$ direction.
The position in $y$ on the UP or DOWN detectors of the arriving electron is given by
\begin{equation}
y = l\frac{p_T}{p},
\end{equation}
where $l$ is the length
from the magnet center to the detector face, and $p$ is the momentum of the electron.

If the $e^{\pm}$ of the pair hitting the upper detector has a fraction of photon energy $z=p/E_\gamma$, the other hitting the lower detector
has a fraction $1-z$. The positions of the pair arriving at the UP and DOWN detectors $y_{\mathrm{up}}$
and $y_{\mathrm{down}}$, $z$ and $E_\gamma$ are related by:
\begin{equation}
zE_\gamma =\frac{lp_T}{y_{\mathrm{up}}},\ \;\;\; (1-z)E_\gamma =\frac{lp_T}{y_{\mathrm{down}}} .
\end{equation}

Both spectrometer detectors cover minimal and maximal positions along $y$, as indicated
in Fig.~\ref{fig:lumi-zeus}. The coincidence requirement then limits 
the range
in $z$ for which the converted photon would be detected by the spectrometer.
The integral of
$\frac{dN}{dz}$ over this range of $z$
then determines the acceptance at a given
$E_{\gamma}$, shown as a solid line in
Fig.~\ref{fig:lumi-spec-acc}.
There is reasonable agreement with the full Geant4 simulation.

The model can be used to determine the magnetic field for the spectrometer dipole magnet in order to optimize the 
rate of coincides. A unique magnetic field will be required for each electron beam energy.

\subsection {Far-Backward Electrons} \label{sec:FB_electrons_main}

\subsubsection {Far-Backward Electron Processes and Measurements}

Downstream of the interaction point the electron beam is accompanied
by a flux of electrons at small angles with respect to the beam
direction and at slightly lower energy.
They are predominantly final state electrons from the bremsstrahlung
process $e+p \longrightarrow e+p+\gamma$,
with an energy distribution the mirror image of the left of
Fig.~\ref{fig:brems_Egthg} with $E_e' = E_e - E_{\gamma}$.
Also, a fraction of the electrons in this region are produced in
quasi-real photoproduction with $Q^2 \approx 0$.

The final state bremsstrahlung electrons provide a powerful tool for
calibrating and verifying the luminosity measurement with photons.
Tagging bremsstrahlung electrons and counting corresponding photons in
the photon detectors provides a direct measure of the luminosity
detector acceptance in the tagged energy range.
This is of paramount importance to precisely determine the pair
conversion probability for the luminosity spectrometer,
which depends on the exit window composition and thickness.
Such measurements will require special runs with low bunch currents to ensure that there is only one bremsstrahlung electron/photon pair in the system per bunch crossing.

Tagging of low-$Q^2$ processes provides an extension of the kinematic
range of DIS processes measured with electrons in the central
detector.
It crosses the transition from DIS to hadronic reactions with
quasi-real photons.
Taggers as depicted in the top left of
Fig.~\ref{fig:rear_layout} provide useful acceptance in the range
$Q^2 <10^{-2}$\,GeV$^2$.
Application of the electron taggers for low-$Q^2$ physics will face a
challenge from the high rate of bremsstrahlung electrons, which can be
addressed by tagger design and correlation with information from the
central detector.

\subsubsection {Far-Backward Electron Detection and Requirements}

Possible locations of detectors for far-backward electrons are shown
in the top left of Fig.~\ref{fig:rear_layout}.
Electrons with energies slightly below the beam are bent out of the
beam by the first lattice dipole after the interaction point.
The beam vacuum chamber must include exit windows for these
electrons.
The windows should be as thin as possible along the electron direction
to minimize energy loss and multiple scattering before the detectors.

The taggers should include calorimeters for triggering and energy
measurements.
They should be finely segmented to disentangle the multiple electron
hits per bunch crossing from the high rate bremsstrahlung process.
The taggers should also have position sensitive detectors to measure the
vertical and horizontal coordinates of electrons.
The combined energy and position measurements allow reconstruction of
the kinematic variable $Q^2$ and $x_{BJ}$.
If the position detectors have multiple layers and are able to
reconstruct the electron direction this will overconstrain the
variable reconstruction and improve their measurement;
this may also provide some measure of background rejection.

\subsubsection {Far-Backward Electron Detector Implementation}

The layout of the backward electron detectors is shown in the top left of
Fig.~\ref{fig:rear_layout}. Beam magnets are shown
in full green, drift space in hatched green and detectors and components 
in red and yellow. Two tagger detectors are proposed, Tagger 1 at $z=-24$~m and Tagger 2 at $z=-37$~m respectively.
The backward electromagnetic calorimeter ECAL, part of the central detector, is located at $z=-3.28$~m.
The rapidity coverage of the ECAL $\sim-4.0<\eta<-1.0$ is implemented in simulations for electron acceptance studies.

\begin{figure}[!ht]
  \centering
  \includegraphics[width=0.9\textwidth]{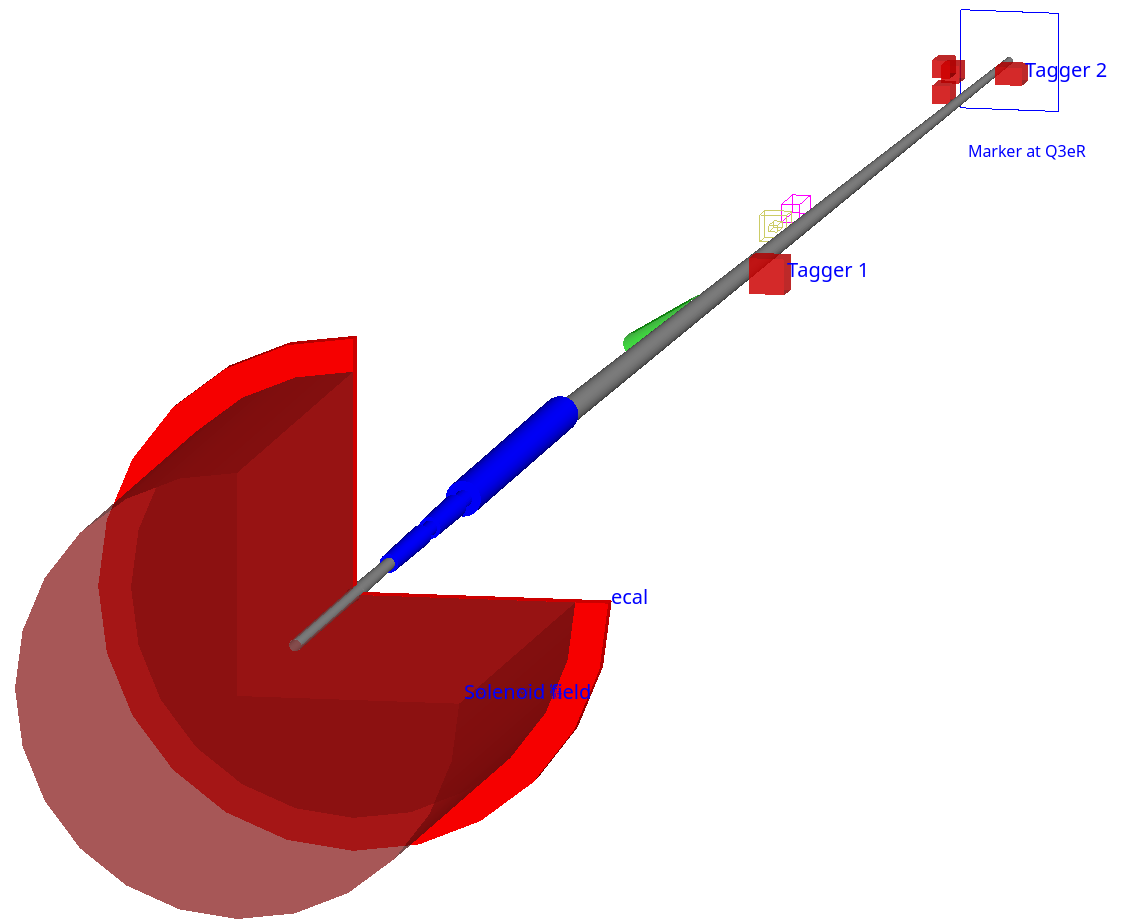}
  \caption{Geant4 model of backward interaction region, on the side of the electron tagger detectors.}
  \label{fig:etag-geant}
\end{figure}

A Geant4 model of the backward side of the interaction region is shown in Fig.~\ref{fig:etag-geant}.
The tagger detectors Tagger 1 and 2 and the backward electromagnetic calorimeter ECAL are implemented
as boxes which register hits by all incoming particles. The solenoid field of the central detector is based in the 3~T BeAST parametrization. Beam magnets eQ1ER, eQ2ER and eB2ER
(Table~\ref{tab:el_ir} ) are shown as blue cylinders. Drift spaces in gray are transparent to all particles. The layout ends
with a marker at the position of the Q3eR magnet.

\begin{figure}[!ht]
  \centering
  \includegraphics[width=0.6\textwidth]{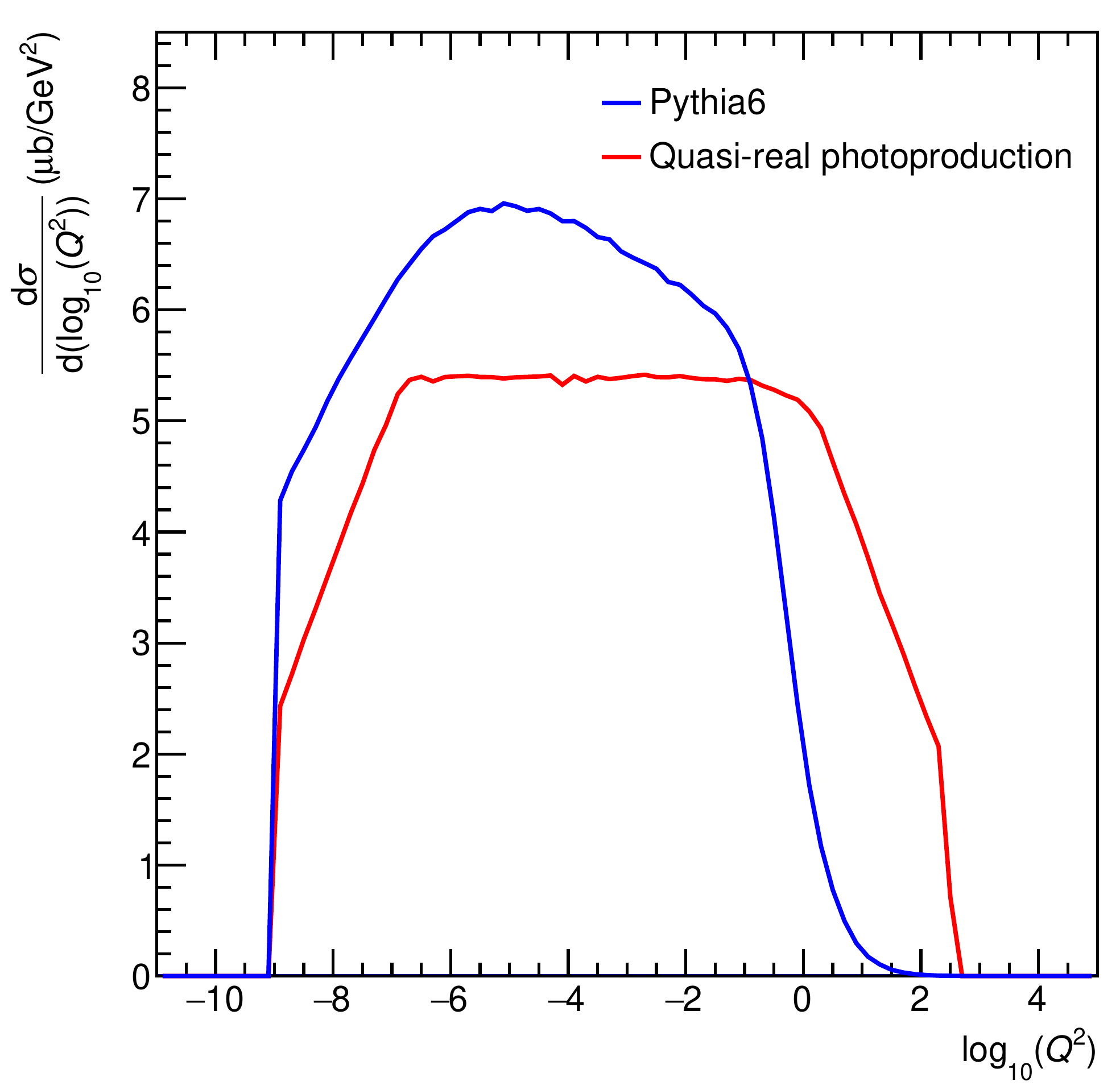}
  \caption{Total cross section as a function of event true $Q^2$ for Pythia~6 and quasi-real photoproduction.}
  \label{fig:etag-sigma}
\end{figure}

Two generated samples, Pythia~6 and quasi-real photoproduction, were used to address acceptance of the taggers at the top energy $18\times 275$~GeV for the electron and proton beams, respectively.
The model of quasi-real photoproduction is based on an approach used in a HERA study \cite{Amaldi:1979qp} and implemented in the eic-lgen event generator \cite{eic-lgen}.
The total cross section for both samples as a function of event true $Q^2$ is shown in Fig.~\ref{fig:etag-sigma}.

\begin{figure}[!ht]
  \centering
  \begin{subfigure}[h]{0.49\textwidth}
    \includegraphics[width=\textwidth]{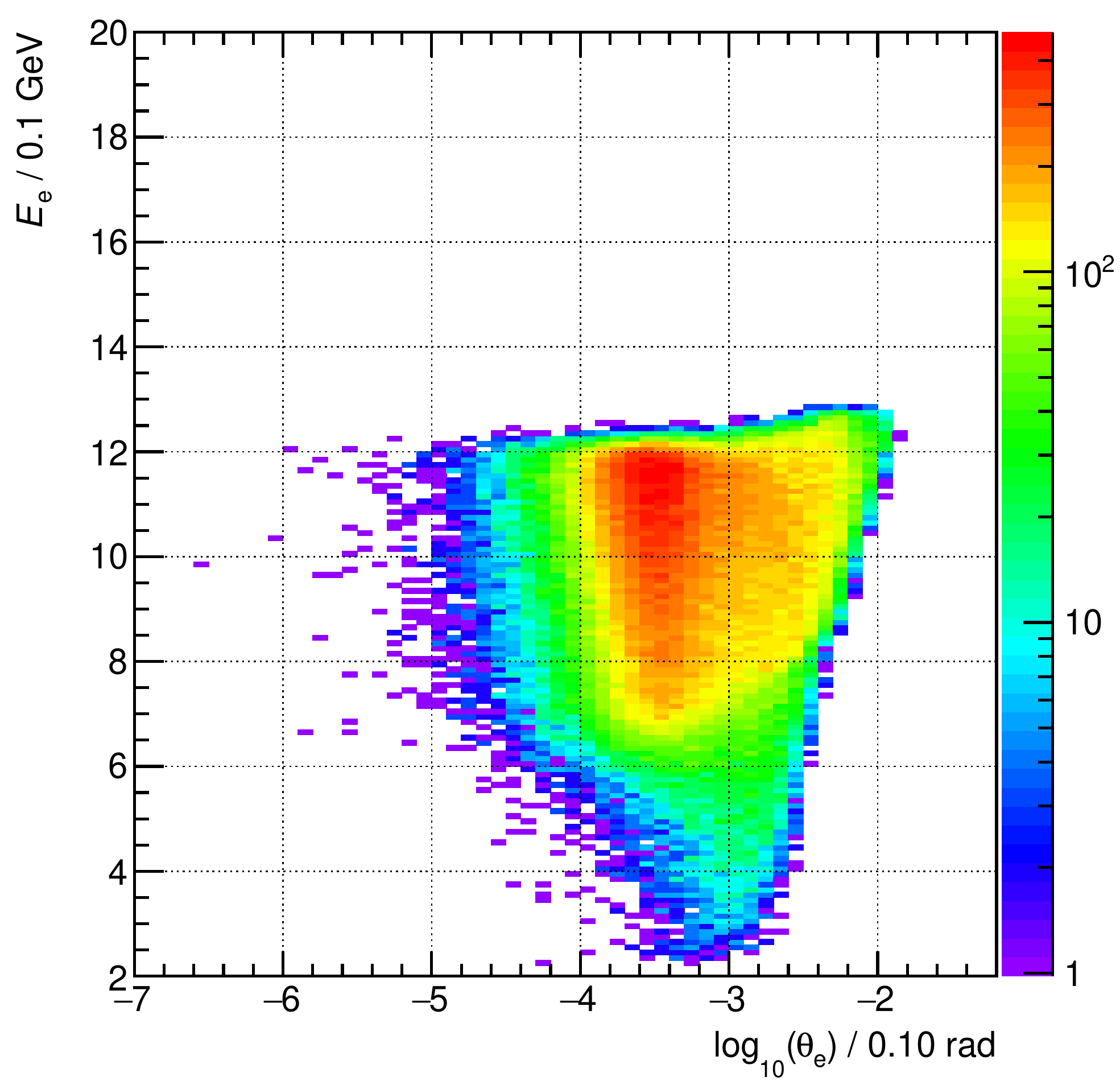}
    \caption{Hit in Tagger 1}
    \label{fig:etag-en-theta-s1}
  \end{subfigure}
  \begin{subfigure}[h]{0.49\textwidth}
    \includegraphics[width=\textwidth]{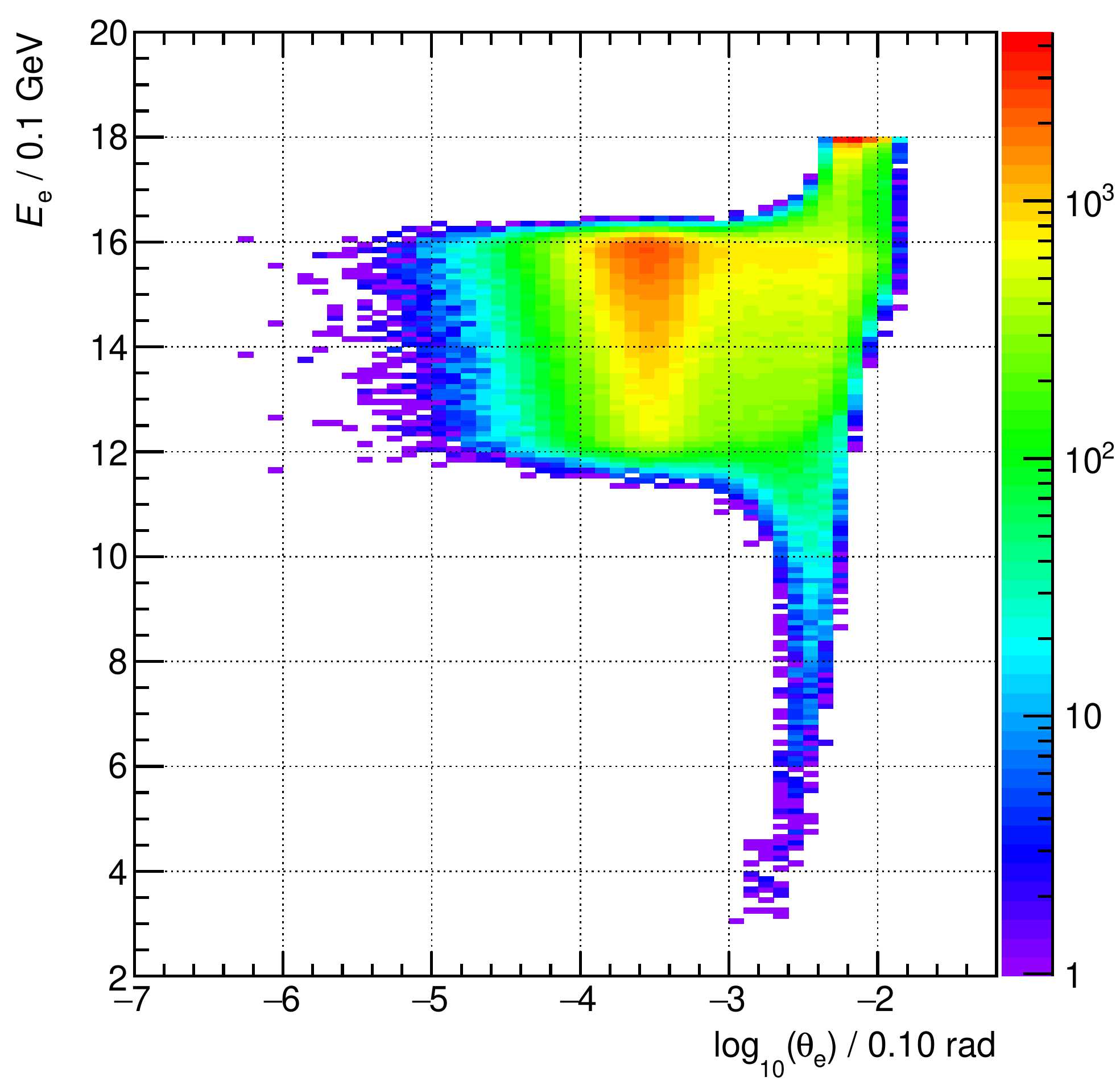}
    \caption{Hit in Tagger 2}
    \label{fig:etag-en-theta-s2}
  \end{subfigure}
  \caption{Scattered electron energy $E_e$ and polar angle $\theta_e$ for quasi-real photoproduction events with a hit in Tagger 1 or 2.}
  \label{fig:etag-en-theta-s12}
\end{figure}

The angular and energy coverage for both tagger detectors is shown in Fig.~\ref{fig:etag-en-theta-s12}. The energy $E_e$ and polar angle $\theta_e$  of scattered electrons is shown for events where the scattered
electron is incident on one of the tagger detectors. The energy coverage is complementary for both tagger detectors.

\begin{figure}[ht]
  \centering
  \includegraphics[width=0.6\textwidth]{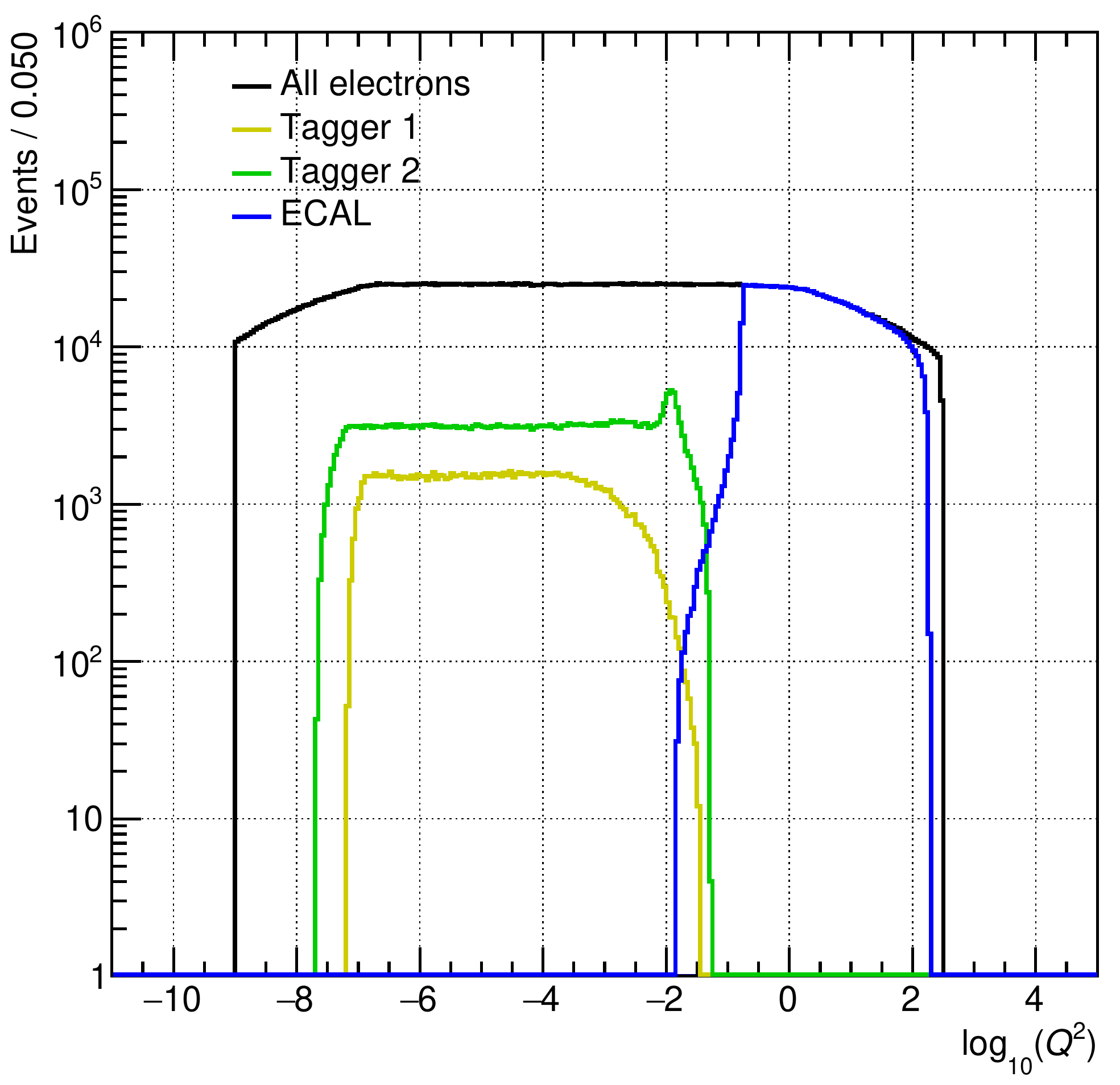}
  \caption{Coverage in $Q^2$ for quasi-real photoproduction events for the tagger detectors and the ECAL.}
  \label{fig:etag-Q2-qr}
\end{figure}

The coverage in $Q^2$ is shown in Fig.~\ref{fig:etag-Q2-qr} for quasi-real photoproduction. Events with a hit in one of the taggers or in the ECAL are shown along with all generated quasi-real events. The taggers cover a similar range in $Q^2$, although as illustrated in Fig.~\ref{fig:etag-en-theta-s12}, the coverage is achieved by different combinations of electron energies and angles. A transition of coverage takes place at the lower reach of the ECAL and the upper reach of the tagger detectors.

\begin{figure}[!ht]
  \centering
  \includegraphics[width=0.6\textwidth]{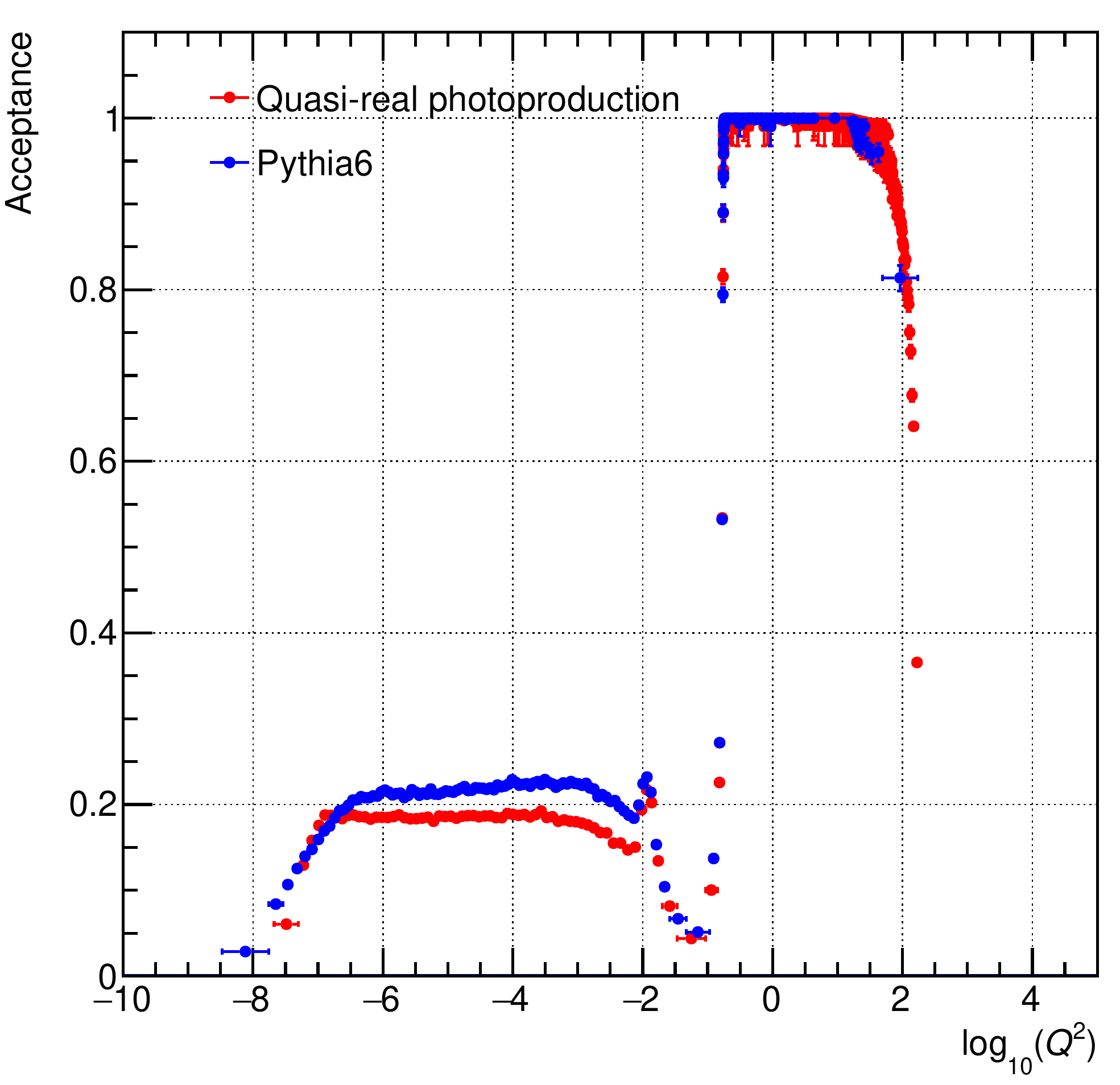}
  \caption{Acceptance versus $Q^2$ for the tagger detectors and the ECAL.}
  \label{fig:etag-lQ2-acc}
\end{figure}

The combined acceptance of the tagger detectors and the ECAL is shown in Fig.~\ref{fig:etag-lQ2-acc} for the Pythia~6 and quasi-real photoproduction samples. The acceptance is obtained as a fraction of all generated
events with a hit in one of the tagger detectors or in the ECAL. A dip occurs at the transition between the ECAL and tagger acceptances, at $Q^2 \sim$ 0.1~GeV$^2$. The magnitude and width of the dip
depends strongly on the inner radius for the ECAL. The acceptance is compatible between the two event generators.

\begin{figure}[!ht]
  \centering
  \begin{subfigure}[ht]{0.49\textwidth}
    \includegraphics[width=\textwidth]{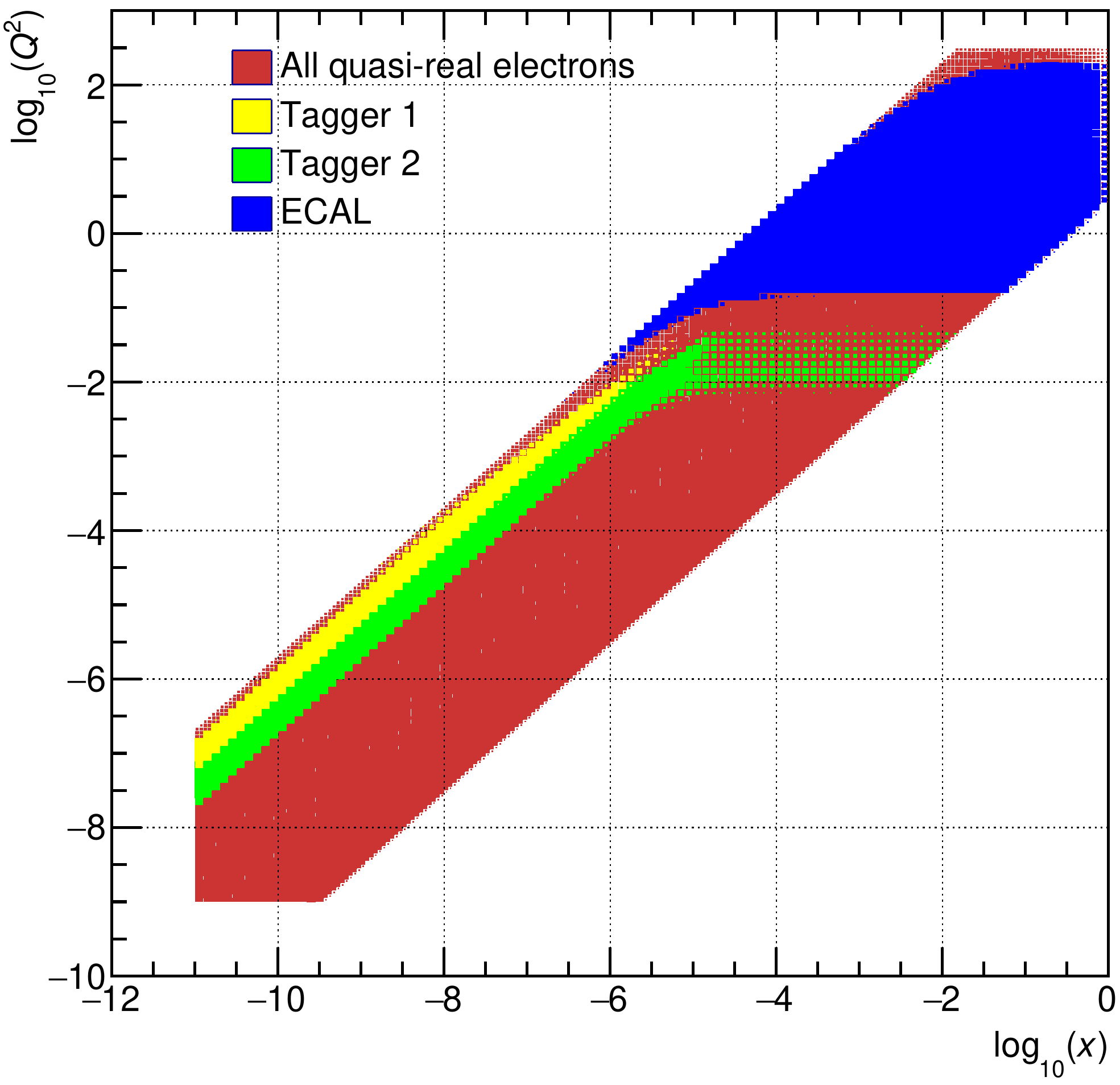}
    \caption{Coverage in $x$ and $Q^2$}
    \label{fig:etag-x-Q2}
  \end{subfigure}
  \begin{subfigure}[ht]{0.49\textwidth}
    \includegraphics[width=\textwidth]{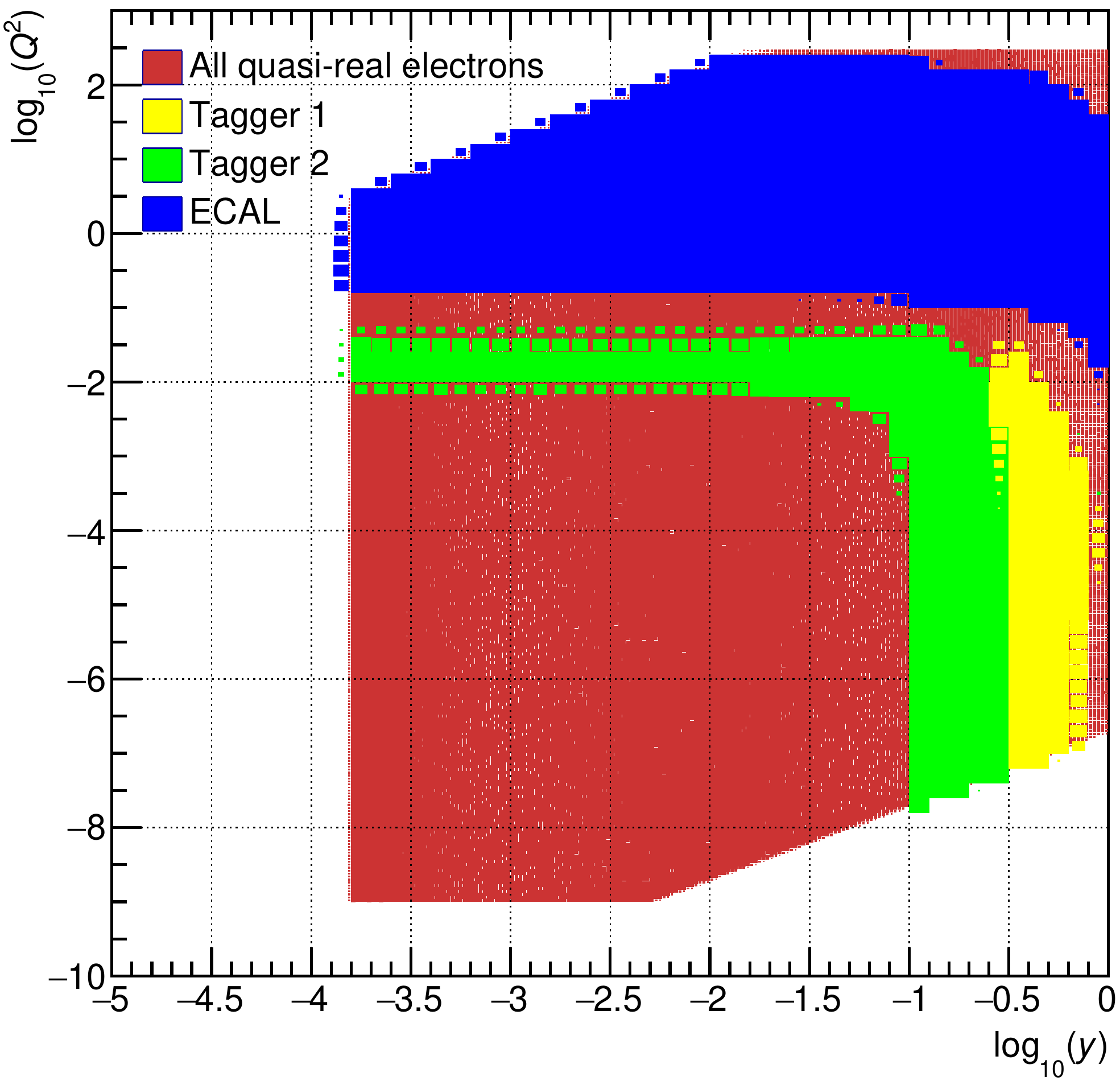}
    \caption{Coverage in $y$ and $Q^2$}
    \label{fig:etag-y-Q2}
  \end{subfigure}
  \caption{Coverage in $x$, $y$ and $Q^2$ for the tagger detectors and the ECAL.}
  \label{fig:etag-xy-Q2}
\end{figure}

Figure~\ref{fig:etag-xy-Q2} shows the tagger and ECAL coverage as functions of the kinematic variables Bjorken-$x$, inelasticity $y$ and 
virtuality $Q^2$.
All generated events are shown as underlying red bands; box diagrams show events with a hit in one of the taggers or in the ECAL.

\begin{figure}[!ht]
  \centering
  \includegraphics[width=0.6\textwidth]
  {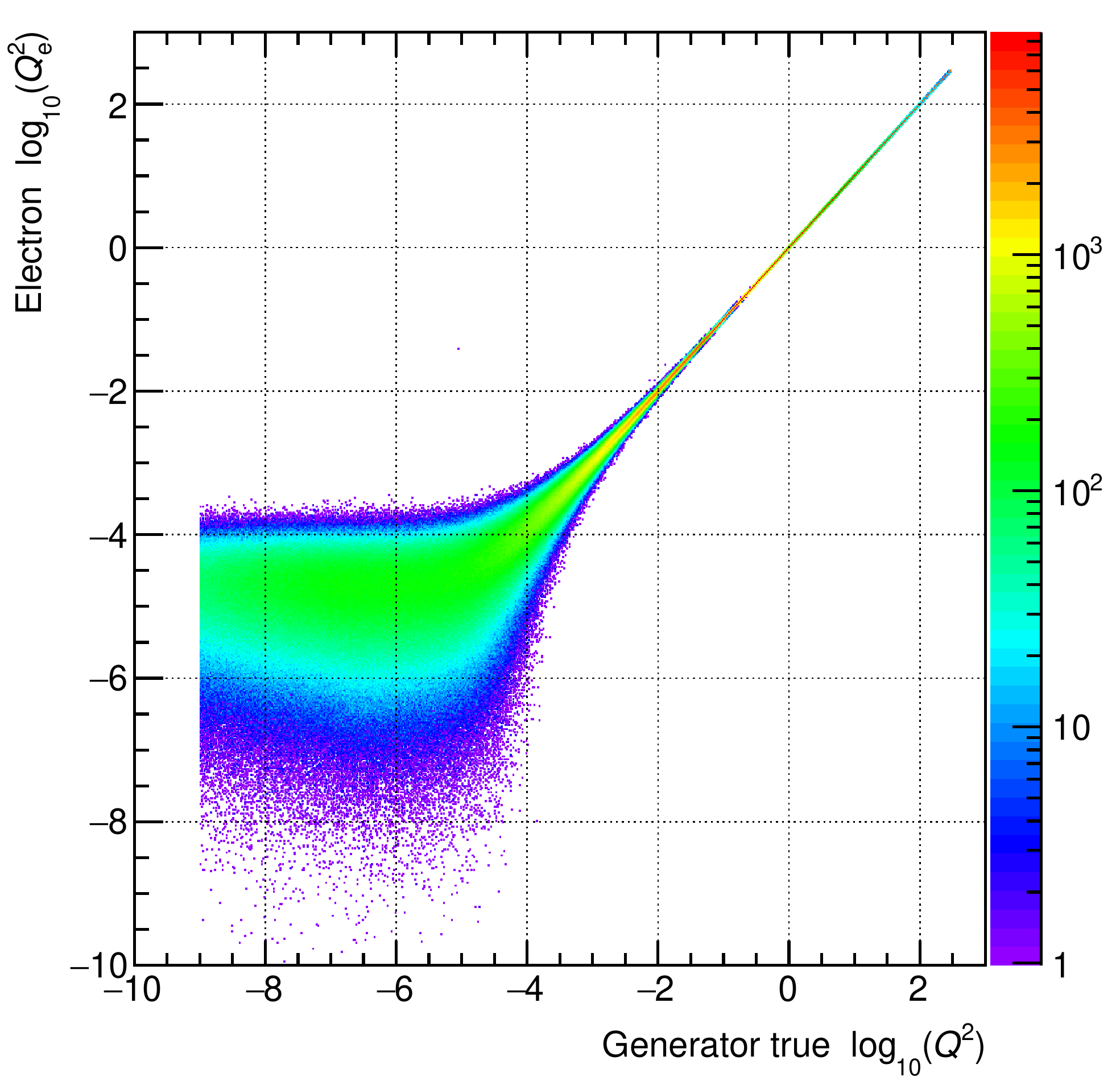}
  \caption{Comparison of generated and reconstructed electron
    $Q^2_e$ with smearing for beam angular divergence.}
  \label{fig:etag_Q2_res}
\end{figure}

The reconstructed versus generated $Q^2$ is shown in
Fig.~\ref{fig:etag_Q2_res}.
The simulation includes smearing from beam angular divergence, which introduces significant errors in the
variable reconstruction
There is reasonable resolution for $Q^2$ as low as $10^{-3}$\,GeV$^2$;
below $10^{-4}$\,GeV$^2$ meaningful reconstruction of $Q^2$ based on the
electron is not possible.

\section{Considered Technologies and Detector Challenges}
\label{part3-sec-Det.Aspects.challenges}
%
The rich panorama of detector technologies illustrated in the previous sections of the present Chapter are summarized in Table~\ref{tab:detector-technologies}.  The corresponding performance according to the present assessment is summarized in the Detector Matrix (https://physdiv.jlab.org/DetectorMatrix/), 
an interactive tool resulting from the effort for the Yellow Report Initiative. Table~\ref{tab:detector-matrix} present a snapshot of the matrix, while the interactive feature gives assess to more details and complementary information: it allows to correlate the listed performance with details about the technology and the simulations used to obtain the performance, as, in particular, the magnet field provided by the central solenoid.

\par 
Some points of tension result from the comparison between the requirements posed to the detector by the EIC physics program, illustrated in Chapter~\ref{part2-chap-DetRequirements} and summarized 
in Sec.~\ref{part3-sec-DetChalReq.PhysReq}, and the performance reported in the Detector Matrix.
They are discussed in the following aiming at raising the community attention to these aspects and urging more dedicated effort, both to understand the impact on the physics reach of non-ideal detector performance and to explore further technological options to overcome the present detector limitations. 
\begin{sidewaystable*}[htb]
\caption{Possible detector technologies for the Central Detector. No technology
ranking between the reference and the alternative options is established, as clarified in Chapter~\ref{part3-chap-Intro}, where the reference detector concept is introduced.}
\label{tab:detector-technologies}
\begin{threeparttable}
\resizebox{\textwidth}{!}{%
\begin{tabular}{|l|l|l|l|l|l|}
\toprule
\textbf{system} &
  \textbf{system   components} &
  \textbf{reference   detectors} &
  \multicolumn{3}{l|}{\textbf{detectors,   alternative options considered by the community}} \\ \hline
\multirow{4}{*}{\textbf{tracking}} &
  vertex &
  MAPS,   20 um pitch &
  MAPS,   10 um pitch &
   &
   \\ \cline{2-6} 
 &
  barrel &
  TPC &
  TPC$^a$ &
  MAPS,   20 um pitch &
  MICROMEGAS$^b$ \\ \cline{2-6} 
 &
  forward \& backward &
  MAPS, 20 um   pitch \& sTGCs$^c$ &
  GEMs &
  GEMs   with Cr electrodes &
   \\ \cline{2-6} 
 &
  very far-forward &
  MAPS, 20 um   pitch \& AC-LGAD$^d$ &
   TimePix (very far-backward) &
   &
\\ 
   & \& far-backward
   &            
   &  
   &
   &
   \\ \hline
\multirow{5}{*}{\textbf{ECal}} &
  barrel &
  W powder/ScFi or Pb/Sc Shashlyk &
  SciGlass &
  W/Sc Shashlyk & \\ \cline{2-6} 
 &
  forward &
  W powder/ScFi &
  SciGlass &
  PbGl &
  Pb/Sc Shashlyk or
  W/Sc Shashlyk \\ \cline{2-6} 
 &
  backward, inner &
  PbWO$_4$ &
  SciGlass &
   &
   \\ \cline{2-6} 
 &
  backward, outer &
  SciGlass &
  PbWO$_4$ &
  PbGl &
  W   powder/ScFi or W/Sc Shashlyk$^e$ \\ \cline{2-6} 
 &
  very far-forward &
  Si/W &
  W powder/ScFi &
  crystals$^f$ &
  SciGlass \\ \hline
\multirow{5}{*}{\textbf{h-PID}} &
  barrel &
  High   performance DIRC \& dE/dx (TPC) &
  reuse   of BABAR DIRC bars &
  fine   resolution TOF &
   \\ \cline{2-6} 
 &
  forward, high p &
  \multirow{2}{*}{double  radiator RICH (fluorocarbon gas, aerogel)} &
  fluorocarbon   gaseous RICH &
  high   pressure Ar RICH &
   \\ \cline{2-2} \cline{4-6} 
 &
  forward, medium p &
   &
  aerogel &
   &
   \\ \cline{2-6} 
 &
  forward, low p &
  TOF &
  dE/dx &
   &
   \\ \cline{2-6} 
 &
  backward &
  modular   RICH (aerogel) &
  proximity   focusing aerogel &
   &
   \\ \hline
\multirow{3}{*}{\textbf{e/h separation}} &
  barrel &
  hpDIRC \&   dE/dx (TPC) &
  very fine resolution TOF &
   &
   \\ \cline{2-6} 
 &
  forward &
  TOF \&   areogel &
   &
   &
   \\ \cline{2-6} 
{\textbf{~~at low p}} &
  backward &
  modular   RICH &
  adding TRD &
  Hadron Blind   Detector &
   \\ \hline
\multirow{4}{*}{\textbf{HCal}} &
  barrel &
  Fe/Sc &
  RPC/DHCAL &
  Pb/Sc &
   \\ \cline{2-6} 
 &
  forward &
  Fe/Sc &
  RPC/DHCAL &
  Pb/Sc &
   \\ \cline{2-6} 
 &
  backward &
  Fe/Sc &
  RPC/DHCAL &
  Pb/Sc &
   \\ \cline{2-6} 
 &
  very far-forward &
  quartz fibers/   scintillators &
   &
   &
   \\ \hline
\bottomrule	
\end{tabular}
} %
\begin{tablenotes}
\small
\item [a] TPC surrounded by a micro-RWELL tracker   
\item [b] set of coaxial cylindrical MICROMEGAS
\item [c] Small-Strip Thin Gas Chamber (sTGC)
\item [d] MAPS for B0 and off-momentum poarticles, LGAD for Roman Pots
\item [e] also Pb/Sc Shashlyk
\item [f] alternative options: PbWO$_4$, LYSO, GSO, LSO
\normalsize
\end{tablenotes}	
\end{threeparttable} 
\end{sidewaystable*}
\begin{sidewaystable}[thb]
\centering
\caption{This matrix summarizes the high level performance of the different subdetectors and a 3~T Solenoid. The interactive version of this matrix can be obtained through the Yellow Report Detector Working Group (https://physdiv.jlab.org/DetectorMatrix/).}
\label{tab:detector-matrix}
\includegraphics[width=1.0\textwidth]{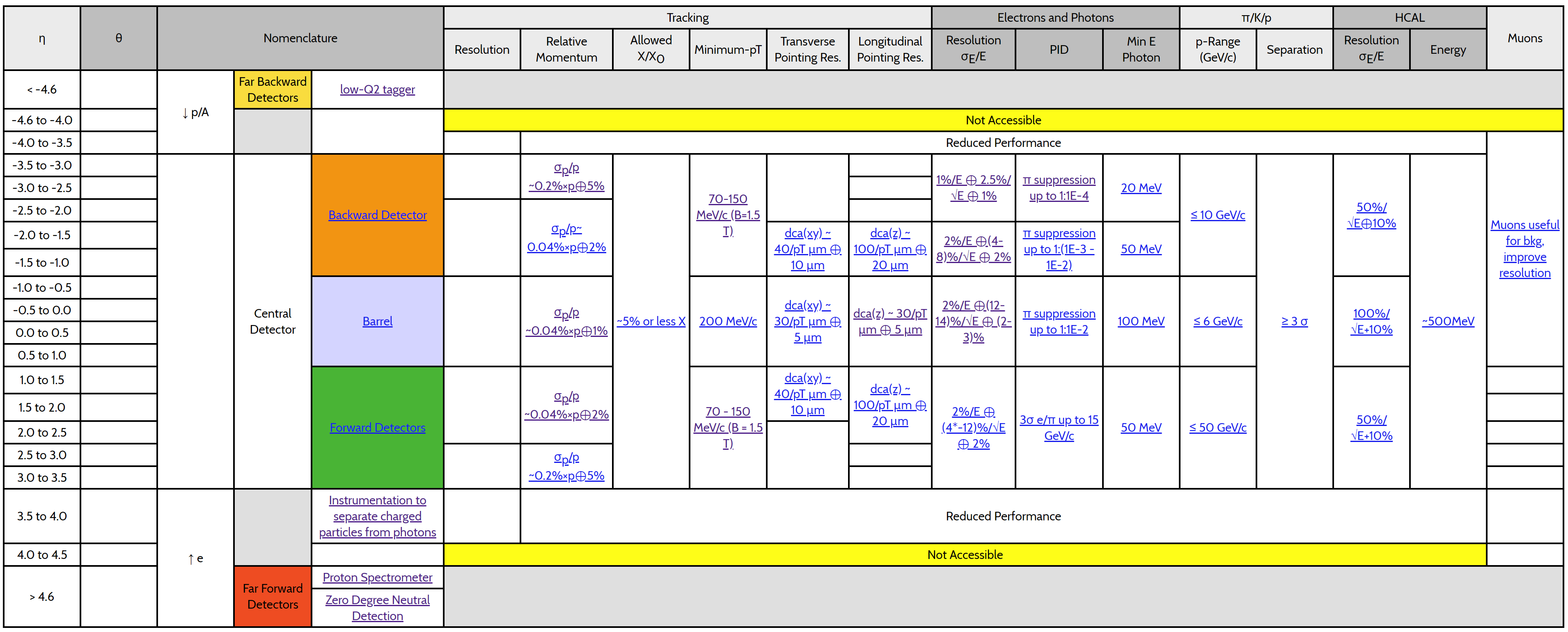} 
\vspace*{3mm}
\end{sidewaystable}
\FloatBarrier

\subsection{Acceptance requirements for the central detector}
Large acceptance in $\eta$ guarantees access to a wide phase-space region in the x—Q$^2$ plane and, therefore, to an accurate study of the evolution of the relevant structure functions and distributions (Sec.~\ref{sec:ePID}, \ref{sec:J-B_rec}, 
\ref{sec:generalSIDIS}, 
\ref{sidis_TMD},
\ref{sec:SIDIS_final-state},
\ref{part2-sec-DetReq.Jets.HQ},
\ref{subsec:bckw_pi0} and
\ref{sec:diff_short-range}).
In particular, it is observed that expanding the coverage in $\eta$ from $\eta$~=~3.5  on the hadron going side extends the region where DIS/SIDIS variables can be reconstructed in the highest x region at low y. For the spectroscopy of XYZ mesons in photoproduction tracking and hadron identification exceeding $\eta$~=~3.5 would be preferable due to the boost of the produced states of interest
(Sec.~\ref{part2-sec-DetReq.SIDIS.Spectroscopy}).
However, it is important to notice that the geometrical acceptance is not the only potential limiting parameter, as other constraints come from the minimum detectable particle momentum and the acceptance in transverse momentum. The reference detector offers complete acceptance in the range $\lvert \eta \rvert <$~3.5, while the calorimeters have more extended acceptance up to $\lvert \eta \rvert =$~4, even if with reduced performance at the detector edges.  These overall features, combined with the further degree of freedom offered by the variable center of mass energy, were found to be adequate for the majority of the physics programme (Chapter~\ref{part2-chap-DetRequirements}). 
\par
Specific requirements are posed by measurements where the detection of very forward particles is needed, as it is the case for exclusive reactions and diffractive scattering. Adequate acceptance is ensured by the very far-forward detectors (B0-spectrometer, off-momentum detectors, Roman pots and ZDC) covering $\eta$-values above 4.6 and optimized for the specific measurements (Sec.~\ref{part3-sec-Det.Aspects.FFDet} ).
\par
Globally, the acceptance offered by the reference detector matches the needs of the overall EIC physics programme.

\subsection{Particle identification challenges in the barrel}
The requirements coming from the physics programme for PID capabilities in the barrel region ($\lvert \eta \rvert  <$1) are not presently matched by the performance of the PID equipment in the reference detector and the alternative options considered by the community. This is the case both concerning  electron/pion separation and hadron identification, discussed separately in the following.
\par
The more demanding figures concerning pion suppression is recalled. In parity-violating asymmetry     A$_{PV}$ measurements in inclusive DIS studies for precision tests of the Standard Model and searches for new physics (Sec.~7.5.2),
a remaining 1\textperthousand  pion contamination in the electron sample is needed to preserve the level of the systematic error due to this background at 10\% of the statistical error. According to the relative abundance of electron and pion production, a pion suppression at the 10$^{-4}$ level is required. The major player in electron identification is electromagnetic calorimetry, where maximum pion rejection is reached for momenta above 2-4~GeV/$c$ and the plateau rejection figure is around 10$^{-2}$ for the technologies foreseen in the barrel (Fig.~\ref{fig:part3-Det.Aspects.ECAL-e-pi-separ1}), a choice imposed by the limited space available and cost considerations. 
PID-dedicated detectors are expected to contribute. In the reference detector, hpDIRC (Sec.~\ref{hpDIRC}) and dE/dx measurements by TPC (Sec.~\ref{dE/dx}) are available. hpDIRC can separate electrons and pions up to approximately 1.3 GeV/$c$ (3$\sigma$ level), dE/dx can cover the higher momenta if adequately supported by TOF measurement (Sec.~\ref{TOF}), which should be added. Due to the short lever arm limited to about 1~m, this option becomes effective if the time resolution is O(10~ps) or better.  
This severe requirement is challenging for the LGAD option; the LAPPD approach is viable, if the LAPPD development will deliver devices that can operate in the high-value magnetic field present in the barrel. 
The addition of a HBD (Sec.~\ref{cherenkov}) would require more space, as it can become available if the all silicon tracking concept is adopted (Sec.~ \ref{tracking_concepts}). Moreover, the concept of an improved HBD providing the required pion rejection rate is, at present, at an initial speculative stage. 
\par
Hadron PID up to 10 GeV/$c$ in the range -1 $< \eta<$ 0 and 15 GeV/$c$ for 0 $< \eta<$ 1 is needed in order to cover the phase space for the whole jet program, in particular for TMD studies from jets (Sec.~8.3.4); 
the impact on physics of reduced performance presently still requires a deeper assessment. Semi-inclusive measurements at mid-x and high Q$^2$ requires, in the barrel,  hadron PID up to 8 GeV/$c$ to have a complete phase space coverage also at the highest center of mass energies (Sec.~8.2.2); 
presently,  the current TMD extraction framework studies indicate that the impact of reduced PID range is not particularly severe.
hpDIRC (Sec.~\ref{hpDIRC}) provides 3~$\sigma$  $\pi$/K separation up to 6-6.5~GeV/$c$, while classical dE/dx (Sec.~\ref{dE/dx}) cannot contribute at high momenta. An intriguing option, presently completely speculative, is by performing dE/dx via cluster counting (Sec.~\ref{dE/dx}), in a dedicated detector to be added if more space becomes available in the barrel, as already mentioned. Preliminary considerations indicate 3~$\sigma$  $\pi$/K separation in the range 1.5-15 GeV/$c$ possible.

\subsection{Hadron calorimetry challenges}
Comparing the physics requirements with the forward hadronic calorimeter performance a discrepancy in the forward region $\eta >$ 3 can be observed. Specifically, there is a desire to have an insert in this region with an energy resolution better than 40\%/$\sqrt{E}$ and a constant term of $\sim$5\% to improve jet energy resolution. For $\eta >$ 3, a constant term of $\sim$ 5\% is needed as jet energies rapidly increase in this region while tracking resolution significantly degrades, enhancing the importance of the HCal energy resolution, which is dominated by the constant term at these energies. As tracking will be absent for $\eta >$ 3.5, good HCal resolution will be imperative for good overall jet energy resolution. Differential TMD measurements with jets, e.g. electron-jet Sivers asymmetry in the valence region 
mid to high $Q^2$ (Sec.~\ref{part2-sec-DetReq.Jets.HQ}). 
Precision calorimetry in the forward direction is also important for large-$x$ processes, which require a resolution $\delta x/x  <$ 0.1, where the HCal energy resolution determines $\delta x/x$ (Sec. ~\ref{part2-sec-DetReq.Diff.Tag}). 
As an example, a 50 GeV hadron/jet energy and 35\%/$\sqrt{E}$ energy resolution results in a resolution $\delta x$=0.05. 
\par
The reference detector HCal in the forward direction is expected to provide an energy resolution of 50\%/$\sqrt{E}$ + 10\%. The reference detector addresses aside from energy resolution the EIC requirements on compactness and mechanical sturdiness, by minimizing the space required for passive mechanical support structures. High resolution calorimetry on the other hand requires a large amount of space and a high sampling frequency. For example, the ZEUS high resolution HCAls required 4 meters, which would be impractical for EIC. A further drawback of these constructions is the large penalty one pays for dead material between ECAL and HCAL. There are possible tradeoffs, e.g., a tail catcher can give the same results as improved sampling frequencies at low energies. Alternative methods for high resolution calorimetry include the dual readout concept in which one identifies an observable that correlates with the number of neutrons and corrects the detected energy event-by-event using this observable. Another alternative option is digital calorimeters. This approach requires significant space for the detector, appropriate design of the magnet and perfect tracking performance at all rapidity.
\par
In summary, EIC requirements up to $\eta \sim$ 3 may be achieved with existing technologies tried by the eRD1 consortium and the STAR forward upgrade with some additional R\&D efforts to improve on performance of STAR-like forward calorimeter system. A high resolution HCal insert for $\eta>$3 will require additional R\&D efforts, e.g. to develop high density fiber calorimeter with SiPM readout.

\section{Polarimetry}
Rapid, precise beam polarization measurements will be crucial for meeting the goals of the EIC physics program as the uncertainty in the polarization propagates directly into the uncertainty for relevant observables (asymmetries, etc.).  In addition, polarimetry will play an important role in facilitating the setup of the accelerator.

The basic requirements for beam polarimetry are:
\begin{itemize}
  \item{Non-destructive with minimal impact on the beam lifetime}
  \item{Systematic uncertainty on the order $\frac{dP}{P}=1\%$ or better}
  \item{Capable of measuring the beam polarization for each bunch in the ring - in particular, the statistical uncertainty of the measurement for a given bunch should be comparable to the systematic uncertainty}
  \item{Rapid, quasi-online analysis in order to provide timely feedback for accelerator setup}
\end{itemize}

\subsection{Electron Polarimetry}\label{sec-Spin}
The most commonly used technique for measuring electron beam polarization in rings and colliders is Compton polarimetry, in which the polarized electrons scatter from 100\% circularly polarized laser photons.  The asymmetry from this reaction is measured via the scattered electrons or high energy backscattered photons.  A brief review and description of several previous Compton polarimeters can be found in~\cite{Aulenbacher:2018weg}.  A particular advantage of Compton polarimetry is that it sensitive to both longitudinal and transverse polarization.

\begin{figure}[tb]
  \centerline{
  \includegraphics[width=0.49\columnwidth]{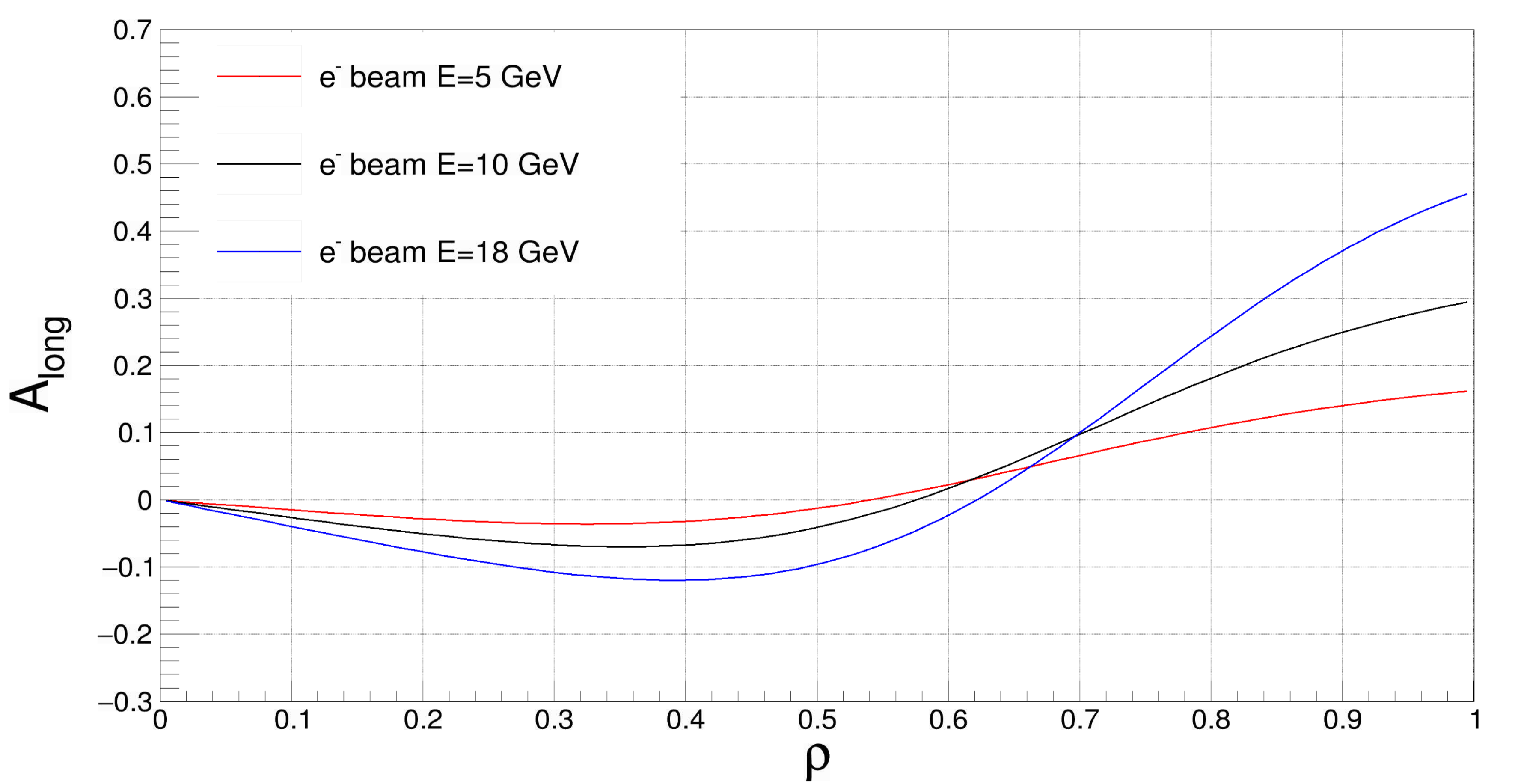}
  \includegraphics[width=0.49\columnwidth]{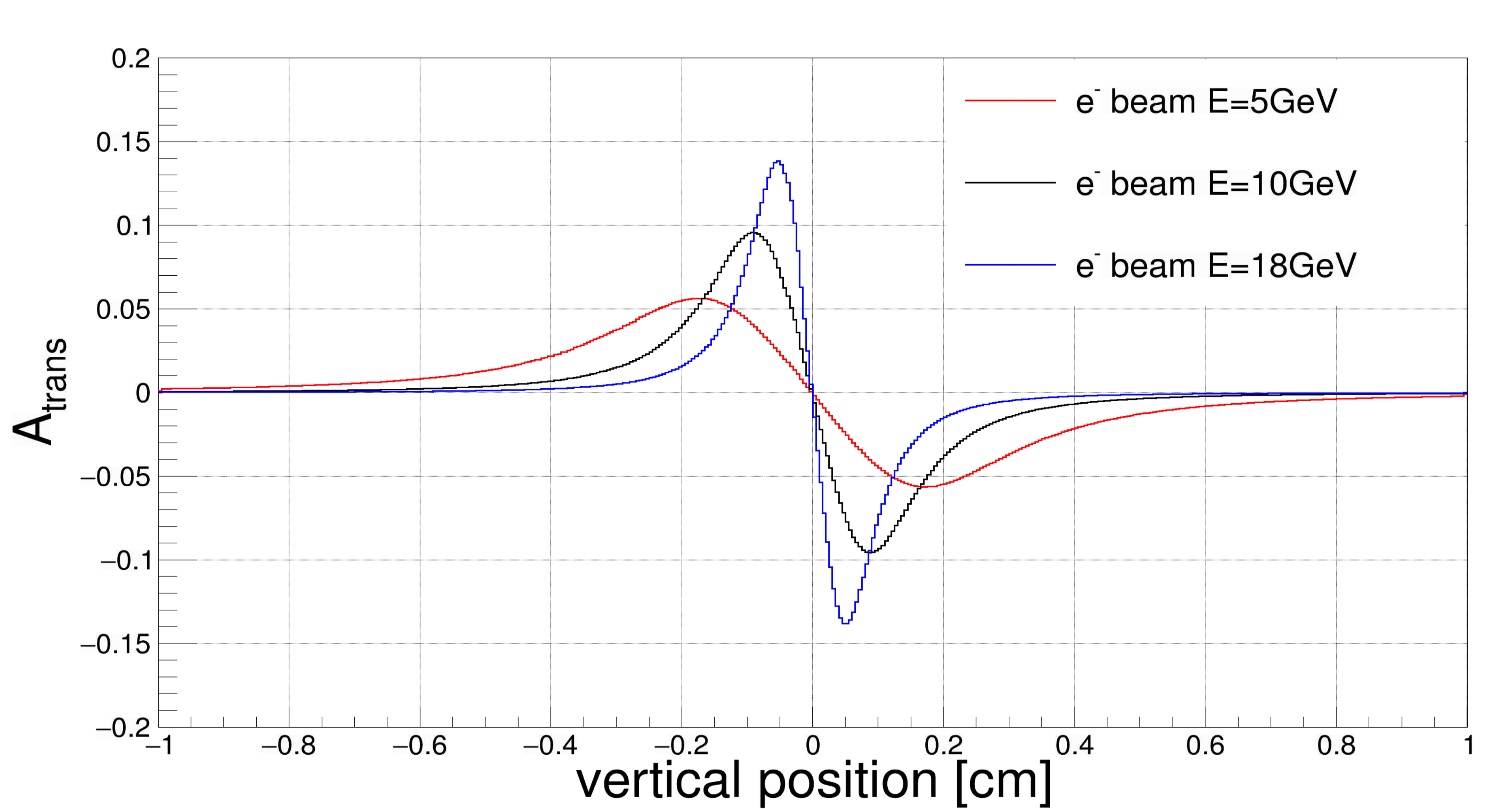}      
  }
  \caption{Longitudinal (left) and transverse (right) analyzing powers assuming a 532~nm wavelength laser colliding with an electron beam at 5\,GeV, 10\,GeV, and 18\,GeV.\label{fig:compton_apower}  The transverse analyzing power is shown for photons projected 25~m from the collision point and plotted vs. the vertical position.}
\end{figure}

The longitudinal analyzing power depends only on the backscattered photon energy and is given by,
\begin{equation}
 A_\textrm{long} = \frac{2 \pi r_o^2 a}{(d\sigma/d\rho)} (1-\rho(1+a)) \left[1-\frac{1}{(1-\rho(1-a))^2}\right],
\end{equation}
where $r_o$ is the classical electron radius, $a=(1+4\gamma E_\textrm{laser} /m_e)^{-1}$ (with the Lorentz factor $\gamma=E_e/m_e$ ), 
 $\rho$ is the backscattered photon energy divided by its kinematic maximum, $E_\gamma/E^{max}_\gamma$, and $d\sigma/d\rho$ is the unpolarized Compton cross section.  In contrast, the transverse analyzing power depends both on the backscattered photon energy and the azimuthal angle ($\phi$) of the photon (with respect to the transverse polarization direction);
\begin{equation}
A_\textrm{tran} = \frac{2 \pi r_o^2 a}{(d\sigma/d\rho)} \cos{\phi} \left[\rho(1-a)
  \frac{\sqrt{4a\rho(1-\rho)}}{(1-\rho(1-a))}\right].
\end{equation}
This azimuthal dependence of the asymmetry results in an ``up-down'' asymmetry (assuming vertically polarized electrons) and requires a detector with spatial sensitivity.  Both the longitudinal and transverse analyzing powers are shown in Fig.~\ref{fig:compton_apower}.

\begin{figure}[tb]
  \centerline{
  \includegraphics[width=\columnwidth]{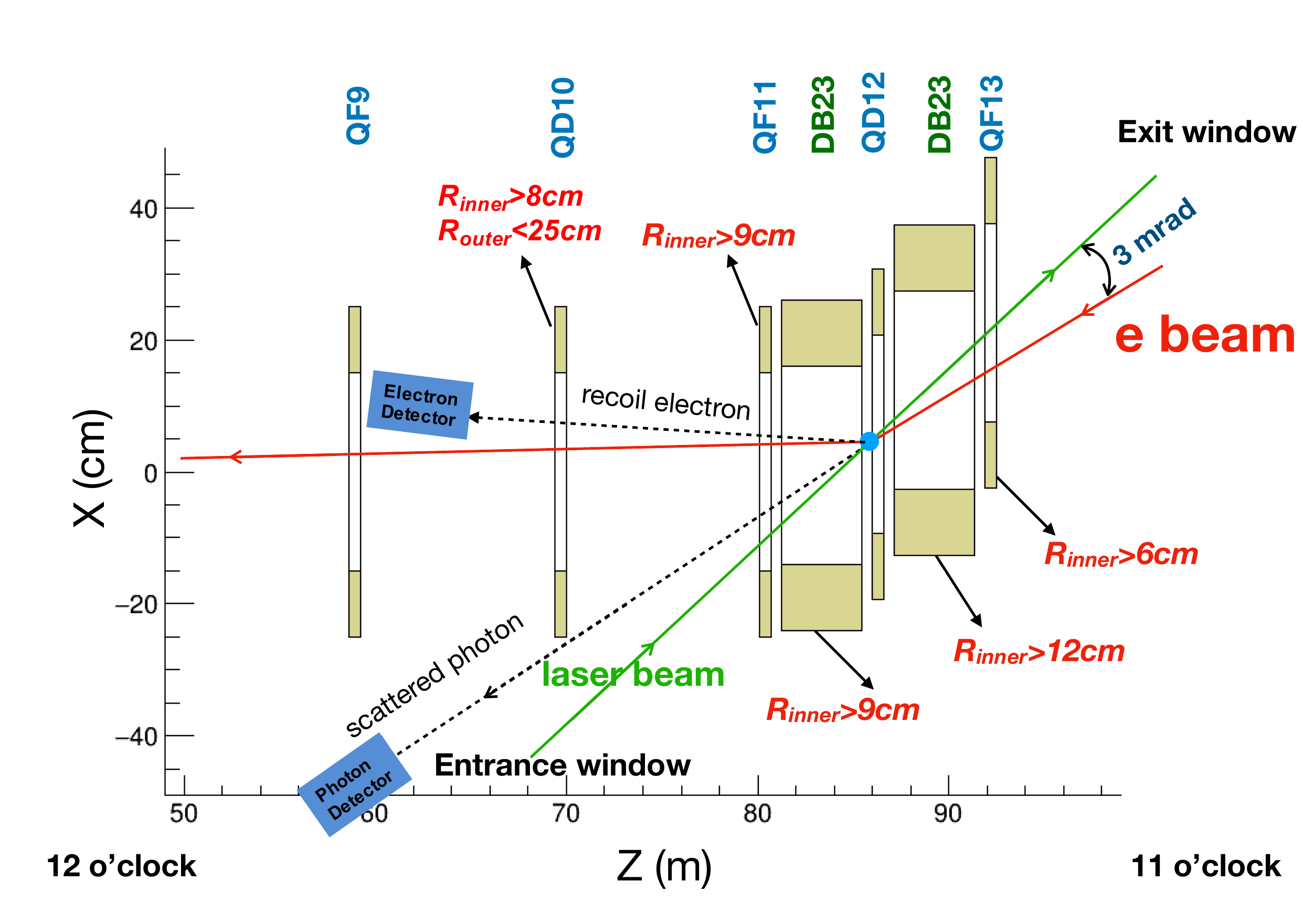}
  }
  \caption{Layout of the Compton polarimeter at IP 12. In this figure the electron beam travels from right to left - the laser beam collides with the electrons just downstream of QD12.  The dipole just downstream of the collision (DB12) steers the unscattered electrons allowing detection of the backscattered photons about 25~m downstream of the collision. DB12 also momentum-analyzes the scattered electrons, facilitating use of a position sensitive electron detector downstream of QD10.  Also noted in the figure are constraints on required apertures of the magnets needed to allow transport of the laser beam, backscattered photons, and scattered electrons.\label{fig:compton_ip12}}
\end{figure}

\begin{figure}[tb]
  \centerline{
  \includegraphics[width=\columnwidth]{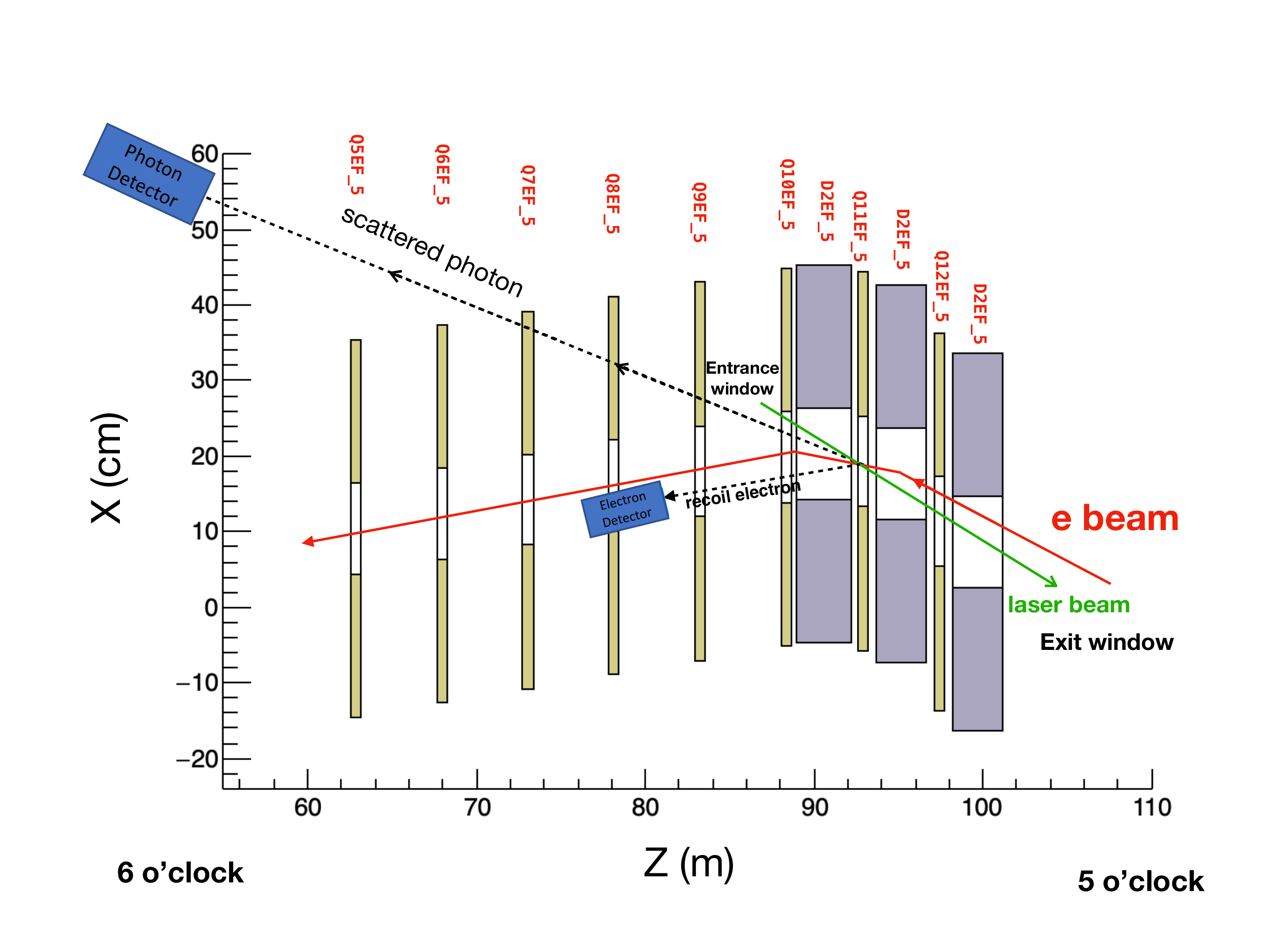}
  }
  \caption{Layout of the Compton polarimeter at IP 6. The laser-electron collision point is at Q11EF.  In contrast to the IP 12 layout, the backscattered photons will not clear the outer aperture of the downstream quadrupoles.  However, an opening for the photons can be accommodated between the quadrupole coils and by creating a small hole in quadrupole steel where needed.\label{fig:compton_ip6}}
\end{figure}

Plans for electron polarimetry at EIC include a Compton polarimeter at IP 12 or near IP 6.  A Compton polarimeter could be easily accommodated at IP 12 (where the electron beam is primarily vertically polarized), however this location is far from IP 6 where the main physics detector will be located. Although the region near IP 6 is crowded, a Compton polarimeter can also be placed in this area with careful attention to integration of the polarimeter with the beamline elements.  It is worth noting that a Compton polarimeter at IP 6, while closer to the main experiment, would measure a mix of longitudinal and transverse polarization ($P_L$=70\% at 18\,GeV and $P_L$=98\% at 5\,GeV), rather than the purely longitudinal polarization expected at the detector IP.  A schematic of the placement of the Compton polarimeter at IP 12 is shown in Fig.~\ref{fig:compton_ip12} and at IP 6 in Fig.~\ref{fig:compton_ip6}.


As noted above, a key requirement of the Compton polarimeter is the ability to make polarization measurements for an individual bunch.  The measurement time to achieve a statistical precision $dP/P$ is given by a combination of the luminosity, Compton cross section,  and analyzing power:

\begin{equation}
    t_{meth} = \left( \mathcal{L}\ \sigma_{\rm{Compton}}\ P_e^2 P_\gamma^2 \ \left(\frac{dP_e}{P_e}\right)^2 \ \rm{A}_{eff}^2\right)^{-1}.
\end{equation}

The effective Compton analyzing power, $\rm{A}_{eff}$, depends on the measurement technique; in order of increasing effective analyzing power, these are integrated, energy-weighted integrated, and differential.  For measurement time estimates here, we will use the smallest analyzing power (i.e., integrated) to be conservative.


Nominal electron beam parameters at IP 12 are provided in Table~\ref{tab:ebeam_ip12}.  Beam properties are identical at the IP 6 Compton location, with the exception of the beam size, which is about 40\% larger horizontally and about a factor of 2 smaller vertically.  Of particular note is the relatively short bunch lifetime at 18\,GeV.  Since measurement of transversely polarized electron beams is generally more time consuming than for longitudinal polarization, we focus on measurements at IP 12 to set a conservative upper limit on the time required for polarization measurements.  Table~\ref{tab:time532} shows the average transverse analyzing power, luminosity, and time required to make a 1\% (statistics) measurement of the beam polarization for an individual bunch, assuming a single Compton-scattered event per crossing.  The constraint of having a single event per crossing is related to the need to make a position sensitive measurement at the photon and electron detectors.  Note that even with this constraint, the measurement times are relatively short and, in particular, shorter than the bunch lifetime in the ring.

\begin{table}[!htb]
\centering
\begin{tabular}{c|ccc}
beam property    & 5~GeV      & 10~GeV     & 18~GeV    \\ \hline
Bunch frequency  & 99~MHz     & 99~MHz     & 24.75~MHz \\
Beam size (x)    & 390~$\mu$m & 470~$\mu$m & 434~$\mu$m \\
Beam size (y)    & 390~$\mu$m & 250~$\mu$m & 332~$\mu$m \\
Pulse width (RMS)& 63.3 ps    & 63.3 ps    & 30 ps \\
Intensity (avg.) & 2.5 A      & 2.5 A      & 0.227 A \\
Bunch lifetime   & $>$30 min  & $>$30 min  & 6 min
\end{tabular}
\caption{Beam parameters at IP12 for the EIC nominal electron beam energies.}
\label{tab:ebeam_ip12}
\end{table}

\begin{table}[!htb]
\centering
\begin{tabular}{c|cccccc}
\multicolumn{1}{l}{beam energy {[}GeV{]}} & \multicolumn{1}{l}{$\sigma_{unpol}$ {[}barn{]}} & \multicolumn{1}{l}{$\langle A_{\gamma}\rangle$} & \multicolumn{1}{l}{t$_{\gamma}${[}s{]}} & \multicolumn{1}{l}{$\langle A_{e}\rangle$} & \multicolumn{1}{l}{t$_{e}${[}s{]}} & \multicolumn{1}{l}{L{[}1/(barn$\cdot$s){]}} \\ \hline
5                                         & 0.569                                           & 0.031                                 & 184                                & 0.029                             & 210                               & 1.37E+05                        \\
10                                        & 0.503                                           & 0.051                                 & 68                                 & 0.050                             & 72                                & 1.55E+05                        \\
18                                        & 0.432                                           & 0.072                                 & 34                                 & 0.075                             & 31                                & 1.81E+05                       
\end{tabular}
\caption{Asymmetries, measurement times needed for a 1\% statistical measurement for one bunch and needed luminosities for three different beam energies for a 532~nm laser.}
\label{tab:time532}
\end{table}

Even for a single electron bunch (circulating through the ring at a frequency of $\approx$75~kHz), the luminosities provided in Table~\ref{tab:time532} can be readily achieved using a single-pass, pulsed laser.
Since the electron beam frequency varies with energy, it would be useful to have a laser with variable pulse frequency.  A laser system based on the gain-switched diode lasers used in the injector at Jefferson Lab~\cite{Hansknecht:2006qq} would provide both the power and flexible pulse frequency desired.  Such a system would make use of a gain-switched diode laser at 1064~nm, amplified to high average power (10-20~W) via a fiber amplifier, and then frequency doubled to 532~nm using a PPLN or LBO crystal.  The repetition rate is set by the applied RF frequency to the gain-switched seed laser.

\begin{figure}[tb]
  \centerline{
  \includegraphics[width=\columnwidth]{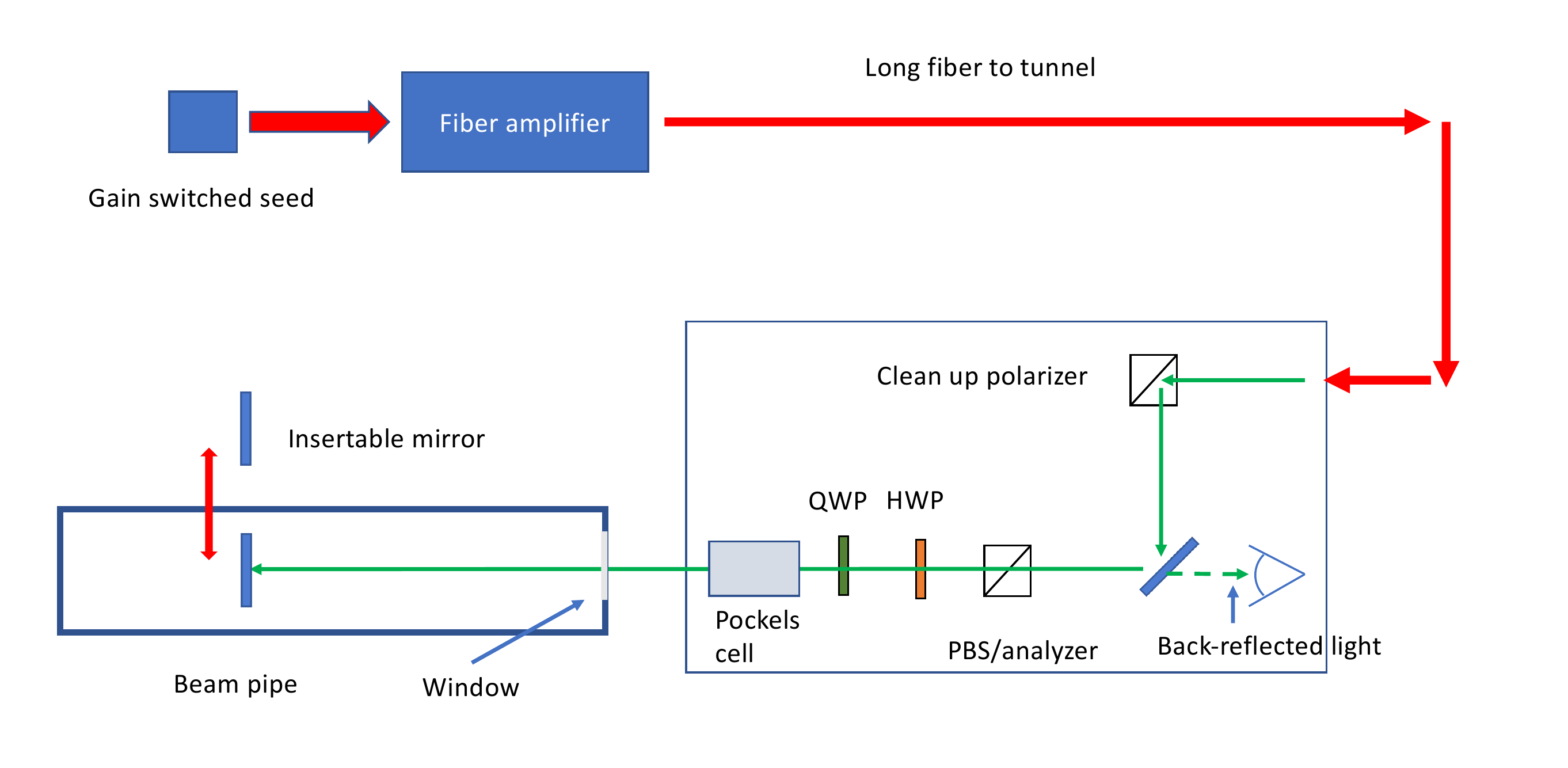}
  }
  \caption{Layout of the Compton polarimeter laser system, including diagnostics to accurately determine the laser polarization at the interaction point.\label{fig:compton_laser}}
\end{figure}

A laser system based on the gain-switched diode lasers used in the injector at Jefferson Lab~\cite{Hansknecht:2006qq} can provide all of the requirements noted above.  The proposed system will make use of a gain-switched diode laser at 1064~nm, amplified to high average power (10-20~W) via a fiber amplifier, and then frequency doubled to 532~nm using a PPLN or LBO crystal.  The repetition rate of the laser is dictated by an applied RF signal and can be readily varied.  In addition to the laser system itself, a system to set up and measure the laser polarization at the interaction point is required.  Determination of the laser polarization in the beamline vacuum is non-trivial due to possible birefringence of the beamline window under mechanical and vacuum stress.  We will employ a technique similar to that used at Jefferson Lab ~\cite{Narayan:2015aua, Vansteenkiste:93} that makes use of optical reversibility theorems to determine the laser polarization inside the vacuum using light reflected backwards through the incident laser transport system.  This polarization monitoring and setup system will require a remotely insertable mirror in the beamline vacuum so will need to be considered in the beamline design.  A schematic of the proposed laser system is shown in Fig.~\ref{fig:compton_laser}.

The detector requirements for the EIC Compton polarimeters are dictated by the requirement to be able to measure the transverse and longitudinal polarization simultaneously.  For longitudinal polarization, this means the detectors will require sensitivity to the backscattered photon and scattered electron energy.  The photon detector can make use of a fast calorimeter, while the electron detector can take advantage of the dispersion introduced by the dipole after the collision point to infer the scattered electron energy from a detector with position sensitivity in the horizontal direction.

To measure transverse polarization, position sensitive detectors are required to measure the up-down asymmetry.  This is particularly challenging given the very small backscattered photon cone at the highest EIC beam energy.  At HERA, the vertical position of the backscattered photon was inferred via shower-sharing between the optically isolated segments of a calorimeter~\cite{Sobloher:2012rc}.  Calibration of the non-linear transformation between the true vertical position and the energy-asymmetry in the calorimeter was a significant source of uncertainty.  The detector for the EIC Compton will measure the vertical position directly via segmented strip detectors, avoiding the calibration issues faced at HERA.

\begin{figure}[tb]
  \centerline{
  \includegraphics[width=\columnwidth]{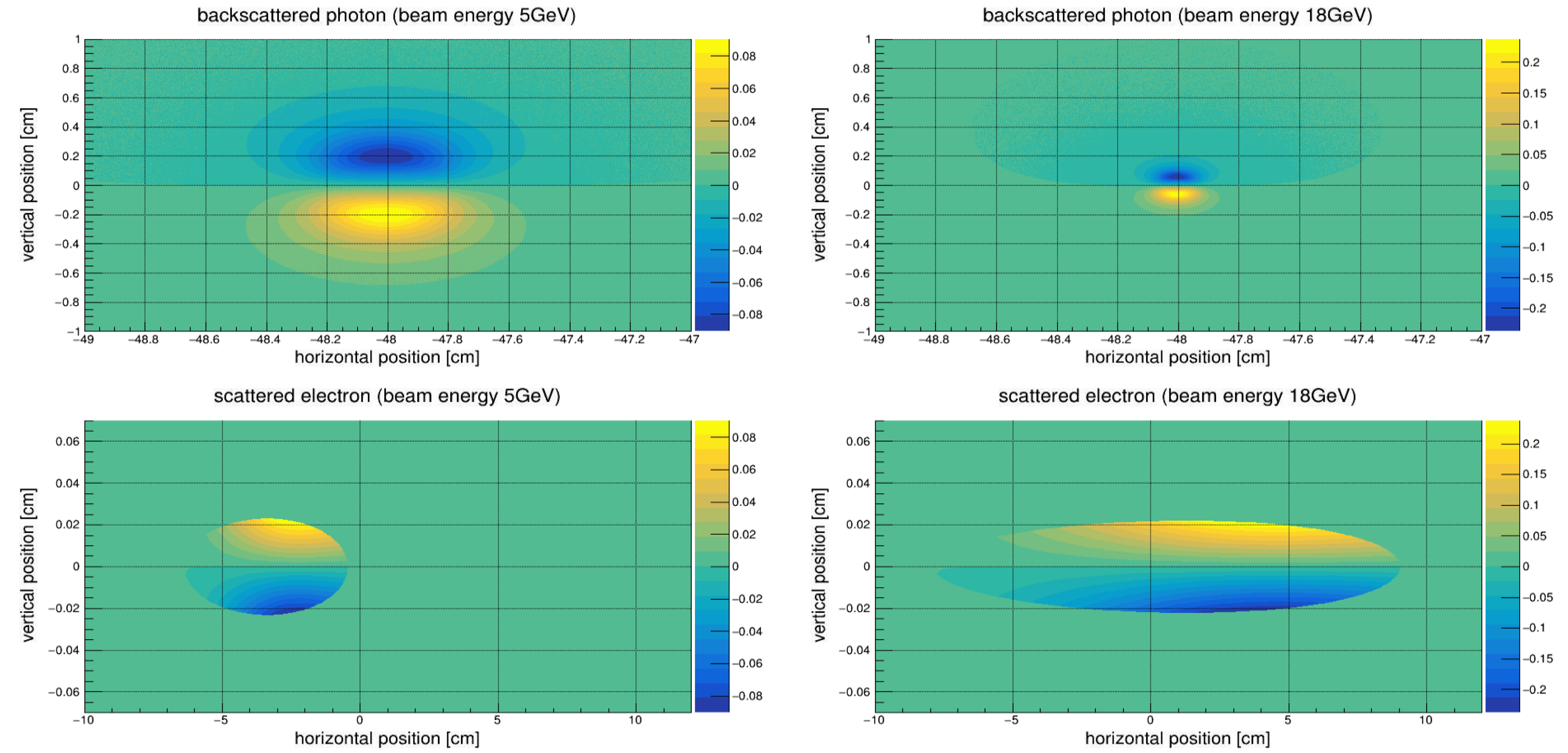}
  }
  \caption{Compton (transverse) analyzing power at the nominal photon and electron detector positions for the IP 12 polarimeter.\label{fig:compton_apower_meas}}
\end{figure}


The transverse Compton analyzing power vs. position at the detector for the IP 12 Compton for the backscattered photons and scattered electrons at 5 and 18\,GeV is shown in Fig.~\ref{fig:compton_apower_meas}.  The backscattered photon cone will be largest at the lowest energy (5\,GeV) - this will determine the required size of the detector. The distribution at 18\,GeV, where the cone is the smallest, sets the requirements for the detector segmentation.  Note that the scattered electrons are significantly more focused than the photons.  Monte Carlo studies indicate that the transverse polarization can be reliably extracted at 18\,GeV with a vertical detector segmentation of 100\,\si{\micro}m for the photon detector and 25\,\si{\micro}m for the electron detector.  The detector size should be at least 16$\times$16\,mm$^2$ for the photons and 10\,cm $\times$ 1\,mm for the scattered electrons.  The horizontal segmentation for the electron detector can be much more coarse due to the large horizontal dispersion introduced by the dipole.  Note that for the IP 6 Compton, the same conclusions apply for the photon detector size and strip-detector segmentation assuming the detector is the same distance from the laser-electron interaction point.  Initial estimates for the IP 6 Compton electron detector suggest that the needed detector size will be similar to the IP 12 detector, about 8\,cm  $\times$ 1\,mm,  and the same general conclusions with respect to the relative vertical and horizontal strip detector pitch will apply. 

Diamond strip detectors are a feasible solution for both the photon and electron detectors.  Diamond detectors are extremely radiation hard and are fast enough to have response times sufficient to resolve the minimum bunch spacing (10~ns) at EIC.  Tests of CVD diamond with specialized electronics have shown pulse widths on the order of 8~ns~\cite{Camsonne:june2017}.  For the photon detector, about 1 radiation length of lead will be placed in front of the strip detectors to convert the backscattered photons.  As an alternative to diamond detectors, HVMAPS detectors are also under consideration.  The radiation hardness and time response of HVMAPS will need to be assessed to determine their suitability for this application.

As noted earlier, the photon detector will also require a calorimeter to be sensitive to longitudinal components of the electron polarization. Only modest energy resolution is needed; radiation hardness and time response are more important requirements for this detector -  a tungsten powder/scintillating fiber calorimeter would meet these requirements.

Backgrounds are an important consideration for Compton polarimetry as well.  The primary processes of interest are Bremsstrahlung and synchrotron radiation.  Monte Carlo studies have shown that the contribution from Bremsstrahlung should be small for a beamline vacuum of 10$^{-9}$ Torr.  Synchrotron radiation, on the other hand, will be a significant concern.  Careful design of the exit window for the backscattered photons will be required to mitigate backgrounds due to synchrotron.  In addition, synchrotron radiation will impact the photon calorimeter linearity and this effect will have to included (along with other effects due to pileup, radiation damage, etc.) when determining the response of the system. On the other hand, the electron detector is not in the direct synchrotron fan, but significant power can be deposited in the detector from one-bounce photons.  This can be mitigated by incorporating tips or a special antechamber in the beampipe between the Compton IP and the detector~\cite{Camsonne:jan2016}.  The electron detector will also be subject to power deposited in the planned Roman Pot housing due to the beam Wakefield.  Preliminary simulations indicate the Wakefield power should not be large enough to cause problems, but this will need to be considered in the detailed Roman Pot design.

In addition to measurements in the EIC electron ring, it is important to be able to determine the electron beam polarization in or just after the Rapid Cycling Synchrotron (RCS) in order to facilitate machine setup and troubleshoot possible issues with the electron beam polarization.  In the RCS, electron bunches of approximately 10 nC are accelerated from 400 MeV to the nominal beam energy (5, 10, or 18\,GeV) in about 100 ms.  These bunches are then injected into the EIC electron ring at 1 Hz. The short amount of time each bunch spends in the RCS, combined with the large changes in energy (and hence polarimeter analyzing power and/or acceptance) make non-invasive polarization measurements, in which the the RCS operates in a mode completely transparent to beam operations, essentially impossible.  However, there are at least two options for making intermittent, invasive polarization measurements.

The first, and perhaps simplest from a polarimetry perspective, would be to operate the RCS in a so-called ``flat-top'' mode~\cite{Meot:2018uoh}.  In this case, an electron bunch in the RCS is accelerated to its full or some intermediate energy, and then stored in the RCS at that energy while a polarization measurement is made.  In this scenario, a Compton polarimeter similar to that described above could be installed in one of the straight sections of the RCS.  The measurement times would be equivalent to those noted in Table~\ref{tab:time532} (since those are for a single stored bunch), i.e., on the order of a few minutes.

Another option would be to make polarization measurements in the transfer line from the RCS to the EIC electron ring.  In this case, one could only make polarization measurements averaged over several bunches.  In addition, the measurement would be much more time consuming due to the low average beam current ($\approx$ 10 nA) since the 10 nC bunches are extracted at 1 Hz.

\begin{figure}[tb]
  \centerline{
  \includegraphics[width=0.7\columnwidth]{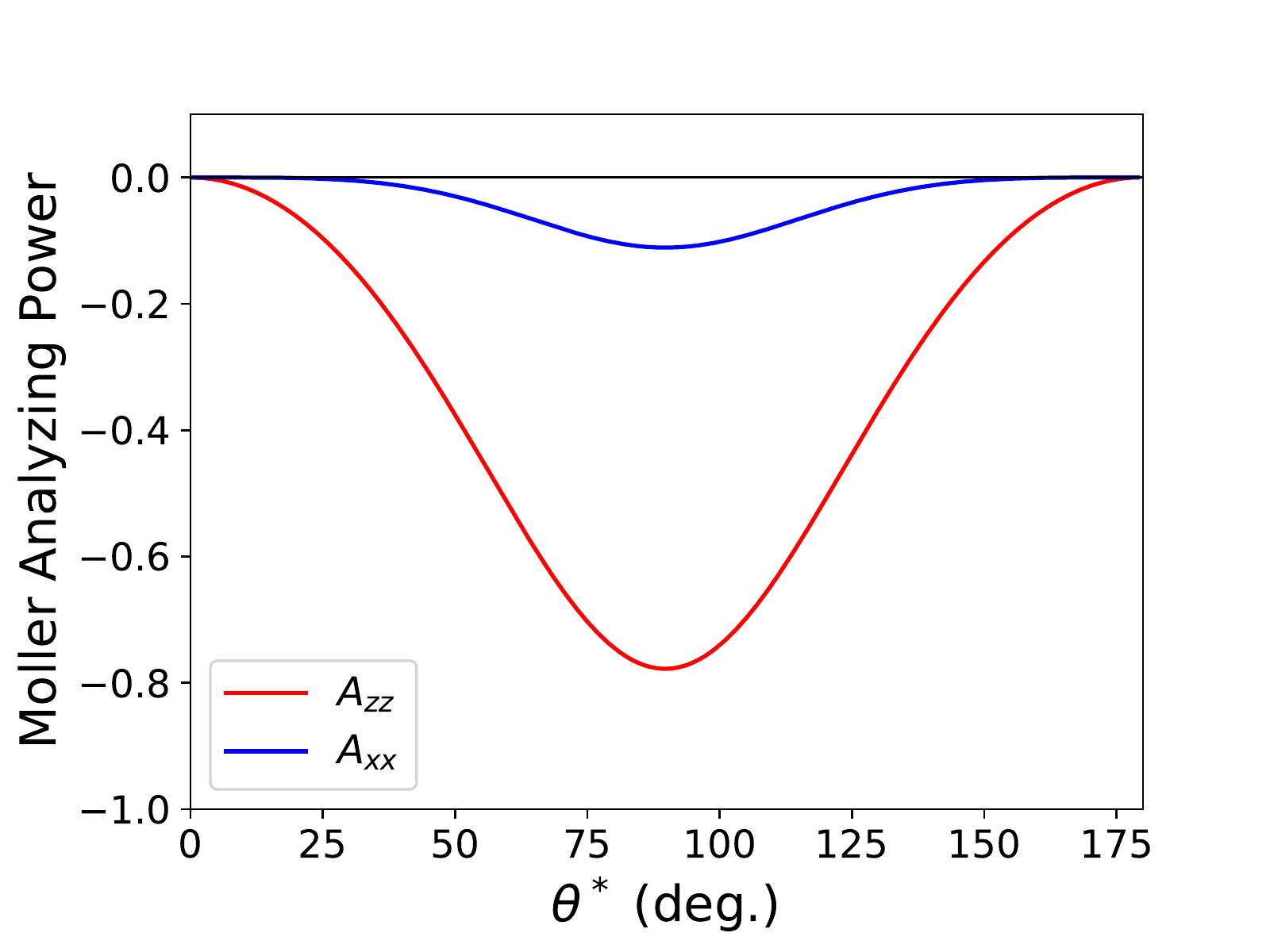}
  }
  \caption{Analyzing power for longitudinally polarized beam and target electrons ($A_{zz}$) and transversely polarized beam and target electrons ($A_{xx}$) vs. center of mass scattering angle, $\theta^*$.  The magnitude for both is largest at $\theta^*$ = $90^{\circ}$; $A_{ZZ}=-7/9$ and $A_{XX}$=-1/9.\label{fig:moller_apower}}
\end{figure}

The measurement time at 10 nA using a Compton polarimeter similar to the one planned for IP12 would take on the order many days.  The IP12 Compton limits the number of interactions to an average of one per crossing to be able to count and resolve the position of the backscattered photons.  A position sensitive detector that could be operated in integrating mode, would allow more rapid measurements. However, the required position resolution (25-100~$\mu$m) would be very challenging for a detector operating in integrating mode.
An alternative to Compton polarimetry would be the use of M\o ller polarimetry.  M\o ller polarimeters can be used to measure both longitudinal and transverse polarization and can make measurements quickly at relatively low currents.  The longitudinal and transverse M\o ller analyzing powers are shown in Fig.~\ref{fig:moller_apower} and are given by,
\begin{eqnarray}
  A_{ZZ} & = & -\frac{\sin^2\theta^* (7+\cos^2\theta^*)} {(3+\cos^2\theta^*)^2},\\
  A_{XX} & = & -\frac{\sin^4\theta^*} {(3 + \cos^2\theta^*)^2},
\end{eqnarray}
where $A_{ZZ}$ is the analyzing power for longitudinally polarized beam and target electrons, $A_{XX}$ for horizontally polarized beam and target electrons, and $\theta^*$ is the center-of-mass scattering angle.  Note that $A_{YY} = -A_{XX}$.  The magnitude of the analyzing power is maximized in both cases at $\theta^*=90^{\circ}$, where $|A_{ZZ}|=7/9$ and $|A_{XX}|=1/9$.

M\o ller polarimeters at Jefferson Lab can make (longitudinal) polarization measurements with a statistical precision of 1\% at average beam currents of 1~$\mu$A with a 4~$\mu$m iron foil target in about 15 minutes.  Electrons from the RCS will be transversely polarized, and the analyzing power will be a factor of 7 smaller, which implies a factor of 50 increase in measurement time for the same precision.  This smaller analyzing power combined with the low average beam current results in very long measurement times.  These long measurements times can be partially mitigated through the use of thicker target foils.  Even then, the measurements still take a significant amount of time - 1.5 hours for a 10\% measurement of the polarization using a 30~$\mu$m target.  While target foil thicknesses of 10-30~$\mu$m have routinely been employed in M\o ller polarimeters, it is possible that even thicker targets (perhaps a factor of 10 thicker) could also be used, reducing the measurement time further.  The maximum useful target thickness would need to be investigated.

\begin{figure}[tb]
  \centerline{
  \includegraphics[width=0.5\columnwidth,angle=-90]{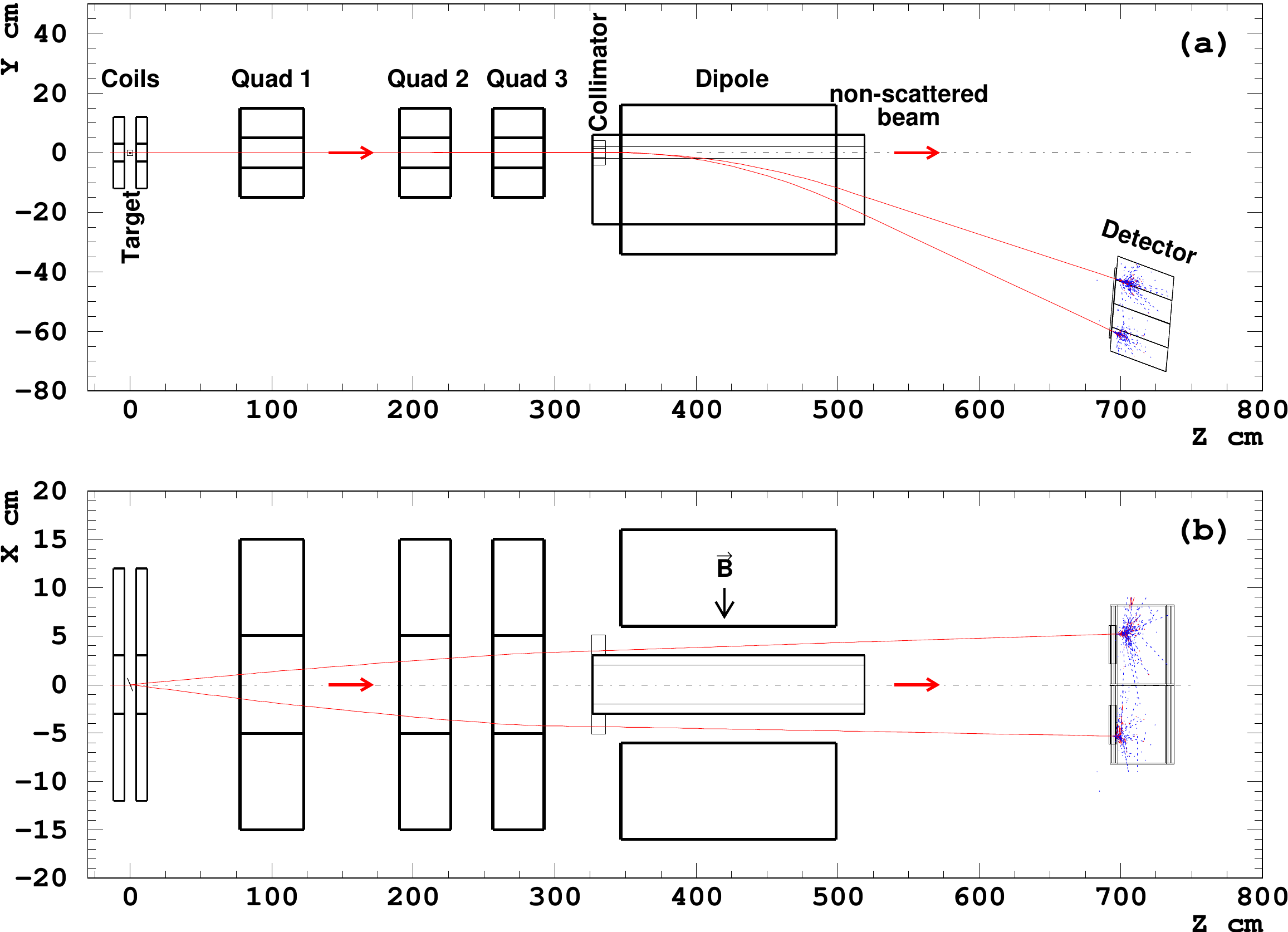}
  }
  \caption{Layout of the M\o ller polarimeter in experimental Hall A at Jefferson Lab.\label{fig:halla_moller}}
\end{figure}

A key drawback of M\o ller polarimetry is that the solid foil targets are destructive to the beam, so cannot be carried out at the same time as normal beam operations.  An additional complication is the requirement for a magneto-optical system to steer the M\o ller electrons to a detector system.  In the experimental Hall A at Jefferson Lab, the M\o ller spectrometer employs several quadrupoles of modest length and aperture, combined with a dipole to deflect the M\o ller electrons into the detector system (see Fig.~\ref{fig:halla_moller}).  The whole system occupies about 7~m of space along the beamline, but the space used by the quadrupoles can also be used for beam transport during normal operations (i.e., when M\o ller measurements are not underway).

The preferred choice for polarimetry at the RCS is a Compton polarimeter in the RCS ring, with measurements taking place during ``flat-top'' mode operation.  However, if this ``flat-top'' mode is not practical, then a M\o ller polarimeter in the RCS transfer line could serve as a reasonable fallback, albeit with reduced precision and a larger impact on the beamline design.

\subsection{Hadron Polarimetry}\label{sec-BD-Spin}

Hadron polarimetry has been successfully performed on RHIC polarized
proton beams for nearly two decades.
Through continual development a relative systematic uncertainty $< 1.5$\%
was achieved for the most recent RHIC polarized proton run.
As the only hadron polarimeter system at a high energy collider it is
the natural starting point for hadron polarimetry at the EIC.

\subsubsection{Proton Polarimetry at RHIC}

\begin{figure*}[ht]
  \centering
  \includegraphics[width=0.5\textwidth]
  {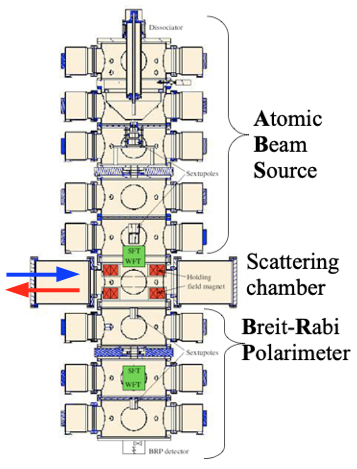}
  \caption{The RHIC polarized hydrogen jet polarimeter. The atomic
    beam source at the top passes polarized hydrogen across the beams (blue
  and read arrows) in the scattering chamber, with detectors left and
  right of the beams. The atomic hydrogen polarization is measured by
  the Breit-Rabi polarimeter at bottom.}
  \label{fig:Hjet_polarim}
\end{figure*}

Hadron polarization is typically measured via a transverse single spin
left right asymmetry: $\epsilon = A_N P$.
Unlike for polarized leptons, the proportionality constant is not
precisely known from theory.
Instead, RHIC employs an absolute polarimeter with a polarized
atomic hydrogen jet target (Hjet),
illustrated in Fig.~\ref{fig:Hjet_polarim}.
Target recoil protons from the elastic process $pp \rightarrow pp$ are
measured in detectors in the scattering chamber left and right of the target.
The hydrogen polarization vector is alternated between vertically up and
down.
The RHIC beam also has bunches with up and down polarization states.
By averaging over the beam states the asymmetry with respect to the
target polarization may be measured, and vice versa:
\begin{equation}
 \epsilon_{\rm target} = A_N P_{\rm target} \; \; \; \;
 \epsilon_{\rm beam} = A_N P_{\rm beam} \; .
\label{eq:Hjet_asyms}
\end{equation}
The equality of $A_N$ in these two relations relations is a result of
time reversal invariance for the purely elastic reaction.
The target polarization is precisely measured with a Breit-Rabi polarimeter.
Combined with the measured asymmetries the beam polarization is determined:
\begin{equation}
  P_{\rm beam} = \frac{\epsilon_{\rm beam}}{\epsilon_{\rm target}} P_{\rm target} \; .
\label{eq:Hjet_polar}
\end{equation}
The absolute polarization measurement is independent of the details of
$A_N$.

The recoil protons are detected in silicon detectors which measure
kinetic energy and time of flight (TOF); segmentation of the detectors
provides the proton scattering angle.
The energy-TOF relation allows identification of protons, separating
backgrounds from inelastic reactions.
The energy-angle measurement of missing mass allows
selection of the purely elastic
reaction required for the validity of Eq.~\ref{eq:Hjet_asyms}.
The resolution is sufficient to distinguish between the
$p$ and $p\pi$ masses, $\sim$140 MeV difference.
The analyzing power $A_N$ is energy dependent; selection of protons in a
fixed energy range defines an effective $A_N$ for the measurement.

The diffuse nature of the polarized jet target provides only a low
rate of interactions, resulting in a measurement limited to a
relative statistical precision of 5-8\% per RHIC fill;
it is not sensitive to
the inevitable decay of beam polarization throughout a fill.
Also, the jet target is wider than the beam and measures
only the average polarization across the beam.
The beam polarization is larger at the center than the edges
transversely; the polarization of colliding beams differs from
the average polarization due to this
effect~\cite{Fischer:2012zzc}.
The polarimeters must measure this transverse polarization profile to
provide correct polarizations for use by collider experiments.

\begin{figure*}[ht]
  \centering
  \includegraphics[width=0.5\textwidth]
  {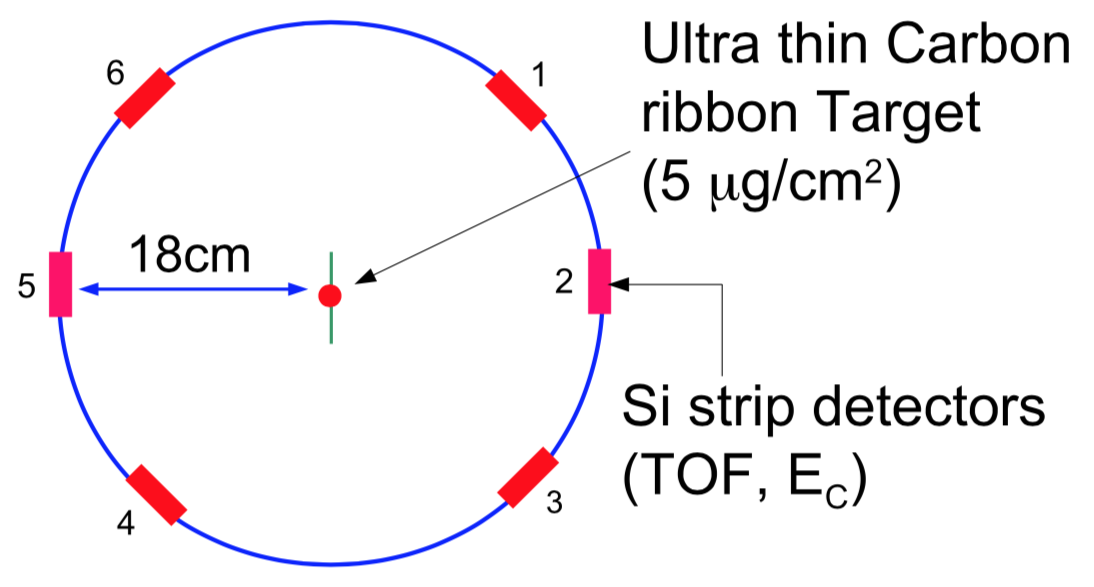}
  \caption{Cross section of the RHIC proton-carbon polarimeter.
    A thin carbon ribbon target is passed across the beam (into page)
    and scattered carbon nuclei are measured in the six detectors.}
  \label{fig:pC_polarim}
\end{figure*}

At RHIC the required finer grained polarization details are provided
by the proton-carbon (pC) relative polarimeter, illustrated in
Fig.~\ref{fig:pC_polarim}.
A thin carbon ribbon target is passed across the beam and scattered
carbon nuclei are measured in detectors arrayed around the beam.
The dense target provides a high interaction rate, allowing an
asymmetry measurement with a few per cent statistical precision in
less than 30 seconds.
Such measurements are made periodically throughout a RHIC fill,
providing a measurement of the beam polarization decay.
The ribbon target is narrower than the beam; thus it is able to
measure asymmetry as a function of position across the beam and
determine the transverse polarization profile.
The absolute polarization scale of the pC polarimeter is set by
normalizing an ensemble of pC measurements to the results from the
Hjet polarimeter for the corresponding RHIC fills.
A pC polarimeter is also installed in the AGS for polarization
measurements during injection.

The recoil carbon nuclei are detected in silicon detectors around the
target which measure kinetic energy and time of flight (TOF).
The energy-TOF relation allows identification of carbon nuclei,
reducing backgrounds from other particles.
A small background remaining under the signal is calibrated along with
the signal in the pC/Hjet normalization.
The analyzing power is energy dependent; selection of carbon nuclei in
a fixed energy range defines an effective analyzing power for the
measurement.

Several effects limit the precision of the polarimetry measurement,
including:
\begin{itemize}
  \item The atomic hydrogen may have a small amount of molecular $H_2$
    contamination not measured by the Breit-Rabi polarimeter,
    affecting $P_{\rm target}$ in Eq.~\ref{eq:Hjet_asyms}.
  \item Non-proton and inelastic proton backgrounds in the Hjet signal
    limit the validity of Eq.~\ref{eq:Hjet_asyms}.
  \item The silicon detectors in the pC polarimeter have a dead layer
    resulting in significant shifts in deposited energies;
    this results in instability of the measured energy range and
    effective analyzing power.
  \item Scattered carbon nuclei lose a significant fraction of energy
    exiting the target; mechanical instability of the targets causes
    this energy loss to vary, resulting in instability of the measured
    energy range and effective analyzing power.
\end{itemize}
Efforts continue to reduce these systematic effects in the RHIC
polarimeters.

For longitudinal spin runs, polarization measurements near the
interaction point and between the spin rotators are required to ensure
that the transverse component of the spin direction is zero.
At RHIC, the collider experiments use the process $pp \rightarrow Xn$,
which has a significant transverse single spin asymmetry for the
produced neutrons.
The neutrons are detected in the experiments' existing zero degree
calorimeters.
The asymmetry is zero when the beam spin is fully longitudinal.

\subsubsection{Proton Polarimetry at the EIC}

Both of the RHIC hadron polarimeters can in principle be used for
proton polarimetry at the EIC.
Beyond the existing limitations at RHIC, two significant difficulties
are presently foreseen, both a consequence of the increased proton beam
currents at the EIC.

First, backgrounds in both polarimeters are observed and lie partially
beneath the signal events.
They are distinguished by energy-TOF distributions different from the
signal allowing separation or estimation of a subtraction from the
signal.
The backgrounds exhibit a small asymmetry with respect to the beam
spin state.
At the EIC, with higher bunch crossing frequency, the backgrounds will
lie under the signal events from adjacent bunches and separation or
subtraction based on energy-TOF will not be possible.
Also, the adjacent bunches may have the same or opposite beam spin as
the central bunch, further convoluting the measurement.
Simulations indicate that the backgrounds are due to fast particles
(pions and protons) which deposit a fraction of their energy in the
silicon and pass entirely through the detector.
A second detector layer may allow vetoing such punch through events,
leaving the signal events which deposit their entire energy in the detector.

Second, materials analysis of the carbon ribbon targets indicates that
the higher proton beam currents
at the EIC will induce heating to temperatures causing the targets to
break after only a few seconds in the beam.
A search for alternative target materials has been initiated.

A possible alternative to the pC relative polarimeter has been
proposed which would mitigate both of these limitations.
It is based on the observation by the PHENIX collaboration of a large
azimuthal asymmetry of forward neutrons in the proton direction in
\pAu ~collisions~\cite{Aidala:2017cnz}.
This effect is well described by an Ultra Peripheral Collision (UPC)
process, in which the high Z Au nucleus emits a photon which produces
neutrons off of the polarized proton~\cite{Mitsuka:2017czj}.
The produced neutrons have a sizable transverse single spin
asymmetry.
A polarimeter based on this process would replace the Au beam with a
high Z fixed target as a source of photons.
A Xe gas jet may be a suitable target and would not suffer from the
heating limitation of the carbon targets.
The neutrons would be detected in a zero degree calorimeter.
The neutrons arrive at the calorimeter promptly and there is no
overlap with adjacent bunches as in the pC polarimeter.

The absolute and relative polarimeters may be situated anywhere in the
EIC hadron ring where the spin vector is transverse.
The Hjet and pC polarimeters each require 1-2 m space along
and transverse to the beam.

The forward neutron process $pp \rightarrow Xn$ used at RHIC to ensure
longitudinal spin at the  interaction points is not available at the EIC.
Instead, one relative polarimeter (pC or alternative) should be
placed near the experimental interaction point between the
spin rotators; a possible location of a pC polarimeter is shown in
Fig.~\ref{fig:overallIRLayout}, it would be located on the rear side between B2APR and Q2pR.
The hadron polarimeters are only sensitive to transverse spin
polarization.
During longitudinal spin runs asymmetry measurements near the
interaction point can verify that the transverse component
of the spin direction is zero.

\subsubsection{\texorpdfstring{$^3{\rm He}$}{} Polarimetry at the EIC}

For $^3{\rm He}$ absolute polarimetry at the EIC, it may be possible
to replace the hydrogen jet target with polarized $^3{\rm He}$.
Such targets have been used in numerous
experiments~\cite{
Anthony:1996mw,Ackerstaff:1997ws,DeSchepper:1998gc,Amarian:2002ar,Krimmer:2009zz,
Krimmer:2011zz,Long:2019iig,Okudaira:2020guz}.
The measurement must select purely elastic scattering events.
The lowest lying excitation of $^3{\rm He}$ is the dissociated
deuteron-proton system, with a mass difference of $\sim$5.5 MeV.
The missing mass measurement from the energy and angle of the recoil
target $^3{\rm He}$ may have insufficient resolution to distinguish
this state.
Detectors downstream of the target can aid tagging of breakup of
target or beam nuclei.

A polarized $^3{\rm He}$ target may only be available on a limited
basis.
In this case it may be used in special runs with an unpolarized proton
beam to determine the asymmetry for the elastic
$^3$He$\uparrow$p process.
During physics operation with polarized $^3{\rm He}$ beams,
the polarization may be continuously measured with an unpolarized
proton target, using the $^3$He$\uparrow$p asymmetry.
To conform with time reversal invariance,
the asymmetry calibration with polarized $^3{\rm He}$ target and
unpolarized proton beam must be performed at the same center of mass
energy as the operational polarized $^3{\rm He}$ beam with
unpolarized proton target.
For both calibration and physics operations, events with breakup
of $^3{\rm He}$ must be excluded to ensure purely elastic scattering.

The pC polarimeter, or an alternative, developed for protons at the EIC
should also provide suitable relative polarimetry for light ions.
A $^3$He\,C polarimeter would require a sufficiently large analyzing power
$A_N$ to provide measurable asymmetries.
Phenomenological analysis indicates its magnitude is $\sim$70\%
(and opposite sign) of
that for pC; this needs to be tested experimentally.

At some point polarized deuteron beams may be developed for the EIC.
The experiences gained from $^3{\rm He}$ will directly inform deuteron
polarimetry.

\subsubsection{R\&D Studies at RHIC}

Polarized proton runs are anticipated for RHIC in 2022 and 2024 before
EIC construction begins.
The polarimeters will be operational for physics measurements during
these runs.
They will also be available for studies to guide further development of
polarimetry for the EIC.

Some polarimeter developments may be tested during normal polarized
proton physics running. These include:
\begin{itemize}
\item Some polarimeter detectors, in both the Hjet and pC
  polarimeters, will have a second layer installed to study tagging of
  punch through events.
\item Any new technologies to replace the carbon targets can be tested
  in the pC polarimeter.
\item A polarimeter based on neutrons from the UPC process could be tested
  during proton running.
  It would require addition of a high Z target, such as a Xe gas jet.
  A zero degree calorimeter would be needed downstream of the target.
\end{itemize}

Unpolarized $^3{\rm He}$ beams have previously been operated in RHIC
for He-Au collisions at 103 GeV/nucleon.
In the next years polarized $^3{\rm He}$ beams will become available,
first at injection energy and later accelerated up to $\sim$58 GeV/nucleon.
These beams may be used for $^3{\rm He}$ polarimetry studies during
regularly scheduled accelerator physics experiments.
Also, the absolute polarimeter may be equipped with a $^3{\rm He}$
target, possibly polarized.
With various combinations of $^3{\rm He}$ beams and targets, numerous
tests for polarimetry may be conducted.
In increasing order of beam and target development, they include:
\begin{itemize}
\item Any $^3{\rm He}$ beam will allow tests of tagging beam nucleus
  breakup with downstream detectors; this will be possible even with
  the existing hydrogen target. The energy-TOF relation for carbon
  nuclei in $^3{\rm He}$+C  scattering may also be verified in the pC
  polarimeter; this has already been shown in the AGS pC polarimeter.
\item With any $^3{\rm He}$ beam and  $^3{\rm He}$ target in the absolute
  polarimeter, the  energy-TOF and energy-angle relations for
  $^3{\rm He}$+$^3{\rm He}$ scattering can be tested.
\item With any $^3{\rm He}$ beam and a polarized $^3{\rm He}$ target
  in the absolute polarimeter, the target asymmetry in
  Eq.~\ref{eq:Hjet_asyms} can be measured, establishing the asymmetry
  for $^3{\rm He}$+$^3{\rm He}$ scattering
\item With a polarized $^3{\rm He}$ beam, the beam asymmetry in
  Eq.~\ref{eq:Hjet_asyms} can be measured with any $^3{\rm He}$ target
  in the absolute polarimeter, establishing the asymmetry for
  $^3{\rm He}$+$^3{\rm He}$ scattering, and with the pC polarimeter,
  establishing the asymmetry for $^3{\rm He}$+C scattering.
\item With both the $^3{\rm He}$ beam and target polarized, the beam
  polarization may be measured as in Eq.~\ref{eq:Hjet_polar}.
\end{itemize}

\section{Readout Electronics and Data Acquisition}
\label{part3-sec-Det.Aspects.DAQ_Electronics}

\subsection{Introduction}

The Readout Electronics and Data Acquisition system is an essential component for the future EIC detectors. The readout electronics is responsible for processing the electric signals from the various detector sensors and converting them into a numerical representation that can be handled by a digital system. The DAQ system, on the other hand, is responsible of collecting, filtering, and storing these data. The overall system must be designed keeping into account the constraints dictated both by the physics program and by the operation environment.

For these reasons, the architecture of the readout system has a very strong impact on the physics program that can be performed at the future EIC experiments. The front-end electronics have to be adapted to the characteristics of the sensors to be equipped, and to the measurements which have to be done with them. In parallel, the DAQ system must offer performance adapted to the data flow coming from these front-end electronics. Filtering features of the DAQ system could be required, in order to maintain the data flow at acceptable level, taken into account the limitation in term of bandwidth of this system. But such a feature would affect directly the EIC physics outcome, since any data discarded at the online level will be lost irretrievably - a careful design, construction, and validation of this system is thus necessary.

This section aims to review the possible solutions on which the readout and DAQ system for the EIC experiments could be built. Hypothesis in term of detector characteristics and data flux are considered, leading to a reflection on the possible architecture on which the DAQ system could be based. Efforts made to validate the proposed architectures are also described. At last a description of the state of the art of the detector front-end electronics is proposed, with a few hypothesis on what could be the possible evolution in this domain.

\subsection{Glossary}

Several terms used in the DAQ and readout electronics domains could be ambiguous or meant differently from one reader to the other. In order to lift up the ambiguities several of these terms are defined below. These definitions are the reference for the whole section.

\subsubsection{Readout electronics terms}
\begin{description}
\item[Front-end electronics (FEE):] The electronics which amplify and put in shape the signals of the detector. After this stage the analog signals are generally digitized using analog-to-digital (ADC), charge-to-digital (QDC), or time-to-digital (TDC) converters \footnote{TDC: digitizer which measures times of amplified signals going above a given threshold.}. FEE is typically associated to data treatment, data bufferization and logic for data transfer to the downstream element in the read-out and DAQ chain. Digitization and data treatment stages are often directly integrated in some of the existing front-end chips.

\item[Amplification stage:] Groups the preamplifier + amplifier/shaper of the detector raw analog signals

\item[Embedded amplification stage:] Preamplifier + amplifier/shaper directly integrated into the detector hardware

\item[Digitization stage / Digitizer:] Transforms amplified signal into digital values (amplitudes, charges and/or times)

\item[Bufferization / data concentration stage:] Setup which concentrates and stores temporarily digital values from several digitizers before to send them to the DAQ, could do data selection and/or reconstruction

\item[Peaking time:] Time between the beginning of the pulse and its maximum after the amplification/shaping stage

\item[Occupation time: ]Time between the beginning and the end of the pulse after the amplification/shaping stage

\item[Analog memory:] Temporary storage of samples of analog signals, generally made of capacitor arrays, before digitization. Allow to select the samples which will be digitized

\item[Amplifier chip:] ASIC which groups the preamplifier and the amplifier/shaper

\item[Digitizing amplifier chip:] ASIC which groups the amplification and the digitization stages
\end{description}
\subsubsection{Data acquisition system terms}

\begin{description}
\item[Triggered readout:] A data acquisition system in which some data from a subset of detectors (“trigger data”) is sent to a dedicated subsystem to produce a trigger decision. This is usually a hardware system, generally based on programmable devices such as FPGAs. The trigger decision is based on a partial elaboration of the “trigger data”. “Trigger primitives” are reconstructed and analyzed to assess whenever all the data from the detector has to be stored for later analysis. In this case, a proper signal is sent back to all the readout elements to control the conversion of detector signals into the digital domain, or to trigger the read-out of a data-window from a continuously filled buffer. A key aspect of a triggered readout system is the fixed latency between the physical event time (FE $\rightarrow$ Trigger system) and the trigger time (Trigger system $\rightarrow$ FE) - in case of systems with multiple trigger levels, this is true for the first-level trigger.

\item[Pipelined/buffered readout: ]A triggered readout system where event data is stored on the front ends and read out asynchronously by the backend when the trigger signal is received.

\item[Second-level / high-level filtering:] In triggered systems, higher-level triggers are often used to reduce deadtime (via a fast clear) or data amount (by dropping the so-far recorded data for that event). Each level in such a system typically has different time constraints and complexity limits. For example: a certain time frame could not be forwarded to the tracker if certain conditions are not met. In certain, complex, triggered setups, the later stages can resemble a streaming system, where a stream of events flows through a network of analysis nodes, and data selection criteria either accept or drop the event. The main remaining difference for this part is then that the data is organized and tagged by an event number instead of time stamps. 

\item[Streaming readout (SRO):] A data acquisition system characterized by a unidirectional data flow from front-end electronics to the storage system.  Each channel, independently, records data over a certain threshold and streams them to a CPU farm for further elaboration. In a streaming readout system there are no dedicated systems to control the conversion into the digital domain or readout of a buffer. Different implementations of streaming readout are possible, depending on the manipulations and filtering applied online to the data.

\item[Unfiltered readout: ]A streaming data acquisition system without any system dedicated to event filtering / building. Only minimal zero suppression at the front-end level is adopted. Data is streamed directly from the front-end electronics to the storage system. Each detector hit is saved together with its time-stamp. 

\item[Zero suppression: ]Removal of data if close to the no-signal level of the detector. For example, in ADC data, removal of the signal digital values below a given threshold.

\item[Noise suppression: ]Removal of data produced by intrinsic or extrinsic detector noise, for example by correlation with neighboring channels or shape analysis.

\item[Feature extraction: ]Calculation of higher-level information. E.g. calculation of hit time and energy from ADC\footnote{ADC: digitizer which measures the amplitude of one or several samples of the amplified signals.} samples, or calculation of track information from hits. Often, but not necessarily, accompanied with the removal of the underlying lower-level data.

\item[Online Physics analysis: ]Analysis of the high-level information provided by the feature extraction steps to produce physics-relevant information (e.g. missing mass).

\item[Data selection: ]In a SRO system, data can be algorithmically selected for further processing and long-term storage. Data not selected are dropped and are not processed further. This is equivalent to the function of first and higher-level triggers in triggered systems but can make use of all detector information and results from further analysis steps including feature extraction and online physics analysis.
\end{description}

\subsection{Overview on DAQ Structure}

Most of the past and currently running particle-physics experiments adopt a DAQ system based on a triggered setup, usually with a multi-layer architecture.
 Usually the first data reduction is achieved by using dedicated boards where a significant filtering is applied by selection algorithms implemented on FPGAs, while the subsequent trigger layers are based on software components: a CPU farm reduces the data stream to a manageable size for storing and off-line processing and applies a second, more sophisticated, level of filtering. 
 The main limitations of a FPGA-based trigger, where FPGAs are actively involved in the events-selection, reside in: the difficulty of implementing algorithms over a certain degree of complexity and sophistication; the difficulty of optimizing the selection criteria that requires reprogramming the boards each time a change is implemented; the partial information accessible at front-end level both in the term of quality (usually it incorporates only basic calibration) and quantity (trigger is usually performed using a limited subset of the full detector).

These limitations may directly affect the ultimate detector performances and the quality of recorded data since only partial information is available at trigger level, when the decision whether to write or not an event to tape has to be taken. Another drawback of this approach consists in the difficulty of changing the FPGA-board in case of unexpected experimental configuration changes or upgrades requiring more trigger resources. 


At the same time, complicated hardware or firmware implementations of trigger logic are hard to characterize, they bias the collected, while the bias assessment can be extremely difficult.

All these issues are largely solved when moving to a full (CPU) software-based system. The FPGA-based system may be replaced by a fully triggerless approach that removes the hardware trigger, performs the full on-line data reconstruction and provides precise selections of (complicated) final states for further high level physics analysis (a similar effort is currently faced at LHC in preparation for the high luminosity upgrade). 

In a triggerless data acquisition scheme, each channel over a threshold implemented on the front-end electronics is transferred after being labeled with a time-stamp, disregarding the status of the other channels. A powerful station of CPUs (usually an on-line farm), connected by a fast network link (usually optical fibers) to the front-end electronics, receives all data from the detector, reorganizes the information ordering hits by time, includes calibration constants, and, at the end, applies algorithms to find specific correlations between reconstructed hits (online event reconstruction), eventually keeping and storing only filtered events. Advantages of this scheme rely in: making use of fully reconstructed hits to define a high-level events selection condition; online algorithms implementation in a high-level programming language; easy reprogramming to upgrade the system configuration and accommodate new requirement. Furthermore, the system can be scaled to match different experimental conditions (unexpected or foreseen in a planned upgrade) by simply adding more computing (CPUs) and/or data transfer (network switches) resources. We underline that FPGAs are still used in a streaming-readout DAQ system, not to take decisions concerning events to select, but to make more ``low-level'' tasks such as adding the time-stamp to the data or canalize the data.

A triggerless option may result in: on-line implementation of calibration parameters, providing a more precise reconstruction of the kinematic quantities; implementation of more sophisticated reconstruction algorithms for a better reconstruction of close-by tracks; improvement in EM/hadron discrimination for a more efficient background rejection. 

These considerations directly apply to the EIC. The EIC physics program will be carried out by measuring different reactions with at least one electron in the final state. Electromagnetic calorimeters will thus play a key role in the online events selection and filtering. For these sub-detectors, a triggerless option may result in: on-line implementation of calibration constants to compensate for longitudinal and transverse EM shower leakage and gain variation, providing a more precise reconstruction of the energy deposition (and therefore an improvement in the ultimate energy resolution); implementation of more sophisticated clustering algorithms for a better reconstruction of close-by tracks allowing to resolve gammas from $\pi^0$ in a wider kinematics; improvement in EM/hadron shower discrimination for a more efficient pion rejection.

A triggerless scheme will facilitate future extensions of the envisaged EIC physics program. For instance, hadron spectroscopy requires  identification of rare exclusive finals states difficult to access experimentally (e.g.\ kaon-rich reactions). This would require implementation at trigger level of  multiple and sophisticated algorithms to select the physics of interest. Same rational is valid for other physics program that will be considered in the future.

The triggerless scheme is also an opportunity to extend the integrated IR-detector design to analysis to optimize physics reach as described above and to streamline workflows. A seamless data processing from DAQ to analysis would allow for a combined software effort for the triggerless scheme, online and offline analysis and to utilize emerging software technologies, e.g. AI / ML, at all levels of the data processing. A near real-time analysis at the EIC with auto-alignment and auto-calibration of the detectors and automated data-quality monitoring would enable significantly faster access to physics results and accelerate science. 
\newline
\newline
For these reasons, we are studying and developing a full streaming-readout DAQ system for the EIC detector, integrating all the sub-detector components.

\subsection{Constraints and Environment}
\label{part3_DetAspects.DaqElectronics.Constraints}


The EIC readout and DAQ system should be designed considering the following constraints, dictated both by the physics program (measurements to be performed) and by the experimental environment. The overall goal for the system, as an integrated component of the EIC detector, is to make it possible to complete the challenging EIC Science program, providing a seamless integration from the DAQ to the physics analysis.

The EIC detector will be made by many sub-components, based on different technologies and with different requirements concerning the values to be measured by them. This translates into specific constraints on each readout solution, in terms of needs and performances. In general, each sub-detectors will introduce its own requirements on the FEE parameters (shaping time, peaking time, gain, $\ldots$). Specific constraints can be mentioned for the different kinds of detector foreseen for the EIC experiments.

\begin{description}
\item[Calorimeters: ]It is anticipated that one of the most realistic options to read out EIC calorimeters will be SiPM (or matrix of SiPM`s) photosensors. The signals coming from these sensors would have a maximum amplitude at a level of 2~V, with a peaking time of 4~ns and a total duration around 60~ns. Several observables should be extracted from the signals: time, amplitude and integral of the pulses, quality of the pulse reconstruction, presence and correction of pile-up, baseline level. The dynamic range of the signals would be also very large, as it is directly correlated to the energy of the detected particles, which would be from 20~MeV to 20~GeV. A 12-bits dynamic range may be too limited, while 13 or 14~bits would allow to have a threshold well below 20~MeV level, corresponding to more than 3 ADC counts, which would let enough dynamic range to take into account the energy resolution and the non-zero baseline level. A compression feature of the readout electronics could also be a solution to compensate a limited number of bits of the ADC. The signal rate would be limited for the electromagnetic calorimeters to 100~kHz/channel over 50 to 100x10³~channels, and for hadronic calorimeters to 500~kHz/channel over around 10x10³~channels. 

To realize the electronics readout chain it would be necessary to develop a novel FE ASIC chip, in which the ADC board would provide the bias voltage to the SiPM and allow for signal amplification, processing and readout. In order to reconstruct the observable a high-rate sampling of the waveform would be necessary in the new chip, with a sampling rate up to 250~MHz during a time gate of 100 to 200~ns. The chip would integrate a digital treatment of the signals in order to extract the observables, or would have to transmit the raw sampled signals to the rest of the electronics chain.

\item[Silicon trackers: ]These pixel detectors will present a very large number of channels as described in the section \ref{part3-sec-Det.Aspects.Tracking}. They will include a front-end directly integrated in the silicon layers, and therefore do not require an external readout electronics. But such a large amount of channels may induce constraints on the DAQ chain in term of data flow, depending on their noise level which is not well determined for the moment.

\item[Gaseous trackers: ]Micro-pattern gaseous detectors (MPGD) are planed to be used for the front tracker and possibly in the barrel tracker. The foreseen number of channels is around 220 thousand, with a rate around 14~Mhit/s for the front tracker. The amplitude of MPGD signals is very low, with a maximum charge at the pC level over a period of a few 10~ns. They require low noise amplifier chips, described in the subsection \ref{readoutElectronicsStateArt}, in order to measure the time and the charge of the signals.

\item[Time projection chamber (TPC): ]The EIC TPC would be using MPGD as ionization electron detectors. The constraints on their read-out would be similar to the other MPGD detectors, excepted the duration of their signals which would be much more longer as they cover the drift time of the ionization electrons over the whole length of the chamber. The readout electronics should then open a time gate up to 20~\(\mu\)s long in order to measure electrons from the whole TPC volume.

\item[RICH-GEM: ]The signals of this detector are similar to the other kinds of MPGD. The only difference would come to the signal amplitude range which would be lower as they are coming from unique photoelectrons. A larger gain of the front-end electronics would be necessary compare to the other MPGD readout.

\item[mRICH, dRICH and DIRC: ]These detectors are read by photosensors, like SiPM, multi-anode PMT or MCP-PMT (cf section \ref{PID-PhotonDetection}). Signals given by these sensors have a quite large, with a maximum of one or a few volts and a pulse length of a few 10~ns. The main observable to be extracted from the signals is the time of the pulse, reconstructed from a threshold applied on the shaped signal, or from a fast digitization of it. A time resolution at the level of 100~ps to 1~ns should be achieved, depending of the kind of detector. Around 500 thousands of channels are foreseen to be read in total.

\item[psTOF: ]These detectors are also read by photosensors like LAPPD or LGAD, with similar signal characteristics compared to RICH and DIRC sensors. However the time resolution required for the readout is an order of magnitude lower than for the other ones, with values around 10 to 20~ps which is very challenging.

\end{description}

Similarly, each sub-detector will be characterized by different radiation levels, affecting the choice of the readout technology.

\begin{figure}
    \centering
    \includegraphics[width=\textwidth]{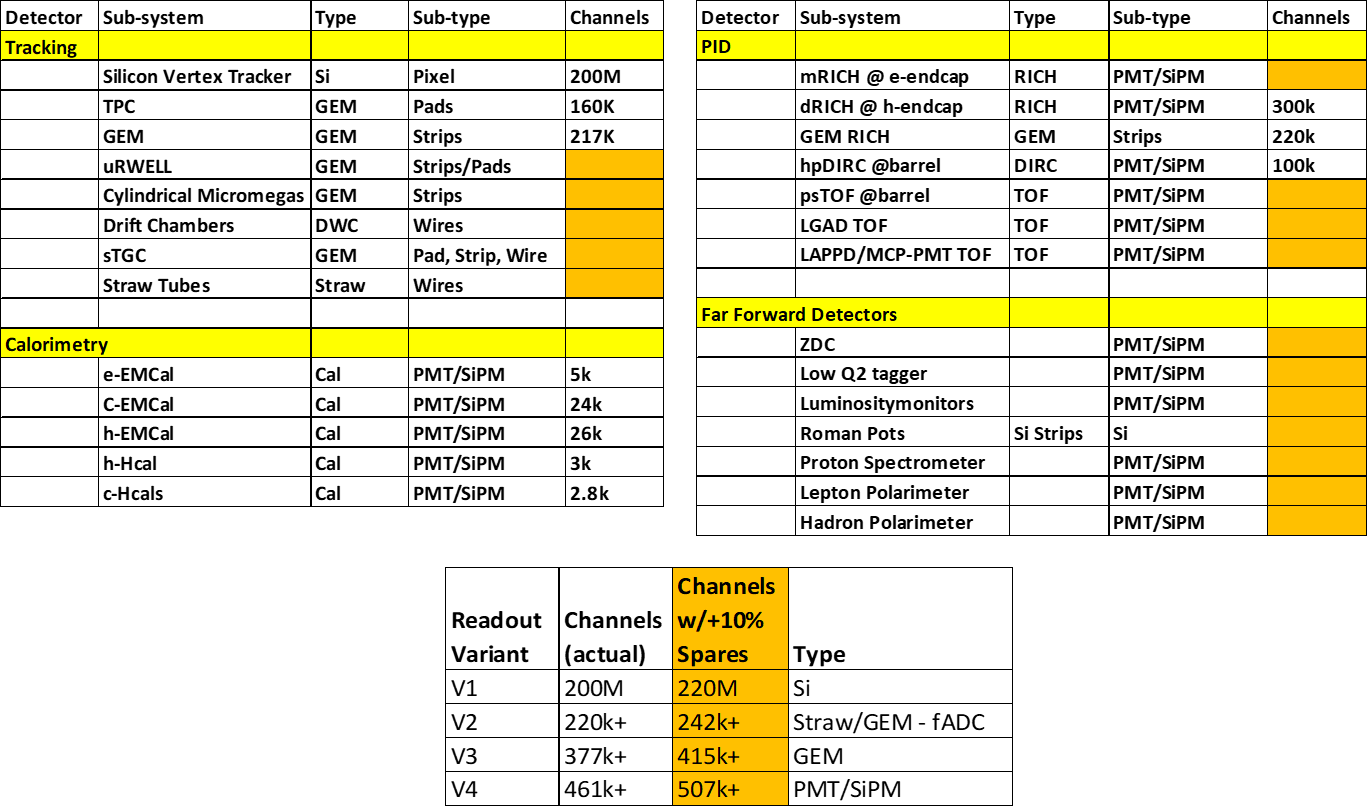}
    \caption{Estimate of the number of different EIC Readout Channels.}
    \label{fig:Part3_Figures.DetAspects.DaqElectronics.EICReadoutChannels}
\end{figure}
\begin{figure}[bh]
    \centering
    \includegraphics[width=\textwidth]{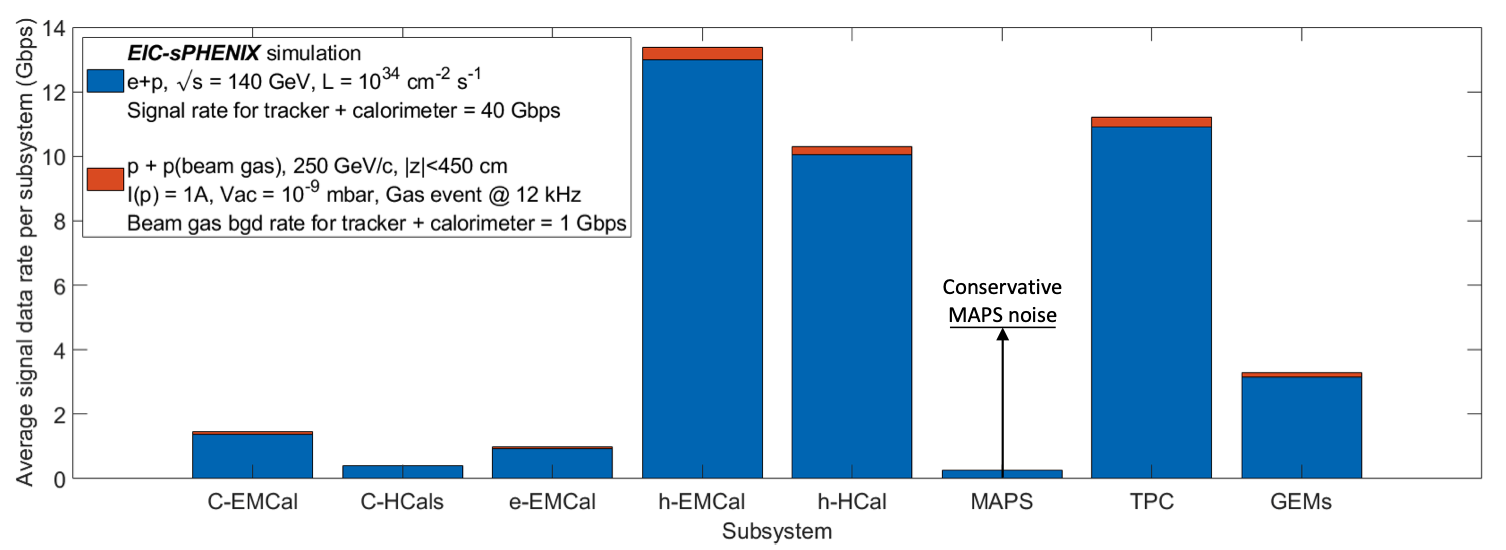}
    \caption{Collision data rate from each detector subsystem for the EIC sPHENIX detector model, at luminosity $\mathcal{L}=10^{34}$ cm$^{-2}$s$^{-1}$~\cite{Adare:2014aaa,sPH-cQCD-2018-001}. The total collision signal is approximately 100 Gbps, including a conservative estimate of the MAPS noise motivated by the recent ALICE ITS2 experience. We note this rate include collision signal only to record down all EIC physics events. In case excessive background rate, e.g. synchrotron photon hits, are observed, further noise and background filtering would be required. }
    \label{fig:Part3_Figures.DetAspects.DaqElectronics.SphenixRate}
\end{figure}
The number of channels anticipated for the EIC detector readout is shown in  Fig.~\ref{fig:Part3_Figures.DetAspects.DaqElectronics.EICReadoutChannels}; this initial estimates includes the large majority of the detectors. It is anticipated that three, and possibly four, different readout solutions will address the front-end readout needs of the various types of detectors.

The main constraint on the DAQ system is the total data rate to be processed, including both the signal (i.e. physics reaction of interest for the EIC physics program) and the background. A preliminary estimate of the total collision signal rate from the EIC detector was discussed in~\cite{sPH-cQCD-2018-001}, assuming $e+p$ collisions at $\mathcal{L}=10^{34}$ cm$^{-2}$s$^{-1}$ and the sPHENIX-based detector concept~\cite{Adare:2014aaa,sPH-cQCD-2018-001}. The calculation includes, for each component of this specific EIC detector model, the signal data rate from $e+p$ collisions and also from $p-p$ (beam-gas) interaction, and also considers a conservative estimate of the MAPS noise motivated by the recent ALICE ITS2 experience.  The result is summarized in Fig.~\ref{fig:Part3_Figures.DetAspects.DaqElectronics.SphenixRate}: a total collision signal rate from the EIC detector of approximately 100 Gb/s is expected.

Further constrains are introduced by the requirement of having, during EIC operations, an immediate online feedback concerning detector performance and data quality. Finally, the engineering requirements related the concrete EIC detector construction and assembly will introduce further constraints on the readout and DAQ system: available space, rating and standards to be satisfied, cooling power availability.

\subsection{Readout Electronics: Present State of the Art}
\label{readoutElectronicsStateArt}




\subsubsection{Introduction}

The role of the readout electronics for an experiment like EIC is crucial. The quality of the data delivered by the data acquisition system will be directly dependent on the performance of each element of the readout chain. The characteristics of these elements should be in accordance with the characteristics of the detectors which will be read by the electronics, as well as with the constraints which are described in the previous section.

The readout chain for a given detector is formed by electronics cards and chips with different functions: signal pre-amplification, amplification and shaping, digitization, data treatment like common mode noise reduction or zero suppression, data bufferization concentration, and transfer to the DAQ system. Side systems like readout trigger system can be also necessary to reduce the flux of data to be treated by the chain. At last several support systems are necessary in order for the readout electronics to work: clock signal distribution to synchronize all the electronics, slow-control to monitor the electronics behavior, power distribution, etc...

In this section a summary of the state of the art for the different elements of the electronics chain are given, with a few examples. Some of the chips described here regroup several functionalities listed above, for instance amplification and digitization.

\subsubsection{Front-end electronics}

The front-end electronics will amplify the signals from the detectors and put them in a shape compatible with the digitization step. The amplification step is important in particular for detectors which deliver very low amplitude signals, like silicon detectors or gaseous detectors. Other detectors like photomultipliers used in some calorimeters deliver larger signals, so the electronics gain should be lower in order to avoid any saturation. Another important parameter is the shaping applied to the signal, which can be characterized by the "peaking time", which is the time taken by the shaped signal to reach its maximum. A short shaping, for instance a few ns of peaking time, enables to get sharp output signals well adapted for fast detectors and fitted to time measurements, but may also induce a non-optimal noise figure. On the other hand slower shapes, in the order of a hundred of ns, are more adapted to slow detectors, for instance gaseous detectors, in order to integrate the totality of their signal and thus to get a more accurate amplitude measurement. A larger peaking time also induce a larger occupancy of the signal in the readout chain, which may limit the signal rate which can be read by the electronics.

In the current designs proposed for the future EIC experiments which are described in the chapter \ref{part1-chap-DetCon} of this document, both silicon and gaseous detectors are considered to measure the trajectory of the secondary particles. However the silicon detectors presently considered, the Monolithic Active Pixel Sensors (MAPS)~\cite{Greiner:2011zz}, integrate directly in the silicon die their own front-end electronics, signal processing and zero-suppression with adjustable threshold. These detectors return addresses of the hit pixels, with typically around 3 pixels per charged particle track. The thresholds are adjusted to give 99\% efficiency and less than $10^{-9}$ fake hit rate.

Several existing chips are dedicated to the readout of small signals coming from gaseous detectors. A few examples of chips used in particle physics experiments are presented below. They are all based on pre-amplifier and amplifier/shaper stages. However the treatment of the signals after these stages vary from one chip to the other, depending on the purpose of these chips. Some of them are more focused on the measurement of the signal time, combining then a fast shaping with a TDC stage, while others are measuring the amplitudes with flash ADCs. 32 to 64 channels are usually read by these chips which are 5 to 15~mm large. Peaking times are usually tunable, with values from 25~ns to 1~\(\mu\)s. Maximum charges accepted by these chips, also usually tunable, cover a range from 50~fC to a few pC. The internal capacitance of the detector channels also plays a role in the behavior of the pre-amplification stage. Depending on the design of this stage, a large capacitance, larger than 100~pF for instance, may alter the gain of the preamplifier and thus, of the whole chip. Some pre-amplification designs prevent this effect, allowing the chip to work with large detector capacitance at the level of several hundreds of pF. It is important to keep the power consumption as low as possible  for highly integrated detectors like the one foreseen for EIC, in order to limit the need of cooling. Power consumption values are typically around 10 to 30~mW/channel.

\subsubsection{Digitization and data treatment}

After amplification and shaping, detector signals are meant to be digitized before to be transmitted to the data acquisition system. Depending on the DAQ structure, signals may be continuously digitized, or the digitization can be triggered only when an interesting event happens. From one kind of chip to the other the digitization strategy can be different. Some chips, like the SAMPA chip (cf below) are indeed able to continuously digitize the signal at a rate of several MHz and to transmit these data to the DAQ. But depending on the kind of detector to be read and the information to be extracted this may or may not be an optimal strategy. That strategy is the most demanding in terms of ADC performance and output data link bandwidth. Present ADC integrated in readout chips are able to read continuously signals with a sampling rate around 10 to 20~MHz, with a ADC dynamics of 10~bits. Data links of a few Gbit/s are also common.
Data treatment may be necessary to reduce the data flux to a scale compatible with the DAQ capacity. Several kinds of algorithms can be applied: common mode reduction which compensate the part of the electronic noise which is common to all channels of a chip, zero suppression which discards the sample measurements below a given thresholds, peak finding, correlation with other detectors which conditions the preservation of the data with data from an other detector, etc. Such data treatments can be performed directly in the chip, for instance in an integrated DSP, or later by specific DSP electronics in the acquisition chain.

Another strategy which may be more adapted to detectors which do not require to store the full signal waveform, for instance trackers, would be to digitize only specific values like signal amplitude, using a sample \& hold (S\&H) circuit like in the VMM chip, or signal time. This strategy produces a much lower data flux. A last strategy is to not include any digitization of the signals in the readout chip, but rather to store them in analog memories which are arrays of capacitances, and to transmit in case of triggers the analog signals to commercial ADCs managed by a FPGA. This strategy is adopted by several chips like the AGET, the DREAM or the AFTER. However this strategy is largely incompatible with the streaming readout structure foreseen for the DAQ.

\subsubsection{Examples of readout chips}


Several chips representing the state of the art of the readout electronics are presented here. They concern mostly gaseous detectors and silicon detector front-end readout but their usage may be extended to other cases. Table~\ref{table:DAQ_Electronics.chip_characteristics} summarizes the characteristics of these chips.

\begin{description}
\item[ATLAS VMM: ]The VMM chip was developed at Brookhaven (BNL) as a 64-channel mixed signal ASIC for the readout of both the ATLAS Micromegas and sTGC detectors, specifically for the ATLAS Muon Spectrometer's New Small Wheel upgrade~\cite{NSW-Kawamoto:2013udg}. Its first version, VMM1, was a simple architecture chip, but a lot of functionality and features were added in the second version, known as VMM2. Another version, VMM3, and its revision VMM3a were also produced. The new versions were designed to contain enhanced features such as deep readout buffer logic, shorter TAC (Time-to-Amplitude Converted) ramps, SEU mitigation circuitry as well as handling of higher input capacitance of the order of 3~nF. The device was fabricated at IBM's foundry with a 130~nm IBM 8RF-DM technology (die size 15.3 x 8.3mm), housing approximately 5.2~million transistors (with nearly 160~k MOSFETs per channel), and packaged in a 1mm pitch 400-BGA (Ball Grid Array) package. It is indeed a state-of-the-art mixed signal ASIC device which aims to achieve the System on Chip (SoC) paradigm.

An excellent feature of the chip is having both time and amplitude (peak) detection circuitry on-board. For each of the 64~channels, a signal obtained from the input pads is amplified by a charge amplifier (CA) and after a shaping circuit (Shaper) is passed over to both a peak detector and time detector working in tandem and giving their respective output to a digitization section. The digitization section is comprised of a novel three-ADC chain in a so-called "Domino Architecture"~\cite{Geronimo4237405}. Output from the peak detector is given to both a 6-bit ADC for a dedicated low-delay output (50~ns delay), to be used for trigger or lower precision measurements, and to a 10-bit ADC for precision read-out, whereas the time detector has its output passed over to an 8-bit ADC for TDC functionality. Outputs from both 8~bit/10~bit ADCs are read-out through a FIFO buffer, which is designed to accommodate 4~MHz data in 16~µs latency window. In addition, a 12-bit coarse code time-stamp is provided to facilitate time measurements, which increments by an external clock providing a cumulative 20-bit timing information. The chip tests claim peak detection digitization process to complete in 250~ns driven by the 10~bits ADC.

The chip also features a novel third-order filter and shaper architecture with a DDF (Delayed Dissipative Feedback) topology. This architecture results into a higher dynamic range, enabling the measurement to achieve a relatively high resolution at very low input capacitance ($<$200~pF). The architecture offers a variable gain in eight values (from 0.5, 1, 3, 4.5, 6, 9, 12, to 16~mV/fC) with four possible shaping time intervals, viz. 25, 50, 100, and 200~ns. The VMM chip's latest versions seem to achieve the promised sub-fC and sub-ns resolutions at 200~pF and 25~ns capacitance and time windows, respectively. They can reach a data flow on the order of 1~Gbit/s. They can be employed for the readout of MPGD detectors requiring time precision at the order of a few ns.

\item[TIGER: ]TIGER is an acronym for the Torino Integrated GEM Electronics for Readout, a mixed signal ASIC chip first developed at INFN Torino. It is a general-purpose chip for readout from gaseous detectors with up to 64 channels, fabricated with a 110nm CMOS technology (fabricated on a die area of 5×5 mm2). While featuring a low-noise level of less than 2000 $e^-$, the chip offers a high input dynamic range of 2.0 to 50.0fC and gains of 12.4 mV/fC for time and 11.9 mV/fC for energy measurements, with time intervals of 60ns and 170ns, respectively. There is a provision of an on-chip calibration circuit which allows injected external pulses to calibrate the amplifiers and exploit the full input dynamic range. The signal conditioning circuitry in the time and energy measurement sections comprise both discriminator and pulse shaper in addition to a Time to Amplitude Converter which works in association with the ADC. A "Channel Controller", running at a clock speed of 200MHZ,  supervises the operation and synchronization of the charge integration, quantization, and time to amplitude conversion sections. The data are readout from the chip using Low-Voltage Differential Signaling (LVDS) standard links.
TIGER is a fine chip, with a simple yet elegant state-of-the-art architecture. Its major advantages include high input dynamic range, two high-resolution (10-bit) Wilkinson ADC's with very low non-linearity at each channel for both time and energy, and fast and a high-speed trigger-less readout, among other features, all offered with a reasonably low-power operation (less than 12~mW per channel while powered with 1.2~V).
The limitations or drawbacks include a bit higher ENC noise, limited value of input capacitance range, no digital processing functions, and possible internal analog signal conditioning structure supporting negative polarity signals only.

\item[SAMPA: ]The SAMPA chip has been designed as a 32-channel device with on-board pre-amplification (CSA charge sensitive amplifier), pulse shaping, quantization (digitizing) and DSP sections, including a high-bandwidth digital interface for computer readout. With the help of its eleven e-links with individual data transfer speed of 320~MB/s, it offers a sufficiently fast bandwidth (~3.4~Gbit/s) to readout all 32 channels, at a sampling rate of 10~MSPS.

The chip is fabricated with 130~nm CMOS technology with a chip area of 9.6 x 9.0~mm² and offered in a 372~Ball Grid Array (BGA) package. A charge-sensitive amplifier amplifies the measured analog signals, followed by a near-Gaussian pulse shaper, a novel element of the design. The 10-bit Successive Approximation ADC digitizes the amplified and shaped signals at a sampling rate of 10~MS/s (which can be configured to up to 20~MS/s), whereas the on-board DSP circuitry filters and carries out signal processing and compression operations on the digitized data. The chip offers a sufficiently high gain of 20-30~mV/fC with a low-noise performance (less than 1000~e\u207b).

SAMPA is a relatively modern chip suitable for high-performance applications. Its superior signal conditioning, digitization and on-board digital signal processing capabilities, as well as fast readout rates, are ideal for applications requiring a high-bandwidth, precision and versatile mixed signal data acquisition architecture. They may be well adapted to the readout of MPGD, in particular in the TPC and in the front trackers

\item[AFTER (ASIC For TPC Electronic Readout): ]The AFTER chip is manufactured with AMS CMOS 0.35~\(\mu\)m technology. The die area is of 7.8 x 7.4~mm² (involving 500,000 transistors). The final chip is produced in a 160-pin LQFP package: (28 x 28 x 1.4~mm). It offers 72~channels which can be preset for a negative/positive polarity by resistor arrays, with a counting rate of up to 0.3~Hz/channel. The chip has a power consumption of less than 10~mW/channel while powered at 3.3~V. This chip has a dynamic range of 120~fC-600~fC with an integral non-linearity of less than 2\% of LSB. However, it does not have an on-board ADC and requires an external ADC (with 20-25~MHz sampling rate). The specified peaking time range, as per the chip's technical sheet, is 100~ns to 2~µs (in 16 denominations). The sampling frequency range spans from 1~MHz to 50~MHz. Input signals sampled in circular analog memory buffers (in the form of a Switched Capacitor Array, SCA, with a depth of 511 time buckets). However, since the chip does not have an on-board ADC it needs an external one to digitize the SCA matrix signals. The SCA can be frozen by an external trigger. The minimum dead-time for the SCA is fixed at 79x40~ns*Number\_Of\_Time\_Buckets (out of 511).

As AFTER chips do not include digitization stage, they should be associated to external ADC ASIC. A suitable commercial or custom low-latency 12-bit ADC ASIC can be employed to work with the chip. A hybrid ASIC chip built by Pacific Microchip Corp. PMCC ADC~\cite{PMCC:32Ch_ADC} is generally employed, as it presents interesting features like 12-bit digitization for up to 32~channels, a 8~ns latency, a 8~Gigabit/s transfer glue-logic on-chip. The company claims to have a fabrication facility down to 7~ns with both CMOS and BiCMOS processes, and have worked with DOE in recent past.

\item[AGET (ASIC for General Electronics for TPC, GET system): ]The AGET chip is the very front-end of the GET system that performs the first concentration of the data from 64 input channels to one analog output connected to an external ADC. Each channel integrates a charge-sensitive pre-amplifier (CSA) with selectable signal polarity, a configurable shaper, a discriminator for multiplicity building and a 512-cell switch capacitor array (SCA). The gain and peaking times are tunable by slow control from 120~fC to 10~pC (4~values) and from 70~ns to 1~${\mu s}$ (16~values) respectively. The filtered signal is sent to an analog memory and discriminator inputs. The SCA for the analog memory is a 512-cell deep circular buffer in which the analog signal from the shaper is continuously sampled and stored. The sampling frequency is adjustable from 1~MHz to 100~MHz depending on the particular requirements of each detector. To process two consecutive events within a time window of 2~ms, such as the implantation of a radioactive ion followed by its decay, the SCA memory can be split into two halves using an adjustable parameter in slow control. The first signal that arrives is sampled and stored in the first half of the SCA memory. This is followed by a switch to the second half of the memory to sample and store the second signal. The system waits for this second signal to arrive for up to 2~ms. The switching from one half of the memory to the other corresponds to 2~sampling times. Sampling is stopped by a trigger decision. In the readout phase, the analog data from the different channels is multiplexed towards a single output and sent to the external 12-bit ADC at a readout frequency of 25~MHz. It is possible to read only a user-defined fraction of the 512~analog cells (1 to 512) beginning from an index defined with a constant offset from the cell corresponding to the trigger arrival. In addition to the 64~input signal channels, the AGET chip has 4 channels that are called fixed-pattern noise (FPN) channels. The inputs of these channels are not connected to the detector but they are treated by the SCA in exactly the same way. The chip is fabricated with 0.35~\(\mu\)m AMS CMOS technology and is 8.5 x 7.6~mm² large. It is housed in a LQFP 160-pin package.

\item[DREAM: ]The DREAM chip is of the same family as the AFTER and AGET chips, and shares most of its characteristics with the AGET. The most noticeable difference is the sensitivity to the input capacitance, as the DREAM is able to maintain its performance with large input capacitance with values up to 200~pF. Its dynamic range is slightly lower compared to the AGET one.

\item[SAMPIC: ]The SAMPIC chip is a 16-channels low depth high-speed digitizer. Each of its 16 channels associates a DLL-based TDC providing a raw time with an ultra-fast analog memory (5~GHz sampling frequency) allowing fine timing extraction as well as other parameters of the pulse. Each channel also integrates a discriminator that can trigger itself independently or participate to a more complex trigger. After triggering, each sample is digitized by an on-chip ADC and only that corresponding to a region of interest is sent serially to the DAQ. The association of the raw and fine timings permits achieving timing resolutions of a few ps rms. This chip accepts input signals up to a level of 1~V, wit a input bandwidth of 1.6~GHz. It is fabricated with the 0.35~\(\mu\)m AMS CMOS technology, and has a power consumption lower than 12~mW/channel.

\item[ALCOR (A Low power Chip for Optical sensor Readout): ]The ALCOR  chip prototype is a first test vehicle for a high-rate digitization back-end for SiPM readout in fast timing applications. It is a 32-channel ASIC that features signal amplification, conditioning and digitization. It features low-power TDCs that provide single-photon tagging with time binning  down to 50~ps and able to work down to cryogenic temperatures. The design of a system-grade ASIC targeting dRICH detector specifications is now being pursued  at INFN. The ALCOR chip is based on a triggerless time-based (time-of-arrival and time-over-threshold) readout and features a SEU-protected logic. A dedicated design of the front-end will allow for integrated cooling and customized decoupling circuits (high pass filter) for possible signal pre-conditioning and count rates well exceeding 500~kHz per channel. The chip architecture and matrix floor-plan will allow for a future version to be assembled chip-on-board with bump-bonding (the first prototype uses wire-bonding padframes), which will be an enabling factor for the design of very compact and robust front-end electronic board.

\end{description}

\begin{sidewaystable}[thbp]
\tiny
 \begin{tabular}{|c|c|c|c|c|c|c|} 
 \hline
                        & SAMPA  & VMM   & TIGER  & DREAM  & AGET & AFTER   \\ [0.5ex]
 \hline\hline
  Architecture     &Front-end + ADC + DSP& Front-end + S\&H + discri + 3xADC& Front-end + S\&H + discri + TDC + ADC &\multicolumn{3}{c|}{Front-end + analog memory}\\ [1ex] 
\hline\hline
   & \multicolumn{6}{c|}{  Analog characteristics} \\ [0.5ex]
 \hline\hline
 Number of channels     & 32       & 64     & 64 & 64  & 64 & 72      \\ 
 Input dynamic range    & 66/500~fC&0.1-2.0pC&2.0-50~fC&50-600~fC& 120~fC - 10~pC & 120-600~fC           \\
 Peaking time range     & 160-300~ns&25, 50, 100 and 200~ns& 60~ns (TDC), 170~ns (ADC)   & 50~ns - 1~\(\mu\)s & 50~ns - 900~ns & 100~ns - 2~\(\mu\)s  \\
 Full signal occupancy  & 550~ns&        &             &    &    &           \\
 Polarity               & +/-      & +/- &             & +/-& +/- & +/- \\
 Detector capacitance range & 18.5~pF/40-80~pF & 200pF& up to 100pF& 200~pF &    &  $<$30pF         \\
 Noise level            & 600/900~e\u207b & 300~e\u207b at 9~mV/fC& up to 2000~e\u207b& 610~e\u207b + 9~e\u207b/pF& 580~e\u207b + 9~e\u207b/pF &   370~e\u207b + 14.6~e\u207b/pF  \\
 Sensitivity/Gain       & 20-30/4~mV/fC&     &      12.4~mV/fC (TDC), 11.9~mV/fC (ADC)       &    &    &    120~fC/mV       \\
 Remarks                &          &        & CR-RC shapers&     &      &           \\ [1ex]
 \hline
   & \multicolumn{6}{c|}{Digital characteristics} \\ [0.5ex]
 \hline
 Sampling frequencies   & 10-20~MHz & 200~MHz & 1-40~MHz & 1-50~MHz& 1-100~MHz & 1-100~MHz          \\
 ADC resolution         & 10-bit   & 10-bit & 10-bit (Wilkinson) &No ADC&No ADC&No ADC\\
 TDC time resolution    &          & 8-bit + 12 global& 5~ns &    &    &           \\
 Remarks                & 10~MS/s  &        &             & Internal trigger & Internal trigger &           \\ [1ex]
 \hline
 Data treatment functions & On-board DSP& none &      none       &    &    &           \\
 Data bandwidth         & 11x320~Mbit/s& $<$1~Gbit/s & 1.28~Gbit/s (triggerless)&    &    &           \\
 Streaming readout capacity & 3.4G~bit/s     &        & Readout on  &    &    &           \\ [1ex]
  & & & internal trig., programable thres. & & &  \\
 \hline
   & \multicolumn{6}{c|}{Other information} \\ [0.5ex]
 \hline
 Die size               & 9.6x9.0~mm²& 15.3x8.3~mm² & 5×5~mm² &    &    &     7.8 x 7.4~mm2      \\
 Package size           & TFBGA 15x15~mm²& 400BGA&             &    &    &     28 x 28~mm      \\
 Power consumption      & 20~mW/ch & 10~mW/ch&        12~mW/ch @ 3.3V       & 10~mW/ch & 10~mW/ch & 10~mW/ch \\ 
 Technology             &130~nm CMOS& 130~nm MOSFET& 110~nm CMOS& 350~nm AMS CMOS   &  350~nm AMS CMOS  &   350~nm AMS CMOS       \\
 Remarks                &          &        &             &    &    &           \\ [1ex]
 \hline
\end{tabular}
\caption{Characteristics of different chips presently available for gaseous detector readout}
\label{table:DAQ_Electronics.chip_characteristics}
\end{sidewaystable}

\subsubsection{Support system}

\subsection{Possible Readout Chip Evolution and Future Technological Constraints}


It seems that a mixed-signal multi-channel (greater than n=64) high-performance ASIC chip architecture is entailed with at least a 10-bit/12-bit resolution SAR ADC (offering a minimum sampling rate of 25~MHz, very low non-linearity, INL $<$2.0$\%$, and a low latency, $<$10.0~ns), working in tandem with a high-speed TDC (with excellent time resolution), preferably with buffer and glue-logic on-board, and a complementary FPGA with intelligent firmware designed, in the form of a total solution for readout. Additional features like an on-board DSP module (for baseline correction, zero suppression, anti-aliasing digital filtering etc.) would be added advantages. Other approaches such as companding ADC can also be explored to make a trade-off for low-resolution (6/8-bit ADC), if resources permit and substantial advantage is expected on the cost of chip component overhead.


There are other ASIC chips available in the HEP community, suitable for GEM and TPC applications. However, detailed evaluation is needed.

Evaluation of radiation, thermal, and magnetic field effects needs to be carried out for all chips, although some of the chips discussed earlier have been through radiation damage tests and seemed to offer satisfactory performance in general, with very little damage. 

So far, unfortunately, one single chip suitable for the prospective experiments at the EIC does not seem to exist. Every chip has some extremely vital feature or necessary benchmark missing.

The most promising places for development of future ASIC and mixed signal devices (and the supporting hardware and firmware/software) seem to be the CEA (France), INFN Torino (Italy) and Brookhaven National Laboratory (USA), etc., where excellent chips have been developed in the past. However, more refinements are needed in existing chip architectures. A collaborative effort with these institutions could be a viable direction for fostering future front-end and readout technologies necessary for endeavors like the EIC.


\subsection{Existing streaming readout DAQ Systems for particle physics experiments}

In the following, we briefly present some existing data acquisition systems for particle physics experiments adopting, completely or in part, a streaming readout approach for data readout. Further examples not reported in this section are the ALICE experiment Online-Offline (O$^2$) system~\cite{Buncic:2015ari} and the new Compass data acquisition system~\cite{Bodlak:2013lla}. 

\subsubsection{LHCb streaming readout DAQ}

The LHCb detector at CERN~\cite{Alves:2008zz} is currently ongoing a major upgrade to replace the current trigger-based DAQ system to a fully streaming DAQ system (see~\cite{Colombo:2018vmp} for a complete description). The new system will allow to acquire and select events at the full 30 MHz rate of proton-proton collisions at the interaction point\footnote{The nominal bunch-crossing frequency at the LHC is 40 MHz, corresponding to one interaction every 25 ns. At the LHCb interaction point, however, one every four collisions is empty, resulting in a 30 MHz physics events rate.}. To reach this goal, all front-end boards in the upgraded LHCb detector will be capable of acquiring signals at the full bunch-crossing frequency. The custom GBT protocol~\cite{Moreira:2009pem} will be used to transport data via optical fibers from the front-end boards to the readout system, with up to 4.5 Gb/s bandwidth per link. Data are then processed by the upgraded LHCb event builder system, capable of aggregating, analyzing, and filtering the events - considering the full 30 MHz collision rate and with a single event size up to 150 kB, the system was scaled to handle a total data rate up to 40 Tb/s. The main components of the event builder system are the readout units and the builder units. Each readout unit is responsible for collecting data from part of the readout board, using point-to-point links, and sends this to a builder unit. For each event, one builder unit receives all the fragments from all the readout units and aggregates them into the event. Each event is then passed to the online processing farm for reconstruction and filtering. A first level filter (HLT1) performs a fast reconstruction and events selection, reducing the input rate from 30 MHz to approximately 1 MHz. A second, more sophisticated, filtering level (HLT2) performs the final event selection, resulting to an output event rate of approximately 100 kHz to be written to the disk.

\subsubsection{sPHENIX Hybrid DAQ}


Construction is ongoing for the sPHENIX triggered-streaming Hybrid
DAQ, which simultaneously reads out the conventionally triggered
calorimeter subsystems and the streaming tracking subsystems~\cite{sPHENIX-TDR}. 
Both the
sPHENIX front-end readout and the back-end DAQ will also serve as an
exercise of a large-scale streaming system that is applicable to
future EIC experiments.

The tracking front-ends consist of the on-detector streaming ASICs for
the readout of the MAPS pixel tracker (ALPIDE), a silicon strip
tracker (PHFX), and GEM-based TPC read out with a new version (V5) of
the ALICE SAMPA chip~\cite{Adolfsson:2017art}. The streaming data
are time-stamped with beam collision clock, aggregated in the
front-end FPGAs, and transported to the back-end DAQ via O(1000)
multi-Gbps fiber links providing O(10) Tbps overall readout
bandwidth. A global timing system provides a low jitter collision
clock, fixed-latency trigger signal, and time-stamp counter to all
front end electronics, which are embedded in the data stream and
serves as the basis for the streaming and hybrid synchronizations.

A fleet of O(50) Front-End Link eXchange
(FELIX)~\cite{Anderson:2016lfn} readout cards hosted in
commodity Linux PCs is used to read out, buffer, and process these
data streams. In the version used by sPHENIX, each FELIX is a
PCI-express card carrying a large FPGA (Xilinx Kintex UltraScale
KU115). It supports 48 bi-directional 10-Gbps optical links to the
front-end and a 100-Gbps PCI-express Gen3 link with the hosting
server’s CPU. It is initially designed for the ATLAS Phase-I upgrade
and continues to be developed utilizing recent parts with a higher
speed for future ATLAS upgrade towards the HL-LHC. The strategy of
using PCIe FPGA cards to bridge the custom front-end and commodity
computing is also used by the LHCb, ALICE, ATLAS, and CBM
experiments. The overall peak disk data rate is designed to
accommodate the RHIC Au+Au collision at the top luminosity that is
orders of magnitude higher charged particle production rate when
compared with the EIC.

While sPHENIX will have a trigger, the overall architecture is 
streaming-oriented and highly parallel. Individual substreams coming 
form the detector are written to different files directly, and 
synchronization will be performed via time stamps, not event numbers. 
The actual event building is moved to the offline analysis, removing
the necessity to build a distributed, fault tolerant, reliable one-shot
online event builder.

The FELIX system also provides the flexibility of throttling the
recorded streaming data corresponding to the calorimeter triggers
(i.e. global zero-suppression) or allows for triggerless recording of
a fraction of or all of the tracker data. The streaming tracker data
are demonstrated to enable a unique set of heavy flavor measurements
that would be otherwise inaccessible, and this streaming DAQ
development is recently commended by the RHIC Program Advisory
Committee.

\subsubsection{The RCDAQ  Data Acquisition System}

sPHENIX uses a powerful but lightweight data acquisition system called
``RCDAQ''~\cite{RCDAQ}. It is currently in use for virtually all
sPHENIX R\&D projects such as test beams, tests in labs, detector
calibrations, and the like. RCDAQ supports all current sPHENIX
front-ends and both triggered and streaming readout modes. It also
supports, by way of plugins, a large variety of commercial or
otherwise available readout electronics, such as the DRS4 Evaluation
Board~\cite{DRS4}, the CERN SRS system~\cite{RD51:SRS}, several CAEN
modules such as the V1742 Waveform Digitizer, and many more.

RCDAQ has long been the de-facto standard data acquisition system for
several EIC R\&D groups, such as eRD1 (calorimetry), eRD6~\cite{eRD6}
(tracking), eRD14 (time-of-flight), and eRD23 (streaming
readout technologies). In addition, RCDAQ is used by dozens of
external groups not connected to the EIC or RHIC R\&D efforts because
of the support for those common readout devices, its built-in support
for ROOT-based online monitoring, comprehensive controls, and small
footprint.

\subsubsection{The ERSAP system}
Development is underway for the Environment for Real-time Streaming, Acquisition and Processing (ERSAP) Streaming Data Readout System at JLab. ERSAP is a backend software system that combines components to form a reactive data flow architecture. This combines software originally developed as part of the CODA data acquisition system and then advanced as part of the CLARA reactive microsevices framework~\cite{Gyurjyan:2011zz} used by the CLAS12 experiment. The system encapsulates each component into a microservice with well defined inputs and outputs that allow for local or remote communications. This allows both horizontal and vertical scaling to make the system highly configurable. It also supports micro-services written in any language (C,C++, Python, Java, ....). Utilizing such a design helps ensure a level future-proofing since individual services can be easily replaced with ones using new syntax, languages, or technology (e.g. heterogeneous hardware components). A prototype of the system was tested successfully in summer 2020 using beam at the CLAS12 Forward Tagger. Also being developed as part of ERSAP is high performance tiered memory or ``Data Lake'' system that allows efficient data cooling (i.e. temporary buffering). The system is scalable enough to be used on a single desktop with other DAQ components in benchtop system or in a dedicated node with a large memory+disk. The Data Lake implements automatic fail-over to disk if its allotted memory resource becomes exhausted.

\subsection{A Progressive Approach toward the EIC DAQ System}

The final goal of the EIC streaming readout system is to reconstruct online all events, adding to the raw-data banks the high-level information from the reconstruction - ideally, four-vectors and PID assignment for all particles in a given interaction, and store all of them to the disk. Eventually, filtering algorithms can run online to tag events according to a certain condition (for example, events belonging to a certain exclusive channel), to speed-up the offline analysis.

\begin{figure}
    \centering
    \includegraphics[width=\textwidth]{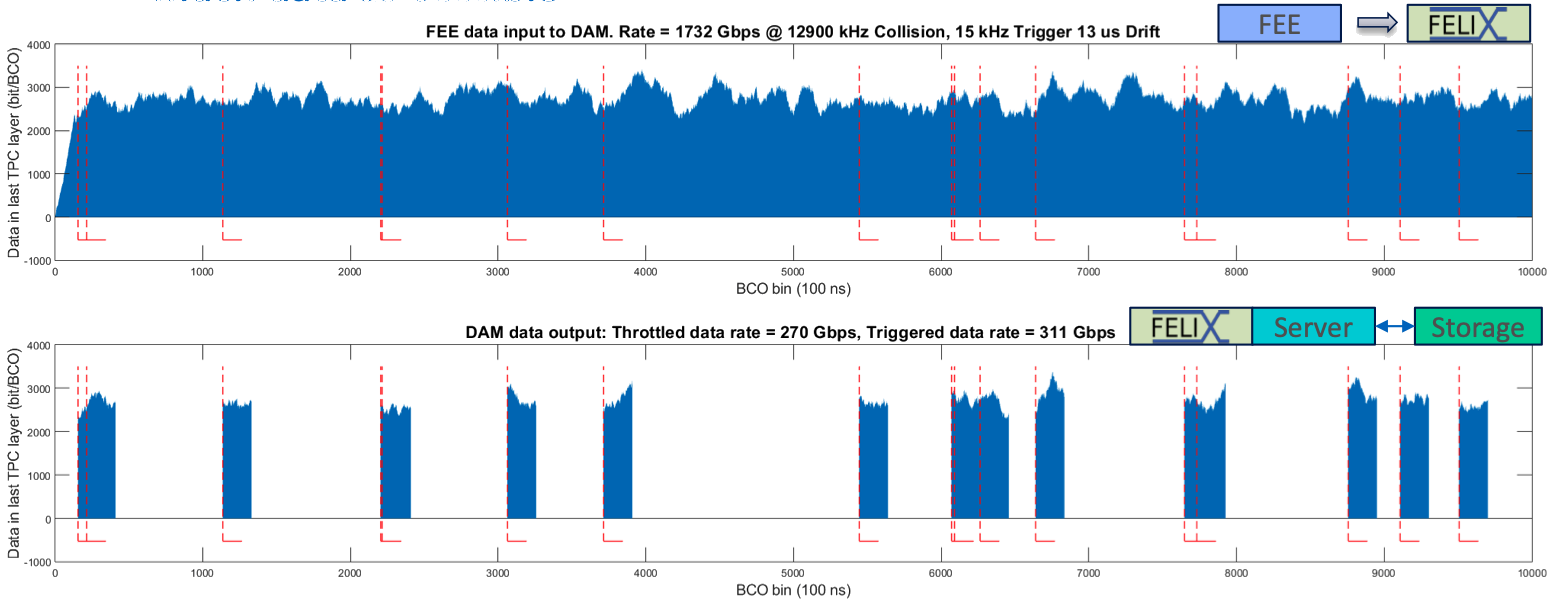}
    \caption{The sPHENIX hybrid DAQ system cross-detector zero suppression. The sPHENIX streaming tracker can use the calorimeter trigger as a data throttle for loss-less data reduction for triggered events + streaming as much data as possible~\cite{sPHENIX-TDR}}
    \label{fig:Part3_Figures.DetAspects.DaqElectronics.SphenixStreaming}
\end{figure}

Based on the preliminary estimates discussed before in Sec.~\ref{part3_DetAspects.DaqElectronics.Constraints}, and considering the technologies that are available already today, the following key arguments concerning the EIC streaming DAQ system can be assessed  (see also Fig.~\ref{fig:Part3_Figures.DetAspects.DaqElectronics.SphenixStreaming}).
\begin{itemize}
    \item In principle, it will be possible to write all raw data directly to the disk, without further online processing. However, unexpected large noise levels could exceed the system capacity, and the system must be prepared for such an event. This is particularly true during the initial phase of EIC operations, when unexpected backgrounds not predicted by simulations and not observed in the preliminary sub-detectors characterization phase could be present, and the machine still needs to be tuned.
    \item High-quality calibration constants are necessary for the online events reconstruction, analysis, and filtering. This requires a depth knowledge of the detector behavior, that may not be available at the beginning of the EIC operations.
\end{itemize}

The solution that we envisage is to design a modular system that will evolve with the experiment. During the first part of the EIC run, a hybrid streaming readout strategy will be adopted, using the so-called ``cross-detector zero suppression technique''. In this scheme, all hits from the detector are streamed to the online computing farm and stored to a temporary buffer. Only ``interesting'' portions of the data stream are further processed, while the others are discarded. 
Technically, this can be achieved both with a parallel hardware system, as in the sPHENIX case, or with a dedicated software component (the sPHENIX hardware-based cross-detector zero suppression system operation is illustrated in Fig.~\ref{fig:Part3_Figures.DetAspects.DaqElectronics.SphenixStreaming}). Online filtering and online reconstruction will be then gradually introduced when the detector will be more under control.

\begin{figure}
    \centering
    \includegraphics[width=\textwidth]{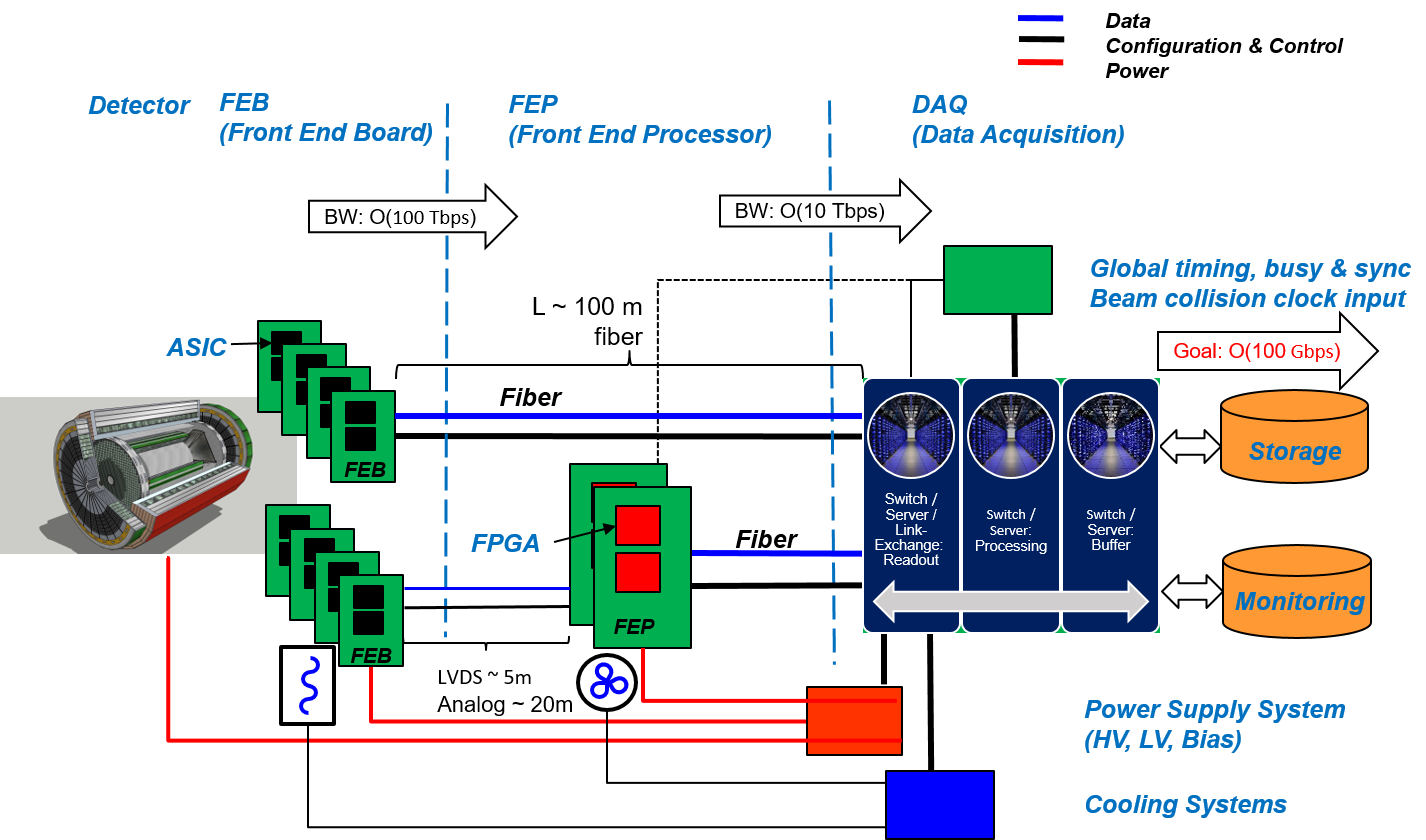}
    \caption{Possible scheme for the EIC Readout Architecture}
    \label{fig:Part3_Figures.DetAspects.DaqElectronics.EICReadoutArchitecture}
\end{figure}

A possible solution for the EIC readout architecture is shown in Fig.~\ref{fig:Part3_Figures.DetAspects.DaqElectronics.EICReadoutArchitecture}. Some front-end boards (FEB) containing ASICs and specific to different sub-detectors will likely require an intermediate stage of processing via FPGAs for data aggregation and reduction by employing front-end processors (FEP).
Data transport to servers or link-exchange cards, such as FELIX, will be made via extensive use of optical fibers. Power supply and cooling systems are planned to be commercial-off-the-shelf (COTS) units. 

An intense R$\&$D program has already started to study and design the EIC readout system, covering all the technical aspects involved with it, including the different FE options compatible with a streaming readout system, the data transport system, the synchronization system, the back-end online processing software (cf sections \ref{part3-sec-DetTechnology.DAQ} and \ref{part3-sec-DetTechnology.Electronics}).

\subsection{Experimental Validation of the Approach}

Despite the conceptual simplicity of a triggerless DAQ, a realistic implementation with the specific detector readout is necessary to validate this solution and demonstrate the expected performances. The sophisticated combination of a suitable front-end electronics, network facilities and CPU algorithms requires a significant effort to identify, or develop in case they are not yet available, the best option for each element, set-up and test the whole scheme and compare results with more traditional approaches. 

A dedicated test and validation program, with complementary experimental efforts, has already started in view of the EIC detector design and construction. In the following, we briefly present these efforts.

\subsubsection{Thomas Jefferson Laboratory efforts}

A first experimental characterization and validation campaign for the new  DAQ approach has started at Jefferson Laboratory in 2020, using a streaming readout solution based on FA250+VTP / Waveboard digitizer boards~\cite{4436457,AMELI2019286} for the front-end readout and on the TriDAS software~\cite{Favaro:2016gvk} interfaced with the JANA2 data analysis framework for the back-end online data reconstruction and filtering~\cite{ Lawrence:2020ikd}.

\begin{figure}[th]
    \centering
    \includegraphics[width=.48\textwidth]{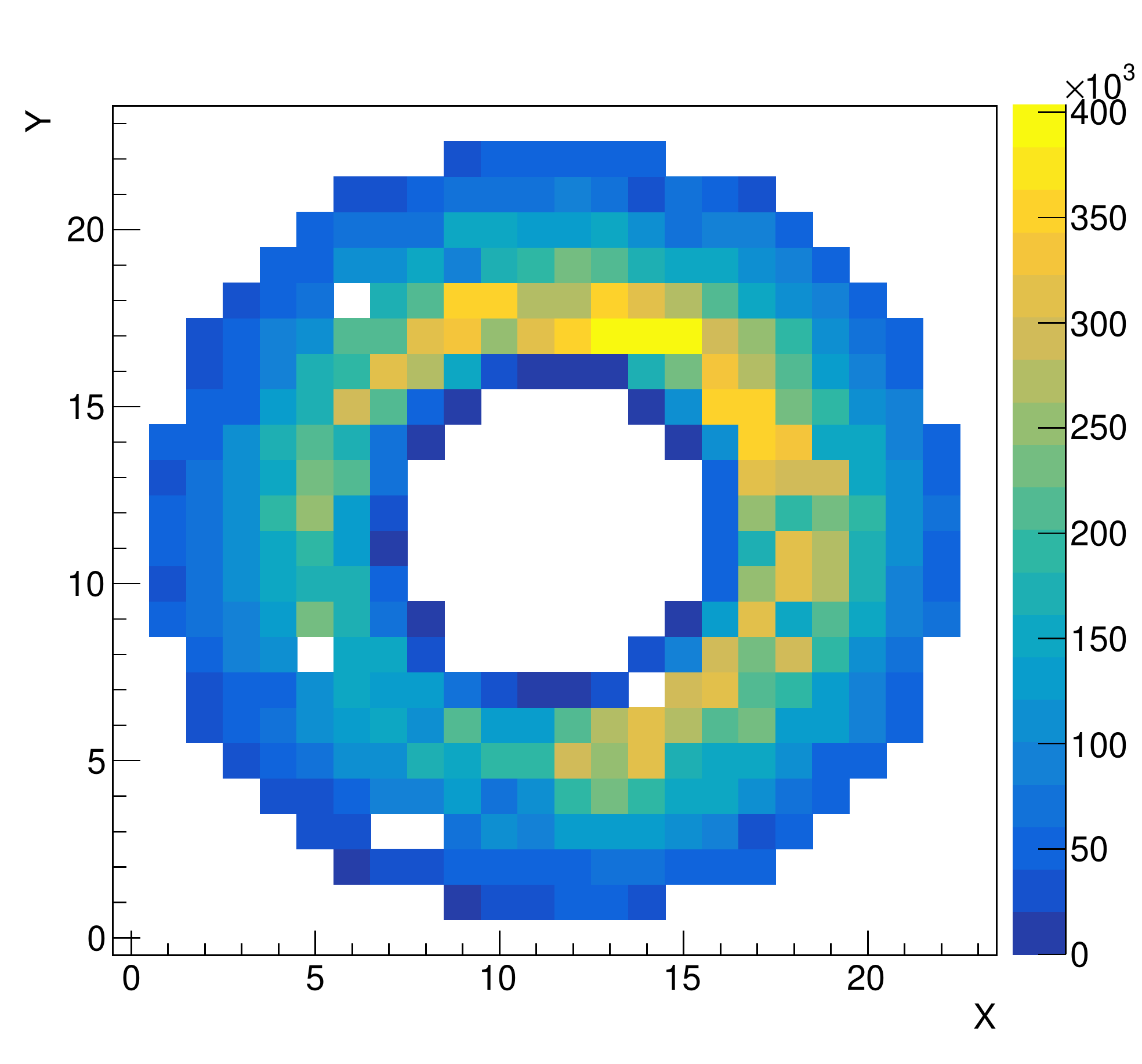}
    \includegraphics[width=.48\textwidth]{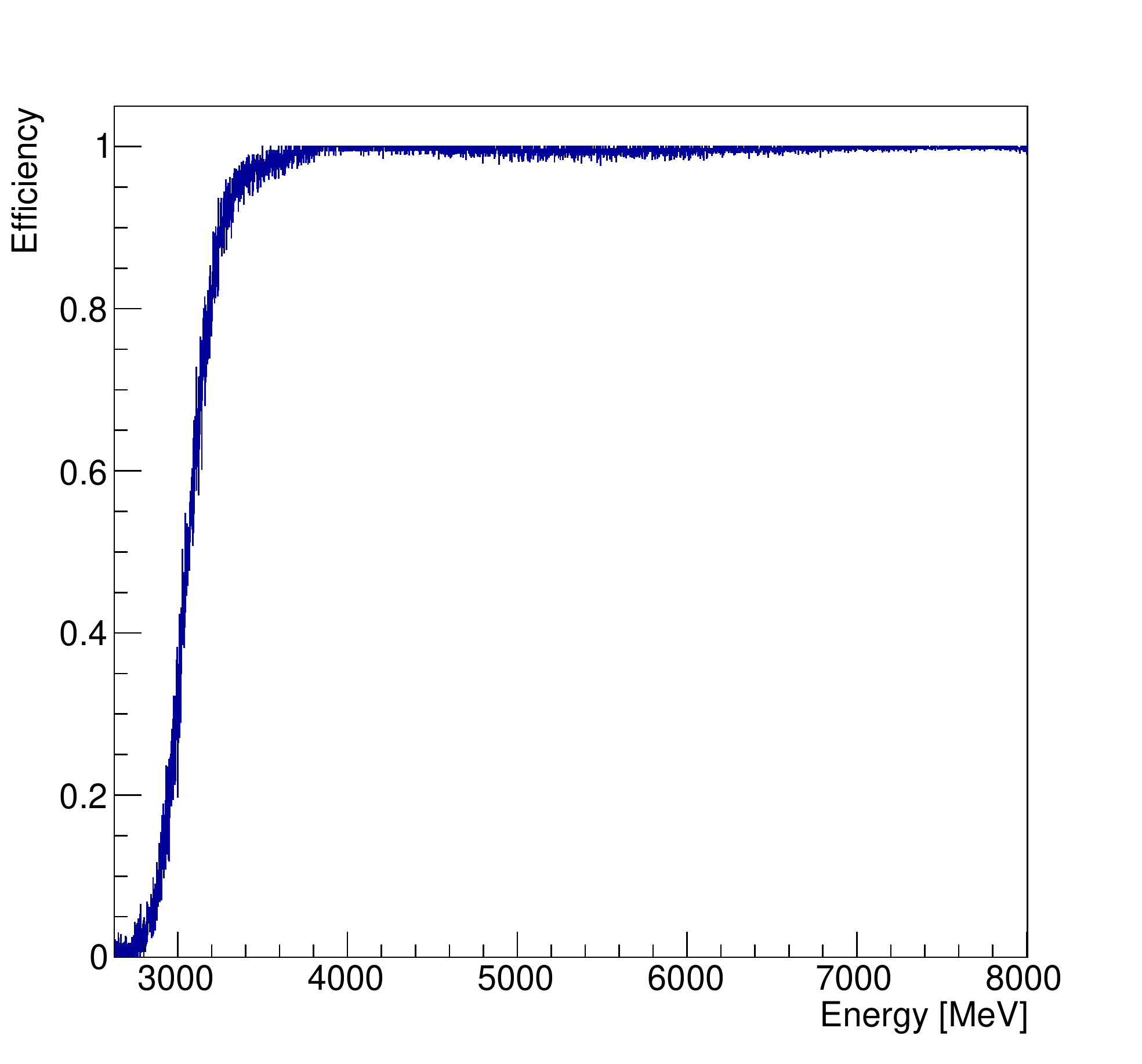}
    \caption{Left: measured FT-Cal hits during the early 2020 Hall-B streaming readout tests. Right: Efficiency of the online clustering algorithm, with a 3.0 GeV cluster threshold applied.}
    \label{fig:Part3_Figures.DetAspects.DAQ_Electronics.HallBtests}
\end{figure}

Due to the comparable luminosity and detector complexity, the CLAS12 detector in Hall B is an ideal study case to characterize and validate the streaming DAQ approach in view of its application for the EIC detector~\cite{Burkert:2020akg}. A first measurement on beam was carried out using the CLAS12-Forward Tagger Calorimeter and Hodoscope detectors~\cite{Acker:2020brf}, with the CEBAF 10.6 GeV electron beam impinging on a lead (early 2020 run) / deuterium (summer 2020 run) target. Some results are reported in Fig.~\ref{fig:Part3_Figures.DetAspects.DAQ_Electronics.HallBtests}, showing the distribution of hits measured with the Forward Tagger Calorimeter, as obtained from the online monitoring, and the corresponding clustering algorithm efficiency as a function of the cluster energy, for an online clustering threshold of 3.0 GeV.

This represented the first attempt to acquire some CLAS12 sub-detectors using streaming readout: the growing interest for this approach is demonstrated by the plans of the CLAS Collaboration to extend it to the full detector in the near future.
During the test, the single $\pi^0$ quasi-real photoproduction reaction was used as a benchmark to assess the performances of the streaming DAQ system. The $\pi^0$ was identified measuring the two photons from the decay in the Forward Tagger Calorimeter, whereas the scattered electron was identified by a combination of an electromagnetic cluster in the Forward Tagger Calorimeter and a geometrically matched signal in the Forward Tagger Hodoscope. Preliminary results show a good agreement between the measured data and the predictions from a Monte Carlo numerical estimate, in terms of the energy distribution and total yield of the measured $\pi^0$. The data analysis is currently in progress, and final results from the test are expected to be published in early 2021.

A pilot beam study was also conducted to test streaming data processing of the CLAS12 Forward Tagger Calorimeter and Hodoscope using ERSAP software package that includes JLAB data acquisition and data processing frameworks, such as CODA, CLARA and JANA. 
Specifically CODA VTP stream firmware was used to stream raw data to stream-aggregator, hit-finder, noise-reduction and event-building micro-services, followed by standard, Forward Tagger reconstruction micro-services from the CLAS12 reconstruction application. CLAS12 reconstruction application is based on the CLARA, which is a reactive micro-services orchestration framework for designing, deploying and scaling data stream processing applications~\cite{Gyurgyan:2016tml,Ziegler:2020gsr}.  The goal of this study was to optimize (both performance and resource utilization) newly developed data-stream curation micro-services, and to estimate existing CLAS12 reconstruction micro-services scaling levels and resource requirements that will keep up with the VTP data stream. Preliminary results were reported at the 22nd IEEE Real Time Conference.

\begin{figure}[th]
    \centering
    \includegraphics[width=\textwidth]{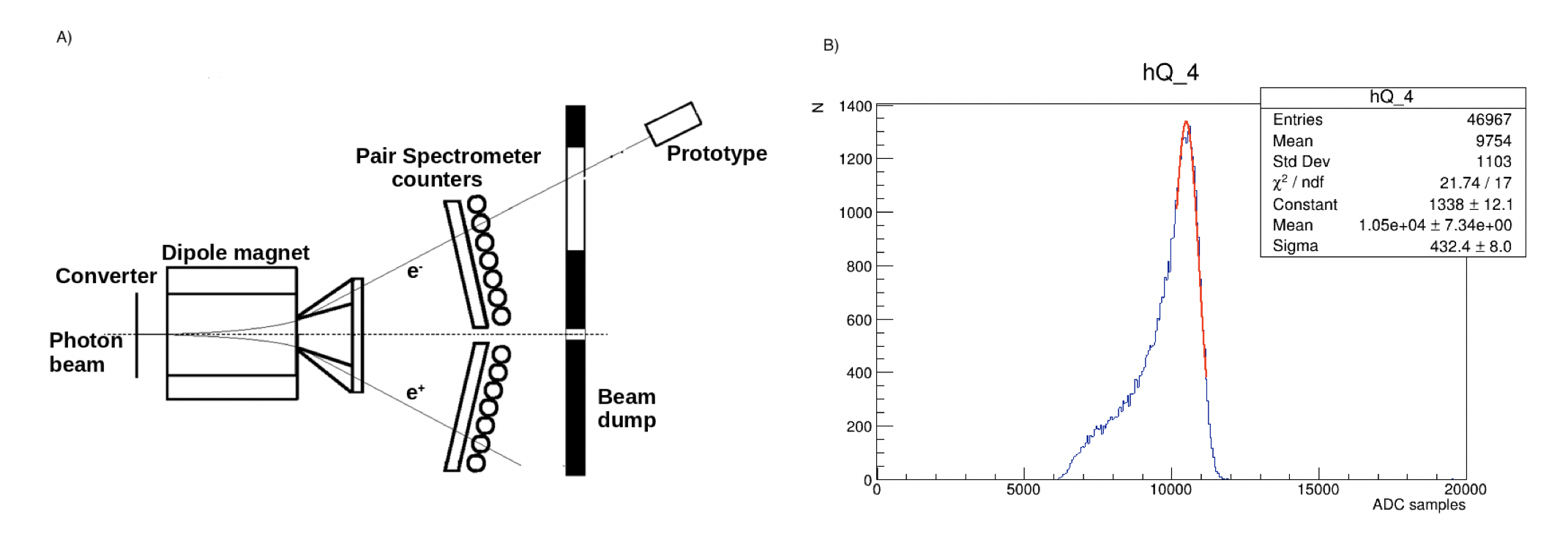}
    \caption{Left: Hall D PS beam test setup scheme; Right: Calorimeter central response (in arbitrary units) to 4.5 GeV impinging electrons.}
    \label{fig:Part3_Figures.DetAspects.DaqElectronics.HallDtests}
\end{figure}
A complementary test was performed in Hall D, at the pair spectrometer (PS) facility~\cite{Barbosa:2015bga}. The general purpose of the beam tests was to study the light yield and the energy resolution of glass-ceramic scintillator bars made in VSL/CUA/Scintilex and new produced PbWO$_4$ crystals made by CRYTUR/SICCAS. A glass-ceramic and a PbWO$_4$ prototype were installed behind the Hall D pair spectrometer and the response to the tagged electrons from the PS was measured (see also Fig.~\ref{fig:Part3_Figures.DetAspects.DaqElectronics.HallDtests}, left panel). The prototypes were also used to test and optimize the entire readout chain: photosensorss (PMT vs SiPM), preamps, fADC or Waveboard digitizers in combination with streaming DAQ system. 
During the spring run 2020 at Jlab HallD a single prototype,  assembled from nine scintillators coupled with R4125-01 Hamamatsu PMTs and active HV dividers with integrated preamplifier, was used. Signals were digitized using a Waveboard device. The SRO tests was performed parasitically during GlueX High Luminosity runs (350 nA photon beam). The waveboard read-out nine calorimeter channels plus two scintillator pads mounted in front of the calorimeter, to tag the impinging electron. The system was operated with a rate up to ~1.5kHz per channel. The full SRO chain (Waveboard+TriDAS+JANA2) was successfully tested, with data collected using different combination of software L2 triggers. The offline data analysis is currently ongoing, and final results from the test are expected to be published in early 2021. Some preliminary results are reported in Fig.~\ref{fig:Part3_Figures.DetAspects.DaqElectronics.HallDtests}, right panel, showing the energy distribution (in arbitrary units) of the center-most calorimeter crystal when hit by 4.5 GeV electrons.

\subsubsection{BNL efforts}

\begin{figure}[hbt!]
\begin{center}
\includegraphics[width=0.85\columnwidth]{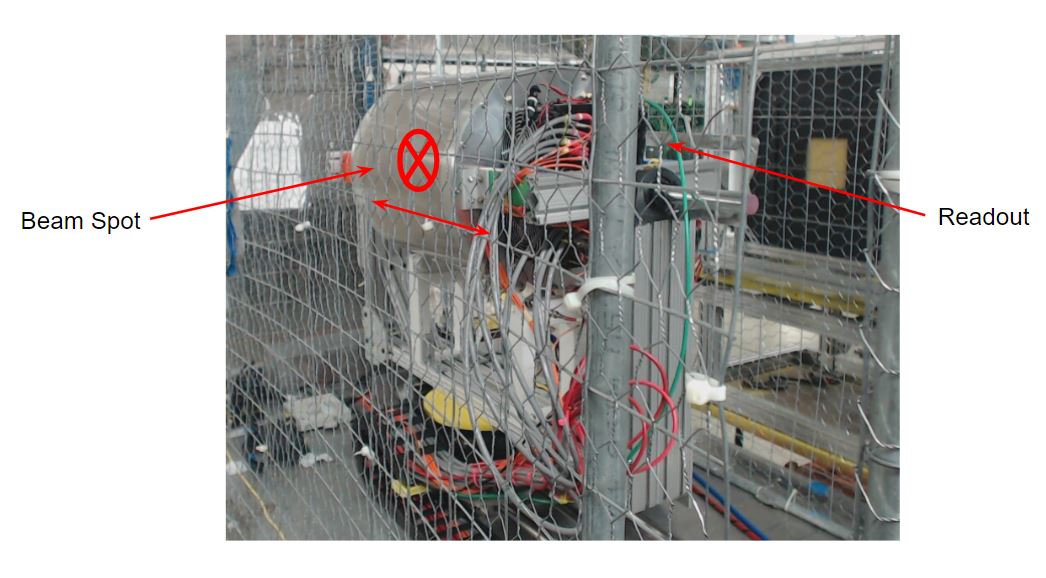}
\includegraphics[width=0.85\columnwidth]{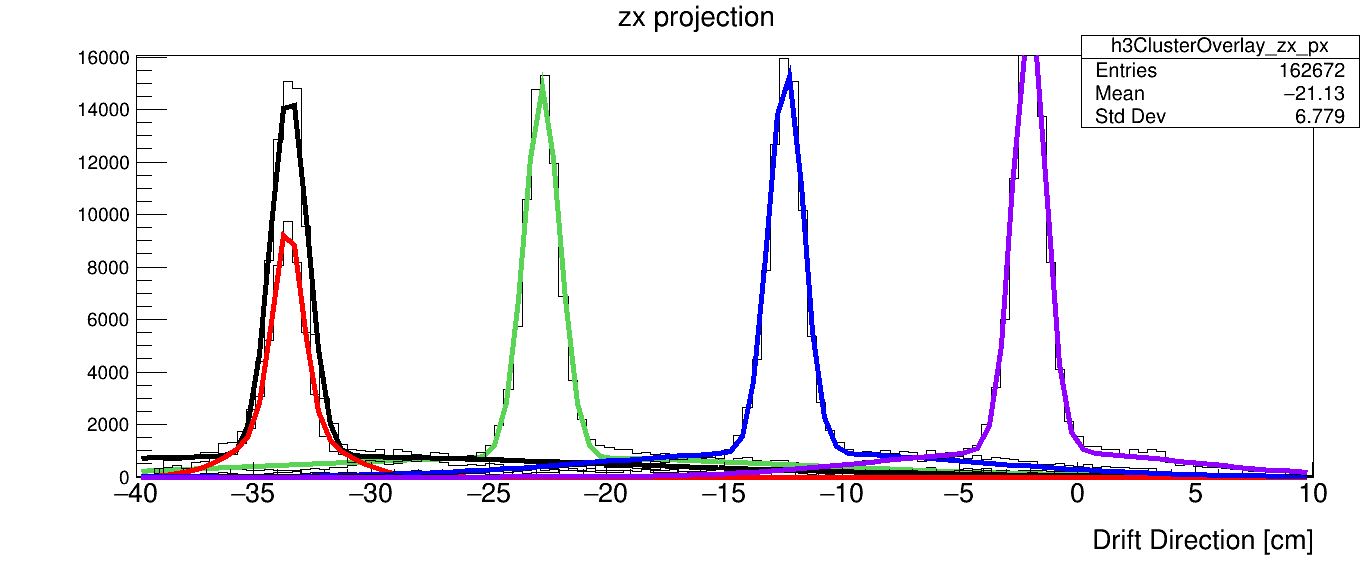}
\end{center}
\caption{\label{fig:Part3_Figures.DetAspects.DaqElectronics.annotated}
  Left: The TPC prototype shown in the test beam which is read out
  with FELIX and the RCDAQ .  The red cross-hair indicates the
  approximate beam position. Right: The reconstructed drift distance
  for 4 different positions of the TPC prototype relative to the
  beam.}
\end{figure}

An example of a detector read out in streaming mode is a prototype of
the sPHENIX TPC that was tested at the FermiLab Test Beam Facility
(FTBF) in 2019. The TPC prototype, shown in
Fig.~\ref{fig:Part3_Figures.DetAspects.DaqElectronics.annotated}, was moved
perpendicular to the beam and rotated with respect to the beam to get
particle tracks at different distances away from the pad plane,
resulting in different drift lengths and angles.

At the test beam, we found that our event rate capability could be
significantly increased by running the FELIX readout in ``streaming
mode''. We still triggered the front-end card with signals from the
FTBF beamline, however, the FELIX cards are oblivious to how the FEE
actually arrived at the decision to send up the data. But by allowing
the FELIX card to format the data as streaming data, one does not need
to wait for all data from a particular beam event to be fully
transmitted. In streaming mode, while data from trigger $n$ are
already arriving from one front-end, other parts can still be
transmitting data from trigger $n-1$, or even $n-2$. In streaming
mode, there is no need to wait for the completion of the data
transmission from a given trigger, as the data parts are later
re-assembled by their embedded clock information. That is what led to
the increased event rate in streaming mode.

This also validated a running mode that sPHENIX is committed to in
production running, combining the streaming data from the trcking
system with triggered data from the calorimeters and the Minimum-Bias
detector. During the test beam we achieved the simultaneous logging of
data from the TPC prototype together with several channels worth of
beamline instrumentation channels read out in ``classic'' triggered
mode.

This also served as an early test of our timing system that provides a
common clock to the various front-end cards, and can on demand also
provide a standard trigger signal to legacy electronics.

\section{Software, Data Analysis and Data Preservation}
\label{part3-sec-Det.Aspects.SDAD}

This section describes the computing needs for the reference detector at the
EIC and discusses the foreseen software developments.

Aside from possible contributions from machine backgrounds, the reconstruction of 
events at the EIC will be easier than the same task at present
LHC or RHIC hadron machines, and, in perspective, much easier than for the High-Luminosity LHC (HL-LHC), which will start operating two years earlier than the EIC, when we may expect a gain in performance for CPUs of about a factor of 10 with respect to now.

Reconstruction time of DIS events at presently running experiments is around
\SI{0.35}{s} (or $\sim$ \SI{5}{HS06 s}) both at COMPASS and at CLAS12, with event sizes of
0.03~MB and 0.02~MB respectively. Filtering out machine background with high 
efficiency will allow to keep the reconstruction time at \SI{5}{HS06 s} also in 2030.
Processing events at the same speed of acquisition, or $500\, 000$ events per 
second, on today nodes will therefore require $200\, 000$ cores or 1500 nodes, 
a computing farm well in the size of the EIC project. The
expected gain in CPU power in the next 10 years, as well as the possible improvement
in the reconstruction software from the use of artificial intelligence and machine learning techniques give a considerable margin to cope with higher event complexity that may come by 
higher backgrounds rates.

Software design and development will constitute an important ingredient for the
future success of the experimental program at the EIC. Moreover, the cost of the
IT related components, from software development to storage systems and to
distributed complex e-Infrastructures can be raised considerably if a proper
understanding and planning is not taken into account from the beginning in the
design of the experiment itself.

A growing group dedicated to this effort already exists. An important step in the
clustering of a core group focusing on computational aspects at an EIC has been
the approval by the EIC Generic R\&D program of the eRD20 proposal, creating in
2016 the EIC Software Consortium (ESC). The ESC has been the backbone to form in
2018 the Software Working Group within the EICUG. The Software Working Group has
supported the Yellow Report initiative and provided the tools for simulations
and helped in the creation of the infrastructure for storage and documentation.

The Software Working Group is starting in parallel a greenfield development that
will focus on different aspects of future needs:
\begin{itemize}
\item Simulations for detector optimization, to cover the more immediate needs
  of the design and integration of the various sub detectors
\item Help in the development of state of the art Monte Carlo event generators
  for the full spectrum of EIC physics. Validation of these generator
  will be possible by using data from HERA and running experiments at CERN and JLab.
\item Development of a full simulation-reconstruction chain allowing to
  benchmark the performances of the reconstruction software.
\item Development of modern ``event reconstruction'' schemes both using standard
  approaches but also exploring novel methods based on artificial intelligence and 
  machine learning.
\end{itemize}   

The reconstruction software will have to cope with the streaming read-out scheme
adopted and will be designed to match the chosen solution.

{\bf Software tools:}
While developing the software for simulation and reconstruction of events from a
detector which will be up and running in 2030, we need to inquire ourselves
about the long term perspective of software used in today HEP experiments
and evaluate different options. Leaving aside for the moment both full purpose
or dedicated Monte Carlo Event Generators (MCEGs) discussed in a separate section, this
means that we have to decide on: how to describe the detector; which program to
use for particle transportation/interaction; reconstruction tools and the data
model.

The choice of LHC experiments for the Run4 and after may guide us in this task.

DD4hep~\cite{Frank:2014zya} is a toolkit for detector description developed
within the AIDA2020 EU program (Advanced European Infrastructures for Detectors
at Accelerators). It can be an interesting option for the EIC as recently the
CMS collaboration announced that it plans to use it for the detector description in all of their data processing applications. Given its use within CMS, it is expected that DD4hep will be supported
over the entire experiment life-time.

It is worth considering DD4hep for the EIC as it is designed to answer a very
common need of experiments, i.e. the development of a coherent set of software
tools for the description of high-energy physics detectors from a single source
of information.  Detector description in general includes not only the geometry
and the materials used in the apparatus, but all parameters describing, e.g., the
detection techniques, constants required by alignment and calibration,
description of the readout structures, conditions data and more. DD4hep reuses existing software components, combines their functionality, and thus minimizes the efforts required by users to leverage their benefits and optimizes flexibility. Reused components include elements of the ROOT geometry package~\cite{Brun:2003xr} and the \textsc{Geant4} simulation tool kit.

\textsc{Geant4}~\cite{Agostinelli:2002hh} is the baseline for detector simulations on all
LHC experiments as well as many NP experiments and is a natural choice for the EIC. We have developed strong connections with the core developer team of \textsc{Geant4} and discussed with them improvements in physics lists and non-standard geometries which may be needed for the EIC. In September of 2019, the Software Working Group organized together with the \textsc{Geant4} Collaboration a technical forum on the EIC. In the technical forum, we reviewed recent updates on \textsc{Geant4}, including the EIC physics list as maintained by the EIC Software Consortium. We requested improvements on photo-nuclear and electro-nuclear reactions that have been included in \textsc{Geant4}version 10.6, and are currently being tested. We will follow the activities of the vector transport R\&D collaboration~\cite{Amadio:2018tnh} that might provide interest techniques for improving \textsc{Geant4}. 

ROOT~\cite{Brun:2003xr} is by today's standards a fundamental ingredient of
virtually all HEP workflows, being used for data persistency, modeling,
graphics, and analysis. It is structured to have excellent, active connections
with the experiments including, at least for LHC, direct contributions from the
experiments.  The ROOT team is investing in future developments for
HL-LHC, and is also assuming interesting approaches to machine learning techniques. It advises the HEP community to not develop its own machine
learning tools but, maybe in a more efficient way, to collaborate with other
sciences on improving and extending existing tool kits. For that they offer a Toolkit for
Multi Variate Analysis (TMVA) to bridge between ROOT and external machine learning
tool( kit)s such as scikit-learn, XGBoost, TensorFlow, Keras, mxnet, or PyTorch.

ACTS~\cite{Ai:2019kze,Ai:2020jbw} (A Common Tracking Software) is an
experiment-independent and framework-independent toolkit for (charged) particle track reconstruction that is being developed for the HL-LHC but is also targeting other HEP and NP experiments, including the EIC. It is designed for modern computing architectures and is agnostic to the details of the detector technologies and magnetic fields configurations. Another important aspect with respect to development are its rigorous unit tests, an essential aspect for the future EIC software.
All these characteristics made this software an interesting option worth
evaluation for the reconstruction software for the EIC reference detector.

Many others codes are under evaluation, e.g., GENFIT~\cite{Rauch:2014wta}, a
generic track-fitting toolkit, GAUDI~\cite{Barrand:2001ny,Clemencic:2010zz}, a
software architecture and framework for building HEP data processing
applications, or JANA2~\cite{Lawrence:2020ikd}, a multi-threaded event
reconstruction.

Finally, following the large worldwide spread, we are moving to the use of tools
facilitating collaborative analysis as presently done at CERN with
SWAN~\cite{Bocchi:2019eis}, as a Service for Web-based ANalysis, built upon the
widely-used Jupyter notebooks.

{\bf Simulations for detector optimization:}
The tools developed for the Yellow Report initiative will be expanded and used for 
extensive full simulations of the reference detector. This is a short term goal
for software developers in order to support with detailed simulation studies the
finalization of the reference detector, thus allowing to move from the CDR stage
toward the full technical design. 

{\bf Monte Carlo event generators for the EIC:}
The Software Working Group, and before the EIC Software Consortium have
initiated a project with the Monte Carlo communities in the US and Europe
(MCnet) to work on MCEGs for the EIC, requiring MCEG for polarized \ep, \eD,
%
and
$^3\mathrm{He}$ as well as \eA\ measurements. The MCEG initiative is connecting the MCEG
efforts in NP and HEP and is encouraging a strong interplay between experiment
and theory already at an early stage of the EIC. As an initial step, we have
started a workshop series on "MCEGs for future \ep\ and \eA\ facilities" where the
third workshop was held in November 2019 at the Erwin Schr\"odinger International
Institute for Mathematics and Physics in Vienna, Austria. During the workshop,
we reviewed the theory for physics with light and heavy ions and discussed the
modifications needed on the general-purpose MCEGs to simulate unpolarized
observables also for \eA\ where a precise treatment of the nucleus and its breakup
is needed. There were presentations about pioneering MCEG projects for \eA\
(BeAGLE, spectator tagging 
in \eD 
, Sartre), as well as on the ongoing development
of the \eA\ adaptation of JETSCAPE and the Mueller dipole formalism in Pythia8. We
also summarized the status of MCEG-data comparisons in HZTool/Rivet that are
critical to tune MCEGs to existing DIS and heavy ion data as well on the ongoing
work of verifying MCEGs for TMDs with TMD theory/phenomenology.  Our current
focus is on benchmarks and validation. We are working with the EICUG on
benchmark MC productions and the validation of MC results. This will facilitate
the adoption of modern MCEGs that have been so far only used by the LHC
community.

As a recent development, the DIRE authors~\cite{Hoche:2015sya,Dulat:2018vuy}
introduced radiative effects in the simulation of DIS. This is an important
step, since a full multidimensional analysis will be needed in the study of TMDs
and GPDs, given the dependence of the cross section over many kinematic
variables. From the experimental point of view, and without entering to much in
detail of the analysis, this means that detectors and radiative effects will have to be
accounted together at simulation level in order to derive matrices transforming
from raw counts in the detector to Born cross sections. Using DIRE in the Pythia or Sherpa general-purpose MCEGs will allow to check the Monte Carlo predictions for radiative effects  both using the data of running DIS experiments (at JLab and COMPASS at CERN) and the outcome of the simulation of
DJANGO~\cite{Schuler:1991yg,Aschenauer:2013iia}, the reference tool for the study of radiative effects at HERA. 

{\bf Data and analysis preservation for the EIC:}
Already during the design of the reference detector, data and analysis preservation (DAP) is an important issue. Decisions on detector design and conclusions on detector performance and physics requirements will be made on the basis of software and data constructs which must be well defined, preserved and documented if these important studies are to be reproducible, and available for use as a well understood basis for progressing to more sophisticated studies. Our DAP activity will include:
\begin{itemize}
    \item Documentation and preservation of simulation and reconstruction tools, analysis code, data products and workflows. 
    \item Documentation and preservation of data and software required for detector development, e.g. test beam experiments. This will include a catalogue of MC data samples for the design of the reference detector, including the event generator data as well as full simulations data.
\end{itemize}
Based on our experience for data and analysis preservation for the design of the reference detector, we will inform the EICUG on possible strategies for data and analysis preservation at the EIC.

\section{Artificial Intelligence for the EIC Detector}
\label{part3-sec-Det.Aspects.AI}

In the world of computing there is growing excitement for what is perceived as the revolution of the new millennium: artificial intelligence (AI). 
In particular the R\&D program of the future EIC could be one of the first programs systematically exploiting AI.  
AI is becoming ubiquitous in nuclear physics \cite{bedaque2020report}.
According to a standard taxonomy \cite{mehta2019high}, AI encompasses all the concepts related to the integration of human intelligence into machines; a subset of AI is machine learning (ML), which is usually grouped into supervised, unsupervised and reinforcement learning; deep learning (DL) is a particular subset of ML based on deep (\textit{i.e.}, made by many hidden layers) neural networks, which is often considered the evolution of ML since it typically outperforms other methods when there is a large amount of data and features, provided sufficient computing resources. In the most frequent applications in our field, features are selected and a model is trained for classification or regression using signal and background examples. 

Experimental particle and nuclear physics is big data \cite{lynch2008big}: 
 the gigantic data volumes produced in modern experiments are typically handled with ``triggers''---a combination of dedicated hardware and software---to decide near-real-time which data to keep for analysis and which to toss out.  
In this respect, AI plays already an important role in experiments like LHCb~\cite{Alves:2008zz},  where machine learning algorithms (see, \textit{e.g.}, topological trigger and ghost probability requirements) make almost 70\% of these decisions, from triggers to higher level analysis of reconstructed data.

Supported by modern electronics able to continuously convert the analog detector signals, new approaches like Streaming Readout \cite{SRO_WS2020} could further the convergence of online and offline analysis: the incorporation of high level AI algorithms in the analysis pipeline can lead to better data quality control during data taking and shorter analysis cycles.  
Recently the Fast Machine Learning workshop \cite{FastML_WS2020} highlighted emerging methods and scientific applications for DL and inference acceleration, with emphasis on ultrafast on-detector inference and real-time systems, hardware platforms, co-processor technologies, and distributed learning.
In this context, AI (used here in a broader sense to embrace different approaches) could foster in the next years significant advances in areas like anomaly detection (see, \textit{e.g.}, \cite{chalapathy2019deep}) and fast calibration/alignment of detectors. 

For tracking detectors, particle tracking is always a computationally challenging step.
Several approaches have been developed recently for tracking based on deep learning \cite{farrell2018novel}, but there are still open questions about the best way to incorporate such techniques. 
 The problem in Nuclear Physics experiments is typically different,
 being characterized by most of the computing cycles spent in propagating the particles through inhomogeneous magnetic fields and material maps. 
 Here AI can contribute to determine the optimal initial track parameters allowing to decrease the number of iterations needed by Kalman-Filter. 
 
As for particle identification and event classification, we have witnessed in the last years a growth of applications based on machine learning both for global particle identification (see, \textit{e.g.}, \cite{Derkach:2018zzv,Derkach:2019amb}) as well as custom novel solutions which combine different architectures for specific detectors (see, \textit{e.g.}, \cite{Fanelli:2019qaq} for imaging Cherenkov detectors). 

The search for rare signatures in large acceptance detectors at high intensities necessitates advanced techniques to filter those events. The GlueX experiment at Jefferson Lab for example is searching for exotic hadrons and demonstrated the utility of machine learning techniques based on BDTs \cite{drucker1996boosting} to achieve the required performance in filtering events with rare reactions~\cite{Dugger:2014xaa}. 

The utilization of jets at the future EIC can be beneficial for a variety of fundamental topics~\cite{Page:2019gbf}, including the gluon Wigner distribution, the gluon Sivers function, the (un)polarized hadronic structure of the photon, the (un)polarized quark and gluon PDFs at moderate to high momentum fraction (x) as well as studies of hadronization and cold nuclear matter properties. 
Machine Learning is having a major impact in jet physics, empowering powerful taggers for boosted jets as well as flavor tagging, and various deep learning applications like recursive neural network which leverage an analogy to natural language processing~\cite{louppe:2017ipp} have been developed. ML4Jets \cite{ML4Jets_2020} is a series of workshop dedicated to these topics. 

Another area where AI can significantly contribute is that of fast simulations. Simulating the detector response of large scale experiments like EIC is typically slow and requires immense computing power. 
One of the most computationally expensive step in the simulation pipeline of a typical experiment is the detailed modeling of the high multiplicity physics processes characterizing the evolution of particle showers inside calorimeters. AI, could speed up simulations and potentially complement the traditional approaches.
Recent advances with generative networks (see, \textit{e.g.}, GAN, VAE, Flow-based models \cite{goodfellow2014generative, doersch2021tutorial, rezende2016variational}) look as a compelling alternative to standard methods with orders of magnitude increase in simulation speed~\cite{Paganini:2017hrr} but so far usually at the cost of reduced accuracy.

Detector design is another fundamental area of research for EIC.   
Advanced detector design often implies performing computationally intensive simulations as part of the design optimization process.   
One of the conclusions from the DOE Town Halls on AI for Science on 2019 \cite{stevens2020ai} was that ``\textit{AI techniques that can optimize the design of complex, large-scale experiments have the potential to revolutionize the way experimental nuclear physics is currently done}''.
There are at present various AI-based  optimization strategies based on, \textit{e.g.}, reinforcement learning or evolutionary algorithm \cite{li2016learning,whitley1994genetic}.   
Among these, Bayesian Optimization (BO) \cite{snoek2012practical, jones1998efficient} has gained popularity for its ability of performing global optimization of black-box functions that are expensive to evaluate and that can be in addition noisy and non-differentiable. 
It consists of a surrogate modelling technique where the regression is typically done through Gaussian processes or decision trees depending on the dimensions of the problem, and a cheap acquisition function is used to suggest which design points to query next, overall minimizing the number of evaluations.  

Recently, an automated, highly-parallelized, and self-consistent procedure has been  developed~\cite{Cisbani:2019xta} and tested for the dual-radiator Ring Imaging Cherenkov (dRICH) design, which has been considered as a case study.
These studies not only showed a statistically significant improvement in performance compared to the existing baseline design but they also provided hints on the relevance of different features of the detector
for the overall performance. 
This procedure can be applied to any detector R\&D, provided that realistic simulations are available.
One example is the optimization of detector materials, e.g. the optimization of large size aerogel composites for aerogel-based detectors in~\cite{AIJLAB_townhall2020}. 

Beyond individual subdetectors AI can be also used to efficiently optimize the design of different sub-detectors combined together, taking into account mechanical and geometrical constraints. 
An interesting approach consists in a multi-objective optimization (see, \textit{e.g.}, \cite{deb2000fast, deb2001multi,feliot2017bayesian}), 
which allows to encode the performance of the detectors as well as other aspects like costs in the design process, to determine the Pareto front~\cite{jin2008pareto}. 
Currently ongoing activities within the EIC R\&D program which are leveraging AI for optimization include the EM/Hadronic Calorimetry, \textit{e.g.}, optimizing the glass/crystal material selection in ``shared rapidity regions'' for best performance of the EM calorimeter.

Even more, AI has the ability to optimize the collection of all subdetectors of a large detector system, using more efficiently the figures of merit we use to evaluate the performance that drive the detector design.  
Remarkably, the design optimization of multiple subdetectors operating together has not been explored yet. This is a high dimensional combinatorial problem that can be solved with AI.

This is undoubtedly a strategic moment to discuss how to fully take advantage of the new opportunities offered by AI to advance research, design and operation of the future EIC.
The interest of the community has been evidenced by the number of contributions and attendance of workshops dedicated to AI in Nuclear Physics, e.g. at the~\cite{bedaque2020report, jointML_WS2020}, and the 2021 AI4EIC-exp workshop~\cite{AI4EICexp_WS2021}, which bring together the communities directly using AI technologies and provide a venue for discussion and identifying the specific needs and priorities for EIC.

\chapter{The Case for Two Detectors}
\label{part3-chap-Two.Detectors}

As documented abundantly elsewhere in this report, the concept of an EIC encapsulates a very broad potential physics program with experimental signatures ranging from exclusive production of single particles in $ep$ scattering to very high multiplicity final states in \eA collisions and potentially spanning a wide range of centre of mass energies. Rarely, if ever, has such a diversity of scientific
output been condensed into a single project. 
The very high target luminosities of the facility imply high statistical precision for some observables, which must be matched with a similar or better level of systematic precision in order to 
realise the full physics potential, correspondingly placing emphasis on carefully optimized instrumentation.    
Each aspect of the physics program can in principle be applied to define its own idealised detector design and configuration. Whilst many aspects align towards the same basic 
needs, it is impossible to optimize for the full breadth of physics ambitions with a single detector. 

All previous high energy colliders have housed more than one interaction point and associated detectors.\footnote{Although not at the highest energies, the lepton-based $B$ factories, Babar and Belle, are an exception to this. In those cases, the basic design precluded having two detectors on the same ring. It is interesting to note that in the absence of multiple detectors at a single site, complementary facilities were built more-or-less concurrently at SLAC and KEK.} In some cases (e.g. LEP, Tevatron), the physics goals of the different experiments have been
similar, or even identical. In others (e.g. RHIC), the focus has been complementary, or completely different. 
Elsewhere, there has been a mixture of the two approaches. For example, at the LHC, ATLAS and CMS aim for the widest survey of physics at the high luminosity energy frontier, whereas ALICE and LHCb have more specialised focus on heavy ion collisions and precision heavy flavour physics, respectively. 
Similarly at HERA, the H1 and ZEUS collider experiments were designed for a general study of energy frontier lepton-hadron physics, whereas the polarised fixed target HERMES experiment had an almost completely orthogonal program.  

Prior to the Yellow Report meeting series and report, a second interaction point was already very much part of the EIC machine design. However, the nature of the experiment that would surround it remained something of a blank page. 
Following extensive discussion among the physics and detector working group communities during the Yellow Report exercise, this chapter assesses complementarities and conflicts among the detector requirements to fulfill the EIC physics aims and explores the opportunities to expand the physics program and mitigate risks through a two detector solution. 
After introducing the boundary conditions imposed by the basic EIC machine design (section~\ref{sec:complement-boundary}), motivations for having two detectors are given in sections~\ref{sec:complement-xcheck} -~\ref{sec:complement-physics}, before a brief discussion of the intrinsic complementarity offered by fixed target mode operation in section~\ref{sec:complement-ft}.

\section{Boundary Conditions and Important Relations}
\label{sec:complement-boundary}

There are several global and IR-specific boundary conditions, which need to be obeyed at both interaction regions and must be kept in mind when designing a two-detector EIC program: 
\begin{itemize}
    \item Both detectors should be able to have optimal performance over the entire EIC center-of-mass energy and luminosity range.
    \item The design of the interaction regions need to be compatible with the current design of the EIC machine.
    \item The Luminosity of a collider like EIC is given by $L=\frac{N_eN_h}{4\pi\sigma_x\sigma_y}H f_{rep}$ with $N_e$ and $N_h$ being the electron and hadron bunch intensities, respectively, and $\sigma_x$ and $\sigma_y$  the horizontal and vertical RMS beam sizes at the interaction point (IP), which are assumed to be identical for the two beams. $H$ is a factor reflecting the impact of the hourglass effect and the crossing angle; $f_{rep} = N_b f_{rev}$ denotes the bunch repetition rate, with $N_b$ the number of bunches per ring and $f_{rev}$ the revolution frequency of the collider rings.
    If one wants to increase luminosity one can increase the number of bunches and as such the bunch repetition rate, the intensity per bunch is reduced. One could decrease the beam emittances by a factor $n$, which immediately leads to increased requirements on the cooling of the beams, like reducing the cooling time by the same factor $n$. There is another issue one needs to deal with, the increased chromaticity. To bring it back to its original value one needs to move the focusing quadrupoles closer to the IP and to do so one needs to increase the crossing angle to have space for the magnets.
    \item The size of the crossing angle directly impacts the acceptance of the detector; a crossing angle of 25 mrad limits the acceptance to a rapidity of $\sim4.2$
    \item The size of the synchrotron radiation fan limits how small one can make the radius of the beam pipe around the IP. To limit the emission of synchrotron radiation the axis of the experimental solenoid is aligned along the electron 
    beam.
    \item The luminosity and the length of the detector are directly coupled. The length of the detector dictates the minimum possible distance between the first focusing quadrupole 
    and the IP, named $L^*$. The luminosity is directly proportional to $1/L^*$, 
    such that larger $L^*$ leads to reduced luminosity.
    \item operating two detectors simultaneously will result in splitting the luminosity between the two interaction regions. 
    \item The need for crab cavities to obtain high luminosity despite a 
    crossing angle limits the longitudinal space a long the beam line for forward detectors. One can not place for example Roman Pots after the crab cavities as the correlation to the scattering at the IR would be completely washed out. 
    Therefore all detectors along the beamline for charged particle detection need to be placed between the crab cavities.
\end{itemize}

\section{Dedicated Detector Designs versus General Purpose Detectors}

In some previous contexts (e.g. LHCb at the LHC), the scientific reach of a collider facility has been enhanced by the addition of a detector dedicated to a particular area of the physics program. The idea that the EIC might be instrumented with one General Purpose Detector 
and one detector strongly optimized towards a particular physics area, for example exclusive
production modes with intact protons, has therefore been explored as part of the Yellow Report exercise. Whilst it would be possible to enhance the output in particular areas with this sort of model, no proposal with a broad enough scope or a compelling enough physics capability to justify the substantial cost of a standalone detector has been identified
within the limitations described in section~\ref{sec:complement-boundary}. 

Here, the realization that the EIC science and detector design are unique in that there has never been a collider detector with both the central and forward (or backward) acceptances maximized in tandem (see section~\ref{part3-sec-DetChalReq.ID.IR}) is equally important. The integration of the detector and Interaction Region allows for choices of two interaction points that place different emphasis on the acceptance of the beam line instrumentation, precision and gaps in various detector regions, and variations of the beam line to emphasize different science processes, possibly exploiting the idea to have the beam line elements act as magnetic spectrometer to provide a secondary focus.

At present the working assumption is therefore that we will aim to produce a pair of General Purpose Detectors that have differences in the details of their physics and technology foci, such that they are designed from the outset to optimize the overall output of the collider as a whole. 
This decision is important in time scheduling for the EIC, since detectors with a dedicated physics focus often operate for a limited period, whereas General Purpose Detectors usually require the maximum possible integrated luminosity and thus ideally operate for the full lifetime of the machine.

\section{Motivation for Two Detectors: Technology Considerations}
\subsection{Cross Checking}
\label{sec:complement-xcheck}

Blind alleys and wrong turns are intrinsically part of the nature of science, whether caused by honest analysis mistakes, instrumental malfunctions or the inevitable statistical fluctuations that appear when searching for new effects beyond our current understanding. 
However, by continuously testing and probing current theories, the scientific method by nature
corrects such diversions when sufficient experimental input becomes available, such that our understanding tends asymptotically towards an ever-improving description of nature. There are many 
famous historical examples of apparently convincing, but wrong, signals for new science, 
from Pons and Fleischmann's cold fusion to 17~keV neutrinos in tritium decays to superluninal neutrinos travelling between CERN and  Gran Sasso \cite{Berlinguette2019,Goodman:2019gin}. 
Collider physics is by no means exempt, with many examples of premature discoveries that disappeared. Examples from lepton-hadron scattering include the HERA signals for leptoquarks \cite{Collaboration:2011qaa,Abramowicz:2012tg} and for at least two different types of pentaquark \cite{Aktas:2007dd,Chekanov:2005at}. These examples illustrate that when embarking on a science program with as much discovery potential as the EIC offers, it is of fundamental importance to be able to cross-check important new results.
This is possible to some extent within a single experiment, for example by collecting larger data sets when we are misled by statistical fluctuations. However, this can be a time-consuming process and 
immense resource can be wasted chasing erroneous signals. Furthermore, there are cases where additional data may even add to the confusion, for example where a false signal is created by a subtle experimental or analytical effect that is never revealed. 

Perhaps the most compelling reason for including two detectors in the EIC design, operated 
independently by different teams of scientists, is thus to ensure a capability to cross check the important, possibly ground-breaking, new results that we expect to obtain. To avoid the possibility of correlated misleading signals in the two detectors, the best approach is to use different instrumentation technologies in the two detectors, as discussed further in section~\ref{sec:complement-tech}. 

\subsection{Technology Redundancy}
\label{sec:complement-tech}

It is natural that any new detector component at a new world-leading facility, particularly one that has a design and construction lead-time of several years, will seek to optimize its performance by employing novel, state-of-the-art technologies. Whilst this in principle ensures the maximum possible return in terms of physics exploitation, it also carries some intrinsic risk. No detector design can be 100\% guaranteed to succeed and, particularly close to the beam, environmental  conditions cannot be perfectly known before operation begins.
There is thus always at least a small risk of failure of any detector sub-component. One means of mitigating this risk when viewing the EIC facility as a whole is to employ different technologies to fulfill similar roles in two complementary detectors. 
The General Purpose Detectors  at the LHC, ATLAS and CMS, are a good example. Whereas CMS is compact and pushes the boundaries of design for example in its all-silicon tracking region, ATLAS has an ambitious outer detector design comprising the one of the world's largest systems of magnets with an unconventional superconducting barrel toroid layout. 

\subsection{Cross Calibration}
\label{sec:complement-xcal}

\begin{figure}[ht]
    \centering
    \includegraphics[width=0.7\textwidth]{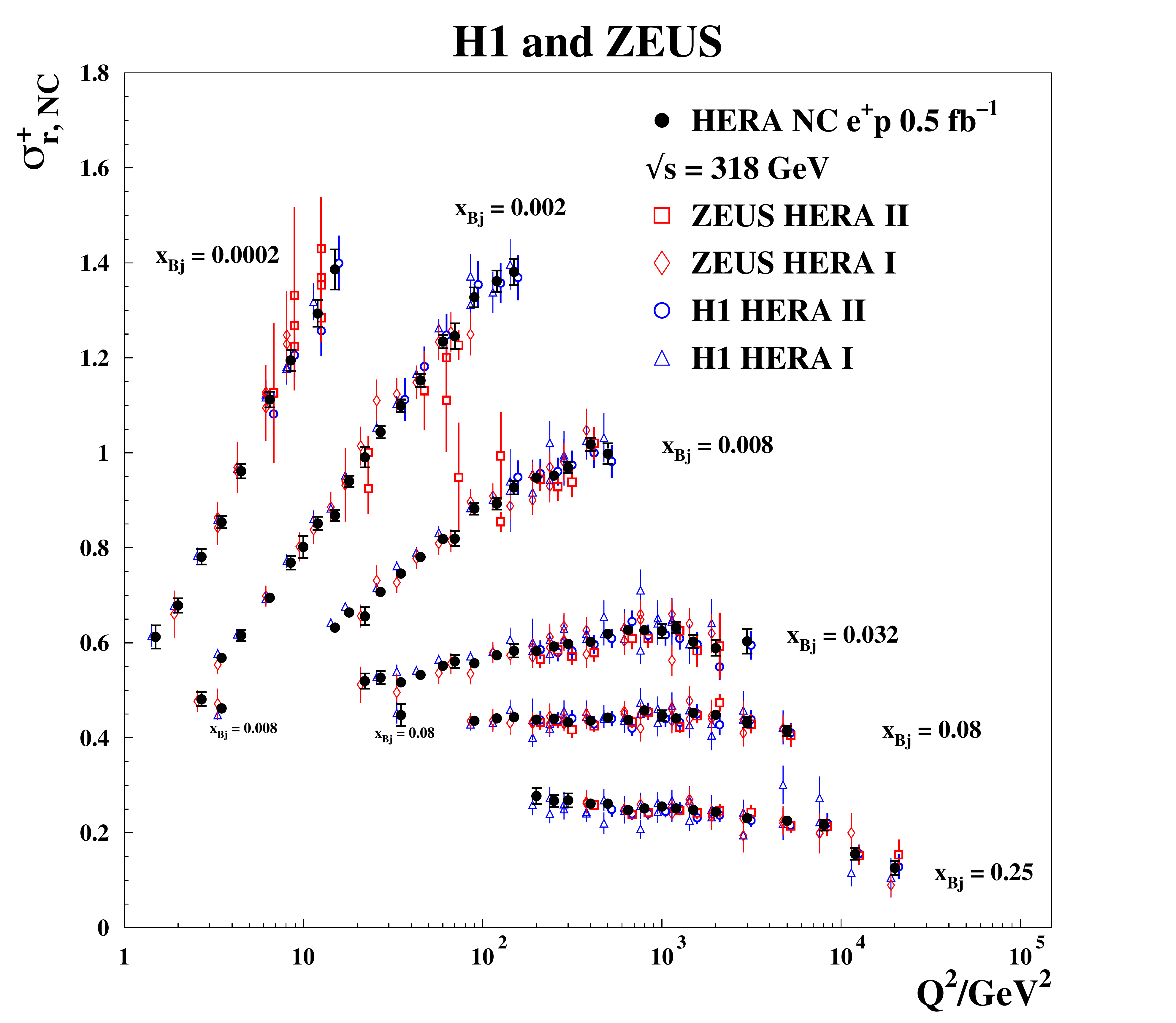}
    \caption{Comparison between input data from H1 and ZEUS and output data from the final combination of HERA data, for selected $x$ values in the $e^+ p$ neutral current cross section. The inner error bars are statistical, whilst the outer error bars show the statistical and systematic uncertainties combined in quadrature. Taken from \cite{Abramowicz:2015mha}.} 
    \label{fig:complement-h1zeus}
\end{figure}

Embedding complementarity in the designs of pairs of detectors fulfilling similar purposes 
as discussed in section~\ref{sec:complement-tech} can offer the opportunity of minimising systematic uncertainties through cross-calibration. A good example is offered by the H1 and ZEUS detectors
at HERA, where the final combination of inclusive DIS data-sets between the two experiments, which created a final combined legacy measurement \cite{Abramowicz:2015mha}, had a largely unexpected positive impact on the combined systematic uncertainties, as illustrated in figure~\ref{fig:complement-h1zeus}.
This arises as a result of the redundancy in the reconstruction of the event kinematics whereby $x$ and $Q^2$ can be obtained either from the scattered electron, the hadronic final state, or a combination of the two. Since, crudely speaking, H1 had the better-performing electromagnetic calorimeter and ZEUS had superior hadronic calorimetry, there was effectively a cross-calibration between the experiments, whereby it became possible to benefit simultaneously from the H1 response to electrons and the ZEUS response to hadrons. 

This example from H1 and ZEUS could be carried forward directly into an EIC context, where $x$ and $Q^2$ reconstruction precision will be of very high importance. There may be other such cases, for 
example in possible trade-offs between particle identification and tracking performance, where space restrictions in the inner detector region may preclude the simultaneous optimization of both. Similarly, 
it may be possible to vary the interaction region design slightly between the two interaction points 
in a way that places different limitations on the acceptance of the beamline instrumentation, or 
exploits the idea of a secondary focus. 
Whilst it may not be possible to optimize the design of all possible components in a single detector, 
it may yet be possible to do so in a carefully pre-prepared combination of two detectors. 

\subsection{Technology Redundancy and Cross Calibration at EIC}

The general requirements for an EIC detector as identified stem from the central goal to cover the entire EIC physics program, the expected event geometries and the constrains coming from the overall collider design. Nevertheless, the detector design and the selected technologies may differ.

The following general characteristics are assumed, the central detector instruments the pseudo-rapidity region $-4 < \eta < 4$. This acceptance range matches the needs of the inclusive, semi-inclusive, jet physics and spectroscopy studies. It is complemented by the very forward and backward detectors (Sec.~\ref{part3-sec-Det.Aspects.FFDet}) ensuring the hermeticity and the forward tagging required by specific items of the physics program; in particular: exclusive reactions and diffractive channels. The main requirements of the central detector dictated by the physics program and the event geometry are related to  (1) tracking and momentum measurements, (2) electron identification, (3) hadron identification and  (4) jet energy measurements, while the overall detector size is imposed by collider design considerations (see section \ref{sec:complement-boundary}):
\begin{enumerate}
\item{very fine vertex resolution, at the 20~\si{\micro}m level for the three coordinates, is needed, while a moderate momentum resolution around 2\% matches the physics requirements;}
\item{the purity requirements for electron/hadron separation are at the 10$^{-4}$ level in the backward and barrel regions and, for this purpose, the figures for the electron energy resolution are very demanding, in particular in the backward region where an r.m.s. of 2\%$\sqrt(E)$ 
is needed; in the same direction, the request is for a light detector, where the material budget should not exceed 5\% X$_0$;}
\item{the identification of the different hadron species in the whole central detector coverage, namely for hadrons with momenta up to 50\,GeV/$c$, is requested with 3$\sigma$ $\pi$/K separation over the whole range as reference figure; }
\item{the measurement of jet energy in the forward direction is a necessity, while moderate resolution of the order of 50\%$\sqrt(E)$ can match the needs;}
\end{enumerate}

Table~\ref{tab:detector-technologies} summarizes the different technologies, which have been identified throughout the Yellow Report initiative to fulfill the science requirements. As can be seen sometimes technologies combine several functions, i.e TRD e-h separation and tracking information, or a AC-LGAD based pre-shower combining tracking, ToF and e-h separation, in one detector versus single function detectors. In addition all of the different technologies will come with different amount of dead material impacting performance and the capability to study systematic detector effects, these are two of the most critical characteristics if one wants to quantify the performance or bias of a measurement.
Such basing the design of the two General Purpose Detectors on different technologies from the beginning will naturally design technological redundancy and the ability for cross calibration into the two General Purpose Detectors and even more importantly both aspects guarantee maximal capability for best cross checking results. To guarantee optimal integration of complementarity in the two General Purpose Detectors excellent communication between both collaborations and as complete as possible testbeam information and GEANT MC tools are needed.

\section{Motivation for Two Detectors: Complementarity of Physics Focus}
\label{sec:complement-physics}

The wide-ranging physics program of the EIC, combined with the physical limitations of the beam and IR design, leads to diverse, sometimes mutually exclusive, demands on the detector design. This section explores some of the examples that have already come to light. Others are expected to emerge
as more detailed simulations of the overall detector design become available and the subtleties involved in combining input from multiple detector components become apparent. 

\subsection{Experimental Solenoid Design}
\label{sec:complement-field}

One area where there is a clear potential conflict between the needs of different physics processes is in the strength of the solenoid magnetic field in the central region of the detector. 
For a single detector, a compromise is required between low field values to avoid loss of acceptance where low transverse momentum charged particles are bent to the extent that they are not reconstructed in the tracking region versus high field values to allow precision tracking measurements of charged particles with large transverse momenta. In the case of the scattered electron in neutral current DIS, the tracking measurement is supplemented by a calorimeter energy measurement that improves in precision with increasing $p_T$.
However, for other cases where hadrons (or perhaps muons) are involved, the tracking measurement 
is the limiting factor. 

\begin{figure*}[ht]
    \centering
    \includegraphics[width=0.95\textwidth]{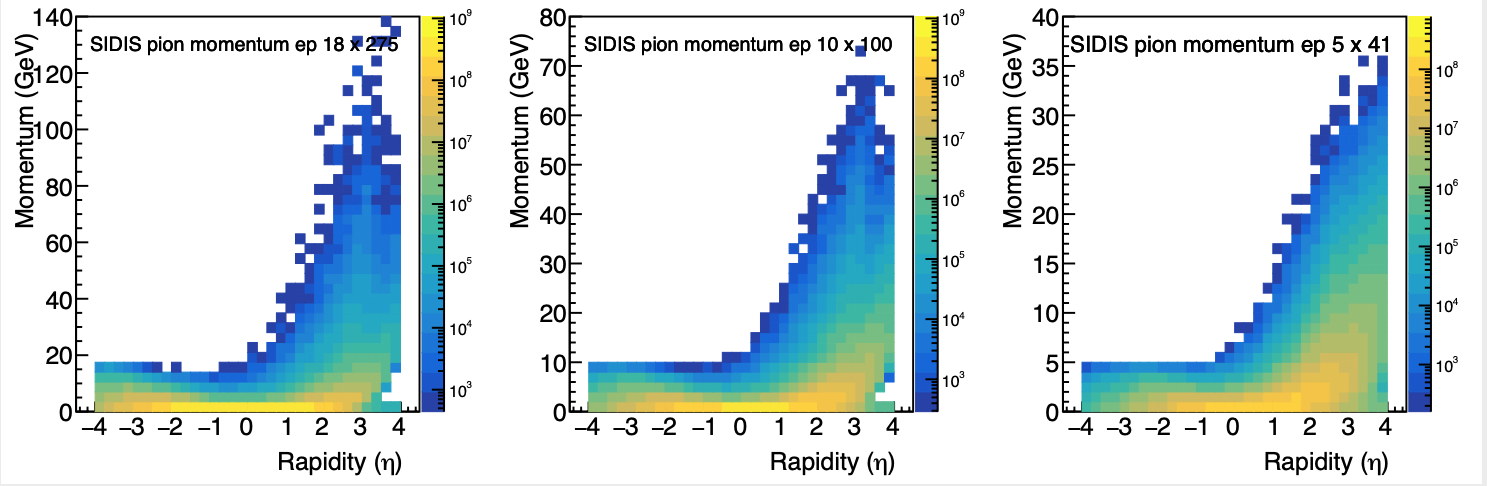}
    \caption{Simulated distribution of momentum versus pseudorapidity for hadron production for different c.o.m energies from left to right of 18\,GeV electrons on 275\,GeV protons, 10\,GeV on 100\,GeV and 5\,GeV on 41\,GeV. The following cuts have been applied: inelasticity $0.01<y<0.9$, momentum transfer $Q^2 > 1$\,GeV, and $W^2>10$\,GeV$^2$, as well as $0.05<z<0.95$ for the pions. The normalisation is arbitrary.} 
    \label{fig:complement-pispectra}
\end{figure*}

As an example, the distribution of kinematics for SIDIS hadron production is illustrated in figure~\ref{fig:complement-pispectra}. Pions are shown here, but other hadrons (K, p and n) 
follow similar distributions.
The distribution is dominated by particles with $p_T$ substantially below $1 \ {\rm GeV}$, which are important for example for precision measurements of semi-inclusive DIS, leading to studies of TMDs and fragmentation functions, and for spectroscopy. 
On the other hand, the distribution extends to the regime of several tens of GeV, where the precision measurements of the highest $p_T$ particles is crucial for methods of reconstructing the event kinematics that rely on hadrons, such as the Jaquet-Blondel method for charged current DIS and mixed hadron-electron methods for neutral current DIS at low $y$.

In the discussion during the Yellow Report the following more quantitative considerations and requirements for the solenoidal  magnet have been identified;
\begin{itemize}
 \item The main advantage of, for example, a 3 Tesla versus a 1.5 T central solenoid field is for the momentum resolution of charged particles as a function of pseudo-rapidity, for detailed studies see section~\ref{part3-sec-Det.Aspects.Magnet}. Doubling the magnetic field can lead to a reduction of the momentum resolution by a factor of $\sim2$ from a leading order 1/B dependence. This is relevant in the central region, but even more so in the forward pseudo-rapidity regions, $\eta > 2$, where the momentum resolutions rapidly worsen. For example, for $\eta \sim3 $, a momentum resolution of $\sim$2-3\% is achievable for pions with momenta up to about 30 GeV/$c$ with a 3 T central field.
 \item The main advantage of accessibility of low central solenoid fields (down to ~0.5 T) is towards 
 the low-$P_T$ acceptance of charged-particle tracks. A central field of 0.5 T roughly equates to a detection capability of charged particles down to transverse momenta of below $\sim0.05$ GeV/$c$. This is for example important in mapping the decay products of heavy-flavor mesons and in measuring inclusive charged particle spectra.
 \item The required low $p_T$ detection threshold strongly depends on one requires particle identification or not. The following table summarizes the achievable lower cut offs for charged pions requiring to reach a PID detector at 1m or only reconstructing the track $p_T$ and its charge with the microvertex tracker.
 \begin{center}
 \begin{tabular}{l|c|c|c}
  lowest $p_T$ & 0.5 Tesla & 1 Tesla & 3 Tesla \\\hline\hline
      with PID @1m  & 75 MeV & 225 MeV & 450 MeV   \\ 
        no PID & 25 MeV & 50 MeV & 100 MeV       \\
 \end{tabular}
 \end{center}
 \item One other critical aspect of the magnet is its bore radius (see also subsection \ref{sec:complement-central}).
 \begin{itemize}
 \item To guarantee sufficient flexibility for detector choices to satisfy EIC science requirements of good tracking and particle identification, and hermetic electromagnetic calorimetry, assuming hadronic calorimetry will be located outside the magnet coil. Note: the barrel region is “defined” as between pseudo-rapidity of -1 and 1. This means that for a certain bore size $x$ (typical assumed bore diameters are of the order of 3.2 m), the coil space will be $\sim1.2*x$, additional space is needed for the cryostat bringing the total length of the magnet to $\sim 3.9m$. The cryostat length defines the separation between the central barrel region and the forward and/or backward end cap regions.
 \item The coil/cryostat length is a delicate balance.\\
  Space in the barrel region for subdetectors is at a much higher premium than in forward and backward
  regions, which can pose limits on reaching the EIC detector requirements. The transition of the barrel to the endcaps  is likely one of the regions where detector infrastructure (support, cabling, etc.) will reside, and detectors such as a RICH will “flare out”, or trade-offs between hermeticity of an electromagnetic calorimeter and a DIRC will occur.
 \end{itemize}
\end{itemize}

It is clearly challenging to simultaneously optimize the magnetic field strength and other features of the magnet to all aspects of the EIC physics program. For a single detector solution, either one physics area has to be prioritised over others, or else a compromise is required, whereby one operates the experiment at different B-fields to optimize for different physics scenarios, but this would be extremely time consuming. On the other hand, in a scenario involving two detectors, the field strengths could be rather different, such that two General Purpose Detector experiments specialize in different physics areas, whilst retaining the ability to cross-check one another.  

\subsection{Tracking versus Particle Identification}
\label{sec:complement-central}

In addition to magnetic field strength considerations, there is also a possible trade-off in the central 
region of the detector between charged particle tracking precision and high performance particle 
identification. Again, optimization for different physics topics leads to different solutions. For example, inclusive DIS requires high quality tracking as discussed in section~\ref{sec:complement-field} together with efficient separation between electrons and pions, whereas semi-inclusive DIS measurements lead to add additional hadron identification requirements. Other topics such as exclusive vector meson production processes, particle spectroscopy and heavy flavour studies all require a mixture of the two with differing emphasis in each case. 

\begin{table*}[ht]
\footnotesize
\begin{tabular}{l|c|c|c|c}
         & \multicolumn{4}{c}{Radial space needs}\\
Function & Minimum & Maximum & Minimum & Maximum \\ \hline\hline
Tracking & \multicolumn{2}{c|}{ All Silicon} & \multicolumn{2}{c}{Silicon + TPC} \\
(includes 5 cm support) & \multicolumn{2}{c|}{\includegraphics[width=0.3\textwidth]{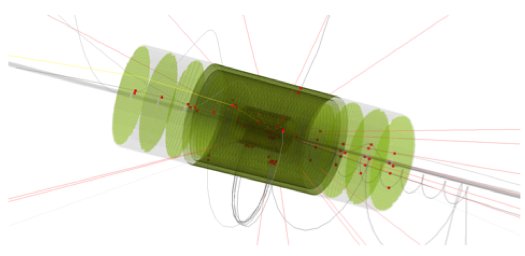}} & \multicolumn{2}{c}{\includegraphics[width=0.2\textwidth]{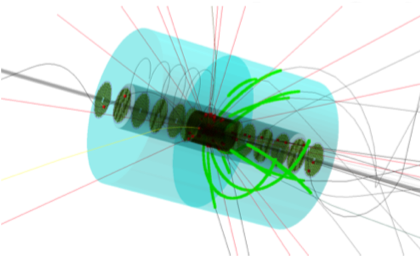}}\\
 & 50 cm & 60 cm & \multicolumn{2}{c}{85 cm}\\ \hline
 Hadron & \multicolumn{2}{c|}{RICH} & \multicolumn{2}{c}{DIRC} \\
 PID & \multicolumn{2}{c|}{50 cm} & \multicolumn{2}{c}{10 cm}\\ \hline
 EM Calorimetry & 30 cm & 50 cm & \multicolumn{2}{c}{High-Resolution to achieve P $<$ 2 GeV}\\
                &      &       & \multicolumn{2}{c}{50 cm} \\ \hline
PID \& EMCal & 10 cm & 15 cm & 10 cm & 15 cm \\
Support Structure & & & & \\ \hline
Total & 140 cm & 175 cm & 155 cm & 160 cm \\ \hline\hline
\end{tabular}
\caption{A high level description of the spatial needs of different subdetector combinations in the barrel.}
\label{two.base.options}
\end{table*}

Much of the space in the central region in the nominal detector design is taken up with a large Time Projection Chamber, which brings the advantages of low material budget and high performance particle identification through specific energy loss ${\rm d} E / {\rm d} x$, with additional particle identification capability through RICH and Transition radiation detectors and only a relatively small silicon vertex detector. It is therefore well matched to a physics program that requires particle ID for a range of particle species and momenta.  

An alternative concept that has been put forward during the Yellow Report discussions is to instrument the central region with an all-silicon tracker, based on CMOS Monalithic Active Pixel Sensors (MAPS). The corresponding device would be significantly more compact (outer radius of $45 \ {\rm cm}$ as compared with $80 \ {\rm cm}$) and would slightly improve transverse momentum measurements and vertexing at the expense of particle identification performance. The space saved might be used for example to implement other solutions to particle identification. These different technological solutions have a direct interplay with the design of the detector solenoid. Table \ref{two.base.options} shows the interplay between subdetector technologies and the requirement for the detector solenoid bore size. Of course this optimization needs to go hand-in-hand with the optimization of the magnitude of the B-field.

Once again, a thorough investigation of the merits of the different designs will become possible when fully integrated simulations of multiple sub-detectors become available. 

\subsection{Hermiticity and Acceptance Gaps}
\label{sec:complement-gaps}

\begin{figure*}[ht]
    \centering
    \includegraphics[width=0.5\textwidth]{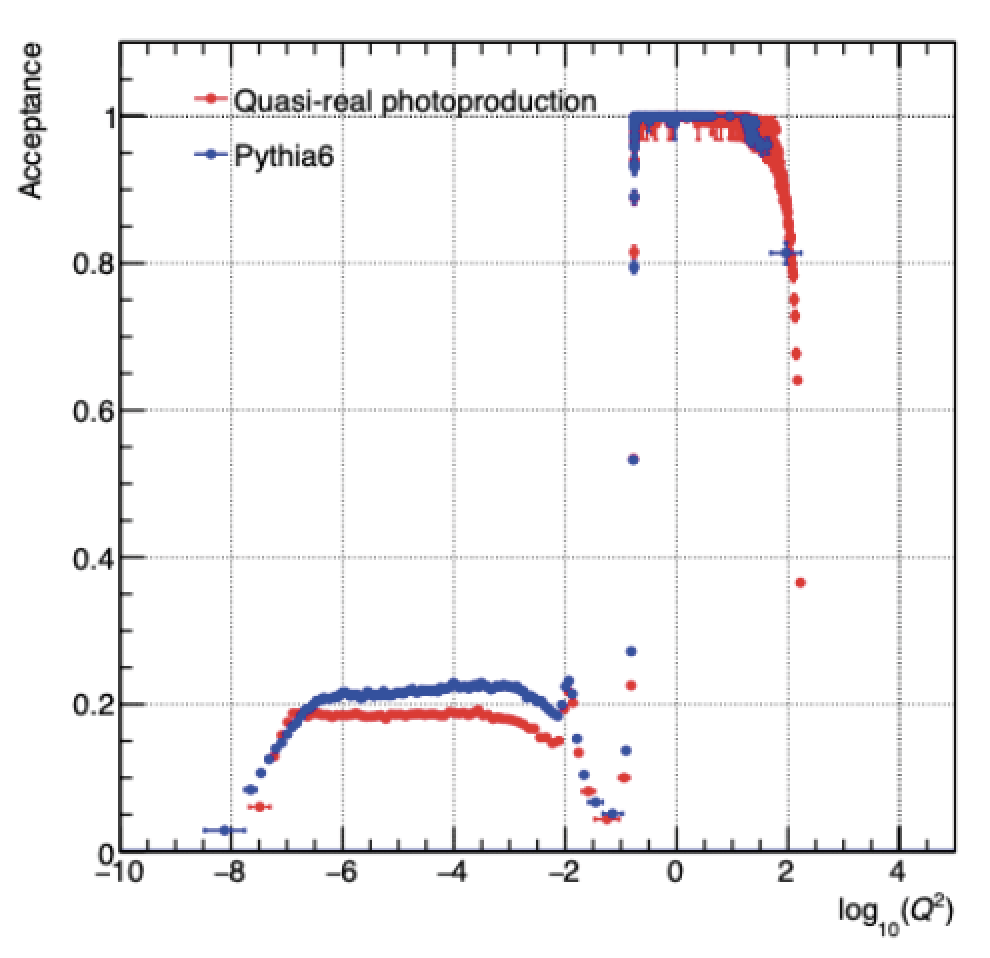}
    \caption{Assessment of the overall acceptance for the reference detector and combined with the low-Q$^2$ taggers placed along the IR for the scattered electron as a function of $Q^2$ when considering both beam-line instrumentation and the central electromagnetic calorimeters. Results are shown from 
    two different physics models.}
    \label{fig:complement-accgap}
\end{figure*}

Whilst the intention of a general purpose detector is to cover the full solid angle for particle production as hermetically as possible, gaps in acceptance and directional peaks in dead material are inevitable in any single design. An example for the case of the scattered electron in inclusive neutral current DIS is shown for the nominal EIC detector design in figure~\ref{fig:complement-accgap}.
There is inevitably a gap in acceptance between the lowest scattering angles where the electron is reconstructed using beam-line tagging and larger scattering angles where the electromagnetic calorimeter in the central detector is used. 
The strong correlation between the scattered electron angle and $Q^2$ leads to a corresponding gap in acceptance as shown in the figure. Although the interaction region design is strongly constrained by machine considerations, it may yet be possible to design the beam pipe and associated instrumentation such that the gap in acceptance arises at a slightly different angle in a second detector compared with that in the first, such that there is complete coverage at all $Q^2$ from the point of view of the overall EIC program. 
Other similar examples will inevitably appear throughout the detector design. One example is the outgoing hadron beamline instrumentation, with the potential to vary the positions of Roman pots and a B0-like detector if a second interaction region design allows suitable variation in gaps between beam elements. In the central detector region, cracks between modules of any sub-detector are unavoidable, as well as at the interface between barrel and end-cap components. Once again, forward planning exploiting the redundancy offered by having two detectors may avoid such issues for the combination. 

\subsection{Optimization to different centre of mass energies}

\begin{figure*}[ht]
    \centering
    \includegraphics[width=0.7\textwidth]{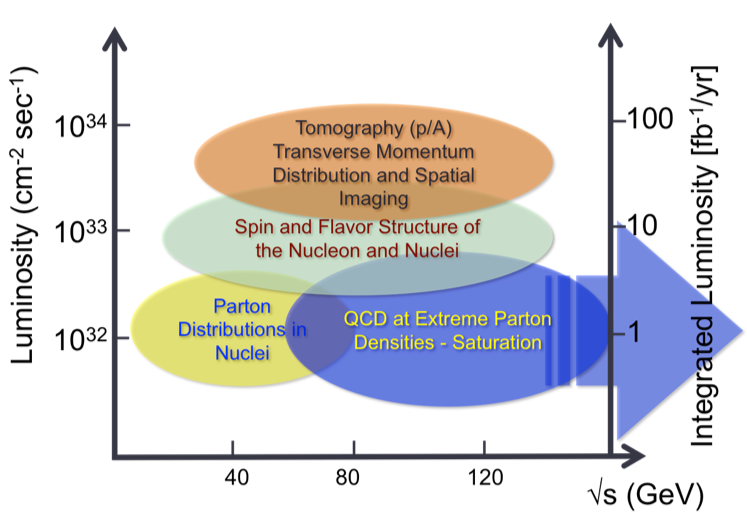}
    \caption{A schematic representation of the luminosity and center-of-mass energy need of the different EIC physics topics.}
    \label{fig:complement-sqrts}
\end{figure*}

The EIC program involves running at least two different centre of mass energies as summarised in figure~\ref{fig:complement-sqrts}.

This gives an other very important opportunity to optimize the two general purpose detectors by also optimizing the two interaction regions for complementarity, i.e. to maximise the instantaneous
luminosity at either high or low $\sqrt s$. If the design of the second interaction region would for example focus on the lower $\sqrt s$ value, then the quadrupole magnets 
might be moved closer to the interaction point, eventually in the acceptance of the detector, the crossing angle and details of the secondary focus may change, which require a correspondingly different detector solution, particularly in terms of the beamline instrumentation. Table \ref{tab:2ndIRreqs} gives an idea how the two different IRs could be optimized taking a complementarity approach.

The slightly different physics focus at large and small $\sqrt s$ also leads to different
optimizations for the central detector components and magnetic fields via the considerations discussed in the previous sections. It may therefore be possible to have two general purpose detectors operating throughout the EIC lifetime, with their performances optimized to different $\sqrt s$ such that all measurements can be cross-checked, but the specialisms
of the two detectors differ. 

\begin{sidewaystable*}[ht]
\begin{center}
\vspace*{8mm}
\caption{Summary of 2nd IR design opportunities and their comparison to the 1st IR. \label{tab:2ndIRreqs}}
\begin{tabular}{c|l|c|c|l}
\toprule
\# & Parameter & EIC IR \#1 & EIC IR \#2 & Impact \\
\midrule
1 & Energy range & & & Facility operation \\
  & ~~~electrons [GeV] & $5-18$ & $5-18$ & \\
  & ~~~protons [GeV] & $41$, $100-275$ & $41$, $100-275$ & \\ \hline
2 & CM energy range  & & & Physics priorities \\
  & of optimum luminosity [GeV] & $80-120$ & $45-80$ & \\  \hline
3 & Crossing angle [mrad] & $\leq 25$ & $\leq 50$ & $p_T$ resolution, acceptance, geometry \\ \hline
4 & Detector space symmetry [m] & $-4.5/+5.0$ & $-(3.5-4.5)/+(5.5-4.5)$ & Forward/rear acceptance balance \\ \hline
5 & Forward angular acceptance [mrad] & 20 & $20-30$ & Spectrometer dipole aperture\\ \hline
6 & Far-forward angular acceptance [mrad] & 4.5 & $5-10$ & Neutron cone, $p_T^{max}$ \\ \hline
7 & Minimum $\Delta(B\rho)/(B\rho)$ allowing for & & & Beam focus with dispersion, \\
  & detection of $p_T=0$ fragments & 0.1 & $0.003-0.01$ & reach in $x_L$ and $p_T$ resolution, \\
  &  & & & reach in $x_B$ for exclusive processes \\ \hline
8 & Angular beam divergence at IP, & & & $p_T^{min}$, $p_T$ resolution \\
  & h/v, rms [mrad] & $0.1/0.2$ & $<0.2$ & \\ \hline
9 & Low $Q^2$ electron acceptance & $<0.1$ & $<0.1$ & Not a hard requirement \\ 
\bottomrule
\end{tabular}
\end{center}
\end{sidewaystable*}

\FloatBarrier
\section{Opportunities from Fixed Target Mode Operation}
\label{sec:complement-ft}

One final consideration is that a considerable range of complementary physics scope can be accessed by running experiments in fixed target mode, either simultaneously with collider mode operation or in dedicated EIC runs. 
This mode of operation has been achieved previously for example at RHIC by the STAR~\cite{Meehan:2017cum} experiment and at the LHC by the LHCb~\cite{Maurice:2017iom} and ALICE experiments~\cite{Barschel:2020drr}. 
It is most easily realizable through the introduction of a gas target which can be spatially separated from the colliding mode interaction point in order to remove ambiguities between beam-beam and beam-target collisions. 
Although the center of mass energy is drastically reduced, the very forward kinematics give access to regions of phase space that are hard to access in colliding beam mode and allow kinematic overlap and hence comparisons with previous fixed target experiments. 

In the EIC context either the electron or the proton / ion beam or both could be used in combination with the fixed target. The \ep and \eA operation mode  naturally accesses high $x$ physics and, depending on target, adds to previous data from HERMES, Compass and JLab. The \pp and \pA modes similarly address novel kinematic regions and interface to previous experiments. This additional physics scope is in principle achievable 'for free', but requires some prior thinking in the detector design.

\section{Summary}

The clear conclusion from the Yellow Report exercise is that the best way to optimize the science output of the EIC is to construct two General Purpose Detectors with associated communities of experimental physicists that operate in friendly competition, as has been the case at most previous collider facilities.
The strongest motivation for this lies in the need for independent cross-checking of important results; the scientific community usually only becomes convinced of exciting new discoveries when two different experiments with different systematics arrive at the same conclusion. 
Studies performed to date already suggest that there is an opportunity to optimize the overall physics output of the EIC in terms of precision and kinematic range through careful complementary choices of basic features of the two general purpose detectors such as  
bore radius and B-field of the Solenoid, as well as sub-detector technologies,
leading to different acceptances,
technology redundancy and cross calibration. 

The strong diversity of EIC science imposes the essential feature that the interaction region and the detector at the EIC are designed so all particles are identified and measured at as close to 100\% acceptance as possible and with the necessary resolutions. Slight variations of the interaction region design between the two interaction points, for example to exploit the idea of a secondary focus, can allow further optimization and enhancement of EIC science reach.
More detailed and precise statements will become possible when a simulation of the full detectors become available, such that the expected performance of different combined sub-detector configurations, or ways to integrate the detector in the interaction region with possible complementary beam line optics, to maximize EIC science reach can be studied in detail and quantitatively.

Perhaps in contrast to previous colliders, the aim is thus to build complementarity into the design of the two general purpose detectors for the EIC from the outset, even before the collaboration formation stage begins.

\chapter{Integrated EIC Detector Concepts}
\label{part3-chap-IntEICDetCon}




The baseline EIC configuration currently includes one fully instrumented Interaction Region (IR) and one general purpose physics detector. 
It is assumed that the detector will be located in IP-6 (the STAR Hall), and that the available infrastructure will either be re-used or will serve as a reference for the future EIC installation. Complementary information about IP-8 (the PHENIX Hall), that can house the second detector, is given where appropriate.

\section{Hall infrastructure}
\begin{figure*}[ht]
\includegraphics[width=\textwidth]{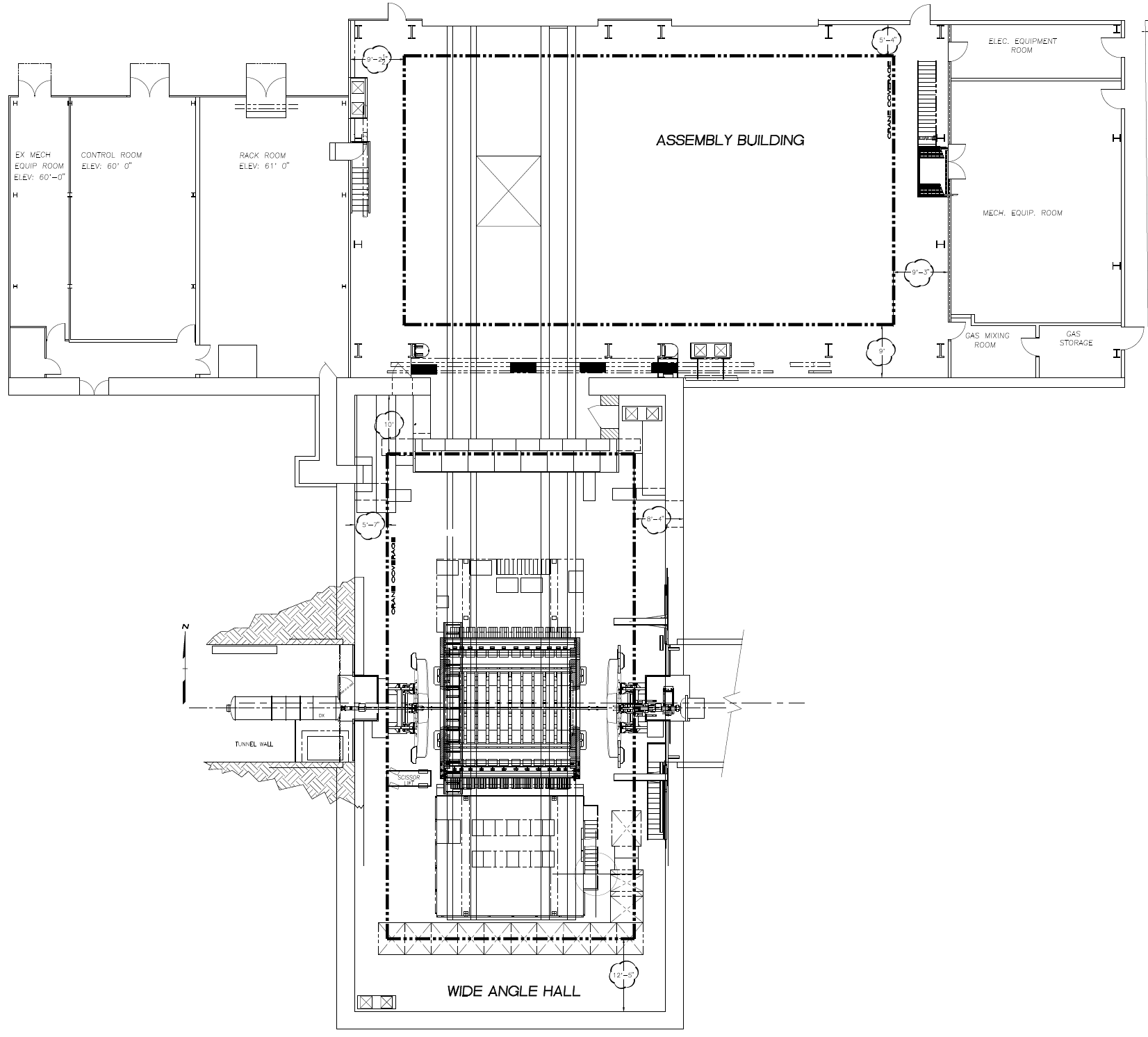}
\caption{ RHIC IP-6 experimental area layout. STAR detector shown schematically in the beam position in the Wide Angle Hall.} 
\label{Integration.fig:IP6}
\end{figure*}

In addition to the beam line area (the Wide Angle Hall), RHIC IP-6 has an Assembly Building with adequate floor space for detector maintenance work, as well as a control room, counting house, office space, electronic/mechanical workshops, gas shed, online computing room and other service areas, as shown in Fig.~\ref{Integration.fig:IP6}.

The general specifications for the IP-6 and IP-8 experimental halls are provided in Tab.~\ref{Integration.tab:IP6}.

\begin{table}[ht]
    \centering
        \caption{IP-6 and IP-8 experimental hall dimensions and related data. Hall width goes parallel to the beam line in this table, see also the Wide Angle Hall boundaries in Fig.~\ref{Integration.fig:IP6}.}
    \label{Integration.tab:IP6}
    \begin{tabular}{lcc}
  	 \toprule
         				& IP-6 & IP-8  \\
	\midrule
         Hall length and width & 3200 cm x 1615 cm & 1737 cm x 1859 cm  \\
         Distance from floor to beam line & 432 cm & 523 cm \\
         Door dimensions (W x H) & 823 cm x 823 cm & 927 cm x 1017 cm \\
         Floor load capacity & 5000 psi & 4000 psi \\
         Crane capacity & 20 ton & 12 ton\\
         \bottomrule
    \end{tabular}
\end{table}

The coordinate system of the EIC experiment is oriented as follows. The z-axis is along the beamline toward the outgoing hadron direction, the y-axis points upward, and the x-axis points toward the EIC-accelerator center.

Detector subsystem infrastructure requirements include various types of cooling, power (clean, utility, generator-backed), cryogenics, cabling, service lines, and gas system specifications for gaseous detectors. These requirements cannot be specifically identified at this stage of planning, but will be developed by the Detector Working Groups for a subset of the EIC detector technologies in the future. 
The following is a list of items which needs to be considered during integration:

\begin{itemize}
    \item {\it Electronics racks and data cables}
    
    The bulk of the electronics cables and service lines for the sub-detectors will be routed through gaps which exist between the barrel and endcap regions. As a result, the installation design for these cables must accommodate the removal/repositioning of the endcaps.
    
    \item {\it Power distribution and grounding }
    
    Electrical requirements for all sub-systems must be defined as part of their functional specification. For power, this includes the voltage and amperage required for each system, as well as any necessary power transformations. Power quality will also need to be identified for each system, e.g. clean power, utility power, generator- backed power, etc.
    
    \item {\it Cooling and gas installation } 
    
    Heat rejection for each system will be identified and quantified. As a best practice, every watt of power that is consumed within or introduced into the experimental hall, must be offset by an equivalent amount of cooling. A cooling assessment will identify which components are cooled by environmental HVAC (Heating, Ventilation, and Air Conditioning), which are cooled by LCW (Low-Conductivity Water), and which are cooled by auxiliary cooling systems. The cumulative load this places on external  heat removal systems, such as chilled water plants and cooling towers, will also be assessed.
    
    The type and volume of gases will be evaluated to determine the best locations for gas storage, the potential risks involved with the various gases, and how gases will be delivered to the sub-systems.

    \item {\it Cryogenic capacity}
    
    The cryogenic demand for each sub-system will be calculated as a collaborative effort between the design engineers and the cryogenic support group. System design will seek to develop a delivery pathway that minimizes losses, and reduces the number of connections and disconnections that are required during normal operations and  maintenance.

    \item {\it Shielding against penetrating particles from the machine  }
    
    The size and configuration of radiation shielding will be calculated as a collaborative effort between the physicists, design engineers, and the radiation control group.
\end{itemize}

\section{Safety and Environmental Protection}

It is assumed that the experimental hall’s safety systems  (sprinklers, oxygen deficiency hazard monitors, smoke alarm) are provided as part of the RHIC infrastructure.  The design and operation of the EIC sub-detector components will follow BNL safety regulations governing radiation controls, interlock  systems,
and hazardous materials or systems such as flammable gases, lasers, cryogenics, etc.

Additionally, the installation area will be equipped with a fast beam dump system which is integrated with the accelerator
controls. This system will prevent radiation damage to the detector in the event that unstable beam conditions occur.

Finally, during the collider commissioning phase, an instrumented beam pipe will be installed in place of the actual detector. This device will be equipped with a robust set of apparatus that
allows the machine performance to be assessed without exposing the actual EIC detector to potentially damaging beam conditions.

\section{Installation} 

At this time, the composition of the EIC central detector is not defined to a level of detail that is sufficient to provide a step-by-step installation procedure. However, the overall detector layout, as well as several boundary conditions, are sufficiently understood to make the following assertions:

\begin{itemize}
    \item In order to maximize luminosity, the beam line final focusing quads must be positioned as close to the IP as possible. The current Interaction Region design provides approximately 9 meters of space for the main physics detector, with accelerator beam line elements installed in the adjacent areas. It is assumed that these elements (quads on the incoming hadron side and B0 magnet on the incoming electron side) WILL NOT be moved for installation or maintenance of the central detector.
    
    \item Having 4$\pi$ coverage in tracking, calorimetry and PID, the general purpose EIC detector is likely to consume 100$\%$ of the available space.
    
    \item The door connecting the assembly area and the installation area is 823 cm wide. Accordingly, the fully assembled, $\sim$9 meter long detector cannot be moved intact between the two areas without making structural modifications. To accommodate this, it is assumed that one or more of the calorimetry endcaps will be placed on independent carriages that allow them to be separated from the main detector before moving.
    
    \item The space in the installation area is not sufficient to perform any significant assembly or maintenance on the central detector (see Fig.~\ref{Integration.fig:IP6}). Consequently, the bulk of assembly and maintenance must be performed in the assembly hall.
\end{itemize}

Although the following considerations do not represent hard constraints, they will impact system design and operation:

\begin{itemize}
    \item The solenoid cryostat chimney must be designed such that it does not need to be disconnected whenever the detector is relocated from the installation area to the assembly area, and vice versa. The current expectation is that the cryo-can will be mounted to the interior wall of the assembly area and will be connected to solenoid using a flexible cryogenic line.  This line will be sufficiently long to remain connected to the solenoid when the central detector is in either room.

    \item In order to minimize the amount of silicon detector cabling in the electron endcap acceptance, the pre-assembled silicon tracker modules must be inserted into their nominal position from the hadron endcap side, with all cabling attached and routed through the ”service gap” between the barrel and the hadron endcaps. This operation cannot be performed with the high-momentum gaseous RICH detector already installed in the hadron endcap, since it will block access to the central area. Additionally, due to space constraints along the beamline, the $\sim$1.5 meter long RICH modules cannot easily be installed into the central detector in the beam position. A possible solution is to pre-assemble the entire central part of the main detector (the barrel, the silicon forward / vertex / backward tracker, and all of the endcap acceptance equipment except for the calorimetry), together with the central piece of the beam pipe, in the assembly area.
\end{itemize}

This set of constraints and supporting considerations provides the foundation for the detector “building block” composition and the installation sequence, described below. A general purpose EIC detector, schematically shown in Fig.~\ref{central-detector-cartoon}, will be used as a reference. Fig.~\ref{central-detector-CAD} shows a perspective view of the EIC detector.

 As illustrated in these figures, the detector can be naturally subdivided into three parts: the central barrel, which is built around the solenoid magnet yoke, and the two endcaps.

\begin{figure}[!ht]
\includegraphics[width=\textwidth]{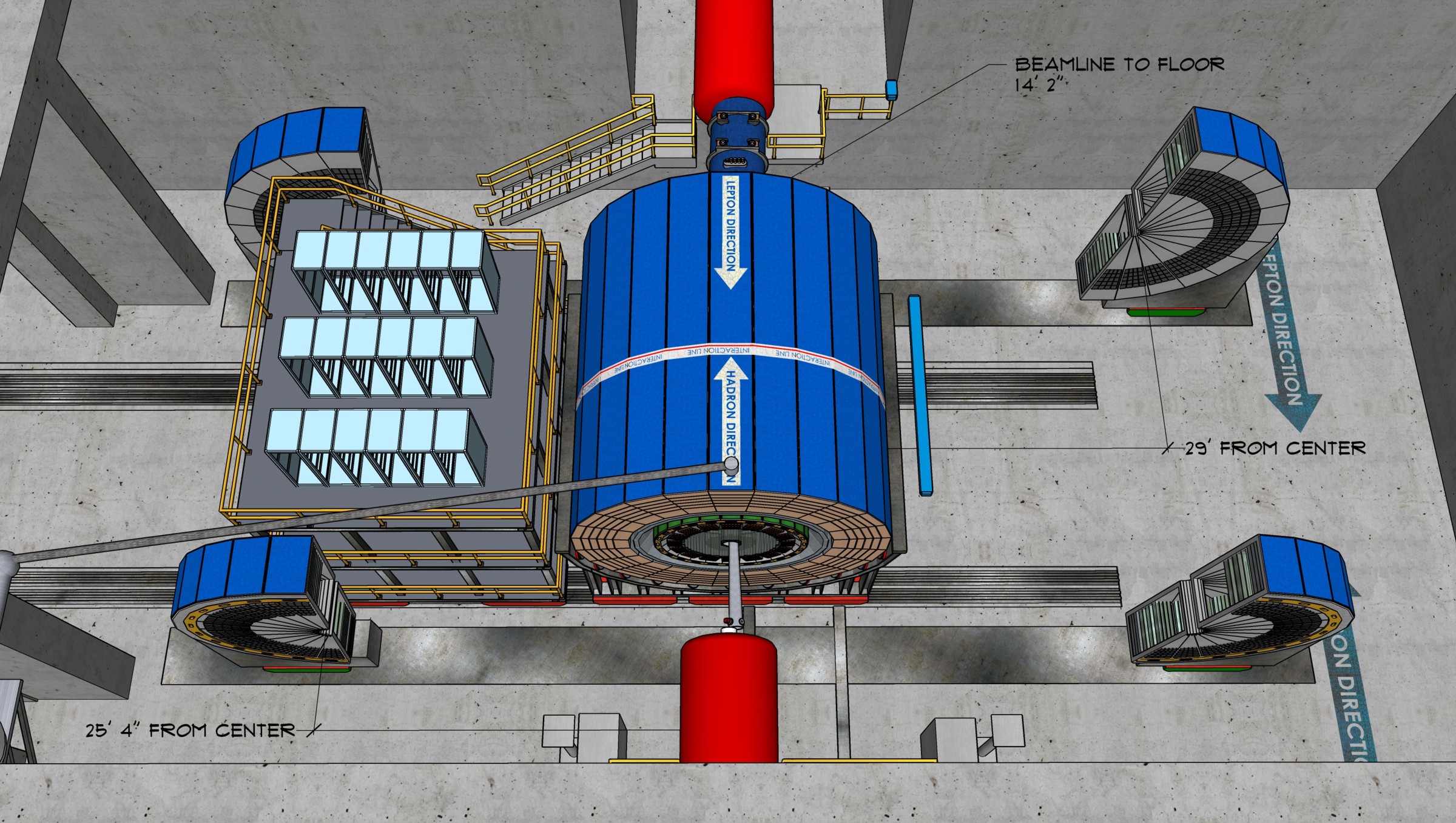}
\caption{ Barrel part of the main detector shown in the beam position. The endcaps shown rolled out to provide the space to access the inner parts of the barrel detectors.} 
\label{Integration.Ass3}
\end{figure}

 
 The endcap hadronic calorimeters are expected to be of an Fe/Sc sandwich type, with the magnetic structural steel used as an absorber. By design, they will be self-supporting, serve as a solenoid flux return, and will be able to provide mechanical support to other subsystems. To optimize construction, as well as the access strategy, it may be beneficial to locate not only the hadronic, but also the electromagnetic calorimetry in the endcap assemblies, as shown in Fig.~\ref{Integration.Ass3} for the hadronic calorimetry. This will certainly be true for the hadron endcap should a spaghetti W/SciFi e/m calorimeter technology be used in a configuration with the photo-sensor electronics installed on the upstream end of the towers. In that case, the hadronic and e/m calorimeter assemblies will likely be physically connected to one another face-to-face, with the barrel hadronic calorimeter to hadron endcap split then also needing to be aligned with the front of the e/m calorimeter.  Once the endcap halves are rolled out, one will have access to both the e/m calorimetry front end electronics, and to the electronics and services of a substantial fraction of the central (barrel) part of the detector. 

In order to meet space constraints, it is expected that the endcap assemblies can be moved laterally relative to the central part of the main detector. This should be achievable with a  few cm of clearance, making it unnecessary to move them a substantial distance outward along the beam line. To accommodate this, it will be required that no part of the central detector is installed in the recess of either of the endcaps, and vice versa.


 In this approach the support frame and the carriage system consist of five independent parts (one for the central piece and two more for each of the endcaps (see Fig.~\ref{Integration.Ass3}), each on their own sets of the heavy duty Hilman rollers. Given the size of the endcap calorimeters, and the expected density of the absorber material, the total weight of each of the four of the endcap halves is estimated to be between 80-100 tons. The weight of the central part of the detector is estimated to be on the order of 500 tons, similar to the fully assembled sPHENIX detector without endcaps.
 
\begin{figure}[ht]
\includegraphics[width=\textwidth]{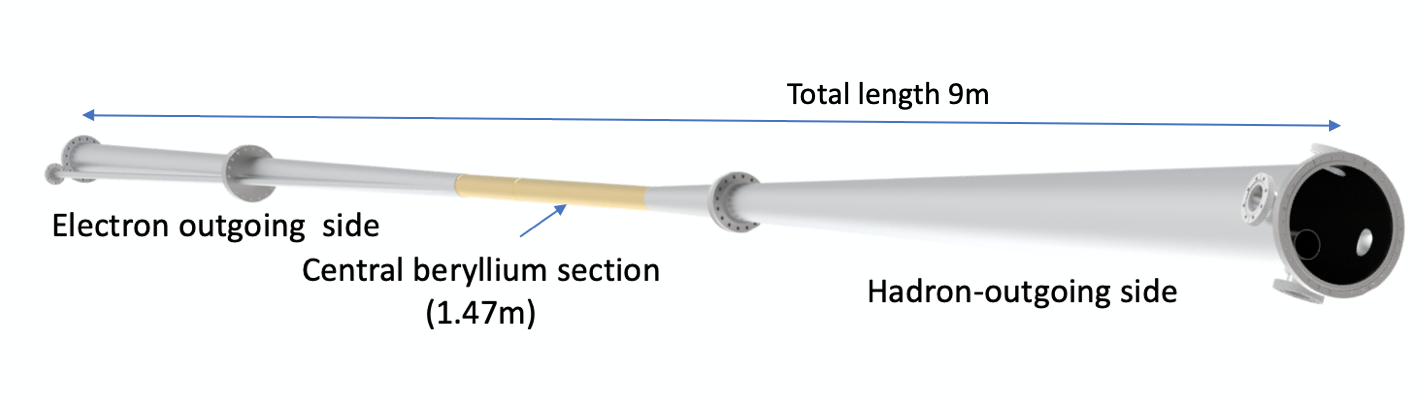}
\caption{Interaction Region vacuum chamber layout.} 
\label{Integration.fig:beampipe}
\end{figure}

 The beam pipe configuration (as shown in Fig.~\ref{Integration.fig:beampipe}), is expected to roughly follow the 1.5 m + 6.0 m + 1.5 m breakdown scheme, matching the main physics detector and consists of a $\sim$6.0 m long central part and two $\sim$1.5 m long endcaps. The central piece may be composed of more than one part. However, the installation procedure described here may be impacted if bulky, permanent flanges are used to interconnect the parts.
 
As shown in Fig.~\ref{central-detector-cartoon}, it is assumed that a clear $\sim$40 cm diameter ”bore” is allocated for the forward / vertex / backward silicon detector assembly installation, and it is not obstructed by any other endcap equipment.

 The pre-assembly sequence of the endcaps is straightforward, and does not require a detailed description at this stage.
    
Starting from the outer parts (the hadronic calorimeter, integrated into the solenoid flux return) the central part of the detector will be assembled on its own support structure. The inner barrel components (the solenoid cryostat, e/m calorimeter modules, PID detectors and the central volume tracker) will be added to the assembly one by one, in sequence, as is typically done for this type of detectors (e.g. BaBar and sPHENIX). Next the central piece of the beam pipe, as well as the two pre-assembled halves of the vertex silicon tracker are installed, with the latter ones connected  to provide minimal clearance to the beam pipe. The endcap tracker and PID detector modules will be installed afterwards, starting from the inner modules.

The installation sequence of the B0 magnet equipment and the pre-assembled main detector blocks can look like this:

\begin{itemize}
    \item The silicon tracker and the e/m calorimeter of the B0 magnet spectrometer are installed in its warm bore.
    \item The approximately 6 m long central part of the main detector, built around the solenoid magnet yoke, is rolled into the beam position, together with the electronics trailer and the pre-installed central piece of the beam pipe. The endcaps are pre-assembled in the experimental hall.  When necessary, they can be split in half and moved away from the beamline, allowing access to the beam pipe for installation or removal.
    \item $\sim$1.5\,m long pieces of the beam pipe are installed, together with the respective pump stands. This operation closes the accelerator Ultra-high vacuum (UHV) volume.
    \item The endcap halves are rolled towards the beam line and bolted together, as well as connected to the solenoid flux return yoke.
\end{itemize}
These actions are performed in the reverse sequence to move the detector from the experimental hall to the assembly area for maintenance.

\begin{figure}[ht]
\includegraphics[width=\textwidth]{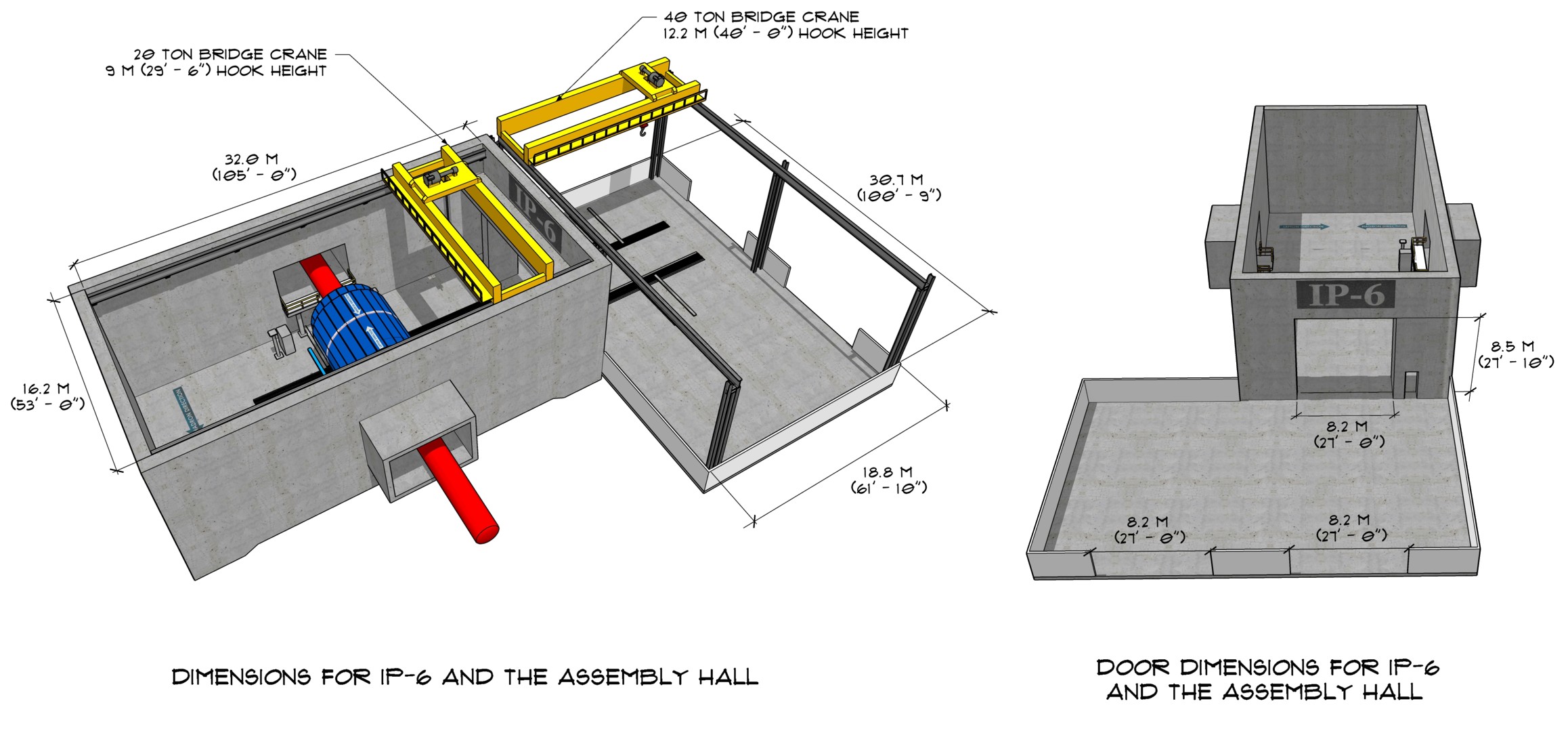}\\
\includegraphics[width=\textwidth]{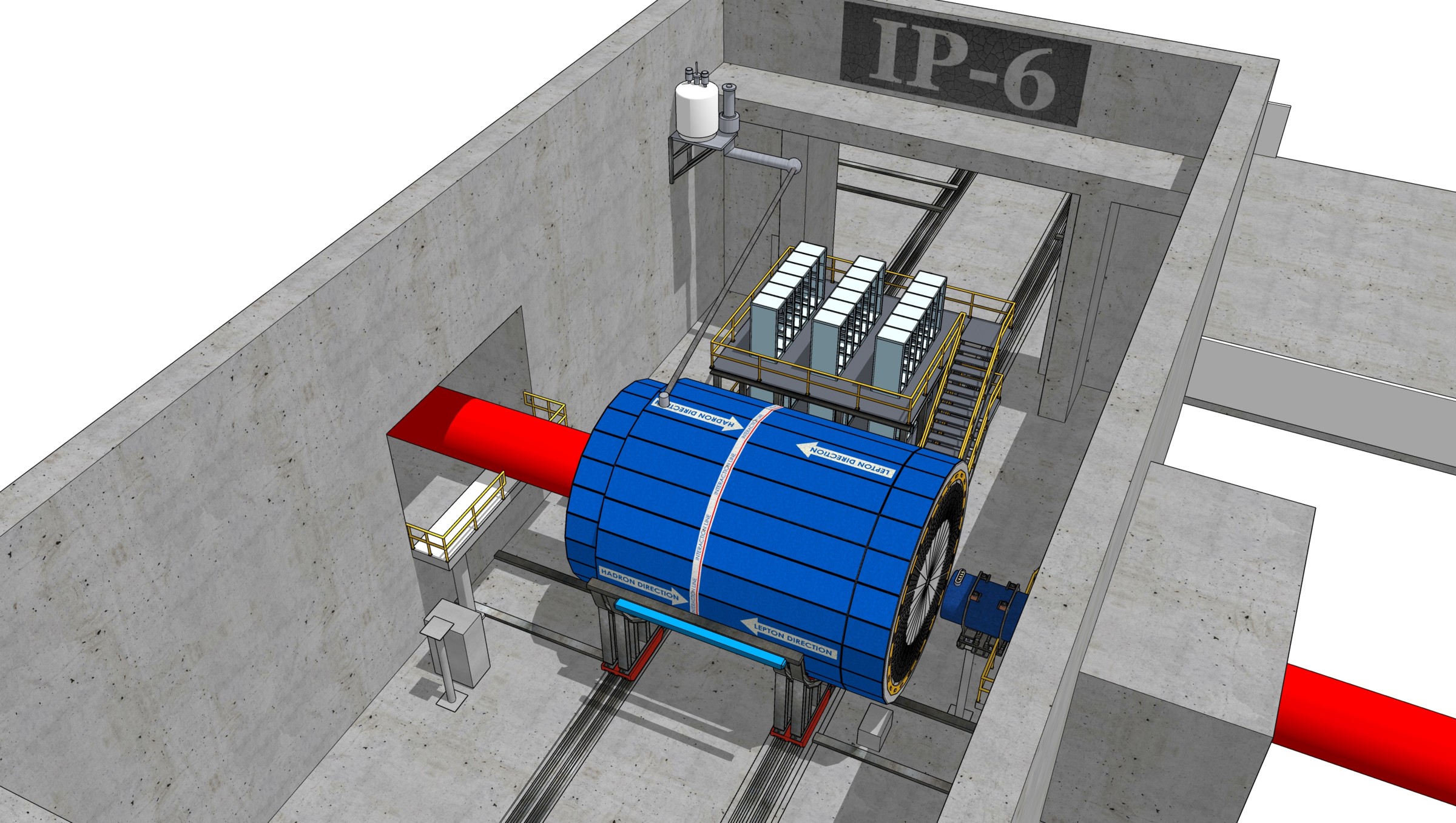}
\caption{ A model of the current detector system design
in the experimental hall with magnets, cryocan, and rear
carriage for the electronics.} 
\label{Integration.fig2}
\end{figure}

Fig.~\ref{Integration.fig2} shows the final installation of the EIC detector in the IP-6 hall.

\section{Detector Alignment}

The internal alignment of the high-precision silicon tracker modules will be done on the bench, prior to installation in the experimental apparatus. It is assumed that the relative alignment of the detector components with respect to one other, to the solenoid magnet and to the beam line elements should be performed to accuracy on the order of  $\sim$100~\si{\micro}m. This level of accuracy can be achieved using modern laser tracker survey apparatus, and by providing a redundant set of alignment marks on the detector frames, which are surveyed together with the network of the permanently mounted 3D points (survey mark nests) in the experimental hall. Maintaining visibility of the detector survey marks within the dense EIC installation environment will be a concern though, particularly for the inner tracker modules. Still, it should be noted that the ultimate alignment on the micron level of accuracy will be performed by software using the real particle tracks.

\section{Access and Maintenance} 

Three different access and / or maintenance scenarios are expected.
A short-term (controlled) access to the detector installation area where there will be no (dis)assembly of the equipment. This scenario would allow access to the electronics trailer, as well as the outer part of the sub-detector components, like front-end electronics (FEE) of the hadronic calorimeters.

\begin{figure}[ht]
\includegraphics[width=\textwidth]{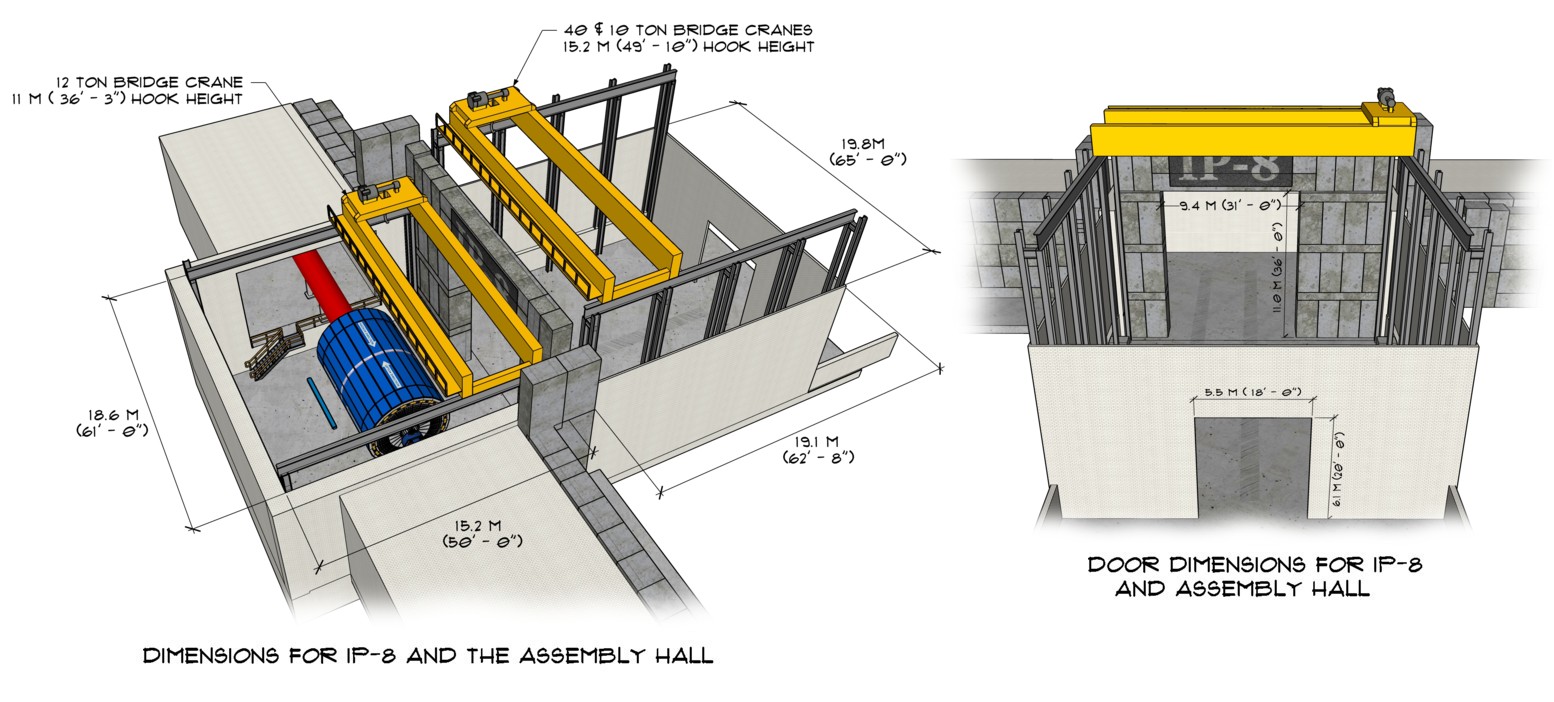}\\
\includegraphics[width=\textwidth]{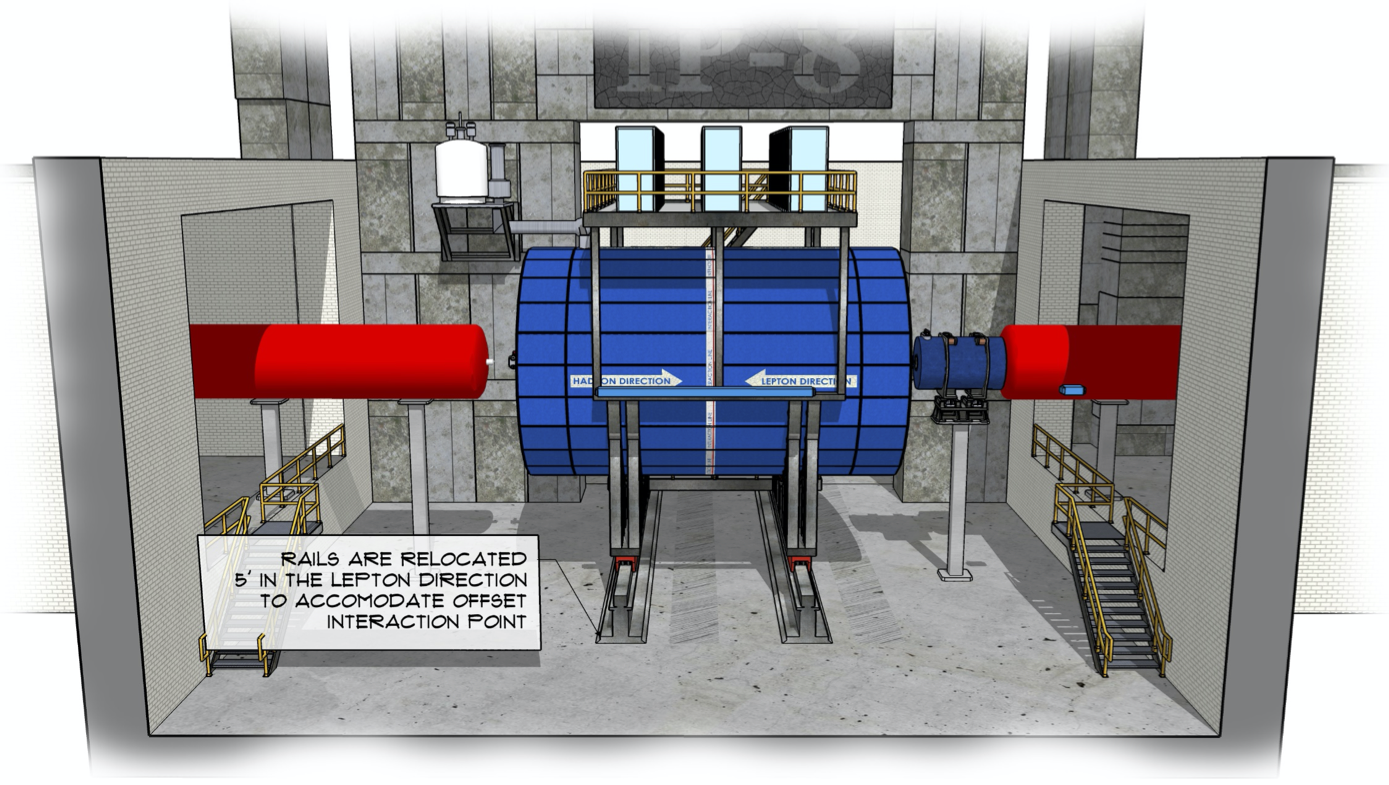}
\caption{ IP-8 (PHENIX Hall) installation.} 
\label{Integration.IP8}
\end{figure}

A short shutdown (typically an emergency event) would allow the detector endcaps to be rolled out as indicated in Fig.~\ref{Integration.Ass3}, providing an access to the endcap e/m calorimeters, outer part of the endcap trackers, beam pipe, as well as to a portion of the barrel part of the detector and the B0 silicon tracker for short maintenance. This procedure will be easier in IP-6 (STAR Hall) than in IP-8 (PHENIX Hall) due to the tighter space constraints in IP-8, see Tab.~\ref{Integration.tab:IP6}, leaving less space to walk or move equipment (Fig.~\ref{Integration.IP8}). 

\begin{figure}[ht]
\includegraphics[width=\textwidth]{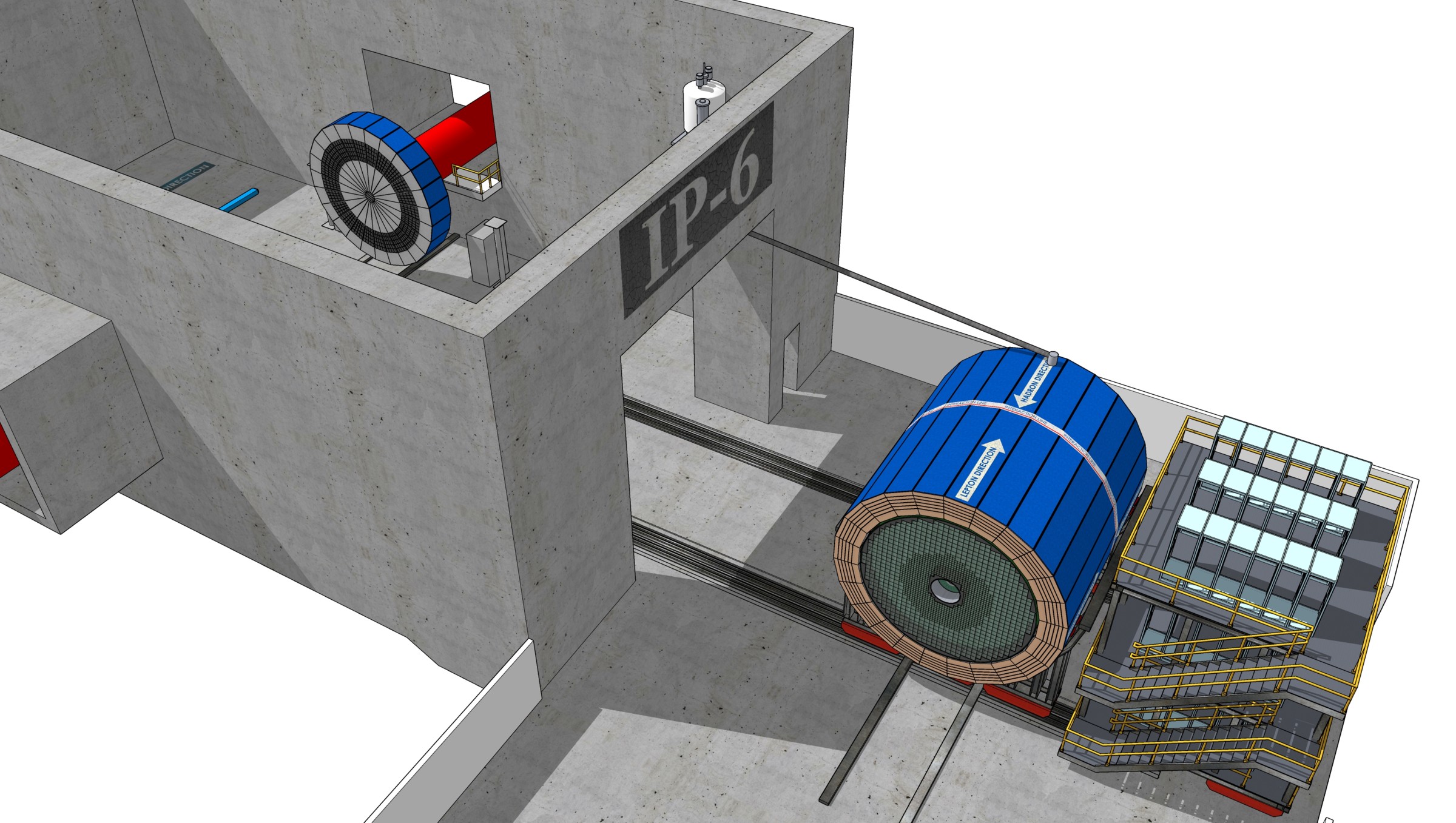}
\caption{The barrel part of the detector and the rear carriage rolled into the maintenance area.}
\label{Integration.Ass2}
\end{figure}

During a long shutdown, the barrel part of the EIC detector could be moved out of the hall completely and sub-components could be disassembled safely. Fig.~\ref{Integration.Ass2}  shows how the barrel part of the detector together with the rear carriage could be rolled into the maintenance area outside of the hall. It is important to keep the readout electronics at the rear carriage next to the detector, to provide an easy way to test sub-components during the shutdown. Such a shutdown involves disassembly of the IP beam pipe section, as well as the beam pipes of the Rapid Cycling Synchrotron (RCS), dismantling of the shielding wall between the installation and assembly halls, and would require several weeks of downtime.

%
%
%
%
%


\chapter{Detector R\&D Goals and Accomplishments}
\label{part3-chap-DetTechnology}

In this report, the EIC community motivates the need for two general-purpose detectors. With this in mind, different specific detector concepts with complementary designs have been developed and studied as described in previous chapters. While significant progress has been reached in developing these concepts, 
work is still needed to ensure that the respective detector technologies reach a viable
state of maturity for construction readiness and EIC science.

The need for R\&D was realized early by the community and laboratories and in January 2011 Brookhaven National Laboratory, in association with Jefferson Lab and the DOE Office of Nuclear Physics, created a generic detector R\&D program to address the scientific requirements for measurements at an EIC. The primary goals of this program were to develop detector concepts and technologies that have particular importance to experiments in an EIC environment and to help ensure that the techniques and resources for implementing these technologies are well established within the EIC user community. It was also meant to stimulate the formation of user groups and collaborations that will be essential for the ultimate design effort and construction of the EIC experiments.

This program is, at the time of writing of this report, supported through R\&D funds provided to BNL by the DOE Office of Nuclear Physics and is open nationally and internationally to the whole EIC community. Funded proposals are selected on the basis of peer review by a standing EIC Detector Advisory Committee consisting of internationally recognized experts in detector technology and collider physics. This committee meets approximately twice per year, to hear and evaluate new proposals, and to monitor progress of ongoing projects\footnote{The web site of the generic R\&D program with a description of the projects and all related documents and presentations is \url{https://wiki.bnl.gov/conferences/index.php/EIC_R\%25D.}}. The program is administered by the BNL Physics Department.

Many of the supported projects, ongoing or completed, developed technologies that are now integral parts of existing detector concepts or are regarded as potential alternatives. The vertex detector R\&D consortium, eRD25, aims to develop new improved MAPS sensors to meet the requirements demanded by the EIC requirements. Various MPGD technologies, such as GEM, Micromega, and μRWELL, have been pursued by the tracking consortium, eRD6, for low material tracking in barrel and forward regions as well as TPC readouts. New concepts like miniTPCs and integrated Cherenkov-TPCs had been developed and tested. Many options for electromagnetic, and recently, hadronic calorimetry have received R\&D effort. From this grew the W-SciFi calorimeter, scintillating fibers embedded in a W-powder composite absorber. In parallel, novel scintillating glasses have been developed with unprecedented quality as an alternative to expensive PbWO$_4$ crystals. The particle ID consortium, eRD14, is pursuing various technologies, such as DIRC, dual RICH with gas and aerogel radiators, and new coating materials like nano-diamonds to replace CsI for RICH photo sensors are under investigation in eRD6. Time-of-Flight detectors, as well as Roman Pots for forward proton detection, require highly segmented AC-LGAD sensors whose development has just started to get supported by the program.

Besides hardware R\&D the program supports various vital projects such as machine background studies and simulation software developments to enable more accurate definition of the physics’ requirements. Sartre and Beagle are two examples of Monte-Carlo event generators whose development was substantially boosted by the program. Both were intensively used in the context of this report.  

The generic R\&D program was and is a vital part of the overall EIC efforts with over 280 participants from 75 institutions. Despite moderate funding, many groups are making excellent progress on many vital technologies needed for an EIC detector. 
The generic R\&D program was not the only source of support for R\&D relevant for an EIC detector. Several National Laboratories, among them BNL, JLab, ANL, and LANL, supported EIC detector R\&D through Laboratory Directed Research \& Development Programs (LDRDs) and many university groups in and outside of the US, active in the many R\&D projects received support from their respective department and/or funding agencies. The EIC also benefited substantially from R\&D conducted for many HEP and NP experiments such as ALICE and LHCb at CERN, Panda at GSI and Belle-II at KEK. 

In the coming years the generic R\&D program will be replaced by a targeted program funded out of the EIC project and guided by a Detector Advisory Committee (DAC). However, the community sees also a need for a continuation of a more generic program to support technologies that go beyond the immediate needs of a day-1 detector.

In the following we discuss the remaining R\&D needs for technologies that are candidates for being deployed in a multi-purpose EIC detector. Here we do not distinguish areas of targeted R\&D, \textit{i.e.}, R\&D needed to ensure a functional baseline EIC detector on day-1 and more generic R\&D, \textit{i.e.},  more future-looking detector concepts and technologies that have the potential to enhance the scope of EIC science in the outyears. The respective timelines
are indicated in the individual section.

\section{Silicon-Vertex Tracking}
\label{part3-sec-DetTechnology.SiliconVertexTracking}

\subsection{Monolithic Active Pixel Sensors (MAPS)}

\begin{figure}[b]
    \centering
    \includegraphics[width=0.9\linewidth]{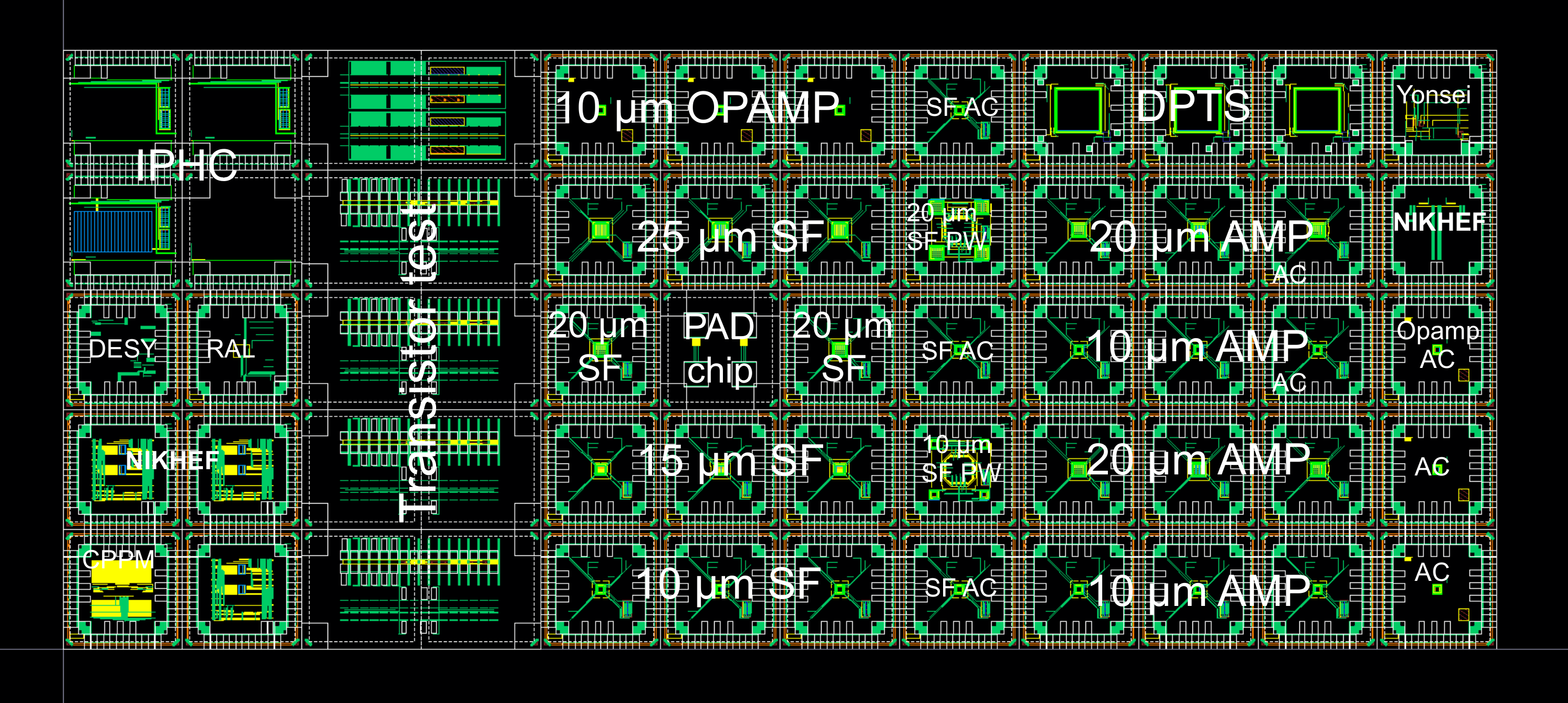}
    \caption{Picture of first MLR run of ITS3 chip submitted 11/2020. The labels indicate the different test areas. IPHC: rolling shutter larger matrices, DESY: pixel test structure, RAL: LVDS/CML receiver/driver, NIKHEF: bandgap, T-sensor, VCO, CPPM: ring-oscillators, Yonsei:
     amplifier structures.}
    \label{fig:MLR1_photo}
\end{figure}

The EIC requires precision tracking with very low $X/X_0$. The goal of MAPS R\&D is to develop sensors that meet the stringent EIC requirements for vertexing and tracking.
The combination of very high single point spatial resolution ($< 5\ \mu$m) and very low mass detector layers makes MAPS technology the most suitable candidate. The need for a new sensor with ITS3 like attributes is demonstrated in Chapter 11. More specifically, work is underway at CERN on a 65 nm MAPS detector for the ALICE Inner Tracking System 3 (ITS3) project \cite{eRD25} and it is suggested that joining this development is the most efficient route to an EIC MAPS detector. The advantage of this route is that the design parameters for the ITS3 based sensor technology closely match EIC needs,  including 10 $\mu$m$^2$ pixels (very precise spatial resolution), low power dissipation (reduced needs for cooling and power delivery leading to reduced infrastructure) and sensors thinned to 30-40 $\mu$m (low $X/X_0$). Furthermore, there are significant advantages in joining a well-funded and staffed existing design effort (high likelihood of success). The ITS3 work is already underway, so funds and support would be needed rapidly to enable full exploitation of this opportunity. An additional consideration is that further effort and funds would be needed to adapt the existing ITS3 design goals to an EIC specific sensor for the barrel and disc layers. 
The needed R\&D is to support the development of a MAPS sensor based on the ITS3 development currently underway at CERN. The work done will follow the path described in eRD-25 R\&D program and transition into an EIC silicon consortium effort. The goal of this consortium is to develop a MAPS sensor and associated powering, cooling, support structures, alignment mechanics and procedures, control and ancillary parts as necessary to produce a detector solution for silicon tracking and vertexing for the central tracking parts of an EIC detector. This will include significant design, testing, prototyping and the groundwork/R\&D to lead to a funded construction project. A more detailed description of the current path that leads to an EIC optimized sensor and associated infrastructure can be found in the eRD25 proposal\footnote{See https://wiki.bnl.gov/conferences/images/6/6d/ERD25-Report-FY21Proposal-Jun20.pdf}.

It is critical that this R\&D program be supported immediately and continuously to allow for the integration of the EIC based design and testing team into the ITS3 effort and to allow for the contribution of the EIC consortium members to the developing design. Looking at the schedule for this development to lead to an EIC optimized sensor in the time-frame needed for detector construction, delay would seriously impact the likelihood of success. We expect this development to take the full ITS3 sensor development time of 4 years plus two additional years to end up with a successful EIC sensor variant.
The product of this effort can be used at either or both interaction points. It is intended as a full silicon based inner tracking and vertexing solution. The overall designs (number of barrel layers and discs, spacing, etc.) may be different for each region, but the need for high precision inner tracking/vertexing is likely to be present at both detectors. The initial timeline is indicated in detail in the FY21 eRD25 proposal. This R\&D program is needed for a day-1 detector and would be very applicable to subsequent/parallel efforts for a detector at the second interaction point.

\subsection{Silicon-Sensor Tracking Fallback}

The goal of sensor tracking fallback R\&D is to monitor developments in 180 nm MAPS, Silicon on Insulator (SOI), and LGAD technologies that can be developed into tracking sensor solutions for EIC tracking. It is prudent to keep abreast of developing sensor technologies and to plan for a fallback solution should the 65 nm MAPS development in collaboration with the ALICE ITS3 project prove to be unsuitable for this purpose.

The combination of very high single point spatial resolution ($< 5\,\mu$m) and very low mass detector layers leads to the selection of silicon based sensors for EIC tracking. While a path to meet the EIC requirements using 65 nm MAPS technology has been identified, production of sensors for construction of an EIC tracking detector should begin in the 2026 time frame in order for a detector to be ready for use in the 2030 time frame. During this development time, contingency plans using other technologies should be developed. The most promising existing technology for a fallback path is 180 nm MAPS based on ALPIDE or Depleted MAPS (DMAPS) sensors such as MALTA. The pros are having an alternative path to success should the existing effort be unsuitable due to technology or schedule considerations. The cons are that general silicon R\&D can be expensive in both material costs and effort and having two parallel path of MAPS development might be prohibitive. While this is not the primary path, this could become the primary path to having a sensor that meets the EIC requirements available in the needed time frame. It is strongly believed that the 65 and 180 nm processes will be available through the expected EIC detector construction period.

The path involving the least amount of additional development is through the adaptation of existing 180 nm designs. At this point it is still prudent to maintain a close watch on the developing technologies of SOI and LGADs as progress is being made in both technologies. 
Doing the baseline R\&D to develop a fallback path is urgent and increases the chances of having a sensor available that meets EIC requirements in the needed time frame. This should be explored in the 180 nm technologies.

This R\&D program should be done in parallel with the timelines developed for the primary ALICE ITS3 based effort on 65 nm MAPS. This development is needed for a day-1 detector and would be very applicable to subsequent/parallel efforts for a detector at the second interaction point or future detector upgrades.
The LGAD technology offers very high temporal resolution (tens of ps) and is also a candidate for TOF and bunch crossing timing at the EIC. While MAPS technologies have proven low mass, low power dissipation and very high single point spatial resolution to match EIC vertex and tracking requirements, these features are not yet available in current state-of-the-art LGAD sensors. R\&D in this direction is however undertaken by a number of HEP and NP groups and progress should be monitored. 

A significant effort in simulation has been made to assess the suitability of additional technologies of HV MAPS and LGADs for use in the hadron endcap region and the B0 tracker. While the current specifications of prototypes in these technologies do not meet the requirements for the central barrel and inner-most discs, the technologies can be adapted for tracking in the outermost discs. Current pixel sizes are 36.4 $\mu$m$^2$ for MALTA and 100 $\mu$m$^2$ for LGADs. The detector design and performance studies are shown in Sec.~\ref{LANL-concept}. A more developed document of the path proposed for this development can be found in \cite{Wong:2020xtc}.

\subsection{Services Reduction – Multiplexing and Serial-Fiber Off Detector Output}
 
The primary goal for this R\&D program is to reduce services loads by reducing the number and volume of the way that data is taken off of the silicon tracking and vertexing detector. This effort will need to balance the reduction in service loads with the risks of losing communication with larger parts of the detector in the event of single point failures. It is possible that even with redundancy, one may be able to reduce the service loads significantly. While this is primarily geared for the silicon tracking barrel layers and discs, the product of this R\&D could be applied to other detectors in the main detector volume. Service reductions would be implemented in a day-one EIC detector, but also could be improved for future detector upgrades.

The envisioned EIC requirement is the need for the reduction of the services loads with the corresponding space and radiation length reduction. This matches the need for very low radiation length of non-active parts of the detector. Most of these services will exist in the acceptance of the tracking detectors and most of the acceptance of the surrounding detectors (PID, Calorimetry, etc.) The advantages may prove to be quite significant. The risks would be related to single point failures, but the hope that redundant paths with higher bandwidth and lower mass connections could ameliorate these yielding net positive results. 

R\&D would be needed in radiation tolerant multiplexing (probably using radiation tolerant FPGAs) and in high speed (5 GHz and above) fiber or multi-fiber optical transmission components. Both of these technologies are complementary and R\&D is urgently needed. 
In general, this R\&D provides the most benefit when it is co-developed with the detector technology (MAPS sensors, GEMS, etc.). This research could also complement and integrate additional efforts in moving some of the early stage analysis onto the detector (providing track candidates, etc.). This R\&D could lag the primary sensor R\&D by up to six months as an estimate, but should be considered as part of the system level approach to developing detector solutions. This is envisioned for a day one detector implementation.

\subsection{Services Reduction - Serial Powering and/or DC-DC Converters for Powering of Detector Components}

This R\&D aims at reducing the services loads by minimizing the number and volume of the primary service load of the silicon tracking and vertexing detector, the power and return cables.  The magnitude of this load in existing architectures has been documented in detail (see section 11.2.11). This effort envisions investigating both possibilities of serial powering, possibly with on-chip regulation and the use of on-detector radiation tolerant DC-DC converters, either or both of which could significantly reduce the required amount of power cabling. While this is primarily geared for the silicon barrel tracking layers and discs, the product of this R\&D could apply to other detectors in the main detector volume.  The product of this R\&D would be envisioned for a day-one EIC detector, but also could be improved for future detector upgrades. 
As with the Multiplexing and Seial Fiber Off Detector Output, the goal is the reduction in services and space needed for services in acceptance of the parts of the tracking detectors and most of the acceptance of the surrounding detectors (PID, Calorimetry, etc.) The advantages may prove to be quite significant. The risk could be related to single point failures in the serial powering chains which, depending on the architecture, could cause loss of powering to larger segments of the detector, and limitation in the current scaling factor for integrated DC-DC converters. The architectural aspects would be a significant part of the R\&D.

This effort envisions investigating both possibilities of serial powering, possibly with on chip regulation and the use of on-detector radiation tolerant DC-DC converters. Both of these technologies are complimentary and R\&D is urgently needed. This R\&D, while initially envisioned for the silicon tracking detector, can be applied to other detector powering systems with commensurate improvements in the powering services loads.
In common with the above multiplexing R\&D program, this R\&D provides the most benefit when it is co-developed with the detector technology (MAPS sensors, GEMS, etc.). This R\&D could lag the primary sensor R\&D by up to six months as an estimate, but should also be considered as part of the system level approach to developing detector solutions. This is envisioned for a day one detector implementation. 

\section{Tracking}
\label{part3-sec-DetTechnology.Tracking}

\subsection{Low-Mass Forward/Backward GEM Detectors}
\label{RD:lowMassGem}
Gas Electron Multipliers (GEMs) are a well-established MPGD detector technology that will soon be operational on a large scale in current NP and HEP experiments, \textit{e.g.},  SBS tracker, ALICE TPC upgrade, and CMS muon upgrade. In a day-one EIC detector, they can provide cost-efficient fast tracking with good spatial resolution in the forward and backward regions because they can cover a large area. For the same reason, GEMs could also be employed as muon detectors at the outside of the detector.

The requirements
on the momentum resolution are summarized in
Sec.~\ref{part3-sec-DetChalReq.PhysReq}.

%
Simulations using EICroot, an early simulation framework, for 10 GeV pions 
showed that for a detector geometry with vertex tracker, TPC, six forward MAPS disks, and with three GEM detector layers each placed in front and behind a RICH vessel, the momentum resolution with a 1.5 T magnetic field
is $\sigma_p/p \leq 1.5\%$ in the GEM acceptance region $1.2 < \eta < 1.7$, which is close to meeting the backwards requirements. In the GEM acceptance region  $1.7 < \eta < 3.1$, the resolution is above 1.5\% rising to about 3\% at $\eta = 3.1$. 
Unlike the requirements, it was observed in the simulation that the resolution does not grow exactly linearly with momentum at higher momenta. For example, for a 40 GeV pion at $\eta = 2.0$, the resolution is $\sigma_p/p = 4\%$, which meets the backwards requirement. 

The available material budget is 5\% of $X_0$. In the active area of one foil-based Triple-GEM detector layer, the material accounts for 0.6\% of $X_0$. Consequently, up to eight layers could be installed in an EIC detector, e.g.\ four in front of a RICH and four behind it. 

To bring low-mass GEM tracker technology to a state where it can be implemented in an EIC detector, some R\&D is still required: Improvements in the simulations and a second beam test.

The simulations need to be repeated and refined in the new fun4all simulation framework. Actually measured spatial resolutions and realistic support materials need to be incorporated properly into the simulation, in particular the materials in the TPC endplates, MAPS support structures, and the GEM support frames. Their impacts on forward/backward tracking performance and RICH seeding need to be fully quantified. The simulations will be used to determine if the material budget used in the GEM design is adequate for controlling multiple scattering and momentum resolution or if it needs to be further adjusted. This is particularly important for the azimuthal overlap region of the detectors where detector frames in a layer overlap with the active detector area in an adjacent layer. The multiple scattering also depends on the material present in front of the GEM layers, e.g.\ TPC endcaps with mechanical support and readout. Consequently, a full simulation with central and forward/backward tracking needs to be run to inform a final design. This should take six months to a year to complete.

\begin{figure}[t]
    \centering
    \includegraphics[width=\linewidth]{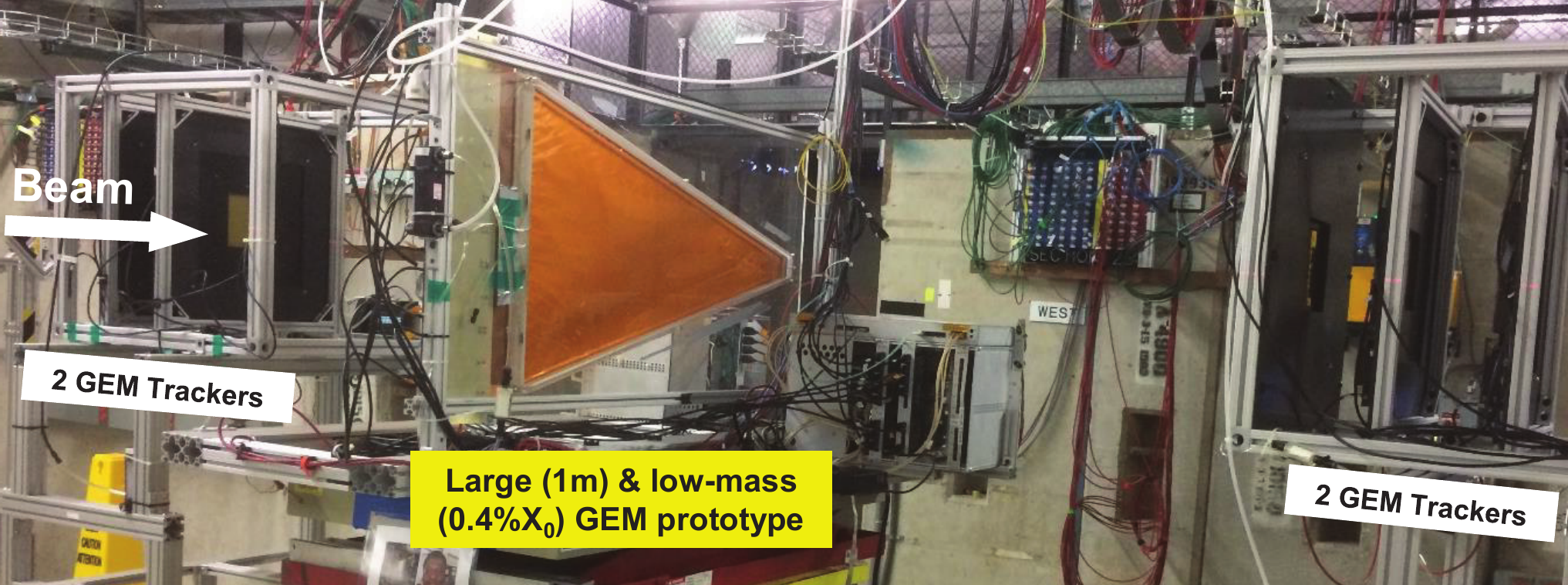}
    \caption{Large low-mass GEM detector prototype with UV strips readout during a beam test at FNAL in 2018.}
    \label{fig:my_label}
\end{figure}

Groups from the eRD6 consortium at the University of Virginia (UVa) and at the Florida Institute of Technology (FIT) constructed prototype detectors and tested them in a beam at Fermilab in 2018 (Fig.~\ref{fig:my_label}).  Detailed results from this multi-year R\&D effort and the current status can be found in the semi-annual reports submitted by the eRD6 consortium to the BNL EIC Generic Detector R\&D reviews~\cite{eRD6cons}. For the glued prototype built at UVa, different types of zebra strip connectorizations need to be tested. The mechanically-stretched FIT prototype developed at FIT with carbon fiber frames has been undergoing major refurbishments of its mechanics and its operation needs to be confirmed. If successful, both prototypes will be evaluated in a second beam test at Fermilab  planned for spring/summer 2021 to finalize the spatial resolution studies and the overall performance characterisation of the prototypes. The performance of the different GEM chamber designs in the beam test will provide additional guidance towards a final design.

\subsection{R\&D Needs for Planar \texorpdfstring{$\mu$}{}RWELL Detectors}\label{planarurwell}
%
%
$\mu$RWELL is a promising MPGD alternative to the well established GEM or Micromegas detectors for tracking in EIC end cap regions. One significant advantage of $\mu$RWELL is that it combines its electron amplification stage ($\mu$RWELL  foil) and the readout plane into a single device, making its fabrication simpler and more cost effective than GEM and Micromegas, specially for large area trackers. In addition,  $\mu$RWELL are expected to be easier to operate and more stable under harsh radiation environment. Thus, large planar $\mu$RWELL an ideal option for EIC end cap trackers, for amplification and readout layer option for TPC end cap readout or for transition radiation detectors (TRDs) required for electron identification.
%

As a relatively new MPGD technology, $\mu$RWELL have never operated in large scale NP or HEP experiments so far. Therefore, several area of R\&D studies both generic and specific to EIC environment are yet to be fully explored  to validate this technology. Below is a list of a few identified R\&D studies needed for EIC.

\textbf{\textit{Generic R\&D: Performances and stabilities of $\mu$RWELL technology}}

\begin{itemize}
\item \textbf{Rate capabilities and spatial resolution studies:} The impact of the uniformity of  $\mu$RWELL resistive layer (DLC), for large area detector on the rate capabilities and spatial resolution performances required detailed R\&D studies. Rate limitation is not expected to be an issue with $\mu$RWELL in the EIC environment because, like the other MPGD technologies such as GEM or micromegas detectors, the rate capabilities of $\mu$RWELL, even in its basic simplest configuration, can reach up to 100 kHz/cm$^2$, exceeding by a few orders of magnitude the expected rate in the EIC end cap regions. The mentioned rate capabilities and a low power consumption of the applied HV are intrinsic features of the detector technology and as such, we do not anticipate any significant benefit in developing a lower-performance version of the detector. The optimization of the rate capabilities in EIC environment will impact the choice of the front end electronics characteristics selected to readout the $\mu$RWELL detectors and not the device itself. However, these rate and spatial resolution studies, performed on small prototypes, require validation in a beam test for large-area detectors.

\item \textbf{Discharge and aging properties of $\mu$RWELL:} With the introduction of the resistive (DLC layer) as one of the key component of $\mu$RWELL, several studies have demonstrated that $\mu$RWELL is, if not spark-free, a robust spark-resistant detector. Several studies also demonstrated the technology robustness against aging in harsh particle environment. Additional R\&D is required to study the best gas mixture for a stable operation of the detector in a wide range of gain for applications at the EIC. The issue to be addressed here is to identify the optimal gas mixture to operate the detector at a stable gain of $\sim \, 5 \, \times\, 10^5$  while minimizing the discharge rate to a level of a "spark-free" device over the lifetime of the EIC operation
\end{itemize}

\textbf{\textit{EIC specific R\&D: Low-mass \& large $\mu$RWELL trackers}}

In addition to the generic R\&D on  $\mu$RWELL technology, high performance tracking with radiation length in the EIC end cap region required dedicated R\&D studies and prototyping for  $\mu$RWELL. The required R\&D, listed below, have strong synergy with the ones described in section \ref{cylurwell}.
\begin{itemize}
\item \textbf{Development of low-mass \& large area $\mu$RWELL:}  R\&D efforts are needed to minimize the material budget of the current standard  $\mu$RWELL to keep the radiation length around 0.4\% per tracking layers. This means the development of a rigid PCB free detector and  lightweight and narrow support structure based on high strength-to-weight ratio materials such as carbon fibers rather than standard G10 fiberglass frames.
\item \textbf{Development of low mass 2D readout plane:} Another important R\&D  area is the development of high resolution low mass and low channel count flexible 2D readout layers to be coupled with the $\mu$RWELL amplification layer. A few new ideas for such readout planes are already being investigated such as the development of "capacitive-sharing 2D strip" or "2D zigzag" readout structures for MPGDs, e.g.\ a 10$\times$10 cm$^2$ $\mu$RWELL with zigzag strip readout is currently being investigated at FIT.
\end{itemize}

\subsection{Large Cylindrical \texorpdfstring{$\mu$}{}RWELL Layer}
\label{cylurwell}
%


Precision tracking is needed in the central region of an EIC detector to 
to provide high angular resolution for barrel PID detectors an help with the PID particle seed reconstruction, leading to better particle separation. This tracking layers needs will be installed near the PID barrel detector and therefore to cover a large acceptance area. In addition, in the case where a hybrid tracker option of TPC + MAPS  is selected as EIC central tracker, we have identified two additional motivations for high-precision and fast-signal tracking detector to overcome the inherent limitations of the TPC + MAPS. The first is to provide a high space point resolution tracking layer to aid in the TPC field distortion corrections and TPC calibrations. The second to serve as a fast (a few ns) tracking layer for bunch crossing tagging, to complement the relatively slow TPC and MAPS detector suite. 

\begin{figure}[t]
    \centering
    \includegraphics[width=0.9\linewidth]{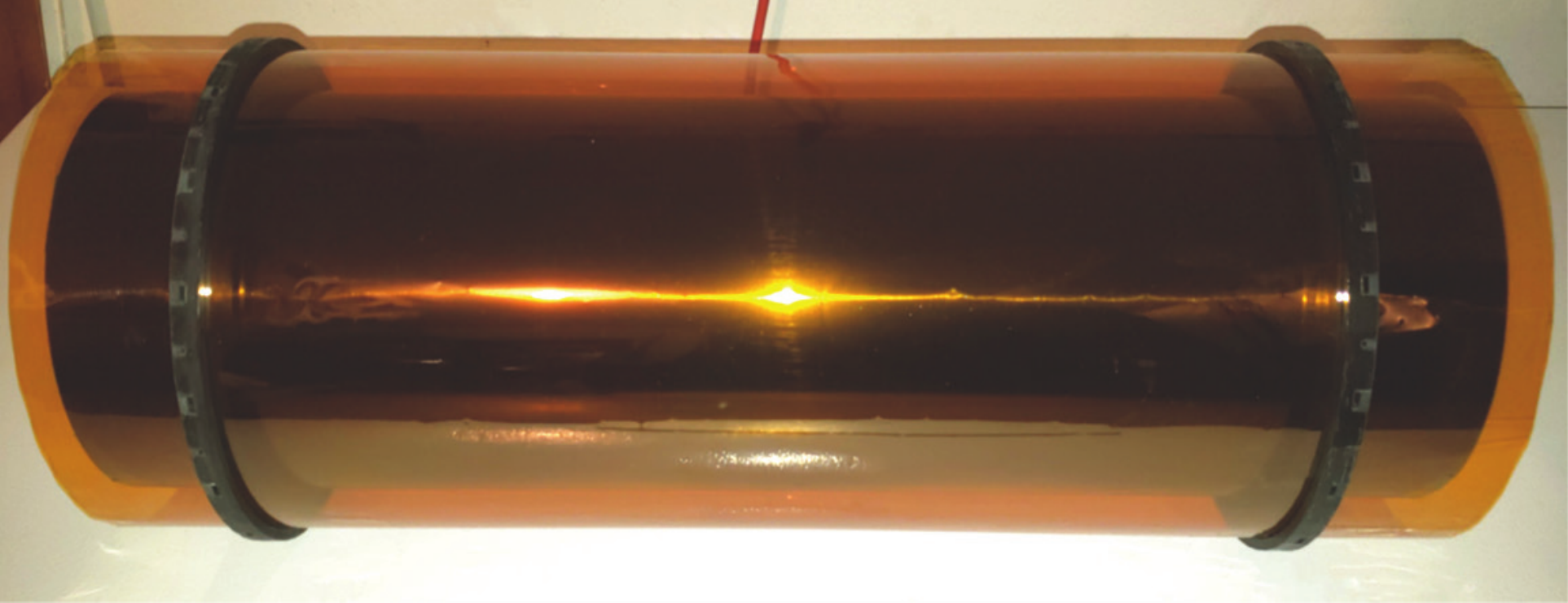}
    \caption{Small mechanical mockup for a $\mu$RWELL layer.}
    \label{fig:CylindricaluRWELLmockupFIT}
\end{figure}
%
To satisfy these specific needs in the central tracking region which combined fast timing high precision tracking,  $\mu$RWELL technology is the ideal detector candidate. The simple construction of a $\mu$RWELL detector compared to triple-GEM or micromegas technologies makes it he best choice to use in a cylindrical geometry. However there are still several R\&D items related to its construction and performance that need to be investigated. The first is related to the $\mu$RWELL technology itself. Efforts are needed to reduce the overall material budget of the current "standard" $\mu$RWELL. This involves the development of low mass amplification and readout structures. Ideally the cylindrical $\mu$RWELL would consist of one large foil and thus have no dead region in the active area. However, as with GEMs, $\mu$RWELL raw foil material is limited to a width of about 50 cm. To provide proper coverage for a barrel PID detector, several $\mu$RWELLs will be needed to form the full cylindrical layer. R\&D is needed to determine the best way to integrate the $\mu$RWELLs into one large cylindrical detector while minimizing dead regions in the active area. In principle, multiple radial layers of $\mu$RWELL are possible, but this will depend on the space available between central tracker and RICH. A single layer with one track stub can already provide the direction of tracks as they impinge on the central RICH as long as the $\mu$RWELL is operated in $\mu$TPC (mini-drift) mode.

Another area of R\&D that is needed is related to the support structure of the cylindrical $\mu$RWELL layer (Fig.~\ref{fig:CylindricaluRWELLmockupFIT}). This involves developing large, high strength and lightweight cylindrical $\mu$RWELL supports to hold the cylindrical shape of the detector. Readout electronics would be plugged into connectors that sit on these cylindrical support frames and connect to the readout strips. Similarly, power connector will also be located on the frames. These interfaces need to be designed and tested. Additionally, end cap structures to hold the cylindrical detector in place need to be designed. Current simulations, which include the $\mu$RWELL, readout structure, drift cathode, detector gas, and cylindrical support rings show a material budget below 1\% $\chi_0$ in the $|\eta|$ < 1 region.

Several performance studies such as rate capabilities, dE/dx, tracking and timing resolutions need to be carried out with a prototype detector operating in $\mu$TPC mode. These results will help to determine the proper readout electronics that are needed. Cylindrical uniformity, discharge rate, and aging properties of the detector will also need to be assessed.  

\subsection{Large Micromegas Barrel Tracker}

The central region of the EIC detector requires very low material budget detectors. Large area Micro-Pattern Gaseous Detectors (MPGD) are a possible solution to complement the silicon vertex detector. In particular, Micromegas detectors have been already successfully employed for building compact and light trackers, such as the Barrel Micromegas Tracker (BMT) of the CLAS12 experiment at the Jefferson Lab. 
Studies conducted within the Yellow Report effort showed that a barrel tracker made of MPGD tiles of a similar technology to the CLAS12 BMT one, would fulfill the requirements in terms of material budget and tracking resolutions. The CLAS12 BMT consists of six concentric layers of curved resistive Micromegas detectors where each layer is composed of three tiles of about 120 degrees width. The material budget of one tile in the active area is about 0.3\% of $X_{0}$. The BMT was designed to withstand ghost hit rates per strip up to 30kHz and its time resolution is about 40 ns.
From the experience of the CLAS12 BMT, the R\&D on the EIC tracker will have two main objectives: reducing even more the material budget and simplifying production and integration.

In the CLAS12 BMT, the thickness of the self-supporting curved detector is determine by the ability to maintain the desired curved shape when constrained at both end by the carbon structure. The  material thickness is $\sim200\ \mu$m FR4 for a radius of $\sim$400 mm. Reducing the thickness further requires R\&D, initially of flat stretched detectors using Micromegas made on a Kapton film of 50 $\mu$m. A detector will consists of two stretched foils (readout and drift) on a carbon frame with pillars to maintain and control the drift distance. Two additional external thin foils (made of 10 $\mu$m polypropylene) will hold the gas pressure instead of the thin electrodes. The R\&D should start with the choice of optimal materials, both for the active region and for the structural components, followed by a full size prototype to demonstrate the integration technique.

Curved detectors impose the use of specific sizes and tools for each curvature radius, thus making the production line more complicate. Excessive large area detector elements require numerous tooling to handle and to control the mechanical uniformity. A modular flat detector that would allow a higher production yield rate and possibly reduce the costs. The necessary R\&D will need study a thin support structure to integrate this modular design.

On most MPGDs, copper is the chosen readout material with a thickness of at least 9 $\mu$m. The use of lower mass material  for the strip readout such as metalized aluminum of about 0.4 $\mu$m requires R\&D. The aluminum strips will have to be protected by a resistive layer to prevent vaporization of the metalized layer due to sparks.

The standard thinnest mesh used for large surface Micromegas detectors is a stainless-steel woven mesh of 18 $\mu$m wires. The alternative solution is electro-formed meshes (i.e Nickel of 10 $\mu$m) which are expensive, very fragile and limited in size ($\sim 30\times30 \text{cm}^{2}$). Since 2018, in parallel with the use of laser techniques for etching ``zigzag'' patterns, a proof of concept has been made of laser etching holes on a small surface with different material (Cu, Al, Steel) of varying thicknesses (10, 15, 20 $\mu$m). R\&D is needed to study this technique on larger surfaces to obtain large thin stretched aluminum foil with millions of holes to be used in the Micromegas bulk process.

Standard connectors made of plastic and brass contacts are quite heavy in term of material budget. If the active area is segmented, the multiplication of connectors can be a problem. Further R\&D will test kapton-kapton connections with metal pixels clamped with light materials (carbon or 3D printed plastic). 

\subsection{MPGD Readout for a Time Projection Chamber}
In general, the TPC for the sPHENIX experiment can serve as a central tracking device in a Day-1 EIC detector. Present configurations for an EIC detector show that the size of the TPC is limited and will be of the dimensions of the sPHENIX-TPC. However, it has to undergo several upgrades and/or modifications in order to be optimized for the EIC program.\newline
The sPHENIX TPC has been optimized for good momentum resolution which requires a good space point resolution for the tracks to be measured. The sPHENIX program does not require PID (dE/dx) to be performed with the TPC. Hence, the optimization for the sPHENIX TPC has concentrated on very good IBF suppression which sacrifices good dE/dx resolution. For the EIC program this feature has to be restored. In the EIC era it is also expected that IBF will not have the same significant impact as during the RHIC program.\newline
With a figure of merit (FOM) according to
\begin{equation*}
    FOM = \frac{Ionization\times Multiplicity\times Rate}{K}\times DVF\times OPF
\end{equation*}
with K: ion mobility; DVF: replacing the dead space factor from 0.1 (sPHENIX) to 1.0 (EIC) to track over the full detector volume; OPV: moving the operating point of the GEM-stack from 0.3\% (sPHENIX) to 2\% IBF for recovering dE/dx resolution; one obtains the following comparison between the sPHENIX environment at RHIC and an EIC detector:
\begin{table}[hbt]
    \centering
    \begin{tabular}{|l|l|l|} 
    \hline
    &Au+Au \@ 200 GeV & EIC (baseline)\\ \hline \hline
    Gas&Ne&Ar\\ \hline
    Ionization (e$^-$/cm)&43&94\\ \hline
    Multiplicity (rel.)&1&10$^{-3}$\\ \hline 
    Rate (rel.)&1&0.69\\ \hline 
    K&6.93&1.96\\ \hline
    DVF&0.1&1\\ \hline 
    OPV&0.3&2\\ \hline\hline
    FOM$^*$&1.00&0.36\\ \hline
    \end{tabular}
    \caption{IBF comparison between the sPHENIX environment and the EIC baseline configuration. The FOM$^*$ is normalized to 1 starting off the sPHENIX case.}
    \label{table::IBF_FOM}
\end{table}
This rough estimation indicates that the IBF will not be of less concern in an EIC-TPC compared to the sPHENIX. It should be noted, though, that the calculation does not account for background events or the case of EIC configuration beyond the baseline which possibly affects the IBF.\newline 
R\&D for a "new" TPC will be considered in the readout electronics section.
\subsubsection{Hybrid and Gating}
A very promising candidate for combining very good IBF suppression and good energy resolution is the hybrid option of combining MicroMegas and GEMs into a single amplification stage. The MicroMegas acts as the main amplification stage and reduces the IBF to a minimum. The GEMs act as pre-amplifiers and provide the necessary field ratios to further suppress IBF. The combination of both technologies provide the robustness needed to operate in a high rate environment. First R\&D projects have been already established and this amplification structure needs continued detailed investigation.\newline
Gating grids that have been used in TPCs based on MWPC cannot be used in an EIC environment. The readout rate would not allow one to cope with the luminosity requirement of the physics program. The requirement is that the TPC will be read out continuously which does not allow a traditional gating grid. Consequently, one has to investigate amplification devices that minimize IBF as described in the previous sections. Thus R\&D is needed to investigate gating innovations that have minimal dead-times. One of the options is to use a passive gating grid which ``naturally'' allows electrons to pass through the structure whereas ions will be attracted to a high degree and eliminated from the gas volume. The investigation of such structures has started and is ongoing.
\subsubsection{GEM and MicroMegas}
Prospects for R\&D are in the restoration of good dE/dx resolution. This requires the investigation of GEM-properties and different gas choices that find the optimum of relatively good IBF suppression and optimum dE/dx resolution.\newline
The MicroMegas technology has the best intrinsic IBF suppression and is a good candidate for good dE/dx resolution. However, stability issues have to be investigated and is an indicator for R\&D needed soon for pursuing the MicroMegas option.  
\subsubsection{Readout Electronics}
A possible issue present for a TPC in an EIC environment is the material budget in the forward/backward region, in particular the electron direction. One can possibly overcome this problem by introducing alternative readout electronics; the readout electronics presently included is the major contribution to the material budget, including all their required infrastructure. It shall be noted that during the design-phase of the sPHENIX TPC emphasis was placed on a possible EIC implementation and therefore the lowest practicable material budget was accounted for.\newline
Possible candidates for improved readout electronics are the TimePix or similar constructed microscopic readout structured front-end electronics. The options are a (a) small sized TPC with microscopically sized readout pads, $\mathcal{O}$($10^{-3}$ mm$^2$) and (b) a regular sized TPC with small sized readout pads, $\mathcal{O}$(0.1 mm$^2$).\newline
Option (a) provides the registration of single electrons from the ionization trail of a track, acting as a form of digital camera. This would allow precise tracking and excellent dE/dx resolution. The R\&D needs on this option are manifold, in particular gas choices and readout capabilities.\newline
Option (b) would provide the registration of single clusters from the ionization trail of a track. This would allow precise tracking and excellent dE/dx resolution. The R\&D needs for this option are focused toward the adaptation of the microscopic readout structure of the front-end electronics and distribution over larger areas.\newline
A further option for decreasing the material budget in the electron-going direction of a TPC would be a single sided readout structure, i.e, having one readout plane whereas the other cap of the TPC consists of a thin cathode. This option would require feasibility studies.\newline
All the above mentioned R\&D topics should be investigated to a mature level until the final design of a possible central tracker in the form of a TPC is established.

\section{Particle Identification}
\label{part3-sec-DetTechnology.PID}

\subsection{A Modular RICH (mRICH) for Particle Identification}
\begin{figure}[bt]
    \centering
    \includegraphics[width=0.9\linewidth]{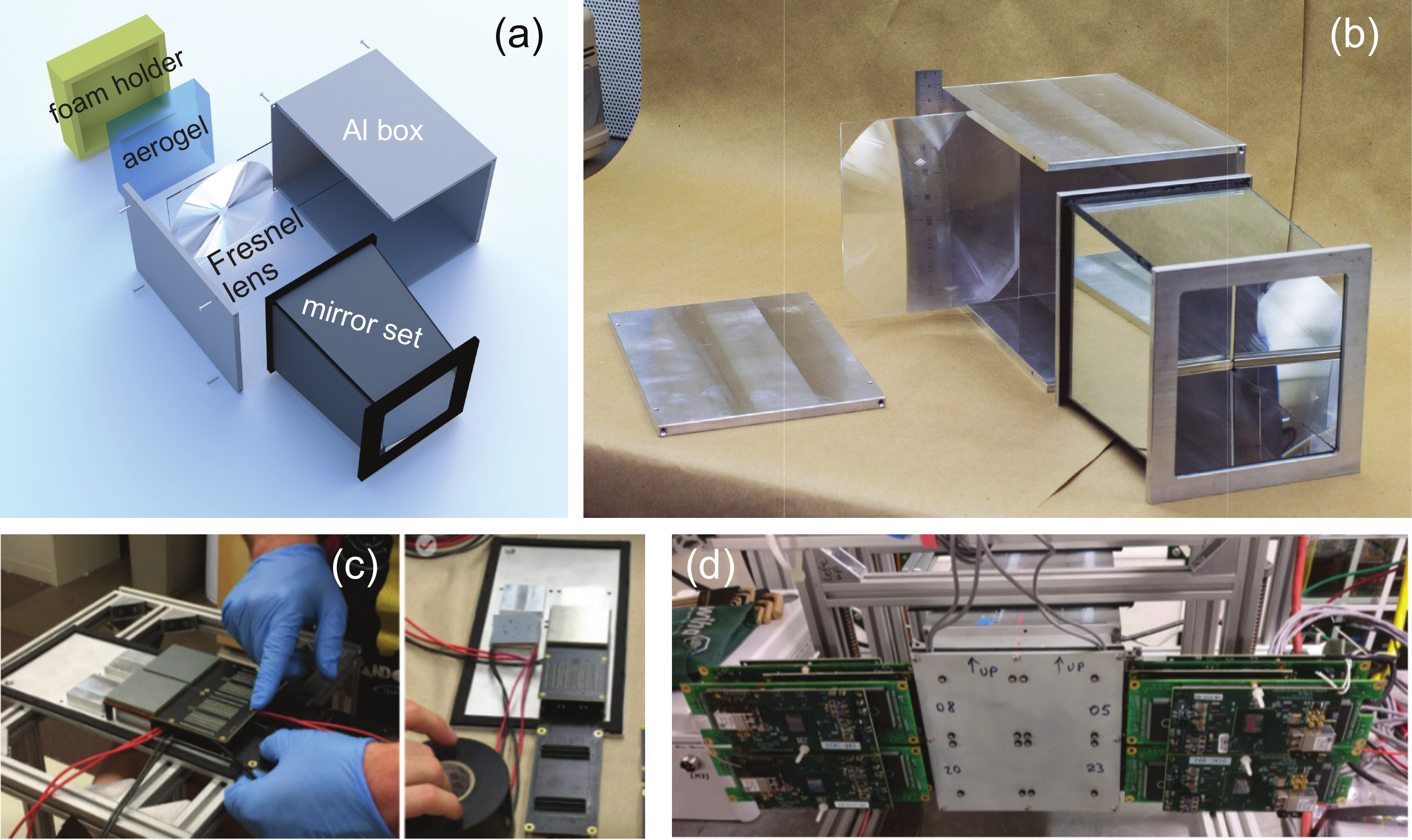}
    \caption{(a) Illustration of mRICH components and (b) prototype assembly for beam tests at JLab and Fermilab. 
    (c) Assembly of SiPM matrices (with cooling) and (d) photo of an mRICH module during beam tests at Fermilab in 2018. }
    \label{fig:mRICH_photos2}
\end{figure}

The mRICH is designed for providing PID capabilities for EIC experiments for kaon and pion separation in momentum coverage between 3 to 10 GeV/$c$ and electron and pion separation around 2 GeV/$c$. 

mRICH detector R\&D has been supported within the EIC eRD14 Consortium since 2015. The key components of an mRICH module include a radiator (Aerogel, $\sim$10 cm $\times$ 10 cm $\times$ 3 cm, n = 1.03), a Fresnel lens (with focal length range from 3” to 6”), a mirror set and a photosensor as shown in Figure~\ref{fig:mRICH_photos2}. The characteristic longitudinal dimension of an mRICH module is from 15 cm to 25 cm depending on the focal length of the lens. A realistic GEANT4-based simulation for mRICH has also been developed and verified with beam test data (see  Figure~\ref{Fig:MRICH-01-Configuration}).

Two rounds of detector prototyping and beam tests were completed with a focus on verifying the detector working principle and performance. The results from the first beam test (in 2016) have been published \cite{Wong:2017ptx}. The second beam test was done in 2018 and the data analysis is still ongoing. Two more beam tests with particle tracking capability are under preparation in order to quantify mRICH PID performance and test new photosensors. One is planned at Fermilab in spring of 2021 for testing the mRICH with a LAPPD readout. The groups involved in this test are BNL, ANL, SBU and GSU. The other test is planned at JLab Hall D in summer of 2021 using secondary electrons in the momentum range $1 < p <  6$ GeV/$c$. The participating groups for this test include DukeU, INFN, JLab, USC and GSU. 

Two key components of an mRICH module are the Aerogel block and a photosensor with single-photon detection capability and fine-segmented pixel size ($< 3$ mm $\times$ 3 mm). The photosensor also needs to be working properly in high magnetic field.

To meet the needs of EIC experiments, a proper photosensor choice is critical. The planned beam test at Fermilab in 2021 will help to evaluate the integration and performance with LAPPD. During the second mRICH beam test in 2018, three SiPM matrices were tested with varying cooling temperature range from $-30^\circ$ C to room temperature as shown in Figure~\ref{fig:mRICH_photos2}(c). The radiation damage effects to SiPM performance is currently under study at INFN. The team also pays close attention to the radiation damage measurements of SiPM sensors from GlueX, STAR and sPHENIX experiments and any new SiPM sensor technology development.

Regarding to the possible kinematic coverage in EIC experiments with mRICH modules, one can envision deployment in (\textit{i}) the electron endcap, (\textit{ii}) the hadron endcap in the range $1 < \eta < 2.5$, and (\textit{iii}) the central barrel region assuming available space in radial direction ($\sim25$ cm) is available. Implementation of mRICH arrays in each of these kinematic regions in the framework of the ePHENIX simulation using GEANT4 are available for acceptance and efficiency studies. The mRICH is considered as a day-1 detector.

Besides the two planned mRICH beam tests in year 2021, there is a longer-term R\&D effort for mRICH toward engineering design which includes: (a) high quality mirror and mirror assembly; (b) mRICH holder box engineering for reducing total weight, easy assembling and projective installation; (c) continued testing with available photosensor options. The efforts related to (a) and (b) were started in summer 2020 with an undergraduate engineering student who designed a new mRICH holder box and a mirror assembly for easy integration and light weight. Assessment of these new designs in comparison with engineering assembling of modular components in other major experiments are going to be the next tasks. At the same time, the physics shop staff at Georgia State University has started looking into machining carbon fiber plates for constructing the next generation of mRICH prototype. 

\subsection{A Dual-Radiator Ring Imaging Cherenkov Detector (dRICH)}
The dual-radiator Ring Imaging Cherenkov (dRICH) detector is designed to provide 
continuous full hadron identification ($\pi/K/p$ separation better than $3\sigma$ apart) 
from $\sim 3$ GeV/$c$ to $\sim 60$ GeV/$c$ in the ion-side end cap of the EIC detector. It also 
offers a remarkable electron and positron identification ($e/\pi$ separation) from few 
hundred MeV up to about 15 GeV/$c$. The baseline geometry covers polar angles from $\sim 5$ 
up to $\sim 25$ degree (pseudorapidity range $\sim 1.5-3$). Achieving such a momentum coverage 
in the forward ion-side region is a key requirement for the EIC physics program. 
Currently, the dRICH is, by design, the only hadron identification detector in EIC 
able to provide continuous coverage in RICH mode over the full momentum range 
required for the forward end-cap.

The dRICH baseline configuration consists of six identical open sectors. Each sector 
has two radiators (aerogel with refractive index $n \approx 1.02$ and gas with 
$n \approx 1.008$) sharing 
the same outward focusing mirror and instrumented area made of highly segmented 
photosensors ($3\times 3$ mm$^2$ pixels). The photosensor tiles are arranged on a curved surface 
in a way that minimises aberrations.  The original benchmark configuration assumed 
$\sim 160$ cm longitudinally long thickness but even a shorter, down to ~$\sim 100$ cm, dRICH 
preliminary version features a performance that fulfills the above mentioned key 
physics requirements, indicating a remarkable flexibility of possible dRICH 
configurations.

To meet the EIC specifications, critical elements are the effective interplay between 
the two radiators and a proper choice of the photosensor, that should preserve 
single-photon detection capability inside a strong magnetic field. The dRICH 
focusing system is designed to keep the detector outside the EIC spectrometer 
acceptance, in a volume with reduced requests in terms of material budget and 
radiation levels. This feature makes dRICH a natural candidate for the exploitation 
of magnetic field tolerant SiPMs with an integrated cooling system to mitigate 
their significant dark count.

The dRICH design and performance has been studied through various means: a full 
Geant4 simulation (including an event based particle reconstruction processor) \cite{Barion:2020iqw}, 
AI-based learning algorithms with Bayesian optimisation to maximise the hadron 
separation \cite{Cisbani:2019xta}, analytic parameterizations taking into account 
the optical properties  of each component and the Geant4 simulated resolutions. 

A small-scale prototype is being developed to investigate critical aspects of the 
proposed dRICH detector, in particular related to the interplay and long-term 
performance of the two radiators and simultaneous imaging. The prototype vessel 
is composed of standard vacuum parts to contain the cost and support pressures 
different from the atmospheric one. This would allow efficient gas exchange and, 
in principle, adjustment of the refractive index and consequent flexibility in the 
gas choice (in the search for alternatives to greenhouse gases). The prototype supports 
the usage of various type of photosensors, in particular SiPM matrices and MCP-PMTs.

A program has been initiated to study the potential of SiPM sensors for Cherenkov 
applications, aiming to an assessment of the use of irradiated SiPM in conjunction 
with the dRICH prototype. Promising SiPM candidates will be irradiated at various 
integrated doses (up to the reference value of $10^{11}$ $n_\mathrm{eq}$ cm$^{-2}$) and will undergo 
controlled annealing cycles at high temperature (up to 180 C). The SiPM response
before and after irradiation will be characterised and their imaging potential will 
be studied with a customised electronics. High-frequency sampling and Time-of-Threshold-based
readouts will be compared. Of particular interest, the ALCOR front-end chip designed
to work down to cryogenics temperatures, features low-power TDCs that provide 
single-photon tagging with binning down to 50 ps and potential counting rate well 
exceeding 500 kHz per channel. The irradiated sensors will be cooled down to the working 
temperature (down to -40 Celsius) to instrument an area suitable for imaging tests 
with the dRICH prototype. After an initial survey of the most promising candidates 
available on the market, a dedicated R\&D could be pursued to meet the EIC 
specifications.

dRICH is considered as a day-1 detector given the EIC physics requirement for PID.

Besides the first SiPM irradiation campaign and the baseline prototype realisation in 
the coming year, there is a longer-term R\&D effort towards engineering design which 
includes: (a) light and stiff support structure in composite materials (b) high quality 
mirror assembly; (c) cost-effective production of high-quality aerogel; (d) 
alternatives to the greenhouse gases; (e) magnetic field tolerant single-photon sensors;\cite{Barion:2020iqw}
(f) dedicated readout electronics and cooling. 

\subsection{High-Performance DIRC}
The high-performance DIRC (hpDIRC) is a proposed hadronic PID system for the barrel region of the central detector, capable of $\pi/K$ separation with $3 \sigma$  or more up to at least 6 GeV/$c$ momentum over a wide angular range. It can also contribute to $e/\pi$ identification at lower momenta and provide a supplemental time-of-flight measurement.
 
\begin{figure}[tb]
    \centering
    \includegraphics[width=0.9\linewidth]{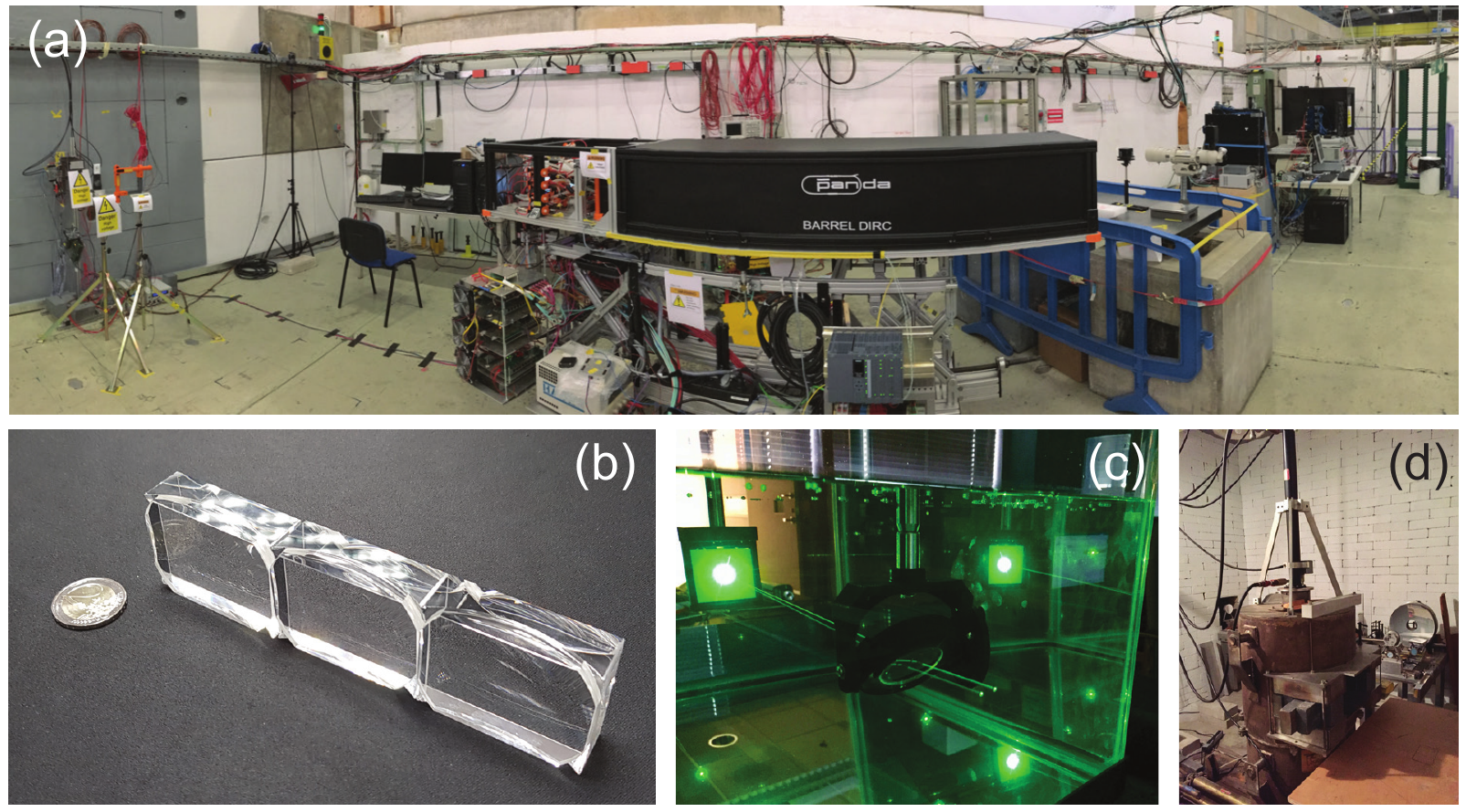}
    \caption{(a) DIRC prototype in T9 beamline at CERN PS. (b) 3-layer lens prototype. (c) Laser setup to characterize 3-layer lens. (d) Radiation hardness setup at BNL's $^{60}$Co chamber.}
    \label{fig:dircPhotos2}
\end{figure}

The hpDIRC is a compact system with a radial thickness of less than 8 cm. 
The design is flexible, the radius and length of the bars can be modified without impact on the PID performance and the shape of the expansion volume prism can be selected for optimum position of the sensors in the magnetic field. 
It has low demands on the detector infrastructure (no cryogenic cooling, no flammable gases) and is easy to operate. 
The R\&D of the hpDIRC is at an advanced stage, Figure~\ref{fig:dircPhotos2} shows examples of ongoing R\&D activities. 
The PID performance estimate is based on test beam results, with excellent agreement between simulation and prototype data. 

Although the conceptual design and performance evaluation, described in detail in section~\ref{hpDIRC}, are advanced, several aspects still require significant R\&D.
Matching the hpDIRC design to the final EIC detector layout, optimizing it for cost efficiency in simulation, and validating the performance with the full system hpDIRC prototype are the most critical items. 
In addition, the hpDIRC requires “external” R\&D by EIC groups working on developing the fast readout electronics for small-pixel MCP-PMTs and on pixelated LAPPD sensors. 
This R\&D is important for several other EIC detectors as well. 
Significant funding is needed soon to upgrade the PANDA DIRC prototype, which is being transferred from GSI to CUA/SBU, to fully equip it with new sensors and electronics, in order to validate the resolution and PID performance with cosmic muons and/or particle beams. 
A new Cosmic Ray Telescope (CRT) facility is being developed for the hpDIRC in collaboration between SBU, ODU, and CUA to study the prototype prior to possible tests in particle beams. 
This CRT will be located at SBU and available for use by other EIC systems.

The feasibility of reusing the BaBar DIRC bars vs. ordering new radiator bars, and of using LAPPDs instead of commercially available MCP-PMTs, have to be determined since they have a large impact on the projected cost. 
R\&D will be required to develop a procedure to transport and safely disassemble the fragile BaBar DIRC bar boxes, extract and separate the radiator bars, and to evaluate the optical quality to ensure that the bars can be used in the hpDIRC.  
The recently discussed potential increase of the PID momentum coverage, required by EIC physics, may require additional design improvements and utilizing possible post-DIRC tracking. 
Since the discussions about higher magnetic field options for the EIC detector are still ongoing, further investigation of a sensor solution for a possible 3T field may be required. 
If the funding for the continuation of the R\&D program is made available, we expect the hpDIRC TDR readiness to be achievable by 2024/2025. 

\subsection{Photosensor: MCP-PMT and LAPPD} 

The choice of photosensors is essential for reaching the cost and performance goals of all EIC PID subsystems. The best possible photosensor solution for each detector component is driven by the detector's specific operational parameters, naturally with cost optimization in mind. Ultimately, it would be preferable to use a common photosensor thus reducing development and procurement costs.

Microchannel-plate photomultipliers (MCP-PMTs) from commercial vendors have shown superior good timing and position resolution as well as high magnetic field tolerance but are generally far too expensive for large area coverage. The recently commercialized new type MCP-PMT using the atomic layer deposition technique as a large area picosecond photodetector (LAPPD) provides a promising cost-effective MCP-PMT for the EIC RICH detectors. Efforts have already been devoted to optimizing the LAPPD as photosensor of choice for EIC Cherenkov detectors (e.g. dRICH, mRICH, DIRC) as well as TOF applications.

\begin{figure}[tb]
    \centering
    \includegraphics[width=0.9\linewidth]{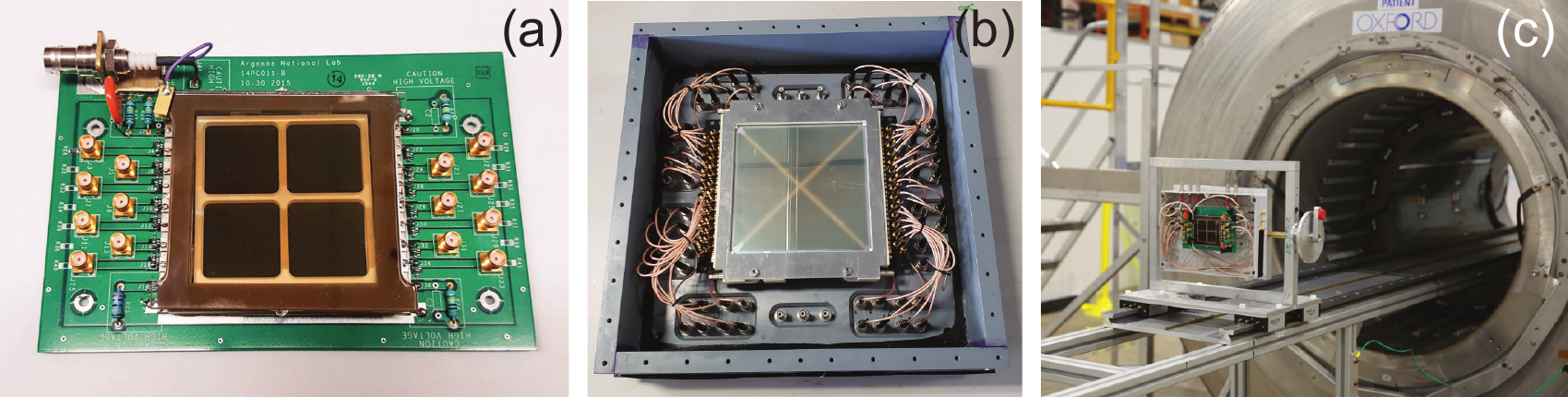}
    \caption{(a) R\&D $6\times6$ cm$^2$ MCP-PMT built at ANL for R\&D test bed. (b) Commercial $20\times20$ cm$^2$ LAPPD for test  at JLab (Hall C). (c) $6\times6$ cm$^2$ MCP-PMT during magnetic field tests.}
    \label{fig:lappdPhotos}
\end{figure}


A list of performance requirements of the photosensors for EIC Cherenkov based detectors is listed in Tab.~\ref{table::photosensorReq}.

\begin{table}[hbt]
    \centering
    \begin{tabular}{l|l|l} 
    \hline
       Parameter  &  gas-RICH, mRICH, dRICH	& DIRC 
       \\ \hline \hline
       Gain	&$\sim10^6$ &	$\sim 10^6$  \\
Timing Resolution &	$\leq 800$ ps&$\leq 100$ ps  \\
Pixel Size&	≤ 3 mm 	& 2–3 mm  \\
Dark Noise &	$\leq$ 1MHz/cm$^2$&	$\leq$ 1kHz/cm$^2$  \\
Radiation Hardness &	Yes &	Yes  \\
Single-photon mode operation&	Yes	& Yes  \\
Magnetic-field tolerance&	Yes (1.5–3 T)&	Yes (1.5–3 T) \\
Photon Detection Efficiency&	$\geq 20$\%	& $\geq 20$\% \\ \hline
    \end{tabular}
    \caption{Performance requirements of photosensors for EIC Cherenkov based detectors.}
    \label{table::photosensorReq}
\end{table}

R\&D at Argonne National Laboratory using the Argonne MCP-PMT ($6\times 6$ cm$^2$), a small format of LAPPD, has demonstrated all the required parameters, including a gain of $10^6-10^7$,fast timing resolution of $\sim 80$ ps (RMS) and of $\sim 20$ ps (SPE), dark noise $\leq$ 1kHz/cm$^2$ etc. Especially a magnetic field tolerance over 1.5 Tesla and less than a 1 mm position resolution with a pixel size of 3$\times$3 mm$^2$ have been demonstrated as well. Generally, MCP-PMT has good radiation hardness, a radiation test on our MCP-PMT will be performed to ensure it. To expedite the application of MCP-PMT for EIC Cherenkov detectors, a 10x10 cm$^2$ MCP-PMT fabrication facility is under construction to produce larger size, high-performance  MCP-PMTs. The commercial available LAPPD module has also achieved almost all the requirements except fine pixel size and magnetic field tolerance. Our industrial partner INCOM has adapted the Argonne MCP-PMT R\&D results to develop low-cost pixelated LAPPDs for EIC Cherenkov detectors. Fine pixel size (3x3 mm$^2$) is the urgent focus for commercial LAPPDs; bench and beam line tests are required for the LAPPD validation. 

The Argonne MCP-PMT/LAPPD R\&D is a generic effort.  These photosensors can be widely used where large areas, low cost and high performance are needed.  The required R\&D is aimed at both near-term and future detector designs. Testing and performance results have already been shared with all EIC Cherenkov and TOF detector design efforts.

Rapid progress has been achieved on the Argonne MCP-PMT/LAPPD. Recently, a Gen-II LAPPD from INCOM was successfully tested at Jefferson Lab in a high rate, high background environment. Furthermore, a Fermilab beam line test of a pixelized MCP-PMT performance is planned for Spring 2021. To validate the LAPPD performance and apply this new technology to the EIC-PID subsystems, critical R\&D is needed in the next two years.  A bench test and multiple beam tests of Cherenkov prototype detectors using the MCP-PMT/LAPPD will need to be performed. For example, an mRICH beam test with LAPPD is mentioned in the mRICH section, and a gaseous RICH detector with Argonne 10x10 cm$^2$ MCP-PMT is under development and planned for a beam line test as well.

\subsection{R\&D Needs for GEM-TRD/Tracker in the Forward Direction}

The identification of secondary electrons in the forward region plays an important role in the physics program of the Electron-Ion Collider (EIC). A high granularity tracker combined with a transition radiation detector for particle identification could become crucial for improving the overall $e/h$ performance of the detector. The scope of the project is to develop a transition radiation detector and tracker capable of providing additional pion rejection in the order of 10-100.   

A low-mass radiator available for mass production is critical and various materials still need to be tested. This includes the optimization of \emph{(i)} a pseudo-regular radiator using thin ($\sim12-15\,\mu$m) Kapton foils with thin net spacers and \emph{(ii)} a detailed test of available fleece and foam materials to improve the photon yield. 

The transition radiation detector readout  is based on well established GEM technology. The main difference to a GEM tracker as discussed in Sec.~\ref{RD:lowMassGem} is the thickness of the drift volume. In order to keep the electric field uniform a special field cage needs to be developed.  This includes the mechanical design and construction of a gas cage  to minimize a Xe-filled gas gap between radiator and the drift cathode.

The anode readout PCB layer of the current GEM-TRD prototype is based on a readout developed for the COMPASS experiment made of X and Y strips with a pitch size of 400 $\mu$m. While this is optimal for a high occupancy environment, the large number of channels does increase the price of the readout electronics. Efforts are under way to develop a new  pad readout that is better 
 suited for GEM-TRD applications. A novel large-pad readout PCB will combine three crucial advantages: large readout pads to reduce the number of readout channels, excellent spatial resolution (despite the large pad size), and improved noise reduction. We also plan to test a zigzag readout board option. 

The GEM TRD will need 2 HV lines, one for the GEM  amplification stage and one to set a uniform drift field. To work in a high occupancy environment, the drift time needs to be minimized, requiring fields of  $\sim$2-3 kV/cm. For a 2 cm drift distance the HV should be at the level of 4-5 kV. Depending on the chosen grounding scheme, the total voltage including GEM stage, could be up to 8-9 kV. Optimization of HV for large drift distances is ongoing. 

In the current tests, the GEM TRD uses readout electronics that was originally developed for the  GlueX wire chambers. It consist of a preamplifier (GAS2 ASIC chip) with shaping times of $\sim$10-12~ns. The flash ADC has a sampling rate of 125 MHz and 12 bit resolution but provides only pipe-lined triggered readout. The total price is about \$50 per channel.  The collected high-resolution data recorded in test beams allow us to estimate 
the needed shaping times of the preamplifier, the FADC sampling rate, and the corresponding resolution.  Development of a new  FADC125 is needed to enable the streaming of zero-suppressed data over fiber links. An alternative would be to adopt other existing readout chip such as SAMPA or VMM3,  for the GEMTRD application. This, however, would require significant improvements of their shaping times. For example, the latest SAMPA version (v5 ) has a peaking time of ~ 80 ns, which is too slow for studying the GEM-TRD performance. Improvements are also needed in the return-to-baseline time for single clusters in order to allow multi-cluster measurements from a single GEM strip/pad. Additionally, a final implementation to the GEM readout based on improvements to the SAMPA or VMM chips will require their compliance to the EICs streaming readout architecture. This  includes off-chip drive enhancements that will provide for better thermal management at the detector while enabling data collection, processing and transport via high speed optical fibers. 

Over the past few years, the price of Xe has gone up significantly requiring the development of a recirculation system to purify, distribute, and recover the gas. This will be based on the design of the ATLAS TRD gas system and therefore will require only moderate R\&D.

\subsection{Gaseous Single Photon Detectors Based on MPGD Technologies}
\label{subsec-gaseous_PD}

Single Photon Detectors (PD) for Cherenkov imaging devices represent a key
challenge at EIC where minimum material budget and operation in high magnetic
field is required. Gaseous PDs, which have played /are playing a major role
in establishing and operating Ring Imaging CHerenkov (RICH) counters,
satisfy these requirements and they represent the most cost-effective
solution when equipping large detector areas. So far, the only photon
converter successfully coupled to gaseous detector is CsI with Quantum Efficiency (QE)
limited to the far UV domain. Optimized detector architecture and operative
conditions have to be established to ensure effective photoelectron extraction
and control of the Ion BackFlow (IBF) to the photocathode. In particular,
Micro Pattern Gaseous Detector (MPGD) technologies offer natural answers
to IBF and photon feedback suppression and fast response, as tested by a number of successful applications:
the PHENIX HBD with triple GEM PDs~\cite{Anderson:2011jw},
the COMPASS RICH upgrade with Hybrid (THGEMS and resistive MICROMEGAS) PDs~\cite{Agarwala:2018wba},
the windowless RICH prototype and test beam with quintuple GEM PDs~\cite{Blatnik:2015bka},
the TPC-Cherenkov (TPCC) tracker prototype with quadruple GEM PDs~\cite{Azmoun:2019doe}.

\begin{figure}[tb]
    \centering
    \includegraphics[width=0.9\linewidth]{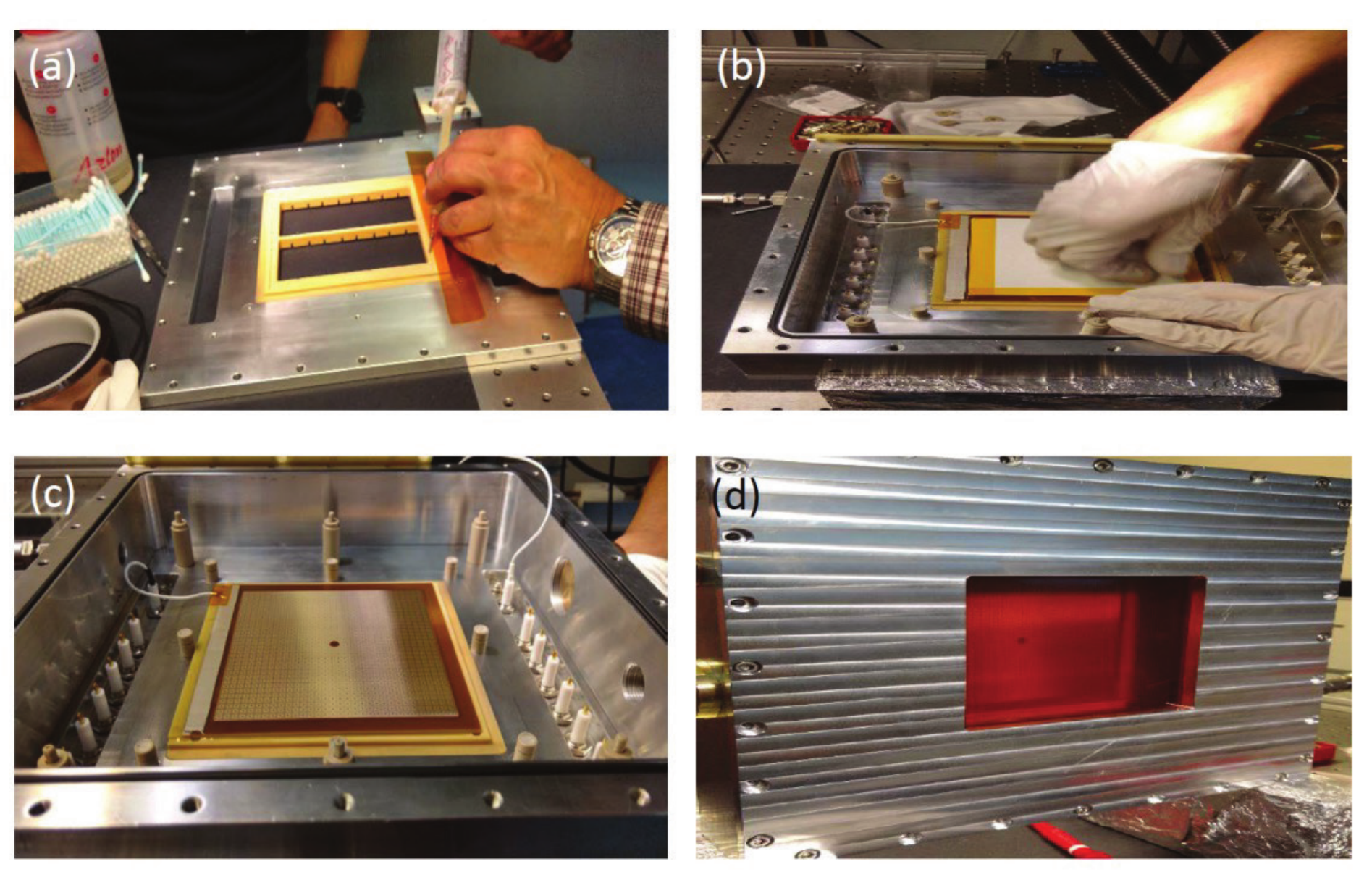}
    \caption{Construction of a prototype of MPGD-based single photon detector with small pad-size. (a) The fiberglass frame supporting the MICROMEGAS (MM) is glued onto the Al chamber structure. (b) The MM is glued onto the fiberglass frame. (c) The MM installed in the chamber and its power lines are visible. (d) The chamber is closed with a mylar window.}
    \label{fig:PhotonDetYR}
\end{figure}

In the EIC context, gaseous PDs represent a valid option for the high momentum RICH with
gaseous radiator.  An R\&D program for further developments of the hybrid approach in operation at COMPASS,
aiming at making it fully adequate for the high momentum RICH at EIC, is ongoing, where the
reduced space availability imposes a compact RICH. The whole program includes:
\begin{enumerate}
\item
Establishing the hybrid PD for a windowless RICH approach to increase the number of detected Cherenkov photons;
\item
Increasing the granularity of the read-out elements for fine resolution with limited lever arm; this item is well advanced (Fig.~\ref{fig:PhotonDetYR});
\item
Comparing the detector performance using either THGEM (as in COMPASS) or GEMs for the first multiplication stages;
\item
Identification of an adequate front-end chip: studies for coupling the hybrid PD with VMM3 ASIC have been initiated;
\item
Coupling of the THGEMs with a novel and more robust photoconverter by Hydrogenated Nano Diamond powder (HND)
to overcome the limitation imposed by the use of CsI. This is due to its chemical fragility in contaminated atmosphere
or under ion bombardment, which imposes gain limitations and complex handing;
very promising initial studies are ongoing.
\end{enumerate}
\par
The R\&D will progress along these lines. The action items 1, 2, 3 and 4 are needed to
make this technology adequate for its use at EIC and they can be completed within a couple of years.
Establishing the novel photoconverter for gaseous PDs will take longer, due to the largely innovative
character of the approach. If converging, it can represent an added value to the project.
It can be selected for the EIC PDs according to its level of maturity
when the detector design is finalized.

\subsection{Fast Timing Silicon Sensor: LGADs} 
\label{part3-sec-DetTechnology.LGADsRD}

The Low Gain Avalanche Detector (LGAD) with internal 
gain~\cite{White:2014oga,Pellegrini:2014lki,Cartiglia:2015iua,Breton:2016zoz,Minafra:2017nqc,Apresyan:2018oln} 
is an ultra-fast silicon sensor technology, which has recently been 
chosen for constructing a fast-timing layer in the
forward rapidity region of the CMS~\cite{CMS:2667167} 
and ATLAS~\cite{Collaboration:2623663} experiments
at the high-luminosity (HL) LHC starting in 2027.
The new timing layers will help the experiments 
mitigate significantly larger pileups of proton-proton
interactions (up to about 200) by providing 4-D vertex 
reconstruction, and serve as a time-of-flight system for 
hadron identification in QCD and heavy-ion physics. 

Traditional $n$-$p$ silicon sensors with gains provided by 
external bias voltages can provide a typical time resolution 
on the order of 150~ps. The LGAD silicon sensors have an 
intrinsic gain of 10--30 provided by a special implant 
layer to generate a strong electric field locally and 
trigger avalanches. This internal gain helps the LGADs to 
achieve a low-jitter fast-rising pulse edge and overcome many 
other noise sources that enable high precision timing measurements 
for MIPs. LGAD sensors of 35--50~$\mu m$ in active area thickness 
can achieve a typical time resolution of about 30~ps. The handling
wafer has a tyical thickness of 150--300$\mu m$.

With excellent timing and position resolutions,
the LGADs provide an attractive option for constructing a 
compact, multi-layer system to simultaneously provide TOF-PID 
and trajectory reconstruction as part of the tracking system. 
In addition, the LGADs have several other key advantages of 
being highly tolerant to strong magnetic fields (up to B$\sim$4~T), radiation-hard (up to $\sim 2 \times 10^{15}$~n$_{\rm eq}$/cm$^{2}$, 
compared to the expected level of radiation of $\sim 10^{11}$~n$_{\rm eq}$/cm$^{2}$ at EIC) and compact (flexible for integration).
To fulfill the requirements for EIC physics, there are three main areas
of R\&D needed, which are discussed below:

\begin{itemize}
    \item \textbf{Time resolution}: while LGAD silicon 
    sensors used by CMS and ATLAS can provide a time 
    resolution of 30--50~ps, particle flight distance 
    at EIC detectors is likely to be much 
    shorter due to tight space constraints. Therefore,
    a total time resolution (including readout electronics) of 20~ps or better per layer is desired to
    meet the PID requirement at low and intermediate momentum regions.
    The jitter contribution to the time resolution 
    is directly related to the signal slew rate, which is
    inversely proportional to the sensor thickness.    
    Reducing the thickness from 50$\mu m$ to 35, 25 and 
    even 20$\mu m$ will not only improve the jitter but 
    can also suppress the Landau noise. Note that to maintain
    the total charge collection for a large signal, both internal
    and external gains applied also need to be optimized.
    Recent R\&D work on 35$\mu m$-thin LGADs shows a time resolution 
    of about 20--25~ps per layer, a promising step toward achieving
    the PID requirements for EIC~\cite{Jadhav:2020ujs}. By stacking several timing layers, the time resolution can be further improved by a factor proportional to the inverse of the square root of the number of layers: as demonstrated in Ref~\cite{Jadhav:2020ujs} by stacking three layers of 35$\mu m$-thin LGADs a time resolution of 14~ps can be reached.  
    
    \item \textbf{Fill factor and position resolution}:
    to serve as (part of) a tracking system, a position resolution 
    much better than the 1~mm pixel size has to be
    accomplished to be competitive to other types of
    silicon pixel and/or strip sensors that are designated 
    for position measurements. The current limitation lies 
    in the approximately 50~$\mu m$ width of the intra-pad no-gain region, which is needed to protect against early breakdowns. Smaller 
    pixel sizes would lead to too low fill factors, or 
    loss of acceptance. The CMS and ATLAS timing layers have 
    a fill factor of 85\% per disk, with the two-disk system 
    compensating for a 100\% acceptance.
    
    To achieve better position resolution (beyond 1~mm pixel
    size), two viable solutions are present. Trench-isolated (TI) LGADs 
    are capable of reducing the no-gain region down to a width of 
    only a few $\mu m$, essentially eliminating it, to achieve 100\% 
    fill factor. All readout schemes can be kept the same as 
    standard LGADs. For AC-coupled LGADs, segmentation is not done 
    on the silicon sensor but at metallic readout contacts sitting 
    on top of a dielectric layer, reading out induced charges. The 
    fill factor is effectively 100\%. The signal pulse is shared 
    among several adjacent pads, further improving its position 
    sensitivity. The metallic readout pads can be fabricated into 
    pixels, strips or any shape desired. The AC-coupled LGADs are 
    also considered as an option for a high precision timing Roman Pots, 
    where R\&D needs are discussed in Sec.~\ref{part3-sec-DetTechnology.RomanPotsRD}.
    
    \item \textbf{ASIC readout chips}: The needs for better timing performance and finer 
    granularity also pose significant challenges to the readout electronics and specifically 
    to the ASIC readout chips. Present ASIC chips designed for CMS and ATLAS timing detectors 
    have a jitter on the order of 20--30~ps, and a pixel granularity of $1.3\times1.3$ mm$^{2}$. The CMS chip uses a 65~nm node technology, while the ATLAS chip uses 130~nm, and their power consumption per ASIC is 1.0 W and 1.2 W, respectively.
    Reduced granularity will make it more difficult to fit all the circuit components within 
    the available space, and is also likely to lead to significantly increased power 
    consumption due to increased total number of channels. However, differently from LHC applications, where low temperatures are needed to minimise the sensor radiation damage, at EIC cooling will only be needed to remove the excessive heat generated by the front-end electronics. Based on architectural designs of CMS
    and ATLAS timing layers, an ASIC chip with a size of $0.5\times0.5$ mm$^{2}$ is feasible to achieve
    and would meet the requirements set by the Roman Pot detector. A finer granularity, likely required for 
    the tracker application, would require dedicated efforts of new architectural designs and adoption of more 
    advanced silicon fabrication processes.
\end{itemize}

\section{Electromagnetic and Hadronic Calorimetry}
\label{part3-sec-DetTechnology.Calorimetry}

\subsection{Tungsten Scintillator Calorimetry}

Tungsten scintillator (W/Scint) calorimetry can play a major role in many of the regions of an EIC detector, covering a rapidity range from $\sim$ -2.0 to 4.0. It offers a very compact design in terms of its short radiation length, thus limiting the total length of the calorimeter, as well as providing a small ($\sim$ few cm) Moliere radius which limits the lateral extent of the shower, therefore allowing good separation between neighboring electromagnetic showers as well as limiting the overlap with hadronic showers. In addition, the energy resolution can be tuned by changing the sampling fraction and sampling frequency to meet the different requirements in the various rapidity regions.

There are primarily two candidates that are being considering for a W/Scint calorimeter for EIC. One is a tungsten scintillating fiber (W/SciFi) SPACAL, which consists of a matrix of tungsten powder and epoxy with embedded scintillating fibers. This technology is used for the sPHENIX barrel EMCAL that consists of more than 6K individual 2D projective absorber blocks. The blocks are read out using SiPMs that are coupled to the blocks using short light guides. This calorimeter is currently under construction and is expected to be completed by the end of 2021. 

\begin{figure}[tb]
    \centering
    \includegraphics[width=0.9\linewidth]{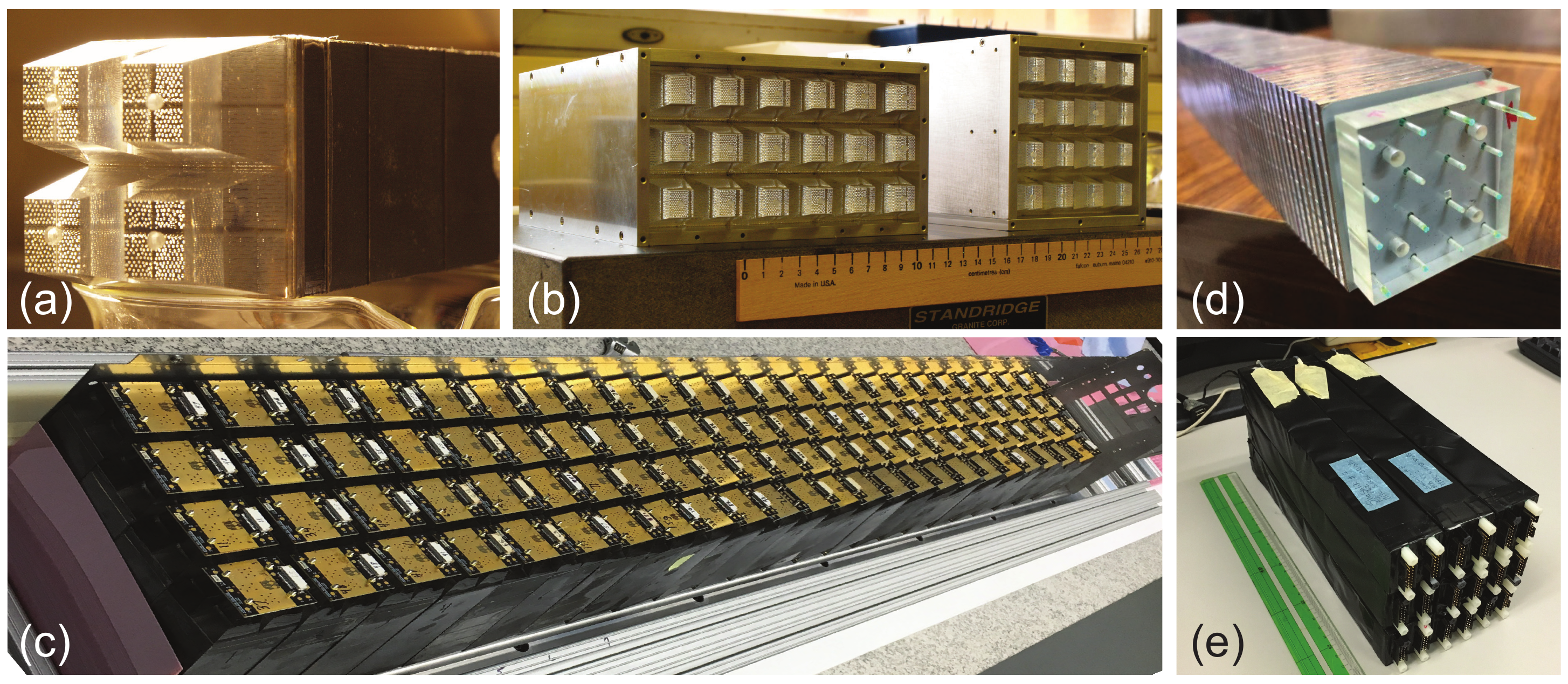}
    \caption{(a)-(b) W-SciFi calorimeter modules assembled during early stages of R\&D. (c) Assembly of a complete sector of the sPHENIX calorimeter based on W-SciFi technology. (d) W/Shashlik calorimeter with 80\% W and 20\% Cu absorber plates between scintillating tiles read out with WLS fibers. (e) Assembly of 9 prototype W/Cu shashlik modules.}
    \label{fig:sciWiFiPhoto2} 
\end{figure}
The technology for producing the blocks was originally developed at UCLA \cite{Tsai:2015bna}. Several of the early prototypes are shown in Fig. ~\ref{fig:sciWiFiPhoto2}(a)\&(b). The technology for producing blocks has now been developed to an industrial scale at the University of Illinois \cite{Aidala:2017rvg} to produce all the blocks for the sPHENIX calorimeter.  Figure ~\ref{fig:sciWiFiPhoto2}(c) shows one of its sectors which consists of 96 blocks. Therefore, no further R\&D is required for producing the blocks, but the method used for reading out the blocks with SiPMs could be improved. This would include the use of large area SiPMs to provide more photocathode coverage and eliminate the boundaries between the light guides which leads to non-uniformities in the energy response. It is planned to refurbish the sPHENIX EMCAL with this type of readout for use as a Day-1 detector at EIC.
   
The second W/Scint technology that is being considered for EIC is a tungsten shashlik (W/Shashlik) design. Many shashlik calorimeters have been built and used by many experiments. A W/Shashlik design offers some distinct advantages but also poses some significant challenges. In addition to being compact and being able to tune the energy resolution as in the W/SciFi, a W/Shashlik offers the possibility of improving the light collection and providing better uniformity by reading out each individual WLS fiber with its own SiPM. This allows a better determination of the shower position and the possibility of using this information to correct for non-uniformities in the energy response. However, the mechanical properties of tungsten make it difficult to machine and requires using a slightly less dense alloy of tungsten, thereby increasing the radiation length and Moliere radius. Also, making a shashlik calorimeter projective makes the mechanical design and assembly more complicated. This technology is currently being studied for EIC by the groups at BNL and Andres Bello University in Chile and a small prototype detector consisting of nine W/Cu shashlik modules, shown in  ~\ref{fig:sciWiFiPhoto2}(d)\&(e), has been constructed and will be tested. 

Both calorimeter technologies use SiPMs as photosensors, but it is well known that these devices are subject to radiation damage, particularly neutrons. The development of more radiation hard SiPMs would be of great benefit for calorimetry at  EIC, as well as for many other detectors, but developing radiation hard SiPMs would take several years of R\&D and require a substantial investment with the manufacturers.

\subsection{SciGlass for Electromagnetic Calorimetry}
Nearly all physics processes require the detection of the scattered electron in the electron endcap (backward rapidities). The requirement of high-precision detection is driven mainly by inclusive DIS where the scattered electron is critical to determine the event kinematics. Excellent electromagnetic calorimeter resolution of better than 2\%/$\sqrt{E}$ is required at small scattering angles, while very good resolution is acceptable at larger angles. For hadron physics measurements with electromagnetic reactions, the most common precision calorimeter material of choice has been lead tungstate, PbWO$_4$ (PWO). However, the production of crystals is slow and expensive. 

\begin{figure}[tb]
    \centering
    \includegraphics[width=0.9\linewidth]{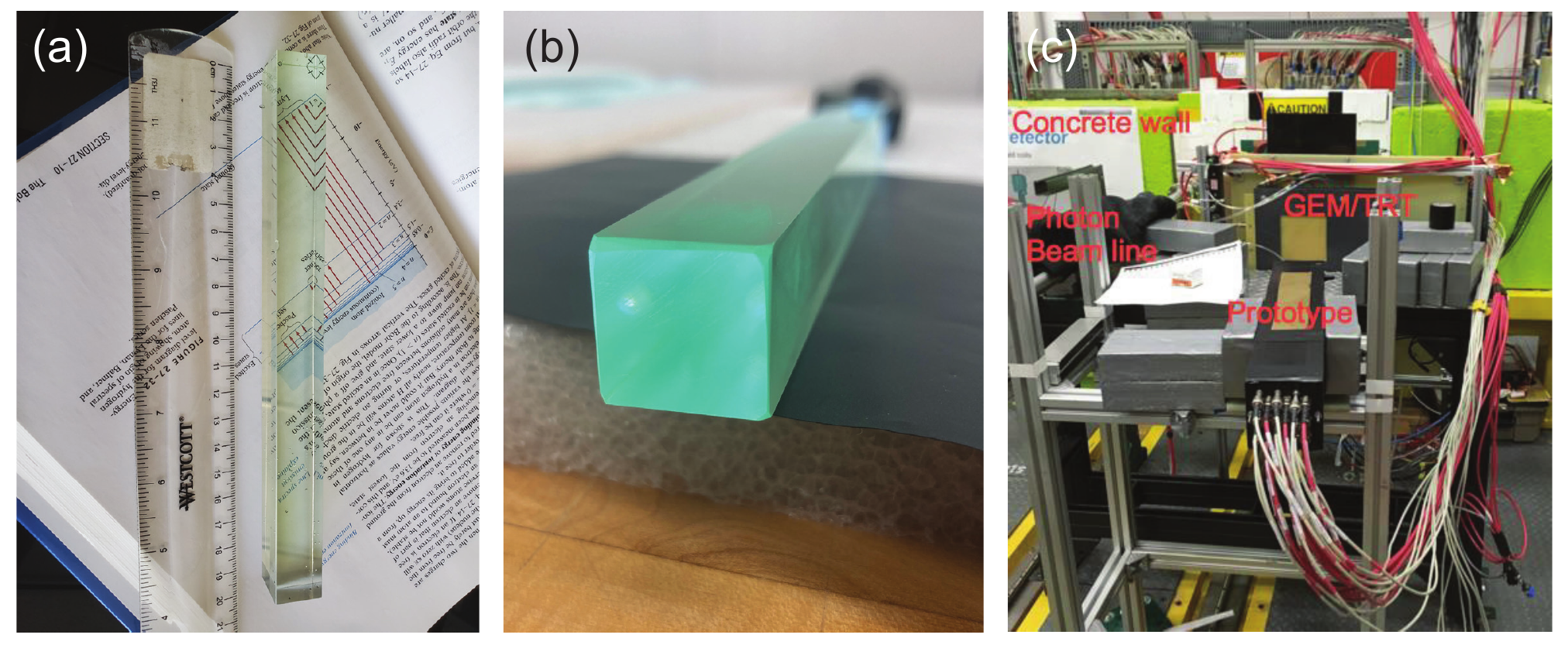}
    \caption{((a,b) Samples of 20 cm long SciGlass towers. (c) Beam test setup at JLab (Hall D) for SciGlass with PMT and SiPM readouts.}
    \label{fig:sciglassPhotos2}
\end{figure}

The technology goal of SciGlass R\&D is to develop a scintillating glass for homogeneous electromagnetic calorimetry. SciGlass is a radiation hard material optimized to provide characteristics similar to or better than PbWO$_4$. SciGlass fabrication is expected to be cheaper, faster, and more flexible than PbWO$_4$ crystals. SciGlass is being developed by Scintilex, LLC in collaboration with the Vitreous State Laboratory at CUA. Tremendous progress has been made in the formulation and production of SciGlass that improves properties and solves the issue of macro defects. Scintilex has demonstrated a successful scale-up method and can now reliably produce glass samples of sizes up to $\sim 10$ radiation lengths (see Figure~\ref{fig:sciglassPhotos2}. Simulations combined with initial beam tests at photon energies of {4-5~GeV} suggest that high resolution competitive with PbWO$_4$ can be reached for $ > 15 X_0$. SciGlass has excellent radiation resistance (no damage up to 1000 Gy electromagnetic and $10^{15}$ n/cm$^2$ hadron irradiation, the highest doses tested to date), response time of 20-50 ns, and good transmittance in the near UV domain (74\% at 440 nm). The SciGlass insensitivity to temperature is also a clear advantage over PbWO$_4$, which has a dependence of about 2-3\%/$^\circ$C and has to be continuously monitored. The present samples have a density up to 5.4 g/cm$^3$, radiation length ($X_0$) of 2.2-2.8 cm and a Moliere radius of 2-3 cm.

The areas of needed R\&D for SciGlass include the final formulation optimization, scale up to block sizes $\gtrsim 15 X_0$, and beam tests to establish characteristics like energy resolution. The most critical items are to demonstrate scale up to block sizes $\gtrsim 15 X_0$ and to establish SciGlass characteristics with beam tests. The evaluation of SciGlass as particle detector has been shared in part with activities on PbWO$_4$ crystals for the electron endcap calorimeter, {\it e.g.}~simulations, radiator characterization and prototype construction, commissioning and beam tests. The approximate timeline for completing the SciGlass R\&D is about one year assuming R\&D funds are available. The goal is to be ready for a day-1 detector. SciGlass could also be available for future detector upgrades.

\subsection{Hadronic Calorimetry}
\par Optimum jet reconstruction will require the use of several detector systems (tracking, EMCAL and HCAL) but is a main driver for hadronic calorimetry. As such, the requirements for the resolution of the hadronic calorimeter are different for the endcaps and the barrel region. The most challenging is the forward region of hadronic endcap where pure calorimetric measurements starts to outperform particle-flow like approaches due to the degradation of tracker performance. For the electron endcap and the barrel region, only modest hadronic energy resolution is required from calorimeter system (ECAL+HCAL). It is believed that these systems can be built using standard construction methods and no additional R\&D efforts are needed. For the hadronic endcap, covering the rapidity range from $\sim$ 1.0 to 2.5 where better energy resolution is required, modest R\&D efforts will be needed to improve the performance of these systems. For example, the STAR Forward Calorimeter, which is currently being constructed using a new and efficient method developed at UCLA \cite{Tsai:2012cpa,Trentalange:2013/12/16ena}, would require improvements for a more efficient light collection scheme due to the relatively low energy of hadrons in this region of hadronic endcap at the EIC. 
\par At more forward rapidities in the hadron endcap, it is important to have the best possible performance of the calorimeter system. The main constraint at the EIC is the lack of space for a high sampling fraction and high sampling frequency calorimetry system, both of which are required to achieve good resolution. Developing a high resolution calorimetry system for this region will require significant R\&D efforts. At present we believe that there is only one technology option that may be suitable for this region, which is a very high density, approximately compensated fiber calorimeter, which would serve as both the ECAL and HCAL with a common readout.

To date, R\&D for hadron calorimetry for EIC has had a low priority and very limited funding. The synergy between the STAR Forward Upgrade and eRD1 R\&D activities lead to construction and testing of two prototypes forward calorimeter systems. One was a compensated system with an ECAL section built with a W/ScFi technique followed by hadronic section made of a lead scintillator sandwich. The other non-compensated version had a lead scintillating shashlyk ECAL and an iron scintillator sandwich HCAL section behind. A later version was a final design prototype for STAR Forward Calorimetry system. Both versions had SiPM readouts and both were tested at FNAL. The performance of both systems led us to believe that the initial requirements for the EIC calorimetry system can be reached with only the modest improvements mentioned above. However, due to lack of funds, both versions of the prototypes had limited size which lead to significant transverse leakage and required an extrapolation of the test results to larger size detectors. This should be avoided for future EIC targeted R\&D. 

A common theme for the R\&D needs for both an ECAL and HCAL at EIC is the readout with SiPM sensors covering a large surface area. This may be challenging at the forward rapidities of the hadron endcap due to the relatively low light yield of hadron calorimeters (compared to EM calorimeters), and the high neutron fluences in this region, which will lead to significant degradations in SiPM performance. Operation of the STAR Forward Calorimetry system in the 2022 500 GeV RHIC run will be very valuable because the conditions at STAR will be very close to those in the EIC hadron endcap in terms of neutron fluxes. Future R\&D is therefore needed in this direction.       

Further R\&D targeted on a reference EIC detector should include the construction and testing of a full-scale prototype
 of a forward hadron calorimeter system. This prototype should include a WScFi electromagnetic-calorimeter section and a Fe/Sc-hadronic section with an integrated tail catcher. These R\&D efforts should also include a preshower detector envisioned for the reference detector design.

\subsection{CSGlass for Hadronic Calorimetry}
Achieving high-quality science at nuclear physics facilities requires the measurement of particle energy with excellent calorimeter energy resolution. Particles that produce EM showers can be detected with high precision. However, there is a need to improve the energy resolution of hadron calorimetry. The technology goal of CSGlass R\&D is to develop a scintillating glass for improving hadronic calorimeter resolution, which is desired for measurements of hadronic jets. 

CSGlass is optimized for the dual readout approach, where one compares the signals produced by Cherenkov and Scintillation light in the same detector. This approach has been a promising method to achieve better performance for hadron calorimeters. Homogeneous crystals are an option, but have to be outfitted with optical filters, which results in insufficient Cherenkov light detection. Crystals are also prone to radiation damage, time consuming to manufacture, and relatively expensive. In comparison, radiation-hard glasses can be tuned for favorable Cherenkov/Scintillation signal ratio, eliminating the need for optical filters, and thus offer great potential for both precision hadron calorimetry and significant cost reductions if competitive performance parameters can be achieved. CSGlass is derived from SciGlass and expected to be similarly resistant to EM and hadron irradiation up to 1000 Gy and $10^{15}$~n/cm$^2$, the highest doses tested so far. The CSGlass interaction length is comparable to crystals and should allow for small tower size. The anticipated space for the homogeneous calorimeter configuration could be similar to the binary system and may provide better resolution. 

The areas of needed R\&D for CSGlass include the demonstration of CSGlass with sufficient UV transparency for Cherenkov light collection, clear separation of Cherenkov and Scintillation light of sufficient intensity (slow scintillation, $ > 500$ nm beneficial), low cost, and characterization of CSGlass in the lab and with test beam R\&D prototypes. The most critical items are the formulation optimization and production of CSGlass test samples. Some of the CSGlass R\&D is shared with SciGlass and PbWO$_4$ crystals for EM calorimeters. The approximate timeline for completing the CSGlass R\&D is around three years assuming R\&D funds are available. CSGlass could be ready for future detector upgrades.

\section{Auxiliary Detectors}
\label{part3-sec-DetTechnology.Aux}

\subsection{Roman Pots and LGAD Technology}
\label{part3-sec-DetTechnology.RomanPotsRD}
A far-forward proton spectrometer, based on the well known technique of Roman Pots, is an integral part of an EIC detector system, essential for the success of its physics program (see 
Secs.~\ref{part2-sec-DetReq.Excl} and ~\ref{part2-sec-DetReq.Diff.Tag}), and thus is envisioned as a subsystem for a day-one EIC detector.
A forward proton spectrometer will provide a critical contribution to the study of inclusive diffractive and exclusive production processes in coherent e+p and e+D collisions. 
Furthermore, it is essential to provide a veto of incoherent background to measurements of exclusive meson production in e+A collisions.

\begin{figure}[tb]
    \centering
    \includegraphics[width=0.85\linewidth]{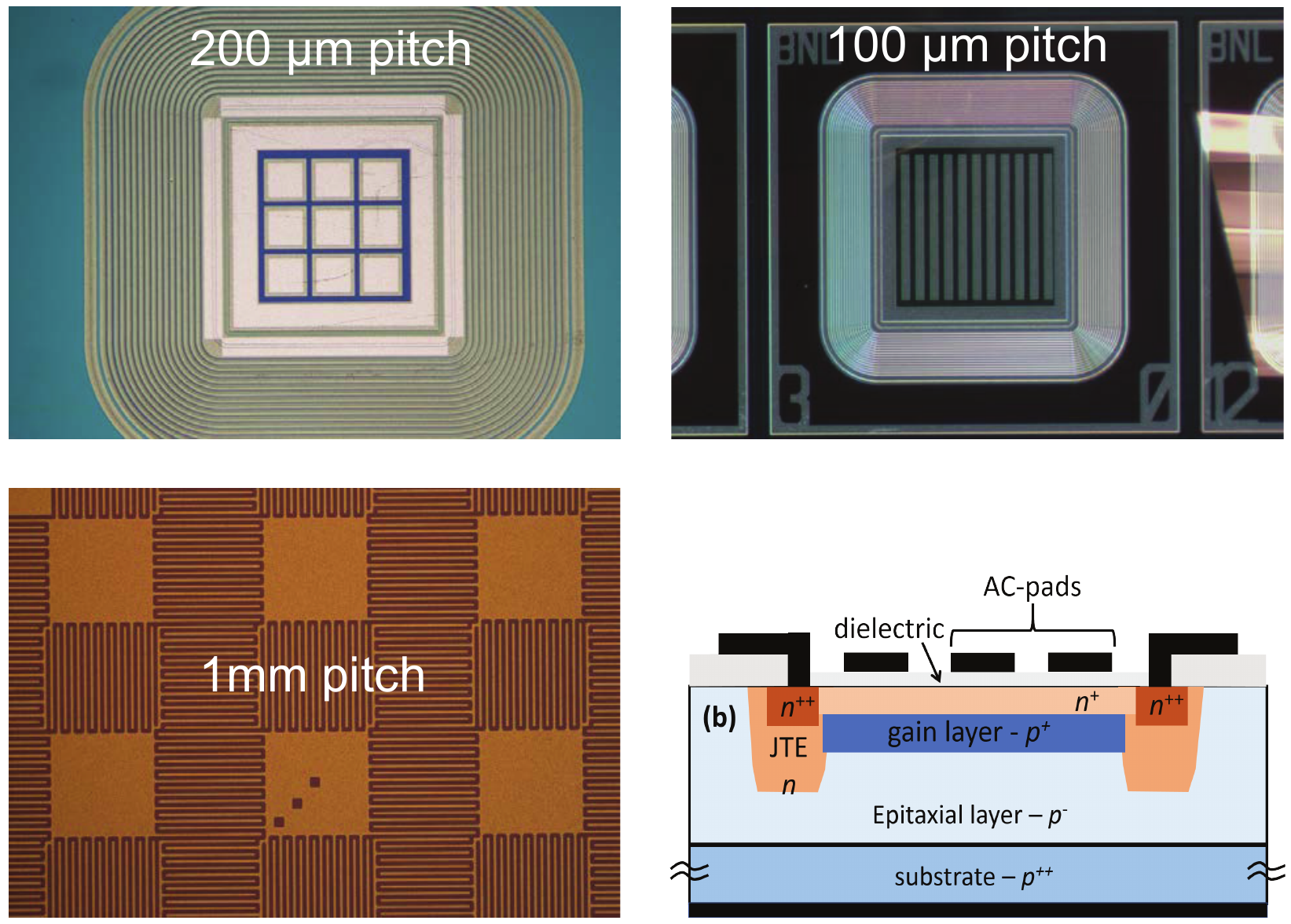}
    \caption{A collage of AC-LGAD sensors that are under development and being tested. The lower right plot shows the schematic design of an AC-LGAD sensor.}
    \label{fig:lgad_photos}
\end{figure}

An innovative silicon technology, based on {\it Low Gain Avalanche Diode (LGAD)}, is proposed to instrument the Roman Pots, as well as other EIC detector subsystems, as it has the potential to combine in a single sensor fine spatial resolution and precise timing. More specifically, by AC-coupling the metal layer (that is connected to the readout electronics) to the active silicon layers of an LGAD ({\it AC-LGAD}), the sensor can be finely pixelated (in the order of few tens of microns) to reach a spatial resolution similar to conventional pixel trackers, and its timing performance can be maintained compatible to the one of standard LGADs, i.e. $\approx30$ ps. A sketch of an AC-LGAD sensor designed at BNL can be seen in Fig.~\ref{fig:lgad_photos} (bottom-right).
While the LGAD technology is established and is being used by the ATLAS and CMS experiments at the LHC for their timing subsystems for the High Luminosity phase (HL-LHC), the AC-LGAD technology is instead under intense development in US, Europe and Japan.
 
Simulations show that $500 ~\mu$m  square pixels and 30--40 ps time resolution are sufficient to achieve the desired physics performance. In more detail, simulations showed that the detector pixels must be at least as small as 500 x 500 $\mu$m$^2$ to make the smearing contribution negligible with respect to the other effects at 275 GeV. Currently available LGAD sensors for the HL-LHC have 1.3 x 1.3 mm$^2$ pixels, which would provide smearing contributions outside of the Roman Pots specifications. The 500 x 500 $~\mu$m$^2$ pixelation can be achieved in AC-LGADs, and, with reasonable effort, in the associated readout electronics, i.e. by small modifications of the ASIC developed for the ATLAS timing detector. It must be noted that a space resolution an order of magnitude smaller than the pixel pitch can be achieved by using the information from the signal sharing between neighboring pixels, with a substantial advantage in power and real estate in the readout electronics. At the same time, in the high acceptance configuration, the impact of the angular divergence on the smearing of the transverse momentum becomes comparable to the contribution from the crab cavity rotation of the beam bunch. To remove the smearing contribution from the crab cavity rotation, in addition to further rejecting the backgrounds, fast timing is required in the range $\approx30$ -- 40 ps. Such timing performance has been demonstrated by the LGAD sensors developed for the HL-LHC, and it has been recently shown to be achievable by AC-LGAD sensors too. In addition, such sensors must be placed as close as possible to the beam, therefore their inactive area at the edge of the sensor must be minimized, and must be $\le100 ~\mu$m. Laboratory tests showed that the inactive edges of LGADs can be reduced to about $50 ~\mu$m, i.e. to values compatible with the Roman Pots specifications.
Several designs of AC-LGAD sensors have been fabricated and tested, as shown in Fig.~\ref{fig:lgad_photos}.
 
In summary, the novel AC-LGAD sensor technology has recently been shown to meet both spatial and timing performance as well as small edge specifications for its application in Roman Pots. However, further work is needed to fully characterize the AC-LGAD performance, test their robustness and optimize their design for the specific implementation in Roman Pots. For instance, the intrinsic sensor gain and thickness can be optimized to improve the time resolution, finer spatial resolution can be achieved by exploiting the signal sharing properties of neighboring pixels, and larger area prototypes with advanced designs need to be fabricated and tested. Most critical at this point in time is the development of an architecture of the readout electronics, and more urgently the ASIC R\&D.  
 
Given the need of fast-timing at EIC and the growing interests in LGAD technology to meet those needs (see time-of-flight detector, 4D tracker, TOPSiDE detector concept, $4\pi$ hybrid LGAD/SOI tracker, preshower), a collaborative effort will be extremely beneficial. An international consortium is being formed to accomplish the above-mentioned R\&D tasks.
 
In a time-frame of two years, thanks to prototyping and laboratory testing, the AC-LGAD can be confirmed as the baseline technology for Roman Pots, while an optimization of the sensor readout can be achieved in a five year time scale.  In a two year timeframe the readout architecture can be developed and its viability demonstrated via simulations as well as laboratory tests based on existing prototypes for the LHC, while in a five year time scale a more detailed design of the ASICs and the readout chain, including initial prototyping, can be achieved.

\subsection {Zero Degree Calorimeter }


The ZDC will serve critical roles for a number
of important physics topics at EIC, such as distinguishing between coherent 
diffractive scattering in which the nucleus remains intact, and incoherent 
scattering in which the nucleus breaks up; measuring geometry of $e + A$ 
collisions, spectator tagging in $e + d / ^3He$, asymmetries of leading 
baryons, and spectroscopy. 
These physics goals require that the ZDCs have high efficiency for neutrons 
and for low-energy photons, excellent energy, $p_T$ and position resolutions, 
large acceptance and sufficient radiation hardness. 





There are several possible approaches to achieve high energy and 
position resolution in a calorimeter. 
For example, the ALICE FoCal \cite{ALICE-PUBLIC-2019-005}, is  
a silicon-tungsten (Si+W) sampling calorimeter with longitudinal 
segmentation. 
Low granularity layers are used for the energy measurement while higher 
granularity layers provide accurate position information. 


From simulations the photon energy resolution for FoCal is estimated to 
be $\sigma_E = 27\% / \sqrt{E} \oplus 1\%$ \cite{ALICECollaboration:2020rog}. 
Other technologies that would provide suitable resolution include 
crystals (PbWO$_4$, LYSO, GSO, LSO), DSB:Ce glass, and W/SciFi. 
PbWO$_4$ crystals and DSB:Ce glass have been developed and 
characterized by the eRD1 Consortium and the Neutral Particle 
Spectrometer project at Jefferson Lab. 
Tests have shown energy resolutions of $\sim 2\%/\sqrt{E}$ for photon 
energies $\sim 4$ GeV \cite{Horn:2015yma}. 

To identify neutrons,  the ZDC needs a hadronic section with a  resolution of $\sigma_E < {50\%}/{\sqrt{E}}$ with an angular 
resolution of at least 3 mrad / $\sqrt{E}$ is desired. 
Cerenkov calorimeters, which measure only the high energy component of 
the showers, give excellent position resolution  and tight containment 
but are non-compensating and so somewhat non-linear. 
Sampling all charged particles produced gives better energy resolution 
at the cost of worse lateral containment. 
We seek to exploit both techniques to maximize both the energy and 
position resolution of the ZDC. 
This could be done by using the quartz fibers developed for the LHC 
ZDCs, \cite{JZCap}, with traditional scintillators. 


In order to detect coherent collisions it is necessary to veto events in which 
 soft photons are emmitted from an excited nucleus
For ${}^{208}$Pb, every bound-state 
decay sequence has at least one photon with an energy of at least 2.6 MeV. 
For a beam momentum of 
275 GeV/$c$, 20\% of these decay photons (with minimum energy 
455 MeV) are detectable in the ZDC aperture of $\sim 4.5$ mrad.
In order to detect such  photons from nuclear excitation 
it is important that the ZDC have the largest possible 
aperture. 
It is possible that a 2nd IR design will allow a larger ZDC acceptance.


The meson structure research for the EIC has shown the need of a tracker, in combination with the ZDC, to be used as a veto detector for $\pi^-$ for an efficient measurement of the $\Lambda \rightarrow n + \pi^0$ channel. Besides this main purpose, adding a tracker could improve the reconstruction of charged particles in the ZDC for other different channels. A non-expensive and feasible option is the use of scintillating fibers (SciFi) as a tracker detector.



The number of spectator neutrons is predicted to be somewhat 
correlated with the collision geometry.  
The required performance of the detector to identify the coherence of 
the collision is under development using the BeAGLE simulation
\cite{Beagle}. 
Some of performance parameters are under ongoing study. 
The optimization of the performance requirements is included in the 
scope of the development based on the requirements known as of now 
as listed below. 


A large acceptance (e.g. 60$\times$60 cm$^2$) to establish good 
identification efficiency between coherent and incoherent collisions is 
necessary for vetoing spectator neutrons from nuclear breakup. 
This large acceptance is also required to determine the collision 
geometry~\cite{EIC:RDHandbook}. 
For studying very forward production and asymmetries of hadrons and 
photons, a large acceptance is also important. 
The EIC aperture of $\pm$4 mrad gives $p_T < 1$ GeV/$c$ coverage for 
275 GeV hadrons and photons, which covers the transition from 
elastic/diffraction to incoherent regime; for low-energy hadron beam 
the acceptance in terms of $p_T$ is more limited e.g. 
$p_T < 0.4$ GeV/$c$ coverage for 100\,GeV beam. 


Due to the strong $\beta$ squeeze $<$ 1 meter for the high luminosity, 
a beam spread of $\sim$20 MeV and $\sim$1 cm of the hadron beam angular 
divergence is induced. 
Thus the position resolution of neutron in sub cm won't help. 
1 cm position resolution provides 300 $\mu$rad angular resolution, which 
can be translated to transverse momentum resolution $p_{\rm T} \sim 30$ 
MeV/$c$ of 100 GeV spectator neutron. 

The minimum energy resolution $\Delta E/E \sim 50\%/\sqrt{E(GeV)}$ to 
distinguish number of spectator neutrons from 20 to 30 for collision 
geometry determination. 
In order to accommodate a single MIP track to 30 spectator neutrons, 
wide dynamic energy range in the readout electronics is required. 

It is anticipated to be a sampling type calorimeter with a sufficient 
longitudinal size of $\sim$10 interaction length\cite{EIC:RDHandbook}. 
It is also required to have a sufficient transverse size of $\sim$2 
interaction length to avoid transverse leakage of the hadron shower and 
to achieve good  hadron energy resolution.

\subsection{Superconducting-Nanowire Particle Detectors}

Superconducting Nanowire Single-Photon Detectors (SNSPDs) have become the 
dominant technology in quantum optics due to their unparalleled timing resolution and quantum efficiency.
The Argonne National Laboratory group, supported by eRD28, is currently investigating
the pathway to transform these sensors into a novel particle detector for the EIC.
The sensors can operate in magnetic fields greater than $5$ T at a high rate with high efficiency,
and with a timing resolution as low as $\lesssim 20$~ps.
The R\&D effort aims to produce a small
(mm$^2$) superconducting nanowire pixel array for detecting high energy particles. 
This first of its kind detector will have the flexibility to be used in multiple far-forward detector systems. It can extend the EIC's scientific reach beyond what is possible with contemporary technology for far-forward detection.

Superconducting nanowire detectors have multiple characteristics that make them a uniquely capable detector technology for applications at the EIC. 
(a) Superconducting nanowire detectors are high-speed detectors and have time resolutions typically on the order of \SI{20}{\pico\second} scale, with a current record of \SI{3}{\pico\second}.
(b) A meandering wire layout allows for small pixel sizes and allows for \si{\micro\meter} position resolution if needed.
(c) Single pixels can operate efficiently at high-rates in strong magnetic fields (up to \SI{5}{\tesla}) \cite{Polakovic:2019wrh}. 
(d) Edgeless sensor configurations are a possibility, with the sensitive element positioned to within a few \SI{100}{\nano\meter} of the substrate edge, eliminating dead material in between the particle beam and the detector.
(e) Wide choice of substrate material -- the detectors can be fabricated on membranes as thin as few \SI{10}{\micro\meter}, further cutting down on material thickness.
(f) Radiation hardness allows for a longer service cycle of detectors operating near the beam and interaction regions.

The EIC R\&D committee identified four applications at the EIC for future R\&D~\cite{eicRDcommittee}. (1) A Roman pot detector in the forward region about 35 meters or more from the interaction point to tag low momentum transfer recoiling ions.
(2) An integrated detector inside the cold bore of superconducting magnets for the forward ion detection would provide tracking in regions of high magnetic fields. 
This would include placing the detector inside the magnet and integrating it with the magnet’s cooling system, eliminating the need for a separate cryogenic system. 
Further applications include (3) placing the detector in front of the ZDC detector and around the forward ion spectrometer section, filling in the detection gaps where radiation hard detectors with excellent position and timing resolution are needed. 
Finally, (4) use in an electron detector for a Compton Polarimeter, because the high rate
capability, allows the nanowire detectors to handle the 100 MHz beam pulse rate to measure the azimuthal asymmetries needed to extract the beam polarization.

Superconducting nanowire sensors are an entirely new technology for high energy particle detection in nuclear physics ~\cite{nanowireReviewPaper}. 
This unique opportunity comes with some R\&D needs to leverage the full potential for applications at the EIC.
Further R\&D includes optimizing the wire parameters or high energy ion detection, developing cryogenic bias and readout ASICs for high channel count tracking detectors, and design integration of superconducting nanowire sensors into the cold bore of superconducting magnets. 
The required R\&D can be completed within the next few years, depending on the specific application.

\section{Data Acquisition}
\label{part3-sec-DetTechnology.DAQ}

\subsection{Streaming-Capable Front-End Electronics, Data Aggregation, and Timing Distribution}

A streaming readout is the likely readout paradigm for the EIC, as it allows easy scaling to the requirements of EIC, enables recording more physics more efficiently, and allows better online monitoring capabilities. The EIC detectors will likely be highly segmented, leading to a large number of readout channels. At the same time, multiplicities and pile-up are likely less demanding than other experiments like sPHENIX. The physics case is very wide, and many analysis will be systematics dominated. It is therefore crucial to minimize systematic effects from the readout, for example trigger biases. Further, minimally biased data recording allows to data-mine for novel physics later in the EIC life-cycle. A streaming readout system further reduces scaling choke-points and allows us to eliminate critical failure points like online event building. See section \ref{part3-sec-Det.Aspects.DAQ_Electronics} for more details.

A working readout system is crucial for any data taking and must be ready at day-1. In fact, ideally, prototypes should be ready for detector tests well ahead of first beam. R\&D is required in multiple areas:
    
    Streaming readout requires the distribution of clock information. While crucial for successful data taking, this is a less demanding task than the distribution of triggers, and a scheme similar to the one at sPHENIX is a likely solution. This approach will be tested by sPHENIX well ahead of EIC completion, and other test beams will likely use other timing systems. Front end electronics need to be read out via some sort of data collection hardware. These will likely be evolutions of already available components like the FELIX cards, and existing hardware can be used during test beams until the final  hardware iterations are available. Both of these research topics are rather low risk.
    
    Of higher risk is the development of suitable front-end electronics. Here, possible front-end readout ASICs have to be matched to the detector requirements. While existing ASICs cover many use cases, it is not clear yet if the requirements of the final detector configurations for the EIC are covered by current capabilities. History tells us that timelines for development of completely new ASICs is 6 or more years, with substantial investment of R\&D personnel, while modifications of existing designs might be done in 3+ years, and possibly smaller teams. It is therefore paramount that cases where new readout ASICs are required are identified soon, and whether a readout is at all feasible within the given constraints. This puts this research into the high-risk and high priority category. We want to note here that this risk is not unique to a streaming readout---in fact, most high-performance ASICs today fit a streaming readout solution better than a triggered one---and is indeed a risk for the readout in general, independent of chosen paradigm. There is further a risk that evolutions of current designs face deprecation of the underlying process nodes, prompting a costly transition to a new node. However current process node timelines and predictions indicate that process nodes used in current-gen chips will be available at the timescale of EIC operation.
    
    The research intrinsically touches upon a wide range of other detector projects. It is very likely that data collection hardware is shared between most detector components. For front-end electronics, designs will be shared as much as possible.

\subsection{Readout Software Architecture, Orchestration and Online Analysis}

In addition to readout hardware, it is important to develop and test protocols and software to provide a stable, high-performing readout. This includes a scalable platform, both in channel count and processing capabilities, and the inclusion of analysis into the online system as much as possible. The system must be resilient against errors in the FEE to enable an overall highly efficient data taking. High quality, high level monitoring will secure the recording of high-quality data, reducing time-to-publication. Similar to the hardware, prototype designs should be ready well in advance to support test beam times, and to collect experience necessary to build the online analysis. 

The development of software and protocol components must go hand in hand with the hardware. As the highest priority, it is important to define a logical protocol for data exchange. This will enable groups to develop interoperable electronics and software components early in the development cycle. The community is actively working on this issue, but revisions will be likely in the years to come.
        
To achieve optimal usage of beam times, techniques must be developed to make the readout resilient against FEE errors (e.g.\ Single Event Upsets) without requiring a full stop and restart of the system.  This issue is exacerbated by the high channel count and density. In a similar fashion, it is an open research question how to best address bandwidth restrictions. Since the data rate is governed by a stochastic process, they will have peak rates substantially above the average data rate, with almost no ceiling. While large memory buffers can mitigate this by smearing out peak rates over time, the system must still be able to handle buffer overflows. For both problems, R\&D is required to develop a framework and control algorithms that react in a predictable and reconstructable way, so that overall detector/DAQ efficiencies can be extracted. Such a system must be available essentially at first beam, with improvements later in the life cycle. 
    
The amount of data collected and the evolving landscape of computing infrastructure from a local to a federated model makes it necessarily to rethink data storage and retrieval to achieve efficient usage of the computing resources. Here, a flexible software layer must be developed to isolate the analysis code from the changing infrastructure. While a first solution is required at first beam, it is likely that this will evolve together with the computing infrastructure during the EIC lifecycle. Connected to this issue is the integration of analysis into the online and near-online processing to maximize data quality. This includes the efficient handling of calibration procedures, and minimization of time delay between analysis and data taking. 
    
The latter points are, to some degree, also required R\&D for other projects like sPHENIX and CLAS-12, and an EIC solution would likely be straight-forward iterative development. On the other hand, even with sophisticated simulations and detector tests, the initial conditions at an EIC in the sense of observed background and dark rates, beam quality etc.\ are hard to predict, and will probably require some time for tuning. The initial rates might overwhelm the readout system and a system to mitigate this risk must be developed. A possible avenue is to include a hard data reduction stage early in the readout system, for example controlled by a trigger, or via software cuts at a very early stage. This capability is the equivalent of raising trigger thresholds or disabling trigger sources in a classical triggered system, and would secure the ability to record data required to understand and calibrate detectors and optimize the machine, at the cost of physics reach during this tune-up period. A possible approach based on hardware signals is essentially realized at sPHENIX, and other implementations are straight forward. Research and development has to show if a pure software-based solution can be implemented, which would allow for more flexibility in the transition to normal operation.

\section{Electronics}
\label{part3-sec-DetTechnology.Electronics}

Development of readout electronics for the EIC Detectors is informed by recent and ongoing development efforts such as LHC upgrades and Belle II.  R\&D efforts are needed on the following key topics.

\subsection{High Precision Timing Distribution Over Large System}
High precision timing distribution is important for sub-detectors like TOF and LGAD based timing detectors. This technology will be used to distribute high precision clocks to the sub-detectors, allowing precision timing of detector signals. It should also support online calibration of the clock phase drift caused by environmental changes, such as temperature drift. As an example, for the proposed TOF in sPHENIX and LGAD detectors in HL-LHC, the required resolution for measurements of individual arrival times of particles is about 25 to 50 ps, to mitigate the relatively short flight lengths, and extreme pile-up and occupancy conditions\cite{DOE:ResearchStudy2019}. Contributions to the resolution comes from both the detector and electronics. In future HEP experiments, the electronics timing requirement may reach picosecond level.  R\&D on phase stabilization within the back-end FPGA is a promising candidate solution, considering a holistic integration of the back-end electronics in the DAQ system with the front-end readout electronics of the sub-detectors.  A benchmark for such a system will be to demonstrate full path transmission of signals and data with the system clock embedded. The most critical aspect of this R\&D is to guarantee the phase stability of the low jitter clock at the sub-detector front-end.  This endeavor leverages of ongoing studies at CERN for the HL-LHC experiments \cite{CERN:HighPrecisionTiming}. Depending on the detailed requirements for EIC, and the maturity of the HL-LHC development efforts, this R\&D may be completed in about 1 year.  
 
\subsection{Usage of COTS Devices in a Radiation Environment}
Depending on the design methodology for the sub-detectors readout electronics, FPGAs may be used in the readout boards exposed to significant radiation. This R\&D will mainly focus on the application of commercial FPGAs in the front-end electronics. The purpose is to provide common FPGA-based solutions for readout electronics in a radiation environment. Similar research has been carried out for existing FPGAs, for example the Xilinx 7 series and Ultrascale series FPGAs in sPHENIX and the LHC experiments. The key outcomes of this study are two-fold. First, the selection of FPGAs (SRAM-based or Flash-based FPGA) depending on the detailed FPGA functional requirements, radiation dose, radiation type and sensitivity to different types of errors caused by expected radiation backgrounds. And second, selecting FPGA firmware design methodologies to mitigate errors such as Single Event Upset (SEU) and Single Event Functional Interrupt (SEFI).  Examples of mitigation methods are the use of TMR (Triple Modular Redundancy) for the firmware logic, data coding with error correction for high-speed serial links, and continuous, background scrubbing of the FPGA configuration memory. Besides FPGA and the memory for configuration, the commercial power regulators should also be qualified. This R\&D should be started as soon as possible. Several radiation test campaigns will be needed, and the whole process may last for 1 to 1.5 years and will inform the selection criteria, as well as, implementation mitigating metrics.

\subsection{Micro-electronics, Opto-Electronics and Powering}
Selecting the most appropriate micro-electronics includes a survey and evaluation of current CMOS technologies, including 65nm and 28nm technology nodes where appropriate.  Not all subdetectors require extreme radiation tolerance and ASICs successfully demonstrated in backgrounds appropriate to experimental needs should be considered.  Where necessary to mitigate radiation effects, the choice of technology, models, cell libraries and IP blocks development for extreme environments \cite{DOE:ResearchStudy2019, CERN:StrateicProg2018} will be considered. Due to the limited available resources within the EIC community, expertise and experience from HEP will be harnessed.  Detailed designs of specific front-end ASICs will depend on the requirements from the various sub-detectors.  In most cases there are multiple, potentially viable ASIC solutions that have already been developed.  However many were not designed for Streaming Readout mode of operation.  Development efforts in coming years will include updating these architectures to support this mode of operation.

To support streaming readout and reduce cabling infrastructure, suitable Opto-electronics are needed.  This means qualifying radiation hard optical link architectures for high speed serial links, including the optical modules and possible common ASICs for data aggregation \cite{CERN:StrateicProg2018}. While this could require a lot of effort, existing designs at 2.5 Gbps, 5 Gbps and 10 Gbps line rates at CERN for LHC and HL-LHC experiments \cite{CERN:GBTProject} should be adequate for EIC. Small revisions may be needed to match EIC machine parameters, such as different base clock frequencies and line rates for the data transmission. 
 
Again to reduce cable infrastructure, power system R\&D may be needed.  This chiefly includes research on radiation tolerant DC-DC converters, following such development activities at CERN \cite{CERN:DCDCProject}, low voltage power distribution and serial powering for tracker modules. 

It is necessary to have regular electronics meetings as early as possible, to discuss the overall electronics development, especially the common ASIC modules, firmware protocols, the FPGA selection and also the PCB procurement. 

To reduce the development and maintenance effort for the readout and data acquisition, it is critical that it be integrated with the detector technology selection, design and prototyping. The detector groups are encouraged to work closely with the readout and DAQ group in considering readout requirements (e.g. noise performance requirement), using the supported readout chips (e.g. streaming compatible chips), and perform tests with the compatible DAQ software (RCDAQ, etc.) at the earliest possible opportunity.

\cleardoublepage

%
%
\chapter*{Acknowledgements}
\addcontentsline{toc}{part}{Acknowledgements}
\markboth{Acknowledgements}{Acknowledgements}

The EIC yellow report was the result of a year long community wide effort.  The authors are indebted to the following colleagues for critical and crucial comments in the preparation of this report: Iris~Abt (MPI-Munich, Germany), Ani~Aprahamyan (AANL, Armenia), Barbara~Badelek (Univ. Warsaw, Poland), Marie~Boer (Virginia Tech., USA), Helen~Caines (Yale Univ., USA), Maria~Chamizo-Llatas (BNL, USA), Janusz~Chwastowski (IFJ PAN, Poland), Dmitry~Denisov (BNL, USA), Markus~Diehl (DESY, Germany), Haiyan~Gao (Duke Univ., USA), Michel~Gar\c con (IRFU CEA-Saclay, France), Donald~Geesaman (ANL, USA), Ed~Kinney (Univ. Colorado-Boulder, USA), Robert~Klanner (Univ. Hamburg, Germany), Christina~Markert (Univ. Texas at Austin, USA), Bob~McKeown (JLab, USA), Hugh~Montgomery (JLab, USA), Piet~Mulders (Vrije Univ. and NIKHEF, Amsterdam, the Netherlands), Eugenio~Nappi (INFN-Bari, Italy), Peter~Petreczky  (BNL, USA), Krzysztof~Piotrzkowski (Cath. Univ. Louvain, Belgium), Oscar~Rondon-Aramayo (Univ. Virginia, USA), James~Symons (LBNL, USA), Cristina~Tuv\`{e} (Univ. and INFN - Catania, Italy), Julia~Velkovskaja (Vanderbilt Univ., USA), and Glenn~Young (BNL, USA). 

We thank the following colleagues for their essential contributions to the simulation studies:
Christian Bierlich (Lund University, Sweden),
Ilkka Helenius (University of Jyväskylä, Finland),
Stefan Hoeche (Fermilab, USA), 
Hannes Jung (DESY, Germany),
Leif Lönnblad  (Lund University, Sweden),
Simon Plätzer (University of Graz / University of Vienna, Austria),
Stefan Prestel (Lund University, Sweden) and
Tobias Toll (Indian Institute of Technology, Delhi, India).

The EICUG Steering Committee is also indebted to Rik Yoshida for conceptualizing
the Yellow Report Initiative in the early stages.

\vspace{0.5cm}
We acknowledge support from the following institutions/agencies:
{\flushleft
\begin{enumerate}

\item DOE, Office of Science, Office of Nuclear Physics (USA)
\begin{description}
\item[Contracts:] DE-AC05-06OR23177, DE-AC02-05CH11231, DE-AC02-06CH11357, DE-AC52-06NA25396, DE-FG02-87ER40365, DE-FG02-88ER40410, DE-FG02-93ER40771, DE-FG02-94ER40818, DE-FG02-03ER41260, DE-FG02-04ER41325, DE-FG02-07ER41460, DE-FG02-09ER41620, DE-SC0004286, DE-SC0008791, DE-SC0010129, DE-SC0020240, DE-SC0020265, DE-SC0012704, DE-SC0016583, DE-SC0020405, DE-SC0011090, DE-SC0013391, DE-SC0018224, DE-SC0019230, DE-SC001012, DE-SC0013405
\end{description}

\item European Union
\begin{description}
\item[Projects:] CA 1521
\item[ERC:] ERC-2018-ADG-835105, ERC-2015-CoG-681707 
\item[Horizon 2020 programme:]  \hfil
\begin{description}
\item[Contracts:] STRONG-2020 n. 824093, AIDA-2020 n. 654168
\item[MSCA:] "RISE" n. 823947, "ParDHonSFF" n. 752748, "SQuHadron" n. 795475, "FELLINI" n. 754496
\end{description}
\end{description}

\item National Science Foundation (USA)
\begin{description}
\item[Awards:] n. 2012826, 2000108, 1812423 
\item[Grants:] n. PHY-1915093, PHY-1714133, PHY-2012430, PHY-2012002, PHY-1945471, DGE-1650604
\end{description}

\item Brookhaven National Laboratory (USA)
\begin{description}
\item[Programs:] LDRD n. 18-037
\end{description}

\item the Netherlands Organization for Scientific Research (NWO - the Netherlands)

\item German Research Foundation (DFG - Germany) 
\begin{description}
\item[Grants:]  KL 1266/9-1, 396021762-TRR257, FOR 2926
\end{description}

\item National Science Center (NCN - Poland) 
\begin{description}
\item[Grants:] n. 2017/26/M/ST2/01074, 2017/27/B/ST2/02755, 2019/33/B/ST2/02588, 2019/34/E/ST2/00186, 2019/35/D/ST2/00272
\end{description}

\item National Council of Science and Technology (CONICET - Argentina)

\item Academy of Finland 
\begin{description}
\item[Projects:] n. 308301, 314764, 321840
\end{description}

\item Italian Ministry of Education, University and Research (MIUR - Italy) 
\begin{description}
\item["FARE" grants:] "3DGLUE" n. R16XKPHL3N
\end{description}

\item Spanish Ministry of Science and Innovation (MICINN - Spain) 
\begin{description}
\item[Grants:] n. FPA2017-83814-P, MDM-2016-0692, PID2019-107844GB-C22, PID2019-106080GB-C21
\end{description}

\item Xunta de Galicia (Spain)
\begin{description}
\item[Projects:] ED431C 2017/07, ED431G 2019/05
\end{description}

\item Junta de Andaluc\'{i}a (Spain)
\begin{description}
\item[Contracts:] n. P18-FRJ-1132
\item[Programs:] "Operativo FEDER 2014-2020" n. UHU-1264517
\end{description}

\item Research Talent Attraction Program (Comunidad Aut\'{o}noma de Madrid - Spain)
\begin{description}
\item[Contracts:] n. 2018-T1/TIC-10313
\end{description}

\item Natural Sciences and Engineering Research Council of Canada (NSERC - Canada) 
\begin{description}
\item[Contracts:] n. SAPIN-2016-00031
\end{description}

\item Los Alamos National Laboratory  (USA)
\begin{description}
\item[Programs:] LDRD n. 20200022DR
\end{description}

\item Science and Technology Facilities Council (STFC - United Kingdom) 
\begin{description}
\item[Grants:] n. ST/T000600/1
\end{description}

\item National Agency of Research and Development (ANID - Chile) 
\begin{description}
\item[Grants:] FONDECyT n. 1191103, 1180232, PIA/APOYO AFB 180002
\end{description}

\item National Autonomous University of Mexico (UNAM - Mexico) 
\begin{description}
\item[Grants:] DGAPA-PAPIIT n. IA 1017120, IN 106921
\end{description}

\item National Council of Science and Technology (CONACyT - Mexico)
\begin{description}
\item[Grants:] "Ciencia de Frontera 2019" n. 51244 (FORDECYT-PRONACES), n. A1-S-21389
\end{description}

\item Czech Ministry of Education, Youth and Sport (MEYS - Czech Republic)
\begin{description}
\item[Grants:] n. LM2015054, LM2018109
\end{description}

\item Chinese University of Hong Kong (CUHK Shenzhen - China) 
\begin{description}
\item[Grants:] n. UDF01001859
\end{description}

\item Guangdong Major Project of Basic and Applied Basic Research (Guangdong - China) 
\begin{description}
\item[Projects:] n. 2020B0301030008
\end{description}

\item National Natural Science Foundation of China (NSFC - China) 
\begin{description}
\item[Grants:] n. 11875112, 11775023, 12022512
\end{description}

\item Hundred Talents Plan for Professionals (Jiangsu Province - China)

\item Simons Foundation (New York City - USA)
\begin{description}
\item[Grants:] "Investigator" n. 327942
\end{description}

\item National Council of Scientific and Technological Involvement (CNPq - Brazil) 
\begin{description}
\item[Grants:] n. 308486/2015-3
\end{description} 

\item The S\~{a}o Paulo Research Foundation (FAPESP - Brazil) 
\begin{description}
\item[Grants:] n. 17/05660-0
\end{description} 

\item National Institute of Science and Technology - Nuclear Physics and Applications (INCT-FNA - Brazil) 
\begin{description}
\item[Grants:] n. 464898/2014-5
\end{description}

\item Argonne National Laboratory (USA)
\begin{description}
\item[Programs:] LDRD
\end{description}

\item Lawrence Berkeley National Laboratory (USA)
\begin{description}
\item[Programs:] LDRD
\end{description}

\item University of California - Office of the President

\item Nuclear Science and Security Consortium (California - USA)

\item Slovenian Research Agency (ARRS - Slovenia)

\end{enumerate}
\vspace{1cm}

A.B, M.D., and T. W. thank EIC-India for their help with benchmarks and validation of the simulation software,
V.V.B. thanks Tanja Horn for the support, 
T.J.H. acknowledges partial support from the JLab EIC Center, and 
S.H.L. thanks Christine Aidala for the support and critical revision of his work.
}

\cleardoublepage

%
%
\begin{appendices}
\phantomsection
\addcontentsline{toc}{part}{\large{Appendices}}
%
%
%
%

\newcommand{\xpom}{x_\mathbb{P}}
\newcommand{\pt}[1]{{\mathbf{#1}}_\perp}

\chapter{Deep Inelastic Scattering Kinematics}
\label{appendix:kinematics}
\section{Structure functions}

In general, the inclusive DIS process can be written as
\begin{equation}
\label{eq:dis_process}
    e(l) + N(p) \to e(l') + X(p_X),
\end{equation}
where $e$ refers to the electron or positron, $N$ is the nucleon in the initial state with momentum $p$, and a system $X$ (which is not measured) is produced with momentum $p_X$. In case of an unpolarized nucleon, the cross-section for this process can be written in terms of the structure functions $F_2$ and $F_L$ in the one photon exchange approximation neglecting electroweak effects as
\begin{equation}
    \frac{\mathrm{d} \sigma}{\mathrm{d} x \mathrm{d} Q^2} = \frac{4\pi \alpha^2}{xQ^4} \left[ \left(1 - y + \frac{y^2}{2}\right) F_2(x,Q^2) - \frac{y^2}{2} F_L(x,Q^2) \right].
\end{equation}
Instead of structure functions, the reduced cross-section $\sigma_r$ is often used
\begin{equation}
    \sigma_r = \frac{\mathrm{d}^2 \sigma}{\mathrm{d}x \mathrm{d}Q^2} \frac{x Q^4}{2\pi \alpha^2 [1+(1-y)^2]} = F_2(x,Q^2) - \frac{y^2}{1+(1-y)^2} F_L(x,Q^2).
\end{equation}

With longitudinally polarized electron and nucleon beams, it is also possible to extract the structure function $g_1$
\begin{equation}
    \frac{1}{2} \left[ \frac{\mathrm{d} \sigma^\rightleftarrows}{\mathrm{d}x \mathrm{d}Q^2} - \frac{\mathrm{d}\sigma^\rightrightarrows}{\mathrm{d}x \mathrm{d}Q^2}  \right] = \frac{4\pi \alpha^2}{Q^4} y(2-y) g_1(x,Q^2).
\end{equation}
Here terms suppressed by $x^2 m_N^2/Q^2$ have been neglected, and $\sigma^\rightleftarrows$ refers to the case where the nucleon and electron spins are opposite (and parallel to the $z$ axis), and $\sigma^\rightrightarrows$ to the scattering process in case of aligned spins. The kinematical variables $x$,$y$ and $Q^2$ are introduced below, and $m_N$ is the nucleon mass and $\alpha$ is the fine structure constant. At large $Q^2$ and to leading order in the strong coupling constant $\alpha_s$ the $F_2$ structure function is proportional to the unpolarized quark and antiquark distributions in the nucleon, and $g_1$ is sensitive to the longitudinally polarized distributions. In this limit $F_L=0$, and it obtains a first contribution at next to leading order in perturbative expansion, and is thus particularly sensitive to the gluon distribution.

\begin{figure}[th]
\subfloat[Inclusive DIS]{%
\includegraphics[width=0.48\textwidth]{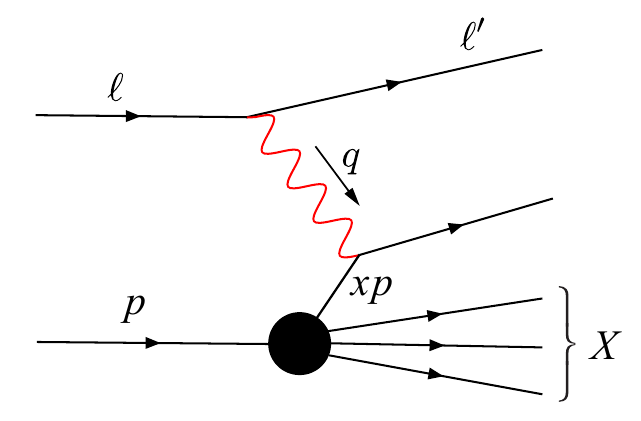}
\label{fig:inclusive}
}
\subfloat[Diffractive scattering ]{%
\includegraphics[width=0.48\textwidth]{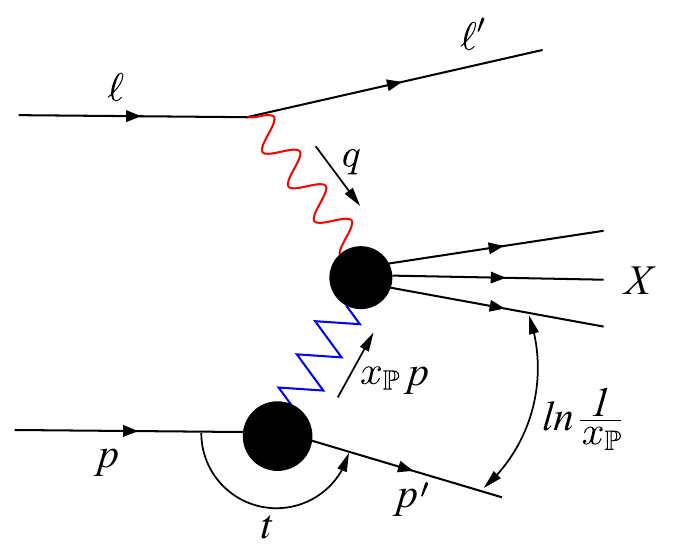}
\label{fig:exclusive}
}
\caption{Kinematical variables of inclusive and exclusive DIS. The blobs correspond to interactions. }
\label{fig:kinematics}
\end{figure}

In diffractive (and also semi-inclusive) scattering, the process becomes
\begin{equation}
\label{eq:ddis_process}
    e(l) + N(p) \to e(l') + N'(p') + X(p_X) ,
\end{equation}
where $N'$ refers to the nucleon or the nucleon remnants in the final state with momentum $p'$ and a specific system  $X$ is produced. The electron mass is neglected in the following discussion, and the nucleon mass $p^2=m_N^2$ is kept non-zero unless otherwise stated. In this appendix, $p$ is a four vector and $\mathbf{p}$ and $\pt{p}$ refer to the three-momentum and the transverse momentum, respectively. The momentum vectors are illustrated in Fig.~\ref{fig:kinematics}.

\section{Invariants}

Let us first consider inclusive scattering where the final state $X$ is not completely determined and the scattered nucleon (nucleon remnants) are not reconstructed. The center-of-mass energy squared for the DIS process can be written using the momenta defined in Eq.~\eqref{eq:dis_process} as
\begin{equation}
    s =(l+p)^2 = m_N^2 + 2 p \cdot l \approx 2\sqrt{E_e E_n}.
\end{equation}
Here $E_e$ is the electron energy and $E_n$ the nucleon energy, and the approximation is valid in the high energy limit where the nucleon mass can be neglected.

As the scattering process is mediated by a virtual photon, the center-of-mass energy $W$ for the photon-nucleon system is generically more useful:
\begin{equation}
    W^2 = (p+q)^2 = m_N^2 - Q^2 + 2p\cdot q.
\end{equation}
Here the virtual photon momentum is $q=l-l'$ and its virtuality $-Q^2=(l-l')^2$. The other useful Lorentz invariant quantities describing the DIS process are

\begin{align}
	x &\equiv \frac{Q^2}{2p \cdot q} = \frac{Q^2}{2m_N\nu} = \frac{Q^2}{Q^2 + W^2 -m_N^2} \\
	y &\equiv \frac{p \cdot q}{p \cdot \ell} = \frac{W^2 + Q^2-m_N^2}{s-m_N^2} 
\end{align}

These invariants have intuitive physical interpretations in particular frames. The Bjorken variable $x$ can be interpreted in the parton model in the infinite momentum frame where the nucleon carries a large longitudinal momentum. In such a frame, $x$ is the fraction of the nucleon momentum carried by the struck parton if the quark masses are neglected. In electron-nucleon collisions, $0<x<1$. 

The variable $y$ is called \emph{inelasticity}. When expressed in the nucleon rest frame, one finds $y=1-\frac{E_l'}{E_l}$, where $E_l$ and $E_l'$ are the energies of the incoming and outgoing leptons in this frame, respectively. Consequently, $0\le y \le 1$, and in particular, the highest possible photon-nucleon center-of-mass energies are reached at the $y\to 1$ limit. A closely related variable $\nu$ also exists: $\nu \equiv \frac{p\cdot q}{m_N}$ describes, in the nucleon rest frame, the electron energy carried away by the virtual photon: $\nu = E_l - E_{l'}$.

The invariants presented above are not independent, and in inclusive scattering the collision kinematics is completely determined by three variables, e.g. $s,Q^2$ and $x$. This becomes apparent when noticing that the invariants defined above satisfy e.g. the following relations:
\begin{align}
    Q^2 &= xy(s-m_N^2),\ \ \text{and} \label{eq:Q2_xy}\\
    W^2 &= \frac{1-x}{x}Q^2 + m_N^2. \label{eq:W2_Q2} 
\end{align}
The smallest kinematically allowed virtuality $Q^2_\text{min}$ can be determined if the electron mass $m_e$ is non-zero: $Q^2_\text{min}=m_e^2\frac{y^2}{1-y}$.


Let us then discuss diffractive production of a system $X$ with an invariant mass $M_X^2$. In the unpolarized case where the cross-section is symmetric in azimuthal angle, we can describe the kinematics by introducing the following new invariants:
\begin{align}
t &\equiv -(p'-p)^2 \\
\xpom &\equiv \frac{(p-p')\cdot q}{p \cdot q} = \frac{M_X^2+Q^2-t}{W^2+Q^2-m_N^2} \label{eq:xpom} \\
\beta &\equiv \frac{Q^2}{2q\cdot(p-p')} = \frac{Q^2}{M_X^2+Q^2-t} \label{eq:beta}
\end{align}
In the infinite momentum frame, $\xpom$ has the interpretation that in the scattering process an exchange of vacuum quantum numbers (a \emph{pomeron} exchange) takes place, and the pomeron carries a fraction of $\xpom$ of the nucleon longitudinal momentum. Similarly, in the partonic language $\beta$ is the longitudinal momentum of the struck parton inside the pomeron. These invariants are not independent, and can be related to the invariants of inclusive DIS discussed above via e.g.
\begin{equation}
    x = \beta \xpom.
\end{equation}
An experimental signature of a diffractive event is the presence of a rapidity gap between the outgoing nucleon (nucleon remnants) and the system $X$. This gap size is $\Delta y \sim \ln 1/\xpom$.

\section{Laboratory frame}
In the laboratory frame the collisions are asymmetric, and the inclusive DIS invariants can be determined by measuring the energy and the scattering angle of the outgoing electron. In the limit of small nucleon mass, the invariants read
\begin{align}
    s &=  4 E_e E_n \\ 
    Q^2 &= 2 E_e E_e' (1-\cos \theta_e) \\
    W^2 &=  4 E_e E_n - 2 E_e'\left[E_n + E_e + (E_n-E_e)\cos \theta_e\right] \\ 
     x &= \frac{E_e E_e' (1-\cos \theta_e)}{2E_e E_n - E_e' E_n(1 + \cos \theta_e)} \\ 
     y &=  \frac{2E_e E_n - E_e' E_n(1+\cos \theta_e)}{2E_e E_n}. 
\end{align}
Here $E_e$ and $E_e'$ are the incoming and outgoing electron energies, and the electron scattering angle is $\theta_e$, with $\theta_e=0$ corresponding to the forward scattering, or photoproduction region $Q^2\approx 0$. Similarly the incoming nucleon energy is $E_n$.   

In exclusive processes it is possible to also measure the momentum of the produced particle and its invariant mass by measuring the decay products. Although the kinematical variables can be reconstructed using the scattered electron only, a common method  to determine $y$ and $Q^2$ is to express these invariants in terms of the scattering angles of both the electron and the produced particle using the double angle method~\cite{Bentvelsen:1992fu}:
\begin{align}
    Q^2 &=  4E_e^2 \frac{\sin \theta_e ( 1-\cos \theta_V)}{\sin \theta_V + \sin \theta_e - \sin(\theta_e + \theta_V)}\\
    y &= \frac{\sin \theta_e(1-\cos \theta_V)}{\sin \theta_V + \sin \theta_e - \sin(\theta_e + \theta_V)}.
\end{align}
Here $\theta_V$ is the scattering angle of the produced particle. These expressions are again valid in the limit where the nucleon mass can be neglected, and other similar methods can be found from Ref.~\cite{Bentvelsen:1992fu}. Note that once $Q^2$ and $y$ are determined, $x$ and $W^2$ can be obtained using Eqs.~\eqref{eq:Q2_xy} and \eqref{eq:W2_Q2}.

The squared momentum transfer $t$ can be written as
\begin{equation}
    t = -\frac{(\pt{p_X} - \pt{l'})^2 + \xpom^2 m_N^2}{1-\xpom} \approx -(\pt{p_X} - \pt{l'})^2.
\end{equation}
Here $\pt{p_X}$ is the transverse momentum of the produced particle and $\pt{l'}$ the transverse momentum of the scattered electron, and the approximation is valid at high energies where $\xpom$ is small and the momentum transfer is approximatively transverse. Note that the kinematical lower bound for $t$ reads
\begin{equation}
   -t > -t_\text{min} = \frac{\xpom^2 m_N^2}{1-\xpom}.
\end{equation}
When $t,Q^2$ and $W^2$ are determined, $\xpom$ can be obtained by using Eq.~\eqref{eq:xpom}.

In exclusive and semi-inclusive processes the particle $X$ is identified by measuring the invariant mass of the decay products.
In inclusive diffraction the invariant mass $M_X^2$ is determined by measuring the total energy $E_X$ and the total momentum $\mathbf{p_X}$ of the produced particles: 
\begin{equation}
    M_X^2= E_X^2 - \mathbf{p_X}^2.
\end{equation}
In these events, it is also possible to construct inelasticity using the hadron method
\begin{equation}
    y_h = \frac{E_X - \mathbf{p_X}_z}{2E_e}.
\end{equation}
The hadron method can also be used to determine inelasticity in exclusive particle production in the photoproduction limit where the scattered electron cannot be detected. For a better experimental accuracy, different methods to construct e.g. inelasticity can be combined (see e.g.~\cite{Adloff:1997sc}). Generically in inclusive diffraction $M_X^2 + Q^2 \gg |t|$, and consequently $t$ can be neglected when determining $\xpom$ and $\beta$ using Eqs.~\eqref{eq:xpom} and~$\eqref{eq:beta}$.

\begin{figure}[ht]
    \centering
    \includegraphics{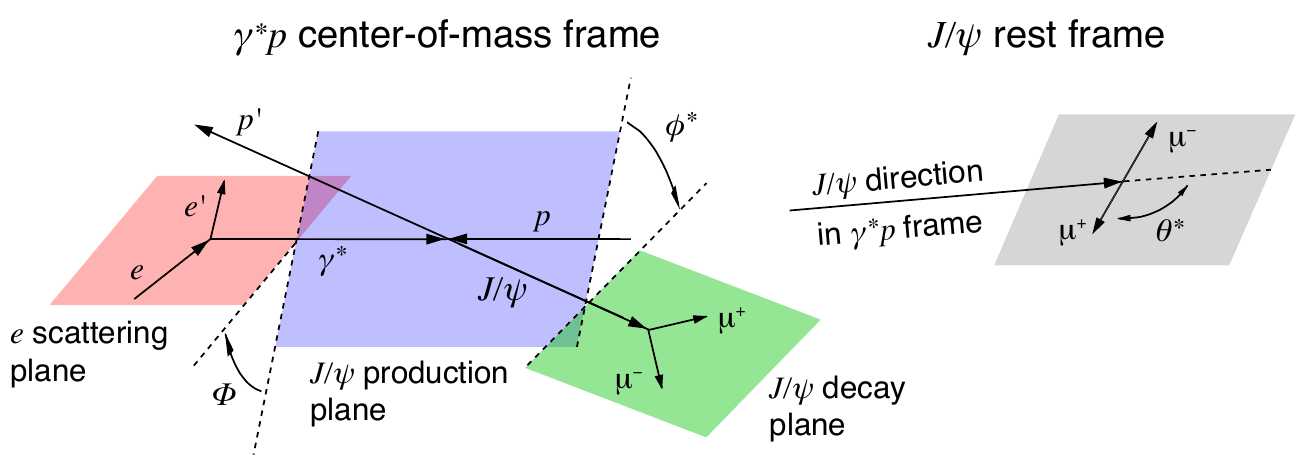}
    \caption{Planes in exclusive vector meson production.}
    \label{fig:planes}
\end{figure}

\section{Breit frame}
A natural frame to describe hard scattering process in DIS is the Breit (or brick wall) frame, where the incoming photon carries no energy, and the parton to which the photon couples to behaves as if it bounced off a brick wall. 
Let us choose that the ultrarelativistic nucleon moves along the positive $z$ axis, and the photon propagates to the $-z$ direction. The nucleon momentum in this frame is $p_z=\frac{1}{2x}Q$, and the parton longitudinal momentum $k_z$ can be written as  $k_z = x p_z = \frac{1}{2}Q$. Similarly, the photon four-momentum reads $q=(0,0,0,-Q)$. Now, after the photon absorption $\mathbf{k'}=-\mathbf{k}$, where $\mathbf{k'}$ is the parton momentum after the scattering. Note that in this frame there is no energy transfer to the proton.

The Breit frame is not the center-of-mass frame for the parton-photon scattering. This is advantageous when separating the produced particles from the beam remnants. In the Breit frame, the produced particles populate the region of negative $z$ momentum, while the beam remnants generically have a positive momentum $z$ component.

\section{Helicity studies}
Studying the helicity structure of exclusive particle production processes requires one to measure the azimuthal angles $\phi^*$ and $\Phi$ defined in Fig.~\ref{fig:planes}. Note that the angles are defined in the frame where the photon and the nucleon momenta are aligned along the same axis (here $z$ axis), so this discussion is valid both in  the Breit frame and in the $\gamma$-nucleon center-of-mass frame. 

The production plane is defined as the plane spanned by the $z$ axis and the momentum of the produced particle. The azimuthal angle between this plane, and the electron scattering plane spanned by the momenta of the incoming and outgoing electron momentum vectors is denoted by $\Phi$ in Fig.~\ref{fig:planes}, where the geometry is illustrated in case of $e^++p \to e^++p+\mathrm{J}/\psi$ scattering. Similarly, we define the decay plane, which is spanned by the momenta of the decay products of the produced particle, and the azimuthal angle between this plane and the production plane is denoted by $\phi^*$. 

The third angle required to specify the geometry $\theta^*$ also shown in Fig.~\ref{fig:planes} is required to determine the polarization state of the produced particle. This angle is defined as the polar angle of the decay particle having the same charge as the incoming lepton in the rest frame of the decaying particle. The $\theta^*=0$ case corresponds to the direction of the produced particle in the photon-nucleon center-of-mass frame.

\cleardoublepage
\chapter{Organizational Structure}

%
%

\centerline{\large\bf Physics working group and sub-working group conveners}
\vspace*{0.5cm}

\begin{itemize}
\item Physics Working Group conveners: Adrian Dumitru (The City University of New York, USA), Olga Evdokimov (University of Illinois Chicago, USA), Andreas Metz (Temple University, USA), and Carlos Mu\~noz Camacho (CNRS Universit\'e Paris-Saclay, France)
\end{itemize}

\vspace*{0.1cm}
\begin{itemize}
\item Inclusive Reactions:  

Conveners: Renee Fatemi (University of Kentucky, USA), Nobuo Sato (JLab, USA), Barak Schmookler (Stony Brook University, USA)
\item Semi-inclusive Reactions:  

Conveners: Ralf Seidl (RIKEN, Japan), Justin Stevens (The College of William\&Mary, USA), Alexey Vladimirov (University of Regensburg, Germany), Anselm Vossen (Duke University, USA), Bowen Xiao (The China University of Hong Kong, China)
\item Jets, Heavy Quarks:  

Conveners: Leticia Mendez (ORNL, USA), Brian Page (BNL, USA), Frank Petriello (ANL \& Northwestern University, USA), Ernst Sichtermann (LBNL, USA), Ivan Vitev (LANL, USA)
\item Exclusive Reactions: 

Conveners: Rapha\"el Dupr\'e (CNRS Universit\'e Paris-Saclay, France), Salvatore Fazio (BNL, USA), Tuomas Lappi (University of Jyv\"askyl\"a, Finland), Barbara Pasquini (University of Pavia, Italy), Daria Sokhan (University of Glasgow, Scotland-UK)
\item Diffractive Reactions \& Tagging:  

Conveners: Wim Cosyn (Florida International University, USA), Or Hen (MIT, USA), Douglas Higinbotham (JLab, USA), Spencer Klein (LBNL, USA), Anna Stasto (Penn State University, USA)
\end{itemize}

\vspace*{0.5cm}

\centerline{\large\bf Detector working group and sub-working group conveners}
\vspace*{0.5cm}

\begin{itemize}
\item Detector Working Group conveners: Ken Barish (UC Riverside, USA), Silvia Dalla Torre (INFN - Trieste, Italy), Tanja Horn (The Catholic University of America, USA), Peter Jones (University of Birmingham, UK), and Markus Diefenthaler, ex-officio (JLab, USA)
\end{itemize}

\vspace*{0.1cm}
\begin{itemize}
\item Tracking (+vertexing)

Conveners: Domenico Elia (INFN - Bari, Italy), Kondo Gnanvo (University of Virginia, USA),  Leo Greiner (LBNL, USA)

\item Particle ID

Conveners: Tom Hemmick (Stony Brook University, USA),  Patrizia Rossi (JLab, USA)

\item Calorimetry (EM and Hadronic)

Conveners: Vladimir Berdnikov  (The Catholic University of America, USA), Eugene Chudakov (JLab, USA)

\item Far-Forward Detectors

Conveners: Alexander Jentsch (BNL, USA),  Michael Murray (University of Kansas, USA)

\item DAQ/Electronics

Conveners: Andrea Celentano (INFN - Genova, Italy), Damien Neyret (CEA Saclay, France)

\item Polarimetry/Ancillary Detectors

Conveners: Elke Aschenauer (BNL, USA), Dave Gaskell (JLab, USA)

\item Central Detector/Integration \& Magnet

Conveners: Willliam Brooks (University of Valparaiso, Chile), Alexander Kiselev (BNL, USA)

\item Forward Detector/IR Integration

Convener: Yulia Furletova (JLab, USA)

\item Detector Complementarity

Conveners: Elke Aschenauer (BNL, USA), Paul Newman (University of Birmingham, UK)
\end{itemize}

\vspace*{0.5cm}
\centerline{\large\bf Software working group conveners}
\vspace*{0.5cm}

\begin{itemize}
\item Andrea Bressan (INFN - Trieste, Italy), Markus Diefenthaler (JLab, USA),  Torre Wenaus (BNL, USA) 
\end{itemize}

\vspace*{0.5cm}
\centerline{\large\bf EIC User Group Steering Committee}
\vspace*{0.5cm}

\begin{itemize}

\item Chair: Bernd Surrow (Temple University, USA), Vice-Chair: Richard Milner (MIT, USA)

\item At Large Members: John Arrington (ANL, USA), Marco Radici (INFN - Pavia, Italy), Barbara Jacak (LBNL and UC Berkeley, USA)

\item Lab Representatives: Thomas Ullrich (BNL, USA), Rolf Ent (JLab, USA)

\item European Representative: Daniel Boer (University Groningen, NL)

\item International Representative: Wouter Deconinck (University of Manitoba, CA)

\item Institutional Board Chair (ex-officio): Christine Aidala (University of Michigan, USA)

\end{itemize}


\cleardoublepage

\chapter{Yellow Report Workshops}
\label{appendix-workshops}

\newlength{\appLongTabWidth}
\newlength{\appTimeTabWidth}
\newlength{\appSpaceTabWidth}
\setlength{\appLongTabWidth}{2.25cm}
\setlength{\appTimeTabWidth}{1.25cm}
\setlength{\appSpaceTabWidth}{1.0cm} 

The Yellow Report initiative entailed a series of four workshops in 2020. They were preceded by a 
kick-off meeting held at MIT on December 12-13, 2019 where most of the organizational structure and strategies of the effort were put in place. The four workshops in 2020 were originally planned to take place at Temple University in Philadelphia (March), University of Pavia in Italy (May), Catholic University of America in Washington D.C.~(September), and LBL in Berkeley (November). Due to the COVID-19 pandemic evolving in early 2020, all of the four workshops were held remote-only, although the names of the original locations were kept as identifiers. In the following we list the agendas of the four workshops for reference; all times are given in EST/EDT. The given presentations and the recording of all sessions are available on the EIC User Group web site \texttt{eicug.org}. 

\section*{1st EIC Yellow Report Workshop at Temple University}
Local organizer (Temple University): Amilkar Quintero, Andreas Metz, Bernd Surrow, Matt Posik 

\subsection*{THURSDAY, 19 MARCH} 
\begin{tabbing}
\hspace*{\appLongTabWidth} \= \kill
\phantom{0}8:30 - 10:40 \> EIC Project Status - Part 1 
\end{tabbing}

\begin{tabbing}
\hspace*{\appSpaceTabWidth} \= \hspace*{\appTimeTabWidth} \= \kill
{} \> \phantom{0}8:30 \> Welcome (10m)\\
{} \> {} \> \footnotesize Speakers: Andreas Metz (Temple), Bernd Surrow (Temple) \\

{} \> \phantom{0}8:40 \> Statement by ALD's BNL and JLab (10m)\\
{} \> {} \> \footnotesize Speakers: Berndt Mueller (BNL), Bob McKeown (JLab) \\

{} \> \phantom{0}8:50 \> EIC project overview (30m)\\
{} \> {} \> \footnotesize Speaker: James Yeck (Wisconsin-Madison and BNL) \\

{} \> \phantom{0}9:20 \> EIC project overview - Discussion / Questions (25m)\\

{} \> \phantom{0}9:45 \> Machine design status and plans (30m)\\
{} \> {} \> \footnotesize Speaker: Ferdinand Willeke (BNL) \\

{} \> 10:15 \> Machine design status and plans - Discussion / Questions (25m)
\end{tabbing}

\begin{tabbing}
\hspace*{\appLongTabWidth} \= \kill
11:00 - 13:00 \> EIC Project Status - Part 2
\end{tabbing}

\begin{tabbing}
\hspace*{\appSpaceTabWidth} \= \hspace*{\appTimeTabWidth} \= \kill
{} \> 11:00 \> IR design status and plans (30m)\\
{} \> {} \> \footnotesize Speaker: Holger Witte (BNL) \\

{} \> 11:30 \> IR design status and plans - Discussion / Questions (15m)\\

{} \> 11:45 \> YR goals and plans (15m)\\
{} \> {} \> \footnotesize Speakers: Rolf Ent (JLab), Thomas Ullrich (BNL) \\

{} \> 12:00 \> YR goals and plans - Discussion / Questions (15m)\\

{} \> 12:15 \> Software WG Overview (30m)\\
{} \> {} \> \footnotesize Speaker: Markus Diefenthaler (JLab) \\

{} \> 12:45 \> Software WG Overview - Discussion / Questions (15m)
\end{tabbing}

\begin{tabbing}
\hspace*{\appLongTabWidth} \= \kill
14:00 - 16:00 \> Detector WG Calorimetry \\
{} \> \footnotesize Conveners: Eugene Chudakov (JLab), Vladimir Berdnikov (CUA)
\end{tabbing}

\begin{tabbing}
\hspace*{\appSpaceTabWidth} \= \hspace*{\appTimeTabWidth} \= \kill
{} \> 14:00 \> Initial considerations for EIC detector EMCal (35m)\\
{} \> {} \> \footnotesize Speaker: Alexander Bazilevsky (BNL) \\

{} \> 14:35 \> Electromagnetic calorimetry technologies for EIC (30m)\\
{} \> {} \> \footnotesize Speaker: Tanja Horn (CUA) \\

{} \> 15:05 \> Jet detection requirements for the EIC calorimeters (25m)\\
{} \> {} \> \footnotesize Speaker: Brian Page (BNL) \\

{} \> 15:30 \> Hadronic calorimetry technologies for EIC (30m)\\
{} \> {} \> \footnotesize Speaker: Oleg Tsai (UCLA)

\end{tabbing}

\begin{tabbing}
\hspace*{\appLongTabWidth} \= \kill
14:00 - 16:00 \> Detector WG Far Forward Detector +\\
{} \> Ancillary detectors/Polarimetry/Luminosity \\
{} \> \footnotesize  Conveners: Alexander Jentsch (BNL), Dave Gaskell, E.~C.~Aschenauer (BNL), \\
{} \> \footnotesize Michael Murray (Kansas), Yulia Furletova (JLab)
\end{tabbing}
\begin{tabbing}
\hspace*{\appSpaceTabWidth} \= \hspace*{\appTimeTabWidth} \= \kill
{} \> 14:00 \> Silicon Sensors for Forward Tracking (30m)\\
{} \> {} \> \footnotesize Speaker: Xuan Li (LANL)\\

{} \> 14:30 \> Sensors for Roman Pots - eRD24 (30m)\\
{} \> {} \> \footnotesize Speaker: Alessandro Tricoli (BNL) \\

{} \> 15:00 \> Zero Degree Calorimetry (30m)\\
{} \> {} \> \footnotesize Speaker: Yuji Goto (RIKEN) \\

{} \> 15:30 \> Open Discussion (30m)\\
{} \> {} \> \footnotesize Speaker: Alexander Jentsch (BNL)
\end{tabbing}

\begin{tabbing}
\hspace*{\appLongTabWidth} \= \kill
14:00 - 16:00 \> Detector WG PID: Introduction \\
{} \> \footnotesize  Conveners: Patrizia Rossi (JLab), Thomas Hemmick (Stony Brook)
\end{tabbing}
\begin{tabbing}
\hspace*{\appSpaceTabWidth} \= \hspace*{\appTimeTabWidth} \= \kill
{} \> 14:00 \> Introduction (10m)\\
{} \> {} \> \footnotesize Speakers: Thomas Hemmick (Stony Brook), Patrizia Rossi (JLab) \\

{} \> 14:10 \> High-Resolution ps TOF for PID at the EIC (20m)\\
{} \> {} \> \footnotesize Speaker: Mickey Chiu (BNL) \\

{} \> 14:30 \> Development of MCP-PMT/LAPPD and exploring their application \\
{} \> {} \> for particle identification (15m)\\
{} \> {} \> \footnotesize Speaker: Junqi Xie (ANL) \\

{} \> 14:45 \> A LGAD-based Time-of-Flight System for EIC - leveraging \\
{} \> {} \> experience from the HL-LHC upgrade (20m)\\
{} \> {} \> \footnotesize Speaker: Wei Li (Rice) \\

{} \> 15:05 \> Evaluation of small photo-sensors in high magnetic fields for EIC PID (15m)\\
{} \> {} \> \footnotesize Speaker: Yordanka Ilieva (South Carolina) \\

{} \> 15:20 \> Dual Ring Imaging Cherenkov status (20m)\\
{} \> {} \> \footnotesize Speakers: Evaristo Cisbani (Rome), Marco Contalbrigo (Ferrara) \\

{} \> 15:40 \> The high-performance DIRC detector for the EIC detector (20m)\\
{} \> {} \> \footnotesize Speaker: Grzegorz Kalicy (ODU) 
\end{tabbing}

\begin{tabbing}
\hspace*{\appLongTabWidth} \= \kill
14:00 - 16:00 \> Detector WG Tracking \\
{} \> \footnotesize  Conveners: Annalisa Mastroserio (Bari), Kondo Gnanvo (UVa), Leo Greiner (LBNL)
\end{tabbing}
\begin{tabbing}
\hspace*{\appSpaceTabWidth} \= \hspace*{\appTimeTabWidth} \= \kill
{} \> 14:00 \> Introduction to YR-Tracking WG and activities (15m)\\
{} \> {} \> \footnotesize Speaker: Kondo Gnanvo (UVa) \\

{} \> 14:15 \> Survey of Silicon Detector Technologies (15m)\\
{} \> {} \> \footnotesize Speaker: Laura Gonella \\

{} \> 14:30 \> ITS3 Technology (15m)\\
{} \> {} \> \footnotesize Speaker: Leo Greiner (LBNL) \\

{} \> 14:45 \> Survey of Gaseous Detector Technologies (10m)\\
{} \> {} \> \footnotesize Speaker: Kondo Gnanvo (UVa) \\

{} \> 14:55 \> Report on eRD6 and eRD22 activities (20m)\\
{} \> {} \> \footnotesize Speaker: Matt Posik (Temple) \\

{} \> 15:15 \> Cylindrical Micromegas for the Central Tracking (10m)\\
{} \> {} \> \footnotesize Speaker: Francesco Bossu (CEA-Saclay) \\

{} \> 15:25 \> Drift Chambers and Straw Tubes for Central Tracking (20m)\\
{} \> {} \> \footnotesize Speaker: Franco Grancagnolo \\

{} \> 15:45 \> sTGCs for the End Cap Tracking (15m)\\
{} \> {} \> \footnotesize Speaker: Daniel Brandenburg (BNL)
\end{tabbing}

\begin{tabbing}
\hspace*{\appLongTabWidth} \= \kill
14:00 - 16:00 \> Physics WG 
\end{tabbing}
\begin{tabbing}
\hspace*{\appSpaceTabWidth} \= \hspace*{\appTimeTabWidth} \= \kill
{} \> 14:00 \> Inclusive reactions WG (10m)\\
{} \> {} \> \footnotesize Speakers: Barak Schmookler (Stony Brook), Renee Fatemi (Kentucky), Nobuo Sato (JLab) \\

{} \> 14:20 \> Semi-inclusive Reactions WG (10m)\\
{} \> {} \> \footnotesize Speakers: Anselm Vossen (Duke), Bowen Xiao (CCNU), Justin Stevens (William \& Mary),\\
{} \> {} \> \footnotesize Ralf Seidl (RIKEN), Vladimirov Alexey (Regensburg) \\

{} \> 14:40 \> Jets, Heavy Quarks WG (10m)\\
{} \> {} \> \footnotesize Speakers: Brian Page (BNL), Ernst Sichtermann (LBNL),\\
{} \> {} \> \footnotesize Frank Petriello (Northwestern), Ivan Vitev (LANL), Leticia Cunqueiro (ORNL) \\

{} \> 15:00 \> Exclusive Reactions WG (10m)\\
{} \> {} \> \footnotesize Speakers: Barbara Pasquini (Pavia), Daria Sokhan, \\
{} \> {} \> \footnotesize Raphael Dupre (IPN Orsay), Salvatore Fazio (BNL), Tuomas Lappi ( Jyvaskyla) \\

{} \> 15:20 \> Diffractive Reactions \& Tagging WG (10m)\\
{} \> {} \> \footnotesize Speakers: Anna Stasto (Penn State), Douglas Higinbotham (JLab), Or Hen (MIT), \\
{} \> {} \> \footnotesize Spencer Klein (LBNL), Wim Cosyn (FIU) \\

{} \> 15:40 \> Discussion (20m)
\end{tabbing}

\begin{tabbing}
\hspace*{\appLongTabWidth} \= \kill
16:00 - 18:00 \> Detector WG Calorimetry \\
{} \> \footnotesize  Conveners: Eugene Chudakov (JLab), Vladimir Berdnikov (CUA) 
\end{tabbing}
\begin{tabbing}
\hspace*{\appLongTabWidth} \= \kill
16:30 - 18:30 \> Detector WG Far Forward Detector + \\
{} \> Ancillary detectors/Polarimetry/Luminosity \\
{} \> \footnotesize Conveners: Alexander Jentsch (BNL), Dave Gaskell, E.~C.~Aschenauer (BNL), \\
{} \> \footnotesize Michael Murray (Kansas), Yulia Furletova (JLab)
\end{tabbing}
\begin{tabbing}
\hspace*{\appSpaceTabWidth} \= \hspace*{\appTimeTabWidth} \= \kill
{} \> 14:30 \> Technical requirements for the luminosity detector and the low-Q2-tagger (30m)\\
{} \> {} \> \footnotesize Speaker: Jaroslav Adam (BNL) \\

{} \> 17:00 \> Technical requirements for the lepton polarimeter (30m)\\
{} \> {} \> \footnotesize Speaker: Alexandre Camsonne (JLAB) \\

{} \> 17:30 \> Technical requirements for the different hadron polarimeters (30m)\\
{} \> {} \> \footnotesize Speaker: Oleg Eyser (BNL) \\

{} \> 18:00 \> Discussion (30m)
\end{tabbing}

\begin{tabbing}
\hspace*{\appLongTabWidth} \= \kill
16:30 - 18:00 \> Detector WG PID \\
{} \> \footnotesize Conveners: Patrizia Rossi (JLab), Thomas Hemmick (Stony Brook)
\end{tabbing}
\begin{tabbing}
\hspace*{\appSpaceTabWidth} \= \hspace*{\appTimeTabWidth} \= \kill
{} \> 16:30 \>mRICH for EIC - past, present and future (20m)\\
{} \> {} \> \footnotesize Speaker: Xiaochun He (Georgia State) \\

{} \> 16:50 \>Quintuple-GEM Based RICH Detector for EIC (20m)\\
{} \> {} \> \footnotesize Speaker: Prakhar Garg (Stony Brook) \\

{} \> 17:10 \>High Momentum PID at EIC (in 10 years from now) (20m)\\
{} \> {} \> \footnotesize Speaker: Silvia Dalla Torre (INFN, Trieste) \\

{} \> 17:30 \>General discussion on the working plan and deliverables \\
{} \> {} \> for the Pavia workshop (30m)
\end{tabbing}

\begin{tabbing}
\hspace*{\appLongTabWidth} \= \kill
16:30 - 18:00 \> Detector WG Tracking \\
{} \> \footnotesize Conveners: Annalisa Mastroserio (Bari), Kondo Gnanvo (UVa), Leo Greiner (LBNL)
\end{tabbing}
\begin{tabbing}
\hspace*{\appSpaceTabWidth} \= \hspace*{\appTimeTabWidth} \= \kill
{} \> 16:30 \> Introduction to YR-Tracking WG Simulation (15m)\\
{} \> {} \> \footnotesize Speaker: Domenico Elia (INFN Bari) \\

{} \> 16:45 \> Overview of Tracking Simulation needs and Plans (30m)\\
{} \> {} \> \footnotesize Speaker: Barbara Jacak \\

{} \> 17:15 \> Including detector services in simulations (15m)\\
{} \> {} \> \footnotesize Speaker: Leo Greiner (LBNL) \\

{} \> 17:30 \> Open Discussion (30m)
\end{tabbing}

\begin{tabbing}
\hspace*{\appLongTabWidth} \= \kill
16:30 - 18:00 \> Physics WG Diffraction and Tagging \\
{} \> \footnotesize Conveners: Anna Stasto (Penn State), Douglas Higinbotham (JLab), Or Hen (MIT), \\
{} \> \footnotesize Spencer Klein (LBNL), Wim Cosyn (FIU)
\end{tabbing}
\begin{tabbing}
\hspace*{\appSpaceTabWidth} \= \hspace*{\appTimeTabWidth} \= \kill
{} \> 16:30 \> Spectator tagging in deuteron breakups and kinematics determination \\
{} \> {} \> using the BeAGLE generator (30m)\\
{} \> {} \> \footnotesize Speaker: Zhoudunming Tu (BNL) \\

{} \> 17:00 \> Update on detection of SRC nucleons in QE kinematics (30m)\\
{} \> {} \> \footnotesize Speaker: Florian Hauenstein (ODU) \\

{} \> 17:10 \> Semi-inclusive DIS measurements on A=3 (30m)\\
{} \> {} \> \footnotesize Speaker: Dien Nguyen (MIT)
\end{tabbing}

\begin{tabbing}
\hspace*{\appLongTabWidth} \= \kill
16:30 - 18:00 \> Physics WG Exclusive \\
{} \> \footnotesize Conveners: Barbara Pasquini (Pavia), Daria Sokhan, \\
{} \> \footnotesize Spencer Raphael Dupre (IPN Orsay), Salvatore Fazio (BNL), Tuomas Lappi (Jyvaskyla)
\end{tabbing}
\begin{tabbing}
\hspace*{\appSpaceTabWidth} \= \hspace*{\appTimeTabWidth} \= \kill
{} \> 16:30 \>Summary of available DVCS and GPDs impact studies in e+p at EIC (20m)\\
{} \> {} \> \footnotesize Speaker: Salvatore Fazio (BNLoratory) \\

{} \> 16:50 \>DVCS and pi0 study (20m)\\
{} \> {} \> \footnotesize Speaker: Francois-Xavier Girod (JLab) \\

{} \> 17:10 \> DVCS Analysis Framework (20m)\\
{} \> {} \> \footnotesize Speaker: Simonetta Liuti (UVa) \\

{} \> 17:30 \>Common discussion: "what we have \& what's next?" (30m)
\end{tabbing}

\begin{tabbing}
\hspace*{\appLongTabWidth} \= \kill
16:30 - 18:00 \> Physics WG Inclusive \\
{} \> \footnotesize Conveners: Barak Schmookler (Stony Brook), Renee Fatemi (Kentucky), Nobuo Sato (JLab)
\end{tabbing}

\begin{tabbing}
\hspace*{\appLongTabWidth} \= \kill
16:30 - 18:00 \> Physics WG Jets, HF \\
{} \> \footnotesize Conveners: Brian Page (BNL), Ernst Sichtermann (LBNL), Frank Petriello (Northwestern), \\
{} \> \footnotesize Ivan Vitev (LANL), Leticia Cunqueiro (ORNL)
\end{tabbing}
\begin{tabbing}
\hspace*{\appSpaceTabWidth} \= \hspace*{\appTimeTabWidth} \= \kill
{} \> 16:30 \>Table of measurements in the Jets and Heavy Flavor Working Group (15m)\\
{} \> {} \> \footnotesize Speaker: Leticia Cunqueiro (ORNL) \\

{} \> 16:45 \>Jets for 3D imaging (20m)\\
{} \> {} \> \footnotesize Speaker: Miguel Arratia (UC Riverside) \\

{} \> 17:05 \>Heavy flavour reconstruction (20m)\\
{} \> {} \> \footnotesize Speaker: Yue Shi Lai (UC Berkeley) \\

{} \> 17:25 \>Jet angularities at the EIC (20m)\\
{} \> {} \> \footnotesize Speaker: Brian Page (BNL) \\

{} \> 17:45 \>Discussion (15m)
\end{tabbing}

\begin{tabbing}
\hspace*{\appLongTabWidth} \= \kill
16:30 - 18:00 \> Physics WG Semi-Inclusive \\
{} \> \footnotesize Conveners: Anselm Vossen (Duke), Bowen Xiao (CCNU), Justin Stevens (William \& Mary), \\
{} \> \footnotesize Ralf Seidl (RIKEN), Vladimirov Alexey (Regensburg)
\end{tabbing}
\begin{tabbing}
\hspace*{\appSpaceTabWidth} \= \hspace*{\appTimeTabWidth} \= \kill
{} \> 16:30 \>Spectroscopy overview/theory (30m)\\
{} \> {} \> \footnotesize Speaker: Alessandro Pilloni (ECT*) \\

{} \> 17:00 \>Spectroscopy experiment (30m)\\
{} \> {} \> \footnotesize Speaker: Justin Stevens (William \& Mary) \\

{} \> 17:30 \>Di-hadron fragmentation update (30m)\\
{} \> {} \> \footnotesize Speaker: Anselm Vossen (Duke)
\end{tabbing}

\subsection*{FRIDAY, 20 MARCH} 

\begin{tabbing}
\hspace*{\appLongTabWidth} \= \kill
\phantom{0}8:30 - 10:30 \> Detector WG Central Detector / Magnet \\
{} \> \footnotesize Conveners: Alexander Kiselev (BNL), William Brooks (UTFSM)
\end{tabbing}
\begin{tabbing}
\hspace*{\appSpaceTabWidth} \= \hspace*{\appTimeTabWidth} \= \kill
{} \> \phantom{0}8:30 \> Introductory remarks (10m)\\
{} \> {} \> \footnotesize Speaker: Alexander Kiselev (BNL) \\

{} \> \phantom{0}8:40 \> Photo-sensors in a strong magnetic field: options for EIC (20m)\\
{} \> {} \> \footnotesize Speaker: Junqi Xie (ANL) \\

{} \> \phantom{0}9:00 \> Forward gaseous RICH performance in EIC-sPHENIX solenoid\\ {} \> {} \> fringe field (15m)\\
{} \> {} \> \footnotesize Speaker: Jin Huang (BNL) \\

{} \> \phantom{0}9:15 \> Low Pt track cutoff implications of a strong solenoid magnetic field (15m)\\
{} \> {} \> \footnotesize Speaker: Yulia Furletova (JLab) \\

{} \> \phantom{0}9:30 \> Momentum resolution and magnetic field strength for an EIC detector (15m)\\
{} \> {} \> \footnotesize Speaker: Nick Lukow (Temple) \\

{} \> \phantom{0}9:45 \> BeAST solenoid magnetic field calculation and accompanying studies (15m)\\
{} \> {} \> \footnotesize Speaker: Alexander Kiselev (BNL) \\

{} \> 10:00 \> Design considerations for the EIC central detector solenoid \\
{} \> {} \> ßdiscussion and QA session) (30m)

\end{tabbing}

\begin{tabbing}
\hspace*{\appLongTabWidth} \= \kill
\phantom{0}8:30 - 10:30 \> Detector WG Electronics/DAQ \\
{} \> \footnotesize Conveners: Andrea Celentano (INFN-Genova), Damien Neyret (CEA Saclay IRFU/DPhN)
\end{tabbing}
\begin{tabbing}
\hspace*{\appSpaceTabWidth} \= \hspace*{\appTimeTabWidth} \= \kill
{} \> \phantom{0}8:30 \> Introduction (30m)\\
{} \> {} \> \footnotesize Speakers: Andrea Celentano (INFN-Genova), Damien Neyret (CEA Saclay IRFU/DPhN) \\

{} \> \phantom{0}9:00 \> Overview of eRD23 activities (30m)\\
{} \> {} \> \footnotesize Speaker: Marco Battaglieri (JLab) \\

{} \> \phantom{0}9:30 \> Discussion (15m)\\

{} \> \phantom{0}9:45 \> EIC data rates and noise estimates for a streaming readout system (30m)\\
{} \> {} \> \footnotesize Speaker: Jin Huang (BNL) \\

{} \> 10:15 \> Discussion (15m) 

\end{tabbing}

\begin{tabbing}
\hspace*{\appLongTabWidth} \= \kill
\phantom{0}8:30 - 10:30 \> Detector WG PID \\
{} \> \footnotesize Conveners: Patrizia Rossi (JLab), Thomas Hemmick (Stony Brook)
\end{tabbing}

\begin{tabbing}
\hspace*{\appLongTabWidth} \= \kill
\phantom{0}8:30 - 10:30 \> Forward Detector/IR integration + \\
{} \> Ancillary detectors/Polarimetry/Luminosity \\
{} \> \footnotesize Conveners: Alexander Jentsch (BNL), Dave Gaskell, E.~C.~Aschenauer (BNL), \\
{} \> \footnotesize Michael Murray (Kansas), Yulia Furletova (JLab)
\end{tabbing}
\begin{tabbing}
\hspace*{\appSpaceTabWidth} \= \hspace*{\appTimeTabWidth} \= \kill
{} \> \phantom{0}9:30 \> Roman Pots at the LHC (30m)\\
{} \> {} \> \footnotesize Speaker: Michael Murray (Kansas) \\

{} \> 10:00 \> Open Discussion (30m)\\
{} \> {} \> \footnotesize Chair: Alexander Jentsch (BNL) 
\end{tabbing}

\begin{tabbing}
\hspace*{\appLongTabWidth} \= \kill
\phantom{0}8:30 - 10:30 \> Physics WG Diffraction and Tagging \\
{} \> \footnotesize Conveners: Anna Stasto (Penn State), Douglas Higinbotham (JLab), \\
{} \> \footnotesize Or Hen (MIT), Spencer Klein (LBNL), Wim Cosyn (FIU)
\end{tabbing}
\begin{tabbing}
\hspace*{\appSpaceTabWidth} \= \hspace*{\appTimeTabWidth} \= \kill
{} \> \phantom{0}8:30 \> Perspectives on diffractive jet production at the EIC (30m)\\
{} \> {} \> \footnotesize Speaker: Michael Klasen (Münster) \\

{} \> \phantom{0}9:00 \> Inclusive diffraction at future EIC (30m)\\
{} \> {} \> \footnotesize Speaker: Anna Stasto (Penn State) \\

{} \> \phantom{0}9:30 \> Coherent vector meson production off heavy nuclei - lessons \\
{} \> {} \> from studies of ultraperipheral collisions at the LHC (30m)\\
{} \> {} \> \footnotesize Speaker: Mark Strikman (PSU) \\

{} \> 10:00 \> Probing quantum fluctuations of the nucleon’s gluon density with \\
{} \> {} \> inelastic diffraction at EIC (30m)\\
{} \> {} \> \footnotesize Speaker: Christian Weiss (JLab)
\end{tabbing}

\begin{tabbing}
\hspace*{\appLongTabWidth} \= \kill
\phantom{0}8:30 - 10:50 \> Physics WG Exclusive \\
{} \> \footnotesize Conveners: Barbara Pasquini (Pavia), Daria Sokhan, \\
{} \> \footnotesize Raphael Dupre (IPN Orsay), Salvatore Fazio (BNL), Tuomas Lappi (Jyvaskyla)
\end{tabbing}
\begin{tabbing}
\hspace*{\appSpaceTabWidth} \= \hspace*{\appTimeTabWidth} \= \kill
{} \> \phantom{0}8:30 \> Summary of available studies on VMP in e+p collisions at EIC (20m)\\
{} \> {} \> \footnotesize Speaker: Sylvester Joosten (ANL) \\

{} \> \phantom{0}8:50 \> Summary of studies and challenges for VMP in e+A collisions at EIC (20m)\\
{} \> {} \> \footnotesize Speaker: Thomas Ullrich (BNL) \\

{} \> \phantom{0}9:10 \> Exclusive di-jet production as an access gluon Wigner fcn. (20m)\\
{} \> {} \> \footnotesize Speaker: Heikki Mantysaari \\

{} \> \phantom{0}9:30 \> TCS with PARTONS (20m)\\
{} \> {} \> \footnotesize Speaker: Jakub Wagner (NCBJ) \\

{} \> \phantom{0}9:50 \> Common discussion: "what we have \& what's next?" (40m)
\end{tabbing}

\begin{tabbing}
\hspace*{\appLongTabWidth} \= \kill
\phantom{0}8:30 - 10:35 \> Physics WG Inclusive \\
{} \> \footnotesize Conveners: Barak Schmookler (Stony Brook), Renee Fatemi (Kentucky), Nobuo Sato (JLab)
\end{tabbing}
\begin{tabbing}
\hspace*{\appSpaceTabWidth} \= \hspace*{\appTimeTabWidth} \= \kill
{} \> \phantom{0}8:30 \> Plan and Workflow (5m)\\
{} \> {} \> \footnotesize Speaker: Renee Fatemi (Kentucky) \\

{} \> \phantom{0}8:35 \> Nuclear shadowing in DIS for future electron-ion colliders (15m)\\
{} \> {} \> \footnotesize Speaker: Michal Krelina (TU Prague) \\

{} \> \phantom{0}8:50 \> EIC impact on unpolarized PDFs: a preliminary study (15m)\\
{} \> {} \> \footnotesize Speaker: Alberto Accardi (Hampton U. and JLab) \\

{} \> \phantom{0}9:05 \> Wish List (15m)\\
{} \> {} \> \footnotesize Speaker: Fredrick Olness (SMU) \\

{} \> \phantom{0}9:20 \> Constraining the unpolarized proton PDFs at EIC through\\ 
{} \> {} \> inclusive measurements (15m)\\
{} \> {} \> \footnotesize Speaker: Xiaoxuan Chu (BNL) \\

{} \> \phantom{0}8:35 \> Kinematic reconstruction methods (15m)\\
{} \> {} \> \footnotesize Speaker: Bernd Surrow (Temple) \\

{} \> \phantom{0}9:50 \> Charm production in CCDIS at EIC (15m)\\
{} \> {} \> \footnotesize Speaker: Jae Nam (Temple) \\

{} \> 10:05 \> First-round studies of the EIC's PDF implications (15m)\\
{} \> {} \> \footnotesize Speaker: Timothy Hobbs (Southern Methodist) \\

{} \> 10:20 \> Testing Lorentz and CPT symmetry at the EIC (15m)\\
{} \> {} \> \footnotesize Speakers: Enrico Lunghi (Indiana), Nathan Sherrill
\end{tabbing}

\begin{tabbing}
\hspace*{\appLongTabWidth} \= \kill
\phantom{0}8:30 - 10:30 \> Physics WG Jets, HF \\
{} \> \footnotesize Conveners: Brian Page (BNL), Ernst Sichtermann (LBNL), Frank Petriello (Northwestern), \\
{} \> \footnotesize Ivan Vitev (LANL), Leticia Cunqueiro (ORNL)
\end{tabbing}
\begin{tabbing}
\hspace*{\appSpaceTabWidth} \= \hspace*{\appTimeTabWidth} \= \kill
{} \> \phantom{0}8:30 \> Table of measurements in the Jets and Heavy Flavor Working Group (15m)\\
{} \> {} \> \footnotesize Speaker: Ivan Vitev (LANL) \\

{} \> \phantom{0}8:45 \> Double longitudinal spin asymmetries with jets (20m)\\
{} \> {} \> \footnotesize Speaker: Frank Petriello (Northwestern) \\

{} \> \phantom{0}9:05 \> The angularity event shapes in DIS at the NNLL accuracy (20m)\\
{} \> {} \> \footnotesize Speaker: Tanmay Maji (Fudan) \\

{} \> \phantom{0}9:25 \> Energy-Energy Correlators for TMD physics and reduction of uncertainties \\
{} \> {} \> due to hadronization (20m)\\
{} \> {} \> \footnotesize Speaker: Haitao Li (LANL) \\

{} \> \phantom{0}9:45 \> LANL plans for heavy flavor, quarkonia and jet studies (20m)\\
{} \> {} \> \footnotesize Speakers: Matt Durham (LANL), Xuan Li (LANL) \\

{} \> 10:05 \> Discussion (25m)
\end{tabbing}

\begin{tabbing}
\hspace*{\appLongTabWidth} \= \kill
\phantom{0}8:30 - 10:30 \> Physics WG Semi-Inclusive \\
{} \> \footnotesize Conveners: Anselm Vossen (Duke), Bowen Xiao (CCNU), Justin Stevens (William \& Mary), \\
{} \> \footnotesize Ralf Seidl (RIKEN), Vladimirov Alexey (Regensburg)
\end{tabbing}
\begin{tabbing}
\hspace*{\appSpaceTabWidth} \= \hspace*{\appTimeTabWidth} \= \kill
{} \> \phantom{0}8:30 \> Lambda fragmentation related measurements (30m)\\
{} \> {} \> \footnotesize Speaker: Jinlong Zhang (Stony Brook) \\

{} \> \phantom{0}9:00 \> (Nuclear) Fragmentation function related measurements (30m)\\
{} \> {} \> \footnotesize Speaker: : Charlotte Van Hulse \\

{} \> \phantom{0}9:30 \> Quark Sivers/TMD related measurements (30m)\\
{} \> {} \> \footnotesize Speaker: Alexei Prokudin (Penn State Berks) \\

{} \> 10:00 \> Parton helicity related measurements (30m)\\
{} \> {} \> \footnotesize Speaker: E.~C.~Aschenauer (BNL) 
\end{tabbing}

\begin{tabbing}
\hspace*{\appLongTabWidth} \= \kill
11:00 - 13:00 \> Detector WG Central Detector / Magnet \\
{} \> \footnotesize Conveners: Alexander Kiselev (BNL), William Brooks (UTFSM)
\end{tabbing}
\begin{tabbing}
\hspace*{\appSpaceTabWidth} \= \hspace*{\appTimeTabWidth} \= \kill
{} \> 11:00 \> A possible method for adding services load to the EIC simulations (20m)\\
{} \> {} \> \footnotesize Speaker: Leo Greiner (LBNL) \\

{} \> 11:20 \> EIC detector infrastructure (25m)\\
{} \> {} \> \footnotesize Speaker: Mark Breitfeller (BNL) \\

{} \> 11:45 \> Synchrotron radiation studies with the present EIC IR vacuum \\
{} \> {} \> system design (30m)\\
{} \> {} \> \footnotesize Speaker: Charles Hetzel (BNL) \\

{} \> 12:15 \> Background sources and studies at the EIC (20m)\\
{} \> {} \> \footnotesize Speaker: Latifa Elouardhiri (JLab) \\

{} \> 12:35 \> Studies of beam-gas background, neutron flux, radiation dose \\
{} \> {} \> at an EIC (15m)\\
{} \> {} \> \footnotesize Speaker: Jin Huang (BNL) \\

{} \> 12:50 \> Summary and concluding remarks (10m) 
\end{tabbing}

\begin{tabbing}
\hspace*{\appLongTabWidth} \= \kill
11:00 - 13:00 \> Detector WG Electronics/DAQ \\
{} \> \footnotesize Conveners: Andrea Celentano (INFN-Genova), Damien Neyret (CEA Saclay IRFU/DPhN)
\end{tabbing}
\begin{tabbing}
\hspace*{\appSpaceTabWidth} \= \hspace*{\appTimeTabWidth} \= \kill
{} \> 11:00 \> Streaming readout simulation: activity at JLab (30m)\\
{} \> {} \> \footnotesize Speaker: Maurizio Ungaro (JLab) 
\end{tabbing}

\begin{tabbing}
\hspace*{\appLongTabWidth} \= \kill
11:00 - 13:00 \> Forward Detector/IR integration + \\
{} \> Ancillary detectors/Polarimetry/Luminosity \\
{} \> \footnotesize Conveners: Alexander Jentsch (BNL), Dave Gaskell, E.~C.~Aschenauer (BNL), \\
{} \> \footnotesize Michael Murray (Kansas), Yulia Furletova (JLab)
\end{tabbing}

\begin{tabbing}
\hspace*{\appLongTabWidth} \= \kill
11:00 - 13:00 \> Physics WG Diffraction and Tagging \\
{} \> \footnotesize Conveners: Anna Stasto (Penn State), Douglas Higinbotham (JLab), \\
{} \> \footnotesize Or Hen (MIT), Spencer Klein (LBNL), Wim Cosyn (FIU)
\end{tabbing}
\begin{tabbing}
\hspace*{\appSpaceTabWidth} \= \hspace*{\appTimeTabWidth} \= \kill
{} \> 11:00 \> Near threshold photoproduction a2(1320) and other topics (30m)\\
{} \> {} \> \footnotesize Speaker: Spencer Klein (LBNL) \\

{} \> 11:30 \> Meson Structure at the EIC (30m)\\
{} \> {} \> \footnotesize Speaker: Richard Trotta (CUA) \\

{} \> 12:00 \> Discussion (1h) 
\end{tabbing}
\begin{tabbing}
\hspace*{\appLongTabWidth} \= \kill
11:00 - 13:00 \> Physics WG Exclusive \\
{} \> \footnotesize Conveners: Barbara Pasquini (Pavia), Daria Sokhan, Raphael Dupre (IPN Orsay), \\
{} \> \footnotesize Salvatore Fazio (BNL), Tuomas Lappi (Jyvaskyla)
\end{tabbing}
\begin{tabbing}
\hspace*{\appSpaceTabWidth} \= \hspace*{\appTimeTabWidth} \= \kill
{} \> 11:00 \> Accessing the transverse force in a nucleon (20m)\\
{} \> {} \> \footnotesize Speaker: Matthias Burkardt (New Mexico State) \\

{} \> 11:20 \> N $\rightarrow$ N$^*$ transition GPDs (20m)\\
{} \> {} \> \footnotesize Speaker: Asli Tandogan (Bochum) \\

{} \> 11:40 \> Common discussion (1h20m)\\
{} \> {} \> \footnotesize Chair: Christian Weiss (JLab) 
\end{tabbing}

\begin{tabbing}
\hspace*{\appLongTabWidth} \= \kill
11:00 - 13:00 \> Physics WG Inclusive \\
{} \> \footnotesize Conveners: Barak Schmookler (Stony Brook), Renee Fatemi (Kentucky), Nobuo Sato (JLab)
\end{tabbing}
\begin{tabbing}
\hspace*{\appSpaceTabWidth} \= \hspace*{\appTimeTabWidth} \= \kill
{} \> 11:00 \> Vertex Level Events (30m)\\
{} \> {} \> \footnotesize Speaker: Nobuo Sato (JLab) \\

{} \> 11:30 \> Detector Level Events (30m)\\
{} \> {} \> \footnotesize Speaker: Barak Schmookler (Stony Brook) \\

{} \> 12:00 \> Reconstruction (30m)\\
{} \> {} \> \footnotesize Chair: Renee Fatemi (Kentucky) 
\end{tabbing}

\begin{tabbing}
\hspace*{\appLongTabWidth} \= \kill
11:00 - 13:00 \> Physics WG Jets, HF \\
{} \> \footnotesize Conveners: Brian Page (BNL), Ernst Sichtermann (LBNL), Frank Petriello (Northwestern),\\ \> \footnotesize Ivan Vitev (LANL), Leticia Cunqueiro (ORNL)
\end{tabbing}
\begin{tabbing}
\hspace*{\appSpaceTabWidth} \= \hspace*{\appTimeTabWidth} \= \kill
{} \> 11:00 \> Gluon Sivers Related Measurements (25m)\\
{} \> {} \> \footnotesize Speaker: Liang Zheng \\

{} \> 11:25 \> TMD measurements in jets (25m)\\
{} \> {} \> \footnotesize Speaker: Felix Ringer (UC Berkeley/LBNL) \\

{} \> 11:50 \> Modification of heavy flavor in e+A collisions at the EIC (25m)\\
{} \> {} \> \footnotesize Speaker: Zelong Liu (LANL) \\

{} \> 12:15 \> Hadrons in jets (25m)\\
{} \> {} \> \footnotesize Speaker: Yiannis Makris (LANL) \\

{} \> 12:40 \> Discussion (20m) 
\end{tabbing}
\begin{tabbing}
\hspace*{\appLongTabWidth} \= \kill
11:00 - 13:00 \> Physics WG Semi-Inc: Joint session with the Jet/HF WG \\
{} \> \footnotesize Conveners: Anselm Vossen (Duke), Bowen Xiao (CCNU), Justin Stevens (William \& Mary), \\ \> \footnotesize Ralf Seidl (RIKEN), Vladimirov Alexey (Regensburg)
\end{tabbing}
\begin{tabbing}
\hspace*{\appSpaceTabWidth} \= \hspace*{\appTimeTabWidth} \= \kill
{} \> 11:00 \> Joint session with the Jet/HF WG (2h)
\end{tabbing}

\begin{tabbing}
\hspace*{\appLongTabWidth} \= \kill
14:00 - 15:30 \> Exclusive/Diffractive - Tagging/DWG
\end{tabbing}
\begin{tabbing}
\hspace*{\appSpaceTabWidth} \= \hspace*{\appTimeTabWidth} \= \kill
{} \> 14:00 \> Separating Coherent and Incoherent Interactions (5m)\\
{} \> {} \> \footnotesize Speaker: Spencer Klein (LBNL) \\

{} \> 14:10 \> Coherent DVCS with light nuclei (5m)\\
{} \> {} \> \footnotesize Speakers: Anna Stasto (Penn State), Barbara Pasquini (Pavia), Daria Sokhan,\\
{} \> {} \> \footnotesize Douglas Higinbotham (JLab), Or Hen (MIT), Raphael Dupre (IPN Orsay),\\
{} \> {} \> \footnotesize Salvatore Fazio (BNL), Spencer Klein (LBNL), Tuomas Lappi (Jyvaskyla) \\

{} \> 14:20 \> Effect of beam smearing on tagging/Accessible t-range (5m)\\
{} \> {} \> \footnotesize Speaker: Christian Weiss (JLab) \\

{} \> 14:30 \> Requirements on tracking from VMP in e+A (5m)\\
{} \> {} \> \footnotesize Speaker: Thomas Ullrich (BNL) \\

{} \> 14:40 \> Meson structure functions - (forward) detector requirements (5m)\\
{} \> {} \> \footnotesize Speaker: Tanja Horn (CUA) \\

{} \> 14:50 \> Current simulation results with IR + detectors (30m)\\
{} \> {} \> \footnotesize Speaker: Alexander Jentsch (BNL) \\

{} \> 15:20 \> Discussion (10m) 
\end{tabbing}

\begin{tabbing}
\hspace*{\appLongTabWidth} \= \kill
14:00 - 15:30 \> Inclusive/SIDIS/Jets/HQ / DWG
\end{tabbing}
\begin{tabbing}
\hspace*{\appSpaceTabWidth} \= \hspace*{\appTimeTabWidth} \= \kill
{} \> 14:00 \> Detector needs for Inclusive Reactions (10m)\\
{} \> {} \> \footnotesize Speakers: Barak Schmookler (Stony Brook), Renee Fatemi (Kentucky), Nobuo Sato (JLab) \\

{} \> 14:15 \> Detector requirements/input for Heavy Flavor (10m)\\
{} \> {} \> \footnotesize Speakers: Ernst Sichtermann (LBNL), Ivan Vitev (LANL) \\

{} \> 14:30 \> Detector requirements/input forJets (10m)\\
{} \> {} \> \footnotesize Speaker: Brian Page (BNL) \\

{} \> 14:45 \> Detector requirements/input for SIDIS (10m)\\
{} \> {} \> \footnotesize Speaker: Justin Stevens (William \& Mary) \\

{} \> 15:00 \> PID Request to PWG / Discussion (30m)\\
{} \> {} \> \footnotesize Speaker: Thomas Hemmick (Stony Brook) 
\end{tabbing}

\begin{tabbing}
\hspace*{\appLongTabWidth} \= \kill
16:00 - 18:30 \> Joint PWG/DWG Session - Detector Complementarity Discussion
\end{tabbing}
\begin{tabbing}
\hspace*{\appSpaceTabWidth} \= \hspace*{\appTimeTabWidth} \= \kill
{} \> 16:00 \> PWG/DWG common questions (1h)\\

{} \> 17:00 \> Detector complementarity discussion (1h30m)\\
{} \> {} \> \footnotesize Speaker: E.~C.~Aschenauer (BNL) 
\end{tabbing}

\subsection*{SATURDAY, 21 MARCH} 

\begin{tabbing}
\hspace*{\appLongTabWidth} \= \kill
\phantom{0}8:30 - 11:00 \> Working Group Summaries
\end{tabbing}
\begin{tabbing}
\hspace*{\appSpaceTabWidth} \= \hspace*{\appTimeTabWidth} \= \kill
{} \> \phantom{0}8:30 \> Inclusive reactions WG summary (10m)\\
{} \> {} \> \footnotesize Speakers: Barak Schmookler (Stony Brook), Renee Fatemi (Kentucky), Nobuo Sato (JLab)\\

{} \> \phantom{0}8:45 \> Semi-inclusive Reactions WG summary (10m)\\
{} \> {} \> \footnotesize Speakers: Anselm Vossen (Duke), Bowen Xiao (CCNU), Justin Stevens (William \& Mary), \\
{} \> {} \> Ralf Seidl (RIKEN), Vladimirov Alexey (Regensburg) \\

{} \> \phantom{0}9:00 \> Jets, Heavy Quarks WG summary (10m)\\
{} \> {} \> \footnotesize Speakers: Brian Page (BNL), Ernst Sichtermann (LBNL), Frank Petriello (Northwestern), \\
{} \> {} \> Ivan Vitev (LANL), Leticia Cunqueiro (ORNL)
 \\

{} \> \phantom{0}9:15 \> Exclusive Reactions WG summary (10m)\\
{} \> {} \> \footnotesize Speakers: Barbara Pasquini (Pavia), Daria Sokhan, Raphael Dupre (IPN Orsay),\\
{} \> {} \> Salvatore Fazio (BNL), Tuomas Lappi (Jyvaskyla) \\

{} \> \phantom{0}9:30 \> Diffractive Reactions \& Tagging WG summary (10m)\\
{} \> {} \> \footnotesize Speakers: Anna Stasto (Penn State), Douglas Higinbotham \\
{} \> {} \> \footnotesize (JLab), Or Hen (MIT),\\
Spencer Klein (LBNL), Wim Cosyn (FIU) \\

{} \> \phantom{0}9:45 \> Tracking WG summary (10m)\\
{} \> {} \> \footnotesize Speakers: Annalisa Mastroserio (Bari), Kondo Gnanvo (UVa), Leo Greiner (LBNL) \\

{} \> \phantom{0}9:55 \> Particle ID WG Summary (10m)\\
{} \> {} \> \footnotesize Speakers: Patrizia Rossi (JLab), Thomas Hemmick (Stony Brook) \\

{} \> 10:05 \> Calorimetry WG Summary (10m)\\
{} \> {} \> \footnotesize Speakers: Eugene Chudakov (JLab), Vladimir Berdnikov (CUA) \\

{} \> 10:15 \> Far-Forward/IR/Polarimetry/Ancillary Detectors Summary (25m)\\
{} \> {} \> \footnotesize Speakers: Alexander Jentsch (BNL), Dave Gaskell, E. C. Aschenauer (BNL),\\
{} \> {} \> Michael Murray (Kansas), Yulia Furletova (JLab) \\

{} \> 10:40 \> DAQ/Electronics WG Summary (10m)\\
{} \> {} \> \footnotesize Speakers: Andrea Celentano (INFN-Genova), Damien Neyret (CEA Saclay IRFU/DPhN) \\

{} \> 10:50 \> Central detector/Integration and Magnet WG Summary (10m)\\
{} \> {} \> \footnotesize Speakers: Alexander Kiselev (BNL), William Brooks
\end{tabbing}

\begin{tabbing}
\hspace*{\appLongTabWidth} \= \kill
11:30 - 13:30 \> Software Discussion / International engagement Discussion 
\end{tabbing}
\begin{tabbing}
\hspace*{\appSpaceTabWidth} \= \hspace*{\appTimeTabWidth} \= \kill
{} \> 11:30 \> Software - Questions / Discussion (1h)\\

{} \> 12:30 \> International engagement - Discussion (1h)
\end{tabbing}
\newpage
\section*{2nd EIC Yellow Report Workshop at Pavia University}
Local organizer: Marco Radici (INFN - Sezione di Pavia)

\subsection*{WEDNESDAY, 20 May} 
\begin{tabbing}
\hspace*{\appLongTabWidth} \= \kill
\phantom{0}8:30 - \phantom{0}9:45 \> EIC project status \\
{} \> \footnotesize Convener: Bernd Surrow (Temple)
\end{tabbing}
\begin{tabbing}
\hspace*{\appSpaceTabWidth} \= \hspace*{\appTimeTabWidth} \= \kill
{} \> \phantom{0}8:30 \> Welcome (10 min)\\
{} \> {} \> \footnotesize Speaker: Marco Radici (INFN - Sezione di Pavia)\\

{} \> \phantom{0}8:40 \> Brief Statement from BNL / JLab ALD's (10 min)\\
{} \> {} \> \footnotesize Speakers: Berndt M\"uller (BNL), Bob McKeown (JLab\\

{} \> \phantom{0}8:50 \> INFN activities in Hadronic Physic (25 min)\\
{} \> {} \> \footnotesize Speaker: Rosario Nania (INFN)\\

{} \> \phantom{0}9:15 \> EIC project overview (30 min)\\
{} \> {} \> \footnotesize Speaker: James Yeck (Wisconsin-Madison and BNL)
\end{tabbing}

\begin{tabbing}
\hspace*{\appLongTabWidth} \= \kill
10:15 - 12:00 \> EIC design and EoI \\
{} \> \footnotesize Convener: Dani\"el Boer (Groningen)
\end{tabbing}
\begin{tabbing}
\hspace*{\appSpaceTabWidth} \= \hspace*{\appTimeTabWidth} \= \kill
{} \> 10:15 \> EIC accelerator and IR design status (40 min)\\
{} \> {} \> \footnotesize Speaker: Ferdinand Willeke (BNL)\\

{} \> 10:55 \> Discussion (20 min)\\

{} \> 11:15 \> EoI process / discussion (45 min)\\
{} \> {} \> \footnotesize Speakers: E.~C.~Aschenauer (BNL), Rolf Ent (JLab)
\end{tabbing}

\begin{tabbing}
\hspace*{\appLongTabWidth} \= \kill
13:30 - 15:30 \> Working Groups Overviews \\
{} \> \footnotesize Convener: Rolf Ent (JLab)
\end{tabbing}
\begin{tabbing}
\hspace*{\appSpaceTabWidth} \= \hspace*{\appTimeTabWidth} \= \kill
{} \> 13:30 \> YR Physics WG Conveners: overview and progress report (1h)\\
{} \> {} \> \footnotesize Speakers: Conveners\\

{} \> 14:30 \> YR Detector WG Conveners: overview and progress report (1h)\\
{} \> {} \> \footnotesize Speakers: Conveners
\end{tabbing}

\begin{tabbing}
\hspace*{\appLongTabWidth} \= \kill
16:00 - 17:30 \> PWG/DWG/SWG workflow \\
{} \> \footnotesize Convener: Thomas Ullrich (BNL)
\end{tabbing}
\begin{tabbing}
\hspace*{\appSpaceTabWidth} \= \hspace*{\appTimeTabWidth} \= \kill
{} \> 16:00 \> Discussion on PWG / DWG / SWG workflow (1h 30m)\\
{} \> {} \> \footnotesize Speakers: Conveners
\end{tabbing}

\subsection*{THURSDAY, 21 May} 

\begin{tabbing}
\hspace*{\appLongTabWidth} \= \kill
\phantom{0}8:30 - 10:00 \> Calorimeter \& Particle ID \& Tracking \\
{} \> \footnotesize  Conveners: Domenico Elia (INFN Bari), Eugene Chudakov (JLab), Kondo Gnanvo (UVa), \\
{} \> \footnotesize Leo Greiner (LBNL), Patrizia Rossi (JLab), Thomas Hemmick (Stony Brook), \\
{} \> \footnotesize Vladimir Berdnikov (CUA)
\end{tabbing}
\begin{tabbing}
\hspace*{\appSpaceTabWidth} \= \hspace*{\appTimeTabWidth} \= \kill
{} \> \phantom{0}8:30 \> Tracking overview (30m)\\
{} \> {} \> \footnotesize Speakers: Conveners\\

{} \> \phantom{0}9:00 \> PID overview (30m)\\
{} \> {} \> \footnotesize Speakers: Conveners\\

{} \> \phantom{0}9:30 \> Calorimeter overview (30m)\\
{} \> {} \> \footnotesize Speaker: Alexander Bazilevsky (BNL)
\end{tabbing}

\begin{tabbing}
\hspace*{\appLongTabWidth} \= \kill
\phantom{0}8:30 - 10:00 \> DAQ \& Electronics \\
{} \> \footnotesize Conveners: Andrea Celentano (INFN-Genova), Damien Neyret (CEA Saclay IRFU/DPhN)
\end{tabbing}
\begin{tabbing}
\hspace*{\appSpaceTabWidth} \= \hspace*{\appTimeTabWidth} \= \kill
{} \> \phantom{0}8:30 \> Introduction and summary of different solutions discussed\\
{} \> {} \> on DAQ structure (30m)\\
{} \> {} \> \footnotesize Speaker: Andrea Celentano (INFN-Genova)\\

{} \> \phantom{0}9:00 \> Discussion and final conclusions about the DAQ structure (1h)
\end{tabbing}

\begin{tabbing}
\hspace*{\appLongTabWidth} \= \kill
\phantom{0}8:30 - 10:20 \> Diffractive Reactions \& Tagging WG \\
{} \> \footnotesize Convener: Anna Maria Stasto (Penn State)
\end{tabbing}

\begin{tabbing}
\hspace*{\appSpaceTabWidth} \= \hspace*{\appTimeTabWidth} \= \kill

{} \> \phantom{0}8:30 \> Inclusive diffraction in DIS. What can we learn beyond HERA? (23m)\\
{} \> {} \> \footnotesize Speaker: Wojtek Slominski (Jagiellonian University)
\\

{} \> \phantom{0}8:53 \> Diffractive dijet production at EIC (23m)\\
{} \> {} \> \footnotesize Speaker: Vadim Guzey (Petersburg Nuclear Physics Institute)\\

{} \> \phantom{0}9:16 \> Pion and Kaon structure studies
 (23m)\\
{} \> {} \> \footnotesize Speaker: Richard Trotta (CUA)\\

{} \> \phantom{0}9:39 \> SRC measurements (23m)\\
{} \> {} \> \footnotesize Speaker: Florian Hauenstein (ODU)
\\

{} \> 10:02 \> 3He measurements (13m)\\
{} \> {} \> \footnotesize Speaker: Ivica Frsicic
\end{tabbing}

\begin{tabbing}
\hspace*{\appLongTabWidth} \= \kill
\phantom{0}8:30 - 10:00 \> Exclusive Reactions WG 
\end{tabbing}

\begin{tabbing}
\hspace*{\appSpaceTabWidth} \= \hspace*{\appTimeTabWidth} \= \kill

{} \> \phantom{0}8:30 \> VM production: electrons and muons (15m)\\
{} \> {} \> \footnotesize Speaker: Sylvester Joosten (ANL)
\\

{} \> \phantom{0}8:45 \> DVCS and pi0 kinematics (15m)\\
{} \> {} \> \footnotesize Speaker: Maxime DEFURNE (CEA)\\

{} \> \phantom{0}9:00 \> CFF extraction from DVCS
 (15m)\\
{} \> {} \> \footnotesize Speaker: Francois-Xavier Girod (JLab)\\

{} \> \phantom{0}9:15 \> Suppression of incoherent breakup in e+A (15m)\\
{} \> {} \> \footnotesize Speaker: Wan Chang (CCNU)\\

{} \> \phantom{0}9:30 \> How kinematics should be assessed / discussed (10m)\\
{} \> {} \> \footnotesize Speaker: Christian Weiss (JLab)
\end{tabbing}

\begin{tabbing}
\hspace*{\appLongTabWidth} \= \kill
\phantom{0}8:30 - 10:30 \> Forward Detectors/IR \& Central Detector \\
{} \> \footnotesize  Conveners: Alexander Jentsch (BNL), Alexander Kiselev (BNL), Michael Murray (Kansas),\\
{} \> \footnotesize William Brooks (Universidad Técnica Federico Santa María), Yulia Furletova (JLab)
\end{tabbing}

\begin{tabbing}
\hspace*{\appSpaceTabWidth} \= \hspace*{\appTimeTabWidth} \= \kill

{} \> \phantom{0}8:30 \> General Discussion of Central Integration/FF Concerns (20m)\\
{} \> {} \> \footnotesize Speaker: Alexander Jentsch (BNL)
\\

{} \> \phantom{0}8:50 \> Beam Pipe in Central Region of IR (30m)\\
{} \> {} \> \footnotesize Speaker: Alexander Kiselev (BNL)\\

{} \> \phantom{0}9:20 \> Low Q2 Tagger Discussion
 (20m)\\
{} \> {} \> \footnotesize Speaker: Jaroslav Adam (BNL)\\

{} \> \phantom{0}9:40 \> Discussion of B0 and Other FF Hadron Detectors (30m)\\
{} \> {} \> \footnotesize Speaker: Alexander Jentsch (BNL) \\

{} \> 10:10 \> Incoherent Veto of Nuclear Breakup (20m)\\
{} \> {} \> \footnotesize Speaker: E.~C.~Aschenauer (BNL)
\end{tabbing}

\begin{tabbing}
\hspace*{\appLongTabWidth} \= \kill
\phantom{0}8:30 - 10:00 \> Inclusive reactions WG 
\end{tabbing}

\begin{tabbing}
\hspace*{\appSpaceTabWidth} \= \hspace*{\appTimeTabWidth} \= \kill

{} \> \phantom{0}8:30 \> Electron PID Studies (15m)\\
{} \> {} \> \footnotesize Speaker: Hanjie Liu (UMass, Amherst)
\\

{} \> \phantom{0}8:45 \> Neutral-Current Cross-sections (15m)\\
{} \> {} \> \footnotesize Speaker: Xiaoxuan Chu (BNL)\\

{} \> \phantom{0}9:00 \> Djangoh NC cross-section comparison with theory (15m)\\
{} \> {} \> \footnotesize Speaker: Matt Posik (Temple)\\

{} \> \phantom{0}9:15 \> Statistical Tests (15m)\\
{} \> {} \> \footnotesize Speaker: Rabah Abdul Khalek\\

{} \> \phantom{0}9:30 \> Update from CT (15m)\\
{} \> {} \> \footnotesize Speaker: Timothy Hobbs (Southern Methodist and EIC Center@JLab)\\

{} \> \phantom{0}9:45 \> Update on "Missing Energy Performance" (15m)\\
{} \> {} \> \footnotesize Miguel Arratia (UC Riverside)
\end{tabbing}

\begin{tabbing}
\hspace*{\appLongTabWidth} \= \kill
\phantom{0}8:30 - 10:00 \> Jets, Heavy Quarks WG 
\end{tabbing}
\begin{tabbing}
\hspace*{\appSpaceTabWidth} \= \hspace*{\appTimeTabWidth} \= \kill

{} \> \phantom{0}8:30 \> Jet substructure and hadronization studies (20m)\\
{} \> {} \> \footnotesize Speaker: Joe Osborn (ORNL)\\

{} \> \phantom{0}8:50 \> Calculations of heavy meson cross sections at the EIC (20m)\\
{} \> {} \> \footnotesize Speaker: Ivan Vitev (LANL)\\

{} \> \phantom{0}9:10 \> Jets for 3D imaging (20m)\\
{} \> {} \> \footnotesize Speaker: Miguel Arratia (UC Riverside)\\

{} \> \phantom{0}9:30 \> Charm and Beauty (20m)\\
{} \> {} \> \footnotesize Speaker: Matthew Kelsey (LBNL)\\

{} \> \phantom{0}9:50 \> Discussion (10m)
\end{tabbing}

\begin{tabbing}
\hspace*{\appLongTabWidth} \= \kill
\phantom{0}8:30 - 10:00 \> Semi-inclusive Reactions WG 
\end{tabbing}

\begin{tabbing}
\hspace*{\appSpaceTabWidth} \= \hspace*{\appTimeTabWidth} \= \kill

{} \> \phantom{0}8:30 \> TMD grids and tools for predictions (20m)\\
{} \> {} \> \footnotesize Speaker: Chiara Bissolotti (Università di Pavia and INFN)\\

{} \> \phantom{0}8:50 \> Di-hadron and Lambda fragmentation (20m)\\
{} \> {} \> \footnotesize Speaker: Christopher Dilks (Duke)\\

{} \> \phantom{0}9:10 \> Spectroscopy at EIC (20m)\\
{} \> {} \> \footnotesize Speaker: Justin Stevens (William \& Mary)
\end{tabbing}

\begin{tabbing}
\hspace*{\appLongTabWidth} \= \kill
10:30 - 11:30 \> Calorimeter \& Particle ID \& Tracking \\
{} \> \footnotesize Conveners: Domenico Elia (INFN Bari), Eugene Chudakov (JLab), Kondo Gnanvo (UVa),\\
{} \> \footnotesize Leo Greiner (LBNL), Patrizia Rossi (JLab), Thomas Hemmick (Stony Brook ),\\
{} \> \footnotesize Vladimir Berdnikov (CUA)
\end{tabbing}

\begin{tabbing}
\hspace*{\appSpaceTabWidth} \= \hspace*{\appTimeTabWidth} \= \kill

{} \> 10:30 \> PID-Tracking-Calorimetry Discussion (1h)
\end{tabbing}

\begin{tabbing}
\hspace*{\appLongTabWidth} \= \kill
10:30 - 12:30 \> Diffractive \& Tagging+Exclusive joint session
\end{tabbing}

\begin{tabbing}
\hspace*{\appSpaceTabWidth} \= \hspace*{\appTimeTabWidth} \= \kill

{} \> 10:30 \> Elastic Hydrogen and Deuteron scattering (5m)\\
{} \> {} \> \footnotesize Speaker: Barak Schmookler (Stony Brook)\\

{} \> 10:35 \> Initial state radiation as a probe of ep and eA scattering (10m)\\
{} \> {} \> \footnotesize Speaker: Prof. Charles Hyde (ODU)\\

{} \> 10:45 \> Discussion (10m)\\

{} \> 10:55 \> Coherent $\gamma^*$ 4He scattering emphasizing desired t-range\\
{} \> {} \> for 4He detection (10m)\\
{} \> {} \> \footnotesize Speaker: Mark Strikman\\

{} \> 11:05 \> Discussion (10m)\\

{} \> 11:15 \> VM production: electrons vs muons (5m)\\
{} \> {} \> \footnotesize Speaker: Sylvester Joosten (ANL)\\

{} \> 11:20 \> Vector meson production simulations (5m)\\
{} \> {} \> \footnotesize Speaker: Sam Heppelman\\

{} \> 11:25 \> Suppression of incoherent background in VM production (5m)\\
{} \> {} \> \footnotesize Speaker: Wan Chang (CCNU)\\

{} \> 11:30 \> Discussion (15m)\\

{} \> 11:45 \> Diffractive dijet photoproduction at the EIC (5m)\\
{} \> {} \> \footnotesize Speaker: Vadim Guzey\\

{} \> 11:50 \> Diffractive dijets in DIS (5m)\\
{} \> {} \> \footnotesize Speaker: Farid Salazar (Stony Brook)\\

{} \> 11:55 \> Discussion (10m)\\

{} \> 12:05 \> U-channel pi0 production (5m)\\
{} \> {} \> \footnotesize Speaker: Wenliang (Bill) Li\\

{} \> 12:10 \> Discussion (20m)
\end{tabbing}

\begin{tabbing}
\hspace*{\appLongTabWidth} \= \kill
10:30 - 12:30 \> Inclusive + SIDIS + Jets \& HQ joint session
\end{tabbing}
\begin{tabbing}
\hspace*{\appSpaceTabWidth} \= \hspace*{\appTimeTabWidth} \= \kill

{} \> 10:30 \> DIS and CC reactions at EIC (20m)\\
{} \> {} \> \footnotesize Speaker: Xiaoxuan Chu (BNL) \\

{} \> 10:50 \> SIDIS summary of inclusive and jet related topics (20m)\\
{} \> {} \> \footnotesize Speaker: Bowen Xiao (CCNU) \\

{} \> 11:10 \> Open heavy flavor and quarkonia at the EIC (20m)\\
{} \> {} \> \footnotesize Speaker: Cheuk-Ping Wong (LANL) \\

{} \> 11:30 \> Summary of EW and BSM physics (20m)\\
{} \> {} \> \footnotesize Speaker: Ciprian Gal (Stony Brook) \\

{} \> 11:50 \> Charm Jet Tagging in Charged-Current Interactions (20m)\\
{} \> {} \> \footnotesize Speaker: Stephen Sekula (Southern Methodist) \\

{} \> 12:10 \> Discussion (20m)
\end{tabbing}

\begin{tabbing}
\hspace*{\appLongTabWidth} \= \kill
10:30 - 12:30 \> Detector Material Budget\\
{} \> \footnotesize Convener: Yulia Furletova (JLab)
\end{tabbing}
\begin{tabbing}
\hspace*{\appSpaceTabWidth} \= \hspace*{\appTimeTabWidth} \= \kill

{} \> 11:30 \> Material Budget requirements (20m)\\
{} \> {} \> \footnotesize Speaker: Yulia Furletova (JLab)\\

{} \> 11:50 \> eRD6 (20m)\\
{} \> {} \> \footnotesize Speaker: Matt Posik (Temple) \\

{} \> 12:10 \> SI material projections (20m)\\
{} \> {} \> \footnotesize Speaker: Leo Greiner (LBNL)
\end{tabbing}

\begin{tabbing}
\hspace*{\appLongTabWidth} \= \kill
13:30 - 15:30 \> Diffractive \& Tagging + Exclusive with DWG
\end{tabbing}

\begin{tabbing}
\hspace*{\appSpaceTabWidth} \= \hspace*{\appTimeTabWidth} \= \kill

{} \> 13:30 \> DVCS and e+D spectator tagging in the FF region (20m)\\
{} \> {} \> \footnotesize Speaker: Alexander Jentsch (BNL) \\

{} \> 13:50 \> Summary of D\&T+Exclusive joint session (1h)\\
{} \> {} \> \footnotesize Speakers: Anna Stasto (Penn State), Barbara Pasquini (Pavia), Daria Sokhan, \\
{} \> {} \> \footnotesize Douglas Higinbotham (JLab), Or Hen (MIT), Raphael Dupre (IPN Orsay), \\
{} \> {} \> \footnotesize Salvatore Fazio (BNL), Spencer Klein (LBNL), Tuomas Lappi (Jyvaskyla), Wim Cosyn (FIU)\\

{} \> 14:50 \> Suppression of incoherent background in VM production (20m)\\
{} \> {} \> \footnotesize Speakers: Barbara Pasquini (Pavia), Daria Sokhan, \\
{} \> {} \> \footnotesize Raphael Dupre (IPN Orsay), Salvatore Fazio (BNL), Tuomas Lappi (Jyvaskyla) \\

{} \> 15:10 \> Magic beam energies for polarized deuteron (20m)\\
{} \> {} \> \footnotesize Speaker: Douglas Higinbotham (JLab)
\end{tabbing}

\begin{tabbing}
\hspace*{\appLongTabWidth} \= \kill
13:30 - 15:30 \> Inclusive + SIDIS + Jets \& HQ with DWG
\end{tabbing}

\begin{tabbing}
\hspace*{\appSpaceTabWidth} \= \hspace*{\appTimeTabWidth} \= \kill

{} \> 13:30 \> Inclusive Reactions Input (15m)\\
{} \> {} \> \footnotesize Speakers: Barak Schmookler (Stony Brook ), Renee Fatemi (Kentucky), \\
{} \> {} \> \footnotesize Nobuo Sato (JLab) \\

{} \> 13:45 \> SIDIS Input (15m)\\
{} \> {} \> \footnotesize Speakers: Anselm Vossen (Duke), Bowen Xiao (CCNU),\\
{} \> {} \> \footnotesize Justin Stevens (William \& Mary), Ralf Seidl
(RIKEN), Vladimirov Alexey (Regensburg) \\

{} \> 14:00 \> Jets+HQ Input (15m)\\
{} \> {} \> \footnotesize Speakers: Brian Page (BNL), Ernst Sichtermann (LBNL), Frank Petriello
(Northwestern),\\
{} \> {} \> \footnotesize Ivan Vitev (LANL), Leticia Cunqueiro (ORNL) \\

{} \> 14:15 \> Tracking+Vertexing Input (15m)\\
{} \> {} \> \footnotesize Speakers: Annalisa Mastroserio (Bari), \\
{} \> {} \> \footnotesize Domenico Elia (Bari), Kenneth Barish (UC Riverside),
Kondo Gnanvo (UVa),\\
{} \> {} \> \footnotesize Leo Greiner (LBNL) \\

{} \> 14:30 \> PID Input (15m)\\
{} \> {} \> \footnotesize Speakers: Kenneth Barish (UC Riverside), Patrizia Rossi (JLab), Thomas Hemmick (Stony Brook) \\

{} \> 14:45 \> Calorimetry Input (15m)\\
{} \> {} \> \footnotesize Speakers: Eugene Chudakov (JLab), Kenneth Barish (UC Riverside), Vladimir Berdnikov (CUA) \\

{} \> 15:00 \> Integration Input (15m)\\
{} \> {} \> \footnotesize Speakers: Alexander Kiselev (BNL), Kenneth Barish (UC Riverside) \\

{} \> 15:15 \> Discussion (15m)
\end{tabbing}

\begin{tabbing}
\hspace*{\appLongTabWidth} \= \kill
16:00 - 17:30 \> Software and PWG/DWG/SWG workflow discussion\\
{} \> \footnotesize Convener: Thomas Ullrich (BNL)
\end{tabbing}

\begin{tabbing}
\hspace*{\appSpaceTabWidth} \= \hspace*{\appTimeTabWidth} \= \kill

{} \> 16:00 \> Software WG: Introduction (20m)\\
{} \> {} \> \footnotesize Speaker: Markus Diefenthaler (JLab) \\

{} \> 16:15 \> Discussion on PWG / DWG / SWG workflow, part 2 (20m)\\
{} \> {} \> \footnotesize Speakers: Conveners
\end{tabbing}

\subsection*{FRIDAY, 22 May} 

\begin{tabbing}
\hspace*{\appLongTabWidth} \= \kill
\phantom{0}8:30 - 10:00 \> Complementarity; Q\&A session\\
{} \> \footnotesize Convener: Christine Aidala (Michigan)
\end{tabbing}

\begin{tabbing}
\hspace*{\appSpaceTabWidth} \= \hspace*{\appTimeTabWidth} \= \kill

{} \> \phantom{0}8:30 \> Complementarity of detectors - Introduction (15m)\\
{} \> {} \> \footnotesize Speakers: E.~C.~Aschenauer (BNL), Paul Newman (Birmingham) \\

{} \> \phantom{0}8:45 \> Open mic session (30m)\\

{} \> \phantom{0}9:15 \> Accelerator and IR design: Q\&A session (45m)
\end{tabbing}

\begin{tabbing}
\hspace*{\appLongTabWidth} \= \kill
\phantom{0}10:30 - 12:30 \> Summaries of PWG/DWG\\
{} \> \footnotesize Convener: John Arrington (ANL)
\end{tabbing}

\begin{tabbing}
\hspace*{\appSpaceTabWidth} \= \hspace*{\appTimeTabWidth} \= \kill

{} \> 10:30 \> Inclusive processes WG summary (10m)\\
{} \> {} \> \footnotesize Speakers: Barak Schmookler (Stony Brook), Renee Fatemi (Kentucky), \\
{} \> {} \> \footnotesize Nobuo Sato (JLab) \\

{} \> 10:45 \> Semi-inclusive processes WG summary (10m)\\
{} \> {} \> \footnotesize Speakers: Anselm Vossen (Duke), Bowen Xiao (CCNU), Justin Stevens (William \& Mary), \\
{} \> {} \> \footnotesize Ralf
Seidl (RIKEN), Vladimirov Alexey (Regensburg) \\

{} \> 11:00 \> Jets, Heavy Quarks WG summary (10m)\\
{} \> {} \> \footnotesize Speakers: Brian Page (BNL), Ernst Sichtermann (LBNL), Frank Petriello
(Northwestern), \\
{} \> {} \> \footnotesize Ivan Vitev (LANL), Leticia Cunqueiro (ORNL) \\

{} \> 11:15 \> Diffractive Reactions \& Tagging WG summary (10m)\\
{} \> {} \> \footnotesize Speakers: Anna Stasto (Penn State), Douglas Higinbotham (JLab), Or Hen (MIT),\\
{} \> {} \> \footnotesize Spencer Klein (LBNL), Wim Cosyn (FIU) \\

{} \> 11:30 \> Exclusive Reactions WG summary (10m)\\
{} \> {} \> \footnotesize Speakers: Barbara Pasquini (Pavia), Daria Sokhan, \\
{} \> {} \> \footnotesize Raphael Dupre (IPN Orsay), Salvatore Fazio
(BNL), Tuomas Lappi (Jyvaskyla) 
\end{tabbing}

\begin{tabbing}
\hspace*{\appLongTabWidth} \= \kill
\phantom{0}14:00 - 15:30 \> Summaries of PWG/DWG\\
{} \> \footnotesize Convener: Ernst Sichtermann (LBNL)
\end{tabbing}

\begin{tabbing}
\hspace*{\appSpaceTabWidth} \= \hspace*{\appTimeTabWidth} \= \kill

{} \> 14:00 \> Tracking WG Summary (10m)\\
{} \> {} \> \footnotesize Speakers: Domenico Elia (INFN Bari), Kondo Gnanvo (UVa), Leo Greiner (LBNL) \\

{} \> 14:10 \> Particle ID Summary (10m)\\
{} \> {} \> \footnotesize Speakers: Patrizia Rossi (JLab), Thomas Hemmick (Stony Brook) \\

{} \> 14:20 \> Calorimeter WG Summary (10m)\\
{} \> {} \> \footnotesize Speakers: Eugene Chudakov (JLab), Vladimir Berdnikov (CUA) \\

{} \> 14:30 \> Forward Detectors WG Summary (10m)\\
{} \> {} \> \footnotesize Speakers: Alexander Jentsch (BNL), Michael Murray 
(Kansas), Yulia Furletova (JLab) \\

{} \> 14:40 \> Polarimetry/Ancillary Detectors WG Summary (10m)\\
{} \> {} \> \footnotesize Speakers: Dave Gaskell, E.~C.~Aschenauer (BNL) \\

{} \> 14:50 \> DAQ/Electronics WG Summary (10m)\\
{} \> {} \> \footnotesize Speakers: Andrea Celentano (INFN-Genova), Damien Neyret (CEA Saclay IRFU/DPhN) \\

{} \> 15:00 \> Central Detector/Magnet WG Summary (10m)\\
{} \> {} \> \footnotesize Speakers: Alexander Kiselev (BNL), William Brooks (UTFSM)
\end{tabbing}

\begin{tabbing}
\hspace*{\appLongTabWidth} \= \kill
\phantom{0}15:30 - 16:30 \> Next steps and plans\\
{} \> \footnotesize Convener: Barbara Jacak (UCB/LBNL)
\end{tabbing}

\newpage
\section*{3rd EIC Yellow Report Workshop at CUA}
Local organizer: Tanja Horn (CUA)

\subsection*{WEDNESDAY, 19 SEPTEMBER} 
\begin{tabbing}
\hspace*{\appLongTabWidth} \= \kill
\phantom{0}9:00 - 12:00 \> EIC Project Status and EoI Information Session \\
{} \> \footnotesize Convener: Tanja Horn (CUA)
\end{tabbing}

\begin{tabbing}
\hspace*{\appSpaceTabWidth} \= \hspace*{\appTimeTabWidth} \= \kill
{} \> \phantom{0}9:00 \> EIC User Group Intro (10 min)\\
{} \> {} \> \footnotesize Speakers: Bernd Surrow (Temple), Richard Milner (MIT)\\

{} \> \phantom{0}9:10 \> Plans for International Engagement (40 min)\\
{} \> {} \> \footnotesize Speaker: Jehanne Gillo (DOE ) \\

{} \> \phantom{0}9:50 \> EIC Project Status (40 min)\\
{} \> {} \> \footnotesize Speaker: James Yeck (Wisconsin-Madison and BNL) \\

{} \> 10:30 \> EoI Information Session (1h 30 min)\\
{} \> {} \> \footnotesize Speakers: 14 speakers, 5 min each
\end{tabbing}

\begin{tabbing}
\hspace*{\appLongTabWidth} \= \kill
13:00 - 16:25 \> OPC Updates and Detector Complementarity \\
{} \> \footnotesize Convener: Grzegorz Kalicy (CUA)
\end{tabbing}

\begin{tabbing}
\hspace*{\appSpaceTabWidth} \= \hspace*{\appTimeTabWidth} \= \kill
{} \> 13:00 \> Current Status of detector solenoid activities (30 min)\\
{} \> {} \> \footnotesize Speaker: Renuka Rajput-Ghoshal (JLab)\\

{} \> 13:30 \> Electronics (15 min)\\
{} \> {} \> \footnotesize Speaker: Fernando Barbosa (JLab) \\

{} \> 13:45 \> Computing (15 min)\\
{} \> {} \> \footnotesize Speaker: Jerome LAURET (BNL)\\

{} \> 14:00 \> Detector Complementarity: Luminosity optimization at low \\
{} \> {} \> COM energies (30 min)\\
{} \> {} \> \footnotesize Speaker: Vadim Ptitsyn (C-AD, BNL)\\

{} \> 14:30 \> Detector Complementarity: Optimization of a 2nd IR (30 min)\\
{} \> {} \> \footnotesize Speaker: Vasiliy Morozov (JLab) \\

{} \> 15:00 \> Detector Complementarity: What we know so far (30 min)\\
{} \> {} \> \footnotesize Speakers: E.~C.~Aschenauer (BNL), Paul Newman (Birmingham) \\

{} \> 15:30 \> Detector Complementarity: Open MIC (30 min)\\
{} \> {} \> \footnotesize Speakers: E.~C.~Aschenauer (BNL), Paul Newman (Birmingham)
\end{tabbing}

\subsection*{THURSDAY, 19 SEPTEMBER} 
\begin{tabbing}
\hspace*{\appLongTabWidth} \= \kill
\phantom{0}9:00 - \phantom{0}9:30 \> YR Overview \\
{} \> \footnotesize  Conveners: Adrian Dumitru (Baruch), Andreas Metz (Temple) , Carlos Munoz Camacho (Orsay), \\
{} \> \footnotesize Olga Evdokimov (UIC), Peter Jones (Birmingham), Silvia Dalla Torre (INFN Trieste), \\
{} \> \footnotesize Tanja Horn (CUA), Kenneth Barish (UC Riverside)
\end{tabbing}

\begin{tabbing}
\hspace*{\appSpaceTabWidth} \= \hspace*{\appTimeTabWidth} \= \kill
{} \> \phantom{0}9:00 \> General Reminders and Workflow (30 min)\\
{} \> {} \> \footnotesize Speakers: Adrian Dumitru (Baruch), Andreas Metz (Temple) , Carlos Munoz Camacho (Orsay), \\
{} \> {} \> \footnotesize Olga Evdokimov (UIC), Peter Jones (Birmingham), Silvia Dalla Torre (INFN Trieste), \\
{} \> {} \> \footnotesize Tanja Horn (CUA), Kenneth Barish (UC Riverside), Rolf Ent (JLab), Thomas Ullrich (BNL)
\end{tabbing}

\begin{tabbing}
\hspace*{\appLongTabWidth} \= \kill
\phantom{0}9:30 - 12:00 \> PWG requirements overview: \\
{} \> what has been established and what still needs work?\\
{} \> \footnotesize  Conveners: Adrian Dumitru (Baruch), Andreas Metz (Temple) , Carlos Munoz Camacho (Orsay), \\
{} \> \footnotesize Olga Evdokimov (UIC)
\end{tabbing}

\begin{tabbing}
\hspace*{\appSpaceTabWidth} \= \hspace*{\appTimeTabWidth} \= \kill
{} \> \phantom{0}9:30 \> Inclusive reactions WG (15m)\\
{} \> {} \> \footnotesize Speakers: Barak Schmookler (Stony Brook ), \\
{} \> {} \> \footnotesize Renee Fatemi (Kentucky), Nobuo Sato (JLab)\\

{} \> 10:00 \> Semi-inclusive Reactions WG (15m)\\
{} \> {} \> \footnotesize Speakers: Anselm Vossen (Duke), Bowen Xiao (CCNU), \\
{} \> {} \> \footnotesize Justin Stevens (William \& Mary), Ralf Seidl (RIKEN), Vladimirov Alexey (Regensburg)\\

{} \> 10:30 \> Jets, Heavy Quarks WG (15m)\\
{} \> {} \> \footnotesize Speakers: Brian Page (BNL), Ernst Sichtermann (LBL), \\
{} \> {} \> \footnotesize Frank Petriello (Northwestern), Ivan Vitev (LANL), Leticia Cunqueiro (ORNL)\\

{} \> 11:00 \> Exclusive Reactions WG (15m)\\
{} \> {} \> \footnotesize Speakers: Barbara Pasquini (Pavia), Daria Sokhan, \\
{} \> {} \> \footnotesize Raphael Dupre (IPN Orsay), Salvatore Fazio (BNL), Tuomas Lappi (Jyvaskyla)\\

{} \> 11:30 \> Diffractive Reactions \& Tagging WG (15m)\\
{} \> {} \> \footnotesize Speakers: Anna Stasto (Penn State), Douglas Higinbotham (JLab), \\
{} \> {} \> \footnotesize Or Hen (MIT), Spencer Klein (LBNL), Wim Cosyn (FIU)
\end{tabbing}

\begin{tabbing}
\hspace*{\appLongTabWidth} \= \kill
13:00 -17:15 \> YR Detailed Discussions: Reference detector, material budget,\\
{} \> integration issues, background level\\
{} \> \footnotesize  Conveners: Peter Jones (Birmingham), Silvia Dalla Torre (INFN Trieste), \\
{} \> \footnotesize Tanja Horn (CUA), Kenneth Barish (UC Riverside)
\end{tabbing}

\begin{tabbing}
\hspace*{\appSpaceTabWidth} \= \hspace*{\appTimeTabWidth} \= \kill
{} \> 13:00 \> DWG detector configurations overview – detector cartoon and \\
{} \> {} \> list of boundary conditions (45m)\\
{} \> {} \> \footnotesize Speaker: Alexander Kiselev (BNL)\\

{} \> 13:45 \> Integration Issues (45m)\\
{} \> {} \> \footnotesize Speaker: Walt Akers (JLab)\\

{} \> 14:30 \> Material Budget (45m)\\
{} \> {} \> \footnotesize Speakers: Jin Huang (BNL), Leo Greiner (LBNL), Matt Posik (Temple) \\

{} \> 15:15 \> Background level (45m)\\
{} \> {} \> \footnotesize Speaker: Marcy Stutzman (JLab)
\end{tabbing}

\subsection*{FRIDAY, 19 SEPTEMBER} 
\begin{tabbing}
\hspace*{\appLongTabWidth} \= \kill
\phantom{0}9:00 - 12:00 \> Discussion on completion tasks for YR and detector testing opportunities
\end{tabbing}

\begin{tabbing}
\hspace*{\appSpaceTabWidth} \= \hspace*{\appTimeTabWidth} \= \kill

{} \> \phantom{0}9:00 \> Overleaf setup and structure (1h 15m)\\
{} \> {} \> \footnotesize Speakers: Adrian Dumitru (Baruch College (CUNY)), Kenneth Barish (UC Riverside) \\

{} \> 10:15 \> Timeline (1h 15m)\\
{} \> {} \> \footnotesize Speakers: Andreas Metz (Temple), Rold Ent (JLab), Thomas Ullrich (BNL) \\

{} \> 11:30 \> Detector testing facilities and opportunities (30m)\\
{} \> {} \> \footnotesize Speaker: Douglas Higinbotham (JLab) 
\end{tabbing}

\begin{tabbing}
\hspace*{\appLongTabWidth} \= \kill
13:00 - 16:00 \> Outlook and next steps
\end{tabbing}

\begin{tabbing}
\hspace*{\appSpaceTabWidth} \= \hspace*{\appTimeTabWidth} \= \kill

{} \> 13:00 \> Overview: Planned activities and new ideas for activities after YR (15m)\\
{} \> {} \> \footnotesize Speaker: Bernd Surrow (Temple) \\

{} \> 13:15 \> EICUG Charter Survey (30m)\\
{} \> {} \> \footnotesize Speaker: Richard Milner (MIT) \\

{} \> 13:45 \> EIC Project Management plans / next steps over the next year (45m)\\
{} \> {} \> \footnotesize Speakers: E.~C.~Aschenauer (JLab), Rolf Ent (JLab) \\

{} \> 14:30 \> CFNS Proposal - IR2 \@ EIC (1h)\\
{} \> {} \> \footnotesize Speaker: Volker Burkert (JLab)
\end{tabbing}
\newpage
\section*{4th EIC Yellow Report Workshop at LBL}
Local organizer (LBNL): Ernst Sichtermann, John Arrington, Barbara Jacak

\subsection*{THURSDAY, 19 NOVEMBER} 

\begin{tabbing}
\hspace*{\appLongTabWidth} \= \kill
11:00 - 14:00 \> EIC Project and Yellow Report Status \\
{} \> \footnotesize  Convener: Bernd Surrow (Temple)
\end{tabbing}

\begin{tabbing}
\hspace*{\appSpaceTabWidth} \= \hspace*{\appTimeTabWidth} \= \kill

{} \> 11:00 \> Welcome and workshop plan (15+5)\\
{} \> {} \> \footnotesize Speakers: Ernst Sichtermann (LBNL), John Arrington (LBNL), \\
{} \> {} \> \footnotesize Barbara Jacak (UC Berkeley, LBNL) \\

{} \> 11:20 \> PWG requirements overview (10+5)\\
{} \> {} \> \footnotesize Speakers: Adrian Dumitru (Baruch College (CUNY)), Andreas Metz (Temple), \\
{} \> {} \> \footnotesize Carlos Munoz Camacho (IJCLab, CNRS/IN2P3), Olga Evdokimov (UIC) \\

{} \> 11:35 \> PWG requirements: Inclusive reactions WG (10+5)\\
{} \> {} \> \footnotesize Speakers: Barak Schmookler (Stony Brook ), Renee Fatemi (Kentucky), \\
{} \> {} \> \footnotesize Nobuo Sato (JLab) \\

{} \> 11:50 \> PWG requirements: Semi-inclusive reactions WG (10+5)\\
{} \> {} \> \footnotesize Speakers: Anselm Vossen (Duke), Bowen Xiao (CCNU),\\
{} \> {} \> \footnotesize Justin Stevens (William \& Mary), Ralf Seidl (RIKEN), Vladimirov Alexey (Regensburg)\\

{} \> 12:05 \> PWG requirements: Jets, Heavy Quarks WG (10+5) \\
{} \> {} \> \footnotesize Speakers: Brian Page (BNL), Ernst Sichtermann (LBNL), \\
{} \> {} \> \footnotesize Frank Petriello (Northwestern), Ivan Vitev (LANL), Leticia Cunqueiro (ORNL) \\

{} \> 12:20 \> PWG requirements: Exclusive Reactions WG (10+5)	\\
{} \> {} \> \footnotesize Speakers: Barbara Pasquini (Pavia), Daria Sokhan, \\
{} \> {} \> \footnotesize Raphael Dupre (IPN Orsay), Salvatore Fazio (BNL), Tuomas Lappi (Jyvaskyla)\\

{} \> 12:35 \> PWG requirements: Diffractive Reactions \& Tagging WG (10+5)\\
{} \> {} \> \footnotesize Speakers: Anna Stasto (Penn State), Douglas Higinbotham (JLab), \\
{} \> {} \> \footnotesize Or Hen (MIT), Spencer Klein (LBNL), Wim Cosyn (FIU)\\

{} \> 13:10 \> CDR Reference Detector (18+7)	\\
{} \> {} \> \footnotesize Speaker: Rolf Ent (JLab)\\

{} \> 12:05 \> Project Status (18+7)\\	
{} \> {} \> \footnotesize Speaker: James Yeck (Wisconsin-Madison and BNL)
\end{tabbing}

\begin{tabbing}
\hspace*{\appLongTabWidth} \= \kill
15:00 - 18:05 \> Yellow Report Updates/Activities \\
{} \> \footnotesize Convener: Ernst Sichtermann (LBNL)
\end{tabbing}

\begin{tabbing}
\hspace*{\appSpaceTabWidth} \= \hspace*{\appTimeTabWidth} \= \kill

{} \> 15:00 \> Boundary conditions on IR / Detector (20+5)	\\
{} \> {} \> \footnotesize Speaker: Alexander Kiselev (BNL)\\

{} \> 15:25 \> Detector Matrix/Updates - Forward region (20+15)	\\
{} \> {} \> \footnotesize Speaker: Michael Murray (Kansas)\\

{} \> 16:00 \> Detector Matrix/Updates - Barrel PID (20+15)\\
{} \> {} \> \footnotesize Speaker: Silvia Dalla Torre (INFN, Trieste)\\

{} \> 16:55 \> Detector Matrix/Updates - Forward HCAL resolution (20+15)\\
{} \> {} \> \footnotesize Speaker: Tanja Horn (CUA)\\

{} \> 17:30 \> Detector Matrix/Updates - Tracking (20+15)\\
{} \> {} \> \footnotesize Speaker: Domenico Elia (INFN Bari)
\end{tabbing}
\subsection*{FRIDAY, 20 NOVEMBER} 
\begin{tabbing}
\hspace*{\appLongTabWidth} \= \kill
11:00 - 14:20 \> Yellow Report Content I \\
{} \> \footnotesize Convener: Christine Aidala (Michigan)
\end{tabbing}

\begin{tabbing}
\hspace*{\appSpaceTabWidth} \= \hspace*{\appTimeTabWidth} \= \kill

{} \> 11:00 \> Yellow Report status/overview (15+10) \\
{} \> {} \> \footnotesize Speaker: Bernd Surrow (Temple) \\

{} \> 11:25 \> Accelerator/IR overview (15+5) \\
{} \> {} \> \footnotesize Speaker: Angelika Drees (BNL) \\

{} \> 11:40 \> Discussion of YR content: Tracking (15+5) \\
{} \> {} \> \footnotesize Speaker: Leo Greiner (LBNL) \\

{} \> 12:05 \> Discussion of YR content: Calorimetry (15+5)	 \\
{} \> {} \> \footnotesize Speaker: Alexander Bazilevsky (BNL) \\

{} \> 12:40 \> Discussion of YR content: PID (15+5)	 \\
{} \> {} \> \footnotesize Speaker: Thomas Hemmick (Stony Brook) \\

{} \> 13:00 \> Discussion of YR content: Magnetic field strength, magnet bore (15+5)	 \\
{} \> {} \> \footnotesize Speakers: E.~C.~Aschenauer (BNL), Renuka Rajput-Ghoshal (JLab), Rolf Ent (JLab) \\

{} \> 13:20 \> Discussion of YR content: Far-forward detectors (15+5)	 \\
{} \> {} \> \footnotesize Speaker: Alexander Jentsch (BNL) \\

{} \> 13:40 \> Discussion of YR content: Readout and DAQ (15+5)	 \\
{} \> {} \> \footnotesize Speakers: Andrea Celentano (INFN-Genova), Damien Neyret (CEA Saclay IRFU/DPhN) \\

{} \> 14:00 \> Discussion of YR content: Software (15+5)	 \\
{} \> {} \> \footnotesize Speaker: Markus Diefenthaler (JLab) 
\end{tabbing}

\begin{tabbing}
\hspace*{\appLongTabWidth} \= \kill
15:20 - 18:20 \> Yellow Report Content II \\
{} \> \footnotesize Convener: John Arrington (LBNL)
\end{tabbing}

\begin{tabbing}
\hspace*{\appSpaceTabWidth} \= \hspace*{\appTimeTabWidth} \= \kill

{} \> 15:20 \> Yellow Report content: section 7.1 (15+5) \\
{} \> {} \> \footnotesize Speakers: Barbara Pasquini (Pavia), Andreas Metz (Temple) \\

{} \> 15:40 \> Yellow Report content: section 7.2 (15+5)	 \\
{} \> {} \> \footnotesize Speaker: Anselm Vossen (Duke) \\

{} \> 16:00 \> Yellow Report content: section 7.3 (15+5) \\
{} \> {} \> \footnotesize Speaker: Spencer Klein (LBNL) \\

{} \> 16:20 \> Yellow Report content: section 7.4 (15+5) \\
{} \> {} \> \footnotesize Speaker: Ivan Vitev (LANL) \\

{} \> 17:00 \> Complementarity: Optimizing luminosity at lower c.m. energy (10+5) \\
{} \> {} \> \footnotesize Speaker: Yuhong Zhang (JLab) \\

{} \> 17:15 \> Complementarity: 2nd IR design considerations (10+5)	 \\
{} \> {} \> \footnotesize Speaker: Vasiliy Morozov (JLab) \\

{} \> 17:30 \> Complementarity: 2nd IR Workshop (10+5)	 \\
{} \> {} \> \footnotesize Speakers: Latifa Elouadrhiri (JLab), Volker Burkert (JLab) \\

{} \> 17:40 \> Complementarity: YR content and discussion (10+20) \\
{} \> {} \> \footnotesize Speaker: E.~C.~Aschenauer (BNL)
\end{tabbing}

\subsection*{SATURDAY, 21 NOVEMBER} 

\begin{tabbing}
\hspace*{\appLongTabWidth} \= \kill
11:00 - 14:00 \> Yellow Report Plans, Future Activities \\
{} \> \footnotesize Convener: Barbara Jacak (UCB and LBNL)
\end{tabbing}

\begin{tabbing}
\hspace*{\appSpaceTabWidth} \= \hspace*{\appTimeTabWidth} \= \kill

{} \> 11:00 \> Detector R\&D needs (15+5) \\
{} \> {} \> \footnotesize Speaker: Patrizia Rossi (JLab) \\

{} \> 11:20 \> Archiving and reproducibility of Yellow Report Studies (20+10)	 \\
{} \> {} \> \footnotesize Speaker: Markus Diefenthaler (JLab) \\

{} \> 11:50 \> Yellow report plans, timeline, discussion (15+15) \\
{} \> {} \> \footnotesize Speaker: Thomas Ullrich (BNL) \\

{} \> 12:40 \> Yellow Report plans, timeline, discussion - finalize plans (20m) \\
{} \> {} \> \footnotesize Speaker: Thomas Ullrich (BNL) \\

{} \> 13:00 \> Expressions of Interest: status, plans (15+5) \\
{} \> {} \> \footnotesize Speakers: E.~C.~Aschenauer (BNL), Rolf Ent (JLab) \\

{} \> 13:20 \> Discussion: Call for detector proposals (10+10)	 \\
{} \> {} \> \footnotesize Speakers: Maria Chamizo (BNL), McKeown Robert (JLab) \\

{} \> 13:40 \> Open discussion (20)
\end{tabbing}

\cleardoublepage

\end{appendices}
\cleardoublepage

%
%
\renewcommand\rightmark{EIC Yellow Report}

\fancyhead{}
\fancyfoot{}

\fancyhead[RE]{\rightmark}
\fancyhead[LE]{\thepage}

\fancyhead[LO]{\leftmark}
\fancyhead[RO]{\thepage}

\renewcommand\bibname{References}
\phantomsection
\addcontentsline{toc}{part}{\large{References}}
\bibliographystyle{elsarticle-num}
\bibliography{EIC_YR}



\afterpage{\blankpage}

\afterpage{\blankpage}

%
%
\newgeometry{textwidth=8.5in,textheight=11.0in}
\includegraphics[width=8.5in,height=10.99in]{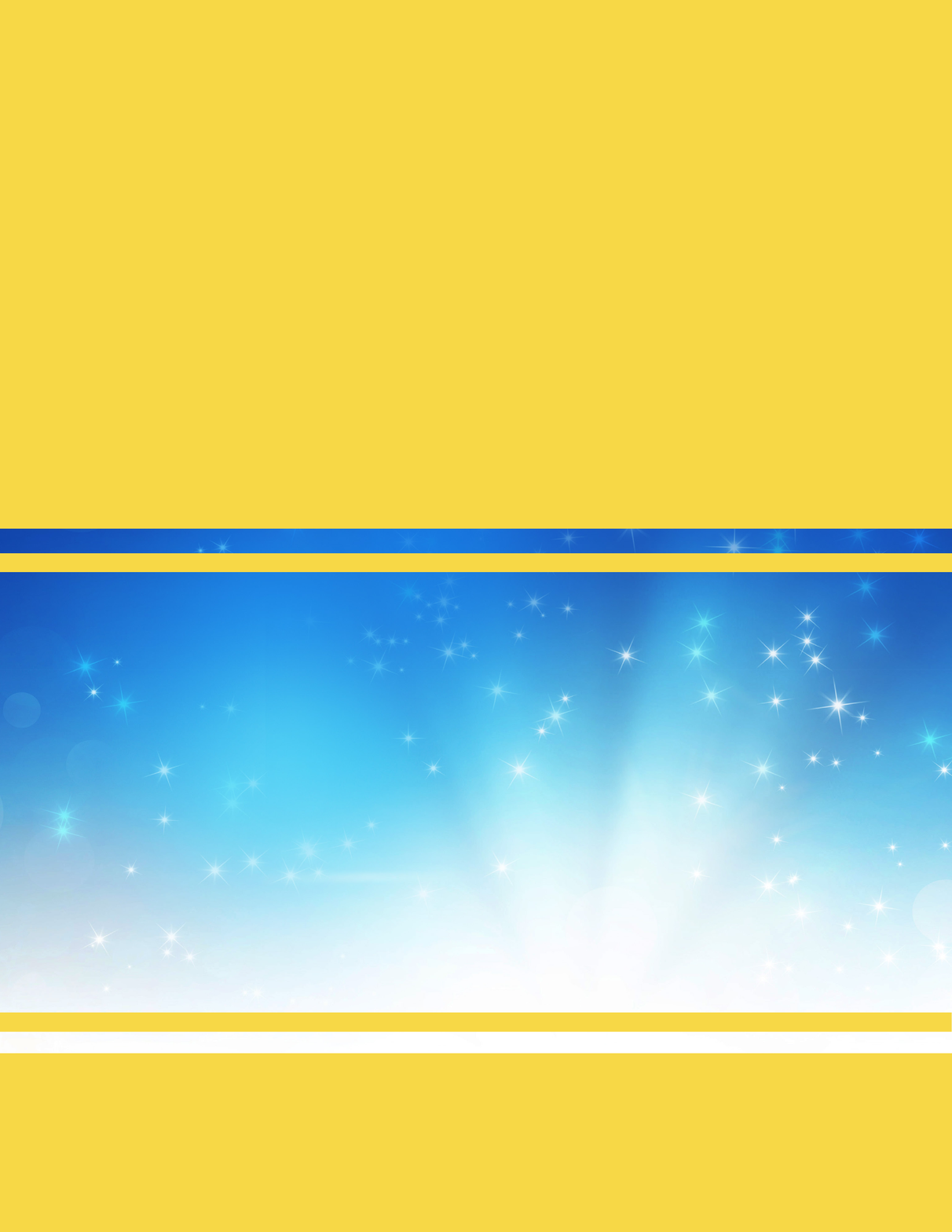}
\restoregeometry

%
%
%
%

\end{document}